%% file: main-arxiv.tex
\documentclass{book} 

\newlength\ys
 
\usepackage{amsmath,amssymb,dsfont,paralist,amsthm,eurosym}
\usepackage[format=default,labelfont=bf]{caption}
\usepackage{placeins}
\usepackage{microtype}
\usepackage[utf8]{inputenc}
\usepackage{scalerel}
\usepackage{stmaryrd}
\usepackage{bbold}
\usepackage{yfonts}
\usepackage{fancyhdr} 

\usepackage{algorithm}
\usepackage{algpseudocode}

\algnewcommand\algorithmicforeach{\textbf{for each}}
\algnewcommand\algorithmicvariables{\textbf{Variables:}}
\algnewcommand\Variables{\item[\algorithmicvariables]}
\algdef{S}[FOR]{ForEach}[1]{\algorithmicforeach\ #1\ \algorithmicdo}
\algrenewcommand\algorithmicrequire{\textbf{Input:}}
\algrenewcommand\algorithmicensure{\textbf{Output:}}
\algnewcommand\Fixedcomment[1]{\hfill\makebox[0.4\textwidth][l]{$\triangleright$ #1}}

\usepackage{graphicx}
\usepackage{tikz}
\usepackage{subcaption}
\usepackage{tikz-qtree}
\usepackage{rotating}

\usetikzlibrary{arrows,matrix}
\usetikzlibrary{matrix,calc}
\usetikzlibrary{shapes.geometric}
\usetikzlibrary{positioning}
\usetikzlibrary{calligraphy}
\usetikzlibrary{fit}

\pgfdeclarelayer{background}
\pgfsetlayers{background,main}

\usetikzlibrary{decorations.pathreplacing}
\tikzset{curlybrace/.style={decoration=brace,decorate}}
\usetikzlibrary{automata}
\usetikzlibrary{calc,positioning}
\usetikzlibrary{decorations.pathmorphing}
\usetikzlibrary{shapes}
\tikzset{trinode/.style={draw,triangle,minimum width=2.0cm}}
\tikzset{snake/.style={decorate,decoration=snake}}
\tikzset{curlybrace/.style={decoration=brace,decorate}}
\tikzset{triangle/.style={regular polygon,regular polygon sides=3}}
\tikzset{edge from parent path={(\tikzparentnode) -- (\tikzchildnode.north)}}
\usetikzlibrary{arrows.meta}

\usepackage{wrapfig}
\usepackage[rightcaption]{sidecap}
\usepackage[colorinlistoftodos]{todonotes}
\usepackage{mathtools}
\usepackage{enumitem}
\usepackage{multicol} 
\usepackage{todonotes}
\usepackage{imakeidx}         

\usepackage{hyperref}

\newtheorem{theorem}{Theorem}[section]
\newtheorem{lemma}[theorem]{Lemma}

\newtheorem{corollary}[theorem]{Corollary}
\newtheorem{observation}[theorem]{Observation}
\newtheorem{example}[theorem]{Example}
\newtheorem{definition}[theorem]{Definition}
\newtheorem{construction}[theorem]{Construction}

\newcommand{\cA}{\mathcal A}
\newcommand{\cB}{\mathcal B}
\newcommand{\cC}{\mathcal C}
\newcommand{\cD}{\mathcal D}

\newcommand{\cF}{\mathcal F}
\newcommand{\cG}{\mathcal G}
\newcommand{\cH}{\mathcal H}
\newcommand{\cK}{\mathcal K}
\newcommand{\cL}{\mathcal L}
\newcommand{\cN}{\mathcal N}
\newcommand{\cM}{\mathcal M}
\newcommand{\cP}{\mathcal P}
\newcommand{\cR}{\mathcal R}
\newcommand{\cS}{\mathcal S}

\newcommand{\cU}{\mathcal U}
\newcommand{\cV}{\mathcal V}
\newcommand{\cW}{\mathcal W}
\newcommand{\MMM}{\mathcal Q}

\newcommand{\C}{\mathrm{C}}
\newcommand{\D}{\mathrm{D}}

\newcommand{\rmH}{\mathrm{H}}
\newcommand{\M}{\mathsf{M}}
\newcommand{\N}{\mathrm{N}}
\newcommand{\Q}{\mathrm{Q}}
\newcommand{\R}{\mathrm{R}}
\newcommand{\T}{\mathrm{T}}
\newcommand{\U}{\mathrm{U}}

\newcommand{\maxrk}{\mathrm{maxrk}}
\newcommand{\Fork}{\mathrm{Fork}}

\newcommand{\QR}[2]{\Q^{\R_#1}_{\not= \mathbb{0}}(#2)}
\newcommand{\Qh}[2]{\Q^{\h_#1}_{\not= \mathbb{0}}(#2)}
\newcommand{\Rv}{\mathrm{R}^{\mathrm{v}}}
\newcommand{\Ra}{\mathrm{R}^{\mathrm{a}}}
\newcommand{\im}{\mathrm{im}}
\newcommand{\charp}{\mathrm{char}}
\newcommand{\murev}{\mu^{\leftarrow}}

\newcommand{\supp}{\mathrm{supp}}
\newcommand{\sgn}{\mathrm{sgn}}

\newcommand{\wt}{\mathrm{wt}}
\newcommand{\past}{\mathrm{past}}

\newcommand{\fl}{\langle \! \langle}
\newcommand{\fr}{\rangle \! \rangle}
\newcommand{\e}{\mathrm{e}} 

\DeclareMathOperator{\n}{-}

\newcommand{\Hom}{\mathrm{Hom}}
\newcommand{\MSO}{\mathrm{MSO}}
\newcommand{\MSOe}{\mathrm{MSO}^{\mathrm{ext}}}
\newcommand{\MSOce}{\mathrm{MSO}^{\mathrm{cext}}}
\newcommand{\Pol}{\mathrm{Pol}}
\newcommand{\Rat}{\mathrm{Rat}}
\newcommand{\Rec}{\mathrm{Rec}}
\newcommand{\Rep}{\mathrm{Rep}}
\newcommand{\RepEx}{\mathrm{RepEx}}
\newcommand{\RecStep}{\mathrm{Rec}\mathrm{Step}}
\newcommand{\budRec}{\mathrm{bud}\mathord{\n}\mathrm{Rec}}
\newcommand{\cdRec}{\mathrm{cd}\mathord{\n}\mathrm{Rec}}

\newcommand{\Reg}{\mathrm{Reg}}
\newcommand{\budwta}{\mathrm{budwta}}
\newcommand{\wta}{\mathrm{wta}}
\newcommand{\rmwta}{\mathrm{wta}}
\newcommand{\mysucc}{\mathrm{succ}}

\newcommand{\RT}{\mathrm{RT}}

\newcommand{\Var}{\mathrm{Var}}

\newcommand{\RQ}{\mathrm{RQ}}
\newcommand{\LQ}{\mathrm{LQ}}
\newcommand{\VRQ}{\mathsf{VRQ}}
\newcommand{\VLQ}{\mathsf{VLQ}}

\newcommand{\sqrun}[1]{\widetilde{#1}}

\newcommand{\A}{\mathsf{A}}
\newcommand{\B}{\mathsf{B}}
\newcommand{\sfA}{\mathsf{A}}
\newcommand{\sfB}{\mathsf{B}}
\newcommand{\sfh}{\mathsf{h}}

\newcommand{\Boole}{\mathsf{Boole}}
\newcommand{\sfC}{\mathsf{C}}
\newcommand{\sfFtwo}{\mathsf{F}_2}
\newcommand{\sfFL}{\mathsf{FL}(2+2)}
\newcommand{\sfG}{\mathsf{G}}
\newcommand{\sfM}{\mathsf{M}}
\newcommand{\sfdM}{\mathsf{dM}}
\newcommand{\Int}{\mathsf{Int}}
\newcommand{\Intminplus}{\mathsf{Int}_{\min,+}}
\newcommand{\Intfour}{\mathsf{Intmod4}}
\newcommand{\Lang}{\mathsf{Lang}}

\newcommand{\TropBM}{\mathsf{TropBM}}
\newcommand{\sfL}{\mathsf{L}}
\newcommand{\Nat}{\mathsf{Nat}}
\newcommand{\Natminplus}{\mathsf{Nat}_{\min,+}}
\newcommand{\Natmaxplus}{\mathsf{Nat}_{\max,+}}
\newcommand{\Natmaxplusn}{\mathsf{Nat}_{\max,+,n}}
\newcommand{\Natmaxmin}{\mathsf{Nat}_{\max,\min}}
\newcommand{\NearSem}{\mathsf{NearSem}}
\newcommand{\Nfive}{\mathsf{N}_5}
\newcommand{\Mthree}{\mathsf{M}_3}
\newcommand{\PP}{\mathsf{PP}}
\newcommand{\Powerset}{\mathsf{PS}}
\newcommand{\Ratnum}{\mathsf{Rat}}
\newcommand{\Ratnumgreaterzero}{\mathsf{Rat}_{\ge 0}}
\newcommand{\Ratminplus}{\mathsf{Rat}_{\min,+}}
\newcommand{\Realnum}{\mathsf{Real}}
\newcommand{\Realnumgreaterzero}{\mathsf{Real}_{\ge 0}}

\newcommand{\St}{\mathsf{S}}
\newcommand{\Sem}{\mathsf{Sem}}
\newcommand{\sfT}{\mathsf{T}}
\newcommand{\Three}{\mathsf{Three}}
\newcommand{\Trunc}{\mathsf{Trunc}}

\newcommand{\UnitIntalg}{\mathsf{UnitInt}_{\mathrm{alg}}}
\newcommand{\UnitIntfuzzy}{\mathsf{UnitInt}}
\newcommand{\UnitIntboundedsum}{\mathsf{UnitInt}_{\mathrm{bs}}}
\newcommand{\V}{\mathsf{V}}
\newcommand{\Viterbi}{\mathsf{Viterbi}}
\newcommand{\aV}{\mathsf{a}\hspace*{-0.23mm}\mathsf{V}}
\newcommand{\0}{\mathbb{0}}
\newcommand{\1}{\mathbb{1}}

\newcommand{\h}{\mathrm{h}}
\newcommand{\LL}{\mathrm{L}}
\newcommand{\Li}{\mathrm{L}_{\mathrm{i}}}
\newcommand{\Lr}{\mathrm{L}_{\mathrm{r}}}

\newcommand{\height}{\mathrm{height}}
\newcommand{\hp}{\mathrm{hp}}
\newcommand{\pos}{\mathrm{pos}}
\newcommand{\prefix}{\mathrm{prefix}}
\newcommand{\size}{\mathrm{size}}
\newcommand{\sub}{\mathrm{sub}}

\newcommand{\yield}{\mathrm{yield}}
\newcommand{\edge}{\mathrm{edge}}
\newcommand{\llabel}{\mathrm{label}}
\newcommand{\Free}{\mathrm{Free}}
\newcommand{\Bound}{\mathrm{Bound}}
\newcommand{\rk}{\mathrm{rk}}

\newcommand{\cut}{\mathrm{cut}}
\newcommand{\Cut}{\mathrm{Cut}}
\newcommand{\lcut}{\mathrm{lcut}}

\newcommand{\inp}{\mathrm{inp}}
\newcommand{\out}{\mathrm{out}}

\newcommand{\sem}[1]{[\![#1]\!]}
\newcommand{\semst}[1]{\llparenthesis #1 \rrparenthesis}
\newcommand{\sema}[1]{\langle #1 \rangle}
\newcommand{\runsem}[1]{[\![#1]\!]^{\mathrm{run}}}
\newcommand{\initialsem}[1]{[\![#1]\!]^{\mathrm{init}}}

\newcommand{\wrtsem}[1]{[\![#1]\!]^{\mathrm{wrt}}}
\newcommand{\ssem}[1]{[\![#1]\!]^{\mathrm{s}}}

\newcommand{\ttsem}[1]{[\![#1]\!]^{\mathrm{tt}}}
\newcommand{\infsum}[3]{\sideset{}{^{#1}}\sum\limits_{#2}#3}

\newcommand{\uh}[1]{\!\!\upharpoonright^{#1,\0}}
\newcommand{\floor}[2]{\lfloor #1 \rfloor_{#2}}

\newcommand{\lm}{\mathrm{l}} 
\newcommand{\lhs}{\mathrm{lhs}}
\newcommand{\rhs}{\mathrm{rhs}}
\newcommand{\first}{\mathrm{first}}
\newcommand{\last}{\mathrm{last}}

\newcommand{\AFwtL}{\mathrm{AFwtL}}
\DeclareMathOperator*{\bigboxplus}{\scalerel*{\boxplus}{\sum}}
\DeclareMathOperator*{\bigplus}{\scalerel*{+}{\sum}}

\newcommand{\psucc}{\succ^+}
\newcommand{\nf}{\mathrm{nf}}
\newcommand{\FL}{\mathrm{FL}(2+2)}
\newcommand{\state}{\mathrm{state}}
\newcommand{\estate}{\mathrm{estate}}
\newcommand{\rhorm}{\mathrm{rho}}
\newcommand{\crhorm}{\mathrm{crho}}

\newcommand{\lM}{\{\!|}
\newcommand{\rM}{|\!\}}

\newcommand{\sfMon}{\mathsf{Mon}}
\newcommand{\rmMon}{\mathrm{Mon}}
\newcommand{\sfnPol}{\mathsf{nPol}}
\newcommand{\sfPol}{\mathsf{Pol}}
\newcommand{\rmnPol}{\mathrm{nPol}}
\newcommand{\eval}{\mathrm{eval}}
\newcommand{\DD}{C}

\newcommand{\id}{\mathrm{id}}

\newcommand{\tree}{\mathrm{tree}}

\newcommand{\ttop}{\mathrm{top}}

\newcommand{\push}{\mathrm{push}}

\newcommand{\CL}{\mathrm{wcl}}

\newcommand{\NXT}{\mathrm{NXT}}
\newcommand{\RRT}{\mathrm{RT}}
\newcommand{\REST}{\mathrm{REST}}

\usepackage[most]{tcolorbox}
\newtheorem{theorem-rect}[theorem]{Theorem}
\newtheorem{corollary-rect}[theorem]{Corollary}
\newtheorem{lemma-rect}[theorem]{Lemma}
\newtheorem{construction-rect}[theorem]{Construction}
\newtheorem{definition-rect}[theorem]{Definition}

\tcolorboxenvironment{theorem-rect}{
		arc=3pt,				
		colback=white,			
		enlarge top by=2mm,		
		enlarge bottom by=-1mm,	
		grow to left by=3mm,	
		grow to right by=3mm,	
		boxrule=0.1mm,			
		left=1.7mm,				
		right=1.7mm				
		}						

\tcolorboxenvironment{corollary-rect}{
		arc=3pt,
		colback=white,
		enlarge top by=2mm,
		enlarge bottom by=-1mm,
		grow to left by=3mm,
		grow to right by=3mm,
		boxrule=0.1mm,
		left=1.7mm,
		right=1.7mm
              }

\tcolorboxenvironment{lemma-rect}{
		arc=3pt,
		colback=white,
		enlarge top by=2mm,
		enlarge bottom by=-1mm,
		grow to left by=3mm,
		grow to right by=3mm,
		boxrule=0.1mm,
		left=1.7mm,
		right=1.7mm
              }

\tcolorboxenvironment{construction-rect}{
		arc=3pt,
		colback=white,
		enlarge top by=2mm,
		enlarge bottom by=-1mm,
		grow to left by=3mm,
		grow to right by=3mm,
		boxrule=0.1mm,
		left=1.7mm,
		right=1.7mm
              }

\tcolorboxenvironment{definition-rect}{
		arc=3pt,
		colback=white,
		enlarge top by=2mm,
		enlarge bottom by=-1mm,
		grow to left by=3mm,
		grow to right by=3mm,
		boxrule=0.1mm,
		left=1.7mm,
		right=1.7mm
              }

\pgfdeclaredecoration{dashsoliddouble}{initial}{
  \state{initial}[width=\pgfdecoratedinputsegmentlength]{
    \pgfmathsetlengthmacro\lw{.7pt+.5\pgflinewidth}
    \begin{pgfscope}
      \pgfpathmoveto{\pgfpoint{0pt}{\lw}}%
      \pgfpathlineto{\pgfpoint{\pgfdecoratedinputsegmentlength}{\lw}}%
      \pgfmathtruncatemacro\dashnum{%
        round((\pgfdecoratedinputsegmentlength-3pt)/6pt)
      }
      \pgfmathsetmacro\dashscale{%
        \pgfdecoratedinputsegmentlength/(\dashnum*6pt + 3pt)
      }
      \pgfmathsetlengthmacro\dashunit{3pt*\dashscale}
      \pgfsetdash{{\dashunit}{\dashunit}}{0pt}
      \pgfusepath{stroke}
      \pgfsetdash{}{0pt}
      \pgfpathmoveto{\pgfpoint{0pt}{-\lw}}%
      \pgfpathlineto{\pgfpoint{\pgfdecoratedinputsegmentlength}{-\lw}}%
      \pgfusepath{stroke}
    \end{pgfscope}
  }
}

\tikzset{small circle/.style={circle, draw=black, inner sep=0pt,outer sep=0pt, minimum size=2.5pt}}

\newcommand{\rhombus}[2][] {%
  \coordinate[#1] (#2);
  \node[small circle] (#2 north) at ($(#2) + (0,0.25)$) {}; 
  \node[small circle] (#2 south) at ($(#2) + (0,-0.25)$) {};
  \node[small circle] (#2 east)  at ($(#2) + (0.5,0)$) {};
  \node[small circle] (#2 west)  at ($(#2) + (-0.5,0)$) {};
  \draw (#2 north) -- (#2 east) -- (#2 south) -- (#2 west) -- (#2 north);
}%

\title{{\bf \LARGE Weighted Tree Automata\\May it be a little more?}\\[0.6cm]
{\large Third Edition}}

\date{\today\endgraf \normalsize
  \vspace{50mm}
  \hspace*{40mm}
  \scalebox{0.8}{
   \begin{tikzpicture}
     \node (1){$\sfM(\cA) =(B^Q,\oplus,\0^Q,\delta_\cA)$};
     \node[right of=1, xshift=25em] (2) {$(\Pol(\Sigma,\B),\oplus,\widetilde{\0},\ttop_\Sigma)$};
     \node[below of=1, yshift=-4em, xshift= 0em] (3) {$\sfM_{\mathrm{im}}(\cA) =(\im(\sfh_\cA),\oplus,\0^Q,\delta_\cA)$};
     \node[right of=3, xshift=30em, yshift=-7em] (4) {$(\Pol(\Sigma,\B),\oplus,\widetilde{\0},\ttop_\Sigma)/_{\approx_{\sem{\cA}}}$};
          \node[right of=3, xshift=6em] (5) {\Large $\cong$};
          \node[right of=5, xshift=5em] (6) {$(\Pol(\Sigma,\B),\oplus,\widetilde{\0},\ttop_\Sigma)/_{\ker(\sfh_\cA)}$};
          \node[below of=5, xshift=8em, yshift=-4.1em] (7) {$\Big((\Pol(\Sigma,\B),\oplus,\widetilde{\0},\ttop_\Sigma)/_{\ker(\sfh_\cA)}\Big)/_{\rho_\cA}$};
          \node[right of=7, xshift=6em] (8) {\Large $\cong$};         
     \draw (2) edge[->,>=stealth] node[fill=white] {$\sfh_\cA$} (1);
        \draw (1) edge[->,>=stealth] node[fill=white] {smallest sub-$(\Sigma,\B)$-semimodule} (3);
        \draw (2) edge[->,>=stealth] node[fill=white] {$\pi_{\approx_{\sem{\cA}}}$} (4);
        \draw (6) edge[->,>=stealth] node[fill=white] {$\pi_{\rho_\cA}$} (7);
     \draw (2) edge[->,>=stealth] node[fill=white] {$\pi_{\ker(\sfh_\cA)}$} (6);
   \end{tikzpicture}
 } 
}

\author{Zolt\'an F\"ul\"op\\University of Szeged\\ Hungary \and
Heiko Vogler\\Technische Universit\"at Dresden\\ Germany}

\makeindex[intoc]

\begin{document}

\frontmatter 

\maketitle
\pagestyle{empty}
$\ $

\newpage

\pagestyle{empty}
Dedicated to the memory of  Magnus Steinby (1941-2021)

\newpage

$\ $

\setcounter{tocdepth}{2}


\include{preface-3rd-edition-by-authors}

\include{preface-2nd-edition-by-authors}

\include{preface}


\tableofcontents

\mainmatter 

\include{introduction}

\include{preliminaries}

\include{basic-wta}

\include{special-wta}

\include{basic-properties}

\include{relationship-initial-run}

\include{support-lang}

\include{pumping-lemmas}

\include{normal-forms}
\include{wcfg}

\include{wrtg-as-wcfg}

\include{closure-properties}

\include{decomposition-results}

\include{crisp-determinization}

\include{determinization-bu-reduced}

\include{support}
\include{Kleene}

\include{Medvedjev}

\include{Buechi}

\include{Myhill-Nerode7}

\include{char-of-rec-by-monoid-rep}

\include{AFwtL}

\include{L-valued-wta}


\cleardoublepage
\bibliographystyle{alpha}
\phantomsection
\addcontentsline{toc}{chapter}{Bibliography} 
\bibliography{book-wta-bib}


\printindex
\end{document}

%% file: preface-3rd-edition-by-authors.tex
\chapter*{Preface to the third edition}

In comparison to the second edition of this book, which was published on June 10, 2024 on arXiv, the following main points have been added or changed.

\begin{compactitem}
  \item We call a tuple $(A,\theta)$ where $A$ is a nonempty set and $\theta$ is a family of operations on $A$ an \emph{algebra} (instead of ``universal algebra'' as in the second edition of this book).
\item We have added Section \ref{sec:free-strong-bimonoid} with the free strong bimonoid of nested polynomials.
\item We have generalized Eilenberg's construction \cite[Thm.~9.1]{eil74} (also cf. \cite[Lm.~3.5.1]{fulvog24} for weighted tree automata over the semiring of natural numbers) to semirings in which the additive monoid is finitely generated (cf. Lemma~\ref{lm:wta-fin-gen-Eilenberg}).
\item We have added Chapter \ref{ch:run-support=initial-algebra-support} on the equality of run support and initial algebra support. This chapter is based on \cite{drovog24}.
\item We have restructured Chapter \ref{ch:Bozapalidis} considerably and have added Section~\ref{sec:B-A-BLs-theorem-for-the-det-case} with a characterization of bu-deterministically recognizable weighted tree languages in terms of m-syntactic congruences. This latter result is based on~\cite{fulvog25}.
  \item We have added Chapter \ref{ch:char-of-rec-by-monoid-rep} which shows a characterization of recognizability in terms of monoids.
  \end{compactitem}

\

\noindent Zolt\'an F\"ul\"op and Heiko Vogler \hfill \today \\
Szeged, Hungary and Dresden, Germany

%% file: preface-2nd-edition-by-authors.tex
\chapter*{Preface to the second edition}

In comparison to the first edition \cite{fulvog22} of this book, which was published on December 13, 2022, we have changed some parts and added further results. A few of the changes and those results are listed below.

\begin{itemize}

\item For definitions by induction and proofs by induction, we now employ the concept of terminating reduction system $(C,\succ)$, instead of well-founded set $(C,\prec)$. Essentially, they are the same, but we think that the new terminology and notations better fit to the presented theory.

\item We have added the definition of universal algebra. Since a universal algebra may contain infinitely many operations, we can consider, e.g., each vector space over the field $\B = (B,\oplus,\otimes,\0,\1)$ (with infinitely many scalar multiplications with $b \in B$) as a universal algebra. A $\Sigma$-algebra is a particular universal algebra with finitely many operations.

  \item By definition (new), each ranked alphabet $\Sigma$ has the property that $\Sigma^{(0)} \not= \emptyset$.

\item We have added two items to Observation 2.6.11 (old) which show how left- and right-distributivity influence the relationships between the sets of bi-locally finite strong bimonoids, weakly locally finite strong bimonoids, and locally finite strong bimonoids (see Observation~\ref{obs:zero-sum-free-property}(6) and (7)).

  \item In Section \ref{sec:rtg} (new) we have recalled the definition of regular tree grammars and reported about a decidability theorem (cf. Theorem~\ref{thm:Koszo22}).

\item We have changed the notion ``wta with identity transition weights'' (old) into ``crisp wta'' (new), because the latter notion seems to be well accepted in the community.

  \item We included a padding lemma (Lemma~\ref{lm:super-big-padding}) which, given a wta over a ranked alphabet with a certain property, can be used to construct another wta  over the same  alphabet with increased ranks such that the latter wta also has that property.  
 
\item Due to \cite{drofultepvog24}, we have strengthened Theorem 3.1.5 (old) by requiring less from the ranked alphabet (cf. Theorem~\ref{thm:closure-of-finite-set-i-recognizable}). By using Lemma~\ref{lm:super-big-padding}, we could even generalize this latter to any branching ranked alphabet (cf. Theorem~\ref{thm:not-monadic-wta-can-compute-closure-of-finite-subset}).

  \item In the (new) Section \ref{sec:images-semantics-restr-sb} we have collected some results on the finiteness of the images of the run semantics and the initial algebra semantics of wta over restricted strong bimonoids.  
    
    \item Theorem 7.4.3 (old) was wrong. So we have dropped it and its consequence Corollary 7.4.4 (old).
      
      \item We have included  new theorems in Chapter 13 (old) ``Elementary operations and M{\'e}dv{\'e}dj{\'e}v's theorem'' and Chapter 14 (old) ``Weighted MSO-logic and B-E-T's theorem''. Their proofs are based on the fact  that the run semantics of a wta over a bi-locally finite strong bimonoid is a recognizable step mapping, which was proved in Chapter 16 (old) ``Crisp-determinization''.
   
 In order to avoid forward references inside the book, we have moved up Chapter 16 (old)  together with the related Chapters 17 and 18 (old) (cf. Chapters~\ref{ch:crisp-determinization},~\ref{ch:determinization}, and \ref{ch:support}). The new theorems are Theorems~\ref{cor:Medvedjev-str-bm} and \ref{thm:Buechi-comm-bi-loc-fin}, respectively.

  \item We have added a new chapter (Chapter \ref{ch:Bozapalidis}) in which we report about the characterization of recognizability in terms of syntactic congruences assuming that the weight algebra is a field (cf.~Theorem~\ref{th:MN-fields}).

  \item We have corrected some typos and small technical mistakes. Also, we have improved the clarity of some of the proofs.

  \end{itemize}

\noindent Zolt\'an F\"ul\"op and Heiko Vogler \hfill June 11, 2024 \\
Szeged, Hungary and Dresden, Germany

%% file: preface.tex
\chapter*{Preface to the first edition}
Weighted automata form a quantitative extension of the traditional automaton model which is fundamental for Computer Science. They can be viewed as classical nondeterministic finite automata in which the transitions are equipped with weights. These weights could model, for instance, the cost, reward or probability of executing the transition. Already in 1961, Marcel Schützenberger investigated weighted automata and described their behaviors on finite words as rational formal power series. This extended Kleene's fundamental theorem on the coincidence of the classes of regular and rational formal languages into a quantitative setting. Subsequently, several decidability problems concerning context-free languages could be solved using weighted automata techniques, and up to now no other proof methods are known for this. This developed into a flourishing field, as described in books by Samuel Eilenberg (1974), Arto Salomaa and Matti Soittola (1978), Wolfgang Wechler (1978), Werner Kuich and Arto Salomaa (1986), Jean Berstel and Christophe Reutenauer (1988), Jacques Sakarovitch (2009) and the ``Handbook of Weighted Automata'' (2009).

Often, the weights employed for weighted automata are taken from a field like the rational numbers, the real numbers or the complex numbers, respectively, with their usual operations. However, we might also consider just the natural numbers which, with their usual operations of addition and multiplication, form the classical example of a semiring. Moreover, in many important applications calculations of the weights combine the operations maximum and addition, or minimum and addition, where the distributivity of the second operation over the first again yields a semiring as weight structure. The weights of the transitions can very conveniently be described as a single matrix (for each letter); the weight at entry  $(i,j)$  is the weight of the transition from state  $i$  to  $j$. Since in semirings multiplication is distributive over addition, matrix multiplication is associative. This enables us to use techniques from algebra to express and analyse the behavior of weighted automata over semirings, and this was one of the reasons for the success of the weighted automaton model. All classical automata can be recast as weighted automata by taking the Boolean semiring  $\{0,1\}$  as weight structure.

With the turn of the century, further quantitative models began to be investigated as weight structure. For instance, we might be interested in average weights, or in discounting of weights. In semiring-weighted automata, the weight of runs is computed by employing the second operation of the semiring, whereas the weights of the (in case of non-deterministic automata, often several) runs are combined into a single weight by using the semiring's first operation. The required distributivity of the second operation over the first is quite a strong mathematical assumption. What happens without this assumption? Such 'semirings without requiring distributivity' were called strong bimonoids. For instance, in lattices, which arise in multi-valued logic, both operations supremum and infimum are associative but in general not distributive over each other, and large parts of lattice theory in Mathematics concern non-distributive lattices. There are many further possibilities how strong bimonoids may arise.
Since for strong bimonoids matrix multiplication is in general not associative, it follows that for weighted automata, techniques from linear algebra can no longer be used. However, fortunately, direct automata constructions can very often still be used to obtain results which previously were derived only for semirings.

In Computer Science, a most important data structure arising, e.g., in programming analysis and from pushdown automata and context-free languages, is given by trees. Therefore, a large part of automata theory concerns automata over trees. Already early on, much research concerned extensions of results from weighted automata over words to weighted tree automata, which becomes often more intricate. Moreover, weighted tree automata have found very interesting recent applications in natural language processing. In view of the above, much research has recently been devoted to weighted tree automata with strong bimonoids as weight structure.

The present book, written by two leading researchers of this area, is the first book on weighted tree automata. It presents large parts of the theory of weighted tree automata over strong bimonoids and semirings in a systematic way. As indicated above, for weighted automata over strong bimonoids two kinds of defining the behavior arise:
\begin{compactitem}
  \item an automata-theoretic oriented approach using runs of the automaton and calculating their weights, and
  \item a universal algebra oriented method similar to the algebraic method mentioned above for words, but adjusted to trees.
  \end{compactitem}
  In contrast to words, trees with tree concatenation do not form a monoid; this is the reason why the algebraic method mentioned above for weighted automata on words now for weighted tree automata has been replaced by a universal algebra approach. Over semirings, the two semantics for weighted tree automata can be shown to coincide, but over strong bimonoids in general they differ. Consequently, general structure results for the possible behaviors of weighted tree automata may have two versions, using the run semantics respectively the initial algebra semantics.
  
The authors show a number of these structure results. These include, among others, descriptions of the behaviors of weighted tree automata by rational expressions, by weighted context-free or regular grammars, by weighted versions of monadic second order logic, by closure results under various operations, and as abstract families of weighted tree languages, as well as weighted versions of classical pumping lemmas and of determinization results. This shows that many structure results for weighted tree automata can also be derived in the case of strong bimonoids. The authors give full mathematical proofs, which are needed as weighted tree automata often have combinatorial intricacies which are easy to overlook. The reader will appreciate that the authors include numerous examples explaining the various aspects of definitions and differences in the results, often illustrating them with pictures.

The field of weighted automata over strong bimonoids is still developing. This systematic presentation of large parts of recent research is therefore very valuable and right in time. I am sure that this book will stimulate much further research in this exciting area.

\

\noindent Leipzig, December 2022\\
Manfred Droste

%% file: introduction.tex
\chapter{Introduction}
\pagestyle{headings}
\setcounter{page}{1}
\label{ch:introduction}
  
The purpose of this book is to present the basic definitions and some of the important results for weighted tree automata in a mathematically rigorous and coherent form with full proofs. The concept of weighted tree automata is part of Automata Theory and it touches the area of Universal Algebra. It originated from two sources: weighted string automata and finite-state tree automata. 

\paragraph{Historical perspective.} 
Weighted string automata were introduced in \cite{sch61}; here we recall briefly \cite[Def.~1']{sch61}. Roughly speaking, a weighted string automaton  $\cA$ is a finite-state automaton in which each transition has a weight, which is an element of the set $\mathbb{Z}$ of integers. Together with summation and multiplication, $\mathbb{Z}$ forms the  \emph{weight algebra} of $\cA$. For each input symbol $a$, the weights of the transitions  are represented as a $Q \times Q$-matrix over $\mathbb{Z}$ where $Q$ is the finite set of states of $\cA$. Due to distributivity of multiplication over summation, the set of $Q \times Q$-matrices over $\mathbb{Z}$ together with matrix multiplication and the unit matrix forms a monoid. Thus, the representation can be extended to a monoid homomorphism from the free monoid over the set of input symbols to the monoid of $Q \times Q$-matrices over~$\mathbb{Z}$. Finally, for a given  input string $w$, its matrix representation is multiplied from the left by an initial weight vector and from the right by a final weight vector; the result is the weight of $w$ computed by $\cA$.  In this way, $\cA$ assigns to each input string a weight in $\mathbb{Z}$; this assignment is called the \emph{formal power series recognized by $\cA$}  or \emph{weighted language recognized by $\cA$}.\footnote{Since we will not deal with convergence questions, we prefer the latter notion.}. We might say that this weighted language is the monoid semantics of $\cA$.
  Although in \cite{sch61} mostly the semiring of natural numbers, the ring of integers, or commutative rings were considered, the idea of \cite{sch61} was general enough to deal with weighted string automata over arbitrary semirings (cf. \cite[p.~58]{wec78}). 
  
Since 1961, weighted string automata and the set of recognizable weighted languages were studied intensively. Different kinds of semantics were investigated (run semantics, free monoid semantics, initial algebra semantics) and various classes of weight algebras were used (e.g.,  bounded lattices, fields, semirings, pairs of t-conorm and t-norm, valuation monoids, strong bimonoids). We note that, neither in bounded lattices nor in valuation monoids or strong bimonoids or pairs of t-conorm and t-norm, the multiplication has to be distributive over the summation.

The development of weighted string automata is witnessed by the following list of books and survey papers: \cite{eil74,salsoi78,wec78,kuisal86,berreu88,kui97,mormal02,sak09,drokuivog09,drokus21}.
The state-of-the-art of this area was also cultivated and extended by the biennial workshops ``Weighted Automata: Theory and Applications'' (WATA) since 2002 (cf. \cite{drovog03,drovog05,drovog07,drovog09,drovog11a,drovog14a,droesilar18,dromalvog21}).

Finite-state tree automata were invented independently by \cite{don65,don70} and \cite{thawri68}. A finite-state tree automaton processes a given input tree over some ranked alphabet $\Sigma$ also by means of transitions. Now a transition on a $k$-ary input symbol $\sigma$ has not only one but $k$ starting states (one for the root of each of the $k$ subtrees below  $\sigma$), and it has one ending state (at the node which is labeled by $\sigma$). Viewed as a bottom-up device\footnote{We draw a tree with its root up and the leaves down, hence bottom-up means: from the leaves towards the root.}, a finite-state tree automaton $\cA$ over an input ranked alphabet $\Sigma$ can be considered as a finite $\Sigma$-algebra, as it is known from Universal Algebra.
Additionally, $\cA$ identifies the set of final states as a subset of the carrier set of that $\Sigma$-algebra. Then $\cA$ recognizes the set of all $\Sigma$-trees which are interpreted in the corresponding $\Sigma$-algebra as some final state.  We might say that this tree language is the initial algebra semantics of $\cA$.

  Also the concept of finite-state tree automata and the set of recognizable tree languages were investigated intensively. Here is a list of books, lecture notes, and survey papers  on finite-state tree automata: \cite{eng75-15,eng80,gecste84,gecste97,comdaugiljaclugtistom08,lodtho21}.
Some more developments on tree automata and tree transducers were published in the series ``International Workshop on Trends in Tree Automata and Tree Transducers'' \cite{man13,fil15,TTATT16}.

It is natural to combine the concepts of weighted string automata and finite-state tree automata, resulting in the concept of  weighted tree automata (wta). In \cite{inafuk75} wta were introduced with the weight algebra being  the real number interval $[0,1]$ in which summation is $\max$   and multiplication is $\min$; such wta were called fuzzy tree automata. In \cite{berreu82}, a wta is a finite-dimensional vector space~$\V$ over some field $\B$ equipped with a $\Sigma$-algebra.  Roughly speaking, each dimension corresponds to a state and each $k$-ary input symbol is interpreted in the $\Sigma$-algebra as a $k$-ary multilinear operation over~$\V$. In the following years, wta over different weight algebras were introduced:  wta over semirings \cite{aleboz87},  wta over distributive multioperator monoids \cite{kui98}, wta over multioperator monoids \cite{fulmalvog09,stuvogfue09}, wta over pairs of t-conorm and t-norm on the unit interval $[0,1]$ \cite{bozlou10}, wta over strong bimonoids \cite{rad10}, and wta over tree valuation monoids \cite{drogoemaemei11}.

Here we list some of the seminal papers and survey papers on wta: \cite{inafuk75,berreu82,aleboz87,kui98,kui99a,boz99,esikui03,esiliu07,fulvog09new,drogoemaemei11}.
We also mention some recent PhD theses: \cite{bor04b,hoe07,mat09,may10,tei16,goe17,her20a,pau20,jon21,doe22}.

\paragraph{Weighted tree automata over strong bimonoids.} In this book we will investigate wta over strong bimonoids. In this paragraph we provide an overview on this automaton model.
A strong bimonoid is an algebra $\B=(B,\oplus,\otimes,\0,\1)$ such that
\begin{compactitem}
\item $(B,\oplus,\0)$ is a commutative monoid (and $\oplus$ is called the summation),
  \item $(B,\otimes,\1)$ is a monoid (and $\otimes$ is called the multiplication), and
\item $\0$ is annihilating with respect to $\otimes$, i.e., $b \otimes \0 = \0 \otimes b = \0$ for each $b \in B$.
\end{compactitem}
For instance, each semiring (and thus each ring, semifield, and field) is a strong bimonoid; also each bounded lattice (and thus each complete lattice and residuated lattice) is a strong bimonoid; we note that a large part of the field of lattice theory refers to non-distributive lattices \cite{bir93,gra03}.
The algebra $(\mathbb{N}_{\infty},+,\min,0,\infty)$ is a strong bimonoid, but it is not a semiring, because, e.g., $\min(a,a+a) \not= \min(a,a) + \min(a,a)$ for each $a \in \mathbb{N}\setminus \{0\}$, and hence the multiplication (here: $\min$) does not distribute over the summation (here:~$+$). Further examples of non-distributive strong bimonoids are the finite lattices $\Nfive$ and $\Mthree$ (cf. Figure \ref{fig:lattices-N5-M3-Graetzer}) and the infinite bounded lattice $\sfFL$ (cf. Figure \ref{fig:FL2+2}). 

A wta over $\Sigma$ and $\B$, for short:  $(\Sigma,\B)$-wta, is a tuple $\cA= (Q,\delta,F)$ which is almost like a  finite-state tree automaton, i.e., $Q$ is a finite set of states, $\delta=(\delta_k\mid k \in \mathbb{N})$ is a family of transition mappings, and $F: Q \to B$ is the root weight mapping. However, now each transition and each final state carries a weight, which is taken from $B$. More precisely, for each $k \in \mathbb{N}$, the mapping $\delta_k$ has the type
\[\delta_k: Q^k \times \Sigma^{(k)} \times Q \to B\]
and it maps each transition $(q_1 \cdots q_k, \sigma,q)$ with starting states $q_1,\ldots,q_k$, a $k$-ary input symbol $\sigma$, and ending state $q$ to a weight, i.e., an element of $B$. Since $\Sigma$ is finite, only finitely many $\delta_k$ are different from the empty mapping. Finally, the mapping  $F: Q\to B$ maps each state to an element of $B$; this is used as root weight (or final weight).

We can define two semantics of $\cA$: the run semantics $\runsem{\cA}$ and the initial algebra semantics~$\initialsem{\cA}$ where each of them is a weighted tree language; more precisely, 
\[
\runsem{\cA}: \T_\Sigma \to B  \ \ \text{ and } \ \  \initialsem{\cA}: \T_\Sigma \to B
\]
where $\T_\Sigma$ is the set of all $\Sigma$-trees.

The run semantics is based on the idea of a run of $\cA$ on a given input tree $\xi$. This is a mapping $\rho$ which maps each position of $\xi$ to a state of $Q$. Thus $\rho$ determines, for each position $w$ of $\xi$, a transition $(q_1 \cdots q_k, \sigma,q)$ where $q_i$ (for $i \in \{1,\ldots,k\}$) and $q$ are the states assigned by $\rho$ to the $i$-th successor of $w$ and to $w$ itself, respectively, and $\sigma$ is the label of $\xi$ at position $w$. By applying $\delta_k$ to $(q_1 \cdots q_k, \sigma,q)$, we obtain the weight of this transition. Then the weight of $\rho$, denoted by $\wt(\xi,\rho)$, is the $\otimes$-product of the weights of the transitions for each position of $\xi$ (where the factors are ordered according to the depth-first post-order of the corresponding positions). Finally, we let
\[
\runsem{\cA}(\xi) = \bigoplus_{\rho\in \R_\cA(\xi)} \wt(\xi,\rho) \otimes F({\rho(\varepsilon)})
\]
where $\R_\cA(\xi)$ is the set of all runs of $\cA$ on $\xi$, and $\rho(\varepsilon)$ is the state assigned by $\rho$ to the root of $\xi$. A weighted tree language $r: \T_\Sigma \to B$ is run recognizable if there exists a $(\Sigma,\B)$-wta $\cA$ such that $r= \runsem{\cA}$. We denote by  $\Rec^{\mathrm{run}}(\Sigma,\B)$ the set of weighted tree languages which are run recognizable  by some $(\Sigma,\B)$-wta.

\sloppy The initial algebra semantics of $\cA$ is based on the concept of vector algebra of $\cA$, which is the $\Sigma$-algebra $(B^Q,\delta_\cA)$ where $\delta_\cA$ associates with each $k$-ary input symbol $\sigma \in \Sigma$ a $k$-ary operation \mbox{$\delta_\cA(\sigma): B^Q \times \cdots \times B^Q \to B^Q$.} 
For each $q\in Q$, the $q$-component of a vector $v\in B^Q$ is denoted by~$v_q$.
The operation $\delta_\cA(\sigma)$ is defined, for every $v_1,\dots,v_k \in B^Q$  and $q \in Q$, by
\[\delta_\cA(\sigma)(v_1,\dots,v_k)_q 
  = \bigoplus_{q_1\cdots q_k \in Q^k} \Big(\bigotimes_{i\in[k]} (v_i)_{q_i}\Big) \otimes \delta_k(q_1\cdots q_k,\sigma,q) \enspace.\]
Then, for every $\xi\in \T_\Sigma$, we let  
\[
  \initialsem{\cA}(\xi) = \bigoplus_{q \in Q} \h_\cA(\xi)_q \otimes F(q)
  \]
where $\h_\cA$ denotes the unique $\Sigma$-algebra homomorphism from the $\Sigma$-term algebra to the $\Sigma$-algebra $(B^Q,\delta_\cA)$. A weighted tree language $r: \T_\Sigma \to B$ is initial algebra recognizable if there exists a $(\Sigma,\B)$-wta $\cA$ such that $r= \initialsem{\cA}$. We denote by  $\Rec^{\mathrm{init}}(\Sigma,\B)$ the set of weighted tree languages which are initial algebra recognizable  by some $(\Sigma,\B)$-wta.

If $\B$ is a semiring, i.e., $\otimes$ distributes over $\oplus$, then $\runsem{\cA}=\initialsem{\cA}$ (as in the case of weighted string automata over semirings, cf. \cite[Lm.~4]{drostuvog10}).
Hence for wta over semirings we drop the attributes ``run'' and ``initial algebra'' and we speak simply about recognizable weighted tree languages. 
For arbitrary strong bimonoids this equality in general  does not hold (cf. \cite[Ex.~25]{drostuvog10} and \cite[Ex.~3.1]{cirdroignvog10}). However, the equality does hold for arbitrary strong bimonoids if $\cA$ is bottom-up-deterministic.

Since the set of trees does not have a monoid structure (in contrast to the set of strings) and distributivity need not hold in a strong bimonoid, there does not exist a semantics of wta over strong bimonoids which directly corresponds to Sch\"utzenberger's semantics of wsa described above. However, for the case that $\B$ is a commutative semiring, the concept of $\B$-$\Sigma$-representation \cite{bozale89} can be viewed as kind of monoid semantics of wta. In that concept it is exploited that the set of contexts over $\Sigma$ (i.e., trees with a variable at exactly one leaf) is free in the set of all monoids. We will elaborate this in Subsection~\ref{sec:monoid-representation-of-wta}.

\paragraph{Topics of the book.} In this book we present a part of the theory of wta over strong bimonoids. Most of the theorems which we present have previously been published in research articles; others are  generalizations of the corresponding published theorems.
Some results are new.

In Chapter \ref{ch:preliminaries} we collect all the technical ingredients for our development such that the book is self-contained and hence also accessible for, e.g., master students of Mathematics or Computer Science. So this chapter is quite long, and of course, it may be consulted on demand.

In Chapter \ref{chapter:basic-wta} we define the concept of wta over some ranked alphabet $\Sigma$ and some  strong bimonoid $\B$, and with each such $(\Sigma,\B)$-wta $\cA$, we associate the run semantics $\runsem{\cA}$ and the initial algebra semantics $\initialsem{\cA}$. We define the restricted versions bottom-up-deterministic wta and crisp-deterministic wta. We give several examples of a wta and both kinds of semantics. We introduce the useful concept of state algebra.

In Chapter \ref{ch:special-cases}, in order to connect the concept of wta with its historical predecessors, we discuss particular cases of wta:
\begin{compactitem}
\item wta over string ranked alphabets (which are equivalent to weighted string automata),
\item wta over the Boolean semiring (which are equivalent to finite-state tree automata),
\item wta over the  semiring of natural numbers (which reflect multiplicities in finite-state tree automata),
\item wta over commutative semirings (which are equivalent to multilinear representations).
 \end{compactitem}

 In Chapter \ref{ch:basic-properties} we prove  basic properties of wta and of their bottom-up and crisp-deterministic subconcepts. The first property concerns the mapping $\h_\cA$ of a wta $\cA$ and a way how to split it.
 The second property is a consequence of the annihilation property of $\0$ for the multiplication $\otimes$. The third property is that, by means of a padding, we can extend the rank of input symbols of a wta without changing its semantics (cf. Lemma~\ref{lm:super-big-padding}). Finally, we consider the images of the semantics of wta and prove a universality property (cf. Theorem~\ref{thm:not-monadic-wta-can-compute-closure-of-finite-subset}) and show sufficient conditions under which these images are finite (cf. Lemmas~\ref{lm:bi-loc-finite-run-image finite} and \ref{lm:loc-finite-init-image finite}). As consequence, we obtain that the initial algebra semantics of wta over weakly locally finite strong bimonoids which are not locally finite, can generate strictly more sets of values than weighted string automata can generate (cf. Corollary \ref{cor:weak-loc-fin-not-loc-fin-weaker-wsa}).

In Chapter \ref{ch:comparison-of-semantics}  we provide two algorithms (cf. Section \ref{sec:complexity-semantics}) which compute in a natural way the values $\runsem{\cA}(\xi)$ and $\initialsem{\cA}(\xi)$ for any given wta $\cA$ and input tree $\xi$. We analyse the complexity of these algorithms and, as is to be expected, the second algorithm is more efficient than the first one. Moreover, we compare the run semantics and the initial algebra semantics  of wta and show some examples of wta for which these semantics are different. We prove that there exist a strong bimonoid $\B$ and a $(\Sigma,\B)$-wta $\cA$ such that $\initialsem{\cA}\not\in \Rec^{\mathrm{run}}(\Sigma,\B)$ (cf. Theorem \ref{thm:init-not-run}).    We show conditions under which, for each wta, the two semantics are equal (cf. Theorems \ref{thm:bu-det:init=run} and  \ref{thm:semiring-run=initial}).

In Chapter \ref{ch:run-support=initial-algebra-support}, we characterize the equality $\supp(\initialsem{\cA})=\supp(\runsem{\cA})$, for each $(\Sigma,\B)$-wta $\cA$, by means of conditions on $\Sigma$ and $\B$ (cf. Theorem~\ref{thm:bi-strongly-zsf-equiv-equ-supp}).

In Chapter \ref{ch:pumping-lemmas}  we prove three pumping lemmas:  one for runs of wta over strong bimonoids (cf. Theorem~\ref{thm:pumping-lemma-for-runs}), another one for supports of weighted tree languages recognizable by wta over positive strong bimonoids  (cf. Corollary \ref{cor:pumping-lemma-for-runs}), and a third one for supports of weighted tree languages recognizable by wta over fields (cf. Theorem \ref{thm:pumping-wta-fields}). We use  these lemmas to show that certain weighted tree languages are not recognizable.

In Chapter \ref{ch:normal-forms} we prove under which conditions a wta can be transformed into an equivalent one which is trim, i.e., each state is ``useful'' (cf. Theorems \ref{thm:t1} and \ref{thm:t2}). Moreover, we show how to normalize the root weight mapping $F$ such that there exists a unique state $q$ with $F(q) = \1$ and $F(p)=\0$ for each state $p$ different from $q$ (cf. Theorem \ref{thm:root-weight-normalization-run}). Finally, we prove that each wta can be transformed into a run equivalent crisp wta (i.e., each transition has weight $\0$ or $\1$) if the multiplication is locally finite \ref{thm:normalizing-transition-weights}.

In Chapter \ref{chapt:wcfg} we recall the concept of weighted context-free grammar from \cite{chosch63} and prove several normal form lemmas (cf. Section \ref{sect:wcfg-normal-forms}).   We prove the fundamental connection between weighted tree languages that are run recognizable by wta and that are generated by weighted context-free grammars (cf. Theorem \ref{thm:yield(wta)=cf}).
This generalizes the well-known theorem \cite{bra69,don70} saying that the yield of a recognizable tree language is a context-free language, and vice versa, each context-free language can be obtained in this way. 

In Chapter \ref{ch:regular-tree-grammars} we introduce weighted regular tree grammars as particular weighted context-free grammars. We show that, under certain conditions, weighted regular tree grammars and the run semantics of wta are equally expressive (cf.  Theorem \ref{thm:rec=reg}).

In Chapter \ref{ch:closure-properties} we prove a number of closure properties of the set of recognizable weighted tree languages, viz.,
closure under sum,  scalar multiplication, Hadamard product, top-concatenations, tree concatenations, Kleene stars, yield-intersection with weighted regular  string languages, strong bimonoid homomorphisms, tree relabelings, linear and nondeleting tree homomorphisms, inverse of linear tree homomorphisms, and weighted projective bimorphisms. A brief summary of these closure results is presented in Table~\ref{fig:table-summary-closure}.

In Chapter \ref{ch:decomposition} we introduce two different kinds of weighted local systems and  generalize another fundamental result from the theory of recognizable tree languages: each recognizable tree language is the image of a local tree language under a tree relabeling (cf. Theorem \ref{thm:decomposition-2}). Moreover, we prove a decomposition of the run semantics of a wta into the inverse of a deterministic tree relabeling, intersection with a local tree language, and a homomorphism into an evaluation algebra  (cf. Theorems \ref{thm:decomposition-1}).

In Chapters \ref{ch:crisp-determinization} and \ref{ch:determinization} we deal with the questions under which conditions it is possible to construct, for a given wta, an equivalent  crisp-deterministic wta and a bottom-up-deterministic wta, respectively. In general, the usual subset construction for unweighted automata leads to an infinite state set and thus  cannot be employed.

In Chapter \ref{ch:support} we deal with the question whether the support of a recognizable weighted tree language is a recognizable tree language. 
We show that, in general, the answer is no, and we give sufficient conditions under which the answer is yes.

In Chapters \ref{ch:Kleene}, \ref{ch:Medvedjev}, \ref{ch:Buechi}, and \ref{ch:Bozapalidis} we show how  well-known results of Kleene, M{\'e}dv{\'e}dj{\'e}v, B\"uchi-Elgot-Trakhtenbrot, and Myhill-Nerode, respectively, can be generalized from the unweighted string case to the weighted tree case.   
 In particular, we show the proofs of the following characterization theorems:
\begin{compactitem}
  \item for commutative semirings: recognizable weighted tree languages in terms of rational weighted tree languages (cf. Theorem \ref{thm:Kleene}),
  \item for semirings: recognizable weighted tree languages in terms of $\times$-restricted representable weighted tree languages (cf. Theorem \ref{thm:Medvedjev}),
    \item for bi-locally finite strong bimonoids: run recognizable weighted tree languages in terms of representable weighted tree languages (cf. Theorem~\ref{cor:Medvedjev-str-bm}),
\item for strong bimonoids: run recognizable weighted tree languages in terms of MSO-definable weighted tree languages (cf.~Theorem~\ref{thm:Buechi}),
\item for commutative semirings: recognizable weighted tree languages in terms of carefully extended MSO-definable weighted tree languages (cf. Theorem \ref{thm:Buechi-extended}),
\item for commutative and bi-locally finite strong bimonoids: run recognizable weighted tree languages in terms of (fully) extended weighted MSO-logic (cf. Theorem \ref{thm:Buechi-comm-bi-loc-fin}), and
\item for fields: recognizable weighted tree languages in terms of syntactic congruences and the corresponding finite-dimensional vector spaces (cf.~Theorems~\ref{th:MN-fields}, \ref{thm:left-quotient-vector-space}, and~\ref{thm:MN-semifield-det}).
 \item for fields: recognizable weighted tree languages in terms of syntactic congruences and the corresponding finite-dimensional vector spaces (cf.~Theorems~\ref{th:MN-fields} and \ref{thm:left-quotient-vector-space}),
  \item for semifields: bottom-up-deterministically recognizable weighted tree languages in terms of m-syntactic congruences and the corresponding finitely generated scalar algebras (cf.~Theorems~\ref{thm:MN-semifield-det}).
  \end{compactitem}

In Chapter \ref{ch:char-of-rec-by-monoid-rep} we recall the concept of $(\Sigma,\B)$-monoid representation and we  prove that, for each $(\Sigma,\B)$-wta $\cA$ over some commutative semiring $\B$ we can construct a consistent $(\Sigma,\B)$-monoid representation which is equivalent to $\cA$ (cf. Theorem~\ref{theo:monoid-characterization-of-run-semantics}). The semantics of the constructed monoid representation can be viewed as the free monoid semantics of $\cA$. Moreover, for each 
consistent $(\Sigma,\B)$-monoid representation, we can construct an equivalent $(\Sigma,\B)$-wta if $\B$ is a field (cf. Corollary~\ref{cor:field-from-consistent-rep-to-wta}).

In Chapter \ref{ch:AFwtL} we prove that, for commutative and $\sigma$-complete semirings, the set of recognizable weighted tree languages is the smallest principal abstract family of weighted tree languages (cf. Theorem~\ref{thm:REGnB-smallest-princ-AFwtL}). This generalizes the well-known situation for the set of recognizable string languages \cite{gin75}.

Finally, in Chapter \ref{ch:L-valued-wta} we collect some of the results of the previous chapters for the special case of wta over bounded lattices. In particular, we prove the inclusion relationship between several sets of recognizable weighted tree languages (cf. Theorem~\ref{thm:Hasse-L-2}). Moreover, we prove the characterizations of $\Rec^{\mathrm{run}}(\Sigma,\sfL)$ in terms of  representations (cf. Corollary \ref{cor:Medvedjev-L-valued-wta-str-bm}) and in terms of extended  weighted MSO formula (cf. Corollary~\ref{thm:BET-bounded-lattices}).

In Figure \ref{fig:overview-models} (at the end of this introduction) we show an overview of the  models of automata and models of grammars which occur in this book, and we indicate their relationship.

\paragraph{Topics not included in this book.}  We have covered some  of the important theorems for wta over strong bimonoids. However, this book is not meant to present the whole theory of wta over strong bimonoids, not even that of wta over semirings. The choice of the presented material is biased by our personal research in this area. In particular, we did not address the topics of the following list. For each topic, we have indicated some of the publications where the reader can start investigating the topic.

{\small
  \begin{compactitem}
\item wta on infinite trees  \cite{rah07,borpen11,lehpen14}
 \item wta on unranked trees \cite{hoemalvog09,drovog11,droheu15,droheuvog15,goe17}
\item (weighted) forest automata \cite{str09,doe19,doe21,doe22}
 \item wta over M-monoids  \cite{kui98,kui99a,mal04,mal05a,fulmalvog09,stuvogfue09,fulstuvog12,teiost15,fulvog18} 
 \item wta over tree valuation monoids \cite{drogoemaemei11,dromei12,teiost15,drofulgoe16,goe17,drofulgoe19}
  \item probabilistic tree automata \cite{magmor70,ell71,wei15}
 \item decidability and undecidability \cite{sei89,sei90,boz91,sei94,bor04,mal04,mal05a,seimankem18,fulkosvog19,pau20,drofulkosvog20b,drofulkosvog21} (except Lemma \ref{lm:fin-der-decidable} and Theorem \ref{thm:run-c-d-decidable})
 \item minimization of wta \cite{hoemalvog09} (except Theorem~\ref{th:minimal-MN} and Theorem~\ref{thm:minimization-theorem-new})
            \item varieties of weighted tree languages \cite{fulste11,stejurcan15}
   \item cut sets of recognizable weighted tree languages over some ordered structure \cite{bormalsestepvog06,sestepvog08}
   \item learning of wta \cite{mal07,drevog07,drehoemal11,balmoh15}
     \item weighted tree transducers \cite{kui99,engfulvog02new,fulvog03,fulgazvog04,mal04,mal05a,mal05b,malvog05,mal06,mal06a,mal06c,fulvog09new,fulmalvog11}
     \item wta  with storage \cite{hervog15,vogdroher16,fulhervog18,fulvog19,hervogdro19,fulvog19a,her20a,doefulvog24}
           \item applications of wta (and weighted tree transducers) in syntactic processing of natural languages \cite{perrilspr94,yamkni02,galhopknimar04,grakni04,knigra05,graknimay08,knimay09,malsat09,fulmalvog10,malsat10,may10,mal11,mal11b,buy13,buevogned12,bue14,bjodrezec15,dregebvog16,mal17,die18,moevog19,moevog19a,moevog21}; also cf. the workshops \cite{drekuh10,drekuh12}.
       \end{compactitem}
     }

     \paragraph{Notes to the reader.}$\ $ 
     
     ... about prerequisites: \\
     The reader is assumed to be familiar with the fundamental concepts, results, and constructions in the theory of tree automata \cite{eng75-15,gecste84,gecste97,comdaugiljaclugtistom08} and context-free languages \cite{har78,hopull79,hopmotull07}.

     \label{page:to-notion-construction}
... about the notion ``construction'':\\
In general, a construction  is an algorithm which takes some objects as input and produces some objects as output in an effective manner.
For instance, \cite[Thm.~2.4.2]{gecste84} states that, for two finite-state tree automata $A_1$ and $A_2$, a finite-state tree automaton $A$ can be constructed which recognizes the intersection $\LL(A_1) \cap \LL(A_2)$ of the tree languages which are recognized by $A_1$ and $A_2$; then the proof of  \cite[Thm.~2.4.2]{gecste84} (essentially) shows an algorithm which, on input $A_1$ and $A_2$, builds up effectively $A$ such that $\LL(A) = \LL(A_1) \cap \LL(A_2)$.

 Also in the present  book we show a number of constructions. However, here the situation is more complicated for those constructions which involve some strong bimonoid $\B$. If the algorithm involves calculations in $\B$, it may happen that the calculations in $\B$ are not effective. This can arise, e.g., because the arguments of the calculation cannot be given effectively or because the operations $\oplus$ and $\otimes$ are not given algorithmically. 
 In this unpleasant situation, we understand by ``construction'' just the definition of the output objects, and not an algorithm for their effective building up process. So, in our understanding a construction
\begin{compactitem}
\item takes some objects as input,
\item if the input objects are given effectively and the operations of the strong bimonoid are effective, then the construction gives the output objects effectively,
\item otherwise the construction merely defines the output objects (in a mathematical sense).
\end{compactitem}
 
Let us illustrate this understanding by considering the ``weighted version'' of \cite[Thm.~2.4.2]{gecste84}. In Theorem~\ref{thm:closure-Hadamard-product}(1) we claim: ``Let $\Sigma$ be a ranked alphabet, $\B$ be a strong bimonoid, and $\cA_1$ and $\cA_2$ be two $(\Sigma,\B)$-wta. If $\B$ is a commutative semiring, then we can construct a $(\Sigma,\B)$-wta $\cA$ such that  $\sem{\cA} = \sem{\cA_1} \otimes  \sem{\cA_2}$.'' If $\B$ is, e.g., the Boolean semiring or the semiring of natural numbers, then our proof shows an algorithm which builds up $\cA$ effectively. However, if $\B$ is, e.g., the field of real numbers and some of the coefficients of $\cA_1$ or $\cA_2$ cannot be given effectively, then our proof just shows the definition of $\cA$.

... about general conventions:\\
In order to avoid many repetitions of the statements ``for each ranked alphabet $\Sigma$'' and ``for each strong bimonoid $\B$'',  we will use the general conventions that $\Sigma$ always stands for an arbitrary, i.e., universally quantified ranked alphabet (cf. page \pageref{page:Sigma0-ne-empty}) and that $\B$  denotes an arbitrary strong bimonoid (cf. page~\pageref{page:general-convention-on-B}). Such a convention is placed inside \verb!\begin{quote}-\end{quote}! in emphasized letter style. In some chapters or sections, we develop results for a particular subset of strong bimonoids, e.g., the set of commutative semirings. Again, in order to avoid repetitions of universal quantifications over this particular subset, we will place a corresponding convention local to these chapters and sections; the scope of each local convention is explicitly indicated. Then,  for each  result (e.g. observation, lemma, theorem, corollary)  within this scope, the general and local conventions and restrictions hold. We hope that this extraction of universal quantifications helps to focus on important argumentation.

... about theorems, lemmas, and corollaries which are in a box:\\
If a reader just wants to look up some result, then checking for local and global conventions and collecting them is a burden and it is better to see *all* the restrictions, requirements, and universal quantifications inside the corresponding latex-environment. To solve the contradicting wishes: avoiding  repetitions of universal quantifications versus fully quantified statements, for some of the main results (lemmas, theorems, and corollaries) we show all the necessary restrictions and requirements inside the latex-environment. In order to ease the search for such main results, we put them into a box.

... about normal forms:\\
In the book we define several instances of terminating reductions systems  and
determine the set of their normal forms. We use such reduction systems for definitions  and
proofs by well-founded induction. For examples of such definitions, see e.g. \eqref{equ:weight-of-run} and the definition of $f$, $g$, and $h$ on page \pageref{page:f-g-h-defined-by-induction}, and for examples of such proofs, see e.g. the proof of Observation \ref{obs:weight-run-explicit} and Lemma \ref{lm:image-monadic-ranked alphabet}. In these cases the normal forms of the reduction system
are used to define or prove the base of the induction.

Besides, we prove several normal form lemmas for weighted tree automata, weighted context-free grammars, and
weighted regular tree grammars (cf., e.g., Theorem \ref{thm:t1}, Lemma \ref{lm:wcfg-in-nonterminal-form}, Theorem \ref{lm:wrtg-chain-free}, and Lemma \ref{lm:rtg-normal-form}); here we use  the phrase ``normal form lemma'' as it is known from the theory of classical (unweighted) automata and formal languages. For instance, a normal form lemma for weighted regular tree grammars states that each weighted regular tree grammar can be transformed into an equivalent one which has a particular property.

Now one may wonder, what the concept of ``normal form of a reduction system'' has in common with the concept of ``normal form lemma'', apart from merely sharing the English words ``normal form''.

We think that, for each normal form lemma, we can define a specific terminating reduction system such that the  grammar or automata which results from the normal form lemma is a normal form of that reduction system. However, several normal forms  of weighted tree automata, weighted context-free grammars, and weighted regular tree grammars are obtained as a result of one complex construction and not as a result of the iteration of some elementary reduction steps; for example, see the construction of a local-trim weighted tree automaton (Theorem \ref{thm:t2}) or of a chain-free weighted context-free grammar (Theorem  \ref{lm:wrtg-chain-free}). In such cases, the construction of the desired normal form can be described in terms of a terminating reduction system in which the reduction relation is the construction itself, hence the reduction terminates in one step. Therefore, in such cases we will not show the specific terminating reduction system explicitly. 

However, for one particular normal form lemma (viz., Lemma \ref{lm:rtg-normal-form}: for each weighted regular tree grammar, we can construct an equivalent one which is alphabetic), the use of terminating reduction systems seems to be very well motivated.

... about possible mistakes:\\
We would like to ask (or: encourage) the readers to share with us their remarks and observations concerning the contents of the book. In particular, the indication of mistakes or of missing references to published results are very welcome. Please, write an email to \verb!fulop@inf.u-szeged.hu! and \verb!heiko.vogler@tu-dresden.de!.

... about an electronic version:\\
This book can be found on arXiv by searching  for the authors or the title.

\paragraph{Acknowledgements.}

In 2017 we started to work on this book. But the seed for this work was laid much earlier. In our academic youth, we both had excellent teachers of the theory of tree automata (Ferenc G\'ecseg and Joost Engelfriet, respectively). After this academic qualification, we met for the first time in the year 1987, when the second author (Heiko) visited the first author (Zolt\'an) and his colleague S\'andor V\'agvölgyi in Szeged. And the result of this start was the publication \cite{fulhervagvog93}. Since then we (the authors) jointly investigated the theory of tree automata and tree transducers. In the beginning, we communicated by hand-written letters as it was usual at that time, or we paid regular visits to each other. When in 1999 Werner Kuich visited Ferenc G\'ecseg, the second author happened to be also in Szeged. Then Werner showed him his fresh paper ``Tree transducers and tree series'' \cite{kui99} and explained its ideas.
From that point on, we (Zoltan and Heiko) worked on the theory of weighted tree automata and weighted tree transducers  (starting with \cite{engfulvog02new}). During the development of that theory,  we benefitted much from the cooperation with Symeon Bozapalidis, Frank Drewes, Manfred Droste, Joost Engelfriet, Andreas Maletti, Mark-Jan Nederhof, George Rahonis, and Magnus Steinby. In the years 2002-2021, the series of  workshops ``Weighted Automata: Theory and Applications'' (WATA), co-organized by Manfred Droste,  gave an excellent infrastructure for presenting our results. This book contains many ideas which were produced by these cooperations during the last 20 years; we are very grateful for this. 

We would like to thank our colleagues for checking some parts of the first edition of the book
and contributing valuable suggestions: Johanna Bj\"orklund, Frank Drewes, Manfred Droste, Zsolt Gazdag, Luisa Herrmann, Eija Jurvanen, Andreas Maletti, George Rahonis, and S\'andor V\'agvölgyi. In particular, we wish to thank Manfred Droste for several fruitful discussions and suggestions, which had a very positive impact on the book. Of course, each remaining mistake is due to the authors.  We are grateful to Felicita Purnama Dewi Gernhardt, Luisa Herrmann, D\'avid K\'osz\'o, and, in particular, Celina Pohl for preparing the figures. In a number of situations, Celina Pohl and Richard M\"orbitz helped us to find the appropriate latex-commands and settings.

\begin{quote}\emph{``... who prologue-like your humble patience pray, gently to hear, kindly to judge, our play.''}\\ W. Shakespeare, Henry V.
  \end{quote}

\

\noindent Zolt\'an F\"ul\"op and Heiko Vogler \hfill \today\\
Szeged, Hungary and Dresden, Germany

\newpage

\begin{figure}[h]
  \hspace*{-15mm}
  \scalebox{0.9}{
\begin{tikzpicture}
\tikzset{transform shape}

\node at (0,1.6) {models of automata};
\node at (12.5,1.6) (ma) {models of grammars};

\node[rectangle,draw,align=left] at (0,0)  (1)
	{$(\Sigma,\B)$-wta \\ (Sec. \ref{sec:basic-defininition-wta})};
\node[rectangle,draw,align=left, below left =1.5cm and 0cm of 1] (2)
	{$(\Gamma,\B)$-wsa \\ (Sec. \ref{sect:string-automata})};
\node[rectangle,draw,align=left, below right =1.5cm and 0cm of 1] (3)
	{$\Sigma$-fta\vphantom{(}\\(Sec. \ref{sec:fta})};
\node[rectangle,draw,align=left, below left =1.5cm and 0cm of 3] (4)
	{$\Gamma$-fsa\vphantom{(} \\ (Sec. \ref{sec:fsa})};

\node[align=left] at (4.5,-5.5) (5)
	{$\strut (\Sigma,\B)$-wta \\ \strut with root \\ \strut root weights};
\node[align=left, baseline=(5.base), right= 2.3cm of 5] (6)
	{$\strut (\Sigma,\B)$-wrtg \\ \strut in tree automata \\ \strut form};

\node[rectangle,draw,align=left] at (12.5,0) (7)
	{$(\Gamma,\B)$-wcfg \\(Sec. \ref{subsect:basic-modell})};
\node[rectangle,draw,align=left, below=2cm of 7] (8)
	{$(\Sigma,\B)$-wrtg \\ (Sec. \ref{sec:grammar-model})};
\node[rectangle,draw,align=left, below right=2cm and 1.5cm of 7] (9)
	{$\Gamma$-cfg \vphantom{(}\\ (Sec. \ref{sec:context-free-grammars})};
\node[rectangle,draw,align=left, below= of 8]  (10) 
	{$(\Sigma,\Psi,\B)$-wpb \\ (Subs. \ref{sect:weighted-proj-bim})};

\draw[shorten <=-1.25cm, shorten >=-3.05cm] ($(1)!0.42!(2)$) edge[loosely dashed] 
	node[below left, pos=2.1] {strings} 
	node[above right, pos=2.1] {trees}
	($(3)!0.58!(4)$);
\draw[shorten <=-2cm, shorten >=-2.5cm] ($(7)!0.5!(8)$) edge[loosely dashed]
	node[below left, pos=2.3] {trees} 
	node[above right, pos=2.3] {strings}
	($(8)!0.5!(9)$);

\draw (1) edge[->,>=stealth] node[fill=white] (m12) {Lm. \ref{lm:wsa=wta-over-string-ra}} (2);
\draw (1) edge[->,>=stealth] node[fill=white] {Cor. \ref{cor:supp-B=fta-1}} (3);
\draw (2) edge[->,>=stealth] node[fill=white] {Obs. \ref{obs:fsa=wsa(B)}} (4);
\draw (3) edge[->,>=stealth] node[align=left, xshift=1.2cm] (m34) 
	{$\Sigma = \Gamma_e$ \\ string r.a.} (4);
	
\draw (7) edge[->,>=stealth] node[align=left, fill=white, left=0.1cm, inner sep=0.1cm]
	{-- $\Gamma = \Sigma^{\Xi}$ \\ -- rhs of \\ \hphantom{--} rules \\ \hphantom{--} are trees} (8);
\draw (7) edge[->,>=stealth] node[fill=white] {Obs. \ref{obs:cfg=wcfg(B)}} (9);
\draw (8) edge[->,>=stealth] (10);

\draw (1) edge[->,>=stealth,out=0,in=90,looseness=1.3]
	node[fill=white, pos=1, yshift=2cm] {Thm. \ref{thm:root-weight-normalization-run}} (5);
\draw (8) edge[->,>=stealth, out=180,in=90,looseness=1]
	node[fill=white] {Lm. \ref{lm:rtg-automata-form-wrtg}} (6.110);
\draw (6) edge[stealth-stealth, shorten <=0.05cm, shorten >=0.05cm]
	node[above=0.1cm] {Lm. \ref{lm:related-semantics}} (5);

\end{tikzpicture}
}

\caption{\label{fig:overview-models} Overview of the models of automata and models of grammars which occur in this book. Their relationship is expressed by arrows. For every pair $(X,Y)$ of models, if $X \rightarrow Y$, then $Y$ is a special case of $X$, and if $X \leftrightarrow Y$, then $X$ and $Y$ are equivalent.\\[2mm]
  wta: weighted tree automaton  \hspace{30mm} wcfg: weighted context-free grammar\\
  wsa: weighted string automaton \hspace{27mm} wrtg: weighted regular tree grammar\\
  fta: \hspace{1.5mm}finite-state tree automaton \hspace{27.1mm} cfg: \hspace{0.69mm} context-free grammar\\
  fsa: \hspace{1.5mm}finite-state string automaton \hspace{23.9mm} wpb:  \hspace{-1.3mm} weighted projective bimorphism\\[3mm]
  $\Sigma$, $\Psi$: ranked alphabets\\
  $\Gamma$:  \hspace{3.3mm} alphabet\\
  $\B$:  \hspace{3mm} strong bimonoid\\
  rhs: \hspace{0.9mm} right-hand side
    }
\end{figure}

%% file: preliminaries.tex
\chapter{Preliminaries}
\label{ch:preliminaries}

\section{Numbers and sets}

\index{N@$\mathbb{N}$}
\index{Np@$\mathbb{N}_+$}
\index{$[n,k]$}
\index{$[n]$}
\index{Z@$\mathbb{Z}$}
\index{Q@$\mathbb{Q}$}
\index{R@$\mathbb{R}$}
We denote the set $\{0,1,2,\ldots\}$ of natural numbers by $\mathbb{N}$ and the set $\mathbb{N}\setminus \{0\}$ by $\mathbb{N}_+$. Let $n,k \in \mathbb{N}$.
Then we denote the set $\{i \in \mathbb{N} \mid n \le i \le k\}$ by $[n,k]$. We abbreviate $[1,k]$ by $[k]$. Thus $[0] = \emptyset$.
We denote the sets of integers, rational numbers, and real numbers by $\mathbb{Z}$, $\mathbb{Q}$, and $\mathbb{R}$, respectively.   

For each $a,b\in \mathbb{R}$, we denote by $\max(a,b)$ and $\min(a,b)$ the maximum and the minimum of  $a$ and $b$ with respect to $\le$, respectively. 
We extend $\max$ and $\min$ to each nonempty, finite subset of $\mathbb{R}$ in a natural way. Later, when $\max$ (or $\min$) is an operation of a monoid,  we will define $\max \emptyset$ (and $\min \emptyset$, respectively) to be the unit element of that monoid (cf. \eqref{eq:extension-odot-finite-set}).

\index{Ninf@$\mathbb{N}_\infty$}
Sometimes we use the set $\mathbb{N} \cup \{\infty\}$. We abbreviate it by $\mathbb{N}_\infty$ and extend the operations $+$ and $\min$ to  $\mathbb{N}_\infty$ in the obvious way: $a + \infty = \infty$ and $\min(a,\infty) = a$ for each $a\in \mathbb{N}_\infty$. In a similar way we proceed with an extension by $-\infty$ and the operation $\max$ (and also with other sets like $\mathbb{Z}$ and~$\mathbb{R}$).

For the set $\mathbb{R}$ of real numbers, we denote its subset $\{r \in \mathbb{R} \mid r \ge 0\}$ by $\mathbb{R}_{\ge 0}$. For the corresponding subsets of integers and rational numbers we use $\mathbb{Z}_{\ge 0}$ and $\mathbb{Q}_{\ge 0}$, respectively.  We denote the set $\mathbb{Q}_{\ge 0} \cup \{\infty\}$ by $\mathbb{Q}_{\ge 0,\infty}$.

Let $A$ be a set. If $A$ is finite, then $|A|$ denotes the number of elements of $A$. We note that if we are given a list $a_1,\ldots,a_n$ of some objects which are not necessarily pairwise different and for some reason we form the set $A=\{a_1,\ldots,a_n\}$, then in general $|A|\le n$. If $A$ is a singleton, i.e., it contains the only element $a$, then sometimes we identify $A$ with $a$, i.e., simply write $a$ for $\{a\}$. The \emph{power-set of $A$}, denoted by $\mathcal{P}(A)$, is the set of all subsets of $A$. We denote by $\cP_{\mathrm{fin}}(A)$ the set of all finite subsets of $A$.

Let $n\in \mathbb{N}$, and $A_1,\ldots, A_n$ be sets. The \emph{Cartesian product of $A_1,\ldots,A_n$} is the set $\{(a_1,\ldots,a_n) \mid a_i \in A_i \text{ for each $i \in [n]$}\}$ and it is denoted by $A_1\times \ldots\times A_n$. The \emph{$n$-fold Cartesian product of a set $A$} is the set $A\times \ldots\times A$, where $A$ appears $n$ times. We abbreviate $A\times \ldots\times A$ by $A^n$. In particular, $A^0 = \{()\}$.

\index{subset}
\index{strict subset}
\index{subset@$\subseteq$}
\index{subset@$\subset$}
Let $A$ and $B$ be sets. If each element of $A$ is also an element of $B$, then $A$ is a \emph{subset} of $B$, denoted by $A \subseteq B$. If $A \subseteq B$ and $A \not= B$, then $A$ is a \emph{strict subset} of $B$, denoted by $A \subset B$.

\section{Strings and languages}
\label{sec:strings}

We recall some notions and notations of strings and (formal) languages from \cite{har78,hopull79,hopmotull14,lin12}.

Let $A$ be a set. A \emph{string} (or: \emph{word}) \emph{over $A$} is a finite sequence $a_1\cdots a_n$ with $n \in \mathbb{N}$ and $a_i \in A$ for each $i \in [n]$. In particular, we denote the sequence $a_1\cdots a_n$ with $n=0$ by $\varepsilon$ and call it the \emph{empty string}. We say that  $a_1\cdots a_n$ has  \emph{length} $n$.  For each $n \in \mathbb{N}$, we denote the \emph{set of strings over $A$ of length $n$} by $A^n$; thus $A^0 = \{\varepsilon\}$. Moreover, we denote by $A^*$ the \emph{set of all strings over $A$}, i.e., $A^* = \bigcup_{n \in \mathbb{N}} A^n$ and we let $A^+=A^*\setminus\{\varepsilon\}$. For each $w \in A^*$, we denote by $|w|$ the length of $w$.

We note that the notation $A^n$ is overloaded in the sense that it denotes two sets: (a) the $n$-fold Cartesian product of $A$ and (b) the set of strings over $A$ of length $n$. Of course,  formally these sets are different. But since there exists a bijection between them, we find it acceptable to use the same notation.

The \emph{concatenation} of the strings $a_1\cdots a_n$ and $b_1\cdots b_m$ is the string $a_1\cdots a_nb_1\cdots b_m$. For every $v,w \in A^*$, we denote the \emph{concatenation of $v$ and $w$} by $v \cdot w$ or simply by $vw$. 

Let $w \in A^*$. For each $n\in \mathbb{N}$, we define $w^n$ such that  $w^0=\varepsilon$ and  $w^n=ww^{n-1}$ for $n\in \mathbb{N}_+$. 
Moreover, we denote by $\mathrm{postfix}(w)$ the set $\{v \in A^* \mid (\exists u\in A^*): w=uv\}$
of \emph{postfixes of $w$}, and  by $\mathrm{prefix}(w)$ the set $\{u\in A^* \mid (\exists v\in A^*): w=uv\}$ of \emph{prefixes of $w$}. Note that $\{\varepsilon, w\} \subseteq \mathrm{postfix}(w) \cap \mathrm{prefix}(w)$.

\index{alphabet}
\index{formal language}
\index{language}
\index{concatenation}
An \emph{alphabet} is a finite and nonempty set. Let $\Gamma$ be an alphabet. Then each subset $L \subseteq \Gamma^*$ is called a \emph{language} (or: \emph{formal language}) \emph{over $\Gamma$}.
 Let $L_1,L_2 \subseteq \Gamma^*$ be two languages. The \emph{concatenation of $L_1$ and $L_2$}, denoted by $L_1 \cdot L_2$ or just $L_1 L_2$, is the language
\(L_1 L_2 = \{w_1w_2 \mid w_1 \in L_1, w_2 \in L_2\}\).

\index{Gamma@$\Gamma$}
\label{p:Gamma-alphabet}
\begin{quote}
  \emph{In the rest of this book, $\Gamma$ will denote an arbitrary alphabet, if not specified otherwise.}
  \end{quote}


\section{Binary relations and mappings}\label{sect:binary-relations}

\index{binary relation}
\index{inverse}
\paragraph{Binary relations.}
Let $A$ and $B$ be sets. A \emph{binary relation (on $A$ and $B$)} is a subset $R \subseteq A \times B$.
As usual, we frequently write $a R b$ for $(a,b)\in R$. For each $A'\subseteq A$, we define $R(A')=\{b\in B\mid (\exists a\in A'): a R b\}$.
If $A=B$, then we call
$R$ a \emph{binary relation on $A$} (or: over $A$). We denote by $R^{-1}$ the binary relation $\{(b,a) \mid aRb\}$, and we call it the \emph{inverse of $R$}. 

\index{composition}
\index{$R_1;R_2$}
Let additionally $C$ be a set and $R_1 \subseteq A \times B$ and $R_2 \subseteq B \times C$ be binary relations. The \emph{composition of $R_1$ and $R_2$}, denoted by $R_1; R_2$, is the binary relation  on $A$ and $C$ defined by 
$R_1;R_2 = \{(a,c) \in A \times C \mid (\exists b \in B): (a,b) \in R_1 \text{ and }  (b,c) \in R_2\}$.

\index{reflexive}
\index{irreflexive}
\index{symmetric}
\index{antisymmetric}
\index{transitive}
\index{strict}
Let $R$ be a binary relation on $A$.
 The relation $R$ is
\begin{compactitem}
\item \emph{reflexive} if $a R a$ for every $a \in A$,
\item \emph{irreflexive} if $(a,a) \not\in R$ for every $a \in A$,
\item \emph{symmetric} if $a R b$ implies that $b R a$ for every $a,b \in A$,
\item \emph{antisymmetric} if $a R b$ and $b R a$ imply that $a=b$ for every $a,b \in A$,
\item \emph{transitive} if $a R b$ and $b R c$ imply that $a R c$ for every $a,b,c \in A$, and
  \item \emph{strict order} if $R$ is irreflexive and transitive.
\end{compactitem}
 The \emph{reflexive, transitive closure of $R$}, denoted by $R^*$, is the binary relation $R^* = \bigcup_{n \in \mathbb{N}} R^n$ where $R^0 = \{(a,a) \mid a \in A\}$ and $R^{n+1} = R^n;R$ for each $n \in \mathbb{N}$. The \emph{transitive closure of $R$}, denoted by $R^+$, is the binary relation $R^+ = \bigcup_{n \in \mathbb{N}_+} R^n$. Then $R^*$ is reflexive and transitive, and $R^+$ is transitive.

\index{equivalence relation}
\index{factor set}
\index{canonical mapping}
If $R$ is reflexive, symmetric, and transitive, then we call it an \emph{equivalence relation (on $A$)}.
Then, for every $a\in A$, the \emph{equivalence class with representative $a$}, denoted by $[a]_R$, is the set $\{b\in A \mid a R b\}$. The \emph{factor set of $A$ modulo $R$}, denoted by $A/\!_R$, is the set $\{ [a]_R \mid a\in A\}$. The \emph{index of $R$} is the cardinality of~$A/\!_R$. The \emph{canonical mapping for $R$}, denoted by $\pi_R$, is the mapping $\pi_R: A \to A/\!_R$ defined, for each $a \in A$, by $\pi_R(a) = [a]_R$.

\index{partial order}
Let $\leq$ be a binary relation on $A$. If $\leq$ is reflexive, antisymmetric, and transitive, then we call it a \emph{partial order (on $A$)}, and we call the pair $(A,\leq)$ a \emph{partially ordered set}.  A partial order $\leq$ on $A$ is a \emph{linear order (on $A$)} if for every $a,b \in A$ we have $a \leq b$ or $b \leq a$. For $a,b \in A$, we write $a < b$ to denote that $a \leq b$ and $a \neq b$.

\index{upper bound}
\index{lower bound}
\index{infimum}
\index{supremum}
Let  $(A,\leq)$ be a partially ordered set and let $B\subseteq A$. An element $a\in A$ is an \emph{upper bound of $B$} if, for each $b\in B$, we have $b \leq a$. Moreover, if $a$ is an upper bound and, for each upper bound $a'$ of $B$, we have $a \leq a'$, then  $a$ is the \emph{least upper bound of $B$}.
If the least upper bound of $B$ exists, then we call it the \emph{supremum of $B$} and denote it by $\sup_\leq B$ (or: $\sup B$). Similarly, we can define \emph{lower bound} and the \emph{greatest lower bound}, i.e.,  the \emph{infimum of $B$}. We denote the latter by $\inf_\leq B$ (or: $\inf B$).

\index{mapping}
\index{domain}
\paragraph{Mappings.} Let $f \subseteq A \times B$ be a binary relation.  We say that $f$ is a \emph{mapping from $A$ to $B$}, denoted by $f: A \rightarrow B$, if for each $a \in A$ there exists a unique $b \in B$ such that $(a,b) \in  f$. In this case we write $f(a)=b$ as usual. The set of all mappings from $A$ to $B$ is denoted by $B^A$. We call $A$ and $B$ the \emph{domain} and \emph{codomain} of $f$, respectively.\index{codomain}
 The \emph{image} of $f$ is the set  $\im(f)=\{f(a)\mid a\in A\}$.\index{image}
A mapping $f: A \to B$ is \emph{injective} if for every $a_1,a_2 \in A$ with $a_1 \ne a_2$ we have that $f(a_1)\ne f(a_2)$.\index{injective}
 It is \emph{surjective} if for each $b \in B$ there exists an $a \in A$ such that $f(a)=b$.\index{surjective}
It is \emph{bijective} if it is injective and surjective. \index{bijective}

In particular, each binary relation $R \subseteq A \times B$ can be considered as a mapping $R: A \to \mathcal{P}(B)$, where $R(a)=\{b \in B \mid aRb\}$ for each $a\in A$.

\index{restriction}
Let $f: A \rightarrow B$ be a mapping and $A'$ and $B'$ be sets such that  $A'\subseteq A$ and $B' \supseteq B$. Then $f$ is also a mapping from $A$ to $B'$, i.e., $f: A \rightarrow B'$. The \emph{restriction of $f$ to $A'$} is the mapping $f|_{A'}:A'\to B$ and defined by $f|_{A'}(a)=f(a)$ for each $a\in A'$.

\index{union of mappings}
Let $A$, $B$, and $C$ be sets such that $A$ and $B$ are disjoint. Moreover, let $f: A \to C$ and $g:B \to C$. We define the \emph{union of $f$ and $g$} as the mapping $(f \cup g): A \cup B \to C$ such that, for each $x \in (A\cup B)$ we let $(f\cup g)(x) = f(x)$ if $x \in A$, and $(f \cup g)(x) = g(x)$ otherwise.
Clearly, $(f \cup g)|_A = f$ and $(f\cup g)|_B = g$.

\index{composition}
\index{$g \circ f$}
Let $C$ be a further set. For two mappings $f: A \rightarrow B$ and $g: B \rightarrow C$, the \emph{composition of $f$ and $g$} is the mapping $(g \circ f) : A \to C$, where $(g \circ f)(a) = g(f(a))$ for every $a \in A$.  
If we view $f$ and $g$ as binary relations, then $g \circ f = f;g$. 

In the usual way we extend a mapping $f: A \to B$ to the mapping $f': \cP(A) \to \cP(B)$ by defining $f'(A') = \{b \in B \mid (\exists a \in A'): f(a)=b\}$ for each $A' \subseteq A$. We denote $f'$ also by $f$.

\index{family}
If we want to emphasize the values of a mapping, then we use the concept of family defined as follows. Let $I$ and $B$ be sets. An \emph{$I$-indexed family over $B$} is a mapping $f: I \rightarrow B$. Such a family is also denoted by $(b_i \mid i \in I)$ where $b_i=f(i)$ for each $i \in I$.  An $I$-indexed family over $B$ is \emph{countable} (\emph{finite}, \emph{nonempty}) if $I$ is countable (finite, nonempty, respectively). We call $I$ the index set of that family.

\index{partitioning}
Let $f= (B_i \mid i \in I)$ be an $I$-indexed family  over $\mathcal{P}(B)$. We call $f$ a {\em family-partitioning of $B$} (with respect to~$I$) \cite[p.197]{hebwei93} if $\bigcup_{i\in I}B_i =B$, and $B_i \cap B_j= \emptyset$ for every $i,j\in I$ with $i\ne j$.
Moreover, a subset $P$ of $\mathcal P(B)$ is a \emph{set-partitioning} of $B$ \cite[p.1]{gra68},\cite[p.29]{wec92},\cite[Def.4.10]{bursan81} if any two different elements of $P$ are disjoint and $B=\bigcup_{H\in P}H$.
The two kinds of partitioning are equivalent in the following sense. If $f$ is a family-partitioning of $B$, then $\{B_i \mid i \in I\}$ is a set-partitioning of $B$ and vice versa, if $P$ is a \emph{set-partitioning} of $B$, then $(H\mid H\in P)$ is a family-partitioning of $B$. In the following we will use both kinds of partitioning under the name \emph{partitioning}.

For each $k \in \mathbb{N}$, a mapping $f: A^k \to A$ is called \emph{$k$-ary operation on $A$} or simply \emph{operation on $A$}. As usual, we identify a nullary operation $f$ on $A$  with the element $f()\in A$. We denote the set of all $k$-ary operations on $A$  by $\mathrm{Ops}^k(A)$, and the set of all operations on $A$ by $\mathrm{Ops}(A)$.

\index{identity mapping}
We define the \emph{identity mapping on $A$} as the mapping $\mathrm{id}_A: A \to A$ such that $\mathrm{id}_A(a) = a$ for each $a \in A$.

Let $A$ and $B$ be sets and $f: A \to \mathrm{Ops}(B)$. Then, for every $a\in A$ and $b_1,\ldots,b_k \in B$ where $k$ is the arity of $f(a)$, we write $f(a)(b_1,\ldots,b_k)$ rather than  $(f(a))(b_1,\ldots,b_k)$.


\section{Hasse diagram, Venn diagram, and Euler diagram}

Let $(A\le)$ be a partially ordered set. There are several ways to represent $(A,\le)$ by means of a diagram in the plane.

\index{Hasse diagram}
\paragraph{Hasse diagram.} Each element $a \in A$ is represented in the diagram by a node which is labeled by $a$. This node is called  $a$-node. The nodes are placed into the diagram in such a way that whenever $a < b$ holds for $a,b \in A$, the $b$-node is drawn above the $a$-node. Moreover, for every $a,b \in A$, the $a$-node and the $b$-node are connected by an edge
if $a < b$ and, for every $c \in A$, the condition $a \le c \le b$ implies that $a = c$ or $b = c$.
From this diagram we can reconstruct $\le$ by noting that $a \le b$ holds if and only if $a = b$ or  the $b$-node  can be reached from the $a$-node via a sequence of
ascending edges.

Figure \ref{fig:Hasse-example} shows an example of a Hasse diagram for $A = (\mathbb{N} \times \mathbb{N})$ and the partial order $\le$ defined by  $(n_1,n_2) \le (k_1,k_2)$ if $n_1 \le k_1$ and $n_2 \le k_2$. Of course, since $A$ is an infinite set, we can only show a finite part of the Hasse diagram.

\begin{figure}
  \centering
  \scalebox{1}{\begin{tikzpicture}
      \matrix (t) [matrix of nodes, column sep=1.0cm, row sep=0.5cm, ampersand replacement=\&]
        {
          \&         \&  $(2,2)$ \& \& \\
          \& $(2,1)$ \&           \& $(1,2)$ \&\\
         $(2,0)$ \&  \&   $(1,1)$ \&  \& $(0,2)$ \\  
	  \& $(1,0)$ \&           \&   $(0,1)$ \&  \\
	         \&         \& $(0,0)$ \&         \&  \\
    }; 
    \draw (t-1-3)--(t-2-2);
    \draw (t-2-2)--(t-3-1);
    \draw (t-1-3)--(t-2-4);
    \draw (t-2-4)--(t-3-5);
    \draw (t-3-1)--(t-4-2);
    \draw (t-2-2)--(t-3-3);
    \draw (t-2-4)--(t-3-3);
    \draw (t-3-5)--(t-4-4);
    \draw (t-3-3)--(t-4-2);
    \draw (t-3-3)--(t-4-4);
    \draw (t-4-2)--(t-5-3);
    \draw (t-4-4)--(t-5-3);
    \draw[dotted] (t-1-3)--+(-1.8,1);
    \draw[dotted] (t-1-3)--+(1.8,1);
    \draw[dotted] (t-2-2)--+(-1.8,1);
    \draw[dotted] (t-2-4)--+(1.8,1);
    \draw[dotted] (t-3-1)--+(-1.8,1);
    \draw[dotted] (t-3-5)--+(1.8,1);
\end{tikzpicture}}

\caption{\label{fig:Hasse-example} A finite part of the Hasse diagram for $A = (\mathbb{N} \times \mathbb{N})$ and $(n_1,n_2) \le (k_1,k_2)$ if $n_1 \le k_1$ and $n_2 \le k_2$.}
  \end{figure}

  \index{Venn diagram}
  \index{Euler diagram}
  \paragraph{Venn diagram and Euler diagram.} These diagrams can be drawn if there exists a finite set $U$ such that $A$ is a collection of subsets of $U$, i.e., $A=\{B_1,\ldots,B_n\}$ for some $n \in \mathbb{N}_+$ and some subsets $B_1,\ldots,B_n$ of $U$; moreover, the partial order $\le$ is the subset relation $\subseteq$. Then each element $a \in U$ is represented by an individual point in the diagram; this point is labeled by $a$.
  Each subset $B_i$ is represented in the diagram by an area which is delimited by an oval or some polygon; this area contains exactly the elements of $B_i$. We call this area the $B_i$-area.

In the Venn diagram, the $B_i$-areas are placed into the diagram in an overlapping manner such that exactly $2^n-1$ overlap areas are created. It is possible that an overlap area does not contain an element of~$\bigcup_{i \in [n]} B_i$.

In the Euler diagram, the $B_i$-areas are placed into the diagram also in an overlapping manner, but now in such a way that each overlap area contains at least one element of $\bigcup_{i \in [n]} B_i$.

Figure \ref{fig:Venn-Euler-example} shows an example of a Venn diagram and an Euler diagram for $U = \{a,b,c\}$ and $A = \{\{a\}, \{a,b\}, \{b,c\}\}$. 

\begin{figure}
  \centering
  \begin{tikzpicture}[scale=1, every node/.style={transform shape}]

\begin{scope}
  \coordinate (ol) at (150:0.8cm); 
  \coordinate (or) at (30:0.8cm);	
  \coordinate (ob) at (270:0.8cm); 
  \draw (ol) circle (1cm);
  \draw (or) circle (1cm);
  \draw (ob) circle (1cm) node[below] {$c$};
  \node at (90:0.55cm) {$a$};
  \node at (330:0.55cm) {$b$};
  \node[above left = 1cm and -0.15cm of ol] {$\{a\}$};
  \node[above right = 0.45cm and 0.85cm of or] {$\{a,b\}$};
  \node[below right = 0.25cm and 0.95cm of ob] {$\{b,c\}$};
\end{scope}
  
\begin{scope}[shift={(6.5cm,0.7cm)}]
  \coordinate (st) at (0.35,0.1); 
  \coordinate (sb) at ([shift=(260:2cm)] st); 
  \draw (st) edge[out=260+90, in=260+90, looseness=0.8] (sb);
  \draw (sb) edge[out=260+270, in=260+270, looseness=0.8] (st);
  \node[circle, minimum size=0.8cm, draw] (a) at (-0.95,-0.45) {$a$};
  \node (b) at ($(st)!0.2!(sb)$) {$b$};
  \node (c) at ($(st)!0.8!(sb)$) {$c$};
  \node[above = 0cm of a] {$\{a\}$};
  \draw (-0.3,-0.15) ellipse (1.8cm and 0.9cm);
  \node at (2.05,0.15) {$\{a,b\}$};
  \node[below = 0.3cm of c] {$\{b,c\}$};
\end{scope}  
\end{tikzpicture}

\caption{\label{fig:Venn-Euler-example} The Venn diagram (left) and the Euler diagram (right) for $U = \{a,b,c\}$ and $A = \{\{a\}, \{a,b\}, \{b,c\}\}$.}
  \end{figure}

  
  \section{Reduction systems}
  \label{sec:reduction-systems}

We recall the concept of reduction system and some of its main properties \cite{hue80,baanip98}.

\index{reduction system}
\index{reduction}
A \emph{reduction system} is a pair $(A,\rightarrow)$ where $A$ is a set and $\rightarrow$ is a binary relation on $A$, called \emph{reduction}.

\index{normal form}
Let $(A,\rightarrow)$ be a reduction system. For every $a,b \in A$, if $a \rightarrow b$, then  \emph{$b$ is a direct successor of $a$}.  An element $a \in A$ is called \emph{in normal form} if $\rightarrow\!(a) = \emptyset$, i.e., there does not exist a $b \in A$ such that $a \rightarrow b$. Let $a,b \in A$. Then $b$ is a \emph{normal form of $a$} if $a \rightarrow^* b$ and $b$ is in normal form.  We denote the set of all normal forms of $a$ by $\nf_\rightarrow(a)$. We say that \emph{$a$ has a normal form} if $\nf_\rightarrow(a) \not= \emptyset$.
If $|\nf_\rightarrow(a)|=1$, then we say that \emph{$a$ has a unique normal form}.

      \index{terminating}
A reduction system  $(A,\rightarrow)$ is called \emph{terminating} if there does not exist an $\mathbb{N}$-indexed family $(a_i \mid i\in \mathbb{N})$ over $A$ such that, for each $i \in \mathbb{N}$, the condition $a_i \rightarrow a_{i+1}$ holds. If a reduction system $(A,\rightarrow)$ is terminating, then we also say that $\rightarrow$ is terminating. 
       
It is easy to see that a terminating binary relation is irreflexive, and that the transitive closure $\rightarrow^+$ of a terminating relation $\rightarrow$ on $A$ is also terminating. 
\index{arrowN@$\rightarrow_\mathbb{N}$}
For instance, the relation 
\[\succ_\mathbb{N} = \{(n+1,n) \mid n \in \mathbb{N}\}
\]
on $\mathbb{N}$ is terminating. Since $> = (\succ_\mathbb{N})^+$ (where $>$ is the usual ``greater than'' relation on $\mathbb{N}$), the relation $>$ is also terminating.

  \begin{observation}\rm \label{obs:terminating-implies-normalizing} \cite[p.~9-10]{baanip98} If the reduction system $(A,\rightarrow)$ is terminating, then each $a \in A$ has a normal form.
  \end{observation}
    
Often we will use a particular basic method in order to show that a reduction system is terminating. This method is based on the concept of monotone measure function. Formally, let $(A,\rightarrow)$  and   $(B,\succ)$ be  reduction systems, and $\varphi: A \to B$ be a mapping. The mapping $\varphi$ is \emph{monotone} if, for every $a_1,a_2 \in A$, the reduction $a_1 \rightarrow a_2$ implies $\varphi(a_1) \succ \varphi(a_2)$.
  \index{monotone measure function}
  \index{monotone embedding}
  If this is the case, then $\varphi$ is called a \emph{monotone measure function}, or: \emph{monotone embedding of $(A,\rightarrow)$ into $(B,\succ)$}.

  \begin{lemma}\rm {\rm\cite[p.~16]{baanip98}}\label{lm:fin-branching-embedding-termination} Let $(A,\rightarrow)$ be a reduction system and $(B,\succ)$ be a terminating reduction system. If there exists a monotone embedding of $(A,\rightarrow)$ into $(B,\succ)$, then $(A,\rightarrow)$ is terminating.
  \end{lemma}
  \begin{proof} Let $\varphi: A \to B$ be a monotone embedding. We prove by contradiction. For this, we assume that $(A,\rightarrow)$ is not terminating. Then there exists an $\mathbb{N}$-indexed family $(a_i \mid i \in \mathbb{N})$ over $A$ such that, for each $i \in \mathbb{N}$, the condition $a_i \rightarrow a_{i+1}$ holds.
    Since $\varphi$ is monotone, for the $\mathbb{N}$-indexed family $(\varphi(a_i) \mid i \in \mathbb{N})$ over $B$ and for each $i \in \mathbb{N}$, the condition $\varphi(a_i) \succ \varphi(a_{i+1})$ holds. This contradicts to the fact that $(B,\succ)$ is terminating.
    \end{proof}

    If in Lemma \ref{lm:fin-branching-embedding-termination} we have that $(A,\rightarrow)$ is finitely branching and $(B,\succ) = (\mathbb{N},>)$, then the implication in the other direction also holds \cite[Lm.~2.3.3]{baanip98}.

    The following lemma shows that the property of being terminating is inherited to subsets.

    \begin{lemma}\rm \label{lm:termination-is-subset-closed} Let $(B,\succ)$ be a terminating reduction system, $A\subseteq B$, and  $\rightarrow {\subseteq} \succ \cap (A\times A)$. Then also $(A,\rightarrow)$ is terminating.
    \end{lemma}
    \begin{proof} The mapping $(\mathrm{id}_B)|_A: A \to B$ is a monotone embedding of $(A,\rightarrow)$ into $(B,\succ)$. Hence, by Lemma~\ref{lm:fin-branching-embedding-termination}, also $(A,\rightarrow)$ is terminating. 
    \end{proof}

    Moreover, the property of being terminating is propagated to Cartesian products.

    \begin{corollary}\rm \label{cor:termination-propagates-to-cartesian-products} Let $(A_1,\rightarrow_1)$ and  $(A_2,\rightarrow_2)$ be two reduction systems and at least one of them is terminating. Moreover, let $\rightarrow$ be the binary relation over $A_1 \times A_2$ such that, for every $a_1,a_2 \in A_1$ and $b_1,b_2 \in A_2$, we let $(a_1,b_1) \rightarrow (a_2,b_2)$ if $a_1 \rightarrow_1 a_2$ and $b_1 \rightarrow_2 b_2$. Then $(A_1 \times A_2, \rightarrow)$ is terminating.
    \end{corollary}
    \begin{proof} Without loss of generality we assume that $(A_1,\rightarrow_1)$ is terminating. We prove by contradiction. Thus we assume that $(A_1 \times A_2, \rightarrow)$ is not terminating. Then there exists an $\mathbb{N}$-indexed family $((a_i,b_i)) \mid i \in \mathbb{N})$ over $A_1 \times A_2$ and for each $i \in \mathbb{N}$, the condition $(a_i,b_i) \rightarrow (a_{i+1},b_{i+1})$ holds. Thus there exists an $\mathbb{N}$-indexed family $(a_i \mid i \in \mathbb{N})$ over $A_1$ and for each $i \in \mathbb{N}$, the condition $a_i \rightarrow_1 a_{i+1}$ holds. This contradicts to the fact that $(A_1,\rightarrow_1)$ is terminating.
    \end{proof}

    \begin{corollary}\rm \label{cor:termination-propagates-to-cartesian-products2} Let $(A,\rightarrow)$ be a reduction system. Moreover, let $\succ$ be the binary relation over $A \times A$ such that, for every $a_1,a_2,b_1,b_2 \in A$, we let $(a_1,b_1) \succ (a_2,b_2)$ if (a) $a_1 \rightarrow a_2$ and $b_1=b_2$ or (b) $a_1=a_2$ and $b_1 \rightarrow b_2$. If $(A,\rightarrow)$ is terminating, then  $(A \times A, \succ)$ is terminating.
      \end{corollary}
      \begin{proof} We prove by contradiction. Assume that $(A \times A, \succ)$ is not terminating, i.e., there exists an $\mathbb{N}$-indexed family $((a_i,b_i) \mid i \in \mathbb{N})$ over $A \times A$ such that, for each $i \in \mathbb{N}$, the condition $(a_i,b_i) \succ (a_{i+1},b_{i+1})$ holds. We consider the 
      $\mathbb{N}$-indexed family $(a_i \mid i \in \mathbb{N})$ over $A$. Since $(A,\rightarrow)$ is terminating, there exists $j\in \mathbb{N}$ such that 
      $a_j=a_{j'}$ for every $j'\ge j$. For similar reasons, there exists $k\in \mathbb{N}$ such that $b_k=b_{k'}$ for every $k'\ge k$. Let $n=\max(j,k)$. Then, for every $n'\ge n$, we have $(a_n,b_n)=(a_{n'},b_{n'})$. This is a contradiction, hence $(A \times A, \succ)$ is terminating.
        \end{proof}
        
          In a number of cases, we will use a combination of Lemmas \ref{lm:fin-branching-embedding-termination} and \ref{lm:termination-is-subset-closed}. This combination concerns reductions to strict substrings. For this scenario the following corollary holds.
          
    \begin{corollary}\rm\label{cor:reduction-to-substring-is-terminating} Let $A$ be a set and $(A^*,\rightarrow)$ be the reduction system such, for every $w_1,w_2 \in A^*$, we let $w_1 \rightarrow w_2$ if $w_2$ is a strict substring of $w_1$, i.e., if there exist $u,v \in A^*$ with $w_1 = uw_2v$ and $uv\not=\varepsilon$. Moreover, let $B\subseteq A^*$, and $\succ {\subseteq} \rightarrow\cap (B\times B)$. Then also $(B,\succ)$ is terminating.
    \end{corollary}
    \begin{proof} First, we prove that $(A^*,\rightarrow)$ is terminating. For this, we consider the length function $\varphi: A^* \to \mathbb{N}$, i.e., for each $w \in A^*$, $\varphi(w)$ is the length of $w$. Obviously, if $w_1 \rightarrow w_2$, then $\varphi(w_1) > \varphi(w_2)$. Hence $\varphi$ is a monotone embedding of the reduction system $(A^*,\rightarrow)$ into the terminating reduction system $(\mathbb{N},>)$. Thus, by Lemma \ref{lm:fin-branching-embedding-termination}, the relation $\rightarrow$ is terminating.

      Second, Lemma \ref{lm:termination-is-subset-closed} implies that $(B,\succ)$ is terminating.
    \end{proof}

    In those cases in which Corollary \ref{cor:reduction-to-substring-is-terminating} applies, we will not show the monotone embedding explicitly, but just refer to this corollary.


  \section{Well-founded induction}
  \label{sec:well-founded-induction}

\index{well-founded induction}
\index{well-founded set}
Let $(C,\succ)$ be a terminating reduction system.  Moreover, let $P \subseteq C$ be a subset, called \emph{property}. We will abbreviate the fact that $c \in P$ by $P(c)$ and say that \emph{$c$ has the property $P$}. Then the following holds:
\begin{equation}\label{equ:well-founded-induction}
\Big( (\forall c \in C): \ [(\forall c' \in C): (c \succ c') \Rightarrow P(c')] \Rightarrow P(c)\Big) \Rightarrow \Big((\forall c \in C): P(c)\Big) \enspace.
\end{equation}
The formula \eqref{equ:well-founded-induction} is called the principle of \emph{proof by well-founded induction on $(C,\succ)$} (for short: \emph{proof by induction on $(C,\succ)$}). We can use this principle to prove the claim that  each $c \in C$ has property $P$. 

Often, the proof of the premise of \eqref{equ:well-founded-induction}, i.e., the formula
\begin{equation*}
\Big((\forall c \in C): \ [(\forall c' \in C): (c \succ c') \Rightarrow P(c')] \Rightarrow P(c)\Big) \enspace,
\end{equation*}
is split into two parts. 
Then the first part is the proof of the \emph{induction base} (for short: I.B.)
\index{I.B.}
\begin{equation}\label{equ:wfi-base}
(\forall c \in \nf_{\succ}(C)):  P(c) 
\end{equation}
and the second part is the proof of the  \emph{induction step} (for short: I.S.)
\index{I.S.}
\begin{equation}\label{eq:wfi-step}
\Big( (\forall c \in C \setminus  \nf_{\succ}(C)): \ [(\forall c' \in C): (c \succ c') \Rightarrow P(c')] \Rightarrow P(c)\Big) \enspace.
\end{equation}
For each $c \in C$, the subformula 
\begin{equation*}
[(\forall c' \in C): (c \succ c') \Rightarrow P(c')]
\end{equation*}
is called \emph{induction hypothesis} (for short: I.H.).
\index{I.H.}

There are several instances of terminating reduction systems  and well-founded induction which are relevant for us. Two instances are based on the set $\mathbb{N}$ of natural numbers and we  present these instances here (later we will  present two more instances which are based on the set of trees):
\begin{itemize}
\item \emph{proof by induction on $(\mathbb{N},\succ_{\mathbb{N}})$} (for short: proof by induction on $\mathbb{N}$): Then $\nf_{\succ_{\mathbb{N}}}(\mathbb{N}) = \{0\}$, and  the induction base \eqref{equ:wfi-base} and the induction step \eqref{eq:wfi-step} read
\[
\text{$P(0)$ \ and \  $\big((\forall n \in \mathbb{N}_+): P(n-1)  \Rightarrow P(n)\big)$ }, \ \text{respectively},
\]

\item \emph{proof by induction on $(\mathbb{N},>)$}: Then $\nf_>(\mathbb{N}) = \{0\}$, and the induction base \eqref{equ:wfi-base} and the induction step \eqref{eq:wfi-step} read
\[
  \text{$P(0)$ \ and \ $\Big((\forall n \in \mathbb{N}_+): \big[(\forall k \in [0,n-1]): P(k)\big]  \Rightarrow P(n)\Big)$}, \ \text{respectively}.
\]
\end{itemize}

We also use the principle of definition of a mapping  by well-founded induction, which is based on the next theorem.
Its historical predecessor is \cite[126. \emph{Satz der Definition durch Induktion}]{ded39} (also cf. \cite[Thm.~1.17]{kla84} and \cite[Thm.~10.19]{win93}) and it is an instance of a more general result in basic set theory, cf.  \cite[Thm.~6.11]{jec03}. Our proof follows the lines of the proof of \cite[p.~1.8, \emph{Rekursionssatz}]{ind76}.  We recall that $\succ\!\!(c)$ denotes the set $\{c' \in C \mid c \succ c'\}$ for each $c\in C$.

\begin{theorem}\label{thm:dedekind} {\rm (cf. \cite[126. Satz]{ded39})} Let $A$ be a set, $(C,\succ)$ be a terminating reduction system, and   $G: \{(c,g) \mid c \in C, g: \ \succ\!\!(c) \to A\} \to A$ a mapping. Then there exists exactly one mapping $f: C \to A$ such that, for each $c \in C$, we have
  \begin{equation} \label{eq:def-by-induction}
    f(c) = G(c,f|_{\succ(c)}) \enspace.
    \end{equation}
  \end{theorem}

\begin{proof} We define the set
\begin{equation}\label{equ:shorter}
  S = \big\{ T \subseteq C \times A \mid
    (\forall c \in C)
    (\forall g: \ \succ\!\!(c) \to A) : (g \subseteq T) \Rightarrow
    ((c,G(c,g)) \in T)  \big\}.
  \end{equation}
 We note that $S\ne \emptyset$ because $(C \times A) \in S$. Moreover, we define
  \[
\rho = \bigcap(T \mid T \in S) \enspace.
\]


Next we show the following three statements.
\begin{compactenum}
\item[(1)] $\rho$  is a mapping of type $\rho: C \to A$,
\item[(2)] for each $c \in C$, we have $\rho(c) = G(c,\rho|_{\succ(c)})$, and
  \item[(3)] for each mapping $h: C \to A$ for which $h(c) = G(c,h|_{\succ(c)})$ for each $c \in C$, we have $h = \rho$.
  \end{compactenum}
  From (1)-(3) it follows that $\rho$ is the desired mapping $f$, and hence we have proved the theorem.
  
  Proof of (1):  We define \[M=\{c\in C \mid  \text{ there exists exactly one $a \in A$ such that $(c,a)\in \rho$}\}\enspace.\]
 
By induction on $(C,\succ)$, we show that $C \subseteq M$.
For this, let $c\in C$.

I.B.: Let $c \in \nf_\succ (C)$. Since $\succ\!\!(c)=\emptyset$, we have $(c,G(c,\emptyset))\in T$ for each $T\in S$. Hence $(c,G(c,\emptyset))\in\rho$. We show that $c\in M$ by contradiction.

We assume that there exists an $a\in A$ such that $a\ne G(c,\emptyset)$ and $(c,a)\in \rho$. Let $\rho'=\rho\setminus \{(c,a)\}$; hence $\rho' \subset \rho$. We show that $\rho'\in S$.
For this, let $c' \in C$ and let $g: \ \succ\!\!(c') \to A$ be such that $g \subseteq \rho'$. Then for this $c'$ and $g$ we also have $g \subseteq \rho$.
Then by the definition of $\rho$ we have $(c',G(c',g))\in \rho$. Now if $c'=c$, then $(c',G(c',g))=(c,G(c,\emptyset))\ne (c,a)$, Otherwise, again $(c',G(c',g))
\ne (c,a)$. Hence in both cases $(c',G(c',g)) \in \rho'$. Then $\rho'\in S$ and thus $\rho\subseteq \rho'$, which contradicts $\rho' \subset \rho$.

I.S.: Let $c \in C\setminus \nf_\succ(C)$. By I.H., for each $c'\in \ \succ\!\!(c)$, we have $c\in M$. Let $g=\{(c',a)\in \rho \mid c' \in \ \succ\!\!(c)\}$.
Then, by the definition of $\rho$, for each $T\in S$, we have $g\subseteq T$ and hence for each $T\in S$, we also have $(c,G(c,g))\in T$. Consequently, $(c,G(c,g))\in 	\rho$. We show that $c\in M$ by contradiction.

We assume that there is an $a\in A$ such that $a\ne G(c,g)$ and $(c,a)\in \rho$. Let $\rho'=\rho\setminus \{(c,a)\}$; hence $\rho' \subset \rho$. We show that $\rho'\in S$.
For this, let $\overline{c} \in C$ and let $g: \ \succ\!\!(\overline{c}) \to A$ be such that $g \subseteq \rho'$. Then for this $\overline{c}$ and $g$ we also have $g \subseteq \rho$. Then by the definition of $\rho$ we have $(\overline{c},G(\overline{c},g))\in \rho$. If $\overline{c}=c$, then $(\overline{c},G(\overline{c},g))=(c,G(c,g))\ne (c,a)$, Otherwise, again $(\overline{c},G(\overline{c},g))
\ne (c,a)$. Hence in both cases $(\overline{c},G(\overline{c},g)) \in \rho'$. Then $\rho'\in S$ and thus $\rho\subseteq \rho'$, which contradicts $\rho' \subset \rho$.

\

Proof of (2): It is obvious that, for each $c\in C$, we have $\rho|_{\succ(c)}\subseteq \rho$. Hence, by the definition of $\rho$, for each $c\in C$, we have $\rho(c)=G(c,\rho|_{\succ(c)})$.

\

Proof of (3): Let $h: C \to A$ be a mapping for which $h(c) = G(c,h|_{\succ(c)})$ for each $c \in C$. By induction on $(C,\succ)$ we show that, for each $c\in C$, we have $h(c)=\rho(c)$. Let $c\in C$. By I.H.  for each $c'\in \succ(c)$, we have $h(c')=\rho(c')$. Hence, $h|_{\succ(c)}=\rho|_{\succ(c)}$ and thus  $h(c) = G(c,h|_{\succ(c)})=G(c,\rho|_{\succ(c)})=\rho(c)$.
 \end{proof}

 We say that the mapping $f$ of Theorem \ref{thm:dedekind}  is \emph{defined by well-founded induction on $(C,\succ)$} (for short: \emph{defined by induction on $(C,\succ)$}).

 \index{schema of primitive recursion}
 The mapping  $G: \{(c,g) \mid c \in C, g: \ \succ\!\!(c) \to A\} \to A$ which determines by Theorem \ref{thm:dedekind} a unique mapping $f: C \to A$, might be called a \emph{schema of primitive recursion}. Such schemata have been investigated, e.g., for $C=\mathbb{N}$ in \cite{pet57}, for $C=\Gamma^*$ in \cite{henindroswei75}, for $C$ being the set of trees over $\Sigma$ in \cite{hup78,engvog91,fulhervagvog93}, and for $C$ being a decomposition algebra in \cite{kla84}.

In many applications of this principle, we will not show the mapping $G$ explicitly, but only implicitly in the definition of $f$. Moreover, we will split \eqref{eq:def-by-induction} into an induction base and an induction step as follows:
\begin{compactenum}
\item[I.B.:] Let $c \in \nf_\succ(C)$. Hence $\succ\!\!(c)=\emptyset$ and thus $f|_\emptyset: \emptyset \to A$.  Then we define $f(c) = G(c,f|_\emptyset)$. That means that $f(c)$ does not depend on elements of $\im(f)$.
\item[I.S.:] Let $c \in C \setminus \nf_\succ(C)$ and we assume that $f(c')$ is defined for each $c' \in \ \succ\!\!(c)$. Then we define $f(c) = G(c,f|_{\succ(c)})$. That means that $f(c)$ may depend on elements of $\{f(c') \mid c' \in \ \succ\!\!(c)\}$.
  \end{compactenum}

Next we demonstrate the principle of definition by well-founded induction. More precisely, given a mapping $h: D \to D$ for some set $D$, we define the $\mathbb{N}$-indexed family of iterated applications $(h^n \mid n \in \mathbb{N})$ of $h$ using this principle.

For this we instantiate the objects $C$, $\succ$, $A$, and $G$ which occur in the principle as follows:
we let $C = \mathbb{N}$, $\succ = \succ_\mathbb{N}$, and  $A=D^D$. Then $G$ has the type
\[
G: \{(n,g) \mid n \in \mathbb{N}, g: \ \succ_\mathbb{N}\!\!(n) \to D^D\} \to D^D \enspace.
\]
We note that $\succ_\mathbb{N}\!\!(0) = \emptyset$ and, for each $n \in \mathbb{N}_+$, the set $\succ_\mathbb{N}\!\!(n) = \{n-1\}$. Now, for every $n \in \mathbb{N}$ and $g: \ \succ_\mathbb{N}\!\!(n) \to D^D$, we define
\[
  G(n,g) = \begin{cases}
    \id_D & \text{ if $n=0$}\\
    h \circ g(n-1) & \text{ if $n\in \mathbb{N}_+$} \enspace.
    \end{cases}
  \]
   By Theorem \ref{thm:dedekind}, there exists a unique mapping $f: \mathbb{N} \to D^D$ such that $f(n) = G(n,f|_{\succ_\mathbb{N}(n)})$. 
Finally, we define the \emph{$\mathbb{N}$-indexed family of iterated applications of a mapping $h: D \to D$}, denoted by $(h^n \mid n \in \mathbb{N})$, by letting $h^n = f(n)$.

Thus, for $n=0$, we have \(h^0 = f(0) = G(0,f|_{\succ_\mathbb{N}(0)}) = \id_D\).
Moreover, for each $n \in \mathbb{N}_+$, we have
\[h^n = f(n) = G(n,f|_{\succ_\mathbb{N}(n)}) = G(n,f|_{\{n-1\}}) = h \circ f|_{\{n-1\}}(n-1) = h \circ f(n-1) = h \circ h^{n-1} \enspace. 
\]

Hence, the $\mathbb{N}$-indexed family of iterated applications of a mapping $h: D \to D$ satisfies the well known equations:
\[h^0 = \id_D \ \text {and} \  h^n =  h \circ h^{n-1} \ \text{ for each $n \in \mathbb{N}_+$} . \text{ In particular, $h^1=h$.} 
\]
 

 \section{Algebraic structures}
 \label{sec:algebraic-structures}

\subsection{Algebras, subalgebras, congruences, and homomorphisms}
\label{ssec:universal-algebras}

We recall some notions from universal algebra \cite{gra68,gogthawagwri77,bursan81,wec92}.

\index{algebra}
An {\em algebra} is a pair $\A=(A,\theta)$ where $A$ is a nonempty set and
$\theta$ is an $I$-indexed family over $\mathrm{Ops}(A)$ for some set $I$.  We call $A$ the \emph{carrier set of $\A$} and we call each element of $\im(\theta)$ an \emph{operation of $\A$}. If $I$ is finite and $\im(\theta) = \{f_1,\ldots,f_n\}$, then we denote $(A,\theta)$ also by $(A,f_1,\ldots,f_n)$ (where we choose an arbitrary ordering among the operations of $\A$).

\index{type of an algebra}
\index{signature of an algebra}
\index{Ktau@$\mathrm{K}(\tau)$}
Let $\A=(A,\theta)$ be an algebra with $\theta: I \to \mathrm{Ops}(A)$. The \emph{type of $\A$} (or: \emph{signature of $\A$}) is the mapping $\tau: I \to \mathbb{N}$ such that, for each $i\in I$, the number $\tau(i)$ is  the arity of the operation $\theta(i)$. Then we also say that \emph{$\A$ is of type~$\tau$}.
Let $\tau: I \to \mathbb{N}$ be a mapping, then we denote the \emph{set of all algebras of type $\tau$} by~$\mathrm{K}(\tau)$.

\label{page:convention-on-algebras}
\begin{quote}\emph{In the rest of this subsection, let $\A=(A,\theta)$ be an arbitrary algebra of type $\tau: I \to \mathbb{N}$ and with $\theta: I \to \mathrm{Ops}(A)$.}
\end{quote}

Let $H\subseteq A$, let $f \in \im(\theta)\cap \mathrm{Ops}^k(A)$ for some $k \in \mathbb{N}$, and let $O\subseteq \im(\theta)$.
\index{closed}
\index{$\langle H \rangle_O$}
We say that $H$ \emph{is closed under $f$} if, for every $a_1,\ldots,a_k \in H$, we have that $f(a_1,\ldots,a_k) \in H$. We say that $H$ \emph{is closed under $O$} if, for each $f \in O$, the set $H$ is closed under $f$. 
We denote by $\langle H \rangle_O$ the smallest subset $B$ of $A$ 
such that $H \subseteq B$ and $B$ is closed under $O$, i.e.\footnote{Since $(\mathcal{P}(A),\cap,\cup,\emptyset,A)$ is a $\sigma$-complete lattice (cf. Example \ref{ex:lattices}(\ref{def:lattice-of-subsets}), the right-hand side is well defined.}, 
\[
\langle H \rangle_O = \bigcap(B \subseteq A \mid H \subseteq B \text{ and $B$ is closed under $O$}) \enspace.
\]

\index{subalgebra}
\index{smallest subalgebra}
Let $B \subseteq A$ and $\B = (B,\theta_B)$ be an algebra of type $\tau$. If, for every $i \in I$ and $b_1,\ldots,b_{\tau(i)}\in B$, the equality 
$\theta_{B}(i)(b_1,\ldots,b_{\tau(i)})=\theta(i)(b_1,\ldots,b_{\tau(i)})$
holds, then we say that $\B$ is a {\em subalgebra of $\A$}.  In order to avoid complex subscripts, we denote $ (B,\theta_B)$ also by $(B,\theta)$.

\index{generating set}
Let $H\subseteq A$ such that $\langle H \rangle_{\im(\theta)}\ne\emptyset$. The subalgebra $(\langle H \rangle_{\im(\theta)},\theta)$ of $\sfA$ is called the \emph{subalgebra of $\A$ generated by $H$}.
By definition, if $(B,\theta)$ is a subalgebra of $\A$ such that $H\subseteq B$, then $\langle H \rangle_{\im(\theta)}\subseteq B$. Therefore $(\langle H \rangle_{\im(\theta)},\theta)$ is the smallest subalgebra of $\A$ which contains $H$.
Moreover, if $H=\emptyset$, then $(\langle \emptyset \rangle_{\im(\theta)},\theta)$ is the smallest subalgebra of $\A$.
If $\langle H \rangle_{\im(\theta)}=A$, then we say that \emph{$\A$ is generated by $H$} or that \emph{$H$ generates $\A$}.  The set $H$ is called \emph{generating set (of $\A$)}.

\index{locally finite}
We say that $\A$ is
\begin{compactitem}
\item \emph{finite} if $A$ is finite,
\item  \emph{locally finite} if, for each finite subset $H \subseteq A$, the set $\langle H\rangle_{\im(\theta)}$ is finite, and
\item \emph{finitely generated} if there exists a finite subset $H \subseteq A$ such that $\langle H\rangle_{\im(\theta)} = A$.\index{finitely generated}
\end{compactitem}

\index{Knaster-Tarski theorem}
\begin{lemma}\rm \label{obs:Knaster-Tarski-applied-to-algebras}  Let $\A=(A,\theta)$ be an algebra and $H \subseteq A$. Let $f: \cP(A) \to \cP(A)$ be defined for each $U \in \cP(A)$ by
  \begin{equation}\label{equ:Knaster-Tarski-mapping}
f(U) = U \cup \{g(u_1,\ldots,u_k) \mid k \in \mathbb{N}, g \in \im(\theta) \cap \mathrm{Ops}^k(A), u_1,\ldots,u_k \in U\} \enspace.
\end{equation}
Then $\langle H\rangle_{\im(\theta)} = \bigcup(f^n(H) \mid n \in \mathbb{N})$. Moreover, if $\langle H\rangle_{\im(\theta)}$ is finite, then we can construct this set.
\end{lemma}
\begin{proof} Obviously, for every $U,U'\in \cP(A)$,
\begin{align}
& \text{for each $n\in \mathbb{N}$, we have $f^n(U)\subseteq f^{n+1}(U)$, and}\label{f-n-f-n+1}\\
& \text{if $U\subseteq U'$ then $f(U)\subseteq f(U')$}\label{f-monotonic}\enspace.
\end{align}
Let us abbreviate $\bigcup(f^n(H) \mid n \in \mathbb{N})$ by $V$ and show that $V=\langle H\rangle_{\im(\theta)}$.

First we show that  $f(V)\subseteq V$ as follows:
\begingroup
\allowdisplaybreaks
\begin{align*}
f(V) = & f\big( \bigcup(f^n(H) \mid n \in \mathbb{N}) \big) \\
= & \bigcup(f^n(H) \mid n \in \mathbb{N}) \cup \big\{g(u_1,\ldots,u_k) \mid k \in \mathbb{N}, g \in \im(\theta) \cap \mathrm{Ops}^k(A), (\forall i\in[k]) : u_i \in  V \big\} \\
= & \bigcup(f^n(H) \mid n \in \mathbb{N}) \cup \{g(u_1,\ldots,u_k) \mid k \in \mathbb{N},  g \in \im(\theta) \cap \mathrm{Ops}^k(A), (\forall i\in[k]) (\exists n_i\in \mathbb{N}) : u_i\in f^{n_i}(H) \}\\
= & \bigcup(f^n(H) \mid n \in \mathbb{N}) \cup \{g(u_1,\ldots,u_k) \mid k \in \mathbb{N},  g \in \im(\theta) \cap \mathrm{Ops}^k(A), (\exists n\in \mathbb{N})(\forall i\in[k])  : u_i\in f^{n}(H) \}\\
& \tag{by (\ref{f-n-f-n+1})}\\
= &   \bigcup \big(f^n(H)\cup \{g(u_1,\ldots,u_k) \mid k \in \mathbb{N},  g \in \im(\theta) \cap \mathrm{Ops}^k(A), (\forall i\in[k])  : u_i\in f^{n}(H) \}\mid n\in \mathbb{N}\big)\big)\\
= & \bigcup \big(f(f^n(H)) \mid n\in\mathbb{N}\big)=\bigcup\big(f^n(H) \mid n \in \mathbb{N}_+\big)\subseteq V\enspace.
\end{align*}
\endgroup
Hence $V$ is closed under the operations of $\im(\theta)$. Since  $\langle H\rangle_{\im(\theta)}$ is the smallest subset of $A$ with this property, we have  $\langle H\rangle_{\im(\theta)}\subseteq V$. 

Next we prove the inclusion $V \subseteq \langle H\rangle_{\im(\theta)}$. By induction on $\mathbb{N}$, we show that $f^n(H) \subseteq \langle H\rangle_{\im(\theta)}$ for each $n \in \mathbb{N}$. The statement is obvious for $n=0$. Then, for each $n \in \mathbb{N}$, we have
\[f^{n+1}(H) = f(f^n(H))\subseteq f(\langle H\rangle_{\im(\theta)})\subseteq \langle H\rangle_{\im(\theta)}\enspace,\]
where the first inclusion follows from the I.H. and (\ref{f-monotonic}), and the second inclusion follows from the definition of $\langle H\rangle_{\im(\theta)}$. Hence $V\subseteq \langle H\rangle_{\theta(\Sigma)}$.

For each $n \in \mathbb{N}$, if $f^n(H)=f^{n+1}(H)$, then $f^{n+1}(H)=f^{n+2}(H)$ obviously.
Now assume that $\langle H\rangle_{\im(\theta)}$ is finite. Then we can find $N\in \mathbb{N}$ such that $f^N(H)=f^{N+1}(H)$ and therefore $\langle H\rangle_{\im(\theta)} = \bigcup(f^n(H) \mid n \in [0,N])$. Hence we can construct the set $\langle H\rangle_{\im(\theta)}$.
\end{proof}

\index{congruence relation}
Let $\sim \,\subseteq A \times A$ be an equivalence relation on $A$. We call $\sim$ a \emph{congruence relation on $\A$} (for short: \emph{congruence on $\A$}) if the following holds: for every $k \in \mathbb{N}$, $f \in \im(\theta) \cap \mathrm{Ops}^k(A)$, $a_1,b_1,\ldots,a_k,b_k \in A$, if $a_1\sim b_1, \ldots, a_k \sim b_k$, then $f(a_1,\ldots,a_k)\sim f(b_1,\ldots,b_k)$.

\index{quotient algebra}
Let $\sim$ a congruence relation on $\A$. The \emph{quotient algebra of $\A$ modulo $\sim$}, denoted by $\A/\!_\sim$, is the algebra $\A/\!_\sim = (A/\!_\sim,\theta/\!_\sim)$ of type $\tau$, where $A/\!_\sim$ is the factor set of $A$ modulo $\sim$ and
$\theta/\!_\sim: I \to \mathrm{Ops}(A/\!_\sim)$ such that, for each $i \in I$ with $\tau(i)=k$ for some $k \in \mathbb{N}$, and for every $a_1,\ldots,a_k \in A$ we have   
\[
(\theta/\!_\sim)(i)([a_1]_\sim,\ldots,[a_k]_\sim) = [\theta(i)(a_1,\ldots,a_k)]_\sim \enspace.
\]
Clearly, this operation is well defined, cf. \cite[p.36]{gra68}. We call
the mapping $\pi_\sim : A \rightarrow A/\!_\sim$ with $\pi_\sim(a) = [a]_\sim$, for each $a \in A$, the \emph{canonical mapping (of $A$ and $\sim$)}.

\index{algebra homomorphism}
Let $\A_1=(A_1,\theta_1)$ and  $\A_2=(A_2,\theta_2)$ be two algebras of the same type $\tau$. Moreover, let $h: A_1 \rightarrow A_2$ be a mapping. Then $h$ is an \emph{algebra homomorphism} (from $\A_1$ to $\A_2$) if, for every 
$i \in I$ with $\tau(i)=k$ for some $k\in \mathbb{N}$, and for every $v_1,\ldots,v_k \in A_1$,
we have
\begin{equation}\label{equ:hom-equality}
h(\theta_1(i)(v_1,\ldots,v_k)) = \theta_2(i)(\sigma)(h(v_1),\ldots,h(v_k))\enspace.
\end{equation}
\index{algebra isomorphism}
\index{$\cong$}
If $h$ is a bijective algebra homomorphism, then it is an {\em algebra isomorphism}. If there exists such an isomorphism, then we say that $\A_1$ and  $\A_2$ are  {\em isomorphic} and we denote this fact by $\A_1 \cong \A_2$.

\begin{theorem} \label{thm:canonical-map-of-congr-is-hom} {\rm \cite[\S~7, Lm.~2]{gra68}} Let $\A=(A,\theta)$ be an algebra and $\sim$ a congruence on $\A$. The canonical mapping $\pi_\sim$ of $\sim$ is an algebra homomorphism from $\A$ to $\A/\!_\sim$. 
  \end{theorem}

Let $B \subseteq A$. We denote by $\theta|_{B}$ the $I$-indexed family such that, for every $i \in I$, we have $\theta|_{B}(i) = \theta(i)|_{B^{\tau(i)}}$.

\begin{lemma}\rm\cite[§7, Lm.~3]{gra68} \label{lm:hom-image=subalgebra} Let $(A_1,\theta_1)$ and  $(A_2,\theta_2)$ be two algebras of the same type $\tau$, and let $h: A_1 \rightarrow A_2$ be an algebra homomorphism. Then $\im(h)$ is closed under the operations in $\im(\theta_2)$. Hence $(\im(h),\theta_2|_{\im(h)})$ is a subalgebra of $(A_2,\theta_2)$.
 \end{lemma}

\begin{lemma}\rm\cite[\S~7, Lm.~4]{gra68}\label{lm:congruence-on-subalgebra} Let $(A_1,\theta_1)$ and  $(A_2,\theta_2)$ be two algebras of the same type $\tau$ such that
$(A_2,\theta_2)$ is a subalgebra of $(A_1,\theta_1)$. Moreover, let $\sim$ be a congruence on $(A_1,\theta_1)$. Then the relation $\sim'=\sim \cap (A_2\times A_2)$ is a congruence on $(A_2,\theta_2)$. Often, we denote $\sim'$ also by $\sim$.
\end{lemma}

\begin{theorem} \label{thm:comp-hom} {\rm \cite[\S 7, Lm. 5]{gra68} } Let $(A_1,\theta_1)$, $(A_2,\theta_2)$, and $(A_3,\theta_3)$ be three algebras. Moreover, let $h: A_1 \to A_2$ and $g: A_2 \to A_3$ be two algebra homomorphisms. Then $g \circ h$ is an algebra homomorphism.
  \end{theorem}

\index{kernel}
Let $h$ be an algebra homomorphism from $(A_1,\theta_1)$ to $(A_2,\theta_2)$. The \emph{kernel of $h$}, denoted by $\mathrm{ker}(h)$, is the equivalence relation on $A_1$ defined by $\mathrm{ker}(h) = \{(a_1,a_2) \in A_1 \times A_1 \mid h(a_1) = h(a_2)\}$. 

\begin{theorem}\label{thm:kernel-is-congruence} {\rm \cite[\S 7, Lm. 6]{gra68}, \cite[\S 11,~Thm.~1]{gra68} }
Let $h$ be an algebra homomorphism from $\A_1=(A_1,\theta_1)$ to $\A_2=(A_2,\theta_2)$.
Then $\mathrm{ker}(h)$ is a congruence relation on $\A_1$. Moreover, if $h$ is surjective, then $\A_1/_{\ker(h)} \cong \A_2$.
\end{theorem}

\begin{corollary} \label{cor:image-of-hom-isomorphic-to-quotient-of-kernel}\rm Let $\A_1=(A_1,\theta_1)$ and $\A_2=(A_2,\theta_2)$ be algebras of the same type. Moreover, let $h$ be an algebra homomorphism from $\A_1$ to $\A_2$. Then $\A_1/_{\ker(h)} \cong (\im(h),\theta_2|_{\im(h)})$.
\end{corollary}
\begin{proof} By Lemma \ref{lm:hom-image=subalgebra}, $(\im(h),\theta_2|_{\im(h)})$ is an algebra of the same type as $\A_2$. Then $h': A_1 \to \im(h)$ defined, for each $a \in A_1$ by $h'(a)=h(a)$ is a homomorphism from $\A_1$ to $(\im(h),\theta_2|_{\im(h)})$. Clearly, $\ker(h)=\ker(h')$. Since $h'$ is surjective, by Theorem \ref{thm:kernel-is-congruence}, we obtain that $\A_1/_{\ker(h)} \cong (\im(h),\theta_2|_{\im(h)})$.
  \end{proof}

Let $\A$ be an algebra. Moreover, let $\sim$ and $\approx$ be congruences on $\A$ such that $\sim \subseteq \approx$. We define the relation $\approx\!/_\sim$ on $A/_\sim$ such that, for every $a,b \in A$, we let
\[
[a]_\sim \ \approx\!/_\sim \  [b]_\sim \ \ \text{ if } \ \ a \approx b \enspace.
\]
The next theorem is also known as the second isomorphism theorem.

\begin{theorem}\label{thm:snd-isom-theorem} {\rm \cite[\S 11, Lm.~2]{gra68} and \cite[\S 11, Thm.~4]{gra68}}
  Let $\A$ be an algebra. Moreover, let $\sim$ and $\approx$ be congruences on $\A$ such that $\sim \subseteq \approx$. Then the following two statements hold.
  \begin{compactenum}
  \item[(1)] The relation $\approx\!/_\sim$ is a congruence on $\A/_\sim$.
  \item[(2)] $(\A/_\sim)/_{(\approx/_\sim)} \cong \A/_\approx$.
    \end{compactenum}
\end{theorem}

\begin{lemma}\rm \cite[Lm.~3.3.1]{baanip98} and \cite[Thm.~II.10.7]{bursan81} \label{lm:baanip-unique-hom}   Let $\sfA$ and $\sfB$ be algebras of the same type $\tau$. Moreover, assume that $\sfA$ is generated by $H$. If $f$ and $g$ are homomorphisms from $\sfA$ to $\sfB$, and if $f$ and $g$ coincide on $H$, then $f=g$.
\end{lemma} 

\index{free algebra in the set $\mathcal{K}$ with generating set $H$}
\index{initial}
Let $\tau: I \to \mathbb{N}$ be a mapping and let  $\mathcal{K} \subseteq \mathrm{K}(\tau)$ be a set of algebras of type $\tau$. Moreover, let $\A=(A,\theta)$ be an algebra and let $H \subseteq A$. The algebra~$\A$ is called \emph{free in~$\mathcal{K}$ with generating set $H$} if the following conditions hold:
\begin{compactenum}
  \item[(a)] $\A \in \mathcal{K}$,
  \item[(b)] $\A$ is generated by $H$, and
\item[(c)] for every algebra $\A'=(A',\theta')$ in $\mathcal{K}$ and mapping $f\colon H \rightarrow A'$, there exists an extension of $f$ to an algebra homomorphism $h\colon A \rightarrow A'$ from $\A$ to $\A'$.
  \end{compactenum}
If the extension $h'$ in (c) exists, then it is unique by Lemma~\ref{lm:baanip-unique-hom}.
  
\label{page:initial-algebra}
If $\A$ is free  in~$\mathcal{K}$ with generating set $H=\emptyset$, then, for each~$\A'=(A',\theta')$ in $\mathcal{K}$, there exists exactly one algebra homomorphism $h\colon A \rightarrow A'$  from $\A$ to $\A'$. In this case $\A$ is called \emph{initial in~$\mathcal{K}$}.

\begin{theorem}{\rm \cite[\S 24, Thm. 1]{gra68}}\label{thm:two-freely-gen-alg-isomorphic}  Let $\tau$ be a type and let  $\mathcal{K} \subseteq \mathrm{K}(\tau)$ be a set of algebras of type $\tau$.  Any two algebras which are free in $\cK$ with the same generating set are isomorphic.
  \end{theorem}

\subsection{Ranked alphabets and $\Sigma$-algebras}
\label{ssec:ranked-alph-Sigma-algebras}

\index{ranked alphabet}
\index{rk@$\rk$}
A {\em ranked alphabet} is a pair $(\Sigma,\rk)$, where
\begin{compactitem}
\item $\Sigma$ is an alphabet and
\item $\rk: \Sigma \rightarrow \mathbb{N}$ is a mapping, called \emph{rank  mapping}, such that $\rk^{-1}(0)\not=\emptyset$.
\end{compactitem}
For each $k \in \mathbb{N}$, we denote the set $\rk^{-1}(k)$ by $\Sigma^{(k)}$. Sometimes we write $\sigma^{(k)}$ to indicate that $\sigma\in\Sigma^{(k)}$. We denote $\max(k \in \mathbb{N} \mid \Sigma^{(k)}\not= \emptyset)$ by $\maxrk(\Sigma)$. Whenever the rank mapping is clear from the context or it is irrelevant, then we abbreviate the ranked alphabet $(\Sigma,\rk)$ by $\Sigma$.

\index{string ranked alphabet}
\index{trivial ranked alphabet}
\index{monadic ranked alphabet}
\index{branching ranked alphabet}
Let $\Sigma$ be a ranked alphabet. We say that $\Sigma$ is
\begin{compactitem}
\item  \emph{trivial} if $\Sigma=\Sigma^{(0)}$,
\item \emph{monadic} if $\Sigma=\Sigma^{(1)} \cup \Sigma^{(0)}$,
\item a \emph{string ranked alphabet} if it is monadic, $\Sigma^{(1)}\not= \emptyset$, and $|\Sigma^{(0)}|=1$,  and
  \item  \emph{branching} if there exists $k \in \mathbb{N}$ such that $k \ge 2$ and $\Sigma^{(k)}\not=\emptyset$.
    \end{compactitem}

   The following table shows some examples of ranked alphabets.

  \

  {\small
  \begin{tabular}{l|c|c|c|c}
    & $\{\alpha^{(0)}, \beta^{(0)}\}$ &  $\{\alpha^{(0)}, \gamma^{(1)}, \delta^{(1)}\}$ &  $\{\alpha^{(0)}, \beta^{(0)},\gamma^{(1)}, \delta^{(1)}\}$ & $\{\alpha^{(0)}, \gamma^{(1)}, \sigma^{(3)}\}$\\\hline
    trivial & yes & no & no & no \\
    string ranked alphabet & no & yes & no & no\\
       monadic & yes & yes &yes & no \\
                                         branching & no & no & no &yes
    \end{tabular}
  }

\index{Sigmaalgebra@$\Sigma$-algebra}
Let $(\Sigma,\rk)$ be a ranked alphabet. A \emph{$\Sigma$-algebra} is an algebra $(A,\theta)$ of type $\rk: \Sigma \to \mathbb{N}$. We denote the set of all $\Sigma$-algebras by $\mathrm{K}(\Sigma)$.

 \index{Sigmaalgebrahomomorhism@$\Sigma$-algebra homomorphism}
 \index{Sigmaalgebraisomorphism@$\Sigma$-algebra isomorphism}
Let $\A_1$ and $\A_2$ be $\Sigma$-algebras and $h$ an algebra homomorphism (algebra isomorphism) from $\A_1$ to $\A_2$. Then we call $h$ a \emph{$\Sigma$-algebra homomorphism} (\emph{$\Sigma$-algebra isomorphism}, respectively).


\subsection{Properties of binary operations}
\label{sec:properties-bin-operations}

\index{extremal}
Let $B$ be a nonempty set.  Let $\odot$ be a binary operation on $B$. The operation $\odot$ is
\begin{compactitem}
\item \emph{associative} if $(a \odot b)\odot c = a \odot (b\odot c)$ for every $a,b,c \in B$,
\item \emph{commutative} if $a \odot b = b \odot a$ for every $a,b \in B$,
\item \emph{idempotent} if $a\odot a=a$ for each $a \in B$,
\item \emph{extremal} if $a\odot b \in \{a,b\}$ for every $a,b \in B$.
\end{compactitem}
Obviously, extremal implies idempotent. The other direction does not hold, e.g., the set union $\cup$ is idempotent but in general not extremal.

An element $e\in B$ is a \emph{unit element (of $\odot$})
if $e\odot a=a\odot e=a$ for every $a\in B$. There is at most one unit element.
Let $e\in B$ be a unit element and $a\in B$. An element $\overline{a} \in B$ is an \emph{inverse of $a$ (with respect to $\odot$)}
if $a \odot \overline{a} = \overline{a} \odot a = e$. If an inverse of $a$ exists, then we also say that $a$ has an inverse. If~$\odot$ is associative, then each element has at most one inverse.

Let $\oplus$ and  $\otimes$ be two binary operations on $B$. The operation $\otimes$ is
\begin{compactitem}
\item  \emph{right-distributive} (with respect to $\oplus$), if $(a\oplus b)\otimes c=(a\otimes c)\oplus (b\otimes c)$ for every $a,b,c \in B$, 
\item \emph{left-distributive} (with respect to $\oplus$), if $a\otimes(b\oplus c)=(a\otimes b)\oplus (a\otimes c)$ for every 
  $a,b,c \in B$,
\item \emph{distributive} (with respect to $\oplus$) if it is both right-distributive and left-distributive (with respect to $\oplus$).
\end{compactitem}
\index{absorption axiom}
Clearly, if $\otimes$  is commutative, then right-distributivity implies left-distributivity and vice versa.
Moreover, $\oplus$ and $\otimes$ satisfy the \emph{absorption axiom} if 
\begin{compactitem}
\item $a \otimes (a \oplus b) = a$ and $a \oplus (a \otimes b)=a$ for  every $a,b \in B$.
\end{compactitem}

\subsection{Semigroups, monoids, and groups}
\label{sec:semigroups}

In this subsection we recall the definitions of semigroups, monoids, and groups; they are particular algebras (cf. Subsection  \ref{ssec:universal-algebras}).
Moreover, we define extensions of the underlying binary operations to finitely many and countably many arguments.
\index{semigroup} \index{monoid} \index{group}
\begin{compactitem}
\item A \emph{semigroup} is an algebra $(B,\odot)$ where $\odot$ is an associative binary operation on $B$.
\item A \emph{monoid} is an algebra $(B,\odot,e)$ where $(B,\odot)$ is a semigroup and $e$ is an element of $B$ which is a unit element of $\odot$.
We note that the unit element $e$ can be considered as a nullary operation on $B$.
\item A \emph{group} is a monoid $(B,\odot,e)$ such that each $a\in B$ has an inverse. We note that the inverses of the elements can be defined in terms of a unary operation on $B$.
\end{compactitem}
The semigroup $(B,\odot)$ is \emph{commutative} if $\odot$ is commutative. Similarly, we define commutative monoids and commutative groups.

\begin{observation}\rm\label{obs:commutative+idempotent-implies-locally finite} Let $(B,\odot)$ be a semigroup such that $\odot$ is commutative and idempotent. Then $(B,\odot)$ is locally finite.
\end{observation}
\begin{proof} Let $H$ be a finite subset of $B$. We note that
$\langle H \rangle_{\{\odot\}}=\{b_1\odot\ldots\odot b_n \mid n\in\mathbb{N}_+, b_1,\ldots, b_n\in H\}$.
Let us abbreviate this set by $\langle H \rangle$. 

By induction on $(\mathbb{N}_+,\succ_{\mathbb{N}_+})$, where $\succ_{\mathbb{N}_+} = \{(n+1,n) \mid n \in \mathbb{N}_+\}$, we prove the following statement.
\begin{eqnarray}\label{eq:induction-com-idem}
  \text{For each $n \in \mathbb{N}_+$ and every $b_1,\ldots,b_n \in  H$, there exist $m\in \mathbb{N}_+$}\\
  \text{ and pairwise different $c_1,\ldots,c_m \in H$ such that $b_1 \odot \ldots \odot b_n = c_1\odot\ldots\odot c_m$.}\nonumber
\end{eqnarray}

I.B.: For $n=1$ the statement is trivially true.

I.S.: Let  $b = b_1 \odot \cdots \odot b_{n+1}$  with  $n \in \mathbb{N}_+$. By I.H. there exist $m\in \mathbb{N}_+$ and pairwise different $c_1,\ldots,c_m \in H$ such that $b_1 \odot \ldots \odot b_n = c_1\odot\ldots\odot c_m$. 

If $b_{n+1}\not \in \{c_1,\ldots,c_m\}$, then \eqref{eq:induction-com-idem} holds with  $m+1$ by letting $c_{m+1} = b_{n+1}$.
Otherwise $b_{n+1}=c_i$ for some $i\in[m]$. Then, by commutativity and idempotency, we have
\[b_1 \odot \cdots \odot b_{n+1} = c_1\odot\ldots\odot c_m \odot b_{n+1} = c_1\odot \ldots\odot c_i \odot b_{n+1} \odot\ldots\odot c_m=c_1\odot\ldots\odot c_m\enspace.\]
Thus \eqref{eq:induction-com-idem} holds with $m$.
This proves \eqref{eq:induction-com-idem}. 
Since $H$ is finite, by \eqref{eq:induction-com-idem} also $\langle H \rangle$ is finite.
\end{proof}

\label{page:finite-summation}
Next we consider a  monoid $(B,\odot,e)$ and formalize two ways of extending $\odot$ to finitely many arguments. Let $I$ be a finite set with $I = \{i_1,\ldots,i_k\}$ for some  $k \in \mathbb{N}$. If (a) $I \subseteq \mathbb{N}$ and $i_1<\ldots<i_k$, where $<$ is the usual total order ``less than'' on natural numbers or (b) $\odot$ is commutative, then we define the operation $\bigodot_I: B^I \rightarrow B$ such that, for each $I$-indexed family $(b_i \mid i\in I)$ of elements in $B$, we have
\begin{equation}\label{eq:extension-odot-finite-set}
\bigodot\nolimits_I (b_i \mid i \in I)=
\left\{
\begin{array}{ll}
b_{i_1}\odot \ldots \odot b_{i_k} & \text{if } I\not=\emptyset\\
e &\text{otherwise}\enspace.
\end{array}
\right.
\end{equation}
We abbreviate $\bigodot_I (b_i \mid i \in I)$ by $\bigodot (b_i \mid i \in I)$ or $\bigodot_{i \in I} b_i$. Thus, in particular, $\bigodot_{i \in [k]} b_i=b_1\odot\ldots\odot b_k$ for each $k\in\mathbb{N}_+$, and $\bigodot_{i \in \emptyset} b_i= e$.

\begin{observation}\rm\label{obs:extremal-sum} Let $I$ be a finite and nonempty index set such that $I \subseteq \mathbb{N}$ or $\odot$ is commutative. If $\odot$ is extremal, then for each $I$-indexed family $(b_i \mid i\in I$) over $B$ there exists a $j\in I$ such that $b_j=\bigodot_{i\in I} b_i$.\hfill $\Box$
\end{observation}

Next let $(B,\odot,e)$ be commutative. We  introduce a notation for the case that the index set is the Cartesian product of two sets as follows. Let $I$ and $J$ be finite sets and $(b_{(i,j)} \mid (i,j ) \in I \times J)$ be an $(I\times J)$-indexed family over $B$. Then it follows from commutativity and associativity that
\begin{equation*}
\bigodot_{ (i,j ) \in I \times J} b_{(i,j)} = \bigodot_{i \in I} \bigodot_{j \in J} b_{(i,j)}\enspace. 
\end{equation*}
Often we will abbreviate the expression $\bigodot_{i \in I} \bigodot_{j \in J} b_{(i,j)}$ by $\bigodot_{i \in I, j \in J} b_{(i,j)}$. We will use this kind of abbreviation also in the case where the index set is the Cartesian product of finitely many sets.

\index{complete}
\index{$\sigma$-complete}
Finally, we consider a commutative monoid $(B,\odot,e)$ and identify algebraic laws such that $\odot$ can be extended to countably  many arguments. We call $(B,\odot,e)$ {\em $\sigma$-complete} if,  for each countable index set $I$, there exists a mapping $\infsum{\odot}{I}\colon B^I \rightarrow B$ such that for each $I$-indexed family $(b_i \mid i\in I)$ over $B$ the equalities \eqref{axiom-i-sum} and \eqref{axiom-ii-sum} below are satisfied (cf.~\cite[p. 124]{eil74}) using $\infsum{\odot}{i\in I}{b_i}$  as abbreviation for $\infsum{\odot}{I} (b_i \mid i \in I)$:\label{page:summation-inf} 
\begin{eqnarray}
  \text{If $I=\{j\}$, then } \infsum{\odot}{i\in I}{b_i} = b_j,\ \
  \text{ and if $I = \{j,j'\}$, then } \infsum{\odot}{i\in  I}{b_i} = b_j  \odot b_{j'}\enspace.\label{axiom-i-sum}\\
  \infsum{\odot}{i\in I}{b_i} = \infsum{\odot}{j \in J}{} (\infsum{\odot}{i\in I_j}{b_i}) \ \text{ for every countable set $J$ and partitioning $(I_j \mid j\in J)$ of $I$}. \label{axiom-ii-sum}
\end{eqnarray}

\label{page:summation-fin}
   \begin{observation}\label{obs:sum-emptyset}\rm Let $(B,\odot,e)$ be a $\sigma$-complete monoid. The following statements hold.
    \begin{compactenum}
    \item[(1)] For each $\emptyset$-indexed family $(b_i \mid i \in \emptyset)$ over $B$, we have $\infsum{\odot}{j \in \emptyset} b_j = e$.

\item[(2)] For each finite $I$-indexed family $(b_i\mid i\in I)$ of elements of $B$, we have $\infsum{\odot}{i \in I}{b_i} = \bigodot_{i\in I}b_i$\enspace. \label{equ:fin0=inf0}
    \item[(3)] For each countable index set $I$, we have \(\infsum{\odot}{j \in I}{e} =e\).
        \end{compactenum}
\end{observation}
\begin{proof} Proof of (1): Let $I=\{1\}$. We consider the $I$-indexed family $(b_i \mid i \in I)$ over $B$ where $b_1$ is an arbitrary element of $B$ (recall that $B \neq \emptyset$). Moreover, let $J=\{1,2\}$, $I_1=\emptyset$, and $I_2=I$. Then we have
   \begingroup
  \allowdisplaybreaks
  \begin{align*}
    b_1 &=  \infsum{\odot}{i \in I} b_i \tag{by \eqref{axiom-i-sum}}\\
        &= \infsum{\odot}{j \in J} \big( \infsum{\odot}{i \in I_j}{b_i}\big)  \tag{by \eqref{axiom-ii-sum}}\\
        &= \big( \infsum{\odot}{i \in I_1}{b_i}\big) \odot \big( \infsum{\odot}{i \in I_2}{b_i}\big) \tag{by \eqref{axiom-i-sum}}\\
          &= \big( \infsum{\odot}{i \in \emptyset}{b_i}\big) \odot b_1 \enspace. \tag{by \eqref{axiom-i-sum}}
  \end{align*}
  \endgroup

Since $\odot$ is commutative, also $b_1 \odot \big(\infsum{\odot}{i \in \emptyset}{b_i}\big) = b_1$. Hence $\infsum{\odot}{i \in \emptyset}{b_i}$ is a unit element, and hence $\infsum{\odot}{i \in \emptyset}{b_i} = e$.

\

  Proof of (2): Let $(b_i\mid i\in I)$ be a finite $I$-indexed family of elements of $B$. If $I = \emptyset$, then the statement follows from Statement (1) and \eqref{eq:extension-odot-finite-set}. If $I \neq \emptyset$, then we can prove by induction on $\mathbb{N}$ the following statement:
  \begin{equation}\label{eq:induction-infsum=finsum}
\text{For every $n \in \mathbb{N}$ and index set $I$ with $|I| = n+1$, we have $\infsum{\odot}{i \in I}{b_i} = \bigodot_{i\in I}b_i$}
\end{equation}

I.B.: For $n=0$, this follows from \eqref{axiom-i-sum} and \eqref{eq:extension-odot-finite-set}.

  I.S.: Now let $n\in \mathbb{N}_+$ and $|I|=n+1$. Moreover, let  $(b_i\mid i\in I)$ be a finite $I$-indexed family of elements of $B$.
  Let $i'\in I$ be an arbitrary index. Using $J = \{1,2\}$ and $I_1 = \{i'\}$ and $I_2 = I\setminus\{i'\}$, we have:
 \begingroup
  \allowdisplaybreaks
  \begin{align*}
  \infsum{\odot}{i\in I}{b_i} 
    &= \infsum{\odot}{j\in J}{}(\infsum{\odot}{i\in I_j}{b_i})
      = (\infsum{\odot}{i\in \{i'\}}{b_i})  \odot (\infsum{\odot}{i\in I\setminus\{i'\}}{b_i}) \tag{by  \eqref{axiom-ii-sum}} \\
  &= b_{i'} \odot \bigodot_{i\in I\setminus\{i'\}}b_i \tag{by \eqref{axiom-i-sum} and I.H.}\\
    &= \bigodot_{i\in I}b_i \enspace.
  \end{align*}
  \endgroup
  Then Statement (2) follows from \eqref{eq:induction-infsum=finsum}.

  \
  
  Proof of (3): Let $I_j=\emptyset$ for every $j\in I$. Then $(I_j \mid j \in I)$ is a partitioning of $\emptyset$. Hence  we have
  \begingroup
  \allowdisplaybreaks
  \begin{align*}
  \infsum{\odot}{j\in I}{e} 
  &= \infsum{\odot}{j\in I}{}(\bigodot_{i\in I_j} e) \tag{by  \eqref{eq:extension-odot-finite-set}} \\
  &= \infsum{\odot}{j\in I}{}(\infsum{\odot}{i\in I_j}{e}) \tag{by Statement (2)}\\
  &= \infsum{\odot}{j\in \emptyset}{e} \tag{because $\bigcup_{j \in I}I_j = \emptyset$ and by (\ref{axiom-ii-sum})} \\
  & = e \tag{by Statement (1)}
    \end{align*}
    \endgroup
  \end{proof}
  
Let $(b_i \mid i \in I)$ be an $I$-indexed family over $B$. Moreover, let $P \subseteq B$ be a property. Then we denote by $(b_i \mid i\in I, b_i \in P)$ the $I'$-indexed family $(b_i \mid i \in I')$ over $B$ where $I' = \{i \in I \mid b_i \in P\}$. Also, we abbreviate $\bigodot(b_i \mid i \in I, b_i \in P)$ by $\bigodot_{\substack{i \in I:\\b_i \in P}}b_i$, and we abbreviate$\sum^{\odot}_I(b_i \mid i \in I, b_i \in P)$ by $\infsum{\odot}{\substack{i \in I:\\b_i \in P}}b_i$.

\index{power}
Let $(B,\odot,e)$ be a monoid and $b \in B$. By induction on $(\mathbb{N},\succ_{\mathbb{N}})$, we define the \emph{family of powers of $b$}, denoted by $(b^n \mid n \in \mathbb{N})$, to be the $\mathbb{N}$-indexed family over $B$ defined such that
\[
b^0 = e \ \ \text{ and, for each $n \in \mathbb{N}$, we let $b^{n+1} = b \odot b^n$}\enspace.
  \]

  \index{finite order}
  \index{order}
  \index{index}
  \index{period}
An element $b \in B$ \emph{has finite order} if $\langle b \rangle_{\{\odot\}}$ is finite.
If this is the case, then there exists a least number $i \in \mathbb{N}_+$ such that $b^i=b^{(i+k)}$ for some $k \in \mathbb{N}_+$, and there exists a least number $p \in \mathbb{N}_+$ such that $b^i = b^{(i+p)}$. We call $i$ and $p$ the {\em index (of $b$)} and the {\em period (of $b$)}, respectively, and denote them by $i(b)$ and $p(b)$, respectively. Moreover, we call $i(b)+p(b)-1$, i.e., the number of elements of $\langle b \rangle_{\{\odot\}}$, the {\em order of $b$} (cf. Figure \ref{fig:period-index}). In particular, the order of $e$ is 1 because $i(e)=p(e)=1$.

Also, if $\odot$ is idempotent, then each $b \in B$ has finite order because $\langle b \rangle_{\{\odot\}} = \{b\}$ is finite. Then, for each $b \in B$, the index of $b$, the period of $b$, and the order of $b$  are $1$.

 \begin{figure}[t]
    \small
    \centering
    \begin{tikzpicture}
        \tikzset{element/.style={circle,fill=black,inner sep=0pt,minimum size=3pt}}
        \node[element,label={below:$b$}] at (0,0) (b) {};
        \node[element,label={below:$b^2$}] at (1,0) (b2) {};
        \node at (2,0) (ld) {$\ldots$};
        \node[element,label={below:$ $}] at (3,0) (ib-1) {};
        \node[element,label={[xshift=1.25cm, yshift=-2em]$b^{i(b)} =b^{(i(b)+p(b))}$}] at (4,0) (ib) {};
        \node[element, shift={(150:.75)}] at (4,.75) (p4) {};
        \node[element, shift={(210:.75)}] at (4,.75) (p5) {};
        \node[element, shift={(330:.75)}] at (4,.75) (p1) {};
        \node[element, shift={(30:.75)}] at (4,.75) (p2) {};
        \node[shift={(90:.75)}] at (4,.75) (p3) {$\cdots$};
        \draw[->, shorten <=1mm, shorten >=1mm] (b) -- (b2);
        \draw[->, shorten <=1mm] (b2) -- (ld);
        \draw[->, shorten >=1mm] (ld) -- (ib-1);
        \draw[->, shorten <=1mm, shorten >=1mm] (ib-1) -- (ib);
        \draw[->, shorten <=1mm, shorten >=1mm] (ib) edge[bend right] (p1) ;
        \draw[->, shorten <=1mm, shorten >=1mm] (p1) edge[bend right] (p2) ;
        \draw[->, shorten <=1mm, shorten >=1mm] (p2) edge[bend right] (p3) ;
        \draw[->, shorten <=1mm, shorten >=1mm] (p3) edge[bend right] (p4) ;
        \draw[->, shorten <=1mm, shorten >=1mm] (p4) edge[bend right] (p5) ;
        \draw[->, shorten <=1mm, shorten >=1mm] (p5) edge[bend right] (ib) ;
        \draw[->] (4.75,0.175) arc (-40:220:1);
        \node at (5.5,1) {$p(b)$};
        \draw[decorate, decoration={brace,amplitude=10pt}] (4,-.65) -- (0,-.65) node[midway,below,yshift=-10pt] {$i(b)$};
    \end{tikzpicture}
    \caption{\label{fig:period-index} Illustration of the index $i(b)$ and the period $p(b)$ of $b$ (cf. \cite[Fig.~1]{fulkosvog19}).}
\end{figure}

\subsection{Strong bimonoids, semirings, rings, and fields}
\label{sec:strong-bimonoid}

\index{strong bimonoid}
A \emph{strong bimonoid} \cite{drostuvog10,rad10,cirdroignvog10,drovog10,drovog12} is an algebra  $\B=(B,\oplus,\otimes,\mathbb{0},\mathbb{1})$ where $(B,\oplus,\mathbb{0})$ is a commutative monoid, $(B,\otimes,\mathbb{1})$ is a monoid, $\mathbb{0}\not=\mathbb{1}$, and $\mathbb{0}$ is an annihilator for $\otimes$, i.e.,  $b \otimes \mathbb{0} = \mathbb{0} \otimes b= \mathbb{0}$ holds for every $b \in B$. 
The operations $\oplus$ and $\otimes$ are called \emph{summation} and \emph{multiplication}, respectively.

\index{strong bimonoid!commutative}
\index{strong bimonoid!right-distributive}
 \index{strong bimonoid!left-distributive}
  \index{strong bimonoid!$\sigma$-complete}
Let $\B=(B,\oplus,\otimes,\mathbb{0},\mathbb{1})$ be a strong bimonoid. It is
\begin{compactitem}
\item \emph{commutative}, if $\otimes$ is commutative, 
\item \emph{left-distributive}, if $\otimes$ is left-distributive (with respect to $\oplus$), 
\item  \emph{right-distributive}, if $\otimes$ is right-distributive (with respect to $\oplus$), 
\item  \emph{distributive}, if $\otimes$ is left-distributive and  right-distributive (with respect to $\oplus$), 
\item \emph{$\sigma$-complete} if $(B,\oplus,\mathbb{0})$ is $\sigma$-complete, and \index{strong bimonoid!$\sigma$-complete},
\item \emph{bi-locally finite} if $(B,\oplus,\mathbb{0})$ and $(B,\otimes,\mathbb{1})$ are locally finite.
  \index{strong bimonoid!bi-locally finite}
\item \emph{weakly locally finite} if, for every finite subset $A\subseteq B$, the set $\CL(A)$, called the \emph{weak closure of $A$}, is finite,  where
$\CL(A)$ is the smallest subset $C\subseteq B$ such that $A\cup\{\0,\1\}\subseteq C$, and $(b\oplus b') \in C$ and $(b\otimes a) \in C$ for every $a \in A$ and $b,b'\in C$.
\index{strong bimonoid!weakly locally finite}
\index{wcl@$\CL(A)$}
\end{compactitem}

We declare that the precedence order of the two operations in expressions as follows:  first $\otimes$, second~$\oplus$.   Using this convention, we can save some parentheses in expressions over $\B$ which use both operations. For instance
\begin{align*}
\bigoplus_{i\in I}{(b_i\otimes b)} \  \text{ can be written as } \  \bigoplus_{i\in I}{b_i\otimes b} 
\end{align*}
for each finite index set $I$, $I$-indexed family $(b_i\mid i\in I)$, and $b\in B$. Similarly, if $\B$ is $\sigma$-complete, then
\begin{align*}
\infsum{\oplus}{i\in I}{(b_i\otimes b)} \  \text{ can be written as } \  \infsum{\oplus}{i\in I}{b_i\otimes b} 
\end{align*}
for each countable index set $I$, $I$-indexed family $(b_i\mid i\in I)$, and $b\in B$.

\index{semiring}
\index{strong bimonoid!semiring}
A \emph{semiring} \cite{hebwei93,gol99} is a strong bimonoid $\B= (B,\oplus,\otimes,\0,\1)$ which is  distributive.
Clearly, if $\B$ is a semiring, then for every finite and nonempty family $(b_i \mid i\in I)$ and $b \in B$ the following equalities hold:
\(
\bigoplus_{i \in I}b \otimes b_i = b \otimes (\bigoplus_{i \in I}b_i)
\)
and
\(
\bigoplus_{i \in I}b_i \otimes b = (\bigoplus_{i \in I}b_i) \otimes b
\).

A semiring $(B,\oplus,\otimes,\0,\1)$ is \emph{$\sigma$-complete} if $(B,\oplus,\mathbb{0})$ is $\sigma$-complete and the following equalities hold for every  countable index set $I$, $I$-indexed family $(b_i \mid i\in I)$, and $b \in B$:
\begin{equation} \label{axiom-iii-sum}
  \infsum{\oplus}{i \in I}{b \otimes b_i} = b \otimes (\infsum{\oplus}{i \in I}{b_i}) \  \text{ and } \
  \infsum{\oplus}{i \in I}{b_i \otimes b} = (\infsum{\oplus}{i \in I}{b_i}) \otimes b \enspace.  
\end{equation}
We note that for semirings $\sigma$-completeness is slightly more general than completeness (cf. e.g.~\cite[p. 125]{eil74} and \cite{esikui03}) because completeness requires the extension of $\oplus$ not only for countable but arbitrary index sets $I$.

\index{ring}
\index{semifield}
\index{field}
Let $(B,\oplus,\otimes,\mathbb{0},\mathbb{1})$ be a semiring.  It is a
\begin{compactitem} 
\item \emph{ring} if $(B,\oplus,\mathbb{0})$ is a commutative group (where we usually denote the inverse of $a \in B$ with respect to $\oplus$ by $-a$),
  \item \emph{semifield} (or: \emph{division semiring}) if  $(B\setminus\{\mathbb{0}\},\otimes,\mathbb{1})$ is a group (where we usually denote the inverse of $a \in B$ with respect to $\otimes$ by $a^{-1}$), and
\item \emph{field} if it is a ring and a commutative semifield.
\end{compactitem}

Next we give a number of examples. We start with examples of semirings because they are more familiar than strong bimonoids.

\begin{example}\label{ex:semirings}\rm Here we show examples of semirings, rings, semifields, and fields. If an example shows a $\sigma$-complete semiring, i.e., there exists a mapping $\infsum{\oplus}{}{}$ which satisfies equalities \eqref{axiom-i-sum}, \eqref{axiom-ii-sum}, and \eqref{axiom-iii-sum}, then we give such a mapping.
\begin{enumerate}
\item 
  \index{Booleansemiring@$\Boole$}
  \index{Boolean semiring}
The \emph{Boolean semiring} $\Boole= (\mathbb{B},\vee,\wedge,0,1)$,
where $\mathbb{B}=\{0,1\}$ (the truth values) and $\vee$ and $\wedge$ denote disjunction and conjunction, respectively. The Boolean semiring $\Boole$ is $\sigma$-complete with the mapping 
\begin{align*}
  \infsum{\vee}{I}{}: \mathbb{B}^I \to \mathbb{B} \ \ \text{ with } \ \ 
  (b_i \mid i \in I) \mapsto \begin{cases} 1 & \text{if there exists $i \in I$ such that $b_i = 1$}\\
    0 &  \text{otherwise}\enspace.\end{cases}
\end{align*}
We note that $\Boole$ is a semifield.

\item \index{Nat@$\Nat$}
The semiring  $\Nat=(\mathbb{N},+,\cdot,0,1)$ of natural numbers, where + and $\cdot$ are the usual addition and multiplication. 

\item 
The semiring  $\Nat_\infty=(\mathbb{N}_\infty,+,\cdot,0,1)$ of natural numbers, where + and $\cdot$ are extended to $\mathbb{N}_\infty$ in the natural way. The semiring $\Nat_\infty$ is $\sigma$-complete with the mapping \index{Natinft@$\Nat_\infty$}
\begin{align*}
  \infsum{+}{I}{}: (\mathbb{N_\infty})^I \to \mathbb{N_\infty} \ \ \text{ with } \ \ 
  (n_i \mid i \in I) &\mapsto \begin{cases} \bigplus_{i \in J} n_j    &\text{if $\{n_i \mid i \in I\} \subseteq \mathbb{N}$ and }\\
    & \text{$J = \{i \in I \mid n_i\ne 0\}$ is finite}\\
     \infty &  \text{otherwise}\enspace.\end{cases}
  \end{align*}
(We recall that $\bigplus$ denotes the extension of $+$ to finite sums in the monoid $(\mathbb{N},+,0)$.)

\index{Int@$\Int$} 
\item  The ring  $\Int=(\mathbb{Z},+,\cdot,0,1)$ of integers.

  \index{Intmodfour@$\Intfour$}
    \item \label{def:ring-Zmod4Z} The ring $\Intfour = (\{0,1,2,3\},+_4,\cdot_4,0,1)$ where $+_4$ and $\cdot_4$ are the usual addition modulo 4 and the usual multiplication modulo 4, respectively.

      \index{Rat@$\Ratnum$}
      \index{Realnum@$\Realnum$}
    \item The field $\Ratnum=(\mathbb{Q},+,\cdot,0,1)$ of rational numbers and the field $\Realnum=(\mathbb{R},+,\cdot,0,1)$ of real numbers.

      \index{Ftwo@$\sfFtwo$}
                \item  \label{ex:two-two-element-sr}
 There exist only two semirings $(\{0,1\},\oplus,\otimes,0,1)$  with exactly two
elements because, for every $a,b\in \{0,1\}$ and $\odot \in \{\oplus,\otimes\}$, the  value of $a \odot b$ is determined by the strong bimonoid axioms, except the value of  $1\oplus 1$.
\begin{compactenum}
  \item If we define $1\oplus 1 = 0$, then $(\{0,1\},\oplus,\otimes,0,1)$ is a field, denoted by $\sfFtwo$; we have $-0 = 0$ and $-1 = 1^{-1} = 1$.
\item If we define $1\oplus 1 = 1$, then $(\{0,1\},\oplus,\otimes,0,1)$ is the Boolean semiring $\Boole$.
\end{compactenum}

\index{Natmaxplus@$\Natmaxplus$}
\index{arctic semiring}
\index{max-plus semiring}
\item The \emph{arctic semiring} $\Natmaxplus =(\mathbb{N}_{-\infty},\max,+,-\infty,0)$. The arctic semiring is often called max-plus semiring.

\index{Natmaxplusn@$\Natmaxplusn$}
\item \label{ex:Nat-max-plus-n}
 The semiring $\Natmaxplusn=([0,n]_{-\infty},\max,\hat{+}_n,-\infty,0)$ where $n \in \mathbb{N}_+$, $[0,n]_{-\infty} = [0,n] \cup \{-\infty\}$, and $b_1 \hat{+}_n b_2 = \min(b_1+b_2,n)$ for every $b_1,b_2 \in [0,n]_{-\infty}$ \cite[Ex.~1.8]{gol99}. The semiring $\Natmaxplusn$ is $\sigma$-complete with the mapping
\begin{align*}
  \infsum{\max}{I}{}: ([0,n]_{-\infty})^I \to [0,n]_{-\infty} \ \ \text{ with } \ \ 
  (n_i \mid i \in I) \mapsto \sup(n_i \mid i \in I) \enspace.
\end{align*}

\item\index{tropical semiring}
\index{Natminplus@$\Natminplus$}
\label{def:Trop}  The \emph{tropical semiring} $\Natminplus =(\mathbb{N}_\infty,\min,+,\infty,0)$ \emph{over $\mathbb{N}$}. The tropical semiring $\Natminplus$ is $\sigma$-complete with the mapping
\begin{align*}
  \infsum{\min}{I}{}: (\mathbb{N}_\infty)^I \to \mathbb{N}_\infty \ \ \text{ with } \ \ 
  (n_i \mid i \in I) \mapsto \inf(n_i \mid i \in I) \enspace.
\end{align*}
\index{min-plus semiring}
The tropical semiring is often called the min-plus semiring.

\index{Natmaxmin@$\Natmaxmin$}
\item \index{max-min semiring}\label{def:max-min-semiring} The semiring $\Natmaxmin = (\mathbb{N}_\infty,\max,\min,0,\infty)$. It is $\sigma$-complete with the mapping
\begin{align*}
  \infsum{\max}{I}{}: (\mathbb{N_\infty})^I \to \mathbb{N_\infty} \ \ \text{ with } \ \ 
  (n_i \mid i \in I) &\mapsto \begin{cases} \sup(n_i \mid i \in I)    &\text{if $\{n_i \mid i \in I\} \subseteq \mathbb{N}$ and }\\
    & \text{it is finite}\\
     \infty &  \text{otherwise}\enspace.\end{cases}
  \end{align*}

  \index{Lang@$\mathsf{Lang}_\Gamma$}
  \index{language semiring}
\item  
  \label{def:formal-lang-semiring}
The \emph{semiring of formal languages}  $\mathsf{Lang}_\Gamma=({\cal P}(\Gamma^*),\cup,\cdot,\emptyset,\{\varepsilon\})$ where $\cdot$ denotes the concatenation of languages. The semiring $\mathsf{Lang}_\Gamma$ is $\sigma$-complete with the mapping
\begin{align*}
  \infsum{\cup}{I}{}: {\cal P}(\Gamma^*)^I \to {\cal P}(\Gamma^*) \ \ \text{ with } \ \ 
  (L_i \mid i \in I) \mapsto \bigcup_{i \in I} L_i \enspace.
\end{align*}

\item
  \label{def:Viterbi-semiring}
  The \emph{Viterbi semiring} $\Viterbi = ([0,1],\max,\cdot,0,1)$ where $[0,1]$ denotes the set $\{r \in \mathbb{R} \mid 0 \le r \le 1\}$ of real numbers\footnote{not to be confused with the set $[0,1]=\{0,1\}$ of natural numbers.}.\index{Viterbi semiring}
\index{Viterbi@$\Viterbi$}
    The Viterbi semiring  is $\sigma$-complete with the  mapping
\begin{align*}
  \infsum{\max}{I}{}: [0,1]^I \to [0,1] \ \ \text{ with } \ \ 
  (r_i \mid i \in I) \mapsto \sup(r_i \mid i \in I) \enspace.
\end{align*}
The Viterbi semiring can be used for calculations with probabilities,
and it is isomorphic to  $(\mathbb{R}_{\ge 0,\infty}, \min, +, \infty, 0)$ where $\mathbb{R}_{\ge 0,\infty}$ denotes the set $\{r \in \mathbb{R} \mid r \ge 0\} \cup \{\infty\}$. The isomorphism  from  $[0,1]$  to  $[0,\infty]$  is  $x \mapsto -\ln(x)$.

    \item
\index{tropical semifield}
\index{Intminplus@$\Intminplus$}
\label{def:TropQ}  The \emph{tropical semifield} $\Intminplus =(\mathbb{Z}_\infty,\min,+,\infty,0)$ \emph{over $\mathbb{Z}$}, and correspondingly the \emph{tropical semifield} $\Ratminplus =(\mathbb{Q}_\infty,\min,+,\infty,0)$ \emph{over $\mathbb{Q}$}.

\index{Ratnumgreaterzero@$\Ratnumgreaterzero$}
\index{Realnumgreaterzero@$\Realnumgreaterzero$}
\item The semifields $\Ratnumgreaterzero= (\mathbb{Q}_{\ge 0},+,\cdot,0,1)$ and $\Realnumgreaterzero= (\mathbb{R}_{\ge 0},+,\cdot,0,1)$.

  \index{log semifield}
\item The \emph{log semifield}  $(\mathbb{R}_{-\infty},\oplus,+,-\infty,0)$  with $x\oplus y= \ln(e^x+e^y)$.

  \index{Sem@$\Sem(\B)$}
\item \label{def:Sem(B)} Let $\B=(B,\otimes,\1)$ be a monoid. In the canonical way, we extend the operation $\otimes$ to finite sets of elements of $B$, i.e., for every $A_1,A_2 \in \cP_{\mathrm{fin}}(B)$ we let $A_1 \otimes A_2 = \{a\otimes b \mid a \in A_1, b \in A_2\}$. Then the algebraic structure  $(\cP_{\mathrm{fin}}(B),\cup,\otimes,\emptyset,\{\1\})$ is a semiring. We denote it by $\Sem(\B)$.

\index{Powerset@$\Powerset_A$}
\item \label{def:semiring-of-posets} Let $A$ be a set. Then $\Powerset_A=(\cP(A),\cup,\cap,\emptyset,A)$ is a semiring. It is $\sigma$-complete with the mapping
  \begin{align*}
  \infsum{\cup}{I}{}: {\cal P}(A)^I \to {\cal P}(A) \ \ \text{ with } \ \ 
  (A_i \mid i \in I) \mapsto \bigcup_{i \in I} A_i \enspace.
  \end{align*}
  Moreover, the monoid $(\cP(A),\cap,A)$ is $\sigma$-complete with the mapping
    \begin{align*}
  \infsum{\cap}{I}{}: {\cal P}(A)^I \to {\cal P}(A) \ \ \text{ with } \ \ 
  (A_i \mid i \in I) \mapsto \bigcap_{i \in I} A_i \enspace.
    \end{align*}
    In the literature (e.g., \cite[p.~4]{gra68}), often the expressions $\sum^\cup_I(A_i \mid i \in I)$ and  $\infsum{\cup}{i \in I}{A_i}$ are written as $\bigcup(A_i \mid i \in I)$. Similarly, $\sum^\cap_I(A_i \mid i \in I)$ and  $\infsum{\cap}{i \in I}{A_i}$ are written as $\bigcap(A_i \mid i \in I)$. In this book, we will also use these notations from the literature.

  \item \label{def:semiring-of-polynomials}\index{Bx@$\B[x_1,\ldots,x_k]$}
    Let $\B = (B,\oplus,\otimes,\0,\1)$ be a semiring and $X_k = \{x_1,\ldots,x_k\}$ for some $k \in \mathbb{N}$. Then
  \[
    \B[x_1,\ldots,x_k] = (B[x_1,\ldots,x_k],\oplus,\cdot,p_0,p_1)
  \]
    is a semiring, called \emph{polynomial semiring (over $k$ variables with coefficients in $B$)}, where $B[x_1,\ldots,x_k]$ is the set of all mappings $p:X_k^*\to B$
    such that the set $\{w \in X_k^* \mid p(w) \not=\0\}$ is finite.
 Each element $p \in B[x_1,\ldots,x_k]$ is called \emph{polynomial (over $k$ variables)}. If $|\{w \in X_k^* \mid p(w) \not=\0\}|\le 1$, then $p$ is called a \emph{monomial}.

Let $p$ and  $q$ be polynomials in $B[x_1,\ldots,x_k]$. We define the \emph{summation of $p$ and $q$}, denoted by $p \oplus q$, for each $u\in X_k^*$, by
 \[
   (p\oplus q)(u) = p(u)\oplus q(u) \enspace.
   \]

   \index{Cauchy product}
We define the \emph{Cauchy product of $p$ and $q$}, denoted by $p \cdot q$, for each $u\in X_k^*$, by 
 \[
   (p\cdot q)(u) = \bigoplus_{\substack{v,w \in X_k^*:\\u=vw}} p(v) \otimes q(w) \enspace.
   \]

   Finally, the polynomials $p_0$ and $p_1$ are defined by
   \begin{compactitem}
   \item $p_0(w)=\0$ for each $w \in X_k^*$ and
   \item $p_1(w)=\1$ if $w=\varepsilon$ and $\0$ otherwise.
   \end{compactitem}
   If $\B$ is commutative, then also $\B[x_1,\ldots,x_k]$ is commutative, and if $\B$ is a ring, then also $\B[x_1,\ldots,x_k]$ is a ring.

   \index{Nx@$\Nat[x_1,\ldots,x_k]$}
   (a) By taking $\B = \Nat$, we obtain the polynomial semiring $\Nat[x_1,\ldots,x_k]$ over $k$ variables with coefficients in $\mathbb{N}$. The semiring $\Nat[x_1,\ldots,x_k]$ is free in the set of all semirings with generating set $X_k$ (cf. e.g. \cite[Ex.~3 on p.166]{wec92}). Clearly, $\Nat[x_1,\ldots,x_k]$ is generated by $X_k$. Moreover, let $\B = (B,\oplus,\otimes,\0,\1)$ be a semiring and $f: X_k \to B$ a mapping. Then $f$ is extended in a unique way to a semiring homomorphism $\widetilde{f}: \mathbb{N}[x_1,\ldots,x_k] \to B$ by letting, for each $p \in \mathbb{N}[x_1,\ldots,x_k]$,
     \[
    \widetilde{f}(p) = \bigoplus_{\substack{w \in X_k^*:\\ p(w) \not= \0}} p(w) \overline{f}(w)
    \]
where $n b = b \oplus \cdots \oplus b$ with $n$ occurrences of $b$, and $\overline{f}$ is the unique extension of $f$ to a monoid homomorphism from the free monoid $(X_k^*,\circ,\varepsilon)$ with generating set $X_k$ to $(B,\otimes,\1)$.

(b) For the special case that $k=1$, we identify $X_1^*$ with $\mathbb{N}$, and we abbreviate $x_1$ by $x$. Then,
 \[
    \B[x] = (B[x],\oplus,\cdot,p_0,p_1)
  \]
is the polynomial semiring over one variable, and each element of $B[x]$ is a mapping $p:\mathbb{N}\to B$. 
   We can denote such a polynomial by $a_n.x^n \oplus \ldots \oplus a_1.x^1 \oplus a_0$ where $n = \max\{k \in \mathbb{N} \mid p(k) \not= \0\}$ and $a_i =p(i)$ for each $i\in[0,n]$.

   The Cauchy product of two polynomials $p$ and $q$ in $\B[x]$ is defined, for each $i\in\mathbb{N}$, by
  \[
    (p\cdot q)(i) =\bigoplus_{j \in [0,i]}p(j) \otimes q(i-j) \enspace.
  \]
          \hfill $\Box$
\end{enumerate}
\end{example}

In the next example, we show a number of strong bimonoids which are not semirings, except the ones given in  Examples \ref{ex:0-1-conorm-norm}, \ref{ex:sb-not-sr}, and \ref{ex:combining-comm-semigroup-semigroup}. For further examples of strong bimonoids, see Observation \ref{obs:bounded-lattice-is-biloc-fin-strong-bimonoid}, where we show that each bounded lattice is a strong bimonoid, as well as the proof of Theorems \ref{thm:comparison-support-theorem-run} and \ref{thm:comparison-support-theorem-initial-algebra}.

\begin{example}\label{ex:strong-bimonoids} \rm In \cite[Ex.~1]{drostuvog10},\cite[Ex.~2.2]{cirdroignvog10}, \cite[Ex.~2.1]{drovog12}, and \cite[Ex.~2.2,~2.3]{drofulkosvog21} a number of examples of  strong bimonoids are given, and in \cite[Ex.~2.1.4]{drovog12} a general construction principle for strong bimonoids  is given, which we recall here. The first two examples are $\sigma$-complete strong bimonoids, and we show the mappings $\infsum{\oplus}{I}{}$ which satisfy equalities \eqref{axiom-i-sum} and \eqref{axiom-ii-sum}.
  \begin{enumerate}
    \item  \index{tropical bimonoid}\label{ex:tropical-bimonoid}
\index{Trop@$\TropBM$}
The algebra $\TropBM = (\mathbb{N}_\infty,+,\min,0,\infty)$ is a commutative strong bimonoid, called the \emph{tropical bimonoid}. However,  it is not bi-locally finite.  Moreover, it is not a semiring, because there exist $a,b,c\in\mathbb N_\infty$ with $\min(a,b+c) \not= \min(a,b) + \min(a,c)$ (e.g., take
$a = b = c\not = 0$).
The strong bimonoid $\TropBM$ is $\sigma$-complete with the mapping
\begin{align*}
  \infsum{+}{I}{}: (\mathbb{N_\infty})^I \to \mathbb{N_\infty} \ \ \text{ with } \ \ 
  (n_i \mid i \in I) &\mapsto \begin{cases} \bigplus_{i \in J} n_j    &\text{if $\{n_i \mid i \in I\} \subseteq \mathbb{N}$ and }\\
    & \text{$J = \{i \in I \mid n_i\ne 0\}$ is finite}\\
     \infty &  \text{otherwise}\enspace.\end{cases}
  \end{align*}

  \index{Trunc@$\Trunc_\lambda$}
\item \label{ex:interval-strong-bimonoid} For each $\lambda \in \mathbb{R}$ with $0 < \lambda< \frac{1}{2}$,  let $\Trunc_\lambda= (B,\oplus,\odot,0,1)$ be the algebra, where
  \begin{compactitem}
    \item $B =\{0\} \cup \{b \in \mathbb{R}\mid \lambda \le b \le 1\}$,
    \item $a \oplus b = \min(a + b, 1)$,  and 
    \item $a\odot b  = a \cdot b$ if $a \cdot b \ge \lambda$, and $0$ otherwise,
    \end{compactitem}
    and where $+$ and $\cdot$ are the usual addition and multiplication of real numbers, respectively.
    Then $\Trunc_\lambda$ is a bi-locally finite and commutative strong bimonoid which can be proved as follows.
    
First we prove that $(B,\oplus,0)$ is locally finite.  By induction on $\mathbb N$, it is easy to show that for every 
 $n\in \mathbb N$ with $n\ge 2$ and  $a_1,\ldots,a_n\in B$, we have
 \begin{equation}\label{eq:oplus-extension}
 \bigoplus_{i \in [n]} a_i =\min(\bigplus_{i \in [n]} a_i,1)\enspace.
    \end{equation}
Now let $A\subseteq B$ be a finite subset. If  $A= \emptyset$ or $A= \{0\}$, then $\langle A\rangle_{\{\oplus,0\}}=\{0\}$, i.e., a finite set. Otherwise, let $m=\min(A\setminus \{0\})$. Then
\begin{equation}\label{eq:oplus=1}
\text{for every $n> \lfloor{\scriptstyle\frac{1}{m}} \rfloor$ and $a_1,\ldots,a_n \in A\setminus \{0\}$, we have  $\bigoplus_{i \in [n]} a_i=1$.}
 \end{equation}
 This follows from  \eqref{eq:oplus-extension} using the approximation
 $\bigplus_{i \in [n]} a_i \ge \bigplus_{i \in [n]} m > 1$.
 Lastly, we obtain that $\langle A\rangle_{\{\oplus,0\}}$ is finite also in this case, because
\begin{align*}
\langle A\rangle_{\{\oplus,0\}}= & \{0\}\cup A \cup \big\{\bigoplus_{i \in [n]} a_i  \mid 2 \le n, a_1,\ldots,a_n \in A\setminus \{0\}\big\} \\
= & \{0,1\}\cup A \cup \big\{\bigoplus_{i \in [n]} a_i  \mid 2\le n \le \lfloor{\scriptstyle\frac{1}{m}} \rfloor, a_1,\ldots,a_n \in A\setminus \{0\}\big\}.\tag{by \eqref{eq:oplus=1}}
\end{align*}

 Next we prove that $(B,\odot,1)$ is locally finite.  By induction on $\mathbb N$, it is easy to show that for every 
 $n\in \mathbb N$ with $n\ge 2$ and  $a_1,\ldots,a_n\in B$, we have
 \begin{equation}\label{eq:odot-extension}
 \text{$\bigodot_{i \in [n]} a_i =\prod_{i \in [n]} a_i$ if  $\prod_{i \in [n]} a_i\ge \lambda$ and 0 otherwise,}
\end{equation}
    where we abbreviate $a_1\cdot\ldots\cdot a_n$ by $\prod_{i \in [n]} a_i$.

    Now let $A\subseteq B$ be a finite subset. If  $A\subseteq \{0,1\}$, then $\langle A\rangle_{\{\odot,1\}}\subseteq\{0,1\}$, hence it is finite.
    Otherwise, let $m=\max(A\setminus \{0,1\})$ and $K=\max(\{k\in \mathbb{N}\mid m^k \ge \lambda\})$. Then
\begin{equation}\label{eq:odot=0}
\text{for every $n> K$ and $a_1,\ldots,a_n \in A\setminus \{0,1\}$, we have  $\bigodot_{i \in [n]} a_i=0$.}
 \end{equation}
 This follows from  \eqref{eq:odot-extension} using the approximation
 $\prod_{i \in [n]} a_i \le \prod_{i \in [n]} m < \lambda$.
 Lastly, we obtain that $\langle A\rangle_{\{\odot,1\}}$ is finite also in this case, because
\begin{align*}
\langle A\rangle_{\{\odot,1\}}= & \ \{1\}\cup A \cup \big\{\bigodot_{i \in [n]} a_i  \mid 2 \le n, a_1,\ldots,a_n \in A\setminus \{0,1\}\big\} \\
= & \ \{1\}\cup A \cup \big\{\bigodot_{i \in [n]} a_i  \mid 2\le n \le K, a_1,\ldots,a_n \in A\setminus \{0\}\big\}.\tag{by \eqref{eq:odot=0}}
\end{align*}

 $\Trunc_\lambda$ is  $\sigma$-complete with the mapping:
\begin{align*}
  \infsum{\oplus}{I}{}: B^I \to B \ \ \text{ with } \ \ 
  (n_i \mid i \in I) &\mapsto \begin{cases} \min\big(\bigoplus_{i \in J} n_j,1\big)     &\text{if $J = \{i \in I \mid n_i\ne 0\}$ is finite}\\
     1 &  \text{otherwise}\enspace.\end{cases}
  \end{align*}
We note that $\Trunc_\lambda$ is not a semiring because $\odot$ is not right-distributive. For instance, for $a=b=0.9$, and $c=\lambda$, we have $(a \oplus b) \odot c=\lambda$, while $(a \odot c) \oplus (b \odot c)=0$ because $a \odot c = b \odot c = 0$.

If $\lambda = \frac{1}{4}$, then $\Trunc_\lambda$ is not weakly locally finite,  because the set $\CL(\{\frac{1}{2}\})$ is infinite. We can show this as follows \cite{dro19}.
Let  $(b_i \mid i \in \mathbb{N})$ such that
\[
b_i = \begin{cases}
  \frac{1}{2} & \text{ if $i =0$}\\
  b_{i-1} \cdot \frac{1}{2}   & \text{ if $i$ is odd}\\
  b_{i-1} + \frac{1}{2} & \text{ if $i$ is even and $i\not=0$}
  \end{cases}
\]
  Then, e.g., $b_0= 1/2$, $b_1=1/4$, $b_2= 1/2+1/4 = 3/4$, $b_3= 3/8$, $b_4= 1/2+3/8 = 7/8$, $b_5= 7/16$, $b_6 = 1/2+7/16 = 15/16$, $b_7 = 15/32$,
$b_8=1/2+15/32 = 31/32$, $b_9=31/64$, $b_{10}= 1/2+31/64 = 63/64$, etc. (In fact, the subsequences $(b_i \mid i \in \mathbb{N}, i \text{ is even})$ and  $(b_i \mid i \in \mathbb{N}, i \text{ is odd})$ converge to $1$ and $\frac{1}{2}$, respectively.)  It is easy to see that $b_i \in \CL(\{\frac{1}{2}\})$ for each $i \in \mathbb{N}$, and that $b_i \not= b_j$ for every $i,j \in \mathbb{N}$ with $i\not=j$. Hence $\CL(\{\frac{1}{2}\})$ is an infinite set and thus $\Trunc_\lambda$ is not weakly locally finite (also cf \cite{tep23}).

  \index{UnitIntfuzzy@$\UnitIntfuzzy_{u,i}$}
\item \label{ex:0-1-conorm-norm} Let $[0,1]=\{r \in \mathbb{R} \mid  0\le r\le 1\}$ and let $i$ be a binary operation on $[0,1]$. We say that $i$ is a \emph{t-norm} (or: \emph{fuzzy intersection}, cf. \cite[p.~62]{kliyua95}) if $i$ satisfies the following conditions: 
  \begin{compactenum}
  \item[(a)] $i$ is commutative and associative,
    \item[(b)] $i(a,1)=a$ for each $a \in [0,1]$ (boundary condition), and
  \item[(c)] $a \le b$ implies $i(c,a) \le i(c,b)$ for every $a,b,c \in [0,1]$ (monotonicity condition).
  \end{compactenum}
  Moreover, let $u$ be a binary operation on $[0,1]$. We say that $u$ is a \emph{t-conorm} (or: \emph{fuzzy union}, cf. \cite[p.~77]{kliyua95}) if $u$ satisfies the the following conditions:
  \begin{compactenum}
  \item[(a)] $u$ is commutative and associative,
    \item[(b)] $u(a,0)=a$ for each $a \in [0,1]$ (boundary condition), and
  \item[(c)]  $a \le b$ implies $u(c,a) \le u(c,b)$ for every $a,b,c \in [0,1]$ (monotonicity condition).
  \end{compactenum}
  Due to the boundary conditions, $([0,1],u,0)$ and $([0,1],i,1)$ are commutative monoids. Since $i(a,0) = i(0,a) \leq i(0,1) = 0$ for each $a \in [0,1]$ (cf. \cite[Thm.~3.10]{kliyua95}), we have that
  \[
    \UnitIntfuzzy_{u,i}= ([0,1],u,i,0,1)
  \]
  is a commutative strong bimonoid for each t-conorm $u$ and each t-norm~$i$.

  There are combinations of t-conorm $u$ and t-norm $i$ such that the strong bimonoid $\UnitIntfuzzy_{u,i}$ is not distributive. This is the case, e.g., for the t-conorm bounded sum (cf. Example \ref{ex:strong-bimonoids}(\ref{ex:0-1-strong-bimonoids})) and the t-norm \emph{bounded difference}
 $i(a,b) = \max(0,a+b-1)$. 
We also refer to \cite[Thm.~3.24]{kliyua95} for sufficient conditions under which $\UnitIntfuzzy_{u,i}$ is not distributive.
  
\index{UnitIntboundedsum@$\UnitIntboundedsum$}
\index{UnitIntalg@$\UnitIntalg$}
\item \label{ex:0-1-strong-bimonoids}
   The algebra $([0,1],\oplus,\cdot,0,1)$ with interval $[0,1]=\{r \in \mathbb{R} \mid  0\le r\le 1\}$ of real numbers and  the usual multiplication $\cdot$ of real numbers is a strong bimonoid for each of the following two definitions of $\oplus$ for every $a,b\in[0,1]$:
\index{algebraic sum}
\index{bounded sum}
\begin{compactitem}
\item $a \oplus b = a +b - a\cdot b$ (called \emph{algebraic sum}) and
\item $a \oplus b = \min(a+b,1)$ (called \emph{bounded sum}).
\end{compactitem}
Both the algebraic sum and the bounded sum are t-conorm (cf. \cite[p.~78]{kliyua95}), hence they are associative. If $\oplus$ is the algebraic sum, then we denote the algebra by $\UnitIntalg$; 
if $\oplus$ is the bounded sum, then we denote the algebra by $\UnitIntboundedsum$. Neither
$\UnitIntalg$ nor $\UnitIntboundedsum$ is a  semiring.

  \index{near semiring}
  \index{NearSem@$\NearSem$}
\item \label{ex:near-semiring}Let $(C,+,0)$ be a commutative monoid. We consider the set $B$ of all mappings $f: C \to C$ such that $f(0)=0$. Moreover, we extend $+$ to $B$ by a pointwise addition on elements of $B$, i.e., for every  $f,g \in B$ and $c \in C$, we define $(f+g)(c) = f(c) + g(c)$. Also, we define the operation $\diamond$ on $B$ such that, for every  $f,g \in B$ and $c \in C$, we have $(f \diamond g)(c) = g(f(c))$. Finally, we denote by $\tilde{0}$ the mapping $\tilde{0}: C \to C$ such that $\tilde{0}(c) = 0$ for each $c \in C$. Then
  \[
    \NearSem_C = (B,+,\diamond,\tilde{0},\id_C)
  \]
is a strong bimonoid. Such an algebra is called a \emph{near semiring (over $C$)} \cite{hooooz67,kri05}. We note that the condition $f(0)=0$ is needed to guarantee that $\tilde{0} \diamond f = \tilde{0}$. Except for trivial cases, the operation $\diamond$ is left-distributive over $+$, but not right-distributive.

As example we consider the commutative monoid $(\cP(Q),\cup,\emptyset)$ for some set $Q$. Then we consider the near semiring $(B,\cup,\diamond,\widetilde{0},\id_{\cP(Q)})$ as described above, i.e.,
\begin{compactitem}
\item $B=\{f \mid f: \cP(Q) \to \cP(Q), f(\emptyset) = \emptyset\}$,
\item $(f\cup g)(U) = f(U) \cup g(U)$ and  $(f \diamond g)(U) = g(f(U))$ for every $f,g \in B$ and $U \in \cP(Q)$, and
  \item $\widetilde{0}(U) = \emptyset$ and $\id_{\cP(Q)}(U)=U$ for each $U  \in \cP(Q)$.
  \end{compactitem}
It is easy to see that the operation $\diamond$ is left-distributive over $\cup$. However, it is not right-distributive over $\cup$ if $|Q|\ge 2$. To show this, let $p,q\in Q$ with $p\ne q$ and let $f,g,h \in B$ such that for every $U \in \cP(Q)$:
\[
f(U) = \begin{cases} \{p\} &\text{ if } U=\{p\} \\\emptyset & \text{otherwise}\end{cases} \enspace,\
g(U) = \begin{cases} \{q\} &\text{ if } U=\{p\} \\\emptyset & \text{otherwise}\end{cases} \enspace, \text{ and } \
h(U) = \begin{cases} \emptyset &\text{ if } |U|=2 \\ U & \text{otherwise.}\end{cases}
  \]
Then $((f\cup g)\diamond h )(\{p\}) =h(\{p,q\})=\emptyset$ and $((f\diamond h) \cup (g\diamond h))(\{p\}) = h(\{p\}) \cup h(\{q\}) = \{p,q\}$.

\item  We consider the strong bimonoid
$(\Gamma^* \cup \{\infty\},\wedge,\cdot,\infty,\varepsilon)$ where
\begin{compactitem}
  \item $\wedge$ is the longest common prefix operation,
  \item $\cdot$ is the usual concatenation of strings, and
  \item $\infty$ is a new element such that $s \wedge \infty = \infty \wedge s = s$ and $s \cdot \infty =
\infty \cdot s = \infty$  for each  $s \in \Gamma^*\cup
\{\infty\}$.
\end{compactitem}
This bimonoid occurs in investigations for natural
language processing, see~\cite{moh00a}. It is clear that $(\Gamma^* \cup
\{\infty\},\wedge,\cdot,\infty,\varepsilon)$ is left-distributive but not right-distributive (consider, e.g., if $\Gamma = \{a,b,c\}$, then  $abc = (a \wedge ab) \cdot bc \not= (a \cdot bc) \wedge (ab \cdot bc) = ab$).

\item \label{ex:sb-not-sr}
  \index{Boolean semiring}
  \index{Three@$\Three$}
 There exist only two strong bimonoids $(\{0,1\},\oplus,\otimes,0,1)$  with exactly two
elements: the Boolean semiring $\Boole$ and the field $\sfFtwo$; in particular, both are semirings (cf. Example \ref{ex:semirings}(\ref{ex:two-two-element-sr})).
However, there exist strong bimonoids with three elements which are not semirings,
take, e.g., $\Three= (\{0,1,2\},\max,\hat\cdot, 0, 1)$ where $a\,\hat\cdot\, b
= (a\cdot b)\bmod 3$ for every $a,b\in\{0,1,2\}$; there $\max(2 \,\hat\cdot\, 2, 2 \,\hat\cdot \,1) \ne 2 \,\hat\cdot\, \max(1,2)$.

\index{PPN@$\PP_{\mathbb{N}}$}
\index{plus-plus strong bimonoid}
\item \label{ex:plus-plus-sb} We consider the algebra $(\mathbb{N}_\0,\oplus,+,\0,0)$ from \cite[Ex.~2.3]{drofulkosvog21}, where
  where $\mathbb{N}_\0 = \mathbb{N} \cup \{ \0 \}$ for some new element $\0 \not\in \mathbb{N}$. The binary operation $\oplus$, if restricted to $\mathbb{N}$, and the binary operation $+$, if restricted to  $\mathbb{N}$,  are the usual addition on natural numbers (e.g. $3 + 2 = 5$). Moreover,
  $\0 \oplus x = x \oplus \0 = x$  and  $\0 + x = x + \0 = \0$  for each $x \in \mathbb{N}_\0$. Thus, $(\mathbb{N}_\0,\oplus,+,\0,0)$ is a strong bimonoid. However, it is not a semiring (e.g., $2+(3\oplus 4) \ne (2+3)\oplus(2+4)$).  We might call this algebra the \emph{plus-plus strong bimonoid of natural numbers} and denote it by $\PP_{\mathbb{N}}$.

\item  \label{ex:Stb-strong-bimonoid} \index{Stb@$\mathsf{Stb}$} We recall the strong bimonoid $\mathsf{Stb}=(\mathbb{N},\oplus,\odot,0,1)$ from \cite[Ex.~25]{drostuvog10}\footnote{$\mathsf{Stb}$ refers to one of the authors}. Intuitively, both operations are commutative and consider their maximal argument, say, $b$. Then, depending on the characteristic of $b$ being even or odd, $\oplus$ delivers $b$ or $b+1$, respectively, and dually, $\odot$ delivers $b+1$ or $b$, respectively.

    Formally, the two commutative operations $\oplus$ and $\odot$ on
        $\mathbb N$ are defined  as follows. First, let $0\oplus a=a$, $0\odot
        a=0$, and $1\odot a = a$ for every $a\in\mathbb N$. If $a,b\in
        \mathbb N\setminus\{0\}$ with $a\leq b$, we put (with $+$
        being the usual addition on $\mathbb{N}$) 
	\begin{align*}
		a\oplus b &=
		\begin{cases}
			b & \text{if $b$ is even}\\
			b+1 & \text{if $b$ is odd.} 
		\end{cases}
	\intertext{If $a,b\in\mathbb N\setminus\{0,1\}$ with $a\leq b$, let}
		a\odot b &=
		\begin{cases}
			b+1 &\text{if $b$ is even}\\
			b & \text{if $b$ is odd.} 
		\end{cases}
	\end{align*}

	We prove that $\oplus$ is associative, i.e., that for every $a,b,c\in \mathbb N$ we have
	\begin{equation}\label{eq:sum-associative}
	a\oplus (b\oplus c)= (a\oplus b)\oplus c\enspace.
	\end{equation}
Since this is obvious if at least one of $a,b$, or $c$ is $0$, we assume that none of them is 0. For each $d\in \{a,b,c\}$, we abbreviate by $d\equiv 0$ that $d$ is an even number and by $d\equiv 1$ that $d$ is odd. Then we can verify that \eqref{eq:sum-associative} holds by making the following case analysis.
\begin{center}
\begin{tabular}{l|l}
  if ... & then both sides of \eqref{eq:sum-associative}\\
  &  are equal to ... \\\hline
$(a \le b) \wedge (b\le c) \wedge (b \equiv 0) \wedge (c \equiv 0)$  &  $c$ \\
$(a \le b) \wedge (b\le c) \wedge (b \equiv 0) \wedge (c \equiv 1)$  & $c+1$ \\
$(a \le b) \wedge (b+1\le c) \wedge (b \equiv 1) \wedge (c \equiv 0)$ \;\;  & $c$ \\
$(a \le b) \wedge (b+1\le c) \wedge (b \equiv 1) \wedge (c \equiv 1)$  & $c+1$ \\
$(a \le b) \wedge (c < b)\wedge (b \equiv 0)$  & $b$ \\
$(a \le b) \wedge (c < b)\wedge (b \equiv 1)$ & $b+1$ \\
etc.
\end{tabular}
\end{center}
Similarly, we can show that $\odot$ associative. In that proof we can assume that none of $a,b$, or $c$ is equal to $0$ or $1$ because otherwise the statement is trivial.

      Then $\mathsf{Stb}=(\mathbb N,\oplus,\odot, 0, 1)$ is a strong bimonoid.  Clearly, it is not a semiring because, e.g., $2 \odot (2 \oplus 3) = 2 \odot 4 = 5$ and $(2\odot 2) \oplus (2\odot 3) = 3 \oplus 3 = 4$.
        It is easy to see that $\mathsf{Stb}$ is bi-locally finite. 
      For this, we show that $(\mathbb N,\oplus,0)$ is locally finite as follows. First, by induction on $\mathbb N$, we show that for every 
 $n\in \mathbb N$ with $n\ge 2$ and  $a_1,\ldots,a_n\in \mathbb{N}_+$ with $a_1\le \ldots\le a_n$, we have
 \begin{equation}\label{eq:sum-biloc-finite}
 \bigoplus_{i \in [n]} a_i =\begin{cases}
 a_n & \text{ if $a_n$ is even}\\
 a_{n}+1 & \text{ otherwise.}
 \end{cases}
   \end{equation}
  I.B.: Let $n=2$. Then the statement follows by the definition of $\oplus$.
  
  I.S.: Next let $n>2$. 
  By I.H., we have that $\bigoplus_{i \in [n-1]} a_i=a_{n-1}$ if $a_{n-1}$ is even, and $a_{n-1}+1$ otherwise. If $a_{n-1}$ is even, then  
  we have $\bigoplus_{i\in [n]} a_i = a_{n-1} \oplus a_n$ and thus the statement follows by the definition of $\oplus$.
  If $a_{n-1}$ is odd, then $\bigoplus_{i\in [n]} a_i = (a_{n-1}+1)\oplus a_n$. If  $(a_{n-1}+1)\le a_n$, then the statement again follows easily.
  If $a_n < a_{n-1}+1$, then by $a_{n-1}\le a_n$, we have $a_{n-1} = a_n$. Then we obtain $\bigoplus_{i\in [n]} a_i = (a_{n-1}+1)\oplus a_n=(a_n +1)+a_n=a_n +1$, where the last equality follows from the fact that $a_n +1$ is even.

  Now let $N \subseteq \mathbb N$ be a finite subset and $K=\max(N)$. It follows from \eqref{eq:sum-biloc-finite}, that for each $a \in \langle N\rangle_{\{\oplus\}}$ we have $a\le K+1$. This proves that $\langle N\rangle_{\{\oplus\}}$ is finite, i.e., $(\mathbb N,\oplus,0)$ is locally finite. The proof of the statement that $(\mathbb N,\odot,1)$ is locally finite is similar.

        However, if we apply $\oplus$ and $\odot$ alternatingly, then the result increases arbitrarily. 
        For instance, let $(b_i \mid n \in \mathbb{N})$ be the family defined by $b_0 = 2$ and, for each $n \in \mathbb{N}$, by
        \[
          b_{n+1} = \begin{cases}
            b_n \oplus 2 & \text{ if $n$ is odd}\\
            b_n \odot 2 & \text{ otherwise}\enspace.
            \end{cases}
          \]
          Then, e.g., $b_0 = 2$, $b_1 = 3$, $b_2 = 4$, $b_3=5$, .... In general we have $\CL(\{2\})=\mathbb{N}$, and hence  $\mathsf{Stb}$ is not weakly locally finite.

        \item \label{ex:rat-number-with-new-multiplication} \cite{dro24} We consider the algebra $(\mathbb{Q},+,\oslash,0,1)$ with the usual addition $+$ on rational numbers, and the binary operation $\oslash$ defined, for every $a,b \in \mathbb{Q}$, by
          \[
            a\oslash b = \begin{cases} \frac{a}{b} & \text{ if  $b \neq 0$,}\\
              0 & \text{ otherwise}
              \end{cases}
\]
The operation $\oslash$ is associative, and $0 \oslash b = b \oslash 0 = 0$ for each $b \in \mathbb{Q}$. Clearly, the algebra is a right-distributive strong bimonoid. Moreover, it is not left-distributive because, e.g., $1 \oslash (2+3) = \frac{1}{2+3} \not= \frac{1}{2} + \frac{1}{3} =  (1 \oslash 2) + (1 \oslash 3)$.

\item  \label{ex:combining-comm-semigroup-semigroup}  Let $(B,+)$ be a commutative semigroup and $(B,\cdot)$ be a semigroup. Combining these two semigroups, we obtain a strong bimonoid structure on $B$ by adding constants $\mathbb{0}$ and $\mathbb{1}$. Formally, let $\mathbb{0},\mathbb{1} \not\in B$ and put $B' = B \cup \{\mathbb{0},\mathbb{1}\}$. Then we define binary operations $\oplus$ and $\odot$ on $B'$ by letting 
  \begin{compactitem}
  \item $\oplus\mid_{B \times B} = +$ and $\mathbb{0} \oplus b = b \oplus \mathbb{0} = b$ and $\mathbb{1} \oplus b = b \oplus \mathbb{1} = b$ if $b\not= \mathbb{0}$, and
  \item $\odot\mid_{B \times B} = \cdot$ and  $\mathbb{0} \odot b = b \odot \mathbb{0} = \mathbb{0}$, and $\mathbb{1} \odot b = b \odot \mathbb{1} = b$ for each $b \in B'$.
    \end{compactitem}
    Then $(B',\oplus,\odot,\mathbb{0},\mathbb{1})$ is a  strong bimonoid.
    \hfill $\Box$
  \end{enumerate}
\end{example}

Next we define some more restrictions on strong bimonoids and prove useful relationships.

Let $\B=(B,\oplus,\otimes,\0,\1)$ be a strong bimonoid. It is
\index{strong bimonoid!additively idempotent}
 \index{strong bimonoid!multiplicatively idempotent} 
 \index{strong bimonoid!extremal} 
 \index{strong bimonoid!zero-sum free}
 \index{strong bimonoid!zero-cancellation free}
 \index{strong bimonoid!zero-divisor free}
 \index{strong bimonoid!positive}
\begin{compactitem}
 \item \emph{additively idempotent} if $\oplus$ is idempotent,
\item \emph{multiplicatively idempotent} if $\otimes$ is idempotent,
\item \emph{zero-sum free} if $a \oplus b = \mathbb{0}$ implies $a = b = \mathbb{0}$ for every $a,b \in B$, 
\item \emph{zero-divisor free} if $a \otimes b = \mathbb{0}$ implies $a = \mathbb{0}$ or $b = \mathbb{0}$ for every $a,b \in B$,
\item \emph{positive} if it is zero-sum free and zero-divisor free,
\item \emph{zero-cancellation free} if $a \otimes b \otimes c \ne \0$ implies $a \otimes c \ne \0$ for every $a,b,c \in B$,
\item \emph{extremal} if $\oplus$ is extremal.
\end{compactitem}

\begin{observation}\label{obs:zero-sum-free-property}\label{obs:zero-div-fre-implies-zero-canc-free}\label{obs:extremal-implies-zero-sum-free}\rm Let $\B=(B,\oplus,\otimes,\0,\1)$ be a strong bimonoid. Then the following statements hold.
  \begin{compactenum}
  \item[(1)] If $\B$ is commutative, then it is zero-cancellation free. Moreover,  there exists a zero-cancellation free semiring which is not commutative.
  \item[(2)]  If $\B$ is zero-divisor free, then it is zero-cancellation free. Moreover,
there exists a zero-cancellation free semiring which is not zero-divisor free.
\item[(3)] If $\B$ is extremal, then it is  additively idempotent.
  \item[(4)] If $\B$ is additively idempotent, then it is  zero-sum free. Moreover,
    there exists a zero-sum free semiring which is not additively idempotent.
  \item[(5)] If $\B$ is locally finite, then it is weakly locally finite. Moreover, if $\B$ is weakly locally finite, then it is bi-locally finite.
  \item[(6)] If $\B$ is right-distributive and bi-locally ﬁnite, then it is weakly locally ﬁnite (cf. \cite[Remark~17]{drostuvog10}).
  \item[(7)]  If $\B$ is left-distributive and weakly locally ﬁnite, then it is locally ﬁnite (cf. \cite[Remark~17]{drostuvog10}).
  \item[(8)] If $\B$ is a bi-locally finite semiring, then it is locally finite.
  \item[(9)] If $\B$ is  $\sigma$-complete, then it is zero-sum free \cite[Prop. 22.28]{gol99}.
     \item[(10)] If $\B$ is  $\sigma$-complete, then, for every countable index set $I$ and family $(b_i\mid i\in B)$ of elements of $B$, we have
\[ \infsum{\oplus}{i\in I}b_i\ne \0 \iff (\exists i\in I): b_i\ne \0\enspace.  \]
\item[(11)]  If $\B$ is zero-sum free, then for every finite index set $I$ and family $(b_i\mid i\in B)$ of elements of $B$ we have
  \[ \bigoplus_{i\in I}b_i\ne \0 \iff (\exists i\in I): b_i\ne \0\enspace.  \]
  \end{compactenum}
\end{observation}
\begin{proof} Proof of (1): For the proof of the first statement, assume that $\B$ is commutative and not zero-cancellation free.
  By the second condition, there are $a,b,c\in B$ such that $a\otimes b \otimes c \ne \0$ and $a\otimes c=\0$. Using the latter and commutativity, we 
  obtain $a\otimes b \otimes c=a\otimes c \otimes b=\0$ which is a contradiction.
   For the second, we can consider the semiring $\mathsf{Lang}_\Gamma=(\cP(\Gamma^*),\cup,\cdot,\emptyset,\{\varepsilon\})$ of  formal languages (cf. Example \ref{ex:semirings} (\ref{def:formal-lang-semiring})), which is zero-cancellation free and not commutative. 

   \
   
Proof of (2): Let us assume that $\B$ is zero-divisor free and not zero-cancellation free. The latter means that there exist $a,b,c \in B$ such that $a \otimes b \otimes c \ne \0$ and $a \otimes c = \0$. Since $\B$ is zero-divisor free, we have $a =\0$ or $c =\0$. Thus $a \otimes b \otimes c = \0$, which contradicts our assumption.

To prove the second statement, we consider the ring $\Intfour = (\{0,1,2,3\},+_4,\cdot_4,0,1)$ as defined in Example \ref{ex:semirings}(\ref{def:ring-Zmod4Z}).  It is easy to see that $\Intfour$ is zero-cancellation free  and not zero-divisor free.

\

Proof of (3): It is obvious.

\

Proof of (4): Let us assume that $\B$ is additively idempotent. Let $a \oplus b= \0$. Then $a = a \oplus \0 = a \oplus(a \oplus b) = (a \oplus a) \oplus b = a \oplus b = \0$. 
In the same way we can prove  that $b=\0$. Hence $\B$ is zero-sum free.
The semiring $\Nat$ of natural numbers is  zero-sum free and not additively idempotent.

\

Proof of (5): The first statement is obvious because for each finite subset $F\subseteq B$ we have $\CL(F) \subseteq \langle F\rangle_{\{\oplus,\otimes,\0,\1\}}$. The second statement follows from the fact that $\langle F\rangle_{\{\oplus,\0\}}\cup \langle F\rangle_{\{\otimes,\1\}} \subseteq \CL(F)$.

\

Proof of (6): Let $A\subseteq B$ be  finite. We show that $\CL(A)\subseteq \langle\langle A\rangle_{\{\otimes,\1\}}\rangle_{\{\oplus,\0\}}$.
 
 By definition, $\CL(A)$ is the smallest subset $C$ of $B$ such that (a) $A \cup\{\0,\1\}\subseteq C$,  (b) $C$ is closed under $\oplus$, and (c) for every
 $a\in C$ and $b\in A$, we have $a\otimes b\in C$. Hence, it is sufficient to show that  $\langle\langle A\rangle_{\{\1,\otimes\}}\rangle_{\{\0,\oplus\}}$ has the properties corresponding to (a), (b), and (c). Obviously,  $A \cup \{\0,\1\}\subseteq \langle\langle A\rangle_{\{\1,\otimes\}}\rangle_{\{\0,\oplus\}}$ and
 $\langle\langle A\rangle_{\{\1,\otimes\}}\rangle_{\{\0,\oplus\}}$ is closed under $\oplus$. Now let $a \in \langle\langle A\rangle_{\{\1,\otimes\}}\rangle_{\{\0,\oplus\}}$ and $b\in A$.
 There are $n\in \mathbb{N}$, $a_1,\ldots,a_n \in \langle A\rangle_{\{\1,\otimes\}}$ such that $a=\bigoplus_{i\in [n]} a_i$. Hence, 
 \[a\otimes b = \big( \bigoplus_{i\in [n]} a_i\big)\otimes b= \bigoplus_{i\in [n]} a_i\otimes b\enspace,\]
 where the second equality follows by right-distributivity. Since $ a_i\otimes b \in  \langle A\rangle_{\{\1,\otimes\}}$ for each $i\in[n]$, we obtain that $a\otimes b \in \langle\langle A\rangle_{\{\1,\otimes\}}\rangle_{\{\0,\oplus\}}$. Hence,  $\langle\langle A\rangle_{\{\1,\otimes\}}\rangle_{\{\0,\oplus\}}$ has also the property corresponding to (c).

 \
 
Proof of (7): The proof can be found in Section \ref{sect:term-tewrite-systems}.

\

Proof of (8): It follows from Statements (6) and (7).

\

Proof of (9): Let $a,b\in B$ such that $a\oplus b=\0$ and let $c=\infsum{\oplus}{i\in \mathbb{N}}{(a\oplus b)}$. Then by Observation \ref{obs:sum-emptyset} we have $c=\0$. Moreover, let $d=\infsum{\oplus}{i\in \mathbb{N}} a$ and $e=\infsum{\oplus}{i\in \mathbb{N}} b$. By renaming indices and using axioms \eqref{axiom-i-sum} and \eqref{axiom-ii-sum}, it is easy to see that 
\[\infsum{\oplus}{i\in \mathbb{N}}{(a\oplus b)}=\infsum{\oplus}{i\in \mathbb{N}} a \oplus \infsum{\oplus}{i\in \mathbb{N}} b\enspace,\] 
hence we have $c=d\oplus e$. Similarly, it is easy to see that 
\[\infsum{\oplus}{i\in \mathbb{N}} a = a\oplus \infsum{\oplus}{i\in \mathbb{N}} a,\]
hence $d=a\oplus d$. Then we obtain
$\0=c=d\oplus e = a\oplus d \oplus e =a$ and $\0=a\oplus b = b$.

\

Proof of (10): Let $\B$ be $\sigma$-complete. First we  prove the implication $\Rightarrow$ in the equivalence. For this we assume that $\infsum{\oplus}{i\in I}b_i\ne \0$ and $(\forall i\in I): b_i= \0$.  By Observation \ref{obs:sum-emptyset}, the second condition implies that $\infsum{\oplus}{i\in I}b_i= \0$, which contradicts the first one.

Lastly we prove the implication $\Leftarrow$. Let $j\in I$ be such that $b_j\ne \0$. Then
\[ \infsum{\oplus}{i\in I}b_i = \infsum{\oplus}{i\in I\setminus\{j\}}b_i \oplus b_j \ne \0,\]
where at the first equality we applied the laws \eqref{axiom-i-sum} and \eqref{axiom-ii-sum}, and the second equality follows from the fact that $\B$ is zero-sum free.

\

Proof of (11): The proof is very similar to the proof of (7). However, instead of Observation \ref{obs:sum-emptyset}, we use the fact that, for every finite index set $I$, we have $\bigoplus_{i\in I}\0 =\0$.
  \end{proof}

  In Subsection \ref{sec:semigroups}, we have defined the concept of powers for a monoid $(B,\odot,e)$. Here we extend this concept to powers in strong bimonoids, and thus we have to distinguish between the additive power and the multiplicative power.

\index{additive power}
\index{multiplicative power}
\index{index in strong bimonoids}
\index{period in strong bimonoids}
  Let $(B,\oplus,\otimes,\0,\1)$ be a strong bimonoid and $b \in B$. By induction of $(\mathbb{N},\succ_{\mathrm{N}})$, we define the \emph{$\mathbb{N}$-indexed family of additive powers of $b$}, denoted by $(nb \mid n \in \mathbb{N})$, and the \emph{$\mathbb{N}$-indexed family of multiplicative powers of $b$}, denoted by $(b^n \mid n \in \mathbb{N})$, to be the $\mathbb{N}$-indexed families over $B$ defined such that
\begin{equation}
0n = \0, \ (n+1)b = b \oplus nb  \ \ \text{ and } \ \  b^0 = \1, \  b^{n+1} = b \otimes b^n \enspace. \label{equ:nb-nexpb}
\end{equation}

\index{order}
\index{finite additive order}
\index{finite multiplicative order}
An element $b$ has \emph{finite additive order} (and \emph{finite multiplicative order}) if $\langle b\rangle_{\oplus}$ is finite (and $\langle b\rangle_{\otimes}$ is finite, respectively).

Of course, now the index $i(b)$ of an element $b$ and its period $p(b)$  refer either to the additive power or to the multiplicative power, and thus, we should notationally distinguish them. However, in each application it will be clear from the context to which sort of power the index and the period refer to. So we will not invent separate notations.


\subsection{The free strong bimonoid of nested polynomials}
\label{sec:free-strong-bimonoid}

Here we define a particular strong bimonoid, the strong bimonoid of nested polynomials, and we prove that it is free in the set of all strong bimonoids. Moreover, we show that the semiring of polynomials is isomorphic to a factor algebra of this strong bimonoid. To give a first impression of nested polynomials, let $\B=(B,\oplus,\otimes,\0,\1)$ be a strong bimonoid and
\[
x_1 \oplus \Big( \big ((\1 \oplus \0)  \oplus  (x_2 \otimes x_3) \big) \otimes (x_2 \otimes \1)   \oplus x_1 \Big)
\]
be an expression over the operations $\oplus$ and $\otimes$ and the unit elements $\0$ and $\1$ of $\B$ and the variables $x_1,x_2,x_3$. Then Figure~\ref{fig:expression1}(a) and (b) illustrate this expression and the corresponding nested polynomial, respectively. In a nested polynomial,
\begin{compactitem}
\item each maximal pattern of occurrences of the summation $\oplus$ is represented by a multiset and
\item each maximal pattern of occurrences of the multiplication $\otimes$ is represented  by the sequence of factors;
\end{compactitem}
in both cases, the unit elements $\0$ and $\1$ are not represented. These building operations may occur nested (cf. \cite{gasmon18} for the case in which the building operators do not occur nested, but multisets of sequences are considered; this case corresponds to semirings).

\index{multiset}
\paragraph{Preliminaries.}
Let $C$ be a set. A \emph{multiset over $C$} is a mapping $M: C \to \mathbb{N}$. The \emph{support of $M$}, denoted by $\supp(M)$, is the set $\supp(M) = \{c \in C \mid M(c)\not=0\}$.
If $\supp(M)$ is finite, then the \emph{size of $M$}, denoted by $\size(M)$, is the sum $\size(M) = \sum_{c \in C} M(c)$.

 We let $\mathrm{M}_0$ denote the multiset over $C$ such that $\supp(\mathrm{M}_0)=\emptyset$. For each $c \in C$, we denote by $\lM c \rM$ the multiset with $\lM c \rM(c') = 1$ if $c=c'$ and $0$ otherwise.
Given two multisets $M_1$ and $M_2$ over $C$, the \emph{summation of $M_1$ and $M_2$}, denoted by $M_1 + M_2$,  is  the multiset over $C$ where $(M_1+M_2)(c) = M_1(c) + M_2(c)$ for each $c \in C$. For each $n \in \mathbb{N}_+$ and multiset  $M$ over $C$, we let $n \cdot M$ denote the multiset $M + \ldots + M$ with $n$ summands.

Given two sequences $s=(s_1,\ldots,s_n)$ and $t=(t_1,\ldots,t_k)$, we denote by $\mathrm{concat}(s,t)$ the sequence $(s_1,\ldots,s_n,t_1,\ldots,t_k)$. In case $n=0$, we have $\mathrm{concat}(s,t)=t$, and symmetrically if $k=0$. For any sequence $s=(s_1,\ldots,s_n)$ we call $n$ the \emph{length of $s$} and denote it by $\mathrm{lg}(s)$, and for each $i \in [n]$, we denote $s_i$ also by $s(i)$.

\begin{figure}
  \centering
  \begin{tikzpicture}

    \begin{scope}

         \node (root1) {$\oplus$}
  child[level distance = 1cm, sibling distance = 30mm] {node  (n1) {$x_1$}}
  child[level distance = 1cm, sibling distance = 30mm] {node (n2) {$\oplus$}
    child[level distance = 1.6cm, sibling distance = 60mm] { node (n21) {$\otimes$}
      child[level distance = 1cm, sibling distance = 40mm] { node (n211) {$\oplus$}
        child[level distance = 1cm, sibling distance = 30mm] { node (n2111) {$\oplus$}
          child[level distance = 1.2cm, sibling distance = 10mm] { node (n21111) {$\1$}}
        child[level distance = 1.2cm, sibling distance = 10mm] { node (n21112) {$\0$}}}
      child[level distance = 1.3cm, sibling distance = 26mm] { node (n2112) {$\otimes$}
        child[level distance = 0.8cm, sibling distance = 10mm] { node (n21121) {$x_2$}}
      child[level distance = 0.8cm, sibling distance = 10mm] {node (n21122) {$x_3$}}}
      }
      child[level distance = 1cm, sibling distance = 30mm] { node (n212) {$\otimes$}
        child[level distance = 1.2cm, sibling distance = 10mm] { node (n2121) {$x_2$}}
      child[level distance = 1.2cm, sibling distance = 10mm] { node (n2122) {$\1$}}}
    }
    child { node (n22) {$x_1$}}};

\node[fit=(root1)(n2),rotate=170,ellipse,dashed,draw] {};
\node[fit=(n21)(n212),rotate=170,ellipse,dashed,draw] {};
\node[fit=(n211)(n2111),rotate=20,ellipse,dashed,draw] {};

\node[xshift=70mm,yshift=-30mm]{$
  \left[ \begin{matrix}
      x_1 \mapsto 2 \\
      \left(
        \left[
          \begin{matrix}
            () \mapsto 1 \\
           (x_2,x_3) \mapsto 1
          \end{matrix}
         \right]
, x_2
      \right) \mapsto 1 
\end{matrix}
\right]$};

\node at (0.25, -6.3) {(a)};
\node at (6.5, -6.3) {(b)};
  \end{scope}
    
 \end{tikzpicture}

\caption{\label{fig:expression1} (a) Example of an expression over the operations $\oplus$ and $\otimes$ and unit elements $\0$ and $\1$  of $\B$ and the variables $x_1,x_2,x_3$; each of the three dashed ellipses stands for a maximal pattern of $\oplus$-occurrences or a maximal pattern of $\otimes$-occurrences; and (b) the corresponding nested polynomial where each nontrivial finite multiset $M$ is denoted by a pair in square brackets with the vertical list of $(\mathrm{value} \mapsto \mathrm{multiplicity})$-relationships of $M$  inside (where the unit element $\0$ is disregarded), and each nontrivial finite sequence $(c_1,\ldots,c_n)$ is presented by itself (except that the unit element $\1$ is disregarded). }
\end{figure}


\paragraph{Definition of nested polynomials.}
Let $\DD$ be a set. We define the \emph{set of nontrivial finite multisets over $\DD$}, denoted by $\cM_{\mathrm{nt}}(\DD)$, to be the set
\[
\cM_{\mathrm{nt}}(\DD) = \{M: \DD \to \mathbb{N} \mid \supp(M) \text{ is finite and } \size(M) \ge 2\}  \enspace.
\]
The condition $\size(M) \ge 2$ prohibits sums with less than two summands. We note that summation of multisets (as defined above) is a commutative binary operation on  $\cM_{\mathrm{nt}}(\DD)$.

We define the \emph{set of nontrivial finite sequences over $\DD$}, denoted by $\cS_{\mathrm{nt}}(\DD)$, to be the set
\[
\cS_{\mathrm{nt}}(\DD) = \{(c_1,\ldots,c_n) \mid n \ge 2, c_1,\ldots,c_n \in \DD\}  \enspace.
\]
Similarly, the condition $n \ge 2$ prohibits products with less than two factors.

Let $X_n=\{x_1,\ldots,x_n\}$ for some $n \in \mathbb{N}$ be a finite set of variables.

\index{nested polynomials}
\index{nPolX@$\rmnPol(X_n)$}
We define \emph{the set of nested polynomials over $X_n$}, denoted by $\rmnPol(X_n)$, to be the smallest set $R$  such that $X_n \subseteq R$, $\mathrm{M}_0 \in R$, $() \in R$,  and $\cM_{\mathrm{nt}}(R) \subseteq R$ and $\cS_{\mathrm{nt}}(R) \subseteq R$. Here ``$\rmnPol$'' stands for \emph{n}ested \emph{pol}ynomials. Intuitively, $\mathrm{M}_0$ and $()$ represent the summation with zero summands and the product with zero factors, respectively.

It is clear that each nested polynomial $r \in \rmnPol(X)$ has exactly one of the following five forms:
\[r =x \ \text{  for some $x\in X_n$ },
\ r=\mathrm{M}_0 \ , 
\ r = () \ ,
\ r \in \cM_{\mathrm{nt}}(\rmnPol(X_n)) \ , \ \text{ or }
  \ r \in \cS_{\mathrm{nt}}(\rmnPol(X_n)) \enspace.
\]

Due to the inductive definition of $\rmnPol(X_n)$, we can use the principle of well-founded induction to define mappings  and to prove properties.
Formally, we define the binary relation $\succ$ on $\rmnPol(X_n)$ as follows. For every $r_1,r_2 \in \rmnPol(X_n)$, we let $r_1 \succ r_2$ if 
\begin{compactitem}
    \item either $r_1 \in \cM_{\mathrm{nt}}(\rmnPol(X_n))$ and $r_2 \in \supp(r_1)$
\item or $r_1 \in \cS_{\mathrm{nt}}(\rmnPol(X_n))$ and  $r_2 = (r_1)(i)$ for some $i \in [\lg(r_1)]$.
\end{compactitem}
Then $\succ$ is a well-founded relation with $\nf_\succ(\rmnPol(X_n)) = X_n \cup \{\mathrm{M}_0, ()\}$. We will use $\succ$ to prove properties of $\rmnPol(X_n)$ by well-founded induction (cf. Section \ref{sec:well-founded-induction})
or to define mappings with $\rmnPol(X_n)$ as domain (using, e.g., Theorem~\ref{thm:dedekind}).

    
    \paragraph{The operations on $\rmnPol(X_n)$.}

    Here we impose a strong bimonoid structure on the set $\rmnPol(X_n)$.
    
      We define the \emph{summation on $\rmnPol(X_n)$} to be the binary operation $\boxplus$ (read: ``boxplus'') as follows. For every $r_1,r_2 \in \rmnPol(X_n)$ we let
   \[
    r_1 \boxplus r_2 =
    \begin{cases}
      r_1 + r_2 & \text{ if $r_1,r_2 \in \cM_{\mathrm{nt}}(\rmnPol(X_n))$}\\
      r_1 & \text{ if $r_1\in \rmnPol(X_n)$ and $r_2 = \mathrm{M}_0$}\\
      r_2 & \text{ if $r_1= \mathrm{M}_0$ and $r_2 \in \rmnPol(X_n)$}\\
      r_1 + \lM r_2 \rM & \text{ if $r_1 \in \cM_{\mathrm{nt}}(\rmnPol(X_n))$ and 
      $r_2 \not\in \cM_{\mathrm{nt}}(\rmnPol(X_n)) \cup \{\mathrm{M}_0\}$}\\
      \lM r_1 \rM + r_2  & \text{ if $r_1 \not\in \cM_{\mathrm{nt}}(\rmnPol(X_n)) \cup \{\mathrm{M}_0\}$ and $r_2 \in \cM_{\mathrm{nt}}(\rmnPol(X_n))$}\\
      \lM r_1, r_2 \rM  & \text{ if $r_1 \not\in \cM_{\mathrm{nt}}(\rmnPol(X_n)) \cup \{\mathrm{M}_0\}$ and  $r_2 \not\in \cM_{\mathrm{nt}}(\rmnPol(X_n)) \cup \{\mathrm{M}_0\}$}\\
      \end{cases}
    \]
\noindent Clearly $r_1 \boxplus r_2 \in \rmnPol(X_n)$. Hence $\boxplus$ is a binary operation on $\rmnPol(X_n)$.

      We define the \emph{multiplication on $\rmnPol(X_n)$} to be the binary operation $\boxtimes$ (read: ``boxtimes'') as follows. For every $r_1,r_2 \in \rmnPol(X_n)$ we let
  \[
    r_1 \boxtimes r_2 =
    \begin{cases}
      \mathrm{concat}(r_1,r_2) & \text{ if $r_1,r_2 \in \cS_{\mathrm{nt}}(\rmnPol(X_n))$}\\
       r_1 & \text{ if $r_1\in \rmnPol(X_n)$ and $r_2 = ()$}\\
      r_2 & \text{ if $r_1= ()$ and $r_2 \in \rmnPol(X_n)$}\\
      \mathrm{M}_0 & \text{ if $r_1\in \rmnPol(X_n)$ and $r_2 = \mathrm{M}_0$}\\
      \mathrm{M}_0 & \text{ if $r_1= \mathrm{M}_0$ and $r_2 \in \rmnPol(X_n)$}\\
      \mathrm{concat}(r_1,(r_2)) & \text{ if $r_1 \in \cS_{\mathrm{nt}}(\rmnPol(X_n))$ and 
      $r_2 \not\in \cS_{\mathrm{nt}}(\rmnPol(X_n)) \cup \{\mathrm{M}_0,()\}$}\\
      \mathrm{concat}((r_1),r_2)  & \text{ if $r_1 \not\in \cS_{\mathrm{nt}}(\rmnPol(X_n)) \cup \{\mathrm{M}_0,()\}$ and $r_2 \in \cS_{\mathrm{nt}}(\rmnPol(X_n))$}\\
      (r_1,r_2) & \text{ if $r_1 \not\in \cS_{\mathrm{nt}}(\rmnPol(X_n)) \cup \{\mathrm{M}_0,()\}$ and  $r_2 \not\in \cS_{\mathrm{nt}}(\rmnPol(X_n)) \cup \{\mathrm{M}_0,()\}$}\\
      \end{cases}
    \]
 \noindent Again it is clear that $r_1 \boxtimes r_2 \in \rmnPol(X_n)$. Hence $\boxtimes$ is a binary operation on $\rmnPol(X_n)$.

\index{nPol(Xn)@$\sfnPol(X_n)$}
    We denote by $\sfnPol(X_n)$ the algebra $(\rmnPol(X_n),\boxplus,\boxtimes,\mathrm{M}_0,())$.
         
    \begin{observation}\rm \label{obs:MS(X_n)-strong-bimonoid} The algebra $\sfnPol(X_n) = (\rmnPol(X_n),\boxplus,\boxtimes,\mathrm{M}_0,())$ is a strong bimonoid.
    \end{observation}
    \begin{proof} 
    It is easy to verify that $\boxplus$ and $\boxtimes$ are associative and that $\boxplus$ is commutative. Moreover, $\mathrm{M}_0$ and $()$ are unit elements of $\boxplus$ and $\boxtimes$, respectively. Finally, $\mathrm{M}_0$ is an annihilator for $\boxtimes$.
      \end{proof}

    \begin{observation}\rm \label{obs:MS(X_n)-generated-by-X_n} The strong bimonoid $\sfnPol(X_n)$ is generated by $X_n$, i.e., $\rmnPol(X_n) = \langle X_n \rangle_{\{\mathrm{M}_0,(),\boxplus,\boxtimes\}}$.
        \end{observation}
        \begin{proof} Obviously, $X_n \cup \{\mathrm{M}_0, ()\} \subseteq \rmnPol(X_n)$, and $\rmnPol(X_n)$ is closed under $\boxplus$ and $\boxtimes$. Since $\langle X_n \rangle_{\{\mathrm{M}_0,(),\boxplus,\boxtimes\}}$ is the smallest set with these properties, we obtain that $\langle X_n \rangle_{\{\mathrm{M}_0,(),\boxplus,\boxtimes\}} \subseteq \rmnPol(X_n)$.

          By well-founded induction on the structure of $\rmnPol(X_n)$, we prove that  $\rmnPol(X_n) \subseteq  \langle X_n \rangle_{\{\mathrm{M}_0,(),\boxplus,\boxtimes\}}$. Clearly, $X_n \cup \{\mathrm{M}_0, ()\} \subseteq \langle X_n \rangle_{\{\mathrm{M}_0,(),\boxplus,\boxtimes\}}$. Now let $r \in \cM_{\mathrm{nt}}(\rmnPol(X_n))$. Then there exist $n \in \mathbb{N}_+$ and $c_1,\ldots,c_n \in \rmnPol(X_n)$ such that $\supp(r) = \{c_1,\ldots,c_n\}$ and $\bigplus_{i\in [n]}~r(c_i)~\ge~2$.  By I.H. we can assume that, for each $i \in [n]$, we have $c_i \in \langle X_n \rangle_{\{\mathrm{M}_0,(),\boxplus,\boxtimes\}}$. Then
          \[
           r = \bigboxplus_{i \in [n]} (\underbrace{c_i \boxplus \ldots \boxplus c_i}_{r(c_i)} )
          \]
and hence $r \in \langle X_n \rangle_{\{\mathrm{M}_0,(),\boxplus,\boxtimes\}}$.

Finally, let $r \in \cS_{\mathrm{nt}}(\rmnPol(X_n))$. Then there exist $n \ge 2$ and $c_1,\ldots,c_n \in \rmnPol(X_n)$ such that $r = (c_1,\ldots,c_n)$.  By I.H. we can assume that $c_j \in \langle X_n \rangle_{\{\mathrm{M}_0,(),\boxplus,\boxtimes\}}$ for each $j \in [n]$. Then we can represent $r$ as follows:
          \[r =  c_1 \boxtimes \cdots \boxtimes c_n \]
and hence $r \in \langle X_n \rangle_{\{\mathrm{M}_0,(),\boxplus,\boxtimes\}}$. 
       \end{proof}
       
 For instance, let $r = \lM x_2,x_2,(\lM x_1,()\rM,x_4)\rM$. Then we have:
 \begingroup
 \allowdisplaybreaks
 \begin{align*}
     r &= x_2 \boxplus x_2 \boxplus (\lM x_1,()\rM,x_4) \\
&= x_2 \boxplus x_2 \boxplus \Big( \lM x_1,() \rM \boxtimes x_4 \Big) \\
&= x_2 \boxplus x_2 \boxplus \Big( \big( x_1 \boxplus () \big) \boxtimes x_4 \Big) \enspace.
 \end{align*} 
 \endgroup

 \begin{quote}{\em In the rest of this subsection, we let $\B=(B,\oplus,\otimes,\0,\1)$ denote an arbitrary strong bimonoid. 
    For each $b \in B$ and $n \in \mathbb{N}$, we let $n b$ denote the sum $b \oplus \ldots \oplus b$ with $n$ summands; in particular, $0b=\0$.}
 \end{quote}

     \index{evalBf@$\eval_{\B,f}$}
Next we show that, for each mapping $f: X_n \to B$, there exists a strong bimonoid homomorphism from $\sfnPol(X_n)$ to $\B$ which extends $f$. Let  $f: X_n \to B$. By induction, we define the mapping $\eval_{\B,f}:~\rmnPol(X_n)~\to~B$ for each $r \in \rmnPol(X_n)$ by
    \[
      \eval_{\B,f}(r) =
      \begin{cases}
        f(r) & \text{ if $r \in X_n$}\\
        \0 & \text{ if $r = \mathrm{M}_0$}\\
                \1 & \text{ if $r = ()$}\\
        \bigoplus\limits_{c \in \supp(r)} r(c) \ \eval_{\B,f}(c) & \text{ if $r \in \cM_{\mathrm{nt}}(\rmnPol(X_n))$}\\
        \bigotimes\limits_{i \in [\mathrm{lg}(r)]} \eval_{\B,f}(r(i)) & \text{ if $r \in \cS_{\mathrm{nt}}(\rmnPol(X_n))$} \enspace.
        \end{cases}
      \]
      We recall that, in the fourth case above,  the expression $r(c) \ \eval_{\B,f}(c)$ denotes $\eval_{\B,f}(c) \oplus \ldots \oplus \eval_{\B,f}(c)$ with $r(c)$ summands.
      
      \begin{lemma}\rm  \label{lm:eval-str-bm-hom} For each mapping $f: X_n \to B$, the mapping  $\eval_{\B,f}$ is a strong bimonoid homomorphism from $\sfnPol(X_n)$ to $\B$.
      \end{lemma}
      \begin{proof} By definition, $\eval_{\B,f}(\mathrm{M}_0) = \0$ and $\eval_{\B,f}(()) = \1$.

        Let $r_1,r_2 \in \rmnPol(X_n)$. We prove that $\eval_{\B,f}(r_1 \boxplus r_2) = \eval_{\B,f}(r_1) \oplus \eval_{\B,f}(r_2)$. We proceed by case analysis.

        \
        
        \noindent
        \underline{$r_1,r_2 \in \cM_{\mathrm{nt}}(\rmnPol(X_n))$:}
        \begingroup
        \allowdisplaybreaks
        \begin{align*}
          \eval_{\B,f}(r_1 \boxplus r_2) &= \eval_{\B,f}(r_1 + r_2)
          \tag{by definition of $\boxplus$}\\
                                          &= \bigoplus\limits_{c \in \supp(r_1+r_2)} (r_1+r_2)(c) \ \eval_{\B,f}(c)
                                            \tag{by definition of $\eval_{\B,f}$}\\
          &= \bigoplus\limits_{c \in \supp(r_1+r_2)} \big(r_1(c) + r_2(c)\big)  \ \eval_{\B,f}(c)
            \tag{by definition of $r_1+r_2$}\\
                                          &= \bigoplus\limits_{c \in \supp(r_1+r_2)} \Big(r_1(c)  \ \eval_{\B,f}(c) \oplus r_2(c) \ \eval_{\B,f}(c)\Big)
                                            \tag{by definition of $n \cdot b$}\\
                                          &= \bigoplus\limits_{c \in \supp(r_1) \cup \supp(r_2)} \Big(r_1(c)  \ \eval_{\B,f}(c) \oplus r_2(c) \ \eval_{\B,f}(c)\Big)\\
                                          &= \bigoplus\limits_{c \in \supp(r_1)} \Big(r_1(c)  \ \eval_{\B,f}(c)\Big) \oplus
                                            \bigoplus\limits_{c \in \supp(r_2)} \Big(r_2(c) \ \eval_{\B,f}(c)\Big)
                                            \tag{by case analysis}\\
                                          &= \eval_{\B,f}(r_1) \oplus \eval_{\B,f}(r_2)
                                            \tag{by definition of $\eval_{\B,f}$} \enspace.
        \end{align*}
        \endgroup

        \
        
        \noindent
        \underline{$r_1\in \rmnPol(X_n)$ and $r_2 = \mathrm{M}_0$:}
        \begingroup
        \allowdisplaybreaks
        \begin{align*}
          \eval_{\B,f}(r_1 \boxplus r_2) &= \eval_{\B,f}(r_1)
                                            = \eval_{\B,f}(r_1) \oplus \0
                                            = \eval_{\B,f}(r_1) \oplus \eval_{\B,f}(r_2)
                                            \tag{by definition of $\boxplus$ and  $\eval_{\B,f}$} \enspace.
        \end{align*}
        \endgroup

        \
        
        \noindent
        \underline{$r_1= \mathrm{M}_0$ and $r_2 \in \rmnPol(X_n)$:} The proof is symmetric to the proof of the previous case.

        \

       \noindent
        \underline{$r_1 \in \cM_{\mathrm{nt}}(\rmnPol(X_n))$ and $r_2 \not\in \cM_{\mathrm{nt}}(\rmnPol(X_n)) \cup \{\mathrm{M}_0\}$:}
        \begingroup
        \allowdisplaybreaks
        \begin{align*}
          \eval_{\B,f}(r_1 \boxplus r_2) &= \eval_{\B,f}(r_1 + \lM r_2 \rM)
          \tag{by definition of $\boxplus$}\\
                                          &= \bigoplus\limits_{c \in \supp(r_1) \cup \{r_2\}} M(c) \  \eval_{\B,f}(c)
                                            \tag{by definition of $\eval_{\B,f}$}
        \end{align*}
        \endgroup
                where $M \in \rmnPol(X_n)$ is defined for each argument $r$ by $M(r) = r_1(r) + 1$ if $r=r_2$, and $M(r) =r_1(r)$ otherwise.
        Then we can continue:
        \begingroup
        \allowdisplaybreaks
        \begin{align*}
          &= \bigoplus\limits_{c \in \supp(r_1) \cup \{r_2\}} M(c) \ \eval_{\B,f}(c)  \\
          &= \Big(\bigoplus\limits_{c \in \supp(r_1)} r_1(c) \ \eval_{\B,f}(c)\Big) \oplus \eval_{\B,f}(r_2)  \\
                                          &= \eval_{\B,f}(r_1) \oplus \eval_{\B,f}(r_2)
                                            \tag{by definition of $\eval_{\B,f}$} \enspace.
        \end{align*}
        \endgroup

        \
        
        \noindent
        \underline{$r_1 \not\in \cM_{\mathrm{nt}}(\rmnPol(X_n)) \cup \{\mathrm{M}_0\}$ and $r_2 \in \cM_{\mathrm{nt}}(\rmnPol(X_n))$:} The proof is symmetric to the proof of the previous case.

        \

        \noindent
        \underline{$r_1 \not\in \cM_{\mathrm{nt}}(\rmnPol(X_n)) \cup \{\mathrm{M}_0\}$ and  $r_2 \not\in \cM_{\mathrm{nt}}(\rmnPol(X_n)) \cup \{\mathrm{M}_0\}$:}
         \begingroup
        \allowdisplaybreaks
        \begin{align*}
          \eval_{\B,f}(r_1 \boxplus r_2) &= \eval_{\B,f}(\lM r_1, r_2\rM)
                                            \tag{by definition of $\boxplus$} \\
                                          &= \bigoplus\limits_{c \in \supp(\lM r_1, r_2\rM)} \lM r_1, r_2\rM(c) \ \eval_{\B,f}(c)
                                            \tag{by definition of $\eval_{\B,f}$}\\
                                          &= \bigoplus\limits_{c \in \{r_1,r_2\}} \lM r_1, r_2\rM(c) \ \eval_{\B,f}(c)\enspace.
                       \end{align*}
                    \endgroup    
We proceed by case analysis.

(a) $r_1=r_2$: Then $\{r_1,r_2\} = \{r_1\}$ and hence
     \begingroup
        \allowdisplaybreaks
        \begin{align*}
        \bigoplus\limits_{c \in \{r_1,r_2\}} \lM r_1, r_2\rM(c) \ \eval_{\B,f}(c)
        &= \lM r_1, r_1\rM(r_1) \ \eval_{\B,f}(r_1)   \\
        &= 2 \ \eval_{\B,f}(r_1) \\
          &= \eval_{\B,f}(r_1) \oplus \eval_{\B,f}(r_1) 
          =  \eval_{\B,f}(r_1) \oplus \eval_{\B,f}(r_2)\enspace.
                    \end{align*}
                    \endgroup

                (b) $r_1\not=r_2$: Then 
     \begingroup
        \allowdisplaybreaks
        \begin{align*}
        \bigoplus\limits_{c \in \{r_1,r_2\}} \lM r_1, r_2\rM(c) \ \eval_{\B,f}(c)
        &= \lM r_1, r_2\rM(r_1) \ \eval_{\B,f}(r_1) \oplus  \lM r_1, r_2\rM(r_2) \ \eval_{\B,f}(r_2)  \\
          &= 1 \ \eval_{\B,f}(r_1) \oplus 1 \ \eval_{\B,f}(r_2)
          = \eval_{\B,f}(r_1) \oplus \eval_{\B,f}(r_2)\enspace.
                    \end{align*}
                    \endgroup    

                    \

                    Finally, we prove that $\eval_{\B,f}(r_1 \boxtimes r_2) = \eval_{\B,f}(r_1) \otimes \eval_{\B,f}(r_2)$. We proceed by case analysis.

     \
        
        \noindent
        \underline{$r_1,r_2 \in \cS_{\mathrm{nt}}(\rmnPol(X_n))$:}
    \begingroup
        \allowdisplaybreaks
        \begin{align*}
          \eval_{\B,f}(r_1 \boxtimes r_2) &= \eval_{\B,f}(\mathrm{concat}(r_1,r_2))
  \tag{by definition of $\boxtimes$}\\
                                           &= \Big(\bigotimes\limits_{i \in [\mathrm{lg}(r_1)]} \eval_{\B,f}(r_1(i))\Big) \otimes
                                             \Big(\bigotimes\limits_{i \in [\mathrm{lg}(r_2)]} \eval_{\B,f}(r_2(i))\Big)
                                             \tag{by definition of $\mathrm{concat}$ and $\eval_{\B,f}$}\\
                                          &= \eval_{\B,f}(r_1) \otimes \eval_{\B,f}(r_2)
                                            \tag{by definition of $\eval_{\B,f}$} \enspace.
     \end{align*}
     \endgroup
     
                                            \
        
        \noindent
        \underline{$r_1\in \rmnPol(X_n)$ and $r_2 = ()$:}
         \begingroup
        \allowdisplaybreaks
        \begin{align*}
          \eval_{\B,f}(r_1 \boxtimes r_2) &= \eval_{\B,f}(r_1) =  \eval_{\B,f}(r_1) \otimes \1 =  \eval_{\B,f}(r_1) \otimes  \eval_{\B,f}(r_2)
  \tag{by definition of $\boxtimes$ and $\eval_{\B,f}$} \enspace.
     \end{align*}
     \endgroup

           \
        
        \noindent
        \underline{$r_1= ()$ and $r_2 \in \rmnPol(X_n)$:} The proof is symmetric to the proof of the previous case.

           \
        
        \noindent
        \underline{$r_1\in \rmnPol(X_n)$ and $r_2 = \mathrm{M}_0$:}
         \begingroup
        \allowdisplaybreaks
        \begin{align*}
          \eval_{\B,f}(r_1 \boxtimes r_2) &= \eval_{\B,f}(\mathrm{M}_0) = \0 = \eval_{\B,f}(r_1) \otimes \0 =  \eval_{\B,f}(r_1) \otimes  \eval_{\B,f}(r_2)
  \tag{by definition of $\boxtimes$ and $\eval_{\B,f}$} \enspace.
     \end{align*}
     \endgroup

           \
        
        \noindent
        \underline{$r_1= \mathrm{M}_0$ and $r_2 \in \rmnPol(X_n)$:} The proof is symmetric to the proof of the previous case.

           \
        
        \noindent
        \underline{$r_1 \in \cS_{\mathrm{nt}}(\rmnPol(X_n))$ and $r_2 \not\in \cS_{\mathrm{nt}}(\rmnPol(X_n)) \cup \{\mathrm{M}_0,()\}$:}
           \begingroup
        \allowdisplaybreaks
        \begin{align*}
          \eval_{\B,f}(r_1 \boxtimes r_2) &= \eval_{\B,f}(\mathrm{concat}(r_1,(r_2)))
                                             \tag{by definition of $\boxtimes$}\\
           &= \Big(\bigotimes\limits_{i \in [\mathrm{lg}(r_1)]} \eval_{\B,f}(r_1(i))\Big) \otimes
                                             \eval_{\B,f}(r_2)
                                             \tag{by definition of $\mathrm{concat}$ and $\eval_{\B,f}$}\\
                                          &= \eval_{\B,f}(r_1) \otimes \eval_{\B,f}(r_2)
                                            \tag{by definition of $\eval_{\B,f}$} \enspace.
        \end{align*}
        \endgroup
        
           \
        
        \noindent
        \underline{$r_1 \not\in \cS_{\mathrm{nt}}(\rmnPol(X_n)) \cup \{\mathrm{M}_0,()\}$ and $r_2 \in \cS_{\mathrm{nt}}(\rmnPol(X_n))$:} The proof is symmetric to the proof of the previous case.

           \
        
        \noindent
        \underline{$r_1 \not\in \cS_{\mathrm{nt}}(\rmnPol(X_n)) \cup \{\mathrm{M}_0,()\}$ and  $r_2 \not\in \cS_{\mathrm{nt}}(\rmnPol(X_n)) \cup \{\mathrm{M}_0,()\}$:}
           \begingroup
        \allowdisplaybreaks
        \begin{align*}
          \eval_{\B,f}(r_1 \boxtimes r_2) &= \eval_{\B,f}((r_1,r_2))
                                             \tag{by definition of $\boxtimes$}\\
          &= \eval_{\B,f}(r_1) \otimes \eval_{\B,f}(r_2)
                                             \tag{by definition of $\mathrm{concat}$ } \enspace.
                                  \end{align*}
        \endgroup             
        \end{proof}

        \begin{lemma}\rm \label{lm:eval-is-unique} For each mapping $f: X_n \to B$, the strong bimonoid homomorphism $\eval_{\B,f}$ is the only strong bimonoid homomorphism from $\sfnPol(X_n)$ to $\B$ which extends $f$.
          \end{lemma}
        \begin{proof} Let $f: X_n \to B$ be a mapping. By Lemma~\ref{lm:eval-str-bm-hom}, the mapping $\eval_{\B,f}$ is a strong bimonoid homomorphism from $\sfnPol(X_n)$ to $\B$.
          By Observation \ref{obs:MS(X_n)-generated-by-X_n}, the strong bimonoid $\sfnPol(X_n)$ is generated by~$X_n$. Then the statement follows by Lemma~\ref{lm:baanip-unique-hom} (for $\A=\sfnPol(X_n)$).
          \end{proof}

         \begin{theorem-rect} \label{thm:MS(X_n)-is-free} For each $n \in \mathbb{N}$, the strong bimonoid $\sfnPol(X_n)$ is free in the set of all strong bimonoids with generating set $X_n$.
          \end{theorem-rect}
             \begin{proof}  It directly follows from  Observation \ref{obs:MS(X_n)-generated-by-X_n} and Lemma~\ref{lm:eval-is-unique}.
          \end{proof}

          \begin{corollary}\rm The following two statements hold.
  \begin{compactenum}
            \item[(1)] For every $n \in \mathbb{N}$ and $f:X_n \to B$, we have $\im(\eval_{\B,f}) = \langle \im(f)\rangle_{\{\oplus,\otimes,\0,\1\}}$. Moreover, $(\im(\eval_{\B,f}),\oplus,\otimes,\0,\1)$ is a sub-strong bimonoid of $\B$.
            \item[(2)] If there exists a finite set $H \subseteq B$ such that $\langle H\rangle_{\{\oplus,\otimes,\0,\1\}}=B$, then $\B \cong \sfnPol(X_n)/_{\ker(\eval_{\B,f})}$ for each bijection $f: X_n \to H$ where $n =|H|$.
              \end{compactenum}
            \end{corollary}
            \begin{proof} Proof of (1): The equality $\im(\eval_{\B,f}) = \langle \im(f)\rangle_{\{\oplus,\otimes,\0,\1\}}$ follows from Observation \ref{obs:MS(X_n)-generated-by-X_n} and Lemma~\ref{lm:eval-str-bm-hom}. By Lemma~\ref{lm:hom-image=subalgebra}, we obtain that $(\im(\eval_{\B,f}),\oplus,\otimes,\0,\1)$ is a sub-strong bimonoid of $\B$.

              \

              Proof of (2): Let $A \subseteq B$ be a finite set such that $\langle H\rangle_{\{\oplus,\otimes,\0,\1\}}=B$, and let $f: X_n \to H$ be a bijection with $n=|H|$. Then, by (1), the mapping $\eval_{\B,f}$ is surjective. Hence, by Lemma~\ref{thm:kernel-is-congruence}, we obtain $\B \cong \sfnPol(X_n)/_{\ker(\eval_{\B,f})}$.
            \end{proof}

            \paragraph{Polynomials.}
\index{polynomials}
\index{PolX@$\Pol(X_n)$}
In polynomials over $X_n$, the building operations $\cM_{\mathrm{nt}}$ and $ \cS_{\mathrm{nt}}$ cannot occur nested; only sequences and multisets of sequences may occur. Formally, 
the \emph{set of polynomials over $X_n$}, denoted by $\Pol(X_n)$, is the set 
\[
\Pol(X_n) = X_n \cup \{\mathrm{M}_0,()\} \cup \cS_{\mathrm{nt}}(X_n) \cup \cM_{\mathrm{nt}}\Big(X_n \cup \{()\} \cup \cS_{\mathrm{nt}}(X_n)\Big) \enspace.
\]
Obviously, $\Pol(X_n) \subseteq \rmnPol(X_n)$ and, e.g., 
\[
(\lM (),()\rM, \lM (),()\rM)\in \rmnPol(X_n) \setminus \Pol(X_n) \enspace,
\]
where, intuitively, $(\lM (),()\rM, \lM (),()\rM)$ corresponds to $(\1 \oplus \1)\otimes (\1 \oplus \1)$.
Thus $\Pol(X_n) \subset \rmnPol(X_n)$.

Next we recall that $\mathbb{N}[x_1,\ldots,x_n] = \{p: X_n^* \to \mathbb{N} \mid \{w \in X_n^* \mid p(w) \not= 0\} \text{ is finite}\}\}$ is the set of polynomials with $n$ variables over $\mathbb{N}$ (cf. Example \ref{ex:semirings}(\ref{def:semiring-of-polynomials})). Obviously, there exists a bijection from $\Pol(X_n)$ to  $\mathbb{N}[x_1,\ldots,x_n]$.
 The following list shows examples of this easy correspondence with $n=3$, where we only show the values for elements in the support of the polynomial and where $p_0 = \emptyset$ and $p_1 = (\varepsilon \mapsto 1)$.

\

{\small
\hspace*{-5mm}
\begin{tabular}{l|c|c|c|c|c|c}
elements in $\Pol(\{x_1,x_2,x_3\})$: & $x_2$ & $\mathrm{M}_0$ & $()$  & $(x_1,x_3,x_3)$
& $\lM (x_1,x_3,x_3), (x_2,x_2), (x_2,x_2), (), ()\rM$\\\hline
elements in $\Nat[x_1,x_2,x_3]$: & $x_2 \mapsto 1$ & $p_0$ & $p_1$ 
& $x_1x_3x_3 \mapsto 1$ & $x_1x_3x_3 \mapsto 1, x_2x_2 \mapsto 2, \varepsilon \mapsto 2$
\end{tabular}
}

\

Clearly, the set $\Pol(X_n)$ of polynomials over $X_n$ is closed under $\boxplus$; in fact, $\boxplus|_{\Pol(X_n)} = \oplus$, i.e., the usual summation of polynomials. However, $\Pol(X_n)$ is not closed under $\boxtimes$, e.g., the product $\lM (),()\rM \boxtimes \lM (),()\rM$ of two polynomials is not a polynomial. Thus $\Pol(X_n)$ is not the carrier set of any sub-strong bimonoid of $\sfnPol(X_n)$.

Finally, we compare
\begin{compactitem}
\item the free strong bimonoid $\sfnPol(X_n) =(\rmnPol(X_n),\boxplus,\boxtimes,\mathrm{M}_0,())$ of nested polynomials and 
\item the semiring $\Nat[x_1,\ldots,x_n] = (\mathbb{N}[x_1,\ldots,x_n],\oplus,\cdot,p_0,p_1)$ of polynomials where the multiplication $\cdot$ is the Cauchy product.
\end{compactitem}
 For the sake of convenience, we abbreviate the strong bimonoid homomorphism $\eval_{\Nat[x_1,\ldots,x_n],\id_{X_n}}$ from $\sfnPol(X_n)$ to $\Nat[x_1,\ldots,x_n]$ by $\eval$. This homomorphism is surjective and hence, by Lemma~\ref{thm:kernel-is-congruence}, we obtain that
            \begin{equation}\label{equ:nPolXn-factorize-modulo-eval=NatXn}
\sfnPol(X_n)/_{\ker(\eval)} \cong \Nat[x_1,\ldots,x_n] \enspace.
\end{equation}
In particular, $\boxtimes/_{\ker(\eval)} = \cdot$, i.e, the Cauchy product on polynomials.
For instance, for every $u \in X_n^*$, we have
\begingroup
\allowdisplaybreaks
\begin{align*}
  \eval\Big((\lM (),()\rM, \lM (),()\rM) \Big)(u)
  &= \Big((p_1 \oplus p_1) \cdot (p_1 \oplus p_1)\Big)(u)\\
  &= \bigplus_{\substack{v,w \in X_n^*:\\u=vw}} (p_1 \oplus p_1)(v) \cdot (p_1 \oplus p_1)(w)\\
  &= \begin{cases}
    (p_1(\varepsilon) + p_1(\varepsilon)) \cdot (p_1(\varepsilon) + p_1(\varepsilon)) & \text{ if $u=\varepsilon$}\\
    0 & \text{ otherwise}
  \end{cases}\\
   &= \begin{cases}
    4 & \text{ if $u=\varepsilon$}\\
    0 & \text{ otherwise} \enspace.
  \end{cases}        
  \end{align*}
\endgroup


\subsection{Lattices and their comparison with strong bimonoids}
\label{sec:lattices} \label{sec:comparison}

\index{lattice}
\index{lattice!bounded}
We recall notions from lattice theory \cite{bir93,gra03} and \cite[Ch.~1]{bursan81}.
A \emph{lattice} is an algebra $(B, \vee, \wedge)$ in which $\vee$ (the \emph{join}) and $\wedge$  (the \emph{meet}) are binary operations,
$(B, \vee)$ and $(B, \wedge)$  are commutative semigroups,
 the operations $\wedge$ and  $\vee$ are idempotent and satisfy the absorption axioms $a\vee(a\wedge b)=a$ and $a\wedge(a\vee b)=a$. The lattice $(B,\vee,\wedge)$ is \emph{bounded} if there exist elements $\mathbb{0}, \mathbb{1} \in B$ such that $\mathbb{0} \vee a = a$ and $\mathbb{1}\wedge a = a$  for every $a\in B$. We denote a bounded lattice also by $(B,\vee,\wedge,\mathbb{0},\mathbb{1})$.

There is an alternative order-theoretic definition of lattice. A  partially ordered set $(B,\le)$ is a \emph{lattice} if for each $a,b \in B$ the elements $\sup_\le\{a,b\}$ and $\inf_\le\{a,b\}$ exist. The lattice $(B,\le)$ is \emph{bounded} if there exist elements $\mathbb{0}, \mathbb{1} \in B$ such that $\inf_{\le}(\0,a) = \0$ and $\sup_{\le}(\1,a) = \1$  for every $a\in B$; in other words, $\0$ and $\1$ are the smallest element and greatest element, respectively, with respect to $\le$. We denote a bounded lattice also by $(B,\le,\0,\1)$.

We recall the well known correspondence between these two definitions.

\begin{theorem} \label{lm:lattice-equivalent-def}  {\rm (cf. \cite[Sect. I.1]{bursan81},  \cite[Ch. I., Thm 8]{bir67}, \cite[Sect. 1.1, Thm 8]{gra78}, \cite[p.~6]{gra03})} Let $B$ be a set. The following two statements hold.
\begin{compactenum}
\item[(1)] Let the algebra $(B,\vee,\wedge)$ be a lattice. We define the partial order $\le$ on $B$ such that
  \[a \le b \ \text{ if } \ a \wedge b = a \ \text{ for every $a,b \in B$} \enspace.
    \]
Then the partially ordered set $(B,\le)$ is a lattice with $\sup\nolimits_\le\{a,b\} =a \vee b$ and $\inf\nolimits_\le\{a,b\} =a \wedge b$ for every $a,b \in B$.

\item[(2)] Let the partially ordered set $(B,\le)$ be a lattice. We define the binary operations $\vee$ and $\wedge$ on $B$ such that
  \[
a \vee b = \sup\nolimits_\le\{a,b\} \ \text{ and } \ a \wedge b = \inf\nolimits_\le\{a,b\}  \ \text{ for every $a,b \in B$} \enspace.
\]
Then the algebra $(B,\vee,\wedge)$ is a lattice. 
\end{compactenum}
\end{theorem}

Clearly, if we  apply the transformation in Theorem \ref{lm:lattice-equivalent-def}(1) to  the lattice $(B,\vee,\wedge)$ and then apply to the resulting lattice $(B,\le)$ the transformation in Theorem \ref{lm:lattice-equivalent-def}(2), then we reobtain the lattice $(B,\vee,\wedge)$. A similar invariant occurs if we apply the transformation of (2) followed by the transformation of (1) to a lattice $(B,\le)$.

Another consequence of Theorem \ref{lm:lattice-equivalent-def} is the following. Let $(B,\vee,\wedge)$ and $(B,\le)$ be bounded lattices such that the transformations of Theorem \ref{lm:lattice-equivalent-def} transform the one into the other. Moreover, let $\0$ and $\1$ be two elements of $B$. Then the following equivalence holds: $(B,\vee,\wedge,\0,\1)$ is a bounded lattice iff $(B,\le,\0,\1)$ is a bounded lattice.

The lattice $(B, \vee, \wedge)$ is \emph{distributive} if $\wedge$ is distributive over $\vee$ and $\vee$ is distributive over~$\wedge$.  If fact, the first of the previous two conditions holds if and only if the second one holds \cite[Lm.~4.3]{davpri12}.   Hence it suffices to require one of them.

\begin{observation}\rm \label{obs:bounded-lattice-is-biloc-fin-strong-bimonoid}
\begin{compactenum} 
   \item[(1)]  For each lattice $(B,\vee,\wedge)$, the semigroups $(B,\vee)$ and $(B,\wedge)$ are locally finite.
  \item[(2)] Each bounded lattice $(B,\vee,\wedge,\0,\1)$ is a bi-locally finite and commutative strong bimonoid.
  \item[(3)] Each distributive bounded lattice $(B,\vee,\wedge,\0,\1)$ is a locally finite and commutative semiring.
    \end{compactenum}
  \end{observation}
  \begin{proof} 
   Proof of (1): Let $(B,\vee,\wedge)$ be a lattice. The statement follows from Observation \ref{obs:commutative+idempotent-implies-locally finite}
  because both $\vee$ and $\wedge$ are commutative and idempotent.

  \
  
  Proof of (2):  Let $\B=(B,\vee,\wedge,\mathbb{0},\mathbb{1})$ be a bounded lattice. Since we have
$\mathbb{0} \wedge a = \mathbb{0} \wedge (\mathbb{0} \vee a)=\mathbb{0}$ for every $a\in B$ (by absorption law), the algebra $\B$ is a strong bimonoid.
  Obviously, it is commutative. Moreover, it follows from  Statement (1) that both $(B,\vee,\0)$ and $(B,\wedge,\1)$ are locally finite.

  \
  
Proof of (3): This follows from  (2) and Observation \ref{obs:zero-sum-free-property}(8).
\end{proof}

  \begin{observation}\rm \label{obs:bounded-lattice=lattice-cap-strong-bimonoid}
    Let $\B= (B,\oplus,\otimes,\mathbb{0},\mathbb{1})$ be an algebra. Then the following two statements hold. 
  \begin{compactenum}
  \item[(1)] $\B$ is a bounded lattice if and only if $(B,\oplus,\otimes)$ is a lattice and $\B$ is a strong bimonoid.
    \item[(2)]  $\B$ is a distributive bounded lattice if and only if $(B,\oplus,\otimes)$ is a lattice and $\B$ is a semiring.\hfill $\Box$
    \end{compactenum}
  \end{observation}
  
We recall that a bounded lattice $(B, \vee, \wedge,\0,\1)$ (which is a particular commutative strong bimonoid) is \emph{$\sigma$-complete} if the monoid $(B,\vee,\0)$ is $\sigma$-complete (cf. Section \ref{sec:strong-bimonoid}).  We note that, for bounded lattices,  our concept is more general than $\sigma$-lattice in \cite[Ch. 9]{bir67} because in a $\sigma$-lattice both the join and the meet exist for arbitrary countable index sets. Moreover, $\sigma$-lattice is more general than complete lattice \cite{bir67,bir93,gra78,gra03} because in a complete lattice the join and the meet exist for arbitrary index sets. Lastly, we note that if meet exists 
for arbitrary index sets, then also join exists  for arbitrary index sets and vice versa \cite[Ch. V, Thm. 3]{bir67} \cite[Sect. 1.3, Lm. 14]{gra78}.

\begin{example}\label{ex:lattices}\rm $\ $
  \begin{enumerate}
    \item \label{def:lattice-of-subsets} Let $A$ be a set. The $\sigma$-complete semiring $\Powerset_A=(\cP(A),\cup,\cap,\emptyset,A)$ is a $\sigma$-complete distributive lattice.

 \item The $\sigma$-complete semiring $\Natmaxmin=(\mathbb{N}_\infty,\max,\min,0,\infty)$ is a  $\sigma$-complete  distributive lattice.
  
\index{Nfive@$\Nfive$}  
\item\label{ex:N-5-label}  \cite[Fig.~2, p.14]{gra03}, \cite[Fig. 5]{bursan81} Let $N_5=\{o,a,b,c,i\}$ be a set with five elements. We consider two binary operations $\vee$ and $\wedge$ such that the following requirements hold:
\begin{align*}
 a \wedge b = b, \ \ b \vee c = i, \ \ \text{ and } \ \ a \wedge c = o. 
\end{align*} 
The definition of the values of $\vee$ and $\wedge$ for other combinations of arguments is given by the unique extension of the above requirements  such that $\Nfive=(N_5,\vee,\wedge,o,i)$
is a bounded lattice (cf. Figure \ref{fig:lattices-N5-M3-Graetzer}). This lattice is not distributive.

\index{Mthree@$\Mthree$}
\item\label{ex:M-3-label} \cite[Fig.~2, p.14]{gra03}, \cite[Fig. 5]{bursan81} Let $M_3=\{o,a,b,c,i\}$ be a set with five elements. We consider two binary operations $\vee$ and $\wedge$ such that the following requirements hold:
\begin{align*}
a \wedge b = a \wedge c = b \wedge c = o \ \ \text{ and } \ \
a \vee b = a \vee c = b \vee c = i. 
\end{align*}
Similarly, the definition of the values of $\vee$ and $\wedge$ for other combinations of arguments is given by the unique extension of the above requirements  such that $\Mthree=(M_3,\vee,\wedge,o,i)$
is a bounded lattice (cf. Figure \ref{fig:lattices-N5-M3-Graetzer}). 
 This lattice is not distributive.

  \index{lattice!orthomodular}
  It  is well known that there exist large sets of lattices that are not distributive \cite{gra03}, e.g., orthomodular lattices. In fact, a lattice is non-distributive if and only if it contains at least one of the two lattices $\Nfive$ and $\Mthree$ as a sublattice \cite[Thm.~1 on p.~80]{gra03}.

\begin{figure}
\begin{center}
 \centering
  \begin{tikzpicture}[scale=0.75, every node/.style={transform shape,scale=1/0.75},
					mycircle/.style={circle,draw,inner sep=0.05cm}]
\node[mycircle, label=left:{$a$}] (a5) {};
\node[mycircle, label=left:{$b$}, below= 1.25cm of a5] (b5) {};
\node[mycircle, label={$i$}, above right= 0.75cm and 1cm of a5] (i5) {};
\node[mycircle, label=below:{$o$}, below right= 0.75cm and 1cm of b5] (o5) {};
\node[mycircle, label=right:{$c$}, right= 1.5cm of $(i5)!0.5!(o5)$] (c5) {};
\draw (a5) -- (b5) -- (o5) -- (c5) -- (i5) -- (a5);
\node[below left= 0.75cm and -0.5cm of c5] {$\Nfive$};

\node[mycircle, label=left:{$a$}, right= 2cm of c5] (a3) {};
\node[mycircle, label=right:{$b$}, right= 1.25cm of a3] (b3) {};
\node[mycircle, label=right:{$c$}, right= 1.25cm of b3] (c3) {};
\node[mycircle, label=above:{$i$}] (i3) at (b3|-i5) {};
\node[mycircle, label=below:{$o$}] (o3) at (b3|-o5) {};
\draw (i3) -- (a3) -- (o3) -- (c3) -- (i3) -- (b3) -- (o3);
\node[below left= 0.75cm and -0.6cm of c3] {$\Mthree$};

\end{tikzpicture}
\end{center}
\caption{\label{fig:lattices-N5-M3-Graetzer} The lattices $\Nfive$ and $\Mthree$ in Figure 2 of \cite[p.14]{gra03}.}
\end{figure}

\item We consider the algebra $(\mathbb{R},\max,\min)$ with maximum and minimum based on the usual $\le$ ordering on the set of real numbers. The algebra $(\mathbb{R},\max,\min)$ is a lattice. It is not bounded (and hence not $\sigma$-complete). Moreover, both $\max$ and $\min$ are extremal.

\item 
\index{lattice!residuated}
A \emph{residuated lattice} is an algebra $\B=(B, \vee,\wedge, \otimes, \rightarrow, \mathbb{0}, \mathbb{1})$ such that
\begin{compactenum}
\item[(L1)] $(B,\vee, \wedge, \mathbb{0}, \mathbb{1})$ is a bounded lattice,
\item[(L2)] $(B, \otimes, \mathbb{1})$ is a commutative monoid with the unit $\mathbb{1}$, and
\item[(L3)] $\rightarrow$ is a binary operation, and $\otimes$ and $\rightarrow$ form an adjoint pair, i.e., they satisfy the adjunction property: for all $a, b, c \in B$,
\(a \otimes b \le c  \ \text{ if and only if } \  a \le (b \rightarrow c)\). (Here $\le$ is the partial order on $B$ defined, as for lattices,  by $a\le b$ iff $a=a\wedge b$ for every $a,b\in B$.) 
\end{compactenum}
If, additionally, $\B$ is a $\sigma$-complete lattice, then $\B$ is called a \emph{$\sigma$-complete residuated lattice}.\index{lattice!$\sigma$-complete residuated}

The operations $\otimes$ and $\rightarrow$ are called \emph{multiplication} and \emph{residuum}, respectively. The algebra $(B, \vee, \otimes, \mathbb{0}, \mathbb{1})$ is a semiring (cf., e.g., \cite[Slide~11]{gal08}),
it is  called the \emph{semiring reduct of $\B$}. Thus, in particular, the semiring reduct of each residuated lattice is a strong bimonoid. 

\item The $\sigma$-complete Boolean algebras of \cite{loo47} are $\sigma$-complete lattices.
  
\item \index{lattice!finite chain}  A \emph{finite chain} is a finite bounded lattice $(L,\le,\0,\1)$ where $\le$ is a linear order. Trivially, each finite chain is distributive. 
 
\item \label{ex:FL2+2} \index{lattice!$\sfFL$}  The bounded lattice
$\sfFL=(\FL,\vee,\wedge,\0,\1)$ (cf. \cite{rol58}, \cite[p.~361]{gra03}, \cite[Fig.~3a]{rivwil09})
is free in the set of all bounded lattices with generating set $\{a,b,c,d\}$, where $a < b$ and $c < d$. Thus, $\sfFL$ is generated by two chains, each of which has two elements (cf. Figure \ref{fig:FL2+2}). The upper bound is $\1= b\vee d$ and the lower bound is $\0= a \wedge c$. This lattice is infinite \cite[Thm.~2]{rol58}, hence it is not locally finite.  Thus,
by Observations \ref{obs:bounded-lattice-is-biloc-fin-strong-bimonoid}(2) and  \ref{obs:zero-sum-free-property}(8), it is not distributive; this also follows from the fact that $\sfFL$ contains $\Nfive$ as sublattice.
    \hfill$\Box$
\end{enumerate}
  \end{example}

\begin{figure}[t]
\centering
	
\begin{tikzpicture}[node distance=0.9cm and 0.2cm, scale=1, every node/.style={transform shape}] 
\coordinate (sw dist) at (-1.5,-0.75); 
\coordinate (se dist) at (1.5,-0.75);  
\coordinate (node dist) at (0,0.9);  
\coordinate (vert dist) at (0,0.65); 
\rhombus[above = 0cm of {(0,0)}]{m0}; 
\rhombus[below = of m0]{m1};
\rhombus[below = of m1]{m2};
\rhombus[below = of m2]{m3};
\rhombus[below = 5.5cm of m3]{m7};
\rhombus[below = of m7]{m8};
\rhombus[below = of m8]{m9};
\rhombus[below = of m9]{m10};

\node[left = 0.62cm  of m0]{};
\node[right = 0.62cm of m0]{};

\node[left = 0.62cm of m1]{};
\node[right = 0.62cm of m1]{};

\node[left = 0.62cm of m2]{};
\node[right = 0.62cm of m2]{};

\node[left = 0.62cm of m3]{};
\node[right = 0.62cm of m3]{};

\node[small circle,fill=black, label=left:${b}$] (l0) at ($(m0 west)+(sw dist)$) {};
\node[small circle] (l1) at ($(m1 west)+(sw dist)$) {};
\node[small circle] (l2) at ($(m2 west)+(sw dist)$) {};
\node[small circle] (l3) at ($(m3 west)+(sw dist)$) {};
\node[small circle,fill=black, label=right:${d}$] (r0) at ($(m0 east)+(se dist)$) {};
\node[small circle] (r1) at ($(m1 east)+(se dist)$) {};
\node[small circle] (r2) at ($(m2 east)+(se dist)$) {};
\node[small circle] (r3) at ($(m3 east)+(se dist)$) {};
\node[small circle] (l7) at ($(m7 west)-(se dist)$) {};
\node[small circle] (l8) at ($(m8 west)-(se dist)$) {};
\node[small circle] (l9) at ($(m9 west)-(se dist)$) {};
\node[small circle,fill=black, label=left:${a}$] (l10) at ($(m10 west)-(se dist)$) {};
\node[small circle] (r7)  at ($(m7 east)-(sw dist)$) {};
\node[small circle] (r8)  at ($(m8 east)-(sw dist)$) {};
\node[small circle] (r9) at ($(m9 east)-(sw dist)$) {};
\node[small circle,fill=black, label=right:${c}$] (r10) at ($(m10 east)-(sw dist)$) {};

\rhombus[above = 0cm of $(l3)!0.5!(l7)$] {l5};
\node[small circle] (l4) at ($(l5)+(vert dist)$) {};
\node[small circle] (l6) at ($(l5)-(vert dist)$) {};
\node[small circle] (m5t) at ($(l5 east)-(sw dist)$) {};
\node[small circle] (m4) at ($(m5t)+(node dist)$) {};
\node[small circle] (m5b) at ($(l5 east)+(se dist)$) {};
\node[small circle] (m6) at ($(m5b)-(node dist)$) {};
\rhombus[above = 0cm of $(r3)!0.5!(r7)$] {r5};
\node[small circle] (r4) at ($(r5)+(vert dist)$) {};
\node[small circle] (r6) at ($(r5)-(vert dist)$) {};

\draw (l0) -- (l1) -- (l2) -- (l3)
      (l4) -- (l5 north)
      (l5 south) -- (l6)
      (l7) -- (l8) -- (l9) -- (l10);
\draw (r0) -- (r1) -- (r2) -- (r3)
      (r4) -- (r5 north)
      (r5 south) -- (r6)
      (r7) -- (r8) -- (r9) -- (r10);
\draw (m0 south) -- (m1 north)
      (m1 south) -- (m2 north)
      (m2 south) -- (m3 north)
      (m4) -- (m5t)
      (m5b) -- (m6)
      (m7 south) -- (m8 north)
      (m8 south) -- (m9 north)
      (m9 south) -- (m10 north);
\draw[dashed] 
      (l3) -- (l4) (r3) -- (r4)
      (m3 south) -- (m4)
      (m6) -- (m7 north)
      (l6) -- (l7) (r6) -- (r7);
\draw (l0) -- (m0 west) (m0 east) -- (r0)
      (l1) -- (m1 west) (m1 east) -- (r1)
      (l2) -- (m2 west) (m2 east) -- (r2)
      (l3) -- (m3 west) (m3 east) -- (r3)
      (l4) -- (m4) -- (r4)
      (l5 east) -- (m5t) -- (r5 west)
      (l5 west) -- (m5b) -- (r5 east)
      (l6) -- (m6) -- (r6)
      (l7) -- (m7 west) (m7 east) -- (r7)
      (l8) -- (m8 west) (m8 east) -- (r8)
      (l9) -- (m9 west) (m9 east) -- (r9)
      (l10) -- (m10 west) (m10 east) -- (r10);

\end{tikzpicture}

  \caption{\label{fig:FL2+2} The bounded lattice $\sfFL$ is free in the set of all bounded lattices with generating set $\{a,b,c,d\}$, where $a < b$ and $c < d$. The upper bound of $\sfFL$ is $b \vee d = \1$, and its lower is $a \wedge c= \0$.}
  \end{figure}

We note that  semiring-reducts of semi-lattice ordered monoids and of $\sigma$-complete residuated lattices, and Brouwerian lattices are commutative semirings.

In Figure \ref{fig:Euler-diagram-extended}, we have placed five sets of algebras into a diagram. The sets are:
\begin{compactitem}
\item the set $\cC_1$ of all lattices,
\item the set $\cC_2$ of all strong bimonoids, 
  \item the set $\cC_3$ of all semirings,
\item the set $\cC_4$ of all bi-locally finite  strong bimonoids, and
\item the set $\cC_5$ of all locally finite strong bimonoids.
  \end{compactitem}

  \begin{observation}\rm \label{obs:Euler-diagram-including-loc-finite} Figure \ref{fig:Euler-diagram-extended} shows the Euler diagram of the sets $\cC_1$, $\cC_2$, $\cC_3$, $\cC_4$, and $\cC_5$. 
  \end{observation}

\usetikzlibrary{patterns}

\begin{figure}[t]
    \begin{center}
\def\firstellipse{(0,0) ellipse (3cm and 4cm)} 
\def\secondellipse{(4.25,0) ellipse (5cm and 2.9cm)} 
\def\thirdellipse{(4.25,0) ellipse (2cm and 1cm)} 
\def\fourthellipse{(2.25,0) ellipse (3cm and 2.3cm)} 
\def\fifthellipse{(2.95,0) ellipse (2.3cm and 1.3cm)} 
\def\legendOne{(-1,-5.5) rectangle (-0.5,-5)} 
\def\legendTwo{(-1,-6.25) rectangle (-0.5,-5.75)} 
\def\legendThree{(-1,-7) rectangle (-0.5,-6.5)} 

\begin{tikzpicture}

	\begin{scope} 
		\clip \firstellipse;
		\fill[black] \secondellipse;
	\end{scope}
	\fill[white] \fourthellipse;

	\begin{scope} 
		\clip \thirdellipse;
		\fill[black] \fourthellipse;
	\end{scope}
	\fill[white] \fifthellipse;

	\begin{scope} 
		\clip \firstellipse;
		\draw[pattern=north west lines] \secondellipse;
	\end{scope}

	\begin{scope} 
		\clip \firstellipse;
		\fill[white]\thirdellipse; 
	\end{scope}
	\begin{scope} 
		\clip \firstellipse;
		\draw[pattern=north east lines]\thirdellipse;
	\end{scope}

	\begin{scope} 
		\clip \thirdellipse;
		\draw[pattern=dots] \fifthellipse;
	\end{scope}

	\draw\firstellipse node[label={[font=\small, text=darkgray, label distance=3.9cm]90:$\mathcal C_1$}, 
	label={[text=blue, label distance=1.5cm]180:$\B_1$}] {}; 
	\draw\secondellipse node[label={[font=\small, text=darkgray, label distance=2.8cm]90:$\mathcal C_2$},
	label={[text=blue, label distance=2cm]80:$\B_9$}] {}; 
	\draw\thirdellipse node[label={[font=\small, text=darkgray, label distance=1.07cm]40:$\mathcal C_3$}, 
	label={[text=blue, inner sep=1pt, fill=white, rounded corners=3pt]0:$\B_6$}, 
	label={[text=blue, label distance=1.18cm, inner sep=1pt, fill=white, rounded corners=3pt]180:$\B_5$}, 
	label={[text=blue, label distance=0.85cm]0:$\B_{10}$}] {}; 
	\draw\fourthellipse node[label={[font=\small, text=darkgray, label distance=2.1cm]85:$\mathcal C_4$},
	label={[text=blue, label distance=1.5cm, inner sep=1pt, fill=white, rounded corners=3pt]140:$\B_2$}, 
	label={[text=blue, label distance=1.7cm]50:$\B_8$}] {}; 
	\draw\fifthellipse node[label={[font=\small, text=darkgray, label distance=1.12cm]87.5:$\mathcal C_5$}, 
	label={[text=blue, label distance=1cm, inner sep=1pt, fill=white, rounded corners=3pt]180:$\B_4$}, 
	label={[text=blue, label distance=0.8cm, inner sep=1pt, fill=white, rounded corners=3pt]150:$\B_3$},
	label={[text=blue, label distance=0.7cm]87.9:$\B_7$}] {};

	\draw[pattern=north west lines]\legendOne node[label={[font=\small, label distance=0.1cm]-10:bounded lattices}] {};
	\draw[pattern=north east lines]\legendTwo node[label={[font=\small, label distance=0.1cm]-10:distributive bounded lattices}] {};
	\draw[pattern=dots]\legendThree node[label={[font=\small, label distance=0.1cm]-10:bi-locally finite semirings = locally finite semirings}] {};
      \end{tikzpicture}
\end{center}
\caption{\label{fig:Euler-diagram-extended} The Euler diagram of the sets $\cC_1$ of all lattices, $\cC_2$ of all strong bimonoids, $\cC_3$ of all semirings,
$\cC_4$ of all bi-locally finite  strong bimonoids, and $\cC_5$ of all locally finite strong bimonoids.  Each black area denotes the empty set. The algebras $\B_1-\B_{10}$ are shown in the proof of Observation \ref{obs:Euler-diagram-including-loc-finite}.}
\end{figure}

\begin{proof} By  Observation \ref{obs:zero-sum-free-property}(8), we have $\cC_3 \cap \cC_4 = \cC_3 \cap \cC_5$.
Moreover, the following subset relationships hold:
\begin{compactitem}
\item $\cC_3 \subseteq \cC_2$,
\item $\cC_5 \subseteq \cC_4$,
\item $\cC_4 \subseteq \cC_2$, 
\item $\cC_1 \cap \cC_3 \subseteq \cC_5$, and 
\item $\cC_1 \cap \cC_2 \subseteq \cC_4$.
\end{compactitem}
Next,  for each nonempty area of the diagram, we show a witness (i.e., an element of that area).

$\cC_1 \setminus \cC_2 \not= \emptyset$: Let $\B_1$ be the lattice $(\mathbb{R},\max,\min)$. Since $\B_1$ is not bounded, it is not a strong bimonoid.

$(\cC_1 \cap \cC_4) \setminus \cC_5 \not= \emptyset$ \cite{cze19}: We let $\B_2$ be the  bounded lattice $\sfFL$ in Example \ref{ex:lattices}(\ref{ex:FL2+2}) (cf. Figure~\ref{fig:FL2+2}).
This lattice is infinite and generated by the two finite chains $a < b$ and $c < d$. Hence it is not locally finite.

$(\cC_1 \cap \cC_5) \setminus \cC_3 \not= \emptyset$: We let $\B_3=\Nfive$ and $\B_4=\Mthree$. These are finite bounded lattices; hence, they are locally finite and commutative strong bimonoids. But neither of them is distributive.

$\cC_1 \cap \cC_3 \not= \emptyset$: We let $\B_5$ be the semiring $\Powerset_A=(\cP(A),\cup,\cap,\emptyset,A)$ for some set $A \not= \emptyset$. This is a bounded lattice (cf. Example \ref{ex:lattices}(1)).

$(\cC_3 \cap \cC_4) \setminus \cC_1 \not= \emptyset$: We let $\B_6$ be the ring $\Intfour= (\{0,1,2,3\},+_4,\cdot_4,0,1)$ defined in Example \ref{ex:semirings}(\ref{def:ring-Zmod4Z}).  This is a finite (hence also locally finite)  semiring but not a lattice, because e.g. $+_4$ is not idempotent.

$\cC_5  \setminus (\cC_1 \cup \cC_3) \not= \emptyset$: We let $\B_7$ be the strong bimonoid $(\{0,1,2,3,\infty\},+_4,\min,0,\infty)$ where $+_4$ is defined as for $\B_6$ (extended in the usual way to $\infty$). Since $\B_7$ is finite, it is also locally finite. 
$\B_7$ is neither a lattice (because $+_4$ is not idempotent) nor a semiring (because, e.g., $\min(1,1+_41) \not= \min(1,1) +_4 \min(1,1)$). 

$\cC_4 \setminus (\cC_1 \cup \cC_3 \cup \cC_5) \not= \emptyset$ \cite{dro19}: Let $\B_8$ be  the strong bimonoid $\Trunc_\lambda=(B,\oplus,\odot,0,1)$ of Example \ref{ex:strong-bimonoids}(\ref{ex:interval-strong-bimonoid}) (also cf. \cite[Ex.~2.1(2)]{drovog12}). Then $\B_8$ is a bi-locally finite and commutative strong bimonoid, but not a locally finite strong bimonoid. Moreover, $\B_8$ is neither a semiring nor a lattice.

$\cC_2 \setminus (\cC_1 \cup \cC_3 \cup \cC_4) \not= \emptyset$: We let $\B_9$ be the tropical bimonoid $\TropBM$. This is neither a lattice (because $+$ is not idempotent) nor a bi-locally finite strong bimonoid  nor a semiring (because, e.g., $\min(1,1+1) \not= \min(1,1) + \min(1,1)$). 

$\cC_3 \setminus (\cC_1 \cup \cC_4)$: We let $\B_{10}$ be the semiring $\Nat=(\mathbb{N},+,\cdot,0,1)$ of natural numbers. This is neither bi-locally finite nor it is a lattice.
\end{proof}

 We finish this subsection with a characterization theorem. It concerns the smallest set which is closed under certain operations.  Such a set can be characterized by using a combination of the fixpoint theorem of Tarski  \cite[Thm.~1]{tar55} and the first fixpoint theorem of Kleene \cite[p.~348]{kle62} (cf. \cite{lasnguson82}). We start with some preparations.
 
  First, we recall some notions and observations from \cite{davpri12}. Let $A$ be a set. The pair $(\cP(A),\subseteq)$ is a $\sigma$-complete lattice in the order-theoretical sense; the corresponding algebraic version is the $\sigma$-complete semiring $\Powerset_A= (\cP(A),\cup,\cap,\emptyset,A)$ where $U \cup U'$ and $U \cap U'$ are the supremum of $U$ and infimum of $U'$, respectively, with respect to $\subseteq$.

Let $D \subseteq \cP(A)$ such that $D \ne \emptyset$. We call $D$ \emph{directed} if, for each $U,U' \in D$, there exists $V \in D$ such that $U \subseteq V$ and $U' \subseteq V$.
  
Let  $f: \cP(A) \to \cP(A)$ be a mapping. We say that
\begin{compactitem}
\item $f$ is \emph{order-preserving} if, for every $U,U' \in \cP(A)$, the inclusion $U \subseteq U'$ implies $f(U) \subseteq f(U')$,
\item $f$ is \emph{continuous} if, for each directed set $D \subseteq \cP(A)$ we have that $f(D)$ is a directed set and $f(\bigcup (U \mid U \in D) ) = \bigcup ( f(U) \mid U \in D)$,
\item $U \in \cP(A)$ is a \emph{fixpoint} of $f$ if $f(U)=U$, and
  \item $U$ is the \emph{least fixpoint} of $f$ if it is a fixpoint  and, for each fixpoint $U'$, the inclusion $U\subseteq U'$ holds.
  \end{compactitem}
We note that each continuous mapping $f: \cP(A) \to \cP(A)$ is order-preserving \cite[Lm.~8.7(ii)]{davpri12}.

\begin{theorem}\label{thm:Knaster-Tarski} {\rm (cf. least fixpoint theorems \cite[Thm.~2.35]{davpri12} and \cite[Thm.~8.15]{davpri12})}  Let $A$ be a set and $f: \cP(A) \to \cP(A)$ be a continuous mapping. Then:
  \begin{eqnarray}\label{eq:Knaster-Tarski}
    \begin{aligned}
  \bigcap(U \in \cP(A) \mid f(U) \subseteq U) = \bigcup(f^n(\emptyset) \mid n \in \mathbb{N}) &\\
  \text{and $\bigcup(f^n(\emptyset) \mid n \in \mathbb{N})$ is the least fixpoint of  $f$.} &
  \end{aligned}
\end{eqnarray}
\end{theorem}
Intuitively, the left-hand side of \eqref{eq:Knaster-Tarski} reflects the definition of a set as the smallest set which satisfies a closure property (formalized  by $f$), and the right-hand side offers a kind of slicing (or stratification) of this set; that slicing can be used in a proof by induction on $\mathbb{N}$.

\label{page:general-convention-on-B}
\begin{quote}\emph{In the rest of this book, $\B$ will denote an arbitrary strong bimonoid $(B, \oplus, \otimes, \0, \1)$,  if not specified otherwise.}
\end{quote}


        \subsection{Semimodules and vector spaces}
        \label{sec:semimodules}

             \index{B@$\B$-semimodule}
        \label{page:general-convention-on-coefficients-for-semimodules}
\begin{quote}\emph{In this subsection, let $\B=(B,\oplus,\otimes,\0,\1)$ be an arbitrary commutative semiring, unless specified otherwise.}
\end{quote}

\index{semimodule}
\paragraph{Semimodules, linear mappings, linear forms, and multilinear operations.}
Let $\V=(V,+,0)$ be a commutative monoid\footnote{We note that the $\0$ of the semiring $B$ can be different from the $0$ of the monoid $V$.}. Moreover, let $\cdot: B\times V \rightarrow V$ be a mapping, called \emph{scalar multiplication}. Then $\V$ is a {\em (left-)$\B$-semimodule (via $\cdot$)} if the following laws hold:
\begin{eqnarray}
(b \otimes b') \cdot v = b \cdot (b' \cdot v)  \label{SM1}\\
b \cdot (v + v') = (b\cdot v) + (b \cdot v')  \label{SM2}\\
(b \oplus b') \cdot v = (b \cdot v) + (b' \cdot v)  \label{SM3}\\
\1 \cdot v = v  \label{SM4}\\
b \cdot 0 = \0 \cdot v = 0 \label{SM5}
\end{eqnarray}
for every $b,b' \in B$ and $v,v' \in V$ (cf. p.149 of \cite{gol99}). 

In particular, $(B,\oplus,\0)$ is a $\B$-semimodule via scalar multiplication $\otimes$. 

In this book, we will deal with several $\B$-semimodules. For each of them, we will use $\cdot$ to denote its scalar multiplication. It will always be clear from the context, which type this scalar multiplication has. Sometimes we drop $\cdot$ and write, e.g., $bv$ instead of $b \cdot v$.

\label{page:B-semimodule-as-universal-algebra}
We can view a $\B$-semimodule  $\V=(V,+,0)$ via $\cdot$ as an algebra (in the sense of Subsection~\ref{ssec:universal-algebras}) as follows. We define the index set $I= \{+,0\} \cup \{(b\cdot)\mid b \in B\}$ and the mapping $\eta: I \to \mathrm{Ops}(V)$ such that
\begin{compactitem}
\item $\eta(+) = +$ and $\eta(0)() = 0$, and
\item for every $b \in B$ and $v \in V$, we let $\eta((b \cdot))(v) = b \cdot v$ for each $b \in B$.
  \end{compactitem}
Then $\V$ is the algebra $(V,\eta)$. Moreover, $\V$ is of type $\tau:I \to \mathbb{N}$ where $\tau(+) = 2$, $\tau(0) = 0$, and $\tau((b\cdot)) = 1$ for each $b \in B$.

Thus, in particular, we can use the concept of ``the $\B$-semimodule $\V$  is generated by some set'' as it is defined for algebras. 

\index{linear mapping}
Let $\V=(V,+,0)$ and $\V'=(V',+',0')$ be $\B$-semimodules. A mapping $f: V\to V'$ is \emph{linear (from $\V$ to $\V'$)} if, for every $b_1,b_2 \in B$ and $u,v\in V$, the equation
\[f(b_1\cdot u+b_2\cdot v)=b_1 \cdot f(u)+'b_2 \cdot f(v)
\]
holds. The semimodules $\V$ and $\V'$  are  \emph{isomorphic} if there exists a bijective linear mapping from $\V$ to $\V'$.

\index{linear form}
Let $\V=(V,+,0)$ be a $\B$-semimodule. A {\em linear form (over $\V$)} is a linear mapping from the $\B$-semimodule $\V=(V,+,0)$ to the $\B$-semimodule  $(B,\oplus,\0)$. That is, a linear form is a mapping $\gamma : V\to B$ such that for every $b,b'\in B$ and $v,v' \in V$, we have $\gamma(b\cdot v + b'\cdot v')= b\otimes \gamma(v) \oplus b'\otimes \gamma(v')$. We note that, by viewing semimodules as algebras, linear mappings and linear forms are nothing else but algebra homomorphisms.

\index{multilinear}
Let $k \in \mathbb{N}_+$. A $k$-ary operation $\omega: V^k \rightarrow V$ is called {\em multilinear (over $\V$)} if
\begin{eqnarray}
  \begin{aligned}\label{ml-Omega}
  &\omega(v_1,\ldots,v_{i-1},\ b \cdot v + b' \cdot v' \ ,v_{i+1},\ldots,v_k) = &\\
  &b \cdot \omega(v_1,\ldots,v_{i-1},v,v_{i+1},\ldots,v_k) + 
    b' \cdot  \omega(v_1,\ldots,v_{i-1},v',v_{i+1},\ldots,v_k) &
    \end{aligned}
\end{eqnarray}
holds for every $i \in [k]$, $b,b' \in B$, and $v,v',v_1,\ldots,v_k \in V$.
\index{linear@${\cal L}(\V^k,\V)$}
We denote by ${\cal L}(\V^k,\V)$ the set of all $k$-ary multilinear operations over $\V$. 

\label{page:why-commutativiy-for-multilinearity}
We explain why commutativity of $\B$ is assumed in the above. For this we consider the expression $\omega(b_1 \cdot v_1, b_2 \cdot v_2)$ and observe that we can evaluate  it in two different ways:
\[
  \begin{array}{cllcll}
    &\omega(b_1 \cdot v_1, b_2 \cdot  v_2)  & & & \omega(b_1 \cdot v_1, b_2 \cdot  v_2) & \\
    = & b_1 \cdot \omega(v_1,b_2\cdot v_2) & \text{(by \eqref{ml-Omega}($i=1$))} \hspace*{16mm}  &
                                                           = & b_2 \cdot  \omega(b_1 \cdot v_1,v_2) & \text{(by \eqref{ml-Omega}($i=2$))}\\
      = & b_1 \cdot (b_2 \cdot  \omega(v_1,v_2)) & \text{(by \eqref{ml-Omega}($i=2$))}  &
                                                                            = & b_2 \cdot (b_1 \cdot  \omega(v_1,v_2)) & \text{(by \eqref{ml-Omega}($i=1$))}\\
    = & (b_1 \otimes b_2) \cdot \omega(v_1,v_2) & \text{(by \eqref{SM1})} &
                                            = & (b_2 \otimes b_1) \cdot  \omega(v_1,v_2) & \text{(by \eqref{SM1})}
    \end{array}
  \]
  Then commutativity of $\B$ guarantees the well-definedness of multilinearity.

\index{endomorphism}
\index{linear@$\cL(\V,\V)$}
We note that each linear mapping from $\V$ to $\V$  is a unary multilinear operation over $\V$.
Thus, $\cL(\V,\V)$ is the set of all linear mappings from $\V$ to $\V$. Each element of $\cL(\V,\V)$ is an \emph{endomorphism}. Since endomorphisms are closed under composition,  $(\cL(\V,\V),\circ,\id_V)$ is a monoid (cf. \cite[p.~384]{mooyaq98}).

\begin{observation}\label{obs:generalization-closure}\rm
Let $D \subseteq V$ and $\omega$ be an operation in ${\cal L}(\V^k,\V)$ for some $k\in \mathbb{N}$. Then $B\cdot D$ is closed under $\omega$ if and only if
\begin{equation}\label{equ:BD-is-closed-iff-D-is-covered}
  \text{for every $d_1,\ldots,d_k \in D$ we have $\omega(d_1,\ldots,d_k) \in B\cdot D$}
  \end{equation}
\end{observation}
\begin{proof} Let $B \cdot D$ be closed under $\omega$. By using \eqref{SM4}, we have $D \subseteq B \cdot D$. Thus, \eqref{equ:BD-is-closed-iff-D-is-covered} holds. To prove the other implication, let  $b_1\cdot d_1,\ldots,b_k\cdot d_k \in B\cdot D$. Then
\begingroup
\allowdisplaybreaks
\begin{align*}
  \omega(b_1\cdot d_1,\ldots,b_k\cdot d_k)
  &= (b_1 \otimes \cdots \otimes b_k) \cdot \omega(d_1,\ldots,d_k)
  \tag{because $\omega$ is multilinear}\\
  &= (b_1 \otimes \cdots \otimes b_k) \cdot (b \cdot d)
  \tag{for some $b\in B$ and $d \in D$, by \eqref{equ:BD-is-closed-iff-D-is-covered}}\\
  &= (b_1 \otimes \cdots \otimes b_k\otimes b) \cdot d
    \tag{ by \eqref{SM1}}\enspace.
\end{align*}
\endgroup
\end{proof}

\label{p:universal-algebra-view-on-semimodules-with-algebra}
  \paragraph{Semimodules equipped with a  $\Sigma$-algebra.}
  \index{SigmaBsemimodules@$(\Sigma,\B)$-semimodules}
  Let $\Sigma$ be a ranked alphabet.
A \emph{$(\Sigma,\B)$-semimodule} is a pair $(\V,\mu)$ where
\begin{compactitem}
\item $\V=(V,+,0)$ be a $\B$-semimodule via $\cdot$,
\item $(V,\mu)$ is a $\Sigma$-algebra such that, for every $k\in \mathbb{N}$ and  $\sigma\in \Sigma^{(k)}$, we have  $\mu(\sigma)\in {\cal L}(\V^k,\V)$, i.e., 
$\mu(\sigma)$ is a $k$-ary multilinear operation over $\V$.
\end{compactitem}
We denote the pair $(\V,\mu)$  also in the form $(V,+,0,\mu)$.

We can view a $(\Sigma,\B)$-semimodule  $(V,+,0,\mu)$ as an algebra (in the sense of Subsection~\ref{ssec:universal-algebras}) as follows. We define the index set $I= \{+,0\} \cup \{(b\cdot)\mid b \in B\} \cup \Sigma$ and the mapping $\eta: I \to \mathrm{Ops}(V)$ such that
\begin{compactitem}

\item $\eta(+) = +$ and $\eta(0)() = 0$,
  \item for every $b \in B$ and $v \in V$, we let $\eta((b \cdot))(v) = b \cdot v$, and
    \item for each $\sigma \in \Sigma$, we let $\eta(\sigma) = \mu(\sigma)$.
    \end{compactitem}
    Then $(V,+,0,\mu)$ is the algebra $(V,\eta)$. Moreover, $(V,+,0,\mu)$ is of type $\tau: I \to \mathbb{N}$ where
    $\tau(+) = 2$, $\tau(0) = 0$, $\tau((b\cdot)) = 1$ for each $b \in B$, and $\tau(\sigma) = \rk(\sigma)$  for each $\sigma \in \Sigma$.
    
 Thus, for $(\Sigma,\B)$-semimodules we can use,  in particular, the concepts ``congruence relation'' and ``initial'' as they are defined for algebras.
Hence, if $\sim$ is a congruence relation on a $(\Sigma,\B)$-semimodule $(V,+,0,\mu)$, then
\begin{compactitem}
\item for every $a,b\in B$ and $v_1,v'_1,v_2,v'_2 \in V$, if $v_1\sim v'_1$ and  $v_2\sim v'_2$, then $(a\cdot v_1+b\cdot v_2) \sim (a\cdot v'_1+b\cdot v'_2)$, and
\item $\sim$ is a congruence relation on the $\Sigma$-algebra $(V,\mu)$.
\end{compactitem}

 Moreover, we call an algebra homomorphism from the $(\Sigma,\B)$-semimodule $(V_1,+_1,0_1,\mu_1)$ to the $(\Sigma,\B)$-semimodule $(V_2,+_2,0_2,\mu_2)$  a \emph{$(\Sigma,\B)$-semimodule homomorphism (from $(V_1,+_1,0_1,\mu_1)$ to $(V_2,+_2,0_2,\mu_2)$)}. Hence, if  $h$ is such a $(\Sigma,\B)$-semimodule homomorphism, then
\begin{compactitem}
\item $h$ is a linear mapping from the $\B$-semimodule $(V_1,+_1,0_1)$ to the $\B$-semimodule $(V_2,+_2,0_2)$ and
  \item $h$ is a $\Sigma$-algebra homomorphism from the $\Sigma$-algebra $(V_1,\mu_1)$ to the $\Sigma$-algebra $(V_2,\mu_2)$.
  \end{compactitem}

  \paragraph{Vector spaces.}
  \index{vector space}
  Let $\B=(B,\oplus,\otimes,\0,\1)$ be a field. A {\em $\B$-vector space} is a $\B$-semimodule  $(V,+,0)$ where $(V,+,0)$ is a commutative group. 
    In particular, $(B,\oplus,\0)$ is a $\B$-vector space via scalar multiplication $\otimes$.
  
  Let $\V=(V,+,0)$ be a $\B$-vector space via $\cdot$. Moreover, let $U\subseteq V$ be a nonempty subset such that for every $u,v\in U$ and $b\in B$, we have $u+v\in U$ and $b\cdot u \in U$. A  \emph{subspace of $\V$} is a subalgebra of the algebra $\V$. Clearly, any subspace of $\V$ is a $\B$-vector space via $\cdot$.
  Two subspaces $(U_1,+,0)$ and $(U_2,+,0)$ of $\V$ are \emph{supplementary} if $U_1\cap U_2 = \{0\}$ and $U_1 + U_2 = V$, where in the latter we denote also by + the natural extension of the binary operation + on $V$ to subsets of $V$.
  
The elements of $V$ are called \emph{vectors}. The vectors $v_1,\ldots, v_n \in V$ are {\em linearly independent} if, for every $b_1, \ldots, b_n\in B$, the equality $b_1\cdot v_1 + \ldots + b_n\cdot v_n =0$ implies that $b_1=\ldots =b_n=\0$. Let  $H \subseteq V$. We say that 
\begin{compactitem}
\item $H$ is \emph{linearly independent} if, for every $n \in \mathbb{N}$ and pairwise different $v_1,\ldots, v_n \in H$, the vectors $v_1,\ldots, v_n$ are linearly independent.
\item $H$ {\em generates} $\V$ (or $H$ {\em is a generating set of} $\V$) if, for every $v\in V$, there exist $n\in \mathbb{N}_+$, $b_1,\ldots,b_n\in B$,  and $v_1,\ldots,v_n\in H$ such that $v=b_1\cdot v_1+\ldots +b_n\cdot v_n$.
\item $H$ is a {\em basis of $V$} if $H$ is linearly independent and $H$ generates $\V$.
\end{compactitem}

A vector space $\V$ that has a finite basis is called {\em finite-dimensional}.
In a finite-dimensional vector space each basis has the same number of elements \cite[p.~341]{mooyaq98}. We say that $\V$ is \emph{$n$-dimensional}, where $n \in \mathbb{N}_+$, if it has a basis consisting of $n$ elements.

\begin{theorem} \label{thm:n-lin-independent-vectors=basis} {\rm  \cite[Thm.~2.38]{axl24}} Let $\V$ be an $n$-dimensional $\B$-vector space for some field $\B$. Then every $n$ linearly independent vectors in $V$ form a basis of $\V$. 
   \end{theorem}

\begin{observation}\rm \label{obs:finite-fields-fin-dim=finite} Let $\B=(B,\oplus,\otimes,\0,\1)$ be a finite field and $\V=(V,+,0)$ be a $\B$-vector space. Then $\V$ is finite-dimensional if and only if  $V$ is finite.
  \end{observation}
  
  \begin{proof} Let $\V$ be finite-dimensional and $H=\{v_1,\ldots,v_n\}$ be a basis for $\V$. Then each element $v$ of $V$ has the form $v = b_1 \cdot v_1 + \ldots + b_n \cdot v_n$ for some $b_1,\ldots,b_n \in B$. Thus, there are at most $|B|^n$ elements of $V$.
     Obviously, if $V$ is finite, then $\V$ is finite-dimensional.
   \end{proof}

   \begin{theorem} \label{thm:subspace-of-finite-dim-vector-space} {\rm  \cite[Thm.~2.25,~2.37]{axl24}} Let $\V$ be finite-dimensional $\B$-vector space for some field $\B$. Then any subspace $\mathsf{U}$ of $\V$ is finite-dimensional and $\dim(\mathsf{U})\le \dim(\V)$. 
   \end{theorem}

Let $\V'=(V',+',0')$ be a $\B$-vector space and let $f: V \to V'$ be a linear mapping from $\V$ to $\V'$.
If $H=\{v_1,\ldots, v_n\}$ is a basis of $\V$, then $f$ is determined by  the vectors $f(v_1),\ldots,f(v_n)$, because for each $v\in V$ there exist $b_1,\ldots,b_n \in B$ such that $v=b_1\cdot v_1+\ldots +b_n\cdot v_n$ and then
\[f(v)=f(b_1\cdot v_1+\ldots +b_n\cdot v_n)=b_1\cdot f(v_1)+'\ldots +'b_n\cdot f(v_n)\enspace.\]

\index{null-space}
The \emph{null-space of $f$},
denoted by $\mathrm{null}(f)$, is the set $\{v\in V\mid f(v)=0'\}$.  It is easy to see that 
$(\mathrm{null}(f),+,0)$ and $(\im(f),+',0')$ are subspaces of $\V$ and $\V'$, respectively.

The following theorem is know as Sylvester's law on nullity.

\begin{theorem} \label{thm:Sylvester-nullity} {\rm \cite[Thm.~5.12]{mooyaq98}, \cite[Thm.~3.21]{axl24}} Let $\V=(V,+,0)$ and $\V'=(V',+',0')$ be $\B$-vector spaces for some field $\B$. Moreover, let $f: V \to V'$ be a linear mapping from $\V$ to $\V'$. If $\V$ is finite-dimensional, then $\dim(\V) = \dim(\mathrm{null}(f)) + \dim(\im(f))$.
\end{theorem}

\index{pseudo-regular}
An endomorphism $f \in \cL(\V,\V)$ is \emph{pseudo-regular} if $(\mathrm{null}(f),+,0)$ and $(\im(f),+,0)$
are supplementary subspaces of $\V$.

\index{SigmaBvectorspace@$(\Sigma,\B)$-vector space}
Now let $\B$ be a field. In this case, we call a $(\Sigma,\B)$-semimodule $(\V,\mu)$ a \emph{$(\Sigma,\B)$-vector space}.  If $\V$ is finite-dimensional then we say that $(\V,\mu)$ is finite-dimensional. Moreover, we call a $(\Sigma,\B)$-semimodule homomorphism a \emph{$(\Sigma,\B)$-vector space homomorphism}. We note that in \cite{bozale89} and \cite{fulste11} $(\Sigma,\B)$-vector spaces are called $\B$-$\Sigma$-algebras and $\B\Sigma$-algebras, respectively.

\begin{observation}\rm \label{obs:finite-fields:sim-finite-index=V-modulo-sim-fin-dim} Let $\B$ be a finite field, $(\V,\mu)$ a $(\Sigma,\B)$-vector space, and $\sim$ a congruence on $(\V,\mu)$. Then $\sim$ has finite index if and only if $(\V,\mu)/_\sim$ is finite-dimensional.
\end{observation}
\begin{proof} Let $\sim$ have finite index. Then $(\V,\mu)/_\sim$ is finite and hence finite-dimensional. Let $(\V,\mu)/_\sim$ be finite-dimensional. Then by Observation~\ref{obs:finite-fields-fin-dim=finite}, the vector space $(\V,\mu)/_\sim$ is finite. Hence $\sim$ has finite index.
\end{proof}


\section{Weighted sets and operations}\label{sect:weighted-sets-languages}

\index{weighted set}
\index{support}
\index{support@$\supp(f)$}
\index{constant}
\index{polynomial}
\index{monomial}
Let $\B=(B,\oplus,\otimes,\mathbb{0},\mathbb{1})$ be a strong bimonoid, $A$ be a set, and $f:A\to B$ be a mapping. We call $f$ a \emph{$\B$-weighted set}.  We will define several notions for $f$ and characteristics of $f$. 
We note that if $\B_1$ and $\B_2$ are different strong bimonoids with the same carrier set $B$, then a mapping $f:A\to B$ is both a $\B_1$-weighted set and a $\B_2$-weighted set. 

Let $f:A\to B$ be a $\B$-weighted set. The {\em support of $f$}, denoted by $\supp_\B(f)$, is defined by 
$\supp_\B(f) = \{a \in A \, | \, f(a) \not= \mathbb{0}\}$. If $\B$ is clear from the context, then we abbreviate $\supp_\B(f)$ by $\supp(f)$.

We call $f$
\begin{compactitem}
\item \emph{polynomial} if $\supp(f)$ is finite.
\item \emph{monomial} if $\supp(f) \subseteq \{a\}$ for some $a \in A$; then we denote $f$ also by $f(a).a$ \enspace.
  \item {\em constant} if there exists a $b\in B$ such that for every $a\in A$, we have $f(a)=b$; then we denote $f$ by~$\widetilde{b}$.
  \end{compactitem}
  \index{weighted set!constant}
  \index{weighted set!polynomial}
  \index{weighted set!monomial}
We note that $\widetilde{\mathbb{0}}$ is a monomial and 
$\widetilde{\mathbb{0}}=\mathbb{0}.a$ for each $a\in A$.

Also, we mention that each polynomial $p: X_k^* \to B$ in $B[x_1,\ldots,x_k]$ (cf. Example~\ref{ex:semirings}\ref{def:semiring-of-polynomials}) is an example of a polynomial $\B$-weighted set.

 \index{characteristic mapping}
\index{chi@$\chi_\B(L)$}
Let $L\subseteq A$. The {\em characteristic mapping of $L$ with respect to  $\B$}, denoted by $\chi_\B(L)$, is the $\B$-weighted set $\chi_\B(L): A \to B$ defined, for each $a \in A$, by $\chi_\B(L)(a)=\mathbb{1}$ if $a \in L$, and $\chi_\B(L)(a)=\mathbb{0}$ otherwise. In particular, $\chi_\B(\emptyset)=\widetilde{\0}$ and $\chi_\B(\{a\})=\1.a$. Certainly, $\supp_\B(\chi_\B(L))=L$ and, if $|B|= 2$, then
  $\chi_\B(\supp_\B(f)) = f$ (i.e., for $\B=\Boole$ and $\B=\sfFtwo$). However, if $|B|>2$, then there exists a $\B$-weighted set $f: A \to B$ such that $\chi_\B(\supp_\B(f)) \ne f$.
If $\B$ is clear from the context, then we abbreviate $\chi_\B$ by~$\chi$.

A \emph{weighted language over $\Gamma$ and $\B$} (for short: $(\Gamma,\B)$-weighted language) is a $\B$-weighted set $r: \Gamma^* \to B$.

Next we introduce some operations on $\B$-weighted sets.
 \index{scalar multiplication}
 Let $r: A \rightarrow B$ and $b \in B$. The \emph{scalar multiplication of $r$ with $b$ from the left (with respect to $\B$)}, denoted by $b\cdot r$, is the $\B$-weighted set  $(b \cdot r): A \rightarrow B$ defined for each $a \in A$ by $(b \cdot r)(a) = b \otimes r(a)$. In a similar way, we define the \emph{scalar multiplication of $r$ with $b$ from the right (with respect to $\B$)} and denote it by $r \cdot b$.

\index{sum}
Let $r_1: A \rightarrow B$ and $r_2: A \rightarrow B$ be two $\B$-weighted sets.   The \emph{sum of $r_1$ and $r_2$ (with respect to $\B$)}, denoted by $(r_1\oplus r_2)$, is the $\B$-weighted set $(r_1 \oplus r_2): A \rightarrow B$ defined for each $a \in A$ by $(r_1\oplus r_2)(a) = r_1(a) \oplus r_2(a)$. It is clear that $\oplus$ is associative and commutative. Thus the algebra $(B^A,\oplus,\widetilde{\0})$ is a commutative monoid. If $\B$ is $\sigma$-complete, then also this monoid is $\sigma$-complete.
If $\B$ is a commutative semiring, then $(B^A,\oplus,\widetilde{\0})$ is a $\B$-semimodule (cf. Section \ref{sec:wta-commutative-semirings} for the case that $A$ is the set of all $\Sigma$-terms).

  \index{Hadamard product}
\index{$\otimes$}
Moreover, the \emph{Hadamard product of $r_1$ and $r_2$ (with respect to $\B$)}, denoted by $(r_1 \otimes r_2)$, is the $\B$-weighted $(r_1 \otimes r_2): A \rightarrow B$ defined for each $a \in A$ by $(r_1 \otimes r_2)(a) = r_1(a) \otimes r_2(a)$. The Hadamard product $\otimes$ is associative and, in general, it is not commutative.
However, for each $L \subseteq A$, the equality $r_1 \otimes \chi(L) = \chi(L) \otimes r_1$ holds.

If $A$ is not empty, then the algebra $(B^A,\otimes,\widetilde{\1})$ is a monoid. Since $\widetilde{\0} \otimes r = r \otimes \widetilde{\0} = \widetilde{\0}$ for each $r \in B^A$, the algebra $(B^A,\oplus,\otimes,\widetilde{\0},\widetilde{\1})$ is a strong bimonoid. If $\B$ is a semiring, then so is  $(B^A,\oplus,\otimes,\widetilde{\0},\widetilde{\1})$. In particular, the semiring $(\mathbb{B}^A,\oplus,\otimes,\widetilde{\0},\widetilde{\1})$ over the Boolean semiring $\Boole = (\mathbb{B},\vee,\wedge,0,1)$ is isomorphic to the semiring $(\cP(A),\cup, \cap, \emptyset,A)$.

Let $r: A \to B$ and $L \subseteq A$ be finite. Then we define $r(L) \in B$ by $r(L) = \bigoplus_{a \in L} r(a)$.


\section{Matrices  and vectors over  a strong bimonoid}
\label{sec:vectors-matrices}

\label{page:conv-Q-finite-set}
\begin{quote}\emph{In this section, we let $Q$ denote an arbitrary finite and nonempty set, unless specified otherwise.}
  \end{quote}

\index{matrix}
\paragraph{Matrices.} A {\em $Q$-square matrix over $B$} is a $\B$-weighted set $M : Q\times Q \to B$, and an entry $M(p,q) \in B$ is denoted by $M_{p,q}$. We recall that the set of all $Q$-square matrices over $B$ is denoted by $B^{Q\times Q}$.

\index{summation}
\index{scalar multiplication}
\index{Hadamard product}
Since each $Q$-square matrix over $B$ is a particular  $\B$-weighted set, the following operations are already defined for $b \in B$ and $Q$-square matrices $M,M_1,M_2$  over $B$
(cf. Section \ref{sect:weighted-sets-languages}).  However, since we use these operations frequently, we recall their definitions, for every $p,q \in Q$:
\begin{compactitem}
\item scalar multiplication from the left: $b \cdot M$, i.e., \ $(b \cdot M)_{p,q}=b \otimes M_{p,q}$,
\item scalar multiplication from the right: $M \cdot b$, i.e., \ $(M \cdot b)_{p,q}=M_{p,q}\otimes b$,
\item summation: $M_1 \oplus M_2$, i.e., \ $(M_1 \oplus M_2)_{p,q}=(M_1)_{p,q} \oplus (M_2)_{p,q}$, and
  \item Hadamard product: $M_1 \otimes M_2$, i.e., \ $(M_1 \otimes M_2)_{p,q}=(M_1)_{p,q} \otimes (M_2)_{p,q}$.
  \end{compactitem}

  \index{matrix0@$\mathrm{M}_\0$}
  From Section \ref{sect:weighted-sets-languages} we also know that $(B^{Q \times Q},\oplus,\otimes,\mathrm{M}_\0,\mathrm{N}_\1)$ is a strong bimonoid, where $\mathrm{M}_\0$ and $\mathrm{N}_\1$ are the matrices with all entries $\0$ and $\1$, respectively.

Also we define another type of multiplication on matrices which might be called the Cauchy-product. Formally, the {\em multiplication of ${Q}$-square matrices  over $B$} is the binary operation $\cdot: B^{Q \times Q} \times B^{Q \times Q} \to B^{Q \times Q}$ by letting for every $M_1,M_2 \in B^{Q \times Q}$ and $p,q \in Q$:
\[
(M_1 \cdot M_2)_{p,q} = \bigoplus_{k \in Q}(M_1)_{p,k} \otimes (M_2)_{k,q} \enspace.
\]
We note that $\cdot$ is overloaded: it is used in $b \cdot M$ where $b \in B$ and $M$ is a $Q$-square matrix, and in $M_1 \cdot M_2$ where $M_1$ and $M_2$ are $Q$-square matrices. But it will always be clear from the context which operation in meant. In general, $\cdot$ is not associative, because $\B$ may lack distributivity.

\index{matrix@$\mathrm{M}_\1$}
 The matrix $\mathrm{M}_\1 \in B^{Q \times Q}$ with
\[
  (\mathrm{M}_\1)_{p,q} = \begin{cases}
    \mathbb{1} & \text{ if } p=q\\
    \mathbb{0} & \text{ otherwise}
    \end{cases}
  \]
  for every $p,q \in Q$, is the unit element for $\cdot$.  Moreover, $\mathrm{M}_\0 \cdot M = M \cdot \mathrm{M}_\0= \mathrm{M}_\0$ for each $M \in B^{Q \times Q}$.

  \index{$n$-th power of a matrix}
  \index{transpose of a matrix}
Let $M \in B^{Q \times Q}$ and $n \in \mathbb{N}$. Then we define the \emph{$n$-th power of $M$}, denoted by $M^n$, as follows: $M^0 =  \mathrm{M}_\1$ and $M^{k+1}=M \cdot M^k$ for each $k \in \mathbb{N}$.
Moreover, we define the \emph{transpose of $M$}, denoted by $M^\mathrm{T}$, as the $Q$-square matrix such that, for every $p,q \in Q$, we let $M^\mathrm{T}_{p,q} = \mathrm{M}_{q,p}$.  It is well-known that
$(M_1 \cdot M_2)^\mathrm{T} = M_2^\mathrm{T} \cdot M_1^\mathrm{T}$.

\index{matrix!semiring of square matrices over $B$}
Let $\B=(B,\oplus,\otimes,\mathbb{0},\mathbb{1})$ be a semiring.  Since the operation $\otimes$ of $\B$ is distributive over $\oplus$, the operation $\cdot$ on $B^{Q\times Q}$ is associative. Thus, $(B^{Q\times Q},\cdot,\mathrm{M}_\1)$ is a monoid. Also $\cdot$ is distributive with respect to $\oplus$. Hence $\B^{Q\times Q}= (B^{Q \times Q},\oplus,\cdot,\mathrm{M}_\0,\mathrm{M}_\1)$ is a semiring, called the \emph{semiring of $Q$-square matrices over $B$}.
If $\B$ is $\sigma$-complete, then so is $\B^{Q \times Q}$ (cf. \cite[Sec.~4]{drokui09}).

We will view the transposition $\mathrm{T}$ of a matrix as a bijective mapping $\mathrm{T}: B^{Q\times Q} \to B^{Q\times Q}$. Then $\mathrm{T}$ is a monoid isomorphism between the monoids $(B^{Q \times Q},\cdot,\mathrm{M}_\1)$ and $(B^{Q \times Q},\diamond,\mathrm{M}_\1)$, where $M_1 \diamond M_2 = M_2 \cdot M_1$ for every $M_1,M_2 \in B^{Q \times Q}$.

 We mention that the semiring of $\{1,2\}$-square matrices over the semiring $\Boole$ (cf. Section~\ref{sec:vectors-matrices}) is not zero-cancellation free, because
\[
 \left( \begin{array}{cc}
          1 & 0\\
          0 & 0
        \end{array}\right)
      \cdot
      \left( \begin{array}{cc}
          1 & 1\\
          1 & 1
             \end{array}\right)
           \cdot
           \left( \begin{array}{cc}
          0 & 0\\
          0 & 1
                  \end{array}\right)
                =
                \left( \begin{array}{cc}
          0 & 1\\
          0 & 0
    \end{array}\right)
\ \ \ \text{ and } \ \ \ 
 \left( \begin{array}{cc}
          1 & 0\\
          0 & 0
        \end{array}\right)
            \cdot
           \left( \begin{array}{cc}
          0 & 0\\
          0 & 1
                  \end{array}\right)
                =
                \left( \begin{array}{cc}
          0 & 0\\
          0 & 0
    \end{array}\right)
\]

\index{vector}
\paragraph{Vectors.} A {\em $Q$-vector $v$ over $B$} is a $\B$-weighted set $v : Q \to B$, and an element $v(q) \in B$ of $v$ is denoted by $v_q$. We recall that the set of all $Q$-vectors over $B$ is denoted by $B^Q$.

\index{summation}
\index{scalar multiplication}
\index{Hadamard product}
Since each $Q$-vector over $B$ is a particular  $\B$-weighted set, for every $b\in B$ and $v,v_1,v_2\in B^Q$, the following operations are already defined (cf. Section \ref{sect:weighted-sets-languages}). Since we use these operations frequently, we recall their definitions, for each $q\in Q$:
\begin{compactitem}
\item scalar multiplication from the left: $b \cdot v$, i.e., \ $(b \cdot v)_{q}=b \otimes v_{q}$,
\item scalar multiplication from the right: $v \cdot b$, i.e., \ $(v \cdot b)_{q}=v_{q}\otimes b$,
\item summation: $v_1 \oplus v_2$, i.e., \  $(v_1 \oplus v_2)_{q}=(v_1)_{q} \oplus (v_2)_{q}$, and
  \item Hadamard product: $v_1 \otimes v_2$, i.e., $(v_1 \otimes v_2)_{q}=(v_1)_{q} \otimes (v_2)_{q}$. 
  \end{compactitem}
\index{zeroQ@$\0^Q$}
\index{oneq@$\1_q$}
\index{obeQ@$\1_Q$}
We denote by $\0^Q$ the $Q$-vector over $B$ which contains $\0$ in each component. Moreover, for each $q\in Q$, we denote by $\1_q$  the \emph{$q$-unit vector in $B^Q$}, i.e., the vector in $B^Q$ defined by $(\1_q)_p=\1$ if $p=q$ and $(\1_q)_p=\0$ otherwise. We denote by $\1_Q$ the set $\{\1_q \mid q \in Q\}$.
The algebra $(B^Q,\oplus,\0^Q)$ is a commutative monoid. Moreover, for each $v\in B^Q$, we have $v=\bigoplus_{q\in Q} v_q\cdot \1_q$.

\begin{observation}\label{obs:B-to-Q-is-semimodule}\rm Let $\B$ be a commutative semiring. For each finite and nonempty set $Q$, the triple $(B^Q,\oplus,\0^Q)$ is a $\B$-semimodule via scalar multiplication.
\end{observation}
    \begin{proof} Obviously, $(B^Q,\oplus,\0^Q)$ is a commutative monoid. Also the properties \eqref{SM1}-\eqref{SM5} are satisfied, where \eqref{SM2} and \eqref{SM3} require that $\B$ is distributive. Hence $(B^Q,\oplus,\0^Q)$ is a $\B$-semimodule.
         \end{proof}

         \index{BQ@$\B^Q$}
\label{p:(BQ,oplus,0)-denotation}
         \begin{quote} \emph{In the rest of the book, we abbreviate $(B^Q,\oplus,\0^Q)$ by $\B^Q$, also we write $\B^Q=(B^Q,\oplus,\0^Q)$.}
           \end{quote}

\index{scalar product}
For vectors, we define another type of multiplication. Formally, the {\em scalar product of $Q$-vectors} is defined for every $v_1, v_2 \in B^Q$ by:
\begin{align*}
v_1\cdot v_2 &= \bigoplus_{q\in Q}(v_1)_q\otimes (v_2)_q \enspace.
\end{align*}

\paragraph{Products.} Here we define products of vectors with matrices.  Let $b\in B$, $v_1,v_2 \in B^Q$, and $M \in B^{Q \times Q}$ be a matrix.

We define  the {\em vector-matrix product} $v_1\cdot  M\in B^Q$ and the {\em matrix-vector product} $M\cdot  v_2\in B^Q$ such that for every $p\in Q$:
\begin{align*}
   (v_1\cdot M)_p &= \bigoplus_{q\in Q} (v_1)_q\otimes M_{q,p} 
  &(M\cdot v_2)_p &= \bigoplus_{q\in Q} M_{p,q} \otimes (v_2)_q\enspace.
\end{align*}

It is easy to calculate that, if $\B$ is left-distributive, then for every $M_1,M_2 \in B^{Q \times Q}$ and $v \in B^Q$, we have
\begin{equation}\label{equ:maxtrix-vector-associative}
  (M_1 \cdot M_2) \cdot v = M_1 \cdot (M_2 \cdot v) \enspace.
\end{equation}

\index{c@$\charp_M$}
\index{characteristic polynomial}
\paragraph{Characteristic polynomials.}\label{page:characteristic-polynomial} Let $(B,\oplus,\otimes,\0,\1)$ be a field.  
Moreover, let $|Q|=n$ and $M \in B^{Q \times Q}$.
The \emph{characteristic polynomial of $M$}, denoted by $\charp_M$, is defined by
\[
  \charp_M(x) = \mathrm{det}(M - x \mathrm{M}_\1)  \]
where $\mathrm{M}_\1 \in B^{Q \times Q}$ is the identity for the matrix multiplication and $\mathrm{det}(M')$ is the determinant of the matrix $M'$ (cf. \cite{mooyaq98}).
Clearly, the degree of $\charp_M$ is $n$, and the coefficient of $x^n$ is $(-1)^n$.
The following shows an example of $M \in \mathbb{R}^{Q\times Q}$ with $|Q|=3$ and $\charp_M(x)$:
\[
  M = \left(\begin{array}{ccc}
              0 & 0  & 1\\
              1 & 3 & 2 \\
              2 & 0 & 2
            \end{array}\right), \ 
           \mathrm{det}(M -x\mathrm{M}_\1) = \left[\begin{array}{ccc}
              -x & 0  & 1\\
              1 & 3-x & 2 \\
              2 & 0 & 2-x
                     \end{array}\right], \
          \charp_M(x) = -x^3 + 5 x^2 - 4x - 6 \enspace.
        \]
        
        The following theorem is due to Cayley and Hamilton. Let  $\charp_M(x) = a_n.x^n \oplus \ldots \oplus a_1.x^1 \oplus a_0$ be a polynomial over $\B$ and $N\in B^{Q \times Q}$. The evaluation of $\charp_M(x)$ in $(B^{Q \times Q},\oplus,\cdot,\mathrm{M}_\0,\mathrm{M}_\1)$ at  $x=N$, denoted by $\charp_M(N)$, is the matrix  $a_n \cdot N^n \oplus \ldots \oplus a_1 \cdot N^1 \oplus a_0 \cdot N^0$ of $B^{Q \times Q}$.
        
        \begin{theorem}{\rm \cite[Ch.~XIV,Thm.~3.1]{lan93}}\label{thm:Cayley-Hamilton} Let $M \in B^{Q \times Q}$. Then $\charp_M(M)=\mathrm{M}_\0$.
        \end{theorem}

\index{unit vector}
\index{zeroQ@$\0^Q$}
\paragraph{Representing endomorphisms by matrices.} Let $\B=(B,\oplus,\otimes,\0,\1)$ be a commutative semiring and $Q=\{q_1,\ldots,q_n\}$ with $n\in \mathbb{N}_+$. By Observation~\ref{obs:B-to-Q-is-semimodule}, the tuple $\B^Q=(B^Q,\oplus,\0^Q)$ is a $\B$-semimodule where $\0^Q$ denotes the $Q$-vector over $\B$ in which each entry is $\0$. We recall that $\1_q$ denotes the $q$-unit vector. 

\label{p:def-psi-prime}
As usual, we can represent each endomorphism $f \in \cL(\B^Q,\B^Q)$ by a $Q$-square matrix over $\B$ (cf. \cite[Ch.~3]{mooyaq98}). For this, we define the mapping $\psi: \cL(\B^Q,\B^Q) \to B^{Q \times Q}$ for each $f \in \cL(\B^Q,\B^Q)$ and $p,q \in Q$ by:
\[
\psi(f)_{p,q} = f(\1_q)_p \enspace,
\]
or, in other words, the $q$th column of $\psi(f)$ is the image of the vector $f(\1_q)$. It is easy to see that $\psi$ is bijective. Moreover, application of a linear mapping to a vector corresponds to the matrix-vector product. That is,
\begin{equation}\label{equ:applcation-of-linear-mapping=matrix-vector-product}
  \text{for every  $f \in \cL(\B^Q,\B^Q)$ and $v\in B^Q$, we have $f(v)=\psi(f)\cdot v$.}
\end{equation}
To see this, let  $v = b_1 \cdot \1_{q_1} \oplus \ldots \oplus b_n \cdot \1_{q_n}$ and $p \in Q$. Then $f(v) = b_1 \cdot f(\1_{q_1}) \oplus \ldots  \oplus b_n \cdot f(\1_{q_n})$ and
\begin{align*}
f(v)_p & = b_1 \otimes f(\1_{q_1})_p \oplus \ldots \oplus b_n \otimes f(\1_{q_n})_p \\
       & = b_1 \otimes \psi(f)_{p,q_1} \oplus \ldots \oplus b_n \otimes \psi(f)_{p,q_n} =\big( \psi(f)\cdot v\big)_p\enspace,
\end{align*}
where the last equality uses commutativity.

The mapping $\psi$ is a monoid homomorphism from the monoid $(\cL(\B^Q,\B^Q),\circ,\id_{B^Q})$ to the monoid $(B^{Q\times Q},\cdot,\mathrm{M}_\1)$, which can be seen as follows.
Clearly, $\psi(\id_{B^Q}) = \mathrm{M}_\1$. Let $f_1,f_2 \in \cL(\B^Q,\B^Q)$. Then, for every $p,q \in Q$, we have
\begingroup
\allowdisplaybreaks
\begin{align*}
  \psi(f_1 \circ f_2)_{p,q} &= (f_1 \circ f_2)(\1_q)_p\\
                             &= f_1(f_2(\1_q))_p\\
                             &= f_1(b_1 \cdot \1_{q_1} \oplus \ldots \oplus b_n \cdot \1_{q_n})_p
                               \tag{where $b_1,\ldots,b_n \in B$ are such that  $f_2(\1_{q}) = b_1 \cdot \1_{q_1} \oplus \ldots \oplus b_n \cdot \1_{q_n}$}\\
                             &= b_1 \otimes f_1(\1_{q_1})_p \oplus \ldots \oplus b_n \otimes f_1(\1_{q_n})_p
                               \tag{since $f_1$ is a linear mapping}\\
                             &=f_2(\1_{q})_{q_1} \otimes f_1(\1_{q_1})_p \oplus \ldots \oplus f_2(\1_{q})_{q_n} \otimes f_1(\1_{q_n})_p
                               \tag{because $b_k = f_2(\1_{q})_{q_k}$ for each $k \in [n]$}\\
                             &= \bigoplus_{k \in [n]} f_2(\1_{q})_{q_k} \otimes f_1(\1_{q_k})_p\\
        &= \bigoplus_{k \in [n]} \psi(f_2)_{q_k,q} \otimes \psi(f_1)_{p,q_k}\\
                             &= \bigoplus_{k \in [n]} \psi(f_1)_{p,q_k} \otimes \psi(f_2)_{q_k,q}
                               \tag{by commutativity}\\
  &= (\psi(f_1) \cdot \psi(f_2))_{p,q} \enspace.
 \end{align*} 
\endgroup
Hence, the two monoids $(\cL(\B^Q,\B^Q),\circ,\id_{B^Q})$ and $(B^{Q\times Q},\cdot,\mathrm{M}_\1)$ are isomorphic.

         \label{page:representing-vector-spaces}
         \paragraph{Representing finite-dimensional vector spaces and linear mappings.}
         \label{p:representing-vector-spaces} Let $\B=(B,\oplus,\otimes,\0,\1)$ be a field and $\V=(V,+,0)$ be a $\B$-vector space via $\cdot$ with the finite basis $Q =  \{q_1,\ldots,q_n\}$. Let $q_1,\ldots, q_n$ be an arbitrary, but fixed  order of the basis vectors.

As usual, each vector of $V$ can be represented by a $Q$-vector over $B$, and vice versa (cf. \cite[Ch.~3]{mooyaq98}).
The key for this representation is the fact that, for each vector $v \in V$, there exist unique $b_1, \ldots,b_n \in B$ such that $v = b_1 \cdot q_1 + \ldots + b_n \cdot q_n$. Then, intuitively, we will represent $v$ by $(b_1,\ldots,b_n)$. Let us formalize this.
 
Obviously, $\B^Q = (B^Q,\oplus,\0^Q)$ is a commutative group. Moreover, it is easy to check that $\B^Q$ is an $n$-dimensional $\B$-vector space via the scalar multiplication $\cdot$. A basis of this vector space is $\1_Q=\{\1_q \mid q \in Q\}$, i.e., the set of all $q$-unit vectors.

Then the $\B$-vector spaces $\V$  and $\B^Q$  are isomorphic via the bijective linear mapping $\psi': V\rightarrow B^Q$ which is determined by the values of the base vectors: for each $q \in Q$, we let
\[\psi'(q)=\1_q \text{ for each } q\in Q.\]
Hence, we identify $\V$ and $\B^Q$.

Also we can represent linear mappings over $\V$ as linear mappings over $\B^Q$ (cf. $\overline{\psi'}$) and, by using $\psi$ from the previous paragraph, as $Q$-square matrices over $\B$ (cf. $\psi''$). For this, we extend $\psi'$ to the mapping $\overline{\psi'}: \cL(\V,\V) \to \cL(\B^Q,\B^Q)$ defined, for each $g \in \cL(\V,\V)$, by
\[
\overline{\psi'}(g) = \psi' \circ g \circ (\psi')^{-1}\enspace.
  \]
For the sake of brevity, we denote $\overline{\psi'}$ also by $\psi'$.

By using $\psi$ from the previous paragraph, each linear mapping $g \in \cL(\V,\V)$ can be represented by a $Q\times Q$-matrix as follows (cf. Figure~\ref{fig:representing-linear-mapping-of-vector-space}). We define the mapping $\psi'': \cL(\V,\V) \to \B^{Q \times Q}$ by
\[
\psi'' = \psi \circ \psi' \enspace,
\]
i.e., for each $g \in \cL(\V,\V)$ and $p,q \in Q$, we have
\[
  \psi''(g)_{p,q} = \psi(\psi'(g))_{p,q} =
  \psi'(g)(\1_q)_p = \psi'(g((\psi')^{-1}(\1_q)))_p = \psi'(g(q))_p\enspace.
\]
Since $\psi'$ is bijective, $\psi''$ inherits from $\psi$ the property of being a bijective monoid homomorphism.

\begin{figure}[t]
\centering
\begin{tikzpicture}
  \node at (-3.1,1.6) (V1) {$\B$-vector space $\V=(V,+,0)$: };
  \node at (-3.5,-0.6) (V2) {$\B$-vector space $\B^Q=(B^Q,\oplus,\0^Q)$: };
\node at (0,1.6) (1) {$V$};
\node at (4,1.6) (2) {$V$};
\node at (0,-0.6) (3) {$B^Q$};
\node at (4,-0.6) (4) {$B^Q$};
\node at (2,-3.0) (5) {$B^{Q \times Q}$};

\draw (1) edge[->,>=stealth] node[fill=white] {$g$} (2);
\draw (3) edge[->,>=stealth] node[fill=white] {$\psi'(g)$} (4);
\draw (1) edge[->,>=stealth] node[fill=white] {$\psi'$} (3);
\draw (2) edge[->,>=stealth] node[fill=white] {$\psi$'} (4);
\draw ($(3)+(2,-0.5)$) edge[->,>=stealth] node[fill=white] {$\psi$} (5);
\draw ($(1)+(2,-0.4)$) edge[->,>=stealth] node[fill=white] {$\psi'$} ($(3)+(2,0.4)$);

    \end{tikzpicture}
  \caption{\label{fig:representing-linear-mapping-of-vector-space} Combination of $\psi': V \to \B^Q$ (and $\psi': \cL(\V,\V) \to \cL(\B^Q,\B^Q)$) and $\psi: \cL(\B^Q,\B^Q) \to B^{Q \times Q}$. }
  \end{figure}

\index{pseudo-regular}
A matrix $M \in B^{Q\times Q}$ is \emph{pseudo-regular}  if the endomorphism $(\psi'')^{-1}(M)$ is pseudo-regular. Equivalent definitions of pseudo-regularity can be found in \cite[Prop. 1]{reu80}.


\section{Trees and tree languages}\label{sect:trees}

We recall some notions from (formal) tree languages \cite{eng75-15,gecste84,gecste97,comdaugiljaclugtistom08}.

\index{Sigma@$\Sigma$}
\index{Delta@$\Delta$}
\label{page:Sigma0-ne-empty}
\begin{quote}{\em In the rest of this book, $\Sigma$ and $\Delta$ will denote arbitrary ranked alphabets, if not specified otherwise.}
\end{quote}

\index{terms}
\index{Sigmaterm@$\Sigma$-terms}
\paragraph{Trees.} 
\index{tsigma@$\T_\Sigma(H)$}
Let $H$ be a set and let $\Xi$ denote the set containing the opening and closing parentheses ``('' and ``)'', respectively, and the comma ``,''.
We assume that $\Sigma$, $H$, and $\Xi$ are pairwise disjoint.
Now we define the $\Sigma$-algebra $((\Sigma \cup H \cup \Xi)^*,\ttop_\Sigma)$ where, for every $k \in \mathbb{N}$, $\sigma \in \Sigma^{(k)}$, and $w_1,\ldots,w_k \in (\Sigma \cup H \cup \Xi)^*$, we let
\[
  \ttop_\Sigma(\sigma)(w_1,\ldots,w_k) = \sigma(w_1,\ldots,w_k)\enspace.
\]

Then the \emph{set of $\Sigma$-terms over $H$}, denoted by $\T_\Sigma(H)$, is the set $\langle H \rangle_{\ttop_\Sigma(\Sigma)}$, i.e., the smallest subset of $(\Sigma \cup H \cup \Xi)^*$ which contains $H$ and is closed under $\ttop_\Sigma(\Sigma)$.
We denote $\T_\Sigma(\emptyset)$ by $\T_\Sigma$ and call it the \emph{set of $\Sigma$-terms}.  If $H$ is finite, then we can view $\T_\Sigma(H)$ as $\T_\Delta$, where the ranked alphabet $\Delta$ is defined as follows:  $\Delta^{(0)}= \Sigma^{(0)} \cup H$ and $\Delta^{(k)}= \Sigma^{(k)}$ for every $k\in \mathbb{N}_+$. 

\index{trees}
\index{Sigmatrees@$\Sigma$-trees}
Since terms can be depicted in a very illustrative way as  trees, i.e., particular graphs, it has become a custom to call $\Sigma$-terms also {\em $\Sigma$-trees} (or just: \emph{trees}). In this book we follow this custom.
Each subset $L \subseteq \T_\Sigma$ is called  a {\em $\Sigma$-tree language} (or just: \emph{tree language}).

Obviously, for each $\xi \in \T_\Sigma$ there exist unique $k\in \mathbb{N}$, $\sigma\in\Sigma^{(k)}$, and $\xi_1,\ldots,\xi_k \in \T_\Sigma$ such that $\xi=\sigma(\xi_1,\ldots,\xi_k)$. (In case $k=0$ we identify $\sigma()$ and $\sigma$.) In order to avoid the repetition of all these quantifications, we henceforth only write that we consider a ``$\xi \in \T_\Sigma$ of the form $\xi=\sigma(\xi_1,\ldots,\xi_k)$'' or ``for every  $\xi = \sigma(\xi_1,\ldots,\xi_k)$''.

As an application of Lemma \ref{obs:Knaster-Tarski-applied-to-algebras},
we obtain the following characterization of the set $\T_\Sigma$.

\begin{observation}\rm \label{ob:TSigma-inductive} Let $f_{\T_\Sigma}: \cP((\Sigma \cup \Xi)^*) \to \cP((\Sigma \cup \Xi)^*)$ be the mapping defined, for each $U \in \cP((\Sigma \cup \Xi)^*)$, by
$f_{\T_\Sigma}(U) =  U \cup \{\sigma(\xi_1,\ldots,\xi_k) \mid k \in \mathbb{N}, \sigma \in \Sigma^{(k)}, \xi_1,\ldots,\xi_k \in U\}$.
    Then, by Lemma \ref{obs:Knaster-Tarski-applied-to-algebras},
    we have
    \(
\T_\Sigma = \bigcup((f_{\T_\Sigma})^n(\emptyset) \mid n \in \mathbb{N})\). 
\hfill $\Box$
\end{observation}


\paragraph{The $\Sigma$-term algebra.}

\index{termalgebra@term algebra}
\index{Sigmatermalgebra@$\Sigma$-term algebra}
\index{topSigma@$\ttop_\Sigma$}
\index{TSigmaH@$\sfT_\Sigma(H)$}
Let $H$ be a set disjoint with $\Sigma$. The \emph{$\Sigma$-term algebra over $H$}, denoted by $\sfT_\Sigma(H)$,  is the subalgebra of the $\Sigma$-algebra $((\Sigma \cup H \cup \Xi)^*,\ttop_\Sigma)$ which is generated by $H$. Thus, since $\langle H \rangle_{\ttop_\Sigma(\Sigma)} = \T_\Sigma(H)$, the $\Sigma$-term algebra over $H$ is the $\Sigma$-algebra
\[\sfT_\Sigma(H) = (\T_\Sigma(H),\ttop_\Sigma)\enspace.
\]
The \emph{$\Sigma$-term algebra}, denoted by $\sfT_\Sigma$, is the $\Sigma$-term algebra over $\emptyset$, i.e., $\sfT_\Sigma = \sfT_\Sigma(\emptyset)$.

Next we prove that $\sfT_\Sigma(H)$ is free in the set of all $\Sigma$-algebras with generating set $H$, by using Theorem~\ref{thm:dedekind}. For this, we use the following auxiliary definitions.  We define the binary relation $\succ$ on $\T_\Sigma(H)$ as follows:   
\begin{align*}
  \text{for every $k \in \mathbb{N}_+$, $\sigma \in \Sigma^{(k)}$,  $\xi_1,\ldots,\xi_k \in \T_\Sigma(H)$, and $i \in [k]$, we let
  $\sigma(\xi_1,\ldots,\xi_k) \succ \xi_i$} \enspace.
\end{align*}
By Corollary \ref{cor:reduction-to-substring-is-terminating}, the relation $\succ$ is terminating and $\nf_{\succ}(\T_\Sigma(H)) =\Sigma^{(0)}\cup H$.
\index{Gthetaf@$G_{\theta,f}$}
Let $\A=(A,\theta)$ be a $\Sigma$-algebra and $f: H \to A$. 
Then we define the mapping
    \[
    G_{\theta,f}: \{(\xi,g) \mid \xi \in \T_\Sigma(H), g: \ \succ\!\!(\xi) \to A \} \to A
  \]
  for every $\xi \in \T_\Sigma(H)$ and $g:\ \succ\!\!(\xi) \to A$ as follows:
  \[
    G_{\theta,f}(\xi,g) = \begin{cases}
      \theta(\sigma)(g(\xi_1),\ldots,g(\xi_k)) & \text{ if } (\exists k \in \mathbb{N}, \sigma \in \Sigma^{(k)}, \xi_1,\ldots,\xi_k \in \T_\Sigma(H)): \xi = \sigma(\xi_1,\ldots,\xi_k)\\
      f(\xi) & \text{ otherwise (i.e., $\xi \in H$)} \enspace.
      \end{cases} 
    \]

    Now we prove that the unique mapping defined by $G_{\theta,f}$ (cf. Theorem~\ref{thm:dedekind})
    is the unique $\Sigma$-algebra homomorphism from $\sfT_\Sigma(H)$  to $\A$ which extends~$f$.

    \begin{lemma} \rm \label{lm:Gthetaf} Let $\A=(A,\theta)$ be a $\Sigma$-algebra and $f: H \to A$. Moreover, let $\phi: \T_\Sigma(H) \to A$ be a mapping. Then the following two statements are equivalent.
\begin{compactenum}
 \item[(A)] For each $\xi \in \T_\Sigma(H)$, the mapping $\phi$ satisfies $\phi(\xi) = G_{\theta,f}(\xi, \phi|_{\succ(\xi)})$.
 \item[(B)] $\phi$ is a  $\Sigma$-algebra homomorphism from $\sfT_\Sigma(H)$ 
   to $\A$, and $\phi$ extends $f$.
\end{compactenum}
\end{lemma}
\begin{proof}  Proof of (A)$\Rightarrow$(B): Let $\xi = \sigma(\xi_1,\ldots,\xi_k)$. Then we can compute as follows:
    \begin{align*}
      \phi(\sigma(\xi_1,\ldots,\xi_k)) &= G_{\theta,f}(\xi,\phi|_{\succ(\xi)})
      \tag{by (A)} \\
                                       &= \theta(\sigma)(\phi|_{\succ(\xi)}(\xi_1),\ldots,\phi|_{\succ(\xi)}(\xi_k))
      \tag{by definition of $G_{\theta,f}$}\\
      &= \theta(\sigma)(\phi(\xi_1),\ldots,\phi(\xi_k)) \enspace.
      \end{align*}
    Hence $\phi$ is a $\Sigma$-algebra homomorphism from $\sfT_\Sigma(H)$ 
    to $\A$.  By definition of $G_{\theta,f}$, the homomorphism $\phi$ extends $f$. 

    \
    
     Proof of (B)$\Rightarrow$(A): We prove by case analysis. Let $\xi \in H$. Then $\phi(\xi) = f(\xi) = G_{\theta,f}(\xi,\phi|_{\succ(\xi)})$.   Let $\xi = \sigma(\xi_1,\ldots,\xi_k)$.  Then we can compute as follows:
    \begin{align*}
      G_{\theta,f}(\xi,\phi|_{\succ(\xi)}) &= \theta(\sigma)(\phi|_{\succ(\xi)}(\xi_1),\ldots,\phi|_{\succ(\xi)}(\xi_k))
      \tag{by definition of $G_{\theta,f}$}\\
                                        &= \theta(\sigma)(\phi(\xi_1),\ldots,\phi(\xi_k))\\
      &= \phi(\xi) \tag{by (B)}\enspace.
    \end{align*}
    \end{proof}

\begin{theorem}\label{thm:initial-iso} {\rm \cite[Prop.~2.3]{gogthawagwri77} and \cite[Thm. 1.3.11]{gecste84}} The $\Sigma$-term algebra $\sfT_\Sigma(H)$ is free in the set of all $\Sigma$-algebras with generating set~$H$. Moreover, the $\Sigma$-term algebra $\sfT_\Sigma$ is initial in the set of all $\Sigma$-algebras.
\end{theorem}
\begin{proof} For the proof of the first statement of the theorem, we proceed along the items (a), (b), and (c) of the definition of a free algebra.
  
  (a) As we saw, the $\Sigma$-term algebra $\sfT_\Sigma(H)$ 
  over $H$ is a $\Sigma$-algebra.

  (b) By definition, we have $\T_\Sigma(H) = \langle H \rangle_{\ttop_\Sigma(\Sigma)}$.

  (c) Let $\A=(A,\theta)$ be a $\Sigma$-algebra and $f:H \to A$ a mapping.
  By Theorem \ref{thm:dedekind}, there exists a unique mapping $h: \T_\Sigma(H) \to A$ such that, for each $\xi \in \T_\Sigma(H)$, we have $h(\xi) = G_{\theta,f}(\xi, h|_{\succ(\xi)})$.  Thus, due to Lemma \ref{lm:Gthetaf}, we obtain that $h$ is the unique $\Sigma$-algebra homomorphism from $\sfT_\Sigma(H)$ 
  to $\A$ which extends~$f$.

  This proves the first statement of the lemma. 
  The second statement follows from the first one with $H=\emptyset$.
  \end{proof}

\index{homA@$\h_\A$}
  \label{p:convention-term-algebra}
\begin{quote}\em
If not specified otherwise, then we denote the unique $\Sigma$-algebra homomorphism from the  $\Sigma$-term algebra $\sfT_\Sigma = (\T_\Sigma,\ttop_\Sigma)$ to some $\Sigma$-algebra $\A$ by $\h_\A$. 
\end{quote}

    \begin{observation}\label{obs:smallest-subalgebra-im} \rm Let $\A=(A,\theta)$ be a $\Sigma$-algebra. Then $\im(\h_\A)=\langle \emptyset \rangle_{\theta(\Sigma)}$ and thus $(\im(\h_\A),\theta)$ is the smallest subalgebra of $\A$.
\end{observation}
\begin{proof} Since $(\langle \emptyset \rangle_{\theta(\Sigma)},\theta)$ is the smallest subalgebra of $\A$,
we have that $\langle \emptyset \rangle_{\theta(\Sigma)}\subseteq \im(\h_\A)$. Moreover, we can show by induction on $\T_\Sigma$ that, for each $\xi\in\T_\Sigma$, we have $\h_\A(\xi)\in \langle \emptyset \rangle_{\theta(\Sigma)}$. Thus $\im(\h_\A)\subseteq \langle \emptyset \rangle_{\theta(\Sigma)}$.
    \end{proof}

    \paragraph{String-like terms.} We can view strings over an alphabet $\Gamma$ as trees. For this, we first define an appropriate ranked alphabet.
    \label{p:string-like-trees-are-strings}

    \index{Gammae@$\Gamma_e$}
Let $e\not\in \Gamma$ be a symbol.  Then $\Gamma$ and $e$ determine the string ranked alphabet  $\Gamma_e=\{a^{(1)} \mid a \in \Gamma\} \cup \{e^{(0)}\}$. Vice versa, each string ranked alphabet $\Sigma$ can be written in the form $\Gamma_e$,  where $\Gamma=\Sigma^{(1)}$ and $e$ is the only element of $\Sigma^{(0)}$. In the following, sometimes we denote a string ranked alphabet by  $\Gamma_e$ without mentioning what $\Gamma$ and $e$ mean.  

Now we define the $\Gamma_e$-algebra\footnote{$(\Gamma^*,\widehat{\Gamma_e})$ should not be confused with the free monoid  $(\Gamma^*,\cdot,\varepsilon)$ with generating set $\Gamma$.} $(\Gamma^*,\widehat{\Gamma_e})$ where $\widehat{\Gamma_e}=(\widehat{b} \mid b \in \Gamma_e)$ is the $\Gamma_e$-indexed family over $\mathrm{Ops}(\Gamma^*)$ such that
\[
  \text{$\widehat{e}=\varepsilon$ \  and \ $\widehat{a}(w)=w a$ for every $w\in \Gamma^*$ and $a\in \Gamma$ \enspace.}
\]
Then we consider the  $\Gamma_e$-term algebra $(\T_{\Gamma_e},\theta_{\Gamma_e})$ and show that
\begin{equation}\label{eq:ismorphism}
  (\Gamma^*,\widehat{\Gamma_e})\cong(\T_{\Gamma_e},\theta_{\Gamma_e})\enspace,
\end{equation}
i.e., the two $\Gamma_e$-algebras $(\Gamma^*,\widehat{\Gamma_e})$ and $(\T_{\Gamma_e},\theta_{\Gamma_e})$ are isomorphic.
\index{succ@$\succ$}
\label{page:def-of-mapping-tree}
For this, we define the mapping
\index{tree@$\tree_e$}
\[
  \tree_e: \Gamma^* \to \T_{\Gamma_e}
\]
as follows. We consider the
reduction system $(\Gamma^*,\succ)$ where $\succ \subseteq \Gamma^* \times \Gamma^*$ is defined by $\succ = \{(wa,w) \mid w\in \Gamma^*, a \in \Gamma\}$.
By Corollary~\ref{cor:reduction-to-substring-is-terminating}, the relation $\succ$ is terminating and $\nf_\succ(\Gamma^*) = \{\varepsilon\}$.
Then we define $\tree_e$ by induction on $(\Gamma^*,\succ)$ by
\begin{compactenum}
  \item[I.B.:] $\tree_e(\varepsilon)=e$ \ and
    \item[I.S.:] $\tree_e(wa)=a(\tree_e(w))$ for every $w\in \Gamma^*$ and $a\in \Gamma$.
\end{compactenum}
It is obvious that $\tree_e$  is a bijection. Lastly, we show that $\tree_e$ is a $\Gamma_e$-algebra homomorphism from $(\Gamma^*,\widehat{\Gamma_e})$ to $(\T_{\Gamma_e},\theta_{\Gamma_e})$: 
\begin{compactitem}
\item[(i)] $\tree_e(\widehat{e})=\tree_e(\varepsilon)=e=\theta_{\Gamma_e}(e)()$, and
\item[(ii)] for every $w\in \Gamma^*$ and $a\in \Gamma$, we have $\tree_e(\widehat{a}(w))=\tree_e(wa)=a(\tree_e(w))= \theta_{\Gamma_e}(a)(\tree_e(w))$.
\end{compactitem}
Hence \eqref{eq:ismorphism} holds.

\paragraph{Proof by induction on $\T_\Sigma(H)$ and proof by induction on $\T_\Sigma$.}
\index{succSigmaH@$\succ_{\Sigma,H}$}
\index{succSigmaHplus@$\succ_{\Sigma,H}^+$}
Let $H$ be a set disjoint with $\Sigma$. Many times we will use the following two instances of well-founded induction to prove that each element of $\T_\Sigma(H)$  has a property:

\label{p:succ-Sigma-H}
\begin{itemize}
  \item \emph{proof by induction on $(\T_\Sigma(H),\succ_{\Sigma,H})$} (for short: \emph{proof by induction on $\T_\Sigma(H)$}):
        we define the binary relation\footnote{This $\succ_{\Sigma,H}$ is exactly the binary relation $\succ$ which we have used to prove that the $\Sigma$-term algebra is free  in the set of all $\Sigma$-algebras with generating set $H$; since we will use it often, we give it a specific denotation.} $\succ_{\Sigma,H}$ on $\T_\Sigma(H)$ as follows:   
\begin{align*}
\text{for every $k \in \mathbb{N}_+$, $\sigma \in \Sigma^{(k)}$,  $\xi_1,\ldots,\xi_k \in \T_\Sigma(H)$, and  $i \in [k]$, we let $\sigma(\xi_1,\ldots,\xi_k) \succ \xi_i$} \enspace.
\end{align*}
By Corollary \ref{cor:reduction-to-substring-is-terminating}, the relation $\succ_{\Sigma,H}$ is terminating and $\mathrm{nf}_{\succ_{\Sigma,H}}(\T_\Sigma(H)) =\Sigma^{(0)}\cup H$.
In this case, the induction base \eqref{equ:wfi-base} and induction step \eqref{eq:wfi-step} read
\begin{align*}
\text{I.B.: } &(\forall \alpha \in \Sigma^{(0)} \cup H): P(\alpha) \\ 
\text{I.S.: }  & \Big(\big(\forall k \in \mathbb{N}_+, \sigma \in \Sigma^{(k)}, \xi_1,\ldots,\xi_k \in \T_\Sigma(H)\big): \big[P(\xi_1)\wedge \ldots \wedge P(\xi_k)\big] \Rightarrow \ P(\sigma(\xi_1,\ldots,\xi_k)) \Big) \enspace. 
\end{align*}
\item \emph{proof by induction on $(\T_\Sigma(H),\psucc_{\Sigma,H})$}, where  $\psucc_{\Sigma,H}=(\succ_{\Sigma,H})^+$, i.e., $\psucc_{\Sigma,H}$ is the transitive closure of $\succ_{\Sigma,H}$.
Again we have $\mathrm{nf}_{\psucc_{\Sigma,H}}(\T_\Sigma(H)) =\Sigma^{(0)} \cup H$. In this case, the induction base \eqref{equ:wfi-base} and the induction step \eqref{eq:wfi-step} read
\begin{align*}
\text{I.B.: }  & (\forall \alpha \in \Sigma^{(0)} \cup H): P(\alpha)\\
\text{I.S.: } & \Big(\big(\forall \xi \in \T_\Sigma(H)\setminus (\Sigma^{(0)} \cup H)\big): \big[\big(\forall \xi'\in \T_\Sigma(H)
  \text{ with $\xi \psucc_{\Sigma,H} \xi'$}\big): P(\xi')\big]  \Rightarrow P(\xi)\Big) \enspace.
\end{align*}
\end{itemize}

\label{page:prec-Sigma}
\index{succSigma@$\succ_\Sigma$}
\index{succSigmaplus@$\psucc_{\Sigma}$}
Often we will use  induction on $\T_\Sigma(H)$ for the case that $H=\emptyset$. We will abbreviate $\succ_{\Sigma,\emptyset}$ and $\psucc_{\Sigma,\emptyset}$ by $\succ_\Sigma$ and $\psucc_{\Sigma}$, respectively; then $\mathrm{nf}_{\succ_\Sigma}(\T_\Sigma) = \mathrm{nf}_{\psucc_\Sigma}(\T_\Sigma)  = \Sigma^{(0)}$. We will call a proof by induction on $\T_\Sigma(\emptyset)$ (i.e., on $(\T_\Sigma,\succ_\Sigma)$) a \emph{proof by induction on $\T_\Sigma$}.
As in the general case, we will use the idioms ``we prove $P$ by induction on $\T_\Sigma(H)$'' and ``we prove $P$ by induction on $\T_\Sigma$'' with their natural meanings.

In some proofs by induction on $\T_\Sigma$, we combine the induction base \eqref{equ:wfi-base} and induction step \eqref{eq:wfi-step} by proving
\begin{equation*}
  (\forall k \in \mathbb{N}, \sigma \in \Sigma^{(k)}, \xi_1,\ldots,\xi_k \in \T_\Sigma):
  [P(\xi_1)\wedge \ldots \wedge P(\xi_k)]
 \Rightarrow \ P(\sigma(\xi_1,\ldots,\xi_k))  \enspace.
\end{equation*}

We will define several mappings on $\T_\Sigma(H)$ by induction on $(\T_\Sigma(H),\succ_{\Sigma,H})$ (cf. \eqref{eq:def-by-induction}). Then we will use the idiom ``we define a mapping by induction on $\T_\Sigma(H)$'' or, if $H=\emptyset$, ``we define a mapping  by induction on $\T_\Sigma$''. 


\index{height@$\height$}
\index{pos@$\pos$}
\index{size@$\size$}
\paragraph{Mappings on trees.}\label{page:algebras-for-height-size-pos}
Let $H$ be a set disjoint with $\Sigma$. 
First, we  define the three mappings 
\[
  \height: \T_\Sigma(H)\rightarrow \mathbb{N} , \ \ \ \size:\T_\Sigma(H)\rightarrow \mathbb{N},
  \ \ \text{ and } \ \ \pos:\T_\Sigma(H)\rightarrow {\cal P}(\mathbb{N_+}^*)
\]
which, intuitively, for each tree $\xi \in \T_\Sigma(H)$ (viewed as graph), deliver the maximal number of edges from the root of $\xi$ to some leaf, the number of nodes of $\xi$, and the set of Gorn-addresses of $\xi$, respectively. For instance, let $\Sigma = \{\sigma^{(2)}, \gamma^{(1)}, \alpha^{(0)}, \beta^{(0)}\}$,  $H = \{a\}$, and $\xi = \sigma(\gamma(\alpha),\sigma(\gamma(a),\beta))$; then
\begin{align*}
   \height(\xi) = 3 \enspace, \ \ 
  \size(\xi) = 7 \enspace, \ \ 
  \pos(\xi) = \{\varepsilon, 1, 11, 2, 21, 211, 22\} \enspace.
  \end{align*}
\label{page:height-size-pos}

Formally, we use Theorem \ref{thm:initial-iso} for the definition of $\height$, $\size$, and $\pos$ as follows.
\begin{enumerate}
\item Let $(\mathbb{N},\theta_1)$ be the $\Sigma$-algebra where $\theta_1$ is defined as follows:
  \begin{compactitem}
  \item for each $\alpha \in \Sigma^{(0)}$, we let $\theta_1(\alpha)() = 0$, and
  \item for each $k \in \mathbb{N}_+$, $\sigma \in \Sigma^{(k)}$, and $n_1,\ldots,n_k \in \mathbb{N}$, we let $\theta_1(\sigma)(n_1,\ldots,n_k) = 1 + \max(n_1,\ldots,n_k)$.
  \end{compactitem}
  Moreover, let $f: H \to \mathbb{N}$ be defined such that, for each $a \in H$, we let $f(a) = 0$. Then
  \[
    \height: \T_\Sigma(H)\rightarrow \mathbb{N}
  \]
  is the unique extension of $f$ to a $\Sigma$-algebra homomorphism from the $\Sigma$-term algebra $\sfT_\Sigma(H)$ 
  over $H$ to the $\Sigma$-algebra $(\mathbb{N},\theta_1)$.  

  \item Let $(\mathbb{N},\theta_2)$ be the $\Sigma$-algebra where $\theta_2$ is defined as follows:
 for each $k \in \mathbb{N}$, $\sigma \in \Sigma^{(k)}$, and $n_1,\ldots,n_k \in \mathbb{N}$, we let $\theta_2(\sigma)(n_1,\ldots,n_k) = 1 + \bigplus_{i \in [k]} n_i$.
  Moreover, let $f: H \to \mathbb{N}$ be defined such that, for each $a \in H$, we let $f(a) = 1$. Then
  \[
    \size: \T_\Sigma(H)\rightarrow \mathbb{N}
  \]
  is the unique extension of $f$ to a $\Sigma$-algebra homomorphism from the $\Sigma$-term algebra $\sfT_\Sigma(H)$ 
  over $H$ to the $\Sigma$-algebra $(\mathbb{N},\theta_2)$.

    \item Let $(\cP({\mathbb{N}_+}^*),\theta_3)$ be the $\Sigma$-algebra where $\theta_3$ is defined as follows:
 for each $k \in \mathbb{N}$, $\sigma \in \Sigma^{(k)}$, and $U_1,\ldots,U_k \in \cP({\mathbb{N}_+}^*)$, we let $\theta_3(\sigma)(U_1,\ldots,U_k) =\{\varepsilon\} \cup \{iv \mid i \in [k], v \in U_i\}$.
  Moreover, let $f: H \to \cP({\mathbb{N}_+}^*)$ be defined such that, for each $a \in H$, we let $f(a) = \{\varepsilon\}$. Then
  \[
    \pos: \T_\Sigma(H)\rightarrow \cP({\mathbb{N}_+}^*)
  \]
  is the unique extension of $f$ to a $\Sigma$-algebra homomorphism from the $\Sigma$-term algebra $\sfT_\Sigma(H)$ 
  over $H$ to the $\Sigma$-algebra $(\cP({\mathbb{N}_+}^*),\theta_3)$.  
\end{enumerate}

  It is easy to verify that the following equalities hold. 
  \begin{compactitem}
  \item[(i)] For each $\alpha \in  \Sigma^{(0)} \cup H$, we have $\height(\alpha) = 0$, $\size(\alpha) = 1$, and $\pos(\alpha) = \{\varepsilon\}$.
\item[(ii)] For each $\xi = \sigma(\xi_1,\ldots,\xi_k)$ with $k \in \mathbb{N}_+$, we have: 
\begin{compactitem}
\item $\height(\xi) = 1 + \max(\height(\xi_i) \mid i \in [k])$,
\item $\size(\xi) = 1 + \bigplus_{i \in [k]}\size(\xi_i)$, and
\item $\pos(\xi) = \{\varepsilon\} \cup \{iv \mid  i \in [k],
v \in \pos(\xi_i)\}$.
\end{compactitem}
\end{compactitem}

\index{root of a tree}
\index{trees!root of a tree}
\index{leaf of a tree}
\index{trees!leaf of a tree}
Let $\xi\in \T_\Sigma(H)$. Sometimes we refer to the position $\varepsilon$ of $\xi$ as the \emph{root of $\xi$}, and to a position $w$ of $\xi$
with $wi\ne \pos(\xi)$ for each $i\in \mathbb{N}_+$ as a  \emph{leaf of $\xi$}.

\begin{observation}\rm \label{ex:TSigma-inductive} For the mapping $f_{\T_\Sigma}: \cP((\Sigma \cup \Xi)^*) \to \cP((\Sigma \cup \Xi)^*)$ defined in Observation \ref{ob:TSigma-inductive},
  there is a connection between the set $f_{\T_\Sigma}^{n+1}(\emptyset)$ and the set of $\Sigma$-trees of height at most $n$. Indeed, we can prove by induction on $\mathbb{N}$ that, for each $n \in \mathbb{N}$, we have
  \(f_{\T_\Sigma}^{n+1}(\emptyset) = \{\xi \in \T_\Sigma \mid \height(\xi) \le n\}\).
\hfill $\Box$
\end{observation}

\begin{figure}
 \centering
\begin{tikzpicture}[scale=1, every node/.style={transform shape},
					level distance= 1cm,
					level 1/.style={sibling distance=20mm},
					level 2/.style={sibling distance=16mm}]
										
\node at (-1.5,0) {$\xi:$}; 
\node (root) {$\sigma$}
  child {node {$\gamma$}
    child {node {$\alpha$}}}
  child {node (pos2) {$\sigma$}
  edge from parent [decoration={dashsoliddouble}, decorate]
   child {node (pos21) {$\gamma$}
   edge from parent [decoration={dashsoliddouble}, decorate]
     child {node {$\beta$}}}
   child {node {$\beta$}}};

\node[right= -0.05cm of pos21] {$\scriptstyle 21$};

\node[anchor=west] at (3.5,0)  {$f(\xi,21)=\gamma=\xi(21)$};
\node[anchor=west] at (3.5,-1) {$g(\xi,21)=\gamma(\beta)=\xi|_{21}$};
\node[anchor=west] at (3.5,-2) {$h(\xi,21)(\zeta)=\sigma(\gamma(\alpha),\sigma(\zeta,\beta))=\xi [\zeta]_{21}$};

\end{tikzpicture}
\caption{\label{fig:label-subtree-replacement} For a given tree $\xi=\sigma(\gamma(\alpha),\sigma(\gamma(\beta),\beta))$ and position $w=21$, the label of $\xi$ at position $w$, the subtree of $\xi$ at position $w$, and the replacement of the subtree of $\xi$ at position $w$ by $\zeta$.}
\end{figure}

\index{succ@$\succ$}
\index{TP@$\mathrm{TP}$}
Next we define three mappings $f$, $g$, and $h$ which show the label at a position of a tree, the subtree at a position of a tree, and the replacement of the subtree at a position of a tree, respectively (cf. Figure~\ref{fig:label-subtree-replacement}).
We define these mappings by well-founded induction. For this, we define the set $\mathrm{TP} = \{(\xi,w) \mid \xi \in \T_\Sigma(H), w \in \pos(\xi)\}$ and the binary relation $\succ$ on $\mathrm{TP}$ as follows:
\begin{align*}
  \text{for every $k \in \mathbb{N}_+$, $\sigma \in \Sigma^{(k)}$, $\xi_1,\ldots,\xi_k \in \T_\Sigma(H)$, $i \in [k]$, and $v \in \pos(\xi_i)$, we let $(\sigma(\xi_1,\ldots,\xi_k),iv) \succ (\xi_i,v)$}\enspace. 
\end{align*}
\label{page:f-g-h-defined-by-induction}
Since the relation $\succ_{\Sigma,H}$ is terminating, by Corollary \ref{cor:termination-propagates-to-cartesian-products}, the relation  $\succ$ is also terminating. Obviously, $\nf_\succ(\mathrm{TP}) = \{(\xi,\varepsilon)\mid \xi \in \T_\Sigma(H)\}$. By induction on $(\mathrm{TP},\succ)$, we define the mappings
\[
  f: \mathrm{TP} \to \Sigma \cup H , \ \ g: \mathrm{TP} \to \T_\Sigma(H), \ \text{ and } \ \
  h: \mathrm{TP} \to \T_\Sigma(H)^{\T_\Sigma(H)}
\]
 as follows:
\begin{compactenum}
\item[I.B.:]  Let $\xi \in \T_\Sigma(H)$. Moreover, let $\zeta \in \T_\Sigma(H)$. Then we define $f(\xi,\varepsilon)$, $g(\xi,\varepsilon)$, and $h(\xi,\varepsilon)(\zeta)$ as follows.
  \begin{compactitem}
    \item If $\xi \in H$, then we define $f(\xi,\varepsilon) = \xi$. If there exist $k \in \mathbb{N}$, $\sigma \in \Sigma^{(k)}$, and $\xi_1,\ldots,\xi_k \in \T_\Sigma(H)$ such that $\xi = \sigma(\xi_1,\ldots,\xi_k)$, then we define $f(\xi,\varepsilon) = \sigma$,

    \item  $g(\xi,\varepsilon) = \xi$, and
      \item $h(\xi,\varepsilon)(\zeta) = \zeta$.  
\end{compactitem}
  
\item[I.S.:] Let $\xi \in \T_\Sigma(H)$ and $iv \in \pos(\xi)$ with $i \in \mathbb{N}$ and $v \in \mathbb{N}^*$. Thus, there exists $k \in \mathbb{N}_+$ with $k \ge i$ and there exist  $\sigma \in \Sigma^{(k)}$ and $\xi_1,\ldots,\xi_k \in \T_\Sigma(H)$ such that $\xi= \sigma(\xi_1,\ldots, \xi_k)$. Moreover, let $\zeta \in \T_\Sigma(H)$. Then  we define $f(\xi,iv)$, $g(\xi,iv)$, and $h(\xi,iv)(\zeta)$ as follows.
\begin{compactitem}
\item $f(\xi,iv) = f(\xi_i,v)$, 
\item $g(\xi,iv) = g(\xi_i,v)$, and 
\item $h(\xi,iv)(\zeta) = \sigma(\xi_1,\ldots,\xi_{i-1},h(\xi_i,v)(\zeta),\xi_{i+1},\ldots,\xi_k)$.
\end{compactitem} 
\end{compactenum}

\index{xiofw@$\xi(w)$}
\label{page:subtree-replacement}
For every $\xi, \zeta \in \T_\Sigma(H)$ and $w \in \pos(\xi)$ we call
\index{label}
\index{subtree}
\index{replacement}
\begin{compactitem}
\item $f(\xi,w)$ the {\em label of $\xi$ at  $w$}, and we denote it by $\xi(w)$,
\item $g(\xi,w)$ the {\em subtree of $\xi$ at  $w$}, and we denote it by $\xi|_w$, and
\item $h(\xi,w)(\zeta)$ the {\em replacement of the subtree of $\xi$ at  $w$ by $\zeta$}, and we denote it by $\xi[\zeta]_w$.
\end{compactitem}

  \index{pos@$\pos_Q(\xi)$}
  \index{sub@$\sub(\xi)$}
For each subset $Q \subseteq \Sigma \cup H$, we define $\pos_Q: \T_\Sigma(H) \rightarrow {\cal P}(\mathbb{N_+}^*)$ by
  \[\pos_Q(\xi) = \{w \in \pos(\xi) \mid \xi(w) \in Q\}\]
  and, for each $\xi \in \T_\Sigma(H)$, we define the \emph{set of subtrees of $\xi$}, denoted by $\sub(\xi)$, as the set
  \[\sub(\xi) = \{\xi|_w \mid w\in \pos(\xi)\} \enspace.\]
In fact, $\sub(\xi) = \{\xi\}\cup\{\xi' \in \T_\Sigma(H) \mid \xi \succ_{\Sigma,H}^+ \xi'\}$.

\index{yield@$\yield_\Gamma$}
Let $\Gamma \subseteq \Sigma^{(0)}$.  We define the mapping $\yield_\Gamma: \T_\Sigma \to \Gamma^*$ by induction on $\T_\Sigma$ as follows.
\begin{compactenum}
\item[I.B.:] Let $\xi \in \Sigma^{(0)}$. If  $\xi \in \Gamma$, then let $\yield_\Gamma(\xi) = \xi$, otherwise let $\yield_\Gamma(\xi) = \varepsilon$.
  
  \item[I.S.:] Let $\xi = \sigma(\xi_1,\ldots,\xi_k)$ with $k\in \mathbb{N}_+$. Then we define 
 $\yield_\Gamma(\xi) = \yield_\Gamma(\xi_1) \cdots \yield_\Gamma(\xi_k)$. 
\end{compactenum}
We abbreviate $\yield_{\Sigma^{(0)}}$ by $\yield$.

   \paragraph{Comparison.} Now we wish to compare two methods for defining mappings  $h: \T_\Sigma(H) \to A$, where $H$ and $A$ are two sets. 
\begin{compactenum}
\item[(1)] We define the mapping $h$ by induction on $\T_\Sigma(H)$, i.e., by using the terminating relation $\succ_{\Sigma,H}$ and Theorem \ref{thm:dedekind}.
  \item[(2)] As additional information, we know that $A$ has a $\Sigma$-algebra structure $(A,\theta)$.
    Then we define $h$ to be the unique extension of some mapping $f: H \to A$ to a $\Sigma$-algebra homomorphism, i.e., by using algebraic methods and, in particular, Theorem \ref{thm:initial-iso}; this theorem guarantees that $h$ is a $\Sigma$-algebra homomorphism form $\sfT_\Sigma(H)$ to $(A,\theta)$.
    \end{compactenum}
    Obviously, Method (1) is more general than Method (2) because the first one can be used to define not only $\Sigma$-algebra homomorphisms.  For instance, let $\Sigma= \{\sigma^{(2)}, \alpha^{(0)}, \beta^{(0)}\}$, $H=\emptyset$, and $A=\T_\Sigma$. Moreover, we define the mapping $G: \{(\xi,g) \mid \xi \in \T_\Sigma, g: \ \succ_{\Sigma}\!\!(\xi) \to \T_\Sigma\} \to \T_\Sigma$ such that for each $(\xi,g)$ we let
    \[
      G(\xi,g) = \begin{cases}
        \sigma(\alpha,g(\xi')) & \text{ if there exists $\xi' \in \T_\Sigma$ such that $\xi = \sigma(\alpha,\xi')$}\\
        \alpha & \text{ otherwise} \enspace.
        \end{cases}
      \]
      According to Theorem \ref{thm:dedekind}, there exists a unique mapping $h: \T_\Sigma \to \T_\Sigma$ such that $h(\xi) = G(\xi,h|_{\succ_{\Sigma}(\xi)})$ for each $\xi \in \T_\Sigma$.
           
      We show that $h$ is not a $\Sigma$-algebra homomorphism from the $\Sigma$-term algebra $\sfT_\Sigma=(\T_\Sigma,\ttop_\Sigma)$ to $\sfT_\Sigma$. (We note that $G \ne G_{\ttop_\Sigma,\emptyset}$.)   We prove this claim by contradiction. For this, we  assume that $h$ is a $\Sigma$-algebra homomorphism from $\sfT_\Sigma$ to  $\sfT_\Sigma$. Then we can calculate as follows: 
             \begingroup
      \allowdisplaybreaks
      \begin{align*}
        \alpha &= G(\sigma(\beta,\alpha),h|_{\{\alpha,\beta\}}) \tag{by definition of $G$}\\
        &= h(\sigma(\beta,\alpha)) \tag{by definition of $h$}\\
        &= \ttop_\Sigma(\sigma)(h(\beta), h(\alpha)) \tag{because $h$ is a $\Sigma$-algebra homomorphism from $\sfT_\Sigma$ to $\sfT_\Sigma$}\\
        &= \sigma(h(\beta), h(\alpha)) \tag{by definition of $\ttop_\Sigma(\sigma$}\\
        &= \sigma(\alpha,\alpha) \tag{by definition of $h$ and by definition of $G$}\enspace.
      \end{align*}
      \endgroup
           This is a contradiction. Hence $h$ is not a $\Sigma$-algebra homomorphism from $\sfT_\Sigma$ to $\sfT_\Sigma$.

    Nevertheless, given that $\A=(A,\theta)$ is some $\Sigma$-algebra, Method (1) can also be used to define a $\Sigma$-algebra homomorphism from $\sfT_\Sigma(H)$ to $\A$. However, in this case we have to prove explicitly that $h$ is a homomorphism. By means of an example, we compare the two methods in the case that $h$ is a $\Sigma$-algebra homomorphism. 
 
By using Method (1):  We define the mapping $\size_1$ by well-founded induction on $(\T_\Sigma(H),\succ_{\Sigma,H})$ with the following induction base and induction step:
\begin{compactenum}
\item[I.B.:] For each $a \in \Sigma^{(0)} \cup H$, we define $\size_1(a) = 1$.
\item[I.S.:] For every $k \in \mathbb{N}_+$,
    $\sigma \in \Sigma^{(k)}$, and $\xi_1,\ldots,\xi_k \in \T_\Sigma(H)$, we define \(\size_1(\sigma(\xi_1,\ldots,\xi_k)) = 1 + \bigplus_{i \in [k]} \size_1(\xi_i)\).
  \end{compactenum}
  By Theorem \ref{thm:dedekind}, there exists exactly one mapping which satisfies these conditions. 
  
  Now we consider the $\Sigma$-algebra $(\mathbb{N},\theta)$ with $\theta(\sigma)(n_1,\ldots,n_k) = 1 + \sum_{i \in [k]} n_i$ for every $k \in \mathbb{N}$, $\sigma \in \Sigma^{(k)}$, and $n_1,\ldots,n_k \in \mathbb{N}$. Then we prove that $\size_1$ is a $\Sigma$-algebra homomorphism from $\sfT_\Sigma(H)$ to $(\mathbb{N},\theta)$. For every $k \in \mathbb{N}$, $\sigma \in \Sigma^{(k)}$, and $\xi_1,\ldots,\xi_k \in \T_\Sigma(H)$, we have
  \[
    \size_1(\ttop_\Sigma(\sigma)(\xi_1,\ldots,\xi_k)) =  \size_1(\sigma(\xi_1,\ldots,\xi_k))
    = 1 + \bigplus_{i \in [k]} \size_1(\xi_i)
    = \theta(\sigma)(\size_1(\xi_1),\ldots,\size_1(\xi_k)) \enspace.
        \]
        Hence $\size_1$ is a $\Sigma$-algebra homomorphism,

By using Method (2): We consider the $\Sigma$-algebra $(\mathbb{N},\theta)$ as above and define the mapping $f: H \to \mathbb{N}$ by $f(a)=1$ for each $a \in H$. Then Theorem \ref{thm:initial-iso} guarantees that there exists a unique extension of $f$, say, $\size_2$, into a $\Sigma$-algebra homomorphism from $\sfT_\Sigma(H)$ to $(\mathbb{N},\theta)$.

Since $\size_1$ is a $\Sigma$-algebra homomorphism and $\size_1(a) = \size_2(a)$  for each $a \in H$, we know that $\size_1 = \size_2$, due to the uniqueness property.

In comparison, for $\Sigma$-algebra homomorphisms, Method (1) seems to be slightly less efficient than Method~(2), because in the first we have to prove explicitly that the mapping $h$ is a $\Sigma$-algebra homomorphism; while in Method (2) this is guaranteed by  Theorem \ref{thm:initial-iso}.

\paragraph{Representing trees by tree domains.}
At the beginning of this section, we have defined the set $\T_\Sigma$ of $\Sigma$-trees as the smallest set of strings which contains the nullary symbols of $\Sigma$ and which is closed under top-concatenations. Here we recall an alternative way, which is  equivalent.

\index{tree domain}
A \emph{tree domain} is a finite and nonempty set $W \subseteq (\mathbb{N}_+)^*$ such that
\begin{compactitem}
\item $W$ is prefix-closed, i.e.,  for each $w\in W$ we have $\prefix(w) \subseteq W$, and 
  \item $W$ is left-closed, i.e., for every $u \in (\mathbb{N}_+)^*$ and $i \in \mathbb{N}_+$, if $ui \in W$ and $i \ge 2$, then $u(i-1) \in W$.
  \end{compactitem}
  Thus $\varepsilon$ is an element of each tree domain. Obviously, for each $\xi \in \T_\Sigma$, the set $\pos(\xi)$ is a tree domain.

  \index{tree mapping}
  \index{Sigmatreemapping@$\Sigma$-tree mapping}
  A \emph{$\Sigma$-tree mapping} is a mapping $t: W \to \Sigma$ such that
  \begin{compactitem}
  \item $W$ is a tree domain and
  \item $t$ is rank preserving, i.e., for each $w \in W$, we have $|\{j \in \mathbb{N}_+ \mid wj \in W\}| = \rk_\Sigma(t(w))$.
  \end{compactitem}
  \index{TFSigma@$\mathrm{TF}_\Sigma$}
    Let us denote the set of $\Sigma$-tree mappings by $\mathrm{TF}_\Sigma$.
    
We define the mapping $\varphi: \T_\Sigma \to \mathrm{TF}_\Sigma$ for each $\xi \in \T_\Sigma$ by
\(\varphi(\xi) = t\) such that $t: \pos(\xi) \to \Sigma$ and for each $w \in \pos(\xi)$ we let $t(w) = \xi(w)$.
For instance, let $\xi$ be the tree in Figure \ref{fig:label-subtree-replacement}. Then $\varphi(\xi) = t$ where $t: W \to \Sigma$ with tree domain $W=\{\varepsilon,1,11,2,21,211,22\}$ and $t(\varepsilon) = t(2) = \sigma$, $t(1)=t(21)=\gamma$, $t(11)=\alpha$, and $t(211)=t(22)=\beta$.

We show that $\varphi$ is bijective.
To show that $\varphi$ is injective, let $\xi_1,\xi_2 \in \T_\Sigma$ be such that $\xi_1\ne \xi_2$ and let $\varphi(\xi_i)=t_i$ for $i\in \{1,2\}$. Then there exists 
$w\in \pos(\xi_1)\cap \pos(\xi_2)$ such that $\xi_1(w)\ne \xi_2(w)$.
By definition of $\varphi$, we have $t_1(w)\ne t_2(w)$, i.e. $t_1\ne t_2$. Hence $\varphi$ is injective.

\index{TD@$\mathrm{TD}$}
For the proof of surjectivity of $\varphi$, we define the reduction system $(\mathrm{TD},\succ)$ where  $\mathrm{TD}$ is the set of all tree domains and, for every $W,W' \in \mathrm{TD}$, we let $W\succ W'$ if there exists $i\in \mathbb{N}_+$ such that $W'=\{w\in (\mathbb{N}_+)^* \mid iw\in W\}$. Then, the cardinality of tree domains is a monotone embedding of $(\mathrm{TD},\succ)$ into the terminating reduction system $(\mathbb{N},>)$. Thus, by Lemma \ref{lm:fin-branching-embedding-termination}, the relation $\succ$ is terminating.
Moreover, $\nf_\succ(\mathrm{TD})=\{\{\varepsilon\}\}$. 

By induction on $(\mathrm{TD},\succ)$, we prove that the following statement holds:
\begin{eqnarray}\label{eq:tree-domain-tree}
  \begin{aligned}
  \text{For every  $W\in \mathrm{TD}$ and $\Sigma$-tree mapping  $t: W \to \Sigma$,} &\\
  \text{there exists a $\xi \in \T_\Sigma$ such that $W=\pos(\xi)$ and  $\varphi(\xi)=t$.} &
  \end{aligned}
\end{eqnarray}
For the proof, let $W \in \mathrm{TD}$ and $t: W \to \Sigma$ be a $\Sigma$-tree mapping. Then there exist $k \in \mathbb{N}$ and $\sigma \in \Sigma$ such that  $t(\varepsilon) = \sigma$. Moreover, for each $i \in [k]$, we define $W_i = \{w \in (\mathbb{N}_+)^* \mid iw \in W\}$ and the $\Sigma$-tree mapping $t_i: W_i \to \Sigma$ by $t_i(w) = t(iw)$ for each $w \in W_i$. Then $W = \{\varepsilon\} \cup \{iw \mid i \in [k], w \in W_i\}$.

Since $\{W' \in \mathrm{TD} \mid W \succ W' \} = \{W_1,\ldots,W_k\}$, the I.H. implies that, for each $i\in [k]$, there exists $\xi_i \in \T_\Sigma$ such that $W_i = \pos(\xi_i)$ and  $\varphi(\xi_i) = t_i$.

Then let $\xi = \sigma(\xi_1,\ldots,\xi_k)$. We claim that $\varphi(\xi) = t$. First, by definition we have that  $\varphi(\xi): \pos(\xi) \to \Sigma$. Moreover,  
\[
 \pos(\xi)  = \{\varepsilon\} \cup \{iw \mid  i \in [k], w \in \pos(\xi_i)\} = \{\varepsilon\} \cup \{iw \mid i \in [k], w \in W_i\} = W \enspace,
\]
where the second equality follows from I.H. Second, for the proof of $\varphi(\xi)=t$ we let $w \in W$. If $w= \varepsilon$, then $\xi(w) = \sigma = t(w)$. If $w = i w'$ for some $i \in \mathbb{N}_+$ and $w' \in W_i$, then
\[
\xi(iw') = \xi_i(w') = t_i(w') = t(iw') \enspace,
\]
where  the second equality follows from I.H.
Hence $\varphi(\xi) = t$.  This proves \eqref{eq:tree-domain-tree}, and hence $\varphi$ is surjective.

Since $\varphi$ is bijective we can specify a tree $\xi \in \T_\Sigma$ also by means of a $\Sigma$-tree mapping (cf., e.g., the proofs of Lemma \ref{lm:wta-wcfg}, of Theorem \ref{thm:closure-tree-concatenation-wrtg}, and of Theorem \ref{thm:decomposition-1}).

\index{prefix order}
\index{smallerpref@$\le_{\mathrm{pref}}$}
\index{lesspref@$\le_{\mathrm{pref}}$}
\label{page:orders-on-positions-of-trees}
\paragraph{Orders on positions of trees.}
The \emph{prefix order on $\pos(\xi)$}, denoted by $\le_{\mathrm{pref}}$, is the partial order defined for every $w,v \in \pos(\xi)$ by
\begin{align*}
w \le_{\mathrm{pref}} v \ \text { iff } w \in \prefix(v) \enspace.
\end{align*}
We let $w <_{\mathrm{pref}} v$ if $(w \le_{\mathrm{pref}} v) \wedge (w\ne v)$.

\index{lexicographic order}
\index{lexi@$\le_{\mathrm{lex}}$}
\index{depth-first post-order}
\index{smallerdp@$\le_{\mathrm{dp}}$}
\index{lessdp@$\le_{\mathrm{dp}}$}
The \emph{lexicographic order on $\pos(\xi)$}, denoted by $\le_{\mathrm{lex}}$, and the \emph{depth-first post-order on $\pos(\xi)$}, denoted by $\le_{\mathrm{dp}}$, are the linear orders on $\pos(\xi)$ defined for every $w,v \in \pos(\xi)$ by
\begin{align*}
w \le_{\mathrm{lex}} v \ &\text{ iff } (w\in \prefix(v)) \vee (w <_{\mathrm{left}\mathord{\n}\mathrm{of}} v)\\
w \le_{\mathrm{dp}} v \ &\text{ iff } (v\in \prefix(w)) \vee (w <_{\mathrm{left}\mathord{\n}\mathrm{of}} v) \enspace, \ \text{where}\\
(w <_{\mathrm{left}\mathord{\n}\mathrm{of}} v) \ &\text{ iff } (\exists u \in \prefix(w) \cap \prefix(v))(\exists i,j \in \mathbb{N}_+)
 (ui \in \prefix(w)) \wedge (uj \in \prefix(v)) \wedge (i < j)\enspace.
\end{align*}
The lexicographic order is also called \emph{depth-first pre-order}.
We let $w <_{\mathrm{lex}} v$ if $(w \le_{\mathrm{lex}} v) \wedge (w\ne v)$, and
 we let $w <_{\mathrm{dp}} v$ if $(w \le_{\mathrm{dp}} v) \wedge (w\ne v)$.

\index{tree relabeling}
\paragraph{Tree relabelings.} A \emph{$(\Sigma,\Delta)$-tree relabeling} (or simply: \emph{tree relabeling}) \cite[Def.~3.1]{eng75-15}  is an $\mathbb{N}$-indexed family $\tau = (\tau_k \mid k \in \mathbb{N})$  of mappings $\tau_k: \Sigma^{(k)} \to \cP(\Delta^{(k)})$.
\index{tree relabeling!non-overlapping}
\index{tree relabeling!deterministic}
We call a tree relabeling $\tau$
\begin{compactitem}
\item \emph{non-overlapping} if $\tau_k(\sigma) \cap \tau_k(\sigma') = \emptyset$ for every $k \in \mathbb{N}$ and  $\sigma,\sigma' \in \Sigma^{(k)}$ with $\sigma \ne \sigma'$. 
\item \emph{deterministic} if for every $k \in \mathbb{N}$ and  $\sigma \in \Sigma^{(k)}$, the set $\tau_k(\sigma)$ contains exactly one element. Then we specify the tree relabeling as an $\mathbb{N}$-indexed family of mappings $\tau_k: \Sigma^{(k)} \to \Delta^{(k)}$. 
    \end{compactitem}

Now let $\tau = (\tau_k \mid k \in \mathbb{N})$ be a $(\Sigma,\Delta)$-tree relabeling. 
We extend $\tau$ to the mapping 
\[
  \tau':\T_\Sigma \rightarrow \cP_{\mathrm{fin}}(\T_\Delta)
\]
by using Theorem \ref{thm:initial-iso} as follows. We define the $\Sigma$-algebra $(\cP_{\mathrm{fin}}(\T_\Delta),\theta_\tau)$ such that, for every $k \in \mathbb{N}$, $\sigma \in \Sigma^{(k)}$, and $U_1,\ldots,U_k \in \cP_{\mathrm{fin}}(\T_\Delta)$, we let
\[
  \theta_\tau(\sigma)(U_1,\ldots,U_k) = \{\gamma(\zeta_1,\ldots,\zeta_k) \, | \, \gamma \in
  \tau_k(\sigma), \zeta_1 \in U_1,\ldots,\zeta_k \in U_k\} \enspace.
\]
Then $\tau':\T_\Sigma \rightarrow \cP_{\mathrm{fin}}(\T_\Delta)$ is the unique $\Sigma$-homomorphism $h$ from the $\Sigma$-term algebra $\sfT_\Sigma$ to the $\Sigma$-algebra $(\cP_{\mathrm{fin}}(\T_\Delta),\theta_\tau)$ (cf. Theorem \ref{thm:initial-iso}). We also call the mapping  $\tau':\T_\Sigma \rightarrow \cP_{\mathrm{fin}}(\T_\Delta)$ a $(\Sigma,\Delta)$-tree relabeling (or simply: tree relabeling) and we abbreviate $\tau'$ by $\tau$. The mapping $\tau$ can also be considered as a binary relation $\tau\subseteq \T_\Sigma \times \T_\Delta$, and thus, in particular, $\tau^{-1}$ is defined.

 We note that $\pos(\zeta) = \pos(\xi)$ for each $\zeta \in \tau(\xi)$. Thus, for every $\zeta \in \T_\Delta$,  the set $\tau^{-1}(\zeta)$ is finite and, if $\tau$ is non-overlapping, then $|\tau^{-1}(\zeta)| \le 1$. If $\tau$ is deterministic, then for every $\xi \in \T_\Sigma$ we have  $|\tau(\xi)| = 1$.

Let $\Omega$ be a ranked alphabet, $\omega=(\omega_k \mid k \in \mathbb{N})$ be a $(\Sigma,\Omega)$-tree relabeling,  and $\tau=(\tau_k \mid k \in \mathbb{N})$ be a $(\Omega,\Delta)$-tree relabeling. The \emph{syntactic composition of $\omega$ and $\tau$}, denoted by $\tau \hat{\circ} \omega$,
is the  $(\Sigma,\Delta)$-tree relabeling
$((\tau \hat{\circ} \omega)_k \mid k \in \mathbb{N})$ such that $(\tau \hat{\circ} \omega)_k =\tau_k \circ \omega_k$, i.e.,  $(\tau \hat{\circ} \omega)_k(\sigma)=\bigcup_{\delta\in \omega_k(\sigma)}\tau_k(\delta)$ for every $k\in \mathbb{N}$ and $\sigma\in \Sigma^{(k)}$. It follows that if $\omega$ and $\tau$ are deterministic, then $\tau \hat{\circ} \omega$ is also deterministic.

\begin{theorem}\label{thm:relabeling-synt-comp}{\rm (cf. \cite[Lm. 3.4]{eng75})} Let $\Omega$ be a ranked alphabet, 
$\omega$ be a $(\Sigma,\Omega)$-tree relabeling,  and $\tau$ be a $(\Omega,\Delta)$-tree relabeling. Then $\tau \hat{\circ} \omega= \tau\circ \omega$, i.e., the syntactic composition of $\omega$ and $\tau$ determines the mapping $\tau\circ \omega$.
\end{theorem}

\index{tree substitution}
\index{xi@$\xi[\xi_1,\ldots,\xi_n]$}
\paragraph{Substitution and contexts.} Now we define {\em tree substitution} (cf. derived operator in \cite[p.~73]{gogthawagwri77}). Let  $Z=\{z_1,z_2,\ldots\}$ be a set of variables, disjoint with $\Sigma$, and let  $Z_n=\{z_1,\ldots,z_n\}$ for every $n \in \mathbb{N}$. Sometimes we write $z$ for $z_1$. Moreover, let $H$ be a set disjoint with $\Sigma$; let $n \in \mathbb{N}$, $\xi \in \T_\Sigma(Z_n)$, and $\xi_1,\ldots,\xi_n \in \T_\Sigma(H)$. Intuitively, we want to define the tree in $\T_\Sigma(H)$  which is obtained from $\xi$ by replacing,  for each $z_i \in Z_n$, each occurrence of $z_i$ by $\xi_i$. Formally, let  $\varphi: Z_n \to \T_\Sigma(H)$ be the mapping  defined by $\varphi(z_i) = \xi_i$ for each $i \in [n]$, and let $\overline{\varphi}$ be the unique extension of $\varphi$ to a $\Sigma$-algebra homomorphism from the $\Sigma$-term algebra $(\T_\Sigma(Z_n),\ttop_\Sigma)$ over $Z_n$ to the $\Sigma$-term algebra $(\T_\Sigma(H),\ttop_\Sigma)$ over $H$. It is easy to see that $\overline{\varphi}(\xi)$ is the desired tree from our informal discussion. In the sequel, we abbreviate  $\overline{\varphi}(\xi)$ by $\xi[\xi_1,\ldots,\xi_n]$.
\index{circz@$\circ_z$}
For $n=1$ and $H=\{z\}$, we also denote $\xi[\xi_1]$ by $\xi \circ_z \xi_1$. Hence, $\circ_z$ can be thought of as a binary operation on $\T_\Sigma(\{z\})$. In fact, $\circ_z$ is associative, hence 
$(\T_\Sigma(\{z\}),\circ_z,z)$ is a monoid. 

\index{contextn@$\C_\Sigma(Z_n)$}
\index{context@$\C_\Sigma$}
\index{context}
\index{Sigmacontext@$\Sigma$-context}
\index{variables}
\label{page:context-and-substitution}
 Next we define \emph{contexts}.  Let $\xi \in \T_\Sigma(Z)$ and $V\subseteq Z$. We say that \emph{$\xi$ is linear in $V$} (\emph{$\xi$ is nondeleting in $V$}) if for each $z\in V$ we have $|\pos_{z}(\xi)|\le 1$ (and $|\pos_{z}(\xi)|\ge 1$, respectively).  We denote by $\C_\Sigma(Z_n)$ the set of all trees $\xi\in\T_\Sigma(Z_n)$ which are both linear and nondeleting in $Z_n$. Since $Z_0=\emptyset$, we have $\C_\Sigma(Z_0) = \T_\Sigma$. We call the elements of $\C_\Sigma(Z_n)$ \emph{contexts over $\Sigma$ and $Z_n$} (or: \emph{$n$-contexts} if $\Sigma$ is clear). We abbreviate $\C_\Sigma(\{z\})$ by $\C_\Sigma$ and call its elements \emph{$\Sigma$-contexts} (or: \emph{contexts}).\label{p:contexts} We also denote by $\circ_z$ the restriction of $\circ_z$ to $\C_\Sigma$.
Since $\C_\Sigma$ is closed under  $\circ_z$, the triple $(\C_\Sigma,\circ_z,z)$ is a monoid.

\index{elementary context}
\index{depth}
For each $c \in \C_\Sigma$, we  denote by $\mathrm{depth}(c)$ the length of the unique $z$-labeled position of $c$, i.e., $\mathrm{depth}(c)=|\pos_z(c)|$. We say that $c$ is \emph{elementary} if $\mathrm{depth}(c)=1$, i.e., there exist $k \in \mathbb{N}_+$, $\sigma \in \Sigma^{(k)}$, $i \in [k]$, and $\xi_1,\ldots,\xi_{i-1},\xi_{i+1},\ldots,\xi_k \in \T_\Sigma$ such that
\begin{align*}
  c =  \sigma(\xi_1,\ldots,\xi_{i-1},z,\xi_{i+1},\ldots,\xi_k)\enspace.
\end{align*}
\index{eC@$\e\C_\Sigma$}
We denote the set of all elementary $\Sigma$-contexts by $\e\C_\Sigma$.

\index{succCSigma@$\succ_{\C_\Sigma}$}
\index{CSigma@$(\C_\Sigma,\succ_{\C_\Sigma})$}
\label{page:order-on-contexts}
In the next lemma we prove that $(\C_\Sigma,\circ_z, z)$ is free in the set of all monoids with generating set $\e\C_\Sigma$. In order to give an inductive proof, we consider the reduction system $(\C_\Sigma,\succ_{\C_\Sigma})$ where
\[
  \succ_{\C_\Sigma} = \ \succ_{\Sigma,\{z\}} \! \cap \ (\C_\Sigma \times \C_\Sigma)\enspace.
\]
Since $\C_\Sigma \subseteq \T_\Sigma(\{z\})$  and $(\T_\Sigma(\{z\}),\succ_{\Sigma,\{z\}})$ is terminating, by Lemma~\ref{lm:termination-is-subset-closed}, also $(\C_\Sigma,\succ_{\C_\Sigma})$ is terminating. Clearly, $\nf_{\succ_{\C_\Sigma}}(\C_\Sigma)=\{z\}$ and we have $c_1\succ_{\C_\Sigma} c_2$ if and only if there exists an elementary context $e\in \e\C_\Sigma$ such that $c_1=e[c_2]$.

\begin{lemma}\rm \cite[Prop.~9.1]{berreu82} \label{lm:freely-generated-context}  The monoid $(\C_\Sigma,\circ_z,z)$ is free in the set of all monoids with generating set $\e\C_\Sigma$.
\end{lemma}
\begin{proof} We abbreviate $(\e\C_\Sigma)^*$ by $\e\C_\Sigma^*$. We prove that the monoid $(\C_\Sigma,\circ_z,z)$ is isomorphic to the monoid $(\e\C_\Sigma^*,\cdot,\varepsilon)$.
Since $(\e\C_\Sigma^*,\cdot,\varepsilon)$ is free  in the set of all monoids with generating set $\e\C_\Sigma$, this implies the statement of the lemma.
  
  We define the mapping $\psi: \e\C_\Sigma \to \C_\Sigma$ with $\psi(e) = e$ for each $e \in \e\C_\Sigma$. Since $(\e\C_\Sigma^*,\cdot,\varepsilon)$ is free in the set of all monoids with generating set $\e\C_\Sigma$, there exists a unique monoid homomorphism $\sem{.}: \e\C_\Sigma^* \to \C_\Sigma$ which extends $\psi$. For each $c\in \e\C_\Sigma^*$, we write $\sem{c}$ instead of $\sem{.}(c)$. Then, for every $n \in \mathbb{N}_+$ and  $e_1, \ldots, e_n \in  \e\C_\Sigma$ we have
  \[\sem{e_1 \cdots e_n} = e_1 \circ_z \cdots \circ_z e_n \ \text{ and } \ \sem{\varepsilon}=z \enspace. 
\]
  \index{$\sem{e_1 \cdots e_n}$}
  
We prove that the mapping $\sem{.}$ is surjective. For this, by induction on $(\C_\Sigma,\succ_{\C_\Sigma})$, we prove that the following statement holds:
  \begin{equation}\label{eq:generated-n}
\text{For each $c \in \C_\Sigma$, we have $c \in  \im(\sem{.})$\enspace. } 
\end{equation}

I.B.: Let $c=z$. Then the statement holds because $\sem{\varepsilon}=z$.

I.S.: Let $c=e[c']$ for some $e\in \e\C_\Sigma$ and $c' \in \C_\Sigma$. By I.H. we have $c' \in \im (\sem{.})$ and thus there exists a $d \in \e\C_\Sigma^*$ such that $\sem{d}=c'$. Since $\sem{ed} = \sem{e}\circ_z \sem{d} =  e \circ_z c' = c$, we have that $c \in \im(\sem{.})$.

Hence \eqref{eq:generated-n} holds and $\sem{.}$ is surjective.

Finally, we prove that $\sem{.}$ is injective. Let $e_1,\ldots, e_n \in \e\C_\Sigma$ and  $e'_1,\ldots, e'_m \in \e\C_\Sigma$ such that $\sem{e_1\cdots e_n} = \sem{e'_1\cdots e'_m}$.
Since $\mathrm{depth}(\sem{e_1 \cdots e_n}) =n$ and $\mathrm{depth}(\sem{e'_1 \cdots e'_m}) =m$, it follows that $n=m$. 
Now assume that there exists $i \in [n]$ such that $e_i \not= e'_i$. Then there exists a $w\in \pos(e_i)\cap \pos(e'_i)$ such that $e_i(w) \ne e'_i(w)$. Using this fact, it is easy to see that $\sem{e_1\cdots e_n} \ne \sem{e'_1\cdots e'_m}$.
 Thus $e_i=e'_i$ for each $i \in [n]$. Hence, $\sem{.}$ is injective.

  So we have proved that $\sem{.}$ is a monoid isomorphism from $(\e\C_\Sigma^*,\cdot,\varepsilon)$ to $(\C_\Sigma,\circ_z,z)$.
\end{proof}

\index{Z@$Z$}
\index{X@$X$}
\label{p:cdot-z}
\begin{quote}\emph{In the rest of the book,  $Z$ and $X$ will denote sets of variables, if not specified otherwise. Moreover, we  let $Z_n=\{z_1,\ldots,z_n\}$ and $X_n=\{x_1,\ldots,x_n\}$ for every $n \in \mathbb{N}$.}
  \end{quote}

  \index{tree homomorphism}
  \index{alphabetic}
\paragraph{Tree homomorphisms.} A \emph{$(\Sigma,\Delta)$-tree homomorphism} (or just \emph{tree homomorphism}) \cite[Def.~3.62]{eng75-15} is a family
$h=(h_k \mid k \in \mathbb{N})$ of mappings $h_k : \Sigma^{(k)} \to \T_\Delta(Z_k)$. We call a tree homomorphism $h$
\begin{compactitem}
\item \emph{linear}  if, for every $k \in \mathbb{N}$ and $\sigma \in \Sigma^{(k)}$, the tree $h_k(\sigma)$ is linear in $Z_k$;
\item \emph{nondeleting} if, for every $k \in \mathbb{N}$ and $\sigma \in \Sigma^{(k)}$, the tree $h_k(\sigma)$ is nondeleting in $Z_k$;
\item \emph{alphabetic} if, for every $k \in \mathbb{N}$ and $\sigma  \in \Sigma^{(k)}$, we have $|\pos_\Delta(h_k(\sigma))| \le 1$;
\item \emph{productive}  if, for every $k \in \mathbb{N}$ and $\sigma  \in \Sigma^{(k)}$, we have $h_k(\sigma) \not\in Z_k$ (this notion is borrowed from \cite{fulmalvog10} where it was called output-productive);
\item \emph{ordered} if $h$ is linear and nondeleting and, for every $k \in \mathbb{N}$ and $\sigma  \in \Sigma^{(k)}$, the sequence of variables occurring in $h_k(\sigma)$ is ordered by increasing indices from left to right, i.e., it is $z_1,\ldots,z_k$;
  \item \emph{simple} if $h$ is alphabetic and ordered (and hence also linear and nondeleting).
  \end{compactitem}

  Now let $H$ be a set  disjoint with $\Sigma$ and $\Delta$. Moreover, let $h= (h_k \mid k \in \mathbb{N})$ be a $(\Sigma,\Delta)$-tree homomorphism. We extend $h$ to  the mapping
  \[
    h : \T_\Sigma(H) \to \T_\Delta(H)
  \]
  by using Theorem \ref{thm:initial-iso} as follows. We define the $\Sigma$-algebra $(\T_\Delta(H),\theta_h)$  such that, for every $k \in \mathbb{N}$, $\sigma \in \Sigma^{(k)}$, and $\xi_1,\ldots,\xi_k \in \T_\Delta(H)$, we let
\[
  \theta_h(\sigma)(\xi_1,\ldots,\xi_k) = h_k(\sigma)[\xi_1,\ldots,\xi_k] \enspace.
\]
Then $h : \T_\Sigma(H) \to \T_\Delta(H)$  is the unique extension of $\id_H$ to a $\Sigma$-algebra homomorphism from the $\Sigma$-term algebra $(\T_\Sigma(H),\ttop_\Sigma)$ over $H$ to the $\Sigma$-algebra $(\T_\Delta(H),\theta_h)$  (cf. Theorem \ref{thm:initial-iso}).

We also call the mapping $h : \T_\Sigma(H) \to \T_\Delta(H)$ a $(\Sigma,\Delta)$-tree homomorphism (or simply: tree homomorphism).

We note that each deterministic tree relabeling can be thought of as a particular tree homomorphism. Indeed, let $(\tau_k \mid k \in \mathbb{N})$ be a deterministic $(\Sigma,\Delta)$-tree relabeling. Then $\tau$ determines the $(\Sigma,\Delta)$-tree homomorphism $(\tau'_k \mid k \in \mathbb{N})$ where, for every $k \in \mathbb{N}$ and $\sigma \in \Sigma^{(k)}$, we have $\tau'_k(\sigma)=\tau_k(\sigma)(z_1,\ldots,z_k)$. It is obvious that $\tau=\tau'$.

Let $\Omega$ be a ranked alphabet, $g=(g_k \mid k \in \mathbb{N})$ be a $(\Sigma,\Omega)$-tree homomorphism,  and $h=(h_k \mid k \in \mathbb{N})$ be an $(\Omega,\Delta)$-tree homomorphism. The \emph{syntactic composition of $g$ and $h$}, denoted by $h \hat{\circ} g$,
is the  $(\Sigma,\Delta)$-tree homomorphism
$((h \hat{\circ} g)_k \mid k \in \mathbb{N})$ such that $(h \hat{\circ} g)_k(\sigma)=h(g_k(\sigma))$ for every $k\in \mathbb{N}$ and $\sigma\in \Sigma^{(k)}$.

\begin{theorem}\label{thm:hom-synt-comp}{\rm (cf. \cite[Lm. 3.4]{eng75}, \cite[Thm. 4.3.7]{gecste84})} Let $\Omega$ be a ranked alphabet, 
$g$ be a $(\Sigma,\Omega)$-tree homomorphism,  and $h$ be an $(\Omega,\Delta)$-tree homomorphism. Then $h \hat{\circ} g= h\circ g$, i.e., the syntactic composition of $g$ and $h$ determines the mapping $h\circ g$.
\end{theorem}


\section{Term rewrite systems}\label{sect:term-tewrite-systems}

We recall some notions and results from \cite{hue80,baanip98}. Moreover, as an application, we prove Observation~\ref{obs:zero-sum-free-property}(7).

Let $\xi \in \T_\Sigma(X)$. We denote by $\Var(\xi)$ the set of all $x \in X$ which occurs in $\xi$.

\index{term rewrite system}
A \emph{term rewrite system (over $\Sigma$)} is a finite set $R$ of term rewrite rules of the form $\ell \to r$ where $\ell,r \in \T_\Sigma(X)$ such that $\ell \not\in X$ and $\Var(r) \subseteq \Var(\ell)$. We call $\ell$ and $r$ the \emph{left-hand side} of the rule and the \emph{right-hand side} of the rule, respectively.

\index{substitution}
A \emph{substitution} is a mapping $\varphi: X \to \T_\Sigma(X)$ such that $\varphi(x)=x$ for almost all $x \in X$. A substitution extends in a unique way to a $\Sigma$-algebra homomorphism $\varphi'$ from $\sfT_\Sigma(X)$ to $\sfT_\Sigma(X)$. In the sequel, we drop the prime from  $\varphi'$. 

The \emph{term rewrite relation induced by $R$}, denoted by  $\Rightarrow_R$, is the binary relation on $\T_\Sigma(X)$ defined as follows. For every $\xi_1,\xi_2 \in \T_\Sigma(X)$, we let $\xi_1 \Rightarrow_R \xi_2$ if there exist a position $w \in \pos(\xi_1)$, a rule $\ell \to r$ in $R$,  a substitution $\varphi: X \to \T_\Sigma(X)$ such that
$\xi_1|_w = \varphi(\ell)$ and $\xi_2 = \xi_1[\varphi(r)]_w$.

In fact, $(\T_\Sigma(X),\Rightarrow_R)$ is a reduction system. We call a term rewrite system $R$ terminating, if $\Rightarrow_R$ is terminating.

Now we turn to the missing proof of Observation \ref{obs:zero-sum-free-property}(7). We will employ the concept of reduction order \cite[Def.~5.2.1]{baanip98} and monotone polynomial interpretation \cite[Def.~5.3.6]{baanip98}, which we will recall here.

Let $\succ$ be a strict order on $\T_\Sigma(X)$. We say that $\succ$ is a \emph{reduction order} if it is
\begin{compactitem}
    \item terminating,
\item \emph{compatible with $\Sigma$-operations}, i.e., for every $k \in \mathbb{N}$, $\sigma \in \Sigma^{(k)}$, $i \in [k]$, and $\xi_1,\ldots,\xi_k,\zeta,\zeta' \in \T_\Sigma(X)$, if $\zeta \succ \zeta'$, then $\sigma(\xi_1,\ldots,\xi_{i-1},\zeta,\xi_{i+1},\ldots,\xi_k) \succ \sigma(\xi_1,\ldots,\xi_{i-1},\zeta',\xi_{i+1},\ldots,\xi_k)$, and
\item \emph{closed under substitution}, i.e., for every $\xi_1,\xi_2 \in \T_\Sigma(X)$ and substitution $\varphi: X \to \T_\Sigma(X)$, if $\xi_1 \succ \xi_2$, then $\varphi(\xi_1) \succ \varphi(\xi_2)$.
\end{compactitem}

  \begin{theorem}\label{thm:5.2.3} {\rm \cite[Thm.~5.2.3]{baanip98} (also cf. \cite{mannes70})} Let $R$ be a term rewrite system over $\Sigma$. Then $R$ is terminating if and only if there exists a reduction order $\succ$ on $\T_\Sigma(X)$ such that, for each $\ell \to r$ in $R$, we have~$\ell  \succ r$. 
  \end{theorem}

  We obtain particular reduction orders by considering monotone polynomial interpretations. For the definition of the latter, we recall that 
  $\mathbb{N}[x_1,\ldots,x_k]$ denotes the set of all polynomials with $k$ variables $x_1,\ldots,x_k$ and coefficients in $\mathbb{N}$ (cf. Example \ref{ex:semirings}(\ref{def:semiring-of-polynomials})).
  
   A \emph{monotone polynomial interpretation of $\Sigma$} is a $\Sigma$-algebra $\cA =(A,\theta)$ such that
  
  \begin{compactitem}
  \item $A \subseteq \mathbb{N} \setminus \{0\}$ and
    \item for every $k \in \mathbb{N}$ and $\sigma \in \Sigma^{(k)}$, $\theta(\sigma)$ is a polynomial in $\mathbb{N}[x_1,\ldots,x_k]$ such that, for each $x_i \in \{x_1,\ldots,x_k\}$, the polynomial $\theta(\sigma)$ contains a monomial with non-zero coefficient in which $x_i$ occurs with exponent at least $1$.
    \end{compactitem}

Let $\cA = (A,\theta)$ be a $\Sigma$-algebra with $A\not=\emptyset$ and $\succ$ a terminating binary relation on $A$. We define the binary relation $\succ_\cA$ on $\T_\Sigma(X)$ as follows. For every $\xi_1,\xi_2 \in \T_\Sigma(X)$, we let $\xi_1 \succ_\cA \xi_2$ if for each $\Sigma$-algebra homomorphism $\pi: \T_\Sigma(X) \to \cA$ we have $\pi(\xi_1) \succ \pi(\xi_2)$. (We recall that each substitution of variables in $\xi_1$ and $\xi_2$ determines uniquely such a homomorphism.)

\begin{corollary} \rm \label{cor:5.3.7+5.3.3}  Let $\cA = (A,\theta)$ be a monotone polynomial interpretation of $\Sigma$ with $A\not=\emptyset$ and let $\succ$ be a terminating binary relation on $A$. Then $\succ_\cA$ is a reduction order on $\T_\Sigma(X)$.
    \end{corollary}
    \begin{proof} This follows from \cite[Lm.~5.3.7]{baanip98} and \cite[Thm.~5.3.3]{baanip98}.
    \end{proof}

\

Now we provide the missing proof of Observation \ref{obs:zero-sum-free-property}(7).
    
\emph{Proof.} Let $A \subseteq B$ be finite and $h: \T_{\{\oplus,\otimes, \0,\1\}}(A) \to B$ be the canonical evaluation homomorphism, i.e., the unique extension of $\mathrm{id}_A$ to a homomorphism from $\sfT_{\{\oplus,\otimes, \0,\1\}}(A)$ to $\B$ (cf. Theorem \ref{thm:initial-iso}). We define the ranked alphabet $\Sigma = \{\oplus^{(2)},\otimes^{(2)},\0^{(0)},\1^{(0)}\} \cup A$ where each element of $A$ has arity~$0$. Then
\[
  \T_{\{\oplus,\otimes, \0,\1\}}(A) = \T_\Sigma \enspace.
  \]

 Clearly, $h( \T_\Sigma) = \langle A \rangle_{\{\oplus,\otimes,\0,\1\}}$. Let $\T_\Sigma^r$ be the set of all $\xi \in \T_\Sigma$ such that, for each $w \in \pos(\xi)$ with $\xi(w) = \otimes$, we have $\xi(w2) \in A\cup \{\0,\1\}$. Clearly, $h(\T_\Sigma^r) = \CL(A)$.

 We prove the following statement.
 \begin{equation}\label{equ:iterated-left-distr}
   \text{For each  $\xi \in \T_\Sigma$, we can construct a  $\zeta \in \T_\Sigma^r$ such that $h(\xi) = h(\zeta)$.}
 \end{equation}
 
 For the construction of $\zeta$ from the given $\xi$,  we will employ a terminating term rewrite system over $\Sigma$; then $\zeta$ will be an arbitrary normal form of $\xi$ with respect to that term rewrite system. (In fact, each $\xi$ has exactly one normal form, but we do not prove this here.) Formally, we consider the set $R$ with the following term rewrite rules:
\begin{align*}
 (1):\ \otimes(x_1,\otimes(x_2,x_3)) &\to  \otimes(\otimes(x_1,x_2),x_3)\\
 (2):\  \otimes(x_1,\oplus(x_2,x_3)) &\to \oplus(\otimes(x_1,x_2),\otimes(x_1,x_3))\enspace,
\end{align*}
which correspond to one direction of associativity of $\otimes$ and of left-distributivity, respectively. Then, $\Rightarrow_R$ is the term rewrite relation induced by $R$ on  $\T_\Sigma(X)$.

It is clear that
\begin{equation}\label{equ:equality-for-T-r}
\nf_{\Rightarrow_R}(\T_\Sigma) = \T_\Sigma^r \enspace.
\end{equation}

Due to associativity and  left-distributivity in $\B$, it is clear that the following statement holds: 
\begin{equation}\label{eq:rediction-preserves-evaluation}
\text{For every $\xi_1,\xi_2 \in \T_\Sigma$ with $\xi_1 \Rightarrow_R \xi_2$, we have $h(\xi_1)=h(\xi_2)$} \enspace.
\end{equation}

Next, by using Corollary \ref{cor:5.3.7+5.3.3}  and Theorem~\ref{thm:5.2.3}, we will show that $\Rightarrow_R$ is terminating\footnote{We are grateful to Stefan Borgwardt for providing us the proof idea and the particular monotone polynomial interpretation.}. 
For this purpose, we define the $\Sigma$-algebra $\cC = (C,\theta)$ \cite{bor24} where
\begin{compactitem}
\item $C = \mathbb{N} \setminus \{0,1\}$ and
  \item $\theta(\oplus)(n_1,n_2) = n_1 + n_2$ for every $n_1,n_2 \in \mathbb{N}$,
  \item $\theta(\otimes)(n_1,n_2) = n_1 \cdot {n_2}^2$ for every $n_1,n_2 \in \mathbb{N}$, and
    \item $\theta(\0)() = \theta(\1)() = \theta(a)() =  2$ for each $a \in A$.
    \end{compactitem}

It is easy to see that $\cC$ is a monotone polynomial interpretation of $\Sigma$. Hence, by Corollary~\ref{cor:5.3.7+5.3.3} (with the usual order $>$ on $\mathbb{N}$, e.g., $5 > 3$),  the relation $>_\cC$ is a reduction order on $\T_\Sigma(X)$.

In order to apply Theorem \ref{thm:5.2.3}, we have to prove that
\[ \ell_1 >_\cC r_1 \ \text{ and } \ \ell_2 >_\cC r_2
\]
where $\ell_i$ and $r_i$ abbreviate the left-hand side and the right-hand side of rule ($i$), respectively, with $i \in \{1,2\}$. 

Let $P_{\ell_1}$, $P_{\ell_2}$, $P_{r_1}$, and $P_{r_2}$ be the polynomials in $\mathbb{N}[x_1,x_2,x_3]$ which are obtained from $\ell_1,\ell_2,r_1,r_2$ by replacing each occurrence of $\oplus$ and $\otimes$ by $\theta(\oplus)$ and $\theta(\otimes)$, respectively, and by substituting these polynomials into each other successively.
Thus
  \begin{align*}
    P_{\ell_1}(x_1,x_2,x_3) &= x_1 \cdot (x_2 \cdot {x_3}^2)^2 = x_1{x_2}^2{x_3}^4\\
    P_{r_1}(x_1,x_2,x_3) &= x_1 {x_2}^2{x_3}^2\\[2mm]
        P_{\ell_2}(x_1,x_2,x_3) &= x_1 \cdot (x_2 + x_3)^2 = x_1{x_2}^2  + 2 \cdot x_1x_2x_3 + x_1{x_3}^2\\
        P_{r_2}(x_1,x_2,x_3) &= x_1{x_2}^2 + x_1{x_3}^2 \enspace.
  \end{align*}

  Then it is easy to see that, for every $n_1,n_2,n_3 \in \mathbb{N} \setminus \{0,1\}$, we have
  \begin{align*}
P_{\ell_1}(n_1,n_2,n_3) > P_{r_1}(n_1,n_2,n_3) \ \text{ and } \ P_{\ell_2}(n_1,n_2,n_3) > P_{r_2}(n_1,n_2,n_3) \enspace.
    \end{align*}
By definition of $>_\cC$, this implies that $\ell_1 >_\cC r_1$ and $\ell_2 >_\cC r_2$ (also cf. \cite[Lm.~5.3.8]{baanip98}). 
Hence, by Theorem \ref{thm:5.2.3}, the term rewrite system $R$ is terminating.

Now we turn to the proof of \eqref{equ:iterated-left-distr}. For this, let $\xi \in \T_\Sigma$. Since $\Rightarrow_R$ is terminating, we can apply Observation \ref{obs:terminating-implies-normalizing}; hence, $\nf_{\Rightarrow_R}(\xi) \not= \emptyset$. Let $\zeta \in \nf_{\Rightarrow_R}(\xi)$. Then $\zeta \in \T_\Sigma^r$ by \eqref{equ:equality-for-T-r} and $h(\xi) = h(\zeta)$ by~\eqref{eq:rediction-preserves-evaluation}. This finishes the proof of \eqref{equ:iterated-left-distr}.

Finally, we can calculate as follows.
\begingroup
\allowdisplaybreaks
\begin{align*}
  \langle A \rangle_{\{\oplus,\otimes,\0,\1\}}
  &= h( \T_\Sigma)\\
  &=  h( \T_\Sigma^r) \tag{the set inclusion $\subseteq$ follows from  \eqref{equ:iterated-left-distr}}\\
  &= \CL(A)
\end{align*}
\endgroup
Since $\B$ is weakly locally finite, the set $\CL(A)$ is finite, and hence $\langle A \rangle_{\{\oplus,\otimes,\0,\1\}}$ is finite. Thus $\B$ is locally finite. \hfill$\Box$


\section{Weighted tree languages and transformations}
\label{sec:weighted-tree-languages-and-transformations}

\subsection{Weighted tree languages and operations}
\label{sec:weighted-tree-languages}

\index{weighted tree language}
\index{SigmaB-weighted tree language@$(\Sigma,\B)$-weighted tree language}
Let $H$ be a set with $\Sigma\cap H=\emptyset$. A {\em weighted tree language over $\Sigma$, $H$, and $\B$}  is a $\B$-weighted set $r: \T_\Sigma(H) \rightarrow B$. If $H=\emptyset$, then we just say \emph{weighted tree language over $\Sigma$ and $\B$} or \emph{$(\Sigma,\B)$-weighted tree language}. A \emph{$\B$-weighted tree language} is a $(\Sigma,\B)$-weighted tree language for some ranked alphabet $\Sigma$. A \emph{weighted $\Sigma$-tree language} is a $(\Sigma,\B)$-weighted tree language for some strong bimonoid~$\B$. Finally, a \emph{weighted tree language} is a $(\Sigma,\B)$-weighted tree language for some ranked alphabet $\Sigma$ and some strong bimonoid~$\B$.

 \index{weighted tree language!constant}
  \index{weighted tree language!polynomial}
  \index{weighted tree language!monomial}
    \index{weighted tree language!scalar multiplication}
  \index{weighted tree language!sum}
    \index{weighted tree language!Hadamard product}
    \index{weighted tree language!application}
      \index{weighted tree language!support}
    Since each $(\Sigma,\B)$-weighted tree language is a particular $\B$-weighted set, by Section \ref{sect:weighted-sets-languages} the following notions and operations are already defined for every $b \in B$, $r: \T_\Sigma \to B$, $r_1: \T_\Sigma \to B$, $r_2: \T_\Sigma \to B$, and  $L\subseteq \T_\Sigma$:
  \begin{compactitem}
  \item polynomial weighted tree language,
  \item monomial weighted tree language,
  \item constant weighted tree language,
    \item support,
  \item characteristic mapping of $L \subseteq \T_\Sigma$ with respect to $\B$,
  \item scalar multiplication from the left ($b \cdot r$) and scalar multiplication from the right ($r \cdot b$),
  \item sum of $r_1$ and $r_2$ ($r_1 \oplus r_2$), 
  \item Hadamard product of $r_1$ and $r_2$ ($r_1 \otimes r_2$), and
    \item application of $r$ to $L$ ($r(L)$).
    \end{compactitem}

    \index{Pol(Sigma,B)@$\Pol(\Sigma,\B)$}
We denote the set of polynomial $(\Sigma,\B)$-weighted tree languages by $\Pol(\Sigma,\B)$. We call each element of $\Pol(\Sigma,\B)$ a \emph{$(\Sigma,\B)$-polynomial}. 

We note that, for each $r\in \Pol(\Sigma,\B)$, there exist $n\in \mathbb{N}_+$, $b_1,\ldots,b_n\in B$, and pairwise different $\xi_1,\ldots,\xi_n \in \T_\Sigma$ such that $r=b_1.\xi_1\oplus \ldots \oplus b_n.\xi_n$. (We recall that, for each $i\in[n]$,  $b_i.\xi_i$ is a monomial such that for each $\xi\in\T_\Sigma$ we have
$(b_i.\xi_i)(\xi)= b_i$ if $\xi= \xi_i$ and $\0$ otherwise). Indeed, if $\supp(r)=\emptyset$, then $r=\0.\alpha$, where $\alpha$ is an arbitrary element of $\Sigma^{(0)}$. Otherwise, $\supp(r)=\{\xi_1,\ldots,\xi_n\}$ for some $n \in \mathbb{N}_+$ and, for each $i\in [n]$, there exists $b_i\in B\setminus \{\0\}$ such that  $r(\xi_i)=b_i$.
Then, by using the definition of monomial and of $\oplus$, we have $r=b_1.\xi_1\oplus \ldots \oplus b_n.\xi_n$.

\index{preimage property}
A $(\Sigma,\B)$-weighted tree language $r$ has the \emph{preimage property} if, for each $b \in B$, the $\Sigma$-tree language $r^{-1}(b)$ is recognizable (in the sense of Section \ref{sec:fta}).

\index{support@$\supp(\cC)$}
We extend the concept of support to sets of weighted tree languages in the natural way: for each set $\cC$ of weighted tree languages, we define $\supp(\cC)=\{\supp(r)\mid r\in \cC\}$.

\paragraph{Evaluation algebras.}
\index{homM@$\h_{\M(\Sigma,\kappa)}$}
\index{evaluation algebra}
We define particular $(\Sigma,\B)$-weighted tree languages; they are the unique $\Sigma$-algebra homomorphisms from $\sfT_\Sigma$ to $\Sigma$-algebras which are based on the evaluation of $\Sigma$-symbols in~$\B$.

Formally,  let $\kappa=(\kappa_k \mid k \in \mathbb{N})$ be an $\mathbb{N}$-indexed family of mappings $\kappa_k: \Sigma^{(k)} \to B$. The \emph{$(\Sigma,\kappa)$-evaluation algebra}, denoted by $\M(\Sigma,\kappa)$, is the $\Sigma$-algebra $(B,\overline{\kappa})$, where
\[
  \overline{\kappa}(\sigma)(b_1,\ldots,b_k) = b_1 \otimes \cdots \otimes b_k \otimes \kappa_k(\sigma)
\]
for every $k \in \mathbb{N}$, $\sigma \in \Sigma^{(k)}$, and $b_1,\ldots,b_k \in B$.
We recall that $\h_{\M(\Sigma,\kappa)}$ denotes the unique $\Sigma$-algebra homomorphism from $\sfT_\Sigma$ to $\M(\Sigma,\kappa)$.
Then, for each $\xi = \sigma(\xi,\ldots,\xi_k)$ in $\T_\Sigma$, we have
\begin{equation}\label{eq:evaluation-of-a-tree} 
  \h_{\M(\Sigma,\kappa)}(\xi)= \overline{\kappa}(\sigma)(\h_{\M(\Sigma,\kappa)}(\xi_1),\ldots,\h_{\M(\Sigma,\kappa)}(\xi_k))
=  \Big(\bigotimes_{i \in [k]} \h_{\M(\Sigma,\kappa)}(\xi_i)\Big) \otimes  \kappa_k(\sigma)\enspace.  
\end{equation}

Obviously, we have 
\begin{equation}\label{eq:hom-kappa}
  \h_{\M(\Sigma,\kappa)}(\xi)= \bigotimes_{\substack{w \in \pos(\xi)\\\text{in $<_{\mathrm{dp}}$ order}}} \kappa_{\rk(\xi(w))}(\xi(w)) \enspace. 
\end{equation}

In particular, $\h_{\M(\Sigma,\kappa)}: \T_\Sigma \to B$ is a $(\Sigma,\B)$-weighted tree language.

\subsection{Weighted tree transformations}\label{sect:weighted-tr-tr}
\label{sec:weighted-tree-transformations}

  \index{weighted tree transformation}
  \index{SigmaDeltaBweightedtreetransf@$(\Sigma,\Delta,\B)$-weighted tree transformation}
\index{supp-i-finite}
\index{supp-o-finite}
Let $A$ be a set and $\tau: \T_\Sigma \times A \rightarrow B$ a $\B$-weighted set. 
We say that
\begin{compactitem}
\item $\tau$ is \emph{supp-i-finite} if, for every $a \in A$, the set  $\{\xi \in \T_\Sigma \mid \tau(\xi,a) \not= \mathbb{0}\}$ is finite;
  \item $\tau$ is \emph{supp-o-finite} if, for every $\xi \in \T_\Sigma$, the set  $\{a \in A \mid \tau(\xi,a) \not= \mathbb{0}\}$ is finite.
  \end{compactitem}
  Here supp-i-finite and supp-o-finite abbreviate ``support input finite'' and ``support output finite'', respectively.
  If $A=\T_\Delta$, then we call $\tau$ a \emph{$(\Sigma,\Delta,\B)$-weighted tree transformation} or simply \emph{weighted tree transformation}.
  
  \index{tausummable@$\tau$-summable}
  Let $r: \T_\Sigma \rightarrow B$ be a weighted tree language and $\tau: \T_\Sigma \times A \rightarrow B$  be a $\B$-weighted set. We say that $r$ is \emph{$\tau$-summable} if $\tau$ is supp-i-finite or $r$ has finite support.
  \index{application}
  If $r$ is $\tau$-summable or $\B$ is $\sigma$-complete, then  the \emph{application of $\tau$ to  $r$}, denoted by $\tau(r)$, is the $\B$-weighted set $\tau(r): A \rightarrow B$ defined, for each $a \in A$, by 
\begin{equation} \label{equ:appl-wtt-to-wtl}
\tau(r)(a) = \infsum{\oplus}{\xi \in \T_\Sigma} r(\xi) \otimes \tau(\xi,a)\enspace.
\end{equation}
Obviously, if $r$ is  $\tau$-summable, then
\begin{equation}\label{equ:appl-wttfinpreim-to-wtl}
  \infsum{\oplus}{\xi \in \T_\Sigma} r(\xi) \otimes \tau(\xi,a) = \bigoplus_{\xi \in \T_\Sigma} r(\xi) \otimes \tau(\xi,a)\enspace. 
  \end{equation}

  We will use this application in three different scenarios.

\paragraph{Scenario 1.} In this scenario, $\tau$ is the characteristic mapping of a binary relation. Formally, let $r:\T_\Sigma \to B$ and $g \subseteq \T_\Sigma \times A$. If $r$ is $\chi(g)$-summable or $\B$ is $\sigma$-complete, then for each $a\in A$ we have
    \begin{equation}\label{obs:app-tree-transf-to-wtl-2}
\chi(g)(r)(a) =  \infsum{\oplus}{\xi \in \T_\Sigma}{r(\xi) \otimes \chi(g)(\xi,a)}
= \infsum{\oplus}{\xi \in g^{-1}(a)}{r(\xi)} \enspace. 
\end{equation}
We will use the following five cases of this scenario. In Case (a) we have $A=\Gamma^*$, and in Cases (b)-(e) we have $A=\T_\Delta$. Moreover, $\tau$ is the characteristic mapping of
\begin{compactitem}
\item[(a)]    a yield mapping $g : \T_\Sigma \to \Gamma^*$, where $\Gamma \subseteq \Sigma^{(0)}$ is an alphabet (Section \ref{sect:yield-of-weighted-cf}), 
\item[(b)]  a tree relabeling $g\subseteq \T_\Sigma \times \T_\Delta$ (Section \ref{sect:tree-relabeling-preserves} and Section \ref{sec:representable-wtl}), 
\item[(c)]  a tree homomorphism $g: \T_\Sigma \to \T_\Delta$ (Section \ref{sect:homomorphism-preserves}), 
\item[(d)]  the inverse of a tree homomorphism $g\subseteq \T_\Delta \times \T_\Sigma$ (Section~\ref{sect:inverse-homomorphism-preserves}), and
\item[(e)] a tree relabeling  $g\subseteq \T_\Sigma \times \T_\Delta$ (Subsection~\ref{subsec:alternative-recog-implies-def}).
\end{compactitem}

  As a consequence of \eqref{obs:app-tree-transf-to-wtl-2}, we obtain the following inclusion and equality: 
  \begin{eqnarray}\label{equ:supp-yield=yieldA}
    \begin{aligned}
             \supp(\chi(g)(r)) \subseteq g(\supp(r)) \  \text{ and,}& \\
             \text{if $\B$ is zero-sum free, then
               $\supp(\chi(g)(r)) = g(\supp(r))$}\enspace.&
             \end{aligned} 
           \end{eqnarray}

First we prove the inclusion. For this, let $a \in A$. Then we have 
\begingroup
\allowdisplaybreaks
\begin{align*}
 a \in \supp(\chi(g)(r))
\  \Longleftrightarrow \ \  &  \Big(\ \infsum{\oplus}{\xi \in g^{-1}(a)}{r(\xi)}\ \Big) \ne \0 \tag{by \eqref{obs:app-tree-transf-to-wtl-2}}\\
\  \Longrightarrow \ \ & (\exists \xi \in g^{-1}(a)): r(\xi) \ne \0 \tag{by Observation \ref{obs:sum-emptyset}}\\[2mm]
   \  \Longleftrightarrow \ \  & a \in g(\supp(r))\enspace.
    \end{align*}
    \endgroup
    If $\B$ is zero-sum free, then the above implication can be turned into an equivalence by Observation \ref{obs:zero-sum-free-property}(10) or (11), respectively.

    The next observation will be useful later.

\begin{observation}\label{obs:inside-out}\rm Let $g \subseteq \T_\Sigma \times \T_\Delta$ be a  $(\Sigma,\Delta)$-tree relabeling, $r: \T_\Sigma \to B$, $L \subseteq \T_\Sigma$, and $\xi \in \T_\Delta$. Then
  \(\chi(g)\big(\chi(L) \otimes r\big)(\xi) = r(g^{-1}(\xi) \cap L)\).
\end{observation}
\begin{proof} 
\begingroup
\allowdisplaybreaks
\begin{align*}
r(g^{-1}(\xi) \cap L)
    &=  \bigoplus_{\zeta \in g^{-1}(\xi) \cap L}r(\zeta)
    \tag{we note that $g^{-1}(\xi) \cap L$ is finite}\\
    &=  \bigoplus_{\zeta \in g^{-1}(\xi)}\chi(L)(\zeta)\otimes r(\zeta)
    =  \bigoplus_{\zeta \in g^{-1}(\xi)}\big(\chi(L)\otimes r\big)(\zeta)\\
    &=\chi(g)\big( \chi(L)\otimes r\big)(\xi) \enspace.  \tag{by \eqref{obs:app-tree-transf-to-wtl-2}}
    \end{align*}
    \endgroup
  \end{proof}

\paragraph{Scenario 2.} \index{diagonalization}
\index{$\overline{r'}$}
\label{page:diagonalization}
In this scenario, $A=\T_\Sigma$ and $\tau$ is the diagonalization of a  $(\Sigma,\B)$-tree language  $r': \T_\Sigma \to B$. We define the \emph{diagonalization of $r'$} \cite[Sect.~2.6]{fulmalvog11} to be the $(\Sigma,\Sigma,\B)$-weighted tree transformation $\overline{r'}: \T_\Sigma \times \T_\Sigma \to B$ such that, for every $\xi,\zeta \in \T_\Sigma$, we let 
  \begin{equation}\label{equ:weighted-tree-lang-as-weighted-tree-transf}
  \overline{r'}(\xi,\zeta) =
  \begin{cases}
    r'(\xi) & \text{ if $\xi =\zeta$}\\
    \0 & \text{ otherwise} \enspace.
    \end{cases} 
  \end{equation}
  Obviously, $\overline{r'}$ is supp-i-finite and hence, for each $r:\T_\Sigma \to B$, the application $\overline{r'}(r)$ is defined. Then the application of the diagonalization of $r'$ to $r$ is the Hadamard product of $r$ and $r'$, i.e., $\overline{r'}(r)=r\otimes r'$, because for  each $\zeta \in \T_\Sigma$ we have
  \begin{equation}\label{obs:application-vs-Hadamard}
    (\overline{r'}(r))(\zeta) = \infsum{\oplus}{\xi \in \T_\Sigma}{r(\xi) \otimes \overline{r'}(\xi,\zeta)} = r(\zeta) \otimes r'(\zeta) =  (r\otimes r')(\zeta)\enspace. 
    \end{equation}
  In the sequel we will sometimes drop the bar from $\overline{r'}$. Then it will be clear from the context whether $r'$ denotes a $(\Sigma,\B)$-weighted tree language or its diagonalization. We will use this type of application, e.g., in the alternative proof of closure of the set of recognizable weighted tree languages under Hadamard product (cf. Subsection \ref{ssec:closure-Had-alternative}).

\paragraph{Scenario 3.} Finally, in this scenario, we have $A = \T_\Delta$, i.e., $\tau$ is a $(\Sigma,\Delta,\B)$-weighted tree transformation. Let $r: \T_\Sigma \to B$ and $\tau: \T_\Sigma \times \T_\Delta \to B$. If $r$ is $\tau$-summable or $\B$ is $\sigma$-complete, then for each $\zeta \in \T_\Delta$ we have
    \begin{equation} \label{equ:appl-wtt-to-wtl-trees}
      \tau(r)(\zeta) =  \infsum{\oplus}{\xi \in \T_\Sigma}{r(\xi) \otimes \tau(\xi,\zeta)} \enspace.
     \end{equation}
We will use this type of application, e.g., in Section \ref{sec:w-projective-bimorphisms} and in Chapter \ref{ch:AFwtL}.

\

\index{composition}
Next we define the composition of weighted tree transformations.
Let $\Omega$ be a ranked alphabet and $\tau: \T_\Sigma \times \T_\Delta \rightarrow B$ and $\tau': \T_\Delta \times \T_\Omega \rightarrow B$ be  weighted tree transformations. If $\tau$ is supp-o-finite, $\tau'$ is supp-i-finite, or $\B$ is $\sigma$-complete, then we define the {\em composition} of $\tau$ and $\tau'$ to be the weighted tree transformation $(\tau;\tau'): \T_\Sigma \times \T_\Omega \rightarrow B$
defined by 
\begin{equation}\label{equ:comp-wtt}
  (\tau;\tau')(\xi,\zeta)=\infsum{\oplus}{\eta\in \T_\Delta}\tau(\xi,\eta)\otimes \tau'(\eta,\zeta) 
  \end{equation}
for every $\xi\in \T_\Sigma$ and $\zeta\in \T_\Omega$. Clearly, if $\tau$ is supp-o-finite or $\tau'$ is supp-i-finite, then
\[\infsum{\oplus}{\eta\in \T_\Delta}\tau(\xi,\eta)\otimes \tau'(\eta,\zeta) = \bigoplus_{\eta\in \T_\Delta}\tau(\xi,\eta)\otimes \tau'(\eta,\zeta) \enspace.\]

We show that, roughly speaking, if $\B$ is a semiring, then the composition of weighted tree transformations is associative. Intuitively, the conditions $P_1$ and $P_2$ of the next observation guarantee that the expressions $(\tau_1;\tau_2);\tau_3$ and $\tau_1;(\tau_2;\tau_3)$, respectively, are well defined.

\begin{observation}\rm\label{obs:composition-semiring-associative} Let $\B$ be a semiring. Moreover, let 
$\Omega$ and $\Psi$ be ranked alphabets and
\(\tau_1: \T_\Sigma \times \T_\Delta \rightarrow B,  \tau_2: \T_\Delta \times \T_\Omega \rightarrow B,  \text{ and } \tau_3: \T_\Omega \times \T_\Psi \rightarrow B\)  be weighted tree transformations such that
the condition $P_1 \wedge P_2$ holds or $\B$ is $\sigma$-complete, where
\begin{compactitem}
\item $P_1$: [$\tau_1$ is supp-o-finite or $\tau_2$ is supp-i-finite] and [$\tau_1;\tau_2$ is supp-o-finite or $\tau_3$ is supp-i-finite] and 
\item $P_2$:  [$\tau_2$ is supp-o-finite or $\tau_3$ is supp-i-finite] and [$\tau_1$ is supp-o-finite or $\tau_2;\tau_3$ is supp-i-finite].
  \end{compactitem}
Then $(\tau_1;\tau_2);\tau_3=\tau_1;(\tau_2;\tau_3)$.
\end{observation}
\begin{proof} Due to the first conjunct (and the second conjunct) of $P_1$ the compositions $\tau_1;\tau_2$ (and $(\tau_1;\tau_2);\tau_3$, respectively) are defined. Moreover, due to the first conjunct (and the second conjunct) of $P_2$ the compositions $\tau_2;\tau_3$ (and $\tau_1;(\tau_2;\tau_3)$, respectively) are defined.

  Let $\xi \in \T_\Sigma$ and $\zeta \in \T_\Psi$. Then
\begingroup
\allowdisplaybreaks
\begin{align*}
\big((\tau_1;\tau_2);\tau_3 \big)(\xi,\zeta)= & \infsum{\oplus}{\eta\in \T_\Omega} (\tau_1;\tau_2)(\xi,\eta) \otimes \tau_3(\eta,\zeta)=  \infsum{\oplus}{\eta\in \T_\Omega} \; \Big( \infsum{\oplus}{\theta\in \T_\Delta}\tau_1(\xi,\theta) \otimes \tau_2(\theta,\eta) \Big) \otimes \tau_3(\eta,\zeta)\\[2mm] = &  \infsum{\oplus}{\eta\in \T_\Omega} \; \Big( \infsum{\oplus}{\theta\in \T_\Delta}\tau_1(\xi,\theta) \otimes \tau_2(\theta,\eta)  \otimes \tau_3(\eta,\zeta) \Big)\tag{\text{by right-distributivity}}\\[2mm]
= & \infsum{\oplus}{\theta\in \T_\Delta}\tau_1(\xi,\theta) \otimes \Big( \infsum{\oplus}{\eta\in \T_\Omega}
\tau_2(\theta,\eta)  \otimes \tau_3(\eta,\zeta) \Big)\tag{\text{by left-distributivity}}\\[2mm]
= &\infsum{\oplus}{\theta\in \T_\Delta}\tau_1(\xi,\theta) \otimes (\tau_2;\tau_3)(\theta,\zeta)=
    \big(\tau_1;(\tau_2;\tau_3)\big)(\xi,\zeta). \qedhere
\end{align*}
\endgroup
\end{proof}

Next we show that, roughly speaking, if $\B$ is a semiring, then the application of the  composition of weighted tree transformations $\tau_1$ and $\tau_2$ to a weighted tree language $r$  can be expressed as the consecutive applications of $\tau_1$ and $\tau_2$ to $r$. Intuitively, the conditions $P_1$ and $P_2$ of the next observation guarantee that the expressions  $(\tau;\tau')(r)$ and $\tau'(\tau(r))$, respectively, are well defined.

\begin{observation} \rm\label{obs:application-of-composition} Let $\B$ be a semiring. Moreover, let $\Omega$ be a ranked alphabet, $\tau: \T_\Sigma \times \T_\Delta \rightarrow B$ and $\tau': \T_\Delta \times \T_\Omega \rightarrow B$ be  weighted tree transformations, and let   $r:\T_\Sigma \to B$ be a weighted tree language such that the condition $P_1 \wedge P_2$ holds or $\B$ is $\sigma$-complete, where
\begin{compactitem}
\item $P_1$:  [$\tau$ is supp-o-finite or $\tau'$ is supp-i-finite] and [$\tau;\tau'$ is supp-i-finite or  $r$ has finite support] and 

\item $P_2$: [$\tau$ is supp-i-finite or $r$ has finite support] and [$\tau'$ is supp-i-finite or $\tau(r)$ has finite support].
\end{compactitem}
Then we have
 $(\tau;\tau')(r) = \tau'(\tau(r))$.
\end{observation}
\begin{proof}  Due to the first conjunct (and the second conjunct) of $P_1$ the composition $\tau;\tau'$ (and the application $(\tau;\tau')(r)$, respectively) are defined. Moreover, due to the first conjunct (and the second conjunct) of $P_2$ the applications $\tau(r)$ (and $\tau'(\tau(r))$, respectively) are defined.

Let $\theta \in \T_\Omega$. Then
\begingroup
\allowdisplaybreaks
\begin{align*}
  (\tau;\tau')(r)(\theta)
  = & \infsum{\oplus}{\xi \in \T_\Sigma}{r(\xi) \otimes (\tau;\tau')(\xi,\theta)}
  =   \infsum{\oplus}{\xi \in \T_\Sigma} r(\xi) \otimes \infsum{\oplus}{\zeta \in \T_\Delta}\tau(\xi,\zeta) \otimes \tau'(\zeta,\theta)\\
  = &  \infsum{\oplus}{\xi \in \T_\Sigma} \infsum{\oplus}{\zeta \in \T_\Delta} r(\xi) \otimes \tau(\xi,\zeta) \otimes \tau'(\zeta,\theta) \tag{\text{by left-distributivity}}\\
  = &  \infsum{\oplus}{\zeta \in \T_\Delta}  \infsum{\oplus}{\xi \in \T_\Sigma} r(\xi) \otimes \tau(\xi,\zeta) \otimes \tau'(\zeta,\theta)\\
  = &  \infsum{\oplus}{\zeta \in \T_\Delta}  \Big(\infsum{\oplus}{\xi \in \T_\Sigma} r(\xi) \otimes \tau(\xi,\zeta)\Big) \otimes \tau'(\zeta,\theta)  \tag{\text{by right-distributivity}}\\
  = &  \infsum{\oplus}{\zeta \in \T_\Delta}  \tau(r)(\zeta) \otimes \tau'(\zeta,\theta) 
  =   \tau'(\tau(r))(\theta)\enspace. \qedhere
\end{align*}
\endgroup
\end{proof}
 

\section{Finite-state string automata}
\label{sec:fsa}

\index{fsa}
\index{finite-state string automaton}
We recall the definition of $\Gamma$-automaton from \cite[p.~12]{eil74}.
A \emph{finite-state string automaton over $\Gamma$} (for short: \emph{$\Gamma$-fsa}) is a quadruple $A=(Q,I,\delta,F)$ where $Q$ is a finite nonempty set of states such that $Q \cap \Gamma = \emptyset$, $I\subseteq Q$ (initial states), $\delta \subseteq Q \times \Gamma \times Q$ (transitions), and $F\subseteq Q$ (final states).

\index{run}
Let $w=a_1\cdots a_n$ be a string in $\Gamma^*$ with $n \in \mathbb{N}$ and $a_1,\ldots,a_n \in \Gamma$. A {\em run of $A$ on $w$} is a string $\rho = q_0 \cdots q_{n}$ in $Q^{n+1}$ such that, for each $i \in [0,n-1]$ we have $(q_i,a_{i+1},q_{i+1}) \in \delta$. We denote $q_0$ and $q_{n}$ by $\first(\rho)$ and $\last(\rho)$, respectively. The \emph{language recognized by $A$}, denoted by $\LL(A)$, is the set
\[
\LL(A) = \{w \in \Gamma^* \mid \text{there exists a run $\rho$ of $A$ on $w$ such that $\first(\rho) \in I$ and $\last(\rho) \in F$}\}\enspace.
\]
\index{recognizable}
Let $L\subseteq \Gamma^*$. We call $L$ \emph{recognizable} if there exists an fsa $A$ such that $\LL(A)=L$.


\section{Context-free grammars}
\label{sec:context-free-grammars}

The theory of context-free grammars is well established; for more details, we refer the reader to, e.g., \cite{har78,hopull79,hopmotull07}. Here we only recall the most basic definitions.

\index{context-free grammar}
\index{cfg}
A \emph{context-free grammar over $\Gamma$} (for short: \emph{$\Gamma$-cfg}) is a triple $G = (N,S,R)$ where $N$ is a finite set (nonterminals) with $N \cap \Gamma = \emptyset$, $S \subseteq N$ with $S \ne \emptyset$ (initial nonterminals), and $R$ is a finite set (rules); each rule has the form $A \to \alpha$ where $A \in N$ and $\alpha\in (N \cup \Gamma)^*$.

Let $r=(A \rightarrow \alpha)$ be a rule. We call $r$ a \emph{chain rule} (an \emph{$\varepsilon$-rule}, a \emph{terminal rule}) if $\alpha \in N$ (and if $\alpha = \varepsilon$, and if $\alpha \in \Gamma^*$,  respectively).  The left-hand side of $r$ is the nonterminal $A$, denoted by $\lhs(r)$. Moreover, the $i$-th occurrence of a nonterminal in $\alpha$ (counted from left to right) is denoted by  $\rhs_{N,i}(r)$. We sometimes want to show the occurrences of elements of $N$ in the right-hand side of rule $r$ more explicitly. Then we will also write $r$ in the form
\[
  A \rightarrow u_0 A_1 u_1 \cdots A_k u_k\enspace,
\]
where $k\in \mathbb{N}$, $u_0,u_1,\ldots,u_k \in \Gamma^*$, and $A_1,\ldots,A_k\in N$. If $S$ contains only one element, say $S_0$, then we denote $G$ by $(N,S_0,R)$.

\index{derivation relation}
\index{Rightarrow@$\Rightarrow_{G}$}
The \emph{derivation relation (of $G$)} is the binary relation $\Rightarrow_{G}$ on $(N \cup \Gamma)^*$ defined as follows. For every $\gamma,\delta \in (N \cup \Gamma)^*$ and rule $A \to \alpha$ in $R$, we have
$\gamma A \delta \Rightarrow_G \gamma \alpha \delta$. If $G$ is clear from the context, then we denote $\Rightarrow_G$ by $\Rightarrow$. We recall that  the reflexive and transitive closure of $\Rightarrow$ is denoted by $\Rightarrow^*$.

\index{derivation}
A \emph{derivation (of $G$)} is a sequence $d= (\gamma_1,\ldots,\gamma_n)$ with $n \in \mathbb{N}_+$, $\gamma_i \in (N \cup \Gamma)^*$, and $\gamma_i \Rightarrow \gamma_{i+1}$ for each $i \in [n-1]$. We denote $d$ also by $\gamma_1 \Rightarrow_G^* \gamma_n$.  
Let $L_1,L_2 \subseteq (N \cup \Gamma)^*$. We denote by $\D_G(L_1,L_2)$ the set of all derivations $\gamma_1\Rightarrow_G^*\gamma_n$ such that $\gamma_1 \in L_1$ and $\gamma_n \in L_2$. 

\index{language generated by $\cG$}
\index{languageG@$\LL(G)$}
The \emph{language generated by $G$} is the set \(\LL(G) = \{w \in \Gamma^* \mid (\exists\, S_0 \in S) : S_0 \Rightarrow_G^* w\}\).
or, equivalently, \(\LL(G) = \{w \in \Gamma^* \mid \D_G(S,w)\ne \emptyset\}\). Let $L \subseteq \Gamma^*$. We call $L$ a \emph{context-free language} if there exists a $\Gamma$-cfg $G$ such that $\LL(G) = L$.

\index{leftmost derivation relation}
\index{Rightarrow@$\Rightarrow_{G,\lm}$}
The \emph{leftmost derivation relation} $\Rightarrow_{G,\lm}$ is defined such that, for every $u\in \Gamma^*$, $\gamma \in (N \cup \Gamma)^*$, and rule $A \to \alpha$ in $R$, we have
$ uA \gamma \Rightarrow_{G,\lm} u\alpha \gamma$.
If $G$ is clear from the context, then we denote $\Rightarrow_{G,\lm}$ by $\Rightarrow_\lm$. The concept of \emph{leftmost derivation} is defined analogously to the concept of derivation. For every $L_1,L_2 \subseteq (N \cup \Gamma)^*$, we denote by $\D_{G,\lm}(L_1,L_2)$ the set of all leftmost derivations $\gamma_1\Rightarrow_{G,\lm}^*\gamma_n$ such that $\gamma_1 \in L_1$ and $\gamma_n \in L_2$.
It is well known that \(\LL(G) = \{w \in \Gamma^* \mid S_0 \in S, S_0 \Rightarrow_{G,\lm}^* w\}\).

Let $G = (N,S,R)$ be a context-free grammar over $\Gamma$ and $A \in N$ be a nonterminal.  
We say that
\begin{compactitem}
\item $A$ is \emph{generating} if there exists $w \in \Gamma^*$ such that $A \Rightarrow^* w$,\index{generating}

\item $A$ is \emph{reachable} if there exist $S_0\in S$ and $\gamma,\delta \in (N \cup \Gamma)^*$ such that $S_0 \Rightarrow^* \gamma A \delta$, and \index{reachable}

\item $A$ is \emph{useful} if it is generating and reachable.\index{useful}

\end{compactitem}
We say that $G$ is \emph{reduced} if $R=\emptyset$ or each nonterminal $A \in N$ is useful.\index{reduced}

\begin{theorem}\label{thm:reduced-cfg}{\rm \cite[Thm.~3.2.3]{har78}} For each $\Gamma$-cfg $G$ with a single initial nonterminal $S_0$, we can  construct a $\Gamma$-cfg $G'$  such that $G'$ has the single initial nonterminal $S_0$, $\LL(G')=\LL(G)$, and $G'$ is reduced.
  \end{theorem}

  Finally, we give a characterization of the languages generated by context-free grammars which have a terminal rule. The characterization uses the concept of  rule trees (cf.  \cite[Def.~3.54]{eng75-15} and \cite[Def.~3.2.8]{gecste84}). Intuitively, each leftmost derivation corresponds to exactly one rule tree, and vice versa.
   Because of our general assumption made at the beginning of Section \ref{sect:trees},
 that each ranked alphabet contains a symbol of rank zero, such a characterization is only possible for cfg which have a terminal rule. Clearly, we can transform each cfg into an equivalent one which has a terminal rule. 

Formally, let $G=(N,S,R)$ be a $\Gamma$-cfg which has a terminal rule. We consider $R$ as ranked alphabet by defining the rank of each rule $r$ to be the number of nonterminals in the right-hand side of $r$. Hence, each terminal rule has rank $0$ and, due to our assumption, $R^{(0)} \ne \emptyset$.

\index{projection}
\index{pi@$\pi_G$}
\label{page:projection}
We define the \emph{projection of $G$}, denoted by $\pi_G$, to be the mapping
\[
  \pi_G: \T_R \to \Gamma^*
\]
defined by induction on $\T_R$ as follows.
For every $k \in \mathbb{N}$, $r \in R^{(k)}$ of the form $r = (A \rightarrow u_0 A_1 u_1 \cdots A_k u_k)$, and $d_1,\ldots,d_k \in \T_R$, we define
\[
  \pi_G(r(d_1,\ldots,d_k)) = u_0 \pi_G(d_1) u_1 \ldots \pi_G(d_k) u_k \enspace.
  \]
If there is no confusion, then we drop the index $G$ from $\pi_G$ and  just write $\pi$.

\index{rule tree}
\index{weighted context-free grammar!rule tree}
Now let $d \in \T_R$. We say that $d$ is a  \emph{rule tree of $G$} if, for every $w\in \pos_R(d)$ and $i \in [\rk_R(d(w))]$,  we have  \(\rhs_{N,i}(d(w))= \lhs(d(wi))\).
  Let $A \in N$, $u \in \Gamma^*$, and $d \in \T_R$ be a rule tree of $G$. We say that $d$ is
  \begin{compactitem}
  \item  an \emph{$A$-rule tree of $G$} if $\lhs(d(\varepsilon))=A$.
  \item a \emph{rule tree of $G$ for $u$} if $\pi_G(d)=u$.
    \end{compactitem}
    \index{RT@$\RT_G(A,u)$}
\index{RT@$\RT_G(N',L)$}
We denote the \emph{set of all $A$-rule trees of $G$ for $u$} by $\RT_G(A,u)$. For every $N'\subseteq N$ and $L \subseteq \Gamma^*$,  we define $\RT_G(N',L) = \bigcup_{A \in N', u \in L} \RT_G(A,u)$, and we abbreviate  $\RT_G(S,L)$ by $\RT_G(L)$. \label{page:rule}
Obviously, for each $u \in \Gamma^*$, we have $\pi_G^{-1}(u) \cap \RT_G(\Gamma^*) = \RT_G(u)$.
Finally, we abbreviate $\RT_G(\Gamma^*)$ by~$\RT_G$.
Then the following is easy to see.

\begin{observation}\rm \label{obs:cfg=pi(rule-trees)} For each $\Gamma$-cfg $G$ which has a terminal rule  we have that $\LL(G) = \pi_G(\RT_G)$.\hfill $\Box$
\end{observation}

We note that $\LL(G) =\emptyset$ for each $\Gamma$-cfg $G$ which does not have a terminal rule. Therefore such grammars are not relevant. On the other hand, $\LL(G)=\emptyset$ does not imply that $G$ does not have a terminal rule.

\begin{example}\rm We consider the alphabet $\Gamma = \{a,b,c\}$ and the $\Gamma$-cfg $G=(N,S,R)$ with $N=\{S,A,B\}$ and the rules
  \[
S \to aAB, \ A \to AaBA, \ A \to B, \ B \to bc, \ A \to \varepsilon \enspace.
\]
Figure \ref{fig:example-rule-tree} shows the $A$-rule tree $d \in \RT_G(A,abcbc)$ of $G$ for $abcbc$.
  \end{example}

  \begin{figure}
    \centering
   \begin{tikzpicture}[scale=1, every node/.style={transform shape},
					node distance=0cm and 0cm,
					level distance= 1.3cm,
					level 1/.style={sibling distance=22mm}]
					
\node at (-2,0.7) {$d\in \RT_{G}(A,abcbc):$}; 
\node (root) {$A \to AaBA$}
  child {node {$A \to \varepsilon$}}
  child {node {$B \to bc$}}
  child {node {$A \to B$}
   child { node {$B \to bc$}}};
\node[below right= 0.1cm and 3.6cm of root] (word) {\strut $abcbc$};
\node[right= of word, baseline=(word.base)] (in) {\strut $\in$};
\node[right= of in, baseline=(word.base)] {\strut $\Gamma^*$};
\draw[->,>=stealth,semithick] ($(word.west)+(-1.4,0)$) -- ($(word.west)+(-0.6,0)$)
  node[midway, above] {$\pi_G$};
\end{tikzpicture}

    \caption{\label{fig:example-rule-tree} The $A$-rule tree $d \in \RT_G(A,abcbc)$ of $G$ for $abcbc$.}
  \end{figure}


  \section{Regular tree grammars}
  \label{sec:rtg}

  \index{SigmaXi@$\Sigma^\Xi$}
  By definition, each tree $\xi$ over $\Sigma$ is a particular string over the alphabet $\Sigma \cup \Xi$ where $\Xi$ contains the opening and closing parentheses and the comma. For convenience, we abbreviate $\Sigma \cup \Xi$ by $\Sigma^\Xi$.  Then, by definition, we have  $\T_\Sigma \subseteq (\Sigma^\Xi)^*$. Of course, $(\Sigma^\Xi)^* \setminus \T_\Sigma \ne \emptyset$.
  
A $\Sigma^\Xi$-cfg $G$ is \emph{tree-generating} if $\LL(G) \subseteq \T_\Sigma$. Next we give an example of a tree-generating $\Sigma^\Xi$-cfg.

\begin{example}\rm\label{ex:tree-generating-cfg} \cite[Ex.~3.9]{kos22}
Let $\Sigma = \{\sigma^{(3)}, \alpha^{(0)}\}$. We consider the $\Sigma^\Xi$-cfg $G=(\{S,A,B,C\}, S, R)$
	where $R$ contains the following rules: \[S \to ASB)\ \ \ \ S \to ACB)\ \ \ \ A \to \sigma( C,\ \ \ \ B \to ,C\ \ \ \ C \to \alpha\enspace.\]
Note that commas in the above rules must be considered as an element of the alphabet $\Sigma^\Xi$.

	Then we have
	\begin{align*}
		S &\Rightarrow ASB) \Rightarrow \sigma( C,SB) \Rightarrow \sigma( \alpha,SB)\\
		&\Rightarrow \sigma( \alpha,ACB) B) \Rightarrow \sigma( \alpha,\sigma( C,CB) B)\\
		&\Rightarrow \sigma( \alpha,\sigma( \alpha,CB) B)\\
		&\Rightarrow \sigma( \alpha,\sigma( \alpha,\alpha B)B)\\
		&\Rightarrow \sigma( \alpha,\sigma( \alpha,\alpha ,C) B)\\
		&\Rightarrow \sigma( \alpha,\sigma( \alpha,\alpha ,\alpha) B)\\
		&\Rightarrow \sigma( \alpha,\sigma( \alpha,\alpha ,\alpha) ,C)\\
		&\Rightarrow \sigma( \alpha,\sigma( \alpha,\alpha ,\alpha) ,\alpha)\enspace.
	\end{align*}
It is easy to see that $\LL(G) \subseteq \T_\Sigma$, and thus, $G$ is tree-generating.\hfill$\Box$
\end{example}
  
  Here we recall the concept of regular tree grammar \cite{bra69} (also cf. \cite{eng75-15,gecste84,comdaugiljaclugtistom08}). A \emph{regular tree grammar over $\Sigma$} (for short: $\Sigma$-rtg) is a $\Sigma^\Xi$-cfg $G = (N,S,R)$ where each rule in $R$ has the form  $A \rightarrow \xi$ with $\xi \in \T_\Sigma(N)$. Obviously, $G$ is $\varepsilon$-free because $\varepsilon \not\in \T_\Sigma(N)$.  Moreover,  each regular tree grammar over $\Sigma$ is a tree generating $\Sigma^\Xi$-cfg.

  A tree language $L \subseteq \T_\Sigma$ is \emph{regular} if there exists a $\Sigma$-rtg $G$ such that $\LL(G)=L$.
 
  Now we give an example of a $\Sigma$-rtg.
  
\begin{example} \label{ex:rtg} \rm \cite[Ex.~3.10]{kos22}
	Let $\Sigma$ be the ranked alphabet defined in Example~\ref{ex:tree-generating-cfg}. We consider the $\Sigma$-rtg $\overline{G}=(\{S\},S,R)$,
	where $R = \{S \to \sigma(\alpha,\alpha,\alpha) ,  S \to \sigma(\alpha,S,\alpha)\}$. Fig.~\ref{fig:derivation} shows, for each $n\in \mathbb{N}_+$, the tree $\xi_n$ and the derivation of $\overline{G}$ for $\xi_n$. In fact, $\LL(\overline{G})=\{\xi_n \mid n\in \mathbb{N}_+\}$.

It is easy to show that, for the tree generating $\Sigma^\Xi$-cfg $G$ defined in Example~\ref{ex:tree-generating-cfg}, we have $\LL(\overline{G}) = \LL(G)$. 
\hfill$\Box$
\end{example} 

The following theorem shows that, in general, for each tree generating $\Sigma^\Xi$-cfg we can construct an equivalent $\Sigma$-rtg.

\begin{figure}[t]
	\centering
	\begin{tikzpicture}
		\node at (0,0) {$S$};
		
		\node at (1,0) {$\Rightarrow_{\overline{G}}$};
		
		\node at (2,0) (s) {$\sigma$};
		\node at (1.5,-.75) (s1) {$\alpha$};
		\node at (2,-.75) (s2) {$S$};
		\node at (2.5, -.75) (s3) {$\alpha$};
		
		\draw	
			(s) -- (s1)
			(s) -- (s2)
			(s) -- (s3)
		;
		
		\node at (3, 0) {$\Rightarrow_{\overline{G}}$};
		
		\node at (4, 0) (t) {$\sigma$};
		\node at (3.5,-.75) (t1) {$\alpha$};
		\node at (4,-.75) (t2) {$\sigma$};
		\node at (4.5, -.75) (t3) {$\alpha$};
		\node at (3.5,-1.5) (t21) {$\alpha$};
		\node at (4,-1.5) (t22) {$S$};
		\node at (4.5,-1.5) (t23) {$\alpha$};
		
		\draw	
			(t) -- (t1)
			(t) -- (t2)
			(t) -- (t3)
			(t2) -- (t21)
			(t2) -- (t22)
			(t2) -- (t23)
		;
		
		\node at (5, 0) {$\Rightarrow^*_{\overline{G}}$};
		
		\node at (6, 0) {$\xi_n=$};
		
		\node at (7, 0) (u) {$\sigma$};
		\node at (6.5,-.75) (u1) {$\alpha$};
		\node at (7,-.75) (u2) {$\sigma$};
		\node at (7.5, -.75) (u3) {$\alpha$};
		\node at (6.5,-1.5) (u21) {$\alpha$};
		\node at (7,-1.4) (u22) {$$};
		\node at (7.5,-1.5) (u23) {$\alpha$};
		\node at (7,-1.5) (u22s) {$\vdots$};
		\node at (7,-2.375) (u222) {$\sigma$};
		\node at (6.5,-3.125) (u2221) {$\alpha$};
		\node at (7, -3.125) (u2222) {$\alpha$};
		\node at (7.5,-3.125) (u2223) {$\alpha$};
		
		\draw	
			(u) -- (u1)
			(u) -- (u2)
			(u) -- (u3)
			(u2) -- (u21)
			(u2) -- (u22)
			(u2) -- (u23)
			(u22s) -- (u222)
			(u222) -- (u2221)
			(u222) -- (u2222)
			(u222) -- (u2223)
		;
		
	\end{tikzpicture}
	\caption{\label{fig:derivation} \cite[Fig.~1]{kos22} A derivation of the $\Sigma$-rtg $\overline{G}$ defined in Example~\ref{ex:rtg} for $n\in \mathbb{N}_+$ and $\xi_n$, where $\xi_n$ is the tree in which the symbol $\sigma$ occurs $n$ times.}
\end{figure}

  \begin{theorem} {\rm \cite[Thm.~V.5 and V.6]{kos22}} \label{thm:Koszo22} Let $G$ be a $\Sigma^\Xi$-cfg. Then the following two statements hold.
    \begin{compactenum}
    \item[(1)] It is decidable whether $G$ is tree generating, i.e., whether $\LL(G) \subseteq \T_\Sigma$.
      \item[(2)] If $\LL(G) \subseteq \T_\Sigma$, then we can construct a $\Sigma$-rtg $G'$ such that $\LL(G')=\LL(G)$.
      \end{compactenum}

    \end{theorem}


\section{Finite-state tree automata}
\label{sec:fta}

We recall the notions of finite-state tree automaton and recognizable tree language from \cite{gecste84,eng75-15,gecste97,comdaugiljaclugtistom08}.

\index{finite-state tree automaton}
\index{fta}
A \emph{finite-state tree automaton over $\Sigma$} (for short: $\Sigma$-fta, or just fta) is a triple $A = (Q,\delta,F)$ where 
\begin{compactitem}
\item $Q$ is a finite nonempty set (states) such that $Q \cap \Sigma = \emptyset$,
\item $\delta=(\delta_k\mid k\in \mathbb{N})$ is a family of relations $\delta_k \subseteq  Q^k\times \Sigma^{(k)} \times Q$\footnote{For each $k \in \mathbb{N}$ with $k > \maxrk(\Sigma)$ we have $\delta_k=\emptyset$, because $\Sigma^{(k)} = \emptyset$. } (transition relations) where we consider $Q^k$ as a set of strings over $Q$ of length $k$, and \index{transition relation}
\item $F \subseteq Q$ (set of root states).
\end{compactitem}
\index{bottom-up-deterministic}
\index{bu-deterministic}
We say that $A$ is {\em bottom-up-deterministic} (for short: bu-deterministic) if for every $k \in \mathbb{N}$, $w\in Q^k$, and $\sigma \in \Sigma^{(k)}$, there exists at most one $q\in Q$ such that $(w,\sigma,q)\in \delta_k$. And we say that $A$ is {\em total} if for every $k \in \mathbb{N}$, $w\in Q^k$, and $\sigma \in \Sigma^{(k)}$, there exists at least one $q\in Q$ such that $(w,\sigma,q)\in \delta_k$.

\index{Sigma@$\Sigma$-algebra associated with fta $A$}
We can associate two semantics with $A$: the initial algebra semantics and the run semantics. Both lead to the same tree language. Since each of the semantics has its benefits we present them both.

\paragraph{Initial algebra semantics.} We define the \emph{$\Sigma$-algebra associated with $A$} to be the $\Sigma$-algebra $({\cal P}(Q),\delta_{A})$  where, for every  $k\in \mathbb{N}$ and $\sigma \in \Sigma^{(k)}$, the mapping $\delta_{A}(\sigma): {\cal P}(Q)^k \to {\cal P}(Q)$ is defined by 
\[
\delta_{A}(\sigma)(P_1,\ldots,P_k)= \{ q \in Q \mid (\exists q_1 \in P_1)\ldots(\exists q_k \in P_k)  : (q_1\cdots q_k,\sigma, q) \in \delta_k\}
\]
for every $P_1,\ldots, P_k\in {\cal P}(Q)$. We denote the unique $\Sigma$-algebra homomorphism from the term algebra $\sfT_\Sigma$ to $(\cP(Q),\delta_A)$ by $\h_A$.
\label{page:algebra-associated-with-fta}
\index{homA@$\h_A$}
\index{languageiA@$\Li(A)$}
The \emph{tree language $i$-recognized by
  ${A}$}, denoted by $\Li(A)$, is defined by
\[\Li(A) =\{ \xi\in \T_\Sigma \mid \h_{A}(\xi)\cap F\neq \emptyset \}\enspace.\]

\index{run}
\paragraph{Run semantics.} Let~$\xi\in\T_\Sigma$. A \emph{run of~$A$ on~$\xi$} is a
mapping $\rho: \pos(\xi) \rightarrow Q$. Let $q \in Q$. Then $\rho$ is called
\begin{compactitem}
\item \emph{$q$-run} if $\rho(\varepsilon)=q$,
\item \emph{valid} if for every~$w\in\pos(\xi)$ it holds that
\(\big(\rho(w1)\cdots\rho(wk),\xi(w),\rho(w)\big)\in\delta_k\)
where~$\xi(w)\in\Sigma^{(k)}$, and
\item \emph{accepting} if $\rho$ is valid and $\rho(\varepsilon) \in F$.
\end{compactitem}

The set of all $q$-runs (all valid $q$-runs, all accepting $q$-runs)
of $A$ on $\xi$ is denoted by $\R_{A}(q,\xi)$ (respectively, $\Rv_{A}(q,\xi)$ and $\Ra_{A}(q,\xi)$). We let
\[
  \R_{A}(\xi) = \bigcup_{q \in Q} \R_{A}(q,\xi) \ \text{ and }\  \Rv_{A}(\xi) = \bigcup_{q \in Q} \Rv_{A}(q,\xi)
  \text{ and }\  \Ra_{A}(\xi) = \bigcup_{q \in F} \Ra_{A}(q,\xi)\enspace.
\]
\index{languagerA@$\Lr(A)$}
The \emph{tree language r-recognized by~$A$}, denoted by $\Lr(A)$, is defined by
\[\Lr(A) = \{\xi\in\T_\Sigma\mid \Ra_{A}(\xi)\not= \emptyset\}\enspace.\]

\begin{lemma}\rm \label{lm:run=init-fta}
Let $A = (Q,\delta,F)$ be a $\Sigma$-fta. Then $\Li(A) = \Lr(A)$.
\end{lemma}
\begin{proof}
Let~$\xi\in\T_\Sigma$. We have
\begin{align*}
  \xi\in \Li(A) 
\  \text{iff} \ \h_A(\xi)\cap F\neq\emptyset
\ \text{iff} \ (\exists q\in F)\colon q\in\h_A(\xi)
\ \text{iff}^{(*)} \ (\exists q\in F)\colon \Rv_{A}(q,\xi)\not= \emptyset 
\ \text{iff} \  \xi\in \Lr(A).
\end{align*}

It remains to prove (*). By induction on $\T_\Sigma$, we prove that the following more general statement holds:

\begin{equation}\label{equ:init=run-for-Boolean-sr}
\text{For every $\xi\in\T_\Sigma$ and $q \in Q$, we have: $q\in\h_A(\xi)$ iff  $\big((\exists \rho \in \R_{A}(q,\xi)): \rho \text{ is valid}\big)$.}
\end{equation}
Let $\xi=\sigma(\xi_1,\ldots,\xi_k)$. Then 
\begingroup
\allowdisplaybreaks
\begin{align*}
  &q\in \h_A(\sigma(\xi_1,\ldots,\xi_k))\\
\text{iff}\quad& (\exists q_1 \cdots q_k \in Q^k): (q_1\in \h_A(\xi_1)) \wedge \ldots \wedge (q_k\in \h_A(\xi_k)) \wedge
((q_1\cdots q_k,\sigma,q)\in\delta_k)\\
  \text{iff}\quad& (\exists q_1\cdots q_k \in Q^k):\\
  &\big((\exists \rho_1 \in \R_{A}(q_1,\xi)): \rho_1 \text{ is valid}\big) \wedge \ldots \wedge \big((\exists \rho_k \in \R_{A}(q_k,\xi)): \rho_k \text{ is valid}\big) \wedge
((q_1\cdots q_k,\sigma,q)\in\delta_k) \tag{I.H.}\\
\text{iff}\quad& (\exists q_1\cdots q_k \in Q^k)(\exists \rho_1 \in \R_{A}(q_1,\xi)) \ldots (\exists \rho_k \in \R_{A}(q_k,\xi)):\\
  &\hspace*{70mm}(\rho_1 \text{ is valid})  \wedge \ldots \wedge  (\rho_k \text{ is valid}) \wedge
    ((q_1\cdots q_k,\sigma,q)\in\delta_k) 
    \tag{by distributivity of $\wedge$ over $\exists$}\\
  \text{iff}\quad& (\exists \rho \in \R_{A}(q,\xi)): \rho \text{ is valid} \enspace. \tag*\qedhere
\end{align*}
\endgroup
\end{proof}

\index{recognizable}
\index{equivalent}
Since $\Li(A)=\Lr(A)$ for each $\Sigma$-fta $A$, we will sometimes  drop the indices $\mathrm{i}$ and $\mathrm{r}$ from $\Li(A)$ and $\Lr(A)$, respectively, and we call the language $\LL(A)$ the \emph{tree language recognized by $A$}.  Two $\Sigma$-fta $A$ and $A'$ are \emph{equivalent} if $\LL(A) = \LL(A')$.
\index{equivalent} A tree language $L \subseteq \T_\Sigma$ is \emph{recognizable} if there exists a $\Sigma$-fta $A$ such that $\LL(A)=L$.
The set of all {\it recognizable $\Sigma$-tree languages} is denoted by $\Rec(\Sigma)$.

\begin{theorem}\label{thm:fta-total-bud-fta} {\rm (cf. \cite[Thm.~2.2.6]{gecste84}\footnote{When citing results of \cite{gecste84} we use the numbers in the arXiv-version of that book.} and \cite[Thm.~3.8]{eng75-15}\footnote{When citing results of \cite{eng75-15}  we use the numbers in the arXiv-version of those lecture notes.})}
 For each $\Sigma$-fta $A$ we can construct a total and bu-deterministic $\Sigma$-fta $B$ such that $\LL(A)=\LL(B)$.
\end{theorem}

\begin{theorem} {\rm (cf. \cite[Thm.~3.25 and 3.32]{eng75-15} and \cite[Thm.~2.4.2]{gecste84}} \label{thm:fta-closure-results} Let $A_1, A_2$ be two $\Sigma$-fta.
  \begin{compactenum}
    \item[(1)] We can construct a $\Sigma$-fta $A$ such that $\LL(A)= \LL(A_1) \cup \LL(A_2)$.
    \item[(2)] We can construct a $\Sigma$-fta $A$ such that $\LL(A)= \LL(A_1) \cap \LL(A_2)$.
    \item[(3)] We can construct a $\Sigma$-fta $A$ such that $\LL(A)= \LL(A_1) \setminus \LL(A_2)$.
    \end{compactenum}
   Thus, the set $\Rec(\Sigma)$ is closed under union, intersection, set subtraction,  and complement.
   \end{theorem}

   \begin{theorem} {\rm (cf. \cite[Thm.~3.25]{eng75-15} and \cite[Thm.~2.3.6]{gecste84}} \label{thm:regular=recognizable-unweighted} Let $L \subseteq \T_\Sigma$. Then $L$ is regular if and only if $L$ is recognizable.
   \end{theorem}
   
For the theory of recognizable tree languages we refer to \cite{eng75-15}, \cite{gecste84,gecste97},\cite{comdaugiljaclugtistom08}.

\section{Recognizable step mappings}
\label{sec:rec-step-mappings}

We recall the concept of recognizable step mapping from \cite{drovog06} (also cf. \cite{drogas05,drogas07,drogas09}). Roughly speaking, these are $(\Sigma,\B)$-weighted tree languages which can be obtained from characteristic mappings of finitely many recognizable $\Sigma$-tree languages by multiplying with elements of $B$ and summing up. 

\index{recognizable step mapping}
Formally, a weighted tree language $r: \T_\Sigma \rightarrow B$ is a \emph{$(\Sigma,\B)$-recognizable step mapping} (or just: \emph{recognizable step mapping}) if there exist $n \in \mathbb{N}_+$, $b_1,\ldots,b_n \in B$, and recognizable $\Sigma$-tree languages $L_1,\ldots,L_n$ such that we have
\begin{equation} \label{equ:def-rec-step-mapping}
r = \bigoplus_{i \in [n]} b_i \cdot \chi(L_i)\enspace,  
\end{equation}
(where we have extended the sum of two weighted tree languages to the sum of finitely many weighted tree languages, cf. page \pageref{page:finite-summation}). In other words, for each $\xi \in \T_\Sigma$, we have
\begin{equation*}
  r(\xi) = \bigoplus_{\substack{i \in [n]:\\\xi \in L_i}} b_i \enspace.
\end{equation*}
\index{step languages}
In particular, if $\xi \not\in L_1 \cup \ldots \cup L_n$, then $r(\xi)=\0$. Moreover, if $n=1$ and $L_1=\emptyset$, then $r = \widetilde{\0}$. The tree languages $L_i$ are called \emph{step languages}.
\index{RecStep@$\RecStep(\Sigma,\B)$}
If $n=1$, then $r$ is called \emph{$(\Sigma,\B)$-recognizable one-step mapping}. We denote by $\RecStep(\Sigma,\B)$  the set of $(\Sigma,\B)$-recognizable step mappings.

Each polynomial $(\Sigma,\B)$-weighted tree language $r=b_1.\xi_1\oplus\ldots\oplus b_n.\xi_n$ is a recognizable step mapping. This is because  also $r=b_1\cdot \chi(\{\xi_1\})\oplus\ldots\oplus b_n\cdot\chi(\{\xi_n\})$ and because singleton $\Sigma$-tree languages are recognizable.

In general, the step languages of a recognizable step mapping need not be disjoint. However, for each recognizable step mapping we can find a characterization in terms of pairwise disjoint step languages.

\begin{observation}\rm \label{obs:partition-step} \cite[Lm.~3.1]{drovog06} Let $r: \T_\Sigma \rightarrow B$ be a  recognizable step mapping. There exist a finite set $F$ and for each $f \in F$, there exist a recognizable $\Sigma$-tree language $U_f$ and an element $a_f \in B$ such that $(U_f \mid f \in F)$ is a partitioning of $\T_\Sigma$ and $r = \bigoplus_{f \in F} a_f \cdot \chi(U_f)$.  
\end{observation}
\begin{proof} Let $n \in \mathbb{N}_+$, $b_1,\ldots,b_n \in B$, and recognizable $\Sigma$-tree languages $L_1,\ldots,L_n$ be such that, for each $\xi \in \T_\Sigma$, Equation \eqref{equ:def-rec-step-mapping} holds.

  Let $F$ be the set   of all mappings of type $\{1,\ldots,n\} \rightarrow \{1,c\}$ (where $1$ and $c$ are just viewed as two distinct symbols). For each mapping $f\in F$, we define the $\Sigma$-tree language $U_f = \bigcap_{i=1}^n L_i^{f(i)}$ where $L_i^1 = L_i$ and $L_i^c = \T_\Sigma \setminus L_i$. We note that the family $(U_f \mid f \in F)$ forms a partitioning of $\T_\Sigma$. Moreover, since the set of recognizable $\Sigma$-tree languages is closed under intersection and complementation \cite{gecste84}, we have that $U_f$ is a recognizable $\Sigma$-tree language. 

  For each  $f \in F$, we define the element $a_f = \bigoplus_{i \in   f^{-1}(1)}b_i$ in $B$. Then clearly,
  for each $\xi \in \T_\Sigma$ we have  $r(\xi) = a_f$ if $\xi \in U_f$, where the $f$ is uniquely determined because $(U_f \mid f \in F)$ forms a partitioning of $\T_\Sigma$. Thus  $r = \bigoplus_{f \in F} a_f \cdot \chi(U_f)$.  
\end{proof}

We mention that recognizable step mappings play an important role 
in the characterization of recognizable weighted languages by weighted MSO-logic (cf. \cite{drogas05,drogas07,drogas09} for strings, and \cite{drovog06,drovog11,fulstuvog12,droheuvog15} for trees; also cf. Section \ref{sec:weighted-conj-univ-quant}). In fact, the semantics of the weighted MSO-formula $\forall x. \varphi$ is a recognizable weighted language if the semantics of $\varphi$ is a recognizable step mapping (\cite[Lm.~4.2]{drogas05} and \cite[Lm.~5.4]{drogas09}, also cf.  \cite[Lm.~5.5]{drovog06} and Lemma \ref{lm:univ-fo-quantification-rec-step} for the tree case); moreover, there exists a weighted MSO-formula $\varphi$ of which the semantics is a recognizable weighted language and the semantics $\forall x. \varphi$ is not recognizable \cite[Ex.~3.6]{drogas09}.

We note that in \cite[Def.~3.4]{ghozah12}, a $(\Sigma,\sfL)$-recognizable step mapping for some $\sigma$-complete residuated lattice $\sfL$ is called an $\sfL$-valued regular tree language.

\section{Fta-hypergraphs}
\index{hypergraph}
\index{fta-hypergraph}
The concept of hypergraph \cite{baucou87,habkre87,coueng12} generalizes that of graphs. In this book we consider particular hypergraphs which we call fta-hypergraphs.
Let $Q$ be a finite set. A \emph{$(Q,\Sigma)$-hypergraph}  is a pair $g = (Q,E)$, where   $E \subseteq \bigcup_{k \in \mathbb{N}} Q^k \times \Sigma^{(k)} \times Q$ (set of hyperedges); the elements of $Q$ are called \emph{nodes}.
An \emph{fta-hypergraph} is a $(Q,\Sigma)$-hypergraph for some finite set $Q$ and some ranked alphabet $\Sigma$.

We can represent  a $(Q,\Sigma)$-hypergraph as a picture as  follows. Each node $q \in Q$ is represented as a circle with $q$ in its center. Each hyperedge $(q_1\cdots q_k,\sigma,q)$ is represented as a box with $\sigma$ in its center and with incoming and outgoing arrows. More specifically, this box has exactly one outgoing arrow, which leads to the representation of the node  $q$. Moreover, it has $k$ incoming arrows, which come from the representations of the nodes $q_1.\ldots,q_k$, respectively. The order among $q_1,\ldots,q_k$ which is determined by the string $q_1 \cdots q_k$, is represented in the picture as follows: starting from the unique outgoing arrow and moving counter-clockwise around the box, the $i$-th incoming arrow comes from the representation of the $i$-th component of the string $q_1 \cdots q_k$.

Figure \ref{fig:hypergraph} illustrates the $(Q,\Sigma)$-hypergraph $(Q,E)$ with $Q = \{\h,0\}$, $\Sigma = \{\sigma^{(2)}, \alpha^{(0)}\}$, and
\[
  E = \{(\varepsilon,\alpha,\h), (\h0,\sigma,\h), (0\h,\sigma,\h), (\varepsilon,\alpha,0), (00,\sigma,0)\}\enspace.
\]
For instance, the hyperedge $(0\h,\sigma,\h)$ is represented by the left-lower box.
In the sequel, we will not distinguish between  an fta-hypergraph and its representation as picture.

\begin{figure}[h]
  \begin{center}
\begin{tikzpicture}
\tikzset{node distance=7em, scale=0.6, transform shape}
\node[state, rectangle] (1) {\Large $\alpha$};
\node[state, right of=1] (2) {\Large h};
\node[state, rectangle, above of=2] (3){\Large $\sigma$};
\node[state, rectangle, below of=2] (4) {\Large $\sigma$};
\node[state, right of=2] (5) [right=5em] {\Large 0};
\node[state, rectangle, below of=5] (6) {\Large $\alpha$};
\node[state, rectangle, right of=5] (7) {\Large $\sigma$};

\tikzset{node distance=2em}
\node[above of=1] (w1) {};
\node[above of=2] (w2) [right=0.05cm] {};
\node[above of=3] (w3) {};
\node[above of=4] (w4) [right=0.05cm]{};
\node[above of=6] (w6) [right=0.05cm] {};
\node[above of=7] (w7) {};

\draw[->,>=stealth] (1) edge (2);
\draw[->,>=stealth] (2) edge (3);
\draw[->,>=stealth] (3) edge[out=-250, in=-210, looseness=1.8] (2);
\draw[->,>=stealth] (2) edge (4);
\draw[->,>=stealth] (4) edge[out=250, in=-150, looseness=1.8] (2);
\draw[->,>=stealth] (5) edge (4);
\draw[->,>=stealth] (5) edge (3);
\draw[->,>=stealth] (6) edge (5);
\draw[->,>=stealth] (5) edge[out=60, in=30, looseness=1.4] (7);
\draw[->,>=stealth] (5) edge[out=-60, in=-30, looseness=1.4] (7);
\draw[->,>=stealth] (7) edge (5);
\end{tikzpicture}
\end{center}
\caption{\label{fig:hypergraph} A $(Q,\Sigma)$-hypergraph.}
\end{figure}

%% file: basic-wta.tex
\addtocontents{toc}{\protect\pagebreak}
\chapter{The~model~of~weighted~tree~automata}
\label{chapter:basic-wta}

In this chapter we present the basic model of weighted tree automata. Intuitively, it results from the concept of fta by applying two steps.

In the first step, for a given $\Sigma$-fta $\cA =(Q,\delta,F)$, we replace each transition relation
\begin{align*}
  \delta_k \subseteq Q^k \times \Sigma^{(k)} \times Q \ \  \ \ \text{ by its characteristic mapping} \ \ \ \  
  \chi_{\Boole}(\delta_k): Q^k \times \Sigma^{(k)} \times Q  \to \mathbb{B} 
\end{align*}
 with respect to the Boolean semiring $\Boole=(\mathbb{B},\vee,\wedge,0,1)$  (where $\mathbb{B}=\{0,1\}$), and similarly, we replace
\begin{align*}
  F \subseteq Q \ \ \ \  \text{ by its characteristic mapping } \ \  \ \ \chi_{\Boole}(F): Q \to \mathbb{B} \enspace.
\end{align*}
For the sake of simplicity, we denote $\chi_{\Boole}(\delta_k)$ and $\chi_{\Boole}(F)$ also by $\delta_k$ and $F$, respectively.

Since there is a bijection between the set of subsets of  $Q^k \times \Sigma^{(k)} \times Q $  and set of mappings of type $Q^k \times \Sigma^{(k)} \times Q \to \mathbb{B}$ (and the same holds for subsets of $Q$ and mappings of type $Q\to \mathbb{B}$), this step does not change the concept of an fta; it is simply another way to specify an fta.

In the second step, we replace the Boolean semiring $\Boole$ by an arbitrary strong bimonoid $\B=(B,\oplus,\otimes,\0,\1)$, i.e., we let
\begin{align*}
  \delta_k : Q^k \times \Sigma^{(k)} \times Q  \to B \ \ \text{ and  } \ \ F: Q \to B \enspace,
\end{align*}
and allow that $\delta_k$ and $F$ can map their arguments to any value of $B$ (i.e., not only to $\0$ and $\1$).

In contrast to the first step, the second one generalizes the concept of fta considerably, because
\begin{compactenum}
\item[(a)] a strong bimonoid $\B$ need not be finite  and
\item[(b)] a number of algebraic laws which hold for the summation $\vee$ and multiplication $\wedge$ of $\Boole$ need not hold anymore in $\B$, e.g., idempotency, commutativity of $\wedge$, zero-sum freeness, zero-divisor freeness, absorption axiom, and distributivity.
\end{compactenum}
Hence, in the proofs of lemmas and theorems on weighted tree automata over some arbitrary strong bimonoid $\B$ we have to be careful not to use the latter properties out of habit.

In Section \ref{sec:basic-defininition-wta} we recall the concept of weighted tree automata and define the run semantics, the initial algebra semantics, and the corresponding notions of recognizable weighted tree language. In Section~\ref{sec:examples} we illustrate these definitions by a number of examples. In Section~\ref{sec:state-algebra-of-wta} we define the concept of state algebra.
Finally, in Section \ref{sect:extension-of-weight-structure} we briefly mention a consequence of extending the weight algebra and we define the concept of Fatou extension. 

For the investigation of wta over finite strong bimonoids, we refer the reader to Chapter \ref{ch:crisp-determinization}. In fact, there we will deal with wta over even more general ``finite-like'' weight algebras:  the initial algebra semantics of wta over locally finite strong bimonoids (cf. Theorem~\ref{thm:loc-finite-rec-step-function}) and the run semantics of wta over bi-locally finite strong bimonoids (cf. Theorem~\ref{thm:bi-loc-finite-rec-step-function}).

\section{Basic definitions}
\label{sec:basic-defininition-wta}

\index{weighted tree automaton}
\index{wta}
\index{root weight vector}
\index{transition mapping}
\index{SigmaBwta@$(\Sigma,\B)$-wta}
A \emph{weighted tree automaton over $\Sigma$ and $\B$} (for short: $(\Sigma,\B)$-wta, or: wta) is a tuple $\cA = (Q,\delta,F)$ where 
\begin{compactitem}
\item $Q$ is a finite nonempty set (\emph{states}) such that $Q \cap \Sigma = \emptyset$,
\item $\delta=(\delta_k\mid k\in\mathbb{N})$ is a family of mappings $\delta_k: Q^k\times \Sigma^{(k)}\times Q \to B$ (\emph{transition mappings})
  where we consider $Q^k$ as set of strings over $Q$ of length $k$, and 
\item $F: Q \rightarrow B$ is a mapping (\emph{root weight vector}).
\end{compactitem}

\index{transition}
For each $k \in \mathbb{N}$, we call each element in $Q^k \times \Sigma^{(k)} \times Q$ a \emph{transition}. For each transition $(w,\sigma,q)$, the element $\delta_k(w,\sigma,q)$ of $B$  is its \emph{weight}, and for each $q \in Q$, the element $F_q$ is its \emph{root weight}. (We recall that $F_q$ denotes $F(q)$.)
\index{wts@$\mathrm{wts}(\cA)$}
We denote the set of all transition weights and root weights occurring in $\cA$ by $\mathrm{wts}(\cA)$, i.e., 
\[\mathrm{wts}(\cA) =  \im(\delta) \cup \im(F) \ \text{ where } \ \im(\delta) = \bigcup_{k \in \mathbb{N}} \im(\delta_k) \enspace.\]

\index{succA@$\mathrm{succ}_\cA(w,\sigma)$}
Let $k \in \mathbb{N}$, $\sigma \in \Sigma^{(k)}$, and $w \in Q^k$. We define the set of successor states of $w$ and $\sigma$ by
\[
\mathrm{succ}_\cA(w,\sigma) = \{p \in Q \mid \delta_k(w,\sigma,p) \ne \0\} \enspace.
\]
\index{esuccA@$\mathrm{esucc}_\cA(w,\sigma)$}
If $\mathrm{succ}_\cA(w,\sigma)$ is a singleton, then we denote its only element by $\mathrm{esucc}_\cA(w,\sigma)$.

Let $\cA = (Q,\delta,F)$ be a $(\Sigma,\B)$-wta. 
\begin{compactitem}
\index{weighted tree automaton!bottom-up-deterministic}
\index{bottom-up-deterministic}
\index{bu-deterministic}
\item  $\cA$ is \emph{bottom-up-deterministic} (for short: bu-deterministic) if for every $k \in \mathbb{N}$, $\sigma \in \Sigma^{(k)}$, and $w \in Q^k$ there exists at most one $q \in Q$ such that $\delta_k(w,\sigma,q) \not = \mathbb{0}$, i.e., $|\mathrm{succ}_\cA(w,\sigma)| \le 1$.
  \index{weighted tree automaton!total}
\item  $\cA$ is \emph{total} if for every $k \in \mathbb{N}$, $\sigma \in \Sigma^{(k)}$, and $w \in Q^k$ there exists at least one state $q$ such that $\delta_k(w,\sigma,q) \not= \mathbb{0}$, i.e., $|\mathrm{succ}_\cA(w,\sigma)| \ge 1$.

\index{weighted tree automaton!crisp}
\index{crisp}
\item $\cA$ is \emph{crisp} if $\im(\delta) \subseteq \{\mathbb{0},\mathbb{1}\}$.
\index{weighted tree automaton!crisp-deterministic}
\index{crisp-deterministic}
\item $\cA$ is \emph{crisp-deterministic} if $\cA$ is bu-deterministic, total, and crisp (cf. \cite[Sec.~5]{cirdroignvog10}). 
\index{unit root weight}
\item $\cA$ has \emph{unit root weights} if $\im(F) \subseteq \{\mathbb{0},\mathbb{1}\}$.

\index{weighted tree automaton!root weight normalized}
\index{root weight normalized}
\item $\cA$ is \emph{root weight normalized} if there exists a state $q \in Q$ such that $\supp(F) = \{q\}$ and $F_q=\mathbb{1}$.
\end{compactitem}

We note that, for each $(\Sigma,\B)$-wta $\cA=(Q,\delta,F)$, we consider the strong bimonoid $\B$ as integral part of the specification of $\cA$. Thus, if $\B'$ is another strong bimonoid which has the same carrier set as $\B$ (but different operations and/or unit elements), then we consider $\cA$ to be different from the $(\Sigma,\B')$-wta $\cA'=(Q,\delta,F)$. In other words, instead of saying that $\cA=(Q,\delta,F)$ is a $(\Sigma,\B)$-wta, we could equivalently say that $\cA=(Q,\Sigma,\B,\delta,F)$ is a wta. And clearly, $(Q,\Sigma,\B,\delta,F)\not= (Q,\Sigma,\B',\delta,F)$.
\label{page:wta-B1-not=-wta-B2}

\paragraph{Representation of wta by fta-hypergraphs.}
We can represent each $(\Sigma,\B)$-wta $\cA=(Q,\delta,F)$ as an fta-hypergraph with extra annotations. For this we first consider the $(Q,\Sigma)$-hypergraph
\[
  g_\cA = (Q,\bigcup_{k \in \mathbb{N}}\supp(\delta_k)) \enspace.
\]
Then we add to $g_\cA$  the root weights of $\cA$ and the weights of transitions as follows. For each $q \in Q$ such that $F_q \not= \mathbb{0}$, we add $F_q$ to the node which represents $q$. If $F_q = \mathbb{0}$, then we do not illustrate $F_q$ in the picture. Moreover, for each transition of $\cA$ with non-$\0$-weight, i.e., element in $\bigcup_{k \in \mathbb{N}}\supp(\delta_k)$, we add its weight to its representing hyperedge.

For instance, let us consider  the $(\Sigma,\Natmaxplus)$-wta $\cA = (Q,\delta,F)$, where
\begin{compactitem}
\item $\Sigma = \{\sigma^{(2)}, \alpha^{(0)}\}$,

\item $\Natmaxplus = (\mathbb{N}_{-\infty},\max,+,-\infty,0)$, the arctic semiring,
 
\item $Q  = \{\h,0\}$, 

\item $\delta_0(\varepsilon,\alpha,\h) = \delta_0(\varepsilon,\alpha,0) = 0$ and for every $q_1,q_2,q\in Q$, 
\[
\delta_2(q_1q_2,\sigma,q) = 
\begin{cases}
1&\text{ if }q_1q_2q\in\{\h0\h, 0\h\h\}\;,\\
0&\text{ if }q_1q_2q=000\;,\\
-\infty&\text{ otherwise\;,}
\end{cases}
\]

\item $F_\h = 0$ and $F_0 = -\infty$.
\end{compactitem}
In Figure \ref{fig:hypgraph-height-syntax} we show how $\cA$ is represented as fta-hypergraph with extra annotations.  
We note that, e.g., the transition $(\h\h,\sigma,0)$ is not represented in the fta-hypergraph, because $\delta_2(\h\h,\sigma,0) = -\infty$ and $-\infty$ is the identity for the summation of
 the arctic semiring $\Natmaxplus$. Also, the value $F_0$ is not represented because $F_0 = -\infty$.

\begin{figure}[t]
  \begin{center}
\begin{tikzpicture}
\tikzset{node distance=7em, scale=0.6, transform shape}
\node[state, rectangle] (1) {\Large $\alpha$};
\node[state, right of=1] (2) {\Large h};
\node[state, rectangle, above of=2] (3){\Large $\sigma$};
\node[state, rectangle, below of=2] (4) {\Large $\sigma$};
\node[state, right of=2] (5) [right=5em] {\Large 0};
\node[state, rectangle, below of=5] (6) {\Large $\alpha$};
\node[state, rectangle, right of=5] (7) {\Large $\sigma$};

\tikzset{node distance=2em}
\node[above of=1] (w1) {0};
\node[above of=2] (w2) [right=0.05cm] {0};
\node[above of=3] (w3) {1};
\node[above of=4] (w4) [right=0.05cm]{1};
\node[above of=6] (w6) [right=0.05cm] {0};
\node[above of=7] (w7) {0};

\draw[->,>=stealth] (1) edge (2);
\draw[->,>=stealth] (2) edge (3);
\draw[->,>=stealth] (3) edge[out=-250, in=-210, looseness=1.8] (2);
\draw[->,>=stealth] (2) edge (4);
\draw[->,>=stealth] (4) edge[out=250, in=-150, looseness=1.8] (2);
\draw[->,>=stealth] (5) edge (4);
\draw[->,>=stealth] (5) edge (3);
\draw[->,>=stealth] (6) edge (5);
\draw[->,>=stealth] (5) edge[out=60, in=30, looseness=1.4] (7);
\draw[->,>=stealth] (5) edge[out=-60, in=-30, looseness=1.4] (7);
\draw[->,>=stealth] (7) edge (5);
\end{tikzpicture}
\end{center}

\vspace{-10mm}

\caption{\label{fig:hypgraph-height-syntax} The fta-hypergraph for the $(\Sigma,\Natmaxplus)$-wta $\cA$.}
\end{figure}

\paragraph{Run semantics.}
\index{run}\label{page:run-sem}
The run semantics of a $(\Sigma,\B)$-wta can be viewed as a generalization of the algebraic path problem (cf. \cite{car71}, \cite[Ch.~8]{zim81}, \cite{mah81}, \cite{rot90}, \cite[Sec.~26.4]{corleiriv90}, and \cite[Ch.~IV.6]{hebwei93}). Roughly speaking, given a finite, directed graph such that each edge is labeled by an element of some $\sigma$-complete strong bimonoid $\B$, the algebraic path problem asks the following: for each pair $q,q'$ of nodes,  how can the value
\[
l_{q,q'} = \infsum{\oplus}{\rho \text{ path from $q$ to $q'$}}{\lambda(\rho)}
\]
in $B$ be computed where, for each path $\rho= (q_0,q_1)(q_1,q_2) \cdots (q_{n-1},q_n)$ from $q_0$ to $q_n$, we let
\[
  \lambda(\rho) = \bigotimes_{i \in [n]} w(q_{i-1},q_{i})
\]
and $w(q_{i-1},q_{i}) \in B$ is the label of the edge $(q_{i-1},q_{i})$.  For instance, if $\B$ is the formal language semiring $\Lang_\Gamma=(\cP(\Gamma^*),\cup,\cdot,\emptyset,\{\varepsilon\})$ and, for every two vertices $q,q'$, we have $w((q,q')) \subseteq \Gamma$, then the graph can be considered as a $\Gamma$-fsa $A$. Then $l_{q,q'} \in \cP(\Gamma^*)$ is the formal language recognized by $A$ starting from $q$ and ending in $q'$. Another example is that $\B$ is the tropical semiring $\Natminplus=(\mathbb{N}_\infty,\min,+,\infty,0)$ and, for every two vertices $q,q'$, the value $w((q,q'))$ is the distance between $q$ and $q'$ (which may be $\infty$). Then the value $l_{q,q'} \in \mathbb{N}_\infty$ is the length of a shortest  path from $q$ to $q'$.

Since a wta $\cA$ corresponds to an fta-hypergraph, the algebraic path problem can be generalized to an ``algebraic fta-hyperpath problem'' in order to capture the recursive structure of $\cA$
(cf. \cite{knu77} where the fta-hypergraph is called superior context-free grammar, and each hyperedge is labeled by a superior function). Given an fta-hypergraph $\cA$ and a node  $q$, a $q$-hyperpath is an unfolding of the fta-hypergraph, starting in node~$q$ and moving in the direction opposite to the direction of the hyperedges; this unfolding results in a $\Sigma$-tree $\xi$ (constituted by protocoling the labels of the visited hyperedges) together with a decoration of each position of $\xi$ by some node of the fta-hypergraph $\cA$; we call this decoration a ``$q$-run of $\cA$ on $\xi$''. We note that the first node of a path in a graph (see above) disappears when generalizing to hyperpaths and fta-hypergraphs, because at each leaf of an fta-hyperpath a transition in $\delta_0$ of $\cA$ is applied, and such transitions do not have source states. 

\index{run}
\index{RAxi@$\R_\cA(\xi)$}
\index{RAqxi@$\R_\cA(q,\xi)$}
Formally, let $\xi \in \T_\Sigma$. A \emph{run of $\cA$ on $\xi$} is a mapping $\rho: \pos(\xi) \rightarrow Q$. If $\rho(\varepsilon)=q$ for some
$q\in Q$, then $\rho$ is also called a \emph{$q$-run}. The \emph{set of all runs of $\cA$ on $\xi$} and the \emph{set of all $q$-runs of $\cA$ on $\xi$} are denoted by $\R_\cA(\xi)$ and $\R_{\cA}(q,\xi)$, respectively. Each run $\rho \in \R_\cA(\xi)$ determines,  for each $w \in \pos(\xi)$, a unique transition, viz., if $k = \rk(\xi(w))$ and $\sigma=\xi(w)$, then $\rho$ determines  the transition $(\rho(w1)\cdots \rho(wk),\sigma,\rho(w))$.  We call this the \emph{transition induced by $\rho$ on $\xi$ at $w$}.
For every $\rho\in \R_\cA(\xi)$ and $w \in\pos(\xi)$, the \emph{run induced by~$\rho$ at position~$w$}, denoted by $\rho|_w$, is the run in $\R_\cA(\xi|_w)$  defined for every $w'\in\pos(\xi|_w)$ by $\rho|_w(w') = \rho(ww')$.

\begin{figure}
  \centering
\begin{tikzpicture}[scale=0.8, every node/.style={transform shape},
					node distance=0.05cm and 0.05cm,
					level distance= 1.25cm,
					level 1/.style={sibling distance=25mm},
			        level 2/.style={sibling distance=17mm},
					mycircle/.style={draw, circle, inner sep=0mm, minimum height=5.5mm},]
					
\node (n0) {$\gamma$}
    child {node (n1) {$\sigma$}
      child {node (n2) {$\alpha$}}
      child {node (n3) {$\beta$}} };
  \node[mycircle, right= of n0] (cn0) {$q$};
  \node[mycircle, right= of n1] (cm1) {$p$};
  \node[mycircle, right= of n2] (cn2) {$q$};
  \node[mycircle, right= of n3] (cn3) {$q$};
  \node[above left= 0cm and 0.2cm of n0] (n) {$\xi|_1 \in \T_\Sigma$};
  \node[right= 1.6cm of n] {$\rho|_1 \in \R_{\cA}(\xi|_1)$};
  \coordinate (topn)  at ($(cn0)!0.5!(n0)+(0.1,0.6)$);
  \coordinate (botrn) at ($(cn3)!0.5!(n3)+(0.9,-0.3)$);
  \coordinate (botln) at ($(cn2)!0.5!(n2)+(-0.7,-0.3)$);
  \draw [bend left=60, looseness=0.60] (topn.center) to (botrn.center);
  \draw [in=-60, out=-120, looseness=0.40] (botrn.center) to (botln.center);
  \draw [bend left=60, looseness=0.60] (botln.center) to (topn.center);

  \node[left= 7cm of n0, yshift=12mm] (m0) {$\sigma$}
 	child {node (mn0) {$\gamma$} 
      child {node (mn1) {$\sigma$}
        child {node (mn2) {$\alpha$}}
        child {node (mn3) {$\beta$}} }}
    child {node (m1) {$\alpha$}};
  \node[mycircle, right= of m0] (cm0) {$q$};
  \node[mycircle, right= of mn0] (cmn0) {$q$};
  \node[mycircle, right= of mn1] (cmn1) {$p$};
  \node[mycircle, right= of mn2] (cmn2) {$q$};
  \node[mycircle, right= of mn3] (cmn3) {$q$};
  \node[mycircle, right= of m1] (cm1) {$p$};
  \node[above left=-0cm and 0.2cm of m0] (m) {$\xi \in \T_\Sigma$};
  \node[right= 1.6cm of m] {$\rho \in \R_{\cA}(\xi)$};
  \coordinate (topm)  at ($(cmn0)!0.5!(mn0)+(0.1,0.6)$);
  \coordinate (botrm) at ($(cmn3)!0.5!(mn3)+(0.9,-0.3)$);
  \coordinate (botlm) at ($(cmn2)!0.5!(mn2)+(-0.7,-0.3)$);
  \draw [bend left=60, looseness=0.60] (topm.center) to (botrm.center);
  \draw [in=-60, out=-120, looseness=0.40] (botrm.center) to (botlm.center);
  \draw [bend left=60, looseness=0.60] (botlm.center) to (topm.center);
  
  \draw[semithick] (-3.6,-1.5) arc (-270:-220:0.8);
  \draw[semithick] (-3.6,-1.5) arc (270:220:0.8);

\end{tikzpicture}
\caption{\label{fig:well-founded-order-run} An example element $((\xi,\rho),(\xi|_1,\rho|_1))$ of $\succ$.}
  \end{figure}

Next we define the weight of a run $\rho \in \R_\cA(\xi)$. For this, we will define a mapping $\wt_\cA: C \to B$ by well-founded induction on $(C,\succ)$ (using Theorem \ref{thm:dedekind} and \eqref{eq:def-by-induction}). One might be tempted to choose $C = \R_\cA(\xi)$. However, this leads to an underspecification, because one can easily imagine a ranked alphabet $\Sigma$ and two trees $\xi_1,\xi_2 \in \T_\Sigma$ such that $\xi_1 \not= \xi_2$ and $\pos(\xi_1) = \pos(\xi_2)$; then $\R_\cA(\xi_1) = \R_\cA(\xi_2)$ and the mapping $\wt_\cA$ is not able to ``see'' the $\Sigma$-labels at the positions; but, of course, the weight of a run depends on these labels.\footnote{For instance, Equations \ref{eq:coding=0} and \ref{eq:unique-run-A} show that the run alone does not determine the value of $\wt$.}  Thus, we let $C$ be the set of pairs where each pair consists of a tree $\xi \in \T_\Sigma$ and  a run $\rho \in \R_\cA(\xi)$; due to this origin, we denote $C$ by $\mathrm{TR}$ (\emph{t}ree-\emph{r}un); and we define an appropriate terminating relation on this set of pairs. 

\index{TR@$\mathrm{TR}$}
\label{page:TR-prec}
Formally, we let
\(
\mathrm{TR} = \{(\xi,\rho) \mid \xi \in \T_\Sigma, \rho \in \R_\cA(\xi)\}
\)
and  define the binary relation $\succ$ on $\mathrm{TR}$ as follows:
\index{succ@$\succ$}
\[  \text{for every $(\xi, \rho) \in \mathrm{TR}$ and $i \in [\rk(\xi(\varepsilon))]$, we let \
$(\xi,\rho) \succ (\xi|_i,\rho|_i)$}  \enspace. \]
Since the relation $\succ_\Sigma$ is terminating, by Corollary \ref{cor:termination-propagates-to-cartesian-products}, the relation  $\succ$ is also terminating. Moreover, we have that $\nf_\succ(\mathrm{TR}) = \{(\alpha,\rho) \mid \alpha \in \Sigma^{(0)}, \rho: \{\varepsilon\} \to Q\}$. The terminating relation $\succ$ is illustrated in Figure~\ref{fig:well-founded-order-run}.

  We define the mapping
  \[
    \wt_\cA: \mathrm{TR} \to B
  \]
  by induction on $(\mathrm{TR},\succ)$ for every $\xi \in \T_\Sigma$ and $\rho \in \R_\cA(\xi)$ by 
\index{wt@$\wt_\cA(\xi,\rho)$}
\begin{equation}\label{equ:weight-of-run}
\wt_\cA(\xi,\rho) = \Big( \bigotimes_{i\in [k]} \wt_\cA(\xi|_i,\rho|_i)\Big) \otimes \delta_k\big(\rho(1) \cdots \rho(k),\sigma,\rho(\varepsilon)\big) \enspace,
\end{equation}
where $k$ and $\sigma$ abbreviate $\rk(\xi(\varepsilon))$ and $\xi(\varepsilon)$, respectively. 
We call $\wt_\cA(\xi,\rho)$ the \emph{weight of $\rho$ by $\cA$ on $\xi$}. If $\xi$ is uniquely determined by the context, then the phrase \emph{weight of $\rho$ by $\cA$} means  the value $\wt_\cA(\xi,\rho)$.
\index{run semantics}
If the wta $\cA$ is clear from the context, then we drop ``by $\cA$'' and the index $\cA$ from the phrase ``weight of $\rho$ by $\cA$ on $\xi$'' and from the denotation $\wt_\cA(\xi,\rho)$, respectively.

\index{semanticA@$\runsem{\cA}$}
Intuitively, the run semantics of $\cA$ on a tree $\xi \in \T_\Sigma$ is the finite summation of the weights of each run $\rho$ on $\xi$, which is additionally multiplied by the root weight $F_{\rho(\varepsilon)}$. Here we use the fact that the binary summation $\oplus$ of $\B$ is extended in a unique way to the summation over the finite $\R_\cA(\xi)$-family $(\wt(\xi,\rho) \otimes F_{\rho(\varepsilon)} \mid \rho \in \R_\cA(\xi))$  over $B$ (cf. page \pageref{page:finite-summation}). Formally,
the {\it run semantics of $\cA$}, denoted by $\runsem{\cA}$, is the weighted tree language $\runsem{\cA}:~\T_\Sigma~\rightarrow~B$ such that,  for each $\xi \in \T_\Sigma$, we let
\begin{equation*}
  \runsem{\cA}(\xi) = \bigoplus_{\rho \in \R_\cA(\xi)}\wt(\xi,\rho) \otimes F_{\rho(\varepsilon)}\enspace. 
  \end{equation*}
Obviously, we have
\begin{equation}\label{equ:runsem-splitted-set-of-runs}
\runsem{\cA}(\xi) = \bigoplus_{q \in Q}\bigoplus_{\rho \in \R_{\cA}(q,\xi)}\wt(\xi,\rho) \otimes F_q\enspace,
\end{equation}
because the $Q$-indexed family $(\R_{\cA}(q,\xi) \mid q \in Q)$ is a partitioning of $\R_\cA(\xi)$.

Next we prove that the weight $\wt_\cA(\xi,\rho)$ of a run $\rho$ is the product of the weights of the transitions induced by $\rho$ on $\xi$.

\begin{observation}\rm \label{obs:weight-run-explicit}  For every $\xi \in \T_\Sigma$ and $\rho\in \R_\cA(\xi)$, we have
  \[
\wt_\cA(\xi,\rho) = \bigotimes_{\substack{u \in \pos(\xi)\\\text{in $<_{\mathrm{dp}}$ order}}} \delta_{\rk(\xi(u))}(\rho(u1) \cdots \rho(u \, \rk(\xi(u))),\xi(u),\rho(u))\enspace.
\]
\end{observation}
\begin{proof} We prove the statement  by induction on $\T_\Sigma$.
    Let $\xi=\sigma(\xi_1,\ldots,\xi_k)$ and $\rho \in \R_\cA(\xi)$. Then we can calculate as follows:
  \begingroup
  \allowdisplaybreaks
  \begin{align*}
    \wt_\cA(\xi,\rho) =&
    \Big( \bigotimes_{i\in [k]} \wt_\cA(\xi_{i},\rho|_{i})\Big) \otimes \delta_k(\rho(1) \cdots \rho(k),\sigma,\rho(\varepsilon))
    \tag{by \eqref{equ:weight-of-run}}\\
                            =&\bigotimes_{\substack{u \in \pos(\xi_{1})\\\text{in $<_{\mathrm{dp}}$ order}}} \delta_{\rk(\xi_1(u))}(\rho|_{1}(u1) \cdots \rho|_{1}(u\, \rk(\xi_1(u))),\xi_1(u),\rho|_{1}(u)) \\[-7mm]
                            &\hspace*{25mm}  \otimes \ldots \otimes \bigotimes_{\substack{u \in \pos(\xi_{k})\\\text{in $<_{\mathrm{dp}}$ order}}} \delta_{\rk(\xi_k(u))}(\rho|_{k}(u1) \cdots \rho|_{k}(u\, \rk(\xi_k(u))),\xi_k(u),\rho|_{k}(u)) \\
                                                       &\hspace*{25mm}  \otimes \delta_k(\rho(1) \cdots \rho(k),\sigma,\rho(\varepsilon)) \tag{by I.H.}\\
                               =& \bigotimes_{\substack{u \in \pos(\xi)\\\text{in $<_{\mathrm{dp}}$ order}}} \delta_{\rk(\xi(u))}(\rho(u1) \cdots \rho(u\, \rk(\xi(u))),\xi(u),\rho(u))\\[-5mm]
  &  \tag{by associativity of $\otimes$ and definition of $<_{\mathrm{dp}}$} 
  \end{align*}
  \endgroup  
\end{proof}

\index{run recognizable}
\index{r-recognizable}
\index{bu-deterministically r-recognizable}
A weighted tree language $r: \T_\Sigma \rightarrow B$ is \emph{run recognizable over $\B$} (for short: run recognizable or r-recognizable) if there exists a $(\Sigma,\B)$-wta $\cA$ such that $r = \runsem{\cA}$. The set of all $(\Sigma,\B)$-weighted tree languages which are run recognizable over $\B$, is  denoted by $\Rec^{\mathrm{run}}(\Sigma,\B)$.  
\index{Rec@$\Rec^{\mathrm{run}}(\Sigma,\B)$}
In the obvious way, we can define the notions of \emph{bu-deterministically r-recognizable over $\B$} and \emph{crisp-deterministically r-recognizable over $\B$}. We denote the corresponding sets of all such  weighted tree languages by $\budRec^{\mathrm{run}}(\Sigma,\B)$ and $\cdRec^{\mathrm{run}}(\Sigma,\B)$, respectively. 
\index{budRec@$\budRec^{\mathrm{run}}(\Sigma,\B)$}
\index{cdRec@$\cdRec^{\mathrm{run}}(\Sigma,\B)$}

\index{r-equivalent}
\index{equivalent}
Two $(\Sigma,\B)$-wta $\cA_1$ and $\cA_2$ are \emph{run equivalent} (r-equivalent) if $\runsem{\cA_1} = \runsem{\cA_2}$.

\paragraph{Initial algebra semantics.}
\index{vector algebra}
\label{page:vector-algebra}
The heart of the initial algebra semantics \cite{gogthawagwri77} of a $(\Sigma,\B)$-wta $\cA$ is a particular $\Sigma$-algebra, called the vector algebra of $\cA$; it is tailormade for $\cA$. Due to the uniqueness of the $\Sigma$-algebra homomorphism from the $\Sigma$-term algebra $\sfT_\Sigma = (\T_\Sigma,\ttop_\Sigma)$ to the vector algebra of $\cA$, each $\Sigma$-tree $\xi$ can be interpreted (or: evaluated) in a unique way in the vector algebra. Eventually, the interpretation of $\xi$ is modified with root weights. On first glance, the definition of the vector algebra of a wta below might look a bit artificial. However, conceptually, this definition results from the straightforward application of the two steps described at the beginning of this chapter to an fta and its initial algebra semantics (the latter being defined in \cite[Def.~2.2.5]{gecste84}).

\index{Vector@$\V(\cA)$}
Formally, the {\em vector algebra of $\cA$} is the $\Sigma$-algebra $\V(\cA)=(B^Q,\delta_\cA)$ where, for every $k \in \mathbb{N}$, $\sigma \in \Sigma^{(k)}$, the $k$-ary operation $\delta_\cA(\sigma): B^Q \times \cdots \times B^Q \to B^Q$ is defined by
\begin{equation}\label{eq:delta-A-definition}
\delta_\cA(\sigma)(v_1,\dots,v_k)_q 
  = \bigoplus_{q_1\cdots q_k \in Q^k} \Big(\bigotimes_{i\in[k]} (v_i)_{q_i}\Big) \otimes \delta_k(q_1\cdots q_k,\sigma,q)
  \end{equation}
  for every $v_1,\dots,v_k \in B^Q$ and $q \in Q$.
   Thus, in particular, for each $\alpha \in \Sigma^{(0)}$, we have
  \[
    \delta_\cA(\alpha)()_q= \bigoplus_{\varepsilon \in Q^0} \mathbb{1} \otimes \delta_0(\varepsilon,\alpha,q) = \delta_0(\varepsilon,\alpha,q) \enspace.
  \]

We abbreviate  $\h_{\V(\cA)}$ by $\h_\cA$, i.e., we denote the unique $\Sigma$-algebra homomorphism from the $\Sigma$-term algebra $\sfT_\Sigma$  to the vector algebra $\V(\cA)$ by $\h_\cA$. Then, for every $\xi = \sigma(\xi_1,\ldots,\xi_k)$ in $\T_\Sigma$ and $q\in Q$, we have
\begin{align*}
\h_\cA(\sigma(\xi_1,\ldots,\xi_k))_q &= \h_\cA(\ttop_\Sigma(\sigma)(\xi_1,\ldots,\xi_k))_q = \delta_{\cA}(\sigma)
(\h_\cA(\xi_1),\ldots, \h_\cA(\xi_k))_q \\
&= \bigoplus_{q_1 \cdots q_k \in Q^k} \Big( \bigotimes_{i\in [k]} \h_\cA(\xi_i)_{q_i}\Big) \otimes \delta_k(q_1\cdots q_k,\sigma,q),
\end{align*}
where $\ttop_\Sigma(\sigma)$ is the operation of the $\Sigma$-term algebra associated to $\sigma$; the second equality holds, because $\h_\cA$ is a $\Sigma$-algebra homomorphism. In the subsequent proofs, we will show only the first and the last expressions of the above calculation. For the particular case that $k=0$, we obtain the following:
\begin{align*}
\h_\cA(\sigma)_q = \h_\cA(\ttop_\Sigma(\sigma)())_q = \delta_{\cA}(\sigma)
()_q = \delta_0(\varepsilon,\sigma,q),
\end{align*}
and thus we will write $\h_\cA(\sigma)_q=\delta_0(\varepsilon,\sigma,q)$ directly.

\index{homA@$\h_\cA$}
\index{initial algebra semantics}
\index{semanticA@$\initialsem{\cA}$}
The \emph{initial algebra semantics of $\cA$}, denoted by $\initialsem{\cA}$,  is the weighted tree language $\initialsem{\cA}: \T_\Sigma \rightarrow B$ defined by  
\begin{equation*}
  \initialsem{\cA}(\xi) =  \h_\cA(\xi) \cdot F\enspace, 
\end{equation*}
where $\cdot$ is the scalar product of two $Q$-vectors over $B$, i.e.,
$\initialsem{\cA}(\xi) = \bigoplus_{q \in Q} \h_\cA(\xi)_q \otimes F_q$.

We note that, in \cite[p.~22]{boz99}, the concept of weighted tree automaton over semirings is defined, where the move function $\mu_\sigma: Q^k \to B^Q$ corresponds to our transition mapping $\delta_k$ on $\sigma$. Then, by induction on $\T_\Sigma$,  he defines the mapping $\mu_\tau: \T_\Sigma \to B^Q$, which corresponds to our $\h_\cA$. The definition of $\mu_\tau$ uses the multilinear extension of the mappings in the  family $(\mu_\sigma \mid \sigma \in \Sigma)$. In view of our discussion on p.~\pageref{page:why-commutativiy-for-multilinearity}, we think that in \cite{boz99} each semiring is assumed to be commutative. In contrast, for each weighted tree automata over a strong bimonoid $\B$, we define the mapping $\delta_\cA(\sigma)$ explicitly and, if $\B$ is a commutative semiring, then we prove that $\delta_\cA(\sigma)$ is multilinear (cf. Lemma~\ref{lm:vector-algebra-is-semimodule-com-sr}).

  \index{accessible subalgebra}
  \index{aV@$\aV(\cA)$}
  The vector algebra $\V(\cA)$ may contain $Q$-vectors which are not accessible by $\cA$, i.e., in general, we have $B^Q \setminus \im(\h_\cA) \ne \emptyset$. Next we define a $\Sigma$-algebra which is a subalgebra of $\V(\cA)$ and has exactly $\im(\h_\cA)$ as carrier set. By Lemma~\ref{lm:hom-image=subalgebra}, the $\Sigma$-algebra \[\aV(\cA) =(\im(\h_\cA),\delta_{\cA})\] is a subalgebra of $\V(\cA)$. 
We call $\aV(\cA)$ the \emph{accessible subalgebra of $\V(\cA)$}.
By  Observation \ref{obs:smallest-subalgebra-im}, $\aV(\cA)$ is the smallest subalgebra of $\V(\cA)$.

We denote the unique $\Sigma$-algebra homomorphism from $\sfT_\Sigma$ to $\aV(\cA)$ by $\h_{\aV(\cA)}$. Since  $\aV(\cA)$ is a subalgebra of $\V(\cA)$, the homomorphism $\h_{\aV(\cA)}$ is a $\Sigma$-algebra homomorphism also from $\sfT_\Sigma$  to $\V(\cA)$, and since $\sfT_\Sigma$ is initial, we have 
\begin{equation}
\h_{\aV(\cA)}(\xi) = \h_\cA(\xi)\text{ for every $\xi \in \T_\Sigma$}\enspace. \label{eq:hNA=hA}
\end{equation}
Then we can express the initial algebra semantics $\initialsem{\cA}$ also in terms of the homomorphism $\h_{\aV(\cA)}$ as follows:
\begin{equation}
\initialsem{\cA} = F' \circ \h_{\aV(\cA)} \label{eq:initial-sem-Nerode}
\end{equation}
where $F': \im(\h_\cA)\to B$ is the mapping defined by $F'(u) = \bigoplus_{q\in Q} u_q\otimes F_q$ for every $u\in \im(\h_\cA)$. Indeed, for each $\xi \in \T_\Sigma$, we have
\[ \initialsem{\cA}(\xi)= \bigoplus_{q\in Q}\h_\cA(\xi)_q\otimes F_q\stackrel{\eqref{eq:hNA=hA}}{=} \bigoplus_{q\in Q}\h_{\aV(\cA)}(\xi)_q\otimes F_q = (F' \circ \h_{\aV(\cA)})(\xi)\enspace.\]

By Theorem \ref{thm:kernel-is-congruence}, the kernel $\ker(\h_\cA)$ is a congruence relation on the $\Sigma$-term algebra $(\T_\Sigma,\ttop_\Sigma)$.\footnote{For the case of wsa, in \cite[Prop.~6.2]{cirdroignvog10} the relation $\ker(\h_\cA)$ was defined and called ``Nerode right congruence''. We refrain from also calling our relation $\ker(\h_\cA)$ ``Nerode congruence relation'', because the Nerode congruence relation from classical formal language theory  means something else: it is based on a language $L \subseteq \Gamma^*$ and defines two strings $w_1$ and $w_2$ to be $L$-equivalent if for each $u \in \Gamma^*$: $w_1u \in L$ iff $w_2u \in L$ (also cf. \cite{koz92} for tree languages). Our $\ker(\h_\cA)$ does not depend on a (weighted) language and hence, it is of a different nature.}

\begin{lemma}\label{lm:Nerode-alg=factor-alg} \rm $\aV(\cA) \cong \sfT_\Sigma/\ker(\h_\cA)$, i.e., the accessible subalgebra of $\V(\cA)$ is isomorphic to  the quotient algebra of the $\Sigma$-term algebra modulo $\ker(\h_\cA)$.
  \end{lemma}
  \begin{proof} Since $\h_{\aV(\cA)}$ is a surjective $\Sigma$-algebra homomorphism from the $\Sigma$-term algebra to $\aV(\cA)$, Theorem \ref{thm:kernel-is-congruence} implies $\aV(\cA) \cong \sfT_\Sigma/\ker(\h_{\aV(\cA)})$. Since \eqref{eq:hNA=hA} implies $\ker(\h_{\cA})=\ker(\h_{\aV(\cA)})$, we obtain $\aV(\cA) \cong \sfT_\Sigma/\ker(\h_{\cA})$.
\end{proof}

\index{initial algebra recognizable}
\index{i-recognizable}
\index{bu-deterministically i-recognizable}
A weighted tree language $r: \T_\Sigma \rightarrow B$ is \emph{initial algebra recognizable over $\B$} (for short: initial algebra recognizable or i-recognizable) if there exists a $(\Sigma,\B)$-wta $\cA$ such that $r = \initialsem{\cA}$. The set of all weighted tree languages over $\Sigma$ and $\B$ which are i-recognizable, is  denoted by $\Rec^{\mathrm{init}}(\Sigma,\B)$.   
\index{Rec@$\Rec^{\mathrm{init}}(\Sigma,\B)$}
In an obvious way, we can define the notion of \emph{bu-deterministically i-recognizable} weighted tree language and \emph{crisp-deterministically i-recognizable} weighted tree language. We denote the corresponding sets of all such  weighted tree languages by $\budRec^{\mathrm{init}}(\Sigma,\B)$ and $\cdRec^{\mathrm{init}}(\Sigma,\B)$, respectively.
\index{budRec@$\budRec^{\mathrm{init}}(\Sigma,\B)$}
\index{cdRec@$\cdRec^{\mathrm{init}}(\Sigma,\B)$}

\index{i-equivalent}
\index{equivalent}
Let $\cA_1$ and $\cA_2$ be $(\Sigma,\B)$-wta. We say that $\cA_1$ and $\cA_2$ are \emph{initial algebra equivalent} (i-equivalent) if $\initialsem{\cA_1} = \initialsem{\cA_2}$. Moreover,  $\cA_1$ and $\cA_2$ are \emph{equivalent} if they are both r-equivalent and i-equivalent.


\section{Examples}
\label{sec:examples}

Here we list a number of weighted tree languages and show how they can be recognized by wta.

\index{$\#_{\Ra_A}$}
\begin{example}{\bf (Number of accepting runs of an fta on a tree.)}\label{ex:valid-runs} \rm
Let $A = (Q,\delta,F)$ be a $\Sigma$-fta. We recall that $\R_A(\xi)$, $\Rv_A(\xi)$, and $\Ra_A(\xi)$ denote the set of runs, valid runs, and accepting runs of $A$ on a tree $\xi$, respectively.
We consider the mapping
\begin{align*}
  \#_{\Ra_A}: \T_\Sigma \to \mathbb{N} \ \ \text{ with } \ \
  \#_{\Ra_A}(\xi) =  |\Ra_A(\xi)| \ \text{ for each $\xi \in \T_\Sigma$}\enspace.
\end{align*}
We call $\#_{\Ra_A}$ the \emph{multiplicity mapping of $A$} (cf. \cite[Sect.~VI.1]{eil74}). \index{multiplicity mapping}

As weight algebra we use the semiring $\Nat=(\mathbb{N},+,\cdot,0,1)$ of natural numbers.
We construct a $(\Sigma,\Nat)$-wta $\cA= (Q,\delta',F')$ which r-recognizes and also i-recognizes $\#_{\Ra_A}$, as follows:
\begin{compactitem}
\item for each $k\in \mathbb{N}$, we have $(\delta')_k=\chi(\delta_k)$ and
\item $F'=\chi(F)$.
\end{compactitem}

Obviously, $\cA$ is crisp and has unit root weights.

Next we show that $\cA$ r-recognizes $\#_{\Ra_A}$, i.e., $\runsem{\cA} = \#_{\Ra_A}$.
We observe that, for each $\xi\in \T_\Sigma$, we have $\R_{\cA}(\xi)=\R_{A}(\xi)$.
By induction on $(\mathrm{TR},\succ)$ (defined on page \pageref{page:TR-prec}), we prove that,  
for each $(\xi,\rho) \in \mathrm{TR}$, we have 
\begin{equation}\label{equ:run=no-runs}
\wt(\xi,\rho)=
\begin{cases}
1 & \text{if $\rho \in \Rv_{A}(\xi)$}\\
0 & \text{otherwise}\enspace. 
\end{cases}
\end{equation}

I.B.: Let $\xi=\alpha$ for some $\alpha \in \Sigma^{(0)}$, and let $\rho: \{\varepsilon\} \to Q$. Then
\(\wt(\xi,\rho) = (\delta')_0(\varepsilon,\alpha,\rho(\varepsilon))\)
by definition of $\wt$. We proceed by case analysis.

\underline{Case (a):} Let $\rho \in \Rv_{A}(\alpha)$. Then $(\varepsilon,\alpha,\rho(\varepsilon)) \in \delta_0$ and by the definition of $(\delta')_0$ we have $(\delta')_0(\varepsilon,\alpha,\rho(\varepsilon))~=~1$. Thus $\wt(\xi,\rho) =1$. 

\underline{Case (b):} Let $\rho \not\in \Rv_{A}(\alpha)$. Then $(\varepsilon,\alpha,\rho(\varepsilon)) \not\in \delta_0$ and we have $(\delta')_0(\varepsilon,\alpha,\rho(\varepsilon)) = 0$, i.e.,  $\wt(\xi,\rho) =0$.

I.S.: Let $(\xi,\rho) \in \mathrm{TR}$ such that $\xi = \sigma(\xi_1,\ldots,\xi_k)$ for some $k \in \mathbb{N}_+$. Then 
\[\wt(\xi,\rho) =
  \Big(\wt(\xi_{1},\rho|_{1}) \cdot \ldots \cdot \wt(\xi_{k},\rho|_{k})\Big) \cdot (\delta')_k(\rho(1) \cdots \rho(k),\sigma,\rho(\varepsilon))\]
by definition of $\wt$. Again, we proceed by case analysis.

\underline{Case (a):} Let $\rho \in \Rv_A(\xi)$. Then, for each $i \in [k]$, we have $\rho|_i \in \Rv_A(\xi_i)$, and $(\rho(1) \cdots \rho(k),\sigma,\rho(\varepsilon)) \in \delta_k$. Hence, by I.H., we have $\wt(\xi_i,\rho|_i) = 1$ for each $i \in [k]$ and, by construction, we have $(\delta')_k(\rho(1) \cdots \rho(k),\sigma,\rho(\varepsilon)) = 1$. Thus $\wt(\xi,\rho) = 1$.

\underline{Case (b1):} Let $\rho \not\in \Rv_A(\xi)$ and $(\forall i\in[k]):\rho|_i \in \Rv_A(\xi_i)$. Then $(\rho(1) \cdots \rho(k),\sigma,\rho(\varepsilon)) \not\in \delta_k$, i.e.,  $(\delta')_k(\rho(1) \cdots \rho(k),\sigma,\rho(\varepsilon)) = 0 $. Hence $\wt(\xi,\rho) = 0$.

\underline{Case (b2):} $\rho \not\in \Rv_A(\xi)$ and $(\exists i\in[k]):\rho|_i \not\in \Rv_A(\xi_i)$. Then, by I.H., $\wt(\xi_i,\rho|_i)=0$ and hence $\wt_\cA(\xi,\rho) =0$.
This finishes the proof of \eqref{equ:run=no-runs}.

Hence, for each $\xi\in \T_\Sigma$,
\begin{align*}
\runsem{\cA}(\xi) & =  \bigplus_{\rho \in \R_{\cA}(\xi)}\wt(\xi,\rho) \cdot F'_{\rho(\varepsilon)}=
\bigplus_{\rho \in \Rv_{A}(\xi)} 1 \cdot F'_{\rho(\varepsilon)} = \bigplus_{\rho \in \Ra_{A}(\xi)} 1 = |\Ra_{A}(\xi)| = \#_{\Ra_A}(\xi)\enspace,
\end{align*}
i.e., $\runsem{\cA}(\xi)$ is the number of accepting runs of $A$ on $\xi$. Hence $\#_{\Ra_A} \in \Rec^{\mathrm{run}}(\Sigma,\Nat)$.

Next we show that $\cA$ also i-recognizes $\#_{\Ra_A}$, i.e., $\initialsem{\cA} = \#_{\Ra_A}$.
By induction on $\T_\Sigma$, we prove that the following statement holds: 
\begin{equation}\label{equ:hom=no-runs}
\text{For every $\xi \in \T_\Sigma$ and $q \in Q$ we have: } \h_{\cA}(\xi)_q = |\Rv_{A}(q,\xi)|.
\end{equation}
Let $\xi = \sigma(\xi_1,\ldots,\xi_k)$. Then we can calculate as follows.
\begingroup
\allowdisplaybreaks
\begin{align*}
  \h_{\cA}(\sigma(\xi_1,\ldots,\xi_k))_q &= \bigplus_{q_1\cdots q_k\in Q^k}  \h_{\cA}(\xi_1)_{q_1} \cdot \ldots \cdot \h_{\cA}(\xi_k)_{q_k}  \cdot (\delta')_k(q_1 \cdots q_k,\sigma,q)\\
                                       &= \bigplus_{q_1\cdots q_k\in Q^k}  |\Rv_{A}(q_1,\xi_1)| \cdot \ldots \cdot |\Rv_{A}(q_k,\xi_k)| \cdot (\delta')_k(q_1 \cdots q_k,\sigma,q) \tag{by I.H.}\\
  &= |\Rv_{A}(q,\sigma(\xi_1,\ldots,\xi_k))| \enspace. \tag{by definitions of $(\delta')_k$ and of $\Rv_{A}(q,\sigma(\xi_1,\ldots,\xi_k))$}
\end{align*}
\endgroup
Then, for each $\xi \in \T_\Sigma$, we have 
\begin{align*}
  \initialsem{\cA}(\xi) & =  \bigplus_{q \in Q}\h_{\cA}(\xi)_q \cdot F'_q\\
  & = \bigplus_{q \in Q} |\Rv_{A}(q,\xi)| \cdot F'_q \tag{by \eqref{equ:hom=no-runs}}\\
  & = \bigplus_{q \in Q} |\Ra_{A}(q,\xi)| \tag{by definitions of $F'$, $\Rv_{A}(q,\xi)$, and  $\Ra_{A}(q,\xi)$}\\
  & = | \bigcup_{q \in Q} \Ra_{A}(q,\xi)| \tag{because $\Ra_{A}(q_1,\xi) \cap \Ra_{A}(q_2,\xi) = \emptyset$ for every $q_1,q_2 \in Q$}\\
  &= |\Ra_{A}(\xi)| \tag{because $\Ra_{A}(\xi)= \bigcup_{q \in Q} \Ra_{A}(q,\xi)$}\\
  & = \#_{\Ra_A}(\xi)\enspace.
\end{align*}
Hence $\#_{\Ra_A} \in \Rec^{\mathrm{init}}(\Sigma,\Nat)$.

There will be a general result which says that, if $\B$ is a semiring, then $\initialsem{\cA} = \runsem{\cA}$ for each $(\Sigma,\B)$-wta $\cA$ (cf. Theorem \ref{thm:semiring-run=initial}).
\hfill $\Box$
\end{example}

\

\begin{example}\rm \cite{dro21}  {\bf (Number of occurrences of transitions of accepting computations of an fta on a tree.)} Let $A = (Q,\delta,F)$ be a $\Sigma$-fta. We consider the mapping
\begin{align*}
  \#_{A}: \T_\Sigma \to \mathbb{N} \ \ \text{ with } \ \ 
  \#_{A}(\xi) = |\Ra_A(\xi)| \cdot \size(\xi) \text{ for each $\xi \in \T_\Sigma$} \enspace.
\end{align*}
As weight algebra we use the plus-plus strong bimonoid $\PP_\mathbb{N} = (\mathbb{N}_\0,\oplus,+,\0,0)$ (cf. Example \ref{ex:strong-bimonoids}(\ref{ex:plus-plus-sb})).

We construct the $(\Sigma,\PP_\mathbb{N})$-wta $\cA= (Q,\delta',F')$ which r-recognizes $\#_{A}$, as follows.
 For every $k\in \mathbb{N}$, $q_1,\ldots,q_k,q \in Q$, and $\sigma \in \Sigma^{(k)}$, we let:
  \[\delta'_k(q_1 \cdots q_k,\sigma,q) =
  \begin{cases}
    1 & \text{ if $(q_1 \cdots q_k,\sigma,q) \in \delta_k$}\\
    \0 & \text{ otherwise}
  \end{cases}  \ \ \text{ and } \ \
 F'_q =
  \begin{cases}
    0 & \text{ if $q \in F$}\\
    \0 & \text{ otherwise} \enspace.
  \end{cases}
  \]
Obviously, for each $\xi\in \T_\Sigma$, we have $\R_{\cA}(\xi)=\R_{A}(\xi)$, and
for each  $\rho\in \R_{\cA}(\xi)$, we have
\begin{equation}
\wt(\xi,\rho)= 
\begin{cases}
\size(\xi) & \text{if $\rho \in \Rv_{A}(\xi)$}\\
\0 & \text{otherwise}\enspace. 
\end{cases}\label{equ:wt-number-transitions}
\end{equation}
Hence, for each $\xi\in \T_\Sigma$,
\begin{align*}
  \runsem{\cA}(\xi) & =  \bigoplus_{\rho \in \R_{\cA}(\xi)}\wt(\xi,\rho) + F'_{\rho(\varepsilon)}\\
                     & = \bigoplus_{\rho \in \Rv_{A}(\xi)} \size(\xi) + F'_{\rho(\varepsilon)}
                       \tag{by \eqref{equ:wt-number-transitions}}\\
                     & = \bigoplus_{\rho \in \Ra_{A}(\xi)} \size(\xi) + 0
  \tag{because $\0$ is the unit element of $\oplus$}\\
                     & = \bigplus_{\rho \in \Ra_{A}(\xi)} \size(\xi)
                       \tag{because $\size(\xi) \in \mathbb{N}$}\\
  & = |\Ra_{A}(\xi)| \cdot \size(\xi) = \#_{A}(\xi)\enspace.
\end{align*}
Hence $\#_{A} \in \Rec^{\mathrm{run}}(\Sigma,\PP_\mathbb{N})$.
\hfill $\Box$
    \end{example}

 \

\index{size@$\size$}
\begin{example}{\bf (Size of trees.)}\label{ex:size} \rm Let $\Sigma = \{\sigma^{(2)}, \alpha^{(0)}\}$. 
  We consider the mapping
  \[\size: \T_\Sigma \to \mathbb{N}\]
  defined on page \pageref{page:height-size-pos}.
  As weight algebra we use the tropical semiring $\Natminplus=(\mathbb{N}_\infty,\min,+,\infty,0)$. Since $\mathbb{N} \subseteq \mathbb{N}_\infty$, by our convention in Section \ref{sect:binary-relations}, $\size$ is also a mapping of type $\size: \T_\Sigma \to \mathbb{N}_\infty$, i.e., a $(\Sigma,\Natminplus)$-weighted tree language.
  
We construct the $(\Sigma,\Natminplus)$-wta $\cA=(Q,\delta,F)$ which r-recognizes $\size$, as follows.
\begin{compactitem}
\item $Q = \{{s}\}$,  (intuitively, the state ${s}$ computes the size of the tree),

\item $\delta_0(\varepsilon,\alpha,{s}) = \delta_2({s}{s},\sigma,{s}) = 1$, and 
\item $F_{s}=0$.
\end{compactitem}
In  Figure \ref{fig:hypgraph-size} we represent $\cA$ as an fta-hypergraph. 
Clearly, $\cA$ is bu-deterministic, total, and root weight normalized; $\cA$ is not crisp-deterministic, because $1 \not\in\{\infty,0\}$.

\begin{figure}[t]
    \begin{center}
\begin{tikzpicture}
\tikzset{node distance=7em, scale=0.6, transform shape}
\node[state, rectangle] (1) {\Large $\alpha$};
\node[state, right of=1] (2) {\Large s};
\node[state, rectangle, right of=2] (3) {\Large $\sigma$};

\tikzset{node distance=2em}
\node[above of=1] (w1) {1};
\node[above of=2] (w2) {0};
\node[above of=3] (w3) {1};

\draw[->,>=stealth] (1) edge (2);
\draw[->,>=stealth] (3) edge (2);
\draw[->,>=stealth] (2) edge[out=60, in=30, looseness=1.4] (3);
\draw[->,>=stealth] (2) edge[out=-60, in=-30, looseness=1.4] (3);
\end{tikzpicture}
\end{center}

\vspace{-5mm}

\caption{\label{fig:hypgraph-size} The fta-hypergraph for the $(\Sigma,\Natminplus)$-wta $\cA=(Q,\delta,F)$ which r-recognizes $\size$.}
\end{figure}

For each $\xi \in \T_\Sigma$, there is exactly one run on $\xi$. We denote it by $\rho^\xi$, hence $\rho^\xi(w) = {s}$ for each $w \in \pos(\xi)$. Obviously, for each $\xi \in \T_\Sigma$:
\[
\wt(\xi,\rho^\xi)= |\pos(\xi)| \enspace.
\]
Then, for each $\xi \in \T_\Sigma$, we have
\[\runsem{\cA}(\xi) = \min( \wt(\xi,\rho) + F_{\rho(\varepsilon)} \mid  \rho \in \R_\cA(\xi)) = \wt(\xi,\rho^\xi) + 0= |\pos(\xi)| = \size(\xi) \enspace.
\]
Hence $\runsem{\cA}=\size$ and thus $\size \in \budRec^{\mathrm{run}}(\Sigma,\Natminplus)$. The construction can easily be generalized to an arbitrary ranked alphabet $\Sigma$.
\hfill $\Box$
\end{example}

\

\begin{example}\rm \label{ex:height}
\index{height}
{\bf (Height of trees.)} Let $\Sigma=\{\sigma^{(2)},\alpha^{(0)}\}$. We consider the mapping  
\[
\height: \T_\Sigma \to \mathbb{N}
\]
defined on page \pageref{page:height-size-pos}.  As weight algebra we use the arctic semiring 
$\Natmaxplus = (\mathbb{N}_{-\infty},\max,+,-\infty,0)$. Thus, the mapping $\height$ is a $(\Sigma,\Natmaxplus)$-weighted tree language.

We construct the $(\Sigma,\Natmaxplus)$-wta $\cA = (Q,\delta,F)$ which i-recognizes $\height$, as follows.
\begin{compactitem}
\item $Q  = \{\h,0\}$, (intuitively, $h$ and $0$ should calculate the height and the natural number $0$, respectively) 

\item $\delta_0(\varepsilon,\alpha,\h) = \delta_0(\varepsilon,\alpha,0) = 0$ and for every $q_1,q_2,q\in Q$, 
\[
\delta_2(q_1q_2,\sigma,q) = 
\begin{cases}
1&\text{ if }q_1q_2q\in\{\h0\h, 0\h\h\}\;,\\
0&\text{ if }q_1q_2q=000\;,\\
-\infty&\text{ otherwise\;.}
\end{cases}
\]

\item $F_\h = 0$ and $F_0 = -\infty$.
\end{compactitem}
In Figure \ref{fig:hypgraph-height} we represent $\cA$ as an fta-hypergraph. 
Clearly, $\cA$ is root weight normalized; $\cA$ is not total, because there does not exist a state $q$ such that $\delta_ 2(\h\h,\sigma,q) \not= -\infty$; $\cA$ is not bu-deterministic, because $\delta_0(\varepsilon,\alpha,\h) = \delta_0(\varepsilon,\alpha,0) = 0 \not= -\infty$.

\begin{figure}[t]
  \begin{center}
\begin{tikzpicture}
\tikzset{node distance=7em, scale=0.6, transform shape}
\node[state, rectangle] (1) {\Large $\alpha$};
\node[state, right of=1] (2) {\Large h};
\node[state, rectangle, above of=2] (3){\Large $\sigma$};
\node[state, rectangle, below of=2] (4) {\Large $\sigma$};
\node[state, right of=2] (5) [right=5em] {\Large 0};
\node[state, rectangle, below of=5] (6) {\Large $\alpha$};
\node[state, rectangle, right of=5] (7) {\Large $\sigma$};

\tikzset{node distance=2em}
\node[above of=1] (w1) {0};
\node[above of=2] (w2) [right=0.05cm] {0};
\node[above of=3] (w3) {1};
\node[above of=4] (w4) [right=0.05cm]{1};
\node[above of=6] (w6) [right=0.05cm] {0};
\node[above of=7] (w7) {0};

\draw[->,>=stealth] (1) edge (2);
\draw[->,>=stealth] (2) edge (3);
\draw[->,>=stealth] (3) edge[out=-250, in=-210, looseness=1.8] (2);
\draw[->,>=stealth] (2) edge (4);
\draw[->,>=stealth] (4) edge[out=250, in=-150, looseness=1.8] (2);
\draw[->,>=stealth] (5) edge (4);
\draw[->,>=stealth] (5) edge (3);
\draw[->,>=stealth] (6) edge (5);
\draw[->,>=stealth] (5) edge[out=60, in=30, looseness=1.4] (7);
\draw[->,>=stealth] (5) edge[out=-60, in=-30, looseness=1.4] (7);
\draw[->,>=stealth] (7) edge (5);
\end{tikzpicture}
\end{center}

\vspace{-10mm}

\caption{\label{fig:hypgraph-height} The fta-hypergraph for the $(\Sigma,\Natmaxplus)$-wta $\cA$ which i-recognizes $\height$.}
\end{figure}

By induction on $\T_\Sigma$, we prove that the following statement holds:
\begin{equation}
\text{For each $\xi \in \T_\Sigma$, we have $\h_\cA(\xi)_{\h} = \mathrm{height}(\xi)$ and $\h_\cA(\xi)_{0} = 0$\enspace.} \label{equ:ex-height}
\end{equation}

I.B.: Let $\xi = \alpha$. Then $\h_\cA(\xi)_{\h} = \delta(\varepsilon,\alpha,\h) = 0$ (and similarly for $\h_\cA(\xi)_{0}$).

I.S.: Now let $\xi = \sigma(\xi_1,\xi_2)$ for some trees $\xi_1,\xi_2 \in \T_\Sigma$. 
Then
\begingroup
\allowdisplaybreaks
\begin{align*}
& \h_\cA(\sigma(\xi_1,\xi_2))_{\h} \\[1mm]
=& \ \max( \h_\cA(\xi_1)_{q_1} +\h_\cA(\xi_2)_{q_2} + \delta_2(q_1q_2,\sigma,\h) \mid q_1,q_2\in Q)\\[2mm]
=& \ \max (\h_\cA(\xi_1)_{\h} +\h_\cA(\xi_2)_{0} + \delta_2(\h0,\sigma,\h),  \h_\cA(\xi_1)_{0} +\h_\cA(\xi_2)_{\h} + \delta_2(0\h,\sigma,\h))
\tag{using that $\delta_2(\h\h,\sigma,\h) = \delta_2(00,\sigma,\h)= -\infty$ and $-\infty$ is neutral for $\max$}\\[2mm]
  =& \ \max(\height(\xi_1) + 0 + 1, \ 0 + \height(\xi_2) + 1)
     \tag{by I.H. and definition of $\delta$}\\[1mm]
=& \ 1 + \max(\height(\xi_1),\height(\xi_2)) = \height(\xi)\enspace.
\end{align*}
\endgroup
Moreover, using similar arguments, we can calculate:
\[
 \h_\cA(\sigma(\xi_1,\xi_2))_{0}
= \h_\cA(\xi_1)_{0} +\h_\cA(\xi_2)_{0} + \delta_2(00,\sigma,0)
= 0 + 0 + 0 = 0\enspace.
\]
This finishes the proof of Statement \eqref{equ:ex-height}.
Then, for each $\xi \in \T_\Sigma$, we have 
\[\initialsem{\cA}(\xi) = \max (\h_\cA(\xi)_q +F_q \mid {q \in Q}) = \max (\h_\cA(\xi)_{\h} + 0, 0+(-\infty)) = \h_\cA(\xi)_{\h} = \height(\xi)\enspace.\] 
Hence $\initialsem{\cA}=\height$, and thus $\height \in \Rec^{\mathrm{init}}(\Sigma,\Natmaxplus)$.

 Next we show that the $(\Sigma,\Natmaxplus)$-wta $\cA$ also r-recognizes the weighted tree language $\height$. First, we observe that for each $\xi \in \T_\Sigma$,
\begin{equation}
  \height(\xi) = \max(|w| \mid w \in \pos_\alpha(\xi)) \enspace. \label{equ:height=max}
    \end{equation}

    Let $\xi \in \T_\Sigma$ be an arbitrary $\Sigma$-tree. We define the run $\rho_0: \pos(\xi) \to Q$ by $\rho_0(w) = 0$ for each $w \in \pos(\xi)$. It is obvious that $\wt(\xi,\rho_0)=0$. 

Moreover, for every $w \in \pos_\alpha(\xi)$, we define the run $\rho_w: \pos(\xi) \to Q$ such that for each $v \in \pos(\xi)$  we let $\rho_w(v) = \h$ if $v$ is a prefix of $w$, and $0$ otherwise. By induction on $\T_\Sigma$, we prove that the following statement holds: 
\begin{equation}
\text{For every $\xi \in \T_\Sigma$ and $w \in  \pos_\alpha(\xi)$, we have }\wt(\xi,\rho_w) = |w| \enspace. \label{equ:length-rhow}
\end{equation}

I.B.: Let $\xi = \alpha$. Then $w = \varepsilon$ and $\wt(\xi,\rho_\varepsilon)=\delta_0(\varepsilon,\alpha,\h)=0$. 

I.S.: Now let $\xi = \sigma(\xi_1,\xi_2)$ and assume that $w=1v$ for some $v \in \pos(\xi_1)$. Then
\begin{align*}
\wt(\sigma(\xi_1,\xi_2),\rho_w)
  &= \wt(\xi_1,\rho_w|_1) +  \wt(\xi_2,\rho_w|_2) +  \delta_2(\rho_w(1)\rho_w(2),\sigma,\rho_w(\varepsilon))\\
  &= \wt(\xi_1,\rho_v) + \wt(\xi_2,\rho_0) + \delta_2(\h0,\sigma,\h)\\
  &= |v| + 0  + 1 \tag{by I.H. and definition of $\delta_2$}\\
  &= |w| \enspace.
\end{align*}
In a similar way we prove $\wt(\xi,\rho_w) = |w|$ if $w = 2v$ for some $v$. This proves \eqref{equ:length-rhow}.
Also it is obvious that
\begin{equation}
\text{ for every $\rho \in \R_\cA(\xi)$ and $w \in \pos_\alpha(\xi)$, if $\rho\not\in \{\rho_0,\rho_w\}$, then $\wt(\xi,\rho) = -\infty$} \enspace. \label{equ:infty}
\end{equation}

  Then we can compute the run semantics of $\cA$ on $\xi$ as follows:
 \begin{align*}
  \runsem{\cA}(\xi) &= \max_{\rho \in \R_\cA(\xi)} \wt(\xi,\rho) + F_{\rho(\varepsilon)}
  = \max(\wt(\xi,\rho) +  F_{\rho(\varepsilon)}  \mid \rho \in \R_\cA(\xi))\\
  &= \max(\wt(\xi,\rho) +  F_\h  \mid \rho \in \R_\cA(\h,\xi)) \tag{\text{because $F_0=-\infty$}}\\
  &= \max(\wt(\xi,\rho_w) +  0  \mid w \in \pos_\alpha(\xi)) \tag{\text{by \eqref{equ:infty}}}\\
                    &= \max(|w|  \mid w \in \pos_\alpha(\xi)) \tag{\text{by \eqref{equ:length-rhow}}}\\
  &= \height(\xi) \tag{\text{by \eqref{equ:height=max}}}\enspace.
\end{align*}
 Hence $\runsem{\cA}=\height$, and thus $\height \in \Rec^{\mathrm{run}}(\Sigma,\Natmaxplus)$. The construction can easily be generalized to an arbitrary ranked alphabet $\Sigma$.
\hfill $\Box$
\end{example}

\

\index{pos@$\pos$}
\begin{example}\rm {\bf (Set of positions.)}\label{ex:set-of-positions} Let  $\Sigma = \{\sigma^{(2)}, \gamma^{(1)}, \alpha^{(0)}\}$.
  We consider the mapping
  \[\pos: \T_\Sigma \to \mathcal{P}(\mathbb{N}_+^*)
  \]
  defined on page \pageref{page:height-size-pos}.

  As weight algebra we use the semiring $\mathsf{Pos} = (\mathcal{P}(\mathbb{N_+}^*),\cup,\circ^R,\emptyset, \{\varepsilon\})$ where $U \circ^R V = V  U$ for every $U,V \in \mathcal{P}(\mathbb{N_+}^*)$. Thus, the mapping $\pos$
is a $(\Sigma,\mathsf{Pos})$-weighted tree language.

We construct the $(\Sigma,\mathsf{Pos})$-wta $\cA = (Q,\delta,F)$ which r-recognizes $\pos$, as follows.
\begin{compactitem}
\item $Q  = \{{e},{p}\}$ (intuitively, ${e}$ and ${p}$ calculate $\varepsilon \in \mathbb{N_+}^*$ and a path in $\mathbb{N_+}^*$, respectively),

\item for every $q_1,q_2,q \in Q$ we define $\delta_0(\varepsilon,\alpha,q) = \{\varepsilon\}$ and 
\[
\delta_1(q_1,\gamma,q) =
\left\{
\begin{array}{ll}
\{\varepsilon\} & \text{if } q_1={e}\\
\{1\} & \text{if } q_1=q={p}\\
\emptyset  & \text{otherwise }
\end{array}
\right.
\ \ \text{ and } \ \
\delta_2(q_1q_2,\sigma,q) =
\left\{
\begin{array}{ll}
\{\varepsilon\} & \text{if } q_1q_2={e}{e}\\
\{1\} & \text{if } q_1q_2q={p}{e} {p}\\
\{2\} & \text{if } q_1q_2q={e} {p} {p}\\
\emptyset  & \text{otherwise }
\end{array}
\right.
\]

\item $F_{e}=\emptyset$ and $F_{p}=\{\varepsilon\}$.
\end{compactitem}
In Figure \ref{fig:ex-positions} we represent $\cA$ as fta-hypergraph.  
Clearly, $\cA$ is root weight normalized; $\cA$  is not bu-deterministic, because e.g. $\delta_1({e},\gamma,{e}) = \delta_1({e},\gamma,{p})=\{\varepsilon\}$, which is different from $\emptyset$;
$\cA$ is not total, because there does not exist a state $q$ such that $\delta_2({p}{p},\sigma,q) \not=\emptyset$.

\begin{figure}
\begin{center}
\begin{tikzpicture}
\tikzset{node distance=7em, scale=0.6, transform shape}
\node[state, rectangle] (1) {\Large $\alpha$};
\node[state, right of=1] (2) {\Large ${e}$};
\node[state, rectangle, above of=2] (3){\Large $\gamma$};
\node[state, rectangle, below of=2] (4) {\Large $\sigma$};
\node[state, rectangle, above right = 0.5em and 13em of 2] (5){\Large $\sigma$};
\node[state, rectangle, below right = 0.5em and 13em of 2] (6){\Large $\gamma$};
\node[state, right of=2] (7)[right=18em]{\Large ${p}$};
\node[state, rectangle, above of=7] (8)[above=2em]{\Large $\sigma$};
\node[state, rectangle, below of=7] (9)[below=2em]{\Large $\sigma$};
\node[state, rectangle, right of=7] (10){\Large $\gamma$};
\node[state, rectangle, right of=7] (11)[below=4em]{\Large $\alpha$};

\tikzset{node distance=2em}
\node[above of=1] (w1) {\{$\varepsilon$\}};
\node[above of=3] (w3)[right=0.05cm] {\{$\varepsilon$\}};
\node[above of=4] (w4)[right=0.05cm] {\{$\varepsilon$\}};
\node[above of=5] (w5) {\{$\varepsilon$\}};
\node[above of=6] (w6) {\{$\varepsilon$\}};
\node[above of=7] (w7)[right=0.05cm] {\{$\varepsilon$\}};
\node[above of=8] (w8)[left=0.05cm] {\{2\}};
\node[above of=9] (w9)[left=0.05cm] {\{1\}};
\node[above of=10] (w10) {\{1\}};
\node[above of=11] (w11) {\{$\varepsilon$\}};

\draw[->,>=stealth] (1) edge (2);
\draw[->,>=stealth] (2) edge (3);
\draw[->,>=stealth] (3) edge[out=-250, in=-210, looseness=1.8] (2);
\draw[->,>=stealth] (2) edge (8);
\draw[->,>=stealth] (8) edge[out=-280, in=-340, looseness=1.4] (7);
\draw[->,>=stealth] (9) edge[out=-65, in=-55, looseness=1.4] (7);
\draw[->,>=stealth] (2) edge[out=365, in=155, looseness=0.3] (5);
\draw[->,>=stealth] (2) edge[out=360, in=205, looseness=0.3] (5);
\draw[->,>=stealth] (2) edge (9);
\draw[->,>=stealth] (4) edge (2);
\draw[->,>=stealth] (2) edge[out=230, in=245, looseness=1.8] (4);
\draw[->,>=stealth] (2) edge[out=315, in=290, looseness=1.8] (4);
\draw[->,>=stealth] (5) edge (7);
\draw[->,>=stealth] (2) edge (6);
\draw[->,>=stealth] (6) edge (7);
\draw[->,>=stealth] (7) edge (8);
\draw[->,>=stealth] (7) edge (9);
\draw[->,>=stealth] (7) edge (10);
\draw[->,>=stealth] (10) edge[out=-30, in=-20, looseness=1.7] (7);
\draw[->,>=stealth] (11) edge (7);
\end{tikzpicture}
\end{center}

\vspace{-10mm}

\caption{\label{fig:ex-positions} The $(\Sigma,\mathsf{Pos})$-wta $\cA = (Q,\delta,F)$ which r-recognizes $\pos$.}
\end{figure}

Let $\xi \in \T_\Sigma$. First we define particular runs on $\xi$ and show their weights. For each $u \in \pos(\xi)$, we define the run $\rho^\xi_u: \pos(\xi) \rightarrow Q$ for each $w \in \pos(\xi)$ by 
\[
\rho^\xi_u(w) =
\left\{
\begin{array}{ll}
{p} & \text{if $w$ is a prefix of $u$}\\
{e} & \text{otherwise.}
\end{array}
\right.
\]
(We recall that $\varepsilon$ and $u$ are prefixes of $u$.)
Also, we define the run $\rho^\xi_{{e}}: \pos(\xi) \rightarrow Q$ by $\rho^\xi_{{e}}(w)={e}$ for each $w \in \pos(\xi)$.

Then we claim that the following three statements hold for each $\xi \in \T_\Sigma$:
\begin{eqnarray}
\wt(\xi,\rho^\xi_{{e}}) = \{\varepsilon\} \label{equ:pos-run0}\\
\text{ for each $\rho \in \R_\cA(\xi) \setminus \Big(\{\rho^\xi_u \mid u \in \pos(\xi)\} \cup \{\rho^\xi_{{e}}\}\Big)$, we have } \wt(\xi,\rho) = \emptyset  \label{equ:pos-other-runs}\\
\text{ for each $u \in \pos(\xi)$, we have }\wt(\xi,\rho^\xi_u) = \{u\}\enspace.  \label{equ:pos-particular-run}
\end{eqnarray}
By direct inspection of the corresponding definitions we get the  proofs of Statements \eqref{equ:pos-run0} and \eqref{equ:pos-other-runs}.

Next we prove Statement \eqref{equ:pos-particular-run}. For this we fix $\xi \in \T_\Sigma$ and $u \in \pos(\xi)$. We define the binary relation $\succ_{\mathrm{d}}$ (\emph{d}irect postfix) on $\mathrm{postfix}(u)$ by letting $v_1 \succ_{\mathrm{d}} v_2$
if there exists an $i \in \mathbb{N}_+$ such that $v_1 = iv_2$.
By Corollary \ref{cor:reduction-to-substring-is-terminating}, the relation $\succ_{\mathrm{d}}$ is terminating. Moreover, we have that $\nf_{\succ_{\mathrm{d}}}(\mathrm{postfix}(u)) = \{\varepsilon\}$.  By induction on $(\mathrm{postfix}(u),\succ_{\mathrm{d}})$, we prove that the following statement holds (where $w$ is determined by $wv=u$.)
\begin{equation}
\text{For each $v \in \mathrm{postfix}(u)$, we have } \wt(\xi|_w,\rho^{\xi|_w}_v) = \{v\}  \label{equ:wfi-pos}\enspace.
\end{equation}

I.B.: Let $v=\varepsilon$. Then $w=u$ and $\rho^{\xi|_u}_\varepsilon: \pos(\xi|_u) \rightarrow Q$ maps $\varepsilon$ to ${p}$ and each other position to ${e}$. Let $\xi(u)=\kappa$ for some $\kappa \in \Sigma^{(k)}$ with $k \in\{0,1,2\}$. Then we calculate as follows (abbreviating $\rho^{\xi|_u}_{{e}}$ by $\nu$):
\begin{align*}
  \wt(\xi|_u,\nu) &= \wt(\xi|_{u1},\nu|_1) \circ^R \ldots \circ^R \wt(\xi|_{uk},\nu|_k) \circ^R \delta_k(\underbrace{{e}\cdots {e}}_{k},\kappa,{p}) \\
                  &= \{\varepsilon\} \circ^R \ldots \circ^R \{\varepsilon\} \circ^R \{\varepsilon\}
                    \tag{by the fact that $\nu|_i = (\rho^{\xi|_u}_{{e}})|_i = \rho_{{e}}^{\xi|_{ui}}$ for each $i \in [k]$, by \eqref{equ:pos-run0}, and by definition of $\delta$}\\
  &=\{\varepsilon\} \enspace.
\end{align*}

I.S.: Let $v \not= \varepsilon$.  Let $\xi(w)=\kappa$ for some $\kappa \in \Sigma^{(k)}$ with $k \in\{1,2\}$. Then there exists a unique $v' \in \mathrm{postfix}(u)$ with $v \succ_{\mathrm{d}} v'$. 

Let $i \in [k]$ be such that $v=iv'$,  then $\wt(\xi|_{wi},\rho^{\xi|_{wi}}_{v'})=\{v'\}$ by I.H.
Then we can calculate as follows (abbreviating $\rho^{\xi|_{w}}_{v}$ by $\nu$):
\begingroup
\allowdisplaybreaks
\begin{align*}
\wt(\xi|_{w},\nu)
=& \ \wt(\xi_{w1},\nu|_1) \circ^R \ldots \circ^R \wt(\xi_{w(i-1)},\nu|_{i-1}) \circ^R \wt(\xi_{wi},\nu|_i)\\
&\qquad  \circ^R \wt(\xi_{w(i+1)},\nu|_{i+1}) \circ^R \ldots \circ^R \wt(\xi_{wk},\nu|_k) \circ^R \delta_k(\underbrace{{e}\cdots {e}}_{i-1} {p} \underbrace{{e}\cdots {e}}_{k -i}, \kappa,{p})\\
  =& \  \wt(\xi_{wi},\nu|_i)  \circ^R \delta_k(\underbrace{{e}\cdots {e}}_{i-1} {p} \underbrace{{e}\cdots {e}}_{k -i}, \kappa,{p})
     \tag{\text{because $\nu|_j = \rho^{\xi|_{wj}}_{{e}}$ for each $j\not=i$; thus $\wt(\xi_{wj},\nu|_j)=\{\varepsilon\}$ by Statement \eqref{equ:pos-run0}}}\\[2mm]
=&  \{v'\}  \circ^R \delta_k(\underbrace{{e}\cdots {e}}_{i-1} {p} \underbrace{{e}\cdots {e}}_{k -i}, \kappa,{p})
 \tag{\text{because $\nu|_i = \rho^{\xi|_{wi}}_{v'}$; by I.H.  $\wt(\xi_{wi},\nu|_i)=\{v'\}$}}\\[2mm]
  =&  \{v'\}  \circ^R \{i\}
     \tag{\text{by definition of $\delta_k(\underbrace{{e}\cdots {e}}_{i-1} {p} \underbrace{{e}\cdots {e}}_{k -i}, \kappa,{p})$}}
= \ \{iv'\} = \{v\}\enspace.
\end{align*}
\endgroup
This finishes the proof of Statement \eqref{equ:wfi-pos}. Choosing $v=u$ in this statement proves Statement \eqref{equ:pos-particular-run}.

Now, for each $\xi \in \T_\Sigma$, we can prove that $\runsem{\cA}(\xi)=\pos(\xi)$ as follows.
\begingroup
\allowdisplaybreaks
\begin{align*}
&\runsem{\cA}(\xi)
= \bigcup (\wt(\xi,\rho) \circ^R F_{\rho(\varepsilon)} \mid \rho \in \R_\cA(\xi)) \\
=& \bigcup (\wt(\xi,\rho) \circ^R F_{\rho(\varepsilon)} \mid \rho \in \{\rho^\xi_u \mid u \in \pos(\xi)\} \cup \{\rho^\xi_{{e}}\} )
\ \  \tag{\text{by \eqref{equ:pos-other-runs}}}\\[2mm]
  =& \bigcup (\wt(\xi,\rho) \circ^R F_{\rho(\varepsilon)} \mid \rho \in \{\rho^\xi_u \mid u \in \pos(\xi)\})
\ \  \tag{\text{because $\rho^\xi_{{e}}(\varepsilon)={e}$ and $F_{e}=\emptyset$}}\\[2mm]
=& \bigcup (\wt(\xi,\rho) \mid \rho \in \{\rho^\xi_u \mid u \in \pos(\xi)\})
\ \ \tag{\text{because $\rho^\xi_u(\varepsilon)={p}$ for each $u \in \pos(\xi)$ and $F_{p}=\{\varepsilon\}$}}\\[2mm]
=& \bigcup (\wt(\xi,\rho^\xi_u)  \mid  u \in \pos(\xi))\\
=& \bigcup (\{u\} \mid u \in \pos(\xi))
\ \  \text{(by \eqref{equ:pos-particular-run})}\\
=& \ \pos(\xi)\enspace.
\end{align*}
\endgroup
Hence $\runsem{\cA} = \pos$ and thus $\pos \in \Rec^{\mathrm{run}}(\Sigma,\mathsf{Pos})$.
\hfill $\Box$
\end{example}

\

\index{yield@$\mathrm{yield}^\cP$}
\begin{example}\rm \label{ex:yield}  {\bf (Yield of trees.)} Let $\Sigma = \{\sigma^{(2)}, \alpha^{(0)}, \beta^{(0)}\}$.
  We consider the mapping   
  \[\yield^\cP: \T_\Sigma \to \cP((\Sigma^{(0)})^*)\]
defined, for each $\xi \in \T_\Sigma$, by $\yield^\cP(\xi) = \{\yield(\xi)\}$. For  the definition of $\yield(\xi)$ we refer to Section~\ref{sect:trees}.  For instance, $\yield^\cP(\sigma(\alpha,\sigma(\alpha,\beta)) = \{\alpha \alpha \beta\}$. 
 
 As weight algebra we use the formal language language semiring $\Lang_{\Sigma} = (\mathcal{P}(\Sigma^*),\cup, \cdot, \emptyset, \{\varepsilon\})$. Thus, the mapping $\mathrm{yield}$ is a $(\Sigma,\Lang_{\Sigma})$-weighted tree language.
 
We define the $(\Sigma,\Lang_{\Sigma})$-wta $\cA=(Q,\delta,F)$ which i-recognizes $\yield^\cP$, as follows.
\begin{compactitem}
\item $Q = \{{y}\}$ (intuitively, ${y}$ calculates the yield),
\item $\delta_0(\varepsilon,\alpha,{y}) = \{\alpha\}$, $\delta_0(\varepsilon,\beta,{y}) = \{\beta\}$, and 
$\delta_2({y}{y},\sigma,{y}) = \{\varepsilon\}$,
\item $F_{y} =\{\varepsilon\}$.
\end{compactitem}
In Figure \ref{fig:yield-ex} we represent $\cA$ as fta-hypergraph.
Clearly, $\cA$ is bu-deterministic, total, and root weight normalized; $\cA$ is not crisp-deterministic because, e.g., $\delta_0(\varepsilon,\alpha,{y}) \not\in \{\emptyset, \{\varepsilon\}\}$.

\begin{figure}[t]
  \begin{center}
\begin{tikzpicture}
\tikzset{node distance=7em, scale=0.6, transform shape}
\node[state] (1) {\Large $y$};
\node[state, rectangle, above left of=1] (2) [below=0.3em] {\Large $\alpha$};
\node[state, rectangle, below left of=1] (3) [above=0.3em] {\Large $\beta$};
\node[state, rectangle, right of=1] (4) {\Large $\sigma$};

\tikzset{node distance=2em}
\node[above of=1] (w1) {\{$\varepsilon$\}};
\node[above of=2] (w2) {\{$\alpha$\}};
\node[above of=3] (w3) {\{$\beta$\}};
\node[above of=4] (w4) {\{$\varepsilon$\}};

\draw[->,>=stealth] (2) edge (1);
\draw[->,>=stealth] (3) edge (1);
\draw[->,>=stealth] (4) edge (1);
\draw[->,>=stealth] (1) edge[out=60, in=30, looseness=1.4] (4);
\draw[->,>=stealth] (1) edge[out=-60, in=-30, looseness=1.4] (4);
\end{tikzpicture}
\end{center}
\caption{\label{fig:yield-ex} The $(\Sigma,\Lang_{\Sigma})$-wta $\cA=(Q,\delta,F)$ which i-recognizes $\yield^\cP$.}
\end{figure}

By induction on $\T_\Sigma$, we prove that the following statement holds:
\begin{equation*}
\text{For each $\xi \in \T_\Sigma$, we have } \h_\cA(\xi)_{y} = \{\yield(\xi)\} \enspace. 
\end{equation*}

I.B.: Let $\xi=\alpha$. Then $\h_\cA(\xi)_{y} = \delta_0(\varepsilon,\alpha,{y}) = \{\alpha\} = \{\yield(\alpha)\}$. For  $\xi=\beta$ the proof is similar.

I.S.: Now let $\xi = \sigma(\xi_1,\xi_2)$ and assume that $\h_\cA(\xi_i)_{y}=\{\yield(\xi_i)\}$ for each $i \in \{1,2\}$.  Then 
\[\h_\cA(\xi)_{y} = \h_\cA(\xi_1)_{y} \ \h_\cA(\xi_2)_{y} \  \delta_2({y}{y},\sigma,{y}) = \{\yield(\xi_1)\} \{\yield(\xi_2)\} \{\varepsilon\}= \{\yield(\xi)\}\enspace.
\]
Then we have $\initialsem{\cA}(\xi) = \h_\cA(\xi)_{y} \ F_{y} = \{\yield(\xi)\} = \yield^\cP(\xi)$. Thus $\yield^\cP \in \budRec^{\mathrm{init}}(\Sigma,\Lang_{\Sigma})$.
\hfill $\Box$
\end{example}

\

\index{transformation monoid}
\begin{example}\rm \label{ex:transformation-monoid} {\bf (Transformation monoid.)} We recall from Section \ref{sec:fsa} that a $\Gamma$-fsa is a tuple $A=(Q,I,\delta,F)$ where $Q$ is the finite set of states, $I \subseteq Q$ and $F \subseteq Q$ are the sets of initial states and final states, respectively, and $\delta\subseteq Q \times \Gamma \times Q$ is the set of transitions. The language $\LL(A)$ recognized by $A$ is defined in terms of runs.

  Alternatively, we wish to describe $\LL(A)$ by using the mapping
  \[
    \bar{\delta}: \Gamma^* \to (\cP(Q) \to \cP(Q)) \enspace.
  \]
as follows.  Intuitively, for every $w \in \Gamma^*$ and $U \subseteq Q$, the set $\bar{\delta}(w)(U)$  is the set of all states which $A$ can enter when starting in some state of $U$ and reading the string $w$.
  \index{succ@$\succ$}
  Formally, we define $\bar{\delta}$ by induction on $(\Gamma^*,\succ)$ where, for every $w_1,w_2 \in \Gamma^*$, we let $w_1 \succ w_2$ if there exists an $a \in \Gamma$ such that $w_1 = w_2a$.
  By Corollary \ref{cor:reduction-to-substring-is-terminating}, the relation $\succ$ is terminating. Moreover, we have that $\succ$ is terminating and $\nf_\succ(\Gamma^*) = \{\varepsilon\}$.
  For every $a \in \Gamma$, $w \in \Gamma^*$, and $U \subseteq Q$, we define
  \[\bar{\delta}(\varepsilon)(U) = U \ \ \text{ and } \ \ \bar{\delta}(wa)(U) = \{p \in Q \mid (\exists r \in \bar{\delta}(w)(U)): (r,a,p) \in \delta\}\enspace.
  \]
 Then it easy to see that  $\LL(A) = \{w \in \Gamma^* \mid \bar{\delta}(w)(I) \cap F \not= \emptyset\}$.

 Let $\Sigma = \{\sigma^{(2)}, \alpha^{(0)}, \beta^{(0)}\}$ and let $A = (Q,I,\delta,F)$ be a $\Gamma$-fsa with $\Gamma = \{\alpha,\beta\}$.
  We define  the mapping
  \begin{align*}
    \mathrm{BPS}_A: \T_\Sigma \to (\cP(Q) \to \cP(Q)) \ \ \text{ with } \ \
    \mathrm{BPS}_A(\xi) =  \bar{\delta}(\yield(\xi)) \ \text{ for each $\xi \in \T_\Sigma$}
    \end{align*}
    where $\yield$ is defined in Section \ref{sect:trees}. Eventually, we will prove that $\mathrm{BPS}_A$ is r-recognizable over a suitable strong bimonoid.
    
  The name ``BPS'' of the mapping refers to Bar-Hillel, Perles, and Shamir \cite{barpersha61}; in that paper they have used a technique to fold the computation of a finite-state automaton onto a derivation tree of a context-free grammar. They used this technique for the proof of the fact that the set of context-free languages is closed under intersection with the set of regular languages. We note that the set of context-free languages and the set of yields of recognizable tree languages are equal, cf. \cite[Thm.~3.20]{bra69} and \cite[Thm.~2.5]{don70} (also cf. Corollary \ref{cor:cfg=yield(fta)-JGC}  and  \cite[Cor.~3.2.4 and Thm.~3.2.5]{gecste84}). Here we illustrate the BPS-technique for the yield of the particular recognizable tree language $\T_\Sigma$ (also cf. Section \ref{sect:yield-of-weighted-cf}). 
    
  We consider the near semiring $\NearSem_{\cP(Q)} = (B,\cup,\diamond,\widetilde{\emptyset},\id)$ over the commutative monoid
    $(\cP(Q),\cup,\emptyset)$ as defined in Example~\ref{ex:strong-bimonoids}(\ref{ex:near-semiring}), where $\id$ abbreviates $\id_{\cP(Q)}$. We recall that $B= \{f \mid f: \cP(Q) \to \cP(Q), f(\emptyset)=\emptyset\}$.

    We note that $(B,\diamond,\id)$ is a monoid. Moreover, it is easy to verify that $\bar{\delta}$ is a monoid homomorphism from $(\Gamma^*,\cdot,\varepsilon)$ to $(B,\diamond,\id)$, i.e., $\bar{\delta}(\varepsilon) = \id$ and,
       \begin{equation}
\text{ for every $v,w \in \Gamma^*$, we have $\bar{\delta}(wv)= \bar{\delta}(w) \diamond \bar{\delta}(v)$. }  \label{eq:bar-delta-mon-hom}
\end{equation}
By induction on $(\Gamma^*,\succ)$, we can easily prove \eqref{eq:bar-delta-mon-hom}.

    Now we construct the $(\Sigma,\NearSem_{\cP(Q)})$-wta $\cA=(\{*\},\delta',F)$ which r-recognizes $\mathrm{BPS}_A$ as follows:
\begin{compactitem}
\item $(\delta')_0(\varepsilon,\alpha,*) = \bar{\delta}(\alpha)$, $(\delta')_0(\varepsilon,\beta,*) = \bar{\delta}(\beta)$, and $(\delta')_2(**,\sigma,*)=\id$ and 
\item $F_*=\id$.
\end{compactitem}
In fact, $\cA$ is bu-deterministic and root weight normalized, and in general $\cA$ is not crisp-deterministic. 

Let $\xi \in \T_\Sigma$ and $\rho_\xi: \pos(\xi) \to Q$ be the run which maps each position of $\xi$ to $*$.
By induction on $\T_\Sigma$, we prove that the following statement holds: 
\begin{equation}
\text{For each $\xi \in \T_\Sigma$, we have }  \wt_\cA(\xi,\rho_\xi) = \bar{\delta}(\yield(\xi)) \enspace. \label{equ:run-BPS}
  \end{equation}

  Let $\xi= \alpha$. Then $\wt_\cA(\xi,\rho_\xi) = (\delta')_0(\varepsilon,\alpha,*) = \bar{\delta}(\alpha) = \bar{\delta}(\yield(\xi)).$
  In the same way we obtain \eqref{equ:run-BPS} for $\xi = \beta$. Now let $\xi=\sigma(\xi_1,\xi_2)$ and assume that \eqref{equ:run-BPS} holds for $\xi_1$ and $\xi_2$. Then
  \begin{align*}
    \wt_\cA(\xi,\rho_\xi) &= \wt_\cA(\xi_1,\rho_{\xi_1}) \diamond \wt_\cA(\xi_2,\rho_{\xi_2}) \diamond (\delta')_2(**,\sigma,*)\\
    &= \bar{\delta}(\yield(\xi_1))  \diamond \bar{\delta}(\yield(\xi_2)) \diamond \id    \tag{by I.H. and construction}\\
                      &= \bar{\delta}(\yield(\xi_1))  \diamond \bar{\delta}(\yield(\xi_2))  \tag{because $\id$ is the identity}\\
                      &= \bar{\delta}(\yield(\xi_1)\ \yield(\xi_2))   \tag{because $\bar{\delta}$ is a monoid homomorphism}\\
                      &= \bar{\delta}(\yield(\sigma(\xi_1,\xi_2))\enspace.
    \end{align*}
        Finally, for each $\xi \in \T_\Sigma$ we have
    \[\runsem{\cA}(\xi) = \bigcup_{\rho \in \R_\cA(\xi)} \wt_\cA(\xi,\rho) = \wt_\cA(\xi,\rho_\xi) = \bar{\delta}(\yield(\xi)) = \mathrm{BPS}_A(\xi)\enspace.\]
    Thus $\mathrm{BPS}_A \in \budRec^{\mathrm{run}}(\Sigma,\NearSem_{\cP(Q)})$.
\hfill $\Box$
\end{example}

\

\begin{example}\rm  
  {\bf (Exponentiation.)}
  \index{exp}
  \label{ex:run-bimonoid} We consider the string ranked alphabet $\Sigma = \{\gamma^{(1)}, \alpha^{(0)}\}$ and the mapping
  \begin{align*}
    \mathrm{exp}: \T_\Sigma \to \mathbb{N} \ \ \text{ with } \ \ 
    \mathrm{exp}(\xi) = 2^{n+1} \ \text{ if $\xi = \gamma^n(\alpha)$ for some $n \in \mathbb{N}$} \enspace.
  \end{align*}
As weight algebra we consider the tropical bimonoid  $\TropBM=(\mathbb{N}_\infty,+,\min,0,\infty)$ from Example~\ref{ex:strong-bimonoids}(\ref{ex:tropical-bimonoid}). Thus $ \mathrm{exp}$ is a $(\Sigma,\TropBM)$-weighted tree language.

    We construct the $(\Sigma,\TropBM)$-wta $\cA =(Q,\delta,F)$ which r-recognizes $\mathrm{exp}$, as follows.  
\begin{compactitem}
\item $Q = \{q_0,q_1\}$,
\item $\delta_0(\varepsilon,\alpha,p) = \delta_1(p,\gamma,q)= 1$ for every $p,q \in Q$, and
\item $F_{q_0}= F_{q_1}=1$.
\end{compactitem}
In Figure \ref{fig:ex-exp-run} we show the wta $\cA$. Obviously, $\cA$ is not bu-deterministic.

\begin{figure}[t]
  \begin{center}

\begin{tikzpicture}
\tikzset{node distance=7em, scale=0.6, transform shape}
\node[state, rectangle] (1) {\Large $\alpha$};
\node[state, right of=1] (2) {\Large $q_0$};
\node[state, rectangle, above of=2] (3) {\Large $\gamma$};
\node[state, rectangle, above right=1em and 8em of 2] (4) {\Large $\gamma$};
\node[state, rectangle, below right=1em and 8em of 2] (5) {\Large $\gamma$};
\node[state, right of=2] (6)[right=12em] {\Large $q_1$};
\node[state, rectangle, above of=6] (7) {\Large $\gamma$};
\node[state, rectangle, right of=6] (8) {\Large $\alpha$};

\tikzset{node distance=2em}
\node[above of=1] (w1) {1};
\node[above of=2] (w2) [right=0.05cm] {1};
\node[above of=3] (w3) {1};
\node[above of=4] (w4) {1};
\node[above of=5] (w5) {1};
\node[above of=6] (w6)[left=0.05cm]  {1};
\node[above of=7] (w7) {1};
\node[above of=8] (w8) {1};

\draw[->,>=stealth] (1) edge (2);
\draw[->,>=stealth] (2) edge (3);
\draw[->,>=stealth] (2) edge (4);
\draw[->,>=stealth] (3) edge[out=-250, in=-210, looseness=1.8] (2);
\draw[->,>=stealth] (4) edge (6);
\draw[->,>=stealth] (5) edge (2);
\draw[->,>=stealth] (6) edge (5);
\draw[->,>=stealth] (6) edge (7);
\draw[->,>=stealth] (7) edge[out=-290, in=30, looseness=1.8] (6);
\draw[->,>=stealth] (8) edge (6);
\end{tikzpicture} 
  \end{center}
  \caption{\label{fig:ex-exp-run} The $(\Sigma,\TropBM)$-wta $\cA$ which r-recognizes $\mathrm{exp}$.}
  \end{figure}

Now let $n \in \mathbb{N}$. We compute $\runsem{\cA}(\gamma^n(\alpha))$. It is easy to see that $\wt(\gamma^n(\alpha),\rho)=1$ for each run $\rho \in \R_\cA(\gamma^n(\alpha))$.  Then for each $n \in \mathbb{N}$ we have 
\[
\runsem{\cA}(\gamma^n(\alpha)) = \hspace{-2mm} \bigplus_{\rho \in \R_\cA(\gamma^n(\alpha))}\hspace{-2mm} {\min(\wt(\gamma^n(\alpha),\rho), \ F_{\rho(\varepsilon)})} = \hspace{-2mm} \bigplus_{\rho \in \R_\cA(\gamma^n(\alpha))}\hspace{-2mm} {1} = |\R_\cA(\gamma^n(\alpha))| = 2^{n+1} =  \mathrm{exp}(\gamma^n(\alpha))\enspace,
\]
where the last but one equality is due to the fact that there exists a bijection between the two finite sets $\R_\cA(\gamma^n(\alpha))$ and $\{q_0,q_1\}^{n+1}$. Hence $\runsem{\cA} = \mathrm{exp}$ and thus  $\mathrm{exp} \in \Rec^{\mathrm{run}}(\Sigma,\TropBM)$.
\hfill $\Box$
\end{example}

\

\begin{example}{\bf (Exponentiation plus one.)} \rm \label{ex:exp+1}
  Let $\Sigma = \{\gamma^{(1)}, \alpha^{(0)}\}$ be a string ranked alphabet.  We define the mapping  
  \begin{align*}
    (\exp +1): \T_\Sigma \rightarrow \mathbb{N} \ \ \text{ with } \ \
    (\exp +1)(\gamma^n(\alpha)) =  2^n +1 \ \text{ for each $n \in \mathbb{N}$} \enspace.
                           \end{align*}

As weight algebra we consider the semiring $\Nat=(\mathbb{N},+,\cdot,0,1)$. Thus, the mapping $\exp +1$ is a $(\Sigma,\Nat)$-weighted tree language.

We construct the $(\Sigma,\Nat)$-wta $\cA = (Q,\delta,F)$ which i-recognizes $(\exp +1)$, as follows: 
\begin{compactitem}
\item $Q  = \{1,{e}\}$ (intuitively, $1$ and ${e}$ calculate $1 \in \mathbb{N}$ and $2^n$, respectively),

\item $\delta_0(\varepsilon,\alpha,1) = \delta_0(\varepsilon,\alpha,{e}) = 1$ and for every $q_1,q\in Q$, 
\[
\delta_1(q_1,\gamma,q) = 
\begin{cases}
1&\text{ if $q_1=q = 1$}\\
2&\text{ if $q_1=q = e$}\\
0&\text{ otherwise\;.}
\end{cases}
\]

\item $F_1 = F_{e} = 1$.
\end{compactitem}
In Figure \ref{fig:ex-exponential} we represent $\cA$ as fta-hypergraph.
\begin{figure}[t]
\begin{center}
\begin{tikzpicture}
\tikzset{node distance=7em, scale=0.6, transform shape}
\node[state, rectangle] (1) {\Large $\alpha$};
\node[state, right of=1] (2) {\Large 1};
\node[state, rectangle, above of=2] (3){\Large $\gamma$};
\node[state, rectangle, right of=2] (4) [right=2em] {\Large $\alpha$};
\node[state, right of=4] (5) {\Large $e$};
\node[state, rectangle, above of=5] (6){\Large $\gamma$};

\tikzset{node distance=2em}
\node[above of=1] (w1) {1};
\node[above of=2] (w2) [right=0.05cm] {1};
\node[above of=3] (w3) {1};
\node[above of=4] (w4) {1};
\node[above of=5] (w5) [right=0.05cm] {1};
\node[above of=6] (w6) {2};

\draw[->,>=stealth] (1) edge (2);
\draw[->,>=stealth] (2) edge (3);
\draw[->,>=stealth] (3) edge[out=-250, in=-210, looseness=1.8] (2);
\draw[->,>=stealth] (4) edge (5);
\draw[->,>=stealth] (5) edge (6);
\draw[->,>=stealth] (6) edge[out=-250, in=-210, looseness=1.8] (5);
\end{tikzpicture}
\end{center}
\caption{\label{fig:ex-exponential}
  The $(\Sigma,\Nat)$-wta $\cA = (Q,\delta,F)$ which i-recognizes $(\exp +1)$.}
\end{figure}
Clearly, $\cA$ is total and has unit root weights; $\cA$ is not bu-deterministic and it is not root weight normalized.

We can prove easily that:
\begin{equation*}
\text{$\h_\cA(\gamma^n(\alpha))_{1} = 1$ and $\h_\cA(\gamma^n(\alpha))_{{e}} = 2^n$  for each $n \in \mathbb{N}$.}
\end{equation*}
Then $\initialsem{\cA}(\gamma^n(\alpha)) = 2^n+1 = (\exp +1)(\gamma^n(\alpha))$ for each $n \in \mathbb{N}$. 
Thus   $(\exp +1) \in \Rec^\mathrm{init}(\Sigma,\Nat)$.
\hfill $\Box$
\end{example}

\

\begin{example} \rm \label{ex:wta-F2-mod2} {\bf (Number of $\gamma$'s modulo 2)}  Let $\Sigma = \{\gamma^{(1)}, \alpha^{(0)}\}$ be a string ranked alphabet.  We define the mapping  
  \begin{align*}
    \mathrm{odd}: \T_\Sigma \rightarrow \{0,1\} \ \ \text{ with } \ \
    \mathrm{odd}(\gamma^n(\alpha)) =  \begin{cases} 1 & \text{ if $n$ is odd}\\
      0 &  \text{ otherwise} 
      \end{cases} \  \ \text{for each $n \in \mathbb{N}$.}
                           \end{align*}

                           As weight algebra we consider the field $\sfFtwo= (\{0,1\},\oplus,\otimes,0,1)$ (cf. Example \ref{ex:strong-bimonoids}(\ref{ex:sb-not-sr})). Thus, the mapping $\mathrm{odd}$ is a $(\Sigma,\sfFtwo)$-weighted tree language.

We construct the $(\Sigma,\sfFtwo)$-wta $\cA = (Q,\delta,F)$ which r-recognizes $\mathrm{odd}$, as follows.
\begin{compactitem}
\item $Q  = \{q_0,q_1\}$ (intuitively, on each input tree $\gamma^{n}(\alpha)$, the wta $\cA$ starts at the leaf in state $q_0$ or $q_1$ and,  at any of the occurrences of $\gamma$,  nondeterministically either stays in $q_0$ or  switches to $q_1$ and remains there),
 
\item for every $p_1,p_2 \in Q$ we define
\[
\delta_0(\varepsilon,\alpha,q_0) = 1  \ \text{ and } \ \delta_0(\varepsilon,\alpha,q_1) = 0
\ \text{ and } \ 
\delta_1(p_1,\gamma,p_2) = 
\left\{
\begin{array}{ll}
1 & \text{if } p_1p_2 \in \{q_0q_0, q_0q_1, q_1q_1\}\\
0 & \text{otherwise}
\end{array}
\right.
\]

\item $F_{q_0}=0$ and $F_{q_1}=1$.
\end{compactitem}
In Figure \ref{fig:odd-F2} we show the $(\Sigma,\sfFtwo)$-wta which r-recognizes $\mathrm{odd}$. Clearly, $\cA$ is crisp and root weight normalized;  $\cA$ is not bu-deterministic.

\begin{figure}[t]
  \begin{center}
\begin{tikzpicture}
\tikzset{node distance=7em, scale=0.6, transform shape}
\node[state, rectangle] (1) {\Large $\alpha$};
\node[state, right of=1] (2) {\Large $q_0$};
\node[state, rectangle, above of=2] (3) {\Large $\gamma$};
\node[state, rectangle, right of=2] (4) {\Large $\gamma$};
\node[state, right of=4] (6) {\Large $q_1$};
\node[state, rectangle, above of=6] (7) {\Large $\gamma$};

\tikzset{node distance=2em}
\node[above of=1] (w1) {1};
\node[above of=3] (w3) {1};
\node[above of=4] (w4) {1};
\node[above of=6] (w6)[left=0.05cm]  {1};
\node[above of=7] (w7) {1};

\draw[->,>=stealth] (1) edge (2);
\draw[->,>=stealth] (2) edge (3);
\draw[->,>=stealth] (2) edge (4);
\draw[->,>=stealth] (3) edge[out=-250, in=-210, looseness=1.8] (2);
\draw[->,>=stealth] (4) edge (6);
\draw[->,>=stealth] (6) edge (7);
\draw[->,>=stealth] (7) edge[out=-290, in=30, looseness=1.8] (6);
\end{tikzpicture} 
  \end{center}
  \caption{\label{fig:odd-F2} The $(\Sigma,\sfFtwo)$-wta $\cA$ which r-recognizes $\mathrm{odd}$.}
  \end{figure}

Let $n \in \mathbb{N}$ and let $\xi = \gamma^n(\alpha)$. Then
\begingroup
\allowdisplaybreaks
\begin{align*}
  \runsem{\cA}(\xi) &= \bigoplus_{\rho \in \R_\cA(\xi)} \wt(\xi,\rho) \otimes F_{\rho(\varepsilon)} 
                                 = \bigoplus_{\rho \in \R_\cA(q_1,\xi)} \wt(\xi,\rho)
                                   = \bigoplus_{\rho \in [n]} 1
                                   = \mathrm{odd}(\xi) \enspace.
\end{align*}
\endgroup

Here we use the additive monoid $(\{0,1\},\oplus,0)$ of $\sfFtwo$ to count modulo 2. At the same time, for each strong bimonoid $\B=(B,\oplus,\otimes,\0,\1)$, we can easily design a crisp-deterministic $(\Sigma,\B)$-wta with two states and unit root weights such that it simulates counting modulo 2 in its state behaviour. Hence, for each strong bimonoid $\B$, the corresponding mapping $\mathrm{odd}_\B: \T_\Sigma \rightarrow \{\0,\1\}$ is r-recognizable by a crisp-deterministic $(\Sigma,\B)$-wta with unit root weights. We note that we use  the state behaviour of a wta for ``counting'' also in Example \ref{ex:recog-step-mapping}.
\hfill $\Box$
  \end{example}

   \begin{example}\rm \label{ex:bu-det+total-wta-det-weo} {\bf (Weighted even-odd.)} Let $\Sigma = \{\sigma^{(2)}, \alpha^{(0)}\}$ be a ranked alphabet. We define the mapping
    \[
      \mathrm{weo}: \T_\Sigma \to \mathbb{Q} \ \ \ \text{ with } \ \ \ \mathrm{weo}(\xi) = \begin{cases} 2 \cdot 2^{\#_\alpha(\xi)} & \text{ if $\#_\alpha(\xi)$ is even}\\
       3 \cdot 2^{\#_\alpha(\xi)} & \text{ otherwise} \end{cases} \ \ \text{ for each $\xi \in \T_\Sigma$}
   \]
      where $\#_\alpha(\xi)$ denotes the number of occurrences of $\alpha$ in $\xi$. The name ``$\mathrm{weo}$'' might be read as ``weighted even-odd''.

      As weight algebra we consider the field $\Ratnum = (\mathbb{Q},+,\cdot,0,1)$ of rational numbers. Thus, $\mathrm{weo}$ is a $(\Sigma,\Ratnum)$-weighted tree language.
      
      Next we define the $(\Sigma,\Ratnum)$-wta $\cA=(Q,\delta,F)$ which i-recognizes $\mathrm{weo}$, as follows. 
    \begin{compactitem}
    \item $Q = \{e,o\}$, where $e$ and $o$ are standing for {\em e}ven and {\em o}dd, respectively,
    \item $\delta_0(\varepsilon,\alpha,o) = 2$,  $\delta_0(\varepsilon,\alpha,e)=0$ and
      \[
        \delta_2(q_1q_2,\sigma,o) = \begin{cases} 1 & \text{ if $q_1\not= q_2$}\\
          0 & \text{ otherwise }
        \end{cases}
       \ \ \text{ and } \ \
         \delta_2(q_1q_2,\sigma,e) = \begin{cases} 1 & \text{ if $q_1= q_2$}\\
          0 & \text{ otherwise }\enspace,
        \end{cases}
      \]
      \item $F_o=3$ and $F_e=2$.
      \end{compactitem}
 In Figure \ref{fig:even-odd-two-three-det} we show $\cA$ as hypergraph. Clearly, $\cA$ is total and bu-deterministic; it is not crisp, because $\delta_0(\varepsilon,\alpha,o) = 2$.  Moreover, it does not have unit root weights.

 \begin{figure}[t]
   \begin{center}
\begin{tikzpicture}
\tikzset{node distance=7em, scale=0.5, transform shape}
\node[state, rectangle] (1) {\Large $\alpha$};
\node[state, right of=1] (2){\Large $o$};
\node[state, rectangle, right of=2] (3)[right=1em]{\Large $\sigma$};
\node[state, rectangle, above of=3] (4)[above=1em]{\Large $\sigma$};
\node[state, rectangle, below of=3] (5)[below=1em]{\Large $\sigma$};
\node[state, right of=3] (6)[right=1em]{\Large $e$};
\node[state, rectangle, right of=6] (7) {\Large $\sigma$};

\tikzset{node distance=2em}
\node[above of=1] (w1)[yshift=0.7em] {\Large 2};
\node[above of=2] (w2)[left=0.1em, yshift=0.7em] {\Large 3};
\node[above of=3] (w3)[yshift=0.7em] {\Large 1};
\node[above of=4] (w4)[yshift=0.7em] {\Large 1};
\node[above of=5] (w5)[yshift=0.7em] {\Large 1};
\node[above of=6] (w6)[yshift=0.7em] {\Large 2};
\node[above of=7] (w7)[yshift=0.7em] {\Large 1};

\draw[->,>=stealth] (1) edge (2);
\draw[->,>=stealth] (2) edge[out=20, in=155, looseness=1.1] (3);
\draw[->,>=stealth] (2) edge[out=-20, in=205, looseness=1.1] (3);
\draw[->,>=stealth] (2) edge (4);
\draw[->,>=stealth] (2) edge (5);
\draw[->,>=stealth] (3) edge (6);
\draw[->,>=stealth] (6) edge (4);
\draw[->,>=stealth] (4) edge[out=110, in=80, looseness=1.4] (2);
\draw[->,>=stealth] (6) edge (5);
\draw[->,>=stealth] (5) edge[out=250, in=-80, looseness=1.4] (2);
\draw[->,>=stealth] (7) edge (6);
\draw[->,>=stealth] (6) edge[out=60, in=30, looseness=2.7] (7);
\draw[->,>=stealth] (6) edge[out=-60, in=-30, looseness=2.7] (7);
\end{tikzpicture} 
\caption{\label{fig:even-odd-two-three-det} The $(\Sigma,\Ratnum)$-wta $\cA=(Q,\delta,F)$ which i-recognizes $\mathrm{weo}$.}
   \end{center}
   \end{figure}
     
      By induction on $\T_\Sigma$, we prove that
      \begin{equation}\label{equ:number-of-alphas-det}
        \begin{aligned}
          &\text{for every $\xi \in \T_\Sigma$ and $q \in Q$, we have:}\\
          &\text{$\h_\cA(\xi)_q = \begin{cases} 2^{\#_\alpha(\xi)} & \text{ if \big($\#_\alpha(\xi)$ is even and $q=e$\big) or
                \big($\#_\alpha(\xi)$ is odd and $q=o$\big)  } \\ 0 & \text{ otherwise} \end{cases}$}
          \end{aligned}
        \end{equation}

        I.B.: Let $\xi = \alpha$. Then $\h_\cA(\xi)_o = \delta_0(\varepsilon,\alpha,o)= 2 = 2^1 = 2^{\#_\alpha(\xi)}$. Moreover, $\h_\cA(\xi)_e= \delta_0(\varepsilon,\alpha,e) = 0$.

        I.S.: Let $\xi = \sigma(\xi_1,\xi_2)$ and $q=o$. Then
        \[
\h_\cA(\xi)_o = \h_\cA(\xi_1)_e \cdot  \h_\cA(\xi_2)_o \cdot \delta_2(eo,\sigma,o)   + \h_\cA(\xi_1)_o \cdot  \h_\cA(\xi_2)_e \cdot \delta_2(oe,\sigma,o) \enspace.
\]
We proceed by case analysis.

$\#_\alpha(\xi_1)$ is even and $\#_\alpha(\xi_2)$ is even: Then $\h_\cA(\xi)_o = 0$.

$\#_\alpha(\xi_1)$ is even and $\#_\alpha(\xi_2)$ is odd: Then
\[
  \h_\cA(\xi)_o = \h_\cA(\xi_1)_e \cdot  \h_\cA(\xi_2)_o \cdot \delta_2(eo,\sigma,o) = 2^{\#_\alpha(\xi_1)} \cdot 2^{\#_\alpha(\xi_2)} = 2^{\#_\alpha(\xi)} \enspace.
  \]

  $\#_\alpha(\xi_1)$ is odd and $\#_\alpha(\xi_2)$ is even: Then
\[
  \h_\cA(\xi)_o = \h_\cA(\xi_1)_o \cdot  \h_\cA(\xi_2)_e \cdot \delta_2(eo,\sigma,o) = 2^{\#_\alpha(\xi_1)} \cdot 2^{\#_\alpha(\xi_2)} = 2^{\#_\alpha(\xi)} \enspace.
  \]

  $\#_\alpha(\xi_1)$ is odd and $\#_\alpha(\xi_2)$ is odd: Then $\h_\cA(\xi)_o = 0$.

\noindent  This proves \eqref{equ:number-of-alphas-det}. Then, for each $\xi \in \T_\Sigma$ we have
  \[
\sem{\cA}(\xi) = \bigplus_{q \in Q} \h_\cA(\xi)_q \cdot F_q = \h_\cA(\xi)_e \cdot 2
+ \h_\cA(\xi)_o \cdot 3 = 
\begin{cases} 2 \cdot 2^{\#_\alpha(\xi)} & \text{ if $\#_\alpha(\xi)$ is even}\\
  3 \cdot 2^{\#_\alpha(\xi)} & \text{ otherwise}  \end{cases}
\ \ = \ \ \mathrm{weo}(\xi) \enspace.\]
Hence $\mathrm{weo} \in \budRec(\Sigma,\B)$.
\hfill $\Box$  
\end{example}

\

\index{$\#_{\sigma(.,\alpha)}$}
\begin{example}\rm \label{ex:number-of-occurrences} {\bf (Number of occurrences of a pattern, semiring $\Nat$.)} \\
  Let $\Sigma = \{\sigma^{(2)}, \omega^{(1)}, \alpha^{(0)}\}$.
 We define the mapping  
\begin{align*}
\#_{\sigma(.,\alpha)}: \T_\Sigma \to \mathbb{N} \ \ \text{ with } \ \  
\#_{\sigma(.,\alpha)}(\xi) =  |U(\xi)| \ \text{ for each $\xi \in \T_\Sigma$}
\end{align*}
where $U(\xi) = \{u \in \pos(\xi) \mid \xi(u) = \sigma, \xi(u2)=\alpha\}$.
Intuitively, $\#_{\sigma(.,\alpha)}(\xi)$ is the number of occurrences of the pattern $\sigma(.,\alpha)$ in $\xi$. For instance, $\#_{\sigma(.,\alpha)}\big(\sigma(\omega(\sigma(\alpha,\alpha)),\sigma(\alpha,\alpha))\big)=2$.

As weight algebra we use the natural number semiring $\Nat=(\mathbb{N}, +,\cdot,0,1)$. Thus, the mapping $\#_{\sigma(.,\alpha)}$ is a $(\Sigma,\Nat)$-weighted tree language.

We construct the $(\Sigma,\Nat)$-wta $\cA = (Q,\delta,F)$ which r-recognizes $\#_{\sigma(.,\alpha)}$, as follows.
\begin{compactitem}
\item $Q  = \{\bot, {a}, {f}\}$ (intuitively, $\bot$ ignores occurrences of the pattern, ${a}$ detects an $\alpha$-labeled leaf, and ${f}$ reports ``pattern found'' up to the root),
 
\item for every $q_1,q_2,q \in Q$ we define
\[
\delta_0(\varepsilon,\alpha,q) = 
\left\{
\begin{array}{ll}
1 & \text{if } q \in \{\bot,{a}\}\\
0 & \text{otherwise}
\end{array}
\right. \ \ 
\delta_1(q_1,\omega,q) = 
\left\{
\begin{array}{ll}
1 & \text{if } q_1q \in \{\bot\bot,{f}{f}\}\\
0 & \text{otherwise}
\end{array}
\right.
\]

\[
\delta_2(q_1q_2,\sigma,q) = 
\left\{
\begin{array}{ll}
1 & \text{if } q_1q_2q \in \{\bot\bot\bot,\bot {a} {f}, \bot {f}{f}, {f}\bot {f}\}\\
0 & \text{otherwise}
\end{array}
\right.
\]

\item $F_\bot=F_{a}=0$ and $F_{{f}}=1$.
\end{compactitem}
Clearly, $\cA$ is root weight normalized; $\cA$ is not total, because there does not exist a state $q$ with $\delta_1({a},\omega,q)\not=0$; $\cA$ is not bu-deterministic.
In Figure \ref{fig:ex-number-patterns} we represent $\cA$ as an fta-hypergraph.

\begin{figure}
  \begin{center}
\begin{tikzpicture}
\tikzset{node distance=7em, scale=0.6, transform shape}
\node[state, rectangle] (1) {\Large $\sigma$};
\node[state, right of=1] (2) {\Large $\bot$};
\node[state, rectangle, above of=2] (3) {\Large $\omega$};
\node[state, rectangle, above right of=2] (4) {\Large $\alpha$};
\node[state, rectangle, below of=1] (5) {\Large $\alpha$};
\node[state, right of=5] (6) {\Large ${a}$};
\node[state, rectangle, right of=2] (7) [right=3em] {\Large $\sigma$};
\node[state, right of=7] (8) [right=3em]{\Large ${f}$};
\node[state, rectangle, above of=8] (9) {\Large $\sigma$};
\node[state, rectangle, below of=8] (10) {\Large $\sigma$};
\node[state, rectangle, right of=8] (11) {\Large $\omega$};

\tikzset{node distance=2em}
\node[above of=1] (w1) {1};
\node[above of=3] (w3) {1};
\node[above of=4] (w4) {1};
\node[above of=5] (w5) {1};
\node[above of=7] (w7) {1};
\node[above of=8] (w8) [left=0.05cm]{1};
\node[above of=9] (w9) [left=0.05cm] {1};
\node[above of=10] (w10) [left=0.05cm] {1};
\node[above of=11] (w11) {1};

\draw[->,>=stealth] (1) edge (2);
\draw[->,>=stealth] (2) edge (3);
\draw[->,>=stealth] (2) edge[out=140, in=-210, looseness=1.8] (1);
\draw[->,>=stealth] (2) edge[out=-140, in=210, looseness=1.8] (1);
\draw[->,>=stealth] (3) edge[out=-250, in=-240, looseness=1.8] (2);
\draw[->,>=stealth] (4) edge (2);
\draw[->,>=stealth] (2) edge (7);
\draw[->,>=stealth] (2) edge (9);
\draw[->,>=stealth] (2) edge (10);
\draw[->,>=stealth] (5) edge (6);
\draw[->,>=stealth] (6) edge (7);
\draw[->,>=stealth] (7) edge (8);
\draw[->,>=stealth] (8) edge (9);
\draw[->,>=stealth] (8) edge (11);
\draw[->,>=stealth] (11) edge[out=-30, in=-30, looseness=1.8] (8);
\draw[->,>=stealth] (8) edge (10);
\draw[->,>=stealth] (9) edge[out=-280, in=40, looseness=1.8] (8);
\draw[->,>=stealth] (10) edge[out=-65, in=-50, looseness=1.8] (8);
\end{tikzpicture}
\end{center}

\vspace{-10mm}

\caption{\label{fig:ex-number-patterns} The $(\Sigma,\Nat)$-wta $\cA = (Q,\delta,F)$ which r-recognizes $\#_{\sigma(.,\alpha)}$.}
  \end{figure}

Let  $\xi \in \T_\Sigma$. We prove that $\runsem{\cA}(\xi) = \#_{\sigma(.,\alpha)}(\xi)$. For this we define, for each $u \in U(\xi)$, the special run  $\rho_u^\xi: \pos(\xi) \rightarrow Q$ for each $w \in \pos(\xi)$ by 
\[
\rho_u^\xi(w) =
\left\{
\begin{array}{ll}
{f} & \text{if $w$ is a prefix of $u$}\\
{a} & \text{if $w=u2$}\\
\bot & \text{otherwise}
\end{array}
\right.
\]

It is obvious that $\wt(\xi,\rho_u^\xi)=1$ and $\rho_u^\xi(\varepsilon)={f}$. Hence $\wt(\xi,\rho_u^\xi)\cdot F_{\rho_u^\xi(\varepsilon)}=1$. Moreover, $\wt(\xi,\rho)\cdot F_{\rho(\varepsilon)}=0$ for each $\rho \in \R_\cA(\xi) \setminus \{\rho_u^\xi \mid u\in  U(\xi)\}$. 

 Then we can calculate as follows.
\[
\runsem{\cA}(\xi) = \bigplus_{\rho \in \R_\cA(\xi)} \wt(\xi,\rho) \cdot F_{\rho(\varepsilon)} = \bigplus_{u \in U(\xi)} \wt(\xi,\rho_u^\xi)\cdot F_{\rho_u^\xi(\varepsilon)} = |U(\xi)|= \#_{\sigma(.,\alpha)}(\xi)\enspace.
\]
Hence $\runsem{\cA}=\#_{\sigma(.,\alpha)}$ and thus $\#_{\sigma(.,\alpha)} \in \Rec^{\mathrm{run}}(\Sigma,\Nat)$.

Next we show that $\cA$ also i-recognizes  $\#_{\sigma(.,\alpha)}$. By induction on $\T_\Sigma$, we prove that the following statement holds:
\begin{eqnarray}
  \begin{aligned}
  & \text{For each $\xi \in \T_\Sigma$, we have $\h_\cA(\xi)_\bot=1$, $\h_\cA(\xi)_a=g(\xi)$, and $\h_\cA(\xi)_f=\#_{\sigma(.,\alpha)}(\xi)$,} \\
  & \text{where $g(\xi) = 1$ if $\xi=\alpha$, and $0$ otherwise.}  
   \end{aligned} \label{eq:h-numberofpatterns-ex}
\end{eqnarray}

I.B.: Let $\xi = \alpha$. Then
\begingroup
\allowdisplaybreaks
\begin{align*}
  \h_\cA(\alpha)_\bot &= \delta_0(\varepsilon,\alpha,\bot) = 1,\\
  \h_\cA(\alpha)_a &= \delta_0(\varepsilon,\alpha,a) = 1 = g(\alpha), \text{ and }\\
  \h_\cA(\alpha)_f &= \delta_0(\varepsilon,\alpha,f) = 0 = \#_{\sigma(.,\alpha)}(\alpha) \enspace.
\end{align*}
\endgroup

I.S.: Let $\xi = \omega(\xi')$.
Then
\begingroup
\allowdisplaybreaks
\begin{align*}
  \h_\cA(\omega(\xi'))_\bot &= \h_\cA(\xi')_\bot \cdot \delta_1(\bot,\omega,\bot) = 1,\\
  \h_\cA(\omega(\xi'))_a &= 0 = g(\omega(\xi')), \text{ and }\\
  \h_\cA(\omega(\xi'))_f &= \h_\cA(\xi')_f \cdot \delta_1(f,\omega,f) = \#_{\sigma(.,\alpha)}(\xi') = \#_{\sigma(.,\alpha)}(\omega(\xi')) \enspace.
\end{align*}
\endgroup
Let $\xi=\sigma(\xi_1,\xi_2)$. 
Then
\begingroup
\allowdisplaybreaks
\begin{align*}
  \h_\cA(\sigma(\xi_1,\xi_2))_\bot &= \h_\cA(\xi_1)_\bot \cdot \h_\cA(\xi_2)_\bot \cdot \delta_2(\bot\bot,\sigma,\bot) = 1,\\
  \h_\cA(\sigma(\xi_1,\xi_2))_a &= 0 = g(\sigma(\xi_1,\xi_2)), \text{ and }\\
  \h_\cA(\sigma(\xi_1,\xi_2))_f &= \h_\cA(\xi_1)_\bot  \cdot \h_\cA(\xi_2)_a  \cdot \delta_2(\bot a,\sigma,f) \ + \\
                     & \ \ \ \  \h_\cA(\xi_1)_\bot  \cdot \h_\cA(\xi_2)_f  \cdot \delta_2(\bot f,\sigma,f) \ + \\
                                          & \ \ \ \   \h_\cA(\xi_1)_f  \cdot \h_\cA(\xi_2)_\bot  \cdot \delta_2(f \bot,\sigma,f)  \\
                                            &= g(\xi_2) +  \#_{\sigma(.,\alpha)}(\xi_2) + \#_{\sigma(.,\alpha)}(\xi_1) \\
  &= \#_{\sigma(.,\alpha)}(\sigma(\xi_1,\xi_2)) \enspace.
\end{align*}
\endgroup
This proves \eqref{eq:h-numberofpatterns-ex}. Then for each $\xi \in \T_\Sigma$ we have
\[
\initialsem{\cA}(\xi) = \bigplus_{q \in Q} \h_\cA(\xi)_q \cdot F_q = \h_\cA(\xi)_f = \#_{\sigma(.,\alpha)}(\xi) \enspace.
\]
Hence $\#_{\sigma(.,\alpha)} \in \Rec^{\mathrm{init}}(\Sigma,\Nat)$.
\hfill $\Box$
 \end{example}

\

\index{$\#_{\sigma(.,\alpha)}$}
\begin{example}\rm \label{ex:number-of-occurrences-arctic} {\bf (Number of occurrences of a pattern, semiring $\Natmaxplus$.)} \sloppy  Again, let $\Sigma = \{\sigma^{(2)}, \omega^{(1)}, \alpha^{(0)}\}$ and let us consider the mapping $\#_{\sigma(.,\alpha)}:\T_\Sigma \to \mathbb{N}$ defined in Example 
\ref{ex:number-of-occurrences}. 

We construct an $(\Sigma,\Natmaxplus)$-wta $\cA=(Q,\delta,F)$ which r-recognizes $\#_{\sigma(.,\alpha)}$, where $\Natmaxplus$ is the arctic semiring. 
For this, we let
\begin{compactitem}
\item $Q=\{\overline{\sigma}, \overline{\omega},\overline{\alpha}\}$,
\item for every $q_1,q_2,q \in Q$ we define
\[
\delta_0(\varepsilon,\alpha,q) = 
\left\{
\begin{array}{ll}
0 & \text{if } q = \overline{\alpha}\\
-\infty & \text{otherwise}
\end{array}
\right. \ \ 
\delta_1(q_1,\omega,q) = 
\left\{
\begin{array}{ll}
0 & \text{if } q=\overline{\omega}\\
-\infty & \text{otherwise}
\end{array}
\right.
\]

\[
\delta_2(q_1q_2,\sigma,q) = 
\left\{
\begin{array}{ll}
1 & \text{if } q_2=\overline{\alpha} \text{ and } q=\overline{\sigma}\\
0 & \text{if } q_2\not=\overline{\alpha} \text{ and } q=\overline{\sigma}\\
-\infty & \text{otherwise}\enspace,
\end{array}
\right.
\]
\item for each $q\in Q$ we define $F_q=0$.
\end{compactitem}
It is clear that $\cA$ is bu-deterministic and total. Since $\delta_2(q_1\overline{\alpha},\sigma,\overline{\sigma}) \not\in \{-\infty,0\}$ for each $q_1 \in Q$, the wta $\cA$ is not crisp-deterministic.

Now let $\xi\in \T_\Sigma$. We define the particular run $\rho^\xi$ on  $\xi$ such that $\rho^\xi(w)=\overline{\xi(w)}$ for each $w\in \pos(\xi)$. Then, for each $w\in \pos(\xi)$, the weight of the transition induced by $\rho^\xi$ on $\xi$ at $w$ is 1 if $\xi(w)=\sigma$ and $\xi(w2)=\alpha$ (i.e., the pattern $\sigma(.,\alpha)$ shows up at $w$)
and 0 otherwise. Hence $\wt(\xi,\rho^\xi)=\#_{\sigma(.,\alpha)}(\xi)$. Moreover, for each $\rho\in \R_\cA(\xi)\setminus \{\rho^\xi\}$ we have $\wt(\xi,\rho)=-\infty$.

Thus we can calculate as follows.
\[
\runsem{\cA}(\xi) = \max\big( \wt(\xi,\rho) + F_{\rho(\varepsilon)} \mid \rho \in \R_\cA(\xi)\big)  = \wt(\xi,\rho^\xi) + F_{\rho^\xi(\varepsilon)} = \#_{\sigma(.,\alpha)}(\xi) + 0 =\#_{\sigma(.,\alpha)}(\xi),\]
hence $\runsem{\cA}(\xi)=\#_{\sigma(.,\alpha)}(\xi)$. Thus $\#_{\sigma(.,\alpha)} \in \budRec^{\mathrm{run}}(\Sigma,\Natmaxplus)$.
\hfill $\Box$
\end{example}

\

\index{diff@$\mathrm{diff}$}
\begin{example}\rm {\bf (Difference of numbers of occurrences of nullary symbols.)}\label{example:diff}
\sloppy Let $\Sigma = \{\sigma^{(2)}, \alpha^{(0)}, \beta^{(0)}\}$ be a ranked alphabet. We define the weighted tree language
\begin{align*}
\mathrm{diff}: \T_\Sigma &\rightarrow \mathbb{Z}_\infty \ \ \text{ such that, for each $\xi \in \T_\Sigma$, we let}\\
\mathrm{diff}(\xi) &=  
\left\{
\begin{array}{ll}
|\pos_\alpha(\xi_1)| - | \pos_\beta(\xi_2)| &\text{if } \xi=\sigma(\xi_1,\xi_2) \text{ for some $\xi_1,\xi_2 \in \T_\Sigma$}\\
\infty & \text{otherwise.}
\end{array}
\right.
\end{align*}
For instance, $\mathrm{diff}(\sigma(\alpha,\sigma(\beta,\beta))) = -1$. As weight algebra we consider the tropical semiring $\Intminplus = (\mathbb{Z}_\infty,\min,+,\infty,0)$ over $\mathbb{Z}$. Thus, the mapping $\mathrm{diff}$ is a $(\Sigma,\Intminplus)$-weighted tree language.

We construct the $(\Sigma,\Intminplus)$-wta $\cA=(Q,\delta,F)$ which i-recognizes $\mathrm{diff}$, as follows:
\begin{compactitem}
\item $Q = \{\#_\alpha,\#_\beta,{d}\}$ (intuitively, $\#_\alpha$  calculates the number of occurrences of $\alpha$,  $\#_\beta$ calculates the negative of the number of occurrences of $\beta$, and ${d}$ calculates the difference),

\item for each  $q \in Q$ we define 
\[
\delta_0(\varepsilon,\alpha,q) =
\left\{
\begin{array}{ll}
1 & \text{if } q=\#_\alpha\\
0 & \text{if } q=\#_\beta\\
\infty & \text{otherwise } 
\end{array}
\right.
\qquad
\delta_0(\varepsilon,\beta,q) =
\left\{
\begin{array}{ll}
0 & \text{if } q=\#_\alpha\\
-1 & \text{if } q=\#_\beta\\
\infty & \text{otherwise } 
\end{array}
\right.
\]
and for each $q_1,q_2,q \in Q$  we define
\[
\delta_2(q_1q_2,\sigma,q) = 
\left\{
\begin{array}{ll}
0 & \text{if } q_1q_2q \in \{\#_\alpha \#_\alpha \#_\alpha, \  \#_\beta \#_\beta \#_\beta,\  \#_\alpha \#_\beta {d}\}\\
\infty & \text{otherwise}
\end{array}
\right.
\]

\item $F_{\#_\alpha}=F_{\#_\beta}=\infty$ and $F_{d}=0$.
\end{compactitem}
Clearly, $\cA$ is root weight normalized; $\cA$ is not total, because there does not exist a state $q$ such that $\delta_2({d}{d},\sigma,q)\not= \infty$; $\cA$ is not bu-deterministic, because $\delta_0(\varepsilon,\alpha,\#_\alpha)\not= \infty \not= \delta_0(\varepsilon,\alpha,\#_\beta)$. In Figure \ref{fig:ex-diff-number-occ} we represent $\cA$ as fta-hypergraph.

\begin{figure}
\begin{center}
\begin{tikzpicture}
\tikzset{node distance=7em, scale=0.6, transform shape}
\node[state, rectangle] (1) {\Large $\sigma$};
\node[state, right of=1] (2) {\Large $\#_\alpha$};
\node[state, rectangle, above of=2] (3) {\Large $\alpha$};
\node[state, rectangle, above right of=2] (4) {\Large $\beta$};
\node[state, rectangle, below of=1] (5) [below=2em]{\Large $\sigma$};
\node[state, right of=5] (6) {\Large $\#_\beta$};
\node[state, rectangle, below of=6] (7) {\Large $\alpha$};
\node[state, rectangle, below right of=6] (8) {\Large $\beta$};
\node[state, rectangle, below right=2em and 9em of 2] (9) {\Large $\sigma$};
\node[state, right of=9] (10) [right=1em] {\Large ${d}$};

\tikzset{node distance=2em}
\node[above of=1] (w1) {0};
\node[above of=3] (w3) {1};
\node[above of=4] (w4) {0};
\node[above of=5] (w5) {0};
\node[above of=7] (w7) [left=0.05cm] {0};
\node[above of=8] (w8) {-1};
\node[above of=9] (w9) {0};
\node[above of=10] (w10) {0};

\draw[->,>=stealth] (1) edge (2);
\draw[->,>=stealth] (2) edge[out=140, in=-210, looseness=1.8] (1);
\draw[->,>=stealth] (2) edge[out=-140, in=210, looseness=1.8] (1);
\draw[->,>=stealth] (3) edge (2);
\draw[->,>=stealth] (4) edge (2);
\draw[->,>=stealth] (5) edge (6);
\draw[->,>=stealth] (6) edge[out=140, in=-210, looseness=1.8] (5);
\draw[->,>=stealth] (6) edge[out=-140, in=210, looseness=1.8] (5);
\draw[->,>=stealth] (7) edge (6);
\draw[->,>=stealth] (8) edge (6);
\draw[->,>=stealth] (2) edge (9);
\draw[->,>=stealth] (6) edge (9);
\draw[->,>=stealth] (9) edge (10);
\end{tikzpicture}
\end{center}
\caption{\label{fig:ex-diff-number-occ} The $(\Sigma,\Intminplus)$-wta $\cA=(Q,\delta,F)$ which i-recognizes $\mathrm{diff}$.}
\end{figure}

First we observe that, for each $\xi \in \T_\Sigma$, we have 
\begin{equation}
  \h_\cA(\xi)_{\#_\alpha} = | \pos_\alpha(\xi)| \  \text{ and } \ \h_\cA(\xi)_{\#_\beta} = - |\pos_\beta(\xi)| \enspace.
  \label{equ:h=number-pos}
\end{equation}

Next we prove that $\initialsem{\cA} = \mathrm{diff}$.
Let $\xi \in \T_\Sigma$. By definition of $F$,  we have:
\[
\initialsem{\cA}(\xi) = \min (\h_\cA(\xi)_{q} + F_q \mid q\in Q)
= \h_\cA(\xi)_{d}+F_d=\h_\cA(\xi)_{d}\enspace.
\]

Let $\xi \in \{\alpha,\beta\}$.  Then 
$\h_\cA(\xi)_{d} = \delta_0(\varepsilon,\xi,{d}) = \infty$. 
Hence $\initialsem{\cA}(\xi) = \infty = \mathrm{diff}(\xi)$.

Now let $\xi = \sigma(\xi_1,\xi_2)$ for some $\xi_1,\xi_2 \in \T_\Sigma$. Then we can calculate as follows:
\begin{align*}
\h_\cA(\sigma(\xi_1,\xi_2))_{d}
=& \ \min \big( \h_\cA(\xi_1)_{q_1} +  \h_\cA(\xi_2)_{q_2} + \delta_2(q_1q_2,\sigma,{d}) \mid q_1,q_2 \in Q\big) \\
= & \  \h_\cA(\xi_1)_{\#_\alpha} +  \h_\cA(\xi_2)_{\#_\beta} + \delta_2(\#_\alpha\#_\beta,\sigma,d)\\
=& \  \h_\cA(\xi_1)_{\#_\alpha} +  \h_\cA(\xi_2)_{\#_\beta} + 0 \\
  =& \  | \pos_\alpha(\xi_1)| - | \pos_\beta(\xi_2)| \tag{by \eqref{equ:h=number-pos}}\\
  =& \ \mathrm{diff}(\sigma(\xi_1,\xi_2)) \enspace.
\end{align*}
Hence $\initialsem{\cA} = \mathrm{diff}$ and thus  $\mathrm{diff} \in \Rec^{\mathrm{init}}(\Sigma,\Intminplus)$.
\hfill $\Box$
\end{example}

\

\index{diff@$\mathrm{diffm}$}
\begin{example}\rm {\bf (Difference of numbers of occurrences of symbols in monadic trees.)}\label{ex:diff-mon}
Let $\Sigma = \{\gamma^{(1)}, \sigma^{(1)},\alpha^{(0)}\}$ be a string ranked alphabet. We define the mapping 
\begin{align*}
\mathrm{diffm}: \T_\Sigma \rightarrow \mathbb{Z} \ \ \text{ with }
\mathrm{diffm}(\xi) =  |\pos_\gamma(\xi)| - | \pos_\sigma(\xi)| \ \text{ for each $\xi \in \T_\Sigma$}\enspace.
\end{align*}
As weight algebra we consider the ring $\Int=(\mathbb{Z},+,\cdot,0,1)$  of integers. Thus, the mapping $\mathrm{diffm}$ is a $(\Sigma,\Int)$-weighted tree language.

We construct the $(\Sigma,\Int)$-wta $\cA=(Q,\delta,F)$ which i-recognizes $\mathrm{diffm}$, as follows. 
\begin{compactitem}
\item $Q=\{1,{d}\}$ (intuitively, the state $1$ calculates the integer $1$ and, at each non-leaf position $u$, the state ${d}$ increases or decreases the difference 
by $1$  depending on the symbol at $u$),

\item $\delta_0(\varepsilon,\alpha,1) = 1$ and $\delta_0(\varepsilon,\alpha,{d}) = 0$, and for every $q_1,q_2 \in Q$ and $\nu \in \{\gamma,\sigma\}$ we define 
\[
\delta_1(q_1,\nu,q_2) = 
\begin{cases}
1 & \text{if }q_1q_2 \in \{11, {d}{d}\}\\
1 &  \text{if }q_1q_2 = 1{d} \text{ and } \nu=\gamma\\
-1 &  \text{if }q_1q_2 = 1{d} \text{ and } \nu=\sigma\\
0 &  \text{if }q_1q_2 = {d}1
\end{cases}
\]
\item $F_1=0$ and $F_{d}=1$.
\end{compactitem} 
Clearly, $\cA$ is root weight normalized; $\cA$ is not bu-deterministic, because $\delta_1(1,\gamma,1) \not=0\not= \delta_1(1,\gamma,{d})$. In Figure \ref{fig:ex-diff-symbols} we represent $\cA$ as fta-hypergraph.

\begin{figure}
\begin{center}
\begin{tikzpicture}
\tikzset{node distance=7em, scale=0.6, transform shape}
\node[state, rectangle] (1) {\Large $\alpha$};
\node[state, right of=1] (2) {\Large 1};
\node[state, rectangle, above of=2] (3) {\Large $\gamma$};
\node[state, rectangle, below of=2] (4) {\Large $\sigma$};
\node[state, rectangle, above right=1em and 8em of 2] (5) {\Large $\gamma$};
\node[state, rectangle, below right=1em and 8em of 2] (6) {\Large $\sigma$};
\node[state, right of=2] (7)[right=12em] {\Large ${d}$};
\node[state, rectangle, above of=7] (8) {\Large $\gamma$};
\node[state, rectangle, below of=7] (9) {\Large $\sigma$};

\tikzset{node distance=2em}
\node[above of=1] (w1) {1};
\node[above of=3] (w3) {1};
\node[above of=4] (w4)[right=0.05cm] {1};
\node[above of=5] (w5) {1};
\node[above of=6] (w6) {-1};
\node[above of=7] (w7)[left=0.05cm] {1};
\node[above of=8] (w8) {1};
\node[above of=9] (w9)[left=0.05cm] {1};

\draw[->,>=stealth] (1) edge (2);
\draw[->,>=stealth] (2) edge (3);
\draw[->,>=stealth] (3) edge[out=-250, in=-210, looseness=1.8] (2);
\draw[->,>=stealth] (2) edge (4);
\draw[->,>=stealth] (4) edge[out=250, in=-150, looseness=1.8] (2);
\draw[->,>=stealth] (2) edge (5);
\draw[->,>=stealth] (2) edge (6);
\draw[->,>=stealth] (5) edge (7);
\draw[->,>=stealth] (6) edge (7);
\draw[->,>=stealth] (7) edge (8);
\draw[->,>=stealth] (8) edge[out=-290, in=30, looseness=1.8] (7);
\draw[->,>=stealth] (7) edge (9);
\draw[->,>=stealth] (9) edge[out=290, in=-30, looseness=1.8] (7);
\end{tikzpicture}
\end{center}

\vspace{-10mm}

\caption{\label{fig:ex-diff-symbols} The $(\Sigma,\Int)$-wta $\cA=(Q,\delta,F)$ which i-recognizes $\mathrm{diffm}$.}
\end{figure}

First, by induction on $\T_\Sigma$, we prove that the following statement holds:
\begin{equation}
\text{For every $\xi\in \T_\Sigma$, we have } \h_\cA(\xi)_1=1 \text{   and  } \h_\cA(\xi)_{d}=|\pos_\gamma(\xi)|-|\pos_\sigma(\xi)| \enspace. \label{equ:support-sigma-gamma}
\end{equation}

I.B.: Let $\xi=\alpha$. Then $\h_\cA(\xi)_1=\delta_0(\varepsilon,\alpha,1)=1$ and $\h_\cA(\xi)_{d}=\delta_0(\varepsilon,\alpha,{d})=0$. Thus \eqref{equ:support-sigma-gamma} holds.

I.S.: Let  $\xi=\gamma(\xi')$ for some $\xi' \in \T_\Sigma$. Then
\begin{align*}\h_\cA(\xi)_1
&=\bigplus_{q\in Q}\h_\cA(\xi')_{q}\cdot \delta_1(q,\gamma,1)\\
  &=\h_\cA(\xi')_{1}\cdot \delta_1(1,\gamma,1) \tag{by definition of $\delta_1$ and by I.H.}\\
  &=1\cdot 1= 1
\end{align*}
and
\begin{align*}
\h_\cA(\xi)_{d}&=\bigplus_{q\in Q}\h_\cA(\xi')_{q}\cdot \delta_1(q,\gamma,{d})
=  \h_\cA(\xi')_{1}\cdot \delta_1(1,\gamma,{d}) + \h_\cA(\xi')_{{d}}\cdot \delta_1({d},\gamma,{d}) \\ 
&=  1\cdot 1 + (|\pos_\gamma(\xi')|-|\pos_\sigma(\xi')|)\cdot 1 \tag{by definition of $\delta_1$ and by I.H.}\\
&=  (|\pos_\gamma(\xi')|+1)-|\pos_\sigma(\xi')|  \\ 
&= |\pos_\gamma(\xi)|-|\pos_\sigma(\xi)|.
\end{align*}
In a similar way we can prove that \eqref{equ:support-sigma-gamma}  holds for $\xi=\sigma(\xi')$.

Using Statement \eqref{equ:support-sigma-gamma},  we obtain for each $\xi\in \T_\Sigma$:
\[\initialsem{\cA}(\xi)= \bigplus_{q\in Q}\h_\cA(\xi)_{q}\cdot F_{q}=\h_\cA(\xi)_{{d}}=|\pos_\gamma(\xi)|-|\pos_\sigma(\xi)| = \mathrm{diffm}(\xi)\enspace.\]
Thus $\mathrm{diffm} \in \Rec^{\mathrm{init}}(\Sigma,\Int)$.
\hfill $\Box$
\end{example}

\

\index{zigzag@$\mathrm{zigzag}$}
\begin{example}\rm {\bf (Length of zigzag-path.)} \label{ex:zigzag} Let $\Sigma = \{\sigma^{(2)}, \alpha^{(0)}\}$ be a ranked alphabet. We define the mapping $\mathrm{zigzag}: \T_\Sigma \rightarrow \mathbb{N}$ which, intuitively, computes for each tree the length of its zigzag-path (where ``zig'' means ``go to the left child'', and ``zag'' means ``go to the right child''). 
Formally, we first define the auxiliary mappings $\mathrm{zz}_l: \T_\Sigma \to (\mathbb{N}_+)^*$ and 
$\mathrm{zz}_r: \T_\Sigma \to (\mathbb{N}_+)^*$ 
by induction on $\T_\Sigma$ as follows. We let
\[
  \mathrm{zz}_l(\alpha)=\mathrm{zz}_r(\alpha) = \varepsilon, 
  \] 
and, for each $\xi = \sigma(\xi_1,\xi_2)$, we define
\[
  \mathrm{zz}_l(\sigma(\xi_1,\xi_2))=1 \ \mathrm{zz}_r(\xi_1) \ \ \text{ and } \ \ 
  \mathrm{zz}_r(\sigma(\xi_1,\xi_2)) =  2 \ \mathrm{zz}_l(\xi_2)\enspace.
\]
Then we let 
\begin{align*}
\mathrm{zigzag}: \T_\Sigma \rightarrow \mathbb{N} \ \ \text{ with } 
\mathrm{zigzag}(\xi) = |\mathrm{zz}_{l}(\xi)| \ \text{ for each $\xi \in \T_\Sigma$} \enspace.
\end{align*}
For instance, for $\xi= \sigma(\sigma(\zeta_1,\alpha),\zeta_2)$ and any $\zeta_1,\zeta_2 \in \T_\Sigma$, we have $\mathrm{zigzag}(\xi) = |\mathrm{zz}_{l}(\xi)| = |12| = 2$.

As weight algebra we use the arctic semiring $\Natmaxplus= (\mathbb{N}_{-\infty},\max,+,-\infty,0)$. Thus, the mapping $\mathrm{zigzag}$ is a $(\Sigma,\Natmaxplus)$-weighted tree language.

We construct the $(\Sigma,\Natmaxplus)$-wta $\cA= (Q,\delta,F)$ which i-recognizes $\mathrm{zigzag}$ as follows.
\begin{compactitem}
\item $Q = \{{l}, {r}, 0\}$ (intuitively, ${l}$ and ${r}$ stand for ``left'' and ``right'', respectively, and $0$ calculates the $0 \in \mathbb{N}$),

\item $\delta_0(\varepsilon,\alpha,{l}) = \delta_0(\varepsilon,\alpha,{r}) = \delta_0(\varepsilon,\alpha,0) = 0$ and 
for every $q_1,q_2,q \in Q$
\[
\delta_2(q_1q_2,\sigma,q) = 
\left\{
\begin{array}{ll}
1 & \text{if } q_1q_2q \in \{{r}0{l}, 0{l}{r}\}\\
0 & \text{if } q_1q_2q = 000\\
-\infty &\text{otherwise}
\end{array}
\right.
\]

\item $F_{l} = 0$, $F_{r} = F_0 = -\infty$.
\end{compactitem}
Clearly, $\cA$ is root weight normalized; $\cA$ is not bu-deterministic.
In Figure \ref{fig:ex-zigzag-hyp} we represent $\cA$ as an fta-hypergraph.
\begin{figure}
  \begin{center}
\begin{tikzpicture}[scale=0.6, transform shape]
\tikzset{node distance=6em}
\node[state, rectangle] (1) {\Large $\alpha$};
\node[state, right of= 1] (2) {\Large ${l}$};
\node[state, rectangle, right= of 2] (3) {\Large $\sigma$};
\node[state, right of= 3] (4) {\Large ${0}$};
\node[state, rectangle, right of= 4] (5) {\Large $\alpha$};
\node[state, right= of 5] (6) {\Large ${r}$};
\node[state, rectangle, right of= 6] (7) {\Large $\alpha$};
\node[state, rectangle, above= of 5] (8) {\Large $\sigma$};
\node[state, rectangle, below= of 3] (9) {\Large $\sigma$};

\tikzset{node distance=2em}
\node[above of=1] (w1) {0};
\node[above of=2, xshift=-0.3cm] (w2) {0};
\node[above of=3] (w3) {0};
\node[above of=5] (w5) {0};
\node[above of=7] (w7) {0};
\node[above of=8, xshift=0.1cm] (w8) {1};
\node[below of=9, xshift=-0.1cm] (w9) {1};

\draw[->,>=stealth] (1) edge (2);
\draw[->,>=stealth] (3) edge (4);
\draw[->,>=stealth] (4) edge[out=140, in=-210, looseness=2.2] (3);
\draw[->,>=stealth] (4) edge[out=-140, in=210, looseness=2.2] (3);
\draw[->,>=stealth] (5) edge (4);
\draw[->,>=stealth] (7) edge (6);
\draw[->,>=stealth] (4) edge[out= 85, in=270, looseness=1] (8.270+30);
\draw[->,>=stealth] (6) edge[out=100, in=270, looseness=1.3] (8.270-30);
\draw[->,>=stealth] (8) edge[out=95, in= 80, looseness=0.6] (2);
\draw[->,>=stealth] (4) edge[out=-95, in=-270, looseness=1.35] (9.-270-30);
\draw[->,>=stealth] (2) edge[out=-80, in=-270, looseness=1] (9.-270+30);
\draw[->,>=stealth] (9) edge[out=-85, in= -100, looseness=0.6] (6);
\end{tikzpicture}
\end{center}

\vspace{-10mm}

\caption{\label{fig:ex-zigzag-hyp} The $(\Sigma,\Natmaxplus)$-wta $\cA= (Q,\delta,F)$ which i-recognizes $\mathrm{zigzag}$.}
\end{figure}

By induction on $\T_\Sigma$, we can prove that the following statement holds: 
\begin{equation}
\text{For each $\xi \in \T_\Sigma$, we have } \h_\cA(\xi)_{l} = |\mathrm{zz}_{l}(\xi)|, \ 
\h_\cA(\xi)_{r} = |\mathrm{zz}_{r}(\xi)|, \text{ and }
\h_\cA(\xi)_0 = 0  \enspace.
\label{equ:zz}
\end{equation}

I.B.: Let $\xi = \alpha$. Then, for each $q \in Q$, we have  $\h_\cA(\xi)_q = \delta_0(\varepsilon,\alpha,q)=0 = |\mathrm{zz}_{l}(\xi)| = |\mathrm{zz}_{r}(\xi)|$.

I.S.: Now let $\xi = \sigma(\xi_1,\xi_2)$ for some $\xi_1,\xi_2 \in \T_\Sigma$. Then
\begin{align*}
\h_\cA(\sigma(\xi_1,\xi_2))_{l}
=& \max ( \h_\cA(\xi_1)_{q_1} + \h_\cA(\xi)_{q_2} + \delta_2(q_1q_2, \sigma,{l}) \mid q_1,q_2\in Q)\\
=&\ \h_\cA(\xi_1)_{{r}} + \h_\cA(\xi_2)_{0} + \delta_2({r}0, \sigma,{l})\\
  =& \ |\mathrm{zz}_{r}(\xi_1)| + 0 +  1 \tag{by I.H.}\\
  =& \ |\mathrm{zz}_{l}(\sigma(\xi_1,\xi_2))|\enspace.
\end{align*}
In a similar way we can prove that $\h_\cA(\sigma(\xi_1,\xi_2))_{r} = |\mathrm{zz}_{{r}}(\sigma(\xi_1,\xi_2))|$,  and $\h_\cA(\sigma(\xi_1,\xi_2))_0 = 0$. This finishes the proof of Statement \eqref{equ:zz}.

Now let $\xi \in \T_\Sigma$. Then, using \eqref{equ:zz} at the fourth equality below, we obtain 
\[
\initialsem{\cA}(\xi) = \max(\h_\cA(\xi)_q + F_q \mid q \in Q)
= \h_\cA(\xi)_{l} + F_l = \h_\cA(\xi)_{l}=|\mathrm{zz}_{l}(\xi)| = \mathrm{zigzag}(\xi)\enspace.
\]

Thus $\initialsem{\cA} = \mathrm{zigzag}$ and hence $\mathrm{zigzag} \in \Rec^{\mathrm{init}}(\Sigma,\Natmaxplus)$. In \cite[Ex.~3.2]{drovog06} and \cite[Ex.~2]{hogmalmay07d}, a $(\Sigma,\Nat)$-wta $N$ (over the semiring of natural numbers) is shown for which $\initialsem{N} = \mathrm{zigzag}$.  
\hfill $\Box$
\end{example}

\

\index{twothree@$\mathrm{twothree}$}
\begin{example} \rm {\bf (Recognizable step mapping twothree.)} \label{ex:recog-step-mapping}\rm Let $\Sigma = \{\sigma^{(2)}, \gamma^{(1)}, \alpha^{(0)}\}$ be a ranked alphabet. We define the mapping $\mathrm{twothree}: \T_\Sigma \rightarrow \mathbb{N}$ for each $\xi \in \T_\Sigma$ by 
\begin{align*}
\mathrm{twothree}: \T_\Sigma \rightarrow \mathbb{N} \ \ \text{ with } \ \ 
  \mathrm{twothree}(\xi) = 
        \begin{cases}
          2 & \text{ if } |\pos(\xi)| \text{ is even }\\
          3 & \text{ otherwise } 
          \end{cases} \ \ \text{for each $\xi \in \T_\Sigma$}\enspace.
\end{align*}
As weight algebra we use the natural number semiring $\Nat= (\mathbb{N},+,\cdot,0,1)$. Thus, the mapping $\mathrm{twothree}$ is a $(\Sigma,\Nat)$-weighted tree language. Since the two $\Sigma$-tree languages
\[L_{{e}} = \{\xi \in \T_\Sigma \mid |\pos(\xi)| \text{ is even } \} \ \ \text{ and } \ \
L_{{o}} = \T_\Sigma \setminus L_{{e}}\]
are recognizable, $\mathrm{twothree}$ is even a recognizable step mapping, because
\[
  \mathrm{twothree} = (2 \cdot \chi_{\Nat}(L_{{e}}) + 3 \cdot \chi_{\Nat}(L_{{o}}))  \enspace.
  \]
We construct the $(\Sigma,\Nat)$-wta  $\cA=(Q,\delta,F)$ which i-recognizes $\mathrm{twothree}$ as follows
\begin{compactitem}
\item $Q = \{{e},{o}\}$ where ${e}$ and ${o}$ stand for ``even'' and ``odd'', respectively,

  \item  $\delta_0(\varepsilon,\alpha,{e}) = 0$ and $\delta_0(\varepsilon,\alpha,{o}) = 1$ and for every $q_1,q_2,q \in Q$ we let
   \[
    \delta_1(q_1,\gamma,q) =
    \begin{cases}
      1 & \text{ if ($q_1={e}$ and $q={o}$) or ($q_1={o}$ and $q={e}$)}\\
      0 & \text{ otherwise},
      \end{cases}
    \]
    
  \[
    \delta_2(q_1q_2,\sigma,q) =
    \begin{cases}
      1 & \text{ if ($q_1=q_2$ and $q={o}$) or ($q_1\not=q_2$ and $q={e}$)}\\
      0 & \text{ otherwise},
      \end{cases}
    \]
    and 
\item $F_{e}=2$ and $F_{o} = 3$.
\end{compactitem}
We note that $\cA$ is crisp-deterministic and consequently also bu-deterministic. Figure \ref{fig:ex-twothree-step-mapping} shows the $(\Sigma,\Nat)$-wta  $\cA$.

Let $\xi \in \T_\Sigma$ and $q \in Q$. It is clear that
\[
  \h_\cA(\xi)_q =
  \begin{cases}
    1 & \text{ if }  \ \big((q={e} \text{ and } |\pos(\xi)| \text{ is even} ) \text{ or }  (q={o} \text{ and } |\pos(\xi)| \text{ is odd} )\big)\\
    0 & \text{ otherwise}
    \end{cases}
\]
Thus
\begin{align*}
  \initialsem{\cA}(\xi) &= \bigplus_{q \in Q} (\h_\cA(\xi)_q \cdot F_q) \\
                      &= (\h_\cA(\xi)_{e} \cdot F_{e}) +  (\h_\cA(\xi)_{o} \cdot F_{o}) = (\h_\cA(\xi)_{e} \cdot 2) + (\h_\cA(\xi)_{o} \cdot 3)\\
  &= \mathrm{twothree}(\xi) \enspace. 
\end{align*}

  \begin{figure}[t]
\begin{center}
\begin{tikzpicture}
\tikzset{node distance=7em, scale=0.7, transform shape}
\node
(1) {};
\node[state, right of=1] (2){\Large ${e}$};
\node[state, rectangle, above of=2] (3){\Large $\sigma$};
\node[state, rectangle, above right=4em and 7em of 2] (4){\Large $\sigma$};
\node[state, rectangle, below of=4] (5)[above=0.5em]{\Large $\gamma$};
\node[state, rectangle, below of=5] (6)[above=0.5em]{\Large $\gamma$};
\node[state, rectangle, below of=6] (7)[above=0.5em]{\Large $\sigma$};
\node[state, right of=2] (8)[right=10em]{\Large ${o}$};
\node[state, rectangle, below of=8] (9){\Large $\sigma$};
\node[state, rectangle, right of=8] (10){\Large $\alpha$};

\tikzset{node distance=2em}
\node[above of=2] (w2)[left=0.3cm] {2};
\node[above of=3] (w3) {1};
\node[above of=4] (w4) {1};
\node[above of=5] (w5) {1};
\node[above of=6] (w6) {1};
\node[above of=7] (w7) {1};
\node[above of=8] (w8)[left=0.2cm] {3};
\node[right of=9] (w9) {1};
\node[above of=10] (w10) {1};

\draw[->,>=stealth] (2) edge (4);
\draw[->,>=stealth] (2) edge (5);
\draw[->,>=stealth] (2) edge (7);
\draw[->,>=stealth] (2) edge[out=110, in=-115, looseness=1.1] (3);
\draw[->,>=stealth] (2) edge[out=70, in=-65, looseness=1.1] (3);
\draw[->,>=stealth] (3) edge[out=60, in=80, looseness=1.4] (8);
\draw[->,>=stealth] (4) edge[out=110, in=50, looseness=1.1] (2);
\draw[->,>=stealth] (5) edge (8);
\draw[->,>=stealth] (6) edge (2);
\draw[->,>=stealth] (7) edge[out=260, in=-100, looseness=1.3] (2);
\draw[->,>=stealth] (8) edge (4);
\draw[->,>=stealth] (8) edge (6);
\draw[->,>=stealth] (8) edge (7);
\draw[->,>=stealth] (8) edge[out=-110, in=115, looseness=1.1] (9);
\draw[->,>=stealth] (8) edge[out=-70, in=65, looseness=1.1] (9);
\draw[->,>=stealth] (9) edge[out=-70, in=-30, looseness=1.8] (8);
\draw[->,>=stealth] (10) edge (8);
\end{tikzpicture}
\end{center}

\vspace{-10mm}

\caption{\label{fig:ex-twothree-step-mapping}  The $(\Sigma,\Nat)$-wta  $\cA$ which i-recognizes $\mathrm{twothree}$.}
\end{figure}

Hence $\mathrm{twothree} \in \budRec^{\mathrm{init}}(\Sigma,\Nat)$.
\hfill $\Box$
\end{example}

\

\index{evaluation algebra}
\begin{example}\rm {\bf (Evaluation algebras.)} \label{ex:h-kappa-crisp-det}  \cite[Lm.~4.3]{fulstuvog12} Let $(\kappa_k \mid k \in \mathbb{N})$ be an $\mathbb{N}$-indexed family of mappings $\kappa_k:~\Sigma^{(k)}~\to~B$.  We recall that $\B=(B,\oplus,\otimes,\0,\1)$ is a strong bimonoid, $\M(\Sigma,\kappa)$ is the $(\Sigma,\kappa)$-evaluation algebra and $\h_{\M(\Sigma,\kappa)}: \T_\Sigma \to B$ is the unique  $\Sigma$-algebra homomorphism, as  defined in Subsection \ref{sec:weighted-tree-languages}.

  We construct the $(\Sigma,\B)$-wta $\cA=(Q,\delta,F)$ which i-recognizes $\h_{\M(\Sigma,\kappa)}$ as follows.
  \begin{compactitem}
  \item $Q=\{{q}\}$ and $F_{q}=\1$ and
    \item for every $k \in \mathbb{N}$, $\sigma \in \Sigma^{(k)}$, we let $\delta_k({q}\cdots {q},\sigma,{q}) = \kappa_k(\sigma)$.
    \end{compactitem}
    Obviously, $\cA$ is bu-deterministic and root weight normalized. Moreover, if there exists a $\sigma$ such that $\kappa_k(\sigma) \not\in \{\0,\1\}$, then $\cA$ is not crisp and hence $\cA$ is not crisp-deterministic.

  It is easy to see that the vector algebra $\V(\cA)=(B^Q,\delta_\cA)$ of $\cA$ and the $(\Sigma,\kappa)$-evaluation algebra $\M(\Sigma,\kappa) = (B,\overline{\kappa})$ are isomorphic $\Sigma$-algebras, by identifying each $Q$-vector over $B$ with its one and only component in $B$, and by verifying that, for every $b_1,\ldots,b_k \in B$, we have 
   \[
     \delta_\cA(\sigma)((b_1),\ldots,(b_k))_{q} = \Big(\bigotimes_{i \in [k]} b_i\Big) \otimes \delta_k({q} \cdots {q},\sigma,{q}) = \Big(\bigotimes_{i \in [k]} b_i\Big) \otimes \kappa_k(\sigma) = \overline{\kappa}(\sigma)(b_1,\ldots,b_k) \enspace.
     \]
     Hence,  for each $\xi \in \T_\Sigma$, we have $\h_\cA(\xi)_{q}=\h_{\M(\Sigma,\kappa)}(\xi)$. Then, for each $\xi \in \T_\Sigma$, we have
    \[
\initialsem{\cA}(\xi) = \bigoplus_{p \in Q} \h_\cA(\xi)_p \otimes F_p = \h_\cA(\xi)_{q} = \h_{\M(\Sigma,\kappa)}(\xi)\enspace. 
\]
Thus, in particular, $\h_{\M(\Sigma,\kappa)} \in  \budRec^{\mathrm{init}}(\Sigma,\B)$.

Above we have proved that, for each  $\mathbb{N}$-indexed family $\kappa = (\kappa_k \mid k \in \mathbb{N})$ of mappings $\kappa_k:~\Sigma^{(k)}~\to~B$, the $\Sigma$-algebra homomorphism $\h_{\M(\Sigma,\kappa)}$ is i-recognizable by a  bu-deterministic $(\Sigma,\B)$-wta. This can be used to show i-recognizability of the mappings $\size$ (cf. Example \ref{ex:size}), $\yield^\cP$   (cf. Example \ref{ex:yield}), and $\mathrm{BPS}_A$  (cf. Example \ref{ex:transformation-monoid}) by instantiating $\B$ and providing $\mathbb{N}$-indexed families of mappings appropriately.

\begin{itemize}
\item We consider the  tropical semiring $\Natminplus=(\mathbb{N}_\infty,\min,+,\infty,0)$ and the $(\Sigma,\Natminplus)$-weighted tree language
  \[\size: \T_\Sigma \to \mathbb{N}\]
of Example \ref{ex:size}  with $\Sigma = \{\sigma^{(2)}, \alpha^{(0)}\}$. We define the $\mathbb{N}$-indexed family $\kappa^{\size} = (\kappa^{\size}_k \mid k \in \mathbb{N})$  of mappings by $\kappa^{\size}_0(\alpha) = \kappa^{\size}_2(\sigma) = 1$. Then $\size = \h_{\M(\Sigma,\kappa^{\size})}$.
  
\item \sloppy We consider the ranked alphabet $\Sigma = \{\sigma^{(2)}, \alpha^{(0)}, \beta^{(0)}\}$ and  the formal language semiring $\mathsf{Lang}_\Sigma=(\cP(\Sigma^*),\cup,\cdot,\emptyset,\{\varepsilon\})$ (where we forget about the rank of symbols).  Moreover, we consider the $(\Sigma,\mathsf{Lang}_\Sigma)$-weighted tree language
  \[\yield^\cP: \T_\Sigma \to \cP((\Sigma^{(0)})^*)\]
of Example \ref{ex:yield}. We define the $\mathbb{N}$-indexed family $\kappa^{\yield} = (\kappa^{\yield}_k \mid k \in \mathbb{N})$  of mappings by $\kappa^{\yield}_0(\alpha) = \{\alpha\}$, $\kappa^{\yield}_0(\beta) = \{\beta\}$, and $\kappa^{\yield}_2(\sigma) = \{\varepsilon\}$. Then $\yield^\cP = \h_{\M(\Sigma,\kappa^{\yield})}$.
  
\item Let $A =(Q,I,\delta,F)$ be an $\{\alpha,\beta\}$-fsa. We consider the  near semiring $\NearSem_{\cP(Q)}=(B,\cup,\diamond,\emptyset,\id_{\cP(Q)})$ with $B = \{f \mid f:\cP(Q) \to \cP(Q), f(\emptyset)=\emptyset\}$ and the $(\Sigma,\NearSem_{\cP(Q)})$-weighted tree language
  \[\mathrm{BPS}_A: \T_\Sigma \to \cP(Q)^{\cP(Q)}\]
of Example \ref{ex:transformation-monoid}  with $\Sigma = \{\sigma^{(2)}, \alpha^{(0)}, \beta^{(0)}\}$. We define the $\mathbb{N}$-indexed family $\kappa^{\mathrm{BPS}} = (\kappa^{\mathrm{BPS}}_k \mid k \in \mathbb{N})$  of mappings such that, for each $U \in \cP(Q)$, we let
  \begin{compactitem}
  \item $\kappa^{\mathrm{BPS}}_0(\alpha)(U) = \{p \in Q \mid (\exists r \in U): (r,\alpha,p) \in \delta\}$,
  \item  $\kappa^{\mathrm{BPS}}_0(\beta)(U) = \{p \in Q \mid (\exists r \in U): (r,\beta,p) \in \delta\}$, and
  \item $\kappa^{\mathrm{BPS}}_2(\sigma)(U) = U$ (i.e., $\kappa^{\mathrm{BPS}}_2(\sigma) = \id_{\cP(Q)}$).
  \end{compactitem}
  Then $\mathrm{BPS}_A = \h_{\M(\Sigma,\kappa^{\mathrm{BPS}})}$.
  \hfill $\Box$
\end{itemize}
\end{example}

\


\begin{example}\rm \label{ex:Sigma-algebra-hom-as-wta} {\bf ($\Sigma$-algebra homomorphism.)} Let $\Sigma$ be a ranked alphabet and $\B=(B,\oplus,\otimes,\0,\1)$ be a  strong bimonoid. Moreover, let $\cB = (B,\theta)$ be a $\Sigma$-algebra such that there exists a family \((b_{\sigma,w} \mid k \in \mathbb{N}, \sigma \in \Sigma^{(k)}, w \in \{0,1\}^k) \) over $B$ and, for every $k \in \mathbb{N}$, $\sigma \in \Sigma^{(k)}$, and $b_1,\ldots,b_k \in B$, the following holds:
  \begin{equation}\label{eq:sigma-algebra-hom=init-sem}
    \theta(\sigma)(b_1,\ldots,b_k) = \bigoplus_{w \in \{0,1\}^k} b_1^{w_1} \otimes \cdots \otimes b_k^{w_k} \otimes b_{\sigma,w}
    \end{equation}
    where $w_i$ denotes the $i$-th symbol of $w$ and $b_i^0 = \1$ and $b_i^1= b_i$. Thus $\theta(\sigma)$ is a polynomial over $k$
    non-commuting  variables with degree at most $1$. Let us denote by $\h_\cB$ the unique $\Sigma$-algebra homomorphism from $\sfT_\Sigma$ to $\cB$.

  We construct the $(\Sigma,\B)$-wta $\cA =(Q,\delta,F)$ such that $\initialsem{\cA} = \h_\cB$, as follows. We let $Q =\{0,1\}$, $F_0=\0$, and $F_1=\1$. Moreover, for every $k \in \mathbb{N}$, $\sigma \in \Sigma^{(k)}$, and $q_1,\ldots,q_k,q \in Q$, we let
  \[
    \delta_k(q_1 \cdots q_k,\sigma,q) =
    \begin{cases}
      b_{\sigma,q_1 \cdots q_k} & \text{ if $q=1$}\\
        \1 & \text{ if $q_1 \cdots q_kq = 0^{k+1}$}\\
        \0 & \text{ otherwise}
      \end{cases}
    \]
In general, $\cA$ is not bu-deterministic, because
    $\delta_k(0^k,\sigma,1) = b_{\sigma,q_1 \cdots q_k}$ and $\delta_k(0^k,\sigma,0) = \1$.

    By induction on $\T_\Sigma$ we  prove the following statement:
    \begin{equation}\label{equ:algebra-hom-turned-into-wta}
\text{For each $\xi \in \T_\Sigma$, we have: $\h_\cA(\xi)_1 = \h_\cB(\xi)$ and $\h_\cA(\xi)_0 = \1$.} 
\end{equation}
Let $\xi = \sigma(\xi_1,\ldots,\xi_k)$. Then we can calculate as follows:
\begingroup\allowdisplaybreaks
\begin{align*}
  \h_\cA(\sigma(\xi_1,\ldots,\xi_k))_1
  &= \bigoplus_{q_1\cdots q_k \in Q^k} \h_\cA(\xi_1)_{q_1} \otimes \cdots \otimes \h_\cA(\xi_k)_{q_k} \otimes \delta_k(q_1 \cdots q_k,\sigma,1)\\
  &= \bigoplus_{q_1\cdots q_k \in Q^k} \h_\cA(\xi_1)_{q_1} \otimes \cdots \otimes \h_\cA(\xi_k)_{q_k} \otimes b_{\sigma,q_1 \cdots q_k} \tag{by construction}\\
  &= \bigoplus_{q_1\cdots q_k \in Q^k} \h_\cB(\xi_1)^{q_1} \otimes \cdots \otimes \h_\cB(\xi_k)^{q_k} \otimes b_{\sigma,q_1 \cdots q_k} \tag{by I.H.}\\
  &= \theta(\sigma)(\h_\cB(\xi_1),\ldots,\h_\cB(\xi_k)) \tag{by definition of $\theta$}\\
  &= \h_\cB(\sigma(\xi_1,\ldots,\xi_k)) \tag{because $\h_\cB$ is a $\Sigma$-algebra homomorphism} \enspace.
\end{align*}
\endgroup
Moreover we can calculate:
\begingroup\allowdisplaybreaks
\begin{align*}
  \h_\cA(\sigma(\xi_1,\ldots,\xi_k))_0
  &= \bigoplus_{q_1\cdots q_k \in Q^k} \h_\cA(\xi_1)_{q_1} \otimes \cdots \otimes \h_\cA(\xi_k)_{q_k} \otimes \delta_k(q_1 \cdots q_k,\sigma,0)\\
  &= \h_\cA(\xi_1)_{0} \otimes \cdots \otimes \h_\cA(\xi_k)_{0} \otimes \1 \tag{by construction}\\
  &= \1 \otimes \cdots \otimes \1 \otimes \1 = \1 \tag{by I.H.} \enspace.
\end{align*}
\endgroup
Then, for each $\xi \in \T_\Sigma$, we have $\initialsem{\cA}(\xi) = \bigoplus_{q \in Q} \h_\cA(\xi)_q \otimes F_q = \h_\cA(\xi)_1 = \h_\cB(\xi)$.

We finish this example with three instances of this general scenario. The first two instances result in wta which we have already constructed in an ad hoc manner, viz., in Example \ref{ex:h-kappa-crisp-det} (evaluation algebras) and in Example \ref{ex:height} (the mapping $\height: \T_\Sigma \to \mathbb{N}$). The third example is related to Theorem~\ref{thm:closure-of-finite-set-i-recognizable} and Corollary \ref{lm:closure-of-finite-set-recognizable}.

\begin{enumerate}
\item Let $(\kappa_k \mid k \in \mathbb{N})$ be an $\mathbb{N}$-indexed family of mappings $\kappa_k:~\Sigma^{(k)}~\to~B$. We consider  the $(\Sigma,\kappa)$-evaluation algebra  $\M(\Sigma,\kappa) = (B,\overline{\kappa})$ and the unique  $\Sigma$-algebra homomorphism $\h_{\M(\Sigma,\kappa)}: \T_\Sigma \to B$ (cf. Subsection \ref{sec:weighted-tree-languages}).
  We note that $\M(\Sigma,\kappa)$ is a particular $\Sigma$-algebra.
  
 Now we define the family \((b_{\sigma,w} \mid k \in \mathbb{N}, \sigma \in \Sigma^{(k)}, w \in \{0,1\}^k) \) over $B$
  such that $b_{\sigma,w} =  \kappa_k(\sigma)$ if $w=1^k$, and $\0$ otherwise, and we define the $\Sigma$-algebra $\cB = (B,\theta)$ as in \eqref{eq:sigma-algebra-hom=init-sem}. 
  Then, for every $b_1,\ldots,b_k \in B$, we have
  \[
    \overline{\kappa}(\sigma)(b_1,\ldots,b_k)= b_1 \otimes \cdots \otimes b_k \otimes \kappa_k(\sigma) = \bigoplus_{w \in \{0,1\}^k} b_1^{w_1} \otimes \cdots \otimes b_k^{w_k} \otimes b_{\sigma,w} =\theta(\sigma)(b_1,\ldots,b_k)\enspace.
  \]
   Hence $\overline{\kappa}=\theta$ and thus $\h_{\M(\Sigma,\kappa)} = \h_{\cB}$. Then, using the general scenario, we can construct a $(\Sigma,\B)$-wta $\cA$ such that $\initialsem{\cA} = \h_{\M(\Sigma,\kappa)}$. In this sense, the construction of the current example subsumes the one given in Example \ref{ex:h-kappa-crisp-det} (where the state $1$ corresponds to state $q$, and state $0$ is superfluous).

\item As strong bimonoid $\B$ we consider the semiring $\Natmaxplus = (\mathbb{N}_{-\infty},\max,+,-\infty,0)$.  Moreover, we extend the $\Sigma$-algebra  $(\mathbb{N},\theta_1)$ defined on page \pageref{page:algebras-for-height-size-pos} to the $\Sigma$-algebra $(\mathbb{N}_{-\infty},\theta_1')$ canonically as follows: 
  \begin{compactitem}
  \item for each $\alpha \in \Sigma^{(0)}$, we let $\theta'_1(\alpha)() = 0$, and
  \item for every $k \in \mathbb{N}_+$, $\sigma \in \Sigma^{(k)}$, and $n_1,\ldots,n_k \in \mathbb{N}_{-\infty}$, we let $\theta_1'(\sigma)(n_1,\ldots,n_k) = 1 + \max(n_1,\ldots,n_k)$. 
\end{compactitem}
Then the unique $\Sigma$-algebra homomorphism from $\sfT_\Sigma$ to $(\mathbb{N}_{-\infty},\theta_1')$ is the mapping $\height: \T_\Sigma \to \mathbb{N}$.

 Now we define the family \((b_{\sigma,w} \mid k \in \mathbb{N}, \sigma \in \Sigma^{(k)}, w \in \{0,1\}^k) \) over $\mathbb{N}_{-\infty}$ such that $b_{\sigma,w} =  1$ if $w$ contains exactly one occurrence of $1$, and $-\infty$ otherwise. Moreover, we define the $\Sigma$-algebra $\cB = (\mathbb{N}_{-\infty},\theta)$ as in~\eqref{eq:sigma-algebra-hom=init-sem}. Then we have
 \begingroup
 \allowdisplaybreaks
 \begin{align*}
   \theta_1'(\sigma)(n_1,\ldots,n_k) &= 1 + \max(n_1,\ldots,n_k) = \max(n_1 +1, \ldots, n_k + 1)\\
                                     &= \max_{w \in \{0,1\}^k} (n_1^{w_1} + \cdots + n_k^{w_k} + b_{\sigma,w}) =\theta(\sigma)(n_1,\ldots,n_k) \enspace.
   \end{align*}
   \endgroup
   Hence $\theta'_1=\theta$ and thus $\height= \h_{\cB}$. Then, using the general scenario,  we can construct a $(\Sigma,\Natmaxplus)$-wta $\cA$ such that $\initialsem{\cA} = \height$. In fact, for $\Sigma=\{\sigma^{(2)},\alpha^{(0)}\}$,  this $(\Sigma,\Natmaxplus)$-wta $\cA$ is the same as the one that we have constructed in Example \ref{ex:height}.

\item In the proof of \cite[Lm.~6.1]{rad10}(2)$\Rightarrow$(1), for each finite subset $A$ of the carrier set of the strong bimonoid $\B$, a  ranked alphabet $\Sigma$  and a $(\Sigma,\B)$-wta $\cA$ are constructed such that $\langle A \rangle_{\{\oplus,\otimes,\0,\1\}} \subseteq \im(\initialsem{\cA})$. Here we use the above general scenario (starting from a particular $\Sigma$-algebra that obeys~\eqref{eq:sigma-algebra-hom=init-sem})  in order to obtain the same $\cA$.\footnote{In fact, the proof of \cite[Ex.~6.1]{rad10}(2)$\Rightarrow$(1) inspired us for this general scenario.}

Let $A \subseteq B$ be a finite subset of $B$.  Clearly, $\langle A \rangle_{\{\oplus,\otimes,\0,\1\}} = \langle A \cup \{\0,\1\} \rangle_{\{\oplus,\otimes\}}$. Let $a_1,\ldots,a_n$ be the elements of $A \cup \{\0,\1\}$, i.e., $A \cup \{\0,\1\} = \{a_1,\ldots,a_n\}$. 

We define the ranked alphabet $\Sigma = \{a^{(0)}_1,\ldots,a^{(0)}_n\} \cup \{\oplus^{(2)}, \otimes^{(2)}\}$. Moreover, we define the family \((b_{\delta,w} \mid k \in \mathbb{N}, \sigma \in \Sigma^{(k)}, w \in \{0,1\}^k) \) over $B$ such that, for every $k \in \mathbb{N}$, $\sigma \in \Sigma^{(k)}$, and $w \in \{0,1\}^k$, we let 
\[
  b_{\delta,w} =
  \begin{cases}
    a_i & \text{ if $\sigma=a_i$ and $w= \varepsilon$}\\
    \1 & \text{ if $\sigma=\oplus$ and $w \in \{10,01\}$}\\
    \1 & \text{ if $\sigma=\otimes$ and $w=11$}\\
    \0 & \text{ otherwise} \enspace.
    \end{cases}
  \]
  Finally, we define the $\Sigma$-algebra $\cB=(B,\theta)$ such that~\eqref{eq:sigma-algebra-hom=init-sem} holds.
  Then we can prove the following statement.
  \begin{eqnarray}
    \begin{aligned}\label{equ:canonical-evaluation-algebra-1}
      &\text{For every $i \in [n]$ and $b_1,b_2 \in B$, we have:}\\
    &       \theta(a_i)() = a_i, \ 
            \theta(\oplus)(b_1,b_2) = b_1 \oplus b_2 \enspace,  \text{ and } 
            \theta(\otimes)(b_1,b_2) = b_1 \otimes b_2 \enspace.
  \end{aligned}
  \end{eqnarray}

   Let $i\in [n]$. Then
\(\theta(a_i)() = \bigoplus_{w \in \{0,1\}^0} b_1^{w_1} \otimes \cdots \otimes b_k^{w_k} \otimes b_{a_i,w} =  b_{a_i,\varepsilon} =  a_i\).
  
Let $b_1,b_2 \in B$. Then:
\begin{align*}
  \theta(\oplus)(b_1,b_2)
  &= \bigoplus_{w \in \{0,1\}^2} b_1^{w_1} \otimes b_2^{w_2} \otimes b_{\oplus,w}\\
  &= (b_1^1 \otimes b_2^0 \otimes b_{\oplus,10}) \oplus (b_1^0 \otimes b_2^1 \otimes b_{\oplus,01})\\
  &= (b_1 \otimes \1 \otimes \1) \oplus (\1 \otimes b_2 \otimes \1)
    =  b_1 \oplus b_2 \ \ \text{ and  }\\
  \theta(\otimes)(b_1,b_2)
  &= \bigoplus_{w \in \{0,1\}^2} b_1^{w_1} \otimes b_2^{w_2} \otimes b_{\otimes,w}\\
  &= b_1^1 \otimes b_2^1 \otimes b_{\otimes,11} = b_1 \otimes b_2 \enspace. 
\end{align*}
This proves \eqref{equ:canonical-evaluation-algebra-1}.
Clearly,  \eqref{equ:canonical-evaluation-algebra-1} and Lemma~\ref{obs:Knaster-Tarski-applied-to-algebras} imply the next statement.
\begin{equation}\label{equ:for-each-a-there-is-a-tree}
\text{For each $a \in \langle A \rangle_{\{\oplus,\otimes,\0,\1\}}$ there exists $\xi \in \T_\Sigma$ such that $\h_\cB(\xi) = a$.}
  \end{equation}
  
  Now, given the $\Sigma$-algebra $\cB=(B,\theta)$, we use our general scenario to construct the $(\Sigma,\B)$-wta $\cA$; then $\initialsem{\cA} = \h_\cB$. Thus, \eqref{equ:for-each-a-there-is-a-tree} implies that $\langle A \rangle_{\{\oplus,\otimes,\0,\1\}} \subseteq \im(\initialsem{\cA})$. 
   \hfill$\Box$
  \end{enumerate}
\end{example}

\section{The state algebra and the support fta of a wta}
\label{sec:state-algebra-of-wta}

It will turn out as a useful perspective on a $(\Sigma,\B)$-wta $\cA=(Q,\delta,F)$, to consider only the state behaviour of $\cA$ on input trees, where the state behaviour is based on the non-zero weighted transitions. We formalize this perspective as the concept of state algebra of $\cA$.

\begin{quote}\emph{In this section, we let $\cA=(Q,\delta,F)$ be an arbitrary $(\Sigma,\B)$-wta, unless specified otherwise.}
  \end{quote}

\index{deltaQ@$\delta^Q$}
\index{state algebra}
\index{StA@$\St(\cA)$}
The \emph{state algebra of $\cA$}, denoted by $\St(\cA)$, is the $\Sigma$-algebra $\St(\cA)=(\cP(Q),\delta_Q)$ where, for every $k \in \mathbb{N}$, $\sigma \in \Sigma^{(k)}$, and $P_1,\ldots,P_k \in \cP(Q)$, we let
\begin{equation*}
\delta_Q(\sigma)(P_1,\ldots,P_k) = \{q \in Q \mid (\exists q_1 \in P_1) \ldots (\exists q_k \in P_k): \delta_k(q_1 \cdots q_k,\sigma,q)\ne \0\} \enspace.
\end{equation*}
  \index{stateA@$\state_\cA$}
We denote the $\Sigma$-algebra homomorphism from $\sfT_\Sigma$ to $\St(\cA)$ by $\state_\cA$. If $\cA$ is clear from the context, then we drop $\cA$ from $\state_\cA$ and simply write $\state$.

  Obviously, for every $k \in \mathbb{N}$, $\sigma \in \Sigma^{(k)}$, and $P_1,\ldots,P_k \in \cP(Q)$, if there exists $i \in [k]$ such that $P_i=\emptyset$, then $\delta_Q(\sigma)(P_1,\ldots,P_k)=\emptyset$. It is easy to check that this annihilation propagates over trees in the following sense.

  \begin{observation}\rm \label{obs:deltaQ-is-strict-in-emptyset}
    For every $\xi \in \T_\Sigma$ and $w \in \pos(\xi)$, we have that $\state(\xi|_w) = \emptyset$ implies $\state(\xi) = \emptyset$.
  \end{observation}

We define the sets
\[\cP_{= 1}(Q)= \{U \in \cP(Q) \mid |U| = 1 \} \ \text{ and }\ \cP_{\le 1}(Q)= \{U \in \cP(Q) \mid |U| \le 1\}.\]
Thus, $\cP_{=1}(Q)= \{\{q\} \mid q \in Q\}$ and  $\cP_{\le 1}(Q) = \cP_{=1}(Q) \cup \{\emptyset\}$.

\begin{observation}\rm \label{obs:subalgebras-of-the-state-algebra} The following two statements hold.
  \begin{compactenum}
  \item[(1)] If $\cA$ is bu-deterministic, then $(\cP_{\le 1}(Q),\delta_Q)$ is a subalgebra of $\St(\cA)$ and $\im(\state) \subseteq \cP_{\le 1}(Q)$.
  \item[(2)] If $\cA$ is total and bu-deterministic, then $(\cP_{=1}(Q),\delta_Q)$ is a subalgebra of~$\St(\cA)$ and $\im(\state) \subseteq \cP_{=1}(Q)$.
    \end{compactenum}
  \end{observation}
  \begin{proof}
    Proof of (1): Since $\cA$ is bu-deterministic, for every $k \in \mathbb{N}$, $\sigma \in \Sigma^{(k)}$, and $w \in Q^k$, we have $|\mathrm{succ}(w,\sigma)| \le 1$. Then it is easy to check that $\cP_{\le 1}(Q)$ is closed under $\delta_Q(\Sigma)$. Hence $(\cP_{\le 1}(Q),\delta_Q)$ is a subalgebra of $\St(\cA)$. By Observation \ref{obs:smallest-subalgebra-im}, the $\Sigma$-algebra  $(\im(\state),\delta_Q)$ is the smallest subalgebra of $\St(\cA)=(\cP(Q),\delta_Q)$. Thus $\im(\state) \subseteq \cP_{\le 1}(Q)$.

    \
    
Proof of (2): It is similar to the proof of Statement (1). However, here we use that $\cA$ is 
total and bu-deterministic and thus, for every $k \in \mathbb{N}$, $\sigma \in \Sigma^{(k)}$, and $w \in Q^k$, we have $|\mathrm{succ}(w,\sigma)| = 1$.
\end{proof}

\index{support fta}
\index{supp@$\supp(\cA)$}
The state algebra of the $(\Sigma,\B)$-wta $\cA$  can also be derived from the concept of the $\Sigma$-algebra associated with the support fta of $\cA$.
The \emph{support fta of $\cA$}, denoted by $\supp_\B(\cA)$, is the $\Sigma$-fta $(Q,\delta',F')$ where, for each $k \in \mathbb{N}$, we let $\delta_k' = \supp_\B(\delta_k)$ and $F' = \supp_\B(F)$. If $\B$ is clear from the context, then we drop $\B$ from $\supp_\B$.
Then it is easy to see that the $\Sigma$-algebra $(\cP(Q),\delta_{\supp(\cA)})$ associated with $\supp(\cA)$ (cf. Section~\ref{sec:fta}) is the same as the state algebra of $\cA$, i.e.,  for every $k \in \mathbb{N}$ and $\sigma \in \Sigma^{(k)}$, we have
\begin{equation}\label{equ:state-algebra=algebra-associated-with-supp-fta}
\delta_{\supp(\cA)}(\sigma) = \delta_Q(\sigma), \ \text{ and thus $\h_{\supp(\cA)} = \state_\cA$}\enspace,
\end{equation}
where $\h_{\supp(\cA)}$ denotes the $\Sigma$-algebra homomorphism from $\sfT_\Sigma$ to  $(\cP(Q),\delta_{\supp(\cA)})$.

\

The state algebra can be used to formulate, for every $\xi \in \T_\Sigma$ and $q \in Q$, necessary and sufficient conditions for $\h_\cA(\xi)_q\ne \0$ and for  the existence of a run $\rho \in \R_\cA(q,\xi)$ with $\wt(\xi,\rho) \ne \0$.

\pagebreak[2]
\begin{lemma}\rm \label{lm:hAxiqne0-implies-qinhQxi} The following four statements hold.
  \begin{compactenum}
  \item[(1)]  For every $\xi \in \T_\Sigma$ and $q \in Q$, if $\h_\cA(\xi)_q \ne \0$, then $q \in \state(\xi)$.
  \item[(2)]  Let $\B$ be positive. For every $\xi \in \T_\Sigma$ and $q \in Q$, if $q \in \state(\xi)$, then $\h_\cA(\xi)_q \ne \0$.
        \end{compactenum}
Moreover,
   \begin{compactenum}
   \item[(3)] For every $\xi \in \T_\Sigma$ and $q \in Q$, if there exists $\rho \in \R_\cA(q,\xi)$ such that $\wt(\xi,\rho) \ne \0$, then $q \in \state(\xi)$.
    \item[(4)] Let $\B$ be zero-divisor free or $\cA$ be crisp. For every $\xi \in \T_\Sigma$ and $q \in Q$, if $q \in \state(\xi)$, then there exists a run $\rho \in \R_\cA(q,\xi)$ such that $\wt(\xi,\rho) \ne \0$.
    \end{compactenum}
  \end{lemma}

  \begin{proof}
Proof of (1): We prove the statement by induction on $\T_\Sigma$. Let $\xi = \sigma(\xi_1,\ldots,\xi_k)$. Then, ignoring the upper indices at $\Rightarrow$ for the time being, we can argue as follows.
    \begingroup
    \allowdisplaybreaks
    \begin{align*}
      & \h_\cA(\xi)_q \not= \0 \\
      \Leftrightarrow \hspace*{6mm} &  \Big(\bigoplus_{q_1 \cdots q_k \in Q^k} \h_\cA(\xi_1)_{q_1} \otimes \ldots \otimes \h_\cA(\xi_k)_{q_k} \otimes \delta_k(q_1\cdots q_k,\sigma,q)\Big) \not= \0\\
        \Rightarrow^{(1)} \hspace*{2mm}  & ( \exists q_1 \cdots q_k \in Q^k):   \h_\cA(\xi_1)_{q_1} \otimes \ldots \otimes \h_\cA(\xi_k)_{q_k} \otimes \delta_k(q_1\cdots q_k,\sigma,q) \not= \0\\
      \Rightarrow^{(2)} \hspace*{2mm}  & ( \exists q_1 \cdots q_k \in Q^k): \Big((\forall i \in [k]): \h_\cA(\xi_i)_{q_i}  \not= \0 \Big)  \text{ and }  \delta_k(q_1\cdots q_k,\sigma,q) \not= \0\\
      \Rightarrow^{(3)} \hspace*{2mm} & ( \exists q_1 \cdots q_k \in Q^k):  \Big( (\forall i \in [k]): q_i \in \state(\xi_i)\Big) 
                           \text{ and }  \delta_k(q_1\cdots q_k,\sigma,q) \not= \0
      \tag{by I.H.}\\
      \Leftrightarrow \hspace*{6mm}  &  q \in \delta_Q(\sigma)\big(\state(\xi_1),\ldots,\state(\xi_k)\big) 
                       \tag{by definition of $\delta_Q(\sigma)$}\\ 
      \Leftrightarrow \hspace*{6mm}  & q \in \state(\xi)
                           \tag{because $\state$ is a $\Sigma$-algebra homomorphism}\enspace.
    \end{align*}
    \endgroup

  \

       Proof of (2): 
 Let $\B$ positive. We prove the statement by induction on $\T_\Sigma$ and by proceeding backwards in the calculation of the proof of Statement (1). Moreover,
 \begin{compactitem}
 \item we replace $\Rightarrow^{(3)}$ by $\Leftarrow$, this is justified by  I.H. of Statement (2),
  \item we replace $\Rightarrow^{(2)}$ by $\Leftarrow$, this is justified  because $\B$ is zero-divisor free, and
    \item We replace $\Rightarrow^{(1)}$ by $\Leftarrow$, this is justified  because $\B$ is zero-sum free.
    \end{compactitem}
    
    \

    Proof of (3): We prove the statement by induction on $\T_\Sigma$. Let $\xi = \sigma(\xi_1,\ldots,\xi_k)$. Then, ignoring the upper indices at $\Rightarrow$ for the time being, we can argue as follows.
    \begingroup
    \allowdisplaybreaks
    \begin{align*}
      & (\exists \rho \in \R_\cA(q,\xi)): \wt(\xi,\rho) \ne \0 \\
      \Leftrightarrow \hspace*{6mm} & (\exists \rho \in \R_\cA(q,\xi)): \Big( \bigotimes_{i \in [k]} \wt(\xi_i,\rho|_i)\Big) \otimes \delta_k(\rho(1) \cdots \rho(k), \sigma,q) \ne \0
                                      \tag{by \eqref{equ:weight-of-run}}\\
      \Rightarrow^{(1)} \hspace*{2mm} & (\exists q_1  \cdots q_k \in Q^k) \Big((\forall i \in [k])(\exists \rho_i \in \R_\cA(q_i,\xi_i)): \wt(\xi_i,\rho_i)\ne \0 \Big) \text{ and } \delta_k(q_1 \cdots q_k, \sigma,q) \ne \0
                                  \tag{by annihilation law in $\B$; choose $q_i=\rho(i)$ and $\rho_i = \rho|_i$ for each $i \in [k]$}\\
  \Rightarrow^{(2)} \hspace*{2mm} & (\exists q_1  \cdots q_k \in Q^k)\Big((\forall i \in [k]): q_i \in \state(\xi_i) \Big)  \text{ and } \delta_k(q_1 \cdots q_k, \sigma,q) \ne \0
                              \tag{by I.H.}\\
      \Leftrightarrow \hspace*{6mm} & q \in \state(\xi)
                                      \tag{as in the proof of Statement (1)} \enspace.
    \end{align*}
    \endgroup

 \

 Proof of (4): Let $\B$ be zero-divisor free. We prove the statement by induction on $\T_\Sigma$ and by proceeding backwards in the calculation of the proof of Statement (3). Moreover,
 \begin{compactitem}
 \item we replace $\Rightarrow^{(2)}$ by $\Leftarrow$, this is justified by  I.H. of Statement (4) and 
 \item we replace $\Rightarrow^{(1)}$ by $\Leftarrow$, this is justified  because $\rho$ can be defined by $\rho(\varepsilon)=q$ and $\rho|_i=\rho_i$ for each $i\in [k]$ and, if $\cA$ is crisp, then $\delta_k(q_1\cdots q_k,\sigma,q) =\1$ and $\wt(\xi_i,\rho_i)=\1$ for each $i \in [k]$ and hence $(\bigotimes_{i \in [k]} \wt(\xi_i,\rho_i)) \otimes \delta_k(q_1\cdots q_k,\sigma,q) =\1$,   otherwise $\B$ is zero-divisor free.
   \qedhere
    \end{compactitem}
      \end{proof}

    Lemma \ref{lm:hAxiqne0-implies-qinhQxi} implies the following equivalence.

    \begin{corollary}\rm \label{cor:hAxiqne0-equiv-qinstatexi-equiv-exists-run}  Let $\B$ be positive. For every  $\xi \in \T_\Sigma$ and $q \in Q$, we have:
      \[
\h_\cA(\xi)_q \ne \0 \ \ \Leftrightarrow \ \ q \in \state(\xi) \ \ \Leftrightarrow \ \ \text{ there exists $\rho \in \R_\cA(q,\xi)$ such that $\wt(\xi,\rho) \ne \0$} \enspace.
        \]
  \end{corollary}

We note that, in general, the other direction of Lemma~\ref{lm:hAxiqne0-implies-qinhQxi}(1) and (3) do not hold already for total and bu-deterministic wta. For this consider, e.g., the ring $\Intfour$ with four elements (cf. Example \ref{ex:semirings}(\ref{def:ring-Zmod4Z})), the ranked alphabet $\Sigma=\{\gamma^{(1)},\alpha^{(0)}\}$, and  the total and bu-deterministic $(\Sigma,\Intfour)$-wta $\cA=(Q,\delta,F)$ with $Q=\{q\}$, $\delta_0(\varepsilon,\alpha,q)=\delta_1(q,\gamma,q)=2$, and $F_q=1$. Then $\state(\gamma(\alpha)) = \{q\}$ and $\h_\cA(\gamma(\alpha))_q = 2 \cdot_4 2 = 0$. Moreover, for the unique run $\rho \in \R_\cA(\gamma(\alpha))$, we also have $\wt_\cA(\gamma(\alpha),\rho) = 2 \cdot_4 2 = 0$.

\section{Extension of the weight algebra}
\label{sect:extension-of-weight-structure}

We finish this chapter with two easy observations on the extension of the weight algebra of a $(\Sigma,\B)$-wta~$\cA$. If $\sfC$ is a strong bimonoid such that $\B$ is a subalgebra of $\sfC$, then we say that $\sfC$ is an \emph{extension} of $\B$. If this is the case, then we can view each $(\Sigma,\B)$-wta~$\cA$ as a $(\Sigma,\sfC)$-wta. Moreover, the run semantics and the initial algebra semantics of the $(\Sigma,\sfC)$-wta $\cA$ are the same as the corresponding semantics of the $(\Sigma,\B)$-wta $\cA$. Thus we obtain the following observation.

\begin{observation}\rm \label{obs:extension-of-weight-structure} Let the strong bimonoid $\sfC$ be an extension of $\B$. Moreover, let $\cA$ be a $(\Sigma,\sfB)$-wta. Then $\initialsem{\cA} \in \Rec^{\mathrm{init}}(\Sigma,\sfC)$ and  $\runsem{\cA} \in \Rec^{\mathrm{run}}(\Sigma,\sfC)$. \hfill $\Box$
\end{observation}

One might also think of the following inverse problem. Let the strong bimonoid $\sfC$ be an extension of $\B$ and let $\cA$ be a $(\Sigma,\sfC)$-wta with $\mathrm{wts}(\cA) \subseteq B$. Since $\B$ is a subalgebra of $\sfC$, we can view $\cA$ as a $(\Sigma,\B)$-wta. Moreover, the run semantics and the initial algebra semantics of the $(\Sigma,\sfC)$-wta $\cA$ are the same as the corresponding semantics of the $(\Sigma,\B)$-wta $\cA$, and hence we obtain the following.

\begin{observation}\rm \label{obs:extension-of-weight-structure-inverse} Let the strong bimonoid $\sfC$ be an extension of $\B$. Moreover, let $\cA$ be a $(\Sigma,\sfC)$-wta with $\mathrm{wts}(\cA) \subseteq B$. Then $\initialsem{\cA} \in \Rec^{\mathrm{init}}(\Sigma,\B)$ and  $\runsem{\cA} \in \Rec^{\mathrm{run}}(\Sigma,\B)$.
\hfill $\Box$
\end{observation}

%% file: special-wta.tex
\chapter{Special cases of weighted tree automata}
\label{ch:special-cases}

In this chapter, we deal with special cases of weighted tree automata where
\begin{compactenum}
\item[(a)] the trees have a special form, viz. they are strings (cf. Section~\ref{sect:string-automata}),
\item[(b)] the weight algebra is the Boolean semiring (cf. Section~\ref{sec:wta-Boolean-semiring}),
\item[(c)] the weight algebra is the semiring of natural numbers (cf. Section~\ref{sect:wta-N}), and
  \item[(d)] the weight algebra is a commutative semiring (cf. Section~\ref{sec:wta-commutative-semirings})
  \end{compactenum}
  These special cases connect the concept of weighted tree automata with the concepts of weighted string automata, finite-state tree automata, multiplicities in finite-state tree automata, and multilinear representations, respectively.

\section{Weighted string automata}
\label{sect:string-automata}

In this section we recall the concept of weighted string automata \cite{sch61,eil74} and show that, essentially, they are equivalent to wta over string ranked alphabets (cf. \cite[p.~324]{fulvog09new}). Moreover, we show that they generalize fsa.
We recall the run semantics and the initial algebra semantics of weighted string automata; in Chapter~\ref{sec:monoid-representation-of-wta} we deal with the free monoid semantics of weighted string automata (cf. Corollary~\ref{cor:cor-VI.6.2-eil74}).

\index{weighted string automaton}
\index{wsa}
A \emph{weighted string automaton (over $\Gamma$ and  $\B$)} 
(for short: $(\Gamma,\B)$-wsa or wsa)   is a tuple
$\cA = (Q,\lambda,\mu,\gamma)$, where $Q$ is a finite nonempty set of
states such that $Q \cap \Gamma = \emptyset$, $\mu: \Gamma \rightarrow B^{Q \times Q}$ is the transition
mapping, and  $\lambda, \gamma \in B^Q$ are the initial weight mapping and the final weight mapping, respectively. We define the run semantics and the initial algebra semantics for $\cA$ as follows. 

\begin{quote}\emph{In the rest of this section, $\cA = (Q,\lambda,\mu,\gamma)$ denotes an arbitrary weighted string automaton over $\Gamma$ and  $\B$ unless specified otherwise.}
\end{quote}

\index{wt@$\wt_\cA(w,\rho)$}
\index{wt@$\wt_\cA^-(w,\rho)$}
\paragraph{Run semantics.} Let $w=a_1\cdots a_n$ be a string in $\Gamma^*$ with $n \in \mathbb{N}$ and $a_1,\ldots,a_n \in \Gamma$. A {\em run of $\cA$ on $w$} is a string $\rho \in Q^{n+1}$. 
Let $\rho = q_0\cdots q_n$ with $q_i \in Q$ for each $i \in [0,n]$.
The {\em weight of $\rho$ for $w$}, denoted by $\wt_\cA(w,\rho)$, is the element of $B$ defined by 
\[\wt_\cA(w,\rho) = \lambda_{q_0}\otimes \wt_\cA^-(w,\rho) \otimes \gamma_{q_n}\]
where
\[\wt_\cA^-(w,\rho) = \mu(a_1)_{q_0,q_1}\otimes \ldots \otimes \mu(a_n)_{q_{n-1},q_n}\enspace.\]
Thus, in particular, we have that $\wt_\cA^-(\varepsilon,\rho)=\mathbb{1}$ and $\wt_\cA(\varepsilon,\rho) = \lambda_{q_0}\otimes \gamma_{q_0}$.

\index{semanticA@$\runsem{\cA}$}
The {\em run semantics of $\cA$} is the weighted language
$\runsem{\cA}:\Gamma^* \to B$ defined by 
\[\runsem{\cA}(w)=\bigoplus_{\rho\in Q^{|w|+1}}\wt_\cA(w,\rho)\] 
for every $w\in \Gamma^*$. In particular, $\runsem{\cA}(\varepsilon) = \bigoplus_{q \in Q} \lambda_{q}\otimes \gamma_{q}$.

\index{run recognizable}
\index{r-recognizable}
A weighted language $r: \Gamma^* \rightarrow B$ is \emph{run recognizable (over $\B)$} (for short: r-recognizable)
if there exists a $(\Gamma,\B)$-wsa $\cA$ such that $r = \runsem{\cA}$.

In Corollary \ref{cor:cor-VI.6.2-eil74}, we will prove the classical monoid representation of wsa \cite[Cor.~VI.6.2]{eil74}.

\index{unit initial weights}
We finish this paragraph with a useful restriction of wsa.
A $(\Gamma,\B)$-wsa $\cA = (Q,\lambda,\mu,\gamma)$ has \emph{unit initial weights} if $\im(\lambda) \subseteq \{\0,\1\}$.

\begin{lemma}\rm \label{lm:wsa-with-identity-initial-weights} \cite{renroslivog23} For each $(\Gamma,\B)$-wsa $\cA$, we can construct a $(\Gamma,\B)$-wsa $\cA'$ such that $\cA'$ has unit initial weights and $\runsem{\cA'} = \runsem{\cA}$.
\end{lemma}
\begin{proof} Let $\cA = (Q,\lambda,\mu,\gamma)$. We construct $\cA'=(Q',\lambda',\mu',\gamma')$ such that
  \begin{compactitem}
  \item $Q' = Q \cup \{\overline{q} \mid q \in Q\}$,
  \item for each $p \in Q'$, we let
    \[\lambda'_p = \begin{cases}\1 & \text{ if $(\exists q \in Q): p = \overline{q}$}\\
        \0 &\text{ otherwise}\end{cases} \ \ \text{ and } \ \
   \gamma'_p = \begin{cases}\lambda_q \otimes \gamma_q & \text{ if $(\exists q \in Q): p = \overline{q}$}\\
        \gamma_p &\text{ otherwise}\end{cases} \]
  \item for every $p_1,p_2 \in Q'$ and $a \in \Gamma$, we let
    \[
      \mu'(p_1,a,p_2) = \begin{cases}
\lambda_q \otimes \mu(q,a,p_2) & \text{ if $(\exists q \in Q): p_1 = \overline{q}$ and $p_2 \in Q$}\\
\mu(p_1,a,p_2) &\text{ if $p_1,p_2 \in Q$}\\
\0 & \text{ otherwise}
        \end{cases}
      \]
    \end{compactitem}
    Obviously, $\cA'$  has unit initial weights.   Next we prove that  $\runsem{\cA'} = \runsem{\cA}$.
    Let $w \in \Gamma^*$. We proceed by case analysis.

    \underline{$w= \varepsilon$:}
    \begin{align*}
      \runsem{\cA'}(w)&=\bigoplus_{p\in Q'}\lambda'_p \otimes \gamma'_p
      =\bigoplus_{p\in Q \cup \{\overline{q} \mid q \in Q\}}\lambda'_p \otimes \gamma'_p\\
                      &= \bigoplus_{p\in \{\overline{q} \mid q \in Q\}}\lambda'_p \otimes \gamma'_p \tag{because $\lambda'_p=\0$ for each $p \in Q$}\\
                      &= \bigoplus_{q \in Q}\1 \otimes (\lambda_q \otimes \gamma_q) \tag{by construction}\\
      &= \bigoplus_{q \in Q} \lambda_q \otimes \gamma_q = \runsem{\cA}(w) \enspace.
    \end{align*}

    \underline{$w= a_1 \cdots a_n$ with $n \ge 1$:}
    \begingroup
    \allowdisplaybreaks
    \begin{align*}
      \runsem{\cA'}(w)&=\bigoplus_{\rho\in (Q')^{|w|+1}}\wt_{\cA'}(w,\rho)\\
                      &=\bigoplus_{p_0\cdots p_n \in (Q')^{|w|+1}} \lambda'_{p_0}\otimes \mu'(a_1)_{p_0,p_1}\otimes \mu'(a_2)_{p_1,p_2}\otimes \ldots \otimes \mu'(a_n)_{p_{n-1},p_n} \otimes \gamma'_{p_n}\\
                      &=\bigoplus_{q_0 \in Q, p_1\cdots p_n \in (Q')^{|w|}} \lambda'_{\overline{q_0}}\otimes \mu'(a_1)_{\overline{q_0},p_1}\otimes \mu'(a_2)_{p_1,p_2}\otimes \ldots \otimes \mu'(a_n)_{p_{n-1},p_n} \otimes \gamma'_{p_n} \tag{because $\lambda'_{p_0}=\0$ for each $p_0 \in Q$}\\
                      &=\bigoplus_{q_0 \in Q, p_1\cdots p_n \in (Q')^{|w|}} \1 \otimes \mu'(a_1)_{\overline{q_0},p_1}\otimes \mu'(a_2)_{p_1,p_2}\otimes \ldots \otimes \mu'(a_n)_{p_{n-1},p_n} \otimes \gamma'_{p_n} \tag{by definition of $\lambda'$}\\
                      &=\bigoplus_{q_0 \in Q, q_1\cdots q_n \in Q^{|w|}} \mu'(a_1)_{\overline{q_0},q_1}\otimes \mu'(a_2)_{q_1,q_2}\otimes \ldots \otimes \mu'(a_n)_{q_{n-1},q_n} \otimes \gamma'_{q_n} \tag{because the values of $\mu'(a_i)$ on other arguments are equal to $\0$}\\
                      &=\bigoplus_{q_0 \in Q, q_1\cdots q_n \in Q^{|w|}} (\lambda_{q_0} \otimes \mu(a_1)_{q_0,q_1})\otimes \mu(a_2)_{q_1,q_2}\otimes \ldots \otimes \mu(a_n)_{q_{n-1},q_n} \otimes \gamma_{q_n} \tag{by construction}\\
      &= \runsem{\cA}(w) \enspace. \qedhere
    \end{align*}
    \endgroup
  \end{proof}

\paragraph{Initial algebra semantics.}
Let  $e\not\in\Gamma$. For the definition of the initial algebra semantics, we use the string ranked alphabet $\Gamma_e$ and the $\Gamma_e$-algebra $(\Gamma^*,\widehat{\Gamma_e})$ defined in Paragraph ``String-like terms'' of Section~\ref{sect:trees}.

\index{semanticA@$\initialsem{\cA}$}
\index{homA@$\h_\cA$}
Also, we define the $\Gamma_e$-algebra $(B^Q,\theta)$ with
\[\theta(e)()=\lambda \ \ \text{ and } \ \ \theta(a)(v)= v\cdot \mu(a) \ \text{ (vector-matrix product, see Section \ref{sec:vectors-matrices})}
\]
for every $v\in B^Q$ and  $a\in \Gamma$.
By \eqref{eq:ismorphism}, $(\Gamma^*,\widehat{\Gamma_e})$ is initial in the set of all $\Gamma_e$-algebras, hence there exists a unique $\Gamma_e$-algebra homomorphism  from $(\Gamma^*,\widehat{\Gamma_e})$ to $(B^Q,\theta)$. Let us denote this by $\h_\cA$. Then the {\em initial algebra semantics of $\cA$} is the weighted language
$\initialsem{\cA}:\Gamma^* \to B$ defined by 
\[\initialsem{\cA}(w)=\bigoplus_{q\in Q}\h_\cA(w)_q\otimes \gamma_q\] 
for every $w\in \Gamma^*$. We note that, using another nullary symbol $e'$ instead of $e$, leads to the same weighted language $\initialsem{\cA}$. 

\index{initial algebra recognizable}
\index{i-recognizable}
A weighted  language $r: \Gamma^* \rightarrow B$ is \emph{initial algebra recognizable (over $\B$)} (for short: i-recognizable) if there exists a $(\Gamma,\B)$-wsa $\cA$ such that $r = \initialsem{\cA}$.

\paragraph{Embedding wsa into wta.}
Now let $\cA = (Q,\lambda,\mu,\gamma)$ be a $(\Gamma,\B)$-wsa, $\Sigma$ a string ranked alphabet, and $\cB=(Q,\delta,F)$
a $(\Sigma,\B)$-wta. We say that  $\cA$ and $\cB$ are {\em related} if
\begin{compactitem}
\item $\Sigma = \Gamma_e$ for some $e \not\in \Gamma$,
\item for every $a\in \Sigma^{(1)}$ and $p,q\in Q$, we have
 $\delta_0(\varepsilon,e,q)=\lambda_q$, and $\delta_1(p,a,q)=\mu(a)_{p,q}$, and 
\item $F=\gamma$.
\end{compactitem}

For the proof of the next lemma, we recall that  $\tree_e$ is an isomorphism from the $\Gamma_e$-algebra $(\Gamma^*,\widehat{\Gamma_e})$ to the 
$\Gamma_e$-term algebra $(\T_{\Gamma_e},\theta_{\Gamma_e})$, cf. Paragraph ``String-like terms'' in Section \ref{sect:trees}.

\begin{lemma}\rm \label{lm:wsa-wta-over-string-ra-related} Let $\Sigma$ be a string ranked alphabet, $\cA = (Q,\lambda,\mu,\gamma)$ a $(\Gamma,\B)$-wsa,  and $\cB=(Q,\delta,F)$ a $(\Sigma,\B)$-wta. If $\cA$ and $\cB$ are related, then
\(\runsem{\cA}=\runsem{\cB}\circ \tree_e\) and \(\initialsem{\cA}=\initialsem{\cB}\circ \tree_e\),
where $e$ denotes the unique nullary element of $\Sigma$.
\end{lemma}
\begin{proof}
To show that $\runsem{\cA}=\runsem{\cB}\circ \tree_e$, let $w\in \Gamma^*$ with $w=a_1\cdots a_n$ for some $n\in \mathbb{N}$. 
It is obvious that there exists a bijection $\mapsto$ between $Q^{|w|+1}$ and $\R_{\cB}(\tree_e(w))$
such that, for $\rho=q_0\cdots q_n$, we have $\rho\mapsto \rho'$, where $\rho'(1^i)=q_{n-i}$ for every $i\in [0,n]$ (note that
$\pos(\tree_e(w))=\{1^i \mid i\in [0,n]\})$. Moreover, if  $\rho\mapsto \rho'$, then due to Observation \ref{obs:weight-run-explicit}  we also have
\[
  \wt_\cA(w,\rho)=\wt_\cB(\tree_e(w),\rho')\otimes F_{q_n}\enspace,
\]
where $\wt_\cB(\tree_e(w),\rho')$ denotes the weight of $\rho'$ on $\tree_e(w)$ for $\cB$. Then
\begin{equation*}
\runsem{\cA}(w)=\bigoplus_{\rho\in Q^{|w|+1}}\wt_\cA(w,\rho)=\bigoplus_{\rho'\in \R_{\cB}(\tree_e(w))}\wt_\cB(\tree_e(w),\rho')\otimes F_{\rho'(\varepsilon)}=\runsem{\cB}(\tree_e(w)),
\end{equation*}
where the second equality follows from the bijection $\mapsto$ described above and the fact that $\rho'(\varepsilon)=q_n$.

Now we show that $\initialsem{\cA}=\initialsem{\cB}\circ \tree_e$.
Since $\tree_e: \Gamma^* \to \T_{\Gamma_e}$ and $\h_\cB: \T_{\Gamma_e} \to B^Q$ are $\Gamma_e$-algebra homomorphisms, it follows from Theorem \ref{thm:comp-hom} that $(\h_\cB \circ \tree_e): \Gamma^* \to B^Q$ is a $\Gamma_e$-algebra homomorphism. Since $\h_\cA$ is the unique $\Gamma_e$-algebra homomorphism from $(\Gamma^*,\widehat{\Gamma_e})$ to $(B^Q,\theta)$, we have $\h_\cA = \h_\cB \circ \tree_e$.

 Then we obtain
 \begin{align*}\initialsem{\cA}(w)=\bigoplus_{q\in Q}\h_\cA(w)_q\otimes \gamma_q= \bigoplus_{q\in Q'}\h_\cB(\tree_e(w))_q\otimes F_q=\initialsem{\cB}(\tree_e(w))\enspace. \hspace{10mm} \qedhere
   \end{align*}
\end{proof}

\begin{lemma-rect}\rm \label{lm:wsa=wta-over-string-ra} Let $\Gamma$ be an alphabet and $\B$ be a strong bimonoid. Then the following two statements hold.
\begin{compactenum} 
\item[(1)] For every $(\Gamma,\B)$-wsa $\cA$ and  $e \not\in \Gamma$, we can construct a $(\Gamma_e,\B)$-wta $\cB$ such that $\runsem{\cA}=\runsem{\cB}\circ \tree_e$ and $\initialsem{\cA}=\initialsem{\cB}\circ \tree_e$.
\item[(2)] For every string ranked alphabet $\Gamma_e$ and $(\Gamma_e,\B)$-wta $\cB$, we can construct a $(\Gamma,\B)$-wsa $\cA$ such that $\runsem{\cA}=\runsem{\cB}\circ \tree_e$ and $\initialsem{\cA}=\initialsem{\cB}\circ \tree_e$.
\end{compactenum}
\end{lemma-rect}
\begin{proof} First we prove  Statement (1). For a given $(\Gamma,\B)$-wsa $\cA$, it is easy to construct a $(\Gamma_e,\B)$-wta $\cB$ such that $\cA$ and $\cB$ are related. Then the statement follows from Lemma \ref{lm:wsa-wta-over-string-ra-related}.
Statement (2) follows analogously.
\end{proof}


\paragraph{Weighted string automata over the Boolean semiring.}
Let $\cA=(Q,\lambda,\mu,\gamma)$ be a $(\Gamma,\Boole)$-wsa and $A=(Q,I,\delta,F)$ be a $\Gamma$-fsa. We say that $\cA$ and $A$ are \emph{related} if $I=\supp(\lambda)$, $F=\supp(\gamma)$, and for every $a \in \Gamma$ and $q,q' \in Q$ we have: $\mu(a)_{q,q'} = 1$ iff $(q,a,q') \in \delta$.

It is easy to see that, if $\cA$ and $A$ are related, then $\supp(\runsem{\cA}) = \LL(A)$. Moreover, for each $(\Gamma,\Boole)$-wsa, we can construct a $\Gamma$-fsa  $A$ such that $\cA$ and $A$ are related; and vice versa, for each $\Gamma$-fsa $A$, we can construct a $(\Gamma,\Boole)$-wsa  $\cA$ such that $\cA$ and $A$ are related. Thus we obtain the following equivalence between $\Gamma$-fsa and $(\Gamma,\Boole)$-wsa. 

\begin{observation}\rm \label{obs:fsa=wsa(B)} Let $L \subseteq \Gamma^*$. Then the following two statements are equivalent.
  \begin{compactenum}
  \item[(A)] We can construct a $\Gamma$-fsa $A$ such that $L = \LL(A)$.
  \item[(B)] We can construct a $(\Gamma,\Boole)$-wsa $\cA$ such that $L = \supp(\runsem{\cA})$.\hfill$\Box$
    \end{compactenum}
  \end{observation}


\section{Weighted tree automata over the Boolean semiring}
\label{sec:wta-Boolean-semiring}

Here we prove that every wta over the Boolean semiring $\Boole=(\mathbb{B},\vee,\wedge,0,1)$ is essentially an fta, and vice versa.
For this we use the concept of support fta (cf. Section~\ref{sec:state-algebra-of-wta}).

\begin{theorem}\label{thm:wta-B=fta} Let $\cA$ be $(\Sigma,\Boole)$-wta. Then $\supp(\initialsem{\cA}) = \supp(\runsem{\cA}) = \LL(\supp(\cA))$. Thus $\initialsem{\cA} = \runsem{\cA}$.
\end{theorem}
\begin{proof} Let $\cA=(Q,\delta,F)$ and $\supp(\cA)=(Q,\delta',F')$. We recall that $\LL(\supp(\cA))=\Li(\supp(\cA)) = \Lr(\supp(\cA))$, cf. Lemma \ref{lm:run=init-fta}.

  First we prove $\supp(\initialsem{\cA}) = \Li(\supp(\cA))$. Since $\Boole$ is positive, by Lemma \ref{lm:hAxiqne0-implies-qinhQxi}, we have:
  \begin{equation}
\text{For every $\xi \in \T_\Sigma$ and $q \in Q$: \ $\h_\cA(\xi)_q \not= 0$ iff $q \in \state(\xi)$} \enspace.\label{equ:init-Bool-fta}
    \end{equation}

Then, for each $\xi \in \T_\Sigma$, we can argue as follows:
   \begingroup
\allowdisplaybreaks
    \begin{align*}
      & \xi \in \supp(\initialsem{\cA})
       \ \text{ iff } \ \Big(\bigvee_{q \in Q} \h_\cA(\xi)_q \wedge F_q\Big) \not= 0 
       \ \text{ iff } \  \big(\bigvee_{q \in F'} \h_\cA(\xi)_q\Big)  \not= 0 \\
      \text{ iff }^{(**)} & (\exists q \in F'): \h_\cA(\xi)_q \not= 0
                    \ \text{ iff } \ (\exists q \in F'): q \in \h_{\supp(\cA)}(\xi)
                     \ \text{ iff } \ \xi \in \Li(\supp(\cA))
    \end{align*}
    \endgroup
    where the implication from right to left at equivalence $(**)$ uses the fact that $\Boole$ is zero-sum free;  the last but one equivalence is due to \eqref{equ:init-Bool-fta} and \eqref{equ:state-algebra=algebra-associated-with-supp-fta}. This proves $\supp(\initialsem{\cA}) = \Li(\supp(\cA))$.

     Next we prove that $\supp(\runsem{\cA}) = \Lr(\supp(\cA))$. We note that $\R_\cA(\xi) = \R_{\supp(\cA)}(\xi)$ for every $\xi \in \T_\Sigma$.  First, by induction on $\T_\Sigma$, we prove that the following statement holds:
  \begin{equation}
    \text{For every $\xi \in \T_\Sigma$ and $\rho \in \R_\cA(\xi)$: \ $\wt_\cA(\xi,\rho) \not= 0$ iff $\rho \in \Rv_{\supp(\cA)}(\xi)$} \enspace. \label{equ:run-Bool-fta}
  \end{equation}
We let $\xi = \sigma(\xi_1,\ldots,\xi_k)$. Then
  \begin{align*}
    &\wt_\cA(\xi,\rho) \not= 0 \
    \text{ iff } \ (\wt_\cA(\xi_1,\rho|_1) \wedge  \ldots \wedge \wt_\cA(\xi_k,\rho|_k) \wedge \delta_k(\rho(1) \cdots \rho(k), \sigma,\rho(\varepsilon))) \not=0\\
    \
    \text{ iff }^{(***)} & \ (\wt_\cA(\xi_1,\rho|_1)  \not=0) \wedge  \ldots \wedge (\wt_\cA(\xi_k,\rho|_k)  \not=0)  \wedge (\delta_k(\rho(1) \cdots \rho(k), \sigma,\rho(\varepsilon)) \not=0)\\
    \text{ iff } & \ (\rho|_1 \in \Rv_{\supp(\cA)}(\xi_1))  \wedge  \ldots \wedge (\rho|_k \in \Rv_{\supp(\cA)}(\xi_k))  \wedge ((\rho(1) \cdots \rho(k), \sigma,\rho(\varepsilon)) \in \delta'_k) \tag{by I.H.}\\
    \text{ iff } & \ \rho \in \Rv_{\supp(\cA)}(\xi)
  \end{align*}
  where at $(***)$ from right to left we have used the fact that $\Boole$ is zero-divisor free.
  This proves \eqref{equ:run-Bool-fta}.
  
   Now let $\xi \in \T_\Sigma$. Then
    \begin{align*}
      & \xi \in \supp(\runsem{\cA})
       \ \text{ iff } \ \Big(\bigvee_{q \in Q}\bigvee_{\rho \in \R_\cA(q,\xi)} \wt_\cA(\xi,\rho) \wedge F_{q} \Big)\not= 0 \\
       \ \text{ iff } &  \  \Big(\bigvee_{q \in F'} \bigvee_{\rho \in \R_\cA(q,\xi)}  \wt_\cA(\xi,\rho)\Big) \not= 0 \\
      \text{ iff }^{(v*)} & (\exists q \in F')(\exists \rho \in \R_\cA(q,\xi)): \wt_\cA(\xi,\rho)  \not= 0\\
                    \ \text{ iff } &  (\exists q \in F')(\exists \rho \in \R_{\supp(\cA)}(q,\xi)):  \rho \in \Rv_{\supp(\cA)}(\xi)
                     \ \text{ iff } \ \xi \in \Lr(\supp(\cA))
    \end{align*}
    where at $(v*)$ from right to left we have used the fact that $\Boole$ is zero-sum free;  the last but one equivalence is due to \eqref{equ:run-Bool-fta}. This proves $\supp(\runsem{\cA}) = \Lr(\supp(\cA))$.
    
Finally we have $\initialsem{\cA} = \chi(\supp(\initialsem{\cA})) = \chi(\LL(\supp(\cA))) = \chi(\supp(\runsem{\cA})) = \runsem{\cA}$, where $\chi$ abbreviates  $\chi_{\Boole}$. 
  \end{proof}

  In Lemma \ref{thm:run-init-positive} we will generalize the first statement of Theorem \ref{thm:wta-B=fta} from the Boolean semiring $\Boole$ to arbitrary positive strong bimonoids. In Theorem \ref{thm:semiring-run=initial} we will generalize the second statement of Theorem \ref{thm:wta-B=fta} from $\Boole$ to arbitrary semirings.

Theorem \ref{thm:wta-B=fta} implies that, intuitively, wta over the Boolean semiring and fta are equivalent devices for specifying recognizable tree languages; this is formalized in Corollary \ref{cor:supp-B=fta-1}.  We mention that Corollary~\ref{cor:supp-F2=fta} contains the same statement as  Corollary \ref{cor:supp-B=fta-1}, except that the semiring $\Boole$ is replaced by the field $\sfFtwo$.

     \begin{corollary-rect}\rm \label{cor:supp-B=fta-1} 
    Let $\Sigma$ be a ranked alphabet. Moreover, let $L \subseteq \T_\Sigma$. Then the following three statements are equivalent.
    \begin{compactenum}
    \item[(A)] We can construct a $\Sigma$-fta $A$ such $L = \LL(A)$.
    \item[(B)] We can construct a $(\Sigma,\Boole)$-wta $\cA$ such that $L = \supp(\runsem{\cA}) = \supp(\initialsem{\cA})$.
      \item[(C)] We can construct a $(\Sigma,\Boole)$-wta $\cA$ such that $\chi(L) = \runsem{\cA} = \initialsem{\cA}$.
    \end{compactenum}
 Moreover, we have  $\supp(\Rec^\mathrm{init}(\Sigma,\Boole)) = \supp(\Rec^\mathrm{run}(\Sigma,\Boole)) = \Rec(\Sigma)$.
\end{corollary-rect}

\begin{proof} Proof of (A)$\Rightarrow$(B):   Let $A=(Q,\delta,F)$ be a $\Sigma$-fta such that $L = \LL(A)$. We construct the $(\Sigma,\Boole)$-wta 
  $\cA=(Q,\delta',F')$ such that  $\supp(\delta'_k)=\delta_k$ for each $k \in \mathbb{N}$, and $\supp(F')=F$ (note that this determines $\delta'$ and $F'$ because $\mathbb{B}=\{0,1\}$). Then $A=\supp(\cA)$ and by Theorem~\ref{thm:wta-B=fta} we obtain  $\LL(A)=\LL(\supp(\cA))= \supp(\runsem{\cA}) = \supp(\initialsem{\cA})$.

  \
  
   Proof of (B)$\Rightarrow$(A): Let $(\Sigma,\Boole)$-wta $\cA$ such that $L = \supp(\runsem{\cA}) = \supp(\initialsem{\cA})$. Then, we construct the $\Sigma$-fta $\supp(\cA)$, and by Theorem~\ref{thm:wta-B=fta}, we have that $\LL(\supp(\cA)) = \supp(\runsem{\cA}) = \supp(\initialsem{\cA})$.

   \
   
   Proof of (B)$\Leftrightarrow$(C): This follows from the equivalence that for every $r: \T_\Sigma \to \mathbb{B}$ and $L\subseteq \T_\Sigma$: $r = \chi(L)$
   iff $L = \supp(r)$. 
   
   Finally, the equality $\supp(\Rec^\mathrm{init}(\Sigma,\Boole)) = \supp(\Rec^\mathrm{run}(\Sigma,\Boole))$ follows from Theorem~\ref{thm:wta-B=fta}; and the equality $\Rec(\Sigma) = \supp(\Rec^\mathrm{run}(\Sigma,\Boole))$  follows from  (A)$\Leftrightarrow$(B).
 \end{proof}


\section{Weighted tree automata over the semiring of natural numbers}
\label{sect:wta-N}

In \cite[Ch.~VI, Thm.~9.1]{eil74} it was proved that, for every alphabet $\Gamma$ and $(\Gamma,\Nat)$-wsa $\cA$, there exists an unambiguous $(\Gamma,\Nat)$-wsa $\cB$ such that $\runsem{\cA} = \runsem{\cB}$. There, unambiguous means that each transition weight and each initial and final weight are elements of  $\{0,1\}$. For wta, we call the corresponding property ``$\cB$ is crisp and it has unit root weights''. In \cite[Lm.~3.5.1]{fulvog24} this result was generalized to wta over $\Nat$. Here we generalize Eilenberg's result even further to wta over semirings with a finitely generated additive monoid.

\begin{quote} \emph{In this section, we let $\B=(B,\oplus,\otimes,\0,\1)$ be a semiring where the additive monoid $(B,\oplus,\0)$ is finitely generated, i.e., there exists a finite set $G \subseteq B$ such that $B = \langle G \rangle_{\{\oplus,\0\}}$. Moreover, we let $\cA=(Q,\delta,F)$ be a $(\Sigma,\B)$-wta.}
\end{quote}

Examples of semirings with a finitely generated additive monoid are the semiring $\Nat$ of natural numbers (with generating set $\{1\}$) and the ring $\Int$ of integers (with generating set $\{1,-1\}$). Moreover, each factor semiring of $\Nat$ and of $\Int$ modulo a congruence relation has this property.
Trivially, other examples are each finite semiring (like $\Boole$, $\Intfour$, and each finite distributive bounded lattice). On the other hand, e.g., the additive monoid $(\mathbb{N}_{-\infty},\max,-\infty)$ of the arctic semiring $\Nat_{\max,+}=(\mathbb{N}_{-\infty},\max,+,-\infty,0)$ is not finitely generated.

Let $G$ be a finite generating set of $(B,\oplus,\0)$. Moreover, let $b \in B$ and $w = g_1 \cdots g_n\in G^*$  for some $n\in\mathbb{N}$. We say that $w$ is a \emph{$G$-path to $b$} if $b=\bigoplus_{j \in [n]}g_j$. In particular, due to \eqref{eq:extension-odot-finite-set}, $\varepsilon$ is a $G$-path to $\0$.
Since $(B,\oplus,\0)$ is finitely generated by $G$, the set $\mathrm{Path}_G(b)$ is not empty for each $b\in B$.
For each $b \in B$, we define the \emph{$G$-length of $b$}, denoted by $|b|_G$, by
\[
|b|_G = \min_{w \in \mathrm{Path}_G(b)} |w| \enspace.
\]
Thus, in particular, $|\0|_G= 0$.
A \emph{family of $G$-paths for $\cA$} is a family $\widetilde{p} = (p(b) \mid b\in  \im(\delta))$ such that, for each $b \in \im(\delta)$, the member $p(b)$ is  a $G$-path to $b$ of length $|b|_G$. We denote the $i$th component of $p(b)$ by $p(b)_i$. Thus, if $p(b)$ has length $n$ for some $n \in \mathbb{N}$, then $b = \bigoplus_{i \in [n]} p(b)_i$.
We let
\[
  m = \max(|b|_G \mid b \in \im(\delta)) \enspace.
  \]

For an arbitrary subset $H\subseteq B$, we say that $\cA$ is \emph{$H$-weighted} if $\im(\delta)\subseteq H \cup \{\0\}$. 
Thus, in particular, $\cA$ is $\{\1\}$-weighted if and only if $\cA$ is crisp.

For each finite generating $G$ of $(B,\oplus,\0)$ and family $\widetilde{p}$ of $G$-paths for $\cA$, we construct the $G$-weighted $(\Sigma,\B)$-wta
\[
  \mathrm{split}_{G,\widetilde{p}}(\cA) = (Q',\delta',F')
\]
according to the following idea (cf. \cite[Thm.~9.1]{eil74}). We supply  $\mathrm{split}_{G,\widetilde{p}}(\cA)$  with $m$ copies of each state of $\cA$; a copy is a pair $(q,\ell)$ with $q \in Q$ and $\ell \in [m]$. If  $\tau=(q_1 \cdots q_k,\sigma, q)$ is  a transition of $\cA$ and  $(q_1,\ell_1),\ldots,(q_k,\ell_k)$ are copies of $q_1,\ldots,q_k$, respectively,  then  $\mathrm{split}_{G,\widetilde{p}}(\cA)$  simulates the weight $\delta_k(\tau)$ by nondeterministically branching from $(q_1,\ell_1)\cdots (q_k,\ell_k)$ while reading $\sigma$ into $(q,1),\ldots, (q,\ell)$. For each $\ell \in [m]$, the weight $\delta'_k(\tau')$ of the transition $\tau' = ((q_1,\ell_1)\cdots (q_k,\ell_k),\sigma,(q,\ell))$ is the following:  if $\ell \in [|\delta_k(q_1 \cdots q_k,\sigma,q)|_G]$, then it is $p(\delta_k(q_1 \cdots q_k,\sigma,q))_\ell$,  and $\0$ otherwise. Thus, in particular, the numbers $\ell_1,\ldots,\ell_k$ do not influence the value of the transition $\tau'$.

Formally, we construct the $(\Sigma,\B)$-wta  $\mathrm{split}_{G,\widetilde{p}}(\cA) = (Q',\delta',F')$ as follows:
    \begin{compactitem}
    \item $Q' = Q \times [m]$,
    \item for each $(q,\ell) \in Q'$, we define $F'_{(q,\ell)} = F_q$,
    \item for every $k \in \mathbb{N}$, $\sigma  \in \Sigma^{(k)}$, $(q_1,\ell_1),\ldots,(q_k,\ell_k), (q,\ell) \in Q'$, we define
         \[
        (\delta')_k\Big((q_1,\ell_1)\cdots (q_k,\ell_k),\sigma, (q,\ell)\Big)  = \begin{cases}
          p(\delta_k(q_1 \cdots q_k,\sigma,q))_\ell & \text{ if $\ell \in [|\delta_k(q_1 \cdots q_k,\sigma, q)|_G]$}\\
          \0 & \text{ otherwise} \enspace.
          \end{cases}
        \]
      \end{compactitem}
      Clearly, $\mathrm{split}_{G,\widetilde{p}}(\cA)$ is $G$-weighted and, if $\cA$ has unit root weights, then $\mathrm{split}_{G,\widetilde{p}}(\cA)$ has unit root weights.

The next lemma generalizes \cite[Ch.~VI, Thm.~9.1]{eil74} from wsa over $\Nat$ to wta over semirings with a finitely generated additive monoid.

\begin{lemma} \label{lm:wta-fin-gen-Eilenberg}\rm Let $\B$ be a semiring, $G \subseteq B$ be a finite generating set of $(B,\oplus,\0)$, and $\widetilde{p} = (p(b) \mid b \in \im(\delta))$ be a family of $G$-paths for $\cA$. Then $\initialsem{\cA}= \initialsem{\mathrm{split}_{G,\widetilde{p}}(\cA)}$.
   \end{lemma}

  \begin{proof} Inside this proof we abbreviate $\mathrm{split}_{G,\widetilde{p}}(\cA)$ by $\cB$.
 The following relation holds between $\delta_k$ and $\delta'_k$:
      \begin{equation}\label{equ:summation-over-generators}
        \begin{aligned}
          &\text{For every  $k \in \mathbb{N}$, $\sigma  \in \Sigma^{(k)}$, $(q_1,\ell_1),\ldots,(q_k,\ell_k) \in Q'$, and $q \in Q$, we have}\\
          &\text{$\bigoplus_{\ell \in [m]} (\delta')_k((q_1,\ell_1) \cdots (q_k,\ell_k),\sigma,(q,\ell)) = \delta_k(q_1 \cdots q_k,\sigma,q)$.}
          \end{aligned}
        \end{equation}
        This can be seen as follows.
        \begingroup
        \allowdisplaybreaks
        \begin{align*}
          &\bigoplus_{\ell \in [m]} (\delta')_k((q_1,\ell_1) \cdots (q_k,\ell_k),\sigma,(q,\ell))\\
          =& \ \ \bigoplus_{\ell \in [|p(\delta_k(q_1\cdots q_k,\sigma,q))|_G]} (\delta')_k((q_1,\ell_1) \cdots (q_k,\ell_k),\sigma,(q,\ell))
          \tag{because for each $\ell \in [|p(\delta_k(q_1\cdots q_k,\sigma,q))|_G+1,m]$, we have $(\delta')_k((q_1,\ell_1) \cdots (q_k,\ell_k),\sigma,(q,\ell))=\0$}\\[3mm]
          =&  \ \ \bigoplus_{\ell \in [|p(\delta_k(q_1\cdots q_k,\sigma,q))|_G]} p(\delta_k(q_1 \cdots q_k,\sigma,q))_\ell
          \tag{by definition of $(\delta')_k$}\\
          =& \ \  \delta_k(q_1 \cdots q_k,\sigma,q) \enspace.
             \tag{by definition of $p(\delta_k(q_1 \cdots q_k,\sigma,q))$}
          \end{align*}
            \endgroup
     This proves \eqref{equ:summation-over-generators}.

      Next, by induction on $\T_\Sigma$, we prove that \eqref{equ:summation-over-generators} propagates to the level of the homomorphisms $\h_\cA$ and $\h_\cB$ as follows:
      \begin{equation}
\text{For every $\xi \in \T_\Sigma$ and $q \in Q$, we have: } \h_\cA(\xi)_q = \bigoplus_{\ell \in [m]} \h_\cB(\xi)_{(q,\ell)} \enspace. \label{equ:wta-N}
\end{equation}
Let $\xi = \sigma(\xi_1,\ldots,\xi_k)$. Then we can calculate as follows.
\begingroup
\allowdisplaybreaks
\begin{align*}
&\h_\cA(\sigma(\xi_1,\ldots,\xi_k))_q \\
  &= \bigoplus_{q_1 \cdots q_k \in Q^k} \h_\cA(\xi_1)_{q_1} \otimes \ldots \cdot \h_\cA(\xi_k)_{q_k} \otimes \delta_k(q_1 \cdots q_k,\sigma,q)\\
  &=\bigoplus_{q_1 \cdots q_k \in Q^k}  \Big(\bigoplus_{\ell_1 \in [m]} \h_\cB(\xi_1)_{(q_1,\ell_1)}\Big) \otimes \ldots \otimes \Big(\bigoplus_{\ell_k \in [m]} \h_\cB(\xi_k)_{(q_k,\ell_k)}\Big) \otimes \delta_k(q_1 \cdots q_k,\sigma,q)
    \tag{by I.H.}\\
  &= \bigoplus_{(q_1,\ell_1) \cdots (q_k,\ell_k) \in (Q')^k} \h_\cB(\xi_1)_{(q_1,\ell_1)} \otimes \ldots \otimes \h_\cB(\xi_k)_{(q_k,\ell_k)} \otimes
    \delta_k(q_1 \cdots q_k,\sigma,q)
    \tag{by right-distributivity}\\
   &= \bigoplus_{(q_1,\ell_1)\cdots (q_k,\ell_k) \in (Q')^k} \h_\cB(\xi_1)_{(q_1,\ell_1)} \otimes \ldots \otimes \h_\cB(\xi_k)_{(q_k,\ell_k)} \otimes
     \Big(\bigoplus_{\ell \in [m]} (\delta')_k((q_1,\ell_1) \cdots (q_k,\ell_k),\sigma,(q,\ell))\Big)
     \tag{by \eqref{equ:summation-over-generators}}\\
  &=\bigoplus_{\ell \in [m]}  \bigoplus_{(q_1,\ell_1)\cdots (q_k,\ell_k) \in (Q')^k} \h_\cB(\xi_1)_{(q_1,\ell_1)} \otimes \ldots \otimes \h_\cB(\xi_k)_{(q_k,\ell_k)} \otimes
    (\delta')_k((q_1,\ell_1) \cdots (q_k,\ell_k),\sigma,(q,\ell))
    \tag{by left-distributivity}\\
  &= \bigoplus_{\ell \in [m]} \h_\cB(\sigma(\xi_1,\ldots,\xi_k))_{(q,\ell)} \enspace.
\end{align*}
\endgroup
This proves \eqref{equ:wta-N}.

 Now let $\xi \in \T_\Sigma$. Then 
\begingroup
\allowdisplaybreaks
\begin{align*}
  \initialsem{\cA}(\xi)&= \bigoplus_{q \in Q} \h_\cA(\xi)_q \otimes F_q\\
                       &= \bigoplus_{q \in Q} \Big(\bigoplus_{\ell \in [m]} \h_\cB(\xi)_{(q,\ell)}\Big) \otimes F_q
  \tag{by \eqref{equ:wta-N}}\\
                       &= \bigoplus_{q \in Q} \bigoplus_{\ell \in [m]} \h_\cB(\xi)_{(q,\ell)} \otimes
                         F_q\tag{by right-distributivity}\\
                       &= \bigoplus_{(q,\ell) \in Q'} \h_\cB(\xi)_{(q,\ell)} \otimes (F')_{(q,\ell)} \tag{by construction}\\
                       & = \initialsem{\cB}(\xi)
                         \enspace. \qedhere
  \end{align*}
\endgroup
  \end{proof}

     We mention that the equality $\initialsem{\cA} = \initialsem{\mathrm{split}_{G,\widetilde{p}}(\cA)}$ can also be proved by using forward bisimulation \cite{hogmalmay07d} as follows.
We define the equivalence relation $\sim$ on the state set $Q'$ of $\mathrm{split}_{G,\widetilde{p}}(\cA)$ such that, for every $(q_1,\ell_1),(q_2,\ell_2) \in Q'$, we let $(q_1,\ell_1) \sim (q_2,\ell_2)$ if $q_1=q_2$. Then $\sim$ is a forward bisimulation on $\mathrm{split}_{G,\widetilde{p}}(\cA)$ in the sense of \cite[Def.~1]{hogmalmay07d} and it determines the $(\Sigma,\B)$-wta $\mathrm{split}_{G,\widetilde{p}}(\cA)/_{\sim}$, see \cite[Def.~3]{hogmalmay07d}. It is easy to see that, apart from state renaming, the $(\Sigma,\B)$-wta $\cA$ and $\mathrm{split}_{G,\widetilde{p}}(\cA)/_{\sim}$ are equal.
Hence $\initialsem{\cA} = \initialsem{\mathrm{split}_{G,\widetilde{p}}(\cA)/_{\sim}}$. Thus, intuitively, the state splitting by Eilenberg's construction can be reversed by a forward bisimulation. Finally, by \cite[Thm.~6]{hogmalmay07d}, we obtain that $\initialsem{\mathrm{split}_{G,\widetilde{p}}(\cA)/_{\sim}} = \initialsem{\mathrm{split}_{G,\widetilde{p}}(\cA)}$. Thus, overall, we have $\initialsem{\cA} = \initialsem{\mathrm{split}_{G,\widetilde{p}}(\cA)}$.

If a family $\widetilde{p}$ of $G$-paths for $\cA$ is given, then we can construct $\mathrm{split}_{G,\widetilde{p}}(\cA)$. Next we show that such a family can be constructed.

\begin{observation}\rm \label{obs:family-G-paths-can-be-constructed} We can construct a family $\widetilde{p} = (p(b) \mid b \in \im(\delta))$ of $G$-paths for $\cA$.
\end{observation}

\begin{proof} We construct $\widetilde{p}$ algorithmically using the following assumptions and data structures. We fix an arbitrary linear order $<$ on $G$. Moreover, we will maintain a counter $i \in \mathbb{N}$, which is initialized with $i=0$. Also, we will create a finite sequence of subsets of $\im(\delta)$; we start with $U_0=\im(\delta)$.

  Now we enumerate the strings in $G^*$ with respect to the lexicographic order on $G^*$ induced by $<$.
Let $w \in G^*$ be the currently enumerated string. 
 If there exists $b \in U_i$ such that $b= \bigoplus_{i \in [|w|]} w_i$, then we let $p(b)=w$ and  define $U_{i+1}=U_i\setminus\{b\}$; moreover, we increment the counter $i$.
 If $U_{i}=\emptyset$, then we stop, and otherwise we continue the enumeration.

 Since $(B,\oplus,\0)$ is generated by $G$, for each $b \in \im(\delta)$ there exists a $w \in G^*$ such that $b=\bigoplus_{i \in [|w|]} w_i$, and hence our algorithm terminates after finitely many steps.
  Since we enumerate the strings in $G^*$ with respect to the lexicographic order, we have that $p(b) = \min_{w \in \mathrm{Path}_G(b)}|w|$; thus, 
$(p(b) \mid b \in \im(\delta))$ is a family of $G$-paths for $\cA$.
\end{proof}

Observation \ref{obs:family-G-paths-can-be-constructed} implies that we can construct a $G$-weighted $(\Sigma,\B)$-wta $\mathrm{split}_{G,\widetilde{p}}(\cA)$, where $\widetilde{p}$ is a family of $G$-paths for $\cA$. Then by Lemma \ref{lm:wta-fin-gen-Eilenberg} we obtain the following corollary.

 \begin{corollary}\rm \label{cor:applying-Eilenberg-to-wta-N-and-wta-Z} The following two statements hold.
   \begin{compactenum} 
   \item[(1)]  For each $(\Sigma,\Nat)$-wta $\cA$, we can construct a crisp $(\Sigma,\Nat)$-wta $\cB$ such that  $\initialsem{\cA}= \initialsem{\cB}$.

     \item[(2)]  For each $(\Sigma,\Int)$-wta $\cA$, we can construct a $\{1,-1\}$-weighted $(\Sigma,\Int)$-wta $\cB$ such that  $\initialsem{\cA}= \initialsem{\cB}$. 
     \end{compactenum}
In both statements the following holds:  if $\cA$ has unit root weights, then $\cB$ has unit root weights.
    \end{corollary}

Assuming that Theorems \ref{thm:semiring-run=initial} and  \ref{thm:root-weight-normalization-run} are already available, Lemma \ref{lm:wta-fin-gen-Eilenberg} implies the following corollary. We note that the proofs of the Theorems \ref{thm:semiring-run=initial} and \ref{thm:root-weight-normalization-run} do not depend on Corollary \ref{cor:wta-N-special-wta-N}.

\begin{corollary}\label{cor:wta-N-special-wta-N} \rm  For each $(\Sigma,\Nat)$-wta $\cA$, we can construct a $(\Sigma,\Nat)$-wta $\cB$ such that $\cB$ is crisp, it has unit root weights, and $\initialsem{\cA}= \initialsem{\cB}$.
 \end{corollary}
\begin{proof} Let $\cA$ be a $(\Sigma,\Nat)$-wta. Since $\Nat$ is distributive, by Theorem \ref{thm:semiring-run=initial} we have $\initialsem{\cA} = \runsem{\cA}$. By Theorem \ref{thm:root-weight-normalization-run} we can construct a root weight normalized $(\Sigma,\Nat)$-wta $\cC$ such that $\runsem{\cA}=\runsem{\cC}$. In particular, $\cC$ has unit root weights.
By Theorem  \ref{thm:semiring-run=initial} we have $\runsem{\cC} = \initialsem{\cC}$. 
  Then, by Lemma \ref{lm:wta-fin-gen-Eilenberg} we can construct a  $(\Sigma,\Nat)$-wta $\cB$ such that $\cB$ is crisp, it has unit root weights, and $\initialsem{\cC} = \initialsem{\cB}$.
 \end{proof}

\index{characteristic wta}
\index{chi@$\chi_\Nat(\cA)$}
Next we define the characteristic wta of a $\Sigma$-fta $A$. For this, let $A=(Q,\delta,F)$ be a  $\Sigma$-fta. The \emph{characteristic wta of $A$}, denoted by $\chi_\Nat(A)$, is the $(\Sigma,\Nat)$-wta $(Q,\delta',F')$ where for every $k \in \mathbb{N}$ we have $\delta_k' = \chi_\Nat(\delta_k)$ and $F' = \chi_\Nat(F)$. Note that $\chi_\Nat(A)$ is crisp and it has unit root weights. (For the definition of the characteristic mapping $\chi_\Nat$, see Section~\ref{sect:weighted-sets-languages}.)

For each crisp $(\Sigma,\Nat)$-wta $\cA$ with unit root weights, we have $\chi_\Nat(\supp(\cA)) = \cA$. (For the definition of $\supp(\cA)$ we refer to Section~\ref{sec:state-algebra-of-wta}.) Moreover, for each $\Sigma$-fta $A$, we have $\supp(\chi_\Nat(A)) = A$.

\label{page:multiplicity-mapping}
      \index{multiplicity mapping}
We recall that, for a  $\Sigma$-fta $A$,  the set of  accepting runs of $A$ on a tree $\xi$ is denoted by $\Ra_A(\xi)$. Moreover, the multiplicity mapping of $A$ is defined as the weighted tree language $\#_{\Ra_A}: \T_\Sigma \to \mathbb{N}$ such that, for each $\xi \in \T_\Sigma$, we have $\#_{\Ra_A}(\xi) =  |\Ra_A(\xi)|$ (cf. Example \ref{ex:valid-runs}).      
      
In Example \ref{ex:valid-runs} we have proved the following result.

\begin{lemma}\label{lm:semantics-chi-A}\rm For each $\Sigma$-fta $A$, we have $\initialsem{\chi_\Nat(A)}=\#_{\Ra_A}$.
\end{lemma}

We also have the following kind of inverse result.

\begin{lemma}\rm \label{lm:special-wta-N-fta} Let $\cA$ be a crisp $(\Sigma,\Nat)$-wta which has unit root weights. Then
  \(\initialsem{\cA} = \#_{\Ra_{\supp(\cA)}}\).
\end{lemma}
\begin{proof} This follows from the fact that $\cA = \chi_\Nat(\supp(\cA))$ and from Lemma~\ref{lm:semantics-chi-A} (by choosing $A=\supp(\cA)$). 
  \end{proof}

  For example, the $(\Sigma,\Nat)$-wta $\cA$ in Example \ref{ex:number-of-occurrences} is crisp and it has unit  root weights. Since $\initialsem{\cA}=\#_{\sigma(.,\alpha)}$, it follows from Lemma \ref{lm:special-wta-N-fta} that $\#_{\sigma(.,\alpha)} = \#_{\Ra_{\supp(\cA)}}$.

 We can conclude the following characterization of $\Rec^{\mathrm{init}}(\Sigma,\Nat)$, which says that the i-recognizable $(\Sigma,\Nat)$-weighted tree languages are exactly the multiplicity mappings of $\Sigma$-fta.

 \begin{theorem-rect} \label{thm:char-wta-N} Let $\Sigma$ be a ranked alphabet. Moreover, let $r: \T_\Sigma \to \mathbb{N}$. Then the following three statements are equivalent.
        \begin{compactenum}
        \item[(A)] We can construct a $(\Sigma,\Nat)$-wta $\cA$ such that $r=\initialsem{\cA}$.
        \item[(B)] We can construct a crisp $(\Sigma,\Nat)$-wta $\cA$ which has unit root weights such that $r=\initialsem{\cA}$.
        \item[(C)] We can construct a $\Sigma$-fta $A$ such that $r = \#_{\Ra_A}$.
          \end{compactenum}
        \end{theorem-rect}
        \begin{proof} (A)$\Rightarrow$(B) follows from Corollary \ref{cor:wta-N-special-wta-N}  and (B)$\Rightarrow$(A) is obvious.
         (B)$\Rightarrow$(C)  follows from  Lemma~\ref{lm:special-wta-N-fta} and (C)$\Rightarrow$(B) follows from Lemma~\ref{lm:semantics-chi-A}.
          \end{proof}

      It will turn out later (cf. Theorem~\ref{thm:semiring-run=initial}) that $\initialsem{\cA} = \runsem{\cA}$ for each $(\Sigma,\Nat)$-wta. Hence, Theorem~\ref{thm:char-wta-N} also shows a characterization of  $\Rec^{\mathrm{run}}(\Sigma,\Nat)$.

Finally, we show that, for each $(\Sigma,\Nat)$-wta $\cA$ and $\xi\in \T_\Sigma$, the value $\runsem{\cA}(\xi)$ is bounded exponentially by $\size(\xi)$.

\begin{lemma}\rm\label{lm:N-wta-upper bound} Let $\cA=(Q,\delta,F)$ be a $(\Sigma,\Nat)$-wta. There exists $K\in \mathbb{N}$ such that, for each $\xi\in \T_\Sigma$,  we have $\runsem{\cA}(\xi)\le K^{\size(\xi)}$.
\end{lemma}
\begin{proof} Let $m=\max(\mathrm{wts}(\cA))$ and $K=|Q|\cdot m^2$. Obviously, for  each $\xi\in \T_\Sigma$, the number of runs on $\xi$ is $|Q|^{\size(\xi)}$. Moreover, for  each $\xi\in \T_\Sigma$ and $\rho\in\R_\cA(\xi)$,  we have $\wt(\xi,\rho)\le m^{\size(\xi)}$. Then
\begin{align*}
\runsem{\cA}(\xi)=\bigplus_{\rho\in \R_\cA(\xi)}\wt(\xi,\rho)\cdot F_{\rho(\varepsilon)}\le \bigplus_{\rho\in \R_\cA(\xi)} m^{\size(\xi)+1}=|Q|^{\size(\xi)}\cdot m^{\size(\xi)+1} \le K^{\size(\xi)}\enspace.
\end{align*}
\end{proof}


    \section{Weighted tree automata over commutative semirings}
    \label{sec:wta-commutative-semirings}

  Here we consider wta over commutative semirings. For each such wta, we represent its initial algebra semantics in terms of a semimodule with multilinear operations and a linear form. The latter combination will be called multilinear representation. This combination is based on the following predecessors:
  \begin{compactitem}
  \item linear representation \cite[p.~119]{berreu82},
  \item finite-dimensional $\B\Sigma$ algebra, \cite[p.~452]{bozale89}, \cite[p.~353]{boz91}, and \cite[p.~738]{fulste11}.
  \end{compactitem}
  Also, vice versa, the semantics of each multilinear representation is the initial algebra semantics of some wta (cf. Theorem~\ref{thm:lin-iff-wta-extended}).

    In the next subsections, we will use notions from Sections \ref{sec:vectors-matrices} and \ref{sec:semimodules}.
 
\label{p:wta-over-fields}
\begin{quote}\emph{In the rest of this section, we let $\B =(B,\oplus,\otimes,\0,\1)$ denote an arbitrary commutative semiring, if not specified otherwise.}
\end{quote}

\subsection{Multilinear representations}
\label{sec:multilinear-representation}

Let $Q$ be a finite and nonempty set. We recall that $\0^Q$ is the $Q$-vector in $B^Q$ which contains $\0$ in each component, and that $\1_q$  is the $q$-unit vector in $B^Q$ for each $q\in Q$. Moreover, $\oplus$ denotes the usual vector addition on $B^Q$ (cf. Section \ref{sec:vectors-matrices}). By Observation~\ref{obs:B-to-Q-is-semimodule}, the tuple $(B^Q,\oplus,\0^Q)$ is a $\B$-semimodule with scalar multiplication.
         
  \index{multilinear representation}
\index{SigmaBmultilinear representation@$(\Sigma,\B)$-multilinear representation}
A {\em $(\Sigma,\B)$-multilinear representation}  is a tuple $\cM=(Q,\mu, \gamma)$ where
  \begin{compactitem}
  \item $Q$ is a finite and nonempty set,
  \item $\mu: \Sigma \to \mathrm{Ops}(B^Q)$ such that, for every $k \in \mathbb{N}$ and $\sigma \in \Sigma^{(k)}$, the operation $\mu(\sigma)$ is a $k$-ary multilinear operation over the $\B$-semimodule $(B^Q,\oplus,\0^Q)$, and
  \item $\gamma: B^Q \to B$ is a linear form over $\B^Q$ (cf. Subsection \ref{sec:semimodules}).  
  \end{compactitem}
    \index{hommu@$\h_\mu$}
Clearly, $(B^Q,\oplus,\0^Q,\mu)$ is a $(\Sigma,\B)$-semimodule (cf. Subsection \ref{sec:semimodules}).
  We call $(B^Q,\mu)$ the \emph{$\Sigma$-algebra associated with $\cM$}, and denote the unique $\Sigma$-algebra homomorphism from the $\Sigma$-term algebra  $\sfT_\Sigma$ to the $\Sigma$-algebra $(B^Q,\mu)$ by~$\h_\mu$.

  \index{semanticM@$\sem{\cM}$}
The weighted tree language \emph{recognized by $\cM=(Q,\mu,\gamma)$} is the weighted tree language $\sem{\cM}:\T_\Sigma \to B$ such that, for every $\xi \in \T_\Sigma$, we have 
  \[
    \sem{\cM}(\xi) = \gamma(\h_\mu(\xi)) \enspace.
    \]
  \index{recognizable by a $(\Sigma,\B)$-multilinear representation}
  A weighted tree language $r:\T_\Sigma \to B$ is \emph{recognizable by a $(\Sigma,\B)$-multilinear representation} if there exists a $(\Sigma,\B)$-multilinear representation $\cM$ with $r=\sem{\cM}$.

  Let $\cM = (Q,\mu,\gamma)$ be a $(\Sigma,\B)$-multilinear representation and let  us consider the set $\1_Q=\{\1_q \mid q \in Q\}$.  We call $(Q,\mu,\gamma)$
\begin{compactitem}
\item \emph{bottom-up-deterministic} (for short: \emph{bu-deterministic}) if the set $B \cdot \1_Q$ is closed under $\mu(\Sigma)$, i.e., for every $k \in \mathbb{N}$, $\sigma \in \Sigma^{(k)}$, and $b_1\cdot \1_{q_1},\ldots,b_k \cdot \1_{q_k} \in B \cdot \1_Q$, we have $\mu(\sigma)(b_1\cdot \1_{q_1},\ldots,b_k \cdot \1_{q_k})  \in B \cdot \1_Q$.
\item \emph{crisp-deterministic} if the set $\1_Q$ is closed under $\mu(\Sigma)$, i.e., for every $k \in \mathbb{N}$, $\sigma \in \Sigma^{(k)}$, and $\1_{q_1},\ldots,\1_{q_k} \in \1_Q$, we have $\mu(\sigma)(\1_{q_1},\ldots,\1_{q_k})  \in \1_Q$.
  \item \emph{$n$-dimensional} if $|Q|=n$.
  \end{compactitem}

\begin{observation}\label{obs:closed-mu-Sigma-simpler}\rm Let $\cM=(Q,\mu,\gamma)$ be a $(\Sigma,\B)$-multilinear representation and let $\1_Q=\{\1_q \mid q \in Q\}$. Then $B \cdot \1_Q$ is closed under $\mu(\Sigma)$ if and only if, for every $k \in \mathbb{N}$, $\sigma \in \Sigma^{(k)}$, and  $\1_{q_1},\ldots,\1_{q_k} \in \1_Q$, we have
\begin{equation}\label{equ:closed-under-mu-simpler}
\mu(\sigma)(\1_{q_1},\ldots,\1_{q_k}) \in B \cdot \1_Q\enspace.
\end{equation}
\end{observation}
\begin{proof} It follows from Observation~\ref{obs:generalization-closure} because $(B^Q,\oplus,\0^Q)$ is a $\B$-semimodule.
\end{proof}


  \begin{example}\rm \label{ex:ml-representation-for-height} We consider the ranked alphabet $\Sigma = \{\sigma^{(2)}, \alpha^{(0)}\}$, the commutative semiring $\Natmaxplus = (\mathbb{N}_{-\infty},\max,+,-\infty,0)$, and the $(\Natmaxplus,\Sigma)$-weighted tree language $\height: \T_\Sigma \to \mathbb{N}$ (as in Example~\ref{ex:height}). In this example, we wish to show a $(\Sigma,\Natmaxplus)$-multilinear representation $\cM = (Q,\mu,\gamma)$ which recognizes $\height$.

    We define $\cM = (Q,\mu,\gamma)$ such that  $Q = \{\h,0\}$, and for every $v_1,v_2 \in (\mathbb{N}_{-\infty})^Q$ we let
    \[
      \begin{array}{ll}
        \mu(\sigma)(v_1,v_2)_\h = \max\{(v_1)_\h + (v_2)_0 + 1 \ , \ (v_1)_0 + (v_2)_\h + 1\} \ \ 
        & \mu(\alpha)()_\h = 0\\
      \mu(\sigma)(v_1,v_2)_0 = (v_1)_0 + (v_2)_0 & 
      \mu(\alpha)()_0 = 0 \enspace.
      \end{array}
    \]
    Finally, we define the linear form $\gamma: (\mathbb{N}_{-\infty})^Q \to \mathbb{N}_{-\infty}$ for each $v \in (\mathbb{N}_{-\infty})^Q$ by
    \(\gamma(v) = v_\h \).

    The mappings $\mu(\sigma)$ and $\mu(\alpha)$ are multilinear. In fact, the $\Sigma$-indexed family $(\mu(\sigma) \mid \sigma \in \Sigma)$ is the same as the $\Sigma$-indexed family $(\delta_\cA(\sigma) \mid \sigma \in \Sigma)$ where $\cA$ is the $(\Sigma,\Natmaxplus)$-wta of Example~\ref{ex:height} with $\initialsem{\cA}=\height$ and $\delta_\cA$ is part of the vector algebra $\V(\cA)=((\Natmaxplus)^Q,\delta_\cA)$; moreover, in Lemma~\ref{lm:vector-algebra-is-semimodule-com-sr}, we will prove that $\delta_\cA(\sigma)$ and $\delta_\cA(\alpha)$ are multilinear. Finally, since $\1_Q=\{0_\h, 0_0\}$, $\mu(\alpha)() = (0,0)$, and $(0,0) \not\in \mathbb{N}_{-\infty}\cdot \1_Q$, Observation~\ref{obs:closed-mu-Sigma-simpler} implies that $\cM$ is not bu-deterministic.
       
Since $\delta_\cA=\mu$, we also have that $\h_\cA=\h_\mu$. Thus, for each $\xi \in \T_\Sigma$, we have $\h_\mu(\xi)_h = \height(\xi)$ and $\h_\mu(\xi)_0 = 0$ (cf. \eqref{equ:ex-height}). Hence, for each $\xi\in \T_\Sigma$, we have
    \[\sem{\cM}(\xi)=\gamma(\h_\mu(\xi))=\h_\mu(\xi)_h = \height(\xi).\] 
    i.e., the weighted tree language recognized by $\cM$ is $\height$.   

It might be tempting to use the tuple $(Q',\mu',\gamma')$ for the recognition of $\height$ where
\begin{compactitem}
\item $Q' = \{q\}$; for the sake of a better comparison with $\cM$ from above, we distinguish $v \in (\mathbb{N}_{-\infty})^Q$ and its only component $v_q$;
   \[
      \begin{array}{ll}
        \mu'(\sigma)(v_1,v_2)_q = 1 + \max\{(v_1)_q, (v_2)_q\} \ \ 
        & \mu'(\alpha)()_q = 0 \enspace,
      \end{array}
    \]
    \item for each $v \in (\mathbb{N}_{-\infty})^{Q'}$ we let $\gamma'(v) = v_q$.
    \end{compactitem}
  Let us denote the unique homomorphism from 
   $\sfT_\Sigma$ to the $\Sigma$-algebra $(Q',\mu')$ by $\h_{\mu'}$. Then for each $\xi\in \T_\Sigma$, we have $\h_{\mu'}(\xi)_q=\height$, hence $\gamma(\h_{\mu'}(\xi))=\height$.
 However, $\mu'(\sigma)$ is not multilinear, as the following two calculations show. Let  $v_{11}=v_{12}=(3)$, $v_2=(6)$, and $b_1=b_2=2$. Then
    \begingroup
        \allowdisplaybreaks
        \begin{align*}
          &\mu'(\sigma)(b_1 \cdot v_{11} \oplus b_2 \cdot v_{12} \ , \ v_2)_q\\
          &= 1 + \max\{\max\{b_1 + (v_{11})_q, b_2 + (v_{12})_q\} \ , \  (v_2)_q\}\\
          &= 1 + \max\{\max\{2+3, 2+3 \} \ , \ 6 \} = 1+ 6 = 7 
        \end{align*}
        \endgroup
        and
         \begingroup
        \allowdisplaybreaks
        \begin{align*}
          &\big(b_1 \cdot \mu'(\sigma)(v_{11}, v_2) \oplus b_2 \cdot  \mu'(\sigma)(v_{12} \ , \ v_2)\big)_q\\
          &= \max\{b_1 + \mu'(\sigma)(v_{11}, v_2)_q \ , \ b_2 +  \mu'(\sigma)(v_{12}, v_2)_q \}\\
          &= \max\{b_1 + 1 + \max\{(v_{11})_q, (v_2)_q\} \ , \ b_2 + 1+ \max\{(v_{12})_q,(v_2)_q\} \}\\
          &= \max\{2 + 1 + \max\{3,6\}\ , \ 2 + 1 + \max\{3,6 \}\} = 9 
            \enspace.
        \end{align*}
        \endgroup
        Hence $\mu'(\sigma)$ is not multilinear, hence $(Q',\mu',\gamma')$ is not a multilinear representation.
    \hfill $\Box$
  \end{example}

  Before showing another example of a multilinear representation, we prove a useful lemma. For this, we note that the algebra $(B^{\T_\Sigma},\oplus,\widetilde{\0})$ is a $\B$-semimodule via  scalar multiplication from left.

\begin{lemma}\rm (cf. \cite[Prop.~3.1]{berreu82}) \label{lm:recognizable-subspace} The set of weighted tree languages which are recognizable by $(\Sigma,\B)$-multilinear representations is a sub-semimodule  of the $\B$-semimodule $(B^{\T_\Sigma},\oplus,\widetilde{\0})$.
  \end{lemma}
  \begin{proof} We show that the subset of $B^{\T_\Sigma}$ which consists of all weighted tree languages which are recognizable by $(\Sigma,\B)$-multilinear representations is closed under (a) scalar multiplication from left and (b) under sum.

  (a) Let $r \in B^{\T_\Sigma}$ and $b\in B$ and assume that $r$ is recognized by the $(\Sigma,\B)$-multilinear representation $(Q,\mu,\gamma)$.
  Then the $(\Sigma,\B)$-weighted tree language $b\cdot r$ is recognizable by the $(\Sigma,\B)$-multilinear representation $(Q,\mu,\gamma')$, where $\gamma'(v)=b\otimes \gamma(v)$ for each $v\in V$.
  
  (b) Let $r_1,r_2\in B^{\T_\Sigma}$ and assume that,  for each $i\in\{1,2\}$, the $(\Sigma,\B)$-multilinear representation $(Q_i,\mu_i,\gamma_i)$ recognizes $r_i$. We assume that $Q_1 \cap Q_2 = \emptyset$. We define the  $(\Sigma,\B)$-multilinear representation $\cM= (Q_1 \cup Q_2,\mu,\gamma)$ which recognizes  $r_1\oplus r_2$ as follows.
  \begin{itemize}
  \item For every $k\in \mathbb{N}$ and $\sigma \in\Sigma^{(k)}$, we define the mapping $\mu(\sigma): (B^{Q_1\cup Q_2})^k \to B^{Q_1\cup Q_2}$ such that, for every $v_1,\ldots,v_k\in B^{Q_1 \cup Q_2}$ and $q \in Q_1 \cup  Q_2$, we let
    \[\mu(\sigma)\big( v_1,\ldots,v_k\big)_{q} =
      \begin{cases}
        \mu_1(\sigma)(v_1|_{Q_1},\ldots,v_k|_{Q_1})_q& \text{ if $q \in Q_1$} \\
        \mu_2(\sigma)(v_1|_{Q_2},\ldots,v_k|_{Q_2})_q& \text{ otherwise}
      \end{cases}
      \enspace,\]
  \item and for each $u \in B^{Q_1\cup Q_2}$, we let $\gamma(u)=\gamma_1(u|_{Q_1})\oplus \gamma_2(u|_{Q_2})$.
  \end{itemize}
  It is easy to see that, for every $k\in \mathbb{N}$ and $\sigma \in\Sigma^{(k)}$, the mapping $\mu(\sigma)$ is multilinear, i.e., $\mu(\sigma) \in \cL((\B^{Q_1 \cup Q_2})^k,\B^{Q_1 \cup Q_2})$,  and that   $\gamma$ is a linear form. Hence $\cM$ is indeed a $(\Sigma,\B)$-multilinear representation.

  By induction on $\T_\Sigma$, it is also easy to show that, for each $q \in Q_1 \cup Q_2$ and $\xi \in \T_\Sigma$, we have
\[
  \h_\mu(\xi)_q=
  \begin{cases}
    \h_{\mu_1}(\xi)_q & \text{ if $q \in Q_1$}\\
    \h_{\mu_2}(\xi)_q & \text{ otherwise} \enspace.
    \end{cases}
  \]
Then, for each $\xi\in \T_\Sigma$, we have $\h_\mu(\xi)|_{Q_1} = \h_{\mu_1}(\xi)$ and $\h_\mu(\xi)|_{Q_2} = \h_{\mu_2}(\xi)$, and hence:
\begin{align*} \sem{\cM}(\xi) = \gamma(\h_\mu(\xi)) = \gamma_1(\h_\mu(\xi)|_{Q_1}) \oplus \gamma_2(\h_\mu(\xi)|_{Q_2})
  = \gamma_1(\h_{\mu_1}(\xi)) \oplus \gamma_2(\h_{\mu_2}(\xi)) =r_1(\xi)\oplus r_2(\xi)\enspace,
  \end{align*}
  i.e., $\cM$ recognizes $r_1\oplus r_2$. 
\end{proof}

In the next example we consider the field  $\Ratnum$ of rational numbers.

  \begin{example}\rm \label{ex-lin-rec} \cite[Ex.~4.1]{berreu82} Here we show an example of a multilinear representation for the field $\Ratnum$ of rational numbers. For this, we consider the weighted tree language $\size: \T_\Sigma \to \mathbb{N}$ (cf. page \pageref{page:algebras-for-height-size-pos}) and view it as a weighted tree language over  $\Ratnum$. We show that $\size$ is recognizable by a $(\Sigma,\Ratnum)$-multilinear representation.

  For each $\delta \in \Sigma$, we define the mapping $\size_\delta: \T_\Sigma \to \mathbb{Q}$ by $\size_\delta(\xi) = |\pos_\delta(\xi)|$. Then we have $\size = \bigplus_{\delta \in \Sigma}\size_\delta$. By Lemma \ref{lm:recognizable-subspace}, a finite sum of weighted tree languages which are recognizable by $(\Sigma,\Ratnum)$-multilinear representation is also recognizable by a $(\Sigma,\Ratnum)$-multilinear representation. Hence,
  it suffices to prove, for an arbitrary but fixed $\delta$ that $\size_\delta$ is recognizable by some $(\Sigma,\Ratnum)$-multilinear representation. From now on let $\delta \in \Sigma$ be arbitrary, but fixed. We assume that $\delta$ has rank $\ell$ for some $\ell \in \mathbb{N}$.

  We define the $(\Sigma,\Ratnum)$-multilinear representation $\cM=(Q,\mu, \gamma)$ with $Q = \{p,q\}$. Clearly, $(\mathbb{Q}^Q,+,0^Q)$ is a $\Ratnum$-vector space with basis $H=\{1_p,1_q\}$, where assuming that $p$ is left of $q$ we have $1_p=(1,0)$ and $1_q=(0,1)$. Certainly, it is sufficient to define each multilinear operation $\mu(\sigma)$ (for $k \in \mathbb{N}$ and $\sigma \in \Sigma^{(k)}$)  and $\gamma$ on $H$.
  For every  $1_{q_1},\ldots,1_{q_\ell} \in \{1_p,1_q\}$, we define
  \[
  \mu(\delta)(1_{q_1},\ldots,1_{q_\ell}) =
  \begin{cases}
    (1,1)  & \mbox{if $q_1=\ldots = q_\ell=p$}\\
(0,1) & \mbox{if there exists exactly one $j \in [\ell]$ with $q_j=q$} \\
(0,0) & \mbox{otherwise},
\end{cases}
\]
and, for every $\sigma \in\Sigma^{(k)}$ with $k \in \mathbb{N}$ and  $\sigma\not=\delta$ and
$1_{q_1},\ldots,1_{q_k} \in \{1_p,1_q\}$, we define
\[
  \mu(\sigma)(1_{q_1},\ldots,1_{q_k}) =
  \begin{cases}
    (1,0) & \mbox{if $q_1=\ldots = q_k=p$}\\
(0,1) & \mbox{if there exists exactly one $j \in [k]$ with $q_j=q$} \\
(0,0) & \mbox{otherwise}.
\end{cases}
\]
Thus, in particular, for $\alpha\in \Sigma^{(0)}$, this gives 
\begin{equation}
  \mu(\alpha)() =
  \begin{cases}
(1,1) & \mbox{if $\alpha=\delta$}\\
(1,0) & \mbox{otherwise}.
\end{cases}\label{equ:size-vector-space}
\end{equation}
Finally, we define $\gamma(1_p)=0$ and $\gamma(1_q)=1$.

First we note that $\cM$ is not bu-deterministic because, e.g., $\mu(\delta)(\overbrace{1_p,\ldots,1_p}^\ell)=(1,1)$ and $(1,1)\not\in \mathbb{Q}\cdot H$.

Next, by induction on $\T_\Sigma$,  we prove that the following statement holds:
\begin{equation}
  \text{For every $\xi \in \T_\Sigma$: \ $\h_\mu(\xi)=1_p+ \size_\delta(\xi) \cdot 1_q$} \enspace.\label{eq:multilinear-ex}
\end{equation}

I.B.: Let $\xi\in\Sigma^{(0)}$. Then \eqref{eq:multilinear-ex} follows from \eqref{equ:size-vector-space}.

I.S.: Now let $\xi=\sigma(\xi_1,\ldots,\xi_k)$. Then we
can calculate as follows:
    \begingroup
      \allowdisplaybreaks
\begin{align*}
\h_\mu(\xi) & =  \mu(\sigma)(\h_\mu(\xi_1),\ldots,\h_\mu(\xi_k))\\[2mm]
& =  \mu(\sigma)(1_p+ \size_\delta(\xi_1) \cdot 1_q,\ldots,1_p+ \size_\delta(\xi_k) \cdot 1_q) \tag{\text{by I.H.}}\\
           & =
\mu(\sigma)(1_p,\ldots,1_p) +
\bigplus_{i\in [k]} \mu(\sigma)(1_p,\ldots,1_p, \size_\delta(\xi_i) \cdot 1_q,1_p,\ldots,1_p)\\
            & \hspace*{8mm}  + \bigplus_{1\leq i < j \leq m} \mu(\sigma)(\ldots, \size_\delta(\xi_i) \cdot 1_q,\ldots, \size_\delta(\xi_j) \cdot 1_q,\ldots)
  \tag{by multilinearity}\\
& =
\mu(\sigma)(1_p,\ldots,1_p) +
\bigplus_{i\in [k]} \size_\delta(\xi_i) \cdot \mu(\sigma)(1_p,\ldots,1_p,1_q,1_p,\ldots,1_p)\\
            &  \hspace*{8mm}   + \bigplus \ldots \size_\delta(\xi_i)\ldots \size_\delta(\xi_j)\ldots \cdot  \mu(\sigma)(\ldots,1_q,\ldots,1_q,\ldots)
  \tag{by multilinearity}\\
& = 
\mu(\sigma)(1_p,\ldots,1_p) +
\bigplus_{i\in [k]} \size_\delta(\xi_i) \cdot 1_q + (0,0)\\
& =  \left\{
\begin{array}{ll}
1_p + 1_q + \bigplus_{i\in [k]} \size_\delta(\xi_i) \cdot 1_q & \mbox{if $\sigma=\delta$}\\
1_p + \bigplus_{i\in [k]} \size_\delta(\xi_i) \cdot 1_q& \mbox{otherwise}.
\end{array}
\right.
\end{align*}
\endgroup
This proves \eqref{eq:multilinear-ex}. Thus,  for each $\xi \in \T_\Sigma$, we have: 
\[
  \sem{\cM}(\xi)= \gamma(\h_\mu(\xi))= \gamma(1_p+ \size_\delta(\xi) \cdot 1_q) = \gamma(1_p) + \size_\delta(\xi) \cdot \gamma(1_q) = \size_\delta(\xi) \enspace.
\]
Thus $\size$ is recognizable by some $(\Sigma,\Ratnum)$-multilinear representation.
\hfill $\Box$
  \end{example}

\subsection{A characterization of wta in terms of multilinear representations}
\label{subs:characterization-with-multilinear-reprs}

  Here we prove that a weighted tree language $r:\T_\Sigma \to \B$ is recognizable by a $(\Sigma,\B)$-multilinear representation if and only if it is initial algebra recognizable by a $(\Sigma,\B)$-wta. For this, let $\cM=(Q,\mu,\gamma)$ be a $(\Sigma,\B)$-multilinear representation and let  $\cA =(Q,\delta,F)$ be a $(\Sigma,\B)$-wta.

  We say that $\cM$ and $\cA$   are {\em related} if 
\begin{compactitem}
\item for every $k \in \mathbb{N}$, $\sigma \in \Sigma^{(k)}$, and
  $q,q_1,\ldots,q_k \in Q$, the equation
  \begin{equation}\label{eq:multilin-repr-and-wta-related}
\mu(\sigma)(\1_{q_1},\ldots,\1_{q_k})_q = \delta_k(q_1\cdots q_k,\sigma,q)
\end{equation}
 holds,
  and
\item for every $q \in Q$, the equation  $\gamma(\1_q) = F_q$ holds.
\end{compactitem}

We recall that  $\V(\cA)= (B^Q,\delta_\cA)$ is the vector algebra of $\cA$.
The intention of the first condition in the definition of relatedness is that $\mu$ and $\delta_\cA$ should play the same role. In order to formalize this, we first prove that each $\delta_\cA(\sigma)$ is a multilinear operation.

\begin{lemma}\label{lm:vector-algebra-is-semimodule-com-sr}\rm Let  $\cA=(Q,\delta,F)$ be a $(\Sigma,\B)$-wta.  For every $k \in \mathbb{N}$ and $\sigma \in \Sigma^{(k)}$, the operation $\delta_\cA(\sigma)$ is multilinear. 
\end{lemma}
\begin{proof} Let $k \in \mathbb{N}$, $\sigma \in \Sigma^{(k)}$, $i \in [k]$, $b,b' \in B$, $v_1,\ldots,v_k, v,v' \in B^Q$, and $q \in Q$. Then we can calculate as follows.
  \begingroup
  \allowdisplaybreaks
  \begin{align*}
    & \delta_\cA(\sigma)(v_1,\ldots,v_{i-1},b\cdot v \oplus b'\cdot v',v_{i+1},\ldots,v_k)_q\\
    =& \bigoplus_{q_1\cdots q_k \in Q^k} \Big(\bigotimes_{j\in [1,i-1]} (v_j)_{q_j}\Big) \otimes
       (b\cdot v \oplus b'\cdot v')_{q_i} \otimes \Big(\bigotimes_{j\in [i+1,k]} (v_j)_{q_j}\Big) \otimes \delta_k(q_1\cdots q_k,\sigma,q)\\
      =& \bigoplus_{q_1\cdots q_k \in Q^k} \Big(\bigotimes_{j\in [1,i-1]} (v_j)_{q_j}\Big) \otimes
         (b \otimes v_{q_i} \oplus b' \otimes (v')_{q_i}) \otimes \Big(\bigotimes_{j\in [i+1,k]} (v_j)_{q_j}\Big) \otimes \delta_k(q_1\cdots q_k,\sigma,q)\\
       =& \bigoplus_{q_1\cdots q_k \in Q^k} \Big[\Big(\bigotimes_{j\in [1,i-1]} (v_j)_{q_j}\Big) \otimes
          b \otimes v_{q_i} \otimes \Big(\bigotimes_{j\in [i+1,k]} (v_j)_{q_j}\Big) \otimes \delta_k(q_1\cdots q_k,\sigma,q)\\
    & \hspace*{15mm} \oplus \Big(\bigotimes_{j\in [1,i-1]} (v_j)_{q_j}\Big) \otimes
      b' \otimes (v')_{q_i} \otimes \Big(\bigotimes_{j\in [i+1,k]} (v_j)_{q_j}\Big) \otimes \delta_k(q_1\cdots q_k,\sigma,q)\Big] \tag{by distributivity of $B$}\\
       =& \bigoplus_{q_1\cdots q_k \in Q^k} \Big[\Big(\bigotimes_{j\in [1,i-1]} (v_j)_{q_j}\Big) \otimes
          b \otimes v_{q_i} \otimes \Big(\bigotimes_{j\in [i+1,k]} (v_j)_{q_j}\Big) \otimes \delta_k(q_1\cdots q_k,\sigma,q)\Big]\\
    & \oplus \bigoplus_{q_1\cdots q_k \in Q^k} \Big[\Big(\bigotimes_{j\in [1,i-1]} (v_j)_{q_j}\Big) \otimes
      b' \otimes (v')_{q_i} \otimes \Big(\bigotimes_{j\in [i+1,k]} (v_j)_{q_j}\Big) \otimes \delta_k(q_1\cdots q_k,\sigma,q)\Big]\\
    =&\ b \otimes \big[ \bigoplus_{q_1\cdots q_k \in Q^k} \Big(\bigotimes_{j\in [1,i-1]} (v_j)_{q_j}\Big) \otimes
          v_{q_i} \otimes \Big(\bigotimes_{j\in [i+1,k]} (v_j)_{q_j}\Big) \otimes \delta_k(q_1\cdots q_k,\sigma,q)\Big]\\
    & \oplus b' \otimes  \Big[\bigoplus_{q_1\cdots q_k \in Q^k} \Big(\bigotimes_{j\in [1,i-1]} (v_j)_{q_j}\Big) \otimes
      (v')_{q_i} \otimes \Big(\bigotimes_{j\in [i+1,k]} (v_j)_{q_j}\Big) \otimes \delta_k(q_1\cdots q_k,\sigma,q)\Big] \tag{by commutativity and distributivity}\\
    &= b \otimes \delta_\cA(\sigma)(v_1,\ldots,v_{i-1},v,v_{i+1},\ldots,v_k)_q
      \oplus b' \otimes \delta_\cA(\sigma)(v_1,\ldots,v_{i-1},v',v_{i+1},\ldots,v_k)_q\\
    &= \big(b \cdot \delta_\cA(\sigma)(v_1,\ldots,v_{i-1},v,v_{i+1},\ldots,v_k)
    \oplus b' \cdot \delta_\cA(\sigma)(v_1,\ldots,v_{i-1},v',v_{i+1},\ldots,v_k)\big)_q\enspace.
    \end{align*}
  \endgroup
 Hence $\delta_\cA(\sigma)$ is multilinear.  
\end{proof}

The following example suggests that distributivity  is a necessary condition for multilinearity of mappings of the form $\delta_\cA(\sigma)$.

\begin{example}\rm \label{ex:wta-deltasigma-not-multilinear} We consider the strong bimonoid $\TropBM=(\mathbb{N}_\infty,+,\min,0,\infty)$ from Example~\ref{ex:strong-bimonoids}(\ref{ex:tropical-bimonoid}), which is commutative and not distributive. Moreover, we consider the string ranked alphabet $\Sigma = \{\gamma^{(1)}, \alpha^{(0)}\}$ and the $(\Sigma,\TropBM)$-wta $\cA =(Q,\delta,F)$ with $Q=\{q_0,q_1\}$  of Example~\ref{ex:run-bimonoid}, which r-recognizes the mapping $\mathrm{exp}$. We will show that the operation $\delta_\cA(\gamma)$ does not satisfy \eqref{ml-Omega} (for $\omega = \delta_\cA(\gamma)$).
  
  First, for every $q \in Q$ and $v \in (\mathbb{N}_\infty)^Q$ we have
  \[
\delta_\cA(\gamma)(v)_q = \min(v_{q_0}, 1) + \min(v_{q_1},1) \enspace.
\]
Hence, for each $v \in (\mathbb{N}_\infty)^Q$ with $v_{q_0} >0$ and $v_{q_1} > 0$, we have $\delta_\cA(\gamma)(v) = \left(\begin{matrix}2 \\ 2\end{matrix}\right)$.

    Second, we let $v_1=v_2=\left(\begin{matrix}2 \\ 2\end{matrix}\right)$ and $b_1=b_2 = 2$  and we compute:
    \begingroup
    \allowdisplaybreaks
    \begin{align*}
      \delta_\cA(\gamma)(b_1 \cdot v_1 \oplus b_2 \cdot v_2)
      &= \delta_\cA(\gamma)(\left(\begin{matrix}\min(2,2) \\ \min(2,2)\end{matrix}\right) +
      \left(\begin{matrix}\min(2,2) \\ \min(2,2)\end{matrix}\right))
      = \delta_\cA(\gamma)(\left(\begin{matrix}4 \\ 4\end{matrix}\right))
      = \left(\begin{matrix}2 \\ 2 \end{matrix}\right)
    \end{align*}
    \endgroup
    and
    \begingroup
    \allowdisplaybreaks
    \begin{align*}
      b_1 \cdot \delta_\cA(\gamma)(v_1) \oplus b_2 \cdot \delta_\cA(\gamma)(v_2)
      &= b_1 \cdot \left(\begin{matrix}2 \\ 2\end{matrix}\right) + b_2 \cdot \left(\begin{matrix}2 \\ 2\end{matrix}\right)
      =  \left(\begin{matrix}\min(2,2) \\ \min(2,2)\end{matrix}\right) + \left(\begin{matrix}\min(2,2) \\ \min(2,2)\end{matrix}\right)
      = \left(\begin{matrix}4 \\ 4\end{matrix}\right) \enspace.
    \end{align*}
    \endgroup
    Thus, the operation $\delta_\cA(\gamma)$ does not satisfy \eqref{ml-Omega}.
    \hfill $\Box$
  \end{example}

In fact, we can obtain the following characterization.

  \begin{corollary-rect}\rm Let $\B$ be a commutative strong bimonoid. Then the following two statements are equivalent.
    \begin{compactenum}
    \item[(A)] $\B$ is distributive.
      \item[(B)] For every ranked alphabet $\Sigma$, $(\Sigma,\B)$-wta $\cA=(Q,\delta,F)$, and $\sigma \in \Sigma$ with $\rk(\sigma) \ge 1$, the operation $\delta_\cA(\sigma)$ satisfies Equation \eqref{ml-Omega} (for $\omega=\delta_\cA(\sigma)$).
      \end{compactenum}
          \end{corollary-rect}

      \begin{proof} Proof of (A)$\Rightarrow$(B): This follows from Lemma~\ref{lm:vector-algebra-is-semimodule-com-sr}.

        \
        
        Proof of (B)$\Rightarrow$(A): We prove the contraposition $\neg$(A)$\Rightarrow \neg$(B). Let $\B = (B,\oplus,\otimes,\0,\1)$ be a commutative strong bimonoid which is  not distributive. Hence there exist $a,b,c \in B$ such that $(a \oplus b) \otimes c \not= a \otimes c \oplus b \otimes c$.

        We define the ranked alphabet $\Sigma = \{\gamma^{(1)}, \alpha^{(0)}\}$ and the $(\Sigma,\B)$-wta $\cA=(Q,\delta,F)$ with $Q=\{q\}$ and $\delta_1(q,\gamma,q)=c$. The other parts of $\cA$ are irrelevant.

        Then for each $v \in B^Q$ we have $\delta_\cA(\gamma)(v)_q = v_q \otimes c$. For $v_1 = (a)$, $v_2 = (b)$, $b_1=b_2=\1$, we can calculate as follows.

        \[
\delta_\cA(\gamma)(b_1 \cdot v_1 \oplus b_2 \cdot v_2)_q = \delta_\cA(\gamma)((a)\oplus (b))_q = (a \oplus b) \otimes c
\]
and
\[
\big(b_1 \cdot \delta_\cA(\gamma)(v_1) \oplus b_2 \cdot \delta_\cA(\gamma)(v_2)\big)_q = a \otimes c \oplus b \otimes c \enspace.
\]
Hence $\delta_\cA(\gamma)$ does not  satisfy \eqref{ml-Omega}.
      \end{proof}

Now we can show the semantic implications of the relatedness between a multilinear representation and a wta.

\begin{lemma}\rm \label{lm:related-linrep-wta} Let $\cM=(Q,\mu,\gamma)$  and $\cA=(Q,\delta,F)$ be  related. The following four statements hold.
\begin{compactitem}
\item[(1)] $\cM$ is bu-deterministic (crisp-deterministic) if and only if $\cA$ is bu-deterministic (crisp-deterministic, respectively).
\item[(2)] For each $\sigma \in \Sigma$, we have that $\mu(\sigma) = \delta_\cA(\sigma)$.
\item[(3)] For every $\xi \in \T_\Sigma$ and $q \in Q$, we have that
  $\h_\mu(\xi)_q = \h_\cA(\xi)_q$.  
\item[(4)] For every $\xi \in \T_\Sigma$ we obtain:
$\sem{\cM}(\xi) = \initialsem{\cA}(\xi)$.
\end{compactitem}
\end{lemma}

\begin{proof} Proof of (1): Let $k\in \mathbb{N}$, $\sigma\in \Sigma^{(k)}$, and $q_1,\ldots,q_k\in Q$.
Let $\cM$ be bu-deterministic. Then, by definition, $B\cdot \1_Q$ is closed under $\mu(\Sigma)$. By the only-if-part of Observation~\ref{obs:closed-mu-Sigma-simpler} we obtain that  there is at most one $q\in Q$ such that $\mu(\sigma)(\1_{q_1},\ldots,\1_{q_k})_q\ne \0$. Hence, using \eqref{eq:multilin-repr-and-wta-related}, there is at most one $q\in Q$ such that $\delta_k(q_1\ldots q_k,\sigma,q)\ne\0$, i.e, $\cA$ is bu-deterministic. Vice versa, if $\cA$ is bu-deterministic, then using~\eqref{eq:multilin-repr-and-wta-related}  and again Observation~\ref{obs:closed-mu-Sigma-simpler},
we obtain that $\cM$ is bu-deterministic. The proof for the case crisp-deterministic is similar and left to the reader.

\

Proof of (2): Let $k \in \mathbb{N}$, $\sigma \in \Sigma^{(k)}$, and $q,q_1,\ldots,q_k \in Q$. Then we have 
\[ \mu(\sigma)(\1_{q_1},\ldots,\1_{q_k})_{q}\ = \ \delta_k(q_1 \cdots q_k,\sigma,q)\  =\ \delta_\cA(\sigma)(\1_{q_1},\ldots,\1_{q_k})_{q}\enspace,\] 
where the first equality holds by \eqref{eq:multilin-repr-and-wta-related} and the second one holds by the definition of $\delta_\cA(\sigma)$. Since $\mu(\sigma)$ is multilinear (by definition) and $\delta_\cA(\sigma)$ is multilinear (due to  Lemma \ref{lm:vector-algebra-is-semimodule-com-sr}) and they coincide on unit vectors, we obtain that $\mu(\sigma) = \delta_\cA(\sigma)$.
This proves (2).

\
  
Proof of (3): It follows from the fact that the two $\Sigma$-algebras $(B^Q,\mu)$ and $(B^Q,\delta_\cA)$ are equal and hence also the two $\Sigma$-algebra homomorphisms $\h_\mu$ and $\h_\cA$ are equal. 

\

Proof of (4): Let $\xi \in \T_\Sigma$. Then 
\begin{align*}
  \initialsem{\cA}(\xi) = \bigoplus_{q\in Q}  \h_\cA(\xi)_q \otimes F_q
                        =  \bigoplus_{q \in Q} \h_\mu(\xi)_q \otimes \gamma(\1_q)
                       &= \gamma(\h_\mu(\xi)) = \sem{\cM}(\xi), 
                          \end{align*}  
                          where the second equality follows from (3); the third equality follows from the fact that $\gamma$ is a linear form over $\B^Q$.
\end{proof}

                        The next theorem contains a characterization of initial algebra  recognizability
in terms of multilinear representations. For the case that the weight algebra $\B$ is a field, the proof was sketched in  \cite[Thm.~3.52]{fulvog09new}. Already in \cite[p.~518]{bor04}, direct constructions have been indicated. There, the author deals with the concepts of $\B$-vector space and basis, and thereby gives the impression that $\B$ is a field. On the background of this understanding, it is a bit puzzling that the author mentions twice the condition that ``the underlying semiring is commutative'', as if the indicated constructions would also hold for commutative semirings. Anyway, the characterization for the case that $\B$ is a commutative semiring has been (re)discovered in \cite{dro22}. However, the statement of preservation of bu-determinism and crisp-determinism is new.

  \begin{theorem-rect} \label{thm:lin-iff-wta-extended} Let $\Sigma$ be a ranked alphabet and $n \in \mathbb{N}$. Moreover, let $\B=(B,\oplus,\otimes,\0,\1)$ be a commutative semiring and let $r:~\T_\Sigma~\to~B$. Then the following two statements are equivalent \cite{dro22}.
    \begin{compactenum}
     \item[(A)] We can construct a $(\Sigma,\B)$-wta $\cA$ with $n$ states such that $\initialsem{\cA} = r$.
    \item[(B)] We can construct an $n$-dimensional $(\Sigma,\B)$-multilinear representation $\cM$ such that $\sem{\cM} = r$.
    \end{compactenum}
    Moreover, if in (A) the  $(\Sigma,\B)$-wta is bu-deterministic (or crisp-deterministic), then so is the $(\Sigma,\B)$-multilinear representation in (B), and vice versa. 
  \end{theorem-rect}
  
\begin{proof} Proof of (A)$\Rightarrow$(B): Let $r: \T_\Sigma \to B$ be i-recognizable by some $(\Sigma,\B)$-wta  $\cA = (Q,\delta,F)$ with $|Q|=n$, i.e., $\initialsem{\cA}(\xi)=r(\xi)$  for every $\xi  \in \T_\Sigma$. We recall that $\V(\cA)=(B^Q,\delta_\cA)$ is the vector algebra of $\cA$. By Lemma \ref{lm:vector-algebra-is-semimodule-com-sr},  for each $\sigma \in \Sigma$, the operation $\delta_\cA(\sigma)$ is multilinear. Moreover, we  consider the linear form $\gamma$ over $\B^Q$ defined, for each $v \in B^Q$, by $\gamma(v) = v \cdot F$.

  Then  $\cM=(Q,\delta_\cA,\gamma)$ is a $(\Sigma,\B)$-multilinear representation.  It is clear that $\cA$ and $\cM$ are related.
By Observation~\ref{lm:related-linrep-wta}(4), we have $\sem{\cM}(\xi) = \initialsem{\cA}(\xi)$  for every $\xi  \in \T_\Sigma$. 
Moreover, by Lemma~\ref{lm:related-linrep-wta}(1), if $\cA$ is bu-deterministic (or crisp-deterministic), then so is~$\cM$. 
  
\

Proof of (B)$\Rightarrow$(A):  Let $r: \T_\Sigma \to B$ be recognizable by the $n$-dimensional $(\Sigma,\B)$-multilinear representation  $\cM=(Q,\mu,\gamma)$.  Hence $r(\xi)= \sem{\cM}(\xi)$ for every $\xi  \in \T_\Sigma$. 

By reading \eqref{eq:multilin-repr-and-wta-related} from right to left and by letting $F_q=\gamma(\1_q)$ for each $q \in Q$, we can construct the $(\Sigma,\B)$-wta  $\cA = (Q,\delta,F)$. Clearly, $\cM$ and $\cA$ are related.
By Lemma~\ref{lm:related-linrep-wta}(4), for every $\xi \in \T_\Sigma$ we obtain: $\sem{\cM}(\xi)$ = $\initialsem{\cA}(\xi)$. Hence $r= \initialsem{\cA}$.  Moreover, by Lemma \ref{lm:related-linrep-wta}(1), if $\cM$ is bu-deterministic (or crisp-deterministic), then so is $\cA$. 
 \end{proof}

%% file: basic-properties.tex
 \chapter{Basic properties of wta}
\label{ch:basic-properties}

This chapter concerns basic properties of each wta. In particular, we prove certain splitting properties of the $\Sigma$-algebra homomorphism $\h_\cA$ for each wta $\cA$ (cf. Section~\ref{sect:splitting-properties-of-hA})
and for each bu-deterministic wta $\cA$ (cf. Lemma~\ref{obs:total-bu-det-wta-calc(new)-H}). We show that the annihilation property of the strong bimonoids propagates over the computation of the semantics of a wta (cf. Section~\ref{sect:properties-of-wta}). We prove a decomposition of the run semantics of crisp wta and a characterization of crisp-deterministic wta (cf. Section~\ref{sec:crisp-wta-dec-and-char}). Moreover, we show how we can extend the arity of input symbols (padding) without changing the semantics (cf. Section~\ref{sect:padding}). Finally, we deal with the images of the semantics of a wta; in particular, we prove a universality property for the initial algebra semantics and we give conditions on the ranked alphabet and the bimonoid which imply the finiteness of the images of the semantics (cf. Section~\ref{sec:images-semantics-restr-sb}).


\section{Splitting properties of the $\Sigma$-algebra homomorphism $\lowercase{\h}_\cA$}
\label{sect:splitting-properties-of-hA}
\label{p:convention-splitting-properties}
\begin{quote}\emph{In this section, we let $\cA=(Q,\delta,F)$ be an arbitrary $(\Sigma,\B)$-wta.}
  \end{quote} 
  Here we consider the $\Sigma$-algebra homomorphism $\h_\cA: \T_\Sigma \to B^Q$ from the $\Sigma$-term algebra $\sfT_\Sigma$ to the vector algebra $\V(\cA)=(B^Q,\delta_\cA)$. We show that the computation of $\h_\cA$ can be split up on input trees of the form $c[\xi]$ where $c \in \C_\Sigma$ and $\xi \in \T_\Sigma$.   For this, we will define the mapping $\h_\cA^\C: \C_\Sigma \to (B^Q \to B^Q)$ and express $\h_\cA(c[\xi])$ in terms of $\h_\cA^\C(c)$ and $\h_\cA(\xi)$, where $B^Q \to B^Q$ denotes the set of all mappings from $B^Q$ to $B^Q$.

  \index{homAcontext@$\h_\cA^\C$}
  First, we define the mapping $g: \e\C_\Sigma \to (B^Q \to B^Q)$ for each elementary context $e \in \e\C_\Sigma$ as follows. Let $e = \sigma(\xi_1,\ldots,\xi_{i-1},z,\xi_{i+1},\ldots,\xi_k)$ for some $k \in \mathbb{N}_+$, $\sigma \in \Sigma^{(k)}$, $i \in [k]$, and  $\xi_1,\ldots,\xi_{i-1},\xi_{i+1},\ldots,\xi_k \in \T_\Sigma$. Then, for each  $v \in B^Q$  we let
  \[
g(e)(v) = \delta_\cA(\sigma)\big(\h_\cA(\xi_1),\ldots,\h_\cA(\xi_{i-1}),v,\h_\cA(\xi_{i+1}),\ldots,\h_\cA(\xi_k)\big) \enspace.
\]
Second, since $(\C_\Sigma,\circ_z,z)$ is free in the set of all monoids with generating set $\e\C_\Sigma$ (cf. Lemma~\ref{lm:freely-generated-context}), there exists a unique monoid homomorphism
\[
\h_\cA^\C: \C_\Sigma \to (B^Q \to B^Q)
\]
from $(\C_\Sigma,\circ_z,z)$ to the monoid $(B^Q \to B^Q,\circ,\id_{B^Q})$ which extends $g$.
Thus, for every $c_1,c_2 \in \C_\Sigma$, we have 
\[
  \h_\cA^\C(c_1[c_2]) = \h_\cA^\C(c_1 \circ_z c_2) = \h_\cA^\C(c_1)\circ \h_\cA^\C(c_2) \enspace.
\]
As usual, function application associates to the left, i.e., $\h_\cA^\C(c)(v)$ stands for $\big(\h_\cA^\C(c)\big)(v)$.

  Now we can prove the mentioned splitting property of $\h_\cA$.

\begin{lemma}\rm \label{lm:hcACchAxi=hAcxi} For every $c \in \C_\Sigma$ and $\xi \in \T_\Sigma$, we have  $\h_\cA(c[\xi]) = \h_\cA^\C(c)\big(\h_\cA(\xi)\big)$. 
\end{lemma}
\begin{proof} We prove the statement by induction on $(\C_\Sigma,\succ_{\C_\Sigma})$  (for the definition of this reduction system cf. page \pageref{page:order-on-contexts}).

  I.B.: Let $c= z$ and $\xi \in \T_\Sigma$. Then $\h_\cA(c[\xi]) = \h_\cA(\xi) = \id_{B^Q}(\h_\cA(\xi)) = \h_\cA^\C(c)(\h_\cA(\xi))$.

  I.S.: Let $c=e[c']$ for some $e \in \e\C_\Sigma$ and $c' \in \C_\Sigma$. We let $e = \sigma(\xi_1,\ldots,\xi_{i-1},z,\xi_{i+1},\ldots,\xi_k)$ for some $k \in \mathbb{N}_+$, $\sigma \in \Sigma^{(k)}$, $i \in [k]$, and  $\xi_1,\ldots,\xi_{i-1},\xi_{i+1},\ldots,\xi_k \in \T_\Sigma$. Then we have
  \begingroup
  \allowdisplaybreaks
  \begin{align*}
   \h_\cA\big((e[c'])[\xi]\big) 
    &= \h_\cA(e[c'[\xi]])
      \tag{by associativity of tree substitution}\\
    &=  \delta_\cA(\sigma)(\h_\cA(\xi_1),\ldots,\h_\cA(\xi_{i-1}),\ \h_\cA(c'[\xi]), \ \h_\cA(\xi_{i+1}),\ldots,\h_\cA(\xi_k)) \\
    &= \delta_\cA(\sigma)(\h_\cA(\xi_1),\ldots,\h_\cA(\xi_{i-1}),\ \h_\cA^\C(c')(\h_\cA(\xi)), \ \h_\cA(\xi_{i+1}),\ldots,\h_\cA(\xi_k))
      \tag{by I.H.} \\
      &= g(e)\big( \h_\cA^\C(c')(\h_\cA(\xi)) \big)
      \tag{by definition of $g$}\\
    &= \h_\cA^\C(e) \big( \h_\cA^\C(c')(\h_\cA(\xi)) \big)
      \tag{because $\h_\cA^\C$ extends $g$}\\
    &=  \big(\h_\cA^\C(e) \circ  \h_\cA^\C(c')\big)(\h_\cA(\xi)) \big)\\
    &=  \h_\cA^\C(e[c'])(\h_\cA(\xi)) \enspace.
      \tag{because $\h_\cA^\C$ is a monoid homomorphism}
  \end{align*}
  \endgroup
  
 \end{proof}

  The following is an immediate consequence of Lemma~\ref{lm:hcACchAxi=hAcxi}. 
  \begin{equation} \label{equ:consistency-for-hAC}
\text{For every $c,c' \in \C_\Sigma$ and $\alpha,\alpha' \in \Sigma^{(0)}$, if $c[\alpha]=c'[\alpha']$, then $\h_\cA^\C(c)(\h_\cA(\alpha)) = \h_\cA^\C(c')(\h_\cA(\alpha'))$} \enspace.
\end{equation}

  The second splitting property concerns $\h_\cA^\C$ and shows the linearity of the mapping $\h_\cA^\C(c)$ for each $c \in \C_\Sigma$ under the assumption that $\B$ is a commutative semiring. Intuitively, if $v$ is a linear combination of vectors, then $\h_\cA^\C(c)(v)$ can be split up along this combination. This splitting extends multilinearity of $\delta_\cA(\sigma)$ (cf. Lemma \ref{lm:vector-algebra-is-semimodule-com-sr}) in an easy way. We recall that $\B^Q=(B^Q,\oplus,\0^Q)$ is a $\B$-semimodule and $\cL(\B^Q,\B^Q)$ is the set of all linear mapping on that semimodule. 

\begin{lemma}\rm \label{lm:haC-linear-mapping}  Let  $\B$ a commutative semiring.  Then, for each $c \in \C_\Sigma$, the mapping $\h_\cA^\C(c)$ is linear, i.e., $\h_\cA^\C: \C_\Sigma \to \cL(\B^Q,\B^Q)$.
\end{lemma}
\begin{proof} By induction on $(\C_\Sigma\succ_{\C_\Sigma})$  (cf. page \pageref{page:order-on-contexts}), we prove the following statement.
  \begin{eqnarray}
    \text{For every $c\in \C_\Sigma$, $b_1,b_2\in B$, and $v_1,v_2\in B ^Q$, we have \ } \label{equ:haC-linear-mapping}\\
    \h_\cA^\C(c)(b_1 \cdot v_1 \oplus  b_2 \cdot v_2)  = b_1 \cdot \h_\cA^\C(c)(v_1) \oplus b_2 \cdot \h_\cA^\C(c)(v_2)\enspace. \notag
    \end{eqnarray}

    I.B.:  Let $c=z$. Then
\[\h_\cA^\C(c)(b_1 \cdot v_1 \oplus b_2 \cdot v_2)= b_1 \cdot v_1 \oplus  b_2 \cdot v_2=b_1 \cdot  \h_\cA^\C(c)(v_1) \oplus b_2 \cdot \h_\cA^\C(c)(v_2)\enspace.\]

I.S.:         Let $c= e[c']$ with $e \in \e\C_\Sigma$ and  $c' \in \C_\Sigma$. Let  $e= \sigma(\xi_1,\ldots,\xi_{i-1},z,\xi_{i+1},\ldots,\xi_k)$. Then we can calculate as follows.
\begingroup
\allowdisplaybreaks
\begin{align*}
  & \h_{\cA}^\C(e[c'])(b_1 \cdot v_1\oplus b_2 \cdot v_2)\\
  &=  \h_{\cA}^\C(e) \Big( \h_\cA^\C(c')(b_1 \cdot v_1\oplus b_2 \cdot v_2) \Big)
  \tag{because $\h_\cA^\C$ is a monoid homomorphism}\\
  &=  \delta_{\cA}(\sigma)\big(\h_{\cA}(\xi_1),\ldots,\h_{\cA}(\xi_{i-1}), \ \h_\cA^\C(c')(b_1 \cdot v_1\oplus b_2 \cdot v_2), \ \h_{\cA}(\xi_{i+1}),\ldots,\h_{\cA}(\xi_k)\big) \\
  &= \delta_{\cA}(\sigma)\big(\h_{\cA}(\xi_1),\ldots,\h_{\cA}(\xi_{i-1}), \ b_1 \cdot \h_{\cA}^\C(c')(v_1) \oplus b_2 \cdot \h_{\cA}^\C(c')(v_2), \ \h_{\cA}(\xi_{i+1}),\ldots,\h_{\cA}(\xi_k)\big) \tag{by I.H.}\\
  &=  b_1 \cdot \delta_{\cA}(\sigma)\big(\h_{\cA}(\xi_1),\ldots,\h_{\cA}(\xi_{i-1}), \ \h_{\cA}^\C(c')(v_1), \ \h_{\cA}(\xi_{i+1}),\ldots,\h_{\cA}(\xi_k)\big) \\
  & \hspace{14mm}  \oplus \ b_2 \cdot \delta_{\cA}(\sigma)\big(\h_{\cA}(\xi_1),\ldots,\h_{\cA}(\xi_{i-1}), \ \h_{\cA}^\C(c')(v_2), \ \h_{\cA}(\xi_{i+1}),\ldots,\h_{\cA}(\xi_k)\big) \tag{by multilinearity of $\delta_\cA(\sigma)$, cf. Lemma \ref{lm:vector-algebra-is-semimodule-com-sr}; this lemma requires that $\B$ is a commutative semiring}\\
  &= b_1 \cdot \h_{\cA}^\C(e)\Big(\h_\cA^\C(c')(v_1)\Big) \ \oplus \ b_2 \cdot \h_{\cA}^\C(e)\Big(\h_\cA^\C(c')(v_2)\Big) \\
  &= b_1 \cdot \h_{\cA}^\C(e[c'])(v_1) \ \oplus \ b_2 \cdot \h_{\cA}^\C(e[c'])(v_2)\enspace.
    \tag{because $\h_\cA^\C$ is a monoid homomorphism}
\end{align*}
\endgroup
    \end{proof}

In the next example we illustrate the evaluation of $\h_\cA^\C(c)(v)$ for a particular $(\Sigma,\Nat)$-wta. We refer the reader to the end of Section~\ref{sec:pos-results-semantics} and, in particular, to Lemma~\ref{lm:connection-between-murev-and-hAC}, where we explain the relationship between the next example and Example \ref{ex:monoid-rep-number-of-occ-of-pattern}.

  \index{$\#_{\sigma(.,\alpha)}$}
  \begin{example}\rm \label{ex:number-of-occ-context-field} Here we consider the  mapping $\#_{\sigma(.,\alpha)}:\T_\Sigma \to \mathbb{N}$ and the $(\Sigma,\Nat)$-wta $\cA$ as defined in Example \ref{ex:number-of-occurrences}.
    Thus, $\Sigma = \{\sigma^{(2)}, \omega^{(1)}, \alpha^{(0)}\}$ and
    \begin{align*}
\#_{\sigma(.,\alpha)}: \T_\Sigma &\to \mathbb{N}\\ 
\xi &\mapsto |U(\xi)| \ \text{ for every $\xi \in T_\Sigma$}
\end{align*}
where $U(\xi) = \{u \in \pos(\xi) \mid \xi(u) = \sigma, \xi(u2)=\alpha\}$. We recall that $\cA=(Q,\delta,F)$, where
\begin{compactitem}
\item $Q  = \{\bot, {a}, {f}\}$ (intuitively, $\bot$ ignores occurrences of the pattern, ${a}$ detects an $\alpha$-labeled leaf, and ${f}$ reports ``pattern found'' up to the root),
 
\item for every $q_1,q_2,q \in Q$ we define
\[
\delta_0(\varepsilon,\alpha,q) = 
\left\{
\begin{array}{ll}
1 & \text{if } q \in \{\bot,{a}\}\\
0 & \text{otherwise}
\end{array}
\right. \ \ 
\delta_1(q_1,\omega,q) = 
\left\{
\begin{array}{ll}
1 & \text{if } q_1q \in \{\bot\bot,{f}{f}\}\\
0 & \text{otherwise}
\end{array}
\right.
\]

\[
\delta_2(q_1q_2,\sigma,q) = 
\left\{
\begin{array}{ll}
1 & \text{if } q_1q_2q \in \{\bot\bot\bot,\bot {a} {f}, \bot {f}{f}, {f}\bot {f}\}\\
0 & \text{otherwise}
\end{array}
\right.
\]

\item $F_\bot=F_{a}=0$ and $F_{{f}}=1$.
\end{compactitem}

For each $c \in \C_\Sigma$, we define the $(Q\times Q)$-matrix :
\begin{equation}\label{equ:matrix-Nc}
  N_c =
  \begin{pmatrix}
    1 & 0 & 0\\
    0 & a(c) & 0\\
    \#_{\sigma(.,\alpha)}(c) & b(c) & 1
    \end{pmatrix}
  \end{equation}
  where 
\begin{compactitem}  
  \item[(a)] the rows and columns are ordered according to the sequence $(\bot,a,f)$, 
\item[(b)] we have extended $\#_{\sigma(.,\alpha)}$ in the obvious way such that it is also applicable to contexts, and 
\item[(c)]  $a(c) = 1$ if $c=z$, and $0$ otherwise; and $b(c) = 1$ if there exists a position $w \in \pos(c)$ such that $z = c(w2)$, and $0$ otherwise. 
\end{compactitem}   
Intuitively, for each $q,p \in Q$, the entry $(N_c)_{q,p}$ is the sum of the weights of all $(q,p)$-runs of $\cA$ on $c$ (for the definition of $(q,p)$-runs cf. page \pageref{page:run-on-contexts}). In particular, $N_z=\mathrm{M}_1$, i.e., the identity $Q\times Q$-matrix for the matrix multiplication.

  In the following, for each $c \in \C_\Sigma$, we  abbreviate the notation $\h_\cA^\C(c)$ by $\semst{c}$.
  In Figure \ref{fig:context-matrix} we illustrate the evaluation of $\semst{c}(v)$ for $c=\sigma(\sigma(\sigma(\alpha,z),\alpha),\alpha)$ and  $v=(1,0,2)$ (where, e.g., $(1,0,2) = \h_\cA(\xi)$ with $\xi = \sigma(\sigma(\alpha,\alpha),\alpha)$).

By induction on the reduction system $(\C_\Sigma,\succ_{\C_\Sigma})$ (cf. page \pageref{page:order-on-contexts}), we prove  that the following statement holds:
\begin{equation}\label{eq:context-matrix}
  \text{For each $c \in \C_\Sigma$ and $v \in \mathbb{N}^Q$, we have $\semst{c}(v) = N_c \cdot v$} \enspace. 
  \end{equation}

\begin{figure}
 \centering 

\scalebox{0.8}{
\rotatebox{0}{
\begin{tikzpicture}[
	transform shape,
	mymatrix/.style={matrix of math nodes,
			ampersand replacement=\&,
			left delimiter  = (,
			right delimiter = )},
	brace/.style={decoration={brace,mirror,amplitude=8pt},
			decorate,
			thick},
	snakearrow/.style={decorate,
			decoration={snake,amplitude=0.7pt,segment length=3mm,pre=lineto,post length=4pt}},
	fixposition/.style={xshift=-3.2cm, yshift=0.45cm}
]

\matrix (M1) [mymatrix]
{ 1 \\ 0 \\ 4 \\ };
\matrix (M2) [mymatrix, below left= 0.75cm and 1.9cm of M1]
{ 1 \& 0 \& 0 \\
  0 \& 0 \& 0 \\
  2 \& 1 \& 1 \\ };
\matrix (M3) [mymatrix, below right= 0.75cm and 2.6cm of M1]
{ 1 \\ 0 \\ 2 \\ };
\matrix (M4) [mymatrix, below left= 0.75cm and 1cm of M2]
{ 1 \& 0 \& 0 \\
  0 \& 0 \& 0 \\
  1 \& 0 \& 1 \\ };
\matrix (M5) [mymatrix, below= 0.75cm of M2]
{ 1 \& 0 \& 0 \\
  0 \& 0 \& 0 \\
  1 \& 0 \& 1 \\ };
\matrix (M6) [mymatrix, below right= 0.75cm and 1cm of M2]
{ 1 \& 0 \& 0 \\
  0 \& 0 \& 0 \\
  0 \& 1 \& 1 \\ };
  
\draw [brace] ($(M3.north east)+(0,0.25)$) -- ($(M2.north west)+(0,0.25)$);
\draw [brace] ($(M6.north east)+(0,0.25)$) -- ($(M4.north west)+(0,0.25)$);

\begin{scope}[
	level 1/.style={node distance=0.4cm and 0.4cm,sibling distance=1.4cm,level distance=1cm},
	rotate=90]
\node[fixposition] at (M4.-90) (s1) {$\sigma$}
	child { node[below= of s1] (z1) {z}}
	child { node[below right= of s1] (a1) {$\alpha$}};
\node[fixposition] at (M5.-90) (s2) {$\sigma$}
	child { node[below= of s2] (z2) {z}}
	child { node[below right= of s2] (a2) {$\alpha$}};
\node[fixposition] at (M6.-90) (s3) {$\sigma$}
	child { node[below= of s3] (z3) {z}}
	child { node[below left= of s3] (a3) {$\alpha$}};
\node[below= 2.5cm of s3] (s4) {$\sigma$}
	child { node (s5) {$\sigma$}
		child { node (a4) {$\alpha$}}
		child { node (a5) {$\alpha$}}}
	child { node (a6) {$\alpha$}};		
\end{scope}

\coordinate[below left = 0.5cm and 0.35cm of s1.base] (cs1);
\coordinate[above right = 0.5cm and 0.1cm of a1.base] (ca1);
\coordinate[below right = 0.5cm and 0.1cm of z1.base] (cz1);
	\draw [in=190, out=-10, looseness=0.65] (cs1) to (cz1);
	\draw [in=-10, out=10,  looseness=0.75] (cz1) to (ca1);
	\draw [in=170, out=170, looseness=0.9] (ca1) to (cs1);
\coordinate[below left = 0.5cm and 0.35cm of s2.base] (cs2);
\coordinate[above right = 0.5cm and 0.1cm of a2.base] (ca2);
\coordinate[below right = 0.5cm and 0.1cm of z2.base] (cz2);
	\draw [in=190, out=-10, looseness=0.65] (cs2) to (cz2);
	\draw [in=-10, out=10,  looseness=0.75] (cz2) to (ca2);
	\draw [in=170, out=170, looseness=0.9] (ca2) to (cs2);
\coordinate[above left = 0.5cm and 0.35cm of s3.base] (cs3);
\coordinate[below right = 0.5cm and 0.1cm of a3.base] (ca3);
\coordinate[above right = 0.5cm and 0.1cm of z3.base] (cz3);
	\draw [in=170, out=10, looseness=0.65] (cs3) to (cz3);
	\draw [in=10, out=-10,  looseness=0.75] (cz3) to (ca3);
	\draw [in=190, out=190, looseness=0.9] (ca3) to (cs3);
\coordinate[below left = 0.2cm and 1.2cm of s1.base] (cs3);
\coordinate[below right = 0.8cm and 0.4cm of a3.base] (ca3);
\coordinate[above right = 2.5cm and 0.4cm of z3.base] (cz3);
	\draw [in=170, out=90, looseness=0.6] (cs3) to (cz3);
	\draw [in=10, out=-10,  looseness=0.7] (cz3) to (ca3);
	\draw [in=270, out=190, looseness=0.5] (ca3) to (cs3);

\draw[snakearrow,->] (M4.-90) to ($(M4.-90)+(0,-1.5)$);
\draw[snakearrow,->] (M5.-90) to ($(M5.-90)+(0,-1.5)$);
\draw[snakearrow,->] (M6.-90) to ($(M6.-90)+(0,-2.3)$);

\node[left = 1cm of M1-1-1.base, anchor=base east] {\strut $N_{c} \cdot h_{\delta_{\cA}}(\xi):$};
\node[left = 1cm of M2-1-1.base, anchor=base east] {\strut $N_{c}:$};
\node[left = 1cm of M3-1-1.base, anchor=base east] {\strut $h_{\delta_{\cA}}(\xi):$};
\node[above left = 1.9cm and 1cm of s1, anchor=north, 
] (xi) {$c:$};
\node[right = 9cm of xi.center, anchor=north, 
] {$\xi:$};

\end{tikzpicture}
} 
} 

  \caption{\label{fig:context-matrix} An illustration of the evaluation of $\semst{c}(v)$ for $c=\sigma(\sigma(\sigma(\alpha,z),\alpha),\alpha)$ and  $v=(1,0,2)$.}
    \end{figure}

I.B.: Let $c = z$ and $v \in \mathbb{N}^Q$. Then $\semst{c}(v) = \semst{z}(v) = \mathrm{id}_{\mathbb{N}^Q}(v) = v = N_c \cdot v$.

I.S.: Let $c = e[c']$ for some $e \in \e\C_\Sigma$ and $c' \in \C_\Sigma$.  Let $e = \sigma(z,\xi)$ for some $\xi \in \T_\Sigma$. Using the I.H., we obtain:
\[
  \semst{c}(v) = \semst{e[c']}(v) = \semst{e}(\semst{c'}(v)) = \semst{e}(N_{c'} \cdot v)
  = \delta_\cA(\sigma)(N_{c'} \cdot v, \h_\cA(\xi)) \enspace.
  \]
 We recall that, by \eqref{eq:h-numberofpatterns-ex}, for each $\xi \in \T_\Sigma$, we have $\h_\cA(\xi)_\bot=1$, $\h_\cA(\xi)_a=g(\xi)$ and $\h_\cA(\xi)_f=\#_{\sigma(.,\alpha)}(\xi)$, where $g(\xi)=1$ if $\xi=\alpha$ and $g(\xi)=0$  otherwise. Then we obtain:
  \begingroup
  \allowdisplaybreaks
  \begin{align*}
    \semst{c}(v)_\bot &= \delta_\cA(\sigma)(N_{c'} \cdot v, \h_\cA(\xi))_\bot
                            = (N_{c'} \cdot v)_\bot \cdot \h_\cA(\xi)_\bot \cdot 1 = v_\bot = (N_c \cdot v)_\bot\\
    \semst{c}(v)_a &= \delta_\cA(\sigma)(N_{c'} \cdot v, \h_\cA(\xi))_a
                              = 0 \tag{the last equality holds because $a(c) = 0$}
                              = (N_c \cdot v)_a\\
    \semst{c}(v)_f &= \delta_\cA(\sigma)(N_{c'} \cdot v, \h_\cA(\xi))_f\\
                          &= (N_{c'} \cdot v)_\bot \cdot \h_\cA(\xi)_a +
                            (N_{c'} \cdot v)_\bot \cdot \h_\cA(\xi)_f +
                            (N_{c'} \cdot v)_f \cdot \h_\cA(\xi)_\bot\\
                          &= (N_{c'} \cdot v)_\bot \cdot (g(\xi) + \#_{\sigma(.,\alpha)}(\xi)) + (N_{c'} \cdot v)_f\\
    &= \left( \begin{pmatrix}
    1 & 0 & 0\\
    0 & 0 & 0\\
    g(\xi) + \#_{\sigma(.,\alpha)}(\xi) & 0 & 1
  \end{pmatrix} \cdot (N_{c'} \cdot v)\right)_f\\
                          &= (N_e \cdot (N_{c'} \cdot v))_f
                            =^{(*)} ((N_e \cdot N_{c'}) \cdot v)_f
                            = (N_{e[c']}\cdot v)_f
                            = (N_c \cdot v)_f \enspace,
  \end{align*}
  \endgroup
  where we have used \eqref{equ:maxtrix-vector-associative} at the equality marked by $(*)$ (note that $\mathbb{N}$ is distributive).

  Now let $e = \sigma(\xi,z)$ for some $\xi \in \T_\Sigma$. Using the I.H., we obtain:
\[
  \semst{c}(v) = \semst{e[c']}(v) = \semst{e}(\semst{c'}(v)) = \semst{e}(N_{c'} \cdot v)
  = \delta_\cA(\sigma)(\h_\cA(\xi), N_{c'} \cdot v) \enspace.
  \]
  Then, using \eqref{eq:h-numberofpatterns-ex}, we obtain:
  \begingroup
  \allowdisplaybreaks
  \begin{align*}
    \semst{c}(v)_\bot &= \delta_\cA(\sigma)(\h_\cA(\xi), N_{c'} \cdot v)_\bot
                            = \h_\cA(\xi)_\bot \cdot (N_{c'} \cdot v)_\bot \cdot 1 = v_\bot = (N_c \cdot v)_\bot\\
    \semst{c}(v)_a &= \delta_\cA(\sigma)(\h_\cA(\xi), N_{c'} \cdot v)_a
                              = 0 \tag{because $a(c) = 0$}
                              = (N_c \cdot v)_a\\
    \semst{c}(v)_f &= \delta_\cA(\sigma)(\h_\cA(\xi), N_{c'} \cdot v)_f\\
                          &= \h_\cA(\xi)_\bot \cdot (N_{c'} \cdot v)_a +
                            \h_\cA(\xi)_\bot \cdot (N_{c'} \cdot v)_f +
                            \h_\cA(\xi)_f \cdot (N_{c'} \cdot v)_\bot\\
                           &= \left( \begin{pmatrix}
    1 & 0 & 0\\
    0 & 0 & 0\\
    \#_{\sigma(.,\alpha)}(\xi) & 1 & 1
  \end{pmatrix} \cdot (N_{c'} \cdot v)\right)_f\\
                          &= (N_e \cdot (N_{c'} \cdot v))_f
                            =^{(*)} ((N_e \cdot N_{c'}) \cdot v)_f
                            = (N_{e[c']}\cdot v)_f
                            = (N_c \cdot v)_f \enspace,
  \end{align*}
  \endgroup
  where, again,  we have used \eqref{equ:maxtrix-vector-associative} at the equality marked by $(*)$.

    Now let $e = \omega(z)$. Using the I.H., we obtain:
\[
  \semst{c}(v) = \semst{e[c']}(v) = \semst{e}(\semst{c'}(v)) = \semst{e}(N_{c'} \cdot v)
  = \delta_\cA(\omega)( N_{c'} \cdot v) \enspace.
  \]
  Then we obtain:
  \begingroup
  \allowdisplaybreaks
  \begin{align*}
    \semst{c}(v)_\bot &= \delta_\cA(\omega)(N_{c'} \cdot v)_\bot
                            =  (N_{c'} \cdot v)_\bot \cdot 1 =  (N_{c'} \cdot v)_\bot \\
    &=\left( \begin{pmatrix}
    1 & 0 & 0\\
    0 & 0 & 0\\
    0 & 0 & 1
  \end{pmatrix} \cdot (N_{c'} \cdot v)\right)_\bot\\
    &= (N_e \cdot (N_{c'} \cdot v))_\bot
                            =^{(*)} ((N_e \cdot N_{c'}) \cdot v)_\bot
                            = (N_{e[c']}\cdot v)_\bot
                            = (N_c \cdot v)_\bot \enspace,\\
    \semst{c}(v)_a &= \delta_\cA(\omega)(N_{c'} \cdot v)_a
                              = 0 \tag{because $a(c) = 0$}
                              = (N_c \cdot v)_a\\
    \semst{c}(v)_f &= \delta_\cA(\omega)(N_{c'} \cdot v)_f
                         = (N_{c'} \cdot v)_f
    = (N_e \cdot (N_{c'} \cdot v))_f = (N_{c} \cdot v)_f,
  \end{align*}
  \endgroup
where, again,  we have used \eqref{equ:maxtrix-vector-associative} at the equality marked by $(*)$.
This proves \eqref{eq:context-matrix}.

Equation \eqref{eq:context-matrix} shows that the vector $\h_\cA^\C(c)(v)$ ($= \semst{c}(v)$) is equal to the matrix-vector product $N_c \cdot v$. In fact, $N_c$ is the matrix representation of the linear mapping $\h_\cA^\C(c)$ under the monoid isomorphism $\psi: \cL(\Nat^Q,\Nat^Q) \to \mathbb{N}^{Q \times Q}$ (cf. page \pageref{p:def-psi-prime}). Formally,
\begin{equation}\label{equ:psi-hAC-c=M-c}
\text{for each $c \in \C_\Sigma$,  we have $\psi(\h_\cA^\C(c)) = N_c$}.
\end{equation}
To see this, let $p,q \in Q$. Then
\[
\psi(\h_\cA^\C(c))_{p,q} = \h_\cA^\C(c)(1_q)_p = (N_c \cdot 1_q)_p = (N_c)_{p,q}
\]
where the first equality follows from the definition of $\psi$ and the second one by  \eqref{eq:context-matrix}.
\hfill $\Box$
\end{example}

\section{Annihilation in wta}
\label{sect:properties-of-wta}

\begin{quote} \emph{In this section, we let $\cA=(Q,\delta,F)$ be an arbitrary $(\Sigma,\B)$-wta.}
  \end{quote}

In this section, we show that the annihilation property of the strong bimonoid $\B$
(i.e., $\mathbb{0} \otimes b = b \otimes \mathbb{0} = \mathbb{0}$) propagates over runs of a $(\Sigma,\B)$-wta $\cA$ on an input tree $\xi \in \Sigma$ and over $Q$-vectors $\h_\cA(\xi)$ (where $Q$ is the set of states of $\cA$; cf. Lemma~\ref{lm:zero-propagation-h}). If $\cA$ is bu-deterministic, then for each input tree $\xi \in \T_\Sigma$, there exists at most one run $\rho \in \R_\cA(\xi)$ with weight different from $\mathbb{0}$, and at most one component $q$ of $\h_\cA(\xi)$ is  different from $\mathbb{0}$; moreover, such a $\rho$ exists if and only if such a component $q$ exists, and if they exist, then $\wt_\cA(\xi,\rho) = \h_\cA(\xi)_q$  (cf. Lemma~\ref{lm:limit-bu-det}). Moreover, by means of the concept of state algebra
we can identify this state and the run (cf. Lemmas~\ref{lm:total-bu-det-sem}).
For each crisp-deterministic $(\Sigma,\B)$-wta  $\cA$,  we prove a characterization result which involves the  congruence relation $\ker(\h_\cA)$  (cf. Theorem~\ref{thm:crisp-wta-final-index}).

\index{Qh@$\Qh{\cA}{\xi}$}
\index{Qh@$\QR{\cA}{\xi}$}
For each $\xi \in \T_\Sigma$, we define the sets
\begin{align*}
\Qh{\cA}{\xi} &= \{q \in Q \mid \h_\cA(\xi)_q\not=\mathbb{0}\}, \ \  \text{ and} \\
\QR{\cA}{\xi}& =\{q\in Q\mid (\exists \rho\in \R_{\cA}(q,\xi)): \wt(\xi,\rho)\ne \mathbb{0}\}\enspace.
\end{align*}

\begin{observation}\rm \label{obs:something-which-is-in-QRxi-and-not-in-Qhxi}
  There exists a string ranked alphabet $\Sigma$, a $(\Intfour,\Sigma)$-wta $\cA$, and a  $\xi \in \T_\Sigma$  such that
  $\QR{\cA}{\xi} \setminus \Qh{\cA}{\xi} \ne \emptyset$.
  \end{observation}

  \begin{proof}
   We consider the string ranked alphabet $\Sigma=\{\gamma^{(1)}, \alpha^{(0)}\}$ and the $(\Intfour,\Sigma)$-wta $\cA=(Q,\delta,F)$ with $Q=\{q_1,q_2,q\}$ and $\delta_0(\varepsilon,\alpha,q_1)=\delta_0(\varepsilon,\alpha,q_2)=2$ and $\delta_1(q_1,\gamma,q)=\delta_1(q_2,\gamma,q)=1$ and each other transition has the weight $0$.

  For the input tree $\xi = \gamma(\alpha)$ we let $\rho \in \R_\cA(\xi)$ such that $\rho(\varepsilon)=q$ and $\rho(1)=q_1$. Then $\wt(\xi,\rho) = \delta_0(\varepsilon,\alpha,q_1) \cdot_4 \delta_1(q_1,\gamma,q) = 2$. Thus $q \in \QR{\cA}{\xi}$.

 Next we calculate $\h_\cA(\xi)_q$.
  \begin{align*}
\h_\cA(\xi)_q = \h_\cA(\alpha)_{q_1} \cdot_4 \delta_1(q_1,\gamma,q) +_4 \h_\cA(\alpha)_{q_2} \cdot_4 \delta_1(q_2,\gamma,q)
    = 2 \cdot_4 1 +_4 2 \cdot_4 1 = 2 +_4 2 = 0 \enspace.
  \end{align*}
Thus $q \in \QR{\cA}{\xi} \setminus \Qh{\cA}{\xi}$. 
\end{proof}

\begin{observation}\rm \label{obs:Qhaxi-is-subset-of-QRAxi} \label{obs:Qhaxi-is-equal-to-QRAxi}$\ $
\begin{compactitem}
\item[(1)] If $\B$ is zero-divisor free or $\cA$ is crisp, then for each $\xi \in \T_\Sigma$, we have $\Qh{\cA}{\xi} \subseteq \state(\xi) = \QR{\cA}{\xi}$.
\item[(2)] If $\B$ is positive, then for each $\xi \in \T_\Sigma$, we have $\Qh{\cA}{\xi} = \state(\xi)= \QR{\cA}{\xi}$.
\end{compactitem}
\end{observation}
\begin{proof} Statement (1)  follows from Lemma~\ref{lm:hAxiqne0-implies-qinhQxi}(1), (3), and (4).
Statement (2)  follows from Corollary~\ref{cor:hAxiqne0-equiv-qinstatexi-equiv-exists-run} immediately.
\end{proof}

\begin{lemma}\label{obs:state-properties-of-wta}\rm For each $\xi\in \T_\Sigma$, the following statements hold.
\begin{compactitem}
\item[(1)] $\Qh{\cA}{\xi}=\emptyset$ if and only if $\h_\cA(\xi)=\0^Q$.
\item[(2)] If $\Qh{\cA}{\xi}=\emptyset$, then $\initialsem{\cA}(\xi)=\0$.
\item[(3)] If $\QR{\cA}{\xi}=\emptyset$, then $\runsem{\cA}(\xi)=\0$.
\item[(4)]  $\Qh{\cA}{\xi} \subseteq \state_\cA(\xi)$ and if $\B$ is positive, then $\Qh{\cA}{\xi} = \state_\cA(\xi)$.
\item[(5)]  $\QR{\cA}{\xi} \subseteq \state_\cA(\xi)$ and if $\B$ is zero-divisor free or $\cA$ is crisp, then 
$\QR{\cA}{\xi} = \state_\cA(\xi)$.
\item[(6)]  If $\state_\cA(\xi)=\emptyset$, then $\initialsem{\cA}(\xi)=\0$ and $\runsem{\cA}(\xi)=\0$.
\end{compactitem}
\end{lemma}
\begin{proof} Statement (1) follows by definition. Statement(2) follows from Statement (1) and the definition of the initial algebra semantics. Statement (3)  follows from the definition of the run semantics immediately. Statement (4) follows from Lemma~\ref{lm:hAxiqne0-implies-qinhQxi}(1) and (2). 
Statement (5) follows from Lemma~\ref{lm:hAxiqne0-implies-qinhQxi}(3) and (4).

To prove Statement (6), let $\state_\cA(\xi)=\emptyset$. Then $\initialsem{\cA}(\xi)=\0$ follows from Statements (4) and (2), and
$\runsem{\cA}(\xi)=\0$ follows from Statements (5) and (3).
\end{proof}

\subsection{Annihilation in arbitrary wta}
\label{subsec:annihilation-arbitrary-wta}

\label{p:convention-annihilation-arbitrary-wta}
\begin{quote}\emph{In this subsection, we let $\cA=(Q,\delta,F)$ be an arbitrary $(\Sigma,\B)$-wta.}
\end{quote}

Since $\mathbb{0}$ is annihilating with respect to the multiplication of $\B$, we obtain the following zero-propagation statements (where the second statement can be compared to \cite[Cor.~3.6]{bor04}).

\begin{lemma} \label{lm:zero-propagation-h} \rm Let $\xi \in \T_\Sigma$ and $w \in \pos(\xi)$. Then the following two statements hold.
\begin{compactenum}
\item[(1)] If $\Qh{\cA}{\xi|_w} = \emptyset$, then $\Qh{\cA}{\xi} = \emptyset$.
\item[(2)]  If $\QR{\cA}{\xi|_w} = \emptyset$, then $\QR{\cA}{\xi} = \emptyset$. 
\end{compactenum}
\end{lemma}
\index{succ@$\succ$}
\begin{proof} Let $\xi \in \T_\Sigma$ and $w \in \pos(\xi)$. In order to allow proof by induction, we define the 
  reduction system $(\prefix(w),\succ)$ where, for every $w_1,w_2 \in \prefix(w)$, we let $w_1 \succ w_2$ if there exists a $j \in \mathbb{N}$ such that $w_2 = w_1 j$. For proving termination, we define the mapping $d: \prefix(v) \to \mathbb{N}$ by $d(w) = |v|-|w|$ for each $w\in \prefix(v)$. Clearly, $d$ is a monotone embedding of the reduction system $(\prefix(v),\succ)$ into the terminating reduction system $(\mathbb{N},\succ_\mathbb{N})$. Hence, by Lemma \ref{lm:fin-branching-embedding-termination}, the reduction system $(\prefix(v),\succ)$ is terminating.
   Moreover, we have that  $\nf_\succ(\prefix(w)) = \{w\}$.

\

Proof of (1): We assume that $\Qh{\cA}{\xi|_w} = \emptyset$. By induction on $(\prefix(w),\succ)$, we prove  that the following statement holds:
\begin{equation}
\text{For each $w'\in \prefix(w)$, we have  $\Qh{\cA}{\xi|_{w'}} = \emptyset\enspace.$} \label{equ:zero-propagation-h}
\end{equation}

I.B.: Let $w'=w$. Then the statement holds by assumption.

I.S.: Now let $w'\in \prefix(w)\setminus \{w\}$ and $\xi(w') = \sigma$ for some $\sigma \in \Sigma^{(k)}$ and $k \in \mathbb{N}$.
Since $w'\in \prefix(w)\setminus \{w\}$, we have that $k \ge 1$ and  there exists a $j\in[k]$ such that $w'j \in \prefix(w)$. 
  Moreover, for each  $q\in Q$, we have 
\begin{align*}
 \h_\cA(\xi|_{w'})_q
  = & \ \h_\cA(\sigma(\xi|_{w'1},\ldots, \xi|_{w'k}))_q
  = \bigoplus_{q_1\cdots q_k \in Q^k}  \Big( \bigotimes_{i \in [k]} \h_\cA(\xi|_{w'i})_{q_i}\Big) \otimes \delta(q_1\cdots q_k,\sigma,q)\enspace.
\end{align*}
Then, by I.H., $\Qh{\cA}{\xi|_{w'j}} = \emptyset$ and thus for each  $q_j \in Q$ we have $\h_\cA(\xi|_{w'j})_{q_j} = \mathbb{0}$. Hence each of the summands is a product in which the $j$-th factor is $\mathbb{0}$. Consequently, each summand is $\mathbb{0}$ and thus  $\h_\cA(\xi|_{w'})_q = \mathbb{0}$. This proves \eqref{equ:zero-propagation-h}.

Finally, $\Qh{\cA}{\xi} = \emptyset$ follows from \eqref{equ:zero-propagation-h} with $w'=\varepsilon$. Thus Statement (1)  holds.

\

Proof of (2): We assume that $\QR{\cA}{\xi|_w} = \emptyset$. Then, by induction on $(\prefix(w),\succ)$,  we prove  that the following statement holds:
\begin{equation}
\text{For each $w'\in \prefix(w)$, we have $\QR{\cA}{\xi|_{w'}} = \emptyset\enspace.$} \label{equ:zero-propagation-r}
\end{equation}
I.B.: Let $w'=w$. Then the statement holds by assumption.

I.S.: Now let $w'\in \prefix(w)\setminus \{w\}$. Let $\xi(w') = \sigma$ for some $\sigma \in \Sigma^{(k)}$ and $k \in \mathbb{N}$. Since  $w'\in \prefix(w)\setminus\{w\}$, we have that $k \ge 1$ and  there exists a $j\in[k]$ such that $w'j \in \prefix(w)$. 

Let $q\in Q$ and $\rho$ be an arbitrary run in $\R_\cA(q,\xi|_{w'})$. The weight of $\rho$, by definition, is
\begin{align*}
\wt(\xi|_{w'},\rho) = \Big( \bigotimes_{i \in [k]} \wt(\xi|_{w'i},\rho|_i)\Big) \otimes \delta(\rho(1)\cdots \rho(k),\sigma,q) 
\enspace.
\end{align*}
By I.H.  $\QR{\cA}{\xi|_{w'j}} = \emptyset$, hence $\wt(\xi|_{w'j},\rho|_{w'j}) = \mathbb{0}$. This implies $\wt(\xi|_{w'},\rho) = \mathbb{0}$ and thus \eqref{equ:zero-propagation-r} holds.

Then $\QR{\cA}{\xi} = \emptyset$ follows from \eqref{equ:zero-propagation-r} for $w'=\varepsilon$. Thus  Statement (2) holds.
\end{proof}

\subsection{Annihilation in bu-deterministic wta}
\label{subsect:annihilation-bu-det-wta}

\label{p:convention-annihilation-bu-det-wta}
\begin{quote} \emph{In this subsection, we let $\cA=(Q,\delta,F)$ be an arbitrary bu-deterministic $(\Sigma,\B)$-wta.}
\end{quote}
\index{estate@$\estate(\xi)$}
Thus, for every $k \in \mathbb{N}$, $\sigma \in \Sigma^{(k)}$, and $w \in Q^k$ there exists at most one $q \in Q$ such that $\delta_k(w,\sigma,q) \not = \mathbb{0}$, or, in other words, $|\mathrm{succ}_\cA(q_1\cdots q_k,\sigma)| \le 1$.
We recall that, by Lemma~\ref{obs:subalgebras-of-the-state-algebra}(2),  we have $\im(\state) \subseteq \cP_{\le 1}(Q)$, where
$\state$ is the unique $\Sigma$-algebra homomorphism from $\sfT_\Sigma$ to the state algebra of $\cA$ (cf. Section~\ref{sec:state-algebra-of-wta}).
To simplify the notation, if $\state(\xi)$ is a singleton, then we denote its one and only element by $\estate(\xi)$.

\index{BQequalone@$B^Q_{=1}$}
\index{BQlessorequalone@$B^Q_{\le 1}$}
Since $\cA$ is bu-deterministic, we can identify a subalgebra of the vector algebra $\V(\cA)$ in which the values of $\h_\cA(\xi)$ can be computed. For this, we define the sets 
\[B^Q_{=1}=\{v \in B^Q \mid |\supp(v)| = 1\} \ \text{ and } \ \ B^Q_{\le 1}= B^Q_{= 1} \cup \{\0^Q\}\enspace.\] 
For each $v \in B^Q_{= 1}$,  we denote by $q_v$ the only state in $\supp(v)$. Hence $v=v_{q_v}\cdot \1_{q_v}$ where $\1_{q_v}$ is the $q_v$-unit vector in $B^Q$. Moreover, for each $\xi \in \T_\Sigma$, we have
\[  |\Qh{\cA}{\xi}|=1 \ \Leftrightarrow \ \h_\cA(\xi)\in B^Q_{=1} \ \Leftrightarrow \ \Qh{\cA}{\xi}=\{q_{\h_\cA(\xi)}\}\enspace.\]

\begin{lemma}\rm \label{lm:properties-sem-of-budet-wta-det-new} The following three statements hold.
    \begin{compactenum}
      \item[(1)] For every $k \in \mathbb{N}$, $\sigma \in \Sigma^{(k)}$, $v_1,\ldots,v_k \in B^Q_{\le 1}$, and $q \in Q$, we have
        \[
          \delta_\cA(\sigma)(v_1,\ldots,v_k)_q =
          \begin{cases}
            (v_1)_{q_{v_1}} \otimes \ldots \otimes (v_k)_{q_{v_k}} \otimes  \delta_k(q_{v_1}\cdots q_{v_k},\sigma,q) 
             &\text{ if $(\forall i \in [k]): v_i \in B^Q_{=1}$}\\
            \0 & \ \ \text{ otherwise}
            \end{cases}
          \]

        \item[(2)]  $B^Q_{\le 1}$ is closed under the operations in $\delta_\cA(\Sigma)$, i.e.,  $(B^Q_{\le 1},\delta_\cA)$ is a $\Sigma$-algebra and it is a subalgebra of the $\Sigma$-algebra $\V(\cA)$.

          \item[(3)] $\im(\h_\cA) \subseteq B_{\le 1}^Q$.
        \end{compactenum}
      \end{lemma}

    \begin{proof} 
 Proof of (1):  If $k=0$, then we are done, because the equality in Statement (1) reduces to $\delta_\cA(\sigma)()_q=\delta_0(\varepsilon,\sigma,q)$, and this holds by the definition of $\delta_\cA$.      
 Thus, let $k \in \mathbb{N}_+$, $\sigma \in \Sigma^{(k)}$, $v_1,\ldots,v_k \in B^Q_{\le 1}$, and $q \in Q$. We recall that 
      \begingroup
      \allowdisplaybreaks
      \begin{align*}
\delta_\cA(\sigma)(v_1,\ldots,v_k)_q = \bigoplus_{q_1 \cdots q_k \in Q^k} (v_1)_{q_1} \otimes \ldots \otimes (v_k)_{q_k} \otimes \delta_k(q_1 \cdots q_k,\sigma,q)\enspace.
          \end{align*}
        \endgroup     
We continue by case analysis.     

\underline{Case (a):} For each $i \in [k]$, we have $v_i \in B^Q_{= 1}$; thus $q_{v_i}$ is defined.  Then
             \begingroup
      \allowdisplaybreaks
      \begin{align*}
        \bigoplus_{q_1 \cdots q_k \in Q^k} (v_1)_{q_1} \otimes \ldots \otimes (v_k)_{q_k} \otimes \delta_k(q_1 \cdots q_k,\sigma,q)
        &= (v_1)_{q_{v_1}} \otimes \ldots \otimes (v_k)_{q_{v_k}} \otimes \delta_k(q_{v_1} \cdots q_{v_k},\sigma,q)
            \end{align*}
            \endgroup
          because all other members of the sum are $\0$. 
          
          \underline{Case (b):} There exists $i \in [k]$ such that $v_i = \0^Q$. Then by Lemma \ref{lm:vector-algebra-is-semimodule-com-sr} we have  $\delta_\cA(\sigma)(v_1,\ldots,v_k)=\0^Q$.

           \
           
         Proof of (2): Let $k \in \mathbb{N}$, $\sigma \in \Sigma^{(k)}$, and $v_1,\ldots,v_k \in B^Q_{\le 1}$. If $v_i\in B^Q_{= 1}$ for each $i\in[k]$, then, by Statement (1) and the fact that $\cA$ is bu-deterministic, 
there exists at most one $q\in Q$ such that $\delta_\cA(\sigma)(v_1,\ldots,v_k)_q\ne \0$. Otherwise, there exists $i\in[k]$ with $v_i=\0^Q$. Then again by Statement (1), we have $\delta_\cA(\sigma)(v_1,\ldots,v_k)=\0^Q$. Hence in both cases  $\delta_\cA(\sigma)(v_1,\ldots,v_k)\in  B^Q_{\le 1}$.

\

Proof of (3): By Observation \ref{obs:smallest-subalgebra-im}, the $\Sigma$-algebra  $(\im(\h_\cA),\delta_\cA)$ is the smallest subalgebra of $\V(\cA)=(B^Q,\delta_\cA)$. By Statement~(2), $(B_{\le 1}^Q,\delta_\cA)$ is a subalgebra of $\V(\cA)=(B^Q,\delta_\cA)$. Thus $\im(\h_\cA) \subseteq B_{\le 1}^Q$.
 \end{proof}

 Due to the fact that $\cA$ is bu-deterministic and by using Lemma \ref{lm:properties-sem-of-budet-wta-det-new}(1), we can now strengthen Lemma~\ref{lm:hAxiqne0-implies-qinhQxi}(1) and (2).

\begin{lemma}\rm \label{lm:hAxiqne0-implies-qinhQxi-for-bu-det-wta} The following two statements hold.
  \begin{compactenum}
  \item[(1)]  For every $\xi \in \T_\Sigma$ and $q \in Q$, if $\h_\cA(\xi)_q \ne \0$, then $\state(\xi)=\{q\}$.
  \item[(2)] Let $\B$ be zero-divisor free. For every $\xi \in \T_\Sigma$ and $q \in Q$, if $\state(\xi)=\{q\}$, then $\h_\cA(\xi)_q \ne \0$.
    \end{compactenum}
  \end{lemma}

  \begin{proof} Proof of (1): Let $\xi \in\T_\Sigma$. By Lemma \ref{obs:subalgebras-of-the-state-algebra}(1), we have $|\state(\xi)|\le 1$. Then the statement follows from Lemma~\ref{lm:hAxiqne0-implies-qinhQxi}(1).

    \

    Proof of (2): We prove the statement by induction on $\T_\Sigma$.
    Let $\xi= \sigma(\xi_1,\ldots,\xi_k)$ and $q \in Q$.
 \begingroup
    \allowdisplaybreaks
    \begin{align*}
      &\state(\xi) =\{q\} \\
    \Leftrightarrow \ \  & \delta_Q(\sigma)\big(\state(\xi_1),\ldots,\state(\xi_k)) = \{q\}
                           \tag{because $\state$ is a $\Sigma$-algebra homomorphism}\\
      \Leftrightarrow \ \ & ( \exists q_1 \cdots q_k \in Q^k):  q_1 \in \state(\xi_1) \text{ and }  \ldots \text{ and }  q_k \in \state(\xi_k)
                            \text{ and }  \delta_k(q_1\cdots q_k,\sigma,q) \not= \0
                            \tag{by definition of $\delta_Q(\sigma)$}\\
       \Rightarrow \ \ & ( \exists q_1 \cdots q_k \in Q^k):  \state(\xi_1)=\{q_1\} \text{ and }  \ldots \text{ and }  \state(\xi_k)=\{q_k\}
                            \text{ and }  \delta_k(q_1\cdots q_k,\sigma,q) \not= \0
      \tag{by Lemma \ref{obs:subalgebras-of-the-state-algebra}(1) }\\
      \Rightarrow \ \   & ( \exists q_1 \cdots q_k \in Q^k):  \h_\cA(\xi_1)_{q_1}  \not= \0 \text{ and } \ldots \text{ and }  \h_\cA(\xi_k)_{q_k} \not= \0 \text{ and }  \delta_k(q_1\cdots q_k,\sigma,q) \not= \0
                          \tag{by I.H.} \\
            \Rightarrow \ \  &  ( \exists q_1 \cdots q_k \in Q^k):  \h_\cA(\xi_1)_{q_1} \otimes \ldots \otimes \h_\cA(\xi_k)_{q_{k}} \otimes \delta_k(q_{1}\cdots q_{k},\sigma,q) \not= \0
                                      \tag{because $\B$ is zero-divisor free}\\
      \Rightarrow \ \ & \delta_\cA(\sigma)(\h_\cA(\xi_1),\ldots,\h_\cA(\xi_k))_q \ne \0
                        \tag{by Lemma~\ref{lm:properties-sem-of-budet-wta-det-new}(1); because by  Lemma~\ref{lm:properties-sem-of-budet-wta-det-new}(3) $\h_\cA(\xi_i) \in B_{=1}^Q$ for each $i \in [k]$}\\
      \Rightarrow \ \ & \h_\cA(\sigma(\xi_1,\ldots,\xi_k))_q \ne \0 \enspace.
                        \qedhere
         \end{align*}
\endgroup
\end{proof}

Lemma~\ref{lm:hAxiqne0-implies-qinhQxi}(1) and Lemma~\ref{lm:hAxiqne0-implies-qinhQxi-for-bu-det-wta} imply the following corollary. It strengthens Corollary~\ref{cor:hAxiqne0-equiv-qinstatexi-equiv-exists-run} for bu-deterministic wta.

\begin{corollary}\rm \label{cor:hAxiqne0-iff-qinhQxi-for-bu-det-wta}  Let $\B$ be zero-divisor free.
 For every $\xi \in \T_\Sigma$ and $q \in Q$, we have:  
 \[\h_\cA(\xi)_q\ne \0  \ \ \Leftrightarrow \ \ \state(\xi)=\{q\} \ \  \Leftrightarrow \ \ \text{ there exists $\rho \in \R_\cA(q,\xi)$ such that $\wt(\xi,\rho)\ne\0$}\enspace.\]
\end{corollary}
\begin{proof} The first equivalence follows from Lemma~\ref{lm:hAxiqne0-implies-qinhQxi-for-bu-det-wta}, the second one follows from 
Lemma~\ref{lm:hAxiqne0-implies-qinhQxi}(3) and (4) and the fact that $|\state(\xi)|\le 1$ (cf. Lemma \ref{obs:subalgebras-of-the-state-algebra}(1)).
\end{proof}

In Figure~\ref{table:summary-relations-between-QhcAxi-statexi-QRcAxi}, we summarize some of the results which relate the sets $\Qh{\cA}{\xi}$, $\state_\cA(\xi)$, and $\QR{\cA}{\xi}$  where $\xi \in \T_\Sigma$ and $q \in Q$.

\begin{figure}
  \footnotesize
  \begin{tabular}{l|c|c|c}
    & $\B$ strong bimonoid & $\B$ zero-divisor free & $\B$ is positive\\[2mm]\hline
    &&& \\[-1mm]
    $(\Sigma,\B)$-wta $\cA$ & $\Qh{\cA}{\xi} \cup \QR{\cA}{\xi} \subseteq \state_\cA(\xi)$ 
                           &  $\Qh{\cA}{\xi} \subseteq \state_\cA(\xi) = \QR{\cA}{\xi}$
                                                    & $\Qh{\cA}{\xi} = \state_\cA(\xi) = \QR{\cA}{\xi}$\\[1mm]
    & (by Lemma \ref{lm:hAxiqne0-implies-qinhQxi}(1) and (3))
    & (by Obs. \ref{obs:Qhaxi-is-subset-of-QRAxi}(1)
     & (by Lemma \ref{obs:Qhaxi-is-equal-to-QRAxi}(2))\\[2mm]\hline
    &&& \\[-1mm]
$(\Sigma,\B)$-wta $\cA$ & as above & $\Qh{\cA}{\xi} = \state_\cA(\xi) = \QR{\cA}{\xi}$ &  as above\\
bu-deterministic & & (by Corollary \ref{cor:hAxiqne0-iff-qinhQxi-for-bu-det-wta})
    \end{tabular}

  \caption{\label{table:summary-relations-between-QhcAxi-statexi-QRcAxi} Summary of results on the relation between the sets $\Qh{\cA}{\xi}$, $\state_\cA(\xi)$, and $\QR{\cA}{\xi}$.}
  \end{figure}

\index{canonical run}
\index{crho@$\crhorm_\xi$}
  Let $\xi \in \T_\Sigma$ such that $|\state(\xi)|=1$.
By Observations~\ref{obs:deltaQ-is-strict-in-emptyset} and \ref{obs:subalgebras-of-the-state-algebra}(1), for each $w \in \pos(\xi)$, we have $|\state(\xi|_w)|=1$ and hence $\estate(\xi|_w)$ exists. We define  the \emph{canonical run of $\cA$ on $\xi$}, denoted by $\crhorm_\xi$, to be the run in $\R_\cA(\xi)$ such that, for each $w \in \pos(\xi)$, we let 
\(\crhorm_\xi(w) = \estate(\xi|_w)\). Hence $\crhorm_\xi \in \R_\cA(\estate(\xi),\xi)$.

In the following lemma we relate the sets $\Qh{\cA}{\xi}$  and $\QR{\cA}{\xi}$ for each $\xi \in \T_\Sigma$.

\begin{lemma-rect}\rm \label{lm:limit-bu-det} \cite[Lm.~3.5]{fulkosvog19} Let $\Sigma$ be a ranked alphabet. Moreover, let $\B=(B,\oplus,\otimes,\0,\1)$ be  a strong bimonoid, let $\cA = (Q,\delta,F)$ be  a bu-deterministic $(\Sigma,\B)$-wta, and let $\xi \in \T_\Sigma$. Then the following three statements hold.
\begin{compactenum}
\item[(1)] $|\Qh{\cA}{\xi}| \le 1$  (cf. \cite[Thm.~3.6]{borvog03}).
\item[(2)] $|\QR{\cA}{\xi}| \le 1$. 
\item[(3)] Either (a) $\Qh{\cA}{\xi}=\emptyset=\QR{\cA}{\xi}$ or (b) $|\state(\xi)|=1$ and $\Qh{\cA}{\xi}=\state(\xi)=\QR{\cA}{\xi}$, and for the canonical run $\crhorm_\xi$ of $\cA$ on $\xi$, we have 
  \begin{compactitem}
  \item $\wt(\xi,\crhorm_\xi)\ne \0$ and  $\h_{\cA}(\xi)_{\estate(\xi)} = \wt(\xi,\crhorm_\xi)$, 
  \item for each $\rho \in \R_{\cA}(\xi) \setminus \{\crhorm_\xi\}$, we have $\wt(\xi,\rho) =\0$,
   \item $\initialsem{\cA}=\h_{\cA}(\xi)_{\estate(\xi)}\otimes F_{\estate(\xi)}$, and
  \item $\runsem{\cA}=\wt(\xi,\crhorm_\xi)\otimes F_{\estate(\xi)}$
    \end{compactitem}
  \end{compactenum}
\end{lemma-rect}

\begin{proof} Proof of (1): This follows from Lemma~\ref{lm:properties-sem-of-budet-wta-det-new}(3).

\

Proof of (2): We prove the statement by induction on $\T_\Sigma$.
Let $\xi = \sigma(\xi_1,\ldots,\xi_k)$. By I.H., the following two cases are possible.
  
  \underline{Case (a):} For each $i \in [k]$, there exists exactly one state $q_i\in Q$, such that there exists a run $\rho_i \in \R_\cA(q_i,\xi_i)$ with $\wt(\xi_i,\rho_i) \ne \0$. Using that $\cA$ is bu-deterministic, we distinguish the following two subcases.
  
  \underline{Case (a1):} There exists exactly one  $q\in Q$ with $\delta_k(q_1\cdots q_k,\sigma,q)\ne \0$. Let $\rho\in \R_\cA(q,\xi)$
  be the run defined by $\rho(iw)=\rho_i(w)$ for every $i\in[k]$ and $w\in \pos(\xi_i)$. Then, for each 
  $\rho'\in  \R_\cA(\xi)\setminus\{\rho\}$ we have $\wt(\xi,\rho')=\0$.
  
  \underline{Case (a2):} For each $q\in Q$ we have $\delta_k(q_1\cdots q_k,\sigma,q)=\0$. Then for each 
  $\rho\in \R_\cA(\xi)$, we have $\wt(\xi,\rho)=\0$.
  
  \underline{Case (b):} There exists $i \in [k]$ such that, for every state $q_i\in Q$ and run $\rho_i \in \R_\cA(q_i,\xi_i)$, we have $\wt(\xi_i,\rho_i) = \0$. Then for each 
  $\rho\in \R_\cA(\xi)$, we have $\wt(\xi,\rho)=\0$.
  
\

Proof of (3): If $\Qh{\cA}{\xi}=\emptyset=\QR{\cA}{\xi}$, then we are done. We prove the other cases by induction on $\T_\Sigma$. Let $\xi = \sigma(\xi_1,\ldots,\xi_k)$.

First, assume that  $\Qh{\cA}{\xi}\not=\emptyset$. Then, by Statement (1), $|\Qh{\cA}{\xi}|=1$ and hence, by Lemma~\ref{lm:hAxiqne0-implies-qinhQxi-for-bu-det-wta}(1),  $\Qh{\cA}{\xi}=\state(\xi)$ and $|\state(\xi)|=1$. Hence 
\[\initialsem{\cA}(\xi) =  \bigoplus_{p \in Q} \h_\cA(\xi)_p \otimes F_p = \h_\cA(\xi)_{\estate_\cA(\xi)} \otimes F_{\estate_\cA(\xi)}\enspace.\]
Moreover, by Lemma~\ref{lm:zero-propagation-h}(1), $\Qh{\cA}{\xi_i}\ne \emptyset$ for each $i\in [k]$. By  I.H., for each $i\in [k]$, we have $|\state(\xi_i)|=1$, 
$\Qh{\cA}{\xi_i}=\state(\xi_i)=\QR{\cA}{\xi_i}$, and $\wt(\xi_i,\crhorm_{\xi_i})\ne \0$ and $\h_{\cA}(\xi_i)_{\estate(\xi_i)} = \wt(\xi_i,\crhorm_{\xi_i})$.

Then we have
\begingroup
\allowdisplaybreaks
\begin{align*}
  \wt(\xi,\crhorm_\xi) &=  \Big(\bigotimes_{i \in [k]}\wt(\xi_i,(\crhorm_\xi)|_i)\Big) \otimes
                              \delta_k(\crhorm_\xi(1)\cdots\crhorm_\xi(k),\sigma,\crhorm_\xi(\varepsilon))
  \tag{by \eqref{equ:weight-of-run}}\\
                            &= \Big(\bigotimes_{i \in [k]}\wt(\xi_i,\crhorm_{\xi_i})\Big) \otimes
                              \delta_k(\crhorm_\xi(1)\cdots\crhorm_\xi(k),\sigma,\crhorm_\xi(\varepsilon))\\
                            &=  \Big(\bigotimes_{i \in [k]} \h_{\cA}(\xi_i)_{\estate(\xi_i)}\Big) \otimes
                              \delta_k(\crhorm_\xi(1)\cdots\crhorm_\xi(k),\sigma,\crhorm_\xi(\varepsilon))
                              \tag{because $\h_{\cA}(\xi_i)_{\estate(\xi_i)} = \wt(\xi_i,\crhorm_{\xi_i})$ for $i\in[k]$}\\
  &=  \Big(\bigotimes_{i \in [k]} \h_{\cA}(\xi_i)_{\estate(\xi_i)}\Big) \otimes
    \delta_k(\estate(\xi_1) \cdots \estate(\xi_k),\sigma, \estate(\xi))
  \tag{by definition of $\crhorm_\xi$}\\
                      &= \h_{\cA}(\xi)_{\estate(\xi)}
                        \tag{by Lemma~\ref{lm:properties-sem-of-budet-wta-det-new}(1)}    \enspace.
\end{align*}
\endgroup
Since $\h_{\cA}(\xi)_{\estate(\xi)}\ne \0$, we have $\wt(\xi,\crhorm_\xi)\ne  \0 $, i.e., $\estate(\xi)\in \QR{\cA}{\xi}$, and thus, by Statement (2),  $\QR{\cA}{\xi}=\state(\xi)$.

Lastly, let $\rho\in \R_\cA(\xi)$ with $\rho\ne \crhorm_\xi$. If $\rho(\varepsilon) \ne \estate(\xi)$, then $\wt(\xi,\rho)=\0$ because $\QR{\cA}{\xi}=\state(\xi)$. Otherwise $\rho\in \R_\cA(\estate(\xi),\xi)$.
Using the definition of $\crhorm_\xi$, from the assumption that $\crhorm_{\xi_i}$ is the only run in  $\R_{\cA}(\xi_i)$  with weight different from $\0$ for each $i\in [k]$, and the fact that $\cA$ is bu-deterministic, it follows easily that $\wt(\xi,\rho)= \0$. Since $\crhorm_{\xi}(\varepsilon)=\estate(\xi)$, we obtain 
\[\runsem{\cA}(\xi)
    =   \bigoplus_{\rho\in \R_\cA(\xi)} \wt_\cA(\xi,\rho) \otimes F_{\rho(\varepsilon)} =  \wt_\cA(\xi,\crhorm_\xi)\otimes F_{\estate_\cA(\xi)}\enspace.\]

\

Second, assume that $\QR{\cA}{\xi}\not=\emptyset$. By Statement (2), there exists $q\in Q$ such that $\QR{\cA}{\xi}=\{q\}$.
By Lemma~\ref{lm:zero-propagation-h}~(2), for each $i \in [k]$, we have $\QR{\cA}{\xi_i}\ne \emptyset$.
Thus, by  I.H., for each $i\in [k]$, we have $|\state(\xi_i)|=1$,
$\Qh{\cA}{\xi_i}=\state(\xi_i)=\QR{\cA}{\xi_i}$ and $\wt(\xi_i,\crhorm_{\xi_i})\ne \0$ and $\h_{\cA}(\xi_i)_{\estate(\xi_i)} = \wt(\xi_i,\crhorm_{\xi_i})$.
Moreover, for each run $\rho_i\in \R_\cA(\xi_i)\setminus\{\crhorm_{\xi_i}\}$, we have  $\wt(\xi_i,\rho_i)= \0$.

Let $\rho$ be the run on $\xi$ defined by $\rho(\varepsilon)=q$ and $\rho|_i=\crhorm_{\xi_i}$ for each $i\in[k]$. Since $\QR{\cA}{\xi}=\{q\}$ and, for each $i\in[k]$,  $\crhorm_{\xi_i}$ is the only run in  $\R_{\cA}(\xi_i)$ with weight different from $\0$, we have $\wt(\xi,\rho)\ne  \0$. Now we calculate as follows:
\begingroup
\allowdisplaybreaks
\begin{align*}
  \wt(\xi,\rho) &=  \Big(\bigotimes_{i \in [k]}\wt(\xi_i,\rho|_i)\Big) \otimes
                              \delta_k(\rho(1)\cdots\rho(k),\sigma,\rho(\varepsilon))
  \tag{by \eqref{equ:weight-of-run}}\\
    &= \Big(\bigotimes_{i \in [k]}\wt(\xi_i,\crhorm_{\xi_i})\Big) \otimes
                               \delta_k(\estate(\xi_1) \cdots \estate(\xi_k),\sigma, q)
                               \tag{by the definition of $\rho$}\\
    &=  \Big(\bigotimes_{i \in [k]} \h_{\cA}(\xi_i)_{\estate(\xi_i)}\Big) \otimes
    \delta_k(\estate(\xi_1) \cdots \estate(\xi_k),\sigma, q) \tag{because $\h_{\cA}(\xi_i)_{\estate(\xi_i)} = \wt(\xi_i,\crhorm_{\xi_i})$ for $i\in[k]$}\\  
                &=  \h_\cA(\xi)_q
                  \tag{by Lemma~\ref{lm:properties-sem-of-budet-wta-det-new}(1), because $\Qh{\cA}{\xi_i}=\state(\xi_i)$ for $i\in[k]$}                 \enspace.    
\end{align*}
\endgroup
Since $\wt(\xi,\rho)\ne \0$, we have $q\in \Qh{\cA}{\xi}$ and thus by Statement (1) we obtain  $\Qh{\cA}{\xi}=\{q\}$. Hence by 
Lemma~\ref{lm:hAxiqne0-implies-qinhQxi-for-bu-det-wta}(1) we have $\state(\xi)=\{q\}$ and thus
$\Qh{\cA}{\xi}=\state(\xi)=\QR{\cA}{\xi}$. Moreover, $\rho=\crhorm_\xi$
and thus $\wt(\xi,\crhorm_\xi)\ne \0$ and  $\h_{\cA}(\xi)_{\estate(\xi)} = \wt(\xi,\crhorm_\xi)$.
Then we can show in the same way as above that,  for each $\rho\in \R_\cA(\xi)$ with $\rho\ne \rhorm_\xi$, we have $\wt(\xi,\rho)\ne \0$.
Finally, it easily follows that $\initialsem{\cA}=\h_{\cA}(\xi)_{\estate(\xi)}\otimes F_{\estate(\xi)}$ and  $\runsem{\cA}=\wt(\xi,\crhorm_\xi)\otimes F_{\estate(\xi)}$.
\end{proof}

\begin{corollary}\rm \label{lm:properties-hA-of-budet-wta-det-new-H} Let $\xi \in \T_\Sigma$. Then
       \[\initialsem{\cA}(\xi)= \begin{cases}
         \h_\cA(\xi)_{\estate(\xi)} \otimes F_{\estate(\xi)} & \text{ if $\h_\cA(\xi)\in B_{=1}^Q$} \\
         \0 & \text{ otherwise,}
    \end{cases}
  \]
    \end{corollary}
 
    \begin{proof} It follows from Lemma \ref{lm:limit-bu-det}(3)(a), Lemma \ref{obs:state-properties-of-wta}(2), and Lemma \ref{lm:limit-bu-det}(3)(b).
     \end{proof}

A further consequence of Lemma \ref{lm:limit-bu-det} is that,  for each bu-deterministic wta, its run semantics coincides with its  initial algebra semantics. We will deal with this topic in Section \ref{sec:pos-results-semantics}. 
Here we present three other easy consequences of Lemma \ref{lm:limit-bu-det}.

\begin{corollary}\label{lm:bu-det-image-multiplication-new}\rm  (1) $\im(\initialsem{\cA}) \subseteq \langle \im(\delta)\rangle_{\{\otimes\}} \otimes \im(F)$ and (2) $\im(\runsem{\cA}) \subseteq  \langle \im(\delta)\rangle_{\{\otimes\}} \otimes \im(F)$.
\end{corollary}
\begin{proof}
  Proof of (1): First, by induction on $\T_\Sigma$, we prove  the following statement.
\begin{equation}\label{eq:bu-det-init-hom-multiplication-new}
\text{For every $\xi\in \T_\Sigma$ and $q\in Q$, we have $\h_\cA(\xi)_q \in \langle\im(\delta)\rangle_{\{\otimes\}}$.}
\end{equation}

For this let $\xi=\sigma(\xi_1,\ldots,\xi_k)$ and $q\in Q$. Since $\h_\cA$ is a $\Sigma$-algebra homomorphism, we have
\begin{align*}
\h_\cA(\sigma(\xi_1,\ldots,\xi_k))_q = \delta_\cA(\sigma)(\h_\cA(\xi_1),\ldots,\h_\cA(\xi_1))_q\enspace.
\end{align*}

By Lemma~\ref{lm:properties-sem-of-budet-wta-det-new}(3), we distinguish two cases.

\underline{Case (a):} For each $i\in[k]$, we have  $\h_\cA(\xi_i)\in B^Q_{= 1}$. Then, by  Lemma~\ref{lm:hAxiqne0-implies-qinhQxi-for-bu-det-wta} (1), we have
\begin{align*}
\h_\cA(\sigma(\xi_1,\ldots,\xi_k))_q=\Big( \bigotimes_{i \in [k]} \h_\cA(\xi_i)_{\estate(\xi_i)} \Big)\otimes \delta_k(\estate(\xi_1)\cdots \estate(\xi_k),\sigma,q)\enspace.
\end{align*}
By I.H.,  for each $i\in[k]$, we have $ \h_\cA(\xi_i)_{\estate(\xi_i)}\in \langle\im(\delta)\rangle_{\{\otimes\}}$ and thus  \eqref{eq:bu-det-init-hom-multiplication-new} holds.

\underline{Case (b):} There exists $i\in[k]$ such that $\h_\cA(\xi_i)=\0^Q$. Thus, by using the I.H., we have  $\0\in \langle\im(\delta)\rangle_{\{\otimes\}}$. Moreover, Lemma~\ref{lm:properties-sem-of-budet-wta-det-new}(1), we have $\h_\cA(\xi)_q=\0$. Hence \eqref{eq:bu-det-init-hom-multiplication-new} holds.
This finishes the proof of \eqref{eq:bu-det-init-hom-multiplication-new}.

In order to prove (1), it suffices show the following.
\begin{align}\label{inclusion-in-multiplicatve-new}
\text{For each  $\xi\in \T_\Sigma$, we have } \initialsem{\cA}(\xi)\in \langle \im(\delta)\rangle_{\{\otimes\}} \otimes \im(F)\enspace. 
\end{align}
Let $\xi\in \T_\Sigma$. If $\h_\cA(\xi)\in B^Q_{=1}$, then \eqref{inclusion-in-multiplicatve-new} follows from  Corollary~\ref{lm:properties-hA-of-budet-wta-det-new-H} and \eqref{eq:bu-det-init-hom-multiplication-new}. Otherwise $\h_\cA(\xi)\in \0^Q$
and thus $\sem{\cA}(\xi)=\0$. Then, by \eqref{eq:bu-det-init-hom-multiplication-new} we have $0\in \langle \im(\delta)\rangle_{\{\otimes\}}$ and thus \eqref{inclusion-in-multiplicatve-new} holds.

\

Proof of (2): This can be shown similarly to the proof of Statement (1), by using Lemma \ref{lm:limit-bu-det}(3).
\end{proof}

A further consequence of Lemma \ref{lm:limit-bu-det}(3) is that both, the run semantics and the initial algebra semantics of a bu-deterministic $(\Sigma,\B)$-wta $\cA$ are determined by the multiplicative monoid of $\B$; the additive part does not play any role (apart from its unit element). 
Hence, if we  replace the addition  by another operation such that the algebra is still a strong bimonoid, then both the run semantics and the initial algebra semantics of $\cA$ remain the same mapping. Below, we prove this statement and show some pairs of strong bimonoids which differ only in the definition of the addition.

\begin{corollary}\label{cor:bu-det-wta-B1-B2}\rm Let $\B_1 = (B,\oplus,\otimes,\0,\1)$ and $\B_2 = (B,+,\otimes,\0,\1)$ be strong bimonoids and $\cA=(Q,\delta,F)$ be a bu-deterministic $(\Sigma,\B_1)$-wta. Then, for the bu-deterministic $(\Sigma,\B_2)$-wta $\cB=(Q,\delta,F)$, we have 
  (1)~$\runsem{\cB}=\runsem{\cA}$ 
  and 
  (2) $\initialsem{\cB}=\initialsem{\cA}$.
  \end{corollary}
\begin{proof} It is obvious that, for every $\xi \in \T_\Sigma$ and $q\in Q$, we have $\R_\cA(q,\xi)=\R_\cB(q,\xi)$. Moreover, since $\delta$ is the family of transition mappings of $\cA$ and of $\cB$, for each $\rho\in \R_\cA(q,\xi)$, we have
\begin{equation}\label{eq:run-weights-equal}
\wt_\cA(\xi,\rho)=\wt_\cB(\xi,\rho)\enspace.
\end{equation}
Hence it follows that
\begin{equation}\label{eq:set-run-states-equal}
\QR{\cA}{\xi}=\QR{\cB}{\xi}\enspace.
\end{equation}

Now let $\xi \in \T_\Sigma$. By Lemma \ref{lm:limit-bu-det}(3) and \eqref{eq:set-run-states-equal}, we distinguish two cases.

\underline{Case (a):} $\Qh{\cA}{\xi}=\emptyset=\QR{\cA}{\xi}=\QR{\cB}{\xi}=\emptyset=\Qh{\cB}{\xi}$. Then by Lemma~\ref{obs:state-properties-of-wta}(3) and (2) we have $\runsem{\cA}(\xi)=\0=\runsem{\cB}(\xi)$ and $\initialsem{\cA}(\xi)=\0=\initialsem{\cB}(\xi)$, respectively.

\underline{Case (b):} $|\state_\cA(\xi)|=|\state_\cB(\xi)|=1$ and
\[\Qh{\cA}{\xi}=\state_\cA(\xi)=\QR{\cA}{\xi}=\QR{\cB}{\xi}=\state_\cB(\xi)=\Qh{\cB}{\xi}\enspace.\]
Then $\estate_\cA(\xi)=\estate_\cB(\xi)$ and by \eqref{eq:run-weights-equal} and Lemma \ref{lm:limit-bu-det}(3)(b) (applied to $\cA$ and $\cB$) we obtain
\begin{align*}
 \runsem{\cA}(\xi)
    =  \wt_\cA(\xi,\crhorm_\xi)\otimes F_{\estate_\cA(\xi)} = \wt_\cB(\xi,\crhorm_\xi)\otimes F_{\estate_\cB(\xi)}
 = \runsem{\cB}(\xi) \enspace.
  \end{align*} 
 By \eqref{eq:run-weights-equal} and Lemma \ref{lm:limit-bu-det}(3)(b) we also have 
 \[\h_{\cA}(\xi)_{\estate_\cA(\xi)} = \wt_\cA(\xi,\crhorm_\xi)=\wt_\cB(\xi,\crhorm_\xi)=\h_{\cB}(\xi)_{\estate_\cB(\xi)}\] 
 and 
 \begin{align*}
\initialsem{\cA}(\xi)
  = \h_\cA(\xi)_{\estate_\cA(\xi)} \otimes F_{\estate_\cA(\xi)} = \h_\cB(\xi)_{\estate_\cB(\xi)} \otimes F_{\estate_\cB(\xi)}  = \initialsem{\cB}(\xi)\enspace. \hspace{4mm} \qedhere
\end{align*}

\end{proof}

Next we show three examples of pairs of strong bimonoids $\B_1 = (B,\oplus,\otimes,\0,\1)$ and $\B_2 = (B,+,\otimes,\0,\1)$:
\begin{compactitem}
\item  the strong bimonoids  $([0,1],\oplus,\cdot,0,1)$ and $([0,1],+,\cdot,0,1)$ where $\oplus$ and $+$ are t-conorms (cf. Example~\ref{ex:strong-bimonoids} (\ref{ex:0-1-strong-bimonoids})) and $\cdot$ is the usual multiplication; examples of a t-conorm $u: [0,1] \times [0,1] \to [0,1]$ are
  \begin{compactitem}
    \item the standard union $u(a,b) = \max(a,b)$,
  \item the algebraic sum $u(a,b) = a +b - a\cdot b$, or
  \item the bounded sum $u(a,b) = \min(a+b,1)$,
    \end{compactitem}
  \item the Boolean semiring $\Boole$ and the field $\sfFtwo$ with two elements (cf. Example \ref{ex:strong-bimonoids} (\ref{ex:sb-not-sr})), and
  \item the plus-plus strong bimonoid of natural numbers $\PP_{\mathbb{N}}=(\mathbb{N}_\0,\oplus,+,\0,0)$ (cf. Example \ref{ex:strong-bimonoids}(\ref{ex:plus-plus-sb})) and the semiring $(\mathbb{N}_\0,\max',+,\0,0)$ where the binary operation $\max'$, if restricted to $\mathbb{N}$,  is the usual operation $\max$ on natural numbers (e.g. $\max(3,2) = 3$). Moreover,  $\max'(\0,x) = \max'(x,\0) = x$ for each $x \in \mathbb{N}_\0$.
  \end{compactitem}

  We mention that Corollary \ref{cor:bu-det-wta-B1-B2} is meaningless if $\B_1 = (B,\oplus,\otimes,\0,\1)$ and $\B_2 = (B,+,\otimes,\0,\1)$ are bounded lattices because in this case $\oplus=+$. The latter can be seen as follows \cite{tep23}. Let $\le$ be the partial order on $B$ defined by $a\le b$ if $a\otimes b=a$ for every $a,b\in B$ (cf. Theorem \ref{lm:lattice-equivalent-def}(1)).  Then, by Theorem \ref{lm:lattice-equivalent-def}(2) and the remark after that theorem, we have $a \oplus b = \sup\nolimits_\le\{a,b\}=a+b$ for every $a,b\in B$.

\

Finally, we show how annihilation propagates over input trees of the form $c[\xi]$ under the assumption that $\B$ is a commutative semiring.

 \begin{lemma}\rm \label{obs:total-bu-det-wta-calc(new)-H} Let $\B$ be a commutative semiring. For every  $\xi \in \T_\Sigma$, $c \in \C_\Sigma$, and $q \in Q$, we have
   \[
     \h_\cA(c[\xi])_q = \begin{cases}
       \h_\cA(\xi)_{\estate(\xi)} \otimes \h_\cA^\C(c)(\1_{\estate(\xi)})_q & \text{ if $\h_\cA(c[\xi]) \in B_{=1}^Q$}\\
       \0 & \text{ otherwise}
       \end{cases}
     \]
 \end{lemma}
 \begin{proof}  Let $\xi \in \T_\Sigma$, $c \in \C_\Sigma$, and $q \in Q$.  We proceed by case analysis.

\underline{$\h_\cA(c[\xi]) \in B_{=1}^Q$:} Thus $\Qh{\cA}{c[\xi]} \ne \emptyset$. Then by Lemma~\ref{lm:zero-propagation-h} we obtain $\Qh{\cA}{\xi} \ne \emptyset$. Hence, by Lemma~\ref{lm:limit-bu-det}(1), also $\h_\cA(\xi) \in B_{=1}^Q$, and thus $\estate(\xi)$ is defined.   Then
   \begingroup
   \allowdisplaybreaks
   \begin{align*}
  \h_\cA(c[\xi])_q &=  \,\h_\cA^\C(c)(\h_\cA(\xi))_q \tag{by Lemma \ref{lm:hcACchAxi=hAcxi}} \\
     &=  \,\h_\cA^\C(c)(\h_\cA(\xi)_{\estate(\xi)}\cdot\,\1_{\estate(\xi)})_q 
\tag{by Lemma~\ref{lm:hAxiqne0-implies-qinhQxi-for-bu-det-wta}(1)}\\
      &=  \, \big(\h_\cA(\xi)_{\estate(\xi)} \cdot \h_\cA^\C(c)(\1_{\estate(\xi)})\big)_q \tag{by Lemma \ref{lm:haC-linear-mapping}; this lemma requires that $\B$ is a commutative semiring }\\
  &=  \, \h_\cA(\xi)_{\estate(\xi)} \otimes \h_\cA^\C(c)(\1_{\estate(\xi)})_q  \enspace.
   \end{align*}
   \endgroup

     \underline{$\h_\cA(c[\xi]) \not\in B_{=1}^Q$:} Then by Lemma~\ref{lm:limit-bu-det}(1) we obtain that $\h_\cA(c[\xi])= \0^Q$.
 \end{proof}

  \subsection{Annihilation in total and bu-deterministic wta}
  \label{sect:properties-total-bu-det-wta}

  \label{p:convention-annihilation-total-bu-det-wta}
\begin{quote} \emph{In this subsection, we let $\cA=(Q,\delta,F)$ be an arbitrary total and bu-deterministic $(\Sigma,\B)$-wta.}
\end{quote}
Thus, for every $k \in \mathbb{N}$, $\sigma \in \Sigma^{(k)}$, and $w \in Q^k$ there exists exactly one $q \in Q$ such that $\delta_k(w,\sigma,q) \not = \mathbb{0}$, or, in other words, $|\mathrm{succ}_\cA(q_1\cdots q_k,\sigma)| = 1$.
As a consequence, for each $\xi \in \T_\Sigma$, for each $\xi \in \T_\Sigma$, the set $\state(\xi)$ contains exactly one element denoted by $\estate(\xi)$.
Hence, for each $\xi \in \T_\Sigma$, the canonical run $\crhorm_\xi$ is defined (cf. Subsection \ref{subsect:annihilation-bu-det-wta}).

\begin{lemma}\rm \label{lm:total-bu-det-sem} Let $\xi \in \T_\Sigma$. Then the following three statements hold.
\begin{compactenum}
\item[(1)] $\h_\cA(\xi)_{\estate(\xi)} = \wt_\cA(\xi,\crhorm_\xi)$.
\item[(2)]  $\initialsem{\cA}(\xi)= \h_\cA(\xi)_{\estate(\xi)} \otimes F_{\estate(\xi)}$,
\item[(3)] $\runsem{\cA}(\xi)= \wt_\cA(\xi,\crhorm_\xi) \otimes F_{\estate(\xi)}$.
\end{compactenum}
\end{lemma}
\begin{proof}  
According to Lemma~\ref{lm:limit-bu-det}(3) and the fact that $|\state(\xi)|=1$ we distinguish two cases. 

\underline{Case (a):} $\Qh{\cA}{\xi} = \emptyset = \QR{\cA}{\xi}$. Then Statement (1) holds because $\h_\cA(\xi)_{\estate(\xi)} = \0 = \wt_\cA(\xi,\crhorm_\xi)$.
Moreover, Statements (2) and (3) also hold because, by Lemma~\ref{obs:state-properties-of-wta}(2) and (3), we have $\initialsem{\cA}(\xi) = \0 =\runsem{\cA}(\xi)$.

\underline{Case (b):}  $\Qh{\cA}{\xi} = \state(\xi) = \QR{\cA}{\xi}$. Then each of Statement (1), (2), and (3) follows from Lemma \ref{lm:limit-bu-det}(3)(b).
\end{proof}

\begin{lemma}\rm \label{cor:total-bu-det-equiv-zero-div-free} Let $\B$ be a strong bimonoid. Then the following two statements are equivalent.
    \begin{compactenum}
    \item[(A)] $\B$ is zero-divisor free.
    \item[(B)] For each $\xi \in \T_\Sigma$, we have  $\Qh{\cA}{\xi}=\state(\xi)=\QR{\cA}{\xi}$.
      \end{compactenum}
    \end{lemma}

    \begin{proof} Proof of (A)$\Rightarrow$(B):  Let $\xi \in \T_\Sigma$. Since $\cA$ is total and bu-deterministic, we have $|\state(\xi)|=1$ by Lemma~\ref{obs:subalgebras-of-the-state-algebra}(2). Moreover, by Corollary~\ref{cor:hAxiqne0-iff-qinhQxi-for-bu-det-wta}, we have that $\Qh{\cA}{\xi}=\state(\xi)$. Then the statement follows from Lemma~\ref{lm:limit-bu-det}(3).

      \
      
      Proof of (B)$\Rightarrow$(A):  Let $a,b \in B$ such that $a\ne\0 \ne b$. We consider the ranked alphabet $\Sigma=\{\gamma^{(1)},\alpha^{(0)}\}$ and the $(\Sigma,\B)$-wta $\cA=(Q,\delta,F)$ with $Q=\{q\}$, $F_q=\1$, and $\delta_0(\varepsilon,\alpha,q)=a$ and $\delta_1(q,\gamma,q)=b$. Then $\cA$ is total and bu-deterministic. Moreover, $\h_\cA(\gamma(\alpha))_q= a \otimes b$ and $\state(\xi)=\{q\}$. By (B) we obtain that $\h_\cA(\gamma(\alpha))_q \ne \0$. Hence $\B$ is zero-divisor free.
     \end{proof}

  \subsection{Annihilation in crisp-deterministic wta}
  \label{sect:properties-crisp-det-wta}
  
\begin{quote}\emph{In this subsection, we let $\cA=(Q,\delta,F)$ be an arbitrary crisp-deterministic $(\Sigma,\B)$-wta.}
\end{quote}

Thus, for every $k \in \mathbb{N}$, $\sigma \in \Sigma^{(k)}$, and $w \in Q^k$ there exists a $q \in Q$ such that $\delta_k(w,\sigma,q) = \mathbb{1}$ and for every $q' \in Q$ with $q'\ne q$ we have $\delta_k(w,\sigma,q') = \mathbb{0}$. We recall that, for each $\xi \in \T_\Sigma$, the set $\state(\xi)$ contains exactly one element denoted by $\estate(\xi)$.

\index{state algebra}
\index{StA@$\St(\cA)$}
\index{homStatealgebra@$\h_{\St(\cA)}$}
Essentially, each crisp-deterministic wta $\cA$ is a finite $\Sigma$-algebra combined with a mapping which reflects the root weights (cf. Lemma~\ref{lm:fin-algebra-crisp-det-wta}) and vice versa (cf. Lemma~\ref{lm:fin-algebra-wta}). 

The next lemma shows how the intial algebra semantics of a crisp-deterministic $(\Sigma,\B)$-wta $\cA$ can be expressed in terms of the mapping $\state$ and the root weight mapping of $\cA$.

\begin{lemma}\rm \label{lm:fin-algebra-crisp-det-wta}  The following three statements hold.
\begin{compactitem}
\item[(1)] For each $\xi \in \T_\Sigma$, we have $\h_\cA(\xi) = \1_{\estate(\xi)}$.
\item[(2)]  For each $\xi \in \T_\Sigma$, we have, $\initialsem{\cA}(\xi)=F_{\estate(\xi)}$.

  \item[(3)]  $\im(\initialsem{\cA})   \subseteq \im(F)$. Hence the set $\im(\initialsem{\cA})$ is finite.
\end{compactitem}
\end{lemma} 

\begin{proof} Proof of (1): First, by induction on $\T_\Sigma$, we prove the following.
\begin{equation}\label{eq:h-xi-state-xi}
\text{For each $\xi \in \T_\Sigma$, we have $\h_\cA(\xi)_{\estate(\xi)}=\1$.}
\end{equation}
Let $\xi=\sigma(\xi_1,\ldots,\xi_k)$. Then
\begingroup
\allowdisplaybreaks
\begin{align*}
 \h_\cA(\xi)_{\estate(\xi)} &= \delta_{\cA}(\sigma)(\h_\cA(\xi_1),\ldots,\h_\cA(\xi_k))_{\estate(\xi)}\\
    &= \h_\cA(\xi_1)_{\estate(\xi_1)} \otimes \ldots \otimes \h_\cA(\xi_k)_{\estate(\xi_k)} \otimes \delta_k(\estate(\xi_1)  \cdots \estate(\xi_k),\sigma,\estate(\xi))
      \tag{by Lemma \ref{lm:properties-sem-of-budet-wta-det-new}(1); note that by I.H., $\h_\cA(\xi_i) \in B_{=1}^Q$ for each $i \in [k]$}\\
                            &= \delta_k(\estate(\xi_1)  \cdots \estate(\xi_k),\sigma,\estate(\xi))
                              \tag{by  I.H.}\\
    &= \1 \tag{because $\estate(\xi)=\delta_Q(\sigma)(\estate(\xi_1),  \ldots, \estate(\xi_k))$}
\end{align*}
\endgroup
Then the statement follows   from \eqref{eq:h-xi-state-xi} and  Lemma \ref{lm:limit-bu-det}(1).

\

Proof of (2): It follows from Statement (1) and Lemma~\ref{lm:total-bu-det-sem}(2).

\

Proof of (3): It follows from Statement (2).
\end{proof}

Now we consider two strong bimonoids $\B_1$ and $\B_2$ which have the same carrier set; note that the unit elements of $\B_1$ and $\B_2$ can be different (examples are given below).
Moreover, we let $\cA$ be a crisp-deterministic $(\Sigma,\B_1)$-wta  and let $\cB$ be the crisp-deterministic $(\Sigma,\B_1)$-wta which we obtain from $\cA$ by  replacing the unit elements of $\B_1$ in the transitions of $\cA$ by the corresponding unit elements of $\B_2$. According to the note on page \pageref{page:wta-B1-not=-wta-B2}, $\cA$ is different from $\cB$. However, due to Lemma \ref{lm:fin-algebra-crisp-det-wta}(2), the initial algebra semantics of $\cA$ and $\cB$  coincide.\footnote{This does not mean that $\cA$ and $\cB$ are i-equivalent because the latter concept is defined for wta over the same strong bimonoid.} In \cite{fulvog23} the next corollary was proved for the special case that $\B_1$ and $\B_2$ are bounded lattices.

\begin{corollary-rect} \rm \label{cor:crisp-det-wta-B1-B2}  Let $\Sigma$ be a ranked alphabet. Moreover, let $\B_1$ and $\B_2$ be strong bimonoids with the same carrier set.  For each crisp-deterministic $(\Sigma,\B_1)$-wta $\cA$, we can construct a crisp-deterministic $(\Sigma,\B_2)$-wta $\cB$ such that $\initialsem{\cB}=\initialsem{\cA}$. In particular, $\cdRec^{\mathrm{init}}(\Sigma,\B_1) = \cdRec^{\mathrm{init}}(\Sigma,\B_2)$.
\end{corollary-rect}
  \begin{proof} Let $\B_1 = (B,\oplus,\otimes,\0,\1)$ and $\B_2 = (B,+,\times,0,1)$ be strong bimonoids. Let $\cA = (Q,\delta,F)$ be a crisp-deterministic $(\Sigma,\B_1)$-wta. We construct the crisp-deterministic $(\Sigma,\B_2)$-wta  $\cB =(Q,\delta',F)$ where for every $k \in \mathbb{N}$, $\sigma \in \Sigma^{(k)}$, and $q_1,\ldots,q_k,q\in Q$ we let
  \[
(\delta')_k(q_1\cdots q_k,\sigma,q) = 
\begin{cases} 1 & \text{ if } \ \delta_k(q_1\cdots q_k,\sigma,q) = \1 \\
0 & \text{ otherwise.}
\end{cases} 
\]
Clearly, $\cB$ is crisp-deterministic.
It is obvious that the state algebra $\St(\cA)$ is equal to the state algebra $\St(\cB)$. As a consequence, the $\Sigma$-algebra homomorphism $\state_\cA$ from $\T_\Sigma$ to $\St(\cA)$ is the same as the $\Sigma$-algebra homomorphism $\state_\cB$ from $\T_\Sigma$ to $\St(\cB)$.    Then, by applying Lemma \ref{lm:fin-algebra-crisp-det-wta}(2) twice, we have $\initialsem{\cA}(\xi) = 
F_{\estate_\cA(\xi)} = F_{\estate_\cB(\xi)} = \initialsem{\cB}(\xi)$ for each $\xi\in\T_\Sigma$.
 \end{proof}

Examples for pairs $(\B_1,\B_2)$ of different strong bimonoids which have the same carrier set (-- and the last three also have the same unit elements --) are
\begin{compactitem}
\item the semiring $(\mathbb{N}_\infty,+,\cdot,0,1)$ and the distributive bounded lattice $(\mathbb{N}_\infty,\max,\min,0,\infty)$,
\item the bounded lattices $\Nfive$ and $\Mthree$ from Examples \ref{ex:lattices}(\ref{ex:N-5-label}) and \ref{ex:lattices}(\ref{ex:M-3-label}), respectively,
\item the tropical semiring $(\mathbb{N}_\infty,\min,+,\infty,0)$ and the tropical bimonoid $(\mathbb{N}_{\infty},+,\min,0,\infty)$,
  \item the semiring of formal languages $(\cP(\Gamma^*),\cup,\cdot,\emptyset,\{\varepsilon\})$ and the semiring $(\cP(\Gamma^*),\cup,\cap,\emptyset, \Gamma^*)$,
\item the two different strong bimonoids with the carrier set $[0,1]$ in Example \ref{ex:strong-bimonoids}(\ref{ex:0-1-strong-bimonoids}) using algebraic sum and bounded sum, respectively,
  \item the strong bimonoid $([0,1],\oplus,\cdot,0,1)$ with bounded sum $\oplus$ (cf. Example \ref{ex:strong-bimonoids}(\ref{ex:0-1-strong-bimonoids})) and the strong bimonoid $([0,1],u,i,0,1)$ with t-conorm $u$ and t-norm $i$ (cf. Example \ref{ex:strong-bimonoids}(\ref{ex:0-1-conorm-norm})), and 
\item the Boolean semiring $\Boole$  and the field $\sfFtwo$ with two elements (cf. Example \ref{ex:strong-bimonoids}(\ref{ex:sb-not-sr}).
\end{compactitem}


\section{Crisp wta: decomposition and characterization}
\label{sec:crisp-wta-dec-and-char}

\begin{quote}\emph{In this section, we let $\cA=(Q,\delta,F)$ be an arbitrary crisp $(\Sigma,\B)$-wta, unless specified otherwise.}
\end{quote}

We denote by  $\cM(Q)$ the set of all  multisets over $Q$.

    \begin{lemma}\rm \label{lm:decomposition-of-crisp-wta} There exists a $\Sigma$-algebra $\sfA=(\cM(Q),\theta)$ and a mapping $F': \cM(Q) \to B$ such that $\runsem{\cA} = F' \circ \h_\sfA$.  
\end{lemma}
\begin{proof} We define the family $\delta' = ((\delta')_k \mid k \in \mathbb{N})$ of mappings $(\delta')_k: Q^k \times \Sigma^{(k)} \times Q \to \mathbb{N}$ such that, for every $k \in \mathbb{N}$, $\sigma \in \Sigma^{(k)}$, and $q_1,\ldots,q_k,q \in Q$, we let  
\begin{align*}
(\delta)'_k(q_1 \cdots q_k,\sigma,q) = 
\begin{cases} 1 & \text{ if  $\delta_k(q_1\cdots q_k,\sigma,q)=\1$}\\
  0 & \text{ otherwise.}
\end{cases}  
\end{align*}
(Note that $\delta_k(q_1\cdots q_k,\sigma,q)\in \{\0,\1\}$.)

  We define the $\Sigma$-algebra $\sfA=(\cM(Q),\theta)$ where for every $k \in \mathbb{N}$, $\sigma \in \Sigma^{(k)}$, $M_1,\ldots,M_k \in \cM(Q)$, and $q \in Q$, we let
    \[
\theta(\sigma)(M_1,\ldots,M_k)(q) = \bigplus_{q_1, \ldots,q_k \in Q} M_1(q_1) \cdot \ldots \cdot M_k(q_k) \cdot (\delta')_k(q_1 \cdots q_k,\sigma,q) \enspace.
\]
Moreover, we define the mapping $F': \cM(Q) \to B$ for each $M \in \cM(Q)$ by
\[
F'(M) = \bigoplus_{q \in Q} M(q)F_q
\]
where $M(q)F_q = F_q \oplus \ldots \oplus F_q$ with $M(q)$ summands.

For each $\xi \in \T_\Sigma$, we let $\#\QR{\cA}{\xi}$ be  the  multiset over $Q$ such that, for each $q \in Q$,
\[
  \#\QR{\cA}{\xi}(q) = |\{\rho \in \R_{\cA}(q,\xi) \mid \wt(\xi,\rho) \ne \0\}| \enspace.
\]
Thus $\QR{\cA}{\xi}=\supp(\#\QR{\cA}{\xi})$.

      By induction on $\T_\Sigma$, we prove the following statement.
    \begin{equation}\label{equ:number-of-non-zero-runs-as-hom}
      \text{For each $\xi \in \T_\Sigma$, we have $\#\QR{\cA}{\xi} = \h_\sfA(\xi)$.}
    \end{equation}
    Let $\xi = \sigma(\xi_1,\ldots,\xi_k)$ and $q \in Q$. Then
    \begingroup
    \allowdisplaybreaks
    \begin{align*}
      \#\QR{\cA}{\xi}(q) &= \bigplus_{q_1, \ldots, q_k \in Q} \#\QR{\cA}{\xi_1}(q_1) \cdot \ldots \cdot \#\QR{\cA}{\xi_k}(q_k) \cdot (\delta')_k(q_1 \cdots q_k,\sigma,q)
      \tag{because $\cA$ is crisp}\\
                                &= \theta(\sigma)\Big(\#\QR{\cA}{\xi_1},\ldots,\#\QR{\cA}{\xi_k}\Big)(q)
      \tag{by definition of $\theta(\sigma)$}\\
                                &= \theta(\sigma)\big(\h_{\sfA}(\xi_1),\ldots,\h_{\sfA}(\xi_k)\big)(q)
      \tag{by I.H.}\\
                                &= \h_\sfA(\xi)(q) \enspace.
      \end{align*}
    \endgroup

    Finally, we can prove that  $\runsem{\cA} = F' \circ \h_\sfA$.
      \begingroup
    \allowdisplaybreaks
    \begin{align*}
      \runsem{\cA}(\xi) &=  \bigoplus_{q \in Q} \bigoplus_{\rho \in \R_\cA(q,\xi)} \wt_\cA(\xi,\rho) \otimes F_{q}
      \tag{by \eqref{equ:runsem-splitted-set-of-runs}}\\
                       &= \bigoplus_{q \in Q}\bigoplus_{\substack{\rho \in \R_\cA(q,\xi):\\\wt_\cA(\xi,\rho)=\1}}  F_{q}
      \tag{because $\wt_\cA(\xi,\rho) \in \{\0,\1\}$}\\
                        &= \bigoplus_{q \in Q} |\{\rho \in \R_\cA(q,\xi) \mid \wt_\cA(\xi,\rho)=\1\}|  F_{q}
                          \tag{because $F_q$ does not depend on $\rho$}\\
                        &= \bigoplus_{q \in Q} \Big(\#\QR{\cA}{\xi}(q)\Big)  F_{q}
                          \tag{by definition of $\#\QR{\cA}{\xi}$}\\
      &= \bigoplus_{q \in Q} \Big(\h_\sfA(\xi)(q)\Big)  F_{q}
        \tag{by \eqref{equ:number-of-non-zero-runs-as-hom}}\\
                       &= F'(\h_\sfA(\xi))
                          \tag{by definition of $F'$}\\
                        &= (F' \circ \h_\sfA)(\xi)
                          \qedhere
  \end{align*}
    \endgroup
  \end{proof}

In the next theorem we characterize the set of weighted tree languages which are i-recognizable by crisp-deterministic wta. As preparation we prove the following lemma which is a kind of inverse of Lemma~\ref{lm:fin-algebra-crisp-det-wta}(2).

\begin{lemma}\label{lm:fin-algebra-wta}\rm $\ $ For every finite  $\Sigma$-algebra $\A=(Q,\theta)$ and mapping $F: Q \to B$, we can construct a crisp-deterministic $(\Sigma,\B)$-wta $\cA$ such that  $F \circ \h_{\A} = \initialsem{\cA}$.
\end{lemma}
\begin{proof} We construct the crisp-deterministic $(\Sigma,\B)$-wta $\cA=(Q,\delta,F)$  where for every $k \in \mathbb{N}$, $\sigma \in \Sigma^{(k)}$, $q\in Q$, and $q_1,\ldots q_k \in Q$ we define
\[
\delta_k(q_1\cdots q_k,\sigma,q) = 
\left\{
\begin{array}{ll}
\mathbb{1} & \text{if } \theta(\sigma)(q_1,\ldots,,q_k)=q\\
\mathbb{0} & \text{otherwise} \enspace.
\end{array}
\right.
\]

Next we prove a relationship between $\h_\A$ and $\state$.
\begin{equation}\label{equ:hA=state-total-bu-det-wta}
\text{ For each $\xi \in \T_\Sigma$, we have $\state(\xi) = \{\h_\A(\xi)\}$.}
\end{equation}
We prove this statement by induction in $\T_\Sigma$. We let $\xi = \sigma(\xi_1,\ldots,\xi_k)$. Then:
\begingroup
\allowdisplaybreaks
\begin{align*}
  \state(\sigma(\xi_1,\ldots,\xi_k))
  &= \{q \in Q\mid (\exists q_1 \in \state(\xi_1))\cdots (\exists q_k \in \state(\xi_k)): \delta_k(q_1 \cdots q_k,\sigma,q) = \1\} \\
  &= \{q \in Q\mid (\exists q_1 \in \state(\xi_1))\cdots (\exists q_k \in \state(\xi_k)): \theta(\sigma)(q_1, \ldots, q_k) = q\}
  \tag{by construction of $\cA$}\\
  &= \{\theta(\sigma)(\h_\A(\xi_1), \ldots, \h_\A(\xi_k))\}
    \tag{by I.H.}\\
  &= \{\h_\A(\sigma(\xi_1,\ldots,\xi_k))\} \enspace.
\end{align*}
\endgroup
This proves \eqref{equ:hA=state-total-bu-det-wta}. Hence $\estate(\xi)=\h_\A(\xi)$.

Now let $\xi \in \T_\Sigma$. Then 
\[\initialsem{\cA}(\xi)=F_{\estate(\xi)}=F_{\h_\A(\xi)}=(F\circ \h_\A)(\xi)\enspace,\] 
where the second equality is justified by Lemma \ref{lm:fin-algebra-crisp-det-wta}(2).
\end{proof}

Now we characterize the weighted tree languages which are i-recognizable by crisp-deterministic wta. Roughly speaking, it shows that each crisp-deterministic $(\Sigma,\B)$-wta consists of a finite $\Sigma$-algebra with carrier set $Q$ and a mapping $F: Q \to B$.  In \cite[Thm.~2]{bozlou10}, the combination of (a) a finite $\Sigma$-algebra with carrier set $Q$ and (b) a mapping $F: Q \to B$ was called ``finite representation''.

\begin{theorem-rect}\label{thm:crisp-wta-final-index} Let $\Sigma$ be a ranked alphabet. Moreover, let $\B=(B,\oplus,\otimes,\0,\1)$ be a strong bimonoid and let $r:~\T_\Sigma~\to~B$. Then the following three statements are equivalent.
  \begin{compactenum}
    \item[(A)] We can construct a finite $\Sigma$-algebra $\A=(Q,\theta)$ and a mapping $F:Q \rightarrow B$ such that $r= F \circ \h_\A$.
    \item[(B)] We can construct a crisp-deterministic $(\Sigma,\B)$-wta $\cA$ such that $r=\initialsem{\cA}$.
    \item[(C)] We can construct a  $(\Sigma,\B)$-wta $\cA$ such that the  congruence relation $\ker(\h_\cA)$  has finite index and $r=\initialsem{\cA}$.
\end{compactenum}
\end{theorem-rect}
\begin{proof}
 Proof of  (A)$\Rightarrow$(B): This follows directly from Lemma \ref{lm:fin-algebra-wta}.

 \
 
Proof of   (B)$\Rightarrow$(C): Let $\cA=(Q,\delta,F)$ be a crisp-deterministic $(\Sigma,\B)$-wta with $r=\initialsem{\cA}$.  The image of $\h_\cA$ is a finite set because, by \eqref{eq:h-xi-state-xi},  $\im(\h_\cA) \subseteq  \{ \mathbb{0},\mathbb{1}\}^Q$. Thus $\ker(\h_\cA)$ has finite index.

\

Proof of (C)$\Rightarrow$(A): Let $\cA=(Q,\delta,F_\cA)$ be a $(\Sigma,\B)$-wta such that $\ker(\h_\cA)$ has finite index and $r=\initialsem{\cA}$. We consider the accessible subalgebra $\aV(\cA) = (\im(\h_\cA),\delta_\cA)$ of $\V(\cA)$ (cf. Section \ref{sec:basic-defininition-wta}). Since $\ker(\h_\cA)$ has finite index, $\im(\h_\cA)$ is finite and hence $\aV(\cA)$ is a finite $\Sigma$-algebra. By Observation~\ref{obs:smallest-subalgebra-im}, we have  $\im(\h_\cA)=\langle \emptyset\rangle_{\delta_\cA(\Sigma)}$, and by Lemma \ref{obs:Knaster-Tarski-applied-to-algebras}, we can construct the set $\langle \emptyset\rangle_{\delta_\cA(\Sigma)}$. Thus $\im(\h_\cA)$ can be constructed. 
 Finally we define the mapping $F:  \im(\h_\cA) \to B$ for every $v \in \im(\h_\cA)$ by $F(v) =  \bigoplus_{q\in Q} v_q\otimes (F_\cA)_q$.
Then by \eqref{eq:initial-sem-Nerode}, we have $\initialsem{\cA}= F \circ \h_{\aV(\cA)}$.
\end{proof}

Finally, we prove that the set of weighted tree languages which are recognized by crisp-deterministic $(\Sigma,\B)$-wta with unit root weights, is exactly the set of characteristic mappings of recognizable $\Sigma$-tree languages. We note that, in the following theorem, the implication (B)$\Rightarrow$(A) is in fact Theorem~\ref{thm:fta-total-bud-fta}.

\begin{theorem}\label{thm:fta-wta} Let $L \subseteq \T_\Sigma$. Then the following three statements are equivalent.
\begin{compactenum}
\item[(A)] We can construct a total and bu-deterministic $\Sigma$-fta $A$ such that $\LL(A)= L$.
\item[(B)] We can construct a $\Sigma$-fta $A$ such that $\LL(A)= L$.
\item[(C)] \sloppy We can construct  a crisp-deterministic $(\Sigma,\B)$-wta $\cA$ with unit root weights such that $\initialsem{\cA}~=~\chi(L)$.
\end{compactenum}
\end{theorem}
\begin{proof} Proof of (A)$\Rightarrow$(B): This is by definition.

  \
  
  Proof of (B)$\Rightarrow$(C): Let $A=(Q,\delta,F_A)$ be a  $\Sigma$-fta such that $L = \LL(A)$. We consider the finite $\Sigma$-algebra $(\cP(Q),\delta_A)$ associated with $A$ (for the definition of this algebra see page \pageref{page:algebra-associated-with-fta}; we recall that the unique $\Sigma$-homomorphism from $\sfT_\Sigma$ to $(\cP(Q),\delta_A)$ is denoted by $\h_A$) and the mapping $F: \cP(Q) \rightarrow B$ defined for each $P \in \cP(Q)$ by 
\[
F(P) = 
\left\{
\begin{array}{ll}
\mathbb{1} & \text{if } P \cap F_A \not= \emptyset\\
\mathbb{0} & \text{otherwise} \enspace.
\end{array}
\right.
\]
Obviously, $\chi(L) = F\circ \h_A$. Then, by Lemma \ref{lm:fin-algebra-wta} we can construct a crisp-deterministic $(\Sigma,\B)$-wta $\cA'=(\cP(Q),\delta',F)$ such that $\initialsem{\cA'} = F \circ \h_A$. Hence $\initialsem{\cA'} = \chi(L)$. Moreover, $\cA'$ has unit root weights and thus (C) holds.

\

Proof of (C)$\Rightarrow$(A): Let $\cA =(Q,\delta,F)$ be a crisp-deterministic $(\Sigma,\B)$-wta with unit root weights and $\initialsem{\cA} = \chi(L)$. 
We consider the support fta $\supp(\cA)=(Q,\delta',F')$ of $\cA$. It is easy to see that $\supp(\cA)$ is total and bu-deterministic. Finally, we prove that $\LL(\supp(\cA))=L$.
Let $\xi \in \T_\Sigma$. Then
\begingroup
\allowdisplaybreaks
\begin{align*}
\xi \in L \ \  \Leftrightarrow  \ \   &  \chi(L)(\xi)= \1 \\
 &  \Leftrightarrow  \ \   \initialsem{\cA}(\xi) = \1
  \tag{by assumption}\\
&  \Leftrightarrow \ \   F_\estate(\xi) = \1
                          \tag{by Lemma~\ref{lm:fin-algebra-crisp-det-wta}(2)} \\
                                      &  \Leftrightarrow \ \    \state_\cA(\xi) \cap F' \ne \emptyset
  \tag{becaue $F'= \supp(F)$ and $\state_\cA(\xi)=\{\estate(\xi)\}$}\\
&  \Leftrightarrow \ \    \h_{\supp(\cA)}(\xi) \cap F' \ne \emptyset
       \tag{by \eqref{equ:state-algebra=algebra-associated-with-supp-fta}}\\
                                      &  \Leftrightarrow \ \  \xi \in \LL(\supp(\cA)) \enspace.
                                        \qedhere
  \end{align*}
  \endgroup
\end{proof}


\section{Padding of wta}
\label{sect:padding}

Given a $(\Sigma,\B)$-wta $\cA$ which has a certain property. Then we might wish to increase the rank of the symbols of $\Sigma$ and to construct a $(\Sigma',\B)$-wta $\cB$ which also has this property and where $\Sigma'$ is the same as $\Sigma$ but with increased ranks. The construction of $\cB$ can be achieved by a kind of padding of $\cA$.  

\index{extension}
Formally, let $(\Sigma,\rk)$ and $(\Sigma,\rk_e)$ be two ranked alphabets over the same set $\Sigma$. We say that  $(\Sigma,\rk_e)$ is an \emph{extension} of  $(\Sigma,\rk)$ if, for every $k \in \mathbb{N}$ and $\sigma \in \Sigma$, the relation $\rk(\sigma) \le \rk_e(\sigma)$ holds and, moreover, $\rk(\sigma)=0$ implies $\rk_e(\sigma)=0$. Then $(\Sigma,\rk)$ is trivial if and only if $(\Sigma,\rk_e)$ is trivial, and if $(\Sigma,\rk)$ is trivial, then $\rk=\rk_e$.  We denote the difference $\rk_e(\sigma) - \rk(\sigma)$ by $\e(\sigma)$.

Now we can prove the following padding lemma.

\begin{lemma}\rm \label{lm:super-big-padding} Let $(\Sigma,\rk)$ and $(\Sigma,\rk_e)$ be two ranked alphabets such that $(\Sigma,\rk_e)$ is an extension of $(\Sigma,\rk)$. Moreover, let $\cA$ be a $((\Sigma,\rk),\B)$-wta. Then we can construct a $((\Sigma,\rk),(\Sigma,\rk_e))$-tree homomorphism $g$ and  a $((\Sigma,\rk_e),\B)$-wta $\cB$ such that the following three statements hold.
  \begin{compactenum}
  \item[(1)] If $(\Sigma,\rk_e)$ is trivial, then $\mathrm{wts}(\cB) = \mathrm{wts}(\cA) \cup \{\1\}$, otherwise $\mathrm{wts}(\cB) = \mathrm{wts}(\cA) \cup \{\0,\1\}$.
  \item[(2)] For each $\xi \in \T_{(\Sigma,\rk)}$ we have $\initialsem{\cB}(g(\xi)) = \initialsem{\cA}(\xi)$ and $\runsem{\cB}(g(\xi)) = \runsem{\cA}(\xi)$ and\\
    for each $\zeta \in \T_{(\Sigma,\rk_e)}\setminus g(\T_{(\Sigma,\rk)})$ we have $\initialsem{\cB}(\zeta) = \runsem{\cB}(\zeta) = \0$. 
  \item[(3)] If $\rk=\rk_e$, then $\im(\initialsem{\cB}) = \im(\initialsem{\cA})$ and $\im(\runsem{\cB}) = \im(\runsem{\cA})$,\\ otherwise $\im(\initialsem{\cB}) = \im(\initialsem{\cA}) \cup\{\0\}$ and $\im(\runsem{\cB}) = \im(\runsem{\cA}) \cup\{\0\}$.
    \end{compactenum}
    \end{lemma}

\begin{proof} Let $\cA=(Q,\delta,F)$. Let $\alpha$ be an arbitrary element of $\Sigma$ with $\rk(\alpha)=0$. Then $\rk_e(\alpha)=0$.

 We define the $((\Sigma,\rk),(\Sigma,\rk_e))$-tree homomorphism $g= (g_k \mid k \in \mathbb{N})$ as follows.
  For every $\sigma \in \Sigma$ and $k\in\mathbb{N}$ such that $\rk(\sigma)=k$, we let
  $g_k(\sigma) = \sigma(z_1,\ldots,z_k,\alpha,\ldots,\alpha)$ with $\e(\sigma)$ occurrences of $\alpha$. This tree homomorphism might be called a  padding.
  Clearly, for each $\xi \in \T_{(\Sigma,\rk)}$, we have $\pos(\xi)\subseteq \pos(g(\xi))$, and for each
  $w\in \pos(g(\xi))$, we have
  \begin{align}\label{eq:how-g-works}
  g(\xi)(w)= \begin{cases}
   \xi(w) & \text{ if $w\in \pos(\xi)$} \\
   \alpha & \text{ otherwise.}
  \end{cases}
  \end{align}

  We construct the $((\Sigma,\rk_e),\B)$-wta $\cB =(Q',\delta',F')$ by letting $Q' = Q \cup \{q_\alpha\}$ where $q_\alpha$ is a new state (i.e., $q_\alpha \not\in Q$), $F'_q = F_q$ for each $q \in Q$ and $F'_{q_\alpha}=\0$. Moreover, for every $\sigma \in \Sigma$ and $q_1',\ldots,q_{\rk_e(\sigma)}',q' \in Q'$, using $k$ as abbreviation for $\rk(\sigma)$, we let
  \begin{align*}
    &(\delta')_{\rk_e(\sigma)}(q_1' \cdots q_{\rk_e(\sigma)}',\sigma,q') =\\
    & \ \ \begin{cases}
      \delta_k(q_1'\cdots q_k',\sigma,q') & \text{ if $q_1',\ldots,q_k' \in Q$, $q_{k+1}' \cdots q_{\rk_e(\sigma)}' = {q_\alpha}^{\e(\sigma)}$, and $q' \in Q$}\\
      \1 & \text{ if $\rk_e=0$, $\sigma=\alpha$, and  $q' = q_\alpha$}\\
      \0 & \text{ otherwise}  \enspace.
      \end{cases}
    \end{align*}

If $\rk=\rk_e$, then the only difference between $\cA$ and $\cB$ is that the later contains the extra state $q_\alpha$ and the transition $(\varepsilon,\alpha,q_\alpha)$ with weight $\1$ (note that $q_\alpha \not\in \supp(F)$).

\

Proof of (1). Obviously, $\mathrm{wts}(\cB) \subseteq \mathrm{wts}(\cA) \cup \{\0,\1\}$ and  $\mathrm{wts}(\cA) \cup \{\1\} \subseteq \mathrm{wts}(\cB)$.

    Assume that $(\Sigma,\rk_e)$ is trivial. Then also $(\Sigma,\rk)$ is trivial and $\rk=\rk_e$.  Hence  $\mathrm{wts}(\cB) = \mathrm{wts}(\cA) \cup \{\1\}$.
    If $(\Sigma,\rk_e)$ is not trivial, then there exists a $\sigma \in \Sigma$ such that $\rk_e(\sigma) \ge 1$. Then, e.g., $(\delta')_{\rk_e(\sigma)}(q_\alpha\cdots q_\alpha,\sigma,q_\alpha) = \0$, and hence $\0 \in \mathrm{wts}(\cB)$. Thus $\mathrm{wts}(\cB) = \mathrm{wts}(\cA) \cup \{\0,\1\}$. This proves (1).

    \

Proof of (2). First, we prove $\initialsem{\cB}(g(\xi)) = \initialsem{\cA}(\xi)$ for each $\xi \in \T_{(\Sigma,\rk)}$.  Obviously the following statement holds:
  \begin{equation}\label{equ:hB-not-alpha=0}
\text{For each $\zeta \in \T_{(\Sigma,\rk_e)}$, if $\zeta \not= \alpha$, then $\h_\cB(\zeta)_{q_\alpha}=\0$.}
\end{equation}

Second, by induction on $\T_{(\Sigma,\rk)}$, we prove the following statement.
    \begin{equation}\label{equ:super-big-padding-init}
\text{For every $\xi \in \T_{(\Sigma,\rk)}$ and $q \in Q$, we have $\h_\cB(g(\xi))_q = \h_\cA(\xi)_q$} \enspace.
\end{equation}

I.B.: Let $\xi = \beta$ with $\beta \in \Sigma$ and $\rk(\beta)=0$, and let $q \in Q$. Then
\[
  \h_\cB(g(\beta))_q = \h_\cB(\beta)_q = (\delta')_0(\varepsilon,\beta,q) = \delta_0(\varepsilon,\beta,q) = \h_\cA(\beta)_q \enspace.
\]

\

I.S.: Let $\xi = \sigma(\xi_1,\ldots,\xi_k)$ with $k \in \mathbb{N}_+$ and  $\xi_1,\ldots,\xi_k \in \T_{(\Sigma,\rk)}$, and let $q \in Q$. Then we can calculate as follows
\begingroup
\allowdisplaybreaks
\begin{align*}
  &\h_\cB(g(\sigma(\xi_1,\ldots,\xi_k)))_q
    = \h_\cB(\sigma(g(\xi_1),\ldots,g(\xi_k),\underbrace{\alpha,\ldots,\alpha}_{\e(\sigma)}))_q
  \tag{by definition of $g$}\\
  &= \bigoplus_{q_1'\cdots q_{\rk_e(\sigma)}' \in (Q')^{\rk_e(\sigma)}}
    \Big(\bigotimes_{j \in [k]} \h_\cB(g(\xi_j))_{q_j'} \Big) \otimes
    \Big(\bigotimes_{j \in [k+1,\rk_e(\sigma)]}  \h_\cB(\alpha)_{q_j'}  \Big) \otimes
     (\delta')_{\rk_e(\sigma)}(q_1'\cdots q_{\rk_e(\sigma)}',\sigma,q)\\[2mm]
  &= \bigoplus_{q_1\cdots q_k\in Q^k}
    \Big(\bigotimes_{j \in [k]} \h_\cB(g(\xi_j))_{q_j} \Big) \otimes
    \Big(\bigotimes_{j \in [k+1,{\rk_e(\sigma)}]} \h_\cB(\alpha)_{q_\alpha}\Big) \otimes
    (\delta')_{\rk_e(\sigma)}(q_1\cdots q_k{q_\alpha}^{\e(\sigma)},\sigma,q)
    \tag{because for each
    $q_1'\cdots q_{\rk_e(\sigma)}' \in (Q')^{\rk_e(\sigma)}\setminus Q^k\{{q_\alpha}^{\e(\sigma)}\}$ we have  $(\delta')_{\rk_e(\sigma)}(q_1'\cdots q_{\rk_e(\sigma)}',\sigma,q) =\0$}\\[3mm]
  &= \bigoplus_{q_1\cdots q_k\in Q^k}
        \Big(\bigotimes_{j \in [k]} \h_\cA(\xi_j)_{q_j} \Big) \otimes
    \delta_k(q_1\cdots q_k,\sigma,q)
    \tag{by I.H. and because $\h_\cB(\alpha)_{q_\alpha}= (\delta')_0(\varepsilon,\alpha,q_\alpha) = \1$}\\[2mm]
  &= \h_\cA(\sigma(\xi_1,\ldots,\xi_k))_q \enspace.
  \end{align*}
\endgroup
This proves \eqref{equ:super-big-padding-init}.

  Now let $\xi \in \T_{(\Sigma,\rk)}$. 
  Then
  \begin{align*}
    \initialsem{\cB}(g(\xi)) &= \bigoplus_{q' \in Q'} \h_\cB(g(\xi))_{q'} \otimes F'_{q'}\\
                             &= \bigoplus_{q \in Q} \h_\cB(g(\xi))_{q} \otimes F'_{q} \tag{because $F'_{q_\alpha}=\0$}\\
                             &=  \bigoplus_{q \in Q} \h_\cB(g(\xi))_{q} \otimes F_{q} \tag{because $F'_q = F_q$ for each $q \in Q$}\\
    &=  \bigoplus_{q \in Q} \h_\cA(\xi)_{q} \otimes F_{q} \tag{by \eqref{equ:super-big-padding-init}}\\
    &= \initialsem{\cA}(\xi) \enspace.
  \end{align*}

  Next we prove $\runsem{\cB}(g(\xi)) = \runsem{\cA}(\xi)$ for each $\xi \in \T_{(\Sigma,\rk)}$.   As preparation, for each  $\xi \in \T_{(\Sigma,\rk)}$, we define the mapping $\varphi_\xi: \R_\cA(\xi) \to \R_\cB(g(\xi))$ such that, for every $\rho \in \R_\cA(\xi)$ and $w \in \pos(g(\xi))$, we let
  \[
    \varphi(\rho)(w) =
    \begin{cases}
      \rho(w) & \text{ if $w \in \pos(\xi)$}\\
      q_\alpha & \text{ otherwise} \enspace,
      \end{cases}
    \]
    cf. \eqref{eq:how-g-works}.  
    Obviously, $\varphi_\xi$ is injective. Moreover,
    \begin{equation}\label{equ:super-big-padding-bad-runs=0}
      \text{ For each $\rho \in \R_\cB(g(\xi)) \setminus \im(\varphi_\xi)$ we have $\wt_\cB(g(\xi),\rho) =\0$.}
      \end{equation}

    By induction on $\T_{(\Sigma,\rk)}$ we prove the following statement.
    \begin{equation}\label{equ:super-big-padding-run} 
\text{For every $\xi \in \T_{(\Sigma,\rk)}$ and $\rho \in \R_\cA(\xi)$ we have $\wt_\cA(\xi,\rho)=\wt_\cB(g(\xi),\varphi_\xi(\rho))$.}
\end{equation}
Let $\xi = \sigma(\xi_1,\ldots,\xi_k)$ and $\rho \in \R_\cA(\xi)$. Then
\begingroup
\allowdisplaybreaks
\begin{align*}
  \wt_\cA(\xi,\rho)&= \Big(\bigotimes_{j \in [k]} \wt_\cA(\xi_j,\rho|_j) \Big) \otimes \delta_k(\rho(1)\cdots \rho(k),\sigma,\rho(\varepsilon))\\
                   &= \Big(\bigotimes_{j \in [k]} \wt_\cB(g(\xi_j),\varphi_{\xi_j}(\rho|_j)) \Big) \otimes \delta_k(\rho(1)\cdots \rho(k),\sigma,\rho(\varepsilon))
                     \tag{by I.H.}\\[2mm]
                   &= \Big(\bigotimes_{j \in [k]} \wt_\cB(g(\xi_j),\varphi_{\xi_j}(\rho|_j)) \Big) \otimes
                     \Big(\bigotimes_{j \in [k+1,\rk_e(\sigma)]} (\delta')_0(\varepsilon,\alpha,q_\alpha) \Big) \otimes
                     \delta_k(\rho(1)\cdots \rho(k),\sigma,\rho(\varepsilon))
                     \tag{because $(\delta')_0(\varepsilon,\alpha,q_\alpha)=\1$ }\\[2mm]
   &= \Big(\bigotimes_{j \in [\rk_e(\sigma)]} \wt_\cB(g(\xi_j),\varphi_{\xi_j}(\rho|_j)) \Big) \otimes
                     \delta_k(\rho(1)\cdots \rho(k),\sigma,\rho(\varepsilon))
                     \tag{because $(\delta')_0(\varepsilon,\alpha,q_\alpha)= \wt_\cB(g(\xi_j),\varphi_{\xi_j}(\rho|_j))$ for each $j \in [k+1,\rk_e(\sigma)]$}\\[2mm]
   &= \Big(\bigotimes_{j \in [\rk_e(\sigma)]} \wt_\cB(g(\xi_j),\varphi_{\xi_j}(\rho|_j)) \Big) \otimes
                     (\delta')_{\rk_e(\sigma)}(\rho(1)\cdots \rho(k) {q_\alpha}^{\e(\sigma)},\sigma,\rho(\varepsilon))
     \tag{by construction}\\[2mm]
    &= \Big(\bigotimes_{j \in [\rk_e(\sigma)]} \wt_\cB(g(\xi)|_j,\varphi_{\xi}(\rho)|_j) \Big) \otimes
      (\delta')_{\rk_e(\sigma)}(\rho(1)\cdots \rho(k) {q_\alpha}^{\e(\sigma)},\sigma,\rho(\varepsilon))
  \tag{by definitions of $g$, $\varphi$, and the restriction $|_j$}\\[2mm]
  &= \wt_\cB(g(\xi),\varphi_\xi(\rho)) \enspace.
  \end{align*}
\endgroup
    This proves \eqref{equ:super-big-padding-run} .

 Now let $\xi \in \T_{(\Sigma,\rk)}$. 
 Then
 \begingroup
 \allowdisplaybreaks
  \begin{align*}
    \runsem{\cB}(g(\xi)) &= \bigoplus_{\rho \in \R_\cB(\xi)} \wt_\cB(g(\xi),\rho) \otimes F'_{\rho(\varepsilon)}\\
    &= \bigoplus_{\rho \in \im(\varphi_\xi)} \wt_\cB(g(\xi),\rho) \otimes F'_{\rho(\varepsilon)}
      \tag{by \eqref{equ:super-big-padding-bad-runs=0}}\\
     &= \bigoplus_{q \in Q'} \bigoplus_{\rho \in \R_\cB(q,g(\xi)) \cap \im(\varphi_\xi)} \wt_\cB(g(\xi),\rho) \otimes F'_q\\
                         &= \bigoplus_{q \in Q} \bigoplus_{\rho \in \R_\cB(q,g(\xi)) \cap \im(\varphi_\xi)} \wt_\cB(g(\xi),\rho) \otimes F_q
    \tag{because $F'_{q_\alpha}=\0$ and $F'_q=F_q$ for each $q\in Q$}\\
                         &=  \bigoplus_{q \in Q}  \bigoplus_{\rho \in \R_\cA(q,\xi)} \wt_\cB(g(\xi),\varphi_\xi(\rho)) \otimes F_q
    \tag{because $\varphi_\xi$ is injective}\\
    &=  \bigoplus_{q \in Q} \bigoplus_{\rho \in \R_\cA(q,\xi)} \wt_\cA(\xi,\rho) \otimes F_{q} \tag{by \eqref{equ:super-big-padding-run}}\\
    &= \runsem{\cA}(\xi) \enspace.
  \end{align*}
  \endgroup
  
  \
  
 Now let $\zeta \in \T_{(\Sigma,\rk_e)} \setminus  g(\T_{(\Sigma,\rk)})$. 
 Then $\rk\ne \rk_e$, thus there exists $w \in \pos(\zeta)$ such that
  $\rk_e(\sigma) > k \not=0$ where $\sigma=\zeta(w)$ and $k = \rk(\sigma)$, and there exists $j_0 \in [k+1,\rk_e(\sigma)]$ such that $\zeta(wj_0) \not= \alpha$, cf. \eqref{eq:how-g-works}.

 First, we prove that $\initialsem{\cB}(\zeta) = \0$. Since $\zeta(wj_0) \not= \alpha$, we have  $\h_\cB(\zeta|_{wj_0})_{q_\alpha}=\0$ by~\eqref{equ:hB-not-alpha=0}. Next we prove that
  \begin{equation}\label{equ:is-0-for-ugly-trees-init}
    Q^{\h_\cB}_{\not=\emptyset}(\zeta|_w) = \emptyset
  \end{equation}
  where $Q^{\h_\cB}_{\not=\emptyset}(\zeta|_w)$ is defined in Section\ref{sect:properties-of-wta}.

  Since $\zeta|_w \not= \alpha$, by \eqref{equ:hB-not-alpha=0}, the state $q_\alpha \not\in Q^{\h_\cB}_{\not=\emptyset}(\zeta|_w)$.
Let $q \in Q$. Then
\begingroup
\allowdisplaybreaks
\begin{align*}
  &\h_\cB(\zeta|_w)_q \\[2mm]
  &= \bigoplus_{q_1'\cdots q_{\rk_e(\sigma)}' \in (Q')^{\rk_e(\sigma)}}
    \Big(\bigotimes_{j \in [k]} \h_\cB(\zeta|_{wj})_{q_j'} \Big) \otimes
    \Big(\bigotimes_{j \in [k+1,\rk_e(\sigma)]}  \h_\cB(\zeta|_{wj})_{q_j'}  \Big) \otimes
    (\delta')_{\rk_e(\sigma)}(q_1'\cdots q_{\rk_e(\sigma)}',\sigma,q)\\[2mm]
  &= \bigoplus_{q_1\cdots q_k\in Q^k}
    \Big(\bigotimes_{j \in [k]} \h_\cB(\zeta|_{wj})_{q_j} \Big) \otimes
    \big(\bigotimes_{j \in [k+1,{\rk_e(\sigma)}]} \h_\cB(\zeta|_{wj})_{q_\alpha}\Big) \otimes
    (\delta')_{\rk_e(\sigma)}(q_1\cdots q_k{q_\alpha}^{\e(\sigma)},\sigma,q)
    \tag{because for each
    $q_1'\cdots q_{\rk_e(\sigma)}' \in (Q')^{\rk_e(\sigma)}\setminus Q^k\{{q_\alpha}^{\e(\sigma)}\}$ we have  $(\delta')_{\rk_e(\sigma)}(q_1'\cdots q_{\rk_e(\sigma)}',\sigma,q) =\0$}\\[3mm] 
  &= \0 \tag{because $\h_\cB(\zeta|_{wj_0})_{q_\alpha}=\0$ and $\bigoplus_{q_1\cdots q_k\in Q^k}\0=\0$} \enspace.
\end{align*}
\endgroup
Hence $q \not\in Q^{\h_\cB}_{\not=\emptyset}(\zeta|_w)$, which means $Q^{\h_\cB}_{\not=\emptyset}(\zeta|_w)=\emptyset$. This proves \eqref{equ:is-0-for-ugly-trees-init}. Then by Lemma~\ref{lm:zero-propagation-h}, also $Q^{\h_\cB}_{\not=\emptyset}(\zeta)=\emptyset$, and hence $\initialsem{\cB}(\zeta) = \0$.

Second we prove that $\runsem{\cB}(\zeta)=\0$. 
Obviously the following statement holds:
  \begin{equation}\label{equ:run-not-alpha=0}
\text{For each $\overline{\zeta} \in \T_{(\Sigma,\rk_e)}$ with $\overline{\zeta} \not= \alpha$ and $\rho \in \R_\cB(q_\alpha,\overline{\zeta})$, we have  $\wt_{\cB}(\overline{\zeta},\rho) =\0$.}
\end{equation}
Since $\zeta(w)\not=\alpha$, by \eqref{equ:run-not-alpha=0} it follows that $q_\alpha \not\in Q^{\R_\cB}_{\not=\emptyset}(\zeta|_w)$. Now let $q\in Q$ and $\rho \in \R_\cB(q,\zeta|_w)$. Then
\begingroup
\allowdisplaybreaks
\begin{align*}
  \wt_{\cB}(\zeta|_w,\rho) = \Big(\bigotimes_{j \in [\rk_e(\sigma)]} \wt_\cB(\zeta|_{wj},\rho|_j)\Big) \otimes
                             (\delta')_{\rk_e(\sigma)}(\rho(1)\cdots \rho(\rk_e(\sigma)),\sigma,q)\enspace.
\end{align*}
\endgroup
Now if $\rho(j_0)\ne q_\alpha$, then $(\delta')_{\rk_e(\sigma)}(\rho(1)\cdots \rho(\rk_e(\sigma)),\sigma,q)=\0$, hence $\wt_{\cB}(\zeta|_w,\rho)=\0$. If $\rho(j_0)= q_\alpha$, then
by $\zeta(wj_0)\ne \alpha$ and \ref{equ:run-not-alpha=0} we obtain $\wt_\cB(\zeta|_{wj_0},\rho|_{j_0})=\0$. Hence again $\wt_{\cB}(\zeta|_w,\rho)=\0$. We conclude that $Q^{\R_\cB}_{\not=\emptyset}(\zeta|_w) =\emptyset$. Then, by Lemma~\ref{lm:zero-propagation-h}, also $Q^{\R_\cB}_{\not=\emptyset}(\zeta)=\emptyset$, and hence $\runsem{\cB}(\zeta) = \0$. This proves (2).

\

Proof of (3). If  $\rk=\rk_e$, then $\T_{(\Sigma,\rk_e)} = g(\T_{\Sigma,\rk)})$, hence (3) follows from (2). 
Otherwise also (3) follows from (2).
\end{proof}


  \section{Images  of the semantics of weighted tree automata}
  \label{sec:images-semantics-restr-sb}

  The semantics of a $(\Sigma,\B)$-wta $\cA$ are mappings $\initialsem{\cA}: \T_\Sigma \to B$ and $\runsem{\cA}: \T_\Sigma \to B$. Their images are subsets of $B$, i.e., $\im(\initialsem{\cA}) \subseteq B$ and  $\im(\runsem{\cA}) \subseteq B$. In this section, we investigate universality and finiteness of images.

  \subsection{Images of initial algebra semantics of wta}

  Here we prove an interesting property of the initial algebra semantics (universality property), which will be applied in several proofs. Roughly speaking it says the following: if $\Sigma$ contains a binary symbol, then for each finite subset $H$ of the strong bimonoid $\B$, we can construct a $(\Sigma,\B)$-wta $\cA$ such that the closure of $H$ is a subset of $\im(\initialsem{\cA})$, see \cite{drofultepvog24}. This result has two weaker predecessors, which are incomparable:
\begin{compactitem}
\item In \cite[Lm.~6.1]{rad10}, the result was proved under the condition that $\Sigma$ contains at least $|H\cup\{\0,\1\}|$ many nullary symbols and at least two binary symbols (also cf. Example~\ref{ex:Sigma-algebra-hom-as-wta}(3)).
\item In \cite[Lm.~12]{fulvog23}, the result was proved under the condition that $\Sigma$ contains a binary symbol and $\B$ is a bounded lattice (exploiting idempotency and commutativity).
\end{compactitem}
Also, we refer to Corollary \ref{lm:closure-of-finite-set-recognizable}, where a weaker result is proved (viz., $\B$ must be a  semiring); its benefit is that its proof is a simple combination of Example~\ref{ex:Sigma-algebra-hom-as-wta} and Corollary~\ref{cor:closure-REC-under-lin-nondel-prod-hom}.

The idea for the construction is that $\cA$ receives as input an encoding $h(\xi)$ of a tree representation $\xi$ of an element $a \in \langle H \rangle_{\{\oplus,\otimes,\0,\1\}}$; the encoding is done by the tree homomorphism $h$, see Figure~\ref{fig:init-can-compute-closure}. Then $\cA$ should satisfy $\initialsem{\cA}(h(\xi)) = \mathrm{eval}(\xi)$ where $\mathrm{eval}$ is the canonical evaluation of $\xi$.

\begin{figure}[t]
          \centering

      \begin{tikzpicture}[transform shape,scale=0.8]
        
        \begin{scope}[level distance=2em, level 1/.style={sibling distance=20mm},
level 2/.style={sibling distance=20mm}]

         \node (root1) {$\otimes$}
  child {node {$a_2$}}
  child {node {$\oplus$}
    child { node {$a_1$}}
    child { node {$a_1$}}};

    \node[above left = 0.1cm and 0.6cm of root1] {$\xi$:};

\end{scope}

\node[xshift=50mm,yshift=-55mm] (exp) {$a = a_2 \otimes (a_1 \oplus a_1)$};

\begin{scope}[xshift=-40mm, yshift=-45mm]

   \node (root2) {$\sigma$}
   child[level distance = 0.6cm, sibling distance = 10mm]  { node (c1) {$\sigma$}
     child[level distance = 1.3cm, sibling distance = 30mm]  { node (c11) {$\sigma$}
       child[level distance = 0.6cm, sibling distance = 5mm] { node (c111) {$\sigma$}
         child[sibling distance = 3mm] { node (c1111) {$\alpha$}}
         child[sibling distance = 3mm] { node (c1112) {$\alpha$}}}
       child[level distance = 0.6cm, sibling distance = 5mm] { node (c112) {$\alpha$}}}
     child[level distance = 1.3cm, sibling distance = 30mm] { node (c12) {$\sigma$}
       child[level distance = 0.6cm, sibling distance = 10mm] { node (c121) {$\alpha$}}
       child[level distance = 0.6cm, sibling distance = 10mm] { node (c122) {$\sigma$}
         child[level distance = 1.3cm, sibling distance = 30mm] { node (c1221) {$\sigma$}
           child[level distance = 0.6cm, sibling distance = 5mm] { node (c12211) {$\alpha$}}
           child[level distance = 0.6cm, sibling distance = 5mm] { node (c12212) {$\alpha$}}}
         child[level distance = 1.3cm, sibling distance = 30mm] { node (c1222) {$\sigma$}
           child[level distance = 0.6cm, sibling distance = 5mm] { node (c12221) {$\alpha$}}
           child[level distance = 0.6cm, sibling distance = 5mm] { node (c12222) {$\alpha$}}}
         }}}
    child[level distance = 0.6cm, sibling distance = 10mm] { node (c2) {$\alpha$}};
    
      \node[above left = 0.1cm and 1.2cm of root2] {$h(\xi)$:};

\node[fit=(root2)(c1)(c2),ellipse,draw] {};
\node[fit=(c12)(c121)(c122),ellipse,draw] {};
\node[fit=(c11)(c1111)(c112),ellipse,draw] {};
\node[fit=(c1221)(c12211)(c12212),ellipse,draw] {};
\node[fit=(c1222)(c12221)(c12222),ellipse,draw] {};

\draw[->] (3.5,-1.0) -- (6.5,-1.0) node[midway,above] {$\initialsem{\cA}$};
\draw[->] (6.0,2.4) -- (8.1,0.3) node[midway,right] {\ \ $\mathrm{eval}$};
\draw[->] (3.0,2.4) -- (1.3,1.0) node[midway,right] {\ \ $h$};

\end{scope}
\end{tikzpicture}
\caption{\label{fig:init-can-compute-closure} The $(\Sigma,\B)$-wta $\cA$ has the property $\initialsem{\cA}(h(\xi)) = \mathrm{eval}(\xi)$.}
\end{figure}

Next we make some formal preparations for the proof of the mentioned result. We assume that $H \cup \{\0,\1\} = \{a_1,\ldots,a_n\}$, and we represent the values in $\langle H \rangle_{\{\oplus,\otimes,\0,\1\}}$ by trees over the ranked alphabet $\Delta =\{\oplus^{(2)}, \otimes^{(2)}\} \cup \{a_1^{(0)},\ldots,a_n^{(0)}\}$. Formally, we define the mapping $\mathrm{eval}: \T_\Delta \to \langle H \rangle_{\{\oplus,\otimes,\0,\1\}}$ by induction, for each $\xi \in \T_\Delta$, as follows:
\[
  \mathrm{eval}(\xi) = \begin{cases} a_i & \text{ if $\xi = a_i$ for some $i \in [n]$}\\
    \mathrm{eval}(\xi_1) \oplus \mathrm{eval}(\xi_2) & \text{ if $\xi = \oplus(\xi_1,\xi_2)$ for some $\xi_1,\xi_2 \in \T_\Delta$}\\
    \mathrm{eval}(\xi_1) \otimes \mathrm{eval}(\xi_2) & \text{ if $\xi = \otimes(\xi_1,\xi_2)$ for some $\xi_1,\xi_2 \in \T_\Delta$}\enspace.
    \end{cases}
  \]
  Clearly, $\mathrm{eval}$ is surjective, i.e., $\langle H \rangle_{\{\oplus,\otimes,\0,\1\}} \subseteq \mathrm{eval}(\T_\Delta)$. 
 \label{page:def-of-eval}

Moreover, we assume that $\Sigma$ contains a binary symbol. Thus, we can fix arbitrary $\alpha \in \Sigma^{(0)}$ and  $\sigma \in \Sigma^{(2)}$. For each $i \in [0,n]$ we define the $\Sigma$-tree $f_i$ by $f_0 = \alpha$ and $f_{i+1} = \sigma(f_i,\alpha)$ for each $i \in [n-1]$.
  We define the $(\Delta,\Sigma)$-tree homomorphism $h = (h_k \mid k \in \mathbb{N})$ by letting
\begin{align*}
  h_0(a_i) &= f_i \ \text{  for each $i \in [n]$,}\\
  h_2(\oplus) &= \sigma(\alpha,\sigma(z_1,z_2)), \ \text{and }\\
    h_2(\otimes) &= \sigma(\sigma(z_1,z_2),\alpha) \enspace.
\end{align*}
Obviously, $h$ is nondeleting, linear, and productive.

Then we construct the $(\Sigma,\B)$-wta $\cA$ such that $\initialsem{\cA}(h(\xi)) = \mathrm{eval}(\xi)$ for each $\xi \in \T_\Delta$. Since   $\mathrm{eval}$ is surjective, each element of $\langle H \rangle_{\{\oplus,\otimes,\0,\1\}}$ occurs in the image of $\initialsem{\cA}$.

\begin{theorem-rect} \label{thm:closure-of-finite-set-i-recognizable}  {\rm \cite[Thm.~5.1]{drofultepvog24}} Let $\Sigma$ be a ranked alphabet such that $\Sigma^{(2)} \not= \emptyset$. Moreover, let $\B=(B,\oplus,\otimes,\0,\1)$ be a strong bimonoid and $H \subseteq B$  be a finite subset. Then we can construct a  $(\Sigma,\B)$-wta  $\cA$  such that $\mathrm{wts}(\cA) = H \cup \{\0,\1\}$ and  $\im(\initialsem{\cA}) = \langle H \rangle_{\{\oplus,\otimes,\0,\1\}}$. In particular, if  $\B$  is generated by  $H$, then we obtain  $\im(\initialsem{\cA}) = B$.
\end{theorem-rect}

\begin{proof} Clearly, $\langle H \rangle_{\{\oplus,\otimes,\0,\1\}} = \langle H \cup \{\0,\1\} \rangle_{\{\oplus,\otimes\}}$. Let $a_1,\ldots,a_n$ be the elements of $H \cup \{\0,\1\}$, i.e., $H \cup \{\0,\1\} = \{a_1,\ldots,a_n\}$. Let $\alpha$ and $\sigma$ be arbitrary elements of $\Sigma^{(0)}$ and $\Sigma^{(2)}$, respectively.
  
  Now we construct the $(\Sigma,\B)$-wta $\cA=(Q,\delta,F)$ as follows (cf. Figure \ref{fig:closure-by-initsem}).
\begin{compactitem}
\item $Q = \{v\} \cup \{q_0,\ldots,q_{n-1}\} \cup \{q_{\mathrm{one}},q_{\oplus}\} \cup \{q_{\otimes}\}$; the intention of the states is as follows:
  \begin{compactitem}
  \item $v$ is the ``main'' state with $\h_\cA(h(\xi))_v = \mathrm{eval}(\xi)$ for each $\xi \in \T_\Delta$ (cf. \eqref{eq:wta-computes-value-on-representations});
  \item the state  $q_i$ with $i \in [0,n-1]$ is used to recognize the tree $f_{i}$ with weight $\1$ (cf. \eqref{eq:usual-values-1}), and in combination with $v$, we will have $\h_\cA(\sigma(f_{i-1},\alpha))_v = a_i$ (cf. I.B. of \eqref{eq:wta-computes-value-on-representations});
  \item the states $q_{\oplus}$ and $q_{\otimes}$ are intermediate states such that $\h_\cA(\sigma(\xi_1,\xi_2)))_\oplus = \mathrm{eval}(\xi_1) \oplus \mathrm{eval}(\xi_2)$ and  $\h_\cA(\sigma(\xi_1,\xi_2)))_\otimes = \mathrm{eval}(\xi_1) \otimes \mathrm{eval}(\xi_2)$, respectively; the state $q_{\mathrm{one}}$ supports $q_\oplus$;
    \item the switch from $q_\oplus$ to $v$ and from $q_\otimes$ to $v$ is triggered by the patterns $\sigma(\alpha,.)$ and $\sigma(.,\alpha)$, respectively,
    \end{compactitem}
\item $F_v=\mathbb{1}$ and, for each $q \in Q \setminus \{v\}$, we let $F_q=\mathbb{0}$, and
\item for each $q \in Q$, we define 
  \[
    \delta_0(\varepsilon,\alpha,q) =
    \begin{cases}
      \1 & \text{if } q \in \{q_0,q_{\mathrm{one}}\}\\
      \0 & \text{otherwise}
      \end{cases}
    \]
    and, for every $p,q,r \in Q$, we define
    \[
\delta_2(pq,\sigma,r) = 
\left\{
\begin{array}{ll}
\mathbb{1} & \text{if there exists $i \in [n-1]$ such that } pqr= q_{i-1}q_0q_i\\
  a_i & \text{if there exists $i \in [n]$ such that } pqr= q_{i-1}q_0v\\
  \1 & \text{if } pqr \in \{vq_{\mathrm{one}}q_{\oplus}, q_{\mathrm{one}}vq_{\oplus},q_0q_{\oplus}v\} \\
  \1 & \text{if } pqr \in \{vvq_{\otimes}, q_{\otimes}q_0v\}\\
    \1 & \text{if } pqr = q_{\mathrm{one}}q_{\mathrm{one}}q_{\mathrm{one}}\\
  \0 & \text{otherwise} \enspace,
\end{array}
\right.
\]
\item for every $k\in \mathbb{N}$, $\eta \in \Sigma^{(k)}$ with $\sigma\ne \eta\ne \alpha$, and $q,q_1,\ldots,q_k\in Q$, we define $\delta_k(q_1\ldots q_k,\eta,q)=\0$.
\end{compactitem}
Clearly, $\mathrm{wts}(\cA) = H \cup \{\0,\1\}$.

\begin{figure}  

\begin{tikzpicture}[scale=1.09]
\tikzset{node distance=7em, scale=0.4, transform shape}
\node[state, rectangle] (b) {\huge $\alpha$};
  \node[state, right of=b] (q0) {\huge $q_0$};
\node[state, rectangle, right of=q0](g1-1) {\huge $\sigma$};
  \node[state, right of=g1-1] (q1) {\huge $q_1$};
\node[state, rectangle, right of=q1] (g1-2) {\huge $\sigma$};
  \node[state, right of=g1-2] (q2) {\huge $q_2$};
\node[state, right of=q2,opacity=0] (empty) {}; 
\node[state, rectangle, right of=empty]  (g1-3) {\huge $\sigma$};
\node[state, right of=g1-3] (qn-1) {\huge $q_{n-1}$};

\node[state, rectangle,above of=q0] (h0) {\huge $\sigma$};
\node[state, rectangle,above of=q1] (h1) {\huge $\sigma$};
\node[state, rectangle,above of=q2] (h2) {\huge $\sigma$};
\node[state, rectangle,above of=qn-1] (hn-1) {\huge $\sigma$};

\tikzset{node distance=3em}
\node[left of=h0]     {\LARGE $a_1$};
\node[left of=h1]     {\LARGE $a_2$};
\node[left of=h2]     {\LARGE $a_3$};
\node[left of=hn-1]     {\LARGE $a_n$};

\draw (q0)    edge[->,>=stealth, out=300, in=350] (h0);
\draw (q0)    edge[->,>=stealth, out=90, in=270] (h0);

\draw (q0)    edge[->,>=stealth, out=290, in=350,looseness=1.8] (h1);
\draw (q1)    edge[->,>=stealth, out=90, in=270] (h1);

\draw (q0)    edge[->,>=stealth, out=280, in=350, looseness=1.5] (h2);
\draw (q2)    edge[->,>=stealth, out=90, in=270] (h2);

\draw (q0)    edge[->,>=stealth, out=270, in=350, looseness=1.3] (hn-1);
\draw (qn-1)    edge[->,>=stealth, out=90, in=270] (hn-1);

\node (dots) at ($(q2.east)!0.5!(empty.east)$) {$\ldots$};

\tikzset{node distance=2em}
\node[below of=b]    (wb)    {\LARGE $\1$};
\node[below of=g1-1] (wg1-1) {\LARGE $\1$};
\node[below of=g1-2] (wg1-2) {\LARGE $\1$};
\node[below of=g1-3] (wg1-3) {\LARGE $\1$};
(
\draw (b)    edge[->,>=stealth] (q0);
\draw (q0)   edge[->,>=stealth, out=0, in=180] (g1-1);
\draw (q0)   edge[->,>=stealth, out=-20, in=210] (g1-1);
\draw (g1-1) edge[->,>=stealth] (q1);
\draw (q1)   edge[->,>=stealth, out=0, in=180] (g1-2);
\draw (q0)   edge[->,>=stealth, out=-30, in=210] (g1-2);
\draw (g1-2) edge[->,>=stealth] (q2);
\draw (empty)edge[->,>=stealth] (g1-3);
\draw (g1-3) edge[->,>=stealth] (qn-1);
\draw (q0)   edge[->,>=stealth, out=-40, in=210] (g1-3);

\node[state, above=5cm of g1-2] (v) {\huge $v$};
\draw (h0)    edge[->,>=stealth] (v);
\draw (h1)    edge[->,>=stealth] (v);
\draw (h2)    edge[->,>=stealth] (v);
\draw (hn-1)    edge[->,>=stealth] (v);
\node[below of=v, yshift=-3mm]     {\LARGE $\1$};

\node[state, rectangle,above=6cm of v,xshift=-6cm] (sigma1) {\huge $\sigma$};
\node[state,above=1cm of sigma1] (state1) {\huge $q_{\mathrm{one}}$};
\draw (state1) edge[->,>=stealth,out=200, in=250,looseness=1.7]  (sigma1);
\draw (state1) edge[->,>=stealth,out=340, in=290,looseness=1.6]  (sigma1);
\draw (sigma1) edge[->,>=stealth]  (state1);
\tikzset{node distance=3em}
\node[below of=sigma1]     {\LARGE $\1$};

\node[state, rectangle,left=2cm of state1] (sigmaleftleft) {\huge $\sigma$};
\draw (v) edge[->,>=stealth,out=140, in=270]  (sigmaleftleft);
\draw (state1) edge[->,>=stealth]  (sigmaleftleft);
\node[left of=sigmaleftleft]     {\LARGE $\1$};

\node[state, rectangle,right=2cm of state1] (sigmaleftright) {\huge $\sigma$};
\draw (v) edge[->,>=stealth,out=110, in=270]  (sigmaleftright);
\draw (state1) edge[->,>=stealth]  (sigmaleftright);
\node[right of=sigmaleftright]     {\LARGE $\1$};

\node[state, rectangle,above=1cm of state1] (alpha1) {\huge $\alpha$};
\draw (alpha1) edge[->,>=stealth] (state1);
\node[above of=alpha1,yshift=-2mm]     {\LARGE $\1$};

\node[state,above=1.5cm of alpha1] (oplus) {\huge $q_\oplus$};
\draw (sigmaleftleft) edge[->,>=stealth,out=90,in=210] (oplus);
\draw (sigmaleftright) edge[->,>=stealth,out=90,in=330] (oplus);

\node[state,rectangle,above=2cm of oplus] (sigmatopleft) {\huge $\sigma$};
\draw (sigmatopleft) edge[->,>=stealth,out=130,in=180, looseness=2.2] (v);
\node[state,left=2cm of oplus] (q0left) {\huge $q_0$};
\draw (q0left) edge[->,>=stealth] (sigmatopleft);
\draw (oplus) edge[->,>=stealth] (sigmatopleft);
\node[above of=sigmatopleft, xshift=0.1cm]     {\LARGE $\1$};

\node[state, rectangle,above=10cm of v,xshift=8cm] (sigma2) {\huge $\sigma$};
\node[right=3mm of sigma2] {\Large $\1$};
\node[state,above=1cm of sigma2] (otimes) {\huge $q_\otimes$};
\draw (v) edge[->,>=stealth,out=60,in=240] (sigma2);
\draw (v) edge[->,>=stealth,out=40,in=280] (sigma2);
\draw (sigma2) edge[->,>=stealth] (otimes);
\node[state, rectangle, above=2cm of otimes] (sigmatopright) {\huge $\sigma$}; 
\draw (otimes) edge[->,>=stealth] (sigmatopright);
\draw (sigmatopright) edge[->,>=stealth,out=50,in=10,looseness=2.1] (v);
\node[right=3mm of sigmatopright] {\Large $\1$};
\node[state,right=1cm of otimes] (q0right) {\huge $q_0$};
\draw (q0right) edge[->,>=stealth] (sigmatopright);
\end{tikzpicture}  

  \caption{\label{fig:closure-by-initsem} The $(\Sigma,\B)$-wta $\cA$ of the proof of Theorem \ref{thm:closure-of-finite-set-i-recognizable}, where the three occurrences of the state $q_0$ have to be identified. The wta $\cA$ might be called ``angle on inline skaters''.}
\end{figure}

\

Next we prove that $\im(\initialsem{\cA}) = \langle H \rangle_{\{\oplus,\otimes,\0,\1\}}$. Since $\mathrm{wts}(\cA) = H \cup \{\0,\1\}$, we have $\im(\initialsem{\cA}) \subseteq \langle H \rangle_{\{\oplus,\otimes,\0,\1\}}$. Thus it remains to prove $\langle H \rangle_{\{\oplus,\otimes,\0,\1\}}\subseteq \im(\initialsem{\cA})$.

For this purpose, we need some auxiliary statements, which are easy to see.
\begin{equation}
  \text{For each $q \in \{v,q_\oplus,q_\otimes\} \cup \{q_1,\ldots,q_{n-1}\}$, we have $\h_\cA(\alpha)_q=\0$\enspace.} \label{eq:usual-values-3}
    \end{equation}
\begin{equation}
\text{For each $i \in [0,n]$, we have $\h_\cA(f_{i})_{q_\otimes}=\0$ \enspace.}
 \label{eq:usual-values-1.5}
\end{equation}
  \begin{equation}
   \text{For every $\zeta_1,\zeta_2 \in \T_\Sigma$, we have $\h_\cA(\sigma(\zeta_1,\zeta_2))_{q_0} = \0$\enspace.} \label{eq:usual-values-4}
 \end{equation}
  \begin{equation}
   \text{For every $\zeta_1,\zeta_2,\zeta_3 \in \T_\Sigma$ and $i \in [0,n-1]$, we have $\h_\cA(\sigma(\zeta_1,\sigma(\zeta_2,\zeta_3)))_{q_i} = \0$ \enspace.} \label{eq:usual-values-5}
 \end{equation}
 \begin{equation}
   \text{For each $\zeta \in \T_\Sigma$, we have $\h_\cA(\zeta)_{q_{\mathrm{one}}}=\1$\enspace.} \label{eq:usual-values-6}
 \end{equation}

As further preparation, we deal with the values of expressions of the form $\h_\cA(f_i)_{q_j}$. 
First, by bounded induction on $i$, we prove the following:
\begin{equation}
  \text{For each $i,j \in [0,n-1]$ with $i \le j$, we have $\h_\cA(f_i)_{q_{j}}=\h_\cA(f_0)_{q_{j-i}}$ \enspace.} \label{eq:usual-values-1.1}
\end{equation}
The I.B. with $i=0$ is trivial. For the I.S., let $i+1 \le j$. We can calculate as follows:
\begin{align*}
\h_\cA(f_{i+1})_{q_{j}} &= \h_\cA(\sigma(f_i,\alpha))_{q_j} = \h_\cA(f_i)_{q_{j-1}} \otimes \h_\cA(\alpha)_{q_0} \otimes \delta_2(q_{j-1}q_0,\sigma,q_j) = \h_\cA(f_i)_{q_{j-1}} = \h_\cA(f_0)_{q_{j-(i+1)}} \enspace,
\end{align*}
where the last equality holds by I.H. This proves \eqref{eq:usual-values-1.1}.

Second, by bounded induction on $i$, we prove the following:
\begin{equation}
  \text{For each $i,j \in [0,n-1]$ with $i \ge j$, we have $\h_\cA(f_i)_{q_{j}}=\h_\cA(f_{i-j})_{q_0}$ \enspace.} \label{eq:usual-values-1.11}
\end{equation}
The I.B. with $i=0$ is trivial. For the I.S., let $i+1 \ge j$. As above we obtain $\h_\cA(f_{i+1})_{q_{j}} =  \h_\cA(f_i)_{q_{j-1}}$. Then, by I.H., we obtain that $\h_\cA(f_i)_{q_{j-1}} = \h_\cA(f_{(i+1)-j})_{q_0}$. This  proves \eqref{eq:usual-values-1.11}.

Since,
\begin{compactitem}
\item for each $i < j$, Equality \eqref{eq:usual-values-1.1} implies that $\h_\cA(f_i)_{q_{j}} = \h_\cA(f_0)_{q_{j-i}} = \h_\cA(\alpha)_{q_{j-i}} = \0$ (because $q_{j-i} \not= q_0$), and
\item for each $i > j$, Equality \eqref{eq:usual-values-1.11} implies that $\h_\cA(f_i)_{q_{j}}=\h_\cA(f_{i-j})_{q_0}= \0$ (because $f_{i-j}\not= \alpha$), and
\item for $i=j$, Equalities \eqref{eq:usual-values-1.1} and \eqref{eq:usual-values-1.11} imply that $\h_\cA(f_i)_{q_{j}}=\h_\cA(f_0)_{q_{j-i}} = \h_\cA(f_{i-j})_{q_0}= \h_\cA(f_0)_{q_0} = \h_\cA(\alpha)_{q_0}= \1$,
\end{compactitem}
the following two statements hold.
\begin{equation}
  \text{For each $i \in [0,n-1]$, we have $\h_\cA(f_i)_{q_{i}}  = \1$ \enspace.} \label{eq:usual-values-1}
\end{equation}
\begin{equation}
  \text{For each $i,j \in [0,n-1]$ with $j \not= i$, we have $\h_\cA(f_i)_{q_{j}}=\0$ \enspace.} \label{eq:usual-values-1.2}
\end{equation}
This finishes the proofs of auxiliary statements.

By induction on $\T_\Sigma$ we prove the following statement:
\begin{equation}\label{eq:wta-computes-value-on-representations}
  \text{For each $\xi \in \T_\Delta$, we have $\h_\cA(h(\xi))_v = \mathrm{eval}(\xi)$} \enspace,
\end{equation}
the ranked alphabet $\Delta$ and the mapping $\mathrm{eval}: \T_\Delta \to \langle H \rangle_{\{\oplus,\otimes,\0,\1\}}$ are defined before this theorem.

I.B.: Let $\xi=a_i$ for some $i \in [n]$. Then $h(a_i)=f_i$ and we can calculate as follows.
\begingroup
\allowdisplaybreaks
\begin{align*}
  \h_\cA(h(a_i))_v=& \ \h_\cA(f_i)_v 
                    = \bigoplus_{p_1p_2\in Q^2} \h_\cA(f_{i-1})_{p_1} \otimes \h_\cA(\alpha)_{p_2} \otimes \delta_2(p_1p_2,\sigma,v)\\[2mm]
  = \ & \Big(\bigoplus_{j \in [0,n-1]} \h_\cA(f_{i-1})_{q_j} \otimes \h_\cA(\alpha)_{q_0} \otimes \delta_2(q_jq_0,\sigma,v) \Big)\\
                  & \oplus \Big( \h_\cA(f_{i-1})_{q_0} \otimes \h_\cA(\alpha)_{q_\oplus} \otimes \delta_2(q_0q_\oplus,\sigma,v)\Big)\\
                    & \oplus \Big( \h_\cA(f_{i-1})_{q_\otimes} \otimes \h_\cA(\alpha)_{q_0} \otimes \delta_2(q_\otimes q_0\sigma,v)\Big)
        \tag{because for each other combination of $p_1p_2$ we have $\delta_2(p_1p_2,\sigma,v) = \0$}\\[2mm]
  = \ & \Big(\bigoplus_{j \in [0,n-1]} \h_\cA(f_{i-1})_{q_j} \otimes \1 \otimes \delta_2(q_jq_0,\sigma,v) \Big)\\
                  & \oplus \Big( \h_\cA(f_{i-1})_{q_0} \otimes \0 \otimes \delta_2(q_0q_\oplus,\sigma,v)\Big)\\
                  & \oplus \Big( \0 \otimes \h_\cA(\alpha)_{q_0} \otimes \delta_2(q_\otimes q_0\sigma,v)\Big)
                    \tag{by \eqref{eq:usual-values-1}(for $i=0$), \eqref{eq:usual-values-3}, and \eqref{eq:usual-values-1.5} }\\[2mm]
   = \ & \bigoplus_{j \in [0,n-1]} \h_\cA(f_{i-1})_{q_j} \otimes \delta_2(q_jq_0,\sigma,v) \\[2mm]
  = \ & \h_\cA(f_{i-1})_{q_{i-1}} \otimes \delta_2(q_{i-1}q_0,\sigma,v)
  \tag{by \eqref{eq:usual-values-1.2}}\\[2mm]
  = \ & \1 \otimes a_i \tag{by \eqref{eq:usual-values-1} and because $ \delta_2(q_{i-1}q_0,\sigma,v)=a_i$}\\[2mm]
        = \ & \mathrm{eval}(a_i) \enspace.
  \end{align*}
  \endgroup

I.S.: We distinguish two cases. First, let $\xi = \oplus(\xi_1,\xi_2)$ for some $\xi_1,\xi_2 \in \T_\Delta$. Then: 
\begingroup
\allowdisplaybreaks
\begin{align*}
  &\h_\cA(h(\oplus(\xi_1,\xi_2)))_v = \h_\cA(\sigma(\alpha,\sigma(h(\xi_1),h(\xi_2))))_v \\[2mm]
  = \  & \bigoplus_{p_1p_2\in Q^2} \h_\cA(\alpha)_{p_1} \otimes \h_\cA(\sigma(h(\xi_1),h(\xi_2)))_{p_2} \otimes \delta_2(p_1p_2,\sigma,v)\\[2mm]
  = \ & \Big(\bigoplus_{j \in [0,n-1]} \h_\cA(\alpha)_{q_j} \otimes \h_\cA(\sigma(h(\xi_1),h(\xi_2)))_{q_0} \otimes \delta_2(q_jq_0,\sigma,v) \Big)\\
  & \oplus \Big( \h_\cA(\alpha)_{q_0} \otimes \h_\cA(\sigma(h(\xi_1),h(\xi_2)))_{q_\oplus} \otimes \delta_2(q_0q_\oplus,\sigma,v)\Big)\\
  & \oplus \Big( \h_\cA(\alpha)_{q_\otimes} \otimes \h_\cA(\sigma(h(\xi_1),h(\xi_2)))_{q_0} \otimes \delta_2(q_\otimes q_0\sigma,v)\Big)
    \tag{because for each other combination $p_1p_2$ we have $\delta_2(p_1p_2,\sigma,v)=\0$}\\[2mm]
   = \ & \Big(\bigoplus_{j \in [0,n-1]} \h_\cA(\alpha)_{q_j} \otimes \0 \otimes \delta_2(q_jq_0,\sigma,v) \Big)\\
  & \oplus \Big( \h_\cA(\alpha)_{q_0} \otimes \h_\cA(\sigma(h(\xi_1),h(\xi_2)))_{q_\oplus} \otimes \delta_2(q_0q_\oplus,\sigma,v)\Big)\\
  & \oplus \Big( \h_\cA(\alpha)_{q_\otimes} \otimes \0 \otimes \delta_2(q_\otimes q_0\sigma,v)\Big)
  \tag{by \eqref{eq:usual-values-4}}\\[2mm]
  = \ & \h_\cA(\alpha)_{q_0} \otimes \h_\cA(\sigma(h(\xi_1),h(\xi_2)))_{q_\oplus} \otimes \delta_2(q_0q_\oplus,\sigma,v)\\[2mm]
  = \ & \1 \otimes \h_\cA(\sigma(h(\xi_1),h(\xi_2)))_{q_\oplus} \otimes \1
        \tag{by \eqref{eq:usual-values-1} (for $i=0$) and definition of $\delta_2$}\\[2mm]
  = \ &  \h_\cA(\sigma(h(\xi_1),h(\xi_2)))_{q_\oplus} \\[2mm]
  = \ & \Big(\h_\cA(h(\xi_1))_v \otimes \h_\cA(h(\xi_2))_{q_{\mathrm{one}}} \otimes \delta_2(q_vq_{\mathrm{one}},\sigma,q_\oplus)\Big) \oplus
        \Big(\h_\cA(h(\xi_1))_{q_{\mathrm{one}}} \otimes \h_\cA(h(\xi_2))_v \otimes \delta_2(q_{\mathrm{one}}q_v,\sigma,q_\oplus)\Big)
  \tag{because for each other combination $p_1p_2$ we have $\delta_2(p_1p_2,\sigma,q_\oplus)=\0$}\\[2mm]
  = \ & (\mathrm{eval}(\xi_1) \otimes \1 \otimes \1) \oplus
        (\1 \otimes \mathrm{eval}(\xi_2) \otimes \1)
   \tag{by I.H., \eqref{eq:usual-values-6}, and definition of $\delta_2$}\\[2mm]
  = \ & \mathrm{eval}(\xi_1) \oplus \mathrm{eval}(\xi_2) = \mathrm{eval}(\oplus(\xi_1,\xi_2))\enspace.
\end{align*}
\endgroup

\

Secondly, let $\xi = \otimes(\xi_1,\xi_2)$ for some $\xi_1,\xi_2 \in \T_\Delta$. Then:
\begingroup
\allowdisplaybreaks
\begin{align*}
  &\h_\cA(\otimes(\xi_1,\xi_2))_v = \h_\cA(\sigma(\sigma(h(\xi_1),h(\xi_2)),\alpha))_v \\[2mm]
  = \  & \bigoplus_{p_1p_2\in Q^2} \h_\cA(\sigma(h(\xi_1),h(\xi_2)))_{p_1} \otimes \h_\cA(\alpha)_{p_2} \otimes \delta_2(p_1p_2,\sigma,v)\\
  = \ & \Big(\bigoplus_{j \in [0,n-1]} \h_\cA(\sigma(h(\xi_1),h(\xi_2)))_{q_j} \otimes \h_\cA(\alpha)_{q_0} \otimes \delta_2(q_jq_0,\sigma,v) \Big)\\[2mm]
   & \oplus \Big( \h_\cA(\sigma(h(\xi_1),h(\xi_2)))_{q_0} \otimes \h_\cA(\alpha)_{q_\oplus} \otimes \delta_2(q_0q_\oplus,\sigma,v)\Big)\\
  & \oplus \Big( \h_\cA(\sigma(h(\xi_1),h(\xi_2)))_{q_\otimes} \otimes \h_\cA(\alpha)_{q_0} \otimes \delta_2(q_\otimes q_0\sigma,v)\Big)
    \tag{because for each other combination $p_1p_2$ we have $\delta_2(p_1p_2,\sigma,v)=\0$}\\[2mm]
   = \ & \Big(\bigoplus_{j \in [0,n-1]} \0 \otimes \h_\cA(\alpha)_{q_0} \otimes \delta_2(q_jq_0,\sigma,v) \Big)\\
  & \oplus \Big( \h_\cA(\sigma(h(\xi_1),h(\xi_2)))_{q_0} \otimes \0 \otimes \delta_2(q_0q_\oplus,\sigma,v)\Big)\\
  & \oplus \Big( \h_\cA(\sigma(h(\xi_1),h(\xi_2)))_{q_\otimes} \otimes \1 \otimes \delta_2(q_\otimes q_0\sigma,v)\Big) \tag{by \eqref{eq:usual-values-5} using that $h(\xi_2)(\varepsilon)=\sigma$, \eqref{eq:usual-values-3}, and \eqref{eq:usual-values-1}(for $i=0$)}\\[2mm]
  = \ & \h_\cA(\sigma(h(\xi_1),h(\xi_2)))_{q_\otimes} \otimes \delta_2(q_\otimes q_0\sigma,v)\\[2mm]
  = \ & \h_\cA(\sigma(h(\xi_1),h(\xi_2)))_{q_\otimes} \tag{by definition of $\delta_2$}\\[2mm]
  = \ & \h_\cA(h(\xi_1))_v \otimes \h_\cA(h(\xi_2))_v \otimes \delta_2(vv,\sigma,{q_\otimes})
  \tag{because for each other combination $p_1p_2$ we have $\delta_2(p_1p_2,\sigma,q_\otimes)=\0$}\\[2mm]
  = \ & \mathrm{eval}(\xi_1) \otimes \mathrm{eval}(\xi_2) \tag{by I.H. and definition of $\delta_2$}\\[2mm]
= \ & \mathrm{eval}(\otimes(\xi_1,\xi_2)) \enspace. 
\end{align*}
\endgroup
This finishes the proof of  \eqref{eq:wta-computes-value-on-representations}.

Now let $a \in \langle A\rangle_{\{\oplus,\otimes,\0,\1\}}$. Since  $\mathrm{eval}$ is surjective, there exists $\xi \in \T_\Delta$ such that $\mathrm{eval}(\xi) = a$.  Then
\begin{equation}\label{equ:A-init-can-compute-the-closure}
\initialsem{\cA}(h(\xi)) = \bigoplus_{p \in Q} \h_\cA(h(\xi))_p \otimes F_p = \h_\cA(h(\xi))_v = \mathrm{eval}(\xi) = a
  \end{equation}
where the last but one equality follows from   \eqref{eq:wta-computes-value-on-representations}.
Hence $\langle H\rangle_{\{\oplus,\otimes,\0,\1\}} \subseteq \im(\initialsem{\cA})$.
\end{proof}

In fact, Theorem \ref{thm:closure-of-finite-set-i-recognizable} holds for each ranked alphabet $\Sigma$ which is branching, i.e., satisfies that $\Sigma^{(k)} \not= \emptyset$ for some $k\ge 2$.

\begin{theorem-rect} \label{thm:not-monadic-wta-can-compute-closure-of-finite-subset} Let $\Sigma$ be a branching ranked alphabet. Moreover, let $\B=(B,\oplus,\otimes,\0,\1)$ be a strong bimonoid and $H \subseteq B$  be a finite subset. Then we can construct a  $(\Sigma,\B)$-wta  $\cB$  such that  $\im(\initialsem{\cB}) = \langle H \rangle_{\{\oplus,\otimes,\0,\1\}}$. In particular, if  $\B$  is generated by  $H$, then we obtain  $\im(\initialsem{\cB}) = B$. 
\end{theorem-rect}

\begin{proof} Let $(\Sigma,\rk)$ be a ranked alphabet which is branching. Hence there exist a $k\in \mathbb{N}$ with $k \ge 2$ and a $\sigma \in \Sigma$ such that $\rk(\sigma)=k$.

  We define the ranked alphabet $(\Sigma,\rk')$ such that for each $\omega \in \Sigma$ we let
 \[
    \rk'(\omega) =
    \begin{cases}
      2 & \text{ if $\omega=\sigma$} \\
      \rk(\sigma) & \text{ if otherwise.}
    \end{cases}
  \]
  Thus, $\rk'(\sigma)=2$ and $(\Sigma,\rk)$ is an extension of $(\Sigma,\rk')$.

  Now let  $H \subseteq B$  be a finite subset.

  First, since $(\Sigma,\rk')$ contains a binary symbol (i.e., $(\rk')^{-1}(2)\not= \emptyset$), by Theorem~\ref{thm:closure-of-finite-set-i-recognizable}, we can construct an $((\Sigma,\rk'),\B)$-wta $\cA$ such that $\im(\initialsem{\cA}) = \langle H \rangle_{\{\oplus,\otimes,\0,\1\}}$.

  Second, by applying Lemma~\ref{lm:super-big-padding} to $(\Sigma,\rk')$, $(\Sigma,\rk)$, and $\cA$, we obtain the $((\Sigma,\rk),\B)$-wta $\cB$. Since $\0 \in \langle H \rangle_{\{\oplus,\otimes,\0,\1\}}$, by Lemma~\ref{lm:super-big-padding}(3) we have $\im(\initialsem{\cB})= \langle H \rangle_{\{\oplus,\otimes,\0,\1\}}$. 
\end{proof}

  \subsection{Image finiteness of the semantics of wta over restricted strong bimonoids}

  In this subsection, we are interested in  sufficient conditions on the strong bimonoid $\B$ such that $\im(\runsem{\cA})$ or $\im(\initialsem{\cA})$ are finite. Moreover, we draw some conclusions about the relationship of the classes $\Rec^{\mathrm{run}}(\Sigma,\B)$ and $\Rec^{\mathrm{init}}(\Sigma,\B)$ under these conditions.

    If $\Sigma$ is trivial, i.e., $\Sigma= \Sigma^{(0)}$, then $\T_\Sigma=\Sigma^{(0)}$ and hence $\T_\Sigma$ is a finite set. Thus  we can observe the following easy fact.

  \begin{observation}\rm \label{lm:trivial-ra-finite-image-prop} Let $\Sigma$ be a trivial ranked alphabet. Then, for each $(\Sigma,\B)$-wta $\cA$, we have that $\im(\runsem{\cA})$ and $\im(\initialsem{\cA})$ are finite sets.
  \end{observation}


  The following result is an obvious consequence of the definition of the run semantics. 

\begin{lemma}\rm \label{lm:bi-loc-finite-run-image finite}
Let $\B$ be bi-locally finite and $\cA$  a $(\Sigma,\B)$-wta $\cA$. Then $\im(\runsem{\cA})$ is finite.
\end{lemma}
\begin{proof} Let $\cA=(Q,\delta,F)$. For each $\xi \in \T_\Sigma$, we have $\wt(\xi,\rho) \otimes F_{\rho(\varepsilon)} \in \langle \mathrm{wts}(\cA)\rangle_{\{\otimes\}}$.
  Thus
  \[
    \runsem{\cA}(\xi) = \bigoplus_{\rho \in \R_\cA(\xi)}\wt(\xi,\rho) \otimes F_{\rho(\varepsilon)} \subseteq \langle \langle \mathrm{wts}(\cA) \rangle_{\{\otimes\}} \rangle_{\{\oplus\}}\enspace.
  \]
  This latter set is finite because $\B$ is bi-locally finite. Hence $\im(\runsem{\cA})$ is also finite.
\end{proof}
 
In Theorem \ref{thm:bi-loc-finite-rec-step-function} we will prove the converse result: if for each $(\Sigma,\B)$-wta $\cA$ the set $\im(\runsem{\cA})$  is finite, then $\B$ is bi-locally finite.

A consequence of Lemma \ref{lm:bi-loc-finite-run-image finite} and Theorem \ref{thm:not-monadic-wta-can-compute-closure-of-finite-subset} is the following result.

\begin{corollary}\rm \label{cor:bilocfin-notlocfin-finitely-gen-inf-set} Let $\B=(B,\oplus,\otimes,\0,\1)$ be bi-locally finite and not locally finite.  If $\Sigma$ is branching, then there exists a $(\Sigma,\B)$-wta $\cA$ such that $\initialsem{\cA} \not\in \Rec^{\mathrm{run}}(\Sigma,\B)$. 
Hence, for such a $\Sigma$, we have $\Rec^{\mathrm{init}}(\Sigma,\B)\setminus \Rec^{\mathrm{run}}(\Sigma,\B)\not= \emptyset$.
\end{corollary}

\begin{proof} Let $H \subseteq B$ be a finite subset such that $\langle H\rangle_{\{\oplus,\otimes,\0,\1\}}$ is an infinite set. Let $\Sigma$ be branching. Then, by Theorem \ref{thm:not-monadic-wta-can-compute-closure-of-finite-subset} we can construct a $(\Sigma,\B)$-wta $\cA$ such that $\im(\initialsem{\cA}) = \langle H \rangle_{\{\oplus,\otimes,0,1\}}$. Thus $\im(\initialsem{\cA})$ is an infinite set. On the other hand, by Lemma \ref{lm:bi-loc-finite-run-image finite}, for each $r\in \Rec^{\mathrm{run}}(\Sigma,\B)$, the set $\im(r)$ is finite. Hence $\initialsem{\cA}\not \in \Rec^{\mathrm{run}}(\Sigma,\B)$.
  \end{proof}

  We show three applications of Corollary \ref{cor:bilocfin-notlocfin-finitely-gen-inf-set}. First, we consider the strong bimonoid $\Trunc_\lambda$ from  Example \ref{ex:strong-bimonoids}(\ref{ex:interval-strong-bimonoid}) which is not a semiring. We recall that $\Trunc_\lambda=(B,\oplus,\odot,0,1)$ where $\lambda \in \mathbb{R}$ and  $B=\{0\} \cup \{b \in \mathbb{R} \mid \lambda \le b \le 1\}$; $\Trunc_\lambda$ is bi-locally finite and not locally finite. In fact, for $\lambda = \frac{1}{4}$, in Example \ref{ex:strong-bimonoids}(\ref{ex:interval-strong-bimonoid}) an infinite family $(b_i \mid \in i \in \mathbb{N})$ of elements of $B$ is given which is generated by $H=\{\frac{1}{2}\}$. Thus, for each branching ranked alphabet $\Sigma$,  Corollary \ref{cor:bilocfin-notlocfin-finitely-gen-inf-set} implies that $\Rec^{\mathrm{init}}(\Sigma,\Trunc_{\frac{1}{4}})\setminus \Rec^{\mathrm{run}}(\Sigma,\Trunc_{\frac{1}{4}})\not= \emptyset$.

    Second, we consider the bounded lattice  $\sfFL$ in Example \ref{ex:lattices}(\ref{ex:FL2+2}). This lattice is infinite  and generated by the two chains $a < b$ and $c < d$, i.e., by four elements (cf. Figure~\ref{fig:FL2+2}). By Observation~\ref{obs:bounded-lattice-is-biloc-fin-strong-bimonoid}, the bounded lattice $\sfFL$ is a bi-locally finite strong bimonoid. Thus, for each branching ranked alphabet $\Sigma$,  Corollary \ref{cor:bilocfin-notlocfin-finitely-gen-inf-set} implies that $\Rec^{\mathrm{init}}(\Sigma,\sfFL)\setminus \Rec^{\mathrm{run}}(\Sigma,\sfFL)\not= \emptyset$.
       
    Third, we consider the strong bimonoid $\mathsf{Stb}=(\mathbb{N},\oplus,\odot,0,1)$ from Example  \ref{ex:strong-bimonoids}(\ref{ex:Stb-strong-bimonoid}).  We saw that  $\mathsf{Stb}$ is bi-locally finite and not locally finite. Moreover, $\langle \{2\} \rangle_{\{\oplus, \odot,0,1\}}=\mathbb{N}$.  Thus, by Corollary \ref{cor:bilocfin-notlocfin-finitely-gen-inf-set}, for each branching ranked alphabet $\Sigma$, we have that $\Rec^{\mathrm{init}}(\Sigma,\mathsf{Stb})\setminus \Rec^{\mathrm{run}}(\Sigma,\mathsf{Stb})\not= \emptyset$. Later a stronger version of this statement will be shown (cf. Theorem \ref{thm:init-not-run} in which we make a weaker assumption on $\Sigma$).

  
The following result is an obvious consequence of the definition of initial algebra semantics. 

\begin{lemma}\rm \label{lm:loc-finite-init-image finite}
Let $\B$ be locally finite and $\cA$ a $(\Sigma,\B)$-wta $\cA$. Then $\im(\initialsem{\cA})$ is finite.
\end{lemma}
\begin{proof} Let $\cA=(Q,\delta,F)$. For each $\xi \in \T_\Sigma$ and $q \in Q$, we have  that $\initialsem{\cA}(\xi) \in \langle \mathrm{wts}(\cA) \rangle_{\{\oplus,\otimes\}}$. This latter set is finite because $\B$ is locally finite. Hence $\im(\initialsem{\cA})$ is also finite.
\end{proof}

In Theorem \ref{thm:loc-finite-rec-step-function} we will prove the converse result: if for each $(\Sigma,\B)$-wta $\cA$ the set $\im(\initialsem{\cA})$  is finite, then $\B$ is locally finite.

If we restrict in Lemma \ref{lm:loc-finite-init-image finite} the ranked alphabet $\Sigma$ to be monadic, then we can relax the condition that $\B$ is locally finite and only require that $\B$ is weakly locally finite.

\begin{lemma}\label{lm:image-monadic-ranked alphabet}\rm Let $\Sigma$ be a monadic ranked alphabet and $\cA$ be a $(\Sigma,\B)$-wta $\cA$. Then the following two statements hold.
  \begin{compactenum}
  \item[(1)] $\im(\initialsem{\cA})\subseteq \CL(\mathrm{wts}(\cA))$.
  \item[(2)] If $\B$ is weakly locally finite, then $\im(\initialsem{\cA})$ is finite.
    \end{compactenum}
\end{lemma}
\begin{proof} Let $\cA=(Q,\delta,F)$.

Proof of (1): First, by induction on $\T_\Sigma$, we prove that the following  statement holds  (where $ \CL(\im(\delta))$ is defined in Subsection~\ref{sec:strong-bimonoid}).
  \begin{equation}  \text{For every $\xi \in \T_\Sigma$ and $q \in Q$, we have that }\h_\cA(\xi)_q \in \CL(\im(\delta)) \enspace. \label{eq:hom-in-closure}
  \end{equation}
The proof for $\xi\in \Sigma^{(0)}$ is trivial. Let $\xi = \sigma(\xi_1)$.
Then  $(\h_\cA(\xi_1)_{p}\otimes \delta_1(p,\sigma,q)) \in \CL(\im(\delta))$ for each $p\in Q$ because $\h_\cA(\xi_1)_{p}\in \CL(\im(\delta))$ by I.H. and  $\delta_1(p,\sigma,q)\in \im(\delta)$. Moreover, $\h_\cA(\xi)_q=\bigoplus_{p\in Q} \h_\cA(\xi_1)_{p} \otimes  \delta_1(p,\sigma,q)$ is also in $\CL(\im(\delta))$ because  $\CL(\im(\delta))$ is closed under $\oplus$. This finishes the proof of \eqref{eq:hom-in-closure}.

Now let $\xi\in \T_\Sigma$. By \eqref{eq:hom-in-closure}, for each $q\in Q$, we have $(\h_\cA(\xi)_q \otimes F_q) \in \CL(\mathrm{wts}(\cA))$. Hence also
$\initialsem{\cA}(\xi) = \bigoplus_{q \in Q} \h_\cA(\xi)_q \otimes F_q$
  is in $\CL(\mathrm{wts}(\cA))$. This proves (1).

  \

  Proof of (2): This follows from (1) and the fact that, if  $\B$ is weakly locally finite, then $\CL(\mathrm{wts}(\cA))$ is finite
\end{proof}

We note that Lemma~\ref{lm:image-monadic-ranked alphabet} can also be obtained from \cite[Lm.~18]{drostuvog10}, where the result was proved for string ranked alphabets, as follows \cite{dro24}. Let $\cA$ be a $(\Sigma,\B)$-wta $\cA$, where $\Sigma$ is monadic, i.e., $\Sigma = \Sigma^{(1)} \cup \Sigma^{(0)}$. If $\Sigma$ is trivial, then  $\im(\initialsem{\cA})$ is finite by Observation~\ref{lm:trivial-ra-finite-image-prop}. Now assume that $\Sigma$ is not trivial. For each $\alpha \in \Sigma^{(0)}$, we let $\Sigma_\alpha = \Sigma^{(1)} \cup \{\alpha\}$. Since $\Sigma$ is not trivial, $\Sigma_\alpha$ is a string ranked alphabet and $\T_\Sigma = \bigcup_{\alpha \in \Sigma^{(0)}} \T_{\Sigma_\alpha}$. Moreover, for each  $\alpha \in \Sigma^{(0)}$, we let $\cA_\alpha$ be the restriction of $\cA$ to $\Sigma_\alpha$. Then, for each $\xi \in \T_{\Sigma_\alpha}$, we have $\initialsem{\cA}(\xi) = \initialsem{\cA_\alpha}(\xi)$, and thus $\im(\initialsem{\cA}) = \bigcup_{\alpha \in \Sigma^{(0)}} \im(\initialsem{\cA_\alpha})$.
By \cite[Lm.~18]{drostuvog10} (using Lemma~\ref{lm:wsa=wta-over-string-ra} as bridge from wsa to wta), for each $\alpha \in \Sigma^{(0)}$, the set $\im(\initialsem{\cA_\alpha}$ is finite. Hence $\im(\initialsem{\cA})$ is finite.

    A consequence of Lemma \ref{lm:image-monadic-ranked alphabet} and Theorem \ref{thm:not-monadic-wta-can-compute-closure-of-finite-subset} is the fact that, for a particular set of strong bimonoids, the possibility of branching increases the computational power of wta with respect to wta over monadic ranked alphabets.
    
\begin{corollary}\rm \label{lm:weak-loc-fin-not-loc-fin-monadic-is-weaker} Let  $\B$ be a weakly locally finite strong bimonoid which is not locally finite. If $\Sigma$ is branching, then we can construct a $(\Sigma,\B)$-wta $\cA$ such that 
  \[\im(\initialsem{\cA}) \not\in \{\im(\initialsem{\cB}) \mid \text{$\Delta$ is a monadic ranked alphabet and  $\cB$ is a $(\Delta,\B)$-wta}\}\enspace.
    \]
    Hence 
    \begin{align*}
             &\ \{\im(\initialsem{\cB}) \mid \text{$\Delta$ is a monadic ranked alphabet and  $\cB$ is a $(\Delta,\B)$-wta}\} \\
      \subset &\ \  \{\im(\initialsem{\cB}) \mid \text{$\Delta$ is a ranked alphabet and  $\cB$ is a $(\Delta,\B)$-wta}\}  \enspace.
\end{align*}
\end{corollary}
\begin{proof} By our assumption, there exists a finite set $H\subseteq B$ such that  $\langle H \rangle_{\{\oplus,\otimes,\0,\1\}}$ is an infinite set.
  By Theorem \ref{thm:not-monadic-wta-can-compute-closure-of-finite-subset} we can construct a $(\Sigma,\B)$-wta $\cA$ such that $\im(\initialsem{\cA}) = \langle H \rangle_{\{\oplus,\otimes,0,1\}}$. Moreover, by Lemma~\ref{lm:image-monadic-ranked alphabet}, for every monadic ranked alphabet $\Delta$ and $(\Delta,\B)$-wta $\cB$, the set  $\im(\initialsem{\cB})$ is finite. 

    Then the second statement, i.e., the strict inclusion follows immediately.
  \end{proof}

In \cite{drofultepvog24}, we defined a strong bimonoid $\sfM(X)$ and proved that it is  weakly locally finite and not locally finite, cf. \cite[Thm.~3.5]{drofultepvog24}.

As a consequence of Lemma~\ref{lm:wsa=wta-over-string-ra} and Corollary~\ref{lm:weak-loc-fin-not-loc-fin-monadic-is-weaker} we obtain the following statement on the comparison of the computational power of wta and wsa: If we consider weakly locally finite strong bimonoids which are not locally finite, and the initial algebra semantics, then weighted tree automata can generate strictly more sets of values  than weighted string automata can generate.

\begin{corollary}\label{cor:weak-loc-fin-not-loc-fin-weaker-wsa} \rm Let  $\B$ be a weakly locally finite strong bimonoid which is not locally finite. Then
 \begin{align*}
      & \ \{\im(\initialsem{\cC}) \mid \text{$\Gamma$ is an alphabet and  $\cC$ is a $(\Gamma,\B)$-wsa}\} \\
       \subset &\ \{\im(\initialsem{\cA}) \mid \text{$\Sigma$ is a ranked alphabet and  $\cA$ is a $(\Sigma,\B)$-wta}\}  \enspace.
\end{align*}
  \end{corollary}
    \begin{proof} The proof is as follows:
     \begin{align*}
      & \ \{\im(\initialsem{\cC}) \mid \text{$\Gamma$ is an alphabet and  $\cC$ is a $(\Gamma,\B)$-wsa}\}  \\
      = &\ \{\im(\initialsem{\cB}) \mid \text{$\Gamma_e$ is a string ranked alphabet and  $\cB$ is a $(\Gamma_e,\B)$-wta}\} \tag{by Lemma~\ref{lm:wsa=wta-over-string-ra}} \\
      \subset &\ \  \{\im(\initialsem{\cA}) \mid \text{$\Sigma$ is a ranked alphabet and  $\cA$ is a $(\Sigma,\B)$-wta}\}  \tag{by Corollary \ref{lm:weak-loc-fin-not-loc-fin-monadic-is-weaker}}
\end{align*}
  \end{proof}

%% file: relationship-initial-run.tex
\chapter[Comparison of the two semantics]{Comparison of \\ initial algebra semantics and \\run semantics}
\label{ch:comparison-of-semantics}

In Section \ref{sec:examples} we gave a number of examples of a $(\Sigma,\B)$-wta $\cA$ and we computed its run semantics or initial algebra semantics. 

In Section \ref{sec:complexity-semantics}, we start with an analysis of the complexity of two algorithms; one
calculates $\initialsem{\cA}(\xi)$ on input $\cA$ and $\xi$, and the other calculates $\runsem{\cA}(\xi)$ on input $\cA$ and $\xi$  (cf. Theorem \ref{lm:compl-run-initial}); both algorithms follow the definition of the respective semantics in a natural way. We refer the reader to \cite{aisgas20} for investigations on the complexity of the evaluation problem in the general framework of weighted tiling systems.

Second, in Section~\ref{sec:neg-results-semantics}, we compare the run semantics and initial algebra semantics of some arbitrary, but fixed wta. In some situations, the semantics are equal. For instance,  for the mapping $\height$, in Example \ref{ex:height} we constructed a $(\Sigma,\Natmaxplus)$-wta $\cA$ and proved that $\initialsem{\cA} = \runsem{\cA} = \height$. The question arises whether $ \initialsem{\cA} = \runsem{\cA}$ for each $(\Sigma,\B)$-wta~$\cA$. In Section~\ref{sec:neg-results-semantics} we will show that the answer is negative, i.e., we show a $(\Sigma,\B)$-wta~$\cA$ such that $\initialsem{\cA} \ne \runsem{\cA}$; in fact, we show four such examples. This stimulates the next question:  is it possible to construct, for each $(\Sigma,\B)$-wta~$\cA$, a $(\Sigma,\B)$-wta $\cB$ such that $\initialsem{\cA}=\runsem{\cB}$ (and similarly with run semantics and initial algebra semantics exchanged)?
In general, the answer to this question is also negative, i.e., we prove that  there exist a strong bimonoid $\B$ and a $(\Sigma,\B)$-wta $\cA$ such that $\initialsem{\cA}\not\in \Rec^{\mathrm{run}}(\Sigma,\B)$ (cf. Theorem \ref{thm:init-not-run}).

  On the positive side, in Section \ref{sec:pos-results-semantics}, we will prove that, under certain conditions (on the wta or the strong bimonoid), the two semantics are equal.

\section[Complexity of calculating the semantics]{Complexity of calculating run semantics and initial algebra semantics} 
\label{sec:complexity-semantics}

In this section we analyse the complexity of two natural algorithms: one calculates $\runsem{\cA}(\xi)$ on input $\cA$ and $\xi$ (cf. Algorithm \ref{alg:computation-run-semantics})  and the other calculates $\initialsem{\cA}(\xi)$ of input $\cA$ and $\xi$ (cf. Algorithm \ref{alg:computation-initial-semantics}). It is obvious that these algorithms are correct.

\begin{algorithm}[h]
    \begin{algorithmic}[1]
        \Require   $(\Sigma,\B)$-wta $\cA=(Q,\delta,F)$ and $\xi \in \T_\Sigma$
        \Variables $b,r : B$; \  $w : \pos(\xi)$
        \Ensure $\runsem{\cA}(\xi)$
        \State $b \gets \0$
        \ForEach{$\mathrm{\rho}: \pos(\xi) \to Q$}\label{alg:run-rho-loop}
        \State $r \gets \1$
        \ForEach{$w \in \pos(\xi)$ in depth-first post-order}
        \State let $k = \rk(\xi(w))$
        \State $r \gets r \otimes \delta_{k}(\rho(w1)\cdots \rho(wk),\xi(w),\rho(w))$
        \EndFor
        \State $b \gets b \oplus (r \otimes F_{\rho(\varepsilon)})$
        \EndFor
        \State \Return $b$
    \end{algorithmic}
    \caption{Computation of the run semantics}\label{alg:computation-run-semantics}
  \end{algorithm}
  \begin{algorithm}[h]
    \begin{algorithmic}[1]
        \Require   $(\Sigma,\B)$-wta $\cA=(Q,\delta,F)$ and $\xi \in \T_\Sigma$
        \Variables $v: \pos(\xi) \to B^Q$; $w : \pos(\xi)$; $b : B$
        \Ensure $\initialsem{\cA}(\xi)$
        \ForEach{$w \in \pos(\xi)$ in depth-first post-order} \label{alg:initial-pos}
              \State let $k = \rk(\xi(w))$
  \ForEach{$q \in Q$} \label{alg:initial-pos-state}
        \State $v(w)_q \gets \0$
        \ForEach{$q_1 \cdots q_k \in Q^k$} \label{alg:initial-pos-state-stateseq}
        \State $v(w)_q \gets v(w)_q \oplus v(w1)_{q_1} \otimes \ldots \otimes v(wk)_{q_k} \otimes \delta_{k}(q_1\cdots q_k,\xi(w),q)$ \label{alg:initial-comp}
        \EndFor
      \EndFor
      \EndFor
              \State $b \gets \0$

        \ForEach{$q \in Q$} \label{alg:initial-root}
        
        \State $b \gets b \oplus (v(\varepsilon)_q \otimes F_{q})$
        \EndFor
        \State \Return $b$
    \end{algorithmic}
    \caption{Computation of the initial algebra semantics}\label{alg:computation-initial-semantics}
\end{algorithm}

\index{$\#_{\mathrm{run}}(\cA,\xi)$}
\index{$\#_{\mathrm{init}}(\cA,\xi)$}
More specifically, we are interested in the number $\#_{\mathrm{run}}(\cA,\xi)$ of occurrences of the strong bimonoid operations $\oplus$ and $\otimes$ that occur in the execution of Algorithm \ref{alg:computation-run-semantics} if the inputs are $\cA$ and $\xi$. Similarly, we are interested in the number $\#_{\mathrm{init}}(\cA,\xi)$ of occurrences of the strong bimonoid operations $\oplus$ and $\otimes$ that occur in the execution of Algorithm \ref{alg:computation-initial-semantics} if the inputs are $\cA$ and $\xi$.

\begin{theorem-rect} \label{lm:compl-run-initial} Let $\Sigma$ be a ranked alphabet. Moreover, let  $\B=(B,\oplus,\otimes,\0,\1)$ be a strong bimonoid and let $\cA=(Q,\delta,F)$ be a $(\Sigma,\B)$-wta. For every $\xi \in \T_\Sigma$, we have
  \begin{compactenum}
  \item[(1)] $\#_{\mathrm{run}}(\cA,\xi) = (\size(\xi)+2)  \cdot |Q|^{\size(\xi)}$, and
  \item[(2)] $\#_{\mathrm{init}}(\cA,\xi) \le \size(\xi) \cdot \max(m+1,2) \cdot (|Q|^{m+1} + |Q|)$ where $m=\maxrk(\Sigma)$.
    \end{compactenum}
\end{theorem-rect}
\begin{proof}  Proof of (1): The loop in line 2 of Algorithm \ref{alg:computation-run-semantics} is executed $|Q|^{\size(\xi)}$ many times. In each execution of the body of this loop, $\size(\xi)+1$ times $\otimes$ is executed and $1$ times $\oplus$ is executed. 

  \
  
 Proof of (2): We abbreviate $\maxrk(\Sigma)$ by $m$. 
 By assuming that each position of $w$ of $\xi$ has $m$ children
 (apart from the leaves of $\xi$) we can approximate $\#_{\mathrm{init}}(\cA,\xi)$ as follows. The loops in lines 1, 3, and 5 are executed $\size(\xi)$ many times, $|Q|$ many times, and maximally $|Q|^m$ many times, respectively. In line 6  there are maximally $m +1$ occurrences of $\oplus$ and $\otimes$. Hence, during execution of the loop in line 1, there are maximally $\size(\xi) \cdot |Q|^{m+1} \cdot (m+1)$ many occurrences of $\oplus$ and $\otimes$. During execution of the loop in line 11 there are  $2 \cdot |Q|$ many occurrences  of $\oplus$ and $\otimes$. Thus in total we obtain
 \begingroup
 \allowdisplaybreaks
  \begin{align*}
    \#_{\mathrm{init}}(\cA,\xi) &\le \size(\xi) \cdot |Q|^{m+1} \cdot (m+1) + 2 \cdot |Q|\\
                                  &\le \size(\xi) \cdot |Q|^{m+1} \cdot \max(m+1,2) + \max(m+1,2) \cdot |Q| \cdot \size(\xi)\\
    &\le \size(\xi) \cdot \max(m+1,2) \cdot  (|Q|^{m+1} + |Q|)\enspace. \qedhere
  \end{align*}
  \endgroup
\end{proof}

    As we can see from Theorem \ref{lm:compl-run-initial}, the effort of calculating $\runsem{\cA}(\xi)$ by Algorithm \ref{alg:computation-run-semantics} is exponential in the size of $\xi$, and the  effort of calculating $\initialsem{\cA}(\xi)$ by Algorithm \ref{alg:computation-initial-semantics} is at most linear in the size of $\xi$. 

    With respect to the discussion on page \pageref{page:run-sem}, 
both algorithms compute the values of the algebraic fta-hyperpath problem of an fta-hypergraph where the weights come from a strong bimonoid. We note that the value computation algorithm in \cite[Alg.~6.1]{moevog21} (also cf. \cite[Alg.~1]{moevog19a}) computes the values of the fta-algebraic hyperpath problem where the weights are calculated in a multioperator monoid.

For Hidden Markov Models (for short: HMM), the efficient calculation of the likelihood of a given observation sequence can be performed by the forward algorithm \cite[App.~A]{jurmar25}. Essentially, the forward algorithm for a HMM and the computation of the initial algebra semantics of a probabilistic wsa are based on the same algorithmic idea of summing up locally.


\section{Negative results for equality of the two semantics}
\label{sec:neg-results-semantics}

In general, the run semantics and the initial algebra semantics of a wta over $\B$ are different, which is witnessed by the following four examples for string ranked alphabets.
 
\begin{example}\rm 
\label{ex:run-not=init} We consider the string ranked alphabet $\Sigma = \{\gamma^{(1)}, \alpha^{(0)}\}$ and the tropical bimonoid  $\TropBM=(\mathbb{N}_\infty,+,\min,0,\infty)$ from Example~\ref{ex:strong-bimonoids}(\ref{ex:tropical-bimonoid}). Moreover, let $\cA =(Q,\delta,F)$ be the $(\Sigma,\TropBM)$-wta from Example~\ref{ex:run-bimonoid}, i.e.,  
\begin{compactitem}
\item $Q = \{q_0,q_1\}$,
\item $\delta_0(\varepsilon,\alpha,p) = \delta_1(p,\gamma,q)= 1$ for every $p,q \in Q$, and
\item $F_{q_0}= F_{q_1}=1$.
\end{compactitem}
From Example~\ref{ex:run-bimonoid} we know that $\runsem{\cA}(\gamma^n(\alpha))= 2^{n+1}$ for every $n \in \mathbb{N}$.
Next we compute $\initialsem{\cA}(\gamma^n(\alpha))$.
\[
\initialsem{\cA}(\gamma^n(\alpha)) = \bigplus_{q \in Q}{\min(\h_\cA(\gamma^n(\alpha))_q, \ F_q)}
= \bigplus_{q \in Q}{\min(\h_\cA(\gamma^n(\alpha))_q, \ 1)}
=^{(*)} \bigplus_{q \in Q}{1} = 2 \enspace,
\] 
where we have to verify $(*)$. For this, by induction on $\mathbb{N}$, we prove that the following statement holds:  
\[
  \text{For every $n \in \mathbb{N}$ and $q \in Q$ we have:  } \h_\cA(\gamma^n(\alpha))_{q}=
\left\{
\begin{array}{ll}
1 &\text{ if } n=0\\
2 &\text{ otherwise}\enspace.
\end{array}
\right.
\]

 I.B.: Let $n=0$. Then $\h_\cA(\gamma^n(\alpha))_q =\h_\cA(\alpha)_q = \delta_0(\varepsilon,\alpha,q) = 1$ for each $q \in Q$.

 I.S.: Let $n \ge 1$. Then for each $q \in Q$:
\begin{align*}
\h_\cA(\gamma^n(\alpha))_q 
=  \bigplus_{p \in Q}{\min(\h_\cA(\gamma^{n-1}(\alpha))_{p}, \ \delta_1(p,\gamma,q))}
= \bigplus_{p \in Q}{1}
= 2\enspace,
\end{align*}
where the second equality follows from (a) the I.H. saying that $1 \le \h_\cA(\gamma^{n-1}(\alpha))_{p} \le 2$ and (b) the fact that $\delta_1(p,\gamma,q)=1$. 
Hence $\runsem{\cA} \not= \initialsem{\cA}$.
\hfill $\Box$
\end{example}

\

\begin{example}\label{ex:run-not=init-2}\rm (adapted from \cite[Ex.~3.1]{cirdroignvog10})
  Let $\UnitIntboundedsum=([0,1],\oplus,\,\cdot\,,0,1)$ be the strong bimonoid given in Example \ref{ex:strong-bimonoids}(\ref{ex:0-1-strong-bimonoids}), where $\oplus$ is the bounded sum, and let $\Sigma=\{\gamma^{(1)},\nu^{(1)},\alpha^{(0)}\}$.  We consider the $(\Sigma,\UnitIntboundedsum)$-wta $\cA=(Q,\delta,F)$, where
\begin{compactitem}
\item $Q=\{{p},{q}\}$,
\item the transition mappings are given as follows:

  $\delta_0(\varepsilon,\alpha,{p})=0.6$, \ $\delta_0(\varepsilon,\alpha,{q})=0$,
  
$\delta_1({p},\gamma,{p})=1$, \
$\delta_1({p},\gamma,{q})=1$, \
$\delta_1({q},\gamma,{p})=0$, \
$\delta_1({q},\gamma,{q})=0$,

$\delta_1({p},\nu,{p})=1$, \
$\delta_1({p},\nu,{q})=0$, \
$\delta_1({q},\nu,{p})=1$, \
$\delta_1({q},\nu,{q})=0$,

\item $F_{p}=0.6$ and $F_{q}=0$.
\end{compactitem}
Figure \ref{fig:Ex3-1:CDIV10} shows the fta-hypergraph of $\cA$.

\begin{figure}
\begin{center}
\begin{tikzpicture}
\tikzset{node distance=7em, scale=0.6, transform shape}
\node[state, rectangle] (1) {\Large $\alpha$};
\node[state, right of=1] (2) {\Large ${p}$};
\node[state, rectangle, above of=2] (3) {\Large $\gamma$};
\node[state, rectangle, below of=2] (4) {\Large $\nu$};
\node[state, rectangle, above right=1em and 8em of 2] (5) {\Large $\gamma$};
\node[state, rectangle, below right=1em and 8em of 2] (6) {\Large $\nu$};
\node[state, right of=2] (7)[right=12em] {\Large ${q}$};

\tikzset{node distance=2em}
\node[above of=1] (w1) {0.6};
\node[above of=3] (w3) {1};
\node[above of=2] (w2)[right=0.05cm] {0.6};
\node[above of=4] (w4)[right=0.05cm] {1};
\node[above of=5] (w5) {1};
\node[above of=6] (w6) {1};

\draw[->,>=stealth] (1) edge (2);
\draw[->,>=stealth] (2) edge (3);
\draw[->,>=stealth] (3) edge[out=-250, in=-210, looseness=1.8] (2);
\draw[->,>=stealth] (2) edge (4);
\draw[->,>=stealth] (4) edge[out=250, in=-150, looseness=1.8] (2);
\draw[->,>=stealth] (2) edge (5);
\draw[->,>=stealth] (6) edge (2);
\draw[->,>=stealth] (5) edge (7);
\draw[->,>=stealth] (7) edge (6);
\end{tikzpicture}
\end{center}

\vspace{-10mm}

\caption{\label{fig:Ex3-1:CDIV10} The $(\Sigma,\UnitIntboundedsum)$-wta $\cA$ of Example \ref{ex:run-not=init-2}.}
  \end{figure}

We consider the tree $\xi = \nu(\gamma(\alpha))$ and the runs $\rho_1$ and $\rho_2$ of $\cA$ on $\xi$ with $\rho_1(\varepsilon)=\rho_1(1)=\rho_1(11)={p}$ and $\rho_2(\varepsilon)={p}$, $\rho_1(1)={q}$, and $\rho_2(11)={p}$. Then $\wt(\xi,\rho_1) = \wt(\xi,\rho_2) = 0.6$. Then we can calculate as follows:
\begin{align*}
  \runsem{\cA}(\xi) &= \bigoplus_{\rho \in \R_\cA(\xi)} \wt(\xi,\rho) \cdot F_{\rho(\varepsilon)}
                      = \bigoplus_{\rho \in \R_\cA({p},\xi)} \wt(\xi,\rho) \cdot 0.6\\
  &= \wt(\xi,\rho_1) \cdot 0.6 \oplus \wt(\xi,\rho_2) \cdot 0.6 = 0.36 \oplus 0.36 = 0.72\enspace.
                        \end{align*}
                        For the calculation of $\initialsem{\cA}(\xi)$ we calculate the following $Q$-vectors where the upper component corresponds to the state ${p}$:
                        \[
\h_\cA(\alpha) = \begin{pmatrix}0.6\\0 \end{pmatrix} \ \ \ \ 
\h_\cA(\gamma(\alpha)) = \begin{pmatrix} \h_\cA(\alpha)_{p} \cdot 1 \\ \h_\cA(\alpha)_{p} \cdot 1 \end{pmatrix} = \begin{pmatrix}0.6\\0.6 \end{pmatrix} \ 
\]
\[
  \h_\cA(\nu(\gamma(\alpha))) = \begin{pmatrix}\h_\cA(\gamma(\alpha))_{p} \cdot 1 \oplus \h_\cA(\gamma(\alpha))_{q} \cdot 1 \\ \h_\cA(\gamma(\alpha))_{p} \cdot 0 \oplus \h_\cA(\gamma(\alpha))_{q} \cdot 0\end{pmatrix} =
  \begin{pmatrix}0.6 \oplus 0.6 \\ 0\end{pmatrix} =
  \begin{pmatrix}1 \\ 0\end{pmatrix}\enspace.
  \]
Then $\initialsem{\cA}(\xi)= \bigoplus_{q' \in Q} \h_\cA(\xi)_{q'} \cdot F_{q'} = \h_\cA(\xi)_{{p}} \cdot 0.6 = 0.6$.
Hence $\runsem{\cA} \not= \initialsem{\cA}$. This inequality is due to the lack of right-distributivity in $\UnitIntboundedsum$:
\begin{align*}
  0.72 &= 0.36 \oplus 0.36 = (0.6 \cdot 0.6) \oplus (0.6 \cdot 0.6) \ne (0.6 \oplus 0.6) \cdot 0.6 = 1 \cdot 0.6 = 0.6\enspace.  \hspace{10mm} \Box
         \end{align*}
\end{example}

As the next two examples show, the run semantics and the initial algebra semantics of a wta can be different even when the strong bimonoid is finite. In the first example the wta is crisp.

\newcommand{\hatcdot}{\ \hat{\cdot}\ }

\begin{example}\rm \label{ex:5-1-3} Let $\Sigma = \{\gamma^{(1)}, \alpha^{(0)}\}$ and $\B=(\{\0,\1,b\},\oplus,\otimes,\0,\1)$ be the commutative strong bimonoid defined by $\1 \oplus \1 = \0$ and $b \oplus c = b$ for each $c \in \{\1,b\}$, and $b \otimes b = b$. The other values follow from the laws for the unit elements. 
Then, e.g., $b \oplus b = b \ne \0 = \0 \otimes b=  (\1 \oplus \1) \otimes b$. Thus $\B$ is not distributive, and hence not a semiring.

We define the $(\Sigma,\B)$-wta $\cA=(Q,\delta,F)$ such that $Q=\{q_1,q_2,q\}$, $F_{q_1}=F_{q_2}=\0$ and $F_q=b$ and
\[
  \delta_0(\varepsilon,\alpha,q_1) = \delta_0(\varepsilon,\alpha,q_2) = \1 \ \ \text{ and } \ \ 
  \delta_1(q_1,\gamma,q) =  \delta_1(q_2,\gamma,q) = \1 
\]
and the weights for the other transitions is $\0$ (cf. Figure~\ref{fig:hypergraph-Ex-5-1-3}). Obviously, $\cA$ is crisp.
   Then
   \begin{align*}
      \initialsem{\cA}(\gamma(\alpha)) &= \bigoplus_{p \in Q} \h_\cA(\gamma(\alpha))_p \otimes F_p
                                         = \h_\cA(\gamma(\alpha))_q \otimes b
      = \Big(\bigoplus_{p \in Q} \h_\cA(\alpha)_p \otimes \delta_1(p,\gamma,q)\Big) \otimes b\\
      &=  \Big((\h_\cA(\alpha)_{q_1}  \otimes \delta_1(q_1,\gamma,q))  \oplus (\h_\cA(\alpha)_{q_2} \otimes \delta_1(q_2,\gamma,q)) \Big) \otimes b\\
                                       &=  \Big((\delta_0(\varepsilon,\alpha,q_1) \otimes \1)  \oplus (\delta_0(\varepsilon,\alpha,q_2) \otimes \1) \Big) \otimes b\\
      &= (\1 \oplus \1) \otimes b = \0 \enspace.
    \end{align*}
    Also we have
      \begin{align*}
        \runsem{\cA}(\gamma(\alpha))
        &= \bigoplus_{p \in Q}\bigoplus_{\rho \in \R_\cA(p,\xi)} \wt(\gamma(\alpha),\rho) \otimes F_p
      = \bigoplus_{\rho \in \R_\cA(q,\xi)} (\wt(\gamma(\alpha),\rho) \otimes b)\\
        &= \big(\delta_0(\varepsilon,\alpha,q_1) \otimes \delta_1(q_1,\gamma,q) \otimes b\Big) \oplus \Big(\delta_0(\varepsilon,\alpha,q_2) \otimes \delta_1(q_2,\gamma,q) \otimes b\Big)\\
            &= (\1 \otimes \1 \otimes b) \oplus (\1 \otimes \1 \otimes b)\\
        &= b \oplus b = b        \enspace.
    \end{align*}
    Hence $\initialsem{\cA}\not= \runsem{\cA}$.
   \hfill $\Box$
\end{example}

\begin{figure}
\begin{center}
\begin{tikzpicture}
\tikzset{node distance=7em, scale=0.6, transform shape}
\node[state, rectangle] (1) {\Large $\alpha$};
\node[state, right of=1] (2) {\Large $q_1$};
\node[state, right=1em and 16em of 2] (3) {\Large $q_2$};
\node[state, rectangle, right of=3] (4) {\Large $\alpha$};
\node[state, rectangle, above right=4em and 2em of 2] (5) {\Large $\gamma$};
\node[state, rectangle, above left=4em and 2em of 3] (6) {\Large $\gamma$};
\node[state, above right=4em and 2.8em of 5] (7) {\Large $q$};

\tikzset{node distance=2em}
\node[above of=1] (w1) {$\1$};
\node[above of=4] (w4) {$\1$};
\node[above of=5] (w5)[left=0.05cm] {$\1$};
\node[above of=6] (w6)[right=0.05cm] {$\1$};
\node[above of=7] (w7) {$b$};

\draw[->,>=stealth] (1) edge (2);
\draw[->,>=stealth] (2) edge (5);
\draw[->,>=stealth] (4) edge (3);
\draw[->,>=stealth] (3) edge (6);
\draw[->,>=stealth] (5) edge (7);
\draw[->,>=stealth] (6) edge (7);
\end{tikzpicture}
\end{center}

\caption{\label{fig:hypergraph-Ex-5-1-3} The fta-hypergraph of the $(\Sigma,\B)$-wta $\cA$ of Example \ref{ex:5-1-3}.}
\end{figure}

  \begin{example}\rm \label{ex:lat-wta-sem-diff}\cite{fulvog23} We consider the bounded lattice $\Nfive=(N_5,\vee,\wedge,o,i)$ shown in Figure \ref{fig:lattices-N5-M3-Graetzer}. Moreover, we consider the string ranked alphabet $\Sigma = \{\gamma^{(1)}, \alpha^{(0)}\}$. Now we let $\cA =(Q,\delta,F)$ be the $(\Sigma,\Nfive)$-wta with $Q=\{q_1,q_2,q\}$, $F_{q_1}=F_{q_2}=o$ and $F_q = a$. Moreover, let
    \begin{compactitem}
  \item    $\delta_0(\varepsilon,\alpha,q_1) = b$, 
      $\delta_0(\varepsilon,\alpha,q_2) = c$, 
      $\delta_1(q_1,\gamma,q) = \delta_1(q_2,\gamma,q) = i$, 
      \item $\delta_0(\varepsilon,\alpha,q) = o$ and, for each $p_1p_2 \in (Q \times Q)\setminus \{q_1q, q_2q\}$, we let $\delta_1(p_1,\gamma,p_2)=o$.
      \end{compactitem}
The fta-hypergraph of $\cA$ is obtained from the one in Figure \ref{fig:hypergraph-Ex-5-1-3} by  adapting the weights appropriately.  

        Then
    \begin{align*}
      \initialsem{\cA}(\gamma(\alpha)) &= \bigvee_{p \in Q} \h_\cA(\gamma(\alpha))_p \wedge F_p
                                         = \h_\cA(\gamma(\alpha))_q \wedge a
      = \Big(\bigvee_{p \in Q} \h_\cA(\alpha)_p \wedge \delta_1(p,\gamma,q)\Big) \wedge a\\
      &=  \Big((\h_\cA(\alpha)_{q_1}  \wedge \delta_1(q_1,\gamma,q))  \vee (\h_\cA(\alpha)_{q_2} \wedge \delta_1(q_2,\gamma,q)) \Big) \wedge a\\
                                       &=  \Big((\delta_0(\varepsilon,\alpha,q_1) \wedge i)  \vee (\delta_0(\varepsilon,\alpha,q_2) \wedge i) \Big) \wedge a\\
      &= (b \vee c) \wedge a = a \enspace.
    \end{align*}
    Also we have
      \begin{align*}
        \runsem{\cA}(\gamma(\alpha))
        &= \bigvee_{p \in Q}\bigvee_{\rho \in \R_\cA(p,\xi)} \wt(\gamma(\alpha),\rho) \wedge F_p
      = \bigvee_{\rho \in \R_\cA(q,\xi)} (\wt(\gamma(\alpha),\rho) \wedge a)\\
        &= \big(\delta_0(\varepsilon,\alpha,q_1) \wedge \delta_1(q_1,\gamma,q) \wedge a\Big) \vee \Big(\delta_0(\varepsilon,\alpha,q_2) \wedge \delta_1(q_2,\gamma,q) \wedge a\Big)\\
            &= (b \wedge i \wedge a) \vee (c \wedge i \wedge a)\\
        &= (b \wedge a) \vee (c \wedge a) = b        \enspace.
    \end{align*}
    Hence $\initialsem{\cA}\not= \runsem{\cA}$.
    \hfill $\Box$
  \end{example}


For the particular string ranked alphabet $\Sigma$ and the $(\Sigma,\TropBM)$-wta $\cA$ of Example~\ref{ex:run-not=init}, it is easy to construct a $(\Sigma,\TropBM)$-wta $\cB$ such that $\initialsem{\cA} = \runsem{\cB}$ (just let $\cB$ have one state $q$, the weight of each transition is 2, and the root weight of $q$ is also 2).
The next theorem shows that, in general, this is not possible. Its proof is a slight adaptation of the proof of  \cite[Ex.~25]{drostuvog10}.

\begin{theorem-rect} {\rm \cite[Ex.~25]{drostuvog10}}\label{thm:init-not-run} Let $\Sigma$ be a ranked alphabet.  The following two statements hold.
  \begin{compactenum}
  \item[(1)] For each $(\Sigma,\mathsf{Stb})$-wta $\cA$, the set $\im(\runsem{\cA})$ is finite.
  \item[(2)] If $\Sigma$ is not trivial, then $\Rec^{\mathrm{init}}(\Sigma,\mathsf{Stb}) \setminus \Rec^{\mathrm{run}}(\Sigma,\mathsf{Stb}) \ne \emptyset$.
    \end{compactenum}
  \end{theorem-rect}
  
  \index{Stb@$\mathsf{Stb}$}
  \begin{proof} Before proving Statements (1) and (2), we recall the strong bimonoid $\mathsf{Stb}=(\mathbb{N},\oplus,\odot,0,1)$ from Example \ref{ex:strong-bimonoids}(\ref{ex:Stb-strong-bimonoid}). The two commutative operations $\oplus$ and $\odot$ on
        $\mathbb N$ are defined  as follows. First, let $0\oplus a=a$, $0\odot
        a=0$, and $1\odot a = a$ for every $a\in\mathbb N$. If $a,b\in
        \mathbb N\setminus\{0\}$ with $a\leq b$, we put (with $+$
        being the usual addition on $\mathbb{N}$) 
	\begin{align*}
		a\oplus b &=
		\begin{cases}
			b & \text{if $b$ is even}\\
			b+1 & \text{if $b$ is odd.} 
		\end{cases}
	\intertext{If $a,b\in\mathbb N\setminus\{0,1\}$ with $a\leq b$, let}
		a\odot b &=
		\begin{cases}
			b+1 &\text{if $b$ is even}\\
			b & \text{if $b$ is odd.} 
		\end{cases}
	\end{align*}

Proof of (1):  It follows from  Lemma \ref{lm:bi-loc-finite-run-image finite} because $\mathsf{Stb}$ is bi-locally finite (cf. Example \ref{ex:strong-bimonoids}(\ref{ex:Stb-strong-bimonoid})).

\

Proof of (2): Let $\Sigma$ be not trivial. Now we consider the particular $(\Sigma,\mathsf{Stb})$-wta $\cA=(Q,\delta,F)$ with $Q=\{{q}_1,{q}_2\}$ and $\delta_k(p_1 \cdots p_k,\sigma,p)= F_p=2$ for every $k \in \mathbb{N}$,  $\sigma\in\Sigma^{(k)}$,  and $p,p_1,\ldots,p_k \in Q$.
By induction on $\T_\Sigma$, we prove that the following statement holds:
\begin{equation}
  \text{For every $\xi \in \T_\Sigma$ and $p \in Q$, we have} : \h_\cA(\xi)_p = 2 \cdot \height(\xi)+2 \enspace. \label{eq:Ex-25}
\end{equation}

I.B.: Let $\xi = \alpha$ be in $\Sigma^{(0)}$. Then $\h_\cA(\xi)_p = \delta_0(\varepsilon,\alpha,p) = 2 = 2 \cdot \height(\xi) +2$.

I.S.: Let $\xi = \sigma(\xi_1,\ldots,\xi_k)$ with $k \ge 1$ and $p \in Q$. Moreover, let $j \in [k]$ be such that $\height(\xi_j) = \max(\height(\xi_i) \mid i \in [k])$.
Then,
\begingroup
\allowdisplaybreaks
\begin{align*}
  \h_\cA(\sigma(\xi_1,\ldots,\xi_k))_{p}
  &= \bigoplus_{p_1 \cdots p_k \in Q} \h_\cA(\xi_1)_{p_1} \odot \ldots \odot \h_\cA(\xi_k)_{p_k} \odot \delta_k(p_1 \cdots p_k,\sigma,p)\\
   &= \bigoplus_{p_1 \cdots p_k \in Q} (2 \cdot \height(\xi_1)+2) \odot \ldots \odot (2 \cdot \height(\xi_k)+2)  \odot 2  \tag{by I.H. and definition of $\delta_k$}\\
                       &= \bigoplus_{p_1 \cdots p_k \in Q} (2 \cdot \height(\xi_j)+3) \tag{by definition of $\odot$ and our choice of $j$; note that $2 \cdot \height(\xi_j)+2$ is even}\\
  &= 2 \cdot \height(\xi_j)+4 \tag{by definition of $\oplus$ and the fact that there are at least two summands due to $k \ge 1$ and $|Q|=2$}\\
  &= 2 \cdot \height(\sigma(\xi_1,\ldots,\xi_k))+ 2\enspace.
\end{align*}
\endgroup
This finishes the proof of \eqref{eq:Ex-25}.

 Hence, for each $\xi \in \T_\Sigma$, we have
 \begingroup
 \allowdisplaybreaks
 \begin{align*}
   \initialsem{\cA}(\xi) = \bigoplus_{p \in Q} \h_\cA(\xi)_q \odot F_p 
                         = \bigoplus_{p \in Q} (2 \cdot \height(\xi) +2) \odot 2
                         =\bigoplus_{p \in Q} (2 \cdot \height(\xi) +3)
   = 2 \cdot \height(\xi) + 4 \enspace.
                           \end{align*}
                           \endgroup
                           Since $\Sigma$ is not trivial, this implies that the set $\mathrm{im}(\initialsem{\cA})$ is infinite.

By using Statement (1), we obtain that
                           \[
                             \initialsem{\cA} \in \Rec^{\mathrm{init}}(\Sigma,\mathsf{Stb}) \setminus \Rec^{\mathrm{run}}(\Sigma,\mathsf{Stb}) \enspace. \qedhere
                             \]
\end{proof}

The result which is dual to Theorem \ref{thm:init-not-run} would be that there exists a ranked alphabet $\Sigma$ and a strong bimonoid $\B$ such that $\Rec^{\mathrm{run}}(\Sigma,\B) \setminus \Rec^{\mathrm{init}}(\Sigma,\B) \ne \emptyset$. The corresponding statement for wsa was shown in \cite[Ex.~26]{drostuvog10}. 
However, we were not able to reproduce that proof, and thus, we cannot include its version for wta into the book.


\section{Positive results for equality of the two semantics}\label{sec:pos-results-semantics}

In this section we show that the run semantics and the initial algebra semantics of a $(\Sigma,\B)$-wta $\cA$ are equal if $\Sigma$ is trivial, $\cA$ is bu-deterministic, or $\B$ is distributive.

\begin{observation}\rm \label{obs:init=run-trivial-ra}
Let $\Sigma$ be trivial. Then for each $(\Sigma,\B)$-wta $\cA$ we have $\initialsem{\cA} = \runsem{\cA}$.
\end{observation}
\begin{proof}  Let $\cA=(Q,\delta,F)$ and  $\alpha \in \Sigma^{(0)}$. Then
\[
  \initialsem{\cA}(\alpha) = \bigoplus_{q \in Q} \h_\cA(\alpha)_q \otimes F_q
  = \bigoplus_{q \in Q} \delta_0(\varepsilon,\alpha,q) \otimes F_q
  = \bigoplus_{q \in Q} \bigoplus_{\rho \in \R_\cA(q,\alpha)} \wt_\cA(\alpha,\rho) \otimes F_q
  = \runsem{\cA}(\alpha) \enspace,
\]
where the third equality follows from that $|\R_\cA(q,\alpha)|=1$ and for the only element $\rho$ of $\R_\cA(q,\alpha)$
we have $\wt_\cA(\alpha,\rho)=\delta_0(\varepsilon,\alpha,q)$.
Since $\T_\Sigma = \Sigma^{(0)}$, we obtain that  $\initialsem{\cA} = \runsem{\cA}$.
\end{proof}

The equality $\initialsem{\cA}=\runsem{\cA}$ holds for each bu-deterministic $(\Sigma,\B)$-wta $\cA$. Roughly speaking, this is due to the fact that the sum
\[
\h_\cA(\sigma(\xi_1,\ldots,\xi_k))_q = \bigoplus_{q_1\cdots q_k \in Q^k} \h_\cA(\xi_1)_{q_1} \otimes \ldots \otimes \h_\cA(\xi_k)_{q_k} \otimes \delta_k(q_1 \cdots q_k,\sigma,q)
\]
contains at most one non-zero summand; and hence, in the computation of $\h_\cA(\xi)_q$, the operations $\oplus$ and $\otimes$ do not alternate.

\label{p:convention-denotation-of-classes}

\begin{theorem-rect}\label{thm:bu-det:init=run} {\rm (cf. \cite[Thm.~3.6]{fulkosvog19})} Let $\Sigma$ be a ranked alphabet. Moreover, let $\B$ be a strong bimonoid. For each bu-deterministic $(\Sigma,\B)$-wta $\cA$ we have $\runsem{\cA} = \initialsem{\cA}$. Thus, in particular, $\budRec^{\mathrm{run}}(\Sigma,\B)=\budRec^{\mathrm{init}}(\Sigma,\B)$ and $\cdRec^{\mathrm{run}}(\Sigma,\B)=\cdRec^{\mathrm{init}}(\Sigma,\B)$.
\end{theorem-rect}
\begin{proof}
 Let  $\cA = (Q,\delta,F)$ and $\xi \in \T_\Sigma$. We proceed by case analysis. Using Lemma  \ref{lm:limit-bu-det}(3) we distinguish the following two cases.
 
 \underline{Case (a):} Let $\QR{\cA}{\xi} =\emptyset=\Qh{\cA}{\xi}$. Then $\runsem{\cA}(\xi) = \0 = \initialsem{\cA}(\xi)$ by Lemma~\ref{obs:state-properties-of-wta}(2) and (3).
 
 \underline{Case (b):} Let $|\state(\xi)|=1$ and $\QR{\cA}{\xi} =\state(\xi) = \Qh{\cA}{\xi}$. Then by Lemma  \ref{lm:limit-bu-det}(3)(b)
 we have 
 \begin{align*}
  \runsem{\cA}(\xi)&=\wt(\xi,\crhorm_\xi) \otimes F_{\estate(\xi)}=\h_{\cA}(\xi)_{\estate(\xi)} \otimes F_{\estate(\xi)}=\initialsem{\cA}(\xi)\enspace.
                     \qedhere
\end{align*}
\end{proof}

\index{bu-deterministically recognizable}
\index{crisp-deterministically recognizable}
\index{cdRec@$\cdRec(\Sigma,\B)$}
\index{budRec@$\budRec(\Sigma,\B)$}
\index{semantic@$\sem{\cA}$}
  \begin{quote}  
    \emph{Due to Theorem \ref{thm:bu-det:init=run}, if $\cA$ is bu-deterministic, then we write $\sem{\cA}$ instead of $\runsem{\cA}$ and $\initialsem{\cA}$.  Moreover, for a weighted tree language $r$, we say that $r$ is bu-deterministically recognizable (instead of bu-deterministically i-recognizable and bu-deterministically r-recognizable).
      As a consequence, we write $\budRec(\Sigma,\B)$ for $\budRec^{\mathrm{run}}(\Sigma,\B)$ (and hence, for $\budRec^{\mathrm{init}}(\Sigma,\B)$). Similarly, we say that $r$ is crisp-deterministically recognizable (instead of crisp-deterministically i-recognizable and crisp-deterministically r-recognizable). Also, we write $\cdRec(\Sigma,\B)$ for $\cdRec^{\mathrm{run}}(\Sigma,\B)$ (and for $\cdRec^{\mathrm{init}}(\Sigma,\B)$).}
  \end{quote}

  
  Also the two semantics are equal if $\B$ is the Boolean semiring (cf. Theorem~\ref{thm:wta-B=fta}).  In Theorem~\ref{thm:semiring-run=initial} we will prove that the two semantics are equal for each $(\Sigma,\B)$-wta if $\B$ is an arbitrary semiring and that even the converse holds.
  The converse  was proved in \cite[Lm.~4]{drostuvog10} for the particular case that $\Sigma$ is a string ranked alphabet; in that case only right-distributivity is needed. It turns out that in the tree case also left-distributivity is needed if $\Sigma$ is branching. Moreover, in \cite[Thm.~3.1(i),(iii)]{liped05}, the implication (A)$\Rightarrow$(B) of the theorem was proved for string ranked alphabets and lattice-ordered monoids in which the unit element of the monoid equals the upper bound of the lattice.

As preparation for the proof of Theorem \ref{thm:semiring-run=initial}, for every $a,b,c \in B$, we define two wta $\cA_r$ and $\cA_l$ where the intention is that, 
\begin{compactitem}
\item $\im(\initialsem{\cA_r}) = \{\0,(a \oplus b) \otimes c\}$ and
  $\im(\runsem{\cA_r}) = \{\0,a \otimes c \oplus b \otimes c\}$, i.e., apart from $\0$ the two sides of the axiom for right-distributivity are computed, and
  \item $\im(\initialsem{\cA_l}) = \{\0,a \otimes (b \oplus c)\}$ and
  $\im(\runsem{\cA_l}) = \{\0,a \otimes b \oplus a \otimes c\}$, i.e., apart from $\0$  the two sides of the axiom for left-distributivity are computed.
\end{compactitem}

\begin{lemma}\rm \label{lm:Ar-right-distributivity} For every $a,b,c \in B$ and  ranked alphabet $\Sigma$ with $\Sigma^{(1)}\not=\emptyset$, we can construct a $(\Sigma,\B)$-wta $\cA_r$ such that  $\im(\initialsem{\cA_r}) = \{\0,(a \oplus b) \otimes c\}$ and  $\im(\runsem{\cA_r}) = \{\0,a \otimes c \oplus b \otimes c\}$.
\end{lemma}

\begin{proof} Let $a,b,c \in B$. Moreover, let  $\alpha \in  \Sigma^{(0)}$ and $\gamma \in \Sigma^{(1)}$.

We construct the $(\Sigma, \B)$-wta $\cA_r = (Q, \delta, F)$ as follows: $Q = \{q_1,q_2,q_3\}$, $F_{q_1} = F_{q_2} = \mathbb{0}$, and $F_{q_3} = c$. 
Moreover, 
\begin{compactitem}
\item $\delta_0(\varepsilon,\alpha, q_1) = a$, $\delta_0(\varepsilon,\alpha, q_2) = b$, and $\delta_0(\varepsilon,\alpha, q_3) = \0$
\item $\delta_1(q_1,\gamma, q_3) = \delta_1(q_2, \gamma, q_3) = \mathbb{1}$ and $\delta_1(p,\gamma,q)=\0$ for each $pq \not\in \{q_1q_3,q_2q_3\}$, and
\item for every $k \in \mathbb{N}$, $\sigma \in \Sigma^{(k)}\setminus\{\alpha,\gamma\}$, and $(w, p) \in Q^k \times Q$, we let $\delta_k(w,\sigma,p)=\0$.
\end{compactitem}
In fact, the state behaviour of $\cA_r$ is very similar to the one of the wta in Example~\ref{ex:lat-wta-sem-diff}.

\begin{figure}
  \centering

\begin{tikzpicture}[level distance=3.5em,
  every node/.style = {align=center}]]
  \pgfdeclarelayer{bg}    
  \pgfsetlayers{bg,main}  

  \newcommand{\mydista}{1.8mm} 
  \newcommand{\mydistaa}{2.2mm} 
  \newcommand{\mydistb}{0.8mm} 
  \tikzstyle{mycircle}=[draw, circle, inner sep=-2mm, minimum height=5mm]
  \tikzstyle{mycirclebig}=[draw, circle, inner sep=-2mm, minimum height=8mm]
  \tikzstyle{mybox}=[draw, ellipse, inner sep=-2mm, minimum height=5mm]

\begin{scope}[xshift=-16mm,level 1/.style={sibling distance=20mm},level 2/.style={sibling distance=15mm}]

 \node at (-0.6, 0.6) {$\xi:$};

 \node (N0) {$\gamma$}
 child {node (N1) {$\alpha$}};

 \node [anchor=west] at ([xshift=0mm]N0.east) {$\h_{\cA_r}(\gamma(\alpha))_{q_3}=a\oplus b$};
 \node [anchor=north] at ([xshift=-3mm,yshift=-13.0mm]N0.east) (bnode) {$\h_{\cA_r}(\alpha)_{q_1}=a$};
 \node [anchor=north] at ([xshift=-11.2mm,yshift=-2.0mm]bnode.east) (bbnode) {$\h_{\cA_r}(\alpha)_{q_2}=b$};
\end{scope}
  
\begin{scope}[xshift=40mm,level 1/.style={sibling distance=20mm},level 2/.style={sibling distance=15mm}]

 \node at (-0.6, 0.6) {$\xi:$};
 \node at (1.3, 0.6) {$\rho_{q_1} \in \R_{\cA_r} (\xi)$};

 \node (N0) {$\gamma$}
 child {node (N1) {$\alpha$}};

 \node [mycircle, anchor=west] at ([xshift=\mydista]N0.east) {$q_3$};
 \node [mycircle, anchor=west] at ([xshift=\mydistb]N1.east) {$q_1$};
\end{scope}

\begin{scope}[xshift=77mm,level 1/.style={sibling distance=20mm},level 2/.style={sibling distance=15mm}]

 \node at (-0.6, 0.6) {$\xi:$};
 \node at (1.3, 0.6) {$\rho_{q_2} \in \R_{\cA_r} (\xi)$};

 \node (N0) {$\gamma$}
 child {node (N1) {$\alpha$}};

 \node [mycircle, anchor=west] at ([xshift=\mydista]N0.east) {$q_3$};
 \node [mycircle, anchor=west] at ([xshift=\mydistb]N1.east) {$q_2$};
\end{scope}
\end{tikzpicture}
  
\caption{\label{fig:semantics-on-gamma-alpha} (left) Computation of $\h_{\cA_r}(\xi)_q$ for some state $q \in Q$ and some subtree $\xi$ of $\gamma(\alpha)$ and\\ (middle and right) runs $\rho_{q_1}$ and $\rho_{q_2}$ of $\cA_r$ on $\xi=\gamma(\alpha)$ for the wta $\cA_r$ in the proof of Lemma~\ref{lm:Ar-right-distributivity}.}
\end{figure}

We consider the particular input tree $\xi = \gamma(\alpha)$ in  $\T_\Sigma$ and calculate $\initialsem{\cA_r}(\xi)$ as follows (cf. Figure~\ref{fig:semantics-on-gamma-alpha}).
\begingroup
\allowdisplaybreaks
\begin{align*}
  \initialsem{\cA_r}(\xi)
  &= \bigoplus_{q \in  Q} \h_{\cA_r}(\xi)_q \otimes F_q 
    = \h_{\cA_r}(\xi )_{q_3} \otimes F_{q_3} \tag{because $F_{q_1} = F_{q_2} = \mathbb{0}$}\\
  &= \Big( \Big( \bigoplus_{q \in  Q} \h_{\cA_r}(\alpha)_{q}\Big)  \otimes  \delta_1(q, \gamma, q_3) \Big) \otimes  c \tag{because $F_{q_3} = c$}\\
  &= \Big( \h_{\cA_r}(\alpha)_{q_1}  \otimes   \delta_1(q_1,\gamma, q_3)
   \ \oplus \ \h_{\cA_r}(\alpha)_{q_2}  \otimes  \delta_1(q_2, \gamma, q_3)\Big) \otimes  c
  \tag{because $\delta_1(q_3,\gamma,q_3)=\0$}\\
  &=(a \oplus b) \otimes  c
    \tag{because $\h_{\cA_r}(\alpha)_{q_1}=a$, $\h_{\cA_r}(\alpha)_{q_2}=b$, and  $\delta_1(q_1,\gamma, q_3) = \delta_1(q_2, \gamma, q_3) = \mathbb{1}$}\enspace.
\end{align*}
\endgroup
Moreover, it is easy to see that, for each $\xi \in \T_\Sigma\setminus \{\gamma(\alpha)\}$, we have $\initialsem{\cA_r}(\xi)=\0$. Hence $\im(\initialsem{\cA_r})=\{\0,(a \oplus b) \otimes c\}$.

\

For the calculation of $\runsem{\cA_r}(\xi )$, where $\xi=\gamma(\alpha)$, we consider the particular runs $\rho_{q_1}$ and $\rho_{q_2}$ of $\cA_r$ on $\xi$ defined by $\rho_{q_1}(\varepsilon) = q_3$, $\rho_{q_1}(1)= q_1$, and  $\rho_{q_2}(\varepsilon)= q_3$, $\rho_{q_2}(1)= q_2$ (cf. Figure~\ref{fig:semantics-on-gamma-alpha}).
Then we can calculate as follows:
\begingroup
\allowdisplaybreaks
\begin{align*}
  \runsem{\cA_r}(\xi )
  &= \bigoplus_{\rho \in \R_\cA(\xi )} \wt(\xi,\rho) \otimes  F_{\rho(\varepsilon )} \\
  &= \bigoplus_{\rho \in  \R_{\cA}(q_3,\xi )} \wt(\xi,\rho) \otimes  c
  \tag{because $F_{q_1} = F_{q_2} = \mathbb{0}$ and $F_{q_3}=c$}\\
  &= \big(\wt(\xi,\rho_{q_1}) \otimes  c\big) \oplus \big(\wt(\xi,\rho_{q_2}) \otimes  c\big)\\
  &= \big(\wt(\alpha,\rho_{q_1}|_1) \otimes  \delta_1(q_1, \gamma, q_3) \otimes  c\big) \oplus
    \big(\wt(\alpha,\rho_{q_2}|_1) \otimes    \delta_1(q_2, \gamma, q_3) \otimes  c \big)
   \tag{by \eqref{equ:weight-of-run}} \\
  &= \big(\delta_0(\varepsilon,\alpha,q_1) \otimes  \delta_1(q_1, \gamma, q_3) \otimes  c\big) \oplus
    \big(\delta_0(\varepsilon,\alpha,q_2) \otimes    \delta_1(q_2, \gamma, q_3) \otimes  c \big) \\
  &= (a \otimes   c) \oplus (b \otimes  c) \enspace.
\end{align*}
\endgroup

Moreover, it is easy to see that, for each $\xi \in \T_\Sigma\setminus \{\gamma(\alpha)\}$, we have $\runsem{\cA_r}(\xi)=\0$. Hence $\im(\runsem{\cA_r})=\{\0,a \otimes c \oplus b \otimes c\}$.
\end{proof}

For later purposes, we construct the wta $\cA_l$ such that it computes slightly more general values. 

\begin{lemma}\rm \label{lm:wta-Abr} (cf. \cite[Ex.~5.10]{drovog24}) Let $a,b,c,d \in B$ and $\Sigma$ be a ranked alphabet with $\Sigma^{(2)}\not= \emptyset$. We can construct  the $(\Sigma,\B)$-wta $\cA_l$ such that $\im(\initialsem{\cA_l}) = \{\0,a \otimes (b \oplus  c) \otimes d\}$ and  $\im(\runsem{\cA_l}) = \{\0, (a \otimes b \otimes d) \oplus (a \otimes c \otimes d)\}$.
  \end{lemma}

\begin{proof} Let $\sigma \in \Sigma^{(2)}$ and $\alpha \in \Sigma^{(0)}$.
 We construct $\cA_l = (Q,\delta,F)$ as follows (cf. Figure \ref{fig:Aleft-wta}).   
\begin{compactitem}
\item $Q  = \{q_1,q_2,q_3,q_4,q_5,q_6\}$,

 \item $\delta_0(\varepsilon,\alpha,q_1) =a$, $\delta_0(\varepsilon,\alpha,q_2) =b$, $\delta_0(\varepsilon,\alpha,q_3) =b'$, $\delta_0(\varepsilon,\alpha,q_4) =\1$, $\delta(\varepsilon,\alpha,q_5)=\delta(\varepsilon,\alpha,q_6)=\0$ and, for every $p_1,p_2,p\in Q$, 
\[
\delta_k(p_1p_2,\sigma,p) = 
\begin{cases}
  \mathbb{1} &\text{ if } p_1p_2 = q_2q_4 \text{ and }  p=q_5 \\
  \mathbb{1} &\text{ if } p_1 p_2 = q_3q_4 \text{ and }  p=q_5,\\
    \mathbb{1} &\text{ if } p_1 p_2 = q_1q_5 \text{ and }  p=q_6,\\
\mathbb{0} &\text{ otherwise.}
\end{cases}
\]
For every $\theta \in \Sigma \setminus \{\alpha, \sigma\}$ with rank $k \in \mathbb{N}$ and every $p_1,\ldots,p_k,p \in Q$, we let $\delta_n(p_1\cdots p_k,\theta,p) = \0$.

\item $F_{q_1}= F_{q_2} = F_{q_3}= F_{q_4}= F_{q_5}=\mathbb{0}$ and $F_{q_6} = c$.
\end{compactitem}

\begin{figure}
  \centering

  \begin{tikzpicture}
  \tikzset{node distance=7em, scale=0.4, transform shape}

  \node[state, circle] (q6) {\huge $q_6$};
  \node[state, rectangle, below= 4em of q6] (sigmat) {\Huge $\sigma$};
  \node[state, circle, xshift=-10em, yshift=-14em] (q1) {\huge $q_1$};
  \node[state, circle, xshift=10em, yshift=-14em] (q5) {\huge $q_5$};
  \node[state, rectangle, below= 4em of q1] (alphal) {\Huge $\alpha$};
  \node[state, rectangle, xshift= 2em, yshift=-21em] (sigmabl) {\Huge $\sigma$};
  \node[state, rectangle, xshift= 18em, yshift=-21em] (sigmabr) {\Huge $\sigma$};
  \node[state, circle, xshift=-2em, yshift=-28em] (q2) {\huge $q_2$};
  \node[state, circle, xshift=5em, yshift=-28em] (q3) {\huge $q_3$};
  \node[state, circle, xshift=24em, yshift=-28em] (q4) {\huge $q_4$};
  \node[state, rectangle, below= 4em of q2] (alphaml) {\huge$\alpha$}; 
  \node[state, rectangle, below= 4em of q3] (alphamr) {\huge$\alpha$}; 
  \node[state, rectangle, below= 4em of q4] (alphar) {\huge$\alpha$};

\draw (sigmat) edge[->,>=stealth] (q6);
\draw (q1) edge[->,>=stealth] (sigmat);
\draw (q5) edge[->,>=stealth] (sigmat);
\draw (alphal) edge[->,>=stealth] (q1);
\draw (sigmabl) edge[->,>=stealth] (q5);
\draw (sigmabl) edge[->,>=stealth] (q5);
\draw (sigmabr) edge[->,>=stealth] (q5);
\draw (q2) edge[->,>=stealth] (sigmabl);
\draw (q3) edge[->,>=stealth] (sigmabr);
\draw (q4) edge[->,>=stealth] (sigmabl);
\draw (q4) edge[->,>=stealth] (sigmabr);
\draw (alphaml) edge[->,>=stealth] (q2);
\draw (alphamr) edge[->,>=stealth] (q3);
\draw (alphar) edge[->,>=stealth] (q4);

\tikzset{node distance=3em}
\node[right of=q6]    (wq6)    {\huge $d$};
\node[right of=sigmat]    (wsigmat)    {\huge $\1$};
\node[right of=alphal]    (walphal)    {\huge $a$};
\node[right of=sigmabl]    (walphal)    {\huge $\1$};
\node[right of=sigmabr]    (walphar)    {\huge $\1$};
\node[right of=alphaml]    (walphaml)    {\huge $b$};
\node[right of=alphamr]    (walphamr)    {\huge $c$};
\node[right of=alphar]    (walphar)    {\huge $\1$};

  \end{tikzpicture}
  \caption{\label{fig:Aleft-wta} The  $(\Sigma,\B)$-wta $\cA_l$ of Lemma \ref{lm:wta-Abr}.}
  \end{figure}

\begin{figure}
  \centering

\begin{tikzpicture}[level distance=3.5em, scale=0.8,
  every node/.style = {align=center}]]
  \pgfdeclarelayer{bg}    
  \pgfsetlayers{bg,main}  

  \newcommand{\mydista}{1.8mm} 
  \newcommand{\mydistaa}{2.2mm} 
  \newcommand{\mydistb}{0.8mm} 
  \newcommand{\mydistc}{-3.2mm} 
  \newcommand{\mydistd}{-8.9mm} 
  \tikzstyle{mycircle}=[draw, circle, inner sep=-2mm, minimum height=5mm]
  \tikzstyle{mybox}=[draw, ellipse, inner sep=-2mm, minimum height=5mm]

\begin{scope}[level 1/.style={sibling distance=30mm},level 2/.style={sibling distance=25mm}]

 \node at (-1.2, 0.2) {$\xi:$};

 \node (N0) {$\sigma$}
 child {node (N1) {$\alpha$}}
         child {node (N3) {$\sigma$}
           child {node (N4) {$\alpha$}}
         child {node (N6) {$\alpha$}}};

 \node [anchor=west] at ([xshift=\mydista]N0.east) {$\h(\xi)_{q_6} = a \otimes (b \oplus c)$};
 \node [anchor=north] at ([xshift=\mydistc]N1.east) {$\h(\alpha)_{q_1}=a$};
 \node [anchor=west] at ([xshift=\mydistb]N3.east) {$\h(\sigma(\alpha,\alpha))_{q_5} = b\oplus c$};
 \node [anchor=north] at ([xshift=\mydistc]N4.east) (bnode) {$\h(\alpha)_{q_2}=b$};
 \node [anchor=north] at ([xshift=\mydistd,yshift=-1.0mm]bnode.east) (bnode) {$\h(\alpha)_{q_3}=c$};
 \node [anchor=north] at ([xshift=\mydistc]N6.east) {$\h(\alpha)_{q_4}=\1$};
\end{scope}
\end{tikzpicture}

\caption{\label{fig:init-2} The illustration of the computation of 
  $\h(\xi)_{q_6}$.}
\end{figure}   

Let us consider the particular tree $\xi=\sigma(\alpha,\sigma(\alpha,\alpha))$.

We calculate $\initialsem{\cA_l}(\xi)$ as follows (cf. Figure~\ref{fig:init-2}) where we abbreviate $\h_{\cA_l}$ by~$\h$:
\begingroup
\allowdisplaybreaks
\begin{align*}
 \initialsem{\cA_l}(\xi) =& \bigoplus_{p \in Q} \h(\xi)_p \otimes F_p=\h(\xi)_{q_6} \otimes d
  \tag{because $F_p = \0$ for each $p \in Q\setminus \{q_6\}$, and $F_{q_6} = d$}\\
  = & \ \Big(\bigoplus_{p_1p_2 \in Q^2}  \h(\alpha)_{p_1} \otimes \h(\sigma(\alpha,\alpha))_{p_2} \otimes \ \delta_2(p_1p_2,\sigma,q_6)\Big)\otimes d\\
   = & \  \h(\alpha)_{q_1} \otimes \h(\sigma(\alpha,\alpha))_{q_5} \otimes \ \delta_2(q_1q_5,\sigma,q_6) \otimes d
       \tag{because  $\delta_2(p_1p_2,\sigma,q_6) =\0$ for each $p_1 p_2 \not= q_1q_5$}\\
  = & \  a \otimes  \h(\sigma(\alpha,\alpha))_{q_5} \otimes \ \1 \otimes d\\
  = & \  a \otimes \h(\sigma(\alpha,\alpha))_{q_5} \otimes d \\
  = & \  a \otimes \Big(\bigoplus_{p_1p_2 \in Q^2} \h(\alpha)_{p_1} \otimes \h(\alpha)_{p_2} \otimes \delta_2(p_1p_2,\sigma,q_5) \Big) \otimes d \\
  = & \  a \otimes \Big(\ \ \ \h(\alpha)_{q_2} \otimes  \h(\alpha)_{q_4} \otimes \delta_2(q_2q_4,\sigma,q)  \\
  & \hspace*{7mm} \  \oplus \ \h(\alpha)_{q_3} \otimes \h(\alpha)_{q_4} \otimes \delta_2(q_3q_4,\sigma,q) \Big) \otimes d
  \tag{because $\delta_2(p_1p_2,\sigma,q_5) = \0$ for each $p_1p_2 \not\in\{q_2q_4, q_3q_4\}$}\\
  = & \  a \otimes \Big(b \otimes \1 \otimes \1  \  \oplus \ c \otimes \1 \otimes \1 \Big) \otimes d\\
  =& \ a \otimes (b \oplus  c) \otimes d \enspace.
\end{align*}
\endgroup

Thus
\begin{equation}\label{equ:wta-specialtree-init-sem}
   \initialsem{\cA_l}(\sigma(\alpha,\sigma(\alpha,\alpha))) = a \otimes (b \oplus  c) \otimes d \enspace.
  \end{equation}

It is easy to see that $\initialsem{\cA_l}(\zeta)=\0$ for each $\zeta \in \T_\Sigma$ with $\zeta\ne \xi$, hence 
\begin{equation}\label{equ:wta-im-init-sem}
   \im(\initialsem{\cA_l}) = \{\0, a \otimes (b \oplus  c) \otimes d\} \enspace.
  \end{equation}

In order to compute $\runsem{\cA_l}(\xi)$, we consider the particular runs $\rho_1,\rho_2 \in \R_{\cA_l}(\xi)$ (cf. Figure~\ref{fig:run-2}) with
\begin{compactitem}
\item $\rho_1(\varepsilon)=q_6$, $\rho_1(1)=q_1$,  $\rho_1(2)=q_5$, $\rho_1(21)=q_2$, $\rho_1(22)=q_4$, and
\item $\rho_2$ is defined in the same way as $\rho_1$ except that $\rho_2(21)=q_3$.
\end{compactitem}

\begin{figure}
  \centering

\begin{tikzpicture}[level distance=3.5em, scale=0.8,
  every node/.style = {align=center}]]
  \pgfdeclarelayer{bg}    
  \pgfsetlayers{bg,main}  

  \newcommand{\mydista}{1.8mm} 
  \newcommand{\mydistaa}{2.2mm} 
  \newcommand{\mydistb}{0.8mm} 
  \tikzstyle{mycircle}=[draw, circle, inner sep=-2mm, minimum height=5mm]
  \tikzstyle{mycirclebig}=[draw, circle, inner sep=-2mm, minimum height=8mm]
  \tikzstyle{mybox}=[draw, ellipse, inner sep=-2mm, minimum height=5mm]

\begin{scope}[level 1/.style={sibling distance=20mm},level 2/.style={sibling distance=20mm}]

 \node at (-0.8, 0.6) {$\xi:$};
 \node at (1.9, 0.6) {$\rho_1 \in \R_{\cA_l} (\xi)$:};

 \node (N0) {$\sigma$}
 child {node (N1) {$\alpha$}}
         child {node (N3) {$\sigma$}
           child {node (N4) {$\alpha$}}
         child {node (N6) {$\alpha$}}};

 \node [mycircle, anchor=west] at ([xshift=\mydista]N0.east) {$q_6$};
 \node [mycircle, anchor=west] at ([xshift=\mydistb]N1.east) {$q_1$};
 \node [mycircle, anchor=west] at ([xshift=\mydistb]N3.east) {$q_5$};
 \node [mycircle, anchor=west] at ([xshift=\mydistb]N4.east) {$q_2$};
 \node [mycircle, anchor=west] at ([xshift=\mydistb]N6.east) {$q_4$};
\end{scope}

\begin{scope}[xshift=70mm,level 1/.style={sibling distance=20mm},level 2/.style={sibling distance=20mm}]

 \node at (-0.8, 0.6) {$\xi:$};
 \node at (1.9, 0.6) {$\rho_2 \in \R_{\cA_l} (\xi)$:};

 \node (N0) {$\sigma$}
 child {node (N1) {$\alpha$}}
         child {node (N3) {$\sigma$}
           child {node (N4) {$\alpha$}}
         child {node (N6) {$\alpha$}}};

 \node [mycircle, anchor=west] at ([xshift=\mydista]N0.east) {$q_6$};
 \node [mycircle, anchor=west] at ([xshift=\mydistb]N1.east) {$q_1$};
 \node [mycircle, anchor=west] at ([xshift=\mydistb]N3.east) {$q_5$};
 \node [mycircle, anchor=west] at ([xshift=\mydistb]N4.east) {$q_3$};
 \node [mycircle, anchor=west] at ([xshift=\mydistb]N6.east) {$q_4$};
\end{scope}
\end{tikzpicture}
  
\caption{\label{fig:run-2} Runs $\rho_1$ and $\rho_2$ of $\cA_l$ on $\sigma(\alpha,\sigma(\alpha,\alpha))$, cf. Lemma~\ref{lm:wta-Abr}.}
\end{figure}
Then we have (where we abbreviate $\wt_{\cA_l}$ by $\wt$):
\begingroup
\allowdisplaybreaks
\begin{align*}
  & \ \runsem{\cA_l}(\xi) = \bigoplus_{\rho \in \R_{\cA_l}(\xi)}\wt(\rho,\xi) \otimes F_{\rho(\varepsilon)}= 
  \bigoplus_{\rho \in \R_{\cA_l,q_6}(\xi)}\wt(\rho,\xi) \otimes d
      \tag{because $F_p = \0$ for each $p \in Q\setminus \{q_6\}$, and $F_{q_6} = d$} \\[2mm]
     =& \ \ \ \big(\wt(\rho_1,\xi) \otimes d\big) \oplus \big(\wt(\rho_2,\xi) \otimes d\big)
     \tag{because $\wt(\rho,\xi)=\0$ for each $\rho \in \R_{\cA_l,q_6}(\xi) \setminus \{\rho_1,\rho_2\}$}\\[2mm]
  =& \ \ \ \Big(\delta_0(\varepsilon,\alpha,q_1) \otimes  \delta_0(\varepsilon,\alpha,q_2) \otimes \delta_0(\varepsilon,\alpha,q_4) \otimes \delta_2(q_2q_4,\sigma,q_5) \otimes \delta_2(q_1q_5,\sigma,q_6) \otimes d\Big) \\
  & \oplus  \Big(\delta_0(\varepsilon,\alpha,q_1) \otimes   \delta_0(\varepsilon,\alpha,q_3) \otimes \delta_0(\varepsilon,\alpha,q_4) \otimes \delta_2(q_3q_4,\sigma,q_5) \otimes \delta_2(q_1q_5,\sigma,q_6) \otimes d\Big) \\[2mm]
  =& \ \ \ \Big(a \otimes  b  \otimes   \1
     \otimes \1 \otimes \1 \otimes d\Big)  \oplus  \Big(a \otimes  c \otimes  \1 \otimes \1 \otimes \1 \otimes d\Big) \\
  =& \ \ \  (a \otimes  b \otimes d) \oplus (a \otimes c\otimes d) \enspace.
\end{align*} 
\endgroup

Thus
\begin{equation}\label{equ:wta-specialtree-run-sem}
   \runsem{\cA_l}(\sigma(\alpha,\sigma(\alpha,\alpha))) = (a \otimes  b \otimes d) \oplus (a \otimes c\otimes d)\enspace.
  \end{equation}

It is easy to see that $\runsem{\cA_l}(\zeta)=\0$ for each $\zeta \in \T_\Sigma$ with $\zeta\ne \xi$, hence
\begin{equation}\label{equ:wta-im-run-sem}
   \im(\runsem{\cA_l}) = \{\0, (a \otimes  b \otimes d) \oplus (a \otimes c\otimes d)\} \enspace.
 \end{equation}
\end{proof}

Now we can prove the characterization of the equality of the two semantics.

\begin{theorem-rect}{\rm (cf. \cite[Thm.~4.1]{rad10}, \cite[Lm. 4.1.13]{bor04b}), and \cite[Thm.~6.3]{drovog24})} \label{thm:semiring-run=initial} Let $\Sigma$ be a ranked alphabet. Moreover, let $\B=(B,\oplus,\otimes,\0,\1)$ be a strong bimonoid. The following three statements are equivalent:
\begin{compactenum}
\item[(A)] If $\Sigma$ is not trivial, then $\B$ is right-distributive, and if $\Sigma$ is branching, then $\B$ is distributive.
\item[(B)] For each $(\Sigma,\B)$-wta $\cA$ we have $\runsem{\cA} = \initialsem{\cA}$.
\item[(C)] For each $(\Sigma,\B)$-wta $\cA$ we have $\im(\runsem{\cA}) = \im(\initialsem{\cA})$.
\end{compactenum}
\end{theorem-rect}
\begin{proof} Proof of (A)$\Rightarrow$(B):
  Let $\cA=(Q,\delta,F)$ and  $\xi\in \T_\Sigma$. Since
\begin{align*}
\runsem{\cA}(\xi) = \bigoplus_{q \in Q}\bigoplus_{\rho \in \R_{\cA}(q,\xi)}\wt(\xi,\rho) \otimes F_q \ \ 
\text{ and } \ \ 
\initialsem{\cA}(\xi) = \bigoplus_{q \in Q} \h_\cA(\xi)_q \otimes F_q\enspace,
\end{align*}
it suffices to show that
\[\text{for each $q\in Q$, we have } \ \h_\cA(\xi)_q\otimes F_q=\bigoplus_{\rho\in \R_{\cA}(q,\xi)}\wt(\xi,\rho)\otimes F_q\enspace.
\]
If $\Sigma$ is trivial, then $|\R_{\cA}(q,\xi)|= 1$. Otherwise
$\B$ is right-distributive. Thus in both cases we have 
\[\bigoplus_{\rho\in \R_{\cA}(q,\xi)}\wt(\xi,\rho)\otimes F_q = \Big(\bigoplus_{\rho\in \R_{\cA}(q,\xi)}\wt(\xi,\rho)\Big)\otimes F_q.\]
Hence, by induction on $\T_\Sigma$, we prove that the following statement holds: 
\begin{equation}
\text{For every $\xi \in \T_\Sigma$ and $q \in Q$, we have }  \h_\cA(\xi)_q=\bigoplus_{\rho\in \R_{\cA}(q,\xi)}\wt(\xi,\rho)\enspace. \label{equ:weight-run=hom}
\end{equation}
For this, let $\xi=\sigma(\xi_1,\ldots,\xi_k)$ and $q\in Q$.
Then
\begingroup
\allowdisplaybreaks
\begin{align*}
&\h_\cA(\xi)_q \\
&= \bigoplus_{q_1\cdots q_k\in Q^k} \Big( \bigotimes_{i\in[k]} \h_\cA(\xi_i)_{q_i}\Big) \otimes\delta_k(q_1\cdots q_k,\sigma,q)\\[2mm]
&= \bigoplus_{q_1\cdots q_k\in Q^k} \Big(\bigoplus_{\rho_1\in \R_{\cA}(q_1,\xi_1)} \mathrm{wt}(\xi_1,\rho_1)\Big)\otimes\ldots\otimes \Big(\bigoplus_{\rho_k\in \R_{\cA}(q_k,\xi_k)} \mathrm{wt}(\xi_k,\rho_k)\Big)\otimes
\delta_k(q_1\cdots q_k,\sigma,q) \ \ \text{(by I.H.)}\\[2mm]
&= \bigoplus_{q_1\cdots q_k\in Q^k} \Big(\bigoplus_{\rho_1\in \R_{\cA}(q_1,\xi_1)} \mathrm{wt}(\xi_1,\rho_1)\otimes \Big(\ldots \otimes \Big(\bigoplus_{\rho_k\in \R_{\cA}(q_k,\xi_k)} \mathrm{wt}(\xi_k,\rho_k)\otimes
\delta_k(\rho_1(\varepsilon)\cdots \rho_k(\varepsilon),\sigma,q)\Big) \ldots\Big)\Big)
\tag{by right-distributivity, in case $k\ge 1$}\\[2mm]
&= \bigoplus_{q_1\cdots q_k\in Q^k} \Big(\bigoplus_{\rho_1\in \R_{\cA}(q_1,\xi_1)}\cdots 
\bigoplus_{\rho_k\in \R_{\cA}(q_k,\xi_k)} \Big( \bigotimes_{i \in [k]} \mathrm{wt}(\xi_i,\rho_i)\Big) 
\otimes \delta_k(\rho_1(\varepsilon)\cdots \rho_k(\varepsilon),\sigma,q)\Big)
\tag{by left-distributivity, in case $k\ge 2$}\\[2mm]
  &= \bigoplus_{q_1\cdots q_k\in Q^k} \Big(\bigoplus_{\substack{\rho\in \R_{\cA}(q,\xi),\\\rho(1)=q_1,\ldots,\rho(k)=q_k}}
  \Big( \bigotimes_{i \in [k]} \mathrm{wt}(\xi_i,\rho|_i)\Big) \otimes \delta_k(\rho(1)\cdots \rho(k),\sigma,\rho(\varepsilon))
\Big)\\[2mm]
&= \bigoplus_{\rho\in \R_{\cA}(q,\xi)}
\Big( \bigotimes_{i \in [k]} \mathrm{wt}(\xi_i,\rho|_i)\Big) \otimes
\delta_k(\rho(1)\cdots \rho(k),\sigma,\rho(\varepsilon))
= \bigoplus_{\rho\in \R_{\cA}(q,\xi)} \mathrm{wt}(\xi,\rho)\;.
\end{align*}
\endgroup

\

Proof of (B)$\Rightarrow$(C): This is trivial.

\

Proof of (C)$\Rightarrow$(A): We have to show:
\begin{compactenum}
\item[(i)]\label{lm:run-and-initial-algebra-semantics:itemi} If $\Sigma$ is not trivial, then $(a \oplus b) \otimes c = a \otimes c \oplus b \otimes c$ for every $a,b,c \in B$, and
\item[(ii)]\label{lm:run-and-initial-algebra-semantics:itemii} if $\Sigma$ is branching, then
  $(a \oplus b) \otimes c = a \otimes c \oplus b \otimes c$
    and $a \otimes (b \oplus c) = a \otimes b \oplus a \otimes c$ for every $a,b,c \in B$.
\end{compactenum}

To prove (i), we assume that $(\Sigma,\rk)$ is not trivial.  Hence there exists a $k\in \mathbb{N}$ with $k \ge 1$ and a $\gamma \in \Sigma$ such that $\rk(\gamma)=k$. Moreover, let  $a,b,c \in B$. 

 First, we define the ranked alphabet $(\Sigma,\rk')$ such that for each $\sigma \in \Sigma$ we let
 \[
    \rk'(\sigma) =
    \begin{cases}
      1 & \text{ if $\sigma=\gamma$} \\
      \rk(\sigma) & \text{ if otherwise.}
    \end{cases}
  \]
Thus, $\rk'(\gamma)=1$ and $(\Sigma,\rk)$ is an extension of $(\Sigma,\rk')$.

  Second, since $(\Sigma,\rk')$ contains a unary symbol, i.e., $(\rk')^{-1}(1) \not= \emptyset$, by Lemma~\ref{lm:Ar-right-distributivity}, we can construct a $((\Sigma,\rk'),\B)$-wta $\cA_r$ such that $\im(\initialsem{\cA_r}) = \{\0,(a \oplus b) \otimes c\}$ and  $\im(\runsem{\cA_r}) = \{\0,a \otimes c \oplus b \otimes c\}$.

  Third, by applying Lemma~\ref{lm:super-big-padding}(3) to $(\Sigma,\rk')$, $(\Sigma,\rk)$, and $\cA_r$,
  we obtain the $((\Sigma,\rk),\B)$-wta $\cB$ such that,
\begin{compactitem}
\item $\im(\initialsem{\cB}) = \im(\initialsem{\cA_r})$ and $\im(\runsem{\cB}) = \im(\runsem{\cA_r})$, or
\item  $\im(\initialsem{\cB}) = \im(\initialsem{\cA_r}) \cup\{\0\}$ and $\im(\runsem{\cB}) = \im(\runsem{\cA_r}) \cup\{\0\}$.
\end{compactitem}
Thus $\im(\initialsem{\cB}) = \{\0,(a \oplus b) \otimes c\}$ and $\im(\runsem{\cB}) = \{\0,a \otimes c \oplus b \otimes c\}$.
By (C) we obtain that $\{\0,(a \oplus b) \otimes c\}=  \{\0,a \otimes c \oplus b \otimes c\}$, and hence $(a \oplus b) \otimes c = a \otimes c \oplus b \otimes c$. Thus $\B$ is right-distributive.

\

To prove (ii), we assume that $(\Sigma,\rk)$ is branching. Let  $a,b,c \in B$.  Since $(\Sigma,\rk)$ is not trivial, by (i), the first law in (ii) holds. So it suffices to prove that the second holds. By our assumption, there exists a $k\in \mathbb{N}$ with $k \ge 2$ and a $\gamma \in \Sigma$ such that $\rk(\gamma)=k$.

First, we define the ranked alphabet $(\Sigma,\rk')$ such that for each $\sigma \in \Sigma$ we let
 \[
    \rk'(\sigma) =
    \begin{cases}
      2 & \text{ if $\sigma=\gamma$} \\
      \rk(\sigma) & \text{ if otherwise.}
    \end{cases}
  \]
Thus, $\rk'(\gamma)=2$ and $(\Sigma,\rk)$ is an extension of $(\Sigma,\rk')$.

  Second, since $(\Sigma,\rk')$ contains a binary symbol,  i.e., $(\rk')^{-2}(1) \not= \emptyset$, by Lemma~\ref{lm:wta-Abr} (with $d=\1$), we can construct a $((\Sigma,\rk'),\B)$-wta $\cA_l$ such that $\im(\initialsem{\cA_l}) = \{\0,a \otimes (b \oplus c) \}$ and  $\im(\runsem{\cA_l}) = \{\0,a \otimes b \oplus a \otimes c\}$.
  
  Third, by applying Lemma~\ref{lm:super-big-padding}(3) to $(\Sigma,\rk')$, $(\Sigma,\rk)$, and $\cA_l$,
  we obtain the $((\Sigma,\rk),\B)$-wta $\cB$ such that,
\begin{compactitem}
\item $\im(\initialsem{\cB}) = \im(\initialsem{\cA_l})$ and $\im(\runsem{\cB}) = \im(\runsem{\cA_l})$, or
\item  $\im(\initialsem{\cB}) = \im(\initialsem{\cA_l}) \cup\{\0\}$ and $\im(\runsem{\cB}) = \im(\runsem{\cA_l}) \cup\{\0\}$.
\end{compactitem}
Thus $\im(\initialsem{\cB}) =  \{\0,a \otimes (b \oplus c) \}$ and $\im(\runsem{\cB}) = \{\0,a \otimes b \oplus a \otimes c\}$.
By (C) we obtain $\{\0,a \otimes (b \oplus c)\}=  \{\0,a \otimes b \oplus a \otimes c \}$, and hence $a \otimes (b \oplus c)= a \otimes b \oplus a \otimes c$. Thus $\B$ is left-distributive.
\end{proof}

In Examples \ref{ex:height} and \ref{ex:number-of-occurrences} we gave a wta over the semiring $\Natmaxplus$ and a wta over the semiring $\Nat$, respectively,  and we have proved that for both wta the run semantics coincides with the initial algebra semantics. The following corollary of Theorem~\ref{thm:semiring-run=initial} shows that this coincidence is valid for each wta over an arbitrary semiring.

\begin{corollary-rect}\label{cor:semiring-run=init} \rm Let $\Sigma$ be a ranked alphabet. Moreover, let $\B=(B,\oplus,\otimes,\0,\1)$ be a semiring. For each $(\Sigma,\B)$-wta $\cA$, we have
  \begin{compactenum}
    \item[(1)]  $\h_\cA(\xi)_q = \bigoplus_{\rho \in \R_\cA(q,\xi)} \wt(\xi,\rho)$ for every $\xi \in \T_\Sigma$ and state $q$ of $\cA$ and
    \item[(2)]    $\runsem{\cA} = \initialsem{\cA}$.
    \end{compactenum}
    Thus, in particular, $\Rec^{\mathrm{run}}(\Sigma,\B)=\Rec^{\mathrm{init}}(\Sigma,\B)$.
     \end{corollary-rect} 
\begin{proof} Since $\B$ is a semiring, Equation \eqref{equ:weight-run=hom} holds, and hence Statement (1) holds. Moreover, Theorem~\ref{thm:semiring-run=initial}((A)$\Rightarrow$(B)) implies that $\runsem{\cA} = \initialsem{\cA}$ and $\Rec^{\mathrm{run}}(\Sigma,\B)=\Rec^{\mathrm{init}}(\Sigma,\B)$. 
\end{proof}

\label{page:convention-sem-wta}
  \begin{quote}
    \emph{Due to Corollary \ref{cor:semiring-run=init}, if $\B$ is a semiring and $\cA$ is a $(\Sigma,\B)$-wta, then we  write $\sem{\cA}$ instead of $\runsem{\cA}$ and $\initialsem{\cA}$.
    Moreover, for an i-recognizable or r-recognizable weighted tree language $r$,  we say that $r$ is recognizable. Also, we denote the set $\Rec^{\mathrm{run}}(\Sigma,\B)$ (and hence, $\Rec^{\mathrm{init}}(\Sigma,\B)$) by $\Rec(\Sigma,\B)$. }
\end{quote}
\index{Rec@$\Rec(\Sigma,\B)$}
\index{semantic@$\sem{\cA}$}

With respect to the complexity of calculating the semantics $\sem{\cA}$ of some $(\Sigma,\B)$-wta $\cA$ where $\B$ is a semiring, Theorem \ref{lm:compl-run-initial} implies the following: in general, it is more efficient to calculate $\sem{\cA}$ according to Algorithm \ref{alg:computation-initial-semantics} (initial algebra semantics) than to calculate it according to  Algorithm \ref{alg:computation-run-semantics} (run semantics).

In Section \ref{sect:string-automata} we have recalled the definition of wsa and the definitions of their run semantics and initial algebra semantics. Moreover, we have shown that wsa can be considered as particular wta (cf. Lemma \ref{lm:wsa-wta-over-string-ra-related}). An easy consequence of the previous results is that the two semantics of each wsa over right-distributive strong bimonoid are equal.

\begin{corollary-rect} \rm \label{cor:wsa-semiring-equal} (cf. \cite[Lm.~4]{drostuvog10}) Let $\Gamma$ be an alphabet and $\B$ be a right-distributive strong bimonoid. For each $(\Gamma,\B)$-wsa $\cA$ we have $\runsem{\cA} = \initialsem{\cA}$.
\end{corollary-rect}
\begin{proof} Let $\cA$ be a $(\Gamma,\B)$-wsa. Moreover, let $e \not\in \Gamma$. By Lemma \ref{lm:wsa-wta-over-string-ra-related}, we can construct  a $(\Gamma_e,\B)$-wta $\cB$ such that $\runsem{\cA} = \runsem{\cB} \circ \tree_e$ and $\initialsem{\cA} = \initialsem{\cB} \circ \tree_e$. We note that $\Gamma \ne \emptyset$ and hence $\Gamma_e$ is not trivial. Then, by Theorem \ref{thm:semiring-run=initial} and again by Lemma \ref{lm:wsa-wta-over-string-ra-related}, we obtain that $\runsem{\cA} = \initialsem{\cA}$.
  \end{proof}
  
\label{page:convention-sem-wsa}
  \begin{quote}
    \emph{Due to Corollary \ref{cor:wsa-semiring-equal}, if $\B$ is a semiring and $\cA$ is a $(\Gamma,\B)$-wsa, then we  write $\sem{\cA}$ instead of $\runsem{\cA}$ and $\initialsem{\cA}$.}
\end{quote}

In fact, Observation \ref{obs:init=run-trivial-ra} is a direct consequence of Theorem \ref{thm:semiring-run=initial}.

\section{Fatou extension}
\label{sec:Fatou-extension}

\index{Fatou extension}
Let $\B$ and $\sfC$ be two semirings such that $\sfC$ is an extension of $\B$, i.e., $\B$ is a subalgebra of $\sfC$. By Observation~\ref{obs:extension-of-weight-structure-inverse}, for each $(\Sigma,\sfC)$-wta $\cA$ with $\mathrm{wts}(\cA) \subseteq B$, we have that $\sem{\cA} \in \Rec(\Sigma,\B)$. We can replace the condition that $\mathrm{wts}(\cA) \subseteq B$ by the weaker condition that $\sem{\cA}: \T_\Sigma \to B$. Then we might ask: is $\sem{\cA} \in \Rec(\Sigma,\B)$ true? In other words,
does the inclusion
\begin{equation}\label{eq:Fatou}
  \Rec(\Sigma,\sfC) \cap B^{\T_\Sigma}     \subseteq \Rec(\Sigma,\B) 
\end{equation}
hold?
If  \eqref{eq:Fatou} holds, then we call $\sfC$ a \emph{Fatou extension of $\B$} (cf. \cite[Ch.~V]{berreu88}).

For instance, for string ranked alphabets, the ring $\Int$ of integers is not a Fatou extension of the semiring $\Nat$ of natural numbers \cite[Ex.~8.1]{kuisal86} (also cf. Lemma \ref{lm:Z-not-Fatou-N}). In contrast, for the case that $\Sigma$ is a string ranked alphabet, the semiring $\Ratnum_{\ge 0}$ is a Fatou extension of the semiring $\Nat$ of natural numbers \cite{fli75} (also cf. \cite[Thm.~3.3]{berreu88}); moreover, for every two fields $\B$ and $\sfC$, if $\sfC$ is an extension of $\B$, then $\sfC$ is a Fatou extension of $\B$ \cite{fli74} (also cf. \cite[Thm.~3.1]{berreu88}).

%% file: support-lang.tex
\chapter{Equality of run support and initial algebra support}
\label{ch:run-support=initial-algebra-support}

\index{run support}
\index{initial algebra support}
In this chapter we consider, for each $(\Sigma,\B)$-wta $\cA$, the $\Sigma$-tree languages $\supp(\runsem{\cA})$ and $\supp(\initialsem{\cA})$, which we call \emph{run support of $\cA$} and \emph{initial algebra support of $\cA$}, respectively. More specifically, we report on the characterization of the statement:
\begin{equation}\label{eq:equality-of-supports}
  \text{for each $(\Sigma,\B)$-wta $\cA$, we have } \ \supp(\runsem{\cA}) = \supp(\initialsem{\cA}) \tag{$\mathrm{Supp}$}
\end{equation}
in terms of properties of $\Sigma$ and of the underlying strong bimonoid $\B$ (cf.~Theorem\ref{thm:bi-strongly-zsf-equiv-equ-supp}). This characterization was proved in \cite{drovog24}. It can be  compared to the characterization of 
the statement:
\begin{equation}\label{eq:equality-of-semantics}
  \text{for each $(\Sigma,\B)$-wta $\cA$, we have } \ \runsem{\cA} = \initialsem{\cA} \tag{$\mathrm{Equ}$}
\end{equation}
which we gave in Theorem~\ref{thm:semiring-run=initial}.

\section{The main result}
\label{sec:the-main-result}

In our characterization result of \eqref{eq:equality-of-supports},  we consider only zero-sum free strong bimonoids. 
For this case Theorem~\ref{thm:semiring-run=initial}(A)$  \Leftrightarrow$(B) says
\begin{align*}
(\forall \B \text{ zero-sum free}):  \  \Big( &  \big(\Sigma \text{ not trivial} \Rightarrow \B \text{ right-distributive}\big) \ \wedge \\
 & \big(\Sigma \text{ branching} \Rightarrow \B \text{ distributive}\big) \Big) \ \ \Leftrightarrow \ \   \mathrm{(Equ)} \enspace.
\end{align*}

Obviously, \eqref{eq:equality-of-semantics} implies \eqref{eq:equality-of-supports}, hence \eqref{eq:equality-of-supports} is a weakening of \eqref{eq:equality-of-semantics}.
In our characterization of \eqref{eq:equality-of-supports},  this will correspond to a weakening of the conditions  ``right-distributive and zero-sum free'' and ``distributive and zero-sum free''. Next we introduce such a weakening for the above conditions.

Let  $\B =(B,\oplus,\otimes,\0,\1)$  be a strong bimonoid.
We say that  $B$  is
\begin{compactitem}
\item \emph{strongly zero-sum free}, if for every $a, b, c \in B$  we have:
  $(a \oplus b) \otimes c = \0$  iff  $a \otimes c = b \otimes c = \0$.
  \item \emph{bi-strongly zero-sum free}, if for every  $a, b, b', c \in B$  we have:
  $a \otimes (b\oplus b') \otimes c = 0$  iff  $a \otimes b \otimes c = a \otimes b' \otimes c = \0$.
\end{compactitem}
It is easy to see that,  for each strong bimonoid $\B$, we have
\begin{compactitem}
\item if $\B$ is right-distributive and  zero-sum free, then $\B$ is strongly zero-sum free and
  \item if $\B$ is left-distributive and strongly zero-sum free, then $\B$ is bi-strongly zero-sum free.
  \end{compactitem}

 We recall that $\B$ is zero-sum free if, for every  $a, b \in B$  we have:  $a \oplus b = \0$  iff  $a = b = \0$. It is zero-divisor free if, for every  $a, b \in B$  we have:  $a \otimes b = \0$  iff  $a = \0$ or $b = \0$. Moreover, $\B$ is positive if $\B$ is zero-sum free and zero-divisor free.
The following implications hold for each strong bimonoid $\B$:
         \begin{center}
$\B$ positive \ $\Rightarrow$ \ $\B$ bi-strongly zero-sum free \ $\Rightarrow$ \ $\B$ strongly zero-sum free \ $\Rightarrow$ \ $\B$ zero-sum free.
\end{center}
We refer to \cite[Sect.~3]{drovog24} for examples of strong bimonoids which witness that none of the above implications holds in the reverse direction.

The characterization of Property \eqref{eq:equality-of-supports} is the following theorem.

\begin{theorem-rect} \label{thm:bi-strongly-zsf-equiv-equ-supp} {\rm \cite[Thm.~5.9]{drovog24}} Let  $\Sigma$  be a ranked alphabet and  $\B$  be a zero-sum free strong bimonoid.
    Then the following two statements are equivalent.
    \begin{compactenum}
    \item[(A)] If $\Sigma$ is not trivial, then  $\B$ is strongly zero-sum free, and \\
       if $\Sigma$ is branching, then $\B$ is bi-strongly zero-sum free.
     \item[(B)] For each $(\Sigma,\B)$-weighted tree automaton $\cA$, we have $\supp(\initialsem{\cA}) = \supp(\runsem{\cA})$.
     \end{compactenum}
     \end{theorem-rect}

     Theorem \ref{thm:bi-strongly-zsf-equiv-equ-supp} follows directly from Lemmas \ref{lm:wta-run-implies-init} and \ref{lm:wta-init-implies-run}.
      Theorem \ref{thm:bi-strongly-zsf-equiv-equ-supp}(A)$\Rightarrow$(B) generalizes \cite[Thm.~4.1]{ghovog23}, where, instead of the condition (A), it is  required that $\B$ is positive, i.e., zero-sum free and zero-divisor free.

    \section{The lemmas and their proofs}
    \label{sect:the-lemmas-and-their-proofs}

       \begin{lemma}\rm \label{lm:wta-run-implies-init}  Let  $\B$ be zero-sum free. Then the following two statements are equivalent.
      \begin{compactenum}
      \item[(A)] If $\Sigma$ is not trivial, then
        $b \otimes c \neq \0$ implies $(b \oplus b') \otimes c \neq \0$ for every $b,b',c \in B$, and\\
        if $\Sigma$ is branching, then  $a \otimes b \otimes c \neq \0$ implies $a \otimes (b \oplus b') \otimes c \neq \0$ for every $a,b,b',c \in B$.
        \item[(B)] For each $(\Sigma,\B)$-wta $\cA$, we have $\supp(\runsem{\cA}) \subseteq \supp(\initialsem{\cA})$.
          \end{compactenum}
        \end{lemma}

Before showing the formal proof (cf. page \pageref{p:proof-of-lemma-run-ne-0-implies-initial-ne-0}), we will explain the proof idea for the implication (A)$\Rightarrow$(B) of Lemma \ref{lm:wta-run-implies-init}, i.e., that under certain conditions on~$\B$, we have $\supp(\runsem{\cA}) \subseteq \supp(\initialsem{\cA})$ for each $(\Sigma,\B)$-wta $\cA$. 

  Let $\cA$ be a $(\Sigma,\B)$-wta and $\xi \in \supp(\runsem{\cA})$. Thus there exists a run $\rho_0 \in \R_\cA(\xi)$ such that $\wt(\xi,\rho_0) \otimes F_{\rho_0(\varepsilon)} \not= \0$. Here we consider the particular tree $\xi$ and run $\rho_0$ as shown in Figure~\ref{fig:0}; thus, in particular, $\Sigma$ is not trivial and branching.

\begin{figure}
    \centering
 \begin{tikzpicture}[scale=0.8, every node/.style={transform shape},
					node distance=0.05cm and 0.05cm,
					mycircle/.style={draw, circle, inner sep=0mm, minimum height=5.5mm},
					mydashed/.style={dash pattern=on 2mm off 1mm, thin},
					mydashedarrow/.style={mydashed,->, shorten >=0.1cm,shorten <=0.1cm}],

                                        \begin{scope}[level 1/.style={sibling distance=50mm},
			  level 2/.style={sibling distance=20mm},
			  level 3/.style={sibling distance=18mm},
			  level 4/.style={sibling distance=16mm}]
			  
  \node (n0) at (0,0) {$\eta$}
  child {node (n1) {$\eta$}
    child {node (n11) {$\alpha$}}
      child {node[inner sep=2.5mm] (n12) {$\sigma$}
        child {node (n121) {$\alpha$}}
        child {node (n122) {$\beta$}
        }}
    child {node (n13) {$\alpha$}}}
    child {node[inner sep=2.5mm] (n2) {$\sigma$}
      child {node (n21) {$\sigma$}
        child {node (n211) {$\beta$}}
        child {node (n212) {$\beta$}}}
      child {node (n22) {$\alpha$}} }
    child {node (n3) {$\alpha$}};
  
  \node [mycircle, right = 0.2cm of n0, yshift=0.3cm] (cn0) {$q_{13}$};
  \node [mycircle, right = 0.1cm of n1] (cn1) {$q_6$};
  \node [mycircle, right = 0.0cm of n11] (cn11) {$q_1$};
  \node [mycircle, right = 0.0cm of n12] (cn12) {$q_4$};
  \node [mycircle, right = 0.0cm of n121] (cn121) {$q_2$};
  \node [mycircle, right = 0.0cm of n122] (cn122) {$q_3$};
  \node [mycircle, right = 0.0cm of n13] (cn13) {$q_5$};
  \node [mycircle, right = 0.0cm of n2] (cn2) {$q_{11}$};
  \node [mycircle, right = 0.0cm of n21] (cn21) {$q_9$};
    \node [mycircle, right = 0.0cm of n211] (cn211) {$q_7$};
  \node [mycircle, right = 0.0cm of n212] (cn212) {$q_8$};
  \node [mycircle, right = 0.0cm of n22] (cn22) {$q_{10}$};
  \node [mycircle, right = 0.0cm of n3] (cn3) {$q_{12}$};

  \node[left= 1.4cm of n0, yshift=0.45cm] {$\xi :$};
  \node[right= 1.4cm of cn0, yshift=0.15cm] {$\rho_0 \in \R_{\cA}(\xi)$};
\end{scope}

\end{tikzpicture}   \caption{\label{fig:0} Tree $\xi$ with run $\rho_0 \in \R_\cA(\xi)$.}
\end{figure}

By definition, the value $\wt(\xi,\rho_0)$ is a product with 13 factors where each factor is a weight of a transition; thus
{\small
  \begin{equation}\label{equ:comp-1}
 \wt(\xi,\rho_0) \otimes F_{q_{13}} = \delta_0(\varepsilon,\alpha,q_1) \otimes \delta_0(\varepsilon,\alpha,q_2) \otimes \delta_0(\varepsilon,\beta,q_3) \otimes \delta_2(q_2q_3,\sigma,q_4)
  \otimes   \delta_0(\varepsilon,\alpha,q_5)
  \otimes \cdots \otimes \delta_3(q_6q_{11}q_{12},\eta,q_{13})
  \otimes F_{q_{13}}  
\end{equation}
}
Again by definition, the transition weights at nullary symbols can be expressed as images of $\h_\cA$, for instance,
\[
  \delta_0(\varepsilon,\alpha,q_1) = \h_\cA(\alpha)_{q_1}, \ \
  \delta_0(\varepsilon,\alpha,q_2) = \h_\cA(\beta)_{q_2}, \ \
  \delta_0(\varepsilon,\beta,q_3) = \h_\cA(\beta)_{q_3},  \text{ and } 
  \delta_0(\varepsilon,\alpha,q_5) = \h_\cA(\alpha)_{q_5} \enspace.
 \]
 Thus \eqref{equ:comp-1} can  be written as
 {\small
   \begin{equation}\label{equ:comp-2}
     \wt(\xi,\rho_0) \otimes  F_{q_{13}} =
     \h_\cA(\alpha)_{q_1} \otimes
   \h_\cA(\alpha)_{q_2} \otimes \h_\cA(\beta)_{q_3} \otimes \delta_2(q_2q_3,\sigma,q_4) \otimes
   \h_\cA(\alpha)_{q_5} \otimes
   \cdots \otimes \delta_3(q_6q_{11}q_{12},\eta,q_{13})
    \otimes F_{q_{13}} \enspace.
\end{equation}
}

Now we replace
\[
  \text{ the subproduct} \ \ \h_\cA(\alpha)_{q_2} \otimes \h_\cA(\beta)_{q_3} \otimes \delta_2(q_2q_3,\sigma,q_4) \ \ \text{ of \eqref{equ:comp-2}}
\]
by
\[
  \text{the sum } \ \ 
\h_\cA(\sigma(\alpha,\beta))_{q_4} = \bigoplus_{p_1p_2 \in Q^2} \h_\cA(\alpha)_{p_1} \otimes \h_\cA(\beta)_{p_2} \otimes \delta_2(p_1p_2,\sigma,q_4)
\]
and obtain the product
\begin{equation}\label{equ:comp-3}
   d = \h_\cA(\alpha)_{q_1} \otimes
   \h_\cA(\sigma(\alpha,\beta))_{q_4} \otimes
   \h_\cA(\alpha)_{q_5} \otimes
   \ldots \otimes \delta_3(q_6q_{11}q_{12},\eta,q_{13})
    \otimes F_{q_{13}} \enspace.
 \end{equation}
 Of course, $\wt(\xi,\rho_0)\otimes F_{q_{13}}$ and $d$ need not be equal. But the fact that  $\Sigma$ is branching implies, by (A), that $\B$ is bi-strongly zero-sum free, i.e., $a \otimes b \otimes c \not= \0$ implies $a \otimes (b \oplus b') \otimes c \not=\0$ for every $a,b,b',c \in B$. Since the subproduct is one of the summands of the sum (viz., the summand for $p_1p_2=q_2q_3$), we obtain that $d \not= \0$, where we instantiate $a,b,b',c$ as follows:

 {\small
\[ \wt(\xi,\rho_0) \otimes F_{q_{13}} \ = \ \underbrace{\h_\cA(\alpha)_{q_1}}_{a} \otimes
   \underbrace{\h_\cA(\alpha)_{q_2} \otimes \h_\cA(\beta)_{q_3} \otimes \delta_2(q_2q_3,\sigma,q_4)}_{b} \otimes
   \underbrace{\h_\cA(\alpha)_{q_5} \otimes
   \ldots \otimes \delta_3(q_6q_{11}q_{12},\eta,q_{13})
   \otimes F_{q_{13}}}_{c}
\]

\[d \ = \ \underbrace{\h_\cA(\alpha)_{q_1}}_{a} \otimes
  \underbrace{\bigoplus_{p_1p_2 \in Q^2} \h_\cA(\alpha)_{p_1} \otimes \h_\cA(\beta)_{p_2} \otimes \delta_2(p_1p_2,\sigma,q_4) }_{b \oplus b'}
  \otimes
   \underbrace{\h_\cA(\alpha)_{q_5} \otimes
   \ldots \otimes \delta_3(q_6q_{11}q_{12},\eta,q_{13})
   \otimes F_{q_{13}}}_{c}
\]

\[
b' = \bigoplus_{p_1p_2 \in Q^2 \setminus \{q_2q_3\}} \h_\cA(\alpha)_{p_1} \otimes \h_\cA(\beta)_{p_2} \otimes \delta_2(p_1p_2,\sigma,q_4) \enspace.
  \]

   }

 Next we can apply the same transformation to 
\[
  \text{ the subproduct} \ \ \h_\cA(\beta)_{q_7} \otimes \h_\cA(\beta)_{q_8} \otimes \delta_2(q_7q_8,\sigma,q_9) \ \ \text{ of \eqref{equ:comp-3}}
\]
and replace it by
\[
  \text{the sum } \ \ 
\bigoplus_{p_1p_2 \in Q^2} \h_\cA(\beta)_{p_1} \otimes \h_\cA(\beta)_{p_2} \otimes \delta_2(p_1p_2,\sigma,q_9) \enspace.
\]
For the resulting product $d'$ we again obtain that $d' \not= \0$.

In this way we keep on to replace each subproduct of the form
\[
  \h_\cA(\xi|_{w1})_{\rho_0(w1)} \otimes \cdots \otimes  \h_\cA(\xi|_{wk})_{\rho_0(wk)} \otimes \delta_k(\rho_0(w1)\cdots \rho_0(wk),\xi(w),\rho_0(w)) 
\]
with $w \in \pos(\xi)$ by the sum 
\[
\h_\cA(\xi|_w)_{\rho_0(w)} = \bigoplus_{p_1\cdots p_k \in Q^k} \h_\cA(\xi|_{w1})_{p_1} \otimes \cdots \otimes \h_\cA(\xi|_{wk})_{p_k} \otimes \delta_k(p_1\cdots p_k,\xi(w),\rho_0(w)) \enspace,
\]
maintaining the invariant that the new product is not equal to $\0$.

As final step, we replace the subproduct
\[
\h_\cA(\xi|_1)_{q_6} \otimes \h_\cA(\xi|_2)_{q_{11}} \otimes \h_\cA(\xi|_3)_{q_{12}} \otimes \delta_3(q_6q_{11}q_{12},\eta,q_{13})
\]
by the sum
\[
  \h_\cA(\xi)_{q_{13}} = \bigoplus_{p_1p_2p_3 \in Q^3} \h_\cA(\xi|_1)_{p_1} \otimes
    \h_\cA(\xi|_2)_{p_2} \otimes \h_\cA(\xi|_3)_{p_3} \otimes  \delta_3(p_1p_2p_3,\eta,q_{13}) \enspace.
  \]
And we obtain that $\h_\cA(\xi)_{q_{13}} \otimes F_{q_{13}} \not= \0$. Thus $\xi \in \supp(\initialsem{\cA})$.

In order to formalize the current form of the product and the change from one form into another, we use the concept of cut through $\xi$ and define a terminating binary relation $\succ$ on the set of all cuts through~$\xi$.

Now let $\xi\in \T_\Sigma$ be an arbitrary tree. A \emph{cut through $\xi$} is a sequence $(w_1,\ldots,w_n)$ such that
    \begin{compactitem}
    \item $n \in \mathbb{N}_+$ and $w_i \in \pos(\xi)$, for each $i \in [n]$,
    \item (independent) for every $i,j \in [n]$ with $i\ne j$, we have $w_i \not\in \prefix(w_j)$, 
    \item (ordered) $w_1 <_{\mathrm{lex}}\ldots <_{\mathrm{lex}}w_n$, and
    \item (complete) for each $w \in \pos_{\Sigma^{(0)}}(\xi)$ there exists $i \in [n]$ such that $w_i \in \prefix(w)$.
    \end{compactitem}
    We denote by $\Cut(\xi)$ the set of all cuts through $\xi$. Clearly, the set $\Cut(\xi)$ is finite and $(\varepsilon) \in \Cut(\xi)$. The \emph{leaves-cut through $\xi$}, denoted by $\lcut(\xi)$, is the cut $(w_1,\ldots,w_n)$ where $n = |\pos_{\Sigma^{(0)}}(\xi)|$. If $\Sigma$ is trivial, i.e., $\Sigma=\Sigma^{(0)}$, then $\lcut(\xi)=(\varepsilon)$ and  $\Cut(\xi) = \{\lcut(\xi)\}$ for each $\xi\in\T_\Sigma$.

Each cut $\kappa \in \Cut(\xi)$ represents a product. 
For instance, the cut $\kappa_1$ shown in Figure~\ref{fig:1}, represents the product 
  \begin{align*}
    &\h_\cA(\xi|_{11})_{q_1} \otimes \h_\cA(\xi|_{12})_{q_4} \otimes \h_\cA(\xi|_{13})_{q_5} \otimes
    \delta_2(q_1q_4q_5,\eta,q_6)\\
    & \ \ \  \otimes \h_\cA(\xi|_{21})_{q_9} \otimes \h_\cA(\xi|_{22})_{q_{10}}
      \otimes \delta_2(q_9q_{10},\sigma,q_{11}) \\
    &\otimes \ \h_\cA(\xi|_3)_{q_{12}} \otimes \delta_3(q_6q_{11}q_{12},\eta,q_{13}) \otimes F_{q_{13}}
  \end{align*}
 and the cut $\kappa_2$ shown in Figure~\ref{fig:2}, represents the product 
  \begin{align*}
    &\h_\cA(\xi|_{11})_{q_1} \otimes \h_\cA(\xi|_{12})_{q_4} \otimes \h_\cA(\xi|_{13})_{q_5} \otimes
    \delta_2(q_1q_4q_5,\eta,q_6)\\
    & \ \ \  \otimes \h_\cA(\xi|_{2})_{q_{11}}  \\
    &\otimes \ \h_\cA(\xi|_3)_{q_{12}} \otimes \delta_3(q_6q_{11}q_{12},\eta,q_{13}) \otimes F_{q_{13}} \enspace.
  \end{align*}
  The cuts $\lcut(\xi)$ and  $(\varepsilon)$ represent the products \eqref{equ:comp-2} and $\h_\cA(\xi)_{q_{13}}$  (with one factor), respectively.
    
    Moreover, we define the binary relation $\succ$ on $\Cut(\xi)$ as follows. For every $\kappa_1,\kappa_2 \in \Cut(\xi)$, we let  $\kappa_1 \succ \kappa_2$ if there exist $n \in \mathbb{N}_+$,  $i \in [n]$,
    $w_i,\ldots,w_{i-1},w_i,w_{i+1},\ldots,w_n \in \pos(\xi)$  such that 
    \begin{compactitem}
    \item $\kappa_1 = (w_1,\ldots,w_{i-1},w_i1,\ldots,w_ik,w_{i+1},\ldots,w_n)$ where $k = \rk(\xi(w_i))$ and
      \item  \(\kappa_2 = (w_1,\ldots,w_{i-1},w_i,w_{i+1},\ldots,w_n)\) \enspace.
  \end{compactitem}
  Intuitively, $\kappa_2$ is obtained from $\kappa_1$ by reducing a sequence $w_i1,\ldots,w_ik$ of children of $w_i$ to $w_i$. The symbol $\succ$ is mnemonic in the following sense: if $\kappa_1 \succ \kappa_2$ for two cuts $\kappa_1,\kappa_2$, then the sequence $\kappa_1$ is \underline{longer} than the sequence $\kappa_2$ because $\kappa_1$ is placed at the side of $\succ$ which has a \underline{longer} vertical extension than the other side.  It is clear that $\succ$ is terminating and $\nf_\succ(\Cut(\xi)) = \{(\varepsilon)\}$.
For instance, for the two particular cuts $\kappa_1$ and $\kappa_2$ shown in Figure \ref{fig:1} and \ref{fig:2}, respectively, we have $\kappa_1 \succ \kappa_2$ .
  
 \begin{figure}
    \centering
 \begin{tikzpicture}[scale=0.8, every node/.style={transform shape},
					node distance=0.05cm and 0.05cm,
					mycircle/.style={draw, circle, inner sep=0mm, minimum height=5.5mm},
					mydashed/.style={dash pattern=on 2mm off 1mm, thin},
					mydashedarrow/.style={mydashed,->, shorten >=0.1cm,shorten <=0.1cm}],

                                        \begin{scope}[level 1/.style={sibling distance=50mm},
			  level 2/.style={sibling distance=20mm},
			  level 3/.style={sibling distance=18mm},
			  level 4/.style={sibling distance=16mm}]
			  
  \node (n0) at (0,0) {$\eta$}
  child {node (n1) {$\eta$}
    child {node (n11) {$\alpha$}}
      child {node[inner sep=2.5mm] (n12) {$\sigma$}
        child {node (n121) {$\alpha$}}
        child {node (n122) {$\beta$}
        }}
    child {node (n13) {$\alpha$}}}
    child {node[inner sep=2.5mm] (n2) {$\sigma$}
      child {node (n21) {$\sigma$}
        child {node (n121) {$\beta$}}
        child {node (n122) {$\beta$}}}
      child {node (n22) {$\alpha$}} }
    child {node (n3) {$\alpha$}};
  
  \node [mycircle, right = 0.2cm of n0, yshift=0.3cm] (cn0) {$q_{13}$};
  \node [mycircle, right = 0.1cm of n1] (cn1) {$q_6$};
  \node [mycircle, right = 0.0cm of n11] (cn11) {$q_1$};
  \node [mycircle, right = 0.0cm of n12] (cn12) {$q_4$};
  \node [mycircle, right = 0.0cm of n13] (cn13) {$q_5$};
  \node [mycircle, right = 0.0cm of n2] (cn2) {$q_{11}$};
  \node [mycircle, right = 0.0cm of n21] (cn21) {$q_9$};
  \node [mycircle, right = 0.0cm of n22] (cn22) {$q_{10}$};
  \node [mycircle, right = 0.0cm of n3] (cn3) {$q_{12}$};

  \begin{scope}[every path/.style= mydashed]
    \node [left = 1cm of n11] (leftofn11) {};
    \draw (leftofn11) -- (cn11.west);
    \draw (cn11.east) -- (cn12.west);
    \draw (cn12.east) -- (cn13.west);
    \draw (cn13.east) -- (cn21.west);
    \draw (cn21.east) -- (cn22.west);
    \draw (cn22.east) to[out=0, in=180] (cn3.west) -- (cn3.west);
    \node [right = 1cm of cn3] (rightofcn3) {};
        \draw (cn3.east) -- (rightofcn3)  node[above] {$\kappa_1$};
  \end{scope}
  
  \node[left= 1.4cm of n0, yshift=0.45cm] {$\xi :$};
  \node[right= 1.4cm of cn0, yshift=0.15cm] {$\rho_0 \in \R_{\cA}(\xi)$};
\end{scope}

\end{tikzpicture}   \caption{\label{fig:1} Tree $\xi$ with run $\rho_0$ and  cut $\kappa_1 = (11,12,13,\underline{21,22},3)$.}
\end{figure}

\begin{figure}
    \centering
 \begin{tikzpicture}[scale=0.8, every node/.style={transform shape},
					node distance=0.05cm and 0.05cm,
					mycircle/.style={draw, circle, inner sep=0mm, minimum height=5.5mm},
					mydashed/.style={dash pattern=on 2mm off 1mm, thin},
					mydashedarrow/.style={mydashed,->, shorten >=0.1cm,shorten <=0.1cm}],

                                        \begin{scope}[level 1/.style={sibling distance=50mm},
			  level 2/.style={sibling distance=20mm},
			  level 3/.style={sibling distance=18mm},
			  level 4/.style={sibling distance=16mm}]
			  
  \node (n0) at (0,0) {$\eta$}
  child {node (n1) {$\eta$}
    child {node (n11) {$\alpha$}}
      child {node[inner sep=2.5mm] (n12) {$\sigma$}
        child {node (n121) {$\alpha$}}
        child {node (n122) {$\beta$}
        }}
    child {node (n13) {$\alpha$}}}
    child {node[inner sep=2.5mm] (n2) {$\sigma$}
      child {node (n21) {$\sigma$}
        child {node (n121) {$\beta$}}
        child {node (n122) {$\beta$}}}
      child {node (n22) {$\alpha$}} }
    child {node (n3) {$\alpha$}};
  
  \node [mycircle, right = 0.2cm of n0, yshift=0.3cm] (cn0) {$q_{13}$};
  \node [mycircle, right = 0.1cm of n1] (cn1) {$q_6$};
  \node [mycircle, right = 0.0cm of n11] (cn11) {$q_1$};
  \node [mycircle, right = 0.0cm of n12] (cn12) {$q_4$};
  \node [mycircle, right = 0.0cm of n13] (cn13) {$q_5$};
  \node [mycircle, right = 0.0cm of n2] (cn2) {$q_{11}$};
  \node [mycircle, right = 0.0cm of n3] (cn3) {$q_{12}$};

  \begin{scope}[every path/.style= mydashed]
    \node [left = 1cm of n11] (leftofn11) {};
    \draw (leftofn11) -- (cn11.west);
    \draw (cn11.east) -- (cn12.west);
    \draw (cn12.east) -- (cn13.west);
    \draw (cn13.east)  to[out=0, in=180] (cn2.west) -- (cn2.west);
    \draw (cn2.east) to[out=0, in=180] (cn3.west) -- (cn3.west);
    \node [right = 1cm of cn3] (rightofcn3) {};
        \draw (cn3.east) -- (rightofcn3)  node[above] {$\kappa_2$};
  \end{scope}
  
  \node[left= 1.4cm of n0, yshift=0.45cm] {$\xi :$};
  \node[right= 1.4cm of cn0, yshift=0.15cm] {$\rho_0 \in \R_{\cA}(\xi)$};
\end{scope}

\end{tikzpicture}   \caption{\label{fig:2} Tree $\xi$ with run $\rho_0$ and  cut $\kappa_2 = (11,12,13,\underline{2},3)$.}
\end{figure}

We have formalized the transformation from $\kappa_1$ to $\kappa_2$  by the binary relation $\succ$. However, for the proof of Lemma~\ref{lm:wta-run-implies-init}, we have to swap its direction for the following reason. We want to prove this lemma by induction on some terminating reduction system $(\Cut(\xi),\rho)$ for which 
$\nf_{\rho}(\Cut(\xi)) = \{\lcut(\xi)\}$ because $\lcut(\xi)$ corresponds to the form of $\wt(\xi,\rho_0) \otimes F_{q_{13}}$ 
shown in \eqref{equ:comp-1}. Thus $\rho = \succ$ is not appropriate because $\nf_{\succ}(\Cut(\xi)) = \{(\varepsilon)\}$.
Therefore, we choose $\rho = \prec$ where  $\prec = \succ^{-1}$, i.e.,
$\prec$ is the inverse of $\succ$. Then  we have $\nf_{\prec}(\Cut(\xi))= \{\lcut(\xi)\}$ and, in the proof of Lemma \ref{lm:wta-run-implies-init}, we perform induction on $(\Cut(\xi),\prec)$.

In order to deal with the set of positions of $\xi$ which are above and on a cut $\kappa$, we truncate $\xi$ at the positions of the cut. Formally, let $\alpha$ be an arbitrary element of $\Sigma^{(0)}$. For each $\kappa =(w_1,\ldots,w_n)$ in $\Cut(\xi)$, we denote by $\xi[\kappa \leftarrow \alpha]$ the tree which is obtained from $\xi$ by replacing, for each $i \in [n]$, the subtree at position $w_i$ by $\alpha$. 
We note that $\{w_1,\ldots,w_n\} \subseteq \pos(\xi[\kappa \leftarrow \alpha])$. Moreover, $\pos(\xi[\kappa \leftarrow \alpha]) \subseteq \pos(\xi)$ and
\begin{eqnarray}\label{equ:order-of-factors}
      &\text{for each $w_1,w_2 \in \pos(\xi[\kappa \leftarrow \alpha])$, we have:}\notag\\
      &\text{$w_1$ occurs left of $w_2$ in the depth-first post-order on $\pos(\xi[\kappa \leftarrow \alpha])$ if and only if}\\
      &\text{$w_1$ occurs left of $w_2$ in the depth-first post-order on $\pos(\xi)$.\notag}
\end{eqnarray}
For instance, for $\xi$ and $\kappa_2$ as shown in Figure~\ref{fig:2}, we have that $\xi[\kappa_2 \leftarrow \alpha]= \eta(\eta(\alpha,\alpha,\alpha),\alpha,\alpha)$.
Equivalence~\ref{equ:order-of-factors} guarantees that the order of the factors after the replacement is the same as before; this is important because $\B$ need not be commutative.

\label{p:proof-of-lemma-run-ne-0-implies-initial-ne-0}
\begin{proof} (Proof of Lemma \ref{lm:wta-run-implies-init})

  (A)$\Rightarrow$(B): If $\Sigma$ is trivial, then for each $(\Sigma,\B)$-wta $\cA$, we have $\runsem{\cA} = \initialsem{\cA}$ by Observation~\ref{obs:init=run-trivial-ra}. This implies (B). Now let $\Sigma$ be not trivial.
          
          Let $\cA$ be a $(\Sigma,\B)$-wta and  $\xi \in \T_\Sigma$ such that $\xi \in \supp(\runsem{\cA})$.
  Hence $\bigoplus_{\rho \in \R_\cA(\xi)} \wt(\xi,\rho) \otimes F_{\rho(\varepsilon)} \not= \0$. Let  $\rho_0 \in \R_\cA(\xi)$ be such that $\wt(\xi,\rho_0) \otimes F_{\rho_0(\varepsilon)} \not= \0$. Such a $\rho_0$ exists, because $\B$ is zero-sum free.
  
  By Observation \ref{obs:weight-run-explicit},
  \begin{equation}\label{equ:wt-rho0-not-zero}
\wt(\xi,\rho_0)  \otimes F_{\rho_0(\varepsilon)} = \big(\bigotimes_{\substack{w \in \pos(\xi)\\\text{in $<_{\mathrm{dp}}$ order}}} \delta_{\rk(\xi(w))}(\rho_0(w1)\cdots \rho_0(w\,\rk(\xi(w)),\xi(w),\rho_0(w))\Big)  \otimes F_{\rho_0(\varepsilon)}
    \end{equation}

  By well-founded induction on $(\Cut(\xi),\prec)$, we prove the following statement: 
  \begin{eqnarray}\label{equ:wta-h-not-equal-0}
    &\text{For every $\kappa \in \Cut(\xi)$ with $\kappa=(w_1,\ldots,w_n)$ and $\alpha\in \Sigma^{(0)}$,}\notag\\
    &\text{we have:} \
      \Big(\bigotimes_{\substack{w \in \pos(\xi[\kappa \leftarrow \alpha])\\\text{in $<_{\mathrm{dp}}$ order}}} b_w\Big)   \otimes F_{\rho_0(\varepsilon)} \not= \0 \enspace,\\
       &\text{where $b_w = \begin{cases}\h_\cA(\xi|_w)_{\rho_0(w)} & \text{if $w \in \{w_1,\ldots,w_n\}$}\\
           \delta_{\rk(\xi(w))}(\rho_0(w1) \cdots \rho_0(w\, \rk(\xi(w))), \xi(w),\rho_0(w)) & \text{otherwise}
        \end{cases}$}\enspace.\notag
\end{eqnarray}

I.B.: We recall that $\nf_\prec(\xi) = \{\lcut(\xi)\}$. So, let $\kappa = \lcut(\xi)$, i.e., $\kappa$ is the leaves-cut through $\xi$.  By the definition of the set of positions of a tree, we have $\pos(\xi) = \pos(\xi[\lcut(\xi) \leftarrow \alpha])$. Moreover, for each $w \in \pos_{\Sigma^{(0)}}(\xi)$, we have $\h_\cA(\xi|_w)_{\rho_0(w)} = \delta_0(\varepsilon,\xi(w),\rho_0(w))$. Hence, by Equation \eqref{equ:wt-rho0-not-zero} and since $\wt(\xi,\rho_0) \otimes F_{\rho_0(\varepsilon)} \not= \0$, we obtain
\begin{align*}
  & \Big(\bigotimes_{\substack{w \in \pos(\xi[\lcut(\xi) \leftarrow \alpha]):\\\text{in $<_\mathrm{dp}$ order}}} \!\! b_w\Big) \otimes F_{\rho_0(\varepsilon)}\\
  = &\ \Big(\bigotimes_{\substack{w \in \pos(\xi):\\\text{in $<_\mathrm{dp}$ order}}} \!\! \h_\cA(\xi|_w)_{\rho_0(w)} \Big) \otimes F_{\rho_0(\varepsilon)}\\
  = &\ \Big(\bigotimes_{\substack{w \in \pos(\xi):\\\text{in $<_\mathrm{dp}$ order}}}  \delta_{\rk(\xi(w))}(\rho_0(w1)\cdots \rho_0(w\,\rk(\xi(w)),\xi(w),\rho_0(w))\Big) \otimes F_{\rho_0(\varepsilon)}\\
  = &\ \wt(\xi,\rho_0) \otimes F_{\rho_0(\varepsilon)} \not=\0 \enspace.
\end{align*}

  I.S.: Let $\kappa \in \Cut(\xi)\setminus \nf_\prec(\xi)$ with $\kappa = (w_1,\ldots,w_n)$.
  Since $\kappa \not\in  \nf_\prec(\xi)$, there exists $\kappa' \in \Cut(\xi)$ with $\kappa \prec \kappa'$. Hence there exists $i \in [n]$ such that
  \[
  \kappa' = (w_1,\ldots,w_{i-1},\underbrace{w_{i}1,\ldots,w_{i}\, \rk(\xi(w_i))}_{\text{successors of $w_i$}}, w_{i+1},\ldots,w_n)
  \]
  We abbreviate $\rk(\xi(w_i))$ by $k$. 
  We assume that  $\Big(\bigotimes_{\substack{w \in \pos(\xi[\kappa' \leftarrow \alpha])\\\text{in $<_\mathrm{dp}$ order}}} b_w\Big) \otimes F_{\rho_0(\varepsilon)} \not= \0$ (I.H.).
Then there exist $a,c \in B$ such that
  \begin{align*}
    & \Big(\bigotimes_{\substack{w \in \pos(\xi[\kappa' \leftarrow \alpha])\\\text{in $<_\mathrm{dp}$ order}}} b_w\Big) \otimes F_{\rho_0(\varepsilon)}\\
    &= a \otimes \Big(\bigotimes_{j \in [k]} b_{w_ij}\Big)
    \otimes b_{w_i} \otimes c \otimes F_{\rho_0(\varepsilon)}\tag{because $w_{i}1 <_\mathrm{dp} \ldots <_\mathrm{dp} w_{i}k$ and  $(\forall j\in[k-1]) :\{v \in \xi[\kappa' \leftarrow \alpha] \mid w_ij <_{\mathrm{dp}} v <_{\mathrm{dp}} w(j+1)\}=\emptyset$}\\
    &= a \otimes \Big(\bigotimes_{j \in [k]} \h_\cA(\xi|_{w_ij})_{\rho_0(w_ij)}\Big)
    \otimes \delta_{k}(\rho_0(w_i1) \cdots \rho_0(w_i\, k), \xi(w_i),\rho_0(w_i))
      \otimes c \otimes F_{\rho_0(\varepsilon)}\enspace. \tag{by definition of $b_{w_ij}$ and $b_{w_i}$; note that each $w_ij$ is on the cut $\kappa'$}
  \end{align*}
  
  If $\Sigma$ is monadic, then $n=k=1$ and $a=\1$.

  By I.H., we know that this quantity is not equal to $\0$.  Then we can compute as follows.
    \begingroup
  \allowdisplaybreaks
  \begin{align*}
    &a \otimes \Big(\Big(\bigotimes_{j \in [k]} \h_\cA(\xi|_{w_ij})_{\rho_0(w_j)}\Big)
      \otimes \delta_{k}(\rho_0(w_i1) \cdots \rho_0(w_i\, k), \xi(w_i),\rho_0(w_i)) \Big)
      \otimes c  \otimes F_{\rho_0(\varepsilon)} \not=\0\\
    \Rightarrow \ \ \  & a \otimes \Big( \bigoplus_{q_1\cdots q_k \in Q^k} \Big(\bigotimes_{j  \in [k]} \h_\cA(\xi|_{w_ij})_{q_j}\Big)
      \otimes \delta_{k}(q_1 \cdots q_k, \xi(w_i),\rho_0(w_i)) \Big)
                         \otimes c  \otimes F_{\rho_0(\varepsilon)} \not=\0 \tag{by Statement (A) of the lemma}\\[2mm]
     \Leftrightarrow \ \ \  & a \otimes \h_\cA(\xi|_{w_i})_{\rho_0(w_i)} \otimes c  \otimes F_{\rho_0(\varepsilon)} \not=\0 \tag{by definition of $\h_\cA(\xi|_{w_i})_{\rho_0(w_i)}$}\\
    \Leftrightarrow \ \ \ & \Big(\bigotimes_{\substack{w \in \pos(\xi[\kappa \leftarrow \alpha])\\\text{in $<_\mathrm{dp}$ order}}} b_w\Big)  \otimes F_{\rho_0(\varepsilon)}\not= \0 \enspace.
  \end{align*}
  \endgroup

  This proves Statement \eqref{equ:wta-h-not-equal-0}. Then we obtain
  \begingroup
  \allowdisplaybreaks
  \begin{align*}
    \0 & \not= \Big(\bigotimes_{\substack{w \in \pos(\xi[(\varepsilon) \leftarrow \alpha])\\\text{in $<_{\mathrm{dp}}$ order}}} b_w\Big)  \otimes F_{\rho_0(\varepsilon)} \tag{by choosing $\kappa = (\varepsilon)$ in \eqref{equ:wta-h-not-equal-0}} \\
       &= \Big(\bigotimes_{\substack{w \in \pos(\alpha)\\\text{in $<_\mathrm{dp}$ order}}} b_w\Big)  \otimes F_{\rho_0(\varepsilon)}
    \tag{because $\xi[(\varepsilon) \leftarrow \alpha] = \alpha$}\\
      &= b_\varepsilon \otimes F_{\rho_0(\varepsilon)} \tag{because $\pos(\alpha) = \{\varepsilon\}$}\\
    & = \h_\cA(\xi)_{\rho_0(\varepsilon)}  \otimes F_{\rho_0(\varepsilon)} \tag{because $\xi|_\varepsilon = \xi$ and by definition of $b_\varepsilon$}\enspace.  
  \end{align*}
  \endgroup
  Since $\B$ is zero-sum free, we obtain
  \(\0 \not= \bigoplus_{q\in Q} \h_\cA(\xi)_q \otimes F_q = \initialsem{\cA}(\xi)\).
Hence $\xi \in \supp(\initialsem{\cA}(\xi))$.

\

(B)$\Rightarrow$(A): We assume that (B) holds. We have to prove:\\
(i) If $\Sigma$ is not trivial, then $b \otimes c \neq \0$ implies $(b \oplus b') \otimes c \neq \0$ for every $b,b',c \in B$, and\\
(ii) if $\Sigma$ is branching, then  $a \otimes b \otimes c \neq \0$ implies $a \otimes (b \oplus b') \otimes c \neq \0$ for every $a,b,b',c \in B$.

To prove (i), we assume that $\Sigma$ is not trivial. Since here the rank mapping will be important, we also show it explicitly and assume that $(\Sigma,\rk)$ is not trivial. Then there exist $\gamma \in \Sigma$ and $k \in \mathbb{N}$ with $k \ge 1$ and $\rk(\gamma)=k$. 
Let $b,b',c \in B$.

 First, we define the ranked alphabet $(\Sigma,\rk')$ such that, for each $\eta \in \Sigma$, we let
 \[
    \rk'(\eta) =
    \begin{cases}
      1 & \text{ if $\eta=\gamma$} \\
      \rk(\eta) & \text{ if otherwise.}
    \end{cases}
  \]
Thus, $\rk'(\gamma)=1$ and $(\Sigma,\rk)$ is an extension of $(\Sigma,\rk')$.

Second, since $(\Sigma,\rk')$ contains a unary symbol, i.e., $(\rk')^{-1}(1) \not= \emptyset$, by Lemma~\ref{lm:Ar-right-distributivity}, we can construct a $((\Sigma,\rk'),\B)$-wta $\cA_r$ such that
\begin{equation}\label{equ:mon-image-run-init}
\im(\runsem{\cA_r}) = \{\0, (b \otimes c) \oplus (b' \otimes c)\} \ \ \text{  and } \ \ 
\im(\initialsem{\cA_r}) = \{\0, (b \oplus b') \otimes c\} \enspace.
\end{equation}

Third, by using the padding lemma (i.e., Lemma~\ref{lm:super-big-padding}), we can construct a $((\Sigma,\rk),\B)$-wta $\cB$ such that, for each $\xi \in \T_{(\Sigma,\rk')}$, we have
\begin{equation}\label{equ:application-padding-lemma-mon-supp-codomain-run-init}
\runsem{\cB}(g(\xi)) = \runsem{\cA_r}(\xi)
\ \ \text{ and } \ \ 
\initialsem{\cB}(g(\xi))) = \initialsem{\cA_r}(\xi)  \enspace, 
\end{equation}
where $g$ is the $((\Sigma,\rk'),(\Sigma,\rk))$-tree homomorphism defined in Lemma~\ref{lm:super-big-padding}.

Now assume that $b \otimes c \not= \0$. Then the following implications hold.
\begingroup
\allowdisplaybreaks
\begin{align*}
& b \otimes c \not= \0 \\
  \Rightarrow \ \ & (b \otimes c) \oplus (b' \otimes c) \not= \0
                \tag{ because $\B$ is zero-sum free}\\
  \Rightarrow \ \ & (\exists \xi \in \T_{(\Sigma,\rk')}): \ \runsem{\cA_r}(\xi)  \not= \0
                    \tag{ by \eqref{equ:mon-image-run-init}}\\
  \Rightarrow \ \ & (\exists \xi \in \T_{(\Sigma,\rk')}): \ \runsem{\cB}(g(\xi)) \not= \0
                    \tag{ by \eqref{equ:application-padding-lemma-mon-supp-codomain-run-init}}\\
  \Rightarrow \ \ & (\exists \xi \in \T_{(\Sigma,\rk')}): \ \initialsem{\cB}(g(\xi))  \not= \0
                    \tag{ by (B): $\supp(\runsem{\cB}) \subseteq \supp(\initialsem{\cB})$}\\
  \Rightarrow \ \ & (\exists \xi \in \T_{(\Sigma,\rk')}): \ \initialsem{\cA_r}(\xi) \not= \0
                    \tag{ by \eqref{equ:application-padding-lemma-mon-supp-codomain-run-init}}\\
  \Rightarrow \ \ & (b \oplus b') \otimes c \not= \0
                    \tag{ by \eqref{equ:mon-image-run-init}} \enspace.
\end{align*}
\endgroup
This proves (i).

\

To prove (ii), we assume that $\Sigma$ is branching.
Again we make the rank mapping visible. By our assumption, there exist $\sigma \in \Sigma$ and $k \in \mathbb{N}$ with $k \ge 2$ and  $\rk(\sigma)=k$.
 Let $a,b,b',c \in B$.

 First, we define the ranked alphabet $(\Sigma,\rk')$ such that for each $\eta \in \Sigma$ we let
 \[
    \rk'(\eta) =
    \begin{cases}
      2 & \text{ if $\eta=\sigma$} \\
      \rk(\eta) & \text{ if otherwise.}
    \end{cases}
  \]
Thus, $\rk'(\sigma)=2$ and $(\Sigma,\rk)$ is an extension of $(\Sigma,\rk')$.

Second, since $(\Sigma,\rk')$ contains a binary symbol, i.e., $(\rk')^{-1}(2) \not= \emptyset$, by Lemma~\ref{lm:wta-Abr}, we can construct a $((\Sigma,\rk'),\B)$-wta $\cA_l$ such that
\begin{equation}\label{equ:br-image-run-init}
  \im(\runsem{\cA_l}) = \{\0, (a \otimes b \otimes c) \oplus (a \otimes b' \otimes c)\} \ \text{ and } \  
  \im(\initialsem{\cA_l}) =  \{\0, a \otimes (b \oplus b') \otimes c\} \enspace.
  \end{equation}

  Third, by using the padding lemma (i.e., Lemma~\ref{lm:super-big-padding}) we can construct a $((\Sigma,\rk),\B)$-wta $\cB$ such that, for each $\xi \in \T_{(\Sigma,\rk')}$, we have
\begin{equation}\label{equ:application-padding-lemma-br-supp-codomain}
\runsem{\cB}(g(\xi)) = \runsem{\cA_l}(\xi)
\ \ \text{ and } \ \ 
\initialsem{\cB}(g(\xi))) = \initialsem{\cA_l}(\xi)  \enspace. 
\end{equation}

 Now assume that $a \otimes b \otimes c \not= \0$.  Then the following implications hold.
\begingroup
\allowdisplaybreaks
\begin{align*}
& a \otimes b \otimes c \not= \0 \\
  \Rightarrow \ \ & (a \otimes b \otimes c) \oplus (a \otimes b' \otimes c) \not= \0
                \tag{ because $\B$ is zero-sum free}\\
  \Rightarrow \ \ & (\exists \xi \in \T_{(\Sigma,\rk')}): \ \runsem{\cA_l}(\xi)  \not= \0
                    \tag{ by \eqref{equ:br-image-run-init}}\\
  \Rightarrow \ \ & (\exists \xi \in \T_{(\Sigma,\rk')}): \ \runsem{\cB}(g(\xi)) \not= \0
                    \tag{ by \eqref{equ:application-padding-lemma-br-supp-codomain}}\\
  \Rightarrow \ \ & (\exists \xi \in \T_{(\Sigma,\rk')}): \ \initialsem{\cB}(g(\xi))  \not= \0
                    \tag{ by (B): $\supp(\runsem{\cB}) \subseteq \supp(\initialsem{\cB})$}\\
  \Rightarrow \ \ & (\exists \xi \in \T_{(\Sigma,\rk')}): \ \initialsem{\cA_l}(\xi) \not= \0
                    \tag{ by \eqref{equ:application-padding-lemma-br-supp-codomain}}\\
  \Rightarrow \ \ & a \otimes (b \oplus b') \otimes c \not= \0
                    \tag{ by \eqref{equ:mon-image-run-init}} \enspace.
\end{align*}
\endgroup
This proves (ii).
\end{proof}

\

          \begin{lemma}\rm  \label{lm:wta-init-implies-run} Let $\B$  be zero-sum free.
    Then the following two statements are equivalent.
    \begin{compactenum}
    \item[(A)] If $\Sigma$ is not trivial, then  $(b \oplus b') \otimes c \neq \0$  implies $\big(b \otimes c \neq \0$ or $b' \otimes c \neq \0\big)$, for all $b,b',c \in B$, and \\
       if $\Sigma$ is branching, then $a \otimes (b \oplus b') \otimes c \neq \0$  implies $\big(a \otimes b \otimes c \neq \0$  or 
       $a \otimes b' \otimes c \neq \0\big)$, for all $a,b,b',c \in B$.
     \item[(B)] For each $(\Sigma,\B)$-wta  $\cA$, we have $\supp(\initialsem{\cA}) \subseteq \supp(\runsem{\cA})$.
         \end{compactenum}
       \end{lemma}

The proof of the implication (A)$\Rightarrow$(B) of Lemma \ref{lm:wta-init-implies-run} proceeds in a similar way, as the proof of the implication (A)$\Rightarrow$(B) of Lemma \ref{lm:wta-run-implies-init} except that we start from the cut $(\varepsilon)$ and proceed by induction towards the cut $\lcut(\xi)$. In each step, a sequence of new states of a run is created (one state for each direct successor of a position) such that the current product is $\not= \0$. 
For this we employ the binary relation $\succ$ on $\Cut(\xi)$ which is the inverse of $\prec$, i.e.,  $\succ = \prec^{-1}$. It is clear that $\succ$ is terminating and $\nf_\succ(\Cut(\xi)) = \{(\varepsilon)\}$.

\begin{proof} (Proof of Lemma \ref{lm:wta-init-implies-run}).

  \underline{(A)$\Rightarrow$(B):} If $\Sigma$ is trivial, then for each $(\Sigma,\B)$-wta $\cA$, we have $\runsem{\cA} = \initialsem{\cA}$ by Observation~\ref{obs:init=run-trivial-ra}. This implies (B). Now let $\Sigma$ be not trivial.

         Let $\cA$ be a weighted tree automaton  over $\Sigma$ and $\B$ and  $\xi \in \T_\Sigma$ such that $\xi \in \supp(\initialsem{\cA})$. Let  $q_\varepsilon\in Q$ be such that $\h_\cA(\xi)_{q_\varepsilon} \otimes F_{q_\varepsilon} \not= \0$. Such a $q_\varepsilon$ exist, because $\B$ is zero-sum free.

  By well-founded induction on $(\Cut(\xi),\succ)$, we prove the following statement: 
  \begin{eqnarray}\label{equ:wta-h-not-equal-0-2}
    &\text{For every $\kappa \in \Cut(\xi)$ with $\kappa=(w_1,\ldots,w_n)$ and 
      $\alpha\in\Sigma^{(0)}$,}\notag\\
    &\text{there exists $\rho \in \R_\cA(\xi[\kappa \leftarrow \alpha])$ such that} \
      \Big(\bigotimes_{\substack{w \in \pos(\xi[\kappa \leftarrow \alpha])\\\text{in $<_\mathrm{dp}$ order}}} b^\rho_w\Big)   \otimes F_{q_\varepsilon} \not= \0 \enspace,\\
       &\text{where $b_w^\rho = \begin{cases}\h_\cA(\xi|_w)_{\rho(w)} & \text{if $w \in \{w_1,\ldots,w_n\}$}\\
           \delta_{\rk(\xi(w))}(\rho(w1) \cdots \rho(w\, \rk(\xi(w))), \xi(w),\rho(w)) & \text{otherwise}
           \end{cases}$}\enspace.\notag
  \end{eqnarray}
  We note that \eqref{equ:wta-h-not-equal-0-2} is similar to \eqref{equ:wta-h-not-equal-0} except that now we have to construct a run $\rho$; in \eqref{equ:wta-h-not-equal-0} the run $\rho_0$ is given from the beginning.

I.B.: We recall that $\nf_\succ(\xi) = \{(\varepsilon)\}$. So, let $\kappa = (\varepsilon)$.  By the definition of the set of positions of a tree, we have  $\pos(\xi[(\varepsilon) \leftarrow \alpha]) = \pos(\alpha) = \{\varepsilon\}$. We define $\rho \in \R_\cA(\xi[(\varepsilon) \leftarrow \alpha])$ such that $\rho(\varepsilon) = q_\varepsilon$ (where $q_\varepsilon$ is specified above). Then  
\begin{align*}
  \Big(\bigotimes_{\substack{w \in \pos(\xi[(\varepsilon) \leftarrow \alpha]):\\\text{in $<_\mathrm{dp}$ order}}} \!\! b^\rho_w\Big) \otimes F_{q_\varepsilon} = b^\rho_\varepsilon \otimes F_{q_\varepsilon} = \h_\cA(\xi|_\varepsilon)_{\rho(\varepsilon)} \otimes F_{q_\varepsilon} = \h_\cA(\xi)_{q_\varepsilon} \otimes F_{q_\varepsilon} \enspace.
\end{align*}
By the above, this quantity is not equal to $\0$.

\

  I.S.: Let $\kappa \in \Cut(\xi) \setminus \nf_\succ(\xi)$.
Since $\kappa \not\in  \nf_\succ(\xi)$, there exists $\kappa' \in \Cut(\xi)$ with $\kappa \succ \kappa'$. Hence
there exist $n \in \mathbb{N}_+$, $i \in [n]$, and  $w_1,\ldots,w_n \in \pos(\xi)$ such that
  \begin{compactitem}
  \item  $\kappa = (w_1,\ldots,w_{i-1},\underbrace{w_{i}1,\ldots,w_{i}\, \rk(\xi(w_i))}_{\text{successors of $w_i$}}, w_{i+1},\ldots,w_n)$ and
    \item  $\kappa' = (w_1,\ldots,w_{i-1},w_{i}, w_{i+1},\ldots,w_n)$ \enspace.
  \end{compactitem}
  We abbreviate $\rk(\xi(w_i))$ by $k$. We assume that there exists $\rho' \in \R_\cA(\xi[\kappa' \leftarrow \alpha])$ such that
  \[
  \Big(\bigotimes_{\substack{w \in \pos(\xi[\kappa' \leftarrow \alpha])\\\text{in $<_\mathrm{dp}$ order}}} b^{\rho'}_w\Big) \otimes F_{q_\varepsilon} \not= \0 \ \ \text{ (I.H.)} \enspace.
  \]
       The first factor of the product on the left-hand side of the inequality can be split as follows: there exist $a,c \in B$ such that
  \begin{align*}
    \Big(\bigotimes_{\substack{w \in \pos(\xi[\kappa' \leftarrow \alpha])\\\text{in $<_\mathrm{dp}$ order}}} b^{\rho'}_w\Big) \otimes F_{q_\varepsilon}
    = a \otimes \h_\cA(\xi|_{w_i})_{\rho'(w_i)} \otimes c \otimes F_{q_\varepsilon} \enspace.
  \end{align*}
  If $\Sigma$ is monadic, then $n=k=1$ and $a=\1$.

          By I.H., we know that this quantity is not equal to $\0$.  Then we can compute as follows.
  \begingroup
  \allowdisplaybreaks
  \begin{align*}
    & a \otimes \h_\cA(\xi|_{w_i})_{\rho'(w_i)} \otimes c  \otimes F_{q_\varepsilon} \not=\0 \\
\Leftrightarrow \ \ \     &a \otimes \Big( \bigoplus_{q_1\cdots q_k \in Q^k}\Big(\bigotimes_{j \in [k]} \h_\cA(\xi|_{w_ij})_{q_j}\Big)
      \otimes \delta_{k}(q_1 \cdots q_k, \xi(w_i),\rho'(w_i)) \Big)
                            \otimes c  \otimes F_{q_\varepsilon} \not=\0
    \tag{by definition of $\h_\cA(\xi|_{w_i})_{\rho'(w_i)}$}\\
    \Rightarrow \ \ \  & a \otimes \Big(\bigotimes_{j \in [k]} \h_\cA(\xi|_{w_ij})_{q_{w_ij}}\Big)
      \otimes \delta_{k}(q_{w_i1} \cdots q_{w_ik}, \xi(w_i),\rho'(w_i))
                         \otimes c  \otimes F_{q_\varepsilon} \not=\0 \tag{by Statement (A) of the lemma such $q_{w_i1},\ldots,q_{w_ik}$ exist}\\
     \Leftrightarrow \ \ \  & a \otimes \Big(\bigotimes_{j \in [k]} b^{\rho}_{w_ij}\Big)
                              \otimes \delta_{k}(q_{w_i1} \cdots q_{w_ik}, \xi(w_i),\rho(w_i)) \otimes c  \otimes F_{q_\varepsilon} \not=\0
                              \tag{where $\rho|_{\pos(\xi[\kappa' \leftarrow \alpha])} = \rho'$ and $\rho(w_ij) = q_{w_ij}$ for each $j \in [k]$ }\\
    \Leftrightarrow \ \ \ & \Big(\bigotimes_{\substack{w \in \pos(\xi[\kappa \leftarrow \alpha])\\\text{in $<_\mathrm{dp}$ order}}} b^{\rho}_w\Big)  \otimes F_{q_\varepsilon}\not= \0 \enspace. \tag{because $w_{i}1 <_\mathrm{dp} \ldots <_\mathrm{dp} w_{i}k$ and  $(\forall j\in[k-1]) :\{v \in \xi[\kappa \leftarrow \alpha] \mid w_ij <_{\mathrm{dp}} v <_{\mathrm{dp}} w(j+1)\}=\emptyset$}
  \end{align*}
  \endgroup
  This proves Statement \eqref{equ:wta-h-not-equal-0-2}.

  By \eqref{equ:wta-h-not-equal-0-2} for $\kappa = \lcut(\xi)$ and an arbitrary $\alpha \in \Sigma^{(0)}$, there exists $\rho \in \R_\cA(\xi[\lcut(\xi) \leftarrow \alpha])$ such that
  \[
\Big(\bigotimes_{\substack{w \in \pos(\xi[\lcut(\xi) \leftarrow \alpha])\\\text{in $<_\mathrm{dp}$ order}}} b^\rho_w\Big)   \otimes F_{q_\varepsilon} \ne \0 \enspace.
\]
Since $\R_\cA(\xi[\lcut(\xi) \leftarrow \alpha]) = \R_\cA(\xi)$, we obtain
  \begingroup
  \allowdisplaybreaks
  \begin{align*}
    \0 &\not= \Big(\bigotimes_{\substack{w \in \pos(\xi[\lcut(\xi) \leftarrow \alpha])\\\text{in $<_\mathrm{dp}$ order}}} b^\rho_w\Big)   \otimes F_{q_\varepsilon} \tag{note that $\rho \in \R_\cA(\xi)$}\\
       &= \Big(\bigotimes_{\substack{w \in \pos(\xi)\\\text{in $<_\mathrm{dp}$ order}}} b^\rho_w\Big)   \otimes F_{q_\varepsilon}\\
       &= \Big(\bigotimes_{\substack{w \in \pos(\xi)\\\text{in $<_\mathrm{dp}$ order}}} \delta_{\rk(\xi(w))}(\rho(w1) \cdots \rho(w\, \rk(\xi(w))), \xi(w),\rho(w)) \Big)   \otimes F_{q_\varepsilon}\\
    &= \wt(\rho,\xi)  \otimes F_{q_\varepsilon} \enspace.
    \end{align*}
  \endgroup

  Since $\B$ is zero-sum free, we also have
  \(\runsem{\cA}(\xi) = \bigoplus_{\rho \in \R_\cA(\xi)} \wt(\rho,\xi) \otimes F_{\rho(\varepsilon)} \not= \0 \).
  Hence $\xi \in \supp(\runsem{\cA})$.
  
\

\underline{(B)$\Rightarrow$(A):} We assume that (B) holds. We have to prove:\\
(i) If $\Sigma$ is not trivial, then $(b \oplus b') \otimes c \neq \0$  implies $\big(b \otimes c \neq \0$ or $b' \otimes c \neq \0\big)$, for every $b,b',c \in B$, and\\
(ii) if $\Sigma$ is branching, then $a \otimes (b \oplus b') \otimes c \neq \0$  implies $\big(a \otimes b \otimes c \neq \0$  or 
       $a \otimes b' \otimes c \neq \0\big)$, for every $a,b,b',c \in B$.

       To prove (i), we assume that $\Sigma$ is not trivial. Since the rank mapping will play a role, we consider the ranked alphabet $(\Sigma,\rk)$ and assume that $(\Sigma,\rk)$ is not trivial.
       Then there exist $\gamma \in \Sigma$ and $k \in \mathbb{N}$ with $k \ge 1$ and $\rk_\Sigma(\gamma)=k$. Let $\alpha \in \Sigma^{(0)}$. Moreover, let $b,b',c \in B$.

 First, we define the ranked alphabet $(\Sigma,\rk')$ such that for each $\eta \in \Sigma$ we let
 \[
    \rk'(\eta) =
    \begin{cases}
      1 & \text{ if $\eta=\gamma$} \\
      \rk(\eta) & \text{ if otherwise.}
    \end{cases}
  \]
Thus, $\rk'(\gamma)=1$ and $(\Sigma,\rk)$ is an extension of $(\Sigma,\rk')$.

Second, since $(\Sigma,\rk')$ contains a unary symbol, i.e., $(\rk')^{-1}(1) \not= \emptyset$,
by Lemma~\ref{lm:Ar-right-distributivity}, we can construct a $((\Sigma,\rk'),\B)$-wta $\cA_r$ such that
\begin{equation}\label{equ:mon-image-init-run}
\im(\runsem{\cA_r}) = \{\0, (b \otimes c) \oplus (b' \otimes c)\} \ \ \text{  and } \ \ 
\im(\initialsem{\cA_r}) = \{\0, (b \oplus b') \otimes c\} \enspace.
\end{equation}

Third, by using the padding lemma (i.e., Lemma~\ref{lm:super-big-padding}), we can construct a $((\Sigma,\rk),\B)$-wta $\cB$ such that, for each $\xi \in \T_{(\Sigma,\rk')}$, we have
\begin{equation}\label{equ:application-padding-lemma-mon-supp-codomain-init-run}
\runsem{\cB}(g(\xi)) = \runsem{\cA_r}(\xi)
\ \ \text{ and } \ \ 
\initialsem{\cB}(g(\xi))) = \initialsem{\cA_r}(\xi) \enspace. 
\end{equation}

Now assume that $(b \oplus b') \otimes c \not= \0$. Then the following implications hold.
\begingroup
\allowdisplaybreaks
\begin{align*}
& (b \oplus b')  \otimes c \not= \0 \\
  \Rightarrow \ \ & (\exists \xi \in \T_{(\Sigma,\rk')}): \ \initialsem{\cA_r}(\xi)  \not= \0
                    \tag{ by \eqref{equ:mon-image-init-run}}\\
  \Rightarrow \ \ & (\exists \xi \in \T_{(\Sigma,\rk')}): \ \initialsem{\cB}(g(\xi)) \not= \0
                    \tag{ by \eqref{equ:application-padding-lemma-mon-supp-codomain-init-run}}\\
  \Rightarrow \ \ & (\exists \xi \in \T_{(\Sigma,\rk')}): \ \runsem{\cB}(g(\xi))  \not= \0
                    \tag{ by (B): $\supp(\initialsem{\cB}) \subseteq \supp(\runsem{\cB})$}\\
  \Rightarrow \ \ & (\exists \xi \in \T_{(\Sigma,\rk')}): \ \runsem{\cA_r}(\xi) \not= \0
                    \tag{ by \eqref{equ:application-padding-lemma-mon-supp-codomain-init-run}}\\
  \Rightarrow \ \ &  (b \otimes c) \oplus (b' \otimes c) \not= \0
                    \tag{ by \eqref{equ:mon-image-init-run}} \\
  \Rightarrow \ \ & \big(b \otimes c \neq \0 \ \text{  or } \ b' \otimes c \neq \0\big)
                    \tag{ because otherwise $b \otimes c \oplus b' \otimes c = \0$ } \enspace.
\end{align*}
\endgroup
This proves (i).

\

To prove (ii), we assume that $(\Sigma,\rk)$ is branching. Then there exist $\sigma \in \Sigma$ and $k \in \mathbb{N}$ with $k \ge 2$ and $\rk(\sigma)=k$. Let $\alpha \in \Sigma^{(0)}$. Moreover, let $a,b,b',c \in B$.

 First, we define the ranked alphabet $(\Sigma,\rk')$ such that for each $\eta \in \Sigma$ we let
 \[
    \rk'(\eta) =
    \begin{cases}
      2 & \text{ if $\eta=\sigma$} \\
      \rk(\eta) & \text{ if otherwise.}
    \end{cases}
  \]
Thus, $\rk'(\sigma)=2$ and $(\Sigma,\rk)$ is an extension of $(\Sigma,\rk')$.

Second, since $(\Sigma,\rk')$ contains a binary symbol, i.e., $(\rk')^{-1}(1) \not= \emptyset$,
by Lemma~\ref{lm:wta-Abr}, we can construct a $((\Sigma,\rk'),\B)$-wta $\cA_l$ such that
\begin{equation}\label{equ:br-image-init-run}
\im(\runsem{\cA_l}) = \{\0, (a \otimes b \otimes c) \oplus (a \otimes b' \otimes c)\} \ \ \text{  and } \ \ 
\im(\initialsem{\cA_l}) = \{\0, a \otimes (b \oplus b') \otimes c\} \enspace.
\end{equation}

Third, by using the padding lemma (i.e., Lemma~\ref{lm:super-big-padding}), we can construct a $((\Sigma,\rk),\B)$-wta $\cB$ such that, for each $\xi \in \T_{(\Sigma,\rk')}$, we have
\begin{equation}\label{equ:application-padding-lemma-br-supp-codomain-init-run}
\runsem{\cB}(g(\xi)) = \runsem{\cA_l}(\xi)
\ \ \text{ and } \ \ 
\initialsem{\cB}(g(\xi))) = \initialsem{\cA_l}(\xi) \enspace. 
\end{equation}

Now assume that $a \otimes (b \oplus b') \otimes c \not= \0$. Then the following implications hold.
\begingroup
\allowdisplaybreaks
\begin{align*}
& a \otimes (b \oplus b')  \otimes c \not= \0 \\
  \Rightarrow \ \ & (\exists \xi \in \T_{(\Sigma,\rk')}): \ \initialsem{\cA_l}(\xi)  \not= \0
                    \tag{ by \eqref{equ:br-image-init-run}}\\
  \Rightarrow \ \ & (\exists \xi \in \T_{(\Sigma,\rk')}): \ \initialsem{\cB}(g(\xi)) \not= \0
                    \tag{ by \eqref{equ:application-padding-lemma-br-supp-codomain-init-run}}\\
  \Rightarrow \ \ & (\exists \xi \in \T_{(\Sigma,\rk')}): \ \runsem{\cB}(g(\xi))  \not= \0
                    \tag{ by (B): $\supp(\initialsem{\cB}) \subseteq \supp(\runsem{\cB})$}\\
  \Rightarrow \ \ & (\exists \xi \in \T_{(\Sigma,\rk')}): \ \runsem{\cA_l}(\xi) \not= \0
                    \tag{ by \eqref{equ:application-padding-lemma-br-supp-codomain-init-run}}\\
  \Rightarrow \ \ &  (a \otimes b \otimes c) \oplus (a \otimes b' \otimes c) \not= \0
                    \tag{ by \eqref{equ:br-image-init-run}} \\
  \Rightarrow \ \ & \big(a \otimes b \otimes c \neq \0 \ \text{  or } \ a \otimes b' \otimes c \neq \0\big)
                    \tag{ because otherwise $(a \otimes b \otimes c) \oplus (a \otimes b' \otimes c) = \0$ } \enspace.
\end{align*}
\endgroup
This proves (ii).
\end{proof}

%% file: pumping-lemmas.tex
\chapter{Pumping lemmas}
\label{ch:pumping-lemmas}

In this chapter we show three pumping lemmas, which have  different flavors: the pumping lemma for runs \cite[Sec.~5]{bor04} (cf. Theorem \ref{thm:pumping-lemma-for-runs}) and, as a corollary, a pumping lemma for wta over positive strong bimonoids (cf. Corollary \ref{cor:pumping-lemma-for-runs}), and  the pumping lemma for wta over fields \cite{berreu82} (cf. Theorem \ref{thm:pumping-wta-fields}).

The pumping lemma in Theorem \ref{thm:pumping-lemma-for-runs} concerns runs of weighted tree automata over $\B$. It says that, given a big tree and a run on that tree, both the tree and the run can be decomposed such that both, the tree and the run can be pumped up along its components, and the weight of a pumped run on the corresponding pumped tree can be constructed as the product of the weights of the components of the run. As a corollary of Theorem \ref{thm:pumping-lemma-for-runs}, we prove a pumping lemma for the support of weighted tree languages recognizable by wta over a positive strong bimonoid $\B$ (cf. Corollary \ref{cor:pumping-lemma-for-runs}). Essentially this is the classical pumping lemma for recognizable $\Sigma$-tree languages \cite[Lm.~2.10.1]{gecste84}.

The pumping lemma in Theorem \ref{thm:pumping-wta-fields} concerns the support of weighted tree languages recognizable by wta over a field $\B$. It says that, given a recognizable weighted tree language $r$ and a big tree in the support of $r$, the tree can be decomposed such that it can be pumped up along its components and infinitely many of the pumped trees belong to the support of $r$.

We mention that in  \cite{mazriv18,chamazmusriv21} five pumping lemmas for wsa  over particular important semirings are shown. They concern weighted languages (a) which are  recognizable by wsa over $\Nat_\infty$, (b) which have the form $\min(r_1,\ldots,r_m)$, where $r_1,\ldots,r_m$ are recognizable by wsa over $\Nat_\infty$, (c) which are recognizable by polynomially-ambiguous wsa over $\Natminplus$, (d) which are recognizable by finitely-ambiguous wsa over $\Natmaxplus$, and (e) which are recognizable by polynomially-ambiguous wsa over $\Natmaxplus$, respectively. Each of these pumping lemmas is used to show that certain weighted languages do not belong to the corresponding set of recognizable weighted languages.

We also mention that  the pumping lemma in \cite{reu80} for wsa over fields (which can be considered as a predecessor of the above mentioned pumping lemma in \cite{berreu82}) was generalized to wsa over Artinian semirings  in \cite{malnue23}.

\section{Pumping lemma for weights of runs of weighted tree automata}
\label{sec:pumping-lemma-for-runs}

The pumping lemma of this section is based on the idea in \cite[Sec.~5]{bor04}.
The form in which we present it is adapted from \cite{drofulkosvog20b,drofulkosvog21}.

\label{p:convention-on-wta-in-pumping}
\begin{quote} \emph{In this section, we let $\cA= (Q,\delta,F)$ be an arbitrary $(\Sigma,\B)$-wta.}
  \end{quote}

For this, let $\xi\in \T_\Sigma(\{z\})$ and $q\in Q$. A \emph{run (of $\cA$) on $\xi$} is a mapping $\rho: \pos(\xi)\to Q$, and it is a  $q$-run if $\rho(\varepsilon)=q$. Also, we denote the set of all runs on $\xi$ by $\R_\cA(\xi)$ and the set of all $q$-runs on $\xi$ by $\R_\cA(q,\xi)$. For every $\xi \in \T_\Sigma(\{z\})$, $\rho\in \R_\cA(\xi)$, and $w \in\pos(\xi)$, the \emph{run induced by~$\rho$ at position~$w$}, denoted by $\rho|_w$, is defined in a similar way as shown on page \pageref{page:TR-prec}. 

  \label{p:par-extension-runs-to-contexts}
\index{TRz@$\mathrm{TR}_z$}
\index{TRzsucc@$(\mathrm{TR}_z,\succ)$}
Next we define the weight of a run on trees in $\T_\Sigma(\{z\})$ in a similar way as shown on page \pageref{page:TR-prec}. 
We consider the reduction system $(\mathrm{TR}_z,\succ)$ where
\(
\mathrm{TR}_z = \{(\xi,\rho) \mid \xi \in \T_\Sigma(\{z\}), \rho \in \R_\cA(\xi)\}
\)
and $\succ$ is the binary relation  on $\mathrm{TR}_z$ defined as follows:
\index{succ@$\succ$}
\[
  \text{for every $(\xi, \rho) \in \mathrm{TR}_z$ and $i \in [\rk(\xi(\varepsilon))]$, we let $(\xi,\rho) \succ (\xi|_i,\rho|_i)$ }\enspace. \]
By Corollaries \ref{cor:reduction-to-substring-is-terminating} and \ref{cor:termination-propagates-to-cartesian-products}, the relation $\succ$ is terminating. Moreover, we have that $\nf_\succ(\mathrm{TR}_z) = \{(\alpha,\rho) \mid \alpha \in \Sigma^{(0)} \cup \{z\}, \rho: \{\varepsilon\} \to Q\}$.
  By induction on $(\mathrm{TR}_z,\succ)$, we define the mapping
  \[
    \wt_\cA: \mathrm{TR}_z \to B
  \]
   for every $\xi \in \T_\Sigma(\{z\})$ and $\rho \in \R_\cA(\xi)$ as follows:
  
  \index{wt@$\wt_\cA(\xi,\rho)$}
  I.B.: If $\xi = z$, then $\wt_\cA(\xi,\rho)=\1$. If $\xi = \alpha$ is in $\Sigma^{(0)}$, then $\wt_\cA(\xi,\rho)=\delta_0(\varepsilon,\alpha,\rho(\varepsilon))$.

  I.S.: Let $\xi = \sigma(\xi_1,\ldots,\xi_k)$ with $k \ge 1$. Then we define  
\begin{equation}\label{eq:weight-of-a-run-context}
\wt_\cA(\xi,\rho) = \Big( \bigotimes_{i\in [k]} \wt_\cA(\xi|_i,\rho|_i)\Big) \otimes \delta_k\big(\rho(1) \cdots \rho(k),\sigma,\rho(\varepsilon)\big) \enspace.
\end{equation}

We call $\wt_\cA(\xi,\rho)$ the \emph{weight of $\rho$ (by $\cA$ on $\xi$)}.

\index{loop}
\index{run on context}
\index{R@$\R_\cA(q,c,p)$}
\index{qtilde@$\widetilde{q}$}
\index{qprunonc@$(q,p)$-run on $c$}
\label{page:run-on-contexts}
Now let $c\in \C_\Sigma$ be a context  with $\pos_z(c)=v$ and let $\rho \in  \R_\cA(q,c)$ for some $q\in Q$. 
If $\rho(v)=p$, then we call $\rho$ a \emph{$(q,p)$-run on $c$} and we denote the set of all such $(q,p)$-runs by $\R_\cA(q,c,p)$. A $(q,q)$-run is called a \emph{loop}. If $\rho \in \R_\cA(z)$ with $\rho(\varepsilon)=q$ for some $q \in Q$, then sometimes we write $\widetilde{q}$ for $\rho$. 

\index{combination of runs}
Let $c_1,c_2 \in \C_\Sigma$, $v = \pos_z(c_1)$, $p,q,r \in Q$,
$\rho_1 \in \R_\cA(q,c_1,r)$, and $\rho_2 \in \R_\cA(r,c_2,p)$.
The {\em combination of $\rho_1$ and $\rho_2$}, denoted by $\rho_1[\rho_2]$, is
the mapping $\rho_1[\rho_2]: \pos(c_1[c_2]) \to Q$ defined for every $u \in \pos(c_1[c_2])$
as follows: if $u=vw$ for some $w$, then we define $\rho_1[\rho_2](u)=\rho_2(w)$,
otherwise we define $\rho_1[\rho_2](u)=\rho_1(u)$. Clearly, $\rho_1[\rho_2] \in \R_\cA(q,c_1[c_2],p)$.

\index{xi@${\xi\mid^v}$}
\index{rho@${\rho\mid^v(w)}$}
Let  $\xi \in \T_\Sigma$ and $v \in \pos(\xi)$. We denote by $\xi|^v$ the context $\xi[z]_v$ obtained by replacing the subtree of $\xi$ at $v$ by $z$ (cf. page \pageref{page:subtree-replacement}).
Moreover, for each $\rho \in \R_\cA(\xi)$, we define the
run $\rho|^v$ on the context $\xi|^v$ such that for every $w \in \pos(\xi|^v)$ we set
$\rho|^v(w) =\rho(w)$. 

\index{lcrho@$l_{c,\rho}$}
\index{rcrho@$r_{c,\rho}$}

Let $c \in \C_\Sigma$, $v =\pos_z(c)$, and $\rho \in \R_\cA(c)$. We define two mappings
\[
  l_{c,\rho}: \prefix(v) \to B \ \ \text{ and } \ \ r_{c,\rho}: \prefix(v) \to B
\]
(cf. \cite[p.~526]{bor04} for deterministic wta). Intuitively, the product \eqref{eq:weight-of-a-run-context} which yields the element $\wt(c,\rho) \in B$, can be split into a left subproduct $l_{c,\rho}(\varepsilon)$ and a right subproduct $r_{c,\rho}(\varepsilon)$, where the border is given by the factor $\mathbb{1}$ coming from the weight of $z$.
Figure \ref{fig:l-and-r} illustrates the mappings $l_{c,\rho}$ and $r_{c,\rho}$.

\index{succ@$\succ$}
Formally, we define  the mappings $l_{c,\rho}$ and $r_{c,\rho}$ by induction on the reduction system $(\prefix(v),\succ)$  where, for every $w_1,w_2 \in \prefix(v)$, we let $w_1 \succ w_2$ if there exists an $i \in \mathbb{N}$ such that $w_2 = w_1 i$. In the proof of Lemma~\ref{lm:zero-propagation-h} we showed that $(\prefix(v),\succ)$ is terminating.
Moreover, we have that  $\nf_\succ(\prefix(v)) = \{v\}$.

  Then we define $l_{c,\rho}$ and $r_{c,\rho}$ by induction on $(\prefix(v),\succ)$ as follows.  Let $w \in \prefix(v)$ and assume that $c(w)=\sigma$ and $\rk_\Sigma(\sigma)=k$. Then we define

\[
  l_{c,\rho}(w) =
  \begin{cases}
 \1   & \text{ if $w=v$}\\
 \bigotimes\limits_{j\in [1,i-1]} \wt(c|_{wj},\rho|_{wj}) \otimes l_{c,\rho}(wi)  & \text{ if $wi \in \prefix(v)$ for some $i \in \mathbb{N}_+$}
    \end{cases}
  \]
\[   r_{c,\rho}(w) =
  \begin{cases}
 \1  & \hspace*{-14mm}\text{ if $w=v$}\\
 r_{c,\rho}(wi) \otimes \bigotimes\limits_{j \in [i+1,k]} \wt(c|_{wj},\rho|_{wj}) \otimes \delta_k(\rho(w1) \cdots \rho(wk),\sigma,\rho(w)) &\\
 &\hspace*{-54mm} \text{ if $wi \in \prefix(v)$ for some $i \in \mathbb{N}_+$}\enspace.
    \end{cases}
  \] 
In the sequel, we abbreviate $l_{c,\rho}(\varepsilon)$ and $r_{c,\rho}(\varepsilon)$ by $l_{c,\rho}$ and $r_{c,\rho}$, respectively.

\begin{sidewaysfigure}
\centering
\begin{tikzpicture}
\footnotesize
\node[anchor=north] at (-7.1,1.5) (t1) {$\wt(c|_{w1},\rho|_{w1})$};
\node[anchor=north] at (-7.1,0.75) {$c|_{w1}$};

\node at (-5.6,1.25) {$\otimes \cdots \otimes$};

\node[anchor=north] at (-3.625,1.5) (ti-1) {$\wt(c|_{w(i-1)},\rho|_{w(i-1)})$};
\node[anchor=north] at (-3.625,0.75) {$c|_{w(i-1)}$};

\node at (-2, 1.25) {$\otimes$};
\node[anchor=north] at (-1.1,1.5) {$l_{c,\rho}(wi)$};
\draw[rounded corners=7pt] (-0.475,1) rectangle (-1.725,1.5); 

\node[anchor=north] at (0,0.75) {$c|_{wi}$};
\draw (0,0) -- (0,-0.25) node [midway,right] {$v$};
\node at (0,-0.375) {$z$};

\node at (2,1.25) {$\otimes$};
\node[anchor=north] at (1.1,1.5) {$r_{c,\rho}(wi)$};
\draw[rounded corners=7pt] (0.475,1) rectangle (1.725,1.5); 

\node[anchor=north] at (3.625,1.5) (ti+1) {$\wt(c|_{w(i+1)},\rho|_{w(i+1)})$};
\node[anchor=north] at (3.625,0.75) {$c|_{w(i+1)}$};

\node at (5.6,1.25) {$\otimes \cdots \otimes$};

\node[anchor=north] at (7.1,1.5) (tk) {$\wt(c|_{wk},\rho|_{wk})$};
\node[anchor=north] at (7.1,0.75) {$c|_{wk}$};

\node at (8.2,1.25) {$\otimes$};

\node at (0,4.5) (s) {$\sigma$};
\node at (-1.125,4.5) {$l_{c,\rho}(w)$};
\draw[rounded corners=7pt] (-0.5,4.25) rectangle (-1.75,4.75); 
\node at (1.125,4.5) {$r_{c,\rho}(w)$};
\draw[rounded corners=7pt] (0.5,4.25) rectangle (1.75,4.75);

\node at (-7.1,1.75) (da11) {};
\node at (-2,4.5) (da12) {};

\draw[-Implies,double distance=2pt] (da11) to [out=90,in=180] (da12);

\node at (6.5,3.25) (da21) {};
\node at (2,4.5) (da22) {};

\draw[-Implies,double distance=2pt] (da21) to [out=90,in=0] (da22);

\draw[-]
  (-8.1,0) -- (-6.1,0) -- (t1) -- (-8.1,0)
  (-2.625,0) -- (-4.625,0) -- (ti-1) -- (-2.625,0)
  (-1,0) -- (1,0) -- (0,1.5) -- (-1,0)
  (2.625,0) -- (4.625,0) -- (ti+1) -- (2.625,0)
  (6.1,0) -- (8.1,0) -- (tk) -- (6.1,0)
  (s) -- (t1) node[midway,xshift=-24pt,anchor=north] {$1$}
  (s) -- (ti-1) node[midway,xshift=-20pt,anchor=north] {$i-1$}
  (s) -- (0,1.5) node[midway,xshift=-8pt,anchor=north] {$i$}
  (s) -- (ti+1) node[midway,xshift=-8pt,anchor=north] {$i+1$}
  (s) -- (tk) node[midway,xshift=-6pt,anchor=north] {$k$}
;

\node at (-4,2) {$\cdots$};
\node at (4,2) {$\cdots$};

\node at (6.5,3) {$\delta_k(\rho(w1) \cdots \rho(wk),\sigma,\rho(w))$};

\draw[dashed,smooth,line cap=round, rounded corners=10pt] (-8.1,0.875) -- (-0.375,0.875) -- (-0.375,1.625) -- (-8.1,1.625) -- cycle;

\draw[dashed,smooth,line cap=round, rounded corners=10pt] (8.5,0.875) -- (0.375,0.875) -- (0.375,1.625) -- (4.5,1.625) -- (4.5,3.25) -- (8.5,3.25) -- cycle;

\draw[decorate, decoration=snake] (0,6.5) -- (s);
\node[anchor=west] at (0,5.5) {$w$};
\draw (0,6.5) -- (-7.75,4.5);
\node at (-4,5.75) {$c \in \C_\Sigma$};
\draw (0,6.5) -- (8.25,4.5);
\node at (4.5,5.75) {$\rho \in \R_\cA(c)$};

\end{tikzpicture}
\caption{\label{fig:l-and-r} Illustration of mappings $l_{c,\rho}$ and $r_{c,\rho}$ (cf. \cite{drofulkosvog20b,drofulkosvog21})}
\end{sidewaysfigure}

\begin{observation}\rm \label{obs:decomp-run-left-right}
Let $c \in \C_\Sigma$ and  $\rho \in \R_\cA(c)$. Then $\wt(c,\rho) = l_{c,\rho} \otimes r_{c,\rho}$.\hfill $\Box$
\end{observation}

\begin{lemma} \label{lm:combining-runs} \rm (cf. \cite[Lm.~5.1]{bor04})
Let $c \in \C_\Sigma$, $\zeta \in \T_\Sigma$, $q', q \in Q$, $\rho \in \R_\cA(q',c,q)$, and $\theta \in \R_\cA(q,\zeta)$. Then $\wt(c[\zeta],\rho[\theta]) = l_{c,\rho} \otimes \wt(\zeta,\theta) \otimes  r_{c,\rho}$.
\end{lemma}
\begin{proof} We prove the statement by induction on the reduction system $(\C_\Sigma,\succ_{\C_\Sigma})$ (cf. page \pageref{page:order-on-contexts}).

  I.B.: Let $c=z$. Then
\begingroup
\allowdisplaybreaks
\begin{align*}
\wt(c[\zeta], \rho[\theta])
&= \wt(\zeta, \theta) \tag{\text{since $c[\zeta]=\zeta$ and $\rho[\theta]=\theta$}}\\
&= \1 \otimes \wt(\zeta,\theta) \otimes \1 \\
&= l_{z,\rho} \otimes \wt(\zeta,\theta) \otimes r_{z,\rho} \tag{\text{since $l_{z,\rho}=\1$ and $r_{z,\rho} = \1$}}
\end{align*}
\endgroup

I.S.: Let $c=\sigma(\xi_1,\ldots,\xi_{i-1},c',\xi_{i+1},\ldots,\xi_k)$ with $k \in \mathbb{N}_+$, $i \in [k]$, $\xi_1,\ldots,\xi_k \in \T_\Sigma$, and $c' \in \C_\Sigma$. Then we have
\begingroup
\allowdisplaybreaks
\begin{align*}
\wt(c[\zeta],\rho[\theta])
&= \Big(\bigotimes_{j\in[1,i-1]} \wt(\xi_j, \rho|_j)\Big) \otimes \wt\big(c'[\zeta], (\rho|_i)[\theta]\big)\\
&\hspace*{10mm}  \otimes \Big(\bigotimes_{j\in[i+1,k]} \wt(\xi_j, \rho|_j)\Big) \otimes \delta_k\big(\rho(1)\cdots \rho(k), \sigma, q'\big) \tag{\text{by the definition of $c[\zeta]$ and $\rho[\theta]$}}\\
&= \Big(\bigotimes_{j\in [1,i-1]} \wt(\xi_j, \rho|_j)\Big) \otimes l_{c',\rho|_i} \otimes \wt(\zeta, \theta) \otimes r_{c',\rho|_i}\\
&\hspace*{10mm} \otimes \Big(\bigotimes_{j\in[i+1,k]} \wt(\xi_j, \rho|_j)\Big) \otimes \delta_k\big(\rho(1) \cdots \rho(k), \sigma, q'\big) \tag{\text{by I.H.}}\\[1.1em]
  &= l_{c,\rho} \otimes \wt(\zeta,\theta) \otimes  r_{c,\rho} \tag{\text{by the definition of $l_{c,\rho}$ and $r_{c,\rho}$}}\\
      & \tag{\text{note that $l_{c|_i,\rho|_i}= l_{c',\rho|_i}$ and $r_{c|_i,\rho|_i}= r_{c',\rho|_i}$}}
\end{align*}
        \endgroup
\end{proof} 

\index{power of $\rho$}
\index{power of $c$}
\index{rho@$\rho^n$}
\index{ctothepower@$c^n$}
Let $c \in \C_\Sigma$. For each $n \in \mathbb{N}$, we define the {\em $n$-th power of $c$}, denoted by $c^n$, by induction on $\mathbb{N}$ as follows: $c^0=z$ and  $c^{n+1}=c[c^n]$. Moreover, let
$q \in Q$ and $\rho \in \R_\cA(q,c,q)$ be a loop. For each $n \in \mathbb{N}$, the {\em $n$-th power of $\rho$}, denoted by $\rho^n$, is the run on $c^n$ defined by induction on $\mathbb{N}$ as follows: $\rho^0=\sqrun{q}$ (note that $c^0=z$) and $\rho^{n+1}=\rho[\rho^n]$. Next we apply the previous results to the weights of powers of loops.

\begin{theorem} \label{thm:decomposition-of-a-run}  {\rm (cf. \cite[Lm.~5.3]{bor04})}
Let $c',c \in \C_\Sigma$ and $\zeta \in \T_\Sigma$, $q',q \in Q$, $\rho' \in \R_\cA(q',c',q)$, $\rho \in \R_\cA(q,c,q)$, and $\theta \in \R_\cA(q,\zeta)$. Then, for each $n \in \mathbb{N}$,  
\begin{equation}\label{equ:pumping-equation}
  \wt(c'\big[c^n[\zeta]\big], \rho'\big[\rho^n[\theta]\big]) =
  l_{c',\rho'} \otimes (l_{c,\rho})^n \otimes \wt(\zeta,\theta) \otimes (r_{c,\rho})^n \otimes r_{c',\rho'}
  \enspace.
\end{equation}
\end{theorem}
\begin{proof}
  First, by induction on $\mathbb{N}$, we prove that the following statement holds:
  \begin{equation}
\text{For each $n \in \mathbb{N}$ we have } \wt(c^n[\zeta],\rho^n[\theta]) = (l_{c,\rho})^n \otimes \wt(\zeta,\theta) \otimes  (r_{c,\rho})^n \enspace. \label{eq:n-context}
    \end{equation}

I.B.:  Let $n=0$.  Then $c^0=z$ and we have
\begin{align*}
\wt(c^0[\zeta],\rho^0[\theta])
&= \wt(\zeta, \theta) \tag{\text{since $c^0[\zeta]=\zeta$ and $\rho^0[\theta]=\theta$}}\\
&= \1 \otimes \wt(\zeta,\theta) \otimes \1\\
&= (l_{c,\rho})^0 \otimes \wt(\zeta,\theta) \otimes (r_{c,\rho})^0 \tag{\text{since $(l_{c,\rho})^0=\1$ and $(r_{c,\rho})^0 = \1$}.}
\end{align*}

I.S.: We assume that the equality holds for $n$. Since $c^{n+1}=c[c^n]$ and $\rho^{n+1}=\rho[\rho^n]$, we have
\begin{align*}
\wt(c^{n+1}[\zeta],\rho^{n+1}[\theta])
&= l_{c,\rho} \otimes \wt(c^n[\zeta],\rho^n[\theta]) \otimes r_{c,\rho} \tag{\text{by Lemma \ref{lm:combining-runs}}}\\
&= l_{c,\rho} \otimes  (l_{c,\rho})^n \otimes \wt(\zeta,\theta) \otimes (r_{c,\rho})^n \otimes r_{c,\rho} \tag{\text{by I.H.}}\\
&= (l_{c,\rho})^{n+1} \otimes \wt(\zeta, \theta) \otimes (r_{c,\rho})^{n+1}. 
\end{align*}
This proves \eqref{eq:n-context}. Now let $n \in \mathbb{N}$. Then  we have
  \begin{align*}
    \wt(c'\big[c^n[\zeta]\big], \rho'\big[\rho^n[\theta]\big])
    &= l_{c',\rho'} \otimes \wt(c^n[\zeta],\rho^n[\theta]) \otimes r_{c',\rho'} \tag{\text{by Lemma \ref{lm:combining-runs}}}\\
    &= l_{c',\rho'} \otimes (l_{c,\rho})^n \otimes \wt(\zeta,\theta) \otimes (r_{c,\rho})^n \otimes r_{c',\rho'} \tag{\text{by \eqref{eq:n-context}}} 
  \end{align*} 
\end{proof}

Next, we prove  our pumping lemma for runs of $\cA$ on trees in $\T_\Sigma$ which are large enough. We note that $\B$ need not be commutative.

\begin{theorem-rect} {\rm \label{thm:pumping-lemma-for-runs} (cf. \cite[Lm.~5.5]{bor04})} Let $\Sigma$ be a ranked alphabet. Moreover, let $\B=(B,\oplus,\otimes,\0,\1)$ be a strong bimonoid and $\cA$ be a $(\Sigma,\B)$-wta.
Let $\xi \in \T_\Sigma$, $q' \in Q$, and $\kappa \in \R_\cA(q',\xi)$. If $\height(\xi) \geq |Q|$, then
there exist $c',c \in \C_\Sigma$, $\zeta \in \T_\Sigma$, $q \in Q$, $\rho' \in \R_\cA(q',c',q)$, $\rho \in \R_\cA(q,c,q)$, and $\theta \in \R_\cA(q,\zeta)$ such that $\xi=c'\big[c[\zeta]\big]$, $\kappa=\rho'\big[\rho[\theta]\big]$,
$\height(c) > 0$, $\height\big(c[\zeta]\big) < |Q|$, and, for each $n \in \mathbb{N}$,
\[
  \wt(c'\big[c^n[\zeta]\big], \rho'\big[\rho^n[\theta]\big]) =
  l_{c',\rho'} \otimes (l_{c,\rho})^n \otimes \wt(\zeta,\theta) \otimes (r_{c,\rho})^n \otimes r_{c',\rho'}
  \enspace.
\]
\end{theorem-rect}
\begin{proof}
  Since $\height(\xi) \geq |Q|$ there exist $u,w \in \mathbb{N}_+^*$ such that $uw \in \pos(\xi)$, $|w| > 0$, $\height(\xi|_u) < |Q|$, and $\kappa(u)=\kappa(uw)$. Then we let $c'=\xi|^u$, $c=(\xi|_u)|^w$, $\zeta=\xi|_{uw}$.  Clearly, \(\xi=c'\big[c[\zeta]\big]\), $\height(c) > 0$ because $|w| > 0$, and $\height\big(c[\zeta]\big) < |Q|$ because $\height(\xi|_u) < |Q|$.
  Moreover, we set $q=\kappa(u)$, $\rho'=\kappa|^u$, $\rho=(\kappa|_u)|^w$ and $\theta=\kappa|_{uw}$.
  Then the statement follows from Theorem \ref{thm:decomposition-of-a-run}. 
\end{proof}

By Example \ref{ex:height}, for $\Sigma = \{\sigma^{(2)}, \alpha^{(0)}\}$, the weighted tree language 
$\height: \T_\Sigma \to \mathbb{N}$ is initial algebra recognizable by a $(\Sigma,\Natmaxplus)$-wta. As an application of Theorem \ref{thm:decomposition-of-a-run}, we show that $\height$ cannot be recognized by a bu-deterministic $(\Sigma,\Natmaxplus)$-wta. In contrast, if we consider an arbitrary string ranked alphabet $\Sigma$, then $\height: \T_\Sigma \to \mathbb{N}$ is in $\budRec(\Sigma,\Natmaxplus)$.

\begin{figure}
        \centering
        \begin{tikzpicture}
            \footnotesize
            \draw (-.3,0) to (1,1.5) to (2.3,0) to (-.3,0);
            \draw
                (-.5,1.5) -- (-.5,0) node[midway,left] {$|Q|$}
                (-.4,1.5) -- (-.6,1.5)
                (-.4,0) -- (-.6,0);
            
            \node at (.5,1.25) {$\zeta_1$};

            \fill[gray!20] (.2,0) to (1,1.1) to (1.8,0);
            \node[circle,fill=black,inner sep=0pt,minimum size=3pt] at (1,1.1) (u) {} node[right = -2pt of u] {$u$} ;

            \draw[decorate, decoration={snake,segment length=6mm}]
 (1,1.5) -- (1,1.1);

            \draw[fill=white,opacity=1] (.6,0) to (1,.6) to (1.4,0) to (.6,0);
            \node[circle,fill=black,inner sep=0pt,minimum size=3pt] at (1,.6) (uv) {} node[right = -2pt of uv] {$uv$};

            \draw[decorate, decoration={snake,segment length=8mm}]
 (1,1.1) -- (1,.6);

            \node[anchor=south,yshift=-2pt] at (0,0) {$c'$};
            \node[anchor=south,yshift=-2pt] at (.5,0) {$c$};
            \node[anchor=south,yshift=-2pt] at (.825,0) {$\zeta$};
            \node[anchor=south,yshift=-2pt] at (1.15,0) {$\theta$};
            \node[anchor=south,yshift=-2pt] at (1.5,0) {$\rho$};
            \node[anchor=south,yshift=-2pt] at (1.9,0) {$\rho'$};

            \draw (2.3,-1.5) to (3.6,1.5) to (4.9,-1.5) to (2.3,-1.5);
            \draw
                (5.1,1.5) -- (5.1,-1.5) node[midway,right] {$2|Q|$}
                (5,1.5) -- (5.2,1.5)
                (5,-1.5) -- (5.2,-1.5);
            
            \node at (4,1.25) {$\zeta_2$};

            \node at (2.3,2) (s) {$\sigma$};

            \draw (s) -- (1,1.5);
            \draw (s) -- (3.6,1.5);

            \node at (2.4,1.1) {$\kappa(u) = \kappa(uv)$};
        \end{tikzpicture}
        \caption{Decomposition of the tree $\xi$ and the run $\kappa$.}\label{fig:pumping-height}
    \end{figure}

\begin{corollary}\label{ex:height-not-bud}\rm \cite[Ex.~5.9]{bor04} For the ranked alphabet $\Sigma = \{\sigma^{(2)}, \alpha^{(0)}\}$, we have that $\height \not\in \budRec(\Sigma,\Natmaxplus)$.
\end{corollary}
\begin{proof} We prove by contradiction. For this, let us assume that there exists a bu-deterministic 
$(\Sigma,\Natmaxplus)$-wta $\cA=(Q,\delta,F)$ such that $\sem{\cA}=\height$. 

Let $\xi=\sigma(\zeta_1,\zeta_2)$ be a tree such that $\height(\zeta_1)=|Q|$ and $\height(\zeta_2)=2|Q|$.
By Lemma \ref{lm:limit-bu-det}(3)(b), there exist a $q\in Q$ and a run $\kappa\in \R_\cA(q,\xi)$ such that
\begin{equation}\nonumber
\sem{\cA}(\xi)=\wt(\xi,\kappa)+F_q \enspace.
\end{equation}
Moreover, there exist $u\in (\mathbb{N}_+)^+$
and $v\in (\mathbb{N}_+)^+$ such that $uv \in \pos(\xi)$ and
$u=1u'$ for some $u'\in \pos(\zeta_1)$ (i.e., $u$ is located in $\zeta_1$), and $\kappa(u)=\kappa(uv)$. Let us introduce 
\begin{compactitem}
\item[-] the contexts $c'=\xi|^{u}$, $c=(\xi|_{u})|^{v}$, the tree $\zeta=\xi|_{uv}$, as well as,
\item[-] the runs $\rho'=\kappa|^{u}$, $\rho=(\kappa|_{u})|^{v}$, and $\theta=\kappa|_{uv}$, see Figure \ref{fig:pumping-height} and
\item[-] let  $\xi_n=c'[c^n[\zeta]]$ and $\kappa_n=\rho'[\rho^n[\theta]]$ for each $n\in \mathbb{N}$.
\end{compactitem}
Note that $\xi_1=\xi$ and $\kappa_1=\kappa$. By Theorem \ref{thm:decomposition-of-a-run}, we have
\begin{equation}\nonumber
  \wt(\xi_n, \kappa_n) =
  l_{c',\rho'} + n\cdot l_{c,\rho} + \wt(\zeta,\theta) + n\cdot r_{c,\rho} + r_{c',\rho'}
\end{equation}
for each $n\in \mathbb{N}$. Since $\height(\xi)=2|Q|+1$, we have
\begin{equation}\nonumber
\sem{\cA}(\xi)= l_{c',\rho'} + l_{c,\rho} + \wt(\zeta,\theta) + r_{c,\rho} + r_{c',\rho'} + F_q =2|Q|+1.
\end{equation}
Thus, each of $l_{c',\rho'}$, $l_{c,\rho}$, $\wt(\zeta,\theta)$, $r_{c,\rho}$, $r_{c',\rho'}$, and  $F_q$ is in $\mathbb{N}$. Therefore $\wt(\xi_n, \kappa_n) \ne -\infty$ for each $n\in \mathbb{N}$
and thus, by Lemma \ref{lm:limit-bu-det}(3)(b), we have
\begin{equation}\nonumber
\sem{\cA}(\xi_n)= l_{c',\rho'} + n\cdot l_{c,\rho} + \wt(\zeta,\theta) + n\cdot r_{c,\rho} +  r_{c',\rho'} + F_q 
\end{equation}
for each $n\in \mathbb{N}$.
It follows that $l_{c,\rho}\ne 0$ or $r_{c,\rho}\ne 0$ because otherwise $\height(\xi_n)$ would be the same number for each $n\in \mathbb{N}$. But then $\sem{\cA}(\xi_0)< \sem{\cA}(\xi_1)$, which is a contradiction because
$\height(\xi_0)= \height(\xi_1)$.
\end{proof}

Contrary to Corollary \ref{ex:height-not-bud}, for each string ranked alphabet $\Sigma$, the weighted tree language $\height: \T_\Sigma \to \mathbb{N}$ is in $\budRec(\Sigma,\Natmaxplus)$. To see this, let $\Sigma$ be a string ranked alphabet and assume that $\Sigma^{(0)}= \{\alpha\}$. Then we consider the bu-deterministic $(\Sigma,\Natmaxplus)$-wta $\cA= (\{q\},\delta,F)$ with $\delta_0(\varepsilon,\alpha,q)=0$, and $\delta_1(q,\gamma,q)=1$ for each $\gamma \in \Sigma^{(1)}$; moreover, we let $F_q=0$. It is easy to see that $\runsem{\cA}=\height$.

As a corollary of Theorem \ref{thm:pumping-lemma-for-runs}, we prove a pumping lemma for the supports of weighted tree languages recognizable by wta over positive strong bimonoids.

\begin{corollary-rect} \rm \label{cor:pumping-lemma-for-runs}
  Let $\Sigma$ be a ranked alphabet. Moreover, let $\B$ be a positive  strong bimonoid and let $L\subseteq \T_\Sigma$. If $L \in \supp(\Rec^{\mathrm{run}}(\Sigma,\B))$, then  there exists a $p \in \mathbb{N}_+$ such that for each $\xi \in L$ with $\height(\xi) \geq p$, there exist $c',c \in \C_\Sigma$, $\zeta \in \T_\Sigma$  such that
\begin{compactitem}
  \item $\xi=c'\big[c[\zeta]\big]$,
  \item $\height(c) > 0$ and $\height\big(c[\zeta]\big) < p$, and
  \item for each $n \in \mathbb{N}$, we have $c'\big[c^n[\zeta]\big] \in L$.
    \end{compactitem}
  \end{corollary-rect}
  \begin{proof} Since $L \in \supp(\Rec^{\mathrm{run}}(\Sigma,\B))$, there exists a  $(\Sigma,\B)$-wta $\cA=(Q,\delta,F)$ such that $L = \supp(\runsem{\cA})$. Let $p=|Q|$.

    Let $\xi \in  L$ with $\height(\xi) \geq p$. There there exist $q' \in Q$ and  $\kappa \in \R_\cA(q',\xi)$ such that $\wt_\cA(\xi,\kappa) \otimes F_{q'} \ne \0$. Thus  $\wt_\cA(\xi,\kappa) \ne \0$ and $F_{q'} \ne \0$.

    By Theorem \ref{thm:pumping-lemma-for-runs}, there exist $c',c \in \C_\Sigma$, $\zeta \in \T_\Sigma$, $q \in Q$, $\rho' \in \R_\cA(q',c',q)$, $\rho \in \R_\cA(q,c,q)$, and $\theta \in \R_\cA(q,\zeta)$ such that $\xi=c'\big[c[\zeta]\big]$, $\kappa=\rho'\big[\rho[\theta]\big]$,
$\height(c) > 0$, $\height\big(c[\zeta]\big) < p$, and, for each $n \in \mathbb{N}$,
\(
  \wt(c'\big[c^n[\zeta]\big], \rho'\big[\rho^n[\theta]\big]) =
  l_{c',\rho'} \otimes (l_{c,\rho})^n \otimes \wt(\zeta,\theta) \otimes (r_{c,\rho})^n \otimes r_{c',\rho'}\). Together with $\wt_\cA(\xi,\kappa) \ne \0$, this implies (for $n=1$) that each of the three values $l_{c',\rho'}$, $\wt(\zeta,\theta)$, and $r_{c,\rho}$ is different from $\0$. Since $\B$ is zero-divisor free, we obtain that 
  \[l_{c',\rho'} \otimes (l_{c,\rho})^n \otimes \wt(\zeta,\theta) \otimes (r_{c,\rho})^n \otimes r_{c',\rho'} \ne \0\]
  for each $n \in \mathbb{N}$, and hence $\wt(c'\big[c^n[\zeta]\big], \rho'\big[\rho^n[\theta]\big]) \ne \0$ (using Theorem \ref{thm:pumping-lemma-for-runs} again). Since $F_{q'} \ne \0$ and $\B$ is zero-divisor free, we obtain that
  \[\wt(c'\big[c^n[\zeta]\big], \rho'\big[\rho^n[\theta]\big])  \otimes F_{q'} \ne \0\enspace.\]
  Since $\B$ is zero-sum free, we obtain:
  \[\runsem{\cA}(c'[c^n[\zeta]]\big]) = \bigoplus_{\nu \in \R_\cA(\xi)} \wt(c'\big[c^n[\zeta]\big], \nu) \otimes F_{\nu(\varepsilon)} \ne \0 \enspace.\]
  Thus, for each $n \in \mathbb{N}$, we have that $c'[c^n[\zeta]] \in L$.
    \end{proof}
    
By Corollary \ref{cor:supp-B=fta-1}, $\supp(\Rec^\mathrm{run}(\Sigma,\Boole)) = \Rec(\Sigma)$. Thus,
    for the case that $\B$ is the Boolean semiring,  Corollary \ref{cor:pumping-lemma-for-runs} shows a slight improvement of the pumping lemma for recognizable tree languages \cite[Lm.~2.10.1]{gecste84} (because \cite[Lm.~2.10.1]{gecste84} does not show the condition $\height\big(c[\zeta]\big) < p$). 
Hence in case $\B=\Boole$, by using the contraposition of Corollary \ref{cor:pumping-lemma-for-runs}, we can give a shorter proof of that 
a certain tree language is not recognizable, than by using \cite[Lm.~2.10.1]{gecste84}. The contraposition  is the following statement:
  \begin{quote} Let $L \subseteq \T_\Sigma$.
    If for each $p \in \mathbb{N}_+$, there exists $\xi \in L$ with $\height(\xi) \ge p$ such that for every  $c',c \in \C_\Sigma$, $\zeta \in \T_\Sigma$ with $\xi = c'[c[\zeta]]$, $\height(c) > 0$, and $\height\big(c[\zeta]\big) < p$, there exists an $n \in \mathbb{N}$ such that $c'\big[c^n[\zeta]\big] \not\in L$,
    then $L$ is not recognizable.
    \end{quote}
        
Let us give an example of this application (cf. \cite[Ex. 4.2]{hopmotull14}).

\begin{example}\rm \label{ex:pumping-lemma-positive} We consider the string ranked alphabet $\Sigma = \{\sigma^{(1)}, \gamma^{(1)}, \alpha^{(0)}\}$ and the $\Sigma$-tree language
  \[L = \{\xi \in \T_\Sigma \mid  |\pos_\gamma(\xi)| = | \pos_\sigma(\xi)|\}\enspace.
    \]

Now we apply the contraposition of Corollary \ref{cor:pumping-lemma-for-runs}  (for $\B=\Boole$). For this let $p \in \mathbb{N}_+$. 
We consider the tree $\xi = \sigma^{p+1}\gamma^{p+1}\alpha$. Clearly, $\xi \in L$ and $\height(\xi) \ge p$.

Now we consider an arbitrary decomposition of $\xi$ of the form $\xi = c'\big[c[\zeta]\big]$ for some $c',c \in \C_\Sigma$, $\zeta \in \T_\Sigma$ such that $\height(c) > 0$ and $\height\big(c[\zeta]\big) < p$. By our choice of $\xi$, we have that  $c = \gamma^k(z)$ for some $k \ge 1$ (we recall that $\height(z) = 0$).

Now we choose $n=2$. Obviously, $c'[c^2[\zeta]]$ contains more $\gamma$s than $\sigma$s, hence $c'[c^2[\zeta]] \not\in L$. Hence, by the contraposition of Corollary \ref{cor:pumping-lemma-for-runs} (for $\B=\Boole$), the tree language $L$ is not recognizable.
\hfill$\Box$
\end{example}

It will turn out later (when Theorem \ref{thm:recognizable=run=init-positive} is available), that $\supp(\Rec^\mathrm{run}(\Sigma,\B)) = \Rec(\Sigma)$ for each positive semiring $\B$, hence Corollary \ref{cor:pumping-lemma-for-runs} is the same slight improvement of the classical pumping lemma for recognizable $\Sigma$-tree languages \cite[Lm.~2.10.1]{gecste84} which we described for the case $\B=\Boole$.

\section[Pumping lemma for recognizable wtl over fields]{Pumping lemma for recognizable weighted tree languages over fields}
\label{sec:pumping-lemma-fields}

Here we deal with weighted tree automata over fields and a pumping lemma which guarantees the existence of infinitely many trees in the support of the semantics of such wta. The pumping lemma is based on \cite{reu80,berreu82}.

\label{p:pumping-for-fields}
\begin{quote}\it In the rest of this section, we let $\B=(B,\oplus,\otimes,\0,\1)$ be an arbitrary field and  $\cA= (Q,\delta,F)$ be an arbitrary $(\Sigma,\B)$-wta with $|Q|=n$. 
\end{quote}

We recall that
\begin{compactitem}
\item $\B^Q=(B^Q,\oplus,\0^Q)$ is a  $\B$-semimodule (cf. Observation \ref{obs:B-to-Q-is-semimodule}),
\item $(\C_\Sigma,\circ_z,z)$ is the free monoid  with generating set $\e\C_\Sigma$ (cf. Lemma \ref{lm:freely-generated-context}), and
\item $\h_\cA^\C: \C_\Sigma \to \cL(\B^Q,\B^Q)$ is the  unique monoid homomorphism from $(\C_\Sigma,\circ_z,z)$  to the monoid $(B^Q \to B^Q,\circ,\id_{B^Q})$ which extends $g$ (cf. Section \ref{sect:splitting-properties-of-hA}), where 
\[
  g(e)(v) = \delta_\cA(\sigma)(\h_\cA(\xi_1),\ldots,\h_\cA(\xi_{i-1}),v,\h_\cA(\xi_{i+1}),\ldots,\h_\cA(\xi_k)) \enspace,
\]
for every elementary context $e \in \e\C_\Sigma$ with $e= \sigma(\xi_1,\ldots,\xi_{i-1},z,\xi_{i+1},\ldots,\xi_k)$ and $v \in B^Q$.
\end{compactitem}


\subsection{Linear recurrence equation}

Weighted tree languages that are recognizable over fields satisfy a particular linear recurrence equation.
Here we do not distinguish between the linear mapping $\h_\cA^\C(c)$ where $c\in \C_\Sigma$ and the matrix $\psi(\h_\cA^\C(c))$, where $\psi$ is the monoid isomorphism defined on page \pageref{p:def-psi-prime}. In particular, $\charp_{\h_\cA^\C(c)}(x)$ denotes the characteristic mapping of the matrix $\psi(\h_\cA^\C(c))$.

\begin{lemma-rect}\rm \label{lm:prop-9.3-berreu82}  Let $\Sigma$ be a ranked alphabet. Moreover, let $\B=(B,\oplus,\otimes,\0,\1)$ be a field and $\cA$ be a $(\Sigma,\B)$-wta with $|Q|=n$.  Moreover, let $c \in \C_\Sigma$ and let $\charp_{\h_\cA^\C(c)}(x) = \bigoplus_{i \in [0,n]} b_{i} . x^i$ where $b_n=(-\1)^n$. Then, for every $c' \in \C_\Sigma$, $\alpha \in \Sigma^{(0)}$, and $\ell \in \mathbb{N}$, we have \\
\[
(-\1)^n \otimes \sem{\cA}(c'[c^{\ell+n}[\alpha]]) \oplus b_{n-1} \otimes  \sem{\cA}(c'[c^{\ell+n-1}[\alpha]]) \oplus \cdots \oplus b_0 \otimes  \sem{\cA}(c'[c^\ell[\alpha]]) = \0 \enspace.\]
           \end{lemma-rect}

           \begin{proof}  For each $c \in \C_\Sigma$, we abbreviate  $\h_\cA^\C(c)$ by $\semst{c}$.

             Now let $c \in \C_\Sigma$ and $\charp_{\semst{c}}(x) = \bigoplus_{i \in [0,n]} b_{i} . x^i$ where $b_n=(-\1)^n$.
             By Theorem \ref{thm:Cayley-Hamilton} (i.e., the theorem of Cayley and Hamilton), we have $\charp_{\semst{c}}(\semst{c}) = \mathrm{M}_\0$.
 
             Now let $c' \in \C_\Sigma$, $\alpha \in \Sigma^{(0)}$, and $\ell \in \mathbb{N}$. Then we can calculate as follows:
             \begingroup
             \allowdisplaybreaks
             \begin{align*}
               & \charp_{\semst{c}}(\semst{c}) = \mathrm{M}_\0\\
               \Rightarrow \ &  \bigoplus_{i \in [0,n]} b_{i} \cdot \semst{c}^i = \mathrm{M}_\0
               \tag{we use $\oplus$ because $b_{i} \cdot \semst{c}^i$ is a matrix}\\
               \Rightarrow \ &  \big(\bigoplus_{i \in [0,n]} b_{i}\cdot \semst{c}^i\big) \cdot \h_\cA(\alpha) = \0^Q \tag{\text{matrix-vector multiplication with $\h_\cA(\alpha)$}}\\
               \Rightarrow \ &  \semst{c'[c^\ell]} \cdot \Big( \big(\bigoplus_{i\in[0,n]} b_{i} \cdot \semst{c}^i\big) \cdot \h_\cA(\alpha)\Big) = \0^Q \tag{matrix-vector multiplication with $\semst{c'[c^\ell]}$}\\
               \Rightarrow \ &  \Big(\semst{c'[c^\ell]} \cdot \big(\bigoplus_{i\in[0,n]} b_{i} \cdot \semst{c}^i\big)\Big) \cdot \h_\cA(\alpha) = \0^Q
               \tag{by associativity in a matrix-matrix-vector product}\\
               \Rightarrow \ &  \Big(\bigoplus_{i\in[0,n]} b_{i} \cdot \semst{c'[c^\ell]} \cdot \semst{c}^i\Big) \cdot \h_\cA(\alpha) = \0^Q  \tag{by distributivity} \\
               \Rightarrow \ &  \Big(\bigoplus_{i \in [0,n]} b_{i} \cdot \semst{c'[c^{\ell +i}]}\Big) \cdot \h_\cA(\alpha) = \0^Q  \tag{because $\h_\cA^\C$ is a monoid homomorphism; recall that $\semst{c'[c^{\ell +i}]} = \h_\cA^\C(c'[c^{\ell +i}])$}\\
               \Rightarrow \ &  \bigoplus_{i \in [0,n]} b_{i} \cdot \semst{c'[c^{\ell +i}]} \cdot \h_\cA(\alpha) = \0^Q  \tag{by distributivity}\\
               \Rightarrow \ &  \bigoplus_{i \in [0,n]} b_{i} \cdot \semst{c'[c^{\ell +i}]}(\h_\cA(\alpha)) = \0^Q
                               \tag{by \eqref{equ:applcation-of-linear-mapping=matrix-vector-product}}\\
  \Rightarrow \ &  \bigoplus_{i \in [0,n]} b_{i} \cdot \h_\cA(c'[c^{\ell +i}[\alpha]]) = \0^Q
                  \tag{\text{by Lemma \ref{lm:hcACchAxi=hAcxi}}}\\
                \Rightarrow \ &  \Big(\bigoplus_{i \in [0,n]} b_{i}\cdot \h_\cA(c'[c^{\ell +i}[\alpha]])\Big) \cdot F = \0
                                \tag{\text{by multiplying the vector $F$ to both sides}}\\
               \Rightarrow \ & \bigoplus_{i\in[0,n]} b_{i}\otimes \Big(\h_\cA(c'[c^{\ell +i}[\alpha]]) \cdot F\Big) = \0
                               \tag{because the mapping $f: B^Q \to B$ defined by $f(v)= v \cdot F$ is linear}\\
               \Rightarrow \ &  \bigoplus_{i\in[0,n]} b_{i}\otimes \sem{\cA}(c'[c^{\ell +i}[\alpha]]) = \0
                               \tag{by definition of $\sem{\cA}$}  \enspace.
             \end{align*}
             \endgroup
                        \end{proof}

\begin{example}\rm \label{ex:continuation-charact-polynomial} We continue with Example \ref{ex:number-of-occ-context-field} and consider the $(\Sigma,\Nat)$-wta $\cA$, which computes the mapping $\#_{\sigma(.,\alpha)}$, i.e., the number of occurrences of the pattern $\sigma(.,\alpha)$ in a given tree. Here we want to show the characteristic polynomial $\charp_{N_c}(x) = \mathrm{det}(N_c - x\mathrm{M}_1)$ of the $(Q\times Q)$-matrix $N_c$ (cf. \eqref{equ:matrix-Nc}). Since this concept is only defined if the weight algebra is a field, we  have to view $\cA$ as a $(\Sigma,\Ratnum)$-wta. This is possible because $\Ratnum$ is an extension of $\Nat$ (cf. Section \ref{sect:extension-of-weight-structure}).

Let $c \in \C_\Sigma$. We show $N_c$ and $N_c - x\mathrm{M}_1$: 
\[
  N_c =
  \begin{pmatrix}
    1 & 0 & 0\\
    0 & a(c) & 0\\
    \#_{\sigma(.,\alpha)}(c) & b(c) & 1
  \end{pmatrix}
  \ \ \text{ and } \ \ 
   N_c - x\mathrm{M}_1 =
  \begin{pmatrix}
    1-x & 0 & 0\\
    0 & a(c)-x & 0\\
    \#_{\sigma(.,\alpha)}(c) & b(c) & 1-x
    \end{pmatrix}\enspace.
  \]
  Then
  \[
\mathrm{det}(N_c - x\mathrm{M}_1) = -x^3 + (a(c)+2)\cdot x^2 -x + a(c) \enspace.
    \]
    Thus, $\charp_{N_c}(x)$  has the following form: 
  \begin{compactitem}
  \item if $c \ne z$, then $a(c)=0$ and $\charp_{N_c}(x) = -x^3 + 2x^2 -x$ and
    \item if $c =z$, then $a(c)=1$ and $\charp_{N_c}(x) = -x^3 + 3x^2 - 3x + 1$.
    \end{compactitem}
    Due to Lemma \ref{lm:prop-9.3-berreu82}, we obtain the following linear recurrence equations for $\#_{\sigma(.,\alpha)}$ for every $c' \in \C_\Sigma$ and every $\ell \in \mathbb{N}$:
    \begin{align*} 
      &\text{if $c\not= z$:} \ \ \  \#_{\sigma(.,\alpha)}(c'[c^{\ell+3}[\alpha]]) = 2 \cdot \#_{\sigma(.,\alpha)}(c'[c^{\ell+2}[\alpha]]) - \#_{\sigma(.,\alpha)}(c'[c^{\ell+1}[\alpha]])\\
     &\text{if $c=z$:} \ \ \   \#_{\sigma(.,\alpha)}(c'[\alpha]) = 3 \cdot \#_{\sigma(.,\alpha)}(c'[\alpha]) - 3 \cdot \#_{\sigma(.,\alpha)}(c'[\alpha]) + 1 \cdot \#_{\sigma(.,\alpha)}(c'[\alpha]) \enspace, 
    \end{align*}
    where the second equality is a tautology.
    \hfill $\Box$
    \end{example}

In the following corollary, $\height: \T_\Sigma \to \mathbb{N}$ is the weighted tree language defined in Example \ref{ex:height}.                        

\index{height}                            
\begin{corollary-rect}\label{cor:height-not-rec-field}\rm \cite[Ex.~9.2]{berreu82} For the ranked alphabet $\Sigma = \{\sigma^{(2)}, \alpha^{(0)}\}$, we have that $\height \not\in \Rec(\Sigma,\Realnum)$.
\end{corollary-rect}
\begin{proof}   We prove by contradiction. For this we assume that $\height \in \Rec(\Sigma,\Realnum)$. Then there exists a $(\Sigma,\Realnum)$-wta $\cA$ such that $\height = \sem{\cA}$. Let $n = |Q|$.

      Let $\xi \in \T_\Sigma$ be a tree with $\height(\xi) = n$. Due to the definition of $\Sigma$ such a $\xi$ exists.   We consider the two elementary $\Sigma$-contexts $e_1= \sigma(z,\xi)$ and $e_2 = \sigma(z,\alpha)$. For each $\ell \in \mathbb{N}$ we define the tree $\xi_\ell = e_1[e_2^\ell[\alpha]]$ in $\T_\Sigma$. Then
      \begin{equation*}
        \height(\xi_\ell) = 1 + \max(\ell,n) \ \  \text{ for every $\ell \in \mathbb{N}$} \enspace. 
        \end{equation*}

By Lemma \ref{lm:prop-9.3-berreu82}, there exist $b_1,\ldots,b_n \in \mathbb{R}$ such that 
   \begin{equation}
(-1)^n \cdot \height(\xi_{\ell+n}) + b_1 \cdot  \height(\xi_{\ell+n-1}) + \cdots + b_n \cdot  \height(\xi_\ell) = 0 \ \  \text{ for every $\ell \in \mathbb{N}$}. \label{eq:height-recurrent-equation}
\end{equation}
By using \eqref{eq:height-recurrent-equation} for $\ell=0$ and $\ell=1$, we obtain
   \begin{equation}
(-1)^n \cdot (1+n)  + b_1 \cdot  (1 + n) + \cdots + b_n \cdot  (1 + n) = 0 \enspace, \text{ and}\label{eq:1}
\end{equation}
   \begin{equation}
(-1)^n \cdot (1+(n+1))  + b_1 \cdot  (1 + n) + \cdots + b_n \cdot  (1 + n) = 0 \enspace, \label{eq:2}
\end{equation}
respectively. Then Equations \eqref{eq:1} and \eqref{eq:2} imply that $n = n +1$ which is a contradiction.
  \end{proof}
  
It follows that, for $\Sigma = \{\sigma^{(2)}, \alpha^{(0)}\}$ and any strong bimonoid $\B$ such that  $\Realnum$ is an extension of $\B$, the weighted tree language $\height$ is not recognizable by any  $(\Sigma,\B)$-wta (cf. Observation \ref{obs:extension-of-weight-structure}). In particular, $\height$ is not  recognizable by any $(\Sigma,\Ratnum)$-wta and  $(\Sigma,\Nat)$-wta. Recall that, at the same time, there is a $(\Sigma,\Natmaxplus)$-wta which recognizes $\height$ (cf. Example \ref{ex:height}).


\subsection{The pumping lemma}

Lemma \ref{lm:freely-generated-context} justifies  to use the following theorem and lemma from \cite{reu80} (using $A = \e\C_\Sigma$). We will use the fact that a matrix $M \in  B^{Q\times Q}$ is pseudo-regular if and only if $x^2$ does not divide its characteristic polynomial $\charp_M$  (cf. \cite[Prop.~1]{reu80}).
We do not show the proof of this statement.

\begin{theorem}{\rm \cite[Thm.~3]{reu80}} \label{thm:Thm3-reu80}  There exists $N \in \mathbb{N}$ such that, for every set $A$, monoid homomorphism $\mu: A^* \to B^{Q \times Q}$, string $w \in A^*$ of length at least $N$, the string $w$ has a factor $v \ne \varepsilon$ such that $\mu(v)$ is a pseudo-regular matrix.
\end{theorem}

\begin{lemma}\rm \cite[Lm.~1]{reu80} \label{lm:Lm1-reu80} Let $M \in  B^{Q \times Q}$, and $\omega, \kappa \in B^Q$.  If $M$ is pseudo-regular and $\omega \cdot M \cdot \kappa \ne \0$, then there exist infinitely many $\ell \in \mathbb{N}$ such that $\omega \cdot M^\ell \cdot \kappa \ne \0$.
  \end{lemma}
\begin{proof}  Let
             \begin{equation}\nonumber
\charp_{M}(x) = (-\1)^n . x^n \oplus b_{n-1} .  x^{n -1} \oplus \cdots \oplus 
b_{1}.  x \oplus b_0 \enspace,
\end{equation}
where $n=|Q|$.

By  our assumption $b_{1}\ne \0$ or $b_0\ne \0$.
Let us abbreviate $\omega \cdot M^\ell \cdot \kappa$ by $a_\ell$ for each $\ell \in \mathbb{N}$. We show that, for each $\ell \in \mathbb{N}$, if $a_\ell \ne \0$, then there exists $m>\ell$ such that $a_m \ne \0$. Since $a_1\ne \0$, this proves the lemma.

Let $\ell\in \mathbb{N}$ and assume that $a_\ell \ne \0$. In addition, assume that $b_0\ne \0$.
By Theorem \ref{thm:Cayley-Hamilton}, we have
\begin{equation}\nonumber
  (-\1)^n \cdot M^n \oplus b_{n-1} \cdot M^{n-1} \oplus \cdots
  \oplus  b_{1}\cdot  M \oplus b_0 \cdot M^0= \mathrm{M}_\0\enspace.
\end{equation}

By multiplying both sides first with $M^\ell$, second from the left with the vector $\omega$, and third from the right with the vector $\kappa$, we obtain
\begin{equation}\nonumber
  (-\1)^n \otimes a_{\ell+n} \oplus b_{n-1} \otimes  a_{\ell+n-1} \oplus \cdots
  \oplus  b_{1}\otimes a_{\ell+1}
\oplus b_0 \otimes a_\ell=\0\enspace.
\end{equation}
Since $b_0\not= \0$ and $a_\ell \not= \0$ and a field is zero-divisor free, we have that $b_0 \otimes a_\ell\not=\0$. Hence there exists $m\in [\ell+1,\ell+n]$ such that $a_m\ne \0$.

Now assume that $b_0= \0$ and 
$b_{1}\ne \0$. By our assumption $n>1$ and by Theorem \ref{thm:Cayley-Hamilton} (as above), we have
\begin{equation}\nonumber
(-\1)^n \otimes a_{\ell+n-1} \oplus b_{n-1} \otimes  a_{\ell+n-2} \oplus \cdots \oplus 
b_{1}\otimes a_{\ell}=\0\enspace.
\end{equation}
Then again $a_m\ne \0$ for some $m\in [\ell+1,\ell+n-1]$.
\end{proof}

 Let $\xi \in \T_\Sigma$, $c \in \C_\Sigma$, and $\alpha \in \Sigma^{(0)}$. The pair $(c,\alpha)$ is a \emph{walk in $\xi$} if $\xi = c[\alpha]$.
The \emph{depth of $(c,\alpha)$} is $\mathrm{depth}(c)$.
Let $c \in \C_\Sigma$. We recall that $c^0=z$, and for every $\ell \in \mathbb{N}$, we have $c^{\ell+1} = c[c^\ell]$.

\begin{theorem-rect}{\rm \cite[Thm.~9.2]{berreu82}}\label{thm:pumping-wta-fields}
 Let $\Sigma$ be a ranked alphabet. Moreover, let $\B=(B,\oplus,\otimes,\0,\1)$ be a field and let $r \in \Rec(\Sigma,\B)$. There exists a constant $N \in \mathbb{N}$ such that, for every  $\xi \in \supp(r)$ and every walk $(c,\alpha)$ in $\xi$ of depth at least $N$, there exist $c_1,c_2,c_3 \in \C_\Sigma$ such that $c_2\ne z$, $c= c_1\circ_z c_2\circ_z c_3$, and the set $\{(c_1 \circ_z c_2^\ell \circ_z c_3)[\alpha]\mid \ell \in \mathbb{N}\} \cap \supp(r)$ is infinite.
\end{theorem-rect}

\begin{proof} Then there exists a $(\Sigma,\B)$-wta $\cA=(Q,\delta,F)$ such that $\sem{\cA}=r$. We let $n=|Q|$. Then 
  \begin{equation}
\text{for each $\xi \in \T_\Sigma$, we have }r(\xi) = \h_\cA(\xi) \cdot F \ \enspace. \label{eq:r=hom}
    \end{equation}
  
    Since the two monoids $(\C_\Sigma,\circ_z,z)$ and $(\e\C_\Sigma^*,\cdot,\varepsilon)$ are isomorphic (cf. proof of Lemma~\ref{lm:freely-generated-context}), we identify them. Also, the two monoids $(B^{Q \times Q},\cdot,\mathrm{M}_\1)$ and $(\cL(\B^Q,\B^Q),\circ,\id_{B^Q})$ are isomorphic (cf Subsection~\ref{sec:semimodules}), and we identify them also.
    
        Then we can apply Theorem \ref{thm:Thm3-reu80}, where we abbreviate $\h_\cA^\C(c)$ by $\semst{c}$ for each context $c \in \C_\Sigma$.  Let $N \in \mathbb{N}$ be the integer from that theorem. Moreover, let $\xi \in \supp(r)$ and $(c,\alpha)$ be a walk of $\xi$ of depth at least $N$. By Theorem \ref{thm:Thm3-reu80}, there exists
      $c_1,c_2,c_3 \in \C_\Sigma$ such that $c = c_1 \circ_z c_2 \circ_z c_3$ with $c_2\ne z$ and $\semst{c_2}$ is a pseudo-regular endomorphism. In particular, we have $\xi = c_1[c_2[c_3[\alpha]]]$.

      For each $\ell \in \mathbb{N}$ we define the element $u_\ell \in B$ by
      \[u_\ell = \gamma(\semst{c_1} ( \semst{c_2}^\ell ( \semst{c_3} (\h_\cA(\alpha) )))\enspace.\]
      In particular,
      \[
        u_1 = \gamma(\semst{c_1} ( \semst{c_2}( \semst{c_3} (\h_\cA(\alpha) ))) = \gamma(\h_\cA( c_1[c_2[c_3[(\alpha)]]] ))) = r(c_1[c_2[c_3[(\alpha)]]]) =r(\xi) \ne \0
        \]
where the second and third equality are due to Lemma \ref{lm:hcACchAxi=hAcxi} and \eqref{eq:r=hom}, respectively.

Thus, we can apply Lemma \ref{lm:Lm1-reu80} (using $\omega = \gamma \cdot \semst{c_1}$, $M=\semst{c_2}$ and $\kappa = \semst{c_3} \cdot \h_\cA(\alpha)$) and hence, there exist infinitely many $\ell$ such that $u_\ell\ne\0$.

Since \(u_\ell = r(c_1[c_2^\ell[c_3[(\alpha)]]])\) (by a similar calculation as for $u_1$), we obtain the statement of the theorem.
    \end{proof}

Theorem \ref{thm:pumping-wta-fields} can be used to show that certain tree languages cannot be the support of recognizable $\B$-weighted tree languages.

\begin{corollary-rect}\rm \label{cor:application-pump-fields} Let $\Sigma=\{\omega^{(1)},\sigma^{(1)},\alpha^{(0)}\}$.   For each field $\B$, the tree language $L=\{\sigma^\ell\omega^\ell(\alpha)\mid \ell\ge 0\}$ is not in $\supp(\Rec(\Sigma,\B))$.
\end{corollary-rect}
\begin{proof}
Let $\B$ be a field. We continue to prove by contradiction. For this, we assume that there exists an $r\in \Rec(\Sigma,\B)$ such that $L=\supp(r)$. Let $N$ be the number for $r$ appearing in Theorem \ref{thm:pumping-wta-fields} and consider the tree $\xi=\sigma^N\omega^N(\alpha)$ and the walk $(c,\alpha)$ in $\xi$, where $c=\sigma^N\omega^N(z)$. 
By Theorem \ref{thm:pumping-wta-fields}, there exist $c_1,c_2,c_3 \in \C_\Sigma$ such that $c_2\ne z$, $c= c_1\circ_z c_2\circ_z c_3$ and the set $\{(c_1 \circ_z c_2^\ell \circ_z c_3)[\alpha]\mid \ell \in \mathbb{N}\} \cap L$ is infinite. On the other hand, for any decomposition  $c= c_1\circ_z c_2\circ_z c_3$ of $c$ with 
$c_2\ne z$ we have $\{(c_1 \circ_z c_2^\ell \circ_z c_3)[\alpha]\mid \ell \in \mathbb{N}\} \cap L\subseteq \{(c_1 \circ_z  c_3)[\alpha],\xi\}$.
This is a contradiction, hence $L\not\in \supp(\Rec(\Sigma,\B))$.
\end{proof}

By similar arguments as  in Corollary \ref{cor:application-pump-fields}, we can show that the tree language $\mathrm{FB}$ is not in $\supp(\Rec(\Sigma,\B))$. The tree language $\mathrm{FB}$ of \emph{fully balanced trees over $\{\sigma^{(2)},\alpha^{(0)}\}$} is  the smallest $\Sigma$-tree language $L$ satisfying (i) $\alpha \in L$, and (ii) $\sigma(\xi,\xi)\in L$ for each $\xi\in L$.

%% file: normal-forms.tex
\addtocontents{toc}{\protect\pagebreak}
\chapter{Normal forms of wta} 
\label{ch:normal-forms}

In this chapter we prove some useful normal forms of wta. More precisely, we define several useful properties which a given wta either has or does not have. Then, for each of these properties, we show that, under some conditions on the weight algebra, any given wta can be transformed into an equivalent wta which has this property. In this sense, the given wta is turned into a normal form. Sometimes normal forms of wta are useful for reducing the complexity of proofs which start from a given wta.

We deal with the following properties: trim, total, root weight normalized, slim, and crisp. However, we start with an easy transformation.

\section{Dropping states which are not useful for the initial algebra semantics}

    \begin{lemma}\label{lm:wta-getting-rid-of-useless-state} \rm
Let $\cA= (Q,\delta,F)$ be a $(\Sigma,\B)$-wta and $q_0 \in Q$ such that, for each $\xi \in \T_\Sigma$, we have $\h_\cA(\xi)_{q_0}=\0$. Moreover, let $\cA_{-q_0}=(Q',\delta',F')$ be the $(\Sigma,\B)$-wta with $Q' = Q \setminus \{q_0\}$, $(\delta')_k = \delta_k|_{(Q')^k \times \Sigma^{(k)} \times Q'}$
for each $k \in \mathbb{N}$ and $F' = F|_{Q'}$.
Then $\initialsem{\cA} = \initialsem{\cA_{-q_0}}$. If $\cA$ is bu-deterministic, then so is $\cA_{-q_0}$. Moreover, if $\cA$ has unit root weights, then $\cA_{-q_0}$ has unit root weights.
\end{lemma}
\begin{proof} First, by induction on $\T_\Sigma$, we can easily show:
  \begingroup
\allowdisplaybreaks
\begin{equation}\label{equ:dropping-a-useless-state-leaves-hA-invariant}
\text{For every $\xi \in \T_\Sigma$ and $q \in Q'$, we have: } \h_{\cA_{-q_0}}(\xi)_q = \h_\cA(\xi)_q\enspace.
\end{equation}
\endgroup
Then, for each $\xi \in \T_\Sigma$, we have:
\begin{align*}
\initialsem{\cA}(\xi) & = \bigoplus_{q \in Q} \h_\cA(\xi)_q \otimes F_q = \bigoplus_{q \in Q'} \h_\cA(\xi)_q \otimes F_q
\tag{because $\h_\cA(\xi)_{q_0} \otimes F_{q_0} = \0$} \\
& = \bigoplus_{q \in Q'} \h_{\cA_{-q_0}}(\xi)_q \otimes (F')_q
       \tag{by \eqref{equ:dropping-a-useless-state-leaves-hA-invariant} and definition of $F'$}\\
       &= \initialsem{\cA_{-q_0}}(\xi) \enspace.
       \qedhere
\end{align*}
\end{proof}

  \section{Trimming a wta}
  \label{sect:trimming-a-wta}

In this section we elaborate two trimming methods for wta. Intuitively, a trimming method takes a wta as input and constructs a run equivalent trim wta as output. A wta is trim if it contains only useful states, i.e., states which occur in a successful run on some input tree.

In the literature trimming methods are known for finite-state automata \cite[Prop.~1.9]{sak09}, context-free grammars \cite[Thm.~3.2.3]{har78}, regular tree grammars \cite[Prop.~2.1.3]{comdaugiljaclugtistom08} and \cite[Lm.~A.2.6]{dre06}, fta \cite[Prop.~1.1]{sei89}, weighted string automata over semirings \cite[p.~408]{sak09},  and wta over strong bimonoids \cite[Thm.~4.2]{drofulkosvog21}. Except for the last two shown references, each of them deals with the unweighted case.

Our first trimming method is applicable to wta over zero-cancellation free strong bimonoids and it leads to a ``strong form'' of trim wta. Our second method is applicable to wta over arbitrary strong bimonoids and it leads to a ``weak form'' of trim wta.  We start by defining some useful concepts.

\subsection{Basic definitions}\label{subsect:basic-definitions-runs}

\index{run!successful}
\index{run!local-successful}
\index{state!useful}
\index{state!local-useful}
\index{useful}
\index{local-useful}
Let $\cA= (Q,\delta,F)$ be a $(\Sigma,\B)$-wta.  Let  $q \in Q$, $\xi \in \T_\Sigma$, and $\rho \in \R_\cA(q,\xi)$.
We say that 
\begin{compactitem}
  \item $\rho$ is \emph{successful} if $\wt(\xi,\rho) \otimes F_q \ne \0$,
 \item $\rho$ is \emph{local-successful} if (a)  for each $v \in \pos(\xi)$ we have $\delta_k(\rho(v1) \cdots \rho(vk),\xi(v),\rho(v)) \ne \0$ where $k = \rk_\Sigma(\xi(v))$ and (b) $F_q \ne\0$.
 \end{compactitem}
 Let $p \in Q$. We say that
 \begin{compactitem}
\item $p$ is \emph{useful} if there exist $\xi\in \T_\Sigma$  and a successful run $\rho\in \R_\cA(\xi)$ such that $p\in \im(\rho)$.
  
\item $p$ is \emph{local-useful} if  there exist $\xi\in \T_\Sigma$  and a local-successful run $\rho\in \R_\cA(\xi)$ such that $p\in\im(\rho)$.

\item $p$ is \emph{local-accessible} if there exist  $\xi\in \T_\Sigma$ and $\rho \in \R_\cA(p,\xi)$ such that, for each $v \in \pos(\xi)$, we have $\delta_k(\rho(v1) \cdots \rho(vk),\xi(v),\rho(v)) \ne \0$ where $k = \rk_\Sigma(\xi(v))$.
    \end{compactitem}

Clearly, useful implies  local-useful. Moreover, if $\B$ is zero-divisor free, then the two notions are equivalent;  thus, in particular, this holds if $\B$ is the Boolean semiring $\Boole$.

\index{trim}
\index{local-trim}
  A  $(\Sigma,\B)$-wta $\cA$ is called \emph{trim} (and  \emph{local-trim}) if each of its states is useful (and local-useful, respectively).  
  We note that the notion of trim
  in \cite{drofulkosvog21} corresponds to the notion
  local-trim.
  In the next observation we formally compare the two different trim properties (cf. Figure \ref{fig:illustration-for-trim-wta}).

  \begin{figure}
    \centering
    \begin{tabular}{lll}
      trim wta  \hspace{4mm}  &$\longrightarrow$  \hspace{4mm}  local-trim wta   &$\longrightarrow$ \hspace{4mm} wta\\[-2mm]
                                     &$\longleftarrow$ \hspace*{-5mm}$\not$       &$\longleftarrow$ \hspace*{-5mm}$\not$
    \end{tabular}
\caption{\label{fig:illustration-for-trim-wta} An illustration of the relationships between trim wta, local-trim wta, and wta where $\longrightarrow$ means ``implies'' (cf. Observation \ref{obs:relationship-trim-defs} (1)-(4)).}
    \end{figure}

  \begin{observation}\rm \label{obs:relationship-trim-defs}  The following five statements hold.
      \begin{compactenum}
      \item[(1)] Each trim $(\Sigma,\B)$-wta  is local-trim.
      \item[(2)] There exist a trivial ranked alphabet $\Sigma$ and  a $(\Sigma,\B)$-wta $\cA$ such that $\runsem{\cA}\ne \widetilde{\0}$ and $\cA$ is not local-trim.
         \item[(3)] There exist a string ranked  alphabet $\Sigma$, a strong bimonoid $\B$, and a local-trim $(\Sigma,\B)$-wta which is not trim.
      \item[(4)]  If $\B$ is zero-divisor free, then for each $(\Sigma,\B)$-wta $\cA$ we have:  $\cA$ is trim iff $\cA$ is local-trim.
       \end{compactenum}
     \end{observation}
     \begin{proof} Statements (1) and (4) are obvious.

       Proof of Statement (2): We consider the ranked alphabet $\Sigma = \{\alpha^{(0)}\}$ and  the $(\Sigma,\B)$-wta $\cA = (Q,\delta,F)$, where  $Q = \{q,p\}$, $\delta(\varepsilon,\alpha,q)=\1$,
       $\delta(\varepsilon,\alpha,p)=\0$, and $F_q =\1$ and $F_p=\0$. Then $\runsem{\cA}(\alpha)=\1$, i.e., $\runsem{\cA}\ne \widetilde{\0}$. We note that $p$ is not local-useful and hence $\cA$ is not local-trim.

       Proof of Statement (3): We consider the string ranked alphabet $\Sigma = \{\gamma^{(1)}, \alpha^{(0)}\}$ and the ring $\Intfour = (\{0,1,2,3\},+_4,\cdot_4,0,1)$ defined in Example \ref{ex:semirings}(\ref{def:ring-Zmod4Z}). This $\Intfour$ is not zero-divisor free because $2\cdot_4 2 =0$.

Then we let $\cA = (Q,\delta,F)$ be the $(\Sigma,\Intfour)$-wta with $Q = \{q,p\}$ and for each $f,g \in Q$ we define
       \[\delta_0(\varepsilon,\alpha,f)=
         \begin{cases}
           1 & \text{if } f = q\\
           0 & \text{otherwise}
         \end{cases}
       \ \ \text{ and } \ \
        \delta_1(f,\gamma,g)=
         \begin{cases}
           2 & \text{if } fg \in \{qp, pq\} \\
           0 & \text{otherwise}
         \end{cases}
       \]
       and $F_q =1$ and $F_p=0$. Then both $p$ and $q$ are local-useful, because for $\xi_1 = \gamma^2(\alpha)$, the run $\rho_1 \in \R_\cA(\xi_1)$ with $\rho_1(\varepsilon)=\rho(11)=q$ and $\rho(1)=p$ is local-successful and $\im(\rho_1)=\{p,q\}$. 
        Hence $\cA$ is local-trim. However, $\cA$ is not trim, because for $\xi_1$ there does not exist a successful run $\rho \in \R_\cA(\xi_1)$ such that $p\in \im(\rho)$.  In particular, $\rho_1$ is not successful because $\wt(\xi_1,\rho_1)=0$.
       \end{proof}

\subsection{Trimming for zero-cancellation free strong bimonoids}
\label{sec:t1}

Here we show our first trimming method. It takes a $(\Sigma,\B)$-wta over some zero-cancellation free strong bimonoid $\B$, and constructs a run equivalent trim $(\Sigma,\B)$-wta. The method employs the pumping lemma to cut out in a careful way contexts from successful runs; due to zero-cancellation freeness of $\B$, this results in smaller successful runs. As preparation we characterize useful states.

\begin{lemma} \rm \label{lm:shortening} Let $\B$ be zero-cancellation free, $\cA=(Q,\delta,F)$ be a $(\Sigma,\B)$-wta, and $p \in Q$. Then $p$ is useful if and only if there exist $q \in Q$, $\xi\in \T_\Sigma$,  and a successful run $\rho\in \R_\cA(q,\xi)$ such that $p\in \im(\rho)$ and $\height(\xi) < 2|Q|$.
\end{lemma}
\begin{proof} The ``if''-direction holds trivially. For the ``only-if''-direction we prove the following.
\begin{eqnarray}
\begin{aligned}
 & \text{For every $q\in Q$, $\xi\in \T_\Sigma$, $v\in \pos(\xi)$, and successful run $\kappa\in \R_\cA(q,\xi)$,}  \\
 & \text{there exist $\widehat{\xi}\in \T_\Sigma$, $\widehat{v}\in \pos(\widehat{\xi})$, and successful run $\widehat{\kappa}\in \R_\cA(q,\widehat{\xi})$}  \\
 & \text{such that $\height(\widehat{\xi})< 2|Q|$,  and $\widehat{\kappa}(\widehat{v})=\kappa(v)$.} 
  \end{aligned} \label{eq:trimming}
  \end{eqnarray}

Let $q \in Q$, $\xi \in \T_\Sigma$, $v \in \pos(\xi)$,  and  $\kappa \in \R_\cA(q,\xi)$  be a successful run.
  We prove \eqref{eq:trimming} in three steps. Intuitively, in Step 1 we cut out repeatedly contexts above $v$ such that eventually the state $\kappa(v)$ occurs at $\widehat{v}$ and $|\widehat{v}|< |Q|$. In Step 2 we cut  out repeatedly contexts below $\widehat{v}$ such that eventually the subtree below $\widehat{v}$ has height smaller than $|Q|$. In Step 3 we cut  out repeatedly contexts aside of  $\widehat{v}$ such that eventually  we obtain the desired result. Since $\B$ is zero-cancellation free, cutting out a context transforms a successful run into a successful run. We illustrate the three steps in Figures \ref{fig:step1}, \ref{fig:step2}, and \ref{fig:step3}, respectively.

  \begin{figure}
            \centering
        \begin{tikzpicture}
            \footnotesize
            \draw (0.5,0.5) to (2.5,2) to (5,0.5);

            \fill[gray!20]
                (1,0) to (2.75,1.25) to (4.5,0);

            \node[circle,fill=black,inner sep=0pt,minimum size=3pt] at (2.75,1.25) (u) {} node[right = -1pt of u] {$u$} ;

            \draw[fill=white,fill opacity=1]
                (0.75,-0.75) to (2.25,0.5) to (4.5,-0.75) to (0.75,-0.75);

            \node[circle,fill=black,inner sep=0pt,minimum size=3pt] at (2.25,0.5) (uw) {} node[right = 1pt of uw] {$uw$};

            \node[circle,fill=black,inner sep=0pt,minimum size=3pt] at (2.5,-0.5) (v) {}  node[right = -2pt of v] {$v=uwx$};

            \draw[decorate, decoration={snake,segment length=12mm}]
 (2.5,2) -- (2.75,1.25);
            \draw[decorate, decoration={snake,segment length=13mm}]
 (2.75,1.25) -- (2.25,0.5);
            \draw[decorate, decoration={snake,segment length=15mm}]
 (2.25,0.5) -- (2.5,-0.5);

            \node at (1.5,1.75) {$\xi$};
            \node at (3.5,1.75) {$\kappa$};

            \node at (2.25,1.15) {$c'$};
            \node at (3.5,1.15) {$\rho'$};

            \node at (2,0.5) {$c$};
            \node at (3.25,0.5) {$\rho$};

            \node at (1,-0.25) {$\zeta$};
            \node at (4,-0.25) {$\theta$};

            \draw (6.5,0.5) to (8.5,2) to (11,0.5);
            \draw  (7.25,0.25) to (8.75,1.25) to (10.75,0.25) to (7.25,0.25);

            \node[circle,fill=black,inner sep=0pt,minimum size=3pt] at (8.75,1.25) (u2) {} node[right = -1pt of u2] {$u$} ;

            \draw[decorate, decoration={snake,segment length=12mm}]
            (8.5,2) -- (8.75,1.25);

            \node[circle,fill=black,inner sep=0pt,minimum size=3pt] at (9,0.5) (v2) {}  node[right = -2pt of v2] {$\hat{v}=ux$};

            \draw[decorate, decoration={snake,segment length=12mm}]
 (8.75,1.25) -- (9,0.5);

            \node at (7.5,1.75) {$\xi'$};
            \node at (9.5,1.75) {$\kappa'$};

            \node at (8.25,1.15) {$c'$};
            \node at (9.5,1.15) {$\rho'$};

            \node at (7.25,0.45) {$\zeta$};
            \node at (10.75,0.45) {$\theta$};

            \draw 
                (11.25,2) -- (11.25,0.5) node[midway,right] {$< |Q|$}
                (11.15,2) -- (11.35,2)
                (11.15,0.5) -- (11.35,0.5);

            \draw[decorate, decoration={snake,amplitude=1pt},->] (5.25,1) -- (6.25,1) node[midway,above] {Step 1$^*$};

        \end{tikzpicture}
    \caption{\label{fig:step1} Cutting out  contexts above $v$ repeatedly.}
  \end{figure}

     \begin{figure}
        \centering
        \begin{tikzpicture}
            \footnotesize
            \draw (0.5,-.25) to (2.5,1.25) to (5,-.25);

            \node at (1.5,1) {$\xi'$};
            \node at (3.5,1) {$\kappa'$};

            \draw (0.75,-1.5) to (2.75,0.5) to (4.75,-1.5) to (0.75,-1.5);
            \node[circle,fill=black,inner sep=0pt,minimum size=3pt] at (2.75,0.5) (v) {} node[right = -1pt of v] {$\hat{v}$} ;
            \fill[gray!20] (1.25,-1.5) to (2.75,0) to (4.25,-1.5);
            \node[circle,fill=black,inner sep=0pt,minimum size=3pt] at (2.75,0) (vu) {} node[right = -1pt of vu] {$\hat{v}u$} ;
            \draw[fill=white,fill opacity=1] (2,-1.5) to (2.75,-0.75) to (3.5,-1.5);
            \node[circle,fill=black,inner sep=0pt,minimum size=3pt] at (2.75,-0.75) (vuw) {} node[right = -1pt of vuw] {$\hat{v}uw$} ;

            \draw[decorate, decoration={snake,segment length=12mm,amplitude=4pt}] (2.5,1.25) -- (2.75,0.5);
            \draw[decorate, decoration={snake,segment length=8mm}] (2.75,0.5) -- (2.75,0);
            \draw[decorate, decoration={snake,segment length=12mm}] (2.75,0) -- (2.75,-0.75);

            \draw 
                (0.5,1.25) -- (0.5,0.5) node[midway,left] {$< |Q|$}
                (0.4,1.25) -- (0.6,1.25)
                (0.4,0.5) -- (0.6,0.5);

            \draw[decorate, decoration={snake,amplitude=1pt},->] (5.25,.5) -- (6.25,.5) node[midway,above] {Step 2$^*$};

            \draw (6.5,-.25) to (8.5,1.25) to (11,-.25);
            \node at (7.5,1) {$\xi''$};
            \node at (9.5,1) {$\kappa''$};

            \draw (7.25,-1) to (8.75,0.5) to (10.25,-1) to (7.25,-1);
            \node[circle,fill=black,inner sep=0pt,minimum size=3pt] at (8.75,0.5) (v2) {} node[right = -1pt of v2] {$\hat{v}$} ;
            \draw (8,-1) to (8.75,-0.25) to (9.5,-1);
            \node[circle,fill=black,inner sep=0pt,minimum size=3pt] at (8.75,-0.25) (vw) {} node[right = -1pt of vw] {$\hat{v}w$} ;

            \draw 
                (11.25,1.25) -- (11.25,0.5) node[midway,right] {$< |Q|$}
                (11.15,1.25) -- (11.35,1.25)
                (11.15,0.5) -- (11.35,0.5)
                (11.25,0.5) -- (11.25,-1) node[midway,right] {$ < |Q|$}
                (11.15,-1) -- (11.35,-1);
            \draw[decorate, decoration={snake,segment length=11mm}] (8.5,1.25) -- (8.75,0.5);
            \draw[decorate, decoration={snake,segment length=11mm}] (8.75,0.5) -- (8.75,-0.25);
        \end{tikzpicture}
        \caption{\label{fig:step2} Cutting out contexts below $\widehat{v}$ repeatedly.}
    \end{figure}

   \begin{figure}
              \centering
        \begin{tikzpicture}
            \footnotesize

            \draw (0.5,-.25) to (2.5,1.25) to (5,-.25);

            \node at (1.5,1) {$\xi''$};
            \node at (3.5,1) {$\kappa''$};
            
            \node at (7.5,1) {$\widehat{\xi}$};
            \node at (9.5,1) {$\widehat{\kappa}$};

            \node[circle,fill=black,inner sep=0pt,minimum size=3pt] at (2.25,.5) (lcp) {} node[right = -1pt of lcp,yshift=2pt] {$\mathrm{lcp}(\hat{v},x)$} ;
            \draw[decorate, decoration={snake,segment length=12mm}] (2.5,1.25) -- (2.25,.5);

            \draw 
                (.25,1.25) -- (.25,-.25) node[midway,left] {$< |Q|$}
                (.15,1.25) -- (.35,1.25)
                (.15,-.25) -- (.35,-.25);

            \draw[dashed] (.4,-.25) -- (3.75,-.25);

            \draw (3,-1) to (3.75,-.25) to (4.5,-1) to (3,-1);
            \node[circle,fill=black,inner sep=0pt,minimum size=3pt] at (3.75,-.25) (v) {} node[right = -1pt of v] {$\hat{v}$} ;

            \draw[decorate, decoration={snake,segment length=13mm}] (2.25,.5) -- (3.75,-.25);

            \draw 
                (4.75,-1) -- (4.75,-.25) node[midway,right] {$< |Q|$}
                (4.65,-1) -- (4.85,-1)
                (4.65,-.25) -- (4.85,-.25);

            \fill[gray!20] (-.25,-2) to (1.25,-.5) to (2.75,-2);
            \node[circle,fill=black,inner sep=0pt,minimum size=3pt] at (1.25,-.5) (u) {} node[right = -1pt of u,xshift=2pt,yshift=-2pt] {$\mathrm{lcp}(\hat{v},x)u$} ;

            \draw[decorate, decoration={snake,segment length=12mm}] (2.25,.5) -- (1.25,-.5);

            \draw[fill=white,fill opacity=1] (.75,-2) to (1.25,-1.25) to (1.75,-2) to (.75,-2);
            \node[circle,fill=black,inner sep=0pt,minimum size=3pt] at (1.25,-1.25) (w) {} node[right = -1pt of w] {$\mathrm{lcp}(\hat{v},x)uw$} ;

            \draw[decorate, decoration={snake,segment length=11mm}] (1.25,-.5) -- (1.25,-1.25);

            \node[circle,fill=black,inner sep=0pt,minimum size=3pt] at (1.1,-1.75) (x) {} node[right = -1pt of x] {$x$} ;
            \draw[decorate, decoration={snake,segment length=8mm}] (1.25,-1.25) -- (1.1,-1.75);

            \draw[decorate, decoration={snake,amplitude=1pt},->] (5.25,.5) -- (6.25,.5) node[midway,above] {Step 3$^*$};

            \draw (6.5,-1.25) to (8.5,1.25) to (11,-1.25) to (6.5,-1.25);
            
            \node[circle,fill=black,inner sep=0pt,minimum size=3pt] at (8.75,0) (v2) {} node[right = -1pt of v2] {$\hat{v}$} ;
            \draw[decorate, decoration={snake,segment length=10mm}] (8.5,1.25) -- (8.75,0);
            \draw 
                (11.25,1.25) -- (11.25,0) node[midway,right] {$< |Q|$}
                (11.15,1.25) -- (11.35,1.25)
                (11.15,0) -- (11.35,0)
                (11.25,0) -- (11.25,-1.25) node[midway,right] {$ < |Q|$}
                (11.15,-1.25) -- (11.35,-1.25);

        \end{tikzpicture}
    \caption{\label{fig:step3} Cutting out  contexts aside of $\widehat{v}$ repeatedly.}
  \end{figure}

 \underline{Step 1:}  We construct a $\xi'\in \T_\Sigma$, a $\widehat{v}\in \pos(\xi')$, and a successful run $\kappa'\in \R_\cA(q,\xi')$ such that $|\widehat{v}|< |Q|$ and $\kappa'(\widehat{v})=\kappa(v)$ as follows.

If $|v|< |Q|$, then we let $\xi'=\xi$, $\widehat{v}=v$, and $\kappa'=\kappa$ and we are ready.

Otherwise, there exist $u,x\in \mathbb{N}_+^*$ and  $w\in \mathbb{N}_+^+$ such that $v=uwx$ and $\kappa(u)=\kappa(uw)$. Let $c'=\xi|^u$, $c=(\xi|_u)|^w$, and $\zeta=\xi|_{uw}$, and let $\rho'=\kappa|^u$, $\rho=(\kappa|_u)|^w$, and $\theta=\kappa|_{uw}$. Then $\xi=c'[c[\zeta]]$ and $\kappa=\rho'[\rho[\theta]]$, where $\rho$ is a loop due to the condition $\rho(u)=\rho(uw)$. Thus, by  Theorem \ref{thm:decomposition-of-a-run} (equality \eqref{equ:pumping-equation} for $n=1$), we have
\[\wt(\xi,\kappa)\otimes F_q=l_{c',\rho'} \otimes l_{c,\rho} \otimes \wt(\zeta,\theta) \otimes r_{c,\rho} \otimes r_{c',\rho'}\otimes F_q.\]
Now we cut out the context $c$ and the corresponding loop $\rho$. Formally, let $\xi'=c'[\zeta]$ and $\kappa'=\rho'[\theta]$. By equality \eqref{equ:pumping-equation} for $n=0$, we obtain 
\[\wt(\xi',\kappa')\otimes F_q=l_{c',\rho'} \otimes \wt(\zeta,\theta)  \otimes r_{c',\rho'}\otimes F_q.\]
Since $\kappa$ is a successful run, i.e., $\wt(\xi,\kappa)\otimes F_q\ne\0$, and $\B$ is zero-cancellation free, we obtain that $\kappa'$ is also successful. Moreover, for $\widehat{v}=ux$, we have $\kappa'(\widehat{v})=\kappa(v)$ by the definition of $\kappa'$. If $|\widehat{v}|\ge |Q|$, then we repeat the above procedure with $\xi'$, $\widehat{v}$, and $\kappa'$. 
After finitely many steps, we obtain $\xi'$, $\widehat{v}$, and $\kappa'$ such that $|\widehat{v}|< |Q|$.

\underline{Step 2:} Given $\xi'\in \T_\Sigma$, $\widehat{v}\in \pos(\xi')$, and the successful run $\kappa'\in \R_\cA(q,\xi')$ constructed in Step 1, we construct $\xi''\in \T_\Sigma$ and successful run $\kappa''\in \R_\cA(q,\xi'')$ such that $\widehat{v}\in \pos(\xi'')$, $\kappa''(\widehat{v})=\kappa'(\widehat{v})$, and $\height(\xi''|_{\widehat{v}})< |Q|$ as follows.

If $\height(\xi'|_{\widehat{v}})< |Q|$, then we let $\xi''=\xi'$ and $\kappa''=\kappa'$ and we are ready.

Otherwise, there exist $u\in \mathbb{N}_+^*$ and $w\in \mathbb{N}_+^+$ such that $\widehat{v}uw\in \pos(\xi')$ and  $\kappa'(\widehat{v}u)=\kappa'(\widehat{v}uw)$.
Similarly to the construction in Step 1, by cutting out the context $(\xi'|_{\widehat{v}u})|^w$ and the loop 
$(\kappa'|_{\widehat{v}u})|^w$ from $\xi'$ and $\kappa'$, respectively, we obtain $\xi''\in \T_\Sigma$ and successful run $\kappa''\in \R_\cA(q,\xi'')$
such that $\widehat{v}\in \pos(\xi'')$, $\kappa''(\widehat{v})=\kappa'(\widehat{v})$ and $\size(\xi''|_{\widehat{v}})<\size(\xi'|_{\widehat{v}})$.
If $\height(\xi''|_{\widehat{v}})\ge |Q|$, then we repeat the above procedure with $\xi''$ and $\kappa''$. 
After finitely many steps, we obtain $\xi''$ and $\kappa''$ such that $\height(\xi''|_{\widehat{v}})< |Q|$.

\underline{Step 3:} Given $\xi''\in \T_\Sigma$, $\widehat{v}\in \pos(\xi'')$, and the successful run $\kappa''\in \R_\cA(q,\xi'')$ constructed in Step~2, we construct $\widehat{\xi}\in \T_\Sigma$ and the successful run $\widehat{\kappa}\in \R_\cA(q,\widehat{\xi})$ such that $\widehat{v}\in \pos(\widehat{\xi})$, 
$\widehat{\kappa}(\widehat{v})=\kappa''(\widehat{v})$, and $\height(\widehat{\xi})<2|Q|$.

Let $x\in \pos(\xi'')$ be such that $x\not\in\prefix(\widehat{v})$.
Let us denote by $\mathrm{lcp}(\widehat{v},x)$ the longest common prefix 
of $\widehat{v}$ and $x$. Moreover, let $\mathrm{rest}(\widehat{v},x)$ be the unique 
string in $\mathbb{N}_+^+$ such that $x=\mathrm{lcp}(\widehat{v},x)\mathrm{rest}(\widehat{v},x)$. We note that if $\widehat{v}\in \prefix(x)$, i.e., $\mathrm{lcp}(\widehat{v},x)=\widehat{v}$, then $|\mathrm{rest}(\widehat{v},x)|<|Q|$ due to the condition $\height(\xi''|_{\widehat{v}})< |Q|$.

If, for each $x\in \pos(\xi'')$ with $x\not\in\prefix(\widehat{v})$ and $\widehat{v}\not\in \prefix(x)$, we also have $|\mathrm{rest}(\widehat{v},x)|<|Q|+1$, then we are ready because $|\widehat{v}|<|Q|$.

Otherwise, there exist $u\in \mathbb{N}_+^+$ and  $w\in \mathbb{N}_+^+$ such that  $\mathrm{lcp}(\widehat{v},x)uw \in \prefix(x)$ and $\kappa''(\mathrm{lcp}(\widehat{v},x)u)=\kappa''(\mathrm{lcp}(\widehat{v},x)uw)$. Similarly to the construction in Step 1, by cutting out the context $(\xi''|_{\mathrm{lcp}(\widehat{v},x)u})|^w$ and the loop $(\kappa''|_{\mathrm{lcp}(\widehat{v},x)u})|^w$ from $\xi''$ and $\kappa''$, respectively, we obtain $\widehat{\xi}\in \T_\Sigma$ and successful run $\widehat{\kappa}\in \R_\cA(q,\widehat{\xi})$ such that $\widehat{v}\in \pos(\widehat{\xi})$, $\widehat{\kappa}(\widehat{v})=\kappa''(\widehat{v})$ and $\size(\widehat{\xi})<\size(\xi'')$. The condition $\widehat{v}\in \pos(\widehat{\xi})$ is ensured by $u\in \mathbb{N}_+^+$.
If there still exists an $x\in \pos(\widehat{\xi})$ such that $x\not\in\prefix(\widehat{v})$, $\widehat{v}\not\in \prefix(x)$, and $|\mathrm{rest}(\widehat{v},x)|\ge|Q|+1$, then we repeat the above procedure with $\widehat{\xi}$ and $\widehat{\kappa}$. 
After finitely many steps, we obtain $\widehat{\xi}$ and $\widehat{\kappa}$ as desired.
\end{proof}

Now we can prove the trimming result for zero-cancellation free strong bimonoids. By Observation~\ref{obs:zero-sum-free-property}(1) this also covers commutative strong bimonoids.

\begin{theorem-rect} \label{thm:t1}
 Let $\Sigma$ be a ranked alphabet. Moreover, let $\B$ be a  zero-cancellation free strong bimonoid and let $\cA$ be a $(\Sigma,\B)$-wta. If $\cA$ contains a useful state, then we can  construct a $(\Sigma,\B)$-wta $\cA'$  such that $\cA'$ is trim and $\runsem{\cA'}=\runsem{\cA}$. 
If $\cA$ is bu-deterministic, then so is $\cA'$.
\end{theorem-rect}
\begin{proof}  Let $\cA=(Q,\delta,F)$. We construct  the set
  \[Q'=\bigcup\big\{\im(\rho)\mid \xi\in \T_\Sigma, \height(\xi) < 2|Q|, \rho\in \R_\cA(q,\xi), \wt(\xi,\rho)\otimes F_{\rho(\varepsilon)}\ne \0\big\}.\]
    By Lemma \ref{lm:shortening}, $Q'$ is the set of all useful states of $\cA$.
  Due to our assumption on  $\cA$, we have $Q' \ne \emptyset$.
Then we construct the  $(\Sigma,\B)$-wta $\cA' = (Q',\delta',F')$ such that
  \begin{compactitem}
\item  $\delta'_k = \delta_k|_{(Q')^k \times \Sigma^{(k)} \times Q'}$ for each  $k \in \mathbb{N}$, and
\item $F' = F|_{Q'}$.
\end{compactitem}
If $\cA$ is bu-deterministic, then so is $\cA'$ because $Q'\subseteq Q$ and each transition mapping $\delta'_k$ is a restriction of $\delta_k$. Clearly,  $\cA'$  is trim.

Let $\xi\in \T_\Sigma$. Then
\begin{align*}
\runsem{\cA}(\xi)& = \bigoplus_{q\in Q}\bigoplus_{\rho\in \R_\cA(q,\xi)}\wt_\cA(\xi,\rho)\otimes F_q = \bigoplus_{q\in Q}\bigoplus_{\substack{\rho\in \R_\cA(q,\xi):\\ \rho \text{ is successful}}}\wt_\cA(\xi,\rho)\otimes F_q \\
& =  \bigoplus_{q\in Q'}\bigoplus_{\rho\in \R_{\cA'}(q,\xi)}\wt_{\cA'}(\xi,\rho)\otimes F'_q
= \runsem{\cA'}(\xi),
\end{align*}
where the third equality follows from the fact that, if $\rho$ is successful, then all states in $\im(\rho)$ are useful and that for every $q\in Q'$ and $\rho\in \R_{\cA'}(q,\xi)$ we have $\wt_{\cA'}(\xi,\rho)=\wt_{\cA}(\xi,\rho)$.
\end{proof}

\subsection{Local-trimming for arbitrary strong bimonoids}
\label{sec:t2}

Here we show our second trimming method. It takes a $(\Sigma,\B)$-wta over an arbitrary strong bimonoid $\B$ as input and constructs a run equivalent local-trim $(\Sigma,\B)$-wta. This trimming method is adapted from \cite[Thm.~3.2.3]{har78} where it is shown how to construct an equivalent  reduced context-free grammar from a given context-free grammar.

\begin{theorem-rect} \label{thm:t2} Let $\Sigma$ be a ranked alphabet. Moreover, let $\B=(B,\oplus,\otimes,\0,\1)$ be a strong bimonoid and
   $\cA$ a $(\Sigma,\B)$-wta. If $\cA$ contains a local-useful state,
  then we can  construct a  $(\Sigma,\B)$-wta $\cA'$  such that $\cA'$ is local-trim and   $\runsem{\cA'}=\runsem{\cA}$. 
 If $\cA$ is bu-deterministic, then so is $\cA'$.
\end{theorem-rect}

\begin{proof} The following proof is based on \cite{dro22}. Let  $\cA = (Q,\delta,F)$. First, in a bottom-up process, we construct the set $Q_1$ of all local-accessible states of $\cA$ (by using a mapping $f$). Second, in a top-down process, we construct the set of all local-useful states (by using a mapping $g$). 

We define the mapping  $f: \cP(Q) \to \cP(Q)$  for each $U \in \cP(Q)$ by
\[
f(U) = U \cup \{q \in Q | \text{ there exist }  k \in \mathbb{N}, q_1,\ldots,q_k \in U, \text{ and } \sigma \in \Sigma^{(k)}  \text{ such that }  \delta_k(q_1\cdots q_k,\sigma,q) \neq \0\}\enspace.
\]
Trivially,  $f$  is order-preserving and hence continuous, as  $Q$  is finite.

Let $Q_1$  be the smallest subset of $Q$ which is closed under $f$. Thus, by Theorem \ref{thm:Knaster-Tarski}, we have that $Q_1 = \bigcup(f^n(\emptyset)\mid n \in \mathbb{N})$.  Since $\bigcup(f^n(\emptyset)\mid n \in \mathbb{N}) = \bigcup(f^n(\emptyset)\mid 0 \le n \le |Q|)$, we can construct $Q_1$.
Moreover, each state in $Q_1$  is local-accessible in $\cA$. Conversely, working up a run, it follows that each local-accessible
state of $\cA$ belongs to $Q_1$. Hence $Q_1$ is the set of all local-accessible states of $\cA$.

Next we define $g: \cP(Q_1) \to \cP(Q_1)$ for each $U \in \cP(Q_1)$  by
\[
g(U) =
\begin{cases}
  \{q \in Q_1 \mid F_q \neq \0\} & \text{ if $U= \emptyset$} \\
  U \cup \{q_i \in Q_1 \mid \text{ there exist } k \in \mathbb{N}_+, q \in U, q_1,\ldots,q_k \in Q_1, i \in [k] \text{ and } \\
   \hspace*{24mm} \sigma \in \Sigma^{(k)} \text{ such that } \delta_k(q_1\cdots q_k,\sigma,q) \neq \0\} & \text{ otherwise} \enspace. 
  \end{cases}
\]
The mapping  $g$  is order-preserving and hence continuous, as  $Q_1$  is finite. Let  $Q'$ be the smallest subset of $Q_1$ which is closed under $g$. Thus, by Theorem \ref{thm:Knaster-Tarski}, we have that $Q' = \bigcup(g^n(\emptyset)\mid n \in \mathbb{N})$. Since $\bigcup(g^n(\emptyset)\mid n \in \mathbb{N}) = \bigcup(g^n(\emptyset)\mid 0 \le n \le |Q_1|)$, we can construct $Q'$.

We claim that  
\begin{equation}\label{eq-Q-prime-all-local-useful}
\text{$Q'$  is the set of all local-useful states of $\cA$.}
\end{equation}

(a) First we show that each $q\in Q'$ is local-useful.  For this, by induction on $\mathbb{N}$, we prove the following statement.
\begin{equation}\label{equ:local-useful-Harrison}
\text{For each $n \in \mathbb{N}_+$, we have that each state in $g^n(\emptyset)$ is local-useful.} 
\end{equation}

I.B.: Clearly, each state  in  $g(\emptyset) = \{q \in Q_1 \mid F_q \neq \0\}$ is local-useful.

I.S.: Let $n \geq 1$ and $p\in g^{n+1}(\emptyset)$. We may assume that $p\in g^{n+1}(\emptyset)\setminus g^{n}(\emptyset)$ because otherwise the statement follows by I.H. immediately. Then there  exist $k \in \mathbb{N}_+$, $j\in[k]$, $q \in g^n(\emptyset)$,  $q_1,\ldots,q_k \in Q_1$, and  $\sigma \in \Sigma^{(k)}$ such that $\delta_k(q_1\cdots q_k,\sigma,q) \neq \0$ and $p=q_j$.

Since $q_1,\ldots,q_k$ are in $Q_1$, i.e., local-accessible, for each $i\in [k]$ there is a tree $\xi_i\in \T_\Sigma$ and run $\rho_i\in \R_\cA(q_i,\xi_i)$ such that for every  $v\in \pos(\xi_i)$, the weight   $\delta_\ell(\rho_i(v1) \cdots \rho_i(v\ell),\xi_i(v),\rho_i(v)) \ne \0$, where $\ell = \rk(\xi(v))$.

By I.H. $q$ is local-useful, hence there is a $\xi\in \T_\Sigma$,  and a local-successful run $\rho\in \R_\cA(\xi)$ such that $q\in\im(\rho)$.
Let $w\in \pos(\xi)$ with $\rho(w)=q$. 

Now consider the tree $\zeta=\sigma(\xi_1,\ldots,\xi_k)$ and define the run $\kappa\in \R_\cA(\zeta)$ such that $\kappa(\varepsilon)=q$ and for each $i\in [k]$ and 
$v \in \pos(\xi_i)$ we have $\kappa(iv)=\rho_i(v)$.

Note that $\kappa(j)=p$ and
\begin{equation}\label{eq:kappa-nonzero-tramsitions}
\text{for every  $v\in \pos(\zeta)$, the weight   $\delta_\ell(\kappa(v1) \cdots \kappa(v\ell),\zeta(v),\kappa(v)) \ne \0$, where $\ell = \rk(\xi'(v))$.}
\end{equation}

Lastly, we consider the tree $\xi'=\xi[\zeta]_w$ and the run $\rho'\in \R_\cA(\xi')$ defined, for each $v \in \pos(\xi')$, by 
\[
\rho'(v)=
\begin{cases}
\kappa(u) & \text{ if } (\exists u\in \mathbb{N}^*): v=wu\\
\rho(v) & \text{ otherwise.} 
\end{cases}
\]
Since $\rho$ is local-successful and $\kappa$ has property \eqref{eq:kappa-nonzero-tramsitions}, the run $\rho'$ is also local-successful.
Moreover, $\rho'(wj)=\kappa(j)=p$, hence $p$ is local-useful. This finishes the proof of \eqref{equ:local-useful-Harrison}.

Since $Q' = \bigcup(g^n(\emptyset)\mid n \in \mathbb{N})$, we obtain that each state of $Q'$ is local-useful in $\cA$.

(b) Next we show that each local-useful state of $\cA$ is in $Q'$. 
For this, let $\xi\in \T_\Sigma$ and $\rho\in \R_\cA(\xi)$ be a local-successful run on $\xi$. By induction on $(\pos(\xi),<_{\mathrm{pref}}^{-1})$ we show that
\begin{equation}\label{eq:loc-succ-in-Q}
\text{for each $w\in \pos(\xi)$, there exists $n\in \mathbb{N}$ such that state $\rho(w)\in g^n(\emptyset)$.}
\end{equation}
We note that, since $\rho$ is local-successful, $\rho(v)\in Q_1$ for each $v\in \pos(\xi)$. 

I.B.: Let $w=\varepsilon$. Again, since $\rho$ is local-successful, we have $F_{\rho(\varepsilon)}\ne \0$. Hence $\rho(\varepsilon)\in g(\emptyset)$.

I.S.: Let $w=vi$ for some $v\in \pos(\xi)$ and $i\in \maxrk(\Sigma)$. By I.H., there exists  $n\in \mathbb{N}$ such that the state $\rho(v)\in g^n(\emptyset)$. Let $\sigma=\xi(v)$ and $k=\rk(\sigma)$. Since $\rho$ is local-successful, we have $\delta_k(\rho(v1)\ldots\rho(vk),\sigma,\rho(v))\ne \0$. Then $\rho(v1),\ldots,\rho(vk) \in g^{n+1}(\emptyset)$ and thus, in particular, $\rho(w)\in g^{n+1}(\emptyset)$.

With this we proved that \eqref{eq-Q-prime-all-local-useful} holds. By our assumption on $\cA$, the set $Q'$ is not empty.

Now we construct the  $(\Sigma,\B)$-wta $\cA' = (Q',\delta',F')$ such that, for each $k\in \mathbb{N}$, $\delta'_k=\delta_k|_{(Q')^k\times \Sigma^{(k)}\times Q'}$, and $F'=F|_{Q'}$. Then  $\cA'$ is local-trim. If $\cA$ is bu-deterministic, then so is $\cA'$ because $Q'\subseteq Q$ and each transition mapping $\delta'_k$ is a restriction of $\delta_k$.

Finally, we prove that $\runsem{\cA} = \runsem{\cA'}$. Let $\xi \in \T_\Sigma$. Obviously, $\R_{\cA'}(\xi) \subseteq \R_\cA(\xi)$ and for each $\rho \in \R_{\cA'}(\xi)$ we have $\wt_{\cA'}(\xi,\rho) = \wt_\cA(\xi,\rho)$. If $\rho \in \R_\cA(\xi)\setminus \R_{\cA'}(\xi)$, then there exists $p\in \im(\rho)$ such that $p$ is not local-useful. Then $\wt_\cA(\xi,\rho)=\0$ or $F_{\rho(\varepsilon)}=\0$ and hence $\wt_\cA(\xi,\rho) \otimes F_{\rho(\varepsilon)}=\0$.  
  Thus we can compute
  \begin{align*}
    \runsem{\cA}(\xi) &= \bigoplus_{\rho \in \R_\cA(\xi)} \wt_\cA(\xi,\rho) \otimes F_{\rho(\varepsilon)}
                  =  \bigoplus_{\rho \in \R_{\cA'}(\xi)} \wt_{\cA'}(\xi,\rho) \otimes F'_{\rho(\varepsilon)}
    =\runsem{\cA'}(\xi). \qedhere
    \end{align*}
\end{proof}

\section{Transforming wta into total wta}

We recall that a $(\Sigma,\B)$-wta $\cA=(Q,\delta,F)$ is total if for every $k \in \mathbb{N}$, $\sigma \in \Sigma^{(k)}$, and $w \in Q^k$ there exists at least one state $q$ such that $\delta_k(w,\sigma,q) \not= \mathbb{0}$.

\begin{theorem-rect}\label{thm:wta-nf-total}  Let $\Sigma$ be a ranked alphabet. Moreover, let $\B=(B,\oplus,\otimes,\0,\1)$ be a strong bimonoid and $\cA$ a $(\Sigma,\B)$-wta. We can construct a $(\Sigma,\B)$-wta $\cA'$ such that $\cA'$ is total,  $\initialsem{\cA}= \initialsem{\cA'}$, and $\runsem{\cA}= \runsem{\cA'}$. Moreover, if $\cA$ is bu-deterministic, then so is~$\cA'$, and if $\cA$ has unit root weights, then so does $\cA'$.
\end{theorem-rect}
\begin{proof} Let $\cA=(Q,\delta,F)$. We construct the $(\Sigma, \B)$-wta $\cA'=(Q',\delta',F')$ as follows.
\begin{compactitem}
\item $Q' = Q \cup \{q_\bot\}$ where $q_\bot \not\in Q$,

\item $\delta' = (\delta'_k \mid k \in \mathbb{N})$ such that for every $k \in \mathbb{N}$, $\sigma \in \Sigma^{(k)}$, and $q \in Q'$, and  $q_1 \cdots q_k \in (Q')^k$:

\[
(\delta')_k(q_1\cdots q_k,\sigma,q) =
\left\{
\begin{array}{ll}
\delta_k(q_1\cdots q_k,\sigma,q) & \text{if } q_1\cdots q_k \in Q^k, q \in Q\\[1mm]
\mathbb{1} & \text{if } q_1 \cdots q_k \in Q^k, q=q_\bot, \text{ and }\\
& (\forall q' \in Q): \delta_k(q_1\cdots q_k,\sigma,q')=\mathbb{0}\\[1mm]
\mathbb{1} & \text{if } q_\bot \in \{q_1,\ldots,q_k\} \text{ and } q=q_\bot \\[1mm]
\mathbb{0} & \text{otherwise }
\end{array}
\right.
\]

\item $F'_q=F_q$ for each $q \in Q$ and $F'_{q_\bot}=\mathbb{0}$.
\end{compactitem}

Obviously, if $\cA$ is bu-deterministic, then so is $\cA'$. The construction also preserves crisp-determinism.

First we prove that $\initialsem{\cA} = \initialsem{\cA'}$. By induction on~$\T_\Sigma$, we prove that the following statement holds:
\begin{equation}
\text{For every $\xi \in \T_\Sigma$ and $q \in Q$: } \h_\cA(\xi)_q = \h_{\cA'}(\xi)_q \enspace.\label{eq:init-h-total}
\end{equation}
Let $\xi = \sigma(\xi_1,\ldots,\xi_k)$ and assume that \eqref{eq:init-h-total} holds for $\xi_1,\ldots,\xi_k$. Let $q \in Q$.  Then we can calculate:
\begingroup
\allowdisplaybreaks
\begin{align*}
\h_{\cA'}(\sigma(\xi_1,\ldots,\xi_k))_q
=& \bigoplus_{q_1 \cdots q_k \in (Q\cup\{q_\bot\})^k} \Big( \bigotimes_{i \in [k]} \h_{\cA'}(\xi_i)_{q_i}\Big)  \otimes \delta'_k(q_1\cdots q_k,\sigma,q)\\
  =& \bigoplus_{q_1 \cdots q_k \in Q^k} \Big( \bigotimes_{i \in [k]} \h_{\cA'}(\xi_i)_{q_i}\Big)  \otimes \delta_k(q_1\cdots q_k,\sigma,q)
     \tag{\text{note that $\delta'_k(q_1\cdots q_k,\sigma,q)=\mathbb{0}$ if one of the $q_i$ is $q_\bot$}}\\
  =& \bigoplus_{q_1 \cdots q_k \in Q^k} \Big( \bigotimes_{i \in [k]} \h_\cA(\xi_i)_{q_i}\Big)  \otimes \delta_k(q_1\cdots q_k,\sigma,q)
     \tag{\text{by I.H.}}\\
=& \ \h_\cA(\sigma(\xi_1,\ldots,\xi_k))_q\enspace.
\end{align*}
\endgroup
Then for each $\xi \in \T_\Sigma$.
\[
\initialsem{\cA'}(\xi) 
= \bigoplus_{q \in Q \cup \{q_\bot\}} \h_{\cA'}(\xi)_q \otimes F'_q
= \bigoplus_{q \in Q} \h_{\cA'}(\xi)_q \otimes F_q
= \bigoplus_{q \in Q} \h_\cA(\xi)_q \otimes F_q  = \initialsem{\cA}(\xi)\enspace,
\]
where the second equality holds because $F'_{q_\bot}=\mathbb{0}$ and $F'|_Q = F$.

Next  we prove that $\runsem{\cA} = \runsem{{\cA'}}$. First, it is easy to see that 
\begin{eqnarray}
& \text{for every $\xi \in \T_\Sigma$ and  $q\in Q$: $\R_\cA(q,\xi)\subseteq \R_{\cA'}(q,\xi)$, and}\nonumber \\
& \text{for each  $\rho \in \R_{\cA'}(q,\xi) \setminus \R_\cA(q,\xi)$:  $\wt_{\cA'}(\xi,\rho)=\mathbb{0}$.} \label{equ:run-init-2}
\end{eqnarray}
 Also it is easy to see that 
\begin{eqnarray}
\text{for each $\xi \in \T_\Sigma$, $q\in Q$, and  $\rho \in \R_\cA(q,\xi)$: $\wt_{\cA'}(\xi,\rho) = \wt_\cA(\xi,\rho)$,} \label{equ:run-init-1}
\end{eqnarray}
where $\wt_{\cA'}$ and $\wt_\cA$ use $\delta'$ and $\delta$, respectively.
Now let $\xi \in \T_\Sigma$. Then we can calculate as follows.
\begingroup
\allowdisplaybreaks
\begin{align*}
\runsem{{\cA'}}(\xi) 
=& \bigoplus_{q\in Q'}\bigoplus_{\rho\in \R_{\cA'}(q,\xi)} \wt_{\cA'}(\xi,\rho) \otimes F'_q\\
  =& \bigoplus_{q\in Q}\bigoplus_{\rho\in \R_{\cA'}(q,\xi)} \wt_{\cA'}(\xi,\rho) \otimes F_q
     \tag{\text{because $F'_{q_\bot}=\mathbb{0}$ and $F'|_Q = F$}}\\
=& \bigoplus_{q\in Q}\bigoplus_{\rho\in \R_\cA(q,\xi)} \wt_{\cA'}(\xi,\rho) \otimes F_q
\tag{\text{by \eqref{equ:run-init-2}}}\\
=& \bigoplus_{q\in Q}\bigoplus_{\rho\in \R_\cA(q,\xi)} \wt_\cA(\xi,\rho) \otimes F_q
\tag{\text{by \eqref{equ:run-init-1}}}\\
=& \ \runsem{\cA}(\xi)\enspace. \qedhere
\end{align*}
\endgroup
\end{proof}

\section{Normalizing root weights of wta}
\label{sec:normalizing-root-weights}

Next we want to show that each wta can be transformed into a run equivalent root weight normalized wta. In \cite[Lm.~6.1.1]{bor04b}  the same statement is proved if $\B$ is a semiring. (We also refer to \cite[Lm.~22]{boz99} and \cite[Lm.~4.8]{dropecvog05} for weaker statements.) Here we present a slightly more complicated construction which allows to prove the statement for an arbitrary strong bimonoid.

\begin{theorem-rect}\label{thm:root-weight-normalization-run} Let $\Sigma$ be a ranked alphabet. Moreover, let $\B$ be a strong bimonoid and let $\cA$ be a $(\Sigma,\B)$-wta. We can construct a $(\Sigma,\B)$-wta ${\cA'}$ such that ${\cA'}$ is  root weight normalized and  $\runsem{\cA}= \runsem{{\cA'}}$.
\end{theorem-rect}
\begin{proof} Let $\cA=(Q,\delta,F)$. The idea for the construction of $\cA'$ is the following. The wta $\cA'$ simulates~$\cA$. Additionally, at each leaf of the given input tree $\xi$, the wta $\cA'$ guesses a state $p$ of $\cA$ and stores it in its states; at each non-nullary symbol of $\Sigma$, the wta $\cA'$ checks whether the guesses in the subtrees are consistent and propagates the guessed state; and at the root, the wta $\cA'$ multiplies the last transition weight with $F_p$ (if $p$ is the state guessed at each leaf) and moves to the final state $q_f$. (We note that the idea for maintaining the guessed final states was used in  \cite[Lm.~9]{hervogdro19} where initial- and final-state normalization was proved for  weighted string automata with storage over unital valuation monoids.)

  We construct the $(\Sigma,\B)$-wta ${\cA'} = (Q',\delta',F')$ as follows:  
\begin{compactitem}
\item $Q' = (Q \times Q) \cup \{q_f\}$ where $q_f \not\in Q$; for each $q' \in Q \times Q$ we let $(q')_i$ denote the $i$-th component of $q'$ for $i \in \{1,2\}$;
\item $\delta' = (\delta'_k \mid k \in \mathbb{N})$ where for each $k \in \mathbb{N}$, $\sigma \in \Sigma^{(k)}$, $q_0' \in Q'$ and $q_1'\cdots q_k' \in (Q')^k$ we define $\delta'_k(q_1'\cdots q_k',\sigma,q_0')$ by case analysis as follows.

\underline{Case (a):} Let $k=0$. Then
    \begin{align*}
\delta'_0(\varepsilon,\sigma,q_0')
= \left\{
\begin{array}{ll}
\delta_0(\varepsilon,\sigma,(q_0')_1) & \text{ if } q_0' \in Q\times Q\\
\bigoplus_{p \in Q} \delta_0(\varepsilon,\sigma,p) \otimes F_p & \text{ if } q_0'=q_f\enspace.
\end{array}
\right. 
\end{align*}

\underline{Case (b):} Let $k \ge 1$. Then 
\begin{align*}
&\delta'_k(q_1'\cdots q_k',\sigma,q_0')\\
= &\left\{
\begin{array}{ll}
\delta_k((q_1')_1\cdots (q_k')_1,\sigma,(q_0')_1) & \text{ if } (\exists p \in Q) \ (\forall i \in [0,k]) \ (\exists q_i \in Q) : q_i' = (q_i,p)\\
\delta_k((q_1')_1\cdots (q_k')_1,\sigma,(q_1')_2) \otimes F_{(q_1')_2} & \text{ if } (\exists p \in Q) \ (\forall i \in [k]) \ (\exists q_i \in Q): q_i' = (q_i,p)  \text{ and } q_0'=q_f\\
\mathbb{0} & \text{ otherwise,}
\end{array}
\right. 
\end{align*}
where in the second case we have used the assumption that $k \ge 1$,
\item $F'_{q_f}=\mathbb{1}$ and $F'_{q'}=\mathbb{0}$ for each $q' \in Q' \setminus \{q_f\}$.
\end{compactitem}

It is clear that, for each position of $\xi \in \T_\Sigma$ except at its root,  ${\cA'}$ behaves exactly as $\cA$. To make this precise, for each $p \in Q$, we define the family $(\psi_{p,\xi} \mid \xi \in \T_\Sigma)$ of  mappings
\[\psi_{p,\xi}: \R_\cA(\xi) \to \R_{\cA'}(\xi)\]
such that,  for every $\rho \in \R_\cA(\xi)$ and $w \in \pos(\xi)$, we let  $\psi_{p,\xi}(\rho)(w) = (\rho(w),p)$. Obviously, $\psi_{p,\xi}$ is  injective.
Also it is obvious that, for every $\xi = \sigma(\xi_1,\ldots,\xi_k)$, $i \in [k]$, and $\rho \in \R_\cA(\xi)$, the following holds:
\begin{equation}
\psi_{p,\xi}(\rho)|_i = \psi_{p,\xi_i}(\rho|_i)\enspace. \label{equ:run-subtree}
  \end{equation}

By induction on $\T_\Sigma$, we prove that the following statement holds:
\begin{equation}
\text{For every $\xi \in \T_\Sigma$, $\rho \in \R_\cA(\xi)$, and $p\in Q$, we have } \wt_\cA(\xi,\rho) = \wt_{\cA'}(\xi,\psi_{p,\xi}(\rho))\enspace. \label{equ:add-p}
\end{equation}
Let $\xi = \sigma(\xi_1,\ldots,\xi_k)$. Then we can calculate as follows:
\begingroup
\allowdisplaybreaks
\begin{align*}
  \wt_\cA(\xi,\rho) &= \Big(\bigotimes_{i \in [k]} \wt_\cA(\xi_i,\rho|_i)\Big) \otimes \delta_k(\rho(1) \cdots \rho(k),\sigma,\rho(\varepsilon))\\
                &=  \Big(\bigotimes_{i \in [k]} \wt_{\cA'}(\xi_i,\psi_{p,\xi_i}(\rho|_i))\Big) \otimes \delta_k'((\rho(1),p) \cdots (\rho(k),p),\sigma,(\rho(\varepsilon),p)) \tag{\text{by I.H. and construction}}\\
                &=  \Big(\bigotimes_{i \in [k]} \wt_{\cA'}(\xi_i,\psi_{p,\xi}(\rho)|_i)\Big) \otimes \delta_k'((\rho(1),p) \cdots (\rho(k),p),\sigma,(\rho(\varepsilon),p)) \tag{\text{by \eqref{equ:run-subtree}}}\\
  &=\wt_{\cA'}(\xi,\psi_{p,\xi}(\rho))\enspace.
\end{align*}
\endgroup

Next we prove that $\runsem{\cA}(\xi)= \runsem{{\cA'}}(\xi)$ for each $\xi \in \T_\Sigma$ by case analysis.

\underline{Case (a):}  Let $\xi = \alpha$ for some $\alpha \in \Sigma^{(0)}$. Since, for each $p \in Q$, we have $\R_\cA(p,\alpha) = \{\rho\}$ with $\rho(\varepsilon)=p$ and $\wt_\cA(\alpha,\rho) =  \delta_0(\varepsilon,\alpha,p)$, and similarly, $\R_{\cA'}(q_f,\alpha) = \{\rho'\}$ with $\rho'(\varepsilon)=q_f$ and $\wt_{\cA'}(\alpha,\rho') =  \delta'_0(\varepsilon,\alpha,q_f)$, we obtain
\begin{align*}
  \runsem{\cA}(\alpha) &= \bigoplus_{p \in Q}  \bigoplus_{\rho \in \R_\cA(p,\xi)} \wt_\cA(\alpha,\rho) \otimes F_{p}
                         = \bigoplus_{p \in Q} \delta_0(\varepsilon,\alpha,p) \otimes F_{p}
                         = \delta'_0(\varepsilon,\alpha,q_f)\\
                       &= \delta'_0(\varepsilon,\alpha,q_f) \otimes F'_{q_f}
                       = \bigoplus_{q' \in Q'}  \bigoplus_{\rho' \in \R_{\cA'}(q',\xi)} \wt_{\cA'}(\alpha,\rho') \otimes F'_{q'}
                         = \runsem{{\cA'}}(\alpha) \enspace.
  \end{align*}

\underline{Case (b):} Let $\xi \in \T_\Sigma \setminus \Sigma^{(0)}$. Let  $p \in Q$. First we relate $p$-runs of $\cA$ on $\xi$ with  $q_f$-runs of ${\cA'}$ on $\xi$ which assign states of the form $(q,p)$ to inner nodes of $\xi$ for some $q \in Q$. Formally, we define the mapping
  \[\varphi_p: \R_\cA(p,\xi) \to \R_{\cA'}^p(q_f,\xi)\]
where $\R_{\cA'}^p(q_f,\xi) = \{\rho' \in \R_{\cA'}(q_f,\xi) \mid (\forall w \in \pos(\xi)\setminus \{\varepsilon\}): \rho'(w) \in Q \times \{p\}\}$, and for every $\rho \in \R_\cA(p,\xi)$ and $w \in \pos(\xi)$ we let
\(  \big(\varphi_p(\rho)\big)(w)  = q_f\) if $w = \varepsilon$, and $(\rho(w),p)$ otherwise.
It is obvious that, for every $p\in Q$, the mapping $\varphi_p$ is bijective
(note that $\xi \not\in \Sigma^{(0)}$) and
\begin{equation}
\text{for every $\rho' \in \R_{\cA'}(q_f,\xi) \setminus \bigcup_{p\in Q}\R_{\cA'}^p(q_f,\xi)$ we have $\wt_{\cA'}(\xi,\rho') = \mathbb{0}$}\enspace. \label{equ:not-p-equal-zero}
\end{equation}
Moreover, the family $(\R_{\cA'}^p(q_f,\xi) \mid p \in Q)$ partitions $\bigcup_{p\in Q}\R_{\cA'}^p(q_f,\xi)$, i.e.,
\begin{equation}
\R_{\cA'}^p(q_f,\xi) \cap \R_{\cA'}^{p'}(q_f,\xi) = \emptyset \ \text{for every $p,p' \in Q$ with $p\not= p'$}\enspace. \label{equ:partition} 
  \end{equation}

  Next we prove a correspondence between the weights of runs of $\cA$ and ${\cA'}$ on $\xi$. We claim:
  \begin{equation}
\text{for every $p\in Q$, and $\rho \in \R_\cA(p,\xi)$, we have $\wt_\cA(\xi,\rho) \otimes F_p = \wt_{\cA'}(\xi,\varphi_p(\rho))$}\enspace. \label{equ:relation-weights}
\end{equation}
For the proof let $\xi = \sigma(\xi_1,\ldots,\xi_k)$ (note that $k \ge 1$). Let $p\in Q$, and $\rho \in \R_\cA(p,\xi)$ (hence $\rho(\varepsilon) =p$). Then we can calculate as follows:
\begingroup
\allowdisplaybreaks
\begin{align*}
  \wt_\cA(\xi,\rho) \otimes F_p &=\Big(\Big(\bigotimes_{i \in [k]} \wt_\cA(\xi_i,\rho|_i)\Big) \otimes \delta_k(\rho(1) \cdots \rho(k),\sigma,p)\Big) \otimes F_p\\
                            &=\Big(\bigotimes_{i \in [k]} \wt_\cA(\xi_i,\rho|_i)\Big) \otimes \Big(\delta_k(\rho(1) \cdots \rho(k),\sigma,p) \otimes F_p\Big)\\
                            &= \Big(\bigotimes_{i \in [k]} \wt_{\cA'}(\xi_i,\psi_{p,\xi_i}(\rho|_i))\Big) \otimes \Big(\delta_k(\rho(1) \cdots \rho(k),\sigma,p) \otimes F_p\Big)
  \tag{\text{by \eqref{equ:add-p}}}\\
                            &= \Big(\bigotimes_{i \in [k]} \wt_{\cA'}(\xi_i,\psi_{p,\xi_i}(\rho|_i))\Big) \otimes \delta_k'((\rho(1),p) \cdots (\rho(k),p),\sigma,q_f)
  \tag{\text{by construction}}\\
                                &= \Big(\bigotimes_{i \in [k]} \wt_{\cA'}(\xi_i,\varphi_p(\rho)|_i)\Big) \otimes \delta_k'(\varphi_p(\rho)(1) \cdots \varphi_p(\rho)(k),\sigma,\varphi_p(\rho)(\varepsilon))
                                  \tag{\text{because $\psi_{p,\xi_i}(\rho|_i)=\varphi_p(\rho)|_i$ for each $i \in [k]$ and $\varphi_p(\rho)(\varepsilon)=q_f$}}\\
                &=   \wt_{\cA'}(\xi,\varphi_p(\rho))\enspace.
\end{align*}
\endgroup

Finally, we can prove that $\sem{\cA}^{\mathrm{run}}(\xi) = \sem{{\cA'}}^{\mathrm{run}}(\xi)$.
\begin{align*}
  \sem{{\cA'}}^{\mathrm{run}}(\xi) &= \bigoplus_{q' \in Q'}  \bigoplus_{\rho' \in \R_{\cA'}(q',\xi)} \wt_{\cA'}(\xi,\rho') \otimes F'_{q'}\\
                                  &=  \bigoplus_{\rho' \in \R_{\cA'}(q_f,\xi)} \wt_{\cA'}(\xi,\rho')
                                 \tag{\text{because $F'_{q_f}=\mathbb{1}$ and $F'_{q'}=\mathbb{0}$ for each $q' \in Q' \setminus \{q_f\}$}}\\
                                &= \bigoplus_{\rho' \in \bigcup_{p\in Q}\R_{\cA'}^p(q_f,\xi)} \wt_{\cA'}(\xi,\rho')
                                  \tag{\text{by \eqref{equ:not-p-equal-zero}}}\\
                                &= \bigoplus_{p\in Q} \bigoplus_{\rho' \in \R_{\cA'}^p(q_f,\xi)} \wt_{\cA'}(\xi,\rho')
                                  \tag{\text{by \eqref{equ:partition}}}\\
                                &= \bigoplus_{p\in Q} \bigoplus_{\rho \in \R_\cA(p,\xi)} \wt_{\cA'}(\xi,\varphi_p(\rho))
  \tag{\text{because $\varphi_p$ is a bijection}}\\
                                &= \bigoplus_{p \in Q}  \bigoplus_{\rho \in \R_\cA(p,\xi)} \wt_\cA(\xi,\rho) \otimes F_{p}
                                  \tag{\text{by \eqref{equ:relation-weights}}}\\
  &= \sem{\cA}^{\mathrm{run}}(\xi)\enspace.   \qedhere
  \end{align*}
\end{proof}

An analysis of the construction in the proof of Theorem~\ref{thm:root-weight-normalization-run} shows that it does not preserve bu-determinism. 
If we restrict the set of weight algebras to the set of commutative semifields, then we can transform each wta into an equivalent one with unit root weights and this transformation preserves bu-determinism. The transformation uses the technique of weight pushing  \cite[Def.~4.1]{hanmalque18} (also cf. \cite[p.~296]{moh97} for the case that $\B$ is the tropical semiring on the set of non-negative real numbers), which is of interest of its own. 

\index{push@$\push_\lambda(\cA)$}
Formally, let $\cA = (Q,\delta,F)$ be a $(\Sigma,\B)$-wta over some commutative semifield $\B$. Moreover, let $\lambda: Q \rightarrow B \setminus \{\mathbb{0}\}$ be a mapping such that $\lambda(q) = F_q$ for each $q \in \supp(F)$. We define the wta $\push_\lambda(\cA)$ to be the $(\Sigma,\B)$-wta $(Q,\delta',F')$ as follows. For every $k \in \mathbb{N}$, $\sigma \in \Sigma^{(k)}$, $q\in Q$, and $q_1\cdots q_k \in Q$: 
\[
\delta'_k(q_1 \cdots q_k,\sigma,q) = \Big(\bigotimes_{i \in [k]} \lambda(q_i)^{-1}\Big) \otimes \delta_k(q_1 \cdots q_k,\sigma,q) \otimes \lambda(q)
\]
and for each $q \in Q$ 
\[
F'_q = 
\left\{
\begin{array}{ll}
\mathbb{1} & \text{ if } q \in \supp(F)\\
\mathbb{0} & \text{ otherwise }\enspace.
\end{array}
\right.
\]

\begin{lemma}\rm (cf. \cite[Prop.~4.2]{hanmalque18}) \label{lm:A=push-lambda(A)} Let  $\cA$ and $\lambda$ be defined as above. Then $\sem{\cA} = \sem{\push_\lambda(\cA)}$.
\end{lemma}
\begin{proof} We abbreviate $\push_\lambda(\cA)$ by $\cA'$. By  induction on $\T_\Sigma$, we prove that the following statement holds:
  \begin{equation}
    \text{For every $\xi \in \T_\Sigma$ and $q \in Q$, we have: } \h_{\cA'}(\xi)_q = \lambda(q) \otimes \h_\cA(\xi)_q \enspace.\label{eq:push-lemma}
  \end{equation}
  Let $\xi = \sigma(\xi_1,\ldots,\xi_k)$. Then
  \begingroup
  \allowdisplaybreaks
\begin{align*}
  &\h_{\cA'}(\sigma(\xi_1,\ldots,\xi_k))_q \\
  &= \bigoplus_{q_1 \cdots q_k \in Q^k} \Big(\bigotimes_{i \in [k]} \h_{\cA'}(\xi_i)_{q_i}\Big) \otimes \delta'(q_1\cdots q_k,\sigma,q)\\
&= \bigoplus_{q_1 \cdots q_k \in Q^k} \Big(\bigotimes_{i \in [k]} \lambda(q_i) \otimes \h_{\cA}(\xi_i)_{q_i}\Big) \otimes 
    \Big(\Big(\bigotimes_{i \in [k]} \lambda(q_i)^{-1}\Big)  \otimes \delta_k(q_1 \cdots q_k,\sigma,q) \otimes \lambda(q)\Big)
    \tag{\text{by I.H. and construction of $\push_\lambda(\cA)$}}\\
&= \lambda(q) \otimes \bigoplus_{q_1 \ldots q_k \in Q^k} \Big(\bigotimes_{i \in [k]}  \h_{\cA}(\xi_i)_{q_i}\Big) \otimes 
    \delta_k(q_1 \cdots q_k,\sigma,q)
    \tag{\text{by commutativity and associativity of $\otimes$ and by distributivity}}\\
&= \lambda(q) \otimes  \h_{\cA}(\sigma(\xi_1,\ldots,\xi_k))_q \enspace.
\end{align*}
\endgroup
Then, for each $\xi \in \T_\Sigma$, we have 
\begin{align*}
  &\sem{\push_\lambda(\cA)}(\xi)\\
  &= \bigoplus_{q \in Q} \h_{\cA'}(\xi)_q \otimes F'_q
                                = \bigoplus_{q \in Q} \h_{\cA}(\xi)_q \otimes \lambda(q) \otimes  F'_q
                                \tag{\text{by \eqref{eq:push-lemma} and commutativity of $\otimes$}}\\
&= \bigoplus_{q \in \supp(F)} \h_{\cA}(\xi)_q \otimes \lambda(q)
\tag{by the definition of $F'$}\\
&= \bigoplus_{q \in \supp(F)} \h_{\cA}(\xi)_q \otimes F_q
\tag{by the definition of $\lambda$}\\
&= \sem{\cA}(\xi)\enspace. \qedhere
\end{align*}
\end{proof}

\begin{theorem-rect}\label{lm:com-semifield-Boolean-root-weights} Let $\Sigma$ be a ranked alphabet. Moreover, let $\B=(B,\oplus,\otimes,\0,\1)$ be a commutative semifield and let $\cA$  be a $(\Sigma,\B)$-wta.
We can construct a $(\Sigma,\B)$-wta ${\cA'}$ such that ${\cA'}$ has unit root weights and $\sem{\cA} = \sem{{\cA'}}$. If $\cA$ has some of the properties  bu-deterministic, total, and trim, then  $\cA'$ has the same properties. 
\end{theorem-rect}

\begin{proof} Let $\cA = (Q,\delta,F)$ and let $\lambda:  Q \rightarrow B \setminus \{\mathbb{0}\}$ be any mapping such that $\lambda(q) = F_q$ for each $q \in \supp(F)$.
  We note that $\push_\lambda(\cA)$ has unit root weights.
By Lemma \ref{lm:A=push-lambda(A)} we have $\sem{\cA} = \sem{\push_\lambda(\cA)}$, and hence we can choose ${\cA'} = \push_\lambda(\cA)$.
  
Let $\push_\lambda(\cA) = (Q,\delta',F')$. Then, for every $k \in \mathbb{N}$, $\sigma \in \Sigma^{(k)}$, $q\in Q$, and $q_1\cdots q_k \in Q^k$, we have $\delta'_k(q_1 \cdots q_k,\sigma,q)\ne \0$ iff $\delta_k(q_1 \cdots q_k,\sigma,q)\ne \0$; and also $F'_q\ne \0$ iff $F_q\ne \0$ for each $q \in Q$. Thus, if $\cA$ is bu-deterministic, then so is $\push_\lambda(\cA)$. Moreover, if $\cA$ is trim, then so is $\push_\lambda(\cA)$.
\end{proof}

\begin{example}\rm We construct $\push_\lambda(\cA)$ for the $(\Sigma,\Ratnum)$-wta $\cA=(Q,\delta,F)$ of Example \ref{ex:bu-det+total-wta-det-weo} and $\lambda: Q \to \mathbb{Q}\setminus \{0\}$ such that $\lambda(o)=3$ and $\lambda(e)=2$. For convenience, we recall that $\Sigma = \{\sigma^{(2)}, \alpha^{(0)}\}$ and 
    \begin{compactitem}
    \item $Q = \{e,o\}$, where $e$ and $o$ are standing for {\em e}ven and {\em o}dd, respectively,
    \item $\delta_0(\varepsilon,\alpha,o) = 2$,  $\delta_0(\varepsilon,\alpha,e)=0$ and
      \[
        \delta_2(q_1q_2,\sigma,o) = \begin{cases} 1 & \text{ if $q_1\not= q_2$}\\
          0 & \text{ otherwise }
        \end{cases}
       \ \ \text{ and } \ \
         \delta_2(q_1q_2,\sigma,e) = \begin{cases} 1 & \text{ if $q_1= q_2$}\\
          0 & \text{ otherwise }\enspace,
        \end{cases}
      \]
      \item $F_o=3$ and $F_e=2$.
      \end{compactitem}
Then $\push_\lambda(\cA)$ is the $(\Sigma,\Ratnum)$-wta $(Q,\delta',F')$ (cf. Figure~\ref{fig:even-odd-two-three-2-det}) where 
  \begin{compactitem}
    \item $(\delta')_0(\varepsilon,\alpha,o) = 2\cdot 3$ and   $(\delta')_0(\varepsilon,\alpha,e)=0 \cdot 2$ and
      \[
        (\delta')_2(q_1q_2,\sigma,o) = \begin{cases} \frac{1}{3}\cdot \frac{1}{2}\cdot 1 \cdot 3  & \text{ if $q_1=o$ and $q_2=e$}\\
          \frac{1}{2}\cdot \frac{1}{3} \cdot  1 \cdot 3 & \text{ if $q_1=e$ and $q_2=o$}\\
          0 & \text{ otherwise }
        \end{cases}
        \ \ \ \
        = \ \ \ \
         \begin{cases} \frac{1}{2}  & \text{ if $q_1\not= q_2$}\\
           0 & \text{ otherwise }
           \end{cases}
        \]
        and
        \[
       (\delta')_2(q_1q_2,\sigma,e) = \begin{cases}  \frac{1}{2}\cdot \frac{1}{2}\cdot 1 \cdot 2 & \text{ if $q_1= q_2=e$}\\
         \frac{1}{3}\cdot \frac{1}{3}\cdot 1 \cdot 2 & \text{ if $q_1= q_2=o$}\\
          0 & \text{ otherwise }
        \end{cases}
        \ \ \ \ = \ \ \ \
        \begin{cases}  \frac{1}{2} & \text{ if $q_1= q_2=e$}\\
         \frac{2}{9} & \text{ if $q_1= q_2=o$}\\
          0 & \text{ otherwise }\enspace,
        \end{cases}
      \]
      \item $(F')_o= (F')_e=1$.
      \end{compactitem}
  Obviously, $\push_\lambda(\cA)$ has unit root weights and it is total and bu-deterministic.    Below, we abbreviate $\push_\lambda(\cA)$ by $\cB$. 

 \begin{figure}[t]
   \begin{center}
\begin{tikzpicture}
\tikzset{node distance=7em, scale=0.5, transform shape}
\node[state, rectangle] (1) {\Large $\alpha$};
\node[state, right of=1] (2){\Large $o$};
\node[state, rectangle, right of=2] (3)[right=1em]{\Large $\sigma$};
\node[state, rectangle, above of=3] (4)[above=1em]{\Large $\sigma$};
\node[state, rectangle, below of=3] (5)[below=1em]{\Large $\sigma$};
\node[state, right of=3] (6)[right=1em]{\Large $e$};
\node[state, rectangle, right of=6] (7) {\Large $\sigma$};

\tikzset{node distance=2em}
\node[above of=1] (w1)[yshift=0.4em] {\Large 6};
\node[above of=2] (w2)[left=0.1em,yshift=0.5em] {\Large 1};
\node[above of=3] (w3)[yshift=0.7em] {\Large $\tfrac{2}{9}$};
\node[above of=4] (w4)[yshift=0.7em] {\Large $\tfrac{1}{2}$};
\node[above of=5] (w5)[yshift=0.7em] {\Large $\tfrac{1}{2}$};
\node[above of=6] (w6)[yshift=0.7em] {\Large 1};
\node[above of=7] (w7)[yshift=0.7em] {\Large $\tfrac{1}{2}$};

\draw[->,>=stealth] (1) edge (2);
\draw[->,>=stealth] (2) edge[out=20, in=155, looseness=1.1] (3);
\draw[->,>=stealth] (2) edge[out=-20, in=205, looseness=1.1] (3);
\draw[->,>=stealth] (2) edge (4);
\draw[->,>=stealth] (2) edge (5);
\draw[->,>=stealth] (3) edge (6);
\draw[->,>=stealth] (6) edge (4);
\draw[->,>=stealth] (4) edge[out=110, in=80, looseness=1.4] (2);
\draw[->,>=stealth] (6) edge (5);
\draw[->,>=stealth] (5) edge[out=250, in=-80, looseness=1.4] (2);
\draw[->,>=stealth] (7) edge (6);
\draw[->,>=stealth] (6) edge[out=60, in=30, looseness=2.7] (7);
\draw[->,>=stealth] (6) edge[out=-60, in=-30, looseness=2.7] (7);
\end{tikzpicture} 
\caption{\label{fig:even-odd-two-three-2-det} The $(\Sigma,\Ratnum)$-wta  $\push_\lambda(\cA) = (Q,\delta',F')$.}
   \end{center}
   \end{figure}

      Using $\cB$ as abbreviation of $\push_\lambda(\cA)$, for every $\xi \in \T_\Sigma$ and $q \in Q$, we have $\h_\cB(\xi)_q = \h_\cA(\xi)_q \cdot F_q$ (cf.~\eqref{eq:push-lemma}). For instance:
      \begingroup
      \allowdisplaybreaks
      \begin{align*}
        \h_\cB(\sigma(\alpha,\alpha))_e &= \h_\cB(\alpha)_e \cdot \h_\cB(\alpha)_e \cdot (\delta')_2(ee,\sigma,e) +
        \h_\cB(\alpha)_o \cdot \h_\cB(\alpha)_o  \cdot (\delta')_2(oo,\sigma,e) \\
                                        &= (\delta')_0(\varepsilon,\alpha,e) \cdot (\delta')_0(\varepsilon,\alpha,e) \cdot (\delta')_2(ee,\sigma,e) +
                                          (\delta')_0(\varepsilon,\alpha,o) \cdot (\delta')_0(\varepsilon,\alpha,o) \cdot (\delta')_2(oo,\sigma,e)\\
                                        &= (0 \cdot 2) \cdot (0 \cdot 2) \cdot \Big(\tfrac{1}{2}\cdot \tfrac{1}{2}\cdot 1 \cdot 2\Big)
                                          + (2 \cdot 3) \cdot (2 \cdot 3) \cdot \Big(\tfrac{1}{3}\cdot \tfrac{1}{3}\cdot 1 \cdot 2\Big) \\
        &= \Big(0  \cdot 0  \cdot 1 +
          2  \cdot 2 \cdot 1 \Big) \cdot 2 \\
        &= \Big(\delta_0(\varepsilon,\alpha,e)  \cdot \delta_0(\varepsilon,\alpha,e)  \cdot \delta_2(ee,\sigma,e) +
        \delta_0(\varepsilon,\alpha,o)  \cdot \delta_0(\varepsilon,\alpha,o)   \cdot \delta_2(oo,\sigma,e)\Big) \cdot 2 \\
        &= \Big(\h_\cA(\alpha)_e \cdot \h_\cA(\alpha)_e \cdot \delta_2(ee,\sigma,e) +
        \h_\cA(\alpha)_o \cdot \h_\cA(\alpha)_o  \cdot \delta_2(oo,\sigma,e)\Big) \cdot 2 \\
                                          &= \h_\cA(\sigma(\alpha,\alpha))_e \cdot F_e \enspace.
        \end{align*}
      \endgroup
     Moreover, $\sem{\push_\lambda(\cA)}(\xi) = \sem{\cA}(\xi)$ for each $\xi \in \T_\Sigma$.
\hfill $\Box$
\end{example}


\section{Transforming a bu-deterministic wta into a slim one}
\label{sect:bu-det-wta-into-simple-wta}

\index{weighted tree automaton!slim}
Here we show that each bu-deterministic $(\Sigma,\B)$-wta $\cA$ with state set $Q$ can be transformed into an equivalent slim $(\Sigma,\B)$-wta. Intuitively, $\cA$ is slim if each state $q \in Q$ can be reached by the support fta $\supp(\cA)$ of $\cA$. This reachability is expressed in the state algebra $\St(\cA)$ of $\cA$ (cf. Section~\ref{sec:state-algebra-of-wta}). 
We recall  that the state algebra of $\cA$ is the $\Sigma$-algebra $\St(\cA) =(\cP(Q),\delta_Q)$ such that, for every $k \in \mathbb{N}$, $\sigma \in \Sigma^{(k)}$, and $P_1,\ldots,P_k \in \cP(Q)$, we have 
\begin{equation*}
\delta_Q(\sigma)(P_1,\ldots,P_k) = \{q \in Q \mid (\exists q_1 \in P_1) \ldots (\exists q_k \in P_k): \delta_k(q_1 \cdots q_k,\sigma,q)\ne \0\} \enspace.
\end{equation*}
The $\Sigma$-algebra homomorphism from $\sfT_\Sigma$ to $\St(\cA)$ is denoted by $\state_\cA$. Since $\cA$ is bu-deterministic,  for each $\xi \in \T_\Sigma$, the set $\state_\cA(\xi)$ contains at most one element (cf. Observation~\ref{obs:subalgebras-of-the-state-algebra}(1)).

Formally, a bu-deterministic $(\Sigma,\B)$-wta $\cA=(Q,\delta,F)$ is \emph{slim} if, for each $q \in Q$, there exists $\xi \in \T_\Sigma$ such that $\state_\cA(\xi) = \{q\}$; or, in other words, $\im(\state_\cA) = \{\{q\} \mid q \in Q\}$.


\begin{theorem}\label{lm:construction-of-simple-det-new} Let  $\cA$ be a bu-deterministic $(\Sigma,\B)$-wta. We can construct a slim $(\Sigma,\B)$-wta $\cA'$ such that $\sem{\cA}=\sem{\cA'}$.
\end{theorem}
\begin{proof}  
First we compute the set $\im(\state_\cA)$. By Observation~\ref{obs:smallest-subalgebra-im}, we have $\im(\state_\cA)= \langle \emptyset\rangle_{\im(\delta_Q)}$. Then by Lemma~\ref{obs:Knaster-Tarski-applied-to-algebras}, we can construct the set $\langle \emptyset\rangle_{\im(\delta_Q)}$, because it is finite.

For the construction of $\cA'$, we distinguish the following two cases.

\underline{$\im(\state_\cA)=\{\emptyset\}$}: By Lemma~\ref{obs:state-properties-of-wta}(6), we have $\sem{\cA}=\widetilde{\0}$. 
We construct the bu-deterministic $(\Sigma,\B)$-wta $\cA'=(\{p\},\delta',F')$ where,  for every $k\in\mathbb{N}$ and $\sigma\in\Sigma^{(k)}$, we let $\delta'_k(p^k,\sigma,p)=\1$, and $F'_p=\0$.
Then $\cA'$ is slim because $\im(\state_\cA) = \{\{p\}\}$,  and it is obvious that $\sem{\cA'}=\widetilde{\0}$.

\underline{$\im(\state_\cA)\ne\{\emptyset\}$}: We construct the $(\Sigma,\B)$-wta $\cA'=(Q',\delta',F')$ where
\begin{compactitem}
\item $Q'= \{q \in Q \mid \{q\} \in \im(\state_\cA)\}$,
\item $\delta'=(\delta'_k \mid k\in \mathbb{N})$ and, for each $k\in \mathbb{N}$, the mapping $\delta'_k$ is the restriction of $\delta_k$ to $(Q')^k \times \Sigma^{(k)}\times Q'$, and 
\item $F'=F|_{Q'}$.
\end{compactitem}

It is obvious that $\cA'$ is bu-deterministic. We show that $\cA'$ is slim. As a first step for this, we show that
\begin{equation}\label{equ:succ-equ-new}
\text{for every $k\in \mathbb{N}$, $w\in (Q')^k$, and $\sigma\in\Sigma^{(k)}$, we have $\mysucc_{\cA'}(w,\sigma)=\mysucc_{\cA}(w,\sigma)$.}
\end{equation}

Let $w=q_1\cdots q_k$ with $q_1,\ldots,q_k\in Q'$.
Obviously, $\mysucc_{\cA'}(w,\sigma)\subseteq\mysucc_{\cA}(w,\sigma)$ by the definition of $\delta'$. 
We show that the inclusion $\mysucc_{\cA}(w,\sigma)\subseteq \mysucc_{\cA'}(w,\sigma)$ also holds. 
This is trivial if $\mysucc_{\cA}(w,\sigma)=\emptyset$, therefore assume that $\mysucc_{\cA}(w,\sigma)=\{p\}$ for some $p\in Q$.
Then $\delta_k(w,\sigma,p)\ne \0$, hence $\delta_Q(\sigma)(\{q_1\},\ldots,\{q_k\})=\{p\}$.
Since $q_i \in Q'$ for each  $i\in[k]$,  there exists $\xi_i\in \T_\Sigma$ such that
$\{q_i\}=\state_\cA(\xi_i)$.
Then $\state_\cA(\sigma(\xi_1,\ldots,\xi_k))=\{p\}$, i.e., $p\in Q'$.
By the definition of $\delta'$, we have $\delta'_k(w,\sigma,p)=\delta_k(w,\sigma,p)\ne \0$, hence 
$\mysucc_{\cA'}(w,\sigma)=\{p\}$.
This proves \eqref{equ:succ-equ-new}.

Next we show the following. 
\begin{equation}\label{equ:theta-equ-new}
  \begin{aligned}
    &\text{For every $k\in \mathbb{N}$, $q_1,\ldots,q_k \in Q'$, and $\sigma\in\Sigma^{(k)}$, we have }\\
    &\delta_Q(\sigma)(\{q_1\},\ldots,\{q_k\})=\delta_{Q'}(\sigma)(\{q_1\},\ldots,\{q_k\}) \enspace.
    \end{aligned}
\end{equation}
We compute as follows:
\begingroup
\allowdisplaybreaks
\begin{align*}
  \delta_{Q'}(\sigma)(\{q_1\},\ldots,\{q_k\})
  &= \mysucc_{\cA'}(q_1\cdots q_k,\sigma)\\
  &= \mysucc_{\cA}(q_1 \cdots q_k,\sigma)
        \tag{by using \eqref{equ:succ-equ-new}}\\
&=  \delta_{Q}(\sigma)(\{q_1\},\ldots,\{q_k\}).
\end{align*}
\endgroup
This proves \eqref{equ:theta-equ-new}.

From \eqref{equ:theta-equ-new} and the fact that, for each $\xi \in \T_\Sigma$, the set $\state_\cA(\xi)$ contains at most one element, it follows that
\begin{equation}\label{equ:state-equ-new}
\text{for every $\xi \in \T_\Sigma$, we have
     $\state_\cA(\xi) = \state_{\cA'}(\xi)$.}
\end{equation}

Then, by \eqref{equ:state-equ-new} and by the definition of $Q'$, we obtain that $\cA'$ is slim.

Lastly, we prove that $\sem{\cA}=\sem{\cA'}$. First we observe that, 
\begin{equation}\label{eq:h-on-Q-minus-Q-prime}
\text{for every $\xi\in \T_\Sigma$ and $q\in (Q\setminus Q')$, we have $\h_{\cA}(\xi)_{q}=\0$.}
\end{equation}
The statement follows from Lemma~\ref{lm:hAxiqne0-implies-qinhQxi-for-bu-det-wta}(1)  because if 
$q\in (Q\setminus Q')$, then $\state_\cA(\xi)\ne \{q\}$.

Next, by induction on $\T_\Sigma$, we show that
\begin{equation}\label{equ:hAxia=hAprimexiq-Z-new}
  \text{for every $\xi \in \T_\Sigma$ and $q\in Q'$, we have
   $\h_\cA(\xi)_{q}=\h_{\cA'}(\xi)_{q}$.}
\end{equation}
Let $\xi=\sigma(\xi_1,\ldots,\xi_k)$ and $q\in Q'$.  Then
\begingroup
\allowdisplaybreaks
\begin{align*}
  \h_{\cA'}(\xi)_{q}
  &= \bigoplus_{q_1 \cdots q_k \in (Q')^k} \Big( \bigotimes_{i\in [k]} \h_{\cA'}(\xi_i)_{q_i}\Big) \otimes \delta'_k(q_1\cdots q_k,\sigma,q)\\
  &= \bigoplus_{q_1 \cdots q_k \in (Q')^k} \Big( \bigotimes_{i\in [k]} \h_{\cA}(\xi_i)_{q_i}\Big) \otimes \delta'_k(q_1\cdots q_k,\sigma,q) \tag{by I.H.}\\
  &= \bigoplus_{q_1 \cdots q_k \in (Q')^k} \Big( \bigotimes_{i\in [k]} \h_{\cA}(\xi_i)_{q_i}\Big) \otimes \delta_k(q_1\cdots q_k,\sigma,q) \tag{by the definition of $\delta'_k$}\\
  &= \bigoplus_{q_1 \cdots q_k \in Q^k} \Big( \bigotimes_{i\in [k]} \h_{\cA}(\xi_i)_{q_i}\Big) \otimes \delta_k(q_1\cdots q_k,\sigma,q) \tag{by \eqref{eq:h-on-Q-minus-Q-prime} }\\
  &= \h_{\cA}(\xi)_{q}
\end{align*}
\endgroup

Now we are ready to show that $\sem{\cA}=\sem{\cA'}$. For this, let $\xi \in \T_\Sigma$. Then
\begingroup
\allowdisplaybreaks
\begin{align*}
  \sem{\cA'}(\xi) & = \bigoplus_{q\in Q'} \h_{\cA'}(\xi)_{q}\otimes F'_q \\
   & =  \bigoplus_{q\in Q'} \h_{\cA}(\xi)_{q}\otimes F_q 
   \tag{by \ref{equ:hAxia=hAprimexiq-Z-new} and the definition of $F'$} \\
   & =  \bigoplus_{q\in Q} \h_{\cA}(\xi)_{q}\otimes F_q 
  \tag{by \eqref{eq:h-on-Q-minus-Q-prime} } \\
   & = \sem{\cA}(\xi)
     \qedhere
\end{align*}
\endgroup
\end{proof}

\begin{example}\rm \label{ex:example-of-slim-wta-for-a-given-bu.det-wta-new}
  Let $\Sigma = \{\alpha^{(0)},\beta^{(0)}\}$ and $\cA=(Q,\delta,F)$ be the bu-deterministic $(\Sigma,\Ratnum)$-wta  with $Q=\{p_1,p_2\}$, and  $\delta_0(\varepsilon,\alpha,p_1)=1$,  $\delta_0(\varepsilon,\alpha,p_2)=\delta_0(\varepsilon,\beta,p_1)=\delta_0(\varepsilon,\beta,p_2)=0$, and $F_{p_1} = F_{p_2}=1$. 

  Then, in the state algebra $\St(\cA) = (\cP(Q),\delta_\cA)$, we have
$\delta_\cA(\alpha)()=\{p_1\}$ and $\delta_\cA(\beta)()=\emptyset$. Thus, $\im(\state_\cA)) = \{\{p_1\}\}$.

By applying the construction described in the proof of Theorem \ref{lm:construction-of-simple-det-new},
    we obtain the bu-deterministic $(\Sigma,\Ratnum)$-wta $\cA'=(Q',\delta',F')$ which is slim and equivalent to $\cA$, where $Q'=\{p_1\}$ and $\delta'_0(\varepsilon,\alpha,p_1) =1$, $\delta'_0(\varepsilon,\beta,p_1)=0$, and $F'_{p_1}=1$.
    \hfill $\Box$
    \end{example}


\section{Normalizing transition weights of wta}
\label{sect:normalizing-transition-weights}

Finally, we want to show that each $(\Sigma,\B)$-wta $\cA$ can be transformed into a run equivalent crisp $(\Sigma,\B)$-wta provided that the set 
\index{HA@$\rmH(\cA)$}
\begin{align}\label{alg:H(A)}
  \rmH(\cA) = \{\wt(\xi,\rho)  \mid \xi \in \T_\Sigma , \rho \in \R_\cA(\xi)\}
\end{align}
is finite. Due to this requirement on $\cA$, we can code the weights of runs into the states and prove the following lemma.

\begin{lemma} \rm \cite[Lm.~6.5]{drofulkosvog21} \label{lm:compute-c(A)-normal-form}
Let $\cA=(Q,\delta,F)$ be a $(\Sigma,\B)$-wta. If $\rmH(\cA)$ is finite, then we can construct the set $\rmH(\cA)$.
\end{lemma}
\begin{proof} Let  $\rmH(\cA)$ be finite.
For every $n \in \mathbb{N}$ and $q \in Q$ let
\[H_{n, q} = \{\wt(\xi,\rho) \mid \xi \in \T_\Sigma, \height(\xi) \leq n, \rho \in \R_\cA(q,\xi)\}\enspace.\]
Clearly, we have $H_{0,q} \subseteq H_{1,q} \subseteq \ldots \subseteq \rmH(\cA)$ for each $q \in Q$.
We claim that
\begin{equation}
  \text{for each $n \in \mathbb{N}$, if for each $q\in Q$: $H_{n,q} = H_{n+1,q}$, then for each $q\in Q$: $H_{n+1,q} = H_{n+2,q}$}.\label{eq:stable}
\end{equation}
To show this, let $n \in \mathbb{N}$, $q \in Q$, and $b \in H_{n+2,q}$. There exist  $\xi \in \T_\Sigma$ and $\rho \in \R_\cA(q,\xi)$ such that $\height(\xi) \leq n+2$ and $\wt(\xi,\rho) = b$. We may assume that $\height(\xi) = n+2$. Hence $\xi=\sigma(\xi_1,\ldots,\xi_k)$  such that $\height(\xi_i)\leq n+1$ for each $i \in [k]$. Clearly, for each $i \in [k]$, we have $\wt(\xi_i,\rho|_i) \in H_{n+1,\rho(i)}$, so by our assumption there exist   $\zeta_i \in \T_\Sigma$ with $\height(\zeta_i)\leq n$ and  run $\theta_i \in \R_\cA\big(\rho(i),\zeta_i\big)$
such that $\wt(\xi_i,\rho|_i) = \wt(\zeta_i,\theta_i)$.

Now let $\zeta=\sigma(\zeta_1,\ldots,\zeta_k)$. Obviously, $\height(\zeta) \le n+1$. Moreover, let $\theta \in \R_\cA(q,\zeta)$ such that $\theta|_i = \theta_i$ for each $i \in [k]$. Clearly, $\wt(\zeta,\theta) \in H_{n+1,q}$, and we calculate
\begin{align*}
    \wt(\zeta,\theta)
    &= \Big( \bigotimes_{i \in [k]} \wt(\zeta_i,\theta|_i) \Big) \otimes \delta_k\big(\theta(1) \cdots \theta(i),\sigma,q\big)\\
    &= \Big( \bigotimes_{i \in [k]} \wt(\xi_i, \rho|_i) \Big) \otimes \delta_k\big(\rho(1) \cdots \rho(i),\sigma,q\big) = \wt(\xi,\rho) = b\enspace. 
\end{align*}
This shows that $b \in H_{n+1,q}$, proving \eqref{eq:stable}.

We recall that $H_{0,q} \subseteq H_{1,q} \subseteq \ldots \subseteq \rmH(\cA)$ for each $q \in Q$.
Obviously,  we can construct $H_{n,q}$ for every $n\in \mathbb{N}$ and $q\in Q$.

Then, since $\rmH(\cA)$ is finite, by constructing $H_{0,q}$ for each $q\in Q$, $H_{1,q}$ for each $q\in Q$, and so on, we can find the least number $n_m\in \mathbb{N}$ such that
$H_{n_m,q}=H_{n_m+1,q}$ for each $q\in Q$ and thus by  \eqref{eq:stable} we have $H_{n_m,q}=H_{j,q}$ for every $q\in Q$ and $j\in \N$ with $j \geq n_m$.

We show that $\rmH(\cA) = \bigcup_{q \in Q} H_{n_m, q}$. For this, let $b\in \rmH(\cA)$,
i.e., $b=\wt(\xi,\rho)$ for some $\xi \in \T_\Sigma$ with $\height(\xi)=j$, $q\in Q$ and $\rho \in \R_\cA(q,\xi)$. Then $b\in H_{j,q}=H_{n_m,q}$. The other inclusion is obvious. Since we can construct the set $\bigcup_{q \in Q} H_{n_m, q}$, the set $\rmH(\cA)$ can be constructed.
\end{proof}

Now we can prove the following normal form theorem.

\begin{theorem-rect}\label{thm:normalizing-transition-weights} Let $\Sigma$ be a ranked alphabet. Moreover, let $\B$ be a strong bimonoid and let $\cA$ be a $(\Sigma,\B)$-wta $\cA$ such that $\rmH(\cA)$ is finite. Then we can construct a crisp $(\Sigma,\B)$-wta $\cB$ such that $\runsem{\cB} = \runsem{\cA}$. Moreover, if $\cA$ is total and bu-deterministic, then $\cB$ is crisp-deterministic.
\end{theorem-rect}
\begin{proof} Let $\B=(B,\oplus,\otimes,\0,\1)$ and $\cA=(Q,\delta,F)$. Since $\rmH(\cA)$ is finite, by Lemma \ref{lm:compute-c(A)-normal-form}, we can construct $\rmH(\cA)$.

  Now we construct the $(\Sigma,\B)$-wta $\cB=(Q',\delta',F')$ as follows.
  \begin{compactitem}
  \item $Q' = Q \times \rmH(\cA)$,
   \item for every $k \in \mathbb{N}$, $\sigma \in \Sigma^{(k)}$, and $(q_1,b_1),\ldots,(q_k,b_k),(q,b) \in Q'$ we define
    \[
      (\delta')_k((q_1,b_1)\cdots (q_k,b_k),\sigma,(q,b)) =
      \begin{cases}
        \1 & \text{ if $b= \Big(\bigotimes_{i \in [k]} b_i\Big) \otimes \delta_k(q_1\cdots q_k,\sigma,q)$}\\
        \0 & \text{ otherwise} 
        \end{cases} \ \ \text{ and }
      \]
        \item for each  $(q,b) \in Q'$, we let $(F')_{(q,b)} = b \otimes F_q$.
        \end{compactitem}
        Obviously, $\cB$ is crisp and  if $\cA$ is total and  bu-deterministic, then $\cB$ is crisp-deterministic.

 As preparation for the proof of $\runsem{\cA} = \runsem{\cB}$, we define a bijection between the set of runs of $\cA$ on an input tree and the set of runs of $\cB$ on that tree. 
Indeed, it is easy to see that for every $q\in Q, \xi \in \T_\Sigma$, and $b\in \rmH(\cA)$, the mapping
\begin{align*}
\varphi: \{ \rho \in \R_\cA(q,\xi) \mid \wt_\cA(\xi,\rho) = b\} \ \to \ \{ \rho' \in \R_\cB((q,b),\xi) \mid \wt_\cB(\xi,\rho') = \1\} 
\end{align*} 
defined, for each $w\in \pos(\xi)$, by $\varphi(\rho)(w) = (\rho(w),\wt_\cA(\xi|_w,\rho|_w))$ is a bijection. 
    
Then we can calculate as follows.
    \begingroup
    \allowdisplaybreaks
    \begin{align*}
      \runsem{\cA}(\xi)
      &=  \bigoplus_{q \in Q}\bigoplus_{\rho \in \R_\cA(q,\xi)} \wt_\cA(\xi,\rho) \otimes F_q\\
      &=  \bigoplus_{q \in Q}\bigoplus_{b \in \rmH(\cA)}\bigoplus_{\substack{\rho \in \R_\cA(q,\xi):\\ \wt_\cA(\xi,\rho) = b}} \wt_\cA(\xi,\rho) \otimes F_q\\
      &=  \bigoplus_{(q,b) \in Q'}\bigoplus_{\substack{\rho \in \R_\cA(q,\xi):\\ \wt_\cA(\xi,\rho) = b}} b \otimes F_q \tag{by the definition of $Q'$}\\
       & =\bigoplus_{(q,b) \in Q'} \bigoplus_{\substack{\rho' \in \R_\cB((q,b),\xi):\\ \wt_\cB(\xi,\rho') = \1}} b \otimes F_q \tag{because $\varphi$ is a bijection}\\
       & =\bigoplus_{(q,b) \in Q'} \bigoplus_{\substack{\rho' \in \R_\cB((q,b),\xi):\\ \wt_\cB(\xi,\rho') = \1}} \wt_\cB(\xi,\rho') \otimes b \otimes F_q\\
       & =\bigoplus_{(q,b) \in Q'} \bigoplus_{\rho' \in \R_\cB((q,b),\xi)} \wt_\cB(\xi,\rho') \otimes b \otimes F_q \tag{because $ \wt_\cB(\xi,\rho')\in \{\0,\1\}$}\\
       & =\bigoplus_{(q,b) \in Q'} \bigoplus_{\rho' \in \R_\cB((q,b),\xi)} \wt_\cB(\xi,\rho')\otimes (F')_{(q,b)} \tag{by definition of $F'$}\\
       & = \runsem{\cB}(\xi) \qedhere
                  \end{align*}
    \endgroup   
  \end{proof} 

%% file: wcfg.tex
\chapter{Weighted context-free grammars}\label{chapt:wcfg}

In the theory of tree languages it is a fundamental theorem that the yield of a recognizable tree language is a context-free language and, vice versa, each context-free language can be obtained in this way \cite{tha67} (also cf. \cite[Thm.~3.28]{eng75-15} and \cite[Sect.~3.2]{gecste84}).  This fundamental theorem has been generalized to the weighted setting in \cite{esikui03} for continuous and commutative semirings and in \cite{fulgaz18} for arbitrary semirings. Here we generalize it for arbitrary strong bimonoids.
 
First, we  recall the concept of weighted context-free grammars (cf. Section \ref{subsect:basic-modell}) and prove a number of normal form lemmas:  nonterminal form, start-separated, local-reduced, chain-free, and
$\varepsilon$-free (cf. Section \ref{sect:wcfg-normal-forms}). Then we prove that a weighted language is context-free if and only if it is the yield of an r-recognizable weighted tree language (cf. Theorem \ref{thm:yield(wta)=cf}).

We note that in \cite{rahtor19} weighted context-free grammars over bimonoids were investigated; roughly speaking, a bimonoid is a strong bimonoid in which the addition need not be commutative and the annihilation law ($b \otimes \0=\0 \otimes b=\0$) need not hold.
Moreover, we note that we will use the concept of weighted context-free grammar to define the concept of weighted regular tree grammar (cf. Chapter \ref{ch:regular-tree-grammars}) and the concept of weighted projective bimorphism (cf. Section~\ref{sec:w-projective-bimorphisms}).

\section{The grammar model }
\label{subsect:basic-modell}

Weighted context-free grammars were introduced in \cite{chosch63} where the weights reflected the degree of ambiguity (also cf \cite{sha67}).
Similar to such grammars, finite systems of algebraic equations were investigated in \cite{salsoi78,kuisal86,esikui03}. We build our definitions along \cite{drovog14}.

\index{weighted context-free grammar}
\index{wcfg}
\index{cfg underlying}
\index{GammaBwcfg@$(\Gamma,\B)$-wcfg}
  A {\em weighted context-free grammar over $\Gamma$ and $\B$} (for short: $(\Gamma,\B)$-wcfg, or: wcfg) is a tuple $\cG = (N,S,R,wt)$, where
\begin{compactitem}
  \item $(N,S,R)$ is a $\Gamma$-cfg which has a terminal rule\footnote{We need this condition because we wish to define the semantics of a wcfg in terms of rule trees (cf. Section \ref{sec:context-free-grammars})} and 
\item  $wt: R \rightarrow B$ is the \emph{weight mapping}. 
\end{compactitem}
 The \emph{cfg underlying $\cG$}, denoted by $\cG^\mathrm{u}$, is the $\Gamma$-cfg $(N,S,R)$.

Since each $(\Gamma,\B)$-wcfg $\cG$ is  an extension of some $\Gamma$-cfg (viz. by some weight mapping $wt$), the concepts and abbreviations which are defined for context-free grammars (cf. Section \ref{sec:context-free-grammars}) are also available for weighted-context free grammars. This concerns, in particular,
\begin{compactitem}
\item the abbreviations $\lhs(r)$ and $\rhs_{N,i}(r)$ for some rule $r$ of $\cG$,
\item the explicit form $A \to u_0A_1u_1 \cdots A_ku_k$ of a rule $r$ of $\cG$,
  \index{pi@$\pi_{\cG^\mathrm{u}}$}
  \index{pi@$\pi_{\cG}$}
  \item the mapping  $\pi_{\cG^\mathrm{u}}: \T_R \to \Gamma^*$, which we will call \emph{projection of $\cG$} and denote by $\pi_\cG$,
\item     \index{RT@$\RT_{\cG^\mathrm{u}}(N',L)$}
the concept of rule tree; we will denote the sets $\RT_{\cG^\mathrm{u}}(A,u)$,  $\RT_{\cG^\mathrm{u}}(N',L)$, $\RT_{\cG^\mathrm{u}}(L)$, and $\RT_{\cG^\mathrm{u}}$ by $\RT_\cG(A,u)$,  $\RT_\cG(N',L)$, $\RT_\cG(L)$, and $\RT_\cG$, respectively.   \index{RT@$\RT_\cG(N',L)$}
  \end{compactitem}

\index{wt@$\wt_\cG$}
Next we define the weight of rule trees by using the concept of evaluation algebra (cf. Section~\ref{sect:trees}).  For this we consider the mapping $wt$ as $\mathbb{N}$-indexed family $(wt_k \mid k \in \mathbb{N})$ of mappings $wt_k: R^{(k)} \to B$ by defining $wt_k = wt|_{R^{(k)}}$. Then, for each $d \in \T_R$, the \emph{weight of $d$} is the value $\h_{\M(R,wt)}(d)$ in $B$, where $\M(R,wt)$ is the $(R,wt)$-evaluation algebra. For convenience, we will abbreviate $\h_{\M(R,wt)}$ by $\wt_\cG$. We note that, for each $d=r(d_1,\ldots,d_k)$ in $\T_R$, we have
\begin{equation}\label{equ:wt-rules-tree-as-evaluation}
  \wt_\cG(d)=\wt_\cG(d_1)\otimes\cdots\otimes\wt_\cG(d_k)\otimes  wt(r)\enspace,
  \end{equation}
by \eqref{eq:evaluation-of-a-tree}.  If there is no confusion, then we drop the index $\cG$ from $\wt_\cG$ and  just write $\wt(d)$ for the weight of $d$.

\label{page:sem-wrtg}
\index{semanticG@$\wrtsem{\cG}$}
\index{weighted context-free grammar!weighted rule tree language}
\index{weighted rule tree language}
We define the \emph{weighted rule tree language of $\cG$}, denoted by $\wrtsem{\cG}$, as the weighted tree language $\wrtsem{\cG}: \T_R \to B$ defined by the Hadamard product
\[
\wrtsem{\cG} = \wt_\cG \otimes \chi(\RT_\cG) \enspace.
\]
\label{page:semwrtg-wls}

\index{finite-derivational}
\index{weighted context-free grammar!finite-derivational}
We say that \emph{$\cG$ is finite-derivational} if the set $\RT_\cG(u)$ is finite for every $u \in \Gamma^*$.
We note that, if $\cG$ is finite-derivational, then $\wrtsem{\cG}$ is $\chi(\pi_\cG)$-summable
(cf. Subsection \ref{sec:weighted-tree-transformations}) because
$\pi_\cG^{-1}(u)\cap \supp(\wrtsem{\cG})\subseteq \RT_\cG(u)$ for each $u\in \Gamma^*$.

\index{semanticG@$\ssem{\cG}$}
\index{weighted  language generated by $\cG$}
If $\cG$ is finite-derivational or $\B$ is $\sigma$-complete, then the {\em weighted  language generated by $\cG$} is the weighted language $\ssem{\cG}: \Gamma^* \rightarrow B$ defined by
\[
  \ssem{\cG} = \chi(\pi_\cG)\big(\wrtsem{\cG}\big) \enspace.
  \]
  Thus, for each $u \in \Gamma^*$, we have
\begin{eqnarray}
\begin{aligned}
 & \ssem{\cG}(u) =  \chi(\pi_\cG)\big(\wrtsem{\cG}\big)(u)
  = \infsum{\oplus}{d \in \pi_\cG^{-1}(u)}{\wrtsem{\cG}(d)} \\
 & =  \infsum{\oplus}{d \in \pi_\cG^{-1}(u)}{\big(\wt_\cG \otimes \chi(\RT_\cG)\big)(d)}
  = \infsum{\oplus}{d \in \RT_\cG(u)}{\wt_\cG(d)}\enspace, 
  \end{aligned} \label{eq:sem-wcfg-der-tree}
\end{eqnarray}
where in the second equality we have used \eqref{obs:app-tree-transf-to-wtl-2}.
In the last expression of \eqref{eq:sem-wcfg-der-tree}, the sum is well defined: if $\B$ is $\sigma$-complete, then the sum is defined on page~\pageref{page:summation-inf} and, if in addition $\cG$ is finite-derivational, then it is equal to $\bigoplus_{d \in \RT_\cG(u)}{\wt(d)}$ by \eqref{equ:fin0=inf0}; if $\B$ is not $\sigma$-complete, then $\cG$ is finite-derivational and the sum denotes $\bigoplus_{d \in \RT_\cG(u)}{\wt(d)}$  by our convention (cf. page \pageref{page:summation-fin}). Thus, the sum is well defined for arbitrary~$\B$. Figure~\ref{fig:rule-tree-weight} illustrates the definition of $\ssem{\cG}(u)$.

\begin{figure}
  \centering
\begin{tikzpicture}[scale=1, every node/.style={transform shape},
					node distance=-0.05cm and -0.05cm,
					mypoint/.style={circle,inner sep=0.75pt,draw,fill}]
				
\node[ellipse, draw, minimum height=3cm, minimum width=6.25cm] (e) {};
\node[above = 0.1cm of e.135] {$\pi_{\cG}^{-1}(u) \subseteq \T_{R}$};
\draw[bend right, out=335, in=-140, looseness=1.1] (e.6) to 
  node[anchor=east,pos=0.3,above left=-0.15cm] {$ \RT_\cG \cap \pi_\cG^{-1}(u) $} 
  coordinate[pos=0.86, right=1.15cm] (a) (e.226);
\node[mypoint,label={$d_1$}] at (a) (cd1) {};
\node[mypoint,label={$d_2$},right=0.5cm of cd1] (cd2) {};
\node[mypoint,label={$d_3$},right=0.5cm of cd2] (cd3) {};
\node[right=0.3cm of cd3] {$\cdots$};

\draw[semithick] (-5.5,-3) to 
  node[midway] (ph) {\phantom{$wt_{\cG}(d_1)$}}
  node[midway, below] {$u \in \Gamma^*$} (-2,-3);
\draw[->,>=stealth, shorten <= 0.5cm] (e) to 
  node[pos=0.4, left=0.4cm] {$\pi_{\cG}$} (ph);

\node at (0,-3) (g) {$\ssem{\cG}(u)=$};
\node[right= of g]   (gd1) {$\mathrm{wt}_{\cG}(d_1)$};
\node[right= of gd1] (ga1) {$\oplus$};
\node[right= of ga1] (gd2) {$\mathrm{wt}_{\cG}(d_2)$};
\node[right= of gd2] (ga2) {$\oplus$};
\node[right= of ga2] (gd3) {$\mathrm{wt}_{\cG}(d_3)$};
\node[right= of gd3] (ga3) {$\oplus$};
\node[right= of ga3]       {\strut $\ldots$};
\draw[->,>=stealth, shorten <= 0.1cm] (cd1) to (gd1);
\draw[->,>=stealth, shorten <= 0.1cm] (cd2) to (gd2);
\draw[->,>=stealth, shorten <= 0.1cm] (cd3) to 
  node[pos=0.4, right=0.5cm] {$\wt_{\cG}$} (gd3);
\end{tikzpicture}
  \caption{\label{fig:rule-tree-weight} An illustration of the definition of $\ssem{\cG}(u)$.}
  \end{figure}

\index{equivalent}
Let $\cG_1$ and $\cG_2$ be two $(\Gamma,\B)$-wcfg such that both $\cG_1$ and $\cG_2$ are finite-derivational or $\B$ is  $\sigma$-complete. Then $\cG_1$ and $\cG_2$  are \emph{equivalent} if $\ssem{\cG_1}=\ssem{\cG_2}$.

\index{weighted context-free language}
A weighted language  $s: \Gamma^* \to B$ is called {\em $(\Gamma,\B)$-weighted context-free language} if there exists a $(\Gamma,\B)$-wcfg $\cG$ which is finite-derivational if $\B$ is not $\sigma$-complete such that $s = \ssem{\cG}$.

\label{page:wcfg-B}
Since the Boolean semiring $\Boole=(\mathbb{B},\vee,\wedge,0,1)$ is $\sigma$-complete, for each $(\Gamma,\Boole)$-wcfg $\cG$, the weighted language $\ssem{\cG}: \Gamma^* \to \mathbb{B}$ is defined, and the support of $\ssem{\cG}$ is a context-free language. Vice versa, each context-free language over $\Gamma$ is the support of some $(\Gamma,\Boole)$-weighted context-free language.
Indeed, the concept of weighted context-free grammars generalizes the concept of context-free grammars in the following sense.

\begin{observation}\rm \label{obs:cfg=wcfg(B)} Let $L \subseteq \Gamma^*$. Then the following two statements are equivalent.
  \begin{compactenum}
  \item[(A)] We can construct a $\Gamma$-cfg $G$ which has a terminal rule such that $\LL(G) = L$.
    \item[(B)] We can construct a $(\Gamma,\Boole)$-wcfg $\cG$ such that $\supp(\ssem{\cG})=L$.    \end{compactenum}
    \end{observation} 
  \begin{proof}  
    Proof of (A)$\Rightarrow$(B): Let $G=(N,S,R)$ be a $\Gamma$-cfg with a terminal rule. Then we construct the $(\Gamma,\Boole)$-wcfg $\cG=(N,S,R,wt)$ such that for each $r \in R$ we let $wt(r) = 1$. Obviously, $\RT_\cG = \RT_G$ and, for each $d \in \RT_\cG$, we have $\wt_\cG(d)=1$. Then we can prove $\LL(G) = \supp(\ssem{\cG})$ as follows.
    \begingroup
    \allowdisplaybreaks
    \begin{align*}
      \supp(\ssem{\cG}) &= \supp(\chi(\pi_\cG)(\wt_\cG \otimes \chi(\RT_\cG))) =  \supp(\chi(\pi_\cG)(\chi(\RT_\cG)))\\
                        &= \pi_\cG(\supp(\chi(\RT_\cG))) \tag{by \eqref{equ:supp-yield=yieldA}}\\
      &=\pi_G(\RT_G) = \LL(G) \tag{because $\pi_\cG=\pi_G$ and $\RT_\cG = \RT_G$}\enspace.
      \end{align*}
      \endgroup

      \

      Proof of (B)$\Rightarrow$(A): Let $\cG=(N,S,R,wt)$ be a $(\Gamma,\Boole)$-wcfg. We distinguish two cases. 

      \underline{Case (a):} For each terminal rule $r \in R$, we have $wt(r) = 0$. Then $\supp(\ssem{\cG}) = \emptyset$. We construct the $\Gamma$-cfg $G=(\{A,S_0\},\{S_0\},R')$ where $R' = \{A \to \varepsilon\}$. Obviously, $\LL(G)= \emptyset$.

      \underline{Case (b):} There exists a terminal rule $r \in R$ for which  $wt(r) = 1$.   Then we construct the $\Gamma$-cfg  $G=(N,S,R')$ where $R' = \supp(wt)$. We can view $R'$ as a ranked alphabet. Hence $\RT_G=\RT_\cG \cap \T_{R'}$. We note that, for each $d \in \RT_\cG$, we have $\wt_\cG(d)=1$ if $d \in \T_{R'}$, and  $\wt_\cG(d) = 0$ otherwise.
      Then we can prove
      \begin{equation}
        \supp(\wt_\cG \otimes \chi(\RT_\cG)) = \RT_G \enspace.\label{equ:wcfg(B)-cfg}
      \end{equation}
      For this let $d \in \T_R$. Then
      \begingroup
    \allowdisplaybreaks
    \begin{align*}
      d \in \supp(\wt_\cG \otimes \chi(\RT_\cG))
      \ \text{ iff } \ \wt_\cG(d) = 1 \wedge d \in \RT_\cG  
      \ \text{ iff } \ d \in \T_{R'} \wedge d \in \RT_\cG
      \ \text{ iff } \ d \in \RT_G\enspace.
  \end{align*}
      \endgroup
      Then we can prove  $\LL(G) = \supp(\ssem{\cG})$ as follows.
        \begingroup
    \allowdisplaybreaks
    \begin{align*}
      \supp(\ssem{\cG}) &= \supp(\chi(\pi_\cG)(\wt_\cG \otimes \chi(\RT_\cG)))\\
                        &= \pi_\cG(\supp(\wt_\cG \otimes \chi(\RT_\cG))) \tag{by \eqref{equ:supp-yield=yieldA}}\\
                        &=\pi_\cG(\RT_G) \tag{by \eqref{equ:wcfg(B)-cfg}}\\
      &=\pi_G(\RT_G) = \LL(G) \tag{because $\pi_\cG=\pi_G$}\enspace.
      \end{align*}
      \endgroup
      \end{proof}


\begin{example}\rm\label{ex:cfg-N-wcf-1} Let  $G = (N,S,R)$ be an arbitrary $\Gamma$-cfg which has a terminal rule.
  \begin{enumerate}
    \item  (cf. \cite[Sec.~2.3]{chosch63}) We consider the $\sigma$-complete semiring $\Nat_\infty = (\mathbb{N}_\infty,+,\cdot,0,1)$ of natural numbers and we define the $(\Gamma,\Nat_\infty)$-wcfg $\cG = (N,S,R,wt)$ such that $wt(r)=1$ for each $r\in R$. Then $\wrtsem{\cG}=\chi(\RT_\cG)$, i.e., the characteristic mapping of $\RT_\cG$. Obviously, the sets $\D_{G,\lm}(S,u)$  and  $\RT_\cG(u)$ are in a one-to-one correspondence for each $u\in \Gamma^*$, and $\LL(G) = \pi_\cG(\RT_\cG)$. Moreover,   for each $u\in \Gamma^*$, we have 
      \[
        \ssem{\cG}(u)=\infsum{+}{d\in \RT_\cG(u)}\wt_\cG(d)=|\RT_\cG(u)|=|\D_{G,\lm}(S,u)|\enspace,
        \]
 where the second equality follows from the fact that   $\wt_\cG(d) = 1$ for each $d \in \RT_\cG(u)$.      
      Hence $\ssem{\cG}(u)$ is the number of leftmost derivations of $u$ in $G$.

    \item We consider the tropical semiring $\Natminplus=(\mathbb{N}_\infty,\min,+,\infty, 0)$, which is $\sigma$-complete.  Moreover, we define the $(\Gamma,\Natminplus)$-wcfg $\cG = (N,S,R,wt)$ such that $wt(r)=1$ for each $r\in R$ (as above). Then, for every $u \in \Gamma^*$ and  $d \in \RT_\cG(u)$, we have $\wt_\cG(d) = \size(d)$ and
      \[
        \ssem{\cG}(u) = \infsum{\min}{d \in \RT_\cG(u)}{\size(d)} \enspace.
      \]
      (We recall that $\infsum{\min}{}{}$ is the extension of min to an arbitrary countable index set.) Thus $\ssem{\cG}(u)$ is the minimal size of a rule tree for $u$.
Since $\size(d)$ is equal to the length of the leftmost derivation which corresponds to $d$, also the number $\ssem{\cG}(u)$ is the minimal length of a leftmost derivation of $u$ in $\cG$.
       \hfill $\Box$
      \end{enumerate}
\end{example}

\begin{example}\rm\label{ex:wcf}  (cf. \cite[Ex. 1]{fulgaz18}) Let $\Gamma=\{a,b\}$. It is known that the language $L=\{w\in\Gamma^*\mid |w|_a=|w|_b\}$ is context-free. It can be generated, for instance, by the $\Gamma$-cfg $G = (\{S\},S,R)$, where $R$ is the set of the following rules:
\begin{align*}
r_1:S\to SS, \; r_2:S\to aSb,  \; r_3:S\to bSa, \text{ and } r_4:S\to\varepsilon\enspace.
\end{align*}
Now we consider the tropical semiring $\Natminplus$ and define the  $(\Gamma,\Natminplus)$-wcfg $\cG = (\{S\},S,R,wt)$, where    $wt(r_1)=wt(r_2)=wt(r_3)=0$ and $wt(r_4)=1$.  

In Figure \ref{fig:ex:cfg:height}, we show two rule trees $d_1$ and $d_2$ of $\cG$, for which the following hold: $\pi_\cG(d_1)=ab$, $\pi_\cG(d_2)=abab$, hence $d_1\in \RT_\cG(ab)$ and $d_2\in \RT_\cG(abab)$.
Clearly, $\wt(d_1)=\wt(d_2)=2$.

Note that $\cG$ is not finite-derivational. In fact, for each $u\in L$, the set $\RT_\cG(u)$ is not finite.
However, the semiring $\Natminplus$ is $\sigma$-complete. Moreover, due to the fact that there exists only one nonterminal, we have $\RT_\cG=\T_R$ (recall that we view $R$ as a ranked alphabet).

It is clear that, for each $d \in \RT_\cG$, the weight $\wt(d)$  is the number of the occurrences of $r_4$ (i.e., the erasing rule) in $d$.
Let us denote this number by $\#_\mathrm{ers}(d)$. 
Hence, for each $d\in \T_R$, we have
\[
  \wrtsem{\cG}(d) =   \#_\mathrm{ers}(d) \enspace.
  \]
Moreover, for each $u\in \Gamma^*$ we have
\(\ssem{\cG}(u)  =  \infsum{\min}{d \in \RT_\cG(u)}{\#_\mathrm{ers}(d)}\).
\hfill $\Box$
\end{example}

\

\begin{figure}
\begin{center}
\begin{tikzpicture}[level distance=4.5em,
  every node/.style = {align=center}]]

  \pgfdeclarelayer{bg}    
  \pgfsetlayers{bg,main}  

\begin{scope}[level 1/.style={sibling distance=25mm},
level 2/.style={sibling distance=13mm}, level 3/.style={sibling distance=15mm}]
 \node {$S\to SS$}
 child {node {$S\to aSb$}  
       child { node[] {$S\to \varepsilon$} }}
 child {node {$S\to \varepsilon$}} ;
 \end{scope}

 \node at (-2.75,-.75) {$d_1$};
 \node at (3.5, -.75) {$d_2$};

\begin{scope}[xshift=55mm,level 1/.style={sibling distance=13mm},
level 2/.style={sibling distance=25mm}, level 3/.style={sibling distance=13mm}]
 \node {$S\to aSb$}
 child {node {$S\to SS$}  
    child { node[] {$S\to bSa$} 
         child { node[] {$S\to \varepsilon$}}}
    child { node {$S\to \varepsilon$} }}

;
 \end{scope}
\end{tikzpicture}
\end{center}
\caption{\label{fig:ex:cfg:height} Rule trees for $ab$ and  $abab$ of Example \ref{ex:wcf}.}
\end{figure}


\section{Normal forms of wcfg}
\label{sect:wcfg-normal-forms}

\index{weighted context-free grammar!nonterminal form}
\index{weighted context-free grammar!start-separated}
\index{weighted context-free grammar!chain-free}
Now we define wcfg which satisfy particular properties.
Let $\cG=(N,S,R,wt)$ be a $(\Gamma,\B)$-wcfg. We say that $\cG$ is
\begin{compactitem}
\item \emph{in nonterminal form} if, for each rule $A \to \alpha$ in $R$, either $\alpha \in \Gamma$ or $\alpha\in N^*$,
\item  \emph{start-separated} if it has exactly one initial nonterminal and this nonterminal does not occur in the right-hand side of a rule, 
\item \emph{chain-free} if $\cG$ does not have chain rules, and
  \item \emph{$\varepsilon$-free} if $\cG$ does not have $\varepsilon$-rules.
\end{compactitem}
Moreover,
\index{local-successful}
\index{local-useful}
\index{local-reduced}
\index{nullable}
\begin{compactitem}
\item a rule tree $d \in \RT_\cG$ is \emph{local-successful} if  $wt(d(w)) \ne \0$ for each $w \in \pos(d)$, 
\item a nonterminal $A \in N$ is \emph{local-useful} if there exists a local-successful $d \in  \RT_\cG$ such that $A$ occurs in $d$, and
\item $\cG$ is \emph{local-reduced} if $\supp(wt)=R$ and each nonterminal in $N$ is local-useful.
\end{compactitem}
Finally, we call a nonterminal $A \in N$ \emph{nullable} if $\RT_\cG(A,\varepsilon) \ne \emptyset$.

We note that $\cG$ has a local-useful nonterminal if and only if it has a local-successful rule tree.

\begin{observation} \rm \label{obs:fin-der+local-red->finite}
Let $\cG=(N,S,R,wt)$ be a finite-derivational and  local-reduced $(\Gamma,\B)$-wcfg. Then, for every $A \in N$ and $u \in \Gamma^*$, the set $\RT_\cG(A,u)$ is finite.
\end{observation}
\begin{proof} We prove by contradiction. We assume that there exist $A\in N$ and $u \in \Gamma^*$ such that $\RT_\cG(A,u)$ is not finite. Since $\cG$ is local-reduced, there exists an $x\in \Gamma^*$ and a  rule tree $d \in \RT_\cG(x)$ such that $A$ occurs in $d$, i.e.,  there exists a $w\in \pos(d)$ with $\lhs(d(w))=A$.
  Then there exists a $y\in \Gamma^*$ such that for each $d' \in \RT_\cG(A,u)$, we have $d[d']_w \in \RT_\cG(y)$. Hence $\RT_\cG(y)$ is not finite which contradicts that $\cG$ is finite-derivational.
 \end{proof}

\begin{observation}\rm \label{obs:bound-on-size-of-rule-trees} If the $(\Gamma,\B)$-wcfg
$\cG$ is chain-free and $\varepsilon$-free, then it is finite-derivational.
  \end{observation}
    \begin{proof} Let $\cG=(N,S,R,wt)$ be chain-free and $\varepsilon$-free.
 By induction on $\T_R$, we prove that the following statement holds:
  \begin{equation}
    \text{For each $d \in \T_R$, we have $\size(d) \le 2^{|\pi_\cG(d)|}$.} \label{eq:size-bound-TR}
    \end{equation}
  For this, let $d = r(d_1,\ldots,d_k)$ be in $\T_R$ with $r = (A \rightarrow u_0 A_1 u_1 \cdots A_k u_k)$. Since $\cG$ is chain-free, in case $k=1$ we have $|u_0u_1|>0$. Moreover, since $\cG$ is $\varepsilon$-free, we have that $|\pi_\cG(d_i)| > 0$ for each $i \in [k]$. Then we can calculate as follows:
    \begingroup
    \allowdisplaybreaks
    \begin{align*}
      \size(r(d_1,\ldots,d_k)) &= 1 + \bigplus_{i\in[k]} \size(d_i)\\
                              &\le 1 + \bigplus_{i\in[k]} 2^{|\pi_\cG(d_i)|} \tag{by I.H.}\\
                               &\le  2^{|u_0u_1 \cdots u_k|} \cdot 2^{|\pi_\cG(d_1)|} \cdot \ldots \cdot 2^{|\pi_\cG(d_k)|} \tag{since $\cG$ is chain-free and $\varepsilon$-free}\\
                              &= 2^{|u_0u_1 \cdots u_k| + {\bigplus\nolimits_{i\in[k]}} |\pi_\cG(d_i)|}
      = 2^{|\pi_\cG(r(d_1,\ldots,d_k))|}\enspace. 
    \end{align*}
    \endgroup
By \eqref{eq:size-bound-TR}, for every $u \in \Gamma^*$ and  $d\in\RT_\cG(u)$, we have that $\size(d) \le 2^{|u|}$, and hence $\cG$ is finite-derivational. \qedhere
    \end{proof}

    Next we consider the following restrictions of a wcfg:
      \begin{compactitem}
    \item nonterminal form,
    \item start-separated,
    \item local-reduced,
    \item chain-free, and
      \item $\varepsilon$-free.
      \end{compactitem}
      We prove that, under appropriate conditions, each wcfg can be transformed into an equivalent one which satisfies one of the mentioned restrictions. As it is common, we call each of these restrictions a normal form of wcfg.

    \begin{lemma}\rm \label{lm:wcfg-in-nonterminal-form}  Let $\cG$ be a $(\Gamma,\B)$-wcfg such that $\cG$ is finite-derivational or $\B$ is $\sigma$-complete. We can construct a $(\Gamma,\B)$-wcfg $\cG'$ such that $\cG'$ is  in nonterminal form and $\ssem{\cG'} = \ssem{\cG}$. Moreover, the construction preserves the properties finite-derivational, start-separated, chain-free, $\varepsilon$-free, and local-reduced. 
  \end{lemma}
  \begin{proof} Let $\cG=(N,S,R,wt)$. We construct the $(\Gamma,\B)$-wcfg $\cG'=(N',S,R',wt')$ such that $N' = N \cup \{A_a \mid a \in \Gamma\}$ and, for each rule $r= (A \to \alpha)$ in $R$, 
\begin{compactitem}
\item if $\alpha \in \Gamma$, then $r$ is in $R'$; we let $wt'(r) = wt(r)$,
\item otherwise the rule $r'= (A \to \alpha')$ is in $R'$ where $\alpha'$ is obtained from $\alpha$ by replacing each symbol $a \in \Gamma$ by $A_a$; we let $wt'(r') = wt(r)$. 
\end{compactitem}
Moreover, for each $a\in \Gamma$, the rule $A_a \to a$ is in $R'$ with $wt'(A_a \to a) = \1$.

Clearly, $\cG'$ is in nonterminal form and the construction preserves the mentioned properties. In particular, if $\cG$ is finite-derivational, then so is $\cG'$. Hence $\ssem{\cG'}$ is defined. It is clear that, for each $u \in \Gamma^*$, the sets $\RT_\cG(u)$ and $\RT_{\cG'}(u)$ are in a one-to-one correspondence, and $\wt_{\cG}(d) = \wt_{\cG'}(d')$ if $d$ and $d'$ correspond to each other. Thus $\ssem{\cG} = \ssem{\cG'}$.
\end{proof}

\begin{lemma}\rm \label{lm:wcfg-start-separated}  Let $\cG$ be a $(\Gamma,\B)$-wcfg such that $\cG$ is finite-derivational or $\B$ is $\sigma$-complete. We can construct a  $(\Gamma,\B)$-wcfg $\cG'$  such that $\cG'$ is start-separated  and $\ssem{\cG'} = \ssem{\cG}$. Moreover, the construction preserves the properties finite-derivational, nonterminal form, $\varepsilon$-free, and local-reduced.  The construction does not preserve the property chain-free. 
  \end{lemma}
\begin{proof} Let  $\cG=(N,S,R,wt)$. We construct the $(\Gamma,\B)$-wcfg $\cG' =(N',S_0,R',wt')$ as follows. We let $N' = N \cup \{S_0\}$ where $S_0 \not\in N \cup \Gamma$.  The set $R'$ contains the following rules.
  \begin{compactitem}
  \item For each $A \in S$, the rule $r = (S_0 \to A)$ is in $R'$ with $wt'(r)=\1$.
  \item Each rule $r \in R$ is in $R'$ with $wt'(r) = wt(r)$.
\end{compactitem}
  It is clear that $\cG'$ is start-separated and that the construction preserves the mentioned properties. In particular, if $\cG$ is finite-derivational, then so is $\cG'$. Hence $\ssem{\cG'}$ is defined. 
  Moreover, it is clear that,
  \begin{eqnarray}
  \begin{aligned}
   & \text{for every $A \in N$ and $u \in \Gamma^*$, we have  $\RT_\cG(A,u) = \RT_{\cG'}(A,u)$ } \\
   & \text{and $\wt_{\cG}(d) = \wt_{\cG'}(d)$ for each $d \in \RT_\cG(A,u)$.}
    \end{aligned} \label{eq:start-separated}
    \end{eqnarray}
  
  We can prove  that $\ssem{\cG} = \ssem{\cG'}$ as follows. Let $u \in \Gamma^*$. Then
  \begingroup
  \allowdisplaybreaks
  \begin{align*}
    \ssem{\cG}(u) &= \infsum{\oplus}{d \in \RT_\cG(S,u)} \wt_\cG(d)
                   \tag{by \eqref{eq:sem-wcfg-der-tree}}\\
                 &= \bigoplus_{A \in S} \ \ \infsum{\oplus}{d \in \RT_\cG(A,u)} \wt_\cG(d)
    \tag{because the family $(\RT_\cG(A,u) \mid A \in S)$ partitions $ \RT_\cG(S,u)$.}\\
                   &=  \bigoplus_{A \in S} \ \ \infsum{\oplus}{d' \in \RT_{\cG'}(A,u)} \wt_{\cG'}(d')  \tag{by \eqref{eq:start-separated}}\\
                   &=  \bigoplus_{A \in S} \ \ \infsum{\oplus}{d' \in \RT_{\cG'}(A,u)} \wt_{\cG'}(d') \otimes wt'(S_0 \to A) \hspace{70mm} \ \\
                 &= \infsum{\oplus}{d'' \in \RT_{\cG'}(S_0,u)} \wt_{\cG'}(d'')
                   \tag{because the family $(\{(S_0\to A)(d') \mid d' \in \RT_{\cG'}(A,u)\}\mid A\in S)$ partitions $\RT_{\cG'}(S_0,u)$}\\
                   &=  \ssem{\cG'}(u). \qedhere
  \end{align*}
  \endgroup
    \end{proof}


Next we will prove that for each wcfg we can construct an equivalent local-reduced wcfg. We follow the outline of the proof of constructing an equivalent local-trim wta from a given wta (cf. Theorem \ref{thm:t2}). 
As an auxiliary tool, we associate to each  $(\Gamma,\B)$-wcfg  $\cG=(N,S,R,wt)$,  the context-free grammar $\mathrm{G}(\cG) = (N\cup\{S_0\},S_0,P)$ over $\Gamma$, where $S_0$ is a new symbol and $P$ is the smallest set of rules such that 
\begin{compactitem}
\item for each $A \in S$, the rule $S_0 \rightarrow A$ is in $P$ and
\item $\supp(wt) \subseteq P$.
\end{compactitem}
Then it is obvious that a nonterminal $A\in N$ is local-useful in $\cG$ if and only if it is useful in 
$\mathrm{G}(\cG)$. This implies that $\cG$ is local-reduced if and only if $\mathrm{G}(\cG)$ is reduced.

\begin{lemma} \rm \label{lm:t2-wrtg}    Let $\cG$ be a $(\Gamma,\B)$-wcfg such that $\cG$ is finite-derivational or $\B$ is $\sigma$-complete. If $\cG$ has a local-successful rule tree,  then we can  construct a  $(\Gamma,\B)$-wcfg $\cG'$  such that $\cG'$ is local-reduced and  $\ssem{\cG'}=\ssem{\cG}$. Moreover, the construction preserves the properties finite-derivational, nonterminal form, start-separated, chain-free, and $\varepsilon$-free.
\end{lemma}
\begin{proof}
Let $\cG = (N,S,R,wt)$. We can construct the context-free grammar $\mathrm{G}(\cG)=(N \cup \{S_0\},S_0,P)$ associated to $\cG$. Due to our assumption on $\cG$ we have $\LL(\mathrm{G}(\cG))\ne \emptyset$.  Thus, by Theorem \ref{thm:reduced-cfg}, a context-free grammar $G'=(N',S_0,P')$ can be constructed such that ${G}'$ is reduced and $\LL(G')= \LL(\mathrm{G}(\cG))$. By the proof of that theorem, we know that $N' = N_u \cup \{S_0\}$, where $N_u$ is the set of all useful nonterminals in $N$.   Moreover, $N_u\ne \emptyset$ by our assumption on $\cG$.

Now let $\cG' = (N_u,S\cap N_u,R',wt')$ be the $(\Gamma,\B)$-wcfg such that
\begin{compactitem}
\item $R' = \{(A \to \alpha) \in R \mid  A \in N_u, \alpha \in (N_u \cup \Gamma)^*\}$ and
\item $wt'(r) = wt(r)$ for each $r \in R'$.
  \end{compactitem}
We note that, since $\cG$ has a local-successful rule tree, there exists a terminal rule $B \to u$ in $R$ and $B$ is useful in $\mathrm{G}(\cG)$. This rule is also in $R'$, hence $R'$ contains a terminal rule.
Moreover, it is obvious that $\cG'$ is local-reduced and the mentioned properties are preserved. Thus, in particular, $\sem{\cG'}$ is defined.
  
Lastly, we prove that $\ssem{\cG} = \ssem{\cG'}$. Let $u \in \Gamma^*$. Obviously, $\RT_{\cG'}(u) \subseteq \RT_\cG(u)$ and for each $d \in \RT_{\cG'}(u)$ we have $\wt_{\cG'}(d) = \wt_\cG(d)$. For each $d \in \RT_\cG(u)\setminus \RT_{\cG'}(u)$, there exists $A' \in N$ such that $A'$ occurs in $d$ and $A'$ is not local-useful. Hence $\wt_\cG(d) = \0$.  
  Thus we can compute
  \begin{align*}
    \hspace*{3cm} \ssem{\cG}(u) = \bigoplus_{d \in \RT_\cG(u)} \wt_\cG(d)
                  =  \bigoplus_{d \in \RT_{\cG'}(u)} \wt_{\cG'}(d)
    =\ssem{\cG'}(u). \hspace{3cm}  \qedhere
    \end{align*}
  \end{proof}


Next we show how a wcfg can be transformed into an equivalent chain-free wcfg.

\begin{theorem-rect} \label{lm:wrtg-chain-free} {\rm (cf. \cite[Thm.~14.2]{kuisal86})} Let $\Gamma$ be an alphabet, $\B$ be a semiring, and $\cG$ be a $(\Gamma,\B)$-wcfg such that (a)~$\cG$ is finite-derivational or (b) $\B$~is $\sigma$-complete. Then there exists a $(\Gamma,\B)$-wcfg $\cG'$ such that $\cG'$ is chain-free and $\ssem{\cG'}=\ssem{\cG}$. If $\cG$ has one of the following properties, then also $\cG'$ has it: finite-derivational, nonterminal form, start-separated, $\varepsilon$-free, and local-reduced. Moreover, if $\cG$ is finite-derivational, then we can even construct $\cG'$.
\end{theorem-rect}

\begin{proof}  Let $\cG=(N,S,R,wt)$. By standard pumping methods, we can decide whether $\cG$ has a local-successful rule tree.
  If the answer is no, then $\ssem{\cG}=\widetilde{\mathbb{0}}$ and thus the statement of the lemma holds obviously. Otherwise, by Lemma \ref{lm:t2-wrtg}  we 
  can construct a $(\Gamma,\B)$-wcfg which is local-reduced and equivalent to $\cG$. So we 
 may assume that $\cG$ is local-reduced. 

 We will prove the theorem by case analysis according to the Cases (a) and (b). Before doing so, we make some common preparations. Since  $\B$ is a semiring, we can consider the semiring $(B^{N \times N},\oplus,\cdot,\mathrm{M}_\0,\mathrm{M}_\1)$ of all $N$-square matrices over $B$ (where we assume an arbitrary but fixed linear order on $N$).\footnote{The multiplication $\cdot$ of matrices is associative only if the multiplication $\otimes$ of $B$ distributes over $\oplus$.}
  We define the $N$-square matrix $M$ over $B$ such that, for every $A,A' \in N$:
  \[
    M_{A',A} = \begin{cases}
      wt(r) & \text{ if } r = (A \rightarrow A') \text{ is in $R$}\\
      \mathbb{0} & \text{ otherwise.}
      \end{cases}
    \]

   We recall that, for each $n \in \mathbb{N}$, the $N$-square matrix $M^n$ over $B$ is defined such that such that
    $M^0 = \mathrm{M}_\1$  and $M^{k+1} = M^k \cdot M$ for each $k \in \mathbb{N}$.
    Since $\otimes$ is distributive with respect to $\oplus$, for each $n\in \mathbb{N}_+$ and $A,A'\in N$, we have
    \begin{equation}
      (M^n)_{A',A} = \bigoplus_{\substack{C_1,\ldots,C_{n+1} \in N: \\ C_1=A, C_{n+1}=A' }} \ \bigotimes_{j\in[0,n-1]} M_{C_{n+1-j},C_{n-j}}\enspace. \label{eq:nth-iteration-matrix}
      \end{equation}
      Hence, $(M^n)_{A',A}$ is the sum of all products 
 \begin{equation}\label{eq:product-n}
 \wt(r_n)\otimes \cdots \otimes \wt(r_1),
 \end{equation} 
 where, for each $i\in [n]$, $r_i$ is a chain-rule of the form $C_i \to C_{i+1}$ such that $C_1=A$ and $C_{n+1}=A'$.

      We start the proof with Case (b), i.e., we assume that $\B$ is $\sigma$-complete.  Then also the semiring  $(B^{N \times N},\oplus,\cdot,\mathrm{M}_\0,\mathrm{M}_\1)$ is $\sigma$-complete and hence the matrix $M^*$ is well defined where $M^* = \infsum{\oplus}{n \in \mathbb{N}}{M^n}$. Hence, $(M^*)_{A',A}$ is the sum of all products \eqref{eq:product-n} for some $n \in \mathbb{N}$.

       Then we eliminate the chain rules from $\cG$  (cf., e.g., \cite[Thm.~3.2]{esikui03}, \cite[Lm.~3.2]{fulmalvog11}, and \cite[Thm.~6.3]{fulhervog18}).
We define the $(\Gamma,\B)$-wcfg $\cG' = (N,S,R',wt')$ as follows. Let
\[R'=\{A\to \alpha \mid A\in N, \alpha \not\in N, (\exists A'\in N): A'\to \alpha \text{ is in $R$}\}.\]
Moreover, for each rule $A \rightarrow \alpha$ in $R'$, we define 
\[
wt'(A \rightarrow \alpha) = \bigoplus_{\substack{A'\in N: \\ (A'\to \alpha)\in R}}  wt(A' \rightarrow \alpha)  \otimes (M^*)_{A',A}\enspace.
\]
Obviously, $\cG'$ is chain-free and each mentioned property is preserved. In particular, $\ssem{\cG'}$ is defined.

In fact, the above definition of $\cG'$ is the same as the corresponding one in the proof of \cite[Thm.~6.3]{fulhervog18} for the case that (i) the storage type is the trivial one
 and (ii) the M-monoid $K$ is the M-monoid associated with the $\sigma$-complete semiring $\B$ (for the concept of ``M-monoid associated with a semiring'' cf. \cite[Def.~8.5]{fulmalvog09} or \cite[p.~261]{fulstuvog12}). By instantiating the correctness proof of \cite[Thm.~6.3]{fulhervog18} to the case specified by (i) and (ii), we obtain a proof of $\ssem{\cG'}=\ssem{\cG}$. For the sake of $\sigma$-completeness, we repeat this correctness proof here.

We note that, in virtue of the definition of $\cG'$, we cannot \underline{construct} $\cG'$. This is due to the fact that the definition of the weights of rules involves the matrix $M^*$, and, although $M^*$ is well defined, in general, we cannot compute it algorithmically.

\index{succ@$\succ$}
For the inductive definition of a mapping that relates rule trees of $\cG$ with rule trees of $\cG'$, we employ the  reduction system  $(\RT_{\cG}(N,\Gamma^*),\succ)$ where we let
   \[
     \succ \ = \ \succ_{R} \cap \ (\RT_{\cG}(N,\Gamma^*) \times \RT_{\cG}(N,\Gamma^*)) 
   \]
   (for the definition of $\succ_R$ we refer to page \pageref{page:prec-Sigma}).
   Since $\succ_{R} $ is terminating, by Lemma \ref{lm:termination-is-subset-closed} also $\succ$ is terminating. Moreover, we have that $\nf_\succ(\RT_{\cG}(N,\Gamma^*))$ is the set of terminal rules of $R$, which is not empty.  
We define the mapping
\(\textrm{eff}: \RT_{\cG}(N,\Gamma^*) \rightarrow \RT_{\cG'}(N,\Gamma^*)\)
 by induction on $(\RT_{\cG}(N,\Gamma^*),\succ)$ as follows. Let  $d\in \RT_{\cG}(N,\Gamma^*)$. Then 
\begin{compactitem}
\item there exist $n\in \mathbb{N}$ and rules $r_1 = (C_1\to C_2)$, \ldots, $r_n=(C_n\to C_{n+1})$, and $C_{n+1}\rightarrow \alpha$ with $\alpha = u_0A_1u_1 \ldots A_ku_k$
in $R$ such that $\alpha \not\in N$ and 
\item for each $i\in[k]$   there exists a rule tree $d_i\in \RT_{\cG}(A_i,\Gamma^*)$
\end{compactitem}
 such that
 \(d=r_1 \ldots r_n (C_{n+1} \rightarrow \alpha)\big(d_1,\ldots,d_k\big)\).
 We define
 \[\textrm{eff}(d)=(C_{1} \rightarrow \alpha)(\textrm{eff}(d_1),\ldots,\textrm{eff}(d_k)).\]
 Then $\lhs(\textrm{eff}(d_i)(\varepsilon))=A_i$ for each $i \in [k]$. Moreover,
 \begin{eqnarray}
 \begin{aligned}
  & \text{for every $A \in N$, $u \in \Gamma^*$, and $d \in \RT_\cG(N,\Gamma^*)$ we have:}  \\ 
  & \text{$d \in \RT_\cG(A,u)$ if and only if  $\mathrm{eff}(d) \in \RT_{\cG'}(A,u)$\enspace.} 
   \end{aligned} \label{eq:Au-equivalent}
   \end{eqnarray}

 Next we will prove a relationship between the weights of rule trees that are related by $\mathrm{eff}$. For this we use the  reduction system $(\RT_{\cG'}(N,\Gamma^*),\succ')$ where we let
 \index{succB@$\succ'$}
   \[
     \succ' \ = \ \succ_{R'} \cap \ (\RT_{\cG'}(N,\Gamma^*) \times \RT_{\cG'}(N,\Gamma^*)) \enspace.
   \]
  Since $\succ_{R'} $ is terminating, by Lemma \ref{lm:termination-is-subset-closed} also $\succ'$ is terminating.
   Moreover, we have that $\nf_{\succ'}(\RT_{\cG'}(N,\Gamma^*))$ is the set of terminal rules of $R'$, which is not empty. Then, by induction on $(\RT_{\cG'}(N,\Gamma^*),\succ')$, we prove that the following statement holds.
\begin{eqnarray}\label{eq:compress}
  \text{For every $d' \in \RT_{\cG'}(N,\Gamma^*)$ we have $\infsum{\oplus}{{\substack{d \in \RT_{\cG}(N,\Gamma^*):\\\textrm{eff}(d)=d'}}}{\wt(d)} = \wt'(d')$}\enspace. \label{eq:sum-weight-chain}
\end{eqnarray}

     Let $d' \in \RT_{\cG'}(N,\Gamma^*)$. Hence there exist $A \in N$ and $u\in \Gamma^*$ such that $d' \in \RT_{\cG'}(A,u)$.  Then
\begin{compactitem}
\item there exists a rule $r' = (A \rightarrow \alpha)$ in $R'$ with $\alpha = u_0A_1u_1 \ldots A_ku_k$, $k\in \mathbb{N}_+$, and $\alpha \not\in N$ and
 \item  for each $i \in [k]$ there exist $v_i \in \Gamma^*$ and  $d_i' \in \RT_{\cG'}(A_i,v_i)$ 
 \end{compactitem}
 such that $u= u_0v_1u_1 \cdots v_ku_k$ and  $d' = r'(d_1',\ldots,d_k')$. Then we can calculate as follows:
 \begingroup
\allowdisplaybreaks
\begin{align*}
  & \infsum{\oplus}{{\substack{d \in \RT_{\cG}(N,\Gamma^*):\\\textrm{eff}(d)=d'}}}{\wt(d)} 
  = \infsum{\oplus}{\substack{d \in \RT_{\cG}(A,u):\\\textrm{eff}(d)=r'(d_1',\ldots,d_k')}}{\wt(d)}
  \tag{by \eqref{eq:Au-equivalent}}\\
&= \infsum{\oplus}{n \in \mathbb{N}}  \bigoplus_{\substack{C_1,\ldots,C_{n+1} \in N: \\ C_1=A }}  \ \
\infsum{\oplus}{\substack{d_1 \in \RT_{\cG}(A_1,v_1),\ldots,d_k \in \RT_{\cG}(A_k,v_k):\\ (\forall i \in [k]): \textrm{eff}(d_i)=d_i' }} \\
  &\hspace*{10mm} \Big(\bigotimes_{i\in[k]} \wt(d_i)\Big)  \otimes  wt(C_{n+1}\rightarrow  u_0A_1u_1 \ldots A_ku_k) \otimes  \bigotimes_{j\in[0,n-1]} M_{C_{n+1-j},C_{n-j}}\\
  %
  &= \infsum{\oplus}{n \in \mathbb{N}}  \bigoplus_{\substack{C_1,\ldots,C_{n+1} \in N: \\ C_1=A }}  \Big(\bigotimes_{i\in[k]} \ \  \infsum{\oplus}{\substack{d_i \in \RT_{\cG}(A_i,v_i):\\ \textrm{eff}(d_i)=d_i' }} \wt(d_i)\Big)
 \\
      &\hspace*{10mm}   \otimes  wt(C_{n+1}\rightarrow  u_0A_1u_1 \ldots A_ku_k) \otimes  \bigotimes_{j\in[0,n-1]} M_{C_{n+1-j},C_{n-j}}
        \tag{\text{because $\B$ is distributive}}\\
   &= \infsum{\oplus}{n \in \mathbb{N}}  \bigoplus_{\substack{C_1,\ldots,C_{n+1} \in N: \\ C_1=A }}   
  \Big(\bigotimes_{i\in[k]} \wt'(d_i')\Big)  \otimes  wt(C_{n+1}\rightarrow   u_0A_1u_1 \ldots A_ku_k) \otimes  \bigotimes_{j\in[0,n-1]} M_{C_{n+1-j},C_{n-j}}
\tag{\text{by I.H.}}\\
    &= \bigoplus_{A'\in N}  \ \infsum{\oplus}{n \in \mathbb{N}} \bigoplus_{\substack{C_1,\ldots,C_{n+1} \in N: \\ C_1=A, C_{n+1}=A' }}   
  \Big(\bigotimes_{i\in[k]} \wt'(d_i')\Big)  \otimes  wt(A'\rightarrow  u_0A_1u_1 \ldots A_ku_k) \otimes  \bigotimes_{j\in[0,n-1]} M_{C_{n+1-j},C_{n-j}}
\tag{\text{by renaming of $C_{n+1}$ by $A'$}}\\
      &=  \Big(\bigotimes_{i\in[k]} \wt'(d_i')\Big)  \otimes \Big(\bigoplus_{A'\in N}  wt(A'\rightarrow  u_0A_1u_1 \ldots A_ku_k)\Big) \otimes  
  \infsum{\oplus}{n \in \mathbb{N}}  \bigoplus_{\substack{C_1,\ldots,C_{n+1} \in N: \\ C_1=A, C_{n+1}=A' }} \bigotimes_{j\in[0,n-1]} M_{C_{n+1-j},C_{n-j}}
\tag{\text{by distributivity}}\\
        &= \Big(\bigotimes_{i\in[k]} \wt'(d_i')\Big)  \otimes  \Big(\bigoplus_{A'\in N}  wt(A'\rightarrow  u_0A_1u_1 \ldots A_ku_k)\Big) \otimes  
  \infsum{\oplus}{n \in \mathbb{N}}  (M^n)_{A',A}
          \tag{\text{by \eqref{eq:nth-iteration-matrix}}}\\
             &= \Big(\bigotimes_{i\in[k]} \wt'(d_i')\Big)  \otimes  \Big(\bigoplus_{A'\in N}  wt(A'\rightarrow  u_0A_1u_1 \ldots A_ku_k)\Big) \otimes  
  \Big(\infsum{\oplus}{n \in \mathbb{N}}  (M^n)\Big)_{A',A}\\
   &= \Big(\bigotimes_{i\in[k]} \wt'(d_i')\Big)  \otimes  \Big(\bigoplus_{A'\in N} wt(A'\rightarrow  u_0A_1u_1 \ldots A_ku_k)\ \otimes  (M^*)_{A',A}\Big)\\
&=  \big(\bigotimes_{i\in[k]} \wt'(d_i')\Big)  \otimes wt'(r') \tag{\text{by the definition of $\cG'$}}\\
&= \wt'(r'(d_1',\ldots,d_k')).
\end{align*}
\endgroup
This proves \eqref{eq:sum-weight-chain}.

Then we can prove for each $u \in \Gamma^*$:
\begingroup
\allowdisplaybreaks
\begin{align*}
\ssem{\cG}(u) 
&= \infsum{\oplus}{d \in \RT_{\cG}(\xi)}{\wt(d)} 
= \bigoplus_{A\in S} \ \ \infsum{\oplus}{d \in \RT_{\cG}(A,u)}{\wt(d)} 
= \bigoplus_{A \in S} \ \ \infsum{\oplus}{d' \in \RT_{\cG'}(A,u)} \ \ \infsum{\oplus}{\substack{d \in \RT_{\cG}(A,u):\\\textrm{eff}(d)=d'}}{\wt(d)}
\\
&= \bigoplus_{A \in S} \ \ \infsum{\oplus}{d' \in \RT_{\cG'}(A,u)} \wt'(d')  \tag{\text{by \eqref{eq:compress}}}\\
&= \infsum{\oplus}{d' \in \RT_{\cG'}(u)} \wt'(d') = \ssem{\cG'}(u)\enspace. 
\end{align*}
\endgroup

\


Now we continue the proof with Case (a), i.e., we assume that $\cG$ is finite-derivational. Since $\cG$ is local-reduced  and finite-derivational, the length of sequences of chain rules in $A$-rule trees of $\cG$ is bounded by $|N|-1$ for each $A\in N$. We show this statement by contradiction. 
For this, let us assume that there exist a string $u  \in \Gamma^*$ and a $d \in \RT_\cG(A,u)$ such that $d$ contains a sequence of chain rules of length at least $|N|$.
Then a nonterminal is repeated in that sequence and, by applying the standard pumping argument, we obtain that $\RT_\cG(A,u)$ is not finite. However, this contradicts Observation 
\ref{obs:fin-der+local-red->finite}.

This boundedness property gives us the possibility to \underline{construct} $\cG'$, because, in order to eliminate the chain rules from $\cG$, now we can use the matrix  $\bigoplus\limits_{n\in[0,|N|-1]}M^{n}$.

 Thus, we construct the $(\Gamma,\B)$-wcfg $\cG' = (N,S,R',wt')$ as follows. As in Case (b), we let
\[R'=\{A\to \alpha\mid A\in N, \alpha\not\in N, (\exists A'\in N): A'\to \alpha \text{ is in $R$}\}.\]
Moreover, for each rule $A \rightarrow \alpha$ in $R'$, we define 
\[
wt'(A \rightarrow \alpha) = \bigoplus_{\substack{A'\in N: \\ (A'\to \alpha)\in R}}  wt(A' \rightarrow \alpha)  \otimes \Big(\bigoplus_{n\in[0,|N|-1]}M^{n}\Big)_{A',A}\enspace.
\]

Obviously,  $\cG'$ is chain-free and, again, all the mentioned properties are preserved. Thus, since $\cG$ is finite-derivational, also $\cG'$ is so, hence $\ssem{\cG'}$ is defined.
\index{succB@$\succ'$}
By induction on $(\RT_{\cG'}(N,\Gamma^*),\succ')$ (where $\succ'$ is the terminating relation defined in the proof of Case (b) above), we prove that the following statement holds:
\begin{eqnarray}\label{eq:compress-finite}
  \text{For every $d' \in \RT_{\cG'}(N,\Gamma^*)$, we have
  $\bigoplus_{\substack{d \in \RT_{\cG}(N,\Gamma^*):\\\textrm{eff}(d)=d'}} \wt(d) = \wt'(d')$.}
 \end{eqnarray}
The proof of \eqref{eq:compress-finite} is the same as the proof of \eqref{eq:compress} except that the four infinite summations
\[
  \infsum{\oplus}{\substack{d \in \RT_{\cG}(A,u):\\\textrm{eff}(d)=r'(d_1',\ldots,d_k')}} \, \text{ and } \
  \infsum{\oplus}{\substack{d_i \in \RT_{\cG}(A_i,v_i):\\ \textrm{eff}(d_i)=d_i' }} \, \text{ and } \
  \infsum{\oplus}{n \in \mathbb{N}} \, \text{ and } \ \infsum{\oplus}{n \in \mathbb{N}} 
\]
are replaced by the finite summations
\[
  \bigoplus_{\substack{d \in \RT_{\cG}(A,u):\\\textrm{eff}(d)=r'(d_1',\ldots,d_k')}} \, \text{ and } \
  \bigoplus_{\substack{d_i \in \RT_{\cG}(A_i,v_i):\\ \textrm{eff}(d_i)=d_i' }} \, \text{ and } \
  \bigoplus_{n \in [0,|N|-1]} \, \text{ and } \ \bigoplus_{n\in [0,|N|-1]}\ \text{, respectively,}
\]
as well as, 
\[(M^*)_{A',A} \text{ is replaced by } \Big(\bigoplus_{n\in[0,|N|-1]}M^{n}\Big)_{A',A}\enspace. \]

By a similar modification of the final calculation of (b) we obtain the proof for $\ssem{\cG}=\ssem{\cG'}$.
\end{proof}


In the next theorem we deal with the transformation of a $(\Gamma,\B)$-wcfg into an equivalent $\varepsilon$-free one. More precisely, for a given   $(\Gamma,\B)$-wcfg $\cG=(N,S,R,wt)$ such that (a) $\B$ is a commutative semiring and (b) $\cG$ is finite-derivational or $\B$~is $\sigma$-complete, we define an $\varepsilon$-free  $(\Gamma,\B)$-wcfg $\cG'$  such that $\ssem{\cG'} = \ssem{\cG} \otimes \chi_\B(\Gamma^+)$. We follow the idea of \cite[Lm.~4.1]{barpersha61} (also cf. \cite[Subsec.~7.1.3]{hopmotull07}) where context-free grammars are transformed into (almost) equivalent $\varepsilon$-free context-free grammars. However, here we have to be careful about the weights and have to cope with two phenomena. 

We explain the first phenomenon by means of an example. Let $\cG$ contain the  initial nonterminal $A$ and the three rules
\[
  \text{$r= (A \to aCC)$, $r' = (C \to b)$, and $r'' = (C \to \varepsilon)$.}
\]
Then $\RT_\cG(A,ab) = \{d_1,d_2\}$ with $d_1 = r(r',r'')$ and  $d_2= r(r'',r')$. Hence
\[\ssem{\cG}(ab) = \wt_\cG(d_1) \oplus \wt_\cG(d_2) \enspace.\]
According to the construction of \cite[Lm.~4.1]{barpersha61}, the wcfg $\cG'$ contains the rules $r$, $r'$, and $\bar{r}=(A \to aC)$, where $\bar{r}$ results from $r$ by erasing either  the first occurrence of $C$ or the second one. Thus $\RT_{\cG'}(A,ab) = \{\bar{r}(r')\}$ and
\[\ssem{\cG'}(ab) = \wt_{\cG'}(\bar{r}(r')) \enspace.\]
We observe that this construction does not take care of the different ways in which a right-hand side 
(e.g. $aC$) results from erasing nonterminals from the right-hand side of the $\cG$-rule (e.g. $aCC$).
In this particular example, we could define the weights of the rules of $\cG'$ such that the weight balance  $\ssem{\cG}(ab) = \ssem{\cG'}(ab)$ is satisfied. But in general, rules of $\cG$ call each other recursively, and then this method of defining weights is not successful. Instead, we will code the occurrences of nonterminals which are selected for erasing into the rules of $\cG'$, and thereby we keep the different ways in which a rule for $\cG'$ was obtained separately.

The second phenomenon is the fact that the empty string $\varepsilon$ can be derived in many different ways, and $\cG'$ has to sum up the weights of all the corresponding rule trees. For instance, if there exists a rule $r= (C \to a A b)$ in $R$ and we have detected that $A$ is nullable,  i.e., $\RT_\cG(A,\varepsilon)\ne\emptyset$, then according to the construction in \cite[Lm.~4.1]{barpersha61} the rule $r' = (C \to ab)$ will be in $\cG'$ and the weight of $r'$ in $\cG'$ will be $wt(r) \otimes W(A)$, where we define
\[
  W(A) = \infsum{\oplus}{d \in \RT_\cG(A,\varepsilon)}{\wt_\cG(d)}\enspace.
\]
By our assumptions on $\cG$ and $\B$, the value $W(A)$ is well defined. We note that, if $\RT_\cG(A,\varepsilon) = \emptyset$, then $W(A) = \0$. If $\cG$ is not finite-derivational (and hence $\B$ is $\sigma$-complete), it is in general not clear how to construct $W(A)$. However, if $\cG$ is finite-derivational, then $W(A)$ can be computed.

  \begin{observation}\label{obs:e-finite} \rm Let  $\cG=(N,S,R,wt)$ be a finite-derivational and local-reduced $(\Gamma,\B)$-wcfg. Then, for each $A \in N$, we can compute $W(A)$.
  \end{observation}
  
  \begin{proof} First we show that 
   \begin{equation}\label{eq:RT(A,e)-finite}
      \RT_\cG(A,\varepsilon) = \{d \in \RT_\cG(A,\varepsilon) \mid \height(d) \le |N|-1\}\enspace. 
      \end{equation}
   We prove \eqref{eq:RT(A,e)-finite} by contradiction and assume that there exists a $d \in \RT_\cG(A,\varepsilon)$ such that $\height(d) \ge |N|$. Then there exist $w \in \pos(d)$ and $v \in \mathbb{N}^+$ such that $\lhs(d(w)) = \lhs(d(wv))$. Then, using $c$ as abbreviation for the context $(d|_w)[z]_v$, we obtain that $d[c^n[d|_{wv}]]_w \in \RT_\cG(A,\varepsilon)$ for each $n \in \mathbb{N}$. Hence $\RT_\cG(A,\varepsilon)$ is not finite, which  contradicts Observation  \ref{obs:fin-der+local-red->finite}. This proves \eqref{eq:RT(A,e)-finite}.

   Then we have
   \begin{align*}
     W(A) &= \infsum{\oplus}{d \in \RT_\cG(A,\varepsilon)}{\wt_\cG(d)}\\
     &= \infsum{\oplus}{\substack{d \in \RT_\cG(A,\varepsilon):\\\height(d) \le |N|-1}}{\wt_\cG(d)}
     \tag{by \eqref{eq:RT(A,e)-finite}} \\
     &=  \bigoplus_{\substack{d \in \RT_\cG(A,\varepsilon):\\\height(d) \le |N|-1}} \wt_\cG(d)
     \tag{by Observation \ref{obs:sum-emptyset}(2)}
     \end{align*}

Obviously, $\bigoplus_{\substack{d \in \RT_\cG(A,\varepsilon):\\\height(d) \le |N|-1}} \wt_\cG(d)$ can be computed. 
  \end{proof}

The next theorem has been achieved in \cite[Thm.~14.6]{kuisal86} for wcfg over a partially ordered, continuous, and commutative semiring $\B$. We follow the construction given in \cite[Thm.~14.6]{kuisal86}.

\begin{theorem-rect} \label{lm:wrtg-e-free} {\rm (cf. \cite[Thm.~14.6]{kuisal86})} Let $\Gamma$ be an alphabet, $\B=(B,\oplus,\otimes,\0,\1)$ be a commutative semiring, and $\cG$ be a $(\Gamma,\B)$-wcfg such that $\cG$ is finite-derivational or $\B$~is $\sigma$-complete. Then there exists an $\varepsilon$-free $(\Gamma,\B)$-wcfg $\cG'$ such that $\ssem{\cG'} = \ssem{\cG} \otimes \chi_\B(\Gamma^+)$. If $\cG$ is finite-derivational, then we can construct $\cG'$.
\end{theorem-rect}
\begin{proof} 
  Let $\cG=(N,S,R,wt)$. By standard pumping methods, we can decide whether $\cG$ has a local-successful rule tree. If the answer is no, then $\ssem{\cG}=\widetilde{\mathbb{0}}$. Hence, the statement of the theorem holds obviously because we can construct an $\varepsilon$-free $(\Gamma,\B)$-wcfg $\cG'$ such that $\ssem{\cG'} = \widetilde{\0}$.

Otherwise, by  Lemma \ref{lm:wcfg-in-nonterminal-form} and Lemma \ref{lm:t2-wrtg}, we can construct a $(\Gamma,\B)$-wcfg which is in nonterminal form, local-reduced, and equivalent to $\cG$. Hence we can assume  
that $\cG$ is in nonterminal form and local-reduced. Then we distinguish two cases.

\underline{Case (a):} Each terminal rule of $\cG$ is an $\varepsilon$-rule. Then $\ssem{\cG} \otimes \chi_\B(\Gamma^+) = \widetilde{\0}$. Again, the statement of the theorem holds obviously.

\underline{Case (b):} There exists a terminal rule of $\cG$, say $A \to u$, such that $u \ne \varepsilon$.
Since $\cG$ is in nonterminal form, we have  $u \in \Gamma$.  We will use this property when we construct $\cG'$. But before this we need some preparations.
  
As in the case of context-free grammars, we construct the set of nullable nonterminal symbols  (cf. \cite[Lm.~4.1]{barpersha61}), i.e., the set
\[E=\{A\in N \mid \RT_\cG(A,\varepsilon)\ne\emptyset\} \enspace.\]
 
In order to prepare for the above mentioned first phenomenon, we introduce the concept of $E$-selection. Intuitively, an $E$-selection is a string over $\{0,1\}$ and it shows, for a given right-hand side $\alpha \in (N\cup \Gamma)^*$ of a rule and each of its nonterminal occurrences, whether that occurrence is selected for erasing (represented by 1) or not (represented by 0). Formally, let $\alpha = u_0A_1 u_1 \ldots A_k u_k$ be in $(N\cup \Gamma)^*$. An \emph{$E$-selection for $\alpha$} is a string $f = f_1\cdots f_k$ with $f_i \in \{0,1\}$ such that, for every $i \in [k]$, if $f_i =1$, then $A_i \in E$, i.e., only occurrences of nullable nonterminals can be selected for erasing. In particular, for each $u\in \Gamma^*$, the only $E$-selection is $f = \varepsilon$.

For instance, if $\alpha = aAbDCC$, $E=\{C,D\}$,  then $f =0110$ is an $E$-selection. It indicates that the occurrence of $D$ and the leftmost occurrence of $C$ are selected for erasing, and the other two occurrences of nonterminals are not. 

      Let $\alpha = u_0A_1u_1 \ldots A_ku_k$ in $(N\cup \Gamma)^*$ and $f$ be an $E$-selection for $\alpha$. Then we define the \emph{application of $f$ to $\alpha$}, denoted by $f(\alpha)$,  to be the string $u_0X_1u_1 \ldots X_ku_k$, where for each $i \in [k]$:
\[
X_i=\begin{cases}
\varepsilon & \text{if } f_i=1\\
A_i & \text{otherwise.}
\end{cases}
\] 
In particular, for each $u \in \Gamma^*$, we have $\varepsilon(u) = u$. For $\alpha$ and $f$ in the above example we have $f(\alpha) =aAbC$.

For each nonterminal which is not selected by $f$ for erasing, we will have to know its target position in $f(\alpha)$.  Formally, let $f = f_1 \cdots f_k$ be an $E$-selection for $\alpha$. We let $\pos_0(f) =\{i \in [k] \mid f_i=0\}$. Then we define the mapping $g_f: \pos_0(f) \to [|\pos_0(f)|]$ for each $i \in \pos_0(f)$ by $g_f(i) =j$ if the $j$th occurrence of $0$ in $f$ (counted from left to right) has position $i$. 
We note that $g_f$ is bijective. For $f$ in the above example, we have $\pos_0(f) = \{1,4\}$ and  $g_f(1)=1$ and $g_f(4)=2$.

      Now we define the $(\Gamma,\B)$-wcfg $\cG' = (N',S,R',wt')$ as follows.
      \begin{compactitem}
      \item $N' = N \cup \{[\alpha,f] \mid \big(\exists (A \to \alpha) \text{ in $R$}\big) : \text{$f$ is an $E$-selection for $\alpha$}\}$ and
      \item $R'$ and $wt'$ are defined as follows.
        For every $r = (A \to \alpha)$ with $\alpha=u_0A_1 u_1 \ldots A_ku_k$ in $R$ and $\alpha\not= \varepsilon$, and $E$-selection  $f$ for $\alpha$ such that $f(\alpha) \ne \varepsilon$,
        \begin{compactitem}
        \item the rule $r' = (A \to [\alpha,f])$ is in $R'$ with $wt'(r')=wt(r)$ and
        \item the rule $r'' = ([\alpha,f] \to f(\alpha))$ is in $R'$ with
          \[
            wt'(r'') = \bigotimes_{\substack{i \in [k]:\\ f_i=1}} W(A_i) \enspace.
          \]
          In particular, if $\{i \in [k] \mid f_i=1\} = \emptyset$, then $wt'(r'') = \1$.
        \end{compactitem}
      \end{compactitem}
      Since $R$ contains a terminal rule $A \to u$ with $u\in \Gamma$, the set $R'$ contains the rule $[u,\varepsilon] \to u$, and hence the $\Gamma$-cfg $(N',S,R')$ contains a terminal rule. Thus, $\cG'$ is a $(\Gamma,\B)$-wcfg. Moreover, if $\cG$ is finite-derivational, then $wt'(r'')$ can be constructed
           by Observation \ref{obs:e-finite}, and hence $\cG'$ can be constructed.

      We note that the construction of $\cG'$ is essentially the same as the construction of the algebraic system $(y_i= q_i \mid 1 \le i \le n)$ in the proof of Theorem \cite[Thm.~14.6]{kuisal86}. For an explanation, we assume that  $N=\{A_1,\ldots,A_n\}$ and $A_j\to \alpha_1$, \ldots, $A_j\to \alpha_\ell$ are all the rules of $\cG$ with left-hand side $A_j$. Then the corresponding algebraic system has the set $Y=N$ as set of variables and it contains the equation
      \[
A_j= p_j  \ \ \text{ where } \ \ p_j = \alpha_1 + \ldots + \alpha_\ell \enspace.
      \]
  The set of all right-hand sides $f(\alpha_i)$ where $1 \le i \le  \ell$ and $f$ is an $E$-selection for $\alpha_i$, is in one-to-one correspondence to the set of all monomials in the polynomial $q_i$ (in the proof of Theorem \cite[Thm.~14.6]{kuisal86}); the latter is obtained by substituting $(\sigma,\varepsilon)\varepsilon + Y$ into $p_j$. The coefficient $(\sigma,\varepsilon)$ is the $N$-vector $(W(A_1),\ldots,W(A_n))$, which is relevant for all those nonterminals which are selected for erasing, and the part $Y$ reflects the case that a nonterminal is not erased.
        
      \

      Next we prove $\ssem{\cG'} = \ssem{\cG} \otimes \chi_\B(\Gamma^+)$.
      As preparation,  we define the binary relation $\succ$ on $\RT_\cG(N,\Gamma^+)$ by
 \index{succ@$\succ$}
      \[\succ = \succ_R \cap  \RT_\cG(N,\Gamma^+) \times \RT_\cG(N,\Gamma^+)\enspace.\]  
 Since $\succ_R$ is terminating, also  $\succ$ has this property (using Lemma \ref{lm:termination-is-subset-closed}).
Moreover, we have $\nf_\succ(\RT_\cG(N,\Gamma^+)) = \{(A\to u) \in R \mid u \in \Gamma^+\}$, which is not empty.
      We define the mapping
\[\varphi: \RT_\cG(N,\Gamma^+) \rightarrow \RT_{\cG'}(N,\Gamma^+)\]
by induction on $(\RT_\cG(N,\Gamma^+),\succ)$ as follows (cf. Figure \ref{fig:e-free}).

Let $d \in \RT_\cG(N,\Gamma^+)$.
Then there exists a rule $(A \to \alpha)$ in $R$ with  $\alpha=u_0A_1 u_1 \ldots A_ku_k$, and for every $i \in [k]$ there exists $d_i \in \RT_\cG(A_i,\Gamma^*)$ such that 
\[
d = (A \to \alpha) \big( d_1,\ldots,d_k\big) \enspace.
\]
We can assume that $\varphi(d_i)$ is defined for each $d_i \in \RT_\cG(N,\Gamma^+)$.

Since $\pi_\cG(d) \not= \varepsilon$, we have $u_0u_1\cdots u_k \not=\varepsilon$ or there  exists $i \in [k]$ such that $d_i \in \RT_\cG(A_i,\Gamma^+)$.
Let $f = f_1 \cdots f_k$ be the $E$-selection for $\alpha$ defined, for each $i \in [k]$, by
\[f_i =
  \begin{cases}
    1 & \text{ if $d_i \in \RT_\cG(A_i,\varepsilon)$}\\
    0 & \text{ otherwise} \enspace.
  \end{cases}
\]
Then the rule $A \to [\alpha,f]$ is in $R'$ and we define 
\[
  \varphi(d) = (A \to [\alpha,f])\Big( ([\alpha,f] \to f(\alpha)) \bar{\zeta} \Big),
\]
where the sequence $\bar{\zeta}$ of rule trees of $\cG'$ is obtained from the sequence
$(d_1,\ldots,d_k)$ by
\begin{compactenum}
\item[(a)] dropping each rule tree $d_i$ with $d_i \in \RT_\cG(A_i,\varepsilon)$ and
\item[(b)] replacing each remaining $d_i$ by $\varphi(d_i)$ (which is defined because $d_i \in \RT_\cG(N,\Gamma^+)$).
\end{compactenum}
For instance, if $r = (S \to aAbCDD)$, $d_1 \in \RT_\cG(A,\Gamma^+)$, $d_2 \in \RT_\cG(C,\varepsilon)$, $d_3 \in \RT_\cG(D,\varepsilon)$,
 and $d_4 \in \RT_\cG(D,\Gamma^+)$, then $f=0110$ and $\bar{\zeta} = (\varphi(d_1),\varphi(d_4))$, cf. Figure \ref{fig:e-free}.

In the particular case that $d_i \in \RT_\cG(A_i,\Gamma^+)$ for each $i\in[k]$, we have $f=0^k$ and
\[
  \varphi(d) = (A \to [\alpha,f])\Big( ([\alpha,f] \to \alpha) (\varphi(d_1),\ldots,\varphi(d_k)) \Big)\enspace,
\]
and in the particular case that $d_i \in \RT_\cG(A_i,\varepsilon)$ for each 
$i\in[k]$, we have $f=1^k$ and
\[
  \varphi(d) = (A \to [\alpha,f])\Big( [\alpha,f] \to u_0u_1 \ldots u_k \Big)\enspace.
\]

\begin{figure}
  \begin{centering}
    \begin{tikzpicture}[every node/.style = {align=center}, scale=0.60, transform shape,line width=0.2pt]

\newcommand{\scalefactor}{0.60}

\begin{scope}[level 1/.style={sibling distance=22mm,level distance=2 cm},
level 2/.style={sibling distance=22mm,level distance=1.5cm}]

 \node (b1) {$S \overset {b_1} {\longrightarrow} aAbCDD$}
  child {node (b2) {$A \overset {b_2} {\longrightarrow} a$}}
  child {node[yshift=-0.6cm,xshift=-0.9cm] (Ce) {}}
  child {node[yshift=-0.6cm, xshift=-0.8cm] (De) {}}
  child {node (b3) {$D \overset {b_3} {\longrightarrow} EF$}
	child {node[xshift=0.3cm] (b4) {$E \overset {b_4} {\longrightarrow} b$}}
	child {node (Fe) {}} };	

 \node[isosceles triangle,draw,rotate=90,anchor=apex,minimum size =1.5cm, isosceles triangle apex angle=55] (TCe) at (Ce.north){};
 \node[isosceles triangle,draw,rotate=90,anchor=apex,minimum size =1.5cm, isosceles triangle apex angle=55] (TDe) at (De.north){};
 \node[isosceles triangle,draw,rotate=90,anchor=apex,minimum size =1.5cm, isosceles triangle apex angle=55] (TFe) at (Fe.north){};
 \node[yshift=-0.25cm] at (TCe.lower side) {$\in \RT_\cG (C, \varepsilon)$};
 \node[yshift=-0.25cm] at (TDe.lower side) {$\in \RT_\cG (D, \varepsilon)$};
 \node[yshift=-0.25cm] at (TFe.lower side) {$\in \RT_\cG (F, \varepsilon)$};

 \draw ($ (b2.north) !0.55! (b2.south) $) ellipse (0.85cm and 0.6cm);
 \draw ($ (b4.north) !0.55! (b4.south) $) ellipse (0.85cm and 0.6cm);

 \node (1) at (-1.55, 0.35) {};
 \node (2) at (1.75, 0.35) {};
 \node (3) at (1.3, -1) {};
 \node (4) at (0.95, -3) {};
 \node (5) at (1.45, -4.75) {};
 \node (6) at (-3.05, -4.75) {};
 \node (7) at (-2, -1.25) {};

 \draw [bend left, looseness=0.75] (1.center) to (2.center);
 \draw [in=45, out=-35] (2.center) to (3.center);
 \draw [in=115, out=-135, looseness=1] (3.center) to (4.center);
 \draw [in=45, out=-75, looseness=0.75] (4.center) to (5.center);
 \draw [bend left=45, looseness=0.50] (5.center) to (6.center);
 \draw [in=-90, out=135, looseness=1] (6.center) to (7.center);
 \draw [in=-145, out=90] (7.center) to (1.center);

 \node (8) at (2.25, -1.65) {};
 \node (9) at (4.1, -1.65) {};
 \node (10) at (5.3, -3.15) {};
 \node (11) at (5.5, -5.5) {};
 \node (12) at (3, -5.5) {};
 \node (13) at (3.5, -3.25) {};
 \node (14) at (2.1, -2.5) {};
				
 \draw [in=165, out=15,looseness=0.8] (8.center) to (9.center);
 \draw [in=115, out=-15] (9.center) to (10.center);
 \draw [in=30, out=-65] (10.center) to (11.center);
 \draw [bend left=15, looseness=0.8] (11.center) to (12.center);
 \draw [in=-60, out=150] (12.center) to (13.center);
 \draw [in=-30, out=120] (13.center) to (14.center);
 \draw [in=195, out=150,looseness=1.5] (14.center) to (8.center);
\end{scope}	

\draw[->] (5.8,-2.2) -- (7,-2.2) node[midway,above] {\large $\varphi$};

\begin{scope}[xshift=12cm,
	level 1/.style={sibling distance=20mm,level distance=1.05cm},
	level 2/.style={sibling distance=55mm,level distance=2cm},
	level 3/.style={sibling distance=20mm,level distance=1.05cm},
	level 4/.style={sibling distance=20mm,level distance=1.8cm},
	level 5/.style={sibling distance=20mm,level distance=1.05cm},
	level 6/.style={sibling distance=20mm,level distance=1.8cm}
	]

 \node (g11) {$S \overset {b_1} {\longrightarrow} [aAbCDD, 0110]$}
	child { node (g12) {$[aAbCDD, 0110] \xrightarrow{W(C)\otimes W(D)} aAbD$}
		child[shorten <=0.4cm*\scalefactor] {node (g21) {$A \overset {b_2} {\longrightarrow} [a, \varepsilon]$}
			child[shorten <=0cm] {node (g22) {$[a, \varepsilon] \overset{\1}{\longrightarrow} a$}}	}
		child[shorten <=0.4cm*\scalefactor] {node (g31) {$D \overset {b_3} {\longrightarrow} [EF, 01]$}
			child[shorten <=0cm] {node (g32) {$[EF, 01] \xrightarrow{W(F)} E$}
				child[shorten <=0.18cm*\scalefactor] {node (g41) {$E \overset {b_4} {\longrightarrow} [b, \varepsilon]$}
					child[shorten <=0cm] {node (g42){$[b, \varepsilon] \overset{\1}{\longrightarrow} b$}}	}}}};
						
 \draw ($ (g11.north) !0.55! (g12.south) $) ellipse (4.95cm and 1.2cm);
 \draw ($ (g21.north) !0.55! (g22.south) $) ellipse (2.2cm and 1.1cm);
 \draw ($ (g31.north) !0.55! (g32.south) $) ellipse (2.2cm and 1.1cm);
 \draw ($ (g41.north) !0.55! (g42.south) $) ellipse (2.2cm and 1.1cm);
\end{scope}

\end{tikzpicture}

\caption{\label{fig:e-free} An illustration of $\varphi(d) =d'$ in the proof of Theorem \ref{lm:wrtg-e-free}.}
\end{centering}
\end{figure}

    Next we prove that $\varphi$ is surjective. 
    More precisely, by induction on $(\RT_{\cG'}(N,\Gamma^+),\succ)$, we prove  that the following statement holds:
    \begin{equation}
\text{For every $d' \in \RT_{\cG'}(N,\Gamma^+)$, there exists a $d \in \RT_{\cG}(A,\Gamma^+)$
such that $\varphi(d)=d'$.} \label{eq:varphi-surjective}
\end{equation}
Let $d' \in \RT_{\cG'}(N,\Gamma^+)$. 
   Then $d'$ has the form
    \[ (A \to [\alpha,f])\Big( ([\alpha,f] \to v_0C_1v_1\ldots C_nv_n)(d'_1,\ldots,d_n') \Big)
      \]
      where $n\in \mathbb{N}$, $f$ is an $E$-selection for $\alpha$; moreover $v_0C_1v_1\ldots C_nv_n = f(\alpha)$ and $d_1' \in \RT_{\cG'}(C_j,\Gamma^+)$, \ldots, $d_n' \in \RT_{\cG'}(C_n,\Gamma^+)$. Let $\alpha = u_0A_1u_1 \ldots A_k u_k$ and $f=f_1\cdots f_k$.  Hence, $C_j = A_{g_f^{-1}(j)}$ for each $j \in [n]$.

      By I.H. we can assume that, for every $j \in [n]$, there exists $e_j \in \RT_\cG(C_j,\Gamma^+)$ such that $\varphi(e_j) = d_j'$. Moreover, for each $i \in [k]$ such that $f_i=1$, we choose an arbitrary rule tree $d_i''$ from $\RT_\cG(A_i,\varepsilon)$. We note that $\RT_\cG(A_i,\varepsilon) \not= \emptyset$ because $f$ is an $E$-selection for $\alpha$. Then we construct the rule tree $d = (A\to \alpha)(d_1,\ldots,d_k)$ in  $\RT_\cG(A,\Gamma^+)$ where for each $i \in [k]$ we let
      \[
        d_i = \begin{cases}
          d''_i & \text{ if } f_i=1\\
          e_{g_f(i)} & \text{ otherwise} \enspace.
          \end{cases}
        \]
        Then $\varphi(d)=d'$. This proves that $\varphi$ is surjective.    
        We mention that, in general, $\varphi$ is not injective because there may be several rule trees for $\varepsilon$ with the same nonterminal on the left-hand side of the rule at their root.

        \

  From the construction of $\cG'$, it is obvious that 
\begin{equation}
\text{for every $A \in N$ and $d \in \RT_\cG(A,\Gamma^+)$, we have $\pi_\cG(d) = \pi_{\cG'}(\varphi(d))$} \label{equ:yield-correspondence}
  \end{equation}
and, by using \eqref{equ:yield-correspondence}, we have that
  \begin{eqnarray}
  \begin{aligned}
   & \text{for every $A \in N$ and $u \in \Gamma^+$, the sets $\RT_\cG(A,u)$ and } \\
   &  \text{$\{(d',d) \mid d' \in \RT_{\cG'}(A,u), d \in \varphi^{-1}(d')\}$ are in a one-to-one correspondence.} 
  \end{aligned} \label{equ:bijectiveness}
  \end{eqnarray}

Finally, by induction on $(\RT_{\cG'}(N,\Gamma^+),\succ)$, we prove that the following statement holds: 
\begin{equation} \label{equ:weight-preservation}
\text{For every $d' \in \RT_{\cG'}(N,\Gamma^+)$ we have $\wt_{\cG'}(d') = \infsum{\oplus}{d \in \varphi^{-1}(d')}{\wt_\cG(d)}$}\enspace.
\end{equation}
Let $d' \in \RT_{\cG'}(N,\Gamma^+)$. Then $d'$ has the form
\[
d' = (A \to [\alpha,f])\Big( ([\alpha,f] \to f(\alpha)) \big(d_1',\ldots,d'_n) \Big)
  \]
  with $\alpha = u_0A_1u_1 \ldots A_ku_k$. Hence $f(\alpha)$ contains $n$ occurrences of $0$. Moreover, for each $j \in [n]$, we have $d'_j \in \RT_{\cG'}(A_{g_f^{-1}(j)},\Gamma^*)$.
  Then we can calculate as follows.
  \begingroup
  \allowdisplaybreaks
  \begin{align*}
    &\wt_{\cG'}(d') \\
    =& \Big( \bigotimes_{j \in [n]} \wt_{\cG'}(d_j') \Big) \otimes wt'([\alpha,f] \to f(\alpha)) \otimes wt'(A \to [\alpha,f])\\
    =& \Big( \bigotimes_{\substack{i \in [k]:\\f_i=0}} \wt_{\cG'}(d_{g_f(i)}') \Big)
    \otimes \Big( \bigotimes_{\substack{i \in [k]:\\f_i=1}} W(A_i) \Big)
    \otimes wt(A \to \alpha)
    \tag{by definition of $f$ and construction}\\
    =& \Big( \bigotimes_{i \in [k]} c_i \Big)     \otimes wt(A \to \alpha)
       \tag{by commutativity}
    \end{align*}
    \endgroup
    where
    \[
      c_i = \begin{cases}
        \wt_{\cG'}(d'_{g_f(i)}) & \text{ if $f_i=0$}\\
        W(A_{i}) & \text{otherwise} \enspace.
        \end{cases}
      \]
      By I.H. and by definition of $W(A_i)$,
           we have
     \[
      c_i = \begin{cases}
        \ \ \infsum{\oplus}{d_i \in \varphi^{-1}(d'_{g_f(i)})}{\wt_\cG(d_i)} & \text{ if $f_i=0$}\\
        \ \  \infsum{\oplus}{d_i \in \RT_{\cG}(A_{i},\varepsilon)}{\wt_\cG(d_i)} & \text{otherwise} \enspace.
        \end{cases}
      \]   
      Then, by distributivity and by surjectivity of $\varphi$ (see \eqref{eq:varphi-surjective}), we obtain
      \begin{align*}
        \Big( \bigotimes_{i \in [k]} c_i \Big)     \otimes wt(A \to \alpha)
          = \infsum{\oplus}{d \in \varphi^{-1}(d')}{\wt_\cG(d)} 
      \end{align*}
      which proves \eqref{equ:weight-preservation}.
  Then for each $u \in \Gamma^+$
  \begingroup
  \allowdisplaybreaks
  \begin{align*}
    \ssem{\cG'}(u) &= \bigoplus_{A \in S} \ \ \infsum{\oplus}{d' \in \RT_{\cG'}(A,u)}{\wt_{\cG'}(d')}\\
                   &= \bigoplus_{A \in S} \ \ \infsum{\oplus}{d' \in \RT_{\cG'}(A,u)} \ \  \infsum{\oplus}{d \in \varphi^{-1}(d')}{\wt_{\cG}(d)}
    \tag{by \eqref{equ:weight-preservation}}\\
                   &= \bigoplus_{A \in S} \ \ \infsum{\oplus}{d \in \RT_{\cG}(A,u)}{\wt_{\cG}(d)} \tag{by \eqref{equ:bijectiveness}}\\
    &= \ssem{\cG}(u) \enspace.
  \end{align*}
  \endgroup

  Since $\ssem{\cG'}(\varepsilon) = \0$, we obtain  $\ssem{\cG'} = \ssem{\cG} \otimes \chi_\B(\Gamma^+)$.
  \end{proof}


    \section{Yields of recognizable weighted tree languages}\label{sect:yield-of-weighted-cf}

    In this section  we state the  main result of this chapter: the yield of an r-recognizable weighted tree language is a weighted context-free language, and vice versa, each weighted context-free language can be obtained in this way. This result generalizes \cite[Thm.~3.20]{bra69} and \cite[Thm.~2.5]{don70} from the unweighted case to the weighted case (also cf. \cite[Thm.~3.2.7]{gecste84}), it generalizes \cite[Thm.~6.8.6]{esikui} (also cf. \cite[Thm.~8.6]{esikui03}) from commutative, continuous semirings to strong bimonoids, and it generalizes Theorem \cite[Thm.1 (1) $\Leftrightarrow$ (5)]{fulgaz18} from semirings to strong bimonoids. (We also refer to \cite[Thm.~8.1 and 8.3]{berreu82} and \cite[Thm.~30]{boz99}.)

\index{yield@$\yield_\Gamma$}
We recall the mapping $\yield_\Gamma$ from Section \ref{sect:trees}. Let $\Gamma \subseteq \Sigma^{(0)}$.  Then the mapping $\yield_\Gamma: \T_\Sigma \to \Gamma^*$ is defined for each $\xi = \sigma(\xi_1,\ldots,\xi_k)$ in $\T_\Sigma$ by  
\[
  \yield_\Gamma(\xi) =
  \begin{cases}
    \yield_\Gamma(\xi_1) \cdots \yield_\Gamma(\xi_k) & \text{ if $ k \ge 1$}\\
    \sigma & \text{ if $k=0$ and $\sigma \in \Gamma$}\\
    \varepsilon  & \text{ otherwise}\enspace.
    \end{cases}
  \]

  Let $r: \T_\Sigma \to B$ be a weighted tree language.   Also we recall from \eqref{obs:app-tree-transf-to-wtl-2} that, if $r$ is $\chi(\yield_\Gamma)$-summable or $\B$ is $\sigma$-complete, then $\chi(\yield_\Gamma)(r): \Gamma^* \to B$ is a  weighted language such that,  for each $u \in \Gamma^*$, we have
\[
\chi(\yield_\Gamma)(r)(u) = \infsum{\oplus}{\xi \in \yield_\Gamma^{-1}(u)}{r(\xi)} \enspace.
  \]

\begin{theorem-rect}{\rm \cite[Thm.~1]{fulgaz18}} \label{thm:yield(wta)=cf} Let $\Gamma$ be an alphabet and $\B$ be a strong bimonoid. For each weighted language $s: \Gamma^* \to B$ the following two statements are equivalent.
    \begin{compactenum}
    \item[(A)] We can construct a $(\Gamma,\B)$-wcfg $\cG$ such that $\cG$ is finite-derivational or $\B$ is $\sigma$-complete, and $s = \ssem{\cG}$.
    \item[(B)] We can construct a ranked alphabet $\Sigma$ with $\Gamma \subseteq \Sigma^{(0)}$ and a $(\Sigma,\B)$-wta $\cA$ such that $\runsem{\cA}$ is $\chi(\yield_\Gamma)$-summable or $\B$ is $\sigma$-complete, and $s = \chi(\yield_\Gamma)\big(\runsem{\cA}\big)$.
           \end{compactenum}
         \end{theorem-rect}
         \begin{proof} This follows from Lemmas \ref{lm:wcfg-wta} and \ref{lm:wta-wcfg}.
           \end{proof}
  
      \begin{lemma}\rm \label{lm:wcfg-wta} Let $\cG$ be a $(\Gamma,\B)$-wcfg such that $\cG$ is finite-derivational or $\B$ is $\sigma$-complete. We can construct a ranked alphabet $\Sigma$ with $\Gamma \subseteq \Sigma^{(0)}$ and a root weight normalized $(\Sigma,\B)$-wta $\cA=(Q,\delta,F)$ such that (a)~$\runsem{\cA}$ is $\chi(\yield_\Gamma)$-summable if $\cG$ is finite-derivational and (b)~$\ssem{\cG} = \chi(\yield_\Gamma)(\runsem{\cA})$.
      \end{lemma}
      \begin{proof} Let $\cG=(N,S,R,wt)$. By Lemmas \ref{lm:wcfg-in-nonterminal-form} and \ref{lm:wcfg-start-separated}, we can construct a $(\Gamma,\B)$-wcfg
      which is in nonterminal form, start-separated, and equivalent to $\cG$. Hence we can assume
      that $\cG$ is in nonterminal form and start-separated.

      We define the $\mathbb{N}$-indexed family $(R_k \mid k \in \mathbb{N})$ such that
      \[
R_k = \{(A\to \alpha) \in R \mid |\alpha|=k\}\enspace.
        \]
We note that, if $\cG$ is $\varepsilon$-free, then $(R_k \mid k \in \mathbb{N})$ is not a ranked alphabet, because $R_0 = \emptyset$. Moreover, a terminal rule $A \to a$ with $a \in \Gamma$ is in $R_1$ (and not in $R^{(0)}$ as defined in Section \ref{subsect:basic-modell}). 

We construct the ranked alphabet $\Sigma$ by $\Sigma^{(k)} = R_k$ for each $k \in \mathbb{N}_+$, and $\Sigma^{(0)} = R_0 \cup \Gamma$.  We construct the $(\Sigma,\B)$-wta $\cA=(Q,\delta,F)$ as follows.

        \begin{compactitem}
        \item $Q = N \cup \overline{\Gamma}$ where $\overline{\Gamma} = \{\overline{a} \mid a \in \Gamma\}$ (note that $Q \cap \Sigma = \emptyset$),

        \item for every $k \in \mathbb{N}$, $\sigma \in \Sigma^{(k)}$, and  $q,q_1,\ldots,q_k \in Q$, we let
          \[
            \delta_k(q_1\cdots q_k,\sigma,q) = \begin{cases}
              wt(\sigma) & \text{ if $q,q_1,\ldots,q_k \in N$ and $\sigma = (q \to q_1 \cdots q_k)$ is in $R$}\\
              wt(\sigma) & \text{ if $k=1$, $q\in N$, $q_1 = \overline{a}$ for some $a \in \Gamma$ such that $\sigma = (q \to a)$ is in $R$}\\
              \1 & \text{ if $k=0$, $q=\overline{\sigma}$, and $\sigma \in \Gamma$}\\
                \0 & \text{ otherwise}\enspace,
              \end{cases}
            \]
      
          \item for each $q \in Q$, we define $F_q=\1$ if $q=S$, and $\0$ otherwise. 
          \end{compactitem}
          Obviously, $\cA$ is root weight normalized.

          For the proof of (a) and (b) we need some preparations. For each $\xi \in \T_\Sigma$, we define the run $\rho_\xi: \pos(\xi) \to Q$ such that, for each $w \in \pos(\xi)$, we let
          \[
            \rho_\xi(w) = \begin{cases}
              \lhs(\xi(w)) & \text{ if $\xi(w) \in R$}\\
              \overline{\xi(w)} & \text{ otherwise} \enspace.
              \end{cases}
            \]
            We define the mapping $\varphi: \RT_\cG \to \{(\xi,\rho) \mid \xi \in \T_\Sigma, \rho \in \R_\cA(\xi)\}$ for every $d \in \RT_\cG$ by $\varphi(d) = (\xi,\rho_\xi)$ where $\xi$ is obtained from $d$ by replacing each leaf which is labeled by a rule of the form $A \to a$ with $a \in \Gamma$ by the tree $(A\to a)(a)$ (cf. Figure \ref{fig:from-wcfg-to-wta}). In particular, for each $(\xi,\rho_\xi) \in \im(\varphi)$, we have that  $\rho_\xi \in \R_\cA(S,\xi)$.
            Obviously, $\varphi$ is injective.

            Next we define $\varphi': \RT_\cG \to \im(\varphi)$ by $\varphi'(d) = \varphi(d)$ for each $d \in \RT_\cG$. Obviously, $\varphi'$ is bijective. Moreover,
            \begin{equation}
              \text{for each $d \in \RT_\cG$,  if $\varphi'(d)=(\xi,\rho_\xi)$, then $\pi(d) = \yield_\Gamma(\xi)$.} \label{eq:pi=yield}
            \end{equation}
            For the proof of (a), assume that $\cG$ is finite-derivational. Then, \eqref{eq:pi=yield} and the fact that $\varphi'$ is bijective imply  that $\yield_\Gamma^{-1}(u) \cap \supp(\runsem{\cA})$ is finite for each $u \in \Gamma^*$, i.e.,  that $\runsem{\cA}$ is $\chi(\yield_\Gamma)$-summable.

            Finally, we prove (b).           
            It is easy to see that
\begin{equation}
\text{for each $d \in \RT_\cG$, we have $\wt_\cG(d) = \wt_\cA(\varphi'(d))$}\enspace. \label{eq:wtG=wtA}
\end{equation}
Moreover, 
            \begin{equation}
\text{for every $\xi \in \T_\Sigma$ and $\rho \in \R_\cA(S,\xi)$: if $(\xi,\rho) \not\in \im(\varphi)$, then $\wt_\cA(\xi,\rho) = \0$}\enspace. \label{eq:not-equal-rho-xi=0}
\end{equation}

Then, for each $u \in \Gamma^*$, we can calculate as follows.
\begingroup
\allowdisplaybreaks
\begin{align*}
  \ssem{\cG}(u) &= \infsum{\oplus}{d \in \RT_\cG(u)}{\wt_\cG(d)}
                 \tag{by \eqref{eq:sem-wcfg-der-tree}}\\
                &= \infsum{\oplus}{\substack{d \in \RT_\cG:\\\pi(d)=u}}{\wt_\cG(d)}\\
               &= \infsum{\oplus}{\substack{(\xi,\rho_\xi) \in \im(\varphi):\\ \yield_\Gamma(\xi) = u}}{\wt_\cG((\varphi')^{-1}((\xi,\rho_\xi)))}
  \tag{because $\varphi'$ is bijective and by \eqref{eq:pi=yield}} \\
   &= \infsum{\oplus}{\substack{(\xi,\rho_\xi) \in \im(\varphi):\\ \yield_\Gamma(\xi) = u}}{\wt_\cA(\xi,\rho_\xi)}
  \tag{by \eqref{eq:wtG=wtA} and because $\varphi((\varphi')^{-1}((\xi,\rho_\xi))) = (\xi,\rho_\xi)$} \\
  %
  %
                &= \infsum{\oplus}{\xi \in \yield_\Gamma^{-1}(u)}{\bigoplus_{\rho \in \R_\cA(S,\xi)} \wt_\cA(\xi,\rho)}
   \tag{by \eqref{eq:not-equal-rho-xi=0}}\\
                &= \infsum{\oplus}{\xi \in \yield_\Gamma^{-1}(u)}{\bigoplus_{q \in Q}\bigoplus_{\rho \in \R_\cA(q,\xi)} \wt_\cA(\xi,\rho) \otimes F_q} 
                  \tag{because $F_S=\1$ and $F_q = \0$ for each $q \in Q\setminus\{S\}$}\\
                &= \infsum{\oplus}{\xi \in \yield_\Gamma^{-1}(u)}{\runsem{\cA}(\xi)}
  \tag{by definition of $\runsem{\cA}$}\\
  &= \Big(\chi(\yield_\Gamma)(\runsem{\cA})\Big)(u)\enspace.  \tag{by \eqref{obs:app-tree-transf-to-wtl-2}} 
\end{align*}
\endgroup
        \end{proof}

                \begin{figure}
          \centering
\begin{tikzpicture}[level distance=3.5em,
  every node/.style = {align=center}]]
  \pgfdeclarelayer{bg}    
  \pgfsetlayers{bg,main}  

  \newcommand\mydist{3mm}
  \tikzstyle{mycircle}=[draw, circle, inner sep=-2mm, minimum height=5mm]

\begin{scope}[level 1/.style={sibling distance=27mm},
level 2/.style={sibling distance=20mm}]

 \node at (-1.2,0.6) {$d\in \RT_{\cG}(ab)$:}; 
 
 \node (T1) {$S \to AB$}
  child {node {$A\to a$} }
  child {node {$B \to BC$}
 	   child { node {$B \to b$}}
 	   child { node {$C \to \varepsilon$}} };
\end{scope}

\begin{scope}[xshift=76mm, level 1/.style={sibling distance=37mm},
level 2/.style={sibling distance=27mm}]

 \node at (-1.3, 0.6) {$\xi \in \T_\Sigma$:};
 \node at (2.5, 0.6) {$\rho_\xi \in \R_\cA (\xi)$:};

 \node (N0) {$S \to AB$}
  child {node (N1) {$A\to a$} 
  	   child { node (N2) {$a$}} }
  child {node (N3) {$B \to BC$}
 	   child { node (N4) {$B \to b$}
 	        child { node (N5) {$b$}} }
 	   child { node (N6) {$C \to \varepsilon$}} };

 \node [mycircle] at ([xshift=\mydist]N0.east) {$S$};
 \node [mycircle] at ([xshift=\mydist]N1.east) {$A$};
 \node [mycircle] at ([xshift=\mydist]N2.east) {$\overline{a}$};
 \node [mycircle] at ([xshift=\mydist]N3.east) {$B$};
 \node [mycircle] at ([xshift=\mydist]N4.east) {$B$};
 \node [mycircle] at ([xshift=\mydist]N5.east) {$\overline{b}$};
 \node [mycircle] at ([xshift=\mydist]N6.east) {$C$};
\end{scope}

\draw [->, thick, xshift=3.3cm, yshift=-1.2cm] (0,0) -- (1,0) node[midway, above] {$\varphi$};

\end{tikzpicture}

\caption{\label{fig:from-wcfg-to-wta} A visualization of $\varphi(d)=(\xi,\rho_\xi)$ in the proof of Lemma \ref{lm:wcfg-wta} with $d \in \R_\cA(ab)$ for $a,b \in\Gamma$, $\xi \in \T_\Sigma$, and $\rho_\xi \in \R_\cA(\xi)$. The states of $\rho_\xi$ are circled.}
          \end{figure}

      \begin{lemma}\rm \label{lm:wta-wcfg} Let $\Gamma \subseteq \Sigma^{(0)}$. Moreover, let $\cA=(Q,\delta,F)$ be a $(\Sigma,\B)$-wta such that $\runsem{\cA}$ is $\chi(\yield_\Gamma)$-summable or $\B$ is $\sigma$-complete. We can construct a $(\Gamma,\B)$-wcfg in nonterminal form  such that (a)~$\cG$~is finite-derivational if $\runsem{\cA}$ is $\chi(\yield_\Gamma)$-summable  and  (b)~$\ssem{\cG} = \chi(\yield_\Gamma)\big(\runsem{\cA}\big)$.
      \end{lemma}

      \begin{proof} Let $\cA=(Q,\delta,F)$. By Theorem \ref{thm:root-weight-normalization-run} we can construct a $(\Sigma,\B)$-wta which is root weight normalized and r-equivalent to $\cA$. Hence, we can
        assume that $\cA$ is root weight normalized. Thus, $\supp(F) = \{q_f\}$ for some $q_f \in Q$ and $F_{q_f}=\1$.

        We construct the $(\Gamma,\B)$-wcfg $\cG=(N,S,R,wt)$ as follows.
        \begin{compactitem}
        \item $N = Q \times \Sigma$ where  each element has the form $[q,\sigma]$ for some $q \in Q$ and $\sigma \in \Sigma$,
          \item $S = \{q_f\} \times \Sigma$,
        \item $R$ is the smallest set such that the following conditions hold; simultaneously, we define $wt$.
          \begin{compactitem}
          \item For every $k \in \mathbb{N}_+$, $\sigma \in \Sigma^{(k)}$, $q,q_1,\ldots,q_k \in Q$, and $\sigma_1,\ldots,\sigma_k \in \Sigma$, the rule \\
            $r=([q,\sigma] \to [q_1,\sigma_1] \cdots [q_k,\sigma_k])$ is in $R$ and $wt(r) = \delta_k(q_1\cdots q_k,\sigma,q)$.
          \item For every $\sigma \in \Gamma$ and $q \in Q$, the rule $r=([q,\sigma]\to \sigma)$ is in $R$ and $wt(r) = \delta_0(\varepsilon,\sigma,q)$.
              \item For every $\sigma \in \Sigma^{(0)}\setminus \Gamma$ and $q \in Q$, the rule $r=([q,\sigma]\to \varepsilon)$ is in $R$ and $wt(r) = \delta_0(\varepsilon,\sigma,q)$.
            \end{compactitem}
          \end{compactitem}

          We define the mapping $\varphi: \{(\xi,\rho) \mid \xi \in \T_\Sigma, \rho\in \R_\cA(q_f,\xi)\} \to \RT_\cG$ for every $(\xi,\rho)$ by \(\varphi(\xi,\rho) = d\) where $d$ is determined by the tree domain $W$ and the  $R$-tree mapping $d': W \to R$ (recall that there is a bijective representation of trees as tree domains and tree mappings, cf. Section \ref{sect:trees}). We define $W = \pos(\xi)$ and for each $w \in W$ (using $k$ as abbreviation for $\rk_\Sigma(\xi(w))$) we let
          \[
            d'(w) =
            \begin{cases}
              [\rho(w),\xi(w)] \to [\rho(w1),\xi(w1)] \cdots  [\rho(wk),\xi(wk)] & \text{ if $k \ge 1$}\\
              [\rho(w),\xi(w)] \to \xi(w) & \text{ if $k=0$ and $\xi(w) \in \Gamma$}\\
                             [\rho(w),\xi(w)] \to \varepsilon  & \text{ if $k=0$ and $\xi(w) \in \Sigma^{(0)} \setminus \Gamma$} \enspace.
              \end{cases}
          \]
          Obviously, $\varphi$ is injective and surjective, hence bijective (cf. Figure \ref{fig:from-wta-to-wcfg}).  Moreover,
            \begin{eqnarray}
            \begin{aligned}
             & \text{for every $\xi \in \T_\Sigma$ and $\rho \in \R_\cA(\xi)$,} \\
             &  \text{if $\varphi(\xi,\rho)=d$, then $\pi(d) = \yield_\Gamma(\xi)$ and $\wt_\cG(d)=\wt_\cA(\xi,\rho)$.} 
              \end{aligned} \label{eq:yield=pi}
            \end{eqnarray}
 Thus, since $\varphi$ is bijective, the assumption that  $\runsem{\cA}$ is $\chi(\yield_\Gamma)$-summable implies that $\cG$ is finite-derivational.

            Then, for each $u \in \Gamma^*$, we can calculate as follows.          
          \begingroup
            \allowdisplaybreaks
\begin{align*}
  \ssem{\cG}(u) &= \infsum{\oplus}{d \in \RT_\cG(u)}{\wt_\cG(d)} 
                = \infsum{\oplus}{\substack{d \in \RT_\cG:\\\pi(d)=u}}{\wt_\cG(d)}\\
                &= \infsum{\oplus}{\substack{\xi \in \T_\Sigma:\\ \yield_\Gamma(\xi) = u}}{
  \bigoplus_{\rho \in \R_\cA(q_f,\xi)} \wt_\cA(\xi,\rho)}
  \tag{because $\varphi$ is bijective and by \eqref{eq:yield=pi}} \\
                &= \infsum{\oplus}{\xi \in \yield_\Gamma^{-1}(u)}{\bigoplus_{\rho \in \R_\cA(q_f,\xi)} \wt_\cA(\xi,\rho)}\\
                &= \infsum{\oplus}{\xi \in \yield_\Gamma^{-1}(u)}{\bigoplus_{\rho \in \R_\cA(\xi)} \wt_\cA(\xi,\rho) \otimes F_{\rho(\varepsilon)}} \tag{because $\supp(F)=\{q_f\}$ and $F_{q_f}=\1$}\\
                &= \infsum{\oplus}{\xi \in \yield_\Gamma^{-1}(u)}{\runsem{\cA}(\xi)}
  \tag{by definition of $\runsem{\cA}$}\\
  &= \chi(\yield_\Gamma)(\runsem{\cA})(u)\enspace.   \tag{by \eqref{obs:app-tree-transf-to-wtl-2}}
\end{align*}
\endgroup
        \end{proof}

        \begin{figure}
          \centering
          \begin{tikzpicture}[level distance=3.5em,
  every node/.style = {align=center}]]
  \pgfdeclarelayer{bg}    
  \pgfsetlayers{bg,main}  

  \newcommand{\mydista}{4mm}
  \newcommand{\mydistb}{3mm}
  \tikzstyle{mycircle}=[draw, circle, inner sep=-2mm, minimum height=5mm]

\begin{scope}[level 1/.style={sibling distance=20mm}]

 \node at (-0.8, 0.6) {$\xi \in \T_\Sigma$:};
 \node at (1.9, 0.6) {$\rho \in \R_\cA (\xi)$:};

 \node (N0) {$\sigma$}
  child {node (N1) {$\gamma$} 
  	   child { node (N2) {$\alpha$}} }
  child {node (N3) {$\beta$} };

 \node [mycircle] at ([xshift=\mydista]N0.east) {$q$};
 \node [mycircle] at ([xshift=\mydistb]N1.east) {$q_1$};
 \node [mycircle] at ([xshift=\mydistb]N2.east) {$q_2$};
 \node [mycircle] at ([xshift=\mydistb]N3.east) {$q_3$};
 
 \node at (0.3,-3.5) {$\Gamma = \{\alpha\}$};
\end{scope}

\begin{scope}[xshift=80mm, level 1/.style={sibling distance=37mm}]

 \node at (-1.5,0.6) {$d\in \RT_{\cG}(\alpha):$}; 
 
 \node (N0) {$[q,\sigma]\to [q_1,\gamma][q_3,\beta]$}
  child {node (N1) {$[q_1,\gamma]\to [q_2,\alpha]$} 
  	   child { node (N2) {$[q_2,\alpha]\to\alpha$}} }
  child {node (N3) {$[q_3,\beta]\to\varepsilon$} };
\end{scope}

\draw [->, thick, xshift=3.3cm, yshift=-1.2cm] (0,0) -- (1,0) node[midway, above] {$\varphi$};

\end{tikzpicture}
\caption{\label{fig:from-wta-to-wcfg} A visualization of $\varphi(\xi,\rho)=d$ in the proof of Lemma \ref{lm:wta-wcfg} with $\xi \in \T_\Sigma$,  $\rho \in \R_\cA(\xi)$, $\Gamma=\{\alpha\}$, and $d \in \RT_\cG(\alpha)$. The states of $\rho$ are circled.}
\end{figure}

Finally, we verify that Theorem \ref{thm:yield(wta)=cf}  generalizes \cite[Thm.~3.20]{bra69} and \cite[Thm.~2.5]{don70}. We achieve this by proving that the latter results are equivalent to Theorem \ref{thm:yield(wta)=cf} for the case that $\B$ is the Boolean semiring $\Boole$.

           \begin{corollary}\label{cor:cfg=yield(fta)-JGC} \rm For each language $L \subseteq \Gamma^*$ the following two statements are equivalent.
    \begin{compactenum}
    \item[(A)] We can construct a context-free grammar $G$ such that $\LL(G) = L$.
    \item[(B)] We can construct a ranked alphabet $\Sigma$ with $\Gamma \subseteq \Sigma^{(0)}$ and a $\Sigma$-fta $A$ such that $L = \yield_\Gamma(\LL(A))$.
           \end{compactenum}
         \end{corollary}
         \begin{proof} Proof of (A)$\Rightarrow$(B): Let $G$ be a $\Gamma$-cfg such that $L= \LL(G)$. By Observation \ref{obs:cfg=wcfg(B)}(A)$\Rightarrow$(B), we can construct a $(\Gamma,\Boole)$-wcfg $\cG$ such that $L=\supp(\ssem{\cG})$. Since $\Boole$ is $\sigma$-complete, Theorem \ref{thm:yield(wta)=cf}(A)$\Rightarrow$(B) implies that we can construct a ranked alphabet $\Sigma$ with $\Gamma \subseteq \Sigma^{(0)}$ and a $(\Sigma,\Boole)$-wta $\cA$ such that $L= \supp(\chi(\yield_\Gamma)(\sem{\cA}))$.  By \eqref{equ:supp-yield=yieldA} (using $r=\sem{\cA}$ and $g=\yield_\Gamma$), we obtain $L = \yield_\Gamma(\supp(\sem{\cA}))$.  By Corollary \ref{cor:supp-B=fta-1}(B)$\Rightarrow$(A), we can construct a $\Sigma$-fta $A$ such that $\supp(\sem{\cA})=\LL(A)$. Thus $L= \yield_\Gamma(\LL(A))$.

           \

 Proof of (B)$\Rightarrow$(A): Let $A$ be a $\Sigma$-fta such that $L=\yield_\Gamma(\LL(A))$. By Corollary \ref{cor:supp-B=fta-1}(A)$\Rightarrow$(B), we can construct a $(\Sigma,\Boole)$-wta $\cA$ such that $L = \yield_\Gamma(\supp(\sem{\cA}))$. Then we can apply \eqref{equ:supp-yield=yieldA} and obtain that $L=\supp(\chi(\yield_\Gamma)(\sem{\cA}))$. Since $\Boole$ is $\sigma$-complete, Theorem \ref{thm:yield(wta)=cf}(B)$\Rightarrow$(A) implies that we can construct a $(\Gamma,\Boole)$-wcfg $\cG$ such that $L=\supp(\ssem{\cG})$. Finally,  Observation \ref{obs:cfg=wcfg(B)}(B)$\Rightarrow$(A), we can construct a $\Gamma$-cfg $G$ such that $L=\LL(G)$. 
\end{proof}

%% file: wrtg-as-wcfg.tex
\chapter{Weighted regular tree grammars}
\label{ch:regular-tree-grammars}

Weighted regular tree grammars were introduced and investigated in \cite{inafuk75} (for $\B = ([0,1],\max,\min,0,1)$) and in \cite{aleboz87} (for $\B$  being an arbitrary semiring).

Here we define weighted regular tree grammars as particular wcfg. This possibility is based on the simple fact that each tree is a particular string. 
We indicate that some of the normal forms for wcfg are also normal forms for weighted regular tree grammars.
 We define two more restricted forms of weighted regular tree grammars: alphabetic  and tree automata form, and prove that they are normal forms (cf. Lemmas \ref{lm:rtg-normal-form} and  \ref{lm:rtg-automata-form-wrtg}, respectively).
 In particular, we prove that weighted regular tree grammars in tree automata form are essentially wta (cf. Theorem \ref{thm:rec=reg}).

\section{The grammar model}
\label{sec:grammar-model}

\index{weighted regular tree grammar}
\index{wrtg}
\index{Xi@$\Xi$}
By definition, each tree $\xi$ over $\Sigma$ is a particular string over the alphabet $\Sigma \cup \Xi$ where $\Xi$ contains the opening and closing parentheses and the comma. For convenience, we abbreviate $\Sigma \cup \Xi$ by $\Sigma^\Xi$.  Then, by definition, we have  $\T_\Sigma \subseteq (\Sigma^\Xi)^*$.
Of course, $(\Sigma^\Xi)^* \setminus \T_\Sigma \ne \emptyset$.

This aspect of the definition of trees (being particular strings) allows one to define particular wcfg, viz. those which generate weighted tree languages. We call them weighted regular tree grammars.  

Formally, a {\em weighted regular tree grammar over $\Sigma$ and $\B$} (for short: $(\Sigma,\B)$-wrtg, or: wrtg) is a $(\Sigma^\Xi,\B)$-wcfg $\cG = (N,S,R,wt)$ where each rule in $R$ has the form  $A \rightarrow \xi$ with $\xi \in \T_\Sigma(N)$. Obviously, $\cG$ is $\varepsilon$-free because $\varepsilon \not\in \T_\Sigma(N)$.
Also here we sometimes want to show the occurrences of elements of $N$ in the right-hand side of a rule more explicitly. Then we will  write a rule in the form
\[
  A \rightarrow \xi[A_1,\ldots,A_k]\enspace,
\]
where $k\in \mathbb{N}$, $\xi \in \C_\Sigma(Z_k)$, and $A_1,\ldots,A_k\in N$ and $\xi[A_1,\ldots,A_k]$ denotes the tree obtained from $\xi$ by replacing each occurrence of $z_i$ by $\xi_i$ (cf. ``Contexts and substitution'' on page \pageref{page:context-and-substitution}).

  \index{pi@$\pi_{\cG}$}
  Let $\cG=(N,S,R,wt)$ be a $(\Sigma,\B)$-wrtg. We recall that we considered $R$ as ranked alphabet (where the rank of a rule is the number of nonterminal occurrences in its right-hand side) and that the projection of $\cG$ has the type $\pi_\cG: \T_R \to (\Sigma^\Xi)^*$.
    Due to the special form of the rules,  we have $\im(\pi_\cG) \subseteq \T_\Sigma$, and thus we can view $\pi_\cG$ as mapping of type
  \[
    \pi_\cG: \T_R \rightarrow \T_\Sigma \enspace,
  \]
    and hence $\RT_\cG = \RT_\cG(\T_\Sigma)$.
  Moreover, it is easy to see that $\pi_\cG$ is determined by the $(R,\Sigma)$-tree homomorphism $\pi_\cG = ((\pi_\cG)_k \mid k \in \mathbb{N})$ defined, for every $k \in \mathbb{N}$ and $r \in R^{(k)}$ of the form $r = (A \rightarrow \xi[A_1,\ldots,A_k])$, by $(\pi_\cG)_k(r) = \xi$.

By Observation \ref{obs:bound-on-size-of-rule-trees} and since $\cG$ is $\varepsilon$-free, if $\cG$ is chain-free, then it is finite-derivational.

\index{semanticG@$\sem{\cG}^{\mathrm{t}}$}
\index{semanticG@$\sem{\cG}^{\mathrm{t}}$}
If $\cG$ is finite-derivational or $\B$ is $\sigma$-complete, then the {\em weighted tree language generated by $\cG$}, denoted by $\sem{\cG}^{\mathrm{t}}$, is the mapping $\sem{\cG}^{\mathrm{t}}: \T_\Sigma \rightarrow B$ defined for each $\xi \in \T_\Sigma$ by
\[
\sem{\cG}^{\mathrm{t}}(\xi) = \ssem{\cG}(\xi) \enspace.
\]
Hence $\sem{\cG}^{\mathrm{t}} = \ssem{\cG}|_{\T_\Sigma}$ and for each $\xi \in \T_\Sigma$, using  \eqref{eq:sem-wcfg-der-tree}, we have
\begin{equation*}
  \sem{\cG}^{\mathrm{t}}(\xi)  
  = \infsum{\oplus}{d \in \RT_\cG(\xi)}{\wt_\cG(d)}\enspace.  \label{eq:sem-wrtg-der-tree}
\end{equation*}
Let  $r$ be a $(\Sigma,\B)$-weighted tree language. It is called {\em regular} if there exists a $(\Sigma,\B)$-wrtg $\cG$ which is finite-derivational if $\B$ is not $\sigma$-complete and for which $r = \sem{\cG}^{\mathrm{t}}$. \index{regular}

Since in this book we will use $\sem{\cG}^{\mathrm{t}}$ many times, we make the following convention.
\label{p:convention-fo-wrtg}
\index{semanticG@$\sem{\cG}$}
\begin{quote}\em In the rest of this book,  for each wrtg $\cG$, we will abbreviate $\sem{\cG}^{\mathrm{t}}$ by $\sem{\cG}$. 
\end{quote}

\index{Reg@$\Reg(\Sigma,\B)$}
\index{Reg@$\Reg_{\mathrm{fd}}(\Sigma,\B)$}
\index{Reg@$\Reg_{\mathrm{nc}}(\Sigma,\B)$}
We denote by (i) $\Reg_{\mathrm{nc}}(\Sigma,\B)$, (ii) $\Reg_{\mathrm{fd}}(\Sigma,\B)$, and (iii) $\Reg(\Sigma,\B)$ the sets of $(\Sigma,\B)$-weighted tree languages  which can be generated by
\begin{compactitem}
\item[(i)] chain-free  $(\Sigma,\B)$-wrtg
\item[(ii)] finite-derivational  $(\Sigma,\B)$-wrtg, and 
\item[(iii)] $(\Sigma,\B)$-wrtg which are finite-derivational if $\B$ is not $\sigma$-complete, respectively. 
\end{compactitem}
Then $\Reg_{\mathrm{nc}}(\Sigma,\B) \subseteq \Reg_{\mathrm{fd}}(\Sigma,\B)\subseteq \Reg(\Sigma,\B)$; moreover, $\Reg_{\mathrm{fd}}(\Sigma,\B)= \Reg(\Sigma,\B)$ if $\B$ is not $\sigma$-complete.


\begin{example} \rm We consider the ranked alphabet $\Sigma = \{\sigma^{(2)},\gamma^{(1)},\alpha^{(0)}\}$ and the set of trees $U \subset \T_\Sigma$ such that each tree $\xi \in U$  has an upper part, which only contains $\sigma$-labeled positions, and a number of lower parts, which only contain $\gamma$- and $\alpha$-labeled positions. Formally,
 \[
    U = \{ \zeta[\zeta_1,\ldots,\zeta_n] \mid  n \in \mathbb{N},  \zeta \in \C_\Sigma(Z_n), \pos_{\{\gamma,\alpha\}}(\zeta)=\emptyset,
    \text{ and } \zeta_1,\ldots,\zeta_n \in \T_{\{\gamma,\alpha\}} \} \enspace.
  \]

Let $\xi \in U$. There are unique $n$, $\zeta$, and $\zeta_1,\ldots,\zeta_n$ such that $\xi = \zeta[\zeta_1,\ldots,\zeta_n]$. Then $\zeta$ is the upper part of $\xi$ and $\zeta_1,\ldots,\zeta_n$ are the lower parts of $\xi$. We denote by $\mathrm{bor}(\xi)$ the set of positions of $\xi$ which are at the border between the upper part and the lower parts, i.e., $\mathrm{bor}(\xi) = \pos_Z(\zeta)$.

Now we wish to determine, for each $\xi \in U$, the minimal height of a lower part of $\xi$. For this we consider the tropical semiring $\Natminplus = (\mathbb{N}_\infty,\min,+,\infty,0)$ and the mapping $f: \T_\Sigma \to \mathbb{N}_\infty$ defined for each $\xi \in \T_\Sigma$ by
\[
  f(\xi) = \begin{cases}
    \min(\height(\xi|_w) \mid w \in \mathrm{bor}(\xi)) & \text{ if $\xi \in U$}\\
    \infty & \text{ otherwise} \enspace.
    \end{cases}
  \]

  We will show that $f$ is a regular $(\Sigma,\Natminplus)$-weighted tree language. For this we construct the following $(\Sigma,\Natminplus)$-wrtg $\cG = (N,\{S\},R,wt)$ with $N=\{S,A,B,C,D\}$ and the following   rules and weights:
  \begin{center}
  \begin{tabular}{lll}
    $S \to \sigma(A,S)$ \ : 0,  & $S \to \sigma(S,A)$ \ : 0,  & $S \to B$ \ : 0,\\
    $A \to \sigma(A,A)$ \ : 0,  & $A \to C$ \ : 0,\\
    $B \to \gamma(B)$ \ : 1,    & $B \to \alpha$ \ : 0,\\
    $C \to \gamma(C)$ \ : 0,    & $C \to \alpha$ : 0.
  \end{tabular} 
\end{center}
We note that $\cG$ contains two chain-rules. However $\cG$ is finite-derivational because of the following. For each $\xi \in U$, there exists a bijection between the sets $\RT_\cG(\xi)$ and $\mathrm{bor}(\xi)$ (in particular, based on $\Sigma$, we can reconstruct from the set $\mathrm{bor}(\xi)$ of positions the tree $\xi$). Thus $|\RT_\cG(\xi)| = |\mathrm{bor}(\xi)|$ for each $\xi \in U$, and $| \RT_\cG(\xi)| = \emptyset$ for each $\xi \in \T_\Sigma\setminus U$. Hence $\cG$ is finite-derivational. Moreover, it is easy to see that
$\supp(\sem{\cG})=U$.

  Now let $\xi \in U$. Then, for every $d \in \RT_\cG(\xi)$, there exists a unique position $w \in \pos(d)$ such that $d(w) = (S\to B)$ and $w \in \mathrm{bor}(\xi)$ (i.e., $\xi|_w$ is a lower part of $\xi$); this is due to the form of the $S$-rules. Let us denote this unique position by $w_d$. Vice versa, for each  $w \in \mathrm{bor}(\xi)$, there exists a $d \in \RT_\cG(\xi)$ such that $d(w) = (S\to B)$. Let us denote this $d$ by $d_w$. Clearly, $w_{d_w}=w$. We refer to Figure \ref{fig:ex-wrtg} for an illustration.

  \begin{figure}
    \centering

\begin{tikzpicture}[level distance=3.5em,
  every node/.style = {align=center}]]
  \pgfdeclarelayer{bg}    
  \pgfsetlayers{bg,main}  
  
  \tikzset{my dbl/.style={double,double distance=1.5pt}}
  \usetikzlibrary{fit}

\begin{scope}[level 1/.style={sibling distance=32mm},
level 2/.style={sibling distance=20mm}]
 
 \node at (-2.75,0.25) {$\xi :$}; 
 
 \node (T1) {$\sigma$}
  child {node {$\sigma$}
  edge from parent [my dbl]  
  	    child { node {$\gamma$}
       		 child { node {$\gamma$} 
       		       child { node {$\alpha$} }}}
       child { node (G1) {$\gamma$}
       edge from parent [my dbl]
  	   		 child { node (A1) {$\alpha$} }}}
  child {node {$\gamma$}
 	   child { node {$\gamma$} 
 	          child { node {$\alpha$} }}} ;
 \node[ellipse,draw,fit=(G1)(A1),minimum height=26mm] {};

 \node at (-0.5,-7) {$12\in \mathrm{bor}(\xi)$};
\end{scope}

\begin{scope}[xshift=80mm,level 1/.style={sibling distance=40mm},
level 2/.style={sibling distance=24mm}]

 \node at (-3, 0.25) {$d\in \RT_{\cG} (S,\xi):$};

 \node {$S\to\sigma (S,A)$}
  child {node {$S\to\sigma (A,S)$}
  edge from parent [my dbl]  
  	    child { node[draw] {$A\to C$}
       		 child { node {$C\to\gamma (C)$} 
       		       child { node {$C\to\gamma (C)$}
       		       		 child { node {$C\to\alpha$} }}}}
       child { node[draw] {$S\to B$}
       edge from parent [my dbl]
  	   		 child { node (G2) {$B\to\gamma (B)$} 
  	   		 	   child { node (A2) {$B\to\alpha$} }}}}
  child {node[draw] {$A\to C$}
 	   child { node {$C\to\gamma (C)$} 
 	          child { node {$C\to\gamma (C)$} 
 	                child {node {$C\to\alpha$} }}}} ;
 \node[fit=(G2)(A2),ellipse,draw,inner sep=-2mm, minimum height=26mm] {};

 \node at (-0.8,-7) {$w_d=12$};
\end{scope}
 
\end{tikzpicture}
   
    \caption{\label{fig:ex-wrtg} An illustration of a rule tree $d \in \RT_\cG(\xi)$ for some $\xi \in U$.}
    \end{figure}

  It is clear that, for each $d \in \RT_\cG(B,\T_{\{\gamma,\alpha\}})$ we have $\wt_\cG(d) = \height(\pi_\cG(d))$.
  Moreover, for each $d \in \RT_\cG(C,\T_{\{\gamma,\alpha\}})$ we have $\wt_\cG(d) = 0$. Since each of the rules
   \begin{align*}
    &S \to \sigma(A,S) \ \ \ S \to \sigma(S,A) \ \ \  S \to B\\
    &A \to \sigma(A,A) \ \ \ A \to C
   \end{align*}
   has weight $0$, we obtain the following:
   \begin{equation}
\text{for each $d \in \RT_\cG(\xi)$ we have  $\wt_\cG(d) = \height(\xi|_{w_d})$} \enspace. \label{equ:wt=height}
\end{equation}
Then we have
\begin{align*}
  \sem{\cG}(\xi) &= \infsum{\min}{d \in \RT_\cG(S,\xi)}{\wt_\cG(d)} = \min(\wt_\cG(d) \mid d \in \RT_\cG(\xi)) \\
                 &= \min(\wt_\cG(d_w) \mid w \in \mathrm{bor}(\xi))\\
                 &= \min( \height(\xi|_{w})  \mid w \in \mathrm{bor}(\xi)) \tag{by \eqref{equ:wt=height}}\\
  &= f(\xi)\enspace.  \hspace{80mm} \Box
\end{align*}
\end{example}


  \begin{lemma}\rm \label{lm:fin-der-decidable} For every $(\Sigma,\B)$-wrtg $\cG$ it is decidable whether $\cG$ is finite-derivational.
  \end{lemma}
\begin{proof} Let $\cG= (N,S,R,wt)$ be a $(\Sigma,\B)$-wrtg. By standard methods, we can decide whether there exists a $\xi \in \T_\Sigma$ such that $\RT_\cG(\xi) \not=\emptyset$. If for every  $\xi \in \T_\Sigma$ we have $\RT_\cG(\xi) =\emptyset$, then $\cG$ is finite-derivational.

  Now let us assume that there exists a $\xi \in \T_\Sigma$ with $\RT_\cG(\xi)\not= \emptyset$. The idea is to transform $\cG$ into a context-free grammar and to apply the construction in the proof of Theorem \ref{thm:reduced-cfg} for reducing that context-free grammar. In fact, we view each tree $\xi \in \T_\Sigma(N)$ as a string over the alphabet $\Sigma^\Xi \cup N$. Formally, we construct the $\Sigma^\Xi$-cfg $G=(N\cup \{S_0\},S_0,P)$, where $S_0$ is a new nonterminal and $P=R\cup \{S_0\to A \mid A\in S\}$. (We note that $G$ is not the context-free grammar associated to $\cG$ in the sense of Chapter \ref{chapt:wcfg}.)  Then $\LL(G) \subseteq \T_\Sigma$ and, due to our assumption on $\cG$, we have  $\LL(G) \ne \emptyset$.

  Then we apply the construction (cf. Theorem \ref{thm:reduced-cfg})  to obtain a reduced $\Sigma^\Xi$-cfg $G'=(N',S_0,P')$ such that $G'$ and $G$ are equivalent. Since $G'$ is reduced, each nonterminal in $N'$ occurs in some rule tree $d \in \RT_{G'}(S_0,\Delta^*)$. 

  Due to the construction, we have that $N'\subseteq N$ and $P'\subseteq P$. Then we construct the $(\Sigma,\B)$-wrtg $\cG'=(N',S',R',wt')$, where $S'=S\cap N'$, $R'=P'\setminus \{S_0\to A \mid A\in S'\}$ (hence $R'\subseteq R$),  and $wt' = wt|_{R'}$. Since $G'$ is reduced, the following property holds for $\cG'$:
  \begin{equation}
    \text{for each $A \in N'$ there exist  $\xi \in \T_\Sigma$ and $d \in \RT_{\cG'}(\xi)$ such that $A$ occurs in $d$.}\label{eq:cfg-reduced}
  \end{equation}
  Moreover, $\RT_{\cG'}(\xi)=\RT_\cG(\xi)$ for each $\xi \in \T_\Sigma$. Thus $\cG'$ is  equivalent to $\cG$, and $\cG'$ is finite-derivational if and only if $\cG$ is finite-derivational.

  Since $\cG'$ satisfies \eqref{eq:cfg-reduced} and it is $\varepsilon$-free, it is finite-derivational iff there do not exist $n\in \mathbb{N}_+$ and a sequence $A_1\to A_2,\ldots, A_{n-1}\to A_n$ of chain rules of $\cG'$ such that $A_1=A_n$. This latter property of $\cG'$ is decidable by a standard algorithm on finite graphs \cite{war62}.
\end{proof}

 \section{Normal forms of wrtg}
 \label{sec:restricted-forms}

 Since each $(\Sigma,\B)$-wrtg $\cG$ is a particular $(\Sigma^\Xi,\B)$-wcfg, we can apply the normal form lemmas proved in Chapter \ref{chapt:wcfg} also to $\cG$. Then it is a question whether the wcfg $\cG'$ constructed in the proof of such a lemma is a wrtg or not. Next we tailor those lemmas for wrtg for which the answer to the question is positive.

 \begin{lemma}\rm \label{lm:normal-form-lemmas-inherited-from-wcfg}
   Let $\cG$ be a $(\Sigma,\B)$-wrtg such that $\cG$ is finite-derivational or $\B$ is $\sigma$-complete. Then the following three statements hold.
  \begin{compactenum}
  \item[(1)] We can construct a start-separated $(\Sigma,\B)$-wrtg $\cG'$ such that $\sem{\cG}=\sem{\cG'}$. The construction preserves the properties finite-derivational and local-reduced.  The construction does not preserve the property chain-free.
   \item[(2)] We can construct a local-reduced $(\Sigma,\B)$-wrtg $\cG'$ such that $\sem{\cG}=\sem{\cG'}$. The construction preserves the properties finite-derivational, start-separated, and chain-free.
    \item[(3)] If $\B$ is a semiring, then there exists a chain-free $(\Sigma,\B)$-wrtg $\cG'$ such that $\sem{\cG}=\sem{\cG'}$. If $\cG$ has one of the properties:  start-separated  and local-reduced, then also $\cG'$ has it. Moreover, if $\cG$ is finite-derivational, then we can construct $\cG'$. 
     \end{compactenum}
  \end{lemma}
  \begin{proof}Statement (1) is a corollary of Lemma \ref{lm:wcfg-start-separated} because if $\cG$ is a wrtg, then the construction in the proof of the lemma yields a  wrtg $\cG'$ with the properties stated in (1). Similarly, Statements (2) and (3) are corollaries of  Lemma \ref{lm:t2-wrtg} and Theorem \ref{lm:wrtg-chain-free}, respectively. 
\end{proof}

  \index{weighted regular tree grammar!tree automata form}
  \index{weighted regular tree grammar!alphabetic}
  \index{alphabetic}
We continue with the definition of two more restricted forms of wrtg. Let $\cG=(N,S,R,wt)$ be a $(\Sigma,\B)$-wrtg. We say that $\cG$ is
    \begin{compactitem}
    \item  \emph{alphabetic} if each rule contains at most one occurrence of a symbol in $\Sigma$.
   \item   in \emph{tree automata form} if
\(
  R = \{A \rightarrow \sigma(A_1,\ldots,A_k) \mid k\in \mathbb{N}, \sigma \in \Sigma^{(k)}, A,A_1,\ldots,A_k \in N\}\enspace.
\)
Thus, there exists a bijection from $R$ to the set $\bigcup_{k \in \mathbb{N }} N \times \Sigma^{(k)} \times N^k$.
\end{compactitem}
Obviously, if $\cG$ is in tree automata form, then it is alphabetic and chain-free. 
  
\begin{lemma}\rm \label{lm:rtg-normal-form} (cf.  \cite[Lm.~3.1]{mezwri67}, \cite[Thm.~1]{inafuk75}, and \cite[Prop.~1.2]{aleboz87}) Let $\cG$ be a $(\Sigma,\B)$-wrtg such that $\cG$ is finite-derivational or $\B$ is $\sigma$-complete. We can construct a $(\Sigma,\B)$-wrtg $\cG'$ such that $\cG'$ is alphabetic  and $\sem{\cG'}=\sem{\cG}$. The construction preserves the properties finite-derivational, start-separated, chain-free, and  local-reduced.
\end{lemma}

\begin{proof} As preparation, we define a measurement for the degree of ``how much'' a $(\Sigma,\B)$-wrtg does not match the condition of being alphabetic. Formally, let $\cG= (N,S,R,wt)$ be a $(\Sigma,\B)$-wrtg.  We define, for each rule $r=(A\to \xi)$, the {\em degree  $\mathrm{deg}(r)$ of $r$} by
\[
  \mathrm{deg}(r) = \begin{cases}
      0 & \text{ if } r \text{ is a chain-rule}\\
      |\pos_\Sigma(\xi)|-1 & \text{ otherwise,}
      \end{cases}
    \]
and we define  the \emph{degree $\mathrm{deg}(\cG)$ of $\cG$} by $\mathrm{deg}(\cG)=\sum_{r\in R} \mathrm{deg}(r)$. Then the  following equivalence holds: $\mathrm{deg}(\cG)=0$ iff each rule of $\cG$ contains at most one occurrence of a symbol in $\Sigma$ (i.e., $\cG$ is alphabetic).

Next we consider a reduction system where its reduction transforms a given wrtg into an equivalent one with a lower degree. Formally, we consider the reduction system $(\mathrm{WRTG}(\Sigma,\B),\succ)$ where $\mathrm{WRTG}(\Sigma,\B)$ denotes the set of all $(\Sigma,\B)$-wrtg and, for every $\cG_1=(N_1,S_1,R_1,wt_1)$ and $\cG_2=(N_2,S_2,R_2,wt_2)$ in $\mathrm{WRTG}(\Sigma,\B)$, we let $\cG_1 \succ \cG_2$ if
\begin{compactitem}
\item there exists a rule $r=(A \rightarrow \sigma(\xi_1,\ldots,\xi_{i-1},\xi_i,\xi_{i+1},\ldots,\xi_k))$ in $R_1$ and 
  \item there exists $i \in [k]$ such that $\xi_i \not\in N_1$
\end{compactitem}
such that $N_2 = N_1 \cup \{[\xi_i]\}$, $S_2= S_1$, $R_2 = (R_1 \setminus \{r\}) \cup \{r_1,r_2\}$ where
\[r_1 = (A \rightarrow \sigma(\xi_1,\ldots,\xi_{i-1},[\xi_i],\xi_{i+1},\ldots,\xi_k)) \ \ \text{ and } \ \ r_2 = ([\xi_i] \to \xi_i)\enspace,
\]
and, for each $r' \in R_2$, we let 
\[
wt_2(r') = \begin{cases}wt_1(r') & \text{ if $r' \in R_1 \setminus \{r\}$}\\
  wt_1(r') & \text{ if $r' = r_1$}\\
  \1 & \text{ if $r' = r_2$} \enspace.
\end{cases}
\]
(This generalizes the construction in \cite[Lm.~2.3.4]{gecste84} and \cite[Thm.~3.22]{eng75-15}.) 
Obviously, if  $\mathrm{deg}(\cG_1)>0$, then we can construct a $\cG_2$ such that $\cG_1 \succ \cG_2$.
In this case $\sem{\cG_1} = \sem{\cG_2}$ and if $\cG_1$ is finite-derivational, start-separated, chain-free, or  local-reduced, then so is $\cG_2$.

Next we show that $\succ$ is terminating. For this, we observe that if $\cG_1 \succ \cG_2$, then $\mathrm{deg}(\cG_1) \succ_{\mathbb{N}} \mathrm{deg}(\cG_2)$, because $\mathrm{deg}(\cG_2)=\mathrm{deg}(\cG_1)-1$.
Hence, $\mathrm{deg}$ is a monotone embedding of $(\mathrm{WRTG}(\Sigma,\B),\succ)$ into the terminating reduction system $(\mathbb{N},\succ_{\mathbb{N}})$ (cf. Section \ref{sec:reduction-systems}). Thus, by Lemma \ref{lm:fin-branching-embedding-termination}, the relation $\succ$ is terminating. Moreover, we have 
\[
\nf_\succ(\mathrm{WRTG}(\Sigma,\B)) = \{\cG \in \mathrm{WRTG}(\Sigma,\B) \mid \cG \text{ is alphabetic}\} \enspace.
\]

Now let $\cG$ be an arbitrary $(\Sigma,\B)$-wrtg. We construct an arbitrary element $\cG'$ of $\nf_\succ(\cG)$. Then $\sem{\cG'}=\sem{\cG}$ and $\cG'$ is alphabetic.
\end{proof}

Next we relate wrtg with wta.

\begin{lemma-rect}\rm \label{lm:rtg-automata-form-wrtg}  Let $\Sigma$ be a ranked alphabet. Moreover, let $\B$ be a strong bimonoid and let $\cG$ be a $(\Sigma,\B)$-wrtg. If (a) $\cG$ is chain-free or  (b) $\B$ is a semiring and  $\cG$ is finite-derivational or  (c) $\B$ is a $\sigma$-complete semiring,  then there exists a $(\Sigma,\B)$-wrtg $\cG'$ such that $\cG'$ is in tree automata form and $\sem{\cG'} = \sem{\cG}$. In Cases (a) and (b) we can even construct $\cG'$.
\end{lemma-rect}

\begin{proof} Let $\cG= (N,S,R,wt)$.  In Case (a), by Lemma \ref{lm:rtg-normal-form} we can construct an alphabetic and chain-free  $(\Sigma,\B)$-wrtg which is equivalent to $\cG$; thus we can assume that $\cG$ is  alphabetic and chain-free.

In Cases (b) and (c), by Lemma \ref{lm:normal-form-lemmas-inherited-from-wcfg}(3), there exists a chain-free  $(\Sigma,\B)$-wrtg which is equivalent to $\cG$. Then, by Lemma \ref{lm:rtg-normal-form}, there exists an alphabetic and chain-free $(\Sigma,\B)$-wrtg which is equivalent to $\cG$. Moreover, in Case (b) we can even assume that $\cG$ is alphabetic and chain-free, because then Lemma \ref{lm:normal-form-lemmas-inherited-from-wcfg}(3) is constructive.

Next we complete the set $R$ of rules by adding rules with weight $\mathbb{0}$. Formally, we construct the $(\Sigma,\B)$-wrtg $\cG' = (N,S,R',wt')$ such that $R'$ is the set of all rules $r= (A \rightarrow \sigma(A_1,\ldots,A_k))$ with $k\in \mathbb{N}$, $\sigma \in \Sigma^{(k)}$, and $A,A_1,\ldots,A_k \in N$. Moreover,  for each $r \in R'$ we let $wt'(r) = wt(r)$ if $r \in R$, and  $wt'(r) =\mathbb{0}$ otherwise. Clearly, $\cG'$ is in tree automata form and $\sem{\cG'} = \sem{\cG}$.
\end{proof}

Let $\cA=(Q,\delta,F)$ be a $(\Sigma,\B)$-wta with unit root weights and $\cG=(N,S,R,wt)$ be a $(\Sigma,\B)$-wrtg in tree automata form. We call $\cA$ and $\cG$ \emph{related} if $Q=N$, for each $q \in Q$ we have $F_q=\1$ if and only if $q \in S$, and
\[
\delta_k(q_1\cdots q_k, \sigma,q) = wt(q \to \sigma(q_1,\ldots,q_k))
\]
for every $k \in \mathbb{N}$, $\sigma \in \Sigma^{(k)}$, $q\in Q$, and $q_1\cdots q_k \in Q^k$. We note that, if $q \not\in S$, then $F_q=\0$ because $\cA$ has unit root weights.
Moreover, if $\cA$ and $\cG$ are related, then $\cA$ is local-trim if and only if $\cG$ is reduced.

\begin{example}\label{ex:wta-wrtg-related}\rm We give an example of a wta and a wrtg which are related. As wta, we choose the root weight normalized $(\Sigma,\Nat)$-wta $\cA=(Q,\delta,F)$ of Example \ref{ex:number-of-occurrences}, which we recall here:
  \begin{compactitem}
    \item $\Sigma = \{\sigma^{(2)}, \gamma^{(1)}, \alpha^{(0)}\}$,
\item $Q  = \{\bot, a, f\}$ and  $F_\bot=F_a=0$ and $F_f=1$,
\item for every $q_1,q_2,q \in Q$ we define
\[
\delta_0(\varepsilon,\alpha,q) = 
\left\{
\begin{array}{ll}
1 & \text{if } q \in \{\bot,a\}\\
0 & \text{otherwise}
\end{array}
\right. \ \ 
\delta_1(q_1,\gamma,q) = 
\left\{
\begin{array}{ll}
1 & \text{if } q_1q \in \{\bot\bot,ff\}\\
0 & \text{otherwise}
\end{array}
\right.
\]

\[
\delta_2(q_1q_2,\sigma,q) = 
\left\{
\begin{array}{ll}
1 & \text{if } q_1q_2q \in \{\bot\bot\bot,\bot a f, \bot ff, f\bot f\}\\
0 & \text{otherwise} \enspace.
\end{array}
\right.
\]
\end{compactitem}

The $(\Sigma,\Nat)$-wrtg $\cG=(Q,f,R,wt)$ in tree automata form which is related to $\cA$ is defined by:
\begin{compactitem}
\item $R=\{q \to \theta(q_1,\ldots,q_k) \mid k \in \mathbb{N}, \theta \in \Sigma^{(k)}, q,q_1,\ldots,q_k \in Q\}$ and
  \item for every $q_1,q_2,q \in Q$
\[
wt(q \to \alpha) = 
\left\{
\begin{array}{ll}
1 & \text{if } q \in \{\bot,a\}\\
0 & \text{otherwise}
\end{array}
\right. \ \ 
wt(q \to \gamma(q_1)) = 
\left\{
\begin{array}{ll}
1 & \text{if } q_1q \in \{\bot\bot,ff\}\\
0 & \text{otherwise}
\end{array}
\right.
\]

\[
wt(q \to \sigma(q_1,q_2)) = 
\left\{
\begin{array}{ll}
1 & \text{if } q_1q_2q \in \{\bot\bot\bot,\bot a f, \bot ff, f\bot f\}\\
0 & \text{otherwise}\enspace.
\end{array}
\right.
\]
\end{compactitem}
\hfill $\Box$
  \end{example}

\begin{lemma-rect}\rm \label{lm:related-semantics} Let $\Sigma$ be a ranked alphabet. Moreover, let $\B$ be a strong bimonoid, let $\cA$ be a $(\Sigma,\B)$-wta with unit root weights, and let $\cG$ be a $(\Sigma,\B)$-wrtg in tree automata form. If $\cA$ and $\cG$ are related, then $\runsem{\cA} =\sem{\cG}$.
  \end{lemma-rect}
     \begin{proof} Let $\cA=(Q,\delta,F)$ and $\cG=(N,S,R,wt)$. Since $\cG$ is in tree automata form,  we have $\pos(d)=\pos(\xi)$ for every $\xi \in \T_\Sigma$ and $d \in \RT_\cG(\xi)$.
       Moreover, since $\cA$ and $\cG$ are related, there exists  a one-to-one correspondence between (a) the rules tree of $\cG$ and (b) pairs of $\Sigma$-trees and runs on them.
       
           For the formalization we recall the set $\mathrm{TR}$ defined in Section \ref{sec:basic-defininition-wta} on p.~\pageref{page:run-sem}. 
We have 
\(
\mathrm{TR} = \{(\xi,\rho) \mid \xi \in \T_\Sigma, \rho \in \R_\cA(\xi)\}.
\)
Then we define the mapping $\varphi: \RT_\cG(N,\T_\Sigma) \to \mathrm{TR}$ such that, for every $A \in N$, $\xi \in \T_\Sigma$, and $d \in \RT_\cG(A,\xi)$, we let
    \[\varphi(d) = (\xi,\rho) \]
    where, for each  $w \in \pos(\xi)$, we let $\rho(w) = \lhs(d(w))$ (cf. Figure \ref{fig:from-wrtg-to-wta}).
Hence, $\rho \in \R_{\cA}(A,\xi)$.  Since $\cA$ and $\cG$ are related, $\varphi$ is bijective.

Next we will prove a relationship between the weights of rule trees of $\cG$ and weights of runs of $\cA$  that are related by $\varphi$. For this we use the reduction system $(\RT_{\cG}(N,\T_\Sigma),\succ)$, where we let
   \[
     \succ \ = \ \succ_{R} \cap \ (\RT_{\cG}(N,\T_\Sigma) \times \RT_{\cG}(N,\T_\Sigma)) \enspace.
   \]
    Since $\succ_R$ is terminating, by Lemma \ref{lm:termination-is-subset-closed} also $\succ$ is terminating and $\nf_{\succ}(\RT_{\cG}(N,\T_\Sigma))$ is the set of terminal rules of $R$, which is not empty. Then, by induction on $(\RT_{\cG}(N,\T_\Sigma),\succ)$,  we prove that the following statement holds:
        \begin{equation}
\text{For each $d \in \RT_\cG(N,\T_\Sigma)$, we have: } \wt_\cG(d) = \wt_\cA(\varphi(d))\enspace. \label{eq:rtg-wta}
\end{equation}
Let $d\in \RT_{\cG}(N,\T_\Sigma)$, i.e., $d \in \RT_\cG(A,\xi)$ for some $A\in N$ and $\xi = \sigma(\xi_1,\ldots,\xi_k)$. Since $\pos(\xi) = \pos(d)$, there exists a rule $r= (A \to \sigma(B_1\ldots,B_k))$ in $R$ and for each $i \in [k]$ there exists a rule tree $d_i \in \RT_\cG(B_i,\xi_i)$ such that $d= r(d_1,\ldots,d_k)$. Then
  \begingroup
\allowdisplaybreaks
\begin{align*}
  \wt_\cG(d) &= \bigotimes_{i \in [k]} \wt_\cG(d_i) \otimes wt(A \to \sigma(B_1\ldots,B_k))
  \tag{by \eqref{equ:wt-rules-tree-as-evaluation}}\\
  &= \bigotimes_{i \in [k]} \wt_\cA(\varphi(d_i)) \otimes \delta_k(B_1\cdots B_k,\sigma,A)
    \tag{by I.H. and definition of related}\\
  &= \wt_\cA(\varphi(d)) \tag{by definition of $\varphi$ and \eqref{equ:weight-of-run}} \enspace.
\end{align*}
\endgroup
This proves \eqref{eq:rtg-wta}. 

           Let $\xi \in \T_\Sigma$. Then
           \begingroup
\allowdisplaybreaks
           \begin{align*}
             \runsem{\cA}(\xi)
             =& \bigoplus_{\rho \in R_\cA(\xi)} \wt_\cA(\xi,\rho) \otimes F_{\rho(\varepsilon)}\\
                               =& \bigoplus_{A \in S} \bigoplus_{\rho \in R_{\cA}(A,\xi)} \wt_\cA(\xi,\rho) \tag{because $\cA$ has unit root weights and $F_A=\1$ iff $A \in S$}\\
=& \bigoplus_{A \in S} \bigoplus_{d \in \RT_{\cG}(A,\xi)} \wt_\cA(\xi,\varphi(d))\\
             &\tag{because $\varphi$ is a bijection and $\varphi(\RT_\cG(A,\xi)) = \R_{\cA}(A,\xi)$ for each $A \in S$}\\
             =& \bigoplus_{A \in S} \bigoplus_{d \in \RT_{\cG}(A,\xi)} \wt_\cG(d)
             \tag{by Equation \eqref{eq:rtg-wta}}\\
             =&\sem{\cG}(\xi)\enspace. \qedhere
           \end{align*}
           \endgroup
         \end{proof}

                \begin{figure}
          \centering
\begin{tikzpicture}[level distance=3.5em,
  every node/.style = {align=center}]]
  \pgfdeclarelayer{bg}    
  \pgfsetlayers{bg,main}  

  \newcommand\mydist{4mm}
  \tikzstyle{mycircle}=[draw, circle, inner sep=-2mm, minimum height=5mm]

\begin{scope}[level 1/.style={sibling distance=27mm},
level 2/.style={sibling distance=20mm}]

 \node at (-1.2,0.6) {$d\in \RT_{\cG}(\xi)$:}; 
 
 \node (T1) {$A \to \sigma(B,C)$}
 child {node {$B\to \gamma(D)$}
            child { node {$D \to \alpha$}} }
  child {node {$C \to \beta$} };
\end{scope}

\begin{scope}[xshift=76mm, level 1/.style={sibling distance=37mm},
level 2/.style={sibling distance=27mm}]

 \node at (-1.3, 0.6) {$\xi \in \T_\Sigma$:};
 \node at (2.5, 0.6) {$\rho \in \R_\cA (\xi)$:};

 \node (N0) {$\sigma$}
  child {node (N1) {$\gamma$} 
  	   child { node (N2) {$\alpha$}} }
  child {node (N3) {$\beta$} };

 \node [mycircle] at ([xshift=\mydist]N0.east) {$A$};
 \node [mycircle] at ([xshift=\mydist]N1.east) {$B$};
 \node [mycircle] at ([xshift=\mydist]N2.east) {$D$};
 \node [mycircle] at ([xshift=\mydist]N3.east) {$C$};

\end{scope}

\draw [->, thick, xshift=3.3cm, yshift=-1.2cm] (0,0) -- (1,0) node[midway, above] {$\varphi$};

\end{tikzpicture}

\caption{\label{fig:from-wrtg-to-wta} A visualization of $\varphi(d)=(\xi,\rho)$ in the proof of Lemma \ref{lm:related-semantics} with $\xi = \sigma(\gamma(\alpha),\beta)$, $d \in \RT_\cG(\xi)$, and $\rho \in \R_\cA(A,\xi)$. The states of $\rho$ are circled.}
          \end{figure}

\begin{lemma}\rm \label{lm:wta-to-wrtg} For each $(\Sigma,\B)$-wta $\cA$, we can construct a $(\Sigma,\B)$-wrtg $\cG$ such that $\cG$ is in tree automata form and $\runsem{\cA} = \sem{\cG}$.
\end{lemma}
\begin{proof}  By Theorem \ref{thm:root-weight-normalization-run}, we can construct a  root weight normalized $(\Sigma,\B)$-wta $\cA'$ such that $\runsem{\cA}=\runsem{\cA'}$. Thus, in particular, $\cA'$ has unit root weights. It is obvious how to construct the $(\Sigma,\B)$-wrtg $\cG$ in tree automata form such that $\cA'$ and $\cG$ are related. By Lemma \ref{lm:related-semantics}, we have $\runsem{\cA'} = \sem{\cG}$.
  \end{proof}

  \begin{corollary} \label{cor:wta-are-wcfg} \rm Each r-recognizable $(\Sigma,\B)$-weighted tree language is a $(\Sigma^\Xi,\B)$-weighted context-free language.
  \end{corollary}
  \begin{proof} By Lemma \ref{lm:wta-to-wrtg}, for each  $(\Sigma,\B)$-wta $\cA$ we can construct a $(\Sigma,\B)$-wrtg $\cG$ such that $\cG$ is in tree automata form and $\runsem{\cA} = \sem{\cG}$. Since $\cG$ is a particular $(\Sigma^\Xi,\B)$-wcfg, we obtain the result.
    \end{proof}

\begin{lemma}\rm  \label{lm:wrtg-to-wta} Let $\cG$ be a $(\Sigma,\B)$-wrtg. If
(a) $\cG$ is chain-free or
(b) $\B$ is a semiring and  $\cG$ is finite-derivational or
(c) $\B$ is a $\sigma$-complete semiring,
  then there exists a $(\Sigma,\B)$-wta $\cA$ such that $\sem{\cG}=\runsem{\cA}$.
In Cases (a) and (b) we can even construct $\cA$.
\end{lemma}
\begin{proof} By Lemma \ref{lm:rtg-automata-form-wrtg}, in each of the Cases (a), (b), and (c) there exists a $(\Sigma,\B)$-wrtg $\cG'$ in tree automata form such that $\sem{\cG'}=\sem{\cG}$. In Cases (a) and (b) we can even construct $\cG'$.  Then, it is obvious how to construct a $(\Sigma,\B)$-wta $\cA$ with unit root weights such that $\cA$ and $\cG'$ are related. Finally, by Lemma \ref{lm:related-semantics}, we have $\sem{\cG'} = \runsem{\cA}$.
\end{proof}

Now we can prove the equivalence of $(\Sigma,\B)$-wta and chain-free $(\Sigma,\B)$-wrtg. For the case that $\B$ is the Boolean semiring, this is \cite[Thm.~2.3.6]{gecste84} and \cite[Thm.~3.25]{eng75-15}.

\begin{theorem-rect}\label{thm:rec=reg} Let $\Sigma$ be a ranked alphabet. Moreover, let $\B$ be a strong bimonoid. Then the following two statements hold.
                       \begin{compactenum}
                       \item[(1)] $\Reg_{\mathrm{nc}}(\Sigma,\B) = \Rec^{\mathrm{run}}(\Sigma,\B)$.
                         \item[(2)] If $\B$ is a  semiring, then $\Reg(\Sigma,\B) = \Rec(\Sigma,\B)$. 
                         \end{compactenum}
                       \end{theorem-rect}
                    
                       \begin{proof} First we prove the inclusions from left to right. In Statement (1), the inclusion  follows from Lemma \ref{lm:wrtg-to-wta}(a).
In Statement (2), if $\B$ is not $\sigma$-complete, then $\Reg(\Sigma,\B) = \Reg_{\mathrm{fd}}(\Sigma,\B)$, hence the inclusion follows from Lemma \ref{lm:wrtg-to-wta}(b). In Statement (2), if $\B$ is $\sigma$-complete, then it follows from Lemma \ref{lm:wrtg-to-wta}(c).
The inclusions from right to left follow from Lemma \ref{lm:wta-to-wrtg} and the fact that, if a wrtg  is in tree automata form, then it is both chain-free and finite-derivational.
\end{proof}

Finally, we mention that, for every $(\Sigma,\B)$-wta $\cA$ with unit root weights and $(\Sigma,\B)$-wrtg $\cG$ in tree automata form, if $\cA$ and $\cG$ are related, then the following equivalence holds:
\[
\text{ $\cA$ is local-trim \ iff \ $\cG$ is local-reduced.}
  \]

  The following result generalizes \cite[Thm.~2]{inafuk75} (also cf. \cite[Thm.~3.57]{eng75-15} and \cite[Thm.~3.2.2]{gecste84} for the unweighted case). It says that the weighted rule tree language of each wcfg is a regular weighted tree language. In Lemma \ref{lm:w-rule-trees-are-local} we will prove that each such weighted rule tree language can be determined by some weighted local system.

  \begin{theorem-rect} Let $\Gamma$ be an alphabet and $\B$ a strong bimonoid. Let $\Gamma$ be an alphabet and $\B$ a strong bimonoid. For each $(\Gamma,\B)$-wcfg $\cG$ with rule set $R$, we can construct an $(R,\B)$-wrtg $\cG'$ such that $\sem{\cG'} = \wrtsem{\cG}$.
  \end{theorem-rect}
    \begin{proof} Let $\cG = (N,S,R,wt)$. We consider $R$ as ranked alphabet such that, for each $r \in R$, the rank of $r$ is the number of nonterminals in the right-hand side of $r$. 

    We construct the $(R,\B)$-wrtg $\cG' = (N,S,R',\wt')$ as follows. If $r=(A \to u_0A_1u_1 \cdots A_ku_k)$ is in $R$, then $r' = (A \to r(A_1,\ldots,A_k))$ is in $R'$, and we let $\wt'(r')=\wt(r)$. In the usual way, we also consider $R'$ as ranked alphabet.

    Let $\xi \in \RT_\cG$. We define $d_\xi \in \RT_{\cG'}$ such that $\pos(d_\xi) = \pos(\xi)$ and, for each $w \in \pos(d_\xi)$, if $\xi(w) = r$ and $r=(A \to u_0A_1u_1 \cdots A_ku_k)$, then we let  $d_\xi(w) = (A \to r(A_1,\ldots,A_k))$. Obviously, $\RT_{\cG'}(\xi) = \{d_\xi\}$ and $\wt_{\cG'}(d_\xi)= \wt_\cG(\xi)$.

    Now let $\xi \in \T_R$. If $\xi \not\in \RT_\cG$, then $\RT_{\cG'}(\xi) = \emptyset$ and 
    \[
\sem{\cG'}(\xi) = \infsum{\oplus}{d \in \RT_{\cG'}(\xi)}{\wt_{\cG'}(d)} = \0 = (\wt_\cG \otimes \chi(\RT_\cG))(\xi) = \wrtsem{\cG}(\xi) \enspace.
\]
If $\xi \in \RT_\cG$, then 
    \[
      \sem{\cG'}(\xi) = \infsum{\oplus}{d \in \RT_{\cG'}(\xi)}{\wt_{\cG'}(d)}
      = \wt_{\cG'}(d_\xi) = \wt_\cG(\xi) = (\wt_\cG \otimes \chi(\RT_\cG))(\xi) = \wrtsem{\cG}(\xi) \enspace.
\]
    \end{proof}

%% file: closure-properties.tex
\chapter{Closure properties}
\label{ch:closure-properties}
\index{Rec@$\Rec^{\mathrm{run}}(\_\,,\B)$}
\index{Rec@$\Rec^{\mathrm{run}}(\Sigma,\_)$}
\index{Rec@$\Rec^{\mathrm{init}}(\_\,,\B)$}
\index{Rec@$\Rec^{\mathrm{init}}(\Sigma,\_)$}
In this chapter we prove that the set of recognizable weighted tree languages is closed under several operations.
For some of the closure results it is technically easier first to prove them in the setting of wrtg and second to instantiate them to wta by  using results on the connection between wrtg and wta (cf. Lemmas \ref{lm:wta-to-wrtg} and \ref{lm:wrtg-to-wta}).

Since some of the operations change the ranked alphabet or the strong bimonoid, we define corresponding sets of recognizable weighted tree languages in which these parameters are left open (using the underscore). Formally, we abbreviate by $\Rec^{\mathrm{run}}(\_\,,\B)$ the set of all r-recognizable $(\Sigma,\B)$-weighted tree languages for some ranked alphabet $\Sigma$. Similarly we define the abbreviations $\Rec^{\mathrm{init}}(\_\,,\B)$ and $\Rec(\_\,,\B)$. Moreover, we abbreviate by $\Rec^{\mathrm{run}}(\Sigma,\_)$ the set of all r-recognizable $(\Sigma,\B)$-weighted tree language for some strong bimonoid $\B$. Similarly we define the abbreviations $\Rec^{\mathrm{init}}(\Sigma,\_)$ and $\Rec(\Sigma,\_)$.

\section{Closure under sum}

In this section we will prove that the sets $\Rec^{\mathrm{run}}(\Sigma,\B)$  and   $\Rec^{\mathrm{init}}(\Sigma,\B)$ are closed under sum. A set $\cL$ of $\B$-weighted tree languages is \emph{closed under sum} if the following holds: for every  $(\Sigma,\B)$-weighted tree languages  $r_1$ and $r_2$, if $r_1,r_2 \in \cL$, then $(r_1\oplus r_2) \in \cL$.

\begin{theorem-rect}\label{thm:closure-sum} Let $\Sigma$ be a ranked alphabet. Moreover, let $\B=(B,\oplus,\otimes,\0,\1)$ be a strong bimonoid and let $\cA_1$ and $\cA_2$ be two $(\Sigma,\B)$-wta. Then the following two statements hold.
\begin{compactenum}
\item[(1)] (cf. \cite[Lm.~5.1(1)]{rad10}) We can construct a $(\Sigma,\B)$-wta $\cB$ such that $\runsem{\cB} = \runsem{\cA_1} \oplus  \runsem{\cA_2}$ and  $\initialsem{\cB} = \initialsem{\cA_1} \oplus  \initialsem{\cA_2}$.
  
\item[(2)] If $\cA_1$ and $\cA_2$ are crisp-deterministic, then we can construct a crisp-deterministic $(\Sigma,\B)$-wta $\cB$ such that $\sem{\cB} = \sem{\cA_1} \oplus  \sem{\cA_2}$.
\end{compactenum}
Thus, in particular,  the sets $\Rec^{\mathrm{run}}(\Sigma,\B)$  and   $\Rec^{\mathrm{init}}(\Sigma,\B)$ are closed under sum. 
\end{theorem-rect}
\begin{proof} Let $\cA_1=(Q_1,\delta_1,F_1)$ and $\cA_2=(Q_2,\delta_2,F_2)$ such that $Q_1 \cap Q_2 = \emptyset$.

Proof of (1): We construct the $(\Sigma,\B)$-wta $\cB = (Q,\delta,F)$ as follows:
\begin{compactitem}
\item $Q = Q_1 \cup Q_2$,
\item for every $k \in \mathbb{N}$, $\sigma\in \Sigma^{(k)}$, $q\in Q$, and $q_1 \cdots q_k \in Q^k$ we define
\[
\delta_k(q_1\cdots q_k,\sigma,q) =
\left\{
\begin{array}{ll}
(\delta_1)_k(q_1\cdots q_k,\sigma,q) & \text{ if } q_1, \ldots, q_k,q \in Q_1\\
(\delta_2)_k(q_1\cdots q_k,\sigma,q) & \text{ if } q_1, \ldots, q_k,q \in Q_2\\
\mathbb{0} & \text{ otherwise}\enspace,
\end{array}
\right.\]

\item for each $q \in Q$ we define $F_q = (F_1)_q$ if $q \in Q_1$, and $F_q=(F_2)_q$ if $q \in Q_2$. 
\end{compactitem}

We prove that $\runsem{\cB} = \runsem{\cA_1} \oplus \runsem{\cA_2}$. Let $\xi \in \T_\Sigma$ and $\rho \in \R_\cB(\xi)$. Obviously, $\R_{\cA_1}(\xi) \cap \R_{\cA_2}(\xi)=\emptyset$ and $\R_{\cA_1}(\xi) \cup \R_{\cA_2}(\xi) \subseteq \R_\cB(\xi)$. Moreover, $\wt_\cB(\xi,\rho) = \wt_{\cA_i}(\xi,\rho)$ if $\rho \in \R_{\cA_i}(\xi)$ for each $i\in \{1,2\}$, and $\wt_\cB(\xi,\rho) = \mathbb{0}$ if $\rho \in \R_\cB(\xi) \setminus (\R_{\cA_1}(\xi) \cup \R_{\cA_2}(\xi))$. 
Thus
\begin{align*}
  \runsem{\cB}(\xi) &= \bigoplus_{\rho \in \R_\cB(\xi)} \wt_\cB(\xi,\rho) \otimes F_{\rho(\varepsilon)} \\
                      &= \bigoplus_{\rho \in \R_{\cA_1}(\xi)} \wt_{\cA_1}(\xi,\rho) \otimes (F_1)_{\rho(\varepsilon)} \oplus \bigoplus_{\rho \in \R_{\cA_2}(\xi)} \wt_{\cA_2}(\xi,\rho) \otimes (F_2)_{\rho(\varepsilon)}\\
                      &= \runsem{\cA_1}(\xi) \oplus \runsem{\cA_2}(\xi_2)\enspace.
  \end{align*}

  Next we prove $\initialsem{\cB} = \initialsem{\cA_1} \oplus \initialsem{\cA_2}$. Let $\xi \in \T_\Sigma$. Obviously, for every $q \in Q$,
  \[
    \h_\cB(\xi)_q =
    \begin{cases}
      \h_{\cA_1}(\xi)_q & \text{ if } q \in Q_1\\
      \h_{\cA_2}(\xi)_q & \text{ if } q \in Q_2 \enspace.     
      \end{cases}
    \]
    Thus,
    \begin{align*}
  \initialsem{\cB}(\xi) &= \bigoplus_{q \in Q} \h_\cB(\xi)_q \otimes F_q
                      = \bigoplus_{q \in Q_1} \h_{\cA_1}(\xi)_q \otimes (F_1)_q  \oplus \bigoplus_{q \in Q_2} \h_{\cA_2}(\xi)_q \otimes (F_2)_q\\
                      &= \initialsem{\cA_1}(\xi) \oplus \initialsem{\cA_2}(\xi)\enspace.
  \end{align*}

  \

Proof of (2):
  Let $\cA_1$ and $\cA_2$ be crisp-deterministic.
We construct the $(\Sigma,\B)$ wta $\cB = (Q,\delta,F)$ as follows:
\begin{compactitem}
\item $Q = Q_1 \times Q_2$;  for each $q \in Q$ we denote its first and second component by $q^1$ and $q^2$, respectively,

\item for every $k \in \mathbb{N}$, $\sigma\in \Sigma^{(k)}$, $q\in Q$ and $q_1 \cdots q_k \in Q^k$ we define
\[
\delta_k(q_1\cdots q_k,\sigma,q) =
(\delta_1)_k(q^1_1\cdots q^1_k,\sigma,q^1) \otimes  (\delta_2)_k(q^2_1\cdots q^2_k,\sigma,q^2)
\]

\item for each $q \in Q$ we define $F_q = (F_1)_{q^1} \oplus (F_2)_{q^2}$.
\end{compactitem}
Obviously, $\cB$ is crisp-deterministic.

We recall that, for each $\xi \in \T_\Sigma$, we have $\state_\cB(\xi) = \{\estate_\cB(\xi)\}$, and similarly for $\cA_1$ and $\cA_2$ (see Subsection~\ref{sect:properties-total-bu-det-wta}). 
First, we prove by induction on $\T_\Sigma$ that,
\begin{equation}\label{eq:crisp-direct-product-estate}
\text{for each $\xi \in \T_\Sigma$, we have $\estate_{\cB}(\xi)=(\estate_{\cA_1}(\xi),\estate_{\cA_2}(\xi))$.}
\end{equation}
Let $\xi = \sigma(\xi_1,\ldots,\xi_k)$. Then we calculate as follows
\begingroup
\allowdisplaybreaks
\begin{align*}
&  \hspace{5mm} \{ \estate_\cB(\xi)\} = \state_\cB(\xi)\\[1mm]
  &= \delta_\Q(\sigma)(\state_\cB(\xi_1),\ldots,\state_\cB(\xi_k))
  \tag{because $\state_\cB$ is a $\Sigma$-algebra homomorphism} \\[1mm]
  &= \delta_\Q\big(\sigma)(\{(\estate_{\cA_1}(\xi_1),\estate_{\cA_2}(\xi_1))\},\ldots, \{(\estate_{\cA_1}(\xi_k),\estate_{\cA_2}(\xi_k))\}\big)
  \tag{by I.H.}\\[2mm]
  &= \{(q_1,q_2) \in Q \mid \delta_k\big((\estate_{\cA_1}(\xi_1),\estate_{\cA_2}(\xi_1)) \cdots (\estate_{\cA_1}(\xi_k),\estate_{\cA_2}(\xi_k)),\sigma,(q_1,q_2)\big) = \1\}
                          \tag{by definition of $\delta_Q(\sigma)$}\\
 &= \{(q_1,q_2) \in Q \mid (\delta_1)_k(\estate_{\cA_1}(\xi_1) \cdots \estate_{\cA_1}(\xi_k),\sigma,q_1) \ \otimes\\
  & \hspace*{27mm}     (\delta_2)_k(\estate_{\cA_2}(\xi_1) \cdots \estate_{\cA_2}(\xi_k),\sigma,q_2)= \1\}
                          \tag{by definition of $\delta_k$}\\[2mm]
&= \{(q_1,q_2) \in Q \mid (\delta_1)_k(\estate_{\cA_1}(\xi_1) \cdots \estate_{\cA_1}(\xi_k),\sigma,q_1)= \1 \text{ and }\\
  & \hspace*{27mm}     (\delta_2)_k(\estate_{\cA_2}(\xi_1) \cdots \estate_{\cA_2}(\xi_k),\sigma,q_2)= \1\}
                          \tag{because $\im(\delta_1) \cup \im(\delta_2) \subseteq \{\0,\1\}$ and $\1 \otimes \1 = \1$}\\
&= \{(\estate_{\cA_1}(\xi),\estate_{\cA_2}(\xi))\} \enspace.                          
\end{align*}
\endgroup
This proves \eqref{eq:crisp-direct-product-estate}.
Then we obtain:
\begin{align*}
\sem{\cB}(\xi)
  &=  F_{\estate_{\cB}(\xi)}
  \tag{by Lemma~\ref{lm:fin-algebra-crisp-det-wta}(2)}\\
  &= F_{(\estate_{\cA_1}(\xi),\estate_{\cA_2}(\xi))}
  \tag{by \eqref{eq:crisp-direct-product-estate} }\\
  &= (F_1)_{\estate_{\cA_1}(\xi)} \oplus (F_2)_{\estate_{\cA_2}(\xi)}
  \tag{by definition of $F$}\\
  &= \sem{\cA_1}(\xi) \oplus \sem{\cA_2}(\xi)
    \tag{by Lemma~\ref{lm:fin-algebra-crisp-det-wta}(2)}
    \enspace. 
\end{align*}
\end{proof}

We note that, in the proof of Theorem \ref{thm:closure-sum}(1), we cannot use the product construction of the proof of Theorem \ref{thm:closure-sum}(2), because that  would require the strong bimonoid $\B$ to be  commutative. Also, vice versa, in the proof of Theorem \ref{thm:closure-sum}(2), we cannot use the union construction of the proof of Theorem \ref{thm:closure-sum}(1), because in general that does not preserve bu-determinism: on a nullary symbol the constructed $(\Sigma,\B)$-wta $\cB$ can start with some $q\in Q_1$ or with some $q' \in Q_2$.

Further, we note that in \cite[Thm.~4.1]{gho22} closure of $\Rec^{\mathrm{init}}(\Sigma,\B)$ under sum was proved for the particular case that $\B$ is a $\sigma$-complete orthomodular lattice. Since each $\sigma$-complete orthomodular lattice is a particular strong bimonoid, \cite[Thm.~4.1]{gho22} follows from Theorem \ref{thm:closure-sum} and from \cite[Lm.~5.1(1)]{rad10}.

For the case of bu-deterministic wta we get the following negative result. (Hence \cite[Lm.~5.1(2)]{rad10} is wrong, because the construction of that lemma does not preserve bu-determinism.)

\begin{theorem}\label{thm:bud-rec-not-closed-under-sum}
  There exists a string ranked alphabet $\Sigma$ such that  $\budRec(\Sigma,\Ratnum)$ is not closed under sum.
\end{theorem}
\begin{proof} We let $\Sigma = \{\gamma^{(1)},\alpha^{(0)}\}$. It is easy to define the bu-deterministic $(\Sigma,\Ratnum)$-wta $\cA_1$ and $\cA_2$ over $\Sigma$ and $\mathbb{Q}$ such that $\sem{\cA_1}(\gamma^n\alpha) = 2^n$ and $\sem{\cA_2}(\gamma^n\alpha)=1$ for each $n \in \mathbb{N}$.  Clearly, $\sem{\cA_1} + \sem{\cA_2} = (\exp +1)$ where the weighted tree language $(\exp +1)$ is defined in Example \ref{ex:exp+1}. However, as Theorem \ref{thm:borchard-not-det} shows, $(\exp +1)$ is not in $\budRec(\Sigma,\Ratnum)$.
  \end{proof}


  \section{Closure under scalar multiplications}

In this section we will prove that the sets $\Rec^{\mathrm{init}}(\Sigma,\B)$ and  $\Rec^{\mathrm{run}}(\Sigma,\B)$ are closed under scalar multiplications from the left and from the right. 
A set $\cL$ of $\B$-weighted tree languages is \emph{closed under scalar multiplications from the left} if for every $b \in B$ and $r \in \cL$, the $\B$-weighted tree language $b \cdot r$ is in $\cL$. Similarly we define the concept \emph{closed under scalar multiplications from the right}. A set $\cL$ is \emph{closed under scalar multiplications} if it is closed under scalar multiplications from the left and closed under scalar multiplications from the right.

  \begin{theorem-rect}\label{thm:closure-scalar} {\rm (cf. \cite[Thm.~5.4]{rad10})} Let $\Sigma$ be a ranked alphabet. Moreover, let $\B=(B,\oplus,\otimes,\0,\1)$ be a strong bimonoid, let $\cA$ be a $(\Sigma,\B)$-wta, and let $b \in B$. Then the following four statements hold.
  \begin{compactenum}
  \item[(1)] If (a) $\B$ is left-distributive or (b) $\cA$ is bu-deterministic, then we can construct a $(\Sigma,\B)$-wta $\cB$ such that $\initialsem{\cB} = b \cdot \initialsem{\cA}$ and $\runsem{\cB} = b \cdot \runsem{\cA}$.
    \item[(2)] If $\B$ is right-distributive, then we can construct a $(\Sigma,\B)$-wta $\cB$ such that  $\initialsem{\cB} = \initialsem{\cA} \cdot b$ and $\runsem{\cB} = \runsem{\cA} \cdot b$.
     \item[(3)] If $\cA$ is bu-deterministic, then we can construct a bu-deterministic $(\Sigma,\B)$-wta $\cB$ such that $\sem{\cB} = \sem{\cA} \cdot b$. Moreover, if $\cA$ is crisp-deterministic, then so is $\cB$.
\item[(4)] If $\cA$ is crisp-deterministic, then we can construct a crisp-deterministic $(\Sigma,\B)$-wta $\cB$ such that $\sem{\cB} = b \cdot \sem{\cA}$.
\end{compactenum}
Thus, in particular, if $\B$ is a semiring, then the set $\Rec(\Sigma,\B)$ is closed under scalar multiplications. 
\end{theorem-rect}
\begin{proof} Proof of (1): Let $\cA=(Q,\delta,F)$. We construct the $(\Sigma,\B)$-wta $\cB=(Q',\delta',F')$ such that, if (a) or (b) holds, then $\initialsem{\cB} = b \cdot \initialsem{\cA}$ and $\runsem{\cB} = b \cdot \runsem{\cA}$, using the following idea.
The wta $\cB$ simulates $\cA$ but, at the leftmost leaf of each input tree, it multiplies the weight of the applied transition with $b$ from the left. The information about this multiplication is propagated up by using states of the form $q^\ell$ for each $q \in Q$. Formally, we let
  \begin{compactitem}
  \item $Q' = Q^0 \cup Q^\ell$ where $Q^0 = \{q^0 \mid q \in Q\}$ and $Q^\ell = \{q^\ell \mid q \in Q\}$,
  \item for every $q \in Q$ and $\alpha \in \Sigma^{(0)}$, we let $\delta'_0(\varepsilon,\alpha,q^0) = \delta_0(\varepsilon,\alpha,q)$ and $\delta'_0(\varepsilon,\alpha,q^\ell) = b \otimes \delta_0(\varepsilon,\alpha,q)$,

    for every $k \in \mathbb{N}_+$, $\sigma \in \Sigma^{(k)}$, and $q_1^{s_1},\ldots,q_k^{s_k},q^{s} \in Q'$, we let
    \[\delta'_k(q_1^{s_1} \cdots q_k^{s_k},\sigma,q^{s}) = 
      \begin{cases}
        \delta_k(q_1 \ldots q_k,\sigma,q) &\text{if } \big(s_1=s=\ell \wedge (\forall i \in [2,k]): s_i=0\big)\\
        & \ \ \  \vee \big(s=0 \wedge(\forall i \in [k]): s_i=0  \big)\\
        \0 & \text{otherwise}
        \end{cases}
      \]
      
      \item for every $q \in Q$ we let $(F')_{q^\ell} = F_q$ and $(F')_{q^0} = \0$.
    \end{compactitem}
    We note that this construction does not preserve bu-determinism.

Now we assume that $\B$ is left-distributive or $\cA$ is bu-deterministic.
    First,  we prove that $\initialsem{\cB} = b \cdot \initialsem{\cA}$.    
    For this, by induction on $\T_\Sigma$,  we can  prove that the following statement holds:
    \begin{equation}
      \text{For every  $\xi \in \T_\Sigma$ and $q \in Q$, we have $\h_\cB(\xi)_{q^0} = \h_\cA(\xi)_q$}\enspace.
      \label{equ:scalar-mult-0}
\end{equation}
Since this proof is straightforward, we do not show it.
Next,  by induction on $\T_\Sigma$, we prove that the following statement holds:
  \begin{equation}
\text{For every  $\xi \in \T_\Sigma$ and $q \in Q$, we have $\h_\cB(\xi)_{q^\ell} = b \otimes \h_\cA(\xi)_q$}\enspace.       \label{equ:scalar-mult-ell}
\end{equation}
Let $\xi = \sigma(\xi_1,\ldots,\xi_k)$. Then
\begingroup
\allowdisplaybreaks
\begin{align*}
  \h_\cB(\xi)_{q^\ell} &= \bigoplus_{\substack{q_1 \cdots q_k \in Q^k,\\s_1,\ldots,s_k \in \{0,\ell\}}}
  \h_\cB(\xi_1)_{q_1^{s_1}} \otimes \ldots \otimes \h_\cB(\xi_k)_{q_k^{s_k}} \otimes
  \delta'_k(q_1^{s_1} \cdots q_k^{s_k},\sigma,q^{\ell})\\
  &= \bigoplus_{q_1 \cdots q_k \in Q^k}
  \h_\cB(\xi_1)_{q_1^{\ell}} \otimes \h_\cB(\xi_2)_{q_2^{0}} \otimes \ldots \otimes \h_\cB(\xi_k)_{q_k^{0}} \otimes
    \delta'_k(q_1^{\ell}q_2^0 \cdots q_k^{0},\sigma,q^{\ell})\\
   &= \bigoplus_{q_1 \cdots q_k \in Q^k}
  b \otimes \h_\cA(\xi_1)_{q_1} \otimes  \h_\cA(\xi_2)_{q_2} \otimes \ldots \otimes \h_\cA(\xi_k)_{q_k} \otimes
     \delta_k(q_1\cdots q_k,\sigma,q)
     \tag{by I.H., \eqref{equ:scalar-mult-0}, and construction}
  \end{align*}
  \endgroup

  We continue with case analysis.

\underline{Case  (a):} Let $\B$ be left-distributive. Then we can continue as follows:
  \begingroup
\allowdisplaybreaks
\begin{align*}
    & \bigoplus_{q_1 \cdots q_k \in Q^k}
  b \otimes \h_\cA(\xi_1)_{q_1}  \otimes \ldots \otimes \h_\cA(\xi_k)_{q_k} \otimes
      \delta_k(q_1\cdots q_k,\sigma,q)\\
    = & \ b \otimes  \bigoplus_{q_1 \cdots q_k \in Q^k}
  \h_\cA(\xi_1)_{q_1} \otimes \ldots \otimes \h_\cA(\xi_k)_{q_k} \otimes
        \delta_k(q_1\cdots q_k,\sigma,q)\\
  =& \  b \otimes  \h_\cA(\xi)_{q} \enspace.
  \end{align*}
  \endgroup

 \underline{Case (b):} Let $\cA$ be bu-deterministic. By Lemma \ref{lm:limit-bu-det}(3), there are two subcases.
    
    \underline{Case (b1):} There exists $i \in [k]$ such that $\Qh{\cA}{\xi_i}= \emptyset$, i.e., for each $q \in Q$ we have $\h_\cA(\xi_i)_q=\0$.
      Then we can continue as follows:
  \begingroup
\allowdisplaybreaks
\begin{align*}
    & \bigoplus_{q_1 \cdots q_k \in Q^k}
  b \otimes \h_\cA(\xi_1)_{q_1} \otimes \ldots \otimes \h_\cA(\xi_i)_{q_i} \otimes \ldots \otimes \h_\cA(\xi_k)_{q_k} \otimes
      \delta_k(q_1\cdots q_k,\sigma,q)\\
  =&\ \bigoplus_{q_1 \cdots q_k \in Q^k}
  b \otimes \h_\cA(\xi_1)_{q_1} \otimes \ldots \otimes \0  \otimes \ldots \otimes  \h_\cA(\xi_k)_{q_k} \otimes
     \delta_k(q_1\cdots q_k,\sigma,q)\\
  =& \ \0\\
  =& \ b \otimes \bigoplus_{q_1 \cdots q_k \in Q^k}
 \h_\cA(\xi_1)_{q_1} \otimes \ldots \otimes \0  \otimes \ldots \otimes  \h_\cA(\xi_k)_{q_k} \otimes
     \delta_k(q_1\cdots q_k,\sigma,q)\\
  =& \ b \otimes \h_\cA(\xi)_{q} \enspace.
\end{align*}
\endgroup

\underline{Case (b2):} For each $i \in [k]$, we have $\Qh{\cA}{\xi_i}=\{\estate_\cA(\xi_i\}$ and $\h_\cA(\xi_i)_{\estate_\cA(\xi_i)} \ne \0$. Then we can continue as follows:
  \begingroup
\allowdisplaybreaks
\begin{align*}
    & \bigoplus_{q_1 \cdots q_k \in Q^k}
  b \otimes \h_\cA(\xi_1)_{q_1} \otimes \ldots \otimes \h_\cA(\xi_k)_{q_k} \otimes
      \delta_k(q_1\cdots q_k,\sigma,q)\\
= & \ b \otimes \h_\cA(\xi_1)_{\estate_\cA(\xi_1)} \otimes \ldots \otimes \h_\cA(\xi_k)_{\estate_\cA(\xi_k)} \otimes
      \delta_k(\estate_\cA(\xi_1)\cdots \estate_\cA(\xi_k),\sigma,q) \\
    &  b \otimes \bigoplus_{q_1 \cdots q_k \in Q^k}
  \h_\cA(\xi_1)_{q_1} \otimes \ldots \otimes \h_\cA(\xi_k)_{q_k} \otimes
      \delta_k(q_1\cdots q_k,\sigma,q)\\
  =& \ b \otimes \h_\cA(\xi)_{q} \enspace.
\end{align*}
\endgroup
This finishes the proof of \eqref{equ:scalar-mult-ell}.

Now we prove that $\initialsem{\cB} = b \cdot \initialsem{\cA}$. For each $\xi \in \T_\Sigma$ we have:
\begin{align*}
  \initialsem{\cB}(\xi) = \bigoplus_{\substack{s\in \{0,\ell\} \\q^s \in Q'}} \h_\cB(\xi)_{q^s} \otimes (F')_{q^s}
  = \bigoplus_{q^\ell \in Q^\ell} \h_\cB(\xi)_{q^\ell} \otimes F_{q^\ell}
   = \bigoplus_{q \in Q} b \otimes \h_\cA(\xi)_{q} \otimes F_{q}\enspace,
\end{align*}
where the last equality follows from \eqref{equ:scalar-mult-ell} and the definition of $\cB$.

We proceed by case analysis as above.

\underline{Case (a):} Let $\B$ be left-distributive. Then we can continue as follows:
\begin{align*}
  \bigoplus_{q \in Q} b \otimes \h_\cA(\xi)_{q} \otimes F_{q}
  =   b \otimes  \bigoplus_{q \in Q}\h_\cA(\xi)_{q} \otimes F_{q}
  = b \otimes \initialsem{\cA}(\xi)
  = (b \cdot \initialsem{\cA})(\xi) \enspace.
\end{align*}

\underline{Case (b):} Let $\cA$ be bu-deterministic. Again, by Lemma \ref{lm:limit-bu-det}(3), there are two cases.

\underline{Case (b1):} $\Qh{\cA}{\xi}=\emptyset$, i.e., for each $q \in Q$ we have $\h_\cA(\xi)_q=\0$. By Lemma~\ref{obs:state-properties-of-wta}(2), we have 
$\initialsem{\cA}(\xi)=\0$, thus we can continue as follows:
\begin{align*}
  \bigoplus_{q \in Q} b \otimes \h_\cA(\xi)_{q} \otimes F_{q}
  = \0 =  b \otimes \initialsem{\cA}(\xi)
   = (b \cdot \initialsem{\cA})(\xi)\enspace.
\end{align*}

\underline{Case (b2):} The set $\Qh{\cA}{\xi}=\{\estate_\cA(\xi)\}$ and $\h_\cA(\xi)_{\estate_\cA(\xi)}\ne\0$.
Then we can continue as follows:
\begingroup
\allowdisplaybreaks
\begin{align*}
  \bigoplus_{q \in Q} b \otimes \h_\cA(\xi)_{q} \otimes F_{q}
  & = b \otimes \h_\cA(\xi)_{\estate_{\cA}(\xi)} \otimes F_{\estate_{\cA}(\xi)}\\
 & =  b \otimes \initialsem{\cA}(\xi) \tag{by Corollary~\ref{lm:properties-hA-of-budet-wta-det-new-H}} \\
  &  = (b \cdot \initialsem{\cA})(\xi)\enspace.
\end{align*}
\endgroup
This finishes the proof of $\initialsem{\cB} = b \cdot \initialsem{\cA}$.

\

Second, we prove that $\runsem{\cB} = b \cdot \runsem{\cA}$. For each $\xi \in \T_\Sigma$ we define the sets
\begin{align*}
  \pos_{\mathrm{left}}(\xi) &= \{w \in \pos(\xi) \mid w = 1^n \text{ for some $n \in \mathbb{N}$}\},\\
  \R_\cB^0(\xi) &= \{\rho \in \R_\cB(\xi) \mid \im(\rho) \subseteq Q^0\}, \text{ and }\\
  \R_\cB^\ell(\xi) &= \{\rho \in \R_\cB(\xi) \mid (\forall w \in \pos_{\mathrm{left}}(\xi)): \rho(w) \in Q^\ell \text{ and } (\forall w \in \pos(\xi)\setminus \pos_{\mathrm{left}}(\xi)): \rho(w) \in Q^0\} \enspace.
\end{align*}
Moreover, for each $\xi \in \T_\Sigma$, we define the mapping $\overline{(.)}: \R_\cB(\xi) \to \R_\cA(\xi)$ such that, for every $\rho \in \R_\cB(\xi)$, we abbreviate $\overline{(.)}(\rho)$ by $\overline{\rho}$ and  the run $\overline{\rho}$ of $\cA$ is obtained from $\rho$ by dropping the upper index (in $\{0,\ell\}$) from each state. 

  By the construction of $\cB$, the following observations are easy to see.
  \begin{equation}
\text{For every $\xi \in \T_\Sigma$ and $\rho \in \R_\cB(\xi)\setminus (\R_\cB^0(\xi) \cup \R_\cB^\ell(\xi))$ we have $\wt_\cB(\xi,\rho)=\0$. } \label{equ:not-left-0}
    \end{equation}
    \begin{equation}
\text{For every $\xi \in \T_\Sigma$ and $\rho \in \R_\cB^0(\xi)$ we have $\wt_\cB(\xi,\rho)=\wt_\cA(\xi,\overline{\rho})$. } \label{equ:0-equal}
\end{equation}
   \begin{equation}
\text{For every $\xi \in \T_\Sigma$ and $\rho \in \R_\cB^\ell(\xi)$ we have $\wt_\cB(\xi,\rho)= b \otimes \wt_\cA(\xi,\overline{\rho})$. } \label{equ:l-equal-b}
\end{equation}
For the proof of \eqref{equ:l-equal-b} we need \eqref{equ:not-left-0}, and \eqref{equ:0-equal}.
Then for each $\xi \in \T_\Sigma$ we can compute as follows.
\begingroup
\allowdisplaybreaks
\begin{align*}
  \runsem{\cB}(\xi)
  &= \bigoplus_{\rho \in \R_\cB(\xi)} \wt_\cB(\xi,\rho) \otimes (F')_{\rho(\varepsilon)}\\
  &= \bigoplus_{\substack{\rho \in \R_\cB(\xi):\\\rho(\varepsilon)\in \Q^\ell}} \wt_\cB(\xi,\rho) \otimes (F')_{\rho(\varepsilon)} \tag{because $(F')_{q^s}=\0$ for each $q^s \in Q^0$}\\
  &= \bigoplus_{\substack{\rho \in \R_\cB(\xi):\\\rho(\varepsilon)\in \Q^\ell}} \wt_\cB(\xi,\rho) \otimes F_{\overline{\rho}(\varepsilon)}
  \tag{by definition of $F'$}\\
    &= \bigoplus_{\rho \in \R^\ell_\cB(\xi)} \wt_\cB(\xi,\rho) \otimes F_{\overline{\rho}(\varepsilon)}
      \tag{by \eqref{equ:not-left-0}}\\
    &= \bigoplus_{\rho \in \R^\ell_\cB(\xi)} b \otimes \wt_\cA(\xi,\overline{\rho}) \otimes F_{\overline{\rho}(\varepsilon)}
      \tag{by \eqref{equ:l-equal-b}}\\
    &= \bigoplus_{\rho \in \R_\cA(\xi)} b \otimes \wt_\cA(\xi,\rho) \otimes F_{\rho(\varepsilon)}
      \tag{because the restriction of $\overline{(.)}$ to  $\R_\cB^\ell(\xi)$ is a bijection}
\end{align*}
\endgroup

We proceed by case analysis.

\underline{Case (a):} Let $\B$ be left-distributive. Then we can continue as follows:
\begin{align*}
\bigoplus_{\rho \in \R_\cA(\xi)} b \otimes \wt_\cA(\xi,\rho) \otimes F_{\rho(\varepsilon)} 
   = b \otimes  \bigoplus_{\rho \in \R_\cA(\xi)} \wt_\cA(\xi,\rho) \otimes F_{\rho(\varepsilon)}      
  =  b \otimes \runsem{\cA}(\xi)
   =  (b \cdot \runsem{\cA})(\xi)\enspace,
\end{align*}
where we use left-distributivity at the first equality.

\underline{Case (b):} Let $\cA$ be bu-deterministic. By Lemma~\ref{lm:limit-bu-det}(3), there are  two cases.

 \underline{Case (b1):} $\QR{\cA}{\xi}=\emptyset$, i.e., for each $\rho \in \R_\cA(\xi)$ we have $\wt_\cA(\xi,\rho)=\0$. 
 By Lemma~\ref{obs:state-properties-of-wta}(3) we have $\runsem{\cA}(\xi)=\0$, thus we can continue as follows:
\begingroup
\allowdisplaybreaks
 \begin{align*}
  \bigoplus_{\rho \in \R_\cA(\xi)} b \otimes \wt_\cA(\xi,\rho) \otimes F_{\rho(\varepsilon)} =  \bigoplus_{\rho \in \R_\cA(\xi)} b \otimes \0 \otimes  F_{\rho(\varepsilon)} =  \0 
      =  b \otimes \runsem{\cA}(\xi) = (b \cdot \runsem{\cA})(\xi) \enspace.
 \end{align*}
 \endgroup
 
\underline{Case (b2):} $\QR{\cA}{\xi}=\{\estate_\cA(\xi)\}$. Thus, the canonical run $\crhorm_\xi$ is the only run on $\xi$ with non-zero weight, and $\wt_\cA(\xi,\crhorm_\xi) = \h_\cA(\xi)_{\estate_\cA(\xi)}$.
 (Note that $\crhorm_\xi(\varepsilon)=\estate_\cA(\xi)$.) Then we continue as follows:
\begin{align*}
  \bigoplus_{\rho \in \R_\cA(\xi)} b \otimes \wt_\cA(\xi,\rho) \otimes F_{\rho(\varepsilon)}
  &= b \otimes \wt_\cA(\xi,\crhorm_\xi) \otimes F_{\estate_\cA(\xi)}\\
  &= b \otimes \runsem{\cA}(\xi)  \tag{by Lemma~\ref{lm:limit-bu-det}(3)(b)} \\
  & = (b \cdot \runsem{\cA})(\xi) \enspace.
\end{align*}
This finishes the proof of $\runsem{\cB} = b \cdot \runsem{\cA}$.

\


Proof of (2): Let $\cA=(Q,\delta,F)$. We construct $\cB=(Q,\delta,F')$ by letting $(F')_q = F_q \otimes b$ for each $q \in Q$.

First, we prove that $\initialsem{\cB} = \initialsem{\cA} \cdot b$.
Obviously,
\begin{equation}\label{equ:ha=hb}
\text{for each $q \in Q$ we have } \h_\cB(\xi)_q = \h_\cA(\xi)_q \enspace. 
\end{equation}
(We note that the proof of \eqref{equ:ha=hb} does not need the assumption that $\B$ is right-distributive.)
Now we assume that $\B$ is right-distributive.

 Then, for each $\xi \in \T_\Sigma$, we can calculate as follows.
  \begingroup
\allowdisplaybreaks
\begin{align*}
  \initialsem{\cB}(\xi) 
  &= \bigoplus_{q \in Q} \h_\cB(\xi)_{q} \otimes (F')_{q}\\
  &= \bigoplus_{q \in Q} \h_\cA(\xi)_{q} \otimes F_{q} \otimes b \tag{by \eqref{equ:ha=hb} and the definition of $\cB$}\\
  &= \big(\bigoplus_{q \in Q} \h_\cA(\xi)_{q} \otimes F_{q}\big) \otimes b \tag{by right-distributivity}\\
  &= \initialsem{\cA}(\xi) \otimes b = (\initialsem{\cA} \cdot b)(\xi) \enspace.
\end{align*}
\endgroup

Second, we prove that $\runsem{\cB} = \runsem{\cA} \cdot b$. Obviously,
\begin{eqnarray}
  \begin{aligned}
  &\text{for each $\xi \in \T_\Sigma$ we have $\R_\cA(\xi)=\R_\cB(\xi)$  and}\\
  &\text{for each $\rho \in \R_\cA(\xi)$ we have $\wt_\cA(\xi,\rho) = \wt_\cB(\xi,\rho)$}\enspace.
  \end{aligned} \label{equ:ha=hb-run}
\end{eqnarray}

Then, for each $\xi \in \T_\Sigma$, we can calculate as follows.
  \begingroup
\allowdisplaybreaks
\begin{align*}
  \runsem{\cB}(\xi) &= \bigoplus_{\rho \in \R_\cB(\xi)} \wt_\cB(\xi,\rho) \otimes (F')_{\rho(\varepsilon)}\\
                    &= \bigoplus_{\rho \in \R_\cA(\xi)} \wt_\cA(\xi,\rho) \otimes F_{\rho(\varepsilon)} \otimes b  \tag{by \eqref{equ:ha=hb-run}}\\
                    &= \big(\bigoplus_{\rho \in \R_\cA(\xi)} \wt_\cA(\xi,\rho) \otimes F_{\rho(\varepsilon)}\big) \otimes b
  \tag{by right-distributivity} \\
  &= \runsem{\cA}(\xi) \otimes b = (\runsem{\cA} \cdot b)(\xi)  \enspace.
\end{align*}
\endgroup

\


Proof of (3): Let $\cA=(Q,\delta,F)$ be bu-deterministic. We construct $\cB=(Q,\delta,F')$ in the same way as in the proof of Statement~(2). It is obvious that this construction preserves bu-determinism and crisp-determinism.
Let $\xi \in \T_\Sigma$. By Lemma \ref{lm:limit-bu-det}(3), there are  two cases.
  
\underline{Case (a):} $\Qh{\cA}{\xi}=\emptyset$, i.e., for each $q \in Q$ we have $\h_\cA(\xi)_q=\0$. By Lemma~\ref{obs:state-properties-of-wta}(2), we have 
$\sem{\cA}(\xi)=\0$. Thus, using  \eqref{equ:ha=hb}, we obtain:
\begin{align*}
      \sem{\cB}(\xi)
      =   \bigoplus_{q \in Q} \h_\cB(\xi)_{q} \otimes (F')_{q}
  = \0 =  \sem{\cA}(\xi) \otimes b = (\sem{\cA} \cdot b)(\xi) \enspace.
    \end{align*}

\underline{Case (b):} $\Qh{\cA}{\xi}= \{\estate_\cA(\xi)\}$ and $\h_\cA(\xi)_{\estate_\cA(\xi)} \ne \0$.  Then  we can continue as follows:
\begingroup
\allowdisplaybreaks
\begin{align*}
  \sem{\cB}(\xi) &=  \h_\cB(\xi)_{\estate_\cB(\xi)} \otimes F'_{\estate_\cB(\xi)} 
\tag{by Corollary~\ref{lm:properties-hA-of-budet-wta-det-new-H}}\\
  &= \h_\cA(\xi)_{\estate_\cA(\xi)} \otimes F_{\estate_\cA(\xi)} \otimes b
\tag{by \eqref{equ:ha=hb}, by $\estate_\cA(\xi)= \estate_\cB(\xi)$, and by definition of $F'$}\\[1mm]
  &= \sem{\cA}(\xi)  \otimes b
\tag{by Corollary~\ref{lm:properties-hA-of-budet-wta-det-new-H}}\\
  &= (\sem{\cA}  \cdot b)(\xi) \enspace.
\end{align*}
\endgroup

\

Proof of (4). Let $\cA=(Q,\delta,F)$ be crisp-deterministic. We construct $\cB=(Q,\delta,F')$  where $F'_q = b \otimes F_q$ for each $q \in Q$. 
  Obviously, $\cB$ is crisp-deterministic and we have $\state_{\cA}=\state_{\cB}$. Thus, by Lemma~\ref{lm:fin-algebra-crisp-det-wta}(2), we obtain
\[
  \sem{\cB}(\xi) =  (F')_{\estate_{\cB}(\xi)}
  = b \otimes F_{\estate_{\cA}(\xi)}
  = b \otimes \sem{\cA}(\xi)
  = (b \cdot \sem{\cA})(\xi)\enspace. \qedhere\]
  \end{proof}


  \section{Characterization of recognizable step mappings}
  \label{sec:recog-step-mapping}

As a kind of application of closure under sum and under scalar multiplications, we show that recognizable step mappings and  crisp-deterministic wta are closely related (cf. Theorem \ref{thm:crisp-det-algebra}). Moreover, we prove a characterization of the set of all recognizable step mappings in terms of characteristic mappings of recognizable tree languages and closure under sum and scalar multiplication (cf. Theorem~\ref{thm:RecStep-smallest}). 

We recall that a weighted tree language $r: \T_\Sigma \rightarrow B$ is a recognizable step mapping  (cf. Section \ref{sec:rec-step-mappings}) if there exist $n \in \mathbb{N}_+$, $b_1,\ldots,b_n \in B$, and recognizable $\Sigma$-tree languages $L_1,\ldots,L_n$ such that 
\begin{equation}\nonumber
r = \bigoplus_{i \in [n]} b_i \cdot \chi(L_i)\enspace.
\end{equation}

In the next theorem, the equivalences (A)$\Leftrightarrow$(B)$\Leftrightarrow$(C)   were proved in \cite[Lm.~8 and Prop. 9]{drostuvog10}, respectively, for the string case. The implication (B)$\Rightarrow$(D) of Theorem~\ref{thm:crisp-det-algebra} was proved in \cite[Lm.~3.1]{drovog06} for the case that $\B$ is a semiring. A similar result was proved in \cite[Thm.~2.1]{li08a} for wsa over a pair of t-conorm and t-norm.
We recall that a weighted tree language $r: \T_\Sigma \to B$ has the preimage property, if $r^{-1}(b)$ is a recognizable $\Sigma$-tree language for each $b \in B$.

\begin{theorem-rect} \label{thm:crisp-det-algebra} Let $\Sigma$ be a ranked alphabet. Moreover, let $\B=(B,\oplus,\otimes,\0,\1)$ be a strong bimonoid and let $r\colon \T_\Sigma \rightarrow B$. Then the following four statements are equivalent. 
\begin{compactenum}
\item[(A)] We can construct a crisp-deterministic $(\Sigma,\B)$-wta $\cA$ such that $r = \sem{\cA}$.
\item[(B)] $r$ is a recognizable step mapping  and we can construct $n\in \mathbb{N}_+$, $b_1,\ldots,b_n \in B$, and $\Sigma$-fta $A_1,\ldots,A_n$ such that $r = \bigoplus_{i\in [n]} b_i \cdot \chi(\LL(A_i))$.
\item[(C)] $\im(r)$ is finite, $r$ has the preimage property, and we can construct $n\in \mathbb{N}_+$ and elements $b_1,\ldots,b_n \in B$ such that
$\im(r)=\{b_1,\ldots,b_n\}$ and, moreover, for each $b\in B$ we can construct  a $\Sigma$-fta which recognizes $r^{-1}(b)$.
\item[(D)] $r$ is a recognizable step mapping with pairwise disjoint step languages and  we can construct $n\in \mathbb{N}_+$, $b_1,\ldots,b_n \in B$, and $\Sigma$-fta $A_1,\ldots,A_n$ such that $r = \bigoplus_{i\in [n]} b_i \cdot \chi(\LL(A_i))$.
\end{compactenum}
Thus, in particular, $\cdRec(\Sigma,\B)=\RecStep(\Sigma,\B)$.
\end{theorem-rect}
\begin{proof}
  Proof of (A)$\Rightarrow$(D): Let $\cA$ be a crisp-deterministic $(\Sigma,\B)$-wta such that $r = \sem{\cA}$. By Theorem~\ref{thm:crisp-wta-final-index}(B)$\Rightarrow$(A), we can construct a finite $\Sigma$-algebra $\A=(Q,\theta)$ and a mapping $F: Q \rightarrow B$ such that $\sem{\cA} = F \circ h_\A$.

Let $n=|Q|$ and let $p_1,\ldots,p_n$ be a list of the elements of $Q$. For each $i \in [n]$ we construct the bu-deterministic and total  $\Sigma$-fta $A_i=(Q,\delta,\{p_i\})$, where for every $k \in \mathbb{N}$, $\sigma \in \Sigma^{(k)}$, and  $q\in Q$ and $q_1\cdots q_k \in Q^k$, we have
\[
\delta_k(q_1\cdots q_k,\sigma)=q  \ \text{ iff } \ \theta(\sigma)(q_1,\ldots,q_k) = q\enspace.
\] 
Since each fta $A_i$ is bu-deterministic, it is obvious that,  for every $i,j \in [n]$  with $i\not= j$, the sets $\LL(A_i)$ and $\LL(A_j)$ are disjoint, i.e.,  $\LL(A_i) \cap \LL(A_j) = \emptyset$.   

Let us abbreviate $F_{p_i}$ by $b_i$ for every $i\in[n]$. Next we prove that $r = \bigoplus_{i \in [n]} b_i \cdot \chi(\LL(A_i))$. Obviously, for every $i \in [n]$, we have  $\h_\A = \h_{A_i}$, where $\h_{A_i}$ is the unique homomorphism from $\sfT_\Sigma$ to the $\Sigma$-algebra $(\cP(Q),\delta_{A_i})$ defined in Section \ref{sec:fta}.

Now let $\xi \in \T_\Sigma$ and let $\h_\A(\xi)=p_j$. Then,  in particular, $\h_\A = \h_{A_{j}}$.
By definition of $A_{j}$ it follows that $\xi \in \LL(A_{j})$, and thus $\chi(\LL(A_{j}))(\xi) = \mathbb{1}$. Moreover, for each $p_{i} \in Q$ with $p_{i}\not= \h_\A(\xi)$, we have $\xi \not\in \LL(A_{i})$ and thus $\chi(\LL(A_{i}))(\xi) = \mathbb{0}$.   Thus  we have
\[
r(\xi) = (F \circ \h_\A)(\xi) = F_{p_j} = b_j \otimes \chi(\LL(A_{j}))(\xi) = \bigoplus_{i \in [n]} b_i \otimes \chi(\LL(A_i))(\xi) = \big(\bigoplus_{i \in [n]} b_i \cdot \chi(\LL(A_i))\big)(\xi)\enspace.
\]
Hence $r$ is a recognizable step mapping with pairwise disjoint step languages.

\

Proof of (D)$\Rightarrow$(C):  For each $i\in[n]$, let us abbreviate $\LL(A_{i})$ by $L_i$. Hence $r = \bigoplus_{i\in [n]} b_i \cdot \chi(L_i)$, where the step languages $L_1,\ldots, L_n$ are pairwise disjoint.
Then $\im(r)=\{b_1,\ldots,b_n\}$ if $\T_\Sigma = \bigcup_{i\in [n]} L_i$ and  $\im(r)=\{b_1,\ldots,b_n,\mathbb{0}\}$ if $\T_\Sigma \setminus  \bigcup_{i\in [n]} L_i\ne \emptyset$; in both cases $\im(r)$ is finite.

Now let $b\in B$. Then
\[
  r^{-1}(b) =
  \begin{cases}
    \bigcup_{i \in [n]: b_i=b}L_i & \text{if } b\ne \0 \\
    \T_\Sigma \setminus \bigcup_{i \in [n]: b_i \ne \0}L_i & \text{ otherwise.}
    \end{cases}
  \]
  Since the set of recognizable $\Sigma$-tree languages is effectively closed under union and complement \cite[Thm.~2.4.2]{gecste84}, in each case we can construct a $\Sigma$-fta which recognizes $r^{-1}(b)$. Thus $r$ has the preimage property.

  \
  
Proof of (C)$\Rightarrow$(B): Let  $\im(r)=\{b_1,\ldots,b_n\}$ for some  $n\in \mathbb{N}_+$ and $b_1,\ldots,b_n\in B$. By our assumption, for each $i\in[n]$, 
we can construct a $\Sigma$-fta $A_i$ with $\LL(A_i)=r^{-1}(b_i)$. Then $r = \bigoplus_{i\in [n]} b_i \cdot \chi(\LL(A_i))$.  Hence $r$ is a recognizable step mapping.

\

Proof of (B)$\Rightarrow$(A):  Let $n \in \mathbb{N}_+$ and  $r = \bigoplus_{i\in [n]} b_i  \cdot \chi(\LL(A_i))$ for some $\Sigma$-fta $A_1,\ldots,A_n$ and coefficients $b_1,\ldots,b_n \in B$.

Let  $i \in [n]$. By Theorem \ref{thm:fta-wta}, we can construct a crisp-deterministic $(\Sigma,\B)$-wta $\cA_i$ with unit root weights such that $\sem{\cA_i}= \chi(\LL(A_i))$. By Theorem \ref{thm:closure-scalar}(4), we can construct a crisp-deterministic $(\Sigma,\B)$-wta $\cA_i'$ such that  $\sem{\cA_i'} = b_i \cdot  \chi(\LL(A_i))$. By Theorem \ref{thm:closure-sum}(2), we can construct a crisp-deterministic $(\Sigma,\B)$-wta $\cA$ such that $\sem{\cA}= \bigoplus_{i\in [n]} b_i  \cdot \chi(\LL(A_i))$.
\end{proof}

We note that Theorem \ref{thm:crisp-det-algebra}((B)$\Rightarrow$(D))  also follows from Observation \ref{obs:partition-step}.

\begin{corollary}\rm \label{cor:RecStep-closed-sum-scalar} $\RecStep(\Sigma,\B)$ is closed under sum and scalar multiplications.
\end{corollary}
\begin{proof} By Theorem \ref{thm:crisp-det-algebra}(A)$\Leftrightarrow$(B), $\RecStep(\Sigma,\B)$ is equal to the set of $(\Sigma,\B)$-weighted tree languages recognizable by crisp-deterministic wta.
By Theorem \ref{thm:closure-sum}(2) the latter is closed under sum. Moreover, by Theorem \ref{thm:closure-scalar}(3) and (4) it is closed under scalar multiplications from the right and the left, respectively.
\end{proof}

We finish this section with a characterization of  $\RecStep(\Sigma,\B)$ which results from several closure properties.

\begin{theorem-rect} \label{thm:RecStep-smallest} Let $\Sigma$ be a ranked alphabet. Moreover, let $\B$ be a strong bimonoid. The set $\RecStep(\Sigma,\B)$ of $(\Sigma,\B)$-recognizable step mappings is the smallest set of $(\Sigma,\B)$-weighted tree languages which contains,  for each recognizable $\Sigma$-tree language~$L$, the characteristic mapping $\chi(L)$ and is closed under sum and scalar multiplication from the left. 
\end{theorem-rect}
\begin{proof} For the sake of convenience, we denote by $\cC$ the smallest set of $(\Sigma,\B)$-weighted tree languages which contains,  for each recognizable $\Sigma$-tree language~$L$, the characteristic mapping $\chi(L)$ and is closed under sum and scalar multiplications from the left. 

  First we prove that $\RecStep(\Sigma,\B) \subseteq \cC$.  
 Let $r\in \RecStep(\Sigma,\B)$ with  $r = \bigoplus_{i\in[n]} b_i \cdot \chi(L_i)$ for some  $n \in \mathbb{N}_+$, $b_1,\ldots,b_n \in B$, and $L_1,\ldots,L_n$ recognizable $\Sigma$-tree languages. Then, for each $i\in [n]$, we have $\chi(L_i) \in \cC$ and $b_i \cdot \chi(L_i) \in \cC$. Since $\cC$ is closed under sum, we have $r \in \cC$.

  Second we prove that $\cC \subseteq \RecStep(\Sigma,\B)$. Obviously, for each recognizable $\Sigma$-tree language $L$, the characteristic mapping $\chi(L)$ is in $\RecStep(\Sigma,\B)$. Moreover, by Corollary 
\ref{cor:RecStep-closed-sum-scalar}, $\RecStep(\Sigma,\B)$  is closed under sum and scalar multiplications from the left. Since $\cC$ is the smallest set with these properties, we obtain that $\cC \subseteq \RecStep(\Sigma,\B)$.
  \end{proof}


  \section{Closure under Hadamard product}
  \label{sec:closure-Hadamard-product}

In this section we will prove that the set $\Rec(\Sigma,\B)$ is closed under Hadamard product if  $\B$ is a commutative semiring. A set $\cL$ of $\B$-weighted tree languages is \emph{closed under Hadamard product} if the following holds: for every  $(\Sigma,\B)$-weighted tree languages  $r_1$ and $r_2$, if $r_1,r_2 \in \cL$, then $(r_1 \otimes r_2) \in \cL$.

\begin{theorem-rect}\label{thm:closure-Hadamard-product} Let $\Sigma$ be a ranked alphabet. Moreover, let $\B=(B,\oplus,\otimes,\0,\1)$ be a strong bimonoid and let $\cA_1$ and $\cA_2$ be two $(\Sigma,\B)$-wta. Then the following three statements hold.
  \begin{compactenum}
  \item[(1)]  (cf. \cite[Cor.~3.9]{bor04}) If $\B$ is a commutative semiring, then we can construct a $(\Sigma,\B)$-wta $\cA$ such that  $\sem{\cA} = \sem{\cA_1} \otimes  \sem{\cA_2}$.
  \item[(2)] (cf. \cite[Lm.~5.3]{rad10}) If $\B$ is commutative and $\cA_1$ and $\cA_2$ are bu-deterministic,  then we can construct a bu-deterministic $(\Sigma,\B)$-wta $\cA$ such that  $\sem{\cA} = \sem{\cA_1} \otimes  \sem{\cA_2}$.
  \item[(3)] If $\cA_1$ and $\cA_2$ are crisp-deterministic,  then we can construct a crisp-deterministic $(\Sigma,\B)$-wta $\cA$ such that  $\sem{\cA} = \sem{\cA_1} \otimes  \sem{\cA_2}$.
  \end{compactenum}
  Thus, in particular, the set $\RecStep(\Sigma,\B)$ is closed under Hadamard product. Moreover, if $\B$ is a commutative semiring, then the set $\Rec(\Sigma,\B)$ is closed under Hadamard product.
\end{theorem-rect}

\begin{proof}  Let $\cA_1=(Q_1,\delta_1,F_1)$ and $\cA_2=(Q_2,\delta_2,F_2)$. We construct the $(\Sigma,\B)$-wta $\cA = (Q,\delta,F)$ which we will use in the proof of each of the Statements (1), (2), and (3).
We define:
\begin{compactitem}
\item $Q = Q_1 \times Q_2$;  for each $q \in Q$ we denote its first and second component by $q^1$ and $q^2$, respectively,

\item for every $k \in \mathbb{N}$, $\sigma\in \Sigma^{(k)}$, $q\in Q$, and $q_1\cdots q_k \in Q^k$ we define
\[
\delta_k(q_1\cdots q_k,\sigma,q) =
(\delta_1)_k(q^1_1\cdots q^1_k,\sigma,q^1) \otimes  (\delta_2)_k(q^2_1\cdots q^2_k,\sigma,q^2)
\]

\item for each $q \in Q$ we define $F_q = (F_1)_{q^1} \otimes (F_2)_{q^2}$.
\end{compactitem}  
Obviously, if $\cA_1$ and $\cA_2$ are bu-deterministic (or crisp-deterministic), then so is $\cA$.

\

Proof of (1): We assume that $\B$ is a commutative semiring.  First, by induction on $\T_\Sigma$, we prove that the following statement holds.
  \begin{equation}
   \text{For every $\xi \in \T_\Sigma$ and $q=(q^1,q^2) \in Q$, we have: }  \h_\cA(\xi)_q = \h_{\cA_1}(\xi)_{q^1} \otimes \h_{\cA_2}(\xi)_{q^2}\enspace. \label{eq:Hadamard}
    \end{equation}

Let $\xi=\sigma(\xi_1,\ldots,\xi_k)$ and $q=(q^1,q^2) \in Q$.
Then:
\begingroup
\allowdisplaybreaks
\begin{align*}
  & \h_\cA(\xi)_q\\
  = & \bigoplus_{q_1 \cdots q_k \in Q^k} \Big(\bigotimes_{i \in [k]} \h_\cA(\xi_i)_{q_i}\Big) \otimes \delta_k(q_1 \cdots q_k,\sigma,q)\\
                     = & \bigoplus_{q_1 \cdots q_k \in Q^k} \Big(\bigotimes_{i \in [k]} (\h_{\cA_1}(\xi_i)_{q_i^1} \otimes \h_{\cA_2}(\xi_i)_{q_i^2})\Big) \otimes \Big(\bigotimes_{j \in [2]} (\delta_j)_k(q_1^j \cdots q_k^j,\sigma,q^j)\Big)\\
  & \tag{by I.H. and construction}\\
  = &  \bigoplus_{q_1^1 \cdots q_k^1 \in (Q_1)^k}\bigoplus_{q_1^2 \cdots q_k^2 \in (Q_2)^k} \Big(\bigotimes_{i \in [k]} (\h_{\cA_1}(\xi_i)_{q_i^1} \otimes \h_{\cA_2}(\xi_i)_{q_i^2})\Big) \otimes \Big(\bigotimes_{j \in [2]} (\delta_j)_k(q_1^j \cdots q_k^j,\sigma,q^j)\Big) \\
  = &  \bigoplus_{q_1^1 \cdots q_k^1 \in (Q_1)^k}\bigoplus_{q_1^2 \cdots q_k^2 \in (Q_2)^k}\\
  & \hspace*{15mm} \Big(\bigotimes_{i \in [k]} \h_{\cA_1}(\xi_i)_{q_i^1}\Big) \otimes (\delta_1)_k(q_1^1 \cdots q_k^1,\sigma,q^1) \ \otimes
       \Big(\bigotimes_{i \in [k]} \h_{\cA_2}(\xi_i)_{q_i^2}\Big) \otimes (\delta_2)_k(q_1^2 \cdots q_k^2,\sigma,q^2) \\
      & \tag{by commutativity}\\
  = &  \bigoplus_{q_1^1 \cdots q_k^1 \in (Q_1)^k} \Big[\Big(\bigotimes_{i \in [k]} \h_{\cA_1}(\xi_i)_{q_i^1}\Big) \otimes (\delta_1)_k(q_1^1 \cdots q_k^1,\sigma,q^1)\Big] \ \otimes\\
  & \hspace*{15mm} 
       \Big(\bigoplus_{q_1^2 \cdots q_k^2 \in (Q_2)^k} \Big(\bigotimes_{i \in [k]} \h_{\cA_2}(\xi_i)_{q_i^2}\Big) \otimes (\delta_2)_k(q_1^2 \cdots q_k^2,\sigma,q^2)\Big) 
       \tag{by left-distributivity}\\
  = &  \Big(\bigoplus_{q_1^1 \cdots q_k^1 \in (Q_1)^k} \Big(\bigotimes_{i \in [k]} \h_{\cA_1}(\xi_i)_{q_i^1}\Big) \otimes (\delta_1)_k(q_1^1 \cdots q_k^1,\sigma,q^1)\Big) \ \otimes\\
  & \hspace*{15mm} 
       \Big(\bigoplus_{q_1^2 \cdots q_k^2 \in (Q_2)^k} \Big(\bigotimes_{i \in [k]} \h_{\cA_2}(\xi_i)_{q_i^2}\Big) \otimes (\delta_2)_k(q_1^2 \cdots q_k^2,\sigma,q^2)\Big) 
       \tag{by right-distributivity}\\
  = & \ \h_{\cA_1}(\xi)_{q^1} \otimes \h_{\cA_2}(\xi)_{q^2}\enspace.
\end{align*}

This proves \eqref{eq:Hadamard}. Then for each $\xi \in \T_\Sigma$:
  \begin{align*}
    \sem{\cA}(\xi) = \bigoplus_{q \in Q} \h_\cA(\xi)_q \otimes F_q
                      &= \bigoplus_{q \in Q} \Big(\h_{\cA_1}(\xi)_{q^1} \otimes \h_{\cA_2}(\xi)_{q^2}\Big) \otimes \Big((F_1)_{q^1} \otimes (F_2)_{q^2}\Big)
                       \tag{\text{by Equation \eqref{eq:Hadamard} and construction of $F$}}\\
                          &= \Big(\bigoplus_{q \in Q_1} \h_{\cA_1}(\xi)_q \otimes (F_1)_{q}\Big) \otimes
                            \Big(\bigoplus_{q \in Q_2} \h_{\cA_2}(\xi)_q \otimes (F_2)_{q}\Big)                        \tag{\text{by commutativity and distributivity of $\B$}}\\
    &= \sem{\cA_1}(\xi) \otimes \sem{\cA_2}(\xi) \enspace.
  \end{align*}
\endgroup

\


Proof of (2): Let $\B$ be commutative and $\cA_1$ and $\cA_2$ be bu-deterministic. Then $\cA$ is bu-deterministic. Here we prove \eqref{eq:Hadamard} as follows. Let $\xi=\sigma(\xi_1,\ldots,\xi_k)$ and $q=(q^1,q^2) \in Q$. As above we obtain 
  \begin{eqnarray}
    \begin{aligned}
  \h_\cA(\xi)_q
  &=   \bigoplus_{q_1^1 \cdots q_k^1 \in (Q_1)^k}\bigoplus_{q_1^2 \cdots q_k^2 \in (Q_2)^k}\\
  &\Big(\bigotimes_{i \in [k]} \h_{\cA_1}(\xi_i)_{q_i^1}\Big) \otimes (\delta_1)_k(q_1^1 \cdots q_k^1,\sigma,q^1) \ \otimes
  \Big(\bigotimes_{i \in [k]} \h_{\cA_2}(\xi_i)_{q_i^2}\Big) \otimes (\delta_2)_k(q_1^2 \cdots q_k^2,\sigma,q^2)\enspace.
  \end{aligned}\label{eq:bu-det-Hadamard}
\end{eqnarray}  
We note that this part uses commutativity and it does not use distributivity. By Lemma \ref{lm:limit-bu-det}(3) (applied to $\cA_1$ and $\cA_2$), there are four cases. 
We can combine the four cases into the following two cases.

\underline{Case (a):} (i) There exists an $i \in [k]$ such that $\Qh{{\cA_1}}{\xi_i}= \emptyset$, i.e.,  for every $q_1 \in Q_1$, we have $\h_{\cA_1}(\xi_i)_{q_1} = \mathbb{0}$ or (ii) there exists a $j \in [k]$ such that $\Qh{{\cA_2}}{\xi_j}= \emptyset$, i.e., for every $q_2 \in Q_2$, we have  $\h_{\cA_2}(\xi_j)_{q_2} = \mathbb{0}$.

Due to \eqref{eq:bu-det-Hadamard} we have $\h_\cA(\xi)_q=\mathbb{0}$. In Case (i), $\h_{\cA_1}(\xi)_{q^1}=\mathbb{0}$ by Lemma \ref{lm:zero-propagation-h}(1). Similarly, in Case~(ii),  $\h_{\cA_2}(\xi)_{q^2}=\mathbb{0}$. In both cases we have $\h_{\cA_1}(\xi)_{q^1} \otimes \h_{\cA_2}(\xi)_{q^2} = \mathbb{0}$.
 
\underline{Case (b):} \sloppy For every $i \in [k]$ we have  $\Qh{{\cA_1}}{\xi_i}=\{\estate_{\cA_1}(\xi_i)\}$ and $\Qh{{\cA_2}}{\xi_i}=\{\estate_{\cA_2}(\xi_i)\}$. Then we have $\h_{\cA_1}(\xi_i)_{\estate_{\cA_1}(\xi_i)} \ne \mathbb{0}$  and $\h_{\cA_2}(\xi_i)_{\estate_{\cA_2}(\xi_i)} \ne \mathbb{0}$.
Then, together with \eqref{eq:bu-det-Hadamard}, we have
\begingroup
\allowdisplaybreaks
\begin{align*}
 & \h_\cA(\xi)_q\\
  &=  
    \Big(\bigotimes_{i \in [k]} \h_{\cA_1}(\xi_i)_{\estate_{\cA_1}(\xi_i)}\Big) \otimes (\delta_1)_k(\estate_{\cA_1}(\xi_1) \cdots \estate_{\cA_1}(\xi_k),\sigma,q^1) \ \otimes \\
  & \hspace*{5mm} \Big(\bigotimes_{i \in [k]} \h_{\cA_2}(\xi_i)_{\estate_{\cA_2}(\xi_i)}\Big) \otimes (\delta_2)_k(\estate_{\cA_2}(\xi_1) \cdots \estate_{\cA_2}(\xi_k),\sigma,q^2)\\
  &= \h_{\cA_1}(\xi)_{q^1} \otimes \h_{\cA_2}(\xi)_{q^2} \enspace.
  \end{align*}
  \endgroup
  This proves \eqref{eq:Hadamard}. 

Now let $\xi \in \T_\Sigma$.  We can prove $\sem{\cA}(\xi) =\sem{\cA_1}(\xi) \otimes \sem{\cA_2}(\xi)$ by case analysis. In a similar way as above, we distinguish two cases.

\underline{Case (a):} Let (i) $\Qh{{\cA_1}}{\xi}=\emptyset$, i.e.,  $\h_{\cA_1}(\xi)_{q_1} = \mathbb{0}$   for every $q_1 \in Q_1$   or (ii) $\Qh{{\cA_2}}{\xi}=\emptyset$, i.e., $\h_{\cA_2}(\xi)_{q_2} = \mathbb{0}$   for every $q_2 \in Q_2$. Then $\sem{\cA_1}(\xi)=\sem{\cA_2}(\xi)=\mathbb{0}$. By \eqref{eq:Hadamard} we have that $\h_\cA(\xi)_{(q_1,q_2)}= \h_{\cA_1}(\xi)_{q_1} \otimes \h_{\cA_2}(\xi)_{q_2}= \mathbb{0}$ for every $(q_1,q_2) \in Q$, and hence $\sem{\cA}(\xi) =\mathbb{0}$.

\underline{Case (b):} \sloppy $\Qh{{\cA_1}}{\xi}=\{\estate_{\cA_1}(\xi)\}$ and $\Qh{{\cA_2}}{\xi}=\{\estate_{\cA_2}(\xi)\}$. Then we have $\h_{\cA_1}(\xi)_{\estate_{\cA_1}(\xi)} \ne \mathbb{0}$  and $\h_{\cA_2}(\xi)_{\estate_{\cA_2}(\xi)} \ne \mathbb{0}$.
Then, by \eqref{eq:Hadamard}, we have that $\h_\cA(\xi)_{(q_1,q_2)}=\mathbb{0}$ for each $(q_1,q_2) \in Q \setminus \{(\estate_{\cA_1}(\xi),\estate_{\cA_2}(\xi))\}$. Then
  \begin{align*}
    \sem{\cA}(\xi) &= \h_\cA(\xi)_{(\estate_{\cA_1}(\xi),\estate_{\cA_2}(\xi))} \otimes F_{(\estate_{\cA_1}(\xi),\estate_{\cA_2}(\xi))}
                     \tag{by Corollary~\ref{lm:properties-hA-of-budet-wta-det-new-H}}\\
    &=  (\h_{\cA_1}(\xi)_{\estate_{\cA_1}(\xi)} \otimes \h_{\cA_2}(\xi)_{\estate_{\cA_2}(\xi)}) \otimes ((F_1)_{\estate_{\cA_1}(\xi)} \otimes (F_2)_{\estate_{\cA_2}(\xi)})
                       \tag{\text{by Equation \eqref{eq:Hadamard} and construction of $F$}}\\
                          &= (\h_{\cA_1}(\xi)_{\estate_{\cA_1}(\xi)} \otimes (F_1)_{\estate_{\cA_1}(\xi)}) \otimes
                            (\h_{\cA_2}(\xi)_{\estate_{\cA_2}(\xi)} \otimes (F_2)_{\estate_{\cA_2}(\xi)})
                            \tag{\text{by commutativity of $\B$}}\\
    &= \sem{\cA_1}(\xi) \otimes \sem{\cA_2}(\xi) \enspace.
  \end{align*}  
  
  \
  
  
  Proof of (3): Now let $\cA_1$ and $\cA_2$ be crisp-deterministic. In the same way as in the proof of Theorem~\ref{thm:closure-sum}(2), we can prove \eqref{eq:crisp-direct-product-estate}:
\begin{equation}\label{equ:state=state-times-state}
\text{for each $\xi \in \T_\Sigma$, we have $\estate_{\cA}(\xi)=(\estate_{\cA_1}(\xi),\estate_{\cA_2}(\xi))$.}
\end{equation}
Then we obtain:
\begin{align*}
\sem{\cA}(\xi)
  &=  F_{\estate_{\cA}(\xi)}
  \tag{by Lemma~\ref{lm:fin-algebra-crisp-det-wta}(2)}\\
  &= F_{(\estate_{\cA_1}(\xi),\estate_{\cA_2}(\xi))}
  \tag{by \eqref{equ:state=state-times-state} }\\
  &= (F_1)_{\estate_{\cA_1}(\xi)} \otimes (F_2)_{\estate_{\cA_2}(\xi)}
  \tag{by definition of $F$}\\
  &= \sem{\cA_1}(\xi) \otimes \sem{\cA_2}(\xi)
    \tag{by Lemma~\ref{lm:fin-algebra-crisp-det-wta}(2)}
    \enspace. 
\end{align*}

 Finally, by Theorem \ref{thm:crisp-det-algebra} (A)$\Leftrightarrow$(B), $\RecStep(\Sigma,\B)$ is equal to the set of $(\Sigma,\B)$-weighted tree languages recognizable by crisp-deterministic wta. By (3) the latter is closed under Hadamard product.
\end{proof}

We note that in \cite[Thm.~4.2(2)]{gho22} closure of $\Rec^{\mathrm{init}}(\Sigma,\B)$ under Hadamard product was proved for the particular case that $\B$ is a $\sigma$-complete distributive  orthomodular lattice. Since each $\sigma$-complete distributive  orthomodular lattice is a particular commutative semiring, \cite[Thm.~4.2(2)]{gho22} follows from Theorem \ref{thm:closure-Hadamard-product} and from \cite[Cor.~3.9]{bor04}.

The next theorem can be compared to \cite[Thm.~5.12.1(b)]{drogoemaemei11}.

\begin{theorem-rect}\label{thm:closure-Hadamard-product-char} Let $\Sigma$ be a ranked alphabet. Moreover, let $\B=(B,\oplus,\otimes,\0,\1)$ be a strong bimonoid, let $\cA$ be a $(\Sigma,\B)$-wta, and let $\cB$ be a crisp-deterministic $(\Sigma,\B)$-wta. Moreover, let $D$ be a $\Sigma$-fta. Then the following four statements hold.
  \begin{compactenum}
  \item[(1)] If $\B$ is right-distributive, then we can construct a $(\Sigma,\B)$-wta $\cC$ such that  $\runsem{\cC} = \runsem{\cA} \otimes \sem{\cB}$. 
    \item[(2)] We can construct a $(\Sigma,\B)$-wta $\cC$ such that  $\runsem{\cC} = \runsem{\cA} \otimes \chi(\LL(D))$. 
          \item[(3)] We can construct a $(\Sigma,\B)$-wta $\cC$ such that  $\runsem{\cC} = \chi(\LL(D)) \otimes \runsem{\cA}$. 
          \item[(4)] \cite{dro22} If $\B$ is left-distributive,
            then  we can construct a $(\Sigma,\B)$-wta $\cC$ such that  $\runsem{\cC} = \sem{\cB} \otimes \runsem{\cA}$.
              \end{compactenum}
  In Statements (1), (2), and (3), if $\cA$ is bu-deterministic (or crisp-deterministic), then so is~$\cC$.
\end{theorem-rect}

\begin{proof} Proof of (1):  Let $\B$ be right-distributive. Moreover, let $\cA=(Q_1,\delta_1,F_1)$ and $\cB=(Q_2,\delta_2,F_2)$.
  We construct the $(\Sigma,\B)$-wta $\cC = (Q,\delta,F)$ in the same way as in the proof of Theorem \ref{thm:closure-Hadamard-product}: 
\begin{compactitem}
\item $Q = Q_1 \times Q_2$;  for each $q \in Q$ we denote its first and second component by $q^1$ and $q^2$, respectively,

\item for every $k \in \mathbb{N}$, $\sigma\in \Sigma^{(k)}$, $q\in Q$, and $q_1\cdots q_k \in Q^k$ we define
\[
\delta_k(q_1\cdots q_k,\sigma,q) =
(\delta_1)_k(q^1_1\cdots q^1_k,\sigma,q^1) \otimes  (\delta_2)_k(q^2_1\cdots q^2_k,\sigma,q^2)
\]

\item for each $q \in Q$ we define $F_q = (F_1)_{q^1} \otimes (F_2)_{q^2}$.
\end{compactitem}  
Obviously, if $\cA$ is bu-deterministic (or crisp-deterministic), then so is $\cC$.

Let $\xi \in \T_\Sigma$. By Lemma \ref{lm:fin-algebra-crisp-det-wta}(2) we have 
\begin{equation}\label{eq:semantics-of-crips}
\sem{\cB}(\xi)= (F_2)_{\estate_{\cB}(\xi)}\enspace.
\end{equation}

Then we define the mapping $\psi: \R_\cA(\xi) \to \R_\cC(\xi)$ for every $\rho \in \R_\cA(\xi)$ and $w \in \pos(\xi)$ by
\[
  \psi(\rho)(w) = (\rho(w),\crhorm_{\xi}(w))
\]
where $\crhorm_\xi$ is the canonical run of $\cB$ on $\xi$ defined in Subsection \ref{subsect:annihilation-bu-det-wta}. Obviously, $\psi$ is injective. Moreover, we define the mapping $\psi': \R_\cA(\xi) \to \im(\psi)$ such that $\psi'(\rho) = \psi(\rho)$ for each $\rho \in \R_\cA(\xi)$. It follows that $\psi'$ is bijective 
(cf. Figure \ref{fig:psi-varphi-Hadamard}).

\begin{figure}
  \centering

\begin{tikzpicture}[level distance=3.5em,
  every node/.style = {align=center}]]
  \pgfdeclarelayer{bg}    
  \pgfsetlayers{bg,main}  

  \newcommand{\mydista}{1.8mm} 
  \newcommand{\mydistaa}{2.2mm} 
  \newcommand{\mydistb}{0.8mm} 
  \tikzstyle{mycircle}=[draw, circle, inner sep=-2mm, minimum height=5mm]
  \tikzstyle{mybox}=[draw, ellipse, inner sep=-2mm, minimum height=5mm]

\begin{scope}[level 1/.style={sibling distance=20mm}]

 \node at (-0.8, 0.6) {$\xi:$};
 \node at (1.9, 0.6) {$\rho \in \R_\cA (\xi)$:};

 \node (N0) {$\sigma$}
  child {node (N1) {$\gamma$} 
  	   child { node (N2) {$\alpha$}} }
  child {node (N3) {$\beta$} };

 \node [mycircle, anchor=west] at ([xshift=\mydista]N0.east) {$q_1$};
 \node [mycircle, anchor=west] at ([xshift=\mydistb]N1.east) {$q_2$};
 \node [mycircle, anchor=west] at ([xshift=\mydistb]N2.east) {$q_3$};
 \node [mycircle, anchor=west] at ([xshift=\mydistb]N3.east) {$q_4$};
\end{scope}

\begin{scope}[xshift=80mm, level 1/.style={sibling distance=30mm}]

 \node at (-0.8, 0.6) {$\xi:$};
 \node at (2.7, 0.6) {$\psi'(\rho) \in \R_\cC(\xi)$:};

 \node (N0) {$\sigma$}
  child {node (N1) {$\gamma$} 
  	   child { node (N2) {$\alpha$}} }
  child {node (N3) {$\beta$} };

 \node [draw, rounded rectangle, anchor=west] at ([xshift=\mydistaa]N0.east) {$q_1,\crhorm_\xi(\varepsilon)$};
 \node [draw, rounded rectangle, anchor=west] at ([xshift=\mydistb]N1.east) {$q_2,\crhorm_\xi(1)$};
 \node [draw, rounded rectangle, anchor=west] at ([xshift=\mydistb]N2.east) {$q_3,\crhorm_\xi(11)$};
 \node [draw, rounded rectangle, anchor=west] at ([xshift=\mydistb]N3.east) {$q_4,\crhorm_\xi(2)$};
\end{scope}

\draw [->, thick, xshift=2.8cm, yshift=-1cm] (0,0) -- (2.2,0) node[midway, above] {$\psi'$};
\draw [->, thick, xshift=2.8cm, yshift=-1.8cm] (2.2,0) -- (0,0) node[midway, above] {$(\psi')^{-1}$};

\end{tikzpicture}

\caption{\label{fig:psi-varphi-Hadamard} An illustration of the mapping $\psi'$.}
  \end{figure}

Since $\cB$ is crisp-deterministic, the following statement is easy to see.
\begin{eqnarray}
  \text{For every $\xi \in \T_\Sigma$ and $\rho \in \R_\cC(\xi)$: } \
  \wt_\cC(\xi,\rho) =
    \begin{cases}
      \wt_\cA(\xi,(\psi')^{-1}(\rho)) & \text{if } \rho \in \im(\psi')\\
      \0 & \text{ otherwise}
      \end{cases}\label{eq:0-for-almost-all}
  \end{eqnarray}
Then, for each $\xi \in \T_\Sigma$, we can compute as follows.
  \begingroup
  \allowdisplaybreaks
 \begin{align*}
   \runsem{\cC}(\xi) &= \bigoplus_{\rho \in \R_\cC(\xi)} \wt_\cC(\xi,\rho) \otimes F_{\rho(\varepsilon)}\\
   &= \bigoplus_{\rho \in \im(\psi')}  \wt_\cA(\xi,(\psi')^{-1}(\rho)) \otimes F_{\rho(\varepsilon)}
     \tag{\text{by \eqref{eq:0-for-almost-all}}}\\
     &= \bigoplus_{\rho \in \im(\psi')}  \wt_\cA(\xi,(\psi')^{-1}(\rho)) \otimes (F_1)_{\rho(\varepsilon)^1} \otimes (F_2)_{\rho(\varepsilon)^2}
   \tag{\text{by the construction of $F$}}\\
                     &= \bigoplus_{\rho \in \im(\psi')}  \wt_\cA(\xi,(\psi')^{-1}(\rho)) \otimes (F_1)_{\rho(\varepsilon)^1} \otimes (F_2)_{\crhorm_\xi(\varepsilon)}  \tag{by the definition of $\psi$}\\
   &= \bigoplus_{\rho' \in \R_\cA(\xi)}  \wt_\cA(\xi,(\psi')^{-1}(\psi'(\rho'))) \otimes (F_1)_{\psi'(\rho')(\varepsilon)^1} \otimes (F_2)_{\crhorm_\xi(\varepsilon)} \tag{because $\psi'$ is bijective}\\
                     &= \bigoplus_{\rho' \in \R_\cA(\xi)}  \wt_\cA(\xi,\rho') \otimes (F_1)_{\rho'(\varepsilon)} \otimes (F_2)_{\crhorm_\xi(\varepsilon)}
   \tag{\text{because $\psi'(\rho')(\varepsilon)^1=\rho'(\varepsilon)$}} \\
     &=^{(*)}   \Big(\bigoplus_{\rho' \in \R_\cA(\xi)} \wt_\cA(\xi,\rho') \otimes (F_1)_{\rho'(\varepsilon)}\Big) \
                        \otimes (F_2)_{\crhorm_\xi(\varepsilon)}
   \tag{\text{by right-distributivity}} \\
    &= \runsem{\cA}(\xi) \otimes \sem{\cB}(\xi) \tag{by \eqref{eq:semantics-of-crips} and $\crhorm_\xi(\varepsilon)=\estate_\cB(\xi)$}\enspace.
  \end{align*}
  \endgroup
The $(*)$ at the last but one equality will be used in the proof of (2).

\

Proof of (2): By Theorem \ref{thm:fta-wta} we can construct  a crisp-deterministic $(\Sigma,\B)$-wta $\cB=(Q_2,\delta_2,F_2)$ with unit root weights (i.e., $\im(F_2) \subseteq \{\0,\1\}$) such that $\sem{\cB} = \chi(\LL(D))$. We construct the $(\Sigma,\B)$-wta~$\cC$ and define the mappings $\psi$ and $\psi'$ as in the proof of Statement (1). Obviously, \eqref{eq:0-for-almost-all} also holds. Then we can prove that  $\runsem{\cC} = \runsem{\cA} \otimes \chi(L(D))$ in the same way as we have proved $\runsem{\cC} = \runsem{\cA} \otimes \sem{\cB}$ in Statement (1) except that the equality $(*)$ is justified because $(F_2)_{\crhorm_\xi(\varepsilon)} \in \{\0,\1\}$. Thus no right-distributivity of $\B$ is used.

\

Proof of (3):  As in the proof of (2), by Theorem \ref{thm:fta-wta} we can construct  a crisp-deterministic $(\Sigma,\B)$-wta $\cB=(Q_2,\delta_2,F_2)$ with unit root weights (i.e., $\im(F_2) \subseteq \{\0,\1\}$) such that $\sem{\cB} = \chi(\LL(D))$. Then, in the same way as in the proof of Statement (1), we construct the $(\Sigma,\B)$-wta $\cC$. We mention that we do not have to reverse the order of $(\delta_1)_k$ and $(\delta_2)_k$ in the right-hand side of the definition of $\delta_k$, because  $\im((\delta_2)_k) \in\{\0,\1\}$ and thus
\[(\delta_2)_k(q^2_1\cdots q^2_k,\sigma,q^2) \otimes  (\delta_1)_k(q^1_1\cdots q^1_k,\sigma,q^1)= (\delta_1)_k(q^1_1\cdots q^1_k,\sigma,q^1) \otimes  (\delta_2)_k(q^2_1\cdots q^2_k,\sigma,q^2) \enspace.\]
Next we define the mappings $\varphi$ and $\psi$ in the same way as in the proof of Statement (1), and of course, Equality \eqref{eq:0-for-almost-all} holds also in this case. Due to the first six steps in the final calculation of the proof of (1), we obtain: 
\begin{align*}
   \runsem{\cC}(\xi) &= \bigoplus_{\rho' \in \R_\cA(\xi)}  \wt_\cA(\xi,\rho') \otimes (F_1)_{\rho'(\varepsilon)} \otimes (F_2)_{\crhorm_\xi(\varepsilon)}\enspace .
\end{align*}
Then we finish the proof as follows:
  \begingroup
  \allowdisplaybreaks
  \begin{align*}
    & \ \ \bigoplus_{\rho' \in \R_\cA(\xi)}  \wt_\cA(\xi,\rho') \otimes (F_1)_{\rho'(\varepsilon)} \otimes (F_2)_{\crhorm_\xi(\varepsilon)}\\
        &=    \bigoplus_{\rho' \in \R_\cA(\xi)}  (F_2)_{\crhorm_\xi(\varepsilon)} \otimes  \wt_\cA(\xi,\rho') \otimes (F_1)_{\rho'(\varepsilon)} \tag{because $(F_2)_{\crhorm_\xi(\varepsilon)} \in \{\0,\1\}$}\\
     &=   (F_2)_{\crhorm_\xi(\varepsilon)} \otimes    \bigoplus_{\rho' \in \R_\cA(\xi)}\wt_\cA(\xi,\rho') \otimes (F_1)_{\rho'(\varepsilon)} \tag{because $(F_2)_{\crhorm_\xi(\varepsilon)} \in \{\0,\1\}$}\\
    &=  \sem{\cB}(\xi) \otimes \runsem{\cA}(\xi) \tag{by \eqref{eq:semantics-of-crips} because $\crhorm_\xi(\varepsilon)=\estate_\B(\xi)$}\enspace.
  \end{align*}
  \endgroup
  Thus no commutativity and no left-distributivity of $\B$ is used.
  
\

Proof of (4): Let $\B$ be left-distributive. By Theorem \ref{thm:crisp-det-algebra}(A)$\Rightarrow$(D),
we can construct $n \in \mathbb{N}_+$, $b_1,\ldots,b_n \in B$, and  $\Sigma$-fta $A_1,\ldots,A_n$ such that $\sem{\cB} = \bigoplus_{i \in [n]} b_i \cdot \chi(\LL(A_i))$ and the step languages $\LL(A_i)$ are pairwise disjoint.
Due to the disjointness of the step languages and by associativity of $\otimes$, we have 
\begingroup
\allowdisplaybreaks
\begin{align*}
  \big( \bigoplus_{i \in [n]} b_i \cdot \chi(\LL(A_i)) \big) \otimes \runsem{\cA}
    = \bigoplus_{i \in [n]} b_i \cdot \big(\chi(\LL(A_i))  \otimes \runsem{\cA}\big) \enspace.
 \end{align*}
 \endgroup
By Theorem \ref{thm:closure-Hadamard-product-char}(3), we can construct a $(\Sigma,\B)$-wta $\cB_i$ such that $\runsem{\cB_i} = \chi(\LL(A_i)) \otimes \runsem{\cA}$. Hence
\begin{align*}
  \bigoplus_{i \in [n]} b_i \cdot \big(\chi(\LL(A_i))  \otimes \runsem{\cA}\big)
  =  \bigoplus_{i \in [n]} b_i \cdot \runsem{\cB_i} \enspace.
 \end{align*}
 Since $\B$ is left-distributive, by Theorem \ref{thm:closure-scalar}(1) we can construct, for each $i \in [n]$, a $(\Sigma,\B)$-wta $\cC_i$ such that $\runsem{\cC_i} = b_i \cdot \runsem{\cB_i}$. Hence
\begin{align*}
  \bigoplus_{i \in [n]} b_i \cdot \runsem{\cB_i}
  = \bigoplus_{i \in [n]} \runsem{\cC_i}\enspace.
 \end{align*}

 By Theorem \ref{thm:closure-sum} we can construct a $(\Sigma,\B)$-wta $\cC$ such that $\runsem{\cC} = \bigoplus_{i \in [n]} \runsem{\cC_i}$.
\end{proof}


The next corollary shows that recognizable step mappings are closed under Hadamard product if the weight structure is a semiring. This result is a special instance of Theorem \ref{thm:closure-Hadamard-product-char} (the proof of which uses Theorem \ref{thm:crisp-det-algebra}). Nevertheless, here we show a proof of the weaker result without referring to Theorem \ref{thm:crisp-det-algebra}; its proof exploits the special structure of recognizable step mappings. The reader may view this proof as an exercise.

\begin{corollary} \label{thm:closure-Hadamard-step} \rm (cf. \cite[Thm.~5.12]{drogoemaemei11} and \cite[Lm.~1.4.22]{her20}) Let $\Sigma$ be a ranked alphabet and $\B=(B,\oplus,\otimes,\0,\1)$ be a semiring. Moreover, let $r: \T_\Sigma \to B$ and $r': \T_\Sigma \to B$ be recognizable $(\Sigma,\B)$-weighted tree languages. If $r$ or $r'$ is a recognizable step mapping, then $r \otimes r'$ is $(\Sigma,\B)$-recognizable.
  \end{corollary}
 \begin{proof}
  First we assume that $r$ is a recognizable step mapping. Hence, there exist $n \in \mathbb{N}_+$, $b_1,\ldots,b_n \in B$, and recognizable $\Sigma$-tree languages $L_1,\ldots,L_n$ such that
  $r = \bigoplus_{i\in[n]} b_i \cdot \chi(L_i)$. Then, for each $\xi \in \T_\Sigma$, we can calculate as follows.
  \begingroup
    \allowdisplaybreaks
    \begin{align*}
       (r \otimes r')(\xi) &= \Big(\bigoplus_{i\in[n]} b_i \otimes \chi(L_i)(\xi)\Big) \otimes r'(\xi)\\
      &= \bigoplus_{i\in[n]} b_i \otimes \chi(L_i)(\xi) \otimes r'(\xi) \tag{by right-distributivity}\\
       &= \Big(\bigoplus_{i\in[n]} b_i \cdot \big(\chi(L_i) \otimes r'\big)\Big)(\xi) \enspace.
      \end{align*}
      \endgroup
      By Theorem \ref{thm:closure-Hadamard-product-char}(3),  the weighted tree language $\chi(L_i) \otimes r'$ is recognizable. Since $\B$ is left-distributive, we can apply Theorem \ref{thm:closure-scalar}(1a) and obtain that the weighted tree language $b_i \cdot \big(\chi(L_i) \otimes r'\big)$
            is recognizable. Finally, by Theorem \ref{thm:closure-sum}, the weighted tree language $\bigoplus_{i\in[n]} b_i \cdot \big(\chi(L_i) \otimes r'\big)$ is recognizable. Hence $r \otimes r'$ is recognizable.

      Now we assume that $r'$ is a recognizable step mapping. Hence, there are $n \in \mathbb{N}_+$, $b_1,\ldots,b_n \in B$, and recognizable $\Sigma$-tree languages $L_1,\ldots,L_n$ such that
  $r' = \bigoplus_{i\in[n]} b_i \cdot \chi(L_i)$. Then, for each $\xi \in \T_\Sigma$, we can calculate as follows.
  \begingroup
    \allowdisplaybreaks
    \begin{align*}
       (r \otimes r')(\xi) &= r(\xi) \otimes \bigoplus_{i\in[n]} b_i \otimes \chi(L_i)(\xi)\\
      &= \bigoplus_{i\in[n]} r(\xi) \otimes b_i \otimes \chi(L_i)(\xi) \tag{by left-distributivity} \\
       &= \Big(\bigoplus_{i\in[n]} \big(r \cdot b_i \big)\otimes \chi(L_i) \Big)(\xi) \enspace.
      \end{align*}
      \endgroup
Since $\B$ is right-distributive, we can apply Theorem \ref{thm:closure-scalar}(2) and obtain that the weighted tree language $r \cdot b_i$ is recognizable. By Theorem \ref{thm:closure-Hadamard-product-char}(2), the weighted tree language $\big(r \cdot b_i\big)\otimes \chi(L_i)$ is recognizable.  Finally, by Theorem \ref{thm:closure-sum}, the weighted tree language $\bigoplus_{i\in[n]} \big(r \cdot b_i \big)\otimes \chi(L_i)$ is recognizable. Hence $r \otimes r'$ is recognizable.
    \end{proof}

Finally, we show an overview of the closure results for the Hadamard product, cf. Figure \ref{fig:overview-closure-results-Hadamard-product}.

        \begin{figure}[h]

          \
          
        \begin{tabular}{lllll}
          Theorem & restriction on  & restriction on  & subset of   & $\runsem{\cA_1} \otimes \runsem{\cA_2}$\\
          &  $(\Sigma,\B)$-wta $\cA_1$ & $(\Sigma,\B)$-wta $\cA_2$ & strong bimonoids & included in \\\hline
          \ref{thm:closure-Hadamard-product}(1) & - & - & commutative semirings & $\Rec(\Sigma,\B)$\\
          \ref{thm:closure-Hadamard-product}(2) & bu-deterministic & bu-deterministic & commutative & $\budRec(\Sigma,\B)$\\
          \ref{thm:closure-Hadamard-product}(3) & crisp-deterministic & crisp-deterministic & all strong bimonoids & $\cdRec(\Sigma,\B)$\\
          \\
          \ref{thm:closure-Hadamard-product-char}(1) & - & crisp-deterministic & right-distributive  & $\Rec^{\mathrm{run}}(\Sigma,\B)$\\
          \ref{thm:closure-Hadamard-product-char}(2) & - & $\runsem{\cA_2}= \chi(L)$ & all strong bimonoids  & $\Rec^{\mathrm{run}}(\Sigma,\B)$\\
                  && for some $L \in \Rec(\Sigma)$ & &\\
          \ref{thm:closure-Hadamard-product-char}(3) & $\runsem{\cA_1}= \chi(L)$ & - & all strong bimonoids  & $\Rec^{\mathrm{run}}(\Sigma,\B)$\\
                  & for some $L \in \Rec(\Sigma)$ &  &&\\
          \ref{thm:closure-Hadamard-product-char}(4) & crisp-deterministic & - & left-distributive & $\Rec^{\mathrm{run}}(\Sigma,\B)$
        \end{tabular}
        
        \caption{\label{fig:overview-closure-results-Hadamard-product} An overview of the closure results for the Hadamard product of $\runsem{\cA_1}$ and $\runsem{\cA_2}$ for two $(\Sigma,\B)$-wta $\cA_1$ and $\cA_2$.}
      \end{figure}


\section{Closure under top-concatenations}
\label{sec:top-concatenation}

\index{top-concatenation}
\index{top@$\ttop_\sigma$}
Let $k \in \mathbb{N}$, $\sigma \in \Sigma^{(k)}$, and $r_1,\ldots,r_k$ be $(\Sigma,\B)$-weighted tree languages. The \emph{top-concatenation of $r_1,\ldots,r_k$ with $\sigma$} is the $(\Sigma,\B)$-weighted tree language $\ttop_\sigma(r_1,\ldots,r_k): \T_\Sigma \to B$  defined, for each $\xi \in \T_\Sigma$, as follows:
\[
  \ttop_\sigma(r_1,\ldots,r_k)(\xi) =
  \begin{cases}
    r_1(\xi_1) \otimes \cdots \otimes r_k(\xi_k) & \text{ if $\xi=\sigma(\xi_1,\ldots,\xi_k)$}\\
    \0 & \text{ otherwise}
    \end{cases}
  \]
  If $\alpha \in \Sigma^{(0)}$, then we call $\ttop_\alpha()$ the \emph{top-concatenation with $\alpha$}.  For each $\xi \in \T_\Sigma$, we have
  \[
  \ttop_\alpha()(\xi) =
  \begin{cases}
    \1 & \text{ if $\xi=\alpha$}\\
    \0 & \text{ otherwise} \enspace,
    \end{cases}
  \]
i.e., $\ttop_\alpha()=\mathbb{1}.\alpha$.
It follows that $\ttop_\sigma(r_1,\ldots,r_k)(\xi) = \0$ for each $\xi \in \T_\Sigma$ with $\xi(\varepsilon)\ne \sigma$.

A set $\cL$ of $\B$-weighted tree languages is \emph{closed under top-concatenations} if for every $k \in \mathbb{N}$, $\sigma \in \Sigma^{(k)}$, and $(\Sigma,\B)$-weighted tree languages $r_1,\ldots,r_k$ in $\cL$, the $(\Sigma,\B)$-weighted tree language $\ttop_\sigma(r_1,\ldots,r_k)$ is in $\cL$. Thus, in particular, a set $\cL$ of $(\Sigma,\B)$-weighted tree languages  which is closed under top-concatenations contains the $(\Sigma,\B)$-weighted tree language $\ttop_\alpha()$ for each $\alpha \in \Sigma^{(0)}$.
  
\begin{theorem} \label{thm:closure-top-concat-wrtg} Let $\B$ be a semiring. Moreover, let $k \in \mathbb{N}$ and $\sigma \in \Sigma^{(k)}$. Also, for each $i \in [k]$, let $\cG_i$ be a $(\Sigma,\B)$-wrtg such that (a) for each $i \in [k]$ the wrtg $\cG_i$ is finite-derivational or (b) $\B$ is $\sigma$-complete. We can construct a $(\Sigma,\B)$-wrtg $\cG$ such that 
          $\sem{\cG} =  \ttop_\sigma(\sem{\cG_1},\ldots,\sem{\cG_k})$.
  
       Thus, in particular, if $\B$ is a semiring, then the set $\Reg(\Sigma,\B)$ is closed under top-concatenations.
      \end{theorem}
      \begin{proof} For each $i \in [k]$, let $\cG_i = (N_i,S_i,R_i,\wt_i)$ be a $(\Sigma,\B)$-wrtg such that $N_i \cap N_j =\emptyset$ for each $i,j \in [k]$ with $i\ne j$. By Lemma \ref{lm:normal-form-lemmas-inherited-from-wcfg}(1), for each  $i \in [k]$, we can construct a wrtg
        which is start-separated and equivalent to $\cG_i$. The construction preserves the property finite-derivational. So we can assume that each $\cG_i$ is start-separated, i.e., $S_i$ is a singleton (and then we assume that $S_i$ denotes the only initial nonterminal).  
         
      We let $S$ be a new nonterminal, i.e., $S \not\in \bigcup_{i \in [k]} N_i$, and construct the $(\Sigma,\B)$-wrtg $\cG=(N,S,R,\wt)$ as follows.
      \begin{compactitem}
      \item $N  = \{S\} \cup \bigcup_{i \in [k]} N_i$ and
      \item $R$ is the smallest set $R'$ of rules which satisfies the following conditions:
        \begin{compactitem}
        \item $r=(S \rightarrow \sigma(S_1,\ldots,S_k))$ is in $R'$  and $\wt(r) = \1$ and
          \item for every $i \in [k]$ and $r  \in R_i$, we let $r \in R'$ and $\wt(r) = \wt_i(r)$.
        \end{compactitem}
             \end{compactitem}
             In case $k=0$ we have $N= \{S\}$ and $R=\{r\}$ with $r=(S \rightarrow \sigma)$ and $\wt(r)=\1$.   
             
It is obvious that, if for each $i \in [k]$ the wrtg $\cG_i$ is finite-derivational, then $\cG$ is finite-derivational. Hence $\sem{\cG}$ is defined.

          Next we prove that
          \begin{equation}
            \text{ $\sem{\cG}(\xi) = \ttop_\sigma(\sem{\cG_1},\ldots,\sem{\cG_k})(\xi)$ for each $\xi \in \T_\Sigma$}\enspace. \label{eq:main-equation-wrtg}
            \end{equation}
            For this, we mention some properties which are easy to see.
          \begin{equation}
\text{For every $\xi \in \T_\Sigma$ with $\xi(\varepsilon)\ne \sigma$ we have $\RT_{\cG}(S,\xi)=\emptyset$.} \label{eq:not-sigma-wrtg}
\end{equation}
\begin{eqnarray}
  \begin{aligned}
     &\text{For every $\xi = \sigma(\xi_1,\ldots,\xi_k)$ in $\T_\Sigma$ and $d \in \RT_{\cG}(S,\xi)$,}\\
     &\text{we have $\lhs(d(i))= S_i$ for each $i \in [k]$.}
   \end{aligned}\label{eq:not-final-states-wrtg}
\end{eqnarray}
\begin{equation}
\text{For every $\xi \in \T_\Sigma$ and $i\in [k]$, we have $\RT_{\cG}(S_i,\xi)= \RT_{\cG_i}(S_i,\xi)$.}\label{eq:weight-0-wrtg}
  \end{equation}

\begin{equation}
\text{For every $\xi \in \T_\Sigma$, $i\in [k]$, and $d \in \RT_{\cG_i}(S_i,\xi)$, we have $\wt_{\cG}(d) = \wt_{\cG_i}(d)$. }\label{eq:weight-equality-wrtg}
  \end{equation}

Now we prove \eqref{eq:main-equation-wrtg}. Let $\xi \in \T_\Sigma$. Then we have:
\begingroup
\allowdisplaybreaks
\begin{align*}
\sem{\cG}(\xi) 
  =  \bigoplus_{d \in \RT_{\cG}(\xi)}\wt_{\cG}(d) = \bigoplus_{d \in \RT_{\cG}(S,\xi)}\wt_{\cG}(\xi) \enspace.
\end{align*}
\endgroup

If $\xi(\varepsilon)\ne \sigma$, then $\sem{\cG}(\xi)= \0$ by \eqref{eq:not-sigma-wrtg}.
Since  $\ttop_\sigma(\sem{\cG_1},\ldots,\sem{\cG_k})(\xi) =\0$, the equality \ref{eq:main-equation-wrtg} is proved for that case.

Now we assume that $\xi = \sigma(\xi_1,\ldots,\xi_k)$. Then we continue with
\begingroup
\allowdisplaybreaks
\begin{align*}
  & \bigoplus_{d \in \RT_{\cG}(S,\xi)}\wt_{\cG}(d)\\
  = & \bigoplus_{\substack{d \in \RT_{\cG}(S,\xi):\\(\forall i \in [k]): \lhs(d(i))=S_i }}\wt_{\cG}(d)
  \tag{\text{by \eqref{eq:not-final-states-wrtg}}}\\
  = & \bigoplus_{\substack{d \in \RT_{\cG}(S,\xi):\\(\forall i \in [k]): \lhs(d(i))=S_i }}
  \wt_{\cG}(d|_1) \otimes \ldots \otimes \wt_{\cG}(d|_k) \otimes \wt(S \rightarrow \sigma(S_1,\ldots, S_k))\\
    = & \bigoplus_{\substack{d \in \RT_{\cG}(S,\xi):\\(\forall i \in [k]): \lhs(d(i))=S_i }}
  \wt_{\cG}(d|_1) \otimes \ldots \otimes \wt_{\cG}(d|_k)
  \tag{\text{since $\wt(S \rightarrow \sigma(S_1,\ldots, S_k))=\1$}}\\
    = & \bigoplus_{d_1 \in \RT_{\cG}(S_1,\xi_1)} \ldots \bigoplus_{d_k \in \RT_{\cG}(S_k,\xi_k)}
        \wt_{\cG}(d_1) \otimes \ldots \otimes \wt_{\cG}(d_k)\\
  = & \bigoplus_{d_1 \in \RT_{\cG_1}(S_1,\xi_1)} \ldots
  \bigoplus_{d_k \in \RT_{\cG_k}(S_k,\xi_k)}  \wt_{\cG}(d_1) \otimes \ldots \otimes \wt_{\cG}(d_k)
  \tag{\text{by \eqref{eq:weight-0-wrtg}}}\\
      = & \bigoplus_{d_1 \in \RT_{\cG_1}(S_1,\xi_1)} \ldots \bigoplus_{d_k \in \RT_{\cG_k}(S_k,\xi_k)}
          \wt_{\cG_1}(d_1) \otimes \ldots \otimes \wt_{\cG_k}(d_k)
  \tag{\text{by \eqref{eq:weight-equality-wrtg}}}\\
  = & \ \Big(\bigoplus_{d_1 \in \RT_{\cG_1}(S_1,\xi_1)} \wt_{\cG_1}(d_1)\Big)  \otimes
     \ldots \otimes  \Big(\bigoplus_{d_k \in \RT_{\cG_k}(S_k,\xi_k)} \wt_{\cG_k}(d_k)\Big)
      \tag{\text{by distributivity}}\\
  = & \ \sem{\cG_1}(\xi_1) \otimes \ldots \otimes \sem{\cG_k}(\xi_k)\\
  = & \ \ttop_\sigma(\sem{\cG_1},\ldots,\sem{\cG_k})(\xi)
      \tag{\text{because $\xi = \sigma(\xi_1,\ldots,\xi_k)$}}
\end{align*}
\endgroup
This proves \eqref{eq:main-equation-wrtg} also in the case that $\xi = \sigma(\xi_1,\ldots,\xi_k)$.
      \end{proof}

      \begin{corollary-rect}\label{cor:closure-under-top-concat-wta} \rm (cf. \cite[Lm.~6.2]{dropecvog05}) Let $\Sigma$ be a ranked alphabet. Moreover, let $\B$ be a semiring, let $k \in \mathbb{N}$, and let $\sigma \in \Sigma^{(k)}$. Also, for each $i \in [k]$, let $\cA_i$ be a $(\Sigma,\B)$-wta. Then we can construct a $(\Sigma,\B)$-wta $\cA$ such that $\sem{\cA} =  \ttop_\sigma(\sem{\cA_1},\ldots,\sem{\cA_k})$.
        Thus, in particular, if $\B$ is a semiring, then the set $\Rec(\Sigma,\B)$ is closed under top-concatenations.
      \end{corollary-rect}
      \begin{proof}  By Lemma \ref{lm:wta-to-wrtg}, for each $i \in [k]$, we can construct $(\Sigma,\B)$-wrtg $\cG_i$ such that $\cG_i$ is in tree automata form and $\sem{\cA_i} = \sem{\cG_i}$.  Then, in particular, $\cG_i$ is  finite-derivational. By Theorem \ref{thm:closure-top-concat-wrtg} we can construct a  finite-derivational $(\Sigma,\B)$-wrtg $\cG$ such that  $\sem{\cG} = \ttop_\sigma(\sem{\cG_1},\ldots,\sem{\cG_k})$. Then,  by Lemma \ref{lm:wrtg-to-wta}, we can construct a $(\Sigma,\B)$-wta $\cA$ such that $\sem{\cG}=\sem{\cA}$. 
        \end{proof}


  \section{Closure under tree concatenations}
  \label{sec:closure-under-tree-concatenation}
  
In this section we recall the definition of tree concatenation and show that $\Rec(\Sigma,\B)$ is closed under tree concatenation. The definition of tree concatenation involves a technical problem which is due to a difference between  the concatenation of strings and tree concatenation \cite[p.~59]{thawri68}. The concatenation of two strings $\xi$ and $\zeta$ is simply the string $\xi\zeta$. This also applies if $\xi$ and $\zeta$ are particular strings, viz. trees. However, $\xi\zeta$ is not a tree. In order to define tree concatenation as operation on trees, in \cite{thawri68} a nullary symbol $\alpha$ is used and each occurrence of $\alpha$ in $\xi$ is replaced by $\zeta$. Hence, tree concatenation consumes the occurrences of $\alpha$ in $\xi$; in contrast, string concatenation does not consume symbols.

Let $r_1$ and $r_2$ be $(\Sigma,\B)$-weighted tree languages  and $\alpha \in \Sigma^{(0)}$. We will recall the definition of the $\alpha$-concatenation of $r_1$ and $r_2$, denoted by $r_1 \circ_{\alpha} r_2$. This is a $(\Sigma,\B)$-weighted tree language and hence, for each $\xi \in \T_\Sigma$, the value $(r_1 \circ_{\alpha} r_2)(\xi)$ has to be defined. Intuitively, $\xi$ has to be decomposed and the resulting parts of $\xi$ have to be evaluated appropriately in $r_1$ or $r_2$.  As preparation for this, we  define the concept of $\alpha$-cut through $\xi$ which helps to identify the mentioned parts of $\xi$.  An $\alpha$-cut through $\xi$ is the same as a cut through $\xi$ as defined in Section~\ref{sect:the-lemmas-and-their-proofs} except that (a) $()$ is an $\alpha$-cut of $\xi$ if $\xi$ does not contain an occurrence of $\alpha$ and (b) in the condition ``complete'' only positions labeled by $\alpha$ are considered.

\index{$\Cut_\alpha(\xi)$}
  \index{cut}
  \index{alphacut@$\alpha$-cut}
Formally, let $\xi\in \T_\Sigma$ and $\alpha \in \Sigma^{(0)}$. An \emph{$\alpha$-cut through $\xi$} is a sequence $(w_1,\ldots,w_n)$ such that 
    \begin{compactitem}
    \item $n \in \mathbb{N}$ and $w_i \in \pos(\xi)$, for each $i \in [n]$,
    \item (independent) for every $i,j \in [n]$ with $i\ne j$, we have $w_i \not\in \prefix(w_j)$, 
    \item (ordered) $w_1 <_{\mathrm{lex}}\ldots <_{\mathrm{lex}}w_n$, and
    \item ($\alpha$-complete) for each $w \in \pos_{\alpha}(\xi)$ there exists $i \in [n]$ such that $w_i \in \prefix(w)$.
    \end{compactitem}
We denote by $\Cut_\alpha(\xi)$ the set of all $\alpha$-cuts through $\xi$. 

In Figure \ref{fig:cut} we show an example tree $\xi$ and illustrate an $\alpha$-cut through $\xi$.

Let $\xi\in \T_\Sigma$. We show that the relation between $\Cut(\xi)$ defined in Section~\ref{sect:the-lemmas-and-their-proofs} and 
$\Cut_\alpha(\xi)$ defined above is $\Cut(\xi) = \bigcap_{\alpha \in \Sigma^{(0)}} \Cut_\alpha(\xi)$. Clearly, $\Cut(\xi)\subseteq \Cut_\alpha(\xi)$ for each $\alpha\in \Sigma^{(0)}$, hence  $\Cut(\xi) \subseteq \bigcap_{\alpha \in \Sigma^{(0)}} \Cut_\alpha(\xi)$. The converse inclusion also holds because,
for each sequence $(w_1,\ldots,w_n) \in \bigcap_{\alpha \in \Sigma^{(0)}} \Cut_\alpha(\xi)$, it holds that 
for every $\alpha\in\Sigma^{(0)}$ and $w \in \pos_{\alpha}(\xi)$ there exists $i \in [n]$ such that $w_i \in \prefix(w)$.
Thus $(w_1,\ldots,w_n) \in \Cut(\xi)$.

Let $\widetilde{w} = (w_1,\ldots,w_n)$ be an element of $\Cut_\alpha(\xi)$. Then we will abbreviate  $\xi[\alpha]_{w_1} \ldots [\alpha]_{w_n}$ by $\xi[\alpha]_{\widetilde{w}}$.
We note that if $\widetilde{w} \in \Cut(\xi)$, then $\xi[\alpha]_{\widetilde{w}}=\xi[\widetilde{w}\leftarrow \alpha]$, where the notation $\xi[\widetilde{w}\leftarrow \alpha]$ is defined in Section~\ref{sect:the-lemmas-and-their-proofs}.
We  point out some properties of $\Cut_\alpha(\xi)$.
\begin{compactitem}
\item The set $\Cut_\alpha(\xi)$ is finite and nonempty; in particular, $(\varepsilon)\in \Cut_\alpha(\xi)$ and $\xi[\alpha]_{(\varepsilon)} = \xi[\alpha]_{\varepsilon} = \alpha$.
\item For each $\widetilde{w} = (w_1,\ldots,w_n)$ in $\Cut_\alpha(\xi)$, we have 
$\pos_\alpha(\xi[\alpha]_{\widetilde{w}})= \{w_1,\ldots,w_n\}$.
\item The following equivalence holds: $\pos_\alpha(\xi)=\emptyset$ if, and only if $() \in \Cut_\alpha(\xi)$; moreover, $\xi[\alpha]_{()} = \xi$. 
\item $\Cut_\alpha(\alpha) = \{(\varepsilon)\}$ and, for each $\beta \in \Sigma^{(0)}\setminus \{\alpha\}$, we have $\Cut_\alpha(\beta) = \{(),(\varepsilon)\}$.
\end{compactitem}

\begin{figure}[t]
   \centering
    $\xi$:
\scalebox{.7}{
    \begin{tikzpicture}[baseline=(t),cut/.style={draw,circle},level 1/.style={sibling distance=3.5cm},level 2/.style={sibling distance=1.5cm}]
        \node (t) {$\eta$}
            child {node {$\eta$}
                child {node {$\beta$}}
                child {node[cut] {$\gamma$}
                    child {node {$\sigma$}
                        child {node {$\beta$}}
                        child {node {$\beta$}}
                    }
                }
                child {node {$\beta$}}
            }
            child {node {$\sigma$}
                child {node[cut] {$\beta$}}
                child {node[cut] {$\eta$}
                    child {node {$\alpha$}}
                    child {node {$\gamma$}
                        child {node {$\alpha$}}
                    }
                    child {node {$\beta$}}
                }
            }
            child {node[cut] {$\alpha$}};
    \end{tikzpicture}
}
    \[
        (12, 21, 22, 3) \in \mathrm{Cut}_\alpha(\xi)
    \]
\caption{\label{fig:cut}An illustration of an $\alpha$-cut through $\xi$. Position $22$ covers the $\alpha$ at positions $221$ and $2221$. Moreover, position $3$ covers $\alpha$ at position $3$.}
\end{figure}

\index{tree concatenation}
\index{alphaconcat@$\alpha$-concatenation}
\index{$r_1 \circ_\alpha r_2$}
Let $r_1: \T_\Sigma \rightarrow B$ and $r_2: \T_\Sigma \rightarrow B$ be weighted tree languages. Moreover, let $\alpha \in \Sigma^{(0)}$. The {\em $\alpha$-concatenation of $r_1$ and $r_2$} \cite{dropecvog05} is the weighted tree language $(r_1 \circ_\alpha r_2): \T_\Sigma \rightarrow B$ defined for every $\xi \in \T_\Sigma$ by 
\[(r_1 \circ_\alpha r_2)(\xi) = \bigoplus\limits_{(w_1,\ldots,w_n) \in \Cut_\alpha(\xi)} 
r_1(\xi[\alpha]_{(w_1,\ldots,w_n)}) \otimes r_2(\xi|_{w_1})\otimes \ldots \otimes r_2(\xi|_{w_n})
\]
Since, for each $\xi\in \T_\Sigma$, the index set $\Cut_\alpha(\xi)$ is finite, we do not need the condition that $\B$ is $\sigma$-complete for the summation. Moreover, we have $(r_1 \circ_\alpha r_2)(\alpha) = r_1(\alpha) \otimes r_2(\alpha)$ and, for each $\beta \in \Sigma^{(0)}\setminus \{\alpha\}$, we have  $(r_1 \circ_\alpha r_2)(\beta) = r_1(\beta) \oplus r_1(\alpha) \otimes r_2(\beta)$. The \emph{tree concatenation of $r_1$ and $r_2$} is the $\alpha$-concatenation of $r_1$ and $r_2$ for some $\alpha \in \Sigma^{(0)}$.

A set $\cL$ of $\B$-weighted tree languages is \emph{closed under tree concatenations} if the following holds: for every $(\Sigma,\B)$-weighted tree languages  $r_1$ and $r_2$ and for each $\alpha \in \Sigma^{(0)}$, if $r_1,r_2 \in \cL$, then $(r_1 \circ_\alpha r_2) \in \cL$.

\begin{theorem}\label{thm:closure-tree-concatenation-wrtg} {\rm (cf. \cite[Thm.~3.35]{eng75-15}, \cite[Thm.~2.4.6]{gecste84}, and  \cite[Lm. 6.6]{fulvog19a})} Let $\B$ be a commutative semiring and let $\alpha \in \Sigma^{(0)}$. Let $\cG_1$ and $\cG_2$ be two $(\Sigma,\B)$-wrtg such that $\cG_1$ and $\cG_2$ are finite-derivational or $\B$ is $\sigma$-complete. Then the following two statements hold.
  \begin{compactenum}
    \item[(1)] There exists a  $(\Sigma,\B)$-wrtg $\cG$ such that $\sem{\cG} = \sem{\cG_1} \circ_\alpha  \sem{\cG_2}$.
    \item[(2)] If $\cG_1$ and $\cG_2$ are finite-derivational, then we can construct a finite-derivational  $(\Sigma,\B)$-wrtg $\cG$ such that $\sem{\cG} = \sem{\cG_1} \circ_\alpha  \sem{\cG_2}$.
\end{compactenum}
    \end{theorem}
  \begin{proof}  Proof of (1): Let $\cG_1=(N_1,S_1,R_1,wt_1)$ and $\cG_2=(N_2,S_2,R_2,wt_2)$ be two $(\Sigma,\B)$-wrtg with $N_1 \cap N_2 = \emptyset$.  By Lemmas \ref{lm:normal-form-lemmas-inherited-from-wcfg}(1) and \ref{lm:rtg-normal-form}, we can assume that $\cG_1$ and $\cG_2$ are start-separated and alphabetic. By Lemma \ref{lm:normal-form-lemmas-inherited-from-wcfg}(3) we can furthermore assume that $\cG_1$ and $\cG_2$ are  chain-free. 
We call a rule $A \to \alpha$ of $R_1$ an \emph{$\alpha$-rule}.

We define the $(\Sigma,\B)$-wrtg $\cG=(N,S_1,R,wt)$ such that $\sem{\cG} = \sem{\cG_1} \circ_\alpha \sem{\cG_2}$ as follows. We let $N = N_1 \cup N_2$. 
Moreover, we let $R = \overline{R_1} \cup R_2$ where $\overline{R_1}$ is obtained from $R_1$ by replacing each $\alpha$-rule  
$r = (A \rightarrow \alpha)$   
by the rule
$r' = (A \rightarrow S_2)$,
and we define $wt(r') = wt_1(r)$. Each rule $r \in R$ which is not of the form $A \rightarrow S_2$ keeps the weight from its original grammar. We call each rule of the form $A \rightarrow S_2$ a $S_2$-rule. It is obvious that, if $\cG_1$ and $\cG_2$ are finite-derivational, then also $\cG$ is finite-derivational. Hence $\sem{\cG}$ is defined.

\begin{figure}[t]
\begin{center}
\begin{tikzpicture}[scale=0.7,level distance=4.5em,
  every node/.style = {align=center}]]

\pgfdeclarelayer{bg}    
\pgfsetlayers{bg,main}  


\begin{scope}[level 1/.style={sibling distance=50mm},
level 2/.style={sibling distance=30mm}, level 3/.style={sibling distance=15mm}]
 \node (root1) {$S_1\to\sigma(A,B)$}
 child {node (ch11){$A\to\alpha$}}
 child {node (ch12){$B\to \delta(B',D)$}
 	child { node (ch13) {$B'\to\beta$} }
    child { node (ch14) {$D\to \alpha$} } };
\end{scope}

\begin{scope}[xshift=-40mm,yshift=-5mm,level 1/.style={sibling distance=13mm}, level 2/.style={sibling distance=25mm}, level 3/.style={sibling distance=13mm}]
 \node {$\cG_1$};
 \end{scope}

\begin{scope}[xshift=-22mm,yshift=-27mm,level 1/.style={sibling distance=13mm}, level 2/.style={sibling distance=25mm}, level 3/.style={sibling distance=13mm}]
 \node (root2){$S_2\to \gamma(E)$}
 child {node (ch21) {$E\to \beta$}};
 \end{scope}

\begin{scope}[xshift=-40mm,yshift=-35mm,level 1/.style={sibling distance=13mm}, level 2/.style={sibling distance=25mm}, level 3/.style={sibling distance=13mm}]
 \node {$\cG_2$};
 \end{scope}

\begin{scope}[xshift=120, yshift=-42mm,level 1/.style={sibling distance=13mm}, level 2/.style={sibling distance=25mm}, level 3/.style={sibling distance=13mm}]
 \node (root3) {$S_2\to \gamma(F)$}
 child {node (ch31) {$F\to \beta$}};
 \end{scope}

\begin{scope}[xshift=21mm,yshift=-48mm,level 1/.style={sibling distance=13mm}, level 2/.style={sibling distance=25mm}, level 3/.style={sibling distance=13mm}]
 \node {$\cG_2$};
 \end{scope}


\setlength{\ys}{-65mm}

\begin{scope}[xshift=0mm,yshift=\ys+0mm,level 1/.style={sibling distance=50mm},
level 2/.style={sibling distance=30mm}, level 3/.style={sibling distance=15mm}]
 \node (root1) {$S_1\to\sigma(A,B)$}
 child {node (ch11){$A\to S_2$}
 child {node (root2){$S_2\to \gamma(E)$}
 child {node (ch21) {$E\to \beta$}}}
}
 child {node (ch12){$B\to \delta(B',D)$}
 	child { node (ch13) {$B'\to\beta$} }
    child { node (ch14) {$D\to S_2$}
 child {node (root3) {$S_2\to \gamma(F)$}
 child {node (ch31) {$F\to \beta$}}}
 } };
\end{scope}

\begin{scope}[xshift=-40mm,yshift=\ys+-5mm,level 1/.style={sibling distance=13mm}, level 2/.style={sibling distance=25mm}, level 3/.style={sibling distance=13mm}]
 \node {$\cG$};
 \end{scope}

\end{tikzpicture}
\end{center}
\caption{\label{fig:concatenation-Joost} Illustration for the proof of $\sem{\cG} = \sem{\cG_1} \circ_\alpha \sem{\cG_2}$.}
\end{figure}

Now we prove that $\sem{\cG} = \sem{\cG_1} \circ_\alpha \sem{\cG_2}$ (see  Figure \ref{fig:concatenation-Joost}).
For this, let $\xi \in \T_\Sigma$ and $\widetilde{w} = (w_1,\ldots,w_n) \in \Cut_\alpha(\xi)$. 
We define
\begin{align*}\RT_{\cG}^{\widetilde{w}}(\xi) = \big\{  d\in \RT_\cG(\xi) \mid  \{w_1,\ldots,w_n\} = \{u \in \pos(d) \mid  \text{ $d(u)$ is a $S_2$-rule}\}\big\},
\end{align*}
Obviously, if $()\in \Cut_\alpha(\xi)$, i.e., $\pos_\alpha(\xi)=\emptyset$, then $\RT_{\cG}^{()}(\xi)=\RT_{\cG_1}(\xi)$. 
The tree in the lower part of Figure \ref{fig:concatenation-Joost} illustrates an element of $\RT_{\cG}^{(1,22)}(\xi)$, where $\xi=\sigma\big(\gamma(\beta),\delta(\beta,\gamma(\beta))\big)$. 
The following is obvious:
\begin{equation}
(\RT_\cG^{\widetilde{w}}(\xi) \mid \widetilde{w} \in \Cut_\alpha(\xi)) \text{ is a partitioning of } \RT_\cG(\xi)\enspace. \label{equ:tree-concat-partitioning}
  \end{equation}

We will show that there exists a bijection
\[
\Phi: \RT_{\cG_1}(\xi[\alpha]_{\widetilde{w}}) \times \RT_{\cG_2}(\xi|_{w_1}) \times \ldots \times \RT_{\cG_2}(\xi|_{w_n}) \rightarrow  \RT_{\cG}^{\widetilde{w}}(\xi).
\]
For this, let  $d \in \RT_{\cG_1}(\xi[\alpha]_{\widetilde{w}})$, $d_1 \in \RT_{\cG_2}(\xi|_{w_1}), \ldots, d_n \in \RT_{\cG_2}(\xi|_{w_n})$. Since $\cG_1$ is alphabetic and chain-free, we have $\pos(d)=\pos(\xi[\alpha]_{\widetilde{w}})$ and, moreover, 
\[\{w_1,\ldots,w_n\}=\{u\in \pos(d)\mid d(u) \text{ is an $\alpha$-rule} \}\enspace.\]
First we define a tree $d'\in \T_R$ by specifying a tree domain and an $R$-tree mapping and then using the bijective representation of trees as tree domains (cf. Section \ref{sect:trees}). 
We define the tree domain
\begin{equation} \label{equat:positions}
W= \pos(d)\cup \bigcup_{i\in[n]}\{w_i1v\mid v\in \pos(d_i)\} 
\end{equation}
and we define the mapping $d': W \to \Sigma$ for each $v \in W$ by 
\begin{equation} \label{equat:labels}
d'(v) = 
\begin{cases}
d(v)    &  \text{if }(\forall i \in [n]) : \neg (w_i \le_{\mathrm{pref}} v) \\
d_i(u)  & \text{if } (\exists i \in [n]) (\exists u \in \mathbb{N}^+):  v=w_i1 u \\
t_i          & \text{if } (\exists i \in [n]) :  v = w_i  \enspace,
\end{cases}
\end{equation}
where $t_i$ is defined as follows: if $d(w_i) = (A\to\alpha)$, then $t_i = (A \rightarrow S_2)$.

Since $W$ is a tree domain and $d'$ is an $R$-tree mapping, we can view $d'$ as a tree over $R$, i.e., $d' \in \T_R$.
It is obvious that $d'\in \RT_{\cG}^{\widetilde{w}}(\xi)$.
Then we  define $\Phi$ by letting
\begin{equation} \label{equat:Phi}
\Phi(d,d_1,\ldots,d_k)=d'.
\end{equation}
It is easy to see that $\Phi$ is bijective. To  show that it is injective, we assume that (\ref{equat:Phi}) holds and that also $\Phi(\hat{d},\hat{d_1},\ldots,\hat{d_k})=d'$ for some
$\hat{d} \in \RT_{\cG_1}(\xi[\alpha]_{\widetilde{w}})$, $\hat{d_1} \in \RT_{\cG_2}(\xi|_{w_1}), \ldots, \hat{d_n} \in \RT_{\cG_2}(\xi|_{w_n})$. We prove that $d=\hat{d}$ by contradiction. If $d\ne\hat{d}$, then there exists an $u\in \pos(d)\cap \pos(\hat{d})$ such that $d(u)\ne \hat{d}(u)$. We note that $d(u),\hat{d}(u)\in R_1$ and that, by (\ref{equat:positions}), also $u\in \pos(d')$. The following four cases are possible:
\begin{compactitem}
\item[-] Neither $d(u)$ nor $\hat{d}(u)$ is an $\alpha$-rule. Then, by  (\ref{equat:labels}), $d'(u)=d(u)$ and $d'(u)=\hat{d}(u)$, which is a contradiction.
\item[-] $d(u)=(A\to \alpha)$  and $\hat{d}(u)$ is not an $\alpha$-rule. Then, by  (\ref{equat:labels}),
$d'(u)=(A \rightarrow S_2)$. On the other hand, $d'(u)=\hat{d}(u)$ is a rule in $R_1$, a contradiction.
\item[-] $d(u)$ is not an $\alpha$-rule and $\hat{d}(u)$ is an $\alpha$-rule. By symmetry, this also  leads to a contradiction.
\item[-] $d(u)=(A\to \alpha)$ and $\hat{d}(u)=(B\to \alpha)$, where $A\ne B$. Then by  (\ref{equat:labels}), $d'(u)=(A \rightarrow S_2)$ and also $d'(u)=(B \rightarrow S_2)$, which is a contradiction.
\end{compactitem}
So we have $d=\hat{d}$. Then, by (\ref{equat:positions}) and (\ref{equat:labels}), $d_i=\hat{d}_i$ which proves that $\Phi$ is injective. Also, it is surjective, because for each $d' \in \RT_{\cG}^{\widetilde{w}}(\xi)$, the positions $w_1,\ldots,w_n$
determine $d \in \RT_{\cG_1}(\xi[\alpha]_{\widetilde{w}})$, $d_1 \in \RT_{\cG_2}(\xi|_{w_1}), \ldots, d_n \in \RT_{\cG_2}(\xi|_{w_n})$ such that $\Phi(d,d_1,\ldots,d_k)=d'$.
Thus $\Phi$ is bijective.

Moreover, if (\ref{equat:Phi}) holds, then
\begin{equation}\label{equat:weights}
\wt_\cG(d') = \wt_{\cG_1}(d) \otimes \bigotimes_{i\in[n]} \wt_{\cG_2}(d_i)
\end{equation}
because $\B$ is commutative.

Then we can compute as follows (keeping in mind that $\cG_1$, $\cG_2$, and $\cG$ are finite-derivational if $\B$ is not $\sigma$-complete):
\begingroup
\allowdisplaybreaks
\begin{align*}
& \hspace*{5mm}(\sem{\cG_1} \circ_\alpha \sem{\cG_2})(\xi) \\
& =  \bigoplus_{\widetilde{w}=(w_1,\ldots,w_n) \in \Cut_\alpha(\xi)} \sem{\cG_1}(\xi[\alpha]_{\widetilde{w}}) \otimes \sem{\cG_2}(\xi|_{w_1})\otimes \ldots \otimes \sem{\cG_2}(\xi|_{w_n})\\
& = \bigoplus_{\widetilde{w}=(w_1,\ldots,w_n) \in \Cut_\alpha(\xi)} \Big(\ \infsum{\oplus}{d \in \RT_{\cG_1}(\xi[\alpha]_{\widetilde{w}})} \wt_{\cG_1}(d) \Big) \ \otimes \ \bigotimes_{i\in[n]} \Big(\ \infsum{\oplus}{d_i \in \RT_{\cG_2}(\xi|_{w_i})} \wt_{\cG_2}(d_i) \Big)\\
  & = \bigoplus_{\widetilde{w}=(w_1,\ldots,w_n) \in \Cut_\alpha(\xi)} \infsum{\oplus}{\substack{d \in \RT_{\cG_1}(\xi[\alpha]_{\widetilde{w} }),\\(\forall i \in [n]): d_i \in \RT_{\cG_2}(\xi|_{w_i})}} \Big(  \wt_{\cG_1}(d) \otimes  \bigotimes_{i\in[n]} \wt_{\cG_2}(d_i)\Big)
  \tag{by distributivity}\\[2mm]
  & = \bigoplus_{\widetilde{w}\in \Cut_\alpha(\xi)}\ \ \infsum{\oplus}{d' \in \RT_{\cG}^{\widetilde{w}}(\xi)} \wt_\cG(d')
    \tag{because $\Phi$ is a bijection and by \eqref{equat:weights}}\\
  & =\infsum{\oplus}{d' \in \RT_{\cG}(\xi)} \wt_\cG(d')
    \tag{by \eqref{equ:tree-concat-partitioning}}\\
  & = \ \sem{\cG}(\xi)\enspace.
\end{align*}
\endgroup
Thus $\sem{\cG} = \sem{\cG_1} \circ_\alpha \sem{\cG_2}$.

\

Proof of (2): Now let $\cG_1$ and $\cG_2$ finite-derivational. Then, by Lemma \ref{lm:normal-form-lemmas-inherited-from-wcfg}(3), we can even \underline{construct} equivalent chain-free wrtg. Consequently, the definition of $\cG$, as it is given in the proof of (1), is constructive. Finally, as pointed out in the proof of (1), the wrtg $\cG$ is finite-derivational. 
\end{proof}

\begin{corollary-rect}\label{cor:closure-tree-concatenation}\rm (cf. \cite[Lm. 6.5]{dropecvog05}) Let $\Sigma$ be a ranked alphabet. Moreover, let $\B$ be a commutative semiring, and let $\cA_1$ and $\cA_2$ be two $(\Sigma,\B)$-wta. Moreover, let  $\alpha \in \Sigma^{(0)}$. Then we can construct a $(\Sigma,\B)$-wta $\cA$ such that  $\sem{\cA} = \sem{\cA_1} \circ_\alpha  \sem{\cA_2}$.
  Thus, in particular, if $\B$ is a commutative semiring, then  the set $\Rec(\Sigma,\B)$ is closed under tree concatenations.
\end{corollary-rect}
\begin{proof}  By Lemma \ref{lm:wta-to-wrtg} we can construct $(\Sigma,\B)$-wrtg $\cG_1$ and $\cG_2$ such that $\cG_1$ and $\cG_2$ are in tree automata form and $\sem{\cA_1} = \sem{\cG_1}$ and $\sem{\cA_2} = \sem{\cG_2}$.  Then, in particular, $\cG_1$ and $\cG_2$ are  finite-derivational. By Theorem \ref{thm:closure-tree-concatenation-wrtg} we can construct a  finite-derivational $(\Sigma,\B)$-wrtg $\cG$ such that  $\sem{\cG} = \sem{\cG_1} \circ_\alpha  \sem{\cG_2}$. Then,  by Lemma \ref{lm:wrtg-to-wta}, we can construct a $(\Sigma,\B)$-wta $\cA$ such that $\sem{\cG}=\sem{\cA}$. 
\end{proof}


\section{Closure under Kleene stars}

In this section we recall the definition of Kleene star and show that $\Rec(\Sigma,\B)$ is closed under Kleene stars. We follow the approach developed in \cite{berreu82,eng03,dropecvog05}.

\index{iteration}
\index{riteration@$r_\alpha^{\ell}$}
Let $r: \T_\Sigma \rightarrow B$ and $\alpha\in\Sigma^{(0)}$. We define the family $(r_\alpha^{\ell} \mid \ell \in \mathbb{N})$ of weighted tree languages $r_\alpha^{\ell}: \T_\Sigma \rightarrow B$ (called the \emph{$\ell$-th iteration of $r$ at $\alpha$}) by induction on $\mathbb{N}$ as follows: 
\begin{compactenum} 
\item[I.B.:] $r_\alpha^{0} = \widetilde{\0}$ and 
\item[I.S.:] $r_\alpha^{\ell+1} =
  (r \circ_\alpha r_\alpha^{\ell}) \oplus \1.\alpha$ for each $\ell \in \mathbb{N}$. 
\end{compactenum} 

We mention that iteration can also be defined in different ways. For instance, if one generalizes the iteration of tree languages as presented in \cite[p.~38]{eng75-15} to the weighted case, then this would read (cf. \cite[Def.~3.7]{dropecvog05}):
\begin{compactenum} 
\item[I.B.:] $r_\alpha^{\mathrm{E},0} = \1.\alpha$ and 
\item[I.S.:] $r_\alpha^{\mathrm{E},\ell+1} =
  r_\alpha^{\mathrm{E},\ell} \circ_\alpha (r \oplus \1.\alpha)$ for each $\ell \in \mathbb{N}$. 
\end{compactenum} 
Another possibility is to generalize the iteration of tree languages as presented in \cite[p.~66]{thawri68} to the weighted case; this reads as follows (cf. \cite[Def.~3.7]{dropecvog05}): 
\begin{compactenum} 
\item[I.B.:] $r_\alpha^{\mathrm{TW},0} = \1.\alpha$ and 
\item[I.S.:] $r_\alpha^{\mathrm{TW},\ell+1} =
  (r  \oplus \1.\alpha) \circ_\alpha r_\alpha^{\mathrm{TW},\ell}$ for each $\ell \in \mathbb{N}$. 
\end{compactenum}
For the Boolean semiring, all three definitions are equivalent. This is due to the facts that disjunction is idempotent and that tree concatenation is associative. If $\B$ is a commutative semiring, then  $r_\alpha^{\mathrm{E},\ell} = r_\alpha^{\mathrm{TW},\ell}$ for each $\ell \in \mathbb{N}$ (cf. \cite[Lm.~3.8]{dropecvog05}). If $\B$ is a commutative and idempotent semiring, then $r_\alpha^\ell \le r_\alpha^{\mathrm{TW},\ell} \le r_\alpha^{\ell+1}$ for each $\ell \in \mathbb{N}$, where $\le$ is defined componentwise: for every $a,b\in B$, we let $a \le b$ if $a \oplus b = b$. For a more detailed comparison we refer the reader to \cite[Sec.~3]{dropecvog05}.

\index{proper}
\index{alphaproper@$\alpha$-proper}
We call $r: \T_\Sigma \to B$ \emph{$\alpha$-proper} if $r(\alpha)=\mathbb{0}$.
For each $\alpha$-proper weighted tree language $r$, the application of the iteration of $r$ to a tree $\xi$ becomes stable after $\height(\xi)+1$ steps. 

\begin{lemma}\rm (cf. \cite[Lm. 3.10]{dropecvog05})\label{lm:iteration-terminates} \rm Let $r: \T_\Sigma \to B$ be $\alpha$-proper. For every $\xi \in \T_\Sigma$ and $\ell \in \mathbb{N}$, if $\ell \geq \height(\xi)+1$, then $r_\alpha^{\ell+1}(\xi) = r_\alpha^\ell(\xi)$. In particular, for every $\ell \in \mathbb{N}_+$ and $\beta \in \Sigma^{(0)} \setminus \{\alpha\}$, we have $r_\alpha^{\ell}(\alpha) = \1$ and $r_\alpha^\ell(\beta) = r(\beta)$.
\end{lemma}

\begin{proof}  We prove the first statement of the lemma by induction on $(\T_\Sigma,\succ_\Sigma^+)$; the second statement of the lemma is proved in the induction base.

I.B.:  Let $\xi \in \Sigma^{(0)}$. We distinguish two cases.

  \underline{Case (a):} Let $\xi = \alpha$. First we prove that $r_\alpha^{\ell}(\alpha) = \1$  for each $\ell \in \mathbb{N}_+$.
  \begingroup
  \allowdisplaybreaks
  \begin{align*}
    r_\alpha^{\ell}(\alpha) &= \Big((r \circ_\alpha r_\alpha^{\ell -1}) \oplus \1.\alpha\Big)(\alpha)
                       = (r \circ_\alpha r_\alpha^{\ell -1})(\alpha) \oplus (\1.\alpha)(\alpha)\\
                       &= r(\alpha) \otimes r_\alpha^{\ell -1}(\alpha) \oplus \1
                       = \0 \otimes r_\alpha^{\ell -1}(\alpha) \oplus \1 
    = \1 \enspace,
  \end{align*}
  \endgroup
  where the third equality is due to the facts that $\Cut_\alpha(\xi)=\{(\varepsilon)\}$ and $(\1.\alpha)(\alpha)=\1$. Since $\height(\xi) +1 = 1$, this implies that, for each $\ell \geq \height(\xi)+1$, we have $r_\alpha^{\ell+1}(\alpha) = r_\alpha^\ell(\alpha)$.

  \underline{Case (b):} Let $\xi \ne \alpha$. First we prove that $r_\alpha^\ell(\xi) = r(\xi)$  for each $\ell \in \mathbb{N}_+$. 
  \begingroup
  \allowdisplaybreaks
  \begin{align*}
    r_\alpha^\ell(\xi) &= \Big((r \circ_\xi r_\alpha^{\ell-1}) \oplus \1.\alpha\Big)(\xi)
                        = (r \circ_\alpha r_\alpha^{\ell-1})(\xi) \oplus (\1.\alpha)(\xi)
    =  (r \circ_\alpha r_\alpha^{\ell-1})(\xi)\\
                       &= r(\xi) \oplus r(\alpha) \otimes r_\alpha^{\ell-1}(\xi)
                       = r(\xi) \oplus \0 \otimes r_\alpha^{\ell-1}(\xi)
    = r(\xi) \enspace,
  \end{align*}
  \endgroup
where the fourth equality is due to the fact that $\Cut_\alpha(\xi)=\{(),(\varepsilon)\}$.  
Since $\height(\xi) +1 = 1$, this implies that, for each $\ell \geq \height(\xi)+1$, we have $r_\alpha^{\ell+1}(\xi) = r_\alpha^\ell(\xi)$.

I.S.: Now let $\xi = \sigma(\xi_1,\ldots,\xi_k)$ with  $k \in \mathbb{N}_+$. Let $\ell \in \mathbb{N}$ with $\ell \ge \height(\xi) +1$.
\begingroup
\allowdisplaybreaks
\begin{align*}
  r_\alpha^{\ell+1}(\xi) &= (r \circ r_\alpha^{\ell})(\xi) \oplus (\1.\alpha)(\xi) = (r \circ r_\alpha^{\ell})(\xi) \tag{because $\xi \ne \alpha$}\\
  &=  \bigoplus\limits_{(w_1,\ldots,w_n) \in \Cut_\alpha(\xi)} 
    r(\xi[\alpha]_{(w_1,\ldots,w_n)}) \otimes r_\alpha^{\ell}(\xi|_{w_1})\otimes \ldots \otimes r_\alpha^{\ell}(\xi|_{w_n})\\[2mm]
                      &= \  r(\xi[\alpha]_{(\varepsilon)}) \otimes r_\alpha^{\ell}(\xi|_{\varepsilon}) \  \oplus\\
  &\ \ \ \ \  \bigoplus\limits_{(w_1,\ldots,w_n) \in \Cut_\alpha(\xi)\setminus \{(\varepsilon)\}} 
    r(\xi[\alpha]_{(w_1,\ldots,w_n)}) \otimes r_\alpha^{\ell}(\xi|_{w_1})\otimes \ldots \otimes r_\alpha^{\ell}(\xi|_{w_n}) \\[2mm]
     &= \  r(\alpha) \otimes r_\alpha^{\ell}(\xi|_{\varepsilon}) \  \oplus\\
  &\ \ \ \ \  \bigoplus\limits_{(w_1,\ldots,w_n) \in \Cut_\alpha(\xi)\setminus \{(\varepsilon)\}} 
    r(\xi[\alpha]_{(w_1,\ldots,w_n)}) \otimes r_\alpha^{\ell}(\xi|_{w_1})\otimes \ldots \otimes r_\alpha^{\ell}(\xi|_{w_n}) \\[3mm]
   &= \bigoplus\limits_{(w_1,\ldots,w_n) \in \Cut_\alpha(\xi)\setminus \{(\varepsilon)\}} 
     r(\xi[\alpha]_{(w_1,\ldots,w_n)}) \otimes r_\alpha^{\ell}(\xi|_{w_1})\otimes \ldots \otimes r_\alpha^{\ell}(\xi|_{w_n})
  \tag{because $r(\alpha)=\0$}\\[3mm]
  &= \bigoplus\limits_{(w_1,\ldots,w_n) \in \Cut_\alpha(\xi)\setminus \{(\varepsilon)\}} 
     r(\xi[\alpha]_{(w_1,\ldots,w_n)}) \otimes r_\alpha^{\ell-1}(\xi|_{w_1})\otimes \ldots \otimes r_\alpha^{\ell-1}(\xi|_{w_n})
  \tag{because, for each $i \in [\ell]$, we have $|w_i| \ge 1$ and hence $\ell-1 \ge \height(\xi|_{w_i})+1$, and then by I.H. }\\[3mm]
  &= \bigoplus\limits_{(w_1,\ldots,w_n) \in \Cut_\alpha(\xi)} 
    r(\xi[\alpha]_{(w_1,\ldots,w_n)}) \otimes r_\alpha^{\ell-1}(\xi|_{w_1})\otimes \ldots \otimes r_\alpha^{\ell-1}(\xi|_{w_n})
    \tag{because $r(\xi[\alpha]_{(\varepsilon)}) \otimes r_\alpha^{\ell-1}(\xi|_{\varepsilon}) = r(\alpha) \otimes r_\alpha^{\ell-1}(\xi|_{\varepsilon}) = \0$ as above} \\[2mm]
  &= (r \circ_\alpha r_\alpha^{\ell-1})(\xi) = (r \circ r_\alpha^{\ell-1})(\xi) \oplus (\1.\alpha)(\xi) =  r_\alpha^{\ell}(\xi)\enspace.   \qedhere
\end{align*}
\endgroup
\end{proof}

\

Lemma \ref{lm:iteration-terminates}  justifies to define the operation Kleene star as follows.
\index{Kleene star}
\index{alphaKleene@$\alpha$-Kleene star}
Let $r: \T_\Sigma \to B$ be $\alpha$-proper. The {\em $\alpha$-Kleene star of~$r$}, denoted by $r_\alpha^*$, is the weighted tree language $r_\alpha^*: \T_\Sigma \rightarrow B$ defined, for every $\xi \in \T_\Sigma$, by 
$r_\alpha^*(\xi) = r_\alpha^{\height(\xi)+1}(\xi)$  \cite{eng03} (cf. \cite[Def.~3.11]{dropecvog05}).  Thus, in particular, by Lemma \ref{lm:iteration-terminates}, we have
\begin{equation}
  r_\alpha^*(\alpha) = r_\alpha^1(\alpha) = \1.  \label{equ:star-alpha=1}
\end{equation}
Moreover, for each $\beta \in \Sigma^{(0)}$ with $\beta \ne \alpha$, by Lemma \ref{lm:iteration-terminates}, we have
\begin{equation}
   r^*_\alpha(\beta) =  r^{1}_\alpha(\beta) = r(\beta). \label{equ:star-beta=r(beta)}
\end{equation}

\begin{example}\label{ex:calculation-star} \rm Let $\Sigma = \{\sigma^{(2)}, \alpha^{(0)}\}$. We consider the $(\Sigma,\Nat)$-weighted tree language $r = 1.\sigma(\alpha,\alpha)$, which is a monomial,  and compute $r_\alpha^*(\sigma(\alpha,\alpha))$:
  \begingroup
  \allowdisplaybreaks
  \begin{align*}
    &r_\alpha^*(\sigma(\alpha,\alpha)) = r_\alpha^2(\sigma(\alpha,\alpha))
                                                = (r \circ_\alpha r_\alpha^1)(\sigma(\alpha,\alpha)) + (1.\alpha)(\sigma(\alpha,\alpha))\\
                                              &= (r \circ_\alpha r_\alpha^1)(\sigma(\alpha,\alpha))\\
  &= r(\alpha) \cdot r_\alpha^1(\sigma(\alpha,\alpha)) \  + 
      \      r(\sigma(\alpha,\alpha)) \cdot r_\alpha^1(\alpha)  \cdot r_\alpha^1(\alpha)
    \tag{because  $\Cut_\alpha(\sigma(\alpha,\alpha)) = \{(\varepsilon),(1,2)\}$}\\
    &= r(\sigma(\alpha,\alpha)) \cdot r_\alpha^1(\alpha)  \cdot r_\alpha^1(\alpha) \tag{because $r(\alpha)=0$}\\
    & = 1  \cdot r_\alpha^1(\alpha)  \cdot r_\alpha^1(\alpha)\\
    & = 1 \tag{by  Lemma \ref{lm:iteration-terminates}} \enspace.
  \end{align*}
  \endgroup

We might wish to compare the calculation of $r_\alpha^2(\sigma(\alpha,\alpha))$ with that of $r_\alpha^3(\sigma(\alpha,\alpha))$:
  \begingroup
  \allowdisplaybreaks
  \begin{align*}
    &r_\alpha^3(\sigma(\alpha,\alpha))
                      = (r \circ_\alpha r_\alpha^2)(\sigma(\alpha,\alpha)) + (1.\alpha)(\sigma(\alpha,\alpha))\\
                                              &= (r \circ_\alpha r_\alpha^2)(\sigma(\alpha,\alpha))\\
  &= r(\alpha) \cdot r_\alpha^2(\sigma(\alpha,\alpha)) \  + 
      \      r(\sigma(\alpha,\alpha)) \cdot r_\alpha^2(\alpha)  \cdot r_\alpha^2(\alpha)
    \tag{because  $\Cut_\alpha(\sigma(\alpha,\alpha)) = \{(\varepsilon),(1,2)\}$}\\
    &= r(\sigma(\alpha,\alpha)) \cdot r_\alpha^2(\alpha)  \cdot r_\alpha^2(\alpha) \tag{because $r(\alpha)=0$}\\
    & = 1  \cdot r_\alpha^2(\alpha)  \cdot r_\alpha^2(\alpha)\\
    & = 1 \tag{by Lemma \ref{lm:iteration-terminates}} \enspace.
  \end{align*}
  \endgroup

Next we compute $r_\alpha^*(\sigma(\alpha,\sigma(\alpha,\alpha)))$:
        \begingroup
  \allowdisplaybreaks
  \begin{align*}
    &r_\alpha^*(\sigma(\alpha,\sigma(\alpha,\alpha))) = r_\alpha^3(\sigma(\alpha,\sigma(\alpha,\alpha)))\\
                                                &= (r \circ_\alpha r_\alpha^2)(\sigma(\alpha,\sigma(\alpha,\alpha))) + (1.\alpha)(\sigma(\alpha,\sigma(\alpha,\alpha)))\\
                                              &= (r \circ_\alpha r_\alpha^2)(\sigma(\alpha,\sigma(\alpha,\alpha)))\\
    &= r(\alpha) \cdot r_\alpha^2(\sigma(\alpha,\sigma(\alpha,\alpha)))\\
    &\ \ \  +  r(\sigma(\alpha,\alpha)) \cdot r_\alpha^2(\alpha)  \cdot r_\alpha^2(\sigma(\alpha,\alpha))\\
    &\ \ \  +   r(\sigma(\alpha,\sigma(\alpha,\alpha))) \cdot r_\alpha^2(\alpha)  \cdot r_\alpha^2(\alpha) \cdot r_\alpha^2(\alpha)
      \tag{because  $\Cut_\alpha(\sigma(\alpha,\sigma(\alpha,\alpha))) = \{(\varepsilon),(1,2),(1,21,22)\}$}\\
 &=   r(\sigma(\alpha,\alpha)) \cdot r_\alpha^2(\alpha)  \cdot r_\alpha^2(\sigma(\alpha,\alpha))
   \tag{because $r(\alpha)=r(\sigma(\alpha,\sigma(\alpha,\alpha)))= 0$}\\
    &= r_\alpha^2(\alpha)  \cdot r_\alpha^2(\sigma(\alpha,\alpha))
      \tag{because $r(\sigma(\alpha,\alpha))= 1$}\\
     &= r_\alpha^2(\sigma(\alpha,\alpha))\enspace.
       \tag{because $r_\alpha^2(\alpha)= 1$}\\
    &=1 \tag{as above}
  \end{align*}
  \endgroup
      \hfill $\Box$
    \end{example}

    \begin{example} \rm \label{ex:abn-as-Kleene-star} We consider the alphabet $\Gamma = \{a,b\}$ and the string ranked alphabet $\Gamma_e = \{a^{(1)}, b^{(1)}, e^{(0)}\}$. Moreover, we consider the arctic semiring $\Nat_{\max,+} = (\mathbb{N}_{-\infty},\max,+,-\infty,0)$  and the monomial $1.a(b(e))$.
    
    We would like to prove that, for each $\xi \in \T_{\Gamma_e}$, we have
     \begin{equation}\label{equ:what-we-want-to-prove}
\big(1.a(b(e))\big)_e^*(\xi) =  \begin{cases} |\pos_a(\xi)| & \text{ if there exists $n \in \mathbb{N}$ such that $\xi = \tree_e((ab)^n)$}\\
                 -\infty &\text{ otherwise,}
                 \end{cases}
               \end{equation}
                 where  $\tree_e: \Gamma^* \to \T_{\Gamma_e}$ is the bijection defined on page \pageref{page:def-of-mapping-tree}.
We note that $1.a(b(e))$ is $e$-proper because $(1.a(b(e)))(e) = -\infty$.
In the rest of this example we abbreviate $1.a(b(e))$ by $r$.

We note that 
\begin{equation}
  \text{ for each $\xi \in \T_{\Gamma_e}$ we have $\cut_e((a(b(\xi))) = \{(\varepsilon), (1)\} \cup \{(11w) \mid (w) \in \cut_e(\xi)\}$} \enspace.
\end{equation}
Moreover, it is easy to see that
\begin{equation}\label{equ:not-11-is-infty}
\text{ for every $\xi \in \T_{\Gamma_e}$ and $(w) \in \cut_e(\xi)\setminus\{(11)\}$, we have $r(\xi[e]_{(w)}) = -\infty$.} 
\end{equation}

In the following, for every $n\in \mathbb{N}$ and $\xi \in \T_{\Gamma_e}$, we abbreviate the tree $\tree_e((ab)^n)[\xi]_{1^{2n}}$ by $(ab)^n\xi$.

First, by induction on $\mathbb{N}$,  we prove that
\begin{equation}\label{abn-iteration}
\text{ for every $n \in \mathbb{N}$ and $\xi \in \T_{\Gamma_e}$, we have $r_e^*((ab)^n\xi) = n + r_e^*(\xi)$} \enspace.
\end{equation}

Let $\xi \in \T_{\Gamma_e}$. 

I.B.: $r_e^*((ab)^0\xi) = r_e^*(e[\xi]_\varepsilon) = r_e^*(\xi) = 0 + r_e^*(\xi)$. 

I.S.: Let $n \in \mathbb{N}$. We abbreviate $\height(\xi)$ by $\ell$.

Then we can calculate as follows.
  \begingroup
    \allowdisplaybreaks
    \begin{align*}
      r_e^*((ab)^{n+1}\xi)
      &= r_e^{2(n+1) + \ell +1}((ab)^{n+1}\xi)  \tag{by definition of $r_e^*((ab)^{n+1}\xi)$}\\
      &= \max\Big( ( r \circ_e r_e^{2(n+1)+\ell} )((ab)^{n+1}\xi) \ , \ (0.e)((ab)^{n+1}\xi) \Big)
      \tag{by definition of $r_e^{2(n+1) + \ell +1}$}\\
      &= \max_{(w) \in \cut_e((ab)^{n+1}\xi)}  r(((ab)^{n+1}\xi)[e]_{(w)}) + r_e^{2(n+1)+\ell}(((ab)^{n+1}\xi)|_w)
\tag{because $(0.e)((ab)^{n+1}\xi) = -\infty$ and by definition of $\circ_e$} \\
&= r(((ab)^{n+1}\xi)[e]_{(11)}) + r_e^{2(n+1)+ \ell}(((ab)^{n+1}\xi)|_{11}) \tag{by \eqref{equ:not-11-is-infty}}\\
      &= r(a(b(e))) + r_e^{2(n+1)+ \ell}((ab)^{n}\xi) \\
      &= 1 + r_e^{2n+\ell+1}((ab)^{n}\xi)  \tag{by definition of $r$ and by Lemma \ref{lm:iteration-terminates}}\\
      &= 1 + r_e^*((ab)^{n}\xi) \tag{by definition of $r_e^*((ab)^{n}\xi)$}\\
      &= (n+1) + r_e^*(\xi)  \tag{by I.H.}\enspace.
\end{align*}
\endgroup
This proves \eqref{abn-iteration}.

Second, we prove that
  \begin{equation} \label{equ:induction-ab*=n}
  \text{for each $n \in \mathbb{N}$, we have $r_e^*(\tree_e((ab)^n)) = n$} \enspace.
  \end{equation}

Let $n \in \mathbb{N}$. Then we can calculate as follows.
\begin{align*}
  r_e^*(\tree_e((ab)^n)) &= r_e^*((ab)^ne)
                           = n + r_e^*(e)
\end{align*}
where the second equality is due to \eqref{abn-iteration} with $\xi = e$. 
Now we evaluate $r_e^*(e)$:
\begingroup
\allowdisplaybreaks
\begin{align*}
  r_e^*(e) &= r_e^{1}(e) \tag{by definition of $r_e^*(e)$}\\
      &= \max\Big( \big[r \circ_e r_e^{0}\big](e)\ , \  (0.e)(e)  \Big)
\tag{by definition of $r_e^{1}$}\\
&= \max\Big( \big[r \circ_e r_e^{0}\big](e)\ , \  0  \Big)\\
& = \max\Big( r(e[e]_{(\varepsilon)}) + r_e^{0}(e|_\varepsilon)\ , \  0  \Big)
\tag{by definition of $\circ_e$ and the fact that $\cut_e(e) =\{(\varepsilon)\}$}\\
& = \max\Big( r(e) + r_e^{0}(e)\ , \  0  \Big)\\
& = \max\Big( -\infty + - \infty \ , \  0  \Big)
\tag{because $r_e^{0} = \widetilde{-\infty}$}\\
&= 0 \enspace.
      \end{align*}
    \endgroup
This proves \eqref{equ:induction-ab*=n}.

Third, we prove that,
\begin{equation}\label{equ:not-abn=infty}
  \text{ for each $\xi \in \T_{\Gamma_e} \setminus \{\tree_e((ab)^n) \mid n\in \mathbb{N}\}$, we have $r_e^*(\xi) = -\infty$.}
\end{equation}
Let $\xi \in \T_{\Gamma_e} \setminus \{\tree_e((ab)^n) \mid n\in \mathbb{N}\}$. Then there exists an $n \in \mathbb{N}$ and a $\zeta \in \T_{\Gamma_e}$ such that
\[\xi = (ab)^n\zeta \  \text{ and } \zeta \in \{a(a(\zeta')) \mid \zeta'\in \T_{\Gamma_e}\} \cup \{a(e)\}
  \cup \{b(\zeta') \mid \zeta'\in \T_{\Gamma_e}\} \enspace.
\]

Thus, by \eqref{abn-iteration} and the definitions of $ r_e^*(\zeta)$ and $\circ_e$, we have
\begin{equation}\label{eq:r-e-star}
  r_e^*(\xi) = n + r_e^*(\zeta) =n+r_e^{\height(\zeta) +1}(\zeta)=n+\max\big((r \circ_e r_e^{\height(\zeta)})(\zeta) \ , \ (0.e)(\zeta) \big)\enspace.
\end{equation}  
 We proceed by case analysis.

\underline{$\zeta = a(a(\zeta'))$ for some $\zeta'\in \T_{\Gamma_e}$:}
Then 
  \begingroup
    \allowdisplaybreaks
\begin{align*}
               & \max\big((r \circ_e r_e^{\height(\zeta)})(\zeta) \ , \ (0.e)(\zeta) \big)\\
             &= \max\big( r(\zeta[e]_{(11)}) + r_e^{\height(\zeta)}(\zeta|_{11}) \ , \ (0.e)(\zeta) \big)
               \tag{by \eqref{equ:not-11-is-infty} and the fact that $(11) \in \cut_e(\zeta)$}\\
  &= \max\big( r(a(a(e))) + r_e^{\height(\zeta)}(\zeta') \ , \ (0.e)(a(a(\zeta'))) \big)\\
  &= \max\big( -\infty + r_e^{\height(\zeta)}(\zeta') \ , \ -\infty   \big) =  -\infty \enspace.
\end{align*}
\endgroup

\

\underline{$\zeta = a(e)$:} Thus $\height(\zeta) = 1$.
Then 
  \begingroup
    \allowdisplaybreaks
\begin{align*}
               & \max\big((r \circ_e r_e^{1})(a(e)) \ , \ (0.e)(a(e)) \big)\\
               &= \max\big((r \circ_e r_e^{1})(a(e)) \ , \ -\infty \big)\\
               &=  (r \circ_e r_e^{1})(a(e))\\
               &= \max\big( r(a(e)[e]_{(\varepsilon)}) + r_e^{1}(a(e)|_{\varepsilon}) \ , \
                 r(a(e)[e]_{(1)}) + r_e^{1}(a(e)|_{1}) \big)
                 \tag{because $\cut_e(a(e)) = \{(\varepsilon),(1)\}$}\\
               &= \max\big( r(e) + r_e^{1}(a(e)) \ , \
                 r(a(e)) + r_e^{1}(e) \big)\\
  &=  \max\big( -\infty + r_e^{1}(a(e))\ , \ -\infty +   r_e^{1}(e)\big)   = -\infty \enspace.
\end{align*}
\endgroup

\

\underline{$\zeta = b(\zeta')$ for some $\zeta'\in \T_{\Gamma_e}$:} 
Then we can proceed as follows.

If $\zeta' = e$, then we have
  \begingroup
    \allowdisplaybreaks
\begin{align*}
  &\max\big((r \circ_e r_e^{\height(\zeta)})(b(e)) \ , \ (0.e)(b(e)) \big)\\
  &=  \max\big( -\infty \ , \ -\infty \big) \tag{because $(11) \not\in \cut_e(b(e)) =\{(\varepsilon),(1)\}$ and by \eqref{equ:not-11-is-infty}}\\
  &= -\infty \enspace.
\end{align*}
\endgroup
  If $\zeta' \not= e$, then there exist $c \in \{a,b\}$ and $\zeta'' \in \T_{\Gamma_e}$ such that $\zeta = b(c(\zeta''))$. Then we have
    \begingroup
    \allowdisplaybreaks
\begin{align*}
  &\max\big((r \circ_e r_e^{\height(\zeta)})(b(c(\zeta''))) \ , \ (0.e)(b(c(\zeta''))) \big)\\
             &=  \max\big( r(b(c(\zeta''))[e]_{(11)}) + r_e^{\height(\zeta)}(b(c(\zeta''))|_{11}) \ , \ (0.e)(b(c(\zeta''))) \big)
               \tag{by \eqref{equ:not-11-is-infty}}\\
  &= \max\big( r(b(c(e))) + r_e^{\height(\zeta)}(\zeta'') \ , \ -\infty \big)\\
  &=  \max\big( -\infty + r_e^{\height(\zeta)}(\zeta'') \ , \ -\infty   \big) = -\infty \enspace.
\end{align*}
\endgroup
Hence, for each possible choice of $\zeta$, we have $\max\big((r \circ_e r_e^{\height(\zeta)})(\zeta) \ , \ (0.e)(\zeta) \big)=-\infty$. This fact, and \eqref{eq:r-e-star}
prove \eqref{equ:not-abn=infty}.
Then \eqref{equ:induction-ab*=n}  and \eqref{equ:not-abn=infty} imply \eqref{equ:what-we-want-to-prove}.
\hfill$\Box$
\end{example}

A set $\cL$ of $\B$-weighted tree languages is \emph{closed under Kleene stars} if the following holds: for each $(\Sigma,\B)$-weighted tree language $r$ and for each $\alpha \in \Sigma^{(0)}$ such that $r$ is $\alpha$-proper, if $r \in \cL$, then $r_\alpha^* \in \cL$.

 \begin{lemma}\rm \label{lm:equation}\cite[Lm. 3.13]{dropecvog05} Let $r: \T_\Sigma \rightarrow B$ be $\alpha$-proper. Then 
$r_\alpha^* = (r \circ_\alpha r_\alpha^*) \oplus  \1.\alpha$.
\end{lemma} 
\begin{proof} Let $\xi \in \T_\Sigma$ and let $\ell = \height(\xi) +1$. Then we can calculate as follows.
    \begingroup
    \allowdisplaybreaks
    \begin{align*}
r_\alpha^*(\xi) = r_\alpha^\ell(\xi) = r_\alpha^{\ell+1}(\xi) =  (r \circ_\alpha r_\alpha^{\ell})(\xi) \oplus (\1.\alpha)(\xi) \enspace,
    \end{align*}
    \endgroup
    where the second equality holds by Lemma \ref{lm:iteration-terminates}.
    Now we proceed with the subexpression $(r \circ_\alpha r_\alpha^{\ell})(\xi)$ as follows.    
       \begingroup
    \allowdisplaybreaks
    \begin{align*}
      &(r \circ_\alpha r_\alpha^{\ell})(\xi) \\
      &= \bigoplus\limits_{(w_1,\ldots,w_n) \in \Cut_\alpha(\xi)} 
    r(\xi[\alpha]_{(w_1,\ldots,w_n)}) \otimes r_\alpha^{\ell}(\xi|_{w_1})\otimes \ldots \otimes r_\alpha^{\ell}(\xi|_{w_n})\\[2mm]
      &= \bigoplus\limits_{(w_1,\ldots,w_n) \in \Cut_\alpha(\xi)\setminus \{(\varepsilon)\}} 
        r(\xi[\alpha]_{(w_1,\ldots,w_n)}) \otimes r_\alpha^{\ell}(\xi|_{w_1})\otimes \ldots \otimes r_\alpha^{\ell}(\xi|_{w_n})
        \tag{because $r(\xi[\alpha]_{(\varepsilon)}) \otimes r_\alpha^{\ell}(\xi|_{\varepsilon}) = r(\alpha) \otimes r_\alpha^{\ell}(\xi|_{\varepsilon}) = \0$}\\[2mm]
       &= \bigoplus\limits_{(w_1,\ldots,w_n) \in \Cut_\alpha(\xi)\setminus \{(\varepsilon)\}} 
         r(\xi[\alpha]_{(w_1,\ldots,w_n)}) \otimes r_\alpha^{\height(\xi|_{w_1})+1}(\xi|_{w_1})\otimes \ldots \otimes r_\alpha^{\height(\xi|_{w_n})+1}(\xi|_{w_n})
         \tag{because, for each $i \in [n]$, we have $|w_i| \ge 1$ and hence $\ell \ge \height(\xi|_{w_i})+1$,}\\
      &\tag{ and then by Lemma \ref{lm:iteration-terminates}.}\\
       &= \bigoplus\limits_{(w_1,\ldots,w_n) \in \Cut_\alpha(\xi)\setminus \{(\varepsilon)\}} 
         r(\xi[\alpha]_{(w_1,\ldots,w_n)}) \otimes r_\alpha^*(\xi|_{w_1})\otimes \ldots \otimes r_\alpha^*(\xi|_{w_n})
      \tag{by definition of $\alpha$-Kleene star}\\
       &= \bigoplus\limits_{(w_1,\ldots,w_n) \in \Cut_\alpha(\xi)} 
         r(\xi[\alpha]_{(w_1,\ldots,w_n)}) \otimes r_\alpha^*(\xi|_{w_1})\otimes \ldots \otimes r_\alpha^*(\xi|_{w_n}) \tag{as above}\\
      &= (r \circ_\alpha r_\alpha^*)(\xi) \enspace.
        \end{align*}
        \endgroup
        Hence $r_\alpha^*(\xi) = (r \circ_\alpha r_\alpha^{\ell})(\xi) \oplus (\1.\alpha)(\xi) = (r \circ_\alpha r_\alpha^*)(\xi) \oplus (\1.\alpha)(\xi)$, which proves the lemma.
  \end{proof}

  In the next theorem we prove that, under certain conditions, the $\alpha$-Kleene star of a regular $(\Sigma,\B)$-weighted tree language is regular. For the Boolean semiring this was proved in \cite[Lm.~12]{thawri68} (also cf.  \cite[Thm.~3.41]{eng75-15} and \cite[Thm.~2.4.8]{gecste84}) and for wta this was proved in \cite[Lm.~6.7]{dropecvog05}. We follow the construction in \cite[Lm. 6.7]{fulvog19a} (for the trivial storage type) and  close a gap in the given correctness proof.

 \begin{theorem}\label{thm:closure-Kleene-star-wrtg}  Let $\B$ be a commutative semiring and let $\alpha \in \Sigma^{(0)}$. Let $\cG$ be a $(\Sigma,\B)$-wrtg such that (a) $\sem{\cG}$ is $\alpha$-proper and (b) $\cG$ is finite-derivational or $\B$ is $\sigma$-complete. Then the following two statements hold.
   \begin{compactenum}
     \item[(1)] There exists a $(\Sigma,\B)$-wrtg $\cG'$ such that  $\sem{\cG'} = \sem{\cG}_\alpha^*$.
     \item[(2)] If $\cG$ is finite-derivational, then we can construct a finite-derivational $(\Sigma,\B)$-wrtg $\cG'$ such that  $\sem{\cG'} = \sem{\cG}_\alpha^*$.
     \end{compactenum}
  \end{theorem}
  \begin{proof} Proof of (1):   Let  $\cG=(N,S,R,wt)$ be a $(\Sigma,\B)$-wrtg and $\alpha \in \Sigma^{(0)}$ such that $\sem{\cG}$ is $\alpha$-proper. By Lemmas  \ref{lm:normal-form-lemmas-inherited-from-wcfg}(1) and \ref{lm:rtg-normal-form}, we can assume that $\cG$ is start-separated and alphabetic. By Lemma \ref{lm:normal-form-lemmas-inherited-from-wcfg}(3) we can furthermore assume that $\cG$ is chain-free. 
  Clearly, each mentioned construction preserves $\alpha$-properness. We call a rule of the form $A \to \alpha$ in $R$ an \emph{$\alpha$-rule}.

    Since $\cG$ is chain-free, for each $\beta \in \Sigma^{(0)}$, we have $\RT_{\cG}(\beta) \subseteq \{S \to \beta\}$, i.e., if $\RT_{\cG}(\beta) \not= \emptyset$, then $d = (S \to \beta)$ is the only rule tree such that $\pi(d) = \beta$. Hence, if $(S \to \alpha)$ is in $R$, then
\[
  wt(S\to \alpha) = \wt_{\cG}(S\to \alpha) = \infsum{\oplus}{d \in \RT_{\cG}(\alpha)}{\wt_{\cG}(d)} = \sem{\cG}(\alpha) = \0 \enspace,
  \]
where the last equality holds because $\cG$ is $\alpha$-proper. Thus, $\sem{\cG} = \sem{\widetilde{\cG}}$ where $\widetilde{\cG}$ is the $(\Sigma,\B)$-wrtg $(N,S,\widetilde{R},\widetilde{wt})$ with $\widetilde{R} = R \setminus \{S \to \alpha\}$ and $\widetilde{wt} = wt|_{\widetilde{R}}$.  Hence we can assume that $\cG$ does not contain the rule $S \to \alpha$. Due to our assumptions on $\cG$, for every $\xi \in \T_\Sigma$ and $d\in  \RT_{\cG}(\xi)$, we have  $\pos(d)=\pos(\xi)$.   

We define the $(\Sigma,\B)$-wrtg $\cG'=(N',\{S,S_0\},R',wt')$, where $S_0 \not\in N$ is a new nonterminal,
$N' = N \cup \{S_0\}$, and $R' = R \cup R_\alpha \cup R_{\text{init}}$, where 
\begin{compactitem}
  \item For each rule $r \in R$ we let $wt'(r) = wt(r)$.
\item $R_\alpha$: If $r=(A \rightarrow \alpha)$ is in $R$, then $r'=(A \rightarrow S)$ is in $R_\alpha$ with $wt'(r') = wt(r)$. 
\item $R_{\text{init}}$: This set contains the rule $S_0 \rightarrow \alpha$ with $wt'(S_0 \rightarrow \alpha)=\mathbb{1}$.
\end{compactitem}
We call each rule of the form $A \to S$ an \emph{$S$-rule}.
 
For each $\xi \in \T_\Sigma$, we have
\begin{equation}\label{eq:RTxi-alpha}
\RT_{\cG'}(\xi)=
\begin{cases}
  \{ S_0 \to \alpha\} & \text{ if } \xi=\alpha \\
\RT_{\cG'}(S,\xi) & \text{ otherwise. }
\end{cases}
\end{equation}
Moreover, since $\cG$ does not contain chain rules and the rule $S\to \alpha$, it is easy to see that $\RT_{\cG'}(S,\beta) = \RT_{\cG}(\beta)$.

Let $\xi \in \T_\Sigma$ and $\widetilde{w} = (w_1,\ldots,w_n) \in \Cut_\alpha(\xi)$.
We define
\begin{align*}\RT_{\cG'}^{\widetilde{w}}(S,\xi) = \{  d'\in \RT_{\cG'}(S,\xi) \mid  &
           (\forall i \in [n]) : w_i \in \pos(d') \wedge  \big( d'(w_i) \text{ is an $\alpha$-rule} \text{ or an $S$-rule} \big) \\
&  \ \ \  \wedge \  \big((\forall v\in \prefix(w_i) \setminus\{w_i\}):  d'(v) \text{ is not an $S$-rule} \big) \} \enspace.
\end{align*}

We have that
\begin{equation}\label{eq:D_G-equ}
(\RT_{\cG'}^{\widetilde{w}}(S,\xi) \mid \widetilde{w}\in \Cut_\alpha(\xi)) \ \text{ is a partitioning of $\RT_{\cG'}(S,\xi)$}\enspace.
\end{equation}
In particular, we have $\RT_{\cG'}(S,\alpha) = \RT_{\cG'}^{(\varepsilon)}(S,\alpha)= \emptyset$.

Let $\{j_1,\ldots,j_\ell\}$ be the set of all $i \in [n]$ such that $\xi(w_i) \not=\alpha$. We denote the set $\{j_1,\ldots,j_\ell\}$ by $\mathrm{iter}(\widetilde{w})$.  

We define the mapping
\[
  \Phi_{\widetilde{w}}: \RT_{\cG}(\xi[\alpha]_{\widetilde{w}}) \times 
  \RT_{\cG'}(S,\xi|_{w_{j_1}}) \times \ldots \times \RT_{\cG'}(S,\xi|_{w_{j_\ell}}) \rightarrow  \RT_{\cG'}^{\widetilde{w}}(S,\xi)
\]
as follows. Let $d \in \RT_{\cG}(\xi[\alpha]_{\widetilde{w}})$ and $d_1 \in \RT_{\cG'}(S,\xi|_{w_{j_1}})$, \ldots, $d_\ell \in \RT_{\cG'}(S,\xi|_{w_{j_\ell}})$. We define the tree
\[
d' = d[(B_1 \to S)(d_1)]_{w_{j_1}} \cdots [(B_\ell \to S)(d_\ell)]_{w_{j_\ell}} \enspace,
\]
where, for each $i \in [\ell]$, $B_i$ is the nonterminal on the left-hand side of the $\alpha$-rule $d(w_{j_i})$. (Since $\pos(\xi[\alpha]_{\widetilde{w}})=\pos(d)$ and $\xi[\alpha]_{\widetilde{w}}(w_{j_i})=\alpha$, the rule $d(w_{j_i})$ is an $\alpha$-rule.)
It is obvious that $yd'\in \RT_{\cG'}^{\widetilde{w}}(S,\xi)$. Now we define
$\Phi_{\widetilde{w}}(d,d_1,\ldots,d_\ell)=d'$.

Also it is easy to see that $\Phi_{\widetilde{w}}$ is a bijection. Moreover, 
\begin{equation}\label{equat:weights-star}
\wt_{\cG'}(\Phi_{\widetilde{w}}(d,d_1,\ldots, d_\ell)) = \wt_\cG(d) \otimes \bigotimes_{i\in[\ell]} \wt_{\cG'}(d_i)
\end{equation}
by the definitions of $R'$ and $\wt_{\cG'}$ and by the fact that $\B$ is commutative.

By induction on $(\T_\Sigma,\psucc_\Sigma)$, we prove that the following statement holds:
\begin{equation}\label{eq:Kleene-star-constr}
  \text{For every $\xi\in \T_\Sigma$, we have: } \sem{\cG'}(\xi) = \sem{\cG}^*_\alpha(\xi)\enspace.
\end{equation}
(We recall that $\cG$ is finite-derivational or $\B$ is $\sigma$-complete).

I.B.: Let $\xi \in \Sigma^{(0)}$. If $\xi \ne \alpha$, then
\begingroup
\allowdisplaybreaks
\begin{align*}
  \sem{\cG'}(\xi) =  \infsum{\oplus}{d' \in \RT_{\cG'}(\xi)} \wt_{\cG'}(d')
                     = \infsum{\oplus}{d' \in \RT_{\cG'}(S,\xi)} \wt_{\cG'}(d')
                    = \infsum{\oplus}{d \in \RT_{\cG}(\xi)} \wt_{\cG}(d)
  = \sem{\cG}(\xi) 
  =\sem{\cG}^*_\alpha(\xi) \enspace,
\end{align*}
\endgroup
where the last equality holds by \eqref{equ:star-beta=r(beta)}.

Now let $\xi = \alpha$.
Then we can calculate as follows:
\begingroup
\allowdisplaybreaks
\begin{align*}
  \sem{\cG'}(\alpha)  = \wt_{\cG'}((S_0 \to \alpha))
  =  \1   = \sem{\cG}^*_\alpha(\alpha) \enspace,
\end{align*}
\endgroup
where the first equality holds by \eqref{eq:RTxi-alpha} and the last equality holds by \eqref{equ:star-alpha=1}.

\

I.S.: Let $\xi \in (\T_\Sigma \setminus \Sigma^{(0)})$. We can calculate as follows.
\begingroup
\allowdisplaybreaks
\begin{align*}
  & \ \sem{\cG'}(\xi) = \infsum{\oplus}{d' \in \RT_{\cG'}(\xi)} \wt_{\cG'}(d')\\
  & = \infsum{\oplus}{d' \in \RT_{\cG'}(S,\xi)} \wt_{\cG'}(d') \tag{by \eqref{eq:RTxi-alpha} because $\xi \ne \alpha$}\\
  & = \bigoplus_{\widetilde{w}\in \Cut_\alpha(\xi)} \ \ \infsum{\oplus}{d' \in \RT_{\cG'}^{\widetilde{w}}(S,\xi)} \wt_{\cG'}(d')
  \tag{\text{by (\ref{eq:D_G-equ})}}\\[2mm]
  & = \bigoplus_{\widetilde{w}=(w_1,\ldots,w_n)\in \Cut_\alpha(\xi)} \ \
    \infsum{\oplus}{\substack{d \in \RT_{\cG}(\xi[\alpha]_{\widetilde{w}}),\\(\forall i \in [\ell]): d_i \in \RT_{\cG'}(S,\xi|_{w_{j_i}})}} \Big(  \wt_\cG(d) \otimes  \bigotimes_{i\in[\ell]} \wt_{\cG'}(d_i)\Big)
  \tag{because $\Phi_{\widetilde{w}}$ is bijective and by \eqref{equat:weights-star}, where $\mathrm{iter}(\widetilde{w}) = \{j_1,\ldots,j_\ell\}$}\\[2mm]
  & = \bigoplus_{\widetilde{w}=(w_1,\ldots,w_n)\in \Cut_\alpha(\xi)} \Big(\ \  \infsum{\oplus}{d \in \RT_{\cG}(\xi[\alpha]_{\widetilde{w}})} \wt_\cG(d) \Big) \ \otimes \
    \bigotimes_{i\in[\ell]} \Big(\ \ \infsum{\oplus}{d_i \in \RT_{\cG'}(S,\xi|_{w_{j_i}})} \wt_{\cG'}(d_i) \Big)
    \tag{\text{by distributivity}}\\[2mm]
  & = \bigoplus_{\widetilde{w}=(w_1,\ldots,w_n)\in \Cut_\alpha(\xi)}    \sem{\cG}(\xi[\alpha]_{\widetilde{w}})        \ \otimes \ \bigotimes_{i\in[\ell]} \Big(\ \ \infsum{\oplus}{d_i \in \RT_{\cG'}(S,\xi|_{w_{j_i}})} \wt_{\cG'}(d_i) \Big)\\[2mm]
  & = \bigoplus_{\widetilde{w}=(w_1,\ldots,w_n)\in \Cut_\alpha(\xi)}    \sem{\cG}(\xi[\alpha]_{\widetilde{w}})        \ \otimes \ \bigotimes_{i\in[\ell]} \Big(\ \ \infsum{\oplus}{d_i \in \RT_{\cG'}(\xi|_{w_{j_i}})} \wt_{\cG'}(d_i) \Big)
  \tag{by \eqref{eq:RTxi-alpha} because $\xi|_{w_{j_i}} \ne \alpha$}\\[2mm]
    & = \bigoplus_{\widetilde{w}=(w_1,\ldots,w_n)\in \Cut_\alpha(\xi)}    \sem{\cG}(\xi[\alpha]_{\widetilde{w}})        \ \otimes \ \bigotimes_{i\in[\ell]} \  \sem{\cG'}(\xi|_{w_{j_i}}) \\[2mm]
  & = \bigoplus_{\widetilde{w}=(w_1,\ldots,w_n)\in \Cut_\alpha(\xi)}    \sem{\cG}(\xi[\alpha]_{\widetilde{w}})        \ \otimes \ \sem{\cG'}(\xi|_{w_1})\otimes \ldots \otimes 
  \sem{\cG'}(\xi|_{w_n}) 
    \tag{because for each $i \in [n]\setminus \mathrm{iter}(\widetilde{w})$ we have $\xi|_{w_i} = \alpha$ and  $\sem{\cG'}(\alpha) = \1$}\\[2mm]
   & = \bigoplus_{\substack{\widetilde{w}=(w_1,\ldots,w_n)\in \Cut_\alpha(\xi):\\\widetilde{w}\ne(\varepsilon)}}    \sem{\cG}(\xi[\alpha]_{\widetilde{w}})        \ \otimes \ \sem{\cG'}(\xi|_{w_1})\otimes \ldots \otimes 
  \sem{\cG'}(\xi|_{w_n})
  \tag{because $\sem{\cG}(\xi[\alpha]_{(\varepsilon)}) = \sem{\cG}(\alpha)=\0$}\\[2mm]
   &   = \bigoplus_{\substack{\widetilde{w}=(w_1,\ldots,w_n)\in \Cut_\alpha(\xi):\\\widetilde{w}\ne(\varepsilon)}} 
  \sem{\cG}(\xi[\alpha]_{\widetilde{w}}) \otimes \sem{\cG}^*_\alpha(\xi|_{w_1})\otimes \ldots \otimes \sem{\cG}^*_\alpha(\xi|_{w_n})
  \tag{\text{by I.H.: since $\widetilde{w}\ne(\varepsilon)$, we have $\xi \succ_\Sigma^+ \xi|_{w_i}$ for every $i\in [n]$}}   \\[2mm]
  &   = \bigoplus_{\widetilde{w}=(w_1,\ldots,w_n)\in \Cut_\alpha(\xi)} \sem{\cG}(\xi[\alpha]_{\widetilde{w}}) \otimes \sem{\cG}^*_\alpha(\xi|_{w_1})\otimes \ldots \otimes \sem{\cG}^*_\alpha(\xi|_{w_n})
  \tag{because  $\sem{\cG}(\alpha)=\0$}\\
  & = (\sem{\cG} \circ_\alpha \sem{\cG}^*_\alpha)(\xi)\\
  & = (\sem{\cG} \circ_\alpha \sem{\cG}^*_\alpha \oplus \1.\alpha)(\xi)
  \tag{because $\xi \ne \alpha$}\\
  & = \sem{\cG}^*_\alpha(\xi) \tag{\text{by Lemma \ref{lm:equation}}} \enspace.
    \end{align*}
    \endgroup
 This finishes the proof of \eqref{eq:Kleene-star-constr}.  Thus $\sem{\cG'}= \sem{\cG}^*_\alpha$.

   \

   Proof of (2): Let $\cG$ be finite-derivational. Then, by Lemma \ref{lm:normal-form-lemmas-inherited-from-wcfg}(3), we can even \underline{construct} an equivalent chain-free wrtg. Consequently, the definition of $\cG'$, as given in the proof of (1), is constructive. Finally, by analysing the construction of $\cG'$, we obtain that $\cG'$ is finite-derivational. 
 \end{proof}

Next we give an example of the concepts which appear in the proof of Theorem \ref{thm:closure-Kleene-star-wrtg}.

\begin{example}\label{ex:example-Kleene-star} \rm
 Let $\Sigma=\{\sigma^{(2)},\alpha^{(0)}\}$ and $\cG$ be the $(\Sigma,\Nat)$-wrtg which has the rules
\begin{align*}
  & S \to \sigma(S,A) : 2 \\
  & S \to \beta : 2 \\
& A \to \alpha : 2 \enspace,
\end{align*}
and the weight of each rule is 2 (indicated by ``: 2'' after the rule). If we apply the construction in the proof of Theorem \ref{thm:closure-Kleene-star-wrtg}, then we obtain the $(\Sigma,\Nat)$-wrtg $\cG'$ with  the rules and weights
\begin{align*}
  & S \to \sigma(S,A): 2  & S_0 \to \alpha : 1 \\
  & S \to \beta : 2  &  A \to S : 2 \\
& A \to \alpha : 2    \enspace.
\end{align*}

Now let us consider the tree $\xi\in \T_\Sigma$, the cut $\widetilde{w} = (12,2)$ in $\Cut_\alpha(\xi)$, and tree $\xi[\alpha]_{\widetilde{w}}$ in Figure \ref{fig:ex:Kleene-star-1}. By definition, we have $\mathrm{iter}(\widetilde{w}) = \{12\}$. 
In the upper part of Figure \ref{fig:ex:Kleene-star-2}, there is an example of a tree $d'\in \RT_{\cG'}^{\widetilde{w}}(S,\xi)$. Moreover, for the trees $d\in \RT_{\cG}(\xi[\alpha]_{\widetilde{w}})$ and $d_1 \in \RT_{\cG'}(S,\xi|_{12})$ in the lower part of Figure \ref{fig:ex:Kleene-star-2}, we have
\[
  d' = \Phi_{\widetilde{w}}(d,d_1) \text{\ \ \ \  and \ \ \ } \wt_{\cG'}(d')=\wt_\cG(d)\cdot \wt_{\cG'}(d_1) \enspace,
\]
where $\wt_{\cG'}(d')=2^8$, $\wt_\cG(d)=2^5$, and $\wt_{\cG'}(d_1)=2^3$. 
\hfill $\Box$
\end{example}

\begin{figure}
\begin{center}
\begin{tikzpicture}[level distance=4.5em,
  every node/.style = {align=center}]]

  \pgfdeclarelayer{bg}    
  \pgfsetlayers{bg,main}  

\begin{scope}[xshift=-17mm,level 1/.style={sibling distance=25mm},
level 2/.style={sibling distance=25mm}, level 3/.style={sibling distance=25mm}]
 \node {$\sigma$}
 child {node {$\sigma$}  
 child {node {$\beta$}}
       child {node{$\sigma$} 
       child {node {$\beta$}}
       child { node {$\alpha$}}}}
       child {node {$\alpha$}} ;
 \end{scope}

 \node at (-3.75,-.75) {$\xi$};
 \node at (1,-.75) {$\widetilde{w}=(12,2)$};
 \node at (3.5, -.75) {$\xi[\alpha]_{\widetilde{w}}$};

 \begin{scope}[xshift=55mm,level 1/.style={sibling distance=25mm},
level 2/.style={sibling distance=25mm}, level 3/.style={sibling distance=25mm}]
 \node {$\sigma$}
 child {node {$\sigma$}  
 child {node {$\beta$}}
       child {node{$\alpha$}}} 
       child {node {$\alpha$}} ;
 \end{scope}

\end{tikzpicture}
\end{center}
\caption{\label{fig:ex:Kleene-star-1} Example of a tree $\xi\in \T_\Sigma$, a cut $\widetilde{w}\in \Cut_\alpha(\xi)$, and a tree $\xi[\alpha]_{\widetilde{w}}$,  cf. Example \ref{ex:example-Kleene-star}.}
\end{figure}

\begin{figure}
\begin{center}
\begin{tikzpicture}[level distance=4.5em,
  every node/.style = {align=center}]]

  \pgfdeclarelayer{bg}    
  \pgfsetlayers{bg,main}  
  
 \begin{scope}[xshift=-17mm,level 1/.style={sibling distance=25mm},
level 2/.style={sibling distance=25mm}, level 3/.style={sibling distance=25mm}]
 \node {$S\to \sigma(S,A)$}
 child {node{$S\to \sigma(S,A)$}
 child {node {$S\to \beta$}}
 child {node {$A\to S$}
 child {node{$S\to \sigma(S,A)$}
 child {node {$S\to \beta$}}
 child {node {$A\to \alpha$}}}}} 
       child {node {$A\to \alpha$}}; 
  \end{scope} 
  
   \node at (-4.85,-.55) {$d'$};
   \node at (2.85,-.55) {$\mathrm{iter}(\widetilde{w}) = \{12\}$};
   
  \begin{scope}[xshift=-37mm,yshift=-80mm,level 1/.style={sibling distance=25mm},
level 2/.style={sibling distance=25mm}, level 3/.style={sibling distance=25mm}]
 \node {$S\to \sigma(S,A)$}
 child {node{$S\to \sigma(S,A)$}
 child {node {$S\to \beta$}}
 child {node {$A\to \alpha$}}}
 child {node {$A\to \alpha$}}; 
 \end{scope} 
 
 \node at (-67mm,-85mm) {$d$};
 \node at (-7mm,-85mm) {$d_1$};

\begin{scope}[xshift=20mm,yshift=-80mm,level 1/.style={sibling distance=25mm},
level 2/.style={sibling distance=25mm}, level 3/.style={sibling distance=25mm}]
 \node {$S\to \sigma(S,A)$}
 child {node {$S\to \beta$}}
 child {node {$A\to \alpha$}}; 
 \end{scope}

\end{tikzpicture}
\end{center}
\caption{\label{fig:ex:Kleene-star-2}  Example of  trees   $d'\in \RT_{\cG'}^{\widetilde{w}}(S,\xi)$, $d\in \RT_{\cG}(\xi[\alpha]_{\widetilde{w}})$, and $d_1 \in \RT_{\cG'}(S,\xi|_{12})$, cf. Figure \ref{fig:ex:Kleene-star-1}  and Example \ref{ex:example-Kleene-star}.}
\end{figure}

 Finally, as a consequence of Theorem \ref{thm:closure-Kleene-star-wrtg}, we show that $\Rec(\Sigma,\B)$ is closed under Kleene stars.

   \begin{corollary-rect}\label{cor:closure-Kleene-star}\rm (cf. \cite[Lm. 6.7]{dropecvog05}) Let $\Sigma$ be a ranked alphabet. Moreover, let  $\alpha \in \Sigma^{(0)}$ and let $\B$ be a commutative semiring. Let $\cA$ be a $(\Sigma,\B)$-wta such that $\sem{\cA}$ is $\alpha$-proper. Then we can construct a $(\Sigma,\B)$-wta $\cA'$ such that  $\sem{\cA'} = \sem{\cA}_\alpha^*$.  Thus, in particular, if $\B$ is a commutative semiring, then  the set $\Rec(\Sigma,\B)$ is closed under Kleene stars.
\end{corollary-rect}
\begin{proof}  By Lemma \ref{lm:wta-to-wrtg} we can construct $(\Sigma,\B)$-wrtg $\cG$ such that $\cG$ is in tree automata form and $\sem{\cA} = \sem{\cG}$.  Then, in particular, $\cG$ is  finite-derivational. By Theorem \ref{thm:closure-Kleene-star-wrtg} we can construct a  finite-derivational $(\Sigma,\B)$-wrtg $\cG'$ such that  $\sem{\cG'} = \sem{\cG}_\alpha^*$. Then,  by Lemma \ref{lm:wrtg-to-wta}, we can construct a $(\Sigma,\B)$-wta $\cA'$ such that $\sem{\cG'}=\sem{\cA'}$. 
  \end{proof}

  
  \section[Closure under yield-intersection]{Closure under yield-intersection with weighted recognizable languages}
\label{sect:Bar-Hillel-Perles-Shamir}

Bar-Hillel, Perles, and Shamir \cite[Thm.~8.1]{barpersha61} proved that the intersection of any context-free language and any recognizable language is again a context-free language.  This theorem is very valuable, in particular, in natural language processing, where one wishes to have a finite description of the set of all derivations of a string given   a context-free grammar \cite{malsat09,malsat10}. Here we extend this result by replacing (a) the context-free language by a recognizable weighted tree language~$r$, (b)~the recognizable language by an r-recognizable weighted language $L$, and (c) the intersection by the Hadamard product. Roughly speaking, we prove  that  the Hadamard product of $r$ and $L \circ \yield$ is a recognizable weighted tree language if $r$ and $L$ take their weights in a commutative semiring (cf. Theorem \ref{thm:BPS}). We refer to \cite{malsat09,malsat10} for even more extended results.

\index{yield-intersection}
A set $\cL$ of $\B$-weighted tree languages is  \emph{closed under yield-intersection  with weighted recognizable languages} if for every  $(\Sigma,\B)$-weighted tree language $r$ in $\cL$, $\Gamma\subseteq \Sigma^{(0)}$, and  r-recognizable $(\Gamma,\B)$-weighted language $L: \Gamma^* \to B$, the $(\Sigma,\B)$-weighted tree language $r \otimes (L \circ \yield_\Gamma)$ is in $\cL$.
    
Since we have already proved that the set $\Rec^{\mathrm{run}}(\Sigma,\B)$ is closed under Hadamard product if $\B$ is a commutative semiring, it remains to prove that $L \circ \yield_\Gamma$ is a recognizable weighted tree language. Indeed, this even holds for  commutative strong bimonoids.
 Our construction is based on the key idea of the proof of \cite[Thm.~8.1]{barpersha61}, cf. also \cite[Thm.~2]{malsat09}. (We also refer to Example \ref{ex:transformation-monoid}.)

\begin{lemma}\label{lm:BPS}
  \rm Let $\Gamma\subseteq \Sigma^{(0)}$ and $\cB$ be a $(\Gamma,\B)$-wsa. Then we can construct a $(\Sigma,\B)$-wta $\cA$ such that $\runsem{\cA} = \runsem{\cB} \circ \yield_\Gamma$.
\end{lemma}
\begin{proof} Let $\cB = (P,\lambda,\mu,\gamma)$.  By  Lemma \ref{lm:wsa-with-identity-initial-weights}, we can assume that $\cB$ has unit initial weights, i.e., $\im(\lambda) \subseteq \{\0,\1\}$. Intuitively, we construct  $\cA$ such that it guesses, at each $\Gamma$-labeled leaf of a given input $\Sigma$-tree $\xi$, a transition of $\cB$ and,  at each $(\Sigma^{(0)}\setminus\Gamma)$-labeled leaf of $\xi$, a pair $(p,p)$ for some $p\in P$. Then, while moving up towards the root of $\xi$, the wta $\cA$ checks whether subruns of $\cB$ can be composed to larger runs of $\cB$. The first state and the last state of such a subrun are coded as a pair and forms a state of $\cA$.
  
Formally, we construct the $(\Sigma,\B)$-wta $\cA = (Q,\delta,F)$ as follows.
  \begin{compactitem}
  \item $Q= P \times P$,
  \item  we let $F_{(p,p')} = \lambda(p) \otimes \gamma(p')$ for every $(p,p') \in Q$,
  \item for every $\alpha \in \Gamma$ and $(p,p') \in Q$
we let 
\[
  \delta_0(\varepsilon,\alpha,(p,p')) =  \mu(p,\alpha,p') \]
and for every $\alpha \in \Sigma^{(0)}\setminus \Gamma$ and $(p,p') \in Q$ we let
\[
  \delta_0(\varepsilon,\alpha,(p,p')) =
  \begin{cases}\1 & \text{ if } p=p'\\ \0 &\text{otherwise}\end{cases}
\]
and for every $k \in \mathbb{N}_+$, $\sigma \in \Sigma^{(k)}$, and $(p_1,p_1'), (p_2,p_2'),\ldots,
(p_k,p_k'), (p,p') \in Q$  we let
\[
\delta_k((p_1,p_1')\cdots (p_k,p_k'), \sigma, (p,p')) =\]
\[
\left\{
\begin{array}{ll}
\mathbb{1} & \hbox{ if } p = p_1, p_i' = p_{i+1} \text{ for each } i \in [k-1], \text{ and } p_k' = p' \\
\mathbb{0} & \hbox{ otherwise.}
\end{array}
\right.
\]
\end{compactitem}
This finishes the construction of $\cA$.

\

 Next we prove that $\runsem{\cA} = \runsem{\cB} \circ \yield_\Gamma$, for which we need some auxiliary tools and statements. Let $\xi \in \T_\Sigma$ and $\rho \in P^{|\yield_\Gamma(\xi)|+1}$. Then there exist $n \in \mathbb{N}$ and $p_1,\ldots,p_n \in P$ such that $\rho = p_0\cdots p_n$. 
 Let
\[u_0,w_1,u_1,\ldots,u_{n-1},w_n, u_n \]
be the list of  elements of $\pos_{\Sigma^{(0)}}(\xi)$ ordered by the lexicographic order $<_{\mathrm{lex}}$ on $\pos(\xi)$, where $\pos_{\Gamma}(\xi) = \{w_1,\ldots,w_n\}$  and, for each $i \in [0,n]$, the $u_i$ is the corresponding list of elements in $\pos_{(\Sigma^{(0)}\setminus\Gamma)}(\xi)$.

\index{succb@$\succ$}
\label{page:prec-for-positions}
We define the binary relation $\succ$ on $\pos(\xi)$ by letting $w_1 \succ w_2$ if there is an $i\in \mathbb{N}$ such that $w_2=w_1i$. For the proof of termination, we let $N=\max(|w| \mid w\in \pos(\xi))$ and we define the mapping $d:\pos(\xi)\to \mathbb{N}$ by $d(w)=N-|w|$ for each $w\in \pos(\xi)$. Clearly, $d$ is a monotone embedding of $(\pos(\xi),\succ)$ into $(\mathbb{N},>)$. Then, by Lemma~\ref{lm:fin-branching-embedding-termination}, the relation $\succ$ is terminating.
Moreover, we have that $\nf_\succ(\pos(\xi))=\pos_{\Sigma^{(0)}}(\xi)$, i.e., it is the set of leaves of $\xi$.

Then we define the mapping
\[
  \varphi_{\xi,\rho}: \pos(\xi) \to Q
\]
by induction on $(\pos(\xi),\succ)$ for each $w \in \pos(\xi)$ as follows.
\[
\varphi_{\xi,\rho}(w)=
\begin{cases}
(p_i,p_i) & \text{ if $w \in u_i$ for some $i\in[0,n]$} \\
(p_{i-1},p_i) & \text{ if $w = w_i$ for some $i\in[n]$} \\
\big(\varphi_{\xi,\rho}(w1)_1, \varphi_{\xi,\rho}(wk)_2 \big) & \text{ otherwise,}
\end{cases}
\]
where $w \in u_i$ abbreviates that $w$ is an element in the list $u_i$ and $k=\rk_\xi(w)$.
Moreover, $\varphi_{\xi,\rho}(w1)_1$ denotes the first component of $\varphi_{\xi,\rho}(w1)$, and $\varphi_{\xi,\rho}(wk)_2$ denotes the second component of $\varphi_{\xi,\rho}(wk)$.
In the particular case that $n=0$, we have $\varphi_{\xi,\rho}(w)=(p_0,p_0)$ for each $w\in \pos(\xi)$. Moreover, $\varphi_{\xi,\rho}(\varepsilon) = (p_0,p_n)$.

We demonstrate the definition of $\rho$  in Figures \ref{fig:Bar-Hillel-a}, \ref{fig:Bar-Hillel-b}, and \ref{fig:Bar-Hillel-c} where $\alpha_i=\xi(w_i)$ for each $i\in [n]$ and hence $\yield_\Gamma(\xi)=\alpha_1 \cdots \alpha_n$.

\begin{figure}
  \centering
\scalebox{0.8}{\begin{tikzpicture}[level distance=2.75em,
  every node/.style = {align=center}]
  \pgfdeclarelayer{bg}    
  \pgfsetlayers{bg,main}  

\node at (2.75,3.75) {$\xi$};

\draw   (5,4) coordinate (a) --
		(0,0) coordinate (c) --
		(10,0) coordinate(b) -- cycle;

\path[draw] (a) to[out = 315, in = 135] (5,0) node[circle,fill,inner sep=1.8pt](dot) {};

\node[xshift=0cm,yshift=0.3cm] at (dot) {$w$};
\node[yshift=-0.35cm] at (dot) {$p_i p_i$};

\node (a1) at (1,-0.85) {$\alpha_1$};
\node[below left of=a1] (p0) {$p_0$};
\node[below right of=a1] (p1) {$p_1$};
\draw (p0) edge[->,bend left, above] (a1);
\draw (a1) edge[->,bend left, above] (p1);

\node (ai) at (8/3+1,-0.85) {$\alpha_i$};
\node[below right of=ai] (pi) {$p_i$};
\draw (ai) edge[->,bend left, above] (pi);

\node (ai1) at (8*2/3+1,-0.85) {$\alpha_{i+1}$};
\node[below left of=ai1] (pi3) {$p_i$};
\draw (pi3) edge[->,bend left, above] (ai1);

\node (an) at (9,-0.85) {$\alpha_n$};
\node[below left of=an] (pn1) {$p_{n-1}$};
\node[below right of=an] (pn) {$p_n$};
\draw (pn1) edge[->,bend left, above] (an);
\draw (an) edge[->,bend left, above] (pn);

\draw let \p{a1}=(a1) in (a1) edge (\x{a1}, 0);
\draw let \p{ai}=(ai) in  (ai) edge (\x{ai},0);
\draw let \p{ai1}=(ai1) in  (ai1) edge (\x{ai1},0);
\draw let \p{an}=(an) in (an) edge (\x{an},0);

\node[xshift=-1cm] at (p0) {$\rho =$};         
\end{tikzpicture}
}
\caption{\label{fig:Bar-Hillel-a} Value of $\varphi_{\xi,\rho}(w)$ if $w \in u_i$ for some $i \in [0,n]$.}
\end{figure}

\begin{figure}
  \centering
  \scalebox{0.8}{
    \begin{tikzpicture}[level distance=2.75em,
  every node/.style = {align=center}]
  \pgfdeclarelayer{bg}    
  \pgfsetlayers{bg,main}  

  \tikzstyle{mycircle}=[draw, circle, inner sep=-2mm, minimum height=5mm]

\node at (2.75,3.75) {$\xi$};

\draw   (5,4) coordinate (a) --
		(0,0) coordinate (c) --
		(10,0) coordinate(b) -- cycle;

\path[draw] (a) to[out = 315, in = 135] (5,0);

\node (a1) at (1,-0.85) {$\alpha_1$};
\node[below left of=a1] (p0) {$p_0$};
\node[below right of=a1] (p1) {$p_1$};
\draw (p0) edge[->,bend left, above] (a1);
\draw (a1) edge[->,bend left, above] (p1);

\node[mycircle] (ai) at (5,-0.85) {$\alpha_i$};
\node[xshift=-8mm] at (ai) {$p_{i-1}$};
\node[xshift=8mm] at (ai) {$p_i$};
\node[xshift=2.2mm, yshift=4mm] at (ai) {$w$};

\node[below left of=ai] (pi1) {$p_{i-1}$};
\node[below right of=ai] (pi) {$p_i$};
\draw (pi1) edge[->,bend left, above] (ai);
\draw (ai) edge[->,bend left, above] (pi);

\node (an) at (9,-0.85) {$\alpha_n$};
\node[below left of=an] (pn1) {$p_{n-1}$};
\node[below right of=an] (pn) {$p_n$};
\draw (pn1) edge[->,bend left, above] (an);
\draw (an) edge[->,bend left, above] (pn);

\draw let \p{a1}=(a1) in (a1) edge (\x{a1}, 0);
\draw let \p{ai}=(ai) in  (ai) edge (\x{ai},0);
\draw let \p{an}=(an) in (an) edge (\x{an},0);

\node[xshift=-1cm] at (p0) {$\rho =$};        
\end{tikzpicture}
}  
\caption{\label{fig:Bar-Hillel-b} Value of $\varphi_{\xi,\rho}(w)$ if $w=w_i$ for some $i \in [n]$.}
\end{figure}

\begin{figure}
  \centering
  \scalebox{0.8}{
\begin{tikzpicture}[level distance=2.75em,
  every node/.style = {align=center}]
  \pgfdeclarelayer{bg}    
  \pgfsetlayers{bg,main}  

\node at (2.75,3.75) {$\xi$};

\draw   (5,4) coordinate (a) --
		(0,0) coordinate (c) --
		(10,0) coordinate(b) -- cycle;
 
\draw	(3.75,1.25) node[circle,fill,inner sep=1.8pt](w1) {} --
        (5,2) node[circle,fill,inner sep=1.8pt](w) {} --
        (6.25,1.25) node[circle,fill,inner sep=1.8pt](wk) {} ;

\node[yshift=0.3cm] at (w) {\strut $w$};
\node[xshift=-0.7cm] at (w) {\strut ${p_1}$};
\node[xshift=+0.7cm] at (w) {\strut ${p_k}'$};

\node[xshift=-0cm,yshift=0.27cm] at (w1) {\strut $w_1$};
\node[xshift=-0.7cm] at (w1) {\strut $p_1$};
\node[xshift=+0.55cm] at (w1) {\strut ${p_1}'$};

\node[xshift=0cm,yshift=0.27cm)] at (wk) {\strut ${w_k}$};
\node[xshift=-0.5cm] at (wk) {\strut $p_k$};
\node[xshift=+0.7cm] at (wk) {\strut ${p_k}'$};
  
\draw let \p{b}=(b), \p{w1}=(w1) in
     (\x{w1}-0.85cm,\y{b}) -- 
     (w1) --
     (\x{w1}+0.85cm,\y{b});
     
\draw let \p{b}=(b), \p{wk}=(wk) in
     (\x{wk}-0.85cm,\y{b}) -- 
     (wk) --
     (\x{wk}+0.85cm,\y{b});

\path[draw] (a) to[out = 315, in = 150] (w);
         
\draw[->, shorten <=0.5cm, shorten >=0.1cm,thick] let \p{w}=(w), \p{w1}=(w1) in 
	(\x{w1}-1cm,\y{w1}) -- 
	(\x{w}-1cm,\y{w});
	
\draw[->, shorten <=0.5cm, shorten >=0.1cm,thick] let \p{w}=(w), \p{wk}=(wk) in
	(\x{wk}+1cm,\y{wk}) -- 
	(\x{w}+1cm,\y{w});

      \end{tikzpicture}
      } 
\caption{\label{fig:Bar-Hillel-c} Value of $\varphi_{\xi,\rho}(w)$ if $w$ is not a leaf of $\xi$.}
\end{figure}

In fact, $\varphi_{\xi,\rho}$ is a run of $\cA$ on $\xi$, i.e., $\varphi_{\xi,\rho} \in \R_\cA(\xi)$. Thus, if
 $\xi = \sigma(\xi_1,\ldots,\xi_k)$, then for each $l \in [k]$, the subrun $(\varphi_{\xi,\rho})|_l$ is defined and we have 
  \begin{equation}
 (\varphi_{\xi,\rho})|_l = \varphi_{\xi_l,\rho|_{(i_l,j_l)}}\enspace,\label{eq:sub-runs}
\end{equation}
where $\rho|_{(i_l,j_l)}$ is the subsequence $p_{i_l} \cdots p_{j_l}$ of $\rho$ determined by the indices
\[ i_l=\max(m\in [n] \mid w_m \in \bigcup_{\kappa =1}^{l-1}\pos_\Gamma(\xi_\kappa)) \text{ and } 
j_l=\max(m\in [n] \mid w_m \in \bigcup_{\kappa =1}^{l}\pos_\Gamma(\xi_\kappa))\]
and the definition $\max(\emptyset)=0$.
Thus $i_1=0$, $j_l=i_{l+1}$ for each $l\in [k-1]$, $j_k=n$, and if $\pos_\Gamma(\xi_l)=\emptyset$, then $i_l=j_l$. (For instance, let $\xi=\sigma(\xi_1,\xi_2,\xi_3)$,
$\pos_\Gamma(\xi_1)=\{w_1,w_2\}$ with  $ w_1 <_{\mathrm{lex}} w_2$, $\pos_\Gamma(\xi_2)= \emptyset$, and 
$\pos_\Gamma(\xi_3)= \{w_3\}$, and let $\rho=p_0p_1p_2p_3$. Then $(i_1,j_1)=(0,2)$, $(i_2,j_2)=(2,2)$, and
$(i_3,j_3)=(2,3)$ and thus the corresponding subsequences are $p_0p_1p_2$, $p_2$, and $p_2p_3$.)

This finishes the preparations.

\

Now, by induction on $\T_\Sigma$,  we  prove that the following statement holds:
\begin{equation}
\text{For every $\xi \in \T_\Sigma$  and $\rho \in P^{|\yield_\Gamma(\xi)|+1}$ we have } \wt_\cB^-(\yield_\Gamma(\xi),\rho) = \wt_\cA(\xi,\varphi_{\xi,\rho}). \label{eq:fold}
\end{equation}

I.B.: Let $\xi \in \Sigma^{(0)}$. We distinguish two cases.

\underline{Case (a):} Let $\xi \in \Sigma^{(0)}\setminus \Gamma$. Now let $\rho \in P^{|\yield_\Gamma(\xi)|+1}$. Since $|\yield_\Gamma(\xi)|+1 = 1$,  there exists a $p \in P$ such that $\rho = p$. Then $\varphi_{\xi,\rho}(\varepsilon)=(p,p)$. Moreover, due to the construction of $\delta$, we have 
\[\wt_\cA(\xi,\varphi_{\xi,\rho}) = \delta_0(\varepsilon,\alpha,(p,p)) = \1\enspace.
\]
Since $\wt_\cB^-(\varepsilon,p)=\1$ by definition, we obtain~\eqref{eq:fold}.

\underline{Case (b):} Let $\xi \in \Gamma$. Now let $\rho \in P^{|\yield_\Gamma(\xi)|+1}$. Since $|\yield_\Gamma(\xi)|+1 = 2$,  there exist $p,p' \in P$ such that $\rho = pp'$. Then $\varphi_{\xi,\rho}(\varepsilon)=(p,p')$. Moreover, due to the construction of $\delta$, we have 
\[\wt_\cA(\xi,\varphi_{\xi,\rho}) = \delta_0(\varepsilon,\alpha,(p,p')) = \mu(p,\alpha,p')\enspace.
\]
Since $\wt_\cB^-(\varepsilon,p)=\mu(p,\alpha,p')$ by definition, we obtain~\eqref{eq:fold}.

I.S.: Let $\xi = \sigma(\xi_1,\ldots,\xi_k)$ with $k \in \mathbb{N}_+$. Then we can calculate as follows:
\begin{align*}
  & \wt_\cB^-(\yield_\Gamma(\xi),\rho)  =   \bigotimes_{l\in [k]} \wt_\cB^-(\yield_\Gamma(\xi_l),\rho|_{(i_l,j_l)})
  \\
   =& \ \bigotimes_{l\in [k]} \wt_\cA(\xi_l,\rho_{\xi_l,\rho|_{(i_l,j_l)}})
      \tag{by  I.H.}\\
  =& \Big(\bigotimes_{l\in[k]} \wt_\cA(\xi_l,(\varphi_{\xi,\rho})|_l)\Big) \otimes
     \delta_k(\varphi_{\xi,\rho}(1) \cdots \varphi_{\xi,\rho}(k),\sigma, \varphi_{\xi,\rho}(\varepsilon))
     \tag{\text{by \eqref{eq:sub-runs} and definition of $\delta$}}\\
  =& \ \wt_\cA(\xi,\varphi_{\xi,\rho})\enspace.
\end{align*}
This finishes the proof of \eqref{eq:fold}.

\

Let $\xi \in \T_\Sigma$ and assume that $\yield_\Gamma(\xi)$ has length $n$.
For each
\(\rho_\cA \in \R_\cA(\xi) \setminus \{\varphi_{\xi,\rho} \mid \rho \in P^{n+1}\}
\)
there exists a $w \in \pos(\xi)$ such that, assuming that $\rk_\Sigma(\xi(w))=k$, we have 
\begin{compactitem}
\item $\rho_\cA(w)_1 \not= \rho_\cA(w1)_1$ or $\rho_\cA(w)_2 \not= \rho_\cA(wk)_2$ or
  \item there exists an $i \in [k-1]$ such that $\rho_\cA(wi)_2\not= \rho_\cA(w(i+1))_1$ or
  \item $k=0$, $\xi(w) \not\in \Gamma$, and $\rho_\cA(w)_1 \not= \rho_\cA(w)_2$.
\end{compactitem}
In this case, by definition of $\delta_k$, we have $\delta_k(\rho_\cA(w1) \cdots \rho_\cA(wk), \xi(w),\rho_\cA(w)) = \mathbb{0}$ and thus $\wt_\cA(\xi,\rho_\cA)=\mathbb{0}$.
Thus we have 
\begin{equation}
\wt_\cA(\xi,\rho_\cA)=\mathbb{0} \text{ for each $\rho_\cA \in \R_\cA(\xi) \setminus  \{\varphi_{\xi,\rho} \mid \rho \in P^{n+1}\}$.}\label{eq:fold-2}
\end{equation}

For each $\rho \in P^{n+1}$ we denote the first state of $\rho$ and the last state of $\rho$  by $\mathrm{fst}(\rho)$ and $\mathrm{lst}(\rho)$, respectively. 
 Then:
\begingroup
\allowdisplaybreaks
\begin{align*}
  &\runsem{\cA}(\xi) =  \bigoplus_{\rho_\cA \in \R_\cA(\xi)} \wt_\cA(\xi,\rho_\cA) \otimes  F_{\rho_\cA(\varepsilon)}\\
  =& \bigoplus_{\rho \in P^{n+1}} \wt_\cA(\xi,\varphi_{\xi,\rho}) \otimes F_{(\mathrm{fst}(\rho),\mathrm{lst}(\rho))} \tag{\text{by \ref{eq:fold-2}} and the fact that $\varphi_{\xi,\rho}(\varepsilon) = (\mathrm{fst}(\rho),\mathrm{lst}(\rho))$}\\
    =& \bigoplus_{\rho \in P^{n+1}}  \wt_\cB^-(\yield_\Gamma(\xi),\rho)     \otimes F_{(\mathrm{fst}(\rho),\mathrm{lst}(\rho))} \tag{by \eqref{eq:fold}}\\
  =& \bigoplus_{(p,p')\in Q'} \bigoplus_{\substack{\rho \in P^{n+1}:\\
  \mathrm{fst}(\rho)=p, \mathrm{lst}(\rho)=p'}}  \wt_\cB^-(\yield_\Gamma(\xi),\rho)     \otimes F_{(p,p')}\\
 =& \bigoplus_{(p,p')\in Q'} \bigoplus_{\substack{\rho \in P^{n+1}:\\
  \mathrm{fst}(\rho)=p, \mathrm{lst}(\rho)=p'}} \wt_\cB^-(\yield_\Gamma(\xi),\rho)     \otimes   \lambda(p)  \otimes \gamma(p')   \tag{\text{by construction}}\\
  &= \bigoplus_{(p,p')\in Q'} \bigoplus_{\substack{\rho \in P^{n+1}:\\
  \mathrm{fst}(\rho)=p, \mathrm{lst}(\rho)=p'}} \lambda(p)  \otimes \wt_\cB^-(\yield_\Gamma(\xi),\rho)\otimes \gamma(p') \tag{because $\im(\lambda)\subseteq \{\0,\1\}$}\\
  =&  \bigoplus_{(p,p')\in Q'} \bigoplus_{\substack{\rho \in P^{n+1}:\\
  \mathrm{fst}(\rho)=p, \mathrm{lst}(\rho)=p'}} \wt_\cB(\yield_\Gamma(\xi),\rho)   \tag{\text{by definition}}\\
 =& \ \runsem{\cB}(\yield_\Gamma(\xi)) \enspace. \qedhere
  \end{align*}
  \endgroup  
\end{proof}

The next theorem was proved in \cite[Thm.~2]{malsat09} via a direct construction. We show a modular proof by  exploiting Lemma~\ref{lm:BPS} and closure under Hadamard product (where the latter requires that $\B$ is a commutative semiring).

\begin{theorem-rect}\label{thm:BPS} {\rm \cite[Thm.~2]{malsat09}} Let $\Sigma$ be a ranked alphabet and $\Gamma$ be an alphabet such that $\Gamma\subseteq \Sigma^{(0)}$. Moreover, let $\B=(B,\oplus,\otimes,\0,\1)$ be a commutative semiring. For every $(\Sigma,\B)$-wta $\cA$ and every $(\Gamma,\B)$-wsa $\cB$, we can construct a $(\Sigma,\B)$-wta $\cA'$ such that
  \[\sem{\cA'} = \sem{\cA} \otimes (\sem{\cB}\circ \yield_\Gamma) \enspace.
  \]
  Thus, in particular, the set $\Rec(\Sigma,\B)$ is closed under yield-intersection.
\end{theorem-rect}
\begin{proof} By Corollary \ref{cor:semiring-run=init} and by convention on page \pageref{page:convention-sem-wta} we have $\sem{\cA}=\runsem{\cA}$ and $\sem{\cA'}=\runsem{\cA'}$. Moreover, by Corollary \ref{cor:wsa-semiring-equal} and by convention on page \pageref{page:convention-sem-wsa} we have $\sem{\cB} = \runsem{\cB}$.  By Lemma \ref{lm:BPS}, we can construct a $(\Sigma,\B)$-wta $\cC$ such that $\sem{\cC} = \sem{\cB} \circ \yield_\Gamma$. Then, by Theorem \ref{thm:closure-Hadamard-product}(1), we can construct a $(\Sigma,\B)$-wta $\cA'$ such that $\sem{\cA'} = \sem{\cA} \otimes \sem{\cC}$.
\end{proof}

Finally, we verify that Theorem \ref{thm:BPS} generalizes the classical result of Bar-Hillel, Perles, and Shamir \cite[Thm.~8.1]{barpersha61}. We achieve this by proving that the latter is equivalent to Theorem  \ref{thm:BPS}  for the case that $\B$ is the Boolean semiring.

\begin{corollary}\rm \label{cor:BPS-classical}  Let $G$ be a $\Gamma$-cfg and $A$ be a $\Gamma$-fsa. Then we can construct a $\Gamma$-cfg $G'$ such that $\LL(G') = \LL(G) \cap \LL(A)$.
\end{corollary}

\begin{proof} By Corollary \ref{cor:cfg=yield(fta)-JGC}, we can construct a ranked alphabet $\Sigma$ with 
$\Gamma \subseteq \Sigma^{(0)}$ and a $\Sigma$-fta $A'$ such that $\LL(G) = \yield_\Gamma(\LL(A'))$. By Corollary \ref{cor:supp-B=fta-1}, we can construct a $(\Sigma,\Boole)$-wta $\cA'$ such that $\LL(G) = \yield_\Gamma(\supp(\sem{\cA'}))$.
  Moreover, by Observation \ref{obs:fsa=wsa(B)}, we can construct a $(\Gamma,\Boole)$-wsa $\cA$ such that $\LL(A)=\supp(\runsem{\cA})$. Then we can calculate as follows.
  \begingroup
  \allowdisplaybreaks
  \begin{align*}
    \LL(G) \cap \LL(A) &= \yield_\Gamma(\supp(\sem{\cA'})) \cap \supp(\runsem{\cA})\\
                       &= \supp(\chi(\yield_\Gamma)(\sem{\cA'})) \cap \supp(\runsem{\cA}) \tag{by \eqref{equ:supp-yield=yieldA}}\\
     &= \supp(\chi(\yield_\Gamma)(\sem{\cA'}) \wedge \runsem{\cA})\\
                       &=^{(*)} \supp(\chi(\yield_\Gamma)(\sem{\cA'} \wedge (\runsem{\cA} \circ \yield_\Gamma)))\enspace,
  \end{align*}
  \endgroup
    where at $(*)$ we have used the following subcalculation for each $w \in \Gamma^*$:
  \begingroup
  \allowdisplaybreaks
  \begin{align*}
    (\chi(\yield_\Gamma)(\sem{\cA'}) \wedge \runsem{\cA})(w)
    &= \chi(\yield_\Gamma)(\sem{\cA'})(w) \wedge \runsem{\cA}(w)\\
    &= \big(\infsum{\vee}{\xi \in \yield_\Gamma^{-1}(w)}{\sem{\cA'}(\xi)}\big) \wedge \runsem{\cA}(w)\\
    &= \infsum{\vee}{\xi \in \yield_\Gamma^{-1}(w)}{\big(\sem{\cA'}(\xi) \wedge \runsem{\cA}(w)\big)}
      \tag{by distributivity}\\
    &= \infsum{\vee}{\xi \in \yield_\Gamma^{-1}(w)}{\big(\sem{\cA'}(\xi) \wedge \runsem{\cA}(\yield_\Gamma(\xi))\big)}
      \\
    &= \infsum{\vee}{\xi \in \yield_\Gamma^{-1}(w)}{\big(\sem{\cA'} \wedge \runsem{\cA} \circ \yield_\Gamma\big)(\xi)}
      \\
    &= \chi(\yield_\Gamma)(\sem{\cA'} \wedge \runsem{\cA} \circ \yield_\Gamma)(w) \enspace.
      \end{align*} 
  \endgroup

  By Theorem \ref{thm:BPS}, we can construct a $(\Sigma,\Boole)$-wta $\cB$ such that $\runsem{\cB} = \sem{\cA'} \wedge (\runsem{\cA} \circ \yield_\Gamma)$. Hence we can continue with:
  \begingroup
  \allowdisplaybreaks
  \begin{align*}
    \supp(\chi(\yield_\Gamma)(\sem{\cA'} \wedge (\runsem{\cA} \circ \yield_\Gamma)))
    &= \supp(\chi(\yield_\Gamma)(\runsem{\cB})) \\
    &= \yield_\Gamma(\supp(\runsem{\cB})) \tag{by \eqref{equ:supp-yield=yieldA}}\enspace.
  \end{align*}
  \endgroup
  By Corollary \ref{cor:supp-B=fta-1}, we can construct a $\Sigma$-fta $C$ such that $\supp(\runsem{\cB}) = \LL(C)$. Moreover, by Corollary~\ref{cor:cfg=yield(fta)-JGC}, we can construct a $\Gamma$-cfg $G'$ such that $\yield_\Gamma(\LL(C)) = \LL(G')$.
\end{proof}


\section{Closure under strong bimonoid homomorphisms}\label{sect:strong-bm-bimorphism}

A set $\cL$ of weighted tree languages is \emph{closed under strong bimonoid homomorphisms} if for every $(\Sigma,\B)$-weighted tree language $r \in \cL$, strong bimonoid $\sfC=(C,+,\times,0,1)$, and strong bimonoid homomorphism $f: B \to C$, the $(\Sigma,\sfC)$-weighted tree language $f\circ r$ is in $\cL$.

\index{f@$f$-image}
\index{image@$f$-image}
\index{f@$f(\cA)$}
\label{p:f(A)}
Let $\cA=(Q,\delta,F)$ be a $(\Sigma,\B)$-wta, $\C$ be a strong bimonoid, and $f: B \rightarrow C$ be a strong bimonoid homomorphism. Then we define the \emph{$f$-image of $\cA$}, denoted by $f(\cA)$, to be the $(\Sigma,\sfC)$-wta $(Q,\delta',F')$ by defining $(\delta')_k = f \circ \delta_k$ for each $k \in \mathbb{N}$, and $F' = f \circ F$. 

Note that,  if $\cA$ is bu-deterministic (or crisp-deterministic), then so is $f(\cA)$. Moreover, if $\cA$ is total and $f^{-1}(0) = \{\0\}$, then $f(\cA)$  is total. The condition $f^{-1}(0) = \{\0\}$ cannot be dropped, which can be seen as follows. Let $\cA=(Q,\delta,F)$ be a total $(\Sigma,\Nat)$-wta (where $\Nat$ is the semiring of natural numbers) such that there exist  $k\in \mathbb{N}$, $\sigma\in \Sigma^{(k)}$, $w\in Q^k$, and $q\in Q$ with $\delta_k(w,\sigma,q)=4$ and for each  $p\in Q\setminus\{q\}$ we have $\delta_k(w,\sigma,p)=0$. We consider the ring $\Intfour=(\{0,1,2,3\},+_4,\cdot_4,0,1)$ of natural numbers modulo 4. Finally, we consider the canonical semiring homomorphism $f:\mathbb{N} \to \{0,1,2,3\}$. Since $f(0)=f(4)=0$, in $f(\cA)$ we have $(\delta')_k(w,\sigma,p)= 0$ for every $p\in Q$. Hence $f(\cA)$ is not total.

Now we can prove the main result of this section.

\begin{lemma}\label{lm:f-image-equivalent} \rm Let $\cA=(Q,\delta,F)$ be a $(\Sigma,\B)$-wta, $(C,+,\times,0,1)$ be a strong bimonoid, and $f: B \rightarrow C$ be a strong bimonoid  homomorphism. Then 
 $f \circ \initialsem{\cA} = \initialsem{f(\cA)}$ and $f \circ \runsem{\cA} = \runsem{f(\cA)}$. 
\end{lemma}
\begin{proof}
First we show that $f \circ \initialsem{\cA} = \initialsem{f(\cA)}$. For this, we define the mapping $\widetilde{f}: B^Q \to C^Q$ for every $v \in B^Q$ and $q \in Q$ by $\widetilde{f}(v)_q = f(v_q)$. We prove that $\widetilde{f}$ is a $\Sigma$-algebra homomorphism from the vector algebra $\V(\cA)=(B^Q,\delta_\cA)$ to the vector algebra $\V(f(\cA))=(C^Q,\delta'_{f(\cA)})$. For this, let $k \in \mathbb{N}$, $\sigma \in \Sigma^{(k)}$, and $v_1,\ldots,v_k \in B^Q$. Then we can calculate as follows.
\begingroup
\allowdisplaybreaks
\begin{align*}
\widetilde{f}(\delta_\cA(\sigma)(v_1,\ldots,v_k))_q
  =& f(\delta_\cA(\sigma)(v_1,\ldots,v_k)_q)\\
  =& f \Big(\bigoplus_{q_1 \cdots q_k \in Q^k} \Big(\bigotimes_{i\in[k]} (v_i)_{q_i}\Big) \otimes \delta_k(q_1\cdots q_k, \sigma, q)\Big)\\
  =& \bigplus_{q_1 \cdots q_k \in Q^k} \Big(\bigtimes_{i\in[k]} f((v_i)_{q_i})\Big) \times f(\delta_k(q_1\cdots q_k, \sigma, q))
  \tag{because $f$ is a strong bimonoid homomorphism}\\
    =& \bigplus_{q_1 \cdots q_k \in Q^k} \Big(\bigtimes_{i\in[k]} \widetilde{f}(v_i)_{q_i}\Big) \times (\delta')_k(q_1\cdots q_k, \sigma, q)
  \tag{be definition of $\widetilde{f}$ and $\delta'$}\\
  =& \delta'_{f(\cA)}(\sigma)(\widetilde{f}(v_1),\ldots,\widetilde{f}(v_k))_q \enspace.
\end{align*}
\endgroup
Hence $\widetilde{f}$ is a $\Sigma$-algebra homomorphism. By Theorem \ref{thm:comp-hom}, $\widetilde{f} \circ \h_\cA$ is a $\Sigma$-algebra homomorphism from the $\Sigma$-term algebra $(\T_\Sigma,\ttop_\Sigma)$ to the vector algebra $\V(f(\cA))=(C^Q,\delta'_{f(\cA)})$. Since $\h_{f(\cA)}$ is also a $\Sigma$-algebra homomorphism of the same type, and $(\T_\Sigma,\ttop_\Sigma)$ is initial, it follows that $\h_{f(\cA)} = \widetilde{f} \circ \h_\cA$. 
Hence, for every $\xi \in \T_\Sigma$ and $q \in Q$, we have $\h_{f(\cA)}(\xi)_q = f(\h_\cA(\xi)_q)$.

Then for each $\xi \in \T_\Sigma$:
\begingroup
\allowdisplaybreaks
\begin{align*}
(f \circ \initialsem{\cA})(\xi) &= f(\initialsem{\cA}(\xi)) 
= f(\bigoplus_{q\in Q} \h_\cA(\xi)_q \otimes F_q)\\
&= \bigplus_{q\in Q} f(\h_\cA(\xi)_q) \times f(F_q)
= \bigplus_{q\in Q} \h_{f(\cA)}(\xi)_q \times F'_q = \initialsem{f(\cA)}(\xi)\enspace.
\end{align*}
\endgroup

Now we show that $f \circ \runsem{\cA} = \runsem{f(\cA)}$. For this, by induction on $\T_\Sigma$, we prove that the following statement holds: 
\begin{equation}
\text{For every $\xi \in \T_\Sigma$ and $\rho \in \R_\cA(\xi)$, we have: } \wt_{f(\cA)}(\xi,\rho) = f(\wt_\cA(\xi,\rho)) \enspace .\label{equ:sr-hom-run}
\end{equation}
Let $\xi = \sigma(\xi_1,\ldots,\xi_k)$. Then
\begingroup
\allowdisplaybreaks
\begin{align*}
  \wt_{f(\cA)}(\xi,\rho) 
 = & \Big(\bigtimes_{i\in[k]} \wt_{f(\cA)}(\xi_i,\rho|_i)\Big) \times \delta'_k(\rho(1)\cdots\rho(k),\sigma,\rho(\varepsilon)) \\
  = & \Big(\bigtimes_{i\in[k]} f(\wt_{\cA}(\xi_i,\rho|_i))\Big) \times f(\delta_k(\rho(1)\cdots\rho(k),\sigma,\rho(\varepsilon)))
      \tag{\text{by I.H. and construction, recall that $\rho|_i\in \R_\cA(\xi_i)$ for $1\le i\le k$}}\\[2mm]
  = & f\Big(\Big(\bigotimes_{i\in[k]} \wt_{\cA}(\xi_i,\rho|_i)\Big)\otimes \delta_k(\rho(1)\cdots\rho(k),\sigma,\rho(\varepsilon))\Big)
      \tag{\text{by the fact that $f$ is a  strong bimonoid homomorphisms}}\\[2mm]
= &  f(\wt_\cA(\xi,\rho)).
\end{align*}
\endgroup
This proves \eqref{equ:sr-hom-run}. Then  for each $\xi \in \T_\Sigma$: 
\begin{align*}
&(f \circ \runsem{\cA})(\xi) = f(\runsem{\cA}(\xi)) 
= f\Big(\bigoplus_{\rho \in \R_\cA(\xi)} \wt_\cA(\xi,\rho) \otimes F_{\rho(\varepsilon)}\Big)\\
&= \bigplus_{\rho \in \R_\cA(\xi)} f(\wt_\cA(\xi,\rho)) \times f(F_{\rho(\varepsilon)})
= \bigplus_{\rho \in \R_{f(\cA)}(\xi)}  \wt_{f(\cA)}(\xi,\rho)\times F'_{\rho(\varepsilon)} = \runsem{f(\cA)}(\xi)\enspace,
\end{align*}
where the third equation uses again that $h$ is a  strong bimonoid homomorphism and the fourth one uses the fact that $\R_\cA(\xi)=\R_{f(\cA)}(\xi)$.
\end{proof}

The next theorem generalizes \cite[Lm. 3]{bormalsestepvog06} and \cite[Thm.~3.9]{fulvog09new} from semirings to strong bimonoids. It uses the following abbreviation. Let $\cC$ be a set of $(\Sigma,\B)$-weighted tree languages, $\sfC=(C,+,\times,0,1)$ be a strong bimonoid and $f: B \rightarrow C$ be a  strong bimonoid  homomorphism. We define $f\circ \cC =\{f\circ r \mid r \in \cC\}$. Obviously, $f\circ \cC$ is a set of $(\Sigma,\sfC)$-weighted tree languages.

\begin{theorem-rect}\label{thm:closure-sr-hom} Let $\Sigma$ be a ranked alphabet. Moreover, let $\B=(B,\oplus,\otimes,\0,\1)$ and $\sfC=(C,+,\times,0,1)$ be strong bimonoids, and let $f: B \rightarrow C$ be a  strong bimonoid  homomorphism. Then the following three statements hold.
  \begin{compactenum}
  \item[(1)] $f \circ \Rec^{\mathrm{init}}(\Sigma,\B)\subseteq \Rec^{\mathrm{init}}(\Sigma,\sfC)$ and $f \circ \Rec^{\mathrm{run}}(\Sigma,\B)\subseteq \Rec^{\mathrm{run}}(\Sigma,\sfC)$. 
  \item[(2)] If $f$ is surjective, then $f \circ \Rec^{\mathrm{init}}(\Sigma,\B)=\Rec^{\mathrm{init}}(\Sigma,\sfC)$ and $f \circ \Rec^{\mathrm{run}}(\Sigma,\B)=\Rec^{\mathrm{run}}(\Sigma,\sfC)$.
  \item[(3)] Statements (1) and (2) also hold for the subsets of
$\Rec^{\mathrm{init}}(\Sigma,\B)$, $\Rec^{\mathrm{init}}(\Sigma,\sfC)$, $\Rec^{\mathrm{run}}(\Sigma,\B)$, and $\Rec^{\mathrm{run}}(\Sigma,\sfC)$ which are recognizable by bu-deterministic wta. The same holds if we replace bu-deterministic by crisp-deterministic.
    \end{compactenum}
    Thus, in particular, the sets $\Rec^{\mathrm{init}}(\Sigma,\_)$, $\Rec^{\mathrm{run}}(\Sigma,\_)$ and $\Rec(\Sigma,\_)$ are closed under strong bimonoid homomorphisms.
  \end{theorem-rect}

\begin{proof} Proof of (1):  Let $\cA$ be a $(\Sigma,\B)$-wta such that $r=\initialsem{\cA}$.  By Lemma \ref{lm:f-image-equivalent} we have that $f \circ \initialsem{\cA} = \initialsem{f(\cA)}$. Hence $f \circ r \in \Rec^{\mathrm{init}}(\Sigma,\sfC)$. By a similar argument we can prove that  $r \in \Rec^{\mathrm{run}}(\Sigma,\B)$ implies $f \circ r \in \Rec^{\mathrm{run}}(\Sigma,\sfC)$. 

 \
  
  Proof of (2):  By (1) we have $f \circ \Rec^{\mathrm{init}}(\Sigma,\B)\subseteq \Rec^{\mathrm{init}}(\Sigma,\sfC)$ and $f \circ \Rec^{\mathrm{run}}(\Sigma,\B)\subseteq \Rec^{\mathrm{run}}(\Sigma,\sfC)$.

  Next we assume that $f$ is surjective. Then we show that also the other inclusions hold. For this, let $\cA'=(Q,\delta',F')$ be an arbitrary $(\Sigma,\sfC)$-wta. We construct the $(\Sigma,\B)$-wta
$\cA=(Q,\delta,F)$ such that
\begin{compactitem}
\item for every $k\in \mathbb{N}$, $\sigma\in \Sigma^{(k)}$, $w\in Q^k$, and $q\in Q$ we let
  $\delta_k(w,\sigma,q)=b$ where $b$ is determined as follows:
  if  $\delta'_k(w,\sigma,q) = 0$, then we let $b = \0$, if $\delta'_k(w,\sigma,q) = 1$,
   then we let $b = \1$, and if $\delta'_k(w,\sigma,q) \not\in \{0,1\}$,
   then we let $b$ be an arbitrary element in $f^{-1}(\delta'_k(w,\sigma,q))$,
and 
\item for each $q\in Q$, we let $F_q=b$, for some $b\in f^{-1}(F'_q)$.
\end{compactitem}
In both items such values  $b$ exist because $f$ is surjective.  We have $\cA'=f(\cA)$ and by Lemma 
\ref{lm:f-image-equivalent} we conclude
$f \circ \initialsem{\cA} = \initialsem{\cA'}$ and $f \circ \runsem{\cA} = \runsem{\cA'}$. Hence the other inclusions also hold.

\

Proof of (3): This statement holds by (1) and (2) and by the fact that the constructions of the $f$-image $f(\cA)$ and of $\cA'$ (in the proof of (2)) preserve bu-determinism and crisp-determinism.
\end{proof}

\newcommand{\free}{\mathrm{free}}
\newcommand{\wts}{\mathrm{wts}}

\index{free variant of $\cA$}
\index{freefA@$\free_f(\cA)$}
Finally, we use the closure under strong bimonoid homomorphism in order to decompose the semantics of a wta over some strong bimonoid $\B$ into the semantics of a wta over the free monoid $\sfnPol(X_n)$ of nested polynomials and the strong bimonoid homomorphism $\eval_{\B,f}$. For the definition of $\sfnPol(X_n)$ and of $\eval_{\B,f}$ we refer to Subsection~\ref{sec:free-strong-bimonoid}. Formally, let $\cA=(Q,\delta,F)$ be a $(\Sigma,\B)$-wta. Moreover, let $|\wts(\cA)|=n$ and $f:X_n \to \wts(\A)$ be a bijection. We define the \emph{free variant of $\cA$ with respect to $f$}, denoted by $\free_f(\cA)$, to be the $(\Sigma,\sfnPol(X_n))$-wta  $\free_f(\cA) =(Q,\delta',F)$ where, for each $k \in \mathbb{N}$, $(\delta')_k = f^{-1} \circ \delta_k$.  In fact, $\cA$ and $\free_f(\cA)$ are identical except that $\cA$ uses elements and operations of the strong bimonoid $\B$ whereas $\free_f(\cA)$ uses the corresponding variables and operations of the free strong bimonoid $\sfnPol(X_n)$ of nested polynomials.

\begin{theorem-rect} \label{thm:MW-theorem} Let $\Sigma$ be a ranked alphabet and  $\B$ be a  strong bimonoid. Moreover, let $\cA$ be a $(\Sigma,\B)$-wta and $f:X_n \to \wts(\A)$ be a bijection where  $n=|\wts(\cA)|$. Then $\runsem{\cA} = \eval_{\B,f} \circ \runsem{\free_f(\cA)}$ and $\initialsem{\cA} = \eval_{\B,f} \circ \initialsem{\free_f(\cA)}$.
\end{theorem-rect}
\begin{proof} Let $\cA$ be a $(\Sigma,\B)$-wta and $f:X_n \to \wts(\A)$ be a bijection where  $n=|\wts(\cA)|$. First we remark that
  \begin{equation}\label{equ:eval-free-A=A}
    \eval_{\B,f}(\free_f(\cA)) = \cA \enspace,
  \end{equation}
 where $\eval_{\B,f}(\free_f(\cA))$ is the $\eval_{\B,f}$-image of $\free_f(\cA)$ defined on p.\pageref{p:f(A)}.

By Lemma~\ref{lm:eval-str-bm-hom}, the mapping  $\eval_{\B,f}$ is a strong bimonoid homomorphism from $\sfnPol(X_n)$ to $\B$. Then, by Lemma \ref{lm:f-image-equivalent} and \eqref{equ:eval-free-A=A}, we have
 \begin{equation}
\eval_{\B,f} \circ \initialsem{\free_f(\cA)} = \initialsem{\eval_{\B,f}(\free_f(\cA))} = \initialsem{\cA}
\end{equation}
and
 \begin{equation}
\eval_{\B,f} \circ \runsem{\free_f(\cA)} = \runsem{\eval_{\B,f}(\free_f(\cA))} = \runsem{\cA}
   \end{equation}
  \end{proof}

  The reader might wish to compare Theorem \ref{thm:MW-theorem} with the approach of abstract semantics in \cite[Sect.~3.5]{gasmon18}.


\section{\ Closure under tree relabelings}\label{sect:tree-relabeling-preserves}

In this section we will prove that $\Rec^{\mathrm{run}}(\_\,,\B)$ is closed under tree relabelings. We recall that a $(\Sigma,\Delta)$-tree relabeling is a family  $\tau = (\tau_k \mid k \in \mathbb{N})$ such that $\tau_k(\sigma) \subseteq \Delta^{(k)}$ for each $\sigma \in \Sigma^{(k)}$. A tree relabeling $\tau$ is non-overlapping if $\tau_k(\sigma) \cap \tau_k(\sigma') = \emptyset$ for every $k \in \mathbb{N}$ and $\sigma,\sigma' \in \Sigma^{(k)}$ with $\sigma\ne \sigma'$. Its extension is the mapping $\tau:  \T_\Sigma \to \cP_{\mathrm{fin}}(\T_\Delta)$, which can be also considered as binary relation $\tau \subseteq \T_\Sigma \times \T_\Delta$. Then its characteristic mapping $\chi(\tau): \T_\Sigma \times \T_\Delta \to B$ is supp-i-finite.

 Since $\chi(\tau)$ is supp-i-finite, for each $r: \T_\Sigma \to B$, the application $\chi(\tau)(r)$ is defined by \eqref{equ:appl-wtt-to-wtl} and  by \eqref{equ:appl-wttfinpreim-to-wtl} and \eqref{obs:app-tree-transf-to-wtl-2}, for each $\zeta \in \T_\Delta$, we have
\[
  \chi(\tau)(r)(\zeta) = \bigoplus_{\xi \in \tau^{-1}(\zeta)} r(\xi) \enspace.
\]

A set $\cL$ of $\B$-weighted tree languages is \emph{closed under tree relabelings} if for every $(\Sigma,\B)$-weighted tree language $r \in \cL$ and every $(\Sigma,\Delta)$-tree relabeling $\tau$, the $(\Delta,\B)$-weighted tree language $\chi(\tau)(r)$ is in~$\cL$.

\begin{theorem-rect} \label{thm:closure-under-tree-relabeling} {\rm (cf. \cite[Lm.~6]{stuvogfue09})} Let $\Sigma$ and $\Delta$ be ranked alphabets. Moreover, let $\B=(B,\oplus,\otimes,\0,\1)$ be a strong bimonoid and let $\cA$ be a $(\Sigma,\B)$-wta. Also, let $\tau$ be a $(\Sigma,\Delta)$-tree relabeling.
  Then we can construct a $(\Delta,\B)$-wta $\cB$ such that
  \[
    \runsem{\cB}=\chi(\tau)\big(\runsem{\cA}\big)\enspace.
  \]
  If, moreover,  $\cA$ is bu-deterministic (or: crisp-deterministic) and  $\tau$ is non-overlapping, then $\cB$ is bu-deterministic (and  crisp-deterministic, respectively). 
  Thus, in particular, the set $\Rec^{\mathrm{run}}(\_\,,\B)$ is closed under tree relabelings.
\end{theorem-rect}

\begin{proof} Let $\cA = (Q,\delta,F)$ be a $(\Sigma,\B)$-wta and $\tau = (\tau_k \mid k \in \mathbb{N})$ be a $(\Sigma,\Delta)$-tree relabeling.

  We construct the $(\Delta,\B)$-wta $\cB = (Q',\delta',F')$ where
  \begin{compactitem}
  \item $Q' = Q \times \Sigma$
    
  \item  for every $k \in \mathbb{N}$, $\gamma \in \Delta^{(k)}$, $(q,\sigma),(q_1,\sigma_1),\ldots,(q_k,\sigma_k) \in Q \times \Sigma$,
    \[
      (\delta')_k((q_1,\sigma_1)\cdots (q_k,\sigma_k),\gamma,(q,\sigma)) =
      \begin{cases}
        \delta_k(q_1\cdots q_k,\sigma,q) & \text{ if $\gamma \in \tau_k(\sigma)$}\\
        \0 & \text{ otherwise,}
        \end{cases}
      \]
        \item $(F')_{(q,\gamma)} = F_q$ for each $(q,\gamma)\in Q'$. 
  \end{compactitem}
  In general, the above construction does not preserve bu-determinism. For instance, let $a,b \in B$ with $a \ne \0 \ne b$, and let $\cA$ be a bu-deterministic $(\Sigma,\B)$-wta with transition mapping $\delta_1$ such that $\delta_1(q,\sigma_1,p)=a$ and $\delta_1(q,\sigma_2,p')=b$. Moreover, let $\tau_1(\sigma_1)=\tau_1(\sigma_2)=\{\gamma\}$ and $\kappa \in \Sigma$. Then $(\delta'_1)((q,\kappa),\gamma,(p,\sigma_1))= a$ and $(\delta'_1)((q,\kappa),\gamma,(p',\sigma_2))= b$ and hence $\cB$ is not bu-deterministic.

However, if $\tau$ is non-overlapping, then this phenomenon cannot occur. Thus, if $\cA$ is bu-deterministic and  $\tau$ is non-overlapping, then $\cB$ is bu-deterministic. Moreover,  if $\cA$ is crisp-deterministic and  $\tau$ is non-overlapping, then $\cB$ is crisp-deterministic.

  Next we prove that $\runsem{\cB}=\chi(\tau)\big(\runsem{\cA}\big)$.
  Let $\xi \in \T_\Sigma$ and $\zeta \in \tau(\xi)$. We define the mapping $\varphi_{\xi,\zeta} : \R_\cA(\xi) \to \R_\cB(\zeta)$ for every $\rho \in \R_\cA(\xi)$ and $w \in \pos(\xi)$ by
  $\varphi_{\xi,\zeta}(\rho)(w) = (\rho(w),\xi(w))$. Obviously, $\varphi_{\xi,\zeta}$ is injective. Moreover, we define the mapping $\varphi_{\xi,\zeta}' : \R_\cA(\xi) \to \im(\varphi_{\xi,\zeta})$ by $\varphi_{\xi,\zeta}'(\rho) = \varphi_{\xi,\zeta}(\rho)$ for each $\rho \in \R_\cA(\xi)$. Clearly, $\varphi_{\xi,\zeta}'$ is bijective.

  The following statements are easy to see.
  \begin{equation}
    \text{For every $\zeta \in \T_\Delta$, $\xi \in \tau^{-1}(\zeta)$, and $\rho \in \R_\cA(\xi)$: \
      $\wt_\cA(\xi,\rho) = \wt_\cB(\zeta,\varphi_{\xi,\zeta}'(\rho))$. } \label{eq:equality-on-runs}
   \end{equation}
    \begin{equation}
\text{For every $\zeta \in \T_\Delta$ and $\rho' \in \R_\cB(\zeta) \setminus \bigcup_{\xi \in \tau^{-1}(\zeta)} \im(\varphi'_{\xi,\zeta}): \  \wt_\cB(\zeta,\rho')=\0$.} \label{eq:run=0-outside}
\end{equation}
  \begin{equation}
    \text{For every $\zeta \in \T_\Delta$ and $\xi_1,\xi_2 \in \tau^{-1}(\zeta)$: \
if $\xi_1\not=\xi_2$, then $\im(\varphi'_{\xi_1,\zeta}) \cap \im(\varphi'_{\xi_2,\zeta}) = \emptyset$.  } \label{eq:disjoint-xi}
\end{equation}

Let $\zeta \in \T_\Delta$. We can calculate as follows.
\begingroup
\allowdisplaybreaks
\begin{align*}
  \runsem{\cB}(\zeta) &= \bigoplus_{\rho' \in \R_\cB(\zeta)} \wt_\cB(\zeta,\rho') \otimes (F')_{\rho'(\varepsilon)}\\
                      &= \bigoplus_{\rho' \in \bigcup_{\xi \in \tau^{-1}(\zeta)} \im(\varphi'_{\xi,\zeta})}
                        \wt_\cB(\zeta,\rho') \otimes (F')_{\rho'(\varepsilon)}
  \tag{\text{by \eqref{eq:run=0-outside}}}\\
                     &= \bigoplus_{\xi \in \tau^{-1}(\zeta)} \bigoplus_{\rho' \in \im(\varphi'_{\xi,\zeta})}
                       \wt_\cB(\zeta,\rho') \otimes (F')_{\rho'(\varepsilon)}
  \tag{\text{by \eqref{eq:disjoint-xi}}}\\
                      &= \bigoplus_{\xi \in \tau^{-1}(\zeta)} \bigoplus_{\rho \in \R_\cA(\xi)} \wt_\cB(\zeta,\varphi_{\xi,\zeta}'(\rho)) \otimes F_{\rho(\varepsilon)}
  \tag{\text{because $\varphi_{\xi,\zeta}'$ is bijective}}\\
                      &= \bigoplus_{\xi \in \tau^{-1}(\zeta)} \bigoplus_{\rho \in \R_\cA(\xi)} \wt_\cA(\xi,\rho) \otimes F_{\rho(\varepsilon)}
  \tag{\text{by \eqref{eq:equality-on-runs}}}\\
  &= \bigoplus_{\xi \in \tau^{-1}(\zeta)} \runsem{\cA}(\xi) = \chi(\tau)\big(\runsem{\cA}\big)(\zeta)\enspace. \qedhere
\end{align*}
\endgroup
\end{proof}

As a consequence of Theorem \ref{thm:closure-under-tree-relabeling} we reobtain the well-known closure of recognizable tree languages under tree relabelings.

\begin{corollary} \rm (cf. \cite[Thm.~3.48]{eng75-15}) \label{cor:recog-closed-under-tree-relabeling} Let $A$ be a $\Sigma$-fta and $\tau$ a $(\Sigma,\Delta)$-tree relabeling. Then we can construct a $\Delta$-fta $B$ such that $\LL(B)= \tau(\LL(A))$.
\end{corollary}

\begin{proof} By Corollary \ref{cor:supp-B=fta-1}(A)$\Rightarrow$(B), we can construct a $(\Sigma,\Boole)$-wta $\cA$ such that $\LL(A) = \supp(\sem{\cA})$. Then we have
  \[
\tau(\LL(A)) = \tau(\supp(\sem{\cA})) = \supp(\chi(\tau)(\sem{\cA}))
    \]
    where the last equality is due to \eqref{equ:supp-yield=yieldA}. By Theorem \ref{thm:closure-under-tree-relabeling} we can construct a $(\Delta,\Boole)$-wta $\cB$ such that $\sem{\cB} = \chi(\tau)(\sem{\cA})$. Finally, by Corollary \ref{cor:supp-B=fta-1}(B)$\Rightarrow$(A), we can construct a $\Delta$-fta $B$ such that $\LL(B) = \supp(\sem{\cB})$.
    Hence $\LL(B) = \tau(\LL(A))$.
  \end{proof}


\section{\ Closure under linear and nondeleting  tree homomorphisms}\label{sect:homomorphism-preserves}

In this section we will prove that $\Rec^{\mathrm{run}}(\_,\B)$ is closed under linear, nondeleting, and productive tree homomorphism. We will even prove a more general closure of regular weighted tree languages under such tree homomorphisms. 

\index{tree homomorphism!productive}
Let $h=(h_k \mid k \in \mathbb{N})$ be a $(\Sigma,\Delta)$-tree homomorphism. We can view its extension $h: \T_\Sigma \to \T_\Delta$ as binary relation $h \subseteq \T_\Sigma \times \T_\Delta$; then its characteristic mapping has type $\chi(h): \T_\Sigma \times \T_\Delta \to B$.  We say that \emph{$h$ is supp-i-finite}  if $\chi(h):\T_\Sigma \times \T_\Delta \to B$ is supp-i-finite, i.e., for each $\zeta \in \T_\Delta$, the set $h^{-1}(\zeta)$ is finite. 

In general, a tree homomorphism $h$ is not supp-i-finite. For instance, if $\Sigma = \{\gamma^{(1)}, \alpha^{(0)}\}$, $\Delta = \{\beta^{(0)}\}$, and  $h(\gamma)=x_1$ and $h(\alpha) = \beta$, then $h^{-1}(\beta) = \{\gamma^n(\alpha) \mid n \in \mathbb{N}\}$ is infinite. However, if $h$ is productive and nondeleting, then $h$ is supp-i-finite.

Now let $h$ be a $(\Sigma,\Delta)$-tree homomorphism and $r: \T_\Sigma \to B$
such that $h$ is supp-i-finite or $\B$ is $\sigma$-complete. Then the application of $\chi(h):\T_\Sigma \times \T_\Delta \to B$  to $r$ is defined by \eqref{equ:appl-wtt-to-wtl}, and by \eqref{obs:app-tree-transf-to-wtl-2}, for each $\zeta \in \T_\Delta$, we have
\[\chi(h)(r)(\zeta)=\infsum{\bigoplus}{\xi \in h^{-1}(\zeta)}{r(\xi)}.\]

Let us denote by $\Hom(\Sigma,\Delta)$ the set of all tree homomorphisms from $\Sigma$ to $\Delta$ and
let $\cC$ be a subset of $\Hom(\Sigma,\Delta)$ such that each $h\in \cC$ is supp-i-finite.
A set $\cL$ of $\B$-weighted tree languages is \emph{closed under tree homomorphisms in $\cC$} if for every $(\Sigma,\B)$-weighted tree language $r \in \cL$ and every $(\Sigma,\Delta)$-tree homomorphism $h$ in  $\cC$, the $(\Delta,\B)$-weighted tree language $\chi(h)(r)$ is in~$\cC$.

The next theorem can be compared to \cite[Thm.~5.3]{fulmalvog11} where the closure of the set of recognizable $\B$-weighted tree languages under linear and nondeleting weighted extended tree transformations was proved if $\B$ is a $\sigma$-complete and commutative semiring. We recall that linear and nondeleting tree homomorphisms are particular linear and nondeleting weighted extended tree transducers. (We also refer to \cite[Prop.~25]{boz99}.)
This result is weaker than what one might expect when looking at \cite[Thm.~3.65]{eng75-15} for the unweighted case. There, it was proved that the set of recognizable tree languages is closed under linear tree homomorphisms (which need not be nondeleting). The underlying construction can be reformulated easily to $\Boole$-weighted tree languages. But we do not know whether it is possible to generalize the latter to $\B$-weighted tree languages because it is not clear how to handle the weights of deleted subtrees.

\begin{theorem}\label{thm:closure-lin-nondel-hom-general}  Let $\cG$ be a $(\Sigma,\B)$-wrtg such that $\cG$ is finite-derivational or $\B$ is $\sigma$-complete. Moreover, let $h$ be a linear and nondeleting tree homomorphism from $\Sigma$ to $\Delta$ such that $h$ is productive or $\B$ is $\sigma$-complete. Then we can construct a $(\Delta,\B)$-wrtg $\cG'$ such that (a) if $\cG$ is finite-derivational (chain-free) and $h$ is productive, then $\cG'$ is finite-derivational (chain-free) and (b)~$\sem{\cG'} =  \chi(h)\big(\sem{\cG}\big)$.
  \end{theorem}
 \begin{proof}  Let $\cG=(N,S,R,wt)$ and $h=(h_k \mid k \in \mathbb{N})$.  By Lemma \ref{lm:rtg-normal-form} we can assume that $\cG$ is alphabetic.
We construct the $(\Delta,\B)$-wrtg $\cG'=(N',S',R',wt')$, where
\begin{compactitem}
\item $N'=N\times \Sigma$,
\item $S'=S\times \Sigma$, and
\item $R'$ is the smallest set of rules satisfying the following conditions:
\begin{compactitem}
\item for each rule $r=(A\to \sigma(A_1,\ldots,A_k))$ in $R$ and for every $\sigma_1,\ldots,\sigma_k\in \Sigma$, the rule $r'=((A,\sigma)\to h_k(\sigma)[(A_1,\sigma_1),\ldots,(A_k,\sigma_k)])$ is in $R'$ and $wt'(r')=wt(r)$, and
\item for each rule $r=(A\to B)$ in $R$ and each $\sigma \in \Sigma$,  the rule $r'=((A,\sigma)\to (B,\sigma))$ in $R'$ and $wt'(r')=wt(r)$.
\end{compactitem}
\end{compactitem}
Since $h$ is linear and nondeleting, the tree $h_k(\sigma)$ is a context for each $\sigma\in \Sigma^{(k)}$ and $k \in \mathbb{N}$. Hence, for each rule of $\cG'$ of the form $(A,\sigma)\to h_k(\sigma)[(A_1,\sigma_1),\ldots,(A_k,\sigma_k)]$,
each nonterminal $(A_i,\sigma_i)$ occurs exactly once in the right-hand side of that rule. Moreover, if 
$\cG$ is chain-free and $h$ is productive, then $\cG'$ is chain-free.

\index{succb@$\succ$}
We define the mapping 
\[\varphi: \RT_\cG(N,\T_\Sigma) \to \RT_{\cG'}(N',\T_\Delta)\]
by induction on the terminating reduction system $(\RT_\cG(N,\T_\Sigma),\succ)$ where $\succ = \succ_R \cap \ (\RT_\cG(N,\T_\Sigma) \times \RT_\cG(N,\T_\Sigma))$.
Then $\nf_\succ(\RT_\cG(N,\T_\Sigma))$ is the set of all rules of $R$ with right-hand side in $\Sigma^{(0)}$.

Let $d \in \RT_\cG(N,\T_\Sigma)$. We can distinguish the following two cases

\underline{Case (a):} Let $d=r(d_1,\ldots,d_k)$ for some rule $r=(A\to \sigma(A_1,\ldots,A_k))$ and $d_i\in \RT_\cG(N,\T_\Sigma)$ for each $i\in[k]$. Then we let $\varphi(d)=r'(d'_1,\ldots,d'_k)$, where $r'=((A,\sigma)\to h_k(\sigma)[(A_1,\sigma_1),\ldots,(A_k,\sigma_k)])$ with $\sigma_i=\pi(d_i)(\varepsilon)$ and $d'_i =\varphi(d_i)$ for each $i\in[k]$.

\underline{Case (b):} Let $d=r(d_1)$ for some rule $r=(A\to B)$ in $R$ and $d_1\in \RT_\cG(N,\T_\Sigma)$. Then we let 
$\varphi(d)=r'(d'_1)$, where $r'=((A,\sigma)\to (B,\sigma))$ and $d'_1 =\varphi(d_1)$.

Due to the construction of $\cG'$,
\begin{equation*}
\text{for every $A \in N$ and $\xi \in \T_\Sigma$, we have } \varphi(\RT_\cG(A,\xi)) = \RT_{\cG'}((A,\xi(\varepsilon)),h(\xi))\enspace,
\end{equation*}
and
\begin{equation}\label{eq:partitioning-homomrphism}
  \text{for every $(A,\sigma)\in N'$ and $\zeta \in \T_\Delta$,  we have } \RT_{\cG'}((A,\sigma),\zeta)=\bigcup_{\substack{\xi\in h^{-1}(\zeta):\\ \xi(\varepsilon)=\sigma}}\varphi(\RT_\cG(A,\xi)) \enspace.\end{equation}
It follows that, if $h$ is productive (and thus it is supp-i-finite) and $\cG$ is finite-derivational, then $\cG'$ is finite-derivational. 
It is easy to see that 
\begin{equation}\label{eq:bijection-homomorphism}
\text{$\varphi$ is a bijection and $\wt_\cG(d)=\wt_{\cG'}(\varphi(d))$ for each 
$d\in \RT_\cG(N,\T_\Sigma)$\enspace.}
\end{equation}
Moreover,
\begin{equation} \label{eq:partitioning-and-proof}
\text{the family $(\varphi(\RT_\cG(A,\xi)) \mid \xi\in h^{-1}(\zeta), \xi(\varepsilon)=\sigma)$ is a partitioning of $\RT_{\cG'}((A,\sigma),\zeta)$ \enspace,}
\end{equation}
because each $d \in \varphi(\RT_\cG(A,\xi))$ encodes $\xi$, thus $\xi\ne \xi'$ implies $\varphi(\RT_\cG(A,\xi))\cap \varphi(\RT_\cG(A,\xi'))=\emptyset$. Hence the union on right-hand side of \eqref{eq:partitioning-homomrphism} is a disjoint union.

Now let $\zeta\in \T_\Delta$. Then we can compute as follows
\begingroup
\allowdisplaybreaks
\begin{align*}
\sem{\cG'}(\zeta)& = \infsum{\oplus}{d'\in \RT_{\cG'}(\zeta)}\wt_{\cG'}(d') = \bigoplus_{(A,\sigma)\in S'}
\ \ \infsum{\oplus}{d'\in \RT_{\cG'}((A,\sigma),\zeta)}\wt_{\cG'}(d') \\[2mm]
                 & =^{(*)}  \bigoplus_{(A,\sigma)\in S'}\;\; \infsum{\oplus}{\substack{\xi \in h^{-1}(\zeta), \xi(\varepsilon) =\sigma\\ d\in \RT_\cG(A,\xi)}} \wt_\cG(d)
  = \bigoplus_{\xi \in h^{-1}(\zeta)}\;\;\infsum{\oplus}{\substack{A\in S\\ d\in \RT_\cG(A,\xi)} } \wt_\cG(d)\\[2mm]
& = \bigoplus_{\xi \in h^{-1}(\zeta)} \sem{\cG}(\xi) = \chi(h)\big(\sem{\cG}\big)(\zeta)\enspace,
\end{align*}
\endgroup
where $(*)$ follows from \eqref{eq:bijection-homomorphism} and \eqref{eq:partitioning-and-proof}.
\end{proof}

\begin{corollary-rect}\rm \label{cor:closure-REC-under-lin-nondel-prod-hom} Let $\Sigma$ and $\Delta$ be ranked alphabets. Moreover, let $\B$ be a strong bimonoid and let $\cA$  be a $(\Sigma,\B)$-wta. Moreover, let $h$ be a linear, nondeleting, and productive tree homomorphism from $\Sigma$ to $\Delta$.  Then we can construct a $(\Delta,\B)$-wta $\cA'$ such that
  \[
    \runsem{\cA'} =  \chi(h)\big(\runsem{\cA}\big) \enspace.
  \]
  Thus, in particular,  the set $\Rec^{\mathrm{run}}(\_,\B)$ is closed under linear, nondeleting, and productive tree homomorphisms.
\end{corollary-rect}
\begin{proof}  By Lemma \ref{lm:wta-to-wrtg} we can construct $(\Sigma,\B)$-wrtg $\cG$ such that $\cG$ is in tree automata form and $\runsem{\cA} = \sem{\cG}$.  In particular, $\cG$ is chain-free.
  By Theorem \ref{thm:closure-lin-nondel-hom-general} we can construct a  chain-free  $(\Sigma,\B)$-wrtg $\cG'$ such that $\sem{\cG'} =  \chi(h)\big(\sem{\cG}\big)$.
  Finally, by Lemma \ref{lm:wrtg-to-wta}(a), we can construct a $(\Sigma,\B)$-wta $\cA'$ such that $\sem{\cG'}=\runsem{\cA'}$.
\end{proof}

In the rest of this section, we show that we can use Corollary \ref{cor:closure-REC-under-lin-nondel-prod-hom} (in combination with Example~\ref{ex:Sigma-algebra-hom-as-wta}(3)) in order to prove Theorem \ref{thm:closure-of-finite-set-i-recognizable}, provided that the underlying strong bimonoid is a semiring. The latter condition is needed in order to get rid of the difference between run semantics (used in  Corollary~\ref{cor:closure-REC-under-lin-nondel-prod-hom}) and initial algebra semantics (used in Example~\ref{ex:Sigma-algebra-hom-as-wta}).

\begin{corollary-rect} \rm (cf. Theorem \ref{thm:closure-of-finite-set-i-recognizable} for a stronger version) \label{lm:closure-of-finite-set-recognizable} Let $\Sigma$ be a ranked alphabet such that $\Sigma^{(2)}\not=\emptyset$. Moreover, let $\B=(B,\oplus,\otimes,\0,\1)$ be a semiring and $A \subseteq B$  be a finite subset. Then we can construct a  $(\Sigma,\B)$-wta  $\cA$  such that  $\im(\sem{\cA}) = \langle A \rangle_{\{\oplus,\otimes,\0,\1\}}$. 
\end{corollary-rect}

\begin{proof}  Clearly, $\langle A \rangle_{\{\oplus,\otimes,\0,\1\}} = \langle A \cup \{\0,\1\} \rangle_{\{\oplus,\otimes\}}$. Let $a_1,\ldots,a_n$ be the elements of $A \cup \{\0,\1\}$, i.e., $A \cup \{\0,\1\} = \{a_1,\ldots,a_n\}$. 

  The idea for the construction of $\cA$ with $\im(\sem{\cA}) = \langle A \rangle_{\{\oplus,\otimes,\0,\1\}}$ is similar to that of the proof of Theorem~\ref{thm:closure-of-finite-set-i-recognizable}. However, we specify the mapping $\mathrm{eval}: \T_\Delta \to  \langle A \rangle_{\{\oplus,\otimes,\0,\1\}}$ in the proof (cf. page \pageref{page:def-of-eval}) by a $(\Delta,\B)$-wta $\cA_1$ here; this is done by using~Example~\ref{ex:Sigma-algebra-hom-as-wta}(3). Then we can apply Corollary~\ref{cor:closure-REC-under-lin-nondel-prod-hom} to $\cA_1$ and the tree homomorphism $h: \T_\Delta \to \T_\Sigma$ defined in the proof of  Theorem~\ref{thm:closure-of-finite-set-i-recognizable}.

  For the sake of convenience, we recall here the definitions of $\Delta$ and $h$.  The ranked alphabet $\Delta$ is $\Delta = \{a^{(0)}_1,\ldots,a^{(0)}_n\} \cup \{\oplus^{(2)}, \otimes^{(2)}\}$. 
  We define the $(\Delta,\Sigma)$-tree homomorphism $h = (h_k \mid k \in \mathbb{N})$ as follows. We fix arbitrary $\alpha \in \Sigma^{(0)}$ and $\sigma \in \Sigma^{(2)}$. Then we let
  \begin{align*}
    h_0(a_i) &= \underbrace{\sigma(\ldots \sigma(}_{i \text{ times}}\alpha \underbrace{,\alpha)\ldots, \alpha)}_{i \text{ times}} \ \text{ for each $i \in [n]$},\\
    h_2(\oplus) &= \sigma(\alpha,\sigma(z_1,z_2)), \text{ and} \\
    h_2(\otimes) &= \sigma(\sigma(z_1,z_2),\alpha)    \enspace.
  \end{align*}
  The tree homomorphism $h: \T_\Delta \to \T_\Sigma$ is linear, nondeleting, and productive.  Moreover, $h$ is injective.  
  
The latter can be seen as follows. First we show that, (a) for every $\xi_1,\xi_2\in \T_\Delta$, if $\xi_1(\varepsilon)\ne \xi_2(\varepsilon)$, then $h(\xi_1)\ne h(\xi_2)$.
Then let $\xi_1,\xi_2\in \T_\Delta$ such that $\xi_1\ne \xi_2$. The latter means that there exists $w\in \pos(\xi_1)\cap \pos(\xi_2)$ such that $(\xi_1|_w)(\varepsilon) \ne (\xi_2|_w)(\varepsilon)$
and $\xi_1(v)=\xi_2(v)$ for each $v\in \prefix(w)\setminus \{w\}$.
Using this fact, statement (a), and that $h$ is nondeleting, we can show that $h(\xi_1)\ne h(\xi_2)$.

  Now we construct the $(\Delta,\B)$-wta $\cA_1$ such that $\sem{\cA_1} = \mathrm{eval}$. For this, we define the $\Delta$-algebra $\cB=(B,\theta)$ exactly as in Example~\ref{ex:Sigma-algebra-hom-as-wta}(3). By~\eqref{equ:for-each-a-there-is-a-tree}, we have:
  \begin{equation}\label{equ:for-each-closure-el-there-exists-a-tree}
\text{For each $a \in \langle A \rangle_{\{\oplus,\otimes,\0,\1\}}$ there exists $\xi \in \T_\Delta$ such that $\h_\cB(\xi) = a$.}
\end{equation}
Clearly, $\h_\cB=\mathrm{eval}$.

  Given the $\Delta$-algebra  $\cB=(B,\theta)$, we construct the $(\Delta,\B)$-wta $\cA_1$  as in Example~\ref{ex:Sigma-algebra-hom-as-wta}(3). Then $\initialsem{\cA_1}=\h_\cB=\mathrm{eval}$ and, by \eqref{equ:for-each-closure-el-there-exists-a-tree}, $\langle A \rangle_{\{\oplus,\otimes,\0,\1\}} \subseteq \im(\initialsem{\cA_1})$. Since $\B$ is a semiring, by Theorem~\ref{thm:semiring-run=initial} we have $\initialsem{\cA_1}= \runsem{\cA_1}$, which by convention we denote by $\sem{\cA_1}$.

  By Corollary \ref{cor:closure-REC-under-lin-nondel-prod-hom}, we can construct a $(\Sigma,\B)$-wta $\cA$ such that
  \(\sem{\cA} =  \chi(h)(\sem{\cA_1})\).  Then we prove that the following statement holds.
  \begin{equation}\label{equ:A2=overline}
    \text{For each $a \in \langle A\rangle_{\{\oplus,\otimes,\0,\1\}}$ there exists $\zeta \in \T_\Sigma$ such that }
    \sem{\cA}(\zeta) =  a \enspace.
  \end{equation}

  Let $a \in \langle A\rangle_{\{\oplus,\otimes,\0,\1\}}$. Since  $\langle A \rangle_{\{\oplus,\otimes,\0,\1\}} \subseteq \im(\sem{\cA_1})$, there exists a $\xi \in \T_\Delta$ such that $a = \sem{\cA_1}(\xi)$. Let us choose an arbitrary such $\xi$ and denote it by $\xi_a$. Then we calculate as follows:
  \begingroup
    \allowdisplaybreaks
    \begin{align*}
      \sem{\cA}(h(\xi_a)) = \chi(h)(\sem{\cA_1})(h(\xi_a))
                           = \infsum{\oplus}{\xi \in h^{-1}(h(\xi_a))} \sem{\cA_1}(\xi)
      = \infsum{\oplus}{\xi \in \{\xi_a\}} \sem{\cA_1}(\xi)
                         = \sem{\cA_1}(\xi_a) = a
    \end{align*}
    \endgroup
where the third equality holds because $h$ is injective. 
     This finishes the proof of \eqref{equ:A2=overline}.
Since $\im(\sem{\cA}) \subseteq \langle A\rangle_{\{\oplus,\otimes,\0,\1\}}$ holds obviously, together with \eqref{equ:A2=overline} we have $\im(\sem{\cA}) = \langle A\rangle_{\{\oplus,\otimes,\0,\1\}}$. 
\end{proof}


  \section{\ Closure under inverse of linear  tree homomorphisms}\label{sect:inverse-homomorphism-preserves}

  Let $h$ be a  $(\Sigma,\Delta)$-tree homomorphism and $r: \T_\Delta \to B$. Since the extension $h: \T_\Sigma \to \T_\Delta$ is a particular binary relation $h \subseteq \T_\Sigma \times \T_\Delta$, its inverse $h^{-1} \subseteq \T_\Delta \times \T_\Sigma$ is well defined. Moreover, $\chi(h^{-1}): \T_\Delta \times \T_\Sigma \to B$ is supp-i-finite, and hence $r$ is $\chi(h^{-1})$-summable. Thus, the application of $\chi(h^{-1})$ to $r$ is defined by \eqref{equ:appl-wtt-to-wtl}, and by \eqref{obs:app-tree-transf-to-wtl-2} we obtain, for each $\xi \in \T_\Sigma$:
\begin{equation*}
  \chi(h^{-1})(r)(\xi) = \infsum{\oplus}{\zeta \in h(\xi)}{r(\zeta)} 
  = r(h(\xi))\enspace. 
\end{equation*}

Again let us denote by $\Hom(\Sigma,\Delta)$ the set of all tree homomorphisms from $\Sigma$ to $\Delta$. Let $\cC$ be a subset of the set $\Hom(\Sigma,\Delta)$.
A set $\cL$ of $\B$-weighted tree languages is \emph{closed under inverse of tree homomorphisms in $\cC$} if for every $(\Delta,\B)$-weighted tree language $r \in \cL$ and every $(\Sigma,\Delta)$-tree homomorphism $h$ in  $\cC$, the $(\Sigma,\B)$-weighted tree language $\chi(h^{-1})(r) = r \circ h$ is in~$\cL$.

In this section we show that, if $\B$ is a commutative semiring, then  $\Rec(\_,\B)$ is closed under the inverse of linear tree homomorphisms.

Let $\cA=(Q,\delta,F)$ be a $(\Sigma,\B)$-wta. We wish to extend the $\Sigma$-algebra homomorphism $\h_\cA: \T_\Sigma \to B^Q$ to a mapping which can also process trees with variables in $Z_n$ (cf. \cite[p.~85]{fulvog15}).
Formally, let $n \in \mathbb{N}$ and $q_1,\ldots, q_n \in Q$. We define the mapping $f: Z_n \to B^Q$ such that, for each $z_i \in Z_n$ and $q \in Q$, we let $f(z_i)_q = \1$ if $q=q_i$, and $\0$ otherwise.  Since the $\Sigma$-term algebra $(\T_\Sigma(Z_n),\ttop_\Sigma)$ over $Z_n$ is free in the set of all $\Sigma$-algebras generated by $Z_n$ (cf. Theorem \ref{thm:initial-iso}), there exists a unique extension of $f$ to a $\Sigma$-algebra homomorphism from $(\T_\Sigma(Z_n),\ttop_\Sigma)$ to the vector algebra $\V(\cA)=(B^Q,\delta_\cA)$. We denote this $\Sigma$-algebra homomorphism by 
$\h_\cA^{q_1 \cdots q_n}$. Thus, in particular, for every $\sigma \in \Sigma^{(k)}$ with $k\in \mathbb{N}$ and $\xi_1,\ldots,\xi_k \in \T_\Sigma(Z_n)$, we have
\[ \h_\cA^{q_1 \cdots q_n}(\sigma(\xi_1,\ldots,\xi_k))_q = \bigoplus_{p_1 \cdots p_k  \in Q^k}
\Big(\bigotimes_{i \in[k]} \h_\cA^{q_1 \cdots q_n}(\xi_i)_{p_i}\Big) \otimes \delta_k(p_1 \cdots p_k, \sigma, q)
\enspace. \]
 Obviously, $\h_\cA^\varepsilon = \h_\cA$.

\newcommand{\seq}{\mathrm{seq}}
\newcommand{\lin}{\mathrm{lin}}

In the next lemma, for every $\zeta \in \T_\Sigma(Z_n)$ which is linear in $Z_n$ and trees $\xi_1,\ldots, \xi_n \in \T_\Sigma$, we wish to express the value $\h_\cA(\zeta[\xi_1,\ldots,\xi_n])_q \in B$ by means of $\h_\cA(\xi_1),\ldots, \h_\cA(\xi_n)$  (cf. Figure \ref{fig:decomp-hA-linear-tree}). 
For this we need some technical tools to identify variables in subterms of $\zeta$.

Let $(m_1,\ldots,m_l)$ be the sequence of indices of variables occurring in $\zeta$ in a left-to-right order;  we denote this sequence by $\seq(\zeta)$. Obviously, $l\le n$ and $m_j \le n$ for each $j \in [l]$. For instance, if $\zeta=\sigma(\sigma(\alpha,z_2),\sigma(z_3,z_1))$, then $\seq(\zeta) = (2,3,1)$. 
We note that, if $\zeta = \sigma(\zeta_1,\ldots,\zeta_k)$, then $\seq(\zeta) = \seq(\zeta_1) \cdots \seq(\zeta_k)$ (by dropping intermediate occurrences of ``)(``).

Moreover, we denote by $\lin(\zeta)$ the tree obtained from $\zeta$ by replacing its variables by  $z_1,\ldots, z_l$ in a left-to-right order. Hence $\seq(\lin(\zeta))=(1,\ldots,l)$.
For instance, $\lin(\sigma(\sigma(\alpha,z_2),\sigma(z_3,z_1))) = \sigma(\sigma(\alpha,z_1),\sigma(z_2,z_3))$.

\begin{figure}
  \begin{tikzpicture}[scale=1, every node/.style={transform shape},
                    node distance=0.05cm and 0cm]

\node (lhs1) {\strut $\h_\cA \scaleobj{1.5}{\Bigg(} $};
\node[base right= 4.75cm of lhs1] (lhs2) {\strut ${\scaleobj{1.5}{\Bigg)}}_{\!\!\!\! q} \;\; =$};
\begin{scope}[level distance= 0.7cm,
      level 1/.style={sibling distance=15mm},
      level 2/.style={sibling distance=8mm}]
  \node (n0) at ($(lhs1.east)!0.5!(lhs2.west)+(0,0.5)$) {$\sigma$}
    child {node (n1) {$\sigma$}
      child {node (n11) {$\alpha$}} 
      child {node[draw, rectangle] {$z_2$}} }
    child {node (n2) {$\sigma$}
      child {node {$z_3$}}
      child {node (n22) {$z_1$}} };
  \coordinate[above= 0.45cm of n0 ] (ttop);
  \coordinate[below left= 0.3cm and 1cm of n11] (tleft);
  \coordinate[below right= 0.3cm and 1cm of n22] (tright);
  \draw (ttop) -- (tleft) -- (tright) -- cycle;
  \coordinate (t1) at ($(tleft)!0.2!(tright)+(0,-0.1)$);
  \coordinate (t2) at ($(tleft)!0.5!(tright)+(0,-0.1)$);
  \coordinate (t3) at ($(tleft)!0.8!(tright)+(0,-0.1)$);
  \node[below= 0.7cm of t1, draw, triangle, inner sep=1pt] (x1) {$\xi_1$};
  \node[below= 0.7cm of t2, draw, triangle,inner sep=1pt] (x2) {$\xi_2$};
  \node[below= 0.7cm of t3, draw, triangle, inner sep=1pt] (x3) {$\xi_3$};
  \node[fit=(x2),draw,inner sep=1pt,minimum width=1.2cm,yshift=-1pt] (rx2) {};
  \draw[->] (x1) -- (t1) node[midway,left] {$z_1$};
  \draw[->] (rx2) -- (t2) node[midway,left] {$z_2$};
  \draw[->] (x3) -- (t3) node[midway,left] {$z_3$};
\end{scope}
\node[above left= of n0] {$\zeta [\xi_1, \xi_2, \xi_3]:$};
\node[above right= 0.2cm and 0.1cm of n2] {$\zeta$};

\node (rhs1) at (5.5,-4.75) {\strut $\ldots \ 
  \oplus \ \ \fboxsep=3pt\fbox{$\h_\cA (\xi_2)_{p_1}$}
  \otimes \h_\cA (\xi_3)_{p_2} 
  \otimes \h_\cA (\xi_1)_{p_3} 
  \otimes \h_\cA^{\, \fboxsep=2pt\fbox{$\scriptstyle{\mkern-1mu p_1 \!}$} \, p_2 p_3} \Bigg( $};
\node[base right= 3.5cm of rhs1] (rhs2) {\strut $\Bigg)_{\!\! q} \ \ \oplus \ \ldots$};
\begin{scope}[scale=0.8,
      level distance= 0.7cm,
      level 1/.style={sibling distance=15mm},
      level 2/.style={sibling distance=8mm}]
  \node (m0) at ($(rhs1.east)!0.5!(rhs2.west)+(0,0.5)$) {$\sigma$}
    child {node (m1) {$\sigma$}
      child {node (m11) {$\alpha$}} 
      child {node[draw, rectangle] {$z_1$}} }
    child {node (m2) {$\sigma$}
      child {node {$z_2$}}
      child {node (m22) {$z_3$}} };
  \coordinate[above= 0.45cm of m0 ] (stop);
  \coordinate[below left= 0.3cm and 1cm of m11] (sleft);
  \coordinate[below right= 0.3cm and 1cm of m22] (sright);
  \draw (stop) -- (sleft) -- (sright) -- cycle;
\end{scope}
\node[above left= of m0] {$\lin(\zeta):$};

\end{tikzpicture}

 \caption{\label{fig:decomp-hA-linear-tree} The value of $\h_\cA(\sigma(\sigma(\alpha,z_2),\sigma(z_3,z_1))[\xi_1,\xi_2,\xi_3])$ where on the right-hand side of the equality only one summand (for $q_1q_2q_3 = p_1p_2p_3$ in $Q^3$) is shown.}
  \end{figure}

\begin{lemma}\rm (cf. \cite[Lm.~5.4]{fulvog15}) \label{lm:cut-tree-hom} Let $\B$ be a  semiring and $\cA=(Q,\delta,F)$ a $(\Sigma,\B)$-wta. Moreover, let $n \in \mathbb{N}$. Then, for every $\zeta \in \T_\Sigma(Z_n)$ which $\zeta$ is linear in $Z_n$, $\xi_1,\ldots,\xi_n \in \T_\Sigma$, and $q \in Q$, we have
  \[
\h_\cA(\zeta[\xi_1,\ldots,\xi_n])_q = \bigoplus_{q_1 \cdots q_l \in Q^l}
\Big(\bigotimes_{j \in[l]} \h_\cA(\xi_{m_j})_{q_j}\Big) \otimes \h_\cA^{q_1 \cdots q_l}(\lin(\zeta))_q \enspace,
\]
where $\seq(\zeta)=(m_1,\ldots,m_l)$.
  \end{lemma}
  \begin{proof} We prove the statement by induction on $\T_\Sigma(Z_n)$. For this, let $\zeta \in \T_\Sigma(Z_n)$ such that $\zeta$ is  linear in $Z_n$.
  
I.B.: Let $\zeta= z_j$ for some $j\in[n]$.  Then $\seq(\zeta)=(j)$ and 
    \[
 \h_\cA(z_j[\xi_1,\ldots,\xi_n])_q   = \h_\cA(\xi_j)_{q}  =    \bigoplus_{q_1 \in Q} \h_\cA(\xi_j)_{q_1} \otimes \h_\cA^{q_1}(z_j)_q  \enspace.
\]
Let $\zeta =\sigma$ for some $\sigma\in \Sigma^{(0)}$. The proof of this case is the proof of case $k=0$ of the I.S.

I.S.: Let $\zeta \in \T_\Sigma(Z_n)\setminus Z_n$.  Then $\zeta = \sigma(\zeta_1,\ldots,\zeta_k)$ for some $k\in\mathbb{N}$, $\sigma\in \Sigma^{(k)}$, and $\zeta_1,\ldots,\zeta_k\in \T_\Sigma(Z_n)$.  
For each $j \in [k]$, we have that $\zeta_j \in \T_\Sigma(Z_n)$ and $\zeta_j$  is linear in $Z_n$.
Moreover, let $\seq(\zeta)=(m_1,\ldots,m_l)$ and let 
$\seq(\zeta_i)=(n^i_1,\ldots,n^i_{\kappa_i})$ for each $i\in [k]$. Hence
\begin{equation}\label{eq:sequences}
l = \sum_{j=1}^k \kappa_j \text{\ \  and\ \ \ } (m_1,\ldots,m_l)=(n^1_1,\ldots,n^1_{\kappa_1}, \ \ldots \ ,n^l_1,\ldots,n^l_{\kappa_l})\enspace .
\end{equation}
Then
\begingroup
\allowdisplaybreaks
\begin{align*}
& \h_\cA(\zeta[\xi_1,\ldots,\xi_n])_q\\[2mm] 
= & \bigoplus_{p_1\cdots p_k\in Q^k}\Big[ \bigotimes_{i\in[k]}
\h_\cA(\zeta_i[\xi_1,\ldots,\xi_n] )_{p_i} \Big]\otimes \delta_k(p_1\cdots p_k,\sigma,q)
\\[2mm]
= & \bigoplus_{p_1\cdots p_k\in Q^k}\Big[\bigotimes_{i\in[k]}
\Big( \bigoplus_{q^i_1 \cdots q^i_{\kappa_i} \in Q^{\kappa_i}}
    \Big(\bigotimes_{j \in[\kappa_i]} \h_\cA(\xi_{n^i_j})_{q^i_j}\Big) \otimes \h_\cA^{q^i_1\cdots q^i_{\kappa_i}}(\lin(\zeta_i))_{p_i}\Big)\Big]\otimes \delta_k(p_1\cdots p_k,\sigma,q)
    \tag{by I.H.}\\
  = & \bigoplus_{p_1\cdots p_k\in Q^k}   \bigoplus_{q^1_1 \cdots q^1_{\kappa_1} \in Q^{\kappa_1}} \cdots  \bigoplus_{q^k_1 \cdots q^k_{\kappa_k} \in Q^{\kappa_k}}\\
  & \hspace*{5mm} \bigotimes_{i\in[k]} \Big( 
    \Big(\bigotimes_{j \in[\kappa_i]} \h_\cA(\xi_{n^i_j})_{q^i_j}\Big) \otimes \h_\cA^{q^i_1\cdots q^i_{\kappa_i}}(\lin(\zeta_i))_{p_i}\Big)\otimes \delta_k(p_1\cdots p_k,\sigma,q)
    \tag{by right-distributivity}\\
  = & \bigoplus_{q^1_1 \cdots q^1_{\kappa_1} \in Q^{\kappa_1}} \cdots  \bigoplus_{q^k_1 \cdots q^k_{\kappa_k} \in Q^{\kappa_k}}
      \Big( \bigotimes_{i\in[k]}     \bigotimes_{j \in[\kappa_i]} \h_\cA(\xi_{n^i_j})_{q^i_j}\Big) \otimes\\
  & \hspace*{5mm} 
     \bigoplus_{p_1\cdots p_k\in Q^k} \Big( \bigotimes_{i\in[k]}  \h_\cA^{q^i_1\cdots q^i_{\kappa_i}}(\lin(\zeta_i))_{p_i} \Big) \otimes \delta_k(p_1\cdots p_k,\sigma,q)
    \tag{by left-distributivity}\\
  = & \bigoplus_{q_1 \cdots q_l \in Q^l}
\Big(\bigotimes_{j \in[l]} \h_\cA(\xi_{m_j})_{q_j}\Big) \otimes \h_\cA^{q_1 \cdots q_l}(\lin(\zeta))_q
\end{align*}
\endgroup
where the last equality uses \eqref{eq:sequences} and the fact that
\[\bigoplus_{p_1\cdots p_k\in Q^k} \Big( \bigotimes_{i\in[k]}  \h_\cA^{q^i_1\cdots q^i_{\kappa_i}}(\lin(\zeta_i))_{p_i} \Big) \otimes \delta_k(p_1\cdots p_k,\sigma,q) = \h_\cA^{q^1 \cdots q^l}(\lin(\zeta))_q,\] where $q^j = q^j_1 \cdots q^j_{\kappa_j}$. 
      \end{proof}
     
    In the next theorem we apply Lemma \ref{lm:cut-tree-hom} to the case where $\zeta = h_k(\sigma)$ for some linear tree homomorphism  $h= (h_k \mid k \in \mathbb{N})$. It is helpful to view $\seq(h_k(\sigma))$ as a set of indices; then we can write something like $i \not\in \seq(h_k(\sigma))$ with the obvious meaning.

    The next theorem can be compared to \cite[Thm.~5.1]{fulmalvog11} where the closure of the set of recognizable $\B$-weighted tree languages under inverse of linear weighted extended tree transformations was proved if $\B$ is a $\sigma$-complete and commutative semiring. We recall that linear  tree homomorphisms are particular linear  weighted extended tree transducers.

    \begin{theorem-rect}\label{thm:closure-under-inverse-tree-hom-2} Let $\Sigma$ and $\Delta$ be ranked alphabets. Moreover, let $\B$ be a commutative semiring and let $\cA$  be a $(\Delta,\B)$-wta.  Moreover, let $h$ be a linear $(\Sigma,\Delta)$-tree homomorphism. Then we can construct a $(\Sigma,\B)$-wta $\cB$ such that
      \[
        \sem{\cB} = \chi(h^{-1})\big(\sem{\cA}\big)\enspace.
      \]
      Thus, in particular, the set $\Rec(\_,\B)$ is closed under the inverse of linear tree homomorphisms.
  \end{theorem-rect}
  \begin{proof} Let $\cA=(Q,\delta,F)$ and $h= (h_k \mid k \in \mathbb{N})$.
   For the construction of $\cB$ we use a technique from \cite[Thm.~2.8]{eng75}: it adds a new state $e$ which takes care of those subtrees which are deleted by the tree homomorphism $h$ (the latter corresponding to the given linear top-down tree transducer in \cite[Thm.~2.8]{eng75}). 

   Formally, we construct $\cB=(Q',\delta',F')$ such that $Q' = Q \cup \{e\}$ and for every $k \in \mathbb{N}$, $\sigma \in \Sigma^{(k)}$, $q_1,\ldots,q_k,q \in Q'$ we let
    \[
      (\delta')_k(q_1 \cdots q_k,\sigma,q) =
      \begin{cases}
        \h_\cA^{w}(\lin(h_k(\sigma)))_q & \text{ if $q \in Q$ and $(\forall i\in [k])$: $q_i = e$ iff $i \not\in \seq(h_k(\sigma))$ }\\
        \1 & \text{ if $q = q_1 = \ldots = q_k = e$}\\
        \0 & \text{ otherwise}
        \end{cases}
      \]
      where $w = q_{m_1} \cdots q_{m_l}$ if $\seq(h_k(\sigma)) = (m_1,\ldots,m_l)$. Moreover, we let $(F')_q = F_q$ for each $q \in Q$, and $(F')_e=\0$.

      The following property is easy to see.
\begin{equation}
\text{For every $\xi \in \T_\Sigma$ we have: $\h_\cB(\xi)_e = \1$.} \label{equ:e=1}
\end{equation}

By induction on $\T_\Sigma$, we prove that the following statement holds:
\begin{equation}
\text{For every $\xi \in \T_\Sigma$ and $q \in Q$ we have: $\h_\cB(\xi)_q = \h_\cA(h(\xi))_q$.} \label{equ:hom-on-image}
\end{equation}
Let $\xi = \sigma(\xi_1,\ldots,\xi_k)$. We assume that $\seq(h_k(\sigma))=(m_1,\ldots,m_l)$.
 Then we can compute as follows.
      \begingroup
      \allowdisplaybreaks
      \begin{align*}
        \h_\cB(\sigma(\xi_1,\ldots,\xi_k))_q
        &= \bigoplus_{q_1 \cdots q_k  \in (Q')^k} \Big(\bigotimes_{i \in[k]} \h_\cB(\xi_i)_{q_i}\Big) \otimes (\delta')_k(q_1 \cdots q_k,\sigma,q) \\
        &=\bigoplus_{q_1 \cdots q_k  \in (Q')^k} \Bigg(\bigotimes_{i \in[k]} \begin{cases}\h_\cA(h(\xi_i))_{q_i} & \text{ if } q_i\in Q \\
\1 & \text{ otherwise}
\end{cases}
\Bigg) \otimes (\delta')_k(q_1 \cdots q_k,\sigma,q)
          \tag{by I.H. and \eqref{equ:e=1}}\\
        &= \bigoplus_{\substack{{q_1 \cdots q_k  \in (Q')^k:}\\{q_i=e \iff i \not\in \seq(h_k(\sigma))}}} \Bigg(\bigotimes_{i \in[k]} \begin{cases}\h_\cA(h(\xi_i))_{q_i} & \text{ if } q_i\in Q \\
\1 & \text{ otherwise}
\end{cases}\Bigg) \otimes \h_\cA^{w}(\lin(h_k(\sigma)))_q
        \tag{by construction where $w = q_{m_1} \cdots q_{m_l}$}\\
        &= \bigoplus_{q_1 \cdots q_l \in Q^l} \Big(\bigotimes_{j \in[l]} \h_\cA(h(\xi_{m_j}))_{q_j}\Big) \otimes \h_\cA^{q_1 \cdots q_l}(\lin(h_k(\sigma)))_q
        \tag{by the commutativity of $\B$}\\
        &= \h_\cA(h_k(\sigma)[h(\xi_1),\ldots,h(\xi_k)])_q  \tag{by Lemma \ref{lm:cut-tree-hom}}\\
         &= \h_\cA(h(\sigma(\xi_1,\ldots,\xi_k))_q \enspace.
           \end{align*}
      \endgroup

      Then, for each $\xi \in \T_\Sigma$ we can compute as follows.
      \begingroup
      \allowdisplaybreaks
      \begin{align*}
        \sem{\cB}(\xi) 
        &= \bigoplus_{q\in Q'} \h_\cB(\xi)_q \otimes (F')_q = \bigoplus_{q\in Q} \h_\cB(\xi)_q \otimes F_q\\
        &= \bigoplus_{q\in Q} \h_\cA(h(\xi))_q \otimes F_q \tag{by \eqref{equ:hom-on-image}}\\
        &= \sem{\cA}(h(\xi)) = \chi(h^{-1})\big(\sem{\cA}\big)(\xi) \enspace. \qedhere
        \end{align*}
      \endgroup
    \end{proof}

    We note that in \cite[Thm.~2.4.18]{gecste84} (also cf. \cite[Cor.~4.60]{eng75-15}) it was proved that the set of recognizable tree languages is closed under inverse of arbitrary tree homomorphisms. The same is true for crisp-deterministically recognizable weighted tree languages over $\UnitIntfuzzy_{u,i}$ \cite[Prop.~6]{bozlou10} (taking Theorem \ref{thm:crisp-wta-final-index}(A)$\Leftrightarrow$(B) into account).
        However, in the arbitrary weighted case, the restriction to linear tree homomorphisms  is important.  In fact, we can give a (nonlinear) tree homomorphism~$h$ and a bu-deterministic wta~${\cA}$ over the semiring~$\Nat$ such that $\chi(h^{-1})(\sem{\cA})$~is not recognizable.

\begin{example} \rm \cite[Ex.~5.1]{fulmalvog11}\label{ex:hom-bw}
  Let~$\Sigma = \{\gamma^{(1)}, \alpha^{(0)}\}$
  and $\Delta = \{\sigma^{(2)}, \alpha^{(0)}\}$.  For each $n \in \mathbb{N}$,
  let~$\zeta_n$ be the fully balanced tree of height~$n$ over~$\Delta$,
  which is defined by~$\zeta_0 = \alpha$ and $\zeta_n = \sigma(\zeta_{n-1},
  \zeta_{n-1})$ for every~$n \geq 1$.  We consider the $(\Sigma,\Delta)$-tree homomorphism
  with $h_1(\gamma) = \sigma(z_1, z_1)$
    and $h_0(\alpha) = \alpha$.  It is clear that $h(\gamma^n(\alpha)) =
  \zeta_n$ for each $n \in \mathbb{N}$ where $\gamma^n(\alpha)$~abbreviates the
  tree~$\gamma(\gamma(\dotsm \gamma(\alpha) \dotsm))$ containing
  $n$~times the symbol~$\gamma$.

  Moreover, we consider the  bu-deterministic $(\Delta,\Nat)$-wta ${\cA} = (\{q\}, \delta, F)$, where $\delta_2(qq, \sigma, q) = 1$, $\delta_0(\varepsilon, \alpha, q) = 2$, and $F_q=1$.  For each $\zeta \in
  \T_\Delta$, we have $\sem{\cA}(\zeta) = 2^n$, where $n$~is the number of
  occurrences of~$\alpha$ in~$\zeta$.  Consequently, we obtain that
  \[\chi(h^{-1})\big(\sem{\cA}\big)(\gamma^n(\alpha)) =  \sem{\cA}(h(\gamma^n(\alpha))) = \sem{\cA}(\zeta_n)=2^{2^n}\] for each $n \in   \mathbb{N}$.

Now we can easily show by contradiction that the weighted
  tree language~$\chi(h^{-1})\big(\sem{\cA}\big)$ is not recognizable.  For this, we assume there
  exists a $(\Delta,\Nat)$-wta $\cB= (Q, \delta', F')$ 
  such that $\sem{\cB} = \chi(h^{-1})\big(\sem{\cA}\big)$.

   By Lemma \ref{lm:N-wta-upper bound}, there exists a $K \in \mathbb{N}$ such that, for each $n \in \mathbb{N}$, we have $\sem{\cB}(\gamma^n(\alpha)) \le K^{n+1}$. Then, for each $n \in \mathbb{N}$, we have
  $2^{2^n} = \chi(h^{-1})\big(\sem{\cA}\big)(\gamma^n(\alpha)) = \sem{\cB}(\gamma^n(\alpha))  \le K^{n+1}$. But this is not true. Hence, there does not exist  a $(\Delta,\Nat)$-wta $\cB= (Q, \delta', F')$ 
  such that $\sem{\cB} = \chi(h^{-1})\big(\sem{\cA}\big)$.
\hfill $\Box$ 
\end{example}


   \section{\ Closure under weighted projective bimorphisms}
   \label{sec:w-projective-bimorphisms}
   
  In this section, we let $\Psi$ be a ranked alphabet.  We define the concept of $(\Sigma,\Psi,\B)$-weighted projective bimorphism  as special weighted regular tree grammar; each such grammar $\cH$  computes a $(\Sigma,\Psi,\B)$-weighted tree transformation, denoted by  $\ttsem{\cH}$; we also call  $\ttsem{\cH}$ a $(\Sigma,\Psi,\B)$-weighted projective bimorphism. Intuitively, a weighted projective bimorphism is a tree relabeling in which, additionally,  unary input symbols can be deleted without producing output and  unary output symbols can be produced without consuming an input symbol.

 Then we prove that, roughly speaking, the sets $\Reg(\_,\B)$ and $\Rec(\_\,,\B)$ are closed under weighted projective bimorphisms if $\B$ is a commutative semiring.
Given an alphabetic $(\Sigma,\B)$-wrtg $\cG$  and a  $(\Sigma,\Psi,\B)$-weighted projective bimorphism $\cH$, we will proceed in four steps as follows: 
\begin{compactenum}
\item we split the semantics of $\cG$ into an  $([R],\Sigma,\B)$-weighted projective bimorphism $\cH_\cG$ and a characteristic mapping $\chi(\T_{[R]})$ such that $\sem{\cG}= \ttsem{\cH_\cG}(\chi(\T_{[R]}))$, where $R$ is the set of rules of $\cG$ viewed as ranked alphabet and $[R]$ is the skeleton alphabet of $R$ (cf. Lemma \ref{lm:decomp-wrtg}),

\item we prove that weighted projective bimorphisms are closed under composition (cf. Theorem \ref{thm:comp-closure-wbim}); thus, in particular, we can construct an $([R],\Psi,\B)$-weighted projective bimorphism $\cH'$ such that $\ttsem{\cH'}=\ttsem{\cH_\cG};\ttsem{\cH}$,
  
\item we merge $\cH'$ and the characteristic mapping $\chi(\T_{[R]})$ into a $(\Psi,\B)$-wrtg $\cG'$ such that $ \sem{\cG'}=\ttsem{\cH'}(\chi(\T_{[R]}))$  (cf. Lemma \ref{lm:comp-wbim}), and

\item finally, we deduce the mentioned closure results for $\Reg(\_,\B)$ and $\Rec(\_\,,\B)$ (cf. Theorem \ref{thm:closure-REG-under-wbim} and  Corollary \ref{cor:closure-REC-under-wbim}).
  \end{compactenum}
  On first glance, this split-compose-merge procedure looks too complicated for the purpose of just showing the closure of recognizable weighted tree languages under weighted projective bimorphisms. However, in Chapter \ref{ch:AFwtL} we will need the result of the second step (i.e., closure of weighted projective bimorphisms under composition), and we want to benefit from this result already here in our current setting.


  \subsection{Weighted projective bimorphisms}
  \label{sect:weighted-proj-bim}

  \index{$[\Sigma\Psi]$}
 We define the ranked alphabet $[\Sigma\Psi]$ such that, for each $k\in \mathbb{N}$, we let $[\Sigma\Psi]^{(k)}=\Sigma^{(k)}\times \Psi^{(k)}$ if $k\ne 1$, and we let $[\Sigma\Psi]^{(1)}=\big((\Sigma^{(1)}\cup \{\varepsilon\})\times (\Psi^{(1)}\cup \{\varepsilon\})\big)\setminus\{(\varepsilon,\varepsilon)\}$.

 \index{pi@$\pi_1$}
 \index{pi@$\pi_2$}
  Moreover, we define the tree homomorphism $\pi_1: \T_{[\Sigma\Psi]} \rightarrow \T_\Sigma$  and the tree homomorphism $\pi_2: \T_{[\Sigma\Psi]} \rightarrow \T_\Psi$ which, intuitively, project to the first component of a symbol $[\sigma,\psi]$ and to the second component, respectively. 
Formally,  the tree homomorphism $\pi_1$ is induced by the family $((\pi_1)_k \mid k \in \mathbb{N})$ of mappings $(\pi_1)_k: [\Sigma\Psi]^{(k)} \to \T_\Sigma(X_k)$ such that, for every $k \in \mathbb{N}$ and $[\sigma,\psi] \in [\Sigma\Psi]^{(k)}$, we let
\[ (\pi_1)_k([\sigma,\psi]) =
\begin{cases}
\sigma(x_1,\ldots,x_k) & \text{ if } \sigma \ne \varepsilon \\
x_1 & \text{ otherwise. }
\end{cases}
\]
We recall that $\sigma =\varepsilon$ is possible only in case $k=1$.
Analogously, we define the family $((\pi_2)_k \mid k \in \mathbb{N})$ of mappings $(\pi_2)_k: [\Sigma\Psi]^{(k)} \to \T_\Psi(X_k)$ which induces the tree homomorphism $\pi_2$. Obviously, $\pi_1$ and $\pi_2$ are linear and nondeleting; moreover, for each $[\sigma,\psi] \in [\Sigma\Psi]^{(k)}$, we have  $\height(\pi_1([\sigma,\psi])) \le 1$ and $\height(\pi_2([\sigma,\psi])) \le 1$.

\index{weighted projective bimorphism}
\index{wpb}
\index{SigmaPsiBweightedprojectivebim@$(\Sigma,\Psi,\B)$-wpb}
A \emph{$\B$-weighted projective bimorphism over $\Sigma$ and $\Psi$} (for short: $(\Sigma,\Psi,\B)$-wpb, or just: wpb) is an alphabetic $([\Sigma\Psi],\B)$-wrtg $\cH =(N,S,R,wt)$.
The alphabets $\Sigma$ and $\Psi$ are called \emph{input alphabet} and \emph{output alphabet} of $\cH$, respectively; correspondingly, trees in $\T_\Sigma$ and $\T_\Psi$ are called   \emph{input trees} and \emph{output trees} of $\cH$, respectively. 

 Obviously, a $(\Sigma,\Psi,\B)$-wpb $\cH$ can contain two types of rules:
\begin{compactenum}
\item $A \to [\sigma,\psi](A_1,\ldots,A_k)$ with $k\in \mathbb{N}$ and 
$[\sigma,\psi] \in [\Sigma,\Psi]^{(k)}$, and
  \item $A \to B$ (chain rule).
  \end{compactenum}

  Let us consider the first type of rules. We call  $\sigma$ the \emph{input of $r$} and $\psi$ the \emph{output of $r$}, and we denote them  by  $\inp(r)$ and $\out(r)$, respectively.
  Moreover, we give the names shown in the second column of the following table to such rules depending on whether they read a symbol from the input  or not and whether they write a symbol to the output  or not; in the last column we show the denotation for the set of all rules of a particular type.
\begin{center}
\begin{tabular}{l|l|l}
  $A \to [\sigma,\psi](A_1,\ldots,A_k)$ in $R$ & type of the rule & denotation for the set\\
                                               & $\mathrm{r}$= read, $\mathrm{w}$=write &  of all rules of that type\\ \hline
\multicolumn{1}{c|}{$\sigma\not=\varepsilon$ and $\psi\not=\varepsilon$} & \multicolumn{1}{c|}{$\mathrm{rw}$-rule} & \multicolumn{1}{c}{$R^{\mathrm{rw}}$}\\
  \multicolumn{1}{c|}{$\sigma=\varepsilon$ and $\psi\not=\varepsilon$} & \multicolumn{1}{c|}{$\varepsilon\mathrm{w}$-rule} & \multicolumn{1}{c}{$R^{\varepsilon\mathrm{w}}$}\\
  \multicolumn{1}{c|}{$\sigma\not=\varepsilon$ and $\psi=\varepsilon$} & \multicolumn{1}{c|}{$\mathrm{r}\varepsilon$-rule} & \multicolumn{1}{c}{$R^{\mathrm{r}\varepsilon}$}
  \end{tabular}
\end{center}
Moreover, we let $R^{\_\mathrm{w}} = R^{\mathrm{rw}} \cup R^{\varepsilon\mathrm{w}}$ and  $R^{\mathrm{r}\_} = R^{\mathrm{rw}} \cup R^{\mathrm{r}\varepsilon}$.
By viewing $R$ as a ranked alphabet, both  $\varepsilon w$-rules and $r\varepsilon$-rules have rank 1, i.e., $R^{\varepsilon\mathrm{w}} \cup R^{\mathrm{r}\varepsilon} \subseteq R^{(1)}$, and each chain rule is in $R^{(1)}$. Finally, we note that, since each wpb is a particular wrtg (and in its turn, each wrtg is a particular wcfg; and in its turn, each wcfg is a cfg equipped with a weight mapping, cf. Figure \ref{fig:overview-models}), we have  the concept of rule tree of $\cH$ available, as well as the abbreviation $\RT_\cH(N',\eta)$ for each $N' \subseteq N$ and $\eta \in \T_{[\Sigma\Psi]}$, cf. page \pageref{page:rule}.

\index{RT@$\RT_\cH(\xi,\zeta)$}
Now we define the weighted tree transformation computed by a $(\Sigma,\Psi,\B)$-wpb $\cH =(N,S,R,wt)$.
For this, let  $N'\subseteq N$, $\xi \in \T_\Sigma$, and $\zeta \in \T_\Psi$. We  define
\begin{equation} \label{eq:bimorphism-def}
\RT_\cH(N',\xi,\zeta) = \bigcup_{\substack{\eta \in \T_{[\Sigma\Psi]}:\\\pi_1(\eta)=\xi, \pi_2(\eta)=\zeta}} \RT_\cH(N',\eta) \enspace
\end{equation}
and we define $\RT_\cH(\xi,\zeta) = \RT_\cH(S,\xi,\zeta)$.
Then the family
\begin{equation}\label{eq:bimorphism-partitioning}
(\RT_\cH(N',\eta) \mid \eta \in \T_{[\Sigma\Psi]}: \pi_1(\eta)=\xi, \pi_2(\eta)=\zeta)
\end{equation}
is a partitioning of the set $\RT_\cH(N',\xi,\zeta)$ because \eqref{eq:bimorphism-def} holds and $\eta\ne \eta'$ implies that $\RT_\cH(N',\eta)\cap \RT_\cH(N',\eta')=\emptyset$.
Moreover,  the set $\{\eta \in \T_{[\Sigma\Psi]} \mid \pi_1(\eta)=\xi, \pi_2(\eta)=\zeta\}$ is finite, hence \eqref{eq:bimorphism-partitioning} is a partitioning into finitely many subsets. Due to this fact, $\cH$ is finite-derivational if and only if, for every 
$\xi \in \T_\Sigma$ and $\zeta \in \T_\Psi$, the set $\RT_\cH(\xi,\zeta)$ is finite.

\index{finite-input}
\index{finite-output}
\index{weighted projective bimorphism!finite-input}
\index{weighted projective bimorphism!finite-output}
We call $\cH$
\begin{compactitem}
\item \emph{finite-input} if for every $\zeta \in \T_\Psi$, the set $\{\xi \in \T_{\Sigma} \mid \RT_\cH(\xi,\zeta) \not=\emptyset\}$ is finite;
\item \emph{finite-output} if for every $\xi \in \T_\Sigma$, the set $\{\zeta \in \T_{\Psi} \mid \RT_\cH(\xi,\zeta) \not=\emptyset\}$ is finite.
  \end{compactitem}
If  $R=R^{\_\mathrm{w}}$, then $\cH$ is finite-input, and if  $R=R^{\mathrm{r}\_}$, then $\cH$ is finite-output. 

\index{semanticH@$\ttsem{\cH}$}
If $\cH$ is finite-derivational or $\B$ is $\sigma$-complete, then we define the {\em weighted projective bimorphism computed by $\cH$} to be the weighted tree transformation $\ttsem{\cH}:~\T_\Sigma~\times~\T_\Psi~\rightarrow~B$ such that, for every $\xi \in \T_\Sigma$ and $\zeta \in \T_\Psi$, we have
\[
\ttsem{\cH}(\xi,\zeta) = \infsum{\oplus}{d \in \RT_\cH(\xi,\zeta)}  \wt_\cH(d) \enspace,
\]
where $\wt_\cH: \T_R \to B$ is the weight mapping defined in \eqref{equ:wt-rules-tree-as-evaluation} (viewing $\cH$ as wcfg).

It is easy to see that, if $\cH$ is finite-input, then $\ttsem{\cH}$ is supp-i-finite, and if $\cH$ is finite-output, then $\ttsem{\cH}$ is supp-o-finite. 

 Let $\tau: \T_\Sigma \times \T_\Psi \to B$ be a weighted tree transformation. We say that $\tau$ is a \emph{$(\Sigma,\Psi,\B)$-weighted projective bimorphism} (or just:  \emph{weighted projective bimorphism}) if there exists a $(\Sigma,\Psi,\B)$-wpb $\cH$  such that $\cH$ is finite-derivational if $\B$ is not $\sigma$-complete, and  $\ttsem{\cH} = \tau$.

A set $\cL$ of $\B$-weighted tree languages is \emph{closed under weighted projective bimorphisms} if for every $(\Sigma,\B)$-weighted tree language $r \in \cL$ and every $(\Sigma,\Psi,\B)$-bimorphism $\cH$, the $(\Psi,\B)$-weighted tree language $\ttsem{\cH}(r)$ is in $\cL$.

In Figure \ref{fig:wpb-illu} we illustrate the relationships between the mappings $\pi_\cH$, $\pi_1$, and $\pi_2$.

\begin{figure}
  \centering
  
\begin{tikzpicture}[scale=0.8, every node/.style={transform shape},
					level distance= 1cm,
					node distance= 0cm and 0cm]									

\begin{scope}[level 1/.style={sibling distance=30mm},level distance= 1.1cm]		  
  \node (droot) {$A \to [\varepsilon , \gamma](B)$}
    child {node (d1) {$B \to [\sigma , \kappa](C,D)$}
      child {node (d11) {$C \to [\gamma , \varepsilon](F)$}
        child {node (d111) {$F \to [\alpha , \beta]$}} }
      child {node (d12) {$D \to E$}
        child {node (d13) {$E \to [\nu , \mu]$}} }};
  \node[above left= of droot] {$d \in \RT_{\cH}(A,\eta) \subseteq \RT_{\cH}(A,\xi,\zeta):$};
  \draw (-4,-4) to[out=65,in=180, looseness=0.8] 
        ($ (droot) + (0,0.85) $) to[out=0, in=115, looseness=0.8]
        (4,-4) to cycle;
\end{scope}

\node (wtd) at (6.1,-1.3) {$\wt_{\cH}(d) \in B$};
\draw[->, >=stealth] ($(wtd.west) + (-1.4,0)$) -- ($(wtd.west) + (-0.4,0)$)
  node[above, midway] {$\wt_{\cH}$};
\draw[->, >=stealth] (0,-4.5) -- (0,-5.3) node[right, midway] {$\pi_{\cH}$};
\draw[->, >=stealth] (-2.25,-7.3) -- (-3.25,-7.3) node[above, midway] {$\pi_{1}$};
\draw[->, >=stealth] (2.25,-7.3) -- (3.25,-7.3) node[above, midway] {$\pi_{2}$};

\begin{scope}[level 1/.style={sibling distance=8mm},level distance= 0.7cm]			     
  \node at (-5,-6.4) (xroot) {$\sigma$}
    child {node (x1) {$\gamma$}
      child {node (x11) {$\alpha$}} }
    child {node (x2) {$\nu$}};
  \node[above left= of xroot] {$\xi \in \T_{\Sigma}:$};
  \draw ($ (xroot) + (-1.5,-1.75) $) to 
        ($ (xroot) + (0,0.65) $) to
        ($ (xroot) + (1.5,-1.75) $) to cycle;
\end{scope}
    
\begin{scope}[level 1/.style={sibling distance=20mm},level distance= 0.9cm]			  
  \node at (-0,-6.4) (eroot) {$[\varepsilon , \gamma]$}
    child {node (e1) {$[\sigma , \kappa]$}
      child {node (e11) {$[\gamma , \varepsilon]$}
        child {node (e111) {$[\alpha , \beta]$}} }
      child {node (e12) {$[\nu , \mu]$}} };
  \node[above left= of eroot] {$\eta \in \T_{[\Sigma \Psi]}:$};
  \draw ($ (e11.west) + (-1,-1.5) $) to[out=70,in=180, looseness=0.65] 
        ($ (eroot) + (0,0.65) $) to[out=0, in=110, looseness=0.65]
        ($ (e12.east) + (1,-1.5) $) to cycle;
\end{scope}
    
\begin{scope}[level 1/.style={sibling distance=13mm},level distance= 0.7cm]   
  \node at (5,-6.4) (zroot) {$\gamma$}
    child {node (z1) {$\kappa$}
      child {node (z11) {$\beta$}} 
      child {node (z12) {$\mu$}} };
  \node[above left= of zroot] {$\zeta \in \T_{\Psi}:$};
  \draw ($ (zroot) + (-1.5,-1.75) $) to
        ($ (zroot) + (0,0.65) $) to
        ($ (zroot) + (1.5,-1.75) $) to cycle;
\end{scope}

\end{tikzpicture}
  \caption{\label{fig:wpb-illu} An illustration of the relationships between the mappings $\pi_\cH$, $\pi_1$, and $\pi_2$.}
  \end{figure}

Up to syntactic modifications, the concept of $(\Sigma,\Psi,\B)$-wpb is the same as the concept of $\B$-weighted generalized relabeling tree transducers over $\Sigma$ and $\Psi$ (for short: $(\Sigma,\Psi,\B)$-transducer) as it was defined in \cite{fulvog19a}. The syntactic modifications are shown in the following table where we assume that $\sigma \in \Sigma^{(k)}$ and $\psi \in \Psi^{(k)}$:

\begin{tabular}{l|l|l}
$k \in \mathbb{N}$ & rule of a  $(\Sigma,\Psi,\B)$-wpb & rule of a  $(\Sigma,\Psi,\B)$-transducer\\\hline
$k \ge 0$ & $A \to [\sigma,\psi](A_1,\ldots,A_k)$ & $(A,\sigma(x_1,\ldots,x_k),A_1 \cdots A_k,\psi(x_1,\ldots,x_k))$\\
$k=1$ & $A \to [\sigma,\varepsilon](A_1)$ & $(A,\sigma(x_1),A_1,x_1)$\\
$k=1$ & $A \to [\varepsilon,\psi](A_1)$ & $(A,x_1,A_1,\psi(x_1))$\\
  $k=1$ & $A \to A_1$ & $(A,x_1,A_1,x_1)$
\end{tabular}

\noindent In \cite{fulvog19a} we have assumed that $\B$ is a commutative and $\sigma$-complete semiring. Here we are more liberal and do not require per se that $\B$ is $\sigma$-complete (and commutative). As a consequence, we have to be more careful with the definedness of the semantics $\ttsem{\cH}$ of a weighted projective bimorphism~$\cH$.

Also, we refer the reader to Theorem \ref{thm:decomp-wpb} where the weighted tree transformation $\ttsem{\cH}$ is decomposed into the inverse of a simple tree homomorphism, followed by the Hadamard product with the recognizable weighted tree language $\wrtsem{\cH}$, followed by a simple tree homomorphism; this decomposition connects our definition of wpb with the classical definition of bimorphisms in \cite{arndau76,arndau82}. For details we refer to Section \ref{sec:decomposition-of-wbp}.

Finally, we note that weighted projective bimorphisms are particular linear and nondeleting weighted extended tree transducers \cite{fulmalvog11} in which each rule processes at most one input symbol and generates at most one output symbol. For a comparison to linear nondeleting recognizable tree transducers \cite{kui99} we refer to \cite{fulmalvog11}.

\begin{example}\label{ex:wbg} \rm We consider the ranked alphabets $\Sigma = \{\sigma^{(2)}, \gamma^{(1)}, \alpha^{(0)}\}$ and  $\Psi = \{\sigma^{(2)}, \gamma'^{(1)}, \alpha^{(0)}\}$. As weight algebra we consider the semiring $\Nat=(\mathbb{N},+,\cdot,0,1)$ of natural numbers.

  We wish to define a $(\Sigma,\Psi,\Nat)$-wpb $\cH$ such that, for each $(\xi, \zeta)\in \supp(\ttsem{\cH})$, there exists a tree $\kappa \in \T_{\{\sigma^{(2)},\alpha^{(0)}\}}$ (called kernel of $(\xi,\zeta)$) and (a) $\xi$ is obtained from $\kappa$ by  inserting  above an arbitrary position an arbitrary number of $\gamma$-labeled positions and (b)  $\zeta$ is obtained from $\kappa$ by inserting  above an arbitrary position an arbitrary number of $\gamma'$-labeled positions; moreover, the insertions of $\gamma$ into $\xi$ and of $\gamma'$ into $\zeta$ can happen interleaved and in an arbitrary ordering. Then, we define $\cH$ such that  $\ttsem{\cH}(\xi,\zeta)$ is the number of all possible orderings of insertions to obtain the pair $(\xi,\zeta)$ from $\kappa$ by this insertion method.

  For this we let $\cH = (N,S,R,wt)$ with $N=S=\{A\}$ and we let $R$ contain the following rules:
  \begin{center}
    \begin{tabular}{lll}
     $A \to [\sigma,\sigma](A,A)$ & $A \to [\alpha,\alpha]$ & (for the  generation of $\kappa$)\\
    $A \to [\gamma,\varepsilon](A)$  & $A \to [\varepsilon,\gamma'](A)$ & (for inserting $\gamma$ and $\gamma'$) \enspace.
    \end{tabular}
    \end{center}
  
  We let $wt(r)=1$ for each $r \in R$. Obviously, $\ttsem{\cH}(\xi,\zeta) = |\RT_\cH(\xi,\zeta)|$ for every $\xi \in \T_\Sigma$ and $\zeta \in \T_\Psi$. 

  For instance, let $\xi = \sigma(\gamma(\alpha),\gamma\gamma(\alpha))$ and $\zeta=\gamma'(\sigma(\gamma'(\alpha),\gamma'(\alpha)))$. The kernel of $(\xi,\zeta)$ is the tree $\kappa= \sigma(\alpha,\alpha)$. Then (a) above position $\varepsilon$ of $\kappa$, $\cH$ inserts no $\gamma$ (for $\xi$) and one $\gamma'$ (for $\zeta$) and there exists one (trivial) ordering of this insertion, (b) above position $1$ of~$\kappa$, $\cH$ inserts one $\gamma$ (for $\xi$) and one $\gamma'$ (for $\zeta$) and there are two orderings of these insertions, and (c) above position $2$ of $\kappa$, $\cH$ inserts two $\gamma$s (for $\xi$) and one $\gamma'$ (for $\zeta$) and there are three orderings of these insertions. Multiplying up the number of all such orderings yields  $\ttsem{\cH}(\xi,\zeta) = 2 \cdot 3 = 6$. 
Figure \ref{fig:ex-wbg2} illustrates two derivations $d_1,d_2 \in \RT_\cH(\xi,\zeta)$, the $[\Sigma\Psi]$-trees  $\eta_1=\pi(d_1)$ and $\eta_2= \pi(d_2)$, and the trees  $\xi$, $\zeta$ and $\kappa$. The brackets indicate the regions where reordering is possible without changing $\xi$ and $\zeta$.  \hfill $\Box$
 \end{example}

  \begin{figure}[t]
    \centering
\scalebox{.7}{ 
\begin{tikzpicture}
\tikzset{
    ncbar angle/.initial=90,
    ncbar/.style={
        to path=(\tikztostart)
        -- ($(\tikztostart)!#1!\pgfkeysvalueof{/tikz/ncbar angle}:(\tikztotarget)$)
        -- ($(\tikztotarget)!($(\tikztostart)!#1!\pgfkeysvalueof{/tikz/ncbar angle}:(\tikztotarget)$)!\pgfkeysvalueof{/tikz/ncbar angle}:(\tikztostart)$)
        -- (\tikztotarget)
    }
}

\tikzset{square left brace/.style={ncbar=-0.25em}}
\tikzset{square right brace/.style={ncbar=0.25em}}

\node (i0) at (0,0) {$A \to [\varepsilon,\gamma'](A)$};
\node[left = 0.25em of i0] {$d_1 \in D_{\cal H}(\eta_1):$};
\node (i1) at (0,-1) {$A \to [\sigma,\sigma](A,A)$};
\node (i11) at (-2,-2) {$A \to [\gamma,\varepsilon](A)$};
\node (i111) at (-2,-3) {$A \to [\varepsilon,\gamma'](A)$};
\node (i1111) at (-2,-4) {$A \to [\alpha,\alpha]$};
\node (i12) at (2,-2) {$A \to [\gamma,\varepsilon](A)$};
\node (i121) at (2,-3) {$A \to [\gamma,\varepsilon](A)$};
\node (i1211) at (2,-4) {$A \to [\varepsilon,\gamma'](A)$};
\node (i12111) at (2,-5) {$A \to [\alpha,\alpha]$};
\draw
  (i0) -- (i1)
  (i1) -- (i11)
  (i1) -- (i12)
  (i11) -- (i111)
  (i111) -- (i1111)
  (i12) -- (i121)
  (i121) -- (i1211)
  (i1211) -- (i12111)
  ;
  
\draw (-3.25,-1.75) to [square left brace] (-3.25,-3.25);
\draw (3.25,-1.75) to [square right brace] (3.25,-4.25);

\node (ii0) at (9,0) {$A \to [\varepsilon,\gamma'](A)$};
\node[left = 0.25em of ii0] {$d_2 \in D_{\cal H}(\eta_2):$};
\node (ii1) at (9,-1) {$A \to [\sigma,\sigma](A,A)$};
\node (ii11) at (7,-2) {$A \to [\varepsilon,\gamma'](A)$};
\node (ii111) at (7,-3) {$A \to [\gamma,\varepsilon](A)$};
\node (ii1111) at (7,-4) {$A \to [\alpha,\alpha]$};
\node (ii12) at (11,-2) {$A \to [\varepsilon,\gamma'](A)$};
\node (ii121) at (11,-3) {$A \to [\gamma,\varepsilon](A)$};
\node (ii1211) at (11,-4) {$A \to [\gamma,\varepsilon](A)$};
\node (ii12111) at (11,-5) {$A \to [\alpha,\alpha]$};
\draw
  (ii0) -- (ii1)
  (ii1) -- (ii11)
  (ii1) -- (ii12)
  (ii11) -- (ii111)
  (ii111) -- (ii1111)
  (ii12) -- (ii121)
  (ii121) -- (ii1211)
  (ii1211) -- (ii12111)
;
\draw (5.75,-1.75) to [square left brace] (5.75,-3.25);
\draw (12.25,-1.75) to [square right brace] (12.25,-4.25);

\draw[->] (0,-5.5) to node[right] {$\pi$} (0,-6.5);

\node (iii0) at (0,-7) {$[\varepsilon,\gamma']$};
\node[left = 1em of iii0] {$\eta_1:$};
\node (iii1) at (0,-8) {$[\sigma,\sigma]$};
\node (iii11) at (-2,-9) {$[\gamma,\varepsilon]$};
\node (iii111) at (-2,-10) {$[\varepsilon,\gamma']$};
\node (iii1111) at (-2,-11) {$[\alpha,\alpha]$};
\node (iii12) at (2,-9) {$[\gamma,\varepsilon]$};
\node (iii121) at (2,-10) {$[\gamma,\varepsilon]$};
\node (iii1211) at (2,-11) {$[\varepsilon,\gamma']$};
\node (iii12111) at (2,-12) {$[\alpha,\alpha]$};
\draw
  (iii0) -- (iii1)
  (iii1) -- (iii11)
  (iii1) -- (iii12)
  (iii11) -- (iii111)
  (iii111) -- (iii1111)
  (iii12) -- (iii121)
  (iii121) -- (iii1211)
  (iii1211) -- (iii12111)
;
\draw (-2.5,-8.75) to [square left brace] (-2.5,-10.25);
\draw (2.5,-8.75) to [square right brace] (2.5,-11.25);

\draw[->] (9,-5.5) to node[right] {$\pi$} (9,-6.5);

\node (iiii0) at (9,-7) {$[\varepsilon,\gamma']$};
\node[left = 1em of iiii0] {$\eta_2:$};
\node (iiii1) at (9,-8) {$[\sigma,\sigma]$};
\node (iiii11) at (7,-9) {$[\varepsilon,\gamma']$};
\node (iiii111) at (7,-10) {$[\gamma,\varepsilon]$};
\node (iiii1111) at (7,-11) {$[\alpha,\alpha]$};
\node (iiii12) at (11,-9) {$[\varepsilon,\gamma']$};
\node (iiii121) at (11,-10) {$[\gamma,\varepsilon]$};
\node (iiii1211) at (11,-11) {$[\gamma,\varepsilon]$};
\node (iiii12111) at (11,-12) {$[\alpha,\alpha]$};
\draw
  (iiii0) -- (iiii1)
  (iiii1) -- (iiii11)
  (iiii1) -- (iiii12)
  (iiii11) -- (iiii111)
  (iiii111) -- (iiii1111)
  (iiii12) -- (iiii121)
  (iiii121) -- (iiii1211)
  (iiii1211) -- (iiii12111)
;
\draw (6.5,-8.75) to [square left brace] (6.5,-10.25);
\draw (11.5,-8.75) to [square right brace] (11.5,-11.25);

\draw[->] (0,-12.5) to node[right] {$\pi_1$} (0,-13.5);
\draw[->] (6,-11) to node[below] {$\pi_1$} (0.5,-13.5);

\node (iiiii0) at (0,-14) {$\sigma$};
\node[left = 1em of iiiii0] {$\xi:$};
\node (iiiii1) at (-1,-15) {$\gamma$};
\node (iiiii11) at (-1,-16) {$\alpha$};
\node (iiiii2) at (1,-15) {$\gamma$};
\node (iiiii21) at (1,-16) {$\gamma$};
\node (iiiii211) at (1,-17) {$\alpha$};
\draw
  (iiiii0) -- (iiiii1)
  (iiiii0) -- (iiiii2)
  (iiiii1) -- (iiiii11)
  (iiiii2) -- (iiiii21)
  (iiiii21) -- (iiiii211)
;

\draw[->] (9,-12.5) to node[right] {$\pi_2$} (9,-13.5);
\draw[->] (3,-11) to node[below] {$\pi_2$} (8.5,-13.5);

\node (iiiiii0) at (9,-14) {$\gamma'$};
\node[left = 1em of iiiiii0] {$\zeta:$};
\node (iiiiii1) at (9,-15) {$\sigma$};
\node (iiiiii11) at (8,-16) {$\gamma'$};
\node (iiiiii111) at (8,-17) {$\alpha$};
\node (iiiiii12) at (10,-16) {$\gamma'$};
\node (iiiiii121) at (10,-17) {$\alpha$};
\draw
  (iiiiii0) -- (iiiiii1)
  (iiiiii1) -- (iiiiii11)
  (iiiiii1) -- (iiiiii12)
  (iiiiii11) -- (iiiiii111)
  (iiiiii12) -- (iiiiii121)
;


\node (iiiiiii0) at (4.5,-15) {$\sigma$};
\node[left = 1em of iiiiiii0] {$\kappa:$};
\node (iiiiiii1) at (3.5,-16) {$\alpha$};
\node (iiiiiii2) at (5.5,-16) {$\alpha$};
\draw
  (iiiiiii0) -- (iiiiiii1)
  (iiiiiii0) -- (iiiiiii2)
  ;

\end{tikzpicture}
}
    \caption{\label{fig:ex-wbg2} Illustration of two rule trees $d_1,d_2 \in \RT_\cH(\xi,\zeta)$, the $[\Sigma\Psi]$-trees  $\eta_1=\pi(d_1)$ and $\eta_2= \pi(d_2)$, and the trees $\xi  = \sigma(\gamma(\alpha),\gamma\gamma(\alpha))$, $\zeta =\gamma'(\sigma(\gamma'(\alpha),\gamma'(\alpha)))$, and $\kappa = \sigma(\alpha,\alpha)$  (cf. Example \ref{ex:wbg}).}
    \end{figure}

    \begin{example}\rm We have noted that, if $\cH$ is finite-input, then $\ttsem{\cH}$ is supp-i-finite. In this example we show that the reverse direction does not hold, even if $\cH$ is chain-free. That means, there exists a chain-free wpb $\cH$ such that $\ttsem{\cH}$ is supp-i-finite and $\cH$ is not finite-input. We let $\Sigma = \{\gamma^{(1)}, \alpha^{(0)}\}$ and $\Psi=\{\beta^{(0)}\}$.
 We consider  the ring  $\Intfour=(\{0,1,2,3\},+_4,\cdot_4,0,1)$ as defined in Example \ref{ex:semirings}(\ref{def:ring-Zmod4Z}) and the chain-free $(\Sigma,\Psi,\Intfour)$-wpb $\cH$ with set $\{A,B\}$ of nonterminals, $A$ as only initial nonterminal, and the rules:
 \[
    r_1= (  A \to [\gamma,\varepsilon](B)) \ \
    r_2= (  B \to [\gamma,\varepsilon](A)) \ \
  r_3 =(  A \to [\alpha,\beta]) \ \
  \]
  and $wt(r_1) = wt(r_2) = 2$ and $wt(r_3)=1$. Obviously, $\T_\Psi = \{\beta\}$ and the set $\{\xi \in \T_\Sigma \mid \ttsem{\cH}(\xi,\beta) \not= 0\} = \{\alpha\}$ is finite. Hence $\ttsem{\cH}$ is supp-i-finite.
 Moreover, $\cH$ is not finite-input, because $\{\xi \in \T_\Sigma \mid \RT_{\cH}(\xi,\beta) \not= \emptyset\} = \{\gamma^n\alpha \mid n \in \mathbb{N}\}$ is infinite.  \hfill $\Box$
      \end{example}

      By definition, each wpb $\cH$ is a particular wrtg. Thus $\cH$ has two semantics: the weighted tree language $\sem{\cH}$ and the weighted tree transformation $\ttsem{\cH}$. The next lemma shows that, under some additional conditions, the equality of the weighted tree languages of two wpb implies the equality of their weighted tree transformations.

\begin{lemma}\rm \label{lm:lift-wbs} Let $\cH$ and $\cH'$ be $(\Sigma,\Psi,\B)$-wpb such that both $\cH$ and $\cH'$ are finite-derivational or $\B$ is $\sigma$-complete. If  $\sem{\cH}=\sem{\cH'}$, then $\ttsem{\cH}=\ttsem{\cH'}$.
\end{lemma}
\begin{proof} Let us assume that $\sem{\cH}=\sem{\cH'}$ and let $\xi \in \T_\Sigma$ and $\zeta \in \T_\Psi$. Then
  \begingroup
  \allowdisplaybreaks
  \begin{align*}
&\ttsem{\cH}(\xi,\zeta) =  \infsum{\oplus}{d \in \RT_\cH(\xi,\zeta)}  \wt_\cH(d) = \bigoplus_{\substack{\eta \in \T_{[\Sigma\Psi]}:\\\pi_1(\eta)=\xi, \pi_2(\eta)=\zeta}} \infsum{\oplus}{d \in \RT_\cH(\eta)}   \wt_\cH(d) = \bigoplus_{\substack{\eta \in \T_{[\Sigma\Psi]}:\\\pi_1(\eta)=\xi, \pi_2(\eta)=\zeta}} \sem{\cH}(\eta) =\\[3mm] & \bigoplus_{\substack{\eta \in \T_{[\Sigma\Psi]}:\\\pi_1(\eta)=\xi, \pi_2(\eta)=\zeta}} \sem{\cH'}(\eta) = 
\bigoplus_{\substack{\eta \in \T_{[\Sigma\Psi]}:\\\pi_1(\eta)=\xi, \pi_2(\eta)=\zeta}} \infsum{\oplus}{d \in \RT_{\cH'}(\eta)}   \wt_{\cH'}(d)= 
\infsum{\oplus}{d \in  \RT_{\cH'}(\xi,\zeta)}  \wt_{\cH'}(d)= \ttsem{\cH'}(\xi,\zeta),
  \end{align*}
  \endgroup
where the second and the sixth equality hold because \eqref{eq:bimorphism-partitioning} is a partitioning (of $\RT_\cH(\xi,\zeta)$ and  $\RT_{\cH'}(\xi,\zeta)$, respectively), the fourth one holds by our assumption, and all other ones hold by definition.
\end{proof}

\FloatBarrier

The next example shows that the reverse of Lemma \ref{lm:lift-wbs} does not hold.

\begin{example}\rm We consider the ranked alphabets $\Sigma = \{\gamma^{(1)}, \alpha^{(0)}\}$ and  $\Psi = \{\gamma'^{(1)}, \alpha^{(0)}\}$. As weight algebra we consider the semiring  $\Boole = (\mathbb{B},\vee,\wedge,0,1)$.

  We consider the two $(\Sigma,\Psi,\Boole)$-wpb $\cH = (N,S,R,wt)$ and  $\cH' = (N,S,R',wt')$ with $N=\{A,B,C\}$, $S=\{A\}$, and the rules shown in the following table.
\begin{center}
  \begin{tabular}{l|l}
    rules of $\cH$: & rules of $\cH'$:\\\hline
    $A \to [\gamma,\varepsilon](B)$ & $A \to [\varepsilon,\gamma'](B)$\\
    $B \to [\varepsilon,\gamma'](C)$ & $B \to [\gamma,\varepsilon](C)$\\
    $C \to [\alpha,\alpha]$ & $C \to [\alpha,\alpha]$
  \end{tabular}
  \end{center}
    and each rule has weight 1. Thus, $\cH$ and $\cH'$ differ in the order in which they read the input symbol $\gamma$ and write the output symbol $\gamma'$.
    For every $\xi \in \T_\Sigma$ and $\zeta \in \T_\Psi$ we have
    \begin{align*}
\ttsem{\cH}(\xi,\zeta) = 1 \ \ \text{ iff } \ \  \xi = \gamma(\alpha) \text{ and } \zeta = \gamma'(\alpha) \ \ \text{ iff } \ \ \ttsem{\cH'}(\xi,\zeta) = 1 \enspace.
    \end{align*}
    Thus $\ttsem{\cH}=\ttsem{\cH'}$. However,
      \begin{align*}
        \supp(\sem{\cH}) = \{[\gamma,\varepsilon][\varepsilon,\gamma']([\alpha,\alpha])\}  \ \text{ and } \
        \supp(\sem{\cH}) = \{[\varepsilon,\gamma'][\gamma,\varepsilon]([\alpha,\alpha])\}
      \end{align*}
      and thus $\sem{\cH} \not= \sem{\cH'}$.
      \hfill $\Box$
  \end{example}


  The next lemma shows that, under certain conditions, we can transform a wpb into an equivalent chain-free wpb  (with respect to the computed weighted tree transformations). This lemma is based on the corresponding lemma for wrtg (cf. Lemma \ref{lm:normal-form-lemmas-inherited-from-wcfg}) which, in its turn, is based on the corresponding lemma for wcfg (cf. Theorem \ref{lm:wrtg-chain-free}).

\begin{lemma}\rm \label{lm:wbim-simple}  (cf. \cite[Lm.~4.1]{fulvog19a}) Let $\B$ be a  semiring and $\cH$ be  a  $(\Sigma,\Psi,\B)$-wpb such that $\cH$ is finite-derivational or $\B$ is $\sigma$-complete. Then the following three statements hold.
  \begin{compactenum}
  \item[(1)] There exists a chain-free $(\Sigma,\Psi,\B)$-wpb $\cH'$ such that $\ttsem{\cH} = \ttsem{\cH'}$.
  \item[(2)] If $\cH$ is finite-derivational, then we can construct a chain-free $(\Sigma,\Psi,\B)$-wpb $\cH'$ such that $\ttsem{\cH} = \ttsem{\cH'}$.
    \item[(3)] In both statements (1) and (2) the following holds: if $\cH$ is finite-input (resp. finite-output), then $\cH'$ is finite-input (resp. finite-output).
    \end{compactenum}
\end{lemma}
\begin{proof} Proof of (1):  By Lemma \ref{lm:normal-form-lemmas-inherited-from-wcfg}(3), there exists an  alphabetic and chain-free $([\Sigma\Psi],\B)$-wrtg $\cH'$ such that $\sem{\cH}=\sem{\cH'}$. Then, by definition of wpb,  $\cH'$ is a chain-free $(\Sigma,\Psi,\B)$-wpb, and by Lemma  \ref{lm:lift-wbs} we have $\ttsem{\cH}=\ttsem{\cH'}$.

  \
  
  Proof of (2): Now let $\cH$ be finite-derivational. Then, Lemma \ref{lm:normal-form-lemmas-inherited-from-wcfg}(3) is constructive and hence we can \underline{construct}  a chain-free $(\Sigma,\Psi,\B)$-wpb $\cH'$ such that $\ttsem{\cH} = \ttsem{\cH'}$.

  \
  
  Proof of (3): By analysing the proof of Theorem \ref{lm:wrtg-chain-free} (on which the proof of Lemma \ref{lm:normal-form-lemmas-inherited-from-wcfg}(3) is based), it is easy to see that, if $\cH$ is finite-input (resp. finite-output), then $\cH'$ is finite-input (resp. finite-output).
\end{proof}


\subsection{Split}
\label{ssec:split}

Next we characterize the weighted tree language $\sem{\cG}$ of a wrtg $\cG$ in terms of the image of a characteristic mapping under a weighted projective bimorphism. For this we introduce the concept of skeleton alphabet as follows. We  define the \emph{skeleton alphabet of $\Sigma$}, denoted by $[\Sigma]$, to be the ranked alphabet with  \label{page:skeleton}
\index{skeleton alphabet}
\index{$[\Sigma]$}
\[
  [\Sigma]^{(k)} = \begin{cases}
\{[k]\} & \text{if } \Sigma^{(k)}\not= \emptyset\\
\emptyset & \text{otherwise}
\end{cases}
\]
for each $k \in \mathbb{N}$ (i.e., $[\Sigma]^{(k)}$ is a singleton or empty and $[\Sigma]^{(0)}=\{[0]\}$).

\label{page:beta-b} Let $\cG=(N,S,R,wt)$ be an arbitrary $(\Sigma,\B)$-wrtg. We define a deterministic tree relabeling $\beta$ which, intuitively,  maps each occurrence of rule $r$ of $\cG$ in a tree $d \in \T_R$ to the symbol of the skeleton alphabet which corresponds to the rank of $r$. Formally, we define $\beta=(\beta_k \mid k \in \mathbb{N})$ with $\beta_k: R^{(k)} \to [R]^{(k)}$ by
\[
  \beta_k(r) = [k] \ \text{ for each $r \in R^{(k)}$} \enspace.
\]
For each $d \in \T_\R$ we call $\beta(d)$ the \emph{skeleton of $d$}.  We note that, for every $\xi \in \T_\Sigma$, the $\T_{[R]}$-indexed family $(\beta^{-1}(b) \cap \RT_\cG(\xi) \mid b \in \T_{[R]})$ is a partitioning of $\RT_\cG(\xi)$. Moreover, for every $\xi \in \T_\Sigma$ and $b \in  \T_{[R]}$, the set $\beta^{-1}(b) \cap \RT_\cG(\xi)$ is finite.
 
\begin{lemma} \rm \label{lm:decomp-wrtg} (cf. \cite[Lm.~4.3]{fulvog19a} for the trivial storage type)  Let $\cG=(N,S,R,wt)$ be an alphabetic $(\Sigma,\B)$-wrtg such that $\cG$ is finite-derivational or $\B$ is  $\sigma$-complete. Then we can construct a chain-free $([R],\Sigma,\B)$-wpb $\cH$ such that (a) $\cH$ is finite-output, (b) $\cH$ is finite-input if $\cG$ is finite-derivational, and (c) $\sem{\cG} = \ttsem{\cH}(\chi(\T_{[R]}))$.
     \end{lemma}
    \begin{proof} We construct the chain-free $([R],\Sigma,\B)$-wpb  $\cH= (N,S,R',wt')$ as follows.
\begin{compactitem}
\item For every $r = (A \rightarrow \sigma(A_1,\ldots,A_k))$ in $R$,\\ 
the rule $r' = (A \to [[k],\sigma](A_1,\ldots, A_k))$ is in $R'$
 and $wt'(r') = wt(r)$.
\item For every $r = (A \rightarrow B)$ in $R$,  the rule $r' = (A \to [[1],\varepsilon](B))$ is in $R'$ and $wt'(r') = wt(r)$.
\end{compactitem}

We continue with introducing  some useful mappings.
We define the deterministic $(R',R)$-tree relabeling
$\varphi=(\varphi_k \mid k \in \mathbb{N})$ for every $k \in \mathbb{N}$ and  $r' \in R'^{(k)}$ by 
\[
  \varphi_k(r') =
  \begin{cases}
   A \rightarrow \sigma(A_1,\ldots,A_k) & \text{ if $r' = (A \to [[k],\sigma](A_1,\ldots, A_k))$}\\
    A \rightarrow B & \text{ if  $r' = (A \to [[1],\varepsilon](B))$} 
    \end{cases}
  \]
  where in the second case $k=1$. 
  Clearly, the deterministic tree relabeling $\varphi: \T_{R'}  \to \T_R$ is bijective.
 Moreover,
  \begin{equation}
    \text{ for each $d \in \RT_{\cH}(N,\T_{[[R]\Sigma]})$, we have: } \wt_\cH(d) = \wt_\cG(\varphi(d))\label{eq:weight-preserving-b}
  \end{equation}
  
  This can be proved easily by induction on $(\RT_\cH(N,\T_{[[R]\Sigma]}), \succ)$ where
  \[\succ = \succ_{R'} \cap \ (\RT_\cH(N,\T_{[[R]\Sigma]}) \times \RT_\cH(N,\T_{[[R]\Sigma]}))\enspace.
  \]
   Since $\succ_{R'}$ is terminating, by Lemma \ref{lm:termination-is-subset-closed} also $\succ$ is terminating. Moreover, we have that  $\nf_\succ(\RT_\cH(N,\T_{[[R]\Sigma]})) = (R')^{(0)}$.

  Also, for every $b \in \T_{[R]}$ and $\xi\in \T_\Sigma$, we have  $\varphi(\RT_\cH(b,\xi)) = \beta^{-1}(b) \cap \RT_\cG(\xi)$. Thus, for every $b \in \T_{[R]}$ and $\xi\in \T_\Sigma$, the mapping $\varphi_{b,\xi}: \RT_\cH(b,\xi) \to  \beta^{-1}(b) \cap \RT_\cG(\xi)$ defined by $\varphi_{b,\xi} = \varphi|_{\RT_\cH(b,\xi)}$ is bijective.  Figure \ref{fig:wrtg-represented-as-wbg} illustrates the connection between the mappings $\beta$ and $\varphi$.

\begin{figure}[h]
  \begin{center}
    \scalebox{.8}{
     \begin{tikzpicture}
\node (a) at (0,0) {$A \to B$};
\node[left = -.1em of a,anchor=east] {$d \in \RT_{\cal G}(\xi)$:};
\node (b) at (0,-1) {$B \to \sigma(C,D)$};
\draw (a) -- (b);
\node (c) at (-1,-2) {$C \to \gamma(E)$};
\draw (b) -- (c);
\node (d) at (1,-1.875) {$\vdots$};
\draw (b) -- (d);

\draw[->] (-2.5,-2.25) -- (-4,-3);
\node[anchor=east,xshift=-9pt] at (-3,-2.5) {$\beta: \mathrm{T}_\mathrm{R} \to \mathrm{T}_\mathrm{[R]}$};

\draw[->] (1.5,-2.25) -- (3.5,-3.25);
\node[anchor=west,xshift=7.5pt] at (2,-2.5) {$\pi: \mathrm{T}_\mathrm{R} \to \mathrm{T}_\Sigma$};

\node (e) at (-4.5, -3.5) {$[1]$};
\node [left of = e] {$b \in \mathrm{T}_\mathrm{[R]}$:};
\node (f) at (-4.5,-4.5) {$[2]$};
\draw (e) -- (f);
\node (g) at (-5.25,-5.5) {$[1]$};
\draw (f) -- (g);
\node[below = -.5em of g,anchor=north] {$\vdots$};
\node (h) at (-3.75,-5.5) {$\vdots$};
\draw (f) -- (h);

\node (i) at (4,-3.5) {$\sigma$};
\node[right =.5em of i,anchor=west] {$\xi \in \mathrm{T}_\Sigma$};
\node (j) at (3.25,-4.25) {$\gamma$};
\draw (i) -- (j);
\node[below = -.5em of j,anchor=north] {$\vdots$};
\node (k) at (4.75,-4.25) {$\vdots$};
\draw (i) -- (k);

\draw[->] (0,-5.5) -- (0,-3.5);
\node[anchor=west] at (0,-4.5) {$\varphi: \mathrm{T}_{\mathrm{R}'} \to \mathrm{T}_\mathrm{R}$};

\node (l) at (0, -6.5) {$A \to \big[[1],\varepsilon\big](B)$};
\node[anchor=east] at (-1.5,-7) {$d' \in \RT_{\cal H}(b,\xi)$:};
\node (m) at (0, -7.5) {$B \to \big[[2],\sigma\big](C,D)$};
\draw (l) -- (m);
\node (n) at (-1,-8.5) {$C \to \big[[1], \gamma\big](E)$};
\draw (m) -- (n);
\node[below = -.5em of n,anchor=north] {$\vdots$};
\node (o) at (1,-8.5) {$\vdots$};
\draw (m) -- (o);
\end{tikzpicture}
}
     \end{center}
\caption{\label{fig:wrtg-represented-as-wbg} An illustration of the mappings $\beta$, $\pi$, and $\varphi$.}
    \end{figure}

  Now we prove (a), (b), and (c).
  Property (a) holds, because $R'=(R')^{\mathrm{r}\_}$ and thus $\cH$ is finite-output (and hence,  $\ttsem{\cH}$ is supp-o-finite).

  To see Property (b), we assume that $\cG$ is finite-derivational and let  $\xi \in \T_\Sigma$. Then the set $\RT_\cG(\xi)$ is finite. Since the family $(\beta^{-1}(b) \cap \RT_\cG(\xi) \mid b \in \T_{[R]})$ is a partitioning of the finite set $\RT_\cG(\xi)$, the set $\{b \in \T_{[R]} \mid \beta^{-1}(b) \cap \RT_\cG(\xi) \not= \emptyset\}$ is also finite. Since $\varphi_{b,\xi}$ is bijective, also the set  $\{b \in \T_{[R]} \mid \RT_\cH(b,\xi) \not= \emptyset\}$ is finite. It means that $\cH$ is finite-input, which proves Property (b).

Now we turn to the proof of Property (c). Obviously,  $\cH$ is chain-free and hence finite-derivational; thus, $\ttsem{\cH}$ is well defined. Moreover, if $\cG$ is finite-derivational, then $\cH$ is finite-input (due to Property (b)), and thus   $\ttsem{\cH}$ is supp-i-finite. Hence $\ttsem{\cH}(\chi(\T_{[R]}))$ is well defined.

Now we prove that $\sem{\cG} = \ttsem{\cH}(\chi(\T_{[R]}))$. For this, let $\xi \in \T_\Sigma$. Then
\begingroup
\allowdisplaybreaks
\begin{align*}
&\big(\ttsem{\cH}(\chi(\T_{[R]}))\big)(\xi) 
                 = \infsum{\oplus}{b \in \T_{[R]}} \chi(\T_{[R]})(b) \otimes  \ttsem{\cH}(b,\xi)
  \tag{\text{because $\ttsem{\cH}$ is supp-i-finite if $\B$ is not $\sigma$-complete}}\\
  & = \infsum{\oplus}{b \in \T_{[R]}} \ttsem{\cH}(b,\xi)\\
  & = \infsum{\oplus}{b \in \T_{[R]}} \bigoplus_{d' \in \RT_\cH(b,\xi)} \wt_\cH(d')
     \tag{\text{because $\cH$ is finite-derivational}}\\
  & = \infsum{\oplus}{b \in \T_{[R]}} \bigoplus_{d' \in \RT_\cH(b,\xi)} \wt_\cG(\varphi_{b,\xi}(d'))
    \tag{\text{by \eqref{eq:weight-preserving-b} and definition of $\varphi_{b,\xi}$}}\\
    & = \infsum{\oplus}{b \in \T_{[R]}} \bigoplus_{d \in \beta^{-1}(b) \cap \RT_\cG(\xi)} \wt_\cG(d)
  \tag{\text{since $\varphi_{b,\xi}$ is bijective}}\\
  &= \infsum{\oplus}{d \in \RT_\cG(\xi)} \wt_\cG(d) \tag{\text{because $(\beta^{-1}(b) \cap \RT_\cG(\xi) \mid b \in \T_{[R]})$ is a partitioning of $\RT_\cG(\xi)$}}\\
  &= \sem{\cG}(\xi). \tag{\text{because $\cG$ is finite-derivational or $\B$ is  $\sigma$-complete}}
\end{align*}
\endgroup
\end{proof}

\subsection{Closure of weighted projective bimorphisms under composition}

In this subsection we prove that weighted projective bimorphisms are closed under composition.
For the proof, we introduce a notation. For every $(\Sigma,\Psi,\B)$-bimorphism $\cH=(N,S,R,\wt)$ and $N'\subseteq N$, we let
\[\RT_\cH(N')=\bigcup_{\eta\in \T_{[\Sigma \Psi]}}\RT_\cH(N',\eta).\]

\begin{theorem-rect}\label{thm:comp-closure-wbim}  {\rm \cite[Thm.~4.2]{fulvog19a}} Let $\Sigma$, $\Psi$, and $\Delta$ be ranked alphabets. Moreover, let $\B$ be a commutative semiring. Let $\cH_1$ be a $(\Sigma,\Psi,\B)$-wpb and $\cH_2$ be a $(\Psi,\Delta,\B)$-wpb such that both $\cH_1$ and $\cH_2$ are finite-derivational or $\B$ is $\sigma$-complete. Moreover, let $\ttsem{\cH_1}$ be supp-o-finite or $\ttsem{\cH_2}$ be supp-i-finite or $\B$ is $\sigma$-complete. Then the following three statements hold.
  \begin{compactenum}
  \item[(1)] There exists a $(\Sigma,\Delta,\B)$-wpb $\cH$ such that $\ttsem{\cH} = \ttsem{\cH_1} ; \ttsem{\cH_2}$.
    \item[(2)] If $\cH_1$ and $\cH_2$ are finite-derivational, then we can construct a finite-derivational $(\Sigma,\Delta,\B)$-wpb $\cH$ such that $\ttsem{\cH} = \ttsem{\cH_1} ; \ttsem{\cH_2}$.

  \item[(3)] In both statements (1) and (2) the following holds. If $\cH_1$ and $\cH_2$ are finite-input, then $\cH$ is finite-input.
    \end{compactenum}
\end{theorem-rect}
\begin{proof} Proof of (1).  Let $\cH_1 = (N_1,S_1,R_1,wt_1)$ and $\cH_2 = (N_2,S_2,R_2,wt_2)$. Since $\cH_1$ and $\cH_2$ are finite-derivational or $\B$ is $\sigma$-complete, the weighted tree transformations $\ttsem{\cH_1}$ and $\ttsem{\cH_2}$ are well defined. Moreover, since $\ttsem{\cH_1}$ is supp-o-finite or $\ttsem{\cH_2}$ is supp-i-finite or $\B$ is $\sigma$-complete, the weighted tree transformation $\ttsem{\cH_1} ; \ttsem{\cH_2}$ is well defined.  By Lemma \ref{lm:wbim-simple}(1) we can assume that  $\cH_1$  and $\cH_2$ are chain-free.

  We define the $(\Sigma,\Delta,\B)$-wpb $\cH = (N,S,R,wt)$ by using a modification of the usual product construction for tree transducers \cite{eng75-15,bak79} and for weighted tree transducers in \cite{engfulvog02new}.
    We have to use a modification for the following reason. Before producing an output symbol $\psi \in \Psi$, the wpb $\cH_1$ can execute a sequence $s_1$ of  $\mathrm{r}\varepsilon$-rules on some input tree $\xi \in \T_\Sigma$. Also, before reading $\psi$, the wpb $\cH_2$ can execute a sequence $s_2$ of $\varepsilon \mathrm{w}$-rules, thereby producing a part of some output tree $\zeta \in \T_\Delta$. If we combine each rule in the sequence $s_1$ with each rule in the sequence $s_2$ into one rule for $\cH$, then we introduce a number of artificial orderings for the applications of the rules in $s_1$ and $s_2$. As a consequence, for one single  summand in $(\ttsem{\cH_1};\ttsem{\cH_2})(\xi,\zeta)$, there will be multiple corresponding summands in $\ttsem{\cH}(\xi,\zeta)$. Thus, in order to establish a weight balance between $\ttsem{\cH}(\xi,\zeta)$ and $(\ttsem{\cH_1};\ttsem{\cH_2})(\xi,\zeta)$,  we have to forbid artificial orderings in the rule trees of $\cH$.

  For this purpose, we introduce auxiliary control nonterminals as follows. Before $\cH$ simulates the combination of a rule $r_1=(A_1 \to[\sigma,\psi](A_{11},\ldots, A_{1k}))$ in $R_1^{\_\mathrm{w}}$ and of a rule $r_2 = (A_2 \to [\psi,\delta](A_{21},\ldots, A_{2k}))$ in $R_2^{\mathrm{r}\_}$, it first simulates rules in $R_1^{\mathrm{r}\varepsilon}$ (using states of the form $\semst{r_1,r_2}$), and second it simulates rules in $R_2^{\varepsilon\mathrm{w}}$ (using states of the form $\sema{r_1,r_2}$).

Formally, we define $\cH$  as follows. We let 
\begin{align*}
N = & \{\semst{r_1,r_2} \mid r_1 \in R_1, r_2 \in R_2\} \ \cup 
 \{\sema{r_1,r_2} \mid r_1 \in R_1^{\_\mathrm{w}}, r_2 \in R_2\} \\
& \cup \{(r_1,r_2) \mid (r_1,r_2) \in R_1^{\_\mathrm{w}} \times R_2^{\mathrm{r}\_}, \out(r_1)= \inp(r_2)\} \text{ and} \\
S = & \{\semst{r_1,r_2} \in Q \mid \lhs(r_1) \in S_1, \lhs(r_2) \in S_2\} \enspace.
\end{align*}
The set $R$ and the weight mapping $\wt$ are defined by the following five cases.
\begin{compactenum}
\item[1.] For every $r_1 = (A_1\to [\sigma,\varepsilon](A_1'))$ in $R_1^{\mathrm{r}\varepsilon}$,
$r_1' \in R_1$ with $\lhs(r_1') = A_1'$, and $r_2 \in R_2$, the rule 
\[r = (\semst{r_1,r_2}\to [\sigma,\varepsilon](\semst{r_1',r_2}))\] 
is in $R$ and $wt(r) = wt_1(r_1)$.
\item[2.] For every $r_1 \in R_1^{\_\mathrm{w}}$ and $r_2 \in R_2$,
the rule  
\[r = (\semst{r_1,r_2}\to\sema{r_1,r_2})\] 
is in $R$ and $wt(r) = \mathbb{1}$.
\item[3.] For every $r_1 \in R_1^{\_\mathrm{w}}$, $r_2 = (A_2\to [\varepsilon,\psi](A_2'))$ in $R_2^{\varepsilon\mathrm{w}}$,
and $r_2' \in R_2$ with $\lhs(r_2') = A_2'$, the rule 
\[r = (\sema{r_1,r_2}\to [\varepsilon,\psi](\sema{r_1,r_2'}))\] 
is in $R$ and $wt(r) = wt_2(r_2)$.
\item[4.] For every $r_1 \in R_1^{\_\mathrm{w}}$ and $r_2 \in R_2^{\mathrm{r}\_}$,
the rule  
\[r = (\sema{r_1,r_2}\to (r_1,r_2))\] 
is in $R$ and $wt(r) = \mathbb{1}$.
\item[5.] For every $r_1 = (A_1\to[\sigma,\psi](A_{11},\ldots, A_{1k}))$ in $R_1^{\_\mathrm{w}}$,  \\
$r_2 = (A_2\to[\psi,\delta](A_{21},\ldots, A_{2k}))$ in $R_2^{\mathrm{r}\_}$, \\
$r_{11},\ldots,r_{1k} \in R_1$ with $\lhs(r_{1j}) = A_{1j}$  ($j \in [k]$), and\\
$r_{21},\ldots,r_{2k} \in R_2$ with $\lhs(r_{2j}) = A_{2j}$ ($j \in [k]$), the rule 
\[r = ((r_1,r_2)\to [\sigma,\delta](\semst{r_{11},r_{21}}, \ldots, \semst{r_{1k},r_{2k}}))\]
is in $R$ and $wt(r) = wt_1(r_1) \otimes wt_2(r_2)$.
\end{compactenum}

Clearly, each rule $r$ produced in Cases 1-4 has rank 1.
Next, we mention that, for each $d \in \RT_\cH(\semst{r_1,r_2},\T_{[\Sigma\Psi]})$ with $r_i\in R_i$ for $i=\{1,2\}$ and each leaf $w \in \pos(d)$ with $w = i_1 \ldots i_m$ for some $m\ge 1$, the sequence of cases which has produced $d(i_1) \ldots d(i_m) \in R^*$, is contained in $( 1^\ell 2 3^u 4 5 )^+$ for some $\ell,u \in \mathbb{N}$ (see Figure \ref{fig:run-schematic}).

\begin{figure}[h]
  \hspace*{40mm}
\scalebox{.8}{
 \begin{tikzpicture}[cut/.style={draw,circle},level 1/.style={sibling distance=1.5cm},level 2/.style={sibling distance=20mm}]
         \node {$\semst{r_1^1,r_2^1}\to [\sigma_1,\varepsilon](\semst{r_1^2,r_2^1})$}
    child {node {$\semst{r_1^{\ell},r_2^1}\to [\sigma_{\ell},\varepsilon](\semst{r_1^{\ell+1},r_2^1})$} edge from parent[dashed]
  child {node  {$\semst{r_1^{\ell+1},r_2^1}\to\langle r_1^{\ell+1},r_2^1\rangle$} edge from parent[solid]
  child {node  {$\langle r_1^{\ell+1},r_2^1\rangle \to [\varepsilon,\psi_1](\langle r_1^{\ell+1},r_2^2\rangle)$}
  child {node  {$\langle r_1^{\ell+1},r_2^{u}\rangle \to [\varepsilon,\psi_u](\langle r_1^{\ell+1},r_2^{u+1}\rangle)$} edge from parent[dashed]
  child {node  {$\langle r_1^{\ell+1},r_2^{u+1}\rangle \to (r_1^{\ell+1},r_2^{u+1})$} edge from parent[solid]
  child {node  {$(r_1^{\ell+1},r_2^{u+1}) \to [\sigma,\delta](\semst{r_{11},r_{21}},\ldots, \semst{r_{1k},r_{2k}})$}
  child {node[trinode] (t1) {$d_1$}}
  child {edge from parent[draw=none] node[draw=none] (ellipsis) {$\ldots$} }
  child {node[trinode] (tk) {$d_k$}}
}}}}}}
;

         \node[xshift=-40mm] {1.}
    child {node {1.} edge from parent[draw=none]
  child {node  {2.}  edge from parent[draw=none]
  child {node  {3.}  edge from parent[draw=none]
  child {node  {3.}  edge from parent[draw=none]
  child {node  {4.}  edge from parent[draw=none]
  child {node  {5.}  edge from parent[draw=none]
  }}}}}}
;

        \node[xshift=40mm] {}
    child {node {$\ell\in\mathbb{N}$} edge from parent[draw=none]
  child {node  {}  edge from parent[draw=none]
  child {node  {}  edge from parent[draw=none]
  child {node  {$u\in\mathbb{N}$}  edge from parent[draw=none]
  child {node  {}  edge from parent[draw=none]
  child {node  {}  edge from parent[draw=none]
  }}}}}}
;

    \end{tikzpicture}
}
\caption{\label{fig:run-schematic} A rule tree $d$ of  $\cH$.}
\end{figure}

We also note that $\cH$ is not chain-free due to rules of type 2 and 4. However, it is easy to see that, if $\cH_1$ and $\cH_2$ are finite-derivational, then  $\cH$ is finite-derivational, because rules of type 2 and 4 cannot occur arbitrarily often in rule trees. Hence  $\ttsem{\cH}$ is well defined.

For the rest of the proof we need some preparation.  We define a mapping $h$ which retrieves from a rule tree in $\RT_\cH(N,\T_{[\Sigma\Delta]})$ the intermediate tree in $\T_\Psi$. Formally, we define the mapping $h: \RT_\cH(N,\T_{[\Sigma\Delta]}) \rightarrow \T_\Psi$ by induction on $(\RT_\cH(N,\T_{[\Sigma\Delta]}),\succ)$ where
\[
  \succ = \succ_R \cap \ (\RT_\cH(N,\T_{[\Sigma\Delta]}) \times \RT_\cH(N,\T_{[\Sigma\Delta]}))\enspace.
\]
  Since $\succ_{R}$ is terminating, by Lemma \ref{lm:termination-is-subset-closed} also
 Obviously, $\succ$ is terminating. Moreover, we have that $\nf_\succ(\RT_\cH(N,\T_{[\Sigma\Delta]}) =  R^{(0)}$.

Let $d = r(d_1,\ldots,d_k)$ be in $\RT_\cH(N,\T_{[\Sigma\Delta]})$ with $r \in R$ and $d_1,\ldots,d_k \in \RT_\cH(N,\T_{[\Sigma\Delta]})$. Then we let
\[
h(r(d_1,\ldots,d_k)) =
\left\{
\begin{array}{ll}
h(d_1) & \text{ if } r \text{ is obtained by Cases 1, 2, 3, or 4}\\
\psi(h(d_1),\ldots,h(d_k)) & \text{ if } r \text{ is obtained by Case 5,}\\
& \text{ $\lhs(r) = (r_1,r_2)$ for some $r_1$ and $r_2$,}\\
& \text{ and $\psi = \out(r_1)$}\enspace.
\end{array}
\right.
\]

  We define the set $N_{\semst{.}} = \{\semst{r_1,r_2} \mid r_1 \in R_1, r_2 \in R_2\}$. Moreover, we define the mapping
  \[
\varphi: \RT_\cH(N_{\semst{.}},\T_{[\Sigma\Delta]}) \to \RT_{\cH_1}(N_1,\T_\Sigma) \times \RT_{\cH_2}(N_2,\T_\Psi)
    \]
    by induction on $(\RT_\cH(N_{\semst{.}},\T_{[\Sigma\Delta]}),\succ)$ where
    \[
      \succ = \psucc_R \cap (\RT_\cH(N_{\semst{.}},\T_{[\Sigma\Delta]}) \times \RT_\cH(N_{\semst{.}},\T_{[\Sigma\Delta]}))
    \]
    as follows.
  Since $\succ_{R}$ is terminating, also $\psucc_R$ is terminating, hence by Lemma \ref{lm:termination-is-subset-closed} also $\succ$ is terminating. Moreover, we have that $\nf_\succ(\RT_\cH(N_{\semst{.}},\T_{[\Sigma\Delta]})) = \{(A \to \xi) \in R \mid A \in N_{\semst{.}}, \xi \in [\Sigma\Delta]^{(0)}\}$.)

    Let $d \in \RT_{\cH}(N_{\semst{.}},\T_{[\Sigma\Delta]})$. Then there exist
$r_1 \in R_1$, $r_2 \in R_2$, $\xi \in \T_\Sigma$, and $\zeta \in \T_\Delta$ such that $d \in \RT_{\cH}(\semst{r_1,r_2},\xi,\zeta)$. 

Due to the definition of rules of $\cH$, the rule tree $d$ has the form shown in Figure~\ref{fig:run-schematic} for some 
$\ell, u \in \mathbb{N}$, rules $r_i^j$, symbols $\sigma_j$ and $\psi_j$ (each of appropriate type which can be read off easily from the definition of $R$), as well as rule $(r_1^{\ell+1},r_2^{u+1}) \to [\sigma,\delta](\semst{r_{11},r_{21}},\ldots, \semst{r_{1k},r_{2k}})$ in $R$, and $d_1,\ldots,d_k \in \RT_{\cH}(N_{\semst{.}},\T_{[\Sigma\Delta]})$; moreover, $r_1 = r^1_1$ and $r_2 = r^1_2$.

Then 
  \begin{compactitem}
  \item there exists a $\psi\in \Psi^{(k)}$ such that $h(d) =\psi(h(d_1),\ldots,h(d_k))$,
  \item $\xi= \sigma(\pi_1(d_1),\ldots,\pi(d_k))$,
    \item $\zeta= \delta(\pi_2(d_1),\ldots,\pi_2(d_k))$, and
    \item   $d_i \in \RT_{\cH}(\semst{r_{1i},r_{2i}},\pi_1(d_i),\pi_2(d_i))$ for each $i \in [k]$.
\end{compactitem}

We define $\varphi(d)= (t_1,t_2)$, where 
$t_1$ and $t_2$ are shown in Figure~\ref{fig:pair-run2} and  where for each $i \in [k]$ we let
$(t_{1i},t_{2i}) = \varphi(d_i)$.

Next we introduce a kind of typing for elements of $\RT_\cH(N_{\semst{.}},\T_{[\Sigma\Delta]})$ and for elements of $\RT_{\cH_1}(N_1,\T_\Sigma) \times \RT_{\cH_2}(N_2,\T_\Psi)$. Formally, let $r_1 \in R_1$, $r_2 \in R_2$, $\xi \in \T_\Sigma$, $\theta \in \T_\Psi$, and $\zeta \in \T_\Delta$. We define the following two sets:
\begin{align*}
    \RT_{\cH}(\semst{r_1,r_2},\xi,\theta,\zeta) &= \{d \in \RT_\cH(\semst{r_1,r_2},\xi,\zeta) \mid 
  h(d) = \theta\}\\
\RT_{\cH_1 \times \cH_2}(r_1,r_2,\xi,\theta,\zeta) &= 
                                                       \{(t_1,t_2) \in  \RT_{\cH_1}(N_1,\xi,\theta) \times  \RT_{\cH_2}(N_2,\theta,\zeta) \mid  t_1(\varepsilon) = r_1, t_2(\varepsilon) = r_2\}.
\end{align*}
Then the families 
\begin{eqnarray*}
  (\RT_{\cH}(\semst{r_1,r_2},\xi,\theta,\zeta) \mid r_1,r_2,\xi,\theta,\zeta \text{ as above})\\
  (\RT_{\cH_1 \times \cH_2}(r_1,r_2,\xi,\theta,\zeta) \mid r_1,r_2,\xi,\theta,\zeta \text{ as above})
  \end{eqnarray*}
  are partitionings of $\RT_\cH(N_{\semst{.}},\T_{[\Sigma\Delta]})$ and of $\RT_{\cH_1}(N_1,\T_\Sigma) \times \RT_{\cH_2}(N_2,\T_\Psi)$, respectively.
One might say that an element $d \in \RT_{\cH}(\semst{r_1,r_2},\xi,\theta,\zeta)$ and an element $(t_1,t_2) \in \RT_{\cH_1 \times \cH_2}(r_1,r_2,\xi,\theta,\zeta)$ have type  $(r_1,r_2,\xi,\theta,\zeta)$.

It is easy to see that
\(\varphi\big(\RT_{\cH}(\semst{r_1,r_2},\xi,\theta,\zeta)\big) \subseteq \RT_{\cH_1 \times \cH_2}(r_1,r_2,\xi,\theta,\zeta)\). Thus, intuitively, $\varphi$ is type preserving. Based on this property, we define the mapping 
\[
\varphi_{r_1,r_2,\xi,\theta,\zeta}: \RT_{\cH}(\semst{r_1,r_2},\xi,\theta,\zeta) \to  \RT_{\cH_1 \times \cH_2}(r_1,r_2,\xi,\theta,\zeta) 
  \]
  for each $d \in \RT_{\cH}(\semst{r_1,r_2},\xi,\theta,\zeta)$ by
  \(\varphi_{r_1,r_2,\xi,\theta,\zeta}(d) = \varphi(d)\).

\begin{figure}[t]
\hspace*{25mm}
\scalebox{.8}{
 \begin{tikzpicture}[cut/.style={draw,circle},level 1/.style={sibling distance=1.5cm},level 2/.style={sibling distance=1.5cm}]
\node[xshift=-25mm] {$t_1$:};
 \node {$r_1^1= (A_1^1\to[\sigma_1,\varepsilon](A_1^2))$}
         child {
           node {$r_1^\ell= (A_1^{\ell}\to [\sigma_{\ell},\varepsilon](A_1^{\ell+1}))$}
           child {
           node {$r_1^{\ell +1}= (A_1^{\ell+1}\to [\sigma,\psi](A_{11},\ldots, A_{1k}))$}
           child {node[trinode] (t1) {$t_{11}$}} 
            child {node{$\cdots$} edge from parent[draw=none]}
            child {node[trinode] (tk) {$t_{1k}$}}  edge from parent[solid]
            } edge from parent[dashed]};  
\node[xshift=25mm] {}
         child {
           node {$\ell\in\mathbb{N}$}edge from parent[draw=none]};
\hspace{70mm}
\node[xshift=-25mm] {$t_2$:};
\node {$r_2^1=(A_2^1\to [\varepsilon,\psi_1](A_2^2))$}
         child {
           node {$r_2^u=(A_2^{u}\to [\varepsilon, \psi_u](A_2^{u+1}))$}
           child {
           node {$r_2^{u+1}= (A_2^{u+1}\to [\psi,\delta](A_{21},\ldots, A_{2k}))$}
           child {node[trinode] (t1) {$t_{21}$}} 
            child {node{$\cdots$} edge from parent[draw=none]}
            child {node[trinode] (tk) {$t_{2k}$}}  edge from parent[solid]
            } edge from parent[dashed]};
\node[xshift=27mm] {}
         child {
           node {$u\in\mathbb{N}$}edge from parent[draw=none]};
    \end{tikzpicture}
}
\caption{\label{fig:pair-run2} A pair $(t_1,t_2)$ of rule trees of $\cH_1$ and $\cH_2$, respectively, where
$\lhs(t_{1j}(\varepsilon))=A_{1j}$ and $\lhs(t_{2j}(\varepsilon))=A_{2j}$ for $j\in [k]$.}
\end{figure}

  We show that each mapping $\varphi_{r_1,r_2,\xi,\theta,\zeta}$ is injective. For this, let $d$ and $d'$ be different trees with   $\varphi_{r_1,r_2,\xi,\theta,\zeta}(d)=(t_1,t_2)$ and $\varphi_{r_1,r_2,\xi,\theta,\zeta}(d')=(t'_1,t'_2)$. Then there exists a position $w\in \pos(d)\cap\pos(d')$ with $d(w)\ne d'(w)$. Since the nonterminals of $\cH$ encode rules of $\cH_1$ and $\cH_2$, this yields that the rules in $t_1$ and  $t'_1$ or the rules in $t_2$ and  $t'_2$ corresponding to $d(w)$ and $d'(w)$, respectively, are different. Hence $(t_1,t_2)\ne (t'_1,t'_2)$, i.e., $\varphi_{r_1,r_2,\xi,\theta,\zeta}$ is injective.

Now let $(t_1,t_2)\in  \RT_{\cH_1 \times \cH_2}(r_1,r_2,\xi,\theta,\zeta)$. Since each $t_1 \in \RT_{\cH_1}(N_1)$ and each  $t_2 \in \RT_{\cH_2}(N_2)$ have the form as shown in Figure~\ref{fig:pair-run2}, we can combine $t_1$ and $t_2$ into a $d \in \RT_{\cH}(\semst{r_1,r_2},\xi,\theta,\zeta)$  shown in Figure~\ref{fig:run-schematic}. For this $d$ we have $\varphi_{r_1,r_2,\xi,\theta,\zeta}(d)= (t_1,t_2)$,
hence each mapping $\varphi_{r_1,r_2,\xi,\theta,\zeta}$ is also surjective.

Moreover, by analysing the weights of the rules shown in Figure~\ref{fig:run-schematic} with those shown in Figure~\ref{fig:pair-run2} and taking the commutativity of $\B$ into account, we easily obtain that
\[\wt_\cH(d)  = \wt_{\cH_1}(t_1) \otimes \wt_{\cH_2}(t_2)\]
for each $d \in \RT_{\cH}(\semst{r_1,r_2},\xi,\theta,\zeta)$,
where $\varphi_{r_1,r_2,\xi,\theta,\zeta}(d)= (t_1,t_2)$.
Thus, each $\varphi_{r_1,r_2,\xi,\theta,\zeta}$ is a weight preserving bijection.

Now we can finish the proof of (a) as follows. For every $\xi \in \T_\Sigma$ and $\zeta \in \T_\Delta$, we have
\begingroup
\allowdisplaybreaks
\begin{align*}
&(\ttsem{\cH_1};\ttsem{\cH_2})(\xi,\zeta) =  \ \ \infsum{\oplus}{\theta \in \T_\Psi} \ttsem{\cH_1}(\xi,\theta) \otimes  \ttsem{\cH_2}(\theta,\zeta)\tag{by \eqref{equ:comp-wtt}}\\
= & \ \infsum{\oplus}{\theta \in \T_\Psi} 
\ \ \infsum{\oplus}{t_1 \in \RT_{\cH_1}(\xi,\theta)} \hspace{2mm}
\ \ \infsum{\oplus}{t_2 \in \RT_{\cH_2}(\theta,\zeta)}
  \wt_{\cH_1}(t_1) \otimes \wt_{\cH_2}(t_2) \tag{\text{by distributivity}}\\
= & \ \infsum{\oplus}{\theta \in \T_\Psi}
\bigoplus_{\substack{r_1 \in R_1:\\\lhs(r_1) \in S_1}} \bigoplus_{\substack{r_2 \in R_2:\\\lhs(r_2) \in S_2}}
\ \ \infsum{\oplus}{\substack{t_1 \in \RT_{\cH_1}(N_1,\xi,\theta):\\t_1(\varepsilon)=r_1}}\hspace{2mm}
\ \ \infsum{\oplus}{\substack{t_2 \in \RT_{\cH_2}(N_2,\theta,\zeta):\\t_2(\varepsilon)=r_2}}
\wt_{\cH_1}(t_1) \otimes \wt_{\cH_2}(t_2)\\
= & \ \infsum{\oplus}{\theta \in \T_\Psi} \hspace{2mm} \bigoplus_{\substack{r_1 \in R_1:\\\lhs(r_1) \in S_1}} \bigoplus_{\substack{r_2 \in R_2:\\\lhs(r_2) \in S_2}} \hspace{3mm}
\infsum{\oplus}{(t_1,t_2) \in \RT_{\cH_1 \times \cH_2}(r_1,r_2,\xi,\theta,\zeta)}
\wt_{\cH_1}(t_1) \otimes \wt_{\cH_2}(t_2)\\
= & \ \infsum{\oplus}{\theta \in \T_\Psi}\hspace{2mm}  \bigoplus_{\substack{r_1 \in R_1:\\\lhs(r_1) \in S_1}} \bigoplus_{\substack{r_2 \in R_2:\\\lhs(r_2) \in S_2}} \hspace{3mm}
\infsum{\oplus}{d \in \RT_{\cH}(\semst{r_1,r_2},\xi,\theta,\zeta)}
  \wt_\cH(d)
  \tag{\text{because  $\varphi_{r_1,r_2,\xi,\theta,\zeta}$ is a weight preserving bijection }}\\
       \tag{\text{
  from $\RT_{\cH}(\semst{r_1,r_2},\xi,\theta,\zeta)$ to $\RT_{\cH_1 \times \cH_2}(r_1,r_2,\xi,\theta,\zeta)$}}\\
  = & \infsum{\oplus}{\theta \in \T_\Psi} \hspace{2mm} \bigoplus_{\substack{r_1 \in R_1:\\\lhs(r_1) \in S_1}} \bigoplus_{\substack{r_2 \in R_2:\\\lhs(r_2) \in S_2}} \hspace{3mm}
\infsum{\oplus}{\substack{d \in \RT_{\cH}(N,\xi,\zeta):\\h(d)=\theta\\\lhs(d(\varepsilon))=\semst{r_1,r_2}}}
\wt_\cH(d)\\
= & \ \infsum{\oplus}{\theta \in \T_\Psi} 
\ \infsum{\oplus}{\substack{d \in \RT_{\cH}(\xi,\zeta):\\h(d)=\theta}}
\wt_\cH(d)
= \ \infsum{\oplus}{d \in \RT_{\cH}(\xi,\zeta)}
\wt_\cH(d)
=  \ttsem{\cH}(\xi,\zeta)\enspace.
\end{align*}
\endgroup

\

Proof of (2): Now let $\cH_1$ and $\cH_2$ be finite-derivational. Then, by Lemma \ref{lm:wbim-simple}(2), we can even \underline{construct} equivalent chain-free wpb. Consequently, the definition of $\cH$, as it is given in the proof of (1), is constructive and $\cH$ is finite-derivational. Moreover, by the proof of (1), we have $\ttsem{\cH} = \ttsem{\cH_1} ; \ttsem{\cH_2}$.

\

Proof of (3).  For this we assume that  $\cH_1$ and $\cH_2$ are finite-input. 
Let $\zeta \in \T_\Delta$. Since $\cH_1$ and $\cH_2$ are finite-input, the set $A= \{\theta \in \T_\Psi \mid \RT_{\cH_2}(\theta,\zeta)\not= \emptyset\}$ is finite, and for every $\theta \in A$ the set $B = \{\xi \in \T_\Sigma \mid \RT_{\cH_1}(\xi,\theta)\not= \emptyset\}$ is finite. Hence the set
\(
  C= \{\xi \in \T_\Sigma \mid (\exists \theta \in \T_\Psi): \RT_{\cH_1}(\xi,\theta)\not= \emptyset \wedge \RT_{\cH_2}(\theta,\zeta)\not= \emptyset\}
\)
is finite.

Next we show that $\{\xi \in \T_\Sigma \mid \RT_{\cH}(\xi,\zeta)\not= \emptyset\} \subseteq C$. For this, let $\xi \in \T_\Sigma$ and $d \in \RT_\cH(\xi,\zeta)$. Then there exists $\semst{r_1,r_2} \in S$ such that $d \in \RT_\cH(\semst{r_1,r_2},\xi,h(d),\zeta)$.
Since $\varphi_{r_1,r_2,\xi,h(d),\zeta}$ is bijective, there exist $t_1 \in \RT_{\cH_1}(N_1,\xi,h(d))$ and $t_2 \in \RT_{\cH_2}(N_2,h(d),\zeta)$. Since $\lhs(r_i) \in S_i$ for $i\in \{1,2\}$, we have $t_1 \in \RT_{\cH_1}(\xi,h(d))$ and $t_2 \in \RT_{\cH_2}(h(d),\zeta)$.
 Hence for  $h(d)\in \T_\Psi$ we have $\RT_{\cH_1}(\xi,h(d))\not= \emptyset$ and $\RT_{\cH_2}(h(d),\zeta)\not= \emptyset$, which means that $\xi \in C$. 

Since $\{\xi \in \T_\Sigma \mid \RT_{\cH}(\xi,\zeta)\not= \emptyset\} \subseteq C$ and $C$ is finite, also $\{\xi \in \T_\Sigma \mid \RT_{\cH}(\xi,\zeta)\not= \emptyset\}$ is finite. Hence $\cH$ is finite-input.
  \end{proof}


\subsection{Merge}

Here we express the application of a weighted projective bimorphism to a characteristic mapping in terms of a weighted regular tree grammar. Since this process is a kind of inverse to the split in Subsection \ref{ssec:split}, we call it merge. We recall that $[\Delta]$ denote the skeleton alphabet of $\Delta$, defined on page \pageref{page:skeleton}.

\begin{lemma} \rm \label{lm:comp-wbim} (cf. \cite[Lm.~4.4]{fulvog19}) Let $\cH$ be a chain-free  $([\Delta],\Psi,\B)$-wpb such that $\cH$ is finite-input or $\B$ is  $\sigma$-complete. Then we can construct an alphabetic $(\Psi,\B)$-wrtg $\cG$ such that (a) $\cG$ is finite-derivational if $\cH$ is finite-input and (b) $\ttsem{\cH}(\chi(\T_{[\Delta]})) = \sem{\cG}$.
 \end{lemma}
\begin{proof} Let  $\cH=(N,S,R,wt)$. We recall that $\cH$ is a chain-free and alphabetic $([[\Delta]\Psi],\B)$-wrtg. We construct the $(\Psi,\B)$-wrtg $\cG = (N',S',R',wt')$ as follows.  We let $N' = R$ and $S' = \{r \in R \mid \lhs(r) \in S\}$. 
The set $R'$ of rules and the weight mapping $wt'$ are defined as follows.
\begin{compactenum}
\item For every $r =(A \to [[k],\psi](A_1,\ldots, A_k))$ in $R$ and every $r_1,\ldots,r_k \in R$ with $\lhs(r_i) =A_i$ for each $i \in [k]$, the rule $r' =(r \rightarrow \psi(r_1,\ldots, r_k))$ is in $R'$ and $wt'(r') = wt(r)$.

  \item For every $r =(A \to [[1],\varepsilon](A_1))$ in $R$ and every $r_1 \in R$ with $\lhs(r_1)=A_1$,\\
 the rule $r' =(r \rightarrow r_1)$ is in $R'$ and $wt'(r') = wt(r)$.

\item  For every $r =(A \to [\varepsilon,\psi](A_1))$ in $R$ and every $r_1 \in R$ with $\lhs(r_1)=A_1$,\\
 the rule $r' =(r \rightarrow \psi(r_1))$ is in $R'$ and $wt'(r') = wt(r)$.
\end{compactenum}
Obviously, $\cG$ is alphabetic.

Before we prove Properties (a) and (b) we need some preparations. We define the mapping $\beta': \T_{R'} \rightarrow \T_{[\Delta]}$
by induction on $\T_{R'}$ for each $r'(d_1',\ldots,d_k') \in \T_{R'}$  as follows:
\[
\beta'(r'(d_1',\ldots,d_k')) =
\left\{
\begin{array}{ll}
[k](\beta'(d_1'),\ldots,\beta'(d_k')) & \text{ if $r'$ is defined by 1.}\\
\ \! \! [1](\beta'(d_1')) & \text{ if $r'$ is defined by 2.}\\
\beta'(d_1') & \text{ if $r'$ is defined by 3.}
\end{array}
\right.
\]
In other words, if the terminal symbol in the right-hand side of $\lhs(r')$ has the form $[[k],\psi]$ or $[[1],\varepsilon]$, then $\beta'$ outputs the terminal symbol $[k]$ and $[1]$, respectively; if the terminal symbol in the right-hand side of $\lhs(r')$ has the form $[\varepsilon,\psi]$, then $\beta'$ does not output any terminal symbol.

(This $\beta'$ is a modification of the mapping $\beta$ defined on page \pageref{page:beta-b}.)
For instance, if $\beta'$ is applied to the right tree in Figure~\ref{fig:bijection-2-b}, then the left tree in Figure~\ref{fig:bijection-1-b} is obtained.

\begin{figure}[t]
  \hspace*{40mm}
\scalebox{.8}{
 \begin{tikzpicture}[cut/.style={draw,circle},level 1/.style={sibling distance=1.5cm},level 2/.style={sibling distance=20mm}]
         \node {$[1]$}
          child { node{$[2]$}
            child {node {$[0]$}}
            child {node {$[0]$}}
                };
\hspace{50mm}
\node {$\psi$}
          child { node{$\omega$}
            child {node {$\gamma$}
            child {node {$\alpha$}} 
                   }
            child {node {$\beta$}}
                };

    \end{tikzpicture}
}
\caption{\label{fig:bijection-1-b} Example trees $b\in \T_{[\Delta]}$ (left) and $\xi \in \T_\Psi$ (right).}
\end{figure}

\begin{figure}[t]
 \hspace*{15mm}
\scalebox{.8}{
 \begin{tikzpicture}[cut/.style={draw,circle},level 1/.style={sibling distance=1.5cm},level 2/.style={sibling distance=40mm}]
      \node{$r_1 : (A \to [\varepsilon,\psi](A'))$}
     child {node {$r_2 :  (A' \to [[1],\varepsilon](C))$}
     child {node {$r_3 :  (C \to [[2],\omega](A_1,A_2))$}
     child {node {$r_4 :  (A_1 \to [\varepsilon,\gamma](D))$}
     child {node {$r_6 :  (D \to [[0],\alpha])$}}
}
     child {node {$r_5 : (A_2 \to [[0],\beta])$}}
}
}
;
\hspace{80mm}
      \node{$r_1 \to \psi(r_2)$}
     child {node {$r_2 \to r_3$}
     child {node {$r_3 \to \omega(r_4,r_5)$}
     child {node {$r_4 \to \gamma(r_6)$}
     child {node {$r_6 \to \alpha$}}
}
     child {node {$r_5 \to \beta$}}
}
}
;

\end{tikzpicture}
}
\caption{\label{fig:bijection-2-b} Example trees $d\in \RT_\cH(N,b,\xi)$ (left) with $b\in \T_{[\Delta]}$ and $\xi\in \T_\Psi$ from Fig. \ref{fig:bijection-1-b}, and $\varphi_{b,\xi}(d)\in \RT_\cG(N',\xi)$ (right). Moreover, $\beta'(\varphi_{b,\xi}(d)) = b$ and $\pi(\varphi_{b,\xi}(d)) = \xi$.}
\end{figure}

\index{succb@$\succ$}
Lastly, we define the relation
\[
  \succ = \succ_R \cap \ \big( \RT_\cH(N,\T_{[[\Delta]\Psi]}) \times  \RT_\cH(N,\T_{[[\Delta]\Psi]})\big)\enspace.
\]
   Since $\succ_{R}$ is terminating, by Lemma \ref{lm:termination-is-subset-closed} also it is easy to see that $\succ$ is terminating. Moreover, we have that $\nf_\succ(\RT_\cH(N,\T_{[[\Delta]\Psi]}))$ is the set of terminal rules of $\cH$. 

Then,
by induction on $(\RT_\cH(N,\T_{[[\Delta]\Psi]}),\succ)$, we define the mapping
\[\varphi: \RT_\cH(N,\T_{[[\Delta]\Psi]}) \rightarrow \RT_\cG(N',\T_\Psi),
\]
as follows.

Let $d \in \RT_\cH(N,\T_{[[\Delta]\Psi]})$. Then there exists $r=(A \to [\kappa,\lambda](A_1,\ldots,A_k))$ in $R$ and $d_1,\ldots,d_k \in \RT_\cH(N,\T_{[[\Delta]\Psi]})$ such that $d = r(d_1,\ldots,d_k)$.
\begin{compactitem}
\item If $\kappa=[k]$  for some $k\ne 1$ and $\lambda \in \Psi$, then we define 
\[\varphi(r(d_1,\ldots,d_k)) =
  (r \rightarrow \lambda(d_1(\varepsilon),\ldots,d_k(\varepsilon))) \Big(\varphi(d_1),\ldots,\varphi(d_k)\Big) \enspace.
\]

\item If $\kappa=[1]$ and $\lambda=\varepsilon$, then  $k=1$ and we define 
  \[
    \varphi(r(d_1)) =
    (r \rightarrow d_1(\varepsilon)) \Big(\varphi(d_1)\Big)\enspace.
    \]

\item If $\kappa=\varepsilon$ and $\lambda \in \Psi$, then $k=1$ and we define

  \[
    \varphi(r(d_1)) =
    (r \rightarrow \psi(d_1(\varepsilon)))\Big(\varphi(d_1)\Big) \enspace.
    \]
\end{compactitem}
The following is easy to see:
\begin{equation*}
\text{for every $b \in \T_{[\Delta]}$ and $\xi \in \T_\Psi$, we have } \varphi\big(\RT_\cH(N,b,\xi)\big) \subseteq \RT_\cG(N',\xi) \cap (\beta')^{-1}(b) \enspace. 
  \end{equation*}
  Thus, for  every $b \in \T_{[\Delta]}$ and $\xi \in \T_\Psi$, we can define the mapping
  \[
\varphi_{b,\xi} : \RT_\cH(N,b,\xi) \to \RT_\cG(N',\xi) \cap (\beta')^{-1}(b)
\]
by letting $\varphi_{b,\xi}(d) = \varphi(d)$ for each $d \in \RT_\cH(N,b,\xi)$.

Next we prove that $\varphi_{b,\xi}$ is bijective for every $b \in \T_{[\Delta]}$ and $\xi \in \T_\Psi$. For this proof let us abbreviate $\varphi_{b,\xi}$, $\RT_\cH(N,b,\xi)$, and $\RT_\cG(N',\xi)\cap (\beta')^{-1}(b)$ by $\varphi$, $\RT_\cH$, and $\RT_\cG$, respectively. To illustrate our arguments we show   examples of $b\in \T_{[\Delta]}$ and  $\xi\in \T_\Psi$ in Figure \ref{fig:bijection-1-b},  and examples of $d \in \RT_\cH$ and  $\varphi(d)\in \RT_\cG$ in Figure \ref{fig:bijection-2-b}. First we observe that $\varphi$ is shape preserving, i.e., for every $d \in \RT_\cH$, we have $\pos(\varphi(d))=\pos(d)$ (see  Figure \ref{fig:bijection-2-b}). Now let $d_1,d_2\in \RT_\cH$ such that $d_1 \ne d_2$. If $\pos(d_1)\ne \pos(d_2)$, then  $\varphi(d_1) \ne \varphi(d_2)$ because $\varphi$ is shape preserving. If 
$\pos(d_1)= \pos(d_2)$, then there exists a $w\in \pos(d_1)$ such that $d_1(w)\ne d_2(w)$. But then $\varphi(d_1)(w)\ne \varphi(d_2)(w)$ because the left-hand side nonterminals in $\varphi(d_1)(w)$ and  $\varphi(d_2)(w)$ are $d_1(w)$ and  $d_2(w)$, respectively (see Figure  \ref{fig:bijection-2-b}). This means that $\varphi$ is injective. Moreover, $\varphi$ is surjective because, given a $d\in \RT_\cG$, we can easily reobtain a $d\in \RT_\cH$ such that $\varphi(d)=d$. In fact, $d$ can be constructed by replacing, at each position $u$ of $d$, the rule $d(u)$ of $\cG$ by the rule $\lhs(d(u))$ of $\cH$  (see again Figure \ref{fig:bijection-2-b}).

Obviously, it also holds that $\varphi_{b,\xi}$ is weight preserving, i.e., $\wt_\cH(d)=\wt_\cG(\varphi_{b,\xi}(d))$ for every $d \in \RT_\cH(N,b,\xi)$. 
Then, for every $b\in \T_{[\Delta]}$ and $\xi \in \T_\Psi$, there exists a weight preserving bijection
$\overline{\varphi}_{b,\xi}: \RT_\cH(b,\xi) \to \RT_\cG(\xi) \cap (\beta')^{-1}(b)$.

Now we prove property (a). Assume that $\cH$ is finite-input and  let $\xi \in \T_\Psi$. Then  the set $\{b \in \T_{[\Delta]} \mid \RT_\cH(b,\xi)\not=\emptyset\}$ is finite. Since $\cH$ is chain-free, it is also  finite-derivational, i.e., the set $\RT_\cH(b,\xi)$ is finite for every $b\in \T_{[\Delta]}$. Since $\varphi_{b,\xi}$ is bijective, we obtain that the set $\RT_\cG(\xi)\cap (\beta')^{-1}(b)$ is finite for each $b \in \T_{[\Delta]}$ and thus the set
$\{b \in \T_{[\Delta]} \mid  \RT_\cG(\xi)\cap (\beta')^{-1}(b)\not=\emptyset\}$ is also finite. Moreover, the family $(\RT_\cG(\xi)\cap (\beta')^{-1}(b) \mid b \in \T_{[\Delta]})$ is a partitioning of $\RT_\cG(\xi)$. Thus, $\RT_\cG(\xi)$ is finite and hence $\cG$ is finite-derivational. This proves (a).

Finally we prove Property (b):  $\ttsem{\cH}(\chi(\T_{[\Delta]})) = \sem{\cG}$. For this, let $\xi \in \T_\Psi$. Then
\begingroup
\allowdisplaybreaks
\begin{align*}
&\big(\ttsem{\cH}(\chi(\T_{[\Delta]}))\big)(\xi) 
                 = \infsum{\oplus}{b \in \T_{[\Delta]}} \chi(\T_{[\Delta]})(b) \otimes  \ttsem{\cH}(b,\xi)
  \tag{\text{by \eqref{equ:appl-wtt-to-wtl-trees} because $\sem{\cH}$ is finite-input or $\B$ is $\sigma$-complete}}\\
& = \infsum{\oplus}{b \in \T_{[\Delta]}} \ttsem{\cH}(b,\xi)\\
& = \infsum{\oplus}{b \in \T_{[\Delta]}} \bigoplus_{d \in \RT_\cH(b,\xi)} 
                                                                                                     \wt_{\cH}(d)
  \tag{\text{because $\cH$ is chain-free and hence finite-derivational}}\\ 
  & = \infsum{\oplus}{b \in \T_{[\Delta]}} \bigoplus_{d \in \RT_\cG(\xi)\cap (\beta')^{-1}(b)} \wt_{\cG}(d)
  \tag{\text{because $\overline{\varphi}_{b,\xi}$ is bijective and weight preserving}}\\
& = \infsum{\oplus}{d \in \RT_\cG(\xi)} \wt_{\cG}(d) = \sem{\cG}(\xi) \tag{\text{because $\cG$ is finite-derivational if $\sem{\cH}$ is finite-input}}\enspace. 
\end{align*}
\endgroup
\end{proof}

\subsection{Closure result for wrtg and wta}

Finally we can prove the closure results for the sets $\Reg(\_,\B)$ and $\Rec(\_,\B)$.

\begin{theorem} \label{thm:closure-REG-under-wbim}  {\rm (cf. \cite[Thm.~6.3]{fulvog19a})} Let  $\B$ be a commutative semiring.
  Let $\cG$ be a $(\Sigma,\B)$-wrtg and $\cH$ be a $(\Sigma,\Psi,\B)$-wpb such that $\big($both $\cG$ and $\cH$ are finite-derivational and $\cH$ is finite-input$\big)$ or $\B$ is $\sigma$-complete. Then the following two statements hold.
  \begin{compactenum}
  \item[(1)] Then there exists a $(\Psi,\B)$-wrtg $\cG'$ such that (a) $\cG'$ is finite-derivational if $\cG$ is finite-derivational and $\cH$ is finite-input and (b) $\ttsem{\cH}(\sem{\cG}) = \sem{\cG'}$.
  \item[(2)] If $\cH$ is finite-derivational, then we can construct a  $(\Psi,\B)$-wrtg $\cG'$ such that (a) $\cG'$ is finite-derivational if $\cG$ is finite-derivational and $\cH$ is finite-input and (b) $\ttsem{\cH}(\sem{\cG}) = \sem{\cG'}$.
    \end{compactenum}
\end{theorem}
\begin{proof} Proof of (1): Let $\cG=(N,S,R,\wt)$. By Lemma \ref{lm:rtg-normal-form}, we may assume that $\cG$ is alphabetic. Then, by Lemma \ref{lm:decomp-wrtg}, we can construct a chain-free $([R],\Sigma,\B)$-wpb $\cH_\cG$ such that $\cH_\cG$ is finite-output, $\cH_\cG$ is finite-input
 if $\cG$ is finite-derivational, and $\sem{\cG}= \ttsem{\cH_\cG}(\chi(\T_{[R]}))$. 
In particular, since $\cH_\cG$ is chain-free, it is also finite-derivational. Then
  \[
\ttsem{\cH}(\sem{\cG}) = \ttsem{\cH}(\ttsem{\cH_\cG}(\chi(\T_{[R]}))) = (\ttsem{\cH_\cG};\ttsem{\cH})(\chi(\T_{[R]}))\enspace,
\]
where the last equality is due to Observation \ref{obs:application-of-composition} (recall that $\B$ is a semiring). Any of the two conditions that (i)~$\ttsem{\cH_\cG}$ is supp-o-finite and (ii) $\ttsem{\cH}$ is supp-i-finite or $\B$ is $\sigma$-complete, assures that by Theorem \ref{thm:comp-closure-wbim}(1) there exists  an $([R],\Psi,\B)$-wpb  $\cH'$ such that $\ttsem{\cH'} = \ttsem{\cH_\cG};\ttsem{\cH}$.
If  $\cH$ and $\cH_\cG$ are finite-input, then by Theorem \ref{thm:comp-closure-wbim}(3), also $\cH'$ is finite-input. By Lemma \ref{lm:wbim-simple}(1) we can assume that $\cH'$ is chain-free and finite-input. 

Finally, by Lemma \ref{lm:comp-wbim}, if $\cH'$ is given, then
we can construct a $(\Psi,\B)$-wrtg $\cG'$ such that $\cG'$ is  finite-derivational $\cH'$ is finite-input and $\ttsem{\cH'}(\chi(\T_{[R]})) = \sem{\cG'}$.

\

Proof of (2): Now assume that $\cH$ is finite-derivational. Then we follow the proof of (1) and, instead of Theorem \ref{thm:comp-closure-wbim}(1) and Lemma \ref{lm:wbim-simple}(1),  we apply Theorem \ref{thm:comp-closure-wbim}(2) and Lemma \ref{lm:wbim-simple}(2), respectively.  Hence, we can even construct $\cH'$. Thus we can also construct $\cG'$ with the mentioned properties.
\end{proof}

\begin{corollary-rect} \label{cor:closure-REC-under-wbim} \rm Let $\Sigma$ and $\Psi$ be ranked alphabets, let $\B$ be a commutative semiring, and let $\cA$ be a $(\Sigma,\B)$-wta. Moreover, let $\cH$ be a $(\Sigma,\Psi,\B)$-wpb such that (a) $\cH$ is finite-derivational and finite-input or (b) $\B$ is $\sigma$-complete.
  Then we can construct a $(\Psi,\B)$-wta $\cA'$ such that
  \[
    \ttsem{\cH}(\sem{\cA}) = \sem{\cA'}\enspace.
  \]
  Thus, in particular, if $\B$ is a commutative semiring, then $\Rec(\_\,,\B)$ is closed under weighted projective bimorphisms.
\end{corollary-rect}
\begin{proof}  By Lemma \ref{lm:wta-to-wrtg} we can construct a $(\Sigma,\B)$-wrtg $\cG$ such that $\cG$ is in tree automata form and $\sem{\cA} = \sem{\cG}$.  Then, in particular, $\cG$ is  finite-derivational. By Theorem \ref{thm:closure-REG-under-wbim} we can construct a  finite-derivational $(\Psi,\B)$-wrtg $\cG'$ such that $\ttsem{\cH}(\sem{\cG}) = \sem{\cG'}$.
Finally, by Lemma \ref{lm:wrtg-to-wta}, we can construct a $(\Sigma,\B)$-wta $\cA'$ such that $\sem{\cG'}=\sem{\cA'}$.
\end{proof}


\subsection{Closure under Hadamard product: an alternative proof}
\label{ssec:closure-Had-alternative}
  
We have proved that the set of recognizable $(\Sigma,\B)$-weighted tree languages is closed under Hadamard product if $\B$ is a commutative semiring (cf. Theorem \ref{thm:closure-Hadamard-product}). This closure also follows from the closure of this set under weighted projective bimorphisms.

\begin{corollary}\rm (cf. Theorem \ref{thm:closure-Hadamard-product})\label{cor:Hadamard-alternative}  Let $\B$ be a commutative semiring. Moreover, let $\cA_1$ and $\cA_2$ be two $(\Sigma,\B)$-wta.  Then we can construct a $(\Sigma,\B)$-wta $\cB$ such that  $\sem{\cB} = \sem{\cA_1} \otimes  \sem{\cA_2}$.
  \end{corollary}
  \begin{proof} Let $\cA_2=(Q,\delta,F)$. By Theorem \ref{thm:root-weight-normalization-run} we can assume that $\cA_2$ is root weight normalized and $\supp(F)=\{q_f\}$. We construct the $(\Sigma,\Sigma,\B)$-wpb $\cH=(N,S,R,wt)$ as follows.
    \begin{compactitem}
    \item $N=Q$ and $S=\{q_f\}$,
      \item For every $k\in \mathbb{N}$, $\sigma \in \Sigma^{(k)}$, $q\in Q$, and $q_1\cdots q_k\in Q^k$, the rule  $r= (q \to [\sigma,\sigma](q_1,\ldots,q_k))$ is in $R$ and $wt(r) = \delta_k(q_1\cdots q_k,\sigma,q)$.
      \end{compactitem}
      We note that $\cH$ is in tree automata form and finite-input.
      Then it is easy to see that $\ttsem{\cH}=\overline{\sem{\cA_2}}$, i.e., $\ttsem{\cH}$ is the diagonalization of $\sem{\cA_2}$ (cf. \eqref{equ:weighted-tree-lang-as-weighted-tree-transf}). Then by Equation \eqref{obs:application-vs-Hadamard} we have $\ttsem{\cH}(\sem{\cA_1})=\sem{\cA_1} \otimes \sem{\cA_2}$ and thus by Corollary  \ref{cor:closure-REC-under-wbim} we can construct a $(\Sigma,\B)$-wta $\cB$ such that $\sem{\cB}= \ttsem{\cH}(\sem{\cA_1})$. 
        \end{proof}


        \subsection[Closure under yield-intersection: an alternative proof]{Closure under yield-intersection with weighted recognizable languages: an alternative proof}\label{ssec:closure-yield-intersection-alternative}
        
        For every $(\Sigma,\B)$-wta $\cA$, $\Gamma \subseteq \Sigma^{(0)}$,  and $(\Gamma,\B)$-wsa $\cB$, we have constructed a $(\Sigma,\B)$-wta $\cA'$ such that  $\sem{\cA'} = \sem{\cA} \otimes (\sem{\cB}\circ \yield_\Gamma)$, if $\B$ is a commutative semiring (cf. Theorem \ref{thm:BPS}).
        Here we give an alternative proof which uses the fact that the set of recognizable $(\Sigma,\B)$-weighted tree languages is closed under weighted projective bimorphisms.

 \begin{corollary}\rm (cf. Theorem \ref{thm:BPS})\label{cor:BPS-alternative}  Let $\B$ be a commutative semiring and $\Gamma \subseteq \Sigma^{(0)}$. For every $(\Sigma,\B)$-wta $\cA$ and every wsa $\cB$ over $\Gamma$ and $\B$, we can construct a $(\Sigma,\B)$-wta $\cA'$ such that  $\sem{\cA'} = \sem{\cA} \otimes (\sem{\cB}\circ \yield_\Gamma)$. Thus, in particular, the set $\Rec^{\mathrm{run}}(\Sigma,\B)$ is closed under yield-intersection.
\end{corollary}
\begin{proof} Let $\cB = (P,\lambda,\mu,\gamma)$   be a $(\Sigma^{(0)},\B)$-wsa.

  The idea is to construct a $(\Sigma,\Sigma,\B)$-wpb $\cH$ such that, for every $\xi,\zeta \in \T_\Sigma$ we have
  \begin{equation} 
    \ttsem{\cH}(\xi,\zeta) =
      \begin{cases}
        \sem{\cB}(\yield_\Gamma(\zeta)) & \text{ if $\xi=\zeta$}\\
        \0 & \text{ otherwise} \enspace. \label{eq:yield-by-wpb}
        \end{cases}
      \end{equation}

  We construct  $\cH = (N,S,R,wt)$ as follows.
  \begin{compactitem}
  \item $N = (P \times P) \cup \{S\}$ where $S$ is a new symbol, and 
  \item $R$ and $wt$ are defined as follows:
    \begin{compactitem}
      \item for each $(p,p') \in N$, the rule $r=(S \to (p,p'))$ is in $R$ and  we let $wt(r) = \lambda(p) \otimes \gamma(p')$,
  \item for each $\alpha \in \Gamma$ and $(p,p') \in P\times P$
    we let \(r=((p,p') \to [\alpha,\alpha])\) be a rule in $R$ with $wt(r) =  \mu(p,\alpha,p')$, 
   \item for each $\alpha \in \Sigma^{(0)} \setminus \Gamma$ and $p \in P$
    we let \(r=((p,p) \to [\alpha,\alpha])\) be a rule in $R$ with $wt(r) =  \1$, and
\item for every $k \in \mathbb{N}_+$, $\sigma \in \Sigma^{(k)}$, and $(p_1,p_1'), (p_2,p_2'),\ldots,
(p_k,p_k'), (p,p') \in N$  we let the rule \(r = ((p,p') \to [\sigma,\sigma]((p_1,p_1'), \ldots,(p_k,p_k')))\) be in $R$ with  
\[
  wt(r) = 
\left\{
\begin{array}{ll}
\mathbb{1} & \hbox{ if } p = p_1, p_i' = p_{i+1} \text{ for each } i \in [k-1], \text{ and } p_k' = p' \\
\mathbb{0} & \hbox{ otherwise.}
\end{array}
\right.
\]
\end{compactitem}
\end{compactitem}
Then $\cH$ is finite-derivational and finite-input, and hence  $\ttsem{\cH}$ is supp-i-finite. Moreover, apart from the chain rules of the form $S \to (p,p')$, the wpb $\cH$ closely corresponds to the wta $\cA$ constructed in the proof of Lemma \ref{lm:BPS}. Indeed, there exists a bijection between $\RT_\cH(\xi,\xi)$ and $\R_\cA(\xi)$, which is defined in a similar way as the bijection in the proof of Lemma~\ref{lm:related-semantics}.

It is quite obvious that \eqref{eq:yield-by-wpb} holds. Since $\ttsem{\cH}$ is supp-i-finite, $\sem{\cA}$ is $\ttsem{\cH}$-summable. Hence $\ttsem{\cH}(\sem{\cA})$ is defined.
       Then, for each $\zeta \in \T_\Sigma$, we have
        \begingroup
        \allowdisplaybreaks
        \begin{align*}
          \ttsem{\cH}(\sem{\cA})(\zeta) 
         &= \infsum{\oplus}{\xi \in \T_\Sigma}{\sem{\cA}(\xi) \otimes \ttsem{\cH}(\xi,\zeta)}\\
          &=\sem{\cA}(\zeta) \otimes \ttsem{\cH}(\zeta,\zeta) \tag{by the second case of  \eqref{eq:yield-by-wpb}}\\
          &=  (\sem{\cA} \otimes (\sem{\cB}\circ\yield_\Gamma))(\zeta)  \tag{by the first case of  \eqref{eq:yield-by-wpb}}
          \end{align*}
          \endgroup
          By Corollary  \ref{cor:closure-REC-under-wbim} we can construct a $(\Sigma,\B)$-wta $\cA'$ such that $\sem{\cA'}= \ttsem{\cH}(\sem{\cA})$. 
        \end{proof}


\subsection{Closure under tree relabelings: an alternative proof}
\label{ssec:closure-tree-relab-alternative}

  Finally, we show that each tree relabeling can be computed by a  particular weighted projective bimorphism. Then, for the case that $\B$ is a commutative semiring, the closure of $\Rec(\_\,,\B)$ under tree relabelings (cf. Theorem \ref{thm:closure-under-tree-relabeling}) can be reobtained as corollary of Corollary \ref{cor:closure-REC-under-wbim}. However, we recall that Theorem \ref{thm:closure-under-tree-relabeling} holds for arbitrary strong bimonoids and not only for commutative semirings.

\begin{observation}\rm \label{obs:tree-relab-as-wgrel} Let $\tau$ be a $(\Sigma,\Delta)$-tree relabeling and $r: \T_\Sigma \to B$. Then we can construct a chain-free, finite-input, and finite-output $(\Sigma,\Delta,\B)$-wpb $\cH$ such that  $\tau(r) = \ttsem{\cH}(r)$.
  \end{observation}
  \begin{proof} Let $\tau= (\tau_k \mid k \in \mathbb{N})$. We construct the  $(\Sigma,\Delta,\B)$-wpb $\cH=  (\{S_0\},S_0,R,\wt)$ as follows. For every $k \in \mathbb{N}$, $\sigma \in \Sigma^{(k)}$, and $\gamma \in \tau_k(\sigma)$, the  set $R$ contains the rule $r= (S_0 \to [\sigma,\gamma](S_0,\ldots,S_0))$  with $k$ occurrences of $S_0$ and $\wt(r) = \mathbb{1}$. Clearly, $\cH$ has the desired properties. In particular, $\ttsem{\cH}$ is supp-i-finite.

    Then for every $\xi \in \T_\Sigma$ and $\zeta \in \T_\Delta$: $\ttsem{\cH}(\xi,\zeta) \in \{\mathbb{0},\mathbb{1}\}$, and $\ttsem{\cH}(\xi,\zeta) = \mathbb{1}$ if and only if $\zeta \in \tau(\xi)$. Now let $\zeta \in \T_\Delta$. Then we can calculate as follows:
    \begin{align*}
      \tau(r)(\zeta) = \bigoplus_{\xi \in \tau^{-1}(\zeta)} r(\xi)
      = \bigoplus_{\xi \in \T_\Sigma} r(\xi) \otimes \ttsem{\cH}(\xi,\zeta)
      = (\ttsem{\cH}(r))(\zeta)
    \end{align*}
    Hence $\tau(r) = \ttsem{\cH}(r)$.
\end{proof}

The following corollary follows from Corollary \ref{cor:closure-REC-under-wbim} and Observation \ref{obs:tree-relab-as-wgrel}.

\begin{corollary} \label{cor:tree-relabeling-alternative}\rm Let $\B$ be a commutative semiring and $\cA$ be a $(\Sigma,\B)$-wta. Moreover, let $\tau$ be a $(\Sigma,\Delta)$-tree relabeling.
 Then we can construct a $(\Delta,\B)$-wta $\cA'$ such that $\tau(\sem{\cA}) = \sem{\cA'}$ (cf.  Theorem \ref{thm:closure-under-tree-relabeling}).
  Thus, in particular, if $\B$ is a commutative semiring, then the set $\Rec(\_\,,\B)$ is closed under tree relabelings.
  \end{corollary}


\section{\ Summary of some of the closure properties}

In the three tables of Figure \ref{fig:table-summary-closure} we summarize some of the closure properties of the set of weighted tree languages recognized by wta. Each entry refers to the theorem or corollary where the precise formulation of the closure property can be found; in the figure we only have indicated the additionally required  properties of the strong bimonoid. For some of the operations in the first table and the third table, we could only prove  the corresponding closure property for the case that $\B$ is a semiring; for a few other operations we did not need distributivity from both sides. In the summary, we did not consider the case of bu-deterministic and crisp-deterministic wta.

\begin{figure}[t]
  \centering
  {\footnotesize
    \begin{tabular}{|l|c|c|c|}
\hline
   & $\Rec^{\mathrm{run}}(\Sigma,\B)$ & $\Rec^{\mathrm{init}}(\Sigma,\B)$ & $\Rec(\Sigma,\B)$\\[1mm]\hline
sum  & Thm. \ref{thm:closure-sum} & Thm. \ref{thm:closure-sum} & Thm. \ref{thm:closure-sum} \\[1mm]\hline
      scalar multiplications & & & \\
      ... from left & Thm. \ref{thm:closure-scalar}(1)  & Thm. \ref{thm:closure-scalar}(1) & Thm. \ref{thm:closure-scalar}(1) \\
      & (left-distr.) &  (left-distr.) & \\
      ... from right &  Thm. \ref{thm:closure-scalar}(2)  & Thm. \ref{thm:closure-scalar}(2) & Thm. \ref{thm:closure-scalar}(2) \\
  & (right-distr.) &  (right-distr.) & \\[1mm]\hline
   Hadamard product &   &    & Thm. \ref{thm:closure-Hadamard-product}, also Cor. \ref{cor:Hadamard-alternative}\\
 &  &   & (commutative)\\[1mm]\hline
top-concatenations &  &   &  Cor. \ref{cor:closure-under-top-concat-wta}\\[1mm]\hline
  tree concatenations &   &   & Cor. \ref{cor:closure-tree-concatenation}  \\ 
&  &   & (commutative)\\[1mm]\hline
Kleene stars &   &   & Cor. \ref{cor:closure-Kleene-star}\\
  &  &   & (commutative)\\[1mm]\hline
  yield-intersection &   &   & Thm. \ref{thm:BPS}, also Cor. \ref{cor:BPS-alternative}\\
&  &   & (commutative)\\
\hline
\multicolumn{4}{c}{}\\[4mm]\hline
 & $\Rec^{\mathrm{run}}(\Sigma,\_)$ & $\Rec^{\mathrm{init}}(\Sigma,\_)$ & $\Rec(\Sigma,\_)$\\\hline
 strong bimonoid hom. & Thm. \ref{thm:closure-sr-hom} & Thm. \ref{thm:closure-sr-hom} & Thm. \ref{thm:closure-sr-hom} \\\hline
  \multicolumn{4}{c}{}\\[4mm]
  \hline
   & $\Rec^{\mathrm{run}}(\_\,,\B)$ & $\Rec^{\mathrm{init}}(\_\,,\B)$ & $\Rec(\_\,,\B)$\\\hline
 tree relabeling   & Thm. \ref{thm:closure-under-tree-relabeling}   &   & Thm. \ref{thm:closure-under-tree-relabeling}, also Cor. \ref{cor:tree-relabeling-alternative}\\[1mm]\hline
linear, nondeleting, and & Cor. \ref{cor:closure-REC-under-lin-nondel-prod-hom} & & Cor. \ref{cor:closure-REC-under-lin-nondel-prod-hom}\\
      productive tree hom.  &  & &  \\[1mm]\hline
      inverse of linear tree hom. & & & Thm. \ref{thm:closure-under-inverse-tree-hom-2}\\
      &&& (commutative) \\[1mm]\hline
weighted projective bimorphisms &   &   & Cor. \ref{cor:closure-REC-under-wbim}  \\
  (finite-derivational, finite-input &  &   & (commutative)\\
  or $\B$ $\sigma$-complete) & & & \\[1mm]\hline
  
    \end{tabular}
    }
\caption{\label{fig:table-summary-closure} A summary of some closure properties of some sets of weighted tree languages recognized by wta,  where ``distr.'' and ``hom.'' abbreviate distributive and homomorphisms, respectively, and in the third column $\B$ is a semiring.}
\end{figure}

%% file: decomposition-results.tex
\addtocontents{toc}{\protect\pagebreak}
\chapter{Characterizations by weighted local systems}
\label{ch:decomposition}

In this chapter we show two characterization theorems for wta. 
The first one is Theorem \ref{thm:decomposition-2}; it is based on a decomposition theorem of wta which is due to \cite{ful15}. The latter result generalizes the fact that each recognizable tree language is the image of a local tree language under a deterministic tree relabeling \cite[Prop.~2]{tha67} (cf. \cite[Thm.~2.9.5]{gecste84} and \cite[Cor.~3.59(i)$\Rightarrow$(ii)]{eng75-15}). 

The second characterization theorem is Theorem \ref{thm:decomposition-1}, which follows the idea of decomposing a bottom-up tree transducer. In \cite[Thm.~3.5]{eng75} it was proved that each bottom-up tree transducer can be decomposed into a relabeling, followed by the intersection with a local tree language, followed by a tree homomorphism. Also the reverse composition result holds \cite[p.220]{eng75}. In Theorem \ref{thm:decomposition-1}(A) $\Rightarrow$ (B) we decompose the run semantics $\runsem{\cA}$ of a $(\Sigma,\B)$-wta $\cA$ into  an inverse deterministic tree relabeling (corresponding to the above relabeling), followed by the intersection with a local tree language, followed by a homomorphism which interprets  trees in some evaluation algebra with carrier set $B$ (corresponding to the above tree homomorphism). Similar decompositions were proved for weighted tree automata over multioperator monoids \cite[Thm.~1]{stuvogfue09}.

 Before proving the decomposition theorems we recall the definitions of local systems, local tree languages, and weighted local system.

  \section{Local tree languages and weighted local systems}
\label{sec:loc-tree-lang-wls}
    
  \index{Fork@$\Fork(\Sigma)$}
  \index{fork}
  We consider the ranked alphabet $\Fork(\Sigma)$  where $\Fork(\Sigma)^{(k)} = \Sigma^k \times \Sigma^{(k)}$ for each $k \in \mathbb{N}$. Each element of $\Fork(\Sigma)^{(k)}$ has the form $(\sigma_1 \cdots \sigma_k,\sigma)$ with $\sigma_1, \ldots, \sigma_k\in \Sigma$
    and $\sigma \in \Sigma^{(k)}$, and it  is called \emph{$k$-fork} or just \emph{fork}.

    \index{local system}
A \emph{$\Sigma$-local system} \cite[Sec.~2.9]{gecste84}  is a pair $(K,H)$ where $K\subseteq\Fork(\Sigma)$ is a set of forks and $H\subseteq \Sigma$. 
The \emph{tree language  generated by $(K,H)$}, denoted by $\LL(K,H)$, is defined as follows. First, we let
\index{languageK@$\LL(K)$}
\[\LL(K)=\{\xi \in \T_\Sigma\mid \text{$(\xi(w1) \cdots \xi(wk), \xi(w))\in K$ for each $w\in \pos(\xi)$, where $k=\rk(\xi(w))$} \} \enspace. \]
Second, we define
\index{languageKH@$\LL(K,H)$}
\[\LL(K,H)=\{\xi \in \LL(K) \mid \xi(\varepsilon)\in H\}.\]
For the particular $\Sigma$-local system $(\Fork(\Sigma),\Sigma)$, we have $\LL(\Fork(\Sigma))=\LL(\Fork(\Sigma),\Sigma) = \T_\Sigma$.
\index{local tree language}
Let $L\subseteq \T_\Sigma$. We call $L$ a \emph{local tree language} if there exists a $\Sigma$-local system  $(K,H)$  such that $L = \LL(K,H)$.

\index{weighted local system}
\index{wls}
A \emph{$(\Sigma,\B)$-weighted local system} (for short: $(\Sigma,\B)$-wls) \cite{ful15} is a tuple $\cS= (g,F)$ such that $g = (g_k \mid   k \in \mathbb{N})$ is an $\mathbb{N}$-indexed family of mappings $g_k: \Fork(\Sigma)^{(k)} \to B$ and $F: \Sigma \to B$. 
We say that $\cS$ has \emph{unit root weights} if $\im(F)\subseteq \{\0,\1\}$.

We define the mapping $g': \T_\Sigma \to B$ by induction on $\T_\Sigma$.
For every $\xi = \sigma(\xi_1,\ldots,\xi_k)$, we let
\[
g'(\xi)= g'(\xi_1) \otimes \cdots \otimes g'(\xi_k) \otimes g_k(\xi_1(\varepsilon) \cdots \xi_k(\varepsilon),\sigma) \enspace.
\]
In the following we drop the prime from $g'$ and simply write $g$ for $g'$. The  \emph{$(\Sigma,\B)$-weighted tree language determined by $\cS$}, denoted by $\sem{\cS}$, is the mapping $\sem{\cS}: \T_\Sigma \to B$ defined for each $\xi \in \T_\Sigma$ by
\index{semanticS@$\sem{\cS}$}
\[
\sem{\cS}(\xi) = g(\xi) \otimes F(\xi(\varepsilon)) \enspace.
\]
Since the value $g(\sigma(\xi_1,\ldots,\xi_k)) \in B$ depends on the root labels of  $\xi_1,\ldots,\xi_k$, in general there does not exist a $\Sigma$-algebra $(B,\lambda)$ such that $g$ is the unique $\Sigma$-algebra homomorphism from $(\T_\Sigma,\ttop_\Sigma)$ to  $(B,\lambda)$.

\begin{example}\label{ex:number-of-occurrences-by-weighted-local} \rm 
We consider the ranked alphabet $\Sigma=\{\sigma^{(2)},\gamma^{(1)},\alpha^{(0)}\}$ and the weighted tree language $\#_{\sigma(.,\alpha)}: \T_\Sigma \to \mathbb{N}$ defined in Example \ref{ex:number-of-occurrences}, where $\#_{\sigma(.,\alpha)}(\xi)$ is the number of occurrences of the pattern $\sigma(.,\alpha)$ in $\xi$ for each $\xi\in \T_\Sigma$. 

We define the $(\Sigma,\Natmaxplus)$-wls ${\cal S}=(g,F)$ such that
\begin{compactitem}
\item for every $\theta_1,\theta_2\in \Sigma$ we let $g_2(\theta_1\theta_2,\sigma)=1$ if $\theta_1\theta_2\in  \{\sigma\alpha,\gamma\alpha,\alpha\alpha\}$ and 0 otherwise; and we let $g_1(\theta_1,\gamma)=0$ and  $g_0(\varepsilon,\alpha)=0$, and 
\item $F(\sigma)=F(\gamma)=F(\alpha)=0$.
\end{compactitem}
It should be clear that $\sem{\cal S}(\xi)=\#_{\sigma(.,\alpha)}(\xi)$ for each $\xi \in \T_\Sigma$.
\hfill $\Box$
\end{example}

Let $(K,H)$ be a $\Sigma$-local system  and $\kappa=(\kappa_k \mid k\in\mathbb{N})$ an $\mathbb{N}$-indexed family of mappings $\kappa_k:~\Sigma^{(k)}~\to~B$. We interpret the trees in $\LL(K,H)$ by the unique $\Sigma$-algebra homomorphism $\h_{\M(\Sigma,\kappa)}: \T_\Sigma \to B$, where $\M(\Sigma,\kappa)$ is the $(\Sigma,\kappa)$-evaluation algebra defined in Section \ref{sect:trees}. Then we obtain the following  $(\Sigma,\B)$-weighted tree language:
\begin{align*}
  (\chi(\LL(K,H)) \otimes \h_{\M(\Sigma,\kappa)}): \T_\Sigma &\to B\\
  \xi &\mapsto
        \begin{cases}\h_{\M(\Sigma,\kappa)}(\xi) & \text{if } \xi\in \LL(K,H) \\
\0 & \text{otherwise} \enspace.
\end{cases}
  \end{align*}
In the next lemma we prove that the weighted tree language $\chi(\LL(K,H)) \otimes \h_{\M(\Sigma,\kappa)}$ can be computed by a $(\Sigma,\B)$-wls.
  
\begin{lemma}\label{lm:local+kappa->weighted-local} \rm Let $(K,H)$ be a $\Sigma$-local system and $\kappa=(\kappa_k \mid k\in\mathbb{N})$ a family of mappings $\kappa_k:~\Sigma^{(k)}~\to~B$. We can construct a $(\Sigma,\B)$-wls $\cS$ which has unit root weights such that 
\(\sem{\cS} = \chi(\LL(K,H)) \otimes \h_{\M(\Sigma,\kappa)}\).
\end{lemma}
\begin{proof} We define the $(\Sigma,\B)$-wls $\cS=(g,F)$ as follows. For every $k\in\mathbb{N}$ and $\sigma_1,\ldots,\sigma_k\in \Sigma$, and $\sigma\in\Sigma^{(k)}$, we have
\begin{align*}
      g_k(\sigma_1\cdots\sigma_k,\sigma) =
      \begin{cases}
        \kappa_k(\sigma) & \text{if } (\sigma_1\cdots\sigma_k,\sigma) \in K \\
        \0 & \text{ otherwise}
        \end{cases} 
      \text{\hspace{10mm} and \hspace{10mm}}
 F(\sigma)=
\begin{cases}
\1 & \text{if } \sigma \in H  \\
        \0 & \text{ otherwise.}
\end{cases} 
\end{align*}

By induction on $\T_\Sigma$, we prove that the following statement holds: 
\begin{equation}\label{eq:g=h}
  \text{For each $\xi\in \T_\Sigma$, we have } g(\xi)=\begin{cases}
\h_{\M(\Sigma,\kappa)}(\xi) & \text{if } \xi \in \LL(K) \\
\0 & \text{ otherwise.} 
\end{cases}
\end{equation}
Let $\xi=\sigma(\xi_1,\ldots,\xi_k)$. Then
\begingroup
\allowdisplaybreaks
\begin{align*}
g(\xi)=& \Big(\bigotimes_{i\in[k]}g(\xi_i)\Big) \otimes g_k(\xi_1(\varepsilon)\cdots \xi_k(\varepsilon),\sigma) \\[2mm]
= & \begin{cases}
\Big(\bigotimes_{i\in[k]}\h_{\M(\Sigma,\kappa)}(\xi_i)\Big) \otimes g_k(\xi_1(\varepsilon)\cdots \xi_k(\varepsilon),\sigma)
& \text{if } (\forall i\in [k]):\xi_i \in \LL(K) \\
\0 & \text{otherwise}
\end{cases}
\tag{by I.H.}\\[2mm]
= & \begin{cases}
\Big(\bigotimes_{i\in[k]}\h_{\M(\Sigma,\kappa)}(\xi_i)\Big) \otimes \kappa_k(\sigma)
& \text{if } (\forall i\in [k]):\xi_i \in \LL(K) \\
& \text{and } (\xi_1(\varepsilon)\cdots \xi_k(\varepsilon),\sigma) \in K \\
\0 & \text{otherwise}
\end{cases}
     \hspace*{5mm}
 \tag{by the definition of $g_k$}\\[2mm]
= & \begin{cases}
\h_{\M(\Sigma,\kappa)}(\xi) & \text{if } \xi \in \LL(K) \\
\0 & \text{ otherwise.} 
\end{cases}
\end{align*}
\endgroup
This proves \eqref{eq:g=h}.
Now let $\xi\in \T_\Sigma$. Then
\begingroup
    \allowdisplaybreaks
\begin{align*}
\sem{\cS}(\xi)=g(\xi)\otimes F(\xi(\varepsilon)) \ = \ 
\begin{cases} g(\xi) & \text{if  $\xi\in \LL(K)$}\\
&  \text{and $\xi(\varepsilon) \in H$}\\
\0 & \text{otherwise}
\end{cases}
\ \ = \ \ 
\begin{cases}\h_{\M(\Sigma,\kappa)}(\xi) & \text{if } \xi\in \LL(K,H) \\
\0 & \text{otherwise}
\end{cases} 
\end{align*}
\endgroup
where the last equality is due to \eqref{eq:g=h}.
\end{proof}

\sloppy It seems that the inverse of Lemma \ref{lm:local+kappa->weighted-local} does not  hold, because mappings of the form $g_k:~\Fork(\Sigma)^{(k)}~\to~B$ cannot be coded by mappings of the form $\kappa_k: \Sigma^{(k)} \to B$.

\index{supportS@$\supp(\cS)$}
In the next lemma, we prove that,  for each $(\Sigma,\Boole)$-wls $\cS=(g,F)$,  the support of $\sem{\cS}$ is a local tree language. As preparation, we define the \emph{support local system of $\cS$}, denoted by $\supp(\cS)$, to be the $\Sigma$-local system $(K,H)$, where $K=\bigcup_{k\in\mathbb{N}}\supp(g_k)$ and $H=\supp(F)$.

\begin{lemma}\label{lm:supp-of-Boolean-weighted-local}\rm Let $\cS=(g,F)$ be a $(\Sigma,\Boole)$-wls. Then $\supp(\sem{\cS})=\LL(\supp(\cS))$.
\end{lemma}
\begin{proof} Let $\supp(\cS) = (K,H)$.
  By induction on $\T_\Sigma$, we prove that the following statement holds:
  \begin{equation}
    \text{For each $\xi\in \T_\Sigma$, we have  $\xi \in \supp(g)$ if and only if $\xi \in \LL(K)$.} \label{equ:supp-wlsB=ls}
    \end{equation}

    For this, let 
$\xi=\sigma(\xi_1,\ldots,\xi_k)$. Then
\begin{align*}
\xi \in \supp(g) \iff & g(\xi_1)\wedge \cdots \wedge g(\xi_k)\wedge g_k(\xi_1(\varepsilon)\cdots\xi_k(\varepsilon),\sigma)=1 \\
\iff & (\forall i\in[k]): \xi_i\in \supp(g_i) \text{ and } g_k(\xi_1(\varepsilon)\cdots\xi_k(\varepsilon),\sigma)=1\\
\iff & (\forall i\in[k]): \xi_i\in \LL(K) \text{ and } (\xi_1(\varepsilon)\cdots\xi_k(\varepsilon),\sigma)\in K\\
& \tag{\text{by I.H. and the definition of $K$}}\\
\iff & \xi \in \LL(K).
\end{align*}
This proves \eqref{equ:supp-wlsB=ls}. 
Then for each $\xi\in \T_\Sigma$ we have
\begin{align*}
\xi\in \supp(\sem{\cS}) &\iff g(\xi)=1 \text{ and } F(\xi(\varepsilon))=1 \iff \xi \in \LL(K) \text{ and } \xi(\varepsilon) \in H \iff \xi\in \LL(K,H). \qedhere
\end{align*}
\end{proof}

The next theorem states that, intuitively, $\Sigma$-local systems and $(\Sigma,\Boole)$-wls are essentially the same.

\begin{theorem}\label{thm:loc=wloc(B)} Let $L\subseteq \T_\Sigma$ be a tree language. The following two statements are equivalent.
\begin{compactenum}
\item[(A)] We can construct a $\Sigma$-local system $(K,H)$ such that $L = \LL(K,H)$.
\item[(B)] We can construct a  $(\Sigma,\Boole)$-wls $\cS$ such that $L=\supp(\sem{\cS})$.
\end{compactenum}
\end{theorem}
\begin{proof} Proof of (A)$\Rightarrow$(B): Let $\kappa=(\kappa_k \mid k\in\mathbb{N})$ be a family of mappings $\kappa_k: \Sigma^{(k)} \to \mathbb{B}$ such that $\kappa_k(\sigma)=1$ for every $k\in \mathbb{N}$ and $\sigma \in \Sigma^{(k)}$. Then
  \[
    L = \LL(K,H) = \supp(\chi(\LL(K,H))) = \supp(\chi(\LL(K,H)) \otimes \h_{\M(\Sigma,\kappa)}) \enspace.
  \]
By Lemma \ref{lm:local+kappa->weighted-local}, we can construct a $(\Sigma,\Boole)$-wls $\cS$ such that  $\chi(L(K,H)) \otimes \h_{\M(\Sigma,\kappa)}=\sem{\cS}$. Thus $L=\supp(\sem{\cS})$. 

\

Proof of (B)$\Rightarrow$(A): This follows from Lemma \ref{lm:supp-of-Boolean-weighted-local}.
\end{proof}


  \section[Characterization by  weighted local systems]{Characterization by  weighted local system and deterministic relabeling}
  \label{sec:char-wls-rel}

In this section we prove the first characterization theorem for recognizable weighted tree languages. Roughly speaking, it says that each recognizable $(\Sigma,\B)$-weighted tree language can be represented as the application of some deterministic tree relabeling to the weighted rule tree language of some  wrtg in tree automata form (cf. Theorem \ref{thm:decomposition-2}).

As preparation, we recall the well known fact that the rule tree language of a context-free grammar is a local tree language (cf. \cite[Prop.~1]{tha67}, also cf. \cite[Thm.~3.57]{eng75-15} and \cite[Thm.~3.2.9]{gecste84}).

\begin{lemma}\label{lm:rule-trees-are-local} \rm Let  $G$ be a $\Gamma$-cfg with rule set $R$. We can construct an $R$-local system $(K,H)$ such that $\LL(K,H) = \RT_G$.
\end{lemma}
\begin{proof}  Let $G = (N,S,R)$. We construct the $R$-local system $(K,H)$ as follows.

We construct the set $K \subseteq \mathrm{Fork}(R)$ such that, for every  $r = (A \to u_0 A_1 u_1 \cdots A_ku_k)$ in $R$  and every $r_1 = (A_1 \to \alpha_1), \ldots, r_k = (A_k \to \alpha_k)$ in $R$, the tuple $(r_1\cdots r_k,r)$ is in $K$.

Moreover, we let $H = \{r \mid \lhs(r) \in S\}$. 
It is obvious that $\LL(K,H) = \RT_G$.
\end{proof}

We note that, due to \cite[Thm.~3.57]{eng75-15}, the inverse of Lemma \ref{lm:rule-trees-are-local} does not hold. Next we generalize Lemma \ref{lm:rule-trees-are-local} to the weighted case.

\begin{lemma} \label{lm:w-rule-trees-are-local} \rm Let $\cG$ be a $(\Gamma,\B)$-wcfg with rule set $R$. The following two statements hold.
\begin{compactenum}
\item[(1)] We can construct an $R$-local system $(K,H)$ such that $\wrtsem{\cG} =  \chi(\LL(K,H)) \otimes \wt_\cG$.
\item[(2)] We can construct an $(R,\B)$-wls $\cS$ with unit root weights such that $\wrtsem{\cG} = \sem{\cS}$.
    \end{compactenum}
\end{lemma}
          \begin{proof} Let $\cG = (N,S,R,wt)$ and $G$ be the $\Gamma$-cfg underlying $\cG$. We construct the $R$-local system $(K,H)$  as in Lemma \ref{lm:rule-trees-are-local} and then, for $(K,H)$ and $wt$, the $(R,\B)$-wls $\cS$ as in  Lemma \ref{lm:local+kappa->weighted-local}. Then we obtain          
            \begingroup
            \allowdisplaybreaks
            \begin{align*}
\wrtsem{\cG} &= \chi(\RT_\cG) \otimes \wt_\cG \tag{by definition and because $ \wt_\cG  \otimes \chi(\RT_\cG) = \chi(\RT_\cG) \otimes \wt_\cG$}\\
&= \chi(\RT_G) \otimes \wt_\cG  \\
&= \chi(\LL(K,H)) \otimes \wt_\cG  
\tag{by Lemma \ref{lm:rule-trees-are-local}}\\
&= \chi(\LL(K,H)) \otimes \h_{\M(R,wt)}
\tag{we recall that $\h_{\M(R,wt)} = \wt_\cG$ holds by our convention}\\
&= \sem{\cS} \enspace.
\tag{by Lemma \ref{lm:local+kappa->weighted-local}}
\end{align*}
\endgroup
          \end{proof}

It is known that each local tree language is recognizable \cite[Thm. 2.9.4]{gecste84}. The following result generalizes this to the weighted case (where \cite[Lm.~1]{ful15} requires that $\B$ is a semiring).

\begin{lemma}\label{lm:weighted-local->bu-recognizable}\rm \cite[Lm.~1]{ful15}
For each $(\Sigma,\B)$-wls $\cS$, we can construct a bu-deterministic $(\Sigma,\B)$-wta such that
$\sem{\cA}=\sem{\cS}$.
\end{lemma}
\begin{proof} Let $\cS=(g,F)$. We construct the $(\Sigma,\B)$-wta ${\cal A} = (Q,\delta,F_\cA)$
as follows:
\begin{compactitem}
\item $Q=\{ \overline{\sigma} \mid \sigma \in \Sigma\}$,
\item for every $k\in \mathbb{N}$, $\sigma_1\ldots \sigma_k \in \Sigma^k$, $\sigma \in \Sigma^{(k)}$, and $\omega\in \Sigma$,
\[
\delta_k(\overline{\sigma_1}\ldots \overline{\sigma_k},\sigma,\overline{\omega}) =
\left\{
\begin{array}{ll}
g_k(\sigma_1\ldots \sigma_k,\sigma) & \text{ if } \omega=\sigma \\
\0 & \text{ otherwise,}
\end{array}
\right.
\]
\item $F_\cA(\overline{\sigma})=F(\sigma)$ for every $\sigma \in \Sigma$.
\end{compactitem}
It is clear that $\cal A$ is bu-deterministic. Next, by induction on $\T_\Sigma$, we prove that the following statement holds: 
\begin{equation}\label{eq:hA=g}
\text{For every $\xi\in T_\Sigma$ and $\omega \in \Sigma$, we have  \ } \h_\cA(\xi)_{\overline{\omega}} =
\left\{
\begin{array}{ll}
g(\xi)& \text{ if } \omega=\xi(\varepsilon)\\
\0 & \text{ otherwise.}
\end{array}
\right.
\end{equation}
Let $\xi=\sigma(\xi_1,\ldots,\xi_k)$. Then
      \begingroup
    \allowdisplaybreaks
\begin{align*}
& \h_\cA(\sigma(\xi_1,\ldots,\xi_k))_{\overline{\omega}} \\[2mm]
  = & \bigoplus_{\overline{\sigma_1}\cdots\overline{\sigma_k} \in Q^k} \h_\cA(\xi_1)_{\overline{\sigma_1}}\otimes \ldots \otimes \h_\cA(\xi_k)_{\overline{\sigma_k}} \otimes
\delta_k(\overline{\sigma_1}\cdots\overline{\sigma_k} ,\sigma,\overline{\omega}) \\[2mm]
    = & \ \h_\cA(\xi_1)_{\overline{\xi_1(\varepsilon)}}\otimes \ldots \otimes \h_\cA(\xi_k)_{\overline{\xi_k(\varepsilon)}} \otimes
        \delta_k(\overline{\xi_1(\varepsilon)}\cdots\overline{\xi_k(\varepsilon)} ,\sigma,\overline{\omega})\\
  & \hspace*{35mm} \text{(because for each $\overline{\sigma_1}\cdots\overline{\sigma_k} \in Q^k$ with
$\overline{\sigma_1}\cdots\overline{\sigma_k} \ne \overline{\xi_1(\varepsilon)}\cdots\overline{\xi_k(\varepsilon)}$}\\
    & \hspace*{35mm} \text{there exists an $i \in [k]$ such that $\h_\cA(\xi_i)_{\overline{\sigma_i}}=\0$ by I.H.)} \\[2mm]
  = & \ g(\xi_1)\otimes\ldots \otimes g(\xi_k) \otimes
    \delta_k(\overline{\xi_1(\varepsilon)}\dots\overline{\xi_k(\varepsilon)} ,\sigma,\overline{\omega})  \tag{\text{by I.H.}}\\[2mm]
  = &  \begin{cases}
    g(\xi_1)\otimes\ldots \otimes g(\xi_k) \otimes 
g_k(\xi_1(\varepsilon)\dots\xi_k(\varepsilon) ,\sigma) & \text{if $\omega =\sigma$} \\
\0 & \text{otherwise}
\end{cases}
 \\[2mm]
  =  & \begin{cases}
    g(\sigma(\xi_1,\ldots,\xi_k) )  & \text{if $\omega =\sigma$}\\
\0 & \text{otherwise.}
\end{cases}
\end{align*}
\endgroup
This proves \eqref{eq:hA=g}.

Finally, let $\xi\in T_\Sigma$. Then we get
\begin{align*}
\sem{{\cal A}}(\xi) = \bigoplus_{\overline{\omega} \in Q} \h_\cA(\xi)_{\overline{\omega}} \otimes F_\cA(\overline{\omega}) = 
g(\xi)\otimes F_\cA(\overline{\xi(\varepsilon)}) = g(\xi)\otimes F(\xi(\varepsilon)) = \sem{{\cal S}}(\xi),
\end{align*}
where the second equality follows from \eqref{eq:hA=g} and the other ones from the corresponding definitions.
\end{proof}

For example, if we apply the construction in Lemma \ref{lm:weighted-local->bu-recognizable} to the $(\Sigma,\Natmaxplus)$-wls ${\cal S}$ of Example \ref{ex:number-of-occurrences-by-weighted-local}, then we obtain the bu-deterministic $(\Sigma,\Natmaxplus)$-wta $\cA$ of Example \ref{ex:number-of-occurrences-arctic}.

Next we verify that Lemma \ref{lm:weighted-local->bu-recognizable} is a generalization of \cite[Thm. 2.9.4]{gecste84}. We achieve this by proving that the latter result is equivalent to Lemma \ref{lm:weighted-local->bu-recognizable} for the case that $\B$ is the semiring $\Boole$.

\begin{corollary}\rm\label{cor:loc-is-recog} For each $\Sigma$-local system $(K,H)$, we can construct a bu-deterministic $\Sigma$-fta $A$ such that $\LL(K,H)=\LL(A)$. Thus, in particular, each local tree language is recognizable.
\end{corollary}
\begin{proof} Let $(K,H)$ be a $\Sigma$-local system. 
  We construct the $(\Sigma,\Boole)$-wls $\cS=(g,F)$ by letting $g_k = \chi(K \cap \Fork(\Sigma)^{(k)})$ for each $k \in \mathbb{N}$, and $F=\chi(H)$. Then $(K,H) = \supp(\cS)$ and, by Lemma \ref{lm:supp-of-Boolean-weighted-local}, we have $\LL(K,H) = \supp(\sem{\cS})$. By Lemma \ref{lm:weighted-local->bu-recognizable}, we can construct a bu-deterministic $(\Sigma,\Boole)$-wta $\cA$ such that $\LL(K,H) = \supp(\sem{\cA})$. By Corollary \ref{cor:supp-B=fta-1}, we can construct a $\Sigma$-fta $A$ such that $\LL(K,H) = \LL(A)$.
  \end{proof}

We can easily demonstrate that bu-deterministic wta are more powerful than weighted local systems.
For instance, let $\Sigma=\{\gamma^{(1)},\alpha^{(0)} \}$ and consider the $(\Sigma,\Boole)$-weighted 
tree language $r$ defined by $r\big(\gamma(\gamma(\alpha))\big)=1$ and $r(\xi)=0$ for every other $\xi\in \T_\Sigma$. It is easy to show that $r \in \budRec(\Sigma,\Boole)$ and  there does not exist a  $(\Sigma,\Boole)$-wls $\cS$ such that $\sem{\cS}=r$.
To see the latter, we assume that there exists a  $(\Sigma,\Boole)$-wls $\cS=(g,F)$ such that $\sem{\cS}=r$.
Thus, using $\xi$ as abbreviation for $\gamma(\gamma(\alpha))$, we have  $\sem{\cS}(\xi) = g(\xi) \wedge F(\gamma) = 1$. Hence $g(\xi)= F(\gamma)= 1$, and thus, in particular, $g_1(\gamma,\gamma) = g_1(\alpha,\gamma) = g_0(\varepsilon,\alpha) = 1$. But this means that, for each $n \in \mathbb{N}_+$, we have $g(\gamma^n(\alpha))= 1$ and hence $\sem{\cS}(\gamma^n(\alpha)) = 1$. This is a contradiction to $\sem{\cS}=r$.

It is also known that each recognizable tree language is the image of a rule tree language of some context-free grammar under a deterministic tree relabeling (cf. \cite[Prop.~2]{tha67}, \cite[Thm.~3.58]{eng75-15}). The next lemma generalizes this to the weighted case.

\begin{lemma}\rm \label{lm:wta-proj-der-tree} Let $\cA$ be a $(\Sigma,\B)$-wta. We can construct a   $(\Sigma,\B)$-wrtg $\cG$ in tree automata form  with rule set $R$ and a deterministic $(R,\Sigma)$-tree relabeling $\tau$  such that  $\runsem{\cA} = \chi(\tau)\big(\wrtsem{\cG}\big)$.
\end{lemma}
\begin{proof}By Lemma \ref{lm:wta-to-wrtg} we can construct a $(\Sigma,\B)$-wrtg $\cG$ in tree automata form such that $\runsem{\cA} = \sem{\cG}$.
Let $R$ be the set of rules of $\cG$. As argued in Section \ref{sec:grammar-model}, the projection $\pi_\cG: \T_R \to \T_\Sigma$ is determined by the $(R,\Sigma)$-tree homomorphism $\pi_\cG = ((\pi_\cG)_k \mid k \in \mathbb{N})$ defined, for every $k \in \mathbb{N}$ and $r \in R^{(k)}$ of the form $r = (A \rightarrow \sigma(A_1,\ldots,A_k))$, by $(\pi_\cG)_k(r) = \sigma(z_1,\ldots,z_k)$ (note that $\cG$ is in tree automata form). 
We construct the deterministic $(R,\Sigma)$-tree relabeling $\tau = (\tau_k \mid k \in \mathbb{N})$ such that, for each rule $A \rightarrow \sigma(A_1,\ldots,A_k)$, we let
\[\tau_k(A \rightarrow \sigma(A_1,\ldots,A_k)) = \sigma \enspace.\]
Obviously, the mappings  $\pi_\cG: \T_R \to \T_\Sigma$ and $\tau: \T_R \to \T_\Sigma$ are equal. Then, by definition, we obtain $\sem{\cG} = \chi(\pi_\cG)(\wrtsem{\cG}) = \chi(\tau)(\wrtsem{\cG})$.
\end{proof}

Now we can prove the first characterization theorem.

\begin{theorem-rect}\label{thm:decomposition-2} Let $\Sigma$ be a ranked alphabet. Moreover, let  $\B$ be a strong bimonoid and let $r: \T_\Sigma \to B$. Then the following three statements are equivalent.
  \begin{compactenum}
  \item[(A)] We can construct a $(\Sigma,\B)$-wta  $\cA$ such that $r=\runsem{\cA}$.
  \item[(B)]  We can construct
    \begin{compactitem}
    \item a ranked alphabet $R$,
    \item a deterministic $(R,\Sigma)$-tree relabeling $\tau$, and
    \item a $(\Sigma,\B)$-wrtg $\cG$ in tree automata form with rule set $R$
    \end{compactitem}
    such that  $r = \chi(\tau)\big(\wrtsem{\cG}\big)$.
  \item[(C)] We can construct
    \begin{compactitem}
    \item a ranked alphabet $R$,
    \item a deterministic $(R,\Sigma)$-tree relabeling $\tau$, and
    \item an $(R,\B)$-wls $\cS$ with unit root weights
    \end{compactitem}
    such that $r= \chi(\tau)\big(\sem{\cS}\big)$.
    \end{compactenum}
  \end{theorem-rect}
 
  \begin{proof} Proof of (A)$\Rightarrow$(B): It follows from Lemma \ref{lm:wta-proj-der-tree}.

    \
    
    Proof of (B)$\Rightarrow$(C): It follows from the fact that each $(\Sigma,\B)$-wrtg is a $(\Sigma^\Xi,\B)$-wcfg and by Lemma~\ref{lm:w-rule-trees-are-local}(2).

    \
    
     Proof of (C)$\Rightarrow$(A): It follows from Lemma \ref{lm:weighted-local->bu-recognizable} and Theorem  \ref{thm:closure-under-tree-relabeling} (closure of $\Rec^{\mathrm{run}}(\_,\B)$ under tree relabelings).
\end{proof}

Next we verify that Theorem \ref{thm:decomposition-2} (A)$\Rightarrow$(C)  generalizes  \cite[Thm. 2.9.5]{gecste84}. We achieve this by proving that the latter result is equivalent to Theorem \ref{thm:decomposition-2} (A)$\Rightarrow$(C) for the case that $\B$ is the  semiring $\Boole$.

\begin{corollary}\rm \label{cor:rec=tau(loc)} Let  $L \subseteq \T_\Sigma$ be recognizable.
 We can construct a ranked alphabet $R$, a deterministic $(R,\Sigma)$-tree relabeling $\tau$, and an $R$-local system $(K,H)$ such that $L= \tau(\LL(K,H))$.
  \end{corollary}
  \begin{proof} Let $A$ be a $\Sigma$-fta such that $L=\LL(A)$. By Corollary \ref{cor:supp-B=fta-1} (A)$\Rightarrow$(B), we can construct a $(\Sigma,\Boole)$-wta $\cA$ such that $L = \supp(\sem{\cA})$.
By Theorem \ref{thm:decomposition-2} (A)$\Rightarrow$(C), we can construct  a ranked alphabet $R$, a deterministic $(R,\Sigma)$-tree relabeling $\tau$, and an $(R,\Boole)$-wls $\cS$ (which has unit root weights by definition) such that $L = \supp(\chi(\tau)\big(\sem{\cS}\big))$. By \eqref{equ:supp-yield=yieldA}, we obtain that $L = \tau(\supp(\sem{\cS}))$. By 
Theorem \ref{thm:loc=wloc(B)}(B)$\Rightarrow$(A), we can construct a $\Sigma$-local system $(K,H)$ such that $L=\tau(\LL(K,H))$.  
  \end{proof}

  If one traces back the constructions involved in the proof of Theorem \ref{thm:decomposition-2}(A)$\Rightarrow$(C) and composes them together, then one can eventually find out how a given wta is decomposed into a wls and a deterministic tree relabeling. Here we compose these constructions and show the resulting direct construction  (where we start  from a wta with unit root weights).

  \begin{construction}\rm \label{const:decomp-wta}  Let $\cA=(Q,\delta,F)$ be a $(\Sigma,\B)$-wta with unit root weights. Then we can construct a ranked alphabet $R$, a deterministic $(R,\Sigma)$-tree relabeling $\tau$, and an $(R,\B)$-wls $\cS$ such that $\runsem{\cA} = \chi(\tau)(\sem{\cS})$ as follows. 
    \begin{compactitem}
    \item For each $k \in \mathbb{N}$, we let $R^{(k)} = \{q \to \sigma(q_1,\ldots,q_k) \mid \sigma \in \Sigma^{(k)}, q,q_1,\ldots,q_k \in Q\}$,
    \item $\tau = (\tau_k \mid k \in \mathbb{N})$ such that, for every $k \in \mathbb{N}$ and $q \to \sigma(q_1,\ldots,q_k)$ in $R^{(k)}$, we let \\$\tau_k(q \to \sigma(q_1,\ldots,q_k))=\sigma$, and
    \item $\cS=(g,F')$ such that $g = (g_k \mid k \in \mathbb{N})$  and for every $k \in \mathbb{N}$, $r_1,\ldots,r_k$ in $R$, and $r \in R^{(k)}$, we let
      \[
        g_k(r_1 \cdots r_k, r) =
        \begin{cases}
          \delta_k(q_1 \cdots q_k,\sigma,q) & \text{ if $r=(q \to \sigma(q_1,\ldots,q_k))$ and }\\
            & \text{ for each $i \in [k]$ we have $\lhs(r_i)=q_i$}\\
          \0 & \text{ otherwise}
          \end{cases}
        \]
        and, for each $r$ in $R$, we let $F'(r) = F(\lhs(r))$.
      \end{compactitem}
    \end{construction}

    In fact, we could even give an arbitrary $(\Sigma,\B)$-wta as input to Construction \ref{const:decomp-wta} (i.e., which does not necessarily have unit root weights) and still we would obtain that $\runsem{\cA} = \chi(\tau)(\sem{\cS})$.


\section[Characterization by local systems]{Characterization by local systems and evaluation algebras}
\label{sec:char-loc-eval-alg}

In the next theorem we prove the second characterization result for recognizable weighted tree languages. In particular, (B) and (C) only differ in the way in which the $\Theta$-tree language is defined: in (B) it is generated by  a $\Theta$-local system and in (C) it is recognized by a bu-deterministic $\Theta$-fta.

\begin{theorem-rect}\label{thm:decomposition-1} Let $\Sigma$ be a ranked alphabet. Moreover, let $\B$ be a strong bimonoid and let $r: \T_\Sigma \to B$. Then the following three statements are equivalent.
\begin{compactenum}
  \item[(A)] We can construct a $(\Sigma,\B)$-wta $\cA$ such that $r=\runsem{\cA}$.
   \item[(B)] We can construct
    \begin{compactitem}
    \item a ranked alphabet $\Theta$,
    \item a deterministic $(\Theta,\Sigma)$-tree relabeling $\tau$,
    \item a $\Theta$-local system $(K,H)$, and
    \item a family $\kappa=(\kappa_k \mid k\in\mathbb{N})$ of mappings $\kappa_k:~\Theta^{(k)}~\to~B$ \end{compactitem}
    such that,  for each $\xi \in \T_\Sigma$, the following holds: $r(\xi) = \h_{\M(\Theta,\kappa)}(\tau^{-1}(\xi) \cap \LL(K,H))$. 
    \item[(C)] We can construct
    \begin{compactitem}
    \item a ranked alphabet $\Theta$,
    \item a deterministic $(\Theta,\Sigma)$-tree relabeling $\tau$,
    \item a bu-deterministic $\Theta$-fta $A$, and
    \item a family $\kappa=(\kappa_k \mid k\in\mathbb{N})$ of mappings $\kappa_k:~\Theta^{(k)}~\to~B$
    \end{compactitem}
    such that,  for each $\xi \in \T_\Sigma$, the following holds:  $r(\xi) = \h_{\M(\Theta,\kappa)}(\tau^{-1}(\xi) \cap \LL(A))$.
\end{compactenum}
  \end{theorem-rect}

  \begin{proof}
Proof of (A)$\Rightarrow$(B): 
  Let $\cA$ be a $(\Sigma,\B)$-wta. By Theorem \ref{thm:decomposition-2} (A)$\Rightarrow$(B), we construct a ranked alphabet $R$, a deterministic $(R,\Sigma)$-tree relabeling $\tau$, and a $(\Sigma,\B)$-wrtg $\cG$ in tree automata form with rule set $R$ such that $\runsem{\cA} = \chi(\tau)(\wrtsem{\cG})$.
Let $wt$ be the weight assignment of $\cG$.
    
Then, as in  Lemma \ref{lm:w-rule-trees-are-local}(1), we can construct an $R$-local system $(K,H)$ such that
\[\runsem{\cA} = \chi(\tau)(\wrtsem{\cG}) = \chi(\tau)(\chi(\LL(K,H)) \otimes \h_{\M(R,wt)}) \enspace.
  \]

  Let $\xi \in \T_\Sigma$. By applying Observation \ref{obs:inside-out} (choosing $\Sigma$, $\Delta$, $L$, and $r$ to be $R$, $\Sigma$, $\LL(K,H)$, and $\h_{\M(\Sigma,wt)}$, respectively), we obtain that
  \[ \runsem{\cA}(\xi) = \h_{\M(R,wt)}(\tau^{-1}(\xi) \cap \LL(K,H)) \enspace.
  \]
  By choosing $\Theta$ and $\kappa$ to be $R$ and $(wt|_{R^{(k)}} \mid k \in \mathbb{N})$, respectively, we have proved (B).

  \
       
Proof of (B)$\Rightarrow$(C): This follows from Corollary \ref{cor:loc-is-recog}, which states that, for each $\Theta$-local system $(K,H)$, a bu-deterministic $\Theta$-fta $A$ can be constructed such that $\LL(A)=\LL(K,H)$.

\
  
Proof of (C)$\Rightarrow$(A):  Let $\Theta$ be a ranked alphabet, $\tau=(\tau_k \mid k \in \mathbb{N})$ be a deterministic $(\Theta,\Sigma)$-tree relabeling, $A = (Q,\delta,F)$ be a bu-deterministic $\Theta$-fta, and a family $\kappa=(\kappa_k \mid k \in \mathbb{N})$ of mappings $\kappa_k: \Theta^{(k)} \to B$. By Theorem \ref{thm:fta-wta}, we can assume that $A$ is total and bu-deterministic. Then, for each $\zeta \in \LL(A)$, we denote by $\rho_\zeta$ the unique valid run of $A$ on $\zeta$. We will abbreviate $\h_{\M(\Theta,\kappa)}$ by $\h$.

    We construct the $(\Sigma,\B)$-wta $\cA$ by coding the preimage of $\tau$ into the states of $\cA$. 
We let   $\cA=(Q',\delta',F')$  with $Q' = Q \times \Theta$ and, for each $(q,\theta) \in Q'$ we let  $F'_{(q,\theta)} = \1$ if $q \in F$, and $F'_{(q,\theta)} = \0$ otherwise. Moreover, for every $k \in \mathbb{N}$, $\sigma \in \Sigma^{(k)}$, and $(q_1,\theta_1),\ldots,(q_k,\theta_k), (q,\theta) \in Q'$ we define
    \[
      (\delta')_k((q_1,\theta_1) \cdots (q_k,\theta_k),\sigma,(q,\theta)) =
      \begin{cases}
        \kappa_k(\theta) & \text{if $(q_1 \cdots q_k, \theta,q) \in \delta_k$ and $\tau_k(\theta)=\sigma$}\\
        \0 & \text{ otherwise.}
        \end{cases} 
      \]

      We let $\R_\cA = \bigcup_{\xi \in \T_\Sigma} \R_\cA(\xi)$, and we define the mapping $\bar{\tau}: \R_\cA \to \T_\Sigma$ such that, for every $\xi \in \T_\Sigma$ and $\rho \in \R_\cA(\xi)$, we let $\bar{\tau}(\rho)$ be the $\Sigma$-tree mapping
      \[
        t: \pos(\xi) \to \Sigma  \text{  with } \ t(w) = \tau_k(\rho(w)_2) \ \text{ for each $w \in \pos(\xi)$ where $k = \rk_\Theta(\rho(w)_2)$} \enspace.
        \]
The reader might wonder why we did not simply define $\bar{\tau}(\rho) = \xi$, because $\rho \in \R_\cA(\xi)$. But this is not well defined because it is possible that there exist $\xi_1,\xi_2 \in \T_\Sigma$ with $\rho \in \R_\cA(\xi_1) \cap \R_\cA(\xi_2)$. We also note that $t$ uniquely determines an element in $\T_\Sigma$ (cf. Section \ref{sect:trees}). 
Moreover, there may exist $\xi \in \T_\Sigma$ and $\rho \in \R_\cA(\xi)$ such that $\bar{\tau}(\rho)\ne \xi$; however, the weight of such a run $\rho$ is $\0$.
Formally,
\begin{equation}
\text{for each $\rho \in \R_\cA(\xi)$: if $\bar{\tau}(\rho) \ne \xi$, then $\wt(\xi,\rho) = \0$}\enspace.\label{eq:coding=0}
\end{equation}

For each $\xi \in \T_\Sigma$, we abbreviate by $\R_\cA(F\times \Theta,\xi)$ the set $\bigcup_{q \in F, \theta \in \Theta}\R_\cA((q,\theta),\xi)$ and by $\R_\cA(F\times \Theta)$ the set $\bigcup_{ \xi \in \T_\Sigma}\R_\cA(F\times \Theta,\xi)$.      
      Then we define the mapping
      \[
\varphi:  \LL(A) \to  \R_\cA(F \times \Theta) 
\]
for each $\zeta \in  \LL(A)$ and $w \in \pos(\xi)$ by $\varphi(\zeta)(w) = (\rho_\zeta(w),\zeta(w))$. (We recall that $\rho_\zeta$ is the unique valid run of $A$ on $\zeta \in \LL(A)$.) It is easy to see that $\varphi$ is a bijection. Moreover, $\varphi(\zeta) \in \R_\cA(F\times \Theta,\tau(\zeta))$.

By induction on $\T_\Theta$, we  prove that the following statement holds:
      \begin{equation}
\text{For every $\zeta \in \LL(A)$, we have } \ \h(\zeta) = \wt(\tau(\zeta),\varphi(\zeta))   \label{eq:unique-run-A}\enspace,
\end{equation}
cf. Figure \ref{fig:composition-wta},
where $\R_A^{\mathrm{v}} (q,\zeta)$ denotes the set of valid $q$-runs on $\zeta$.

Let $\zeta = \theta(\zeta_1,\ldots,\zeta_k)$. Then
\begingroup
\allowdisplaybreaks
\begin{align*}
   \h(\theta(\zeta_1,\ldots,\zeta_k))
  &= \h(\zeta_1) \otimes \ldots \otimes \h(\zeta_k) \otimes \kappa_k(\theta)\\
  &= \Big(\bigotimes_{i \in [k]} \wt(\tau(\zeta_i),\varphi(\zeta_i)) \Big) \otimes \kappa_k(\theta)
  \tag{by I.H.}\\
  &= \Big(\bigotimes_{i \in [k]} \wt(\tau(\zeta_i), \varphi(\zeta_i)) \Big) \otimes (\delta')_k((\rho_{\zeta_1}(\varepsilon),\zeta_1(\varepsilon)) \cdots (\rho_{\zeta_k}(\varepsilon),\zeta_k(\varepsilon)), \tau_k(\theta), (\rho_{\zeta}(\varepsilon),\zeta(\varepsilon)))
  \tag{by definition of $\delta'$}\\
  &= \Big(\bigotimes_{i \in [k]} \wt(\tau(\zeta_i), \varphi(\zeta)|_i) \Big) \otimes (\delta')_k(\varphi(\zeta_1)(\varepsilon) \cdots \varphi(\zeta_k)(\varepsilon), \tau_k(\theta), \varphi(\zeta)(\varepsilon))
    \tag{by definition of $\varphi$}\\
  &= \Big(\bigotimes_{i \in [k]} \wt(\tau(\zeta_i), \varphi(\zeta)|_i) \Big) \otimes (\delta')_k(\varphi(\zeta)(1) \cdots \varphi(\zeta)(k), \tau_k(\theta), \varphi(\zeta)(\varepsilon)) \\
  &=  \wt(\tau_k(\theta)(\tau(\zeta_1),\ldots,\tau(\zeta_k)), \varphi(\zeta))
  \tag{by definition of $\wt$}\\
  &= \wt(\tau(\theta(\zeta_1,\ldots,\zeta_k)), \varphi(\zeta)) \enspace.
  \end{align*}
\endgroup

       \begin{figure}
      \centering
     
\begin{tikzpicture}[level distance=2.75em,
  every node/.style = {align=center},scale=0.9]
  \pgfdeclarelayer{bg}    
  \pgfsetlayers{bg,main}  

  \newcommand{\mydista}{2.1mm} 
  \newcommand{\mydistaa}{2.5mm} 
  \newcommand{\mydistb}{0.8mm} 
  \tikzstyle{mycircle}=[draw, circle, inner sep=-2mm, minimum height=5mm]

\begin{scope}[xshift=43mm,level 1/.style={sibling distance=22mm}]

 \node at (0, 0.8) {$\zeta\in\tau^{-1}(\xi)\cap \LL(A):$};
 \node at (2.5,0) {$\rho_\zeta \in \R_A^{\mathrm{v}} (q,\zeta)$};

 \node (N0) {$\theta_1$}
  child {node (N1) {$\theta_2$} 
  	   child { node (N2) {$\theta_3$}} }
  child {node (N3) {$\theta_4$} };
  
 \node [mycircle, anchor=west] at ([xshift=\mydistb]N0.east) {$q$};
 \node [mycircle, anchor=west] at ([xshift=\mydistb]N1.east) {$q_2$};
 \node [mycircle, anchor=west] at ([xshift=\mydistb]N2.east) {$q_3$};
 \node [mycircle, anchor=west] at ([xshift=\mydistb]N3.east) {$q_4$};

\end{scope}

\begin{scope}[yshift=-35mm, level 1/.style={sibling distance=15mm}]

 \node at (-1.5,0) {$\xi:$};

 \node (N0) {$\sigma$}
  child {node (N1) {$\gamma$} 
  	   child { node (N2) {$\alpha$}} }
  child {node (N3) {$\beta$} };
  
\end{scope}

\begin{scope}[xshift=105mm, yshift=-35mm]

 \node {$\h(\zeta)=\wt(\xi,\varphi(\zeta))$};
\end{scope}

\begin{scope}[xshift=38mm,yshift=-64mm, level 1/.style={sibling distance=27mm}]

 \node at (2.7,0.8){$\rho\in \R_\cA((q,\theta_1),\xi)$};

 \node (N0) {$\sigma$}
  child {node (N1) {$\gamma$} 
  	   child { node (N2) {$\alpha$}} }
  child {node (N3) {$\beta$} };
 
 \node [draw, rounded rectangle, anchor=west] at ([xshift=\mydistaa]N0.east) {$q,\theta_1$};
 \node [draw, rounded rectangle, anchor=west] at ([xshift=\mydistb]N1.east) {$q_2,\theta_2$};
 \node [draw, rounded rectangle, anchor=west] at ([xshift=\mydistb]N2.east) {$q_3,\theta_3$};
 \node [draw, rounded rectangle, anchor=west] at ([xshift=\mydistb]N3.east) {$q_4,\theta_4$};
\end{scope}

\draw [->, thick, xshift=0.5cm, yshift=-2.85cm] (2,1) -- (0,0) node[midway, above left] {$\tau$};
\draw [->, thick, xshift=3.8cm, yshift=-2.85cm] (0,0) -- (0,-2.9) node[pos=0.3,right] {$\varphi$};
\draw [->, thick, xshift=7cm, yshift=-2.85cm] (-0.4,1.3) -- (1.6,0) node[midway, above right] {$\h$};

\draw [->, thick, xshift=1.53cm, yshift=-5cm] (4,-0.2) -- (0,0.5) node[pos=0.2,above right] {$\overline{\tau}$};
\draw [->, thick, xshift=6.25cm, yshift=-5cm] (1.2,-0.2) -- (2.3,1) node[pos=0.5,above left] {$\wt$};
\end{tikzpicture}

\caption{\label{fig:composition-wta} An illustration of \eqref{eq:unique-run-A}: $\h(\zeta) = \wt(\tau(\zeta),\varphi(\zeta))$, where $\tau(\zeta) = \xi$.}
\end{figure}

 This proves \eqref{eq:unique-run-A}.  Next we prove that $\runsem{\cA}(\xi) = \h(\tau^{-1}(\xi) \cap \LL(A))$ for each $\xi \in \T_\Sigma$. Let $\xi \in \T_\Sigma$.
          \begingroup
    \allowdisplaybreaks
    \begin{align*}
      \h(\tau^{-1}(\xi) \cap \LL(A)) &= \bigoplus_{\zeta \in \tau^{-1}(\xi) \cap \LL(A)} \h(\zeta) \tag{\text{by definition}}\\
      &= \bigoplus_{\zeta \in \tau^{-1}(\xi) \cap \LL(A)} \wt(\xi,\varphi(\zeta)) \tag{\text{by \eqref{eq:unique-run-A}}}\\
                                            &= \bigoplus_{q \in F,\theta \in \Theta}\ \  \bigoplus_{\substack{\rho \in \R_\cA((q,\theta),\xi):\\\bar{\tau}(\rho)=\xi}} \wt(\xi,\rho)  \tag{\text{because $\varphi$ is bijective}}\\
      &= \bigoplus_{q \in F,\theta \in \Theta}\ \  \bigoplus_{\rho \in \R_\cA((q,\theta),\xi)} \wt(\xi,\rho)  \tag{\text{by \eqref{eq:coding=0}}}
      \\
      &= \bigoplus_{\rho \in \R_\cA(\xi)} \wt(\xi,\rho) \otimes F'_{\rho(\varepsilon)}
      \tag{by definition of $F'$}\\
      &= \runsem{\cA}(\xi) \qedhere
    \end{align*} 
    \endgroup
    \end{proof}

   As we have seen, the proof of Theorem \ref{thm:decomposition-1}(A)$\Rightarrow$(B) is based on Theorem \ref{thm:decomposition-2} (A)$\Rightarrow$(B). Since Observation \ref{obs:inside-out} is symmetric, one might wonder whether the proof of Theorem \ref{thm:decomposition-2} (A)$\Rightarrow$(B) could be based on Theorem \ref{thm:decomposition-1}(A)$\Rightarrow$(B). However, this is not possible because there exist local tree languages which are not rule tree languages, see our remark after Lemma \ref{lm:rule-trees-are-local}.

%% file: crisp-determinization.tex
\chapter{Crisp-determinization}
\label{ch:crisp-determinization}

We recall that, in a crisp-deterministic $(\Sigma,\B)$-wta $\cB$, the weight of each transition is one of the unit weights $\0$ and $\1$ of the strong bimonoid $\B$; weights different from these units may only occur as root weights. Also, since crisp-determinism implies bu-determinism, the two semantics of $\cB$ coincide, i.e., $\initialsem{\cB} = \runsem{\cB}$ (cf. Theorem \ref{thm:bu-det:init=run}). We denoted the semantics of $\cB$ by $\sem{\cB}$.

In this chapter we will investigate the question under which conditions a given wta can be transformed into an equivalent crisp-deterministic wta. More precisely,
\index{initial algebra crisp-determinizable}
\index{run crisp-determinizable}
\index{crisp-determinizable}
a $(\Sigma,\B)$-wta $\cA$ is \emph{initial algebra crisp-determinizable} if there exists an i-equivalent crisp-deterministic $(\Sigma,\B)$-wta, i.e., there exists a crisp-deterministic $(\Sigma,\B)$-wta $\cB$ such that $\initialsem{\cA} = \sem{\cB}$. It is \emph{run crisp-determinizable} if there exists an r-equivalent crisp-deterministic $(\Sigma,\B)$-wta, i.e.,  there exists a crisp-deterministic $(\Sigma,\B)$-wta $\cB$ such that $\runsem{\cA} = \sem{\cB}$. 
It follows from the previous definitions that, if $\B$ is semiring, then $\cA$ is initial algebra crisp-determinizable if and only if it is run crisp-determinizable (because  by Corollary \ref{cor:semiring-run=init} $\initialsem{\cA}=\runsem{\cA}$).

In the sequel we will identify conditions under which a given $(\Sigma,\B)$-wta $\cA$ is initial algebra crisp-determinizable and under which conditions it is run crisp-determinizable. 
We approach the answers by recalling that each  crisp-deterministic $(\Sigma,\B)$-wta $\cB$ recognizes a $(\Sigma,\B)$-recognizable step mapping, and vice versa, each $(\Sigma,\B)$-recognizable step mapping can be obtained in this way (cf. Theorem \ref{thm:crisp-det-algebra}). Hence, $\cA$ is initial algebra crisp-determinizable if and only if $\initialsem{\cA}$ is a recognizable step mapping, i.e., $\im(\initialsem{\cA})$ is finite and $\initialsem{\cA}$ has the preimage property (and similarly for run crisp-determinizable).

Thus, Theorem \ref{thm:crisp-det-algebra} provides a blue-print for the answers to the crisp-determinization problems. On the negative side, if we consider a $(\Sigma,\B)$-wta $\cA$ for which $\initialsem{\cA}$ is not a recognizable step mapping, then $\cA$ is not initial algebra crisp-determinizable, and similarly for $\runsem{\cA}$ and run crisp-determinizability. For instance, let $\cA$ be the $(\Sigma,\Natmaxplus)$-wta from Example \ref{ex:height} for which  $\initialsem{\cA} = \height$. Since $\im(\height)$ is not finite,  $\initialsem{\cA}$ is not a recognizable step mapping (by definition). Hence, by Theorem  \ref{thm:crisp-det-algebra} and the fact that $\Natmaxplus$ is a semiring, $\cA$ is neither initial algebra crisp-determinizable nor run crisp-determinizable.
On the positive side, each $(\Sigma,\B)$-wta $\cA$ for which $\initialsem{\cA}$ is a recognizable step mapping, is initial algebra crisp-determinizable, and similarly for the run semantics and run crisp-determinizability.

In Section \ref{sec:i-crisp-determinization} we will deal with  initial algebra crisp-determinization and in Section \ref{sec:r-crisp-determinization} with run crisp-determinization.
This chapter is based on results and constructions in \cite{fulkosvog19}, \cite{drofulkosvog20b}, and \cite{drofulkosvog21}.

\section{Initial algebra crisp-determinization}
\label{sec:i-crisp-determinization}

In this section we will show an equivalence between local finiteness and initial algebra crisp-determinizability (cf. Theorem \ref{thm:loc-finite-rec-step-function}).

If we recall the way in which a $(\Sigma,\B)$-wta $\cA$ computes $\h_\cA(\xi)$ for some input tree $\xi \in \T_\Sigma$, then we see that the summation and multiplication of $\B$ are used in an alternating way. One possibility to guarantee that these computations only produce finitely many values in $B$, is to assume that $\B$ is locally finite. Indeed, for each locally finite strong bimonoid $\B$, the set $\Rec^{\mathrm{init}}(\Sigma,\B)$ contains only recognizable step mappings (cf.  \cite[Thm.~9]{bormalsestepvog06}, \cite[Lm.~6.1]{drovog06},  \cite[Cor.~3.16]{fulvog09new},  \cite[p.~9]{cirdroignvog10}, \cite[Lm.~18]{drostuvog10}, and \cite[Cor.~6.10]{fulkosvog19}). Since each distributive bounded lattice is a particular locally finite semiring (cf. Subsection \ref{sec:comparison}), the next lemma can also be compared to \cite[Thm.~3.8]{rah09}.

\begin{lemma}\rm \label{lm:loc-finite-weakly-loc-fin-rec-step-function} Let (a) $\B$ be locally finite or (b)~let $\B$ be weakly locally finite and $\Sigma$ is monadic.  Moreover, let $\cA$ be a $(\Sigma,\B)$-wta. Then $\im(\h_\cA)$ is finite and $\cA$ is initial algebra crisp-determinizable.
\end{lemma}
\begin{proof}   Let $\cA = (Q,\delta,F)$.  We recall that $\im(\delta)=\bigcup_{k\in \mathbb{N}}\im(\delta_k)$.

  First we show that the set \(H=\{\h_\cA(\xi)_q \mid \xi \in \T_\Sigma, q \in Q\}\)
  is finite.

  \underline{Case (a):} Let  $\B$ be locally finite.  Since $H\subseteq \langle\im(\delta)\rangle_{\{\oplus,\otimes\}}$ and $ \langle\im(\delta)\rangle_{\{\oplus,\otimes\}}$ is finite, the set $H$ is finite.

  \underline{Case (b):} Let $\B$ be weakly locally finite and $\Sigma$ be monadic. 
Since $\Sigma$ is monadic, for every $\xi \in \T_\Sigma$ and $q \in Q$, we have $\h_\cA(\xi)_q \in \CL(\im(\delta))$,
by \eqref{eq:hom-in-closure} in the proof of Lemma \ref{lm:image-monadic-ranked alphabet}.  This means that $H\subseteq \CL(\im(\delta))$, and since $\B$ is weakly locally finite, we obtain that $H$ is finite.

Hence, in both Cases (a) and (b), the set $H$ is finite. Then the set $\im(\h_\cA)$ is also finite because  $\h_\cA(\xi)\in H^Q$ for every $\xi\in \T_\Sigma$. Hence the  congruence relation $\ker(\h_\cA)$ of $\cA$ has finite index. Thus, by Theorem~\ref{thm:crisp-wta-final-index}(C)$\Rightarrow$(B),
the $(\Sigma,\B)$-wta $\cA$ is initial algebra crisp-determinizable. 
\end{proof}

Lemma \ref{lm:loc-finite-weakly-loc-fin-rec-step-function} states that each $(\Sigma,\B)$-wta is initial algebra crisp-determinizable if one of its conditions (a) and (b) holds. In fact, if one traces back the proofs of Lemma \ref{lm:loc-finite-weakly-loc-fin-rec-step-function}, Theorem \ref{thm:crisp-wta-final-index}, and Lemma \ref{lm:fin-algebra-wta}, then one can construct the corresponding crisp-deterministic wta. Since its construction is rather hidden, we want to show this explicitly here (cf. Definition \ref{constr:subset}).

In fact, the construction of the corresponding crisp-deterministic wta generalizes the well-known subset method which transforms an fta into an equivalent bu-deterministic fta (cf. \cite[Thm.~1]{thawri68}, \cite[Thm.~3.8]{eng75-15}, and \cite[Thm.~2.2.6]{gecste84}).  The set of states of the bu-deterministic fta is the set of all subsets of the state set of the given fta.  Here, for a given wta $\cA = (Q,\delta,F)$, we represent the states of the equivalent bu-deterministic wta by vectors in $B^Q$. However, we do not use all vectors in $B^Q$, but only those which are images of trees under $\h_{\cA}$. This corresponds to using the reachable subsets in the subset method. We generalize the subset method for any wta and strong bimonoid $\B$, and
we keep the name ``subset method'' also for the generalized version.

A special case of this subset method was given in the proof of \cite[Thm.~2.1]{bel02} for the case of string ranked alphabets (i.e., weighted automata) and complete distributive lattices. (We note that distributivity is not assumed in \cite{bel02},  but it is  needed because the many-fold composition of matrices in the definition of the acceptance degree, i.e., run semantics, assumes associativity of matrix multiplication, and in its turn, this assumes distributivity \cite{bel08}). In fact, the notion of deterministic fuzzy automata in \cite{bel02}  corresponds to our notion of crisp-deterministic wta over some string ranked alphabet and some complete distributive lattice due to Lemma \ref{lm:wsa=wta-over-string-ra} and Theorem \ref{thm:crisp-wta-final-index}(A)$\Leftrightarrow$(B).
 In \cite[Thm.~5 and 8]{bozlou10} the subset method was presented for  wta over $\UnitIntfuzzy_{u,i}$ where $(u,i)$ are particular pairs of t-conorm $u$ and t-norm $i$. \label{page:Belohlavek}
Another special case of the subset method is given in \cite[Thm.~3.5]{mogzahame11} for wta over $([0,1],\nabla,\Delta,0,1)$ where $\nabla$ and $\Delta$ are some t-conorm and t-norm, respectively, such that $(\nabla,\Delta)$ are finite range. The latter property implies that $([0,1],\nabla,\Delta,0,1)$ is locally finite (but not vice versa).

      \begin{definition-rect}\rm \label{constr:subset} Let $\Sigma$ be a ranked alphabet. Moreover, let $\B=(B,\oplus,\otimes,\0,\1)$ be a strong bimonoid and let $\cA=(Q,\delta,F)$ be a $(\Sigma,\B)$-wta. The~\emph{subset method} transforms $\cA$ into the triple $\sub(\cA)=(Q',\delta',F')$, where
	\begin{compactitem}
		\item $Q' = \im(\h_{\cA})$,
		
		\item $\delta'=(\delta'_k\mid k\in\mathbb{N})$ is the family of mappings $\delta'_k: (Q')^k\times \Sigma^{(k)}\times Q' \to B$ defined by
                  \[
                    \delta'_k(u_1\ldots u_k,\sigma,u)=
                    \begin{cases}\1 & \text{ if $\delta_\cA(\sigma)(u_1,\ldots,u_k)=u$}\\
                      \0 & \text{ otherwise}
                      \end{cases}
                  \]
                  for every~$k\in\mathbb{N}$, $\sigma\in\Sigma^{(k)}$, $u_1,\ldots,u_k\in Q'$, and~$u\in Q'$,
		
		\item $F': Q'\to B$ is a mapping defined by $(F')_u = \bigoplus_{q\in Q} u_q\otimes F_q$ for every $u\in Q'$.
                \end{compactitem}
      \end{definition-rect}
      \index{subset method}
\index{sub@$\sub(\cA)$}

Since the set $Q'$ can be infinite, the triple $\sub(\cA)$, in general, is not a wta.

      \begin{lemma}\rm \label{lm:image-ha-finite-subA-crisp-det-wta} Let $\cA$ be a $(\Sigma,\B)$-wta. If $\im(\h_\cA)$ is finite, then the triple $\sub(\cA)$ can be constructed and, moreover, $\sub(\cA)$ is a  crisp-deterministic $(\Sigma,\B)$-wta and $\initialsem{\cA} = \sem{\sub(\cA)}$.
        \end{lemma}
\begin{proof} By Observation \ref{obs:smallest-subalgebra-im}, we have  $\im(\h_\cA) = \langle \emptyset\rangle_{\delta_\cA(\Sigma)}$ and by our assumption the set $\im(\h_{\cA})$ is finite. Hence,  by Lemma \ref{obs:Knaster-Tarski-applied-to-algebras}, $\im(\h_\cA)$  can be constructed.  Then $\sub(\cA)$ is a crisp-deterministic $(\Sigma,\B)$-wta.
                                      
Moreover, the subset method is the composition
of the constructions in the proofs of Theorem \ref{thm:crisp-wta-final-index}(C)~$\Rightarrow$~(A) and Lemma \ref{lm:fin-algebra-wta}. Indeed, in Theorem  \ref{thm:crisp-wta-final-index}(C) $\Rightarrow$ (A) we construct the finite algebra $\sfA=(\im(\h_\cA), \delta_\cA)$ and the mapping $F'$ such that $\initialsem{\cA}=F' \circ \h_\sfA$. Then, using $\sfA$ and $F'$ as input, Lemma~\ref{lm:fin-algebra-wta} constructs the crisp-deterministic $(\Sigma,\B)$-wta $\cB=(\im(\h_\cA),\delta',F')$ such that $F' \circ \h_\sfA = \sem{\cB}$. Since $\cB= \sub(\cA)$, we obtain $\sem{\cB}= \sem{\sub(\cA)}$. Hence it follows that $\initialsem{\cA} = \sem{\sub(\cA)}$.
\end{proof}

  Each condition in Lemma \ref{lm:loc-finite-weakly-loc-fin-rec-step-function} guarantees finiteness of $\im(\h_\cA)$. Thus we obtain the following theorem.      

      \begin{theorem-rect} \label{thm:subset-method-lf-strong-bm} Let $\Sigma$ be a ranked alphabet. Moreover, let $\B=(B,\oplus,\otimes,\0,\1)$ be a strong bimonoid such that (a)~$\B$ is locally finite or (b) $\B$ is weakly locally finite and $\Sigma$ is monadic. Then the following two statements hold.
        \begin{compactenum}
        \item[(1)] For each $(\Sigma,\B)$-wta $\cA$, the triple $\sub(\cA)$ can be constructed; moreover, $\sub(\cA)$ is a  crisp-deterministic $(\Sigma,B)$-wta and $\initialsem{\cA} = \sem{\sub(\cA)}$.
          \item[(2)] $\Rec^{\mathrm{init}}(\Sigma,\B) \subseteq \Rec^{\mathrm{run}}(\Sigma,\B)$.
          \end{compactenum}
        \end{theorem-rect}
        \begin{proof} Proof of (1): Lemma \ref{lm:loc-finite-weakly-loc-fin-rec-step-function}, the set $\im(\h_\cA)$ is finite. Then the statement follows from Lemma~\ref{lm:image-ha-finite-subA-crisp-det-wta}.

\

Proof of (2): 
Let $\cA$ be a $(\Sigma,\B)$-wta. By (1) we have  $\initialsem{\cA} = \sem{\sub(\cA)}$, where $\sem{\sub(\cA)}$ is a crisp-deterministic $(\Sigma,\B)$-wta.  Since, by Theorem \ref{thm:bu-det:init=run}, $\sem{\sub(\cA)}= \runsem{\sub(\cA)}$, we have $\Rec^{\mathrm{init}}(\Sigma,\B) \subseteq \Rec^{\mathrm{run}}(\Sigma,\B)$.
\end{proof}

\index{preimage theorem}
As a consequence of Theorem \ref{thm:subset-method-lf-strong-bm}, we show a preimage theorem. 
In general, a \emph{preimage theorem} states, for some modifier $x \in \{\mathrm{run}, \text{init}\}$, under which conditions on $\B$, for each  $(\Sigma,\B)$-wta $\cA$, the weighted tree language $\sem{\cA}^x$  has the preimage property (i.e., for each $b \in B$ we have that $(\sem{\cA}^x)^{-1}(b)$ is a recognizable tree language, cf. Subsection \ref{sec:weighted-tree-languages}). Clearly, if $\B$ is a semiring, then we can disregard the modifier $x$ (cf. Corollary~\ref{cor:semiring-run=init}). The next preimage theorem concerns the initial algebra semantics (i.e., $x = \text{init}$). 

\begin{corollary}\rm \label{cor:preimage-fin-loc-finite-weakly-loc-fin-rec-step-function}
Let (a) $\B$ be locally finite or (b)~let $\B$ be weakly locally finite and $\Sigma$ monadic.  Moreover, let $\cA$ be a $(\Sigma,\B)$-wta. Then $\initialsem{\cA}$ has the preimage property. Moreover, for each $b \in B$, we can construct a $\Sigma$-fta $A$ such that $\LL(A)=(\initialsem{\cA})^{-1}(b)$.
\end{corollary}
\begin{proof} 
By Theorem \ref{thm:subset-method-lf-strong-bm} we can construct the crisp-deterministic $(\Sigma,\B)$-wta $\sub(\cA)$ such that $\initialsem{\cA} = \sem{\sub(\cA)}$. Then our statements follow from Theorem \ref{thm:crisp-det-algebra}(A)$\Rightarrow$(C).
\end{proof} 

A special case of Corollary  \ref{cor:preimage-fin-loc-finite-weakly-loc-fin-rec-step-function}(a)  was proved in  \cite[Lm.~6.1]{drovog06} for  locally finite commutative semirings (taking into account that for wta over semirings the run semantics and the initial algebra semantics coincide cf. Corollary \ref{cor:semiring-run=init}(2)). Hence we obtain the following equivalence between local finiteness, initial algebra crisp-determinizability, and finiteness of
    initial algebra semantics (where the equivalence with Statement (C) is due to \cite{dro22}).

\begin{theorem-rect}\label{thm:loc-finite-rec-step-function} Let  $\B$ be a strong bimonoid. Then the following four statements are equivalent.
\begin{compactenum}
\item[(A)]  $\B$ is locally finite.
\item[(B)]  For each ranked alphabet $\Sigma$ and for each $(\Sigma,\B)$-wta $\cA$, we can construct a  crisp-deterministic $(\Sigma,\B)$-wta~$\cB$ such that $\initialsem{\cA}=\sem{\cB}$.
\item[(C)] For each ranked alphabet $\Sigma$ and for each $(\Sigma,\B)$-wta $\cA$, the set $\im(\initialsem{\cA})$  is finite.
  \item[(D)] For each branching ranked alphabet $\Sigma$ and for each $(\Sigma,\B)$-wta $\cA$, the set $\im(\initialsem{\cA})$  is finite.
\end{compactenum}
\end{theorem-rect}
\begin{proof} Proof of (A)$\Rightarrow$(B): This follows from Theorem \ref{thm:subset-method-lf-strong-bm}(1).

  \
  
  Proof of (B)$\Rightarrow$(C): This follows from Theorem \ref{thm:crisp-det-algebra}(A)$\Rightarrow$(C).

  \
  
  Proof of (C)$\Rightarrow$(D): This holds trivially.

  \
  
  Proof of (D)$\Rightarrow$(A): Let $A\subseteq B$ be a finite set. Let $\Sigma$ be any ranked alphabet which is branching. By Theorem \ref{thm:not-monadic-wta-can-compute-closure-of-finite-subset} we can construct a $(\Sigma,\B)$-wta $\cB$ such that $\im(\initialsem{\cB}) = \langle A \rangle_{\oplus,\otimes,\0,\1\}}$. Hence, by assumption~(D), the set $\langle A \rangle_{\oplus,\otimes,\0,\1\}}$ is finite. Thus $\B$ is locally finite.
\end{proof}

A special case of Theorem \ref{thm:loc-finite-rec-step-function} was proved in \cite[Thm.~2.1]{bel02} for the case of string ranked alphabets (i.e., weighted automata) and complete distributive lattices (cf. the remark on p.\pageref{page:Belohlavek}).
Moreover, it is easy to see that the Viterbi semiring $([0,1],\max,\cdot,0,1)$ is not locally finite. Hence, by Theorem \ref{thm:loc-finite-rec-step-function}, there exists a ranked alphabet $\Sigma$ and a $(\Sigma, \Viterbi)$-wta $\cA$ such that $\im(\sem{\cA})$ is infinite.  The wsa in \cite[Ex.~3.1]{liped05} can be thought of as such a wta.
Also we note that \cite[Thm.~3.4]{liped05} is a special case of Theorem \ref{thm:loc-finite-rec-step-function}(A)$\Leftrightarrow$(B) for lattice-ordered monoids.

In Theorem \ref{thm:wta-B=fta} and Corollary \ref{cor:supp-B=fta-1} we proved that $\Rec(\Sigma,\Boole)$ and $\Rec(\Sigma)$ are essentially the same. One might wonder, how the set $\Rec(\Sigma,\sfFtwo)$ of weighted tree languages can be characterized, where $\sfFtwo=(\{0,1\},\oplus,\otimes,0,1)$ is the field from Example \ref{ex:semirings}(\ref{ex:two-two-element-sr}), i.e., the other strong bimonoid with two elements.
The answer is given in the following corollary, which corresponds to Corollary \ref{cor:supp-B=fta-1}.

\begin{corollary-rect}\rm \label{cor:supp-F2=fta} 
    Let $\Sigma$ be a ranked alphabet. Moreover, let $L \subseteq \T_\Sigma$. Then the following three statements are equivalent.
    \begin{compactenum}
    \item[(A)] We can construct a $\Sigma$-fta $A$ such $L = \LL(A)$.
    \item[(B)] We can construct a $(\Sigma,\sfFtwo)$-wta $\cA$ such that $L = \supp(\sem{\cA})$.
      \item[(C)] We can construct a $(\Sigma,\sfFtwo)$-wta $\cA$ such that $\chi(L) = \sem{\cA}$.
    \end{compactenum}
Moreover, we have  $\supp(\Rec(\Sigma,\sfFtwo)) = \Rec(\Sigma)$.
\end{corollary-rect}

  \begin{proof} Proof of (A)$\Rightarrow$(C): It follows from Theorem \ref{thm:fta-wta} (B)$\Rightarrow$(C).

    \
    
  Proof of (C)$\Leftrightarrow$(B): This follows from the equivalence that for every $r: \T_\Sigma \to \{0,1\}$ and $L\subseteq \T_\Sigma$: $r = \chi(L)$
   iff $L = \supp(r)$. 

   \
   
  Proof of (C)$\Rightarrow$(A):  Let $\cA$ be a $(\Sigma,\sfFtwo)$-wta with $\sem{\cA}=\chi(L)$. By Theorem~\ref{thm:loc-finite-rec-step-function},  we can construct a crisp-deterministic $(\Sigma,\sfFtwo)$-wta $\cB$ such that $\sem{\cA} = \sem{\cB}$. Since the carrier set of $\mathsf{F}_2$ is the set $\{0,1\}$, $\cB$ has unit root weights. Then, by Theorem~\ref{thm:fta-wta} (C)$\Rightarrow$(B), we can construct a $\Sigma$-fta $A$ such that $\LL(A) = L$. 
   
Finally, the equality $\supp(\Rec(\Sigma,\sfFtwo)) = \Rec(\Sigma)$ follows from (A)$\Leftrightarrow$(B).
  \end{proof}

    \begin{corollary-rect}\rm \label{cor:F2=Boole} 
      Let $\Sigma$ be a ranked alphabet. Then $\Rec(\Sigma,\Boole) = \Rec(\Sigma,\sfFtwo)$.
    \end{corollary-rect}
    \begin{proof} Let $r \in \Rec(\Sigma,\Boole)$. Since $\Boole$ is locally finite, by Theorem \ref{thm:loc-finite-rec-step-function}, we can construct a crisp-deterministic $(\Sigma,\Boole)$-wta
       $\cB$ such that $r = \sem{\cB}$. Hence $r \in \cdRec(\Sigma,\Boole)$. Since the semirings $\Boole$ and $\sfFtwo$ have the same carrier set, by Corollary \ref{cor:crisp-det-wta-B1-B2}, we have that $r \in \cdRec(\Sigma,\sfFtwo)$. Hence $r \in \Rec(\Sigma,\sfFtwo)$. The proof of the other inclusion is very similar.
    \end{proof}


\section{Run crisp-determinization}
\label{sec:r-crisp-determinization}

In this section we will prove a sufficient condition under which a $(\Sigma,\B)$-wta $\cA$ is run crisp-determinizable (cf. Theorem \ref{thm:bounded-finaddorder-r-c-det}). We will continue with two characterizations of this property  (cf. Theorems \ref{thm:bi-loc-finite-rec-step-function} and \ref{thm:past-finite-mon-char-cd}), where the first one deals with arbitrary strong bimonoids and the second one with past-finite monotonic strong bimonoids.

\subsection{Sufficient condition for run crisp-determinizability}
\label{subsec:sufficient}

Let $\cA=(Q,\delta,F)$ be a $(\Sigma,\B)$-wta. We recall that, for each $\xi \in \T_\Sigma$, we have
\[
\runsem{\cA}(\xi) = \bigoplus_{\rho \in \R_\cA(\xi)} \wt(\xi,\rho) \otimes F_{\rho(\varepsilon)}\enspace.
\]
In contrast to the initial algebra semantics, multiplication and summation are not used alternatingly but first multiplication is used (for the computation of $\wt(\xi,\rho) \otimes F_{\rho(\varepsilon)}$) and second summation is used.  In the following we will formulate a sufficient condition for run crisp-determinizability of $\cA$ which is based on the values which occur in the computation of $\runsem{\cA}$ and on structural properties of $\cA$.

For the following discussion we recall the definition $ \rmH(\cA) = \{\wt(\xi,\rho)  \mid \xi \in \T_\Sigma , \rho \in \R_\cA(\xi)\}$ (cf. (\ref{alg:H(A)} of Section \ref{sect:normalizing-transition-weights}) and define the set
\index{HA@$\rmH(\cA)$}
\index{CA@$\C(\cA)$}
\begin{align}\label{alg:H(A)-and-C(A)}
\C(\cA) = \{\wt(\xi,\rho) \otimes F_{\rho(\varepsilon)} \mid \xi \in \T_\Sigma , \rho \in \R_\cA(\xi)\} \enspace.
\end{align}
Clearly, if the set $\rmH(\cA)$ is finite, then $\C(\cA)$ is also finite, because $\C(\cA)\subseteq \rmH(\cA)\otimes \im(F)$.

\begin{lemma} \rm \cite[Lm.~6.5]{drofulkosvog21} \label{lm:compute-c(A)}
Let $\cA=(Q,\delta,F)$ be a $(\Sigma,\B)$-wta. If $\rmH(\cA)$ is finite, then we can construct the set $\C(\cA)$.
\end{lemma}
\begin{proof} Let $\rmH(\cA)$ be finite. By Lemma \ref{lm:compute-c(A)-normal-form}, we construct the set $\rmH(\cA)$. We recall that in the proof of that lemma, for every $n \in \mathbb{N}$ and $q \in Q$, we construct the set
  \[H_{n, q} = \{\wt(\xi,\rho) \mid \xi \in \T_\Sigma, \height(\xi) \leq n, \rho \in \R_\cA(q,\xi)\}\enspace,\]
and denote by $n_m\in \mathbb{N}$  the least number such that $H_{n_m,q}=H_{n_m+1,q}$ for each $q\in Q$.
  
Now we prove that the set $\C(\cA)$ can be constructed. Let $n_m$ be the number as before and
\[C=\{\wt(\xi,\rho)\otimes F_{\rho(\varepsilon)}\mid \xi \in \T_\Sigma, \height(\xi) \leq n_m, \rho \in \R_\cA(\xi)\}\enspace.\]
  It suffices to show that $C=\C(\cA)$ because we can construct the set $C$. 
It is obvious that  $C\subseteq \C(\cA)$. For the proof of the other inclusion,   let $b\in \C(\cA)$,
i.e., $b=\wt(\xi,\rho)\otimes F_q$ for some $\xi \in \T_\Sigma$, $q\in Q$, and  $\rho \in \R_\cA(q,\xi)$. Since $\wt(\xi,\rho)\in \rmH(\cA)$, by the proof of constructing the set $\rmH(\cA)$ (cf. Lemma \ref{lm:compute-c(A)-normal-form}), we have $\wt(\xi,\rho)\in H_{n_m, q}$, i.e., there exist $\xi' \in \T_\Sigma$ with  $\height(\xi') \leq n_m$, and  $\rho' \in \R_\cA(q,\xi')$ such that $\wt(\xi,\rho)=\wt(\xi',\rho')$. Hence $b\in C$.
\end{proof}

We continue with an analysis of $\im(\runsem{\cA})$. We can achieve that $\im(\runsem{\cA})$ is finite if  we guarantee that (a) the set $\C(\cA)$ is finite and (b) there exists an upper bound $K$ such that each $b\in \C(\cA)$ is summed up at most $K$ times.
\index{complete run number mapping}
To express this more precisely,  for each $b \in \C(\cA)$, we define the mapping $f_{\cA,b}: \T_\Sigma \to \mathbb{N}$, called {\em complete run number mapping of $b$}, such that for each $\xi \in \T_\Sigma$  we let
\index{fAb@$f_{\cA,b}(\xi)$}
\[f_{\cA,b}(\xi) = |\{\rho \in \R_\cA(\xi) \mid \wt(\xi,\rho) \otimes F_{\rho(\varepsilon)} = b\}|\enspace.\]
We recall from \eqref{equ:nb-nexpb} that, for each $n \in \mathbb{N}$, the value $nb$ is the sum $b \oplus \ldots \oplus b$ with $n$ summands.
Finally, we define the $(\Sigma,\B)$-weighted tree language
\index{rAb@$r_{\cA,b}$}
\[r_{\cA,b}: \T_\Sigma \to B \ \text{ with } \ r_{\cA,b}(\xi)= (f_{\cA,b}(\xi))b \ \text{ for each $\xi \in \T_\Sigma$}\enspace.\]
Thus, if $\C(\cA)$ is finite, then for each $\xi \in \T_\Sigma$, we have
\begin{equation}\label{eq:semantics=sum-f}
\runsem{\cA}(\xi)=\bigoplus_{b\in \C(\cA)}(f_{\cA,b}(\xi))b= \bigoplus_{b\in \C(\cA)} r_{\cA,b}(\xi)\enspace.
\end{equation}
Overall, if $\C(\cA)$ is finite and $r_{\cA,b}$ is a recognizable step mapping for each $b \in \C(\cA)$, then $\runsem{\cA}$ is a recognizable step mapping (because recognizable step mappings are closed under sum, cf. Corollary~\ref{cor:RecStep-closed-sum-scalar}).

In order to find out under which conditions $r_{\cA,b}$ is a recognizable step mapping (cf. Lemma \ref{lm:bounded-finite-order=>rec-step-mapping}), we first prove that $f_{\cA,b}$ is an r-recognizable  $(\Sigma,\mathbb{N})$-weighted tree language. For this we use the set 
$\rmH(\cA)$.
The next lemma generalizes \cite[Thm.~11]{drostuvog10} and \cite[Thm.~6.2(a)]{drogoemaemei11}
from the string case to the tree case.

\begin{lemma}\rm \cite[Thm.~6.6]{drofulkosvog21} \label{lm:f_(a,b)-recog} Let $\cA$ be a $(\Sigma,\B)$-wta. If $\rmH(\cA)$ is finite, then for each $b \in \C(\cA)$ we can construct a crisp $(\Sigma,\Nat)$-wta $\cA'_b$ with unit root weights such that $\sem{\cA'_b}=f_{\cA,b}$.
  \end{lemma}
  \begin{proof} Let $\rmH(\cA)$ be finite. By Lemma \ref{lm:compute-c(A)-normal-form}, we construct $\rmH(\cA)$. Let $b \in \C(\cA)$. We define the $(\Sigma,\Nat)$-wta  $\cA'_b=(Q',\delta',F'_b)$ as follows:
    \begin{compactitem}
    \item $Q' = Q \times \rmH(\cA)$,
      \item for every $k\in \mathbb{N}$, $\sigma \in \Sigma^{(k)}$ and $(q_1,y_1),\ldots,(q_k,y_k), (q,y) \in Q'$, let 
  \[\delta'_k\big((q_1,y_1)\cdots(q_k,y_k),\sigma,(q,y)\big) = 
  \begin{cases}
      1 &\text{ if $\big(\bigotimes_{i\in [k]} y_i\big) \otimes \delta_k(q_1 \cdots q_k, \sigma, q) = y$}\\
      0 &\text{ otherwise,} 
  \end{cases} \ \text{ and}\]
\item for each $(q,y) \in Q$, we let $(F'_b)_{(q,y)} = 1$ if $y \otimes F_q = b$, and $0$ otherwise.
\end{compactitem}
  Let $\xi \in \T_\Sigma$ and $b \in \C(\cA)$. It is obvious that,
  \begin{equation}
    \text{for each $\rho \in \R_{\cA'_b}(\xi)$, we have } \wt_{\cA_b'}(\xi,\rho) \cdot (F'_b)_{\rho'(\varepsilon)}\in \{0,1\}\enspace.  \label{eq:weight-in-01}
    \end{equation}
 Next we observe that there exists a bijection between the two sets
  \[\{ \rho \in \R_\cA(\xi) \mid \wt_\cA(\xi,\rho)\otimes F_{\rho(\varepsilon)} = b \} \ \ \text{ and } \ \  \{ \rho' \in \R_{\cA'_b}(\xi)\big)\mid \wt_{\cA'_b}(\xi,\rho') \cdot (F'_b)_{\rho'(\varepsilon)} = 1\}\enspace.\]
  Then it follows that
  \begin{align*}
    f_{\cA,b}(\xi) &= |\{ \rho \in \R_\cA(\xi) \mid \wt_\cA(\xi,\rho)\otimes F_{\rho(\varepsilon)} = b \}| \\
                   &= |\{ \rho' \in \R_{\cA'_b}(\xi)\mid \wt_{\cA'_b}(\xi,\rho') \cdot (F'_b)_{\rho'(\varepsilon)} = 1\}| \\
                   &= \bigplus_{\substack{\rho' \in \R_{\cA'_b}(\xi):\\\wt_{\cA'_b}(\xi,\rho') \cdot (F'_b)_{\rho'(\varepsilon)} = 1}} 1\\
                   &= \bigplus_{\rho' \in \R_{\cA'_b}(\xi)} \wt_{\cA_b'}(\xi,\rho') \cdot (F_b')_{\rho'(\varepsilon)}
    \tag{by \eqref{eq:weight-in-01}}\\
    &= \sem{\cA'_b}(\xi) \qedhere
    \end{align*}
\end{proof}

Next we analyse the preimage of $f_{\cA,b}=\sem{\cA'_b}$, where $\cA'_b$ is the $(\Sigma,\Nat)$-wta of Lemma \ref{lm:f_(a,b)-recog}. For this we prove two preimage theorems for an arbitrary $(\Sigma,\Nat)$-wta instead of the specific $\cA'_b$ (cf. Lemmas \ref{lm:preimage-N-1} and \ref{lm:preimage-N-2}), where the first one is a generalization of \cite[Cor.~III.2.5]{berreu88}).

\begin{lemma}\rm \label{lm:preimage-N-1} (cf. \cite[Lm.~6.3(2)]{drovog06} and \cite[Lm.~6.3]{drofulkosvog21}) Let  $\cA$ be a $(\Sigma,\Nat)$-wta. Then $\sem{\cA}$ has the preimage property and,  for each $n  \in \mathbb{N}$, we can construct a $\Sigma$-fta $A$ such that $\LL(A)=\sem{\cA}^{-1}(n)$.
\end{lemma}
\begin{proof}   Let $n \in \mathbb{N}$ and $M = \{k \in \mathbb{N} \mid k > n\}$. Moreover, we let $\sim$ be the equivalence relation on the set $\mathbb{N}$ defined such that its $n+2$ classes are the singleton sets $\{k\}$ for each $k \in [0,n]$ and the set $M$.

  As is well known, $\sim$  is a congruence relation on the semiring of natural numbers, which can be seen as follows. Let $n_1,n_2,n_1',n_2' \in \mathbb{N}$ such that $n_1 \sim n_1'$ and $n_2\sim n_2'$.  Since,  for each $k \in \mathbb{N}$ with $k \le n$, the equivalence class $\{k\}$ is a singleton and $\Nat$ is commutative, the only interesting case arises if $n_1,n_1' \in M$. So we assume that $n_1,n_1' \in M$.
  For the summation, we obviously have $n_1 \leq n_1 + n_2$ and $n_1' \leq n_1' + n_2'$, and hence $n_1 + n_2 \in M$ and $n_1' + n_2' \in M$.
  For the multiplication, if $n_2 \neq 0 \neq n_2'$, then similarly we obtain  that $n_1 \cdot n_2\in M$ and $n_1' \cdot n_2' \in M$. If $n_2 = 0 =n_2'$, then $n_1 \cdot n_2 \in \{0\}$ and $n'_1 \cdot n'_2 \in \{0\}$.
Hence $\sim$ is a congruence relation on $\Nat$.

Since $\sim$ has finite index, the quotient semiring $\Nat/_\sim$ is finite. Let $h: \mathbb{N} \to \mathbb{N}/_\sim$ be the canonical semiring homomorphism, i.e., $h(n) = [n]_\sim$ for each $n \in \mathbb{N}$.  Then, by Theorem  \ref{thm:closure-sr-hom}, the $(\Sigma,\Nat/_\sim)$-weighted tree language $h \circ \sem{\cA}$ is recognizable, and by Corollary \ref{cor:preimage-fin-loc-finite-weakly-loc-fin-rec-step-function}
we have that $(h \circ \sem{\cA})^{-1}(\{n\})$ is a $\Sigma$-recognizable tree language for each $n \in \mathbb{N}$. Since $(h \circ \sem{\cA})^{-1}(\{n\})= \sem{\cA}^{-1}(n)$ for each $n \in \mathbb{N}$, we obtain that $\sem{\cA}$ has the preimage property. 

Clearly, we can give effectively the congruence classes of $\sim$, i.e., the elements of $\mathbb{N}/_\sim$, by choosing only one representative for each congruence class.  By Lemma \ref{lm:f-image-equivalent} and the obvious fact that $h(\cA)$ is constructible,
we can construct the $(\Sigma,\Nat/_\sim)$-wta $h(\cA)$ such that $\sem{h(\cA)} = h \circ \sem{\cA}$. Since $\mathbb{N}/_\sim$ is finite, by Corollary \ref{cor:preimage-fin-loc-finite-weakly-loc-fin-rec-step-function},
 we can construct a finite-state $\Sigma$-tree automaton $A$ which recognizes $\sem{h(\cA)}^{-1}(\{n\})= \sem{\cA}^{-1}(n)$. 
\end{proof}

In the next lemma we deal with the preimage of a set under a recognizable weighted tree language over the semiring of natural numbers. For the sake of simplicity, we call this theorem also preimage theorem. It is a generalization of \cite[Ch.~VI, Thm.~10.1]{eil74} (also cf. \cite[Cor.~III.2.4]{berreu88}). For every $m\in \mathbb{N}$ and $n\in \mathbb{N}_+$, we define
\[m+n\mathbb{N}=\{m+n\cdot j \mid j\in \mathbb{N}\}\enspace.\]
Moreover, for each $m \in \mathbb{N}$, we let  $\overline{m} = m+n\mathbb{N}$, the equivalence class of $m$ modulo $n$. The semiring of natural numbers modulo $n$ is the semiring $\Nat/n\Nat=(\{\overline{0},\ldots,\overline{n-1}\},+_n,\cdot_n,\overline{0},\overline{1}\})$, where for every  $k,\ell\in [0,n-1]$ we define $\overline{k} +_n\overline{\ell} = \overline{k+\ell \!\mod(n)}$ and $\overline{k} \cdot_n \overline{\ell} = \overline{k\cdot\ell \!\mod(n)}$.

\begin{lemma}\rm \label{lm:preimage-N-2} (cf. \cite[Lm.~6.3(2)]{drovog06} and \cite[Lm.~6.4]{drofulkosvog21}) Let  $\cA$ be a $(\Sigma,\Nat)$-wta. Moreover, let $m \in \mathbb{N}$ and $n \in \mathbb{N}_+$. Then (a) the $\Sigma$-tree language $\sem{\cA}^{-1}(m + n\cdot \mathbb{N})$ is recognizable and (b) we can construct a $\Sigma$-fta $A$ such that $\LL(A)=\sem{\cA}^{-1}(m + n\cdot \mathbb{N})$.
\end{lemma}
\begin{proof} Proof of (a): Let $m \in \mathbb{N}$ and $n \in \mathbb{N}_+$. Let $m < n$. By Theorem~\ref{thm:closure-sr-hom}, $(h \circ \sem{\cA}) \in \mathrm{Rec}(\Sigma,\Nat/n\Nat)$, where
  $h: \mathbb{N} \to \mathbb{N}/n\mathbb{N}$ is the canonical semiring homomorphism.
  Moreover, $\sem{\cA}^{-1}(m + n \cdot \mathbb{N}) = \sem{\cA}^{-1}\big(h^{-1}(\overline{m})\big) = (h \circ \sem{\cA})^{-1}(\overline{m})$, where the first equality holds because $m < n$. Since $\Nat/n\Nat$ is a finite semiring, by  Corollary \ref{cor:preimage-fin-loc-finite-weakly-loc-fin-rec-step-function}, 
the $\Sigma$-tree language $(h \circ \sem{\cA})^{-1}(\overline{m})$ is recognizable. Now assume that $m \geq n$. Then there exist  $m' \in [0,n-1]$ and $k \in \mathbb{N}_+$ such that $m = m' + n \cdot k$. Then
\begin{equation}\label{eq:cA-minus-1-equal-stdiff}
  \sem{\cA}^{-1}(m + n \cdot \mathbb{N}) = \sem{\cA}^{-1}(m' + n \cdot \mathbb{N}) \setminus \bigcup_{j=0}^{k-1} \sem{\cA}^{-1}(m' + n \cdot j)\enspace.
  \end{equation}
As we saw, the $\Sigma$-tree language $\sem{\cA}^{-1}(m' + n \cdot \mathbb{N})$ is recognizable because $m'<n$. Moreover, by Lemma~\ref{lm:preimage-N-1}, for each $j \in [0,k-1]$, the $\Sigma$-tree language $\sem{\cA}^{-1}(m' + n \cdot j)$ is also recognizable. Finally, $\Sigma$-tree languages are closed under union and subtraction (cf. Theorem \ref{thm:fta-closure-results}). Thus, also in this case, the $\Sigma$-tree language $\sem{\cA}^{-1}(m + n \cdot \mathbb{N})$ is recognizable.

Proof of (b): We follow the proof of (a). Let $m \in \N$. Assume that $m < n$. Obviously, we can  give effectively the equivalence classes modulo $n$, i.e., the elements of $\mathbb{N}/n\mathbb{N}$, by choosing only one representative for each equivalence class.  Then we can construct the $(\Sigma,\Nat/n\Nat)$-wta $h(\cA)$ (cf. page \pageref{p:f(A)}). By  Lemma \ref{lm:f-image-equivalent} we have $\sem{h(\cA)}=h \circ \sem{\cA}$.

  Since $\Nat/n\Nat$ is a  finite semiring, by Corollary \ref{cor:preimage-fin-loc-finite-weakly-loc-fin-rec-step-function}, we can construct  a $\Sigma$-fta which recognizes $(h \circ \sem{\cA})^{-1}(\overline{m}) = \sem{\cA}^{-1}(m + n \cdot \mathbb{N})$.

Now assume that $m \geq n$. Since $m' < n$, by the above, we can construct a $\Sigma$-fta which recognizes $\sem{\cA}^{-1}(m' + n \cdot \mathbb{N})$. Moreover, by Lemma~\ref{lm:preimage-N-1}, for each $j \in [0,k-1]$, we can also construct  a $\Sigma$-fta which recognizes $\sem{\cA}^{-1}(m' + n \cdot j)$. Thus, for each tree language which occurs on the right-hand side of \eqref{eq:cA-minus-1-equal-stdiff} a $\Sigma$-fta can be constructed. Hence, by Theorem \ref{thm:fta-closure-results}, we can construct a $\Sigma$-fta which recognizes $\sem{\cA}^{-1}(m + n \cdot \mathbb{N})$.
\end{proof}

Now we have collected the necessary preimage theorems to show that, for each $(\Sigma,\B)$-wta $\cA$ and $b\in \C(\cA)$, the mapping $r_{\cA,b}$ is a recognizable step mapping if $f_{\cA,b}$ is bounded or $b$ has finite additive order (cf. Lemma \ref{lm:bounded-finite-order=>rec-step-mapping}).

\index{bounded}
We call the mapping $f_{\cA,b}$  {\em bounded} if there exists $K \in \mathbb{N}$ such that $f_{\cA,b}(\xi) \leq K$ for each $\xi \in \T_\Sigma$.

\begin{lemma}\rm \cite[Thm.~6.6]{drofulkosvog21}\label{lm:bounded-finite-order=>rec-step-mapping} Let $\cA$ be a $(\Sigma,\B)$-wta such that $\rmH(\cA)$ is finite and let $b \in \C(\cA)$. If the mapping  $f_{\cA,b}$ is bounded or $b$ has finite additive order, then we can construct a crisp-deterministic $(\Sigma,\B)$-wta $\cB_b$ such that $\sem{\cB_b} = r_{\cA,b}$.
  \end{lemma}
 
\begin{proof} By performing Algorithm  \ref{alg:construction-crisp-det-wta}, a crisp-deterministic $(\Sigma,\B)$-wta $\cB_b$ is constructed.

\begin{algorithm}[h]
    \begin{algorithmic}[1]
        \Require (a) $(\Sigma,\B)$-wta $\cA$ such that $\rmH(\cA)$ is finite and (b) $b \in \C(\cA)$ such that $f_{\cA,b}$ is bounded or $b$ has finite additive order
        \Ensure cd $(\Sigma,\B)$-wta $\cB_b$
        
        \State by applying Lemma \ref{lm:f_(a,b)-recog} to~$\cA$, we construct the $(\Sigma,\Nat)$-wta $\cA'_b$ s.t. $\sem{\cA'_b}=f_{\cA,b}$; \label{l:faB=cAb'}

        \
        
            \ForEach{$i=0,1,2,\ldots$} \label{l:for-each-i-begin}
                \State by applying Lemma~\ref{lm:preimage-N-1} to $\cA'_b$, we construct the $\Sigma$-fta $A_{b,i}$ s.t. $\LL(A_{b,i}) = \sem{\cA'_b}^{-1}(i)$; \label{l:lm:preimage-N-1}
                \If {$\T_\Sigma \subseteq \bigcup_{j \in [0,i]} \LL(A_{b,j})$} let $K \gets i$ and exit the for-loop;
                \Comment{$f_{b,\cA}$ is bounded by $K$} \label{l:fbA-bounded}
                     \EndIf
                \If {$ib = jb$ for some $j < i$}  \label{l:b-fin-add-index?}
                \State $i(b) \gets \min(j \in \mathbb{N}_+ \mid j < i \text{ and } jb=ib)$ and  $p(b)\gets i-i(b)$; \Comment{construct the index and period of $b$}
                \ForEach{$j \in [i(b),i(b)+p(b)-1]$} \label{l:update-begin}
                       \State by applying  Lemma \ref{lm:preimage-N-2} to $\cA'_b$, we construct the $\Sigma$-fta $A_{b,j}$ s.t. $\LL(A_{b,j}) = \sem{\cA'_b}^{-1}(j + p(b)\cdot \mathbb{N})$
                       \EndFor \label{l:update-end}
                       \State  let $K \gets i(b)+p(b)-1$   and  exit the for-loop \label{l:exit-because-fin-add-index}
                       \Comment{$b$ has additive order $K$}
                       \EndIf       
                       \EndFor \label{l:for-each-i-end}

                       \

                 \ForEach{$j \in [0,K]$} \label{l:constr-D-begin}
                       \State by applying Theorem \ref{thm:fta-wta}(B)$\Rightarrow$(C) to $A_{b,j}$, we construct the cd $(\Sigma,\B)$-wta $\cC_{b,j}$ s.t.
        \[\sem{\cC_{b,j}} = \chi(\LL(A_{b,j}))\enspace;\]
        \State by applying Theorem \ref{thm:closure-scalar}(4) to $jb$ and $\cC_{b,j}$, we construct the cd $(\Sigma,\B)$-wta $\cD_{b,j}$ s.t.
        \[\sem{\cD_{b,j}}= (jb)\cdot \sem{\cC_{b,j}}\enspace;\]
                        \EndFor \label{l:constr-D-end}
                        \State by applying iteratively Theorem~\ref{thm:closure-sum}(2) to the members of  the finite family $(\cD_{b,j} \mid j \in [0,K])$, we construct the cd $(\Sigma,\B)$-wta $\cB_{b}$ s.t.  \[\sem{\cB_{b}}= \bigoplus_{j \in  [0,K]}\sem{\cD_{b,j}}\enspace. \] \label{l:constr-Bb}
           \State  \Return $\cB_b$
    \end{algorithmic}
    \caption{Construction of the cd $(\Sigma,\B)$-wta $\cB_b$ such that $\sem{\cB_{b}} = r_{\cA,b}$ (cd=crisp-deterministic)}\label{alg:construction-crisp-det-wta}
  \end{algorithm}

We prove that $\sem{\cB_b} = r_{\cA,b}$.  
  If, before execution of line 3, the variable $i$ has the value $\ell$, then the family $(A_{b,j}\mid j \in [0,\ell-1])$ of $\Sigma$-fta has already been constructed by using Lemma \ref{lm:preimage-N-1}; in particular, if $\ell = 0$, then this family is empty. Thus, after execution of line 3, the family $(A_{b,j}\mid j \in [0,\ell])$ of $\Sigma$-fta has been constructed (where $A_{b,\ell}$ is also constructed by Lemma \ref{lm:preimage-N-1}).

  Then, line 4 asks whether the mapping $f_{\cA,b}$ is bounded by $\ell$ (i.e., $\T_\Sigma \subseteq \bigcup_{j \in [0,\ell]} \LL(A_{b,j})$; due to \cite[Thm.~2.10.3]{gecste84} this is decidable). If so, then the for-loop is exited with $K=\ell$. 
  
  If not, then line 6 asks whether $b$ has finite additive order $\ell$ (i.e., $\ell b = jb$ for some $j < \ell$). If so, then $\ell = \min(\ell' \mid \exists j < \ell': \ell'b = jb)$ and the index $i(b)$ of $b$ and the period $p(b)$ of $b$ (cf. Fig.~\ref{fig:period-index}) are constructed and, due to lines 8-10, the upper part $(A_{b,j}\mid j \in [i(b),i(b)+p(b)-1])$ of the family of $\Sigma$-fta (generated in line 3  is reconstructed by using Lemma \ref{lm:preimage-N-2}. Thereafter the for-loop is exited with $K=\ell$.  (We note that $K = i(b)+p(b)-1$, i.e., $K$ is the finite additive order of $b$.) 
  
   Since, by our assumption, the mapping $f_{\cA,b}$ is bounded or $b$ has finite additive order, the for-loop in lines~2-13, will eventually terminate.

  If the for-loop in  lines 2-13 was terminated due to the exit in line 4, then the family $(A_{b,j}\mid j \in [0,K])$ of $\Sigma$-fta was constructed such that $\LL(A_{b,j}) = \sem{\cA'_b}^{-1}(j)$ for each $j \in [0,K]$. Thus, due to line 1, for each $j \in [0,K]$ and $\xi \in \T_\Sigma$, we have:
$\xi \in \LL(A_{b,j})$ iff $f_{\cA,b}(\xi)=j$. Hence $ \LL(A_{b,j})\cap \LL(A_{b,j'}) = \emptyset$ for every $j,j' \in [0,K]$ with $j \neq j'$, and for each $\xi \in \T_\Sigma$ there exists $j \in [0,K]$ such that $\xi \in \LL(A_{b,j})$; that means that the family  $(\LL(A_{b,j}) \mid j \in [0,K])$ is a partitioning of $\T_\Sigma$. Since $r_{\cA,b}(\xi)= (f_{\cA,b}(\xi))b$ for each $\xi \in \T_\Sigma$, we finally obtain
  \[r_{\cA,b} = \bigoplus_{j \in [0,K]} (jb) \cdot \chi(\LL(A_{b,j}))\enspace.\]

If the for-loop in  lines 2-13 was terminated due to the exit in line 11, then the family $(A_{b,j}\mid j \in [0,K])$ of $\Sigma$-fta was constructed such that
\begin{compactitem}
\item $\LL(A_{b,j}) = \sem{\cA'_b}^{-1}(j)$ for each $j \in [0,i(b)-1]$ and
\item $\LL(A_{b,j}) = \sem{\cA'_b}^{-1}(j + p(b)\cdot \mathbb{N})$  for each $j \in [i(b),K]$.
\end{compactitem}
Thus, due to line 1,
\begin{compactitem}
\item for each $j \in [0,i(b)-1]$ and $\xi \in \T_\Sigma$, we have:
$\xi \in \LL(A_{b,j})$ iff $f_{\cA,b}(\xi)=j$.
\item for each $j \in [i(b),K]$ and $\xi \in \T_\Sigma$, we have:
$\xi \in \LL(A_{b,j})$ iff $f_{\cA,b}(\xi) \in (j + p(b) \cdot \mathbb{N})$.
  \end{compactitem}
Hence also in this case the family $(\LL(A_{b,j}) \mid j \in [0,K])$ is a partitioning of $\T_\Sigma$ and as above we obtain
\[r_{\cA,b} = \bigoplus_{j \in [0,K]} (jb) \cdot \chi(\LL(A_{b,j}))\enspace.\]

In lines 14-17, the family $(\cD_{b,j} \mid j \in [0,K])$ of crisp-deterministic $(\Sigma,\B)$-wta is constructed such that 
\[\sem{\cD_{b,j}} =  (jb)\cdot \chi(\LL(A_{b,j})) \enspace.
\]
Finally, in line 18, the crisp-deterministic $(\Sigma,\B)$-wta $\cB_b$ is constructed such that $\sem{\cB_b} = \bigoplus_{j \in [0,K]} \sem{\cD_{b,j}}$. Hence 
\[
\sem{\cB_b} = \bigoplus_{j \in [0,K]} (jb)\cdot \chi(\LL(A_{b,j})) = r_{\cA,b}\enspace.
\]
\end{proof}  

Next we can show the main theorem of this section. It originated from \cite[Thm.~11]{drostuvog10}.

\begin{theorem-rect} {\rm \cite[Thm.~6.6]{drofulkosvog21}}\label{thm:bounded-finaddorder-r-c-det} Let $\Sigma$ be a ranked alphabet. Moreover, let $\B$ be a strong bimonoid and let $\cA$ be a $(\Sigma,\B)$-wta such that $\rmH(\cA)$ is finite. If, for each $b \in \C(\cA)$, the mapping $f_{\cA,b}$ is bounded or $b$ has finite additive order, then we can construct a crisp-deterministic $(\Sigma,\B)$-wta $\cB$ such that $\runsem{\cA} = \sem{\cB}$.
\end{theorem-rect}
\begin{proof} Since $\rmH(\cA)$ is finite, also $\C(\cA)$ is finite.  By Lemma \ref{lm:compute-c(A)}, we construct $\C(\cA)$. Then, by Lemma \ref{lm:bounded-finite-order=>rec-step-mapping}, for each $b \in \C(\cA)$, we construct  a crisp-deterministic $(\Sigma,\B)$-wta $\cB_b$ such that $\sem{\cB_b} = r_{\cA,b}$.  
Then by \eqref{eq:semantics=sum-f}, we have $\runsem{\cA} = \bigoplus_{b \in\C(\cA)} \sem{\cB_b}$.
Hence, by Theorem \ref{thm:closure-sum}(2), we can construct a crisp-deterministic $(\Sigma,\B)$-wta $\cB$ such that $\runsem{\cA} = \sem{\cB}$.
\end{proof}

There is a slight alternative to the construction in the proof of Theorem \ref{thm:bounded-finaddorder-r-c-det}  if $\B$ is bi-locally finite. Instead of computing the sets $\rmH(\cA)$ and  $\C(\cA)$ (by Lemmas \ref{lm:compute-c(A)-normal-form} and \ref{lm:compute-c(A)}), we can construct the set $\langle \mathrm{wts}(\cA)\rangle_{\{\otimes\}}$ (by Lemma \ref{obs:Knaster-Tarski-applied-to-algebras}). Then we could replace, in Lemmas \ref{lm:f_(a,b)-recog} and \ref{lm:bounded-finite-order=>rec-step-mapping}, the sets $\rmH(\cA)$ and  $\C(\cA)$ by the set $\langle \mathrm{wts}(\cA)\rangle_{\{\otimes\}}$, and we are also able to construct a crisp-deterministic wta which is run-equivalent to $\cA$. However, this modified construction has the disadvantage that, in the construction of $\cA_b'$ in Lemma \ref{lm:f_(a,b)-recog}, its state space contains, in general, too many useless states, because many of the elements of $\langle \mathrm{wts}(\cA)\rangle_{\{\otimes\}}$ cannot be generated by~$\cA$.
Note, however, that bi-local finiteness is a very strong restriction.
In contrast, Theorem~\ref{thm:bounded-finaddorder-r-c-det} holds for \emph{all} strong bimonoids and gives a structural condition
for the wta  $\cA$  which ensures its crisp-determinizability.

Clearly, Theorem \ref{thm:bounded-finaddorder-r-c-det} also applies to strong bimonoids which are additively idempotent and multiplicatively locally finite. However, in that case we can show a considerably easier construction because we can use the constructions in the proof of Theorem \ref{thm:normalizing-transition-weights} and Lemma \ref{lm:add-idemp-unit-trans-cd}. 
As preparation, we show a crisp-determinization of crisp wta over additively idempotent strong bimonoids.

  \begin{lemma} \rm \label{lm:add-idemp-unit-trans-cd} Let $\B$ be additively idempotent. Let $\cA$ be a crisp $(\Sigma,\B)$-wta. Then we can construct a crisp-deterministic $(\Sigma,\B)$-wta $\cB$ such that $\runsem{\cA} = \sem{\cB}$.
  \end{lemma}

 \begin{proof} Let $\cA=(Q,\delta,F)$. We consider the state algebra $\St(\cA)=(\cP(Q),\delta_Q)$ of $\cA$ (cf. Section~\ref{sec:state-algebra-of-wta}) and we define the mapping $F': \cP(Q) \to B$ for each $U \in \cP(Q)$ by
   \[
(F')_U = \bigoplus_{q \in U} F_q \enspace.
     \]
Then we prove that $\runsem{\cA} = F' \circ \state$. Let $\xi \in \T_\Sigma$. We recall that, for each $n \in \mathbb{N}$ and $b \in B$, the expression $nb$ abbreviates the sum $b \oplus \ldots \oplus b$ with $n$ summands.
  \begingroup
    \allowdisplaybreaks
    \begin{align*}
      \runsem{\cA}(\xi) &=  \bigoplus_{q \in Q} \bigoplus_{\rho \in \R_\cA(q,\xi)} \wt_\cA(\xi,\rho) \otimes F_{q}
      \tag{by \eqref{equ:runsem-splitted-set-of-runs}}\\
                        &=  \bigoplus_{q \in \QR{\cA}{\xi}} \bigoplus_{\rho \in \R_\cA(q,\xi)} \wt_\cA(\xi,\rho) \otimes F_{q}
      \tag{for the definition of $\QR{\cA}{\xi}$ cf. Section \ref{sect:properties-of-wta}}\\
                        &= \bigoplus_{q \in \QR{\cA}{\xi}}\bigoplus_{\substack{\rho \in \R_\cA(q,\xi):\\\wt_\cA(\xi,\rho)=\1}}  F_{q}
      \tag{because $\wt_\cA(\xi,\rho) \in \{\0,\1\}$}\\
                        &= \bigoplus_{q \in \QR{\cA}{\xi}} |\{\rho \in \R_\cA(q,\xi) \mid \wt_\cA(\xi,\rho)=\1\}|  F_{q}
                          \tag{because $F_q$ does not depend on $\rho$}\\
                   &= \bigoplus_{q \in \QR{\cA}{\xi}}  F_{q}
                     \tag{by idempotency of $\oplus$}\\
                        &= (F')_{\QR{\cA}{\xi}}
      \tag{by definition of $F'$}\\
                        &= (F')_{\state(\xi)}
                          \tag{by Lemma~\ref{obs:state-properties-of-wta}(5)}
                         \enspace.
   \end{align*}
   \endgroup
   Since $\St(\cA)$ is a finite $\Sigma$-algebra, by  Lemma~\ref{lm:fin-algebra-wta}, we can construct a crisp-deterministic $(\Sigma,\B)$-wta $\cB$ such that $\sem{\cB} = F' \circ \state$. Hence $\sem{\cB} = \runsem{\cA}$.
   \end{proof}
 
 Then the next corollary follows from  Theorem \ref{thm:normalizing-transition-weights} and Lemma \ref{lm:add-idemp-unit-trans-cd}. 
 

 \begin{corollary-rect}\rm \label{thm:cd-for-l-valued-wta} Let $\Sigma$ be a ranked alphabet.  Let $\B$ be an additively idempotent and multiplicatively locally finite strong bimonoid. Moreover, let $\cA$ be a $(\Sigma,\B)$-wta. We can construct a crisp-deterministic $(\Sigma,\B)$-wta $\cB$ such that $\runsem{\cA} = \sem{\cB}$.
 \end{corollary-rect}
 
\begin{proof} Let $\cA$ be a $(\Sigma,\B)$-wta. Since $\B$ is multiplicatively locally finite, the set $\rmH(\cA)$ (cf. \eqref{alg:H(A)}) is finite. Then, by Theorem \ref{thm:normalizing-transition-weights}, we can construct a crisp $(\Sigma,\B)$-wta $\cB$ such that $\runsem{\cB}=\runsem{\cA}$.

   Since $\B$ is additively idempotent, by Lemma \ref{lm:add-idemp-unit-trans-cd} we can construct a crisp-deterministic $(\Sigma,\B)$-wta $\cC$ such that $\runsem{\cB} = \sem{\cC}$.
   \end{proof}

Next we show some examples of  additively idempotent and multiplicatively locally finite strong bimonoid:
\begin{compactitem}
      \item each bounded lattice.
\item the semiring $\Natmaxplusn=([0,n]_{-\infty},\max,\hat{+}_n,-\infty,0)$ from Example \ref{ex:semirings}(\ref{ex:Nat-max-plus-n}).
\item the semiring $\Natmaxmin = (\mathbb{N}_\infty,\max,\min,0,\infty)$ from Example \ref{ex:semirings}(\ref{def:max-min-semiring}).
\item the strong bimonoid   $(B,\max,\odot,0,1)$  where
  \begin{compactitem}
    \item $B =\{0\} \cup \{b \in \mathbb{R}\mid \lambda \le b \le 1\}$ for some $\lambda \in \mathbb{R}$ with $0 < \lambda< \frac{1}{2}$  and 
    \item $a\odot b  = a \cdot b$ if $a \cdot b \ge \lambda$, and $0$ otherwise,
    \end{compactitem}
    and  $\cdot$ is  the usual multiplication of real numbers (thus, $\odot$ is the same as the multiplication of $\Trunc_\lambda$ of  Example~\ref{ex:strong-bimonoids}(\ref{ex:interval-strong-bimonoid})).
  \end{compactitem}

\subsection{Equivalence of bi-local finiteness and run crisp-determinizability}
\label{subsec:equv-run-cr-det}

First we recall, for a $(\Sigma,\B)$-wta $\cA=(Q,\delta,F)$, the definitions of $\rmH(\cA)$ and of $\C(\cA)$:
\begin{align*}
\rmH(\cA) = \{\wt(\xi,\rho)  \mid \xi \in \T_\Sigma , \rho \in \R_\cA(\xi)\} \ \text{ and } \
\C(\cA) = \{\wt(\xi,\rho) \otimes F_{\rho(\varepsilon)} \mid \xi \in \T_\Sigma , \rho \in \R_\cA(\xi)\} \enspace.
\end{align*}
Then we can prove the following characterization of bi-locally finiteness in terms of crisp-determinization, where the equivalence with Statement (C) is due to \cite{dro22}. It can be  compared directly to Theorem~\ref{thm:loc-finite-rec-step-function}. For the comparison we recall that each locally finite strong bimonoid is bi-locally finite, but not vice versa (cf. the Euler diagram in Figure \ref{fig:Euler-diagram-extended}). We note that in Lemma \ref{lm:bi-loc-finite-run-image finite} we have already proved that (A)$\Rightarrow$(C).

\begin{theorem-rect} \label{thm:bi-loc-finite-rec-step-function} {\rm (cf. \cite[Cor.~7.7]{fulkosvog19})} Let $\B=(B,\oplus,\otimes,\0,\1)$ be a strong bimonoid. Then the following four statements are equivalent.
\begin{compactenum}
\item[(A)]  $\B$ is bi-locally finite.
\item[(B)]  For each ranked alphabet $\Sigma$ and for each $(\Sigma,\B)$-wta $\cA$, we can construct a  crisp-deterministic $(\Sigma,\B)$-wta $\cB$ such that $\runsem{\cA}=\sem{\cB}$.  
\item[(C)] For each ranked alphabet $\Sigma$ and for each $(\Sigma,\B)$-wta $\cA$, the set $\im(\runsem{\cA})$  is finite.
\item[(D)] For each string ranked alphabet $\Sigma$ and for each $(\Sigma,\B)$-wta $\cA$, the set $\im(\runsem{\cA})$  is finite.
\end{compactenum}
Thus, for each ranked alphabet $\Sigma$, if $\B$ is bi-locally finite, then $\Rec^{\mathrm{run}}(\Sigma,\B) \subseteq \Rec^{\mathrm{init}}(\Sigma,\B)$, and if $\B$ is locally finite, then $\Rec^{\mathrm{run}}(\Sigma,\B) = \Rec^{\mathrm{init}}(\Sigma,\B)$.
\end{theorem-rect}
\begin{proof} Proof of (A)$\Rightarrow$(B): Since $\B$ is bi-locally finite, the set $\rmH(\cA)$ is finite and each $b\in \C(\cA)$ has finite additive order. Hence the statement follows from Theorem \ref{thm:bounded-finaddorder-r-c-det}.

  \
  
Proof of (B)$\Rightarrow$(C): It follows from Theorem \ref{thm:crisp-det-algebra}(A)$\Rightarrow$(C).

\

Proof of (C)$\Rightarrow$(D): It is obvious.

\

Proof of  (D)$\Rightarrow$(A): This can be obtained by an easy adaptation of \cite[Lm.~12]{drostuvog10} from weighted string automata to wta over string ranked alphabets as follows. 

  First we show that the additive monoid $(B,\oplus,\0)$ is locally finite. Since $\oplus$ is commutative and associative, it suffices to show that, for each $b \in B$, the monoid $\langle b \rangle_{\{\oplus\}}$ is finite. Thus, let $b \in B$ and $\Sigma = \{\gamma^{(1)}, \alpha^{(0)}\}$. We construct the $(\Sigma,\B)$-wta $\cA=(Q,\delta,F)$, where $Q=\{p,q\}$, $F_p=\0$, $F_q=\1$, and $\delta$ is defined as follows (cf. Figure \ref{fig:add-loc-finite}):
  \begin{compactitem}
  \item $\delta_0(\varepsilon,\alpha,p) = \delta_1(p,\gamma,p)= \delta_1(q,\gamma,q)= \1$,
  \item $\delta_1(p,\gamma,q)=b$, and
  \item $\delta_0(\varepsilon,\alpha,q) = \delta_1(q,\gamma,p)= \0$.
  \end{compactitem}
  Then, for each $n \in \mathbb{N}$, we have $\runsem{\cA}(\gamma^n(\alpha)) = nb$ (cf. \eqref{equ:nb-nexpb}). Hence, $\langle b \rangle_{\{\oplus\}} \subseteq \im(\runsem{\cA})$.
  By (D), the set $\im(\runsem{\cA})$ is finite. Hence $\langle b \rangle_{\{\oplus\}}$ is also finite. 

Next we prove that the multiplicative monoid $(B,\otimes,\1)$ is locally finite.  Let $n \in \mathbb{N}$ and $b_1, \ldots, b_n \in B$. We show that the set
\[B'=\{b_{l_1} \otimes\ldots\otimes b_{l_m}\mid m\in\mathbb{N}, l_1,\ldots,l_m \in [n]\}\]
is finite. Let $\Sigma = \{\gamma_1^{(1)}, \gamma_2^{(1)}, \beta^{(0)}\}$.  We construct the $(\Sigma,\B)$-wta $\cA'=(Q',\delta',F')$ with $Q'=\{q_0,\ldots,q_n\}$, $F'_{q_0}=\1$ and $F'_{q}=\0$ for every $q \in Q \setminus \{q_0\}$, and $\delta'$ is defined as follows (cf. Figure~\ref{fig:mult-loc-fin}):
\begin{compactitem}
\item $\delta'_0(\varepsilon,\beta,q_0)=\1$,
\item $\delta'_1(q_{i-1},\gamma_1,q_i)=\1$ for each $i \in [n]$,
\item $\delta'_1(q_i,\gamma_2,q_0)=b_i$ for each $i \in [n]$, and
\item $\delta'_k(q'_1 \cdots q'_k,\sigma,q')=\0$ for each other combination of $k\in\mathbb{N}$, $\sigma\in\Sigma^{(k)}$ and $q'_1\cdots q'_k\in Q^k$ and $q'\in Q'$.
\end{compactitem}
Hence, $\runsem{\cA'}(\gamma_2\gamma_1^{l_m}\gamma_2\gamma_1^{l_{m-1}}\gamma_2\ldots\gamma_1^{l_1}\beta)=b_{l_1} \otimes \ldots  \otimes b_{l_m}$ for every $m \in \mathbb{N}$ and $l_1,\ldots,l_m \in [n]$. Thus $B' \subseteq \im(\runsem{\cA'})$. By (D),
 the set $\im(\runsem{\cA'})$ is finite, and therefore $B'$ is finite. Hence $(D) \Rightarrow (A)$.
   
 For the proof of the next statement, let  $\cA$ be a $(\Sigma,\B)$-wta. If $\B$ is bi-locally finite, then by (A)$\Rightarrow$(B) we can construct a crisp-deterministic $(\Sigma,\B)$-wta $\cB$ such that $\runsem{\cA}=\sem{\cB}$. Since, in particular, $\sem{\cB}= \initialsem{\cB}$, we have $\Rec^{\mathrm{run}}(\Sigma,\B) \subseteq \Rec^{\mathrm{init}}(\Sigma,\B)$. 

If $\B$ is locally finite, then this inclusion together with Theorem \ref{thm:subset-method-lf-strong-bm}(2)
imply that $\Rec^{\mathrm{run}}(\Sigma,\B) = \Rec^{\mathrm{init}}(\Sigma,\B)$.
\end{proof}

\begin{figure}
  \centering
\begin{tikzpicture}
\tikzset{node distance=7em, scale=0.6, transform shape}
\node[state, rectangle]  (1) {\Large $\alpha$};
\node[state, right of=1] (2) {\Large $p$};
\node[state, rectangle, above of=2, yshift=-0.7cm] (3) {\Large $\gamma$};
\node[state, rectangle, right of=2] (4) {\Large $\gamma$};
\node[state, right of=4] (5) {\Large $q$};
\node[state, rectangle, above of=5, yshift=-0.7cm] (6) {\Large $\gamma$};

\tikzset{node distance=2em}
\node[above of=1] (w1) {$\1$};
\node[above of=3] (w3) {$\1$};
\node[above of=4] (w4) {$b$};
\node[below of=5] (w5) [right=0.05cm] {$\1$};
\node[above of=6] (w6) {$\1$};

\draw (1) edge[->,>=stealth] (2);
\draw (2) edge[->,>=stealth] (4);
\draw (4) edge[->,>=stealth] (5);

\draw (2.45)   edge[->,>=stealth, out=40,   in=-5,    looseness=1.4] (3.east);
\draw (3.west) edge[->,>=stealth, out=180+5,in=180-40,looseness=1.4] (2.135);
\draw (5.45)   edge[->,>=stealth, out=40,   in=-5,    looseness=1.4] (6.east);
\draw (6.west) edge[->,>=stealth, out=180+5,in=180-40,looseness=1.4] (5.135);
\end{tikzpicture}
  
\caption{\label{fig:add-loc-finite} The $(\Sigma,\B)$-wta $\cA$ with $\runsem{\cA}(\gamma^n(\alpha)) = nb$.}
\end{figure}

\begin{figure}
\centering
\begin{tikzpicture}
\tikzset{node distance=7em, scale=0.6, transform shape}
\node[state, rectangle] (b) {\Large $\beta$};
\node[state, right of=b] (q0) {\Large $q_0$};
\node[state, rectangle, right of=q0](g1-1) {\Large $\gamma_1$};
\node[state, right of=g1-1] (q1) {\Large $q_1$};
\node[state, rectangle, right of=q1] (g1-2) {\Large $\gamma_1$};
\node[state, right of=g1-2] (q2) {\Large $q_2$};
\node[state,right of=q2,opacity=0] (empty) {}; 
\node[state, rectangle, right of=empty]  (g1-3) {\Large $\gamma_1$};
\node[state, right of=g1-3] (qn) {\Large $q_n$};

\coordinate (q0-qn) at ($(q0)!0.5!(qn)$);

\node[state, rectangle] (g2-1) at (g1-1  |- 0,5.5em) {\Large $\gamma_2$};
\node[state, rectangle] (g2-2) at (q1    |- 0,11em) {\Large $\gamma_2$};
\node[state, rectangle] (g2-3) at (q0-qn |- 0,16.5em) {\Large $\gamma_2$};

\node (dots) at ($(q2.east)!0.5!(empty.east)$) {$\ldots$};

\tikzset{node distance=2em}
\node[above of=b]    (wb)    {$\1$};
\node[above of=g1-1] (wg1-1) {$\1$};
\node[above of=g1-2] (wg1-2) {$\1$};
\node[above of=g1-3] (wg1-3) {$\1$};
\node[above of=g2-1] (wg2-1) {$b_1$};
\node[above of=g2-2] (wg2-2) {$b_2$};
\node[above of=g2-3] (wg2-3) {$b_n$};
\node[below of=q0] (wq0) {$\1$};

\draw (b)    edge[->,>=stealth] (q0);
\draw (q0)   edge[->,>=stealth] (g1-1);
\draw (g1-1) edge[->,>=stealth] (q1);
\draw (q1)   edge[->,>=stealth] (g1-2);
\draw (g1-2) edge[->,>=stealth] (q2);
\draw (empty)edge[->,>=stealth] (g1-3);
\draw (g1-3) edge[->,>=stealth] (qn);

\draw (q1.90) edge[->,>=stealth, out=90, in=0]  (g2-1.0);
\draw (g2-1)  edge[->,>=stealth, out=180,in=80] (q0.60);
\draw (q2)   edge[->,>=stealth, out=90, in=0] (g2-2);
\draw (g2-2) edge[->,>=stealth, out=180,in=85] (q0.80);
\draw (qn)   edge[->,>=stealth, out=90, in=0] (g2-3);
\draw (g2-3) edge[->,>=stealth, out=180,in=90] (q0.100);
\end{tikzpicture}

\caption{\label{fig:mult-loc-fin} The $(\Sigma,\B)$-wta $\cA'$ with $\runsem{\cA'}(\gamma_2\gamma_1^{l_m}\gamma_2\gamma_1^{l_{m-1}}\gamma_2\ldots\gamma_1^{l_1}\beta)=b_{l_1} \otimes \ldots  \otimes b_{l_m}$.}
\end{figure}

As a consequence of Theorem \ref{thm:bi-loc-finite-rec-step-function}(A)$\Rightarrow$(B) and Theorem \ref{thm:crisp-det-algebra} (A)$\Rightarrow$(C), we show one more preimage theorem.

\begin{corollary}\rm \label{cor:preimage-fin-stb}\cite[Cor.~6.9]{drofulkosvog21} Let $\B$ be bi-locally finite and $\cA$ a $(\Sigma,\B)$-wta. Then, for each $b \in B$, we can construct a $\Sigma$-fta $A$ such that $\LL(A) = (\runsem{\cA})^{-1}(b)$.
\end{corollary}

Another consequence of Theorem \ref{thm:bi-loc-finite-rec-step-function} (in combination with Theorem~\ref{thm:not-monadic-wta-can-compute-closure-of-finite-subset})  is the following corollary.

\begin{corollary}\rm \label{cor:run-cr-det-and-not-init-al-cr-det} Let $\Sigma$ be an arbitrary branching ranked alphabet. Moreover, let $\B$ be bi-locally finite and not locally finite. Then there exists a $(\Sigma,\B)$-wta $\cA$ which is run crisp-determinizable and not initial algebra crisp-determinizable.
  \end{corollary}

  \begin{proof} Since $\B$ is bi-locally finite, by Theorem~\ref{thm:bi-loc-finite-rec-step-function}(A)$\Rightarrow$(B), $\cA$ is run crisp-determinizable.  On the other hand, let $H \subseteq B$ be a finite set such that $\langle H \rangle_{\{\oplus,\otimes,\0,\1\}}$ is infinite.
    By Theorem~\ref{thm:not-monadic-wta-can-compute-closure-of-finite-subset}, we can construct a $(\Sigma,\B)$-wta $\cA$ such that $\im(\initialsem{\cA})=\langle H \rangle_{\{\oplus,\otimes,\0,\1\}}$. Since $\im(\initialsem{\cA})$ is infinite, by Lemma~\ref{lm:fin-algebra-crisp-det-wta}(3), $\cA$ is not initial algebra crisp-determinizable.
    \end{proof}

\subsection[Run crisp-determinizability for  past-finite monotonic strong bimonoids]{\sloppy Characterization of run crisp-determinizability for  past-finite monotonic strong bimonoids}
\label{sec:char-past-finite-crisp}

In this subsection we show a characterization of run crisp-determinizability for wta over past-finite monotonic strong bimonoids (cf. Theorem \ref{thm:past-finite-mon-char-cd}). Each past-finite monotonic strong bimonoid shares many properties with the semiring of natural numbers, like (a) having a partial order on its carrier set (which is not necessarily total), (b) being zero-sum free and zero-divisor free,  (c) monotonicity of its operations with respect to that partial order, and (d) a strong kind of well-foundedness (called past-finiteness). However, in general, distributivity is not required and also does not follow from the axioms. The idea of past-finite strong bimonoids and the characterization stems from \cite{drofulkosvog21}.

\index{past-finite}
First we define these strong bimonoids. 
Let $(B,\preceq)$ be a partially  ordered set. We say that $\preceq$ is \emph{past-finite} if for each $b \in B$ the set $\past(b) = \{a \in B \mid a \preceq b\}$ is finite. In the sequel, we write $a \prec b$ if $a \preceq b$ and $a \ne b$. 

Thus, for each past-finite partial order $\preceq$, the relation $\prec$ is well-founded. However, there exists a well-founded relation $\prec$ such that the relation $\preceq$ (defined by $a \preceq b$ if  $a \prec b$ or $a = b$) is a partial order and it is not past-finite. For instance, let $B=\{b,a_1,a_2,\ldots\}$ and $\prec \,= \{(a_1,b), (a_2,b),\ldots\}$.

\index{monotonic}
Let us consider a strong bimonoid $\B$ and let $\preceq$  be a partial order on $B$.  We say that $\B$  is \emph{monotonic with respect to $\preceq$} (cf. \cite[Def.~5]{borfulgazmal05} and \cite{drofulkosvog21}) if
\begin{eqnarray}
& \text{for every $a,b \in B$, we have $a \preceq a \oplus b$ and } \label{MON1}\\
& \text{for every  $a,b,c \in B \setminus \{\0\}$  with  $b \ne \1$  we have:  $a \otimes c \prec a \otimes b \otimes c$\enspace.\label{MON2}}
\end{eqnarray}
From \eqref{MON2} we easily  obtain that
\begin{equation}
\text{for every  $a,b \in B \setminus \{\0\}$  with  $b \ne \1$  we have:  $a  \prec a \otimes b$ and $a  \prec b \otimes a$\enspace.\label{MON3}}
\end{equation}
If $\B$ is monotonic with respect to $\preceq$, then we also write that $(\B,\preceq)=(B,\oplus,\otimes,\0,\1,\preceq)$
is a \emph{monotonic strong bimonoid}.
A monotonic strong bimonoid $(\B,\preceq)$ has several properties \cite[Lm.~14]{borfulgazmal05}.
\index{one-product free}

\begin{lemma}\label{lm:monotonic-properties} \rm If $(\B,\preceq)$ is a monotonic strong bimonoid, then
\begin{compactitem}
\item[(1)] $\0 \preceq b$ for each $b\in B$, and $\1 \preceq b$ for each $b\in B\setminus\{\0\}$,
\item[(2)] $\B$ is positive, i.e., zero-sum free and zero-divisor free,
\item[(3)] $\B$ is \emph{one-summand free}, i.e., for every $a,b \in B$ if $a \oplus b = \1$, then $a,b\in\{\0, \1\}$, and
\item[(4)] $\B$ is \emph{one-product free}, i.e., for every $a,b \in B$ if $a \otimes b = \1$, then $a= b= \1$.
\end{compactitem}
\end{lemma}
\begin{proof}
Let $a, b \in B$.

Proof of (1): By \eqref{MON1}, we
      obtain $\mathbb{0} \preceq \mathbb{0} \oplus b = b$. If $b \neq \mathbb{0}$,
      then by \eqref{MON3},
      $\mathbb{1} \preceq \mathbb{1} \otimes b = b$.
      
         \
     
 Proof of (2):   First we prove zero-sum freeness. Let $b \in B\setminus\{\0\}$. Then, by Item~(1) and \eqref{MON1}, we have $\mathbb{0} \prec b \preceq a \oplus b$. Hence $\B$ is zero-sum free.

     Next we show zero-divisor freeness. Let $a, b \in B\setminus\{\0\}$. Then, by Item~(1) and \eqref{MON3}, $\mathbb{0} \prec a \preceq a \otimes b$. Thus $\B$ is zero-divisor free.

     \
     
Proof of (3):  Assume that $a \notin \{\mathbb{0},\mathbb{1}\}$.
      Then, by~(1) and \eqref{MON1}, we obtain $\mathbb{1} \prec a \preceq a \oplus b$. Hence
      $\B$ is one-summand free.
      
     \
 
 Proof of (4):  We show by contradiction that $a \otimes b = \mathbb{1}$
      implies $a = \mathbb{1}$ and $b = \mathbb{1}$. Assume that
      $a \neq \mathbb{1}$ or $b \neq \mathbb{1}$ and $a \otimes b =
      \mathbb{1}$. Apparently, $a, b \in B\setminus\{\0\}$. Hence by~(1)
      and \eqref{MON3} we obtain
      $\mathbb{1} \preceq b \prec a \otimes b$ or $\mathbb{1} \preceq a \prec a \otimes
      b$. This contradicts to $a \otimes b =
      \mathbb{1}$. Consequently, $\B$ is one-product free.
    \end{proof}
    
Lemma \ref{lm:monotonic-properties}(1) and \eqref{MON3} imply that, for each $b \in B \setminus \{\0,\1\}$ and $n \in \mathbb{N}$, we have $\0 \prec \1 \prec  b^n \prec b^{n+1}$, i.e., 
\begin{equation*}
\0 \prec \1 \prec b \prec b\otimes b \prec b \otimes b \otimes b \ \cdots \enspace.
\end{equation*}
Hence, if  $(\B,\preceq)$ is a finite monotonic strong bimonoid, then  $B$ has only two elements $\0$ and $\1$, we have  $\0 \prec \1$ and $\1 \oplus \1 = \1$. Thus $\B$  is isomorphic to the Boolean semiring  $\Boole$.

\index{monotonic!past-finite}
We call a monotonic strong bimonoid $(B,\oplus,\otimes,\0,\1,\preceq)$ \emph{past-finite} if $(B,\preceq)$ is past-finite. 
 In \cite[Ex.~2.2-2.4]{drofulkosvog21} a number of past-finite monotonic strong bimonoids are shown. Here we only mention the following examples:
\begin{compactenum}
\item[(a)] the semiring $(\Boole,\le)$, where $0 \le 1$,
\item[(b)] the semiring $(\Nat,\le)$ of natural numbers, where $\le$ is the common relation ``less than or equal to'' on  $\mathbb{N}$,
\item[(c)] the semiring $(\Natmaxplus,\le)$, where $\le$ is the natural extension of $\le$ to $\mathbb{N}_{-\infty}$, 

\item[(d)] the semiring $(\cP_{\mathrm{f}}(\Gamma^*),\cup,\cdot,\emptyset,\{\varepsilon\},\preceq)$  where $\cP_{\mathrm{f}}(\Gamma^*)$ is the set of finite subsets of $\Gamma^*$ and $L_1 \preceq L_2$ if there exists an injective mapping $f: L_1 \to L_2$ such that $w$ is a substring of $f(w)$ for each $w \in L_1$,
  
\item[(e)] the semiring $(\cP_{\mathrm{f}}(\mathbb{N}),\cup,+,\emptyset,\{0\},\preceq)$  where $\cP_{\mathrm{f}}(\mathbb{N})$ is the set of finite subsets of $\mathbb{N}$, the operation $+$ is extended in the usual way to sets, and $N_1 \preceq N_2$ if there exists an injective mapping $f: N_1 \to N_2$ such that $n \le f(n)$ for each $n \in N_1$, and 
\item[(f)]  the plus-plus strong bimonoid $(\PP_\mathbb{N},\le) = (\mathbb{N}_\0,\oplus,+,\0,0,\leq)$ of natural numbers (cf. Example \ref{ex:strong-bimonoids}(\ref{ex:plus-plus-sb})), where $\leq$  is the usual order on  $\mathbb{N}$  together with
  $\0 \le x$  for each  $x \in \mathbb{N}$. Then  $(\PP_\mathbb{N},\le)$  is a past-finite monotonic strong bimonoid which is not a semiring.
  \end{compactenum}

The set of past-finite monotonic strong bimonoids shares another property with the semiring of natural numbers (cf. Lemma \ref{lm:preimage-N-1}).

\begin{theorem}\label{thm:preimage-past-finite-monotonic} {\rm \cite[Thm.~6.10]{drofulkosvog21}} Let $(\B,\preceq)$ be past-finite monotonic strong bimonoid and $\cA$ a $(\Sigma,\B)$-wta and $b\in B$. 
\begin{compactenum}
\item[(1)] The tree language $(\runsem{\cA})^{-1}(b)$ is recognizable.  Hence $\runsem{\cA}$ has the preimage property.
\item[(2)] If  the set $\past(b)$ can be constructed, then we can construct  a finite-state $\Sigma$-tree automaton which recognizes $\sem{\cA}^{-1}(b)$.
\end{compactenum}
\end{theorem}

\begin{proof}  Let $b \in B$. 

Proof of (1): We define the set $C = B \setminus \past(b) = \{a \in B\mid a \npreceq b\}$. Moreover, we define the equivalence relation $\sim$ on the set $B$ such that its classes are the singleton sets $\{a\}$ for each $a \in \past(b)$ and the set $C$.

  We claim that $\sim$  is a congruence relation. To show this let $c,c' \in C$  and  $d \in B$. Since $(\B,\preceq)$ is monotonic, we have $c \preceq c \oplus d$ and $c' \preceq c' \oplus d$. Hence $c \oplus d  \in C$ and $c' \oplus d \in C$. By commutativity of $\oplus$ we also have $d \oplus c \in C$ and $d \oplus c' \in C$. Thus $\oplus$ obeys the structure of the equivalence classes. Now let $d \ne \0$. Then we obtain $c \preceq c \otimes d$  and  $c \preceq d\otimes c$, showing that $c\otimes d \in C$ and $d\otimes c \in C$,  and similarly  $c'\otimes d \in C$ and $d\otimes c' \in C$. Hence $\sim$ is a congruence relation on the strong bimonoid $\B$.

  By definition, the quotient strong bimonoid of $\B$ modulo $\sim$ is finite. Let $h: B \to B/_\sim$ be the canonical strong bimonoid homomorphism, i.e., for each $c \in B$ we let $h(c) = [c]_\sim$. By Lemma \ref{thm:closure-sr-hom}, we obtain $(h \circ \runsem{\cA}) \in \Rec^{\mathrm{run}}(\Sigma,\B/_\sim)$. Moreover $(\runsem{\cA})^{-1}(b) = (h \circ \runsem{\cA})^{-1}(\{b\})$.  Since $B/_\sim$ is finite, by Corollary \ref{cor:preimage-fin-stb}, the $\Sigma$-tree language $(h\circ \runsem{\cA})^{-1}(\{b\})$ is recognizable. Thus  $\runsem{\cA}$ has the preimage property.
  
  \

  Proof of (2):  Assume that $\past(b)$ is constructed. Then we can construct the congruence relation $\sim$ as defined in the proof of Statement 1 and the canonical strong bimonoid homomorphism $h: B \to B/_\sim$. 
  We can construct the $(\Sigma,\B/_\sim)$-wta $h(\cA)$, and by Lemma \ref{lm:f-image-equivalent}, we have $\runsem{h(\cA)}= h \circ \runsem{\cA}$.
  Since $\B/_\sim$ is finite,  by Corollary \ref{cor:preimage-fin-stb}, we can construct   a finite-state $\Sigma$-tree automaton which recognizes  $(\runsem{h(\cA)})^{-1}(\{b\})$.
 \end{proof}

By Theorems \ref{thm:crisp-det-algebra} and \ref{thm:preimage-past-finite-monotonic}(1), $\cA$ is run crisp-determinizable if and only if $\im(\runsem{\cA})$ is finite. As first step towards our characterization theorem we analyse the impact of this finiteness property on the structure of a $(\Sigma,\B)$-wta $\cA$ where 
$(\B,\preceq)$ is a past-finite monotonic strong bimonoid.  In particular, we consider local-non-zero runs and  small loops of $\cA$. (We recall that the definition of a run on a context and the weight of such a run can be found on page \pageref{p:par-extension-runs-to-contexts}.)

For each $\xi \in \T_\Sigma(\{z\})$, we call a run $\rho \in  \R_\cA(\xi)$ \emph{local-non-zero} if for each $w \in \pos_\Sigma(\xi)$ we have $\delta_k(\rho(w1) \cdots \rho(wk), \sigma, \rho(w)) \ne \0$ where $k = \rk(\xi(w))$ and $\sigma = \xi(w)$. We note that, for every $\xi \in \T_\Sigma$ and $\rho \in  \R_\cA(\xi)$, the run $\rho$ is local-successful (cf. Subsection \ref{subsect:basic-definitions-runs}) if it is local-non-zero and $F_{\rho(\varepsilon)}\neq \0$.

We say that \emph{small loops of $\cA$ have weight~$\1$} if, for every $q \in Q$, $c \in \C_\Sigma$, and local-non-zero $\rho \in \R_\cA(q,c,q)$, if $\height(c) < |Q|$, then $\wt(c,\rho) = \1$. For the definition of $\R_\cA(q,c,q)$ we refer to page \pageref{p:par-extension-runs-to-contexts}.

\begin{theorem} \label{thm:past-mon+imA-fin->small-loops} {\rm \cite[Thm.~7.1]{drofulkosvog21}}
Let $(\B,\preceq)$ be a past-finite monotonic strong bimonoid and $\cA$ be a local-trim $(\Sigma,\B)$-wta.   If $\im(\runsem{\cA})$ is finite, then small loops of $\cA$ have weight~$\1$.
\end{theorem}
\begin{proof}
  We prove by contraposition.  Suppose there exist  $q \in Q$, $c \in \C_\Sigma$, and local-non-zero $\rho \in \R_\cA(q,c,q)$ such that $ \height(c) < |Q|$ and $\wt(c,\rho) \ne \1$. Since $\rho$ is local-non-zero and $(\B,\preceq)$ is monotonic, we have that $\1 \prec \wt(c,\rho)$. Since $\cA$ is local-trim, the state $q$ is local-useful and thus there exist  $\xi \in \T_\Sigma$, $\theta \in \R_\cA(q,\xi)$, $c' \in \C_\Sigma$, $q' \in Q$ with $F_{q'}\ne\0$, and $\rho' \in \R_\cA(q',c',q)$, and $\theta$ and $\rho'$ are local-non-zero. By Theorem \ref{thm:decomposition-of-a-run}, for each $n \in \mathbb{N}$, we have
\[
 \wt(c'\big[c^{n}[\xi]\big], \rho'\big[\rho^{n}[\theta]\big]) =
  l_{c',\rho'} \otimes (l_{c,\rho})^n \otimes \wt(\xi,\theta) \otimes (r_{c,\rho})^n \otimes r_{c',\rho'}\enspace.
\]
Since $\1 \prec \wt(c,\rho) = l_{c,\rho}\otimes r_{c,\rho}$ , we have $\1 \prec l_{c,\rho}$  or $\1 \prec r_{c,\rho}$, because $\B$ is one-product free. Thus, by monotonicity of $(\B,\preceq)$, we obtain
\begin{equation}\label{eq:prec-sequence}
  \wt  (c'\big[c^0[\xi]\big], \rho'\big[\rho^0[\theta]\big]  ) \prec \wt  (c'\big[c^1[\xi]\big], \rho'\big[\rho^1[\theta]\big]  ) \prec \ldots \enspace.
\end{equation}

Next we define a sequence $\xi_1, \xi_2, \xi_3, \ldots$ 
of trees in $\T_\Sigma$ such that the elements $\runsem{\cA}(\xi_1)$, $\runsem{\cA}(\xi_2)$, $\runsem{\cA}(\xi_3)$, $\ldots$ are pairwise different as follows.
We let $\xi_1 = c'\big[c[\xi]\big]$. Since $(\B,\preceq)$ is past-finite, the set $P_1 = \past(\runsem{\cA}(\xi_1))$ is finite. By \eqref{eq:prec-sequence} we choose $n_2$ such that $\wt(c'\big[c^{n_2}[\xi]\big], \rho'\big[\rho^{n_2}[\theta]\big]) \not\in P_1$ and let $\xi_2=c'\big[c^{n_2}[\xi]\big]$. Since $\rho'\big[\rho^{n_2}[\theta]\big] \in \R_\cA(q',\xi_2)$ and $(\B,\preceq)$ is monotonic, we have
\[\wt(\xi_2, \rho'\big[\rho^{n_2}[\theta]\big]) \preceq \wt(\xi_2, \rho'\big[\rho^{n_2}[\theta]\big])\otimes F_{q'}
  \preceq \bigoplus_{\kappa \in \R_\cA(\xi_2)} \wt(\xi_2,\kappa) \otimes F_{\kappa(\varepsilon)}
  = \runsem{\cA}(\xi_2).
\]
(Note that $F_{q'}$ may be $\1$.) Hence $\runsem{\cA}(\xi_2) \notin P_1$. Put $P_2 = \past(\runsem{\cA}(\xi_2))$. 
Then we choose $n_3 \in \mathbb{N}$ such that  $\wt(c'\big[c^{n_3}[\xi]\big], \rho'\big[\rho^{n_3}[\theta]\big]) \not\in P_1 \cup P_2$ and let $\xi_3=c'\big[c^{n_3}[\xi]\big]$. As before,  we have
\(\runsem{\cA}(\xi_3) \notin P_1 \cup P_2\). Continuing this process, we obtain the desired sequence of trees. It means that $\im(\runsem{\cA})$ is infinite.
\end{proof}

We recall, for a $(\Sigma,\B)$-wta $\cA=(Q,\delta,F)$, the definitions
\begin{align*}
\rmH(\cA) = \{\wt(\xi,\rho)  \mid \xi \in \T_\Sigma , \rho \in \R_\cA(\xi)\} \text{ and }
\C(\cA) = \{\wt(\xi,\rho) \otimes F_{\rho(\varepsilon)} \mid \xi \in \T_\Sigma , \rho \in \R_\cA(\xi)\} \enspace.
\end{align*}

Also a kind of inverse of Theorem \ref{thm:past-mon+imA-fin->small-loops} holds.

\begin{lemma} \label{lm:weight-of-reduced-tree} \rm \cite[Lm.~5.5,~6.7]{drofulkosvog21}
  Let $\B$ be a one-product free strong bimonoid and $\cA$ be a $(\Sigma,\B)$-wta.
  If small loops of $\cA$ have weight $\1$, then
  \begin{compactenum}
  \item[(1)] for every $\xi \in \T_\Sigma$, $q' \in Q$, and local-non-zero $\kappa \in \R_\cA(q',\xi)$, there exist $\xi' \in \T_\Sigma$ and $\kappa' \in \R_\cA(q',\xi')$ such that $\height(\xi') < |Q|$ and $\wt(\xi,\kappa) = \wt(\xi', \kappa')$ and
  \item[(2)] the set $\rmH(\cA)$ is finite.
    \end{compactenum}
  \end{lemma}
   \begin{proof} Proof of (1): The idea is to cut out iteratively local-non-zero, small loops with weight $\1$. Formally, let $\xi \in \T_\Sigma$, $q' \in Q$, and $\kappa \in \R_\cA(q',\xi)$ be a local-non-zero run. We may assume that $\height(\xi) \geq |Q|$.   Applying Theorem~\ref{thm:pumping-lemma-for-runs} (for $n=1$ and $n=0$),
  there exist  $c,c' \in \C_\Sigma$, $\zeta \in \T_\Sigma$, $q \in Q$, $\rho' \in \R_\cA(q',c',q)$, $\rho \in \R_\cA(q,c,q)$, and $\theta \in \R_\cA(q,\zeta)$ 
  such that   $\xi=c'\big[c[\zeta]\big]$, $\kappa = \rho'\big[\rho[\theta]\big]$,  $\height(c) > 0$, $\height\big(c[\zeta]\big) < |Q|$, and
  \begin{align*}
  \wt(\xi,\kappa) = & \wt(c'\big[c[\zeta]\big], \rho'\big[\rho[\theta]\big]) =
    l_{c',\rho'} \otimes  l_{c,\rho} \otimes \wt(\zeta,\theta) \otimes  r_{c,\rho} \otimes r_{c',\rho'} \ \text{ (for $n=1$)}\enspace,\\
  & \wt(c'[\zeta],\rho'[\theta]) = l_{c',\rho'} \otimes  \wt(\zeta,\theta) \otimes   r_{c',\rho'} \ \text{ (for $n=0$)}\enspace.
  \end{align*}
Since $\kappa$ is local-non-zero, also the runs $\rho'$, $\rho$, and $\theta$ are local-non-zero. Thus $\wt(c,\rho) = \1$ by our assumption.

  Moreover, by Observation \ref{obs:decomp-run-left-right}, we have 
  $\wt(c,\rho) = l_{c,\rho} \otimes r_{c,\rho}$. Since $\B$ is one-product free, we have $l_{c,\rho} =  r_{c,\rho}= \1$. Hence  we have $\wt(\xi,\kappa)=\wt(c'[\zeta],\rho'[\theta])$.
  
  Obviously,  $\rho'[\theta] \in \R_\cA(q',c'[\zeta])$  and  $\size(c'[\zeta]) < \size (\xi)$. If $\height(c'[\zeta])<|Q|$, then we are ready. 
  Otherwise we continue with  $c'[\zeta]$, $q'$, and  $\rho'[\theta]$ as before.
  After finitely many steps, we obtain  $\xi' \in T_\Sigma$  and $\kappa' \in \R_\cA(q',\xi')$
  with  $\height(\xi') < |Q|$  as required.

  \

 Proof of (2): If small loops of $\cA$ have weight  $\1$, then by Statement 1 of this lemma  we have
    \[\rmH(\cA) = \{\wt(\xi,\rho) \mid \xi \in \T_\Sigma, \height(\xi) < |Q|, \text{ and } \rho \in \R_\cA(\xi)\}\enspace.\]
     Hence $\rmH(\cA)$ is finite. 
\end{proof}

Now we can prove the main result of this subsection.

\begin{theorem-rect} \label{thm:past-finite-mon-char-cd} {\rm \cite[Thm.~6.10,~7.1]{drofulkosvog21}} Let $\Sigma$ be a ranked alphabet. Moreover, let $(\B,\preceq)=(B,\oplus,\otimes,\0,\1,\preceq)$ be a past-finite monotonic strong bimonoid and let $\cA$ be a local-trim $(\Sigma,\B)$-wta. Then the following two statements are equivalent.
  \begin{compactenum}
  \item[(A)] $\cA$ is run crisp-determinizable.
        \item[(B)] Small loops of $\cA$ have weight $\1$ and, for each $b \in \C(\cA)$, the mapping $f_{\cA,b}$ is bounded or $b$ has finite additive order.
    \end{compactenum}
\end{theorem-rect}
\begin{proof}
Proof of (A)$\Rightarrow$(B): Since $\cA$ is run crisp-determinizable, the set $\im(\runsem{\cA})$ is finite by Theorem \ref{thm:crisp-det-algebra}. Thus, by Theorem  \ref{thm:past-mon+imA-fin->small-loops}, small loops of $\cA$ have weight $\1$.

Now let $b\in \C(\cA)$. If the mapping $f_{\cA,b}$ is not bounded, then there exists an infinite sequence $\xi_1,\xi_2,\ldots$ of trees in $\T_\Sigma$ such that $f_{\cA,b}(\xi_1) < f_{\cA,b}(\xi_2) < \ldots$. By Equality \eqref{eq:semantics=sum-f}, we have $\big(f_{\cA,b}(\xi_i)\big)b \preceq \runsem{\cA}(\xi_i)$ for each $i\in \mathbb{N}$. Thus $\big(f_{\cA,b}(\xi_i)\big)b \in P$, where
$P=\bigcup_{a\in \im(\runsem{\cA})}\past(a)$. Since  $\im(\runsem{\cA})$ is finite and $(\B,\preceq)$ is past-finite, the set $P$ is also finite. 
Hence $\big(f_{\cA,b}(\xi_i)\big)b=\big(f_{\cA,b}(\xi_j)\big)b$ for some $i,j\in \mathbb{N}$ with $i<j$, which implies that $b$ has finite additive order. 

\

Proof of (B)$\Rightarrow$(A): This implication follows from Lemma \ref{lm:weight-of-reduced-tree} (we recall that each monotonic strong bimonoid is one-product free)  and Theorem \ref{thm:bounded-finaddorder-r-c-det}.
\end{proof}

Finally we show an application of Theorem \ref{thm:past-finite-mon-char-cd}.

\begin{theorem} \label{thm:run-c-d-decidable} {\rm \cite[Thm.~10]{drofulkosvog20b}} Let $(\B,\preceq)$ be an  additively locally finite and past-finite monotonic strong bimonoid. Moreover, let $\cA$ be a $(\Sigma,\B)$-wta which contains at least one  local-useful state. It is decidable whether $\cA$ is run crisp-determinizable.
\end{theorem}
\begin{proof} Let $\cA=(Q,\Sigma,\delta)$. By Theorem \ref{thm:t2}, we  can  construct a  $(\Sigma,\B)$-wta $\cA'$  such that $\cA'$ is local-trim and   $\runsem{\cA'}=\runsem{\cA}$. Hence we may assume that $\cA$ is local-trim. Since $\B$ is additively locally finite, each $b \in \C(\cA)$ has finite additive order. Thus, by Theorem~\ref{thm:past-finite-mon-char-cd}, $\cA$ is run crisp-determinizable if and only if small loops of $\cA$ have weight $\1$. The latter property is decidable because (a) there exist  only finitely many $c \in \C_\Sigma$  such that  $\height(c) < |Q|$, and (b) since  $(\B,\preceq)$ is monotonic, for every  $c \in \C_\Sigma$, $q \in Q$, and  $\rho \in \R_\cA(q,c,q)$  we have $\wt(c,\rho) = \1$ if and only if for each  $v \in \pos(c)$  we have $\delta_k(\rho(v1)\cdots\rho(vk),\sigma,\rho(v)) = \1$  where $\sigma = c(v)$  and  $k = \rk(\sigma)$, and (c) this is decidable because $\B$ has an effective test for $\1$.
\end{proof}

The decidability problem addressed in Theorem \ref{thm:run-c-d-decidable} is meaningful, because there exists an additively locally finite and past-finite monotonic semiring and a wta over that semiring which is not run crisp-determinizable. Such a semiring and wta is the arctic semiring $\Natmaxplus$ and the $(\Sigma,\Natmaxplus)$-wta $\cA$ in Example \ref{ex:height}, respectively. We recall that $\initialsem{\cA} = \runsem{\cA}=\height$ (cf. Theorem \ref{thm:bu-det:init=run}). As we mentioned $\cA$ is not run crisp-determinizable because $\im(\height)$ is not finite.

We refer the reader to  \cite{drofulkosvog20b,drofulkosvog21} for further decidability and undecidability results for wta and wsa
over past-finite monotonic strong bimonoids.

%% file: determinization-bu-reduced.tex
\chapter{Determinization of wta over semirings}
\label{ch:determinization}

\index{bu-determinizable}
In this chapter, we consider as weight algebras only semirings. A $(\Sigma,\B)$-wta $\cA$ over some semiring $\B$ is \emph{bu-determinizable} (or just: determinizable) if there exists an equivalent bu-deterministic $(\Sigma,\B)$-wta $\cB$, i.e., $\sem{\cA} = \sem{\cB}$.
 \footnote{We recall that $\sem{\cA} = \initialsem{\cA} = \runsem{\cA}$ because $\B$ is a semiring.}

If $\B$ is locally finite (e.g., finite), then we can apply the subset method of Definition \ref{constr:subset} to $\cA$ and obtain an i-equivalent crisp-deterministic $(\Sigma,\B)$-wta $\cB$. In Section \ref{sec:subset}, it will turn out that, in general, the  subset method will not yield a bu-deterministic wta, i.e., it is not able to determinize each wta. One might think that there exist other, successful approaches. However, there are recognizable weighted tree languages which are not bu-deterministically recognizable (as already in the string case, cf. \cite[pp.354]{klilommaipri04}).
Hence, if $r$ is such a weighted tree language, then each wta $\cA$ with $\sem{\cA}=r$ is not bu-determinizable.
For example, the  mapping $\height: \T_\Sigma \to \mathbb{N}$ is recognizable by a $(\Sigma,\Natmaxplus)$-wta (cf. Example~\ref{ex:height}) but it is not bu-deterministically recognizable (cf. Corollary~\ref{ex:height-not-bud}). In Section \ref{sec:negative-results-determinization} we show two further recognizable weighted tree languages which are not bu-deterministically recognizable.  

 \sloppy In Section \ref{sec:extremal-commutative-zero-divisor-free-semirings-determinization} we show an advanced version of the  subset method which is based on factorization and it is called determinization by factorization (cf.  Definition \ref{constr:det-A}). We  elaborate sufficient conditions under which  determinization by  factorization transforms a given wta into a bu-deterministic wta which is equivalent to the given wta.  More precisely, if $\B$ is an extremal and commutative semiring for which there exists a maximal factorization, and the given $(\Sigma,\B)$-wta $\cA$ has the twinning property, then $\cA$ is bu-determinizable (cf. Theorem \ref{thm:extremal-twins-determinizable}) and a bu-deterministic wta equivalent to $\cA$ can be obtained by applying determinization by factorization to $\cA$.
The approach of bu-determinization of wta by factorization was published in \cite{buemayvog10} (also cf. \cite[Sec.~5]{bue14}), and it is based on the determinization by factorization of wsa \cite{kirmae05}. In its turn, the latter is a generalization of determinization of  wsa over the tropical semiring \cite{moh97}.

 We mention that in  \cite[Sect.~4.5]{pau20} the decidability of the determinization of $(\Sigma,\Realnum_{\max,+})$-wta was considered, where $\Realnum_{\max,+}$ is the semifield $(\mathbb{R}_{-\infty},\max,+,-\infty,0)$. One of the main results is \cite[Thm. 4.33]{pau20} which states the following. For a finitely ambiguous $(\Sigma,\Realnum_{\max,+})$-wta $\cA$ it is decidable whether there exists a deterministic $(\Sigma,\Realnum_{\max,+})$-wta $\cA'$ such that $\sem{\cA}=\sem{\cA'}$. If such an automaton $\sem{\cA'}$ exists, then it can be constructed. Here finitely ambiguous means that there exists an integer $M\ge 1$ such that, for each tree $\xi\in \T_\Sigma$, there are at most $M$ accepting runs of $\cA$ on $\xi$. We also mention that the base of the construction of $\cA'$ is not the factorization as in this chapter.

Also, we mention that in \cite{doefelsti21} wta over particular semirings were considered which are based on groups. Let $\sfG=(G,\otimes,\1)$ be a group. Then $\Sem(\sfG)=(\cP_{\mathrm{fin}}(G),\cup,\otimes,\emptyset, \{\1\})$ is a semiring where $\otimes$ is lifted to finite subsets of $G$ in the straightforward way (cf. Example \ref{ex:semirings}(\ref{def:Sem(B)})). A \emph{group-weighted  tree automaton over $\Sigma$ and $\sfG$} (for short: $(\Sigma,\sfG)$-gwta) is a $(\Sigma,\Sem(\sfG))$-wta. In \cite[Thm.~1]{doefelsti21} it is proved that each  $(\Sigma,\sfG)$-gwta  is sequentializable if it has the twinning property. A  $(\Sigma,\sfG)$-gwta is sequential if it is bu-deterministic and for each transition $(q_1 \cdots q_k,\sigma,q)$ we have that $|\delta_k(q_1 \cdots q_k,\sigma,q)| \le 1$. It is easy to see that each crisp-deterministic $(\Sigma,\sfG)$-gwta is sequential, and each sequential $(\Sigma,\sfG)$-gwta is bu-deterministic. Moreover, both inclusions are strict.

Finally, we mention that in \cite{doe22} a general framework for the determinization was established, which is based on a theory of factorizations in (multiplicative) monoids \cite[Sec.~3.3]{doe22}). Roughly speaking, the main determinization result \cite[Thm.~3.78]{doe22} says the following. Let $\B=(B,\oplus,\otimes,\0,\1)$ be a semiring; moreover, let $\M = (M,\otimes,\1)$ be a finitely generated submonoid of $(B,\otimes,\1)$ which satisfies a certain monotonicity property and admits centering factorizations; then $\B' = (\langle M \rangle_\oplus,\oplus,\otimes,\0,\1)$ is a semiring. Moreover, let $\cA$ be a $(\Sigma,\B')$-wta  which satisfies a certain twinning property. Then \cite[Thm.~3.78]{doe22} says that, if $\cA$ is finitely $M$-ambiguous or $\B'$ is additively idempotent, then $\cA$ is $\M$-sequentializable, which means sequential and the transition weights are in $M$. As illustrated in \cite[Ex.~3.95]{doe22},  the twinning properties of \cite{buemayvog10} and of \cite[Thm.~3.78]{doe22} are incomparable.

\label{p:convention-in-determinization}
\begin{quote}\em In this chapter, we let $\B=(B,\oplus,\otimes,\0,\1)$ be an arbitrary semiring and $\cA=(Q,\delta,F)$ be an arbitrary $(\Sigma,\B)$-wta, unless specified otherwise.
\end{quote}

\section{Applying the subset method to wta over arbitrary semirings}
\label{sec:subset}

One can be tempted to apply the subset method (cf. Definition \ref{constr:subset}) to a $(\Sigma,\B)$-wta $\cA$ where $\B$  is an arbitrary  semiring. Then one obtains the triple $\sub(\cA)=(Q',\delta',F')$.  We note that $Q'$ is finite if and only if $\sub(\cA)$ is a crisp-deterministic wta.

\begin{lemma}\rm \label{lm:subset:finite-implies-rec-step-map} If  $\sub(\cA)$ is a crisp-deterministic $(\Sigma,\B)$-wta, then $\sem{\cA}$ is a recognizable step mapping.
    \end{lemma}
    \begin{proof} Let  $\sub(\cA)$ be a crisp-deterministic $(\Sigma,\B)$-wta. Then by Theorem \ref{thm:subset-method-lf-strong-bm}(1) we have $\sem{\sub(\cA)}=\sem{\cA}$. Hence the statement follows from Theorem \ref{thm:crisp-det-algebra}.
\end{proof}

This lemma shows the limitation of the  subset method: if $\sem{\cA}$ is not a recognizable step mapping, then  $\sub(\cA)$ is not a crisp-deterministic wta, i.e.,  the  subset method does not yield a wta. In the following we give an example of a (not bu-deterministic) $(\Sigma,\B)$-wta $\cA$ which cannot be determinized by the subset method because $\sem{\cA}$ is not a recognizable step mapping (also cf. \cite[Ex.~4.1]{buemayvog10}).

\begin{example}\rm\label{ex:gen-subset-not-applicable}
   We consider  the tropical semifield $\Ratminplus=(\mathbb{Q}_{\infty},\min,+,\infty,0)$  and the ranked alphabet $\Sigma = \{\sigma^{(2)}, \alpha^{(0)}\}$. Moreover, we let $\cA =(Q,\delta,F)$ be the $(\Sigma,\Ratminplus)$-wta given by
    \begin{compactitem}
    \item $Q= \{p,p'\}$ and $F_p=0$ and $F_{p'}=\infty$ and
      \item $\delta_0(\varepsilon,\alpha,p)=1$, $\delta_2(pp,\sigma,p)=1$, $\delta_2(pp,\sigma,p')=0.5$, and $\delta_2(p'p,\sigma,p)=1.5$, and all other values of $\delta_0$ and $\delta_2$ are~$\infty$. 
      \end{compactitem}
      Figure \ref{fig:ex-unfolding} shows the fta-hypergraph of $\cA$.
      Obviously, $\cA$ is not bu-deterministic. Moreover, for each $\xi \in \T_\Sigma$, we have \[
        \h_\cA(\xi)_{p'} = \begin{cases} \infty & \text{ if $\xi = \alpha$}\\
          \size(\xi) -0.5 & \text{ otherwise}   
        \end{cases}
        \ \ \text{ and } \ \
        \h_\cA(\xi)_p = \size(\xi)     \enspace.
      \]
      Thus, for each $\xi \in \T_\Sigma$, we have  $\sem{\cA}(\xi) = \sem{\cA}(\xi) = \size(\xi)$.\footnote{Indeed, we obtained  $\cA$ by starting from the  bu-deterministic wta shown in Example~\ref{ex:size} and by  ``unfolding'' the transition on $\sigma$.} Obviously, $\size$ is not a recognizable step mapping, because $\im(\size)$ is an infinite set. Thus, by the contraposition of Lemma \ref{lm:subset:finite-implies-rec-step-map} $\sub(\cA) = (Q',\delta',F')$ is not a crisp-deterministic wta. In particular, the set $Q'$ is not finite.  We obtain that, in general, the  subset method is not appropriate to determinize wta. 
\hfill $\Box$
\end{example}

      \begin{figure}[t]
\begin{center}
\begin{tikzpicture}
\tikzset{node distance=7em,scale=0.7, transform shape}
\node[state, rectangle] (1) {\Large $\alpha$};
\node[state, right of=1] (2) {\Large $p$};
\node[state, rectangle, right of=2] (3) {\Large $\sigma$};
\node[state, rectangle, below of=2] (4) {\Large $\sigma$};
\node[state, rectangle, left of=4] (5)[right=0.3em] {\Large $\sigma$};
\node[state, below of=4] (6) {\Large $p'$};

\tikzset{node distance=2em}
\node[above of=1] (w1)[right=-0.7em] {1};
\node[above of=2] (w2) {0};
\node[above of=3] (w3) {1};
\node[right of=4] (w4)[right] {0.5};
\node[above of=5] (w5) {1.5};

\draw[->,>=stealth] (1) edge (2);
\draw[->,>=stealth] (3) edge (2);
\draw[->,>=stealth] (2) edge[out=60, in=30, looseness=1.4] (3);
\draw[->,>=stealth] (2) edge[out=-60, in=-30, looseness=1.4] (3);
\draw[->,>=stealth] (2) edge (5);
\draw[->,>=stealth] (2) edge[out=-105, in=115, looseness=1.3] (4);
\draw[->,>=stealth] (2) edge[out=-75, in=65, looseness=1.3] (4);
\draw[->,>=stealth] (4) edge (6);
\draw[->,>=stealth] (6) edge (5);
\draw[->,>=stealth] (5) edge[out=170, in=-150, looseness=1.5] (2);
\end{tikzpicture}
\end{center}
\caption{\label{fig:ex-unfolding} The $(\Sigma,\Ratminplus)$-wta $\cA$.}
\end{figure}

Finally, we note that the reverse of Lemma \ref{lm:subset:finite-implies-rec-step-map} does not hold.
For instance, consider the $(\Sigma,\Natmaxplus)$-wta $\cA$ of Example \ref{ex:height} which recognizes the weighted tree language $\height: \T_\Sigma \to \mathbb{N}$. In the definition of $\cA$, let us change the mapping $F$ by defining $F_h=-\infty$,  and let us denote the obtained wta by $\cA'$. Obviously, $\sem{\cA'}=\widetilde{-\infty}$,
i.e., a recognizable step mapping. 
On the other hand, $Q'=\{\big(\begin{smallmatrix}n\\0\end{smallmatrix}\big)\mid n\in\mathbb{N}\}$, i.e., an infinite set. Hence $\sub(\cA)$ is not a crisp-deterministic $(\Sigma,\Natmaxplus)$-wta.


\section{Negative results for determinization}
\label{sec:negative-results-determinization}

We have seen that the  subset method does not yield the desired determinization. Actually, there is no determinization which works for each $(\Sigma,\B)$-wta because there exist recognizable weighted tree languages which are not bu-deterministically recognizable. For instance, the weighted tree language   $\height$ is not bu-deterministically recognizable (cf. Corollary \ref{ex:height-not-bud}). In this section we show two further examples of such weighted tree languages.

\begin{theorem}\label{thm:zigzag-not-bu-det}{\rm (\cite{den17}, also cf. \cite[Ex.~7]{mal08c})} The weighted tree language $\mathrm{zigzag}: \T_\Sigma\to \mathbb{N}$ defined in  Example \ref{ex:zigzag} is in $\Rec(\Sigma,\Natmaxplus)$, but not in $\budRec(\Sigma,\Natmaxplus)$. Hence, each $(\Sigma,\Natmaxplus)$-wta which recognizes $\mathrm{zigzag}$ is not bu-determinizable.  
\end{theorem}
\begin{proof} We recall that $\Sigma = \{\sigma^{(2)}, \alpha^{(0)}\}$ and  $\Natmaxplus = (\mathbb{N}_{-\infty},\max,+,-\infty,0)$.  We prove the claim by contradiction.
      For this we assume that there exists a bu-deterministic $(\Sigma, \Natmaxplus)$-wta $\cA = (Q,  \delta, F)$ such that $\sem{\cA} = \mathrm{zigzag}$. 

By Lemma \ref{lm:limit-bu-det}(1),  for every $\xi \in \T_\Sigma$ we have $|\Q^{\h_\cA}_{\not=-\infty}(\xi)| \le 1$. If there exists a tree $\xi \in \T_\Sigma$ such that $|\Q^{\h_\cA}_{\not=-\infty}(\xi)| = 0$, then $\sem{\cA}(\xi) = -\infty \not= \mathrm{zigzag}(\xi)$.
      Thus  for every $\xi \in \T_\Sigma$  we have  $|\Q^{\h_\cA}_{\not=-\infty}(\xi)| = 1$. Let us denote,   for each $\xi \in \T_\Sigma$, the unique element in $\Q^{\h_\cA}_{\not=-\infty}(\xi)$ by $q_\xi$.

      If there exists a $\xi \in \T_\Sigma$ such that $F_{q_\xi} = -\infty$, then $\sem{\cA}(\xi) = -\infty \not= \mathrm{zigzag}(\xi)$. Thus $F_{q_\xi} \not= -\infty$ for every $\xi \in \T_\Sigma$.
      In conclusion we know:
      \begin{equation}
        (\forall \xi \in \T_\Sigma): \big(\sem{\cA}(\xi) = \h_\cA(\xi)_{q_\xi} + F_{q_\xi} = \mathrm{zigzag}(\xi)\big) \wedge \big((\forall q \in Q): \text{if } q \not= q_\xi, \text{ then } \h_\cA(\xi)_q = -\infty\big)\text{.} \label{eq:zigzag:obs}
      \end{equation}

      Since $Q$ is finite and $\mathrm{zigzag}(\T_\Sigma)$ is not, there exist trees $\xi_1, \xi_2 \in \T_\Sigma$ such that $q_{\xi_1} = q_{\xi_2}$ and $\mathrm{zigzag}(\xi_1) \not= \mathrm{zigzag}(\xi_2)$.
      Since $\cA$ is bu-deterministic, we have $q_{\sigma(\alpha, \xi_1)} = q_{\sigma(\alpha, \xi_2)}$. Let us abbreviate $q_{\xi_1}$ by $q$ and
      $q_{\sigma(\alpha, \xi_1)}$ by $q'$, respectively.  Due to the definition of $\mathrm{zigzag}$ we have that $\mathrm{zigzag}(\sigma(\alpha, \xi_1)) = \mathrm{zigzag}(\sigma(\alpha, \xi_2)) = 1$.

      Then we can derive the following sequence of implications:
      \begingroup
       \allowdisplaybreaks
\[
      \begin{array}{lrcl}
  &           \mathrm{zigzag}(\sigma(\alpha, \xi_1))    & = &\mathrm{zigzag}(\sigma(\alpha, \xi_2))\\[2mm]
\Rightarrow &
              \h_\cA(\sigma(\alpha, \xi_1))_{q'} + F_{q'}      & = & \h_\cA(\sigma(\alpha, \xi_2))_{q'} + F_{q'} \\
        &&&\text{(by \eqref{eq:zigzag:obs} and because $q_{\sigma(\alpha, \xi_1)} = q_{\sigma(\alpha, \xi_2)}$)} \\[2mm]
\Rightarrow &
               \h_\cA(\sigma(\alpha, \xi_1))_{q'}       & = & \h_\cA(\sigma(\alpha, \xi_2))_{q'} \ \ \      \text{(because $F_{q'} \not= -\infty$)} \\[2mm]
\Rightarrow &
               \h_\cA(\alpha)_{q_\alpha} + \h_\cA(\xi_1)_q + \delta_2(q_\alpha q,\sigma, q')      & = &\h_\cA(\alpha)_{q_\alpha} + \h_\cA(\xi_2)_q + \delta_2(q_\alpha q,\sigma, q')\\
&&&          \text{(by definition  of $\h_\cA$ and because $q= q_{\xi_1} = q_{\xi_2}$)} \\[2mm]
\Rightarrow &
               \h_\cA(\xi_1)_q                                                        & = & \h_\cA(\xi_2)_q \\
&&&          \text{(because $\h_\cA(\alpha)_{q_\alpha} \not= -\infty \not= \delta_2(q_\alpha q,\sigma, q')$)} \\[2mm]
\Rightarrow &
               \h_\cA(\xi_1)_q + F_q                                                 & = & \h_\cA(\xi_2)_q + F_q  \ \ \       \text{(because $F_q \not= -\infty$)} \\[2mm]
\Rightarrow &
               \mathrm{zigzag}(\xi_1)                                             & = & \mathrm{zigzag}(\xi_2) \ \ \       \text{(by \eqref{eq:zigzag:obs})}
      \end{array}
    \]
    \endgroup
    This is a contradiction. Thus $\mathrm{zigzag} \not\in \budRec(\Sigma,\Natmaxplus)$.
As a consequence, each $(\Sigma,\Natmaxplus)$-wta which recognizes $\mathrm{zigzag}$ (including the one in Example \ref{ex:zigzag}) is not bu-determinizable.
\end{proof}

\begin{theorem}{\rm (cf. \cite[Thm.~6.3]{borvog03})}\label{thm:borchard-not-det} The weighted tree language $(\mathrm{exp}+1): \T_\Sigma \rightarrow \mathbb{N}$ defined in Example \ref{ex:exp+1} is in $\Rec(\Sigma,\Ratnum)$, but not in $\budRec(\Sigma,\Ratnum)$, where $\Ratnum=(\mathbb{Q},+,\cdot,0,1)$ is the field over the rational numbers. Hence, each $(\Sigma,\Ratnum)$-wta which recognizes $(\exp + 1)$ is not bu-determinizable.
\end{theorem}
\begin{proof} We recall that $\Sigma = \{\gamma^{(1)}, \alpha^{(0)}\}$. In  Example \ref{ex:exp+1}, we gave a $(\Sigma,\Nat)$-wta over the semiring $\Nat=(\mathbb{N},+,\cdot,0,1)$ of natural numbers which recognizes $(\exp + 1)$.    We can view this wta as a $(\Sigma,\Ratnum)$-wta (cf. Section \ref{sect:extension-of-weight-structure}), which we denote by $\cA_\mathbb{Q}$ in this proof. Hence $(\exp + 1) \in \Rec(\Sigma,\Ratnum)$.

   Next we prove that $(\exp + 1) \not\in \budRec(\Sigma,\Ratnum)$, and we prove this by contradiction. We assume that $(\mathrm{exp}+1) \in \budRec(\Sigma,\Ratnum)$. Then, by Theorem \ref{lm:com-semifield-Boolean-root-weights} there exists a bu-deterministic $(\Sigma,\Ratnum)$-wta $\cA=(Q,\delta,F)$ with unit root weights such that $\sem{\cA}=(\mathrm{exp}+1)$.

By Lemma \ref{lm:limit-bu-det}(1),  for every $n \in \mathbb{N}$ we have $|\Q^{\h_\cA}_{\ne 0}(\gamma^n(\alpha))| \le 1$. If $|\Q^{\h_\cA}_{\ne 0}(\gamma^n(\alpha))| =0$ or $\Q^{\h_\cA}_{\ne 0}(\gamma^n(\alpha)) \cap \supp(F) = \emptyset$, then $\sem{\cA}(\gamma^n(\alpha))=0$ which is a contradiction. Thus $|\Q^{\h_\cA}_{\ne 0}(\gamma^n(\alpha)) \cap \supp(F)|=1$. We denote the state in the set $\Q^{\h_\cA}_{\ne 0}(\gamma^n(\alpha)) \cap \supp(F)$ by $q_n$. Thus 
\begin{equation}
\sem{\cA}(\gamma^n(\alpha)) = \h_\cA(\gamma^n(\alpha))_{q_n} = 2^n+1 \enspace.\label{equ:A=exp}
\end{equation}
It is an easy exercise to show that, for every $q \in Q$ and $n \ge 1$:
\[
\delta_0(\varepsilon,\alpha,q) =
\left\{
\begin{array}{ll}
2 & \text{ if } q=q_0\\
0 & \text{ otherwise }
\end{array}
\right.
\]
and 
\[
\delta_1(q_{n-1},\gamma,q) =
\left\{
\begin{array}{ll}
(2^{n-1}+1)^{-1} \cdot (2^n+1) & \text{ if } q=q_n\\
0 & \text{ otherwise \enspace.}
\end{array}
\right.
\]

Since $Q$ is finite, there exist $n,i \in \mathbb{N}$ such that $0 \le i < n$ and $q_i=q_n$. Then we can calculate as follows:
\begin{align*}
 \sem{\cA}(\gamma^{n+1}(\alpha))
&=  \h_\cA(\gamma^n(\alpha))_{q_n} \cdot \delta_1(q_n,\gamma,q_{n+1})
= (2^n+1) \cdot \delta_1(q_n,\gamma,q_{n+1})\\
&= (2^n+1) \cdot \delta_1(q_i,\gamma,q_{i+1})
= (2^n+1) \cdot(2^{i}+1)^{-1} \cdot (2^{i+1}+1)
\not= 2^{n+1} +1
\enspace,
\end{align*}
where the inequality can be proved by contradiction as follows.
\begin{align*}
&(2^n+1) \cdot(2^{i}+1)^{-1} \cdot (2^{i+1}+1) = 2^{n+1} +1\\
\text{ iff \ } & (2^n+1) \cdot (2^{i+1}+1) = (2^{i}+1)\cdot (2^{n+1} +1)\\
\text{ iff \ } & 2^{n+i+1} + 2^n + 2^{i+1} +1 =  2^{n+i+1} + 2^{i} + 2^{n+1} +1\\
\text{ iff \ } & 2^i =2^n \\
\text{ iff \ } & i=n.
\end{align*}
This a contradiction because $i<n$.

However, $ \sem{\cA}(\gamma^{n+1}(\alpha)) \not= 2^{n+1} +1$ which contradicts Equation \eqref{equ:A=exp}.  Hence $(\mathrm{exp}+1) \not\in \budRec(\Sigma,\Ratnum)$.
As a consequence, each $(\Sigma,\Ratnum)$-wta which recognizes $(\exp +1)$ (including $\cA_\mathbb{Q}$ from the beginning of the proof) is not bu-determinizable.
\end{proof}


\section{Positive result for determinization}
\label{sec:extremal-commutative-zero-divisor-free-semirings-determinization}

In this section we recall the approach of \cite{buemayvog10} (based on \cite{kirmae05}) and adapt it to our notations.

\label{p:convention-in-determinization-positive}
\begin{quote}\emph{We recall that  $\V(\cA)=(B^Q,\delta_\cA)$ is the vector algebra of $\cA$, and that $\0^Q \in B^Q$ is the $Q$-vector over $\B$ which contains $\0$ in each component (cf. Section~\ref{sec:vectors-matrices}).   We assume that there exists $\alpha \in \Sigma^{(0)}$ such that $\delta_\cA(\alpha)() \ne \0^Q$.}
\end{quote}

In this section we use the addition of $B^Q$-vectors and the multiplication of a $Q$-vector with a scalar both from the left and  the right (cf. Section \ref{sec:vectors-matrices}). Also we recall that, for every $\xi \in \T_\Sigma$ and $q \in Q$ we have $\h_\cA(\xi)_q= \bigoplus_{\rho \in \R_\cA(q,\xi)} \wt(\xi,\rho)$ (cf. \eqref{equ:weight-run=hom}).  

In Subsection \ref{ssec:determinization-by-factorization}, we start by defining the concepts of factorization and twinning property, we show the construction of the triple~$\det_{(f,g)}(\cA)$\footnote{The notation $\det_{(f,g)}(\cA)$ should not be confused with the determinant of a matrix.} for some wta $\cA$ and some factorization $(f,g)$, and we show the main determinization theorem of this section (cf. Theorem \ref{thm:extremal-twins-determinizable}) and an application to a particular wta.
In Subsections \ref{ssec:properties-of-factorizations} and \ref{ssec:properties-of-twins-property}, we elaborate properties and examples of factorizations and of the twinning property, respectively. 
Finally, in Subsection \ref{ssec:proof-of-main-theorem}, we show the proof of Theorem \ref{thm:extremal-twins-determinizable}.


\subsection{Determinization by factorization}
\label{ssec:determinization-by-factorization}

  \index{factorization}
  \index{maximal factorization}
    \index{trivial factorization}
    The idea of factorization is to represent each $Q$-vector $u \in B^Q$ as a scalar product $b \cdot u'$ where $b \in B$ and $u' \in B^Q$; moreover, the way in which $Q$-vectors are split into elements of $B$ and remaining $Q$-vectors in $B^Q$ should be uniform.   Formally, a \emph{factorization over $B^Q$} (or: \emph{factorization}) is a pair $(f,g)$ of mappings, where 
    \[f:B^Q\setminus\{\0^Q\}\rightarrow B^Q \text{ and } g:B^Q\setminus\{\0^Q\}\rightarrow B\] such that $u=g(u)\cdot f(u)$ for every $u\in B^Q\setminus\{\0^Q\}$. Thus $\0^Q \not\in \im(f)$ and $\0 \not\in \im(g)$. We call $g(u)$ the \emph{common factor of $u$} and $f(u)$ the \emph{remainder of $u$}. The factorization $(f,g)$ is called 
 \begin{compactitem}   
\item[-]    \emph{maximal} if, for every $u\in B^Q\setminus\{\0^Q\}$ and $b\in B$ with $b\cdot u\neq \0^Q$, we have $f(u)=f(b\cdot u)$, and
\item[-] \emph{trivial} if $f(u)=u$ and $g(u)=\mathbb{1}$ for every $u \in B^Q\setminus\{\0^Q\}$.
\end{compactitem}   
We recall from page \pageref{p:par-extension-runs-to-contexts} that, for every $c \in \C_\Sigma$ and $p,q\in Q$, we denote the set of all runs of $\cA$ on $c$ and the set of all $(q,p)$-runs of $\cA$ on $c$ by $\R_\cA(c)$ and $\R_{\cA}(q,c,p)$, respectively. Also the weight of a run on a context is defined on page \pageref{p:par-extension-runs-to-contexts}.  Moreover, for every $\xi \in \T_\Sigma$, $\rho\in \R_\cA(\xi)$, and $w\in \pos(\xi)$, we defined the run $\rho|^w$ on the context $\xi|^w$.

    Additionally, for each $\xi \in \T_\Sigma\cup \C_\Sigma$ and $R \subseteq \R_\cA(\xi)$, we define
\[
  \wt(R) = \bigoplus_{\rho \in R} \wt(\xi,\rho)\enspace.
\]
\index{victorious}
By Observation \ref{obs:extremal-sum}, if $\B$ is  extremal, then there exists $\rho \in R$ such that $\wt(\xi,\rho) = \wt(R)$. We call such a run \emph{victorious in $R$}.

\index{twinning property}
The wta~$\cA$ has the \emph{twinning property} if,  for every $p,q\in Q$, $\xi\in \T_\Sigma$, and~$c\in \C_\Sigma$, the following holds:
\begin{align*}
  &\text{If } \ \wt(\R_\cA(p,c,p))\neq \mathbb{0} \text{ and  } \wt(\R_\cA(p,\xi))\neq \mathbb{0} \text{ and } \\
  & \ \ \ \ \wt(\R_\cA(q,c,q))\neq \mathbb{0} \text{ and } \wt(\R_\cA(q,\xi))\neq \mathbb{0}\enspace,\\
&\text{then } \wt(\R_\cA(p,c,p)) = \wt(\R_\cA(q,c,q)) \enspace.
\end{align*}

The following determinization by factorization is based on \cite[Sect.~3.3]{kirmae05} and \cite[p.~11]{buemayvog10}. Roughly speaking, it differs from the  subset method as follows. In the  subset method, for every $u_1,\ldots,u_k \in B^Q$ and $\sigma \in \Sigma^{(k)}$, the transition
\[
  (u_1\cdots u_k, \sigma, \delta_\cA(\sigma)(u_1,\ldots,u_k)) \ \ \text{with weight} \ \ \1
\]
is constructed. In the determinization by factorization, the $Q$-vector $\delta_\cA(\sigma)(u_1,\ldots,u_k)$ is factorized as
\[
\delta_\cA(\sigma)(u_1,\ldots,u_k) = \underbrace{g(\delta_\cA(\sigma)(u_1,\ldots,u_k))}_{\text{common factor in $B$}} \cdot \underbrace{f(\delta_\cA(\sigma)(u_1,\ldots,u_k))}_{\text{remainder in $B^Q$}}
\]
and the transition
\[(u_1\cdots u_k,\sigma,f(\delta_\cA(\sigma)(u_1,\ldots,u_k))) \ \ \text{with weight} \ \  g(\delta_\cA(\sigma)(u_1,\ldots,u_k))
\]
is constructed. Thus, the remainder $f(\delta_\cA(\sigma)(u_1,\ldots,u_k))$ is the target state of the transition and the common factor $g(\delta_\cA(\sigma)(u_1,\ldots,u_k))$ is its weight. We do this because, as we will see,  in certain cases the set of remainders will be a finite set, while the set of vectors produced by the  subset method is infinite.

In the following we give the definition of determinization by factorization. Like the  subset method,  it is not a construction in the sense described in the Introduction,  because the set $Q'$ might be infinite and in this case the triple $(Q',\delta',F')$ is not a bu-deterministic wta.
 We will need further conditions to assure that $(Q',\delta',F')$ is a bu-deterministic wta (cf. Theorem \ref{thm:extremal-twins-determinizable}).

     \index{determinization}
      \begin{definition-rect}\rm \label{constr:det-A}  \cite[p.~11]{buemayvog10} Let $\Sigma$ be a ranked alphabet. Moreover, let $\B=(B,\oplus,\otimes,\0,\1)$ be a semiring, let $\cA=(Q,\delta,F)$ be a $(\Sigma,\B)$-wta, and let $(f,g)$ be a factorization over $B^Q$. The \emph{determinization by factorization} transforms $\cA$ and $(f,g)$ into the triple~$\det_{(f,g)}(\cA)=(Q',\delta',F')$ where
	\begin{compactitem}
		\item $Q'$ is the smallest set $P\subseteq B^Q$ such that, for every~$k\in\mathbb{N}$, $\sigma\in\Sigma^{(k)}$, $u_1,\ldots,u_k\in P$: if $\delta_\cA(\sigma)(u_1,\ldots,u_k)\neq\0^Q$, then $f(\delta_\cA(\sigma)(u_1,\ldots,u_k))\in P$,
		
		\item $\delta'=(\delta'_k\mid k\in\mathbb{N})$ is the family of mappings $\delta'_k: (Q')^k\times \Sigma^{(k)}\times Q' \to B$ defined for every~$k\in\mathbb{N}$, $\sigma\in\Sigma^{(k)}$, $u_1,\ldots,u_k\in Q'$, and~$u\in Q'$ by
                  \[\delta'_k(u_1\ldots u_k,\sigma,u)=
		\begin{cases}
			g(\delta_\cA(\sigma)(u_1,\ldots,u_k))&\text{if }\delta_\cA(\sigma)(u_1,\ldots,u_k)\neq\0^Q\text{ and}\\
			& \quad u=f(\delta_\cA(\sigma)(u_1,\ldots,u_k))\;,\\
		\mathbb{0}&\text{otherwise}\;,
              \end{cases}
            \]
			
		\item $F'_u = \bigoplus_{q\in Q} u_q\otimes F_q$ for every $u\in Q'$. 
                \end{compactitem}
We call the triple $\det_{(f,g)}(\cA)$ the~\emph{determinization of~$\cA$ by~$(f,g)$}. 
      \end{definition-rect}
   
In the following, we list some properties of determinization by factorization.

\begin{compactitem}

\item[(1)] Since $\0^Q \not\in \im(f)$, we have $\0^Q \not\in Q'$.
     
\item[(2)]  If $Q'$ is finite, then $\det_{(f,g)}(\cA)$ is a bu-deterministic $(\Sigma,\B)$-wta. However, even in this case $\det_{(f,g)}(\cA)$ is not necessarily crisp-deterministic. 

\item[(3)] If $(f,g)$ is the trivial factorization, then the determinization by factorization is the same as the  subset method except that the latter may generate the state $\0^Q$.

\item[(4)] If  $(f,g)$ is the trivial factorization and $\B$ is locally finite, then the determinization by factorization is the same as the subset method except that the latter may generate the state $\0^Q$; in particular, the state set $Q'$ is finite. Hence  $\det_{(f,g)}(\cA)$ is a bu-deterministic $(\Sigma,\B)$-wta.  Moreover, $\sub(\cA)$ and $\det_{(f,g)}(\cA)$ are equivalent, because the state $\0^Q$ is useless for $\sub(\cA)$. More precisely, (a)  the operation $\delta_\cA(\sigma)$ yields $\0^Q$ if there exists an argument which is $\0^Q$ (cf. \eqref{eq:delta-A-definition}) and (b) $F'(\0^Q)=\0$. Thus, if $\0^Q \in \im(\h_\cA)$, then we can also drop $\0^Q$ from the state set without changing the semantics of $\sub(\cA)$. 
 Then, by applying the construction of Theorem \ref{thm:wta-nf-total} to $\det_{(f,g)}(\cA)$ (which transforms a given wta into an equivalent total one such that bu-determinism is preserved), we obtain a crisp-deterministic wta which is equivalent to $\det_{(f,g)}(\cA)$. This construction can be compared to the construction of the Nerode wsa in \cite[Sec.~6]{cirdroignvog10}. 
  
\item[(5)] If $\B$ is the semiring $\Boole$, then there exists exactly one factorization $(f,g)$, viz. the trivial and maximal factorization defined by  $g(u) = 1$ and $f(u)=u$. Since  $\mathbb{B}$ is finite,  determinization by factorization is a particular case of the one described in (4). 
\end{compactitem}

The main result of this chapter will be the following theorem.

\begin{theorem-rect}{\rm \cite[Thm.~5.2]{buemayvog10}}\label{thm:extremal-twins-determinizable} Let $\Sigma$ be a ranked alphabet. Moreover, let $\B$ be an extremal and commutative semiring, let $\cA$ be a $(\Sigma,\B)$-wta with the twinning property, and let $(f,g)$ be a maximal factorization over $B^Q$. Then
  \begin{compactenum}
  \item[(1)] $\det_{(f,g)}(\cA)$ is a bu-deterministic $(\Sigma,\B)$-wta (finiteness),
  \item[(2)] $\sem{\det_{(f,g)}(\cA)}= \sem{\cA}$ (correctness), and
  \item[(3)] $\det_{(f,g)}(\cA)$ is minimal with respect to the number of states  among all bu-deterministic $(\Sigma,\B)$-wta which are obtained from $\cA$ by determinization by factorization (minimality).
    \end{compactenum}
  \end{theorem-rect}

  \begin{proof} It follows from Theorems~\ref{thm:finiteness}, \ref{thm:correctness}, and \ref{thm:minimality}, which are proved in the next subsections. 
  \end{proof}

We show six examples of extremal and commutative semirings:
    \begin{compactitem}
    \item the semiring $\Boole = (\mathbb{B},\vee,\wedge,0,1)$,
    \item the arctic semiring $\Natmaxplus = (\mathbb{N}_{-\infty},\max,+,-\infty,0)$,
    \item the semiring $\Natmaxplusn= ([0,n]_{-\infty},\max,\hat{+}_n,-\infty,0)$  where $[0,n]_{-\infty}$ is an abbreviation of $[0,n]\cup\{-\infty\}$ (defined in Example \ref{ex:semirings}(\ref{ex:Nat-max-plus-n})),
    \item the tropical semifield $\Ratminplus = (\mathbb{Q}_{\infty},\min,+,\infty,0)$,
           \item the semifield $(\mathbb{R}_{\ge 0},\max,\cdot,0,1)$, and
      \item the  semiring $\Viterbi=([0,1],\max,\cdot,0,1)$.
      \end{compactitem}

We note that, due to Lemma \ref{lm:maxzd} (below), Theorem \ref{thm:extremal-twins-determinizable} cannot be used to determinize  a $(\Sigma,\B)$-wta $(Q,\delta,F)$ with at least two states if $\B$ is not zero-divisor free, because in this case there does not exist a maximal factorization of $B^Q$.

      \begin{example} \rm \label{ex:factorization} Again, we consider 
             the $(\Sigma,\Ratminplus)$-wta  $\cA =(Q,\delta,F)$ given in Example \ref{ex:gen-subset-not-applicable}. For convenience, we recall that
      \begin{compactitem}
    \item $Q= \{p,p'\}$ and $F_p=0$ and $F_{p'}=\infty$ and
      \item $\delta_0(\varepsilon,\alpha,p)=1$, $\delta_2(pp,\sigma,p)=1$, $\delta_2(pp,\sigma,p')=0.5$, and $\delta_2(p'p,\sigma,p)=1.5$, and all other values of $\delta_0$ and $\delta_2$ are~$\infty$. 
      \end{compactitem}
      
As it turned out in Example \ref{ex:gen-subset-not-applicable}, the  subset method is not applicable to determinize  $\cA$.  Now we determinize $\cA$ by factorization according to Definition \ref{constr:det-A}.
      
 For this, we define the mappings $f: (\mathbb{Q}_{\infty})^Q \setminus \{\infty_Q\} \to (\mathbb{Q}_{\infty})^Q$ and $g: (\mathbb{Q}_{\infty})^Q \setminus \{\infty_Q\} \to \mathbb{Q}_{\infty}$  for each $u=\big(\begin{smallmatrix}u_p\\u_{p'}\end{smallmatrix}\big)$ in $(\mathbb{Q}_{\infty})^Q \setminus \{\infty_Q\}$ by 
\begin{align*}
f(u)_q = u_q - g(u)\  \text{ for each $q \in Q$} \ \ \ \text{ and } \ \ \ g(u) =\min(u_p,u_{p'})\enspace.
    \end{align*}
    Then $(f,g)$ is a maximal  factorization.
    
Then  $\det_{(f,g)}(\cA)=(Q',\delta',F')$, which we will abbreviate by $\cA'$ in the sequel, is defined by (cf. Figure~\ref{fig:ex-det-unfolding})
    \begin{compactitem}
    \item $Q' = \{\big(\begin{smallmatrix}0\\\infty\end{smallmatrix}\big), \big(\begin{smallmatrix}0.5\\0\end{smallmatrix}\big) \}$ and
      $F'(\big(\begin{smallmatrix}0\\\infty\end{smallmatrix}\big)) = 0$ and  $F'(\big(\begin{smallmatrix}0.5\\0\end{smallmatrix}\big)) = 0.5$

  \item $\delta'_0(\varepsilon,\alpha,\big(\begin{smallmatrix}0\\\infty\end{smallmatrix}\big)) = 1$ and 
    \begin{align*}
          \delta'_2(\big(\begin{smallmatrix}0\\\infty\end{smallmatrix}\big)\big(\begin{smallmatrix}0\\\infty\end{smallmatrix}\big),\sigma,\big(\begin{smallmatrix}0.5\\0\end{smallmatrix}\big)) = 0.5, \hspace{5mm}
   &\delta'_2(\big(\begin{smallmatrix}0\\\infty\end{smallmatrix}\big)\big(\begin{smallmatrix}0.5\\0\end{smallmatrix}\big),\sigma,\big(\begin{smallmatrix}0.5\\0\end{smallmatrix}\big)) = 1,\\
          \delta'_2(\big(\begin{smallmatrix}0.5\\0\end{smallmatrix}\big)\big(\begin{smallmatrix}0\\\infty\end{smallmatrix}\big),\sigma,\big(\begin{smallmatrix}0.5\\0\end{smallmatrix}\big)) = 1,  \hspace{5mm} 
   &\delta'_2(\big(\begin{smallmatrix}0.5\\0\end{smallmatrix}\big)\big(\begin{smallmatrix}0.5\\0\end{smallmatrix}\big),\sigma,\big(\begin{smallmatrix}0.5\\0\end{smallmatrix}\big)) = 1.5\enspace,
    \end{align*}
    and $\delta'_0(\varepsilon,\alpha,p)=\delta'_2(pq,\sigma,r)=\infty$ for every other combination of $p,q,r\in Q'$.
 \end{compactitem}

 \begin{figure}[t]
   \begin{center}
\begin{tikzpicture}
\tikzset{node distance=7em, scale=0.6, transform shape}
\node[state, rectangle] (1) {$\alpha$};
\node[state, right of=1] (2){$\binom{0}{\infty}$};
\node[state, rectangle, right of=2] (3)[right=1em]{$\sigma$};
\node[state, rectangle, above of=3] (4)[above=1em]{$\sigma$};
\node[state, rectangle, below of=3] (5)[below=1em]{$\sigma$};
\node[state, right of=3] (6)[right=1em]{$\binom{0.5}{0}$};
\node[state, rectangle, right of=6] (7) {$\sigma$};

\tikzset{node distance=2em}
\node[above of=1] (w1) {1};
\node[above of=2] (w2) {0};
\node[above of=3] (w3) {0.5};
\node[above of=4] (w4) {1};
\node[above of=5] (w5) {1};
\node[above of=6] (w6)[left=0.1em] {0.5};
\node[above of=7] (w7) {1.5};

\draw[->,>=stealth] (1) edge (2);
\draw[->,>=stealth] (2) edge[out=20, in=155, looseness=1.1] (3);
\draw[->,>=stealth] (2) edge[out=-20, in=205, looseness=1.1] (3);
\draw[->,>=stealth] (2) edge (4);
\draw[->,>=stealth] (2) edge (5);
\draw[->,>=stealth] (3) edge (6);
\draw[->,>=stealth] (6) edge (4);
\draw[->,>=stealth] (4) edge[out=-290, in=80, looseness=1.4] (6);
\draw[->,>=stealth] (6) edge (5);
\draw[->,>=stealth] (5) edge[out=290, in=-80, looseness=1.4] (6);
\draw[->,>=stealth] (7) edge (6);
\draw[->,>=stealth] (6) edge[out=60, in=30, looseness=1.4] (7);
\draw[->,>=stealth] (6) edge[out=-60, in=-30, looseness=1.4] (7);
\end{tikzpicture} 
\caption{\label{fig:ex-det-unfolding} The $(\Sigma,\Ratminplus)$-wta $\det_{(f,g)}(\cA)=(Q',\delta',F')$.}
   \end{center}
   \end{figure}
 
Since $Q'$ is finite, $\det_{(f,g)}(\cA)$ is a bu-deterministic $(\Sigma,\Ratminplus)$-wta.
  For two  transitions of $\det_{(f,g)}(\cA)$, we illustrate the way in which they are constructed. We abbreviate vectors of the form
  $\big(\begin{smallmatrix}a-c\\ b-c\end{smallmatrix}\big)$  by $\big(\begin{smallmatrix}a\\ b\end{smallmatrix}\big)-c$
    for every $a,b,c \in \mathbb{Q}_{\ge 0}$.
  
\underline{$\delta'_0(\varepsilon,\alpha,\big(\begin{smallmatrix}0\\\infty\end{smallmatrix}\big))$:} We have $\delta_\cA(\alpha)() = \big(\begin{smallmatrix}1\\\infty\end{smallmatrix}\big) \not= \big(\begin{smallmatrix}\infty\\\infty\end{smallmatrix}\big)$ and $g(\big(\begin{smallmatrix}1\\\infty\end{smallmatrix}\big))=1$. Thus
\[
f(\delta_\cA(\alpha)()) = f(\big(\begin{smallmatrix}1\\\infty\end{smallmatrix}\big)) = \big(\begin{smallmatrix}1\\\infty\end{smallmatrix}\big) -g(\big(\begin{smallmatrix}1\\\infty\end{smallmatrix}\big))  =  \big(\begin{smallmatrix}1\\\infty\end{smallmatrix}\big) -1 = \big(\begin{smallmatrix}0\\\infty\end{smallmatrix}\big)\enspace.
\]
Thus 
\[
\delta'_0(\varepsilon,\alpha,\big(\begin{smallmatrix}0\\\infty\end{smallmatrix}\big)) = g(\delta_\cA(\alpha)()) = g(\big(\begin{smallmatrix}1\\\infty\end{smallmatrix}\big)) = 1\enspace.
\]

\underline{$\delta'_2(\big(\begin{smallmatrix}0\\\infty\end{smallmatrix}\big)\big(\begin{smallmatrix}0.5\\0\end{smallmatrix}\big),\sigma,\big(\begin{smallmatrix}0.5\\0\end{smallmatrix}\big))$:} We have
    \begin{align*}
      \delta_\cA(\sigma)(\big(\begin{smallmatrix}0\\\infty\end{smallmatrix}\big)\big(\begin{smallmatrix}0.5\\0\end{smallmatrix}\big)) =
      \big(\begin{smallmatrix}\min(0+0.5+1, \infty+0.5+1.5)\\0+0.5+0.5\end{smallmatrix}\big)= \big(\begin{smallmatrix}1.5\\1\end{smallmatrix}\big)\enspace.
    \end{align*}
    Since $g(\big(\begin{smallmatrix}1.5\\1\end{smallmatrix}\big))=1$, we have 
\[
f(\delta_\cA(\sigma)(\big(\begin{smallmatrix}0\\\infty\end{smallmatrix}\big),\big(\begin{smallmatrix}0.5\\0\end{smallmatrix}\big))) 
=f(\big(\begin{smallmatrix}1.5\\1\end{smallmatrix}\big))
=\big(\begin{smallmatrix}1.5\\1\end{smallmatrix}\big)  -g(\big(\begin{smallmatrix}1.5\\1\end{smallmatrix}\big)) 
= \big(\begin{smallmatrix}1.5\\1\end{smallmatrix}\big)  -1
=\big(\begin{smallmatrix}0.5\\0\end{smallmatrix}\big)\enspace.
\]
Thus we obtain 
\[
\delta'_2(\big(\begin{smallmatrix}0\\\infty\end{smallmatrix}\big)\big(\begin{smallmatrix}0.5\\0\end{smallmatrix}\big),\sigma,\big(\begin{smallmatrix}0.5\\0\end{smallmatrix}\big)) 
= g(\delta_\cA(\sigma)(\big(\begin{smallmatrix}0\\\infty\end{smallmatrix}\big)\big(\begin{smallmatrix}0.5\\0\end{smallmatrix}\big)))
= g(\big(\begin{smallmatrix}1.5\\1\end{smallmatrix}\big))
= 1\enspace.
\]
Finally we show how the value of $F'(\big(\begin{smallmatrix}0.5\\0\end{smallmatrix}\big))$ is obtained. 
    \[
      F'(\big(\begin{smallmatrix}0.5\\0\end{smallmatrix}\big)) = \big(\begin{smallmatrix}0.5 &0\end{smallmatrix}\big) \cdot \big(\begin{smallmatrix}0\\\infty\end{smallmatrix}\big) =
      \min(0.5+0, 0+\infty) = 0.5\enspace. 
      \] 

Indeed, $\sem{\det_{(f,g)}(\cA)}(\xi) = \size(\xi)$ for each $\xi \in \T_\Sigma$ which can be seen as follows. Let us abbreviate $\h_{\det_{(f,g)}(\cA)}$ by $\h$, and the states $\big(\begin{smallmatrix}0\\\infty\end{smallmatrix}\big)$ and $\big(\begin{smallmatrix}0.5\\0\end{smallmatrix}\big)$ by $q_1$ and $q_2$, respectively. Then for each $\xi \in \T_\Sigma$ we have 
\[
\h(\xi)_{q_1} = 
\begin{cases}
1 & \text{ if } \xi = \alpha\\
\infty & \text{ otherwise}
\end{cases}
\ \text{ and } \ 
\h(\xi)_{q_2} = 
\begin{cases}
\infty & \text{ if } \xi = \alpha\\
\size(\xi) -0.5 & \text{ otherwise}\enspace.
\end{cases}
\]
This can be proved easily by induction on $\T_\Sigma$. Then
\[
\sem{\det\nolimits_{(f,g)}(\cA)}(\xi) = \min(\h(\xi)_{q_1} + F'(q_1), \h(\xi)_{q_2} + F'(q_2) ) = \size(\xi)
\]
where the last equality follows from case analysis $\xi=\alpha$ or $\xi\not=\alpha$. \hfill $\Box$
    \end{example}

In \cite[Ex.~3.1]{buemayvog10} another example of a wta $\cA$ is shown for which $\sub(\cA)$ is not finite, but $\det_{(f,g)}(\cA)$ is a bu-deterministic wta which is equivalent to $\cA$.
    

    \subsection{Examples and properties of factorizations}
    \label{ssec:properties-of-factorizations}

    We start with some examples of factorizations for extremal and  commutative semirings.

    \begin{example}\rm \label{ex:sr-max-factorization} In the following list we show pairs where each pair consists of an extremal and commutative semiring $\B$ and a maximal factorization $(f,g)$ over $B^Q$. We assume that $u \in B^Q \setminus \{\0^Q\}$.
\begin{enumerate}
\item the semiring $\Boole = (\mathbb{B},\vee,\wedge,0,1)$ and the trivial factorization 

\item  $\Natmaxplus = (\mathbb{N}_{-\infty},\max,+,-\infty,0)$ and $g(u) = \min(u_q \mid q \in Q, u_q \ne -\infty)$ and $f(u)_q = u_q - g(u)$ for each $q \in Q$  

\item $\Natmaxplusn =([0,n]_{-\infty},\max,\hat{+}_n,-\infty,0)$ where $n \in \mathbb{N}_+$ and $g(u) = \min(u_q \mid q \in Q, u_q \ne -\infty)$ and $f(u)_q=u_q - g(u)$ for every $q\in Q$

\item  $\Ratminplus = (\mathbb{Q}_{\infty},\min,+,\infty,0)$ and $g(u) = \min(u_q\mid q\in Q)$ and  $f(u)_q = u_q - g(u)$ for each $q \in Q$

  \item  $(\mathbb{R}_{\ge 0},\max,\cdot,0,1)$ and  $g(u) = \max( u_q \mid q \in Q)$ and  $f(u) = \frac{1}{g(u)}\cdot u$ 

\item the semiring $\Viterbi=([0,1],\max,\cdot,0,1)$ and $g(u) = \max( u_q\mid q\in Q)$ and $f(u) = \frac{1}{g(u)}\cdot u$ \hfill $\Box$

\end{enumerate}
\end{example}

Next we will show some properties of factorizations.

The last two examples in Example \ref{ex:sr-max-factorization} show particular semifields with maximal factorizations. In fact, for each zero-sum free semifield, we can show a general construction of a maximal factorization.

\begin{lemma}\rm \label{lm:semifield-factorization}\cite[Lm.~4.2]{buemayvog10}  Let $\B$ be a zero-sum free semifield. Then $(f,g)$ is a maximal factorization where  $g(u)=\bigoplus_{q \in Q} u_q$ and $f(u)= g(u)^{-1} \cdot  u$ for each $u \in B^Q \setminus \{\0^Q\}$.  
\end{lemma}
\begin{proof} First we show that $(f,g)$ is a factorization. Let $u \in B^Q \setminus \{\0^Q\}$. Since $\B$ is zero-sum free, $g(u) \ne \0$ and hence $g(u) \cdot f(u)  = g(u) \cdot (g(u)^{-1} \cdot u) = (g(u) \otimes g(u)^{-1}) \cdot u = u$.

  Second we show that $(f,g)$ is maximal. Let $b \in B$ such that $b \cdot u \ne \0^Q$. Moreover let $q \in Q$. Then
  \begingroup
  \allowdisplaybreaks
  \begin{align*}
    \big(f(b\cdot u)\big)_q =& \big(g(b \cdot u)^{-1} \cdot (b \cdot u)\big)_q = (\bigoplus_{q'\in Q} b\otimes u_{q'})^{-1} \otimes b \otimes u_q\\
    =& (b \otimes \bigoplus_{q'\in Q} u_{q'})^{-1} \otimes b \otimes u_q
       = (\bigoplus_{q'\in Q} u_{q'})^{-1} \otimes b^{-1} \otimes b \otimes u_q\\
    =& g(u)^{-1} \otimes u_q = \big(f(u)\big)_q \enspace. \qedhere
  \end{align*}
\endgroup
  \end{proof}

  The next lemma shows that, for a commutative semiring  with zero-divisors and $|Q|\ge 2$, there does not exist a  maximal factorization (also cf. \cite[Lm.~1]{gergon20} where, however, $Q$ is an infinite countable set).

  \begin{lemma}\rm \cite[Lm.~4.4]{buemayvog10} \label{lm:maxzd}
  If~$\B$ is commutative and $(f,g)$ is a maximal factorization, then $|Q| = 1$ or~$\B$ is zero-divisor free.
\end{lemma}

\begin{proof} If $|Q| = 1$, then the statement holds.  So assume that $|Q| \ge 2$ and let  $a_1\in B\setminus\{\mathbb{0}\}$ and $a_2\in B$ be such that $a_1\otimes a_2=\mathbb{0}$.
	We choose a pair $q_1,q_2\in Q$ such that $q_1\neq q_2$.  For each $i \in \{1,2\}$, we define the $Q$-vector $u_i\in B^Q$ as the $Q$-vector whose $q_i$-component is~$\mathbb{1}$ while the other components are~$\mathbb{0}$. Since $(f,g)$ is maximal and $a_1\otimes a_2=\mathbb{0}$, we have that
	\begin{align*}
		f(u_1) &= f(a_1\cdot u_1) = f(a_1\cdot u_1 \oplus (a_1\otimes a_2)\cdot u_2) = f(a_1\cdot(u_1\oplus a_2\cdot u_2))\\
		&= f(u_1\oplus a_2\cdot u_2)\;.
	\end{align*}
 Let $u = u_1\oplus a_2\cdot u_2$.   Since~$f(u_1)= f(u)$ and $(f,g)$ is a factorization, we obtain the equalities
	\begin{align*}
		g(u_1)\otimes f(u_1)_{q_1} &= (u_1)_{q_1} = \mathbb{1}\tag{I}\\
		g(u_1)\otimes f(u_1)_{q_2} &= (u_1)_{q_2} = \mathbb{0}\tag{II}\\
		g(u)\otimes f(u_1)_{q_1} &= u_{q_1} = \mathbb{1}\tag{III}\\
		g(u)\otimes f(u_1)_{q_2} &= u_{q_2} = a_2.\tag{IV}
	\end{align*}
     	By (I) and (III), and using commutativity, we derive
	\begin{align*}
		g(u_1) = g(u_1)\otimes \bigl(g(u) \otimes f(u_1)_{q_1} \bigr) = \bigl(g(u_1)\otimes f(u_1)_{q_1}\bigr)\otimes g(u) = g(u)\;.
	\end{align*}
	Then by (II) and (IV) we obtain that  $a_2= \0$. Hence $\B$ is zero-divisor free.
      \end{proof}
      
If $|Q|=1$, then $\cA$ is already bu-deterministic and there is no need to determinize it. Thus, as a consequence of Lemma~\ref{lm:maxzd}, if we want to determinize by means of maximal factorization a $(\Sigma,\B)$-wta which is not bu-deterministic, then $\B$ has to be zero-divisor free.

The next lemma shows that, e.g., in the representation 
\[f(\delta_\cA(\sigma)( f(\delta_\cA(\alpha)()),f(\delta_\cA(\alpha)())))\]
of a state of $Q'$, the inner two occurrences of $f$ can be dropped without changing the value of the expression, i.e.,
\[f(\delta_\cA(\sigma)( f(\delta_\cA(\alpha)()),f(\delta_\cA(\alpha)())))
   =
  f(\delta_\cA(\sigma)(\delta_\cA(\alpha)(),\delta_\cA(\alpha)())) = f(\h_\cA(\sigma(\alpha,\alpha)))\enspace.
\]
The next lemma also says that dropping the $f$ from  $\delta_\cA(\sigma)(\ldots,f(u_i),\ldots) \ne \0^Q$ preserves the non-zero property  if $\B$ is zero-divisor free.

\begin{lemma} \label{lm:l1} \rm \cite[Lm.~5.5]{buemayvog10}
Let $\B$ be commutative and $(f,g)$ be a maximal factorization. Furthermore, let~$k\in\mathbb{N}$, $\sigma\in\Sigma^{(k)}$, and $u_1,\ldots,u_k\in B^Q\setminus \{\0^Q\}$. Then the following two statements hold.
  \begin{compactenum}
  \item[(1)] If $\delta_\cA(\sigma)(u_1,\ldots,u_k)\neq\0^Q$, then $\delta_\cA(\sigma)(f(u_1),\ldots,f(u_k))\neq\0^Q$ and\\ $f(\delta_\cA(\sigma)(u_1,\ldots,u_k)) = f(\delta_\cA(\sigma)(f(u_1),\ldots,f(u_k)))$.
    \item[(2)] If $\B$ is zero-divisor free and $\delta_\cA(\sigma)(f(u_1),\ldots,f(u_k))\neq\0^Q$, then $\delta_\cA(\sigma)(u_1,\ldots,u_k)\neq\0^Q$.
        \end{compactenum}
\end{lemma}

\begin{proof}
	Clearly we have  ($\star$)
	\begin{align*}
		\delta_\cA(\sigma)(u_1,\ldots,u_k)
		&= \delta_\cA(\sigma)(g(u_1)\cdot f(u_1),\ldots,g(u_k)\cdot f(u_k)) \tag{$(f,g)$ factorization}\\
		&= \big(g(u_1)\otimes\cdots\otimes g(u_k)\big)\cdot\delta_\cA(\sigma)(f(u_1),\ldots,f(u_k))\;. \tag{Lemma \ref{lm:vector-algebra-is-semimodule-com-sr}}
	\end{align*}
	Proof of (1). Let $\delta_\cA(\sigma)(u_1,\ldots,u_k)\neq\0^Q$. By~($\star$) also $\delta_\cA(\sigma)(f(u_1),\ldots,f(u_k))\neq\0^Q$. Applying~$f$ to~($\star$) and using that~$(f,g)$ is maximal, we obtain that $f(\delta_\cA(\sigma)(u_1,\ldots,u_k)) = f(\delta_\cA(\sigma)(f(u_1),\ldots,f(u_k)))$.

Proof of (2). We assume that $\delta_\cA(\sigma)(f(u_1),\ldots,f(u_k))\neq\0^Q$. Since  $g(u_i)\neq \mathbb{0}$ for every~$i\in[k]$, and since~$\B$ is zero-divisor free,  ($\star$) yields that $\delta_\cA(\sigma)(u_1,\ldots,u_k)\neq\0^Q$.
\end{proof}


Let $\B$ be commutative and $(f,g)$ a maximal factorization. Then the set $Q'$ of states of $\det_{(f,g)}(\cA)$ can be enumerated by enumerating $\T_\Sigma$ in the sense that $Q'\subseteq f(\h_\cA(\T_\Sigma)\setminus\{\0^Q\})$ (cf. Lemma \ref{lm:l2}). As preparation we have the following easy observation.

       \begin{lemma}\rm \cite[Obs.~4.5]{buemayvog10}\label{obs:strat} Let $\det_{(f,g)}(\cA)=(Q',\delta',F')$. Moreover, let $(Q'_n\mid n\in\mathbb{N})$ be the family defined by
$Q_0' = \emptyset$, and for every~$n\in\mathbb{N}$ we let
         \[
        Q_{n+1}' = Q_{n}'\cup \{f(\delta_\cA(\sigma)(u_1,\ldots,u_k)) \mid  
        k\in\mathbb{N}, \sigma\in\Sigma^{(k)}, u_1,\ldots,u_k\in Q_n', \delta_\cA(\sigma)(u_1,\ldots,u_k)\neq\0^Q\} \enspace.
        \]
 Then  $Q' = \bigcup (Q_n' \mid n\in\mathbb{N})$.
 Moreover, $Q'$ is finite iff there exists an~$N\in\mathbb{N}$ with $Q'_{N+1} = Q_N'$.
 \hfill$\Box$
\end{lemma}

\begin{proof} We define the mapping
  \(
h: \cP(B^Q) \to \cP(B^Q)
\)
for each $U \in \cP(B^Q)$ by
\[
  h(U) = U \cup \{f(\delta_\cA(\sigma)(u_1,\ldots,u_k)) \mid  
  k\in\mathbb{N}, \sigma\in\Sigma^{(k)}, u_1,\ldots,u_k\in U, \delta_\cA(\sigma)(u_1,\ldots,u_k)\neq\0^Q\}\enspace.
  \]
  It is easy to see that $h$ is continuous. Moreover, we have
  \begingroup
  \allowdisplaybreaks
  \begin{align*}
    Q' &= \bigcap (U \mid U \in \cP(B^Q), h(U) \subseteq U)
         \tag{by definition of $Q'$ and $h$}\\
       &= \bigcup (h^n(\emptyset) \mid n \in \mathbb{N})
         \tag{by Theorem \ref{thm:Knaster-Tarski}}\\
    &= \bigcup (Q_n' \mid n \in \mathbb{N})
      \tag{by definition of $Q_n'$ and $h$} \enspace.
  \end{align*}
  \endgroup

  The second statement of the lemma is obvious.
  \end{proof}

Then we can show the following enumeration lemma (which is based on \cite[Lm.~2]{kirmae05}). Intuitively, it is obtained by iterating Lemma~\ref{lm:l1}, thereby pulling out $f$ from $Q'_1$, then from $Q'_2$, then from $Q'_3$, and so on.

\begin{lemma} \rm \cite[Lm.~5.8]{buemayvog10} \label{lm:l2}
	Let $\B$ be commutative, $(f,g)$ a maximal factorization, and $\det_{(f,g)}(\cA) = (Q',\delta',F')$. Then $Q'\subseteq f(\h_\cA(\T_\Sigma)\setminus\{\0^Q\})$.
\end{lemma}
\begin{proof}
  Again, we prove by case analysis on the cardinality of $Q$.
  
\underline{Case (a): }  Let $|Q| = 1$. We can identify~$B^Q$ with~$B$. Then,
\begin{equation}\label{eq:number-of-states=1-then}
  \text{for each $u\in B\setminus \{\mathbb{0}\}$, we have $f(u) = f(\mathbb{1})$.}
\end{equation}
To see this we compute:
\begingroup
\allowdisplaybreaks
\begin{align*}
  f(u)&=f(u\cdot \mathbb{1})
  \tag{where on the left-hand side $u \in B^Q$ and on the right-hand side $u \in B$}\\
      &= f(\mathbb{1}) \enspace.
        \tag{because $(f,g)$ is maximal}
\end{align*}
\endgroup

By our assumption on~$\cA$, there exists an~$\alpha\in\Sigma^{(0)}$ such that $\delta_\cA(\alpha)()\neq \mathbb{0}$. Hence \[\h_\cA(\alpha) = \delta_\cA(\alpha)()\neq \mathbb{0}\] and
thus, by \eqref{eq:number-of-states=1-then}, we have  $f(\h_\cA(\alpha)) = f(\mathbb{1})$ and $f(\h_\cA(\T_\Sigma)\setminus\{\0\})=\{f(\mathbb{1})\}$.

Next, by induction on $\mathbb{N}_+$, we show that $Q_{n}' = \{f(\1)\}$ for each $n \in \mathbb{N}_+$. The I.B. follows from $f\big(\delta_\cA(\alpha)()\big)= f(\mathbb{1})$ and \eqref{eq:number-of-states=1-then}; then the I.S. can be proved by using \eqref{eq:number-of-states=1-then}.

Thus $Q'=\{f(\mathbb{1})\}$ and we obtain $Q'= f(\h_\cA(\T_\Sigma)\setminus\{\0\})$.

\underline{Case (b): }  Let $|Q| > 1$. By Lemma~\ref{lm:maxzd}, $\B$ is zero-divisor free. Using Lemma~\ref{obs:strat}, it suffices to prove the following statement by induction on~$\mathbb{N}$:
\begin{equation}
  \text{for each~$n\in\mathbb{N}$, we have $Q'_n\subseteq f(\h_\cA(\T_\Sigma)\setminus\{\0^Q\})$.} \label{eq:enumeration-Q'}
\end{equation}

I.B.: Let $n=0$. Then \eqref{eq:enumeration-Q'} is trivially true.

I.S.: Let~$n=n'+1$ for some $n' \in\mathbb{N}$. We assume that \eqref{eq:enumeration-Q'} holds for $n'$. Let $u\in Q'_{n}$. If $u\in Q'_{n'}$, then we are done. Otherwise, there exist~$k\in\mathbb{N}$, $\sigma\in\Sigma^{(k)}$, and~$u_1,\ldots,u_k\in Q'_{n'}$ such that $\delta_\cA(\sigma)(u_1,\ldots,u_k) \ne \0^Q$ and  $u=f(\delta_\cA(\sigma)(u_1,\ldots,u_k))$. By I.H., for each~$i\in [k]$ there exists $\xi_i \in \T_\Sigma$ such that $u_i=f(\h_\cA(\xi_i))$.  Hence
	\begin{align*}
		u
		&= f(\delta_\cA(\sigma)(u_1,\ldots,u_k)) 
           = f(\delta_\cA(\sigma)(f(\h_\cA(\xi_1)),\ldots,f(\h_\cA(\xi_k))))\enspace.
        \end{align*}
By assumption $\delta_\cA(\sigma)(u_1,\ldots,u_k) \ne \0^Q$, i.e., $\delta_\cA(\sigma)(f(\h_\cA(\xi_1)),\ldots,f(\h_\cA(\xi_k))) \ne \0^Q$. By Lemma~\ref{lm:l1}(2) (using zero-divisor freeness), $\delta_\cA(\sigma)(\h_\cA(\xi_1),\ldots,\h_\cA(\xi_k)) \ne \0^Q$. Thus, we can continue with:
        \begin{align*}
		f(\delta_\cA(\sigma)(f(\h_\cA(\xi_1)),\ldots,f(\h_\cA(\xi_k)))) &= f(\delta_\cA(\sigma)(\h_\cA(\xi_1),\ldots,\h_\cA(\xi_k))) \tag{Lemma~\ref{lm:l1}(1)}\\
		&= f(\h_\cA(\sigma(\xi_1,\ldots,\xi_k))) 
		\in f(\h_\cA(\T_\Sigma)\setminus\{\0^Q\})\;.  \qedhere
	\end{align*}
        
\end{proof} 
   

    \subsection{On the twinning property}
    \label{ssec:properties-of-twins-property}
    
    Obviously, there exist $(\Sigma,\B)$-wta which do not have the twinning property. For example, consider the $(\Sigma,\Natmaxplus)$-wta $\cA =(Q,\delta,F)$ with $\Sigma = \{\gamma^{(1)}, \alpha^{(0)}\}$, $Q=\{p,q\}$, and
    \begin{align*}
      \delta_0(\varepsilon,\alpha,p) = 2 \ \ \  & \delta_1(p,\gamma,p) =5\\
            \delta_0(\varepsilon,\alpha,q) = 3 \ \ \ & \delta_1(q,\gamma,q) =7 
    \end{align*}
    and the weight of each  other transition is $-\infty$. 
    Then, for $\xi = \alpha$ and context $c = \gamma(z)$, we have
    \begin{align*}
      \wt(\R_\cA(p,\xi)) = 2 \ne -\infty \ \ \ & \wt(\R_\cA(p,c,p)) = 5 \ne  -\infty\\
      \wt(\R_\cA(q,\xi)) = 3 \ne  -\infty \ \ \ & \wt(\R_\cA(q,c,q)) = 7 \ne  -\infty
    \end{align*}
    and $\wt(\R_\cA(p,c,p)) \ne \wt(\R_\cA(q,c,q))$.

    But, in particular, each bu-deterministic wta has the twinning property.
    
\begin{observation}\label{obs:bu-deterministic-twins}\rm
The conditions $\wt(\R_\cA(p,\xi))\neq \0$ and $\wt(\R_\cA(q,\xi))\neq \0$ imply that
$p,q\in \Q_{\not= \0}^{\R_\cA}(\xi)$ (cf. Section \ref{sect:properties-of-wta}). If $\cA$ is bu-deterministic, 
then by Lemma \ref{lm:limit-bu-det}(2) it follows that $p=q$. Hence, each bu-deterministic wta
has the twinning property. \hfill $\Box$
\end{observation}

In the next example we illustrate that proving the twinning property can be a tedious task even for a small wta.

\begin{example} \rm \label{ex:factorization2} We prove that the wta $\cA$ given in Example \ref{ex:gen-subset-not-applicable} has the twinning property. For the sake of convenience, we repeat the definition of that $(\Sigma,\Ratminplus)$-wta.  

  We recall that $\Ratminplus=(\mathbb{Q}_{\infty},\min,+,\infty,0)$  is the tropical semifield. We consider the ranked alphabet $\Sigma = \{\sigma^{(2)}, \alpha^{(0)}\}$ and the $(\Sigma,\Ratminplus)$-wta $\cA =(Q,\delta,F)$ given by
    \begin{itemize}
    \item $Q= \{p,p'\}$ and $F_p=0$ and $F_{p'}=\infty$ and
      \item $\delta_0(\varepsilon,\alpha,p)=1$, $\delta_2(pp,\sigma,p)=1$, $\delta_2(pp,\sigma,p')=0.5$, and $\delta_2(p'p,\sigma,p)=1.5$, and all other values of $\delta_0$ and $\delta_2$ are~$\infty$. 
      \end{itemize}

      Next we show that $\cA$ has the twinning property.
First, by induction on $\T_\Sigma$, we prove that the following statement holds:
\begin{eqnarray}
  \begin{aligned}
    &\text{For every $\xi\in \T_\Sigma$, $\rho \in \R_\cA(p,\xi)$, and $\rho' \in \R_\cA(p',\xi)$ we have}\\
  &\text{(1) $\wt(\xi,\rho)=\infty$ or $\wt(\xi,\rho)=\size(\xi)$ and}\\
  &\text{(2) $\wt(\xi,\rho')=\infty$ or $\wt(\xi,\rho')=\size(\xi)-0.5$.}
  \end{aligned}\label{equ:p-tree-1}
\end{eqnarray}

I.B.: Let $\xi=\alpha$. Then (1) holds because $\R_\cA(p,\xi)=\{\rho\}$, where $\rho(\varepsilon)=p$. Hence $\wt(\xi,\rho)=\delta_0(\varepsilon,\alpha,p)=1=\size(\xi)$. Also, (2) holds because $\R_\cA(p',\xi)=\{\rho\}$, where $\rho(\varepsilon)=p'$. Hence $\wt(\xi,\rho)=\delta_0(\varepsilon,\alpha,p')=\infty$.

I.S.: Let $\xi=\sigma(\xi_1,\xi_2)$.  For the proof, let $\rho \in \R_\cA(p,\xi)$. Then
\begin{equation}\label{equ:p-p-run}
\wt(\xi,\rho)=\wt(\xi_1,\rho|_1)+\wt(\xi_2,\rho|_2)+\delta_2(\rho(1)\rho(2),\sigma,p),
\end{equation} 
where $\rho|_i$ is the run induced by $\rho$ at position $i$  for each $i\in [2]$; in particular, $\rho|_i\in \R_\cA(\rho(i),\xi_i)$.

For the proof of \eqref{equ:p-tree-1}(1), we  distinguish three cases as follows.

\underline{Case (a):} Let $\rho(1)\rho(2)=pp$. By  I.H., part (1), $\wt(\xi_i,\rho|_i)=\infty$ or $\wt(\xi_i\rho|_i)=\size(\xi_i)$ for each $i\in [2]$. If $\wt(\xi_i,\rho|_i)=\infty$ for some $i\in [2]$, then by \eqref{equ:p-p-run} we have $\wt(\xi,\rho)=\infty$.
Otherwise, again by \eqref{equ:p-p-run}, we have $\wt(\xi,\rho)=\size(\xi_1)+\size(\xi_2)+\delta_2(pp,\sigma,p)=\size(\xi_1)+\size(\xi_2)+1=\size(\xi)$.

\underline{Case (b):} Let $\rho(1)\rho(2)=p'p$. By  I.H., part (2), $\wt(\xi_1,\rho|_1)=\infty$ or $\wt(\xi_1,\rho|_1)=\size(\xi_1)-0.5$, and by I.H., part (1), $\wt(\xi_2,\rho|_2)=\infty$ or $\wt(\xi_2,\rho|_2)=\size(\xi_2)$.
If $\wt(\xi_i\rho|_i)=\infty$ for some $i\in [2]$, then by \eqref{equ:p-p-run} we have $\wt(\xi,\rho)=\infty$.
Otherwise, also by \eqref{equ:p-p-run}, we have $\wt(\xi,\rho)=\size(\xi_1)-0.5+\size(\xi_2)+\delta_2(p'p,\sigma,p)=\size(\xi_1)-0.5+\size(\xi_2)+1.5=\size(\xi)$.

\underline{Case (c):} Let $\rho(1)\rho(2)\not\in \{pp,p'p\}$.  Then $\delta_2(\rho(1)\rho(2),\sigma,p)=\infty$ and thus by 
\eqref{equ:p-p-run} we have $\wt(\xi,\rho)=\infty$.

For the proof of \eqref{equ:p-tree-1}(2), we  distinguish two cases.

\underline{Case (a):} Let $\rho(1)\rho(2)=pp$. By  I.H., part (1), $\wt(\xi_i,\rho|_i)=\infty$ or $\wt(\xi_i,\rho|_i)=\size(\xi_i)$ for each $i\in [2]$. If $\wt(\xi_i,\rho|_i)=\infty$ for some $i\in [2]$, then by \eqref{equ:p-p-run} we have $\wt(\xi,\rho)=\infty$.
Otherwise, by \eqref{equ:p-p-run}, we have $\wt(\xi,\rho)=\size(\xi_1)+\size(\xi_2)+\delta_2(pp,\sigma,p')=\size(\xi_1)+\size(\xi_2)+0.5=\size(\xi)-0.5$.

\underline{Case (b):} Let $\rho(1)\rho(2)\ne pp$. Then $\delta_2(\rho(1)\rho(2),\sigma,p')=\infty$ and thus by 
\eqref{equ:p-p-run} we have $\wt(\xi,\rho)=\infty$.

This finishes the proof of \eqref{equ:p-tree-1}.

\

Next we prove the following statement for contexts. For every $c\in \C_\Sigma\setminus\{z\}$:
\begin{eqnarray}
  \begin{aligned}
&\text{(1) for every $\rho \in \R_\cA(p,c,p)$:  $\wt(c,\rho)=\infty$ or $\wt(c,\rho)=\size(c)-1$ and}\\
&\text{(2) for every $\rho \in \R_\cA(p,c,p')$: $\wt(c,\rho)=\infty$ or $\wt(c,\rho)=\size(c)-0.5$ and}\\
&\text{(3) for every $\rho \in \R_\cA(p',c,p)$: $\wt(c,\rho)=\infty$ or $\wt(c,\rho)=\size(c)-1.5$, and}\\
&\text{(4) for every $\rho \in \R_\cA(p',c,p')$: $\wt(c,\rho)=\infty$ or $\wt(c,\rho)=\size(c)-1$.}
\end{aligned}\label{equ:p-p-context}
\end{eqnarray}

\index{succC@$\succ_{\C_\Sigma}$}
For the proof, we consider the relation $\succ_{\C_\Sigma}$ on $\C_\Sigma$ defined on page \pageref{page:order-on-contexts}.
We denote the restriction of $\succ_{\C_\Sigma}$ to $\C_\Sigma\setminus\{z\}$ also by $\succ_{\C_\Sigma}$ and we note that $\succ_{\C_\Sigma}$ is terminating and $\nf_{\succ_{\C_\Sigma}}(\C_\Sigma\setminus\{z\})=\e\C_\Sigma$, i.e., the set of elementary contexts. 

We  prove \eqref{equ:p-p-context} by induction on $(\C_\Sigma\setminus\{z\},\succ_{\C_\Sigma})$

I.B.: We distinguish the two cases that (a) $e=\sigma(z,\xi)$ and (b) $e=\sigma(\xi,z)$ for some $\xi\in \T_\Sigma$. Then we proceed as follows.

\begin{enumerate}
\item[(1)] Let $\rho \in  \R_\cA(p,e,p)$. If $\wt(e,\rho)\ne \infty$, then the following two cases are possible:
\begin{itemize}
\item $e$ has the form  (a), $\rho(1)=\rho(2)=p$, and $\rho|_2\in \R_\cA(p,\xi)$.
By \eqref{equ:p-tree-1}(1), we have $\wt(\xi,\rho|_2)=\size(\xi)$. Hence $\wt(e,\rho)=0+\wt(\rho|_2)+\delta_2(pp,\sigma,p)=\size(\xi)+1=\size(e)-1$.
\item $e$ has the form  (b),  $\Big(\rho(1)=\rho(2)=p$ and $\rho|_1\in \R_\cA(p,\xi)\Big)$  or $\Big(\rho(1)=p'$, $\rho(2)=p$, and $\rho|_1\in \R_\cA(p',\xi)\Big)$.
In the first case, by  symmetry,  we get as before $\wt(e,\rho)=\size(e)-1$.
In the second case, by \eqref{equ:p-tree-1}(2), we have $\wt(\xi,\rho|_1)=\size(\xi)-0.5$. Hence $\wt(e,\rho)=\wt(\rho|_1)+0+\delta_2(p'p,\sigma,p)=\size(\xi)-0.5+1.5=\size(\xi)+1=\size(e)-1$.
\end{itemize}
\item[(2)] Let $\rho \in  \R_\cA(p,e,p')$. If $\wt(e,\rho)\ne \infty$, then
\begin{itemize}
\item $e$ has the form  (a),  $\rho(1)=p'$, $\rho(2)=p$ and $\rho|_2\in \R_\cA(p,\xi)$.
By \eqref{equ:p-tree-1}(1), we have $\wt(\xi,\rho|_2)=\size(\xi)$. Hence $\wt(e,\rho)=0+\wt(\rho|_2)+\delta_2(p'p,\sigma,p)=\size(\xi)+1.5=\size(e)-0.5$.
\end{itemize}
\item[(3)] Let $\rho \in  \R_\cA(p',e,p)$. If $\wt(e,\rho)\ne \infty$, then the following two cases are possible:
\begin{itemize}
\item $e$ has the form  (a), $\rho(1)=\rho(2)=p$, and $\rho|_2\in \R_\cA(p,\xi)$.
By \eqref{equ:p-tree-1}(1), we have $\wt(\xi,\rho|_2)=\size(\xi)$. Hence $\wt(e,\rho)=0+\wt(\rho|_2)+\delta_2(pp,\sigma,p')=\size(\xi)+0.5=\size(e)-1.5$.
\item $e$ has the form  (b), $\rho(1)=\rho(2)=p$, and $\rho|_1\in \R_\cA(p,\xi)$. By symmetry with the previous case, we obtain $\wt(e,\rho)=\size(e)-1.5$.
\end{itemize}
\item[(4)] Let $\rho \in  \R_\cA(p',e,p')$. We have $\wt(e,\rho)=\infty$ because $\delta_2(p'q,\sigma,p)=\delta_2(qp',\sigma,p)=\infty$ for each $q\in Q$.
\end{enumerate}

I.S.: We distinguish the two cases that (a) $c=\sigma(c',\xi)$ and (b) $c=\sigma(\xi,c')$ for some $c'\in \C_\Sigma$ and $\xi\in \T_\Sigma$. Then we proceed as follows.

\begin{enumerate}
\item[(1)] Let $\rho \in  \R_\cA(p,c,p)$. If $\wt(c,\rho)\ne \infty$, then the following two cases are possible:
\begin{itemize}
\item $c$ has the form  (a), $\rho(1)=\rho(2)=p$, $\rho|_1\in \R_\cA(p,c',p)$, and $\rho|_2\in \R_\cA(p,\xi)$.
By I.H., part (1), we have $\wt(c',\rho|_1)=\size(c')-1$ and by \eqref{equ:p-tree-1}(1), we have $\wt(\xi,\rho|_2)=\size(\xi)$. Hence $\wt(c,\rho)=\wt(c',\rho|_1)+\wt(\xi,\rho|_2)+\delta_2(pp,\sigma,p)=\size(c')-1 +\size(\xi)+1=\size(c)-1$.
\item $c$ has the form  (a), $\rho(1)=p'$, $\rho(2)=p$, $\rho|_1\in \R_\cA(p',c',p)$, and $\rho|_2\in \R_\cA(p,\xi)$.
By I.H., part (3) we have $\wt(c',\rho|_1)=\size(c')-1.5$ and by \eqref{equ:p-tree-1}(1), we have $\wt(\xi,\rho|_2)=\size(\xi)$. Hence $\wt(c,\rho)=\wt(c',\rho|_1)+\wt(\xi,\rho|_2)+\delta_2(p'p,\sigma,p)=\size(c')-1.5 +\size(\xi)+1.5=\size(c)-1$.
\item $c$ has the form  (b),  $\rho(1)=\rho(2)=p$, $\rho|_1\in \R_\cA(p,\xi)$, and $\rho|_2\in \R_\cA(p,c',p)$. By  symmetry with the previous case, we obtain $\wt(c,\rho)=\size(c)-1$.
\item $c$ has the form  (b),   $\rho(1)=p'$, $\rho(2)=p$, $\rho|_1\in \R_\cA(p',\xi)$ and $\rho|_2\in \R_\cA(p,c',p)$.
By \eqref{equ:p-tree-1}(2), we have $\wt(\xi,\rho|_1)=\size(\xi)-0.5$
and by I.H., part (1), we have $\wt(c',\rho|_2)=\size(c')-1$. 
Hence $\wt(c,\rho)=\wt(\xi,\rho|_1)+\wt(c',\rho|_2)+\delta_2(p'p,\sigma,p)=\size(\xi)-0.5+\size(c')-1+1.5=\size(c)-1$.
\end{itemize}
\item[(2)] Let $\rho \in  \R_\cA(p,c,p')$. If $\wt(c,\rho)\ne \infty$, then
we have the following cases.
\begin{itemize}
\item $c$ has the form  (a),  $\rho(1)=p'$, $\rho(2)=p$, $\rho|_1\in \R_\cA(p',c',p')$, and $\rho|_2\in \R_\cA(p,\xi)$. 
By I.H., part (4) and by \eqref{equ:p-tree-1}(1),  we have $\wt(c',\rho|_1)=\size(c')-1$ and $\wt(\xi,\rho|_2)=\size(\xi)$, respectively. Hence $\wt(c,\rho)=\wt(c',\rho|_1)+\wt(\xi,\rho|_2)+\delta_2(p'p,\sigma,p)=\size(c')-1+\size(\xi)+1.5=\size(c)-0.5$.
\item $c$ has the form  (a),  $\rho(1)=\rho(2)=p$, $\rho|_1\in \R_\cA(p,c',p')$, and $\rho|_2\in \R_\cA(p,\xi)$.  By I.H., part (2) and by \eqref{equ:p-tree-1}(1),  we have $\wt(c',\rho|_1)=\size(c')-0.5$ and $\wt(\xi,\rho|_2)=\size(\xi)$, respectively. Hence $\wt(c,\rho)=\wt(c',\rho|_1)+\wt(\xi,\rho|_2)+\delta_2(pp,\sigma,p)=\size(c')-0.5+\size(\xi)+1=\size(c)-0.5$. 
\item $c$ has the form  (b),  $\rho(1)=p'$, $\rho(2)=p$, $\rho|_1\in \R_\cA(p',\xi)$, and $\rho|_2\in \R_\cA(p,c',p')$. By  \eqref{equ:p-tree-1}(2) and by I.H., part (2)  
$\wt(\xi,\rho|_1)=\size(\xi)-0.5$ and $\wt(c',\rho|_2)=\size(c')-0.5$, respectively. Hence $\wt(c,\rho)=\wt(\xi,\rho|_1)+\wt(c',\rho|_2)+\delta_2(p'p,\sigma,p)=\size(\xi)-0.5+\size(c')-0.5+1.5=\size(c)-0.5$. 
\item $c$ has the form  (b),  $\rho(1)=\rho(2)=p$, $\rho|_1\in \R_\cA(p,\xi)$, and $\rho|_2\in \R_\cA(p,c',p')$. By symmetry, we have $\wt(c,\rho)=\size(c)-0.5$.
\end{itemize}
\item[(3)] Let $\rho \in  \R_\cA(p',c,p)$. If $\wt(c,\rho)\ne \infty$, then the following two cases are possible:
\begin{itemize}
\item $c$ has the form  (a), $\rho(1)=\rho(2)=p$, $\rho|_1\in \R_\cA(p,c',p)$, and $\rho|_2\in \R_\cA(p,\xi)$. By I.H., part (1) and by \eqref{equ:p-tree-1}(1), we have $\wt(c',\rho|_1)=\size(c')-1$ and  $\wt(\xi,\rho|_2)=\size(\xi)$, respectively. Hence $\wt(c,\rho)=\wt(c',\rho|_1)+\wt(\xi,\rho|_2)+\delta_2(pp,\sigma,p')=\size(c')-1+\size(\xi)+0.5=\size(c)-1.5$.
\item $c$ has the form  (b), $\rho(1)=\rho(2)=p$, $\rho|_1\in \R_\cA(p,\xi)$, and $\rho|_2\in \R_\cA(p,c',p)$. By symmetry with the previous case, we obtain $\wt(c,\rho)=\size(c)-1.5$.
\end{itemize}
\item[(4)] Let $\rho \in  \R_\cA(p',c,p')$. If $\wt(c,\rho)\ne \infty$, then the following two cases are possible:
\begin{itemize}
\item $c$ has the form  (a), $\rho(1)=\rho(2)=p$, $\rho|_1\in \R_\cA(p,c',p')$, and $\rho|_2\in \R_\cA(p,\xi)$. By I.H., part (2) and by \eqref{equ:p-tree-1}(1), we have 
$\wt(c',\rho|_1)=\size(c')-0.5$ and $\wt(\xi,\rho|_2)=\size(\xi)$, respectively. 
Hence $\wt(c,\rho)=\wt(c',\rho|_1)+\wt(\xi,\rho|_2)+\delta_2(pp,\sigma,p')=\size(c')-0.5+\size(\xi)+0.5=\size(c)-1$.
\item $c$ has the form  (b), $\rho(1)=\rho(2)=p$, $\rho|_1\in \R_\cA(p,\xi)$, and $\rho|_2\in \R_\cA(p,c',p')$. By symmetry with the previous case, we obtain $\wt(c,\rho)=\size(c)-1$.
\end{itemize}
\end{enumerate}
This finishes the proof of the statement \ref{equ:p-p-context}.

Now we prove that $\cA$ has the twinning property. In fact, we show that even the following stronger property holds: for every context $c \in \C_\Sigma$, if $\wt(\R_\cA(p,c,p))\ne\infty$ and $\wt(\R_\cA(p',c,p'))\ne\infty$, then
$\wt(\R_\cA(p,c,p))=\wt(\R_\cA(p',c,p'))$. The proof is as follows. If $c=z$, then $\R_\cA(p,c,p) =\{\rho\}$ with $\rho(\varepsilon)=p$ and $\R_\cA(p',c,p') =\{\rho'\}$ with $\rho'(\varepsilon)=p'$. Hence $\wt(\R_\cA(p,c,p))=0=\wt(\R_\cA(p',c,p'))$. If $c\ne z$, then the condition $\wt(\R_\cA(p,c,p))\ne\infty$ implies that there exists a run $\rho\in \R_\cA(p,c,p)$ with $\wt(c,\rho)\ne\infty$. By  \eqref{equ:p-p-context}(1) we know that $\wt(c,\rho)=\size(c)-1$ for every such $\rho$, hence by the fact that summation of $\Ratminplus$ is $\min$, we obtain that $\wt(\R_\cA(p,c,p))=\size(c)-1$. Similarly, the conditions $\wt(\R_\cA(p',c,p'))\ne\infty$ and \eqref{equ:p-p-context}(4) imply that $\wt(\R_\cA(p',c,p'))=\size(c)-1$.

This ends the proof of the fact that $\cA$ has the twinning property.
\hfill $\Box$
\end{example}

Thus, the $(\Sigma,\Ratminplus)$-wta $\cA$ given in Example \ref{ex:gen-subset-not-applicable} has the twinning property. Also the other requirements of Theorem \ref{thm:extremal-twins-determinizable} are satisfied, because  $\Ratminplus = (\mathbb{Q}_{\infty},\min,+,\infty,0)$ is an extremal semiring and the pair $(f,g)$ is a maximal  factorization with $f: (\mathbb{Q}_{\infty})^Q \setminus \{\infty_Q\} \to (\mathbb{Q}_{\infty})^Q$ and $g: (\mathbb{Q}_{\infty})^Q \setminus \{\infty_Q\} \to \mathbb{Q}_{\infty}$ defined for each $u=\big(\begin{smallmatrix}u_p\\u_{p'}\end{smallmatrix}\big)$ in $\mathbb{Q}_{\infty}^Q \setminus \{\infty_Q\}$ by 
\begin{align*}
f(u)_q = u_q - g(u)\  \text{ for each $q \in Q$} \ \ \text{ and } \ \ g(u) =\min(u_p,u_{p'})\enspace.
    \end{align*}
   Hence, for the bu-deterministic $(\Sigma,\Ratminplus)$-wta $\det_{(f,g)}(\cA)$ (cf. Figure \ref{fig:ex-det-unfolding}), Theorem \ref{thm:extremal-twins-determinizable} guarantees that
    \begin{compactitem}
      \item $\sem{\det_{(f,g)}(\cA)}= \sem{\cA} = \size$  and
      \item $\det_{(f,g)}(\cA)$ is minimal with respect to the number of states  among all bu-deterministic $(\Sigma,\Ratminplus)$-wta which are equivalent to $\cA$ and  obtained by determinization by factorization.
      \end{compactitem}
      
    We note that the bu-deterministic $(\Sigma,\Natminplus)$-wta of Example \ref{ex:size}, which also recognizes $\size$, has only one state, and hence it is smaller than $\det_{(f,g)}(\cA)$ (where, in this comparison, we disregard the difference between $\Natminplus$ and $\Ratminplus$). However, by means of two transformations, viz. weight pushing and forward bisimulation, we can transform $\det_{(f,g)}(\cA)$ into the $(\Sigma,\Ratminplus)$-wta of Example \ref{ex:size} (cf. Figure \ref{fig:pipeline}). Let us briefly explain these transformations.
  
         Since $\Ratminplus$ is a commutative semifield,
        we can apply weight pushing to $\det_{(f,g)}(\cA)$ (cf. Section~\ref{sec:normalizing-root-weights}). More precisely, we define the mapping $\lambda(\binom{0}{\infty})=0$ and $\lambda(\binom{0.5}{0})=0.5$. Then Figure \ref{fig:pipeline} shows the $(\Sigma,\Ratminplus)$-wta $\push_\lambda(\det_{(f,g)}(\cA))$. By Lemma \ref{lm:A=push-lambda(A)} we have that $\sem{\det_{(f,g)}(\cA)} = \sem{\push_\lambda(\det_{(f,g)}(\cA))}$.
      
\begin{figure}

\vspace{2cm}

         \begin{tikzpicture}
        \node at (-0.8, 1.2)  {$\cA$:};
\tikzset{node distance=7em,scale=0.7, transform shape}
\node[state, rectangle] (1) {$\alpha$};
\node[state, right of=1] (2) {$p$};
\node[state, rectangle, right of=2] (3) {$\sigma$};
\node[state, rectangle, below of=2] (4) {$\sigma$};
\node[state, rectangle, left of=4] (5)[right=0.3em] {$\sigma$};
\node[state, below of=4] (6) {$p'$};

\tikzset{node distance=2em}
\node[above of=1] (w1)[right=-0.7em] {1};
\node[above of=2] (w2) {0};
\node[above of=3] (w3) {1};
\node[right of=4] (w4)[right] {0.5};
\node[above of=5] (w5) {1.5};

\draw[->,>=stealth] (1) edge (2);
\draw[->,>=stealth] (3) edge (2);
\draw[->,>=stealth] (2) edge[out=60, in=30, looseness=1.4] (3);
\draw[->,>=stealth] (2) edge[out=-60, in=-30, looseness=1.4] (3);
\draw[->,>=stealth] (2) edge (5);
\draw[->,>=stealth] (2) edge[out=-105, in=115, looseness=1.3] (4);
\draw[->,>=stealth] (2) edge[out=-75, in=65, looseness=1.3] (4);
\draw[->,>=stealth] (4) edge (6);
\draw[->,>=stealth] (6) edge (5);
\draw[->,>=stealth] (5) edge[out=170, in=-150, looseness=1.5] (2);
\end{tikzpicture}

\vspace{-63mm}

\hspace*{65mm}
\begin{tikzpicture}
        \node at (0.8, 2.2)  {$\det_{(f,g)}(\cA)$:};
\tikzset{node distance=7em, scale=0.6, transform shape}
\node[state, rectangle] (1) {$\alpha$};
\node[state, right of=1] (2){$\binom{0}{\infty}$};
\node[state, rectangle, right of=2] (3)[right=1em]{$\sigma$};
\node[state, rectangle, above of=3] (4)[above=1em]{$\sigma$};
\node[state, rectangle, below of=3] (5)[below=1em]{$\sigma$};
\node[state, right of=3] (6)[right=1em]{$\binom{0.5}{0}$};
\node[state, rectangle, right of=6] (7) {$\sigma$};

\tikzset{node distance=2em}
\node[above of=1] (w1) {1};
\node[above of=2] (w2) {0};
\node[above of=3] (w3) {0.5};
\node[above of=4] (w4) {1};
\node[above of=5] (w5) {1};
\node[above of=6] (w6)[left=0.1em] {0.5};
\node[above of=7] (w7) {1.5};

\draw[->,>=stealth] (1) edge (2);
\draw[->,>=stealth] (2) edge[out=20, in=155, looseness=1.1] (3);
\draw[->,>=stealth] (2) edge[out=-20, in=205, looseness=1.1] (3);
\draw[->,>=stealth] (2) edge (4);
\draw[->,>=stealth] (2) edge (5);
\draw[->,>=stealth] (3) edge (6);
\draw[->,>=stealth] (6) edge (4);
\draw[->,>=stealth] (4) edge[out=-290, in=80, looseness=1.4] (6);
\draw[->,>=stealth] (6) edge (5);
\draw[->,>=stealth] (5) edge[out=290, in=-80, looseness=1.4] (6);
\draw[->,>=stealth] (7) edge (6);
\draw[->,>=stealth] (6) edge[out=60, in=30, looseness=1.4] (7);
\draw[->,>=stealth] (6) edge[out=-60, in=-30, looseness=1.4] (7);
\end{tikzpicture}

\vspace{10mm}

\hspace*{4mm}
\begin{tikzpicture}
          \node at (0.8, 2)  {$(\push_\lambda(\det_{(f,g)}(\cA)))/R$:};
\tikzset{node distance=7em, scale=0.6, transform shape}
\node[state, rectangle] (1) {$\alpha$};
\node[state, right of=1] (2) {$\{\binom{0}{\infty},\binom{0.5}{0}\}$};
\node[state, rectangle, right of=2] (3) {$\sigma$};

\tikzset{node distance=2em}
\node[above of=1] (w1) {1};
\node[above of=2, yshift=8mm] (w2) {0};
\node[above of=3] (w3) {1};

\draw[->,>=stealth] (1) edge (2);
\draw[->,>=stealth] (3) edge (2);
\draw[->,>=stealth] (2) edge[out=60, in=30, looseness=1.4] (3);
\draw[->,>=stealth] (2) edge[out=-60, in=-30, looseness=1.4] (3);
\end{tikzpicture}

\vspace{-49mm}

\hspace*{60mm}
\begin{tikzpicture}
        \node at (0.8, 2)  {$\push_\lambda(\det_{(f,g)}(\cA))$:};
\tikzset{node distance=7em, scale=0.6, transform shape}
\node[state, rectangle] (1) {$\alpha$};
\node[state, right of=1] (2){$\binom{0}{\infty}$};
\node[state, rectangle, right of=2] (3)[right=1em]{$\sigma$};
\node[state, rectangle, above of=3] (4)[above=1em]{$\sigma$};
\node[state, rectangle, below of=3] (5)[below=1em]{$\sigma$};
\node[state, right of=3] (6)[right=1em]{$\binom{0.5}{0}$};
\node[state, rectangle, right of=6] (7) {$\sigma$};

\tikzset{node distance=2em}
\node[above of=1] (w1) {1};
\node[above of=2] (w2) {0};
\node[above of=3] (w3) {1};
\node[above of=4] (w4) {1};
\node[above of=5] (w5) {1};
\node[above of=6] (w6)[left=0.1em] {0};
\node[above of=7] (w7) {1};

\draw[->,>=stealth] (1) edge (2);
\draw[->,>=stealth] (2) edge[out=20, in=155, looseness=1.1] (3);
\draw[->,>=stealth] (2) edge[out=-20, in=205, looseness=1.1] (3);
\draw[->,>=stealth] (2) edge (4);
\draw[->,>=stealth] (2) edge (5);
\draw[->,>=stealth] (3) edge (6);
\draw[->,>=stealth] (6) edge (4);
\draw[->,>=stealth] (4) edge[out=-290, in=80, looseness=1.4] (6);
\draw[->,>=stealth] (6) edge (5);
\draw[->,>=stealth] (5) edge[out=290, in=-80, looseness=1.4] (6);
\draw[->,>=stealth] (7) edge (6);
\draw[->,>=stealth] (6) edge[out=60, in=30, looseness=1.4] (7);
\draw[->,>=stealth] (6) edge[out=-60, in=-30, looseness=1.4] (7);
\end{tikzpicture}

\vspace{-150mm}

\hspace*{35mm}
\begin{tikzpicture}
\draw[thick,->,>=stealth] ([shift=(123:6.5cm)]0,0) arc (123:67:6.5cm)
  node[midway, above] {$\mathrm{\det}_{(f,g)}$};
\draw[thick,->,>=stealth] ([shift=(11.5:6.5cm)]0,0) arc (11.5:-25:6.5cm)
  node[midway, right] {$\mathrm{push}_\lambda$};
\draw[thick,->,>=stealth] ([shift=(-75:6.5cm)]0,0) arc (-75:-119:6.5cm)
  node[midway, below] { $(.)/R$};
\end{tikzpicture}

\vspace{5mm}

\caption{\label{fig:pipeline} The following
$(\Sigma,\Ratminplus)$-wta are shown clockwise, starting
up-left:  (a) $\cA$ of Figure~\ref{fig:ex-unfolding},
(b)~$\det_{(f,g)}(\cA)$,  (c) $\push_\lambda(\det_{(f,g)}(\cA))$, and
(d) $(\push_\lambda(\det_{(f,g)}(\cA)))/R$. The wta in (b), (c), and (d)
are bu-deterministic. }
         \end{figure}

For the second transformation (viz., forward bisimulation, cf. \cite{hogmalmay07d}), we define the equivalence relation $R = Q \times Q$. Thus, the factor set $Q/R$ contains one equivalence class, namely $Q$. One can easily check that $R$ is a forward bisimulation on $\push_\lambda(\det_{(f,g)}(\cA))$ in the sense of \cite[Def.~1]{hogmalmay07d}. Thus we can construct the forward aggregate $(\Sigma,\Ratminplus)$-wta $(\push_\lambda(\det_{(f,g)}(\cA)))/R$ (cf. \cite[Def.~3]{hogmalmay07d}). By \cite[Thm.~6]{hogmalmay07d} we have $\sem{\push_\lambda(\det_{(f,g)}(\cA))} = \sem{(\push_\lambda(\det_{(f,g)}(\cA)))/R}$.
(We note that the relation $R$ is \underline{not} a forward bisimulation on $\det_{(f,g)}(\cA)$ because $F(\binom{0}{\infty}) = 0 \ne 0.5 = F(\binom{0.5}{0})$.)
It is obvious  that the forward aggregate $(\Sigma,\Ratminplus)$-wta $(\push_\lambda(\det_{(f,g)}(\cA)))/R$ is exactly the $(\Sigma,\Ratminplus)$-wta $\cA$ of Example~\ref{ex:size} (modulo state renaming and the change of the weight algebra from $\Natminplus$ to $\Ratminplus$). 

We also obtain the bu-deterministic $(\Sigma,\Ratminplus)$-wta of Example \ref{ex:size} if we apply the minimization of Theorem \ref{thm:minimization-theorem-new} to the bu-deterministic $(\Sigma,\Ratminplus)$-wta $\det_{(f,g)}(\cA)$ in Figure \ref{fig:ex-det-unfolding}, followed by weight pushing (cf. Example \ref{ex:applying-MN-to-det}).

Finally, we mention that in \cite[Sect. 5.5]{buemayvog10}, \cite{buefis12}, and \cite[Sect.~5.4]{bue14} the question of deciding the twinning property is investigated.


\subsection{Proof of Theorem \ref{thm:extremal-twins-determinizable}}
\label{ssec:proof-of-main-theorem}

We organize the proof of Theorem \ref{thm:extremal-twins-determinizable} as follows:
\begin{compactenum}
\item[(1)] finiteness of $\det_{(f,g)}(\cA)$ (cf. Theorem~\ref{thm:finiteness}),
\item[(2)] correctness of $\det_{(f,g)}(\cA)$ (cf. Theorem~\ref{thm:correctness}), and
\item[(3)] minimality of $\det_{(f,g)}(\cA)$ (cf. Theorem~\ref{thm:minimality}).
  \end{compactenum}

\subsubsection{Finiteness of $\det_{(f,g)}(\cA)$ (proof of Theorem~\ref{thm:extremal-twins-determinizable}(1))}

Here we show sufficient conditions under which $\det_{(f,g)}(\cA)$ is a $(\Sigma,\B)$-wta (cf. Theorem~\ref{thm:finiteness}).

Let $\xi\in \T_\Sigma$, $w\in\pos(\xi)$, and $\rho\in \R_\cA(\xi)$ such that $\rho(w) = \rho(\varepsilon)$ and assume that $\B$ is extremal and commutative. Then the following kind of Bellman optimality property holds (where ``optimality'' has to be replaced by ``victory''): If $\rho$ is victorious in $\R_\cA(\rho(\varepsilon),\xi))$, then $\rho|^w$ is  victorious for $\R_\cA(\rho(\varepsilon),\xi|^w,\rho(w))$ (modulo right multiplication with $\wt(\xi|_w,\rho|_w)$). 
The property is based on the following trivial fact.
\begin{align}
  & \text{Let $\B$ be extremal, $b \in B$, and $B_1,B_2 \subseteq B$ be finite subsets.}\nonumber\\
  & \text{If $b \in B_1$, $B_1 \subseteq B_2$, and $b = \bigoplus_{b' \in B_2}b'$, then $b = \bigoplus_{b' \in B_1}b'$.}\label{eq:fact-extremal}
  \end{align}

\begin{observation}\rm \label{ob:bellman} \cite[Obs.~5.12]{buemayvog10}
Let~$\B$ be extremal and commutative. Moreover, let $\xi\in \T_\Sigma$, $w\in\pos(\xi)$, and $\rho\in \R_\cA(\xi)$ such that $\rho(w) = \rho(\varepsilon)$ and  $\wt(\xi,\rho) = \wt(\R_\cA(\rho(\varepsilon),\xi))$. Then
\[
  \wt(\xi,\rho) = \wt(\xi|^w,\rho|^w)\otimes \wt(\xi|_w,\rho|_w) = \wt(\R_\cA(\rho(\varepsilon),\xi|^w,\rho(\varepsilon)))\otimes \wt(\xi|_w,\rho|_w)\enspace.
\]
\end{observation}

\begin{proof}
	We have
	\begin{align*}
           &\wt(\R_\cA(\rho(\varepsilon),\xi|^w,\rho(\varepsilon)))\otimes \wt(\xi|_w,\rho|_w)\\
		&= \textstyle \bigl(\bigoplus_{\nu\in \R_\cA(\rho(\varepsilon),\xi|^w,\rho(\varepsilon))} \wt(\xi|^w,\nu)\bigr)\otimes \wt(\xi|_w,\rho|_w)\\
		&= \textstyle \bigoplus_{\nu\in \R_\cA(\rho(\varepsilon),\xi|^w,\rho(\varepsilon))} \wt(\xi|^w,\nu)\otimes\wt(\xi|_w,\rho|_w) \tag{by distributivity}\\
		&= \wt(\xi,\rho)\tag{$\star$}\\
		&= \wt(\xi|^w,\rho|^w)\otimes\wt(\xi|_w,\rho|_w)\;.\tag{by commutativity}
	\end{align*}
	        At $(\star)$ we have used \eqref{eq:fact-extremal} with
        \begin{compactitem}
        \item $b = \wt(\xi,\rho)$,
        \item $B_1 = \{\wt(\xi|^w,\nu)\otimes\wt(\xi|_w,\rho|_w) \mid \nu\in \R_\cA(\rho(\varepsilon),\xi|^w,\rho(\varepsilon))\}$, and
          \item $B_2 = \{\wt(\xi,\nu)\mid \nu\in \R_\cA(\rho(\varepsilon),\xi)\}$.
          \end{compactitem}
          Obviously, (a) $b \in B_1$, (b) by commutativity we have $B_1 \subseteq B_2$, and (c) by assumption (viz. $\wt(\xi,\rho) = \wt(\R_\cA(\rho(\varepsilon),\xi))$)  we have $b = \bigoplus_{b' \in B_2} b'$.
\end{proof}

The next definitions are taken from \cite[Def.~5.13]{buemayvog10}.
  Let $S\subseteq Q$. Then we define
  \[
    \cC'(S) = \{(\xi,\kappa) \mid \xi\in \T_\Sigma, \kappa: S \to \R_\cA(\xi), (\forall q\in S)\colon \kappa_q\in \R_\cA(q,\xi), \wt(\xi,\kappa_q)\neq \mathbb{0}\}\enspace.
  \]
  We set $\cC' = \bigcup_{S\subseteq Q} \cC'(S)$.

  In the following, we define (a) the mapping $\overline{\wt}:\cC'\rightarrow B^Q$ and (b) for each $S \subseteq Q$, the family $(U(\xi,\kappa)\mid(\xi,\kappa)\in\cC'(S))$ and (c) the sets $\cC(S) \subseteq \cC'(S)$ and $\cC$. To this end, let $S \subseteq Q$ and  $(\xi,\kappa)\in\cC'(S)$.
    Then
	\begin{itemize}
		\item for each~$q\in Q$ we set $\overline{\wt}(\xi,\kappa)_q = \wt(\xi,\kappa_q)$ if $q\in S$, otherwise we set $\overline{\wt}(\xi,\kappa)_q = \mathbb{0}$, 
		
		\item we define $U(\xi,\kappa)$ to be the set of all pairs $(w_1,w_2)\in\pos(\xi)\times\pos(\xi)$ such that $w_1 <_{\mathrm{pref}} w_2$ and for every~$q\in S$ we have $\kappa_q(w_1)=\kappa_q(w_2)$, and 

		\item we have $(\xi,\kappa) \in\cC(S)$ iff for every $(w_1,w_2)\in U(\xi,\kappa)$ and $q\in S$ we have
                  \[
                    \wt(\xi|_{w_1},\kappa_q|_{w_1}) = \wt(\R_\cA(\kappa_q(w_1),\xi|_{w_1}))
                  \]
                  i.e., $\kappa_q|_{w_1}$ is victorious in $\R_\cA(\kappa_q(w_1),\xi|_{w_1})$). We set $\cC = \bigcup_{S\subseteq Q} \cC(S)$.
	\end{itemize}

\begin{lemma} \label{lm:init}\rm \cite[Lm.~5.14]{buemayvog10}
  Let~$\B$ be commutative and  extremal. 
  For each $\xi\in \T_\Sigma$ there exists $(\xi,\kappa) \in\cC$ such that, for each $q \in Q$, we have $\overline{\wt}(\xi,\kappa)_q= \wt(\R_\cA(q,\xi))$.
  \end{lemma}

\begin{proof}
	We begin by showing the following statement: 
\begin{align}\label{equ:P-statement}
& \text{for each~$\xi\in  \T_\Sigma$ there exists a $\kappa: Q\to \R_\cA(\xi)$ such that} \\
& \text{for each $q\in Q$ we have $\kappa_q\in \R_\cA(q,\xi)$ and $P(\xi,\kappa,q)$ holds,}\nonumber
\end{align}
 where $P(\xi,\kappa,q)$ is the abbreviation of the  statement
\begin{equation*}
\text{for every  
 $w\in \pos(\xi): \wt(\xi|_w,\kappa_q|_w)=\wt(\R_\cA(\kappa_q(w),\xi|_{w}))$}.
\end{equation*}
We prove \eqref{equ:P-statement} by induction on $\T_\Sigma$. Let $\xi=\sigma(\xi_1,\ldots,\xi_k)$. By I.H., for each $i\in[k]$, there exists a $\kappa_i: Q\to \R_\cA(\xi_i)$ such that,
for each $q\in Q$, we have $(\kappa_i)_q\in \R_\cA(q,\xi_i)$ and the statement $P(\xi_i,\kappa_i,q)$ holds. We define $\kappa$ as follows: for every $q\in Q$, we let $\kappa_q\in \R_\cA(\xi)$
be such that 
\begin{compactitem}
\item $\kappa_q(\varepsilon)= q$,
\item for every $i\in [k]$ and $w\in \pos(\xi_i)$, we let $\kappa_q(iw)=(\kappa_i)_{p_i}(w)$, where the states $p_1,\ldots,p_k$ are defined such that
\begin{equation}\label{eq:victorious}
\Big(\bigotimes_{i\in [k]} \wt(\xi_i,(\kappa_i)_{p_i})\Big)\otimes \delta_k(p_1\cdots p_k,\sigma,q) =\bigoplus_{q_1,\ldots q_k\in Q}\Big(\bigotimes_{i\in [k]} \wt(\xi_i,(\kappa_i)_{q_i})\Big)\otimes \delta_k(q_1\cdots q_k,\sigma,q),
\end{equation}
where, for each $i \in [k]$, the state $p_i$ is defined such that
\begin{equation}\label{eq:victorious-1}
\bigoplus_{q_i \in Q} \wt(\xi_i,(\kappa_i)_{q_i}) = \wt(\xi_i,(\kappa_i)_{p_i}) \enspace.
  \end{equation}
We recall that such a choice of $p_i$ exists by  Observation \ref{obs:extremal-sum}.

\end{compactitem}

Let $q\in Q$.  We show that $P(\xi,\kappa,q)$ holds. For this, let $w\in \pos(\xi)$. We proceed by case analysis.

\underline{Case (a):} Let $w=\varepsilon$. Then
\begingroup
\allowdisplaybreaks
\begin{align*}
  \wt(\xi, \kappa_q)& = \Big(\bigotimes_{i\in [k]} \wt(\xi|_i,\kappa_q|_i)\Big)\otimes\delta_k(\kappa_q(1)\cdots \kappa_q(k),\sigma,\kappa_q(\varepsilon)) \\
& = \Big(\bigotimes_{i\in [k]} \wt(\xi_i,(\kappa_i)_{p_i})\Big)\otimes\delta_k(p_1\cdots p_k,\sigma,q)  \tag{by definition of $\kappa$}\\
& = \Big(\bigotimes_{i\in [k]} \bigoplus_{q_i \in Q} \wt(\xi_i,(\kappa_i)_{q_i})\Big)\otimes\delta_k(p_1\cdots p_k,\sigma,q)  \tag{by \eqref{eq:victorious-1}}\\
& = \bigoplus_{q_1,\ldots,q_k\in Q} \Big(\bigotimes_{i\in [k]} \wt(\xi_i,(\kappa_i)_{q_i})\Big)\otimes \delta_k(q_1\cdots q_k,\sigma,q)
\tag{by distributivity}  \\
& = \bigoplus_{q_1,\ldots,q_k\in Q} \Big(\bigotimes_{i\in [k]}\wt(\R_\cA(q_i,\xi_i))\Big) \otimes\delta_k(q_1\cdots q_k,\sigma,q) 
\tag{by $P(\xi_i,\kappa_i,q_i)$ for  $w=\varepsilon$}\\
                    & = \bigoplus_{q_1,\ldots,q_k\in Q} \bigoplus_{\rho_1 \in \R_\cA(q_1,\xi_1)} \ldots \bigoplus_{\rho_k \in \R_\cA(q_k,\xi_k)} \Big(\bigotimes_{i\in [k]} \wt(\xi_i,\rho_i)\Big)\otimes \delta_k(q_1\cdots q_k,\sigma,q)
  \tag{by distributivity}\\
& =\bigoplus_{\rho \in \R_\cA(q,\xi)} \wt(\xi,\rho)=\wt(\R_\cA(q,\xi))=\wt(\R_\cA(\kappa_q(\varepsilon),\xi)) \enspace.
\end{align*}
\endgroup

\underline{Case (b):} Let $w=iv$. Then $\wt(\xi|_{iv},\kappa_q|_{iv})= \wt(\xi_i|_v,(\kappa_i)_{p_i}|_v))= \wt(\R_\cA((\kappa_i)_{p_i}(v),\xi_i|_v))= \wt(\R_\cA(\kappa_q(iv),\xi|_{iv}))$,
where the second equality holds by $P(\xi_i,\kappa_i,p_i)$ for  $v$.

This finishes the proof of \eqref{equ:P-statement}.

Now we prove the statement of our lemma. Let $\xi\in \T_\Sigma$. By   (\ref{equ:P-statement}) there exists a $\kappa : Q \to  \R_\cA(\xi)$ such that for every $q\in Q$ we have $\kappa_q\in \R_\cA(q,\xi)$ and $P(\xi,\kappa,q)$. Let $S=\{q\in Q \mid \wt(\R_\cA(q,\xi))\ne \mathbb{0}\}$ and $\kappa'=\kappa|_{S}$.
We note that $(\xi,\kappa') \in \cC(S)$ (and hence, $(\xi,\kappa') \in \cC$), because
\begin{compactitem}
\item for every $q \in S$: $\kappa'_q \in \R_\cA(q,\xi)$ and $\wt(\xi,\kappa'_q) = \wt(\R_\cA(q,\xi)) \not=\mathbb{0}$  (hence $(\xi,\kappa') \in \cC'(S)$) and
  \item for every $w \in \pos(\xi)$: $\wt(\xi|_w,\kappa'_q|_w) = \wt(\R_\cA(\kappa'_q(w),\xi|_w))$ by  (\ref{equ:P-statement}) (hence $(\xi,\kappa') \in \cC(S)$).
  \end{compactitem}
Then, for each $q\in S$, we have $\kappa'_q\in \R_\cA(q,\xi)$ and $P(\xi,\kappa',q)$. Moreover, for each $q\in S$: 
\begin{align*}
  \overline{\wt}(\xi,\kappa')_q  =  \wt(\xi,\kappa'_q) = & \wt(\R_\cA(q,\xi)).
\end{align*}
If $q\in Q\setminus S$, then also $\overline{\wt}(\xi,\kappa')_q =\mathbb{0} = \wt(\R_\cA(q,\xi))$ by  definitions of $\overline{\wt}$ and of $S$.
\end{proof}

The next lemma shows how one slice can be cut out.

\begin{lemma} \label{lm:step}\rm  \cite[Lm.~5.15]{buemayvog10}
	Let~$\B$ be extremal and commutative and~$\cA$ have the twinning property. Moreover, let $S\subseteq Q$ and $(\xi,\kappa)\in \cC(S)$ such that $U(\xi,\kappa)\neq\emptyset$. Then there exist $(\xi',\kappa')\in \cC(S)$ and~$b\in B$ such that $\overline{\wt}(\xi,\kappa)=b\cdot\overline{\wt}(\xi',\kappa')$ and $|U(\xi,\kappa)| > |U(\xi',\kappa')|$.
\end{lemma}

\begin{proof}   Since $U(\xi,\kappa)\neq\emptyset$, there exists a pair $(w_1,w_2)\in U(\xi,\kappa)$ such that for every $(w_1',w_2')\in U(\xi,\kappa)$, if $w_1'\le_{\mathrm{pref}} w_1$, then $w_1'=w_1$. (Thus, intuitively,  there does not exist a repetition of states above $w_1$.)  We construct $\xi' = \xi[\xi|_{w_2}]_{w_1}$ and, for every~$q\in S$ and $w\in\pos(\xi')$, we set $\kappa'_q(w) = \kappa_q(w_2v)$ if $w=w_1v$ and $\kappa'_q(w) = \kappa_q(w)$ otherwise. Before defining $b\in B$, we show that $\kappa$  has the following property: 
\begin{equation}\label{equ:invariance}
(\forall q',\overline{q}\in S) : \wt(\R_\cA(\kappa_{q'}(w_1),(\xi|^{w_2})|_{w_1},\kappa_{q'}(w_2)))=\wt(\R_\cA(\kappa_{\overline{q}}(w_1),(\xi|^{w_2})|_{w_1},\kappa_{\overline{q}}(w_2))).
\end{equation}

We prove \eqref{equ:invariance} by using the twinning property of $\cA$. Roughly speaking, the twinning property is an implication where the premise has two conditions, viz.,  (i)  certain states repeat and (ii)  certain weights are not $\mathbb{0}$.
For (i) we remark that $(w_1,w_2)\in U(\xi,\kappa)$ and thus $\kappa_{q'}(w_1)=\kappa_{q'}(w_2)$ and $\kappa_{\overline{q}}(w_1)=\kappa_{\overline{q}}(w_2)$.
For (ii), by definition, we have $\wt(\xi,\kappa_{q'})\neq \mathbb{0}$ and $\wt(\xi,\kappa_{\overline{q}})\neq \mathbb{0}$. 
Thus, using commutativity,
\begin{compactitem}
\item $\0 \ne \wt(\xi,\kappa_{q'}) = \wt(\xi_{w_1}, \kappa_{q'}|_{w_1}) \otimes \wt((\xi|^{w_2})|_{w_1}, (\kappa_{q'}|^{w_2})|_{w_1}) \otimes \wt(\xi|_{w_2},\kappa_{q'}|_{w_2})$ and
\item $\0 \ne \wt(\xi,\kappa_{\overline{q}}) = \wt(\xi_{w_1}, \kappa_{\overline{q}}|_{w_1}) \otimes  \wt((\xi|^{w_2})|_{w_1}, (\kappa_{\overline{q}}|^{w_2})|_{w_1}) \otimes \wt(\xi|_{w_2},\kappa_{\overline{q}}|_{w_2})$
\end{compactitem}
and hence 
\begin{compactitem}
\item $\wt((\xi|^{w_2})|_{w_1}, (\kappa_{q'}|^{w_2})|_{w_1}) \otimes \wt(\xi|_{w_2},\kappa_{q'}|_{w_2})\neq \mathbb{0}$ and
\item $\wt((\xi|^{w_2})|_{w_1}, (\kappa_{\overline{q}}|^{w_2})|_{w_1}) \otimes \wt(\xi|_{w_2},\kappa_{\overline{q}}|_{w_2})\neq \mathbb{0}$.
\end{compactitem}
Since $\mathbb{0}$ is annihilating, each of the four values
\[
  \wt((\xi|^{w_2})|_{w_1}, (\kappa_{q'}|^{w_2})|_{w_1}), \ \
  \wt(\xi|_{w_2},\kappa_{q'}|_{w_2}), \ \
  \wt((\xi|^{w_2})|_{w_1}, (\kappa_{\overline{q}}|^{w_2})|_{w_1}), \ \ \text{ and } \ \
  \wt(\xi|_{w_2}, \kappa_{\overline{q}}|_{w_2})
\]
is different from $\mathbb{0}$.
Since~$\B$ is extremal, and hence by Observation \ref{obs:extremal-implies-zero-sum-free}(3,4) it is zero-sum free, we obtain that
\begin{compactitem}
\item $\wt(\R_\cA(\kappa_{q'}(w_1),(\xi|^{w_2})|_{w_1},\kappa_{q'}(w_2))) \neq \mathbb{0}$ and  $\wt(\R_\cA(\kappa_{q'}({w_2}),\xi|_{w_2}))\neq \mathbb{0}$ and 
\item $\wt(\R_\cA(\kappa_{\overline{q}}(w_1),(\xi|^{w_2})|_{w_1},\kappa_{\overline{q}}(w_2)))\neq \mathbb{0}$ and $\wt(\R_\cA(\kappa_{\overline{q}}({w_2}),\xi|_{w_2}))\neq \mathbb{0}$.
  \end{compactitem}
  Hence 
(\ref{equ:invariance}) follows by the twinning property of $\cA$.

Next we define $b$ by case analysis as follows:
\[ b= \begin{cases}
\0 & \text{ if $S=\emptyset$} \\
\wt(\R_\cA(\kappa_{q}(w_1),(\xi|^{w_2})|_{w_1},\kappa_{q}(w_2))) \text{ for some } q\in S & \text{ otherwise.}
\end{cases}
\]
We note that, by (\ref{equ:invariance}), $b$ does not depend on the choice of $q$. Now we show that
the following three statements hold:
  \begin{compactitem}
  \item[(1)] $(\xi',\kappa')\in \cC(S)$,
  \item[(2)] $\overline{\wt}(\xi,\kappa)=b\cdot \overline{\wt}(\xi',\kappa')$, and
  \item[(3)] $|U(\xi,\kappa)| >  |U(\xi',\kappa')|$.
    \end{compactitem}

    	We begin with the proof of Statement (1). It is easy to see that $(\xi',\kappa')\in \cC'(S)$. Now let $(w_1',w_2')\in U(\xi',\kappa')$ and~$q\in S$. We show that $\wt(\xi'|_{w_1'}, \kappa_q'|_{w_1'}) = \wt(\R_\cA(\kappa_q'({w_1'}),\xi'|_{w_1'}))$. Note that $w_1'\not<_{\mathrm{pref}} w_1$. We distinguish two cases.

	\begin{figure}
          \centering
		\begin{tikzpicture}[label distance=-0.1cm]

\definecolor{mygray}{rgb}{0.9,0.9,0.9}

\newcommand\xtwo{0.18}
\newcommand\ytwo{-0.75}
\newcommand\xthree{0.05}
\newcommand\ythree{-1.5}

\begin{scope}

	\coordinate (pos2) at (\xtwo,\ytwo);
	\coordinate (pos3) at (\xthree,\ythree);
	
	\coordinate (al) at (-1.5,-3);
	\coordinate (bl) at (1.5,-3);
	\coordinate (cl) at (0,0);
	\coordinate (am) at ($(-1.25,-2.5)	+(pos2)$);
	\coordinate (bm) at ($(1.25,-2.5)	+(pos2)$);
	\coordinate (cm) at ($(0,0)			+(pos2)$);
	\coordinate (as) at ($(-1,-2)		+(pos3)$);
	\coordinate (bs) at ($(1,-2)		+(pos3)$);
	\coordinate (cs) at ($(0,0)			+(pos3)$);

	\draw (al) -- (cl) -- (bl) -- cycle;		

	\fill[mygray] (am) -- (bm) -- (cm) -- cycle;
	\draw (am) -- (cm) -- (bm) -- cycle;		

	\fill[white] (as) -- (bs) -- (cs) -- cycle;
	\draw (as) -- (cs) -- (bs) -- cycle;
	
	\coordinate (w2v1) at ($(cs)+(0.32,-1.15)$);
	\coordinate (w2v2) at ($(cs)+(0.52,-1.65)$);
	\path[draw] (cl) to[out = 250, in = 80] (cm) 
		node[circle,fill,inner sep=0.8pt,label=left:\small $w_1$] {};
	\path[draw] (cm) to[out = 290, in = 100] (cs) 
		node[circle,fill,inner sep=0.8pt,label=right:\small $w_2$] {};
	\path[draw] (cs) to[out = 255, in = 95] (w2v1) 
		node[circle,fill,inner sep=0.8pt,label=left:\small $w_2v_1$] {};
	\path[draw] (w2v1) to[out = 275, in = 85] (w2v2) 
		node[circle,fill,inner sep=0.8pt,label=left:\small $w_2v_2$] {};
		
	\node at (0,-3.95) {\small $(\xi,\kappa)$};

\end{scope}

\begin{scope}[xshift=3.5cm]

	\coordinate (pos2) at (\xtwo,\ytwo);
	
	\coordinate (al) at (-1.5,-3);
	\coordinate (bl) at (1.5,-3);
	\coordinate (cl) at (0,0);
	\coordinate (as) at ($(-1,-2)		+(pos2)$);
	\coordinate (bs) at ($(1,-2)		+(pos2)$);
	\coordinate (cs) at ($(0,0)			+(pos2)$);
	
	\coordinate (pa) at (\xtwo -1.5 -\ytwo/2,-3);
	\coordinate (pb) at (\xtwo +1.5 +\ytwo/2 ,-3);

	\draw (as) -- (pa) -- (al) -- (cl) -- (bl) -- (pb) -- (bs);			

	\fill[white] (as) -- (bs) -- (cs) -- cycle;
	\draw (as) -- (cs) -- (bs) -- cycle;
	
	\coordinate (w2v1) at ($(cs)+(0.32,-1.15)$);
	\coordinate (w2v2) at ($(cs)+(0.52,-1.65)$);
	\path[draw] (cl) to[out = 250, in = 80] (cs) 
		node[circle,fill,inner sep=0.8pt,label=left:\small $w_1$] {};
	\path[draw] (cs) to[out = 255, in = 95] (w2v1) 
		node[circle,fill,inner sep=0.8pt,label=left:\small $w_1v_1$] {};
	\path[draw] (w2v1) to[out = 275, in = 85] (w2v2) 
		node[circle,fill,inner sep=0.8pt,label=left:\small $w_1v_2$] {};
		
	\node at (0,-3.55) {\small $(\xi',\kappa')$};
	
\end{scope}

\begin{scope}[xshift=7.5cm]

	\coordinate[shift={(0.125,-0.25cm)}] (pos2) at (\xtwo,\ytwo);
	\coordinate[shift={(0.125,-0.25cm)}] (pos3) at (\xthree,\ythree);
	
	\coordinate (al) at (-1.5,-3);
	\coordinate (bl) at (1.5,-3);
	\coordinate (cl) at (0,0);
	\coordinate (am) at ($(-1.125,-2.25)+(pos2)$);
	\coordinate (bm) at ($(1.125,-2.25)	+(pos2)$);
	\coordinate (cm) at ($(0,0)			+(pos2)$);
	\coordinate (as) at ($(-0.875,-1.75)+(pos3)$);
	\coordinate (bs) at ($(0.875,-1.75)	+(pos3)$);
	\coordinate (cs) at ($(0,0)			+(pos3)$);

	\draw (al) -- (cl) -- (bl) -- cycle;		

	\fill[mygray] (am) -- (bm) -- (cm) -- cycle;
	\draw (am) -- (cm) -- (bm) -- cycle;		

	\fill[white] (as) -- (bs) -- (cs) -- cycle;
	\draw (as) -- (cs) -- (bs) -- cycle;
	
	\coordinate (w2v1) at ($(cl)+(-0.52,-1.35)$);
	\coordinate (w2v2) at ($(cl)+(-0.72,-1.85)$);
	\path[draw] (cl) to[out = 260, in = 85] (cm) 
		node[circle,fill,inner sep=0.8pt,label=left:\small $w_1$] {};
	\path[draw] (cm) to[out = 290, in = 100] (cs) 
		node[circle,fill,inner sep=0.8pt,label=right:\small $w_2$] {};
	\path[draw] (cl) to[out = 255, in = 80] (w2v1) 
		node[circle,fill,inner sep=0.8pt,label=right:\small $w'_1$] {};
	\path[draw] (w2v1) to[out = 270, in = 70] (w2v2) 
		node[circle,fill,inner sep=0.8pt,label=right:\small $w'_2$] {};
		
	\node at (0,-3.95) {\small $(\xi,\kappa)$};		
	
\end{scope}

\begin{scope}[xshift=11cm]  

	\coordinate[shift={(0.125,-0.25cm)}] (pos2) at (\xtwo,\ytwo);
	
	\coordinate (al) at (-1.5,-3);
	\coordinate (bl) at (1.5,-3);
	\coordinate (cl) at (0,0);
	\coordinate (as) at ($(-0.875,-1.75)+(pos2)$);
	\coordinate (bs) at ($(0.875,-1.75)	+(pos2)$);
	\coordinate (cs) at ($(0,0)			+(pos2)$);
	
	\coordinate (pa) at (\xtwo -1.5 + 0.25 -\ytwo/2,-3);
	\coordinate (pb) at (\xtwo +1.5 +\ytwo/2 ,-3);

	\draw (as) -- (pa) -- (al) -- (cl) -- (bl) -- (pb) -- (bs);			

	\fill[white] (as) -- (bs) -- (cs) -- cycle;
	\draw (as) -- (cs) -- (bs) -- cycle;
	
	\coordinate (w2v1) at ($(cl)+(-0.52,-1.35)$);
	\coordinate (w2v2) at ($(cl)+(-0.72,-1.85)$);
	\path[draw] (cl) to[out = 260, in = 85] (cs) 
		node[circle,fill,inner sep=0.8pt,label=left:\small $w_1$] {};
	\path[draw] (cl) to[out = 255, in = 80] (w2v1) 
		node[circle,fill,inner sep=0.8pt,label=right:\small $w'_1$] {};
	\path[draw] (w2v1) to[out = 270, in = 70] (w2v2) 
		node[circle,fill,inner sep=0.8pt,label=right:\small $w'_2$] {};
		
	\node at (0,-3.55) {\small $(\xi',\kappa')$};
	
\end{scope}

\text (a) and (b)
\node at (3.25 - 1.5, -4.3) {\textbf{(a)}};
\node at (10.75 - 1.5, -4.3) {\textbf{(b)}};

\end{tikzpicture}
		\caption{The Cases (a) and (b) for $(w_1',w_2')\in U(\kappa')$ (cf. \cite[Fig.~7]{buemayvog10}).}
		\label{fig:step}
	\end{figure}

	\underline{Case~(a):} There exist~$v_1,v_2\in\mathbb{N}^*$ such that $w_1' = w_1v_1$ and $w_2' = w_1 v_2$ (cf.\ Fig.~\ref{fig:step}(a)). Since $\kappa_q'|_{w_1} = \kappa_q|_{w_2}$ for every~$q\in S$, we obtain that $(w_2 v_1,w_2 v_2) \in U(\xi,\kappa)$. Hence
	\begin{align*}
		\wt(\xi'|_{w_1'},\kappa'_q|_{w_1'})
		= \wt(\xi|_{w_2v_1},\kappa_q|_{w_2v_1})
		= \wt(\R_\cA(\kappa_q({w_2v_1}),\xi|_{w_2v_1}))
		= \wt(\R_\cA(\kappa_q'({w_1'}),\xi'|_{w_1'}))\;.
	\end{align*}
	
	\underline{Case~(b):} Otherwise, $\kappa_q'|_{w_1'} = \kappa_q|_{w_1'}$ (cf.\ Fig.~\ref{fig:step}(b)). Thus $(w_1',w_2')\in U(\xi,\kappa)$ and
	\begin{align*}
		\wt(\xi'|_{w_1'},\kappa'_q|_{w_1'})
		= \wt(\xi|_{w_1'},\kappa_q|_{w_1'})
		= \wt(\R_\cA(\kappa_q({w_1'}) , \xi|_{w_1'}))
		= \wt(\R_\cA(\kappa_q'({w_1'}),\xi'|_{w_1'}))\;.
	\end{align*}
	
	Now we prove Statement (2) for the non-trivial case of $b$, i.e., $S\neq\emptyset$. Let~$q\in S$. Then
	\begin{align*}
		&\wt(\xi,\kappa_q)\\
		&= \wt(\xi|^{w_1}, \kappa_q|^{w_1})\otimes \wt((\xi|^{w_2})|_{w_1},(\kappa_q|^{w_2})|_{w_1}) \otimes \wt(\xi|_{w_2},\kappa_q|_{w_2})\tag{because $\B$ is commutative}\\
		&= \wt(\xi|^{w_1}, \kappa_q|^{w_1})\otimes \wt(\R_\cA(\kappa_q(w_1),(\xi|^{w_2})|_{w_1},\kappa_q(w_2))) \otimes \wt(\xi|_{w_2},\kappa_q|_{w_2}) \tag{Observation~\ref{ob:bellman}}\\
		&= b\otimes \wt(\xi|^{w_1},\kappa_q|^{w_1}) \otimes\wt(\xi|_{w_2},\kappa_q|_{w_2}) \tag{because $\B$ is commutative}\\
		&= b\otimes \wt(\xi',\kappa'_q)\;.
	\end{align*}
	
	Finally, for the proof of Statement (3), we remark that~$\kappa'$ is obtained from~$\kappa$ by removing the cycle $(w_1,w_2)$, and that this process does not introduce new cycles. Hence $\lvert U(\xi,\kappa)\rvert >  \lvert U(\xi',\kappa')\rvert$.
\end{proof}

The following lemma is used for the proof that our cutting process can only end in a finite set of trees. For this we define the finite set
\[
  \cF= \{ (\xi,\kappa)\in\cC'\mid \height(\xi)  < | Q|^{| Q|} \} \enspace.
\]
Next we will use $U^{-1}(\emptyset)$ as shorthand for the set $\{(\xi,\kappa) \mid U(\xi,\kappa) = \emptyset\}$.

\begin{lemma} \label{lm:done} \rm \cite[Lm.~5.16]{buemayvog10}
	$U^{-1}(\emptyset)\subseteq\cF$.
\end{lemma}

\begin{proof} Let $(\xi,\kappa)\in\cC'\setminus\cF$. We show that $U(\xi,\kappa) \neq \emptyset$ follows. First of all, there exists an $S\subseteq Q$ such that $(\xi,\kappa)\in \cC'(S)$. Since $(\xi,\kappa)\not\in\cF$, there exists a~$w\in\pos(\xi)$ such that $| w| \geq | Q|^{| S|}$. Hence, there exist $k\in\mathbb{N}$, $w_1,\ldots,w_k\in\mathbb{N}^*$, and~$u_1,\ldots,u_k\in Q^{S}$ such that $k > | Q|^{| S|}$, $w_i\in\pos(\xi)$ for every~$i\in[k]$, $w_1 <_{\mathrm{pref}} w_2 <_{\mathrm{pref}} \ldots <_{\mathrm{pref}} w_k$, and $\kappa_q|_{w_i}\in \R_\cA((u_i)_q,\xi)$ for every~$i\in [k]$ and~$q\in S$. By the pigeon-hole principle, there exist $i,j\in[k]$ such that $i<j$ and $u_i = u_j$. Hence $\kappa_q(w_i)=(u_i)_q=(u_j)_q=\kappa_q(w_j)$ for every $q\in S$, which means that $(w_i,w_j)\in U(\xi,\kappa)$.
\end{proof}

We recall that  $\det_{(f,g)}(\cA) = (Q',\delta',F')$, and by Lemma \ref{lm:l2} we have that  $Q' \subseteq f(\h_\cA(\T_\Sigma)\setminus \{\0^Q\})$ if $\B$ is commutative and  $(f,g)$ is a maximal factorization. If we could show that $\h_\cA(\T_\Sigma)$ is finite, then we have shown that $Q'$ is finite. In the next lemma we prove that  $\h_\cA(\T_\Sigma)$ is finite under the conditions that $\B$ is commutative and extremal and~$\cA$ has the twinning property. The idea of the proof is the following. Let us consider a large tree $\xi \in \T_\Sigma$ for which we want to compute $\h_\cA(\xi)$, which is nothing else but the $Q$-vector $(\wt(\R_\cA(q,\xi)) \mid q \in Q)$. Then, by Lemma \ref{lm:init}, there exists $\kappa: Q \to \R_\cA(\xi)$ such that, for each $q \in Q$, we have $\wt(\R_\cA(q,\xi)) = \wt(\xi,\kappa_q) = \overline{\wt}(\xi,\kappa)_q$ (which needs extremality), i.e., $\kappa_q$ is victorious for $\R_\cA(q,\xi)$. If $\xi$ is large enough, then there exists a pair $(w_1,w_2) \in \pos(\xi) \times \pos(\xi)$ such that $\kappa_q(w_1) = \kappa_q(w_2)$. Now we can cut out the slice of $\xi$ which is determined by $w_1$ and $w_2$. By Lemma \ref{lm:step}, we obtain a pair $(\xi',\kappa')$ such that $\overline{\wt}(\xi,\kappa) = b \cdot \overline{\wt}(\xi',\kappa')$ where $b$ is the weight of the victorious run on the slice which we cut out, and $\xi'$ contains one less pair of positions with repeating states. After repeating this cutting process, we end up in a pair $(\xi'',\kappa'')$ such that $\xi''$ does not contain a pair of positions with repeating states, i.e., $(\xi'',\kappa'') \in U^{-1}(\emptyset)$. By Lemma \ref{lm:done}, the set $U^{-1}(\emptyset)$ is finite.

The following lemma is based on \cite[Thm.~5]{kirmae05}. 

\begin{lemma}\rm \cite[Lm.~5.9]{buemayvog10}\label{thm:extremal-twins-finite-set-of-vectors}  	If~$\B$ is commutative and extremal and~$\cA$ has the twinning property, then there exists a finite set~$P\subseteq B^Q$ with $\h_\cA(\T_\Sigma) \subseteq \{ b\cdot u\mid b\in B, u\in P \}$.
\end{lemma}
\begin{proof}
	Let $\xi\in \T_\Sigma$.  We set  $P=\{ \overline{\wt}(\xi,\kappa) \mid (\xi,\kappa) \in \cF)$.
	
        Let $n\in\mathbb{N}$ be maximal such that there exist $(\xi_1,\kappa_1),\ldots,(\xi_n,\kappa_n)\in \cC$ and $b_1,\ldots,b_n\in B$ such that
        \begin{compactitem}
          \item $\xi = \xi_1$, 
        \item      $\h_\cA(\xi)=\overline{\wt}(\xi_1,\kappa_1)$ and $\overline{\wt}(\xi_1,\kappa_1) = b_i\cdot\overline{\wt}(\xi_i,\kappa_i)$ for every~$i\in[n]$, and
        \item $|U(\xi_{i+1},\kappa_{i+1})| < |U(\xi_i,\kappa_i)|$ for every~$i\in[n-1]$.
        \end{compactitem}
        We claim that 
\begin{equation}\label{equ:equ1}
n>0
\end{equation}
and
\begin{equation}\label{equ:equ2}
(\xi_n,\kappa_n)\in \cF,
\end{equation}
which allows us to derive
	\begin{align*}
		\h_\cA(\xi)
		= \overline{\wt}(\xi_1,\kappa_1)
		= b_n\cdot\overline{\wt}(\xi_n,\kappa_n)
		\in \{ b\cdot u\mid b\in B, u\in P \}\;.
	\end{align*}
	Claim  (\ref{equ:equ1}) follows from Lemma~\ref{lm:init} if we set $b_1=1$.
	Finally, we prove (\ref{equ:equ2}). Assume that $U(\xi_n,\kappa_n)\neq\emptyset$. By Lemma~\ref{lm:step}, there exist~$(\xi',\kappa')$ and~$b'$ such that $\overline{\wt}(\xi_n,\kappa_n)=b'\cdot \overline{\wt}(\xi',\kappa')$ and $|U(\xi',\kappa')|<|U(\xi_n,\kappa_n)|$. Using $\kappa_{n+1} = \kappa'$ and $b_{n+1} = b'\cdot b_n$, we see that~$n$ was not maximal. Hence, $U(\xi_n,\kappa_n)=\emptyset$, and by Lemma~\ref{lm:done}, $(\xi_n,\kappa_n)\in\cF$.
\end{proof}

\begin{theorem}{\rm \cite[Thm.~5.10]{buemayvog10}}\label{thm:finiteness} Let $\B$ be commutative and extremal and let $(f,g)$ be a maximal factorization of $B^Q$. Moreover, let $\cA$ have the twinning property. Then $\det_{(f,g)}(\cA)$ is a $(\Sigma,\B)$-wta.  
\end{theorem}

\begin{proof}
	Let $\det_{(f,g)}(\cA) = (Q',\delta',F')$. By Lemma~\ref{lm:l2}, we have $Q'\subseteq f(\h_\cA(\T_\Sigma)\setminus\{\0^Q\})$. We show that $Q'$ is finite. Lemma~\ref{thm:extremal-twins-finite-set-of-vectors} yields that there exists a finite set~$P\subseteq B^Q$ such that $\h_\cA(\T_\Sigma) \subseteq \{ b\cdot u\mid b\in B, u\in P\}$. We calculate
	\[
          Q' \subseteq f(\h_\cA(\T_\Sigma)\setminus\{\0^Q\}) \subseteq f(\{ b\cdot u\mid b\in B, u\in P\}\setminus\{\0^Q\}) \subseteq f(P\setminus\{\0^Q\})
        \]
        because $(f,g)$ is maximal. Hence, $Q'$ is finite.
\end{proof}


\subsubsection{Correctness  of $\det_{(f,g)}(\cA)$ (proof of Theorem~\ref{thm:extremal-twins-determinizable}(2))}

\begin{theorem} \label{thm:correctness} {\rm \cite[Thm.~5.4]{buemayvog10}}
	Let $(f,g)$ be a factorization and $\B$ be commutative. If~$\det_{(f,g)}(\cA)$ is a wta, then $\sem{\cA}=\sem{\det_{(f,g)}(\cA)}$.
\end{theorem}
\begin{proof} Let $\det_{(f,g)}(\cA) =(Q',\delta',F')$ be a wta, i.e.,  $Q'$ be finite. We abbreviate  $\det_{(f,g)}(\cA)$ by $\cA'$. By induction on~$\T_\Sigma$, we prove that the following statement holds:
  \begin{equation} 
    \text{For every~$\xi\in \T_\Sigma$, we have: } \h_\cA(\xi) = \bigoplus_{u\in Q'} \h_{\cA'}(\xi)_u\cdot u \enspace. \label{eq:det-correct}
  \end{equation}

	Let $\xi=\sigma(\xi_1,\ldots,\xi_k)$. Since $\cA'$ is bu-deterministic, by Lemma~\ref{lm:limit-bu-det}(1) there are $u_1',\ldots,u_k'\in Q'$ such that, for every $i\in[k]$ and $u\in Q'\setminus\{u_i'\}$, we have $\h_{\cA'}(\xi_i)_u=\mathbb{0}$.
	We derive ($\star$):
        	\begin{align*}
		\h_\cA(\xi) &= \delta_\cA(\sigma)(\h_\cA(\xi_1),\ldots,\h_\cA(\xi_k))\\
                                  &= \delta_\cA(\sigma)(\bigoplus_{u_1 \in Q'}\h_{\cA'}(\xi_1)_{u_1}\cdot u_1,\ldots,
                              \bigoplus_{u_k \in Q'} \h_{\cA'}(\xi_k)_{u_k}\cdot u_k)\tag{by I.H.}\\
	& = \delta_\cA(\sigma)(\h_{\cA'}(\xi_1)_{u_1'}\cdot u_1',\ldots,\h_{\cA'}(\xi_k)_{u_k'}\cdot u_k')\\
                      	&= (\h_{\cA'}(\xi_1)_{u_1'}\otimes \ldots\otimes  \h_{\cA'}(\xi_k)_{u_k'})\cdot \delta_\cA(\sigma)(u_1',\ldots, u_k'),\tag{Lemma \ref{lm:vector-algebra-is-semimodule-com-sr}}
                \end{align*}
                where Lemma \ref{lm:vector-algebra-is-semimodule-com-sr} uses the assumption that $\B$ is commutative.
                
	Moreover, we derive ($\dagger$): for every $u\in Q'$
	\begin{align*}
		\h_{\cA'}(\xi)_u &= \delta_{\cA'}(\sigma)(\h_{\cA'}(\xi_1),\ldots \h_{\cA'}(\xi_k))_u\\
		&= \textstyle \bigoplus_{u_1,\ldots,u_k\in Q'} \h_{\cA'}(\xi_1)_{u_1}\otimes \ldots\otimes  \h_{\cA'}(\xi_k)_{u_k}\otimes \delta'_k(u_1\ldots u_k,\sigma,u)\\
		&= \h_{\cA'}(\xi_1)_{u_1'}\otimes\ldots\otimes \h_{\cA'}(\xi_k)_{u_k'}\otimes \delta'_k(u_1'\ldots u_k',\sigma,u)\;.
	\end{align*}
	Now we distinguish two cases.

\underline{Case (a):} Let  $\delta_\cA(\sigma)(u_1',\ldots, u_k')=\0^Q$. Then ($\star$) implies $\h_\cA(\xi)=\0^Q$. By definition of $\delta'$, we have $\delta'_k(u_1'\ldots u_k',\sigma,u) = \mathbb{0}$ for every~$u\in Q'$. Hence, ($\dagger$) implies $\h_{\cA'}(\xi)=\0^Q$.
	
\underline{Case (b):} Let $\delta_\cA(\sigma)(u_1',\ldots, u_k')\neq\0^Q$. Then we set $u' = f(\delta_\cA(\sigma)(u_1',\ldots, u_k'))$ and derive
        \begingroup
        \allowdisplaybreaks
	\begin{align*}
		\h_\cA(\xi) &= (\h_{\cA'}(\xi_1)_{u_1'}\otimes \ldots\otimes \h_{\cA'}(\xi_k)_{u_k'}) \cdot \delta_\cA(\sigma)(u_1',\ldots, u_k')\tag{by $(\star)$}\\
		&= (\h_{\cA'}(\xi_1)_{u_1'}\otimes \ldots\otimes \h_{\cA'}(\xi_k)_{u_k'}\otimes g(\delta_\cA(\sigma)(u_1',\ldots, u_k')))\cdot u'\tag{$(f,g)$ is a factorization}\\
		&= \h_{\cA'}(\xi)_{u'}\cdot u'\tag{by definition of $\delta_k'(u_1' \cdots u_k',\sigma,u')$ and by $(\dagger)$}\\
		&=  \bigoplus_{u\in Q'} \h_{\cA'}(\xi)_u\cdot u\;.\tag{Lemma~\ref{lm:limit-bu-det}(1)}
	\end{align*}
\endgroup
        
	Now we show that $\sem{\cA}(\xi)=\sem{\cA'}(\xi)$ for every $\xi \in \T_\Sigma$.
 \begingroup
 \allowdisplaybreaks
 \begin{align*}
		\sem{\cA'}(\xi)
		&=  \bigoplus_{u\in Q'} \h_{\cA'}(\xi)_u\otimes F'_u\\
		&=  \bigoplus_{u\in Q'} \h_{\cA'}(\xi)_u\otimes \big(\bigoplus_{q\in Q} u_q\otimes F_q\big) \tag{by construction of $\det_{(f,g)}(\cA)$}\\
                &= \bigoplus_{u\in Q'} \bigoplus_{q\in Q} \big(\h_{\cA'}(\xi)_u\otimes u_q\otimes F_q \big)\tag{by left-distributivity}\\
    &= \bigoplus_{q\in Q} \bigoplus_{u\in Q'} \big(\h_{\cA'}(\xi)_u\otimes u_q  \otimes F_q\big) \tag{by commutativity of $\oplus$}\\
      &= \bigoplus_{q\in Q} \Big[\big( \bigoplus_{u\in Q'} \h_{\cA'}(\xi)_u\otimes u_q \big) \otimes F_q \Big]\tag{by right-distributivity}\\
		&= \bigoplus_{q\in Q} \Big[\big(\bigoplus_{u\in Q'}  \h_{\cA'}(\xi)_u\cdot u\big)_q \otimes F_q\Big] \\
		&= \bigoplus_{q\in Q} \h_{\cA}(\xi)_q\otimes F_q \tag{by \eqref{eq:det-correct}}\\
                &= \sem{\cA}(\xi)\;. \qedhere
	\end{align*}
        \endgroup
      \end{proof}


\subsubsection{Minimality of $\det_{(f,g)}(\cA)$ (proof of Theorem~\ref{thm:extremal-twins-determinizable}(3))}

The following theorem corresponds to \cite[Thm.~3]{kirmae05}. 

\begin{theorem} \label{thm:minimality} {\rm \cite[Thm.~5.6]{buemayvog10}}
	Let~$\B$ be commutative and let $(f,g)$ and $(\tilde{f},\tilde{g})$ be factorizations such that $(f,g)$ is maximal. Moreover, let $\det_{(f,g)}(\cA)=(Q',\delta',F')$ and $\det_{(\tilde{f},\tilde{g})}(\cA) = (\tilde{Q},\tilde{\delta},\tilde{F})$. Then~$Q'=f(\tilde{Q})$; hence $\lvert Q'\rvert\leq \lvert\tilde{Q}\rvert$, and if $\det_{(\tilde{f},\tilde{g})}(\cA)$ is a wta, then so is~$\det_{(f,g)}(\cA)$.
\end{theorem}

\begin{proof} We prove by case analysis on the cardinality of $Q$.

\underline{Case (a):} Let $\lvert Q\rvert=1$. Then we can identify~$B^Q$ with~$B$. Since $(f,g)$ is maximal, we have that $f(B\setminus\{\mathbb{0}\})=\{f(\mathbb{1})\}$. Since $Q'\neq\emptyset$, $\tilde{Q}\neq\emptyset$, and $\tilde{Q}\subseteq B\setminus\{\mathbb{0}\}$, we obtain that $Q' = \{f(\1)\} = f(\tilde{Q})$.
	
\underline{Case (b):} Let $\lvert Q\rvert>1$. By Lemma~\ref{lm:maxzd}, $\B$ is zero-divisor free.
	Note that ($\star$) for every~$u\in B^Q\setminus\{\0^Q\}$, we have
		$\tilde{g}(u)\cdot\tilde{f}(u) = u = g(u)\cdot f(u)$,
	and by applying~$f$ we obtain
		$f(\tilde{f}(u)) = f(u) = f(f(u))$
	because~$(f,g)$ is maximal.
	
	We begin with the proof of $f(\tilde{Q})\subseteq Q'$. Using Lemma~\ref{obs:strat}, it suffices to prove the following statement by induction on~$\mathbb{N}$:
        \begin{equation*}
          \text{for every~$n\in\mathbb{N}$, we have $f(\tilde{Q}_n)\subseteq Q'$.} \label{eq:strat-ind}
        \end{equation*}

        I.B.: Since $\tilde{Q}_0 = \emptyset$, the statement holds trivially.

        I.S.: Let~$n\in\mathbb{N}$ and $\tilde{u}\in f(\tilde{Q}_{n+1})$.
If $\tilde{u}\in f(\tilde{Q}_{n})$, then we are ready. Otherwise, there exist $k\in\mathbb{N}$, $\sigma\in\Sigma^{(k)}$, and~$\tilde{u}_1,\ldots,\tilde{u}_k\in \tilde{Q}_n$ such that $\tilde{u}=f(\tilde{f}(\delta_\cA(\sigma)(\tilde{u}_1,\ldots,\tilde{u_k})))$.
	Hence
        \begingroup
        \allowdisplaybreaks
	\begin{align*}
		\tilde{u}
		&= f(\tilde{f}(\delta_\cA(\sigma)(\tilde{u}_1,\ldots,\tilde{u}_k)))\\
		&= f(\delta_\cA(\sigma)(\tilde{u}_1,\ldots,\tilde{u}_k))\tag{$\star$}\\
		&= f(\delta_\cA(\sigma)(f(\tilde{u}_1),\ldots,f(\tilde{u}_k))) \tag{Lemma~\ref{lm:l1}(2) and Lemma~\ref{lm:l1}(1)}\\
		&\in Q'\;. \tag{I.H. and def. of~$Q'$}
	\end{align*}
        \endgroup
	
	Now we prove $Q'\subseteq f(\tilde{Q})$. Using Lemma~\ref{obs:strat} again, it suffices to prove the following statement by induction on~$\mathbb{N}$:
	\begin{equation*}
	 \text{for every~$n\in\mathbb{N}$, we have $Q'_n\subseteq f(\tilde{Q})$.} 
	 \end{equation*}
	 
	  I.B.: Since $Q'_0 = \emptyset$, the statement holds trivially.
	 
	 I.S.: Let~$n\in\mathbb{N}$ and $u'\in Q_{n+1}'$. Then there are~$k\in\mathbb{N}$, $\sigma\in\Sigma^{(k)}$, and~$u'_1,\ldots,u'_k\in Q'_n$ such that $u'=f(\delta_\cA(\sigma)(u'_1,\ldots,u'_k))$. By I.H., there exist $\tilde{u}_1,\ldots,\tilde{u}_k\in \tilde{Q}$ such that $u_i'=f(\tilde{u}_i)$ for every~$i\in[k]$.
	Hence
	\begin{align*}
		u'
		&= f(\delta_\cA(\sigma)(f(\tilde{u}_1),\ldots,f(\tilde{u}_k)))\\
		&= f(\delta_\cA(\sigma)(\tilde{u}_1,\ldots,\tilde{u}_k)) \tag{Lemma~\ref{lm:l1}(2) and Lemma~\ref{lm:l1}(1)}\\
		&= f(\tilde{f}(\delta_\cA(\sigma)(\tilde{u}_1,\ldots,\tilde{u}_k)))\tag{$\star$}\\
		&\in f(\tilde{Q})\;. \qedhere
	\end{align*}
\end{proof}

\subsection{\sloppy Applying determinization by factorization to bu-deterministic wta}
\label{sec:det-of-bu-det}

In this subsection let $\cA=(Q,\delta,F)$ be a bu-deterministic $(\Sigma,\B)$-wta. Then of course there is no need to determinize $\cA$. However, in order to show the robustness of determinization by factorization, we apply it to $\cA$ and a maximal factorization $(f,g)$. We prove that the resulting triple $\det_{(f,g)}(\cA)=(Q',\delta',F')$ is indeed a bu-deterministic wta (i.e., $Q'$ is finite) by showing that $|Q'| \le |Q|$. This is due to the fact that $\h_\cA(\T_\Sigma)$  consists of single-valued elements of $B^Q$ and that $f$ takes single-valued $Q$-vectors to $q$-unit vectors. We note that we do not require that $\B$ is extremal (which we do in Theorem \ref{thm:finiteness}).

Formally, for every $b \in B$ and $q \in Q$, we let $b_q$ be the element of  $B^Q$  such that $(b_q)_q=b$ and $(b_q)_p = \0$ for each $p \in Q\setminus \{q\}$. We recall that 
\[
 B_{\le 1}^Q = \{b_q \in B^Q \mid q \in Q, b \in B\} \enspace.
 \]
 By Lemma \ref{lm:limit-bu-det}(1), the following is obvious.

\begin{observation} \label{ob:budet-single-value} $\h_\cA(\T_\Sigma) \subseteq B_{\le 1}^Q$.\hfill $\Box$
\end{observation}

Then we can prove the following result.

\begin{corollary}\rm \label{lm:determinization-bu-det} \cite[Lm. 5.3.11]{bue14} Let $\B$ be commutative and $\cA=(Q,\delta,F)$ be a bu-deterministic $(\Sigma,\B)$-wta. Moreover, let $(f,g)$ be a maximal factorization and $\det_{(f,g)}(\cA)=(Q',\delta',F')$.  Then we have $|Q'| \le |Q|$.
\end{corollary}
\begin{proof} By Lemma \ref{lm:l2} and Observation \ref{ob:budet-single-value}, we have 
\[Q'\subseteq f(\h_\cA(\T_\Sigma)\setminus\{\0^Q\})\subseteq f(B_{\le 1}^Q\setminus\{\0^Q\})\enspace.\]
Since $(f,g)$ is maximal, for each $q\in Q$ and $b_q \in B_{\le 1}^Q$, we have $f(b_q)=f(b\cdot \1_q)=f(\1_q)$. Hence $|f(B_{\le 1}^Q\setminus\{\0^Q\})|\le |f(\{\1_q\mid q\in Q\})| \le |Q|$, which proves our lemma.
\end{proof}

%% file: support.tex
\chapter{Support of recognizable weighted tree languages}
\label{ch:support}

\index{support theorem}
In this chapter we will consider the question whether the support of a recognizable $(\Sigma,\B)$-weighted tree language is a recognizable $\Sigma$-tree language. We will start with a negative result by showing a string ranked alphabet and a $(\Sigma,\Int)$-wta $\cA$ (where $\Int$ is the ring of integers) such that $\supp(\sem{\cA})$ is not recognizable. We will continue with positive results and show \emph{support theorems}; a support theorem states conditions under which the support of a recognizable weighted tree language is a recognizable tree language. The support theorems  concern the run semantics or the initial algebra semantics; moreover, they are based on additional requirements on the wta or on the strong bimonoid.

\section{Negative result for support}
\label{sec:negative-results-for-support}

\begin{lemma}\label{lem:berreu88-Ex-III-3-1} \rm \cite[Ex. III. 3.1]{berreu88} For the $(\Sigma,\Int)$-wta $\cA$  in Example \ref{ex:diff-mon}, we have $\supp(\sem{\cA}) \not\in \Rec(\Sigma)$. \end{lemma}
\begin{proof} We recall that
 $\Sigma= \{\sigma^{(1)}, \gamma^{(1)}, \alpha^{(0)}\}$ and  that   $\sem{\cA}(\xi) = |\pos_\gamma(\xi)| - |\pos_\sigma(\xi)|$ for each $\xi \in \T_\Sigma$. Hence 
\[\supp(\sem{\cA})=\{\xi\in \T_\Sigma\mid |\pos_\gamma(\xi)|\ne|\pos_\sigma(\xi)|\}.\]
By contradiction, we can easily prove that $\supp(\sem{\cA})$ is not recognizable. For this, we assume that $\supp(\sem{\cA})\in \Rec(\Sigma)$. The complement of $\supp(\sem{\cA}))$ is the tree language
\[L=\{\xi\in \T_\Sigma\mid |\pos_\gamma(\xi)|=|\pos_\sigma(\xi)|\}\enspace.\]
Since, by Theorem \ref{thm:fta-closure-results}, recognizable $\Sigma$-tree languages are closed under complement, $L$ is also recognizable. However, this contradicts to Example \ref{ex:pumping-lemma-positive} in which we proved that $L$ is not a recognizable $\Sigma$-tree language. Hence $\supp(\sem{\cA})$ is not recognizable.
\end{proof}

The following result immediately follows from Lemma \ref{lem:berreu88-Ex-III-3-1}.

\begin{theorem}\label{thm:supp-not-rec}  There exists a string ranked alphabet $\Sigma$ such that $\supp(\Rec(\Sigma,\Int))\setminus \Rec(\Sigma) \ne \emptyset$.
\end{theorem}

\section{Positive results for support}

In this section we elaborate some support theorems. By a support theorem we mean a result which, for a certain set of recognizable weighted tree languages, guarantees that the support of each element of that set is a recognizable tree language. We start with a support theorem which can be derived from the preimage theorems. Then we consider (arbitrary) wta and classify the support theorem according to two criteria: the type of semantics of the wta (initial algebra semantics or run semantics) and the set of strong bimonoids. Finally, we compare some of the sets of strong bimonoids for which we have a support theorem.

\subsection{Consequences of preimage theorems}\label{subsect:conseq-preimage-theorems}

In Chapter \ref{ch:crisp-determinization} we have proved some preimage theorems. Roughly speaking, each of them states conditions such that the following implication holds: if a recognizable weighted tree language $r: \T_\Sigma \to B$ satisfies the conditions, then for each $b \in B$, the $\Sigma$-tree language $r^{-1}(b)$ is recognizable. In a straightforward way, each of these preimage theorems implies a support theorem. In the next corollary we merely collect these support theorems (without dealing with their relationship).

\begin{corollary-rect} \rm \label{thm:support-thm-from-preimage} Let $\Sigma$ be a ranked alphabet. Moreover, let $\B=(B,\oplus,\otimes,\0,\1)$ be a strong bimonoid and let $r:\T_\Sigma \to B$. If one of the following conditions is satisfied:
  \begin{compactenum}
  \item[(1)] $r$ is a recognizable step mapping,
  \item[(2)] $r \in \Rec^{\mathrm{init}}(\Sigma,\B)$ and (a)~$\B$ is locally finite or (b)~$\B$ is weakly locally finite and $\Sigma$ is monadic,
        \item[(3)] $r \in \Rec^{\mathrm{run}}(\Sigma,\B)$ and $\B$ is bi-locally finite,
      \item[(4)] $r \in \Rec^{\mathrm{run}}(\Sigma,\B)$ and $(\B,\preceq)$ is a past-finite monotonic strong bimonoid,
    \end{compactenum}
 then $\supp(r) \in \Rec(\Sigma)$.   Moreover, if $r$ is given effectively, then in each case we can construct a $\Sigma$-fta $A$ such that $\supp(r)= \LL(A)$.
\end{corollary-rect}
  \begin{proof} Since $\supp(r) = \T_\Sigma \setminus r^{-1}(\0)$ and the set of $\Sigma$-recognizable tree languages is closed under complement (cf. Theorem~\ref{thm:fta-closure-results}), the following equivalence holds:
    \begin{equation}
\text{$r^{-1}(\0)$ is recognizable if and only if  $\supp(r)$ is recognizable.} \label{equ:r-1-recog=>supp-recog}
      \end{equation}
Then, each of the four implications  is justified by equivalence \eqref{equ:r-1-recog=>supp-recog} and the corresponding preimage theorem:
(1) Theorem \ref{thm:crisp-det-algebra}$(B)\Rightarrow (C)$,
(2) Corollary \ref{cor:preimage-fin-loc-finite-weakly-loc-fin-rec-step-function},
 (3) Corollary \ref{cor:preimage-fin-stb}, and (4) Theorem~\ref{thm:preimage-past-finite-monotonic}(1).
 
Now let $r$ be given effectively. Then, in each of the Cases (1)-(3),  by the corresponding preimage theorem and Theorem~\ref{thm:fta-closure-results}  we can construct a $\Sigma$-fta $A$ such that $\supp(r)= \LL(A)$. In Case(4) we use Theorem~\ref{thm:preimage-past-finite-monotonic}(2)  and Theorem~\ref{thm:fta-closure-results} to construct the $\Sigma$-fta $A$; note that $\past(\0)=\{\0\}$.
 \end{proof}

    \subsection{Both semantics and  positive strong bimonoids}
    \label{ssec:both-semantics-positive-bm}

Corollary \ref{cor:supp-B=fta-1} is  a support theorem for the Boolean semiring. 
If we analyse the proof of the underlying Theorem \ref{thm:wta-B=fta}, then we realize that we have used the fact that the semiring  $\Boole$ is positive, i.e., zero-sum free and zero-divisor free. 
In this subsection we generalize Corollary \ref{cor:supp-B=fta-1}  from $\Boole$ to an arbitrary positive strong bimonoid.

We define the particular mapping $\sgn: B \to \mathbb{B}$ for each $b \in B$ by
\[
  \sgn(b) =
  \begin{cases}
    1 & \text{ if $b \ne \0$}\\
    0 & \text{ otherwise}
    \end{cases}
\]

\begin{lemma}\rm \label{lm:positive-hom}  Let $\B$ be positive. Then $\sgn: B \to \mathbb{B}$ is a strong bimonoid homomorphism from $\B$ to $\Boole$. 
\end{lemma}
\begin{proof} 
Let $b_1,b_2\in B$. Then 
\begin{align*}
\sgn(b_1\oplus b_2) = 1 &\text{ iff } b_1\oplus b_2\not=\mathbb{0}  \text{ iff}^* \ b_1\not= \mathbb{0} \vee b_2\not= \mathbb{0}\\
&\text{ iff } (\sgn(b_1)=1) \vee (\sgn(b_2)=1) \text{ iff } (\sgn(b_1) \vee \sgn(b_2)) = 1
\end{align*}
where at equivalence iff$^*$ from right to left we have used the fact that $\B$ is zero-sum free. Thus $\sgn(b_1\oplus b_2) =  \sgn(b_1) \vee \sgn(b_2)$. Also
\begin{align*}
\sgn(b_1\otimes b_2) = 1 &\text{ iff } b_1\otimes b_2\not=\mathbb{0}  \text{ iff}^* \ b_1\not= \mathbb{0} \wedge b_2\not= \mathbb{0}\\
&\text{ iff } (\sgn(b_1)=1) \wedge (\sgn(b_2)=1) \text{ iff } (\sgn(b_1) \wedge \sgn(b_2)) = 1
\end{align*}
where at equivalence iff$^*$ from right to left we have used the fact that $\B$ is zero-divisor free. Hence $\sgn(b_1\otimes b_2) = \sgn(b_1) \wedge \sgn(b_2)$. Moreover, $\sgn(\0)=0$ and $\sgn(\1)=1$. Thus $\sgn$ is a strong bimonoid homomorphism.
\end{proof}

Our next support theorem can be compared to  \cite[Thm.~3.12]{fulvog09new}.
In Lemma \ref{thm:run-init-positive} we prove that the run support and the initial algebra support of a wta $\cA$ are recognizable tree languages  if the strong bimonoid is positive. Since each monotonic strong bimonoid is positive (cf. Subsection \ref{sec:char-past-finite-crisp}),  Lemma~\ref{thm:run-init-positive} implies Corollary~\ref{thm:support-thm-from-preimage}(4). Moreover, since each positive strong bimonoid is bi-strongly zero-sum free, by Theorem~\ref{thm:bi-strongly-zsf-equiv-equ-supp}, the tree languages $\supp(\runsem{\cA})$ and $\supp(\initialsem{\cA})$ are equal. Finally, the fta $\supp(\sgn(\cA))$ recognizes the support languages. (We recall that in Section~\ref{sect:strong-bm-bimorphism} the wta $f(\cA)$ is defined for  each strong bimonoid homomorphism $f$).

\begin{lemma}\label{thm:run-init-positive} \rm Let  $\B$ be positive and $\cA$ be a $(\Sigma,\B)$-wta. Then $\LL(\supp(\sgn(\cA)))=\supp(\initialsem{\cA})=\supp(\runsem{\cA})$.
Hence, both $\supp(\runsem{\cA})$ and $\supp(\initialsem{\cA})$ are recognizable $\Sigma$-tree languages.
\end{lemma}
\begin{proof} First we prove that $\LL(\supp(\sgn(\cA)))=\supp(\runsem{\cA})$.
  For this, we consider the mapping $\sgn: B \to \mathbb{B}$. Since $\B$ is positive, $\sgn$ is a strong bimonoid homomorphism by Lemma \ref{lm:positive-hom}. We note that $\sgn(\cA)$ is a $(\Sigma,\Boole)$-wta.  Then we have
\begin{align*}
\LL(\supp(\sgn(\cA))) = & \ \supp(\runsem{\sgn(\cA)}) \tag{by Theorem \ref{thm:wta-B=fta}}\\
= & \ \supp(\sgn \circ \runsem{\cA}) \tag{by Lemma \ref{lm:positive-hom} and Lemma \ref{lm:f-image-equivalent}} \\
= & \ \supp(\runsem{\cA})\enspace.  \tag{by definition of $\supp$ and $\sgn$}
\end{align*}
Using the same results as in the above calculation, we can also prove that $\LL(\supp(\sgn(\cA)))=\supp(\initialsem{\cA})$. Since $\supp(\sgn(\cA))$ is a $\Sigma$-fta, both $\supp(\runsem{\cA})$ and $\supp(\initialsem{\cA})$ are recognizable $\Sigma$-tree languages.
\end{proof}

It is clear that Lemma \ref{thm:run-init-positive} generalizes also Theorem \ref{thm:wta-B=fta}, because $\Boole$ is positive. However, we have used Theorem \ref{thm:wta-B=fta}  for the proof of Lemma \ref{thm:run-init-positive}.

We note that, in spite of Lemma \ref{thm:run-init-positive}, there exist a positive strong bimonoid $\B$ and a $(\Sigma,\B)$-wta $\cA$ such that $\initialsem{\cA} \not= \runsem{\cA}$; witnesses of this fact are shown in Examples \ref{ex:run-not=init} and \ref{ex:run-not=init-2}, as well as in the proof of  Theorem~\ref{thm:init-not-run}.

The following theorem is a generalization of \cite[Thm. 3.12]{fulvog09new}
from positive semirings to positive strong bimonoids.

\index{support theorem} 
\begin{theorem-rect}\label{thm:recognizable=run=init-positive} Let $\Sigma$ be a ranked alphabet. Moreover, let  $\B$ be a positive strong bimonoid. Then
\[\Rec(\Sigma)=\supp(\Rec^{\mathrm{init}}(\Sigma,\B))=\supp(\Rec^{\mathrm{run}}(\Sigma,\B))\enspace.\]
\end{theorem-rect}
\begin{proof} The inclusion $\Rec(\Sigma)\subseteq \supp(\Rec^{\mathrm{init}}(\Sigma,\B))$ follows from (B) $\Rightarrow$ (C) of Theorem \ref{thm:fta-wta} (without using that $\B$ is positive).
The inclusion $\supp(\Rec^{\mathrm{init}}(\Sigma,\B))\subseteq \Rec(\Sigma)$ and the equality $\supp(\Rec^{\mathrm{init}}(\Sigma,\B))=\supp(\Rec^{\mathrm{run}}(\Sigma,\B))$ follow from Lemma \ref{thm:run-init-positive}.
\end{proof}

At this point we can prove that $\Int$ is not a Fatou extension of $\Nat$ (cf. Section \ref{sec:Fatou-extension}).

\index{Fatou extension}
\begin{lemma}\label{lm:Z-not-Fatou-N}\rm \cite[Ex. 8.1 on p.~129]{kuisal86} The ring $\Int$ is not a Fatou extension of the semiring $\Nat$.
\end{lemma}
\begin{proof} Let $r=\sem{\cA}$, where $\cA$ is the $(\Sigma,\Int)$-wta  in Example \ref{ex:diff-mon}. By Theorem \ref{thm:closure-Hadamard-product} , $r'=r\otimes r$ is also recognizable by a $(\Sigma,\Int)$-wta.
      Moreover, $\im(r') \subseteq \mathbb{N}$, hence $r'\in \Rec(\Sigma,\Int) \cap \mathbb{N}^{\T_\Sigma}$.
Next we observe that $\supp(r')=\supp(r)$ and hence, by Lemma \ref{lem:berreu88-Ex-III-3-1}, the $\Sigma$-tree language $\supp(r')$ is not recognizable. 
Since the semiring $\Nat$ is positive, by Theorem \ref{thm:recognizable=run=init-positive}, we obtain that $r' \not \in \Rec(\Sigma,\Nat)$.
\end{proof}


\subsection{Both semantics and  commutative semirings which are not rings}

Here we show a support theorem  which deals with commutative semirings which are not rings. It is based on the following result. 

\begin{theorem} \label{thm:Wang-Thm.2.1} {\rm \cite[Thm.~2.1]{wan97} (also cf. \cite[Lm.~9.3]{drokus21})} Let $\B$ be a commutative semiring which is not a ring. Then there exists a semiring homomorphism from $\B$ onto the Boolean semiring.
\end{theorem}

\index{support theorem}
\begin{theorem-rect}\label{thm:support-not-ring} Let $\Sigma$ be a ranked alphabet. Moreover, let $\B=(B,\oplus,\otimes,\0,\1)$ be a commutative semiring which is not a ring, and let $L \subseteq \T_\Sigma$ be a tree language. If $\chi_\B(L) \in \Rec(\Sigma,\B)$, then $L$ is  in $\Rec(\Sigma)$.
\end{theorem-rect}

\begin{proof} Let $\chi_\B(L) \in \Rec(\Sigma,\B)$. By Theorem \ref{thm:Wang-Thm.2.1}, there exists a semiring homomorphism $h: B \rightarrow \mathbb{B}$. Clearly, $h(\chi_\B(L)) = \chi_{\Boole}(L)$.
  Then, by Theorem \ref{thm:closure-sr-hom}, we have $\chi_{\Boole}(L) \in \Rec(\Sigma,\Boole)$. Thus, since $L = \supp(\chi_{\Boole}(L))$, we have $L \in \supp(\Rec(\Sigma,\Boole))$. By Corollary \ref{cor:supp-B=fta-1} we obtain that $L \in \Rec(\Sigma)$.
\end{proof}


\subsection{Run semantics and commutative, zero-sum free strong bimonoids}
\label{sect:Kirsten}

In \cite[Thm.~3.1]{kir11} (also cf. \cite[Thm.~1]{kir09}) it was proved that the support of every recognizable $(\Gamma,\B)$-weighted string language is a recognizable string language provided that the weight algebra $\B$ is a commutative zero-sum free semiring. This generalizes \cite[Cor.~5.2]{wan98} where this result was proved for commutative quasi-positive semirings, because each commutative quasi-positive semiring is a commutative zero-sum free semiring by definition.
Here we show how \cite[Thm.~3.1]{kir11} can be generalized to the tree case by adapting the proof technique from \cite{kir11} (cf. \cite{droheu15}, \cite[Thm.~4.6]{goe17}, and  \cite[Thm.~4.3(1)]{fulhervog18}).
Distributivity is not needed. 

\label{p:convention-support}
\begin{quote}\emph{In this subsection, we assume that $\B$ is commutative.}
\end{quote}

Let us consider a $(\Sigma,\B)$-wta $\cA$, a tree $\xi \in \T_\Sigma$, and a run $\rho \in \R_\cA(\xi)$. By definition, $\wt_\cA(\xi,\rho)$ is the product of some occurrences of elements of $\im(\delta)$ (cf. Observation \ref{obs:weight-run-explicit}). Since $\im(\delta)$ is finite, we can sort its elements in a vector of $n$ components, where $n=|\im(\delta)|$. Then,  since $\otimes$ is commutative, we can represent $\wt_\cA(\xi,\rho)$ as a vector $\bar z \in \mathbb{N}^n$ of the multiplicities of these occurrences. Next we will formalize this representation.

Let $\vec{b}=(b_1,\ldots,b_n)$ be an arbitrary, but fixed enumeration of $\im(\delta)$. Then $b_i \not= b_j$ for $i\not=j$.   Let $(z_1,\ldots,z_n), (z_1',\ldots,z_n') \in \mathbb{N}^n$. Then we define
\[
(z_1,\ldots,z_n) + (z_1',\ldots,z_n') = (z_1+z_1',\ldots,z_n+z_n') \enspace.
\]
It is clear that the binary operation $+$ is associative.
By identifying each $b_j \in \im(\delta)$ with $(\underbrace{0,\ldots,0}_{j-1},1,\underbrace{0,\ldots,0}_{n-j})$, we have
\[
  (z_1,\ldots,z_n) + b_j = (z_1,\ldots,z_{j-1},z_j+1,z_{j+1},\ldots,z_n)
\enspace.
\]

For the inductive definition of the representation, we use the terminating reduction system
\((\mathrm{TR},\succ)\) defined in Section \ref{sec:basic-defininition-wta} on p.~\pageref{page:TR-prec}. For the sake of convenience we recall that
\(
\mathrm{TR} = \{(\xi,\rho) \mid \xi \in \T_\Sigma, \rho \in \R_\cA(\xi)\}
\)
and that the terminating relation $\succ$ on $\mathrm{TR}$ is the binary relation defined as follows:
\index{succb@$\succ$}
\[\text{for every $(\xi, \rho) \in \mathrm{TR}$ and $i \in [\rk(\xi(\varepsilon))]$, we let $(\xi,\rho) \succ (\xi|_i,\rho|_i)$} \enspace. \]
Clearly $\nf_\succ(\mathrm{TR}) = \{(\alpha,\rho) \mid \alpha \in \Sigma^{(0)}, \rho: \{\varepsilon\} \to Q\}$.
Then we define the mapping $\overline{(.)}: \mathrm{TR} \to \mathbb{N}^n$ by induction on $(\mathrm{TR},\succ)$ for every $\xi \in \T_\Sigma$ and $\rho \in \R_\cA(\xi)$ by  
\[
  \overline{(\xi,\rho)} = (0,\ldots,0) +  \overline{(\xi|_1,\rho|_1)} + \ldots + \overline{(\xi|_k,\rho|_k)} + \delta_k(\rho(1) \cdots \rho(k),\xi(\varepsilon),\rho(\varepsilon))
\]
where $k = \rk(\xi(\varepsilon))$.
We have added the vector $(0,\ldots,0)$ in order to have $\overline{(\xi,\rho)}$ well defined for the case $k=0$.

Moreover, each vector $\vec{z} =(z_1,\ldots,z_n) \in \mathbb{N}^n$
represents an element of $B$. Formally, we define the mapping $\sem{.} \colon \mathbb{N}^n \to B$ for each $\vec z=(z_1,\dotsc,z_n) \in \mathbb{N}^n$ by
\[
\sem{\vec z\ } = b_1^{z_1} \otimes \ldots \otimes b_n^{z_n} \enspace.
\]
Since $\B$ is commutative,
\begin{equation}
  \text{for every $\xi \in \T_\Sigma$ and $\rho \in \R_\cA(\xi)$, we have $\wt_\cA(\xi,\rho) = \sem{\overline{(\xi,\rho)}}$  \enspace.} \label{equ:weight-tuple} 
\end{equation}
In other words, $\wt_\cA = \sem{.} \circ \overline{(.)}$. In this sense, the vector $\overline{(\xi,\rho)}$ represents $\wt_\cA(\xi,\rho)$.

\


Next we will analyse the set $\sem{.}^{-1}(\0)$ of vectors $\vec{z} \in \mathbb{N}^n$ which are mapped to $\0$. (We note that $(0,\ldots,0) \not\in \sem{.}^{-1}(\0)$, because $b_1^0 \otimes \ldots \otimes b_n^0 = \1$.)

For this, we consider the partially ordered set $(\mathbb{N}^{n},\le)$, where the binary relation $\le$ is defined as follows: for all vectors $\vec{z}=(z_1,\dotsc,z_n)\in\mathbb{N}^{n}$ and $\vec{y} = (y_1,\dotsc,y_n) \in\mathbb{N}^{n}$, we define $\vec{z} \le \vec{y}$ if $z_i \le y_i$ for each  $i\in [n]$. We note that, for every $\vec{z}\,,\vec{y}\, \in \mathbb{N}^n$, if $\sem{\vec{z}\,} = \0$ and $\vec{z}\, \le \vec{y}\,$, then $\sem{\vec{y}\,} = \0$ (because $\0$ is annihilating).
Let $M \subseteq \mathbb{N}^n$. An element $\vec{z}\in M$ is \emph{minimal in $M$} if,  for each  $\vec{y} \in M$, the assumption $\vec{y} \le \vec{z}$ implies $\vec{y}=\vec{z}$. We denote by $\min(M)$ the set of all minimal elements in $M$.
Thus, in particular, $\min(\emptyset) = \emptyset$.
The following result is called Dickson's Lemma.

\begin{lemma}\label{lm:Dickson} \rm \cite{dic13} (cf. also \cite[Lm. 2.1]{kir11} and \cite{krerob08}) For each $M \subseteq \mathbb{N}^n$, the set $\min(M)$ is finite.
  \end{lemma}

By Lemma \ref{lm:Dickson}, the set  $\min(\sem{.}^{-1}(\0))$ is finite. Hence there exists a smallest number $m \in \mathbb{N}$ satisfying $\min(\sem{.}^{-1}(\0)) \subseteq \{0,\dotsc,m\}^n$, and we  call this number $m$ the \emph{degree of $\vec{b}$} and denote it by $\operatorname{dg}(\vec{b}\,)$. (We note that if $\B$ is zero-divisor free and $\0 \not\in \im(\delta)$, then $\sem{.}^{-1}(\0)=\emptyset$, hence $\operatorname{dg}(\vec{b}\,)=0$. We also note that, for any other enumeration $\vec{b'}$ of $\im(\delta)$, we have $\mathrm{dg}(\vec{b}) = \mathrm{dg}(\vec{b'})$.)

The next lemma states that, if $\sem{\vec{z}\,\,} = \0$ for some $\vec{z}\,$, then also the evaluation of $\vec{z}\,'$ results in $\0$ where $\vec{z}\,'$ is obtained from $\vec{z}\,$ by restricting the components to $\operatorname{dg}(\vec{b})$. Formally, for every $\vec{z}\, \in \mathbb{N}^n$, we define the \emph{cut of $\vec{z}\,$}, denoted by $\lfloor \vec{z}\,\rfloor_{\operatorname{dg}(\vec{b})}$, to be the vector $\lfloor \vec{z}\,\rfloor_{\operatorname{dg}(\vec{b})} \in \mathbb{N}^n$ with 
\[( \lfloor \vec{z}\, \rfloor_{\operatorname{dg}(\vec{b})})_i = \min \{z_i, \operatorname{dg}(\vec{b})\}\]
as $i$-th component for each $i \in [n]$.

\begin{lemma}\rm \label{Kirsten2}(\cite[Lm. 4.1]{kir11}) For every $\vec{z}\, \in \mathbb{N}^n$, we have $\sem{\vec{z}\,} = \0$ iff $\sem{\lfloor \vec{z}\, \rfloor_{\operatorname{dg}(\vec{b})}} = \0$.
\end{lemma}
\begin{proof} Let $\sem{\floor{\vec{z}\,}{\operatorname{dg}(\vec{b})}} = \0$. Since $\lfloor \vec{z}\, \rfloor_{\operatorname{dg}(\vec{b})} \le \vec{z}\,$ and $\0$ is an annihilator for $\otimes$, we have $\sem{\vec{z}\,} = \0$. 

  Now let $\sem{\vec{z}\,} = \0$. Then there exists a $\vec y \in \min(\sem{.}^{-1}(\0))$ such that $\vec{y} \le \vec{z}\,$. We prove that $\bar y \le \floor{\vec{z}\,}{\operatorname{dg}(\vec{b})}$. For this let  $\bar y = (y_1,\ldots,y_n)$, $\vec{z}\, = (z_1,\ldots,z_n)$, and $i \in [n]$. If $z_i \le \operatorname{dg}(\vec{b})$, then $y_i \le z_i = (\floor{\vec{z}\,}{\operatorname{dg}(\vec{b})})_i$. If $z_i > \operatorname{dg}(\vec{b})$, then $y_i \le \operatorname{dg}(\vec{b}) = (\floor{\vec{z}\,}{\operatorname{dg}(\vec{b})})_i$. Hence $\vec y \le \floor{\vec{z}\,}{\operatorname{dg}(\vec{b})}$ and thus $\sem{\lfloor \vec{z}\, \rfloor_{\operatorname{dg}(\vec{b})}} = \0$.
  \end{proof}

In Figure \ref{fig:support-tuples-Hasse-diagram} we indicate the partial order on $\mathbb{N}^3$.
We assume that $\min(\sem{.}^{-1}(\0))=\{\vec{z_1},\vec{z_2},\vec{z_3}\}$. For each $i \in \{1,2,3\}$ we indicate the set $\{\vec y \mid \vec{z_i} \le \vec y\,\}$ by a shaded filter. Clearly, for each $\vec y$ in one of these filters we have $\sem{\vec{y}\,}=\0$. We also indicate the set $\{0,\ldots,j\}^n$ for $j=0$, $j=1$, and $j=\operatorname{dg}(\vec{b})$.

\begin{figure}[t]
  \begin{center}
\begin{tikzpicture}[triangle/.style = {bottom color=black!20, top color=white, isosceles triangle,shape border rotate=90, draw=white, minimum height=3cm, isosceles triangle apex angle=90, anchor=north, scale=0.9},
    node rotated/.style = {rotate=180},
    border rotated/.style = {shape border rotate=180}]
\matrix (m) [matrix of math nodes, row sep=2.5em,
column sep=0.3em, text height=1.5ex, text depth=0.25ex]
{ & & & &&& \\
(2,0,0)&\qquad(1,1,0)\qquad&(1,0,1)&&(0,2,0)&\qquad(0,1,1)\qquad&(0,0,2)&&\\
&(1,0,0) & & (0,1,0)&  &(0,0,1)&& \\
&&&(0,0,0)&&&& \\};

\path[-] 	(m-4-4) edge (m-3-2)
			(m-4-4) edge (m-3-4)
			(m-4-4) edge (m-3-6)
			(m-3-2) edge (m-2-1)
			(m-3-2) edge (m-2-2)
			(m-3-2) edge (m-2-3)
			(m-3-4) edge (m-2-2)
			(m-3-4) edge (m-2-5)
                        (m-3-4) edge (m-2-6)
			(m-3-6) edge (m-2-3)
			(m-3-6) edge (m-2-6)
			(m-3-6) edge (m-2-7);
			
\draw[dashed, rounded corners, black!40] ($(m.south west)+(1,0)$) -- +(0,1) -- ($(m.south east)+(-1,1)$) -- ($(m.south east)+(-1,0)$);

\draw[dashed, rounded corners, black!40] ($(m.south west)+(0,0.5)$) -- +(0,1.5) -- +(1,1.5) -- +(2.3,3.2) -- +(6.5,3.2) -- +(7,2.2) -- +(9,2.2) -- +(9.5,3.2) -- +(10.8,3.2) -- ($(m.south east)+(-1.5,2)$) -- ($(m.south east)+(0,2)$) -- ($(m.south east)+(0,0.5)$);

\node[black!40] at ($(m.south east)+(-2,0.5)$) {$\subseteq\{0\}^n$};
\node[black!40] at ($(m.south east)+(-1,1.5)$) {$\subseteq\{0,1\}^n$};
\node[black!40] at ($(m.south east)+(-1.0,7.2)$) {$\subseteq \{0,\ldots,\mathrm{dg}(\vec{b})\}^n$};

\node[rotate=120] at ($(m-2-1)+(-1,1.5)$) {$\ldots$};
\node[rotate=60] at ($(m-2-7)+(1,1.5)$) {$\ldots$};
\node[rotate=90] at ($(m-3-4)+(0,3)$) {$\ldots$};

\node (b1) at ($(m-2-1)+(1,3)$) {$\bullet$};
\node (b2) at ($(m-3-4)+(0,6)$) {$\bullet$};
\node (b3) at ($(m-2-6)+(0,2)$) {$\bullet$};

\node at ($(b1)+(0.3,-0.2)$) {$\vec{z}\,_1$};
\node at ($(b2)+(0.3,-0.2)$) {$\vec{z}\,_2$};
\node at ($(b3)+(0.3,-0.2)$) {$\vec{z}\,_3$};

\draw[dashed, rounded corners, black!40] ($(b2)+(-7.5,-2)$) -- ($(b2)+(-7.5,0)$) -- ($(b2)+(7.5,0)$) -- ($(b2)+(7.5,-2)$);

\begin{pgfonlayer}{background}
\node[triangle, rotate=180, fill] at (b1) {};
\node[triangle, rotate=180, fill] at (b2) {};
\node[triangle, rotate=180, fill] at (b3) {};
\node[] at ($(b1)+(0,1.2)$) {$\{\vec y \mid \vec{z_1} \le \vec y\,\}$};
\node[] at ($(b2)+(0,1.4)$) {$\{\vec y \mid \vec{z_2} \le \vec y\,\}$};
\node[] at ($(b3)+(0,1.2)$) {$\{\vec y \mid \vec{z_3} \le \vec y\,\}$};
\end{pgfonlayer}
\end{tikzpicture}
  \end{center}
\caption{\label{fig:support-tuples-Hasse-diagram} The Hasse diagram of tuples $\mathbb{N}^3$ with $\min(\sem{.}^{-1}(\0))=\{\vec{z_1},\vec{z_2},\vec{z_3}\}$.}
  \end{figure}

  Now we can prove the main support theorem of this subsection.  We follow the proof  of the corresponding result \cite[Thm.~3.1]{kir11} for weighted string automata over semirings. This result has been generalized to (a)  weighted unranked tree automaton over  zero-sum free, commutative strong bimonoids \cite{droheu15} and (b)~weighted unranked tree automata over zero-sum free, commutative, zero-preserving
  tv-monoids \cite[Thm.~4.7]{goe17}. We also refer to \cite[Thm.~4.4]{fulhervog18} where the support theorem was proved for weighted tree automata over $\sigma$-complete and commutative strong bimonoids\footnote{(We note that, by Observation \ref{obs:zero-sum-free-property}(9), $\sigma$-completeness implies zero-sum freeness.)}.

\index{support theorem}
\begin{theorem-rect} \label{thm:Kirsten} Let $\Sigma$ be a ranked alphabet. Moreover, let $\B=(B,\oplus,\otimes,\0,\1)$ be a  commutative strong bimonoid and let $\cA$ be a $(\Sigma,\B)$-wta. If (a) $\B$ is zero-sum free or (b) $\cA$~is bu-deterministic and root weight normalized, then $\supp(\runsem{\cA}) \in \Rec(\Sigma)$.
\end{theorem-rect}
\begin{proof}
  Let $\cA=(Q,\delta,F)$ be a $(\Sigma,\B)$-wta.
  If $\cA$ is not root weight normalized, then by Theorem~\ref{thm:root-weight-normalization-run} we can construct a run equivalent and root weight normalized 
  $(\Sigma,\B)$-wta.
  Hence, in both cases (a) and (b) we may assume that $\cA$ is root weight normalized.
  Thus, there exists $q_f \in Q$ such that $F_{q_f}= \1$ and $F_{q}=\0$ for each $q \in Q\setminus \{q_f\}$. Let $\vec{b} = (b_1,\dotsc,b_n) \in B^n$ be an arbitrary but fixed  enumeration of $\im(\delta)$. 
  
Then we define  the $\Sigma$-fta $A = (Q', \delta', F')$ which counts, up to the threshold $\operatorname{dg}(\vec{b\,})$, the number of occurrences of the $b_j$'s in the runs of $\cA$. Formally,  we let $T=\{0,\dotsc,\operatorname{dg}(\vec{b}\,)\}$ and  we define 
  \begin{compactitem}
  \item $Q' = Q \times T^n$, 
  \item $F' = \{(q_f,\vec{z}\,) \in Q' \mid \sem{\vec{z}\,} \ne \0\}$,
  \item we define  for each $k \in \mathbb{N}$
       \begin{align*}
      \delta_k' = \{((q_1,\vec{z}_1) \cdots (q_k,\vec{z}_k),\sigma,(q,\vec{z})) \mid \
      & \sigma \in \Sigma^{(k)}, ((q_1,\vec{z}_1) \cdots (q_k,\vec{z}_k))\in (Q')^k,(q,\vec{z}) \in Q', \\
      &\vec{z}= \floor{(0,\ldots,0) + \vec{z}_1 + \ldots + \vec{z}_k + \delta_k(q_1 \ldots q_k,\sigma,q)}{\operatorname{dg}(\vec{b})} \}\enspace.
      \end{align*}
  \end{compactitem}

  Now we prove that $\supp(\runsem{\cA}) = \LL(A )$. First, for each  $\xi \in \T_\Sigma$, we define the mapping
  \[
    \varphi: \R_\cA(\xi) \to \Rv_{A}(\xi)
  \]
  as follows. For every $\rho \in \R_\cA(\xi)$, we define the family $(\varphi(\rho)(w) \mid w \in \pos(\xi))$ of states $\varphi(\rho)(w) \in Q'$.

  For this definition, we employ the reduction system $(\pos(\xi),\succ)$ which was defined and proved to be terminating in the proof of  Lemma \ref{lm:BPS}.
  We recall that, for every $w_1,w_2 \in \pos(\xi)$, we have $w_1 \succ w_2$ if there exists $i\in \mathbb{N}$ such that $w_2 = w_1i$, and that $\nf_\succ(\pos(\xi))=\pos_{\Sigma^{(0)}}(\xi)$, i.e., it is the set of leaves of $\xi$.
    
  Now we define $\varphi(\rho)$ by induction on $(\pos(\xi),\succ)$ as follows.  
 Let $w \in \pos(\xi)$ with  $\xi(w) = \sigma$ for some $\sigma \in \Sigma^{(k)}$ with $k \in \mathbb{N}$. We assume that $\varphi(\rho)(wi)= (q_i,\vec{z_i})$ for each $i \in [k]$. Then we define
  \[
    \varphi(\rho)(w) = \big(\rho(w), \floor{(0,\ldots,0) +  \vec{z}_1 + \ldots + \vec{z}_k + \delta_k(q_1 \ldots q_k,\sigma,\rho(w))}{\operatorname{dg}(\vec{b})}\big)\enspace.
  \]
  It is clear that the run $\varphi(\rho)$ of the $\Sigma$-fta $A$  is valid and that $\varphi$ is bijective. Moreover, by induction on $\T_\Sigma$, we can prove the following:

\begin{equation}
  \text{ for every $\xi \in \T_\Sigma$, $(q,\vec{z}) \in Q'$,  and  $\rho' \in \Rv_{A}((q,\vec{z}),\xi)$: \
    $\vec{z} = \floor{\overline{(\xi,\varphi^{-1}(\rho'))}}{\operatorname{dg}(\vec{b})}$ } \label{equ:weight-truncated}
\end{equation}
Let $\xi = \sigma(\xi_1,\ldots,\xi_k)$,  $(q,\vec{z}) \in Q'$,  and $\rho' \in \Rv_{A}((q,\vec{z}),\xi)$. Then, for each $i \in [k]$, there exist $(q_i,\vec{z_i}) \in Q'$ such that  $\rho'|_i \in \Rv_{A}((q_i,\vec{z_i}),\xi_i)$ and
\begin{equation}\label{eq:induction-weight}
\vec{z} = \floor{(0,\ldots,0) +  \vec{z_1} + \ldots + \vec{z_k} + \delta_k(q_1\cdots q_k,\sigma,q)}{\operatorname{dg}(\vec{b})} \enspace.
  \end{equation}
Then we can calculate as follows (where we abbreviate $\floor{.}{\operatorname{dg}(\vec{b})}$ by $\floor{.}{}$):
\begingroup
\allowdisplaybreaks
\begin{align*}
  &\ \floor{\overline{(\sigma(\xi_1,\ldots,\xi_k),\varphi^{-1}(\rho'))}}{}\\
  = &\ \floor{(0,\ldots,0) +  \overline{(\xi_1,\varphi^{-1}(\rho'|_1))} + \ldots + \overline{(\xi_k,\varphi^{-1}(\rho'|_k)} + \delta_k(q_1\cdots q_k,\sigma,q)}{} \tag{by definition of $\overline{(.)}$}\\
  = & \ \floor{(0,\ldots,0) +  \floor{\overline{(\xi_1,\varphi^{-1}(\rho'|_1))}}{} + \ldots + \floor{\overline{(\xi_k,\varphi^{-1}(\rho'|_k)}}{} + \delta_k(q_1\cdots q_k,\sigma,q)}{}\tag{because $\floor{\vec{y_1}+\vec{y_2}}{} = \floor{\vec{y_1}+\floor{\vec{y_2}}{}}{} =\floor{\floor{\vec{y_1}}{}+\floor{\vec{y_2}}{}}{}$ for every $\vec{y_1},\vec{y_2} \in T^n$} \\
  = & \ \floor{(0,\ldots,0) +  \vec{z_1} + \ldots + \vec{z_k} + \delta_k(q_1\cdots q_k,\sigma,q)}{} \tag{by I.H.}\\
  = & \ \vec{z} \tag{by \eqref{eq:induction-weight}} \enspace.
 \end{align*}
 \endgroup
 This proves \eqref{equ:weight-truncated}.

  Now let $\xi \in \T_\Sigma$. We note that, if $\B$ is not zero-sum free, then $\cA$ is bu-deterministic. Then
  \begingroup
  \allowdisplaybreaks
  \begin{align*}
    & \xi \in \supp(\runsem{\cA})
      \ \text{ iff } \  \runsem{\cA}(\xi) \not= \0 \\
      \ \text{ iff } &   \bigoplus_{\rho \in \R_\cA(q_f,\xi)} \wt_\cA(\xi,\rho) \not= \0 \tag{because $\supp(F)=\{q_f\}$ and $F_{q_f}=\1$}\\
    \text{iff } & (\exists \rho \in \R_\cA(q_f,\xi))\colon \ \wt_\cA(\xi,\rho) \not= \0
                  \tag{\text{because $\B$ is zero-sum free or $\cA$ is bu-deterministic}}\\
    \text{iff } & (\exists \vec{z} \in T^n) (\exists \rho' \in \Rv_{A}((q_f,\vec{z}),\xi))\colon \ \wt_\cA(\xi,\varphi^{-1}(\rho')) \not= \0 \ \
      \tag{\text{because $\varphi$ is bijective}}\\
    \text{iff } & (\exists \vec{z} \in T^n) (\exists \rho' \in \Rv_{A}((q_f,\vec{z}),\xi))\colon \
                  \sem{    \overline{(\xi,\varphi^{-1}(\rho'))}} \not= \0
    \tag{by \eqref{equ:weight-tuple}}\\
    \text{iff } & (\exists \vec{z} \in T^n) (\exists \rho' \in \Rv_{A}((q_f,\vec{z}),\xi))\colon \
                  \sem{    \floor{\overline{(\xi,\varphi^{-1}(\rho'))}}{\operatorname{dg}(\vec{b}\,)}} \not= \0 \ \
                   \tag{\text{Lemma \ref{Kirsten2}}}\\
    \text{iff } & (\exists \vec{z} \in T^n) (\exists \rho' \in \Rv_{A}((q_f,\vec{z}),\xi))\colon \
                  \sem{\vec{z}\,} \not= \0  \ \ \tag{by \eqref{equ:weight-truncated} }\\
    \text{iff } & (\exists q' \in F')\colon \ \Rv_{A}(q',\xi) \not= \emptyset
    \tag{by construction of $F'$}\\
    \text{iff } &  \xi \in \LL(A)\enspace.  \qedhere
    \end{align*} 
    \endgroup
  \end{proof}

As a direct consequence of Theorem \ref{thm:Kirsten} we obtain a support theorem for wta over each strong bimonoid in which the summation is a t-conorm (cf. Example \ref{ex:strong-bimonoids}(\ref{ex:0-1-conorm-norm}) and thus, in particular, for wta over $\UnitIntfuzzy_{u,i}$ where $u$ is a t-conorm and $i$ is a t-norm.  The latter  consequence generalizes the case $\lambda=0$ of \cite[Thm.~7]{san68} from $\UnitIntfuzzy_{\max,\min}$ to $\UnitIntfuzzy_{u,i}$ for each tuple $(u,i)$ of t-conorm $u$ and t-norm $i$.

  \begin{corollary}\rm Let $\B=([0,1],u,\otimes,0,1)$ be a commutative strong bimonoid such that $u$ is a t-conorm. Moreover, let $\cA$ be a $(\Sigma,\B)$-wta. Then $\supp(\runsem{\cA}) \in \Rec(\Sigma)$.
  \end{corollary}
  \begin{proof}  We show that $\B=([0,1],u,\otimes,0,1)$ is zero-sum free. Then the statement follows from  Theorem~\ref{thm:Kirsten}.

    Let $a,b \in [0,1]$ and $u(a,b)=0$. Using the boundary condition and monotonicity condition of $u$ (cf. Example \ref{ex:strong-bimonoids}(\ref{ex:0-1-conorm-norm})), we have
    \begingroup
    \allowdisplaybreaks
    \begin{align*}
      a &= u(a,0) \tag{by the boundary condition for $u$}\\
        &\le u(a,b) \tag{by monotonicity condition for $u$} \\
      &=0 \tag{by assumption}\enspace.
    \end{align*}
    \endgroup
    This implies that $a=0$. In a similar way we can derive that $b=0$. Hence $\B=([0,1],u,\otimes,0,1)$ is zero-sum free. 
  \end{proof}

  In Theorem \ref{thm:Kirsten}(a), the strong bimonoid $\B$ is assumed to be commutative and zero-sum free. The theorem is not constructive because, in order to construct an fta which accepts the support of a run recognizable weighted tree language, we should construct  $\mathrm{dg}(\vec{b})$. However,    it is not clear how to construct $\mathrm{dg}(\vec{b})$. We mention that in \cite{kir11}  the zero generation problem (ZGP)  is considered, and the decidability of the ZGP implies that  $\operatorname{dg}(\vec{b})$ can be constructed and hence the fta in the support theorem can be constructed. 
   
  In the following theorem, we require additionally that $\B$
    is multiplicatively idempotent. Then we do not need to compute $\mathrm{dg}(\vec{b})$ and we can construct the desired fta.   We note that our assumption on $\B$  implies that the ZGP is decidable. However, we do not wish to include the  ZGP in our proof. For more details on ZGP we refer to, e.g., \cite{kir11,droheu15,goe17,fulhervog18}.

  \begin{theorem-rect}\label{thm:Kirsten-comm-idemp} Let $\Sigma$ be a ranked alphabet. Moreover, let  $\B=(B,\oplus,\otimes,\0,\1)$ be a zero-sum free, commutative, and multiplicatively idempotent strong bimonoid. Then, for each $(\Sigma,\B)$-wta $\cA$, we can construct a $\Sigma$-fta $A$ such that $\LL(A)=\supp(\runsem{\cA})$.
  \end{theorem-rect}
  \begin{proof} Let $\cA=(Q,\delta,F)$ be a $(\Sigma,\B)$-wta. By Theorem~\ref{thm:root-weight-normalization-run} we can assume that $\cA$ is root weight normalized and $F_{q_f} =\1$ and $F_q=\0$ for each $q \in Q \setminus \{q_f\}$.  Let $\vec{b} = (b_1,\ldots,b_n) \in B^n$ be an arbitrary but fixed  enumeration of $\im(\delta)$.

    Since $\B$ is commutative and  multiplicatively idempotent, it is clear that $\min(\sem{.}^{-1}(\0)) \subseteq \{0,1\}^n$. Hence, instead of $\mathrm{dg}(\vec{b})$, we can use 1  as threshold for the frequency with which an element of $\im(\delta)$ can occur in the weight of a run. More precisely,  we construct the $\Sigma$-fta $A$ in the same way as in the proof of Theorem \ref{thm:Kirsten} except that we define the set $Q'$ of states of $A$ by $Q' = Q \times \{0,1\}^n$. We obtain  $\LL(A)=\supp(\runsem{\cA})$.
  \end{proof}

Finally we mention that in \cite[Thm. 3.5]{kir14} the following characterization result was shown: for each semiring $\B$, the supports of all recognizable $\B$-weighted string languages are recognizable if and only if in every finitely generated subsemiring of $\B$, there exists a congruence relation of finite index such that $\{\0\}$ is a singleton congruence class. We refer the reader for further  support theorems for wsa to \cite{berreu88,drokus21}.

\subsection{Comparison of support theorems}
\label{sec:comp-supp-theorems}

Finally, we will make two comparisons of the support theorems shown in Figure \ref{fig:table-classes-vs-semantics}. First we compare some support theorems concerning run semantics, then we do the same for
initial algebra semantics. Since each such support theorem requires a particular subset of strong bimonoids, we compare these support theorems by comparing the corresponding sets of strong bimonoids with respect to set inclusion.  We denote the involved sets of strong bimonoids as follows: 
\begin{compactenum}
  \item[$\cC$:] set of all strong bimonoids,
\item[$\cC_1$:] set of all positive strong bimonoids,
  \item[$\cC_2$:]  set of all commutative semirings which are not rings,
\item[$\cC_3$:]  set of all commutative, zero-sum free strong bimonoids,
\item[$\cC_4$:]  set of all locally finite strong bimonoids,
\item[$\cC_5$:]  set of all bi-locally finite strong bimonoids.
\end{compactenum}
Hence, first we compare the sets $\cC$, $\cC_1$, $\cC_2$, $\cC_3$, and $\cC_5$ (cf. Theorem \ref{thm:comparison-support-theorem-run}), and second the sets $\cC$, $\cC_1$, $\cC_2$, and $\cC_4$ (cf. Theorem \ref{thm:comparison-support-theorem-initial-algebra}).

\begin{figure}[t]
  \begin{center}
    \begin{tabular}{l|c|c}
  subsets of strong bimonoids  & \multicolumn{2}{c}{support theorems for}\\
  & run semantics & initial algebra semantics \\\hline
    $\cC$  set of all strong bimonoids & Corollary  \ref{thm:support-thm-from-preimage}(1) & Corollary  \ref{thm:support-thm-from-preimage}(1)\\
       & (for crisp-deterministic wta) & (for crisp-deterministic wta) \\\hline
  $\cC_1$  positive & Theorem  \ref{thm:recognizable=run=init-positive} & Theorem \ref{thm:recognizable=run=init-positive} \\
    $\cC_2$ commutative semirings, not rings & Theorem~\ref{thm:support-not-ring} & Theorem~\ref{thm:support-not-ring}\\
$\cC_3$  commutative, zero-sum free & Theorem \ref{thm:Kirsten} & \\
  $\cC_4$  locally finite & & Corollary \ref{thm:support-thm-from-preimage}(2)  \\
$\cC_5$  bi-locally finite & Corollary \ref{thm:support-thm-from-preimage}(3) & \\
\end{tabular}
\caption{\label{fig:table-classes-vs-semantics} Some support theorems proved in this chapter.}
\end{center}
\end{figure}

The following observation is trivial, but useful.

  \begin{observation}\rm \label{lm:12-3}  $(\cC_1 \cap \cC_2) \subseteq \cC_3$.\hfill $\Box$
  \end{observation}

    \begin{theorem} \label{thm:comparison-support-theorem-run} {\rm \cite{doemoe20,dro22}}
Figure~\ref{fig:Venn-diagram-support-run} shows the Euler diagram of the sets $\cC_1$, $\cC_2$, and $\cC_3$. Moreover, for each of the shown seven nonempty regions $X$, we show two strong bimonoids $\B_i$ and $\B_j$ in the form of the fraction $\frac{\B_i}{\B_j}$ such that $\B_i \in X \cap \cC_5$ and $\B_j \in X \setminus \cC_5$.
\end{theorem}

\begin{figure}[t]
  \begin{center}
    \begin{tikzpicture}[scale=0.9]
\draw (-0.9, 0) circle (1.5);
\draw (0.9, 0) circle (1.5);

\draw (0.0,-0.4) ellipse (4cm and 3cm);

\draw[yshift=1.2cm -3.8cm]
	(1.5,1.5)
		.. controls (1.5,0.67) and (0.83,0) ..
	(0,0)
		.. controls (-0.83,0) and (-1.5,0.67) ..
	(-1.5,1.5)
		.. controls (-1.5,2.33) and (-0.83,3.8) ..
	(0,3.8)
		.. controls (0.83,3.8) and (1.5,2.33) ..
                (1.5,1.5);

\node at (0,0.2) 	{{\Large $\frac{\B_1}{\B_4}$}};
\node at (0.86,-0.7){{\Large $\frac{\B_2}{\B_7}$}};
\node at (-0.86,-0.7){{\Large $\frac{\B_3}{\B_9}$}};
\node at (0,-1.9) 	{{\Large $\frac{\B_5}{\B_{12}}$}};
\node at (1.7,0.3) 	{{\Large $\frac{\B_6}{\B_{11}}$}};
\node at (-1.7,0.3) {{\Large $\frac{\B_8}{\B_{10}}$}};
\node at (3.2,-0.5) {{\Large $\frac{\B_{13}}{\B_{14}}$}};
\node at (-3.2,2.0) {$\cC$};
\node at (-2.5,1.0){{$\cC_1$}};
\node at (2.5,1.0) {{$\cC_2$}};
\node at (1.4,-2.35) 	{{$\cC_3$}};

\end{tikzpicture}
  \end{center}
  \caption{\label{fig:Venn-diagram-support-run}  The Euler diagram of the sets $\cC_1$, $\cC_2$, and $\cC_3$.}
\end{figure}

\begin{proof} For the sake of a more complete picture, we also show the set $\cC$ of all strong bimonoids in Figure~\ref{fig:Venn-diagram-support-run}. In the following, we prove that each of the seven regions $X$ which are shown in the figure is not empty. Together with Observation \ref{lm:12-3} this proves that Figure \ref{fig:Venn-diagram-support-run} is the Euler diagram of $\cC_1$, $\cC_2$, and $\cC_3$. Interleaved,  we prove the existence of $\B_i$ and $\B_j$ with the desired properties. 

  We organize the proof as follows. Since there are four sets involved (viz. $\cC_1,\cC_2,\cC_3$, and $\cC_5$) there exist 16 Boolean combinations of these sets. Each Boolean combination is identified by an \underline{expression} of the form $D \setminus E$ where $D$ is the intersection of some of the sets in  $\{\cC_1,\cC_2,\cC_3,\cC_5\}$ and $E$ is the union of those sets in  $\{\cC_1,\cC_2,\cC_3,\cC_5\}$ which do not occur in $D$. We order the 16 Boolean combinations by increasing number of sets occurring in $E$, starting with $E=\emptyset$.

\underline{$\cC_1 \cap \cC_2 \cap \cC_3 \cap \cC_5 \not= \emptyset$:}  By Observation \ref{lm:12-3}, $\cC_1 \cap \cC_2 \cap \cC_3 \cap \cC_5=\cC_1 \cap \cC_2 \cap \cC_5$. The Boolean semiring $B_1= \Boole = (\mathbb{B},\vee,\wedge,0,1)$ is in $\cC_1 \cap \cC_2 \cap  \cC_5$.
  
\underline{$(\cC_2 \cap \cC_3 \cap \cC_5) \setminus \cC_1 \not= \emptyset$:}  Let $A =\{a,b\}$ be a set. The commutative semiring
\[\B_2= \Powerset_A=(\mathcal{P}(A),\cup,\cap,\emptyset,A)
\]
is not zero-divisor free, because, e.g., $\{a\} \cap \{b\} =\emptyset$ and $\{a\}\not=\emptyset$ and $\{b\}\not=\emptyset$. Hence $\B_2 \not\in \cC_1$. Since $\B_2$ is zero-sum free, finite, and not a ring, we have $\B_2 \in \cC_2 \cap \cC_3 \cap \cC_5$.

\underline{$(\cC_1 \cap \cC_3 \cap \cC_5) \setminus \cC_2 \not= \emptyset$:} We consider the algebra
 \[
\B_3 = (\{0,1,2,3,4\},+',\min,0,4)
   \] 
where $a +' b = \min(a+b,4)$. Obviously $\B_3$ is a commutative strong bimonoid. Since $\B_3$ is also finite and positive, it is in $\cC_1 \cap \cC_3 \cap \cC_5$. $\B_3$ is not distributive, because $\min(2, 2+'2) = 2$ and $\min(2,2) +' \min(2,2) = 2+'2 = 4$. Hence $\B_3 \not\in \cC_2$.

\underline{$(\cC_1 \cap \cC_2 \cap \cC_5) \setminus \cC_3 \not= \emptyset$:} This follows from Observation \ref{lm:12-3}.

\underline{$(\cC_1 \cap \cC_2 \cap \cC_3) \setminus \cC_5 \not= \emptyset$:} By Observation \ref{lm:12-3}, $(\cC_1 \cap \cC_2 \cap \cC_3) \setminus \cC_5=(\cC_1 \cap \cC_2) \setminus \cC_5$. The semiring 
\[\B_4 = \Nat =  (\mathbb{N},+,\cdot,0,1)\]
of natural numbers is positive and commutative, it is not a ring and not bi-locally finite.

\underline{$(\cC_3 \cap \cC_5) \setminus (\cC_1 \cup \cC_2) \not= \emptyset$:}  We consider the algebra
\[\B_5 = (\{0,1,2,3\},+',\cdot_4,0,1)\]
where $a +' b= \min(a+b,3)$ and $\cdot_4$ is multiplication modulo $4$. Obviously, $\B_5$ is a commutative strong bimonoid. Since $\B_5$ is commutative, finite, and zero-sum free, we have $\B_5 \in \cC_3 \cap \cC_5$. Since $2 \cdot_4 2 = 0$, $\B_5 \not\in \cC_1$. The strong bimonoid $\B_5$ is not distributive, because $2 \cdot_4 (3 +' 1) = 2$ and $2 \cdot_4 3 +' 2 \cdot_4 1 = 3$. Thus $\B_5 \not\in \cC_2$.

\underline{$(\cC_2  \cap \cC_5) \setminus (\cC_1 \cup \cC_3) \not= \emptyset$:} The semiring
\[\B_6 = \Powerset_{\{a,b\}} = (\cP(\{a,b\}), \cap, \cup, \{a,b\}, \emptyset)\]
is commutative, not a ring, and finite; hence $\B_6 \in \cC_2 \cap \cC_5$. Moreover, $\B_6$ is not zero-sum free; hence $\B_6 \not\in \cC_1 \cup \cC_3$.

\underline{$(\cC_2 \cap \cC_3) \setminus (\cC_1 \cup \cC_5) \not= \emptyset$:} We consider the commutative semiring
\[
\B_7 = (\mathbb{N} \times \mathbb{N}, + , \times, (0,0), (1,1))
\]
with componentwise addition and multiplication. This is not a ring (hence $\B_7 \in \cC_2$), and it is zero-sum free (hence $\B_7 \in \cC_3$). Moreover, $\B_7$ is not zero-divisor free because, e.g., $(1,0) \times (0,1) = (0,0)$ (hence $\B_7 \not\in \cC_1$) and not bi-locally finite (hence $\B_7 \not\in \cC_5$).

\label{p:B-8}
\underline{$(\cC_1 \cap \cC_5) \setminus (\cC_2 \cup \cC_3) \not= \emptyset$:} By Observation \ref{lm:12-3}, $(\cC_1  \cap \cC_5) \setminus (\cC_2 \cup \cC_3)=(\cC_1\cap \cC_5)\setminus \cC_3$.  We consider the algebra
\[
\B_8 = (\{0,a,b,1\},+,\cdot,0,1)
\]
where $+$ is defined as supremum with respect to the partial order $0 < a < b < 1$; moreover, the operation~$\cdot$ is determined by $x\cdot y = x$ for every $x,y \in \{a,b\}$. Then  $(\{0,a,b,1\},+,0)$ and $(\{0,a,b,1\},\cdot,1)$ are commutative monoids  and $x \cdot 0 = 0$. Hence $\B_8$ is a strong bimonoid.

Moreover, $\B_8$ is zero-sum free and zero-divisor free (hence $\B_8 \in \cC_1$) and $\B_8$ is  finite (hence $\B_8 \in \cC_5$). Since $\cdot$ is not commutative ($a\cdot b \not= b \cdot a$), we have $\B_8 \not\in \cC_3$.

\underline{$(\cC_1 \cap \cC_3) \setminus (\cC_2 \cup \cC_5) \not= \emptyset$:} The tropical bimonoid
\[\B_9 = \TropBM = (\mathbb{N}_\infty,+,\min,0,\infty)\]
is positive and commutative, hence $\B_9 \in \cC_1 \cap \cC_3$. But  $\B_9$ is neither distributive nor bi-locally finite.

\underline{$(\cC_1 \cap \cC_2) \setminus (\cC_3 \cup \cC_5) \not= \emptyset$:} This follows from Observation \ref{lm:12-3}.

\underline{$\cC_1 \setminus (\cC_2 \cup \cC_3 \cup \cC_5) \not= \emptyset$:} By Observation \ref{lm:12-3},
$\cC_1 \setminus (\cC_2 \cup \cC_3 \cup \cC_5) = \cC_1\setminus (\cC_3\cup \cC_5)$.
We consider the formal language semiring
\[\B_{10} = \Lang_\Gamma = (\cP(\Gamma^*),\cup, \circ, \emptyset,\{\varepsilon\})\enspace.\]
This is a positive, non-commutative, and not bi-locally finite  semiring.

\underline{$\cC_2 \setminus (\cC_1 \cup \cC_3 \cup \cC_5) \not= \emptyset$:}  By Observation \ref{lm:12-3},
$\cC_2 \setminus (\cC_1 \cup \cC_3 \cup \cC_5) = \cC_2\setminus (\cC_3\cup \cC_5)$. We consider the commutative semiring
\[
  \B_{11} = (\mathbb{Z}\times \mathbb{N},+,\cdot,(0,0),(1,1))
\]
with pointwise addition and pointwise multiplication. $\B_{11}$ is not a ring, not zero-sum free, and not bi-locally finite.

\underline{$\cC_3 \setminus (\cC_1 \cup \cC_2 \cup \cC_5) \not= \emptyset$:}  We consider the strong bimonoid
\[\B_{12} = (\mathbb{N},+,\cdot_4,0,1)
\]
with \(a \cdot_4 b = (a \cdot b) \mathrm{mod} 4\).
Then $\B_{12}$ is commutative and zero-sum free, i.e., $\B_{12} \in \cC_3$. Since $2 \cdot_4 2 = 0$, $\B_{12}$ contains zero-divisors, and hence $\B_{12} \not\in \cC_1$. $\B_{12}$ is not a semiring, because, e.g., $2 \cdot_4 (3+1) = 0$ and $2 \cdot_4 3 + 2 \cdot_4 1 = 2 + 2 = 4$). Hence $\B_{12} \not\in \cC_2$. Also $\B_{12}$ is not bi-locally finite (because $\langle 1 \rangle_+$ is not finite). Thus $\B_{12} \in C_3 \setminus (\cC_1 \cup  \cC_2 \cup \cC_5)$.

\underline{$\cC_5 \setminus (\cC_1 \cup \cC_2 \cup \cC_3) \not= \emptyset$:}  We consider the ring
\[\B_{13} = \Intfour = (\{0,1,2,3\}, +_4,\cdot_4,0,1)
\]
as defined in Example \ref{ex:semirings}(\ref{def:ring-Zmod4Z}).  Since $\B_{13}$ is finite, $\B_{13} \in \cC_5$. Since $\B_{13}$ is  a ring and is not zero-sum free, $\B_{13} \not\in \cC_1 \cup \cC_2 \cup \cC_3$.

\underline{$\cC\setminus (\cC_1\cup \cC_2 \cup \cC_3 \cup \cC_5)$:} The ring
\[\B_{14} = \Int = (\mathbb{Z},+,\cdot, 0,1)\]
is neither zero-sum free nor bi-locally finite. 
\end{proof}

In particular, it follows from the proof of Theorem \ref{thm:comparison-support-theorem-run}  that each of the support theorems Theorem \ref{thm:recognizable=run=init-positive}, Theorem \ref{thm:support-not-ring}, Theorem \ref{thm:Kirsten}, and Corollary \ref{thm:support-thm-from-preimage}(3), for $\cC_1$, $\cC_2$, $\cC_3$, and $\cC_5$ respectively, has its own benefit with respect to run semantics.
More precisely, let $i \in \{1,2,3,5\}$. Then there exists $\B \in \cC_i \setminus (\cC_j \cup \cC_k \cup \cC_\ell)$ with pairwise different $j,k,\ell \in \{1,2,3,5\} \setminus\{i\}$ such that the support theorem  for $\cC_i$ implies the property $\supp(\runsem{\cA}) \in \Rec(\Sigma)$ for each $(\Sigma,\B)$-wta $\cA$,  and this property does not follow from the support theorems for $\cC_j$, $\cC_k$, and $\cC_\ell$.


Next we compare the sets  $\cC_1$, $\cC_2$, and $\cC_4$.
  
\begin{theorem}\label{thm:comparison-support-theorem-initial-algebra} {\rm \cite{doemoe20,dro22}}
Figure~\ref{fig:Venn-diagram-support-run-2} shows the Euler diagram of the sets $\cC_1$, $\cC_2$, and $\cC_4$. Thus, for every $\cD_1 \in \{\cC_1, \cC\setminus \cC_1\}$, $\cD_2 \in \{\cC_2, \cC\setminus \cC_2\}$, and $\cD_4 \in \{\cC_4, \cC\setminus \cC_4\}$, we have $\cD_1\cap \cD_2\cap \cD_4\ne\emptyset$. 
\end{theorem}
\begin{proof}
\underline{$\cC_1 \cap \cC_2  \cap \cC_4 \not= \emptyset$:} The Boolean semiring
  \[\B_1 = \Boole = (\mathbb{B},\vee,\wedge,0,1)
  \]
  is in $\cC_1 \cap \cC_2 \cap \cC_4$.

\underline{$(\cC_1 \cap \cC_2) \setminus \cC_4 \not= \emptyset$:} The commutative semiring
  \[\B_2 = \Nat = (\mathbb{N},+,\cdot,0,1)
  \]
  is positive and not a ring and not locally finite.

\underline{$(\cC_1 \cap \cC_4) \setminus \cC_2 \not= \emptyset$:}
We consider the strong bimonoid
\[
\B_3 = (\{0,a,b,1\},+,\cdot,0,1)
\]
which is the same as $\B_8$ on page \pageref{p:B-8}, i.e., $+$ is defined as supremum with respect to the partial order $0 < a < b < 1$; moreover, the operation~$\cdot$ is determined by $x\cdot y = x$ for every $x,y \in \{a,b\}$. 

This $\B_3$ is zero-sum free and zero-divisor free (hence $\B_3 \in \cC_1$) and $\B_3$ is  finite (hence $\B_3 \in \cC_4$). Since $\cdot$ is not commutative ($a\cdot b \not= b \cdot a$), we have $\B_3 \not\in \cC_2$.

\underline{$(\cC_2 \cap \cC_4) \setminus \cC_1 \not= \emptyset$:} We consider the algebra
\[ \B_4 = (\{-1,0,1\} \times \{0,1\}, +',\cdot,(0,0),(1,1))\]
where $(a_1,a_2) +' (b_1,b_2) = (   \max(-1,\min(a_1+b_1,1)), \min(a_2+b_2,1))$ (i.e., $+'$ is defined componentwise by using the usual addition except that  the first component is truncated by $1$ from above and by $-1$ from below, and the second component is truncated by $1$ from above). The multiplication is defined componentwise. Obviously, $\B_4$ is a finite commutative semiring.

Since there does not exist an element $(a_1,a_2)$ such that $(0,1) +' (a_1,a_2) = (0,0)$, $\B_4$ is not a ring. Hence $\B_4 \in \cC_2$. $\B_4$ is not zero-sum free, because $(-1,0) +' (1,0) = (0,0)$. Hence $\B_4 \not\in \cC_1$.

\underline{$\cC_1  \setminus (\cC_2 \cup \cC_4) \not= \emptyset$:}  We consider the formal language semiring 
\[\B_5 = \Lang_\Gamma=(\cP(\Gamma^*),\cup, \circ, \emptyset,\{\varepsilon\})\enspace.
\]
 This is a positive, non-commutative, and not locally finite  semiring.

\underline{$\cC_2  \setminus (\cC_1 \cup \cC_4) \not= \emptyset$:} We consider the commutative semiring 
\[\B_6= (\mathbb{Z}\times \mathbb{N},+,\cdot,(0,0),(1,1))
\]
 with pointwise addition and pointwise multiplication. $\B_6$ is not a ring, not zero-sum free, and not locally finite.

\underline{$\cC_4  \setminus (\cC_1 \cup \cC_2) \not= \emptyset$:}  The ring 
\[\B_7= \Intfour = (\{0,1,2,3\}, +_4,\cdot_4,0,1)
\]
as defined in Example \ref{ex:semirings}(\ref{def:ring-Zmod4Z}), is in $\cC_4 \setminus (\cC_1  \cup \cC_2)$.

\underline{$\cC \setminus (\cC_1 \cup \cC_2 \cup \cC_4)$:} The ring 
\[\B_8= \Int = (\mathbb{Z},+,\cdot, 0,1)\]
 is neither zero-sum free nor locally finite. 
\end{proof}

\begin{figure}[t]
  \begin{center}
    \begin{tikzpicture}[scale=0.9]
\draw (-0.9, 0) circle (1.5);
\draw (0.9, 0) circle (1.5);
\draw (0, -1.6) circle (1.5);

\draw (0.0,-0.5) ellipse (4.0 and 3.0);
                
\node at (0,-0.6) 	{$\B_1$};
\node at (0,0.5) 	{$\B_2$};
\node at (0.86,-0.94){$\B_4$};
\node at (-0.86,-0.94){$\B_3$};
\node at (0,-2.3) 	{$\B_7$};
\node at (1.5,0.3) 	{$\B_6$};
\node at (-1.5,0.3) {$\B_5$};
\node at (2.8,-1.5) {$\B_8$};
\node at (-3.5,2.0) {$\cC$};
\node at (-2.5,1.0){{$\cC_1$}};
\node at (2.5,1.0) {{$\cC_2$}};
\node at (1.6,-2.65) 	{{$\cC_4$}};

\end{tikzpicture}
  \end{center}
  \caption{\label{fig:Venn-diagram-support-run-2}  The Euler diagram of the sets $\cC_1$, $\cC_2$, and $\cC_4$.}
\end{figure}

In particular, it follows from the proof of Theorem \ref{thm:comparison-support-theorem-initial-algebra} that each of the support theorems  Theorem \ref{thm:recognizable=run=init-positive}, Theorem \ref{thm:support-not-ring},   and Corollary \ref{thm:support-thm-from-preimage}(2), for $\cC_1$, $\cC_2$, and $\cC_4$ respectively, has its own benefit with respect to initial algebra semantics.
More precisely, let $i \in \{1,2,4\}$. Then there exists $\B \in \cC_i \setminus (\cC_j \cup \cC_k )$ with different $j,k\in \{1,2,4\} \setminus\{i\}$ such that the support theorem  for $\cC_i$ implies the property $\supp(\initialsem{\cA}) \in \Rec(\Sigma)$ for each $(\Sigma,\B)$-wta $\cA$,  and this property does not follow from the support theorems for $\cC_j$ and~$\cC_k$.

\section{Collection of results on tree languages generated by wta}

We can view each $(\Sigma,\B)$-wta $\cA$ as an automaton which accepts the tree languages $\supp(\initialsem{\cA})$, $\supp(\runsem{\cA})$, and $(\initialsem{\cA})^{-1}(b)$ and  $(\initialsem{\cA})^{-1}(b)$ for each $b \in B$. We have proved some results which show sufficient conditions on $\B$ under which some of these tree languages are recognizable, and some conditions under which the tree languages $\supp(\initialsem{\cA})$ and $\supp(\runsem{\cA})$ are equal. Each result requires that the underlying strong bimonoid $\B$ is zero-sum free or even more restricted.

Here we collect these results in a list.
We recall from Lemma \ref{lm:monotonic-properties}(2) and Section \ref{sec:the-main-result} that the following sequence of conditions gives rise to a strictly increasing chain of sets of strong bimonoids with respect to set inclusion.

\begin{center}
  past-finite monotonic \ $\subset$ \   positive \  $\subset$ \  bi-strongly zero-sum-free 
\end{center}

Figure \ref{fig:collection-support} shows the list of these results, where the first three follow the order prescribed by this chain.

\begin{figure}[t]

\hspace*{-5mm}{\small
  \begin{tabular}{lcl}
    strong bimonoid  $\B$ is  ... & & \\[2mm]\hline\\[-1mm]
(1)  past-finite monotonic & $\Longrightarrow$ & for every $(\Sigma,\B)$-wta $\cA$ and $b \in B$: $(\runsem{\cA})^{-1}(b)$ is recognizable \\
  & & (cf. Theorem \ref{thm:preimage-past-finite-monotonic} \\[2mm]\hline\\[-1mm]
(2) positive  &  $\Longrightarrow$ & for every $(\Sigma,\B)$-wta $\cA$ and $x \in \{\mathrm{run}, \mathrm{init}\}$: $\supp(\sem{\cA}^x)$ is recognizable \\
  && (cf. Lemma \ref{thm:run-init-positive})\\[2mm]\hline\\[-1mm]
(3)   bi-strongly zero-sum-free &  $\iff$ & for every branching $\Sigma$ and $(\Sigma,\B)$-wta $\cA$: $\supp(\initialsem{\cA}) = \supp(\runsem{\cA})$ \\
                                  && (by Theorem \ref{thm:bi-strongly-zsf-equiv-equ-supp})\\[2mm]\hline\\[-1mm]
 (4)   zero-sum free, commutative  &  $\Longrightarrow$ & for every $(\Sigma,\B)$-wta $\cA$: $\supp(\runsem{\cA})$ is recognizable \\
  && (cf. Theorem \ref{thm:Kirsten})
\end{tabular}
}
\caption{\label{fig:collection-support} A list of results of tree languages accepted by wta, where in Entries (1)-(3) the sets of strong bimonoids form a strictly increasing chain with respect to set inclusion. }
\end{figure}

%% file: Kleene.tex
\chapter{Rational operations and Kleene's theorem}
\label{ch:Kleene}

In Chapter \ref{ch:closure-properties} we have proved that, in particular, for every commutative semiring $\B$,  the set $\Rec(\Sigma,\B)$ of recognizable $(\Sigma,\B)$-weighted tree languages is closed under the rational operations, i.e.,  sum, tree concatenations, and Kleene stars. 

In Theorem \ref{thm:rec-implies-rat}, we will show that each recognizable weighted tree language can be expressed by polynomial weighted tree language and rational operations (cf. \cite[Thm.~5.2]{dropecvog05}). This generalizes the analysis part of Kleene's result \cite{kle56} (recognizable implies rational), and the corresponding result for tree languages \cite[Thm.~9]{thawri68} (cf. \cite[Thm.~3.43]{eng75-15} and \cite[Thm.~2.5.7]{gecste84}) and weighted string languages (cf. \cite{sch61}). We recall that, in the tree case, extra symbols for tree concatenation were supplied; we will deal with this issue in detail below.

Taking together the closure results from Chapter \ref{ch:closure-properties} and the analysis result from this chapter, we obtain Kleene's result for recognizable weighted tree languages in Theorem  \ref{thm:Kleene} (cf. \cite[Thm.~2.3]{aleboz87} and \cite[Thm.~7.1]{dropecvog05}). It generalizes the corresponding results for the unweighted case (cf. \cite[Thm.~8]{thawri68}, \cite[Thm.~3.43]{eng75-15}, and \cite[Thm.~2.5.8]{gecste84}).

Kleene's result has been extended both, to structures different from finite trees and to algebras different from semirings, e.g.,
formal tree series over additive $\B$-$\Sigma$-algebras for a commutative, $\sigma$-complete semiring~$\B$ \cite[p.28]{boz99} (also cf. \cite[Thm.~6.14]{kui98}),
formal power series in partially commuting variables (traces) and semirings \cite{drogas99},
infinite strings and max-plus semiring \cite{drokus03,drokus06}, strings and lattice-ordered monoids \cite{liped05},
pictures and commutative semirings   \cite{bozgra05,mae05,mae07},
strings and Conway semirings \cite[Sect. 1.3]{esikui},
trees over infinite ranked alphabets and continuous and commutative semirings \cite[Thm. 4.3]{esikui03},
\cite[Thm. 6.4.3]{esikui},
trees and $\sigma$-complete distributive lattices  \cite{esiliu07},
 trees and distributive multioperator monoids (which satisfy some closure properties) \cite{fulmalvog09},
trees and tree-valuation monoids \cite{drofulgoe19},
 and 
weighted tree automata with storage over commutative, $\sigma$-complete semirings \cite{fulvog19,doefulvog24}.
 
In this chapter we recall from \cite{dropecvog05} the proof of the equivalence of recognizable weighted tree languages and rational weighted tree languages for commutative semirings.

\label{p:convention-for-Kleene}
  \begin{quote}
\emph{In the rest of this chapter, we let $\B=(B,\oplus,\otimes,\0,\1)$ be an arbitrary commutative semiring.}
    \end{quote}

    
    \section{Rational weighted tree languages}
    \label{sec:rational-wtl-def}

Here we formally define the concept of rational operations and rational weighted tree languages.

\index{rational operation}
An operation on the set of $(\Sigma,\B)$-weighted tree languages is a \emph{rational operation} if it is the sum, a tree concatenation, or a Kleene star.
\index{Rat@$\Rat(\Sigma,\B)$}
The \emph{set of  rational $(\Sigma,\B)$-weighted tree languages}, denoted by $\Rat(\Sigma,\B)$, is the smallest set of $(\Sigma,\B)$-weighted tree languages which contains each polynomial $(\Sigma,\B)$-weighted tree language and is closed under the rational operations. We call each weighted tree language in $\Rat(\Sigma,\B)$ \emph{rational}.

For our main result, we will have to prove, in  particular, that the semantics $\sem{\cA}$ of a $(\Sigma,\B)$-wta $\cA=(Q,\delta,F)$ is rational. We follow the idea of \cite[Sect.~3]{thawri68} where extra symbols for the tree concatenations are used, and these extra symbols are the states of $\cA$. In order to have such extra nullary symbols available for tree concatenations and \underline{not} to change the type of $\sem{\cA}$ (which is $\T_\Sigma \to B$ and not $\T_{\Sigma\cup Q} \to B$) we use the concept of $0$-extension of ranked alphabet and $\0$-extension of a weighted tree language (cf. \cite[Def.~1]{fulvog19} and \cite[Def.~2]{doefulvog24}).

\index{zero@$0$-extension of $\Sigma$}
Formally, a \emph{$0$-extension of $\Sigma$} is a ranked alphabet $\Theta$ such that $\Sigma \subseteq \Theta$,  
$\rk_\Theta(\sigma)=\rk_\Sigma(\sigma)$ for each $\sigma\in\Sigma$, and $\rk_\Theta(\sigma)=0$ for each $\sigma\in \Theta\setminus \Sigma$. If $\Theta$ is a $0$-extension of $\Sigma$, then we write $\Theta \ge_0 \Sigma$.

\index{r@$r\uh{\Theta}$}
\index{zero@$\0$-extension}
Now let $\Theta \ge_0 \Sigma$ and $r: \T_\Sigma \to B$. The \emph{$\0$-extension of $r$ to $\T_\Theta$}, denoted by $r\uh{\Theta}$, is the weighted tree language $r\uh{\Theta}: \T_\Theta \rightarrow B$ such that
\begin{compactenum}
\item[(a)]  $(r\uh{\Theta})|_{\T_\Sigma} = r$ and
  \item[(b)] $(r\uh{\Theta})(\xi)=\0$ for every $\xi\in \T_\Theta\setminus \T_\Sigma$.
  \end{compactenum}

These concepts have the following transitivity property.

  \begin{observation}\rm \label{obs:transitivity-extension} Let $\Theta$ and $\Delta$ be ranked alphabets such that $\Delta \ge_0 \Theta$ and $\Theta \ge_0 \Sigma$. Then $\Delta \ge_0  \Sigma$. Moreover, let $r: \T_\Sigma \to B$. Then $(r\uh{\Theta})\uh{\Delta} = r \uh{\Delta}$.\hfill $\Box$
    \end{observation}

\index{Ratext@$\Rat(\Sigma,\B)^{\mathrm{ext}}$}
    The \emph{set of extended rational $(\Sigma,\B)$-weighted tree languages}, denoted by $\Rat(\Sigma,\B)^{\mathrm{ext}}$, contains each $(\Sigma,\B)$-weighted tree language $r$ such that $r\uh{\Theta} \in \Rat(\Theta,\B)$ for some $0$-extension $\Theta$ of $\Sigma$. 
  
\sloppy Similarly, the \emph{set of extended recognizable $(\Sigma,\B)$-weighted tree languages}, denoted by $\Rec(\Sigma,\B)^{\mathrm{ext}}$, contains each $(\Sigma,\B)$-weighted tree language $r$ such that  $r\uh{\Theta} \in \Rec(\Theta,\B)$ for some $0$-extension $\Theta$ of $\Sigma$.

In the next two sections, we will prove that $\Rec(\Sigma,\B) \subseteq \Rat(\Sigma,\B)^{\mathrm{ext}}$ (cf. Theorem \ref{thm:rec-implies-rat}) and $\Rat(\Sigma,\B) \subseteq \Rec(\Sigma,\B)$ (cf. Theorem \ref{thm:rat-implies-rec}). These theorems imply the following main result of this chapter.

 \begin{theorem-rect}\label{thm:Kleene} Let $\Sigma$ be a ranked alphabet. Moreover, let $\B$ be a commutative semiring. Then $\Rec(\Sigma,\B)^{\mathrm{ext}} = \Rat(\Sigma,\B)^{\mathrm{ext}}$.
 \end{theorem-rect}
 \begin{proof} 
   \underline{$\Rec(\Sigma,\B)^{\mathrm{ext}} \subseteq  \Rat(\Sigma,\B)^{\mathrm{ext}}$:} Let $r \in \Rec(\Sigma,\B)^{\mathrm{ext}}$. Then $r\uh{\Theta} \in \Rec(\Theta,\B)$ for some $0$-extension $\Theta$ of $\Sigma$.  
   By Theorem \ref{thm:rec-implies-rat} (with $\Sigma=\Theta$), we have that $r\uh{\Theta} \in \Rat^{\mathrm{ext}}(\Theta,\B)$. Hence $(r\uh{\Theta})\uh{\Delta} \in \Rat(\Delta,\B)$ for some $0$-extension $\Delta$ of $\Theta$. Since $\Delta \ge_0 \Theta$ and $\Theta \ge_0 \Sigma$, we obtain by Observation \ref{obs:transitivity-extension} that $(r\uh{\Theta})\uh{\Delta} = r \uh{\Delta}$. Hence  $r\uh{\Delta} \in \Rat(\Delta,\B)$. Since $\Delta \ge_0 \Sigma$, we obtain $r \in \Rat(\Sigma,\B)^{\mathrm{ext}}$.
   
   \underline{$\Rat(\Sigma,\B)^{\mathrm{ext}} \subseteq  \Rec(\Sigma,\B)^{\mathrm{ext}}$:} Let $r \in \Rat(\Sigma,\B)^{\mathrm{ext}}$. Then  $r\uh{\Theta} \in \Rat(\Theta,\B)$ for some $0$-extension $\Theta$ of $\Sigma$.
By Theorem \ref{thm:rat-implies-rec} (with $\Sigma = \Theta$), we have that $r\uh{\Theta} \in \Rec(\Theta,\B)$.  Hence we have $r \in \Rec(\Sigma,\B)^{\mathrm{ext}}$.
\end{proof}

We note that Theorem \ref{thm:Kleene} is similar to  \cite[Thm.~7.1]{dropecvog05}, but slightly different in the following way. Theorem 7.1 of \cite{dropecvog05} says that $\B^{\mathrm{rec}}\fl \T_\Sigma(Q_\infty)\fr = \B^{\mathrm{rat}}\fl \T_\Sigma(Q_\infty)\fr$, where 
\begin{align*}
  \B^{\mathrm{rec}}\fl \T_\Sigma(Q_\infty)\fr \ \ \text{ is defined to be } \ \bigcup (\B^{\mathrm{rec}}\fl \T_\Theta\fr \mid  \text{ranked alphabet $\Theta$ such that $\Theta \ge_0 \Sigma$})  \   \text{ and } \\
 \B^{\mathrm{rat}}\fl \T_\Sigma(Q_\infty)\fr \ \ \text{ is defined to be } \ \bigcup (\B^{\mathrm{rat}}\fl \T_\Theta\fr \mid  \text{ranked alphabet $\Theta$ such that $\Theta \ge_0 \Sigma$})  \   \text{ and } 
    \end{align*}
\index{Brec@$\B^{\mathrm{rec}}\fl \T_\Sigma(Q_\infty)\fr$}
\index{Brat@$\B^{\mathrm{rat}}\fl \T_\Sigma(Q_\infty)\fr$}
\index{Brat@$\B^{\mathrm{rat}}\fl \T_\Theta\fr$}
$\B^{\mathrm{rat}}\fl \T_\Theta\fr$ is the smallest set of $(\Theta,\B)$-weighted tree languages which is closed under scalar multiplication, top-concatenations, and rational operations. Moreover, $\B^{\mathrm{rec}}\fl \T_\Theta\fr$ is just another denotation of the set $\Rec(\Theta,\B)$.
  
However, this setting yields the following type conflict when trying to prove that $\B^{\mathrm{rec}}\fl \T_\Theta\fr \subseteq \B^{\mathrm{rat}}\fl \T_\Sigma(Q_\infty)\fr$ for some $0$-extension $\Theta$ of $\Sigma$. For each $(\Theta,\B)$-wta $\cA =(Q,\delta,F)$, its semantics has the type $\sem{\cA}: \T_\Theta\to B$. In \cite[Thm.~5.2]{dropecvog05} it is proved that there exists an $r \in \B^{\mathrm{rat}}\fl \T_{\Theta \cup Q}\fr$ (where $\Theta \cup Q$ is a $0$-extension of $\Theta$) such that, for each $\xi \in \T_\Theta$, the equality $r(\xi)=\sem{\cA}(\xi)$ holds  and, for each $\xi \in \T_{\Theta \cup Q}\setminus \T_\Theta$, we have $r(\xi)=\0$.
Nevertheless, by the definition of $\B^{\mathrm{rat}}\fl \T_{\Theta \cup Q}\fr$, $r$ has the type $r: \T_{\Theta \cup Q} \to B$ which is different from the type of $\sem{\cA}$. Hence $\sem{\cA}\not = r$ and, in fact, there does not exist an $r \in \B^{\mathrm{rat}}\fl \T_{\Theta \cup Q}\fr$ such that $\sem{\cA} = r$.

We solve this type conflict by the concepts of $0$-extension and $\0$-extension. Such a type conflict does not occur in the proof of the analysis theorem  \cite[Thm.~9]{thawri68} because (unweighted) tree languages are not mappings.

In \cite{dro22} the natural question was posed whether $\Rec(\Sigma,\B) \setminus \Rat(\Sigma,\B) \ne \emptyset$, i.e., whether the extra nullary symbols used in  $\Rat(\Sigma,\B)^{\mathrm{ext}}$ are really necessary (or just comfortable to use). We claim that, for the ranked alphabet $\Sigma = \{\sigma^{(2)}, \sigma'^{(2)}, \alpha^{(0)}\}$, the weighted tree language $\chi_\Boole(L(K,H)) \in \Rec(\Sigma,\Boole) \setminus \Rat(\Sigma,\Boole)$ where $(K,H)$ is the $\Sigma$-local system with $K = \{(\sigma \sigma',\sigma), (\alpha\alpha,\sigma), (\alpha\alpha,\sigma')\}$ and $H=\{\sigma\}$.

Finally, we note that there were investigations to overcome the difference between string concatenation and tree concatenation (cf. the discussion at the beginning of Section \ref{sec:closure-under-tree-concatenation}) by employing forests and forest concatenation \cite{str09,doe19,doe21}. However, since we want to deal with trees, we will not report on these investigations.


 \section{From recognizable to rational}

 In this section we prove Theorem \ref{thm:rec-implies-rat}, i.e., for each $(\Sigma,\B)$-recognizable weighted tree language $r$ there exists a $0$-extension $\Theta$ of $\Sigma$ such that $r\uh{\Theta} \in \Rat(\Theta,\B)$. 

Intuitively, this result shows that the semantics of a wta can be computed in a dynamic programming style. This computation is organized in the same way as the computation of the transitive closure of the edge relation of graphs in \cite{war62}, of the all-pairs shortest-path problem in \cite{flo62}, and of the algebraic path problem for idempotent $\sigma$-complete  semirings \cite[Alg.~5.5]{ahohopull74}. We follow the lines in \cite{dropecvog05} (which was crucially inspired by \cite{eng03}).

As preparation, we extend the concept of run for trees which may contain states of a wta. If for a position $w$ of a tree $\xi$ the symbol $\xi(w)$ is a state, then the extended run assigns the state $\xi(w)$ to that position. Formally, let $\cA=(Q,\delta,F)$ be a $(\Sigma,\B)$-wta and $\xi \in \T_\Sigma(Q)$. A \emph{run of $\cA$ on $\xi$} is a mapping $\rho: \pos(\xi) \to Q$ such that $\rho(w) = \xi(w)$ for each $w \in \pos_Q(\xi)$.
   In the same way as for runs on trees in $\T_\Sigma$, we define the restriction $\rho|_i$ for a run of $\cA$ on $\xi \in \T_\Sigma(Q)$ and $i \in [\rk(\xi(\varepsilon))]$. Moreover, we denote the set of all runs of $\cA$ on $\xi$ by $\R_\cA(\xi)$.

Then we define, for each $\xi \in \T_\Sigma(Q)$, the weight of each run in $\R_\cA(\xi)$. For this we extend the definition of $\wt_\cA: \mathrm{TR} \to B$ given in \eqref{equ:weight-of-run}.  Formally, we let
\(
\mathrm{TR}_Q = \{(\xi,\rho) \mid \xi \in \T_\Sigma(Q), \rho \in \R_\cA(\xi)\}
\)
and we define the binary relation $\succ$ on $\mathrm{TR}_Q$ as follows:
\index{succb@$\succ$}
\[
  \text{for every $(\xi,\rho) \in \mathrm{TR}_Q$ and $i \in [\rk(\xi(\varepsilon)]$, we let $(\xi,\rho) \succ (\xi|_i,\rho|_i)$} \enspace.
\]
  By Corollaries \ref{cor:reduction-to-substring-is-terminating} and \ref{cor:termination-propagates-to-cartesian-products}, the relation $\succ$ is terminating. Moreover, we have  
  \[
    \nf_\succ(\mathrm{TR}_Q) = \{(\alpha,\rho) \mid \alpha \in \Sigma^{(0)}, \rho: \{\varepsilon\} \to Q\} \ \cup \
    \{(q,\rho) \mid q \in Q, \rho = \{(\varepsilon,q)\}\}\enspace.
    \]
  We define the mapping
  \[
    \wt_\cA': \mathrm{TR}_Q \to B
  \]
  by induction on $(\mathrm{TR}_Q,\succ)$ as follows. Let $(\xi,\rho)  \in \mathrm{TR}_Q$. 

  I.B.: If $\xi \in Q$ (and hence $\rho = \{(\varepsilon,\xi)\}$), then we  let $\wt_\cA'(\xi,\rho) = \1$.

  I.S.: If $\xi = \sigma(\xi_1,\ldots,\xi_k)$, then  we let
  \index{wt@$\wt_\cA(\xi,\rho)$}
\begin{equation}\label{equ:weight-of-run-Q}
\wt_\cA'(\xi,\rho) = \Big( \bigotimes_{i\in [k]} \wt_\cA'(\xi|_i,\rho|_i)\Big) \otimes \delta_k\big(\rho(1) \cdots \rho(k),\sigma,\rho(\varepsilon)\big) \enspace,
\end{equation}
where $k$ and $\sigma$ abbreviate $\rk(\xi(\varepsilon))$ and $\xi(\varepsilon)$, respectively. 
We call $\wt_\cA'(\xi,\rho)$ the \emph{weight of $\rho$ (by $\cA$ on $\xi$)}.
If there is no confusion, then we drop the index $\cA$ from $\wt_\cA'$ and write just  $\wt'(\xi,\rho)$ for the weight of $\rho$. Since, for each $\xi \in \T_\Sigma$, we have  $\wt'(\xi,\rho) = \wt(\xi,\rho)$, we drop the prime from $\wt'$ and simply write $\wt$.

Next we define restrictions on (generalized) runs in the way that we restrict the set of states which may occur at positions different from the root and different from $Q$-labeled positions. Formally,  let $q \in Q$, $P \subseteq Q$, and $\xi \in \T_\Sigma(Q)$. A \emph{$q$-run of $\cA$ on $\xi$ using $P$} is a run $\rho: \pos(\xi) \to Q$ such that
 \begin{compactitem}
   \item $\rho(\varepsilon) = q$, 
 \item $\rho(w) \in P$ for each $w \in \pos(\xi) \setminus (\{\varepsilon\}\cup \pos_Q(\xi))$.
 \end{compactitem}
       The set of all $q$-runs of $\cA$ on $\xi$ using $P$ is denoted by $\R_\cA^P(q,\xi)$. We denote the set $\bigcup_{q \in \Q} \R_\cA^P(q,\xi)$ by $\R_\cA^P(\xi)$.

  
Next we recall a particular weighted tree language \cite{eng03,dropecvog05} with a slight modification. We consider the $0$-extension $\Sigma \cup Q$ of $\Sigma$, i.e., $(\Sigma \cup Q)^{(0)} = \Sigma^{(0)} \cup Q$. For every $P \subseteq Q$ and $q \in Q$, we define the weighted tree language $S_\cA(P,q): \T_\Sigma(Q) \to B$ for each $\xi \in \T_\Sigma(Q)$ by
\index{SAPq@$S_\cA(P,q)(\xi)$}
\[
  S_\cA(P,q)(\xi) =
  \begin{cases}
    \bigoplus_{\rho \in \R_\cA^P(q,\xi)} \wt(\xi,\rho) & \text{ if $\xi \in \T_\Sigma(Q) \setminus Q$}\\
      \0 & \text{ otherwise\enspace.}
    \end{cases}
  \]
Then, for each $q' \in Q$, the weighted tree language $S_\cA(P,q)$ is $q'$-proper. We recall that in \cite{eng03,dropecvog05} the language $S_\cA(Q',P,q)$ was defined where $Q' \subseteq Q$. But since we will only use $S_\cA(Q,P,q)$, we refrain from the first parameter and keep it fixed with value $Q$. 


Now we prove that each $S_\cA(P,q)$ is a rational weighted tree language. For this, we will have to decompose runs of $\cA$ at particular positions. 
Formally, let $P \subseteq Q$, $q \in Q$, and $p \in Q \setminus P$. Let $\xi \in \T_\Sigma(Q) \setminus Q$. We define the mapping
\[
  \varphi: \R_\cA^{P \cup \{p\}}(q,\xi) \to \U^{P \cup \{p\}}(q,\xi) 
  \]
  where
  \begin{align*}
   \U^{P \cup \{p\}}(q,\xi) =  \{(\widetilde{v},\rho',\rho_1,\ldots,\rho_n) \mid \ &
                               n \in \mathbb{N}, \widetilde{v}=(v_1,\ldots,v_n) \text{ in $\cut_p(\xi) \setminus \{(\varepsilon)\}$}, \rho' \in \R_\cA^P(q,\xi[p]_{\widetilde{v}}),\\
    & \text{and } \xi|_{v_i} \not\in Q\setminus \{p\} \text{ and } \rho_i \in \R_\cA^{P\cup \{p\}}(p,\xi|_{v_i}) \text{ for each $i \in [n]$}\}
  \end{align*}
    and for every $\rho \in \R_\cA^{P \cup \{p\}}(q,\xi)$  we define
  \[\varphi(\rho) = ((v_1,\ldots,v_n),\rho',\rho_1,\ldots,\rho_n)\]
  such that the following conditions hold (cf. Figure \ref{fig:decomposition-run-Kleene}):
\begin{compactenum}
\item $\{w \in \rho^{-1}(p) \setminus \{\varepsilon\} \mid (\forall v \in \prefix(w) \setminus \{w\}): \rho(v) \ne p \}  = \{v_1,\ldots,v_n\}$,
\item $\rho': \pos(\xi[p]_{\widetilde{v}}) \to Q$ is such that $\rho' = \rho|_{\pos(\xi[p]_{\widetilde{v}})}$, and
  \item for each $i \in [n]$, we have $\rho_i = \rho|_{v_i}$.
  \end{compactenum}
  By the first condition we have $\xi|_{v_i} \not\in Q\setminus \{p\}$, because otherwise $\rho(v_i) \in Q\setminus \{p\}$, and hence $v_i \not\in \rho^{-1}(p)$.

  \begin{figure}
    \centering
\begin{tikzpicture}[scale=1, every node/.style={transform shape},
					node distance=0.05cm and 0.05cm,
					mycircle/.style={draw, circle, inner sep=0mm, minimum height=5.5mm},
					mydashed/.style={dash pattern=on 2mm off 1mm, thin},
					mydashedarrow/.style={mydashed,->, shorten >=0.1cm,shorten <=0.1cm}],

\begin{scope}[level 1/.style={sibling distance=50mm},
			  level 2/.style={sibling distance=20mm},
			  level 3/.style={sibling distance=18mm},
			  level 4/.style={sibling distance=16mm}]
			  
  \node (n0) at (0,0) {$\sigma$}
    child {node (n1) {$\sigma$}
      child {node[inner sep=2.5mm] (q11) {\phantom{$\sigma$}}} 
      child {node[inner sep=2.5mm] (n12) {$\sigma$}
        child {node (n121) {$q_2$}}
        child {node (n122) {$\sigma$}
          child {node (n1221) {$\beta$}}
          child {node (n1222) {$p$}} }}}        
    child {node[inner sep=2.5mm] (n2) {$\sigma$}
      child {node (n21) {$\alpha$}}
      child {node (n22) {$\beta$}} };
  \node[inner sep=2.5mm] at (q11) (n11) {$p$};
  
  \node [mycircle, right = 0.3cm of n0] (cn0) {$q$};
  \node [mycircle, left = of n1] (cn1) {$q_1$};
  \node [mycircle, left = of n11] (cn11) {$p$};
  \node [mycircle, right = of n12] (cn12) {$p$};
  \node [mycircle, left = of n121] {$q_2$};
  \node [mycircle, right = of n122] (cn122) {$q_1$};
  \node [mycircle, left = of n1221] (cn1221) {$q_1$};
  \node [mycircle, right = of n1222] {$p$};
  \node [mycircle, right = of n2] (cn2) {$p$};
  \node [mycircle, left = of n21] {$q_1$};
  \node [mycircle, right = of n22] (cn22) {$q_1$};
  
  \node[inner sep=2.5mm] at (cn11) (p11) {\phantom{$\sigma$}};
  \node[inner sep=2.5mm] at (cn12) (p12) {\phantom{$\sigma$}};
  \node[inner sep=2.5mm] at (cn2) (p2) {\phantom{$\sigma$}};
  \node[fit=(q11)(p11),rectangle,draw,inner sep=0mm] (r11) {};
  \node[fit=(n12)(p12),rectangle,draw,inner sep=0mm] (r12) {};
  \node[fit=(n2)(p2),rectangle,draw,inner sep=0mm] (r2) {};
  
  \node[left= 0.4cm of n0, yshift=0.55cm] {$\xi :$};
  \node[right= 0.4cm of cn0, yshift=0.55cm] {$\rho \in R_{\cA}^{P\cup \{p\}}(q,\xi)$};
  
  \begin{scope}[every path/.style= mydashed]
    \draw ($(r11.west)+(-1,0)$) -- (r11.west);
    \draw (r11.east) -- (r12.west);
    \draw (r12.east) to[out=0, in=180] ($(r2.west)+(-0.5,0)$) -- (r2.west);
    \draw (r2.east) -- ($(r2.east)+(2,0)$) node[above] {cut$_p$};
  \end{scope}
  
  \begin{scope}[node distance=0.1cm and 0.1cm,
                every path/.style={dashed, opacity=0.5}]
    \node[above= of cn2, xshift=0.3cm, yshift=0.2cm] (0) {};
    \node[below= of cn22, xshift=0.2cm, yshift=-0.3cm] (1) {};
    \node[below right= of cn122] (2) {};
    \node[below right= of n1221] (3) {};
    \node[above left= of cn1221, xshift=-0.1cm, yshift=0.1cm] (4){};
    \node[] at ($(n11)!0.5!(n12)$) (5) {};
    \node[below= of cn1, xshift=-0.2cm, yshift=-0.2cm] (6) {};
    \node[above= of cn1, xshift=-0.2cm, yshift=0.2cm] (7) {};
    \draw (6.center)
      to[bend right=270, looseness=1.75] (7.center)
      to[in=180, out=0, looseness=0.25] (0.center)
      to[in=15, out=0, looseness=1.25] (1.center) 
      to (2.center)
      to[in=50.5, out=-165, looseness=0.75] (3.center)
      to[in=232.5, out=232.5, looseness=1.75] (4.center)
      to[in=310, out=50.5, looseness=1.4] (5.center)
      to[in=0, out=130, looseness=0.6] (6.center);
  \end{scope}
\end{scope}

\begin{scope}[yshift=-9.5cm,
              level 1/.style={sibling distance=33mm},
			  level 2/.style={sibling distance=17mm},
			  level distance= 1.3cm] 
  \node[level 1/.style={sibling distance=42mm}] (a0) {$\sigma$}
    child {node (a1) {$\sigma$}
      child {node (a11) {$p$}}
      child {node (a12) {$p$}} }
    child {node (a2) {$p$}};
  \node[mycircle, right= of a0, xshift=0.1cm] (ca0) {$q$};
  \node[mycircle, left= of a1] {$q_1$};
  \node[mycircle, left= of a11] {$p$};
  \node[mycircle, right= of a12] {$p$};
  \node[mycircle, right= of a2] {$p$};
  \node[left= 0.3 of a0] {$\xi [p]_{\tilde{v}}:$};
  \node[right= 0.4 of ca0] {$\rho'$};
\end{scope}

\begin{scope}[level 1/.style={sibling distance=15mm},
			  level 2/.style={sibling distance=12mm},
			  level distance=1.2cm]
  
  \node[below left = 1.5cm and 1cm of a11] (b0) {$p$};
  \node[mycircle, right= of b0] (cb0) {$p$};
  \draw[mydashedarrow] (b0) -- (a11);
  \node[left= 0.2 of b0] {$\xi \vert _{11} :$};
  \node[right= 0.3 of cb0] {$\rho_1$};
  
  \node[right = 4cm of b0] (c0) {$\sigma$}
    child {node (c1) {$q_2$}}
    child {node (c2) {$\sigma$} 
      child {node (c21) {$\beta$}}
      child {node (c22) {$p$}} };
  \node[mycircle, right= of c0] (cc0) {$p$};
  \node[mycircle, left= of c1] {$q_2$};
  \node[mycircle, right= of c2] {$q_1$};
  \node[mycircle, left= of c21] {$q_1$};
  \node[mycircle, right= of c22] {$p$};
  \draw[mydashedarrow] (c0) -- (a12);
  \node[left= 0.2 of c0] {$\xi \vert _{12} :$};
  \node[right= 0.3 of cc0] {$\rho_2$};
    
  \node[right = 4cm of c0] (d0) {$\sigma$}
    child {node (d1) {$\alpha$}}
    child {node (d2) {$\beta$}}; 
  \node[mycircle, right= of d0](cd0) {$p$};
  \node[mycircle, left= of d1] {$q_1$};
  \node[mycircle, right= of d2] {$q_1$};
  \draw[mydashedarrow] (d0) -- (a2.300);
  \node[left= 0.2 of d0] {$\xi \vert _2 :$};
  \node[right= 0.3 of cd0] {$\rho_3$};
\end{scope}

\node[anchor=west] (W) at (2.8,-4.7) {$W=\pos(\xi)\setminus (\{\varepsilon\} \cup \pos_Q(\xi))$};
\node[below= 0.5cm of W.west, anchor=west] {$\rho(W) \subseteq P\cup\{p\}$};
\draw[|->] (0,-7) to node[midway,right] (phi) {$\varphi$} (0,-8.5) ;
\node[anchor=west] at (W.west |- phi) {$\varphi(\rho) = (\tilde{v} = (11,12,2),\rho',\rho_1,\rho_2,\rho_3)$};

\end{tikzpicture}
    \caption{\label{fig:decomposition-run-Kleene} An example of $\rho \in \R_\cA^{P \cup \{p\}}(q,\xi)$ and $\varphi(\rho)$ with $Q = \{p,q_1,q_2,q\}$, $P=\{q_1\}$.} 
    \end{figure}

  Obviously, $\varphi$ is bijective and, if $\varphi(\rho) = ((v_1\ldots,v_n),\rho',\rho_1,\ldots,\rho_n)$, then 
  \begin{equation}
\wt(\xi,\rho) = \wt(\xi[p]_{\widetilde{v}},\rho') \otimes \bigotimes_{i\in [n]} \wt(\xi|_{v_i},\rho_i)  \enspace. \label{equ:decomposition-of-runs} 
    \end{equation}
We note that \eqref{equ:decomposition-of-runs} uses \eqref{equ:weight-of-run-Q},  Observation \ref{obs:weight-run-explicit}, and the assumption that $\B$ is commutative.


  The next lemma shows how the weighted tree languages in the family
  \[
    (S_\cA(P,q) \mid P \subseteq Q, q \in Q)
  \]
  can be computed in a dynamic programming style (cf. Figure \ref{fig:Kleene-dyn-prog})).

  \begin{figure}
    \centering
\begin{tikzpicture}[level distance=2.75em,
  every node/.style = {align=center}]
  \pgfdeclarelayer{bg}    
  \pgfsetlayers{bg,main}  

  \tikzstyle{mycircle}=[draw, circle, inner sep=-2mm, minimum height=5mm]

\draw   (7,7) coordinate (a) --
		(0,0) coordinate (c) --
		(14,0) coordinate(b) -- cycle;
		
\node at (5,7) {$\xi\in \T_{\Sigma \cup Q} \backslash Q:$};
\node at (9.75,7) {$S(P \cup \{p\},q)$};

\node[mycircle] at (7.5,7) {$q$};
\node at (6,4.65) {\small $S(P,p)$};
		
\def\a{2}
\def\b{5.5}
\def\c{6.25}
\def\d{8.125}
\def\e{8.625}
		
\begin{scope}[xshift=\a cm]
\node[mycircle] (p1) at (1.5,2.5) {$p$};
\node[yshift=-1.2cm] at (p1) {\small $S(P,p)$};
\draw[dashed] (0,0) -- (p1) -- (3,0);
\end{scope}
		
\begin{scope}[xshift=\b cm]
\node[mycircle] (p2) at (2.5,4.7) {$p$};
\node[yshift=-1cm, xshift=0.12cm] at (p2) {\small $S(P,p)$};
\draw[dashed] (0,0) -- (p2) -- (6.5,0);
\end{scope}
		
\begin{scope}[xshift=\c cm]
\node[mycircle] (p3) at (1,2) {$p$};
\node[yshift=-1.5cm,xshift=-0.18cm] at (p3) {\small $S(P,p)$};
\draw[dashed] (0,0) -- (p3) -- (1.5,0);
\end{scope}
		
\begin{scope}[xshift=\d cm]
\node[mycircle] (p4) at (1.4,2) {$p$};
\node[yshift=-0.85cm,xshift=0.05cm] at (p4) {\small $S(P,p)$};
\draw[dashed] (0,0) -- (p4) -- (3,0);
\end{scope}
		
\begin{scope}[xshift=\e cm]
\node[mycircle] (p5) at (0.9,0.7) {$p$};
\node[yshift=-0.5cm,xshift=0.05cm] at (p5) {\small $S(P,p)$};
\draw[dashed] (0,0) -- (p5) -- (2,0);
\end{scope}			
 
\end{tikzpicture}
    \caption{\label{fig:Kleene-dyn-prog} An illustration of the equality $S_\cA(P\cup \{p\},q) = S_\cA(P,q) \circ_p S_\cA(P,p)_p^* $, where $S_\cA$ is abbreviated by $S$.}
    \end{figure}

  \begin{lemma}\rm \label{lm:dyn-prog} (cf. \cite[Lm.~12]{eng03} and \cite[Lm.~5.1]{dropecvog05}) For every $P \subseteq Q$, $p \in Q\setminus P$, and $q \in Q$, we have
    \[
S_\cA(P\cup \{p\},q) = S_\cA(P,q) \circ_p S_\cA(P,p)_p^* \enspace.
      \]
    \end{lemma}
    \begin{proof} 
Let $P \subseteq Q$ and $p \in Q\setminus P$.  By  induction on $\T_{\Sigma}(Q)$, we prove that the following statement holds: 
      \begin{equation*}
        \text{For every $\xi \in \T_{\Sigma}(Q)$ and $q \in Q$, we have } S_\cA(P\cup \{p\},q)(\xi) = \Big(S_\cA(P,q) \circ_p S_\cA(P,p)_p^*\Big)(\xi)  \enspace.
      \end{equation*}
      
  I.B.:    Let $\xi \in Q$. Then $S_\cA(P\cup \{p\},q)(\xi) = \0 =  (S_\cA(P,q) \circ_p S_\cA(P,p)_p^*)(\xi)$, because $S_\cA(P,q)$ is $q'$-proper for each $q'\in Q$. The case that $\xi \in \Sigma^{(0)}$ is covered in the I.S.
       
  I.S.:    Let $\xi = \sigma(\xi_1,\ldots,\xi_k)$. For each $\widetilde{v} = (v_1,\ldots,v_n)$ in $\cut_p(\xi)$,  we denote $v_i$ also by $\widetilde{v}_i$.
      \begingroup
      \allowdisplaybreaks
      \begin{align*}
        & S_\cA(P\cup \{p\},q)(\xi)\\
        = & \bigoplus_{\rho \in \R_\cA^{P\cup \{p\}}(q,\xi)} \wt(\xi,\rho)
        \tag{\text{by definition and the fact that $\xi \not\in Q$}}\\
        = & \bigoplus_{(\widetilde{v},\rho',\rho_1,\ldots,\rho_n) \in \U^{P \cup \{p\}}(q,\xi)}
            \wt(\xi[p]_{\widetilde{v}} ,\rho')
            \otimes \bigotimes_{i\in [n]} \wt(\xi|_{\widetilde{v}_i},\rho_i)
        \tag{\text{because $\varphi$ is a weight preserving bijection}}\\[3mm]
         =& \bigoplus\limits_{\substack{\widetilde{v}=(v_1,\ldots,v_n) \in \cut_p(\xi)\setminus \{(\varepsilon)\}:\\
         \ \xi[p]_{\widetilde{v}} \not\in Q\\
        (\forall i \in [n]): \xi|_{v_i} \not\in Q \setminus \{p\}}} \ \
        \bigoplus_{\rho' \in \R_\cA^{P}(q,\xi[p]_{\widetilde{v}})}
\bigoplus_{\rho_1 \in \R_\cA^{P\cup \{p\}}(p,\xi|_{v_1})} \!\!
        \cdots \!\!
        \bigoplus_{\rho_n \in \R_\cA^{P\cup \{p\}}(p,\xi|_{v_n})} \wt(\xi[p]_{\widetilde{v}} ,\rho')
          \otimes \bigotimes_{i\in [n]} \wt(\xi|_{v_i},\rho_i)
          \tag{by definition of $\U^{P \cup \{p\}}(q,\xi)$}
        \\[4mm]
         =& \bigoplus\limits_{\substack{\widetilde{v}=(v_1,\ldots,v_n) \in \cut_p(\xi)\setminus \{(\varepsilon)\}:\\
        \xi[p]_{\widetilde{v}} \not\in Q\\
        (\forall i \in [n]): \xi|_{v_i} \not\in Q \setminus \{p\}}}
         \Big(\bigoplus_{\rho' \in \R_\cA^{P}(q,\xi[p]_{\widetilde{v}})} \wt(\xi[p]_{\widetilde{v}} ,\rho')\Big)
        \otimes
           \bigotimes_{i\in [n]} \Big(\bigoplus_{\rho_i \in \R_\cA^{P\cup \{p\}}(p,\xi|_{v_i})}  \wt(\xi|_{v_i},\rho_i)\Big)
           \tag{\text{by distributivity}}
        \\[4mm]
        =& \bigoplus\limits_{\substack{\widetilde{v}=(v_1,\ldots,v_n) \in \cut_p(\xi)\setminus \{(\varepsilon)\}:\\
        \xi[p]_{\widetilde{v}} \not\in Q,\\
        (\forall i \in [n]): \xi|_{v_i} \not\in Q \setminus \{p\}}}
           S_\cA(P,q)(\xi[p]_{\widetilde{v}}) \otimes
        \bigotimes_{i\in [n]} \big((S_\cA(P\cup \{p\},p) \oplus \1.p\big)(\xi|_{v_i})
      \end{align*}
      \endgroup
            In the previous equality we have used the following argument (for the second factor up to the $(n+1)$-st factor). 
      If, for some $i \in [n]$, we have $\xi|_{v_i}=p$, then $\R_\cA^{P\cup \{p\}}(p,\xi|_{v_i})$ contains exactly one run, say $\rho$, and $\rho(\varepsilon)=p$. Then $\bigoplus_{\rho_i \in \R_\cA^{P\cup \{p\}}(p,\xi|_{v_i})}  \wt(\xi|_{v_i},\rho_i) =\1$. On the other hand, then we have $S_\cA(P\cup \{p\},p)(\xi|_{v_i}) =\0$. Thus, we have to add $\1.p$.

      Then we can continue as follows:
            \begingroup
      \allowdisplaybreaks
      \begin{align*}
        & \bigoplus\limits_{\substack{(v_1,\ldots,v_n) \in \cut_p(\xi)\setminus \{(\varepsilon)\}:\\
        \xi[p]_{(v_1,\ldots,v_n)} \not\in Q,\\
        (\forall i \in [n]): \xi|_{v_i} \not\in Q \setminus \{p\}}}
           S_\cA(P,q)(\xi[p]_{(v_1,\ldots,v_n)}) \otimes
        \bigotimes_{i\in [n]} \big((S_\cA(P\cup \{p\},p) \oplus \1.p\big)(\xi|_{v_i})\\
        =& \bigoplus\limits_{\substack{(v_1,\ldots,v_n) \in \cut_p(\xi)\setminus \{(\varepsilon)\}:\\\xi[p]_{(v_1,\ldots,v_n)} \not\in Q,\\
        (\forall i \in [n]): \xi|_{v_i} \not\in Q \setminus \{p\}}} 
           S_\cA(P,q)(\xi[p]_{(v_1,\ldots,v_n)}) \otimes
        \bigotimes_{i\in [n]} \big((S_\cA(P,p) \circ_p S_\cA(P,p)_p^*) \oplus \1.p\big)(\xi|_{v_i})
      \end{align*}
      \endgroup
         The previous equality uses, for each $i \in [n]$, the following facts. If $\xi|_{v_i} \not\in Q$, then we have used the I.H. with $q=p$.  This is possible because $\xi|_{v_i}$ is a subtree of $\xi$ due to the condition $\xi[p]_{(v_1,\ldots,v_n)} \ne p$. If $\xi|_{v_i}=p$, then we have used that
      \[\big((S_\cA(P\cup \{p\},p) \oplus \1.p\big)(\xi|_{v_i})
= \1 =  \big((S_\cA(P,p) \circ_p S_\cA(P,p)_p^*) \oplus \1.p\big)(\xi|_{v_i})\enspace.
\]

  Then we can continue as follows:
         \begingroup
      \allowdisplaybreaks
      \begin{align*}
        & \bigoplus\limits_{\substack{(v_1,\ldots,v_n) \in \cut_p(\xi)\setminus \{(\varepsilon)\}:\\\xi[p]_{(v_1,\ldots,v_n)} \not\in Q,\\
        (\forall i \in [n]): \xi|_{v_i} \not\in Q \setminus \{p\}}} 
           S_\cA(P,q)(\xi[p]_{(v_1,\ldots,v_n)}) \otimes
        \bigotimes_{i\in [n]} \big((S_\cA(P,p) \circ_p S_\cA(P,p)_p^*) \oplus \1.p\big)(\xi|_{v_i})\\
        =& \bigoplus\limits_{\substack{(v_1,\ldots,v_n) \in \cut_p(\xi)\setminus \{(\varepsilon)\}:\\\xi[p]_{(v_1,\ldots,v_n)} \not\in Q,\\
        (\forall i \in [n]): \xi|_{v_i} \not\in Q \setminus \{p\}}} 
           S_\cA(P,q)(\xi[p]_{(v_1,\ldots,v_n)}) \otimes
           \bigotimes_{i\in [n]} S_\cA(P,p)_p^*(\xi|_{v_i})
        \tag{\text{by Lemma \ref{lm:equation}}}\\
        =& \bigoplus\limits_{(v_1,\ldots,v_n) \in \cut_p(\xi)} 
           S_\cA(P,q)(\xi[p]_{(v_1,\ldots,v_n)}) \otimes
           \bigotimes_{i\in [n]} S_\cA(P,p)_p^*(\xi|_{v_i})
        \tag{\text{by definition of $S_\cA(P,q)$ and because $S_\cA(P,p)_p^*(q)=\0$ for each $q \in Q \setminus\{p\}$}}\\
        =& \Big(S_\cA(P,q) \circ_p S_\cA(P,p)_p^*\Big)(\xi)
           \tag{\text{by definition of $\circ_p$}}
        \end{align*}
      \endgroup
      \end{proof}

    With the help of Lemma \ref{lm:dyn-prog} we can show that weighted languages of the form $S_\cA(P,q)$ are rational. 

    \begin{lemma}\rm \label{lm:S-rat} For every $P \subseteq Q$ and $q \in Q$ we have $S_\cA(P,q) \in \Rat(\Sigma \cup Q,\B)$.
    \end{lemma}
    \begin{proof} We prove the statement by induction on $(\mathcal{P}(Q),\succ)$, where for every $P,P' \in \cP(Q)$  we define $P\succ P'$ if there exists a $p\in Q\setminus P'$ such that $P=P'\cup\{p\}$.
            Since the cardinality of sets is a monotone embedding from $(\mathcal{P}(Q),\succ)$ into the terminating reduction system $(\mathbb{N},>)$, Lemma \ref{lm:fin-branching-embedding-termination} implies that  $\succ$ is terminating. Moreover, we have that $\nf_\succ(\cP(Q))=\{\emptyset\}$.
      
   I.B.: Let $P= \emptyset$. For every $k \in \mathbb{N}$, $\sigma \in \Sigma^{(k)}$, $q,q_1,\ldots,q_k \in Q$, we define the run $\rho_{q_1\cdots q_k,q}^\sigma: \pos(\sigma(q_1,\ldots,q_k)) \to Q$ by $\rho_{q_1\cdots q_k,q}^\sigma(\varepsilon) = q$ and $\rho_{q_1\cdots q_k,q}^\sigma(i)=q_i$ for each $i \in [k]$. Then for each $\xi \in \T_\Sigma(Q)$ we have
      \[
        \R_\cA^\emptyset(q,\xi) =
        \begin{cases}
          \{\rho_{q_1\cdots q_k,q}^\sigma\} & \text{ if $\xi=\sigma(q_1,\ldots,q_k)$}\\
          \emptyset & \text{ otherwise}\enspace.
          \end{cases}
        \]
        Let $\Sigma(Q) = \{\sigma(q_1,\ldots,q_k) \mid k \in \mathbb{N}, \sigma \in \Sigma^{(k)},  q_1,\ldots,q_k \in Q\}$. Then
        \[
          S_\cA(\emptyset,q) = \bigoplus_{\sigma(q_1,\ldots,q_k) \in \Sigma(Q)} \delta_k(q_1 \cdots q_k,\sigma,q). \sigma(q_1,\ldots,q_k)\enspace.
        \]
        Since $S_\cA(\emptyset,q)$ is polynomial, we have  that $S_\cA(\emptyset,q)$ is rational.
        
   I.S.: Let  $P = P'\cup\{p\}$ for some $p \in Q\setminus P'$. 
        For the induction step, we assume that $S_\cA(P',q)$ is rational for each $q \in Q$. By Lemma \ref{lm:dyn-prog}, we have $S_\cA(P'\cup \{p\},q) = S_\cA(P',q) \circ_p S_\cA(P',p)_p^*$. Thus, by I.H. and the definition of rational weighted tree languages, we obtain that $S_\cA(P'\cup \{p\},q)$ is rational.
      \end{proof}

 \begin{theorem} \label{thm:rec-implies-rat} {\rm (cf. \cite[Thm.~5.2]{dropecvog05})} $\Rec(\Sigma,\B) \subseteq \Rat(\Sigma, \B)^{\mathrm{ext}}$.
 \end{theorem}
 \begin{proof} Let $\cA = (Q,\delta,F)$ be a $(\Sigma,\B)$-wta with $Q = \{q_1,\ldots,q_n\}$. By Theorem \ref{thm:root-weight-normalization-run} we can assume that $\cA$ is root weight normalized. Let $q_f \in Q$ such that $\supp(F) = \{q_f\}$ and $F(q_f)=\1$.

   We will prove that $\sem{\cA} \in \Rat(\Sigma,\B)^{\mathrm{ext}}$. 
   For this we want to employ Lemma \ref{lm:S-rat} (for $P=Q$ and $q= q_f$). We observe that $S_\cA(Q,q_f)(\xi)=\sem{\cA}(\xi)$ for each $\xi\in \T_\Sigma$. Since $\supp(S_\cA(Q,q_f))$ may contain trees in $\T_\Sigma(Q)\setminus \T_\Sigma$, the weighted tree language $S_\cA(Q,q_f)$ may not be the $\0$-extension of $\sem{\cA}$ to $\T_{\Sigma\cup Q}$. In order to make it so, we  annihilate those trees from $\supp(S_\cA(Q,q_f))$.

      Therefore, we define $r: \T_\Sigma(Q) \to B$ for each $\xi \in \T_\Sigma(Q)$ by
      \[
r(\xi) = (\cdots (S_\cA(Q,q_f) \circ_{q_1} \widetilde{\0}) \cdots \circ_{q_n} \widetilde{\0})(\xi) \enspace.
        \]

      It is obvious that, for each $\xi \in \T_\Sigma(Q)$, we have
      \[
        r(\xi) = \begin{cases}
          \sem{\cA}(\xi) &\text{ if $\xi \in \T_\Sigma$}\\
          \0 & \text{otherwise}
        \end{cases}
      \]      
      i.e., $r = \sem{\cA}\uh{\Sigma \cup Q}$.

      It remains to prove that $r \in \Rat(\Sigma \cup Q,\B)$.      
        By Lemma \ref{lm:S-rat} we have that $S_\cA(Q,q_f) \in \Rat(\Sigma \cup Q,\B)$. By definition, $\widetilde{\0}$ is a polynomial, hence $\widetilde{\0} \in \Rat(\Sigma \cup Q,\B)$. Since $\Rat(\Sigma \cup Q,\B)$ is closed under $q_i$-concatenation, we obtain that $r \in \Rat(\Sigma \cup Q,\B)$. Hence $\sem{\cA} \in \Rat(\Sigma,\B)^{\mathrm{ext}}$.
\end{proof}

We finish this section by an informal comparison. 
If we analyse the proofs of Theorem \ref{thm:rec-implies-rat} and of Lemmas \ref{lm:dyn-prog} and \ref{lm:S-rat}, then we realize that, for each wta $\cA$, we can represent the weighted tree language $\sem{\cA}$ in terms of polynomial weighted tree languages
and the operations tree concatenation and Kleene stars; in particular, the summation $\oplus$ is not needed. This is different from the situation for (unweighted) string languages. In the following we explain the reason of this difference.

In the string case, assuming that the set $Q$ of states of some finite-state string automaton $A$ is $\{1,\ldots,n\}$, the dynamic programming equality which we use in the analysis of the behaviour of $A$ is
\begin{equation}
L_{i,j}^{(k+1)} = L_{i,j}^{(k)} \cup L_{i,k+1}^{(k)}(L_{k+1,k+1}^{(k)})^*L_{k+1,j}^{(k)} \label{equ:dyn-prog-strings}
\end{equation}
\cite[p.~58]{har78}, where $i$, $j$, and $k$ are states; and $L_{i,j}^{(k)}$ is the set of strings which lead $A$ from state $i$ to state $j$ and,  intermediately,  may only visit states in $\{1,\ldots,k\}$. 

Generalizing this scenario to the (unweighted) tree case means to turn the input string 90$^{\mathrm{o}}$ counter-clockwise, i.e., the starting state is at the leaf, and to extend this monadic tree into a ``real'' tree by allowing arbitrary ranks. This implies that the computation does not only start at one point (leaf) in one state $i$ but at each leaf, and  hence $i$ must be replaced by a whole set $P$ of \underline{possible} starting states. Then \eqref{equ:dyn-prog-strings} turns into 
\begin{equation}
L_{P,j}^{(k+1)} = L_{P,j}^{(k)} \cup L_{P \cup\{k+1\},j}^{(k)} \circ_{k+1} (L_{P \cup\{k+1\},k+1}^{(k)})^*_{k+1}  \circ_{k+1}  L_{P,k+1}^{(k)} \label{equ:dyn-prog-trees}
  \end{equation}
where we assume that $k +1 \not\in P$  (and by keeping in mind that the order of arguments of tree concatenation is reversed with respect to that order in string concatenation).  This corresponds to $(*)$ on page 78 of \cite{gecste84}. Now we can realize that the first part of the union, i.e., the set $L_{P,j}^{(k)}$, is a subset of $L_{P \cup\{k+1\},j}^{(k)}$ and, since the state $k+1$ does not occur in trees in $L_{P,j}^{(k)}$, we obtain that
  \[
 L_{P,j}^{(k)} \subseteq L_{P \cup\{k+1\},j}^{(k)} \circ_{k+1} (L_{P \cup\{k+1\},k+1}^{(k)})^*_{k+1}  \circ_{k+1}  L_{P,k+1}^{(k)} \enspace.
\]
Hence \eqref{equ:dyn-prog-trees} turns into
\begin{equation}
L_{P,j}^{(k+1)} = L_{P \cup\{k+1\},j}^{(k)} \circ_{k+1} (L_{P \cup\{k+1\},k+1}^{(k)})^*_{k+1}  \circ_{k+1}  L_{P,k+1}^{(k)} \label{equ:dyn-prog-trees-2}
  \end{equation}
and the union disappeared. Then \eqref{equ:dyn-prog-trees-2} corresponds to the equality in \cite[p.~21]{eng75-15} where we have to identify the set $\{1,\ldots,k\}$ of states with the set  $Q$ of nonterminals, the state $k+1$ with the nonterminal $\B$, and the state $j$ with the nonterminal $A$. Hence, we can represent each recognizable tree language in terms of finite tree languages and the operations tree concatenation and Kleene stars; in particular, the union is not needed. And we generalized this scenario to recognizable weighted tree languages.


\section{From rational to recognizable}

Here we prove Theorem \ref{thm:rat-implies-rec}, i.e., $\Rat(\Sigma,\B) \subseteq \Rec(\Sigma,\B)$, or in words: each rational $(\Sigma,\B)$-weighted tree language is recognizable.

 \begin{theorem} \label{thm:rat-implies-rec} $\Rat(\Sigma,\B) \subseteq \Rec(\Sigma,\B)$.
 \end{theorem}

\begin{proof} Let $r$ be a polynomial $(\Sigma,\B)$-weighted tree language. Then there exist $n \in \mathbb{N}_+$, $b_1,\ldots,b_n \in B$, and $\xi_1,\ldots,\xi_k \in \T_\Sigma$ such that $r= b_1.\xi_1 \oplus \ldots \oplus b_n.\xi_n$. Since  for each $i \in [n]$ the singleton $\{\xi_i\}$ is a recognizable $\Sigma$-tree language, the weighted tree language $r$ is a recognizable step mapping. By Theorem \ref{thm:crisp-det-algebra}, there exists a crisp-deterministic $(\Sigma,\B)$-wta $\cA$ such that $\sem{\cA} = r$. Hence $r \in \Rec(\Sigma,\B)$.
   
   By Theorem \ref{thm:closure-sum} and  Corollaries \ref{cor:closure-tree-concatenation}  and \ref{cor:closure-Kleene-star}, the set $\Rec(\Sigma,\B)$ is closed under the rational operations. Since $\Rat(\Sigma,\B)$ is the smallest set which contains each polynomial $(\Sigma,\B)$-weighted tree language and is closed under the rational operations, the statement follows.
\end{proof}


\section{Alternative definition of rational weighted tree languages}\label{sect:alternative-definition}

\begin{figure}[t]
  \centering
  {\small
\begin{tabular}{lcccc}
 &   $\Pol(\Sigma,\B)$ & scalar multipl. & top-concat.  & rational operations\\\hline
  $\B^{\mathrm{rat}}\fl \T_\Sigma\fr$ \cite{dropecvog05}:  &  & closed & closed & closed \\[2mm]
  $\B^{\mathrm{rat}}_{+\Pol}\fl \T_\Sigma\fr$ \cite{dropecvog05}:  & includes  & closed & closed & closed \\[2mm]
  $\Rat(\Sigma,\B)$:  & includes & & &  closed\\[3mm]
  %
    %
  $\B^{\mathrm{rat}}\fl \T_\Sigma(Q_\infty)\fr$ \cite{dropecvog05} &
     \multicolumn{4}{l}{$= \ \bigcup (\B^{\mathrm{rat}}\fl \T_\Theta\fr \mid \text{ ranked alphabet $\Theta$ such that } \Theta \ge_0 \Sigma)$} \\[3mm]
  $\B^{\mathrm{rat}}\fl \T_\Sigma\fr^{\mathrm{ext}}$ &
  \multicolumn{4}{l}{$= \{r: \T_\Sigma \to B \mid (\exists \Theta \ge_0 \Sigma): r \uh{\Theta} \in \B^{\mathrm{rat}}\fl \T_\Theta\fr\} $} \\[3mm]
    $\Rat(\Sigma,\B)^{\mathrm{ext}}$ &
  \multicolumn{4}{l}{$= \{r: \T_\Sigma \to B \mid (\exists \Theta \ge_0 \Sigma): r \uh{\Theta} \in \Rat(\Theta,\B)\} $}
\end{tabular}

\caption{\label{fig:overview-rational} An overview of the definition of several sets of rational weighted tree language. The denotation $\B^{\mathrm{rat}}_{+\Pol}\fl \T_\Sigma\fr$ does not occur in \cite{dropecvog05}, but we invented this here for the sake of brevity.}
}
\end{figure}

\index{Bratpol@$\B^{\mathrm{rat}}_{+\Pol}\fl \T_\Sigma\fr$}
In Figure \ref{fig:overview-rational}
we give an overview of the relevant sets of rational  weighted tree languages where the sets $\B^{\mathrm{rat}}_{+\Pol}\fl \T_\Sigma\fr$ and $\B^{\mathrm{rat}}\fl \T_\Sigma\fr^{\mathrm{ext}}$ will be defined below.

In this section we will prove the following: including polynomial weighted tree languages into $\B^{\mathrm{rat}}\fl \T_\Sigma\fr$ (yielding the set $\B^{\mathrm{rat}}_{+\Pol}\fl \T_\Sigma\fr$) does not enrich the set $\B^{\mathrm{rat}}\fl \T_\Sigma\fr$ (cf. Observation \ref{obs:Brat=BratPol} and  \cite[Obs.~3.19]{dropecvog05}).  Moreover, we prove that the extensions of the sets $\B^{\mathrm{rat}}\fl \T_\Sigma\fr$ and $\Rat(\Sigma,\B)$  are equal when using our concept of $\0$-extension (cf. Theorem \ref{thm:two-equal-classes-rational}).


As preparation, we characterize the set $\Pol(\Sigma,\B)$ of polynomial $(\Sigma,\B)$-weighted tree languages in terms of closure properties.

\begin{lemma}\rm \label{lm:poly-are-rational} The set $\Pol(\Sigma,\B)$ is the  smallest set of $(\Sigma,\B)$-weighted tree languages which is closed under sum, scalar multiplications, and top-concatenations. Hence, in particular, $\Pol(\Sigma,\B) \subseteq \B^{\mathrm{rat}}\fl\T_\Sigma\fr$.
\end{lemma}
\begin{proof}
  For convenience, we denote by $\cC$ the  smallest set of $(\Sigma,\B)$-weighted tree languages which is closed under sum, scalar multiplications, and top-concatenations.

  First we prove that each polynomial $(\Sigma,\B)$-weighted tree language is in  $\cC$. As preparation we prove by induction on $\T_\Sigma$ that the monomial $\1.\xi$ is in $\cC$ for each $\xi \in \T_\Sigma$. Let $\xi = \sigma(\xi_1,\ldots,\xi_k)$ and assume that $\1.\xi_i$ is in $\mathrm{Pol}(\Sigma,\B)$ for each $i \in [k]$. Obviously,
  \[\1.\xi =
    \ttop_\sigma(\1.\xi_1,\ldots,\1.\xi_k)\]
  and hence $\1.\xi$ is in $\cC$ because $\cC$ is closed under  top-concatenations.

  Now let $r: \T_\Sigma \to B$ be polynomial. We distinguish two cases.
 
  \underline{Case (a):} Let $\supp(r) = \emptyset$. Let $\alpha \in \Sigma^{(0)}$ (recall that $\Sigma^{(0)} \ne \emptyset$ by definition). Then $r = \0 \otimes \1.\alpha$, and since $\1.\alpha$ is in $\cC$ and $\cC$ is  closed under scalar multiplications, we have that $r$ is in $\cC$.

  \underline{Case (b):} Let $\supp(r) = \{\xi_1,\ldots,\xi_n\}$ for some $n \in \mathbb{N}_+$.  Then $r = \bigoplus_{i\in [n]} r(\xi_i) \otimes (\1.\xi_i)$. By the above, $\1.\xi_i$ is in~$\cC$. Since $\cC$ is closed under  scalar multiplications and sum, we have that $r$ is in $\cC$.

Next we prove that each $r \in \cC$ is polynomial, i.e., $\supp(r)$ is finite. This is easy to see, because the application of top-concatenations, scalar multiplications, and sum preserve the property of finite support.

Thus $\Pol(\Sigma,\B) =\cC$. 
The inclusion $\Pol(\Sigma,\B) \subseteq \B^{\mathrm{rat}}\fl\T_\Sigma\fr$ follows from the definition of $\B^{\mathrm{rat}}\fl\T_\Sigma\fr$.
\end{proof}

We denote by $\B^{\mathrm{rat}}_{+\Pol}\fl \T_\Sigma\fr$ the smallest set of $(\Sigma,\B)$-weighted tree languages which contains $\Pol(\Sigma,\B)$ and is closed under scalar multiplications, top-concatenations,  and the rational operations. 

\begin{observation}\rm \label{obs:Brat=BratPol}\cite[Obs.~3.19]{dropecvog05} $\B^{\mathrm{rat}}\fl \T_\Sigma\fr = \B^{\mathrm{rat}}_{+\Pol}\fl \T_\Sigma\fr$.
  \end{observation}
  \begin{proof}
    The inclusion $\B^{\mathrm{rat}}\fl \T_\Sigma\fr \subseteq \B^{\mathrm{rat}}_{+\Pol}\fl \T_\Sigma\fr$ is obvious. The inclusion $\B^{\mathrm{rat}}_{+\Pol}\fl \T_\Sigma\fr \subseteq \B^{\mathrm{rat}}\fl \T_\Sigma\fr$ holds because $\Pol(\Sigma,\B) \subseteq \B^{\mathrm{rat}}\fl \T_\Sigma\fr$ (by Lemma \ref{lm:poly-are-rational}), the fact that  $\B^{\mathrm{rat}}\fl \T_\Sigma\fr$ is closed under scalar multiplications, top-concatenations,  and the rational operations, and the fact that $\B^{\mathrm{rat}}_{+\Pol}\fl \T_\Sigma\fr$ is the smallest such set.
    \end{proof}


Next we prove that the extensions of the sets $\B^{\mathrm{rat}}\fl \T_\Sigma\fr$ and $\Rat(\Sigma,\B)$ (using our concepts of $\0$-extension) are equal.

\index{Bratext@$\B^{\mathrm{rat}}\fl \T_\Sigma \fr^{\mathrm{ext}}$}
The \emph{set of extended DPV-rational $(\Sigma,\B)$-weighted tree languages}, denoted by $\B^{\mathrm{rat}}\fl \T_\Sigma \fr^{\mathrm{ext}}$, contains each $(\Sigma,\B)$-weighted tree language $r$ such that $r\uh{\Theta} \in \B^{\mathrm{rat}}\fl \T_\Theta \fr$ for some $0$-extension $\Theta$ of $\Sigma$.

Thus, we wish to show that $\Rat(\Sigma,\B)^{\mathrm{ext}} = \B^{\mathrm{rat}}\fl \T_\Sigma \fr^{\mathrm{ext}}$. Intuitively, this means that, for the extended sets, closure under  scalar multiplications and top-concatenations can be traded for polynomials, and vice versa. We prove the two directions in separate lemmas.

\begin{lemma}\rm \label{lm:RAT<=RATDPV} $\Rat(\Sigma,\B)^{\mathrm{ext}} \subseteq \B^{\mathrm{rat}}\fl \T_\Sigma\fr^{\mathrm{ext}}$.
  \end{lemma}
  \begin{proof} Let $r \in \Rat(\Sigma,\B)^{\mathrm{ext}}$. Hence $r\uh{\Theta} \in \Rat(\Theta,\B)$ for some $0$-extension $\Theta$ of $\Sigma$. 

    As next step we prove that $r\uh{\Theta} \in \B^{\mathrm{rat}}\fl \T_\Theta\fr$.
    By Lemma \ref{lm:poly-are-rational}, we have that $\Pol(\Theta,\B) \subseteq \B^{\mathrm{rat}}\fl \T_\Theta\fr$.  Hence $\B^{\mathrm{rat}}\fl \T_\Theta\fr$ is a set of $(\Theta,\B)$-weighted tree languages which contains $\Pol(\Theta,\B)$ and is closed under the rational operations. Since $\Rat(\Theta,\B)$ is the smallest such set, we obtain $\Rat(\Theta,\B) \subseteq \B^{\mathrm{rat}}\fl \T_\Theta\fr$. Hence $r\uh{\Theta} \in \B^{\mathrm{rat}}\fl \T_\Theta\fr$. Thus, by definition, we obtain that $r \in \B^{\mathrm{rat}}\fl \T_\Sigma\fr^{\mathrm{ext}}$.
  \end{proof}

For the other direction, i.e.,  $\B^{\mathrm{rat}}\fl \T_\Sigma\fr^{\mathrm{ext}} \subseteq  \Rat(\Sigma,\B)^{\mathrm{ext}}$, we first prove an auxiliary technical lemma.

  \begin{lemma} \rm \label{lm:ext-of-rat-is-rat} Let $\Delta$ be a $0$-extension of $\Sigma$ and let $r: \T_\Sigma \to B$. If $r \in \Rat(\Sigma,\B)$, then   $r\uh{\Delta} \in \Rat(\Delta,\B)$.
  \end{lemma}
    \begin{proof} First, we characterize $\Rat(\Sigma,\B)$ by means of Theorem \ref{thm:Knaster-Tarski} such that we can prove a statement on $\Rat(\Sigma,\B)$ by induction on $\mathbb{N}$. 
For this, we consider the $\sigma$-complete lattice $(\cP(B^{\T_\Sigma}),\subseteq)$  and  the mapping $f: \cP(B^{\T_\Sigma}) \to \cP(B^{\T_\Sigma})$ defined for each $C \in \cP(B^{\T_\Sigma})$ by
\begin{align*}
f(C) = & \ C \cup \Pol(\Sigma,\B) \cup \{ r_1 \oplus r_2 \mid r_1,r_2 \in C \} \cup  \{ r_1 \circ_\alpha r_2 \mid r_1, r_2 \in C, \alpha\in \Sigma^{(0)}\} \ \cup \\
& \ \{(r)^*_\alpha \mid r\in C, \alpha\in \Sigma^{(0)}, \text{ and $r$ is $\alpha$-proper} \}\enspace. 
\end{align*}
The mapping $f$ is continuous and, by definition,
\[
  \Rat(\Sigma,\B) = \bigcap(C \mid C \in \cP(B^{\T_\Sigma})  \text{ such that }f(C) \subseteq C)\enspace.
  \]
Then, by Theorem \ref{thm:Knaster-Tarski}, we have
\begin{equation}
  \Rat(\Sigma,\B) = \bigcup( f^n(\emptyset) \mid n \in \mathbb{N})\enspace.
  \label{eq:inductive-char-Rat}
  \end{equation}
Now,  by induction on $\mathbb{N}$,  we prove that the following statement holds:
  \begin{equation}
\text{For every  $n \in \mathbb{N}$ and $r: \T_\Sigma \to B$:  if $r \in f^n(\emptyset)$, then $r\uh{\Delta} \in \Rat(\Delta,\B)$.} \label{eq:induction}
\end{equation}

I.B.: Let $n=0$. Since $f^n(\emptyset) = f^0(\emptyset) = \emptyset$, statement \eqref{eq:induction}  trivially holds.

I.S.: Let $n = n' + 1$ for some $n' \in \mathbb{N}$. Moreover, let $r \in f^{n}(\emptyset)$. We assume that \eqref{eq:induction} holds for each $r' \in f^{n'}(\emptyset)$. We proceed by case analysis.

\underline{Case (a):} Let $r \in f^{n'}(\emptyset)$. Then our statement holds by I.H.

\underline{Case (b):}  Let $r \in \Pol(\Sigma,\B)$. Then obviously $r\uh{\Delta}$ is a $(\Delta,\B)$-polynomial, hence $r\uh{\Delta}\in \Rat(\Delta,\B)$.

\underline{Case (c):} Let $r=r_1\oplus r_2$ for some $(\Sigma,\B)$-weighted tree languages $r_1$ and $r_2$ in $f^{n'}(\emptyset)$.
By  I.H., we have $r_1\uh{\Delta} \in \Rat(\Delta,\B)$ and $r_2\uh{\Delta} \in \Rat(\Delta,\B)$. Obviously, $r\uh{\Delta} = r_1\uh{\Delta} \oplus r_2\uh{\Delta}$. Since $\Rat(\Delta,\B)$ is closed under sum, we have that $r\uh{\Delta} \in \Rat(\Delta,\B)$.

\underline{Cases (d) and (e):} Let $r=r_1\circ_\alpha r_2$ and $r=(r_1)^{*}_\alpha$, respectively,  for some $r_1$ and $r_2$ in $f^{n'}(\emptyset)$ and $\alpha\in \Sigma$.
The proofs of the cases  are similar to the proof of Case (c).

This finishes the proof of \eqref{eq:induction}.
The statement of the lemma follows from \eqref{eq:induction} and \eqref{eq:inductive-char-Rat}.
      \end{proof}

      In the proof of $\B^{\mathrm{rat}}\fl \T_\Sigma\fr^{\mathrm{ext}} \subseteq  \Rat(\Sigma,\B)^{\mathrm{ext}}$  we will use the fact that, for each $0$-extension $\Theta$ of $\Sigma$, the set $\Rat(\Theta,\B)^{\mathrm{ext}}$ is closed under  scalar-multiplications, top-concatenations,  and the rational operations. One way to prove this is to use $\Rat(\Theta,\B)^{\mathrm{ext}} \subseteq \Rec(\Theta,\B)^{\mathrm{ext}}$ (by Theorem~\ref{thm:Kleene}), to exploit closure properties of $\Rec(\Theta,\B)^{\mathrm{ext}}$, and to use $\Rec(\Theta,\B)^{\mathrm{ext}} \subseteq \Rat(\Theta,\B)^{\mathrm{ext}}$ (again by Theorem~\ref{thm:Kleene}). But here we show an alternative proof, in which we stay inside the area of rational weighted tree languages.

  \begin{lemma}\rm \label{lm:RAText-closed-abc} Let $\Theta$ be a ranked alphabet. Then $\Rat(\Theta,\B)^{\mathrm{ext}}$  is closed under (a)~scalar-multiplications, (b)~top-concatenations,  and the (c) rational operations.
  \end{lemma}
  \begin{proof}   (a)  Here we prove that $\Rat(\Theta,\B)^{\mathrm{ext}}$ is closed under scalar-multiplications. Let $r \in \Rat(\Theta,\B)^{\mathrm{ext}}$ and $b \in B$. We have to prove that $b \otimes r \in \Rat(\Theta,\B)^{\mathrm{ext}}$.

By our assumption, $r\uh{\Delta} \in \Rat(\Delta,\B)$ for some $0$-extension $\Delta$ of $\Theta$.
    Let $\alpha \in \Delta^{(0)}$ be an arbitrary nullary symbol (recall that $\Delta^{(0)} \ne \emptyset$ by definition of ranked alphabets.) We  define the $(\Delta,\B)$-weighted tree language $s$ by
    \[
      s = (b.\alpha )\circ_\alpha (r\uh{\Delta}) \enspace.
    \]
      Since $b.\alpha \in \Rat(\Delta,\B)$  and $\Rat(\Delta,\B)$ is closed under tree concatenations, we have that $s \in \Rat(\Delta,\B)$. Moreover, for each $\xi \in \T_{\Delta}$, we have
      \[
        s(\xi) = b \otimes \big(r\uh{\Delta}(\xi)\big) \enspace .
        \]
        Hence $s = b \otimes r\uh{\Delta}$, and since $b \otimes r\uh{\Delta}=(b \otimes r)\uh{\Delta}$ we obtain  that $b \otimes r \in \Rat(\Theta,\B)^{\mathrm{ext}}$.        

        \
        
        (b) We prove that $\Rat(\Theta,\B)^{\mathrm{ext}}$ is closed under top-concatenations. Let $k \in \mathbb{N}$,  $\sigma \in \Theta^{(k)}$, and $r_1,\ldots,r_k$ be $(\Theta,\B)$-weighted tree languages in $\Rat(\Theta,\B)^{\mathrm{ext}}$. We have to prove that $\ttop_\sigma(r_1,\ldots,r_k) \in \Rat(\Theta,\B)^{\mathrm{ext}}$.

    For each $i\in [k]$ let $r_i\uh{\Delta_i} \in \Rat(\Delta_i,\B)$ for some $0$-extension $\Delta_i$ of $\Theta$. Without loss of generality we can assume that $(\Delta_i\setminus \Theta) \cap (\Delta_j \setminus \Theta)=\emptyset$ for every $i,j \in [k]$ with $i\ne j$. We define the ranked alphabet $\Delta = \bigcup_{i \in[k]} \Delta_i$. Obviously, $\Delta \ge_0 \Delta_i$ for each $i \in [k]$.
    By Observation \ref{obs:transitivity-extension} we have that $(r_i\uh{\Delta_i})\uh{\Delta} = r_i \uh{\Delta}$ and hence, by Lemma \ref{lm:ext-of-rat-is-rat}, we have that $r_i\uh{\Delta} \in \Rat(\Delta,\B)$. 

    Let $P = \{p_1,\ldots,p_k\}$ be a set disjoint with $\Delta$. We view $\Delta \cup P$ as $0$-extension of $\Delta$. We define the $(\Delta\cup P,\B)$-weighted tree language $r$ by
    \[
r = (\ldots(\1.\sigma(p_1,\ldots,p_k) \circ_{p_1} r_1\uh{\Delta\cup P}) \ldots ) \circ_{p_k} r_k\uh{\Delta\cup P} \enspace.
      \]
      Since $\1.\sigma(p_1,\ldots,p_k) \in \Rat(\Delta\cup P,\B)$ and $r_i\uh{\Delta\cup P} \in \Rat(\Delta \cup P,\B)$ (for each $i \in [k]$ by Lemma \ref{lm:ext-of-rat-is-rat}) and $\Rat(\Delta\cup P,\B)$ is closed under tree concatenations, we have that $r \in \Rat(\Delta \cup P,\B)$.

      We will show that, for each $\xi \in \T_{\Delta\cup P}$, we have
      \begin{equation}\label{eq:r-is-extension}
        r(\xi) = \begin{cases}
          \ttop_\sigma(r_1,\ldots,r_k)(\xi) & \text{if $\xi=\sigma(\xi_1,\ldots,\xi_k)$ for some $\xi_1,\ldots,\xi_k \in \T_\Theta$}\\
          \0 & \text{otherwise}\enspace.
          \end{cases}
        \end{equation}
Hence $r = \ttop_\sigma(r_1,\ldots,r_k)\uh{\Delta\cup P}$. Since $\Delta \cup P \ge_0   \Theta$, we obtain that $\ttop_\sigma(r_1,\ldots,r_k) \in \Rat(\Theta,\B)^{\mathrm{ext}}$.

For the proof of \eqref{eq:r-is-extension}, for each $i\in[0,k]$ we  define
\[s_i=(\ldots(\1.\sigma(p_1,\ldots,p_k) \circ_{p_1} r_1\!\!\upharpoonright) \ldots ) \circ_{p_i} r_i\!\!\upharpoonright\enspace,\]
where $r_j\!\!\upharpoonright$ is an abbreviation of $r_j\uh{\Delta\cup P}$ for each $j \in[i]$.
(We note that $s_0=\1.\sigma(p_1,\ldots,p_k)$.)
Then, by induction on $([0,k],>)$, we prove that the following statement holds:
\begin{eqnarray}
  \begin{aligned}
& \text{For each $i\in[0,k]$ and $\xi \in \T_{\Delta\cup P}$, we have}\\
& s_i(\xi) = \begin{cases}
          \bigotimes_{j \in [i]} r_j(\xi_j) & \text{if $\xi=\sigma(\xi_1,\ldots,\xi_i,p_{i+1},\ldots,p_k)$ for some $\xi_1,\ldots,\xi_i \in \T_{\Theta}$}\\
          \0 & \text{otherwise}\enspace.
        \end{cases}
        \end{aligned} \label{eq:general formula}
\end{eqnarray}
Let $\xi \in \T_{\Delta\cup P}$. 

I.B.: Let $i=0$. The statement is obvious because $s_0=\1.\sigma(p_1,\ldots,p_k)$ and the product of an $\emptyset$-indexed family over $B$ is defined to be $\1$.

I.S.: Let $i = i'+1$: Then we have
\begin{align}\label{eq:induction-formula}
s_{i}(\xi)& =(s_{i'} \circ_{p_{i}} r_{i}\!\!\upharpoonright )(\xi)
    = \bigoplus\limits_{(w_1,\ldots,w_n) \in \cut_{p_{i}}(\xi)} 
s_{i'}(\xi[p_{i}]_{(w_1,\ldots,w_n)}) \otimes r_{i}\!\!\upharpoonright(\xi|_{w_1})\otimes \ldots \otimes r_{i}\!\!\upharpoonright(\xi|_{w_n})\enspace.
\end{align}
By I.H. we obtain
\[
s_{i'}(\xi[p_{i}]_{(w_1,\ldots,w_n)}) = \begin{cases}
          \bigotimes_{j \in [i']} r_j(\xi_j) & \text{if $\xi[p_{i}]_{(w_1,\ldots,w_n)}=\sigma(\xi_1,\ldots,\xi_{i'},p_{i},\ldots,p_k)$} \\
          &  \text{for some $\xi_1,\ldots,\xi_{i'} \in \T_{\Theta}$}\\
          \0 & \text{otherwise}\enspace.
\end{cases}
\]
Now we observe that the condition $\xi[p_{i}]_{(w_1,\ldots,w_n)}=\sigma(\xi_1,\ldots,\xi_{i'},p_{i},\ldots,p_k)$ for some $\xi_1,\ldots,\xi_{i'} \in \T_{\Theta}$ implies that $n=1$, $w_1=i$ and there exists $\xi_{i} \in \T_{\Delta\cup P}$ such that $\xi=\sigma(\xi_1,\ldots,\xi_{i'},\xi_{i},p_{i},\ldots,p_k)$.

Hence \eqref{eq:induction-formula} continues as
\begin{align*}
= \begin{cases}
\Big(\bigotimes_{j \in [i']} r_j(\xi_j)\Big)\otimes r_{i}\!\!\upharpoonright(\xi|_{i}) & \text{if $\xi=\sigma(\xi_1,\ldots,\xi_{i'},\xi_{i},p_{i+1},\ldots,p_k)$} \\
& \text{for some $\xi_1,\ldots,\xi_{i'} \in \T_{\Theta}$ and $\xi_{i} \in \T_{\Delta\cup P}$} \\
\0 & \text{otherwise}\enspace.
\end{cases}
\end{align*}
Since $\xi=\sigma(\xi_1,\ldots,\xi_{i'},\xi_{i},p_{i+1},\ldots,p_k)$ implies that $\xi|_{i}=\xi_{i}$,
and  $r_{i}\!\!\upharpoonright$ is the $\0$-extension of $r_{i}$ to  $\T_{\Delta\cup P}$, we obtain
\begin{align*}
= \begin{cases}
\bigotimes_{j \in [i]} r_j(\xi_j) & \text{if $\xi=\sigma(\xi_1,\ldots,\xi_{i'},\xi_{i},p_{i+1},\ldots,p_k)$} \\
& \text{for some $\xi_1,\ldots,\xi_{i} \in \T_{\Theta}$} \\
\0 & \text{otherwise}\enspace.
\end{cases}
\end{align*}
This proves \eqref{eq:general formula}. From this latter, in case $i=k$ and using the definition of $\ttop_\sigma(r_1,\ldots,r_k)$ we obtain \eqref{eq:r-is-extension}.

        (c)  We prove that $\Rat(\Theta,\B)^{\mathrm{ext}}$ is closed under the rational operations, i.e., sum, tree concatenations, and Kleene stars.

        \underline{Closure under sum:} Let $r_1, r_2 \in \Rat(\Theta,\B)^{\mathrm{ext}}$. We show that $r_1 \oplus r_2 \in \Rat(\Theta,\B)^{\mathrm{ext}}$.

        Let $r_1\uh{\Delta_1} \in \Rat(\Delta_1,\B)$ for some $0$-extension $\Delta_1$ of $\Theta$. Also let  $r_2\uh{\Delta_2} \in \Rat(\Delta_2,\B)$ for some $0$-extension $\Delta_2$ of $\Theta$. Without loss of generality we can assume that $(\Delta_1\setminus \Theta) \cap (\Delta_2 \setminus \Theta)=\emptyset$. We define the ranked alphabet $\Delta = \Delta_1 \cup \Delta_2$. Let $i \in [2]$. Obviously, $\Delta \ge_0 \Delta_i$, and hence $\Delta \ge_0 \Theta$. By Observation \ref{obs:transitivity-extension}, we have that $(r_i\uh{\Delta_i})\uh{\Delta} = r_i \uh{\Delta}$ and hence, by Lemma \ref{lm:ext-of-rat-is-rat}, we have that $r_i\uh{\Delta} \in \Rat(\Delta,\B)$.

        Since $\Rat(\Delta,\B)$ is closed under sum, we obtain that $r_1\uh{\Delta} \oplus r_2\uh{\Delta} \in \Rat(\Delta,\B)$. Since  \((r_1 \oplus r_2)\uh{\Delta}(\xi) = r_1\uh{\Delta}(\xi) \oplus r_2\uh{\Delta}(\xi)\) for each $\xi \in \T_\Delta$, we have $(r_1 \oplus r_2)\uh{\Delta} \in \Rat(\Delta,\B)$. Hence $r_1 \oplus r_2 \in \Rat(\Theta,\B)^{\mathrm{ext}}$.

          \underline{Closure under tree concatenations:} Let $r_1, r_2 \in \Rat(\Theta,\B)^{\mathrm{ext}}$ and $\alpha \in \Theta^{(0)}$. We show that $r_1 \circ_\alpha r_2 \in \Rat(\Theta,\B)^{\mathrm{ext}}$.
  In the same way as above we have that $r_1\uh{\Delta} \in \Rat(\Delta,\B)$ and $r_2\uh{\Delta} \in \Rat(\Delta,\B)$ for some $0$-extension $\Delta$ of $\Theta$.

          Since $\Rat(\Delta,\B)$ is closed under tree concatenations, we obtain that $r_1\uh{\Delta} \circ_\alpha r_2\uh{\Delta} \in \Rat(\Delta,\B)$. Since  \((r_1 \circ_\alpha r_2)\uh{\Delta}(\xi) = r_1\uh{\Delta}(\xi) \circ_\alpha r_2\uh{\Delta}(\xi)\) for each $\xi \in \T_\Delta$, we have $(r_1 \circ_\alpha r_2)\uh{\Delta} \in \Rat(\Delta,\B)$. Hence $r_1 \circ_\alpha r_2 \in \Rat(\Theta,\B)^{\mathrm{ext}}$.

            \underline{Closure under Kleene stars:} Let $\alpha \in \Theta^{(0)}$ and $r \in \Rat(\Theta,\B)^{\mathrm{ext}}$ be an $\alpha$-proper weighted tree language. We show that $r_\alpha^* \in \Rat(\Theta,\B)^{\mathrm{ext}}$.
          By definition, we have that $r\uh{\Delta} \in \Rat(\Delta,\B)$ for some $0$-extension $\Delta$ of~$\Theta$.

          Since $\Rat(\Delta,\B)$ is closed under Kleene stars, we obtain that $(r\uh{\Delta})_\alpha^* \in \Rat(\Delta,\B)$. Since  $(r_\alpha^*)\uh{\Delta}(\xi) = (r\uh{\Delta})_\alpha^*(\xi)$  for each $\xi \in \T_\Delta$, we have $(r_\alpha^*)\uh{\Delta} \in \Rat(\Delta,\B)$. Hence $r_\alpha^* \in \Rat(\Theta,\B)^{\mathrm{ext}}$.
    \end{proof}

  \begin{lemma}\rm \label{lm:RATDPV<=RAT} $\B^{\mathrm{rat}}\fl \T_\Sigma\fr^{\mathrm{ext}} \subseteq \Rat(\Sigma,\B)^{\mathrm{ext}}$.
  \end{lemma}
  \begin{proof} Let $r \in \B^{\mathrm{rat}}\fl \T_\Sigma\fr^{\mathrm{ext}}$. Hence $r\uh{\Theta} \in \B^{\mathrm{rat}}\fl \T_\Theta\fr$ for some $0$-extension $\Theta$ of $\Sigma$. 
  
By Lemma \ref{lm:RAText-closed-abc}, the set $\Rat(\Theta,\B)^{\mathrm{ext}}$  is closed under (a) scalar-multiplications, (b) top-concatenations,  and the (c) rational operations.
 Since  $\B^{\mathrm{rat}}\fl \T_\Theta\fr$ is the smallest set which has these closure properties, we obtain that $\B^{\mathrm{rat}}\fl \T_\Theta\fr \subseteq \Rat(\Theta,\B)^{\mathrm{ext}}$, and hence $r\uh{\Theta} \in \Rat(\Theta,\B)^{\mathrm{ext}}$. Thus $(r\uh{\Theta})\uh{\Delta} \in \Rat(\Delta,\B)$ for some $0$-extension $\Delta$ of $\Theta$. Since $\Delta \ge_\0 \Theta$ and $\Theta \ge_0 \Sigma$ and $(r\uh{\Theta})\uh{\Delta} = r\uh{\Delta}$ by Observation \ref{obs:transitivity-extension}, we obtain that $r\uh{\Delta} \in \Rat(\Delta,\B)$. Thus, by definition, we have $r \in \Rat(\Sigma,\B)^{\mathrm{ext}}$.
      \end{proof}

Then the next theorem follows from Lemmas \ref{lm:RAT<=RATDPV} and \ref{lm:RATDPV<=RAT}.

\begin{theorem-rect} \label{thm:two-equal-classes-rational} Let $\Sigma$ be a ranked alphabet. Moreover, let $\B$ be a commutative semiring. Then $\Rat(\Sigma,\B)^{\mathrm{ext}} = \B^{\mathrm{rat}}\fl \T_\Sigma\fr^{\mathrm{ext}}$.
\end{theorem-rect}

%% file: Medvedjev.tex
\chapter{Elementary operations and M{\'e}dv{\'e}dj{\'e}v's theorem} 
\label{ch:Medvedjev}

In his paper \cite{med56} (cf. also \cite{bli65}), M{\'e}dv{\'e}dj{\'e}v proved a characterization of the set of regular string languages as the smallest set of string languages which contains some elementary sets and is closed under some elementary operations. He considered this as an alternative to Kleene's characterization in terms of finite languages and the rational operations. This characterization has been generalized to the tree case in \cite{cos72}
(cf. \cite[Thm.~2.8.6]{gecste84}) and to the weighted case for semirings in \cite{her17} and \cite[Ch.~5]{her20a}.

In this chapter, we will prove two main theorems: one for wta over semirings and restricted representations (cf. Theorem \ref{thm:Medvedjev}; this closely corresponds to \cite[Ch.~5]{her20a}) and one for wta over bi-locally finite strong bimonoids and arbitrary representations (cf. Theorem~\ref{cor:Medvedjev-str-bm} and, for the particular case of bounded lattices cf. Corollary~\ref{cor:Medvedjev-L-valued-wta-str-bm}). We recall that the set of semirings and the set of bi-locally finite strong bimonoids are incomparable (cf. Figure~\ref{fig:Euler-diagram-extended}).

  \section{Representable weighted tree languages}
  \label{sec:representable-wtl}
  
  In this section we define the generalizations of the mentioned elementary sets and elementary operations to the weighted tree case. First we define the set of $\B$-representations. It is a family of sets $\RepEx(\Sigma,\B)$ of expressions where $\Sigma$ is an arbitrary ranked alphabet; this provides the type of these expressions.

  \index{Repex@$\RepEx(\Sigma,\B)$}
  Formally, the family of \emph{$\B$-representations}, denoted by $\RepEx(\B)$, is the family
  \[\RepEx(\B) = (\RepEx(\Sigma,\B) \mid \Sigma \text{ ranked alphabet})
  \]
defined as the smallest family $R=(R(\Sigma) \mid \Sigma \text{ ranked alphabet})$ of expressions such that Conditions (1)-(6) hold. (We recall that $\Delta$ is an arbitrary ranked alphabet.)
\begin{compactenum}
  \index{RRT@$\RRT_{\Sigma,\sigma,b}$}
\item[(1)] For every  $\sigma \in \Sigma$ and $b \in B$, the expression $\RRT_{\Sigma,\sigma,b}$ is in $R(\Sigma)$.

\index{Next@$\NXT_{\Sigma,\widetilde{\gamma},b}$}
\item[(2)]   For every $k \in \mathbb{N}_+$, element $\widetilde{\gamma} = \gamma_1\cdots\gamma_k$  of $\Sigma^k$, and $b \in B$, the expression $\NXT_{\Sigma,\widetilde{\gamma},b}$ is in $R(\Sigma)$.

\item[(3)] If $e_1,e_2 \in R(\Sigma)$, then $e_1 + e_2 \in R(\Sigma)$. 

\item[(4)]  If $e_1,e_2 \in R(\Sigma)$, then $e_1 \times e_2 \in R(\Sigma)$.

\item[(5)] If $e \in R(\Delta)$ and $\tau$ is a  $(\Delta,\Sigma)$-tree relabeling, then $\tau(e) \in R(\Sigma)$.

  \index{Rest@$\REST(e)$}
  \item[(6)]  If $e \in R(\Sigma)$, then $\REST(e) \in R(\Sigma)$.  
\end{compactenum}
We note that in Case (5) the type of the representation changes from $\Delta$ to $\Sigma$ according to the type of the tree relabeling. Moreover, in \cite{her17} and \cite[Ch.~5]{her20a}, only deterministic tree relabelings are allowed. But since the set of recognizable weighted tree languages is closed under arbitrary tree relabelings (cf. Theorem \ref{thm:closure-under-tree-relabeling}), we can prove the main theorem also for a larger set of representations.

Informally, we might understand the set $\RepEx(\B)$ as the set of expressions generated by the following EBNF; it uses the infinite family $(e_\Sigma \mid \Sigma \text{ ranked alphabet})$ of nonterminals, and for each nonterminal $e_\Sigma$ it has the following rule:
\begin{equation}
e_\Sigma ::= \RRT_{\Sigma,\sigma,b} \mid \NXT_{\Sigma,\widetilde{\gamma},b} \mid e_\Sigma + e_\Sigma \mid e_\Sigma \times e_\Sigma \mid \tau(e_\Delta) \mid \REST(e_\Sigma)  \label{eq:EBNF-representations}
  \end{equation}
where $\sigma \in \Sigma$, $\widetilde{\gamma} \in \Sigma^k$ for some $k \in \mathbb{N}_+$, $b \in B$, and $\tau$ is a  $(\Delta,\Sigma)$-tree relabeling. 

\index{succb@$\succ$}
 In order to perform inductive proofs or to define objects by induction, we will consider the  reduction system
        \[(\RepEx(\B),\succ)
        \]
        where  $\succ$ is the binary relation on $\RepEx(\B)$ defined as follows. 
        For every $e_1,e_2 \in \RepEx(\B)$ we let  $e_1 \succ e_2$ if either of the following three cases hold.

        Case (a): There exist $f_1,f_2 \in \RepEx(\Sigma,\B)$ such that $e_1$ has the form $f_1 +f_2$ or $f_1 \times f_2$ and $e_2 \in \{f_1,f_2\}$.

        Case (b): There exist $f \in \RepEx(\Delta,\B)$ and a $(\Delta,\Sigma)$-tree relabeling $\tau$ such that $e_1=\tau(f)$ and $e_2 = f$.

        Case (c): There exist $f \in \RepEx(\Sigma,\B)$ such that $e_1=\REST(f)$ and $e_2 = f$.

        By Corollary \ref{cor:reduction-to-substring-is-terminating}, the relation $\succ$ is terminating. Moreover, $\nf_{\succ}(\RepEx(\B))$ is the set of all representations of the form $\RRT_{\Sigma,\sigma,b}$ or  $\NXT_{\Sigma,\widetilde{\gamma},b}$.
        For every $e_1,e_2 \in \RepEx(\B)$, we say that \emph{$e_2$ is a subexpression of $e_1$} if $e_1 \succ^* e_2$. We note that the relation ``is a subexpression of'' is reflexive.

\index{semantice@$\sem{e}$}
Next we define the semantics of $\B$-expressions by induction on $(\RepEx(\B),\succ)$. In particular, for each $e \in \RepEx(\Sigma,\B)$, the semantics of $e$  is a weighted tree language $\sem{e}:   \T_\Sigma \to B$.

I.B.: We distinguish two cases.

(a) Let $\sigma \in \Sigma$ and $b \in B$. 
  We define $\sem{\RRT_{\Sigma,\sigma,b}}: \T_\Sigma \to B$ by 
  \[\sem{\RRT_{\Sigma,\sigma,b}} =  b \cdot \chi(L_{\Sigma,\sigma})
\]
where
\( L_{\Sigma,\sigma} = \{\xi \in \T_\Sigma \mid \xi(\varepsilon)=\sigma\}\).
That is, for every $\xi \in \T_\Sigma$, we have
\[
  \sem{\RRT_{\Sigma,\sigma,b}}(\xi) = \begin{cases}
    b & \text{ if $\xi \in  L_{\Sigma,\sigma}$}\\
    \0 & \text{ otherwise} \enspace.
    \end{cases}
  \]
A weighted tree language of the form $\sem{\RRT_{\Sigma,\sigma,b}}$ is called \emph{$(\Sigma,\sigma,b)$-root mapping} or simply \emph{root mapping}.

(b)  Let $k \in \mathbb{N}_+$, $\widetilde{\gamma} = \gamma_1\cdots\gamma_k$ be an element of $\Sigma^k$, and $b \in B$. 
  We define  $\sem{\NXT_{\Sigma,\widetilde{\gamma},b}}: \T_\Sigma \to B$ by
\[\sem{\NXT_{\Sigma,\widetilde{\gamma},b}} = b \cdot \chi(L_{\Sigma,\widetilde{\gamma}})\enspace,
\]
where
\( L_{\Sigma,\widetilde{\gamma}} = \{\xi \in \T_\Sigma \mid \rk(\xi(\varepsilon))=k \wedge (\forall i \in [k]): \xi(i) = \gamma_i\}\).
That is, for every $\xi \in \T_\Sigma$, we have
\[
  \sem{\NXT_{\Sigma,\widetilde{\gamma},b}}(\xi) =
  \begin{cases}
    b & \text{ if $\xi \in L_{\Sigma,\widetilde{\gamma}}$}\\
    \0 & \text{ otherwise} \enspace.
    \end{cases}
  \]
In particular, if $\Sigma^{(k)}=\emptyset$, then $ L_{\Sigma,\widetilde{\gamma}} =\emptyset$ and thus $\sem{\NXT_{\Sigma,\widetilde{\gamma},b}}=\widetilde{\0}$.  
  \index{next mapping}
 A weighted tree language of the form $\sem{\NXT_{\Sigma,\widetilde{\gamma},b}}$ is called \emph{$(\Sigma,\widetilde{\gamma},b)$-next mapping} or simply \emph{next mapping}.

 \

 I.S.: We distinguish four cases.

(a) Let $e_1,e_2 \in \RepEx(\Sigma,\B)$.
  We define $\sem{e_1 + e_2}: \T_\Sigma \to B$ by  $\sem{e_1 + e_2} = \sem{e_1} \oplus \sem{e_2}$.

(b)  Let $e_1,e_2 \in \RepEx(\Sigma,\B)$.
We define $\sem{e_1 \times e_2}: \T_\Sigma \to B$ by $\sem{e_1 \times  e_2} = \sem{e_1} \otimes \sem{e_2}$.

(c) Let $e \in \RepEx(\Delta,\B)$
and $\tau$ is a  $(\Delta,\Sigma)$-tree relabeling. We define $\sem{\tau(e)}: \T_\Sigma \to B$ by $\sem{\tau(e)} = \chi(\tau)(\sem{e})$.

(d)  Let $e \in \RepEx(\Sigma,\B)$.
    We define $\sem{\REST(e)}: \T_\Sigma \to B$  for each $\xi \in \T_\Sigma$ by
  \[
    \sem{\REST(e)}(\xi) = \bigotimes_{\substack{w \in \pos(\xi)\\ \text{in $<_{\mathrm{dp}}$ order }}}  \sem{e}(\xi|_w) \enspace,
  \]
  i.e., the factors of the product are ordered by depth-first post-order $<_{\mathrm{dp}}$ on $\pos(\xi)$.
  \index{restriction}
  The weighted tree language $\sem{\REST(e)}$ is called the \emph{restriction of $\sem{e}$}.


We extend the binary operation $+$ on the set of $\B$-expressions in the natural way to an operation with a finite set $I$ of arguments $e_i$ with $i \in I$, and we denote this expression by $\bigplus_{i \in I} e_i$. The weighted tree language $\sem{\bigplus_{i \in I} e_i}$  is well defined, because $\oplus$ is associative and commutative.

\index{elementary}
  Let $r: \T_\Sigma \to B$ be a weighted tree language. We say that $r$ is \emph{elementary} if $r$ is a root mapping or a next mapping. Since each of the $\Sigma$-tree languages $L_{\Sigma,\sigma}$ and $L_{\Sigma,\widetilde{\gamma}}$ is recognizable,  each elementary $(\Sigma,\B)$-weighted tree language is a $(\Sigma,\B)$-recognizable one-step mapping. The operations sum, Hadamard product, tree relabeling, and restriction (cf. (3)-(6), respectively) are called \emph{elementary operations}.

  \index{representable}
A weighted tree language $r: \T_\Sigma \to B$ is \emph{representable} if there exists an $e \in \RepEx(\Sigma,\B)$  such that $r = \sem{e}$. The set of all representable $(\Sigma,\B)$-weighted tree languages is denoted by $\Rep(\Sigma,\B)$. Moreover, $\Rep(\_,\B)$ denotes the set of all representable $(\Sigma,\B)$-weighted tree languages for some ranked alphabet $\Sigma$. Hence, $\Rep(\_,\B)$ is  the smallest set of $\B$-weighted tree languages which contains the elementary weighted tree languages and is closed under the elementary operations.

Next we give two examples of recognizable $(\Sigma,\Natmaxplus)$-weighted tree languages which are representable.

\begin{example}\label{ex:recognizable-representable}\rm We consider the ranked alphabet $\Sigma=\{\sigma^{(2)},\gamma^{(1)},\alpha^{(0)}\}$, the arctic semiring $\Natmaxplus=(\mathbb{N}_{-\infty},\max,+,-\infty,0)$, and the weighted tree language $\#_{\sigma(.,\alpha)}: \T_\Sigma \to \mathbb{N}$ defined in Example~\ref{ex:number-of-occurrences}. We recall that,  for each $\xi\in \T_\Sigma$, the value $\#_{\sigma(.,\alpha)}(\xi)$ is the number of occurrences of the pattern $\sigma(.,\alpha)$ in $\xi$. In Example \ref{ex:number-of-occurrences-arctic} we gave a 
bu-deterministic $(\Sigma,\Natmaxplus)$-wta $\cA$ such that $\sem{\cA}(\xi)=\#_{\sigma(.,\alpha)}(\xi)$ for each $\xi \in \T_\Sigma$. 

Now we give a $\Natmaxplus$-representation $e$ such that $\sem{e}=\sem{\cA}$. We define 
\[
e = \REST\Big(\Big(\bigplus\nolimits_{\kappa \in \Sigma}\NXT_{\Sigma,\kappa\alpha,0}\Big) \times \RRT_{\Sigma,\sigma,1} + \bigplus\nolimits_{\kappa \in \Sigma} \RRT_{\Sigma,\kappa,0}\Big)
\]
where we have assumed that $\times$ has higher binding priority than $+$.

First, we observe that, for each $\xi \in \T_\Sigma$, we have
\(\sem{\bigplus\limits_{\kappa \in \Sigma} \RRT_{\Sigma,\kappa,0}}(\xi) = 0\).
Second, by using the  abbreviation
\[e' = \Big(\bigplus\nolimits_{\kappa \in \Sigma}\NXT_{\Sigma,\kappa\alpha,0}\Big) \times \RRT_{\Sigma,\sigma,1} \enspace,
\]
we prove  by case analysis that, 
\begin{equation}\label{equ:number-of-patterns-infty}
  \text{ for each $\xi \in \T_\Sigma$, we have } \ \sem{e'}(\xi) =
  \begin{cases}
    1 & \text{ if $\xi=\sigma(\xi',\alpha)$ for some $\xi' \in \T_\Sigma$}\\
    -\infty & \text{ otherwise}\enspace.
    \end{cases}
\end{equation}

\underline{Case (a):} Let  $\xi(\varepsilon) = \alpha$. Then 
\begin{align*}
 \sem{e'}(\xi) &= \sem{\bigplus\nolimits_{\kappa \in \Sigma}\NXT_{\Sigma,\kappa\alpha,0}}(\xi) + \sem{\RRT_{\Sigma,\sigma,1}}(\xi)\\
&=  \sem{\bigplus\nolimits_{\kappa \in \Sigma}\NXT_{\Sigma,\kappa\alpha,0}}(\xi) + -\infty  \tag{because $\xi(\varepsilon) = \alpha$} \\
&= -\infty \enspace.
\end{align*}

\underline{Case (b):} Let $\xi(\varepsilon) = \gamma$. In a similar way as in Case (a), we can calculate that $\sem{e'}(\xi) = -\infty$. 

\underline{Case (c):} Let $\xi(\varepsilon) = \sigma$ and $\xi(2)\ne\alpha$. Then
\begin{align*}
 \sem{e'}(\xi) &= \max(\sem{\NXT_{\Sigma,\sigma\alpha,0}}(\xi), \sem{\NXT_{\Sigma,\gamma\alpha,0}}(\xi), \sem{\NXT_{\Sigma,\alpha\alpha,0}}(\xi)) + \sem{\RRT_{\Sigma,\sigma,1}}(\xi)\\
&= \max(-\infty,-\infty,-\infty) + 1  
= -\infty \enspace.
\end{align*}

\underline{Case (d):} Let $\xi(\varepsilon) = \sigma$, $\xi(2)=\alpha$, and $\xi(1)=\sigma$. Then
\begin{align*}
 \sem{e'}(\xi) &= \max(\sem{\NXT_{\Sigma,\sigma\alpha,0}}(\xi), \sem{\NXT_{\Sigma,\gamma\alpha,0}}(\xi), \sem{\NXT_{\Sigma,\alpha\alpha,0}}(\xi)) + \sem{\RRT_{\Sigma,\sigma,1}}(\xi)\\
               &= \max(0,-\infty,-\infty) + 1
                 = 0 +1 
= 1 \enspace.
\end{align*}
  
\underline{Case (e):} Let $\xi(\varepsilon) = \sigma$, $\xi(2)=\alpha$, and $\xi(1)=\gamma$. Then
\begin{align*}
 \sem{e'}(\xi) &= \max(\sem{\NXT_{\Sigma,\sigma\alpha,0}}(\xi), \sem{\NXT_{\Sigma,\gamma\alpha,0}}(\xi), \sem{\NXT_{\Sigma,\alpha\alpha,0}}(\xi)) + \sem{\RRT_{\Sigma,\sigma,1}}(\xi)\\
               &= \max(-\infty,0,-\infty) + 1
                 = 0 +1 
                 = 1 \enspace.
                 \end{align*}

                 \underline{Case (f):} Let $\xi(\varepsilon) = \sigma$, $\xi(2)=\alpha$, and $\xi(1)=\alpha$. Then
\begin{align*}
 \sem{e'}(\xi) &= \max(\sem{\NXT_{\Sigma,\sigma\alpha,0}}(\xi), \sem{\NXT_{\Sigma,\gamma\alpha,0}}(\xi), \sem{\NXT_{\Sigma,\alpha\alpha,0}}(\xi)) + \sem{\RRT_{\Sigma,\sigma,1}}(\xi)\\
               &= \max(-\infty,-\infty,0) + 1
                 = 0 +1 
= 1 \enspace.
\end{align*}
This proves \eqref{equ:number-of-patterns-infty}.

Hence,  for each $\xi \in \T_\Sigma$, we have
\begin{equation}
  \sem{e' + \bigplus\nolimits_{\kappa \in \Sigma} \RRT_{\Sigma,\kappa,0}}(\xi) =
  \begin{cases}
    1 & \text{ if $\xi=\sigma(\xi',\alpha)$ for some $\xi' \in \T_\Sigma$}\\
    0 & \text{ otherwise}  \enspace.
    \end{cases} \label{eq:net-effect}
\end{equation}

Finally, for each $\xi \in \T_\Sigma$, we can calculate as follows:
\begingroup
\allowdisplaybreaks
\begin{align*}
  \sem{e}(\xi) &= \sem{\REST(e' +  \bigplus\nolimits_{\kappa \in \Sigma} \RRT_{\Sigma,\kappa,0})}(\xi)
               = \bigplus_{w \in \pos(\xi)} \sem{e' +  \bigplus\nolimits_{\kappa \in \Sigma} \RRT_{\Sigma,\kappa,0}}(\xi|_w)\\
               &= \bigplus_{\substack{w \in \pos(\xi):\\ \xi(w)=\sigma, \xi(w2)=\alpha}} 1 \tag{by \eqref{eq:net-effect}}\\
  &= \#_{\sigma(.,\alpha)}(\xi) \enspace.   \hspace{70mm} \Box
\end{align*}
\endgroup
\end{example}

\begin{example}\label{ex:recognizable-representable-height}\rm We consider the ranked alphabet $\Sigma=\{\sigma^{(2)},\alpha^{(0)}\}$, the arctic semiring $\Natmaxplus=(\mathbb{N}_{-\infty},\max,+,-\infty,0)$, and the weighted tree language $\height: \T_\Sigma \to \mathbb{N}$. In Example \ref{ex:height} we gave a $(\Sigma,\Natmaxplus)$-wta $\cA$ such that $\runsem{\cA}=\height$. The definition of $\cA$ was based on the fact that
  \[
    \text{for each $\xi \in \T_\Sigma$, we have $\height(\xi) = \max_{w \in \pos_\alpha(\xi)} |w|$.}
  \]
  Then, roughly speaking, the $\max$ on the right-hand side matches with the operation $\max$ of the semiring, which is used in the definition of the run semantics. 
  
  Now we give a $\Natmaxplus$-representation $e$ such that $\sem{e}=\height$. For this, we will use the fact that
  \[
    \text{for each $\xi \in \T_\Sigma$, we have $\height(\xi) = \max_{w \in \pos(\xi)} |w|$.}
  \]
Here the $\max$ is taken over all the positions of $\xi$ (and not only over the $\alpha$-labeled ones).
  
Intuitively, for a given position $w \in \pos(\xi)$, we will mark the positions on the path from the root 
of $\xi$ down to $w$ by an additional symbol $\#$. 
For this, we define
  \begin{compactitem}
  \item the ranked alphabet $\Sigma_{\#}$ for each $k \in \mathbb{N}$  by $(\Sigma_{\#})^{(k)} = \{(\sigma,\#) \mid \sigma \in \Sigma^{(k)}\}$ and
    \item the ranked alphabet $\Delta$ for each $k \in \mathbb{N}$ by $\Delta^{(k)} = \Sigma^{(k)} \cup (\Sigma_{\#})^{(k)}$.
    \end{compactitem}

    In the set $\T_\Delta$ there are particular trees which are of interest for us: the $\#$-marked symbols occur exactly on a path from the root to some position. Formally, let $\zeta \in\T_\Delta$. We say that \emph{$\zeta$ shows a path} if, there exists $w \in \pos(\zeta)$ such that, for each $v \in \prefix(w)$, we have $\zeta(v) \in \Sigma_{\#}$ and, for each $v \in \pos(\zeta) \setminus \prefix(w)$, we have $\zeta(v) \in \Sigma$. If $\zeta$ shows a path, then the position $w$ which satisfies the above condition is uniquely determined. We denote this position by $w_\zeta$.

    The marking is done by the inverse of a tree relabeling. We define the $(\Delta,\Sigma)$-tree relabeling $\tau =(\tau_k \mid k \in \mathbb{N})$ for each $k \in \mathbb{N}$ and $\kappa \in \Delta^{(k)}$ as follows:
\[\tau_k(\kappa) =
\begin{cases}
\sigma & \text{ if $\kappa = (\sigma,\#)$ for some $\sigma \in \Sigma^{(k)}$} \\
\kappa & \text{ if $\kappa \in \Sigma^{(k)}$.}
\end{cases}
\]
Intuitively, $\tau$ drops $\#$.
It is easy to see that, for each $\xi \in \T_\Sigma$,  
\begin{equation}\label{equ:position-hash-marked-sequence}
  \text{the sets $\pos(\xi)$ and $\{\zeta \mid \zeta \in \tau^{-1}(\xi), \text{ $\zeta$ shows a path}\}$ are in a one-to-one correspondence.}
  \end{equation}

  We define the $\Natmaxplus$-representation
\[
  e = \tau\big(\REST(f) + g\big)
\]
where
\begin{compactitem}
\item $f = f_1 + f_2 + f_3$
  \item $f_1 = \bigplus\nolimits_{k\in \mathbb{N}, \sigma \in \Sigma^{(k)}, \widetilde{\gamma} \in \Sigma^k}
    (\RRT_{\Delta,\sigma,0} \times \NXT_{\Delta,\widetilde{\gamma},0})$,

    \item $f_2= \bigplus\nolimits_{k\in \mathbb{N}, \sigma \in \Sigma^{(k)}, i \in [k], \widetilde{\gamma} \in \Sigma^k} (\RRT_{\Delta,(\sigma,\#),1} \times \NXT_{\Delta,\widetilde{\gamma},0})$
    
    \item $f_3 = \bigplus\nolimits_{k\in \mathbb{N}, \sigma \in \Sigma^{(k)}, i \in [k], \widetilde{\gamma} \in \Sigma^{i-1}(\Sigma_{\#}) \Sigma^{k-i}} (\RRT_{\Delta,(\sigma,\#),1} \times \NXT_{\Delta,\widetilde{\gamma},0})$
  \item $g= \bigplus_{\kappa \in \Delta} \RRT_{\Delta,\kappa,0}$
  \end{compactitem}
  Intuitively, the combinations of root mappings and next mappings in $f_1$, $f_2$, and $f_3$ guarantee that $\supp(\sem{\REST(f)}) \subseteq \{\zeta \in \T_\Delta \mid \text{$\zeta$ shows a path}\}$.

  In particular, for each $\zeta \in \T_\Delta$, we have:
  \begin{compactitem}
  \item $\sem{f_1}(\zeta) = 0$ iff $\zeta(\varepsilon) \in \Sigma$ and, for each $j \in [\rk_\Sigma(\sigma)]$, we have $\zeta(j) \in \Sigma$,
  \item $\sem{f_2}(\zeta) = 1$ iff $\zeta(\varepsilon) \in \Sigma_\#$ and, for each $j \in [\rk_\Sigma(\sigma)]$, we have $\zeta(j) \in \Sigma$, and
 \item $\sem{f_3}(\zeta) = 1$ iff $\zeta(\varepsilon) \in \Sigma_\#$ and there exists exactly one $i \in [\rk_\Sigma(\sigma)]$ such that $\zeta(i) \in \Sigma_\#$ and, for each $j \in [\rk_\Sigma(\sigma)]\setminus \{i\}$, we have $\zeta(j) \in \Sigma$.
\end{compactitem}

  Then the following statements are easy to see.
  \begin{equation}\label{eq:height-representation-1}
    \text{For each $\zeta \in \T_\Delta$, we have 
      $\sem{\REST(f)}(\zeta) = \begin{cases}|w_\zeta|  &\text{if } \text{$\zeta$ shows a path}\\ -\infty & \text{otherwise}
        \end{cases}$ }
    \end{equation}
   
     \begin{equation}\label{eq:height-representation-3}
    \text{For each $\zeta \in \T_\Delta$,  we have $\sem{g}(\zeta) = 0$.}
  \end{equation}

  As combination of \eqref{eq:height-representation-1} and \eqref{eq:height-representation-3}, we obtain:
    \begin{equation}\label{eq:height-representation-2}
    \text{For each $\zeta \in \T_\Delta$, we have 
      $\sem{\REST(f)+g}(\zeta) = \begin{cases}|w_\zeta|  &\text{if } \text{$\zeta$ shows a path}\\ 0 & \text{otherwise}
        \end{cases}$ }
    \end{equation}

  Then we can calculate, for each $\xi \in \T_\Sigma$, as follows:
  \begingroup
  \allowdisplaybreaks
  \begin{align*}
    \sem{e}(\xi) &= \max_{\zeta \in \tau^{-1}(\xi)} \sem{\REST(f)+g}(\zeta)\\
                 &= \max_{\substack{\zeta \in \tau^{-1}(\xi):\\\text{ $\zeta$ shows a path}}} |w_\zeta|  \tag{by \eqref{eq:height-representation-2} and the fact that $\tau^{-1}(\xi)$ contains a tree which shows a path}\\
                 &= \max_{w \in \pos(\xi)} |w|  \tag{by \eqref{equ:position-hash-marked-sequence}} \\
    & = \height(\xi) \enspace.
    \end{align*}
    \endgroup
\hfill $\Box$
\end{example}

It turns out that each recognizable $(\Sigma,\B)$-weighted tree language is representable, but not vice versa as the next example shows.

\begin{example}\rm \label{ex:representable-not-recognizable} \cite[Ex.~5.2.1]{her20a} We consider the string ranked alphabet $\Sigma = \{\gamma^{(1)}, \alpha^{(0)}\}$ and the semiring $\Nat=(\mathbb{N},+,\cdot,0,1)$ of natural numbers. Moreover, we let $r: \T_\Sigma \to \mathbb{N}$ such that $r(\gamma^n(\alpha)) = 2^{(n+1)^2}$ for each $n \in \mathbb{N}$. 

We show that there does not exist a $(\Sigma,\Nat)$-wta $\cA$ such that $\sem{\cA} = r$. For this, we assume that there exists such a $(\Sigma,\Nat)$-wta. Then, by Lemma \ref{lm:N-wta-upper bound}, there exists $K \in \mathbb{N}$ such that, for each $n \in \mathbb{N}$, we have $\sem{\cA}(\gamma^n(\alpha)) = \sem{e}(\gamma^{n}(\alpha)) = 2^{(n+1)^2} \le K^{n+1}$. However, this is a contradiction, because if $n$ is big enough, then $2^{(n+1)^2} > K^{n+1}$.

  Next we consider the $\mathbb{N}$-representation
  \[
e = \REST(\RRT_{\Sigma,\gamma,2} + \RRT_{\Sigma,\alpha,2}) \times \REST(\REST(\RRT_{\Sigma,\gamma,4} + \RRT_{\Sigma,\alpha,1}))
\]
Let $n \in \mathbb{N}$. Then we can calculate as follows (where we denote by $\prod$ the extension of the binary operation $\cdot$ to finite numbers of arguments):
\begingroup
\allowdisplaybreaks
\begin{align*}
  & \sem{\REST(\RRT_{\Sigma,\gamma,2} + \RRT_{\Sigma,\alpha,2})}(\gamma^n(\alpha))\\
  =& \prod_{w \in \pos(\xi)} \sem{\RRT_{\Sigma,\gamma,2} + \RRT_{\Sigma,\alpha,2}}(\gamma^n(\alpha)|_w)
  \tag{note that $\cdot$ is commutative}\\
  =&  \sem{\RRT_{\Sigma,\gamma,2} + \RRT_{\Sigma,\alpha,2}}(\alpha) \cdot \prod_{w \in \pos(\xi)\setminus\{1^n\}} \sem{\RRT_{\Sigma,\gamma,2} + \RRT_{\Sigma,\alpha,2}}(\gamma^n(\alpha)|_w)\\
  =&  \sem{\RRT_{\Sigma,\alpha,2}}(\alpha) \cdot \prod_{w \in \pos(\xi)\setminus\{1^n\}} \sem{\RRT_{\Sigma,\gamma,2}}(\gamma^n(\alpha)|_w)\\
  =& 2 \cdot \prod_{w \in \pos(\xi)\setminus\{1^n\}} 2 = 2^{n+1} \enspace.
 \end{align*}
 \endgroup
 In a similar way we can calculate that $\sem{\REST(\RRT_{\Sigma,\gamma,4} + \RRT_{\Sigma,\alpha,1})}(\gamma^n(\alpha)) = 4^n$. Using Gaussian sum, we have
  \begingroup
\allowdisplaybreaks
\begin{align*}
  \REST(\REST(\RRT_{\Sigma,\gamma,4} + \RRT_{\Sigma,\alpha,1}))(\gamma^n(\alpha))
  &= \prod_{w \in \pos(\xi)} \REST(\RRT_{\Sigma,\gamma,4} + \RRT_{\Sigma,\alpha,1})(\gamma^n(\alpha)|_w)\\
  &= 1 \cdot 4^1 \cdot 4^2 \cdot \ldots \cdot 4^n = 4^{\frac{n^2+n}{2}} = 2^{n^2+n}\enspace.
  \end{align*}
  \endgroup
  Finally, $\sem{e}(\gamma^n(\alpha)) = 2^{n+1}  \cdot 2^{n^2+n} = 2^{(n+1)^2}$.
  \hfill$\Box$
\end{example}

Hence, for a characterization of $\Rec(\Sigma,\B)$ in terms of representations, we have to restrict the set $\Rep(\Sigma,\B)$ (cf. \cite[Sec.~5.2]{her20a}). We define the restriction in a purely syntactic way (in contrast to \cite{her17,her20a}). In fact, we define two restrictions: restricted representations and $\times$-restricted representations.  Intuitively,
\begin{compactitem}
\item in a restricted representation, (a) $\REST$ must not occur nested
  and (b) each tree relabeling occurring in a subexpression of the form $\REST(e)$ must be non-overlapping, and
  \item a $\times$-restricted representation is a restricted representation such that, in each subexpression of the form $e_1 \times e_2$, there exists at least one $i \in \{1,2\}$ such that $e_i$ does not contain $\REST$ and each tree relabeling in $e_i$ is non-overlapping.
  \end{compactitem}

\index{RepExB@$\RepEx^{r}(\B)$}
Formally, let $\RepEx^r(\B)$ be the subset of $\RepEx(\B)$ generated by the following EBNF; it uses the infinite family $(e_\Sigma^r \mid \Sigma \text{ ranked alphabet}) \cup (f_\Sigma^r \mid \Sigma \text{ ranked alphabet})$ of nonterminals, each $e_\Sigma^r$ is an initial nonterminal, and for every nonterminals $e_\Sigma^r$ and  $f_\Sigma^r$ it has the following rules:
\begin{eqnarray}
  \begin{aligned}
  e_\Sigma^r ::= &\ \RRT_{\Sigma,\sigma,b} \mid \NXT_{\Sigma,\widetilde{\gamma},b} \mid e_\Sigma^r + e_\Sigma^r \mid e_\Sigma^r \times e_\Sigma^r \mid \tau(e_\Delta^r) \mid \REST(f_\Sigma^r) \\
  f_\Sigma^r ::= &\ \RRT_{\Sigma,\sigma,b} \mid \NXT_{\Sigma,\widetilde{\gamma},b} \mid f_\Sigma^r + f_\Sigma^r \mid f_\Sigma^r \times f_\Sigma^r \mid \tau'(f_\Delta^r)
  \end{aligned}\label{eq:syntax-restricted-representation}
\end{eqnarray}
where $\sigma \in \Sigma$, $\widetilde{\gamma} \in \Sigma^k$ for some $k \in \mathbb{N}_+$, $b \in B$, $\tau$ is a  $(\Delta,\Sigma)$-tree relabeling, and  $\tau'$ is a  non-overlapping  $(\Delta,\Sigma)$-tree relabeling. We call each  element of  $\RepEx^r(\B)$ a \emph{restricted representation}.
\index{restricted representation}

\index{RepExtimesB@$\RepEx^{\times r}(\B)$}
Let $\RepEx^{\times r}(\B)$ be the subset of $\RepEx^r(\B)$ generated by the following EBNF; it uses the infinite family $(e_\Sigma^{\times r} \mid \Sigma \text{ ranked alphabet}) \cup (f_\Sigma^r \mid \Sigma \text{ ranked alphabet})$ of nonterminals, each $e_\Sigma^r$ is an initial nonterminal, and for every nonterminals $e_\Sigma^{\times r}$ and  $f_\Sigma^r$ it has the following rules:
\begin{eqnarray}
  \begin{aligned}
  e_\Sigma^{\times r} ::= & \ \RRT_{\Sigma,\sigma,b} \mid \NXT_{\Sigma,\widetilde{\gamma},b} \mid e_\Sigma^{\times r} + e_\Sigma^{\times r} \mid e_\Sigma^{\times r} \times f_\Sigma^r \mid f_\Sigma^r \times e_\Sigma^{\times r} \mid f_\Sigma^r \times f_\Sigma^r \mid \tau(e_\Delta^{\times r}) \mid \REST(f_\Sigma^r) \\
  f_\Sigma^r ::= & \ \RRT_{\Sigma,\sigma,b} \mid \NXT_{\Sigma,\widetilde{\gamma},b} \mid f_\Sigma^r + f_\Sigma^r \mid f_\Sigma^r \times f_\Sigma^r \mid \tau'(f_\Delta^r)
                   \end{aligned}\label{eq:syntax-x-restricted-representation}
\end{eqnarray}
where $\sigma \in \Sigma$, $\widetilde{\gamma} \in \Sigma^k$ for some $k \in \mathbb{N}_+$, $b \in B$, $\tau$ is a  $(\Delta,\Sigma)$-tree relabeling, and  $\tau'$ is a non-overlapping  $(\Delta,\Sigma)$-tree relabeling.  We call each  element of  $\RepEx^{\times r}(\B)$ a \emph{$\times$-restricted representation}.
\index{$\times$-restricted representation}
We note that the rules for the nonterminals $f_\Sigma^r$ in \eqref{eq:syntax-restricted-representation} and \eqref{eq:syntax-x-restricted-representation} are identical. Clearly, $\RepEx^{\times r}(\B) \subset \RepEx^{r}(\B) \subset \RepEx(\B)$.

\index{restricted representable}
\index{$\times$-restricted representable}
  A weighted tree language $r: \T_\Sigma \to B$ is \emph{restricted representable} (and \emph{$\times$-restricted representable}) if there exists an $e \in \RepEx^r(\Sigma,\B)$  (and an $e \in \RepEx^{\times r}(\Sigma,\B)$, respectively) such that $r = \sem{e}$.

  We remark that, in \cite{her17,her20a}, restricted representations are defined in a less strict way as here: a representation $e$ is restricted in the sense of \cite{her17,her20a} if, whenever $e$ contains a subexpression of the form $\REST(e')$, then $\sem{e'}$ is a recognizable step mapping (and similarly for $\times$-restricted representations). Thus, in particular, such a restricted representation may contain nested occurrences of $\REST$ (which is not possible in our definition of restricted representation). The expressive power of both forms of restricted representations is equal: they characterize the set of run recognizable weighted tree languages. The variant which we defined above has the advantage that the corresponding characterization result (cf. Theorem~\ref{thm:Medvedjev}) is constructive. The constructivity is guaranteed by the next corollary (and by the constructivity of a number of closure results).

\begin{corollary}\rm \label{cor:char-restricted-representable} Let $e \in \RepEx^r(\Sigma,\B)$ be generated  by the nonterminal $f_\Sigma^r$ using the rules in~\eqref{eq:syntax-restricted-representation}. Then we can construct $n \in \mathbb{N}_+$, $b_1,\ldots,b_n \in B$, and  $\Sigma$-fta $A_1,\ldots,A_n$ such that $\sem{e} = \bigoplus_{i \in [n]} b_i \cdot \chi(\LL(A_i))$. Thus, in particular, $\sem{e}$ is a recognizable step mapping. 
\end{corollary}
\begin{proof} We prove the statement by induction on $(\RepEx(\B)',\succ')$, where $\RepEx(\B)'$ is the set of all representations which can be generated  by nonterminals of the form $f_\Sigma^r$ using the rules in~\eqref{eq:syntax-restricted-representation}, and $\succ' = \succ|_{\RepEx(\B)' \times \RepEx(\B)'}$.

I.B.: For each $e \in \RepEx(\B)'$ of the form $e= \RRT_{\Sigma,\sigma,b}$ or $e = \NXT_{\Sigma,\widetilde{\gamma},b}$,  we can easily construct the $\Sigma$-fta $A_{\Sigma,\sigma}$ and $A_{\Sigma,\widetilde{\gamma}}$ such that $\LL(A_{\Sigma,\sigma})= L_{\Sigma,\sigma}$ and $\LL(A_{\Sigma,\widetilde{\gamma}}) = L_{\Sigma,\widetilde{\gamma}}$. 
Then $\sem{\RRT_{\Sigma,\sigma,b}} = b \cdot \chi(\LL(A_{\Sigma,\sigma}))$ and $\sem{\NXT_{\Sigma,\widetilde{\gamma},b}} = b \cdot \chi(\LL(A_{\Sigma,\widetilde{\gamma}}))$.

I.S.:  It follows from I.H., Theorem \ref{thm:crisp-det-algebra}(B)$\Rightarrow$(A), and
\begin{compactitem}
\item Theorem \ref{thm:closure-sum}(2) (for the sum), 
\item Theorem~\ref{thm:closure-Hadamard-product}(3) (for the Hadamard product), and  
\item Theorem \ref{thm:closure-under-tree-relabeling} (for the non-overlapping tree relabeling).  \qedhere
\end{compactitem}
\end{proof}

For instance, the $\Natmaxplus$-representation
\[
e = \REST\Big(\Big(\bigplus\nolimits_{\kappa \in \Sigma}\NXT_{\Sigma,\kappa\alpha,0}\Big) \times \RRT_{\Sigma,\sigma,1} + \bigplus\nolimits_{\kappa \in \Sigma} \RRT_{\Sigma,\kappa,0}\Big)
\] shown in  Example \ref{ex:recognizable-representable} is $\times$-restricted.

For each semiring $\B$, we will prove that the set
of $(\Sigma,\B)$-recognizable weighted tree languages can be characterized by $\times$-restricted representations (cf. Theorem~\ref{thm:Medvedjev}). On the other hand, for each commutative semiring $\B$, the set $\Rec(\Sigma,\B)$ is characterized by polynomial weighted tree languages
and rational operations (cf. Theorem~\ref{thm:Kleene}). Thus, one might ask how the Kleene stars are expressed in terms of $\times$-restricted representations. In the next example, we illustrate this possibility in an ad hoc way.

\begin{example}\rm \label{ex:abn-arctic} We consider the alphabet $\Gamma = \{a,b\}$ and the string ranked alphabet $\Gamma_e = \{a^{(1)}, b^{(1)}, e^{(0)}\}$. Moreover, we consider the arctic semiring $\Nat_{\max,+} = (\mathbb{N}_{-\infty},\max,+,-\infty,0)$. We would like to construct a restricted representation $f \in \RepEx(\Sigma,\Nat_{\max,+})$ such that
  \[
    \sem{f} = \big(\ 1. a(b(e)) \ \big)_e^* \enspace,
  \]
  i.e., $\sem{f}$ is the $e$-Kleene star of the polynomial $(\Gamma_e,\Nat_{\max,+})$-weighted tree language  $1.a(b(e))$.
  Recall that, by \eqref{equ:what-we-want-to-prove}, 
   for each $\xi \in \T_{\Gamma_e}$, we have 
  \[
\big(\ 1. a(b(e)) \ \big)_e^*(\xi) =  \begin{cases} |\pos_a(\xi)| & \text{ if there exists $n \in \mathbb{N}$ such that $\xi = \tree_e((ab)^n)$}\\
                 -\infty &\text{ otherwise,}
                 \end{cases}
               \]
               where  $\tree_e: \Gamma^* \to \T_{\Gamma_e}$ is the bijection defined on page \pageref{page:def-of-mapping-tree}.
              In particular,  $\big(\ 1. a(b(e)) \ \big)_e^*(\tree_e((ab)^n)) = n$ for each $n \in \mathbb{N}$ and, e.g., $\big(\ 1. a(b(e)) \ \big)_e^*(a(a(e))) = -\infty$.

  For this purpose, we propose the $\Nat_{\max,+}$-representation
  \begin{align*}
    f &= (\RT_{\Gamma_e,a,0}+ \RT_{\Gamma_e,e,0}) \times \REST(h) \ \text{ with }\\
    h &= \big(\RRT_{\Gamma_e,a,1} \times \NXT_{\Gamma_e,b,0}\big) \  + \ \big( \RRT_{\Gamma_e,b,0} \times \NXT_{\Gamma_e,a,0}\big) \  + \ \big( \RRT_{\Gamma_e,b,0} \times \NXT_{\Gamma_e,e,0}\big)\ + \  \RRT_{\Gamma_e,e,0}\enspace.
\end{align*}
Clearly, $f$ is $\times$-restricted. We claim that $\sem{f} = (1. a(b(e)) )_e^*$.

Let $\xi \in \T_{\Gamma_e}$.
\begingroup
\allowdisplaybreaks
\begin{align*}
  \sem{f}(\xi) &= \max\big(\sem{\RT_{\Gamma_e,a,0}}(\xi) \ , \ \sem{\RT_{\Gamma_e,e,0}}(\xi)\big) + \bigplus_{w \in \pos(\xi)} \sem{h}(\xi|_w)\\
               &= \max\big(\sem{\RT_{\Gamma_e,a,0}}(\xi) \ , \ \sem{\RT_{\Gamma_e,e,0}}(\xi)\big) + \bigplus_{w \in \pos_a(\xi)} \sem{h}(\xi|_w) \ + \bigplus_{w \in \pos_b(\xi)} \sem{h}(\xi|_w) \ + \bigplus_{w \in \pos_e(\xi)} \sem{h}(\xi|_w) \tag{since $+$ is commutative}\\
  &= \begin{cases} 0 & \text{ if $\xi(\varepsilon)  \in \{a,e\}$}\\
                 -\infty & \text{ otherwise}
               \end{cases} \\
  &\ \ \  + \begin{cases} |\pos_a(\xi)| & \text{ if, for every $w \in \pos_a(\xi)$: $\xi(w1)=b$}\\
                 -\infty &\text{ otherwise}
               \end{cases}\\
  &\ \ \ + \begin{cases} 0 & \text{ if, for every $w \in \pos_b(\xi)$: $\xi(w1)=a$ or $\xi(w1)=e$}\\
                 -\infty &\text{ otherwise}
               \end{cases}\\
  &\ \ \ + 0\\[3mm]
               &= \begin{cases} |\pos_a(\xi)| & \text{ if $\xi(\varepsilon) \in \{a,e\}$ and for every $w \in \pos(\xi)$:}\\
                   & \text{ \big($\xi(w)=a$ implies $\xi(w1)=b$\big) and }\\
                 & \text{ \big($\xi(w)=b$ implies  ($\xi(w1)=a$ or $\xi(w1)=e$)\big)}\\
                 -\infty &\text{ otherwise}
                 \end{cases}\\[2mm]
               &= \begin{cases} |\pos_a(\xi)| & \text{ if there exists $n \in \mathbb{N}$ such that $\xi = \tree_e((ab)^n)$}\\
                 -\infty&\text{ otherwise} \enspace.
                 \end{cases}
\end{align*}
\endgroup
Hence $f$ is the desired $\Nat_{\max,+}$-representation.
\hfill $\Box$
  \end{example}

\section[The first main result for $\RepEx(\Sigma,\B)$]{The main result for $\RepEx(\Sigma,\B)$ over semirings}

  Now we can formulate the first main result of this chapter: the M{\'e}dv{\'e}dj{\'e}v characterization of wta over semirings.

  \begin{theorem-rect} \label{thm:Medvedjev} {\rm (cf. \cite[Thm.~6]{her17} and \cite[Thm.~5.2.2]{her20a})} Let $\Sigma$ be a ranked alphabet. Moreover, let  $\B=(B,\oplus,\otimes,\0,\1)$ be a semiring and let $r:~\T_\Sigma~\to~B$. Then the following two statements are equivalent.
    \begin{compactenum}
    \item[(A)] We can construct a $(\Sigma,\B)$-wta $\cA$ such that $\sem{\cA} = r$.
      \item[(B)] We can construct an $e \in \RepEx^{\times r}(\Sigma,\B)$ such that $\sem{e} = r$. 
      \end{compactenum}
      Moreover, if $\B$ is commutative, then (A) and (B) are equivalent to:
       \begin{compactenum}
    \item[(C)] We can construct an $e \in \RepEx^{r}(\Sigma,\B)$ such that $\sem{e} = r$. 
      \end{compactenum}
    \end{theorem-rect}
    The proof of this theorem follows from Lemmas \ref{lm:rec-rep} and \ref{lm:rep-rec}, which will be shown in the next two sections. 

        We note that in \cite{her17} and \cite[Sect.~5.3]{her20}, a comparison between representable weighted tree languages and weighted tree languages defined by the weighted MSO-logic of \cite{drogas05,drogas07,drogas09} has been started.


      \subsection{From recognizable to restricted representable}
    \label{sec:Med-rec-rep}

      Since each wta can be decomposed into a deterministic tree relabeling and a weighted local system (cf. Theorem \ref{thm:decomposition-2}) and the set $\Rep(\_,\B)$ is closed under tree relabelings (by definition), we first prove that each weighted tree language determined by a weighted local system is $\times$-restricted representable.

      \begin{lemma}\rm \label{lm:wls-representable} Let $\cS$ be a $(\Sigma,\B)$-wls. Then we can construct an $e \in \RepEx^{\times r}(\Sigma,\B)$ such that $\sem{e}=\sem{\cS}$. 
        \end{lemma}
        \begin{proof} Let $\cS=(g,F)$ be a $(\Sigma,\B)$-wls.  The idea for the construction of $e \in \RepEx^{\times r}(\Sigma,\B)$ is to simulate the weight $g_k((\sigma_1 \cdots \sigma_k,\sigma))$ of the fork $(\sigma_1 \cdots \sigma_k,\sigma)$ by the representation 
          $\NXT_{\Sigma,\sigma_1\ldots\sigma_k,\1}  \times \RRT_{\Sigma,\sigma,g_k(\sigma_1 \cdots \sigma_k,\sigma)}$ if $k \ge 1$, and by the representation $\RRT_{\Sigma,\sigma,g_0(\varepsilon,\sigma)}$ if $k=0$.
          
Formally, we define the $\B$-representation \(e = \REST(e_1 + e_2) \times e_3 \) where
          \[
            e_1 = \bigplus\nolimits_{\sigma \in \Sigma^{(0)}} \RRT_{\Sigma,\sigma,g_0(\varepsilon,\sigma)}
            \ \ \text{ and } \ \
            e_2 = \bigplus\nolimits_{\substack{k\ge 1, \sigma \in \Sigma^{(k)},\\\sigma_1,\ldots,\sigma_k \in \Sigma}} (\NXT_{\Sigma,\sigma_1\cdots\sigma_k,\1}  \times \RRT_{\Sigma,\sigma,g_k(\sigma_1 \cdots \sigma_k,\sigma)})
           \]
           \[
              \text{ and }  \ \ 
  e_3 = \bigplus\nolimits_{\sigma \in \Sigma} \RRT_{\Sigma,\sigma,F(\sigma)} \enspace.
             \]
            It is easy to see that $e$ is $\times$-restricted.
             It remains to show that $\sem{\cS} = \sem{e}$.
          Obviously, 
          \begin{equation}
\text{ for every $\xi \in \T_\Sigma$, we have $\sem{e_3}(\xi) = F(\xi(\varepsilon))$.} \label{eq:root-wls}
            \end{equation}
            Moreover,
            \begin{eqnarray}
              \begin{aligned}
      &\text{for every $\xi \in \T_\Sigma$ and $w \in \pos(\xi)$, we have}\\
      &\text{$\sem{e_1 +e_2}(\xi|_w) = g_{\rk(\xi(w))}(\xi(w1) \cdots \xi(w\, \rk(\xi(w))),\xi(w))$.}
      \end{aligned}\label{eq:inner-wls}
            \end{eqnarray}

            Then, for each $\xi \in \T_\Sigma$, we can calculate as follows.
          \begingroup
          \allowdisplaybreaks
          \begin{align*}
            \sem{\REST(e_1 + e_2) \times e_3}(\xi) 
            &= \Big(\bigotimes_{\substack{w \in \pos(\xi),\\\text{in $<_\mathrm{dp}$ order}}} \sem{e_1 +e_2}(\xi|_w)\Big)    \otimes \sem{e_3}(\xi)\\
            &= \Big(\bigotimes_{\substack{w \in \pos(\xi),\\\text{in $<_\mathrm{dp}$ order}}} g_{\rk(\xi(w))}\big(\xi(w1) \cdots \xi(w\, \rk(\xi(w))),\xi(w)\big) \Big)    \otimes \sem{e_3}(\xi) \tag{by \eqref{eq:inner-wls}}\\
            &= g(\xi) \otimes F(\xi(\varepsilon)) \tag{by definition of $g$ and \eqref{eq:root-wls}}\\
            &=  \sem{\cS}(\xi) \enspace. \qedhere
          \end{align*} 
          \endgroup
          \end{proof}

        \begin{lemma} \label{lm:rec-rep}\rm (cf. \cite[Lm.~12]{her17} and \cite[Thm.~5.2.8]{her20a}) Let $\cA$ be a $(\Sigma,\B)$-wta. Then we can construct an $e \in \RepEx^{\times r}(\Sigma,\B)$ such that $\runsem{\cA} = \sem{e}$.
        \end{lemma}
        \begin{proof} By Theorem \ref{thm:decomposition-2}, we can construct  a ranked alphabet $\Theta$, a deterministic $(\Theta,\Sigma)$-tree relabeling $\tau$, and a $(\Theta,\B)$-weighted local system $\cS$ with unit root weights
          such that $\runsem{\cA} = \chi(\tau)(\sem{\cS})$. By Lemma~\ref{lm:wls-representable}, we can construct an $e' \in \RepEx^{\times r}(\Theta,\B)$ such that $\sem{e'} = \sem{\cS}$. Since $e'$ is a $\times$-restricted representation, also $\tau(e')$ is a $\times$-restricted representation. Moreover, $\runsem{\cA}  = \chi(\tau)(\sem{\cS})  = \chi(\tau)(\sem{e'}) = \sem{\tau(e')}$. Thus we can choose $e= \tau(e')$.
        \end{proof}

    
    \subsection{From restricted representable to recognizable}
    \label{sec:Med-rep-rec}

    Here we prove that each $\times$-restricted representable weighted tree language is recognizable; moreover, if $\B$ is commutative, then each  restricted representable weighted tree language is recognizable (cf. Lemma~\ref{lm:rep-rec}). As to be expected, the proof is by induction on the structure of representations. 

    We start with a lemma which shows the effect of the restriction of recognizable step mappings. 
One might compare the construction of this lemma with the one of \cite[Lm.~5.5]{drovog06} (based in \cite[Lm.~4.4]{drogas07}) in which the following is proved for a  formula $\varphi$ of weighted monadic second order logic: if $\sem{\varphi}$ is a recognizable step mapping, then $\sem{\forall x.\varphi}$ is recognizable (cf. Lemma~\ref{lm:univ-fo-quantification-rec-step}).

\begin{lemma}\rm \label{lm:rest-rec-step-mapping} (cf. \cite[Lm.~10]{her17} and \cite[Lm.~5.2.5]{her20a})
Let $e \in \RepEx(\Sigma,\B)$ such that there exists a crisp-deterministic $(\Sigma,\B)$-wta $\cA$ with $\sem{\cA}=\sem{e}$.
We can construct a bu-deterministic $(\Sigma,\B)$-wta $\cB$ such that $\runsem{\cB} = \sem{\REST(e)}$.
\end{lemma}
\begin{proof} Let $\cA=(Q,\delta,F)$.  By Lemma \ref{lm:fin-algebra-crisp-det-wta}(2), we have
\begin{equation}
  \text{for every $\xi \in \T_\Sigma$ and $w \in \pos(\xi)$, we have } \sem{\cA}(\xi|_w) = F_{\crhorm_\xi(w)} \enspace. \label{eq:value-of-single-run}
\end{equation}
We recall that $\estate(\xi|_w) = \crhorm_\xi(w)$.
By the definition of the canonical run (cf. Subsection \ref{subsect:annihilation-bu-det-wta}) and of the state algebra for a crisp-deterministic wta (cf. Section~\ref{sec:state-algebra-of-wta}) and by Lemmas \ref{lm:fin-algebra-crisp-det-wta}(1) and \ref{lm:total-bu-det-sem}(1), the following holds: 
    \begin{eqnarray}
      \begin{aligned}
        &\text{for every $\xi \in \T_\Sigma$ and $w \in \pos(\xi)$, we have}\\
        &\text{$\delta_{\rk(\xi(w))}(\crhorm_\xi(w1) \cdots \crhorm_\xi(w \, \rk(\xi(w))),\xi(w),\crhorm_\xi(w))=\1$ and  $\wt_\cA(\xi|_w,(\crhorm_\xi)|_w) = \1$.}
        \end{aligned} \label{eq:transition-weight=1}
\end{eqnarray}

      We construct the $(\Sigma,\B)$-wta $\cB$ such that $\runsem{\cB} = \sem{\REST(e)}$ using the following idea. On a given input tree $\xi$, the wta $\cB$ simulates the state behaviour of $\cA$ and, at each position $w$ of $\xi$, the weight of the transition $t$ which $\cA$ applies at $w$, is the product of $\delta_k(t) \in \{\0,\1\}$ (where $k = \rk(\xi(w))$) and the root weight of the target state of $t$. 
Formally, we let $\cB = (Q,\delta',F')$ with $(F')_q = \1$ for each $q \in Q$, and for every $k \in \mathbb{N}$, $\sigma \in \Sigma^{(k)}$, and $q,q_1,\ldots,q_k \in Q$ we let
      \[
(\delta')_k(q_1 \cdots q_k,\sigma,q) = \delta_k(q_1 \cdots q_k,\sigma,q) \otimes F_q\enspace.
\]
We recall that $\delta_k(q_1 \cdots q_k,\sigma,q) \in \{\0,\1\}$.
Obviously, $\cB$ is bu-deterministic, but it is not crisp-deterministic, and not even total. 
 Moreover, for each $\xi \in \T_\Sigma$, we have $\R_\cA(\xi) = \R_\cB(\xi)$ and
        \begin{equation}
          \text{for each $\rho \in \R_\cB(\xi)$: if $\rho \ne \crhorm_\xi$, then $\wt_\cB(\xi,\rho) =\0$.}\label{eq:can-get-worse}
        \end{equation}
Then, for each $\xi \in \T_\Sigma$, we can calculate as follows:
        \begingroup
        \allowdisplaybreaks
        \begin{align*}
          \runsem{\cB}(\xi) &= \wt_\cB(\xi,\crhorm_\xi) \tag{by \eqref{eq:can-get-worse} and the definition of $F'$}\\
                            &= \bigotimes_{\substack{w \in \pos(\xi)\\ \text{in $<_{\mathrm{dp}}$ order }}} (\delta')_{\rk(\xi(w))}(\crhorm_\xi(w1) \cdots \crhorm_\xi(w \, \rk(\xi(w))),\xi(w),\crhorm_\xi(w))
          \tag{by Observation \ref{obs:weight-run-explicit}}\\
                            &= \bigotimes_{\substack{w \in \pos(\xi)\\ \text{in $<_{\mathrm{dp}}$ order }}} \delta_{\rk(\xi(w))}(\crhorm_\xi(w1) \cdots \crhorm_\xi(w \, \rk(\xi(w))),\xi(w),\crhorm_\xi(w)) \otimes F_{\crhorm_\xi(w)}
          \tag{by the definition of $\delta'$}\\
&= \bigotimes_{\substack{w \in \pos(\xi)\\ \text{in $<_{\mathrm{dp}}$ order }}}  F_{\crhorm_\xi(w)}
          \tag{by \eqref{eq:transition-weight=1}}\\
                         &= \bigotimes_{\substack{w \in \pos(\xi)\\ \text{in $<_{\mathrm{dp}}$ order }}}  \sem{\cA}(\xi|_w) \tag{by \eqref{eq:value-of-single-run}}\\
          &=  \bigotimes_{\substack{w \in \pos(\xi)\\ \text{in $<_{\mathrm{dp}}$ order }}}  \sem{e}(\xi|_w)
          = \sem{\REST(e)}(\xi) \enspace. \qedhere
        \end{align*}
        \endgroup
      \end{proof}

      In the next lemma, we will need distributivity, because the Hadamard product is one of the elementary operations, and distributivity is needed in order to guarantee closure of the set of recognizable weighted tree languages under Hadamard product.

    \begin{lemma} \label{lm:rep-rec}\rm (cf. \cite[Lm.~11]{her17} and \cite[Thm.~5.2.7]{her20a}) Let $\B$ be a semiring. Then the following two statements hold.
    \begin{compactenum}
    \item[(1)] For every ranked alphabet $\Sigma$ and $e \in \RepEx^{\times r}(\Sigma,\B)$, we can construct a $(\Sigma,\B)$-wta $\cA$ such that $\sem{\cA} = \sem{e}$.
    \item[(2)] Let $\B$ be commutative.  For  every ranked alphabet $\Sigma$ and  $e \in \RepEx^{r}(\Sigma,\B)$, we can construct a $(\Sigma,\B)$-wta $\cA$ such that $\sem{\cA} = \sem{e}$.
      \end{compactenum}
    \end{lemma}
\begin{proof} Proof of (1): We prove the statement  by induction on the terminating reduction system $(\RepEx^{\times r}(\B),\succ')$ which is defined in an obvious way as restriction of $(\RepEx(\B),\succ)$.  
      
      I.B.: Let $e= \RRT_{\Sigma,\sigma,b}$ or $e=\NXT_{\Sigma,\widetilde{\gamma},b}$.  Then the statement follows from Corollary \ref{cor:char-restricted-representable} and Theorem~\ref{thm:crisp-det-algebra}(B)$\Rightarrow$(A).

I.S.: We distinguish four cases.

      \underline{Case (a):} Let  $e = e_1 + e_2$  for some $e_1,e_2 \in \RepEx^{\times r}(\Sigma,\B)$. By I.H. we can construct $(\Sigma,\B$-wta $\cA_1$ and $\cA_2$ such that $\sem{e_1}=\sem{\cA_1}$ and $\sem{e_2}=\sem{\cA_2}$.  Then by Theorem \ref{thm:closure-sum}(1) we can construct a $(\Sigma,\B)$-wta $\cA$ such that $\sem{\cA} = \sem{\cA_1} \oplus \sem{\cA_2}$. Thus $\sem{\cA} = \sem{\cA_1} \oplus \sem{\cA_2} = \sem{e_1} \oplus \sem{e_2} = \sem{e_1 + e_2} = \sem{e}$.
      
      \underline{Case (b):} Let $e = e_1 \times e_2$  for some $e_1,e_2 \in \RepEx^{\times r}(\Sigma,\B)$. We distinguish three cases.

      Case (b1): Let  $e_1$ be generated by the nonterminal $e_\Sigma^r$ and let $e_2$ be generated by the nonterminal $f_\Sigma^r$ using the rules in~\eqref{eq:syntax-restricted-representation}. By I.H. we can construct a $(\Sigma,\B)$-wta $\cA_1$ such that $\sem{\cA_1}=\sem{e_1}$. 
      By Corollary \ref{cor:char-restricted-representable}, we can construct $n \in \mathbb{N}_+$, $b_1,\ldots,b_n \in B$, and  $\Sigma$-fta $A_1,\ldots,A_n$ such that $\sem{e_2} = \bigoplus_{i \in [n]} b_i \cdot \chi(\LL(A_i))$.
      By  Theorem \ref{thm:crisp-det-algebra}(B)$\Rightarrow$(A), we can construct a crisp-deterministic $(\Sigma,\B)$-wta $\cA_2$ such that $\sem{\cA_2} = \sem{e_2}$.
      Since $\B$ is right-distributive, by Theorem \ref{thm:closure-Hadamard-product-char}(1), we can construct a $(\Sigma,\B)$-wta $\cA$ such that $\sem{\cA} = \sem{\cA_1} \otimes \sem{\cA_2}$. Then  $\sem{\cA} = \sem{\cA_1} \otimes \sem{\cA_2} = \sem{e_1} \otimes \sem{e_2} = \sem{e_1 \times e_2} =\sem{e}$.

Case (b2): Let $e_1$ be generated by the nonterminal $f_\Sigma^r$ and  let $e_2$ be generated by the nonterminal $e_\Sigma^r$  using the rules in~\eqref{eq:syntax-restricted-representation}. By I.H. we can construct  a $(\Sigma,\B)$-wta $\cA_2$  such that $\sem{\cA_2}= \sem{e_2}$. 
As for $e_2$ in  Case (b1), we can construct a crisp-deterministic $(\Sigma,\B)$-wta $\cA_1$ such that $\sem{\cA_1} = \sem{e_1}$. Then, by using Theorem \ref{thm:closure-Hadamard-product-char}(4) (which assumes left-distributivity)
instead of  Theorem \ref{thm:closure-Hadamard-product-char}(1), we can finish the proof in the same way as in Case (b1).

Case (b3): Let $e_1$ and $e_2$ be generated by the nonterminal $f_\Sigma^r$ using the rules in~\eqref{eq:syntax-restricted-representation}.
As indicated in the previous cases, we can construct crisp-deterministic $(\Sigma,\B)$-wta $\cA_1$ and $\cA_2$ such that $\sem{e_1} = \sem{\cA_1}$ and $\sem{e_2} = \sem{\cA_2}$. By Theorem \ref{thm:closure-Hadamard-product}(3), we can construct a crisp-deterministic $(\Sigma,\B)$-wta $\cA$ such that $\sem{\cA} = \sem{\cA_1} \otimes \sem{\cA_2}$, and then we can finish as in Case (b1).

\underline{Case (c):} Let $e = \tau(e')$ for some $e' \in \RepEx^{\times r}(\Delta,\B)$ and  $(\Delta,\Sigma)$-tree relabeling $\tau$. By I.H. we can construct a $(\Delta,\B)$-wta $\cA'$ such that $\sem{\cA'} = \sem{e'}$.
Then  by Theorem  \ref{thm:closure-under-tree-relabeling}, we can construct a $(\Sigma,\B)$-wta $\cA$ such that $\sem{\cA} = \chi(\tau)(\sem{\cA'})$. Then $\sem{\cA} = \chi(\tau)(\sem{\cA'}) = \chi(\tau)(\sem{e'}) =  \sem{\tau(e')} = \sem{e}$.

\underline{Case (d):} Let $e=\REST(e')$  for some $e' \in \RepEx^{\times r}(\Sigma,\B)$
which is generated by the nonterminal $f_\Sigma^r$ using the rules in~\eqref{eq:syntax-restricted-representation}. By Corollary \ref{cor:char-restricted-representable} and Theorem \ref{thm:crisp-det-algebra}$(B) \Rightarrow (A)$, we can construct a crisp-deterministic $(\Sigma,\B)$-wta $\cB$ such that $\sem{\cB} = \sem{e'}$. Then, by Lemma \ref{lm:rest-rec-step-mapping},  we can construct a $(\Sigma,\B)$-wta $\cA$ such that $\sem{\cB} = \sem{\REST(e')}$.

\

Proof of (2): The proof of this statement proceeds in the same way as the proof of (1) except that, in the case that $e= e_1 \times e_2$, we  use  Theorem \ref{thm:closure-Hadamard-product}(1) (which needs commutativity) instead of Theorems \ref{thm:closure-Hadamard-product-char}(1) and   \ref{thm:closure-Hadamard-product}(3) for closure under Hadamard product.
\end{proof}

         \section[The second main result for $\RepEx(\Sigma,\B)$]{The main result for $\RepEx(\Sigma,\B)$ over bi-locally finite strong bimonoids}

         Since each run recognizable weighted tree language over a bi-locally finite strong bimonoid is a recognizable step mapping (cf. Theorems~\ref{thm:bi-loc-finite-rec-step-function} and \ref{thm:crisp-det-algebra}), we can show that the notions of representable, restricted representable, and $\times$-restricted representable coincide. This is  the second main result of this chapter: the M{\'e}dv{\'e}d{\'e}v's characterization of wta over bi-locally finite strong bimonoids. 

 \begin{theorem-rect} \label{cor:Medvedjev-str-bm} Let $\Sigma$ be a ranked alphabet. Moreover, let $\B=(B,\oplus,\otimes,\0,\1)$ be a bi-locally finite strong bimonoid and let $r: \T_\Sigma \to B$.  Then the following four statements are equivalent.
  \begin{compactenum}
  \item[(A)] We can construct a $(\Sigma,\B)$-wta $\cA$ such that $r=\runsem{\cA}$.
    \item[(B)] We can construct an $e \in \RepEx(\Sigma,\B)$ such that $r=\sem{e}$.
  \item[(C)] We can construct an $e \in \RepEx^r(\Sigma,\B)$ such that $r=\sem{e}$.
  \item[(D)] We can construct an $e \in \RepEx^{\times r}(\Sigma,\B)$ such that $r=\sem{e}$.
    \end{compactenum}
  \end{theorem-rect}
  \begin{proof} The implication (A)$\Rightarrow$(D) holds by Lemma \ref{lm:rec-rep}. 
    The implications (D)$\Rightarrow$(C) and (C)$\Rightarrow$(B) hold by definition.

    \
    
    Proof of (B)$\Rightarrow$(A): By induction on $(\RepEx(\B),\succ)$ we prove the following statement:
    \begin{eqnarray}\label{eq:RepEx-rec-step-m-comm-bi-loc-fin}
      \text{For every ranked alphabet $\Sigma$ and every $e \in \RepEx(\Sigma,\B)$,}\\
      \text{we can construct a crisp-deterministic $(\Sigma,\B)$-wta $\cA$ such that $\sem{e} = \runsem{\cA}$.}\notag
    \end{eqnarray}

    I.B.: For the case that $e = \RRT_{\Sigma,\sigma,b}$ or $e = \NXT_{\Sigma,\widetilde{\gamma},b}$, the statement follows from Corollary \ref{cor:char-restricted-representable} and Theorem \ref{thm:crisp-det-algebra}(B)$\Rightarrow$(A).

    I.S.: We distinguish four cases.

    \underline{Case (a):} Let $e = e_1 + e_2$ where $e_1,e_2 \in \RepEx(\Delta,\B)$. By I.H. we can construct crisp-deterministic $(\Sigma,\B)$-wta $\cA_1$ and $\cA_2$ such that $\sem{\cA_1} = \sem{e_1}$ and $\sem{\cA_2} = \sem{e_2}$. By Theorem \ref{thm:closure-sum}(2) we can construct a crisp-deterministic $(\Sigma,\B)$-wta $\cA$ such that $\sem{\cA} = \sem{\cA_1} \oplus \sem{\cA_2}$. Thus $\sem{\cA} = \sem{\cA_1} \oplus \sem{\cA_2} = \sem{e_1} \oplus \sem{e_2} = \sem{e_1 + e_2} = \sem{e}$.

    \underline{Case (b):} Let $e = e_1 \times e_2$. The proof is the same as the one for Case (a) except that we use Theorem~\ref{thm:closure-Hadamard-product}(3) (instead of Theorem \ref{thm:closure-sum}(2)).

    \underline{Case (c):} Let $e = \tau(e')$ where $\tau$ is a $(\Delta,\Sigma)$-tree relabeling and $e' \in \RepEx(\Delta,\B)$. By I.H.  we can construct a crisp-deterministic $(\Delta,\B)$-wta $\cB$  such that $\sem{e'} = \sem{\cB}$.
    By Theorem~\ref{thm:closure-under-tree-relabeling} we can construct $(\Sigma,\B)$-wta $\cC$ such that $\runsem{\cC} = \chi(\tau)(\sem{\cB})$. By Theorem~\ref{thm:bi-loc-finite-rec-step-function}(A)$\Rightarrow$(B), we can construct a crisp-deterministic $(\Sigma,\B)$-wta $\cA$ such that $\sem{\cA} = \runsem{\cC}$. Hence $\sem{\cA} = \runsem{\cC} = \chi(\tau)(\sem{\cB}) = \chi(\tau)(\sem{e'}) = \sem{\tau(e')} = \sem{e}$.
    
    \underline{Case (d):} Let $e = \REST(e')$ where $e' \in \RepEx(\Sigma,\B)$. By I.H. we can construct a crisp-deterministic $(\Sigma,\B)$-wta $\cB$ such that $\sem{\cB} = \sem{e'}$. By  Lemma~\ref{lm:rest-rec-step-mapping},  we can construct a $(\Sigma,\B)$-wta $\cC$ such that $\runsem{\cC} = \sem{\REST(e')}$. By Theorem~\ref{thm:bi-loc-finite-rec-step-function}(A)$\Rightarrow$(B), we can construct a crisp-deterministic $(\Sigma,\B)$-wta $\cA$ such that $\sem{\cA} = \runsem{\cC}$. Thus $\sem{\cA} = \runsem{\cC}  = \sem{\REST(e')} = \sem{e}$.

    This proves \eqref{eq:RepEx-rec-step-m-comm-bi-loc-fin}. Clearly, \eqref{eq:RepEx-rec-step-m-comm-bi-loc-fin} implies (B)$\Rightarrow$(A).
  \end{proof}

%% file: Buechi.tex
\chapter[Weighted MSO-logic and B-E-T's theorem]{Weighted MSO-logic and B\"uchi-Elgot-Trakhtenbrot's theorem}
\label{ch:Buechi}

The  B\"uchi-Elgot-Trakhtenbrot's theorem  (B-E-T's theorem\footnote{where ``'s'' distributes over ``-''})
\cite{buc60,elg61,tra61}, cf. also \cite{str94}, states that regular string languages and string languages definable in monadic second-order logic coincide. In this chapter we prove the corresponding B-E-T's theorem for the set $\Rec^{\mathrm{run}}(\Sigma,\B)$ of run recognizable $(\Sigma,\B)$-weighted tree languages
for an arbitrary strong bimonoid $\B$ (cf. Theorem \ref{thm:Buechi}).
The theorem states that, for every $(\Sigma,\B)$-weighted tree language $r$ the following equivalence holds: $r$ is r-recognizable if and only if $r$ is definable by a sentence of $\MSO(\Sigma,\B)$-logic. 

The first approach for a weighted version of B-E-T's theorem appeared in \cite{drogas05,drogas07,drogas09} where they generalized from the unweighted version (i.e., $\mathbb{B}$-weighted string languages) to the weighted version (i.e., $\B$-weighted string languages where $\B$ is an arbitrary semiring). Then the interpretation of a formula~$\varphi$ on a string  $w$ does not deliver the Boolean-valued answer to the question whether $w$ is a model of~$\varphi$, but  it delivers  an arbitrary element in $B$ as truth value. A first problem in the design of the weighted logic  was to give a semantics to the negation of a formula, because in a semiring (or strong bimonoid) it is not clear whether a value has a kind of complement. The solution published in \cite{drogas05,drogas07,drogas09} can be understood as follows:
start from the usual MSO-logic with atomic formulas (i.e., tests $\mathrm{P}_a(x)$ on labels, next relation $x \le y$, and membership $x \in X$), negation, disjunction, first-order existential quantification, and second-order existential quantification; turn each formula into a semantically equivalent formula in which the negations are applied to atomic formulas only; this can be achieved by introducing conjunction, first-order universal quantification, and second-order universal quantification as part of the logic and by replacing
\begin{compactitem}
  \item $\neg(\varphi \vee \psi)$ \ by \  $\neg \varphi \wedge \neg \psi$,
\item $\neg \exists x. \varphi$ \ by \ $\forall x. \neg \varphi$, and
\item $\neg \exists X. \varphi$ \ by \ $\forall X. \neg \varphi$.
  \end{compactitem}
  As a part of the solution, the truth value of each atomic formula can only be $\0$ or $\1$ (i.e., the unit elements of $\B$); thus, its truth value can be complemented. Moreover, each $b \in B$ is added as new atomic formula, and this has the semantics $b$. Disjunction and existential quantification are semantically expressed by the summation of the semiring $\B$, and conjunction and universal quantification are expressed by the multiplication of $\B$. Overall, the following EBNF shows the syntax of the weighted MSO-logic (where now $\varphi$ is a nonterminal and not a formula):
\begin{eqnarray}
   \begin{aligned}
    \varphi ::= & \ b \mid \mathrm{P}_a(x) \mid x \le y \mid x \in X \mid \neg \mathrm{P}_a(x) \mid \neg(x \le y) \mid \neg(x \in X) \mid \\
    &\ \varphi \vee \varphi \mid \varphi \wedge \varphi \mid \exists x. \varphi \mid \forall x. \varphi \mid \exists X. \varphi \mid \forall X. \varphi \enspace.
    \end{aligned}\label{equ:syntax-DG-MSO}
  \end{eqnarray}

  In fact, the first-order fragment of \eqref{equ:syntax-DG-MSO} is a subset of the fuzzy predicate calculus \cite[Sec.~1.2.2]{wec78}. Since Wechler uses fuzzy algebras \cite[Def.~1.2]{wec78} as weight algebras, atomic formulas can have any truth value (not only $\0$ and $\1$) and the negation can be applied to any formula. The interpretation of the propositional connectives and quantifiers as it is defined in \cite{drogas05,drogas07,drogas09} is essentially the same as in the fuzzy predicate calculus \cite[Def.~1.8 and 1.10]{wec78}.

It turned out that the weighted MSO-logic generated by \eqref{equ:syntax-DG-MSO} is too strong for B-E-T's theorem. Hence, in \cite{drogas05,drogas07,drogas09} restrictions are defined which, roughly speaking, amount to (a) having a kind of commutativity of the semiring multiplication (for exchanging values which appear in conjunctions), (b) restricting the application of  first-order universal quantification to formulas of which the semantics is a recognizable step mapping, and (c) dropping second-order universal quantification. Indeed, this restricted weighted MSO-logic characterizes recognizable weighted string languages \cite[Thm.~4.7]{drogas09}. In the proof (at least) two technical difficulties were mastered.

(A) The proof of ``recognizable implies definable'' follows the well-known idea of expressing runs of an automaton by formulas.
However, since the resulting  formulas just check structural properties, i.e., have semantic value $\0$ or $\1$, it was important to disambiguate subformulas which involve disjunction or existential quantification, because the underlying semiring is not necessarily additively idempotent.

(B) Since conjunction and first-order universal quantification are part of the weighted logic, preservation of recognizability under these constructs had to be proven. The main problem showed up with first-order universal quantification, because there exists a non recognizable  weighted language which is definable by a formula of the form $\forall x. \varphi$ in that weighted logic \cite[Ex.~3.4]{drogas05}. Thus, appropriate restrictions on the form of the body formula $\varphi$ had to be found (cf. restriction (b) above), and then a  new automaton construction had to be invented. This problem did not arise in classical unweighted MSO logic, because in classical logic universal quantification can be expressed using existential quantification and negations.

This approach to weighted MSO-logic has been generalized to a number of structures (different from strings) and weight algebras (different from semirings), e.g. for
Mazurkiewicz traces and semirings \cite{mei06},
ranked and unranked trees and semirings \cite{drovog06,drovog11},
finite and infinite strings with discounting and semirings \cite{drorah07},
nested words and semirings \cite{mat08,mat10},
timed words and semirings \cite{qua09},
pictures and semirings \cite{fic11},
strings and infinite strings and valuation monoids \cite{dromei12},
ranked and unranked trees and tree-valuation monoids \cite{drogoemaemei11,droheuvog15}. 
In \cite{bolgasmonzei14} (see also \cite{bolgasmonzei10}) an approach different to \cite{drogas09} was taken: there, not the logic was restricted but the automaton model was  extended. Indeed, the set of weighted string languages recognizable by pebble two-way weighted automata is equal to the set of weighted string languages definable by, roughly speaking, a first-order fragment of the unrestricted weighted MSO-logic of \cite{drogas09} enriched by a transitive closure operator. In \cite{fulvog15} semiring-weighted tree automata were characterized by a weighted transitive closure logic. For a survey we refer to \cite{gasmon18}. 
The disambiguation reported in (A) above is cumbersome. This has been overcome in \cite{bolgasmonzei10,fulstuvog12,bolgasmonzei14} by allowing explicitly (Boolean-valued) MSO-formulas as guards of weighted formulas. Finally, we mention that, in \cite{li08}, B-E-T's theorem has been proved for mv-algebra. 

In this chapter we follow the approach of \cite{fulstuvog12} in which B-E-T's theorem was proved for recognizable weighted tree languages over absorptive multioperator monoids \cite[Thm.~4.1]{fulstuvog12}. In particular, each strong bimonoid (and hence, each semiring) is an absorptive multioperator monoid. In the MSO-logic proposed in that paper, formulas of classical MSO-logic on trees can be used to guard weighted formulas. Moreover, a universal quantification over some recognizable step mapping (as it appears in the restricted MSO-logic of \cite{drogas05}) is represented by
a guarded atomic formula where the guard simulates the step languages and the atomic formula is interpreted as a unique algebra homomorphism (cf. Lemma \ref{lm:getting-rid-of-universal-quant}). For semirings, each formula of the restricted weighted MSO-logic of \cite{drovog06} (which is a straightforward generalization of \cite{drogas09}) can be transformed syntactically into an equivalent formula of the logic in \cite{fulstuvog12} (cf. \cite[Lm.~5.10]{fulstuvog12}), and vice versa (cf. \cite[Lm.~5.12]{fulstuvog12}). 

We note that each strong bimonoid is a particular tree-valuation monoid of \cite{drogoemaemei11}. In \cite[Sec. 6]{drogoemaemei11} this is made explicit and from the main result of  \cite{drogoemaemei11} a B-E-T's theorem has been derived for wta over strong bimonoids. We also note that  the approach of \cite{fulstuvog12} has been generalized in \cite{fulhervog18} to weighted tree grammars with storage over complete M-monoids, and B-E-T's theorem was proved in \cite[Thm.~7.4]{fulhervog18} (for the string case cf. \cite[Thm.~8]{vogdroher16}).

The goal of this chapter is to report on B-E-T's theorem \cite[Thm.~4.1]{fulstuvog12} for the particular case of strong bimonoids (cf. Theorem \ref{thm:Buechi}).  Moreover, we prove that extending carefully the logic by weighted conjunction and weighted first-order universal quantification on recognizable step formulas does not increase its expressive power if the underlying weight algebra is a commutative semiring (cf. Theorem \ref{thm:Buechi-extended}). In Subsection \ref{sec:MSO-ext-commutative-bi-locally finite}, we prove that this also holds for the full extension  if the underlying weight algebra is a commutative and bi-locally finite strong bimonoid (cf. Theorem \ref{thm:Buechi-comm-bi-loc-fin}); this is closely related to \cite[Thm.~5.3]{drovog12}.
  Finally, in Section \ref{sec:relationship-decomp-rec-implies-def} we show a strong relationship between a decomposition result of wta (proved in Chapter \ref{ch:decomposition}) and the fact that, for weighted tree languages, r-recognizable implies definable (cf. Theorems \ref{thm:Buechi-2} and \ref{thm:decomposition-1-alternative}).

  \section{The MSO-logic $\MSO(\Sigma)$}
  \label{sect:MSO-definition}

  We recall the monadic second-order logic on trees (cf. \cite{thawri68,don70,gecste97}) and make several formal definitions, constructions, and proofs explicit.

\index{MSO@$\MSO(\Sigma)$}
As first-order variables we use small letters from the end of the Latin alphabet, e.g.,  $x,x_1,x_2,\ldots,y,z$, and as second-order variables we use capital letters, like $X,X_1,X_2,\ldots,Y,Z$. The \emph{set of monadic second-order formulas over $\Sigma$}, denoted by $\MSO(\Sigma)$, is the set of all expressions generated by the following EBNF with nonterminal $\varphi$:
	\begin{equation}\label{equ:syntax-MSO-unweighted}
		\varphi\,\,::=\llabel_\sigma(x) \mid \edge_i(x,y) \mid x\in X\mid \neg \varphi \mid (\varphi\vee\varphi) \mid
		\exists x.\varphi \mid \exists X.\varphi\enspace,
	\end{equation}
	where $\sigma\in\Sigma$ and $i \in [\maxrk(\Sigma)]$.

\index{succMSO@$\succ_{\MSO(\Sigma)}$}
        In order to perform inductive proofs or to define objects for $\MSO(\Sigma)$ by induction, we will consider the reduction system
        \[(\MSO(\Sigma),\succ_{\MSO(\Sigma)})
        \]
        where  $\succ_{\MSO(\Sigma)}$ is the binary relation on $\MSO(\Sigma)$ defined as follows. 
        For every $\varphi_1,\varphi_2 \in \MSO(\Sigma)$,  we let  $\varphi_1 \succ_{\MSO(\Sigma)} \varphi_2$ if either of the following two cases holds.

  Case (a) : There exists $\psi \in \MSO(\Sigma)$ such that $\varphi_1$ has the form $\neg \psi$, $\exists x.\psi$, or $\exists X.\psi$ and $\varphi_2 = \psi$. 
        
  Case (b) : There exist $\psi_1,\psi_2 \in \MSO(\Sigma)$ such that $\varphi_1= (\psi_1 \vee \psi_2)$ and $\varphi_2 \in \{\psi_1,\psi_2\}$.

  By Corollary \ref{cor:reduction-to-substring-is-terminating}, the relation $\succ_{\MSO(\Sigma)}$ is terminating.
    Moreover, we have that  $\nf_{\succ_{\MSO(\Sigma)}}(\MSO(\Sigma))$ is the set of formulas of the form $\llabel_\sigma(x)$, $\edge_i(x,y)$, and $x\in X$.
For every  $\varphi,\psi \in \MSO(\Sigma)$, we say that \emph{$\psi$ is a subformula of $\varphi$} if $\varphi \succ_{\MSO(\Sigma)}^* \psi$. We note that the relation ``is a subformula of'' is reflexive.

Next we prepare the definition of the semantics of  $\MSO(\Sigma)$-formulas.
\index{assignment@$\cV$-assignment for $\xi$}
	Let $\cV$ be  a finite set of variables; we abbreviate by $\cV^{(1)}$ and $\cV^{(2)}$ the set of first-order variables in $\cV$ and the set of second-order variables in $\cV$, respectively.
	Let $\xi\in\T_\Sigma$. A \emph{$\cV$-assignment for $\xi$} is a mapping $\eta$ with domain $\cV$ which maps each first-order variable in $\cV$ to a position of $\xi$ and each second-order variable to a subset of $\pos(\xi)$. By $\Phi_{\cV, \xi}$ we denote the set of all $\cV$-assignments for $\xi$. Let $\eta\in\Phi_{\cV,\xi}$, $x$ be a first-order variable, and $w\in\pos(\xi)$. By $\eta[x\mapsto w]$ we denote the $(\cV\cup\{x\})$-assignment for $\xi$ that agrees with $\eta$ on $\cV\setminus \{x\}$ and that satisfies $\eta[x\mapsto w](x)=w$. Similarly, if $X$ is a second-order variable and $W\subseteq \pos(\xi)$, then $\eta[X\mapsto W]$ denotes the $(\cV\cup\{X\})$-assignment for $\xi$ that agrees with $\eta$ on $\cV\setminus \{X\}$ and that satisfies $\eta[X\mapsto W](X)=W$.

For the definition of the semantics of $\MSO(\Sigma)$-formulas, we need the notion of free variable; for later purpose, we also define the notion of bound variables. Formally, for each $\varphi\in\MSO(\Sigma)$, we define the set $\Free(\varphi)$ of \emph{free variables of $\varphi$} and the set $\Bound(\varphi)$ of \emph{bound variables of $\varphi$} by induction on $(\MSO(\Sigma),\succ_{\MSO(\Sigma)})$  as follows:
  
  I.B.: We distinguish three cases.
  	\begin{compactitem}
	\item $\Free(\llabel_\sigma(x))=\{x\}$ and $\Bound(\llabel_\sigma(x))=\emptyset$,
	\item $\Free(\edge_i(x,y))=\{x,y\}$ and $\Bound(\edge_i(x,y))=\emptyset$,
	\item $\Free(x\in X)=\{x,X\}$ and $\Bound(x\in X)=\emptyset$
	\end{compactitem}
	
I.S.: We distinguish four cases.
	\begin{compactitem}
	\item $\Free(\neg \varphi)=\Free(\varphi)$ and $\Bound(\neg \varphi)=\Bound(\varphi)$,
       \item $\Free(\varphi_1\vee\varphi_2)=\Free(\varphi_1)\cup\Free(\varphi_2)$ and $\Bound(\varphi_1\vee\varphi_2)=\Bound(\varphi_1)\cup\Bound(\varphi_2)$, 
       \item $\Free(\exists x. \varphi)=\Free(\varphi)\setminus\{x\}$  and $\Bound(\exists x. \varphi)=\Bound(\varphi)\cup \{x\}$, and
         \item $\Free(\exists X. \varphi)=\Free(\varphi)\setminus\{X\}$ and $\Bound(\exists X. \varphi)=\Bound(\varphi)\cup\{X\}$.
	\end{compactitem}

 If a formula $\varphi$ has the free variables, say, $x$, $y$, and $X$ and no others,
then we also write $\varphi(x,y,X)$.
As usual, we abbreviate formulas like $\exists x.\exists y.\exists z. \varphi$ by $\exists x,y,z.\varphi$.  A formula $\varphi \in \MSO(\Sigma)$ is called \emph{$\MSO(\Sigma)$-sentence} (or just: \emph{sentence}) if $\Free(\varphi) = \emptyset$.

Let $\varphi\in\MSO(\Sigma)$ and $\cV$ be a finite set such that $\Free(\varphi) \subseteq \cV$. For every $\xi\in\T_\Sigma$ and $\eta\in \Phi_{\cV,\xi}$, the relation ``$(\xi,\eta)$ satisfies $\varphi$'', denoted by $(\xi,\eta)\models\varphi$, is defined by induction on $(\MSO(\Sigma),\succ_{\MSO(\Sigma)})$  as follows.

I.B.:  We distinguish three cases.

\begin{tabular}{lcl}
$(\xi,\eta) \models \llabel_{\sigma}(x)$ & iff & $\sigma = \xi(\eta(x))$\\

$(\xi,\eta) \models \edge_i(x,y)$ & iff & $\eta(y) = \eta(x).i$\\

$(\xi,\eta) \models (x\in X)$ & iff & $\eta(x) \in \eta(X)$
\end{tabular}

I.S.:  We distinguish four cases.

\begin{tabular}{lcl}
$(\xi,\eta) \models (\varphi_1\vee\varphi_2)$ & iff &  $(\xi,\eta) \models \varphi_1$ or~$(\xi,\eta)
\models\varphi_2$\\

$(\xi,\eta) \models (\neg \varphi)$ & iff & $(\xi,\eta) \models\varphi$ is not true\\

$(\xi,\eta) \models (\exists x. \varphi)$ & iff & there exists  a~$w\in\pos(\xi)$ such
that~$(\xi,\eta[x \rightarrow w]) \models \varphi$\\

$(\xi,\eta) \models (\exists X. \varphi)$ & iff & there exists a set~$W\subseteq
\pos(\xi)$ such that~$(\xi,\eta[X \rightarrow W]) \models \varphi$.
\end{tabular}

\index{MSO definable}
We denote by~$\LL_{\cV}(\varphi)$ the set $\{(\xi,\eta) \mid \xi \in \T_{\Sigma}, \eta \in \Phi_{\cV,\xi}, (\xi,\eta) \models \varphi\}$, and we will simply write~$\LL(\varphi)$ instead of~$\LL_{\Free(\varphi)}(\varphi)$. If $\varphi$ is a sentence, i.e., $\Free(\varphi)=\emptyset$, then $\Phi_{\Free(\varphi),\xi}$ contains exactly one assignment, which is the $\emptyset$; then we can identify $(\xi,\emptyset)$ with $\xi$, and hence $\LL(\varphi)$ can be thought of as a subset of $\T_\Sigma$.
A~tree language~$L\subseteq \T_\Sigma$ is called \emph{MSO~definable} if there exists a
sentence~$\varphi\in\MSO(\Sigma)$ such that~$L=\LL(\varphi)$.

For every formula $\varphi$, finite set $\cV$ of variables with $\Free(\varphi) \subseteq \cV$, and $(\xi,\eta) \in \LL_\cV(\varphi)$, the membership of $(\xi,\eta)$ in $ \LL_\cV(\varphi)$ is independent from the values assigned to variables outside of $\Free(\varphi)$. This is formally expressed in the following consistency lemma.

\begin{lemma}\label{lm:msocons} \rm
	Let $\varphi\in\MSO(\Sigma)$. For each finite set $\cV$ of variables containing $\Free(\varphi)$ and for every $\xi \in \T_\Sigma$ and $\eta \in \Phi_{\cV,\xi}$, the following equivalence holds:
$(\xi,\eta)\in \LL_\cV(\varphi)\text{ if and only if }(\xi,\eta|_{\Free(\varphi)})\in \LL(\varphi)$.
              \end{lemma}             
              \begin{proof}
                We prove the statement by induction on $(\MSO(\Sigma),\succ_{\MSO(\Sigma)})$. Let $\xi \in \T_\Sigma$ and $\eta \in \Phi_{\cV,\xi}$.

I.B.: For the induction base we distinguish three cases.

\underline{Case (a):} Let $\varphi=\llabel_{\sigma}(x)$. Then $\Free(\varphi)=\{x\}$ and we have
\begin{align*}
 (\xi,\eta) \models \varphi \ \text{ iff } \ \sigma = \xi(\eta(x))
\ \text{ iff } \ \sigma = \xi(\eta|_{\{x\}}(x)) \ \text{ iff } \ (\xi,\eta|_{\{x\}})\models \varphi \ \text{ iff } \ 
& (\xi,\eta|_{\Free(\varphi)})\models \varphi  \enspace.
\end{align*}

\underline{Case (b):} Let $\varphi=\edge_i(x,y)$. Then $\Free(\varphi)=\{x,y\}$ and we have
\begin{align*}
(\xi,\eta) \models \varphi \ \text{ iff } \ \eta(y)= \eta(x).i
\ \text{ iff } \ \eta|_{\{x,y\}}(y)= \eta|_{\{x,y\}}(x).i
\ \text{ iff } \ (\xi,\eta|_{\{x,y\}})\models \varphi  \ \text{ iff } \ 
& (\xi,\eta|_{\Free(\varphi)})\models \varphi  \enspace.
\end{align*}

\underline{Case (c):} Let $\varphi=(x\in X)$. Then $\Free(\varphi)=\{x,X\}$ and the proof of this case is very similar to the proof of Case (b).

\

I.S.: For the induction step we distinguish four cases.

\underline{Case (a):} Let $\varphi = \varphi_1 \vee \varphi_2$. Then Then $\Free(\varphi)=\Free(\varphi_1)\cup \Free(\varphi_2)$ and we have
\begingroup
\allowdisplaybreaks
\begin{align*}
  (\xi,\eta) \models \varphi & \ \text{ iff } \ (\xi,\eta) \models \varphi_1 \text{ or } (\xi,\eta) \models \varphi_2 \\
                             & \ \text{ iff } \   (\xi,\eta|_{\Free(\varphi_1)}) \models \varphi_1 \text{ or } (\xi,\eta|_{\Free(\varphi_2)}) \models \varphi_2  \tag{by I.H.}\\
                             & \ \text{ iff } \   (\xi,\eta|_{\Free(\varphi)}) \models \varphi_1 \text{ or } (\xi,\eta|_{\Free(\varphi)}) \models \varphi_2  \tag{by I.H.}\\  
                             & \ \text{ iff } \   (\xi,\eta|_{\Free(\varphi)}) \models \varphi\enspace.                               
\end{align*}
\endgroup

\underline{Case (b):} Let $\varphi = \neg \varphi'$. This is similar to Case (a). 

\underline{Case (c):} Let $\varphi=(\exists x. \varphi')$. Then $\Free(\varphi)=\Free(\varphi')\setminus \{x\}$ and we have
\begingroup
\allowdisplaybreaks
\begin{align*}
  (\xi,\eta) \models \varphi & \ \text{ iff } \  (\exists w\in \pos(\xi)): (\xi,\eta[x\to w]) \models \varphi'\\
                             & \ \text{ iff } \ (\exists w\in \pos(\xi)): (\xi,(\eta[x\to w])|_{\Free(\varphi')}) \models \varphi' \tag{by I.H.}\\
                            & \ \text{ iff } \ (\exists w\in \pos(\xi)): (\xi,((\eta|_{\Free(\varphi')})[x\to w])|_{\Free(\varphi')}) \models \varphi' \tag{because $(\eta[x\to w])|_{\Free(\varphi')}=\big((\eta|_{\Free(\varphi')})[x\to w]\big)|_{\Free(\varphi')}$}\\
                             & \ \text{ iff } \  (\exists w\in \pos(\xi)): (\xi,(\eta|_{\Free(\varphi')})[x\to w]) \models \varphi'   \tag{by I.H.}\\
                             & \ \text{ iff } \  (\exists w\in \pos(\xi)): (\xi,(\eta|_{\Free(\varphi)})[x\to w]) \models \varphi'   \tag{by (*)}\\
 & \ \text{ iff } \ (\xi,\eta|_{\Free(\varphi)})\models \varphi   \enspace.
\end{align*}
\endgroup
where at (*) we have used that either (i) $x\not \in \Free(\varphi')$ and thus $\Free(\varphi')=\Free(\varphi)$ or (ii) $x\in \Free(\varphi')$ and thus  $\Free(\varphi')=\Free(\varphi)\cup\{x\}$. In both cases we have $(\eta|_{\Free(\varphi')})[x\to w] = (\eta|_{\Free(\varphi)})[x\to w]$.

\underline{Case (d):} Let $\varphi = (\exists X.\varphi')$. This case is similar to Case (c).
\end{proof}

              We will use the following macros:
              \begin{align*}
                             \varphi \wedge \psi &:= \neg(\neg\varphi \vee \neg \psi)\\
   \varphi \to \psi &:= \neg \varphi \vee \psi\\
                \varphi \leftrightarrow \psi &:= (\varphi \to \psi) \wedge (\psi \to \varphi)\\
                \forall x. \varphi &:= \neg(\exists x. \neg \varphi)\\
                \forall X. \varphi &:= \neg(\exists X. \neg \varphi)\\
                \bigvee\limits_{i \in I} \varphi_i & := \varphi_{i_1} \vee \cdots \vee \varphi_{i_n} \text{ and}\\
                \bigwedge\limits_{i \in I} \varphi_i & := \varphi_{i_1} \wedge \cdots \wedge \varphi_{i_n} \text{ for each finite family $(\varphi_i \mid i \in I)$ with $I=\{i_1,\ldots,i_n\}$}\\
                \llabel_\Delta(x) &:= \bigvee_{\sigma \in \Delta} \llabel_\sigma(x) \ \text{ for each $\Delta \subseteq \Sigma$}\\
                \edge(x,y) &:= \bigvee_{i \in [\maxrk(\Sigma)]}  \edge_i(x,y)\\
                (x=y) &:= \forall X. (x \in X) \leftrightarrow (y \in X)\\
                \mathrm{root}(x) &:= \neg \exists y. \edge(y,x) 
              \end{align*}
              Since  disjunction and conjunction are associative, the placement of parentheses in the expressions $\varphi_{i_1} \vee \cdots \vee \varphi_{i_n}$ and $\varphi_{i_1} \wedge \cdots \wedge \varphi_{i_n}$ is irrelevant and hence not shown.

                            \begin{example}\label{ex:path}\rm (cf. e.g. \cite[Sec.~1.3.1]{coueng12}) We consider the following formulas in $\MSO(\Sigma)$:
                  \begin{align*}
                    \mathrm{closed}(X) &= \forall z_1. \forall z_2. (\edge(z_1,z_2) \wedge (z_1 \in X)) \rightarrow (z_2 \in X)\\
                    \mathrm{path}(x,y) &= \forall X. (\mathrm{closed}(X) \wedge (x \in X)) \rightarrow (y \in X) \enspace.
                  \end{align*}

                  We claim that $\mathrm{path}(x,y)$ specifies a descending path from $x$ to $y$. Formally, we claim that 
\begin{equation}
\LL_{\{x,y\}}(\mathrm{path}(x,y)) = \{(\xi,\{(x,u),(y,v)\}) \mid \xi \in \T_\Sigma, u,v \in \pos(\xi) \text{ such that } u \in \prefix(v)\}\enspace. \label{equ:set-of-models-of-path}
\end{equation}

(a) First we prove the inclusion from left to right.
Let $\xi \in \T_\Sigma$, $u,v \in \pos(\xi)$, and $\eta = \{(x,u),(y,v)\}$ such that $(\xi,\eta) \in \LL_{\{x,y\}}(\mathrm{path}(x,y))$.  Then $\xi \in \T_\Sigma$ and $(\xi,\eta) \models \mathrm{path}(x,y)$.
  
Moreover, we have:
                      \begin{equation}
                        \text{for each set $U \subseteq \pos(\xi)$: if $(\xi,\eta \cup \{(X,U)\}) \models \mathrm{closed}(X)$ and $u \in U$, then $v \in U$. } \label{eq:path-U}
                      \end{equation}
                      Moreover, we have $(\xi,\eta \cup \{(X,\pos(\xi|_u)\}) \models \mathrm{closed}(X)$ and $u \in \pos(\xi|_u)$. Hence, by \eqref{eq:path-U} with $U = \pos(\xi|_u)$, we have that $v \in \pos(\xi|_u)$. Thus $u \in \mathrm{prefix}(v)$.

 (b) Next we prove the inclusion from right to left.
 For this, let $\xi \in \T_\Sigma$, $u,v \in \pos(\xi)$, and $\eta = \{(x,u),(y,v)\}$ such that $u \in \mathrm{prefix}(v)$.
 Thus there  exist $n \in \mathbb{N}$, $j_1,\ldots,j_n \in \mathbb{N}_+$ such that $v = uj_1 \cdots j_n$. Now let $U \subseteq \pos(\xi)$ such that  $(\xi,\{(X,U)\}) \models \mathrm{closed}(X)$ and $u \in U$.

                      First, by induction on $([0,n],>)$, we prove that the following statement holds:
                      \begin{equation}
\text{For each $\ell \in [0,n]$, we have $u j_1 \cdots j_\ell \in U$.} \label{eq:ind-path}
                        \end{equation}
                        
                        I.B.: Let $\ell=0$. The statement trivially holds, because $u \in U$.

                        I.S.: Let $\ell= \ell' +1$ for some $\ell' \in \mathbb{N}$. We assume that  \eqref{eq:ind-path} holds for $\ell'\in [0,n-1]$. By I.H.,  we have  $u j_1 \cdots j_{\ell'} \in U$. Since $(\xi,\{(X,U)\}) \models \mathrm{closed}(X)$ and
                        \[
(\xi,\{(X,U),(z_1,u j_1 \cdots j_{\ell'}),(z_2,u j_1 \cdots j_{\ell'} j_{\ell})\}) \models (\edge(z_1,z_2) \wedge z_1 \in X)\enspace,
\]
we have that 
                \[
(\xi,\{(X,U),(z_1,u j_1 \cdots j_{\ell'}),(z_2,u j_1 \cdots j_{\ell'} j_{\ell})\}) \models (z_2 \in X)
\]
and hence $u j_1 \cdots j_{\ell'} j_{\ell} \in U$. This proves \eqref{eq:ind-path}.

\sloppy Then \eqref{eq:ind-path} implies (by choosing $\ell = n$) that
\(u j_1 \cdots j_n \in U\), and hence $v \in U$.
Thus  $(\xi,\eta) \models \mathrm{path}(x,y)$ and hence $(\xi,\eta) \in \LL_{\{x,y\}}(\mathrm{path}(x,y))$.
This finishes the proof of  \eqref{equ:set-of-models-of-path}.

We note that the formula $\exists X. \big((\mathrm{closed}(X) \wedge x \in X) \rightarrow y \in X\big)$ does not specify a descending path from $x$ to $y$. To see this, we consider the tree $\xi = \sigma(\gamma(\alpha),\beta)$ and the assignment $\eta(x)=1$ and $\eta(y)=2$. Obviously, there does not exist a descending path from $1$ to $2$.
 However, with the assignment $\eta'$ with $\eta'(x)=\eta(x)$, $\eta'(y)=\eta(y)$, and  $\eta'(X) = \pos(\xi)$  we have 
\begin{align*}
(\xi,\eta') \models (\mathrm{closed}(X) \wedge (x \in X)) \rightarrow (y \in X)
\end{align*}
and hence 
\begin{align*}
(\xi,\eta) \models \exists X. \big((\mathrm{closed}(X) \wedge (x \in X)) \rightarrow (y \in X)\big)\enspace.
\end{align*}
\hfill $\Box$
                  \end{example}

                  Finally, we recall B-E-T's theorem for recognizable tree languages. It follows from Lemmas \ref{lm:thawri-fta-to-MSO} and~\ref{lm:thawri-MSO-to-fta}.

\begin{theorem-rect} \label{thm:that_wright}  {\rm \cite{thawri68,don70}}
Let $\Sigma$ be a ranked alphabet. Moreover, let $L\subseteq \T_{\Sigma}$.
                    Then the following two statements are equivalent.
		\begin{compactenum}
			\item[(A)] We can construct an fta $A$ over $\Sigma$ such that $\mathrm L(A) =L$.
			\item[(B)] We can construct a sentence $\varphi\in\MSO(\Sigma)$ such that $\LL(\varphi)=L$.
                        \end{compactenum}
                      \end{theorem-rect}

\begin{lemma} \label{lm:thawri-fta-to-MSO} \rm For each $\Sigma$-fta $A$, we can construct a sentence $\varphi \in \MSO(\Sigma)$ such that $\LL(A) = \LL(\varphi)$.
\end{lemma}
\begin{proof} Let $A = (Q,\delta,F)$ be a $\Sigma$-fta. Without loss of generality we can assume that there
  exists an $n\in\mathbb{N}$ such that~$Q = [n]$.   We let $\cU = \{X_1,\ldots,X_n\}$.

 Now we construct a sentence $\varphi \in \MSO(\Sigma)$ such that $\LL(\varphi) = \LL(A)$. The following relationship will be the key for this construction. Let~$\xi\in\T_\Sigma$. We can relate 
\begin{compactenum}
\item[(a)] mappings from~$\pos(\xi)$ into~$Q$ (i.e., runs of $A$ on $\xi$) and 
\item[(b)] $Q$-indexed partitionings over~$\cP(\pos(\xi))$
\end{compactenum}
 in the following way. Let $\rho\colon\pos(\xi)\rightarrow Q$ and $P=(P_q\mid q\in Q)$ be a $Q$-indexed partitioning over $\cP(\pos(\xi))$. Then~$\rho$ and~$P$ are related if and only if $P_q = \rho^{-1}(q)$
 for every~$q\in Q$. Obviously, this relationship is a one-to-one correspondence.

We define the sentence $\varphi \in \MSO(\Sigma)$ by 
\[
\varphi = \exists X_1.\, \ldots \exists X_n.\
\varphi_{\mathrm{part}} \wedge \varphi_{\mathrm{valid}} \wedge \varphi_{\mathrm{final}}
\] 
where 
\begin{itemize}
\item $\varphi_{\mathrm{part}} = \forall x. \bigvee\limits_{q\in Q}\Big( (x \in X_q) \wedge \bigwedge\limits_{\substack{p\in Q:\\p\neq q}} \neg (x \in X_p)\Big)$,

\item $\varphi_{\mathrm{valid}} = \varphi_{\mathrm{valid},\neq 0} \wedge \varphi_{\mathrm{valid},=0}$ with\\[3mm]
  $\varphi_{\mathrm{valid},\neq 0}  = \forall x. \bigwedge\limits_{\substack{k\in\mathbb{N}_+,\sigma\in \Sigma^{(k)}}} \llabel_\sigma(x) \rightarrow$\\[-2mm]
\hspace*{3.5cm} $\forall y_1.\, \ldots \forall y_k. \;\Bigg( \edge_1(x,y_1) \wedge\ldots\wedge \edge_k(x,y_k) \rightarrow$\\[-2mm]
\hspace*{5.3cm} $\bigvee\limits_{\substack{q,q_1,\ldots,q_k \in Q: \\(q_1 \ldots q_k,\sigma, q) \in \delta_k}} \bigg( (y_1 \in X_{q_1}) \wedge \ldots \wedge (y_k \in X_{q_k}) \wedge (x \in X_q) \bigg) \Bigg)$ \ \  and\\[4mm]
$\varphi_{\mathrm{valid},= 0}  = \forall x. \bigwedge\limits_{\substack{\alpha\in \Sigma^{(0)}}} \llabel_\alpha(x) \rightarrow
\Big(\bigvee\limits_{\substack{q \in Q: \\(\varepsilon,\alpha, q) \in \delta_0}}  (x \in X_q) \Big)$, \ \ and\\

\item $\varphi_{\mathrm{final}} = \forall x. \big(\mathrm{root}(x) \rightarrow \bigvee\limits_{q \in F} (x \in X_q)    \big)$.
\end{itemize}

For every $\xi \in \T_\Sigma$ and $\eta \in\Phi_{\cU,\xi}$, the following statement is obvious.
\begin{align*}
  (\xi,\eta) \models \varphi_{\mathrm{part}} \wedge \varphi_{\mathrm{valid}} \wedge \varphi_{\mathrm{final}}  \ \text{ iff } \ &\text{ the $Q$-indexed family $\Psi = (\eta(X_q) \mid q \in Q)$ is a partitioning of $\pos(\xi)$}\\
 \ &\text{ and the run related to $\Psi$ is accepting \enspace.}
\end{align*}
Finally, we can calculate as follows:
\begingroup
\allowdisplaybreaks
\begin{align*}
  \xi \in \LL(\varphi) \
  \text{ iff } \  & \ \text{there exists a $\eta \in \Phi_{\cU,\xi}$ such that }\\
  & \ \text{the $Q$-indexed family $\Psi = (\eta(X_q) \mid q \in Q)$ is a partitioning of $\pos(\xi)$}\\
  \ & \ \text{and the run related to $\Psi$ is accepting}\\[2mm]
\text{ iff }  &\ \R^{\mathrm{a}}_{\mathcal{A}}(\xi) \not= \emptyset\\[2mm]
\text{ iff }  &\ \xi \in \LL(A)\enspace. \qedhere
  \end{align*}
\endgroup
\end{proof}


  Next we will prove that definable implies recognizable. For this we need a few preparations.

  Since we will prove that definable implies recognizable by induction on $(\MSO(\Sigma),\succ_{\MSO(\Sigma)})$, we will also have to construct an fta also for formulas which contains free variables. However, a $\Sigma$-fta recognizes $\Sigma$-trees and not pairs $(\xi,\eta)$ with $\xi \in \T_\Sigma$ and $\eta \in \Phi_{\cV,\xi}$. Thus, by using an extended ranked alphabet, we code $\cV$-assignment for $\xi$ into the labels of $\xi$. Formally, we define the ranked alphabet $\Sigma_\cV$ by letting
        \[
          (\Sigma_\cV)^{(k)}=\Sigma^{(k)}\times \cP(\cV) \ \text{ for each  $k\in\mathbb{N}$} \enspace.
        \]
   Instead of $(\Sigma_\cV)^{(k)}$ we will write $\Sigma_\cV^{(k)}$.     We identify the sets $\Sigma$ and $\Sigma_\emptyset$.
	A tree $\zeta\in \T_{\Sigma_\cV}$ is called \emph{valid} if for each $x\in \cV^{(1)}$ there exists a unique $w\in\pos(\zeta)$ such that $x$ is contained in the second component of $\zeta(w)$. We denote by $\T_{\Sigma_\cV}^\mathrm v$ the set of all valid trees in $\T_{\Sigma_\cV}$.

        	The two sets $\{(\xi,\eta)\mid\xi\in\T_\Sigma, \eta\in\Phi_{\cV,\xi}\}$ and $\T_{\Sigma_\cV}^\mathrm v$ are in a one-to-one correspondence via $(\xi,\eta)\mapsto \zeta$, where $\pos(\zeta)=\pos(\xi)$ and 
	\begin{align*}
		 \zeta(w)=\bigl(\xi(w),\{x\in \cV^{(1)}\mid w=\eta(x)\}\cup \{X\in \cV^{(2)}\mid w\in\eta(X)\}\bigr)\;.
	\end{align*}
        In the sequel, we will not distinguish between the sets $\{(\xi,\eta)\mid\xi\in\T_\Sigma, \eta\in\Phi_{\cV,\xi}\}$ and $\T_{\Sigma_\cV}^\mathrm v$. In particular, sometimes we call pairs $(\xi,\eta)$ in the first set also trees.

\begin{example} \rm\label{ex:example-MSO-otation} Let $\Sigma=\{\sigma^{(2)},\alpha^{(0)}\}$ and $\cV=\cV^{(1)}\cup \cV^{(2)}$ with $\cV^{(1)}=\{x,y\}$ and
$\cV^{(2)}=\{X,Y\}$.

Let $\xi=\sigma(\sigma(\alpha,\alpha),\alpha)$. We define an element $\eta\in \Phi_{\cV, \xi}$ by
\[ \eta(x)=\varepsilon, \ \eta(y)=1, \  \eta(X)=\{\varepsilon,1\}, \text{ and } \ \eta(Y)=\{12\}
\enspace.\]

Moreover, e.g., $(\sigma,\{x,y,X\})\in \Sigma^{(2)}_\cV$ and $(\alpha,\emptyset), (\alpha,\{Y\})\in \Sigma^{(0)}_\cV$. According to the one-to-one correspondence defined above, we have
\begin{align*}
(\xi,\eta) \ \mapsto \ & \ \zeta=(\sigma,\{x,X\})\Big((\sigma,\{y,X\})\Big((\alpha,\emptyset),(\alpha,\{Y\}) \Big),(\alpha,\emptyset) \Big)\enspace,
\end{align*}
see Figure \ref{fig:one-to-one}.
Then $\zeta\in \T_{\Sigma_\cV}^\mathrm v$. Moreover, for $\zeta' = (\sigma,\{x\})\Big((\sigma,\{x,y,X\})\Big(\alpha,((\alpha,\emptyset),\{Y\}) \Big),(\alpha,\emptyset) \Big)$, we have $\zeta' \in \T_{\Sigma_\cV}\setminus \T_{\Sigma_\cV}^\mathrm v$ because $x$ occurs in the second component of both $\zeta'(\varepsilon)=(\sigma,\{x\})$ and $\zeta'(1)=(\sigma,\{x,y,X\})$.
Hence there does not exist $\eta'\in \Phi_{\cV, \xi}$ such that $(\xi,\eta') \mapsto \zeta'$.
\hfill$\Box$
\end{example}        
   

\begin{figure}
\begin{center}

\begin{align*}
&\Sigma=\{\sigma^{(2)},\alpha^{(0)}\}, \  \cV=\cV^{(1)}\cup \cV^{(2)}, \ \cV^{(1)}=\{x,y\}, \  \text{ and } \
\cV^{(2)}=\{X,Y\}\\[.5em]
&\Sigma_\mathcal{V}: (\sigma,\{x,X\})^{(2)},\ (\sigma,\{x,y\})^{(2)}, \ (\alpha,\{Y\})^{(0)}, \ (\alpha,\emptyset)^{(0)},\ \ldots
\end{align*}

\vspace*{1.5em}

\begin{tikzpicture}

\node at (0,0) (e) {$\sigma$};
\node at (-1,-1) (1) {$\sigma$};
\node at (1,-1) (2) {$\alpha$};
\node at (-2,-2) (11) {$\alpha$};
\node at (0,-2) (12) {$\alpha$};

\draw
  (e) -- (1)
  (e) -- (2)
  (1) -- (11)
  (1) -- (12)
;

\node at (-1,-3.2) {$\xi \in \mathrm{T}_\Sigma$};
\node at (1.5,-3.2) {$\eta$};

\node at (2.5,0) (x) {$x$};
\node at (2.5,-.675) (y) {$y$};
\node at (2.5,-1.375) (X) {$X$};
\node at (2.5,-2) (Y) {$Y$};

\draw[dashed,->] (x) -- (e);
\draw[dashed,->] (y)  edge   [bend left=20]   node {} (1);
\draw[dashed,->] (X)  edge   [bend left=20]   node {} (1);
\draw[dashed,->] (X)  -- (e);
\draw[dashed,->] (Y) -- (12);

\node at (7,0) (e') {$(\sigma,\{x,X\})$};
\node at (6,-1) (1') {$(\sigma,\{y,X\})$};
\node at (8,-1) (2') {$(\alpha,\emptyset)$};
\node at (5,-2) (11') {$(\alpha,\emptyset)$};
\node at (7,-2) (12') {$(\alpha,\{Y\})$};

\draw
  (e') -- (1')
  (e') -- (2')
  (1') -- (11')
  (1') -- (12')
;

\node at (6,-3.2) {$\zeta\in \mathrm{T}_{\Sigma_\mathcal{V}}^\mathrm{v}$};

\end{tikzpicture}
\end{center}
\caption{\label{fig:one-to-one} The one-to-one correspondence $(\xi,\eta) \mapsto \zeta$.}
\end{figure}

We ``lift'' the update of a variable assignment to trees in $\T_{\Sigma_\cV}^\mathrm v$ as follows. Let $(\xi,\eta)\in \T_{\Sigma_\cV}^\mathrm v$, $x$ be a first-order variable, and $w\in \pos(\xi)$. By $(\xi,\eta)[x\mapsto w]$ we denote the valid tree $(\xi,\eta[x\mapsto w])$ over $\Sigma_{\cV\cup\{x\}}$.
                Similarly, if $X$ is a second-order variable and $W\subseteq\pos(\zeta)$, then $(\xi,\eta)[X\mapsto W]$ denotes the valid tree $(\xi,\eta[X\mapsto W])$ over $\Sigma_{\cV\cup\{X\}}$. Moreover, let $\cU \subseteq \cV$. Then $\eta|_\cU$ is a $\cU$-assignment and we denote
  the tree~$(\xi,\eta|_\cU)$  also by $(\xi,\eta)|_\cU$.

  
As second preparation, we observe that, for each finite set $\cV$ of variables, the set  $\T_{\Sigma_\cV}^\mathrm{v}$ is a recognizable $\Sigma_\cV$-tree language.

\begin{lemma}\rm \label{lm:TSigma-valid-fta}
Let~$\Sigma$ be a ranked alphabet and~$\cV$ be a finite set of variables. Then we can construct a total and bu-deterministic  $\Sigma_\cV$-fta $A$ such that $\LL(A) = \T_{\Sigma_\cV}^{\mathrm{v}}$.
\end{lemma}
\begin{proof} Let $x_1,\ldots,x_m$ be the first-order variables in $\cV$, i.e., $\cV^{(1)} = \{x_1,\ldots,x_m\}$.  Obviously, $\T_{\Sigma_\cV}^{\mathrm{v}} = \bigcap_{j \in [m]} L_j$ where
  \[
    L_j = \{\xi \in \T_{\Sigma_\cV} \mid x_j \text{ occurs exactly once in } \xi\}\enspace.
    \]
    First, for each $j \in [m]$, we  construct a total and bu-deterministic  $\Sigma_\cV$-fta $B_j = (Q,\delta,F)$ such that $L_j = \LL(B_j)$. The idea for this is to count the number of occurrences of $x_j$ up to 2 while traversing the tree from the leaves towards the root. If $x_j$ is encountered twice, then the tree automaton reaches a non-final state which is propagated towards the root.

    Formally, we let $Q = \{0,1,2\}$ and $F = \{1\}$. 
 For every $k \in \mathbb{N}$, $(\sigma ,U) \in \Sigma_\cV^{(k)}$, and $q_1,\ldots,q_k \in Q$ we let
\[
\delta_k(q_1\cdots q_k,(\sigma,U))=
\begin{cases}
2 & \text{ if }  \  \big(\bigplus_{i \in [k]}q_i \ge 2\big) \vee  \big((\bigplus_{i \in [k]}q_i = 1) \wedge (x_j \in U)\big)\\[2mm]
1 & \text{ if } \
\big((\bigplus_{i \in [k]}q_i = 1) \wedge (x_j \not\in U)\big) \vee
                                             \big((\bigplus_{i \in [k]}q_i = 0) \wedge (x_j \in U)\big)\\[2mm]
                                             0 & \text{ otherwise} \enspace,  
\end{cases}
\]
where $\bigplus$ denotes the extension of $+$ to finite sums in the monoid $(\mathbb{N},+,0)$.
It is clear that $\LL(A) = L_j$.

Then, by Theorems \ref{thm:fta-total-bud-fta} and \ref{thm:fta-closure-results},  we can construct a total and bu-deterministic  $\Sigma_\cV$-fta $A$ such that $\LL(A) =  \bigcap_{j \in [m]} \LL(B_j)$. Then
\[
  \LL(A) =  \bigcap_{j \in [m]} \LL(B_j) =  \bigcap_{j \in [m]} L_j = \T_{\Sigma_\cV}^{\mathrm{v}} \enspace.
  \]
    \end{proof}

As third preparation, we prove that two particular tree languages are recognizable.

\begin{lemma} \label{Gamma-a} \rm
Let~$\Delta$ be a ranked alphabet, $\Gamma\subseteq\Delta$, and~$L_\Gamma = \{\xi\in\T_\Delta\mid (\exists w \in \pos(\xi)): \xi(w) \in \Gamma\}$. Then we can construct a total and bu-deterministic  $\Delta$-fta $A$ such that $\LL(A) = L_\Gamma$.
\end{lemma}
\begin{proof} We construct the 
  $\Delta$-fta $A =
(Q,\delta,F)$ with $Q = \{0,1\}$ (and each state is viewed as natural number), $F = \{1\}$, and for every $k \ge 0$, $\sigma \in \Delta^{(k)}$, $q_1,\ldots,q_k \in Q$ we define
\[
\delta_k(q_1\cdots q_k,\sigma) = 
\left\{
\begin{array}{ll}
1 & \hbox{ if } \big(\bigplus\limits_{i \in [k]}q_i \ge 1\big) \vee (\sigma \in \Gamma) \\[2mm]
0 & \hbox{ otherwise} \enspace.
\end{array}
\right.
\]
It is obvious that $A$ is total and deterministic, and that $\LL(A) = L_{\Gamma}$.
\end{proof}

\begin{lemma} \label{Gamma1,2-a} \rm
Let~$\Delta$ be a ranked alphabet, $\Gamma_1,\Gamma_2\subseteq\Delta$, $j \in \mathbb{N}$ with $j \geq 1$, and~$L_{\Gamma_1,\Gamma_2,j} = \{\xi\in\T_\Delta\mid (\exists w_1,w_2 \in\pos(\xi)): w_2 = w_1j, \xi(w_1)\in\Gamma_1, \xi(w_2)\in\Gamma_2\}$. Then we can construct a $\Delta$-fta $A$ such that $\LL(A) = L_{\Gamma_1,\Gamma_2,j}$.
\end{lemma}
\begin{proof} We construct the $\Delta$-fta $A = (Q,\delta,F)$ with $Q = \{0,1,2\}$, $F = \{2\}$, and for every $k \ge 0$,  we define $\delta_k$ to be the smallest set $\delta'_k$ such that $(q_1\cdots q_k, \sigma, q) \in \delta_k'$ if and only if (at least) one of the following four conditions is satisfied:
\begin{compactitem}
\item $q=0$,
\item $\sigma \in \Gamma_2$ and $q=1$,
\item $j \le k$, $q_j = 1$, $\sigma \in \Gamma_1$, and $q=2$, or
\item $(\exists l \in [k]): q_l = 2$ and $q = 2$.
\end{compactitem}
It is obvious that $L(\mathcal A) = L_{\Gamma_1,\Gamma_2,j}$.
\end{proof}

As final preparation, we prove a consistency lemma for fta.

\begin{lemma}\rm \label{lm:consistency-fta} Let $\varphi \in \MSO(\Sigma)$, let $\cV$ abbreviate $\Free(\varphi)$, and let $A$ be a $\Sigma_\cV$-fta such that $\LL(A) = \LL(\varphi)$. Moreover, let $V$ be a finite set of variables. Then we can construct a $\Sigma_{\cV \cup V}$-fta $A'$ such that $\LL(A') = \LL_{\cV \cup V}(\varphi)$.
\end{lemma}
\begin{proof}  \underline{Case (a):} Let $V \subseteq \cV$.  Then we can choose $A' = A$ and we are ready.

  \underline{Case (b):} Let $V \not\subseteq \cV$. Moreover, let $A=(Q,\delta,F)$. First we construct the $\Sigma_{\cV \cup V}$-fta $B= (Q,\delta',F)$ such that, for every $ k \in \mathbb{N}$, $(\sigma,\cW) \in (\Sigma_{\cV \cup V})^{(k)}$, and $q_1,\ldots,q_k,q \in Q$ we define
  \[
\delta'_k(q_1 \cdots q_k,(\sigma,\cW),q) = \delta_k(q_1 \cdots q_k,(\sigma,\cW \cap \cV),q) \enspace.
\]
Obviously, for each  $\xi \in \T_{\Sigma_{\cV \cup V}}$, we have $\pos(\xi) = \pos(\xi|_\cV)$ and $\R_B^{\mathrm{a}}(\xi) = \R_A^{\mathrm{a}}(\xi|_\cV)$. 
Hence, we have
\begin{equation}\label{eq:auxiliary}
  \xi\in \LL(B) \ \ \text{ iff } \ \ \R_B^{\mathrm{a}}(\xi) \ne \emptyset \ \ \text{ iff } \ \ \R_A^{\mathrm{a}}(\xi|_\cV) \ne \emptyset \ \ \text{ iff } \ \ \xi|_\cV\in \LL(A)
   \enspace.
  \end{equation}

Next we show that
\begin{equation} \label{eq:adding-variable-to-fta}
\LL(B) \cap \T^{\mathrm{v}}_{\Sigma_{\cV \cup V}} = \LL_{\cV \cup V}(\varphi)\enspace. 
\end{equation}
For this, let $\xi \in \T_{\Sigma_{\cV \cup V}}$. Then we have
\begin{align*}
\xi \in \LL(B) \cap \T^{\mathrm{v}}_{\Sigma_{\cV \cup V}} & \text{ iff }  (\xi|_\cV \in \LL(A)) \wedge (\xi \in \T^{\mathrm{v}}_{\Sigma_{\cV \cup V}}) \tag{by \eqref{eq:auxiliary}}\\
& \text{ iff } (\xi|_\cV \in \LL(\varphi)) \wedge (\xi \in \T^{\mathrm{v}}_{\Sigma_{\cV \cup V}}) \\
& \text{ iff } \xi \in \LL_{\cV \cup V}(\varphi) \tag{by Lemma \ref{lm:msocons}}
\end{align*}

Now, by Lemma \ref{lm:TSigma-valid-fta}, we can construct a $\Sigma_{\cV \cup V}$-fta $C$ such that $\LL(C)=\T^{\mathrm{v}}_{\Sigma_{\cV \cup V}}$. 
By Theorem~\ref{thm:fta-closure-results}, we can construct and $\Sigma_{\cV \cup V}$-fta $A'$ with $\LL(A')= \LL(B) \cap \LL(C)$. Then, by \eqref{eq:adding-variable-to-fta} we have $\LL(A') = \LL_{\cV \cup V}(\varphi)$.
\end{proof}

Now we can prove that definable implies recognizable. 

\begin{lemma}\label{lm:thawri-MSO-to-fta} \rm For every finite set $\cV$ of variables and formula $\varphi \in \MSO(\Sigma)$ with $\Free(\varphi) \subseteq \cV$, we can construct a $\Sigma_\cV$-fta $A$ such that $\LL(A)= \LL_\cV(\varphi)$.
  \end{lemma}
  \begin{proof} First, by induction on $(\MSO(\Sigma),\succ_{\MSO(\Sigma)})$, we prove the following.
    \begin{equation}\label{eq:def-rec-with-free-var-only}
      \text{For each $\varphi \in \MSO(\Sigma)$, we can construct a $\Sigma_{\Free(\varphi)}$-fta $B$ such that $\LL(B)=\LL(\varphi)$.}
      \end{equation}
  
 I.B.: For the induction base we distinguish three cases.
   
    \underline{Case (a):} Let $\varphi = \llabel_{\sigma}(x)$. Then $\Free(\varphi)=\{x\}$. Moreover,  let~$\Gamma = \{(\sigma,\{x\})\}$. Clearly~$\LL(\varphi) = L_{\Gamma}\cap \T_{\Sigma_{\{x\}}}^{\mathrm{v}}$. By Lemma \ref{Gamma-a} we can construct a $\Sigma_{\{x\}}$-fta $C$ such that $\LL(C) = L_{\Gamma}$.  By Lemma~\ref{lm:TSigma-valid-fta}  we can construct a $\Sigma_{\cV}$-fta $D$ such that  $\LL(D) = \T_{\Sigma_{\{x\}}}^{\mathrm{v}}$. 
By Theorem \ref{thm:fta-closure-results}, we can construct a $\Sigma_{\{x\}}$-fta $B$ such that  $\LL(B) =  \LL(C) \cap \LL(D)$.

\underline{Case (b):} Let $\varphi = \edge_j(x,y)$. Then $\Free(\varphi)=\{x,y\}$.  Moreover,  let~$\Gamma_1 = \{ (\sigma,\{x\}) \mid \sigma \in \Sigma\}$ and~$\Gamma_2 = \{ (\sigma,\{y\}) \mid \sigma \in \Sigma\}$.  
Clearly~$\LL(\varphi) = L_{\Gamma_1,\Gamma_2,j}\cap \T_{\Sigma_{\{x,y\}}}^{\mathrm{v}}$.
By Lemma~\ref{Gamma1,2-a}, we can construct a $\Sigma_{\{x,y\}}$-fta $C$ such that $\LL(C) = L_{\Gamma_1,\Gamma_2,j}$.  Then we proceed as in Case (a).

\underline{Case (c):} Let $\varphi = (x \in X)$.  Then $\Free(\varphi)=\{x,X\}$.  Moreover,  let~$\Gamma = \{(\sigma,\{x,X\})\mid \sigma \in \Sigma\}$. Clearly~$\LL(\varphi) = L_{\Gamma}\cap \T_{\Sigma_{\{x,X\}}}^{\mathrm{v}}$.  Then we proceed as in Case (a).
  
I.S.:  For the induction step we distinguish four cases.

\underline{Case (a):} Let $\varphi = (\neg \psi)$.  We note  that~$\LL(\varphi) = \T_{\Sigma_{\Free(\varphi)}}^{\mathrm{v}} \setminus \LL(\psi)$. By Lemma~\ref{lm:TSigma-valid-fta}  we can construct a $\Sigma_{\Free(\varphi)}$-fta $D$ such that  $\LL(D) = \T_{\Sigma_{\Free(\varphi)}}^{\mathrm{v}}$.
By I.H. we can construct a $\Sigma_{\Free(\varphi)}$-fta $C$ such that $\LL(C)= \LL(\psi)$.  By Theorem \ref{thm:fta-closure-results}, we can construct a $\Sigma_{\Free(\varphi)}$-fta $B$ such that  $\LL(B) =  \LL(D) \setminus \LL(C)$.

\underline{Case (b):} Let $\varphi = \psi_1 \vee \psi_2$. We note that~\(\LL(\varphi) = \LL_{\Free(\varphi)}(\psi_1) \cup \LL_{\Free(\varphi)}(\psi_2)\).
By I.H. we can construct $\Sigma_{\Free(\psi_i)}$-fta $C_i$ such that  $\LL(C_i) = \LL(\psi_i)$ for $i\in\{1,2\}$. 
Then, by Lemma \ref{lm:consistency-fta} we can construct $\Sigma_{\Free(\varphi)}$-fta $D_i$ such that 
$\LL(D_i) = \LL_{\Free(\varphi)}(\psi_i)$ for $i\in\{1,2\}$. Lastly,
by Theorem \ref{thm:fta-closure-results}, we can construct a $\Sigma_{\Free(\varphi)}$-fta $B$ such that  $\LL(B) =  \LL(D_1) \cup \LL(D_2)$.

\underline{Case (c):} Let $\varphi = (\exists x.\psi)$.   By I.H. we can construct a $\Sigma_{\Free(\psi)}$-fta $C$ such that $\LL(C) = \LL(\psi)$. Then, 
by Lemma \ref{lm:consistency-fta}, we can construct a $\Sigma_{\Free(\psi) \cup \{x\}}$-fta $D= (Q,\delta,F)$ such that $\LL(D) = \LL_{\Free(\psi) \cup \{x\}}(\psi)$. 

Lastly, we construct the $\Sigma_{\mathrm{Free}(\varphi)}$-fta $B =(Q',\delta',F')$ such that~$Q' =
Q\times\{0,1\}$ (where $0$ and $1$ are understood as natural numbers, which can be summed up),
$F'=F\times\{1\}$, and for each~$k\in\mathbb{N}$, we have
\begin{align*}
\delta_k' =\;& \{ \bigl((q_1,0)\cdots(q_k,0),(\sigma,U), (q, 0)\bigr) \mid (\sigma,U) \in \Sigma_{\mathrm{Free}(\varphi)}^{(k)}, (q_1\cdots q_k,(\sigma,U),q)\in \delta_k\}\\
\cup\;& \{ \bigl((q_1,0)\cdots(q_k,0), (\sigma,U), (q, 1)\bigr) \mid (\sigma,U) \in \Sigma_{\mathrm{Free}(\varphi)}^{(k)}, (q_1\ldots
q_k, (\sigma,U \cup \{x\}), q)\in \delta_k\}\\
\cup\;& \{ \bigl((q_1,p_1)\cdots(q_k,p_k), (\sigma,U), (q, 1)\bigr) \mid (\sigma,U) \in \Sigma_{\mathrm{Free}(\varphi)}^{(k)},(q_1\cdots q_k, (\sigma,U), q)\in \delta_k, \textstyle\bigplus_{i\in [k]}p_i=1\}.
\end{align*}

Now we prove that~$\LL(\varphi) = \LL(B)$. For this, let $\zeta = (\xi,\eta)$ be in $\T_{\Sigma_{\Free(\varphi)}}^{\mathrm{v}}$. For each $\rho: \pos(\xi) \to Q'$, we define $\pi_1(\rho): \pos(\xi) \to Q$ and $\pi_2(\rho): \pos(\xi) \to \{0,1\}$ such that, for each $w \in \pos(\xi)$, we let $\pi_1(\rho)(w)$ and $\pi_2(\rho)(w)$ be the first component of $\rho(w)$ and the second component of $\rho(w)$, respectively. 
\begingroup
\allowdisplaybreaks
\begin{align*}
  & \zeta \in \LL(\varphi) \\
  \text{iff }\ &\zeta \models (\exists x. \psi)\\
  \text{iff } \ & (\exists w \in \pos(\xi)): (\xi,\eta[x \mapsto w]) \models \psi\\
  \text{iff } \ & (\exists w \in \pos(\xi)): (\xi,\eta[x \mapsto w]) \in \LL(D)\\
  \text{iff } \ & (\exists w \in \pos(\xi), \rho_1 \in \R_D^{\mathrm{v}}((\xi,\eta[x \mapsto w]))): \rho_1(\varepsilon) \in F\\ 
  \text{iff } \ & (\exists w \in \pos(\xi), \rho \in \Rv_B(\zeta)): \pi_1(\rho) \in \Rv_D((\xi,\eta[x \mapsto w])) \wedge  \pi_1(\rho)(\varepsilon) \in F \wedge \\
  & \hspace*{5mm}  (\forall v \in \prefix(w)): \pi_2(\rho)(v) = 1 \wedge (\forall v \in \pos(\xi) \setminus \prefix(w)): \pi_2(\rho)(v) = 0\\
  \text{iff } \ & (\exists \rho \in \Rv_B(\zeta)): \rho(\varepsilon) \in F'\\
  \text{iff } \ & \zeta \in \LL(B) \enspace.
  \end{align*}
  \endgroup

  We can also give an alternative proof as follows (using closure under tree relabelings). We observe that $\LL(\varphi) = \tau(\LL_{\Free(\psi) \cup \{x\}}(\psi))$ where $\tau= (\tau_k \mid k \in \mathbb{N})$ is the deterministic $(\Sigma_{\mathrm{Free}(\psi)\cup \{x\}}, \Sigma_{\mathrm{Free}(\varphi)})$-tree relabeling defined, for each $k\in \mathbb{N}$ and $(\sigma,U) \in \Sigma_{\mathrm{Free}(\psi)\cup \{x\}}^{(k)}$ by $\tau((\sigma,U)) = (\sigma,U \cap \Free(\varphi))$. This can be seen as
follows:
\begin{align*}
	(\xi,\rho) \in \LL(\varphi)
	& \ \text{iff } \ (\xi,\rho) \models \varphi \\
	&\ \text{iff } \ (\exists w \in \pos(\xi))\colon (\xi,\rho[x \rightarrow w]) \models \psi \\
	&\ \text{iff } \ (\exists \zeta' \in \T_{\Sigma_{\mathrm{Free}(\psi) \cup \{x\}}}^{\mathrm{v}})\colon \tau(\zeta') = (\xi,\rho) \text{ and }  \zeta' \models \psi \\
	&\ \text{iff } \ (\exists \zeta' \in \LL_{\mathrm{Free}(\psi) \cup \{x\}}(\psi))\colon \tau(\zeta') = (\xi,\rho) \\
	&\ \text{iff } \ (\xi,\rho) \in \tau(\LL_{\mathrm{Free}(\psi)\cup \{x\}}(\psi))\text{.}
\end{align*}
By I.H. we can construct a $\Sigma_{\Free(\psi)}$-fta $C$ such that $\LL(C) = \LL(\psi)$. Hence, by Lemma~\ref{lm:consistency-fta}, we can also construct a $\Sigma_{\Free(\psi) \cup \{x\}}$-fta $D$ such that   $\LL(D) = \LL_{\Free(\psi) \cup \{x\}}(\psi)$. Then, by the fact that $\LL(\varphi) = \tau(\LL_{\Free(\psi) \cup \{x\}}(\psi))$ and by Corollary \ref{cor:recog-closed-under-tree-relabeling}, we can construct a $\Sigma_{\Free(\varphi)}$-fta $B$ such that $\LL(B) = \LL(\varphi)$.

\underline{Case (d):} Let $\varphi = (\exists X.\psi)$.   By I.H. we can construct a $\Sigma_{\Free(\psi)}$-fta $C$ such that $\LL(C) = \LL(\psi)$. Then,  by Lemma \ref{lm:consistency-fta}, we can construct a $\Sigma_{\Free(\psi) \cup \{X\}}$-fta $C'=(Q,\delta,F)$ such that $\LL(C') = \LL_{\Free(\psi) \cup \{X\}}(\psi)$.

Lastly, we construct the $\Sigma_{\Free(\varphi)}$-fta~$B= (Q,\delta',F)$ such that, for
every~$k\in\mathbb{N}$, we let
\begin{align*}
  \delta_k' = \{  (q_1 \cdots q_k,(\sigma,U),q) \mid &(\sigma,U) \in \Sigma_{\Free(\varphi)}^{(k)} \text{ and }\\
  &\big( (q_1 \cdots q_k,(\sigma,U),q) \in \delta_k \text{ or } (q_1 \cdots q_k,(\sigma,U \cup \{X\}),q) \in \delta_k\big)\}
\end{align*}
Then it is easy to see that $\LL(B) = \LL(\varphi)$.

Also here we can give an alternative proof as follows. Now we observe that $\LL(\varphi) = \tau(\LL_{\Free(\psi) \cup \{X\}}(\psi))$ where $\tau= (\tau_k \mid k \in \mathbb{N})$ is the $(\Sigma_{\mathrm{Free}(\psi)\cup \{X\}}, \Sigma_{\mathrm{Free}(\varphi)})$-tree relabeling defined, for every $k\in \mathbb{N}$ and $(\sigma,U) \in \Sigma_{\mathrm{Free}(\psi)\cup \{X\}}^{(k)}$ by $\tau((\sigma,U)) = (\sigma,U \cap \Free(\varphi))$. Then the proof can be finished similarly as in Case~(c).

This finishes the proof of \eqref{eq:def-rec-with-free-var-only}.
Finally, by applying Lemma \ref{lm:consistency-fta} to the fta $B$,
we obtain the desired $\Sigma_\cV$-fta $A$ with $\LL(A) = \LL_\cV(\varphi)$.
  \end{proof}


                    \section{The weighted MSO-logic $\MSO(\Sigma,\B)$}
                  \label{sec:adding-weights}

                  Now we add weights to $\MSO(\Sigma)$ in the following way. As atomic formulas of the weighted logic $\MSO(\Sigma,\B)$ we use formulas of the form $\rmH(\kappa)$, where $\kappa$ is an $\mathbb{N}$-indexed family of mappings $\kappa_k: (\Sigma_\cU)^{(k)} \to B$. This introduces the weights into the logic. Roughly speaking, the semantics of $\rmH(\kappa)$ is the unique $\Sigma$-algebra homomorphism $\h_{\M(\Sigma,\kappa)}$ from the $\Sigma_\cU$-term algebra to the $(\Sigma,\kappa)$-evaluation algebra (cf. Section \ref{sect:trees} and Equation \eqref{eq:hom-kappa}). In the setting of the weighted logics of \cite{drogas05}, the formula $\rmH(\kappa)$ can be simulated by a weighted first-order universal quantification over a formula of which the semantics is a recognizable step mapping (cf. \cite[Lm. 5.12]{fulstuvog12} and Lemma \ref{lm:forall-rec-step-formula=H(kappa)}).
                  
\index{MSO@$\MSO(\Sigma,\B)$}
	We define the set of \emph{MSO formulas over $\Sigma$ and $\B$}, denoted by  $\MSO(\Sigma,\B)$, by the following EBNF with nonterminal $e$:
	\begin{equation}
		\textstyle e\,\, ::= \mathrm H(\kappa)\mid (\varphi\rhd e)\mid (e+e)\mid \bigplus\nolimits_x e\mid \bigplus\nolimits_X e\;, \label{eq:MSO-weighted}
              \end{equation}
              where
        \begin{compactitem}
        \item there exists a finite set $\cU$ of variables such that $\kappa = (\kappa_k \mid k \in \mathbb{N})$ is an $\mathbb{N}$-indexed family of mappings $\kappa_k: (\Sigma_\cU)^{(k)} \to B$,
                   \item $\varphi\in\MSO(\Sigma)$.
        \end{compactitem}
        
        \index{weighted first-order existential quantification}
        \index{weighted second-order existential quantification}
        \index{guarded formula}
        We will drop parentheses whenever no confusion arises. We call formulas of the form $\mathrm H(\kappa)$ \emph{atomic formulas}, formulas of the form $\varphi \rhd e$ \emph{guarded formulas}, formulas of the form $(e_1 + e_2)$ \emph{weighted disjunction}, and  formulas of the form $\bigplus\nolimits_x e$ and $\bigplus\nolimits_X e$ the \emph{weighted first-order existential quantification (of $e$)} and \emph{weighted second-order existential quantification (of $e$)}, respectively.

\index{succMSO@$\succ_{\MSO(\Sigma,\B)}$}
       As for $\MSO(\Sigma)$-formulas in Section \ref{sect:MSO-definition}, in order to perform inductive proofs or to define objects by induction, we will consider the reduction system
        \[(\MSO(\Sigma,\B),\succ_{\MSO(\Sigma,\B)})
        \]
        where  $\succ_{\MSO(\Sigma,\B)}$ is the binary relation on $\MSO(\Sigma,\B)$ defined as follows. For every $e_1,e_2 \in \MSO(\Sigma,\B)$, we let  $e_1 \succ_{\MSO(\Sigma,\B)} e_2$ if either of the following three cases holds. 

        Case (a) There exist $\varphi \in \MSO(\Sigma)$ and $e \in \MSO(\Sigma,\B)$ such that $e_1 = (\varphi \rhd e)$ and $e_2 = e$.

        Case (b) There exist $f_1,f_2 \in \MSO(\Sigma,\B)$ such that $e_1 = f_1+ f_2$ and $e_2 \in \{f_1,f_2\}$.

                Case (c) There exists $f \in \MSO(\Sigma,\B)$ such that $e_1$ has the form $\bigplus_x f$ or $\bigplus_X f$ and $e_2 = f$.

                                By Corollary \ref{cor:reduction-to-substring-is-terminating}, the relation $\succ_{\MSO(\Sigma,\B)}$ is terminating. Moreover, we have that $\nf_{\succ_{\MSO(\Sigma,\B)}}(\MSO(\Sigma,\B))$ is the set of formulas of the form $\rmH(\kappa)$.

                                For every $e \in \MSO(\Sigma,\B)$ and $w \in \MSO(\Sigma,\B) \cup \MSO(\Sigma)$, we say that \emph{$w$ is a subformula of $e$} if $w$ is a substring of $e$. We note that the relation ``is a subformula of'' is reflexive.

                                Next we define the semantics of MSO formulas over $\Sigma$ and $\B$. For this purpose, we need the notion of free variable; for later purpose, we also define the notion of bound variables. Formally, for each $e\in\MSO(\Sigma,\B)$, we define the set $\Free(e)$ of \emph{free variables of $e$} and the set $\Bound(e)$ of \emph{bound variables of $e$} by induction on $(\MSO(\Sigma,\B),\succ_{\MSO(\Sigma,\B)})$  as follows:
                   
I.B.: If $\cU$ is a finite set of variables and $\kappa = (\kappa_k \mid n \in \mathbb{N})$ with $\kappa_k: (\Sigma_\cU)^{(k)} \to B$, then $\Free(\mathrm H(\kappa))=\cU$ and $\Bound(\mathrm H(\kappa))=\emptyset$.
	
I.S.: We distinguish four cases.
	
\begin{compactitem}	
          	\item $\Free(\varphi\rhd e)=\Free(\varphi)\cup\Free(e)$ and  $\Bound(\varphi\rhd e)=\Bound(\varphi)\cup\Bound(e)$, 
		\item $\Free(e_1+e_2)=\Free(e_1)\cup\Free(e_2)$ and $\Bound(e_1+e_2)=\Bound(e_1)\cup\Bound(e_2)$, and 
		\item $\Free(\bigplus\nolimits_x e)=\Free(e)\setminus\{x\}$ and $\Bound(\bigplus\nolimits_x e)=\Bound(e)\cup \{x\}$, and
                  \item $\Free(\bigplus\nolimits_X e)=\Free(e)\setminus\{X\}$ and $\Bound(\bigplus\nolimits_X e)=\Bound(e)\cup\{X\}$.
	\end{compactitem}
	If $\Free(e)=\emptyset$, then we call $e$ a \emph{sentence}.

        Since the semantics of an atomic formula uses the concept of evaluation algebra, we briefly recall it from Section \ref{sect:trees}. For a given $\mathbb{N}$-indexed family $\kappa = (\kappa_k \mid k \in \mathbb{N})$ with $\kappa_k: \Sigma^{(k)} \to B$, the $(\Sigma,\kappa)$-evaluation algebra $(\B,\overline{\kappa})$ with
        \(
          \overline{\kappa}(\sigma)(b_1,\ldots,b_k) = b_1 \otimes \cdots \otimes b_k \otimes \kappa_k(\sigma)
        \),
        is denoted by $\M(\Sigma,\kappa)$, and the $\Sigma$-algebra homomorphism from the $\Sigma$-term algebra to $\M(\Sigma,\kappa)$ is denoted by $\h_{\M(\Sigma,\kappa)}$. We will employ this concept for various ranked alphabets when defining the semantics of atomic formulas.
        \index{homM@$\h_{\M(\Sigma,\kappa)}$}
        \index{homkappa@$\h_\kappa$}
        \label{p:convention-denotation-of-h}
        \begin{quote} \em In the sequel we will abbreviate  $\h_{\M(\Sigma,\kappa)}$ by $\h_\kappa$.
\end{quote}

\index{$\kappa[\cU\leadsto\cV]$}
Now let $\cU$ and $\cV$ be finite sets of variables such that  $\cU\subseteq\cV$. Moreover, let $\kappa = (\kappa_k \mid k \in \mathbb{N})$ with $\kappa_k: \Sigma_\cU^{(k)} \to B$.  We define the $\mathbb{N}$-indexed family
\(\kappa[\cU\leadsto\cV] = (\kappa[\cU\leadsto\cV]_k \mid k \in \mathbb{N})\)
with $\kappa[\cU\leadsto\cV]_k: \Sigma_\cV^{(k)} \to B$ such that, for every $\sigma\in\Sigma^{(k)}$ and $\cW\subseteq\cV$, we have
\[
  \kappa[\cU\leadsto\cV]_k(\sigma,\cW)=\kappa_k(\sigma,\cU\cap \cW) \enspace.
\]

Let $e \in \MSO(\Sigma,\B)$ and $\cV$ be a finite set of variables such that $\cV \supseteq \Free(e)$. We define the \emph{semantics of $e$ with respect to $\cV$}, denoted by 
$\sem{e}_\cV$, to be a $(\Sigma_\cV,\B)$-weighted tree language.
In other words, for each $e \in \MSO(\Sigma,\B)$, we define the family  $(\sem{e}_\cV \mid \cV \supseteq \Free(\e),  \cV \text{ finite})$. Since  $\sem{e}_\cV$ depends on the semantics of the subformulas of $e$ (each with respect to its own set of variables), we define the family
\[
  \Big(\big(\sem{e}_\cV \mid \cV \supseteq \Free(\e), \cV \text{ finite} \big) \mid e \in \MSO(\Sigma,\B) \Big) 
\]
by induction on  $(\MSO(\Sigma,\B),\succ_{\MSO(\Sigma,\B)})$ as follows.

I.B.: Let $\cU$ be a finite set of variables and $\kappa = (\kappa_k \mid k \in \mathbb{N})$ be an $\mathbb{N}$-indexed family with $\kappa_k:~(\Sigma_\cU)^{(k)}~\to~B$. Let $\cV \supseteq \Free(\rmH(\kappa))$ be a finite set of variables. Then, for each $\zeta \in \T_{\Sigma_\cV}$, we define
                  \\\mbox{}\hspace*{15mm} $\sem{\mathrm H(\kappa)}_\cV(\zeta)=
                  \begin{cases}\h_{\kappa[\cU\leadsto\cV]}(\zeta) & \text{if } \zeta \in  \T_{\Sigma_\cV}^\mathrm v\\
                    \0 & \text{otherwise}\enspace.
                  \end{cases}$

I.S.: We distinguish four cases.       
\begin{itemize}
                      \item Let  $\varphi\in \MSO(\Sigma)$ and $e\in \MSO(\Sigma,\B)$.  Let $\cV \supseteq \Free(\varphi\rhd e)$ be a finite set of variables. Then, for each $\zeta \in \T_{\Sigma_\cV}$,  we define
              \\\mbox{}\qquad
		$\sem{\varphi\rhd e}_\cV(\zeta) = \begin{cases}
		       	 \sem{e}_\cV(\zeta) &\text{if $\zeta\in \LL_\cV(\varphi)$}\\
                              	 \0 &\text{otherwise} \enspace.
              \end{cases}$
              
		\item Let $e_1,e_2\in \MSO(\Sigma,\B)$.  Let $\cV \supseteq \Free(e_1 + e_2)$ be a finite set of variables. Then, for each $\zeta \in \T_{\Sigma_\cV}$,  we define \\\mbox{}\qquad
                  $\sem{e_1+e_2}_\cV(\zeta) =
                  \begin{cases} \bigl(\sem{e_1}_\cV(\zeta) \oplus \sem{e_2}_\cV(\zeta)&  \text{if } \zeta \in  \T_{\Sigma_\cV}^\mathrm v\\
                    \0 & \text{otherwise}\enspace.
                  \end{cases}$
                  
		\item Let $x$ be a first-order variable and $e\in \MSO(\Sigma,\B)$. Let $\cV \supseteq \Free(\bigplus\nolimits_x e)$ be a finite set of variables.  Then, for each $\zeta \in \T_{\Sigma_\cV}$,  we define
                  \\\mbox{}\qquad
                  $\sem{\bigplus\nolimits_x e}_\cV(\zeta) =
                  \begin{cases}\bigoplus_{w\in \pos(\zeta)} \sem{e}_{\cV\cup\{x\}}(\zeta[x\mapsto w]) &  \text{if }  \zeta \in  \T_{\Sigma_\cV}^\mathrm v\\
                    \0 & \text{otherwise}\enspace.
                  \end{cases}$

            \item Let $X$ be a second-order variable and $e\in \MSO(\Sigma,\B)$.  Let $\cV \supseteq \Free(\bigplus\nolimits_X e)$ be a finite set of variables. Then, for each $\zeta \in \T_{\Sigma_\cV}$,  we define
              \\[2mm]\mbox{}\qquad
              $\sem{\bigplus\nolimits_X e}_\cV(\zeta) =
              \begin{cases}\bigoplus_{W\subseteq \pos(\zeta)} \sem{e}_{\cV\cup\{X\}}(\zeta[X\mapsto W]) & \text{if } \zeta \in  \T_{\Sigma_\cV}^\mathrm v\\
                    \0 & \text{otherwise}\enspace.
                  \end{cases}$
             
                       \end{itemize}
 
By definition, it is clear that the following holds:
\begin{eqnarray}
\begin{aligned}
  &\text{for every finite set $\cV$ of variables, $\varphi \in \MSO(\Sigma)$, and $e \in \MSO(\Sigma,\B)$}\\
  &\text{such that $\cV \supseteq \Free(\varphi \rhd e)$, we have $\sem{\varphi\rhd e}_\cV=\chi(\LL_\cV(\varphi))\otimes \sem{e}_\cV$.}
\end{aligned}\label{eq:guarded-e}
  \end{eqnarray}

\index{definable}
We abbreviate $\sem{e}_{\Free(e)}$ by $\sem{e}$. We say that a $(\Sigma,\B)$-weighted tree language $r$ is \emph{definable by a formula in $\MSO(\Sigma,\B)$-logic} (or: \emph{definable})  if there exists a sentence $e\in\MSO(\Sigma,\B)$ with $\sem{e} = r$.

Next we show a consistency lemma for formulas in $\MSO(\Sigma,\B)$.

\begin{lemma}\label{lm:consistency-MSO} \rm (cf. \cite[Lm.~3.8]{fulstuvog12})
	Let $e\in\MSO(\Sigma,\B)$ and let $\cV$ and $\cW$ be finite sets of variables with $\Free(e)\subseteq \cW\subseteq\cV$. Then, for each $(\xi,\eta)\in \T_{\Sigma_\cV}^{\mathrm v}$, we have
	$\sem{e}_\cV(\xi,\eta) = \sem{e}_\cW(\xi,\eta|_\cW)$.
      \end{lemma}
      \begin{proof} We prove the statement by induction on $(\MSO(\Sigma,\B),\succ_{\MSO(\Sigma,\B)})$.

I.B.: Let $e=\mathrm H(\kappa)$. Let $\cU$ be a finite set of variables and $\kappa = (\kappa_k \mid k \in \mathbb{N})$ an $\mathbb{N}$-indexed family of mappings $\kappa_k: (\Sigma_\cU)^{(k)} \to B$. Moreover, let $\cV$ and $\cW$ be finite sets of variables with $\cU\subseteq \cW\subseteq\cV$ and $(\xi,\eta)\in \T_{\Sigma_\cV}^{\mathrm v}$.   Then
\begin{align*}
\sem{\mathrm H(\kappa)}_\cV (\xi,\eta) = \h_{\kappa[\cU\leadsto \cV]}(\xi,\eta)=^{(*)} \h_{\kappa[\cU\leadsto \cW]}(\xi,\eta|_\cW)=\sem{\mathrm H(\kappa)}_\cW(\xi,\eta|_\cW)\enspace,
\end{align*}
where the equality $(*)$ can be proved as follows. Let $\zeta\in \T^{\mathrm v}_{\Sigma_\cV}$ and $\zeta'\in \T^{\mathrm v}_{\Sigma_\cW}$ be the trees which 
correspond to $(\xi,\eta)$ and $(\xi,\eta|_\cW)$, respectively. We have to show that 
\begin{equation}\label{eq:two-kappas-1}
\h_{\kappa[\cU\leadsto \cV]}(\zeta)= \h_{\kappa[\cU\leadsto \cW]}(\zeta')\enspace.
\end{equation}
Obviously, $\zeta'$ can be obtained from $\zeta$ by replacing each symbol $(\sigma,\cV')$ in $\zeta$ by $(\sigma,\cW\cap\cV')$. By \ref{eq:hom-kappa}, $\h_{\kappa[\cU\leadsto \cV]}(\zeta)$ is the product of the weights of the symbols of $\zeta$, where the factors are ordered according to the depth-first post-order of the positions. The product $\h_{\kappa[\cU\leadsto \cW]}(\zeta')$ is defined analogously. Since $\pos(\zeta)=\pos(\zeta')$, there exists a bijection between the lists of the factors of the two products. Moreover, the factors which correspond to each other are equal, because
for every $k\in \mathbb{N}$, $\sigma\in \Sigma^{(k)}$, and $\cV' \subseteq \cV$, by definition we have
\begin{align*}
\kappa[\cU\leadsto \cV]_k(\sigma,\cV')=\kappa_k(\sigma,\cU\cap \cV')=\kappa_k(\sigma,\cU\cap(\cW\cap \cV'))=\kappa[\cU\leadsto \cW]_k(\sigma,\cW\cap \cV')\enspace .
\end{align*}
This proves \eqref{eq:two-kappas-1} and hence $(*)$.

I.S.: For each form of $e$, the induction step is straightforward, hence we do not show the details. We note that in case $e = (\varphi \rhd e')$ we use Lemma \ref{lm:msocons}.
\end{proof}

\begin{example}\rm \label{ex:MSO-easy} Let $b \in B$ and $x$ be a first-order variable.
 \begin{enumerate}
\item  We define the family  $\kappa(b) = (\kappa(b)_k \mid k \in \mathbb{N})$ of mappings $\kappa(b)_k: \Sigma_{\{x\}}^{(k)} \to B$ such that, for each $(\sigma,U) \in \Sigma_{\{x\}}^{(k)}$, we let
  \index{kappa@$\kappa(b)$}
\[
  \kappa(b)_k((\sigma,U)) = \begin{cases}
    b & \text{ if $U=\{x\}$}\\
   \1 & \text{ otherwise} \enspace.
    \end{cases}
  \]
  Then, 
  \begin{equation}\label{equ:H(kappa(b))=b}
   \text{for every $\xi \in \T_{\Sigma}$ and $w \in \pos(\xi)$, we have }  \sem{\mathrm{H}(\kappa(b))}_{\{x\}}(\xi,[x \mapsto w]) = b  \enspace.
 \end{equation}
 This can be seen as follows:
 \begingroup
 \allowdisplaybreaks
 \begin{align*}
   \sem{\mathrm{H}(\kappa(b))}_{\{x\}}(\xi,[x \mapsto w])
   &= \h_{\kappa(b)[\{x\} \leadsto \{x\}]}(\xi,[x \mapsto w])
   = \h_{\kappa(b)}(\xi,[x \mapsto w])\\
   &= \bigotimes_{\substack{v \in \pos(\xi)\\\text{in $<_{\mathrm{dp}}$ order}}} \kappa(b)_{\rk(\xi(v))}((\xi,[x \mapsto w])(v)) \tag{by \eqref{eq:hom-kappa}}\\
   &= \underbrace{\1 \otimes \cdots \otimes \1}_{n-1} \  \otimes \ b \otimes \1 \otimes \cdots \otimes \1 = b \enspace,
   \end{align*}
 \endgroup
where we assume that $w$ occurs at the $n$-th position in the $<_{\mathrm{dp}}$ order of $\pos(\xi)$.

 \item  Let us consider the $\MSO(\Sigma,\B)$-sentence $\langle b \rangle$ defined by
\[
\langle b \rangle = \bigplus\nolimits_x (\mathrm{root}(x) \rhd \mathrm{H}(\kappa(b)))
\]
where the family $\kappa(b)$ is defined in the first part of this example.
It is clear that $\LL_{\{x\}}(\mathrm{root}(x))= \{(\xi,[x \mapsto \varepsilon])\mid \xi \in \T_\Sigma\}$.
  Then, for each  $\xi \in \T_\Sigma$, we have 
  \begingroup
  \allowdisplaybreaks
\begin{align*}
    \sem{\langle b \rangle}(\xi) &= \sem{\bigplus\nolimits_x (\mathrm{root}(x) \rhd \mathrm{H}(\kappa(b)))}(\xi)
                   = \bigoplus_{w\in \pos(\xi)} \sem{\mathrm{root}(x) \rhd \mathrm{H}(\kappa(b))}_{\{x\}}(\xi,[x\mapsto w])\\
                  & = \bigoplus_{w\in \pos(\xi)} \chi\big(\LL_{\{x\}}(\mathrm{root}(x))\big)(\xi,[x\mapsto w])\otimes \sem{\mathrm{H}(\kappa(b))}_{\{x\}}(\xi,[x\mapsto w]) \tag{by \eqref{eq:guarded-e}} \\
                                 &= \bigoplus_{\substack{w\in \pos(\xi):\\(\xi,[x\mapsto w])\in \LL_{\{x\}}(\mathrm{root}(x))}}\sem{\mathrm{H}(\kappa(b))}_{\{x\}}(\xi,[x\mapsto w])  = \sem{\mathrm{H}(\kappa(b))}_{\{x\}}(\xi,[x\mapsto \varepsilon])\\
  &= b  \tag{by \eqref{equ:H(kappa(b))=b}}\enspace.
  \end{align*} 
  \endgroup  
  Hence $\sem{\langle b \rangle}=\widetilde{b}$.
   \item Now let $\varphi$ be an $\MSO(\Sigma)$-formula. We consider the $\MSO(\Sigma,\B)$-formula
$\varphi \rhd \langle b\rangle$.
Then, for each finite set $\cV$ of variables with $\Free(\varphi)\subseteq \cV$ and  $\xi \in \T_{\Sigma_\cV}$, we have 
  \begingroup
  \allowdisplaybreaks
\begin{align*}
     \sem{\varphi \rhd \langle b \rangle }_\cV(\xi) 
                &   = \chi(\LL_\cV(\varphi))(\xi) \otimes \sem{\langle b \rangle}_\cV (\xi) \tag{by \eqref{eq:guarded-e}} \\
                  & = \sem{\langle b \rangle}_\cV (\xi) \otimes \chi(\LL_\cV(\varphi))(\xi) \tag{because $\chi(\LL(\varphi))(\xi) \in \{\0,\1\}$} \\
                  & = \sem{\langle b \rangle} (\xi) \otimes \chi(\LL_\cV(\varphi))(\xi) \tag{by Lemma  \ref{lm:consistency-MSO}}\\
                  & = \big(b \cdot \chi(\LL_\cV(\varphi))\big)(\xi)  \tag{because $\sem{\langle b \rangle}(\xi)=b$} \enspace .
  \end{align*} 
  \endgroup  
  Hence $\sem{\varphi \rhd \langle  b\rangle}_\cV = b \cdot \chi(\LL_\cV(\varphi))$.\hfill$\Box$
\end{enumerate}      
\end{example}


Now we will define particular $\MSO(\Sigma,\B)$-formulas called recognizable step formulas.  We show that the semantics of recognizable step sentences are recognizable step mappings, and vice versa, each recognizable step mapping is the semantics of such a recognizable step sentence.

  \index{recognizable step formula}
A \emph{$(\Sigma,\B)$-recognizable step formula} is a formula in $\MSO(\Sigma,\B)$ of the form 
\begin{equation}\label{eq:rec-step-formula}
(\varphi_1 \rhd \langle b_1\rangle) + \ldots + (\varphi_n \rhd \langle b_n\rangle)\enspace,
\end{equation}
where $\varphi_i$ is an $\MSO(\Sigma)$-formula and $\langle b_i \rangle$ is an $\MSO(\Sigma,\B)$-sentence defined in Example \ref{ex:MSO-easy}(2) for each $i\in [n]$. If, in addition, $\varphi_1,\ldots,\varphi_n$ are sentences, then \eqref{eq:rec-step-formula} is called a \emph{$(\Sigma,\B)$-recognizable step sentence}.

By Example \ref{ex:MSO-easy}(3),  for each finite set $\cV$ of variables such that $\cV \supseteq \bigcup_{i \in [n]} \Free(\varphi_i)$, we have
\begin{equation}\label{equ:semantics-rec-step-sentence}
\sem{(\varphi_1 \rhd \langle b_1\rangle) + \ldots + (\varphi_n \rhd \langle b_n\rangle)}_\cV = \bigoplus_{i\in [n]} b_i \cdot \chi(\LL_\cV(\varphi_i))\enspace.
\end{equation}

  \begin{lemma} \label{lm:rec-step-formula-lemma}\rm Let $r: \T_{\Sigma} \to B$ be a weighted tree language, $n \in \mathbb{N}_+$, and $b_1,\ldots,b_n \in B$. Then the following two statements are equivalent.
    \begin{compactenum}
    \item[(A)] We can construct $\Sigma$-fta  $A_1,\ldots, A_n$ such that $r = \bigoplus_{i \in [n]} b_i \cdot \chi(\LL(A_i))$.
      \item[(B)] We can construct sentences $\varphi_1,\ldots,\varphi_n$ in $\MSO(\Sigma)$ such that $r = \sem{(\varphi_1 \rhd \langle b_1\rangle) + \ldots + (\varphi_n \rhd \langle b_n\rangle)}$.
      \end{compactenum}
    \end{lemma}
    \begin{proof}  The proof follows from Theorem \ref{thm:that_wright}(A)$\Leftrightarrow$(B) and  \eqref{equ:semantics-rec-step-sentence}.
      \end{proof}


\begin{example}\label{ex:MSO-easy-2}\rm In this example, the weight algebra is the semiring $\Nat=(\mathbb{N},+,\cdot,0,1)$.

\begin{enumerate} 

\item We consider  the $\MSO(\Sigma,\Nat)$-sentence
\[ e= \bigplus\nolimits_x \mathrm{H}(\kappa(1))\enspace,\]
where  the $\mathbb{N}$-indexed  family $\kappa(1)$ of mappings is defined in Example \ref{ex:MSO-easy}.

Then, for each  $\xi \in \T_\Sigma$, we have 
  \begingroup
  \allowdisplaybreaks
\begin{align*}
    \sem{e}(\xi) &= \sem{\bigplus\nolimits_x \mathrm{H}(\kappa(1))}(\xi)
                   = \bigplus_{w\in \pos(\xi)} \sem{\mathrm{H}(\kappa(1))}_{\{x\}}(\xi,[x\mapsto w])\\
                 & =     \bigplus_{w\in \pos(\xi)} 1 \tag{by \eqref{equ:H(kappa(b))=b}}\\
  &= \ \size(\xi)         \enspace.
  \end{align*} 
  \endgroup
Hence $\sem{e} = \size$.

\item Now let $\Sigma = \{\sigma^{(2)}, \gamma^{(1)}, \alpha^{(0)}\}$ and consider the $\MSO(\Sigma,\Nat)$-sentence
\[ e= \bigplus\nolimits_x (\varphi(x) \rhd \mathrm{H}(\kappa(1)))\enspace,\]
where $\varphi(x)= \big(\llabel_\sigma(x) \wedge \forall y. (\edge_2(x,y) \to \llabel_\alpha(y))\big)$.
It is easy to see that \[\LL_{\{x\}}(\varphi(x))=\{(\xi,[x \mapsto w]) \mid \xi \in \T_\Sigma, w\in \pos(\xi), \xi(w)=\sigma \text{ and } \xi(w2)=\alpha \}\enspace.\]
Then for each  $\xi \in \T_\Sigma$, we have 
  \begingroup
  \allowdisplaybreaks
\begin{align*}
    \sem{e}(\xi) &= \sem{\bigplus\nolimits_x (\varphi(x) \rhd \mathrm{H}(\kappa(1)))}(\xi)
                   = \bigplus_{w\in \pos(\xi)} \sem{\varphi(x) \rhd \mathrm{H}(\kappa(1))}_{\{x\}}(\xi,[x\mapsto w])\\
                 & = \bigplus_{w\in \pos(\xi)} \chi\big(\LL_{\{x\}}(\varphi(x))\big)(\xi,[x\mapsto w])\cdot \sem{\mathrm{H}(\kappa(1))}_{\{x\}}(\xi,[x\mapsto w]) \tag{by \eqref{eq:guarded-e}}\\%
                   &= \bigplus_{\substack{w\in \pos(\xi): \\ \xi(w)=\sigma, \xi(w2)=\alpha}}\sem{\mathrm{H}(\kappa(1))}_{\{x\}}(\xi,[x\mapsto w])  \\
                 & =     \bigplus_{\substack{w\in \pos(\xi): \\ \xi(w)=\sigma, \xi(w2)=\alpha}} 1 \tag{by \eqref{equ:H(kappa(b))=b}} \\
  &=\#_{\sigma(.,\alpha)}(\xi)    \tag{for $\#_{\sigma(.,\alpha)}$, see Example \ref{ex:number-of-occurrences}}     \enspace.
  \end{align*} 
  \endgroup  
  Hence $\sem{e} = \#_{\sigma(.,\alpha)}$. \hfill $\Box$
\end{enumerate}
\end{example}

\begin{example}\rm \label{ex:height-MSO} Here we show that the weighted tree language $\height: \T_\Sigma \to \mathbb{N}$ is definable over the arctic semiring $\Natmaxplus=(\mathbb{N}_{-\infty},\max,+,-\infty,0)$.
 We let $\varphi$ be the following $\MSO(\Sigma)$-formula with $\Free(\varphi) = \{x,X\}$:
\begin{align*}
  \varphi(x,X) &= \mathrm{leaf}(x) \wedge \mathrm{Path}(X,x), \text{ where}\\
  \mathrm{leaf}(x) &= \neg \exists y. \edge(x,y),\\
  \mathrm{Path}(X,x) &= \forall y. (y \in X) \leftrightarrow \mathrm{path}(y,x),
  \end{align*}
and $\mathrm{path}(y,x)$ is the formula defined in Example \ref{ex:path}.

 The following is obvious: 
   \begin{equation}\label{eq:phi-height}
    \LL_{\{x,X\}}(\varphi) = \{(\xi,[x \mapsto w,X \mapsto \prefix(w)]) \mid  \xi \in \T_\Sigma, w \in \pos(\xi), \xi(w) \in \Sigma^{(0)}\}
    \end{equation}

Then we consider the $\MSO(\Sigma,\Natmaxplus)$-formula
\[
e_{\height} = \bigplus\nolimits_x\bigplus\nolimits_X (\varphi(x,X) \rhd \mathrm{H}(\kappa))
\]
where $\kappa = (\kappa_k \mid k \in \mathbb{N})$ and $\kappa_k: (\Sigma_{\{x,X\}})^{(k)} \to B$ defined, for each $(\sigma,U) \in (\Sigma_{\{x,X\}})^{(k)}$, by
\[
  \kappa_k((\sigma,U)) = \begin{cases}
    1 & \text{ if $X \in U$ and $k \ge 1$}\\
    0 & \text{ otherwise} \enspace.
    \end{cases}
  \]
  It is clear that, for every $(\xi,[x \mapsto w,X \mapsto \prefix(w)]) \in \LL_{\{x,X\}}(\varphi)$, we have
  \begin{equation}\label{eq:H-constant-height}
    \sem{\mathrm{H}(\kappa)}_{\{x,X\}}(\xi,[x \mapsto w,X \mapsto \prefix(w)])  = |w|  \enspace.
  \end{equation}

   Then, for each  $\xi \in \T_\Sigma$, we have 
  \begingroup
  \allowdisplaybreaks
  \begin{align*}
        \sem{e_{\height}}(\xi) &= \sem{\bigplus\nolimits_x \bigplus\nolimits_X (\varphi(x,X) \rhd \mathrm{H}(\kappa))}(\xi)\\
                   &= \max\Big( \max\big( \sem{\varphi(x,X) \rhd \mathrm{H}(\kappa)}_{\{x,X\}}(\xi[x\mapsto w, X \mapsto W]) \mid W \subseteq \pos(\xi)\big) \mid w\in \pos(\xi)\Big)\\
                   &= \max\Big(\sem{\varphi(x,X) \rhd \mathrm{H}(\kappa)}_{\{x,X\}}(\xi[x\mapsto w, X \mapsto W]) \mid W \subseteq \pos(\xi), w\in \pos(\xi)\Big)\\
                   &= \max\Big( \sem{\mathrm{H}(\kappa)}_{\{x,X\}}(\xi[x\mapsto w, X \mapsto \prefix(w)]) \mid  w\in \pos(\xi),\xi(w) \in \Sigma^{(0)}\Big) \tag{by \eqref{eq:phi-height}}\\
                               &= \max( |w| \mid w\in \pos(\xi), \xi(w) \in \Sigma^{(0)})  \tag{by \eqref{eq:H-constant-height}}\\
    &= \height(\xi) \enspace.  \hspace{90mm} \Box
  \end{align*} 
  \endgroup
\end{example}


We finish this section with two easy properties. Roughly speaking, they say that weighted existential quantification can be expressed by deterministic tree relabelings, where in the first-order case we have to take care of preserving validity.

\begin{lemma}\rm \label{lm:existential-quant=det-tree-relab}  
  Let $e \in \MSO(\Sigma,\B)$.

  \begin{compactenum}
    \item[(1)] Let $\cV = \Free(\bigplus_x e)$. Then 
      \(
        \sem{\bigplus\nolimits_x e} = \chi(\tau)(\sem{e}_{\cV\cup \{x\}}) \otimes \chi(\T_{\Sigma_\cV}^{\mathrm{v}})\),
      where $\tau = (\tau_k \mid k \in \mathbb{N})$ is the  deterministic $(\Sigma_{\cV \cup \{x\}},\Sigma_\cV)$-tree relabeling such that, for every $k \in \mathbb{N}$, $\sigma \in \Sigma^{(k)}$, and $\cW \subseteq \cV \cup \{x\}$, we let $\tau_k((\sigma,\cW)) = \{(\sigma,\cW \setminus \{x\})\}$.

      \item[(2)] Let $\cV = \Free(\bigplus_X e)$. Then
         \(
        \sem{\bigplus\nolimits_X e} = \chi(\tau)(\sem{e}_{\cV\cup \{X\}})\),
      where  $\tau = (\tau_k \mid k \in \mathbb{N})$ is the  deterministic $(\Sigma_{\cV \cup \{X\}},\Sigma_\cV)$-tree relabeling such that, for every $k \in \mathbb{N}$, $\sigma \in \Sigma^{(k)}$, and $\cW \subseteq \cV \cup \{X\}$, we let $\tau_k((\sigma,\cW)) = \{(\sigma,\cW \setminus \{X\})\}$.
          \end{compactenum}
      \end{lemma}
      \begin{proof}
         Proof of (1):  Let $\cV = \Free(\bigplus_x e)$ and $\xi \in \T_{\Sigma_\cV}$.
 We note that $x \not\in \cV$.  We distinguish the following two cases.

\underline{Case (a):} Let $\xi \not\in \T_{\Sigma_\cV}^{\mathrm{v}}$. Then
\(
  \sem{\bigplus\nolimits_x e}(\xi) = \0 = \chi(\tau)(\sem{e}_{\cV\cup \{x\}})(\xi) \otimes \chi(\T_{\Sigma_\cV}^{\mathrm{v}})(\xi) 
  \).

\underline{Case (b):} Let $\xi \in \T_{\Sigma_\cV}^{\mathrm{v}}$.  Then we can calculate as follows:
\begingroup
\allowdisplaybreaks
\begin{align*}
  \sem{\bigplus\nolimits_x e}(\xi)
 &= \bigoplus_{w\in \pos(\xi)} \sem{e}_{\cV \cup \{x\}}(\xi[x\mapsto w])\\
  &= \bigoplus_{\zeta \in \{\xi[x\mapsto w] | w \in \pos(\xi)\}} \sem{e}_{\cV \cup \{x\}}(\zeta)
  \tag{because $\xi[x\mapsto w] \ne \xi[x\mapsto w']$ for $w\ne w'$}\\
  &= \bigoplus_{\zeta \in \tau^{-1}(\xi)\cap \T_{\Sigma_{\cV \cup \{x\}}}^{\mathrm{v}}} \sem{e}_{\cV \cup \{x\}}(\zeta)
    \tag{by definition of $\tau$} \\
  &= \bigoplus_{\zeta \in \tau^{-1}(\xi)} \sem{e}_{\cV\cup \{x\}}(\zeta)
\tag{because $\sem{e}_{\cV\cup \{x\}}(\zeta)=\0$ for $\zeta \in \tau^{-1}(\xi)\setminus \T_{\Sigma_{\cV\cup \{x\}}}^{\mathrm{v}}$}  \\
  &= \chi(\tau)(\sem{e}_{\cV\cup \{x\}})(\xi) \tag{by \eqref{obs:app-tree-transf-to-wtl-2}}\\
  &= \big(\chi(\tau)(\sem{e}_{\cV\cup \{x\}})\otimes  \chi(\T_{\Sigma_\cV}^{\mathrm{v}})\big)(\xi) \enspace.\tag{because $\chi(\T_{\Sigma_\cV}^{\mathrm{v}})(\xi)=\1$}
  \end{align*}
  \endgroup

  \
  
  Proof of (2):  Let $\cV = \Free(\bigplus_X e)$ and  $\xi \in \T_{\Sigma_\cV}$. We note that $X \not\in \cV$.
Since $\cU^{(1)}=\cV^{(1)}$ (i.e., the set of first-order variables of $\cU$ is the same as the set of first-order variables of $\cV$), we have
\begin{equation}\label{eq:validity-invariant-under-adding-X}
\text{for every $\xi \in \T_{\Sigma_\cV}$ and $\zeta \in \tau^{-1}(\xi)$: $\xi \in \T_{\Sigma_\cV}^{\mathrm{v}}$ iff $\zeta \in \T_{\Sigma_{\cV \cup \{X\}}}^{\mathrm{v}}$.}
  \end{equation}

  We distinguish the following two cases.

\underline{Case (a):} Let $\xi \not\in \T_{\Sigma_\cV}^{\mathrm{v}}$. Then
\(\sem{\bigplus\nolimits_X e}(\xi) = \0 = \chi(\tau)(\sem{e}_{\cV\cup \{x\}})(\xi)\)
where the second equality holds by~\eqref{eq:validity-invariant-under-adding-X}.

\underline{Case (b):} Let $\xi \in \T_{\Sigma_\cV}^{\mathrm{v}}$.  Then we can calculate as follows:
\begingroup
\allowdisplaybreaks
\begin{align*}
  \sem{\bigplus\nolimits_X e}(\xi)
  &= \bigoplus_{W\subseteq \pos(\xi)} \sem{e}_{\cV\cup\{X\}}(\xi[X\mapsto W])\\
  &= \bigoplus_{\zeta \in \{\xi[X\mapsto W] | W \subseteq \pos(\xi)\}} \sem{e}_{\cV\cup\{X\}}(\zeta)
  \tag{because $\xi[X\mapsto W] \ne \xi[X\mapsto W']$ for $W\ne W'$}\\
  &= \bigoplus_{\zeta \in \tau^{-1}(\xi)} \sem{e}_{\cV\cup \{X\}}(\zeta)
  \tag{by definition of $\tau$} \\
  &= \chi(\tau)(\sem{e}_{\cV\cup \{X\}})(\xi) \tag{by \eqref{obs:app-tree-transf-to-wtl-2}}
  \end{align*}
  \endgroup
         \end{proof}


        \section{The main result for $\MSO(\Sigma,\B)$}
        \label{sec:the-main-result:BET}

       The main theorem of this chapter will be
                the following B-E-T theorem for weighted tree languages over strong bimonoids.

            \begin{theorem-rect}\label{thm:Buechi} Let $\Sigma$ be a ranked alphabet. Moreover, let $\B=(B,\oplus,\otimes,\0,\1)$ be a strong bimonoid and let $r: \T_\Sigma \to B$. Then the following two statements are equivalent.
    \begin{compactenum}
    \item[(A)] We can construct a $(\Sigma,\B)$-wta $\cA$ such that $r=\runsem{\cA}$.
    \item[(B)] We can construct a sentence $e \in \MSO(\Sigma,\B)$ such that $r=\sem{e}$.
      \end{compactenum}
    \end{theorem-rect}

    This theorem follows from Theorems \ref{thm:wta-MSO} and \ref{thm:MSO-wta}, which we will prove in the next two subsections.

  \subsection{From recognizable to definable}

  We prove that, for each $(\Sigma,\B)$-wta $\cA$, we can construct a sentence $e \in \MSO(\Sigma,\B)$ such that the run semantics $\runsem{\cA}$ is equal to $\sem{e}$. In the usual way, we use second-order quantifications to label positions of the given tree $\xi \in \T_\Sigma$ by transitions of $\cA$; by means of a formula of $\MSO(\Sigma)$-logic we can check whether this labeling corresponds to a run on $\xi$. Finally, we define a mapping $\kappa$ which translates each transition into its weight.

  \begin{theorem} {\rm \cite[Lm.~4.2]{fulstuvog12}} \label{thm:wta-MSO} For each $(\Sigma,\B)$-wta $\cA$, we can construct a sentence $e \in \MSO(\Sigma,\B)$ such that $\sem{e} = \runsem{\cA}$.
  \end{theorem}

  \begin{proof} Let $\cA=(Q,\delta,F)$ be a $(\Sigma,\B)$-wta. By Theorem \ref{thm:root-weight-normalization-run} we can assume that $\cA$ is root weight normalized. Thus $\supp(F)$ contains exactly one element, say, $q_f$, and $F_{q_f}=\1$.

  We define the set \[\cU = \bigcup(Q^k \times  Q \mid k \in \mathbb{N} \text{ such that } \rk^{-1}(k) \not= \emptyset)\enspace.\] Let $n \in \mathbb{N_+}$ be the cardinality of $\cU$. Then we choose an arbitrary bijection $\nu: \cU \to \{X_1,\ldots,X_n\}$ and fix it for the rest of the proof. In the sequel, we will not distinguish between a state behaviour $(q_1 \cdots q_k,q) \in \cU$ and the second-order variable $\nu((q_1 \cdots q_k,q))$. Thus, it is legitimate to consider each $(q_1 \cdots q_k,q) \in \cU$ as a second-order variable. Then, in particular, $\cU=\cU^{(2)}$ and hence $\T_{\Sigma_\cU} = \T_{\Sigma_\cU}^{\mathrm{v}}$.
    
    Moreover, for every $\xi \in \T_\Sigma$ and $\rho \in \R_\cA(\xi)$, we define the $\cU$-assignment $\eta_{\xi,\rho} \in \Phi_{\cU,\xi}$ as follows. For each $(q_1 \cdots q_k,q) \in \cU$, we let
    \[
\eta_{\xi,\rho}((q_1 \cdots q_k,q)) = \{w \in \pos(\xi) \mid \rho(w) = q, \rk(\xi(w))=k, (\forall i \in [k]): \rho(wi)=q_i\} \enspace.
      \]

Now we define the formula $\varphi \in \MSO(\Sigma)$ where $\Free(\varphi)=\cU$ with the following intention:
\begin{eqnarray}
  \begin{aligned}\label{eq:corr-run-assignment}
  &\text{The mapping $f: \R_\cA(q_f,\xi) \to \{\eta \in \Phi_{\cU,\xi} \mid (\xi,\eta) \in \LL_\cU(\varphi)\}$ which is}\\
  &\text{defined for each $\rho \in \R_\cA(q_f,\xi)$ by $f(\rho) = \eta_{\xi,\rho}$, is bijective.}
  \end{aligned}
         \end{eqnarray}       

         For this, we let 
 \begin{equation}\label{eq:formula-phi}       
        \varphi = \varphi_{\mathrm{part}} \wedge \varphi_{\mathrm{run}} \wedge \varphi_{\mathrm{suc}}
 \end{equation}       
        where
\begin{compactitem}
\item $\varphi_{\mathrm{part}}$ checks whether the family $(\eta(X)\mid X \in \cU)$ forms a partitioning of $\pos(\xi)$, i.e., whether each position of $\xi$ is assigned to exactly one transition,
\item $\varphi_{\mathrm{run}}$ checks whether $\eta$ codes a run in $\R_\cA(\xi)$, and 
\item $\varphi_{\mathrm{suc}}$ checks whether the target state of the transition associated to the root of $\xi$ is $q_f$.
        \end{compactitem}        
 Formally, we define     
                \begingroup
        \allowdisplaybreaks
        \begin{align*}
           \varphi_{\mathrm{part}}       &= \forall x. \Big(\bigvee_{X \in \cU} \big( x \in X \wedge \bigwedge_{Y \in \cU \setminus\{X\}} \neg(x \in Y)\big)  \Big)\\
          \varphi_{\mathrm{run}} &= \forall x. \bigwedge_{(q_1 \cdots q_k,q) \in \cU} \Bigg[(x \in (q_1 \cdots q_k,q)) \to   \bigwedge_{i \in [k]} \Big( \forall y. \edge_i(x,y) \to
            \bigvee_{(q_1' \cdots q_\ell',q_i) \in \cU} y \in (q_1' \cdots q_\ell',q_i)  \Big) \Bigg]\\[3mm]
          \varphi_{\mathrm{suc}} &= \forall x. \mathrm{root}(x) \to \bigvee_{(q_1 \cdots q_k,q_f)  \in \cU} x \in (q_1 \cdots q_k,q_f)
        \end{align*}
        \endgroup
        It is easy to check that \eqref{eq:corr-run-assignment} holds.

       Now we define the $\mathbb{N}$-indexed family $(\kappa_k \mid k \in \mathbb{N})$ with $\kappa_k: \Sigma_\cU^{(k)} \to B$ which supplies the weights. For every $k \in \mathbb{N}$, $\sigma \in \Sigma^{(k)}$, and $\cW \subseteq \cU$ we let
        \begin{equation}\label{eq:kappa-in-rec-def}
          \kappa_k((\sigma,\cW))=
          \begin{cases}
            \delta_k(q_1 \cdots q_k,\sigma,q) & \text{ if $\cW =\{(q_1 \cdots q_k,q)\}$}\\
              \0 &\text{ otherwise}
            \end{cases}
          \end{equation}

Let $\xi \in \T_\Sigma$ and $\rho \in \R_\cA(\xi)$ be arbitrary but fixed.
We consider $(\xi,\eta_{\xi,\rho})$ to be an element of  $\T_{\Sigma_\cU}^{\mathrm{v}}$ (by the identification which we discussed in Section \ref{sect:MSO-definition}).

\index{succb@$\succ$}
For the inductive proof of the next statement, we employ  the reduction system \((\pos(\xi),\succ)\) which was defined and proved to be terminating in the proof of Lemma~\ref{lm:BPS}.
We recall that,  for every  $w_1,w_2 \in \pos(\xi)$, we have $w_1 \succ w_2$ if there exists $i\in \mathbb{N}$ such that $w_2 = w_1i$, and that we have  $\nf_\succ(\pos(\xi))=\pos_{\Sigma^{(0)}}(\xi)$, i.e., it is the set of leaves of $\xi$.  Then we can prove the next statement by induction on $(\pos(\xi),\succ)$.
                  \begin{equation}
              \text{For every $w \in \pos(\xi)$:} \ \
              \h_\kappa((\xi,\eta_{\xi,\rho})|_w) = \wt_\cA(\xi|_w,\rho|_w) \enspace. \label{eq:h=wt1}
            \end{equation}
 For this, let  $w \in \pos(\xi)$ with $\sigma = \xi(w)$, and $k = \rk_\Sigma(\sigma)$. Then we have
          \[
(\xi,\eta_{\xi,\rho})(w) = (\sigma, \{X \in \cU \mid w \in \eta_{\xi,\rho}(X)\}) = (\sigma, \{(\rho(w1) \cdots \rho(wk),  \rho(w))\}) \enspace.
\]
We obtain:
\begingroup
\allowdisplaybreaks
\begin{align*}
  \h_\kappa((\xi,\eta_{\xi,\rho})|_w) &= \Big( \bigotimes_{i \in [k]} \h_\kappa((\xi,\eta_{\xi,\rho})|_{wi}) \Big) \otimes
                                        \kappa_k((\sigma, \{(\rho(w1) \cdots \rho(wk),  \rho(w))\}))  \tag{by \eqref{eq:evaluation-of-a-tree}}\\
                                      &= \Big( \bigotimes_{i \in [k]} \wt_\cA(\xi|_{wi},\rho|_{wi})) \Big) \otimes
                                        \delta_k(\rho(w1) \cdots \rho(wk),\sigma,  \rho(w))
                                        \tag{by I.H.  and the definition of $\kappa$}\\
  &= \wt_\cA(\xi|_{w},\rho|_{w})) \tag{by Observation \ref{obs:weight-run-explicit}} \enspace.
\end{align*}
\endgroup

Now we define the sentence $e \in \MSO(\Sigma,\B)$ and prove that it simulates $\cA$ (we recall that we do not distinguish between elements of $\cU$ and elements of $\{X_1,\ldots,X_n\}$):
\begin{equation}\label{eq:main-MSO-formula}
e = \bigplus\nolimits_{X_1} \cdots \bigplus\nolimits_{X_n} (\varphi \rhd \mathrm{H}(\kappa)) \enspace.
\end{equation}
Then, for each $\xi \in \T_\Sigma$, we have:
\begingroup
\allowdisplaybreaks
\begin{align*}
  \sem{e}(\xi) &= \bigoplus_{W_1,\ldots,W_n \subseteq \pos(\xi)} \sem{\varphi \rhd \mathrm{H}(\kappa)}_{\cU}(\xi[X_1\mapsto W_1,\ldots, X_n \to W_n])\\
  &
      = \bigoplus_{\eta \in \Phi_{\cU,\xi}}  \sem{\varphi \rhd \mathrm{H}(\kappa)}_{\cU}(\xi, \eta)
                 =  \bigoplus_{\substack{\eta \in \Phi_{\cU,\xi}:\\ (\xi,\eta) \in \LL_\cU(\varphi)}} \sem{\mathrm{H}(\kappa)}_{\cU}(\xi, \eta)\\
               &=  \bigoplus_{\substack{\eta \in \Phi_{\cU,\xi}:\\ (\xi,\eta) \in \LL_\cU(\varphi)}} \h_\kappa((\xi, \eta))
  \tag{because $\kappa[\cU \leadsto \cU]= \kappa$}\\
                           &= \bigoplus_{\rho \in \R_\cA(q_f,\xi)} \h_\kappa((\xi,\eta_{\xi,\rho}))
   \tag{by \eqref{eq:corr-run-assignment}}\\
               &= \bigoplus_{\rho \in \R_\cA(q_f,\xi)} \wt_\cA(\xi,\rho)
                 \tag{by \eqref{eq:h=wt1} for $w= \varepsilon$}\\
  &= \runsem{\cA}(\xi) \enspace. \qedhere
  \end{align*}
\endgroup
\end{proof}

 \begin{example} \rm \label{ex:wta-MSO} We illustrate the definition of the $\MSO(\Sigma,\B)$-formula \eqref{eq:main-MSO-formula} in the proof of Theorem \ref{thm:wta-MSO}. For this we consider the root weight normalized $(\Sigma,\Natmaxplus)$-wta $\cA = (Q,\delta,F)$ from Example \ref{ex:height}, which i-recognizes the $(\Sigma,\Natmaxplus)$-weighted tree language $\height$. We recall that $\Sigma=\{\sigma^{(2)},\alpha^{(0)}\}$, $\Natmaxplus=(\mathbb{N}_{-\infty},\max,+,-\infty,0)$, $Q = \{\h,0\}$, $\delta_0(\varepsilon,\alpha,\h) = \delta_0(\varepsilon,\alpha,0) = 0$ and for every $q_1,q_2,q\in Q$, 
\[
\delta_2(q_1q_2,\sigma,q) = 
\begin{cases}
1&\text{ if }q_1q_2q\in\{\h0\h, 0\h\h\}\;,\\
0&\text{ if }q_1q_2q=000\;,\\
-\infty&\text{ otherwise} \enspace,
\end{cases}
\]
and $F_\h=0$ and $F_0=-\infty$.

The set $\cU$ of second-order variables is the following:
\begin{align*}
  \cU = \{&(\varepsilon,\alpha,\h), \ (\varepsilon,\alpha,0), \\
          & (00,\sigma,0), \ (0\h,\sigma,0), \ (\h 0,\sigma,0), \ (\h\h,\sigma,0),\\
  & (00,\sigma,\h), \ (0\h,\sigma,\h), \ (\h 0,\sigma,\h), \ (\h\h,\sigma,\h) \}\enspace.
  \end{align*}
  
Let $\varphi \in \MSO(\Sigma)$ be the instance of the MSO formula \eqref{eq:formula-phi} for the particular $(\Sigma,\Natmaxplus)$-wta $\cA$ above.  We will not  give further details of $\varphi$. Instead, in Figure \ref{fig:ex-wta-to-MSO}, we show the tree $(\xi,\eta_2) \in \T_{\Sigma_\cU}^{\mathrm{v}}$ which satisfies $\varphi$, where $\xi = \sigma(\sigma(\alpha,\alpha),\alpha)$ and the assignment $\eta_2$ is defined in Figure \ref{fig:table-ex}.

 Next we instantiate the family of \eqref{eq:kappa-in-rec-def} to the particular $(\Sigma,\Natmaxplus)$-wta $\cA$ above. This provides the following $\mathbb{N}$-indexed family $\kappa = (\kappa_k \mid k \in \mathbb{N})$ with $\kappa_k: (\Sigma_\cU)^{(k)} \to \mathbb{N}_{-\infty}$:
  \begingroup
  \allowdisplaybreaks
  \begin{align*}
\kappa_0((\alpha,\{(\varepsilon,\alpha,\h)\}) &= \delta_0(\varepsilon,\alpha,\h) = 0\\
   \kappa_0((\alpha,\{(\varepsilon,\alpha,0)\}) &= \delta_0(\varepsilon,\alpha,0) = 0\\
    \kappa_2((\sigma,\{(\h 0,\sigma,\h)\}) &= \delta_2(\h 0, \sigma,\h) = 1\\
    \kappa_2((\sigma,\{(0 \h,\sigma,\h)\}) &= \delta_2(0 \h, \sigma,\h) = 1\\
    \kappa_2((\sigma,\{(0 0,\sigma,0)\}) &= \delta_2(00, \sigma,0) = 0
  \end{align*}
  \endgroup
  and for each other argument, $\kappa_0$ and $\kappa_2$ yield $-\infty$.
  Then we instantiate the formula \eqref{eq:main-MSO-formula} to be the $\MSO(\Sigma,\Natmaxplus)$-formula $e$ is defined by
  \[
    e = \bigplus\nolimits_{(\varepsilon,\alpha,\h)} \bigplus\nolimits_{(\varepsilon,\alpha,0)} \bigplus\nolimits_{(00,\sigma,0)} \cdots \ \bigplus\nolimits_{ (\h\h,\sigma,\h)} \ \ \varphi \rhd \mathrm H(\kappa) \enspace.
  \]

  Next we compute $\sem{e}(\xi)$ for the tree $\xi = \sigma(\sigma(\alpha,\alpha),\alpha)$. 
  First, we note that 
  \[\{(\xi,\eta) \in \T_{\Sigma_\cU} \mid (\xi,\eta) \models \varphi\} = \{(\xi,\eta_1), (\xi,\eta_2), (\xi,\eta_3)\}\]
  where the $\cU$-assignments $\eta_1$, $\eta_2$, and $\eta_3$ for $\xi$ are defined in the table shown in Figure \ref{fig:table-ex} (where $X \in \cU$). 

\begin{figure}[t]
  \[ 
     \begin{array}[t]{r|c|c|c}
       X    & \eta_1(X) & \eta_2(X) & \eta_3(X) \\\hline
       (\varepsilon,\alpha,\h) & \{11\} & \{12\} & \{2\}\\
       (\varepsilon,\alpha,0) & \{12,2\} & \{11,2\} & \{11,12\} \\
       (00,\sigma,0) & \emptyset & \emptyset & \{1\}\\
       (0\h,\sigma,0) & \emptyset & \emptyset & \emptyset\\
       (\h 0,\sigma,0) & \emptyset & \emptyset & \emptyset\\
       (\h\h,\sigma,0) & \emptyset & \emptyset & \emptyset \\
       (00,\sigma,\h) & \emptyset & \emptyset & \emptyset \\
       (0\h,\sigma,\h) & \emptyset & \{1\} & \{\varepsilon\} \\
       (\h 0,\sigma,\h) & \{\varepsilon, 1\} & \{\varepsilon\} & \emptyset\\
       (\h\h,\sigma,\h) & \emptyset & \emptyset & \emptyset\\  
    \end{array}
  \]
  \caption{\label{fig:table-ex} The $\cU$-assignments $\eta_1$, $\eta_2$, and $\eta_3$ for $\xi$.}
  \end{figure}
  
 By applying the unique $\Sigma_\cU$-algebra homomorphism $\h_{\kappa}$ to $(\xi,\eta_2)$, we obtain 
   \begingroup
   \allowdisplaybreaks
   \begin{align*}
     \h_{\kappa}((\xi,\eta_2)) &= \h_{\kappa}((\xi,\eta_2)|_1) + \h_{\kappa}((\xi,\eta_2)|_2) + \kappa_2((\sigma,\{(\h 0,\sigma,h)\}))\\
     &= \h_{\kappa}((\xi,\eta_2)|_1) + \h_{\kappa}((\xi,\eta_2)|_2) + 1\\
                          &= \Big(\h_{\kappa}((\xi,\eta_2)|_{11}) + \h_{\kappa}((\xi,\eta_2)|_{12}) + \kappa_2((\sigma,\{(0 \h,\sigma,h)\}))\Big) + \h_{\kappa}((\xi,\eta_2)|_2) + 1\\
          &= \Big(\h_{\kappa}((\xi,\eta_2)|_{11}) + \h_{\kappa}((\xi,\eta_2)|_{12}) + 1\Big) + \h_{\kappa}((\xi,\eta_2)|_2) + 1\\
                          &= \Big(\kappa_0((\alpha,\{(\varepsilon,\alpha,0)\}) + \kappa_0((\alpha,\{(\varepsilon,\alpha,\h)\}) + 1\Big) + \kappa_0((\alpha,\{(\varepsilon,\alpha,0)\}) + 1\\
        &= \Big(0 + 0 + 1\Big) + 0 + 1 = 2 \enspace.
    \end{align*}
    \endgroup
    By similar computations we obtain $\h_{\kappa}((\xi,\eta_1)) = 2$ and $\h_{\kappa}((\xi,\eta_3)) = 1$.  Finally,
    \begingroup
   \allowdisplaybreaks
   \begin{align*}
     \sem{e}(\xi) &= \max( \sem{\mathrm H(\kappa)}_\cU(\xi,\eta_1),  \ \sem{\mathrm H(\kappa)}_\cU(\xi,\eta_2), \ \sem{\mathrm H(\kappa)}_\cU(\xi,\eta_3))\\
     &= \max( \h_{\kappa}((\xi,\eta_1)), \ \h_{\kappa}((\xi,\eta_2)), \  \h_{\kappa}((\xi,\eta_3))  )\\
                  &= \max(2,2,1) = 2 = \height(\xi)\enspace. \hspace{70mm} \Box
   \end{align*}
   \endgroup
\end{example}

    \begin{figure}
      \centering
      \begin{tikzpicture}[level distance=4.5em,
  every node/.style = {align=center}]]

  \pgfdeclarelayer{bg}    
  \pgfsetlayers{bg,main}  

\begin{scope}[level 1/.style={sibling distance=25mm},
level 2/.style={sibling distance=26mm}, level 3/.style={sibling distance=15mm}]
 \node {$(\sigma,\{(\h 0,\sigma,\h)\})$}
 child {node {$(\sigma,\{(0 \h,\sigma,\h)\})$}  
   child { node[] {$(\alpha,\{(\varepsilon,\alpha,0)\})$} }
   child { node[] {$(\alpha,\{(\varepsilon,\alpha,\h)\})$} }
 }
 child {node {$(\alpha,\{(\varepsilon,\alpha,0)\})$}} ;
 \end{scope}

\end{tikzpicture}
\caption{\label{fig:ex-wta-to-MSO} The $\Sigma_\cU$-tree $(\xi,\eta_2)$; it satisfies $\varphi$.}
\end{figure}

    \subsection{From definable to recognizable}
    \label{ssec:definable-implies-recognizable}

By induction on $(\MSO(\Sigma,\B),\succ_{\MSO(\Sigma,\B)})$, we prove that, for each formula $e \in \MSO(\Sigma,\B)$, we can construct a $(\Sigma_{\Free(e)},\B)$-wta $\cA$ such that $\runsem{\cA} = \sem{e}$. 

For this we make some preparation. As first preparation, we prove a consistency lemma for wta.

    \begin{lemma}\rm \label{lm:consistency-wta} Let $e \in \MSO(\Sigma,\B)$, let $\cV$ abbreviate $\Free(e)$, and let $\cA$ be a $(\Sigma_\cV,\B)$-wta such that $\runsem{\cA} = \sem{e}$. Moreover, let $V$ be a finite set of variables. Then we can construct a $(\Sigma_{\cV \cup V},\B)$-wta $\cA'$ such that $\runsem{\cA'} = \sem{e}_{\cV \cup V}$.
\end{lemma}

\begin{proof}  
  \underline{Case (a):} Let $V \subseteq \cV$.  Then we can choose $\cA' = \cA$ and we are ready.

  \underline{Case (b):} Let $V \not\subseteq \cV$. Moreover, let $\cA=(Q,\delta,F)$. First we construct the $(\Sigma_{\cV \cup V},\B)$-wta $\cB= (Q,\delta',F)$ such that, for every $ k \in \mathbb{N}$, $(\sigma,\cW) \in (\Sigma_{\cV \cup V})^{(k)}$, and $q_1,\ldots,q_k,q \in Q$ we define
  \[
\delta'_k(q_1 \cdots q_k,(\sigma,\cW),q) = \delta_k(q_1 \cdots q_k,(\sigma,\cW \cap \cV),q) \enspace.
\]
 
Obviously, for each  $\xi \in \T_{\Sigma_{\cV \cup V}}$, we have $\pos(\xi) = \pos(\xi|_\cV)$ and $\R_\cB(\xi) = \R_\cA(\xi|_\cV)$. 
Moreover, for each $\rho \in \R_\cB(\xi)$, we have $\wt_\cB(\xi,\rho) = \wt_\cA(\xi|_\cV,\rho)$.
Then
\[
  \runsem{\cB}(\xi) = \bigoplus_{\rho \in \R_\cB(\xi)} \wt_\cB(\xi,\rho) \otimes F_{\rho(\varepsilon)}
  =  \bigoplus_{\rho \in \R_\cA(\xi|_\cV)} \wt_\cA(\xi|_\cV,\rho) \otimes F_{\rho(\varepsilon)}
  = \runsem{\cA}(\xi|_\cV) \enspace.
  \]

Now we prove the following statement.
\begin{equation} \label{eq:adding-variable-to-wta}
\text{For each $\xi \in \T_{\Sigma_{\cV \cup V}}$ we have $\runsem{\cB}(\xi) \otimes \chi(\T^{\mathrm{v}}_{\Sigma_{\cV \cup V}})(\xi) = \sem{e}_{\cV \cup V}(\xi)$. }
\end{equation}
For this, let $\xi \in \T_{\Sigma_{\cV \cup V}}$.

\underline{Case (i):} Let $\xi$ not be valid. Then both sides of \eqref{eq:adding-variable-to-wta} evaluate to $\0$.

\underline{Case (ii):} Let $\xi$ be valid. 
  Then we obtain
  \begingroup
  \allowdisplaybreaks
  \begin{align*}
    \runsem{\cB}(\xi) \otimes \chi(\T^{\mathrm{v}}_{\Sigma_{\cV \cup V}})(\xi)
    &= \runsem{\cB}(\xi) \tag{because $\xi$ is valid}\\
    &= \runsem{\cA}(\xi|_\cV)\\
    &= \sem{e}(\xi|_\cV)\\
    &=  \sem{e}_{\cV \cup V}(\xi) \tag{by Lemma \ref{lm:consistency-MSO}}
  \end{align*}
  \endgroup
  This finishes the proof of \eqref{eq:adding-variable-to-wta}.
  
  By Lemma \ref{lm:TSigma-valid-fta}, we can construct a $\Sigma_{\cV \cup V}$-fta $A$ such that $\LL(A)=\T^{\mathrm{v}}_{\Sigma_{\cV \cup V}}$. Finally, by Theorem \ref{thm:closure-Hadamard-product-char}, we can construct a $(\Sigma_{\cV \cup V},\B)$-wta $\cA'$ such that $\runsem{\cA'} =  \runsem{\cB} \otimes \chi(\T^{\mathrm{v}}_{\Sigma_{\cV \cup V}})$. Hence $\runsem{\cA'} = \sem{e}_{\cV \cup V}$.
    \end{proof}
    
    Now we proceed by transforming atomic formulas, guarded formulas, and weighted existential quantifications into a wta. (For formulas of the form $e_1 + e_2$ we will exploit a closure result.)

    \begin{lemma} \rm \label{lm:H-kappa-bu-det} \cite[Lm.~4.5]{fulstuvog12} For every finite sets $\cV$ of variables and  family $\kappa = (\kappa_k \mid k \in \mathbb{N})$ of mappings $\kappa_k: (\Sigma_\cV)^{(k)} \to B$, we can construct a bu-deterministic $(\Sigma_\cV,\B)$-wta $\cB$ such that $\sem{\mathrm{H}(\kappa)} = \sem{\cB}$.
    \end{lemma}
    
    \begin{proof} We recall that $\Free(\mathrm{H}(\kappa)) = \cV$. It is easy to see that $\sem{\mathrm{H}(\kappa)} = \h_{\kappa} \otimes \chi(\T_{\Sigma_\cV}^\mathrm{v})$. 
   In Example \ref{ex:h-kappa-crisp-det} we have constructed a bu-deterministic $(\Sigma_\cV,\B)$-wta $\cA$ such that $\sem{\cA} = \h_{\kappa}$. Then, by Lemma \ref{lm:TSigma-valid-fta} and by Theorem \ref{thm:closure-Hadamard-product-char}(2), we can construct a bu-deterministic $(\Sigma_\cV,\B)$-wta $\cB$ such that $\sem{\cB} = \h_{\kappa} \otimes \chi(\T_{\Sigma_\cV}^\mathrm{v})$.
    \end{proof}

  In the next lemma we prove that adding a guard to a  formula preserves r-recognizability.
  
      \begin{lemma} \label{lm:rec-closed-under-guards} \rm (cf. \cite[Lm.~4.10]{fulstuvog12})  Let $\varphi \in \MSO(\Sigma)$ and $e \in \MSO(\Sigma,\B)$. We let $\cU = \Free(\varphi)$ and $\cV = \Free(e)$. If there exists a $(\Sigma_\cV,\B)$-wta $\cA$ such that $\runsem{\cA} = \sem{e}$, then we can construct a $(\Sigma_{\cU \cup \cV},\B)$-wta $\cB$ such that $\runsem{\cB} = \sem{\varphi \rhd e}$.
\end{lemma}
      \begin{proof} Starting with $\cA$, we can apply Lemma \ref{lm:consistency-wta};
        thereby  we can construct a  $(\Sigma_{\cU \cup \cV},\B)$-wta $\cA'$ such that $\runsem{\cA'} = \sem{e}_{\cU \cup \cV}$. 
Moreover, by Lemma \ref{lm:thawri-MSO-to-fta}, we can construct a $\Sigma_{\cU\cup \cV}$-fta $D$ which recognizes $\LL_{\cU \cup \cV}(\varphi)$, i.e., $\LL(D) = \LL_{\cU \cup \cV}(\varphi)$.
Then, by Theorem  \ref{thm:closure-Hadamard-product-char}(2), we can construct a $(\Sigma_{\cU \cup \cV},\B)$-wta $\cB$ such that  $\runsem{\cB} = \chi(\LL(D)) \otimes \runsem{\cA'}$.
Finally,  we can calculate as follows.
\begingroup
\allowdisplaybreaks
\begin{align*}
  \sem{\varphi \rhd e}
  &= \chi(\LL_{\cU \cup \cV}(\varphi)) \otimes \sem{e}_{\cU \cup \cV}  \tag{by \eqref{eq:guarded-e}}\\
  &= \chi(\LL(D)) \otimes \runsem{\cA'}
  = \runsem{\cB} \enspace. \qedhere
\end{align*}
\endgroup
\end{proof}

    The next lemma shows that weighted existential quantification preserves r-recognizability.
    
    \begin{lemma}\label{lm:rec-closed-under-ex-quant}  \rm \cite[Lm.~4.9]{fulstuvog12}   Let $e \in \MSO(\Sigma,\B)$. Then the following two statements hold.
      \begin{compactenum}
      \item[(1)] If $\cV = \Free(\bigplus_x e)$, $\cU=\Free(e)$, and there exists a $(\Sigma_{\cU},\B)$-wta $\cA$ with $\runsem{\cA} = \sem{e}$, then we can construct a $(\Sigma_{\cV},\B)$-wta $\cB$ such that $\runsem{\cB} = \sem{\bigplus_x e}$.
      \item[(2)] If $\cV = \Free(\bigplus_X e)$, $\cU=\Free(e)$, and there exists a $(\Sigma_{\cU},\B)$-wta $\cA$ with $\runsem{\cA} = \sem{e}$, then we can construct a $(\Sigma_{\cV},\B)$-wta $\cB$ such that $\runsem{\cB} = \sem{\bigplus_X e}$.
          \end{compactenum}
      \end{lemma}

      \begin{proof} Proof of (1). Assume that $\cV = \Free(\bigplus_x e)$, $\cU=\Free(e)$, and there exists a  $(\Sigma_{\cU},\B)$-wta $\cA$ such that $\runsem{\cA} = \sem{e}$.  We note that $\cV=\cU \setminus \{x\}$, thus  $x \not\in \cV$.

We define the deterministic $(\Sigma_{\cV \cup \{x\}},\Sigma_\cV)$-tree relabeling $\tau = (\tau_k \mid k \in \mathbb{N})$ such that, for every $k \in \mathbb{N}$, $\sigma \in \Sigma^{(k)}$, and $\cW \subseteq \cV \cup \{x\}$ we let $\tau_k((\sigma,\cW)) = \{(\sigma,\cW \setminus \{x\})\}$.

Then, starting from the $(\Sigma_{\cU},\B)$-wta $\cA$, by Lemma \ref{lm:consistency-wta}, we can construct a $(\Sigma_{\cU \cup \{x\}},\B)$-wta $\cA'$ such that $\runsem{\cA'} = \sem{e}_{\cU \cup \{x\}}$. Since $\cU \cup \{x\} = (\cU \setminus\{x\}) \cup \{x\} = \cV \cup \{x\}$, we obtain that $\runsem{\cA'} = \sem{e}_{\cV \cup \{x\}}$.
  By using $\cA'$ and  Theorem \ref{thm:closure-under-tree-relabeling}, we can construct a $(\Sigma_{\cV},\B)$-wta which r-recognizes $\chi(\tau)(\sem{e}_{\cV\cup \{x\}})$.
By Lemma \ref{lm:TSigma-valid-fta},  we can construct a $\Sigma_{\cV}$-fta which recognizes $\T_{\Sigma_\cV}^{\mathrm{v}}$.
By Theorem \ref{thm:closure-Hadamard-product-char}(2), we can construct a $(\Sigma_\cV,\B)$-wta $\cB$ such that $\runsem{\cB} = \chi(\tau)(\sem{e}_{\cV\cup \{x\}})\otimes  \chi(\T_{\Sigma_\cV}^{\mathrm{v}})$.
Finally, by Lemma \ref{lm:existential-quant=det-tree-relab}(1), we obtain $\runsem{\cB}= \sem{\bigplus\nolimits_x e}$.

Proof of (2). Assume that $\cV = \Free(\bigplus_X e)$, $\cU = \Free(e)$, and there exists a $(\Sigma_{\cU},\B)$-wta $\cA$ such that $\runsem{\cA} = \sem{e}$. We note that $\cV=\cU \setminus\{X\}$, this $X \not\in \cV$.

We define the $(\Sigma_{\cV \cup \{X\}},\Sigma_\cV)$-tree relabeling $\tau = (\tau_k \mid k \in \mathbb{N})$ such that, for every $k \in \mathbb{N}$, $\sigma \in \Sigma^{(k)}$, and $\cW \subseteq \cV \cup \{X\}$ we let $\tau_k((\sigma,\cW)) = \{(\sigma,\cW \setminus \{X\})\}$.

Similarly as above, starting from the $(\Sigma_{\cU},\B)$-wta $\cA$, by Lemma \ref{lm:consistency-wta}, we can construct a $(\Sigma_{\cU \cup \{X\}},\B)$-wta $\cA'$ such that $\runsem{\cA'} = \sem{e}_{\cU \cup \{X\}}$. Since $\cU \cup \{X\} = (\cU \setminus\{X\}) \cup \{X\} = \cV \cup \{X\}$, we obtain that $\runsem{\cA'} = \sem{e}_{\cV \cup \{X\}}$.
By using $\cA'$ and  Theorem \ref{thm:closure-under-tree-relabeling}, we can construct a $(\Sigma_{\cV},\B)$-wta $\cB$ such that $\runsem{\cB} = \chi(\tau)(\sem{e}_{\cV\cup \{X\}})$. Finally, by Lemma \ref{lm:existential-quant=det-tree-relab}(2), we obtain $\runsem{\cB}= \sem{\bigplus\nolimits_X e}$.
  \end{proof}

Now we can prove that definability implies recognizability.

\begin{theorem}\label{thm:MSO-wta} Let $e\in\MSO(\Sigma,\B)$ and let $\cV$ abbreviate $\Free(e)$. Then we can construct a $(\Sigma_\cV,\B)$-wta $\cB$ such that   $\runsem{\cB} = \sem{e}$.
\end{theorem}
\begin{proof} By induction  on $(\MSO(\Sigma,\B),\succ_{\MSO(\Sigma,\B)})$, we prove  the statement of the theorem.

I.B.:
Let $e = \mathrm{H}(\kappa)$ where  $\kappa: \Sigma_\cV \to B$ for some finite set $\cV$ of variables.   By Lemma \ref{lm:H-kappa-bu-det}, we can construct a bu-deterministic $(\Sigma_\cV,\B)$-wta $\cB$ such that $\sem{\cB}= \sem{\mathrm{H}(\kappa)}$.
  
I.S.:
\underline{Case (a):} Let $e = (\varphi \rhd e')$ and  $\cV = \Free(\varphi \rhd  e')$. By I.H., we can construct a $(\Sigma_{\Free(e')},\B)$-wta $\cA$ such that $\runsem{\cA}=\sem{e'}$. By Lemma \ref{lm:rec-closed-under-guards}, we can construct a $(\Sigma_\cV,\B)$-wta $\cB$ such that $\runsem{\cB}=\sem{e}$.

\underline{Case (b):} Let $e = e_1 + e_2$ and let $\cV = \Free(e_1 + e_2)$. By I.H., we can construct a $(\Sigma_{\Free(e_1)},\B)$-wta $\cA_1$ and a $(\Sigma_{\Free(e_2)},\B)$-wta  $\cA_2$ such that $\runsem{\cA_1}=\sem{e_1}$ and $\runsem{\cA_2}=\sem{e_2}$. By Lemma \ref{lm:consistency-wta},
 we can construct  $(\Sigma_\cV,\B)$-wta $\cA_1'$  and $\cA_2'$ such that $\runsem{\cA_1'} = \sem{e_1}_\cV$ and $\runsem{\cA_2'} = \sem{e_2}_\cV$.
By Theorem \ref{thm:closure-sum} we can construct a $(\Sigma_\cV,\B)$-wta $\cB$ such that $\runsem{\cB}= \runsem{\cA_1'} + \runsem{\cA_2'}$. Then $\runsem{\cB}= \runsem{\cA_1'} + \runsem{\cA_2'} = \sem{e_1}_\cV + \sem{e_2}_\cV = \sem{e_1 + e_2}_\cV$.

\underline{Case (c):} Let $e = \bigplus_x e'$ and  let $\cV = \Free(\bigplus_x e')$.
By I.H., we can construct a $(\Sigma_{\Free(e')},\B)$-wta $\cA$ such that $\runsem{\cA} = \sem{e'}$. By Lemma~\ref{lm:rec-closed-under-ex-quant}(1), we can construct a $(\Sigma_\cV,\B)$-wta $\cB$ such that $\runsem{\cB}=\sem{e}$.
    
\underline{Case (d):} Let $e = \bigplus_X e'$ and  let $\cV = \Free(\bigplus_X e')$. We can finish the proof as in Case (d), except that we use Lemma~\ref{lm:rec-closed-under-ex-quant}(2).
 \end{proof}


\section[Adding weighted conjunction and  universal quantification]{Adding weighted conjunction and weighted universal quantification}
\label{sec:weighted-conj-univ-quant}

In this section we extend $\MSO(\Sigma,\B)$ by weighted conjunction, weighted first-order universal quantification, and  weighted second-order universal quantification \cite{drogas05,drogas07,drogas09}; the resulting logic is denoted by $\MSOe(\Sigma,\B)$. Since $\MSOe(\Sigma,\B)$ is more powerful than run recognizability, we identify a fragment of $\MSOe(\Sigma,\B)$, called carefully extended weighted MSO-logic and denoted by $\MSOce(\Sigma,\B)$,
which characterizes the set of run recognizable $(\Sigma,\B)$-weighted tree languages, if $\B$ is a commutative semiring (cf. Theorem \ref{thm:Buechi-extended}).
Roughly speaking, in the fragment $\MSOce(\Sigma,\B)$,
 (a)~the  weighted first-order universal quantification may only be applied to a recognizable step formula, and (b)~the weighted second-order universal quantification is forbidden. The latter restriction  corresponds to the restriction (c) of the weighted MSO-logic in \cite{drogas05,drogas07,drogas09} which is  discussed in the preface of this chapter. Our  restriction (a) is more severe than the corresponding restriction (b) there, because it forbids nesting of weighted first-order universal quantification. 
In Subsection \ref{sec:MSO-ext-commutative-bi-locally finite} we will prove that, for each commutative and bi-locally finite strong bimonoid $\B$, the fully extended weighted logic $\MSOe(\Sigma,\B)$ characterizes the set of run recognizable $(\Sigma,\B)$-weighted tree languages (cf. Theorem~\ref{thm:Buechi-comm-bi-loc-fin}).
\label{p:convention-commutative-Buechi}
\begin{quote} \emph{In this section, we assume that $\B$ is commutative.}
\end{quote}

\subsection{The extended weighted $\MSO$-logic $\MSOe(\Sigma,\B)$}

\index{MSO@$\MSOe(\Sigma,\B)$}
        	We define the set of \emph{extended MSO formulas over $\Sigma$ and $\B$}, denoted by  $\MSOe(\Sigma,\B)$, by the following EBNF with nonterminal $e$:
	\begin{equation}
		\textstyle e\,\, ::= \mathrm H(\kappa) \mid (\varphi\rhd e) \mid (e + e)\mid  (e \times e) \mid \bigplus\nolimits_x e\mid \bigplus\nolimits_X e\mid \bigtimes_x e \mid \bigtimes_X e\;, \label{eq:syntax-MSO-ext}
              \end{equation}
              	where
        \begin{compactitem}
        \item there exists a finite set $\cU$ of variables such that $\kappa = (\kappa_k \mid k \in \mathbb{N})$ is an $\mathbb{N}$-indexed family of mappings $\kappa_k: (\Sigma_\cU)^{(k)} \to B$,
            \item $\varphi\in\MSO(\Sigma)$.
     \end{compactitem}
     
     \index{weighted conjunction}
     \index{weighted first-order universal quantification}
     \index{weighted second-order universal quantification}
        Formulas of the form $(e_1 \times e_2)$, $\bigtimes_x e$, and $\bigtimes_X e$ are called \emph{weighted conjunction}, \emph{weighted first-order universal quantification}, and \emph{weighted second-order universal quantification}, respectively. Obviously, $\MSO(\Sigma,\B) \subset \MSOe(\Sigma,\B)$.

\index{succMSOext@$\succ_{\MSOe(\Sigma,\B)}$}
       As for $\MSO(\Sigma,\B)$-formulas in Section \ref{sec:adding-weights}, in order to perform inductive proofs or to define objects by induction, we will consider the reduction system
        \[(\MSOe(\Sigma,\B),\succ_{\MSOe(\Sigma,\B)})
        \]
        where  $\succ_{\MSOe(\Sigma,\B)}$ is the binary relation on $\MSOe(\Sigma,\B)$ defined as follows. For every $e_1,e_2 \in \MSOe(\Sigma,\B)$, we let  $e_1 \succ_{\MSOe(\Sigma,\B)} e_2$ if either of the following three cases holds. 

        Case (a) There exist $\varphi \in \MSO(\Sigma)$ and $e \in \MSOe(\Sigma,\B)$ such that $e_1 = (\varphi \rhd e)$ and $e_2 = e$.

        Case (b) There exist $f_1,f_2 \in \MSOe(\Sigma,\B)$ such that $e_1$ has the form $f_1+ f_2$ or $f_1 \times f_2$ and $e_2 \in \{f_1,f_2\}$.

                Case (c) There exists $f \in \MSOe(\Sigma,\B)$ such that $e_1$ has the form $\bigplus_x f$, $\bigplus_X f$, $\bigtimes_x f$, or $\bigtimes_X f$ and $e_2 = f$.

                                \sloppy By Corollary \ref{cor:reduction-to-substring-is-terminating}, the relation $\succ_{\MSOe(\Sigma,\B)}$ is terminating. Moreover, we have that $\nf_{\succ_{\MSOe(\Sigma,\B)}}(\MSO(\Sigma,\B))$ is the set of formulas of the form $\rmH(\kappa)$.
   For every $e \in \MSOe(\Sigma,\B)$ and $w \in \MSOe(\Sigma,\B) \cup \MSO(\Sigma)$, we say that \emph{$w$ is a subformula of $e$} if $w$ is a substring of $e$. We note that the relation ``is a subformula of'' is reflexive.

   For each $e\in\MSOe(\Sigma,\B)$, the set $\Free(e)$ of \emph{free variables of $e$} and set $\Bound(e)$ of \emph{bound variables of $e$} are defined in the same way as for $\MSO(\Sigma,\B)$-formulas and additionally we have the following three cases in the I.S.:
   	\begin{compactitem}
		\item $\Free(e_1 \times e_2)=\Free(e_1)\cup\Free(e_2)$ and $\Bound(e_1 \times e_2)=\Bound(e_1)\cup\Bound(e_2)$,
		\item $\Free(\bigtimes \nolimits_x e)=\Free(e)\setminus\{x\}$ and $\Bound(\bigtimes \nolimits_x e)=\Bound(e)\cup \{x\}$, and
                  \item   $\Free(\bigtimes \nolimits_X e)=\Free(e)\setminus\{X\}$ and $\Bound(\bigtimes \nolimits_X e)=\Bound(e)\cup \{X\}$.
                \end{compactitem}     
\index{semantice@$\sem{e}_\cV$}        

By induction on  $(\MSOe(\Sigma,\B),\succ_{\MSOe})$, we define the $\MSOe(\Sigma,\B)$-indexed family
\[
  \Big((\sem{e}_\cV \mid \cV \supseteq \Free(\e),  \cV \text{ finite}) \mid e \in \MSOe(\Sigma,B) \Big) 
\]
  as follows,  where $\sem{e}_\cV$ is a $(\Sigma_\cV,\B)$-weighted tree language which  is called the \emph{semantics of $e$ with respect to $\cV$}. 
  
I.B.:  Let $e = \mathrm{H}(\kappa)$. Let $\cV \supseteq \Free(e)$ be a finite set of variables. The definition of $\sem{e}_\cV$ is the same as in Section \ref{sec:adding-weights}.

I.S.: We distinguish four cases.
\begin{itemize}
\item Let $e= (\varphi \rhd e')$, $e = e_1 + e_2$, $e = \bigplus_x e'$, or $e = \bigplus_X e'$. Let $\cV \supseteq \Free(e)$ be a finite set of variables. The definition of $\sem{e}_\cV$ is the same as in Section \ref{sec:adding-weights}. 

		\item Let $e_1,e_2\in \MSOe(\Sigma,\B)$. Let $\cV \supseteq \Free(e_1 \times e_2)$ be a finite set of variables.  Then, for each $\zeta \in \T_{\Sigma_\cV}$, we define \\\mbox{}\qquad
                  $\sem{e_1 \times e_2}_\cV(\zeta) =
                  \begin{cases} \bigl(\sem{e_1}_\cV(\zeta) \otimes \sem{e_2}_\cV(\zeta)&  \text{if } \zeta \in  \T_{\Sigma_\cV}^\mathrm v\\
                    \0 & \text{otherwise}\enspace.
                  \end{cases}$

	\item Let $x$ be a first-order variable and $e \in \MSOe(\Sigma,\B)$.  Let $\cV \supseteq \Free(\bigtimes \nolimits_x e)$ be a finite set of variables. Then, for each $\zeta \in \T_{\Sigma_\cV}$, we define
                  \\\mbox{}\qquad
                  $\sem{\bigtimes \nolimits_x e}_\cV(\zeta) =
                  \begin{cases}\bigotimes_{w\in \pos(\zeta)} \sem{e}_{\cV\cup\{x\}}(\zeta[x\mapsto w]) & \text{if $\zeta \in  \T_{\Sigma_\cV}^\mathrm v$}\\
                    \0 & \text{otherwise} \enspace.
                  \end{cases}$

                  \item Let $X$ be a second-order variable and $e \in \MSOe(\Sigma,\B)$.  Let $\cV \supseteq \Free(\bigtimes \nolimits_X e)$ be a finite set of variables. Then, for each $\zeta \in \T_{\Sigma_\cV}$, we define
                  \\\mbox{}\qquad
                  $\sem{\bigtimes \nolimits_X e}_\cV(\zeta) =
                  \begin{cases}\bigotimes_{W \subseteq \pos(\zeta)} \sem{e}_{\cV\cup\{X\}}(\zeta[X\mapsto W]) & \text{if $\zeta \in  \T_{\Sigma_\cV}^\mathrm v$}\\
                    \0 & \text{otherwise} \enspace.
                    \end{cases}$
                  \end{itemize}
                  We note that the definitions of $\sem{\bigtimes \nolimits_x e}_\cV$ and $\sem{\bigtimes \nolimits_X e}_\cV$ use the fact that $\otimes$ is commutative (cf. page \pageref{page:finite-summation}).

 Next we show  that the extension of an assignment does not change the semantics of extended MSO formulas.

\begin{lemma}\label{obs:consistency} \rm (cf. Lemma \ref{lm:consistency-MSO})
	Let $e\in\MSOe(\Sigma,\B)$ and let $\cV$ and $\cW$ be finite sets of variables with $\Free(e)\subseteq \cW\subseteq\cV$. Then for every $(\xi,\eta)\in \T_{\Sigma_\cV}^{\mathrm v}$, we have
	$\sem{e}_\cV(\xi,\eta) = \sem{e}_\cW(\xi,\eta|_\cW)$.
      \end{lemma}

      \begin{proof} We prove the statement by induction on $(\MSOe(\Sigma,\B),\succ_{\MSOe})$. The proof is very similar to that of Lemma \ref{lm:consistency-MSO} and thus it is dropped.
      \end{proof}

We will prove the main theorem of this section (cf. Theorem \ref{thm:Buechi-extended}) by replacing subformulas of a formula in $\MSOe(\Sigma,\B)$ by equivalent formulas in $\MSO(\Sigma,\B)$. Next we will formalize this replacement.

Let $\varphi$ be an $\MSO(\Sigma)$-formula and $V$ and $W$ be two first-order variables or two second-order variables.  Intuitively, we let $\varphi[V/W]$ be the $\MSO(\Sigma)$-formula obtained from $\varphi$ by replacing each free occurrence of $V$ 
by $W$  (similar to $\alpha$-conversion in $\lambda$-calculus). Formally, we define the $\MSO(\Sigma)$-formula $\varphi[V/W]$ by induction on $(\MSO(\Sigma),\succ_{\MSO(\Sigma)})$ (cf.  \eqref{equ:syntax-MSO-unweighted}) as follows:

I.B.: We distinguish three cases.
      \[
      \llabel_\sigma(x)[V/W] =
      \begin{cases}
        \llabel_\sigma(W) & \text{ if $x=V$}\\
        \llabel_\sigma(x) & \text{ otherwise}
      \end{cases}
      \hspace{10mm}
      \edge_i(x,y)[V/W] =
      \begin{cases}
        \edge_i(W,y) & \text{ if $x=V$, $y\not=V$}\\
                \edge_i(x,W) & \text{ if $y=V$, $x\not=V$}\\
        \edge_i(W,W) & \text{ if $x=y=V$}\\
        \edge_i(x,y) & \text{ otherwise}
        \end{cases}
      \]
      \[
      (x \in X)[V/W] =
      \begin{cases}
        (W \in X) & \text{ if $x=V$}\\
        (x \in W) & \text{ if $X=V$}\\
        (x \in X) & \text{ otherwise}
        \end{cases}
      \]

I.S.: We distinguish four cases:
      \[
      (\neg \varphi)[V/W] = \neg (\varphi[V/W])
      \hspace{10mm}
      (\varphi_1 \vee \varphi_2)[V/W] = \varphi_1[V/W] \vee \varphi_2[V/W]
      \]
      \[
      (\exists x. \varphi)[V/W] =
      \begin{cases}
        (\exists x. \varphi) & \text{ if $x=V$}\\
        \exists x. (\varphi[V/W]) & \text{ otherwise}
      \end{cases}
      \hspace{10mm}
      (\exists X. \varphi)[V/W] =
      \begin{cases}
        (\exists X. \varphi) & \text{ if $X=V$}\\
        \exists X. (\varphi[V/W]) & \text{ otherwise.}
      \end{cases}
      \]

Next by induction on  $(\MSOe(\Sigma,\B),\succ_{\MSOe})$ we define $e[V/W]$ for each $\MSOe(\Sigma,\B)$-formula $e$ and first-order or second order variables $V$ and $W$.

I.B.: Let $e=\mathrm{H}(\kappa)$ and $\kappa=(\kappa_k\mid k\in \mathbb{N})$ with $\kappa_k:\Sigma_\cU^{(k)} \to B$. Then
\[\mathrm{H}(\kappa)[V/W] =
\begin{cases}
\mathrm{H}(\kappa) & \text{ if $V\not\in \cU$}\\
\mathrm{H}(\kappa') & \text{ otherwise,}
\end{cases}
\] 
where, for each $k\in \mathbb{N}$, the mapping $\kappa'_k:\Sigma_\cV^{(k)} \to B$ is defined as follows: $\cV=(\cU\setminus \{V\})\cup\{W\}$ and for every $(\sigma,\cW) \in \Sigma_\cV^{(k)}$, we have
\[
\kappa'_k((\sigma,\cW))=
\begin{cases}
\kappa_k((\sigma,\cW)) & \text{ if $W\not\in \cW$}\\
\kappa_k\big((\sigma,(\cW\setminus\{W\})\cup\{V\})\big) & \text{ otherwise.}\\
\end{cases}
\]

I.S.: We distinguish seven cases.

\begin{itemize}
\item Let  $e = (\varphi \rhd e')$. Then $e[V/W] = (\varphi[V/W] \rhd e'[V/W])$.

\item The definition of $e[V/W]$ for the other six cases of $e$ is left to the reader.
\end{itemize}

\

We observe that renaming a bound variable in a formula does not change the semantics of that formula.

\begin{observation}\label{lm:renaming-bounded-variables}\rm  The following two statements hold.
\begin{enumerate}
\item[(1)] Let $\exists V. \varphi$ be an $\MSO(\Sigma)$-formula, where $V$ is a  variable, and $\cV$ be a finite set of variables with $\Free(\exists V. \varphi)\subseteq \cV$. Moreover, let $W$ be a variable of the same type as $V$ with $W\not\in \cV$. Then $\LL_\cV(\exists V. \varphi)=\LL_\cV(\exists W. \varphi[V/W])$. 
\item[(2)]  Let $Q_V e$ be an $\MSOe(\Sigma,\B)$-formula for some $Q\in\{\bigplus,\bigtimes\}$, where $V$ is a  variable, and $\cV$ be a finite set of variables with $\Free(Q_V e)\subseteq \cV$. Moreover, let $W$ be a variable of the same type as $V$ with $W\not\in \cV$. Then $\sem{Q_V  e}_\cV=\sem{Q_W e[V/W]}_\cV$.  \hfill $\Box$
\end{enumerate}
\end{observation}

Moreover, we formalize three variants of the observation that, for each formula, the replacement of a subformula by an equivalent formula does not change the semantics of the original formula.

\begin{observation}\rm \label{obs:replacing-equiv-subformula-1} Let $\varphi\in\MSO(\Sigma)$ and let $\cV$ be a finite set of variables such that $\cV \supseteq \Free(\varphi)$.
Moreover, let $\psi$ be a subformula of $\varphi$. Also let $\psi' \in \MSO(\Sigma)$ such that $\Free(\psi') = \Free(\psi)$ and $\LL(\psi') = \LL(\psi)$. Then $\LL_\cV(\varphi) = \LL_\cV(\varphi[\psi/\psi'])$, where $\varphi[\psi/\psi']$ is the formula in $\MSO(\Sigma)$ obtained from $\varphi$ by replacing each occurrence of the subformula $\psi$  by $\psi'$.\hfill $\Box$
  \end{observation}

\begin{observation}\rm \label{cor:replacing-equiv-subformula-2} Let $e\in\MSOe(\Sigma,\B)$ and let $\cV$ be a finite set of variables such that $\cV \supseteq \Free(e)$. 
Moreover, let $\psi$ be an $\MSO(\Sigma)$-subformula of $e$. Also let $\psi' \in \MSO(\Sigma)$ such that $\Free(\psi') = \Free(\psi)$ and $\LL(\psi') = \LL(\psi)$. 
Then $\sem{e}_\cV = \sem{e[\psi/\psi']}_\cV$, where $e[\psi/\psi']$ is the formula in $\MSOe(\Sigma,\B)$ obtained from $e$ by replacing each occurrence of the subformula $\psi$  by $\psi'$.\hfill $\Box$
  \end{observation}

\begin{observation}\rm \label{obs:replacing-equiv-subformula-3} 	Let $e\in\MSOe(\Sigma,\B)$ and let $\cV$ be a finite set of variables such that $\cV \supseteq \Free(e)$. 
Moreover, let $t$ be an $\MSOe(\Sigma,\B)$ subformula of $e$. Also let $t' \in \MSOe(\Sigma,\B)$ such that $\Free(t') = \Free(t)$ and $\sem{t'} = \sem{t}$. Then $\sem{e}_\cV = \sem{e[t/t']}_\cV$, where $e[t/t']$ is the formula in $\MSOe(\Sigma,\B)$ obtained from $e$ by replacing each occurrence of the subformula $t$  by $t'$.
\hfill $\Box$
  \end{observation}
   
Let $e$ be an $\MSOe(\Sigma,\B)$-formula. We say that $e$ is  \emph{variable separated} if $\Free(e)\cap \Bound(e)= \emptyset$.  

Next we show that each $\MSOe(\Sigma,\B)$-formula can be transformed into an equivalent one which is variable separated (cf. rectified formula in \cite{gal87} and \cite{sch89}).
  
\begin{lemma}\label{lm:normal-form-weighted-MSO-ext} \rm Let $e$ be an $\MSOe(\Sigma,\B)$-formula and $\cV$ be a finite set of variables with $\Free(e)\subseteq \cV$. We can construct an $\MSOe(\Sigma,\B)$-formula $e'$ such that $e'$ is variable separated, $\Free(e)=\Free(e')$, and $\sem{e}_\cV= \sem{e'}_\cV$.
\end{lemma}  
\begin{proof}  Let $\cV$ be a finite set of variables and $\MSOe(\Sigma,\B)_\cV$ be the set of all those $e \in \MSOe(\Sigma,\B)$ such that $\Free(e) \subseteq \cV$.

Intuitively, we define the reduction relation $\succ$ on $\MSOe(\Sigma,\B)_\cV$ such that, in each reduction step, the variable $V$ of an (unweighted or weighted)  quantification where $V$ also occurs freely somewhere else in $e$, is replaced by a fresh variable $W$. Formally, for every $e_1,e_2 \in \MSOe(\Sigma,\B)_\cV$, we let $e_1 \succ e_2$ if either (a) or (b) holds.

Case (a): \begin{compactitem}
\item $e_1$ contains a subformula of the form $\varphi=\exists V.\psi$ such that $V\in \Free(e_1)$ and
\item $e_2 = e_1[\varphi/\varphi']$ where $\varphi'= \exists W. \psi[V/W]$ for some variable $W$ of the same type as $V$  with $W\not \in \Free(e_1)\cup \Bound(e_1)$.
\end{compactitem}

Case (b): \begin{compactitem}
\item $e_1$ contains a subformula of the form $t=Q_V u$ for some $Q\in\{\bigplus,\bigtimes\}$  such that $V\in \Free(e_1)$ and
\item $e_2 = e_1[t/t']$ where  $t'= Q_W u[V/W]$ for some  variable $W$ of the same type as $V$   with $W\not \in \Free(e_1)\cup \Bound(e_1)$.
\end{compactitem}

Next we prove that $\succ$ is terminating. For each $e \in \MSOe(\Sigma,\B)_\cV$, we let $\#(e)$ denote the number of occurrences of subformulas of $e$ of the form $\exists V.\psi$ or of the form $Q_V u$, where $Q\in\{\bigplus,\bigtimes\}$ and $V\in \Free(e)$. Obviously, for each $e_1\in \MSOe(\Sigma,\B)$ with $\#(e_1)>0$, we can construct an  $e_2\in \MSOe(\Sigma,\B)$ such that $e_1\succ e_2$.
Moreover, for every $e_1, e_2 \in \MSOe(\Sigma,\B)_\cV,\succ)$ we have that, if $e_1 \succ e_2$, then $\#(e_1) \succ_{\mathbb{N}} \#(e_2)$. Thus, $\#$ is a monotone embedding of $(\MSOe(\Sigma,\B)_\cV,\succ)$ into $(\mathbb{N},\succ_{\mathbb{N}})$, and hence, by  Lemma \ref{lm:fin-branching-embedding-termination}, the reduction system $(\MSOe(\Sigma,\B)_\cV,\succ)$ is terminating.
Obviously, we have
\begin{equation}\label{equ:nf-variable-separation}
\nf_\succ(\MSOe(\Sigma,\B)_\cV) = \{e \in \MSOe(\Sigma,\B)_\cV \mid e \text{ is variable separated}\} \enspace.
  \end{equation}

  Next we prove that, for every $e_1, e_2 \in \MSOe(\Sigma,\B)_\cV$ we have the following: if $e_1 \succ e_2$, then $\sem{e_1}_\cV= \sem{e_2}_\cV$. Let $e_1, e_2 \in \MSOe(\Sigma,\B)_\cV,\succ)$ with $e_1 \succ e_2$. 
In case (a), we have $\Free(\varphi')=\Free(\varphi)$ and by Observation \ref{lm:renaming-bounded-variables}(1), we have $\LL(\varphi')=\LL(\varphi)$. Hence, by Observation \ref{cor:replacing-equiv-subformula-2}, we have $\sem{e_1}_\cV = \sem{e_1[\varphi/\varphi']}_\cV= \sem{e_2}_\cV$.
In case (b), we have $\Free(t')=\Free(t)$ and by Observation \ref{lm:renaming-bounded-variables}(2), we have $\sem{t'}=\sem{t}$. Hence, by Observation \ref{obs:replacing-equiv-subformula-3}, we have $\sem{e_1}_\cV = \sem{e_1[t/t']}_\cV = \sem{e_2}_\cV$.

Finally, we can prove the statement of the lemma as follows. Let $e$ be an $\MSOe(\Sigma,\B)$-formula and $\cV$ be a finite set of variables with $\Free(e)\subseteq \cV$, i.e., $e \in \MSOe(\Sigma,\B)_\cV$. Then we construct an $f \in \nf_\succ(e)$. By the above, $\sem{f}_\cV= \sem{e}_\cV$,  and, by \eqref{equ:nf-variable-separation}, $f$ is variable separated.
\end{proof}

The next example shows the usefulness of weighted conjunction and weighted universal first-order quantification as specification tool.

\begin{example}\rm In this example we show that, for each $(\Sigma,\B)$-weighted local system  $\cS$ (cf. Section~\ref{sec:loc-tree-lang-wls}) we can construct a sentence $e_\cS \in \MSOe(\Sigma,\B)$ such that (a)  $\sem{e_\cS} = \sem{\cS}$ and (b) $e_\cS$ does not contain weighted second-order universal or existential quantification. For the definition of the macro $\langle b \rangle$ (with $b \in B$) we refer to Example \ref{ex:MSO-easy}. In an obvious way, we extend the syntax of $\MSOe(\Sigma,\B)$ by allowing that $+$ can combine finitely many formulas (and not just two). Hence, for each finite family $(e_i \mid i \in I)$ of formulas $e_i \in \MSOe(\Sigma,\B)$ we may write
  \[
\bigplus_{i \in I} e_i  \enspace.
\]
This  formula has the obvious semantics.

  Let $\cS=(g,F)$. We construct the sentence $e_\cS = e_1 \times e_2$ in $\MSOe(\Sigma,\B)$ where the intuition behind $e_1$ and  $e_2$ is that they simulate $g$ and $F$, respectively.
  Formally, we let
  \begin{align*}
    e_1 &= \bigtimes\nolimits_x  \Big(\bigplus_{\substack{k \in \maxrk(\Sigma),\\f \in \Fork(\Sigma)^{(k)}}} \big(\varphi_f(x) \rhd \langle g(f)\rangle\big)\Big)\\
    e_2 &= \bigplus\nolimits_x  \Big( \bigplus_{\sigma \in \Sigma} \big(\mathrm{root}(x) \wedge \llabel_\sigma(x) \rhd \langle F(\sigma)\rangle \big)\Big) 
  \end{align*}
  and, for each fork $(\sigma_1 \cdots \sigma_k,\sigma) \in \Fork(\Sigma)^{(k)}$, the formula $\varphi_{(\sigma_1 \cdots \sigma_k,\sigma)}(x)$ in $\MSO(\Sigma_{\{x\}})$ is defined by
  \begin{align*}
    \varphi_{(\sigma_1 \cdots \sigma_k,\sigma)}(x) &= \llabel_\sigma(x) \wedge \forall y. \bigwedge_{i \in [k]} \big(\edge_i(x,y) \rightarrow \llabel_{\sigma_i}(y)\big) \enspace.
  \end{align*}
  Clearly, the subformulas
  \[
    \bigplus_{\substack{k \in \maxrk(\Sigma),\\f \in \Fork(\Sigma)^{(k)}}} \big(\varphi_f(x) \rhd \langle g(f)\rangle\big) \ \  \text{ and } \ \
    \bigplus_{\sigma \in \Sigma} \big(\mathrm{root}(x) \wedge \llabel_\sigma(x) \rhd \langle F(\sigma)\rangle \big) 
  \]
of $e_1$ and $e_2$, respectively,  are recognizable step formulas. 
 
Now we consider the semantics of $e_\cS$.  The following is easy to see:
\begin{eqnarray}
  \begin{aligned}
      &\text{For every $\xi \in \T_{\Sigma}$ and $w \in \pos(\xi)$, we have that} \\ 
      &\text{ $(\xi,[x \mapsto w]) \models \varphi_{(\sigma_1 \cdots \sigma_k,\sigma)}(x)$ \ \  iff }\\
      &\text{\Big($\xi(w) = \sigma$ and $\xi(wi) = \sigma_i$ for each $i \in [\rk(\xi(w))]$\Big).}
      \end{aligned} \label{eq:varphi-fork}
    \end{eqnarray}
 Since for every $\xi \in \T_\Sigma$ and $w \in \pos(\xi)$, there exists exactly one fork $f$ which occurs in $\xi$ at $w$, we can also see the following easily (using \eqref{eq:varphi-fork}).
 \begin{eqnarray}
   \begin{aligned}
      &\text{For every $\xi \in \T_{\Sigma}$ and $w \in \pos(\xi)$, we have that}\\
      &\text{$\sem{\bigplus_{\substack{k \in \maxrk(\Sigma),\\f \in \Fork(\Sigma)^{(k)}}} \big(\varphi_f(x) \rhd \langle g(f)\rangle\big)}(\xi,[x \mapsto w]) = g\big((\xi(w1)\cdots \xi(w \ \rk(\xi(w))),\xi(w))\big)$.}
      \end{aligned}\label{eq:fork-weight}
      \end{eqnarray}

      Now let $\xi \in \T_\Sigma$. Then we can calculate as follows.
      \begingroup
      \allowdisplaybreaks
      \begin{align*}
        \sem{e_1}(\xi) &= \sem{ \bigtimes\nolimits_x  \Big(\bigplus_{\substack{k \in \maxrk(\Sigma),\\f \in \Fork(\Sigma)^{(k)}}} \big(\varphi_f(x) \rhd \langle g(f)\rangle\big)\Big)}(\xi)\\
                       &= \bigotimes\limits_{w \in \pos(\xi)}\sem{ \Big(\bigplus_{\substack{k \in \maxrk(\Sigma),\\f \in \Fork(\Sigma)^{(k)}}} \big(\varphi_f(x) \rhd \langle g(f)\rangle\big)\Big)}_{\{x\}}(\xi,[x \mapsto w])\\
                       &= \bigotimes\limits_{w \in \pos(\xi)} g\big((\xi(w1)\cdots \xi(w \ \rk(\xi(w))),\xi(w))\big)
        \tag{by \eqref{eq:fork-weight}}\\
        &= g(\xi) \enspace.
        \end{align*}
        \endgroup

        Moreover, we can calculate as follows.
            \begingroup
      \allowdisplaybreaks
      \begin{align*}
        \sem{e_2}(\xi) &=
  \sem{\bigplus\nolimits_x\Big( \bigplus_{\sigma \in \Sigma} \big(\mathrm{root}(x) \wedge \llabel_\sigma(x) \rhd \langle F(\sigma)\rangle \big)\Big) }(\xi)\\
                       &= \bigoplus\limits_{w \in \pos(\xi)}\sem{ \bigplus_{\sigma \in \Sigma} \big(\mathrm{root}(x) \wedge \llabel_\sigma(x) \rhd \langle F(\sigma)\rangle \big)}_{\{x\}}(\xi,[x \mapsto w])\\
                       &= \sem{\langle F(\xi(\varepsilon))\rangle} \tag{because $(\xi,[x \mapsto w])\models (\mathrm{root}(x) \wedge \llabel_\sigma(x))$ iff $w=\varepsilon$ and $\sigma = \xi(w)$}\\
        &= F(\xi(\varepsilon))\enspace.
        \end{align*}
        \endgroup
        Hence $\sem{e}(\xi) = \sem{e_1 \times e_2}(\xi) = \sem{e_1}(\xi) \otimes \sem{e_2}(\xi) = g(\xi) \otimes F(\xi(\varepsilon)) = \sem{\cS}(\xi)$.
        \hfill $\Box$
  \end{example}

\subsection{The carefully extended weighted $\MSO$-logic $\MSOce(\Sigma,\B)$}
\label{ssec:carefully-extended-weighted-MSO}

Similar to the situation of the weighted $\MSO$ formulas defined in \cite{drovog06} (also cf. \cite[Ex.~3.4]{drogas05}), there exists an extended MSO formula $e \in \MSOe(\Sigma,\Nat)$ such that $\sem{e}$ is not recognizable.
In fact, we can give an atomic formula such that its weighted first-order universal quantification is not recognizable. The same holds for the weighted second-order universal quantification.

  \begin{example}\rm \label{ex:fo-universal-quant-too strong} Let $\Sigma = \{\gamma^{(1)}, \alpha^{(0)}\}$. We consider the semiring  $\Nat$ of natural numbers and the formula $e \in \MSOe(\Sigma,\Nat)$, where
  \[
e = \bigtimes\nolimits_x \rmH(\kappa) 
    \]
    and $\kappa = (\kappa_k \mid k \in \mathbb{N})$ with $\kappa_k: \Sigma_{\{x\}}^{(k)} \to \mathbb{N}$ and
    $\kappa_k((\sigma,\cU)) = 2$ for each $(\sigma,\cU) \in \Sigma_{\{x\}}^{(k)}$.

    Then, for each $n \in \mathbb{N}$, we have (using $\prod$ as notation for the generalization of $\cdot$ to a finite number of arguments)
    \begingroup
    \allowdisplaybreaks
    \begin{align*}
      \sem{\bigtimes\nolimits_x \rmH(\kappa)}(\gamma^n(\alpha))
      &= \prod_{w \in \pos(\gamma^n(\alpha))} \sem{\rmH(\kappa)}_{\{x\}}(\gamma^n(\alpha)[x \mapsto w])\\
      &= \prod_{w \in \pos(\gamma^n(\alpha))} \h_\kappa(\gamma^n(\alpha)[x \mapsto w])
        = \prod_{w \in \pos(\gamma^n(\alpha))} 2^{n+1}
      = 2^{(n+1)^2} \enspace.     
      \end{align*}
\endgroup

    In Example \ref{ex:representable-not-recognizable} we have shown that there does not exist a $(\Sigma,\Nat)$-wta $\cA$ such that $\sem{\cA} = \sem{e}$.
 
  Also weighted second-order universal quantification grows too fast for being recognizable; it grows even faster than weighted first-order universal quantification. To see this, we consider the formula $f \in \MSOe(\Sigma,\Nat)$, where
  \[
f = \bigtimes\nolimits_X \rmH(\kappa) 
    \]
    and $\kappa = (\kappa_k \mid k \in \mathbb{N})$ with $\kappa_k: \Sigma_{\{X\}}^{(k)} \to \mathbb{N}$ and
    $\kappa_k((\sigma,\cU)) = 2$ for each $(\sigma,\cU) \in \Sigma_{\{X\}}^{(k)}$.

    Then, for each $n \in \mathbb{N}$, we have
    \begingroup
    \allowdisplaybreaks
    \begin{align*}
      \sem{\bigtimes\nolimits_X \rmH(\kappa)}(\gamma^n(\alpha))
      = \prod_{W \subseteq \pos(\gamma^n(\alpha))} \sem{\rmH(\kappa)}_{\{X\}}(\gamma^n(\alpha)[X \mapsto W])
        = \prod_{W \subseteq  \pos(\gamma^n(\alpha))} 2^{n+1}
      = 2^{(n+1) \cdot 2^{n+1} } \enspace.     
      \end{align*}
      \endgroup  
      As in Example \ref{ex:representable-not-recognizable}, we can prove that there does not exist a $(\Sigma,\Nat)$-wta $\cA$ such that $\sem{\cA} = \sem{f}$.
      \hfill $\Box$
  \end{example}

    \index{$\MSOce(\Sigma,\B)$} 
    \index{carefully extended MSO formulas}
    \label{page:carefully-extended-MSO-formulas}
In order to decrease the computational power, we define the above mentioned fragment of $\MSOe(\Sigma,\B)$. Formally, we define the set of \emph{carefully extended MSO formulas over $\Sigma$ and $\B$}, denoted by $\MSOce(\Sigma,\B)$, to be the set of all formulas $e\in \MSOe(\Sigma,\B)$ such that the following two conditions hold:
\begin{compactitem}
\item if $e$ contains a subformula of the form $\bigtimes_x e'$, then $e'$ is a recognizable step formula, i.e., a formula in $\MSO(\Sigma,\B)$ of the form
  \(
    (\varphi_1 \rhd \langle b_1\rangle) + \ldots + (\varphi_n \rhd \langle b_n\rangle)
  \)
  as specified in \eqref{eq:rec-step-formula}   and

  \item $e$  does not have a subformula of the form $\bigtimes_X e'$.
\end{compactitem}
Thus, in particular, weighted first-order universal quantification cannot be nested. Moreover, obviously, $\MSO(\Sigma,\B) \subset \MSOce(\Sigma,\B) \subset \MSOe(\Sigma,\B)$.


In the extension $\MSOce(\Sigma,\B)$ of $\MSO(\Sigma,\B)$, formulas of the form $\rmH(\kappa)$ are obsolete; more precisely, they can be expressed by weighted first-order universal quantification over recognizable step formulas; we recall that recognizable step formulas are elements of $\MSO(\Sigma,\B)$.

  \begin{lemma}\rm \label{lm:forall-rec-step-formula=H(kappa)} (cf. \cite[Lm.~5.12]{fulstuvog12}) Let $\cU$ be a finite set of variables and $\kappa = (\kappa_k \mid k \in \mathbb{N})$ be an $\mathbb{N}$-indexed family  with $\kappa_k: \Sigma_\cU^{(k)} \to B$. Moreover, let $x \not\in \cU$. We can construct a recognizable step formula $e$ such that $\Free(e)=\cU\cup \{x\}$ and $\sem{\rmH(\kappa)} = \sem{\bigtimes_x e}$.
  \end{lemma}

          \begin{proof} Let $\Sigma_\cU = \{(\sigma_1,U_1), \ldots, (\sigma_n,U_n)\}$.  We define the recognizable step formula
            \[
              e = \Big(\varphi_{(\sigma_1,U_1)}(\{x\} \cup U_1) \rhd \langle b_{(\sigma_1,U_1)}\rangle\Big) + \ldots + \Big(\varphi_{(\sigma_n,U_n)}(\{x\} \cup U_n) \rhd \langle b_{(\sigma_n,U_n)}\rangle\Big)
            \]
            where, for each $i \in [n]$, we let
            \begin{compactitem}
            \item $b_{(\sigma_i,U_i)} = \kappa_{\rk((\sigma_i,U_i))}((\sigma_i,U_i))$ and
            \item $\varphi_{(\sigma_i,U_i)}(\{x\} \cup U_i)$ is the $\MSO(\Sigma)$-formula
\begin{align*}
  \llabel_{\sigma_i}(x)\wedge \Big(\bigwedge_{y\in U_i}(x=y)\Big)\wedge \Big(\bigwedge_{y\in \cU\setminus U_i}\neg (x=y)\Big)\wedge \Big(\bigwedge_{X\in U_i}(x\in X)\Big) \wedge \Big(\bigwedge_{X\in \cU\setminus U_i}\neg(x\in X)\Big) \enspace.
\end{align*}
              \end{compactitem}

              Obviously, for every $\zeta \in \T_{\Sigma_\cU}^{\mathrm{v}}$, $w \in \pos(\zeta)$, and $(\sigma_i,U_i) \in \cU$, we have
              \begin{equation}\label{eq:step-formula-semantics}
                \zeta[x \mapsto w] \in \LL_{\cU\cup\{x\}}(\varphi_{(\sigma_i,U_i)}) \ \text{ iff } \ \zeta(w) = (\sigma_i,U_i) \enspace,
              \end{equation}
              and hence $( \LL_{\cU\cup\{x\}}(\varphi_{(\sigma_i,U_i)}) \mid i \in [n])$ is a partitioning of $\T_{\Sigma_{\cU \cup \{x\}}}^{\mathrm{v}}$.
Thus we have  
\begin{equation}\label{eq:one-summand-semantics}
  \sem{\varphi_{(\sigma_i,U_i)}(\{x\} \cup U_i)  \rhd \langle b_{(\sigma_i,U_i)}\rangle}(\zeta[x \mapsto w]) =
  \begin{cases}
    b_{(\sigma_i,U_i)} & \text{ if $\zeta(w) = (\sigma_i,U_i)$}\\
    \0 & \text{ otherwise} \enspace.
    \end{cases}
              \end{equation}

              Let $\zeta \in  \T_{\Sigma_\cU}$. If $\zeta$ is not valid, then $\sem{\rmH(\kappa)}(\zeta) = \0 = \sem{\bigtimes_x e}(\zeta)$.
               Now let $\zeta$ be valid. Then  we have
                \begingroup
                \allowdisplaybreaks
                \begin{align*}
                  \sem{\bigtimes\nolimits_x e}(\zeta) 
                  =& \bigotimes_{w \in \pos(\zeta)} \sem{e}_{\cU \cup \{x\}}(\zeta[x \mapsto w])\\
                  =&  \bigotimes_{w \in \pos(\zeta)} b_{\zeta(w)} \tag{by \eqref{eq:one-summand-semantics}}\\
                  =&  \bigotimes_{w \in \pos(\zeta)} \kappa_{\rk(\zeta(w))}(\zeta(w)) \tag{by definition of $b_{\zeta(w)}$}\\
                  =& \ \sem{\rmH(\kappa)}(\zeta) \tag{by \eqref{eq:hom-kappa}} 
                  \end{align*}
                \endgroup
         \end{proof}

          In  Lemma \ref{lm:getting-rid-of-universal-quant} we will show how formulas of the form $\bigtimes_x e$, where $e$ is a recognizable step formula, can be simulated by a formula in $\MSO(\Sigma,\B)$ (cf. $t$ in \eqref{eq:simulation-forall}). Here we prove that atomic formulas alone are not powerful enough to simulate recognizable step formulas.

          \begin{lemma}\rm \label{lm:exists-rec-step-formula-not=H(kappa)} For the ranked alphabet $\Sigma=\{\alpha^{(2)},\gamma^{(1)},\alpha^{(0)}\}$ we can construct a recognizable step sentence $e$ over $\Sigma$ and $\Nat$ such that $\sem{e}\not=\sem{\rmH(\kappa)}$ for any family $\kappa=(\kappa_k\mid k\in \mathbb{N}$) of mappings $\kappa_k:\Sigma^{(k)}\to \mathbb{N}$.
\end{lemma}

\begin{proof} By Theorem  \ref{thm:crisp-det-algebra}(A)$\Rightarrow$(B) and Lemma  \ref{lm:rec-step-formula-lemma}(A)$\Rightarrow$(B),  it is sufficient to construct a crisp-deterministic $(\Sigma,\Nat)$-wta $\cA$ such that $\sem{\cA}\not=\sem{\rmH(\kappa)}$ for any family $\kappa=(\kappa_k\mid k\in \mathbb{N}$) of mappings $\kappa_k:\Sigma^{(k)}\to \mathbb{N}$.

We consider the crisp-deterministic $(\Sigma,\Nat)$-wta $\cA$ constructed in Example \ref{ex:recog-step-mapping}. There it was shown that
$\sem{\cA}=\mathrm{twothree}$, where the weighted tree language $\mathrm{twothree}: \T_\Sigma \rightarrow \mathbb{N}$ is defined by
\begin{align*}
  \mathrm{twothree}(\xi) = 
        \begin{cases}
          2 & \text{ if } |\pos(\xi)| \text{ is even }\\
          3 & \text{ otherwise } 
          \end{cases} 
\end{align*}
for each $\xi \in \T_\Sigma$.

We show by contradiction that $\cA$ has the desired property. For this, we assume that there is a  family $\kappa=(\kappa_k\mid k\in \mathbb{N}$) of mappings $\kappa_k:\Sigma^{(k)}\to \mathbb{N}$ with $\sem{\rmH(\kappa)}=\sem{\cA}$. By definition,
\[\sem{\rmH(\kappa)}(\gamma^n(\alpha))=\kappa_0(\alpha)\cdot (\kappa_1(\gamma))^n\]
for each $n\in \mathbb{N}$. Since $\im(\sem{\cA})=\{2,3\}$, i.e., a finite set, it follows that $\kappa_1(\gamma)=0$ or $\kappa_1(\gamma)=1$. In the first case $\sem{\rmH(\kappa)}(\gamma^n(\alpha))=0$ for each $n \in \mathbb{N}_+$. In the second case $\sem{\rmH(\kappa)}(\gamma^n(\alpha))=\kappa_0(\alpha)$ for each $n\in \mathbb{N}$. Thus, in both cases, we have $\sem{\rmH(\kappa)}(\gamma^2(\alpha))=\sem{\rmH(\kappa)}(\gamma(\alpha))$. This contradicts  $\sem{\rmH(\kappa)}=\sem{\cA}$, because $\sem{\cA}(\gamma^2(\alpha)) = 2 \not= 3 = \sem{\cA}(\gamma(\alpha))$.
\end{proof}


\subsection[The main result for $\MSOce(\Sigma,\B)$]{The main result for $\MSOce(\Sigma,\B)$ over commutative semirings}
\label{ssec:main-result-MSOcext}

In this subsection we prove that the two weighted logics $\MSO(\Sigma,\B)$ and $\MSOce(\Sigma,\B)$ are equivalent (cf. Theorem \ref{thm:Buechi-extended}), if $\B$ is a commutative semiring. This theorem follows directly from Lemmas \ref{lm:getting-rid-of-universal-quant} and Lemma \ref{lm:getting-rid-of-conj}.
For each bounded lattice $\B$, we will prove that even the two weighted logics $\MSO(\Sigma,\B)$ and $\MSOe(\Sigma,\B)$ are equivalent (cf. Corollary \ref{thm:BET-bounded-lattices}).

\begin{theorem-rect} \label{thm:Buechi-extended} Let $\Sigma$ be a ranked alphabet. Moreover, let $\B$ be a commutative semiring, let $e \in \MSOce(\Sigma,\B)$, and let $\cV$ be a finite set of variables such that $\cV \supseteq \Free(e)$. Then we can construct an $\MSO(\Sigma,\B)$-formula $f$ such that $\Free(e) = \Free(f)$ and  $\sem{e}_\cV=\sem{f}_\cV$.
\end{theorem-rect}

In the next lemma we get rid of weighted first-order universal quantification. We follow the proof of \cite[Lm.~5.10]{fulstuvog12}; in its turn, the construction in the proof of the latter lemma was inspired by the construction in the proof of \cite[Lm.~5.4]{drogas09}. In the proof we will use that $\B$ is commutative.

\begin{lemma}\rm (cf. \cite[Lm.~5.10]{fulstuvog12})\label{lm:getting-rid-of-universal-quant} Let $e \in \MSOce(\Sigma,\B)$ and let $\cV$  be a finite set of variables such that  $\cV \supseteq \Free(e)$. Then we can construct an $\MSOce(\Sigma,\B)$-formula $f$ such that (a) $\Free(e) = \Free(f)$, (b)~$\sem{e}_\cV=\sem{f}_\cV$, and (c) $f$ does not contain a subformula of the form~$\bigtimes_x e'$.
\end{lemma}
\begin{proof} Intuitively, we define a reduction system on $\MSOce(\Sigma,\B)$ such that, in each reduction step, the semantics of the formula is preserved and  the number of occurrences of subformulas of the form $\bigtimes_x e'$ decreases by one.

Formally, we define the binary relation $\succ \subseteq \MSOce(\Sigma,\B) \times \MSOce(\Sigma,\B)$ such that, for every $e,f \in \MSOce(\Sigma,\B)$, we let $e \succ f$ if
 \begin{compactitem}
\item  $e$ has a subformula of the form $\bigtimes_x e'$, where $e'$ is a recognizable step formula, i.e., there exist $n \in \mathbb{N}_+$, $b_1,\ldots,b_n \in B$, and $\varphi_1,\ldots,\varphi_n \in \MSO(\Sigma)$ such that
  \(e' = (\varphi_1 \rhd \langle b_1\rangle) + \ldots + (\varphi_n \rhd \langle b_n\rangle)\), we abbreviate by $\cV'$ the set $\Free(\bigtimes_x e')$, and
\item $f$ is obtained from $e$ by replacing an occurrence of $\bigtimes_x e'$ by the $\MSO(\Sigma,\B)$-formula
   \begin{equation}
   t =  \bigplus\nolimits_{X_{1}}\ldots\bigplus\nolimits_{X_{n}}\Bigl(\bigl(\bigwedge\nolimits_{i\in[n]}\forall x. \big((x\in X_{i}) \leftrightarrow  \varphi_i\big)\bigr)\rhd \mathrm H(\kappa)\Bigr) 
    \label{eq:simulation-forall} 
   \end{equation}
  where
  \begin{compactitem}
    \item $X_{1},\ldots,X_{n}$ are second-order variables which do not occur in $\Free(e) \cup \Bound(e)$ and 

      \item $\kappa$ is the $\mathbb{N}$-indexed family $(\kappa_k\mid k \in \mathbb{N})$ of mappings $\kappa_k:\Sigma_\cU^{(k)}\to B$, where $\cU=\cV'\cup\{X_{1},\ldots,X_{n}\}$, such that, for each $(\sigma,\cW) \in \Sigma_\cU^{(k)}$, we define
\[
  \kappa_k((\sigma,\cW)) = \bigoplus_{i\in[n]:X_{i}\in \cW} b_i \enspace.
\]
\end{compactitem}
\end{compactitem}
We make the following observations:
   
  \begin{compactenum}
   
   \item Intuitively, the $\MSO(\Sigma)$-formula $\bigl(\bigwedge\nolimits_{i\in[n]}\forall x.\big((x\in X_i) \leftrightarrow  \varphi_i\big)\bigr)$  (which is a part of $t$) guarantees that, for each $i \in [n]$, the set of positions assigned to $X_i$ is exactly the set of positions for which $\varphi_i$ is true (cf.~\eqref{eq:connection-X-x}).
  
\item On purpose, we use the variable $x$, which might occur in $\varphi_i$, also as variable for the universal quantification; thereby we ``synchronize'' the sequence $X_{1},\ldots,X_{n}$ with the formulas $\varphi_1,\ldots,\varphi_n$.
Obviously, $\Free(t) = \cV'$ and hence  $\Free(\bigtimes_x e') = \Free(t)$ and thus $\Free(e) = \Free(f)$.
\end{compactenum}

Next we prove that $\succ$ is terminating. For each $e \in \MSOce(\Sigma,\B)$, we let $\#_{\bigtimes}(e)$ denote the number of occurrences of a subformula of the form $\bigtimes_x e'$.
Obviously, for each $e\in \MSOce(\Sigma,\B)$ with $\#_{\bigtimes}(e)>0$, we can construct an  $f\in \MSOce(\Sigma,\B)$ such that $e\succ f$.
    Moreover, the mapping $\#_{\bigtimes}$ is a monotone embedding of  $(\MSOce(\Sigma,\B),\succ)$ into $(\mathbb{N},\succ_\mathbb{N})$  because $e\succ f$  implies that  $\#_{\bigtimes}(e) = \#_{\bigtimes}(f)+1$. Hence, by Lemma \ref{lm:fin-branching-embedding-termination}, the reduction system $(\MSOce(\Sigma,\B),\succ)$ is terminating and, obviously,
 \begin{equation}\label{eq:set-of-normal forms}
\nf_\succ(\MSOce(\Sigma,\B)) = \{e \in \MSOce(\Sigma,\B) \mid e \text{ does not have a subformula of the form $\bigtimes\nolimits_x e'$}\} \enspace.
\end{equation}   

Next we prove that, for every  $e,f\in \MSOce(\Sigma,\B)$, if  $e\succ f$, then  $\sem{e}_\cV = \sem{f}_\cV$. 
For this purpose, we will prove that 
\begin{equation} \label{equ:formula-which-simulates-rec-step-map}
   \sem{t} = \sem{\bigtimes\nolimits_{x} e'}\enspace,
 \end{equation}
where $t$ and $e'$ are specified as in the definition of $\succ$, and then we will use Observation \ref{obs:replacing-equiv-subformula-3}. We start with the analysis of the semantics of the $\MSO(\Sigma)$-subformula $\bigl(\bigwedge\nolimits_{i\in[n]}\forall x.((x\in X_i) \leftrightarrow  \varphi_i)\bigr)$. Let $(\xi,\mu) \in \T_{\Sigma_{\mathcal V'}}^\mathrm{v}$ and let $V_1,\ldots,V_n \subseteq \pos(\xi)$. Obviously, the following equivalence holds:
\begin{eqnarray}
  \begin{aligned}
  &(\xi,\mu[X_{1} \mapsto V_1,\ldots,X_{n} \mapsto V_n]) \models \bigl(\bigwedge\nolimits_{i\in[n]}\forall x.((x\in X_i) \leftrightarrow  \varphi_i)\bigr) \ \text{ iff }\\
  &(\forall i \in [n]): \ V_i = \{w\in \pos(\xi)\mid (\xi,\mu[x\mapsto w]) \in \LL_{\cV'\cup \{x\}}(\varphi_i)\}\enspace.
\end{aligned} \label{eq:connection-X-x}
  \end{eqnarray}
 For every $i\in[n]$, we abbreviate the set $\{w\in \pos(\xi)\mid (\xi,\mu[x\mapsto w]) \in \LL_{\cV'\cup \{x\}}(\varphi_i)\}$ by $W(\xi,\mu,i)$.

  Now we prove \eqref{equ:formula-which-simulates-rec-step-map}.
For this, let $(\xi,\mu) \in \T_{\Sigma_{\cV'}}$. If $(\xi,\mu)$ is not valid, then both sides of \eqref{equ:formula-which-simulates-rec-step-map} evaluate to $\0$. Now let $(\xi,\mu) \in \T_{\Sigma_{\cV'}}^{\mathrm{v}}$. Then we obtain
  \begingroup
  \allowdisplaybreaks
\begin{align*}
  \sem{t}(\xi,\mu) &= \bigoplus_{V_1,\ldots,V_n \subseteq \pos(\xi)} \Bigl(\bigl(\bigwedge\nolimits_{i\in[n]}\forall x.((x\in X_i) \leftrightarrow  \varphi_i)\bigr)\rhd \mathrm H(\kappa)\Bigr)(\xi,\mu[X_1 \mapsto V_1,\ldots,X_n \mapsto V_n])\\
                   &= \sem{\mathrm H(\kappa)}_\cU(\xi,\mu[X_1\mapsto W(\xi,\mu,1),\ldots, X_n\mapsto W(\xi,\mu,n)])
   \tag{by \eqref{eq:connection-X-x}}\\
                      &=  \mathrm h_{\kappa[\cU \leadsto \cU]}((\xi,\mu[X_1\mapsto W(\xi,\mu,1),\ldots, X_n\mapsto W(\xi,\mu,n)]))\\
                      &=  \mathrm h_\kappa((\xi,\mu[X_1\mapsto W(\xi,\mu,1),\ldots, X_n\mapsto W(\xi,\mu,n)]))\\
	&=\bigotimes_{w\in \pos(\xi)}\kappa_{k}\Big((\xi(w),\cW\cup \{X_j \mid w \in W(\xi,\mu,j)\})\Big) \tag{by \eqref{eq:hom-kappa} and the fact that $\B$ is commutative}
\end{align*}
\endgroup
where $k=\rk(\zeta(w))$ and $\cW=\{y\in (\cV')^{(1)}\mid w=\mu(y)\}\cup \{Y\in (\cV')^{(2)}\mid w\in \mu(Y)\}$ (note that $\cW\cap \{X_{1},\ldots,X_{n}\}=\emptyset$).

Let $w\in \pos(\xi)$. Clearly,
\begingroup
\allowdisplaybreaks
\begin{align*}
  \kappa_{k}((\xi(w),\cW\cup \{X_j \mid w \in W(\xi,\mu,j)\})) &= \bigoplus\limits_{\substack{i\in[n]:\\ X_i\in \cW\cup \{X_j \mid w \in W(\xi,\mu,j)\}}} b_i \\
  &= \ \bigoplus\limits_{\substack{i\in[n]:\\ w \in W(\xi,\mu,i)}} b_i \tag{because $X_i\not\in \cW$}\\
 &= \bigoplus\limits_{\substack{i\in[n]:\\ (\xi,\mu[x\mapsto w])\in \LL_{\mathcal V' \cup \{x\}}(\varphi_i)}}  \hspace{-4mm} b_i  \\ 
  &= \bigoplus_{i \in [n]} b_i \otimes \chi(\LL_{\mathcal V' \cup \{x\}}(\varphi_i))((\xi,\mu[x\mapsto w]))\\
  &=\sem{e'}_{\cV'\cup\{x\}}(\xi,\mu[x\mapsto w])\;.
\end{align*}
\endgroup

Hence,
\begingroup
\allowdisplaybreaks
\begin{align*}
  \sem{t}(\xi,\mu) =\bigotimes_{w\in \pos(\xi)}\sem{e'}_{\cV' \cup\{x\}}(\xi,\mu[x\mapsto w])
                   =\sem{\bigtimes\nolimits_x e'}_{\cV'}(\xi,\mu) 
   =\sem{\bigtimes\nolimits_x e'}(\xi,\mu) \enspace.
 \end{align*}
\endgroup 
This proves \eqref{equ:formula-which-simulates-rec-step-map}.

Now let $e,f \in \MSOce(\Sigma,\B)$ such that $e \succ f$ and $e'$ and $t$ are specified as in the definition of $\succ$. Then, by \eqref{equ:formula-which-simulates-rec-step-map} and Observation \ref{obs:replacing-equiv-subformula-3},  we obtain $\sem{e}_\cV = \sem{f}_\cV$.

Now let $e \in \MSOce(\Sigma,\B)$ be an arbitrary formula. Let us construct an $f\in \nf_\succ(e)$.  Then $\sem{e}_\cV = \sem{f}_\cV$ and, by \eqref{eq:set-of-normal forms},  $f$ does not contain a subformula of the form $\bigtimes\nolimits_x e'$.
\end{proof}

\begin{example}\rm In this example we want to illustrate the construction defined in the proof of Lemma~\ref{lm:getting-rid-of-universal-quant}. We consider the string ranked alphabet $\Sigma = \{\gamma^{(1)}, \alpha^{(0)}\}$ and the semiring $\Nat=(\mathbb{N},+,\cdot,0,1)$.
  Moreover, we let $e = (\bigtimes\nolimits_x \llabel_\alpha(x) \rhd \langle 2\rangle)$. Obviously, $\Free(e) = \emptyset$; we let $\cV = \emptyset$.

 Then the formula $t$ in $\MSO(\Sigma,\Nat)$ (cf. \eqref{eq:simulation-forall})  has the following form (with $n=1$ and $\varphi_1 = \llabel_\alpha(x)$):
  \begin{equation*}
   t =  \bigplus\nolimits_{X_1}\Bigl(\bigl(\forall x. \big((x\in X_1) \leftrightarrow  \llabel_\alpha(x)\big)\bigr)\rhd \mathrm H(\kappa)\Bigr) \enspace,
  \end{equation*}
  where $\kappa$ is the $\mathbb{N}$-indexed family of mappings $\kappa_k :\Sigma^{(k)}_{\{X_1\}} \to \mathbb{N}$ such that
  \begin{compactitem}
    \item $\kappa_0((\alpha,\emptyset))=\kappa_1((\gamma,\emptyset)) = \bigplus_{X_1\in \emptyset} 2 = 0$ and
 \item $\kappa_0((\alpha,\{X_1\}))=\kappa_1((\gamma,\{X_1\}) = \bigplus_{X_1\in \{X_1\}} 2 = 2$ .
\end{compactitem}
  For the tree $\gamma(\alpha) \in \T_{\Sigma_\emptyset}^{\mathrm{v}}$, we can calculate as follows:
  \begingroup
  \allowdisplaybreaks
  \begin{align*}
    &\sem{(\bigtimes\nolimits_x \llabel_\alpha(x) \rhd \langle 2\rangle)}_\emptyset (\gamma(\alpha)) \\
    =\ & \sem{\llabel_\alpha(x) \rhd \langle 2\rangle}_{\{x\}} (\gamma(\alpha),[x \mapsto 1]) \cdot \sem{\llabel_\alpha(x) \rhd \langle 2\rangle}_{\{x\}} (\gamma(\alpha),[x \mapsto \varepsilon])\\
    =\ & \sem{\langle 2\rangle}_{\{x\}} (\gamma(\alpha),[x \mapsto 1]) \cdot 0 
    =  2 \cdot 0 = 0 \enspace.
    \end{align*}
  \endgroup
    Using the above mentioned instance of $t$, we can calculate as follows:
    \begingroup
  \allowdisplaybreaks
  \begin{align*}
    &\sem{\bigplus\nolimits_{X_1}\Bigl(\bigl(\forall x. \big((x\in X_1) \leftrightarrow  \llabel_\alpha(x)\big)\bigr)\rhd \mathrm H(\kappa)\Bigr)}_\emptyset (\gamma(\alpha)) \\
    = \ & \bigplus_{W \subseteq \{\varepsilon,1\}} \sem{\underbrace{\bigl(\forall x. \big((x\in X_1) \leftrightarrow  \llabel_\alpha(x)\big)\bigr)\rhd \mathrm H(\kappa)}_{g}}_{\{X_1\}} (\gamma(\alpha),[X_1 \mapsto W])\\
    = \ & \sem{g}_{\{X_1\}} (\gamma(\alpha),[X_1 \mapsto \{\varepsilon, 1\}])
    + \sem{g}_{\{X_1\}} (\gamma(\alpha),[X_1 \mapsto \{1\}]) \\
    & + \sem{g}_{\{X_1\}} (\gamma(\alpha),[X_1 \mapsto \{\varepsilon\}])
    +  \sem{g}_{\{X_1\}} (\gamma(\alpha),[X_1 \mapsto \emptyset]) \enspace.
    \end{align*}
  \endgroup
  Due to \eqref{eq:connection-X-x}, we have
  \begin{eqnarray}
    \begin{aligned}
  &(\gamma(\alpha),[X_1 \mapsto V_1']) \models \bigl(\forall x.((x\in X_1) \leftrightarrow  \llabel_\alpha(x)\bigr) \ \text{ iff }\\
   &V_1' = \{w\in \pos(\gamma(\alpha))\mid (\gamma(\alpha),[x\mapsto w]) \in \LL_{\{x\}}(\llabel_\alpha(x))\}= \{1\}\enspace.
\end{aligned}
  \end{eqnarray}
Since $V_1' = \{1\}$, we can continue as follows: 
   \begingroup
  \allowdisplaybreaks
  \begin{align*}
    & \sem{g}_{\{X_1\}} (\gamma(\alpha),[X_1 \mapsto \{\varepsilon, 1\}])
    + \sem{g}_{\{X_1\}} (\gamma(\alpha),[X_1 \mapsto \{1\}]) \\
    & \ \ \ \  + \sem{g}_{\{X_1\}} (\gamma(\alpha),[X_1 \mapsto \{\varepsilon\}])
    +  \sem{g}_{\{X_1\}} (\gamma(\alpha),[X_1 \mapsto \emptyset]) \\
    = \ &  \sem{\rmH(\kappa)}_{\{X_1\}} (\gamma(\alpha),[X_1 \mapsto \{1\}])\\
    = \ &   \kappa_0((\alpha,\{X_1\}) \cdot \kappa_1((\gamma,\emptyset))\\
     = \ & 2 \cdot 0 = 0 \enspace.     \hspace{80mm} \Box
    \end{align*}
  \endgroup
  \end{example}


Next we eliminate weighted conjunction in formulas which do not contain weighted first-order universal quantification. Informally, we define a mapping $\hp$ which, on a pair $(e_1,e_2)$ of $\MSO(\Sigma,\B)$-formulas as argument, shows an $\MSO(\Sigma,\B)$-formula as result such that $e_1 \times e_2$ and $\hp(e_1,e_2)$ are equivalent (cf. proof of \cite[Lm.~5.7]{fulstuvog12}). For technical purposes, we have to assume that $e_1 \times e_2$ is variable separated. Moreover, we will assure that $\Free(\hp(e_1,e_2)) = \Free(e_1) \cup \Free(e_2)$ and  $\Bound(\hp(e_1,e_2)) = \Bound(e_1) \cup \Bound(e_2)$.

Since we will define $\hp$ by induction, we first have to introduce a terminating reduction system. Formally, we let $\mathrm{vs}\MSO^2(\Sigma,\B)$ denote the set
\[
\{(e_1,e_2) \in \MSO(\Sigma,\B) \times \MSO(\Sigma,\B) \mid e_1 \times e_2 \text{ is variable separated}\} 
  \]
and we define the reduction system
  \[
(\mathrm{vs}\MSO^2(\Sigma,\B), \succ)
\]
such that, for every $(e_1,e_2),(e_1',e_2') \in \mathrm{vs}\MSO^2(\Sigma,\B)$, we let $(e_1,e_2) \succ (e_1',e_2')$ if
\begin{compactitem}
\item $e_1 \succ_{\MSO(\Sigma,\B)} e_1'$ (i.e., $e_1'$ is a direct subformula of $e_1$, cf. Section \ref{sec:adding-weights})) and $e_2=e_2'$ or
  \item $e_1 = e_1'= \mathrm H(\kappa)$ for some $\mathbb{N}$-indexed family  $\kappa=(\kappa_k \mid k \in \mathbb{N})$ and $e_2 \succ_{\MSO(\Sigma,\B)} e_2'$.
\end{compactitem}
By Corollary \ref{cor:termination-propagates-to-cartesian-products2} and Lemma \ref{lm:termination-is-subset-closed}, the relation $\succ$ is terminating and $\nf_{\succ}(\mathrm{vs}\MSO^2(\Sigma,\B))$ is the set of all pairs of the form $(\mathrm H(\kappa_1),\mathrm H(\kappa_2))$ where for each $i \in [2]$, $\kappa_i = ((\kappa_i)_k \mid k \in \mathbb{N})$ is an  $\mathbb{N}$-indexed family  of mappings $(\kappa_i)_k:~\Sigma_{\cU_i}^{(k)}~\to~B$ for some finite sets $\cU_i$. In Figure \ref{fig:well-founded-order-on-pairs-of-MSO} we illustrate the well-founded order $\succ$ by means of an example.

\begin{figure}
 \centering
\begin{tikzpicture}[scale=0.8, every node/.style={transform shape},
					level distance= 1.5cm,
					level 1/.style={sibling distance=60mm},
					level 2/.style={sibling distance=80mm},
                                        level 3/.style={sibling distance=40mm}]
										 
                                        \node (root) {$\Big(\ \varphi \rhd (\rmH(\kappa_1) + \rmH(\kappa_2)) \ , \ (\psi \rhd \rmH(\kappa_3)) + \bigplus_x \rmH(\kappa_4)\ \Big)$}
                                        child {node {$\Big(\ \rmH(\kappa_1) + \rmH(\kappa_2) \ , \ (\psi \rhd \rmH(\kappa_3)) + \bigplus_x \rmH(\kappa_4)\ \Big)$}
  child {node {$\Big(\ \rmH(\kappa_1) \ , \ (\psi \rhd \rmH(\kappa_3)) + \bigplus_x \rmH(\kappa_4)\ \Big)$}
    child {node {$\Big(\ \rmH(\kappa_1) \ , \ \psi \rhd \rmH(\kappa_3)\ \Big)$}
    child {node  {$\Big(\ \rmH(\kappa_1) \ , \ \rmH(\kappa_3)\ \Big)$}
      }}
    child {node {$\Big(\ \rmH(\kappa_1) \ , \ \bigplus_x \rmH(\kappa_4)\ \Big)$}
    child {node {$\Big(\ \rmH(\kappa_1) \ , \ \rmH(\kappa_4)\ \Big)$}}}}
  child {node {$\Big(\ \rmH(\kappa_2) \ , \ (\psi \rhd \rmH(\kappa_3)) + \bigplus_x \rmH(\kappa_4)\ \Big)$}
   child {node {$\Big(\ \rmH(\kappa_2) \ , \ \psi \rhd \rmH(\kappa_3)\ \Big)$}
     child {node {$\Big(\ \rmH(\kappa_2) \ , \ \rmH(\kappa_3)\ \Big)$}}}
   child {node {$\Big(\ \rmH(\kappa_2) \ , \ \bigplus_x \rmH(\kappa_4)\ \Big)$}
   child {node {$\Big(\ \rmH(\kappa_2) \ , \ \rmH(\kappa_4)\ \Big)$}}}}};

\end{tikzpicture}
\caption{\label{fig:well-founded-order-on-pairs-of-MSO} Illustration of $\succ$ where moving from a node labeled by $A$ to its direct successor labeled by $B$ means that $A \succ B$.}
\end{figure}

By induction on $(\mathrm{vs}\MSO^2(\Sigma,\B), \succ)$, we define the family $(\hp(e_1,e_2) \mid e_1,e_2 \in \mathrm{vs}\MSO^2(\Sigma,\B))$ as follows. Let $(e_1,e_2) \in \mathrm{vs}\MSO^2(\Sigma,\B)$.

I.B.:  Let $e_1= \mathrm H(\kappa_1)$ and $e_2= \mathrm H(\kappa_2)$ where for each $i \in [2]$, the members of the  $\mathbb{N}$-indexed family  $\kappa_i=((\kappa_i)_k \mid k \in \mathbb{N})$ have the type $(\kappa_i)_k:~\Sigma_{\cU_i}^{(k)}~\to~B$. Then we define
  $\hp(e_1,e_2)=\mathrm H(\kappa)$,
  where $\kappa = (\kappa_k \mid k \in \mathbb{N})$ and for every $k \in \mathbb{N}$ we define  the mapping $\kappa_k: \Sigma_{\cU_1 \cup \cU_2}^{(k)} \to B$ for each $(\sigma,V) \in \Sigma_{\cU_1 \cup \cU_2}^{(k)}$ by
  \(\kappa_k((\sigma,V)) = (\kappa_1)_k((\sigma,V \cap \cU_1)) \otimes (\kappa_2)_k((\sigma,V\cap \cU_2))\).

I.S.:  We distinguish eight cases.
  
(a) Let $e_1 = g_1 + g_2$. We define  $\hp((g_1+g_2), e_2)=\hp(g_1,e_2)+\hp(g_2,e_2)$.

(b) Let $e_1 = \bigplus_x g$.  We define  $\hp(\bigplus_x g,e_2)= \bigplus_x \hp(g,e_2)$.

(c) Let $e_1 = \bigplus_X g$.  We define  $\hp(\bigplus_X g,e_2)= \bigplus_X \hp(g,e_2)$.
  
(d) Let $e_1= (\varphi \rhd g)$. We  define $\hp(\varphi \rhd g, e_2)= (\varphi \rhd \hp(g,e_2))$.

In each of the following cases, let $e_1= \mathrm H(\kappa_1)$ with $\kappa_1=((\kappa_1)_k \mid k \in \mathbb{N})$ and $(\kappa_1)_k:~\Sigma_{\cU_1}^{(k)}~\to~B$.

  (e) Let $e_2 = g_1 + g_2$. We define  $\hp(\mathrm H(\kappa_1), g_1 + g_2))=\hp(\mathrm H(\kappa_1),g_1)+\hp(\mathrm H(\kappa_1),g_2)$.

(f)    Let  $e_2 = \bigplus_x g$. We define  $\hp( \mathrm H(\kappa_1),\bigplus_x g)= \bigplus_x \hp( \mathrm H(\kappa_1),g)$.

(g) Let $e_2 = \bigplus_X g$. We define  $\hp(\mathrm H(\kappa_1),\bigplus_X g)= \bigplus_X \hp(\mathrm H(\kappa_1),g)$.
  
(h) Let  $e_1= (\varphi \rhd g)$. We  define $\hp(\mathrm H(\kappa_1),\varphi \rhd g)= (\varphi \rhd \hp(\mathrm H(\kappa_1),g))$.

It is easy to see that in each case $\Free(\hp(e_1,e_2)) = \Free(e_1) \cup \Free(e_2)=\Free(e_1 \times e_2)$ and  $\Bound(\hp(e_1,e_2)) = \Bound(e_1) \cup \Free(e_2)=\Bound(e_1 \times e_2)$.

\begin{example}\rm Here we show how to compute $\hp(e_1,e_2)$ for the pair $(e_1,e_2)$ which is the label of the root of the tree in Figure\ref{fig:well-founded-order-on-pairs-of-MSO}. We do not specify the families $\kappa_1,\kappa_2,\kappa_3,\kappa_4$, and hence we do not compute $\hp(\rmH(\kappa_1),\rmH(\kappa_3))$, $\hp(\rmH(\kappa_1),\rmH(\kappa_4))$, $\hp(\rmH(\kappa_2),\rmH(\kappa_3))$, and $\hp(\rmH(\kappa_2),\rmH(\kappa_4))$.
    \begingroup
  \allowdisplaybreaks
  \begin{align*}
    &\hp\Big(\ \varphi \rhd (\rmH(\kappa_1) + \rmH(\kappa_2)) \ , \ (\psi \rhd \rmH(\kappa_3)) + \bigplus\nolimits_x \rmH(\kappa_4)\ \Big)\\
    =& \ \varphi \rhd \hp\Big(\ \rmH(\kappa_1) + \rmH(\kappa_2) \ , \ (\psi \rhd \rmH(\kappa_3)) + \bigplus\nolimits_x \rmH(\kappa_4)\ \Big) \tag{by (d)}\\
    =& \ \varphi \rhd \big(\ \hp\Big(\rmH(\kappa_1)\ , \ (\psi \rhd \rmH(\kappa_3)) + \bigplus\nolimits_x \rmH(\kappa_4)\Big) \ + \\
     & \hspace*{11mm} \hp\Big(\rmH(\kappa_2) \ , \ (\psi \rhd \rmH(\kappa_3)) + \bigplus\nolimits_x \rmH(\kappa_4)\Big)\ \big) \tag{by (a)}\\
= & \ \varphi \rhd \big(\  \hp\Big(\rmH(\kappa_1)\ , \ \psi \rhd \rmH(\kappa_3)\Big) + \hp\Big(\rmH(\kappa_1)\ , \ \bigplus\nolimits_x \rmH(\kappa_4)\Big) \ +\\
&  \hspace*{11mm} \hp\Big(\rmH(\kappa_2)\ , \ \psi \rhd \rmH(\kappa_3)\Big) + \hp\Big(\rmH(\kappa_2)\ , \ \bigplus\nolimits_x \rmH(\kappa_4)\Big)\big) \tag{by (e)}\\
= & \ \varphi \rhd \big(\ (\psi \rhd  \hp\Big(\rmH(\kappa_1)\ , \ \rmH(\kappa_3)\Big)) + (\bigplus\nolimits_x \hp\Big(\rmH(\kappa_1)\ , \ \rmH(\kappa_4)\Big)) \ +\\
    &  \hspace*{11mm} (\psi \rhd  \hp\Big(\rmH(\kappa_2)\ , \ \rmH(\kappa_3)\Big)) + (\bigplus\nolimits_x \hp\Big(\rmH(\kappa_2)\ , \ \rmH(\kappa_4)\Big))\big) \tag{by (f)} \enspace.
  \end{align*}
  \endgroup
  \hfill $\Box$
  \end{example}

Next we prove that, roughly speaking, $e_1\times e_2$ and $\hp(e_1,e_2)$ are equivalent, if $\B$ is distributive.

\begin{lemma}\label{lm:e1xe2-equivalent-hp(e1,e2)} \rm (cf. \cite[Lm.~5.7]{fulstuvog12}) Let $\B$ be distributive, let $e_1,e_2\in\MSO(\Sigma,\B)$ such that $e_1\times e_2$ is variable separated, and let $\cU_1$ and $\cU_2$ abbreviate $\Free(e_1)$ and $\Free(e_2)$, respectively.
  For each finite set $\cV$ of variables which contains $\cU_1 \cup \cU_2$ we have \(\sem{e_1 \times e_2}_\cV = \sem{\hp(e_1,e_2)}_\cV\).
  \end{lemma}
  \begin{proof} We prove the statement of the lemma by induction on $(\mathrm{vs}^2\MSO(\Sigma,\B), \succ)$ (see above).
    In the proof we refer to I.B. and Cases (a)--(h) of I.S. of the definition of $\hp(e_1,e_2)$. Let $\zeta \in \T_{\Sigma_\cV}^{\mathrm{v}}$.
    
    I.B.: Let $e_1$ and $e_2$ be as in I.B. of of the definition of $\hp(e_1,e_2)$. Then we can calculate as follows:
\begingroup
\allowdisplaybreaks
  \begin{align*}
        \sem{\mathrm H(\kappa_1) \times \mathrm H(\kappa_2)}_\cV(\zeta)
    &=\sem{\mathrm H(\kappa_1)}_\cV(\zeta)\otimes \sem{\mathrm H(\kappa_2)}_\cV(\zeta) 
    = \mathrm h_{\kappa_1[\cU_1 \leadsto\cV]}(\zeta) \otimes \mathrm h_{\kappa_2[\cU_2 \leadsto\cV]}(\zeta)\\
    &= \mathrm h_{\kappa[\cU_1 \cup \cU_2\leadsto\cV]}(\zeta)
      =\sem{\mathrm H(\kappa)}_\cV(\zeta)
      = \sem{\hp(\mathrm H(\kappa_1),\mathrm H(\kappa_2))}_\cV(\zeta)\enspace,
  \end{align*}
  \endgroup
  where the third equality holds because 
  \begin{align*}
    &\bigotimes_{\substack{w\in \pos(\zeta),\\ \zeta(w)=(\sigma,V)}} \kappa_1((\sigma,\cU_1\cap V) \otimes \bigotimes_{\substack{w\in \pos(\zeta),\\ \zeta(w)=(\sigma,V)}} \kappa_2((\sigma,\cU_2\cap V)) \\
    =&  \bigotimes_{\substack{w\in \pos(\zeta),\\ \zeta(w)=(\sigma,V)}} \kappa_1((\sigma,\cU_1\cap V)) \otimes \kappa_2((\sigma,\cU_2\cap V))
    \end{align*}
where the latter equality holds because $\B$ is commutative.

I.S.: We distinguish eight cases.

Let $e_1$ be as in Case (a). Then our statement holds due to right-distributivity of $\B$.

Let $e_1$ be as in Case (b). Then
\begingroup
\allowdisplaybreaks
\begin{align*}
  \sem{(\bigplus\nolimits_x g) \times e_2)}_\cV(\zeta)
    &=  \sem{\bigplus\nolimits_x g}_{\cV}(\zeta) \otimes  \sem{e_2}_{\cV}(\zeta)\\
  &=  \Big(\bigoplus_{w\in \pos(\zeta)}\sem{g}_{\cV\cup\{x\}}(\zeta[x\mapsto w])\Big) \otimes \sem{e_2}_{\cV}(\zeta)\\
  &=  \bigoplus_{w\in \pos(\zeta)}\Big(\sem{g}_{\cV\cup\{x\}}(\zeta[x\mapsto w]) \otimes \sem{e_2}_{\cV}(\zeta)\Big)
      \tag{by right-distributivity}\\
    &= \bigoplus_{w\in \pos(\zeta)} \Big(\sem{g}_{\cV\cup\{x\}}(\zeta[x\mapsto w]) \otimes  \sem{e_2}_{\cV\cup \{x\}}(\zeta[x\mapsto w])\Big)
  \tag{by Lemma \ref{obs:consistency}; we note that $e_1\times e_2$ is variable separated}\\
  &=  \bigoplus_{w\in \pos(\zeta)}\Big(\sem{g}_{\cV\cup\{x\}}\otimes \sem{e_2}_{\cV\cup\{x\}}\Big)(\zeta[x\mapsto w])\\
  &=  \bigoplus_{w\in \pos(\zeta)}\sem{\hp(g,e_2)}_{\cV\cup\{x\}}(\zeta[x\mapsto w])
   \tag{by I.H.}\\
    &= \sem{\bigplus\nolimits_x \hp(g,e_2)}_\cV(\zeta)
      \tag{by definition of semantics}\\
      &= \sem{\hp(\bigplus\nolimits_x g,e_2)}_\cV(\zeta)\enspace.
       \tag{by definition of $\hp$}
\end{align*}
\endgroup

Let $e_1$ be as in Case (c). Then the corresponding equation for $\hp(\bigplus_X g,e_2)$ can be verified in a similar way as for $\hp(\bigplus_x g,e_2)$.

Let $e_1$ be as in Case (d). Then
\begingroup
\allowdisplaybreaks
\begin{align*}
  \sem{(\varphi \rhd g) \times e_2}_\cV(\zeta)
  &=  \sem{\varphi \rhd g}_\cV(\zeta) \otimes \sem{e_2}_\cV(\zeta) 
  =  \begin{cases}
    \sem{g}_\cV(\zeta) \otimes \sem{e_2}_\cV(\zeta) & \text{if $\zeta \in \LL_\cV(\varphi)$}\\
    \0 & \text{otherwise}
  \end{cases}
       \\
    &= \begin{cases}
      \sem{\hp(g,e_2)}_\cV(\zeta) & \text{if $\zeta \in \LL_\cV(\varphi)$}\\
      \0 & \text{otherwise}
    \end{cases}
    \tag{by I.H.}\\
  &    =\sem{\varphi \rhd \hp(g,e_2)}_\cV(\zeta)
    = \sem{\hp(\varphi \rhd g, e_2)}_\cV(\zeta) \enspace.
  \end{align*}
  \endgroup

  Let $e_1$ and $e_2$ be as in Case (e). Then our statement holds by left-distributivity. 

  Let $e_1$ and $e_2$ be as in Case (f). Then the proof is analogous to the proof of Case (b) except that we use left-distributivity instead of right-distributivity.

  Let $e_1$ and $e_2$ be as in Case (g). Then the proof is analogous to the proof of Case (c)  except that we use left-distributivity instead of right-distributivity.

  Let $e_1$ and $e_2$ be as in Case (h). Then the proof is analogous to the proof of Case (d).
 \end{proof}


In  Lemma \ref{lm:e1xe2-equivalent-hp(e1,e2)}, we assume that $e_1 \times e_2$ is variable separated.
We have used this condition in the proof  for the case that $e_1=\bigplus\nolimits_x g$ (cf. Case (b)
of the definition of $\hp(e_1,e_2)$). More precisely, it is used in the justification of the equality:
\begin{equation}\label{equ:equ-which-needs-prop-on-bounded-vars}
\bigoplus_{w\in \pos(\zeta)}\Big(\sem{g}_{\cV\cup\{x\}}(\zeta[x\mapsto w]) \otimes \sem{e_2}_{\cV}(\zeta)\Big)
  = \bigoplus_{w\in \pos(\zeta)} \Big(\sem{g}_{\cV\cup\{x\}}(\zeta[x\mapsto w]) \otimes  \sem{e_2}_{\cV\cup \{x\}}(\zeta[x\mapsto w])\Big)
\end{equation}
Here we show that, in general, Equation \eqref{equ:equ-which-needs-prop-on-bounded-vars} is wrong if we drop the condition that $\big(\bigplus\nolimits_x g\big) \times e_2$ is variable separated. For this we let $\Sigma = \{\gamma^{(1)}, \alpha^{(0)}\}$, $\cV = \{x\}$, $\xi = \gamma(\alpha)$, $\eta(x) = \varepsilon$, $\zeta = (\xi,\eta)$ in $\T_{\Sigma_\cV}^{\mathrm{v}}$ (i.e., $\zeta = (\gamma,\{x\})((\alpha,\emptyset))$), and
\begin{compactitem}
\item $\kappa = (\kappa_k \mid k \in \mathbb{N})$ with $\kappa_1((\gamma,\{x\})) = \kappa_0((\alpha,\emptyset)) = \1$ and  $\kappa_1((\gamma,\emptyset)) = \kappa_0((\alpha,\{x\})) = \0$ and
  \item $\nu = (\nu_k \mid k \in \mathbb{N})$ with $\nu_1((\gamma,\{x\})) = \nu_0((\alpha,\emptyset))= \0$ and $\nu_1((\gamma,\emptyset) = \nu_0((\alpha,\{x\})) = \1$.
  \end{compactitem}
  We let $g = \rmH(\nu)$ and $e_2= \rmH(\kappa)$. We note that $\Bound(\bigplus_x g) \cap \Free(e_2) = \{x\}$, hence $\Free\big(\big(\bigplus\nolimits_x g\big) \times e_2\big)\cap \Bound\big(\big(\bigplus\nolimits_x g\big) \times e_2\big)\ne\emptyset$.
  We can calculate the left-hand side of \eqref{equ:equ-which-needs-prop-on-bounded-vars} as follows:
  \begingroup
  \allowdisplaybreaks
  \begin{align*}
    &\bigoplus_{w\in \pos(\zeta)}\Big(\sem{g}_{\cV\cup\{x\}}(\zeta[x\mapsto w]) \otimes \sem{e_2}_{\cV}(\zeta)\Big)\\
    = & \bigoplus_{w\in \{\varepsilon,1\}}\Big(\sem{\rmH(\nu)}_{\{x\}}(\zeta[x\mapsto w]) \otimes \sem{\rmH(\kappa)}_{\{x\}}(\zeta)\Big)\\
          = & \bigoplus_{w\in \{\varepsilon,1\}}\Big(\sem{\rmH(\nu)}_{\{x\}}(\zeta[x\mapsto w]) \otimes \1\Big)\\
    = &\ \sem{\rmH(\nu)}_{\{x\}}(\zeta[x\mapsto \varepsilon]) \oplus \sem{\rmH(\nu)}_{\{x\}}(\zeta[x\mapsto 1])\\%
    = & \ \sem{\rmH(\nu)}_{\{x\}}((\gamma,\{x\})((\alpha,\emptyset))) \oplus \sem{\rmH(\nu)}_{\{x\}}((\gamma,\emptyset)((\alpha,\{x\})))=  \0 \oplus \1  = \1\enspace.
    \end{align*}
    \endgroup
 And we can calculate the right-hand side of \eqref{equ:equ-which-needs-prop-on-bounded-vars} as follows:
  \begingroup
  \allowdisplaybreaks
  \begin{align*}
    &\bigoplus_{w\in \pos(\zeta)}\Big(\sem{g}_{\cV\cup\{x\}}(\zeta[x\mapsto w]) \otimes \sem{\rmH(\kappa)}_{\cV\cup \{x\}}(\zeta[x \mapsto w)\Big)\\
    = &\bigoplus_{w\in \{\varepsilon, 1\}}\Big(\sem{\rmH(\nu)}_{\{x\}}(\zeta[x\mapsto w]) \otimes \sem{\rmH(\kappa)}_{\{x\}}(\zeta[x \mapsto w)\Big)\\
    = & \Big(\sem{\rmH(\nu)}_{\{x\}}((\gamma,\{x\})((\alpha,\emptyset))) \otimes \sem{\rmH(\kappa)}_{\{x\}}((\gamma,\{x\})((\alpha,\emptyset)))\Big) \ \oplus \ \\
      &  \Big(\sem{\rmH(\nu)}_{\{x\}}((\gamma,\emptyset)((\alpha,\{x\}))) \otimes \sem{\rmH(\kappa)}_{\{x\}}((\gamma,\emptyset)((\alpha,\{x\})))\Big)\\
     = & (\0 \otimes \1) \oplus (\1 \otimes \0) = \0     \enspace.
    \end{align*}
  \endgroup
Thus, without the mentioned property, Equation \eqref{equ:equ-which-needs-prop-on-bounded-vars} does not hold in general.


In the next lemma we get rid of weighted conjunction in formulas which do not contain weighted first-order universal quantification. Here we will need commutativity and distributivity of~$\B$. We repeatedly apply Lemma
\ref{lm:e1xe2-equivalent-hp(e1,e2)}.

  \begin{lemma}\rm \label{lm:getting-rid-of-conj}  Let $\B$ be distributive, let $e \in \MSOce(\Sigma,\B)$, and let $\cV$  be a finite set of variables such that $\cV \supseteq \Free(e)$. If $e$ does not contain a subformula of the form $\bigtimes_x e'$, then we can construct an $\MSO(\Sigma,\B)$-formula $f$ such that  $\Free(e) = \Free(f)$ and  $\sem{e}_\cV=\sem{f}_\cV$.
  \end{lemma}
  
   \begin{proof} Intuitively, we will define a terminating reduction system on a subset of $\MSOce(\Sigma,\B)$-formulas such that each reduction replaces a weighted conjunction $e_1 \times e_2$ by the formula which is denoted by $\hp(e_1,e_2)$.

   Formally, we define the set 
     \begin{align*}
       \mathrm{res}\MSOce(\Sigma,\B) = \{e \in \MSOce(\Sigma,\B) \mid & \text{ $e$ is variable separated and }\\
       & \text{ $e$ does not contain a subformula of the form $\bigtimes\nolimits_x e'$}\} \enspace.
     \end{align*}
    
     We define the reduction system $(\mathrm{res}\MSOce(\Sigma,\B),\succ)$ where, for every $e,f \in \mathrm{res}\MSOce(\Sigma,\B)$, we let
     \begin{align*}
       e \succ f \  \text{ if } \ &\text{ $e$ contains a subformula of the form $e_1 \times e_2$ such that $e_1,e_2\in  \MSO(\Sigma,\B)$ and }\\
                                  & \text{ $f$ is obtained from $e$ by replacing this subformula by the formula which is}\\
       & \text{ denoted by $\hp(e_1,e_2)$}\enspace.
     \end{align*}
  We note that, since $\Free(e_1 \times e_2) = \Free(\hp(e_1,e_2))$ and $\Bound(e_1 \times e_2) = \Bound(\hp(e_1,e_2))$, also the formula $f$ is variable separated, and hence $f$ is indeed an element of the set $\mathrm{res}\MSOce(\Sigma,\B)$.
   
     Next we prove the following statement.
     \begin{equation}\label{equ:replacing-subformula-wconj-preserves-semantics} 
 \text{Let  $e,f \in \mathrm{res}\MSOce(\Sigma,\B)$. If $e \succ f$, then  $\sem{e}_\cV= \sem{f}_\cV$ for each $\cV$ which contains $\Free(e)$.}
       \end{equation}
 Since $\Free(e_1 \times e_2)=\Free(\hp(e_1,e_2))$, we have that $\Free(e) = \Free(f)$. Since $e$ is variable separated, also $e_1 \times e_2$ is variable separated. Then, by Lemma~\ref{lm:e1xe2-equivalent-hp(e1,e2)}, we have  $\sem{e_1 \times e_2}=\sem{\hp(e_1,e_2)}$, and thus, by  Observation \ref{obs:replacing-equiv-subformula-3}, we have $\sem{e}_\cV= \sem{f}_\cV$ for each $\cV$ which contains $\Free(e)$. This finishes the proof of~\eqref{equ:replacing-subformula-wconj-preserves-semantics}.

 For proving the fact that $\succ$ is terminating, we define the mapping $\#_\times: \mathrm{res}\MSOce(\Sigma,\B) \to \mathbb{N}$ such that, for each $e \in \mathrm{res}\MSOce(\Sigma,\B)$, we let $\#_\times(e)$ be the number of occurrences of~$\times$. It is clear that, if $e \succ f$, then $\#_\times(e) = \#_\times(f) +1$. Hence $\#_\times$ is a monotone embedding of   $(\mathrm{res}\MSOce(\Sigma,\B),\succ)$ into the terminating reduction system $(\mathbb{N},>)$. Thus, by Lemma \ref{lm:fin-branching-embedding-termination}, also $(\mathrm{res}\MSOce(\Sigma,\B),\succ)$ is terminating. Clearly,
 \begin{equation}\label{equ:nf-wconj}
   \nf_\succ(\mathrm{res}\MSOce(\Sigma,\B)) \subseteq \MSO(\Sigma,\B) \enspace.
 \end{equation}

   Now we can prove the statement of the lemma. Let $e \in \MSOce(\Sigma,\B)$ and  $\cV \supseteq \Free(e)$ be a finite set of variables. Moreover, $e$ does not contain a subformula of the form $\bigtimes_x e'$. By Lemma \ref{lm:normal-form-weighted-MSO-ext}, we can assume that $e$ is variable separated, i.e., $e\in \mathrm{res}\MSOce(\Sigma,\B)$. Now we compute an arbitrary element $f$ in the set $\nf_\succ(e)$. By \eqref{equ:nf-wconj}, the formula $f$ is an element of 
   $\MSO(\Sigma,\B)$. And by \eqref{equ:replacing-subformula-wconj-preserves-semantics}, we obtain that  $\sem{e}_\cV= \sem{f}_\cV$ for each $\cV$ which contains $\Free(e)$. Moreover, we have that  $\Free(e) = \Free(f)$. Thus, $f$ is the required formula.
 \end{proof}


\subsection{Weighted  first-order universal quantification of recognizable step mappings}\label{subs:univ-fo-quantification-rec-step}

In Example \ref{ex:fo-universal-quant-too strong} we gave an atomic formula such that its weighted first-order universal quantification is not recognizable.
Here we show that, if the semantics of a weighted $\MSOe$-formula is a recognizable step mapping, then its weighted first-order universal quantification is run recognizable.

\begin{lemma}\rm  (cf. \cite[Lm.~4.2]{drogas05}, \cite[Lm.~4.4]{drogas07}, and  \cite[Lm.~5.5]{drovog06}) \label{lm:univ-fo-quantification-rec-step}  Let $e \in \MSOe(\Sigma,\B)$ and let $\cU$ abbreviate $\Free(e)$. Moreover, let $n\in \mathbb{N}_+$, $b_1,\ldots,b_n \in B$, and $\Sigma_{\cU\cup \{x\}}$-fta $A_1,\ldots,A_n$  such that $\sem{e}_{\cU \cup \{x\}}=\bigoplus_{j \in [n]} b_j \cdot \chi(\LL(A_j))$. Then we can construct a $(\Sigma_{\cV},\B)$-wta $\cB$, where $\cV = \Free(\bigtimes_x e)$, such that  $\runsem{\cB} = \sem{\bigtimes_x e}$.
\end{lemma}
\begin{proof} Since $\cV = \cU \setminus \{x\}$, we have $\cV \cup \{x\} = \cU \cup \{x\}$.
  By Theorem \ref{thm:crisp-det-algebra}(B)$\Rightarrow$(D) we can assume that $(\LL(A_j)\mid j\in[n])$ is a partitioning of $\T_{\Sigma_{\cV \cup \{x\}}}$.

  First, for each $\xi \in \T_{\Sigma_\cV}$, we  define the relation $\nu_\xi \subseteq  \pos(\xi) \times [n]$ by
  \[
\nu_\xi = \{(w,j) \in \pos(\xi) \times [n] \mid  \xi[x\mapsto w] \in \LL(A_j)\} \enspace.
\]
Since $(\LL(A_j)\mid j\in[n])$ is a partitioning of $\T_{\Sigma_{\cV \cup \{x\}}}$, this relation is a mapping $\nu_\xi: \pos(\xi) \to [n]$.

In order to understand the idea for the construction of $\cB$, let us perform a few  steps in the calculation of $\sem{\bigtimes_x e}(\xi)$ where $\xi \in \T_{\Sigma_\cV}$.
\begingroup
\allowdisplaybreaks
\begin{align*}
  \sem{\bigtimes\nolimits_x e}(\xi) = \bigotimes_{w \in \pos(\xi)} \sem{e}_{\cV \cup \{x\}}(\xi[x \mapsto w])
                                    = \bigotimes_{w \in \pos(\xi)} \Big(\bigoplus_{j \in [n]} b_j \cdot \chi(\LL(A_j))\Big) (\xi[x \mapsto w])
  = \bigotimes_{w \in \pos(\xi)}  b_{\nu_\xi(w)} \enspace.
   \end{align*}
   \endgroup

   Before constructing $\cB$,  we wish to construct a bu-deterministic wta $\cC$ which computes $\bigotimes_{w \in \pos(\xi)}  b_{\nu_\xi(w)}$; as input, $\cC$ receives the information $\xi$ and $\nu_\xi$. Since $\cC$ can only read trees (and not pairs of the form $(\xi,\nu)$), we encode each pair $(\xi,\nu)$ where $\xi \in \T_{\Sigma_\cV}$ and $\nu: \pos(\xi) \to [n]$ into one tree. For this purpose, we introduce the new ranked alphabet $n{\Sigma_\cV}$ by letting
 \[
   (n{\Sigma_\cV})^{(k)}=  \Sigma_\cV^{(k)} \times [n] \  \text{ for each $k \in \mathbb{N}$}\enspace.
 \]
Instead of $(n{\Sigma_\cV})^{(k)}$ we write $n{\Sigma_\cV}^{(k)}$. Moreover, for  $((\sigma,U),j) \in n{\Sigma}_\cV$ we also write $(\sigma,U,j)$.

  Now it is clear that the two sets $\T_{n{\Sigma}_\cV}$ and $\T_{\Sigma_\cV}^{+\mathrm{steps}}$,  where $\T_{\Sigma_\cV}^{+\mathrm{steps}} = \{(\xi,\nu) \mid \xi \in \T_{\Sigma_\cV}, \nu: \pos(\xi) \to [n]\}$, are in a one-to-one correspondence. Indeed,
  \begin{compactitem}
    \item each tree $\zeta \in \T_{n{\Sigma}_\mathcal{V}}$ corresponds to the pair
      $(\xi,\nu)$, where $\pos(\xi) = \pos(\zeta)$ and, for each $w \in \pos(\xi)$, if $\zeta(w) = (\sigma,U,j)$, then $\xi(w) = (\sigma,U)$ and  $\nu(w) = j$ and
    \item each pair $(\xi,\nu) \in \T_{\Sigma_\cV}^{+\mathrm{steps}}$ corresponds to the tree $\zeta \in \T_{n{\Sigma}_\mathcal{V}}$ where $\pos(\zeta)=\pos(\xi)$ and, for each $w \in \pos(\zeta)$, we let  $\zeta(w) = (\xi(w),\nu(w))$.
    \end{compactitem}
    Due to this one-to-one correspondence, we can assume that elements of $\T_{n{\Sigma}_\mathcal{V}}$ have the form $(\xi,\nu)$ and, vice versa, that each pair $(\xi,\nu)$ with $\xi \in \T_{\Sigma_\cV}$ and $\nu: \pos(\xi) \to [n]$  is an element of~$\T_{n{\Sigma}_\cV}$.

Now we can turn back to the expression $\bigotimes_{w \in \pos(\xi)}  b_{\nu_\xi(w)}$ above and can be more precise with respect to the type and the semantics of $\cC$.  We will construct a bu-deterministic $(n{\Sigma}_{\cV},\B)$-wta $\cC$ such that, for each $(\xi,\nu) \in \T_{n{\Sigma}_\cV}$, we have
   \begin{equation}\label{eq:wta-extended-ra-with-steps}
     \sem{\cC}((\xi,\nu)) =  \begin{cases}\bigotimes_{w \in \pos(\xi)}  b_{\nu_\xi(w)} &\text{ if $\nu= \nu_\xi$}\\
       \0 &\text{ otherwise}\enspace.
       \end{cases}
     \end{equation}

For the time being, let us assume that we have constructed such a $\cC$ already.     Then, by using the deterministic tree relabeling $\tau: n{\Sigma}_\cV \to \Sigma_\cV$ defined, for each $(\sigma,U,j)$ by $\tau((\sigma,U,j)) = (\sigma,U)$, we  can continue the calculation as follows: for each $\xi \in \T_{\Sigma_\cV}$, we have 
\begingroup
\allowdisplaybreaks
\begin{align*}
  \bigotimes_{w \in \pos(\xi)}  b_{\nu_\xi(w)}
  = \sem{\cC}((\xi,\nu_\xi))
  = \bigoplus_{(\xi,\nu) \in \tau^{-1}(\xi)} \sem{\cC}((\xi,\nu))
  = \chi(\tau)\big( \sem{\cC} \big)(\xi)
   \end{align*}
   \endgroup
   where the first and the second equality hold due to \eqref{eq:wta-extended-ra-with-steps}.
   
   Lastly, by Theorem \ref{thm:closure-under-tree-relabeling} (closure of $\Rec^{\mathrm{run}}(\_,\B)$ under tree relabelings),  we can construct the desired $(\Sigma_\cV,\B)$-wta $\cB$ such that $\runsem{\cB}=\chi(\tau)\big(\sem{\cC}\big)$ and we have proved the lemma. (We note that, since $\tau$ is overlapping, in general $\cB$ is not bu-deterministic.)
   
Thus it remains to construct the bu-deterministic $(n{\Sigma}_{\cV},\B)$-wta $\cC$ such that \eqref{eq:wta-extended-ra-with-steps} holds. For this purpose we first construct an (unweighted) total and bu-deterministic $n{\Sigma}_{\cV}$-fta $C$ such that $\LL(C) = L$ where 
    \[
L = \{(\xi,\nu_\xi) \mid \xi \in \T_{\Sigma_\cV}\} \enspace.
      \]
      Assume that $C = (\widetilde{Q},\widetilde{\delta},\widetilde{F})$ is such an $n{\Sigma}_{\cV}$-fta. Then we  construct the bu-deterministic  $(n{\Sigma}_{\cV},\B)$-wta $\cC= (\widetilde{Q},{\delta},{F})$ such that, for every $k \in \mathbb{N}$, $(\sigma,U,j) \in n{\Sigma}^{(k)}$, and $q_1,\ldots,q_k,q \in \widetilde{Q}$, we let
      \[
        \delta_k(q_1\cdots q_k,(\sigma,U,j),q) =
        \begin{cases}
          b_j & \text{ if $\widetilde{\delta}_k(q_1\cdots q_k,(\sigma,U,j)) = q$}\\
          \0 & \text{ otherwise} 
          \end{cases}
        \]
and $F_q = \1$ if $q \in \widetilde{F}$ and $\0$ otherwise.
It is obvious that \eqref{eq:wta-extended-ra-with-steps} holds.

Thus it remains to construct a total and bu-deterministic $n{\Sigma}_{\cV}$-fta $C$ such that $\LL(C) = L$.

We observe the following:
\begin{align*}
  L &= \{(\xi,\nu_\xi) \mid \xi \in \T_{\Sigma_\cV}\}\\
    &=\{(\xi,\nu) \in \T_{n{\Sigma}_\cV} \mid (\forall j \in [n], w \in \mathrm{pos}(\xi)): \text{ if } \nu(w) = j, \text{ then } \xi[x \rightarrow w] \in L(A_j)\}\\
    &= \bigcap_{j \in [n]} \{(\xi,\nu) \in \T_{n{\Sigma}_\cV} \mid (\forall w \in \mathrm{pos}(\xi)): \text{ if } \nu(w) = j, \text{ then } \xi[x \rightarrow w] \in L(A_j)\}\enspace.
\end{align*}
Thus,  due to Theorems \ref{thm:fta-closure-results}(2) and \ref{thm:fta-total-bud-fta}, it suffices to construct, for each $j \in [n]$, a total and  bu-deterministic $n{\Sigma}_{\cV}$-fta $C_j$ such that $\LL(C_j) = L_j$ where
\[
L_j =  \{(\xi,\nu) \in \T_{n{\Sigma}_\mathcal{V}} \mid (\forall
w \in \mathrm{pos}(\xi)): \text{ if } \nu(w) = j, \text{ then }
\xi[x \rightarrow w] \in L(A_j)\}\enspace.
\]

Let $j\in[n]$. The idea behind the construction of $C_j$  follows the one in the proof of \cite[Lm.~4.4]{drogas07} and it is roughly described as follows. On an input tree $\zeta = (\xi,\nu)$, the $n{\Sigma}_{\cV}$-fta  $C_j$ simulates the work of $A_j$ on $\xi$ and, whenever a position $w$ of $\zeta$ is encountered for which $\nu(w) = j$ holds, then, additionally,  $C_j$  splits off a copy of $A_j$; this copy  behaves as if at $w$ the $x$ would occur (or, in other words: as if $A_j$ would have $\xi[x \mapsto w]$ as input tree). Moreover, since we have to guarantee that the placement of $x$ is done at most once in $\xi$, we maintain a bit $d$ in every state of $C_j$ which indicates whether the $x$ was placed ($d = 1$) or not ($d = 0$). 

In the sequel, we will reuse the standard notations $\delta$, $F$, $\widetilde{Q}$, $\widetilde{\delta}$, and $\widetilde{F}$, which occurred already in the specification of $C$ and $\cC$, for other purposes. This will be harmless because below we will not deal with $C$ and $\cC$ anymore.

Formally,  let $A_j=(Q,\delta,F)$ be the $\Sigma_{\cV \cup \{x\}}$-fta given in the statement of the lemma. By Theorem~\ref{thm:fta-total-bud-fta}, we may assume that $A_j$ is total and bu-deterministic.
We construct the total and bu-deterministic $n{\Sigma}_\mathcal{V}$-fta
$C_j =(\widetilde{Q},\widetilde{\delta},\widetilde{F})$, where
\begin{itemize}
\item $\widetilde{Q} = \mathcal{P}(Q \times \{0,1\})$, 
\item $\widetilde{F} = \{P \subseteq Q \times \{0,1\} \mid P \cap (Q\times\{1\}) \subseteq F \times \{1\}\}$, and 
\item $\widetilde{\delta}$ is defined such that, for every $k\in \mathbb{N}$, 
$(\sigma,U,l) \in n{\Sigma}_\mathcal{V}^{(k)}$,  and $P_1,\ldots,P_k \in \widetilde{Q}$, we let
\[
  \widetilde{\delta}_{k}(P_1\cdots P_k, (\sigma,U,l)) = P \ \ 
  \text{ where } \
P = 
\begin{cases}
P'  & \text{ if } l \not= j\\
P' \cup P'' & \text{ otherwise }
\end{cases}
\]
and 
\begin{align*}
P' = \Big\{\Big(\delta_{k}(p_1\cdots p_k,(\sigma,U)),\bigplus\nolimits_{i\in[k]} d_i\Big)  \mid (\forall i \in [k]): (p_i,d_i) \in  P_i   \text{ and } \bigplus\nolimits_{i\in[k]} d_i \le 1\Big\}
\end{align*}
 and 
\begin{align*}
P'' = \Big\{\Big(\delta_{k}(p_1\cdots p_k,(\sigma,U \cup \{x\})),1\Big) \mid (\forall i \in [k]): (p_i,0) \in P_i \Big\}\enspace.
\end{align*}
\end{itemize}
Since $A_j$ is total and bu-deterministic, the states $\delta_{k}(p_1\cdots p_k,(\sigma,U))$ and $\delta_{k}(p_1\cdots p_k,(\sigma,U \cup \{x\}))$ are defined. Obviously, $P'$ formalizes the propagation of the bit value $0$ if $\bigplus_{i\in[k]} d_i = 0$, and the propagation of the bit value 1 if $\bigplus_{i\in[k]} d_i = 1$.  If $l = j$, then additionally to this propagation, $C_j$ can change from the bit vector $(0,\ldots,0)$ to the bit value 1, thereby placing the $x$ to this position; the union with $P''$ can be described as ``splitting off a new copy of $A_j$''. Also it is clear that $C_j$ is total and bu-deterministic. Now we will show that $C_j$ recognizes $L_j$.

Since $A_j$ and $C_j$ are total and bu-deterministic, for each $(\xi,\nu) \in \T_{n{\Sigma}_\cV}$ and $w \in \pos(\xi)$, the sets of valid runs $\R_{A_j}^\mathrm{v}(\xi)$, $\R_{A_j}^\mathrm{v}(\xi[x \rightarrow w])$, and $\R_{C_j}^\mathrm{v}((\xi,\nu))$ are singletons. We denote these runs by letting
\[
  \R_{A_j}^\mathrm{v}(\xi) = \{\rho_\xi\}, \
  \R_{A_j}^\mathrm{v}(\xi[x \rightarrow w]) = \{\rho_{\xi[x\rightarrow w]}\}, \ \text{  and }  \ \R_{C_j}^\mathrm{v}((\xi,\nu)) = \{\rho_{(\xi,\nu)}\} \enspace.
  \]
By induction on $\T_{n{\Sigma}_\mathcal{V}}$, we prove the following statement.
\begin{equation}\label{eq:all}
\text{For each $(\xi,\nu) \in \T_{n{\Sigma}_\mathcal{V}}$, we have } \rho_{(\xi,\nu)}(\varepsilon) =
\left\{\left(\rho_{\xi}(\varepsilon),0\right)\right\} \cup \{(\rho_{\xi[x \rightarrow w]}(\varepsilon),1) \mid w \in \mathrm{pos}(\xi), \nu(w)
= j\}\enspace.
\end{equation}

Let $\zeta = (\sigma,U,l)(\zeta_1,\ldots,\zeta_k)$ be in $\T_{n{\Sigma}_\mathcal{V}}$ with $\zeta = (\xi,\nu)$ and $\zeta_i = (\xi_i,\nu_i)$ for every $i \in [k]$. Thus $\xi = (\sigma,U)(\xi_1,\ldots,\xi_k)$.
By I.H., \eqref{eq:all} holds for $\zeta_1,\ldots,\zeta_k$. First we prove that
\begin{eqnarray}
  \begin{aligned}\label{eq:two}
&\Big\{\Big(\delta_k(p_1\ldots p_k,(\sigma,U)),
 \bigplus\nolimits_{i\in[k]} d_i\Big) \mid (\forall i\in[k]):(p_i,d_i) \in \rho_{(\xi_i,\nu_i)}(\varepsilon) \text{ and }
 \bigplus\nolimits_{i\in[k]} d_i \le 1\Big\}\\
&=  \{(\rho_{\xi}(\varepsilon),0)\} \cup 
\{(\rho_{\xi[x \rightarrow uw]}(\varepsilon),1) \mid 
u \in [k], w \in \mathrm{pos}(\xi_u), \text{ and } \nu_u(w) = j\}
\end{aligned}
\end{eqnarray}
as follows:
\begingroup
\allowdisplaybreaks
\begin{align*}
  & \Big\{\Big(\delta_k(p_1\cdots p_k,(\sigma,U)), \bigplus\nolimits_{i\in[k]} d_i\Big) \mid (\forall i\in[k]): (p_i,d_i) \in \rho_{(\xi_i,\nu_i)}(\varepsilon) \hbox{ and } \bigplus\nolimits_{i\in[k]} d_i \le 1\Big\}\\[3mm]
   = \ & \Big\{\Big(\delta_k(p_1\cdots p_k,(\sigma,U)), \bigplus\nolimits_{i\in[k]} d_i\Big) \mid (\forall i\in[k]): \\
    & \big[(p_i,d_i) = (\rho_{\xi_i}(\varepsilon),0) \hbox{ or } (p_i,d_i) = (\rho_{\xi_i[x \rightarrow w]}(\varepsilon),1),
     w \in \mathrm{pos}(\xi_i), \nu_i(w) = j \big] \hbox{ and } \bigplus\nolimits_{i\in[k]} d_i \le 1\Big\}
     & \tag{by I.H.}\\
  =\ & \left\{\left(\delta_{k}\big(\rho_{\xi_1}(\varepsilon) \cdots \rho_{\xi_k}(\varepsilon),(\sigma,U)\big),0\right)\right\}\  \\[2mm]
& \cup \left\{\left(\delta_{k}\big(\rho_{\xi_1}(\varepsilon)\cdots \rho_{\xi_u[x \rightarrow w]}(\varepsilon) \cdots \rho_{\xi_k}(\varepsilon),(\sigma,U)\big),1\right)\mid  u \in[k], w \in \mathrm{pos}(\xi_u), \nu_u(w) = j\right\}\\[3mm]
= \ & \left\{\left(\rho_{\xi}(\varepsilon),0\right)\right\} \cup 
\left\{\left(\rho_{\xi[x \rightarrow uw]}(\varepsilon),1\right) \mid u\in[k], w \in \mathrm{pos}(\xi_u), \nu_u(w) = j\right\}\enspace.
   \end{align*}
   \endgroup
   This finishes the proof of  \eqref{eq:two}. Now we can complete the proof of \eqref{eq:all}.

\underline{Case (i):} Let $l\ne j$. Then 
\begin{align*}
  \rho_{(\xi,\nu)}(\varepsilon) &=  \widetilde{\delta}_{k}(\rho_{(\xi,\nu)}(1) \ldots \rho_{(\xi,\nu)}(k),(\sigma,U,l)) = \widetilde{\delta}_{k}(\rho_{(\xi_1,\nu_1)}(\varepsilon) \ldots \rho_{(\xi_k,\nu_k)}(\varepsilon),(\sigma,U,l))\\[2mm]
  &= \Big\{\Big(\delta_{k}\big(p_1\ldots p_k,(\sigma,U)\big), \bigplus\nolimits_{i\in[k]} d_i\Big) \mid (\forall i\in[k]):(p_i,d_i) \in \rho_{(\xi_i,\nu_i)}(\varepsilon) \text{ and } \bigplus\nolimits_{i\in[k]} d_i \le 1\Big\}\\[3mm]
&= \{(\rho_{\xi}(\varepsilon),0)\} \cup 
    \{(\rho_{\xi[x \rightarrow uw]}(\varepsilon),1) \mid u\in[k],  w \in \mathrm{pos}(\xi_u),\nu_u(w) = j\} \tag{by \eqref{eq:two}}\\[3mm]
&= \{(\rho_{\xi}(\varepsilon),0)\} \cup 
\{(\rho_{\xi[x \rightarrow w]}(\varepsilon),1) \mid w \in \mathrm{pos}(\xi), \nu(w) = j\} \tag{because $\nu(\varepsilon) = l \not= j$}\enspace.
\end{align*}

\underline{Case (ii):} Let $l = j$. Then
\begingroup
\allowdisplaybreaks
\begin{align*}
  \rho_{(\xi,\nu)}(\varepsilon) &=  \widetilde{\delta}_{k}(\rho_{(\xi_1,\nu_1)}(\varepsilon) \ldots \rho_{(\xi_k,\nu_k)}(\varepsilon),(\sigma,U,l))\\[3mm]
&= \Big\{\Big(\delta_{k}(p_1\ldots p_k,(\sigma,U)), \bigplus\nolimits_{i\in[k]} d_i\Big) \mid (\forall i\in[k]): (p_i,d_i) \in \rho_{(\xi_i,\nu_i)}(\varepsilon) \text{ and } \bigplus\nolimits_{i\in[k]} d_i \le 1\Big\} \ \\
                                & \hspace*{5mm} \cup  \{(\delta_{k}(\rho_{\xi_1}(\varepsilon)\ldots \rho_{\xi_k}(\varepsilon),(\sigma,U\cup\{x\})),1)\}
  \tag{by definition of $\widetilde{\delta}$ and  \eqref{eq:two} for reducing $P''$} \\[4mm]
  &= \{(\rho_{\xi}(\varepsilon),0)\} \cup 
    \{(\rho_{\xi[x \rightarrow uw]}(\varepsilon),1) \mid u\in[k],  w \in \mathrm{pos}(\xi_u),\nu_u(w) = j\} 
\cup \{(\rho_{\xi[x \rightarrow \varepsilon]}(\varepsilon),1)\}\\
& \tag{by  \eqref{eq:two} and definition of run}\\[2mm]
&= \{(\rho_{\xi}(\varepsilon),0)\} \cup 
\{(\rho_{\xi[x \rightarrow w]}(\varepsilon),1) \mid w \in \mathrm{pos}(\xi), \nu(w) = j\}\enspace.
\end{align*}
\endgroup
This finishes the proof of \eqref{eq:all}.

Finally, let $(\xi,\nu) \in \T_{n{\Sigma}_\cV}$. Then we can calculate as follows.
\begingroup
\allowdisplaybreaks
\begin{align*}
  (\xi,\nu) \in L_j
  & \text{ iff } \   (\forall w \in \mathrm{pos}(\xi)): \nu(w) = j \text{ implies } \xi[x\rightarrow w] \in \LL(A_j)\\
&  \text{ iff } \   (\forall w \in \mathrm{pos}(\xi)): \nu(w) = j \text{ implies } \rho_{\xi[x\rightarrow
                        w]}(\varepsilon) \in F\\
  & \text{ iff } \ \{(\rho_{\xi[x\mapsto w]}(\varepsilon),1) \mid w \in \pos(\xi), \nu(w) = j\} \subseteq F \times \{1\}\\
&  \text{ iff } \ \rho_{(\xi,\nu)}(\varepsilon) \in \widetilde{F} \tag{by \eqref{eq:all}} \\
&  \text{ iff } \ (\xi,\nu) \in \LL(C_j) \enspace.
  \end{align*}
\endgroup
Hence $\LL(C_j) = L_j$, and this finishes the proof of the lemma.
\end{proof}

Finally, we point out the close  relationship between Lemma \ref{lm:univ-fo-quantification-rec-step} and Lemma \ref{lm:getting-rid-of-universal-quant}: both show how to eliminate weighted first-order universal quantification. The difference is that in Lemma \ref{lm:univ-fo-quantification-rec-step} we end up in a wta, whereas in Lemma \ref{lm:getting-rid-of-universal-quant} we end up in an $\MSO(\Sigma,\B)$-formula.

\subsection[The main result for $\MSOe(\Sigma,\B)$]{The main result for $\MSOe(\Sigma,\B)$ over commutative and bi-locally finite strong bimonoids}
\label{sec:MSO-ext-commutative-bi-locally finite}

In this subsection we prove the B-E-T-theorem for commutative and bi-locally finite strong bimonoids (cf. Theorem \ref{thm:Buechi-comm-bi-loc-fin}) where the involved weighted MSO-logic is  $\MSOe(\Sigma,\B)$ (i.e., the fully extended $\MSO(\Sigma,\B)$ without any restrictions). This theorem can be compared to \cite[Thm.~5.3]{drovog12} if one drops negation from the latter. We note that we use commutativity in order to handle weighted second-order universal quantification (cf. Lemma \ref{lm:rec-closed-under-univ-second-quant-cbl}); on the contrary, for M{\'e}dv{\'e}dj{\'e}v's theorem for wta over bi-locally finite strong bimonoids (cf. Theorem~\ref{cor:Medvedjev-str-bm}) commutativity is not necessary.

                \begin{quote}\emph{In this subsection, we let $\B= (B,\oplus,\otimes,\0,\1)$ be an arbitrary commutative and bi-locally finite strong bimonoid.}
                \end{quote}

                For the proof of the B-E-T-theorem, in particular, we have to prove that weighted first-order universal quantification and  weighted second-order universal quantification preserve run recognizability. However, let us start with a consistency lemma (cf. Lemma \ref{lm:consistency-wta}).

\begin{lemma}\rm \label{lm:consistency-wta-MSOe-2}  Let $e \in \MSOe(\Sigma,\B)$, let $\cU$ abbreviate $\Free(e)$, and let $\cA$ be a $(\Sigma_\cU,\B)$-wta such that $\runsem{\cA} = \sem{e}$. Moreover, let $V$ be a finite set of variables. Then we can construct a $(\Sigma_{\cU \cup V},\B)$-wta $\cA'$ such that $\runsem{\cA'} = \sem{e}_{\cU \cup V}$.
\end{lemma}

\begin{proof} The proof is the same as the one for Lemma \ref{lm:consistency-wta} except that in Case (b)(ii) we have to use the consistency lemma for $\MSOe(\Sigma,\B)$ (i.e., Lemma \ref{obs:consistency}) instead of Lemma \ref{lm:consistency-MSO}.
  \end{proof}

Next we prove that  weighted first-order universal quantification preserves run recognizability.

            \begin{lemma}\label{lm:rec-closed-under-univ-first-quant-2}  \rm Let $e \in \MSOe(\Sigma,\B)$ and  $x$  any first-order variable. Moreover, let $\cU$ and $\cV$ abbreviate $\Free(e)$ and $\Free(\bigtimes_x e)$, respectively. If  there exists a $(\Sigma_{\cU},\B)$-wta $\cA$ with $\runsem{\cA} = \sem{e}$, then we can construct a $(\Sigma_{\cV},\B)$-wta $\cB$ such that   $\runsem{\cB} = \sem{\bigtimes_x e}$. 
   \end{lemma}
   \begin{proof} Let $\cA$ be a $(\Sigma_\cU,\B)$-wta such that $\runsem{\cA} = \sem{e}$. Since $\cU \subseteq \cV\cup\{x\}$, by Lemma \ref{lm:consistency-wta-MSOe-2}, we can construct a $(\Sigma_{\cV \cup \{x\}},\B)$-wta $\cA'$ such that $\runsem{\cA'} = \sem{e}_{\cV\cup\{x\}}$.
        
 Since $\B$ is bi-locally finite, by Theorem \ref{thm:bi-loc-finite-rec-step-function}(A)$\Rightarrow$(B) and Theorem \ref{thm:crisp-det-algebra} $\runsem{\cA'}$ is a recognizable step mapping and we can construct $n \in \mathbb{N}$, $b_1,\ldots,b_n \in L$, and $\Sigma_{\cV \cup \{x\}}$-fta $A_1,\ldots,A_n$ such that
              \[
\runsem{\cA'} = \bigoplus_{j \in [n]} b_j \cdot \chi(\LL(A_j)) \enspace.
\]
Then by Lemma \ref{lm:univ-fo-quantification-rec-step} we can construct a $(\Sigma_{\cV},\B)$-wta $\cB$ such that  $\runsem{\cB} = \sem{\bigtimes_x e}$.
     \end{proof}

     Next we prove that  weighted second-order universal quantification preserves run recognizability.

\begin{lemma}\label{lm:rec-closed-under-univ-second-quant-cbl}  \rm (cf. \cite[Lm.~5.14]{drovog12}, \cite[Prop.~6.4]{drovog06}) Let $e \in \MSOe(\Sigma,\B)$  and  $X$ any second-order variable. Moreover, let $\cU=\Free(e)$ and $\cV = \Free(\bigtimes_X e)$.
              If there exists a $(\Sigma_{\cU},\B)$-wta $\cA$ with $\runsem{\cA} = \sem{e}$, then we can construct a $(\Sigma_{\cV},\B)$-wta $\cB$ such that  $\runsem{\cB} = \sem{\bigtimes_X e}$.
            \end{lemma}
             \begin{proof} Let $\zeta \in \T_{\Sigma_\cV}$. By definition, we have that 
              \[
\sem{\bigtimes \nolimits_X e}(\zeta) =
                  \begin{cases}\bigotimes\limits_{W \subseteq \pos(\zeta)} \sem{e}_{\cV\cup\{X\}}(\zeta[X\mapsto W]) & \text{if $\zeta \in  \T_{\Sigma_\cV}^\mathrm v$}\\
                    \0 & \text{otherwise} \enspace.
                    \end{cases}
                \]
which is equivalent to
                \[
\sem{\bigtimes \nolimits_X e}(\zeta) = \Big(\bigotimes_{W \subseteq \pos(\zeta)} \sem{e}_{\cV\cup\{X\}}(\zeta[X\mapsto W])\Big)  \otimes \chi(\T_{\Sigma_\cV}^{\mathrm{v}})(\zeta) \enspace.
                  \]

              As auxiliary tool, we define the deterministic $(\Sigma_{\cV\cup \{X\}},\Sigma_\cV)$-tree relabeling $\tau=(\tau_k \mid k \in \mathbb{N})$ with $\tau_k((\sigma,\cW)) = (\sigma,\cW\setminus\{X\})$ for each $(\sigma,\cW) \in \Sigma_{\cV \cup \{X\}}$. Then we have  
  \begin{align*}
\sem{\bigtimes \nolimits_X e}(\zeta) = & \Big(\bigotimes_{\xi \in \tau^{-1}(\zeta)} \sem{e}_{\cV\cup\{X\}}(\xi)\Big)  \otimes \chi(\T_{\Sigma_\cV}^{\mathrm{v}})(\zeta) \enspace.
  \end{align*}

  Let $\cA$ be a $(\Sigma_{\cU},\B)$-wta with $\runsem{\cA} = \sem{e}$. By Lemma  \ref{lm:consistency-wta-MSOe-2}, we can construct a $(\Sigma_{\cU \cup \{X\}},\B)$-wta $\cA'$ such that $\runsem{\cA'} = \sem{e}_{\cU\cup\{X\}}$.
                  Since $\cU \cup \{X\} = \cV \cup \{X\}$, we have
 \[
\sem{\bigtimes \nolimits_X e}(\zeta) = \Big(\bigotimes_{\xi \in \tau^{-1}(\zeta)} \runsem{\cA'}(\xi)\Big)  \otimes \chi(\T_{\Sigma_\cV}^{\mathrm{v}})(\zeta) \enspace.
                  \]

                  Since $\B$ is bi-locally finite, $\runsem{\cA'}$ is a recognizable step mapping. More precisely, by Theorem \ref{thm:bi-loc-finite-rec-step-function} (A)$\Rightarrow$(B) and Theorem  \ref{thm:crisp-det-algebra} (A)$\Rightarrow$(B),
we can construct $n \in \mathbb{N}_+$, $b_1,\ldots,b_n \in B$, and $\Sigma_{\cV \cup \{X\}}$-fta $A_1,\ldots,A_n$ such that
              \[
\runsem{\cA'} = \bigoplus_{j \in [n]} b_j \cdot \chi(\LL(A_j)) \enspace.
                \]
 Moreover, by Observation \ref{obs:partition-step}, we can assume that the family $(\LL(A_j) \mid j \in [n])$ is a partitioning of $\T_{\Sigma_{\cV \cup \{X\}}}$.     Thus
\[
\sem{\bigtimes \nolimits_X e}(\zeta) = \Big(\bigotimes_{\xi \in \tau^{-1}(\zeta)} \big(\bigoplus_{j \in [n]} b_j \otimes \chi(\LL(A_j))(\xi) \big) \Big)  \otimes \chi(\T_{\Sigma_\cV}^{\mathrm{v}})(\zeta) \enspace.
                  \]

          Since the tree languages $\LL(A_1),\ldots,\LL(A_n)$ are pairwise disjoint and $\B$ is commutative, we obtain
\begin{equation}\label{eq:product-equation}
\sem{\bigtimes \nolimits_X e}(\zeta) = \Big(\bigotimes_{j \in [n]} b_j^{m_j(\zeta)} \Big)  \otimes \chi(\T_{\Sigma_\cV}^{\mathrm{v}})(\zeta) 
                  \end{equation}
                  where, for each $j \in [n]$, the weighted tree language $m_j: \T_{\Sigma_{\cV}} \to \mathbb{N}$ is defined, for each $\theta \in \T_{\Sigma_\cV}$, by 
                                   $m_j(\theta) = |\tau^{-1}(\theta) \cap \LL(A_j)|$.

We show that $m_j$ is $(\Sigma_{\cV},\Nat)$-recognizable as follows. Since $\LL(A_j)$ is a recognizable $\Sigma_{\cV \cup \{X\}}$-tree language, by Theorem \ref{thm:fta-wta}(B)$\Rightarrow$(C), the weighted tree language $\chi_{\Nat}(\LL(A_j))$ is $(\Sigma_{\cV \cup \{X\}},\Nat)$-recognizable. Thus, by Theorem \ref{thm:closure-under-tree-relabeling}, the weighted tree language $\tau(\chi_{\Nat}(\LL(A_j)))$ is $(\Sigma_{\cV},\Nat)$-recognizable. Moreover, for each $\theta \in \T_{\Sigma_{\cV}}$, we have
\[\tau(\chi_{\Nat}(\LL(A_j)))(\theta)=\bigplus\nolimits_{\xi \in \tau^{-1}(\theta)}\chi_{\Nat}(\LL(A_j))(\xi)= |\tau^{-1}(\theta) \cap \LL(A_j)| =  m_j(\theta)\enspace.\]                 
                                   
Next we analyse the product $\bigotimes_{j \in [n]} b_j^{m_j(\zeta)}$. Since $\B$ is, in particular, multiplicatively locally finite, each element of $B$ has finite multiplicative order. Thus, for each element there exists an index and a period (cf. Subsection \ref{sec:strong-bimonoid}).

We recall that, for each $j\in [n]$, the index of $b_j$ and the period of $b_j$ are denoted by $i(b_j)$ and $p(b_j)$, respectively. Then, for each $j\in[n]$, there exists a uniquely determined $d_j(\zeta) \in [0, i(b_j)+p(b_j)-1]$ such that $b_j^{m_j(\zeta)}=b_j^{d_j(\zeta)}$ and $m_j(\zeta)\in d_j(\zeta)+p(j)\mathbb{N}$. Moreover, for every $d\in [0,i(b_j)]$, we have
$b_j^{m_j(\zeta)}=b_j^d$ if and only if $m_j(\zeta)=d$, and for every $d\in [i(b_j),i(b_j)+p(b_j)-1]$, we have $b_j^{m_j(\zeta)}=b_j^d$ if and only if $m_j(\zeta)\in d+p(b_j)\mathbb{N}$. Then 

\begin{equation}\label{eq:this-is-recstep-mapping} \bigotimes_{j \in [n]} b_j^{m_j(\zeta)}  = \bigoplus_{\substack{d_1,\ldots,d_n:\\ (\forall j\in[n]): 0\le d_j < i(j)+p(j)}} b_1^{d_1}\otimes \ldots \otimes b_n^{d_n}\otimes \chi(L^1_{d_1}\cap\ldots\cap L^n_{d_n})(\zeta),
\end{equation}
where, for every $d\in[0,i(b_j)+p(b_j)]$ and $j\in [n]$, we have $L_d^j=\{\theta \in \T_{\Sigma_\cV}\mid b_j^{m_j(\zeta)}=b_j^d\}$. 

Hence 
\begin{align*}
L_d^j=& \{\theta \in \T_{\Sigma_\cV}\mid m_j(\zeta)=d\}= m_j^{-1}(d), \text{ if $ 0\le d <i(b_j)$, and}\\
L_d^j=& \{\theta \in \T_{\Sigma_\cV}\mid m_j(\zeta)\in d+p(b_j)\mathbb{N}\}= m_j^{-1}(d+p(b_j)\mathbb{N}), \text{ if $i(b_j)\le d < i(b_j)+p(b_j)$}.
\end{align*}
Since $m_j$ is $(\Sigma_{\cV},\Nat)$-recognizable, by Lemmas \ref{lm:preimage-N-1} and \ref{lm:preimage-N-2}, in both cases $L_d^j$ is a recognizable $\Sigma_\cV$-language. Hence, the weighted tree language on the right-hand side of \eqref{eq:this-is-recstep-mapping} is a recognizable step mapping (because the set $\Rec(\Sigma_\cV)$ is closed under intersection, cf. Theorem \ref{thm:fta-closure-results}(2)). Thus, by  Theorem  \ref{thm:crisp-det-algebra} (B)$\Rightarrow$(A), we can construct a crisp-deterministic $(\Sigma_{\cV},\B)$-wta $\cA''$ such that $\bigotimes_{j \in [n]} b_j^{m_j(\zeta)}=\sem{\cA''}(\zeta)$. Then, by \eqref{eq:product-equation}, we obtain
\[\sem{\bigtimes \nolimits_X e}(\zeta) = \sem{\cA''}(\zeta)  \otimes \chi(\T_{\Sigma_\cV}^{\mathrm{v}})(\zeta) 
                  \]
Lastly, by Lemma \ref{lm:TSigma-valid-fta} and Theorem \ref{thm:closure-Hadamard-product-char}(2), we can construct a $(\Sigma_{\cV},\B)$-wta $\cB$ such that $\runsem{\cB} = \sem{\bigtimes_X e}$. (We note that $\cB$ is even crisp-deterministic.)                      
               \end{proof}

     Now we can prove the B-E-T-theorem for wta over commutative and bi-locally finite strong bimonoids. It can be compared to  \cite[Cor. 6.5]{drogoemaemei11}.

                \begin{theorem-rect} \label{thm:Buechi-comm-bi-loc-fin} {\rm (cf. \cite[Thm.~5.3]{drovog12})} Let $\Sigma$ be a ranked alphabet. Moreover, let $\B=(B,\oplus,\otimes,\0,\1)$ be a commutative and bi-locally finite strong bimonoid and let $r: \T_\Sigma \to B$. Then the following four statements are equivalent.
                  \begin{compactenum}
                      \item[(A)] We can construct a $(\Sigma,\B)$-wta $\cB$ such that $r=\runsem{\cB}$.
        \item[(B)] We can construct a $(\Sigma,\B)$-recognizable step formula $e$ such that $\Free(e) = \emptyset$ and  $r = \sem{e}$.
        \item[(C)] We can construct a sentence $e \in \MSO(\Sigma,\B)$ such that $r = \sem{e}$
        \item[(D)] We can construct a sentence $e \in \MSOe(\Sigma,\B)$ such that $r = \sem{e}$.
\end{compactenum}
            \end{theorem-rect}
            \begin{proof} Proof of (A)$\Rightarrow$(B): Let $\cB$ be a $(\Sigma,\B)$-wta such that $r= \runsem{\cB}$. By Theorem \ref{thm:bi-loc-finite-rec-step-function}(A)$\Rightarrow$(B),  and Theorem \ref{thm:crisp-det-algebra}, $r$ is a recognizable step mapping and we can construct $n \in \mathbb{N}_+$, $b_1,\ldots,b_n \in L$, and $\Sigma$-fta $A_1,\ldots,A_n$ such that $r = \bigoplus_{i \in [n]} b_i \cdot \chi(\LL(A_i))$.   By Lemma~\ref{lm:rec-step-formula-lemma},  we can construct sentences $\varphi_1,\ldots,\varphi_n$ in $\MSO(\Sigma)$ such that $r = \sem{(\varphi_1 \rhd \langle b_1\rangle) + \ldots + (\varphi_n \rhd \langle b_n\rangle)}$. Then we let $e = (\varphi_1 \rhd \langle b_1\rangle) + \ldots + (\varphi_n \rhd \langle b_n\rangle)$.

              \
              
                  Proof of (B)$\Rightarrow$(C): Since each $(\Sigma,\B)$-recognizable step formula is in $\MSO(\Sigma,\B)$, this is trivial.

                  \
                  
               Proof of (C)$\Rightarrow$(D): Since $\MSO(\Sigma,\B) \subset \MSOe(\Sigma,\B)$, this is trivial.

               \
               
               Proof of  (D)$\Rightarrow$(A): We follow the proof of  Theorem  \ref{thm:MSO-wta} and extend it appropriately. Formally, by induction  on $(\MSOe(\Sigma,\B),\succ_{\MSOe(\Sigma,\B)})$, we prove  the following statement.
                 \begin{eqnarray}
                 \begin{aligned}
                 &\text{For every $e\in\MSOe(\Sigma,\B)$ and $\cV = \Free(e)$,}\\
                 &\text{we can construct a $(\Sigma_\cV,\B)$-wta $\cB$ such that   $\runsem{\cB} = \sem{e}$.}
                 \end{aligned}
               \end{eqnarray}

               I.B.: The proof for the case that  $e = \mathrm{H}(\kappa)$ is the same as in the proof of   Theorem  \ref{thm:MSO-wta}.

               I.S.: The proofs for the cases that  $e = (\varphi \rhd e')$, $e = e_1 + e_2$,  $e = \bigplus_x e'$, or $e = \bigplus_X e'$ are the same as in the proof of   Theorem  \ref{thm:MSO-wta} (except that for $e_1 + e_2$ we use Lemma \ref{lm:consistency-wta-MSOe-2} instead of Lemma~\ref{lm:consistency-wta}).     

     Let $e = e_1 \times e_2$. Let $\cV_1=\Free(e_1)$ and $\cV_2= \Free(e_2)$. By I.H.  we can construct a $(\Sigma_{\cV_1},\B)$-wta $\cA_1$ such that $\runsem{\cA_1} = \sem{e_1}$. By Lemma \ref{lm:consistency-wta-MSOe-2}, we can construct a $(\Sigma_{\cV},\B)$-wta $\cA_1'$ such that $\runsem{\cA_1'} = \sem{e_1}_\cV$.
By Theorem \ref{thm:bi-loc-finite-rec-step-function}(A)$\Rightarrow$(B), we can construct a crisp-deterministic  $(\Sigma_{\cV},\B)$-wta $\cA_1''$ such that $\sem{\cA_1''} = \sem{e_1}_\cV$. In the same way we can construct 
a crisp-deterministic  $(\Sigma_{\cV},\B)$-wta $\cA_2''$ such that $\sem{\cA_2''} = \sem{e_2}_\cV$.
By Theorem \ref{thm:closure-Hadamard-product}(3), we can construct a crisp-deterministic $(\Sigma_\cV,\B)$-wta $\cB$ such that $\sem{\cB} = \sem{\cA_1''} \otimes \sem{\cA_2''}$. Thus $\sem{\cB} = \sem{e_1 \times e_2}$.

Let $e = \bigtimes_x e'$. By I.H. and Lemma  \ref{lm:rec-closed-under-univ-first-quant-2}, we can construct a 
$(\Sigma_\cV,\B)$-wta $\cB$ such that $\runsem{\cB} = \sem{e}$.

Let $e = \bigtimes_X e'$. By I.H. and Lemma \ref{lm:rec-closed-under-univ-second-quant-cbl}, we can construct a $(\Sigma_\cV,\B)$-wta $\cB$ such that $\runsem{\cB} = \sem{e}$.
          \end{proof}

\section[Relationship between decompositions and B-E-T's theorem]{Relationship between a decomposition theorem of wta and  B-E-T's theorem}
\label{sec:relationship-decomp-rec-implies-def}

In this final section we want to show a strong relationship between the following two results which we have proved already:
\begin{enumerate}
\item[(1)] the decomposition of a $(\Sigma,\B)$-wta $\cA$ (cf. Theorem \ref{thm:decomposition-1}(A) $\Rightarrow$ (B), roughly):
  we can construct $\tau$, $(G,H)$, $\Theta$, and $\kappa$ such that, for each $\xi \in \T_\Sigma$: \ \
  $\runsem{\cA}(\xi) = \h_{\M(\Theta,\kappa)}(\tau^{-1}(\xi) \cap \LL(G,H))$,

\item[(2)] the definability of $\runsem{\cA}$ for each $(\Sigma,\B)$-wta $\cA$ (cf. Theorem \ref{thm:wta-MSO}): we can construct a sentence $e \in \MSO(\Sigma,\B)$ such that $\runsem{\cA} = \sem{e}$.
  \end{enumerate}
 We will show that (1) can be used to prove (2) (cf. Theorem \ref{thm:Buechi-2}) and vice versa (cf. Theorem \ref{thm:decomposition-1-alternative}). 
 Thus Theorem \ref{thm:Buechi-2} can be seen  as an alternative proof of ``r-recognizable implies definable'' (cf. Subsection \ref{subsec:alternative-recog-implies-def}); also,  Theorem \ref{thm:decomposition-1-alternative} can be seen as an alternative proof of the decomposition result Theorem \ref{thm:decomposition-1}(A) $\Rightarrow$ (B) (cf. Subsection  \ref{subsec:alternative-decomposition}).

This section elaborates the ideas of \cite[Sec.~7.5]{hervogdro19} for the particular case of the trivial storage type and strong bimonoids as weight algebras.

          \subsection{Alternative proof of ``r-recognizable implies definable''}
          \label{subsec:alternative-recog-implies-def}

          We start with an auxiliary lemma which states that  the set of definable weighted tree languages is closed under  tree relabelings. Actually, we have proved this result already: each definable weighted tree language is r-recognizable (by Theorem \ref{thm:MSO-wta}), r-recognizable weighted tree languages are closed under tree relabelings (cf. Theorem \ref{thm:closure-under-tree-relabeling}), and each r-recognizable weighted tree language is definable (by Theorem  \ref{thm:wta-MSO}). But we do \underline{not} want to use Theorem  \ref{thm:wta-MSO} for this closure result. So we show a proof with a direct construction. 

          Essentially the next lemma has been proved in \cite[Lm.~15]{her17} and \cite[Lm.~5.3.2]{her20}. We follow this proof, but we mention that the notion of ``definable'' in \cite{her17} and \cite{her20} refers to the weighted logic in \cite{drogas05,drovog06}, and this is different from the one in the present chapter. 

          \begin{lemma}\rm \label{lm:closure-under-tree-relab-alternative} Let $\Theta$ be a ranked alphabet, $e \in \MSO(\Theta,\B)$ be a sentence, and $\tau = (\tau_k \mid k \in \mathbb{N})$ be a $(\Theta,\Sigma)$-tree relabeling. Then we can construct a sentence $f \in \MSO(\Sigma,\B)$ such that $\sem{f} = \chi(\tau)(\sem{e})$.
          \end{lemma}
          \begin{proof} Let $(\theta_1,\ldots,\theta_n)$ be an arbitrary, but fixed enumeration of $\Theta$. For each $i\in [n]$, we let $k_i = \rk_\Theta(\theta_i)$.
            Moreover, we let $\cV = \{X_\theta \mid \theta \in \Theta\}$ be a set of second-order variables. The idea is
            \begin{compactitem}
            \item to code, for each $\zeta \in \T_\Sigma$, each preimage $\xi \in \tau^{-1}(\zeta)$ as a tree in $\T_{\Sigma_\cV}$ where the second order variable $X_\theta$ represents the symbol $\theta \in \Theta$, and
            \item to evaluate $e'$ on $\xi$ where $e'$ is obtained from $e$ by replacing each subformula of the form $\llabel_\theta(x)$ by $(x \in X_\theta)$.
              \end{compactitem}
              
              Formally, we construct the  $\MSO(\Sigma,\B)$-formula
              \[
f = \Big(\bigplus\nolimits_{X_{\theta_1}} \cdots \bigplus\nolimits_{X_{\theta_n}} (\psi_{\mathrm{part}} \wedge \psi_{\mathrm{relab}}) \rhd e' \Big)
                \]
                where
                \begin{compactitem}
                \item $\psi_{\mathrm{part}} = \forall x. \Big( \bigvee_{i\in [n]} \big( (x \in X_{\theta_i}) \wedge \bigwedge_{\substack{j \in [n]:\\i\ne j}} \neg(x \in X_{\theta_j})\big)\Big)$,

                \item $\psi_{\mathrm{relab}} = \forall x. \Big( \bigwedge_{i\in [n]}
                  \big( \neg(x \in X_{\theta_i}) \vee \llabel_{\tau_{k_i}(\theta_i)}(x) \big)\Big)$, and

                  \item $e'$ is obtained from $e$ by replacing each subformula of the form $\llabel_\theta(x)$ by $(x \in X_\theta)$.
                  \end{compactitem}
                  We note that $\psi_{\mathrm{part}}$ and   $\psi_{\mathrm{relab}}$ are in $\MSO(\Sigma)$ and $e'$ is in $\MSO(\Sigma,\B)$. Moreover, $\Free(\psi_{\mathrm{part}})=\Free(\psi_{\mathrm{relab}})= \cV$ and $\Free(e')\subseteq \cV$, hence the formula $f$ is a sentence.                  

                  \
                  
                  First we analyse the semantics of $\psi_{\mathrm{part}}$ and  $\psi_{\mathrm{relab}}$. 
                  The following two statements are obvious:
                  \begin{eqnarray}
                    \begin{aligned}
                    &\text{for every $\zeta \in \T_\Sigma$ and $\eta \in \Phi_{\zeta,\cV}$}: (\zeta,\eta) \models \psi_{\mathrm{part}} \ \ \text{ iff }\\
                    &\ \ \{\eta(X_{\theta_1}),\ldots,\eta(X_{\theta_n})\} \text{ is a partitioning of $\pos(\zeta)$}
                    \end{aligned} \label{eq:partitioning}
                  \end{eqnarray}
                  and
                  \begin{eqnarray}
                    \begin{aligned}
                    &\text{for every $\zeta \in \T_\Sigma$ and $\eta \in \Phi_{\zeta,\cV}$}: (\zeta,\eta) \models \psi_{\mathrm{relab}} \ \ \text{ iff }\\
                    &\ \ (\forall w \in \pos(\zeta), i \in [n]): (w \in \eta(X_{\theta_i})) \to (\zeta(w) \in \tau_{k_i}(\theta_i)) \enspace.
                    \end{aligned} \label{eq:relabeling}
  \end{eqnarray}

In the following, for every $\zeta \in \T_\Sigma$ and $U_1,\ldots,U_n \subseteq \pos(\zeta)$, we abbreviate the $\cV$-assignment $[X_{\theta_1}\mapsto U_1,\ldots,X_{\theta_n}\mapsto U_n]$ for $\zeta$ by $[X_{\theta_i}\mapsto U_i \mid i\in [n]]$. Then, putting \eqref{eq:partitioning} and \eqref{eq:relabeling} together, we obtain the statement:
\begin{eqnarray}
  \begin{aligned}
                    &\text{for every $\zeta \in \T_\Sigma$ and $\eta \in \Phi_{\zeta,\cV}$}: (\zeta,\eta) \models (\psi_{\mathrm{part}}  \wedge \psi_{\mathrm{relab}}) \ \ \text{ iff } \\
                    &\ \ (\exists \xi \in \T_\Theta) : (\zeta \in \tau(\xi)) \wedge (\eta = [X_{\theta_i}\mapsto \pos_{\theta_i}(\xi)\mid i\in[n]])\enspace.
                    \end{aligned}\label{eq:image-relab}
                     \end{eqnarray}
We note that $\pos(\zeta)=\pos(\xi)$.                   
                                          Moreover, by construction of $e'$, we have
                       \begin{eqnarray}
                    \text{for every $\zeta \in \T_\Sigma$ and $\xi \in \tau^{-1}(\zeta)$}: \sem{e}(\xi) = \sem{e'}(\zeta,[X_{\theta_i}\mapsto \pos_{\theta_i}(\xi)\mid i\in[n]]) \label{eq:simulation-tau} \enspace.                    \end{eqnarray}
                 
                  Now we prove that $\sem{f} = \chi(\tau)(r)$.
                                    Let $\zeta \in \T_\Sigma$. Then
\begingroup
\allowdisplaybreaks
\begin{align*}
  &\sem{f}(\zeta) \\
  &=  \sem{\bigplus\nolimits_{X_{\theta_1}} \cdots \bigplus\nolimits_{X_{\theta_n}} (\psi_{\mathrm{part}} \wedge \psi_{\mathrm{relab}}) \rhd e'}(\zeta)\\
  &=  \bigoplus_{W_1,\ldots,W_n \subseteq \pos(\zeta)} \sem{(\psi_{\mathrm{part}} \wedge \psi_{\mathrm{relab}}) \rhd e'}(\zeta,[X_{\theta_i}\mapsto W_i \mid i\in [n]])\\
  &=  \begin{cases}
    \bigoplus_{W_1,\ldots,W_n \subseteq \pos(\zeta)} 
    \sem{e'}(\zeta,[X_{\theta_i}\mapsto W_i \mid i\in [n]]) & \text{ if $(\zeta,[X_{\theta_i}\mapsto W_i \mid i\in [n]]) \models (\psi_{\mathrm{part}} \wedge \psi_{\mathrm{relab}})$}\\
    \0 & \text{ otherwise}
  \end{cases}
  \\
  &=  \bigoplus_{\xi \in \tau^{-1}(\zeta)} \sem{e'}(\zeta,[X_{\theta_i}\mapsto \pos_{\theta_i}(\xi)\mid i\in[n]])
  \tag{by \eqref{eq:image-relab}}\\
  &= \bigoplus_{\xi \in \tau^{-1}(\zeta)} \sem{e}(\xi)
    \tag{by \eqref{eq:simulation-tau}}\\
  &= \chi(\tau)(\sem{e})(\zeta) \tag{by \eqref{obs:app-tree-transf-to-wtl-2} and the fact that $\tau^{-1}(\zeta)$ is finite}
\end{align*}
\endgroup
\end{proof}

              Now we can show the alternative proof of the fact that r-recognizable implies definable. 
              
  \begin{theorem}\label{thm:Buechi-2} For each $(\Sigma,\B)$-wta $\cA$, we can construct a sentence $f \in \MSO(\Sigma,\B)$ such that $\sem{f}=\runsem{\cA}$.
  \end{theorem}
  \begin{proof}  By Theorem \ref{thm:decomposition-1}(A) $\Rightarrow$ (B), we can construct
a ranked alphabet $\Theta$, a deterministic $(\Theta,\Sigma)$-tree relabeling $\tau$,  a $\Theta$-local system $(K,H)$, and
    a family $\kappa=(\kappa_k \mid k\in\mathbb{N})$ of mappings $\kappa_k:~\Theta^{(k)}~\to~B$ 
    such that, for each $\xi \in \T_\Sigma$, we have $\runsem{\cA}(\xi) = \h_{\M(\Theta,\kappa)}(\tau^{-1}(\xi) \cap \LL(K,H))$.
  By Observation \ref{obs:inside-out} we have
  \begin{equation}\label{eq:wta-decomp-repl}
    \runsem{\cA}(\xi) = \chi(\tau)(\chi(\LL(K,H)) \otimes \h_{\M(\Theta,\kappa)}) \enspace.
  \end{equation}
  
  By Corollary \ref{cor:loc-is-recog}, we can construct a bu-deterministic $\Sigma$-fta $A$ such that $\LL(K,H)=\LL(A)$.
  By Lemma \ref{lm:thawri-fta-to-MSO}, we can construct a sentence $\varphi \in \MSO(\Theta)$ such that $\LL(A)=\LL(\varphi)$.
  Then
  \begin{align*}
    \chi(\LL(K,H)) \otimes \h_{\M(\Theta,\kappa)}
    &= \chi(\LL(A)) \otimes \h_{\M(\Theta,\kappa)}
      = \chi(\LL(\varphi)) \otimes \h_{\M(\Theta,\kappa)}\\
    &=\chi(\LL(\varphi)) \otimes \sem{\rmH(\kappa)} \tag{because $\rmH(\kappa)$ is a sentence}\\
                                    &= \sem{\varphi \rhd \rmH(\kappa)}\enspace.   \tag{by \eqref{eq:guarded-e}}
    \end{align*}

    By Lemma \ref{lm:closure-under-tree-relab-alternative}, we can construct a sentence $f \in \MSO(\Sigma,\B)$ such that $\sem{f} = \chi(\tau)(\sem{\varphi \rhd \rmH(\kappa)})$. Finally, by \eqref{eq:wta-decomp-repl}, we obtain $\sem{f} = \runsem{\cA}$. 
    \end{proof}

          \subsection{Alternative proof of the  decomposition Theorem \ref{thm:decomposition-1}(A) $\Rightarrow$ (B) }
          \label{subsec:alternative-decomposition}

          The alternative proof uses the fact that, for each $(\Sigma,\B)$-wta $\cA$, we can construct a sentence $e \in \MSO(\Sigma,\B)$ such that $\runsem{\cA}=\sem{e}$  (cf. Theorem \ref{thm:wta-MSO}) and two easy technical lemmas. 
The first  lemma claims that the set of models of the $\MSO(\Sigma)$-formula $\varphi$  in \eqref{eq:formula-phi} is a local tree language.

\begin{lemma}\rm \label{lm:varphiwta-local} Let $\cA=(Q,\delta,F)$ be a root weight normalized  $(\Sigma,\B)$-wta with $\supp(F)=\{q_f\}$  and   $\cU = \bigcup(Q^k \times Q \mid k \in \mathbb{N} \text{ such that } \rk^{-1}(k) \not= \emptyset)$. Moreover, let $\varphi$ be the $\MSO(\Sigma)$-formula $\varphi$  in \eqref{eq:formula-phi}. Then we can construct a $\Sigma_\cU$-local system $(G,H)$ such that  $\LL(G,H) = \LL_\cU(\varphi)$.
    \end{lemma}
    \begin{proof} We define the $\Sigma_\cU$-local system $(G,H)$ such that $G$ is the set of all forks 
      \[
\Big((\sigma_1,\{(w_1,q_1)\}) \cdots (\sigma_k,\{(w_k,q_k)\}), \ (\sigma,\{(w,q)\}) \Big) 
\]
in $\mathrm{Fork}(\Sigma_\cU)$ such that
\begin{compactitem}
\item $\sigma\in \Sigma^{(k)}$ and $\sigma_1,\ldots,\sigma_k \in \Sigma$ for some $k\in\mathbb{N}$,
\item $q,q_1,\ldots,q_k\in Q$, and
\item $w= q_1 \cdots q_k$ and for each $i \in [k]$ we have $w_i \in Q^{\rk(\sigma_i)}$.
  \end{compactitem}
Moreover, $H = \{(\sigma,\{(w,q_f)\}) \mid  k \in \mathbb{N}, \sigma \in \Sigma^{(k)}, w \in Q^k\}$.
Then it is easy to see that $\LL(G,H) = \LL_\cU(\varphi)$.
\end{proof}

Now we can give an alternative proof of Theorem \ref{thm:decomposition-1}(A)$\Rightarrow$(B).

    \begin{theorem}\label{thm:decomposition-1-alternative}  Let $\cA$ be a $(\Sigma,\B)$-wta. Then we can construct
    \begin{compactitem}
    \item a ranked alphabet $\Theta$,
    \item a deterministic $(\Theta,\Sigma)$-tree relabeling $\tau$,
    \item a $\Theta$-local system $(G,H)$, and
    \item a family $\kappa=(\kappa_k \mid k\in\mathbb{N})$ of mappings $\kappa_k:~\Theta^{(k)}~\to~B$ \end{compactitem}
    such that,  for each $\xi \in \T_\Sigma$, the following holds: $\runsem{\cA}(\xi) = \h_{\M(\Theta,\kappa)}(\tau^{-1}(\xi) \cap \LL(G,H))$. 
  \end{theorem}
  
  \begin{proof} Let $\cA=(Q,\delta,F)$. By Theorem \ref{thm:root-weight-normalization-run} we can assume that $\cA$ is root weight normalized, i.e.,  $\supp(F)$ contains exactly one element $q_f$ and $F(q_f)=\1$.
    
   We define $\cU = \bigcup(Q^k \times Q \mid k \in \mathbb{N} \text{ such that } \rk^{-1}(k) \not= \emptyset)$ and $\Theta = \Sigma_\cU$.
   We recall the $\MSO(\Sigma,\B)$-formula defined in \eqref{eq:main-MSO-formula}: 
\[e = \bigplus\nolimits_{X_1} \cdots \bigplus\nolimits_{X_n} (\varphi \rhd \mathrm{H}(\kappa)) 
\]
and $\kappa = (\kappa_k \mid k \in \mathbb{N})$ with $\kappa_k:(\Sigma_\cU)^{(k)} \to B$.
By the proof of Theorem \ref{thm:wta-MSO}, we have  $\runsem{\cA} = \sem{e}$.

For each $i \in [n]$, we consider the subformula
\[
  e_i = \bigplus\nolimits_{X_i} \cdots \bigplus\nolimits_{X_n} (\varphi \rhd \mathrm{H}(\kappa))\enspace.
\]
Obviously, $\Free(e_i)= \{X_1,\ldots,X_{i-1}\}$ and $X_i \not\in \{X_1,\ldots,X_{i-1}\}$. Thus we can apply Lemma \ref{lm:existential-quant=det-tree-relab}(2) (with $\cV=\Free(e_i)$) and thereby construct the deterministic $(\Sigma_{\{X_1,\ldots,X_i\}},\Sigma_{\{X_1,\ldots,X_{i-1}\}})$-tree relabeling $\tau_i$ such that 
\[
  \sem{\bigplus\nolimits_{X_i} \cdots \bigplus\nolimits_{X_n} (\varphi \rhd \mathrm{H}(\kappa))}_{\{X_1,\ldots,X_{i-1}\}} = \chi(\tau_i)\Big( \sem{\bigplus\nolimits_{X_{i+1}} \cdots \bigplus\nolimits_{X_n} (\varphi \rhd \mathrm{H}(\kappa))}_{\{X_1,\ldots,X_{i}\}}\Big)\enspace.
\]

Hence, by the $n$-fold application of Lemma \ref{lm:existential-quant=det-tree-relab}(2), we can construct deterministic tree relabelings $\tau_1,\ldots,\tau_n$ such that 
           \begingroup
          \allowdisplaybreaks
          \begin{align*}
            \sem{e} = \sem{\bigplus\nolimits_{X_1} \cdots \bigplus\nolimits_{X_n} (\varphi \rhd \mathrm{H}(\kappa))}
            = \chi(\tau_1)\Big( \ldots \chi(\tau_n)(\sem{\varphi \rhd \mathrm{H}(\kappa)}_\cU) \ldots \Big) \enspace.
          \end{align*}
          \endgroup
Then we construct the deterministic tree relabeling  $\tau$ such that $\tau=\tau_1 \hat{\circ} \ldots \hat{\circ} \tau_n$ (cf. Section \ref{sect:trees}) and, by Theorem \ref{thm:relabeling-synt-comp}, we have
          \begin{align*}
            \chi(\tau_1)\Big( \ldots \chi(\tau_n)(\sem{\varphi \rhd \mathrm{H}(\kappa)}_\cU) \ldots \Big)
            = \chi(\tau)\Big( \sem{\varphi \rhd \mathrm{H}(\kappa)}_\cU \Big) \enspace.
          \end{align*}
          
          Now let $\xi \in \T_\Sigma$. We can calculate as follows (where we abbreviate  the $\Theta$-algebra homomorphism $\h_{\M(\Theta,\kappa)}$ by $\h_\kappa$).
          
                                                \begingroup
          \allowdisplaybreaks
          \begin{align*}
            \chi(\tau)\Big( \sem{\varphi \rhd \mathrm{H}(\kappa)}_\cU \Big)(\xi)
            &  = \chi(\tau)\big(\chi(\LL_\cU(\varphi))\otimes \sem{\mathrm{H}(\kappa)}_\cU \big)(\xi)
              \tag{by \eqref{eq:guarded-e}}\\
            &  = \chi(\tau)\big(\chi(\LL_\cU(\varphi))\otimes \h_{\kappa[\cU\leadsto \cU]}  \otimes \chi(\T_{\Sigma_\cU}^\mathrm{v}) \big)(\xi)  \tag{by  definition of $\sem{\mathrm{H}(\kappa)}_\cU$}\\
            &  = \chi(\tau)\big(\chi(\LL_\cU(\varphi))\otimes \h_\kappa  \otimes \chi(\T_{\Sigma_\cU}^\mathrm{v}) \big)(\xi)  \tag{because $\h_{\kappa[\cU\leadsto \cU]} = \h_\kappa$}\\
            &  = \chi(\tau)\big(\chi(\LL_\cU(\varphi))\otimes \h_\kappa \big)(\xi)
            \tag{because $\LL_\cU(\varphi) \subseteq \T_{\Sigma_\cU}^\mathrm{v}$}\\
            & = \h_\kappa(\tau^{-1}(\xi) \cap \LL_\cU(\varphi)) \tag{by Observation \ref{obs:inside-out}}\enspace.
          \end{align*}
          \endgroup   
By Lemma \ref{lm:varphiwta-local}, we can construct a $\Sigma_\cU$-local system $(G,H)$ such that $\LL_\cU(\varphi) = \LL(G,H)$. Hence we obtain \(\runsem{\cA}(\xi) = \h_\kappa(\tau^{-1}(\xi) \cap \LL(G,H)) \).
          \end{proof}

%% file: Myhill-Nerode7.tex
\chapter{Syntactic congruences and B-A's theorem}
\label{ch:Bozapalidis}

For string languages, the Myhill-Nerode theorem characterizes the set of recognizable languages in terms of index-finiteness of their syntactic congruences \cite{myh57}, \cite{ner58} (cf. also \cite{hopull79, koz97}). This theorem has been generalized to recognizable tree languages \cite{magmor70,ste92,koz92} and \cite[Thm.~2.7.1]{gecste97}. In this chapter we show three algebraic characterizations of the set of (bu-deterministically) recognizable weighted tree languages.

In Section~\ref{sec:B-A-BLs-theorem-section} we recall a characterization of the set of recognizable weighted tree languages
over fields in terms of  finite dimensionality of vector spaces which result from factorizing the vector space of polynomial weighted tree languages by syntactic congruences of weighted tree languages (cf. Theorem~\ref{th:MN-fields}). Originally, this result was proved  in \cite[Prop.~2]{bozale89} where recognizability was defined in terms of multilinear representations; due to Theorem~\ref{thm:lin-iff-wta-extended}, this refers to the same set of weighted tree languages. We call this result the Bozapalidis-Alexandrakis-theorem, for short: B-A's theorem.

In Section~\ref{sec:characterization-rec-by-left-quotients}, we recall the characterization of  the set of recognizable weighted tree languages over fields
in terms of finite dimensionality of vector spaces of their left quotients \cite[Thm.~2.1]{bozlou83}  and in terms of finite dimensionality of vector spaces of their right quotients \cite[Thm.~2.1 and~3.1]{bozlou83} (cf. Theorem~\ref{thm:left-quotient-vector-space}). We note that, also in \cite{bozlou83}, recognizability was defined in terms of multilinear representations (cf. Theorem~\ref{thm:lin-iff-wta-extended}).

Finally, in Section~\ref{sec:B-A-BLs-theorem-for-the-det-case}, we prove a characterization of the set of weighted tree languages which are recognizable by bu-deterministic wta over commutative semifields \cite[Thm.~7.3]{fulvog25}; this characterization follows the idea of B-A's theorem but it uses finitely generated and cancellative $\B$-scalar algebras instead of finite-dimensional $\B$-vector spaces (cf. Theorem~\ref{thm:MN-semifield-det}). We compare this result with the characterization result in \cite{bor03} (cf. Corollary~\ref{lm:comparison-Bjorn-result-det} and Lemma~\ref{lm:comparison-degree-index}).

\section{B-A's theorem for the general case}
\label{sec:B-A-BLs-theorem-section}

In this section we recall B-A's theorem (cf. Theorem~\ref{th:MN-fields}). Its proof uses the fact that the $(\Sigma,\B)$-semimodule of polynomial weighted tree languages is initial in the set of all $(\Sigma,\B)$-semimodules (cf. Theorem~\ref{lm:Pol-SigmaB-initial}).

\subsection{The $(\Sigma,\B)$-semimodule of polynomial weighted tree languages}
\label{sec:SigmaB-vector-space-of-polynomials}

\label{page:semimodule-polynomial-wtl}

\begin{quote}\emph{In this subsection, we let $\B=(B,\oplus,\otimes,\0,\1)$ be an arbitrary commutative semiring, unless specified otherwise.}
\end{quote}

Here we define the $(\Sigma,\B)$-semimodule of polynomial $(\Sigma,\B)$-weighted tree languages, for short: $(\Sigma,\B)$-polynomials (cf. Subsection~\ref{sec:weighted-tree-languages}). We will prove that it is initial in the set of all $(\Sigma,\B)$-semimodules. 

\index{Pol(Sigma,B)@$\mathsf{Pol}(\Sigma,\B)$}
Formally, we start from  the $\B$-semimodule $\mathsf{Pol}(\Sigma,\B)$ defined by
\[
  \mathsf{Pol}(\Sigma,\B) = (\Pol(\Sigma,\B),\oplus,\widetilde{\0})
\]
of $(\Sigma,\B)$-polynomials via the scalar multiplication $\cdot$; the latter is defined such that, for every $b \in B$ and $s \in \Pol(\Sigma,\B)$, we let $(b \cdot s)(\xi) = b \otimes s(\xi)$ for each $\xi \in \T_\Sigma$. Moreover, we recall that the sum $s_1 \oplus s_2$ of two polynomials $s_1,s_2 \in \Pol(\Sigma,\B)$ is defined by $(s_1 \oplus s_2)(\xi) = s_1(\xi) \oplus s_2(\xi)$ for each
$\xi \in \T_\Sigma$. If $\B$ is a field, then $\mathsf{Pol}(\Sigma,\B)$ is a $\B$-vector space with basis $\{\1.\xi \mid \xi \in \T_\Sigma\}$.

The next theorem was shown in \cite[p.~451]{bozale89} and \cite[p.~352]{boz91} for fields.

\index{oneTSigma@$\1.\T_\Sigma$}
\begin{theorem-rect}\rm \label{lm:Pol-SigmaB-free}  Let $\Sigma$ be a ranked alphabet and $\B$ be a commutative semiring. The $\B$-semimodule $\mathsf{Pol}(\Sigma,\B)$ is free in the set of all $\B$-semimodules with generating set $\1.\T_\Sigma=\{\1.\xi \mid \xi \in \T_\Sigma\}$.
\end{theorem-rect}
\begin{proof} First we recall from page \pageref{page:B-semimodule-as-universal-algebra} that $\sfPol(\Sigma,\B)=(\Pol(\Sigma,\B),\oplus,\widetilde{\0})$ is the algebra $(\Pol(\Sigma,\B),\eta_{\mathrm{Pol}})$ where
\begin{compactitem}
\item $I= \{\oplus,\widetilde{\0}\} \cup \{(b\cdot)\mid b \in B\}$,
  \item $\eta_{\mathrm{Pol}}: I \to \mathrm{Ops}(\Pol(\Sigma,\B))$ such that
\begin{compactitem}

\item $\eta_{\mathrm{Pol}}(\oplus) = \oplus$ and $\eta_{\mathrm{Pol}}(\widetilde{\0})() = \widetilde{\0}$ and
  \item for every $b \in B$ and $r \in \Pol(\Sigma,\B)$, we let $\eta_{\mathrm{Pol}}((b \cdot))(r) = b \cdot r$.
    \end{compactitem}
  \end{compactitem}

Now we prove the three properties (a), (b), and (c) of the definition of free algebra on page~\pageref{page:initial-algebra}.
    
(a) Clearly, $\mathsf{Pol}(\Sigma,\B)$ is a member of the set of all $\B$-semimodules.

    (b) It is obvious that the algebra $(\Pol(\Sigma,\B),\eta_{\mathrm{Pol}})$ is generated by $\1.\T_\Sigma$, i.e., $\langle \1.\T_\Sigma \rangle_{\im(\eta_{\mathrm{Pol}})} = \Pol(\Sigma,\B)$.  

(c) Let $(V,+,0)$ be an arbitrary $\B$-semimodule and $f: \1.\T_\Sigma \to V$ be a mapping. We will prove that there exists a $\B$-semimodule homomorphism from $\sfPol(\Sigma,\B)$ to $(V,+,0)$ which extends $f$.

For this, we define the mapping $h:\Pol(\Sigma,\B) \to V$ such that, for each $n\in \mathbb{N}$, $b_1,\ldots,b_n \in B$, and $\xi_1,\ldots,\xi_n\in \T_\Sigma$, we let
\begin{equation}\label{eq:from-h-to-hV-free}
  h(b_1.\xi_1 \oplus\ldots\oplus b_n.\xi_n) = b_1\cdot f(\1.\xi_1) +\ldots +  b_n\cdot f(\1.\xi_n)\enspace.
  \end{equation}
Clearly, $h$ extends $f$.
Now we prove that $h$ is a $\B$-semimodule homomorphism, i.e., a linear mapping, from $\mathsf{Pol}(\Sigma,\B)$ to $(V,+,0)$.  
For the proof, we let $d_1,d_2 \in B$ and  $s_1,s_2 \in \Pol(\Sigma,\B)$.
By using \eqref{SM1}, \eqref{SM2}, and the definition of $h$, we can easily prove that $h(d_1 \cdot s_1 \oplus d_2 \cdot s_2)= d_1 \cdot h(s_1) \oplus d_2 \cdot h(s_2)$. Thus $h$ is a linear mapping.
\end{proof}

Next we are going to combine $\mathsf{Pol}(\Sigma,\B)$ and $\ttop_\Sigma$ where, for each $\sigma \in \Sigma$, the operation $\ttop_\Sigma(\sigma)$ is the restriction of the top-concatenation $\ttop_\sigma$ (cf. Section \ref{sec:top-concatenation}) to polynomials.
\footnote{In fact, the notation $\ttop_\Sigma$ is overloaded because it is used already in the $\Sigma$-term algebra $\sfT_\Sigma=(\T_\Sigma,\ttop_\Sigma)$. However, since the present use of $\ttop_\Sigma$  straightforwardly generalizes its former use, we keep the notation $\ttop_\Sigma$. The type of $\ttop_\Sigma$ will always be clear from the context.} Thus, the type of $\ttop_\Sigma(\sigma)$ is
\[
\ttop_\Sigma(\sigma): \Pol(\Sigma,\B)^k \to B^{\T_\Sigma}\enspace.
\]
  This combination of $\mathsf{Pol}(\Sigma,\B)$ and $\ttop_\Sigma$ is a $(\Sigma,\B)$-semimodule if
  \begin{compactenum}
  \item[(a)]  $(\Pol(\Sigma,\B),\ttop_\Sigma)$ is a $\Sigma$-algebra, i.e.,  for each $\sigma \in \Sigma$: $\im(\ttop_\Sigma(\sigma)) \subseteq \mathsf{Pol}(\Sigma,\B)$  and
  \item[(b)] for each $\sigma \in \Sigma$, the operation $\ttop_\Sigma(\sigma)$ is multilinear.
  \end{compactenum}
 
 In the next two lemmas and corollary we prove that these requirements are indeed satisfied.

\begin{lemma}\rm \label{polynomials-closed-under-top(sigma)} Let $k\in \mathbb{N}$,  $\sigma \in \Sigma^{(k)}$, and $s_1,\ldots,s_k\in \Pol(\Sigma,\B)$ such that,  for each $i\in [k]$ there exist $n_i\in\mathbb{N}$, $b_{i1},\ldots, b_{in_i}\in B$, and $\xi_{i1},\ldots, \xi_{in_i}\in \T_\Sigma$ with 
$s_i=b_{i1}.\xi_{i1} \oplus\cdots\oplus b_{in_i}.\xi_{in_i}$. Then we have
\[ \ttop_\Sigma(\sigma)(s_1,\ldots,s_k)=\bigoplus_{j_1\in[n_1],\ldots,j_k\in[n_k]} (b_{1j_1}\otimes\ldots\otimes b_{kj_k}).\sigma(\xi_{1j_1},\ldots,\xi_{kj_k})\enspace.\]
Thus, in particular, $\ttop_\Sigma(\sigma)$ is an operation on $\Pol(\Sigma,\B)$.
\end{lemma}
\begin{proof} Let $\xi \in \T_\Sigma$. Then we can calculate as follows.
\begingroup
  \allowdisplaybreaks
  \begin{align*}
    & \ttop_\Sigma(\sigma)(s_1,\ldots,s_k)(\xi)\\
    &= \begin{cases}
      s_1(\xi_1) \otimes \ldots \otimes s_k(\xi_k) & \text{ if $\xi = \sigma(\xi_1,\ldots,\xi_k)$}\\
      \0 & \text{ otherwise} 
    \end{cases}
    \tag{by the definition of $\ttop_\Sigma(\sigma)$} \\
    & = \begin{cases}
        b_{1j_1}\otimes\ldots\otimes b_{kj_k} & \text{ if $(\exists j_1\in[n_1]): \xi_1=\xi_{1j_1},\ldots,\text{ and }(\exists j_k\in[n_k]): \xi_k=\xi_{kj_k}$}\\
      \0 & \text{ otherwise} 
    \end{cases} \\
    & = \bigoplus_{j_1\in[n_1],\ldots,j_k\in[n_k]} \big((b_{1j_1}\otimes\ldots\otimes b_{kj_k}).\sigma(\xi_{1j_1},\ldots,\xi_{kj_k})\big)(\xi)\\
    & = \Big(\bigoplus_{j_1\in[n_1],\ldots,j_k\in[n_k]} (b_{1j_1}\otimes\ldots\otimes b_{kj_k}).\sigma(\xi_{1j_1},\ldots,\xi_{kj_k})\Big)(\xi) \qedhere
    \end{align*} 
  \endgroup
\end{proof}

\begin{lemma}\rm \label{top(sigma)-multilinear} For each $\sigma \in \Sigma$, the operation $\ttop_\sigma$  defined in Section \ref{sec:top-concatenation} is multilinear.
\end{lemma}

\begin{proof} Let $k \in \mathbb{N}$ and  $\sigma \in \Sigma^{(k)}$. Moreover, let $s_1,\ldots,s_k,s,s' \in B^{\T_\Sigma}$ and $b,b' \in B$. Also let  $i \in [k]$ and $\xi \in \T_\Sigma$. We proceed by case analysis.

\underline{$\xi(\varepsilon) = \sigma$:} For each $i \in [k]$, we abbreviate $\xi|_i$ by $\xi_i$.  Then we can calculate as follows.
  \begingroup
  \allowdisplaybreaks
  \begin{align*}
    & \ttop_\sigma(s_1,\ldots,s_{i-1},b \cdot s \oplus b' \cdot s',s_{i+1},\ldots,s_k)(\xi)\\[2mm]
    &= s_1(\xi_1) \otimes \ldots \otimes s_{i-1}(\xi_{i-1}) \otimes (b\cdot s \oplus b'\cdot s')(\xi_i) \otimes s_{i+1}(\xi_{i+1}) \otimes \ldots \otimes s_k(\xi_k)
    \tag{by the definition of $\ttop_\sigma$} \\[2mm]
    &= s_1(\xi_1) \otimes \ldots \otimes s_{i-1}(\xi_{i-1}) \otimes \big((b\cdot s)(\xi_i) \oplus
      (b' \cdot s')(\xi_i)\big) \otimes s_{i+1}(\xi_{i+1}) \otimes \ldots \otimes s_k(\xi_k) \\[2mm]
    &= \big(s_1(\xi_1) \otimes \ldots \otimes s_{i-1}(\xi_{i-1}) \otimes (b \cdot s)(\xi_i) 
      \otimes s_{i+1}(\xi_{i+1}) \otimes \ldots \otimes s_k(\xi_k)\big) \ \oplus\\
      &\hspace*{5mm} \big(s_1(\xi_1) \otimes \ldots \otimes s_{i-1}(\xi_{i-1}) \otimes (b' \cdot s')(\xi_i)
      \otimes s_{i+1}(\xi_{i+1}) \otimes \ldots \otimes s_k(\xi_k)\big)
      \tag{by distributivity}\\[2mm]
    &= b \otimes \big(s_1(\xi_1) \otimes \ldots \otimes s_{i-1}(\xi_{i-1}) \otimes s(\xi_i) 
      \otimes s_{i+1}(\xi_{i+1}) \otimes \ldots \otimes s_k(\xi_k)\big) \ \oplus\\
      &\hspace*{5mm} b' \otimes \big(s_1(\xi_1) \otimes \ldots \otimes s_{i-1}(\xi_{i-1}) \otimes s'(\xi_i)
      \otimes s_{i+1}(\xi_{i+1}) \otimes \ldots \otimes s_k(\xi_k)\big) 
           \tag{by associativity and commutativity of $\otimes$}\\[2mm]
        &= b \otimes \ttop_\sigma(s_1,\ldots,s_{i-1},s,s_{i+1},\ldots,s_k)(\xi) \ \oplus \ b' \otimes \ttop_\sigma(s_1,\ldots,s_{i-1},s',s_{i+1},\ldots,s_k)(\xi) \\[2mm]
    & = \Big(b \cdot \ttop_\sigma(s_1,\ldots,s_{i-1},s,s_{i+1},\ldots,s_k) \ \oplus \ b' \cdot \ttop_\sigma(s_1,\ldots,s_{i-1},s',s_{i+1},\ldots,s_k\Big)(\xi)
      \enspace.
  \end{align*}
  \endgroup
  
  \

  \underline{$\xi(\varepsilon) \ne \sigma$:} Then
  \begingroup
  \allowdisplaybreaks
\begin{align*}
  &\ttop_\sigma(s_1,\ldots,s_{i-1},b \cdot s \oplus b' \cdot s',s_{i+1},\ldots,s_k)(\xi)\\
    &=\0 \tag{by definition of $\ttop_\sigma$}\\
    &= b \otimes \0 \oplus b' \otimes \0\\
    & =  b \otimes \Big( \ttop_\sigma(s_1,\ldots,s_{i-1},s,s_{i+1},\ldots,s_k)(\xi)\Big) \ \oplus \ b' \otimes \Big(\ttop_\sigma(s_1,\ldots,s_{i-1},s',s_{i+1},\ldots,s_k\Big(\xi)) \tag{by definition of  $\ttop_\sigma$}\\
    & = \Big(b \cdot \ttop_\sigma(s_1,\ldots,s_{i-1},s,s_{i+1},\ldots,s_k) \ \oplus \ b' \cdot \ttop_\sigma(s_1,\ldots,s_{i-1},s',s_{i+1},\ldots,s_k\Big)(\xi)
      \tag{by the definitions of scalar multiplication and of sum}
\end{align*}
\endgroup
\end{proof}

By Lemmas \ref{polynomials-closed-under-top(sigma)} and \ref{top(sigma)-multilinear} we immediately obtain the following result.

\begin{corollary}\rm \label{cor:top(sigma)-multilinear} For each $\sigma \in \Sigma$, the mapping $\ttop_\Sigma(\sigma)$ is a multilinear operation on $\Pol(\Sigma,\B)$.
  \end{corollary}

By Lemma \ref{polynomials-closed-under-top(sigma)} and Corollary \ref{cor:top(sigma)-multilinear}, the algebra 
\[
(\Pol(\Sigma,\B),\oplus,\widetilde{\0},\ttop_\Sigma)
\]
is a $(\Sigma,\B)$-semimodule. We call it the \emph{$(\Sigma,\B)$-semimodule of polynomials}.
\index{SigmaBsemimodule@$(\Sigma,\B)$-semimodule of polynomials}
\index{semimodule of polynomials}
The next theorem shows the significance of this $(\Sigma,\B)$-semimodule.
It generalizes \cite[p.451]{bozale89} and \cite[p.352]{boz91} from fields to commutative semirings.

\begin{theorem-rect}\rm \label{lm:Pol-SigmaB-initial}  Let $\Sigma$ be a ranked alphabet and $\B$ be a commutative semiring. The $(\Sigma,\B)$-semimodule $(\Pol(\Sigma,\B),\oplus,\widetilde{\0},\ttop_\Sigma)$ is initial in the set of all $(\Sigma,\B)$-semimodules.
\end{theorem-rect}
\begin{proof} First we recall from page \pageref{p:universal-algebra-view-on-semimodules-with-algebra} that $(\Pol(\Sigma,\B),\oplus,\widetilde{\0},\ttop_\Sigma)$ is the algebra $(\Pol(\Sigma,\B),\eta_{\mathrm{Pol},\Sigma})$ where
\begin{compactitem}
\item $I= \{\oplus,\widetilde{\0}\} \cup \{(b\cdot)\mid b \in B\} \cup \Sigma$,
  \item $\eta_{\mathrm{Pol},\Sigma}: I \to \mathrm{Ops}(\Pol(\Sigma,\B))$ such that
\begin{compactitem}

\item $\eta_{\mathrm{Pol},\Sigma}(\oplus) = \oplus$ and $\eta_{\mathrm{Pol},\Sigma}(\widetilde{\0})() = \widetilde{\0}$,
  \item for every $b \in B$ and $r \in \Pol(\Sigma,\B)$, we let $\eta_{\mathrm{Pol},\Sigma}((b \cdot))(r) = b \cdot r$, and
  \item for each $\sigma \in \Sigma$, we let $\eta_{\mathrm{Pol},\Sigma}(\sigma) = \ttop_\Sigma(\sigma)$.
    \end{compactitem}
  \end{compactitem}
  
Now we prove the three properties (a), (b), and (c) of the definition of initial algebra on page \pageref{page:initial-algebra}.
    
(a) As proven above, $(\Pol(\Sigma,\B),\oplus,\widetilde{\0},\ttop_\Sigma)$ is a member of the set of all $(\Sigma,\B)$-semimodules.

    (b) We prove that the algebra $(\Pol(\Sigma,\B),\eta_{\mathrm{Pol},\Sigma})$ is generated by $\emptyset$, i.e., that $\langle \emptyset\rangle_{\im(\eta_{\mathrm{Pol},\Sigma})} = \Pol(\Sigma,\B)$. By Lemma \ref{lm:poly-are-rational}, we have that $\langle \emptyset\rangle_{\im(\eta_{\mathrm{Pol},\Sigma})\setminus \{\widetilde{\0}\}} = \Pol(\Sigma,\B)$. Then $\langle \emptyset\rangle_{\im(\eta_{\mathrm{Pol},\Sigma})}=  \Pol(\Sigma,\B)\cup\{\widetilde{\0}\}=  \Pol(\Sigma,\B)$.

(c) Let $(V,+,0,\mu)$ be an arbitrary $(\Sigma,\B)$-semimodule. We will prove that there exists a $(\Sigma,\B)$-semimodule homomorphism from $(\Pol(\Sigma,\B),\oplus,\widetilde{\0},\ttop_\Sigma)$ to $(V,+,0,\mu)$.    

For this, we define the mapping $h:\Pol(\Sigma,\B) \to V$ such that, for each $n\in \mathbb{N}$, $b_1,\ldots,b_n \in B$, and $\xi_1,\ldots,\xi_n\in \T_\Sigma$, we let
\begin{equation}\label{eq:from-h-to-hV}
  h(b_1.\xi_1 \oplus\ldots\oplus b_n.\xi_n) = b_1\cdot f(\1.\xi_1) +\ldots +  b_n\cdot f(\1.\xi_n)\enspace,
  \end{equation}
where $f$ denotes the unique  $\Sigma$-algebra homomorphism from $(\1.\T_\Sigma,\ttop_\Sigma)$ to $(V,\mu)$. (Clearly, $(\1.\T_\Sigma,\ttop_\Sigma)$ is initial in the set of all $\Sigma$-algebras.)
Now we prove that $h$ is a $(\Sigma,\B)$-semimodule homomorphism from $(\Pol(\Sigma,\B),\oplus,\widetilde{\0},\ttop_\Sigma)$ to $(V,+,0,\mu)$. Since $h$ is defined in the same way as in \eqref{eq:from-h-to-hV-free}, by Theorem~\ref{lm:Pol-SigmaB-free}, 
$h$ is a $\B$-semimodule homomorphism, i.e., a linear mapping, from $\mathsf{Pol}(\Sigma,\B)$ to $(V,+,0)$.  Thus it remains to prove that $h$ is a  $\Sigma$-algebra homomorphism from $(\Pol(\Sigma,\B),\ttop_\Sigma)$ to $(V,\mu)$.

For this, for each $i\in [k]$, let $s_i=b_{i1}.\xi_{i1} \oplus\cdots\oplus b_{in_i}.\xi_{in_i}$ for some $n_i\in\mathbb{N}$, $b_{i1},\ldots, b_{in_i}\in B$ and $\xi_{i1},\ldots, \xi_{in_i}\in \T_\Sigma$.
Then we can calculate as follows.
\begingroup
\allowdisplaybreaks
\begin{align*}
  & h\big(\ttop_\Sigma(\sigma)(s_1,\ldots,s_k)\big)\\[2mm]
&= h\big(\bigoplus_{j_1\in[n_1],\ldots,j_k\in[n_k]} (b_{1j_1}\otimes\ldots\otimes b_{kj_k}).\sigma(\xi_{1j_1},\ldots,\xi_{kj_k})\big) \tag{by Lemma \ref{polynomials-closed-under-top(sigma)}}\\[2mm]
  &= \bigplus_{j_1\in[n_1],\ldots,j_k\in[n_k]} (b_{1j_1}\otimes\ldots\otimes b_{kj_k})\cdot f(\1.\sigma(\xi_{1j_1},\ldots,\xi_{kj_k}))
  \tag{by \eqref{eq:from-h-to-hV}}\\[2mm]
  &= \bigplus_{j_1\in[n_1],\ldots,j_k\in[n_k]} (b_{1j_1}\otimes\ldots\otimes b_{kj_k})\cdot \mu(\sigma)(f(\1.\xi_{1j_1}),\ldots,f(\1.\xi_{kj_k}))
  \tag{because $f$ is a $\Sigma$-algebra homomorphism}\\[2mm]
  &= \mu(\sigma)\big(\bigplus_{j_1\in[n_1]}b_{1j_1}\cdot f(\1.\xi_{1j_1}),\ldots, \bigplus_{j_k\in[n_k]}b_{kj_k}\cdot f(\1.\xi_{kj_k})  \big)
  \tag{because $\mu(\sigma)$ is multilinear}\\[2mm]
  &= \mu(\sigma)\big(h(s_1),\ldots,h(s_k)\big)
    \tag{by \eqref{eq:from-h-to-hV}}
    \enspace.
\end{align*}
\endgroup
This proves that $h$ is a $\Sigma$-algebra homomorphism. Hence $h$ is a $(\Sigma,\B)$-semimodule homomorphism. 
\end{proof}

\begin{lemma}\rm \label{lm:congruences-can-be-lifted-to-contexts} Let $\approx$ be a congruence on $(\Pol(\Sigma,\B),\oplus,\widetilde{\0},\ttop_\Sigma)$. Let $s = a_1.\xi_1 \oplus \ldots \oplus a_n.\xi_n$ and $t=b_1.\zeta_1 \oplus \ldots \oplus b_\ell.\zeta_\ell$ be in $\Pol(\Sigma,\B)$ such that $s \approx t$. Then, for each $c \in \C_\Sigma$, we have
  \[
a_1.c[\xi_1] \oplus \ldots \oplus a_n.c[\xi_n] \approx b_1.c[\zeta_1] \oplus \ldots \oplus b_\ell.c[\zeta_\ell] \enspace.
    \]
  \end{lemma}
  \begin{proof} We prove by induction on $(\C_\Sigma,\succ_{\C_\Sigma})$.

    I.B.: For $c=z$ the statement is the assumption.

    I.S.: Let $c =e[c']$ for some elementary context $e = \sigma(\theta_1,\ldots,\theta_{i-1},z,\theta_{i+1},\ldots,\theta_k)$  in $\e\C_\Sigma$ and some context $c '\in \C_\Sigma$. We assume that the statement holds for $c'$ (I.H.); thus $s' \approx t'$ with
    \begin{align*}
      s' = a_1.c'[\xi_1] \oplus \ldots \oplus a_n.c'[\xi_n] \ \ \text{ and } \ \
      t' = b_1.c'[\zeta_1] \oplus \ldots \oplus b_\ell.c'[\zeta_\ell] \enspace.
    \end{align*}
    Since $\approx$ is reflexive, for each $j \in [k]\setminus \{i\}$ we have $\1.\theta_j \approx \1.\theta_j$.  
    Since $\approx$ is a congruence we have
    \[
 \ttop_\Sigma(\sigma)\Big( \1.\theta_1,\ldots,\1.\theta_{i-1},s',\1.\theta_{i+1},\ldots,\1.\theta_k)  \Big)
      \ \approx \ \ttop_\Sigma(\sigma)\Big( \1.\theta_1,\ldots,\1.\theta_{i-1},t',\1.\theta_{i+1},\ldots,\1.\theta_k)  \Big) \enspace.
    \]
    Moreover, the following equivalences hold:
    \begingroup
    \allowdisplaybreaks
    \begin{align*}
      &\ttop_\Sigma(\sigma)\Big( \1.\theta_1,\ldots,\1.\theta_{i-1},s',\1.\theta_{i+1},\ldots,\1.\theta_k)  \Big)
      \ \approx \ \ttop_\Sigma(\sigma)\Big( \1.\theta_1,\ldots,\1.\theta_{i-1},t',\1.\theta_{i+1},\ldots,\1.\theta_k)  \Big)\\[3mm]
      \Leftrightarrow \ \   & \bigoplus_{j \in [n]} a_j. \sigma(\theta_1,\ldots,\theta_{i-1},c'[\xi_j],\theta_{i+1},\ldots,\theta_n) \ \approx \
                              \bigoplus_{j \in [\ell]} b_j. \sigma(\theta_1,\ldots,\theta_{i-1},c'[\zeta_j],\theta_{i+1},\ldots,\theta_n)
        \tag{by Lemma \ref{polynomials-closed-under-top(sigma)}}\\[3mm]
      \Leftrightarrow \ \   & \bigoplus_{j \in [n]} a_j. c[\xi_j]  \ \approx \
      \bigoplus_{j \in [\ell]} b_j. c[\zeta_j] \qedhere
    \end{align*}
    \endgroup

   \end{proof}


  \subsection{The syntactic congruences and syntactic $(\Sigma,\B)$-semimodules}
\label{sec:Syntactic-congruences-and-syntactic-SigmaB-vector-spaces}

\begin{quote}\emph{In this subsection, we let $\B=(B,\oplus,\otimes,\0,\1)$ be an arbitrary commutative semiring, unless specified otherwise. }
\end{quote}

Here we will factorize the $(\Sigma,\B)$-semimodule $(\Pol(\Sigma,\B),\oplus,\widetilde{\0},\ttop_\Sigma)$ of $(\Sigma,\B)$-polynomials by particular congruences each of which depends on a weighted tree language.
\index{simr@$\approx_r$}
Formally, for each $r: \T_\Sigma \to B$, we define the relation $\approx_r$ on $\Pol(\Sigma,\B)$ as follows. Let $s = b_1.\xi_1 \oplus \ldots \oplus b_m.\xi_m$ and $t=a_1.\zeta_1\oplus\ldots\oplus a_n.\zeta_n$ be elements of $\mathrm{Pol}(\Sigma,\B)$. Then we let  $s \approx_r t$ if, for each $c \in \C_\Sigma$, we have
      \begin{equation*}
        b_1 \otimes r(c[\xi_1]) \oplus \ldots \oplus b_m \otimes r(c[\xi_m]) = a_1 \otimes r(c[\zeta_1]) \oplus \ldots \oplus a_n \otimes r(c[\zeta_n])  \enspace.
      \end{equation*}
We note that, for the case that $\B$ is a field, the congruence $\approx_r$ is the kernel of the unique homomorphism from the $(\Sigma,\B)$-vector space $(\mathrm{Pol}(\Sigma,\B),\oplus,\widetilde{\0},\ttop_\Sigma)$ to the $(\Sigma,\B)$-vector space $(\VLQ(r),\theta_r)$ (cf. the proof of Lemma~\ref{lm:VLQ-simorphism-lemma}); the $(\Sigma,\B)$-vector space $(\VLQ(r),\theta_r)$ is defined in \cite{bozlou83} (also cf.~\cite{bozale89}, \cite{boz91}, \cite[Sec.~3.7]{fulvog09new}, \cite[Sec.~18.1]{fulvog24}, and Section \ref{sec:characterization-rec-by-left-quotients}).

      \begin{lemma}\rm \label{lm:simr-is-a-congruence} The relation $\approx_r$ is  a congruence relation on $(\mathrm{Pol}(\Sigma,\B),\oplus,\widetilde{\0},\ttop_\Sigma)$.
      \end{lemma}
      \begin{proof} It is obvious that $\approx_r$ is an equivalence relation. It remains to show that the relation $\approx_r$ is compatible with scalar multiplication, sum, and each operation in $\ttop_\Sigma(\Sigma)$.

      \underline{Scalar multiplication:} Let $s,t\in \Pol(\Sigma,\B)$ with $s=\bigoplus_{j\in [m]} b_j.\xi_j$ and $t=\bigoplus_{j \in [n]} a_j.\zeta_j$ such that $s \approx_r t$. Moreover, let $b\in B$. By definition, for each $c \in \C_\Sigma$, we have
      \[ \bigoplus_{j \in [m]} b_j\otimes r(c[\xi_j]) = \bigoplus_{j \in [n]} a_j\otimes r(c[\zeta_j]).\] 
Hence, for each $c \in \C_\Sigma$, we also have 
     \[ \bigoplus_{j \in [m]} (b\otimes b_j)\otimes r(c[\xi_j]) = \bigoplus_{j \in [n]}  (b\otimes a_j)\otimes r(c[\zeta_j]).\] 
     This proves that $b.s \approx_r b.t$.
      
     \underline{Sum:}  For $i\in\{1,2\}$, let $s_i,t_i\in \Pol(\Sigma,\B)$ with $s_i=\bigoplus_{j\in [m_i]} b_{ij}.\xi_{ij}$ and $t_i=\bigoplus_{j \in [n_i]} a_{ij}.\zeta_{ij}$ such that $s_1\approx_r t_1$ and 
     $s_2\approx_r t_2$. By the definition of $\approx_r$, for each  $c \in \C_\Sigma$, we have
     \[ \bigoplus_{j \in [m_1]} b_{1j}\otimes r(c[\xi_{1j}]) = \bigoplus_{j \in [n_1]} a_{1j}\otimes r(c[\zeta_{1j}]) \text{\ \ \ and\ \ \ } \bigoplus_{j \in [m_2]} b_{2j}\otimes r(c[\xi_{2j}]) = \bigoplus_{j \in [n_2]} a_{2j}\otimes r(c[\zeta_{2j}]).\] 
     Hence, for each  $c \in \C_\Sigma$, we also have
      \[ \bigoplus_{j \in [m_1]} b_{1j}\otimes r(c[\xi_{1j}]) \oplus  \bigoplus_{j \in [m_2]} b_{2j}\otimes r(c[\xi_{2j}]) =
      \bigoplus_{j \in [n_1]} a_{1j}\otimes r(c[\zeta_{1j}])      
      \oplus \bigoplus_{j\in [n_2]} a_{2j}\otimes r(c[\zeta_{2j}]).\] 
      This proves that $s_1\oplus s_2 \approx_r t_1\oplus t_2$.
      
     \underline{Operations in $\ttop_\Sigma(\Sigma)$:} Let $k\in \mathbb{N}$ and  $\sigma\in \Sigma^{(k)}$.  Moreover, for each $i\in[k]$,  let $s_i,t_i\in \Pol(\Sigma,\B)$ with $s_i=\bigoplus_{j\in [m_i]} b_{ij}.\xi_{ij}$ and $t_i=\bigoplus_{j \in [n_i]} a_{ij}.\zeta_{ij}$ such that  $s_i\approx_r t_i$. By the definition of $\approx_r$, for every $i\in[k]$ and $c \in \C_\Sigma$, we have
     \begin{equation}\label{eq:synt-congruence-top-sigma-1} \bigoplus_{j\in [m_i]} b_{ij}\otimes r(c[\xi_{ij}]) = \bigoplus_{j \in [n_i]} a_{ij}\otimes r(c[\zeta_{ij}]).
     \end{equation}
     Since, by \eqref{polynomials-closed-under-top(sigma)}, we have 
 \[ \ttop_\Sigma(\sigma)(s_1,\ldots,s_k)=\bigoplus_{j_1\in[m_1],\ldots,j_k\in[m_k]} (b_{1j_1}\otimes\ldots\otimes b_{kj_k}).\sigma(\xi_{1j_1},\ldots,\xi_{kj_k})\]     
 and
 \[ \ttop_\Sigma(\sigma)(t_1,\ldots,t_k)=\bigoplus_{j_1\in[n_1],\ldots,j_k\in[n_k]} (a_{1j_1}\otimes\ldots\otimes a_{kj_k}).\sigma(\zeta_{1j_1},\ldots,\zeta_{kj_k}),\]  
 we have to show that, for each $c\in \C_\Sigma$, the equality
 \begin{equation}\label{eq:synt-congruence-top-sigma-2}\bigoplus_{\substack{j_1\in[m_1],\\ \ldots,\\ j_k\in[m_k]}} b_{1j_1}\otimes\ldots\otimes b_{kj_k}\otimes r\big(c[\sigma(\xi_{1j_1},\ldots,\xi_{kj_k})]\big)=
 \bigoplus_{\substack{j_1\in[n_1],\\ \ldots,\\ j_k\in[n_k]}} a_{1j_1}\otimes\ldots\otimes a_{kj_k}\otimes r\big(c[\sigma(\zeta_{1j_1},\ldots,\zeta_{kj_k})]\big)
 \end{equation}   
 holds. For this, let $c\in\C_\Sigma$. We compute as follows:
 \begingroup
 \allowdisplaybreaks
   \begin{align*}
  & \bigoplus_{\substack{j_1\in[m_1],\\ \ldots,\\ j_k\in[m_k]}} b_{1j_1}\otimes\ldots\otimes b_{kj_k}\otimes r\big(c[\sigma(\xi_{1j_1},\ldots,\xi_{kj_k})]\big)\\
  & =  \bigoplus_{\substack{j_2\in[m_2],\\ \ldots,\\ j_k\in[m_k]}} \Big( \bigoplus_{j_1 \in [m_1]} b_{1j_1}\otimes r\big(c[\sigma(\xi_{1j_1},\xi_{2j_2}\ldots,\xi_{kj_k})]\big) \Big)\otimes b_{2j_2}\otimes\ldots\otimes b_{kj_k} \\
  & =  \bigoplus_{\substack{j_2\in[m_2],\\ \ldots,\\ j_k\in[m_k]}} \Big( \bigoplus_{j_1 \in [n_1]} a_{1j_1}\otimes r\big(c[\sigma(\zeta_{1j_1},\xi_{2j_2}\ldots,\xi_{kj_k})]\big) \Big)\otimes b_{2j_2}\otimes\ldots\otimes b_{kj_k} \tag{by \eqref{eq:synt-congruence-top-sigma-1} for $i=1$ and context $c'=c[\sigma(z,\xi_{2j_2},\ldots,\xi_{kj_k})]$}\\
  & = \bigoplus_{\substack{j_1\in[n_1],\\ j_2\in[m_2],\\\ldots,\\ j_k\in[m_k]}} a_{1j_1}\otimes b_{2j_2}\otimes \ldots\otimes b_{kj_k}\otimes r\big(c[\sigma(\zeta_{1j_1},\xi_{2j_2},\ldots,\xi_{kj_k})]\big)\\
  & \ldots \\
  & = \bigoplus_{\substack{j_1\in[n_1],\\ \ldots,\\ j_{k-1}\in[n_{k-1}],\\j_k\in[m_k]}} a_{1j_1}\otimes  \ldots\otimes a_{k-1j_{k-1}}\otimes  b_{kj_k}\otimes r\big(c[\sigma(\zeta_{1j_1},\ldots,\zeta_{k-1j_{k-1}},\xi_{kj_k})]\big)\\
  & = \bigoplus_{\substack{j_1\in[n_1],\\ \ldots,\\ j_{k-1}\in[n_{k-1}]}} a_{1j_1}\otimes  \ldots\otimes a_{k-1j_{k-1}}\otimes \Big( \bigoplus_{j_k\in[m_k]} b_{kj_k}\otimes r\big(c[\sigma(\zeta_{1j_1},\ldots,\zeta_{k-1j_{k-1}},\xi_{kj_k})]\big)\Big)\\
  & = \bigoplus_{\substack{j_1\in[n_1],\\ \ldots,\\ j_{k-1}\in[n_{k-1}]}} a_{1j_1}\otimes  \ldots\otimes a_{k-1j_{k-1}}\otimes \Big( \bigoplus_{j_k\in[n_k]} a_{kj_k}\otimes r\big(c[\sigma(\zeta_{1j_1},\ldots,\zeta_{k-1j_{k-1}},\zeta_{kj_k})]\big)\Big)
  \tag{by \eqref{eq:synt-congruence-top-sigma-1} for $i=k$ and context $c'=c[\sigma(\zeta_{1j_1},\ldots,\zeta_{k-1j_{k-1}},z)]$}\\
  & =  \bigoplus_{\substack{j_1\in[n_1],\\ \ldots,\\ j_k\in[n_k]}} a_{1j_1}\otimes\ldots\otimes a_{kj_k}\otimes r\big(c[\sigma(\zeta_{1j_1},\ldots,\zeta_{kj_k})]\big) \enspace.
   \end{align*}
   \endgroup
 This finishes the proof of  \eqref{eq:synt-congruence-top-sigma-2}.    
     \end{proof}

\index{syntactic congruence}
\index{syntactic $(\Sigma,\B)$-semimodule}
\index{syntactic $\B$-semimodule}
\index{syntactic $(\Sigma,\B)$-vector space}
\index{syntactic $\B$-vector space}
We call $\approx_r$ the \emph{syntactic congruence of $r$}, and we call
\[(\mathrm{Pol}(\Sigma,\B),\oplus,\widetilde{\0},\ttop_\Sigma)/_{\approx_r} = (\Pol(\Sigma,\B)/_{\approx_r},\oplus/_{\approx_r},[\widetilde{\0}]_{\approx_r},\ttop_\Sigma/_{\approx_r})
\]
the \emph{syntactic $(\Sigma,\B)$-semimodule of $r$}. (It is easy to prove that, for each $\sigma \in \Sigma$, the operation $\ttop_\Sigma/_{\approx_r}(\sigma)$ is multilinear; this also follows from Lemma~\ref{lm:simr-is-a-congruence} and \cite[Th.~II.6.10 and Lm.~II.11.3]{bursan81}.)

If $\B$ is a field, then $(\mathrm{Pol}(\Sigma,\B),\oplus,\widetilde{\0},\ttop_\Sigma)/_{\approx_r}$ is called \emph{syntactic $(\Sigma,\B)$-vector space}.
\sloppy We recall that a congruence  on the $(\Sigma,\B)$-semimodule $(\mathrm{Pol}(\Sigma,\B),\oplus,\widetilde{\0},\ttop_\Sigma)$ is a congruence both on the $\B$-semimodule $\mathsf{Pol}(\Sigma,\B)$ and the $\Sigma$-algebra $(\mathrm{Pol}(\Sigma,\B),\ttop_\Sigma)$ (cf. Subsection  \ref{sec:semimodules}). We call $\mathsf{Pol}(\Sigma,\B)/_{\approx_r}$ the  \emph{syntactic $\B$-semimodule of $r$}; if $\B$ is a field, then we call it \emph{syntactic $\B$-vector space}.

     \begin{example}\rm \label{ex:for-congruence} We consider the ranked alphabet $\Sigma=\{\sigma^{(2)},\gamma^{(1)},\alpha^{(0)}\}$ and the $(\Sigma,\Ratnum)$-weighted tree language $\#_\gamma: \T_\Sigma \to \mathbb{Q}$ such that, for each $\xi \in \T_\Sigma$, we let $\#_\gamma(\xi) = | \pos_\gamma(\xi)|$, i.e., the number of occurrences of $\gamma$ in $\xi$. We recall that $\Ratnum = (\mathbb{Q},+,\cdot,0,1)$ denotes the field of rational numbers.
       Here we show  the congruence $\approx_{\#_\gamma}$.

       For the general description of $\approx_{\#_\gamma}$, we define the mappings 
    \begin{align*}
      \mathrm{sumc}: \Pol(\Sigma,\Ratnum) \to \mathbb{Q} \ \text{ and } \ 
      \mathrm{sum}_\gamma: \Pol(\Sigma,\Ratnum) \to \mathbb{Q}
    \end{align*}
    such that,
    for each $(\Sigma,\Ratnum)$-polynomial weighted tree language
  \(s = b_1 . \xi_1 + \ldots + b_n . \xi_n\), we let 
  \[
    \mathrm{sumc}(s) = \bigplus_{i \in [n]} b_i  \ \ \text{ and } \ \ \mathrm{sum}_\gamma(s) =  b_{1}\cdot\#_\gamma(\xi_{1}) + \ldots + b_{n}\cdot\#_\gamma(\xi_{n}) \enspace.
  \]

Then we can prove the following.
  \begin{equation}\label{equ:form-of-ker-Varphi-for-num-gamma}
    \begin{aligned}
    &\text{For every $s_1,s_2 \in \Pol(\Sigma,\Ratnum)$, we have $s_1 \approx_{\#_\gamma} s_2$ iff }\\
    &\mathrm{sumc}(s_1) = \mathrm{sumc}(s_2)\ \text{ and } \ \mathrm{sum}_\gamma(s_1) = \mathrm{sum}_\gamma(s_2)
      \end{aligned}
\end{equation}
For this, let
\begin{align*}
s_1 = b_{11}.\xi_{11} + \ldots + b_{1n_1}.\xi_{1n_1} \ \ \text{ and } \ \
s_2 = b_{21}.\xi_{21} + \ldots + b_{2n_2}.\xi_{2n_2}
\end{align*}
be two $(\Sigma,\Ratnum)$-polynomial weighted tree languages. Then we can calculate as follows (by using the obvious extension of $\#_\gamma$ to contexts). 
\begingroup
\allowdisplaybreaks
\begin{align*}
  &s_1 \approx_{\#_\gamma} s_2\\[2mm]
  \text{iff } \ & (\forall c \in \C_\Sigma): b_{11} \cdot \#_\gamma(c[\xi_{11}]) + \ldots + b_{1n_1} \cdot \#_\gamma(c[\xi_{1n_1}])\\
 & \hspace*{15mm} = b_{21} \cdot \#_\gamma(c[\xi_{21}]) + \ldots + b_{2n_2} \cdot \#_\gamma(c[\xi_{2n_2}]) \\[2mm]
  \text{iff } \ & (\forall c \in \C_\Sigma): b_{11} \cdot (\#_\gamma(c) + \#_\gamma(\xi_{11})) + \ldots + b_{1n_1} \cdot (\#_\gamma(c) + \#_\gamma(\xi_{1n_1}))\\
  & \hspace*{15mm} = b_{21} \cdot (\#_\gamma(c) + \#_\gamma(\xi_{21})) + \ldots + b_{2n_2} \cdot (\#_\gamma(c) + \#_\gamma(\xi_{2n_2})) \\[2mm]
  \text{iff } \ & (\forall c \in \C_\Sigma): \mathrm{sumc}(s_1) \cdot \#_\gamma(c) + \mathrm{sum}_\gamma(s_1) 
                  = \mathrm{sumc}(s_2) \cdot \#_\gamma(c) + \mathrm{sum}_\gamma(s_2)\\[2mm]
  \text{iff } \ & (\forall n \in \mathbb{N}): f(n) = g(n)
                  \tag{where $f(n)= \mathrm{sumc}(s_1) \cdot n + \mathrm{sum}_\gamma(s_1)$ and
                  $g(n) = \mathrm{sumc}(s_2) \cdot n + \mathrm{sum}_\gamma(s_2)$}\\[2mm]
  \text{iff } \ & \mathrm{sumc}(s_1)  = \mathrm{sumc}(s_2) \ \text{ and } \
                  \mathrm{sum}_\gamma(s_1) =  \mathrm{sum}_\gamma(s_2)\enspace,
  \end{align*}
  \endgroup
  where the last equivalence holds, because the two linear mappings $f$ and $g$ are equal if and only if their two parameters coincide pairwise. This proves \eqref{equ:form-of-ker-Varphi-for-num-gamma}.
  
  An example of a pair $(s_1,s_2)$ in $\approx_{\#_\gamma}$ is
  \begin{align*}
    s_1 =  4.\sigma(\gamma(\alpha,\alpha)) + 5.\alpha \ \ \text{ and } \ \
    s_2= 2. \sigma(\gamma(\alpha),\gamma(\alpha)) + 7.\alpha \enspace,
  \end{align*}
  because $\mathrm{sumc}(s_1) = \mathrm{sumc}(s_2) = 9$ and $\mathrm{sum}_\gamma(s_1)= \mathrm{sum}_\gamma(s_2) =4$.

The congruence $\approx_{\#_\gamma}$ has infinitely many equivalence classes, which can be seen as follows. For each  $n\in \mathbb{N}$ there exists a $\xi_n\in \T_\Sigma$ with $\mathrm{sum}_\gamma(1.\xi_n)=n$. Thus, for every $n,m \in \mathbb{N}$ with  $n\ne m$, the monomial $(\Sigma,\Ratnum)$-weighted tree languages $1.\xi_n$ and $1.\xi_m$ are in different classes of $\approx_{\#_\gamma}$.
 \hfill $\Box$
\end{example}

\index{saturates}
Intuitively, we can retrieve from $\approx_r$ the weighted tree language $r$ by means of a linear form. To
formalize this, we consider an arbitrary $r : \T_\Sigma \rightarrow B$ and an arbitrary congruence $\approx$ on $(\Pol(\Sigma,\B),\oplus,\widetilde{\0},\ttop_\Sigma)$. We say that $\approx$ \emph{saturates $r$} if there exists a linear form $\gamma: \Pol(\Sigma,\B)/_\approx \to B$ such that $r(\xi) = \gamma([\1.\xi]_{\approx})$ for each $\xi \in \T_\Sigma$. If this is the case, then we say that $\approx$ saturates $r$ via $\gamma$.

 \begin{lemma}\rm \label{lm:uniqueness-of-linear-form} Let $r: \T_\Sigma \to B$ and $\approx$ be a congruence on $(\mathrm{Pol}(\Sigma,\B),\oplus,\widetilde{\0},\ttop)$ such that $\approx$ saturates~$r$. Then there exists exactly one linear form $\gamma: \Pol(\Sigma,\B)/_\approx \to B$ such that $\approx$ saturates $r$ via $\gamma$.
    \end{lemma}
    \begin{proof} Since $\approx$ saturates $r$ there exists a linear form $\gamma: \Pol(\Sigma,\B)/_\approx \to B$ such that  $\approx$ saturates $r$ via $\gamma$. Now let  $\gamma': \Pol(\Sigma,\B)/_\approx \to B$ be a linear form such that  $\approx$ saturates $r$ via $\gamma'$. Then, for each $s = b_1.\xi_1 \oplus \ldots \oplus b_m.\xi_m$ in $\Pol(\Sigma,\B)$, we can calculate as follows.
      \begingroup
      \allowdisplaybreaks
      \begin{align*}
        \gamma(s) &= \gamma\Big(\Big[ \bigoplus_{i \in [m]} b_i.\xi_i \Big]_\approx\Big)\\
        &= \bigoplus_{i\in [m]} b_j \otimes \gamma([\1.\xi_i]_\approx)
                               \tag{because $\gamma$ is a scalar-linear form}\\
                  &= \bigoplus_{i\in [m]} b_j \otimes r(\xi_i)
                    \tag{because $\approx$ saturates $r$ via $\gamma$}\\
        &= \bigoplus_{i\in [m]} b_j \otimes \gamma'([\1.\xi_i]_\approx)
                            \tag{because $\approx$ saturates $r$ via $\gamma'$}\\
                             &=  \gamma'\Big(\Big[ \bigoplus_{i \in [m]} b_i.\xi_i \Big]_\approx\Big) = \gamma'(s)
          \tag{because $\gamma'$ is a scalar-linear form} \enspace.
      \end{align*}
      \endgroup
      \end{proof}

The next lemma shows a property of $\approx_r$.
  
 \begin{lemma}\rm \label{lm:approx-r-is-coarsest} Let $r:\T_\Sigma \to \B$. Then $\approx_r$ is the coarsest congruence among the congruences on $(\Pol(\Sigma,\B),\oplus,\widetilde{\0},\ttop_\Sigma)$ which saturate $r$.
\end{lemma}

\begin{proof} First, we prove that $\approx_r$ saturates $r$. We define the mapping $\gamma: \Pol(\Sigma,\B)/_{\approx_r} \ \to B$ for each $(\Sigma,\B)$-polynomial $s = b_1.\xi_1 \oplus \ldots \oplus b_m.\xi_m$ by $\gamma([s]_{\approx_r}) = b_1\otimes r(\xi_1) \oplus \ldots \oplus b_m \otimes r(\xi_m)$. We show that  $\gamma$ is well defined. For this, let $t= d_1.\zeta_1 \oplus \ldots \oplus d_n.\zeta_n$ be a $(\Sigma,\B)$-polynomial such that $s \approx_r t$. For $c=z$, this yields
   \[b_1\otimes r(\xi_1) \oplus \ldots \oplus b_m \otimes r(\xi_m)= d_1\otimes r(\zeta_1) \oplus \ldots \oplus d_n \otimes r(\zeta_n)\enspace,\]
   i.e., $\gamma([s]_{\approx_r}) = \gamma([t]_{\approx_r})$; hence $\gamma$ is well defined.
   Obviously, $\gamma$ is a linear form. Moreover, for each $\xi \in \T_\Sigma$, we have $\gamma([\1.\xi]_{\approx_r})=r(\xi)$. That means that $\approx_r$ saturates $r$ via $\gamma$.

   Now let $\approx$ be a congruence on $(\Pol(\Sigma,\B),\oplus,\widetilde{\0},\ttop_\Sigma)$ which saturates $r$  via the scalar-linear form $\gamma': \Pol(\Sigma,\B)/_{\approx}  \to B$.
   Moreover, let $s,t \in \Pol(\Sigma,\B)$ with $s=a_1.\xi_1 \oplus \ldots \oplus a_n.\xi_n$ and $t=b_1.\zeta_1 \oplus \ldots \oplus b_\ell.\zeta_\ell$ and $s \approx t$. Then
  \begingroup
  \allowdisplaybreaks
\begin{align*}
& s \approx t \\
  \Rightarrow \ \ & (\forall c\in \C_\Sigma): a_1.c[\xi_1] \oplus \ldots \oplus a_n.c[\xi_n] \approx
                    b_1.c[\zeta_1] \oplus \ldots \oplus b_\ell.c[\zeta_\ell]
                    \tag{by Lemma \ref{lm:congruences-can-be-lifted-to-contexts}} \\
\Leftrightarrow \ \ & (\forall c\in \C_\Sigma):  [a_1.c[\xi_1] \oplus \ldots \oplus a_n.c[\xi_n]]_\approx =
                      [b_1.c[\zeta_1] \oplus \ldots \oplus b_\ell.c[\zeta_\ell]]_\approx\\
  \Rightarrow \ \ & (\forall c\in \C_\Sigma): \gamma'\Big([a_1.c[\xi_1] \oplus \ldots \oplus a_n.c[\xi_n]]_\approx\Big) =
                    \gamma'\Big([b_1.c[\zeta_1] \oplus \ldots \oplus b_\ell.c[\zeta_\ell]]_\approx\Big) \\
  \Rightarrow \ \ & (\forall c\in \C_\Sigma): a_1 \otimes \gamma'\big([\1.c[\xi_1]]_\approx\big) \oplus \ldots \oplus a_n \otimes \gamma'\big([\1.c[\xi_n]]_\approx\big) =
                    b_1 \otimes \gamma'\big([\1.c[\zeta_1]]_\approx\big) \oplus \ldots \oplus b_\ell \otimes \gamma'\big([\1.c[\zeta_\ell]]_\approx\big)
                    \tag{because $\gamma'$ is a scalar-linear form}\\
  \Rightarrow \ \ & (\forall c\in \C_\Sigma): a_1 \otimes r(c[\xi_1]) \oplus \ldots \oplus a_n \otimes r(c[\xi_n]) =
                    b_1 \otimes r(c[\zeta_1]) \oplus \ldots \oplus b_\ell \otimes r(c[\zeta_\ell])
                    \tag{because $\approx$ saturates $r$ via $\gamma'$}\\
\Rightarrow \ \ & s \approx_r t  \enspace. 
\qedhere
\end{align*}
\endgroup
\end{proof}


\subsection{The $(\Sigma,\B)$-semimodule associated to a wta}
\label{ssec:(Sigma,B)-semimodules-induced-by-wta}

\begin{quote}\emph{In this subsection, we let $\B=(B,\oplus,\otimes,\0,\1)$ be an arbitrary commutative semiring, unless specified otherwise. Moreover, we let $\cA=(Q,\delta,F)$ be an arbitrary $(\Sigma,\B)$-wta. }
\end{quote}

\index{PsiA@$\Psi_\cA$}
\index{MA@$\sfM(\cA)$}
\index{hA@$\sfh_\cA$}
In a natural way, $\cA$ induces a $(\Sigma,\B)$-semimodule which consists of the vector algebra $\V(\cA)=(B^Q,\delta_\cA)$ of $\cA$ enhanced to a $(\Sigma,\B)$-semimodule. Formally, we define the $(\Sigma,\B)$-semimodule $\sfM(\cA)$ by 
\[
  \sfM(\cA) = (B^Q,\oplus,\0^Q,\delta_\cA)
\]
and we denote by $\sfh_\cA$ the unique $(\Sigma,\B)$-semimodule homomorphism from $(\mathrm{Pol}(\Sigma,\B),\oplus,\widetilde{\0},\ttop_\Sigma)$ to $\sfM(\cA)$ (cf. Figure~\ref{fig:overview-congruence-relations-isos-semimodules}). In particular, $\sfh_\cA$ is a linear mapping from the $\B$-semimodule $\mathsf{Pol}(\Sigma,\B)$ to $(B^Q,\oplus,\0^Q)$.   
By \eqref{eq:from-h-to-hV}, for each $s = b_1.\xi_1 \oplus \ldots \oplus b_m.\xi_m$ in $\Pol(\Sigma,\B)$, we have
\begin{equation}\label{equ:pol-sfh-related-to-h}
  \sfh_\cA(s) = b_1 \cdot \h_\cA(\xi_1) \oplus \ldots \oplus b_m \cdot \h_\cA(\xi_m) \enspace,
  \end{equation}
  where $\h_\cA: \T_\Sigma \to B$ is the unique $\Sigma$-algebra homomorphism from $\sfT_\Sigma$ to $\V(\cA)$. By Theorem~\ref{thm:kernel-is-congruence}, the kernel $\ker(\sfh_\cA)$ of $\sfh_\cA$ is a congruence on $(\Pol(\Sigma,\B),\oplus,\widetilde{\0},\ttop)$.
  
\index{MAim@$\sfM_{\mathrm{im}}(\cA)$}
By Observation~\ref{obs:smallest-subalgebra-im}, the semimodule $\sfM_{\mathrm{im}}(\cA)$, defined by
\[
  \sfM_{\mathrm{im}}(\cA) = (\im(\sfh_\cA),\oplus,\0^Q,\delta_\cA)
\]
is the smallest sub-$(\Sigma,\B)$-semimodule  of $\sfM(\cA)$.
By Corollary~\ref{cor:image-of-hom-isomorphic-to-quotient-of-kernel}, we have
\begin{equation}\label{eq:isomorphism-Polkerh-Mim}
(\Pol(\Sigma,\B),\oplus,\widetilde{\0},\ttop_\Sigma)/_{\ker(\sfh_\cA)} \cong  \sfM_{\mathrm{im}}(\cA) \enspace.
\end{equation}
In a part of Figure \ref{fig:overview-congruence-relations-isos-semimodules} we visualize \eqref{eq:isomorphism-Polkerh-Mim}.

    \begin{figure}[t]
 \begin{center}
   \begin{tikzpicture}
     \node (1){$\sfM(\cA) =(B^Q,\oplus,\0^Q,\delta_\cA)$};
     \node[right of=1, xshift=25em] (2) {$(\Pol(\Sigma,\B),\oplus,\widetilde{\0},\ttop_\Sigma)$};
     \node[below of=1, yshift=-4em, xshift= 0em] (3) {$\sfM_{\mathrm{im}}(\cA) =(\im(\sfh_\cA),\oplus,\0^Q,\delta_\cA)$};
     \node[right of=3, xshift=30em, yshift=-7em] (4) {$(\Pol(\Sigma,\B),\oplus,\widetilde{\0},\ttop_\Sigma)/_{\approx_{\sem{\cA}}}$};
          \node[right of=3, xshift=6em] (5) {\Large $\cong$};
          \node[right of=5, xshift=5em] (6) {$(\Pol(\Sigma,\B),\oplus,\widetilde{\0},\ttop_\Sigma)/_{\ker(\sfh_\cA)}$};
          \node[below of=5, xshift=8em, yshift=-4.1em] (7) {$\Big((\Pol(\Sigma,\B),\oplus,\widetilde{\0},\ttop_\Sigma)/_{\ker(\sfh_\cA)}\Big)/_{\rho_\cA}$};
          \node[right of=7, xshift=6em] (8) {\Large $\cong$};

     \draw (2) edge[->,>=stealth] node[fill=white] {$\sfh_\cA$} (1);
        \draw (1) edge[->,>=stealth] node[fill=white] {smallest sub-$(\Sigma,\B)$-semimodule} (3);
        \draw (2) edge[->,>=stealth] node[fill=white] {$\pi_{\approx_{\sem{\cA}}}$} (4);
        \draw (6) edge[->,>=stealth] node[fill=white] {$\pi_{\rho_\cA}$} (7);
     \draw (2) edge[->,>=stealth] node[fill=white] {$\pi_{\ker(\sfh_\cA)}$} (6);

   \end{tikzpicture}
 \end{center}
 \caption{\label{fig:overview-congruence-relations-isos-semimodules} Overview of the relation between several $(\Sigma,\B)$-semimodules  for an arbitrary $(\Sigma,\B)$-wta~$\cA$. }
      \end{figure}

      \begin{example}\rm \label{ex:illustration-of-kerPsiA} In Example \ref{ex:for-congruence}, we have considered the $(\Sigma,\Ratnum)$-weighted tree language $\#_\gamma: \T_\Sigma \to \mathbb{Q}$ for $\Sigma=\{\sigma^{(2)},\gamma^{(1)},\alpha^{(0)}\}$, and we have computed the congruence $\approx_{\#_\gamma}$. Here we will illustrate the congruence $\ker(\sfh_\cA)$ for a $(\Sigma,\Ratnum)$-wta $\cA$ with $\sem{\cA} = \#_\gamma$. For this,  we define $\cA=(Q,\delta,F)$, where  
  \begin{compactitem}
  \item $Q = \{q_1,q_2\}$,
  \item for every $k \in \mathbb{N}$, $p_1,\ldots,p_k,p \in Q$ we let
    $\delta_0(\varepsilon,\alpha,q_1) = 1$, $\delta_0(\varepsilon,\alpha,q_2)=0$,\\
    $\delta_1(p_1,\gamma,p) =
    \begin{cases}
      1 & \text{ if $p=q_2$ or $p_1=p=q_1$}\\
      0 & \text{ otherwise}
    \end{cases}$, \ \  and \\
    $\delta_2(p_1p_2,\sigma,p) =
    \begin{cases}
      1 & \text{ if ($p=q_2$ and $p_1\ne p_2$) or ($p_1=p_2=p=q_1$)}\\
      0 & \text{ otherwise}
    \end{cases}$
    \item $F_{q_1}= 0$ and $F_{q_2}=1$.
    \end{compactitem}
    Since $\delta_1(q_1,\gamma,q_1)=\delta_1(q_1,\gamma,q_2)= 1$, the wta $\cA$ is not bu-deterministic. Moreover, $\cA$ is crisp and has unit root weights, see Figure \ref{fig:ex-wta-for-number-of-gammas}.

    By induction on $\T_\Sigma$, it is easy to prove that, for each $\xi \in \T_\Sigma$, we have $\h_\cA(\xi)_{q_1}=1$ and $\h_\cA(\xi)_{q_2} = \#_\gamma(\xi)$, where $\#_\gamma(\xi) = | \pos_\gamma(\xi)|$, i.e., the number of occurrences of $\gamma$ in $\xi$. Thus $\sem{\cA}=\#_\gamma$.

      For instance, for the polynomial weighted tree languages $s_1=4.\sigma(\gamma(\alpha),\alpha)$ and
    $s_2= 2.\alpha + 2.\sigma(\gamma(\alpha),\gamma(\alpha))$, we have $s_1 \ker(\sfh_\cA) s_2$, because
    \begin{align*}
      \sfh_\cA(s_1) &= 4 \cdot \h_\cA(\sigma(\gamma(\alpha),\alpha)) = 4 \cdot \left(\begin{matrix}1\\1\end{matrix}\right) = \left(\begin{matrix}4\\4\end{matrix}\right)\\
      \sfh_\cA(s_2) &= 2 \cdot \h_\cA(\alpha) + 2 \cdot \h_\cA(\sigma(\gamma(\alpha),\gamma(\alpha)))
                      = 2 \cdot \left(\begin{matrix}1\\0\end{matrix}\right) + 2 \cdot \left(\begin{matrix}1\\2\end{matrix}\right)
                      = \left(\begin{matrix}2\\0\end{matrix}\right) + \left(\begin{matrix}2\\4\end{matrix}\right)
                          = \left(\begin{matrix}4\\4\end{matrix}\right) \enspace.
    \end{align*}

    For the general description of $\ker(\sfh_\cA)$, we use the mappings $\mathrm{sumc}: \Pol(\Sigma,\Ratnum) \to \mathbb{Q}$ and $\mathrm{sum}_\gamma: \Pol(\Sigma,\Ratnum) \to \mathbb{Q}$ as they are defined in Example~\ref{ex:for-congruence}. 
  Then, for each $s= b_1.\xi_1 + \ldots + b_n.\xi_n$ in $\Pol(\Sigma,\Ratnum)$ we have
  \[
    \sfh_\cA(s) = b_1 \cdot \left( \begin{matrix} 1  \\ \#_\gamma(\xi_1) \end{matrix}\right) + \ldots + b_n \cdot \left( \begin{matrix} 1 \\ \#_\gamma(\xi_n) \end{matrix}\right)
    = \left( \begin{matrix} \mathrm{sumc}(s) \\ \mathrm{sum}_\gamma(s) \end{matrix}\right) \enspace.
  \]
  Thus we obtain:
   \begin{equation}\label{equ:form-of-PsiA-for-num-gamma}
    \begin{aligned}
    &\text{For every $s_1,s_2 \in \Pol(\Sigma,\Ratnum)$, we have $s_1 \ker(\sfh_\cA) s_2$ iff }\\
    &\mathrm{sumc}(s_1) = \mathrm{sumc}(s_2)\ \text{ and } \ \mathrm{sum}_\gamma(s_1) = \mathrm{sum}_\gamma(s_2)
      \end{aligned}
    \end{equation}
    Hence, in this example, we have $\ker(\sfh_\cA)=\approx_{\sem{\cA}}$.
    \hfill $\Box$
  \end{example}
  
  
  \begin{figure}
  \centering
  \begin{tikzpicture}
\tikzset{node distance=7em, scale=0.6, transform shape}
\node[state, rectangle] (1) {\Large $\alpha$};
\node[state, right of=1] (2) {\Large $q_1$};
\node[state, rectangle, above of=2] (3) {\Large $\gamma$};
\node[state, rectangle, below of=2] (4) {\Large $\sigma$};
\node[state, rectangle, right= 10em of 2] (5) {\Large $\gamma$};
\node[state, rectangle, above of=5] (6) {\Large $\sigma$};
\node[state, rectangle, below of=5] (7) {\Large $\sigma$};
\node[state, right= 10em  of 5] (8) {\Large $q_2$};
\node[state, rectangle, right of=8] (9) {\Large $\gamma$};

\tikzset{node distance=2em}
\node[above of=1] (w1) {$1$};
\node[above of=3] (w3) {$1$};
\node[above of=4,  right=0.05cm] (w4) {$1$};
\node[above of=5] (w5) {$1$};
\node[above of=6, left=0.05cm] (w6) {$1$};
\node[above of=7] (w7) {$1$};
\node[right of=8] (w3) {$1$};
\node[above of=9, right=0.05cm] (w9) {$1$};

\draw[->,>=stealth] (1) edge (2);
\draw[->,>=stealth, out=180, in=135, looseness=1.2] (3) edge (2);
\draw[->,>=stealth, out=45, in=0, looseness=1.2] (2) edge (3);

\draw[->,>=stealth] (4) edge (2);
\draw[->,>=stealth, out=225, in=260, looseness=2.3] (2) edge (4);
\draw[->,>=stealth, out=315, in=280, looseness=2.3] (2) edge (4);

\draw[->,>=stealth] (2) edge (6);
\draw[->,>=stealth] (2) edge (5);
\draw[->,>=stealth] (2) edge (7);
\draw[->,>=stealth] (5) edge (8);
\draw[->,>=stealth] (8) edge (6);
\draw[->,>=stealth] (8) edge (7);
\draw[->,>=stealth, out=90, in=110, looseness=1.1] (6) edge (8);
\draw[->,>=stealth, out=270, in=250, looseness=1.1] (7) edge (8);
\draw[->,>=stealth, out=300, in=270, looseness=1.5] (8) edge (9);
\draw[->,>=stealth, out=90, in=60, looseness=1.5] (9) edge (8);
\end{tikzpicture}
\caption{\label{fig:ex-wta-for-number-of-gammas} The $(\Sigma,\Ratnum)$-wta $\cA$ with $\sem{\cA} = \#_\gamma$.}
  \end{figure}


  The next lemma relates $\ker(\sfh_{\cA})$ and the syntactic congruence $\approx_{\sem{\cA}}$.

  \begin{lemma}\rm \label{lm:kerPsiA-subseteeq-sim-semA} (cf. \cite{bozale89}) We have $\ker(\sfh_{\cA}) \subseteq \,\approx_{\sem{\cA}}$.
      \end{lemma}

      \begin{proof}  By Lemma \ref{lm:approx-r-is-coarsest}, $\approx_{\sem{\cA}}$ is the coarsest congruence on $(\Pol(\Sigma,\B),\oplus,\widetilde{\0},\ttop_\Sigma)$ which saturates $\sem{\cA}$. Thus it suffices to prove that $\ker(\sfh_\cA)$ saturates $\sem{\cA}$. For this we define the mapping $\gamma: \Pol(\Sigma,\B)/_{\ker(\sfh_\cA)} \to B$ for each $s = b_1.\xi_1 \oplus \ldots \oplus b_n.\xi_n$ in $\Pol(\Sigma,\B)$ by
  \[
\gamma\Big([b_1.\xi_1 \oplus \ldots \oplus b_n.\xi_n]_{\ker(\sem{\cA})}  \Big) = b_1 \otimes \sem{\cA}(\xi_1) \oplus \ldots \oplus b_n \otimes \sem{\cA}(\xi_n) \enspace.
\]
We prove that $\gamma$ is well defined. For this let $s = b_1.\xi_1 \oplus \ldots \oplus b_n.\xi_n$ and $t=a_1.\zeta_1 \oplus \ldots \oplus a_\ell.\zeta_\ell$ be in $\Pol(\Sigma,\B)$. Then we can calculate as follows.
\begingroup
\allowdisplaybreaks
\begin{align*}
  &s \ker(\sfh_\cA) t\\
  \Leftrightarrow & \ \ b_1 \otimes \h_\cA(\xi_1) \oplus \ldots \oplus b_n \otimes \h_\cA(\xi_n) = a_1 \otimes \h_\cA(\zeta_1) \oplus \ldots \oplus a_\ell \otimes \h_\cA(\zeta_\ell)
  \tag{by \eqref{equ:pol-sfh-related-to-h}}\\
  \Rightarrow & \ \ \big(b_1 \otimes \h_\cA(\xi_1) \oplus \ldots \oplus b_n \otimes \h_\cA(\xi_n)\big)\cdot F = \big(a_1 \otimes \h_\cA(\zeta_1) \oplus \ldots \oplus a_\ell \otimes \h_\cA(\zeta_\ell)\big)\cdot F\tag{where $\cdot$ is the scalar product defined in Section~\ref{sec:vectors-matrices}}\\
  \Leftrightarrow & \ \ \big(b_1 \otimes \h_\cA(\xi_1)\big)\cdot F \oplus \ldots \oplus \big(b_n \otimes \h_\cA(\xi_n)\big)\cdot F = \big(a_1 \otimes \h_\cA(\zeta_1)\big)\cdot F \oplus \ldots \oplus \big(a_\ell \otimes \h_\cA(\zeta_\ell)\big)\cdot F\\
  \Leftrightarrow & \ \ b_1 \otimes \big(\h_\cA(\xi_1)\cdot F\big) \oplus \ldots \oplus b_n \otimes \big(\h_\cA(\xi_n)\cdot F\big) = a_1 \otimes \big(\h_\cA(\zeta_1)\cdot F\big) \oplus \ldots \oplus a_\ell \otimes \big(\h_\cA(\zeta_\ell)\cdot F\big)\\
  \Leftrightarrow & \ \ b_1 \otimes \sem{\cA}(\xi_1) \oplus \ldots \oplus b_n \otimes \sem{\cA}(\xi_n) = a_1 \otimes \sem{\cA}(\zeta_1) \oplus \ldots \oplus a_\ell \otimes \sem{\cA}(\zeta_\ell)\\
  \Leftrightarrow & \ \ \gamma([s]_{\ker(\sfh_\cA)})=\gamma([t]_{\ker(\sfh_\cA)}) \enspace.
\end{align*}
\endgroup
Hence $\gamma$ is well defined. It is easy to see that $\gamma$ is a linear form. Since, for each $\xi \in \T_\Sigma$, we have $\gamma([\1.\xi]_{\ker(\sfh_\cA)})= \sem{\cA}(\xi)$, the congruence $\ker(\sfh_\cA)$ saturates $\sem{\cA}$ via $\gamma$.
    \end{proof}

  \index{rhoA@$\rho_\cA$}
As a kind of bridge between the congruences $\ker(\sfh_\cA)$ and $\approx_{\sem{\cA}}$,  we define the binary relation $\rho_\cA$ on $\Pol(\Sigma,\B)/_{\ker(\sfh_\cA)}$ such that, for every $s_1, s_2\in \Pol(\Sigma,\B)$, we let
       \begin{equation*}
      [s_1]_{\ker(\sfh_\cA)} \ \rho_\cA \  [s_2]_{\ker(\sfh_\cA)} \ \text{ if and only if } \
    s_1 \ \approx_{\sem{\cA}} \  s_2 \enspace.
    \end{equation*}
    By Lemma~\ref{lm:kerPsiA-subseteeq-sim-semA}, the relation $\rho_\cA$ is well defined.
        For a better understanding of the next theorem, the reader is  advised to consult with  Figure~\ref{fig:overview-congruence-relations-isos-semimodules}.

 \begin{samepage}
    \begin{theorem-rect}\label{lm:congruences-modulo-congruences} {\rm (cf. \cite{bozale89})} Let $\B$ be a commutative semiring. Moreover, let $\cA$ be a $(\Sigma,\B)$-wta. Then the following statements hold.
      \begin{compactenum}
    \item[(1)]  The relation $\rho_\cA$ is a congruence on the $(\Sigma,\B)$-semimodule $(\mathrm{Pol}(\Sigma,\B),\oplus,\widetilde{\0},\ttop_\Sigma)/_{\ker(\sfh_\cA)}$.
    \item[(2)] $\Big((\mathrm{Pol}(\Sigma,\B),\oplus,\widetilde{\0},\ttop_\Sigma)/_{\ker(\sfh_\cA)}\Big)/_{\rho_\cA} \cong (\mathrm{Pol}(\Sigma,\B),\oplus,\widetilde{\0},\ttop_\Sigma)/_{\approx_{\sem{\cA}}}$.
    \end{compactenum}
  \end{theorem-rect}
\end{samepage}

\begin{proof} By Lemma~\ref{lm:kerPsiA-subseteeq-sim-semA}, we have $\ker(\sfh_\cA) \subseteq \approx_{\sem{\cA}}$. Then the statements follow from
 Theorem~\ref{thm:snd-isom-theorem} with
 $\A=(\Pol(\Sigma,\B),\oplus,\widetilde{\0},\ttop_\Sigma)$, $\sim \,= \,\ker(\sfh_\cA)$, $\approx \,=\, \approx_{\sem{\cA}}$, and $\approx\!/_\sim = \rho_\cA$.
\end{proof}

\begin{example} \rm From Examples \ref{ex:for-congruence} and \ref{ex:illustration-of-kerPsiA} we obtain that $\ker(\sfh_\cA) = \approx_{\#_\gamma}$. Hence the relation $\rho_\cA$ is the identity on $\mathrm{Pol}(\Sigma,\B)/_{\ker(\sfh_\cA)}$.
  \hfill $\Box$
\end{example}

\begin{example} \rm \label{ex:wta-for-which-sim-is-different-from-kerPsi} Here we show an example of a wta $\cA$ for which $\ker(\sfh_\cA) \subset \, \approx_{\sem{\cA}}$, i.e., 
$\ker(\sfh_\cA)$ is a proper subset of $\approx_{\sem{\cA}}$. For this, we let $\Sigma = \{\gamma^{(1)},\alpha^{(0)},\beta^{(0)}\}$ and consider the $(\Sigma,\Ratnum)$-wta $\cA=(Q,\delta,F)$ with
  \begin{compactitem}
  \item $Q= \{p,q,\bot\}$; in the vectors below we follow the order $p$ is above $q$ is above $\bot$,
  \item $\delta_0(\varepsilon,\alpha,p) = \delta_0(\varepsilon,\beta,q) = \delta_1(p,\gamma,\bot) =\delta_1(q,\gamma,\bot) = \delta_1(\bot,\gamma,\bot) = 1$, and  each other transition has weight $0$, and
    \item $F_p=F_q=1$ and $F_\bot=0$.
    \end{compactitem}
    The wta $\cA$ is crisp-deterministic and it has unit root weights.
    
    Obviously, for every $n \in \mathbb{N}_+$ and $x \in \{\alpha,\beta\}$, we have
    \[
      \h_\cA(\alpha) =
      \left(\begin{matrix}1\\0\\0\end{matrix}\right) \ , \
      \h_\cA(\beta) =
      \left(\begin{matrix}0\\1\\0\end{matrix}\right) \ , \ \text{ and } \ 
      \h_\cA(\gamma^n(x)) =
\left(\begin{matrix}0\\0\\1\end{matrix}\right)
\]
and, for each $\xi \in \T_\Sigma$, we have  $\sem{\cA}(\xi) = 1$ if $\xi \in \{\alpha,\beta\}$, and $0$ otherwise.
Clearly, each $s \in \Pol(\Sigma,\Ratnum)$ has the form
\begin{equation}\label{equ:form-of-polynomial}
s = b_1.\alpha + b_2.\beta + c_1.\gamma^{n_1}(\alpha)+\ldots + c_k.\gamma^{n_k}(\alpha)+d_1.\gamma^{m_1}(\beta)+\ldots + d_\ell.\gamma^{m_\ell}(\beta)
\end{equation}
for some $k,\ell \in \mathbb{N}$, $b_1,b_2,c_i,d_j \in \mathbb{Q}$ and $n_i,m_j \in \mathbb{N}_+$.
For each such $s$ we have
\[
\sfh_\cA(s) = \left(\begin{matrix}e\\f\\g\end{matrix}\right) 
  \]
  where $e = b_1$, $f = b_2$, and $g = \sum_{i \in [k]} c_i +  \sum_{j \in [\ell]} d_j$.
Moreover, for each $c \in \C_\Sigma$, we have 
  \begin{align}
    &b_1.\sem{\cA}(c[\alpha]) + b_2.\sem{\cA}(c[\beta])\nonumber\\
    &+ c_1.\sem{\cA}(c[\gamma^{n_1}(\alpha)]+\ldots + c_k.\sem{\cA}(c[\gamma^{n_k}(\alpha)])\nonumber\\
    &+d_1.\sem{\cA}(c[\gamma^{m_1}(\beta)])+\ldots + d_\ell.\sem{\cA}(c[\gamma^{m_\ell}(\beta)]) \nonumber\\
    & =
    \begin{cases}
      b_1 + b_2 & \text{ if $c=z$}\\
      0 & \text{ otherwise} 
      \end{cases} \ \ 
  = \ \ \begin{cases}
      e + f & \text{ if $c=z$}\\
      0 & \text{ otherwise} 
      \end{cases} \enspace.\label{align:for-syntactic-congruence}
    \end{align}

Now let $s_1,s_2 \in \Pol(\Sigma,\Ratnum)$ with
  \[
  \sfh_\cA(s_1) = \left(\begin{matrix}e_1\\f_1\\g_1\end{matrix}\right) \ \
    \text{ and } \ \
      \sfh_\cA(s_2) = \left(\begin{matrix}e_2\\f_2\\g_2\end{matrix}\right) \enspace.
        \]
        Then we have 
\[
  s_1 \ker(\sfh_\cA) \ s_2 \ \text{ iff } \ \ e_1=e_2 \  \text{and} \   f_1=f_2 \ \text{and} \ g_1=g_2 
  \]
and, by (\ref{align:for-syntactic-congruence}),
    \[
      s_1 \approx_{\sem{\cA}} s_2 \ \ \text{ iff } \ \ e_1 + f_1  = e_2  + f_2 \enspace.
     \]
     Thus, for example, for $s_1= 3.\alpha + 4.\beta + 5.\gamma(\alpha)$ and $s_2 = 2.\alpha + 5.\beta + 6.\gamma^2(\alpha)$, we have 
     \[
       \neg \Big(s_1 \ker(\sfh_\cA) \ s_2\Big)
       \ \ \ \text{ and } \ \ \ s_1  \approx_{\sem{\cA}} s_2
       \ \ \ \text{ and } \ \ \ [s_1]_{\ker(\sfh_\cA)} \ \rho_\cA \ [s_2]_{\ker(\sfh_\cA)} \enspace.
       \]
     \hfill$\Box$
   \end{example}

The next lemma shows that, for each recognizable $(\Sigma,\B)$-weighted tree language $r$ where $\B$ is a field, the number of states of any $\Sigma,\B)$-wta which recognizes $r$, is at least the dimension of $\sfPol(\Sigma,\B)/_{\approx_{\sem{\cA}}}$. The steps of its proof can be tracked on Figure~\ref{fig:overview-congruence-relations-isos-semimodules}

  \begin{lemma}\rm\label{lm:dimension-synt-vector-space-bounded} Let $\B$ be a field. Then the $\B$-vector space $\sfPol(\Sigma,\B)/_{\approx_{\sem{\cA}}}$ is finite-dimensional and $\dim\big(\sfPol(\Sigma,\B)/_{\approx_{\sem{\cA}}}\big)\le |Q|$.
  \end{lemma} 
\begin{proof} Clearly, the $\B$-vector space $(B^Q,\oplus,\0^Q)$ is finite-dimensional and its dimension is $|Q|$, i.e.,
  \[
|Q| = \dim\big( (B^Q,\oplus,\0^Q) \big) \enspace.
  \]
  By \cite[$\S$~7, Lm.~3]{gra68} $(\im(\sfh_\cA),\oplus,\0^Q)$ is a sub-$\B$-vector space of $(B^Q,\oplus,\0^Q)$, and by Theorem~\ref{thm:subspace-of-finite-dim-vector-space},  it is also finite-dimensional and
  \[
\dim\big( (B^Q,\oplus,\0^Q) \big) \ge  \dim\big( (\im(\sfh_\cA),\oplus,\0^Q)\big) \enspace.
    \]

By \eqref{eq:isomorphism-Polkerh-Mim}, $(\im(\sfh_\cA),\oplus,\0^Q) \cong \sfPol(\Sigma,\B)/_{\ker(\sfh_\cA)}$ and thus
  \[
  \dim\big( (\im(\sfh_\cA),\oplus,\0^Q)\big) = \dim\big( \sfPol(\Sigma,\B)/_{\ker(\sfh_\cA)}\big) \enspace.
    \]
    By Theorem \ref{lm:congruences-modulo-congruences}(1), $\rho_\cA$ is a congruence on $\sfPol(\Sigma,\B)/_{\ker(\sfh_\cA)}$ and thus, by Theorem \ref{thm:canonical-map-of-congr-is-hom}, the canonical mapping $\pi_{\rho_\cA}$ is a linear mapping. Hence, Theorem~\ref{thm:Sylvester-nullity},  we have
  \[
\dim\big( \sfPol(\Sigma,\B)/_{\ker(\sfh_\cA)}\big) \ge \dim\Big(\big(\sfPol(\Sigma,\B)/_{\ker(\sfh_\cA)}\big)/_{\rho_\cA}\Big) \enspace.
\]
Lastly, by Theorem \ref{lm:congruences-modulo-congruences}(2), we obtain
$\big(\sfPol(\Sigma,\B)/_{\ker(\sfh_\cA)}\big)/_{\rho_\cA} \cong \sfPol(\Sigma,\B)/_{\approx_{\sem{\cA}}}$ and thus 
 \[
\dim\Big(\big(\sfPol(\Sigma,\B)/_{\ker(\sfh_\cA)}\big)/_{\rho_\cA}\Big) = \dim\big( \sfPol(\Sigma,\B)/_{\approx_{\sem{\cA}}} \big) \enspace.
    \]
     \end{proof}


\subsection[Construction of a wta from a wtl, a congruence, and a finite basis]{Construction of a wta from a weighted tree language, a congruence, and a finite basis over some field}
\label{ssec:construction-wta-from-wta-congruence-basis}

\begin{quote} {\em In this subsection,  we let $\B=(B,\oplus,\otimes,\0,\1)$ be a field.  }
\end{quote}

As preparation for the proof of Theorem~\ref{th:MN-fields}(B)$\Rightarrow$(A), here, for every 
\begin{compactitem}
\item weighted tree language $r: \T_\Sigma \to B$,
\item congruence $\approx$ on the $(\Sigma,\B)$-vector space
  $(\Pol(\Sigma,\B),\oplus,\widetilde{\0},\ttop_\Sigma)$ which saturates $r$, and
\item finite basis $H$ of the $\B$-vector space $\sfPol(\Sigma,\B)/_{\approx}$ with $H \subseteq \{[\1.\zeta]_{\approx} \mid \zeta \in \T_\Sigma\}$,
\end{compactitem}
we define a $(\Sigma,\B)$-wta $\wta(r,\approx,H) = (H,\delta,F)$ and show that it recognizes $r$.
Moreover, if, for each $s \in \Pol(\Sigma,\B)$, the coefficients in the linear combination of $[s]_\approx$ over the basis $H$ are  computable,
then we can even construct $\wta(r,\approx,H)$ (cf. Lemma \ref{lm:semantics-constructed-wta}(2)).

\begin{quote} {\em In the rest of this subsection,  we abbreviate $[\1.\xi]_\approx$ by $[\1.\xi]$ for each $\xi\in \T_\Sigma$. Moreover, for every $s \in \Pol(\Sigma,\B)$ and base vector $[\1.\zeta] \in H$, we denote by $[s]_{[\1.\zeta]}$ the coefficient of $[\1.\zeta]$ in the linear combination of $[s]$ over the basis $H$. }
\end{quote}

\begin{definition}\rm \label{def:wta-from-finite-dimensional} \cite[p.353]{boz91} Let $r$, $\approx$, and $H$ be given as in the above list of objects. Moreover, let $\gamma: \mathrm{Pol}(\Sigma,\B)/_{\approx}\to B$ be the linear form uniquely determined by $\approx$ and $r$ (cf. Lemma~\ref{lm:uniqueness-of-linear-form}). 

We define the $(\Sigma,\B)$-wta $\wta(r,\approx,H)=(H,\delta,F)$ where 
\begin{compactitem}
\item $\delta=(\delta_k: H^k \times \Sigma^{(k)} \times H \to B \mid k \in \mathbb{N})$ and for every $k \in \mathbb{N}$, $\sigma \in \Sigma^{(k)}$, and $[\1.\zeta_{i_1}],\ldots,[\1.\zeta_{i_k}],[\1.\zeta] \in H$, we let
  \[
\delta_k\big([\1.\zeta_{i_1}]\cdots[\1.\zeta_{i_k}],\sigma,[\1.\zeta]\big) = [\1.\sigma(\zeta_{i_1},\ldots,\zeta_{i_k})]_{[\1.\zeta]} \enspace,
    \]
  \item for each $[\1.\zeta] \in H$, we define $F_{[\1.\zeta]} = \gamma([\1.\zeta])$.
    \hfill$\Box$
\end{compactitem}
\end{definition}

\begin{lemma}\rm \label{lm:semantics-constructed-wta}
  Let $r$, $\approx$, and $H$ be given as in the list at the beginning of this subsection.
    Then the following two statements hold.
  \begin{compactenum}
  \item[(1)] $\sem{\wta(r,\approx,H)}=r$.
    \item[(2)] If, for each $s \in \Pol(\Sigma,\B)$, the coefficients in the linear combination of $s$ on the basis $H$ are computable, then we can even construct $\wta(r,\approx,H)$.
      \end{compactenum}
  \end{lemma}
\begin{proof} Proof of (1):  We abbreviate $\wta(r,\approx,H)$ by $\cA$.  By induction on $\T_\Sigma$, we can prove the following statement.
    \begin{equation}\label{equ:hom-is-canonical-vector-new}
\text{For each $\xi \in \T_\Sigma$ and $[\1.\zeta] \in H$, we have $\h_{\cA}(\xi)_{[\1.\zeta]} = [\1.\xi]_{[\1.\zeta]}$.}
\end{equation}

Let $\xi = \sigma(\xi_1,\ldots,\xi_k)$ and $[\1.\zeta] \in H$.  Then we can calculate as follows.
\begingroup
\allowdisplaybreaks
\begin{align*}
  & \ \ \h_{\cA}(\sigma(\xi_1,\ldots,\xi_k))_{[\1.\zeta]}\\
  &= \bigoplus_{[\1.\zeta_1]\cdots [\1.\zeta_k] \in H^k}  \h_{\cA}(\xi_1)_{[\1.\zeta_1]} \otimes \ldots \otimes \h_{\cA}(\xi_k)_{[\1.\zeta_k]} \otimes \delta_k([\1.\zeta_1] \cdots [\1.\zeta_k], \sigma, [\1.\zeta])
  \tag{by the definition of $\h_\cA$}\\
   &=\bigoplus_{[\1.\zeta_1]\cdots [\1.\zeta_k] \in H^k}  [\1.\xi_1]_{[\1.\zeta_1]} \otimes \ldots \otimes [\1.\xi_k]_{[\1.\zeta_k]} \otimes
    [\1.\sigma(\zeta_1, \ldots,\zeta_k)]_{[\1.\zeta]} 
    \tag{by I.H. and the definition of $\delta_k$}\\
    &=  \bigoplus_{[\1.\zeta_1]\cdots [\1.\zeta_k] \in H^k}  \big([\1.\xi_1]_{[\1.\zeta_1]} \otimes \ldots \otimes [\1.\xi_k]_{[\1.\zeta_k]}\big) \cdot [\1.\sigma(\zeta_1,\ldots,\zeta_k)]_{[\1.\zeta]}\\
    &=  \bigoplus_{[\1.\zeta_1]\cdots [\1.\zeta_k] \in H^k}  \big([\1.\xi_1]_{[\1.\zeta_1]} \otimes \ldots \otimes [\1.\xi_k]_{[\1.\zeta_k]}\big) \cdot 
    (\ttop_\Sigma/_{\approx_r})(\sigma)\Big([\1.\zeta_1],\ldots,[\1.\zeta_k]\Big)_{[\1.\zeta]}\\
    &= \bigoplus_{[\1.\zeta_1]\cdots [\1.\zeta_k] \in H^k}(\ttop_\Sigma/_{\approx_r})(\sigma)\Big([\1.\xi_1]_{[\1.\zeta_1]}\cdot [\1.\zeta_1],\ldots, [\1.\xi_k]_{[\1.\zeta_k]}\cdot [\1.\zeta_k]\Big)_{[\1.\zeta]}\\
    &= (\ttop_\Sigma/_{\approx_r})(\sigma)\Big(\bigoplus_{[\1.\zeta] \in H} [\1.\xi_1]_{[\1.\zeta]}\cdot [\1.\zeta],\ldots,\bigoplus_{[\1.\zeta] \in H} [\1.\xi_k]_{[\1.\zeta]}\cdot [\1.\zeta]\Big)_{[\1.\zeta]}\\
    &= (\ttop_\Sigma/_{\approx_r})(\sigma)\big([\1.\xi_1],\ldots,[\1.\xi_k]\big)_{[\1.\zeta]}\\
    &= [\1.\sigma(\xi_1,\ldots,\xi_k)]_{[\1.\zeta]} \enspace.
\end{align*}
\endgroup
This proves \eqref{equ:hom-is-canonical-vector-new}.

Then we can calculate for each $\xi \in \T_\Sigma$ as follows.
\begingroup
\allowdisplaybreaks
\begin{align*}
  \sem{\cA}(\xi) &= \bigoplus_{[\1.\zeta] \in H}  \h_\cA(\xi)_{[\1.\zeta]} \otimes F_{[\1.\zeta]}\\
  &= \bigoplus_{[\1.\zeta] \in H}  [\1.\xi]_{[\1.\zeta]} \otimes \gamma([\1.\zeta])
  \tag{by \eqref{equ:hom-is-canonical-vector-new} and by the definition of $F_{[\1.\zeta]}$}\\
                 &= \bigoplus_{[\1.\zeta] \in H}   \gamma\big([\1.\xi]_{[\1.\zeta]} \cdot [\1.\zeta]\big)
  \tag{because $\gamma$ is a linear form; note that $\otimes$ is a the scalar multiplication in the $\B$-vector space $(B,\oplus,\0)$}\\
                 &= \gamma\big( \bigoplus_{[\1.\zeta] \in H}  [\1.\xi]_{[\1.\zeta]} \cdot [\1.\zeta] \big)
  \tag{because $\gamma$ is a linear form}\\
                 &= \gamma([\1.\xi]) = r(\xi)
                   \tag{because $\approx$ saturates $r$ via $\gamma$}\enspace.
  \end{align*}
  \endgroup

  \

Proof of (2): This follows from the assumption on the computability of the coefficients in the linear combination of a vector over a basis.
\end{proof}

\begin{example}\label{ex:contruction-of-wta-in-MN7} \rm Here we consider the ranked alphabet $\Sigma$ and the $(\Sigma,\Ratnum)$-weighted tree language $\#_\gamma:~\T_\Sigma~\to~\mathbb{Q}$ of Example~\ref{ex:for-congruence} and construct the $(\Sigma,\Ratnum)$-wta $\wta(\#_\gamma,\approx_{\#_\gamma},H)$ according to Definition~\ref{def:wta-from-finite-dimensional}. First however, we have to show that $\mathsf{Pol}(\Sigma,\Ratnum)/\approx_{\#_\gamma}$ is finite dimensional and give its basis. We abbreviate $[s]_{\approx_{\#_\gamma}}$ by $[s]$ for each $s \in \mathrm{Pol}(\Sigma,\Ratnum)$.

 We claim that $H=\{[1.\alpha], [1.\gamma(\alpha)]\}$ is such a basis. For this, we have to prove that $H$ generates $\mathsf{Pol}(\Sigma,\Ratnum)/\approx_{\#_\gamma}$  and $H$ is linearly independent. 
  
We show  that  $H$ generates $\mathsf{Pol}(\Sigma,\Ratnum)/\approx_{\#_\gamma}$ by proving the following statement.
\begin{eqnarray}
  \begin{aligned}\label{equ:representation-of-VLQr-vectors-in-term-of-base-vectors}
&\text{For each  $[b_1.\xi_1+\ldots+ b_n.\xi_n]$  in $\mathsf{Pol}(\Sigma,\Ratnum)/\approx_{\#_\gamma}$, the equality}\\
&\text{$[b_1.\xi_1+\ldots+ b_n.\xi_n]=a_1\cdot[1.\alpha] + a_2\cdot[1.\gamma(\alpha)]$ holds where }\\
& \hspace*{5mm}  a_1 = b_1 \cdot (1-\#_\gamma(\xi_1)) + \ldots + b_n\cdot(1-\#_\gamma(\xi_n)) \ \text{ and }\\
& \hspace*{5mm} a_2 = b_1\cdot\#_\gamma(\xi_1)+\ldots+ b_n\cdot\#_\gamma(\xi_n).
  \end{aligned}
\end{eqnarray}
(We note that both the multiplication in $\Ratnum$ and the scalar multiplication with elements of $\mathbb{Q}$ are denoted by $\cdot$.)

Let $s_1=b_1.\xi_1+\ldots+ b_n.\xi_n$ and $s_2=a_1.\alpha+a_2.\gamma(\alpha)$. Since $\approx_{\#_\gamma}$ is a congruence, we have
\[[b_1.\xi_1+\ldots+ b_n.\xi_n]=a_1\cdot[1.\alpha] + a_2\cdot[1.\gamma(\alpha)] \text{ if and only if } [s_1]=[s_2].\]
We prove the latter by using \eqref{equ:form-of-ker-Varphi-for-num-gamma}. Since
\begin{align*}
&\mathrm{sumc}(s_2)=a_1+a_2= b_1+\ldots +b_n= \mathrm{sumc}(s_1) \text{ and } \\
&\mathrm{sum}_\gamma(s_2)=a_1\cdot 0 +a_2\cdot 1 = a_2 =  b_1\cdot\#_\gamma(\xi_1)+\ldots+ b_n\cdot\#_\gamma(\xi_n) = \mathrm{sum}_\gamma(s_1),
\end{align*}
by \eqref{equ:form-of-ker-Varphi-for-num-gamma} we obtain $[s_1]=[s_2]$. This proves \eqref{equ:representation-of-VLQr-vectors-in-term-of-base-vectors}, i.e., $H$ generates $\mathsf{Pol}(\Sigma,\Ratnum)/\approx_{\#_\gamma}$.

Next we prove that $[1.\alpha]$ and $[1.\gamma(\alpha)]$ are linearly independent. For this, let $b_1,b_2 \in \mathbb{Q}$ and assume that
\[
  b_1 \cdot[1. \alpha] +  b_2\cdot[1.\gamma(\alpha)] = [\widetilde{0}],\text{ i.e., that } [b_1 . \alpha +  b_2.\gamma(\alpha)] = [\widetilde{0}]\enspace.
\]
By \eqref{equ:form-of-ker-Varphi-for-num-gamma}, we obtain
\[b_1+b_2=\mathrm{sumc}(b_1 . \alpha +  b_2.\gamma(\alpha))=\mathrm{sumc}(\widetilde{0})=0 \text{ and } b_2=\mathrm{sum}_\gamma(b_1 . \alpha +  b_2.\gamma(\alpha))=\mathrm{sum}_\gamma(\widetilde{0})=0\enspace. \]
It follows that $b_1=b_2=0$,
i.e., $H$ is linearly independent. Hence $H=\{[1.\alpha], [1.\gamma(\alpha)]\}$ is a basis of $\mathsf{Pol}(\Sigma,\Ratnum)/\approx_{\#_\gamma}$.

  By using \eqref{equ:form-of-ker-Varphi-for-num-gamma}, for each $\xi\in\T_\Sigma$, we can compute $a_1,a_2\in \mathbb{Q}$ for which 
  $[1.\xi]=a_1\cdot[1.\alpha] + a_2\cdot[1.\gamma(\alpha)]$. For instance, we have  $[1.\gamma^2(\alpha)]=(-1)\cdot[1.\alpha] + 2\cdot[1.\gamma(\alpha)]$. Moreover, by Lemma~\ref{lm:approx-r-is-coarsest}, the linear form $\gamma: \mathsf{Pol}(\Sigma,\Ratnum)/\approx_{\#_\gamma} \to \mathbb{Q}$ defined by
  $\gamma([1.\alpha])=\#_\gamma(\alpha)=0$ and $\gamma([1.\gamma(\alpha)])=\#_\gamma(\gamma(\alpha))=1$ saturates $\approx_{\#_\gamma}$.
  Thus we can construct the $(\Sigma,\Ratnum)$-wta $\wta(\#_\gamma,\approx_{\#_\gamma},H)=(H,\delta,F)$ in Definition~\ref{def:wta-from-finite-dimensional} as follows:
\begin{compactitem}
\item the transition mappings are defined by:
  
  \begin{tabular}{ll}
    $\delta_0(\varepsilon,\alpha,[1.\alpha]) = 1$   &  $\delta_0(\varepsilon,\alpha,[1.\gamma(\alpha)])=0$\\[2mm]
  $\delta_1([1.\alpha],\gamma,[1.\alpha]) = 0$    & $\delta_1([1.\alpha],\gamma,[1.\gamma(\alpha)]) =1$\\
  $\delta_1([1.\gamma(\alpha)],\gamma,[1.\alpha]) = -1$ & $\delta_1([1.\gamma(\alpha)],\gamma,[1.\gamma(\alpha)]) =2$\\[2mm]
  $\delta_2([1.\alpha][1.\alpha],\sigma,[1.\alpha])= 1$ &    $\delta_2([1.\alpha][1.\alpha],\sigma,[1.\gamma(\alpha)])= 0$\\
   $\delta_2([1.\alpha][1.\gamma(\alpha)],\sigma,[1.\alpha])= 0$ & $\delta_2([1.\alpha][1.\gamma(\alpha)],\sigma,[1.\gamma(\alpha)])= 1$\\
 $\delta_2([1.\gamma(\alpha)][1.\alpha],\sigma,[1.\alpha])= 0$ &  $\delta_2([1.\gamma(\alpha)][1.\alpha],\sigma,[1.\gamma(\alpha)])= 1$\\
    $\delta_2([1.\gamma(\alpha)][1.\gamma(\alpha)],\sigma,[1.\alpha])= -1$ &  $\delta_2([1.\gamma(\alpha)][1.\gamma(\alpha)],\sigma,[1.\gamma(\alpha)])= 2$
  \end{tabular}
  
  and
 
  \item $F_{[1.\alpha]} = 0$ and $F_{[1.\gamma(\alpha)]} = 1$.
  \end{compactitem}

  For each $\xi \in \T_\Sigma$, we have
  \[
\h_{\wta(\#_\gamma,\approx_{\#_\gamma},H)}(\xi) = \left(\begin{matrix} 1 -\#_\gamma(\xi) \\ \#_\gamma(\xi) \end{matrix}\right) \enspace.
\]
This can be compared with
\[
\h_\cA(\xi) = \left(\begin{matrix} 1 \\ \#_\gamma(\xi) \end{matrix}\right)
\]
where $\cA$ is the $(\Sigma,\Ratnum)$-wta from Example~\ref{ex:illustration-of-kerPsiA}. 
\hfill $\Box$
\end{example}

Alternatively to Definition \ref{def:wta-from-finite-dimensional}, we can obtain a $(\Sigma,\B)$-wta $\cA$ from $r$, $\approx$, and $H$ (as above) with $\sem{\cA}=r$ by using the concept of $(\Sigma,\B)$-multilinear representation and Lemma~\ref{thm:lin-iff-wta-extended} (B)$\Rightarrow$(A). 

Formally, let $\gamma: \Pol(\Sigma,\B)/_{\approx} \to B$ be the linear form over $\sfPol(\Sigma,\B)/_\approx$ such that $r(\xi) = \gamma([\1.\xi]_\approx)$ for each $\xi \in \T_\Sigma$ (by Lemma~\ref{lm:uniqueness-of-linear-form} this $\gamma$ is unique). Moreover, let $\psi: \Pol(\Sigma,\B)/_\approx \to B^H$ be the isomorphism from the $\B$-vector space $\sfPol(\Sigma,\B)/_\approx$ to the $\B$-vector space $(B^H,\oplus,\widetilde{\0})$ as defined on  page~\pageref{page:representing-vector-spaces}. 
Then we define the $(\Sigma,\B)$-multilinear representation $\cM=(H,\eta^*,\gamma^*)$ as follows:
\begin{compactitem}
  \item for every $k \in \mathbb{N}$, $\sigma \in \Sigma^{(k)}$, and $v_1,\ldots,v_k \in B^H$, we let 
\[
\eta^*(\sigma)(v_1,\ldots,v_k) = \psi\big(\ttop_\Sigma/_\approx(\sigma)(\psi^{-1}(v_1),\ldots,\psi^{-1}(v_k))\big) \enspace,
\]
and
\item $\gamma^*: B^H \to B$ is a mapping  defined by $\gamma^*(v) =   \gamma(\psi^{-1}(v))$ for each $v \in B^H$.
\end{compactitem}
We can show that $\sem{\cM}=r$. Then, by  Theorem~\ref{thm:lin-iff-wta-extended}(A)$\Rightarrow$(B), we can construct a $(\Sigma,\B)$-wta $\cA$ such that $\sem{\cA} = \sem{\cM}$.


\subsection{B-A's-theorem and minimality}\label{sect:B-A-LB}

\begin{quote}\emph{In this subsection, we let $\B=(B,\oplus,\otimes,\0,\1)$ denote an arbitrary field, unless specified otherwise}
  \end{quote}

 Now we can prove  B-A's-theorem.

\begin{theorem-rect} \label{th:MN-fields} {\rm \cite[Prop.~2]{bozale89}} Let $\Sigma$ be a ranked alphabet. Moreover, let $\B=(B,\oplus,\otimes,\0,\1)$ be a field and let $r: \T_\Sigma \to B$. Then the following three statements are equivalent.
  \begin{compactenum}
  \item[(A)] $r \in \Rec(\Sigma,\B)$.
  \item[(B)] There exists a congruence relation $\approx$ on the $(\Sigma,\B)$-vector space $(\Pol(\Sigma,\B),\oplus,\widetilde{\0},\ttop_\Sigma)$ such that the $\B$-vector space $\mathsf{Pol}(\Sigma,\B)/_\approx$ is finite-dimensional and $\approx$ saturates $r$.
    \item[(C)] The $\B$-vector space $\mathsf{Pol}(\Sigma,\B)/_{\approx_r}$ is finite-dimensional.
    \end{compactenum}
  \end{theorem-rect}
  \begin{proof}      Proof of  (A)$\Rightarrow$(C): Let $\cA=(Q,\delta,F)$ be a $(\Sigma,\B)$-wta such that $r= \sem{\cA}$. Then the statement follows from Lemma~\ref{lm:dimension-synt-vector-space-bounded}. 

    \
    
Proof of (C)$\Rightarrow$(B): We choose $\approx \ = \approx_r$. Then $\mathsf{Pol}(\Sigma,\B)/_\approx$ is finite-dimensional and, by Lemma~\ref{lm:approx-r-is-coarsest}, $\approx$ saturates $r$.

\

Proof of (B)$\Rightarrow$(A): Let $\approx$ be a congruence relation $\approx$ on the $(\Sigma,\B)$-vector space $(\Pol(\Sigma,\B),\oplus,\widetilde{\0},\ttop_\Sigma)$ such that the $\B$-vector space $\mathsf{Pol}(\Sigma,\B)/_\approx$ is finite-dimensional and $\approx$ saturates $r$.

By Lemma \ref{lm:semantics-constructed-wta}, the $(\Sigma,\B)$-wta $\wta(r,\approx,H)$ recognizes $r$, i.e., 
 $\sem{\rmwta(r,\approx,H)}=r$. Hence $r \in \Rec(\Sigma,\B)$.
    \end{proof}

The next lemma shows that, for each $(\Sigma,\B)$-wta $\cA$, the wta $\wta(\sem{\cA},\approx_{\sem{\cA}},H)$ is an equivalent minimal one where $H$ a finite basis of $\sfPol(\Sigma,\B)/_{\approx_{\sem{\cA}}}$. 
Let $\cA=(Q,\delta,F)$ be a $(\Sigma,\B)$-wta. We say that $\cA$ is \emph{minimal} if, for each $(\Sigma,\B)$-wta $\cB = (Q',\delta',F')$ with $\sem{\cA} = \sem{\cB}$, we have $|Q| \le |Q'|$.
\index{weighted tree automaton!minimal}

\begin{lemma}\rm \label{lm:wta(r,approxr,H)-is-minimal} Let $\cA=(Q,\delta,F)$ be a $(\Sigma,\B)$-wta and $H$ a finite basis of $\sfPol(\Sigma,\B)/_{\approx_{\sem{\cA}}}$. Then $\wta(\sem{\cA},\approx_{\sem{\cA}},H)$ is minimal.
\end{lemma}
\begin{proof}  Let $\cB=(Q',\delta',F')$ be an arbitrary $(\Sigma,\B)$-wta. Then
   \[|Q| = \dim(\mathsf{Pol}(\Sigma,\B)/_{\approx_{\sem{\cA}}})\le |Q'|,\]
   where the equality holds by Definition~\ref{def:wta-from-finite-dimensional} and the approximation follows from  Lemma~\ref{lm:dimension-synt-vector-space-bounded}. 
\end{proof}

\begin{corollary}\rm \label{th:minimal-MN} {\rm \cite[Cor.~on~p.458]{bozale89}, \cite[Prop.~03]{boz91}} Let $\Sigma$ be a ranked alphabet and $\B$ be a field. Moreover, let $\cA$ be a $(\Sigma,\B)$-wta with $r=\sem{\cA}$ and let
$H \subseteq \{[\1.\zeta]_{\approx_r} \mid \zeta \in \T_\Sigma\}$ be a basis of $\sfPol(\Sigma,\B)/_{\approx_r}$.
If, for each $[\1.\xi] \in \Pol(\Sigma,\B)/_{\approx_r}$, the coefficients of the linear combination of $[\1,\xi]$ with respect to $H$ can be computed, then we can construct a minimal $(\Sigma,\B)$-wta which is equivalent to $\cA$.
\end{corollary}

\begin{proof} It follows from Lemmas \ref{lm:semantics-constructed-wta} and \ref{lm:wta(r,approxr,H)-is-minimal}.
\end{proof}


\section[Characterization in terms of vector spaces of left quotients]{Characterization of recognizability in terms of vector spaces of left quotients}
\label{sec:characterization-rec-by-left-quotients}

In this section we recall from \cite{bozlou83} the definitions of left quotient of a weighted tree language $r$ and right quotient of~$r$ for the case that $\B$ is a field. Moreover, we define the $(\Sigma,\B)$-vector space $\VLQ(r)$ of left quotients of $r$ and the $(\Sigma,\B)$-vector space $\VRQ(r)$  of right quotients of $r$. We prove that $\VLQ(r)$ is finite-dimensional if and only if $\VRQ(r)$ is finite-dimensional (cf. Theorem~\ref{lm:Fr=fVr}). Moreover, we give a characterization of recognizability of $r$ in terms of finite dimensionality of $\VLQ(r)$ and of $\VRQ(r)$ (cf. Theorem~\ref{thm:left-quotient-vector-space}).

\begin{quote}\emph{In this section, we let $\B=(B,\oplus,\otimes,\0,\1)$ be an arbitrary field, unless specified otherwise. }
\end{quote}

\index{left quotient}
Let $r: \T_\Sigma \to B$ be a weighted tree language.
For each tree $\xi \in \T_\Sigma$, we define the \emph{left quotient of $r$ with respect to $\xi$}, denoted by $\xi^{-1}r$, to be the $\B$-weighted set 
  \[
\xi^{-1}r: \C_\Sigma \to B \ \ \text{ such that, for each } c \in \C_\Sigma,  \text{ we let } \xi^{-1}r(c) = r(c[\xi]) \enspace.
\]

\index{right quotient}
    Moreover, for each context $c \in \C_\Sigma$, we define the \emph{right quotient of $r$ with respect to $c$}, denoted by $rc^{-1}$, to be the weighted tree language 
  \[
rc^{-1}: \T_\Sigma \to B \ \ \text{ such that, for each } \xi \in \T_\Sigma, \text{ we let } rc^{-1}(\xi) = r(c[\xi]) \enspace.
    \]

Hence, for every $\xi \in \T_\Sigma$ and $c \in \C_\Sigma$, we have $rc^{-1}(\xi) = \xi^{-1}r(c)$ and, in particular, $r z^{-1}(\xi) = r(\xi) = \xi^{-1}r(z)$.

We consider the $\B$-weighted sets $r: \C_\Sigma \to B$, $r_1: \C_\Sigma \to B$, and $r_2: \C_\Sigma \to B$, and the value $b \in B$. In Section \ref{sect:weighted-sets-languages}, we have defined
\begin{compactitem}
\item the \emph{scalar multiplication of $r$ with $b$ (from the left)}, denoted by $b\cdot r$, and 
\item the \emph{sum of $r_1$ and $r_2$},  denoted by $r_1 \oplus  r_2$.
  \end{compactitem}
We note that $\oplus$ is overloaded: $\oplus$ denotes an operation on $B^{\C_\Sigma}$ and an operation on $B^{\T_\Sigma}$. But we think that this notational ambiguity is preferable against a heavy notational discrimination.

  Obviously, the algebras $(B^{\C_\Sigma},\oplus,\widetilde{\0})$ and $(B^{\T_\Sigma},\oplus,\widetilde{\0})$ are commutative monoids; moreover, they are $\B$-semimodules via the scalar multiplication $\cdot$.

\index{LQr@$\LQ(r)$}
  \index{RQr@$\RQ(r)$}
  Let $r: \T_\Sigma \to B$. Moreover, let
\[
   \LQ(r) = \{\xi^{-1}r \in B^{\C_\Sigma} \mid \xi \in \T_\Sigma\} \ \text{ and } \
  \RQ(r) = \{rc^{-1} \in B^{\T_\Sigma} \mid c \in \C_\Sigma\} \enspace.
  \]
  \index{VLQr@$\VLQ(r)$}
  \index{VRQr@$\VRQ(r)$}
  We denote 
  \begin{compactitem}
  \item by $\VLQ(r)$  the $\B$-vector space $(\langle \LQ(r)\rangle_{\{\oplus,\widetilde{\0}\}\cup\{b \cdot  \mid b\in B\}},\oplus,\widetilde{\0})$ of  $(B^{\C_\Sigma},\oplus,\widetilde{\0})$ generated by $\LQ(r)$ and 

\item by $\VRQ(r)$   the $\B$-vector space $(\langle \RQ(r)\rangle_{\{\oplus,\widetilde{\0}\}\cup\{b\cdot \mid b\in B\}},\oplus,\widetilde{\0})$ of  $(B^{\T_\Sigma},\oplus,\widetilde{\0})$ generated by $\RQ(r)$. 
\end{compactitem}

In the following we abbreviate $\langle \LQ(r)\rangle_{\{\oplus,\widetilde{\0}\}\cup\{b \cdot \mid b\in B\}}$ and 
$\langle \RQ(r)\rangle_{\{\oplus,\widetilde{\0}\}\cup\{b \cdot \mid b\in B\}}$ 
by $\langle \LQ(r)\rangle$ and $\langle \RQ(r)\rangle$, respectively.

Next we give an example of $\VLQ(r)$ for a recognizable weighted tree language $r$.

\begin{example}\rm \label{ex:du-det-recog-infinite-index} We consider the string ranked alphabet $\Sigma = \{\gamma^{(1)}, \alpha^{(0)}\}$ and the mapping
  \begin{align*}
    \mathrm{exp}: \T_\Sigma \to \mathbb{N} \ \ \text{ with } \ \ 
    \mathrm{exp}(\xi) = 2^{n+1} \ \text{ if $\xi = \gamma^n(\alpha)$ for some $n \in \mathbb{N}$} 
  \end{align*}
  from Example \ref{ex:run-bimonoid}. Then $\mathrm{exp}$ can be thought of as a $(\Sigma,\Ratnum)$-weighted tree language. Moreover, it is easy to see that the bu-deterministic 
$(\Sigma,\Ratnum)$-wta $\cA = (Q,\delta,F)$ given by
\begin{compactitem}
\item $Q  = \{q\}$, \ $\delta_0(\varepsilon,\alpha,q) = \delta_1(q,\gamma,q) = 2$, and \ $F_q = 1$
\end{compactitem}
recognizes $\exp$, i.e., $\sem{\cA}=\exp$. 

Let us define the mapping $\mathrm{e}: \C_\Sigma \to \mathbb{Q} $ for each $c\in \C_\Sigma$ by $\mathrm{e}(c) = 2^{m}$, where $m \in \mathbb{N}$ is defined by  $c = \gamma^m(z)$. Now let $\xi \in \T_\Sigma$ with $\xi=\gamma^n(\alpha)$ for some 
$n \in \mathbb{N}$. Then, for each $c \in \C_\Sigma$, we have 
\[\xi^{-1}\mathrm{exp}(c)=\mathrm{exp}(c[\xi])=2^{n+1}\cdot\mathrm{e}(c)\enspace.\]
Hence the set $\mathrm{LQ}(\exp)=\{\xi^{-1}\mathrm{exp}\mid  \xi\in \T_\Sigma\}$ is infinite and thus also the $\mathbb{Q}$-vector space $\VLQ(\mathrm{exp})$ is infinite. 

As an exercise, we show that the dimension of $\VLQ(\mathrm{exp})$ is 1 by proving that it is generated by $\{\alpha^{-1}\mathrm{exp}\}$. For this, let 
$b_1\cdot \xi_1^{-1}\mathrm{exp} + \ldots + b_m\cdot \xi_m^{-1}\mathrm{exp}$ be an arbitrary element of $\langle \mathrm{LQ}(\exp)\rangle$ with $\xi_i=\gamma^{n_i}(\alpha)$ for each $i\in[m]$.
We show that
\[b_1\cdot \xi_1^{-1}\mathrm{exp} + \ldots + b_m\cdot \xi_m^{-1}\mathrm{exp}=a\cdot\alpha^{-1}\mathrm{exp},\]
where $a=b_1\cdot 2^{n_1}+\ldots + b_m\cdot 2^{n_m}$. Let $c\in\C_\Sigma$. Then
\begin{align*}
\big(b_1\cdot \xi_1^{-1}\mathrm{exp} + \ldots + b_m\cdot \xi_m^{-1}\mathrm{exp}\big)(c)&\;= b_1\cdot \xi_1^{-1}\mathrm{exp}(c) + \ldots + b_m\cdot \xi_m^{-1}\mathrm{exp}(c) \\
& \,=b_1\cdot 2^{n_1+1}\mathrm{e}(c) + \ldots + b_m\cdot 2^{n_m+1}\mathrm{e}(c) \\
                                                                                       & \,=\big(b_1\cdot 2^{n_1} + \ldots + b_m\cdot 2^{n_m}\big)\cdot 2\cdot\mathrm{e}(c) \\
  & \,= a \cdot \mathrm{exp}(c[\alpha]) \\
& \,= \big(a\cdot \alpha^{-1}\mathrm{exp}\big)(c)\enspace.
\end{align*}\hfill$\Box$
\end{example}

\begin{theorem-rect} {\rm \cite[Thm.~3.1]{bozlou83}} \label{lm:Fr=fVr} Let $\Sigma$ be a ranked alphabet. Let $B$ be a field and $r: \T_\Sigma \to B$. Then the $\B$-vector space  $\VLQ(r)$ is finite-dimensional if and only if  the $\B$-vector space $\VRQ(r)$ is finite-dimensional. Moreover, if $\VLQ(r)$ and $\VRQ(r)$ are finite-dimensional, then $\dim(\VLQ(r)) = \dim(\VRQ(r))$.  
\end{theorem-rect}

\begin{proof} Let us assume that $\VLQ(r)$ is $n$-dimensional for some $n \in \mathbb{N}$ and let $Q_1=\{\xi^{-1}_1r,\ldots,\xi^{-1}_nr\}$ be a basis of $\VLQ(r)$; moreover, let $\xi^{-1}_1r,\ldots,\xi^{-1}_nr$ be an arbitrary but fixed order of the base vectors.
  Then, by the representation of finite-dimensional vector spaces discussed on page \pageref{page:representing-vector-spaces}, the $\B$-vector space $\VLQ(r)$ is isomorphic to the $\B$-vector space 
  $\B^{Q_1}=(B^{Q_1},\oplus,\0^{Q_1})$ using the linear extension of the bijective mapping $\psi: Q_1 \to B^{Q_1}$ defined by $\psi(\xi_i^{-1}r) = \1_{\xi^{-1}_ir}$ for each $i \in [n]$ (note that the $i$th base vector of $\VLQ(r)$ is $\xi_i^{-1}r$, and that $\1_{\xi^{-1}_ir}$ is the $\xi_i^{-1}r$-unit vector of $B^{Q_1}$).  Thus, we are allowed to identify the two $\B$-vector spaces $\VLQ(r)$ and $\B^{Q_1}$, which we will do in the sequel.
  
Next we define the mapping $\psi_1: \RQ(r) \to B^{Q_1}$ such that, for every $c\in \C_\Sigma$ and $i \in [n]$, we let
  \[
    \psi_1(rc^{-1})_{\xi_i^{-1}r} = r(c[\xi_i])\enspace.
  \]
Then we consider its linear extension $\psi'_1: \langle \RQ(r)\rangle \to B^{Q_1}$ from the $\B$-vector space $\VRQ(r)$ to the $\B$-vector space $\B^{Q_1}$, and write $\psi_1$ instead of $\psi'_1$.

   Now we  prove that $\psi_1$ is injective. For this, let $s_1,s_2 \in \langle\RQ(r)\rangle$. We show that $\psi_1(s_1) = \psi_1(s_2)$ implies that $s_1 = s_2$. Since $\psi_1$ is linear, this is equivalent to the implication:
  \[
    \text{if $\psi_1(s_1 \oplus  s_2') = {\0}^{Q_1}$, then $s_1 \oplus s_2' = \widetilde{\0}\enspace,$}
    \]
    where $s_2'$ is obtained from $s_2$ by replacing each coefficient by its inverse in $\B$ (with respect to $\oplus$). We prove this implication by showing that
  \begin{equation}
\text{for each $s \in \langle\RQ(r)\rangle$: if }  \psi_1(s) = {\0}^{Q_1}, \text{ then } s = \widetilde{\0} \enspace. \label{eq:inj-psi}
    \end{equation}

 Let  $s: \T_\Sigma \to B$ be an element of $\langle\RQ(r)\rangle$.
 Thus there exist $m \in \mathbb{N}$, $b_1,\ldots,b_m \in B$, and $c_1, \ldots,c_m \in \C_\Sigma$ such that
    \begin{equation}\label{eq:s-in-detail}
    s=b_1 \cdot rc^{-1}_1 \oplus  \cdots \oplus  b_m \cdot  rc^{-1}_m = \bigoplus_{j \in [m]} b_j \cdot rc_j^{-1} \enspace.
  \end{equation}

  Now let $\xi \in \T_\Sigma$. Then we have: 
  \begin{equation}
    s(\xi)=\bigoplus_{j\in[m]} b_j\otimes rc_j^{-1}(\xi)
    = \bigoplus_{j\in[m]} b_j\otimes \xi^{-1}r(c_j) \enspace. \label{eq:sxi-help}
  \end{equation}

  Since $Q_1$ is a basis of $\VLQ(r)$, there exist $a_1\ldots,a_n \in B$ such that
  \begin{equation}\label{equ:xircj-as-vector}
\xi^{-1}r = a_1 \cdot \xi_1^{-1}r \oplus \ldots \oplus a_n \cdot \xi_n^{-1}r \enspace.
\end{equation}

Then we can calculate as follows.
\begingroup
\allowdisplaybreaks
\begin{align*}
  s(\xi) &= \bigoplus_{j\in[m]} b_j\otimes \xi^{-1}r(c_j) \tag{by \eqref{eq:sxi-help}}\\
         &= \bigoplus_{j\in[m]} b_j\otimes (a_1 \cdot \xi_1^{-1}r \oplus \ldots \oplus a_n \cdot \xi_n^{-1}r)(c_j) \tag{by \eqref{equ:xircj-as-vector}}\\
  &= \bigoplus_{j\in[m]} \Big((b_j \otimes a_1) \otimes \xi_1^{-1}r(c_j) \oplus \ldots \oplus (b_j \otimes a_n) \otimes \xi_n^{-1}r(c_j)\Big) \\
         &= \Big(\bigoplus_{j\in[m]} (b_j \otimes a_1) \otimes \xi_1^{-1}r(c_j)\Big) \oplus \ldots \oplus \Big(\bigoplus_{j\in[m]}(b_j \otimes a_n) \otimes \xi_n^{-1}r(c_j)\Big) \tag{by commutativity and associativity of $\oplus$} \\
         &= a_1 \otimes \Big(\bigoplus_{j\in[m]} b_j\otimes  \xi_1^{-1}r(c_j)\Big) \oplus \ldots \oplus
           a_n \otimes \Big(\bigoplus_{j\in[m]}b_j \otimes \xi_n^{-1}r(c_j)\Big) \tag{by commutativity of $\otimes$ and distributivity} \enspace.
\end{align*}
\endgroup
Since, for each $i \in [n]$, we have $\xi_i^{-1}r(c_j) = \psi_1(rc_j^{-1})_{\xi_i^{-1}r}$, we obtain the following:
\begin{equation}\label{equ:representation-of-sxi}
     s(\xi) = a_1 \otimes \Big(\bigoplus_{j\in[m]} b_j\otimes \psi_1(rc_j^{-1})_{\xi_1^{-1}r} \Big) \oplus \ldots \oplus a_n \otimes \Big(\bigoplus_{j\in[m]}b_j \otimes \psi_1(rc_j^{-1})_{\xi_n^{-1}r}\Big) \enspace.
  \end{equation}

  Now let  $\psi_1(s)=\0^{Q_1}$. Since $\psi_1$ is linear, by \eqref{eq:s-in-detail}, we have
 \[
    \psi_1(s)
    = \bigoplus_{j\in [m]} b_j \cdot \psi_1(rc^{-1}_j) = \0^{Q_1}
    \]
and hence, for each $i \in [n]$, we have
  \[
    \0 =  (\psi_1(s))_{\xi_i^{-1}r} = \left(\bigoplus_{j\in [m]} b_j \cdot \psi_1(rc^{-1}_j)\right)_{\xi_i^{-1}r}=
    \bigoplus_{j\in [m]} b_j \otimes \psi_1(rc^{-1}_j)_{\xi_i^{-1}r} \enspace
  \]
  By \eqref{equ:representation-of-sxi}, it follows that $s = \widetilde{\0}$.
This proves \eqref{eq:inj-psi} and hence that $\psi_1$ is injective.
 
By Sylvester's law of nullity (cf. Theorem \ref{thm:Sylvester-nullity}), we have
\[
  \dim(\VRQ(r)) = \dim(\ker(\psi_1)) + \dim(\im(\psi_1)) \enspace.
\]
Since $\psi_1$ is injective, the kernel of $\psi_1$ only contains the vector $\widetilde{\0}$, and hence $\dim(\ker(\psi_1)) = 0$. Thus
\[
  \dim(\VRQ(r)) = \dim(\im(\psi_1)) \enspace.
\]
Since $\im(\psi_1)$ is a subspace of $\B^{Q_1}$, we obtain that
\[
 \dim(\VRQ(r)) =  \dim(\im(\psi_1)) \le \dim(\B^{Q_1}) = \dim(\VLQ(r))\enspace.
  \]

  For the other direction we can proceed in a similar way as above. First, we assume that $\VRQ(r)$ is finite-dimensional, say $\ell$-dimensional, and with basis $Q_2=\{rc_1^{-1},\ldots,rc_\ell^{-1}\}$; moreover, let $rc_1^{-1},\ldots,rc_\ell^{-1}$ be an arbitrary but fixed order of the base vectors. We identify $\VRQ(r)$ with the $\B$-vector space $\B^{Q_2}=(B^{Q_2},\oplus,\0^{Q_2})$.

Second, we define a linear mapping $\psi_2: \langle \LQ(r) \rangle \to B^{Q_2}$ from  $\VLQ(r)$ to  $\B^{Q_2}$ such that, for every $\xi \in \T_\Sigma$ and $i \in [\ell]$, we let
  \[
    \psi_2(\xi^{-1}r)_{rc_i^{-1}} = r(c_i[\xi])\enspace.
  \]

In a similar way as for $\psi_1$ above, we can prove that $\psi_2$ is injective. Hence $\dim(\VLQ(r)) \le \dim(\VRQ(r))$.

Altogether we obtain the statements of the theorem.
\end{proof}


Let $r: \T_\Sigma \to B$. Now we extend the $\B$-vector space $\VLQ(r) = (\langle \LQ(r)\rangle,\oplus,\widetilde{\0})$  into the particular $(\Sigma,\B)$-vector space $(\langle \LQ(r)\rangle,\oplus,\widetilde{\0},\theta_r)$ by using the following observation  (cf. Subsection~\ref{sec:semimodules} for the concept of $(\Sigma,\B)$-vector space). Since each $\B$-vector space has a basis (assuming Zorn's lemma), also $\VLQ(r)$ has a basis, say $H$ (which need not be finite). Clearly, $H \subseteq \LQ(r)$.
Then, for every $k \in \mathbb{N}$ and $\sigma \in \Sigma^{(k)}$, we define the mapping $\theta_r(\sigma): H^k \to \LQ(r)$ such that, for every $\eta_1^{-1}r,\ldots, \eta_k^{-1}r \in H$, we let\[
\theta_r(\sigma)(\eta_1^{-1}r,\ldots, \eta_k^{-1}r) = \sigma(\eta_1,\ldots,\eta_k)^{-1}r \enspace.
\]

In the canonical way, we extend $\theta_r(\sigma)$  to a $k$-ary multilinear operation on $\langle \LQ(r)\rangle$, and we denote this extension also by $\theta_r(\sigma)$. Formally, let $s_1,\ldots,s_k \in \langle \LQ(r)\rangle$. For every $i \in [k]$, there exist unique $m_i \in \mathbb{N}$, $b_{i1},\ldots,b_{im_i} \in B$, and $\eta_{i1}^{-1}r,\ldots,\eta_{im_i}^{-1}r \in H$ such that
  \[
s_i = \bigoplus_{j \in [m_i]} b_{ij}\cdot (\eta_{ij}^{-1}r) \enspace.
  \]
Then
\begin{align*}
  \theta_r(\sigma)\Big( \bigoplus_{j \in [m_1]} b_{1j}\cdot (\eta_{1j}^{-1}r), \ldots, \bigoplus_{j \in [m_k]} b_{kj}\cdot (\eta_{kj}^{-1}r) \Big) 
  = \bigoplus_{\substack{j_1 \in [m_1]\\\ldots \\ j_k \in [m_k]}} 
  \big( b_{1j_1} \otimes \ldots \otimes b_{kj_k}\big) \cdot
  \theta_r(\sigma)(\eta_{1j_1}^{-1}r,\ldots, \eta_{kj_k}^{-1}r)\enspace.
\end{align*}
\index{homthetar@$\h_{\theta_r}$}
Hence $(\langle \LQ(r)\rangle,\oplus,\widetilde{\0},\theta_r)$ is a $(\Sigma,\B)$-vector space. We denote the unique $\Sigma$-algebra homomorphism from the $\Sigma$-term algebra to the $\Sigma$-algebra $(\langle \LQ(r)\rangle,\theta_r)$ by $\h_{\theta_r}$.


By Theorem~\ref{lm:Pol-SigmaB-initial}, the $(\Sigma,\B)$-vector space $(\Pol(\Sigma,\B),\oplus,\widetilde{\0},\ttop_\Sigma)$ is initial in the set of all $(\Sigma,\B)$-vector spaces. Hence, there exists a unique  $(\Sigma,\B)$-vector space homomorphism from $(\Pol(\Sigma,\B),\oplus,\widetilde{\0},\ttop_\Sigma)$ to $(\langle \LQ(r)\rangle,\oplus,\widetilde{\0},\theta_r)$. We denote this homomorphism by $\Phi_r$, i.e.,
\[
\Phi_r: \Pol(\Sigma,\B) \to \langle \LQ(r) \rangle \ \text{ is a $(\Sigma,\B)$-vector space homomorphism}\enspace.
  \]
  
  The next lemma shows some properties of  $\Phi_r$.
  
\index{Phir@$\Phi_r$}
\begin{lemma}\rm \label{lm:Phir-surjective} Let $r:\T_\Sigma \to \B$. Then the following two statements hold. 
    \begin{compactenum}
    \item[(1)] For  each $s \in \Pol(\Sigma,\B)$ with $s=b_1.\xi_1 \oplus\ldots\oplus b_n.\xi_n$, we have $\Phi_r(s) = b_1\cdot \xi_1^{-1}r \oplus\ldots\oplus b_n\cdot \xi_n^{-1}r$.
    \item[(2)]  $\Phi_r$ is surjective. 
         \end{compactenum}
  \end{lemma}
  
  \begin{proof} 
Proof of (1):  First, by induction on $\T_\Sigma$, we prove the following statement.
\begin{equation}\label{equ:hom-can-do-quotients}
\text{For each $\xi \in \T_\Sigma$ we have: \ $\h_{\theta_r}(\xi) = \xi^{-1}r$.}
  \end{equation}
  Let $\xi = \sigma(\xi_1,\ldots,\xi_k)$. Then we can calculate as follows.
  \begingroup
  \allowdisplaybreaks
  \begin{align*}
    \h_{\theta_r}(\sigma(\xi_1,\ldots,\xi_k))
    &= \theta_r(\sigma)\big(\h_{\theta_r}(\xi_1),\ldots,\h_{\theta_r}(\xi_k)\big)\\
    &= \theta_r(\sigma)\big(\xi_1^{-1}r,\ldots,\xi_k^{-1}r\big)
      \tag{by I.H.}\\
    &= (\sigma(\xi_1,\ldots,\xi_k))^{-1}r  \enspace. \tag{by definition of $\theta_r$}
    \end{align*}
    This proves \eqref{equ:hom-can-do-quotients}.
    
    Let $s = b_1.\xi_1 \oplus\ldots\oplus b_n.\xi_n$ be a $(\Sigma,\B)$-polynomial. Then we can calculate as follows.
    \begin{align*}
      \Phi_r(s) &= \Phi_r(b_1.\xi_1 \oplus\ldots\oplus b_n.\xi_n)\\
                &= b_1\cdot \h_{\theta_r}(\xi_1) \oplus\ldots\oplus b_n\cdot \h_{\theta_r}(\xi_n)\tag{by \eqref{eq:from-h-to-hV}}\\
                &= b_1\cdot \xi_1^{-1}r \oplus\ldots\oplus b_n\cdot \xi_n^{-1}r
                  \tag{by \eqref{equ:hom-can-do-quotients}}
    \end{align*}
    \endgroup

    \
    
    Proof of (2): We let $t= b_1\cdot \xi_1^{-1}r \oplus \ldots \oplus b_n\cdot \xi_n^{-1}r$ be an arbitrary element of $\VLQ(r)$. Then $\Phi_r(b_1.\xi_1 \oplus \ldots \oplus b_n.\xi) =t$. Hence $\Phi_r$ is surjective.
\end{proof}

We recall that for each $(\Sigma,\B)$-weighted tree language $r$, we denote by $\approx_r$ the syntactic congruence of $r$.

\begin{lemma}\label{lm:VLQ-simorphism-lemma}\rm \cite[Prop.~3]{bozale89}, \cite[p.~354]{boz91} For each $r:\T_\Sigma \to B$, we have 
\begin{equation}\label{eq:isomorphism-VLQ}
(\Pol(\Sigma,\B),\oplus,\widetilde{\0},\ttop_\Sigma)/_{\approx_r} \cong (\langle \LQ(r)\rangle,\oplus,\widetilde{\0},\theta_r)\enspace.
\end{equation}
\end{lemma}
\begin{proof} By Theorem \ref{thm:kernel-is-congruence}, we have $(\Pol(\Sigma,\B),\oplus,\widetilde{\0},\ttop_\Sigma)/_{\ker(\Phi_r) } \cong (\langle \LQ(r)\rangle,\oplus,\widetilde{\0},\theta_r)$. Hence it suffices to show that $\ker(\Phi_r)=\approx_r$. For this, let  $s = b_1.\xi_1 \oplus \ldots \oplus b_m.\xi_m$ and $t=a_1.\zeta_1\oplus\ldots\oplus a_n.\zeta_n$ be elements of $\mathrm{Pol}(\Sigma,\B)$.
Then
\begin{align*}
& \Phi_r(s)=\Phi_r(t) \\
  \text{iff \ } & b_1\cdot \xi_1^{-1}r \oplus\ldots\oplus b_m\cdot \xi_m^{-1}r
                  = a_1\cdot \zeta_1^{-1}r \oplus\ldots\oplus a_n\cdot \zeta_n^{-1}r\\
  \text{iff \ } & (\forall c\in \C_\Sigma) : \big( b_1\cdot \xi_1^{-1}r \oplus\ldots\oplus b_m\cdot \xi_m^{-1}r\big)(c)
                  =\big( a_1\cdot \zeta_1^{-1}r \oplus\ldots\oplus a_n\cdot \zeta_n^{-1}r\big)(c)\\
  \text{iff \ } & (\forall c\in \C_\Sigma) : b_1\otimes (\xi_1^{-1}r)(c) \oplus\ldots\oplus b_m\otimes (\xi_m^{-1}r)(c)
  =  a_1\otimes (\zeta_1^{-1}r)(c) \oplus\ldots\oplus a_n\otimes (\zeta_n^{-1})(c)\\                
  \text{iff \ } & (\forall c\in \C_\Sigma) : b_1 \otimes r(c[\xi_1]) \oplus \ldots \oplus b_m \otimes r(c[\xi_m])
                  = a_1 \otimes r(c[\zeta_1]) \oplus \ldots \oplus a_n \otimes r(c[\zeta_n])\\
\text{iff \ } & s \approx_r t\enspace. \qedhere
\end{align*}
\end{proof}

Let $s \in \Pol(\Sigma,\B)$ with $s=b_1.\xi_1 \oplus\ldots\oplus b_n.\xi_n$. In the following we abbreviate $b_1\cdot \xi_1^{-1}r \oplus\ldots\oplus b_n\cdot \xi_n^{-1}r$
by $s^{-1}r$. Using this abbreviation we can write $\Phi_r(s) = s^{-1}r$.

In particular, the isomorphism in \eqref{eq:isomorphism-VLQ} is the mapping
\begin{equation}\label{eq:explicit-isomorphism}
  \psi: \Pol(\Sigma,\B)/_{\approx_r} \to \langle \LQ(r) \rangle \ \text{ with } \
  \psi([s]_{\approx_r}) = s^{-1}r \enspace.
\end{equation}
We recall that the canonical homomorphism from $\sfPol(\Sigma,\B)$ to $\sfPol(\Sigma,\B)/_{\approx_r}$ is denoted by $\pi_{\approx_r}$, i.e.,
\[
\pi_{\approx_r}: \Pol(\Sigma,\B) \to \Pol(\Sigma,\B)/_{\approx_r} \ \text{defined by $\pi_{\approx_r}(s) = [s]_{\approx_r}$ for each $s \in \Pol(\Sigma,\B)$.}
  \]
Thus, using the isomorphism $\psi$ in \eqref{eq:explicit-isomorphism}, we have
\(
\Phi_r = \psi \circ \pi_{\approx_r}.
  \)
Figure~\ref{fig:illustration-Sigma-B-vector-spaces} illustrates the relationship between the involved $(\Sigma,\B)$-vector spaces.

\begin{figure}
  \centering
  \begin{tikzpicture}
    \node at (2,2) (Pol) {$(\Pol(\Sigma,\B),\oplus,\widetilde{\0},\ttop_\Sigma)$};
    \node[below left=2em and 0.01em of Pol] (VLQ) {$(\langle \LQ(r)\rangle,\oplus,\widetilde{\0},\theta_r)$};
    \node[below right=2em and 0.01em of Pol] (PolPhi) {$(\Pol(\Sigma,\B),\oplus,\widetilde{\0},\ttop_\Sigma)/_{\approx_r}$};

    \draw[->] (PolPhi) edge (VLQ);
     \draw[->] (Pol) edge (VLQ);
    \draw[->] (Pol) edge (PolPhi);

     \node at (0.4, 1.5) {$\Phi_r$};
     \node at (4.1, 1.5) {$\pi_{\approx_r}$};
     \node at (2,1.02) {$\psi$};

     \node at (10,2.2) {$\Phi_r(s) = s^{-1}r$};
     \node at (10.45,1.7) {$\approx_r\ = \ker(\Phi_r)$};
     \node at (10.1,1.2) {$\pi_{\approx_r}(s) = [s]_{\approx_r}$};
     \node at (9.86,0.7) {$\psi([s]_{\approx_r}) = s^{-1}r$};
  \end{tikzpicture}
  \caption{\label{fig:illustration-Sigma-B-vector-spaces} Illustration of the relationship between the $(\Sigma,\B)$-vector spaces $(\Pol(\Sigma,\B),\oplus,\widetilde{\0},\ttop_\Sigma)$, $(\langle \LQ(r)\rangle,\oplus,\widetilde{\0},\theta_r)$, and $(\Pol(\Sigma,\B),\oplus,\widetilde{\0},\ttop_\Sigma)/_{\approx_r}$.}
  \end{figure}

We finish this section with a characterization of recognizability in terms of finite dimensionality of left quotients and finite dimensionality of right quotients.

        \begin{theorem-rect} \label{thm:left-quotient-vector-space} {\rm \cite[Thm.~2.1 and 3.1]{bozlou83}} Let $\Sigma$ be a ranked alphabet and $\B=(B,\oplus,\otimes,\0,\1)$ be a field. Moreover, let $r: \T_\Sigma \to B$. The following three statements are equivalent.
          \begin{compactenum}
          \item[(A)] $r \in \Rec(\Sigma,\B)$.
          \item[(B)] $\VLQ(r)$ is finite-dimensional.
          \item[(C)] $\VRQ(r)$ is finite-dimensional.
            \end{compactenum}
          \end{theorem-rect}
          \begin{proof} The proof immediately follows from Theorem \ref{th:MN-fields}, Lemma \ref{lm:VLQ-simorphism-lemma}, and Theorem  \ref{lm:Fr=fVr}.
                    \end{proof}


          \section{B-A's theorem for the deterministic case}
          \label{sec:B-A-BLs-theorem-for-the-det-case}

    In this section, we will recall a deterministic version of B-A's theorem from \cite{fulvog25}: it characterizes the set $\budRec(\Sigma,\B)$ of all bu-deterministically recognizable weighted tree languages over some commutative semifield $\B$ in terms of quotients of appropriate algebras. In principle, we follow the proof of Theorem~\ref{th:MN-fields}.
When analysing this proof under the assumption that $r$ is recognized by some bu-deterministic wta, then one realizes that the addition of the involved $\B$-vector space is not needed. 
This corresponds to Corollary~\ref{cor:bu-det-wta-B1-B2}, which tells us that the summation $\oplus$ of the weight algebra $\B$ is irrelevant when we deal with bu-deterministic $(\Sigma,\B)$-wta. For this reason, we define the concepts of $\B$-scalar algebra and $(\Sigma,\B)$-scalar algebra. Roughly speaking, they  are reducts of the concept of  $\B$-semimodule and $(\Sigma,\B)$-semimodule, respectively, in which the summation $\oplus$ is dropped.

In analogy to the $(\Sigma,\B)$-semimodule $(\Pol(\Sigma,\B),\oplus,\widetilde{\0},\ttop_\Sigma)$ of polynomial weighted tree languages, we define the $(\Sigma,\B)$-scalar algebra $(\mathrm{Mon}(\Sigma,\B),\widetilde{\0},\ttop_\Sigma)$ of monomial weighted tree languages, and we prove that $(\mathrm{Mon}(\Sigma,\B),\widetilde{\0},\ttop_\Sigma)$ is initial in the set of all $(\Sigma,\B)$-scalar algebras (cf. Theorem~\ref{lm:Mon-SigmaB-initial-det}). In the first main result of this section, we will prove that a weighted tree language $r: \T_\Sigma \to B$ over a commutative semifield $\B$  is in $\budRec(\Sigma,\B)$  if and only if the $\B$-scalar algebra $\sfMon(\Sigma,\B)/_{\sim_r}$ is finitely generated, where $\sim_r$ is the restriction of $\approx_r$ to monomials  (cf. Theorem~\ref{thm:MN-semifield-det}). In the second main result of this section, we prove that, for each bu-deterministic wta we can construct an equivalent d-minimal bu-deterministic wta under some reasonable decidability assumptions (cf. Theorem~\ref{thm:minimization-theorem-new}).
We note that an alternative d-minimization algorithm for bu-deterministic wta over commutative semifields was given in \cite{mal08e}.

In contrast to \cite{fulvog25}, we try to keep the development as general as possible. Hence, the first subsections use an arbitrary commutative semiring.

    \subsection{Scalar algebras}
    \label{subsect:scalar-algebras}

    \begin{quote} \emph{In this subsection, we let $\B =(B,\oplus,\otimes,\0,\1)$ be an arbitrary commutative semiring, unless specified otherwise.}
    \end{quote}

Let $\V=(V,0)$ where $V$ is a set and $0\in V$. Moreover, let $\cdot: B\times V \rightarrow V$ be a mapping, called \emph{scalar multiplication} such that the following laws hold for every $b,b' \in B$ and $v \in V$:
    \begin{eqnarray}
&(b \otimes b') \cdot v = b \cdot (b' \cdot v)  \label{SM1-det}\\
&\1 \cdot v = v  \label{SM4-det} \\
&      b \cdot 0 = \0 \cdot v = 0 \label{SM5-det} \enspace.
\end{eqnarray}
    Then we call $\V$ a \emph{$\B$-scalar algebra (via the scalar multiplication $\cdot$)}.
    
We note that $0$ is the unique element of $V$ for which (\ref{SM5-det}) holds. To see this, let $0' \in V$ for which (\ref{SM5-det}) holds. Then, for every $v\in V$, we have $0'=\0 \cdot v= 0$. 
In the sequel we will always assume that the scalar multiplication is denoted by $\cdot$ if not specified otherwise. In particular, $(B,\0)$ is a $\B$-scalar algebra via scalar multiplication $b\cdot v=b\otimes v$ for every $b,v\in B$.
Moreover, for each $\B$-semimodule $(V,+,0)$ via scalar multiplication $\cdot$, the reduct $(V,0)$ is a $\B$-scalar algebra via the same scalar multiplication.

A \emph{$(\Sigma,\B)$-scalar algebra} is a triple $(V,0,\mu)$ where
\begin{compactitem}
\item $(V,0)$ is a $\B$-scalar algebra and 
\item  $(V,\mu)$ is a $\Sigma$-algebra such that, for every $k \in \mathbb{N}_+$, $\sigma \in \Sigma^{(k)}$, $i \in [k]$, $b \in B$, and $v,v_1,\ldots,v_k \in V$, we have
\begin{eqnarray}
  \begin{aligned}
  &\mu(\sigma)\big(v_1,\ldots,v_{i-1},\ b \cdot v \ ,v_{i+1},\ldots,v_k\big) = 
  b \cdot \mu(\sigma)\big(v_1,\ldots,v_{i-1},v,v_{i+1},\ldots,v_k\big) \enspace.
    \end{aligned}\label{ml-Omega-det}
\end{eqnarray}
\end{compactitem}
In particular, (\ref{ml-Omega-det})  (with $b=\0$) and (\ref{SM5-det}) imply that, for every $k \in \mathbb{N}_+$, $\sigma \in \Sigma^{(k)}$, $i \in [k]$,  and $v_1,\ldots,v_k \in V$, we have
\begin{eqnarray}
  \begin{aligned}
  &\mu(\sigma)(v_1,\ldots,v_{i-1},0 ,v_{i+1},\ldots,v_k) = 0\enspace.
    \end{aligned}\label{ml-absorbtive-det}
\end{eqnarray}
Clearly, for each $(\Sigma,\B)$-semimodule $(V,+,0,\mu)$ via scalar multiplication $\cdot$, the reduct $(V,0,\mu)$ is a $(\Sigma,\B)$-scalar algebra via the same scalar multiplication. 

We can represent each $(\Sigma,\B)$-scalar algebra $(V,0,\mu)$ as an algebra $(V,\eta)$ by viewing the scalar multiplications as unary operations on $V$.
For this, we define the index set  $I= \{0\} \cup \{(b\cdot)\mid b \in B\} \cup \Sigma$ and the mapping $\eta: I \to \mathrm{Ops}(V)$ such that
\begin{compactitem}
\item $\eta(0)()=0$,
  \item for every $b \in B$ and $v \in V$, we let $\eta((b \cdot))(v) = b \cdot v$, and
    \item for each $\sigma \in \Sigma$, we let $\eta(\sigma) = \mu(\sigma)$.
\end{compactitem}
And similarly, each $\B$-scalar algebra can be represented as algebra.
Thus, all the concepts like subalgebra,
        congruence, homomorphism, and finitely generated, and all the  results of universal algebra
        are available for the concepts of $\B$-scalar algebra and  $(\Sigma,\B)$-scalar algebra.

        We call a  $\B$-scalar algebra homomorphism a \emph{scalar-linear mapping} (in analogy to the concept of linear mappings between vector spaces).
        
Let $(V,0)$ be a $\B$-scalar algebra. A \emph{scalar-linear} form is a scalar-linear mapping from $(V,0)$ to the $\B$-scalar algebra $(B,\0)$, i.e., it is a mapping $\gamma: V \to B$ such that $\gamma(b\cdot v) = b \otimes \gamma(v)$ for every $b\in B$ and $v \in V$. 

For each $H \subseteq V$ with $H\ne \emptyset$, we let $B \cdot H$ denote the set $\{b \cdot v \mid b \in B, v \in H\}$.

\begin{observation}\rm \label{obs:scalar-algebra-fin-gen-subalg-hom-1} 
  Let $(V,0)$ be a $\B$-scalar algebra. Then the following two statements hold.
  \begin{compactenum}
  \item[(1)] Let $H \subseteq V$ with $H\ne \emptyset$. Then 
  $H$ generates $\V$  (i.e, $\langle H \rangle_{\{0\} \cup \{\eta(b\cdot) \mid b \in B\}}=V$)  if and only if 
  $B\cdot H =V$.
  
  \item[(2)] Let $(V',0')$ be a $\B$-scalar algebra and  $h: V \to V'$ a scalar-linear mapping from  $(V,0)$ to $(V',0')$. Then $(\im(V),0')$ is a sub-$\B$-scalar algebra of $(V',0')$.
    \end{compactenum}
\end{observation}
\begin{proof} Proof of (1): This is obvious.

  \
  
  Proof of (2): This follows from Lemma~\ref{lm:hom-image=subalgebra} and the fact that $(V,0)$ and $(V',0')$ are particular algebras.
  \end{proof}

\subsection{The $(\Sigma,\B)$-scalar algebra  of monomial weighted tree languages}
\label{subsec:sc-alg-general-and-monomials}

    \begin{quote} \emph{In this subsection, we let $\B =(B,\oplus,\otimes,\0,\1)$ be an arbitrary commutative semiring, unless specified otherwise.}
    \end{quote}

Here we  define a particular $\B$-scalar algebra and a particular $(\Sigma,\B)$-scalar algebra, both are based on monomial weighted tree languages. We recall that a monomial $(\Sigma,\B)$-weighted tree language is a mapping $r:\T_\Sigma \to B$ such that there exist a $b \in B$ and a tree $\xi\in\T_\Sigma$ with $r(\xi)=b$ and $r(\zeta)=\0$ for each $\zeta\in\T_\Sigma$ with $\zeta \ne \xi$. We denote this monomial by $b.\xi$.

Now let
 \[
\sfMon(\Sigma,\B) = (\rmMon(\Sigma,\B),\widetilde{\0})
\]
where
\begin{compactitem}
\item $\rmMon(\Sigma,\B) = \{b.\xi \mid b \in B, \xi \in \T_\Sigma\}$ (i.e., $\rmMon(\Sigma,\B)$ is the set of monomial $(\Sigma,\B)$-weighted tree languages) and
\item $\widetilde{\0}: \T_\Sigma \to B$ is the constant zero $(\Sigma,\B)$-weighted tree language (i.e., $\widetilde{\0}(\xi) = \0$ for each $\xi \in \T_\Sigma$),
 \end{compactitem}
 and let
 \[\cdot: B\times \rmMon(\Sigma,\B) \rightarrow \rmMon(\Sigma,\B)\] be the scalar multiplication of $(\Sigma,\B)$-weighted tree languages, i.e.,  for each $a \in B$ and $b.\xi \in \rmMon(\Sigma,\B)$, we have $a \cdot (b.\xi) = (a \otimes b).\xi$.

 It is easy to check that the laws \eqref{SM1-det}, \eqref{SM4-det}, and \eqref{SM5-det} hold, hence $\sfMon(\Sigma,\B)$ is a  $\B$-scalar algebra via the scalar multiplication $\cdot$. We call it the \emph{$\B$-scalar algebra  of monomial $(\Sigma,\B)$-weighted tree languages}.

Clearly,  $\sfMon(\Sigma,\B)$ is generated by the set $\{\1.\xi\mid \xi\in \T_\Sigma\}$.
 However, it is not finitely generated.
 To show this, assume that there exists a finite subset $H\subseteq \rmMon(\Sigma,\B)$ which generates $\sfMon(\Sigma,\B)$. 
We may assume without lost of  generality that $H\subseteq \{\1.\xi\mid \xi\in \T_\Sigma\}$. 
Since $H$ is finite, there exists $\zeta \in \T_\Sigma$ such that $\1.\zeta \not \in H$. Since $H$ is a generating set, there exists $b\in B$ and $\1.\xi\in H$ with $\1.\zeta=b\cdot(\1.\xi)=b.\xi$. This is a contradiction because $\zeta\ne \xi$.

The relationship between the $\B$-scalar algebra $(\mathrm{Mon}(\Sigma,\B),\widetilde{\0})$ and the $\B$-semimodule $(\mathrm{Pol}(\Sigma,\B),\oplus,\widetilde{\0})$ is the following: the $\B$-scalar algebra $(\rmMon(\Sigma,\B),\widetilde{\0})$ is a subalgebra of the  $\B$-scalar algebra $(\Pol(\Sigma,\B),\widetilde{\0})$; in its turn, the $\B$-scalar algebra $(\Pol(\Sigma,\B),\widetilde{\0})$ is a reduct of the $\B$-semimodule $(\Pol(\Sigma,\B),\oplus,\widetilde{\0})$.

\begin{theorem-rect}\rm \label{lm:Mon-SigmaB-free}  Let $\Sigma$ be a ranked alphabet and $\B$ be a commutative semiring. The $\B$-scalar algebra $\sfMon(\Sigma,\B)$ is free in the set of all $\B$-scalar algebras with generating set $\1.\T_\Sigma=\{\1.\xi \mid \xi \in \T_\Sigma\}$.
\end{theorem-rect}
\begin{proof} The proof is the same as the one of Theorem \ref{lm:Pol-SigmaB-free} except that we have to replace $\sfPol(\Sigma,\B)$ by $\sfMon(\Sigma,\B)$ and we have to deal with monomials instead of polynomials.
\end{proof}

In Section \ref{sec:SigmaB-vector-space-of-polynomials} we have seen that, for each $\sigma \in \Sigma$, the top-concatenation $\ttop_\Sigma(\sigma)$ is an operation on $\Pol(\Sigma,\B)$. In particular, by Lemma~\ref{polynomials-closed-under-top(sigma)}, 
for every $k\in \mathbb{N}$,  $\sigma \in \Sigma^{(k)}$, and $b_1.\xi_1,\ldots,b_k.\xi_k\in \rmMon(\Sigma,\B)$, we have
\begin{equation}\label{equ:top-for-monomials}
  \ttop_\Sigma(\sigma)(b_1.\xi_1,\ldots,b_k.\xi_k)= (b_1\otimes\ldots\otimes b_k).\sigma(\xi_1,\ldots,\xi_k)\enspace.
  \end{equation}
Since the right-hand side of \eqref{equ:top-for-monomials} is a monomial,  $\ttop_\Sigma(\sigma)$ is an operation also on $\rmMon(\Sigma,\B)$.

  \begin{lemma}\rm \label{lm:monomials-are-smallest-closed-under-scalar-m-top-det} (cf. \cite[Lm.~15.4.1]{fulvog24}) The set $\rmMon(\Sigma,\B)$ is the  smallest set of $(\Sigma,\B)$-weighted tree languages which is closed under scalar multiplications and top-concatenations. 
\end{lemma}
\begin{proof}
 
  For convenience, we denote by $\cC$ the  smallest set of $(\Sigma,\B)$-weighted tree languages which is closed under scalar multiplications and top-concatenations.
  
  Since $\rmMon(\Sigma,\B)$ is closed under scalar multiplications and top-concatenations, we have $\cC \subseteq \rmMon(\Sigma,\B)$.  
  
  For  the proof of the other inclusion, first we show that the monomial $\1.\xi$ is in $\cC$ for each $\xi \in \T_\Sigma$. We use induction on $\T_\Sigma$. Let $\xi = \sigma(\xi_1,\ldots,\xi_k)$ and assume that $\1.\xi_i$ is in $\cC$ for each $i \in [k]$. Since $\cC$ is closed under  top-concatenations, we obtain that 
  \(\1.\xi =
    \ttop_\Sigma(\sigma)(\1.\xi_1,\ldots,\1.\xi_k)\) is also in $\cC$. (In case $k=0$, we have $\1.\xi =\1.\sigma$.)
  
  Now let $b.\xi \in \rmMon(\Sigma,\B)$ for some $b\in B$ and $\xi\in\T_\Sigma$. Since  $\1.\xi \in \cC$ and $\cC$ is closed under scalar multiplications,
  we obtain that  $b.\xi=b\cdot (\1.\xi)$ is also in $\cC$.
    \end{proof}

By \eqref{equ:top-for-monomials} and commutativity of $\otimes$, for each $\sigma \in \Sigma$, the operation  $\ttop_\Sigma(\sigma)$ on $\rmMon(\Sigma,\B)$ satisfies the law (\ref{ml-Omega-det}). Therefore
\[
    (\rmMon(\Sigma,\B),\widetilde{\0},\ttop_\Sigma) 
  \]
is a $(\Sigma,\B)$-scalar algebra. We call it the \emph{$(\Sigma,\B)$-scalar algebra  of monomial $(\Sigma,\B)$-weighted tree languages}.

The importance of $(\mathrm{Mon}(\Sigma,\B),\widetilde{\0},\ttop_\Sigma)$ is shown in the next theorem: it is initial in the set of all $(\Sigma,\B)$-scalar algebras. This corresponds to the fact that $(\mathrm{Pol}(\Sigma,\B),\oplus,\widetilde{\0},\ttop_\Sigma)$ is initial in the set of all $(\Sigma,\B)$-semimodules (cf. Theorem~\ref{lm:Pol-SigmaB-initial}).

\begin{theorem-rect} \label{lm:Mon-SigmaB-initial-det} The $(\Sigma,\B)$-scalar algebra $(\mathrm{Mon}(\Sigma,\B),\widetilde{\0},\ttop_\Sigma)$ is initial in the set of all $(\Sigma,\B)$-scalar algebras.
\end{theorem-rect}
\begin{proof} Again the proof can be easily derived from the proof of Theorem \ref{lm:Pol-SigmaB-initial}. When proving that $(\rmMon(\Sigma,\B),\widetilde{\0},\ttop_\Sigma)$ is generated by $\emptyset$, we use Lemma~\ref{lm:monomials-are-smallest-closed-under-scalar-m-top-det}.
\end{proof}


\subsection{The m-syntactic congruences and m-syntactic $(\Sigma,\B)$-scalar algebras}
\label{subsec:syntactic-congruence-det}

    \begin{quote} \emph{In this subsection, we let $\B =(B,\oplus,\otimes,\0,\1)$ be an arbitrary commutative semiring, unless specified otherwise.}
    \end{quote}

In Section \ref{sec:Syntactic-congruences-and-syntactic-SigmaB-vector-spaces},  the syntactic congruence $\approx_r$
of a $(\Sigma,\B)$-weighted tree language $r: \T_\Sigma \to B$  over the commutative semiring $\B$ is defined as a congruence on the $(\Sigma,\B)$-semimodule $(\Pol(\Sigma,\B),\oplus,\widetilde{\0},\ttop_\Sigma)$ of polynomial weighted tree languages. Here we consider the restriction of  $\approx_r$ to monomials, i.e., we let
\[\sim_r \, = \,\approx_r \cap \,\big(\rmMon(\Sigma,\B) \times \rmMon(\Sigma,\B)\big).\]
In more detail, for every $b_1.\xi_1, b_2.\xi_2 \in \rmMon(\Sigma,\B)$, we let
\[
b_1.\xi_1 \sim_r b_2.\xi_2 \ \text{ iff } \ (\forall c \in \C_\Sigma): b_1 \otimes r(c[\xi_1]) = b_2 \otimes r(c[\xi_2]) \enspace.
  \]
  
  Since, by Lemma~\ref{lm:simr-is-a-congruence}, $\approx_r$ is a congruence on the $(\Sigma,\B)$-semimodule $(\Pol(\Sigma,\B),\oplus,\widetilde{\0},\ttop_\Sigma)$,  it is also a congruence on the $(\Sigma,\B)$-scalar algebra $(\Pol(\Sigma,\B),\widetilde{\0},\ttop_\Sigma)$.
  Moreover, since $(\mathrm{Mon}(\Sigma,\B),\widetilde{\0},\ttop_\Sigma)$ is a subalgebra of $(\Pol(\Sigma,\B),\widetilde{\0},\ttop_\Sigma)$, by Lemma \ref{lm:congruence-on-subalgebra},
  the relation $\sim_r$ is a congruence on $(\mathrm{Mon}(\Sigma,\B),\widetilde{\0},\ttop_\Sigma)$.
   
  Since $\sim_r$ is the restriction of $\approx_r$ to \underline{m}onomials, we call it the \emph{m-syntactic congruence of $r$}. Moreover, we call the quotient algebras 
  \begin{align*}
     \sfMon(\Sigma,\B)/_{\sim_r} &= (\rmMon(\Sigma,\B)/_{\sim_r},[\widetilde{\0}]_{\sim_r}) \ \ \text{and}\\
    (\rmMon(\Sigma,\B),\widetilde{\0},\ttop_\Sigma)/_{\sim_r} &= (\rmMon(\Sigma,\B)/_{\sim_r},[\widetilde{\0}]_{\sim_r},\ttop_\Sigma/_{\sim_r})
  \end{align*}
  the \emph{m-syntactic $\B$-scalar algebra} and the \emph{m-syntactic $(\Sigma,\B)$-scalar algebra (of $r$)}, respectively.

  \begin{example}\rm \label{ex-syntactic-congruence} Let $\Sigma=\{\sigma^{(2)}, \alpha^{(0)}\}$ be a ranked alphabet. We consider the $(\Sigma,\Ratnum)$-weighted tree language $\mathrm{weo}: \T_\Sigma \to \mathbb{Q}$ in Example \ref{ex:bu-det+total-wta-det-weo}  and compute the congruence $\sim_{\mathrm{weo}}$. By definition, for each $\xi \in \T_\Sigma$, we have
    \[
      \mathrm{weo}(\xi) = \begin{cases} 2 \cdot 2^{\#_\alpha(\xi)} & \text{ if $\#_\alpha(\xi)$ is even}\\
       3 \cdot 2^{\#_\alpha(\xi)} & \text{ otherwise,} \end{cases}
   \]
   where $\#_\alpha(\xi)$ denotes the number of occurrences of $\alpha$ in $\xi$.

 Then, for every $b_1.\xi_1,b_2.\xi_2 \in \rmMon(\Sigma,\Ratnum)$, we have 
\begin{align*} 
& \hspace*{5mm} b_1.\xi_1 \sim_r b_2.\xi_2   \\
&  \text{iff } (\forall c \in \C_\Sigma): b_1 \otimes \mathrm{weo}(c[\xi_1]) = b_2 \otimes \mathrm{weo}(c[\xi_2])\\
&  \text{iff }   \Big[(\forall c\in \C_\Sigma) : \Big( \big(b_1\cdot 2 \cdot 2^{\#_\alpha(c[\xi_1])} = b_2\cdot 2 \cdot 2^{\#_\alpha(c[\xi_2])} \text{ and } \#_\alpha(c[\xi_1]) \text{ and } \#_\alpha(c[\xi_2]) \text{ are even} \big) \text{ or }\\
& \hspace*{25mm} \big( b_1\cdot 3 \cdot 2^{\#_\alpha(c[\xi_1])} = b_2\cdot 3 \cdot 2^{\#_\alpha(c[\xi_2])} \text{ and } \#_\alpha(c[\xi_1]) \text{ and } \#_\alpha(c[\xi_2]) \text{ are odd}\big)\Big)\Big]\\
&  \hspace*{25mm} \text{or \ } b_1=b_2=0\\[2mm]
&  \text{iff } \Big(\big( b_1\cdot 2^{\#_\alpha(\xi_1)} = b_2\cdot  2^{\#_\alpha(\xi_2)} \text{ and } \#_\alpha(\xi_1) \text{ and } \#_\alpha(\xi_2) \text{ are even }\big) \text{ or }\\
 & \hspace*{5mm} \big(  b_1\cdot 2^{\#_\alpha(\xi_1)} = b_2\cdot  2^{\#_\alpha(\xi_2)} \text{ and } \#_\alpha(\xi_1) \text{ and } \#_\alpha(\xi_2) \text{ are odd } \big)\Big)\\
 & \hspace*{5mm} \text{or \ } b_1=b_2=0\\[2mm]
 &   \text{iff } \Big(b_1 \cdot 2^{\#_\alpha(\xi_1)} =  b_2 \cdot 2^{\#_\alpha(\xi_2)} \text{\  and\  } \big(\#_\alpha(\xi_1) \! \mod(2)\big) = \big(\#_\alpha(\xi_2) \! \mod (2)\big)\Big)\text{ or \ } b_1=b_2=0\enspace,
\end{align*}
where at the third equivalence we extend $\#_\alpha$ to contexts in the obvious way and use the fact that $\#_\alpha(c[\xi])=\#_\alpha(c)+\#_\alpha(\xi)$ for each $\xi\in \T_\Sigma$. Clearly,  dependency in $\sfMon(\Sigma,\B)/_{\sim_{\mathrm{weo}}}$ is decidable.
\hfill$\Box$
\end{example}

\begin{lemma}\rm \label{lm:sim-r-is-coarsest} Let $r:\T_\Sigma \to \B$. Then $\sim_r$ is the coarsest congruence among the congruences on $(\rmMon(\Sigma,\B),\widetilde{\0},\ttop_\Sigma)$ which saturate $r$.
\end{lemma}

\begin{proof} The proof of the statement that $\sim_r$ saturates $r$ is similar to the proof of the statement that $\approx_r$ saturates $r$ (cf. Lemma~\ref{lm:approx-r-is-coarsest}).

  Let $\sim$ be a congruence on $(\rmMon(\Sigma,\B),\widetilde{\0},\ttop_\Sigma)$ which saturates $r$  via the scalar-linear form 
$\gamma: \rmMon(\Sigma,\B)/_{\sim}  \to B$. Moreover, let $b_1.\xi_1, b_2.\xi_2 \in \rmMon(\Sigma,\B)$ such that
$b_1.\xi_1 \sim b_2.\xi_2$. Then
\begingroup
\allowdisplaybreaks
\begin{align*}
& b_1.\xi_1 \sim b_2.\xi_2 \\
\Rightarrow \ \ & (\forall c\in \C_\Sigma): b_1.c[\xi_1] \sim b_2.c[\xi_2] \tag{because $\sim$ is a congruence} \\
\Leftrightarrow \ \ & (\forall c\in \C_\Sigma): [b_1.c[\xi_1]]_\sim = [b_2.c[\xi_2]]_\sim \\
\Rightarrow \ \ & (\forall c\in \C_\Sigma): \gamma\big([b_1.c[\xi_1]]_\sim\big) = \gamma\big([b_2.c[\xi_2]]_\sim\big) \\
\Rightarrow \ \ & (\forall c\in \C_\Sigma): b_1\cdot\gamma\big([\1.c[\xi_1]]_\sim\big) = b_2\cdot\gamma\big([\1.c[\xi_2]]_\sim\big) \tag{because $\gamma$ is a scalar-linear form}\\
\Rightarrow \ \ & (\forall c\in \C_\Sigma): b_1\cdot r(c[\xi_1]) =  b_2\cdot r(c[\xi_2]) \tag{because $\sim$ saturates $r$ via $\gamma$}\\
  \Rightarrow \ \ &  b_1.\xi_1 \sim_r b_2.\xi_2 \enspace.
                    \qedhere
\end{align*}
\endgroup
\end{proof}

    Intuitively, we can retrieve from $\sim_r$ the weighted tree language $r$ by means of a scalar-linear form. To formalize this, we consider an arbitrary $r: \T_\Sigma \to B$ and an arbitrary congruence $\sim$ on $(\rmMon(\Sigma,\B),\widetilde{\0},\ttop_\Sigma)$. We say that $\sim$ \emph{saturates $r$} if there exists a scalar-linear form $\gamma: \rmMon(\Sigma,\B)/_\sim \to B$ such that, for every $\xi \in \T_\Sigma$, we have that $r(\xi) = \gamma([\1.\xi]_\sim)$. If this is the case, then we say that  \emph{$\sim$ saturates $r$ via $\gamma$}.
    
    \begin{lemma}\rm \label{lm:uniqueness-of-scalar-linear-form} Let $r: \T_\Sigma \to B$ and $\sim$ be a congruence on $(\rmMon(\Sigma,\B),\widetilde{\0},\ttop_\Sigma)$ such that $\sim$ saturates $r$. Then there exists exactly one scalar-linear form $\gamma: \rmMon(\Sigma,\B)/_\sim \to B$ such that $\sim$ saturates $r$ via $\gamma$.
    \end{lemma}
    \begin{proof} The proof is similar to the one of Lemma~\ref{lm:uniqueness-of-linear-form}.
      \end{proof}


\subsection{The $(\Sigma,\B)$-scalar algebra associated to  a bu-deterministic wta}
\label{subsec:sc-alg-wta}

\begin{quote} {\em In this subsection, we let $\B=(B,\oplus,\otimes,\0,\1)$ denote an arbitrary commutative semiring unless specified otherwise. Moreover, we let  $\cA=(Q,\delta,F)$ denote an arbitrary bu-deterministic $(\Sigma,\B)$-wta. 
  }
\end{quote}

In a natural way, $\cA$  induces a $(\Sigma,\B)$-scalar algebra as follows. We recall that 
\[B^Q_{=1}=\{v \in B^Q \mid |\supp(v)| = 1\} \ \text{ and } \ \ B^Q_{\le 1}= B^Q_{= 1} \cup \{\0^Q\}\enspace.\] 
For each $v \in B^Q_{= 1}$, we denote by $q_v$ the only state of $\supp(v)$. 
Then $(B^Q_{\le 1},\0^Q)$ is a $\B$-scalar algebra via the scalar multiplication $\cdot: B \times B^Q_{\le 1} \to B^Q_{\le 1}$ defined in the obvious way.

\index{dMA@$\sfdM(\cA)$}
\index{hA@$\sfh_\cA$}
By Lemma~\ref{lm:properties-sem-of-budet-wta-det-new}(2), $(B^Q_{\le 1},\delta_\cA)$ is a $\Sigma$-algebra and, 
by Lemma~\ref{lm:properties-sem-of-budet-wta-det-new}(1), for each $\sigma \in \Sigma$, the operation $\delta_\cA(\sigma)$ satisfies the law \eqref{ml-Omega-det}.
Hence the triple 
\[
  \sfdM(\cA)=(B^Q_{\le 1},\0^Q,\delta_\cA)
\]
is a $(\Sigma,\B)$-scalar algebra via scalar multiplication $\cdot$.

Since, by Theorem~\ref{lm:Mon-SigmaB-initial-det}, $(\mathrm{Mon}(\Sigma,\B),\widetilde{\0},\ttop_\Sigma)$ is initial in the set of all $(\Sigma,\B)$-scalar algebras, there exists a unique scalar-linear mapping from $(\mathrm{Mon}(\Sigma,\B),\widetilde{\0},\ttop)$ to $\sfdM(\cA)$. We denote it by~$\sfh_\cA$. By \eqref{eq:from-h-to-hV} restricted to monomials, for each $b.\xi \in \rmMon(\Sigma,\B)$, we have
\begin{equation}\label{equ:sfh-related-to-h}
\sfh_\cA(b.\xi) = b \cdot \h_\cA(\xi) \enspace,
\end{equation}
because, by \eqref{eq:bu-det-init-hom-multiplication-new} we have that $\im(\h_\cA) \subseteq B_{\le 1}^Q$, and thus $\h_\cA$ is the unique $\Sigma$-algebra homomorphism from $\sfT_\Sigma$ also to the $\Sigma$-algebra $(B^Q_{\le 1},\delta_\cA)$.

\index{dMAim@$\sfdM_{\mathrm{im}}(\cA)$}
By Observation~\ref{obs:smallest-subalgebra-im}, the algebra $\sfdM_{\mathrm{im}}(\cA)$, defined by
\[
  \sfdM_{\mathrm{im}}(\cA) = (\im(\sfh_\cA),\0^Q,\delta_\cA) \enspace,
\]
is the smallest sub-$(\Sigma,\B)$-scalar algebra of $\sfdM(\cA)$.
By Corollary~\ref{cor:image-of-hom-isomorphic-to-quotient-of-kernel}, we have
\begin{equation}\label{eq:isomorphism}
(\rmMon(\Sigma,\B),\widetilde{\0},\ttop_\Sigma)/_{\ker(\sfh_\cA)} \cong  \sfdM_{\mathrm{im}}(\cA) \enspace,
\end{equation}
where $\ker(\sfh_\cA)$ is the kernel of $\sfh_\cA$.
In a part of Figure \ref{fig:overview-congruence-relations-isos} we visualize \eqref{eq:isomorphism}.

    \begin{figure}[t]
 \begin{center}
   \begin{tikzpicture}
     \node (1){$\sfdM(\cA) =(B_{\le 1}^Q,\0^Q,\delta_\cA)$};
     \node[right of=1, xshift=25em] (2) {$(\rmMon(\Sigma,\B),\widetilde{\0},\ttop_\Sigma)$};
     \node[below of=1, yshift=-4em] (3) {$\sfdM_{\mathrm{im}}(\cA) =(\im(\sfh_\cA),\0^Q,\delta_\cA)$};
     \node[right of=3, xshift=13em] (4) {$(\rmMon(\Sigma,\B),\widetilde{\0},\ttop_\Sigma)/_{\ker(\sfh_\cA)}$};
     \node[below of=4, yshift=-3em] (5) {$\Big((\rmMon(\Sigma,\B),\widetilde{\0},\ttop_\Sigma)/_{\ker(\sfh_\cA)}\Big)/_{\rho_\cA}$};
     \node[right of=5, xshift=15em] (6) {$(\rmMon(\Sigma,\B),\widetilde{\0},\ttop_\Sigma)/_{\sim_{\sem{\cA}}}$};

     \draw (2) edge[->,>=stealth] node[fill=white] {$\sfh_\cA$} (1);
        \draw (1) edge[->,>=stealth] node[fill=white] {smallest sub-$(\Sigma,\B)$-scalar algebra} (3);
     \draw (2) edge[->,>=stealth] node[fill=white] {$\pi_{\ker(\sfh_\cA)}$} (4);
     \draw (2) edge[->,>=stealth] node[fill=white] {$\pi_{\sim_{\sem{\cA}}}$} (6);
     \draw (4) edge[->,>=stealth] node[fill=white] {$\pi_{\rho_\cA}$} (5);

          \node[right of=3, xshift=5.2em] (7) {\Large $\cong$};
          \node[right of=5, xshift=7em] (8) {\Large $\cong$};

   \end{tikzpicture}
 \end{center}
 \caption{\label{fig:overview-congruence-relations-isos} Overview of the relation among several $(\Sigma,\B)$-scalar algebras which occur in Theorem~\ref{lm:congruences-modulo-congruences-det}. }
      \end{figure}

\begin{example}\rm \label{ex:kernel-congr-of-running-example-det} We determine the congruence  $\ker(\sfh_\cA)$ for the total and bu-deterministic $(\Sigma,\Ratnum)$-wta $\cA$ in Example~\ref{ex:bu-det+total-wta-det-weo}. By using \eqref{equ:number-of-alphas-det}, for every $b_1.\xi_1,b_2.\xi_2 \in \rmMon(\Sigma,\Ratnum)$, we obtain
  \begin{align*}
  b_1.\xi_1 \ \ker(\sfh_\cA) \  b_2.\xi_2  \ \  \text{ iff } \ \ &
   \Big(b_1 \cdot 2^{\#_\alpha(\xi_1)} =  b_2 \cdot 2^{\#_\alpha(\xi_2)} \text{\  and\  } \big(\#_\alpha(\xi_1)\!\mod (2)\big) = \big(\#_\alpha(\xi_2)\!\mod (2)\big)\Big)\\ & \text{or } b_1=b_2=0\enspace.
  \end{align*}
  For instance
\[
  8.\alpha 
   \ \ker(\sfh_\cA) \  
  2. \sigma(\sigma(\alpha,\alpha),\alpha) \hspace{7mm} 
  4.\alpha 
   \ \ker(\sfh_\cA) \  
  1. \sigma(\sigma(\alpha,\alpha),\alpha)  \hspace{7mm} 
  1.\alpha 
   \ \ker(\sfh_\cA) \  
  \tfrac{1}{4}. \sigma(\sigma(\alpha,\alpha),\alpha) \ \ \text{ and }
\]
\[
  16.\sigma(\alpha,\alpha) \  \ker(\sfh_\cA) \  
  4.\sigma(\sigma(\alpha,\alpha),\sigma(\alpha,\alpha)) \hspace{7mm}
   2.\sigma(\alpha,\alpha) \ \ker(\sfh_\cA) \  
  \tfrac{1}{2}.\sigma(\sigma(\alpha,\alpha),\sigma(\alpha,\alpha)) \ \ \text{ and }
\]
\[
  \neg(4.\alpha \ \ker(\sfh_\cA) \  2.\sigma(\alpha,\alpha)) \hspace{7mm}
   \neg(7.\alpha \ \ker(\sfh_\cA) \ 3.\sigma(\sigma(\alpha,\alpha),\alpha))  \hspace{7mm}
   \neg(3.\alpha \  \ker(\sfh_\cA) \  4.\alpha)
  \enspace.
\]
In general, for each $b.\xi \in \rmMon(\Sigma,\B)$ we have
\[
  [b.\xi]_{\ker(\sfh_\cA)} = \begin{cases}\{(b \cdot 2^{\#_\alpha(\xi) -\#_\alpha(\zeta)}).\zeta \mid \zeta \in \T_\Sigma, \big(\#_\alpha(\xi) \!\mod (2)\big) = \big(\#_\alpha(\zeta) \! \mod (2)\big)\} & \text{ if $b\ne 0$}\\
  \{b.\zeta \mid \zeta \in \T_\Sigma\} = \{\widetilde{0}\}& \text{ otherwise.}
 \end{cases} 
\]
Since, e.g., for every $b_1,b_2 \in \mathbb{Q}$ with $b_1 \ne b_2$, we have $\neg(b_1.\alpha \ \ker(\sfh_\cA) \  b_2.\alpha)$, the index of $\ker(\sfh_\cA)$ is not finite.
\hfill $\Box$
   \end{example}

   Next we consider the congruence $\sim_{\sem{\cA}}$. We relate $\sfdM_{\mathrm{im}}(\cA)$ and $(\rmMon(\Sigma,\B),\widetilde{\0},\ttop_\Sigma)/_{\sim_{\sem{\cA}}}$ (cf. Figure~\ref{fig:overview-congruence-relations-isos}). First we prove the following lemma.

\begin{lemma}\rm \label{lm:kerPsiA-subseteeq-sim-semA-det} Let $\cA$ be a bu-deterministic $(\Sigma,\B)$-wta. Then $\ker(\sfh_\cA) \subseteq \sim_{\sem{\cA}}$.
\end{lemma}
\begin{proof} The proof is a special case of the proof of Lemma~\ref{lm:kerPsiA-subseteeq-sim-semA}.
\end{proof}

As in the general case, we define the binary relation $\rho_\cA$ on $\rmMon(\Sigma,\B)/_{\ker(\sfh_\cA)}$ such that, for every $[b_1.\xi_1]_{\ker(\sfh_\cA)}, [b_2.\xi_2]_{\ker(\sfh_\cA)} \in \rmMon(\Sigma,\B)/_{\ker(\sfh_\cA)}$, we let
       \begin{equation*}
      [b_1.\xi_1]_{\ker(\sfh_\cA)} \ \rho_\cA \  [b_2.\xi_2]_{\ker(\sfh_\cA)} \ \text{ if  } \
    b_1.\xi_1 \ \sim_{\sem{\cA}} \  b_2.\xi_2 \enspace.
  \end{equation*}
 The relation $\rho_\cA$ is well defined because $\ker(\sfh_\cA) \subseteq \sim_{\sem{\cA}}$ (cf. Lemma \ref{lm:kerPsiA-subseteeq-sim-semA-det}).

    \begin{theorem-rect}\label{lm:congruences-modulo-congruences-det} (cf. Figure \ref{fig:overview-congruence-relations-isos}.) Let $\B$ be a commutative semiring. Moreover, let $\cA$ be a bu-deterministic $(\Sigma,\B)$-wta. Then the following statements hold.
      \begin{compactenum}
    \item[(1)]  The relation $\rho_\cA$ is a congruence on the $(\Sigma,\B)$-scalar algebra $(\rmMon(\Sigma,\B),\widetilde{\0},\ttop_\Sigma)/_{\ker(\sfh_\cA)}$.
    \item[(2)] $\Big((\rmMon(\Sigma,\B),\widetilde{\0},\ttop_\Sigma)/_{\ker(\sfh_\cA)}\Big)/_{\rho_\cA} \cong (\rmMon(\Sigma,\B),\widetilde{\0},\ttop_\Sigma)/_{\sim_{\sem{\cA}}}$.
    \end{compactenum}
  \end{theorem-rect}
  
  \begin{proof} Proof of (1): By Lemma \ref{lm:kerPsiA-subseteeq-sim-semA-det}, we have $\ker(\sfh_\cA) \subseteq \sim_{\sem{\cA}}$.  Then  Statement (1) follows from Theorem~\ref{thm:snd-isom-theorem}(1)  (with $\A=(\rmMon(\Sigma,\B),\widetilde{\0},\ttop_\Sigma)$, $\approx = \sim_{\sem{\cA}}$, $\sim = \ker(\sfh_\cA)$, and $\approx\!/\!\sim \ = \rho_\cA$).

    \

    Proof of (2): It follows Theorem~\ref{thm:snd-isom-theorem}(2).
    \end{proof}

\subsection{Scalar-basis of cancellative scalar algebras over commutative semifields}
\label{subsec:prep-constr-budet-wta-semifield}

We would like to follow the lines of Subsection \ref{ssec:construction-wta-from-wta-congruence-basis} and define a bu-deterministic $(\Sigma,\B)$-wta in a similar way as $\rmwta(r,\approx,H)$ (cf. Definition~\ref{def:wta-from-finite-dimensional}). We recall that, for the definition of $\rmwta(r,\approx,H)$,  we have used the following fact: in the finite-dimensional syntactic $\B$-vector space $\sfPol(\Sigma,\B)/_{\approx}$,
each vector $[s]_{\approx}$ can be written in a unique way as a linear combination
\[
  [s]_{\approx} = \bigoplus_{\1.\zeta \in H} [s]_{[\1.\zeta]}. [\1.\zeta]_{\approx} 
  \]
  over the basis $H \subseteq \{[\1.\zeta]_{\approx} \mid \zeta \in \T_\Sigma\}$. In this linear combination, $[s]_{[\1.\zeta]} \in B$ is the coefficient of the base vector $[\1.\zeta]_{\approx}$. In order to have a similar unique representation for quotients of monomials (cf. Lemma~\ref{lm:unique-decomposition-det}), we develop the concept of \emph{scalar-basis of cancellative scalar algebras} and we use multiplicative inverses.

  \begin{quote} \emph{In this subsection, we let $\B=(B,\oplus,\otimes,\0,\1)$ denote an arbitrary commutative semifield. Moreover, we let $B^{-\0}$ denote the set $B \setminus \{-\0\}$.}
    \end{quote}

    Since $(B\setminus \{\0\},\otimes,\0)$ is a commutative group, we can view each $\B$-scalar algebra as a left group action \cite{ker99,hum04} extended by an annihilating zero.

The existence of inverses implies a kind of zero-divisor freeness for the scalar multiplication.

\begin{observation}\rm \label{obs:product=0-implies-vector=0-det} Let $\V=(V,0)$ be a $\B$-scalar algebra.  For each $v \in V\setminus\{0\}$ and $b \in B$, if $b \cdot v = 0$, then $b=\0$.
\end{observation}
\begin{proof} We prove by contradiction.  Assume that there exist  $v \in V\setminus\{0\}$ and $b \in B$ such that $b \cdot v = 0$ and $b\ne\0$. Then $b^{-1} \cdot (b \cdot v)=b^{-1}\cdot 0 =0$. On the other hand, $b^{-1} \cdot (b \cdot v)=(b^{-1}\otimes b) \cdot v= \1\cdot v =v \ne 0$, a contradiction.
\end{proof}

A $\B$-scalar algebra  $\V=(V,0)$ is \emph{cancellative} if, for every $b_1,b_2\in B$ and $v \in V\setminus \{0\}$ we have that $b_1 \cdot v = b_2\cdot v$ implies $b_1=b_2$.
Trivially, if $V=\{0\}$, then $\V$ is cancellative. Moreover, the $\B$-scalar algebra $(B,\0)$ is cancellative, because $\B$ is a semifield.

We note that, for each $\B$-vector space $(V,+,0)$ with field $\B$ and scalar multiplication $\cdot$ \cite{axl24,mooyaq98}, the reduct $(V,0)$ is a cancellative $\B$-scalar algebra with the same scalar multiplication. The cancellativity follows from the fact that, in this case, $b_1 \cdot v = b_2\cdot v$ implies that $(b_1+(-b_2))\cdot v =0$, where $-b_2$ is the additive inverse of $b_2$. By Observation \ref{obs:product=0-implies-vector=0-det}, the latter implies that $b_1=b_2$.

\begin{lemma}\rm \label{lm:Mon-mod-simr-is-cancellative} Let $r: \T_\Sigma \to B$. The $\B$-scalar algebra $(\rmMon(\Sigma,\B),\widetilde{\0})/_{\sim_r}$ is cancellative.
\end{lemma}
\begin{proof} Let $b_1,b_2 \in B$ and $[b.\xi] \in (\rmMon(\Sigma,\B)/_{\sim_r})\setminus \{[\widetilde{\0}]_{\sim_r}\}$. Then, by definition of $\sim_r$,  there exists $c_0 \in \C_\Sigma$ such that $b \otimes r(c_0[\xi]) \ne \0$. 

  Now assume that $b_1 \cdot [b.\xi]_{\sim_r} = b_2 \cdot [b.\xi]_{\sim_r}$. Then
  \[
(\forall c \in \C_\Sigma): (b_1 \otimes b) \otimes r(c[\xi]) = (b_2 \otimes b) \otimes r(c[\xi]) \enspace.
\]
By considering the instance $c=c_0$ of this equation and dividing both sides by $b \otimes r(c_0[\xi])$, we obtain 
$b_1=b_2$. Hence $(\rmMon(\Sigma,\B),\widetilde{\0})/_{\sim_r}$ is cancellative.
  \end{proof}


  For a $\B$-vector space, the elements of a basis are independent. For $\B$-scalar algebras we define the appropriate analogue. Formally, let $(V,0)$ be a $\B$-scalar algebra. Two elements $u,v\in V$ are \emph{dependent (in $(V,0)$)} if there exists a $b\in B$ with $u=b\cdot v$ or $v=b \cdot u$. If $u$ and $v$ are not dependent, then we say that they are \emph{independent}. This definition of dependency coincides with the concept of linear dependency in vector spaces in the following sense. If $(V,+,0)$ is a $\B$-vector space for some field $\B$, then
  \begin{align}
    &\text{for every $u,v \in V$ we have: $u,v$ are dependent in the $\B$-scalar algebra $(V,0)$}\label{equ:dependent-linearly-dependent}\\[-1mm]                                                                            
    &\text{if and only if $u,v$ are linearly dependent in the $\B$-vector space $(V,+,0)$.\notag}
      \end{align}
     We say that \emph{dependency in $(V,0)$ is decidable} if, for every $u,v \in V$, it is decidable whether $u$ and $v$ are dependent or not. 

\index{pair-independent}
A subset $H\subseteq V$ is called \emph{pair-independent} if any two different elements of $H$ are independent. In particular,
the empty set and each singleton subset of $V$ is pair-independent. Moreover, if $|H| \ge 2$ and $H$ is pair-independent,  then $0 \not\in H$. This is because if $0\in H$ and there exists $a\in H$ with $a\ne 0$, then we have $0= \0\cdot a$, i.e.,  $H$ is not pair-independent.

\index{scalar-basis}
  Let $(V,0)$ be a  $\B$-scalar algebra and $H$ a pair-independent generating set of $(V,0)$. Then we call $H$ a \emph{scalar-basis} (of $(V,0)$). For instance $\{\1\}$ is a scalar-basis of the $\B$-scalar algebra $(B,\0)$;
    and $\{\1.\xi \mid \xi \in \T_\Sigma\}$ is a scalar-basis of the $\B$-scalar algebra $(\sfMon(\Sigma,\B),\widetilde{\0})$.

  \begin{lemma}\rm \label{lm:crucial-canc-pair-ind-imply-uniqueness-det-1}
     Let $\V=(V,0)$ be a $\B$-scalar algebra, and $H \subseteq V$ be a finite generating set of $\V$. Then there exists a finite scalar-basis  $H' \subseteq H$. Moreover, if dependency in $\V$ is decidable, then we can construct a finite scalar-basis $H' \subseteq H$.
  \end{lemma}
     
 \begin{proof} If $H$ is pair-independent, then we let $H'=H$ and we are ready.

Otherwise, by induction, we define a finite sequence $H_0,\ldots,H_m$ of subsets of $H$ for some  $m \ge 1$
  such that (a) for each $i \in [0,m]$, the set $H_i$ generates $\V$, (b) for each $i \in [0,m-1]$, $H_i$ is not pair-independent, and (b) $H_m$ is pair-independent.
    
Let $H_0=H$. Now assume that $H_i$ is already defined and it generates $\V$. If $H_i$ is pair-independent, then let $H'=H_i$ and we stop; then $m=i$.
Otherwise, there exist two different $u,v \in H_i$ and some $b \in B$ such that $u =b \cdot v$ or $v =b \cdot u$. 
Assume that $u =b \cdot v$. We define the set $H_{i+1} = H_i \setminus\{u\}$; since $u\ne v$ we have  $v \in H_{i+1}$. We show that $H_{i+1}$ generates $\V$.
    For this, let $v'\in V$. Since $H_i$ generates $\V$, there exist $u'\in H_i$ and $a\in B$ with $v'=a\cdot u'$. If $u'\ne u$, then $u'\in H_{i+1}$ and we are done. Otherwise, we have $v'=a\cdot u' =a\cdot u = a\cdot (b \cdot v) = (a\otimes b)\cdot v$. The proof of the case that $v =b \cdot u$ is similar by symmetry.
    
Then (a), (b), and (c) hold.

To prove the second statement, we assume that dependency in $\V$ is decidable. Then, for each $i \in \mathbb{N}$, we can decide if $H_i$ is pair-independent and thus, if this is not the case, we can construct $H_{i+1}$. Hence we can find the smallest $i \in \mathbb{N}$ with  $H_i$ such that $H_i$ is pair-independent.
\end{proof}

Next we will show that a scalar-basis of a $\B$-scalar algebra shares relevant properties with a basis of a $\B$-vector space.

 \begin{lemma}\rm \label{lm:unique-decomposition-det}
     Let $\V=(V,0)$ be a cancellative $\B$-scalar algebra, and $H \subseteq V$ be a scalar-basis of $\V$.  Then, for each $v \in V\setminus \{0\}$, there exist unique $d \in B^{-\0}$ and $u \in H\setminus\{0\}$ such that $v=d\cdot u$.
\end{lemma}
    
  \begin{proof} Let $v \in V\setminus\{0\}$. Since $H$ generates $\V$, there exist $d\in B$ and $u \in H$ such that $v=  d\cdot u$. Since $v \ne 0$, we have $d \ne \0$ and $u \ne 0$. We prove the uniqueness by contradiction. For this, we assume that there exist $d,d' \in B^{-\0}$ and $u,u' \in H\setminus \{0\}$ such that
$v=d\cdot u=d'\cdot u'$ and $d\ne  d'$ or $u\ne u'$. 
If $u \ne u'$, then $u =(d^{-1}\otimes d')\cdot u'$ contradicts that $H$ is pair-independent.
Thus, we have $u=u'$ and  $d\cdot u=d'\cdot u$ where $d\not=d'$. This contradicts the fact that $\V$ is cancellative. Thus $d=d'$.
\end{proof}

\index{scal@$\mathrm{scal}(v)$}
\index{gen@$\mathrm{gen}(v)$}
\index{dec@$\mathrm{dec}(v)$}
Given $\V$ and $H$ as in Lemma \ref{lm:unique-decomposition-det}, then we define the mapping
\begin{equation}\label{equ:scalar-vector-det-new}
\mathrm{dec}: V\setminus \! \{0\} \to B^{-\0} \times \big(H \setminus\! \{0\}\big) \enspace,
\end{equation}
 called \emph{decomposition mapping for $V\setminus \! \{0\}$},
 such that for each $v \in V\setminus \! \{0\}$ we let $\mathrm{dec}(v) = (d,u)$ if $v = d \cdot u$.
By Lemma \ref{lm:unique-decomposition-det}, the mapping $\mathrm{dec}$ is well defined. We denote the first component and second component of $\mathrm{dec}(v)$ by $\mathrm{scal}(v)$ and $\mathrm{gen}(v)$, respectively. Thus, for each $v \in V\setminus \{0\}$, we have 
\[
  v = \mathrm{scal}(v) \cdot \mathrm{gen}(v) \enspace.
\]

Next we show that the cardinality of a scalar-basis is unique.

\begin{lemma}\rm\label{dimension-of-B-scalar-algebra} Let $\V=(V,0)$ be a finitely generated $\B$-scalar algebra. Moreover, let $H_1$ and $H_2$ be finite scalar-bases of $\V$. Then $|H_1|=|H_2|$.
\end{lemma}
\begin{proof} We give a proof by contradiction. For this, assume that $|H_1|<|H_2|$. Since $H_1$ generates $\V$, there exist $u\in H_1$, $v_1,v_2\in H_2$ with $v_1\ne v_2$, and
  $b_1,b_2\in B^{-\0}$ such that $v_1=b_1\cdot u$ and $v_2=b_2\cdot u$.  Hence
\[b_2\cdot v_1=b_2\cdot (b_1 \cdot u)=(b_2\otimes b_1)\cdot u= (b_1\otimes b_2)\cdot u=b_1\cdot( b_2\cdot u)=b_1\cdot v_2\enspace\]
Then we have $v_1=(b_2^{-1}\otimes b_1)\cdot v_2$, which contradicts the assumption that $H_2$ is pair-independent.
\end{proof}

\index{degree}
For each  finitely generated  $\B$-scalar algebra $\V$, we define the \emph{degree of $\V$}, denoted by $\deg(\V)$, to be the cardinality of a finite  scalar-basis  of $\V$. By Lemma \ref{dimension-of-B-scalar-algebra}, $\deg(\V)$ is well defined.

In the next three lemmas, we will show how the degree changes under taking sub-$\B$-scalar algebras, isomorphic images, and quotients.

\begin{lemma}\rm \label{obs:scalar-algebra-fin-gen-subalg-hom-book} Let $(V,0)$ be a  $\B$-scalar algebra and $(V',0)$ a sub-$\B$-scalar algebra of it.  If $(V,0)$ is finitely generated, then $(V',0)$ is finitely generated and $\deg((V',0))\le \deg((V,0))$.
\end{lemma}
\begin{proof} The statement is obvious if $V'=\{0\}$. Therefore, we assume that $V'\not =\{0\}$. By Lemma~\ref{lm:crucial-canc-pair-ind-imply-uniqueness-det-1}, there exists a finite scalar-basis $H$ of $(V,0)$.
Let $H'=\{u\in H \mid (\exists b\in B^{-\0}): b\cdot u\in V'\}$. 
  Then $H'\ne \emptyset$ because $H$ generates $(V,0)$ and $\{0\}\ne V'\subseteq V$. We prove that $H'\subseteq V'$ and that $B\cdot H'=V'$ (cf. Observation~\ref{obs:scalar-algebra-fin-gen-subalg-hom-1}(1)).
  
  For the proof of the inclusion, let $u\in H'$. By definition, there exist $u\in H$ and  $b\in B^{-\0}$ such that $b\cdot u\in V'$. Since $(V',0)$  is a $\B$-scalar algebra, we have that $b^{-1}\cdot(b\cdot u)\in V'$. But $b^{-1}\cdot(b\cdot u)=(b^{-1}\otimes b)\cdot u=\1\cdot u =u$, hence  $u\in V'$. 
  
  Now we prove the equality. First we note that $B\cdot H'\subseteq V'$ because $H'\subseteq V'$ and  $(V',0)$ is a $\B$-scalar algebra. For the proof of $ V' \subseteq B\cdot H'$, let $v\in V'$. If $v=0$, then $v\in B\cdot V'$ obviously, so assume that $v\ne 0$. Since $H$ generates $(V,0)$, there exists $u\in H$ and $b\in B^{-\0}$ such that $v=b\cdot u$. Then  by definition, $u\in H'$, i.e., $v\in B\cdot H'$.
  
  Hence $H'$ generates $(V',0)$. By Lemma \ref{lm:crucial-canc-pair-ind-imply-uniqueness-det-1}, there exists a pair-independent subset $H''\subseteq H'$ which generates 
  $(V',0)$. Then we obtain $\deg((V',0))=|H''| \le |H'| \le |H| = \deg((V,0))$.
 \end{proof}

\begin{observation}\rm \label{obs:isomorphism-preserves-degree-book} Let $(V,0)$ and $(V',0')$ be  $\B$-scalar algebras, $(V,0)$ be finitely generated, and $f: V \to V'$ be a $\B$-scalar algebra isomorphism. Then $(V',0')$ is finitely generated and  $\deg((V,0)) = \deg((V',0'))$.
\end{observation}

Let $(V,0,\mu)$ be a $(\Sigma,\B)$-scalar algebra and let $\sim$ be a congruence on $(V,0,\mu)$. We recall that the quotient algebra of $(V,0,\mu)$ with respect to $\sim$ is the $(\Sigma,\B)$-scalar algebra
\[(V,0,\mu)/_\sim =(V/_\sim,[0]_\sim,\mu/_\sim) \enspace.\]
In this quotient algebra the scalar multiplication is the mapping $\cdot/_\sim : B\times V/_\sim \to V/_\sim$ defined for every $b\in B$ and $[v]_\sim \in V/_\sim$ by $b \cdot\!/_\sim [v]_\sim = [b\cdot v]_\sim$. In the sequel we write $\cdot$ for  $\cdot/_\sim$. Moreover, $\mu/_\sim$ is defined, for every $k\in\mathbb{N}$, $\sigma\in\Sigma^{(k)}$, and $[v_1]_\sim,\ldots,[v_k]_\sim \in V/_\sim$, by
\[\mu/_\sim(\sigma)([v_1]_\sim,\ldots,[v_k]_\sim)=[\mu(\sigma)(v_1,\ldots,v_k)]_\sim\enspace.\]
The canonical mapping $\pi_\sim: V\to V/_\sim$, defined by $\pi_\sim(v)=[v]_\sim$ for each $v\in V$,  is a homomorphism from $\V$ to  $(\V,\mu)/_\sim$ (cf. Lemma~\ref{thm:canonical-map-of-congr-is-hom}).

\begin{lemma}\rm\label{dimension-of-quotient-B-scalar-algebra-book} Let $\V=(V,0)$ be a finitely generated  $\B$-scalar algebra and let $\sim$ be a congruence on $\V$.
Then the quotient $\B$-scalar algebra $\V/_\sim=(V/_\sim,[0]_\sim)$ is also finitely generated and $\deg(\V/_\sim)\le \deg(\V)$.
\end{lemma}
\begin{proof} By Lemma~\ref{lm:crucial-canc-pair-ind-imply-uniqueness-det-1}, there exists a finite scalar-basis $H$ of $(V,0)$.
 Then the set $H/_\sim=\{[u]_\sim\mid u\in H\}$ generates $\V/_\sim$. To see this, let 
$[v]_\sim \in V/_\sim$. Since $H$ generates $\V$, there are $u\in H$ and $b\in B$ such that $v=b\cdot u$. Then $ [v]_\sim=b\cdot [u]_\sim$ because $\sim$ is a congruence.
Moreover 
\[\deg(\V/_\sim) \le |H/_\sim| \le |H|= \deg(\V).\]
\end{proof}

\subsection[Construction of a bu-det wta from a wtl, a congruence, and a finite scalar-basis]{Construction of a bu-deterministic wta from a weighted tree language, a congruence, and a finite scalar-basis over some commutative semifield}
\label{subsec:constr-budet-wta-semifield}

  \begin{quote} \emph{In this subsection, we let $\B=(B,\oplus,\otimes,\0,\1)$ denote an arbitrary commutative semifield.}
    \end{quote}

As a preparation for the proof of Theorem~\ref{thm:MN-semifield-det}(B)$\Rightarrow$(A),  for every
  \begin{compactitem}
  \item weighted tree language $r:\T_\Sigma \to B$,
  \item congruence $\sim$ on the $(\Sigma,\B)$-scalar algebra $(\rmMon(\Sigma,\B),\widetilde{\0},\ttop_\Sigma)$ such that $\sim$ saturates $r$ and $\sfMon(\Sigma,\B)/_\sim$ is cancellative,  and
  \item finite scalar-basis $H$ of the $\B$-scalar algebra $\sfMon(\Sigma,\B)/_\sim$ such that $H \subseteq  \{[\1.\zeta]_\sim \mid \zeta \in \T_\Sigma\}$,
       \end{compactitem}
we define a bu-deterministic $(\Sigma,\B)$-wta  $\budwta(r,\sim,H)=(H,\delta,F)$ and show that it recognizes $r$ (cf. Definition~\ref{def:wta(r,sim,H)} and Lemma \ref{lm:wta(r,sim,H)}(1) and (2)). Moreover, we show that, if  the decomposition mapping for $\big(\rmMon(\Sigma,\B)/_\sim\big) \setminus \{[\widetilde{\0}]_\sim\}$ is computable,
then we can even construct $\budwta(r,\sim,H)$ (cf. Lemma \ref{lm:wta(r,sim,H)}(3)).  

\begin{quote} {\em In the rest of this subsection,  we abbreviate $[\1.\xi]_\sim$ by $[\1.\xi]$ for each $\xi\in \T_\Sigma$.}
\end{quote}

Using the decomposition mapping 
\[
\mathrm{dec}: \big(\rmMon(\Sigma,\B)/_\sim\big) \setminus \{[\widetilde{\0}]\} \to B^{-\0} \times \big(H \setminus \{[\widetilde{\0}]\}\big)
\]
for $V=\rmMon(\Sigma,\B)/_\sim$ as it is defined in \eqref{equ:scalar-vector-det-new},
we have for each $[\1.\xi]\ne[\widetilde{\0}]$:
\begin{equation}\label{equ:decomposition-of-class}
  [\1.\xi] = \mathrm{scal}([\1.\xi]) \cdot \mathrm{gen}([\1.\xi]) \enspace.
  \end{equation}
Then, for every $\xi \in \T_\Sigma$ and $[\1.\zeta] \in H$, we introduce the notation  $[\1.\xi]_{[\1.\zeta]}$ for the value in $B$ defined by
\begin{equation}\label{equ:definition-of-coefficient-det-new-new}
  [\1.\xi]_{[\1.\zeta]}=\begin{cases} \mathrm{scal}([\1.\xi]) & \text{ if $[\widetilde{\0}]\ne[\1.\xi]$ and $\mathrm{gen}([\1.\xi]) = [\1.\zeta]$}\\
\0 & \text{ otherwise.}
\end{cases}
\end{equation}
We note that the well-definedness of the mapping $\mathrm{dec}$, and hence of the notation  $[\1.\xi]_{[\1.\zeta]}$, is due to Lemma~\ref{lm:unique-decomposition-det}.

\begin{definition}\rm \label{def:wta(r,sim,H)} Let $r$, $\sim$, and $H$ be given as in the above list of objects.  Moreover, let $\gamma: \rmMon(\Sigma,\B)/_\sim \to B$ be the scalar-linear form which is uniquely determined by $\sim$ and $r$ (cf. Lemma~\ref{lm:uniqueness-of-scalar-linear-form}).

  We define the $(\Sigma,\B)$-wta $\budwta(r,\sim,H)=(H,\delta,F)$ where
  \begin{compactitem}
\item $\delta = (\delta_k:H^k\times \Sigma^{(k)}\times H \to B\mid k \in \mathbb{N})$ and for every $k \in \mathbb{N}$, $\sigma \in \Sigma^{(k)}$, $[\1.\zeta_{1}],\ldots,[\1.\zeta_{k}],[\1.\zeta]\in H$,  we define
  \begin{equation}\label{eq:delta-k-definition-det-new}
   \delta_k\big([\1.\zeta_{1}]\cdots[\1.\zeta_{k}],\sigma,[\1.\zeta]\big) = [\1.\sigma(\zeta_{1},\ldots,\zeta_{k})]_{[\1.\zeta]}
          \end{equation}
and
        \item $F: H \to B$ such that, for each $[\1.\zeta] \in H$, we define $F_{[\1.\zeta]} = \gamma([\1.\zeta])$. 
        \end{compactitem}
                        \hfill $\Box$
  \end{definition}

\begin{samepage}
   \begin{lemma}\rm \label{lm:wta(r,sim,H)}  Let $r$, $\sim$, and $H$ be given as in Definition \ref{def:wta(r,sim,H)}. Then the following statements hold.
    \begin{compactenum}
    \item[(1)] The $(\Sigma,\B)$-wta $\budwta(r,\sim,H)$ is bu-deterministic.
      \item[(2)] $\sem{\budwta(r,\sim,H)}=r$.
      \item[(3)] If  the decomposition mapping for $\big(\rmMon(\Sigma,\B)/_\sim\big) \setminus \{[\widetilde{\0}]\}$ is computable, then we can  construct $\budwta(r,\sim,H)$.
              \end{compactenum}
            \end{lemma}
            \end{samepage}
  \begin{proof} Proof of (1):  The statement follows from the fact that, due to \eqref{equ:definition-of-coefficient-det-new-new}, in the definition~\eqref{eq:delta-k-definition-det-new} there exists at most one $[\1.\zeta] \in H$ with $[\1.\sigma(\zeta_{1},\ldots,\zeta_{k})]_{[\1.\zeta]}\ne\0$.

    \
    
      Proof of (2): Let $\budwta(r,\sim,H) = (H,\delta,F)$ and let us abbreviate $\budwta(r,\sim,H)$ by $\cA$.  By induction on $\T_\Sigma$, we prove the following statement.
    \begin{equation}\label{equ:hom-is-canonical-det-new}
\text{For each $\xi \in \T_\Sigma$ and $[\1.\zeta] \in H$, we have $\h_{\cA}(\xi)_{[\1.\zeta]} = [\1.\xi]_{[\1.\zeta]}$.}
\end{equation}

Let $\xi = \sigma(\xi_1,\ldots,\xi_k)$ and $[\1.\zeta] \in H$.  Then we can calculate as follows.
\begingroup
\allowdisplaybreaks
\begin{align*}
  & \ \ \h_{\cA}(\sigma(\xi_1,\ldots,\xi_k))_{[\1.\zeta]}\\
  &= \bigoplus_{[\1.\zeta_1]\cdots [\1.\zeta_k] \in H^k}  \h_{\cA}(\xi_1)_{[\1.\zeta_1]} \otimes \ldots \otimes \h_{\cA}(\xi_k)_{[\1.\zeta_k]} \otimes \delta_k([\1.\zeta_1] \cdots [\1.\zeta_k], \sigma, [\1.\zeta])\\
  &=\bigoplus_{[\1.\zeta_1]\cdots [\1.\zeta_k] \in H^k}  [\1.\xi_1]_{[\1.\zeta_1]} \otimes \ldots \otimes [\1.\xi_k]_{[\1.\zeta_k]} \otimes
    \delta_k([\1.\zeta_1] \cdots [\1.\zeta_k], \sigma, [\1.\zeta])
    \tag{by I.H.}\\
   &=\bigoplus_{[\1.\zeta_1]\cdots [\1.\zeta_k] \in H^k}  [\1.\xi_1]_{[\1.\zeta_1]} \otimes \ldots \otimes [\1.\xi_k]_{[\1.\zeta_k]} \otimes
    [\1.\sigma(\zeta_1, \ldots,\zeta_k)]_{[\1.\zeta]} \enspace.
    \tag{by the definition of $\delta_k$}\\
\end{align*}
\endgroup
If $k=0$, then we are ready. For $k\ge 1$ we continue by case analysis.

\underline{case (a):} There exists $i \in [k]$ such that $[\1.\xi_i]= [\widetilde{\0}]$, i.e.,  for each $[\1.\zeta] \in H$, we have $[\1.\xi_i]_{[\1.\zeta]} = \0$. Then  we can continue as follows:
 \begingroup
\allowdisplaybreaks
\begin{align*}
  &\bigoplus_{[\1.\zeta_1]\cdots [\1.\zeta_k] \in H^k}  [\1.\xi_1]_{[\1.\zeta_1]} \otimes \ldots \otimes [\1.\xi_k]_{[\1.\zeta_k]} \otimes [\1.\sigma(\zeta_1, \ldots,\zeta_k)]_{[\1.\zeta]}\\[2mm]
  &= \0\\
  &= \Big(\ttop/\!_\sim(\sigma)([\1.\xi_1],\ldots,[\1.\xi_{i-1}],[\widetilde{\0}], [\1.\xi_{i+1}],\ldots,[\1.\xi_k])\Big)_{[\1.\zeta]}\tag{by (\ref{ml-absorbtive-det}) for $\ttop/\!_\sim(\sigma)$}\\
      &= \Big(\ttop/\!_\sim(\sigma)([\1.\xi_1],\ldots,[\1.\xi_k])\Big)_{[\1.\zeta]}\\
  &=[\1.\sigma(\xi_1,\ldots,\xi_k)]_{[\1.\zeta]} \enspace.
\end{align*}
\endgroup
  
\underline{case (b):}  For each $i \in [k]$ we have $[\1.\xi_i]\ne [\widetilde{\0}]$; thus, by \eqref{equ:decomposition-of-class}, we have $[\1.\xi_i] = \mathrm{scal}([\1.\xi_i]) \cdot \mathrm{gen}([\1.\xi_i])$. Since $\mathrm{gen}([\1.\xi_i]) \in H$, there exists $\widehat{\zeta_i} \in \T_\Sigma$ such that $\mathrm{gen}([\1.\xi_i]) = [\1.\widehat{\zeta_i}]$. Then, by \eqref{equ:definition-of-coefficient-det-new-new},  $[\1.\xi_i] = [\1.\xi_i]_{[\1.\widehat{\zeta_i}]}\cdot [\1.\widehat{\zeta_i}]$ for each $i \in [k]$. Now we can continue as follows:
   \begingroup
\allowdisplaybreaks
\begin{align*}
  &\bigoplus_{[\1.\zeta_1]\cdots [\1.\zeta_k] \in H^k}  [\1.\xi_1]_{[\1.\zeta_1]} \otimes \ldots \otimes [\1.\xi_k]_{[\1.\zeta_k]} \otimes [\1.\sigma(\zeta_1, \ldots,\zeta_k)]_{[\1.\zeta]}\\[2mm]
  &=   [\1.\xi_1]_{[\1.\widehat{\zeta_1}]} \otimes \ldots \otimes [\1.\xi_k]_{[\1.\widehat{\zeta_k}]} \otimes
    [\1.\sigma(\widehat{\zeta}_1,\ldots,\widehat{\zeta}_k)]_{[\1.\zeta]}
    \tag{because $ [\1.\xi_i]_{[\1.\zeta']}=\0 $ if $[\1.\zeta']\ne [\1.\widehat{\zeta}_i]$ for $i\in [k]$}\\
   &= [\1.\xi_1]_{[\1.\widehat{\zeta_1}]} \otimes \ldots \otimes [\1.\xi_k]_{[\1.\widehat{\zeta_k}]} \otimes
     \big(\ttop/\!_\sim(\sigma)([\1.\widehat{\zeta}_1],\ldots,[\1.\widehat{\zeta}_k])\big)_{[\1.\zeta]}
  \tag{by definition of $\ttop/\!_\sim(\sigma)$}\\
   &= \Big( \big([\1.\xi_1]_{[\1.\widehat{\zeta_1}]} \otimes \ldots \otimes [\1.\xi_k]_{[\1.\widehat{\zeta_k}]}\big) \cdot
     \ttop/\!_\sim(\sigma)([\1.\widehat{\zeta}_1],\ldots,[\1.\widehat{\zeta}_k]) \Big)_{[\1.\zeta]}
  \tag{by axioms for scalar multiplication of weighted tree languages}\\
  &= \Big( \ttop/\!_\sim(\sigma)([\1.\xi_1]_{[\1.\widehat{\zeta_1}]}\cdot [\1.\widehat{\zeta_1}],\ldots,
    [\1.\xi_k]_{[\1.\widehat{\zeta_k}]}\cdot [\1.\widehat{\zeta_k}]\Big)_{[\1.\zeta]}
  \tag{by \eqref{ml-Omega-det} for $\ttop/\!_\sim(\sigma)$}\\
  &= \Big(\ttop/\!_\sim(\sigma)([\1.\xi_1],\ldots,[\1.\xi_k])\Big)_{[\1.\zeta]}
    \tag{because  $[\1.\xi_i] = [\1.\xi_i]_{[\1.\widehat{\zeta_i}]}\cdot [\1.\widehat{\zeta_i}]$ for each $i \in [k]$}\\
  &= [\1.\sigma(\xi_1,\ldots,\xi_k)]_{[\1.\zeta]} \enspace.
     \tag{by definition of $\ttop/_\sim(\sigma)$}
  \end{align*}
  \endgroup
  This finishes the proof of \eqref{equ:hom-is-canonical-det-new}.

   Now let $\xi \in \T_\Sigma$. By Lemma~\ref{lm:limit-bu-det}(1) we distinguish the following two cases.                
                           
    \underline{$\h_\cA(\xi) \not\in B_{=1}^H$:} Then $\h_\cA(\xi)=\0^H$ and, by \eqref{equ:hom-is-canonical-det-new}, we have $[\1.\xi]=[\widetilde{\0}]$. Thus
    \begingroup
    \allowdisplaybreaks
    \begin{align*}
      \sem{\cA}(\xi) = \h_\cA(\xi) \cdot F =  \0  = \gamma([\widetilde{\0}]) = \gamma([\1.\xi]) = r(\xi)
    \end{align*}
    \endgroup
    where the last equality holds because $\sim$ saturates $r$ via the scalar-linear form $\gamma$.

    \underline{$\h_\cA(\xi) \in B_{=1}^H$:}   Similarly as in case (b) above, there exists $\widehat{\zeta} \in \T_\Sigma$ such that $[\1.\widehat{\zeta}]\in H$ and $[\1.\xi] = [\1.\xi]_{[\1.\widehat{\zeta}]}\cdot [\1.\widehat{\zeta}]$.
Then $\estate_{\cA}(\xi) = [\1.\widehat{\zeta}]$ and we can continue as follows:
     \begingroup
    \allowdisplaybreaks
    \begin{align*}
      \sem{\cA}(\xi) &= \h_{\cA}(\xi)_{[\1.\widehat{\zeta}]} \otimes F_{[\1.\widehat{\zeta}]}
      \tag{by Corollary~\ref{lm:properties-hA-of-budet-wta-det-new-H}}\\
                        &= [\1.\xi]_{[\1.\widehat{\zeta}]} \otimes \gamma([\1.\widehat{\zeta}]) 
      \tag{by \eqref{equ:hom-is-canonical-det-new} and definition of $F$}\\
                     &= \gamma\Big([\1.\xi]_{[\1.\widehat{\zeta}]} \cdot [\1.\widehat{\zeta}]\Big) 
                           \tag{because $\gamma$ is a scalar-linear form}\\
      &= \gamma([\1.\xi]) 
      \tag{by the above}\\
                          &= r(\xi)
                            \tag{because $\sim$ saturates $r$ via the scalar-linear form $\gamma$}\enspace. 
    \end{align*}
    \endgroup

\
    
    Proof of (3): Let $k \in \mathbb{N}$, $\sigma \in \Sigma^{(k)}$, and $[\1.\zeta_{1}],\ldots,[\1.\zeta_{k}],[\1.\zeta]\in H$. We show that we can compute the right-hand side of \eqref{eq:delta-k-definition-det-new}. For this, we decide if  $[\1.\sigma(\zeta_{1},\ldots,\zeta_{k})]=[\widetilde{\0}]$, i.e., if $\1.\sigma(\zeta_{1},\ldots,\zeta_{k})\sim \widetilde{\0}$. (This is possible because, by our general convention, $\sim$ is given effectively.)  If the answer is yes, then $[\1.\sigma(\zeta_{1},\ldots,\zeta_{k})]_{[\1.\zeta]}=\0$. Otherwise,
  we compute $\mathrm{scal}([\1.\sigma(\zeta_{1},\ldots,\zeta_{k})])$ and $\mathrm{gen}([\1.\sigma(\zeta_{1},\ldots,\zeta_{k})])$.
  Now if $\mathrm{gen}([\1.\sigma(\zeta_{1},\ldots,\zeta_{k})])=[\1.\zeta]$, then 
  $[\1.\sigma(\zeta_{1},\ldots,\zeta_{k})]_{[\1.\zeta]}=\mathrm{scal}([\1.\sigma(\zeta_{1},\ldots,\zeta_{k})])$,
  otherwise again  $[\1.\sigma(\zeta_{1},\ldots,\zeta_{k})]_{[\1.\zeta]}=\0$.
  
  Moreover, let $[\1.\zeta] \in H$. Then we can compute $F_{[\1.\zeta]}$ because $\gamma([\1.\zeta])=r(\zeta)$ and by our convention 
  the mapping $r$ is computable.
          \end{proof}

\begin{example}\rm \label{ex:bu-det+total-wta3-det} Here we illustrate the construction of $\budwta(r,\sim,H)$.
For this, we consider the $(\Sigma,\Ratnum)$-weighted tree language
$r: \T_\Sigma \to \mathbb{Q}$ of Example \ref{ex:bu-det+total-wta-det-weo} and the m-syntactic congruence $\sim_r$ (cf. Example~\ref{ex-syntactic-congruence}). By Lemma~\ref{lm:sim-r-is-coarsest}, $\sim_r$ saturates $r$ via the scalar-linear form $\gamma: \rmMon(\Sigma,\Ratnum)/_{\sim_r} \to \mathbb{Q}$
  defined by
  \[
\gamma([b.\xi]_{\sim_r}) = b \cdot r(\xi) \ \text{ for each $b.\xi \in \rmMon(\Sigma,\Ratnum)$} \enspace.
\]
In the sequel, we abbreviate $[b.\xi]_{\sim_r}$ by $[b.\xi]$.

    By Lemma~\ref{lm:Mon-mod-simr-is-cancellative}, the $\B$-scalar algebra $\sfMon(\Sigma,\Ratnum)/_{\sim_r}$ is cancellative.
In order to show that $\sfMon(\Sigma,\Ratnum)/_{\sim_r}$ is finitely generated, we consider the set
  \[
H = \{[1.\alpha], \ [1.\sigma(\alpha,\alpha)] \} 
\]
and show that it is a generating set of $\sfMon(\Sigma,\Ratnum)/_{\sim_r}$. By using the characterization of $\sim_r$ given in  Example  \ref{ex-syntactic-congruence}, 
for every $\xi \in \T_\Sigma$ and $b\in \mathbb{Q}$, we have
\begin{align}\label{eq:congruence-classes}
[b.\xi] = 
\begin{cases}
[(b\cdot 2^{\#_\alpha(\xi)-1}).\alpha]=(b\cdot 2^{\#_\alpha(\xi)-1})\cdot[1.\alpha] & \text{ if \ $\big(\#_\alpha(\xi)\!\mod(2)\big)=1$}\\
[(b\cdot 2^{\#_\alpha(\xi)-2}).\sigma(\alpha,\alpha)]=(b\cdot 2^{\#_\alpha(\xi)-2})\cdot[1.\sigma(\alpha,\alpha)] & \text{ if \ $\big(\#_\alpha(\xi)\!\mod(2)\big)=0$}.
\end{cases}
\end{align}
This shows that $H$ generates $\sfMon(\Sigma,\Ratnum)/_{\sim_r}$. By \eqref{eq:congruence-classes}, we also obtain that there does not exist $b \in \mathbb{Q}$ such that $[1.\alpha] = b \cdot [1.\sigma(\alpha,\alpha)]$ because $\big(\#_\alpha(\alpha) \! \mod(2)\big) \not = \big(\#_\alpha(\sigma(\alpha,\alpha)) \! \mod (2)\big)$. Thus  $H$ is pair-independent, and hence $H$ is a finite scalar-basis.
Hence $r$, $\sim_r$, and $H$ satisfy all the conditions of  Definition \ref{def:wta(r,sim,H)}.

Moreover, \eqref{eq:congruence-classes} also  shows that  the decomposition mapping $\mathrm{dec}$ for $\big(\rmMon(\Sigma,\Ratnum)/_{\sim_r}\big) \setminus \{[\widetilde{\0}]\}$ is computable. In particular, for each $\xi\in\T_\Sigma$, we have
\begin{align}\label{eq:dec-mapping}
\mathrm{dec}([1.\xi])=
\begin{cases} (2^{\#_\alpha(\xi)-1}, [1.\alpha]) & \text{ if \ $\big(\#_\alpha(\xi)\!\mod(2)\big)=1$}\\
(2^{\#_\alpha(\xi)-2}, [1.\sigma(\alpha,\alpha)]) & \text{ if \ $\big(\#_\alpha(\xi)\!\mod(2)\big)=0$}.
\end{cases}
\end{align}

Hence, by  Lemma \ref{lm:wta(r,sim,H)}(3), we can construct the $(\Sigma,\Ratnum)$-wta $\budwta(r,\sim_r,H) =(H,\delta,F)$ as follows (cf. Figure~\ref{fig:even-odd-two-three-3-det}).
 We have
\begin{compactitem}
\item $\delta_0(\varepsilon,\alpha,[1.\alpha]) = 1$, \
  $\delta_0(\varepsilon,\alpha,[1.\sigma(\alpha,\alpha)]) = 0$, and for every $q_1q_2 \in H$ we have
  \[
    \delta_2(q_1q_2,\sigma,[1.\alpha]) = 
    \begin{cases} 4  & \text{ if $q_1\not= q_2$}\\
           0 & \text{ otherwise }
           \end{cases}
    \]
        and
        \[
       \delta_2(q_1q_2,\sigma,[1.\sigma(\alpha,\alpha)]) = 
        \begin{cases} 4 & \text{ if $q_1= q_2= [1.\sigma(\alpha,\alpha)]$}\\
         1 & \text{ if $q_1= q_2= [1.\alpha]$}\\
          0 & \text{ otherwise }\enspace.
        \end{cases}
        \]
       For instance $\delta_2([1.\sigma(\alpha,\alpha)][1.\alpha],\sigma,[1.\alpha])=4$ because
       $[1.\sigma(\sigma(\alpha,\alpha),\alpha)]_{[1.\alpha]}=\mathrm{scal}([1.\sigma(\sigma(\alpha,\alpha),\alpha)])=2^{3-1}=4$.
       
        \item $F_{[1.\alpha]} = \gamma([1.\alpha]) = r(\alpha) = 6$ and
  $F_{[1.\sigma(\alpha,\alpha)]} = \gamma([1.\sigma(\alpha,\alpha)]) = r(\sigma(\alpha,\alpha)) = 8$.
\end{compactitem}
By Lemma \ref{lm:wta(r,sim,H)}(1), $\budwta(r,\sim_r,H)$ is a bu-deterministic $(\Sigma,\B)$-wta; it is even total. 
By Lemma~\ref{lm:wta(r,sim,H)}(2), we have $\sem{\budwta(r,\sim_r,H)}=r$.

 \begin{figure}[t]
   \begin{center}
\begin{tikzpicture}
\tikzset{node distance=7em, scale=0.5, transform shape}
\node[state, rectangle] (1) {\Large $\alpha$};
\node[state, right of=1] (2){\Large $[1.\alpha]$};
\node[state, rectangle, right of=2] (3)[right=1em]{\Large $\sigma$};
\node[state, rectangle, above of=3] (4)[above=1em]{\Large $\sigma$};
\node[state, rectangle, below of=3] (5)[below=1em]{\Large $\sigma$};
\node[state, right of=3] (6)[right=1em]{\Large $[1.\sigma(\alpha,\alpha)]$};
\node[state, rectangle, right of=6] (7) {\Large $\sigma$};

\tikzset{node distance=2em}
\node[above of=1] (w1)[yshift=0.7em] {\Large 1};
\node[above of=2] (w2)[left=0.1em,yshift=1em] {\Large 6};
\node[above of=3] (w3)[yshift=0.7em] {\Large 1};
\node[above of=4] (w4)[yshift=0.7em] {\Large 4};
\node[above of=5] (w5)[yshift=0.7em] {\Large 4};
\node[above of=6] (w6)[above=1.8em] {\Large 8};
\node[above of=7] (w7)[yshift=0.7em] {\Large 4};

\draw[->,>=stealth] (1) edge (2);
\draw[->,>=stealth] (2) edge[out=20, in=155, looseness=1.1] (3);
\draw[->,>=stealth] (2) edge[out=-20, in=205, looseness=1.1] (3);
\draw[->,>=stealth] (2) edge (4);
\draw[->,>=stealth] (2) edge (5);
\draw[->,>=stealth] (3) edge (6);
\draw[->,>=stealth] (6) edge (4);
\draw[->,>=stealth] (4) edge[out=110, in=80, looseness=1.4] (2);
\draw[->,>=stealth] (6) edge (5);
\draw[->,>=stealth] (5) edge[out=250, in=-80, looseness=1.4] (2);
\draw[->,>=stealth] (7) edge (6);
\draw[->,>=stealth] (6) edge[out=60, in=30, looseness=2.7] (7);
\draw[->,>=stealth] (6) edge[out=-60, in=-30, looseness=2.7] (7);
\end{tikzpicture} 
\caption{\label{fig:even-odd-two-three-3-det} The $(\Sigma,\Ratnum)$-wta $\budwta(r,\sim_r,H) = (H,\delta,F)$.}
   \end{center}
   \end{figure}

   In the following we demonstrate this equality, where we abbreviate $\h_{\budwta(r,\sim_r,H)}$ by $\h$.
      \begin{equation}\label{equ:Atilde-h-det}
        \begin{aligned}
        \text{ For each $\xi \in \T_\Sigma$, we have}  \ \ \ 
        &\h(\xi)_{[1.\alpha]} =
          \begin{cases}
            2^{\#_\alpha(\xi)-1} & \text{ if $\#_\alpha(\xi)$ is odd}\\
            0 & \text{ otherwise}
          \end{cases} \ \ \  \text{ and }\\
        & \h(\xi)_{[1.\sigma(\alpha,\alpha)]} =
          \begin{cases}
            2^{\#_\alpha(\xi)-2} & \text{ if $\#_\alpha(\xi)$ is even}\\
            0 & \text{ otherwise} \enspace.
          \end{cases}
          \end{aligned}
        \end{equation}
        We prove this statement by induction on $\T_\Sigma$. In the following we abbreviate $\delta_{\budwta(r,\sim_r,H)}$ by $\delta$.

        I.B.: $\h(\alpha)_{[1.\alpha]} = \delta(\alpha)()_{[1.\alpha]} = \delta_0(\varepsilon,\alpha,[1.\alpha])= 1 = 2^{1-1}$ and similarly $\h(\alpha)_{[1.\sigma(\alpha,\alpha)]} = 0$.

        I.S.: Let $\xi = \sigma(\xi_1,\xi_2)$. First we consider the target state $[1.\alpha]$. Then
        \[
          \h(\sigma(\xi_1,\xi_2))_{[1.\alpha]}
          = \delta(\sigma)\Big(\h(\xi_1),\h(\xi_2)\Big)_{[1.\alpha]} \enspace.
          \]
        We proceed by case analysis.
        
                \underline{$\#_\alpha(\xi_1)$ is odd and $\#_\alpha(\xi_2)$ is odd:}
\begin{align*}
          \delta(\sigma)\Big(\h(\xi_1),\h(\xi_2)\Big)_{[1.\alpha]}
          &= \h(\xi_1)_{[1.\alpha]} \cdot \h(\xi_2)_{[1.\alpha]} \cdot \delta_2([1.\alpha][1.\alpha],\sigma,[1.\alpha])\\
          &= \h(\xi_1)_{[1.\alpha]} \cdot \h(\xi_2)_{[1.\alpha]} \cdot 0 = 0 \enspace.
        \end{align*}

       \underline{$\#_\alpha(\xi_1)$ is odd and $\#_\alpha(\xi_2)$ is even:}
        \begin{align*}
          \delta(\sigma)\Big(\h(\xi_1),\h(\xi_2)\Big)_{[1.\alpha]}
          &= \h(\xi_1)_{[1.\alpha]} \cdot \h(\xi_2)_{[1.\sigma(\alpha,\alpha)]} \cdot \delta_2([1.\alpha][1.\sigma(\alpha,\alpha)],\sigma,[1.\alpha])\\
          &=  2^{\#_\alpha(\xi_1)-1} \cdot  2^{\#_\alpha(\xi_2)-2} \cdot 4 = 2^{\#_\alpha(\xi)-1}  \enspace.
        \end{align*}

        \underline{$\#_\alpha(\xi_1)$ is even and $\#_\alpha(\xi_2)$ is odd:} This case is similar to the previous one.

               \underline{$\#_\alpha(\xi_1)$ is even and $\#_\alpha(\xi_2)$ is even:}
 \begin{align*}
          \delta(\sigma)\Big(\h(\xi_1),\h(\xi_2)\Big)_{[1.\alpha]}
          &= \h(\xi_1)_{[1.\sigma(\alpha,\alpha)]} \cdot \h(\xi_2)_{[1.\sigma(\alpha,\alpha)]} \cdot \delta_2([1.\sigma(\alpha,\alpha)][1.\sigma(\alpha,\alpha)],\sigma,[1.\alpha])\\
          &= \h(\xi_1)_{[1.\sigma(\alpha,\alpha)]} \cdot \h(\xi_2)_{[1.\sigma(\alpha,\alpha)]} \cdot 0 = 0 \enspace.
 \end{align*}

 Second we consider the target state $[1.\sigma(\alpha,\alpha)]$. Then
        \[
          \h(\sigma(\xi_1,\xi_2))_{[1.\sigma(\alpha,\alpha)]}
          = \delta(\sigma)\Big(\h(\xi_1),\h(\xi_2)\Big)_{[1.\sigma(\alpha,\alpha)]} \enspace.
          \]
        We proceed by case analysis.
        
                \underline{$\#_\alpha(\xi_1)$ is odd and $\#_\alpha(\xi_2)$ is odd:}
\begin{align*}
          \delta(\sigma)\Big(\h(\xi_1),\h(\xi_2)\Big)_{[1.\sigma(\alpha,\alpha)]}
          &= \h(\xi_1)_{[1.\alpha]} \cdot \h(\xi_2)_{[1.\alpha]} \cdot \delta_2([1.\alpha][1.\alpha],\sigma,[1.\sigma(\alpha,\alpha)])\\
          &= 2^{\#_\alpha(\xi_1)-1} \cdot 2^{\#_\alpha(\xi_2)-1} \cdot 1 = 2^{\#_\alpha(\xi)-2} \enspace.
        \end{align*}

       \underline{$\#_\alpha(\xi_1)$ is odd and $\#_\alpha(\xi_2)$ is even:}
        \begin{align*}
          \delta(\sigma)\Big(\h(\xi_1),\h(\xi_2)\Big)_{[1.\sigma(\alpha,\alpha)]}
          &= \h(\xi_1)_{[1.\alpha]} \cdot \h(\xi_2)_{[1.\sigma(\alpha,\alpha)]} \cdot \delta_2([1.\alpha][1.\sigma(\alpha,\alpha)],\sigma,[1.\sigma(\alpha,\alpha)])\\
          &=  \h(\xi_1)_{[1.\alpha]} \cdot \h(\xi_2)_{[1.\sigma(\alpha,\alpha)]} \cdot 0 = 0\enspace.
        \end{align*}

        \underline{$\#_\alpha(\xi_1)$ is even and $\#_\alpha(\xi_2)$ is odd:} This case is similar to the previous one.

               \underline{$\#_\alpha(\xi_1)$ is even and $\#_\alpha(\xi_2)$ is even:}
 \begin{align*}
          \delta(\sigma)\Big(\h(\xi_1),\h(\xi_2)\Big)_{[1.\sigma(\alpha,\alpha)]}
          &= \h(\xi_1)_{[1.\sigma(\alpha,\alpha)]} \cdot \h(\xi_2)_{[1.\sigma(\alpha,\alpha)]} \cdot \delta_2([1.\sigma(\alpha,\alpha)][1.\sigma(\alpha,\alpha)],\sigma,[1.\sigma(\alpha,\alpha)])\\
          &= 2^{\#_\alpha(\xi_1)-2} \cdot 2^{\#_\alpha(\xi_2)-2} \cdot 4 = 2^{\#_\alpha(\xi_1)-2} \enspace.
 \end{align*}
 This proves \eqref{equ:Atilde-h-det}. 

 Now let $\xi \in \T_\Sigma$. By \eqref{equ:Atilde-h-det} and Lemma~\ref{lm:hAxiqne0-implies-qinhQxi-for-bu-det-wta}(1), we have
\begin{align*}
\state(\xi)=
\begin{cases}
\{[1.\alpha]\} & \text{ if $\#_\alpha(\xi)$ is odd}\\
\{[1.\sigma(\alpha,\alpha)]\} & \text{ otherwise.}
\end{cases}
\end{align*}
Then, by Corollary~\ref{lm:properties-hA-of-budet-wta-det-new-H}, we have
\begin{align*}
   \sem{\budwta(r,\sim_r,H)}(\xi) &=
   \begin{cases}
  \h(\xi)_{[1.\alpha]} \cdot F_{[1.\alpha]}    & \text{ if $\#_\alpha(\xi)$ is odd}\\
  \h(\xi)_{[1.\sigma(\alpha,\alpha)]} \cdot F_{[1.\sigma(\alpha,\alpha)]}    & \text{ otherwise}
\end{cases}\\[3mm]
   &=  \begin{cases}
     2^{\#_\alpha(\xi)-1} \cdot 6 = 3\cdot 2^{\#_\alpha(\xi)-1}  & \text{ if $\#_\alpha(\xi)$ is odd}\\
    2^{\#_\alpha(\xi)-2} \cdot 8 = 4\cdot 2^{\#_\alpha(\xi)-2} & \text{ otherwise}
  \end{cases}
                             \ \ \       = r(\xi) \enspace.
\end{align*}
\hfill $\Box$
\end{example}

\subsection{The deterministic version of B-A's theorem}
\label{subsect:det-version-BA-theorem}

As final preparation for the proof of our first main result, we prove that, for each bu-deterministically recognizable $(\Sigma,\B)$-weighted tree language~$r$, the number of states of any bu-deterministic $(\Sigma,\B)$-wta which recognizes $r$, is at least the degree of $\sfMon(\Sigma,\B)/_{\sim_r}$. The steps of its proof can be tracked on Figure~\ref{fig:overview-congruence-relations-isos}.  This lemma can be compared to \cite[Cor.~3]{bor03}.
    
  \begin{lemma}\rm\label{lm:dimension-synt-factor-algebra-bounded} Let $\cA=(Q,\delta,F)$ be a  bu-deterministic $(\Sigma,\B)$-wta. Then the $\B$-scalar algebra $\sfMon(\Sigma,\B)/_{\sim_{\sem{\cA}}}$ is finitely generated and $\deg\big( \sfMon(\Sigma,\B)/_{\sim_{\sem{\cA}}}\big)\le |Q|$.
  \end{lemma} 
\begin{proof} Clearly, the $\B$-scalar algebra $(B^Q_{\le 1},\0^Q)$ is finitely generated and its degree is $|Q|$, i.e.,
  \[
|Q| = \deg\big( (B^Q_{\le 1},\0^Q) \big) \enspace.
\]
We recall that $\sfh_\cA$ denotes the unique scalar-linear mapping
from $(\mathrm{Mon}(\Sigma,\B),\widetilde{\0},\ttop)$ to $\sfdM(\cA)=(B^Q_{\le 1},\0^Q,\delta_\cA)$.
  By Lemma \ref{obs:scalar-algebra-fin-gen-subalg-hom-1}(2), $(\im(\sfh_\cA),\0^Q)$ is a sub-$\B$-scalar algebra of $(B^Q_{\le 1},\0^Q)$, and by Lemma \ref{obs:scalar-algebra-fin-gen-subalg-hom-book},  it is also finitely generated and
  \[
\deg\big( (B^Q_{\le 1},\0^Q) \big) \ge  \deg\big( (\im(\sfh_\cA),\0^Q)\big) \enspace.
    \]
 
    By \eqref{eq:isomorphism}, $(\im(\sfh_\cA),\0^Q) \cong \sfMon(\Sigma,\B)/_{\ker(\sfh_\cA)}$ and thus, by Observation~\ref{obs:isomorphism-preserves-degree-book}, we have
  \[
\deg\big( (\im(\sfh_\cA),\0^Q)\big) = \deg\big( \sfMon(\Sigma,\B)/_{\ker(\sfh_\cA)}\big) \enspace.
\]
 Then, by Theorem \ref{lm:congruences-modulo-congruences-det}(1), $\rho_\cA$ is a congruence on $\sfMon(\Sigma,\B)/_{\ker(\sfh_\cA)}$ and, by Lemma \ref{dimension-of-quotient-B-scalar-algebra-book},  we have
  \[
\deg\big( \sfMon(\Sigma,\B)/_{\ker(\sfh_\cA)}\big) \ge \deg\Big(\big(\sfMon(\Sigma,\B)/_{\ker(\sfh_\cA)}\big)/_{\rho_\cA}\Big) \enspace.
    \]
    Lastly, by Theorem \ref{lm:congruences-modulo-congruences-det}(2), we obtain
$\big(\sfMon(\Sigma,\B)/_{\ker(\sfh_\cA)}\big)/_{\rho_\cA} \cong \sfMon(\Sigma,\B)/_{\sim_{\sem{\cA}}}$ and thus, by Observation~\ref{obs:isomorphism-preserves-degree-book}, we have 
 \[
\deg\Big(\big(\sfMon(\Sigma,\B)/_{\ker(\sfh_\cA)}\big)/_{\rho_\cA}\Big) = \deg\big( \sfMon(\Sigma,\B)/_{\sim_{\sem{\cA}}} \big) \enspace.
    \]
     \end{proof}

Now we present our first main result, which might be called the deterministic version of B-A's theorem. 

\begin{samepage}
  \begin{theorem-rect}\label{thm:MN-semifield-det}  Let $\B$  a commutative semifield. Moreover, let $r: \T_\Sigma \to B$. The following three statements are equivalent.
  \begin{compactenum}
  \item[(A)] $r \in \budRec(\Sigma,\B)$.
  \item[(B)] There exists a congruence $\sim$ on the $(\Sigma,\B)$-scalar algebra $(\rmMon(\Sigma,\B),\widetilde{\0},\ttop_\Sigma)$ such that  $\sim$ saturates~$r$ and the $\B$-scalar algebra $\sfMon(\Sigma,\B)/_\sim$ is finitely generated and cancellative.
  \item[(C)]  The $\B$-scalar algebra $\sfMon(\Sigma,\B)/_{\sim_r}$ is finitely generated.
  \end{compactenum}
  \end{theorem-rect}
\end{samepage}
\begin{proof}  Proof of  (A)$\Rightarrow$(C): Let $\cA$ be a bu-deterministic $(\Sigma,\B)$-wta with $\sem{\cA} = r$.  Then the implication follows from Lemma \ref{lm:dimension-synt-factor-algebra-bounded}.

  \
  
  Proof of (C)$\Rightarrow$(B): We choose $\sim\,=\,\sim_r$.
  Then  $\sim$ is a congruence on $(\rmMon(\Sigma,\B),\widetilde{\0},\ttop_\Sigma)$  and, by Lemma~\ref{lm:sim-r-is-coarsest}, it saturates $r$. By Lemma~\ref{lm:Mon-mod-simr-is-cancellative}, the $\B$-scalar algebra $\sfMon(\Sigma,\B)/_{\sim}$ is cancellative, and by our choice it is finitely generated.

  \
  
Proof of (B)$\Rightarrow$(A): Let $\B=(B,\oplus,\otimes,\0,\1)$. Moreover, let $\sim$ be a congruence on  $(\rmMon(\Sigma,\B),\widetilde{\0},\ttop_\Sigma)$ such that $\sim$ saturates $r$ via the scalar-linear form $\gamma: \rmMon(\Sigma,\B)/_{\sim}\to B$, i.e., for each $\xi \in \T_\Sigma$, we have $r(\xi) = \gamma([\1.\xi]_\sim)$. Moreover, let  $\sfMon(\Sigma,\B)/_{\sim}$ be cancellative and generated by the finite set $H$.  Without loss of generality, we can assume that $H \subseteq \{[\1.\zeta]_\sim \mid \zeta \in \T_\Sigma\}$.

By Lemma \ref{lm:crucial-canc-pair-ind-imply-uniqueness-det-1} there exists a finite scalar-basis $H' \subseteq H$ of $\sfMon(\Sigma,\B)/_\sim$.
Then, by Lemma \ref{lm:wta(r,sim,H)}, the $(\Sigma,\B)$-wta $\budwta(r,\sim,H')$ (defined in Definition \ref{def:wta(r,sim,H)}) is bu-deterministic  and  $\sem{\budwta(r,\sim,H')}=r$.  
    \end{proof}

\begin{example}\rm\label{ex:congruence} We give a weighted tree language $r$ and a congruence $\sim$ which satisfies the conditions in Theorem \ref{thm:MN-semifield-det}(B) and, moreover,
$\sim\,\subset \sim_r$. For this, let $r$ be the $(\Sigma,\Ratnum)$-weighted tree language from Example~\ref{ex-syntactic-congruence}. Moreover, we define the relation $\sim$ on 
$\mathrm{Mon}(\Sigma,\Ratnum)$ such that,
 for every $b_1.\xi_1,b_2.\xi_2 \in \rmMon(\Sigma,\Ratnum)$, we let
\begin{align*}
  & b_1.\xi_1 \ \sim \  b_2.\xi_2  \ \text{ iff } 
   \Big(b_1 \cdot 2^{\#_\alpha(\xi_1)} =  b_2 \cdot 2^{\#_\alpha(\xi_2)} \text{ and } 
   \big(\#_\alpha(\xi_1)\!\mod(4)\big) =\big(\#_\alpha(\xi_2)\!\mod(4)\big)\Big)\\
   & \hspace*{28mm}\text{or \ } b_1=b_2=0\enspace.
\end{align*}
Obviously, we have $\sim\,\subset \sim_r$ (cf. Example  \ref{ex-syntactic-congruence}). It is easy to see that $\sim$ is a congruence on $(\mathrm{Mon}(\Sigma,\Ratnum),\widetilde{0},\ttop)$. 

To show that  $\sim$ saturates $r$, we define the mapping   $\gamma: \mathrm{Mon}(\Sigma,\Ratnum)/_\sim \to \mathbb{Q}$ by $\gamma([b.\xi]_\sim)=b\cdot r(\xi)$ for every $b\in \mathbb{Q}$ and $\xi\in\T_\Sigma$. It is easy to see that $\gamma$ is a scalar-linear form and it saturates $r$ because $\gamma([1.\xi]_\sim)=r(\xi)$ for each $\xi\in\T_\Sigma$.

Moreover, $\sfMon(\Sigma,\Ratnum)/_\sim$ is generated by 
\[H = \{[1.\alpha]_{\sim}, \ [1.\sigma(\alpha,\alpha)]_{\sim}, \ [1.\sigma(\alpha,\sigma(\alpha,\alpha))]_{\sim}, \  [1.\sigma(\sigma(\alpha,\alpha),\sigma(\alpha,\alpha))]_{\sim}\} \enspace.\]
We demonstrate this by an example. Let $b\in \mathbb{Q}$ and $\xi \in \T_\Sigma$ with $\big(\#_\alpha(\xi)\!\mod(4)\big)=3$. Then, we have
\[[b.\xi]_\sim = [(b\cdot 2^{\#_\alpha(\xi)-3}).\sigma(\alpha,\sigma(\alpha,\alpha))]_\sim=(b\cdot 2^{\#_\alpha(\xi)-3})\cdot[1.\sigma(\alpha,\sigma(\alpha,\alpha))]_\sim\enspace,\]
where the first equality holds by the definition of $\sim$ and the second holds because $\sim$ is a congruence.

Lastly, we show that $\sfMon(\Sigma,\B)/_\sim$ is cancellative. For this, let $b,b_1,b_2\in \mathbb{Q}$ with $b\ne 0$ and $\xi\in \T_\Sigma$ such that $b_1\cdot [b.\xi]_\sim = b_2\cdot [b.\xi]_\sim$. Then we argue as follows:
\begin{align*}
& \hspace*{3mm} b_1\cdot [b.\xi]_\sim = b_2\cdot [b.\xi]_\sim \\
\text{iff } &  \hspace*{3mm} [(b_1\cdot b).\xi]_\sim =  [(b_1\cdot b).\xi]_\sim \\
\text{then } &  \hspace*{3mm} [(b_1\cdot b).\xi]_{\sim_r} =  [(b_1\cdot b).\xi]_{\sim_r}\tag{because $\sim\subset \sim_r$}\\
\text{iff } &  \hspace*{3mm} b_1\cdot [b.\xi]_{\sim_r} = b_2\cdot [b.\xi]_{\sim_r}\\
\text{then } &  \hspace*{3mm} b_1 = b_2\enspace. \tag{by Lemma \ref{lm:Mon-mod-simr-is-cancellative} }
\end{align*}
\hfill$\Box$
\end{example}

\begin{example}\rm \label{ex:number-of-gammas}  We consider the ranked alphabet $\Sigma$ and the $(\Sigma,\Ratnum)$-weighted tree language $\#_\gamma$ of Example~\ref{ex:for-congruence}. We prove that the $\Ratnum$-scalar algebra $\sfMon(\Sigma,\Ratnum)/_{\sim_{\#_\gamma}}$ is not finitely generated. This means  $\#_\gamma\not\in \budRec(\Sigma,\Ratnum)$  (cf. Theorem  \ref{thm:MN-semifield-det} (A)$\Leftrightarrow$(C)).
In the following we abbreviate $\sim_{\#_\gamma}$ by~$\sim_{\gamma}$.

First we give the following characterization of $\sim_{\gamma}$:
\begin{equation}\label{equ:char-of-sim-for-num-gamma}
    \begin{aligned}
    &\text{for every $b_1.\xi_1, b_2.\xi_2\in \mathrm{Mon}(\Sigma,\Ratnum)$, we have $b_1.\xi_1 \sim_{\gamma} b_2.\xi_2$ }\\
    &\text{if and only if }(b_1=b_2\ \text{ and } \ \#_\gamma(\xi_1)=\#_\gamma(\xi_2)) \text{ or } b_1=b_2=0\enspace.
      \end{aligned}
\end{equation} 
Let $b_1.\xi_1, b_2.\xi_2\in \mathrm{Mon}(\Sigma,\Ratnum)$
such that $b_1.\xi_1 \sim_{\gamma} b_2.\xi_2$. The proof of the if-part is obvious. 

We prove the only-if-part by case analysis.   

\underline{$b_1=0$ or $b_2=0$:} It follows from the definition of $ \sim_{\gamma}$ that
$b_1=b_2=0$, and hence the right-hand side of \eqref{equ:char-of-sim-for-num-gamma} holds. 

\underline{ $b_1\ne 0\ne b_2$:} Then
\begingroup
\allowdisplaybreaks
\begin{align*}
  &b_1.\xi_1 \sim_{\gamma} b_2.\xi_2\\[2mm]
  \text{iff } \ & (\forall c \in \C_\Sigma): b_1\cdot \#_\gamma(c[\xi_1]) =  b_2\cdot \#_\gamma(c[\xi_2])\\[2mm]
  \text{iff } \ & (\forall c \in \C_\Sigma): b_1\cdot \#_\gamma(c) + b_1\cdot \#_\gamma(\xi_1)=b_2\cdot \#_\gamma(c) + b_2\cdot \#_\gamma(\xi_2)\\[2mm]
  \text{iff } \ & (\forall n \in \mathbb{N}): b_1\cdot n + a_1 = b_2\cdot n + a_2 \tag{where $a_i=b_i\cdot \#_\gamma(\xi_i)$ for $i\in\{1,2\}$}\\[2mm]
  \text{iff } \ &  b_1=b_2 \text{ and }  a_1=a_2\\[2mm]
  \text{iff } \ &  b_1=b_2 \text{ and } \#_\gamma(\xi_1)=\#_\gamma(\xi_2)\enspace.
 \end{align*}
  \endgroup 
  This finishes the proof of \eqref{equ:char-of-sim-for-num-gamma}. It follows from \eqref{equ:char-of-sim-for-num-gamma} that, for every $b\in \mathbb{Q}\setminus\{0\}$ and $\xi\in\T_\Sigma$, we have 
  \begin{equation}\label{eq:eq-class-of-sim-gamma}
  [b.\xi]_{\sim_{\gamma}}=\{b.\zeta\mid \xi\in\T_\Sigma, \#_\gamma(\xi)=\#_\gamma(\zeta)\}.
  \end{equation}
  
  Now we show by contradiction that $\sfMon(\Sigma,\Ratnum)/_{\sim_{\gamma}}$ is not finitely generated. For this, assume that
  there exists an $n\in\mathbb{N}$ and a finite subset $H=\{[b_1.\xi_1]_{\sim_{\gamma}},\ldots,[b_n.\xi_n]_{\sim_{\gamma}}\}$ of $\mathrm{Mon}(\Sigma,\Ratnum)/_{\sim_{\gamma}}$ such that $H$ generates $\sfMon(\Sigma,\Ratnum)/_{\sim_{\gamma}}$. We may assume without loss of generality that $b_i\ne 0$ for each $i\in[n]$. Now let $b\in \mathbb{Q}\setminus\{0\}$ and $\zeta\in \T_\Sigma$ such that
  $\#_\gamma(\zeta)\not \in \{\#_\gamma(\xi_1),\ldots,\#_\gamma(\xi_n)\}$. Then it follows from \eqref{eq:eq-class-of-sim-gamma}
  that there does not exist $a\in \mathbb{Q}$ and $i\in[n]$ such that $[b.\zeta]_{\sim_{\gamma}}=a\cdot [b_i.\xi_i]_{\sim_{\gamma}}$. This contradicts the assumption that $H$ generates $\sfMon(\Sigma,\Ratnum)/_{\sim_{\gamma}}$. Hence the $\Ratnum$-scalar algebra 
$\sfMon(\Sigma,\Ratnum)/_{\sim_{\gamma}}$ is not finitely generated.
 \hfill $\Box$
\end{example}


\subsection[Comparison with Borchardt's characterization theorem]{Comparison of the deterministic version of B-A's theorem and Borchardt's characterization theorem}

  \begin{quote} \emph{In this subsection, we let $\B=(B,\oplus,\otimes,\0,\1)$ denote an arbitrary commutative semifield.}
  \end{quote}

Here we present a short comparison of Theorem~\ref{thm:MN-semifield-det}(A)$\Leftrightarrow$(C) and the characterization result of $\budRec(\Sigma,\B)$ in terms of index-finiteness of the Myhill-Nerode congruence $\equiv_r$ defined in \cite{bor03,bor04b}. For this we recall the definition of $\equiv_r$ from \cite[p.~153]{bor03} and \cite[p.~139]{bor04b} and the characterization result itself. 

Let $r:\T_\Sigma \to B$ be a $(\Sigma,\B)$-weighted tree language. The Myhill-Nerode congruence  $\equiv_r$ of $r$  is a congruence on the $\Sigma$-algebra $\mathsf{T}_\Sigma$ defined  as follows:
    for every $\xi_1,\xi_2 \in \T_\Sigma$,
    \begin{equation} \label{equ:relation-between-equivr-simr}
      \xi_1 \equiv_r \xi_2 \ \      \text{ if } \ \ (\exists b_1,b_2 \in B^{-\0}): b_1.\xi_1 \sim_r b_2.\xi_2\enspace.
      \end{equation}

To give an example of the Myhill-Nerode congruence, we consider the weighted tree language $\mathrm{weo}$ in Example \ref{ex:bu-det+total-wta-det-weo}.  We recall that the m-syntactic congruence $\sim_{\mathrm{weo}}$ (cf. Example~\ref{ex-syntactic-congruence}) is the following: for every $b_1.\xi_1,b_2.\xi_2 \in \rmMon(\Sigma,\Ratnum)$, we have
\begin{align*}
  b_1.\xi_1  \sim_{\mathrm{weo}} b_2.\xi_2  \ \ &\text{ iff } \ \  \Big(b_1 \cdot 2^{\#_\alpha(\xi_1)} =  b_2 \cdot 2^{\#_\alpha(\xi_2)} \text{\  and\  }  \big(\#_\alpha(\xi_1) \! \mod (2)\big) = \big(\#_\alpha(\xi_2) \! \mod (2)\big)\Big)\\
  & \hspace*{8mm}\text{or \ } b_1=b_2=0\enspace.
  \end{align*}
  By the definition of $\equiv_{\mathrm{weo}}$ we obtain
  \begin{align*}
  \xi_1  \equiv_{\mathrm{weo}} \xi_2 \ \ \text{ if and only if } \ \ \big(\#_\alpha(\xi_1) \! \mod (2)\big) = \big(\#_\alpha(\xi_2) \! \mod (2)\big) \enspace.
  \end{align*}
  Hence $\equiv_{\mathrm{weo}}$ has two congruence classes.
    
    We note that in \cite{bor03} the author considers restricted $(\Sigma,\B)$-wta, viz. those with unit root weights. In such restricted wta, the root weight mapping $F: Q \to B$ obeys the condition that $\im(F) \subseteq \{\0,\1\}$. In \cite[Lm.~6.1.4]{bor04b}\footnote{The last inclusion of \cite[Lm.~6.1.4]{bor04b} should read: $\cA_f^{d,bu}\langle\!\langle T_\Sigma \rangle\!\rangle \subseteq \cA^{d,bu}\langle\!\langle T_\Sigma \rangle\!\rangle$.} (also cf. \cite[Prop.~4.2]{hanmalque18} and \cite[Lm.~7.3.2]{fulvog24}), it is proved that this is only a restriction of the syntax because, for every $(\Sigma,\B)$-wta $\cA$ there is an equivalent $(\Sigma,\B)$-wta $\cA'$ with unit root weights; if $\cA$ is bu-deterministic, then $\cA'$ is so. Hence, our set $\budRec(\Sigma,\B)$ is equal to the set $\cA^{d,bu}\langle\!\langle T_\Sigma \rangle\!\rangle$ in \cite{bor03,bor04b}

    \begin{samepage}
    \begin{theorem}{\rm (\cite[Thm.~2]{bor03} and \cite[Thm.~7.3.1]{bor04b})} \label{thm:borchardt-MN-det}  Let $\B$ a commutative semifield. Moreover, let $r: \T_\Sigma \to B$. Then the following two statements are equivalent.
      \begin{compactenum}
      \item[(A)] $r \in \budRec(\Sigma,\B)$.
        \item[(B)] The congruence $\equiv_r$ on the $\Sigma$-term algebra $\sfT_\Sigma$ has finite index.
        \end{compactenum}
      \end{theorem}
\end{samepage}

From Theorem~\ref{thm:MN-semifield-det}(A)$\Leftrightarrow$(C) and  Theorem~\ref{thm:borchardt-MN-det}
we obtain the following corollary.

 \begin{corollary}\label{lm:comparison-Bjorn-result-det}\rm  Let $\B$ a commutative semifield. Moreover, let $r: \T_\Sigma \to B$. Then the following two statements are equivalent.
 \begin{compactitem}
 \item[(A)] The congruence $\equiv_r$ on the $\Sigma$-term algebra $\sfT_\Sigma$ has finite index.
 \item[(B)]  The $\B$-scalar algebra $\sfMon(\Sigma,\B)/_{\sim_r}$ is finitely generated.
  \end{compactitem}
\end{corollary}

  \begin{lemma}\label{lm:comparison-degree-index}\rm  Let $r: \T_\Sigma \to B$. If the $\B$-scalar algebra $\sfMon(\Sigma,\B)/_{\sim_r}$ is finitely generated and the congruence $\equiv_r$ on the $\Sigma$-term algebra $\sfT_\Sigma$ has finite index, then 
 \begin{align*}
 \deg(\sfMon(\Sigma,\B)/_{\sim_r}) =  \begin{cases} |\T_\Sigma/_{\equiv_r}| & \text{ if $[\widetilde{\0}]_{\sim_r}=\{\widetilde{\0}\}$}\\
                            |\T_\Sigma/_{\equiv_r}| - 1& \text{ otherwise.}
              \end{cases}
 \end{align*}
 \end{lemma}
 \begin{proof} In the proof we abbreviate $\sim_r$ and $\equiv_r$ by $\sim$ and $\equiv$, respectively. First we define the set
   \[
   T_\0=\{\zeta\in \T_\Sigma \mid (\exists a\in B^{-\0}) : \widetilde{\0}\sim a.\zeta\} \enspace.
   \]
 We note that  $[\widetilde{\0}]_{\sim}=\{\widetilde{\0}\}$ if and only if  $T_\0=\emptyset$. 
 We show that if $T_\0\not=\emptyset$, then $T_\0$  constitutes a congruence class of $\equiv_r$. Formally,
 \begin{equation}\label{equ:T0-char}
   \text{if  $T_\0\not=\emptyset$, then for each $\zeta \in T_\0$, we have $T_\0=[\zeta]_{\equiv}$.}
 \end{equation}
 For this, let $\zeta \in T_\0$. By definition, there exists an $a\in B^{-\0}$ such that $\widetilde{\0}\sim a.\zeta$.
 
 To show that $T_\0\subseteq[\zeta]_{\equiv}$, let $\zeta'\in T_\0$. By definition, there exists an $a'\in B^{-\0}$ such that $\widetilde{\0} \sim a'.\zeta'$. Since $\sim$ is transitive, we have $a.\zeta\sim a'.\zeta'$. Hence,  by \eqref{equ:relation-between-equivr-simr}, we have $\zeta'\in [\zeta]_{\equiv}$.
 
 To show that $[\zeta]_{\equiv}\subseteq T_\0$, let $\zeta'\in [\zeta]_{\equiv}$. By \eqref{equ:relation-between-equivr-simr}, there exist $b,b'\in B^{-\0}$ with
 $b.\zeta \sim b'.\zeta'$. Then $\1.\zeta\sim (b^{-1}\otimes b').\zeta'$, and thus $\widetilde{\0}\sim a.\zeta\sim (a\otimes b^{-1}\otimes b').\zeta'$. Hence $\zeta'\in T_\0$. This proves \eqref{equ:T0-char}.
 
 Now we prove the statement of the lemma by case analysis.

 \underline{$T_\0=\T_\Sigma$:} Let $\xi,\zeta \in \T_\Sigma$. Then $\xi,\zeta \in T_\0$ and by \eqref{equ:T0-char} we obtain that $\xi \equiv \zeta$. Hence $|\T_\Sigma/_{\equiv}|=1$. 

 Next we prove that $\rmMon(\Sigma,\B)/_{\sim}=\{\rmMon(\Sigma,\B)\} =\{[\0]_\sim\}$.
 For this, let $b.\xi \in \rmMon(\Sigma,\B)$. Since $\xi \in T_\0$, there exists $a \in B^{-\0}$ such that $\widetilde{\0} \sim a.\xi$. Then we can calculate as follows:
 \[
 \widetilde{\0} = (b \otimes a^{-1}) \cdot \widetilde{\0} \sim (b \otimes a^{-1}) \cdot a.\xi =  (b \otimes a^{-1} \otimes a).\xi = b.\xi \enspace,\]
 where the first and second equality hold by (\ref{SM5-det}) and the definition of the scalar multiplication,  respectively, in $\sfMon(\Sigma,\B)$.
Thus $\sfMon(\Sigma,\B)/_{\sim}=\{\rmMon(\Sigma,\B)\} =\{[\0]_\sim\}$. Since $\{[\0]_\sim\}$ is generated by $\emptyset$ (which is pair-independent), we have $\deg(\sfMon(\Sigma,\B)/_{\sim}) = 0$. Since $T_\0 = \T_\Sigma \ne \emptyset$, the statement of the lemma holds.

 \underline{$T_\0\subset \T_\Sigma$:} Let \[f :  \big(\rmMon(\Sigma,\B)/_{\sim}\setminus \{[\widetilde{\0}]_{\sim}\}\big)\to \big(\T_\Sigma/_{\equiv}\setminus \{T_\0\}\big)\] 
be the mapping defined, for every $b\in  B^{-\0}$ and $\xi\in \T_\Sigma$, by $f([b.\xi]_{\sim}) = [\xi]_{\equiv}$. (Note that if $T_\0=\emptyset$, then then $\T_\Sigma/_{\equiv}\setminus \{T_\0\} = \T_\Sigma/_{\equiv}$ because each equivalence class is non-empty.)
The mapping $f$ is well defined because, (a) if $[b.\xi]_{\sim}\ne [\widetilde{\0}]_{\sim}$, then $[\xi]_{\equiv}\ne T_\0$, and (b)
if there exist $a\in B^{-\0}$ and $\zeta \in \T_\Sigma$ with $b.\xi \sim a.\zeta$,  we have $\xi \equiv \zeta$ by \eqref{equ:relation-between-equivr-simr}.  Moreover, $f$ is surjective.

 Next let $g: \big(\T_\Sigma/_{\equiv}\setminus \{T_\0\}\big) \to   \big(\rmMon(\Sigma,\B)/_{\sim}\setminus \{[\widetilde{\0}]_{\sim}\}\big)$ be an injective mapping such that, for each $[\xi]_{\equiv} \in  \big(\T_\Sigma/_{\equiv}\setminus \{T_\0\}\big)$, we have
 $f(g([\xi]_{\equiv} ))=[\xi]_{\equiv}$. Such a mapping $g$ exists because $f$ is surjective. 
 
 Now let $H=\{g([\xi]_{\equiv})\mid \xi \in \T_\Sigma, [\xi]_{\equiv}\ne T_\0\}$. Since $T_\0\subset \T_\Sigma$. we have that $H\ne\emptyset$. Moreover, since $g$ is injective, we have 
 \begin{align*}
 |H| =  \begin{cases} |\T_\Sigma/_{\equiv}| & \text{ if $T_\0=\emptyset$}\\
                            |\T_\Sigma/_{\equiv}| - 1& \text{ otherwise,}
              \end{cases}
 \end{align*}
where in the second case we use the fact that $T_\0$ is a congruence class of $\equiv$ (cf. \eqref{equ:T0-char}).
 Hence it suffices to prove that $H$ is a pair-independent generating set of $\sfMon(\Sigma,\B)/_{\sim}$.

 First we prove that $H$ generates $\sfMon(\Sigma,\B)/_{\sim}$, i.e., that $B\cdot H=\rmMon(\Sigma,\B)/_{\sim}$.
 Since $B\cdot H\subseteq \rmMon(\Sigma,\B)/_{\sim}$ is obvious, it suffices to prove that $\rmMon(\Sigma,\B)/_{\sim} \subseteq B\cdot H$.
 For this, let $[a.\xi]_{\sim}$ be an element  of $\rmMon(\Sigma,\B)/_{\sim}$  for some $a\in B$ and some $\xi \in \T_\Sigma$. We distinguish the following three cases.

  \underline{$a=\0$, i.e., $[a.\xi]_{\sim}=[\widetilde{\0}]_{\sim}$}: Then $[a.\xi]_{\sim}=\0\cdot [b.\xi]_{\sim}$ for each 
  $[b.\xi]_{\sim}\in H$.

     \underline{$a\ne\0$ and $[a.\xi]_{\sim} \in H$:} Then we are done because $[a.\xi]_{\sim} = \1 \cdot [a.\xi]_{\sim}  \in B \cdot H$.
     
     \underline{$a\ne\0$ and $[a.\xi]_{\sim} \not\in H$:} Let $g([\xi]_{\equiv})=[b.\xi]_{\sim}$
   for some $b\in B^{-\0}$. Then we have $[a.\xi]_{\sim}=a\cdot[\1.\xi]_{\sim}=(a\otimes b^{-1} \otimes b)\cdot[\1.\xi]_{\sim} =(a\otimes b^{-1})\cdot[b.\xi]_{\sim}$, i.e., $[a.\xi]_{\sim}\in B\cdot H$.
     
This proves that $B\cdot H=\rmMon(\Sigma,\B)/_{\sim}$. 

 Lastly we show that $H$ is pair-independent. For this, let $[b.\xi]_{\sim}$ and $[c.\zeta]_{\sim}$ be two different elements of 
 $H$ for some $b,c\in  B^{-\0}$ and $\xi,\zeta \in  \T_\Sigma$. 
 
 We begin with showing that $\xi \not\equiv \zeta$. Since $[b.\xi]_{\sim}\in H$, there exists an $\eta\in \T_\Sigma$ such that
  $g([\eta]_{\equiv})=[b.\xi]_{\sim}$. Then $f([b.\xi]_{\sim}) = f(g([\eta]_{\equiv}))=[\eta]_{\equiv}$. On the other hand 
  $f([b.\xi]_{\sim}) = [\xi]_{\equiv}$. Hence $[\xi]_{\equiv} = [\eta]_{\equiv}$ and thus $g([\xi]_{\equiv})=[b.\xi]_{\sim}$.
   Similarly, we can show that $g([\zeta]_{\equiv})=[c.\zeta]_{\sim}$. Since $g$ is a mapping, we obtain that
   $[\xi]_{\equiv}\ne [\zeta]_{\equiv}$. 

   Then we show that $[b.\xi]_{\sim}$ and $[c.\zeta]_{\sim}$ are independent by contradiction. For this, assume that there exists $a\in B$ with $[b.\xi]_{\sim}=a\cdot [c.\zeta]_{\sim}$. Then we have $[b.\xi]_{\sim}=[(a\otimes c).\zeta]_{\sim}$ and since $a=\0$ is not possible and $\B$ is zero-divisor free, we have $\xi \equiv \zeta$. This is a contradiction.
  \end{proof}

\subsection{d-Minimal bu-deterministic wta and its construction}
\label{subs:minimization}

Here we show how to construct, for a given bu-deterministic $(\Sigma,\B)$-wta an equivalent one which is d-minimal.

  \begin{quote}{\em In this subsection, we let $\B=(B,\oplus,\otimes,\0,\1)$ denote an arbitrary commutative semifield. Moreover, $\cA=(Q,\delta,F)$ denotes an arbitrary bu-deterministic $(\Sigma,\B)$-wta, if not specified otherwise.}
  \end{quote}

\index{bu-deterministic!d-minimal}
\index{dminimal@d-minimal}
We say that $\cA$ is \emph{d-minimal} if, for each bu-deterministic $(\Sigma,\B)$-wta  $\cB=(Q',\delta',F')$ with $\sem{\cA}=\sem{\cB}$, we have $|Q|\le |Q'|$. Thus each minimal bu-det wta is d-minimal.
The next lemma can be compared to \cite[Thm.~3(a)]{bor03}.

 \begin{lemma}\label{cor:min-du-det-wta-exists}\rm For each finite scalar-basis  $H$ of $\sfMon(\Sigma,\B)/_{\sim_{\sem{\cA}}}$, the $(\Sigma,\B)$-wta $\budwta(\sem{\cA},\sim_{\sem{\cA}},H)$ is defined and it is d-minimal.
 \end{lemma}
 \begin{proof}  Let us abbreviate $\sem{\cA}$ by $r$. By Lemma \ref{lm:sim-r-is-coarsest} the congruence $\sim_r$ saturates $r$ and, by Lemma \ref{lm:Mon-mod-simr-is-cancellative},  the $\B$-scalar algebra $\sfMon(\Sigma,\B)/_{\sim_r}$ is  cancellative. Now let $H$ be a finite scalar-basis of  $\sfMon(\Sigma,\B)/_{\sim_{r}}$.
 Then $\budwta(r,\sim_r,H)$ is defined (cf. Definition \ref{def:wta(r,sim,H)}) and, by Lemma \ref{lm:wta(r,sim,H)}, it is a bu-deterministic $(\Sigma,\B)$-wta and it is equivalent to $\cA$.

Next we show that $\budwta(r,\sim_r,H)$ is d-minimal.  We recall that $H$ is the set of states of $\budwta(r,\sim_r,H)$ and $|H|=\deg\big( \mathsf{Mon}(\Sigma,\B)/_{\sim_r}\big)$.  Let $\cB=(Q',\delta',F')$ be an arbitrary bu-deterministic $(\Sigma,\B)$-wta with  $r=\sem{\cB}$. By Lemma~\ref{lm:dimension-synt-factor-algebra-bounded}, we have $|H|\le |Q'|$.  Thus $\budwta(r,\sim_r,H)$ is d-minimal.
\end{proof}

In the following we show that we can construct a finite generating set $\mathrm{H}_{\cA}$ of $\sfMon(\Sigma,\B)/_{\sim_{\sem{\cA}}}$ (cf. Corollary \ref{cor:construct-finite-scalar-basis}).
Hence, if dependency in  $\sfMon(\Sigma,\B)/_{\sim_{\sem{\cA}}}$ is decidable, then, by Lemma \ref{lm:crucial-canc-pair-ind-imply-uniqueness-det-1},  we can construct a finite scalar-basis $H \subseteq \mathrm{H}_{\cA}$ of $\sfMon(\Sigma,\B)/_{\sim_{\sem{\cA}}}$.
Since $\sfMon(\Sigma,\B)/_{\sim_{\sem{\cA}}}$ is cancellative, the decomposition mapping
\[\mathrm{dec}: \big(\rmMon(\Sigma,\B)/_{\sim_{\sem{\cA}}}\big) \setminus \{[\widetilde{\0}]_{\sim_{\sem{\cA}}}\} \to B^{-\0} \times \big(H \setminus \{[\widetilde{\0}]_{\sim_{\sem{\cA}}}\}\big)\]
for $\big(\rmMon(\Sigma,\B)/_{\sim_{\sem{\cA}}}\big) \setminus \{[\widetilde{\0}]_{\sim_{\sem{\cA}}}\}$
exists (cf. \eqref{equ:scalar-vector-det-new}). 
If $\mathrm{dec}$ is computable, then by Lemma~\ref{lm:wta(r,sim,H)}(3) we can construct the d-minimal $(\Sigma,\B)$-wta $\budwta(\sem{\cA},\sim_{\sem{\cA}},H)$.
The construction  of the d-minimal $(\Sigma,\B)$-wta  proceeds in three steps and it is summarized in Theorem \ref{thm:minimization-theorem-new}

\

\underline{Step 1:} We construct a slim $(\Sigma,\B)$-wta which is equivalent to $\cA$ (cf. Theorem~\ref{lm:construction-of-simple-det-new}).
We recall  that $\cA$ is slim if, for each $q \in Q$, there exists $\xi \in \T_\Sigma$ such that $\state_\cA(\xi) = \{q\}$.

    In fact, for each bu-deterministic wta, being slim is a necessary condition for being d-minimal.

\begin{lemma}\label{lm:minimal-implies-slim}\rm  If $\cA$ is d-minimal, then $\cA$ is slim.
\end{lemma}
\begin{proof} We prove the contraposition of the statement. Let $\cA$ be not slim. Then there exists a $q_0 \in Q$ such that, for each $\xi \in \T_\Sigma$, we have $\state_\cA(\xi)\ne \{q_0\}$. By Lemma~\ref{lm:hAxiqne0-implies-qinhQxi-for-bu-det-wta}(1) we have that $\h_\cA(\xi)_{q_0}=\0$ for each $\xi\in\T_\Sigma$. 

Now we consider the $(\Sigma,\B)$-wta $\cA_{-q_0}=(Q',\delta',F')$ as it is defined in Lemma~\ref{lm:wta-getting-rid-of-useless-state}. Clearly, $\cA_{-q_0}$ is bu-deterministic and $|Q'| = |Q|-1$. Since $\sem{\cA_{-q_0}} = \sem{\cA}$ by Lemma~\ref{lm:wta-getting-rid-of-useless-state}, we obtain that $\cA$ is not d-minimal.
\end{proof}

\

\underline{Step 2:} For each slim $(\Sigma,\B)$-wta $\cA$, we define the concept of a candidate set induced by $\cA$ for $\sfMon(\Sigma,\B)/_{\sim_{\sem{\cA}}}$. We prove that each such candidate set generates $\sfMon(\Sigma,\B)/_{\sim_{\sem{\cA}}}$.
Essentially, this is due to the fact that $\cA$ is slim. Then we show how to construct such a candidate set.

Let $\cA$ be slim with $Q= \{q_1,\ldots,q_n\}$ and let us abbreviate $\sem{\cA}$ by $r$.  Since $\cA$ is slim, none of the sets $\state_\cA^{-1}(\{q_1\}),\ldots,\state_\cA^{-1}(\{q_n\})$ is empty. 
For every $\zeta_1 \in \state_\cA^{-1}(\{q_1\}), \ldots, \zeta_n \in \state_\cA^{-1}(\{q_n\})$, we call the set
\[
\{[\1.\zeta_1]_{\sim_r}, \ldots,  [\1.\zeta_n]_{\sim_r}\},
\]
a \emph{candidate set induced by $\cA$ for $\sfMon(\Sigma,\B)/_{\sim_r}$}.
(Hence for each slim  $(\Sigma,\B)$-wta  there exists such a candidate set.)

We recall that $\cP_{=1}(Q)= \{\{q\} \mid q \in Q\}$. Moreover, for each $\xi \in \T_\Sigma$,  if $\state_\cA(\xi) \in \cP_{=1}(Q)$, then $\estate_\cA(\xi)$ denotes the unique element in the set $\state_\cA(\xi)$.

\begin{lemma} \label{thm:bu-det-r-construction-ofVLQ(r)-det} \rm \sloppy Let  $\cA$
  be slim  with $|Q|= n$ and let us abbreviate $\sem{\cA}$ by $r$. Moreover, let $H = \{[\1.\zeta_1]_{\sim_r}, \ldots,  [\1.\zeta_n]_{\sim_r}\}$ be a candidate set induced by $\cA$ for $\sfMon(\Sigma,\B)/_{\sim_r}$. The following two statements hold.
\begin{itemize}
\item[(1)] For every $b \in B$ and  $\xi \in \T_\Sigma$, we have:
 
 -  if $\state_\cA(\xi) \in \cP_{=1}(Q)$, then  there exists $j \in [n]$ such that\\
  \hspace*{3mm} $[b.\xi]_{\sim_r} = d \cdot [\1.\zeta_j]_{\sim_r}$ where
     $d = b \otimes \h_\cA(\xi)_{\estate_\cA(\xi)} \otimes \big(\h_\cA(\zeta_j)_{\estate_\cA(\zeta_j)}\big)^{-1}$,

 - otherwise $[b.\xi]_{\sim_r} = \0\cdot [\1.\zeta_j]_{\sim_r}$ for each $j\in [n]$.
 
\item[(2)]  $H$ generates the $\B$-scalar algebra $\sfMon(\Sigma,\B)/_{\sim_r}$.
  \end{itemize}
\end{lemma}

\begin{proof} First we show that 
\begin{equation*}
\text{$\h_\cA(\zeta_j)_{\estate_\cA(\zeta_j)} \ne \0$ for each $j \in [n]$.}
\end{equation*}
Let $j \in [n]$. By definition $\zeta_j \in \state_\cA^{-1}(\{q_j\})$, i.e., $\state_\cA(\zeta_j) = \{q_j\}=\{\estate_\cA(\zeta_j)\}$.
 Since $\B$ is zero-divisor free,  the statement follows by Lemma \ref{lm:hAxiqne0-implies-qinhQxi-for-bu-det-wta}(2).
 Hence $d$ in Statement (1) is well defined.

 \
 
Proof of (1):   Let $b \in B$ and $\xi \in \T_\Sigma$. 
  We distinguish the following two cases.

    \underline{$\state_\cA(\xi) \in \cP_{=1}(Q)$:}  Let $j \in [n]$ be the unique number such that $\state_\cA(\xi) = \state_\cA(\zeta_j)$.  We prove that $b.\xi \sim_r d.\zeta_j$. Let $c \in \C_\Sigma$. By Corollary \ref{cor:hAxiqne0-iff-qinhQxi-for-bu-det-wta}, we have that $\h_\cA(c[\xi]) \in B_{=1}^Q$ if and only if $\state_\cA(c[\xi]) \in \cP_{=1}(Q)$. Then
          \begingroup
  \allowdisplaybreaks
\begin{align*}
  & b \otimes r(c[\xi]) \\
  = & \
  \begin{cases} b \otimes \h_\cA(c[\xi])_{\estate_\cA(c[\xi])} \otimes F_{\estate_\cA(c[\xi])}
    & \text{ if $\state_\cA(c[\xi]) \in \cP_{=1}(Q)$}\\
    \0  & \text{ otherwise}
    \end{cases}
      \tag{by Corollary~\ref{lm:properties-hA-of-budet-wta-det-new-H}}\\[2mm]
      = &\ \begin{cases}
      b \otimes  \h_\cA(\xi)_{\estate_\cA(\xi)} \otimes \h_\cA^\C(c)(\1_{\estate_\cA(\xi)})_{\estate_\cA(c[\xi])} \otimes F_{\estate_\cA(c[\xi])} & \text{ if $\state_\cA(c[\xi]) \in \cP_{=1}(Q)$}\\
    \0  & \text{ otherwise}
    \end{cases}
  \tag{by Lemma~\ref{obs:total-bu-det-wta-calc(new)-H}}\\[2mm]
  = &\ \begin{cases}
  b \otimes  \h_\cA(\xi)_{\estate_\cA(\xi)} \otimes \h_\cA^\C(c)(\1_{\estate_\cA(\zeta_j)})_{\estate_\cA(c[\zeta_j])} \otimes F_{\estate_\cA(c[\zeta_j])}& \text{ if $\state_\cA(c[\zeta_j]) \in \cP_{=1}(Q)$}\\
    \0  & \text{ otherwise}
    \end{cases}
  \tag{because  $\estate_\cA(\xi) = \estate_\cA(\zeta_j)$ and thus $\estate_\cA(c[\xi]) = \estate_\cA(c[\zeta_j])$} \\[2mm]
  = &\ \begin{cases}
  d \otimes \h_\cA(\zeta_j)_{\estate_\cA(\zeta_j)} \otimes \h_\cA^\C(c)(\1_{\estate_\cA(\zeta_j)})_{\estate_\cA(c[\zeta_j])} \otimes F_{\estate_\cA(c[\zeta_j])}& \text{ if $\state_\cA(c[\zeta_j]) \in  \cP_{=1}(Q)$}\\
    \0  & \text{ otherwise}
    \end{cases}
  \tag{where $d$ is defined in Statement (1) of the theorem}\\[2mm]
  = &\ \begin{cases}
  d \otimes \h_\cA(c[\zeta_j])_{\estate_\cA(c[\zeta_j])} \otimes F_{\estate_\cA(c[\zeta_j])}& \text{ if $\state_\cA(c[\zeta_j]) \in  \cP_{=1}(Q)$}\\
    \0  & \text{ otherwise}
    \end{cases}
      \tag{by Lemma~\ref{obs:total-bu-det-wta-calc(new)-H}}\\[2mm]
  = &\ d \otimes r(c[\zeta_j])
      \tag{by Corollary~\ref{lm:properties-hA-of-budet-wta-det-new-H}}
      \enspace.
\end{align*}
\endgroup
Thus $[b.\xi]_{\sim_r} =  d \cdot [\1.\zeta_j]_{\sim_r}$.

\

\underline{$\state_\cA(\xi) \not\in \cP_{=1}(Q)$:} Then $\state_\cA(\xi)=\emptyset$.
  We prove that $b.\xi \sim_r \widetilde{\0}$. Let $c \in \C_\Sigma$.
  \begin{align*}
    b \otimes r(c[\xi]) = b \otimes \big( \h_\cA(c[\xi]) \cdot F \big)
    = b \otimes (\0^Q \cdot F)
    = \0 \enspace,
  \end{align*}
  where the second equality follows by  Lemma~\ref{lm:hAxiqne0-implies-qinhQxi}(1).
Thus $[b.\xi]_{\sim_r} =  [\widetilde{\0}]_{\sim_r}$ and hence $[b.\xi]_{\sim_r} = \0\cdot [\1.\zeta_j]_{\sim_r}$ for each $j\in [n]$.

\

Proof of (2): By Statement (1), we have  $\rmMon(\Sigma,\B)/_{\sim_r} \subseteq B \cdot H$. Since $B \cdot H \subseteq \rmMon(\Sigma,\B)/_{\sim_r}$ and $H \ne \emptyset$, Statement(2) follows from Observation \ref{obs:scalar-algebra-fin-gen-subalg-hom-1}(1).
\end{proof}

\begin{example}\rm \label{ex:candidate-set-det} The $(\Sigma,\Ratnum)$-wta $\cA$ of Example \ref{ex:bu-det+total-wta-det-weo} is slim; it has the state set $Q=\{o,e\}$. An example of a candidate set induced by $\cA$ for $\sfMon(\Sigma,\Ratnum)/_{\sim_r}$ (with $r= \sem{\cA}$) is
  \[
H = \{[1.\alpha]_{\sim_r}, \ [1.\sigma(\alpha,\alpha)]_{\sim_r} \}
\]
because $\state_\cA(\alpha) = \{o\}$ and $\state_\cA(\sigma(\alpha,\alpha))=\{e\}$.
(The set $\{[1.\sigma(\sigma(\alpha,\alpha),\alpha)]_{\sim_r}, \ [1.\sigma(\alpha,\alpha)]_{\sim_r} \}$ is another example of a candidate set induced by $\cA$ for $\sfMon(\Sigma,\Ratnum)/_{\sim_r}$ because also $\state_\cA\big(\sigma(\sigma(\alpha,\alpha),\alpha)\big)=o$.) We demonstrate Lemma \ref{thm:bu-det-r-construction-ofVLQ(r)-det}(1) as follows:
for each $b \in \mathbb{Q}$, we have
\begin{align*}
  [b.\sigma(\sigma(\alpha,\alpha),\alpha)]_{\sim_r}
  = & \ d \cdot [1.\alpha]_{\sim_r} \ \text{ with }\\
  & \ d = b \cdot \h_\cA(\sigma(\sigma(\alpha,\alpha),\alpha))_o \cdot (\h_\cA(\alpha)_o)^{-1} = b \cdot 8 \cdot 2^{-1} = b \cdot 4 \ \text{ and }\\[2mm]
   [b.\sigma(\sigma(\alpha,\alpha),\sigma(\alpha,\alpha))]_{\sim_r}
  = &\ d' \cdot [1.\sigma(\alpha,\alpha)]_{\sim_r} \ \text{ with }\\
  &\ d' = b \cdot \h_\cA(\sigma(\sigma(\alpha,\alpha),\sigma(\alpha,\alpha)))_o \cdot (\h_\cA(\sigma(\alpha,\alpha))_o)^{-1} = b \cdot 16 \cdot 4^{-1} = b \cdot 4 \enspace.
    \end{align*}
    By Lemma \ref{thm:bu-det-r-construction-ofVLQ(r)-det}(2), the set $H$ generates $\sfMon(\Sigma,\Ratnum)/_{\sim_r}$.
    \hfill $\Box$
  \end{example}
  
  The question arises whether, for each slim  $(\Sigma,\B)$-wta, each candidate set induced by $\cA$ is pair-independent.   By the following example we demonstrate that the answer is negative.

  \begin{example}\rm \label{ex:candidate-set-not-independent} Let $\Sigma=\{\alpha^{(0)},\beta^{(0)}\}$.  Then $\T_\Sigma=\{\alpha,\beta\}$ and $\C_\Sigma=\{z\}$.
  Let $\cA=(Q,\delta,F)$ be the $(\Sigma,\Ratnum)$-wta where 
  \begin{compactitem}
  \item $Q=\{q_0,q_1\}$,
  \item  $\delta_0(\varepsilon,\alpha,q_0)=1$ and $\delta_0(\varepsilon,\beta,q_1)=1$,
  \item $F_{q_0}=2$ and $F_{q_1}=1$.
 \end{compactitem}
 Then $\sem{\cA}(\alpha)=2$ and $\sem{\cA}(\beta)=1$.

 The only candidate set is $\{[1.\alpha]_{\sim_{\sem{\cA}}},[1.\beta]_{\sim_{\sem{\cA}}}\}$.
We have $[1.\alpha]_{\sim_{\sem{\cA}}}\ne[1.\beta]_{\sim_{\sem{\cA}}}$ because (with the only context $c=z$) \[1\cdot \sem{\cA}(c[\alpha]) =1\cdot \sem{\cA}(\alpha) =2 \ne 1 = 1\cdot \sem{\cA}(\beta)= 1\cdot \sem{\cA}(c[\beta]).\]
 Moreover, $\{[1.\alpha]_{\sim_{\sem{\cA}}},[1.\beta]_{\sim_{\sem{\cA}}}\}$ is not pair-independent because $[1.\alpha]_{\sim_{\sem{\cA}}}=2\cdot[1.\beta]_{\sim_{\sem{\cA}}}$.
 This is because $2\cdot[1.\beta]_{\sim_{\sem{\cA}}}=[2.\beta]_{\sim_{\sem{\cA}}}$ and $1.\alpha \sim_{\sem{\cA}} 2.\beta$ because for every context (we have only $c=z$) we have
 \[1\cdot \sem{\cA}(c[\alpha])=1\cdot \sem{\cA}(\alpha)=1\cdot 2 = 2 = 2\cdot 1 = 2\cdot \sem{\cA}(\beta)=2\cdot \sem{\cA}(c[\beta]).\]
   \hfill $\Box$
 \end{example}
 

  \begin{lemma}\rm \label{lm:tq-has-minimal-height} Let  $\cA$ be slim. We can construct a finite generating set $\mathrm{H}_{\cA}$ of  $\sfMon(\Sigma,\B)/_{\sim_{\sem{\cA}}}$.
  \end{lemma}
  \begin{proof} Let us abbreviate $\sem{\cA}$ by $r$. First we construct a mapping $\mathrm{t}_\cA: Q \to \T_\Sigma$ such that $\state(\mathrm{t}_\cA(q))=\{q\}$ for each $q\in Q$. The construction is as follows.

We assume that a linear order $\xi_1,\xi_2,\ldots$ on $\T_\Sigma$ is given.
We construct a finite prefix of a sequence $t_0,t_1,t_2,\ldots$ of partial mappings from $Q$ to $\T_\Sigma$ as follows. We let $t_0=\emptyset$. Let $t_i$ be already constructed.
If $t_i$ is defined for each $q\in Q$,  then we stop the construction and we let  $\mathrm{t}_\cA=t_i$.
Otherwise we proceed by case analysis. If $\state(\xi_{i+1})=\emptyset$ or  $\state(\xi_{i+1})\in \cP_{=1}(Q)$ and $t_i(\estate(\xi_{i+1}))$ is already defined, then let $t_{i+1}=t_i$.
Otherwise we let $t_{i+1}=t_i \cup \{(\estate(\xi_{i+1}),\xi_{i+1})\}$.
Since $\cA$ is slim, and hence $\im(\state) = \cP_{=1}(Q)$, there exists an $i \in \mathbb{N}$ such that $t_i$ is a mapping, and hence the construction eventually stops.
  
Then the set $\mathrm{H}_{\cA} = \{[\1.\mathrm{t}_\cA(q)]_{\sim_{r}} \mid q \in Q\}$
is a candidate set induced by $\cA$ for $\sfMon(\Sigma,\B)/_{\sim_{r}}$.
By Lemma \ref{thm:bu-det-r-construction-ofVLQ(r)-det}(2), the set $\mathrm{H}_\cA$ generates $\sfMon(\Sigma,\B)/_{\sim_{r}}$.
\end{proof}

\begin{corollary}\rm \label{cor:construct-finite-scalar-basis} Let $\B$ be a commutative semifield and $\cA$  a bu-deterministic $(\Sigma,\B)$-wta. We can construct a finite generating set $\mathrm{H}_{\cA}$ of $\sfMon(\Sigma,\B)/_{\sim_{\sem{\cA}}}$.
\end{corollary}
\begin{proof} Let $\cA=(Q,\delta,F)$. By Theorem~\ref{lm:construction-of-simple-det-new}, we can construct a slim $(\Sigma,\B)$-wta $\cA'$ such that $\sem{\cA}=\sem{\cA'}$. Then, by Lemma~\ref{lm:tq-has-minimal-height}
we can construct a finite generating set $\mathrm{H}_{\cA'}$ of  $\sfMon(\Sigma,\B)/_{\sim_{\sem{\cA'}}}$. We let  $\mathrm{H}_{\cA}= \mathrm{H}_{\cA'}$.
  \end{proof}

\

 \underline{Step 3:} 
Now we show how to construct, for a given  bu-deterministic $(\Sigma,\B)$-wta, an equivalent d-minimal bu-deterministic $(\Sigma,\B)$-wta.
  
 \begin{theorem-rect}\label{thm:minimization-theorem-new} Let $\B$ be a commutative semifield. Moreover, let $\cA$ be a bu-deterministic $(\Sigma,\B)$-wta and let us abbreviate $\sem{\cA}$ by $r$. We assume that dependency in $\sfMon(\Sigma,\B)/_{\sim_r}$ is decidable and the decomposition mapping $\mathrm{dec}$ for $\big(\rmMon(\Sigma,\B)/_{\sim_r}\big)\setminus \{[\widetilde{\0}]_{\sim_r}\}$ is computable.
   Then we can construct a finite scalar-basis $H$ of $\sfMon(\Sigma,\B)/_{\sim_r}$
 and we can construct the d-minimal  bu-deterministic $(\Sigma,\B)$-wta $\budwta(r,\sim_r,H)$.
     \end{theorem-rect} 
     \begin{proof} By Corollary~\ref{cor:construct-finite-scalar-basis}, we can construct a finite generating set  $\mathrm{H}_{\cA}$ of $\sfMon(\Sigma,\B)/_{\sim_{r}}$.
    Since dependency in $\sfMon(\Sigma,\B)/_{\sim_r}$ is decidable, by Lemma~\ref{lm:crucial-canc-pair-ind-imply-uniqueness-det-1}, we can construct a finite scalar-basis $H\subseteq \mathrm{H}_{\cA}$ of $\sfMon(\Sigma,\B)/_{\sim_r}$.
  By Lemma \ref{cor:min-du-det-wta-exists} the $(\Sigma,\B)$-wta $\budwta(r,\sim_r,H)$ is defined and d-minimal.
       Since the decomposition mapping $\mathrm{dec}$ is computable, we can construct $\budwta(r,\sim_r,H)$ (cf. Lemma \ref{lm:wta(r,sim,H)}(3)).
   \end{proof}

Finally, we give an example of the construction of the d-minimal bu-deterministic $(\Sigma,\B)$-wta. 

  \begin{example}\rm \label{ex:B-implies-A} Here we show an example of the construction of a d-minimal bu-deterministic wta as it is described in the proof of Theorem \ref{thm:minimization-theorem-new}.

We consider the ranked alphabet  $\Sigma=\{\gamma^{(1)},\alpha^{(0)}\}$. Each tree in $\T_\Sigma$ has the form $\gamma(\ldots\gamma(\alpha)\ldots)$ with $n$ occurrences of $\gamma$ for some $n\in\mathbb{N}$. We denote this tree by $\gamma^n(\alpha)$.
    Let us consider the  $(\Sigma,\Ratnum)$-wta $\cA=(Q,\delta,F)$ (cf. Figure~\ref{fig:making-wta-minimal}(left)) where
\begin{compactitem}
\item $Q=\{q_1,q_2,q_3\}$,
\item $\delta_0(\varepsilon,\alpha,q_1)=1$,  $\delta_1(q_1,\gamma,q_2)=\delta_1(q_2,\gamma,q_3)=\delta_1(q_3,\gamma,q_2)=1$, and $\delta_1(p,\gamma,q)=0$ for any other combination $p,q\in Q$, and
\item $F_{q_1}=F_{q_3}=2$ and $F_{q_2}=3$.
\end{compactitem}
Obviously, $\cA$ is total and bu-deterministic.  Moreover, $\cA$ is slim, because, e.g.,  $\state(\alpha)=q_1$, $\state(\gamma(\alpha))=q_2$, and $\state(\gamma^2(\alpha))=q_3$.
Also, for each $n \in \mathbb{N}$, we have
\[\sem{\cA}(\gamma^n(\alpha))=\begin{cases}
2 & \text{ if $n$ is even,} \\
3 & \text{ otherwise.} 
\end{cases}
\]

 \begin{figure}[t]
   \centering
     \begin{tikzpicture}

       \hspace*{-30mm}
       \begin{scope}
\tikzset{node distance=7em, scale=0.5, transform shape}
\node[state, rectangle] (alpha) {\Large $\alpha$};
\node[state, right of=alpha] (q1) {\Large $q_1$};
\node[state, rectangle, right of=q1] (left-gamma){\Large $\gamma$};
\node[state, right of=left-gamma] (q2){\Large $q_2$};
\node[state, rectangle, right of=q2] (right-gamma) {\Large $\gamma$};
\node[state, right of=right-gamma] (q3){\Large $q_3$};
\node[state, rectangle, above of=right-gamma] (up-gamma-right) {\Large $\gamma$};

\tikzset{node distance=2em}
\node[above of=q1] (w0)[yshift=-5em] {\Large 2};
\node[above of=alpha] (w1)[yshift=0.7em] {\Large 1};
\node[above of=q2] (w2)[yshift=-5em] {\Large 3};
\node[above of=left-gamma] (w3)[yshift=0.7em] {\Large 1};
\node[above of=right-gamma] (w4)[yshift=0.7em] {\Large 1};
\node[above of=q3] (w5)[yshift=-5em] {\Large 2};
\node[above of=up-gamma-right] (w7)[yshift=0.7em] {\Large 1};

\draw[->,>=stealth] (alpha) edge (q1);
\draw[->,>=stealth] (q1) edge (left-gamma);
\draw[->,>=stealth] (q2) edge (right-gamma);
\draw[->,>=stealth] (left-gamma) edge (q2);
\draw[->,>=stealth] (q2) edge (right-gamma);
\draw[->,>=stealth] (right-gamma) edge (q3);
\draw[->,>=stealth] (q3) edge[out=120, in=-20, looseness=1] (up-gamma-right);
\draw[->,>=stealth] (up-gamma-right) edge[out=200, in=60, looseness=1] (q2);
       \end{scope}

\hspace*{80mm}
       
          \begin{scope}
\tikzset{node distance=7em, scale=0.5, transform shape}
\node[state, rectangle] (alpha){\Large $\alpha$};
\node[state, right of=alpha] (1alpha){\Large $[1.\alpha]_{\sim_r}$};
\node[state, rectangle, right of=1alpha] (gamma) {\Large $\gamma$};
\node[state, right of=gamma] (1gammaalpha)[xshift=2em] {\Large $[1.\gamma(\alpha)]_{\sim_r}$};
\node[state, rectangle, above of=gamma] (up-gamma) {\Large $\gamma$};

\tikzset{node distance=2em}
\node[above of=alpha] (w1)[yshift=0.7em] {\Large 1};
\node[above of=1alpha] (w2)[yshift=-6em] {\Large 2};
\node[above of=gamma] (w3)[yshift=0.7em] {\Large 1};
\node[above of=up-gamma] (w4)[yshift=0.7em] {\Large 1};
\node[above of=1gammaalpha] (w5)[yshift=-7em] {\Large 3};

\draw[->,>=stealth] (alpha) edge (1alpha);
\draw[->,>=stealth] (1alpha) edge (gamma);
\draw[->,>=stealth] (gamma) edge (1gammaalpha);
\draw[->,>=stealth] (1gammaalpha) edge[out=130, in=-20, looseness=1] (up-gamma);
\draw[->,>=stealth] (up-gamma) edge[out=200, in=60, looseness=1] (1alpha);
       \end{scope}
       
\end{tikzpicture} 
\caption{\label{fig:making-wta-minimal} The $(\Sigma,\Ratnum)$-wta $\cA$ of Example \ref{ex:B-implies-A} (left) and $\budwta(r,\sim_r,\mathrm{H}_{\cA})$ (right).}
   \end{figure}

Let us abbreviate $\sem{\cA}$ by $r$. It is easy to show that, for every $m,n \in \mathbb{N}$ and $b_1,b_2\in \mathbb{Q}$, we have
\[b_1.\gamma^m(\alpha) \sim_r b_2.\gamma^n(\alpha) \ \text{ iff } \ \Big(b_1=b_2 \text{\ \ and \ \ } \big( m\!\!\!\mod (2) = n\!\!\!\mod (2)\big)\Big) \text{ or } b_1=b_2=0.\]
 Hence, for every $n\in \mathbb{N}$ and $b\in \mathbb{Q}$, we have
 \begin{align}\label{ex:what-is-bgamman}
   [b.\gamma^n(\alpha)]_{\sim_r}=\begin{cases}  \{b.\gamma^m(\alpha) \mid m \in \mathbb{N} \ \text{ and } \  m\!\!\!\!\mod (2) = n\!\!\!\!\mod (2)\} & \text{ if $b\ne 0$}\\
 \{\widetilde{0}\} & \text{otherwise.}
  \end{cases}
  \end{align}
Using the above, it is easy to see that dependency in $\sfMon(\Sigma,\B)/_{\sim_r}$ is decidable. 

By Lemma \ref{lm:tq-has-minimal-height}, we construct the mapping $\mathrm{t}_\cA: Q \to \T_\Sigma$ with 
\[\mathrm{t}_\cA(q_1)=\alpha, \ \ \mathrm{t}_\cA(q_2)=\gamma(\alpha), \text{ and} \ \ \mathrm{t}_\cA(q_3)=\gamma^2(\alpha) \enspace.
\]

We have $1.\alpha \sim_r 1.\gamma^2(\alpha)$, i.e., $[1.\alpha]_{\sim_r}=[1.\gamma^2(\alpha)]_{\sim_r}$. Moreover, there does not exist $b\in \mathbb{Q}$ with $1.\alpha \sim_r b.\gamma(\alpha)$, hence the elements  $[1.\alpha]_{\sim_r}$ and $[1.\gamma(\alpha)]_{\sim_r}$ are independent. We obtain that $\mathrm{H}_{\cA}=\{[1.\alpha]_{\sim_r}, [1.\gamma(\alpha)]_{\sim_r}\}$, and this set is pair-independent. Thus there is no need to apply Lemma~\ref{lm:crucial-canc-pair-ind-imply-uniqueness-det-1} in this example.
  
By using the characterization in \eqref{ex:what-is-bgamman}, we can compute the mapping $\mathrm{dec}$. In fact, for every $b\in \mathbb{Q}$ and $n \in \mathbb{N}$, we have
\begin{align}\label{eq:congruence-classes-another}
[b.\gamma^n(\alpha)]_{\sim_r} = 
\begin{cases}
[b.\alpha]_{\sim_r}=b \cdot[1.\alpha]_{\sim_r} & \text{ if \ $\big(n\!\!\mod(2)\big)=0$}\\
[b.\gamma(\alpha)]_{\sim_r}= b\cdot [1.\gamma(\alpha)]_{\sim_r} & \text{ if \ $\big(n\!\!\mod(2)\big)=1$}.
\end{cases}
\end{align}
In particular, for each $n \in \mathbb{N}$, we have
\[
  \mathrm{dec}([1.\gamma^n(\alpha)]_{\sim_r}) =
  \begin{cases}
(1,[1.\alpha]_{\sim_r}) & \text{ if \ $\big(n\!\!\mod(2)\big)=0$}\\
(1,[1.\gamma(\alpha)]_{\sim_r}) & \text{ if \ $\big(n\!\!\mod(2)\big)=1$}.
    \end{cases}
  \]

Finally, we construct the bu-deterministic $(\Sigma,\Ratnum)$-wta $\budwta(r,\sim_r,\mathrm{H}_{\cA})=(\mathrm{H}_{\cA},\delta',F')$ (cf. Figure~\ref{fig:making-wta-minimal}(right)), where
\begin{compactitem}
  \item $\mathrm{H}_{\cA}=\{[1.\alpha]_{\sim_r}, [1.\gamma(\alpha)]_{\sim_r}\}$,
\item $(\delta')_0(\varepsilon,\alpha,[1.\alpha]_{\sim_r}) = (\delta')_1([1.\alpha]_{\sim_r},\gamma,[1.\gamma(\alpha)]_{\sim_r})
  = (\delta')_1([1.\gamma(\alpha)]_{\sim_r},\gamma,[1.\alpha]_{\sim_r}) = 1$ and\\
  $(\delta')_0(\varepsilon,\alpha,[1.\gamma(\alpha)]_{\sim_r}) = 0$, and
  \item $(F')_{[1.\alpha]_{\sim_r}} = 2$ and $(F')_{[1.\gamma(\alpha)]_{\sim_r}} = 3$.
\end{compactitem}
By Lemma \ref{lm:wta(r,sim,H)}, we have $\sem{\budwta(r,\sim_r,\mathrm{H}_{\cA})}=r$.
By Lemma~\ref{cor:min-du-det-wta-exists}, the wta $\budwta(r,\sim_r,\mathrm{H}_{\cA})$ is d-minimal.
\hfill$\Box$
  \end{example}


  \subsection{A characterization of d-minimality for bu-deterministic wta}\label{subsect:characterization-of-minimlity}

Here we characterize the property of a bu-deterministic wta $\cA$ of being d-minimal by the properties that $\cA$ is slim and the degrees of $\sfMon(\Sigma,\B)/_{\ker(\sfh_\cA)}$ and $\sfMon(\Sigma,\B)/_{\sim_{\sem{\cA}}}$ coincide. We recall that, for each bu-deterministic wta $\cA$ (i.e., also for a non-d-minimal one), we have $\deg(\sfMon(\Sigma,\B)/_{\ker(\sfh_\cA)}) \ge \deg(\sfMon(\Sigma,\B)/_{\sim_{\sem{\cA}}})$.

  \begin{quote}{\em In this subsection, we let $\B=(B,\oplus,\otimes,\0,\1)$ denote an arbitrary commutative semifield. Moreover, we let  $\cA=(Q,\delta,F)$ denote an arbitrary bu-deterministic $(\Sigma,\B)$-wta.}
  \end{quote}

We recall that, for every $b \in B$ and $q \in Q$, we let $b_q$ denote the element of  $B^Q$  such that $(b_q)_q=b$ and $(b_q)_p = \0$ for each $p \in Q\setminus \{q\}$. Then $B_{\le 1}^Q = \{\0^Q\} \cup \{b_q \mid b \in B^{-\0}, q \in Q\}$.

\begin{lemma}\rm\label{lm:slim-then-h-A-surjective} If $\cA$ is slim, then $\im(\sfh_\cA)=B_{\le 1}^Q$ and $\deg(\sfMon(\Sigma,\B)/_{\ker(\sfh_\cA)})=|Q|$.
\end{lemma}
\begin{proof} Let $\cA$ be slim. Since $\im(\sfh_\cA) \subseteq B_{\le 1}^Q$, we prove $B_{\le 1}^Q \subseteq \im(\sfh_\cA)$.
  For each $\xi \in \T_\Sigma$, we have $\sfh_\cA(\widetilde{\0}) = \0^Q$. Hence $\0^Q \in \im(\sfh_\cA)$. Now let $b \in B^{-\0}$ and $q \in Q$. We prove that $b_q \in \im(\sfh_\cA)$.  Since $\cA$ is slim, there exists $\xi \in \T_\Sigma$ such that $\state_\cA(\xi)=\{q\}$. By Lemma~\ref{lm:hAxiqne0-implies-qinhQxi-for-bu-det-wta}(2), we have $\h_\cA(\xi)_q \ne \0$. Then
  \begingroup
  \allowdisplaybreaks
  \begin{align*}
    b_q &= b \cdot \1_q \\
        &= (c \otimes \h_\cA(\xi)_q) \cdot \1_q
          \tag{with $c=b \otimes (\h_\cA(\xi)_q)^{-1}$}\\
    &= c \cdot (\h_\cA(\xi)_q \cdot \1_q)\\
        &= c \cdot \h_\cA(\xi)
          \tag{by Lemma~\ref{lm:properties-sem-of-budet-wta-det-new}(1) and $\h_\cA(\xi)_q \ne \0$}\\
        &= c \cdot \sfh_\cA(\1.\xi)
          \tag{by \eqref{equ:sfh-related-to-h}}\\
    &= \sfh_\cA(c.\xi) \tag{because $\sfh_\cA$ is a scalar-linear mapping}\enspace.
  \end{align*}
  \endgroup
    Hence $b_q \in \im(\sfh_\cA)$.  Thus $B_{\le 1}^Q \subseteq \im(\sfh_\cA)$.

Since $\im(\sfh_\cA)=B_{\le 1}^Q$, we have $\sfdM_{\mathrm{im}}(\cA) =(B_{\le 1}^Q,\0^Q,\delta_\cA)$.
By~\eqref{eq:isomorphism}, this implies $\deg(\sfMon(\Sigma,\B)/_{\ker(\sfh_\cA)})=\deg((B_{\le 1}^Q,\0^Q))=|Q|$.
\end{proof}

\begin{lemma}\rm\label{lm:minimal-then-degrees-equal}
 If $\cA$ is d-minimal, then   $\deg(\sfMon(\Sigma,\B)/_{\ker(\sfh_\cA)})=\deg(\sfMon(\Sigma,\B)/_{\sim_{\sem{\cA}}})$.
 \end{lemma}
 \begin{proof} Let $\cA$ be d-minimal. By Lemma~\ref{lm:minimal-implies-slim}, $\cA$ is slim and hence,  by Lemma~\ref{lm:slim-then-h-A-surjective},   $\deg(\sfMon(\Sigma,\B)/_{\ker(\sfh_\cA)})=|Q|$.
Moreover, we also have $\deg(\sfMon(\Sigma,\B)/_{\sim_{\sem{\cA}}})=|Q|$ because, by   Lemma \ref{cor:min-du-det-wta-exists}, 
$ \deg(\sfMon(\Sigma,\B)/_{\sim_{\sem{\cA}}})$ is the number of the states of a d-minimal bu-deterministic $(\Sigma,\B)$-wta which recognizes $\sem{\cA}$.
\end{proof}

\begin{lemma}\rm\label{lm:degrees-equal-then-minimal}
 If $\cA$ is slim and $\deg(\sfMon(\Sigma,\B)/_{\ker(\sfh_\cA)})=\deg(\sfMon(\Sigma,\B)/_{\sim_{\sem{\cA}}})$, then $\cA$ is d-minimal.
  \end{lemma}
\begin{proof} Let $\cA$ be slim. By Lemma~\ref{lm:slim-then-h-A-surjective},   $\deg(\sfMon(\Sigma,\B)/_{\ker(\sfh_\cA)})=|Q|$. Then, by our assumption, we have 
$\deg(\sfMon(\Sigma,\B)/_{\sim_{\sem{\cA}}})=|Q|$. Hence $\cA$ is d-minimal because, by Lemma \ref{cor:min-du-det-wta-exists},
$\deg(\sfMon(\Sigma,\B)/_{\sim_{\sem{\cA}}})$ is the number of the states of a d-minimal bu-deterministic wta which recognizes $\sem{\cA}$.
\end{proof}

Now by Lemmas~\ref{lm:minimal-implies-slim} and \ref{lm:minimal-then-degrees-equal}   and Lemma \ref{lm:degrees-equal-then-minimal} we obtain the following characterization of d-minimality.

\begin{corollary}\rm\label{minimal-iff-degrees-equal} Let $\cA$ be a  bu-deterministic $(\Sigma,\B)$-wta. Then $\cA$  is d-minimal if and only if $\cA$ is slim and $\deg(\sfMon(\Sigma,\B)/_{\ker(\sfh_\cA)})=\deg(\sfMon(\Sigma,\B)/_{\sim_{\sem{\cA}}})$.
\end{corollary}

\subsection[An example of state reduction for bu-deterministic wta]{An example on the comparison of two state reduction methods for bu-deterministic wta}

Let $\B$ be an extremal and commutative semifield and $\cA=(Q,\delta,F)$ be a bu-deterministic $(\Sigma,\B)$-wta. For this scenario, we have elaborated two methods for trying to reduce the number of states of $\cA$.

\begin{enumerate}
\item Method using a maximal factorization (cf. Subsection~\ref{sec:det-of-bu-det}): Let  $(f,g)$ be a maximal factorization over $B^Q$. By Observation \ref{obs:bu-deterministic-twins}, $\cA$ has the twinning property. Thus, by Theorem~\ref{thm:extremal-twins-determinizable}, for
  \[
    \text{the bu-deterministic $(\Sigma,\B)$-wta $\det\nolimits_{(f,g)}(\cA)=(Q^{\mathrm{fac}},\delta^{\mathrm{fac}},F^{\mathrm{fac}})$ (cf. Definition~\ref{constr:det-A})}
  \]
  we have  $\sem{\cA}=\sem{\det_{(f,g)}(\cA)}$.  By Observation~\ref{lm:determinization-bu-det} we have $|Q^{\mathrm{fac}}| \le |Q|$.

\item Method using the m-syntactic congruence (cf. Subsection~\ref{subs:minimization}): By Theorem~\ref{thm:MN-semifield-det}(A)$\Rightarrow$(C), the $\B$-scalar algebra $\sfMon(\Sigma,\B)/_{\sim_{\sem{\cA}}}$ is finitely generated; let $H$ be a scalar-basis of $\sfMon(\Sigma,\B)/_{\sim_{\sem{\cA}}}$. 
  By Lemma~\ref{lm:wta(r,sim,H)}, for
  \[
    \text{the  bu-deterministic $(\Sigma,\B)$-wta $\budwta(r,\sim,H)=(Q^{\mathrm{syn}},\delta^{\mathrm{syn}},F^{\mathrm{syn}})$ (cf.  Definition \ref{def:wta(r,sim,H)})]
    }
  \]
  we have $\sem{\cA}= \sem{\budwta(r,\sim,H)}$.  By Observation~\ref{cor:min-du-det-wta-exists}, $\budwta(r,\sim,H)$ is d-minimal; in particular, $|Q^{\mathrm{syn}}| \le |Q|$.
  \end{enumerate}
  By Observation~\ref{cor:min-du-det-wta-exists} we also have that $|Q^{\mathrm{syn}}| \le |Q^{\mathrm{fac}}|$ (because  $\sem{\cA}=\sem{\det_{(f,g)}(\cA)}$).

Next we show an example for the application of the two methods, where the second method yields a smaller bu-deterministic wta than the first method, i.e., $|Q^{\mathrm{syn}}| < |Q^{\mathrm{fac}}|$.

\begin{example}\label{ex:applying-MN-to-det} \rm We consider the extremal and commutative semifield $\Ratminplus= (\mathbb{Q}_\infty,\min,+,\infty,0)$ and the bu-deterministic $(\Sigma,\Ratminplus)$-wta $\cA=(Q,\delta,F)$ in Figure~\ref{fig:ex-det-unfolding-repeated}. Apart from state renaming, $\cA$ is the same as the bu-deterministic $(\Sigma,\Ratminplus)$-wta in Figure~\ref{fig:ex-det-unfolding}; thus $\sem{\cA} = \size$ (cf. Example~\ref{ex:factorization}). We apply the two methods mentioned above.

 \begin{figure}[t]
   \begin{center}
\begin{tikzpicture}
\tikzset{node distance=7em, scale=0.6, transform shape}
\node[state, rectangle] (1) {\large $\alpha$};
\node[state, right of=1] (2){$p$};
\node[state, rectangle, right of=2] (3)[right=1em]{\large $\sigma$};
\node[state, rectangle, above of=3] (4)[above=1em]{\large $\sigma$};
\node[state, rectangle, below of=3] (5)[below=1em]{\large $\sigma$};
\node[state, right of=3] (6)[right=1em]{\large $p'$};
\node[state, rectangle, right of=6] (7) {\large $\sigma$};

\tikzset{node distance=2em}
\node[above of=1] (w1) {1};
\node[above of=2] (w2) {0};
\node[above of=3] (w3) {0.5};
\node[above of=4] (w4) {1};
\node[above of=5] (w5) {1};
\node[above of=6] (w6)[left=0.1em] {0.5};
\node[above of=7] (w7) {1.5};

\draw[->,>=stealth] (1) edge (2);
\draw[->,>=stealth] (2) edge[out=20, in=155, looseness=1.1] (3);
\draw[->,>=stealth] (2) edge[out=-20, in=205, looseness=1.1] (3);
\draw[->,>=stealth] (2) edge (4);
\draw[->,>=stealth] (2) edge (5);
\draw[->,>=stealth] (3) edge (6);
\draw[->,>=stealth] (6) edge (4);
\draw[->,>=stealth] (4) edge[out=-290, in=80, looseness=1.4] (6);
\draw[->,>=stealth] (6) edge (5);
\draw[->,>=stealth] (5) edge[out=290, in=-80, looseness=1.4] (6);
\draw[->,>=stealth] (7) edge (6);
\draw[->,>=stealth] (6) edge[out=60, in=30, looseness=1.4] (7);
\draw[->,>=stealth] (6) edge[out=-60, in=-30, looseness=1.4] (7);
\end{tikzpicture} 
\caption{\label{fig:ex-det-unfolding-repeated} The $(\Sigma,\Ratminplus)$-wta $\cA=(Q,\delta,F)$.}
   \end{center}
   \end{figure}

\underline{Method using a maximal factorization:}\\
We consider the maximal factorization $(f,g)$ with 
$f: (\mathbb{Q}_{\infty})^Q \setminus \{\infty_Q\} \to (\mathbb{Q}_{\infty})^Q$ and $g: (\mathbb{Q}_{\infty})^Q \setminus \{\infty_Q\} \to \mathbb{Q}_{\infty}$  for each $u=\big(\begin{smallmatrix}u_p\\u_{p'}\end{smallmatrix}\big)$ in $(\mathbb{Q}_{\infty})^Q \setminus \{\infty_Q\}$ by 
\begin{align*}
f(u)_q = u_q - g(u)\  \text{ for each $q \in Q$} \ \ \ \text{ and } \ \ \ g(u) =\min(u_p,u_{p'})\enspace.
    \end{align*}
  We can construct the bu-deterministic $(\Sigma,\B)$-wta $\det_{(f,g)}(\cA)=(Q^{\mathrm{fac}},\delta^{\mathrm{fac}},F^{\mathrm{fac}})$ according to Definition~\ref{constr:det-A}. In fact, apart from state renaming ($p \to \binom{0}{\infty}$ and $p' \to \binom{\infty}{0}$), the bu-deterministic wta $\det_{(f,g)}(\cA)$ is the same as $\cA$. In particular, $|Q^{\mathrm{fac}}| = 2$.

\

\underline{Method using the m-syntactic congruence:}\\
  We abbreviate $\sim_\size$ by $\sim$ and each equivalence class $[b.\xi]_{\sim}$ by $[b.\xi]$. 

   First, we characterize $\sim$. Let $b_1.\xi_1, b_2.\xi_2 \in \rmMon(\Sigma,\Ratminplus)$. Then:
   \begingroup
   \allowdisplaybreaks
   \begin{align*}
     b_1,\xi_1 \sim b_2.\xi_2\\
     \text{iff } \ & \ (\forall c \in \C_\Sigma): b_1 + \size(c[\xi_1]) = b_2 + \size(c[\xi_2])\\
     \text{iff } \ & \ (\forall c \in \C_\Sigma): b_1 + \size(c) + \size(\xi_1) = b_2 + \size(c) + \size(\xi_2)\\
     \text{iff } \ & \  b_1 + \size(\xi_1) = b_2 + \size(\xi_2)\enspace.
   \end{align*}
   \endgroup
   
   Obviously, $\cA$ is slim.

   We consider the candidate set $H = \{[1.\alpha], [1.\sigma(\alpha,\alpha)]\}$ induced by $\cA$ for $\sfMon(\Sigma,\Ratminplus)/_\sim$. By Lemma \ref{thm:bu-det-r-construction-ofVLQ(r)-det}(3), the set $H$ generates the $\Ratminplus$-scalar algebra $\sfMon(\Sigma,\Ratminplus)/_\sim$. The two elements $[1.\alpha]$ and $[1.\sigma(\alpha,\alpha)]$ are dependent, because  $ [1.\sigma(\alpha,\alpha)] = 2 + [1.\alpha]$. The latter can be seen as follows. 
\begingroup
   \allowdisplaybreaks
   \begin{align*}
     & [1.\sigma(\alpha,\alpha)] = 2 + [1.\alpha]\\
     \text{iff } \ & \    [1.\sigma(\alpha,\alpha)] = [3.\alpha]\\
     \text{iff } \ & \  1 + \size(\sigma(\alpha,\alpha)) = 3 + \size(\alpha)
                     \tag{by the above characterization of $\sim$}\\
     \text{iff } \ & \ 4 = 4 \enspace.
     \end{align*}
     \endgroup
     Thus $H = \{[1.\alpha]\}$ is a finite scalar-basis of $\sfMon(\Sigma,\Ratminplus)/_\sim$.

We can construct the  bu-deterministic $(\Sigma,\B)$-wta $\budwta(\size,\sim,H)=(Q^{\mathrm{syn}},\delta^{\mathrm{syn}},F^{\mathrm{syn}})$ according to Definition~\ref{def:wta(r,sim,H)},  as follows.
     \begin{compactitem}
     \item $Q^{\mathrm{syn}} = H = \{[1.\alpha]\}$,
       \item 
   \begin{align*}
  \delta^{\mathrm{syn}}_0(\varepsilon,\alpha,[1.\alpha]) &= [1.\alpha]_{[1.\alpha]} = 0 \\
  \delta^{\mathrm{syn}}_2([1.\alpha][1.\alpha],\sigma,[1.\alpha]) &= [1.\sigma(\alpha,\alpha)]_{[1.\alpha]} = 2
   \end{align*} and,
 \item $F^{\mathrm{syn}}_{[1.\alpha]} = \gamma([1.\alpha]) = \size(\alpha) = 1$.
   \end{compactitem}

   \
   
   We observe that $|Q^{\mathrm{syn}}|=1 < 2 = |Q^{\mathrm{fac}}|$.
Moreover, we note that, if we apply weight pushing to   $\budwta(\size,\sim,H)$, then we obtain the bu-deterministic $(\Sigma,\Ratminplus)$-wta of Example~\ref{ex:size} except that the state is renamed.
\hfill $\Box$
  \end{example}

\section[Dimension and degree for a fixed weighted tree language]{A study on dimension and degree for a fixed bu-deterministically recognizable weighted tree language} 

In this section we show that there exists a ranked alphabet $\Sigma$, a field $\B$, and a bu-deterministically recognizable $(\Sigma,\B)$-weighted tree language $r$ such that the dimension of the $\B$-vector space $\sfPol(\Sigma,\B)/_{\approx_r}$ is less than the degree of the $\B$-scalar algebra $\sfMon(\Sigma,\B)/_{\sim_r}$. 

We consider the ranked alphabet $\Sigma = \{0^{(1)}, 1^{(1)}, \#^{(0)}\}$. We drop parentheses from trees in $\T_\Sigma$, e.g., we write $01\#$ instead of $0(1(\#))$. Also, we consider  the $(\Sigma,\Ratnum)$-weighted tree language $r: \T_\Sigma \to \mathbb{Q}$ such that, for each $\xi \in \T_\Sigma$, we let
  \[
    r(\xi) =
    \begin{cases}
      1 & \text{if there exist $\gamma \in \Sigma^{(1)}$ and $\zeta \in \T_\Sigma$ such that $\xi= \gamma1\zeta$}\\
      0 & \text{otherwise} \enspace.
      \end{cases}
    \]

In Figure \ref{fig:one-below-top}, a crisp (nondeterministic) $(\Sigma,\Ratnum)$-wta $\cA$ and a crisp-deterministic $(\Sigma,\Ratnum)$-wta $\cB$ are shown; both recognize $r$, which can be seen as follows.

    For the proof of $\sem{\cA}=r$, by induction on $\T_\Sigma$, we can easily show the following. For each $\xi \in \T_\Sigma$ we have
    \begin{align*}
      & \h_\cA(\xi)_{q_1}= 1 \\
      & \h_\cA(\xi)_{q_2}= \begin{cases} 1 & \text{if there exists $\zeta \in \T_\Sigma$ such that $\xi=1\zeta$}\\
        0 & \text{otherwise}
        \end{cases}\\
& \h_\cA(\xi)_{q_3}= \begin{cases} 1 & \text{if there exist $\gamma \in \Sigma^{(1)}$ and $\zeta \in \T_\Sigma$ such that $\xi= \gamma1\zeta$}\\
        0 & \text{otherwise}\enspace.
      \end{cases}
    \end{align*}
    Then, for each $\xi \in \T_\Sigma$, we have
    \(
\sem{\cA}(\xi) = \h_\cA(\xi) \cdot F = \h_\cA(\xi)_{q_3} = r(\xi)\).

    For the proof of $\sem{\cB}=r$, we observe that
    \[
      \im(\h_\cA)= \Big\{\left(\begin{matrix}1\\ 0\\ 0\end{matrix}\right),
      \left(\begin{matrix}1\\1\\0\end{matrix}\right),
      \left(\begin{matrix}1\\0\\1\end{matrix}\right),
      \left(\begin{matrix}1\\1\\1\end{matrix}\right)\Big\} \enspace
    \]
    where the ordering is: $q_1$ at the top, $q_2$ in the middle, and $q_3$ at the bottom.
    Thus $\im(\h_\cA)$ is finite. Then, by Lemma~\ref{lm:image-ha-finite-subA-crisp-det-wta}, the crisp-deterministic $(\Sigma,\Ratnum)$-wta $\sub(\cA)$ can be constructed; moreover, $\sem{\cA} = \sem{\sub(\cA)}$. It is easy to see that $\sub(\cA)=\cB$. Thus $\sem{\cB}=r$.

    \begin{figure}[t]
  \centering
  \begin{tikzpicture}
\tikzset{node distance=6em, scale=0.6, transform shape}
\node[state, rectangle] (1) {\Large $\#$};
\node[state, right of=1] (2) {\Large $q_1$};
\node[state, rectangle, above of=2] (3) {\Large $0$};
\node[state, rectangle, below of=2] (4) {\Large $1$};
\node[state, rectangle, right= 3em of 2] (5) {\Large $1$};
\node[state, right= 3em  of 5] (6) {\Large $q_2$};
\node[state, rectangle, right of=6, yshift=3em] (7) {\Large $1$};
\node[state, rectangle, below of=7] (8) {\Large $0$};
\node[state, right of=6, xshift=6em] (9) {\Large $q_3$};

\tikzset{node distance=2em}
\node[above of=1] (w1) {$1$};
\node[above of=3] (w3) {$1$};
\node[above of=4] (w4) {$1$};
\node[above of=5] (w5) {$1$};
\node[above of=9] (w3) {$1$};
\node[above of=8] (w3) {$1$};
\node[above of=7] (w9) {$1$};

\draw[->,>=stealth] (1) edge (2);
\draw[->,>=stealth, out=180, in=135, looseness=1.2] (3) edge (2);
\draw[->,>=stealth, out=45, in=0, looseness=1.2] (2) edge (3);

\draw[->,>=stealth, out=315, in=0, looseness=1.2] (2) edge (4);
\draw[->,>=stealth, out=180, in=225, looseness=1.2] (4) edge (2);

\draw[->,>=stealth] (2) edge (5);
\draw[->,>=stealth] (5) edge (6);
\draw[->,>=stealth, out=45, in=180] (6) edge (7);
\draw[->,>=stealth, out=315, in=180] (6) edge (8);
\draw[->,>=stealth, out=0, in=135] (7) edge (9);
\draw[->,>=stealth, out=0, in=225] (8) edge (9);
\end{tikzpicture}

\vspace{8mm}


 \begin{tikzpicture}
\tikzset{node distance=6em, scale=0.6, transform shape}
\node[state, rectangle] (1) {\Large $\#$};
\node[state, right of=1, xshift=2em] (2) { $\left(\begin{matrix}1\\ 0\\ 0\end{matrix}\right)$};
\node[state, rectangle, above of=2, yshift=1em] (3) {\Large $0$};
\node[state, rectangle, right= 3em of 2] (5) {\Large $1$};
\node[state, right= 3em  of 5] (6) { $\left(\begin{matrix}1\\ 1\\ 0\end{matrix}\right)$};
\node[state, rectangle, right of=6, yshift=5em, xshift=1.5em] (7) {\Large $1$};
\node[state, right of=6, xshift=9em] (9) { $\left(\begin{matrix}1\\ 1\\ 1\end{matrix}\right)$};
\node[state, rectangle, below of=7, yshift=-15em] (10) {\Large $0$};
\node[state,  below of=6, yshift=-10em] (11) { $\left(\begin{matrix}1\\ 0\\ 1\end{matrix}\right)$};
\node[state, rectangle, below of=6, xshift=3em, yshift=-2em] (12) {\Large $0$};
\node[state, rectangle, below of=6, xshift=-3em, yshift=-2em] (13) {\Large $1$};
\node[state, rectangle, below of=2, xshift=3em, yshift=-4em] (14) {\Large $0$};
\node[state, rectangle, right of=9, xshift=2em] (15) {\Large $1$};

\tikzset{node distance=2em}
\node[above of=1] (w1) {$1$};
\node[above of=3] (w3) {$1$};
\node[above of=5] (w5) {$1$};
\node[above of=9, yshift=2.4em] (w9) {$1$};
\node[above of=7] (w7) {$1$};
\node[above of=10] (w10) {$1$};
\node[above of=11, yshift=2.4em] (w11) {$1$};
\node[above of=12, xshift=1em] (w12) {$1$};
\node[above of=13, xshift=1em] (w13) {$1$};
\node[above of=14] (w14) {$1$};
\node[above of=15, xshift=1em] (w15) {$1$};

\draw[->,>=stealth] (1) edge (2);
\draw[->,>=stealth, out=180, in=135, looseness=1.2] (3) edge (2);
\draw[->,>=stealth, out=45, in=0, looseness=1.2] (2) edge (3);

\draw[->,>=stealth] (2) edge (5);
\draw[->,>=stealth] (5) edge (6);
\draw[->,>=stealth, out=60, in=180] (6) edge (7);
\draw[->,>=stealth, out=0, in=120] (7) edge (9);
\draw[->,>=stealth, out=270, in=0] (9) edge (10);
\draw[->,>=stealth, out=180, in=0] (10) edge (11);
\draw[->,>=stealth, out=300, in=90] (6) edge (12);
\draw[->,>=stealth, out=270, in=60] (12) edge (11);
\draw[->,>=stealth, out=120, in=270] (11) edge (13);
\draw[->,>=stealth, out=90, in=240] (13) edge (6);
\draw[->,>=stealth, out=180, in=300] (11) edge (14);
\draw[->,>=stealth, out=120, in=280] (14) edge (2);
\draw[->,>=stealth, out=45, in=90] (9) edge (15);
\draw[->,>=stealth, out=270, in=315] (15) edge (9);

\end{tikzpicture}

\caption{\label{fig:one-below-top} A crisp $(\Sigma,\Ratnum)$-wta $\cA$ (top) and a crisp-deterministic $(\Sigma,\Ratnum)$-wta $\cB$ (bottom); both recognize~$r$.}
\end{figure}

\

Since $r$ is recognizable, by Theorem~\ref{th:MN-fields}(A)$\Rightarrow$(C), the $\Ratnum$-vector space $\mathsf{Pol}(\Sigma,\Ratnum)/_{\approx_r}$ is finite dimensional. Next we determine its dimension. As a first step, we characterize the congruence $\approx_r$ on the $\Ratnum$-vector space $\sfPol(\Sigma,\Ratnum)$. For this, we define the tree languages
\begin{align*}
  L_1 &= \{\#\}\cup \{0\#\} \cup \{00\xi\mid \xi \in \T_\Sigma\}\\
  L_2 &= \{1\#\}\cup\{10\xi\mid \xi \in \T_\Sigma\}\\
    L_3 &= \{01\xi\mid \xi \in \T_\Sigma\} \cup \{11\xi\mid \xi \in \T_\Sigma\} \enspace.
\end{align*}
It is clear that $(L_1,L_2,L_3)$ is a partitioning of $\T_\Sigma$. We choose the following representatives:
\[
\# \in L_1, \ \ 1\# \in L_2, \ \ \text{ and } \ \ 01\# \in L_3 \enspace. 
\]
Moreover, it is easy to see that we have:
\begin{equation}\label{equ:i-determines-characteristic}
  \begin{aligned}
    &\text{For every $i,j \in [3]$, $\xi_1 \in L_i$, and $\xi_2 \in L_j$, the following holds:}\\
    &i=j \ \ \Leftrightarrow \ \ (\forall c \in \C_\Sigma): r(c[\xi_1])= r(c[\xi_2]) \enspace.
    \end{aligned}
\end{equation}
For instance, for  $\xi_1= \#$ and $\xi_2= 0\#$ in $L_1$ and $c_1=z$ and $c_2=01z$ we have
\[
r(c_1[\xi_1])=0=r(c_1[\xi_2]) \ \ \text{ and } \ \ r(c_2[\xi_1])=1=r(c_2[\xi_2]) \enspace.
\]
And for $\xi_1= \#$ in $L_1$ and $\xi_2= 1\#$ in $L_2$ and $c=0z$ we have
\[
r(c[\xi_1]) = r(0\#) = 0 \ \ \text{ and } \ \ r(c[\xi_2]) = r(01\#) = 1 \enspace.
  \]

For each $s = b_1.\xi_1 + \ldots + b_n.\xi_n$ in $\Pol(\Sigma,\Ratnum)$ and $i \in [3]$, we define the rational number
\[
\mathrm{coeff}(s)_i = \bigplus_{\substack{j \in [n]:\\\xi_j \in L_i}} b_j \enspace.
  \]

Now we can prove the following representation of polynomials modulo $\approx_r$.
\begin{equation}\label{equ:representation-of-a-polynomial}
  \begin{aligned}
    &\text{For each $s = b_1.\xi_1 + \ldots + b_n.\xi_n$ in $\Pol(\Sigma,\Ratnum)$, we have}\\
    &s \approx_r \mathrm{coeff}(s)_1.\# + \mathrm{coeff}(s)_2.1\# + \mathrm{coeff}(s)_3.01\# \enspace.
  \end{aligned}
\end{equation}

Let $c \in \C_\Sigma$.
\begingroup
\allowdisplaybreaks
\begin{align*}
  & b_1 \cdot r(c[\xi_1]) + \ldots + b_n\cdot r(c[\xi_n])\\
  &= \Big(\bigplus_{\substack{j \in [n]:\\\xi_j \in L_1}} b_j \cdot r(c[\xi_j])\Big)
  + \Big(\bigplus_{\substack{j \in [n]:\\\xi_j \in L_2}} b_j \cdot r(c[\xi_j])\Big)
  + \Big(\bigplus_{\substack{j \in [n]:\\\xi_j \in L_3}} b_j \cdot r(c[\xi_j])\Big)
  \tag{because $(L_1,L_2,L_3)$ is a partitioning}\\
  &= \Big(\bigplus_{\substack{j \in [n]:\\\xi_j \in L_1}} b_j \cdot r(c[\#])\Big)
  + \Big(\bigplus_{\substack{j \in [n]:\\\xi_j \in L_2}} b_j \cdot r(c[1\#])\Big)
  + \Big(\bigplus_{\substack{j \in [n]:\\\xi_j \in L_3}} b_j \cdot r(c[01\#])\Big)
  \tag{by \eqref{equ:i-determines-characteristic}} \\
  &= \Big(\bigplus_{\substack{j \in [n]:\\\xi_j \in L_1}} b_j\Big) \cdot r(c[\#])
  + \Big(\bigplus_{\substack{j \in [n]:\\\xi_j \in L_2}} b_j\Big) \cdot r(c[1\#])
  + \Big(\bigplus_{\substack{j \in [n]:\\\xi_j \in L_3}} b_j\Big) \cdot r(c[01\#])
  \tag{because $r(c[\#])$, $r(c[1\#])$, and $r(c[01\#])$ do not depend on $j$}\\
  &= \mathrm{coeff}(s)_1 \cdot r(c[\#])
  +  \mathrm{coeff}(s)_2 \cdot r(c[1\#])
  +  \mathrm{coeff}(s)_3 \cdot r(c[01\#]) \enspace.
\end{align*}
\endgroup
This proves \eqref{equ:representation-of-a-polynomial}.

By \eqref{equ:representation-of-a-polynomial} we obtain that, for every $s_1,s_2 \in \Pol(\Sigma,\Ratnum)$, we have:
\begin{equation}\label{charaterization-of-approx-r-01}
s_1 \approx_r s_2 \ \ \text{ iff } \ \ \text{ for each $i \in [3]$ we have } \ \mathrm{coeff}(s_1)_i=\mathrm{coeff}(s_2)_i \enspace.
\end{equation}

Next we show that $H = \{[1.\#]_{\approx_r}, [1.1\#]_{\approx_r}, [1.01\#]_{\approx_r}\}$ is a basis of $\sfPol(\Sigma,\Ratnum)/_{\approx_r}$.
For this we drop $\approx_r$ from $[s]_{\approx_r}$ and simply write $[s]$.

Equation~\ref{equ:representation-of-a-polynomial} shows that $H$ generates $\sfPol(\Sigma,\Ratnum)/_{\approx_r}$, because for each $s \in \sfPol(\Sigma,\Ratnum)$, we have
\begin{equation}
[s] = \mathrm{coeff}(s)_1 . [1.\#] + \mathrm{coeff}(s)_2. [1.1\#] + \mathrm{coeff}(s)_3. [1.01\#] \enspace.
\end{equation}

Next we prove that $H$ are linearly independent. For this let $b_1,b_2,b_3 \in \mathbb{Q}$ and
\[
b_1.[1.\#] + b_2. [1.1\#] + b_3. [1.01\#] = [0], \ \text{ i.e. } \ [b_1.\# + b_2.1\# + b_3.01\#] = [0]  \enspace.
  \]
  By \eqref{charaterization-of-approx-r-01}, we obtain that $b_1=b_2=b_3=0$. Thus $H$ is linearly independent. Hence
  $H$ is a basis of $\sfPol(\Sigma,\Ratnum)/_{\approx_r}$ and the $\Ratnum$-vector space $\sfPol(\Sigma,\Ratnum)/_{\approx_r}$  has dimension $3$.

Thus, by Lemma~\ref{lm:wta(r,approxr,H)-is-minimal}, a minimal $(\Sigma,\Ratnum)$-wta which recognizes $r$ has 3 states. Therefore, the crisp $(\Sigma,\Ratnum)$-wta $\cA$ shown in Figure \ref{fig:one-below-top} is a minimal $(\Sigma,\Ratnum)$-wta which recognizes $r$.

\

Since $r$ is bu-deterministically recognizable, by Theorem\ref{thm:MN-semifield-det}(A)$\Rightarrow$(C), the $\Ratnum$-scalar algebra $\mathsf{Mon}(\Sigma,\Ratnum)/_{\sim_r}$ is finitely generated. Thus, by Lemma~\ref{lm:crucial-canc-pair-ind-imply-uniqueness-det-1}, it has a finite scalar-basis. Next we determine the degree of $\mathsf{Mon}(\Sigma,\Ratnum)/_{\sim_r}$.

First we characterize the congruence $\sim_r$ on the $\Ratnum$-scalar algebra $\sfMon(\Sigma,\Ratnum)$.
Let us partition $\T_\Sigma$ into four classes as follows
  \begin{align*}
 L_{1}&=\{\#\}\cup \{0\#\} \cup \{00\xi\mid \xi \in \T_\Sigma\} &  L_{3/1}&=\{11\xi\mid \xi \in \T_\Sigma\}\\
L_{2}&=\{1\#\}\cup\{10\xi\mid \xi \in \T_\Sigma\} &  L_{3/0}&=\{01\xi\mid \xi \in \T_\Sigma\}.
  \end{align*}
The partitions $L_1$ and $L_2$ are the same as in the characterization of $\approx_r$ and  $L_3= L_{3/1} \cup L_{3/0}$. 
We denote by $\equiv$ the equivalence relation on $\T_\Sigma$ determined by the above partitioning.

We prove the following statement:
\begin{equation}\label{eq:sim-r-characterization}
\begin{aligned}
&\text{For every $b_1.\xi_1,b_2.\xi_2 \in \mathrm{Mon}(\Sigma,\Ratnum)$ we have} \\
&\text{$b_1.\xi_1\sim_r b_2.\xi_2 \ \ \Leftrightarrow \ \ b_1=b_2=0$ or ($b_1\ne 0$, $b_1=b_2$ and $\xi_1 \equiv \xi_2$). }
\end{aligned}
\end{equation}
By definition, we have $b_1.\xi_1\sim_r b_2.\xi_2$ if for each $c \in \C_\Sigma$ we have 
\begin{equation}\label{eq:sim-r-definition}
b_1\cdot r(c[\xi_1])= b_2\cdot r(c[\xi_2]).
\end{equation}

\underline{Proof of $\Leftarrow$:} We proceed by case analysis. If $b_1=b_2=0$, then the implication holds obviously. Next, for instance,
let $b_1=b_2$ with $b_1\ne 0$ and $\xi_1,\xi_2\in L_{2}$. Then, for the context $c_1=z$, we have $r(c_1[\xi_1])=  r(c_1[\xi_2]) =0$; 
and for $c_2\in \{0z,1z\}$ we have  $r(c_2[\xi_1])=  r(c_2[\xi_2]) =1$ and thus \eqref{eq:sim-r-definition} holds. Furthermore, for any other $c \in \C_\Sigma$, both $r(c[\xi_1])$ and $r(c[\xi_2])$ depend only on $c$, hence again \eqref{eq:sim-r-definition} holds.
The other cases can be handled similarly.

\underline{Proof of $\Rightarrow$:} Let us assume that $b_1.\xi_1\sim_r b_2.\xi_2$.
If $b_1=b_2=0$, then we are done. Therefore let $b_1\ne 0$. Then for the context $c=01z$, we have $r(c[\xi_1])=  r(c[\xi_2]) =1$, hence by \eqref{eq:sim-r-definition} we obtain $b_1=b_2$.

We prove $\xi_1 \equiv \xi_2$ by contradiction. For this, we assume that $\xi_1 \not\equiv \xi_2$ and proceed by case analysis. 
For instance, let $\xi_1\in L_{1}$ and $\xi_2\in L_{2}$. Then, for the context $c=0z$, we have $r(c[\xi_1])=0$ and $r(c[\xi_2])=1$, which contradicts \eqref{eq:sim-r-definition}.
The other cases can be handled similarly.

This finishes the proof of \eqref{eq:sim-r-characterization}.

Now let $H = \{[1.\#]_{\sim_r}, [1.1\#]_{\sim_r}, [1.11\#]_{\sim_r}, [1.01\#]_{\sim_r}\}$. We show that $H$ is a scalar-basis of $\sfMon(\Sigma,\Ratnum)/_{\sim_r}$.
In the rest we drop $\sim_r$ from $[b.\xi]_{\sim_r}$ and simply write $[b.\xi]$.

First we show that $H$ generates $\sfMon(\Sigma,\Ratnum)/_{\sim_r}$. For this, let $b.\xi \in \rmMon(\Sigma,\Ratnum)/_{\sim_r}$.
Let $\zeta$ be the unique element of $\{\#,1\#,11\#,01\#\}$ with $\zeta \equiv \xi$. Then $b\cdot [1.\zeta] = [b.\zeta] = [b.\xi]$, where the last equality follows by \eqref{eq:sim-r-characterization}. 

Then we show by contradiction that any two different elements of $H$ are independent. Assume, for instance, that $[1.1\#]$ and  $[1.11\#]$ are dependent. Then there exists $b\in \mathbb{Q}$ such that  $[1.1\#] =b\cdot [1.11\#]$, i.e., that
$[1.1\#] =[b.11\#]$. Then by \eqref{eq:sim-r-characterization} we obtain that $b=1$ and $1.1\#\equiv 1.11\#$, which is a contradiction.
For each other pair, the proof of independency is similar.

Hence the degree of $\sfMon(\Sigma,\Ratnum)/_{\sim_r}$ is 4.

Thus, by Lemma~\ref{cor:min-du-det-wta-exists}, a d-minimal bu-deterministic $(\Sigma,\Ratnum)$-wta which recognizes $r$ has 4 states. Therefore, the crisp-deterministic $(\Sigma,\Ratnum)$-wta $\cB$ shown in Figure \ref{fig:one-below-top} is a d-minimal bu-deterministic $(\Sigma,\Ratnum)$-wta which recognizes $r$.

Then $r$ is an example of a bu-deterministically
recognizable $(\Sigma,\Ratnum)$-weighted tree language, for which the number of states of a minimal 
$(\Sigma,\Ratnum)$-wta which recognizes $r$ is less than the number of states of a d-minimal bu-deterministic
$(\Sigma,\Ratnum)$-wta which recognizes $r$.

%% file: char-of-rec-by-monoid-rep.tex
\chapter[Characterization in terms of monoid representations]{Characterization of recognizability in terms of consistent monoid representations}
    \label{ch:char-of-rec-by-monoid-rep}

            The monoid representation of recognizable weighted tree languages was invented in \cite{boz94,bozale89} (cf. \cite[p.~356]{fulvog09new}); there it was required that the weight algebra $\B$ is a field.  Under this restriction they could prove that consistent $(\Sigma,\B)$-monoid representations and $(\Sigma,\B)$-wta are equally powerful \cite[Thm.~1 and 2]{bozale89}.
For the proof, they used the concepts of syntactic $(\Sigma,\B)$-vector spaces and of vector spaces of left quotients  of weighted tree languages, cf. Subsection~\ref{sec:Syntactic-congruences-and-syntactic-SigmaB-vector-spaces} and Section~\ref{sec:characterization-rec-by-left-quotients}, and related characterization results \cite[Props.~2 and 3]{bozale89}, cf.  Theorem~\ref{th:MN-fields} and  Lemma~\ref{lm:VLQ-simorphism-lemma}, respectively.

Here we show via a direct construction the following: for every ranked alphabet $\Sigma$ and semiring $\B$ such that $\B$ is commutative if $\Sigma$ is not monadic, the semantics of each $(\Sigma,\B)$-wta is equal to the semantics of a consistent $(\Sigma,\B)$-monoid representation (cf. Theorem~\ref{theo:monoid-characterization-of-run-semantics}). The latter semantics might be called the \emph{free monoid semantics} of the wta. For the special case in which $\Sigma$ is string ranked, we reobtain the well-known characterization of the run semantics (or: behaviour) of a wsa over semirings, cf. Theorem~\ref{theo:string-ranked:wta=monoid-rep} (\cite[Cor.~VI.6.2]{eil74}; also cf. \cite[Ch.~III, Sect.~2, Cor.~2.12]{sak09}, \cite[Lm.~5]{drostuvog10}, and \cite[Thm:~3.2]{drokus21}; and \cite[Ch.~I, Sect.~5]{berreu88} where $\B$ is commutative).
We will also prove the converse of Theorem~\ref{theo:monoid-characterization-of-run-semantics} for the case that $\B$ is a field (cf. Corollary~\ref{cor:field-from-consistent-rep-to-wta}). The characterization result is Corollary~\ref{cor:run-wta-mon-rep-initit-wta}. We refer to Figure \ref{fig:overview-monoid-representations-new} for an overview of the main results of this chapter.

\begin{quote}\emph{In this chapter, we let $\B=(B,\oplus,\otimes,\0,\1)$ denote an arbitrary semiring, unless stated otherwise.}
  \end{quote}

\section{$(\Sigma,\B)$-monoid representation}
\label{sect:Sigma-B-monoid-representation}

The central definition uses the set $\e\C_\Sigma$ of all elementary $\Sigma$-contexts (cf. p.~\pageref{page:context-and-substitution}).

\index{SigmaBmonoidrepresentation@$(\Sigma,\B)$-monoid representation}
A \emph{$(\Sigma,\B)$-monoid representation} is a tuple $\cR= (Q,\lambda,\mu,\gamma)$ where
\begin{compactitem}
  \item $Q$ is a finite and nonempty set,
\item $\lambda: \Sigma^{(0)} \to B^Q$ is a mapping (\emph{leaves weight mapping}),
\item $\mu: \e\C_\Sigma \to B^{Q \times Q}$ is an $\e\C_\Sigma$-indexed family over $B^{Q \times Q}$ (\emph{transition matrices}), and
  \item $\gamma \in B^Q$ is mapping (\emph{root weight vector}).
  \end{compactitem}
  \index{murev@$\murev$}
  
  We recall that the monoid $(\C_\Sigma,\circ_z,z)$ is free in the set of all monoids with generating set $\e\C_\Sigma$  (cf. Lemma~\ref{lm:freely-generated-context}). Now we extend $\mu$ in the unique way to the monoid homomorphism
  \[\murev: \C_\Sigma \to B^{Q \times Q}\]
  from the monoid $(\C_\Sigma,\circ_z,z)$ to the monoid $(B^{Q \times Q},\diamond,\mathrm{M}_\1)$ where, for every $M_1,M_2 \in B^{Q \times Q}$, we let $M_1 \diamond M_2 = M_2 \cdot M_1$ (where $\cdot$ is the usual matrix multiplication and $\mathrm{M}_\1$ is the $Q$-square matrix over $B$ in which the entries on the diagonal are $\1$ and the other entries are $\0$). Since $\B$ is distributive, the multiplication of matrices is associative.

  \label{page:def-consistent-mon-rep}
\index{consistent}
  A $(\Sigma,\B)$-monoid representation $(Q,\lambda,\mu,\gamma)$ is called
  \begin{compactitem}
  \item \emph{$n$-dimensional} if $|Q|=n$.
  \item \emph{consistent} if, for every $c,c' \in \C_\Sigma$, and $\alpha,\alpha' \in \Sigma^{(0)}$, the following holds: if $c[\alpha]=c'[\alpha']$, then $\lambda(\alpha) \cdot \murev(c) = \lambda(\alpha') \cdot \murev(c')$.
    \end{compactitem}

   \index{semanticR@$\sem{\cR}$}
    Let $\cR= (Q,\lambda,\mu,\gamma)$ be a consistent $(\Sigma,\B)$-monoid representation. The $(\Sigma,\B)$-weighted tree language \emph{recognized by $\cR$}, denoted by $\sem{\cR}$, is the weighted tree language $\sem{\cR}: \T_\Sigma \to B$ such that, for each $\xi \in \T_\Sigma$, we have
    \[
      \sem{\cR}(\xi) = \lambda(\alpha) \cdot \murev(c) \cdot \gamma
    \]
    for every $\alpha \in \Sigma^{(0)}$ and $c \in \C_\Sigma$ with $\xi=c[\alpha]$.
    The consistency condition guarantees the well-definedness of $\sem{\cR}$.

    For proofs of properties of $\murev$, we often use the terminating reduction system $(\C_\Sigma,\succ_{\C_\Sigma})$ which is defined on page \pageref{page:order-on-contexts}.

     \begin{example}\label{ex:monoid-rep-number-of-occ-of-pattern}\rm  Let $\Sigma = \{\sigma^{(2)}, \omega^{(1)}, \alpha^{(0)}\}$. We consider the mapping $\#_{\sigma(.,\alpha)}: \T_\Sigma \to \mathbb{N}$ from Example~\ref{ex:number-of-occurrences}: for each tree $\xi \in \T_\Sigma$ it returns the number of occurrences of the pattern $\sigma(.,\alpha)$ in $\xi$. 
Here we wish to define a $(\Sigma,\Nat)$-monoid representation $\cR$ such that $\sem{\cR} = \#_{\sigma(.,\alpha)}$. 

We let $\cR=(Q,\lambda,\mu,\gamma)$ as follows.
\begin{compactitem}
\item $Q = \{\bot,a,f\}$,
  
\item since each elementary context $e \in \e\C_\Sigma$ has the form either $e=\sigma(z,\xi)$ or $e=\sigma(\xi,z)$ for some $\xi \in \T_\Sigma$ or $e=\omega(z)$, it suffices to define the $(Q \times Q)$-matrices $\mu(\sigma(z,\xi))$ and $\mu(\sigma(\xi,z))$ for each $\xi \in \T_\Sigma$ and $\mu(\omega(z))$:
  \[
  \mu(\sigma(z,\xi)) =
  \begin{pmatrix}
    1 & 0 & \#_{\sigma(.,\alpha)}(\sigma(z,\xi))\\
    0 & 0 & 0\\
    0 & 0 & 1
  \end{pmatrix}
 \ \  \text{ and } \ \ 
  \mu(\sigma(\xi,z)) =
  \begin{pmatrix}
    1 & 0 & \#_{\sigma(.,\alpha)}(\xi)\\
    0 & 0 & 1\\
    0 & 0 & 1
  \end{pmatrix}
\]
 \[\text{ and } \ \ 
  \mu(\omega(z)) =
  \begin{pmatrix}
    1 & 0 & 0\\
    0 & 0 & 0\\
    0 & 0 & 1
  \end{pmatrix}
  \]
  where 
\begin{compactitem}  
  \item[(a)] the rows and columns are ordered according to the sequence $(\bot,a,f)$ and
\item[(b)] we have extended $\#_{\sigma(.,\alpha)}$ in the obvious way such that it is also applicable to contexts,
\end{compactitem}   

\item $\lambda(\alpha) = \begin{pmatrix}1 & 1 & 0 \end{pmatrix}$ and $\gamma =  \begin{pmatrix}0 \\ 0 \\ 1 \end{pmatrix}$.
\end{compactitem}

\begin{figure}
 \centering 

\scalebox{0.8}{
\rotatebox{0}{
\begin{tikzpicture}[
	transform shape,
	mymatrix/.style={matrix of math nodes,
			ampersand replacement=\&,
			left delimiter  = (,
			right delimiter = )},
	brace/.style={decoration={brace,mirror,amplitude=8pt},
			decorate,
			thick},
	snakearrow/.style={decorate,
			decoration={snake,amplitude=0.7pt,segment length=3mm,pre=lineto,post length=4pt}},
	fixposition/.style={xshift=3.2cm, yshift=0.45cm}
]

\matrix (M2) [mymatrix] 
{ 1 \& 0 \& 2 \\
  0 \& 0 \& 0 \\
  0 \& 0 \& 1 \\ };
\matrix (M4) [mymatrix, below left= 0.75cm and 1cm of M2]
{ 1 \& 0 \& 1 \\
  0 \& 0 \& 0 \\
  0 \& 0 \& 1 \\ };
\matrix (M5) [mymatrix, below= 0.75cm of M2]
{ 1 \& 0 \& 1 \\
  0 \& 0 \& 0 \\
  0 \& 0 \& 1 \\ };
\matrix (M6) [mymatrix, below right= 0.75cm and 1cm of M2]
{ 1 \& 0 \& 0 \\
  0 \& 0 \& 1 \\
  0 \& 0 \& 1 \\ };
  
\draw [brace] ($(M6.north east)+(0,0.25)$) -- ($(M4.north west)+(0,0.25)$);

\begin{scope}[
	level 1/.style={node distance=0.4cm and 0.4cm,sibling distance=1.4cm,level distance=1cm},
	rotate=-90]
\node[fixposition] at (M4.-90) (s1) {$\sigma$}
	child { node[below= of s1] (z1) {z}}
	child { node[below right= of s1] (a1) {$\alpha$}};
\node[fixposition] at (M5.-90) (s2) {$\sigma$}
	child { node[below= of s2] (z2) {z}}
	child { node[below right= of s2] (a2) {$\alpha$}};
\node[fixposition] at (M6.-90) (s3) {$\sigma$}
	child { node[below= of s3] (z3) {z}}
	child { node[below left= of s3] (a3) {$\alpha$}};
\end{scope}
        
\draw[snakearrow,<-] (M4.-90) to ($(M4.-90)+(0,-2.3)$);
\draw[snakearrow,<-] (M5.-90) to ($(M5.-90)+(0,-2.3)$);
\draw[snakearrow,<-] (M6.-90) to ($(M6.-90)+(0,-1.5)$);

\node[left = 1cm of M2-1-1.base, anchor=base east] {\strut $M_{c}:$};
\begin{scope}[rotate=-90]
\node[above right = 6cm and -1cm of s1, anchor=north] (xi) {$c:$};
\end{scope}
\end{tikzpicture}
} 
} 

  \caption{\label{fig:context-matrix-monoid-rep} An illustration of the evaluation of $\murev(c)$ for $c= \sigma(\alpha,z)\circ_z \sigma(z,\alpha) \circ_z \sigma(z,\alpha) =\sigma(\alpha,\sigma(\sigma(z,\alpha),\alpha))$.}
    \end{figure}
     
Next we prove that $\cR$ is consistent. For this purpose we  define, for each context $c$, a matrix $M_c$ and we prove a property of $\murev$ in terms of $M_c$. For each $c \in \C_\Sigma$, we define the $(Q\times Q)$-matrix :
\[
  M_c =
  \begin{pmatrix}
    1 & 0 & \#_{\sigma(.,\alpha)}(c)\\
    0 & a(c) & b(c)\\
    0 & 0 & 1
    \end{pmatrix}
  \]
  where
  \begin{compactitem}
    \item[(a)] $a(c) = 1$ if $c=z$, and $0$ otherwise, and 
    \item[(b)] $b(c) = 1$ if there exists a position $w \in \pos(c)$ such that $z = c(w2)$, and $0$ otherwise.
    \end{compactitem}
    In particular, $M_z=\mathrm{M}_1$, because $a(z)=1$ and $\#_{\sigma(.,\alpha)}(z)=b(z)=0$.

By induction on $(\C_\Sigma,\succ_{\C_\Sigma})$, we prove that the following statement holds (cf. Figure~\ref{fig:context-matrix-monoid-rep}).
\begin{equation}\label{eq:context-matrix-monoid-rep}
  \text{For each $c \in \C_\Sigma$, we have $\murev(c) = M_c$} \enspace. 
  \end{equation}
    
I.B.: Let $c = z$. Then $\murev(c) = \murev(z) = \mathrm{M}_1 = M_c$.

I.S.: Let $c = e[c']$ for some $e \in \e\C_\Sigma$ and $c' \in \C_\Sigma$.
Using the I.H., we obtain:
\[
  \murev(c) = \murev(e[c']) = \murev(e \circ_z c') = \murev(c') \cdot  \murev(e)  = M_{c'} \cdot \mu(e)
 \enspace.
  \]

 We proceed by case analysis. First let $e = \sigma(z,\xi)$ for some $\xi \in \T_\Sigma$.
Then
\[
M_{c'} \cdot \mu(\sigma(z,\xi))= 
 \begin{pmatrix}
    1 & 0 & \#_{\sigma(.,\alpha)}(c')\\
    0 & a(c') &  b(c')\\
    0 & 0 & 1
  \end{pmatrix}
  \cdot
  \begin{pmatrix}
    1 & 0 & \#_{\sigma(.,\alpha)}(\sigma(z,\xi))\\
    0 & 0 &  0\\
    0 & 0 & 1
  \end{pmatrix}
=
 \begin{pmatrix}
    1 & 0 & r\\
    0 & 0 & t\\
    0 & 0 & 1
    \end{pmatrix}
  \]
where $r= \#_{\sigma(.,\alpha)}(\sigma(z,\xi)) + \#_{\sigma(.,\alpha)}(c')$ and $t= b(c')$.
Since 
\begin{compactitem}
\item $a(c)=0$,
\item  $r= \#_{\sigma(.,\alpha)}(\sigma(z,\xi)) + \#_{\sigma(.,\alpha)}(c') = \#_{\sigma(.,\alpha)}(e[c']) = \#_{\sigma(.,\alpha)}(c)$, and 
\item $t=  b(c')= b(\sigma(c',\xi))= b(c)$
\end{compactitem}
 we have $ M_{c'} \cdot \mu(\sigma(z,\xi)) = M_c$. Hence $\murev(c) = M_c$.

\

  Second let $e = \sigma(\xi,z)$ for some $\xi \in \T_\Sigma$. Then
\[
M_{c'} \cdot \mu(\sigma(\xi,z)) = 
 \begin{pmatrix}
    1 & 0 & \#_{\sigma(.,\alpha)}(c')\\
    0 & a(c') & b(c')\\
    0 & 0 & 1
  \end{pmatrix}
  \cdot
  \begin{pmatrix}
    1 & 0 & \#_{\sigma(.,\alpha)}(\xi)\\
    0 & 0 &  1\\
    0 & 0 & 1
  \end{pmatrix}
=
 \begin{pmatrix}
    1 & 0 & r'\\
    0 & 0 & t'\\
    0 & 0 & 1
    \end{pmatrix}
  \]
where $r'= \#_{\sigma(.,\alpha)}(\xi) + \#_{\sigma(.,\alpha)}(c')$ and 
$t'= a(c') + b(c')$.
Since 
\begin{compactitem}
\item $a(c)=0$,
\item  $r'= \#_{\sigma(.,\alpha)}(\xi) + \#_{\sigma(.,\alpha)}(c') = \#_{\sigma(.,\alpha)}(e[c']) = \#_{\sigma(.,\alpha)}(c)$, and 
\item $t'= a(c')  + b(c') = b(\sigma(\xi,c')) = b(c)$
\end{compactitem}
we have $M_{c'} \cdot \mu(\sigma(\xi,z)) = M_c$. Hence $\murev(c) = M_c$.

\

Third and finally, let $e=\omega(z)$. Then
\[
M_{c'} \cdot \mu(\omega(z)) = 
 \begin{pmatrix}
    1 & 0 & \#_{\sigma(.,\alpha)}(c')\\
    0 & a(c') & b(c')\\
    0 & 0 & 1
  \end{pmatrix}
  \cdot
  \begin{pmatrix}
    1 & 0 & 0\\
    0 & 0 & 0 \\
    0 & 0 & 1
  \end{pmatrix}
  =
  \begin{pmatrix}
    1 & 0 & \#_{\sigma(.,\alpha)}(c')\\
    0 & 0 & b(c')\\
    0 & 0 & 1
  \end{pmatrix}
  = M_c \enspace.
\]
Hence $\murev(c) = M_c$.
This proves \eqref{eq:context-matrix-monoid-rep}.

\

Now we prove that $\cR$ is consistent. Let $c,c' \in \C_\Sigma$ such that $c[\alpha] = c'[\alpha]$. We proceed by case analysis. 

If $c=z$, then $c'=z$, and trivially $\lambda(\alpha) \cdot \murev(c) = \lambda(\alpha) \cdot \murev(c')$ because $\murev(c) = \murev(c')= \mathrm{M}_1$.

Now let $c\not= z$; hence $a(c)=0$. Using~\eqref{eq:context-matrix-monoid-rep} and the fact that $a(c)=0$, we have
\begin{equation}\label{equ:lambda-murev-for-example-mon-rep}
  \lambda(\alpha) \cdot \murev(c)
  =
  \begin{pmatrix}1 & 1 & 0 \end{pmatrix}
  \cdot
  \begin{pmatrix}
    1 & 0 & \#_{\sigma(.,\alpha)}(c)\\
    0 & 0 &  b(c)\\
    0 & 0 & 1
    \end{pmatrix}
=
\begin{pmatrix}1 & 0 & \#_{\sigma(.,\alpha)}(c) + b(c) \end{pmatrix} \enspace.
\end{equation}
Since also $c'\not= z$ by analogy we have
\[
  \lambda(\alpha) \cdot \murev(c') = 
\begin{pmatrix}1 & 0 & \#_{\sigma(.,\alpha)}(c') + b(c') \end{pmatrix} \enspace.
\]
Since  $c[\alpha] = c'[\alpha]$,  we have
\[\#_{\sigma(.,\alpha)}(c) + b(c) = \#_{\sigma(.,\alpha)}(c[\alpha]) =\#_{\sigma(.,\alpha)}(c'[\alpha]) = \#_{\sigma(.,\alpha)}(c') + b(c')\enspace.\]
Hence  $\lambda(\alpha) \cdot \murev(c) = \lambda(\alpha') \cdot \murev(c')$. Thus  $\cR$ is consistent. 

It is instructive to analyse the consistency condition for the particular case in which $\xi=\sigma(\alpha,\alpha)$. Then $\xi$ can be decomposed in two different ways:
$\xi = \sigma(z,\alpha)[\alpha]$ and $\xi= \sigma(\alpha,z)[\alpha]$.
Then 
\[\#_{\sigma(.,\alpha)}(\sigma(z,\alpha)) + b(\sigma(z,\alpha))
= 1 + 0 = 1 
= 0 +1 
= \#_{\sigma(.,\alpha)}(\sigma(\alpha,z)) + b(\sigma(\alpha,z)) \enspace.
\]
The analysis for this particular case can be generalized as follows.  
\begin{equation}\label{equ:number-of-occ-of-pattern-in-context-alpha}
\text{For each $c \in \C_\Sigma$, we have $\#_{\sigma(.,\alpha)}(c) + b(c) = \#_{\sigma(.,\alpha)}(c[\alpha])$.}
\end{equation}
We prove the statement by induction on $(\C_\Sigma,\prec_{\C_\Sigma})$.

I.B.: Let $c=z$. Then $\#_{\sigma(.,\alpha)}(c) + b(c) = 0 + 0 = 0 = \#_{\sigma(.,\alpha)}(\alpha)$.

I.S.: Let $c=e[c']$ for some $e \in \e\C_\Sigma$ and $c \in \C_\Sigma$. Then we can calculate as follows.
\begingroup
\allowdisplaybreaks
\begin{align*}
  \#_{\sigma(.,\alpha)}(e[c']) + b(e[c'])
  &= \begin{cases}
    1 + \#_{\sigma(.,\alpha)}(c') + b(c') & \text{ if $e=\sigma(z,\alpha)$}\\
    \#_{\sigma(.,\alpha)}(e|_2) + \#_{\sigma(.,\alpha)}(c') + b(c') & \text{ if $e=\sigma(z,\xi)$ for some $\xi \in \T_\Sigma\setminus\{\alpha\}$}\\
    \#_{\sigma(.,\alpha)}(e|_1) + \#_{\sigma(.,\alpha)}(c') + b(c') & \text{ if $e=\sigma(\xi,z)$ for some $\xi \in \T_\Sigma$}
  \end{cases}\\[3mm]
  &= \begin{cases}
    1 + \#_{\sigma(.,\alpha)}(c[\alpha]) & \text{ if $e=\sigma(z,\alpha)$}\\
    \#_{\sigma(.,\alpha)}(e|_2) + \#_{\sigma(.,\alpha)}(c'[\alpha]) & \text{ if $e=\sigma(z,\xi)$ for some $\xi \in \T_\Sigma\setminus\{\alpha\}$}\\
    \#_{\sigma(.,\alpha)}(e|_1) + \#_{\sigma(.,\alpha)}(c'[\alpha]) & \text{ if $e=\sigma(\xi,z)$ for some $\xi \in \T_\Sigma$}
  \end{cases}
                                                                     \tag{by I.H.}\\[3mm]
  &= \#_{\sigma(.,\alpha)}(e[c'[\alpha]]) = \#_{\sigma(.,\alpha)}((e[c'][\alpha]) \enspace.
\end{align*}
\endgroup
This proves \eqref{equ:number-of-occ-of-pattern-in-context-alpha}.

Finally we prove that $\sem{\cR} = \#_{\sigma(.,\alpha)}$. Let $\xi \in \T_\Sigma$. 
Moreover, let $c \in \C_\Sigma$ with $\xi=c[\alpha]$, and let $w = \pos_z(c)$. Then , we have:
\begingroup
\allowdisplaybreaks
\begin{align*}
  \sem{\cR}(\xi) &= \lambda(\alpha) \cdot \murev(c) \cdot \gamma \\
                 &= (\lambda(\alpha) \cdot M_c)_f
  \tag{by \eqref{eq:context-matrix-monoid-rep} and by definition of $\gamma$}\\
                 &= \Big((\begin{matrix}1& 1 & 0\end{matrix}) \cdot
                                          \begin{pmatrix}
    1 & 0 & \#_{\sigma(.,\alpha)}(c)\\
    0 & a(c) & b(c)\\
    0 & 0 & 1
  \end{pmatrix}\Big)_f
  \tag{by definitions of $\lambda(\alpha)$ and $M_c$}\\
  &= (\begin{matrix}1& a(c) & \#_{\sigma(.,\alpha)} + b(c)\end{matrix})_f \\
                 &=\#_{\sigma(.,\alpha)}(c) + b(c) \\
                 &= \#_{\sigma(.,\alpha)}(\xi)
                   \tag{by \eqref{equ:number-of-occ-of-pattern-in-context-alpha}}\enspace.
\end{align*}
\endgroup
\hfill $\Box$
  \end{example}

  \section{From recognizability to consistent  monoid representations}
  \label{sec:monoid-representation-of-wta}

  Here we show that the semantics of a wta can be expressed in terms of a consistent monoid representation.
For this, we use the concepts of runs on contexts and weights and combinations of such runs as defined on page \pageref{p:par-extension-runs-to-contexts}.

We  associate with each $(\Sigma,\B)$-wta $\cA$ a consistent $(\Sigma,\B)$-monoid representation $\cR$ and prove that the semantics of $\cA$ is equal to the semantics of $\cR$. The latter semantics might be called the \emph{free monoid semantics} of $\cA$. We recall that a ranked alphabet $\Sigma$ is monadic if $\Sigma=\Sigma^{(0)} \cup \Sigma^{(1)}$. The next theorem generalizes \cite[Thm.~1]{bozale89} from fields to commutative semirings.
 \index{free monoid semantics}

\begin{theorem-rect} \label{theo:monoid-characterization-of-run-semantics} Let $\Sigma$ be a ranked alphabet and $\B=(B,\oplus,\otimes,\0,\1)$ a  semiring such that, if $\Sigma$ is not monadic, then $\B$ is commutative. Moreover, let $\cA$ be a $(\Sigma,\B)$-wta. We can construct a consistent $(\Sigma,\B)$-monoid representation $\cR$ such that $\sem{\cA}=\sem{\cR}$.
  \end{theorem-rect}

  \begin{proof} Let  $\cA=(Q,\delta,F)$. We construct the $(\Sigma,\B)$-monoid representation $\cR=(Q,\lambda,\mu,\gamma)$ with
    \begin{compactitem}
    \item $\lambda: \Sigma^{(0)} \to B^Q$ such that, for every $\alpha \in \Sigma^{(0)}$ and $q \in Q$, we let $\lambda(\alpha)_q = \delta_0(\varepsilon,\alpha,q)$.

\item $\mu: \e\C_\Sigma \to B^{Q \times Q}$  such that, for each $e \in \e\C_\Sigma$ and $p,q \in Q$, we let
  \[\mu(e)_{p,q} = \bigoplus_{\rho \in \R_\cA(q,e,p)} \wt_\cA(e,\rho) \enspace,\]
  and

\item $\gamma = F$.
\end{compactitem}

Next we prove that $\cR$ is consistent. For this, we first prove the following statement by induction on~$(\C_\Sigma,\succ_{\C_\Sigma})$.
\begin{equation}\label{equ:murev-c1c2=weight-c1c2}
  \text{For every $c \in \C_\Sigma$ and $p,q \in Q$, we have
  $\murev(c)_{p,q} = \bigoplus_{\rho \in \R_\cA(q,c,p)} \wt_\cA(c,\rho)$.}
\end{equation}

I.B.: Let $c=z$ and $p,q \in Q$. Then $\murev(c)_{p,q}= (\mathrm{M}_\1)_{p,q} = \bigoplus_{\rho \in \R_\cA(q,z,p)} \wt_\cA(z,\rho)$.

I.S.: Let $c=e[c']$ for some $e\in \e\C_\Sigma$ and $c' \in \C_\Sigma$. Then we can calculate as follows (where we use the depth-first post-order $<_{\mathrm{dp}}$ defined in the paragraph ``Orders on positions of trees'', p.~\pageref{page:orders-on-positions-of-trees}).
\begingroup
\allowdisplaybreaks
\begin{align*}
  \murev(e[c'])_{p,q}
  &= \murev(e \circ_z c')_{p,q} = \big(\murev(c') \cdot \murev(e)\big)_{p,q}\\
  &= \bigoplus_{r \in Q}  \murev(c')_{p,r} \otimes \murev(e)_{r,q}
  = \bigoplus_{r \in Q}  \murev(c')_{p,r} \otimes \mu(e)_{r,q}\\
  &= \bigoplus_{r \in Q}\Big(\bigoplus_{\rho_1 \in \R_\cA(r,c',p)} \
    \wt_\cA(c',\rho_1)\Big)  \otimes
    \Big(\bigoplus_{\rho_2 \in \R_\cA(q,e,r)} \wt_\cA(e,\rho_2)\Big)
    \tag{by I.H. and definition of $\mu$}\\
  &= \bigoplus_{r \in Q}\bigoplus_{\rho_1 \in \R_\cA(r,c',p)}\bigoplus_{\rho_2 \in \R_\cA(q,e,r)}
    \wt_\cA(c',\rho_1) \otimes \wt_\cA(e,\rho_2)
    \tag{by distributivity}\\
  &= \bigoplus_{r \in Q}\bigoplus_{\rho_1 \in \R_\cA(r,c',p)}\bigoplus_{\rho_2 \in \R_\cA(q,e,r)}
    \wt_\cA(e[c'],\rho_2[\rho_1])
      \tag{$v <_{\mathrm{dp}} \varepsilon$ for each $v \in \pos(e[c'])\setminus \{\varepsilon\}$; }\\
  &\hspace*{4mm}\text{(if $\Sigma$ is monadic, then $e=\omega(z)$ for some $\omega \in \Sigma^{(1)}$, and then $\wt_\cA(e,\rho_2) = \delta_1(r,\omega,q)$} \\ 
  &\hspace*{4mm}\text{ is the last factor in the product $\wt_\cA(e[c'],\rho_2[\rho_1])$; otherwise $\B$ is commutative.)}\\[2mm]
  &= \bigoplus_{\rho \in \R_\cA(q,e[c'],p)}
    \wt_\cA(e[c'],\rho) \enspace.
\end{align*}
\endgroup
This proves \eqref{equ:murev-c1c2=weight-c1c2}.
Next we prove:
\begin{equation}\label{equ:h(c[alpha])=sum-delta0-times-h1c}
  \text{For every $c \in \C_\Sigma$, $\alpha \in \Sigma^{(0)}$, and $q \in Q$, we have
    $(\lambda(\alpha) \cdot \murev(c))_q = \bigoplus_{\rho \in \R_\cA(q,c[\alpha])} \wt_\cA(c[\alpha],\rho)$.}
\end{equation}
For the proof,  we use $u_z$ as abbreviation for the position $\pos_z(c)$:
\begingroup
\allowdisplaybreaks
\begin{align*}
  &\big(\lambda(\alpha) \cdot \murev(c)\big)_q \\
  & = \bigoplus_{p \in Q} \lambda(\alpha)_p \otimes \murev(c)_{p,q}\\
  &= \bigoplus_{p \in Q}
    \delta_0(\varepsilon,\alpha,p) \otimes
    \bigoplus_{\rho \in \R_\cA(q,c,p)} \wt_\cA(c,\rho)
  \tag{by definition of $\lambda$ and by \eqref{equ:murev-c1c2=weight-c1c2}}\\[2mm]
  &= \bigoplus_{p \in Q} \bigoplus_{\rho \in \R_\cA(q,c,p)}
    \delta_0(\varepsilon,\alpha,p) \otimes \wt_\cA(c,\rho)
                                     \tag{by distributivity}\\[2mm]
  &= \bigoplus_{p \in Q} \bigoplus_{\rho \in \R_\cA(q,c,p)}
    \delta_0(\varepsilon,\alpha,p) \otimes
    \bigotimes_{\substack{u \in \pos(c)\setminus \{u_z\}\\\text{in $<_{\mathrm{dp}}$ order}}}
  \delta_{\rk(c(u))}(\rho(u1) \cdots \rho(u \, \rk(c(u))),c(u),\rho(u))
  \tag{by an extension of Observation \ref{obs:weight-run-explicit} to weights of runs on contexts;} \\
  &       \tag{note that $\wt_\cA(z,\rho|_{u_z})=\1$} \\[2mm]
  &= \bigoplus_{p \in Q} \bigoplus_{\rho \in \R_\cA(q,c,p)}
    \delta_0(\varepsilon,\alpha,p) \otimes
    \bigotimes_{\substack{u \in \pos(c[\alpha])\setminus \{u_z\}\\\text{in $<_{\mathrm{dp}}$ order}}}
    \delta_{\rk(c[\alpha](u))}(\rho(u1) \cdots \rho(u \, \rk(c[\alpha](u))),c[\alpha](u),\rho(u))
    \tag{because $\pos(c[\alpha])\setminus \{u_z\}=\pos(c)\setminus \{u_z\}$}\\[2mm]
  &= \bigoplus_{p \in Q} \bigoplus_{\rho \in \R_\cA(q,c,p)}   
    \Big[\bigotimes_{\substack{u \in \pos(c[\alpha])\\\text{in $<_{\mathrm{dp}}$ order}}}
  \delta_{\rk(c[\alpha](u))}(\rho(u1) \cdots \rho(u \, \rk(c[\alpha](u))),c[\alpha](u),\rho(u))\Big]
    \tag{note that $\delta_0(\varepsilon,c[\alpha](u_z),\rho(u_z)) = \delta_0(\varepsilon,\alpha,p)$;}\\
  &   \tag{in case that $\Sigma$ is monadic, we have that $u_z <_{\mathrm{dp}} v$ for each $v \in \pos(c[\alpha])\setminus \{u_z\}$ and}\\
  &\tag{hence, $\delta_0(\varepsilon,\alpha,p)$ is the first factor of the embraced product; otherwise $\B$ is commutative}\\[2mm]
  &= \bigoplus_{\rho \in \R_\cA(q,c[\alpha])}
    \bigotimes_{\substack{u \in \pos(c[\alpha])\\\text{in $<_{\mathrm{dp}}$ order}}}
    \delta_{\rk(c[\alpha](u))}(\rho(u1) \cdots \rho(u \, \rk(c[\alpha](u))),c[\alpha](u),\rho(u))
    \tag{because the $Q$-indexed family $(\R_\cA(q,c,p) \mid p \in Q)$ is a partitioning of $\R_\cA(q,c[\alpha])$}\\[2mm]
  &= \bigoplus_{\rho \in \R_\cA(q,c[\alpha])} \wt_\cA(c[\alpha],\rho) \enspace.
 \tag{by Observation \ref{obs:weight-run-explicit}}
\end{align*}
\endgroup
This proves \eqref{equ:h(c[alpha])=sum-delta0-times-h1c}.

Now let $c,c' \in \C_\Sigma$ and $\alpha,\alpha' \in \Sigma^{(0)}$ such that $c[\alpha]=c'[\alpha']$. Then, by \eqref{equ:h(c[alpha])=sum-delta0-times-h1c}, we have  $\lambda(\alpha) \cdot \murev(c) = \lambda(\alpha') \cdot \murev(c')$. Hence $\cR$ is consistent.

Finally, we prove that  $\sem{\cA}= \sem{\cR}$. Let $\xi \in \T_\Sigma$ and $c \in \C_\Sigma$ and $\alpha \in \Sigma^{(0)}$ such that $\xi = c[\alpha]$. Then:
\begingroup
\allowdisplaybreaks
\begin{align*}
  \sem{\cA}(\xi) &= \bigoplus_{\rho \in \R_\cA(\xi)} \wt_\cA(\xi,\rho) \otimes F_{\rho(\varepsilon)}
                 = \bigoplus_{q \in \Q} \bigoplus_{\rho \in \R_\cA(q,\xi)} \wt_\cA(\xi,\rho) \otimes F_q\\
                 &= \bigoplus_{q \in \Q} \Big(\bigoplus_{\rho \in \R_\cA(q,\xi)} \wt_\cA(\xi,\rho)\Big) \otimes \gamma_q
  \tag{by distributivity and definition of $\gamma$}\\
                 &=  \bigoplus_{q \in Q} \big(\lambda(\alpha) \cdot \murev(c)\big)_q \otimes \gamma_q
  \tag{by \eqref{equ:h(c[alpha])=sum-delta0-times-h1c}}\\
                 &= \lambda(\alpha) \cdot \murev(c) \cdot \gamma = \sem{\cR}(\xi) \enspace.
                   \qedhere
\end{align*}
\endgroup
\end{proof}

\begin{observation}\label{obs:ex:monoid-rep-number-of-occ-of-pattern-results-from-construction}
  \rm  The monoid representation of Example \ref{ex:monoid-rep-number-of-occ-of-pattern} results from applying the construction in the proof of  Theorem~\ref{theo:monoid-characterization-of-run-semantics} to the $(\Sigma,\Nat)$-wta $\cA$ of Example~\ref{ex:number-of-occurrences}.
\end{observation}

\paragraph{The case of string ranked alphabets.}
Here we analyse the concept of $(\Sigma,\B)$-monoid representation in the case where $\Sigma$ is a string ranked alphabet, say, $\Sigma = \Gamma_e$ for some alphabet $\Gamma$ and some nullary symbol $e$. Then $\Gamma_e$-trees are essentially $\Gamma$-strings (cf. page \pageref{p:string-like-trees-are-strings}). Moreover, elementary contexts have a special form, viz.,  $\e\C_{\Gamma_e} = \{a(z) \mid a \in \Gamma\}$. This implies that each $(\Gamma_e,\B)$-monoid representation is consistent as follows.
    
  \begin{observation}\rm \label{obs:string-ranked-alphabet-implies-consistency-of-monoid-rep} Let $\Gamma_e$  be a string ranked alphabet and $\cR$ be a $(\Gamma_e,\B)$-monoid representation. Then $\cR$ is consistent.
  \end{observation}
  \begin{proof} Since  $\Gamma_e$ is monadic, each $c \in \C_{\Gamma_e}$ has exactly one leaf.  Hence, for every $c,c' \in \C_{\Gamma_e}$, the following holds: if $c[e]=c'[e]$, then $c=c'$. Since $\Gamma_e^{(0)}= \{e\}$, it follows that $\cR$ is consistent.
  \end{proof}

  As corollary of Theorems \ref{theo:monoid-characterization-of-run-semantics} and \ref{lm:wsa=wta-over-string-ra}(1), we obtain the well-known characterization of the run semantics (or: behaviour) of a wsa \cite[Cor.~VI.6.2]{eil74} (also cf. \cite[Ch.~III, Sect.~2, Cor.~2.12]{sak09}, \cite[Lm.~5]{drostuvog10}, and \cite[Thm.~3.2]{drokus21}; and \cite[Ch.~I,~Sect.~5]{berreu88} where $\B$ is commutative).

  \begin{corollary-rect}\label{cor:cor-VI.6.2-eil74} \rm \cite[Cor.~VI.6.2]{eil74} Let $\Gamma$ be an alphabet and $\B=(B,\oplus,\otimes,\0,\1)$ a semiring. Moreover, let $\cA=(Q,\lambda,\mu,\gamma)$ be a $(\Gamma,\B)$-wsa. Then, for each $w \in \Gamma^*$, we have $\sem{\cA}(w) = \lambda \cdot \mu(w) \cdot \gamma$.
  \end{corollary-rect}
  \begin{proof} We recall that $\mu: \Gamma \to B^{Q \times Q}$. We also denote by $\mu$ the extension of $\mu$ to a monoid homomorphism from the free monoid $(\Gamma^*,\cdot,\varepsilon)$ to the monoid $(B^{Q \times Q},\cdot,\mathrm{M}_\1)$. 
  
  By Theorem \ref{lm:wsa=wta-over-string-ra}(1), we can construct a $(\Gamma_e,\B)$-wta $\cB=(Q,\delta,F)$ such that $\sem{\cA} = \sem{\cB} \circ \tree_e$. We recall that $\delta_0(\varepsilon,e,q) = \lambda_q$ for each $q \in Q$, $\delta_1(p,a,q) = \mu(a)_{p,q}$ for every $a \in \Gamma$ and $p,q \in Q$, and $F=\gamma$.

    Since $\Gamma_e$ is monadic, by Theorem \ref{theo:monoid-characterization-of-run-semantics}, we can construct a consistent $(\Gamma_e,\B)$-monoid representation $\cR=(Q,\lambda',\mu',\gamma')$ such that $\sem{\cR}=\sem{\cB}$. We recall that $\lambda'(e)_q= \delta_0(\varepsilon,e,q)$ for each $q \in Q$, $\mu'(a(z))_{p,q}= \bigoplus_{\rho \in \R_\cB(q,a(z),p)} \wt_\cB(a(z),\rho)$ for every $a \in \Gamma$ and  $p,q\in Q$, and $\gamma'=F$. Hence, by combining the above two constructions, we have that $\lambda'(e)=\lambda$, and  $\mu'(a(z))_{p,q}=\delta_1(p,a,q) = \mu(a)_{p,q}$ for  every $p,q\in Q$, and $\gamma'=\gamma$.

    By induction on $\Gamma^*$, we define the auxiliary mapping $\tree_z: \Gamma^* \to \C_{\Gamma_e}$ by $\tree_z(\varepsilon)=z$ and $\tree_z(wa) = a(\tree_z(w))$ for every $a \in \Gamma$ and $w \in \Gamma^*$. Clearly, for each $w \in \Gamma^*$, we have $\tree_e(w) = \tree_z(w)[e]$.
(We note that, although $(\Gamma^*,\cdot,\varepsilon)$ and $(\C_{\Gamma_e},\circ_z,z)$ are monoids, the mapping $\tree_z$ is not a monoid homomorphism, because it reverses the order of the arguments. To see this, let $a,b \in \Gamma$ with $a \not= b$. Then $\tree_z(ab) = b(a(z)) = b(z) \circ_z a(z) = \tree_z(b) \circ_z \tree_z(a)$; if $\tree_z$ was a monoid homomorphisms, then we could continue with $\tree_z(b) \circ_z \tree_z(a) = \tree_z(ba)$; but $\tree_z(ab) = b(a(z)) \not=  a(b(z)) = \tree_z(ba)$.)

Then, by induction on $\Gamma^*$, we prove the following statement (cf. Figure \ref{fig:roles-of-rows-columns}).
  \begin{equation}\label{equ:murev-treee=mu}
\text{For each $w \in \Gamma^*$, we have $(\mu')^{\leftarrow}(\tree_z(w)) = \mu(w)$.}
\end{equation}

  \begin{figure}[t]
    \centering
\begin{tikzpicture}[
	scale=1,
	transform shape,
	zigzag/.style={decorate,decoration={zigzag,amplitude=0.7pt,segment length=1.6mm,pre=lineto,pre length=4pt}},
	myellipse/.style={draw,ellipse,minimum height=1.1cm,minimum width=0.4cm,outer sep=1mm}]

\node[anchor=base west] at (0,3)
	{\strut $(\Gamma,\B)$-wsa \ $\cA =(Q,\lambda,\mu,\tau)$};
\node[anchor=base west] at (0,2) (term1) 
	{\strut $\mu(abb) = \mu(a) \cdot \mu(b) \cdot \mu(b)$};
\node[anchor=base west] at (6,3)
	{\strut $(\Gamma_e,\B)$-monoid representation $\cR=(Q,\lambda',\mu',\gamma')$};
\node[anchor=base west] at (6,2) (term2)
	{\strut $\mu'(b(z))
	\diamond \mu'(b(z)) 
	\diamond \mu'(a(z))
        = (\mu')^{\leftarrow}(\tree_z(abb))$};

\node[anchor=base west] at (5,2) (equality)
	{\strut $=$};      

\coordinate (coordl1) at (1.15,-0.3);
\coordinate[below=1.2cm of coordl1] (coordl2);
\coordinate[below=1.2cm of coordl2] (coordl3);
\coordinate[below=1.2cm of coordl3] (coordl4);
\node at (coordl1) (bl1) {b};
\node at (coordl2) (bl2) {b};
\node at (coordl3) (al3) {a};
\node at (coordl4) (el4) {z};
\node[right=0.7cm of coordl1, anchor=north, rotate=90] (ml1) {$\mu(b)$};
\node[right=0.7cm of coordl2, anchor=north, rotate=90] (ml2) {$\mu(b)$};
\node[right=0.7cm of coordl3, anchor=north, rotate=90] (ml3) {$\mu(a)$};
\node[right=0.7cm of coordl4, anchor=north, rotate=90] (ml4) {$\mathrm{M}_\1$};
\node at ($(ml1)!0.5!(ml2)$) {$\cdot$};
\node at ($(ml2)!0.5!(ml3)$) {$\cdot$};
\node at ($(ml3)!0.6!(ml4)$) {$\cdot$};
\draw (bl1) to (bl2);
\draw (bl2) to (al3);
\draw (al3) to (el4);
\draw[<-,zigzag] (bl1) -- (ml1);
\draw[<-,zigzag] (bl2) -- (ml2);
\draw[<-,zigzag] (al3) -- (ml3);
\draw[<-,zigzag] (el4) -- (ml4);

\coordinate[right=7cm of coordl1] (coordr1);
\coordinate[right=7cm of coordl2] (coordr2);
\coordinate[right=7cm of coordl3] (coordr3);
\coordinate[right=7cm of coordl4] (coordr4);
\node[yshift= 0.8cm] at (coordr1) (br1) {b};
\node[yshift= 0.2cm] at (coordr1) (zr1) {z};
\node[yshift= 0.3cm] at (coordr2) (br2) {b};
\node[yshift=-0.3cm] at (coordr2) (zr2) {z};
\node[yshift=-0.2cm] at (coordr3) (ar3) {a};
\node[yshift=-0.8cm] at (coordr3) (zr3) {z};
\node[myellipse,yshift=0.5cm] at (coordr1) (er1) {};
\node[myellipse] at (coordr2) (er2) {};
\node[myellipse,yshift=-0.5cm] at (coordr3) (er3) {};
\node[right=1.5 of coordr1, anchor=south, yshift=-1.2cm, rotate=-90] (mr1) {$\mu'(b(z)) \ \diamond \ \mu'(b(z)) \ \diamond \ \mu'(a(z))$};
\draw (br1) to (zr1);
\draw (br2) to (zr2);
\draw (ar3) to (zr3);

\draw [->] ($(coordl2)+(2.7,0)$) to ($(coordl2)+(5.7,0)$) node[xshift=-1.5cm,above] {$\T_{\Gamma_z} \rightarrow (\e\C_\Sigma)^*$};
\draw [->, out=225, in=150, looseness=1.4] ($(term1)+(-0.1,-0.5)$) to ($(coordl2)+(-1,0)$) node[midway, xshift=-1.0cm] {tree$_z$};
\draw [->, out=20, in=-45, looseness=1.3] ($(coordr2)+(3,0)$) to ($(term2.-10)+(0.4,-0.3)$);

\end{tikzpicture}

    \caption{\label{fig:roles-of-rows-columns} An illustration of Equation \eqref{equ:murev-treee=mu} with $a,b \in \Gamma$ and $w=abb$.}
    \end{figure}

I.B.: Let $w = \varepsilon$. Then $(\mu')^{\leftarrow}(\tree_z(\varepsilon)) = (\mu')^{\leftarrow}(z) = \mathrm{M}_\1=\mu(\varepsilon)$.

I.S.: Let $a \in \Gamma$ and $w \in \Gamma^*$. We can calculate as follows.  
  \begingroup
  \allowdisplaybreaks
  \begin{align*}
    (\mu')^{\leftarrow}(\tree_z(wa)) &= (\mu')^{\leftarrow}(a(\tree_z(w))) = (\mu')^{\leftarrow}(a(z) \circ_z \tree_z(w))\\
                                     &= (\mu')^{\leftarrow}(a(z)) \diamond (\mu')^{\leftarrow}(\tree_z(w))
    \tag{because $(\mu')^{\leftarrow}$ is a monoid homomorphism}\\
                                     &= (\mu')^{\leftarrow}(\tree_z(w)) \cdot (\mu')^{\leftarrow}(a(z))
    \tag{by definition of $\diamond$}\\
                        &= (\mu')^{\leftarrow}(\tree_z(w)) \cdot \mu'(a(z))
    \tag{because $a(z) \in \e\C_\Sigma$}\\
                        &= \mu(w) \cdot \mu(a)
                          \tag{by I.H. and by the above}\\
                        &= \mu(wa)
                          \tag{because $\mu$ is a monoid homomorphism}\enspace.
  \end{align*}
  \endgroup
  This proves \eqref{equ:murev-treee=mu}.

  Let $w \in \Gamma^*$. Then we can calculate as follows.
  \begingroup
  \allowdisplaybreaks
  \begin{align*}
    \sem{\cA}(w) &= (\sem{\cB} \circ \tree_e)(w) = \sem{\cB}(\tree_e(w))\\
                    &= \sem{\cR}(\tree_e(w))\\
                    &= \lambda'(e) \cdot (\mu')^{\leftarrow}(\tree_z(w)) \cdot \gamma'
    \tag{because $\tree_e(w) = \tree_z(w)[e]$}\\
                    &= \lambda \cdot \mu(w) \cdot \gamma \enspace.
                      \tag{by constructions and by  \eqref{equ:murev-treee=mu}}
  \end{align*}
  \endgroup
    \end{proof}

    In fact, if $\Sigma$ is  string ranked, then the run semantics of $(\Sigma,\B)$-wta can be characterized by $(\Sigma,\B)$-monoid representations (cf. \cite[p.~120]{berreu82}).

     \begin{theorem-rect}\label{theo:string-ranked:wta=monoid-rep} Let $\Sigma$ be a string ranked alphabet and $\B=(B,\oplus,\otimes,\0,\1)$ be a semiring. Moreover, let $r: \T_\Sigma \to B$. Then the following two statements are equivalent.
      \begin{compactenum}
        \item[(A)] We can construct a $(\Sigma,\B)$-wta $\cA$ such that $\sem{\cA} = r$.
        \item[(B)] We can construct a consistent $(\Sigma,\B)$-monoid representation $\cR$ such that $\sem{\cR} = r$.
        \end{compactenum}
        \end{theorem-rect}
        \begin{proof} Proof of (A)$\Rightarrow$(B): This follows from Theorem~\ref{theo:monoid-characterization-of-run-semantics}.

          \
          
           Proof of (B)$\Rightarrow$(A): Let $\cR=(Q,\lambda,\mu,\gamma)$ be a consistent $(\Sigma,\B)$-monoid representation 
           and $\Sigma = \Gamma_e$ for some alphabet $\Gamma$ such that $\sem{\cR}=r$.

           We construct the $(\Sigma,\B)$-wta $\cA=(Q,\delta,\gamma)$ such that, for every $p,q \in Q$ and $\omega \in \Gamma$ we let
           \[
\delta_0(\varepsilon,e,q) = \lambda_q  \ \text{ and } \ \delta_1(p,\omega,q) = \mu(\omega(z))_{p,q} \enspace.
\]
If we apply the construction in the proof of Theorem~\ref{theo:monoid-characterization-of-run-semantics} to $\cA$, then we reobtain $\cR$. This is due to the fact that 
\[
\mu(\omega(z))_{p,q} = \bigoplus_{\rho \in \R_\cA(q,\omega(z),p)} \wt_\cA(\omega(z),\rho) = \delta_1(p,\omega,q) \enspace.
\]
Thus we have $r= \sem{\cR}=\sem{\cA}$ where the second equality holds by Theorem~\ref{theo:monoid-characterization-of-run-semantics}.
\end{proof}

\paragraph{Relationship between the monoid homomorphisms $\h_\cA^\C$ and $\murev$.}

The consistency condition for monoid representations (defined on page \pageref{page:def-consistent-mon-rep}) has the form:
\[\text{for every $c,c' \in \C_\Sigma$, and $\alpha,\alpha' \in \Sigma^{(0)}$: if $c[\alpha]=c'[\alpha']$, then $\lambda(\alpha) \cdot \murev(c) = \lambda(\alpha') \cdot \murev(c')$}
\]
where we assume that $\B$ is a commutative semiring.
This looks very similar to \eqref{equ:consistency-for-hAC}, i.e.,
\[
  \text{\text{For every $c,c' \in \C_\Sigma$ and $\alpha,\alpha' \in \Sigma^{(0)}$, if $c[\alpha]=c'[\alpha']$, then $\h_\cA^\C(c)(\h_\cA(\alpha)) = \h_\cA^\C(c')(\h_\cA(\alpha'))$} }
\]
which holds for an arbitrary strong bimonoid as weight algebra.
Obviously, $\h_\cA^\C(c)$ and $\h_\cA(\alpha)$ correspond to $\murev(c)$ and $\lambda(\alpha)$, respectively. 
We recall that \eqref{equ:consistency-for-hAC} is a consequence of Lemma \ref{lm:hcACchAxi=hAcxi}. In Lemma \ref{lm:hcACchAxi=hAcxi}, we do not need commutativity, because the mapping $\h_\cA^\C(c)$ places its argument $\h_\cA(\alpha)$ to the correct place, which does not hold for $\murev(c)$. Moreover, in Lemma \ref{lm:hcACchAxi=hAcxi}, we do not need distributivity, because there is no matrix multiplication involved. 
Here we show a formal relationship between $\h_\cA^\C$ and $\murev$ for the case in which the weight algebra is a commutative semiring.

For this, we consider an arbitrary $(\Sigma,\B)$-wta $\cA=(Q,\delta,F)$ for a commutative semiring $\B$. Moreover, we consider the $(\Sigma,\B)$-monoid representation $\cR=(Q,\lambda,\mu,\gamma)$ 
which is the result of applying the construction in the proof of Theorem~\ref{theo:monoid-characterization-of-run-semantics} to $\cA$. We recall that $\mu: \e\C_\Sigma \to B^{Q \times Q}$ is defined, for each $e \in \e\C_\Sigma$ and $p,q \in Q$, by $\mu(e)_{p,q} = \bigoplus_{\rho \in \R_\cA(q,e,p)} \wt_\cA(e,\rho)$. We also recall that $\B^Q=(B^Q,\oplus,\1_Q)$ is a $\B$-semimodule (cf. Observation \ref{obs:B-to-Q-is-semimodule}). Here we  prove the precise relationship between the monoid homomorphism
\[
  \h_\cA^\C: \C_\Sigma \to \cL(\B^Q,\B^Q) \enspace,
\]
as defined in Section~\ref{sect:splitting-properties-of-hA} and by Lemma \ref{lm:haC-linear-mapping}, and the monoid homomorphism
\[
  \murev: \C_\Sigma \to B^{Q \times Q}  \enspace.
\]
For this purpose, we use the monoid isomorphism $\psi: \cL(\B^Q,\B^Q) \to B^{Q \times Q}$ as defined on page \pageref{p:def-psi-prime}, and the transposition mapping $\mathrm{T}$ as defined in Section~\ref{sec:vectors-matrices}.

\begin{lemma}\rm \label{lm:connection-between-murev-and-hAC} Let $\Sigma$ be a ranked alphabet and $\B$ a commutative semiring. Moreover, let $\cA$ be a $(\Sigma,\B)$-wta and let $\cR=(Q,\lambda,\mu,\gamma)$ be the $(\Sigma,\B)$-monoid representation which is the result of applying the construction in the proof of Theorem~\ref{theo:monoid-characterization-of-run-semantics} to $\cA$. Then, for each $c \in \C_\Sigma$, we have $\psi(\h_\cA^\C(c)) = \murev(c)^\mathrm{T}$ (cf. Figure \ref{fig:overview-mappings-on-contexts}).
  Since $\mathrm{T}$ and $\psi$ are bijective, we have
  \begin{compactitem}
  \item $\h_\cA^\C = \psi^{-1} \circ \mathrm{T} \circ \murev$ and
    \item $\murev = \mathrm{T} \circ \psi \circ \h_\cA^\C$.
    \end{compactitem}
  \end{lemma}
  \begin{proof}  Since $\psi: \cL(\B^Q,\B^Q) \to B^{Q \times Q}$ and $\mathrm{T}: B^{Q \times Q} \to B^{Q \times Q}$ are monoid homomorphisms,  the mappings $\psi \circ \h_\cA^\C$ and $\mathrm{T} \circ \murev$ are monoid homomorphisms from $(\C_\Sigma,\circ_z,z)$ to $(B^{Q \times Q},\cdot,\mathrm{M}_\1)$. Since $(\C_\Sigma,\circ_z,z)$ is free in the set of all monoids with generating set $\e\C_\Sigma$, for each mapping $g: \e\C_\Sigma \to B^{Q \times Q}$ there exists a unique monoid homomorphism from $(\C_\Sigma,\circ_z,z)$ to $(B^{Q \times Q},\cdot,\mathrm{M}_\1)$ which extends $g$. Thus $\psi \circ \h_\cA^\C=\mathrm{T} \circ \murev$, if $\psi \circ \h_\cA^\C$ and $\mathrm{T} \circ \murev$ coincide on the generating set $\e\C_\Sigma$.
  
    For the proof of the latter, let $e \in \e\C_\Sigma$ with $e= \sigma(\xi_1,\ldots,\xi_{i-1},z,\xi_{i+1},\ldots,\xi_k)$ for some $k \in \mathbb{N}$, $\sigma \in \Sigma^{(k)}$, $\xi_1,\ldots,\xi_{i-1},\xi_{i+1},\ldots,\xi_k \in \T_\Sigma$, and $p,q \in Q$. Then, for every $q,p \in Q$, we can calculate as follows.
\begingroup
\allowdisplaybreaks
\begin{align*}
  &\Big((\psi \circ \h_\cA^\C)(e)\Big)_{q,p} = \psi(\h_\cA^\C(e))_{q,p}\\
  &= \h_\cA^\C(e)(\1_p)_q
  \tag{by definition of $\psi$}\\
  &= \delta_\cA(\sigma)\big(\h_\cA(\xi_1),\ldots,\h_\cA(\xi_{i-1}),\1_p,\h_\cA(\xi_{i+1}),\ldots,\h_\cA(\xi_k)\big)_q
  \tag{by definition of $\h_\cA^\C$} \\
  &= \bigoplus_{q_1 \cdots q_k \in Q}
    \Big(\bigotimes_{j \in [1,i-1]} \h_\cA(\xi_j)_{q_j}\Big)
    \otimes
    (\1_p)_{q_i}
    \otimes
    \Big(\bigotimes_{j \in [i+1,k]} \h_\cA(\xi_j)_{q_j}\Big)
    \otimes \delta_k(q_1 \cdots q_k,\sigma,q)  \\
  &= \bigoplus_{q_1 \cdots q_k \in Q^k}
    \Big(\bigotimes_{j \in [1,i-1]} \bigoplus_{\rho_j \in \R_\cA(q_j,\xi_j)} \wt(\xi_j,\rho_j)\Big)
    \otimes
    (\1_p)_{q_i}
    \otimes
    \Big(\bigotimes_{j \in [i+1,k]} \bigoplus_{\rho_j \in \R_\cA(q_j,\xi_j)} \wt(\xi_j,\rho_j)\Big)
    \otimes \delta_k(q_1 \cdots q_k,\sigma,q)
  \tag{by Corollary \ref{cor:semiring-run=init}(1)}\\
  &= \bigoplus_{q_1 \cdots q_k \in Q^k} \bigoplus_{\rho_1 \in \R_\cA(q_1,\xi_1)} \ldots \bigoplus_{\rho_{i-1} \in \R_\cA(q_{i-1},\xi_{i-1})}
    \bigoplus_{\rho_{i+1} \in \R_\cA(q_{i+1},\xi_{i+1})} \ldots \bigoplus_{\rho_k \in \R_\cA(q_k,\xi_k)}\\
  & \hspace*{9mm}  \Big(\bigotimes_{j \in [1,i-1]}  \wt(\xi_j,\rho_j)\Big)
    \otimes
    (\1_p)_{q_i}
    \otimes
    \Big(\bigotimes_{j \in [i+1,k]}  \wt(\xi_j,\rho_j)\Big)
    \otimes \delta_k(q_1 \cdots q_k,\sigma,q)
  \tag{by distributivity}\\
  &= \bigoplus_{\substack{q_1 \cdots q_{i-1} \in Q^{i-1}\\q_{i+1} \cdots q_k \in Q^{k-i}}} \bigoplus_{\rho_1 \in \R_\cA(q_1,\xi_1)} \ldots \bigoplus_{\rho_{i-1} \in \R_\cA(q_{i-1},\xi_{i-1})}
    \bigoplus_{\rho_{i+1} \in \R_\cA(q_{i+1},\xi_{i+1})} \ldots \bigoplus_{\rho_k \in \R_\cA(q_k,\xi_k)}\\
  & \hspace*{9mm}  \Big(\bigotimes_{j \in [1,i-1]}  \wt(\xi_j,\rho_j)\Big)
    \otimes
    \1
    \otimes
    \Big(\bigotimes_{j \in [i+1,k]}  \wt(\xi_j,\rho_j)\Big)
    \otimes \delta_k(q_1 \cdots q_{i-1} p  q_{i+1},\ldots, q_k,\sigma,q)\\
  &= \bigoplus_{\rho \in \R_\cA(q,e,p)} \wt(e,\rho)
  \tag{due to the form of runs on the elementary context $e$}\\
  &= \mu(e)_{p,q}
    =  \murev(e)_{p,q}
    =  \mathrm{T}(\murev(e))_{q,p}
    = \Big((\mathrm{T} \circ \murev)(e)\Big)_{q,p} \enspace.\qedhere
  \end{align*}
  \endgroup
    \end{proof}

  \begin{figure}[t]
    \centering
  \scalebox{0.9}{
\begin{tikzpicture}
\tikzset{transform shape}

\node at (0,1.6) (1) {$(\cL(\B^Q,\B^Q),\circ,\id_{B^Q})$};
\node at (6,1.6) (4) {$(B^{Q \times Q},\diamond,\mathrm{M}_\1)$};
\node at (3,0) (3) {$(B^{Q \times Q},\cdot,\mathrm{M}_\1)$};
\node at (3,3.2) (2) {$(\C_\Sigma,\circ_z,z)$};

\draw (2) edge[->,>=stealth] node[fill=white] {$\h_\cA^\C$} (1);
\draw (1) edge[->,>=stealth] node[fill=white] {$\psi$} (3);
\draw (2) edge[->,>=stealth] node[fill=white] {$\murev$} (4);
\draw (4) edge[->,>=stealth] node[fill=white] {$\mathrm{T}$} (3);

\end{tikzpicture}
}
\caption{\label{fig:overview-mappings-on-contexts} Relationship between four the monoids $(\C_\Sigma,\circ_z,z)$, $(\cL(\B^Q,\B^Q),\circ,\id_{B^Q})$, $(B^{Q \times Q},\diamond,\mathrm{M}_\1)$, and $(B^{Q \times Q},\cdot,\mathrm{M}_\1)$.} 
\end{figure}

In particular, Lemma \ref{lm:connection-between-murev-and-hAC} clarifies the relationship between Example \ref{ex:monoid-rep-number-of-occ-of-pattern} and Example \ref{ex:number-of-occ-context-field}.
\begin{compactitem}
  \item In Example \ref{ex:monoid-rep-number-of-occ-of-pattern}, we have $\murev(c) = M_c$ for each $c \in \C_\Sigma$ (cf. \eqref{eq:context-matrix-monoid-rep}).
\item In Example \ref{ex:number-of-occ-context-field}, we have $\psi(\h_\cA^\C(c)) = N_c$ for each $c \in \C_\Sigma$ (cf. \eqref{equ:psi-hAC-c=M-c}).
\end{compactitem}
Since the monoid representation of Example \ref{ex:monoid-rep-number-of-occ-of-pattern} results from applying the construction in the proof of  Theorem~\ref{theo:monoid-characterization-of-run-semantics} to the $(\Sigma,\Nat)$-wta $\cA$ (cf. Observation~\ref{obs:ex:monoid-rep-number-of-occ-of-pattern-results-from-construction}), if follows from Lemma \ref{lm:connection-between-murev-and-hAC} that
\[
  M_c = \murev(c) = \psi(\h_\cA^\C(c))^{\mathrm{T}} = (N_c)^\mathrm{T} \enspace.
\]


\section{From consistent monoid representations to recognizability}
\label{sect:from-consistent-monoid-representations-to-recognizability}

Here we show the opposite direction: for every field $\B$ and  consistent $(\Sigma,\B)$-monoid representation $\cR$, we can construct an equivalent $(\Sigma,\B)$-wta  (cf. Theorem~\ref{theo:from-monoid-rep-to-wta-comm-semiring}). Corollary~\ref{cor:field-from-consistent-rep-to-wta} is a restatement of \cite[Thm.~2]{bozale89}.

\begin{theorem-rect} \label{theo:from-monoid-rep-to-wta-comm-semiring} Let $\Sigma$ be a ranked alphabet and  $\B$ be a commutative semiring. Let $\cR=(Q,\lambda,\mu,\gamma)$ be a consistent $(\Sigma,\B)$-monoid representation. Then the following two statements hold.
\begin{compactenum}
\item[(1)] We can construct (a)~a~$(\Sigma,\B)$-semimodule $\V_\cR=(\V,\eta)$ where $\V$ is a $\B$-sub-semimodule of $(B^Q,\oplus,\widetilde{\0})$  and (b)~a~linear form $\gamma'$ over the $\B$-semimodule $\V$ such that $\sem{\cR}= \gamma' \circ \h_\eta$.
 
\item[(2)] Let $\V_\cR=(\V,\eta)$ and $\gamma'$ be the $\B$-sub-semimodule of $(B^Q,\oplus,\widetilde{\0})$  and the linear form, respectively, as constructed in the proof of (1). If $\B$ is a field, then  we can construct a $(\Sigma,\B)$-wta $\cA$ such that $\sem{\cA}= \gamma' \circ \h_\eta$.
\end{compactenum}
\end{theorem-rect}

\begin{proof}
Proof of (1): Let $\cR=(Q,\lambda,\mu,\gamma)$.
Let $\1.\T_\Sigma = \{\1.\xi \mid \xi \in \T_\Sigma\}$. 
  We define the mapping $\h_\cR: \1.\T_\Sigma \to B^Q$ such that, for each $\xi \in \T_\Sigma$, we let $\h_\cR(\1.\xi) = \lambda(\alpha) \cdot \murev(c)$ for $\alpha \in \Sigma^{(0)}$ and $c \in \C_\Sigma$ with $\xi=c[\alpha]$. Due to the consistency of $\cR$, the mapping $\h_\cR$ is well defined.
  
  By Theorem~\ref{lm:Pol-SigmaB-free} there exists a linear mapping $\h_\cR'$ from the $\B$-semimodule $\mathsf{Pol}(\Sigma,\B)$ to the $\B$-semimodule $\B^Q=(B^Q,\oplus,\widetilde{\0})$ which extends $\h_\cR$, i.e.,
  \[
\h_\cR': \Pol(\Sigma,\B) \to B^Q \ \ \text{with $\h_\cR'(\1.\xi) = \h_\cR(\1.\xi)$ for each $\xi \in \T_\Sigma$}\enspace. 
  \]
  By Lemma~\ref{lm:baanip-unique-hom}, it is the unique extension of $\h_\cR$. By the proof of Theorem~\ref{lm:Pol-SigmaB-free} 
  \[
\h_\cR'(s) = b_1 \cdot \h_\cR(\1.\xi_1) \oplus \ldots \oplus b_\ell \cdot \h_\cR(\1.\xi_\ell) 
\]
for each $s= b_1.\xi_1 \oplus \ldots \oplus b_\ell.\xi_\ell$ (cf. \eqref{eq:from-h-to-hV-free}).
Then $(\im(\h_\cR'),\oplus,\widetilde{\0})$ is a sub-semimodule of the $\B$-semimodule $\B^Q$.
We define the desired $\B$-sub-semimodule $\V$ (see (a) of the lemma)  to be $\V=(\im(\h_\cR'),\oplus,\widetilde{\0})$.

\

Next we will define the $\Sigma$-algebra $(\im(\h_\cR'),\eta)$ such that, for each $\xi \in \T_\Sigma$, we have  $\h_\cR(\1.\xi)=\h_\eta(\xi)$.
For the definition of $\eta$ we need some prerequisites.

We let $\Pol_\C(\Sigma,\B)$ be the set of polynomial $\B$-weighted context languages, i.e., $\Pol_\C(\Sigma,\B) = \{ r \in B^{\C_\Sigma} \mid \supp(r) \text{ is finite}\}$. We extend $\murev$ to the mapping $(\mu^{\leftarrow})': \Pol_\C(\Sigma,\B) \to B^{Q\times Q}$ such that, for each $t= d_1.c_1 \oplus \ldots \oplus d_m.c_m$ in $\Pol_\C(\Sigma,\B)$, we let
  \[
(\mu^{\leftarrow})'(t) = d_1 \cdot \murev(c_1) \oplus \ldots \oplus d_m \cdot \murev(c_m) \enspace.
\]
In the sequel, we abbreviate $(\mu^{\leftarrow})'$ by $\murev$.
Moreover, for every $r' = d_1.c_1 \oplus \ldots \oplus d_\ell.c_\ell$ in $\Pol_\C(\Sigma,\B)$ and $r= b_1.\xi_1 \oplus \ldots \oplus b_m.\xi_m$ in $\Pol(\Sigma,\B)$, we let
\[
r' \circ_z r = \bigoplus_{i\in [\ell], j \in [m]} (d_i \otimes b_j). (c_i \circ_z \xi_j) \enspace.
\]

Next we define two algebras $(\Pol(\Sigma,\B),\theta)$ and $(B^Q,\nu)$ of the same type and 
prove that $\h_\cR'$ is an algebra homomorphism from $(\Pol(\Sigma,\B),\theta)$ to $(B^Q,\nu)$. The mappings $\theta$ and $\nu$ are $\Pol_\C(\Sigma,\B)$-indexed families over $\mathrm{Ops}^1(\Pol(\Sigma,\B))$ and $\mathrm{Ops}^1(B^Q)$, respectively, and they are defined, for each $r' \in \Pol_\C(\Sigma,\B)$, such that
\begin{align*}
  \theta(r') \in \mathrm{Ops}^1(\Pol(\Sigma,\B)) \ &\text{ with } \theta(r')(r) = r' \circ_z r \ \text{ for each $r \in \Pol(\Sigma,\B)$} \ \ \text{ and } \ \\
  \nu(r') \in \mathrm{Ops}^1(B^Q) \ &\text{ with } \nu(r')(v) = v \cdot \murev(r') \ \text{ for each $v \in B^Q$} \enspace.
\end{align*}
In order to show that $\h_\cR'$ is an algebra homomorphism from $(\Pol(\Sigma,\B),\theta)$ to $(B^Q,\nu)$, we prove  the following statement.
\begin{equation}\label{equ:consistency-for-extension-of-lambda-mu}
  \text{For every $r' \in \Pol_\C(\Sigma,\B)$, and $r \in \Pol(\Sigma,\B)$, we have
    $\h_\cR'(r' \circ_z r) = \h_\cR'(r) \cdot \murev(r')$.}
\end{equation}
For this, first let $c,c' \in \C_\Sigma$, $\xi \in \T_\Sigma$, and $\alpha \in \Sigma^{(0)}$ such that $\xi=c'[\alpha]$. Then:
\begingroup
\allowdisplaybreaks
\begin{align*}
  \h_\cR\Big(\1.(c \circ_z \xi)\Big) &=  \h_\cR\Big(\1.(c \circ_z c' \circ_z \alpha)\Big)
                           \tag{by associativity of $\circ_z$}\\
                         &= \lambda(\alpha) \cdot \murev(c \circ_z c')
                           \tag{by definition of $\h_\cR$}\\
  &= \lambda(\alpha) \cdot \murev(c') \cdot \murev(c)
                           \tag{because $\murev$ is a monoid homomorphism from $(\C_\Sigma,\circ_z,z)$ to $(B^{Q\times Q},\diamond,\mathrm{M}_\1)$}\\
                         &= \h_\cR(\1.c'[\alpha]) \cdot \murev(c)
                           \tag{by definition of $\h_\cR$}\\
  &= \h_\cR(\1.\xi) \cdot \murev(c) \enspace.
\end{align*}
\endgroup

Now let $r' \in \Pol_\C(\Sigma,\B)$ and $r \in \Pol(\Sigma,\B)$ with $r' = d_1.c_1 \oplus \ldots \oplus d_\ell.c_\ell$ and $r= b_1.\xi_1 \oplus \ldots \oplus b_m.\xi_m$. Then
\begingroup
\allowdisplaybreaks
\begin{align*}
  \h_\cR'(r' \circ_z r) &= \h_\cR'\Big( \bigoplus_{i\in [\ell], j \in [m]} (d_i \otimes b_j). (c_i \circ_z \xi_j)\Big)
  \tag{by definition of $\circ_z$}\\
                       &= \bigoplus_{i\in [\ell], j \in [m]} (d_i \otimes b_j) \cdot \h_\cR\Big(\1.(c_i \circ_z \xi_j)\Big)
   \tag{by definition of $\h_\cR'$}\\
                        &= \bigoplus_{i\in [\ell], j \in [m]} (d_i \otimes b_j) \cdot \Big(\h_\cR(\1.\xi_j) \cdot \murev(c_i)\Big)
 \tag{by the above calculation}\\
                        &= \bigoplus_{i\in [\ell], j \in [m]} (b_j \cdot \h_\cR(\1.\xi_j) \cdot (d_i \cdot \murev(c_i))
  \tag{by associativity and commutativity}\\
                        &= \Big(\bigoplus_{j\in [m]} b_j \cdot \h_\cR(\1.\xi_j)\Big) \cdot \Big(\bigoplus_{i\in [\ell]} d_i \cdot \murev(c_i)\Big)
                           \tag{by distributivity} \\
                        &= \h_\cR'(\bigoplus_{j\in [m]} b_j.\xi_j) \cdot \murev(\bigoplus_{i\in [\ell]} d_i.c_i)
  \tag{by definition of $\h_\cR'$ and the extension of $\murev$}\\
  &=  \h_\cR'(r) \cdot \murev(r') \enspace.
\end{align*}
\endgroup
This proves \eqref{equ:consistency-for-extension-of-lambda-mu}, i.e., $\h_\cR'$ is an algebra homomorphism from $(\Pol(\Sigma,\B),\theta)$ to $(B^Q,\nu)$. By Theorem~\ref{thm:kernel-is-congruence}, $\ker(\h_\cR')$ is a congruence relation on $(\Pol(\Sigma,\B),\theta)$.

 Now we define the $\Sigma$-algebra $(\im(\h_\cR'),\eta)$ such that,
  for every $k \in \mathbb{N}$, $\sigma \in \Sigma^{(k)}$, and $v_1,\ldots,v_k \in \im(\h_\cR')$, we let
    \begin{equation}\label{equ:definition-of-eta-field-is-needed}
\eta(\sigma)(v_1,\ldots,v_k) = \h_\cR'(\ttop_\Sigma(\sigma)(s_1,\ldots,s_k))
\end{equation}
where $s_1,\ldots,s_k \in \Pol(\Sigma,\B)$ are chosen such that $\h_\cR'(s_i) = v_i$ for each $i \in [k]$. Since $v_i \in \im(\h_\cR')$ such an $s_i$ exists.
Next we prove that $\eta$ is well defined (i.e., independent from the choice of the $s_i$'s)
by proving that $\ker(\h_\cR')$ is a congruence relation on the $\Sigma$-algebra $(\Pol(\Sigma,\B),\ttop_\Sigma)$.
To prove this, let $s_1,s'_1,\ldots,s_k,s'_k \in \Pol(\Sigma,\B)$ such that $\h_\cR'(s_i)=\h_\cR'(s'_i)$ for each $i\in[k]$. We show the following statement.
\begin{equation}\label{equ:one-by-one}
  \begin{aligned}
    &\text{For every $j \in [k]$, we have}\\
  &\h_\cR'\Big(\ttop_\Sigma(\sigma)(s_1',\ldots,s_{j-1}',s_j,s_{j+1},\ldots,s_k)\Big)=
  \h_\cR'\Big(\ttop_\Sigma(\sigma)(s'_1,\ldots,s'_{j-1},s_j',s_{j+1},\ldots,s_k)\Big) \enspace.
  \end{aligned}
\end{equation}
Let $j \in [k]$. Then we compute as follows:
\begingroup
\allowdisplaybreaks
\begin{align*}
  & \h_\cR'\Big(\ttop_\Sigma(\sigma)(s_1',\ldots,s_{j-1}',s_j,s_{j+1},\ldots,s_k)\Big)\\
  &=  \h_\cR'\Big(\ttop_\Sigma(\sigma)(s_1',\ldots,s_{j-1}',z,s_{j+1},\ldots,s_k) \circ_z s_j \Big)
  \tag{by an obvious generalization of the definition of $\ttop_\Sigma(\sigma)$; and by using commutativity}\\
  &=  \h_\cR'\Big(\theta\big( \ttop_\Sigma(\sigma)\big(s_1',\ldots,s_{j-1}',z,s_{j+1},\ldots,s_k)\big)(s_j) \Big)
  \tag{by definition of $\theta$}\\
  &= \h_\cR'\Big(\theta \big( \ttop_\Sigma(\sigma)\big(s_1',\ldots,s_{j-1}',z,s_{j+1},\ldots,s_k)\big)(s'_j )\Big)
  \tag{because $\ker(\h_\cR')$ is a congruence relation on the $\Sigma$-algebra  $(\Pol(\Sigma,\B),\theta)$}\\
  &= \h_\cR'\Big(\ttop_\Sigma(\sigma)(s_1',\ldots,s_{j-1}',s_j',s_{j+1},\ldots,s_k)\Big)
    \tag{as above}\enspace.
\end{align*}
\endgroup
This proves \eqref{equ:one-by-one}.
Using \eqref{equ:one-by-one} and the transitivity of equality, we obtain
\[
\h_\cR'(\ttop_\Sigma(\sigma)(s_1,\ldots,s_k))=
  \h_\cR'(\ttop_\Sigma(\sigma)(s'_1,\ldots,s_k')) \enspace.
  \]
Hence $\ker(\h_\cR')$ is a congruence relation on $(\Pol(\Sigma,\B),\ttop_\Sigma)$ and thus, $\eta(\sigma)$ is well defined (cf. Figure~\ref{fig:illustration-top-theta-nu-eta}).

  \begin{figure}[t]
  \begin{tabular}{ccc}
    $\Sigma$-algebras: & & algebras with $\Pol_\C(\Sigma,\B)$-indexed\\
    && families of operations: \\\hline 
    $(\Pol(\Sigma,\B),\ttop_\Sigma)$  & & $(\Pol(\Sigma,\B),\theta)$\\
    \\
    $\downarrow$ $\h_\cR'$ & & $\downarrow$  $\h_\cR'$\\
    \\
    $(\im(\h_\cR'),\eta)$ & & $(B^Q,\nu)$\\[5mm]
    $\h_\cR'\Big(\ttop_\Sigma(\sigma)(s_1',\ldots,s_{j-1}',s_j,s_{j+1},\ldots,s_k)\Big)$ & =
                         & $\h_\cR'\Big(\theta\big( \ttop_\Sigma(\sigma)\big(s_1',\ldots,s_{j-1}',z,s_{j+1},\ldots,s_k)\big)(s_j) \Big)$\\
    && $\downarrow$ = \\
                       && because $\ker(\h_\cR')$ is a congruence on $(\Pol(\Sigma,\B),\theta)$\\
                       && $\downarrow$ = \\
   $\h_\cR'\Big(\ttop_\Sigma(\sigma)(s_1',\ldots,s_{j-1}',s_j',s_{j+1},\ldots,s_k)\Big)$  & =
                         & $\h_\cR'\Big(\theta\big( \ttop_\Sigma(\sigma)\big(s_1',\ldots,s_{j-1}',z,s_{j+1},\ldots,s_k)\big)(s_j') \Big)$\\
    \end{tabular}
\caption{\label{fig:illustration-top-theta-nu-eta} Illustration of the proof of \eqref{equ:one-by-one}.}
  \end{figure}

  \

Now we can prove the following.
\begin{equation} \label{equ:etasigma-is-multilinear}
  \text{For each $\sigma \in\Sigma$, the operation $\eta(\sigma)$ is multilinear.}
\end{equation}
Let $k \in \mathbb{N}$, $\sigma \in \Sigma^{(k)}$, $i \in [k]$, $v_1,\ldots,v_k \in \im(\h_\cR')$, $b,b' \in B$, and $v,v' \in \im(\h_\cR')$. Then we can calculate as follows.
\begingroup
\allowdisplaybreaks
\begin{align*}
  &\eta(\sigma)(v_1,\ldots,v_{i-1},b\cdot v \oplus b' \cdot v',v_{i+1},\ldots,v_k)\\[2mm]
  &=\eta(\sigma)(\h_\cR'(s_1),\ldots,\h_\cR'(s_{i-1}),b\cdot \h_\cR'(s) \oplus b' \cdot \h_\cR'(s'),\h_\cR'(s_{i+1}),\ldots,\h_\cR'(s_k))
  \tag{where $\h_\cR'(s_j)=v_j$ for each $j \in [k]\setminus \{i\}$ and $\h_\cR'(s)=v$ and $\h_\cR'(s')=v'$}\\[2mm]
  &= \eta(\sigma)(\h_\cR'(s_1),\ldots,\h_\cR'(s_{i-1}),\h_\cR'(b\cdot s \oplus b' \cdot s'),\h_\cR'(s_{i+1}),\ldots,\h_\cR'(s_k))
  \tag{by linearity of $\h_\cR'$}\\[2mm]
  &= \h_\cR'\Big(\ttop_\Sigma(\sigma)(s_1,\ldots,s_{i-1}, b\cdot s \oplus b' \cdot s',s_{i+1},\ldots,s_k)\Big)
   \tag{by definition of $\eta(\sigma)$}\\
  &= \h_\cR'\Big(b \cdot \ttop_\Sigma(\sigma)(s_1,\ldots,s_{i-1}, s,s_{i+1},\ldots,s_k)
    \oplus b' \cdot \ttop_\Sigma(\sigma)(s_1,\ldots,s_{i-1}, s',s_{i+1},\ldots,s_k) \Big)
   \tag{by multilinearity of $\ttop_\Sigma(\sigma)$, cf. Corollary \ref{cor:top(sigma)-multilinear}}\\
  &= b \cdot \h_\cR'\Big(\ttop_\Sigma(\sigma)(s_1,\ldots,s_{i-1}, s,s_{i+1},\ldots,s_k)\Big)
    \oplus b' \cdot \h_\cR'\Big(\ttop_\Sigma(\sigma)(s_1,\ldots,s_{i-1}, s',s_{i+1},\ldots,s_k) \Big)
  \tag{by linearity of $\h_\cR'$}\\[2mm]
  &= b \cdot  \eta(\sigma)(v_1,\ldots,v_{i-1},v,v_{i+1},\ldots,v_k) \oplus
    b' \cdot \eta(\sigma)(v_1,\ldots,v_{i-1},v',v_{i+1},\ldots,v_k)
    \tag{by definition of $\h_\cR'$} \enspace.
\end{align*}
\endgroup
This proves \eqref{equ:etasigma-is-multilinear}, i.e., $\eta(\sigma)$ is multilinear.

So far we have obtained the following.

\begin{equation} \label{equ:imhR-Sigma-B-vector-space}
  \text{The tuple $\V_\cR= (\im(\h_\cR'),\oplus,\widetilde{\0},\eta)$ is a $(\Sigma,\B)$-semimodule.}
\end{equation}

\

As final preparation, we prove the following statement.
\begin{equation} \label{equ:hR-Sigma-algebra-hom}
  \begin{aligned}
    &\text{The mapping $\h_\cR'$ is the unique $(\Sigma,\B)$-semimodule homomorphism}\\
  &\text{from the initial $(\Sigma,\B)$-semimodule $(\Pol(\Sigma,\B),\oplus,\widetilde{\0},\ttop_\Sigma)$ }\\
  &\text{to the $(\Sigma,\B)$-semimodule $\V_\cR=(\im(\h_\cR'),\oplus,\widetilde{\0},\eta)$.}
\end{aligned}
\end{equation}
Since $\h_\cR'$ is linear, it suffices to prove that it is a $\Sigma$-algebra homomorphism from $(\Pol(\Sigma,\B),\ttop_\Sigma)$ to $(\im(\h_\cR'),\eta)$. This follows from the definition of 
$\eta(\sigma)$ because, for every $k\in \mathbb{N}$, $\sigma \in \Sigma^{(k)}$, and $s_1,\ldots,s_k \in \Pol(\Sigma,\B)$, we have 
\begingroup
\allowdisplaybreaks
\begin{align*}
  \h_\cR'(\ttop_\Sigma(\sigma)(s_1,\ldots,s_k)) = \eta(\sigma)(\h_\cR'(s_1),\ldots,\h_\cR'(s_k)) \enspace.
  \end{align*}
  \endgroup
  This proves \eqref{equ:hR-Sigma-algebra-hom}.

  Then,  for every $k \in \mathbb{N}$, $\sigma \in \Sigma^{(k)}$, and $\xi_1,\ldots,\xi_k \in \T_\Sigma$ we have:
 \begingroup
\allowdisplaybreaks
\begin{align*}
  &\h_\cR(\ttop_\Sigma(\sigma)(\1.\xi_1,\ldots,\1.\xi_k)) = \h_\cR(\1.\sigma(\xi,\ldots,\xi_k))\\
  &= \h'_\cR(\1.\sigma(\xi,\ldots,\xi_k)) = \h_\cR'(\ttop_\Sigma(\sigma)(\1.\xi_1,\ldots,\1.\xi_k))\\
  &= \eta(\sigma)(\h_\cR'(\1.\xi_1),\ldots,\h_\cR'(\1.\xi_k))
  \tag{by \eqref{equ:hR-Sigma-algebra-hom}}\\
      &= \eta(\sigma)(\h_\cR(\1.\xi_1),\ldots,\h_\cR(\1.\xi_k)) \enspace.
    \end{align*}
  \endgroup  
Thus, $\h_\cR: \1.\T_\Sigma \to B^Q$ is a $\Sigma$-algebra homomorphism from the $\Sigma$-algebra $(\1.\T_\Sigma,\ttop_\Sigma)$ to the $\Sigma$-algebra $(\im(\h_\cR'),\eta)$.

We recall that $\h_\eta$ is the unique $\Sigma$-algebra homomorphism from $\sfT_\Sigma$ to $(\im(\h_\cR'),\eta)$. Since the $\Sigma$-algebras $\sfT_\Sigma$ and  $(\1.\T_\Sigma,\ttop_\Sigma)$ are isomorphic,  we have $\h_\cR(\1.\xi) = \h_\eta(\xi)$ for each $\xi \in \T_\Sigma$. Thus, for each $\xi \in \T_\Sigma$ with $\xi=c[\alpha]$, we have
\[
  \sem{\cR}(\xi) = \lambda(\alpha) \cdot \murev(c) \cdot \gamma = \h_\cR(\1.\xi) \cdot \gamma
  = \h_\eta(\xi) \cdot \gamma \enspace.
\]

We define the linear form $\gamma'$ over $\V$ by $\gamma'(v) = v \cdot \gamma$ for each $v \in \im(\h_\cR')$ (scalar multiplication of the vectors $v$ and $\gamma$). Then $\sem{\cR}(\xi) = \gamma'(\h_\eta(\xi))$.

\

Proof of (2): Let $\V_\cR=(\V,\eta)$ be the $(\Sigma,\B)$-semimodule as constructed in the proof of (1). We recall that $\V$  is a sub-semimodule of $(B^Q,\oplus,\widetilde{\0})$. Since $\B$ is a field, $(B^Q,\oplus,\widetilde{\0})$ is a $|Q|$-dimensional $\B$-vector space and, by Theorem~\ref{thm:subspace-of-finite-dim-vector-space}, $\V$ is a finite-dimensional $\B$-vector space. Let $H=\{u_1,\ldots,u_m\}$ be a  basis of $\V$. (We note that $H \subseteq V = \im(\h_\cR') \subseteq B^Q$.) Thus, for each $v \in V$, there exist unique $b_1,\ldots,b_m \in B$ such that $v = b_1 \cdot u_1 \oplus \ldots \oplus b_m \cdot u_m$. We abbreviate $b_i$ by $v_{u_i}$ for each $i \in [m]$. Also, let  $\gamma'$ be the linear form over $\V$ as constructed in the proof of (1).

We construct the $(\Sigma,\B)$-wta $\cA=(H,\delta,F)$ where
\begin{compactitem}
  \item for every $k \in \mathbb{N}$, $\sigma \in \Sigma^{(k)}$, and $v_1,\ldots,v_k,v \in H$, we let 
\[
\delta_k(v_1 \cdots v_k,\sigma,v) = \eta(\sigma)(v_1,\ldots,v_k)_{v} \enspace,
\]
and
\item $F : H \to B$ is the mapping  defined by $F_v = \gamma'(v)$ for each $v \in H$.
\end{compactitem}

\
  
By induction on $\T_\Sigma$, we prove the following statement.
\begin{equation}\label{obs:fin-dim-vector-space-ml-representation-semantics-H}
\text{For every $\xi \in \T_\Sigma$ and $v\in H$, we have $\h_{\cA}(\xi)_v= \h_\eta(\xi)_v$.}
  \end{equation}

  Let $\xi = \sigma(\xi_1,\ldots,\xi_k)$. Then we can calculate as follows.
  \begingroup
  \allowdisplaybreaks
  \begin{align*}
    & \h_\cA(\sigma(\xi_1,\ldots,\xi_k))_v\\
    &=  \bigoplus_{v_1\cdots v_k\in H^k}\Big(\bigotimes_{i\in[k]}\h_{\cA}(\xi_i)_{v_i}\Big) \otimes \delta_k(v_1\cdots v_k,\sigma,v)\\
    &=  \bigoplus_{v_1\cdots v_k\in H^k}\Big(\bigotimes_{i\in[k]}\h_\eta(\xi_i)_{v_i}\Big) \otimes \eta(\sigma)(v_1,\ldots,v_k)_{v}
    \tag{by I.H. and the definition of $\delta_k$ }\\
    &= \bigoplus_{v_1\cdots v_k\in H^k} \eta(\sigma)\big(\h_\eta(\xi_1)_{v_1}\cdot v_1,\ldots,\h_\eta(\xi_k)_{v_k}\cdot v_k\big)_v
    \tag{by \eqref{equ:etasigma-is-multilinear} $\eta(\sigma)$ is multilinear} \\
    &= \eta(\sigma)\big(\bigoplus_{v_1\in H}\h_\eta(\xi_1)_{v_1}\cdot v_1,\ldots,\bigoplus_{v_k\in H}\h_\eta(\xi_k)_{v_k}\cdot v_k\big)_v
    \tag{by \eqref{equ:etasigma-is-multilinear} $\eta(\sigma)$ is multilinear}\\
    &= \eta(\sigma)\big(\h_\eta(\xi_1),\ldots,\h_\eta(\xi_k)\big)_v
    \\
    &= \h_\eta(\xi)_v
    \tag{$\h_\eta$ is a homomorphism}\enspace.
      \end{align*}
  \endgroup 
  This proves \eqref{obs:fin-dim-vector-space-ml-representation-semantics-H}. Now let $\xi \in \T_\Sigma$.   Then 
    \begingroup
    \allowdisplaybreaks
    \begin{align*}
      \sem{\cA}(\xi) &= \bigoplus_{v\in H}\h_\cA(\xi)_v\otimes F_v\\
      &= \bigoplus_{v\in H} \h_\eta(\xi)_v \otimes \gamma'(v)
      \tag{by \ref{obs:fin-dim-vector-space-ml-representation-semantics-H} and the definition of $F$} \\
      &=\bigoplus_{v\in H} \gamma'(\h_\eta(\xi)_v \cdot v)
      \tag{because $\gamma'$ is a linear form over $\V$}\\
      &= \gamma'\big( \bigoplus_{v\in H} \h_\eta(\xi)_v \cdot v\big)
      \tag{because $\gamma'$ is a linear form over $\V$}\\
      &= \gamma'\big( \h_\eta(\xi)\big)\enspace.
    \end{align*}
    \endgroup
\end{proof}

We note that, in the proof of Theorem \ref{theo:from-monoid-rep-to-wta-comm-semiring}(2), we could have taken an alternative way which, informally, consiststs of two steps. First, a $(\Sigma,\B)$-multilinear representation $\cM$ is constructed such that $\sem{\cM}=\gamma' \circ \h_\eta$. Second, by using Theorem~\ref{thm:lin-iff-wta-extended}(A)$\Rightarrow$(B), a $(\Sigma,\B)$-wta $\cA$ can be constructed such that $\sem{\cA} = \sem{\cM}$.

More precisely, let $H= \{u_1,\ldots,u_m\}$ be a finite basis of $\V=(\im(\h_\cR'),\oplus,\widetilde{\0})$. Moreover, let $\psi : \im(\h_\cR') \to B^H $ be the isomorphism from $\V$ to the $\B$-vector space  $(B^H,\oplus,\widetilde{\0})$ as defined on page~\pageref{page:representing-vector-spaces}.
Then we construct the $(\Sigma,\B)$-multilinear representation $\cM=(H,\eta^*,\gamma^*)$, where
\begin{compactitem}
  \item for every $k \in \mathbb{N}$, $\sigma \in \Sigma^{(k)}$, and $v_1,\ldots,v_k \in B^H$, we let 
\[
\eta^*(\sigma)(v_1,\ldots,v_k) = \psi\big(\eta(\sigma)(\psi^{-1}(v_1),\ldots,\psi^{-1}(v_k))\big) \enspace,
\]
and
\item $\gamma^*: B^H \to B$ is a mapping  defined by $\gamma^*(v) =   \gamma'(\psi^{-1}(v))$ for each $v \in B^H$.
\end{compactitem}
We can show that $\sem{\cM}=\gamma'\circ \h_\eta$. Then, by  Theorem~\ref{thm:lin-iff-wta-extended}(A)$\Rightarrow$(B), we can construct a $(\Sigma,\B)$-wta $\cA$ such that $\sem{\cA} = \sem{\cM}$.

\begin{example}\label{ex:example-for-not-finitely-generated}\rm We consider the consistent $(\Sigma,\Nat)$-monoid representation $\cR=(Q,\lambda,\mu,\gamma)$ with $Q= \{\bot,a,f\}$ of Example~\ref{ex:monoid-rep-number-of-occ-of-pattern} and we view $\cR$ as a $(\Sigma,\Ratnum)$-monoid representation.
  We will construct a $(\Sigma,\Ratnum)$-wta $\cB=(H,\delta,F)$ with  $\sem{\cR}= \sem{\cB}$, according to the proof of Theorem \ref{theo:from-monoid-rep-to-wta-comm-semiring}.
  
First,  for an arbitrary $\xi\in \T_\Sigma$, we give $\h'_\cR(1.\xi) \in \mathbb{Q}^Q$
in an explicit form. For this, let $\xi\in \T_\Sigma$ and 
  $c \in \C_\Sigma$ such that $\xi=c[\alpha]$. Since $\murev(z) = \mathrm{M}_1$, by \eqref{equ:lambda-murev-for-example-mon-rep} and \eqref{equ:number-of-occ-of-pattern-in-context-alpha}, we have
  \begin{align}
  & \h'_\cR(1.\xi)= \h'_\cR(1.c[\alpha]) = \h_\cR(1.c[\alpha]) = \lambda(\alpha) \cdot \murev(c) \label{equ:h-R}\\[2mm] 
  &  = 
  \begin{cases}
    (1 \ 1 \ 0) & \\
    ( 1 \  0 \  \#_{\sigma(.,\alpha)}(c) + b(c)) & 
    \end{cases}
    =
    \begin{cases}
    (1 \ 1 \ 0) & \\
    ( 1 \ 0 \ \#_{\sigma(.,\alpha)}(c[\alpha])) & 
    \end{cases}
    =
    \begin{cases}
    (1 \ 1 \ 0) & \text{ if $c=z$}\\
    (1 \  0 \  \#_{\sigma(.,\alpha)}(\xi) ) & \text{ otherwise,}\nonumber
    \end{cases}
  \end{align}
  where the components are ordered by the sequence $(\bot,a,f)$ and $b(c) = 1$ if there exists a position $w \in \pos(c)$ such that $z = c(w2)$, and $0$ otherwise. Hence, for each $\xi \in \T_\Sigma$, we have $\h'_\cR(1.\xi)= \h_\cA(\xi)$ where $\cA$ is the $(\Sigma,\Nat)$-wta of Example~\ref{ex:number-of-occurrences}.

Next we give a basis of the $\Ratnum$-vector space $\V=(\im(\h_\cR'),+,0^Q)$. 
Formula \eqref{equ:h-R} for trees $\xi=\alpha$, $\xi=\omega(\alpha)$, and $\xi=\sigma(\alpha,\alpha)$
shows that the vectors  $(1 \ 1 \ 0)$, $(1 \ 0 \ 0)$, and $(1 \ 0 \ 1)$, respectively, are in $\im(\h_\cR')$. These three vectors are linearly independent. To see this, let $b_1,b_2,b_3\in \mathbb{Q}$ such that
\[b_1\cdot(1 \ 0 \ 0) + b_2\cdot(1 \ 1 \ 0) + b_3\cdot (1 \ 0 \ 1)= ( 0 \ 0 \ 0).\]
This implies $b_1+b_2+b_3=0$, $b_2=0$, and $b_3=0$, i.e., that $b_1=b_2=b_3=0$. Moreover,  $\dim(\V)\le 3$ because 
$\V$ is a subspace of the 3-dimensional vector space $(\mathbb{Q}^Q,+,0_Q)$ (cf. Theorem~\ref{thm:subspace-of-finite-dim-vector-space}). Hence $\dim(\V)= 3$
and, by Theorem~\ref{thm:n-lin-independent-vectors=basis}, $H=\{(1 \ 0 \ 0), (1 \ 1 \ 0), (1 \ 0 \ 1)\}$ is a basis of $\V$.

Next we do not construct the $\Sigma$-algebra $(\im(\h_\cR'),\eta)$ but the $(\Sigma,\Ratnum)$-wta $\cB=(H,\delta,F)$ immediately.
By the definition of $\delta$ and by using \eqref{equ:definition-of-eta-field-is-needed} and \eqref{equ:h-R}, we give $\delta$ for each $v\in H$ as follows.

For $\alpha$ we have:
\begin{align*}
  \delta_0(\varepsilon,\alpha,v)
  = \h_\cR'(1.\alpha)_v
  = (1 \ 1 \ 0)_v
  = \begin{cases}
    1 & \text{ if $v=  (1 \ 1 \ 0)$}\\
    0 & \text{ otherwise} \enspace.
    \end{cases}
\end{align*}

For $\omega$ we have:
\begin{align*}
  \delta_1((1 \ 0 \ 0),\omega,v)
  = \h_\cR'(1.\omega(\omega( \alpha)))_v
  = (1 \ 0 \ 0)_v
= \begin{cases}
    1 & \text{ if $v=  (1 \ 0 \ 0)$}\\
       0 & \text{ otherwise} \enspace.
    \end{cases}
\end{align*}

\begin{align*}
  \delta_1((1 \ 1 \ 0),\omega,v)
  = \h_\cR'(1.\omega(\alpha))_v
  = (1 \ 0 \ 0)_v
 =  \begin{cases}
    1 & \text{ if $v=  (1 \ 0 \ 0)$}\\
       0 & \text{ otherwise} \enspace.
    \end{cases}
\end{align*}

\begin{align*}
  \delta_1((1 \ 0 \ 1),\omega,v)
  = \h_\cR'(1.\omega(\sigma(\alpha,\alpha)))_v
  = (1 \ 0 \ 1)_v
= \begin{cases}
    1 & \text{ if $v=  (1 \ 0 \ 1)$}\\
       0 & \text{ otherwise} \enspace.
    \end{cases}
\end{align*}

For $\sigma$ we have:
\begin{align*}
  \delta_2((1 \ 0 \ 0) (1 \ 0 \ 0), \sigma,v)
  = \h_\cR'(1.\sigma(\omega(\alpha),\omega(\alpha)))_v
  = (1 \ 0 \ 0)_v
  =   \begin{cases}
    1 & \text{ if $v=  (1 \ 0 \ 0)$}\\
     0 & \text{ otherwise} \enspace.
    \end{cases}
\end{align*}

\begin{align*}
  \delta_2((1 \ 0 \ 0) (1 \ 1 \ 0), \sigma,v)
  = \h_\cR'(1.\sigma(\omega(\alpha),\alpha))_v
  = (1 \ 0 \ 1)_v
  =   \begin{cases}
    1 & \text{ if $v=  (1 \ 0 \ 1)$}\\
     0 & \text{ otherwise} \enspace.
    \end{cases}
\end{align*}

\begin{align*}
  \delta_2((1 \ 0 \ 0) (1 \ 0 \ 1), \sigma,v)
  = \h_\cR'(1.\sigma(\omega(\alpha),\sigma(\alpha,\alpha)))_v
  = (1 \ 0 \ 1)_v
  =   \begin{cases}
    1 & \text{ if $v=  (1 \ 0 \ 1)$}\\
     0 & \text{ otherwise} \enspace.
    \end{cases}
\end{align*}

\begin{align*}
  \delta_2((1 \ 1 \ 0) (1 \ 0 \ 0), \sigma,v)
  = \h_\cR'(1.\sigma(\alpha,\omega(\alpha)))_v
  = (1 \ 0 \ 0)_v
  =   \begin{cases}
    1 & \text{ if $v=  (1 \ 0 \ 0)$}\\
     0 & \text{ otherwise} \enspace.
    \end{cases}
\end{align*}

\begin{align*}
 \delta_2((1 \ 1 \ 0) (1 \ 1 \ 0), \sigma,v) = \h_\cR'(1.\sigma(\alpha,\alpha))_v = (1 \ 0 \ 1)_v =
  \begin{cases}
    1 & \text{ if $v=  (1 \ 0 \ 1)$}\\
     0 & \text{ otherwise} \enspace.
    \end{cases}
\end{align*}

\begin{align*}
  \delta_2((1 \ 1 \ 0) (1 \ 0 \ 1), \sigma,v)
  = \h_\cR'(1.\sigma(\alpha,\sigma(\alpha,\alpha)))_v
  = (1 \ 0 \ 1)_v
  =   \begin{cases}
    1 & \text{ if $v=  (1 \ 0 \ 1)$}\\
     0 & \text{ otherwise} \enspace.
    \end{cases}
\end{align*}

\begin{align*}
  \delta_2((1 \ 0 \ 1) (1 \ 0 \ 0), \sigma,v)
  = \h_\cR'(1.\sigma(\sigma(\alpha,\alpha),\omega(\alpha)))_v
  = (1 \ 0 \ 1)_v
  =   \begin{cases}
    1 & \text{ if $v=  (1 \ 0 \ 1)$}\\
     0 & \text{ otherwise} \enspace.
    \end{cases}
\end{align*}

\begin{align*}
  \delta_2((1 \ 0 \ 1) (1 \ 1 \ 0), \sigma,v)
  = \h_\cR'(1.\sigma(\sigma(\alpha,\alpha),\alpha))_v
  = ( 1 \ 0 \ 2)_v
  =   \begin{cases}
    -1 & \text{ if $v=  (1 \ 0 \ 0)$}\\
    2 & \text{ if $v=  (1 \ 0 \ 1)$}\\
     0 & \text{ otherwise} \enspace.
    \end{cases}
\end{align*}

\begin{align*}
  \delta_2((1 \ 0 \ 1) (1 \ 0 \ 1), \sigma,v)
  = \h_\cR'(1.\sigma(\sigma(\alpha,\alpha),\sigma(\alpha,\alpha)))_v
  = ( 1 \ 0 \ 2)_v
  =   \begin{cases}
    -1 & \text{ if $v=  (1 \ 0 \ 0)$}\\
    2 & \text{ if $v=  (1 \ 0 \ 1)$}\\
     0 & \text{ otherwise} \enspace.
    \end{cases}
\end{align*}

We recall that $\gamma = ( 0 \ 0 \ 1)$ and that $\gamma'(v)=v\cdot \gamma$ for each $v\in \im(\h'_\cR)$.
Thus, we obtain $F_{(1 \ 0 \ 0)}= \gamma'\big((1 \ 0 \ 0)\big)=(1 \ 0 \ 0)\cdot (0 \ 0 \ 1) =0$. Similarly, we have $F_{(1 \ 1 \ 0)}= 0$ and $F_{(1 \ 0 \ 1)}= 1$.

This finishes the definition of the $(\Sigma,\Ratnum)$-wta $\cB$. Obviously, we can view $\cB$ also as a $(\Sigma,\Int)$-wta (cf. Observation \ref{obs:extension-of-weight-structure-inverse}).

\

By induction on $\sfT_\Sigma$ we prove the following statement:
\begin{equation}\label{equ:hA-for-wta-constructed-from-monoid-rep}
  \begin{aligned}
    \text{For each $\xi \in \T_\Sigma$, we have }  \ \
    \h_\cB(\xi) = \begin{matrix} {\tiny (1 \ 0 \ 0)} \\ {\tiny (1 \ 1 \ 0)} \\ {\tiny (1 \ 0 \ 1)} \end{matrix}
\left(
\begin{matrix} f(\xi) \\  g(\xi) \\ \#_{\sigma(.,\alpha)}(\xi)\end{matrix}
\right)
  \end{aligned}
  \end{equation}
where
\[
g(\xi) = \begin{cases}
  1 & \text{ if $\xi = \alpha$}\\
  0 & \text{ otherwise}
  \end{cases} \  \text{ and } \ \ 
f(\xi) =
\begin{cases}
  0 & \text{ if $\xi = \alpha$}\\
  1 -\#_{\sigma(.,\alpha)}(\xi) & \text{ otherwise.}
\end{cases} 
\]

Let $\xi \in \T_\Sigma$.

I.B.: Let $\xi=\alpha$. Then $\h_\cB(\xi)= \left(\begin{matrix} 0 \\  1 \\ 0 \end{matrix}\right)$.

I.S.: We proceed by case analysis.

\underline{Case (a):} Let $\xi= \omega(\xi_1)$ for some $\xi_1 \in \T_\Sigma$. Then 
$\h_\cB(\xi)
= \left(\begin{matrix} f(\xi_1)+g(\xi_1) \\ 0 \\  \#_{\sigma(.,\alpha)}(\xi_1) \end{matrix}\right)
=\left(\begin{matrix} f(\xi_1)+g(\xi_1)  \\ g(\xi) \\  \#_{\sigma(.,\alpha)}(\xi) \end{matrix}\right)$ and
\[
f(\xi_1)+g(\xi_1)
= \begin{cases}
  1 & \text{ if $\xi_1=\alpha$}\\
  1- \#_{\sigma(.,\alpha)}(\xi_1) & \text{ otherwise}
  \end{cases} \ \ 
= 1-\#_{\sigma(.,\alpha)}(\xi_1)
= 1-\#_{\sigma(.,\alpha)}(\xi)
= f(\xi)\enspace.
\]

\underline{Case (b):} Let $\xi= \sigma(\xi_1,\xi_2)$ for some $\xi_1,\xi_2 \in \T_\Sigma$. We show that $\h_\cB(\xi)_{(1 \ 0 \ 0)}=f(\xi)$.
\begingroup
\allowdisplaybreaks
\begin{align*}
  \h_\cB(\xi)_{(1 \ 0 \ 0)} =& \ \ \
  \h_\cB(\xi_1)_{(1 \ 0 \ 0)} \cdot \h_\cB(\xi_2)_{(1 \ 0 \ 0)} \cdot \delta_2((1 \ 0 \ 0)(1 \ 0 \ 0),\sigma,(1 \ 0 \ 0)) \\
  &    +  \h_\cB(\xi_1)_{(1 \ 1 \ 0)} \cdot \h_\cB(\xi_2)_{(1 \ 0 \ 0)} \cdot \delta_2((1 \ 1 \ 0)(1 \ 0 \ 0),\sigma,(1 \ 0 \ 0))\\
  &   + \h_\cB(\xi_1)_{(1 \ 0 \ 1)} \cdot \h_\cB(\xi_2)_{(1 \ 1 \ 0)} \cdot \delta_2((1 \ 0 \ 1)(1 \ 1 \ 0),\sigma,(1 \ 0 \ 0))\\
  &   + \h_\cB(\xi_1)_{(1 \ 0 \ 1)} \cdot \h_\cB(\xi_2)_{(1 \ 0 \ 1)} \cdot \delta_2((1 \ 0 \ 1)(1 \ 0 \ 1),\sigma,(1 \ 0 \ 0))\\[2mm]
  =& \ \ \ f(\xi_1) \cdot f(\xi_2) \cdot 1 \\
  &   + g(\xi_1) \cdot f(\xi_2) \cdot 1\\
  &   + \#_{\sigma(.,\alpha)}(\xi_1) \cdot g(\xi_2) \cdot (-1)\\
  &   + \#_{\sigma(.,\alpha)}(\xi_1) \cdot \#_{\sigma(.,\alpha)}(\xi_2) \cdot (-1)
  \enspace.
\end{align*}
\endgroup
We proceed by case analysis.

\underline{Case (b1):} Let $\xi_1=\alpha$. Then $g(\xi_1)=1$, $f(\xi_1)=0$, and $\#_{\sigma(.,\alpha)}(\xi_1)=0$. Hence 
\(\h_\cB(\xi)_{(1 \ 0 \ 0)} = f(\xi_2)\). 

We continue with distinguishing subcases.

\underline{Case (b11):} Let $\xi_2=\alpha$. Then $f(\xi_2) = 0 = 1- \#_{\sigma(.,\alpha)}(\sigma(\alpha,\alpha)) = f(\sigma(\alpha,\alpha)) = f(\xi)$.

\underline{Case (b12):} Let $\xi_2\ne \alpha$. Then $f(\xi_2)= 1- \#_{\sigma(.,\alpha)}(\xi_2) = 1- \#_{\sigma(.,\alpha)}(\sigma(\alpha,\xi_2)) = f(\sigma(\alpha,\xi_2)) = f(\xi)$.

This finishes the proof for Case (b1).

\

\underline{Case (b2):} Let $\xi_1\ne \alpha$. Then $g(\xi_1)=0$ and $f(\xi_1)=1-\#_{\sigma(.,\alpha)}(\xi_1)$.
Then
\[\h_\cB(\xi)_{(1 \ 0 \ 0)} = (1-\#_{\sigma(.,\alpha)}(\xi_1)) \cdot f(\xi_2)
+ \#_{\sigma(.,\alpha)}(\xi_1) \cdot g(\xi_2) \cdot (-1)
    + \#_{\sigma(.,\alpha)}(\xi_1) \cdot \#_{\sigma(.,\alpha)}(\xi_2) \cdot (-1)
\enspace.
\]
We continue again with subcases.

\underline{Case (b21):} Let $\xi_2=\alpha$. Then $g(\xi_2)=1$, $f(\xi_2)=0$, and $\#_{\sigma(.,\alpha)}(\xi_2)=0$ and thus
\[\h_\cB(\xi)_{(1 \ 0 \ 0)}
= - \#_{\sigma(.,\alpha)}(\xi_1)
= 1 - (1 + \#_{\sigma(.,\alpha)}(\xi_1))
= 1 - \#_{\sigma(.,\alpha)}(\sigma(\xi_1,\alpha))
= f(\sigma(\xi_1,\alpha))= f(\xi)
 \enspace.
\]

\underline{Case (b22):} Let $\xi_2\ne \alpha$. Then $g(\xi_2)=0$ and $f(\xi_2) = 1 - \#_{\sigma(.,\alpha)}(\xi_2)$ and thus
\begingroup
\allowdisplaybreaks
\begin{align*}
  \h_\cB(\xi)_{(1 \ 0 \ 0)} &= (1-\#_{\sigma(.,\alpha)}(\xi_1)) \cdot (1 - \#_{\sigma(.,\alpha)}(\xi_2))
  + \#_{\sigma(.,\alpha)}(\xi_1) \cdot \#_{\sigma(.,\alpha)}(\xi_2) \cdot (-1)\\
  &= 1
  - \#_{\sigma(.,\alpha)}(\xi_1)
  - \#_{\sigma(.,\alpha)}(\xi_2)
  + \#_{\sigma(.,\alpha)}(\xi_1) \cdot \#_{\sigma(.,\alpha)}(\xi_2)
  -\#_{\sigma(.,\alpha)}(\xi_1) \cdot \#_{\sigma(.,\alpha)}(\xi_2)\\
  &=  1
  - \#_{\sigma(.,\alpha)}(\xi_1)
  - \#_{\sigma(.,\alpha)}(\xi_2)\\
  &=  1
  - \#_{\sigma(.,\alpha)}(\xi) = f(\xi) 
  \enspace.
\end{align*}
\endgroup
This finishes the proof for Case (b2). Hence $\h_\cB(\xi)_{(1 \ 0 \ 0)}=f(\xi)$.

\

Next we show that $\h_\cB(\xi)_{(1 \ 1 \ 0)}=g(\xi)$ as follows.
Since there do not exist states $q_1,q_2 \in H$ such that $\delta_2(q_1q_2,\sigma,(1 \ 1 \ 0)) \ne 0$, we have that 
$\h_\cB(\xi)_{(1 \ 1 \ 0)}=0$. Since $\xi=\sigma(\xi_1,\xi_2)$, we have $g(\xi)=0$.
Thus $\h_\cB(\xi)_{(1 \ 1 \ 0)}=0=g(\xi)$.

\

Lastly, we show that $\h_\cB(\xi)_{(1 \ 0 \ 1)}=\#_{\sigma(.,\alpha)}(\xi)$ as follows.
\begingroup
\allowdisplaybreaks
\begin{align*}
  \h_\cB(\xi)_{(1 \ 0 \ 1)} =& \ \ \
   \h_\cB(\xi_1)_{(1 \ 0 \ 0)} \cdot \h_\cB(\xi_2)_{(1 \ 1 \ 0)} \cdot \delta_2((1 \ 0 \ 0)(1 \ 1 \ 0),\sigma,(1 \ 0 \ 1))\\  
  &+ \h_\cB(\xi_1)_{(1 \ 0 \ 0)} \cdot \h_\cB(\xi_2)_{(1 \ 0 \ 1)} \cdot \delta_2((1 \ 0 \ 0)(1 \ 0 \ 1),\sigma,(1 \ 0 \ 1))\\
  &+  \h_\cB(\xi_1)_{(1 \ 1 \ 0)} \cdot \h_\cB(\xi_2)_{(1 \ 1 \ 0)} \cdot \delta_2((1 \ 1 \ 0)(1 \ 1 \ 0),\sigma,(1 \ 0 \ 1))\\
  &+ \h_\cB(\xi_1)_{(1 \ 1 \ 0)} \cdot \h_\cB(\xi_2)_{(1 \ 0 \ 1)} \cdot \delta_2((1 \ 1 \ 0)(1 \ 0 \ 1),\sigma,(1 \ 0 \ 1))\\
  &+ \h_\cB(\xi_1)_{(1 \ 0 \ 1)} \cdot \h_\cB(\xi_2)_{(1 \ 0 \ 0)} \cdot \delta_2((1 \ 0 \ 1)(1 \ 0 \ 0),\sigma,(1 \ 0 \ 1))\\  
  &+ \h_\cB(\xi_1)_{(1 \ 0 \ 1)} \cdot \h_\cB(\xi_2)_{(1 \ 1 \ 0)} \cdot \delta_2((1 \ 0 \ 1)(1 \ 1 \ 0),\sigma,(1 \ 0 \ 1))\\  
  &+ \h_\cB(\xi_1)_{(1 \ 0 \ 1)} \cdot \h_\cB(\xi_2)_{(1 \ 0 \ 1)} \cdot \delta_2((1 \ 0 \ 1)(1 \ 0 \ 1),\sigma,(1 \ 0 \ 1))\\[2mm]
  =& \ \ \ f(\xi_1) \cdot g(\xi_2) \cdot 1\\  
  &+ f(\xi_1) \cdot \#_{\sigma(.,\alpha)}(\xi_2) \cdot 1\\
  &+ g(\xi_1) \cdot g(\xi_2) \cdot 1\\
  &+ g(\xi_1) \cdot \#_{\sigma(.,\alpha)}(\xi_2) \cdot 1\\
  &+ \#_{\sigma(.,\alpha)}(\xi_1) \cdot f(\xi_2) \cdot 1\\  
  &+ \#_{\sigma(.,\alpha)}(\xi_1) \cdot g(\xi_2) \cdot 2\\  
  &+ \#_{\sigma(.,\alpha)}(\xi_1) \cdot \#_{\sigma(.,\alpha)}(\xi_2) \cdot 2 \enspace.
\end{align*}
\endgroup
We proceed by case analysis.

\underline{Case (b1):} Let $\xi_1=\alpha$. Then $g(\xi_1)=1$, $f(\xi_1)=0$, and $\#_{\sigma(.,\alpha)}(\xi_1)=0$. Hence 
\(\h_\cB(\xi)_{(1 \ 0 \ 1)} = g(\xi_2) + \#_{\sigma(.,\alpha)}(\xi_2) = \#_{\sigma(.,\alpha)}(\xi)\).

\underline{Case (b2):} Let $\xi_1\ne \alpha$. Then $g(\xi_1)=0$ and $f(\xi_1)=1-\#_{\sigma(.,\alpha)}(\xi_1)$.
Hence 
\begin{align*}
  \h_\cB(\xi)_{(1 \ 0 \ 1)} &= (1-\#_{\sigma(.,\alpha)}(\xi_1)) \cdot g(\xi_2)
  + (1-\#_{\sigma(.,\alpha)}(\xi_1)) \cdot \#_{\sigma(.,\alpha)}(\xi_2)
  + \#_{\sigma(.,\alpha)}(\xi_1) \cdot f(\xi_2)\\
  & + \#_{\sigma(.,\alpha)}(\xi_1) \cdot g(\xi_2) \cdot 2 
  + \#_{\sigma(.,\alpha)}(\xi_1) \cdot \#_{\sigma(.,\alpha)}(\xi_2) \cdot 2   \enspace.
  \end{align*}

We continue with two subcases. 

\underline{Case (b21):} Let $\xi_2= \alpha$. Then $g(\xi_2)=1$, $f(\xi_2)=0$, and $\#_{\sigma(.,\alpha)}(\xi_2)=0$. Thus
\begin{align*}
  \h_\cB(\xi)_{(1 \ 0 \ 1)} &=  (1-\#_{\sigma(.,\alpha)}(\xi_1))
  + \#_{\sigma(.,\alpha)}(\xi_1) \cdot 2 
  = \#_{\sigma(.,\alpha)}(\xi_1) + 1
  = \#_{\sigma(.,\alpha)}(\xi)\enspace.
\end{align*}

\underline{Case (b22):}  Let $\xi_2\ne \alpha$. Then $g(\xi_2)=0$ and $f(\xi_2)=1-\#_{\sigma(.,\alpha)}(\xi_2)$. Thus
\begin{align*}
  \h_\cB(\xi)_{(1 \ 0 \ 1)} &=  (1-\#_{\sigma(.,\alpha)}(\xi_1)) \cdot \#_{\sigma(.,\alpha)}(\xi_2)
  + \#_{\sigma(.,\alpha)}(\xi_1) \cdot (1-\#_{\sigma(.,\alpha)}(\xi_2))\\
  & \ \ + \#_{\sigma(.,\alpha)}(\xi_1) \cdot \#_{\sigma(.,\alpha)}(\xi_2) \cdot 2  \\[2mm]
  &=  \#_{\sigma(.,\alpha)}(\xi_1)+\#_{\sigma(.,\alpha)}(\xi_2) \\
  &= \#_{\sigma(.,\alpha)}(\xi)\enspace.
\end{align*}
This finishes the proof for Case (b2). Hence $\h_\cB(\xi)_{(1 \ 0 \ 1)}=\#_{\sigma(.,\alpha)}(\xi)$.

Now we have proved \eqref{equ:hA-for-wta-constructed-from-monoid-rep}.

\

Then, for each $\xi\in \T_\Sigma$, we have
\[\sem{\cB}(\xi)
=\bigplus_{v\in H}\h_\cB(\xi)_v\cdot F_v
=\h_\cB(\xi)_{(1 \ 0 \ 1)}\cdot F_{(1 \ 0 \ 1)}
=\#_{\sigma(.,\alpha)}(\xi)\cdot 1
=\sem{\cR}(\xi)\]
where the second equality follows from $F_{(1 \ 0 \ 0)}=F_{(1 \ 1 \ 0)}=0$ and the third equality follows from \eqref{equ:hA-for-wta-constructed-from-monoid-rep}.

 Since the consistent $(\Sigma,\Nat)$-monoid representation $\cR=(Q,\lambda,\mu,\gamma)$ (with which we started the present example) is the monoid representation which results from applying the construction in the proof of  Theorem~\ref{theo:monoid-characterization-of-run-semantics} to the $(\Sigma,\Nat)$-wta $\cA$ of Example~\ref{ex:number-of-occurrences} (cf. Observation \ref  {obs:ex:monoid-rep-number-of-occ-of-pattern-results-from-construction}
), we can consider the following pipeline:
    \begin{align*}
    &\text{ $(\Sigma,\Nat)$-wta $\cA$}
      \tag{of Example~\ref{ex:number-of-occurrences}}\\
      & \leadsto \text{consistent $(\Sigma,\Nat)$-monoid representation $\cR$ with $\sem{\cA}=\sem{\cR}$}
      \tag{by Theorem \ref{theo:monoid-characterization-of-run-semantics}}\\
      &  \leadsto \text{ $\cR$ is a consistent $(\Sigma,\Ratnum)$-monoid representation}\\
      & \leadsto \text{$(\Sigma,\Ratnum)$-wta $\cB$ with $\sem{\cR}=\sem{\cB}$}
      \tag{by Theorem \ref{theo:from-monoid-rep-to-wta-comm-semiring}}\\
      & \leadsto \text{$\cB$ is a $(\Sigma,\Int)$-wta}
      \tag{by Observation \ref{obs:extension-of-weight-structure-inverse}, because $\mathrm{wts}(\cB) \subseteq \mathbb{Z}$ and $\Int$ is a sub-semiring of $\Ratnum$ }
         \end{align*}
      Thus, it makes sense to compare $\cA$ and $\cB$.
  Both wta $\cA$ and $\cB$ have three states. By \eqref{eq:h-numberofpatterns-ex} and \eqref{equ:hA-for-wta-constructed-from-monoid-rep}, for each $\xi \in \T_\Sigma$, we have
    \begin{equation}\label{equ:hA-hB-comparison}
  \begin{aligned}
\h_\cA(\xi) = \begin{matrix} {\tiny \bot} \\ {\tiny a} \\ {\tiny f} \end{matrix}
\left(
\begin{matrix} 1 \\ g(\xi) \\ \#_{\sigma(.,\alpha)}(\xi) \end{matrix}
\right)
\ \ \   \text{ and } \ \ \
    \h_\cB(\xi) = \begin{matrix} {\tiny (1 \ 0 \ 0)} \\ {\tiny (1 \ 1 \ 0)} \\ {\tiny (1 \ 0 \ 1)} \end{matrix}
\left(
\begin{matrix} f(\xi) \\ g(\xi) \\ \#_{\sigma(.,\alpha)}(\xi) \end{matrix}
\right) \enspace.
  \end{aligned}
  \end{equation}
    It is easy to check that, for each $\xi \in \T_\Sigma$, we have
    \[
\h_\cB(\xi)_{(1 \ 0 \ 0)} = \h_\cA(\xi)_\bot - \h_\cA(\xi)_a - \h_\cA(\xi)_f  \enspace.
    \]
 
\hfill $\Box$
\end{example}

The combination of Theorem \ref{theo:from-monoid-rep-to-wta-comm-semiring}(1) and (2) yields the result  \cite[Thm.~2]{bozale89}.

\begin{corollary-rect}\label{cor:field-from-consistent-rep-to-wta} \cite[Thm.~2]{bozale89}
  Let $\Sigma$ be a ranked alphabet and $\B$ be a field. Let $\cR=(Q,\lambda,\mu,\gamma)$ be a consistent $(\Sigma,\B)$-monoid representation. Then we can construct a $(\Sigma,\B)$-wta $\cA$ such that $\sem{\cA}=\sem{\cR}$.
\end{corollary-rect}

\begin{proof} By Theorem \ref{theo:from-monoid-rep-to-wta-comm-semiring}(1)
  we construct  (a)~a~$(\Sigma,\B)$-semimodule $\V_\cR=(\V,\eta)$ where $\V$ is a $\B$-sub-semimodule of $(B^Q,\oplus,\widetilde{\0})$  and (b)~a~linear form $\gamma'$ over the $\B$-semimodule $\V$ such that $\sem{\cR}= \gamma' \circ \h_\eta$.

  Since $\B$ is a field, by Theorem \ref{theo:from-monoid-rep-to-wta-comm-semiring}(2),  we can construct a $(\Sigma,\B)$-wta $\cA$ such that $\sem{\cA}= \gamma' \circ \h_\eta$.
\end{proof}

\section{The characterization result}

Theorem \ref{theo:monoid-characterization-of-run-semantics} and Corollary \ref{cor:field-from-consistent-rep-to-wta} imply the following  corollary (cf. Figure~\ref{fig:overview-monoid-representations-new}).

\begin{corollary-rect}\label{cor:run-wta-mon-rep-initit-wta} \rm \cite[Thms.~1 and 2]{bozale89} (see also \cite[Thm.~3.54]{fulvog09new}) Let $\Sigma$ be a ranked alphabet and $\B$ be a field. Moreover, let $r: \T_\Sigma \to B$. Then the following two statements are equivalent:
  \begin{compactenum}
  \item[(A)] We can construct a $(\Sigma,\B)$-wta $\cA$ such that $\sem{\cA}=r$.
  \item[(B)] We can construct a consistent $(\Sigma,\B)$-monoid representation $\cR$ such that $\sem{\cR}=r$.
      \end{compactenum}
\end{corollary-rect}

\begin{figure}
  {\small
\begin{tikzpicture}

  \node (n1)  {$(\Sigma,\B)$-wta $\cA=(Q,\delta,F)$};

  \node[below of= n1, yshift=-120] (n2)
  {\begin{tabular}{ll}consistent $(\Sigma,\B)$-monoid\\ representation $\cR=(Q,\lambda,\mu,\gamma)$\end{tabular}};

  \node[right of= n2, xshift=180, yshift=-50] (n3)
  {\begin{tabular}{ll}consistent $(\Sigma,\B)$-monoid\\ representation $\cR=(Q,\lambda,\mu,\gamma)$ \end{tabular}};

 \node[above of= n3, yshift=100]  (n4)
  {\begin{tabular}{ll}$(\Sigma,\B)$-semimodule $\V_\cR=(\V,\eta)$\\
     where $\V \le (B^Q,\oplus,\widetilde{\0})$\\
     and $\gamma'$ is a linear form over $\V$ \end{tabular}};


 \node[above of= n4, yshift=100]  (n6)
      {$(\Sigma,\B)$-wta $\cA$};

\draw (n1) edge[->,>=stealth] node[fill=white]
{\fbox{\begin{tabular}{ll} Theorem \ref{theo:monoid-characterization-of-run-semantics}\\ $\B$ commutative semiring\\ $\sem{\cA}=\sem{\cR}$  \end{tabular}}} (n2);
 
\draw (n3) edge[->,>=stealth] node[fill=white]
{\fbox{\begin{tabular}{ll} Theorem \ref{theo:from-monoid-rep-to-wta-comm-semiring}(1)\\ $\B$ commutative semiring\\ $\sem{\cR}=\gamma' \circ \h_\eta$  \end{tabular}}} (n4);

\draw (n4) edge[->,>=stealth] node[fill=white]
{\fbox{\begin{tabular}{ll} Theorem \ref{theo:from-monoid-rep-to-wta-comm-semiring}(2)\\ $\B$ field\\ $\gamma' \circ \h_\eta = \sem{\cA}$  \end{tabular}}} (n6);

\draw (n3) edge[->,>=stealth,out=10, in=350] node[fill=white]
      {\fbox{\begin{tabular}{ll} Corollary \ref{cor:field-from-consistent-rep-to-wta}\\ $\B$  field\\ $\sem{\cA}=\sem{\cR}$  \end{tabular}}} (n6);

\end{tikzpicture}
} 

\caption{\label{fig:overview-monoid-representations-new} Overview of the main results of this chapter, where $\le$ means ``$\B$-sub-semimodule of'' and $\Sigma$ denotes an arbitrary ranked alphabet.}
  \end{figure}

  \

  When comparing Corollary \ref{cor:run-wta-mon-rep-initit-wta} and Theorem \ref{theo:string-ranked:wta=monoid-rep}, we observe a kind of trade-off between the type of ranked alphabets (arbitrary / string) and the class of weight algebras (fields / semirings) in the sufficient conditions for the equivalence (A) $\Leftrightarrow$ (B):
  
  (A) $\Leftrightarrow$ (B) if
  \begin{compactitem}
  \item $\Sigma$ is a ranked alphabet and $\B$ is a field (cf. Corollary \ref{cor:run-wta-mon-rep-initit-wta}) or
    \item $\Sigma$ is a string ranked alphabet and $\B$ is a semiring (cf. Theorem \ref{theo:string-ranked:wta=monoid-rep}).
    \end{compactitem}
  Moreover, it would be interesting to characterize the equivalence (A) $\Leftrightarrow$ (B) in terms of sufficient and necessary conditions.
    
  \

Finally, we present a new characterization result for the class $\Rec(\Sigma)$ of recognizable $\Sigma$-tree languages in terms of consistent $(\Sigma,\sfFtwo)$-monoid representations. It follows directly from  Corollary \ref{cor:run-wta-mon-rep-initit-wta} and Corollary~\ref{cor:supp-F2=fta} .

  \begin{corollary}\rm Let $\Sigma$ be ranked alphabet and $L\subseteq \T_\Sigma$. Then the following two statements are equivalent.
    \begin{compactenum}
    \item[(A)] We can construct a $\Sigma$-fta $A$ such that $L = \LL(A)$.
      \item[(B)] We can construct a  consistent $(\Sigma,\sfFtwo)$-monoid representation $\cR$ such that $L = \supp(\sem{\cR})$.
      \end{compactenum}
    \end{corollary}

%% file: AFwtL.tex
\chapter{Abstract families of weighted tree languages}
\label{ch:AFwtL}

In the 60's and 70's of the previous century, Ginsburg, Greibach, and Hopcroft  proposed a unifying concept for the study of closure properties of sets of formal languages \cite{gingre67,gingre69,gingrehop69,gin75}. This is the concept of \emph{abstract family of languages} (AFL). Roughly speaking, an AFL is a set of formal languages which is closed under intersection with regular languages, inverse homomorphisms, $\varepsilon$-free homomorphisms, union, concatenation, and Kleene star. An AFL is \emph{full} if it is closed under arbitrary homomorphisms. Of particular interest are principal AFL; an  AFL is \emph{principal} if it is generated from one formal language by using the mentioned closure properties. For instance, the sets of regular languages, context-free languages, stack languages, nested-stack languages, and  recursively enumerable languages are full principal AFL \cite[Sec.~2]{gingre70}.

The importance of principal AFL shows up in one of the main theorems of  AFL-theory \cite[Thm.~5.2.1]{gin75}: a set $\cL$ is a full principal AFL if and only if there exists a finitely encoded abstract family of acceptors (AFA) $\cD$ such that $\cL$ is set of all formal languages accepted by $\cD$. Roughly speaking, an acceptor $\cD$ is a one-way nondeterministic finite-state automaton which uses an additional storage (indicated by the type $\cD$) such as, e.g. counter, pushdown, stack, or nested-stack. Thus, in other words, ``each family of languages defined by a set of 'well-behaving' one-way nondeterministic acceptors of a same 'type' is a full principal AFL'' (cited from \cite[p.~109]{engmeilee80}. 

In this chapter  we define the concept of abstract family of weighted tree languages. It is inspired by the definitions of abstract family of languages \cite{gingre67,gingre69,gingre70,gingrehop69,gin75}, abstract family of formal power series \cite{kuisal86}, abstract family of fuzzy languages \cite{asv03} (also cf. \cite[Ch.~3]{wec78}),  abstract family of tree series \cite[Thm.~3.5]{kui99c} and \cite[Sect.~7]{esikui03}, and sheaf of forests~\cite{bozrah94}.

 \label{p:convention-for-AFwtL}
\begin{quote}\em In the rest of this chapter, $\B$ denotes an arbitrary commutative and $\sigma$-complete semiring. 
  \end{quote}

  Thus, by Theorem \ref{thm:rec=reg}, we have that $\Rec(\Sigma,\B) = \Reg(\Sigma,\B)$, and, in particular, the closure results developed in Chapter \ref{ch:closure-properties} for $\Rec(\Sigma,\B)$ also hold for $\Reg(\Sigma,\B)$.
  
In this chapter we follow the notions and definitions in \cite{fulvog19a}. The main theorem is the following (cf. Theorem \ref{thm:REGnB-smallest-princ-AFwtL}): for each $n \in \mathbb{N}$, the set $\Reg(n,\B)$ of all $(n,\B)$-weighted tree languages is the smallest principal abstract family of $(n,\B)$-weighted tree languages.

\section{The basic definitions}

\index{weighted tree language!$(n,\B)$-weighted tree language}
For each $n \in \mathbb{N}$, an \emph{$(n,\B)$-weighted tree language} is a $(\Sigma,\B)$-weighted tree language for some ranked alphabet $\Sigma$ with $\maxrk(\Sigma) \le n$.

In this chapter  we will often consider a set $\cL$ of $(n,\B)$-weighted tree languages and write that ``$\cL$ is closed under some operation'', like scalar multiplication, sum, tree concatenations, or weighted projective bimorphisms (wpb). If we consider a binary operation, then we always assume that the two arguments of the operation are based on the same ranked alphabet. For instance, if we write that a set $\cL$ of $(n,\B)$-weighted tree languages is closed under sum, then this means that, for every ranked alphabet $\Sigma$ with $\maxrk(\Sigma) \le n$, and $r_1: \T_\Sigma \to B$ and $r_2: \T_\Sigma \to B$ in $\cL$, the weighted tree language $r_1 \oplus r_2$ is in~$\cL$.

\index{Reg@$\Reg(n,\B)$}
As a consequence of the definition of an $(n,\B)$-weighted tree language, we have that a regular $(n,\B)$-weighted tree language is a regular $(\Sigma,\B)$-weighted tree language for some ranked alphabet $\Sigma$ with $\maxrk(\Sigma) \le n$. We denote the set of all  regular $(n,\B)$-weighted tree languages by $\Reg(n,\B)$. That is
\[
\Reg(n,\B) = \bigcup (\Reg(\Sigma,\B) \mid  \text{ $\Sigma$ is a ranked alphabet  such that $\maxrk(\Sigma) \le n$}) \enspace.
  \]

  \index{family of $(n,\B)$-weighted tree languages}
Similarly to the unweighted string case, we require that a family of weighted tree languages contains at least one element. Here we let this element be a recognizable step mapping which is  unit-valued, one-step, and over a very simple ranked alphabet. Formally, let $\cL$ be a set of $(n,\B)$-weighted tree languages. We call $\cL$ a \emph{family of $(n,\B)$-weighted tree languages} if there exists a ranked alphabet $\Delta$ such that 
\begin{compactitem}
\item  $\maxrk(\Delta) = n$,
\item 
$|\Delta^{(k)}|=1$  for each $k \in [0,n]$, and 
\item $\chi(\T_\Delta) \in \cL$. 
\end{compactitem} 
If $\Delta$ is such a ranked alphabet, then for each ranked alphabet $\Sigma$ with $\maxrk(\Sigma) \le n$ and $\xi \in \T_\Sigma$, there exists a unique $\zeta \in \T_\Delta$ such that $\pos(\zeta) = \pos(\xi)$. We will exploit this obvious fact later.

\index{tree cone}
\index{nbtreecone@$(n,\B)$-tree cone}
Let $\cL$ be a family of $(n,\B)$-weighted tree languages. 
We call $\cL$ an \emph{$(n,\B)$-tree cone} if it is closed under wpb.  (For the definition of wpb, we refer to cf. Section \ref{sec:w-projective-bimorphisms}. For related concepts, we refer to \cite[p.~136]{ber79} for rational cones, to \cite[p.~8]{kui99c} for recognizable tree cones, and to \cite[Sect.~7]{esikui03} for recognizable tree series cones.)

 \index{AFwtL@$\AFwtL$}
 \index{nBAFwtL@$(n,\B)$-$\AFwtL$}
 \index{abstract family of wtls}
 We call $\cL$  an \emph{abstract family of $(n,\B)$-weighted tree languages} (for short: $(n,\B)$-$\AFwtL$, or just: AFwtL) if $\cL$ is an $(n,\B)$-tree cone which is closed under the rational operations, i.e., under sum, tree concatenations, and Kleene stars.

 The above definition of $(n,\B)$-$\AFwtL$ is equivalent to the one in \cite{fulvog19a} where additionally closure under scalar multiplication was required. However, scalar multiplication of some $(\Sigma,\B)$-weighted tree language $r$ with some $b \in B$ can be simulated as the Hadamard product of $r$ and the constant weighted tree language $\widetilde{b}$. In its turn, the Hadamard product of $r$ and $\widetilde{b}$, can be expressed as application $\ttsem{\cH}(r)$ for some $(\Sigma,\Sigma,\B)$-wpb $\cH$ such that $\ttsem{\cH}(\xi,\zeta) =b$ if $\xi=\zeta$, and $\0$ otherwise (cf. \eqref{equ:appl-wtt-to-wtl}). It is obvious how to construct such an $\cH$. Thus, scalar multiplication can be simulated by wpb, and hence closure under scalar multiplication can be neglected in the definition of $(n,\B)$-AFwtL. We note that the concept of full abstract family of languages \cite{gingre67,gingre69,gingrehop69,gin75} is a particular instance of the concept of AFwtL (cf. \cite[Sec.~7]{fulvog19a}).

\index{FS@$\cF(\cS)$}
Let $\cS$ be a set of $(n,\B)$-weighted tree languages. We denote the smallest $(n,\B)$-$\AFwtL$ which contains $\cS$ by $\cF(\cS)$. 

\index{principal}
\index{generator}
Now assume that $\cL$ is an $(n,\B)$-$\AFwtL$. Then it   is \emph{principal} if there exists an $(n,\B)$-weighted tree language $g \in \cL$ such that $\cL = \cF(\{g\})$; and if this is the case, then $g$ is called \emph{generator (of $\cL$)}.

\section{Characterization of tree cones}
\label{sec:decomposition-of-wbp}

As defined above, a family of $(n,\B)$-weighted tree languages is an $(n,\B)$-tree cone if it is closed under wpb. Here we characterize the property of being an $(n,\B)$-tree cone by decomposing each wpb into (a) the inverse of a linear and nondeleting tree homomorphism, (b) the Hadamard product with a recognizable weighted tree language, and (c) a linear and nondeleting tree homomorphism. In other words, we make the two homomorphisms and the center language of a wpb explicit (cf. Corollary \ref{cor:char-tree-cones}). The characterization resembles Nivat's decomposition theorem of gsm \cite{niv68}, Arnold's and Dauchet's bimorphism \cite{arndau76,arndau82}, Ginsburg's decomposition of a-transducers \cite[Lm.~3.2.2]{gin75}, and Engelfriet's decomposition of bottom-up tree transducers \cite[Thm.~3.5]{eng75}. Moreover, it is an instance of \cite[Thm.~4.2]{fulmalvog11}.

Let $\cH = (N,S,R,wt)$ be a $(\Sigma,\Psi,\B)$-wpb. We recall that $\cH$ is a particular $([\Sigma\Psi],\B)$-wrtg and hence, we have defined the $(R,[\Sigma\Psi])$-tree homomorphism $\pi: \T_\R \to \T_{[\Sigma\Psi]}$,
which extracts from a rule tree the derived $[\Sigma\Psi]$-tree. Moreover, for the wpb $\cH$, we have defined the $([\Sigma\Psi],\Sigma)$-tree homomorphism $\pi_1$ and the $([\Sigma\Psi],\Psi)$-tree homomorphism $\pi_2$, which project a $[\Sigma\Psi]$-tree to its first and second component, respectively. Each of the three tree homomorphisms $\pi$, $\pi_1$, and $\pi_2$ is linear and nondeleting. Now we compose (a) $\pi$ and $\pi_1$ and (b) $\pi$ and $\pi_2$, and thereby obtain two linear and nondeleting tree homomorphisms $h_1^{\cH}$ and $h_2^{\cH}$.

Formally, we define $h_1^\cH = \pi_1 \hat{\circ} \pi$ and $h_2^\cH = \pi_2 \hat{\circ} \pi$, where $\hat{\circ}$ is the syntactic composition of tree homomorphisms defined in Section \ref{sect:trees}. For the convenience of the reader, we recall that
$h_1^{\cH} = ((h_1^{\cH})_k \mid k \in \mathbb{N})$ is the $(R,\Sigma)$-tree homomorphism and $h_2^{\cH} = ((h_2^{\cH})_k \mid k \in \mathbb{N})$ is an $(R,\Psi)$-tree homomorphism such that the following holds. 
\begin{compactitem}
\item For every $k \in \mathbb{N}$ and rule  $r = (A \to [\sigma,\psi](A_1,\ldots,A_k))$ with $[\sigma,\psi] \in [\Sigma\Psi]^{(k)}$, we have
\[(h_1^{\cH})_k(r) =
  \begin{cases}
    \sigma(x_1,\ldots,x_k) & \text{if $\sigma\not=\varepsilon$}\\
    x_1 & \text{otherwise}
  \end{cases}
 \ \ \text{ and } \ \
  (h_2^{\cH})_k(r) =
    \begin{cases}
    \psi(x_1,\ldots,x_k) & \text{if $\psi\not=\varepsilon$}\\
    x_1 & \text{otherwise} \enspace.
  \end{cases}
\]
\item For each chain-rule $r = (A \to B)$, we have
\((h_1^{\cH})_1(r) = x_1 \ \text{ and } \ (h_2^{\cH})_1(r) = x_1\).
\end{compactitem}
In fact, $h_1^{\cH}$ and $h_2^{\cH}$ are simple, i.e., linear, nondeleting, alphabetic, and ordered tree homomorphisms. Moreover, by Theorem \ref{thm:hom-synt-comp}, we have $h_1^\cH = \pi_1 \circ \pi$ and $h_2^\cH = \pi_2 \circ \pi$.

  Now we can prove the characterization of wpb reported above. We mention that the composition of weighted tree transformations involved in the following theorem is associative due to Observation \ref{obs:composition-semiring-associative}. Moreover, we will use the concepts of characteristic mapping and diagonalization as follows (cf. Subsections~\ref{sec:weighted-tree-languages} and \ref{sect:weighted-tr-tr}). Since $h_1^{\cH}$ and $h_2^{\cH}$ are particular binary relations, i.e.,  $h_1^{\cH} \subseteq \T_R \times \T_\Sigma$ and $h_2^{\cH} \subseteq \T_R \times \T_\Psi$, we can consider the characteristic mappings $\chi((h_1^{\cH})^{-1}): \T_\Sigma \times \T_R \to B$ and $\chi(h_2^{\cH}): \T_R \times \T_\Psi \to B$. Also we will use the diagonalization $\overline{\wrtsem{\cH}}: \T_R \times \T_R \to B$ of $\wrtsem{\cH}$.

  \begin{theorem}\label{thm:decomp-wpb}   Let $\cH$ be a $(\Sigma,\Psi,\B)$-wpb. Then $\ttsem{\cH} = \chi((h_1^{\cH})^{-1}) ; \overline{\wrtsem{\cH}}; \chi(h_2^{\cH})$.
  \end{theorem}
  \begin{proof} Let $\xi \in \T_\Sigma$ and $\zeta \in \T_\Psi$. 
  By the definition of $\RT_\cH(\T_{[\Sigma\Psi]})$ in Section \ref{subsect:basic-modell} (notice the fact that each wpb in a particular weighted context-free grammar) and by \eqref{eq:bimorphism-def}, we have
    \begin{equation}
\RT_\cH(\xi,\zeta) = \{d \in \T_R \mid (\xi,\zeta)=(h_1^{\cH}(d),h_2^{\cH}(d)), d \in \RT_\cH(\T_{[\Sigma\Psi]})\} \enspace. \label{equ:split-index-set}
      \end{equation}
    Then we can calculate as follows.
    \begingroup
    \allowdisplaybreaks
    \begin{align*}
      \ttsem{\cH}(\xi,\zeta) &= \infsum{\oplus}{d \in \RT_\cH(\xi,\zeta)}{\wt_\cH(d)}\\
                             &= \infsum{\oplus}{\substack{d \in \T_R:\\(\xi,\zeta)=(h_1^{\cH}(d),h_2^{\cH}(d))}}{\wt_\cH(d) \otimes \chi(\RT_\cH(\T_{[\Sigma\Psi]}))(d)}
      \tag{by \eqref{equ:split-index-set}}\\
                             &= \infsum{\oplus}{\substack{d \in \T_R:\\(\xi,\zeta)=(h_1^{\cH}(d),h_2^{\cH}(d))}}{\wrtsem{\cH}(d)}
      \tag{by definition of $\wrtsem{\cH}$}\\
                             &= \infsum{\oplus}{d \in \T_R}{\chi(h_1^{\cH})(d,\xi) \otimes \overline{\wrtsem{\cH}}(d,d) \otimes \chi(h_2^{\cH})(d,\zeta)}\\
                             &= \infsum{\oplus}{d,d' \in \T_R}{\chi(h_1^{\cH})(d,\xi) \otimes \overline{\wrtsem{\cH}}(d,d') \otimes \chi(h_2^{\cH})(d',\zeta)}
      \tag{because $\overline{\wrtsem{\cH}}(d,d') = \0$ if $d\not=d'$}\\
               &= \infsum{\oplus}{d,d' \in \T_R}{\chi((h_1^{\cH})^{-1})(\xi,d) \otimes \overline{\wrtsem{\cH}}(d,d') \otimes \chi(h_2^{\cH})(d',\zeta)}\\
                             &= (\chi((h_1^{\cH})^{-1}); \overline{\wrtsem{\cH}}; \chi(h_2^{\cH}))(\xi,\zeta) \enspace.
                               \tag{by definition of composition, see Subsection \ref{sect:weighted-tr-tr}}
    \end{align*}
    \endgroup
  \end{proof}

  An \emph{$n$-tree homomorphism} is a $(\Sigma,\Psi)$-tree homomorphism where $\Sigma$ and $\Psi$ are ranked alphabets and $\maxrk(\Sigma) \le n$ and $\maxrk(\Psi) \le n$. The closure under $n$-tree homomorphisms and closure under inverse $n$-tree homomorphisms are defined in a way analogous to the corresponding closure under tree homomorphisms and inverse tree homomorphisms (cf. Sections \ref{sect:homomorphism-preserves} and \ref{sect:inverse-homomorphism-preserves}).
  
A set $\cL$ of $(n,\B)$-weighted tree languages is \emph{closed under Hadamard product with recognizable $(n,\B)$-weighted tree languages} if, for every  $(\Sigma,\B)$-weighted tree language $r$ in $\cL$ and for each recognizable $(\Sigma,\B)$-weighted tree language $r'$, we have $r\otimes r'$ is in $\cL$.

  \begin{corollary-rect}\rm \label{cor:char-tree-cones} Let $\B=(B,\oplus,\otimes,\0,\1)$ be a commutative and $\sigma$-complete semiring. Let $n \in \mathbb{N}$ and  $\cL$ be a family of $(n,\B)$-weighted tree languages. Then the following three statements are equivalent.
    \begin{compactenum}
    \item[(A)] $\cL$ is an $(n,\B)$-tree cone.
    \item[(B)] $\cL$ is closed under (a) simple $n$-tree homomorphisms, (b) inverse of simple $n$-tree homomorphisms, and (c) Hadamard product with recognizable $(n,\B)$-weighted tree languages.
    \item[(C)] $\cL$ is closed under (a) simple $n$-tree homomorphisms, (b) inverse of simple $n$-tree homomorphisms, and (c) Hadamard product with bu-deterministically recognizable $(n,\B)$-weighted tree languages.
      \end{compactenum}
    \end{corollary-rect}

    \begin{proof} Proof of (A)$\Rightarrow$(B):
      (a) Let $h = (h_k \mid k \in \mathbb{N})$ be a simple $(\Sigma,\Psi)$-tree homomorphism with $\maxrk(\Sigma) \le n$ and $\maxrk(\Psi) \le n$. Then we construct the $(\Sigma,\Psi,\B)$-wpb $\cH= (N,S,R,wt)$ as follows. We let $N=S=\{A\}$ and, for every $k \in \mathbb{N}$ and $\sigma \in \Sigma^{(k)}$, we have the following.

      If $h_k(\sigma) = \psi(x_1,\ldots,x_k)$, then $r = (A \to [\sigma,\psi](A,\ldots,A))$ is in $R$ and $wt(r) = \1$. 

      If $k= 1$ and $h_k(\sigma) = x_1$, then $r = (A \to [\sigma,\varepsilon](A))$ is in $R$ and $wt(r) = \1$.

      It is obvious that $\ttsem{\cH} = \chi(h)$. 
           Since $\cL$ is closed under $(n,\B)$-wpb, we obtain that $\cL$ is closed under simple $n$-tree homomorphisms.

      (b) This can be proved in the same way as (a) except that the order of $\sigma$ and $\psi$ (and $\sigma$ and $\varepsilon$) in the rules of $R$ are exchanged.
      
      (c) Let $r: \T_\Sigma \to B$ be a weighted tree language in $\cL$ and $\cA$ be a $(\Sigma,\B)$-wta. By definition, $\maxrk(\Sigma) \le n$.  By the proof of Corollary \ref{cor:Hadamard-alternative}, we can construct a $(\Sigma,\Sigma,\B)$-wpb $\cH$ such that $\ttsem{\cH} = \overline{\sem{\cA}}$, i.e., $\ttsem{\cH}$ is the diagonalization of $\sem{\cA}$ (cf. page \pageref{page:diagonalization}).
Then by Equation \eqref{obs:application-vs-Hadamard} we have  $\ttsem{\cH}(r) = r \otimes \sem{\cA}$. Since $\cL$ is closed under wpb, we can conclude that $\cL$ is closed under Hadamard product with recognizable $(n,\B)$-weighted tree languages.

\

      Proof of (B)$\Rightarrow$(C): It is obvious.

      \
      
      Proof of (C)$\Rightarrow$(A): Let $r: \T_\Sigma \to B$ be in $\cL$ and let $\cH$ be a $(\Sigma,\Psi,\B)$-wpb  with $\maxrk(\Psi) \le n$. Let $R$ be the set of rules of $\cH$. By Theorem \ref{thm:decomp-wpb}, we have
      \[
        \ttsem{\cH}(r) =  \Big(\chi((h_1^{\cH})^{-1}) ; \overline{\wrtsem{\cH}}; \chi(h_2^{\cH})\Big)(r)
      \]
      and thus, by Observation \ref{obs:application-of-composition}, we have
      \[
        \ttsem{\cH}(r) =  \chi(h_2^{\cH})\Big(\overline{\wrtsem{\cH}}\Big(\chi((h_1^{\cH})^{-1})\Big(r\Big)\Big)\Big)  \enspace.
      \]
      Let us denote the weighted tree language  $\chi((h_1^{\cH})^{-1})(r)$ by $r'$. 
      Since $\cL$ is closed under inverse of simple $n$-tree homomorphism, $r'$ is in $\cL$. Then, by Equation \eqref{obs:application-vs-Hadamard}, we have
      \[
\overline{\wrtsem{\cH}}(r') = r' \otimes \wrtsem{\cH} \enspace.
        \]
  By Lemma \ref{lm:w-rule-trees-are-local}(2) and Lemma \ref{lm:weighted-local->bu-recognizable}, the weighted tree language $\wrtsem{\cH}$  is bu-deterministically recognizable. Since $\cL$ is closed under Hadamard product with bu-deterministically recognizable $n$-weighted tree languages, we obtain that $\overline{\wrtsem{\cH}}(r')$ is in $\cL$. Let us denote this weighted tree language by $r''$. Finally, since $\cL$ is closed under simple $n$-tree homomorphism, the weighted tree language  $\chi(h_2^{\cH})(r'')$ is in $\cL$. Hence $\ttsem{\cH}(r)$ is in~$\cL$.
            \end{proof}


\section{$\Reg(n,\B)$ is an AFwtL}

Here we prove that the set of regular $(n,\B)$-weighted tree languages is an AFwtL.
  For each $n \in \mathbb{N}$, we define the ranked alphabet $\Delta_n$ by
  \[
    \Delta_n = \bigcup_{k \in [0,n]} \Delta_n^{(k)} \ \text{ and } \ \Delta_n^{(k)} = \{[k]\} \ \text{ for each $k \in [0,n]$} \enspace.\]
  
We note that, for each ranked alphabet $\Sigma$ with $\maxrk(\Sigma) =n$ for some $n \in \mathbb{N}$, the skeleton alphabet of $\Sigma$ is a subset of  $\Delta_n$, i.e., $[\Sigma] \subseteq \Delta_n$. For the definition of the skeleton alphabet we refer to page~\pageref{page:skeleton}.
  
\begin{lemma}\rm \label{lm:RegnB-is-family} \cite[Lm.~6.2,6.9]{fulvog19a}  The $(\Delta_n,\B)$-weighted tree language $\chi(\T_{\Delta_n})$ is in $\Reg(n,\B)$. Moreover, $\Reg(n,\B)$ is a family of $(n,\B)$-weighted tree languages.  
\end{lemma}
\begin{proof} We construct the $(\Delta_n,\B)$-wrtg $\cG = (\{*\},\{*\},R,\wt)$ such that, for each $k \in [0,n]$, $R$ contains the rule $r = (* \rightarrow [k](*,\ldots,*))$ with $k$ occurrences of $*$ in the right-hand side and  $\wt(r) = \1$. Since  $\sem{\cG} = \chi(\T_{\Delta_n})$, we have $\chi(\T_{\Delta_n}) \in \Reg(n,\B)$.

  Since $\maxrk(\Delta_n) = n$, $|\Delta_n^{(k)}|=1$  for each $k \in [0,n]$, and 
$\chi(\T_{\Delta_n}) \in  \Reg(n,\B)$, we have that $\Reg(n,\B)$ is a family of $(n,\B)$-weighted tree languages.  
\end{proof}

\begin{theorem} {\rm \cite[Thm.~6.8]{fulvog19a}} \label{thm:RegnB-is-AFwtL} $\Reg(n,\B)$ is an AFwtL.
\end{theorem}
\begin{proof}  By Lemma \ref{lm:RegnB-is-family}, the set $\Reg(n,\B)$ is a family of $(n,\B)$-weighted tree languages.

  By Theorem \ref{thm:closure-REG-under-wbim}, $\Reg(n,\B)$ is closed under wpb, and hence $\Reg(n,\B)$ is a tree cone. We note that the image of an $(n,\B)$-weighted tree language under a wpb is again an $(n,\B)$-weighted tree language. 

  Let $r,r_1,r_2,s \in \Reg(\Sigma,\B)$ for some ranked alphabet $\Sigma$ with $\maxrk(\Sigma) \le n$. Moreover, let $\alpha \in \Sigma^{(0)}$. We require that $s$ is $\alpha$-proper. By Theorems \ref{thm:closure-sum}, \ref{thm:closure-tree-concatenation-wrtg}, and \ref{thm:closure-Kleene-star-wrtg}, the $(\Sigma,\B)$-weighted tree languages $r_1 \oplus r_2$, $r_1 \circ_\alpha r_2$, and $s^*_\alpha$ are in $\Reg(\Sigma,\B)$, respectively. (We recall that $\Rec(\Sigma,\B) = \Reg(\Sigma,\B)$ because we assumed that $\B$ is a $\sigma$-complete semiring.)
      Hence $\Reg(n,\B)$ is closed under the rational operations, i.e., under sum, tree concatenations, and Kleene stars. Hence,  $\Reg(n,\B)$ is an AFwtL. 
  \end{proof}

  \section{The AFwtL $\Reg(n,\B)$ is principal}

 We will prove that the $(\Delta_n,\B)$-weighted tree language $\chi(\T_{\Delta_n})$ is the generator of $\Reg(n,\B)$. 

\begin{lemma}\rm \label{lm:technical} \cite[Lm.~6.10]{fulvog19a} Let $\cG$ be a $(\Sigma,\B)$-wrtg with $\maxrk(\Sigma) \le n$. Then there exists a $(\Delta_n,\Sigma)$-wpb $\cH$ such that $\sem{\cG} = \ttsem{\cH}(\chi(\T_{\Delta_n}))$.
\end{lemma}
\begin{proof} Let $\cG=(N,S,R,wt)$ with $\maxrk(\Sigma) \le n$. By Lemma \ref{lm:rtg-automata-form-wrtg}, we can assume that $\cG$ is in tree automata form. Thus $[R]=[\Sigma]$ and hence $[R] \subseteq \Delta_n$ (also for $n=0$). 

  By Lemma \ref{lm:decomp-wrtg}, there exists a chain-free $([R],\Sigma,\B)$-wpb $\cH'$ such that  $\sem{\cG} = \ttsem{\cH'}(\chi(\T_{[R]}))$.  Since $[R] \subseteq \Delta_n$, we can view $\cH'$ as a $(\Delta_{n},\Sigma,\B)$-wpb $\cH$. Obviously, $\sem{\cH'}=\sem{\cH}|_{\T_{[R]} \times \T_\Sigma}$ and, for each $(b,\xi) \in (\T_{\Delta_{n}} \times \T_\Sigma) \setminus (\T_{[R]} \times \T_\Sigma)$, we have that $\sem{\cH}(b,\xi) =\0$.  Thus $\sem{\cH'}(\chi(\T_{[R]}))
= \sem{\cH}(\chi(\T_{\Delta_n}))$. Hence $\sem{\cG} = \ttsem{\cH}(\chi(\T_{\Delta_n}))$.
\end{proof}

\begin{lemma}\rm \label{lm:Reg=principal-AFwtL} \cite[Lm.~6.11]{fulvog19a} Let $\maxrk(\Sigma) \le n$ and  $r: \T_\Sigma \to B$. The following statements are equivalent.
  \begin{compactenum}
  \item[(A)]  $r \in \Reg(n,\B)$.
  \item[(B)] There exists a $(\Delta_n,\Sigma)$-wpb $\cH$ such that $r = \ttsem{\cH}(\chi(\T_{\Delta_n}))$.
    \item[(C)] $r \in \cF(\{\chi(\T_{\Delta_n})\})$.
    \end{compactenum}
  \end{lemma}
  \begin{proof} Proof of (A)$\Rightarrow$(B): This follows from Lemma \ref{lm:technical}.

    Proof of (B)$\Rightarrow$(C): This holds by definition of $\cF(\{\chi(\T_{\Delta_n})\})$.

    Proof of (C)$\Rightarrow$(A): By Theorem \ref{thm:RegnB-is-AFwtL}, the set $\Reg(n,\B)$ is an AFwtL. By Lemma \ref{lm:RegnB-is-family}, we have that $\chi(\T_{\Delta_n})$ is in $\Reg(n,\B)$. Since  $\cF(\{\chi(\T_{\Delta_n})\})$ is the smallest AFwtL containing  $\chi(\T_{\Delta_n})$, we obtain that  $r \in \Reg(n,\B)$.
    \end{proof}

  The equivalence of (A) and (C) of Lemma \ref{lm:Reg=principal-AFwtL}  implies the following theorem.

  \begin{theorem} \label{thm:Reg-is-principal}{\rm \cite[Thm.~6.12]{fulvog19a}}  The AFwtL $\Reg(n,\B)$ is principal with generator $\chi(\T_{\Delta_n})$.
    \end{theorem}


 Next we prove that the set $\Reg(n,\B)$ is contained in each $(n,\B)$-tree cone.

    \begin{corollary}\rm\cite[Cor.~6.13]{fulvog19a}\label{cor:tree-cone-in-RegnB} For each $(n,\B)$-tree cone $\cL$ we have $\Reg(n,\B) \subseteq \cL$.   
    \end{corollary}

    \begin{proof} Let $\cG$ by a $(\Sigma,\B)$-wrtg with $\maxrk(\Sigma) \le n$. By Lemma~\ref{lm:technical}, there exists a $(\Delta_n,\Sigma)$-wpb $\cH$ such that
      \[\sem{\cG} = \ttsem{\cH}(\chi(\T_{\Delta_n})) \enspace.
      \]

Now let $\cL$ be an $(n,\B)$-tree cone. Since $\cL$ is a family of $(n,\B)$-weighted tree languages, there exists a ranked alphabet $\Delta$ such that 
 $\maxrk(\Delta) =n$,
 $|\Delta^{(k)}|=1$ for each $k \in [0,n]$, and 
 $\chi(\T_\Delta)$ is in $\cL$.  The only difference between $\Delta$ and $\Delta_n$ is the fact that the symbols look different.
  Let $\sema{k}$ denote the unique element in $\Delta^{(k)}$.

We construct the $(\Delta,\Delta_{n},\B)$-wpb $\cN =(\{*\},\{*\},R,\wt)$ such that for each $k \in [0,n]$, the rule
 \[* \to [\sema{k},[k]](*, \ldots, *)
 \]
 is in $R$ with $k$ occurrences of $*$. Obviously, $\chi(\T_{\Delta_n}) = \ttsem{\cN}(\chi(\T_{\Delta}))$.   Hence,
 \[
   \sem{\cG} = \ttsem{\cH}(\ttsem{\cN}(\chi(\T_{\Delta}))) \enspace.
   \]

   By Theorem  \ref{thm:comp-closure-wbim}, wpb are closed under composition. Hence there exists a $(\Delta,\Sigma)$-wpb $\cH'$ such that
    \[
   \sem{\cG} = \ttsem{\cH'}(\chi(\T_{\Delta})) \enspace.
   \]
 
   Since $\chi(\T_{\Delta})$ is in $\cL$ and $\cL$ is an $(n,\B)$-tree cone, we obtain that
   $\sem{\cG}$ is in $\cL$.
\end{proof}

\begin{theorem-rect}\label{thm:REGnB-smallest-princ-AFwtL} Let $\B$ be a commutative and $\sigma$-complete semiring and $n \in \mathbb{N}$. Then $\Reg(n,\B)$ is the smallest principal abstract family of $(n,\B)$-weighted tree languages.
\end{theorem-rect}
\begin{proof} By Theorem \ref{thm:Reg-is-principal}, the set $\Reg(n,\B)$ is a principal $(n,\B)$-AFwtL.
Now let $\cL$ be an arbitrary  principal $(n,\B)$-AFwtL. Since, in particular, $\cL$ is an $(n,\B)$-tree cone, by Corollary \ref{cor:tree-cone-in-RegnB}, we have $\Reg(n,\B) \subseteq \cL$.
\end{proof}

Other examples of principal abstract families of $(n,\B)$-weighted tree languages are the members of the infinite family $(\Reg(\mathrm{P}^\ell,n,\B) \mid \ell \in \mathbb{N})$ (cf. \cite[Thm.~8.1]{fulvog19a}). Here $\Reg(\mathrm{P}^\ell,n,\B)$ is the set of all $(n,\B)$-weighted tree languages which are generated by weighted regular tree grammars with additional storage $\mathrm{P}^\ell$, the $\ell$-iterated pushdown storage \cite{gre70,mas76,eng86}. The particular  instance $(\Reg(\mathrm{P}^\ell,n,\Boole) \mid \ell \in \mathbb{N})$ is known as the OI-hierarchy \cite{wan74,dam82,engsch77,engsch78,damgoe82,eng91c}, which starts with the set of regular tree languages ($\ell=0$) and the set of OI context-free tree languages ($\ell=1$).

%% file: L-valued-wta.tex
\chapter{Corollaries and theorems for wta over bounded lattices}
\label{ch:L-valued-wta}

In the literature, $(\Sigma,\sfL)$-wta have been investigated where $\sfL$ is a bounded lattice
(or a bounded lattice with some further restrictions, like distributivity), e.g., \cite{inafuk75,mormal02,esiliu07,bozlou10,mogzahame11,ghozah12,ghozahame12,ghozah16,ghozah17,gho18,gho22} and for the string case \cite{drovog10,drovog12};
for a survey we refer to \cite{rah09}. We also refer to \cite{asv96} for a bibliography on fuzzy automata, grammars, and languages. In this chapter, we will present some part of the theory of wta over bounded lattices which follows from the results of the previous chapters.
This attempt is reasonable, due to Observation \ref{obs:bounded-lattice-is-biloc-fin-strong-bimonoid}(2) and~(3), i.e.,
\begin{compactitem}
\item each bounded lattice is a particular bi-locally finite and commutative strong bimonoid, and
\item each distributive bounded lattice is a particular locally finite and commutative semiring,
\end{compactitem}
respectively. 
We will use the above facts without reference in the proofs of the corollaries and theorems in this chapter. For each lattice-oriented result that we present here, we will only refer to the covering strong bimonoid-oriented result of some previous chapter; there the reader can find references to the literature.

\label{p:convention-L-wta}
\begin{quote}\emph{In this chapter, we let $\sfL = (L,\vee,\wedge,\0,\1)$ be an arbitrary bounded lattice.\footnote{We use the capital letter ``L'' for the following different meanings: $\sfL$ for some bounded lattice with carrier set $L$, $L_i$ for some tree language, and $\LL(A)$ for the tree language recognized by some fta $A$. It will always be clear from the context which meaning is appropriate.}}
\end{quote}

\section{Definition of wta over bounded lattices}

Since a bounded lattice is a particular strong bimonoid, Section \ref{sec:basic-defininition-wta} provides the definition of a $(\Sigma,\sfL)$-wta. This definition coincides with the definition of weighted tree automata over bounded lattices as it is given in the literature, e.g. \cite{esiliu07} (apart from notational variations).
In the same way, the notions of crisp-deterministic $(\Sigma,\sfL)$-wta and  bu-deterministic $(\Sigma,\sfL)$-wta are defined. The definitions of the run semantics and initial algebra semantics of a $(\Sigma,\sfL)$-wta are also given in Section \ref{sec:basic-defininition-wta}. For the sake of convenience, we repeat their main parts here.

Let $\cA=(Q,\delta,F)$ be a $(\Sigma,\sfL)$-wta. 
We note that, due to \eqref{equ:weight-of-run}, for each $\xi = \sigma(\xi_1,\ldots,\xi_k)$ and $\rho \in \R_\cA(\xi)$, we have
\begin{equation}\label{equ:weight-of-run-B}
\wt_\cA(\xi,\rho) = \Big( \bigwedge_{i\in [k]} \wt_\cA(\xi_i,\rho|_i)\Big) \wedge \delta_k\big(\rho(1) \cdots \rho(k),\sigma,\rho(\varepsilon)\big) \enspace,
\end{equation}
and the run semantics of $\cA$ is defined, for each $\xi \in \T_\Sigma$, by
\begin{equation*}
  \runsem{\cA}(\xi) = \bigvee_{\rho \in \R_\cA(\xi)}\wt(\xi,\rho) \wedge F_{\rho(\varepsilon)}\enspace. \label{equ:runsem-lattice} 
  \end{equation*}

  Moreover, due to \eqref{eq:delta-A-definition}, the interpretation function $\delta_\cA$ of the vector algebra $\V(\cA) = (L^Q,\delta_\cA)$ is defined, for every $k \in \mathbb{N}$, $\sigma \in \Sigma^{(k)}$, $v_1,\ldots,v_k \in L^Q$, and $q \in Q$, by
  \begin{equation}
\delta_\cA(\sigma)(v_1,\dots,v_k)_q 
  = \bigvee_{q_1\cdots q_k \in Q^k} \Big(\bigwedge_{i\in[k]} (v_i)_{q_i}\Big) \wedge \delta_k(q_1\cdots q_k,\sigma,q) \enspace.
  \end{equation}

We recall that the unique $\Sigma$-algebra homomorphism from $\T_\Sigma$ to  $\V(\cA)$ is denoted by $\h_\cA$. Then the initial algebra semantics of $\cA$ is defined, for each $\xi \in \T_\Sigma$, by  
\begin{equation*}
  \initialsem{\cA}(\xi) = \bigvee_{q\in Q}\h_\cA(\xi)_q \wedge F_{q}\enspace.  
  \end{equation*}

 For instance, in \cite{asv03,esiliu07} and in \cite{gho22} the initial algebra semantics was used (where $\sfL$ is a $\sigma$-complete distributive lattice and a $\sigma$-complete orthomodular lattice, respectively). In \cite{bel02} the run semantics was used (for wsa where $\sfL$ is a complete distributive lattice, cf. the remark on page \pageref{page:Belohlavek}).

As we have seen in Example \ref{ex:lat-wta-sem-diff}, there exists a  $(\Sigma,\Nfive)$-wta $\cA$ such that $\runsem{\cA} \not= \initialsem{\cA}$. We recall that $\cA$ is not bu-deterministic and that $\Nfive$ (cf. Figure \ref{fig:lattices-N5-M3-Graetzer}) is not distributive. However, in the following cases 
run semantics and initial algebra semantics coincide.

\begin{corollary-rect}\rm  \label{cor:bu-det-lattice:init=run}  Let $\Sigma$ be a ranked alphabet. Moreover, let $\sfL$ be a bounded lattice. For each $(\Sigma,\sfL)$-wta~$\cA$  the following two statements hold.
  \begin{compactenum}
  \item[(1)] If $\cA$ is bu-deterministic, then $\runsem{\cA} = \initialsem{\cA}$.
  \item[(2)] If $\sfL$ is distributive, then $\runsem{\cA} = \initialsem{\cA}$.
    \end{compactenum}
  \end{corollary-rect}
  \begin{proof} Proof of (1) and (2) follows from Theorem \ref{thm:bu-det:init=run} and Corollary \ref{cor:semiring-run=init}(2), respectively.
    \end{proof}

\label{page:convention-L-wta-cd}
\begin{quote}
    \emph{Due to Corollary \ref{cor:bu-det-lattice:init=run}, for each $(\Sigma,\sfL)$-wta $\cA$, if  $\cA$ is bu-deterministic or
    $\sfL$ is distributive, then we  write $\sem{\cA}$ instead of $\runsem{\cA}$ and $\initialsem{\cA}$. Moreover, if $\sfL$ is distributive, then for an i-recognizable or r-recognizable weighted tree language $r$,  we say that $r$ is recognizable and we denote the set $\Rec^{\mathrm{run}}(\Sigma,\sfL)$ (and hence, $\Rec^{\mathrm{init}}(\Sigma,\sfL)$) by $\Rec(\Sigma,\sfL)$.}
\end{quote}

\section{Crisp-determinization}

\begin{corollary-rect}\rm  \label{cor:B-valued-wta-rec-crisp-deterministic} Let $\Sigma$ be a ranked alphabet. Moreover, let $\sfL$ be a bounded lattice. For each $(\Sigma,\sfL)$-wta~$\cA$ the following statements hold.
\begin{compactitem}
\item[(1)] We can construct a crisp-deterministic $(\Sigma,\sfL)$-wta $\cB$ such that $\runsem{\cA}=\sem{\cB}$.
\item[(2)] If $\sfL$ is locally finite, then we can construct a crisp-deterministic $(\Sigma,\sfL)$-wta $\cB$ such that $\initialsem{\cA}=\sem{\cB}$.
\end{compactitem}
Thus, in particular, $\Rec^{\mathrm{run}}(\Sigma,\sfL) = \cdRec(\Sigma,\sfL)$ and, if $\sfL$ is locally finite, then $\Rec^{\mathrm{init}}(\Sigma,\sfL) = \cdRec(\Sigma,\sfL)$.
\end{corollary-rect}
\begin{proof} Proof of (1): Since $\sfL$ is an additively idempotent  and multiplicatively locally finite strong bimonoid, we can apply Corollary \ref{thm:cd-for-l-valued-wta} and obtain the desired crisp-deterministic $(\Sigma,\sfL)$-wta $\cB$ with  $\runsem{\cA}=\sem{\cB}$.

  Proof of (2): It follows from Theorem \ref{thm:subset-method-lf-strong-bm}(1).
  
Finally, the equality of the sets follow from Statements (1) and (2) and the trivial fact that $\cdRec(\Sigma,\sfL) \subseteq \Rec^{\mathrm{run}}(\Sigma,\sfL) \cap \Rec^{\mathrm{init}}(\Sigma,\sfL)$.
\end{proof}

Corollary \ref{cor:B-valued-wta-rec-crisp-deterministic}(2)  can be applied, e.g., to the locally finite bounded lattice $(\mathbb{N}_\infty, \max,\min,0,\infty)$. Moreover, it can be applied to  each finite lattice and to each distributive bounded lattice.

A direct consequence of Corollary \ref{cor:B-valued-wta-rec-crisp-deterministic} and Corollary \ref{cor:crisp-det-wta-B1-B2} is the stability of the set of run recognizable weighted tree languages under changing the bounded lattice $\sfL_1$ into the bounded lattice $\sfL_2$ if $\sfL_1$ and $\sfL_2$ have the same carrier set (as, e.g., $\Nfive$ and $\Mthree$) and similarly for initial algebra recognizable weighted tree languages over locally finite bounded lattices. Moreover, the expressive power of wta with run semantics over bounded lattices is the same as that of wta with initial algebra semantics over locally finite bounded lattices \cite{dro22}.

\begin{corollary}\rm \label{cor:bounded-lattice-stability-run-recog} Let $\sfL_1$ and $\sfL_2$ be bounded lattices with the same carrier set. Then the following three statements hold.
  \begin{compactenum}
  \item[(1)] $\Rec^{\mathrm{run}}(\Sigma,\sfL_1) = \cdRec(\Sigma,\sfL_1) = \cdRec(\Sigma,\sfL_2) = \Rec^{\mathrm{run}}(\Sigma,\sfL_2)$.
  \item[(2)] If $\sfL_1$ and $\sfL_2$ are locally finite, then $\Rec^{\mathrm{init}}(\Sigma,\sfL_1) = \cdRec(\Sigma,\sfL_1) = \cdRec(\Sigma,\sfL_2) = \Rec^{\mathrm{init}}(\Sigma,\sfL_2)$.

    \item[(3)] If $\sfL_2$ is locally finite, then $\Rec^{\mathrm{run}}(\Sigma,\sfL_1)  = \Rec^{\mathrm{init}}(\Sigma,\sfL_2)$.
    \end{compactenum}
      \end{corollary}
         
\begin{proof} Proof of (1):  The equalities $\Rec^{\mathrm{run}}(\Sigma,\sfL_1) = \cdRec(\Sigma,\sfL_1)$ and $\cdRec(\Sigma,\sfL_2) = \Rec^{\mathrm{run}}(\Sigma,\sfL_2)$ follow from Corollary \ref{cor:B-valued-wta-rec-crisp-deterministic}(1). The equality $\cdRec(\Sigma,\sfL_1) = \cdRec(\Sigma,\sfL_2)$ follows from  Corollary \ref{cor:crisp-det-wta-B1-B2}  (and Theorem \ref{thm:bu-det:init=run}, and the convention on page \pageref{p:convention-denotation-of-classes}).

        Proof of (2): The proof is analogous to the proof of (1) but now we use Corollary \ref{cor:B-valued-wta-rec-crisp-deterministic}(2).

        Proof of (3): By (1) we have $\Rec^{\mathrm{run}}(\Sigma,\sfL_1) = \cdRec(\Sigma,\sfL_2)$. Then, by (2), we obtain  $\Rec^{\mathrm{run}}(\Sigma,\sfL_1) = \Rec^{\mathrm{init}}(\Sigma,\sfL_2)$. 
      \end{proof}

We recall from Section \ref{sec:rec-step-mappings} that a $(\Sigma,\sfL)$-weighted tree language $r$ is a recognizable step mapping if there exist $n \in \mathbb{N}_+$, $b_1,\ldots,b_n \in L$, and recognizable $\Sigma$-tree languages $L_1,\ldots,L_n$ such that 
\begin{equation*}
r = \bigvee_{i \in [n]} b_i \cdot \chi(L_i)\enspace. 
\end{equation*}
The expression $b_i \cdot \chi(L_i)$ denotes the scalar multiplication of $b_i$ and $\chi(L_i)$ (see Section \ref{sect:weighted-sets-languages}) and is defined, for each $\xi \in \T_\Sigma$, by $(b_i \cdot \chi(L_i))(\xi) = b_i \wedge \chi(L_i)(\xi)$.

\begin{corollary-rect}\rm \label{cor:B-valued-wta-rec-step-mapping}  Let $\Sigma$ be a ranked alphabet. Moreover, let  $\sfL=(L,\vee,\wedge,\0,\1)$ be a bounded lattice. For each $(\Sigma,\sfL)$-wta $\cA$, the following statements hold.
\begin{compactitem}
\item[(1)] $\runsem{\cA}$ is a recognizable step mapping.
\item[(2)] If $\sfL$ is locally finite, then $\initialsem{\cA}$ is a recognizable step mapping.
\end{compactitem}
Moreover, in (1) we can construct $n\in \mathbb{N}_+$, $b_1,\ldots,b_n\in L$, and $\Sigma$-fta $A_1,\ldots,A_n$ such that $\runsem{\cA}=\bigvee_{i \in [n]} b_i \cdot \chi(\LL(A_i))$. The same holds for $\initialsem{\cA}$ in (2).
\end{corollary-rect}
\begin{proof} The proof follows from Corollary \ref{cor:B-valued-wta-rec-crisp-deterministic} and Theorem \ref{thm:crisp-det-algebra} (A)$\Leftrightarrow$(B). 
  \end{proof}

\begin{corollary-rect}\label{cor:crisp-det=rec-step} \rm Let $\Sigma$ be a ranked alphabet. Moreover, let $\sfL=(L,\vee,\wedge,\0,\1)$ be a bounded lattice and let $r: \T_\Sigma \to L$. Then the following two statements are equivalent.
  \begin{compactenum}
  \item[(A)]  We can construct a $(\Sigma,\sfL)$-wta $\cA$ such that $r=\runsem{\cA}$.
  \item[(B)] $r$ is a recognizable step mapping and we can construct $n\in\mathbb{N}_+$, $b_1,\ldots,b_n\in L$ and $\Sigma$-fta $A_1,\ldots,A_n$ such that $r=\bigvee_{i=1}^n b_i\cdot \chi(\LL(A_i))$.
  \end{compactenum}
  Moreover, if $\sfL$ is distributive, then $\runsem{\cA}$  in condition (A)  can be replaced by $\sem{\cA}$.
\end{corollary-rect}
\begin{proof} Proof of (A)$\Rightarrow$(B): This follows from Corollary \ref{cor:B-valued-wta-rec-crisp-deterministic}(1) and Theorem \ref{thm:crisp-det-algebra} (A)$\Rightarrow$(B).

  \
  
    Proof of (B)$\Rightarrow$(A): By Theorem \ref{thm:crisp-det-algebra}(B)$\Rightarrow$(A), we can construct a crisp-deterministic $(\Sigma,\sfL)$-wta $\cA$ such that $r = \sem{\cA}$. By the convention on page \pageref{page:convention-L-wta-cd}, $\sem{\cA}$ is an abbreviation of  $\runsem{\cA}$.

Let  $\sfL$ be distributive. Then the statement follows from Corollary~\ref{cor:bu-det-lattice:init=run}(2).
 \end{proof}

 \section[Relationship between several sets of recognizable wtl]{Relationship between several sets of recognizable weighted tree languages}
\label{sec:relationship-L-wta}

Here we study the relationship between several sets of r-recognizable, i-recognizable, or crisp-deterministically recognizable weighted tree languages where the weight algebra is a bounded lattice,  locally finite bounded lattice, distributive bounded lattice, or finite chain. By combining these modes of recognizability and  weight algebras, we obtain twelve sets of recognizable  weighted tree languages.

It will turn out that eleven of the twelve sets are equal; let us denote this set by $C$. The exception is the set  of i-recognizable weighted tree languages  over bounded lattices. Moreover, we show that $C$ is a subset of the latter set and, if the ranked alphabet is large enough, then the inclusion is strict. 
Thus, roughly speaking, using initial algebra semantics over bounded lattices, one can specify strictly more weighted tree languages than using another combination.
The results of this section are taken from~\cite{fulvog23}.

More precisely, we consider the set
          \[
\MMM = \{\Rec^y(\Sigma,Z) \mid y \in \{\mathrm{run},\mathrm{init}\}, Z \in \{\mathrm{BL},\mathrm{lfBL},\mathrm{dBL},\mathrm{FC}\}\}\cup \{ \cdRec(\Sigma,Z) \mid Z \in \{\mathrm{BL},\mathrm{lfBL},\mathrm{dBL},\mathrm{FC}\}\},
\]
where
\index{RecinitSigmaBL@$\Rec^{\mathrm{init}}(\Sigma,\mathrm{BL}$)}
\index{RecrunSigmaBL@$\Rec^{\mathrm{run}}(\Sigma,\mathrm{BL}$)}
\index{RecinitSigmalfBL@$\Rec^{\mathrm{init}}(\Sigma,\mathrm{lfBL}$)}
\index{RecrunSigmaBL@$\Rec^{\mathrm{run}}(\Sigma,\mathrm{lfBL}$)}
\index{RecinitSigmadBL@$\Rec^{\mathrm{init}}(\Sigma,\mathrm{dBL}$)}
\index{RecrunSigmadBL@$\Rec^{\mathrm{run}}(\Sigma,\mathrm{dBL}$)}
\index{RecinitSigmadBL@$\Rec^{\mathrm{init}}(\Sigma,\mathrm{FC}$)}
\index{RecrunSigmadBL@$\Rec^{\mathrm{run}}(\Sigma,\mathrm{FC}$)}
\index{cdRecSigmaBL@$\cdRec(\Sigma,\mathrm{BL})$}
\index{cdRecSigmalfBL@$\cdRec(\Sigma,\mathrm{lfBL})$}
\index{cdRecSigmadBL@$\cdRec(\Sigma,\mathrm{dBL})$}
\index{cdRecSigmadBL@$\cdRec(\Sigma,\mathrm{FC})$}
\begin{compactitem}
\item $\mathrm{BL}$, $\mathrm{lfBL}$, $\mathrm{dBL}$, $\mathrm{FC}$ denote the sets of bounded lattices, locally finite bounded lattices, distributive bounded lattices, and finite chains,
\item $\Rec^y(\Sigma,Z)= \bigcup_{\sfL \in Z} \Rec^y(\Sigma,\sfL)$, i.e., the set of all mappings $r:\T_\Sigma \to L$ where $L$ is the carrier set of some $\sfL \in Z$ and $r$ is a $y$-recognizable $(\Sigma,\sfL)$-weighted tree language,  and
\item $\cdRec(\Sigma,Z)= \bigcup_{\sfL \in Z} \cdRec(\Sigma,\sfL)$, i.e., the set of all mappings $r:\T_\Sigma \to L$ where $L$ is the carrier set of some $\sfL \in Z$ and $r$ is a crisp-deterministically recognizable $(\Sigma,\sfL)$-weighted tree language.
  \end{compactitem}

  Our goal is to present the inclusion diagram of $\MMM$ (cf. Theorem \ref{thm:Hasse-L-2}), i.e., the Hasse diagram of $\MMM$ with set inclusion as partial order. 
    
  By definition we have $\mathrm{FC}\subset \mathrm{dBL}$ and by Observation \ref{obs:Euler-diagram-including-loc-finite}, we have $\mathrm{dBL} \subset \mathrm{lfBL} \subset \mathrm{BL}$. Thus
  \begin{equation}\label{eq:y-run-upwards-2} 
    \text{ for each $y \in  \{\mathrm{run},\mathrm{init}\}$, we have $\Rec^y(\Sigma,\mathrm{FC}) \subseteq \Rec^y(\Sigma,\mathrm{dBL}) \subseteq \Rec^y(\Sigma,\mathrm{lfBL}) \subseteq \Rec^y(\Sigma,\mathrm{BL})$}
  \end{equation}
and  we have
    \begin{equation}\label{eq:y-run-upwards-2-cd}
    \text{ $\cdRec(\Sigma,\mathrm{FC}) \subseteq \cdRec(\Sigma,\mathrm{dBL}) \subseteq \cdRec(\Sigma,\mathrm{lfBL}) \subseteq \cdRec(\Sigma,\mathrm{BL})$.}
  \end{equation}

  In Figure \ref{fig:relationship-between-nine-sets} (ignoring the ovals for the time being) we show all the twelve elements of $\MMM$, organized according to their known subset relationships; these relationships follow from (a) Equations \eqref{eq:y-run-upwards-2} and \eqref{eq:y-run-upwards-2-cd},  (b) the trivial fact that each crisp-deterministic  wta is a wta, and (c) Theorem \ref{thm:bu-det:init=run}. We note that this is not a Hasse diagram; in particular, the edges do not show strict inclusions but merely inclusions.

\begin{figure}
\centering
\begin{tikzpicture}[scale=1, every node/.style={transform shape},
					node distance=1cm and 4.5cm,
					rounded/.style={rounded corners=0.65cm*1}] 

  \node[anchor=center] (l1) {$\Rec^{\mathrm{run}}(\Sigma,\mathrm{BL})$};
  \node[below = 2cm of l1.center,anchor=center] (l2) {$\Rec^{\mathrm{run}}(\Sigma,\mathrm{lfBL})$};
  \node[below = 2cm of l2.center,anchor=center] (l3) {$\Rec^{\mathrm{run}}(\Sigma,\mathrm{dBL})$};
  \node[below = 2cm of l3.center,anchor=center] (l4) {$\Rec^{\mathrm{run}}(\Sigma,\mathrm{FC})$};
  \node[below right = of l1.center,anchor=center] (c1) {$\cdRec(\Sigma,\mathrm{BL})$};
  \node[below right = of l2.center,anchor=center] (c2) {$\cdRec(\Sigma,\mathrm{lfBL})$};
  \node[below right = of l3.center,anchor=center] (c3) {$\cdRec(\Sigma,\mathrm{dBL})$};
  \node[below right = of l4.center,anchor=center] (c4) {$\cdRec(\Sigma,\mathrm{FC})$};
  \node[above right = of c1.center,anchor=center] (r1) {$\Rec^{\mathrm{init}}(\Sigma,\mathrm{BL})$};
  \node[above right = of c2.center,anchor=center] (r2) {$\Rec^{\mathrm{init}}(\Sigma,\mathrm{lfBL})$};
  \node[above right = of c3.center,anchor=center] (r3) {$\Rec^{\mathrm{init}}(\Sigma,\mathrm{dBL})$};
  \node[above right = of c4.center,anchor=center] (r4) {$\Rec^{\mathrm{init}}(\Sigma,\mathrm{FC})$};
  
  \draw (c1) to (c2) to (r2) to (r1) to (c1) to (l1) to (l2) to (c2) to (c3) to (l3) to (l2);
  \draw (c3) to (r3) to (r2);
  \draw (l3) to (l4) to (c4) to (r4) to (r3);
  \draw (c3) to (c4);
    
  \begin{scope}[node distance=0.55cm and 0.4cm]	
    \coordinate[above left  = of l1, xshift=0.25cm] (l1a);
    \coordinate[above right = of c1, yshift=-0.4cm] (c1a);
    \coordinate[below right = of c1, xshift=-0.25cm] (c1b);
    \coordinate[below left = of l1, yshift=0.4cm] (l1b);
    \draw[rounded] (l1a)--(c1a)--(c1b)--(l1b)--cycle;
    \coordinate[above left  = of l2, xshift=0.25cm] (l2a);
    \coordinate[above = of c2, yshift=-0.2cm] (c2a);
    \coordinate[above right = of r2, xshift=-0.25cm] (r2a);
    \coordinate[below right = of r2, yshift=0.4cm] (r2b);
    \coordinate[below = of c2, yshift=0.2cm] (c2b);
    \coordinate[below left = of l2, yshift=0.4cm] (l2b);
    \draw[rounded] (l2a)--(c2a)--(r2a)--(r2b)--(c2b)--(l2b)--cycle;
    \coordinate[above left  = of l3, xshift=0.25cm] (l3a);
    \coordinate[above = of c3,yshift=-0.2cm] (c3a);
    \coordinate[above right = of r3, xshift=-0.25cm] (r3a);
    \coordinate[below right = of r3, yshift=0.4cm] (r3b);
    \coordinate[below = of c3, yshift=0.2cm] (c3b);
    \coordinate[below left = of l3, yshift=0.4cm] (l3b);
    \draw[rounded] (l3a)--(c3a)--(r3a)--(r3b)--(c3b)--(l3b)--cycle;
    \coordinate[above left  = of l4, xshift=0.25cm] (l4a);
    \coordinate[above = of c4,yshift=-0.2cm] (c4a);
    \coordinate[above right = of r4, xshift=-0.25cm] (r4a);
    \coordinate[below right = of r4, yshift=0.4cm] (r4b);
    \coordinate[below = of c4, yshift=0.2cm] (c4b);
    \coordinate[below left = of l4, yshift=0.4cm] (l4b);
    \draw[rounded] (l4a)--(c4a)--(r4a)--(r4b)--(c4b)--(l4b)--cycle;
   \end{scope}

\end{tikzpicture}
\caption{\label{fig:relationship-between-nine-sets} Illustration of some of the set inclusions (vertical and diagonal lines) and some of the equalities (ovals) of the elements of $\MMM$.}
\end{figure}
 
From the trivial fact that $\cdRec(\Sigma,\mathrm{BL}) \subseteq \Rec^{\mathrm{run}}(\Sigma,\mathrm{BL}) \cap \Rec^{\mathrm{init}}(\Sigma,\mathrm{lfBL})$ and from Corollary \ref{cor:B-valued-wta-rec-crisp-deterministic}, we directly obtain four equalities (corresponding to the four ovals in Figure \ref{fig:relationship-between-nine-sets})

\begin{corollary}\rm \label{cor:three-ovals} \cite{fulvog23} The following four statements hold.
  \begin{compactenum}
  \item[(1)] $\Rec^{\mathrm{run}}(\Sigma,\mathrm{BL}) = \cdRec(\Sigma,\mathrm{BL})$.
  \item[(2)] $\Rec^{\mathrm{run}}(\Sigma,\mathrm{lfBL}) = \cdRec(\Sigma,\mathrm{lfBL}) = \Rec^{\mathrm{init}}(\Sigma,\mathrm{lfBL})$.
    \item[(3)] $\Rec^{\mathrm{run}}(\Sigma,\mathrm{dBL}) = \cdRec(\Sigma,\mathrm{dBL}) = \Rec^{\mathrm{init}}(\Sigma,\mathrm{dBL})$.  
     \item[(4)] $\Rec^{\mathrm{run}}(\Sigma,\mathrm{FC}) = \cdRec(\Sigma,\mathrm{FC}) = \Rec^{\mathrm{init}}(\Sigma,\mathrm{FC})$.  \hfill$\Box$
    \end{compactenum}
\end{corollary}

Next we prove that $\cdRec(\Sigma,\mathrm{BL}) \subseteq \cdRec(\Sigma,\mathrm{FC})$. 

\begin{lemma}\rm  \label{lm:cdRec-in-both-2} \cite{dro22}  \cite{fulvog23} For every  crisp-deterministic $(\Sigma,\sfL)$-wta $\cA$ we can construct a finite chain $\sfL'$ with carrier set $\mathrm{wts}(\cA)\cup\{\0\}$ and a crisp-deterministic $(\Sigma,\sfL')$-wta $\cB$ such that $\sem{\cA}=\sem{\cB}$. Thus, in particular, $\cdRec(\Sigma,\mathrm{BL}) \subseteq \cdRec(\Sigma,\mathrm{FC})$.
\end{lemma}
\begin{proof} Let $\cA=(Q,\delta,F)$. We note that $\1 \in \mathrm{wts}(\cA)$ because $\Sigma^{(0)} \not= \emptyset \not= Q$ and $\cA$ is crisp-deterministic.  
  We construct the finite chain $\sfL' =(\mathrm{wts}(\cA)\cup\{\0\},\le,\0,\1)$ where $\le$ is an arbitrary linear order on $\mathrm{wts}(\cA)\cup \{\0\}$.
   It is obvious that, for the crisp-deterministic $(\Sigma,\sfL')$-wta $\cB=(Q,\delta,F)$, we have $\sem{\cA}=\sem{\cB}$. This also proves that $\cdRec(\Sigma,\mathrm{BL}) \subseteq \cdRec(\Sigma,\mathrm{FC})$.
\end{proof}

From the previous results, we obtain an intermediate answer to our goal.

\begin{corollary}\label{cor:five-sets-2}\rm \cite{fulvog23}  Let $\MMM'=\MMM\setminus \{\Rec^{\mathrm{init}}(\Sigma,\mathrm{BL}) \}$. The set $\MMM'$ contains exactly one element, i.e., the eleven sets:
  \begin{compactitem}
  \item  $\Rec^{\mathrm{run}}(\Sigma,\mathrm{BL})$, $\Rec^{\mathrm{run}}(\Sigma,\mathrm{lfBL})$, $\Rec^{\mathrm{run}}(\Sigma,\mathrm{dBL})$, $\Rec^{\mathrm{run}}(\Sigma,\mathrm{FC})$, 
  \item $\Rec^{\mathrm{init}}(\Sigma,\mathrm{lfBL})$, 
    $\Rec^{\mathrm{init}}(\Sigma,\mathrm{dBL})$, $\Rec^{\mathrm{init}}(\Sigma,\mathrm{FC})$, and
  \item $\cdRec(\Sigma,\mathrm{BL})$, $\cdRec(\Sigma,\mathrm{lfBL})$, $\cdRec(\Sigma,\mathrm{dBL})$, and $\cdRec(\Sigma,\mathrm{FC})$
      \end{compactitem}
  are equal.
 \end{corollary}
  \begin{proof} By  Lemma \ref{lm:cdRec-in-both-2}, $\cdRec(\Sigma,\mathrm{BL})\subseteq \cdRec(\Sigma,\mathrm{FC})$. Hence, by the inclusions shown in Figure \ref{fig:relationship-between-nine-sets} and Corollary \ref{cor:three-ovals}, all elements of $\MMM'$ are equal.
\end{proof}

  Next we prove that $\Rec^{\mathrm{init}}(\Sigma,\mathrm{BL}) \setminus \Rec^{\mathrm{init}}(\Sigma,\mathrm{lfBL}) \not= \emptyset$ under some mild conditions on $\Sigma$. For this, we consider the bounded lattice $\sfFL$ in  Example \ref{ex:lattices}(\ref{ex:FL2+2}).

\begin{lemma}\label{lm:not-in-BL-2} \rm (cf. \cite{fulvog23}) If $\Sigma$ is branching, then we can construct  a $(\Sigma,\sfFL)$-wta $\cA$ such that $\initialsem{\cA} \not\in \Rec^{\mathrm{init}}(\Sigma,\mathrm{lfBL})$. In particular, for such a $\Sigma$ we have  $\Rec^{\mathrm{init}}(\Sigma,\mathrm{BL}) \setminus \Rec^{\mathrm{init}}(\Sigma,\mathrm{lfBL}) \not= \emptyset$.
\end{lemma}
\begin{proof} We consider the finite generating set $A = \{a,b,c,d\}$  of $\sfFL$ (cf. Figure \ref{fig:FL2+2}), where $a<b$ and $c<d$. Since $b \vee d = \1$ and $a \wedge c = \0$ and $\FL$ is generated by $A$, we have $\langle A \rangle_{\{\vee,\wedge,\0,\1\}} = \langle A \rangle_{\{\vee,\wedge\}} =  \FL$, which is an infinite set. By Theorem~\ref{thm:not-monadic-wta-can-compute-closure-of-finite-subset}, we can construct  a $(\Sigma,\sfFL)$-wta $\cA$ such that $\im(\initialsem{\cA}) = \langle A \rangle_{\{\vee,\wedge,\0,\1\}}$, i.e., $\im(\initialsem{\cA})$ is an infinite set.

On the other hand, by Corollary \ref{cor:B-valued-wta-rec-step-mapping}, each element  $r \in \Rec^{\mathrm{init}}(\Sigma,\mathrm{lfBL})$ is a recognizable step mapping, and hence $\im(r)$ is finite. Thus $\initialsem{\cA} \not\in \Rec^{\mathrm{init}}(\Sigma,\mathrm{lfBL})$.
\end{proof}

From Corollary \ref{cor:five-sets-2}, the fact that $\Rec^{\mathrm{init}}(\Sigma,\mathrm{lfBL}) \subseteq \Rec^{\mathrm{init}}(\Sigma,\mathrm{BL})$, and Lemma \ref{lm:not-in-BL-2} we obtain the final picture of the Hasse diagram for $\MMM$.

\begin{theorem-rect}\label{thm:Hasse-L-2}{\rm (cf. \cite{fulvog23})} Let $\Sigma$ be a ranked alphabet. Moreover, let
  \[\MMM = \{\Rec^y(\Sigma,Z) \mid y \in \{\mathrm{run},\mathrm{init}\}, Z \in \{\mathrm{BL},\mathrm{lfBL},\mathrm{dBL},\mathrm{FC}\}\}\cup \{ \cdRec(\Sigma,Z) \mid Z \in \{\mathrm{BL},\mathrm{lfBL},\mathrm{dBL},\mathrm{FC}\}\}\enspace.
  \]
  Then the following three statements hold.
  \begin{compactenum}
    \item[(1)] $\Rec^{\mathrm{run}}(\Sigma,\mathrm{BL}) =\Rec^{\mathrm{run}}(\Sigma,\mathrm{lfBL}) =\Rec^{\mathrm{run}}(\Sigma,\mathrm{dBL})=\Rec^{\mathrm{run}}(\Sigma,\mathrm{FC}) =\Rec^{\mathrm{init}}(\Sigma,\mathrm{lfBL})=\Rec^{\mathrm{init}}(\Sigma,\mathrm{dBL})=\Rec^{\mathrm{init}}(\Sigma,\mathrm{FC})=\cdRec(\Sigma,\mathrm{BL})=\cdRec(\Sigma,\mathrm{lfBL})=\cdRec(\Sigma,\mathrm{dBL}) =\cdRec(\Sigma,\mathrm{FC})$.
       We denote this set by $C$.
    \item[(2)] $C \subseteq \Rec^{\mathrm{init}}(\Sigma,\mathrm{BL})$.
    \item[(3)] If  $\Sigma$ is branching, then we have $C \subset \Rec^{\mathrm{init}}(\Sigma,\mathrm{BL})$. In this case, the Hasse diagram for $\MMM$ with set inclusion as partial order consists of the two nodes $C$ and~$\Rec^{\mathrm{init}}(\Sigma,\mathrm{BL})$.
      \end{compactenum}
\end{theorem-rect}

\section{Support}

  \begin{corollary-rect}\rm  \label{cor:B-valued-wta-support}  Let $\Sigma$ be a ranked alphabet. Moreover, let $\sfL$ be a bounded lattice. For each $(\Sigma,\sfL)$-wta $\cA$ the following statements hold.
\begin{compactitem}
\item[(1)] The $\Sigma$-tree language $\supp(\runsem{\cA})$ is recognizable.
\item[(2)] If $\sfL$ is locally finite, then  the  $\Sigma$-tree language $\supp(\initialsem{\cA})$ is recognizable.
\end{compactitem}
Moreover, for the tree language $\supp(\runsem{\cA})$ in (1), we can construct a $\Sigma$-fta which recognizes that tree language.
The same holds for the tree languages $\supp(\initialsem{\cA})$ in (2). 
\end{corollary-rect}  
\begin{proof} Proof of (1) and (2): They follow from Corollary \ref{thm:support-thm-from-preimage}(3) and (2), respectively. We note that (1) also follows from Theorem \ref{thm:Kirsten}(a) because $\sfL$ is zero-sum free.

Now consider the tree language $\supp(\runsem{\cA})$ in (1). By Corollary \ref{cor:B-valued-wta-rec-crisp-deterministic}(1), we can construct a crisp-deterministic $(\Sigma,\sfL)$-wta $\cB$ such that $\runsem{\cA}=\sem{\cB}$.
Then $\supp(\runsem{\cA})=\T_\Sigma \setminus \sem{\cB}^{-1}(\0)$. By Theorem \ref{thm:crisp-det-algebra}(C), we can construct a $\Sigma$-fta which recognizes $\sem{\cB}^{-1}(\0)$. Finally, it is easy to construct a $\Sigma$-fta which recognizes the complement tree language $\T_\Sigma \setminus \sem{\cB}^{-1}(\0)$ (cf. e.g \cite[Thm. 2.4.2]{gecste84}).

By using Theorem \ref{thm:Kirsten-comm-idemp}, we can give an alternative proof for the constructability  of a $\Sigma$-fta which recognizes $\supp(\runsem{\cA})$ in (1). Moreover, by using Corollary \ref{cor:B-valued-wta-rec-crisp-deterministic}(2) and Theorem~\ref{thm:crisp-det-algebra}(C), we can give an alternative proof for the constructability a $\Sigma$-fta which recognizes 
$\supp(\initialsem{\cA})$ in (2).
\end{proof}

  \section{Closure results}

  Here we instantiate the closure results of Chapter \ref{ch:closure-properties} to the particular case where $\sfL$ is a (distributive) bounded lattice (cf. Figure \ref{fig:table-summary-closure} for the more general cases).

  \begin{corollary-rect}\label{cor:closure-Rec-bounded-lattices} \rm 
  Let $\Sigma$ be a ranked alphabet. Moreover, let $\sfL$ be a bounded lattice.   The following statements hold.
    \begin{compactenum}
      \item[(1)] The sets $\Rec^{\mathrm{run}}(\Sigma,\sfL)$ and  $\Rec^{\mathrm{init}}(\Sigma,\sfL)$ are closed under sum  (cf. Theorem \ref{thm:closure-sum}).

      \item[(2)] If $\sfL$ is distributive, then the set $\Rec(\Sigma,\sfL)$ is closed under
        \begin{compactitem}
        \item scalar multiplications (cf. Theorem \ref{thm:closure-scalar}),
          \item  Hadamard product (cf. Theorem \ref{thm:closure-Hadamard-product}),
          \item  top-concatenations (cf. Corollary \ref{cor:closure-under-top-concat-wta}),
\item tree concatenations (cf. Corollary \ref{cor:closure-tree-concatenation}),
\item    Kleene stars (cf. Corollary \ref{cor:closure-Kleene-star}), and
\item yield-intersection (cf. Theorem \ref{thm:BPS}).
  \end{compactitem}

\item[(3)] The sets $\Rec^{\mathrm{run}}(\Sigma,\_)$ and  $\Rec^{\mathrm{init}}(\Sigma,\_)$ are closed under homomorphisms between two bounded lattices (cf. Theorem \ref{thm:closure-sr-hom}).
  
\item[(4)] The set  $\Rec^{\mathrm{run}}(\_,\sfL)$ is closed under
  \begin{compactitem}
    \item tree relabelings  (cf. Theorem \ref{thm:closure-under-tree-relabeling}) and 
    \item linear, nondeleting, and productive tree homomorphisms (cf. Corollary \ref{cor:closure-REC-under-lin-nondel-prod-hom}).
  \end{compactitem}
  
\item[(5)] If $\sfL$ is distributive, then the set $\Rec(\_,\sfL)$ is closed under
  \begin{compactitem}
  \item inverse of linear tree homomorphisms (cf. Theorem \ref{thm:closure-under-inverse-tree-hom-2}),
    \item weighted projective bimorphisms (cf. Corollary \ref{cor:closure-REC-under-wbim}).
\end{compactitem}
      \end{compactenum}
    \end{corollary-rect}

    \section{Characterization by rational operations}

    We recall that an  operation on the set of $(\Sigma,\sfL)$-weighted tree languages is a rational operation if it is the sum, a tree concatenation, or a Kleene star.
Moreover, the set of  rational $(\Sigma,\sfL)$-weighted tree languages, denoted by $\Rat(\Sigma,\sfL)$, is the smallest set of $(\Sigma,\sfL)$-weighted tree languages which contains each polynomial $(\Sigma,\sfL)$-weighted tree language and is closed under the rational operations.

Due to the need for extra symbols at which tree concatenation can take place (cf. \cite[Sect.~3]{thawri68}), we have introduced in Section \ref{sec:rational-wtl-def} the concept of $\0$-extension and defined the sets of extended rational $(\Sigma,\sfL)$-weighted tree languages, denoted by $\Rat(\Sigma,\sfL)^{\mathrm{ext}}$, and the set of extended recognizable $(\Sigma,\sfL)$-weighted tree languages, denoted by $\Rec(\Sigma,\sfL)^{\mathrm{ext}}$, which formalize these extensions.

Then Theorem \ref{thm:Kleene} implies the following Kleene theorem for distributive bounded lattices.

\begin{corollary-rect} \label{cor:thm:Kleene-distr-bounded-lattices} \rm Let $\Sigma$ be a ranked alphabet. Moreover, let $\sfL$ be a distributive bounded lattice. Then 
\[\Rec(\Sigma,\sfL)^{\mathrm{ext}}=\Rat(\Sigma,\sfL)^{\mathrm{ext}}\enspace.\]
  \end{corollary-rect}

  \section{Characterization by elementary operations}
  \label{sec:Medvedjev-for-bounded-lattices}

Since $\sfL$ is a bounded lattice, we can show that,  the notions of representable, restricted representable, and $\times$-restricted representable (as defined in Chapter~\ref{ch:Medvedjev} for an arbitrary strong bimonoid) coincide. This results in M{\'e}dv{\'e}d{\'e}v's theorem for wta over bounded lattices (cf. Corollary \ref{cor:Medvedjev-L-valued-wta-str-bm}).
Its proof follows from Theorem \ref{cor:Medvedjev-str-bm} and the fact that
each bounded lattice is a bi-locally finite strong bimonoid.
                   
\begin{corollary-rect} \label{cor:Medvedjev-L-valued-wta-str-bm} \rm Let $\Sigma$ be a ranked alphabet. Moreover, let $\sfL=(L,\vee,\wedge,\0,\1)$ be a  bounded lattice and let $r: \T_\Sigma \to L$.  Then the following four statements are equivalent.
  \begin{compactenum}
  \item[(A)] We can construct a $(\Sigma,\sfL)$-wta $\cA$ such that $r=\runsem{\cA}$.
    \item[(B)] We can construct an $e \in \RepEx(\Sigma,\sfL)$ such that $r=\sem{e}$.
  \item[(C)] We can construct an $e \in \RepEx^r(\Sigma,\sfL)$ such that $r=\sem{e}$.
  \item[(D)] We can construct an $e \in \RepEx^{\times r}(\Sigma,\sfL)$ such that $r=\sem{e}$.
    \end{compactenum}
  \end{corollary-rect}

    \section{Characterization by weighted MSO-logic}
    \label{sec:L-wta-BET}

          In this situation where $\sfL$ is a bounded lattice, we can prove that run recognizability of a $(\Sigma,\sfL)$-weighted tree language is equivalent to its definability in $\MSOe(\Sigma,\sfL)$, and the transformations from wta to formula and vice versa are constructive (cf. Corollary \ref{thm:BET-bounded-lattices}). We recall the syntax of the formulas in $\MSOe(\Sigma,\sfL)$ from Section \ref{sec:weighted-conj-univ-quant}. It was given by the following EBNF with nonterminal $e$:
	\begin{equation}
		\textstyle e\,\, ::= \mathrm H(\kappa)\mid (\varphi\rhd e) \mid (e + e) \mid  (e \times e) \mid \bigplus\nolimits_x e\mid \bigplus\nolimits_X e\mid \bigtimes_x e \mid \bigtimes_X e \enspace. \label{eq:syntax-MSO-ext-L}
              \end{equation}
              In particular, no restriction is needed on the use of weighted conjunction, weighted first-order universal  quantification, and  weighted second-order universal quantification. 

              Since each bounded lattice is a commutative and bi-locally finite strong bimonoid, the proof of the next corollary follows from  Theorem \ref{thm:Buechi-comm-bi-loc-fin}.

          \begin{corollary-rect}\label{thm:BET-bounded-lattices}\rm Let $\Sigma$ be a ranked alphabet. Moreover, let $\sfL=(L,\vee,\wedge,\0,\1)$ be a bounded lattice and let $r: \T_\Sigma \to L$. Then the following four statements are equivalent.
        \begin{compactenum}
        \item[(A)] We can construct a $(\Sigma,\sfL)$-wta $\cB$ such that $r=\runsem{\cB}$.
        \item[(B)] We can construct a $(\Sigma,\sfL)$-recognizable step formula $e$ such that $\Free(e) = \emptyset$ and  $r = \sem{e}$.
        \item[(C)] We can construct a sentence $e \in \MSO(\Sigma,\sfL)$ such that $r = \sem{e}$
        \item[(D)] We can construct a sentence $e \in \MSOe(\Sigma,\sfL)$ such that $r = \sem{e}$.
\end{compactenum}
            \end{corollary-rect}

\section{Abstract families of weighted tree languages}

In Chapter \ref{ch:AFwtL} we have defined and investigated abstract families of weighted tree languages over commutative and $\sigma$-complete semirings. Since each $\sigma$-complete lattice is a  commutative and $\sigma$-complete semiring, we can immediately apply the main result of that chapter (cf. Theorem~\ref{thm:REGnB-smallest-princ-AFwtL}) to the case of $\sigma$-complete lattices.

For this, we recall that, for each $n \in \mathbb{N}$, an \emph{$(n,\sfL)$-weighted tree language} is a $(\Sigma,\sfL)$-weighted tree language for some ranked alphabet $\Sigma$ with $\maxrk(\Sigma) \le n$, and $\Reg(n,\sfL)$ denotes the set of all  regular $(n,\sfL)$-weighted tree languages. Moreover,  a family $\cL$ of $(n,\sfL)$-weighted tree languages is an abstract family of $(n,\sfL)$-weighted tree languages if $\cL$ is an $(n,\sfL)$-tree cone which is closed under the rational operations, i.e., under sum, tree concatenations, and Kleene stars.

\begin{corollary-rect}\label{thm:L-REGnB-smallest-princ-AFwtL}\rm Let $\sfL$ be a $\sigma$-complete lattice and $n \in \mathbb{N}$. Then $\Reg(n,\sfL)$ is the smallest principal abstract family of $(n,\sfL)$-weighted tree languages.
\end{corollary-rect}

%% file: book-wta-bib.bib
@book{ahohopull74,
author= {A.V. Aho and J.E. Hopcroft and J.D. Ullman},
title = {The Design and Analysis of Computer Algorithms},
publisher = {Addison-Wesley},
year = {1974}
}

@InProceedings{aisgas20,
  author =	{C. Aiswarya and P. Gastin},
  title =	{Weighted Tiling Systems for Graphs: Evaluation Complexity},
  booktitle =	{40th IARCS Annual Conference on Foundations of Software Technology and Theoretical Computer Science (FSTTCS 2020)},
  pages =	{34:1--34:17},
  series =	{Leibniz International Proceedings in Informatics (LIPIcs)},
  ISBN =	{978-3-95977-174-0},
  ISSN =	{1868-8969},
  year =	{2020},
  volume =	{182},
  editor =	{N. Saxena and S. Simon},
  publisher =	{Schloss Dagstuhl--Leibniz-Zentrum f{\"u}r Informatik},
  address =	{Dagstuhl, Germany},
  URL =		{https://drops.dagstuhl.de/opus/volltexte/2020/13275},
  URN =		{urn:nbn:de:0030-drops-132753},
  doi =		{10.4230/LIPIcs.FSTTCS.2020.34}
}

@Article{aleboz87,
  Title                    = {Weighted grammars and {K}leene's theorem},
  Author                   = {A. Alexandrakis and S. Bozapalidis},
  Journal                  = {Inform. Process. Lett.},
  Year                     = {1987},
  Number                   = {1},
  Pages                    = {1--4},
  Volume                   = {24}
}

@InProceedings{arndau76,
  Title                    = {Bi-transductions de for\^ets},
  Author                   = {A. Arnold and M. Dauchet},
  Booktitle                = {3rd Int. Coll. Automata, Languages and Programming (ICALP 1976)},
  Year                     = {1976},
  Editor                   = {S. Michaelson and R. Milner},
  Pages                    = {74--86},
  Publisher                = {Edinburgh University Press}
}

@Article{arndau82,
  Title                    = {Morphismes et bimorphismes d'arbres},
  Author                   = {A. Arnold and M. Dauchet},
  Journal                  = {Theoret. Comput. Sci.},
  Year                     = {1982},
  Number                   = {1},
  Pages                    = {33--93},
  Volume                   = {20}
}

@article{asv96,
author = {P.R.J. Asveld},
year = {1996},
title = {A bibliography on fuzzy automata, grammars and languages},
journal = {Bulletin of the European Association for Theoretical Computer Science},
volume = {58},
pages = {187--196}
}

@Article{asv03,
  Title                    = {Algebraic aspects of families of fuzzy languages},
  Author                   = {P.R.J. Asveld},
  Journal                  = {Theoret. Comput. Sci.},
  Year                     = {2003},
  Pages                    = {417--445},
  Volume                   = {293(2)}
}

@book{axl24,
title = {Linear Algebra Done Right},
author = {S. Axler},
year = {2024},
publisher = {Springer},
series = {Undergraduate Texts in Mathematics},
note = {4th edition}
}

@Book{baanip98,
  Title                    = {Term Rewriting and All That},
  Author                   = {F. Baader and T. Nipkow},
  Publisher                = {Cambridge University Press},
  Year                     = {1998}
}

@Article{bak79,
  Title                    = {Composition of top-down and bottom-up tree transductions},
  Author                   = {B.S. Baker},
  Journal                  = {Inf. Control},
  Year                     = {1979},
  Number                   = {2},
  Pages                    = {186--213},
  Volume                   = {41}
}

@inproceedings{balmoh15,
title  = {Learning Weighted Automata},
pages = {1--21},
series = {Lecture Notes in Computer Science},
author = {B. Balle and M. Mohri},
booktitle =  {6th Int. Conf. on Algebraic Informatics (CAI 2015)},
publisher = {Springer},
volume = {9270},
editor = {A. Maletti},
year = {2015}
}

@Article{barpersha61,   
  Title                    = {On {F}ormal {P}roperties of {S}imple {P}hrase {S}tructure {G}rammars},
  Author                   = {Y. {Bar--Hillel} and M. Perles and E. Shamir},
  Journal                  = {Z. Phonetik. Sprach. Komm.},
  Year                     = {1961},
  Pages                    = {143--172},
  Volume                   = {14}
}

@Article{baucou87,
author= {M. Bauderon and B. Courcelle}, 
title= {Graph expressions and graph rewritings},
journal= {Math. Syst. Theory},
year="1987",
volume="20",
number="1",
pages="83--127"
}

@article{bel02,
author = {R. B{\v{e}}lohl\'avek},
title = {Determinism and fuzzy automata},
journal = {Information Sciences},
Volume = {143},
Issues = {1–4},
Pages = {205--209},
year = {2002}
}

@miscellaneous{bel08,
author = {R. B{\v{e}}lohl\'avek},
title = {personal communication},
year = {2008}
}

@Book{ber79,
  Title                    = {Transductions and Context-Free Languages},
  Author                   = {J. Berstel},
  Publisher                = {Teubner Stuttgart},
  Year = {1979}
}

@book{bir93,
author = {G. Birkhoff},
title = {Lattice Theory},
publisher = {American Mathematical Society},
year = {1993},
series = {Colloquium Publications},
volume = {25},
note = {3rd edition}
}

@book{bir67,
author = {G. Birkhoff},
title = {Lattice Theory},
publisher = {American Mathematical Society},
year = {1967},
series = {Colloquium Publications},
volume = {25},
note = {3rd edition}
}

@inproceedings{bor03,
author = {B. Borchardt},
title = {The {M}yhill-{N}erode Theorem for Recognizable Tree Series},
booktitle = {7th Int. Conf. on Developments in Language Theory (DLT)},
series = {LNCS},
volume = {2710},
pages = {146--158},
year = {2003}
}

@Article{bormalsestepvog06,
  Title                    = {Cut sets as recognizable tree languages},
  Author                   = {B. Borchardt and A. Maletti and B. {\v{S}}e\v{s}elja and A. Tepav\v{c}evic and H. Vogler},
  Journal                  = {Fuzzy Sets and Systems},
  Year                     = {2006},
  Pages                    = {1560--1571},
  Volume                   = {157},
  number = {11}
}

@inproceedings{borpen11,
  address = {Debrecen, Hungary},
  author = {S. {Borgwardt} and R. {Pe{\~n}aloza}},
  booktitle = {13th Int. Conf. on Automata and Formal Languages (AFL 2011)},
  editor = {P. {D{\"o}m{\"o}si} and I. Szabolcs},
  pages = {108--122},
  publisher = {College of Ny{\'i}regyh{\'a}za},
  title = {The {I}nclusion {P}roblem for {W}eighted {A}utomata on {I}nfinite {T}rees},
  year = {2011},
}

@Article{boz94,
  Title                    = {Representable tree series},
  Author                   = {S. Bozapalidis},
  Journal                  = {Fundamenta Informaticae},
  Year                     = {1994},
  Pages                    = {367--389},
  Volume                   = {21}
}

@Article{boz99,
  Title                    = {Equational elements in additive algebras},
  Author                   = {S. Bozapalidis},
  Journal                  = {Theory Comput. Syst.},
  Year                     = {1999},
  Number                   = {1},
  Pages                    = {1--33},
  Volume                   = {32}
}

@Article{bozale89,
  Title                    = {Repr{\'e}sentation matricielles de s{\'e}ries d'arbre reconnaissables},
  Author                   = {S. Bozapalidis and A. Alexandrakis},
  Journal                  = {Informatique th{\'e}orique et applications},
  Year                     = {1989},
  Number                   = {4},
  Pages                    = {449--459},
  Volume                   = {23}
  }

@article{bozgra05,
author = {S. Bozapalidis and A. Grammatikopoulou},
title = {Recognizable Picture Series},
journal = {J. Autom. Lang. Comb.},
volume = {10},
number = {2--3},
pages = {159--183},
year= {2005}
}

@article{bozlou83,
author = {S. Bozapalidis and O. Louscou-Bozapalidou},
title = {The rank of a formal tree power series},
journal = {Theoret. Comput. Sci.},
volume = {27},
pages = {211--215},
year = {1983}
}

@article{bozlou10,
author = {S. Bozapalidis and O. Louscou-Bozapalidou},
title = {Fuzzy tree language recognizability},
journal = {Fuzzy Sets and Systems},
volume = {161},
pages = {716--734},
year = {2010}
}

@article{bra69,
author = {W.S. Brainerd},
title = {Tree generating regular systems},
journal = {Inf. Control},
volume = {14},
number = {2},
pages = {217--231},
year = {1969}
}

@Article{buc60,  
  Title                    = {Weak Second-Order Arithmetic and Finite Automata},
  Author                   = {J.R. B{\"u}chi},
  Journal                  = {Zeitschr. f. math. Logik und Grundlagen der Math.},
  Year                     = {1960},
  Pages                    = {66--92},
  Volume                   = {6}
}

@phdthesis{bue14,
title = {Algebraic decoder specification: coupling formal-language theory and statistical machine translation},
author = {M. B{\"u}chse},
year = {2014},
school = {Technische Universit{\"at} Dresden, Germany}
}

@inproceedings{buefis12,
    title = {Deciding the Twins Property for Weighted Tree Automata over Extremal Semifields},
    booktitle = {Workshop on Applications of Tree Automata Techniques in Natural Language Processing (ATANLP 2012)},
    year = {2012},
    author = {M. B\"uchse and A. Fischer},
        publisher = "Association for Computational Linguistics"
}

@Article{buemayvog10,
  Title                    = {Determinization of Weighted Tree automata using Factorizations},
  Author                   = {M. B{\"u}chse and H. Vogler and J. May},
  Journal                  = {J. Autom. Lang. Comb.},
  Year                     = {2010},
  Pages                    = {229--254},
  Volume                   = {15(3-4)}
}

@Book{berreu88,
  Title                    = {Rational Series and Their Languages},
  Author                   = {J. Berstel and Ch. Reutenauer},
  Publisher                = {Springer-Verlag},
  Year                     = {1988},
  Series                   = {EATCS Monographs on Theoretical Computer Science},
  Volume                   = {12},

  ISSN                     = {3-540-18626-3}
}

@Article{berreu82,
  Title                    = {Recognizable formal power series on trees},
  Author                   = {J. Berstel and C. Reutenauer},
  Journal                  = {Theoret. Comput. Sci.},
  Year                     = {1982},
  Number                   = {2},
  Pages                    = {115--148},
  Volume                   = {18},

  Location                 = {vogler-2004}
}

@inproceedings{bjodrezec15,
title = {An Efficient Best-Trees Algorithm for Weighted Tree Automata over the Tropical Semiring},
author = {J. Bj{\"o}rklund and F. Drewes and N. Zechner},
year = {2015},
booktitle = {9th Int. Conf. on Language and Automata Theory and Applications (LATA 2015)},
editor={A.-H. Dediu and E. Formenti and C. Martin-Vide and B. Truthe},
publisher = {Springer},
series = {Lecture Notes in Computer Science},
volume = {8977},
pages = {97--108}
}

@article{bli65,
title={{\'U}.{T}. {M}{\'e}dv{\'e}d{\'e}v. {O} klass{\'e} sobytij, dopuska{\'u}\v{s}ih pr{\'e}dstavl{\'e}ni{\'e} v kon{\'e}\v{c}nom avtomat{\'e} ({O}n a class of events representable in a finite automaton). {A}vtomaty, {I}zdat{\'e}l'stvo {I}nostrannoj {L}it{\'e}ratury, {M}oscow 1956, pp. 385–401.},
volume={30},
DOI={10.2307/2269654},
number={3},
journal={J.  Symb. Logic},
publisher={Cambridge University Press},
author={A.J. Blikle},
year={1965},
pages={381–382}}

@article{bolgasmonzei14,
  publisher = {ACM Press},
  journal = {ACM Transactions on Computational Logic},
  author = {B. Bollig and P. Gastin and B. Monmege and M. Zeitoun},
  title = {Pebble Weighted Automata and Weighted Logics},
  volume = {15},
  number = {2},
  year = 2014
}

@inproceedings{bolgasmonzei10,
  year = 2010,
  volume = 6199,
  series = {Lecture Notes in Computer Science},
  publisher = {Springer},
  editor = {S. Abramsky and F. Meyer{ }auf{ }der{ }Heide and P. Spirakis},
  booktitle = {37th {I}nt. 
               {C}olloquium on {A}utomata, {L}anguages and 
               {P}rogramming ({ICALP} 2010)~-- {P}art~{II}},
  author = {B. Bollig and P. Gastin and B. Monmege and M. Zeitoun},
  title = {Pebble Weighted Automata and Transitive Closure Logics},
  pages = {587-598}
}

@Book{bor04b,
  Title                    = {The Theory of Recognizable Tree Series},
  Author                   = {B. Borchardt},
  Publisher                = {Verlag f{\"u}r {W}issenschaft und {F}orschung},
  Year                     = {2005},
  Note                     = {(PhD thesis, 2004, Technische Universit{\"a}t Dresden, Germany)}
}

@Article{bor04,
  Title                    = {A Pumping Lemma and Decidability Problems for Recognizable Tree Series},
  Author                   = {B. Borchardt},
  Journal                  = {Acta Cybern.}, 
  Year                     = {2004},
  Number                   = {4},
  Pages                    = {509--544},
  Volume                   = {16}
}

@Article{borfulgazmal05,
  Title                    = {Bounds for Tree Automata with Polynomial Costs}, 
  Author                   = {B. Borchardt and Z. F{\"u}l{\"o}p and Z. Gazdag and A. Maletti},
  Journal                  = {J. Autom. Lang. Comb.},
  Year                     = {2005},
  Pages                    = {107--157},
  Volume                   = {10},
  Issue                    = {2/3}
}

@Article{borvog03,
  Title                    = {Determinization of Finite State Weighted Tree Automata},
  Author                   = {B. Borchardt and H. Vogler},
  Journal                  = {J. Autom. Lang. Comb.},
  Year                     = {2003},
  Number                   = {3},
  Pages                    = {417--463},
  Volume                   = {8},

  Location                 = {vogler-2187}
}

@unpublished{bor24,
author = {S. Borgwardt},
title = {personal communication},
note = {Technische Universit{\"a}t Dresden},
year = {2024}
}

@Article{boz91,
  Title                    = {Effective construction of the syntactic algebra of a recognizable series on trees},
  Author                   = {S. Bozapalidis},
  Journal                  = {Acta Inform.},
  Year                     = {1991},
  Pages                    = {351--363},
  Volume                   = {28},

  Location                 = {vogler-2026}
}

@Article{bozrah94,
  Title                    = {On two families of forests},
  Author                   = {S. Bozapalidis and G. Rahonis},
  Journal                  = {Acta Inform.},
  Year                     = {1994},
  Pages                    = {235--260},
  Volume                   = {31},
  number = {3}
}

@article{buevogned12,
author = {M. B{\"u}chse and  H. Vogler and M. Nederhof},
title = {Tree parsing for tree-adjoining machine translation},
journal = {J. Logic and Comput.},
volume = {24(2)},
pages = {351--373},
year = {2014}
}

@book{bursan81,
	address = {New York},
	publisher = {Springer-Verlag},
	title = {A Course in Universal Algebra},
	isbn = {0-387-90578-2},
	note = {Corrected version available at http://www.thoralf.uwaterloo.ca/htdocs/ualg.html},
	volume = {78},
	author = {S. Burris and H.P. Sankappanavar},
	editor = {P.R. Halmos and F.W. Gehring and C.C. Moore},
	series = {Graduate Texts in Mathematics},
	year = {1981},
	edition = {first},
	location = {maletti-alg-1}
}

@masterthesis{buy13,
title = {Probabilistic Tree Transducers for Grammatical Error Correction},
author = {J.M. Buys}, 
year = {2013},
note = {Master thesis, Computer Science Division, Department of Mathematical Sciences, University of Stellenbosch,  South Africa}
}

@misc{chamazmusriv21,
      title={Pumping lemmas for weighted automata}, 
      author={A. Chattopadhyay and F. Mazowiecki and A. Muscholl and C. Riveros},
      year={2021},
      note = {arXiv.2001.06272}
}

@InCollection{chosch63,
  Title                    = {The Algebraic Theory of Context-Free Languages},
  Author                   = {N. Chomsky and M.P. Sch{\"u}tzenberger},
  Booktitle                = {Computer Programming and Formal Systems},
  Publisher                = {North-Holland, Amsterdam},
  Year                     = {1963},
  series = {Studies in Logic and the Foundations of Mathematics},
Volume = {35},
  Editor                   = {P. Braffort and D. Hirschberg},
  Pages                    = {118--161}
}

@Article{car71,
  Title                    = {An Algebra for Network Routing Problems},
  Author                   = {B. A. Carr{\'e}},
  Journal                  = {IMA J. Applied Math.},
  Year                     = {1971},
  Pages                    = {273--294},
  Volume                   = {7},
  number = {3}
}

@Article{cirdroignvog10,
  Title                    = {Determinization of weighted finite automata over strong bimonoids},
  Author                   = {M. \'Ciri\'c and M. Droste and J. Ignjatovi\'c and H. Vogler},
  Journal                  = {Inform. Sci.},
  Year                     = {2010},
  Pages                    = {3497--3520},
  Volume                   = {180(18)}
}

@Misc{comdaugiljaclugtistom08,
  Title                    = {Tree Automata Techniques and Applications},
  Author                   = {H. Comon and M. Dauchet and R. Gilleron and F. Jacquemard and D. Lugiez and S. Tison and M. Tommasi},
  Year                     = {2007}
}

@Book{corleiriv90,
  Title                    = {Introduction to Algorithms},
  Author                   = {T.H. Cormen and C.E. Leiserson and R.L. Rivest},
  Publisher                = {The MIT Press},
  Year                     = {1990}
}

@article{cos72,
author = {O.L. Costich},
title = {A {M}edvedev Characterization of Sets Recognized by Generalized Finite Automata},
journal = {Mathematical Systems Theory},
volume = {6(1-2)},
pages = {263--267},
year = {1972}
}

@book{coueng12,
title = {Graph Structure and Monadic Second-Order Logic: 
A Language-Theoretic Approach},
Author = {B. Courcelle and J. Engelfriet}, 
publisher = {Cambridge University Press},
Series = {Encyclopedia of Mathematics and its Applications},
Volume = {138},
year = {2012}
}

@unpublished{cze19,
author = {G. Cz{\'e}dli},
title  = {personal communication},
year = {2019}
}

@Article{dam82,
  Title                    = {The {IO}- and {OI}-hierarchies},
  Author                   = {W. Damm},
  Journal                  = {Theoret. Comput. Sci.},
  Year                     = {1982},
  Pages                    = {95--208},
  Volume                   = {20}
}

@Inproceedings{damgoe82,
  Title                    = {An automata-theoretical characterization of the {OI}-hierarchy},
  Author                   = {W. Damm and A. Goerdt},
editor = {M. Nielsen and E.M. Schmidt},
booktitle = {Automata, Languages and Programming. ICALP 1982},
series = {Lecture Notes in Computer Science}, 
volume = 140,
publisher = {Springer, Berlin, Heidelberg},
year = 1982,
pages = {141--153}
}

@book{davpri12,
author = {B.A. Davey and H.A. Priestley},
title = {Introduction to Lattices and Order},
publisher = {Cambridge University Press},
year = {2012}
}

@book{ded39,
title = {Was sind und was sollen die Zahlen?},
author = {R. Dedekind},
publisher = {Friedr. Vieweg $\&$ Sohn, Braunschweig},
year = {1939}
}

@Unpublished{den17,
  Title                    = {zigzag is not bottom-up-deterministically recognizable},
  Author                   = {T. Denkinger},
  Note                     = {personal communications},
  Year                     = {2017}
}

@article{dic13,
author = {L.~E. Dickson},
title = {Finiteness of the Odd Perfect and Primitive Abundant Numbers with n Distinct Prime Factors},
journal = {American Journal of Mathematics}, 
volume = {35},
number = {4},
pages = {413--422}, 
year = 1913
}

@phdthesis{die18,
author = {T. Dietze},
title = {A Formal View on Training of Weighted Tree Automata by Likelihood-Driven State Splitting and Merging},
school = {Technische Universit{\"a}t Dresden},
year = {2018}
}

@mastersthesis{doe19,
author = {F. D{\"o}rband},
title = {A {K}leene theorem for weighted forest automata},
school = {Technische Universit{\"a}t Dresden},
year = {2019}
}

@article{doe21,
author = {F. D{\"o}rband},
title = {Kleene and {B}{\"u}chi Theorems for Weighted Forest Languages over {M}-Monoids},
journal = {Inf. Comput.},
volume = {281},
note = {Article 104765},
year = {2021}
}

@phdthesis{doe22,
author = {F. D{\"o}rband},
title = {Characterisation Theorems for
Weighted Tree Automaton Models},
school = {Technische Universit{\"a}t Dresden, Germany},
year = {2022}
}

@inproceedings{doefelsti21,
author = {F. D{\"o}rband and K. Stier and T. Feller},
title = {Sequentiality of Group-Weighted Tree Automata},
booktitle = {Int. Conf. on Language and Automata Theory and Applications (LATA 2021)},
year = {2021},
editor = {A. Leporati and C. Martin-Vide and D. Shapira and C. Zandron},
series = {Lecture Notes in Computer Science},
volume = {12638},
pages = {267--278}
}

@article{doefulvog24,
author = {F. D{\"o}rband and Z. F{\"u}l{\"o}p and H. Vogler},
title = {Rational Weighted Tree Languages with Storage},
journal = {Inf. Comput.},
volume={301},
pages={105205},
note = {special issue for CAI 2019},
year = {2024}
}

@unpublished{doemoe20,
author = {F. D{\"o}rband and R. M\"orbitz},
title = {Incomparability of classes for support theorems},
note = {personal communication},
year = 2020
}

@Article{don65,
  Title                    = {Decidability of the weak second-order theory of two successors},
  Author                   = {J.E. Doner},
  Journal                  = {Notices Amer. Math. Soc.},
  Year                     = {1965},
  Pages                    = {819},
  Volume                   = {12},
  note = {Abstract No. 65T 648, 819}
}

@article{don70,
author = {J. Doner},
title = {Tree acceptors and some of their applications},
journal = {J. Comput. Syst. Sci.},
volume = {4},
pages = {406--451},
year = {1970}
}

@book{dre06,
title = {Grammatical Picture Generation: A Tree-Based Approach},
author = {F. Drewes},
year = {2006},
publisher = {Springer},
series = {Texts in Theoretical Computer Science. An EATCS Series}
}

@inproceedings{dregebvog16,
author = {F. Drewes and K. Gebhardt and H. Vogler},
title = {{EM}-training for weighted aligned hypergraph bimorphisms},
booktitle = {Proceedings of the ACL Workshop on Statistical NLP and Weighted automata},
publishe = {Association of Computational Linguistics},
year = {2016},
pages = {60--69}
}

@article{drehoemal11,
title = {{MAT} learners for tree series: an abstract data type and two realizations},
author = {F. Drewes and J. H{\"o}gberg and A. Maletti},
year = {2011},
journal = {Acta Inform.},
Volume = {48(3)},
pages = {165--189}
}

@proceedings{drekuh10,
editor = {F. Drewes and M. Kuhlmann},
title = {Proceedings of the Workshop on Applications of Tree Automata Techniques in Natural Language Processing},
publisher = {Association of Computational Linguistics},
year = {2010}
}

@proceedings{drekuh12,
editor = {F. Drewes and M. Kuhlmann},
title = {Proceedings of the Workshop on Applications of Tree Automata Techniques in Natural Language Processing},
publisher = {Association of Computational Lingusitics},
year = {2012}
}

@Article{drevog07,
  Title                    = {Learning Deterministically Recognizable Tree Series},
  Author                   = {F. Drewes and H. Vogler},
  Journal                  = {J. Autom. Lang. Comb.},
  Year                     = {2007},
  Pages                    = {333--354},
  Volume                   = {12},
number = {3}
}

@unpublished{dro19,
author = {M. Droste},
title = {Personal communication},
year = {2019}
}

@unpublished{dro21,
author = {M. Droste},
title = {Weighted Automata over Monotonic Strong Bimonoids: Decidability and Undecidability of Finite Image},
note = {Talk at the International Colloquium
Recent Advances of Quantitative Models in Computer Science, June 22-23, 2021, online},
year = {2021}
}

@unpublished{dro22,
author = {M. Droste},
year = {2022},
title = {Personal communication}
}

@unpublished{dro24,
author = {M. Droste},
year = {2024},
title = {Personal communication}
}

@proceedings{droesilar18,
editor = {M. Droste and Z. {\'E}sik and K. Larsen},
title = {Special issue on Quantitative Models and Weighted Automata},
note = {Soft Computing, (22)4, 2018},
publisher = {Springer},
year = {2018}
}

@inproceedings{drofulgoe16,
author = {M. Droste and Z. F{\"u}l{\"o}p and D. G{\"o}tze},
title = {A {K}leene Theorem for Weighted Tree Automata over Tree Valuation Monoids},
booktitle = {Proceedings of Language Automata Theory and Applications (LATA 2016)},
series = {Lecture Notes in Computer Science},
editor = {A.-H. Dediu and  J. Janou\v{s}ek and C. Martin-Vide and  B. Truthe},
volume = {9618},
pages = {452--463},
publisher = {Springer, Cham},
year = {2016}
}

@article{drofulgoe19,
author = {D. G{\"o}tze and  Z. F{\"u}l{\"o}p and M. Droste},
title = {A {K}leene theorem for weighted tree automata over tree valuation monoids},
journal = {Inf. Comput.},
volume = {269},
year = {2019},
note = {Article 104445}
}

@article{drofulkosvog21,
author = {M. Droste and Z. F{\"u}l{\"o}p and D. K{\'o}sz{\'o} and H. Vogler},
title = {Finite-image property of weighted tree automata over past-finite monotonic strong bimonoids},
journal =  {Theoret. Comput. Sci.},
year = {2022},
volume = {919},
pages = {118--143}
}

@inproceedings{drofulkosvog20b,
author = {M. Droste and Z. F{\"u}l{\"o}p and D. K{\'o}sz{\'o} and H. Vogler},
title = {Crisp-Determinization of Weighted Tree Automata over Additively Locally Finite and Past-Finite Monotonic Strong Bimonoids Is Decidable},
year = {2020},
series = {Lecture Notes in Computer Science},
volume = {12442},
pages = {39--51},
publisher = {Springer Nature Switzerland},
booktitle = {Descriptional Complexity of Formal Systems (DCFS 2020)},
editor = {G. Jir\'askov\'a and G. Pighizzini}
}

@unpublished{drofultepvog24,
author = {M. Droste and Z. F{\"u}l{\"o}p and A. Tepav\v{c}evi{\'c} and H. Vogler},
title = {The generating power of weighted tree automata with initial algebra semantics},
year = {2024},
note = {submitted for publication, see also: ar{X}iv 2405.20753}
}

@article{drogas99,
author = {M. Droste and P. Gastin},
title = {The {K}leene-{S}ch{\"u}tzenberger theorem for formal power series in partially
commuting variables},
journal = {Inf. Comput.},
volume = {153},
pages = {47--80},
year = {1999},
note = {Extended abstract in: 24th ICALP, Lecture Notes in Computer Science,
vol. 1256, pp. 682–692. Springer (1997)}
}

@InProceedings{drogas05,
  Title                    = {Weighted Automata and Weighted Logics},
  Author                   = {M. Droste and P. Gastin},
  Booktitle                = {32nd Int. Colloquium Automata, Languages and Programming (ICALP 2005)},
  Year                     = {2005},
  Editor                   = {L.~Caires and G.~F. Italiano and L.~Monteiro and C.~Palamidessi and M.~Yung},
  Pages                    = {513--525},
  Publisher                = {Springer, Berlin, Heidelberg},
  Series                   = {Lecture Notes in Computer Science},
  Volume                   = {3580}
}

@Article{drogas07,
  Title                    = {Weighted automata and weighted logics},
  Author                   = {M. Droste and P. Gastin},
  Journal                  = {Theoret. Comput. Sci.},
  Year                     = {2007},
  Number                   = {1-2},
  Pages                    = {69--86},
  Volume                   = {380},
  Publisher                = {Elsevier Science Publishers Ltd.}
}

@InCollection{drogas09,
  Title                    = {Weighted Automata and Weighted Logics},
  Author                   = {M. Droste and P. Gastin},
  Booktitle                = {Handbook of {W}eighted {A}utomata},
  Publisher                = {Springer-Verlag},
  Year                     = {2009},
  Chapter                  = {5},
  series = {Monographs in Theoretical Computer Science. An EATCS Series},
  Editor                   = {M. Droste and W. Kuich and H. Vogler}
}

@Inproceedings{drogoemaemei11,
author="M. Droste 
and D. G{\"o}tze
and S. M{\"a}rcker
and I. Meinecke",
editor="W. Kuich and G. Rahonis",
title="Weighted Tree Automata over Valuation Monoids and Their Characterization by Weighted Logics",
bookTitle="Algebraic Foundations in Computer Science",
year="2011",
series = {Lecture Notes in Computer Science},
volume = {7020},
publisher="Springer, Berlin, Heidelberg",
pages="30--55"
}

@article{droheu15,
author = {M.~Droste and D.~Heusel},
title = {The supports of weighted unranked tree automata},
journal = {Fundam. Inform.}, 
volume = {136(1-2)},
pages = {37--58}, 
year = 2015
}

@InProceedings{droheuvog15,
  Title                    = {Weighted Unranked Tree Automata over Tree Valuation Monoids and Their Characterization by Weighted Logics},
  Author                   = {M. Droste and D. Heusel and H. Vogler},
  Booktitle                = {6th Int. Conf. on Algebraic Informatics (CAI 2015)},
  Year                     = {2015},
  Pages                    = {90--102},
  Series                   = {Lecture Notes in Computer Science},
  Volume                   = {9270},
editor = {A. Maletti},
publisher = {Springer, Cham}
}

@incollection{drokus21,
author = {M. Droste and D. Kuske},
title = {Weighted automata},
chapter = "4",
booktitle = {Handbook of Automata Theory},
editor = {J.-E. Pin},
pages = {113--150},
publisher = {European Mathematical Society},
year = {2021}
}

@Article{dropecvog05,
  Title                    = {A {K}leene Theorem for Weighted Tree Automata},
  Author                   = {M. Droste and Chr. Pech and H. Vogler},
  Journal                  = {Theory Comput. Syst.},
  Year                     = {2005},
  Pages                    = {1--38},
  Volume                   = {38},

  Location                 = {vogler-2168}
}

@InCollection{drokui09,
  Title                    = {Semirings and Formal Power Series},
  Author                   = {M. Droste and W. Kuich},
  Booktitle                = {Handbook of Weighted Automata},
  Publisher                = {Springer-Verlag},
  Year                     = {2009},
  Chapter                  = {1},
  Editor                   = {M. Droste and W. Kuich and H. Vogler},
    series = {Monographs in Theoretical Computer Science. An EATCS Series},
  Pages                    = {3--28}
}

@inproceedings{drokus03,
author = {M. Droste and D. Kuske},
title = {Skew and Infinitary Formal Power Series},
booktitle = {13th Int. Colloquium on Automata, Languages and Programming (ICALP 2003)},
volume = {2719},
series = {Lecture Notes in Computer Science},
pages = {426--438},
publisher = {Springer, Berlin, Heidelberg},
year = {2003}
}

@article{drokus06,
title = {Skew and infinitary formal power series},
journal = {Theoret. Comput. Sci.},
volume = {366},
author = {M. Droste and D. Kuske},
number = {3},
pages = {199--227},
year = {2006}
}

@Book{drokuivog09,
  Title                    = {Handbook of Weighted Automata},
  Editor                   = {M. Droste and W. Kuich and H. Vogler},
  Publisher                = {Springer-Verlag},
  Year                     = {2009},
  Series                   = {EATCS Monographs in Theoretical Computer Science}
}

@proceedings{dromalvog21,
editor = {M. Droste and A. Maletti and H. Vogler},
title = {Special issue on 9th International Workshop Weighted Automata: Theory and Applications (May 22–26, 2018, Leipzig, Germany)}, 
publisher = {Information and Computation},
volue = {282},
year = {2022}
}

@article{dromei12,
title = {Weighted automata and weighted {MSO} logics for average and long-time behaviors},
journal = {Inf. Comput.},
volume = {220--221},
pages = {44--59},
year = {2012},
issn = {0890-5401},
author = {M. Droste and I. Meinecke}
}

@InProceedings{drorah07,
author="Droste, M. and Rahonis, G.",
editor="J. Holub
and J. {\v{Z}}{\v{d}}{\'a}rek",
title="Weighted Automata and Weighted Logics with Discounting",
booktitle="Implementation and Application of Automata (CIAA 2007)",
year="2007",
series = {Lecture Notes in Computer Science},
volume = {4783},
publisher="Springer, Berlin, Heidelberg",
pages="73--84"
}

@manuscript{drovog24,
  Title                    = {Run supports and initial algebra supports
of weighted automata},
  Author                   = {M. Droste and H. Vogler},
  year = {2024},
      url={https://arxiv.org/abs/2409.08727},
  note = {ar{X}iv, cs.FL, 2409.08727}
  }

@Article{drostuvog10,
  Title                    = {Weighted finite automata over strong bimonoids},
  Author                   = {M. Droste and T. St\"uber and H. Vogler},
  Journal                  = {Inform. Sci.},
  Year                     = {2010},
  Pages                    = {156--166},
  Volume                   = {180},
  number = {1}
}

@inproceedings{drovog10,
  Author                   = {M. Droste and H. Vogler},
title = {Kleene and {B}{\"u}chi Theorems for Weighted Automata and Multi-valued Logics over Arbitrary Bounded Lattices},
editor = {Y. Gao and H. Lu and S. Seki and S. Yu},
booktitle = {Developments in Language Theory (DLT 2020)},
series = {Lecture Notes in Computer Science},
volume = {6224},
pages = {160--172},
year = {2010},
publisher = {Springer}
}

@Article{drovog14,
  Title                    = {The {C}homsky-{S}ch\"utzenberger theorem for quantitative context-free languages},
  Author                   = {M. Droste and H. Vogler},
  Journal                  = {Int. J. Found. Computer Science},
  Year                     = {2014},
  Volume                   = {25(8)},
  pages = {955--969}, 
  Page                     = {955--969}
}

@Article{drovog12,
  Title                    = {Weighted automata and multi-valued logics over arbitrary bounded lattices},
  Author                   = {M. Droste and H. Vogler},
  Journal                  = {Theoret. Comput. Sci.},
  Year                     = {2012},
  Volume                   = {418},
  Pages                     = {14-–36}
}

@Article{drovog11,
  Title                    = {Weighted Logics for Unranked Tree Automata},
  Author                   = {M. Droste and H. Vogler},
  Journal                  = {Theory of Computing Systems},
  Year                     = {2011},
  Note                     = {published online first, 29. June 2009},
  Number                   = {1},
  Pages                    = {23--47},
  Volume                   = {48}
}

@Article{drovog06,
  Title                    = {Weighted tree automata and weighted logics},
  Author                   = {M. Droste and H. Vogler},
  Journal                  = {Theoret. Comput. Sci.},
  Year                     = {2006},
  Pages                    = {228--247},
  Volume                   = {366},
number = {3}
}

@proceedings{drovog03,
editor = {M. Droste and H. Vogler},
title = {Selected Papers of the Workshop “Weighted Automata: Theory and Applications” Dresden University of Technology (Germany), March 4-8, 2002},
publisher = {J. Autom. Lang. Comb.},
volume = {8(2)}, 
year ={2003}
}

@proceedings{drovog05,
editor = {M. Droste and H. Vogler},
title =  {Selected Papers of the Workshop “Weighted Automata: Theory and Applications” Dresden University of Technology (Germany), June 1-5, 2004},
publisher = {J. Autom. Lang. Comb.},
volume = {10(2/3)},
year = {2005}
}

@proceedings{drovog07,
editor = {M. Droste and H. Vogler},
title = {Selected Papers of the Workshop “Weighted Automata: Theory and Applications” University of Leipzig (Germany), March 27-31, 2006},
publisher ={J. Autom. Lang. Comb.},
volume = {12(4)},
year = {2007}
}

@proceedings{drovog09,
editor = {M. Droste and H. Vogler},
title = {Selected Papers of the Workshop “Weighted Automata: Theory and Applications" Technische Universität Dresden (Germany), May 13-16, 2008},
publisher  = {Acta Cybernetica},
volume = {19(2)},
year = {2009}
}

@proceedings{drovog11a,
editor = {M. Droste and H. Vogler},
title = {Selected Papers of the Workshop “Weighted Automata: Theory and Applications” University of Leipzig (Germany), May 3 - 7, 2010},
publisher = {Acta Cybernetica},
volume = {20(2)}, 
year = {2011}
}

@proceedings{drovog14a,
editor = {M. Droste and H. Vogler},
title = {Selected Papers of the Workshop “Weighted Automata: Theory and Applications", WATA 2012, May 29 – June 2, 2012, Dresden, Germany},
publisher = {Theoret. Comput. Sci.},
volume = {534(1)},
year = {2014}
}

@Book{eil74,
  Title                    = {Automata, Languages, and Machines -- Volume A},
  Author                   = {S. Eilenberg},
  Publisher                = {Academic Press},
  Year                     = {1974},
  Series                   = {Pure and Applied Mathematics},
  Volume                   = {59}
}

@Article{ell71,
  Title                    = {Probabilistic tree automata},
  Author                   = {C.A. Ellis},
  Journal                  = {Inform. and Control},
  Year                     = {1971},
  Number                   = {5},
  Pages                    = {401--416},
  Volume                   = {19}
}

@Article{eng75,
  Title                    = {Bottom-up and top-down tree transformations - a comparison},
  Author                   = {J. Engelfriet},
  Journal                  = {Math. Systems Theory},
  Year                     = {1975},
  Number                   = {3},
  Pages                    = {198--231},
  Volume                   = {9}
  }

@TechReport{eng75-15,
  Title                    = {Tree automata and tree grammars},
  Author                   = {J. Engelfriet},
  Institution              = {Inst. of Mathematics, University of Aarhus, Department of Computer Science, Ny Munkegade, 8000 Aarhus C, Denmark},
  Year                     = {1975},
  Number                   = {{DAIMI FN-10}},
  note = {ar{X}iv:1510.02036}
}

@incollection{eng80,
title = {Some open questions and recent results on tree transducers and tree languages},
editor = {R.V.  Book},
booktitle = {Formal Language Theory},
publisher = {Academic Press},
pages = {241--286},
year = {1980},
isbn = {978-0-12-115350-2},
doi = {https://doi.org/10.1016/B978-0-12-115350-2.50014-2},
url = {https://www.sciencedirect.com/science/article/pii/B9780121153502500142},
author = {J. Engelfriet}
}

@TechReport{eng86,
  Title                    = {Context--free grammars with storage},
  Author                   = {J. Engelfriet},
  Institution              = {University of Leiden},
  Year                     = {1986},
  Note                     = {see also: arXiv:1408.0683 [cs.FL], 2014},
  Number                   = {86-11}
}

@Article{eng91c,
author = {J. Engelfriet},
title = {Iterated stack automata and complexity classes},
journal = {Inform. Control},
volume = {95},
pages = {21--75},
year = {1991}
}

@Unpublished{eng03,
  Title                    = {Alternative {K}leene theorem for weighted automata},
  Author                   = {J. Engelfriet},
  Note                     = {personal communication},
  Year                     = {2003}
}

@Article{engfulvog02new,
  Title                    = {Bottom-up and Top-down Tree Series Transformations},
  Author                   = {J. Engelfriet and Z. F{\"u}l{\"o}p and H. Vogler},
  Journal                  = {J.\ Autom.\ Lang.\ Comb.},
  Year                     = {2002},
  Number                   = {1},
  Pages                    = {11--70},
  Volume                   = {7}
}

@Article{engmeilee80,
  Title                    = {Stack Machines and Classes of Nonnested Macro Languages},
  Author                   = {J. Engelfriet and E. M. Schmidt and J. van Leeuwen},
  Journal                  = {J. Assoc. Comput. Machinery},
  Year                     = {1980},
  Number                   = {1},
  Pages                    = {96--117},
  Volume                   = {27}
}

@Article{engsch77,
  Title                    = {{IO} and {OI}.{I}},
  Author                   = {J. Engelfriet and E.M. Schmidt},
  Journal                  = {J. Comput. System Sci.},
  Year                     = {1977},
  Number                   = {3},
  Pages                    = {328--353},
  Volume                   = {15}
}

@Article{engsch78,
  Title                    = {{IO} and {OI}.{II}},
  Author                   = {J. Engelfriet and E.M. Schmidt},
  Journal                  = {J. Comput. System Sci.},
  Year                     = {1978},
  Number                   = {1},
  Pages                    = {67--99},
  Volume                   = {16}
}

@Article{engvog91,
  Title                    = {Modular tree transducers},
  Author                   = {J. Engelfriet and H. Vogler},
  Journal                  = {Theoret. Comput. Sci.},
  Year                     = {1991},
  Pages                    = {267--303},
  Volume                   = {78}
}

@Article{elg61,
  Title                    = {{D}ecision Problems of Finite Automata Design and Related Arithmetics},
  Author                   = {C.C. Elgot},
  Journal                  = {Trans. Amer. Math. Soc.},
  Year                     = {1961},
  Pages                    = {21--52},
  Volume                   = {98},
  number = {1}
  }

@Article{esikui03,
  Title                    = {Formal Tree Series},
  Author                   = {Z. {\'E}sik and W. Kuich},
  Journal                  = {J. Autom. Lang. Comb.},
  Year                     = {2003},
  Number                   = {2},
  Pages                    = {219--285},
  Volume                   = {8}
}

@unpublished{esikui,
title = {Modern Automata Theory},
author = {Z. {\'E}sik and W. Kuich}
}

@article{esiliu07,
author = {Z.~{\'E}sik and G.~Liu},
title = {Fuzzy tree automata},
journal = {Fuzzy Sets and Systems},
volume = {158},
number = {113},
pages = {1450--1460},
year = 2007
}

@article{fic11,
author = {I. Fichtner},
year = {2011},
title = {Weighted Picture Automata and Weighted Logics},
journal = {Theory Comput. Syst.},
volume = {48},
pages = {48--78}
}

@article{fli74,
author = {M. Fliess},
title = {Matrices de {H}ankel},
journal = {J. Math. Pures Appl.},
volume = {(9)53},
pages = {197--222},
year = {1974}
}

@article{fli75,
author = {M. Fliess},
title = {S{\'e}ries rationelles positives et processus stochastiques},
journal = {Ann. Inst. H. Poincar{\'e} Set. B (N.S.)},
volue = {11},
pages = {1--21},
year = {1975}
}

@article{flo62,
author = {R.N. Floyd},
title = {Algorithm 97: {S}hortest path},
journal = {Comm. ACM},
volume = {5(6)},
pages = {345},
year = {1962}
}

@Article{ful15,
  Title                    = {Local Weighted tree languages},
  Author                   = {Z. F{\"u}l{\"o}p},
  Journal                  = {Acta Cybernetica},
  Year                     = {2015},
  Pages                    = {393--402},
  Volume                   = {22(2)}
}

@article{fulgaz18,
journal = {Acta Cybernetica},
volume = {23(3)},
year = {2018},
pages = {867--886},
title = {Weighted Languages Recognizable by Weighted Tree Automata},
author = {Z. F{\"u}l{\"o}p and Z. Gazdag}
}

@article{fulgazvog04,
title = {Hierarchies of tree series transformations},
journal = {Theoret. Comput. Sci.},
volume = {314},
number = {3},
pages = {387-429},
year = {2004},
author = {Z. F{\"u}l{\"o}p and Z. Gazdag and H. Vogler}
}

@Article{fulhervagvog93,
  Title                    = {Tree transducers with external functions},
  Author                   = {Z. F{\"u}l{\"o}p and F. Herrmann and S. V\'agv{\"o}lgyi and H. Vogler},
  Journal                  = {Theoret. Comput. Sci.},
  Year                     = {1993},
  Pages                    = {185--236},
  Volume                   = {108}
}

@article{fulkosvog19,
    title={Crisp-determinization of weighted tree automata over strong bimonoids},
    author={Z. F{\"u}l{\"o}p and D. K{\'o}sz{\'o} and H. Vogler},
    journal ={Discrete Math. Theoret. Comput. Sci.},
    volume = {23},
    number = {1},
    note = {$\#$18, see also arxiv.org:1912.02660v2, 2019},
    year = {2021}
}

@Article{fulmalvog11,
  Title                    = {Weighted {E}xtended {T}ree {T}ransducers},
  Author                   = {Z. F{\"u}l{\"o}p and A. Maletti and H. Vogler},
  Journal                  = {Fundamenta Informaticae},
  Year                     = {2011},
  Pages                    = {1--39},
  Volume                   = {112}
}

@inproceedings{fulmalvog10,
  Author                   = {Z. F{\"u}l{\"o}p and A. Maletti and H. Vogler},
title = {Preservation of Recognizability for Synchronous Tree Substitution Grammars},
booktitle = {Workshop on Applications of Tree Automata in Natural Language Processing, ACL 2010  (ATANLP 2010)},
pages = {1--9}, 
year = 2010,
publisher = {Association for Computational Linguistics}
}

@Article{fulmalvog09,
  Title                    = {A {K}leene {T}heorem for {W}eighted {T}ree {A}utomata over {D}istributive {M}ultioperator {M}onoids},
  Author                   = {Z. F{\"u}l{\"o}p and A. Maletti and H. Vogler},
  Journal                  = {Theory Comput. Syst.},
  Year                     = {2009},
  Pages                    = {455--499},
  Volume                   = {44}
}

@Article{fulstuvog12,
  Title                    = {A {B}\"uchi-Like Theorem for Weighted Tree Automata over Multioperator Monoids},
  Author                   = {Z. F{\"u}l{\"o}p and T. St\"uber and H. Vogler},
  Journal                  = {Theory Comput. Syst.},
  note = {Submitted on 14 October 2009, published online 28 October 2010},
  Year                     = {2012},
  Pages                    = {241--278},
  Volume                   = {50}
}

@article{fulvog03,
author = {Z. F{\"u}l{\"o}p and H. Vogler},
title = {Tree Series Transformations that Respect Copying},
journal = {Theory Comput. Syst.},
volume = {36},
pages = {247--293},
year = {2003}
}

@InCollection{fulvog09new,
  Title                    = {Weighted Tree Automata and Tree Transducers},
  Author                   = {Z. F{\"u}l{\"o}p and H. Vogler},
  Booktitle                = {Handbook of Weighted Automata},
  series = {Monographs in Theoretical Computer Science. An EATCS Series},
  Publisher                = {Springer-Verlag},
  Year                     = {2009},
  Chapter                  = {9},
  Editor                   = {M. Droste and W. Kuich and H. Vogler},
  Pages                    = {313--403}
}

@Article{fulvog15,
  Title                    = {Characterizing Weighted {MSO} for Trees by Branching Transitive Closure Logics},
  Author                   = {Z. F{\"u}l{\"o}p and H. Vogler},
  Journal                  = {Theoret. Comput. Sci.},
  Year                     = {2015},
  Pages                    = {82--105},
  Volume                   = {594}
}

@Article{fulvog18,
  Title                    = {Characterizations of recognizable weighted tree languages by logic and bimorphisms},
  Author                   = {Z. F{\"u}l{\"o}p and H. Vogler},
  Journal                  = {Soft Computing},
  Year                     = {2018},
  volume = {22},
  pages = {1035--1046}
}

@inproceedings{fulvog19,
Title = {Rational Weighted Tree Languages with Storage and the {K}leene-{G}oldstine Theorem}, 
  Author = {Z. F{\"u}l{\"o}p and H. Vogler},
  booktitle = {Int. Conf. on Algebraic Informatics (CAI 2019)},
pages = {138--150},
editor= {M. \'Ciri\'c and M. Droste and J.-\'E. Pin},
year= {2019},
series = {Lecture Notes in Computer Science},
volume = {11545},
publisher={Springer International Publishing}
}

@article{fulvog19a,
  Author                   = {Z. F{\"u}l{\"o}p and H. Vogler},
  title = {Principal abstract families of weighted tree languages},
journal = {Inf. Comput.},
volume= {282},
note = {Available online 16 November 2020, Article 104657},
year = {2022}
}

@article{fulhervog18,
  Title                    = {Weighted Regular Tree Grammars with Storage},
  Author                   = {Z. F\"ul\"op and L. Herrmann and H. Vogler},
  journal = {Discrete Math. Theoret. Comput. Sci.},
  volume = {20},
  number = {1},
  note = {$\#$26},
  Year                     = {2018}
}

@book{fulvog22,
  Author                   = {Z. F\"ul\"op  and H. Vogler},
title = {Weighted Tree Automata -- May it be a little more?},
publisher ={arXiv:2212.05529v1 [cs.FL]},
year = {2022}
 }

@book{fulvog24,
  Author                   = {Z. F\"ul\"op  and H. Vogler},
title = {Weighted Tree Automata -- May it be a little more? Second edition.},
publisher ={arXiv:2212.05529v2 [cs.FL]},
year = {2024}
 }

@manuscript{fulvog25,
author = {Z. F\"ul\"op and H. Vogler},
title = {Characterization of deterministically recognizable weighted tree languages over commutative semifields by finitely generated and cancellative scalar algebras},
year = {2025},
note = {ar{X}iv, 2509.14914, cs.FL},
      url={https://arxiv.org/abs/2509.14914}, 
}

@article{fulvog23,
  Author                   = {Z. F\"ul\"op  and H. Vogler},
title = {A comparison of sets of recognizable weighted tree
languages over specific sets of bounded lattices},
journal = {International Journal of Foundations of Computer Science},
year = {2024},
volume = {35, Nos. 1$\&$2},
pages = {51--76},
note = {DOI: 10.1142/S0129054123480039}
 }

@Article{fulste11,
  Title                    = {Varieties of recognizable tree series over fields},
  Author                   = {Z. F\"ul\"op and M. Steinby},
  Journal                  = {Theoret. Comput. Sci.},
  Year                     = {2011},
  Pages                    = {736--752},
  Volume                   = {412}
}

@InProceedings{galhopknimar04,
  Title                    = {What's in a translation rule?},
  Author                   = {M. Galley and M. Hopkins and K. Knight and D. Marcu},
  Booktitle                = {Proc. HLT-NAACL 2004},
  Year                     = {2004},
  Editor                   = {S. Dumais, D. Marcu and S. Roukos},
  Pages                    = {273--280},
  Publisher                = {Association for Computational Linguistics},

  Location                 = {vogler-2412}
}

@unpublished{gal08,
author = {N. Galatos},
title = {Residuated lattices},
note = {Tutorial at "Summer School on General Algebra and Ordered Sets" (SSAOS 2008)},
year = {2008}
}

@Book{gal87,
  Title                    = {Logic for {C}omputer {S}cience: {F}oundations of {A}utomatic {T}heorem {P}roving},
  Author                   = {J. H. Gallier},
  Publisher                = {John Wiley \& Sons},
Year                     = {1987}
}

@article{gasmon18,
  publisher = {Springer},
  journal = {Soft Computing},
  author = {P. Gastin and B. Monmege},
  title = {A unifying survey on weighted logics and weighted automata},
  volume = {22},
  number = {4},
  year = {2018},
  pages = {1047-1065}
}

@InCollection{gecste97,
  Title                    = {Tree Languages},
  Author                   = {F. G{\'e}cseg and M. Steinby},
  Booktitle                = {Handbook of Formal Languages},
  Publisher                = {Springer-Verlag},
  Year                     = {1997},
  Chapter                  = {1},
  Editor                   = {G. Rozenberg and A. Salomaa},
  Pages                    = {1--68},
  Volume                   = {3}
}

@Book{gecste84,
  Title                    = {Tree Automata},
  Author                   = {F. G{\'e}cseg and M. Steinby},
  Publisher                = {Akad{\'e}miai Kiad{\'o}, Budapest},
  note = {ar{X}iv:1509.06233},
  Year                     = {1984}
}

@article{gergon20,
journal = {Fuzzy Sets and Systems},
Volume = {397},
Pages = {186--196},
title = {Conditions for the existence of maximal factorizations},
author = {S. Gerdjikova and J.R.G. de Mend{\'i}vil},
year = {2020}
}

@article{gho18,
author = {M. Ghorani},
title = {Tree automata on complete residuated lattice-valued logic: reduction algorithms and decision problems},
journal = {Iranian Journal of Fuzzy Systems},
volume = {15(7)},
pages = {103--119},
year = {2018}
}

@article{gho22,
author = {M. Ghorani},
title = {Characterization of tree automata based on quantum logic},
journal = {Int. J. of Theoretical Physics},
volume = {61:13},
pages = {1--24},
year = {2022}
}

@article{ghovog23,
title = {A theorem on supports of weighted tree automata over strong bimonoids},
author = {M. Ghorani and H. Vogler},
journal = {New Mathematics and Natural Computation},
year = {2023},
note = {online ready, https://doi.org/10.1142/S1793005723500230}
}

@Article{ghozah12,
  Title                    = {Characterizations of complete residuated lattice-valued finite tree automata},
  Author                   = {M. Ghorani and M.M. Zahedi},
  Journal                  = {Fuzzy Sets and Systems},
  Year                     = {2012},
  Pages                    = {28--46},
  Volume                   = {199}
  }

@Article{ghozahame12,
  Title                    = {Algebraic properties of complete residuated lattice valued tree automata},
  Author                   = {M. Ghorani and M.M. Zahedi and R. Ameri},
  Year                     = {2012},
  Note                     = {DOI 10.1007/s00500-012-0856-y},
  Pages                    = {1723-–1732},
  journal = {Soft Computing},
  Volume                   = {16}
}

@article{ghozah16,
author = {M. Ghorani and M.M. Zahedi},
title = {Alternating Regular Tree Grammars in the Framework of Lattice-Valued Logic},
journal = {Iranian Journal of Fuzzy Systems},
volume = {13(2)},
pages = {71--94},
year = {2016}
}

@article{ghozah17,
author = {M. Ghorani and M.M. Zahedi},
title = {Coding tree languages based on lattice-valued logic},
journal = {Soft Computing},
volume = {21(14)},
pages = {3815--3825},
year = {2017}
}

@Book{gin75,
  Title                    = {Algebraic and automata-theoretic properties of formal languages},
  Author                   = {S. Ginsburg},
  Publisher                = {North-Holland},
  series = {Fundamental studies in computer science},
  volume = {2},
  Year                     = {1975}
}

@inproceedings{gingre67,
  Title                    = {Abstract families of languages},
  Author                   = {S. Ginsburg and S.A. Greibach},
booktitle = {8th Annual Symposium on Switching and Automata Theory (SWAT 1967)},
Year = {1967},
Volume = {1},
Pages = {128--139}
}

@incollection{gingre69,
  Title                    = {Abstract families of languages},
  Author                   = {S. Ginsburg and S.A. Greibach},
  booktitle                 = {Studies in Abstract Families of Languages},
  Year                     = {1969},
  editor = {S. Ginsburg and S. Greibach and J. Hopcroft},
  Pages                    = {1--32},
  publisher= {American Mathematical Society},
  Volume                   = {87}
}

@Article{gingre70,
  Title                    = {Principal {AFL}},
  Author                   = {S. Ginsburg and S. Greibach},
  Journal                  = {J. Comput. Syst. Sci.},
  Year                     = {1970},
  Pages                    = {308--338},
  Volume                   = {4},
  number = {4}
}

@Book{gingrehop69,
  Title                    = {Studies in Abstract Families of Languages},
  Author                   = {S. Ginsburg and S. Greibach and J. Hopcroft},
  Publisher                = {American Mathematical Society},
  Year                     = {1969},
  number                   = {87},
  Journal                  = {Memoirs of the American Mathematical Society}
}

@phdthesis{goe17,
author = {D. G{\"o}tze},
title = {Weighted Unranked Tree Automata over Tree Valuation Monoids},
school = {University of Leipzig},
year = {2017}
}

@Article{gogthawagwri77,
  Title                    = {Initial Algebra Semantics and Continuous Algebras},
  Author                   = {J.A. Goguen and J.W. Thatcher and E.G. Wagner and J.B. Wright},
  Journal                  = {J. Assoc. Comput. Machinery},
  Year                     = {1977},
  Pages                    = {68--95},
  Volume                   = {24},
number = {1}
}

@Book{gol99,
  Title                    = {Semirings and their Applications},
  Author                   = {J.S. Golan},
  Publisher                = {Kluwer Academic Publishers},
  Year                     = {1999},
  Address                  = {Dordrecht}
}

@Article{gre70,
author = {S. Greibach},
title = {Full {AFLs} and nested iterated substitution},
journal = {Inform. Control},
year = {1970},
volume = {16},
pages = {7--35}
}

@Book{gra68,
  Title                    = {Universal Algebra},
  Author                   = {G. Gr{\"a}tzer},
  Publisher                = {D. van Nostrand Comp.},
  Year                     = {1968}
}

@Book{gra03,
  Title                    = {General Lattice Theory},
  Author                   = {G. Gr{\"a}tzer},
  Publisher                = {Birkh{\"a}user Verlag},
  Year                     = {2003},
  Edition                  = {2nd}
}

@Book{gra78,
  Title                    = {General Lattice Theory},
  Author                   = {G. Gr{\"a}tzer},
  Publisher                = {Birkh{\"a}user Verlag},
  Year                     = {1978}
}

@InProceedings{grakni04,
  Title                    = {Training tree transducers},
  Author                   = {J. Graehl and K. Knight},
  Booktitle                = {Proc.\ Human Language Technologies and North American ACL},
  Year                     = {2004},
  Address                  = {Boston, Massachusetts, USA, May 2 - 7},
  Pages                    = {105--112},
  Publisher                = {Association for Computational Linguistics}
}

@Article{graknimay08,
  Title                    = {Training tree transducers},
  Author                   = {J. Graehl and K. Knight and J. May},
  Journal                  = {Computational Linguistics},
  Year                     = {2008},
  Number                   = {3},
  Pages                    = {391--427},
  Volume                   = {34}
}

@InProceedings{habkre87,
author= {A. Habel and H.-J. Kreowski}, 
title= {Some structural aspects of hypergraph languages generated by hyperedge replacement},
booktitle={ Annual Symposium on Theoretical Aspects of Computer Science
(STACS 1987)},
year="1987",
series = {Lecture Notes in Computer Science},
volume = {247},
editor = {F.J. Brandenburg and G. Vidal-Naquet and M. Wirsing},
publisher="Springer Berlin Heidelberg",
pages="207--219"
}

@ARTICLE{hanmalque18,
  AUTHOR = {T. Hanneforth and A. Maletti and D. Quernheim},
  TITLE = {Pushing for weighted tree automata},
  JOURNAL = {Log.\@ Methods Comput.\@ Sci.},
  VOLUME = 14,
  NUMBER = {1},
  PAGES = {1--16},
  YEAR = 2018}

@Book{har78,
  Title                    = {Introduction to Formal Language Theory},
  Author                   = {M. A. Harrison},
  Publisher                = {Addison Wesley},
  Year                     = {1978}
}

@book{hebwei93,
author = {U. Hebisch and H.J. Weinert},
title = {Semirings - Algebraic Theory and Applications in Computer Science},
publisher = {World Scientific},
year = {1993}
}

@Article{henindroswei75,
  Title                    = {On primitive recursive wordfunctions},
  Author                   = {F.W. von Henke and K. Indermark and G. Rose and K. Weihrauch},
  Journal                  = {Computing},
  Year                     = {1975},
  Pages                    = {217--234},
  Volume                   = {15}
}

@InProceedings{her17,
  Title                    = {A {M}edvedev Characterization of Recognizable Tree Series},
  Author                   = {L. Herrmann},
  Booktitle                = {21th Int. Conf. on Developments in Language Theory (DLT 2017)},
  Year                     = {2017},
  Pages                    = {210--221},
  Publisher                = {Springer},
  Series                   = {Lecture Notes in Computer Science},
  Volume                   = {10396},

  Doi                      = {doi:10.1007/978-3-319-62809-7_15},
  Journal                  = {Proceedings of 21th Int. Conf. on Developments in Language Theory (DLT 2017)}
}

@unpublished{her20,
author = {L. Herrmann},
title = {personal communication},
year = {2020}
}

@phdthesis{her20a,
author = {L. Herrmann},
title = {Weighted Tree Automata with Storage},
school = {Technische Universit{\"a}t Dresden},
year = {2020}
}

@InProceedings{hervog15,
  Title                    = {A {C}homsky-{S}ch\"utzenberger Theorem for Weighted Automata with Storage},
  Author                   = {L. Herrmann and H. Vogler},
  Booktitle                = {6th Int. Conf. on Algebraic Informatics (CAI 2015)},
  editor = {A. Maletti},
  Year                     = {2015},
  Pages                    = {115--127},
  Series                   = {Lecture Notes in Computer Science},
  Volume                   = {9270}
}

@article{hervogdro19,
title = {Weighted automata with storage},
author = {L. Herrmann and H. Vogler and M. Droste},
journal = {Inf. Comput.},
volume = {269},
year = {2019},
note = {Article 104447}
}

@phdthesis{hoe07,
author = {J. H{\"o}gberg},
title = {Contributions to the Theory and Applications of Tree Languages},
school = {Ume{\aa} University},
year = {2007}
}

@Article{hoemalvog09,
  Title                    = {Bisimulation minimisation of weighted automata on unranked trees},
  Author                   = {J. H{\"o}gberg and A. Maletti and H. Vogler},
  Journal                  = {Fundam. Inform.},
  Year                     = {2009},
  Number                   = {1-2},
  Pages                    = {103--130},
  Volume                   = {92}
}

@INPROCEEDINGS{hogmalmay07d,
  AUTHOR = {J. H{\"o}gberg and A. Maletti and J. May},
  TITLE = {Bisimulation Minimisation For Weighted Tree Automata},
  BOOKTITLE = {11th Int.\@ Conf.\@ Developments in Language Theory (DLT 2007)},
  PAGES = {229--241},
  EDITOR = {T. Harju and J. Karhum{\"a}ki and A. Lepist{\"o}},
  VOLUME = 4588,
  SERIES = {Lecture Notes in Computer Science},
  PUBLISHER = {Springer},
  YEAR = 2007}

@Article{hooooz67,
  Title                    = {Fundamental notions in the theory of seminearrings},
  Author                   = {W.G. van Hoorn and B. van Rootselaar},
  Journal                  = {Compositio Math.},
  Year                     = {1967},
  Pages                    = {65--78},
  Volume                   = {18},
  number = {1--2}
}

@book{hopmotull07,
author = {J.E. Hopcroft and R. Motawi and J.D. Ullman},
title = {Introduction to Automata Theory, Languages, and Computation},
publisher = {Pearson, Addison-Wesley},
year = {2007}
}

@Book{hopull79,
  Title                    = {Introduction to automata theory, languages, and computation},
  Author                   = {J. E. Hopcroft and J. D. Ullman},
  Publisher                = {Addison--Wesley},
  Year                     = {1979},

  Bibedited                = {vogler 10.12.96},
  Srcfile                  = {biblio.bib}
}

@Book{hopmotull14,
  Title                    = {Introduction to Automata Theory, Languages, and Computation},
  Author                   = {J. E. Hopcroft and R. Motwani and J. D. Ullman},
  Publisher                = {Pearson Education Limited},
Year                     = {2014}
}

@article{hue80,
title = {Confluent Reductions: Abstract Properties and Applications to Term Rewriting Systems},
author = {G. Huet},
journal = {Journal of the ACM},
Volume = {27(4)},
pages = {797--821},
year = {1980}
}

@Book{hum04,
  Title                    = {A course in group theory},
  Author                   = {J. F. Humphreys},
  Publisher                = {Oxford University Press},
  Year                     = {2004},
  Note                     = {}
}

@Article{hup78,
  Title                    = {Rekursive {F}unktionen in mehrsortigen {P}eano-{A}lgebren},
  Author                   = {U. Hupbach},
  Journal                  = {Elektron. Informationsverarb. Kybern.},
  Year                     = {1978},
  Pages                    = {491--506},
  Volume                   = {14}
  }

@incollection{inafuk75,
author = {Y. Inagaki and T. Fukumura},
title = {On the description of fuzzy meaning of context-free languages},
editor = {L.A. Zadeh},
booktitle = {Fuzzy Sets and Their Applications to Cognitive and Decision Processes},
pages = {301--328},
publisher = {Academic Press, New York},
year = {1975}
}

@booklet{ind76,
author = {K. Indermark},
title = {Berechenbarkeit und {A}lgorithmen},
year = {1976},
note = {Lecture notes, {RWTH} {A}achen}
}

@book{jec03,
author= {T. Jech},
title = {Set Theory, The Third Millennium Edition, revised and expanded},
publisher = {Springer Berlin, Heidelberg},
year = {2003}
}

@phdthesis{jon21,
author = {A. Jonsson},
title = {Best Trees Extraction and Contextual Grammars for Language Processing},
school = {Ume{\aa} University, Sweden},
year = {2021}
}

@Book{jurmar25,
  author =       "Daniel Jurafsky and James H. Martin",
  title =        "Speech and Language Processing: An Introduction to Natural Language Processing, 
  		  Computational Linguistics, and Speech Recognition,
		   with Language Models",
  year =         "2025",
  url = {https://web.stanford.edu/~jurafsky/slp3/},
  note = "Online manuscript released August 24, 2025",
  edition =         "3rd",
  }

@Book{ker99,
  Title                    = {Applied finite group actions},
  Author                   = {A. Kerber},
  Publisher                = {Springer-Verlag},
  Year                     = {1999},
  Note                     = {}
}

@article{kir11,
author = {D.~Kirsten},
title = {The support of a recognizable series over a zero-sum free, commutative semiring is recognizable},
journal = {Acta Cybern.}, 
volume = {20(2)},
pages = {211--221}, 
year = {2011}
}

@inproceedings{kir09,
author = {D.~Kirsten},
title = {The support of a recognizable series over a zero-sum free, commutative semiring is recognizable},
booktitle = {9th Int. Conf. on Developments in Language Theory (DLT)},
editor = {V. Diekert and D. Nowotka},
series = {Lecture Notes in Computer Science},
volume = {5583},
publisher = {Sprigner},
pages = {326--333},
year = {2009}
}

@article{kir14,
author = {D.~Kirsten},
title = {An algebraic characterization of semirings for which the support of every recognizable series is recognizable},
journal = {Theoret. Comput. Sci.}, 
volume = {534},
pages = {43--52}, 
year = {2014}
}

@article{kirmae05,
author = {D.~Kirsten and I.~M{\"a}urer},
title = {On the Determinization of Weighted Automata},
journal = {J. Autom. Lang. Comb.}, 
volume = {10},
number = {2--3},
pages = {287--312}, 
year = {2005}
}

@article{klilommaipri04,
author = {I. Klimann and S. Lombardy and J. Mairesse and C. Prieur},
title = {Deciding unambiguity and sequentiality from a finitely ambiguous max-plus automaton},
journal = {Theoret. Comput. Sci.}, 
volume = {327},
pages = {349--373}, 
year = {2004}
}

@article{kla84,
author = {H. Klaeren},
title = {A constructive method for abstract algebraic software specification},
journal = {Theoret. Comput. Sci.},
volume = {30},
pages = {139--204},
year = {1984}
}

@InCollection{kle56,
  Title                    = {Representation of events in nerve nets and finite automata},
  Author                   = {S.E. Kleene},
  Booktitle                = {Automata Studies},
  volume = {34},
  Publisher                = {Princeton University Press, Princeton, N.J.},
  Year                     = {1956},
  Editor                   = {C.E. Shannon and J. McCarthy},
  Pages                    = {3--42}
}

@book{kle62,
  Author                   = {S.E. Kleene},
title = {Introduction to Meta Mathematics},
publisher = {Van Nostrand, New York},
year = {1962}
}

@Book{kliyua95,
  Title                    = {Fuzzy Sets and Fuzzy Logic, Theory and Application},
  Author                   = {G.J. Klir and B. Yuan},
  Publisher                = {Prentice-Hall, Englewood Cliffs, NJ},
  Year                     = {1995}
}

@InProceedings{knigra05,
author={K. Knight and J. Graehl},
editor="A. Gelbukh",
title="An Overview of Probabilistic Tree Transducers for Natural Language Processing",
booktitle=" Int. Conf. on Intelligent Text Processing and Computational Linguistics (CICLing 2005)",
year="2005",
series = {Lecture Notes in Computer Science},
volume = {3406},
publisher="Springer Berlin Heidelberg",
address="Berlin, Heidelberg",
pages="1--24"
}

@incollection{knimay09,
author = {K. Knight and J. May},
title = {Applications of Weighted Automata in Natural Language Processing},
booktitle = {Handbook of Weighted Automata},
series = {Monographs in Theoretical Computer Science. An EATCS Series},
publisher = {Springer-Verlag},
year = {2009},
editor = {M. Droste and W. Kuich and H. Vogler},
chapter = {14}
}

@Article{knu77,
  Title                    = {A generalization of {D}ijkstra's algorithm},
  Author                   = {D.E. Knuth},
  Journal                  = {Inform. Process. Lett.},
  Year                     = {1977},
 Number                   = {1},
  Pages                    = {1--5},
  Volume                   = {6}
}

@Article{koz92,
  Title                    = {On the {M}yhill-{N}erode Theorem for Trees},
  Author                   = {D. Kozen},
  Journal                  = {Bull. Eur. Assoc. Theoret. Comput. Sci. EATCS},
  Year                     = {1992},
  Pages                    = {170--173},
  Volume                   = {47},
  Address                  = {Cornell University},
  Location                 = {vogler-2013, Bormann-52}
}

@book{koz97,
author = {D. Kozen},
title = {Automata and Computability},
publisher = {Springer},
year = {1997}
}

@article{kos22,
author = {D. K{\'o}sz{\'o}},
title = {Tree generating context-free grammars and regular tree grammars are equivalent},
journal = {Annales Mathematicae et Informaticae},
year = {2022},
volume = {56},
pages = {58--70}
}

@book{krerob08,
author = {M.~Kreuzer and L.~Robbiano},
title = {Computational Commutative Algebra 1},
publisher = {Springer Publishing Company, Incorporated}, 
year = 2008
}

@PhdThesis{kri05,
  Title                    = {Near-Semirings: Theory and Applications},
  Author                   = {K.V. Krishna},
  School                   = {IIT Delhi, New Delhi, India},
  Year                     = {2005}
}

@InCollection{kui97,
  Title                    = {Semirings and Formal Power Series: Their Relevance to Formal Languages and Automata},
  Author                   = {W. Kuich},
  Booktitle                = {Handbook of Formal Languages},
  Volume = {1},
  Publisher                = {Springer-Verlag},
  Year                     = {1997},
  Chapter                  = {9},
  Editor                   = {G. Rozenberg and A. Salomaa},
  Pages                    = {609--677}
}

@InProceedings{kui98,
  Title                    = {Formal {P}ower {S}eries over {T}rees},
  Author                   = {W. Kuich},
  Booktitle                = {3rd Int. Conf. Developments in Language Theory (DLT 1997), Thessaloniki, Greece},
  Year                     = {1998},
  Editor                   = {S. Bozapalidis},
  Pages                    = {61--101},
  Publisher                = {Aristotle University of Thessaloniki}
}

@InProceedings{kui99a,
  Title                    = {Linear systems of equations and automata on distributive multioperator monoids},
  Author                   = {W. Kuich},
  Booktitle                = {Contributions to General Algebra 12 - Proceedings of the 58th Workshop on General Algebra ``58. Arbeitstagung Allgemeine Algebra'', Vienna University of Technology. June 3-6, 1999},
  Year                     = {1999},
  Pages                    = {1--10},
  editor = {D. Dorninger and G. Eigenthaler and M. Goldstern and H. K. Kaiser and W. More and W. B M{\"u}ller},
  Publisher                = {Verlag Johannes Heyn}
}

@Article{kui99,
  Title                    = {Tree transducers and formal tree series},
  Author                   = {W. Kuich},
  Journal                  = {Acta Cybernet.},
  Year                     = {1999},
  Pages                    = {135--149},
  Volume                   = {14},
  number = {1}
}

@InCollection{kui99c,
  Title                    = {Full abstract families of tree series {I}},
  Author                   = {W. Kuich},
  Booktitle                = {Jewels are forever},
  Publisher                = {Springer-Verlag},
  Year                     = {1999},
  Editor                   = {J. Karhum{\"a}ki and H. Maurer and G. Paun and G. Rozenberg},
  Pages                    = {145--156}
}

@Book{kuisal86,
  Title                    = {Semirings, Automata, Languages},
  Author                   = {W. Kuich and A. Salomaa},
  Publisher                = {Springer-Verlag},
  Year                     = {1986},
  Series                   = {EATCS Monographs on Theoretical Computer Science EATCS Ser.},
  Volume                   = {5}
}

@book{lan93,
author = {S. Lang},
title = {Algebra},
note = {3rd edition},
publisher ={Addison–Wesley, Reading},
year = {1993}
}

@article{lasnguson82,
author = {J. Lassez and V. Nguyen and L. Sonenberg},
title = {Fixed point theorems and semantics: a folk tale},
journal = {Inf. Proc. Letters},
volume = {14},
pages = {112--116},
year = {1982}
}

@article{lehpen14,
journal = {Theoret. Comput. Sci.},
volume = {534},
year = {2014},
pages = {53--68},
title = {The complexity of computing the behaviour of lattice automata on infinite trees},
author = {K. Lehmann and R. Pe{\~n}aloza}
}

@inproceedings{li08,
author = {Y. Li},
title = {Fuzzy Finite Automata and Fuzzy Monadic Second-Order Logic},
booktitle = {2008 IEEE International Conference on Fuzzy Systems (IEEE World Congress on Computational Intelligence)},
year = {2008},
note = {DOI: 10.1109/FUZZY.2008.4630353}
}

@article{li08a,
author = {Y. Li},
title = {Approximation and robustness of fuzzy ﬁnite automata},
journal = {International Journal of Approximate Reasoning},
volume = {47},
year = {2008},
pages = {247--257}
}

@Article{liped05,
  Title                    = {Fuzzy finite automata and fuzzy regular expressions with membership values in lattice-ordered monoids},
  Author                   = {Y. Li and W. Pedrycz},
  Journal                  = {Fuzzy Sets and Systems},
  Year                     = {2005},
  Pages                    = {68--92},
  Volume                   = {156}
}

@Book{lin12,
  Title                    = {An Introduction to Formal Languages and Automata},
  Author                   = {P. Linz},
  Publisher                = {Jones \& Bartlett Publishers},
  Year                     = {2012}
}

@incollection{lodtho21,
author = {C. L{\"o}ding and W. Thomas},
title = {Automata on Finite Trees},
chapter = "7",
booktitle = {Handbook of Automata Theory},
editor = {J.-E. Pin},
pages = {113--150},
publisher = {European Mathematical Society},
year = {2021}
}

@article{loo47,
author = {L. H. Loomis},
title = {On the representation of $\sigma$-complete  {B}oolean algebras},
journal = {Bull. Amer. Math. Soc.}, 
year = {1947},
volume = {53},
number = {8},
pages = {757--760}
}

@article{magmor70,
author = {M. Magidor and G. Moran},
title = {Probabilistic tree automata and context free languages},
journal = {Israel Journal of Mathematics},
volume = {8},
year = {1970},
pages = {340--348}
}

@inproceedings{mah81,
author = {B. Mahr},
year = {1981},
title = {A birds eye view to path problems},
editor = {H. Noltemeier},
booktitle = {Graphtheoretic Concepts in Computer Science. WG 1980},
series = {Lecture Notes in Computer Science},
volume = {100},
publisher = {Springer, Berlin, Heidelberg}
}

@inproceedings{mal04,
  Title                    = {Relating Tree Series Transducers and Weighted Tree Automata},
  Author                   = {A. Maletti},
  editor = {C.S. Calude and E. Calude and M.J.  Dinneen},
  booktitle = {Int. Conf. Developments in Language Theory (DLT 2004)},
  series = {Lecture Notes in Computer Science},
  volume = {3340},
  pages = {321--333},
  publisher = {Springer, Berlin, Heidelberg},
  year = {2004}
  }

@Article{mal05a,
  Title                    = {Relating tree series transducers and weighted tree automata},
  Author                   = {A. Maletti},
  Journal                  = {Internat. J. Found. Comput. Sci.},
  Year                     = {2005},
  Pages                    = {723--741},
  Volume                   = {16},
  number = {4}
}

@inproceedings{mal05b,
author = {A. Maletti},
year = {2005},
title = {The Power of Tree Series Transducers of Type I and II},
editor = {C. De Felice and A. Restivo},
booktitle = {Int.  Conf. Developments in Language Theory (DLT 2005)},
series = {Lecture Notes in Computer Science},
volume = {3572},
publisher = {Springer, Berlin, Heidelberg},
pages = {338--349}
}

@INPROCEEDINGS{mal06,
  AUTHOR = {A. Maletti},
  TITLE = {Hierarchies of Tree Series Transformations Revisited},
  BOOKTITLE = {10th Int.\@ Conf.\@ Developments in Language Theory (DLT 2006)},
  PAGES = {215--225},
  EDITOR = {O. H.~Ibarra and Z. Dang},
  VOLUME = "4036",
  SERIES = {Lecture Notes in Computer Science},
  PUBLISHER = {Springer},
  YEAR = "2006"}

@PHDTHESIS{mal06c,
  AUTHOR = {A. Maletti},
  TITLE = {The Power of Tree Series Transducers},
  SCHOOL = {Technische Universit{\"a}t Dresden},
  YEAR = 2006}

@ARTICLE{mal06a,
  AUTHOR = {A. Maletti},
  TITLE = {Compositions of tree series transformations},
  JOURNAL = {Theoret. Comput. Sci.},
  VOLUME = 366,
  NUMBER = 3,
  PAGES = {248--271},
  YEAR = 2006}

@inproceedings{mal07,
    title = {Learning Deterministically Recognizable Tree Series -- Revisited},
    booktitle  = {2nd Int. Conf. Algebraic Informatics (CAI 2007)},
    pages = {218--235}, 
    publisher = {Springer-Verlag},
    editor = {S. Bozapalidis and G. Rahonis},
    year = {2007},
    series  = {Lecture Notes in Computer Science},
    volume = {4728},
    author = {A. Maletti}
}

@INPROCEEDINGS{mal08c,
  AUTHOR = {A. Maletti},
  TITLE = {Myhill {N}erode Theorem for Recognizable Tree Series --- Revisited},
  BOOKTITLE = {8th Int.\@ Conf.\@ Latin American Theoretical Informatics (LATIN 2008)},
  PAGES = {106--120},
  EDITOR = {E. Laber and C. Bornstein and L. Nogueira and L. Faria},
  VOLUME = 4957,
  SERIES = {Lecture Notes in Computer Science},
  PUBLISHER = {Springer},
  YEAR = 2008}

@ARTICLE{mal08e,
  TITLE = {Minimizing deterministic weighted tree automata},
  JOURNAL = {Inf. Comput.},
  NUMBER = 11,
  VOLUME = 207,
  PAGES = {1284--1299},
  AUTHOR = {A. Maletti},
  YEAR = 2009}

@ARTICLE{mal11,
  AUTHOR = {A. Maletti},
  TITLE = {Survey: Weighted Extended Top-down Tree Transducers --- Part {I}: Basics and Expressive Power},
  JOURNAL = {Acta Cybernet.},
  VOLUME = 20,
  NUMBER = 2,
  PAGES = {223--250},
  YEAR = 2011}

@ARTICLE{mal11b,
  AUTHOR = {A. Maletti},
  TITLE = {Survey: {W}eighted {E}xtended {T}op-{D}own {T}ree {T}ransducers --- Part {II}: Application in Machine Translation},
  JOURNAL = {Fundam.\@ Inform.},
  VOLUME = 112,
  NUMBER = {2--3},
  PAGES = {239--261},
  YEAR = 2011}

@article{mal17,
journal = {Theoret. Comput. Sci.},
Volume = {679},
Pages = {2--17},
year = {2017},
title = {Survey: Finite-state technology in natural language processing},
author = {A. Maletti}
}

@INPROCEEDINGS{malnue23,
  AUTHOR = {A. Maletti and  N. O. Nuernbergk},
  TITLE = {Pumping Lemmata for Recognizable Weighted Languages
over {A}rtinian Semirings},
  BOOKTITLE = {16th International Conference on Automata and Formal Languages (AFL 2023)},
  PAGES = {155–169},
  EDITOR = {Zs. Gazdag and Sz. Iv\'an and G. Kov\'asznai},
  VOLUME = 386,
  SERIES = {Electronic Proceedings in Theoretical Computer Science},
  PUBLISHER = {},
  YEAR = 2023}

@INPROCEEDINGS{malsat09,
  AUTHOR = {A. Maletti and G. Satta},
  TITLE = {Parsing Algorithms based on Tree Automata},
  BOOKTITLE = {Proceedings of the 11th Int. Conf. on Parsing Technologies (IWPT 2009)},
  PAGES = {1--12},
  PUBLISHER = {Association for Computational Linguistics},
  YEAR = {2009}
}

@INPROCEEDINGS{malsat10,
  AUTHOR = {A. Maletti and G. Satta},
  TITLE = {Parsing and Translation Algorithms Based on Weighted Extended Tree Transducers},
  BOOKTITLE = {1st Workshop Applications of Tree Automata in Natural Language Processing (ATANLP 2010)},
  PAGES = {19--27},
  EDITOR = {F. Drewes and M. Kuhlmann},
  PUBLISHER = {Association for Computational Linguistics},
  YEAR = 2010}

@ARTICLE{malvog05,
  AUTHOR = {A. Maletti and H. Vogler},
  TITLE = {Incomparability Results for Classes of  Polynomial Tree Series Transformations},
  JOURNAL = {J.~Autom.\@ Lang.\@ Comb.},
  VOLUME = 10,
  NUMBER = 4,
  PAGES = {535--568},
  YEAR = 2005}

@inproceedings{mat08,
booktitle = {Int. Colloquium on Automata, Languages, and Programming (ICALP 2008)},
pages = {221--232},
title = {Weighted Logics for Nested Words and Algebraic Formal Power Series},
author = {C. Mathissen},
editor = {L. Aceto and I. Damg{\aa}rd and L.A. Goldberg and M.M Halld{\'o}rsson and A. Ing{\'o}lfsd{\'o}ttir and I. Walukiewicz},
year = {2008},
series = {Lecture Notes in Computer Science},
volume = {5126},
publisher = {Springer, Berlin, Heidelberg}
}

@Article{mas76,
  Title                    = {Multilevel stack automata},
  Author                   = {A. N. Maslov},
  Journal                  = {Problems Inform. Transmission},
  Year                     = {1976},
  Pages                    = {38--43},
  Volume                   = {12}
}

@article{mat10,
author = {C. Mathissen},
year = {2010},
title = {Weighted Logics for Nested Words and Algebraic Formal Power Series},
journal = {Logical Methods Comput. Sci.},
volume = {6},
number = {1}
}

@inproceedings{mae05,
author = {I. M{\"a}urer},
title = {Rational and recognizable picture series},
booktitle = {Conf. on Algebraic Informatics (CAI 2005), Thessaloniki},
pages = {141--155},
publisher = {Aristotle University of Thessaloniki Press},
year = {2005}
}

@article{mae07,
author = {I. M{\"a}urer},
title = {Characterizations of recognizable picture series},
journal = {Theoret. Comput. Sci.},
volume = {374},
pages = {214--228},
year = {2007}
}

@inproceedings{mannes70,
author = {Z. Manna and S. Ness},
title = {On termination of Markov algorithms},
booktitle = {Proc. 3rd Hawaii International Conference on System  Sciences},
pages = {798-792},
publisher = {},
year = {1970}
}

@phdthesis{mat09,
title = {Weighted Automata and Weighted Logics over Tree-like Structures},
author = {C. Mathissen},
school = {University of Leipzig},
year = {2009}
}

@phdthesis{may10,
author = {J.D.L. May},
title = {Weighted tree automata and transducers for syntactic natural language processing},
school = {University of Southern California},
year = {2010}
}

@InProceedings{mazriv18,
  author =	{F. Mazowiecki and C. Riveros},
  title =	{Pumping Lemmas for Weighted Automata},
  booktitle =	{35th Symposium on Theoretical Aspects of Computer Science (STACS 2018)},
  pages =	{50:1--50:14},
  series =	{Leibniz International Proceedings in Informatics (LIPIcs)},
  year =	{2018},
  volume =	{96},
  editor =	{R. Niedermeier and B. Vall{\'e}e},
  publisher =	{Schloss Dagstuhl--Leibniz-Zentrum fuer Informatik},
}

@miscellaneous{med56,
author = {{\'U}. T. M{\'e}dv{\'e}d{\'e}v},
title = {O klass{\'e} sobytij, dopuska{\'u}\v{s}ih pr{\'e}dstavl{\'e}ni{\'e} v kon{\'e}\v{c}nom avtomat{\'e}},
publisher = {Avtomaty, Izdat{\'e}l'stvo Inostrannoj Lit{\'e}ratury, Moscow},
year = {1956},
pages = {385--401}
}

@inproceedings{mei06,
author = {I. Meinecke},
year = {2006},
title = {Weighted Logics for Traces},
booktitle = {1st Int. Computer Science Conf. on Theory and Applications (CSR 2006)},
series = {Lecture Notes in Computer Science},
editor = {D. Grigoriev and J. Harrison E.A. Hirsch},
volume = {3967},
publisher = {Springer, Berlin, Heidelberg},
pages = {235--246}
}

@Article{mezwri67,
  Title                    = {Algebraic automata and context-free sets},
  Author                   = {J. Mezei and J.B. Wright},
  Journal                  = {Inf. Control},
  Year                     = {1967},
  number = {1--2},
  Pages                    = {3--29},
  Volume                   = {11}
}

@book{mormal02,
author = {J.N. Mordeson and D.S. Malik},
year = {2002},
title = {Fuzzy Automata and Languages: Theory and Applications},
publisher = {Chapman and Hall}
}

@misc{moevog19,
author = {R. M{\"o}rbitz and H. Vogler},
title = {Weighted Parsing for Grammar-Based Language Models over Multioperator Monoids},
   year={2019},
   note = {arxiv.1911.06585},
      eprint={1911.06585},
      archivePrefix={arXiv},
      primaryClass={cs.FL}
}

@inproceedings{moevog19a,
    title = {Weighted parsing for grammar-based language models},
    author = {R. M{\"o}rbitz  and H. Vogler},
    editor = {H. Vogler and A. Maletti},
    booktitle = "Proc. 14th Int. Conf. on Finite-State Methods and Natural Language Processing (FSMNLP 2019)",
    publisher = "Association for Computational Linguistics",
    url = "https://www.aclweb.org/anthology/W19-3108",
    doi = "10.18653/v1/W19-3108",
    pages = "46--55",
    year = {2019},
}

@article{moevog21,
author = {R. M{\"o}rbitz and H. Vogler},
title = {Weighted Parsing for Grammar-Based Language Models over Multioperator Monoids},
journal = {Inf. Comput.},
volume = {281},
note = {Article 104774},
   year={2021},
   }

@Article{mogzahame11,
  Title                    = {New direction in fuzzy tree automata},
  Author                   = {S. Moghari and M.M. Zahedi and R. Ameri},
  Journal                  = {Iranian J. Fuzzy Systems},
  Year                     = {2011},
  Pages                    = {59--68},
  Volume                   = {8(5)}
}

@Article{moh97,
  Title                    = {Finite-State Transducers in Language and Speech Processing},
  Author                   = {M. Mohri},
  Journal                  = {Computational Linguistics},
  Year                     = {1997},
  Number                   = {2},
  Pages                    = {269--311},
  Volume                   = {23}
}

@Article{moh00a,
  Title                    = {Minimization algorithms for sequential transducers},
  Author                   = {M. Mohri},
  Journal                  = {Theoret. Comput. Sci.},
  Year                     = {2000},
  Pages                    = {177--201},
  Volume                   = {234},
  number = {1--2}
}

@book{mooyaq98,
author = {H.G. Moore and A. Yaqub},
title = {A first course in linear algebra with applications},
publisher ={Academic Press},
year = {1998}
}

@techreport{myh57,
author = {J. Myhill},
title = {Finite automata and the representation of events},
note = {WADD Technical Report 57--624},
institution = {Wright-Patterson Air Force Abase},
year = {1957}
}

@Article{ner58,
  Title                    = {Linear automaton transformations},
  Author                   = {A. Nerode},
  Journal                  = {Proc. of the American Mathematical Society},
  Year                     = {1958},
  Pages                    = {541--544},
  Volume                   = {9},

  Location                 = {vogler-2377}
}

@Article{niv68,
  Title                    = {Transduction des langages de {C}homsky},
  Author                   = {M. Nivat},
  Journal                  = {Ann. de l'Inst. Fourier},
  Year                     = {1968},
  Pages                    = {339--455},
  Volume                   = {18}
}

@phdthesis{pau20,
author = {E. Paul},
title = {Expressiveness and Decidability of Weighted Automata and Weighted Logics},
year = {2020},
school = {University of Leipzig} 
}

@inproceedings{perrilspr94,
  title={Weighted rational transductions and their application to human language processing},
  author={F. Pereira and M. Riley  and R. Sproat},
  booktitle={Human Language Technology: Proceedings of a Workshop held at Plainsboro, New Jersey, March 8-11, 1994},
  year={1994}
}

@Book{pet57,
  Title                    = {Rekursive Funktionen},
  Author                   = {R. P{\'e}ter},
  Publisher                = {Akad\'emiai Kiad\'o, Budapest},
  Year                     = {1957},
  Note                     = {English translation: Academic Press, New York, 1967}
  }

@InProceedings{qua09,
author="K. Quaas",
editor="V. Diekert and D. Nowotka",
title="Weighted Timed {MSO} Logics",
booktitle="Int. Conf. on Developments in Language Theory (DLT 2009)",
year="2009",
series = {Lecture Notes in Computer Science},
volume = {5583},
publisher="Springer, Berlin, Heidelberg",
address="Berlin, Heidelberg",
pages="419--430"
}

@Article{rad10,
  Title                    = {Weighted tree automata over strong bimonoids},
  Author                   = {D. Radovanovi\'c},
  Journal                  = {Novi Sad J. Math.},
  Year                     = {2010},
  Number                   = {3},
  Pages                    = {89--108},
  Volume                   = {40}
}

@article{rah07,
author = {G. Rahonis},
title = {Weighted {M}uller Tree Automata and Weighted Logics},
journal = {J. Autom. Lang. Comb.},
year = {2007},
Volume = {12},
number =  {4},
pages = {455--483}
}

@incollection{rah09,
 author = {G. Rahonis},
 title = {Fuzzy Languages},
  Booktitle                = {Handbook of Weighted Automata},
  Publisher                = {Springer-Verlag},
  Year                     = {2009},
  series = {EATCS Monographs in Theoretical Computer Science},
  Chapter                  = {12},
  Editor                   = {M. Droste and W. Kuich and H. Vogler},
  Pages                    = {481--517}
}

@article{rahtor19,
journal = {Scientific Annals of Computer Science},
volume = {29(1)},
year = {2019},
pages = {59--80},
title = {Weighted Context-Free Grammars Over Bimonoids},
author = {G. Rahonis and F. Torpari}
}

@unpublished{renroslivog23,
author= {J. Rennert and M. Rossel and Y. Li and H. Vogler},
title = {Discussion during a lecture on weighted tree automata at {TU} {D}resden},
year = {2023}
}

@Article{reu80,
  Title                    = {{A}n {O}dgen-Like Iteration Lemma for Rational Power Series},
  Author                   = {Ch. Reutenauer},
  Journal                  = {Acta Inform.},
  Year                     = {1980},
  Pages                    = {189--197},
  Volume                   = {13}
}

@article{rivwil09,
author = {I. Rival and R. Wille},
title = {Lattices freely generated by partially ordered sets: which can be "drawn"},
journal = {Journal für die reine und angewandte Mathematik},
volume = {310},
pages = {56--80},
year = {2009}
}

@article{rol58,
author = {H.L. Rolf},
title = {The free lattice generated by a set of chains},
journal = {Pacific J. Math.},
volume = {8},
year = {1958},
pages = {585--595}
}

@incollection{rot90,
  Title                    = {Path Problems in Graphs},
  Author                   = {G. Rote},
  editor = {G. Tinhofer  and E. Mayr and H. Noltemeier and M.M. Syslo},
  booktitle                  = {Computational Graph Theory. Computing Supplementum},
  Year                     = {1990},
  Pages                    = {155--189},
  Volume                   = {7},
  publisher = {Springer, Vienna}
}

@book{sak09,
title = {Elements of Automata Theory},
author = {J. Sakarovitch},
year = {2009},
publisher = {Cambridge University Press}
}
